北京大學

# Postdoctoral Research Report

## Research on the Power Transport Theorem Based Decoupling Mode Theory for Transceiving Systems

Author: **Renzun Lian**

**rzlian@vip.163.com (Email)**

Supervisor: **Prof. Mingyao Xia**

**myxia@pku.edu.cn (Email)**

Institution: **Peking University (PKU)**

**School of Electronics Engineering and Computer Science**

**Department of Electronics**

# Research on the Power Transport Theorem Based Decoupling Mode Theory for Transceiving Systems

(Peking University Postdoctoral Research Report)

Author: **Renzun Lian**

**rzlian@vip.163.com (Email)**

Supervisor: **Prof. Mingyao Xia**

**myxia@pku.edu.cn (Email)**

Institution: **Peking University (PKU)**

**School of Electronics Engineering and Computer Science**

**Department of Electronics**

# 版权声明

There are two ways to do calculations. The first way, which I prefer, is to have a clear *physical picture*. The second way is to have a rigorous mathematical formalism.

—— Prof. Enrico Fermi

The physical pictures of *eigen-mode theory* (*EMT*) and the conventional *characteristic mode theory* (*CMT*) reveal a fact that: the EMT and CMT are the modal theories for electromagnetic wave-guiding and scattering (for details, please see the Appendices E~H) systems respectively, rather than for electromagnetic transceiving systems. **This Postdoctoral Research Report is devoted to establishing a novel modal theory — *decoupling mode theory* (*DMT*) — for transceiving systems, and constructing the *energy-decoupled modes* (*DMs*) of objective transceiving system.**

This Postdoctoral Research Report is a companion volume of the author's Doctoral Dissertation "RESEARCH ON THE WORK-ENERGY PRINCIPLE BASED CHARACTERISTIC MODE THEORY FOR SCATTERING SYSTEMS". The English and Chinese versions of the dissertation can be downloaded from the following links.

English Version: https://arxiv.org/abs/1907.11787

Chinese Version: https://kns.cnki.net/KCMS/detail/detail.aspx?dbcode=CDFD&dbname=CDFDLAST2020&filename=1019851418.nh&v=MTMzOThSOGVYMUx1eFlTN0RoMVQzcVRyV00xRnJDVVI3cWZZK1p0Rnlqa1U3M1BWRjI2Rjd1OUg5WE5wNUViUEk=

Renzun Lian & Mingyao Xia

Peking University (PKU) · Yan Yuan

February 2021

# ABSTRACT

TRANSCEIVING system, scattering system, and wave-guiding system widely exist in electromagnetic (EM) engineering domain, and they are essentially different from each other. The transceiving system transmits and/or receives EM energies, while the scattering system and wave-guiding system scatters and guids EM energies respectively.

For a certain transceiving/scattering/wave-guiding system, its all physically realizable working modes constitute a linear space — modal space. It is unrealistic to study all the modes in the space, because the number of the modes is infinite. In fact, it is sufficient to study the "basis of modal space" — fundamental modes. For a certain modal space, its fundamental mode set is not unique. Among the various fundamental mode sets, the energy-decoupled mode set, whose elements don't have net energy exchange in integral period, is usually more desired, because the energy-decoupled modes don't exchange the informations carried by them due to the fact that energy is the carrier of information.

Characteristic mode theory (CMT) and eigen-mode theory (EMT) are just the ones used to construct energy-decoupled modes. After many years of development, the theoretical foundations of CMT and EMT have been greatly developed, and CMT and EMT have been frequently used to approximately analyze transceiving system (especially the transmitting antenna part). Unfortunately, the work-energy principle (WEP) based physical picture of CMT clearly reveals the fact that: CMT is a rigorous modal theory for scattering systems rather than for transceiving systems. In addition, EMT is also not a rigorous modal theory for transceiving systems but for wave-guiding systems. In fact, how to establish a rigorous modal theory for transceiving systems is just one of important challenges existing in the realm of electromagnetism. Focusing on this challenge, this research report does some novel studies in the following 8 main aspects.

**Aspect I. Divisions for Transceiving System and Surrounding Environment**

The transceiving system placed in a surrounding environment (simply called environment) is divided into two sub-systems: transmitting system (simply called transmitter, which is used to generate EM energy and release the energy into environment) and receiving system (simply called receiver, which is used to collect EM energy from environment and convert the energy into perceptible signals).

The transmitter is further divided into three sub-structures: generating structure (simply called generator, which is used to generate EM energy), transmitting antenna structure (simply called tra-antenna, which is used to modulate the energy to be released into environment), and guiding structure (simply called tra-guide, which is used to guide the energy from generator to tra-antenna and then to environment).

The receiver is further divided into three sub-structures: receiving antenna structure (simply called rec-antenna, which is used to collect EM energy from environment), absorbing structure (simply called absorber, which is used to convert the energy into perceptible signals), and guiding structure (simply called rec-guide, which is used to guide the energy from rec-antenna to absorber).

In addition, the environment is also divided into two sub-structures: infinity (which is used to receive some EM energies radiated by transmitter and scattered by receiver) and propagation medium (simply called medium, which is used as the area for propagating the EM energies radiated by transmitter and scattered by receiver).

**Aspect II. Division for Three-dimensional Euclidean Space**

Evidently, the transceiving system and its surrounding environment occupy whole three-dimensional Euclidean space $\mathbb{E}^3$. Thus the above divisions for transceiving system and environment can also be viewed as a division for $\mathbb{E}^3$, from the viewpoint of point set topology. Specifically, the $\mathbb{E}^3$ is divided into three one-port regions (corresponding to generator, absorber, and infinity respectively) and five two-port regions (corresponding to tra-guide, tra-antenna, medium, rec-antenna, and rec-guide respectively).

The topological structures of the various regions are rigorously depicted using the language of point set topology. Taking two-port region as an example, the region has a cavity enclosed by its boundary; the cavity doesn't include the boundary, such that the EM fields in the cavity satisfy the differential form of Maxwell's equations; the boundary includes three parts — a penetrable input port, a penetrable output port, and an impenetrable electric wall used to separate the input and output ports (also used to partially separate the cavity from its surrounding environment).

**Aspect III. Power Transport Theorem (PTT) for Transceiving Systems**

Due to the existence of the ports, the different structures/regions will exchange EM energies with each other, i.e., there exist some powers flowing from one to another. The physical law governing the power passing through any single structure/region is just famous Poynting's theorem (PtT). This report proves that the PtT for a single structure/region is equivalent to the WEP for the structure/region, and this conclusion is consistent with the relation between the PtT and WEP for any single scatterer. In addition, this report also derives the physical law governing the transporting process of the power flowing in whole {transceiving system, environment} / whole $\mathbb{E}^3$ (which is the cascade of a series of structures/regions), and calls it power transport theorem (PTT).

Just like that the WEP for scattering problem contains a source term — driving power — used to drive a steady work-energy transformation for scattering systems, the PTT for transceiving problem also contains a source term — input power — used to sustain a steady power transport for transceiving systems.

**Aspect IV. PTT-Based Decoupling Mode Theory (DMT)**

It had been exhibited previously that: the CMT for scattering systems can be effectively established in WEP framework, and the deriving power operator (DPO) can be selected as the generating operator of characteristic modes (specifically, the modes can be evaluated from orthogonalizing DPO). Similarly, this report under PTT framework establishes a novel modal theory — decoupling mode theory (DMT) — for transceiving systems, and constructs a set of fundamental modes — input-power-decoupled mode (IP-DMs) — for any objective transceiving system. The IP-DMs are calculated from orthogonalizing IPO (i.e., IPO is selected as the generating operator of the modes), so they are input-power-decoupled, and then they are energy-decoupled in any integral period. This is just the reason why the modal theory and fundamental modes are called decoupling mode theory and input-power-decoupled modes respectively.

The general process used to calculate the IP-DMs is summarized as follows:

Step 1. Mathematically describing the topological structure of the two-port region occupied by objective system

Step 2. Deriving the source-field relationships satisfied related to the two-port region

Step 3. Mathematically depicting the modal space of the objective system by employing the source-field relationships and some necessary EM field boundary conditions

Step 4. Deriving the PTT governing the transporting process of the power passing through the two-port region

Step 5. Recognizing the source term — IPO (which is used to sustain a steady transporting of power) — contained in the PTT

Step 6. Constructing IP-DMs by orthogonalizing the frequency-domain IPO defined on the modal space

The above processes are applicable to any kind of transceiving system and any sub-structure contained in the transceiving system.

**Aspect V. PTT-Based DMT for Some Separate Structures and Modal Matching**

Using the PTT-based DMT, this report constructs the IP-DMs of the previously mentioned various two-port sub-structures (i.e. tra-guide, tra-antenna, medium, rec-antenna, and rec-guide) separately. For a practical transceiving system, there exist the interactions among the IP-DMs of different sub-structures, because the modal EM fields cannot be confined within the corresponding sub-structure (due to the existence of the penetrable ports).

This report provides two different ways to consider the interactions. The first way, which this report doesn't recommend, is to apply modal matching technique to the interfaces among the interactive sub-structures. The second way, which this report prefers, is to treat the interactive sub-structures as a combined system first and then to establish the DMT for the system. As exhibited in this report, the second way has both clearer physical picture and simpler mathematical process than the first way. In fact, this report only employs the first way as an introduction for the second way.

**Aspect VI. PTT-Based DMT for Combined Systems**

Based on the ideology of the second way, this report treats the tra-antenna and environment as a whole — augmented tra-antenna, and constructs the IP-DMs of various augmented tra-antennas (such as augmented metallic, material, and metal-material composite tra-antennas); this report treats the environment and rec-antenna as a whole — augmented rec-antenna, and constructs the IP-DMs of various augmented rec-antennas; this report treats the tra-guide and augmented tra-antenna as a whole — augmented tra-guide-tra-antenna combined system (simply called TGTA system), and constructs the IP-DMs of various TGTA systems; this report treats the augmented tra-antenna and rec-antenna as a whole — augmented tra-antenna-rec-antenna combined system (simply called TARA system), and constructs the IP-DMs of various TARA systems.

In fact, the IP-DMs of other kinds of combined systems can also be constructed similarly.

**Aspect VII. Modal Decomposition for Any Working Mode of Transceiving System**

Imitating the modal decomposition for the working mode of scattering system, the modal decomposition for any physically realizable working mode of a certain transceiving system is done in this report. Specifically, any working mode of a certain transceiving system is decomposed into three energy-decoupled components — purely inductive component (which can be expanded in terms of the inductive IP-DMs), purely resonant component (which can be expanded in terms of the resonant IP-DMs), and purely capacitive component (which can be expanded in terms of the capacitive IP-DMs).

**Aspect VIII. Electric-Magnetic Decoupling Factor of Any Working Mode of Transceiving System**

Employing the modal decomposition, this report introduces a novel concept of "electric-magnetic decoupling factor (simply called decoupling factor or Θ-factor)" to the working mode of transceiving system. The Θ-factor can be viewed as a generalization for the traditional Q-factor (i.e. quantity factor). The Θ-factor has ability to quantitatively depict the decoupling degree between the electric energy and magnetic energy carried by the working mode. Specifically, the larger Θ-factor is, the stronger electric-magnetic decoupling is (i.e., the weaker electric-magnetic coupling is).

Employing the Θ-factor, this report clearly reveals another important physical meaning (i.e. the third physical meaning) possessed by modal significance (MS). The three current known physical meanings of MS are

Physical Meaning 1. a quantitative description for the weight of a certain IP-DM contained in whole IP-DM-based modal expansion formulation,

Physical Meaning 2. a quantitative description for the weight of the modal active power contained in the whole modal power of the IP-DM, and

Physical Meaning 3. a quantitative description for the coupling degree of the electric energy and magnetic energy carried by the IP-DM.

"The three somewhat different physical meanings are collected in a single body" reflects that: the MS is a very important physical quantity in modal theory.

**SUMMARY.** In fact, this postdoctoral research report "RESEARCH ON THE POWER TRANSPORT THEOREM BASED DECOUPLING MODE THEORY FOR TRANSCEIVING SYSTEMS" can be viewed as a companion volume of the author's doctoral dissertation "RESEARCH ON THE WORK-ENERGY PRINCIPLE BASED CHARACTERISTIC MODE THEORY FOR SCATTERING SYSTEMS". This report is devoted to establishing the PTT-based DMT for transceiving systems, and to constructing the corresponding energy-decoupled modes — IP-DMs — by orthogonalizing frequency-domain IPO.

The DMT has both similarity and differences with conventional CMT and EMT as follows:

Similarity. DMT, CMT, and EMT are the modal theories focusing on constructing energy-decoupled modes.

Difference 1. DMT, CMT, and EMT focus on different objective EM systems. Specially, DMT focuses on transceiving systems; CMT focuses on scattering systems; EMT focuses on wave-guiding systems.

Difference 2. DMT, CMT, and EMT are supported by different framework. Specially, DMT is supported by power transport theorem framework; CMT is supported by work-energy principle framework; EMT is supported by Sturm-Liouville theory framework.

Difference 3. DMT, CMT, and EMT utilize different modal generating operators. Specially, DMT utilizes input power operator; CMT utilizes driving power operator; EMT utilizes Sturm-Liouville operator / Helmholtz's operator.

The usages of the CMT-based and EMT-based modal analysis for various antennas are only approximate, but by no means rigorous. As exhibited in this report, the wave-guiding system is only a part contained in transceiving system, so the DMT is also applicable to the modal analysis for wave-guiding systems.

# Main Symbol Table

For the convenience of expressions, this report utilizes some abbreviations frequently. Now, we list the abbreviations (in alphabetical order) and their full names in the following table:

| **Abbreviations** | **Full Names** |
|---|---|
| Abs | absorbing structure in receiving system |
| absorber | absorbing structure in receiving system |
| augmented rec-antenna | combined system constituted by receiving antenna and the grounding structure related to receiving system |
| augmented tra-antenna | combined system constituted by transmitting antenna and the grounding structure related to transmitting system |
| AV | all variables |
| BS | basic solutions |
| BV | basic variables |
| CM | characteristic mode |
| CMT | characteristic mode theory |
| DDC | definition domain compression |
| decoupling equation | modal decoupling equation, i.e. the equation used to calculate energy-decoupled modes |
| decoupling factor | electric-magnetic decoupling factor |
| DM | energy-decoupled mode |
| DMT | decoupling mode theory |
| DoJ | definition of J, where J is the equivalent electric current |
| DoM | definition of M, where M is the equivalent magnetic current |
| DP | driving power |
| DPO | driving power operator |
| DRA | dielectric resonator antenna |
| DVE | dependent variable elimination |
| DVE-DoJ | dependent variable elimination based on the definition of J |
| DVE-DoM | dependent variable elimination based on the definition of M |
| DW | driving work |
| EM | electromagnetic |
| EMT | eigen-mode theory |
| Env | surrounding environment |
| environment | surrounding environment |
| ES | equivalent surface (current) |
| FCE | field continuation equation |
| feed | feeding structure for transmitting antenna |
| Gen | generating structure in transmitting system |
| generator | generating structure in transmitting system |
| grounding structure | electric wall used to partially separate transceiving system from its surrounding environment |
| HM | magnetic field – magnetic current interaction |
| HM-DoM-based IPO | input power operator expressed in terms of the interaction between magnet field and magnetic current and employing the definition of equivalent magnetic current to depict modal space |
| IE | integral equation |
| IE-ComSca-CMT | integral equation based characteristic mode theory for composite scattering systems |
| IE-MatSca-CMT | integral equation based characteristic mode theory for material scattering systems |
| IE-MetSca-CMT | integral equation based characteristic mode theory for metallic scattering systems |
| IE-ScaSys-CMT | integral equation based characteristic mode theory for scattering systems |
| IMO | impedance matrix operator |
| Inf | infinity |
| IP-DMs | input-power-decoupled modes, i.e. the energy-decoupled modes calculated from orthogonalizing input power operator |
| IPO | input power operator |
| IS | induced surface current |
| IVM | intermediate variable method |
| JE | electric current – electric field interaction |
| JE-DoJ-based IPO | input power operator expressed in terms of the interaction between electric current and electric field and employing the definition of equivalent electric current to depict modal space |
| JM | electric current – magnetic current interaction |
| Med | propagation medium |
| medium | propagation medium |
| MH | magnetic current – magnetic field interaction |
| MIA | modal input admittance |
| MII | modal input impedance |
| MS | modal significance |
| OEMS | objective electromagnetic structure/system |
| OPO | output power operator |
| PIFA | planar inverted-F antenna |
| PMCHWT | Poggio-Miller-Chang-Harrington-Wu-Tsai (integral equation) |
| PML | perfectly matched load |
| PMO | perturbation matrix operator |
| PTT | power transport theorem |
| PtT | Poynting's theorem |
| PTT-ComGuid-DMT | power transport theorem based decoupling mode theory for composite guiding structures |
| PTT-ComRecAnt-DMT | power transport theorem based decoupling mode theory for augmented composite receiving antennas |
| PTT-ComSys-DMT | power transport theorem based decoupling mode theory for combined systems |
| PTT-ComTraAnt-DMT | power transport theorem based decoupling mode theory for augmented composite transmitting antennas |
| PTT-DMT | power transport theorem based decoupling mode theory |
| PTT-FreeSpace-DMT | power transport theorem based decoupling mode theory for free space |
| PTT-Guid-DMT | power transport theorem based decoupling mode theory for guiding structures |
| PTT-MatGuid-DMT | power transport theorem based decoupling mode theory for material guiding structures |
| PTT-MatRecAnt-DMT | power transport theorem based decoupling mode theory for augmented material receiving antennas |
| PTT-MatTraAnt-DMT | power transport theorem based decoupling mode theory for augmented material transmitting antennas |
| PTT-MetGuid-DMT | power transport theorem based decoupling mode theory for metallic guiding structures |
| PTT-MetRecAnt-DMT | power transport theorem based decoupling mode theory for augmented metallic receiving antennas |
| PTT-MetTraAnt-DMT | power transport theorem based decoupling mode theory for augmented metallic transmitting antennas |
| PTT-RecAnt-DMT | power transport theorem based decoupling mode theory for augmented receiving antennas |
| PTT-TARA-DMT | power transport theorem based decoupling mode theory for the combined systems constituted by augmented transmitting antennas and augmented receiving antennas |
| PTT-TGTA-DMT | power transport theorem based decoupling mode theory for the combined systems constituted by guiding structures and augmented transmitting antennas |
| PTT-TraAnt-DMT | power transport theorem based decoupling mode theory for augmented transmitting antennas |
| PTT-TRSys-DMT | power transport theorem based decoupling mode theory for transceiving systems |
| Q-factor | quantity factor |
| Rec-ant | receiving antenna |
| rec-antenna | receiving antenna |
| receiver | receiving system |
| rec-ground | electric wall used to partially separate receiving system from its surrounding environment |
| Rec-guid | guiding structure in receiving system |
| rec-guide | guiding structure in receiving system |
| SDC | solution domain compression |
| SDC-DoJ | solution domain compression based on the definition of electric current |
| SDC-DoM | solution domain compression based on the definition of magnetic current |
| SFR | source-free region |
| SIE | surface integral equation |
| SLO | Sturm-Liouville operator |
| SLT | Sturm-Liouville theory |
| SLT-SFReg-EMT | Sturm-Liouville theory based eigen-mode theory for source-free regions |
| SLT-WGSys-EMT | Sturm-Liouville theory based eigen-mode theory for wave-guiding systems |
| SM | scattering matrix |
| SM-ScaSys-CMT | scattering matrix based characteristic mode theory for scattering systems |
| SVIE | surface-volume integral equation |
| TARA system | combined system constituted by augmented transmitting antenna and augmented receiving system |
| TE | transverse electric (wave) |
| TGTA system | combined system constituted by guiding structure and augmented transmitting antenna |
| TM | transverse magnetic (wave) |
| TPR | two-port region |
| Tra-ant | transmitting antenna |
| tra-antenna | transmitting antenna |
| tra-ground | electric wall used to partially separate transmitting system from its surrounding environment |
| Tra-guid | guiding structure in transmitting system |
| transmitter | transmitting system |
| VIE | volume integral equation |
| WEP | work-energy principle |
| WEP-ComSca-CMT | work-energy principle based characteristic mode theory for composite scattering systems |
| WEP-MatSca-CMT | work-energy principle based characteristic mode theory for material scattering systems |
| WEP-MetSca-CMT | work-energy principle based characteristic mode theory for metallic scattering systems |
| WEP-ScaSys-CMT | work-energy principle based characteristic mode theory for scattering systems |
| Θ-factor | electric-magnetic decoupling factor |

# Contents

# Chapter 1 Introduction

**CHAPTER MOTIVATION:** By a brief review for the *characteristic mode theory* (*CMT*), this chapter reveals the fact that the CMT is for scattering systems rather than for transceiving systems, and then introduces the main topic focused on by this postdoctoral research report — "how to establish a rigorous modal theory for transceiving systems".

This chapter is organized as: research background and significance of this report (Sec. 1.1) → research history and state related to this report (Sec. 1.2) → important problems and challenges focused on by this report (Sec. 1.3) → main contributions and innovations of this report (Sec. 1.4) → study outline and roadmap of this report (Sec. 1.5).

## 1.1 Research Background and Significance

*Energy* is one of important *information carriers*. In electromagnetic (EM) engineering, it has had a long history to transmit and receive information by loading the information to EM energy. Just by controlling the *space-time distribution* of the energy, *EM device* (alternatively called *EM system*, or simply called *system*) finally realizes the control for the information. The different space-time distributions of the energy correspond to the different *working states* (alternatively called *working modes*, or simply called *modes*) of the system. For a linear system, its all physically realizable modes constitute a linear space — *modal space*.

In the modal space, some modes can exchange energies with each other, but some modes can not. The modes without energy exchange are called *energy-decoupled modes*, or simply called *decoupled modes* (*DMs*). Because of the absence of energy coupling, the DMs can work independently, and then the DMs can carry independent informations. Due to the energy-decoupling feature and the information-decoupling feature, the DMs have many important engineering applications. Thus, to construct a set of DMs in modal space has become one of the important topics in mathematical physics, applied mathematics and engineering electromagnetics, etc.

## 1.2 Research History and State

Some theoretical and practical methods related to DMs have been established, and the methods can be collectively referred to as *modal theory*. The studies for modal theory have had a relatively long history, and the earliest studies can be dated back to some famous mathematicians and physicists, such as Bernoulli family, D'Alembert, Euler, Fourier, Sturm, and Liouville *et al*.

### 1.2.1 Eigen-Mode Theory

The most classical modal theory used in electromagnetism is *electromagnetic eigen-mode theory*, or simply called *eigen-mode theory* (*EMT*), and it has been used to construct the *eigen-modes* of various closed EM systems, such as wave-guiding structures and wave-oscillating cavities etc., for many years. Under famous *Sturm-Liouville theory* (*SLT*) framework, **electromagnetic EMT focuses on constructing a set of eigen-modes propagating or oscillating in a region with perfectly electric wall**, by solving *Sturm-Liouville equation*. The operator contained in Sturm-Liouville equation — *Sturm-Liouville operator* (*SLO*) — can be viewed as the *generating operator* of the eigen-modes. Some detailed discussions for electromagnetic EMT can be found in Refs. [1~3], and some detailed discussions for SLT can be found in Ref. [4].

In fact, it is not difficult to prove that **the classical EM eigen-modes of metallic waveguides and metallic cavities are energy-decoupled**.

### 1.2.2 From Eigen-Mode Theory to Characteristic Mode Theory

In the 1960s[5,6] and 1970s[7~12], an alternative modal theory — *characteristic mode theory* (*CMT*) — was built for open EM scattering systems, and the modes derived from CMT are now called *characteristic modes* (*CMs*).

The concept of CM was introduced by Prof. Garbacz[5] in 1965 for the first time. Under *scattering matrix* (*SM*) framework, Prof. Garbacz constructed a kind of far-field-decoupled CMs — *Garbacz's CMs* — for lossless scattering systems only, by orthogonalizing *perturbation matrix operator* (*PMO*). A systematical and detailed discussion for Prof. Garbacz's *SM-based CMT for scattering systems* (*SM-ScaSys-CMT*) can be found in his Ph.D. Dissertation [6], and a relatively simplified discussion for SM-ScaSys-CMT can be found in the Sec. 2.2 of Ref. [13].

### 1.2.3 Characteristic Mode Theory — From Scattering Matrix Framework to Integral Equation Framework

Following Prof. Garbacz's pioneering works, Prof. Harrington *et al.*[8~12], under *integral equation* (*IE*) framework, constructed "another" kind of CMs — *Harrington's CMs* — for both lossless and lossy scattering systems, by orthogonalizing *impedance matrix operator* (*IMO*). For the lossless scattering systems, it is very clear that Harrington's CMs are far-field-decoupled. Thus, for lossless scattering systems, Harrington's CMs are equivalent (but not necessarily identical) to Garbacz's CMs in the sense of decoupling modal far fields.

From Prof. Garbacz's SM-ScaSys-CMT to Prof. Harrington's *IE-ScaSys-CMT* (*IE-based CMT for scattering systems*), it is not only a transformation for the carrying framework of CMT — from *SM framework* to *IE framework*, but also a transformation for the construction method of CMs — from *orthogonalizing PMO method* to *orthogonalizing IMO method*, as shown in Tab. 1-1. The transformations significantly extend the application range of CMT①. In addition, perhaps more importantly, the transformations significantly simplify the calculation process for CMs, because to obtain the IMO is much easier than to obtain the PMO. Thus, IE-ScaSys-CMT has a wider spread in EM engineering than SM-ScaSys-CMT. Some valuable reviews for the engineering applications of IE-ScaSys-CMT can be found in Refs. [14~16].

Table 1-1 The evolutions of CMT and the comparisons from aspects of carrying framework, construction method, and physical picture

| ASPECTS / THEORIES | Carrying Framework | Construction Method | Physical Picture |
|---|---|---|---|
| **SM-ScaSys-CMT (1960s)[5~7]** | SM framework | Orthogonalizing PMO method | To construct a set of *far-field-decoupled modes* |
| | ↓ | | |
| **IE-ScaSys-CMT (1970s)[8~12]** | IE framework | Orthogonalizing IMO method | Not clarified by its founders |
| | ↓ | | |
| **WEP-ScaSys-CMT (2019)[13]** | WEP framework | Orthogonalizing DPO method | To construct a set of *energy-decoupled modes* |

After half a century progress, IE-ScaSys-CMT has had a great development in many aspects[14~16]. But, at the same time, some problems are also exposed[13,17], for example:

① Because: IE-ScaSys-CMT[8~12] is applicable to both lossless and lossy scattering systems, but SM-ScaSys-CMT[5~7] is applicable to lossless scattering systems only.

Problem I: When the objective scattering system is placed in a non-free-space *background* (or called *environment*), if the *background Green's functions* are employed to establish CM calculation formulations, then the calculated CMs seemingly don't satisfy a reasonable energy-decoupling relation. Then, how to reasonablely calculate the CMs of the objective scattering system, which is placed in non-free-space environment? In addition, how to reasonablely calculate the CMs of "the objective scattering system which is placed in non-free-space environment"?

Problem II: When the objective scattering system is lossy, the far-field-decoupling feature and the energy-decoupling feature cannot be simultaneously satisfied by the CMs of the system usually. Then, which feature is the preferred/indispensable one for CMs, and which feature is the subordinate/dispensable one for CMs, and why?

Problem III: When the objective scattering system is magnetodielectric, i.e., the system is both magnetic and dielectric, the modal *characteristic values* don't have a clear physical meaning as the characteristic values of metallic scattering systems. Then, how to physically explain this phenomenon, and whether or not the characteristic values of any kind of systems must have the same physical meaning as the ones of metallic systems?

Problem IV: To normalize *modal real power* to 1 has already become a consensus in the realm of CMT since Prof. Harrington's classical Ref. [9], but there has not been a physical explanation for the normalization way up until now. Then, what is the physical reason to normalize CMs like that?

Problem V: For material scattering systems, the modes outputted from *surface integral equation* (*SIE*) based and *volume integral equation* (*VIE*) based CM formulas are not always consistent with each other. In addition, for metal-material composite scattering systems, the modes outputted from SIE-based and *surface-volume integral equation* (*SVIE*) based formulas are also not always consistent with each other. Then, what are the reasons leading to the inconsistencies, and how to resolve the problem?

### 1.2.4 Characteristic Mode Theory — From Integral Equation Framework to Work-Energy Principle Framework

To resolve the above Problems I~V and some other problems listed in the Chap. 1 of Ref.

[13], the *work-energy principle* (*WEP*) for EM scattering systems was derived, and then the concept of *driving power* (*DP*) was introduced, by some researchers[13,18,19] recently. Employing the WEP and DP, the above-mentioned Problems I~V were resolved to a certain degree. Taking the material objective scattering system $\mathbb{V}$ shown in Fig. 1-1 as an example, the WEP, DP, and the resolution schemes are simply reviewed as below.

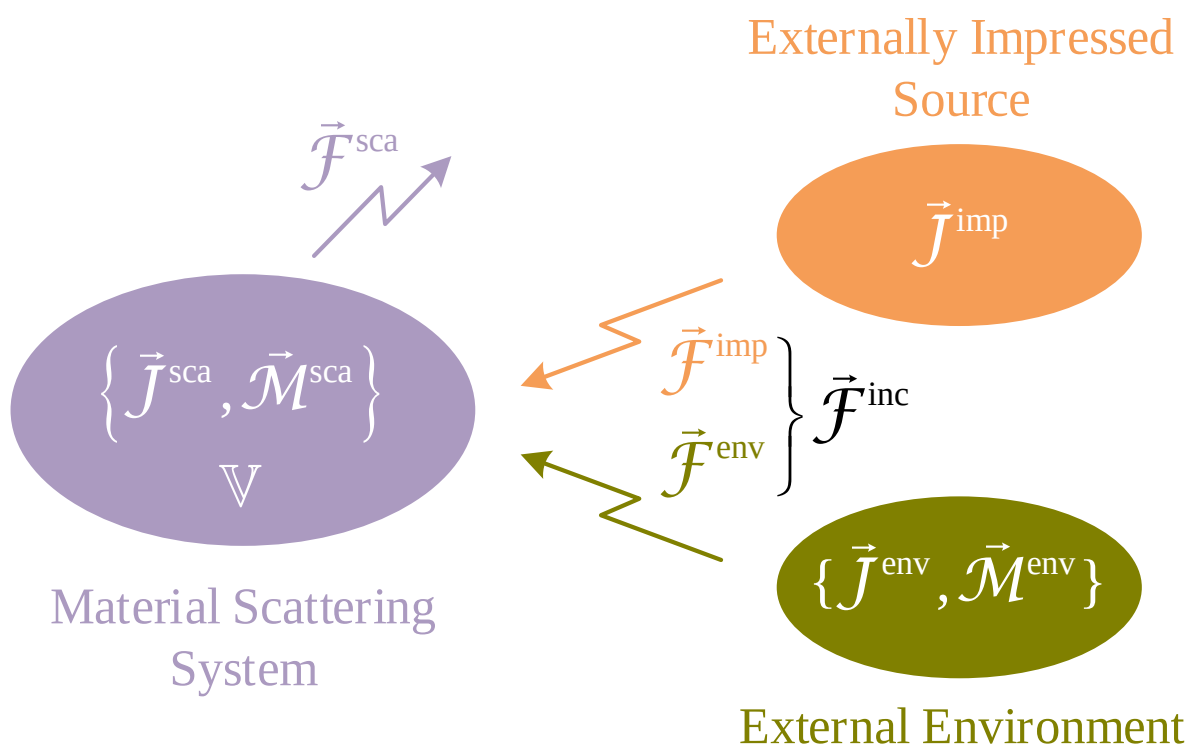

Figure 1-1 A magnetodielectric and lossy material scattering system $\mathbb{V}$, which is placed in a non-free-space environment, is excited by an externally impressed source

The scattering system $\mathbb{V}$ is magnetodielectric and lossy, and its *magnetic permeability*, *dielectric permittivity*, and *electric conductivity* are $\ddot{\mu}$, $\ddot{\varepsilon}$, and $\ddot{\sigma}$ respectively, and the material parameters are restricted to real and symmetrical two-order tensors①. In the following discussions, the time-domain field and current are denoted as $\vec{\mathcal{F}}$ and $\vec{\mathcal{C}}$ respectively, and the frequency-domain field and current are denoted as $\vec{F}$ and $\vec{C}$ respectively, where $\vec{\mathcal{F}}=\vec{\mathcal{E}}\,/\,\vec{\mathcal{H}}$ & $\vec{\mathcal{C}}=\vec{\mathcal{J}}\,/\,\vec{\mathcal{M}}$ and $\vec{F}=\vec{E}\,/\,\vec{H}$ & $\vec{C}=\vec{J}\,/\,\vec{M}$.

Under the excitation of *externally impressed fields* $\{\vec{\mathcal{E}}^{imp},\vec{\mathcal{H}}^{imp}\}$, some *scattered electric and magnetic currents* $\{\vec{\mathcal{J}}^{sca},\vec{\mathcal{M}}^{sca}\}$ are induced on $\mathbb{V}$[20,21], and, at the same time, some *environment electric and magnetic currents* $\{\vec{\mathcal{J}}^{env},\vec{\mathcal{M}}^{env}\}$ are also induced on environment, as shown in Fig. 1-1. Then, some *scattered fields* $\{\vec{\mathcal{E}}^{sca},\vec{\mathcal{H}}^{sca}\}$ and *environmental fields* $\{\vec{\mathcal{E}}^{env},\vec{\mathcal{H}}^{env}\}$ distributing on whole *three-dimensional Euclidean space* $\mathbb{E}^3$ are generated by $\{\vec{\mathcal{J}}^{\rm sca},\vec{\mathcal{M}}^{\rm sca}\}$ and $\{\vec{\mathcal{J}}^{\rm env},\vec{\mathcal{M}}^{\rm env}\}$ respectively, as shown in Fig. 1-1. In Refs. [13] and [18], the summation of the $\vec{\mathcal{F}}^{\rm imp}$ and $\vec{\mathcal{F}}^{\rm env}$ is called *externally resultant field* or alternatively called *incident field*, and denoted as $\vec{\mathcal{F}}^{\rm inc}$, i.e., $\vec{\mathcal{F}}^{\rm inc}=\vec{\mathcal{F}}^{\rm imp}+\vec{\mathcal{F}}^{\rm env}$②. In addition, the summation of the $\vec{\mathcal{F}}^{\rm inc}$ and $\vec{\mathcal{F}}^{\rm sca}$ is the *total field*, and it is denoted as $\vec{\mathcal{F}}^{\rm tot}$, i.e., $\vec{\mathcal{F}}^{\rm tot}=\vec{\mathcal{F}}^{\rm inc}+\vec{\mathcal{F}}^{\rm sca}$.

① The reason for this restriction can be found in the App. B of Ref. [13].
② A *work-energy principle* based explanation and a *Huygens-Fresnel principle* based explanation for this treatment had been given in Ref. [13]; another *Poynting's theorem* based explanation had been done by our research group, and will be simply reviewed in the Sec. 1.2.4.4 of this report.

#### 1.2.4.1 Carrying Framework of WEP-ScaSys-CMT

The *energy conservation law* tells us that the action of $\{\vec{\mathcal{E}}^{\text{inc}},\vec{\mathcal{H}}^{\text{inc}}\}$ on $\{\vec{\mathcal{J}}^{\text{sca}},\vec{\mathcal{M}}^{\text{sca}}\}$ will lead to a transformation between work and energy, and the *work-energy transformation* can be quantitatively expressed as follows[13,18]:

$$\mathcal{W}^{\text{Driv}} = \mathcal{E}_{\mathbb{S}_\infty}^{\text{rad}} + \mathcal{E}_{\mathbb{V}}^{\text{dis}} + \Delta\left(\mathcal{E}_{\mathbb{E}^3}^{\text{mag}} + \mathcal{E}_{\mathbb{E}^3}^{\text{ele}}\right) + \Delta\left(\mathcal{E}_{\mathbb{V}}^{\text{mag}} + \mathcal{E}_{\mathbb{V}}^{\text{pol}}\right) \tag{1-1}$$

in which

$$\mathcal{W}^{\text{Driv}} = \int_{t_0}^{t_0+\Delta t}\left[\left\langle \vec{\mathcal{J}}^{\text{sca}},\vec{\mathcal{E}}^{\text{inc}}\right\rangle_{\mathbb{V}} + \left\langle \vec{\mathcal{M}}^{\text{sca}},\vec{\mathcal{H}}^{\text{inc}}\right\rangle_{\mathbb{V}}\right]dt \tag{1-2a}$$

$$\mathcal{E}_{\mathbb{S}_\infty}^{\text{rad}} = \int_{t_0}^{t_0+\Delta t}\left[\oiint_{\mathbb{S}_\infty}\left(\vec{\mathcal{E}}^{\text{sca}}\times\vec{\mathcal{H}}^{\text{sca}}\right)\cdot\hat{n}_{\mathbb{S}_\infty}^{+}\, dS\right]dt \tag{1-2b}$$

$$\mathcal{E}_{\mathbb{V}}^{\text{dis}} = \int_{t_0}^{t_0+\Delta t}\left\langle \vec{\vec{\sigma}}\cdot\vec{\mathcal{E}}^{\text{tot}},\vec{\mathcal{E}}^{\text{tot}}\right\rangle_{\mathbb{V}} dt \tag{1-2c}$$

$$\begin{aligned}\Delta\left(\mathcal{E}_{\mathbb{E}^3}^{\text{mag}} + \mathcal{E}_{\mathbb{E}^3}^{\text{ele}}\right) = &\left[(1/2)\left\langle \vec{\mathcal{H}}^{\text{sca}},\mu_0\vec{\mathcal{H}}^{\text{sca}}\right\rangle_{\mathbb{E}^3} + (1/2)\left\langle \varepsilon_0\vec{\mathcal{E}}^{\text{sca}},\vec{\mathcal{E}}^{\text{sca}}\right\rangle_{\mathbb{E}^3}\right]_{t=t_0+\Delta t}\\ &-\left[(1/2)\left\langle \vec{\mathcal{H}}^{\text{sca}},\mu_0\vec{\mathcal{H}}^{\text{sca}}\right\rangle_{\mathbb{E}^3} + (1/2)\left\langle \varepsilon_0\vec{\mathcal{E}}^{\text{sca}},\vec{\mathcal{E}}^{\text{sca}}\right\rangle_{\mathbb{E}^3}\right]_{t=t_0}\end{aligned} \tag{1-2d}$$

$$\begin{aligned}\Delta\left(\mathcal{E}_{\mathbb{V}}^{\text{mag}} + \mathcal{E}_{\mathbb{V}}^{\text{pol}}\right) = &\left[(1/2)\left\langle \vec{\mathcal{H}}^{\text{tot}},\Delta\vec{\vec{\mu}}\cdot\vec{\mathcal{H}}^{\text{tot}}\right\rangle_{\mathbb{V}} + (1/2)\left\langle \Delta\vec{\vec{\varepsilon}}\cdot\vec{\mathcal{E}}^{\text{tot}},\vec{\mathcal{E}}^{\text{tot}}\right\rangle_{\mathbb{V}}\right]_{t=t_0+\Delta t}\\ &-\left[(1/2)\left\langle \vec{\mathcal{H}}^{\text{tot}},\Delta\vec{\vec{\mu}}\cdot\vec{\mathcal{H}}^{\text{tot}}\right\rangle_{\mathbb{V}} + (1/2)\left\langle \Delta\vec{\vec{\varepsilon}}\cdot\vec{\mathcal{E}}^{\text{tot}},\vec{\mathcal{E}}^{\text{tot}}\right\rangle_{\mathbb{V}}\right]_{t=t_0}\end{aligned} \tag{1-2e}$$

In the above equations, *time interval* $\Delta t$ is a positive real number; the *inner product* is defined as that $<\vec{f},\vec{g}>_\Omega=\int_\Omega \vec{f}^\dagger\cdot\vec{g}\, d\Omega$, where superscript "$\dagger$" is the *transpose conjugate operation*; integral domain $\mathbb{S}_\infty$ is a spherical surface with infinite radius; $\hat{n}_{\mathbb{S}_\infty}^{+}$ is the *normal direction* of $\mathbb{S}_\infty$, and points to *infinity*; symbols $[\text{quantity}]_{t=t_0}$ and $[\text{quantity}]_{t=t_0+\Delta t}$ denote the values of the physical quantities at time points $t_0$ and $t_0+\Delta t$ respectively; $\Delta\vec{\vec{\mu}}=\vec{\vec{\mu}}-\vec{\vec{I}}\mu_0$, and $\Delta\vec{\vec{\varepsilon}}=\vec{\vec{\varepsilon}}-\vec{\vec{I}}\varepsilon_0$, and $\vec{\vec{I}}=\hat{x}\hat{x}+\hat{y}\hat{y}+\hat{z}\hat{z}$ is *unit dyad*.

The above Eq. (1-1) has a very clear physical meaning: **in time interval $t_0 \sim t_0+\Delta t$, the work $\mathcal{W}^{\text{Driv}}$ done by the $\{\vec{\mathcal{E}}^{\text{inc}},\vec{\mathcal{H}}^{\text{inc}}\}$ acting on $\{\vec{\mathcal{J}}^{\text{sca}},\vec{\mathcal{M}}^{\text{sca}}\}$ is transformed into the following four parts**

**Part I. the *radiated energy* $\mathcal{E}_{\mathbb{S}_\infty}^{rad}$ carried by $\{\vec{\mathcal{E}}^{\text{sca}},\vec{\mathcal{H}}^{\text{sca}}\}$ from scattering system to infinity,**

**Part II. the *Joule heating energy* $\mathcal{E}_{\mathbb{V}}^{dis}$ dissipated in $\mathbb{V}$,**

**Part III. the increment of the *magnetic field energy* $\mathcal{E}_{\mathbb{E}^3}^{mag}$ and *electric field energy* $\mathcal{E}_{\mathbb{E}^3}^{ele}$ stored in the fields distributing on $\mathbb{E}^3$, and**

**Part IV. the increment of the *magnetization energy* $\mathcal{E}_{\mathbb{V}}^{mag}$ and *polarization energy* $\mathcal{E}_{\mathbb{V}}^{pol}$ stored in the matter occupying $\mathbb{V}$.**

Equation (1-1) is very similar to the *work-energy principle for the particles in mechanism*[22,23], so Refs. [13,18,19] called it the *work-energy principle (WEP) for the scattering systems in electromagnetism*. Based on the WEP, a *WEP-based CMT for scattering systems* (*WEP-ScaSys-CMT*) was established in Refs. [13,18,19], such that the carrying framework for CMT was transformed again — from *IE framework* to *WEP framework* — as shown in Tab. 1-1.

### 1.2.4.2 Generating Operator of CMs

**It is clear that the term $\mathcal{W}^{\rm Driv}$ is the source to drive the work-energy transformation and to sustain a stationary working of the scattering system**, so it is called *driving work* (*DW*), and the corresponding power is called *driving power* (*DP*) and denoted as $\mathcal{P}^{\rm Driv}$, i.e.,[13,18,19]

$$\mathcal{P}^{\rm Driv} = \left\langle \vec{\mathcal{J}}^{\rm sca}, \vec{\mathcal{E}}^{\rm inc} \right\rangle_{\mathbb{V}} + \left\langle \vec{\mathcal{M}}^{\rm sca}, \vec{\mathcal{H}}^{\rm inc} \right\rangle_{\mathbb{V}} \tag{1-3}$$

The above time-domain *driving power operator* (*DPO*) has two different frequency-domain versions as follows[13,18]:

$$P^{\rm driv} = (1/2)\left\langle \vec{J}^{\rm sca}, \vec{E}^{\rm inc} \right\rangle_{\mathbb{V}} + (1/2)\left\langle \vec{M}^{\rm sca}, \vec{H}^{\rm inc} \right\rangle_{\mathbb{V}} \tag{1-4}$$

$$P^{\rm DRIV} = (1/2)\left\langle \vec{J}^{\rm sca}, \vec{E}^{\rm inc} \right\rangle_{\mathbb{V}} + (1/2)\left\langle \vec{H}^{\rm inc}, \vec{M}^{\rm sca} \right\rangle_{\mathbb{V}} \tag{1-5}$$

where the coefficient 1/2 originates from the time average of *time-harmonic fields*[24,25].

Recently, under the WEP framework, Refs. [13,18,19] effectively constructed the CMs by orthogonalizing the *frequency-domain DPO* $P^{driv}$ (the $P^{\rm driv}$ can be viewed as the *generating operator of CMs*①), and then the construction method for CMs was transformed again — from *orthogonalizing IMO method* to *orthogonalizing DPO method* — as shown in Tab. 1-1. In addition, as proved in Refs. [13,26], there also exists the following surface formulation

$$P^{\rm driv} = -(1/2)\left\langle \vec{J}^{\rm equ}, \vec{E}^{\rm inc} \right\rangle_{\partial\mathbb{V}} - (1/2)\left\langle \vec{M}^{\rm equ}, \vec{H}^{\rm inc} \right\rangle_{\partial\mathbb{V}} \tag{1-6}$$

for the frequently-domain DPO $P^{\rm driv}$ given in Eq. (1-4), where $\vec{J}^{\rm equ}$ and $\vec{M}^{\rm equ}$ are the *equivalent surface electric and magnetic currents* distributing on $\partial\mathbb{V}$ (which is the

① The reason not selecting $P^{\rm DRIV}$ as the generating operator will be given in the following Sec. 1.2.4.3, and the reason can also be found in Refs. [13,18].

boundary of $\mathbb{V}$), and they are defined as that $\vec{J}^{\text{equ}} = \hat{n}_{\partial\mathbb{V}}^{-} \times \vec{H}_{-}^{\text{tot}}$ and $\vec{M}^{\text{equ}} = \vec{E}_{-}^{\text{tot}} \times \hat{n}_{\partial\mathbb{V}}^{-}$, where $\hat{n}_{\partial\mathbb{V}}^{-}$ is the *inner normal direction* of $\partial\mathbb{V}$ and $\{\vec{E}_{-}^{\text{tot}}, \vec{H}_{-}^{\text{tot}}\}$ are the total fields distributing on the inner surface of $\partial\mathbb{V}$.

### 1.2.4.3 Physical Picture of WEP-ScaSys-CMT

The CMs derived from orthogonalizing $P^{\text{driv}}$ satisfy the following *decoupling relation* (or alternatively called *orthogonality relation*)[13,18,26]

$$(1/2)\left\langle \vec{J}_{\xi}^{\text{sca}}, \vec{E}_{\zeta}^{\text{inc}} \right\rangle_{\mathbb{V}} + (1/2)\left\langle \vec{M}_{\xi}^{\text{sca}}, \vec{H}_{\zeta}^{\text{inc}} \right\rangle_{\mathbb{V}} = 0 \quad , \quad \text{if } \xi \neq \zeta \tag{1-7}$$

However, the modes derived from orthogonalizing $P^{\text{DRIV}}$ satisfy the following intertwining relation[13,18,27]

$$(1/2)\left\langle \vec{J}_{\xi}^{\text{sca}}, \vec{E}_{\zeta}^{\text{inc}} \right\rangle_{\mathbb{V}} + (1/2)\left\langle \vec{H}_{\xi}^{\text{inc}}, \vec{M}_{\zeta}^{\text{sca}} \right\rangle_{\mathbb{V}} = 0 \quad , \quad \text{if } \xi \neq \zeta \tag{1-8}$$

Obviously, the CMs satisfying Eq. (1-7) are completely decoupled, but the modes satisfying Eq. (1-8) are not! In fact, this is just the reason to utilize frequency-domain DPO $P^{\text{driv}}$ as the generating operator of CMs rather than utilizing $P^{\text{DRIV}}$ [13,18].

The *frequency-domain power-decoupling relation* (1-7) implies the following *time-domain energy-decoupling relation* (or called *time-average DP-decoupling relation*)

$$\frac{1}{T}\int_{t_0}^{t_0+T}\left[\left\langle \vec{\mathcal{J}}_{\xi}^{\text{sca}}, \vec{\mathcal{E}}_{\zeta}^{\text{inc}} \right\rangle_{\mathbb{V}} + \left\langle \vec{\mathcal{M}}_{\xi}^{\text{sca}}, \vec{\mathcal{H}}_{\zeta}^{\text{inc}} \right\rangle_{\mathbb{V}}\right] dt = 0 \quad , \quad \text{if } \xi \neq \zeta \tag{1-9}$$

where $T$ is the *time period* of the *time-harmonic EM field*. Equation (1-9) clearly reveals the fact that: **the $\zeta$-th *modal incident fields* $\{\vec{\mathcal{E}}_{\zeta}^{inc}, \vec{\mathcal{H}}_{\zeta}^{inc}\}$ don't supply net energy to the $\xi$-th *modal scattered currents* $\{\vec{\mathcal{J}}_{\xi}^{sca}, \vec{\mathcal{M}}_{\xi}^{sca}\}$ in any integral period, if $\zeta \neq \xi$.** This above gives CMT a very clear *physical picture* — **to construct a set of modes without net energy exchange in any integral period (i.e. to construct a set of *energy-decoupled modes*) for objective scattering system**. Thus, the most essential feature of CMs is that: **the CMs can be independently excited, and then the CMs can carry independent informations.** In fact, this is just the most core charm of CMs.①

For visually exhibiting the above conclusions, we provide an example here. We consider the material elliptical cylinder shown in Fig. 1-2. Its X-directional and Y-directional radiuses are 4mm and 6mm respectively, and its Z-directional height is 5mm.

① There were always someones to ask the following question: **what characters of the objective scattering system does characteristic modes depict?** In fact, **if the modes were called *energy-decoupled modes* at the beginning of their introduction, there wouldn't be the question!** However, this report will still call the modes (which were constructed by Prof. Garbacz *et al.*[5~7] and Prof. Harrington *et al.*[8~12] for scattering systems) *characteristic modes* to conform the traditional custom.

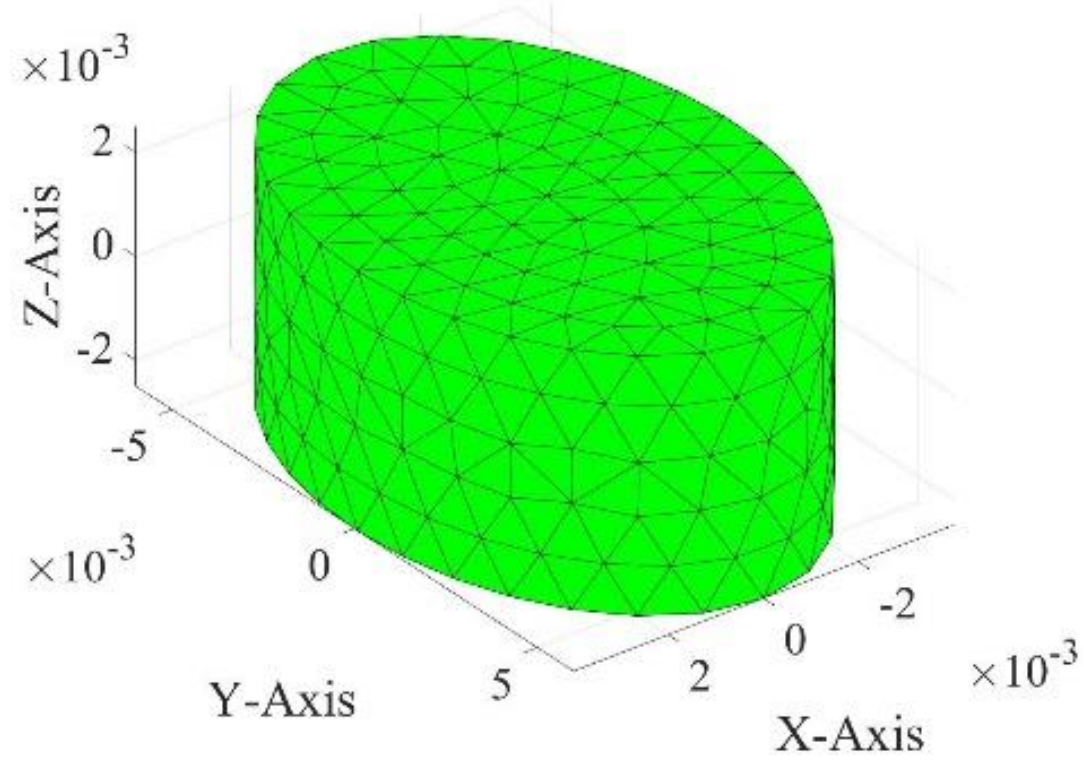

Figure 1-2 Geometry of a material elliptical cylinder. Its X-directional and Y-directional radiuses are 4mm and 6mm respectively, and its Z-directional height is 5mm

When its permeability, permittivity, and conductivity are $\overset{\leftrightarrow}{\mu} = \overset{\leftrightarrow}{I}2\mu_0$, $\overset{\leftrightarrow}{\varepsilon} = \overset{\leftrightarrow}{I}10\varepsilon_0$, and $\overset{\leftrightarrow}{\sigma} = \overset{\leftrightarrow}{I}0$ respectively, we calculate its CMs by orthogonalizing $P^{\text{driv}}$ and $P^{\text{DRIV}}$ respectively, and the associated characteristic values in decibel (dB) are shown in the following Figs. 1-3(a) and 1-3(b) respectively.

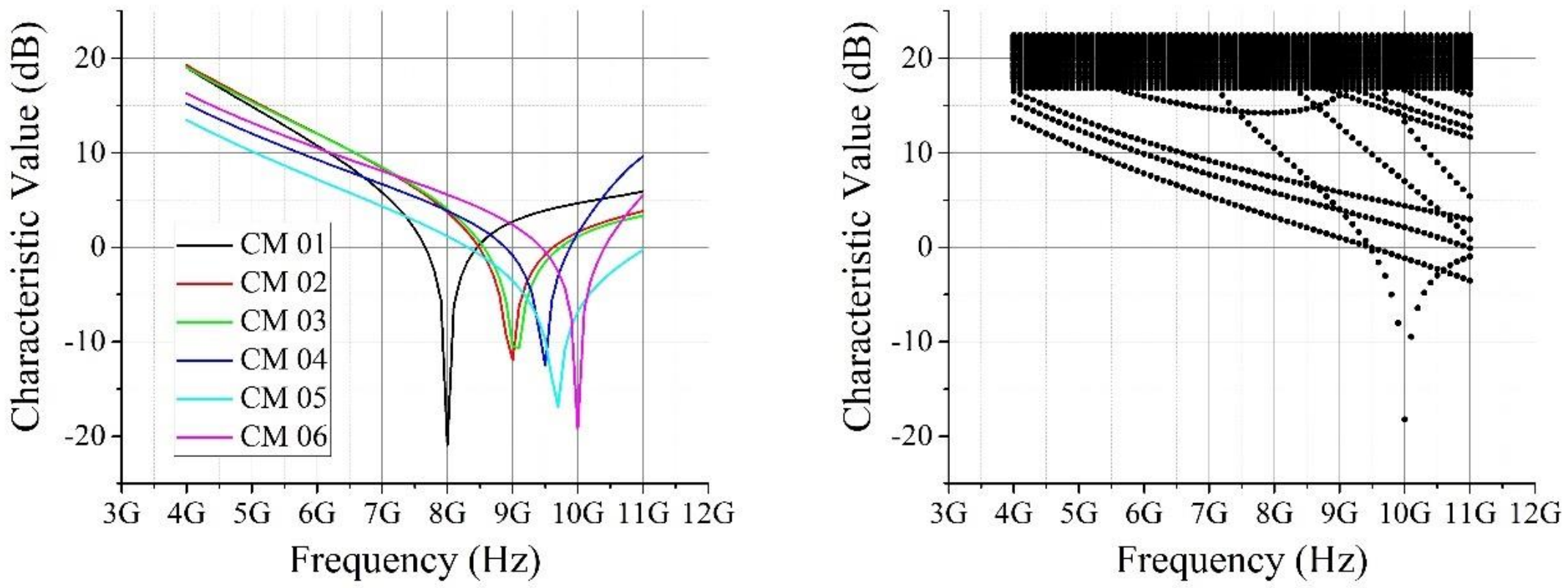

(a) (b)

Figure 1-3 Characteristic values of the CMs of a material elliptical cylinder, whose topological structure is shown in Fig. 1-2 and whose material parameters are $\overset{\leftrightarrow}{\mu} = \overset{\leftrightarrow}{I}2\mu_0$, $\overset{\leftrightarrow}{\varepsilon} = \overset{\leftrightarrow}{I}10\varepsilon_0$, and $\sigma = \overset{\leftrightarrow}{I}0$. (a) Characteristic values calculated from orthogonalizing $P^{\text{driv}}$ (using the volume formulation given in Ref. [11]); (b) characteristic values calculated from orthogonalizing $P^{\text{DRIV}}$ (using the volume formulation given in Ref. [27])

From the above Figs. 1-3(a) and 1-3(b), it is obvious that the CMs calculated from orthogonalizing $P^{\text{driv}}$ and the CMs calculated from orthogonalizing $P^{\text{DRIV}}$ are different from each other. The *time-average DP orthogonality matrices* corresponding to the CMs given in Figs. 1-3(a) and 1-3(b) are visually shown in the following Figs. 1-4(a) and 1-4(b) respectively.

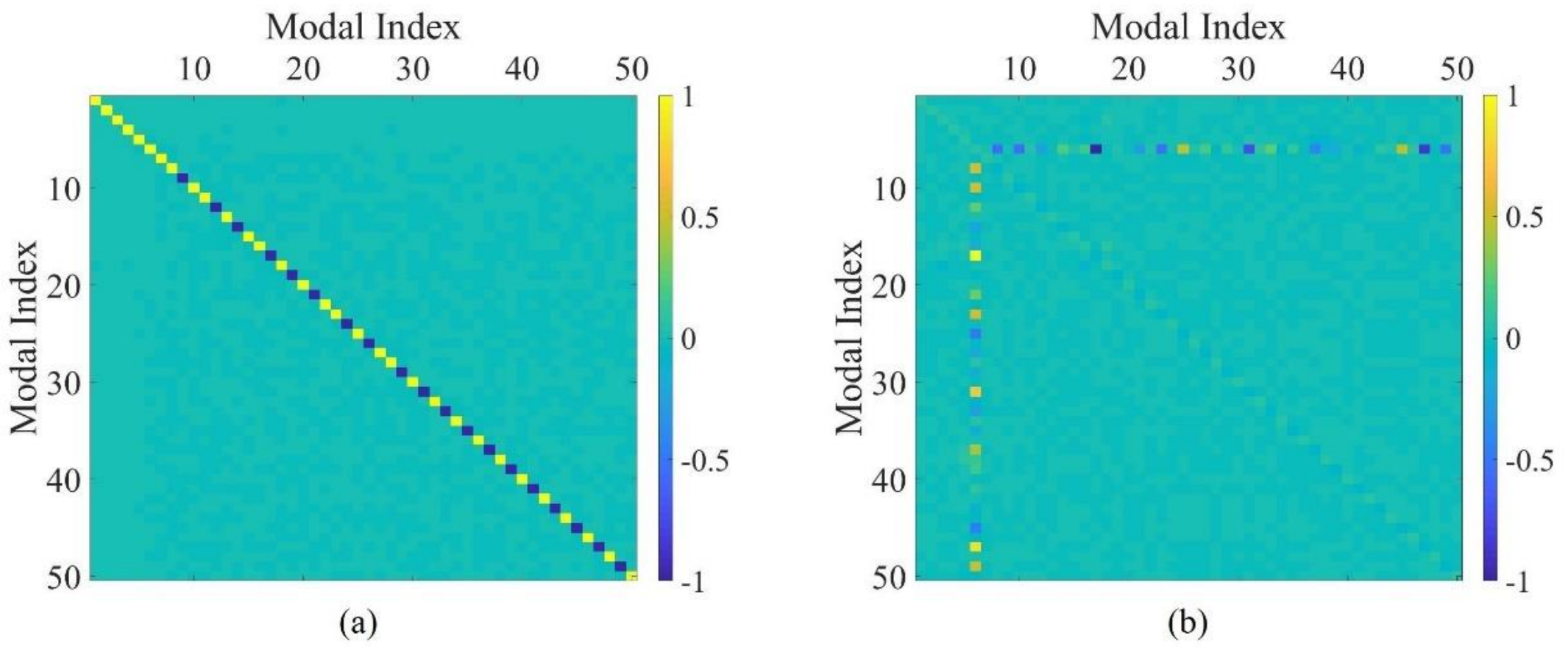

Figure 1-4 Time-average DP orthogonality matrices of the CMs (at 9 GHz) shown in Fig. 1-3. (a) Orthogonality matrix corresponding to the CMs (at 9 GHz) shown in Fig. 1-3(a); (b) orthogonality matrix corresponding to the CMs (at 9 GHz) shown in Fig. 1-3(b)

Evidently, the $P^{\text{driv}}$-based CMs are indeed energy-decoupled completely, but the $P^{\text{DRIV}}$-based CMs are not. The reasons leading to some $-1$ distributing on the diagonal of the matrix shown in Fig. 1-4(a) can be found in Ref. [18].

#### 1.2.4.4 On the Surrounding Environment of the Objective Scattering System

The Problem I mentioned in previous Sec. 1.2.3 had been studied in Ref. [13], and the core viewpoints and conclusions are simply reviewed here.

**An Explicit Expose on the Problem**

If the exterior of $\mathbb{V}$ is denoted as $\text{ext}\,\mathbb{V}$, and $\text{ext}\,\mathbb{V}$ is with parameters $\{\vec{\mu}_{\text{env}},\vec{\varepsilon}_{\text{env}},\vec{\sigma}_{\text{env}}\}$, then the impressed fields $\{\vec{E}^{\text{imp}},\vec{H}^{\text{imp}}\}$, environmental fields $\{\vec{E}^{\text{env}},\vec{H}^{\text{env}}\}$, and scattered fields $\{\vec{E}^{\text{sca}},\vec{H}^{\text{sca}}\}$ satisfy the following *Maxwell's equations*

$$\nabla\times\vec{H}^{\text{imp}} = j\omega\varepsilon_0\vec{E}^{\text{imp}}+\vec{J}^{\text{imp}} \quad , \quad \vec{r}\in\text{ext}\,\mathbb{V} \tag{1-10a}$$
$$-\nabla\times\vec{E}^{\text{imp}} = j\omega\mu_0\vec{H}^{\text{imp}} \quad , \quad \vec{r}\in\text{ext}\,\mathbb{V} \tag{1-10b}$$

and

$$\nabla\times\vec{H}^{\text{env}} = j\omega\varepsilon_0\vec{E}^{\text{env}}+\vec{J}^{\text{env}} \quad , \quad \vec{r}\in\text{ext}\,\mathbb{V} \tag{1-11a}$$
$$-\nabla\times\vec{E}^{\text{env}} = j\omega\mu_0\vec{H}^{\text{env}}+\vec{M}^{\text{env}} \quad , \quad \vec{r}\in\text{ext}\,\mathbb{V} \tag{1-11b}$$

and

$$\nabla\times\vec{H}^{\text{sca}} = j\omega\varepsilon_0\vec{E}^{\text{sca}} \quad , \quad \vec{r}\in\text{ext}\,\mathbb{V} \tag{1-12a}$$
$$-\nabla\times\vec{E}^{\text{sca}} = j\omega\mu_0\vec{H}^{\text{sca}} \quad , \quad \vec{r}\in\text{ext}\,\mathbb{V} \tag{1-12b}$$

In Eq. (1-10a), $\vec{J}^{\text{imp}}$ is the *externally impressed electric current* as shown in Fig. 1-1. Equations (1-12a) and (1-12b) are due to that $\text{ext}\,\mathbb{V}$ is a source-free region for scattered fields $\{\vec{E}^{\text{sca}},\vec{H}^{\text{sca}}\}$.

The summation of the above Maxwell's equations (1-10)~(1-12) is as follows:

$$\nabla\times\vec{H}^{\text{tot}} = j\omega\varepsilon_0\vec{E}^{\text{tot}} + \vec{J}^{\text{env}} + \vec{J}^{\text{imp}} \quad , \quad \vec{r}\in\text{ext}\,\mathbb{V} \tag{1-13a}$$

$$-\nabla\times\vec{E}^{\text{tot}} = j\omega\mu_0\vec{H}^{\text{tot}} + \vec{M}^{\text{env}} \quad , \quad \vec{r}\in\text{ext}\,\mathbb{V} \tag{1-13b}$$

because of that $\vec{F}^{\text{tot}} = \vec{F}^{\text{imp}} + \vec{F}^{\text{env}} + \vec{F}^{\text{sca}}$. Substituting *volume equivalence principle* $\{\vec{J}^{env} = j\omega\Delta\overleftrightarrow{\varepsilon}_{env}^{c}\cdot\vec{E}^{tot}, \vec{M}^{env} = j\omega\Delta\overleftrightarrow{\mu}_{env}\cdot\vec{H}^{tot}\}$ [13,28] (where $\Delta\overleftrightarrow{\mu}_{\text{env}} = \overleftrightarrow{\mu}_{\text{env}} - \overleftrightarrow{I}\mu_0$ and $\Delta\overleftrightarrow{\varepsilon}_{\text{env}}^{\text{c}} = \overleftrightarrow{\varepsilon}_{\text{env}}^{\text{c}} - \overleftrightarrow{I}\varepsilon_0$ and $\overleftrightarrow{\varepsilon}_{\text{env}}^{\text{c}} = \overleftrightarrow{\varepsilon}_{\text{env}} + \overleftrightarrow{\sigma}_{\text{env}}/j\omega$) into the above Maxwell's equations (1-13), it is immediately obtained that

$$\nabla\times\vec{H}^{\text{tot}} = j\omega\overleftrightarrow{\varepsilon}_{\text{env}}^{\text{c}}\cdot\vec{E}^{\text{tot}} + \vec{J}^{\text{imp}} \quad , \quad \vec{r}\in\text{ext}\,\mathbb{V} \tag{1-14a}$$

$$-\nabla\times\vec{E}^{\text{tot}} = j\omega\overleftrightarrow{\mu}_{\text{env}}\cdot\vec{H}^{\text{tot}} \quad , \quad \vec{r}\in\text{ext}\,\mathbb{V} \tag{1-14b}$$

where the relations $\overleftrightarrow{\varepsilon}_{\text{env}}^{\text{c}} = \Delta\overleftrightarrow{\varepsilon}_{\text{env}}^{\text{c}} + \overleftrightarrow{I}\varepsilon_0$ and $\overleftrightarrow{\mu}_{\text{env}} = \Delta\overleftrightarrow{\mu}_{\text{env}} + \overleftrightarrow{I}\mu_0$ have been utilized. Maxwell's equations (1-14) imply the following *convolution integral formulation*[29]

$$\vec{F}^{\text{tot}}(\vec{r}) = \underbrace{\overleftrightarrow{G}_{\text{env}}^{JF} * \left(-\vec{J}^{\text{equ}}\right) + \overleftrightarrow{G}_{\text{env}}^{MF} * \left(-\vec{M}^{\text{equ}}\right)}_{\vec{F}^1} + \underbrace{\overleftrightarrow{G}_{\text{env}}^{JF} * \vec{J}^{\text{imp}}}_{\vec{F}^2} \quad , \quad \vec{r}\in\text{ext}\,\mathbb{V} \tag{1-15}$$

if the fields $\vec{F}^{\text{tot}}$ and $\{\overleftrightarrow{G}_{\text{env}}^{JF},\overleftrightarrow{G}_{\text{env}}^{MF}\}$ satisfy the *Sommerfeld's radiation condition* corresponding to environment parameters $\{\overleftrightarrow{\mu}_{\text{env}},\overleftrightarrow{\varepsilon}_{\text{env}},\overleftrightarrow{\sigma}_{\text{env}}\}$. Here, $\{\overleftrightarrow{G}_{\text{env}}^{JF},\overleftrightarrow{G}_{\text{env}}^{MF}\}$ are the *environmental Green's functions* (or called *background Green's functions*) corresponding to environmental parameters $\{\overleftrightarrow{\mu}_{\text{env}},\overleftrightarrow{\varepsilon}_{\text{env}},\overleftrightarrow{\sigma}_{\text{env}}\}$, and the *convolution integral operator* "$*$" is defined as that $\overleftrightarrow{G} * \vec{C} = \int_\Omega \overleftrightarrow{G}(\vec{r},\vec{r}\,')\cdot\vec{C}(\vec{r}\,')d\Omega'$.

Traditionally, the term $\vec{F}^1 = \overleftrightarrow{G}_{\text{env}}^{JF} * (-\vec{J}^{\text{equ}}) + \overleftrightarrow{G}_{\text{env}}^{MF} * (-\vec{M}^{\text{equ}})$ in Eq. (1-15) is treated as scattered field $\vec{F}^{\text{sca}}$. But, this treatment may not be reasonable, because a correct convolution integral expression for $\vec{F}^{\text{sca}}$ is that $\vec{F}^{\text{sca}} = \overleftrightarrow{G}_0^{JF} * (-\vec{J}^{\text{equ}}) + \overleftrightarrow{G}_0^{MF} * (-\vec{M}^{\text{equ}})$[13], where $\{\overleftrightarrow{G}_0^{JF},\overleftrightarrow{G}_0^{MF}\}$ are the *free-space dyadic Green's functions* corresponding to free-space parameters $\{\mu_0,\varepsilon_0\}$.

To construct CMs, the classical PMCHWT (Poggio-Miller-Chang-Harrington-Wu-Tsai) based scheme focuses on orthogonalizing the following IMO

$$\mathcal{Z}^{\text{PMCHWT}} = -(1/2)\left\langle \vec{J}^{\text{equ}}, \vec{E}_-^{\text{tot}} - \vec{E}_+^1 \right\rangle_{\partial\mathbb{V}} - (1/2)\left\langle \vec{M}^{\text{equ}}, \vec{H}_-^{\text{tot}} - \vec{H}_+^1 \right\rangle_{\partial\mathbb{V}} \tag{1-16}$$

where $\{\vec{E}_+^1,\vec{H}_+^1\}$ are the $\{\vec{E}^1,\vec{H}^1\}$ distributing on the outer surface of $\partial\mathbb{V}$. Substituting Eq. (1-15) into the above Eq. (1-16), we immediately have that

$$\begin{aligned}\mathcal{Z}^{\text{PMCHWT}} &= -(1/2)\left\langle \vec{J}^{\text{equ}}, \vec{E}^{2} \right\rangle_{\partial\mathbb{V}} - (1/2)\left\langle \vec{M}^{\text{equ}}, \vec{H}^{2} \right\rangle_{\partial\mathbb{V}} \\ &= (1/2)\left\langle \vec{J}^{\text{sca}}, \vec{E}^{2} \right\rangle_{\mathbb{V}} + (1/2)\left\langle \vec{M}^{\text{sca}}, \vec{H}^{2} \right\rangle_{\mathbb{V}}\end{aligned} \tag{1-17}$$

where $\vec{E}^{2} = \overleftrightarrow{G}_{\text{env}}^{JE} * \vec{J}^{\text{imp}}$ and $\vec{H}^{2} = \overleftrightarrow{G}_{\text{env}}^{JH} * \vec{J}^{\text{imp}}$, and the second equality can be rigorously proved by employing a method similar to the one used in the Sec. 4.3.2 of Ref. [13].

Obviously, the right-hand side of the second equality in Eq. (1-17) is the power done by fields $\{\vec{E}^{2}, \vec{H}^{2}\}$ on scattered currents $\{\vec{J}^{\text{sca}}, \vec{M}^{\text{sca}}\}$. However, what are the fields $\{\vec{E}^{2}, \vec{H}^{2}\}$? We don't know what they are, but we know that: **they are neither the incident fields $\{\vec{E}^{\text{inc}}, \vec{H}^{\text{inc}}\}$ nor the impressed fields $\{\vec{E}^{\text{imp}}, \vec{H}^{\text{imp}}\}$!**

Because: It had been pointed out previously that $\vec{F}^{1} \neq \vec{F}^{\text{sca}}$. Thus, $\vec{F}^{2}$ must not be $\vec{F}^{\text{inc}}$ due to Eq. (1-15) and the fact that $\vec{F}^{\text{tot}} = \vec{F}^{\text{sca}} + \vec{F}^{\text{inc}}$.

The impressed field $\vec{F}^{\text{imp}}$ can be written as that $\vec{F}^{\text{imp}} = \overleftrightarrow{G}_{0}^{JF} * \vec{J}^{\text{imp}}$ [13]. Thus, $\vec{F}^{2}$ must not be $\vec{F}^{\text{imp}}$ due to that $\vec{F}^{2} = \overleftrightarrow{G}_{\text{env}}^{JF} * \vec{J}^{\text{imp}}$ and $\overleftrightarrow{G}_{0}^{JF} \neq \overleftrightarrow{G}_{\text{env}}^{JF}$.

Hence, **it is not a reasonable scheme to utilize non-free-space environmental Green's functions to establish CM formulations**. Then,

Question A: How to effectively establish the CM formulations for the objective scattering systems, which are placed in non-free-space environment?

Question B: How to effectively establish the CM formulations for "the objective scattering systems which are placed in non-free-space environment"?

Below, we answer the Question A first, and then answer the Question B.

**Answer to Question A**

On the Question A, Refs. [13,18,19] pointed out that the generating operator of CMs is frequency-domain DPO $P^{\text{driv}}$, and the operator is the complex power done by the *externally resultant field*, i.e. *incident field* $\vec{F}^{inc}$ (which is the summation of *externally impressed field* $\vec{F}^{imp}$ and *externally environmental field* $\vec{F}^{env}$), acting on the scattered current $\vec{C}^{\text{sca}}$ as exhibited in Eqs. (1-4) and (1-6), and it is neither the power done by $\vec{F}^{\text{imp}}$ on $\vec{C}^{\text{sca}}$ nor the power done by $\vec{F}^{\text{env}}$ on $\vec{C}^{\text{sca}}$. Based on this viewpoint, **there is no need to utilize non-free-space environmental Green's functions to establish the CM formulation for the objective scattering system, which is placed in a non-free-space environment**.

In fact, the above viewpoint can be naturally derived from WEP[13]. Here, we simply review a *Huygens-Fresnel principle* based explanation (E1)[13], and also provide an alternative *Poynting's theorem* based explanation (E2), for supporting the above viewpoint.

**(E1) A Huygens-Fresnel Principle Based Explanation for Treating $\vec{F}^{\text{imp}}$ and $\vec{F}^{\text{env}}$ as a Whole[13]**

As shown in Fig. 1-1, *Huygens' surface* $\partial\mathbb{V}$ (i.e. the boundary of $\mathbb{V}$) separates the whole three-dimensional Euclidean space $\mathbb{E}^3$ into two parts *the interior of* $\mathbb{V}$ (denoted as $\text{int}\,\mathbb{V}$) and *the exterior of* $\mathbb{V}$ (denoted as $\text{ext}\,\mathbb{V}$). Because both the source $\vec{J}^{\text{imp}}$ generating $\{\vec{E}^{\text{imp}},\vec{H}^{\text{imp}}\}$ and the sources $\{\vec{J}^{\text{env}},\vec{M}^{\text{env}}\}$ generating $\{\vec{E}^{\text{env}},\vec{H}^{\text{env}}\}$ distribute on $\text{ext}\,\mathbb{V}$ only, and the sources $\{\vec{J}^{\text{sca}},\vec{M}^{\text{sca}}\}$ generating $\{\vec{E}^{\text{sca}},\vec{H}^{\text{sca}}\}$ distribute on $\text{int}\,\mathbb{V}$ only, then *Huygens-Fresnel principle* tells us that

$$\vec{G}_0^{JF} * \left[\hat{n}_{\partial\mathbb{V}}^{-} \times \left(\vec{H}^{\text{imp}} + \vec{H}^{\text{env}}\right)\right] + \vec{G}_0^{MF} * \left[\left(\vec{E}^{\text{imp}} + \vec{E}^{\text{env}}\right) \times \hat{n}_{\partial\mathbb{V}}^{-}\right] = \begin{cases} \vec{F}^{\text{imp}} + \vec{F}^{\text{env}} , & \vec{r} \in \text{int}\,\mathbb{V} \\ 0 , & \vec{r} \in \text{ext}\,\mathbb{V} \end{cases} \tag{1-18}$$

and

$$\vec{G}_0^{JF} * \left[\left(-\hat{n}_{\partial\mathbb{V}}^{-}\right) \times \vec{H}^{\text{sca}}\right] + \vec{G}_0^{MF} * \left[\vec{E}^{\text{sca}} \times \left(-\hat{n}_{\partial\mathbb{V}}^{-}\right)\right] = \begin{cases} 0 , & \vec{r} \in \text{int}\,\mathbb{V} \\ \vec{F}^{\text{sca}} , & \vec{r} \in \text{ext}\,\mathbb{V} \end{cases} \tag{1-19}$$

where $\hat{n}_{\partial\mathbb{V}}^{-} \times (\vec{H}^{\text{imp}} + \vec{H}^{\text{env}})$ and $(\vec{E}^{\text{imp}} + \vec{E}^{\text{env}}) \times \hat{n}_{\partial\mathbb{V}}^{-}$ are the *Huygens' second sources corresponding to externally resultant fields* $\{\vec{E}^{imp} + \vec{E}^{env},\vec{H}^{imp} + \vec{H}^{env}\}$, and $(-\hat{n}_{\partial\mathbb{V}}^{-}) \times \vec{H}^{\text{sca}}$ and $\vec{E}^{\text{sca}} \times (-\hat{n}_{\partial\mathbb{V}}^{-})$ are the *Huygens' second sources corresponding to scattered fields* $\{\vec{E}^{sca},\vec{H}^{sca}\}$.

The difference between the above Huygens-Fresnel principles (1-18) and (1-19) gives the following *surface equivalence principle*

$$\vec{G}_0^{JF} * \vec{J}^{\text{equ}} + \vec{G}_0^{MF} * \vec{M}^{\text{equ}} = \begin{cases} \vec{F}^{\text{imp}} + \vec{F}^{\text{env}} , & \vec{r} \in \text{int}\,\mathbb{V} \\ -\vec{F}^{\text{sca}} , & \vec{r} \in \text{ext}\,\mathbb{V} \end{cases} \tag{1-20}$$

where the relations $\vec{F}^{\text{imp}} + \vec{F}^{\text{env}} + \vec{F}^{\text{sca}} = \vec{F}^{\text{tot}}$, $\vec{J}^{\text{equ}} = \hat{n}_{\partial\mathbb{V}}^{-} \times \vec{H}_{-}^{\text{tot}}$, and $\vec{M}^{\text{equ}} = \vec{E}_{-}^{\text{tot}} \times \hat{n}_{\partial\mathbb{V}}^{-}$ have been utilized. It is not difficult to find out that: **to treat $\vec{F}^{\text{imp}}$ and $\vec{F}^{\text{env}}$ as a whole is a natural selection for deriving DPO (1-6) from surface equivalence principle (1-20) (i.e. from Huygens-Fresnel principles (1-18) and (1-19)).** The above-mentioned viewpoint had been pointed out in Ref. [13].

**(E2) An Alternative Poynting's Theorem Based Explanation for Treating $\vec{F}^{\text{imp}}$ and $\vec{F}^{\text{env}}$ as a Whole**

For generality, we discuss in time domain here. The corresponding conclusion in frequently domain is completely similar.

As everyone knows, there exists the following time-domain *Poynting's theorem* for the region $\mathbb{V}$.

$$\begin{aligned} 0 \;=\; & \oiint_{\partial\mathbb{V}} \left(\vec{\mathcal{E}}^{\text{tot}} \times \vec{\mathcal{H}}^{\text{tot}}\right) \cdot \hat{n}_{\partial\mathbb{V}}^{+} dS + \left\langle \ddot{\sigma} \cdot \vec{\mathcal{E}}^{\text{tot}}, \vec{\mathcal{E}}^{\text{tot}} \right\rangle_{\mathbb{V}} \\ & + \frac{d}{dt}\left[ (1/2)\left\langle \vec{\mathcal{H}}^{\text{tot}}, \ddot{\mu} \cdot \vec{\mathcal{H}}^{\text{tot}} \right\rangle_{\mathbb{V}} + (1/2)\left\langle \ddot{\varepsilon} \cdot \vec{\mathcal{E}}^{\text{tot}}, \vec{\mathcal{E}}^{\text{tot}} \right\rangle_{\mathbb{V}} \right] \end{aligned} \tag{1-21}$$

because of the time-domain Maxwell's equations satisfied by the $\{\vec{\mathcal{E}}^{\text{tot}}, \vec{\mathcal{H}}^{\text{tot}}\}$ on $\mathbb{V}$. By some simple operations, we can derive the following two relations

$$\begin{aligned} & \oiint_{\partial\mathbb{V}} \left(\vec{\mathcal{E}}^{\text{tot}} \times \vec{\mathcal{H}}^{\text{tot}}\right) \cdot \hat{n}_{\partial\mathbb{V}}^{+} dS \\ = & - \left\langle \vec{J}^{\text{sca}}, \vec{\mathcal{E}}^{\text{imp}} + \vec{\mathcal{E}}^{\text{env}} \right\rangle_{\mathbb{V}} - \left\langle \vec{\mathcal{M}}^{\text{sca}}, \vec{\mathcal{H}}^{\text{imp}} + \vec{\mathcal{H}}^{\text{env}} \right\rangle_{\mathbb{V}} \\ & + \oiint_{\mathbb{S}_\infty} \left(\vec{\mathcal{E}}^{\text{sca}} \times \vec{\mathcal{H}}^{\text{sca}}\right) \cdot \hat{n}_{\mathbb{S}_\infty}^{+} dS \\ & + \frac{d}{dt}\left[ (1/2)\left\langle \vec{\mathcal{H}}^{\text{sca}}, \mu_0 \vec{\mathcal{H}}^{\text{sca}} \right\rangle_{\mathbb{E}^3 \setminus \mathbb{V}} + (1/2)\left\langle \varepsilon_0 \vec{\mathcal{E}}^{\text{sca}}, \vec{\mathcal{E}}^{\text{sca}} \right\rangle_{\mathbb{E}^3 \setminus \mathbb{V}} \right] \\ & - \frac{d}{dt}\left[ \frac{1}{2}\left\langle \vec{\mathcal{H}}^{\text{tot}} + \vec{\mathcal{H}}^{\text{sca}}, \mu_0 \left(\vec{\mathcal{H}}^{\text{imp}} + \vec{\mathcal{H}}^{\text{env}}\right) \right\rangle_{\mathbb{V}} + \frac{1}{2}\left\langle \varepsilon_0 \left(\vec{\mathcal{E}}^{\text{tot}} + \vec{\mathcal{E}}^{\text{sca}}\right), \vec{\mathcal{E}}^{\text{imp}} + \vec{\mathcal{E}}^{\text{env}} \right\rangle_{\mathbb{V}} \right] \end{aligned} \tag{1-22}$$

and

$$\begin{aligned} & \frac{d}{dt}\left[ (1/2)\left\langle \vec{\mathcal{H}}^{\text{tot}}, \ddot{\mu} \cdot \vec{\mathcal{H}}^{\text{tot}} \right\rangle_{\mathbb{V}} + (1/2)\left\langle \ddot{\varepsilon} \cdot \vec{\mathcal{E}}^{\text{tot}}, \vec{\mathcal{E}}^{\text{tot}} \right\rangle_{\mathbb{V}} \right] \\ = \; & \frac{d}{dt}\left[ (1/2)\left\langle \vec{\mathcal{H}}^{\text{tot}}, \Delta\ddot{\mu} \cdot \vec{\mathcal{H}}^{\text{tot}} \right\rangle_{\mathbb{V}} + (1/2)\left\langle \Delta\ddot{\varepsilon} \cdot \vec{\mathcal{E}}^{\text{tot}}, \vec{\mathcal{E}}^{\text{tot}} \right\rangle_{\mathbb{V}} \right] \\ & + \frac{d}{dt}\left[ (1/2)\left\langle \vec{\mathcal{H}}^{\text{sca}}, \mu_0 \vec{\mathcal{H}}^{\text{sca}} \right\rangle_{\mathbb{V}} + (1/2)\left\langle \varepsilon_0 \vec{\mathcal{E}}^{\text{sca}}, \vec{\mathcal{E}}^{\text{sca}} \right\rangle_{\mathbb{V}} \right] \\ & + \frac{d}{dt}\left[ \frac{1}{2}\left\langle \vec{\mathcal{H}}^{\text{tot}} + \vec{\mathcal{H}}^{\text{sca}}, \mu_0 \left(\vec{\mathcal{H}}^{\text{imp}} + \vec{\mathcal{H}}^{\text{env}}\right) \right\rangle_{\mathbb{V}} + \frac{1}{2}\left\langle \varepsilon_0 \left(\vec{\mathcal{E}}^{\text{tot}} + \vec{\mathcal{E}}^{\text{sca}}\right), \vec{\mathcal{E}}^{\text{imp}} + \vec{\mathcal{E}}^{\text{env}} \right\rangle_{\mathbb{V}} \right] \end{aligned} \tag{1-23}$$

In Eqs. (1-21) and (1-22), $\hat{n}_{\partial\mathbb{V}}^{+}$ is the *outer normal direction* of $\partial\mathbb{V}$, and it is obvious that $\hat{n}_{\partial\mathbb{V}}^{+} = -\hat{n}_{\partial\mathbb{V}}^{-}$ on whole $\partial\mathbb{V}$.

Substituting Eqs. (1-22) and (1-23) into Poynting's theorem (1-21), we immediately obtain the following relation

$$\begin{aligned} & \left\langle \vec{J}^{\text{sca}}, \overbrace{\vec{\mathcal{E}}^{\text{imp}} + \vec{\mathcal{E}}^{\text{env}}}^{\vec{\mathcal{E}}^{\text{inc}}} \right\rangle_{\mathbb{V}} + \left\langle \vec{\mathcal{M}}^{\text{sca}}, \overbrace{\vec{\mathcal{H}}^{\text{imp}} + \vec{\mathcal{H}}^{\text{env}}}^{\vec{\mathcal{H}}^{\text{inc}}} \right\rangle_{\mathbb{V}} \\ = \; & \oiint_{\mathbb{S}_\infty} \left(\vec{\mathcal{E}}^{\text{sca}} \times \vec{\mathcal{H}}^{\text{sca}}\right) \cdot \hat{n}_{\mathbb{S}_\infty}^{+} dS \\ & + \left\langle \ddot{\sigma} \cdot \vec{\mathcal{E}}^{\text{tot}}, \vec{\mathcal{E}}^{\text{tot}} \right\rangle_{\mathbb{V}} \\ & + \frac{d}{dt}\Big\{ \left[ (1/2)\left\langle \vec{\mathcal{H}}^{\text{sca}}, \mu_0 \vec{\mathcal{H}}^{\text{sca}} \right\rangle_{\mathbb{E}^3} + (1/2)\left\langle \varepsilon_0 \vec{\mathcal{E}}^{\text{sca}}, \vec{\mathcal{E}}^{\text{sca}} \right\rangle_{\mathbb{E}^3} \right] \\ & \qquad + \left[ (1/2)\left\langle \vec{\mathcal{H}}^{\text{tot}}, \Delta\ddot{\mu} \cdot \vec{\mathcal{H}}^{\text{tot}} \right\rangle_{\mathbb{V}} + (1/2)\left\langle \Delta\ddot{\varepsilon} \cdot \vec{\mathcal{E}}^{\text{tot}}, \vec{\mathcal{E}}^{\text{tot}} \right\rangle_{\mathbb{V}} \right] \Big\} \end{aligned} \tag{1-24}$$

Clearly, Eq. (1-24) is just the *time-differential form* of the WEP given in Eq. (1-1). From WEP (1-24), it is not difficult to find out that: **it is a very natural selection to treat $\vec{\mathcal{F}}^{\rm imp}$ and $\vec{\mathcal{F}}^{\rm env}$ as a whole — *externally resultant field / incident field*, i.e.,**

$$\vec{\mathcal{F}}^{\rm inc} = \vec{\mathcal{F}}^{\rm imp} + \vec{\mathcal{F}}^{\rm env} \tag{1-25}$$

**as done in Refs. [13,18,19].**

In fact, the above process "from Eq. (1-21) to Eq. (1-24)" also implies the following fact: **for EM scattering problem, work-energy principle (1-1) and Poynting's theorem (1-21) are equivalent to each other.**

**Answer to Question B**

When we want to calculate the CMs of "an objective scattering system placed in non-free-space environment", we treat the original objective scattering system and the non-free-space environment as a whole — *augmented objective scattering system*, and we calculate the CMs of the augmented system, and then we obtain the modal currents distributing on the original system and the modal currents distributing on environment.

In fact, the above-obtained modal currents distributing on the original objective scattering system are just the characteristic currents of "the objective scattering system placed in non-free-space environment".

### 1.2.4.5 On the Energy Decoupling and Far-field Decoupling of CMs

For the CMs, which are from simultaneously orthogonalizing the *positive and negative Hermitian parts of* $P^{driv}$, time-average DP-decoupling relation (1-9) is equivalent to

$$(1/2)\left\langle \frac{1}{\eta_0}\cdot \vec{E}_\xi^{\rm sca}, \vec{E}_\zeta^{\rm sca} \right\rangle_{\mathbb{S}_\infty} + (1/2)\left\langle \vec{\vec{\sigma}}\cdot \vec{E}_\xi^{\rm tot}, \vec{E}_\zeta^{\rm tot} \right\rangle_{\mathbb{V}} = 0 \quad , \quad \text{if } \xi \neq \zeta \tag{1-26}$$

where $\eta_0$ is the *wave impedance of free space* and $\eta_0 = \sqrt{\mu_0/\varepsilon_0}$. Evidently, when the objective scattering system is lossless (i.e., $\vec{\vec{\sigma}} = 0$), the above decoupling relation further implies the following *far-field-decoupling relation*①

$$\left\langle \vec{E}_\xi^{\rm sca}, \vec{E}_\zeta^{\rm sca} \right\rangle_{\mathbb{S}_\infty} = 0 \quad , \quad \text{if } \xi \neq \zeta \text{ and } \vec{\vec{\sigma}} = 0 \tag{1-27}$$

But, when the objective scattering system is lossy (i.e., $\vec{\vec{\sigma}} \neq 0$), the above far-field-decoupling relation doesn't exist as exhibited in the following examples.

We still consider the material elliptical cylinder shown in Fig. 1-2, but now it is a

① In fact, this is just the reason why Harrington's CMs are equivalent to (but not necessarily identical to) Garbacz's CMs in the lossless case, in the aspect of decoupling modal far fields.

lossy one rather than the lossless one considered in the previous Sec. 1.2.4.3. The lossy material elliptical cylinder is with permeability $\vec{\mu} = \vec{I}\,2\mu_0$, permittivity $\vec{\varepsilon} = \vec{I}\,10\varepsilon_0$, and conductivity $\vec{\sigma} = \vec{I}\,1$. We calculate the CMs of the lossy elliptical cylinder, and show the associated *modal significances* (*MSs*) in the following Fig. 1-5.

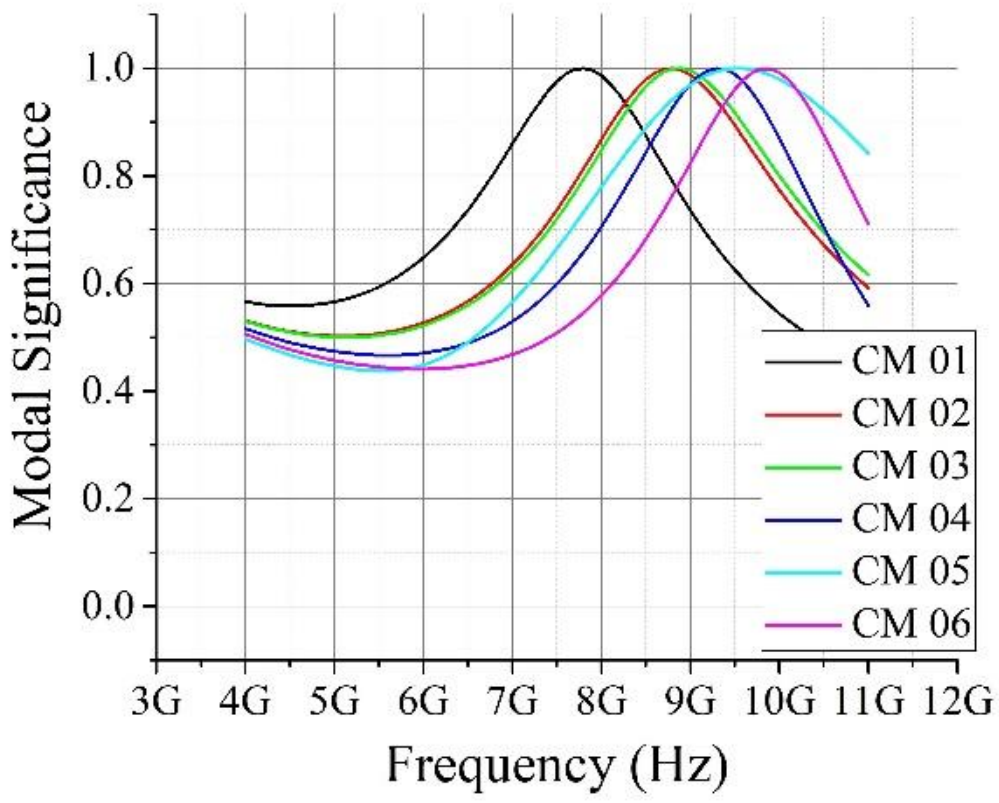

Figure 1-5 MSs of the CMs of a lossy material elliptical cylinder, whose topological structure is shown in Fig. 1-2 and whose material parameters are $\vec{\mu} = \vec{I}\,2\mu_0$, $\vec{\varepsilon} = \vec{I}\,10\varepsilon_0$, and $\vec{\sigma} = \vec{I}\,1$

For the CMs shown in Fig. 1-5, their *time-average DP orthogonality matrix* and *far-field orthogonality matrix* are visually illustrated in Figs. 1-6(a) and 1-6(b) respectively.

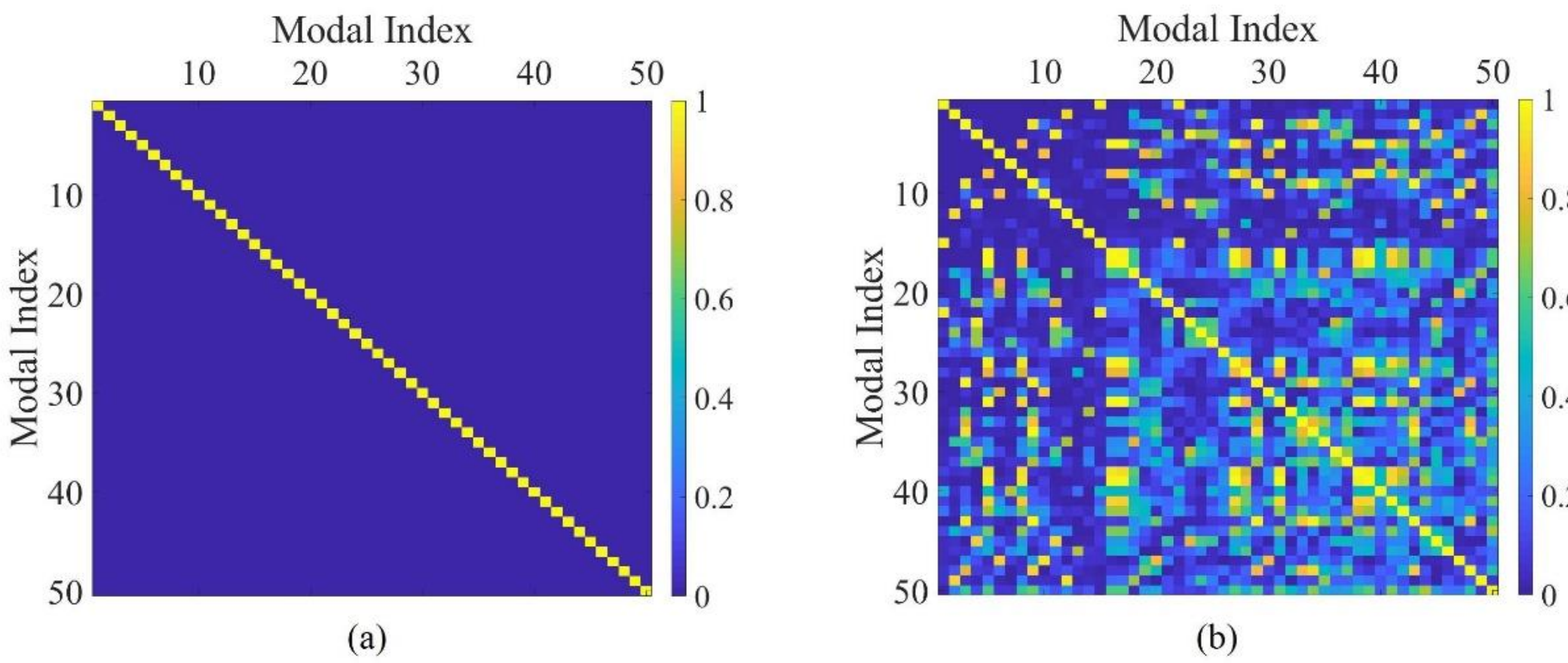

Figure 1-6 Orthogonality matrices of the CMs (at 9 GHz) shown in Fig. 1-5. (a) Time-average DP orthogonality matrix; (b) far-field orthogonality matrix

Evidently, for the lossy material elliptical cylinder, the time-average DP-decoupling relation (1-9)/(1-26) indeed holds, but the far-field-decoupling relation (1-27) doesn't.

Of course, if the operator $P^{\text{driv}} - P^{\text{dis}}$ is selected to generate CMs, then the outputted modes can satisfy far-field-decoupling relation (1-27), where $P^{\text{dis}}$ is the *dissipated power* of system $\mathbb{V}$ due to the non-zero $\vec{\sigma}$ (and $P^{\text{dis}} = (1/2) < \vec{\sigma} \cdot \vec{E}^{\text{tot}}, \vec{E}^{\text{tot}} >_{\mathbb{V}}$).

But, this may not be a good selection, because the modes generated by $P^{\rm driv} - P^{\rm dis}$ cannot guarantee energy-decoupling relation (1-9). Here, we want to emphasize that: **it is not the fundamental goal to decouple modal far fields; it should not sacrifice the energy-decoupling feature for satisfying the far-field-decoupling feature.** The reasons leading to this viewpoint are mainly the following two.

Reason 1: Now, we consider two modes $\{\vec{J}_1^{\rm sca}, \vec{M}_1^{\rm sca}\}$ and $\{\vec{J}_2^{\rm sca}, \vec{M}_2^{\rm sca}\}$, which are driven by fields $\{\vec{E}_1^{\rm inc}, \vec{H}_1^{\rm inc}\}$ and $\{\vec{E}_2^{\rm inc}, \vec{H}_2^{\rm inc}\}$ respectively and generate fields $\{\vec{E}_1^{\rm sca}, \vec{H}_1^{\rm sca}\}$ and $\{\vec{E}_2^{\rm sca}, \vec{H}_2^{\rm sca}\}$ respectively. The modes have decoupled far fields, but don't satisfy energy-decoupling relation (1-9). Driving fields $\{\vec{E}_1^{\rm inc}, \vec{H}_1^{\rm inc}\}$ carry information 1, and driving fields $\{\vec{E}_2^{\rm inc}, \vec{H}_2^{\rm inc}\}$ carry information 2.

Obviously, under the driving of $\{\vec{E}_1^{\rm inc}, \vec{H}_1^{\rm inc}\} / \{\vec{E}_2^{\rm inc}, \vec{H}_2^{\rm inc}\}$, fields $\{\vec{E}_1^{\rm sca}, \vec{H}_1^{\rm sca}\}$ / $\{\vec{E}_2^{\rm sca}, \vec{H}_2^{\rm sca}\}$ carry the information 1 / information 2 from *source zone*① to *far zone*. But, at the same time, fields $\{\vec{E}_1^{\rm sca}, \vec{H}_1^{\rm sca}\} / \{\vec{E}_2^{\rm sca}, \vec{H}_2^{\rm sca}\}$ also carry the information 2 / information 1 from source zone to far zone, because $\{\vec{E}_2^{\rm inc}, \vec{H}_2^{\rm inc}\}$ / $\{\vec{E}_1^{\rm inc}, \vec{H}_1^{\rm inc}\}$ also provide energy to $\{\vec{J}_1^{\rm sca}, \vec{M}_1^{\rm sca}\}$ / $\{\vec{J}_2^{\rm sca}, \vec{M}_2^{\rm sca}\}$ due to the absence of energy-decoupling relation (1-9). In other words, **the modes not satisfying energy-decoupling relation (1-9) cannot be driven independently, and then cannot achieve a real sense of decoupling for the information in far zone**.

Reason 2: Far-field-decoupling relation $< \vec{E}_\xi^{\rm sca}, \vec{E}_\zeta^{\rm sca} >_{\mathbb{S}_\infty} = 0$ is a global decoupling relation rather than a local decoupling relation. Specifically, the product $(\vec{E}_\xi^{\rm sca})^\dagger \cdot \vec{E}_\zeta^{\rm sca}$ may not be zero on the everywhere of $\mathbb{S}_\infty$, though the integral of $(\vec{E}_\xi^{\rm sca})^\dagger \cdot \vec{E}_\zeta^{\rm sca}$ on whole $\mathbb{S}_\infty$ is indeed zero.

This fact implies that: for any receiver not occupying whole $\mathbb{S}_\infty$, it cannot be guaranteed that $< \vec{E}_\xi^{\rm sca}, \vec{E}_\zeta^{\rm sca} >_{\rm receiver} = 0$. In other words, **global decoupling relation $< \vec{E}_\xi^{\rm sca}, \vec{E}_\zeta^{\rm sca} >_{\mathbb{S}_\infty} = 0$ may not be able to provide a real sense of decoupling for the fields on any finite-sized receiver at far zone**.

Thus energy-decoupling feature is more desired than far-field-decoupling feature.

#### 1.2.4.6 On the Physical Meaning of Characteristic Values

For a metallic objective scattering system, whose permeability and permittivity are $\mu_0$ and $\varepsilon_0$, the characteristic values $\lambda_\xi$ satisfy the following relation[9]

① Source zone is the region in where sources $\{\vec{J}^{\rm sca}, \vec{M}^{\rm sca}\}$ exist.

$$\lambda_{\xi} = \frac{2\omega\Big[(1/4)\big\langle \vec{H}_{\xi}^{\text{sca}}, \mu_0 \vec{H}_{\xi}^{\text{sca}} \big\rangle_{\mathbb{E}^3} - (1/4)\big\langle \varepsilon_0 \vec{E}_{\xi}^{\text{sca}}, \vec{E}_{\xi}^{\text{sca}} \big\rangle_{\mathbb{E}^3}\Big]}{(1/2) \oiint_{\mathbb{S}_\infty} \Big[\vec{E}_{\xi}^{\text{sca}} \times \big(\vec{H}_{\xi}^{\text{sca}}\big)^{\dagger}\Big] \cdot \hat{n}_{\mathbb{S}_\infty}^{+} dS} \tag{1-28}$$

Thus, it has been thought for a long time that the characteristic values of metallic systems have a physical explanation — the ratio of *modal imaginary power* to *modal real power*.

For a magnetodielectric objective scattering system, whose permittivity and permeability are not $\varepsilon_0$ and $\mu_0$, there exist *scattered electric current* and *scattered magnetic current* simultaneously due to the existence of both *polarization* and *magnetization* effects. If the operator $P^{\text{DRIV}}$ is used as the generating operator of CMs, then the derived $\lambda_\xi$ satisfy the following relation[27]

$$\lambda_{\xi} = \frac{2\omega\left\{\left[\frac{1}{4}\big\langle \vec{H}_{\xi}^{\text{sca}}, \mu_0 \vec{H}_{\xi}^{\text{sca}} \big\rangle_{\mathbb{E}^3} - \frac{1}{4}\big\langle \varepsilon_0 \vec{E}_{\xi}^{\text{sca}}, \vec{E}_{\xi}^{\text{sca}} \big\rangle_{\mathbb{E}^3}\right] + \left[\frac{1}{4}\big\langle \vec{H}_{\xi}^{\text{tot}}, \Delta\vec{\vec{\mu}} \cdot \vec{H}_{\xi}^{\text{tot}} \big\rangle_{\mathbb{V}} - \frac{1}{4}\big\langle \Delta\vec{\vec{\varepsilon}} \cdot \vec{E}_{\xi}^{\text{tot}}, \vec{E}_{\xi}^{\text{tot}} \big\rangle_{\mathbb{V}}\right]\right\}}{(1/2) \oiint_{\mathbb{S}_\infty} \Big[\vec{E}_{\xi}^{\text{sca}} \times \big(\vec{H}_{\xi}^{\text{sca}}\big)^{\dagger}\Big] \cdot \hat{n}_{\mathbb{S}_\infty}^{+} dS + (1/2)\big\langle \vec{\vec{\sigma}} \cdot \vec{E}_{\xi}^{\text{tot}}, \vec{E}_{\xi}^{\text{tot}} \big\rangle_{\mathbb{V}}} \tag{1-29}$$

Seemingly, Eq. (1-29) is "similar" to Eq. (1-28). **However, the CMs derived from orthogonalizing operator $P^{\text{driv}}$ don't satisfy a similar relation to Eq. (1-28) and Eq. (1-29)[13,18]!**

Then, a very natural question is that: since operator $P^{\text{DRIV}}$ can lead to the characteristic values with a energy relation (1-29) "similar" to energy relation (1-28), why don't we construct CMs from orthogonalizing operator $P^{\text{DRIV}}$?

The reason to this question is as follows: the CMs derived from orthogonalizing $P^{\text{driv}}$ are completely energy-decoupled as shown in Eqs. (1-7)&(1-9) and Figs. 1-4(a)&1-6(a), but the modes derived from orthogonalizing $P^{\text{DRIV}}$ are not completely energy-decoupled as shown in Eq. (1-8) and Fig. 1-4(b). Previously, we have clarified that the central purpose of CMT is to construct a set of energy-decoupled modes. **In fact, the characteristic values are only the by-products during the process to construct the energy-decoupled modes[13,18]. Thus, it should not sacrifice the energy-decoupling feature for the so-called physical explanation of characteristic values.**

### 1.2.4.7 On Modal Normalization

Any working mode under the driving of a previously known incident field can be linearly expanded in terms of the CMs as follows:

$$\text{A Working Mode} = \sum\nolimits_{\xi} c_{\xi} \cdot \text{CM}_{\xi} \tag{1-30}$$

with expansion coefficients

$$c_\xi = \frac{1}{P_\xi^{\text{driv}}}\Big[(1/2)\big\langle \vec{J}_\xi^{\text{sca}}, \vec{E}^{\text{inc}}\big\rangle_{\mathbb{V}} + (1/2)\big\langle \vec{M}_\xi^{\text{sca}}, \vec{H}^{\text{inc}}\big\rangle_{\mathbb{V}}\Big] \tag{1-31}$$

where $P_\xi^{\text{driv}}$ is the $\xi$-th *modal driving power*, and $\{\vec{E}^{\text{inc}}, \vec{H}^{\text{inc}}\}$ is the previously known incident field.

If we want $c_\xi$ to have ability to quantitatively reflect the *weight of a CM in whole modal expansion formulation*, it is necessary to appropriately normalize all of the CMs. To find a reasonable *normalization factor*, we employ the time-domain energy-decoupling relation (1-9) due to its clear physical meaning. From the time-domain energy-decoupling relation, it is not difficult to derive the following *expansion formulation for time-average DP*[18]

$$\begin{aligned}\frac{1}{T}\int_{t_0}^{t_0+T}\Big[\big\langle \vec{\mathcal{J}}^{\text{sca}}, \vec{\mathcal{E}}^{\text{inc}}\big\rangle_{\mathbb{V}} + \big\langle \vec{\mathcal{M}}^{\text{sca}}, \vec{\mathcal{H}}^{\text{inc}}\big\rangle_{\mathbb{V}}\Big]dt &= \text{Re}\big\{P^{\text{driv}}\big\} \\ &= \sum\nolimits_\xi |c_\xi|^2\,\text{Re}\big\{P_\xi^{\text{driv}}\big\} \qquad (1\text{-}32)\end{aligned}$$

where the first equality is evident[24,25], and the second equality is due to Eqs. (1-7) and (1-30). From Eq. (1-32), it is easy to conclude that: **a reasonable modal normalization way is to normalize $\text{Re}\{P_\xi^{\text{driv}}\}$ to 1, i.e., a reasonable modal normalization factor is the square root of time-average modal DP $\text{Re}\{P_\xi^{\text{driv}}\}$.** In fact, this normalization way is just the classical normalization way selected by Prof. Harrington *et al.*[8~12].

By using $\sqrt{\text{Re}\{P_\xi^{\text{driv}}\}}$ as the modal normalization factor, it traditionally has the following relation

$$P_\xi^{\text{driv}} = \underbrace{\text{Re}\big\{P_\xi^{\text{driv}}\big\}}_{1} + j\,\underbrace{\text{Im}\big\{P_\xi^{\text{driv}}\big\}}_{\lambda_\xi} \tag{1-33}$$

and hence we obtain the following famous *Parseval's identity*[18,19]

$$\frac{1}{T}\int_{t_0}^{t_0+T}\Big[\big\langle \vec{\mathcal{J}}^{\text{sca}}, \vec{\mathcal{E}}^{\text{inc}}\big\rangle_{\mathbb{V}} + \big\langle \vec{\mathcal{M}}^{\text{sca}}, \vec{\mathcal{H}}^{\text{inc}}\big\rangle_{\mathbb{V}}\Big]dt = \sum\nolimits_\xi |c_\xi|^2 \tag{1-34}$$

In fact, from the viewpoint of geometry, Parseval's identity (1-34) can also be viewed as the *Pythagorean theorem in inner product space*.

#### 1.2.4.8 On the Unwanted Modes Outputted From Characteristic Equation

The Problem V mentioned in previous Sec. 1.2.3 had been studied in Refs.

[13,16,18,19,26,30~36,etc.], and the core viewpoints and conclusions of Refs. [13,18,19,26,33,34] are simply reviewed here.

**An Explicit Expose on the Problem**

Because both $\vec{F}^{\rm tot}$ and $\vec{F}^{\rm sca}$ are tangentially continuous on $\partial\mathbb{V}$, then the tangential $\vec{F}^{\rm inc}$ on $\partial\mathbb{V}$ is equal to the difference between the tangential $\vec{F}_{-}^{\rm tot}$ and the tangential $\vec{F}_{+}^{\rm sca}$ (where $\vec{F}_{-}^{\rm tot}$ is the total field on the inner surface of $\partial\mathbb{V}$ and $\vec{F}_{+}^{\rm sca}$ is the scattered field on the outer surface of $\partial\mathbb{V}$), so the classical PMCHWT-based CMT[12] rewrites the DPO $P^{\rm driv}$ given in Eq. (1-6) as follows[13,18]:

$$\begin{aligned} P^{\rm driv} &= -(1/2)\left\langle \vec{J}^{\rm equ}, \vec{E}_{-}^{\rm tot} - \vec{E}_{+}^{\rm sca} \right\rangle_{\partial\mathbb{V}} - (1/2)\left\langle \vec{M}^{\rm equ}, \vec{H}_{-}^{\rm tot} - \vec{H}_{+}^{\rm sca} \right\rangle_{\partial\mathbb{V}} \\ &= -(1/2)\left\langle \vec{J}^{\rm equ}, {\rm P.V.}\,\mathcal{E}_0\left(\vec{J}^{\rm equ}, \vec{M}^{\rm equ}\right)\right\rangle_{\partial\mathbb{V}} - (1/2)\left\langle \vec{M}^{\rm equ}, {\rm P.V.}\,\mathcal{H}_0\left(\vec{J}^{\rm equ}, \vec{M}^{\rm equ}\right)\right\rangle_{\partial\mathbb{V}} \\ &\quad -(1/2)\left\langle \vec{J}^{\rm equ}, {\rm P.V.}\,\mathcal{E}_{\rm m}\left(\vec{J}^{\rm equ}, \vec{M}^{\rm equ}\right)\right\rangle_{\partial\mathbb{V}} - (1/2)\left\langle \vec{M}^{\rm equ}, {\rm P.V.}\,\mathcal{H}_{\rm m}\left(\vec{J}^{\rm equ}, \vec{M}^{\rm equ}\right)\right\rangle_{\partial\mathbb{V}} \\ &= \mathcal{Z}^{\rm PMCHWT}\left(\vec{J}^{\rm equ}, \vec{M}^{\rm equ}\right) \end{aligned} \tag{1-35}$$

In Eq. (1-35), operator $\mathcal{F}_{0/{\rm m}}(\vec{J},\vec{M})$ is defined as that $\mathcal{F}_{0/{\rm m}}(\vec{J},\vec{M}) = \overleftrightarrow{G}_{0/{\rm m}}^{JF} * \vec{J} + \overleftrightarrow{G}_{0/{\rm m}}^{MF} * \vec{M}$, where $\mathcal{F}_{0/{\rm m}} = \mathcal{E}_{0/{\rm m}} / \mathcal{H}_{0/{\rm m}}$ and correspondingly $F = E / H$, and $\overleftrightarrow{G}_{0/{\rm m}}^{JF}$ & $\overleftrightarrow{G}_{0/{\rm m}}^{MF}$ are the *dyadic Green's functions* with parameters $\{\mu_0, \varepsilon_0\} / \{\vec{\vec{\mu}}, \vec{\vec{\varepsilon}}, \vec{\vec{\sigma}}\}$; ${\rm P.V.}\,\mathcal{F}_{0/{\rm m}}$ denotes the *principal value* of operator $\mathcal{F}_{0/{\rm m}}$.

For the lossless material elliptical cylinder considered in Sec. 1.2.4.3 and the lossy material elliptical cylinder considered in Sec. 1.2.4.5, we calculate their CMs by employing $\mathcal{Z}^{\rm PMCHWT}$ as generating operator, and show the associated MSs in Fig. 1-7(a) for the lossless case and in Fig. 1-7(b) for the lossy case.

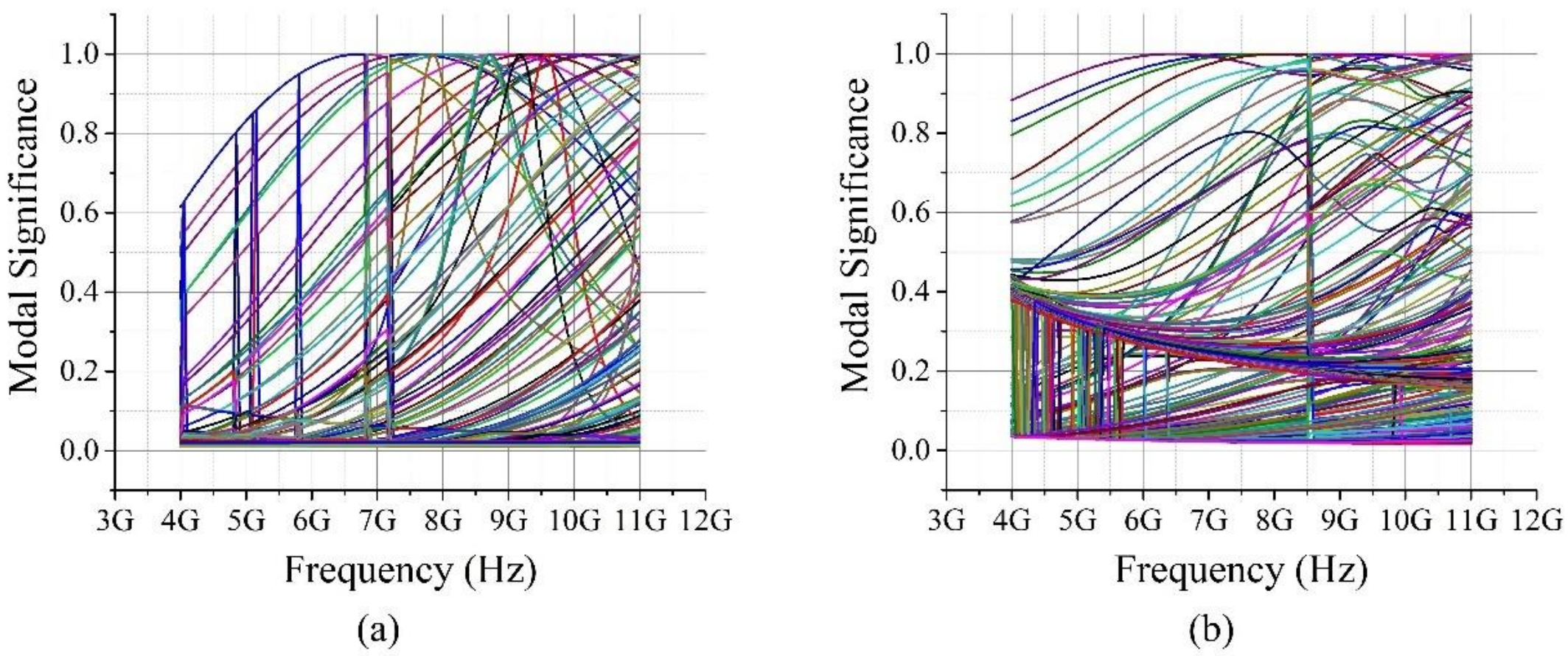

Figure 1-7 MSs of the CMs derived from orthogonalizing $\mathcal{Z}^{\rm PMCHWT}$. (a) MSs of the lossless material elliptical cylinder considered in Sec. 1.2.4.3; (b) MSs of the lossy material elliptical cylinder considered in Sec. 1.2.4.5

By comparing the above Figs. 1-7(a)&1-7(b) with the previous Figs. 1-3(a)&1-5①, it is not difficult to find out that: a direct usage of operator $\mathcal{Z}^{\mathrm{PMCHWT}}$ outputs many *unwanted modes*.

The reasons leading to the unwanted modes are mainly the following two[13,18,19,26,33,34]:

Reason A. An improperly enlarged space is used as "the modal space of the objective scattering system" (as shown in Fig. 1-8) / "the *definition domain* of generating operator" / "the *solution domain* of *characteristic equation*";

Reason B. An improper operator is used as the generating operator of CMs.

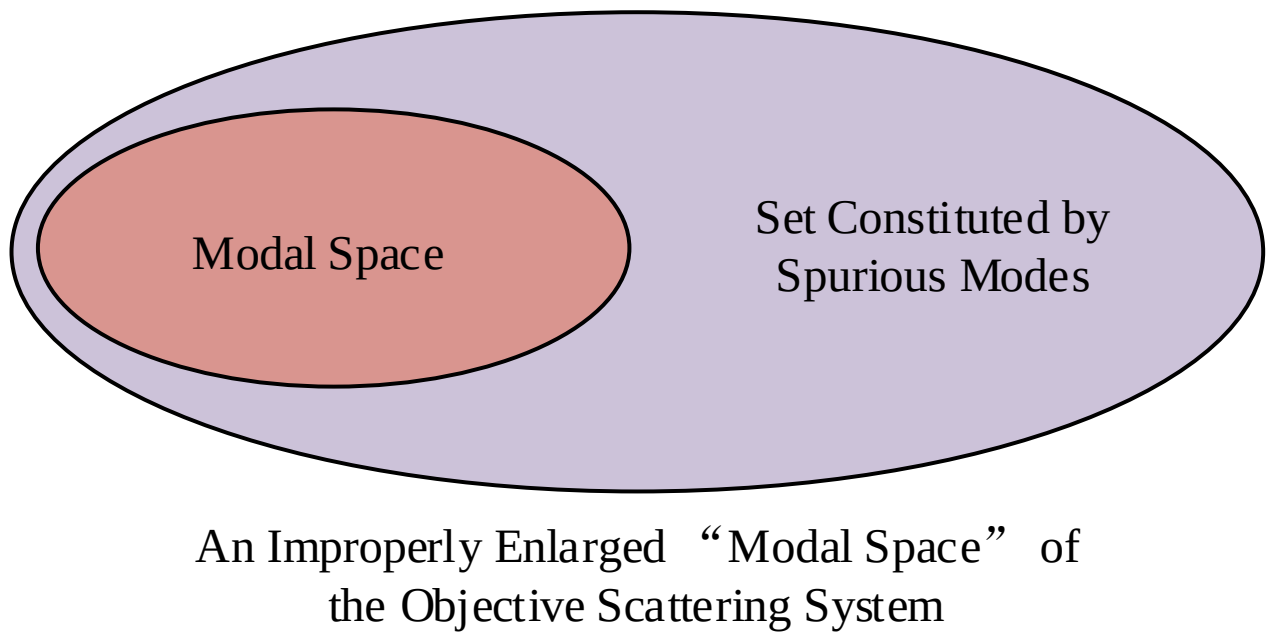

Figure 1-8 An improperly enlarged space is used as the modal space of the objective scattering system

The unwanted modes resulted from the reason A are usually called *spurious modes*②, and the corresponding problem is called *spurious mode problem* in this report. The unwanted modes resulted from the reason B are called *inaccurate modes*③ in this report, and the corresponding problem is called *inaccurate mode problem*.

**Resolution Schemes for Spurious Mode Problem — DVE, SDC/DDC, and IVM**

The above-mentioned reason A originates from that: the $\vec{J}^{\mathrm{equ}}$ and $\vec{M}^{\mathrm{equ}}$ contained in operator $\mathcal{Z}^{\mathrm{PMCHWT}}$ are not independent, but they are not properly related to each other. By properly establishing the dependence relation between $\vec{J}^{\mathrm{equ}}$ and $\vec{M}^{\mathrm{equ}}$, we can obtain the proper modal space of the objective scattering system. By applying the dependence relation to operator $\mathcal{Z}^{\mathrm{PMCHWT}}$, Refs. [13,26,33] made "the operator $\mathcal{Z}^{\mathrm{PMCHWT}}$ with both $\vec{J}^{\mathrm{equ}}$ and $\vec{M}^{\mathrm{equ}}$" become "the $\tilde{\mathcal{Z}}^{\mathrm{PMCHWT}}$ with only $\vec{J}^{\mathrm{equ}}$"④ or "the $\underset{\sim}{\mathcal{Z}}^{\mathrm{PMCHWT}}$

① The characteristic values $\lambda_\xi$ shown in Fig. 1-3(a) can be easily transformed to the associated MSs by using the well known relation $\mathrm{MS}_\xi = 1/|1+j\lambda_\xi|$.
② In Chinese, "虚假模式" or "伪模".
③ In Chinese, "误差模式" or "误差模".
④ The dependence relation for obtaining $\tilde{\mathcal{Z}}^{\mathrm{PMCHWT}}$ is established by using the tangential scattered magnetic field continuity on material boundary[13,26] (or equivalently using the definition of $\vec{J}^{\mathrm{equ}}$ [13,33]).

with only $\vec{M}^{\text{equ}}$ "①. Here, the tilde "~" is to distinguish the operators defined on the proper modal space from the one defined on the improperly enlarged space. The MSs calculated from orthogonalizing operator $\underset{\sim}{\mathcal{Z}}^{\text{PMCHWT}}$ are shown in Fig. 1-9.

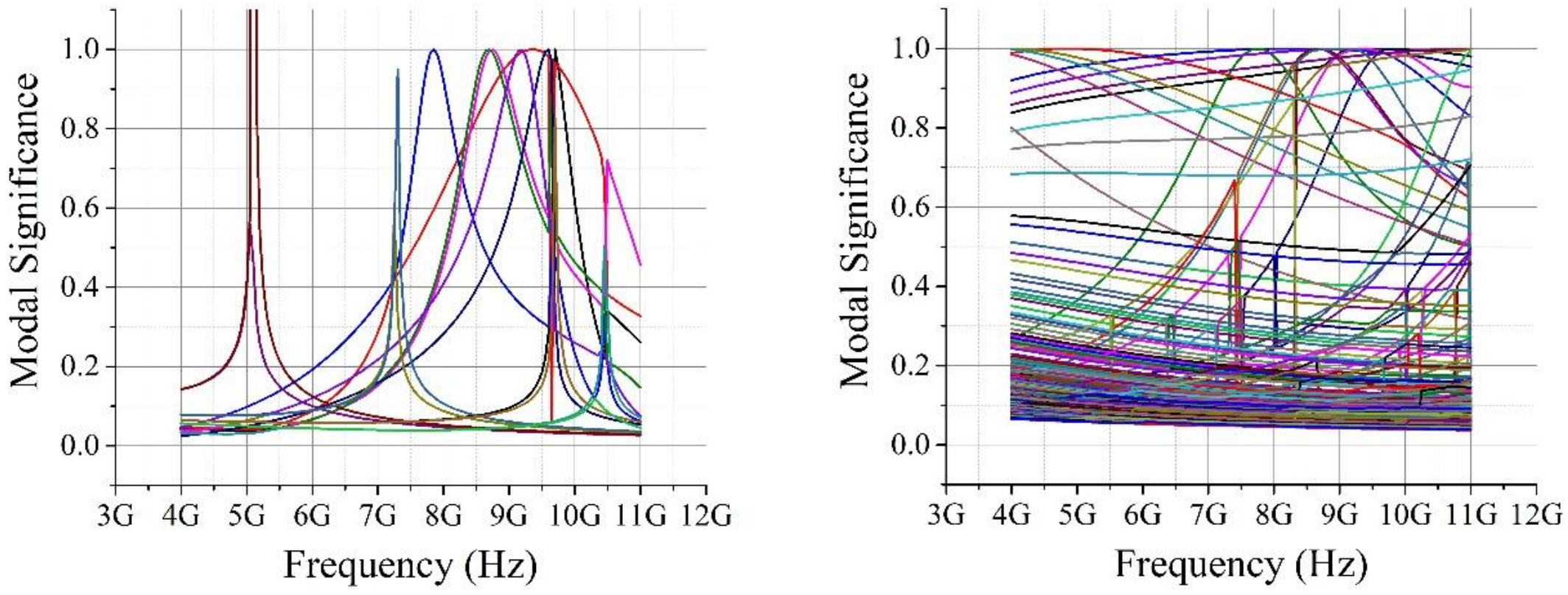

Figure 1-9 MSs calculated from orthogonalizing $\underset{\sim}{\mathcal{Z}}^{\text{PMCHWT}}$. (a) MSs of the lossless material elliptical cylinder considered in Sec. 1.2.4.3; (b) MSs of the lossy material elliptical cylinder considered in Sec. 1.2.4.5

Obviously, the above *dependent variable elimination* (*DVE*) scheme improves the calculating results. But, there still exist some other *unwanted extra modes*, which are just the inaccurate modes mentioned previously.

In addition, by compressing the definition domain of $\mathcal{Z}^{\text{PMCHWT}}$, Refs. [18,19] made the operator $\mathcal{Z}^{\text{PMCHWT}}$ with an improper definition domain become the one with a proper definition domain, and then the characteristic equation with an improper solution domain naturally becomes the one with a proper solution domain, and, at the same time, the modal space of the objective scattering system also becomes a proper one. A detailed and complete discussion for this *definition domain compression* (*DDC*) / *solution domain compression* (*SDC*) scheme can be found in Refs. [18,19]. The CMs generated by the operator with a proper definition domain are almost the same as the ones shown in Fig. 1-9.

Recently, Dr. Guo and Prof. Xia *et al.*[34] proposed another scheme for relating $\vec{J}^{\text{equ}}$ and $\vec{M}^{\text{equ}}$ by employing an intermediate variable — *effective current*, such that the operator $\mathcal{Z}^{\text{PMCHWT}}$ becomes the one with only the intermediate variable. A detailed discussion and some typical examples for this *intermediate variable method* (*IVM*) can be found in Ref. [34].

① The dependence relation for obtaining $\underset{\sim}{\mathcal{Z}}^{\text{PMCHWT}}$ is established by using the tangential scattered electric field continuity on material boundary[13,26] (or equivalently using the definition of $\vec{M}^{\text{equ}}$ [13,33]).

The *spurious modes* shown in Fig. 1-7 don't satisfy Maxwell's equations, so they are also alternatively called *unphysical modes* or *non-physical modes* sometimes. The spurious modes are mainly originated from the improper description for modal space[13,18,19]. However, the origination of the *incorrect modes* shown in Fig. 1-9 is mainly the improper generating operator (which leads to some considerable numerical inaccuracy during calculating CMs), so this kind of modes are called *inaccurate modes* in this report to be distinguished from the *spurious modes*. The resolution for the inaccurate mode problem needs to employ some proper generating operators as simply reviewed below.

**Resolution Scheme for Inaccurate Mode Problem — Proper Generating Operator**

Driving power $P^{\text{driv}}$ can be decomposed into the *dissipated power* $P^{dis}$ (where $P^{\text{dis}} = (1/2) < \vec{\vec{\sigma}} \cdot \vec{E}^{\text{tot}}, \vec{E}^{\text{tot}} >_{\mathbb{V}}$) and a *non-dissipated power* $P^{non\text{-}dis}$ as follows:

$$P^{\text{driv}} = \underbrace{\left(P^{\text{driv}} - P^{\text{dis}}\right)}_{P^{\text{non-dis}}} + P^{\text{dis}} \tag{1-36}$$

Based on the conclusions obtained in Refs. [13,18,19], it is easy to prove that the *non-dissipation term* $P^{non\text{-}dis}$ has the following operator form

$$P^{\text{non-dis}} = \underbrace{-\frac{1}{2}\left\langle \vec{J}^{\text{equ}}, \text{P.V.}\,\mathcal{E}_0\left(\vec{J}^{\text{equ}}, \vec{M}^{\text{equ}}\right)\right\rangle_{\partial\mathbb{V}} - \frac{1}{2}\left\langle \vec{M}^{\text{equ}}, \text{P.V.}\,\mathcal{H}_0\left(\vec{J}^{\text{equ}}, \vec{M}^{\text{equ}}\right)\right\rangle_{\partial\mathbb{V}}}_{\mathcal{P}^{\text{non-dis}}\left(\vec{J}^{\text{equ}}, \vec{M}^{\text{equ}}\right)} \tag{1-37}$$

and the operator form of the *dissipation term* $P^{dis}$ has the following multiple choices

$$P^{\text{dis}} = \begin{cases} \underbrace{\text{Re}\left\{-(1/2)\left\langle \vec{J}^{\text{equ}}, \hat{n}_- \times \vec{M}^{\text{equ}}\right\rangle_{\partial\mathbb{V}}\right\}}_{\mathcal{P}^{\text{dis}}_{\text{JM}}\left(\vec{J}^{\text{equ}}, \vec{M}^{\text{equ}}\right)} \\ \underbrace{\text{Re}\left\{-\left\langle \vec{J}^{\text{equ}}, \text{P.V.}\,\mathcal{E}_{\text{m}}\left(\vec{J}^{\text{equ}}, \vec{M}^{\text{equ}}\right)\right\rangle_{\partial\mathbb{V}}\right\}}_{\mathcal{P}^{\text{dis}}_{\text{JE}}\left(\vec{J}^{\text{equ}}, \vec{M}^{\text{equ}}\right)} \\ \underbrace{\text{Re}\left\{-\left\langle \vec{M}^{\text{equ}}, \text{P.V.}\,\mathcal{H}_{\text{m}}\left(\vec{J}^{\text{equ}}, \vec{M}^{\text{equ}}\right)\right\rangle_{\partial\mathbb{V}}\right\}}_{\mathcal{P}^{\text{dis}}_{\text{MH}}\left(\vec{J}^{\text{equ}}, \vec{M}^{\text{equ}}\right)} \\ \underbrace{\text{Re}\left\{-(1/2)\left\langle \vec{J}^{\text{equ}}, \text{P.V.}\,\mathcal{E}_{\text{m}}\left(\vec{J}^{\text{equ}}, \vec{M}^{\text{equ}}\right)\right\rangle_{\partial\mathbb{V}} - (1/2)\left\langle \vec{M}^{\text{equ}}, \text{P.V.}\,\mathcal{H}_{\text{m}}\left(\vec{J}^{\text{equ}}, \vec{M}^{\text{equ}}\right)\right\rangle_{\partial\mathbb{V}}\right\}}_{\mathcal{Z}^{\text{dis}}_{\text{pmchwt}}\left(\vec{J}^{\text{equ}}, \vec{M}^{\text{equ}}\right)} \\ \underbrace{-(1/2)\left\langle \vec{J}^{\text{equ}}, \text{P.V.}\,\mathcal{E}_{\text{m}}\left(\vec{J}^{\text{equ}}, \vec{M}^{\text{equ}}\right)\right\rangle_{\partial\mathbb{V}} - (1/2)\left\langle \vec{M}^{\text{equ}}, \text{P.V.}\,\mathcal{H}_{\text{m}}\left(\vec{J}^{\text{equ}}, \vec{M}^{\text{equ}}\right)\right\rangle_{\partial\mathbb{V}}}_{\mathcal{Z}^{\text{dis}}_{\text{PMCHWT}}\left(\vec{J}^{\text{equ}}, \vec{M}^{\text{equ}}\right)} \end{cases} \tag{1-38}$$

In Eq. (1-38), the operator form $\mathcal{P}_{\text{JM}}^{\text{dis}}$ is expressed as the interaction between electric current and magnetic curre, so it is correspondingly called *JM interaction form*, and this is just the reason to use subscript “JM”; the operator form $\mathcal{P}_{\text{JE}}^{\text{dis}}$ is expressed as the interaction between electric current and electric field, so it is correspondingly called *JE interaction form*, and this is just the reason to use subscript “JE”; the operator form $\mathcal{P}_{\text{MH}}^{\text{dis}}$ is expressed as the interaction between magnetic current and magnetic field, so it is correspondingly called *MH interaction form*, and this is just the reason to use subscript “MH”; the operator form $\mathcal{Z}_{\text{PMCHWT}}^{\text{dis}}$ is just the one selected by the tradition PMCHWT-based CMT, so the subscript “PMCHWT” is used; the operator form $\mathcal{Z}_{\text{pmchwt}}^{\text{dis}}$ is an improved version for $\mathcal{Z}_{\text{PMCHWT}}^{\text{dis}}$, and the subscript “pmchwt” is to distinguish $\mathcal{Z}_{\text{pmchwt}}^{\text{dis}}$ from $\mathcal{Z}_{\text{PMCHWT}}^{\text{dis}}$.

As concluded in Refs. [13,18,19], after applying the DVE or DDC/SDC scheme to $\mathcal{P}^{\text{driv}} = \mathcal{P}^{\text{non-dis}} + \mathcal{P}^{\text{dis}}$, **the different choices for $\mathcal{P}^{\text{dis}}$ have different numerical performances**, and

- when the objective scattering system is lossless, $\tilde{\mathcal{P}}^{\text{dis}} \equiv 0$ and $\underset{\sim}{\mathcal{P}}^{\text{dis}} \equiv 0$ theoretically[13], so it has a more satisfactory numerical performance to select $\tilde{\mathcal{P}}_0^{\text{driv}} = \tilde{\mathcal{P}}^{\text{non-dis}} + 0$ (if $\mu_r > \varepsilon_r$) / $\underset{\sim}{\mathcal{P}}_0^{\text{driv}} = \underset{\sim}{\mathcal{P}}^{\text{non-dis}} + 0$ (if $\varepsilon_r > \mu_r$) as the generating operator of CMs[13,18];
- when the material parameters of the objective scattering system satisfy relation $\varepsilon_r > \mu_r$, the operator forms $\underset{\sim}{\mathcal{P}}_{\text{JM}}^{\text{driv}} = \underset{\sim}{\mathcal{P}}^{\text{non-dis}} + \underset{\sim}{\mathcal{P}}_{\text{JM}}^{\text{dis}}$ and $\underset{\sim}{\mathcal{P}}_{\text{JE}}^{\text{driv}} = \underset{\sim}{\mathcal{P}}^{\text{non-dis}} + \underset{\sim}{\mathcal{P}}_{\text{JE}}^{\text{dis}}$ are more desired than the other operator forms[13,18];
- when the material parameters of the objective scattering system satisfy relation $\mu_r > \varepsilon_r$, the operator forms $\tilde{\mathcal{P}}_{\text{JM}}^{\text{driv}} = \tilde{\mathcal{P}}^{\text{non-dis}} + \tilde{\mathcal{P}}_{\text{JM}}^{\text{dis}}$ and $\tilde{\mathcal{P}}_{\text{MH}}^{\text{driv}} = \tilde{\mathcal{P}}^{\text{non-dis}} + \tilde{\mathcal{P}}_{\text{MH}}^{\text{dis}}$ are more desired than the other operator forms[13,18];
- when $\varepsilon_r > \mu_r$ / $\mu_r > \varepsilon_r$, the numerical performance of $\underset{\sim}{\mathcal{Z}}^{\text{pmchwt}} = \underset{\sim}{\mathcal{P}}^{\text{non-dis}} + \underset{\sim}{\mathcal{Z}}_{\text{pmchwt}}^{\text{dis}}$ / $\tilde{\mathcal{Z}}^{\text{pmchwt}} = \tilde{\mathcal{P}}^{\text{non-dis}} + \tilde{\mathcal{Z}}_{\text{pmchwt}}^{\text{dis}}$ is usually worse than $\underset{\sim}{\mathcal{P}}_{\text{JE}}^{\text{driv}}$ / $\tilde{\mathcal{P}}_{\text{MH}}^{\text{driv}}$, and is usually better than $\underset{\sim}{\mathcal{P}}_{\text{MH}}^{\text{driv}}$ / $\tilde{\mathcal{P}}_{\text{JE}}^{\text{driv}}$;
- for all cases, the numerical performance of $\tilde{\mathcal{Z}}^{\text{PMCHWT}} = \tilde{\mathcal{P}}^{\text{non-dis}} + \tilde{\mathcal{Z}}_{\text{PMCHWT}}^{\text{dis}}$ / $\underset{\sim}{\mathcal{Z}}^{\text{PMCHWT}} = \underset{\sim}{\mathcal{P}}^{\text{non-dis}} + \underset{\sim}{\mathcal{Z}}_{\text{PMCHWT}}^{\text{dis}}$ is usually worse than $\tilde{\mathcal{Z}}^{\text{pmchwt}}$ / $\underset{\sim}{\mathcal{Z}}^{\text{pmchwt}}$.

For the lossless elliptical cylinder considered in the previous sections, the MSs calculated from the above-mentioned various generating operators are shown in the following figures. (The corresponding characteristic-value-based benchmark is shown in Fig. 1-3(a), and the characteristic values can be easily transformed to MSs.)

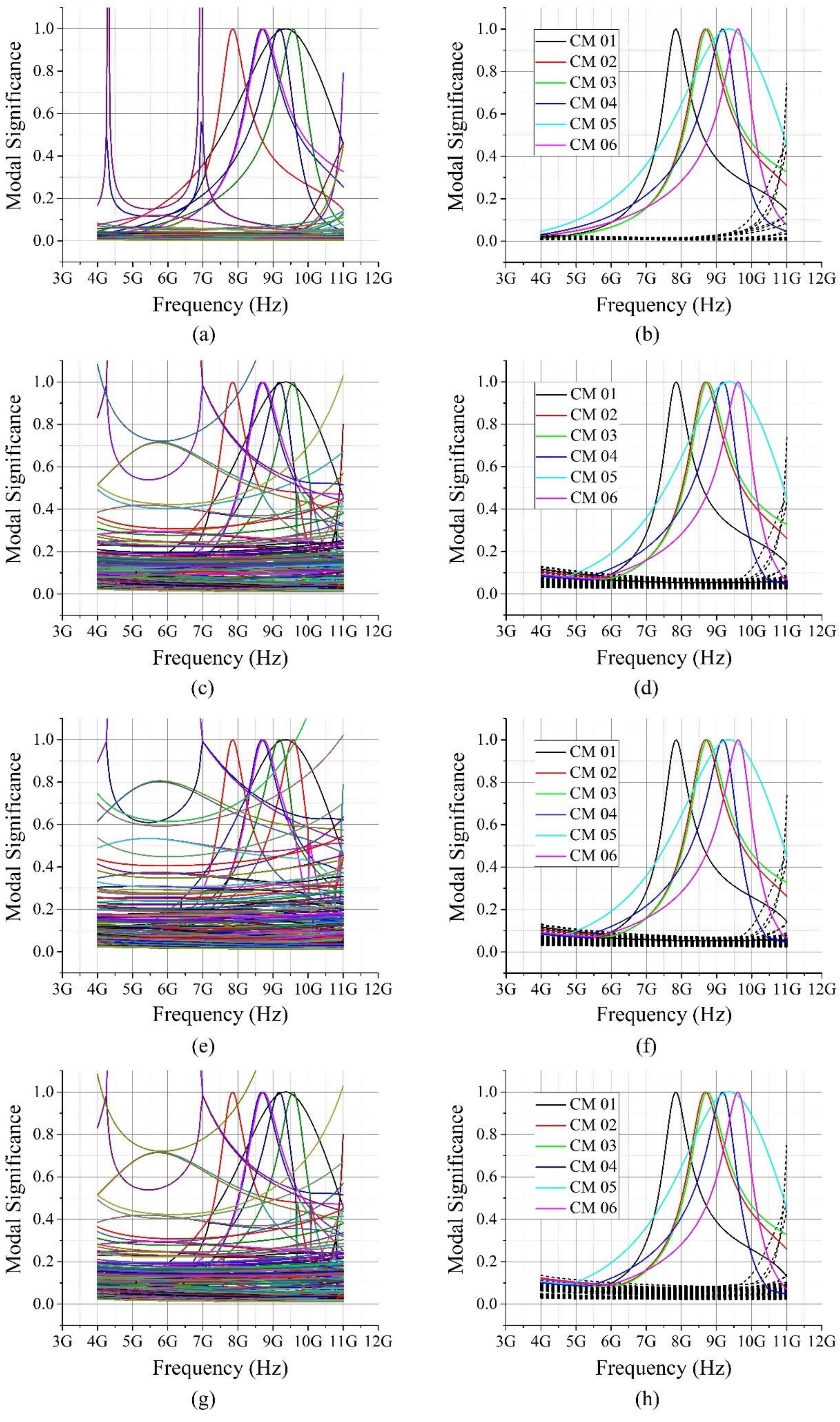
Modal Significance
Frequency (Hz)
3G 4G 5G 6G 7G 8G 9G 10G 11G 12G
0.0 0.2 0.4 0.6 0.8 1.0
CM 01
CM 02
CM 03
CM 04
CM 05
CM 06
(a)
(b)
(c)
(d)
(e)
(f)
(g)
(h)

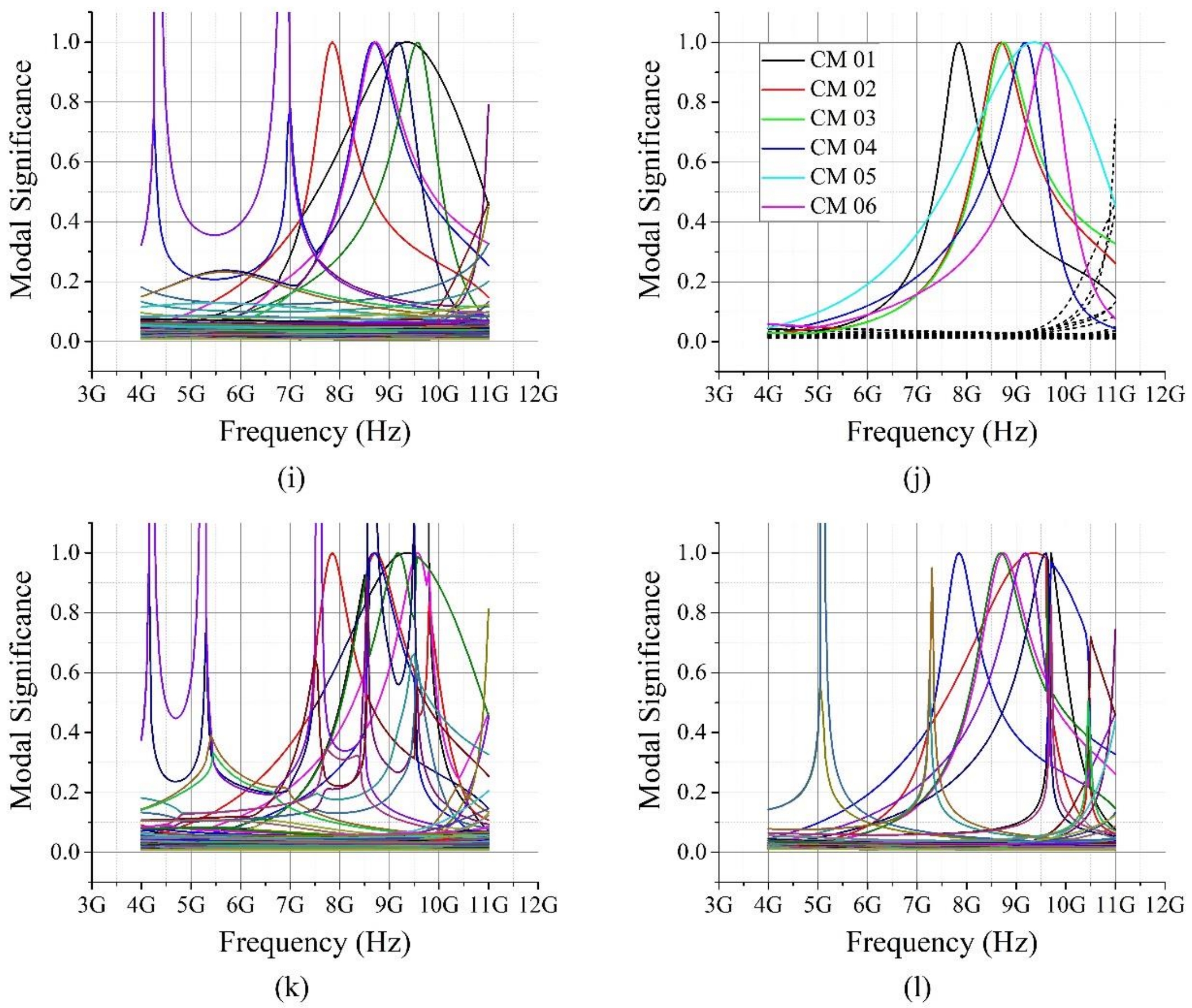

Figure 1-10 MSs of the CMs of the lossless material elliptical cylinder considered in Sec. 1.2.4.3. (a) $\tilde{\mathcal{P}}_{0}^{\text{driv}}$ -based results; (b) $\underset{\sim}{\mathcal{P}}_{0}^{\text{driv}}$ -based results; (c) $\tilde{\mathcal{P}}_{\text{JM}}^{\text{driv}}$ -based results; (d) $\underset{\sim}{\mathcal{P}}_{\text{JM}}^{\text{driv}}$ -based results; (e) $\tilde{\mathcal{P}}_{\text{JE}}^{\text{driv}}$ -based results; (f) $\underset{\sim}{\mathcal{P}}_{\text{JE}}^{\text{driv}}$ -based results; (g) $\tilde{\mathcal{P}}_{\text{MH}}^{\text{driv}}$ -based results; (h) $\underset{\sim}{\mathcal{P}}_{\text{MH}}^{\text{driv}}$ -based results; (i) $\tilde{\mathcal{Z}}^{\text{pmchwt}}$ -based results; (j) $\underset{\sim}{\mathcal{Z}}^{\text{pmchwt}}$ -based results; (k) $\tilde{\mathcal{Z}}^{\text{PMCHWT}}$ -based results; (l) $\underset{\sim}{\mathcal{Z}}^{\text{PMCHWT}}$ -based results

From the figures exhibited above, it is not difficult to find out that: the transformation from $\vec{M}^{\text{equ}}$ to $\vec{J}^{\text{equ}}$ is more desired than the transformation from $\vec{J}^{\text{equ}}$ to $\vec{M}^{\text{equ}}$, and an experiential explanation for this phenomenon can be found in Ref. [13]; among the figures using the more desired transformation, the Fig. 1-10(b) (corresponding to the operator $\underset{\sim}{\mathcal{P}}_{0}^{\text{driv}}$, in which the dissipated power is directly written as 0) is the best one, and the Fig. 1-10(l) (corresponding to the operator $\underset{\sim}{\mathcal{Z}}^{\text{PMCHWT}}$, which is just the PMCHWT operator) is the worst one, and the explanations for these phenomena can be found in Ref. [13].

For the lossy elliptical cylinder considered in the previous sections, the MSs calculated from the above-mentioned various operators are shown in the following figures. (The corresponding benchmark is shown in Fig. 1-5.)

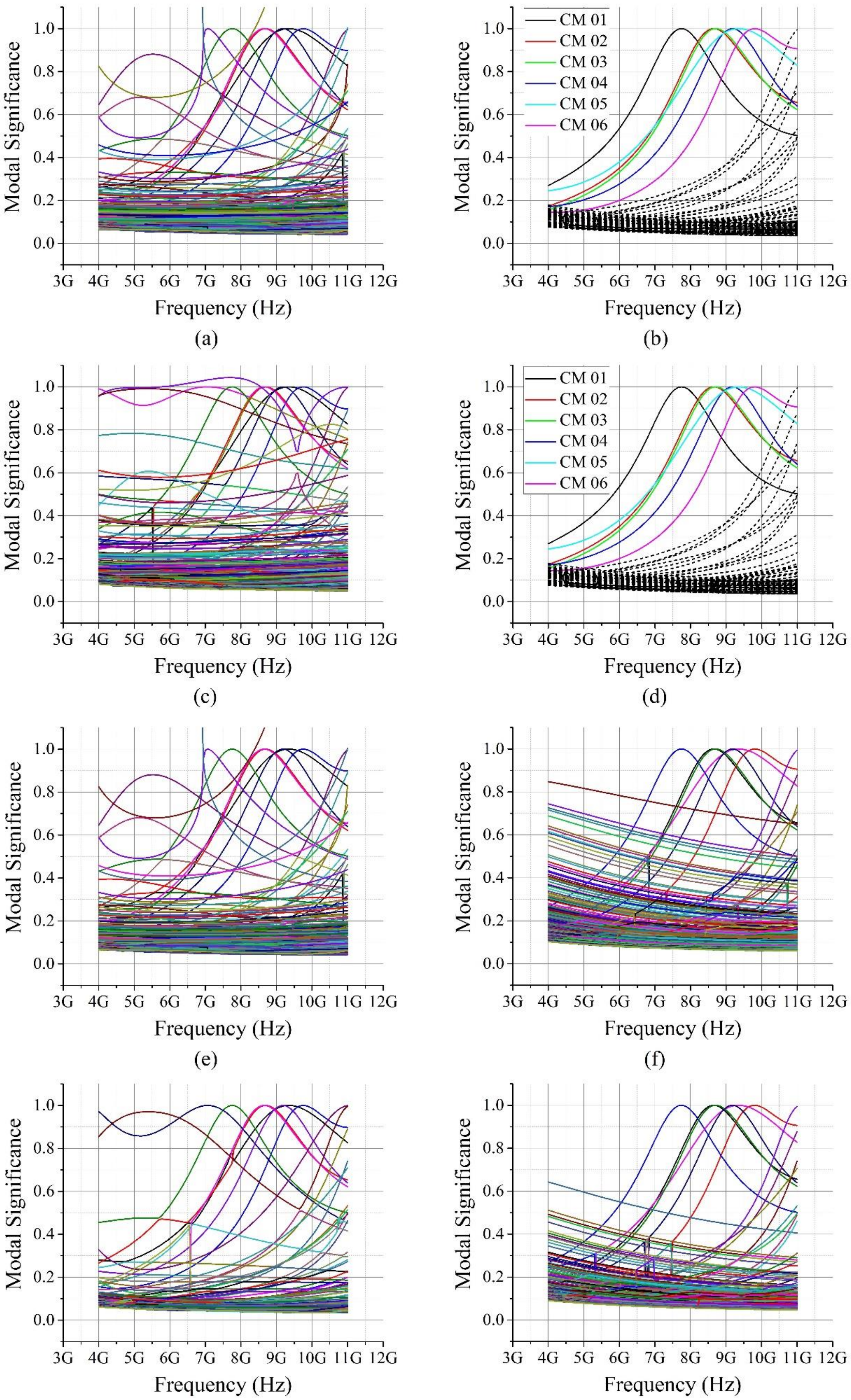
Modal Significance
Frequency (Hz)
3G 4G 5G 6G 7G 8G 9G 10G 11G 12G
0.0 0.2 0.4 0.6 0.8 1.0
CM 01
CM 02
CM 03
CM 04
CM 05
CM 06
(a)
(b)
(c)
(d)
(e)
(f)
(g)
(h)

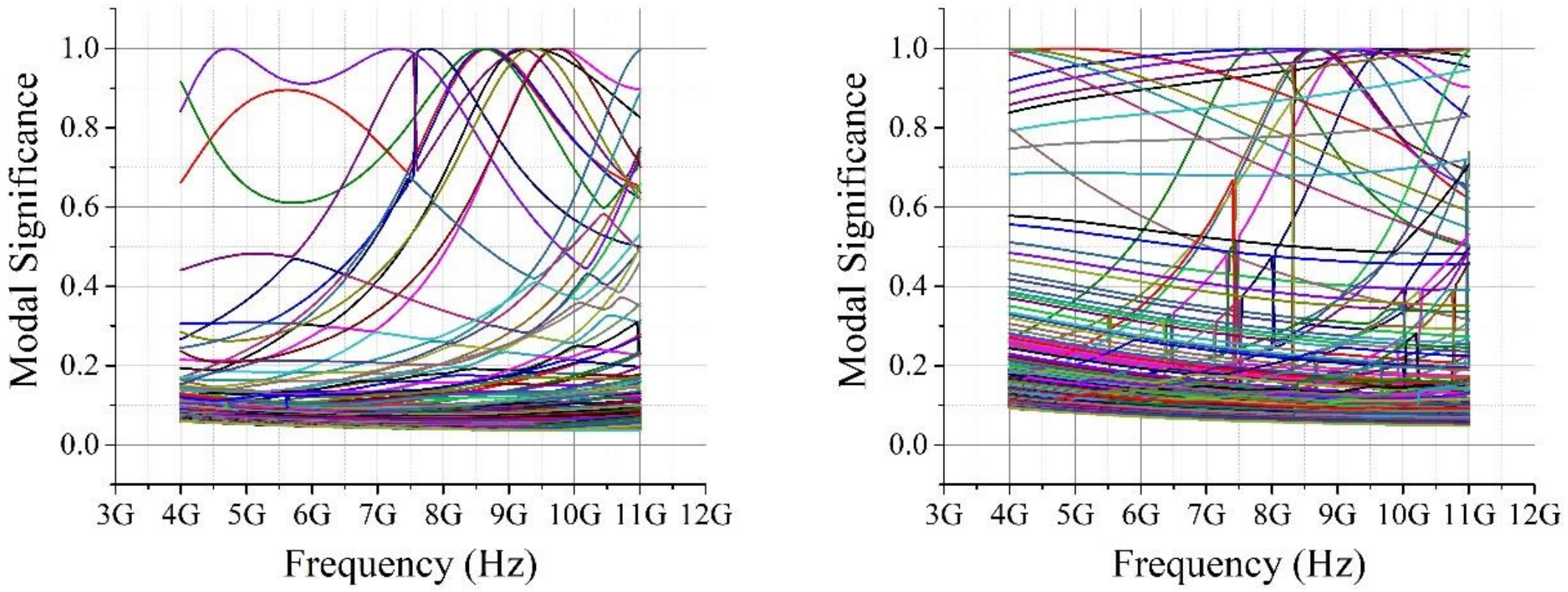

(i) (j)

Figure 1-11 MSs of the CMs of the lossy material elliptical cylinder considered in Sec. 1.2.4.5. (a) $\tilde{\mathcal{P}}_{\text{JM}}^{\text{driv}}$ -based results; (b) $\underset{\sim}{\mathcal{P}}_{\text{JM}}^{\text{driv}}$ -based results; (c) $\tilde{\mathcal{P}}_{\text{JE}}^{\text{driv}}$ -based results; (d) $\underset{\sim}{\mathcal{P}}_{\text{JE}}^{\text{driv}}$ -based results; (e) $\tilde{\mathcal{P}}_{\text{MH}}^{\text{driv}}$ -based results; (f) $\underset{\sim}{\mathcal{P}}_{\text{MH}}^{\text{driv}}$ -based results; (g) $\tilde{\mathcal{Z}}^{\text{pmchwt}}$ -based results; (h) $\underset{\sim}{\mathcal{Z}}^{\text{pmchwt}}$ -based results; (i) $\tilde{\mathcal{Z}}^{\text{PMCHWT}}$ -based results; (j) $\underset{\sim}{\mathcal{Z}}^{\text{PMCHWT}}$ -based results

Evidently, the above Figs. 1-10 and 1-11 support the previous conclusions.

Based on the above-mentioned conclusions, the following operator can be introduced

$$\text{Generating Operator of CMs} = \begin{cases} \tilde{\mathcal{P}}_{0}^{\text{driv}} & , \quad \text{if } \mu_r \ge \varepsilon_r \text{ and } \sigma = 0 \\ \underset{\sim}{\mathcal{P}}_{0}^{\text{driv}} & , \quad \text{if } \mu_r \le \varepsilon_r \text{ and } \sigma = 0 \\ \tilde{\mathcal{P}}_{\text{JM}}^{\text{driv}} \text{ or } \tilde{\mathcal{P}}_{\text{MH}}^{\text{driv}} & , \quad \text{if } \mu_r \ge \varepsilon_r \text{ and } \sigma \ne 0 \\ \underset{\sim}{\mathcal{P}}_{\text{JM}}^{\text{driv}} \text{ or } \underset{\sim}{\mathcal{P}}_{\text{JE}}^{\text{driv}} & , \quad \text{if } \mu_r \le \varepsilon_r \text{ and } \sigma \ne 0 \end{cases} \tag{1-39}$$

as the generating operator of CMs, and the operator has a satisfactory numerical performance for both lossless and lossy systems[13,18,19].

## 1.3 Important Problems and Challenges — How to Establish a Rigorous Modal Theory for Transceiving Systems?

In the above Sec. 1.2, some important CMT-related problems existing in IE framework and their resolution schemes in WEP framework are simply reviewed. It is not difficult to find out many advantages of the WEP framework. In fact, besides the above-listed advantages, the WEP framework has another crucial advantage.

From the physical picture revealed under WEP framework, it can be found out that: **the CMT is a modal theory for *scattering systems*① (shuch as the one shown in Fig.**

① *Electromagnetic scattering* is the phenomenon that "an electromagnetic field resulting from currents induced in a **1-1) rather than for *transceiving systems*① (such as the one shown in Fig. 1-12). The existed CMT-based modal analysis for various antennas (which work at transmitting state) are only approximate, but by no means rigorous. "How to effectively establish a rigorous modal theory for various transceiving systems?" is a very important problem and challenge, and this is just the central issue focused on by this report.**

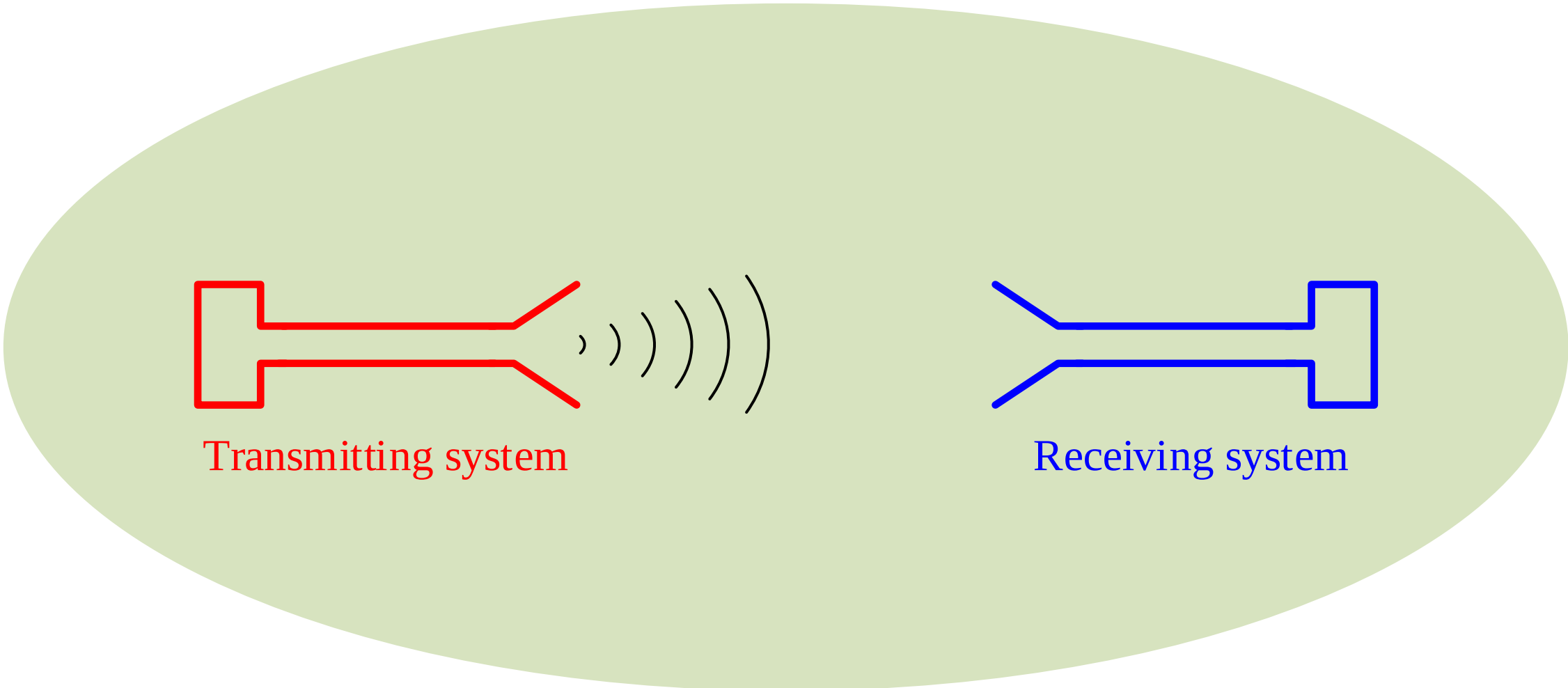

Figure 1-12 A typical transceiving system

## 1.4 Main Contributions and Innovations of This Research Report — Achieving a Leap From WEP-ScaSys-CMT to PTT-TRSys-DMT

This report is devoted to establishing a novel modal theory for transceiving systems. Because the novel modal theory is different from both the conventional EMT and CMT, thus this report will give it another name to avoid confusion. **Because the novel modal theory aims at constructing a set of energy-decoupled modes, thus this report calls it *decoupling mode theory* (*DMT*). Because this report establishes the DMT under a novel framework — *power transport theorem* (*PTT*) *framework*, then the DMT is called *PTT-based DMT for transceiving systems* (*PTT-TRSys-DMT*).**

### 1.4.1 Energy-decoupled Modes and Decoupling Mode Theory

The fundamental principle of PTT-TRSys-DMT is given in this subsection, and a detailed comparison among CMT, DMT and EMT will be provided in the subsequent Sec. 1.4.2.

secondary, conducting or dielectric object by electromagnetic waves incident on that object from one or more primary sources"[37].

① *Transmitting system* is "a device or circuit that generates high-frequency electric energy, controled or modulated, which can be radiated by an antenna"[37]. *Receiving system* is "a device for converting radio-frequency power into perceptible signals"[37]. *Transceiving system* is "a device that both transmits and receives data"[37].

#### 1.4.1.1 Region Division for Transceiving System

*Transceiving system* is an EM device that both transmits and/or receives EM energy[37]. Clearly, the whole transceiving system can be divided into two sub-systems *transmitting system* and *receiving system*, as shown in Fig. 1-13.

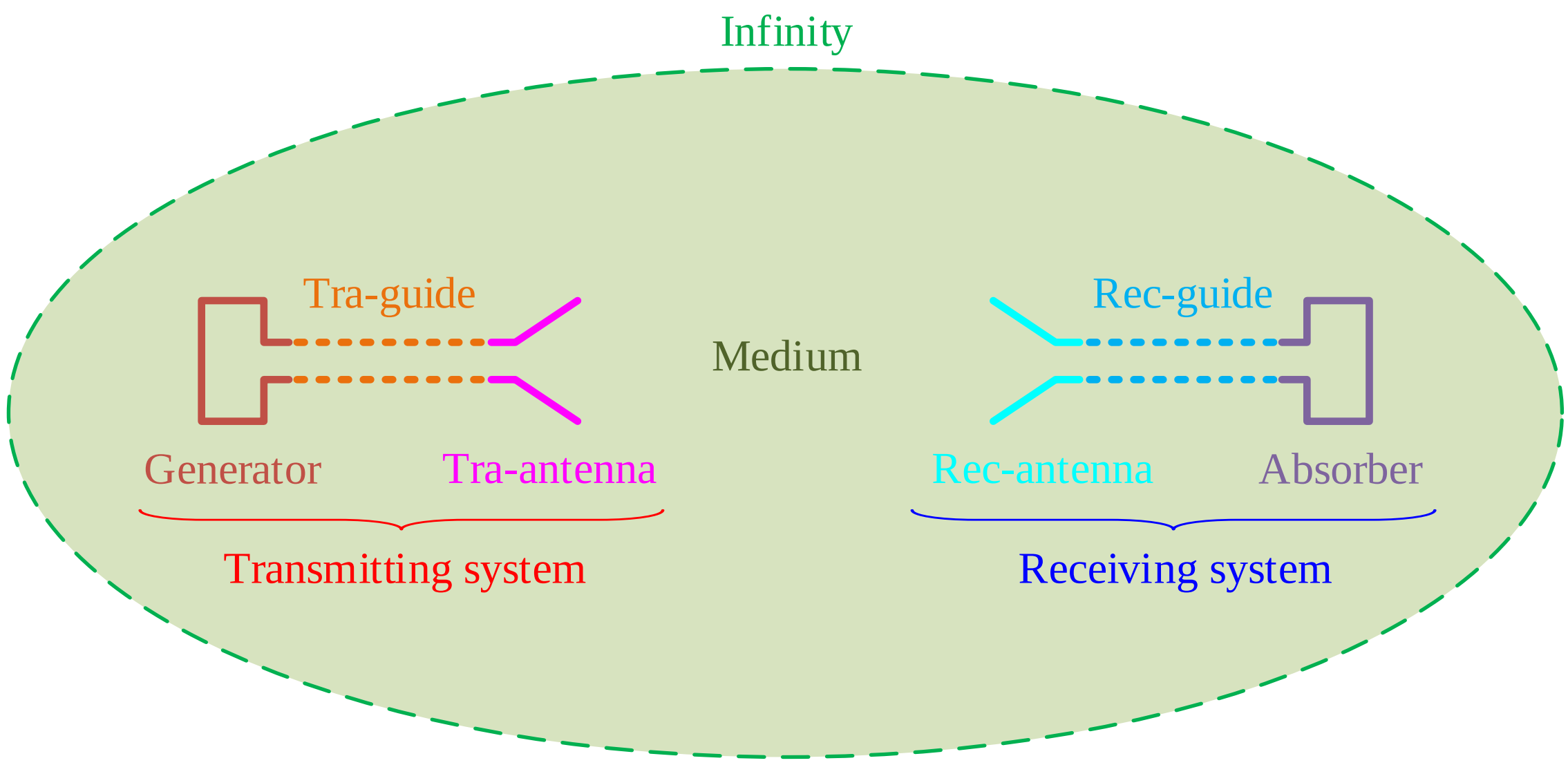

Figure 1-13 Region division for the transceiving system shown in Fig. 1-12

*Transmitting system* is an EM device used to generate EM energy and to release the energy into *surrounding environment* (simply called *environment*)[37]. In general, transmitting system consists of three main sub-structures: *generating structure* (simply called *generator*) used to generate EM energy, *transmitting antenna structure* (simply called *transmitting antenna*) used to modulate the energy to be released into environment, and *wave-guiding structure* (simply called *guide*) used to guide the energy from generator to transmitting antenna and then to environment.

*Receiving system* is an EM device used to collect EM energy from environment and to convert the energy into perceptible signals[37]. In general, receiving system consists of three main sub-structures: *receiving antenna structure* (simply called *receiving antenna*) used to collect EM energy from environment, *absorbing structure* (simply called *absorber*) used to convert the energy into perceptible signals, and *wave-guiding structure* (simply called *guide*) used to guide the energy from receiving antenna to absorber.

In addition, environment generally consists of two main sub-structures: "*infinity* used to receive some energies radiated by transmitting system and scattered by receiving system" and "*propagation medium* (simply called *medium*) used as the area for propagating the radiated and scattered energies.

In this report, the transmitting antenna and receiving antenna are simply denoted as *tra-antenna* and *rec-antenna* respectively; the guide in transmitting system and the guide in receiving system are simply denoted as *tra-guide* and *rec-guide* respectively, to be distinguished from each other. Sometimes, we also simply call the transmitting system and receiving system *transmitter* and *receiver* respectively.

### 1.4.1.2 Power Transport Theorem for Transceiving Systems

Along tra-guide, EM power/energy flows from generator to tra-antenna; with the modulation of tra-antenna, the power is released into medium; by flowing through medium, the power finally reaches {infinity, receiving system}; with the collection of rec-antenna, the power is inputted into receiving system; along rec-guide, the power flows from rec-antenna to absorber. At the same time, there also exist some EM powers scattered by receiving system, and these scattered powers flow from receiving system to {transmitting system, infinity} by passing through medium.

In the above processes, the carrier of EM power is EM field. During carrying the power from one structure to another, the field acts on the structures, and the actions will lead to some induced currents on the structures, and the induced currents will correspondingly generate some EM fields. The EM fields generated by the induced currents will further react on the structures. In summary, generator, tra-guide, tra-antenna, medium, rec-antenna, rec-guide, and absorber interact with each others by the EM fields generated by the currents on them.

Under time-harmonic excitation, the interactions among the EM fields and the structures will finally reach a *stationary state* (i.e. a *dynamic equilibrium*). At the stationary state, the EM power flows in the structures are illustrated in the following power flow graph where symbols "Gen", "Tra-guid", "Tra-ant", "Med", "Inf", "Env", "Rec-ant", "Rec-guid", and "Abs" denote the generator, tra-guide, tra-antenna, medium, infinity, environment, rec-antenna, rec-guide, and absorber respectively; arrows “ $\xrightleftharpoons{\text{power}}$ ” mean that: the power mainly flows from the left to the right, but there also exists a “reflected power” from the right to the left; the *Sommerfeld’s radiation condition* at infinity is just the reason why there doesn’t exist any reflected power from infinity. How to analyze and design the flowing way of the power is an important topic in EM engineering, and this type of problem is called *transceiving problem* in this report.

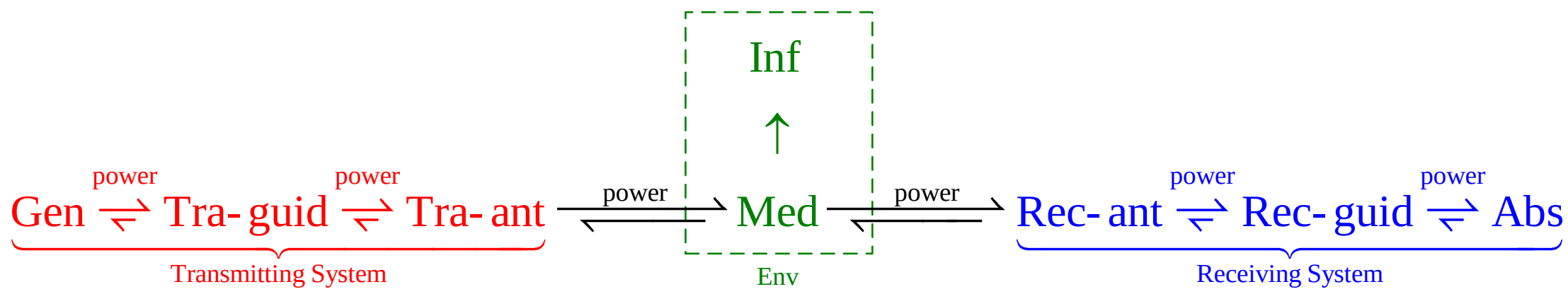

Figure 1-14 Power flow graph for the transceiving problem

Based on the Poynting’s theorem corresponding to a region (for details please see Sec. 2.2), the above power flow process can be quantitatively expressed as the following *power transport theorem* (for details please see Sec. 2.4)

$$P^{\mathrm{O}\rightleftharpoons\mathrm{G}} = \left(P_{\mathrm{dis}}^{\mathrm{G}} + jP_{\mathrm{sto}}^{\mathrm{G}}\right) + \underbrace{\left(P_{\mathrm{dis}}^{\mathrm{A}} + jP_{\mathrm{sto}}^{\mathrm{A}}\right) + \underbrace{\left(P_{\mathrm{Mdis}}^{\mathrm{Mdis}} + jP_{\mathrm{Msto}}^{\mathrm{Msto}}\right) + P_{\mathrm{sca}}^{\mathrm{rad}} + \underbrace{\left(P_{\mathrm{A}}^{\mathrm{dis}} + jP_{\mathrm{A}}^{\mathrm{sto}}\right) + P_{\mathrm{A}\rightleftharpoons\mathrm{G}}}_{P_{\mathrm{M}\rightleftharpoons\mathrm{A}}}}_{P^{\mathrm{A}\rightleftharpoons\mathrm{M}}}}_{P^{\mathrm{G}\rightleftharpoons\mathrm{A}}} \tag{1-40}$$

where the meanings of $P^{\mathrm{O}\rightleftharpoons\mathrm{G}}$, $P^{\mathrm{G}\rightleftharpoons\mathrm{A}}$, $P^{\mathrm{A}\rightleftharpoons\mathrm{M}}$, and $P_{\mathrm{M}\rightleftharpoons\mathrm{A}}$ are as follows:

$P^{\mathrm{O}\rightleftharpoons\mathrm{G}}$ is the net power inputted into tra-guide, and it is called the *input power of tra-guide*;

$P^{\mathrm{G}\rightleftharpoons\mathrm{A}}$ is the net power inputted into tra-antenna, and it is called the *input power of tra-antenna*;

$P^{\mathrm{A}\rightleftharpoons\mathrm{M}}$ is the net power inputted into propagation medium, and it is called the *input power of propagation medium*;

$P_{\mathrm{M}\rightleftharpoons\mathrm{A}}$ is the net power inputted into rec-antenna, and it is called the *input power of rec-antenna*;

$P_{\mathrm{A}\rightleftharpoons\mathrm{G}}$ is the net power outputted from rec-antenna, and it is called the *output power of rec-antenna*.

The meanings of the other powers appeared in Eq. (1-42) will be carefully explained in the future Sec. 2.4.

In the Chap. 3 of this report, the *energy-decoupled modes* (*DMs*) of tra-guide will be constructed by orthogonalizing *input power operator* (*IPO*) $P^{\mathrm{O}\rightleftharpoons\mathrm{G}}$; in the Chap. 6 of this report, the DMs of tra-antenna will be constructed by orthogonalizing IPO $P^{\mathrm{G}\rightleftharpoons\mathrm{A}}$; in the Chap. 7 of this report, the DMs of rec-antenna will be constructed by orthogonalizing IPO $P_{\mathrm{M}\rightleftharpoons\mathrm{A}}$; in the Chap. 8 of this report, the DMs of *tra-guide-tra-antenna system* (*TGTA system*) and *tra-antenna-rec-antenna system* (*TARA system*) will be constructed by orthogonalizing the corresponding IPOs.

## 1.4.2 Comparisons Among CMT, DMT, and EMT

Here, we simply compare the above-mentioned various modal theories from the aspects of carrying framework, generating operator, and physical picture, as summarized in the following Tab. 1-2.

Table 1-2 Comparisons among various modal theories from the aspects of carrying framework, generating operator, and physical picture

| VARIOUS MODAL THEORIES | | | CARRYING FRAMEWORK | GENERATING OPERATOR | PHYSICAL PICTURE | REFERENCES |
|---|---|---|---|---|---|---|
| **EMT** | SLT-SFReg-EMT | | Sturm-Liouville Theory (SLT) | Sturm-Liouville Operator (SLO) | To Construct a Set of Energy-decoupled Modes | [1~3] |
| **CMT** | SM-ScaSys-CMT | | Scattering Matrix (SM) for a Scatterer | Perturbation Matrix Operator (PMO) | To Construct a Set of Far-field-decoupled Modes | [5~7] |
| | IE-ScaSys-CMT | IE-MetSca-CMT | Integral Equation (IE) for a Scatterer | Impedance Matrix Operator (IMO) | Not Clarified by Its Founders | [8~10] |
| | | IE-MatSca-CMT | | | | [11,12] |
| | | IE-ComSca-CMT | | | | [38~43] |
| | WEP-ScaSys-CMT | WEP-MetSca-CMT | Work-Energy Principle (WEP) for a Scatterer | Driving Power Operator (DPO) | To Construct a Set of Energy-decoupled Modes | Chap. 3 in [13] |
| | | WEP-MatSca-CMT | | | | [18], Chap. 4 in [13] |
| | | WEP-ComSca-CMT | | | | [19], Chap. 5 in [13] |
| | WEP-CMT for Wireless Power Transfer (WPT) Systems | | WEP for a WPT System | | | [349], App. G in this report |
| | WEP-CMT for Yagi-Uda Arrays | | WEP for a Yagi-Uda Array | | | [350], App. H in this report |
| **DMT** | PTT-Guid-DMT | PTT-MetGuid-DMT | Power Transport Theorem (PTT) for a Region | Input Power Operator (IPO) | | Secs. 3.2 in this report |
| | | PTT-MatGuid-DMT | | | | Sec. 3.3 in this report |
| | | PTT-ComGuid-DMT | | | | Secs. 3.4&3.5 in this report |
| | | PTT-FreeSpace-DMT | | | | Sec. 3.6 in this report |
| | PTT-TraAnt-DMT | PTT-MetTraAnt-DMT | | | | Sec. 6.2 in this report |
| | | PTT-MatTraAnt-DMT | | | | Sec. 6.3 in this report |
| | | PTT-ComTraAnt-DMT | | | | Secs. 6.4~6.6 in this report |
| | PTT-RecAnt-DMT | PTT-MetRecAnt-DMT | | | | Secs. 7.2&7.3 in this report |
| | | PTT-MatRecAnt-DMT | | | | |
| | | PTT-ComRecAnt-DMT | | | | Sec. 7.4 in this report |
| | PTT-ComSys-DMT | PTT-TGTA-DMT | | | | Sec. 8.2 in this report |
| | | PTT-TARA-DMT | | | | Sec. 8.3 in this report |

It is not difficult to find out that: the *SLT-based EMT for source-free region* (*SLT-SFReg-EMT*) has a same physical purpose — to construct a set of energy-decoupled modes — as the WEP-ScaSys-CMT and PTT-TRSys-DMT. In fact, it will be exhibited in the Chap. 3 of this report that the SLT-SFReg-EMT can also be classified into the special case of PTT-TRSys-DMT, just like having classified Prof. Harrington's IE-ScaSys-CMT into the special case of WEP-ScaSys-CMT in Ref. [13].

## 1.5 Study Outline and Symbolic System of This Research Report

In this section, we give the research outline and roadmap of this report, and briefly introduce the symbolic system used in this report.

### 1.5.1 Outline

This report is devoted to establishing the DMT for transceiving systems by orthogonalizing the frequency-domain IPO derived in PTT framework.

**Chapter 1** is devoted to exposing the core problem focused on by this report, and sketches out the outline of this report. Section 1.1 briefly mentions the importance and research value of modal theory in engineering applications. Section 1.2 briefly reviews several transformations for modal theory during the development of modal theory — from SLT-SFReg-EMT to SM-ScaSys-CMT (Sec. 1.2.2), from SM-ScaSys-CMT to IE-ScaSys-CMT (Sec. 1.2.3), and from IE-ScaSys-CMT to WEP-ScaSys-CMT (Sec. 1.2.4). The carrying framework, generating operator, physical picture, and core advantages of WEP-ScaSys-CMT are highlighted in Sec. 1.2.4, and the physical picture clearly reveals the fact that: **CMT is a custom-made modal theory for scattering systems.** Then, Sec. 1.3 exposes one of the most central challenges faced by modal theory is that: **how to establish a custom-made modal theory for transceiving systems?** To respond to the challenge, **a novel modal theory — PTT-TRSys-DMT — is established in this report for transceiving systems**, and the basic strategies for establishing PTT-TRSys-DMT are briefly discussed in Sec. 1.4.1. At the same time, some brief comparisons among SLT-SFReg-EMT, SM-ScaSys-CMT, IE-ScaSys-CMT, WEP-ScaSys-CMT, and PTT-TRSys-DMT are done in Sec. 1.4.2 from the aspects of carrying framework, generating operator, and physical picture. Here, we provide the logical relationship diagram for the chapters and sections contained in this report, and also exhibit the research roadmap of this report, for the convenience of readers.

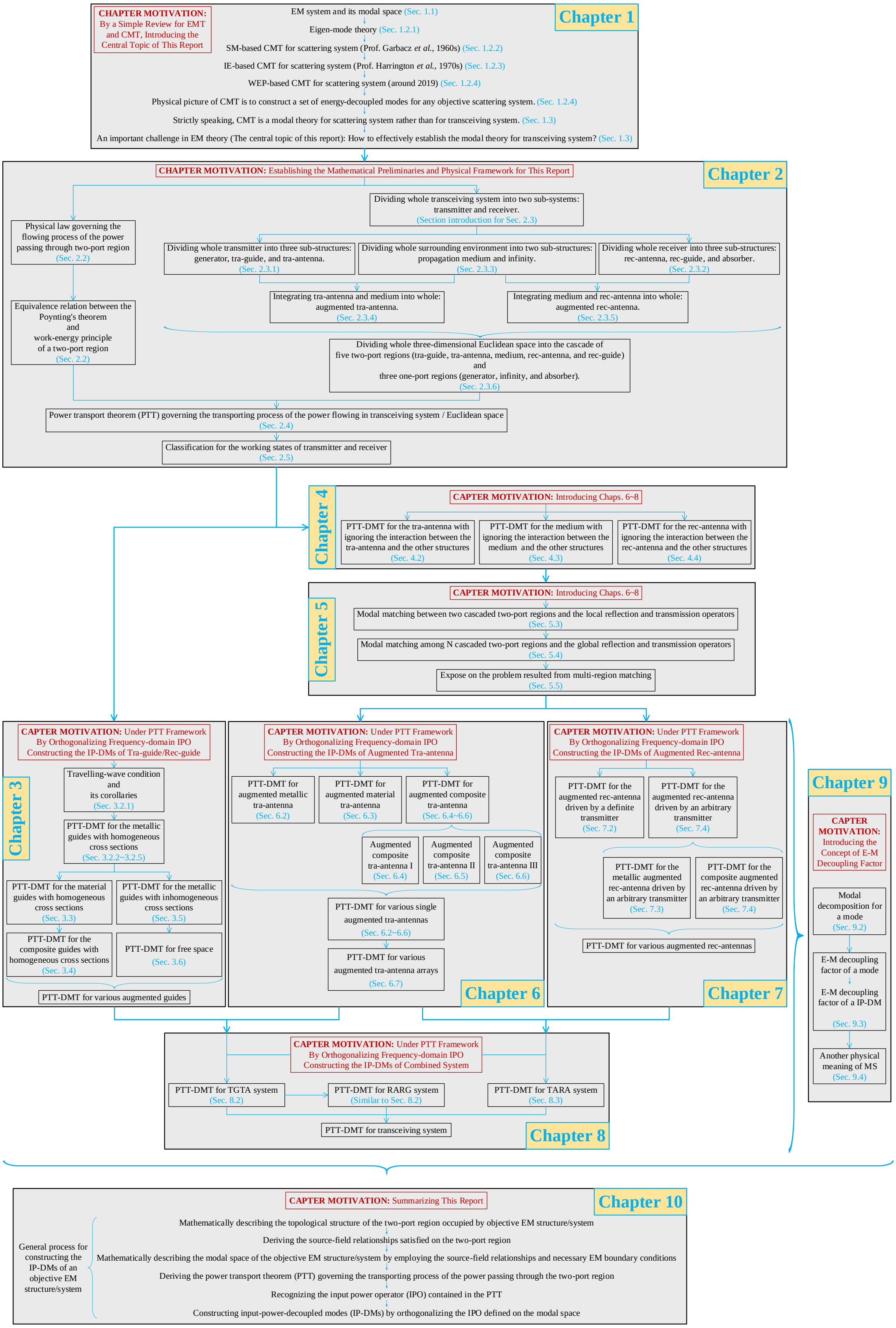

Figure 1-15 Logical relationship diagram and research roadmap for the chapters and sections contained in this report

**Chapter 2** is dedicated to deriving the PTT governing the power exchanges among the various structures in transceiving systems, and providing a quantitative description for the transporting process of the powers flowing in transceiving systems. From Maxwell's equations, Sec. 2.2 derives the Poynting's theorem (PtT) corresponding to any two-port region (which has a penetrable input port, a penetrable output port, and an impenetrable electric wall), and exhibits the fact that the PtT is equivalent to the WEP corresponding to the two-port region. In Sec. 2.3, by employing the language of *point set topology*, whole transceiving system is divided into transmitting sub-system and receiving sub-system, and then the two sub-systems and surrounding environment are respectively further divided into a series of sub-structures, and all the obtained sub-structures (except {generator, absorber, infinity}, which are not interested in this report) are two-port regions. Specifically, the transmitting sub-system is divided into generating structure (an one-port region, which is not interested in this report), guiding structure, and transmitting antenna structure; the receiving sub-system is divided into receiving antenna structure, guiding structure, and absorbing structure (an one-port region, which is not interested in this report); the environment is divided into propagation medium and infinity (an one-port region, which is not interested in this report). By applying the PtT obtained in Sec. 2.2 to the two-port regions obtained in Sec. 2.3, Sec. 2.4 derives a mathematical formulation — *power transport theorem* (*PTT*) — used to quantitatively describe the transporting way of the powers flowing in whole {transceiving system, surrounding environment}. Based on the PTT, Sec. 2.5 classifies the working states of transmitting sub-system and receiving sub-system into some typical ones. The results obtained in Chap. 2 are crucial to this report, because they constitute the theoretical foundation for establishing the DMT for various sub-structures contained in transceiving system.

**Chapter 3** is dedicated to establishing the *PTT-based DMT for guiding structures* (*PTT-Guid-DMT*) and constructing the corresponding DMs by orthogonalizing IPO. In Sec. 3.2, some important conclusions related to the *traveling-wave modes* of metallic guiding structures are derived, and, by employing the conclusions, the DMs of the metallic guiding structures are obtained. In Sec. 3.3, the DMs of the material guiding structures are similarly obtained. In Sec. 3.4, the DMs of the metal-material composite guiding structures are also obtained. In Sec. 3.5, the results obtained in Secs. 3.2~3.4 are further generalized to some more complicated guiding structures. In Sec. 3.6, the DMs of the free space are also constructed.

**Chapters 4 and 5** are two transitional chapters used to introduce the concepts of *augmented tra-antenna* and *augmented rec-antenna* etc., and then used for leading to the subsequent Chaps. 6~8. In Sec. 4.2, by treating tra-antenna (not including *grounding structure*) as a two-port region, the DMs of the tra-antenna are constructed. In Sec. 4.3, by treating propagation medium as a two-port region, the DMs of the propagation medium are constructed. In Sec. 4.4, by treating rec-antenna (not including grounding structure) as a two-port region, the DMs of the rec-antenna are constructed. In Sec. 5.2, the field distributing in any two-port region is decomposed into two components — the *incident component* excited by input port and the *reflected component* excited by output port. In Sec. 5.3, the interaction between two cascaded two-port regions is analysed quantitatively. In Sec. 5.4, the results obtained in Sec. 5.3 are further generalized to multiple cascaded two-port regions. In Sec. 5.5, by combining the results obtained in Secs. 4.2~4.4 and 5.2~5.4, it is not difficult to observe that: **whole modal analysis process will be significantly simplified, if the grounding structure is integrated to tra-antenna and rec-antenna (i.e., if the tra-antenna and medium are treated as a combined system, and the medium and rec-antenna are treated as a combined system).** In fact, the above observation is just the main origination for the subsequent Chaps. 6~8.

**Chapter 6** is dedicated to establishing the *PTT-based DMT for augmented tra-antennas (which include grounding structures) — PTT-TraAnt-DMT* — and constructing the corresponding DMs by orthogonalizing frequency-domain IPO. In Sec. 6.2, the DMs of the augmented metallic tra-antennas are obtained. In Sec. 6.3, the DMs of the augmented material tra-antennas are similarly obtained. In Secs. 6.4~6.6, the results obtained in Secs. 6.2 and 6.3 are generalized to augmented metal-material composite tra-antennas. Section 6.7 gives a further generalization from a single tra-antenna to a two-element tra-antenna array.

**Chapter 7** is dedicated to establishing the *PTT-based DMT for augmented rec-antennas (which include grounding structures) — PTT-RecAnt-DMT* — and constructing the corresponding DMs by orthogonalizing frequency-domain IPO. In Sec. 7.2, the DMs of the augmented metallic rec-antennas driven by definite transmitting systems are obtained. In Sec. 7.3, the DMs of the augmented metallic rec-antennas driven by arbitrary transmitting systems are obtained. In Sec. 7.4, the results obtained in Sec. 7.3 are further generalized to augmented metal-material composite rec-antennas.

**Chapter 8** is dedicated to establishing the *PTT-based DMT for various combined*

*systems* — *PTT-ComSys-DMT*, such as the *PTT-based DMT for augmented tra-guide-tra-antenna and tra-antenna-rec-antenna systems (which include grounding structures)* — *PTT-TGTA-DMT* and *PTT-TARA-DMT*, and deriving the corresponding DMs from orthogonalizing frequency-domain IPO, by generalizing the results obtained in Chaps. 6 and 7.

**Chapter 9** is dedicated to discussing the DM-based *modal decomposition* for any working mode of the objective transceiving system (Sec. 9.2). Based on the modal decomposition, the chapter introduces a novel concept of *electric-magnetic decoupling factor* used to quantitatively depict the coupling degree between the electric energy and magnetic energy carried by any working mode (Sec. 9.3).

**Chapter 10** systematically summarizes the central problems focused on by this report, the fundamental principle established in this report, the main methods used by this report, and the important conclusions derived in this report.

**Appendices** list some relatively detailed formulations related to the main body of this report for readers' references.

### 1.5.2 Symbolic System

The $e^{j\omega t}$ convention is used in this report. For the time-domain physical quantities, this report explicitly exhibits their time variable $t$, for example: time-domain quantity $Q(t)$. The quantities not explicitly containing time variable $t$ are in frequency domain, for example: frequency-domain quantity $Q$. If $\{Q(t), Q\}$ correspond to a linear physical quantity, then $Q(t) = \mathrm{Re}\{Qe^{j\omega t}\}$.

To assist readers to understand the symbols with superscripts or subscripts, this report provides the meanings of the superscripts or subscripts when the symbols first appear. For all of the abbreviations used in this report, their full names are provided when they first appear, such as power transport theorem (PTT), input power operator (IPO), and energy-decoupled mode (DM), etc. We also list all of the abbreviations in the Main Symbol Table after the Abstract of this report, for the convenience of readers' references.

# Chapter 2 Power Transport Theorem for Transceiving System

**CHAPTER MOTIVATION:** The central purpose of this chapter is to derive the *power transport theorem* (*PTT*) governing the transporting process of the electromagnetic power flowing in an arbitrary *transceiving system* by employing the *Poynting's theorem* (*PtT*) / *work-energy principle* (*WEP*) for some *one-port and two-port regions*.

## 2.1 Chapter Introduction

The electromagnetic (EM) energy carried by a non-static EM field usually has flowability. The physical law governing the flowing process is *power transport theorem* (*PTT*). This chapter focuses on deriving the PTT for *transceiving systems*. In fact, the PTT can also be viewed as the generalization for conventional *Poynting's theorem* (*PtT*) and *work-energy principle* (*WEP*).

Firstly, Sec. 2.2 simply reviews the PtT corresponding to a general *two-port region* (which has a penetrable *input port* and a penetrable *output port* separated by an impenetrable *electric wall*), and exhibites the equivalence between the PtT and WEP for the two-port region. Secondly, Sec. 2.3 divides the whole transceiving system into two sub-systems *transmitting sub-system* and *receiving sub-system*, and also divides the sub-systems into some sub-structures (which are some one-port or two-port regions), and then realizes the *region division* for the transceiving system. By applying the PtT to the obtained sub-structures, Sec. 2.4 derives a mathematical expression — PTT — for quantitatively depicting the power flowing way in whole transceiving system and the power exchanging way among the sub-structures. Finally, the working states of transceiving system, transmitting sub-system, receiving sub-system, and various sub-structures are respectively classified in Sec. 2.5.

By utilizing the results obtained in this chapter, we will focus on establishing the *decoupling mode theory* (*DMT*) and constructing the *input-power-decoupled modes* (*IP-DMs*) for the sub-structures and sub-systems in the other chapters of this report.

## 2.2 Poynting's Theorem / Work-Energy Principle for a Two-port Region

In this section, we consider the interaction between the EM fields and EM currents distributing in a region $\mathbb{V}$. The *boundary* of $\mathbb{V}$ is denoted as $\partial\mathbb{V}$, and $\partial\mathbb{V}$ is not contained in $\mathbb{V}$, i.e., $\mathbb{V} = \text{int}\,\mathbb{V}$①. Here, $\text{int}\,\mathbb{V}$ is the *interior* of $\mathbb{V}$, and it is a commonly used concept in *point set topology*[44]. For the generality and convenience of the following discussions, we consider the case that $\partial\mathbb{V}$ is constituted by three sub-boundaries $\mathbb{S}_{\text{in}}$ (a penetrable sub-boundary), $\mathbb{S}_{\text{out}}$ (a penetrable sub-boundary), and $\mathbb{S}_{\text{ele}}$ (an impenetrable electric wall), i.e., $\partial\mathbb{V} = \mathbb{S}_{\text{in}} \cup \mathbb{S}_{\text{out}} \cup \mathbb{S}_{\text{ele}}$, and the three sub-boundaries are pairwise disjoint. The above-mentioned $\mathbb{V}$ and various boundaries are shown in Fig. 2-1. In the figure, $\hat{n}_{\mathbb{S}_{\text{in}}}$ is the *normal direction* of $\mathbb{S}_{\text{in}}$, and points to the interior of $\mathbb{V}$; $\hat{n}_{\mathbb{S}_{\text{out}}}$ is the normal direction of $\mathbb{S}_{\text{out}}$, and points to the *exterior* of $\mathbb{V}$; $\hat{n}_{\partial\mathbb{V}}^{+}$ is the *outer normal direction* of whole $\partial\mathbb{V}$, and it is obvious that $\hat{n}_{\mathbb{S}_{\text{in}}} = -\hat{n}_{\partial\mathbb{V}}^{+}$ (on $\mathbb{S}_{\text{in}}$) and $\hat{n}_{\mathbb{S}_{\text{out}}} = \hat{n}_{\partial\mathbb{V}}^{+}$ (on $\mathbb{S}_{\text{out}}$) as shown in the figure.

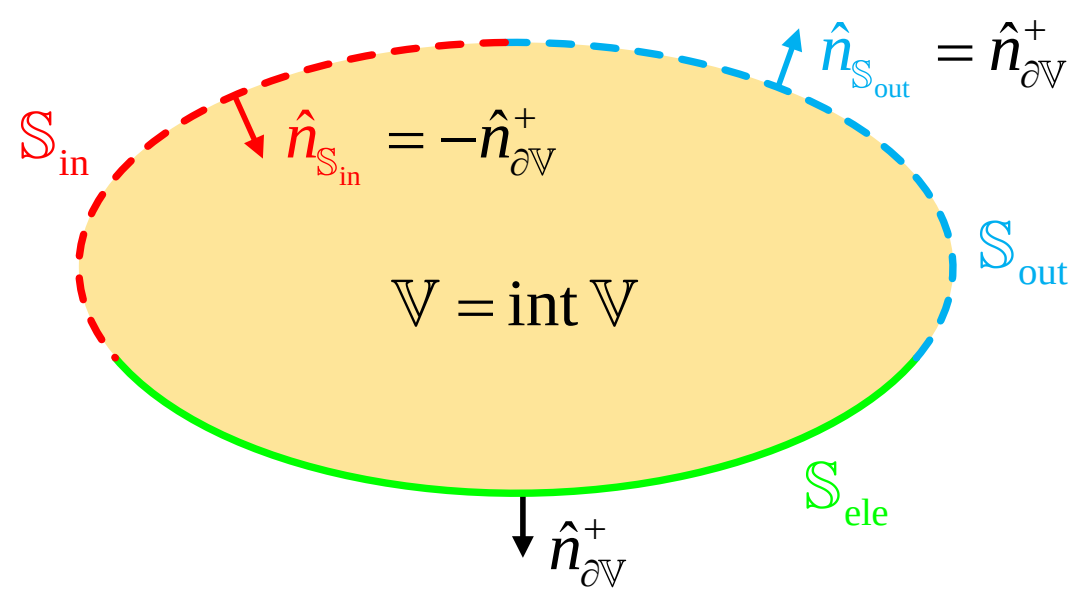

Figure 2-1 Sub-boundaries of a region and their normal direction vectors

If there doesn't exist any *impressed source* distributing in $\mathbb{V}$, and *polarization electric current model*, *Drude model*, and *magnetization magnetic current model* are respectively used to depict the *polarization phenomenon*, *conduction phenomenon*, and *magnetization phenomenon* of the matter under the action of EM field, then we have the following time-domain *Maxwell's equations*

$$\nabla \times \vec{\mathcal{H}} = \varepsilon_0 \frac{\partial \vec{\mathcal{E}}}{\partial t} + \vec{J}^{\text{PV}} + \vec{J}^{\text{CV}} \quad , \quad \vec{r} \in \mathbb{V} \tag{2-1a}$$

$$-\nabla \times \vec{\mathcal{E}} = \mu_0 \frac{\partial \vec{\mathcal{H}}}{\partial t} + \vec{\mathcal{M}}^{\text{MV}} \quad , \quad \vec{r} \in \mathbb{V} \tag{2-1b}$$

In the above Maxwell's equations, $\{\vec{\mathcal{E}}, \vec{\mathcal{H}}\}$ are the EM fields distributing in $\mathbb{V}$; $\varepsilon_0$ and $\mu_0$ are the free-space *dielectric permittivity* and *magnetic permeability* respectively; $\vec{J}^{\text{PV}}$ is the *polarization volume electric current*, and $\vec{J}^{\text{PV}} = \Delta\ddot{\varepsilon} \cdot \partial\vec{\mathcal{E}}/\partial t$, where $\Delta\ddot{\varepsilon} = \ddot{\varepsilon} - \ddot{I}\varepsilon_0$ and $\ddot{\varepsilon}$ is the dielectric permittivity of $\mathbb{V}$ and $\ddot{I}$ is the *unit dyad* used in previous Chap. 1; $\vec{J}^{\text{CV}}$ is the *conduction volume electric current*, and $\vec{J}^{\text{CV}} = \ddot{\sigma} \cdot \vec{\mathcal{E}}$,

① The restriction $\mathbb{V} = \text{int}\,\mathbb{V}$ is not a mathematical game! If boundary $\partial\mathbb{V}$ is included in $\mathbb{V}$, the differential form of Maxwell's equations will not hold in whole $\mathbb{V}$, because the EM fields are not differentiable on $\partial\mathbb{V}$.

where $\vec{\sigma}$ is the *electric conductivity* of $\mathbb{V}$; $\vec{\mathcal{M}}^{\text{MV}}$ is the *magnetization volume magnetic current*, and $\vec{\mathcal{M}}^{\text{MV}} = \Delta\vec{\mu} \cdot \partial\vec{\mathcal{H}} / \partial t$, where $\Delta\vec{\mu} = \vec{\mu} - \vec{I}\mu_0$ and $\vec{\mu}$ is the magnetic permeablity of $\mathbb{V}$.

By doing some simple operations, the above Maxwell's equations (2-1) leads to the following relation

$$0 = \oiint_{\partial\mathbb{V}} \left(\vec{\mathcal{E}} \times \vec{\mathcal{H}}\right) \cdot \hat{n}_{\partial\mathbb{V}}^{+} dS + \left\langle \vec{J}^{\text{CV}}, \vec{\mathcal{E}} \right\rangle_{\mathbb{V}}$$
$$+ \frac{d}{dt}\left[\frac{1}{2}\left\langle \vec{\mathcal{H}}, \mu_0\vec{\mathcal{H}} \right\rangle_{\mathbb{V}} + \frac{1}{2}\left\langle \varepsilon_0\vec{\mathcal{E}}, \vec{\mathcal{E}} \right\rangle_{\mathbb{V}}\right] + \left[\left\langle \vec{\mathcal{H}}, \vec{\mathcal{M}}^{\text{MV}} \right\rangle_{\mathbb{V}} + \left\langle \vec{J}^{\text{PV}}, \vec{\mathcal{E}} \right\rangle_{\mathbb{V}}\right] \quad (2\text{-}2)$$

where the *inner product* is defined as the one used in the previous Chap. 1. Because $\partial\mathbb{V} = \mathbb{S}_{\text{in}} \cup \mathbb{S}_{\text{out}} \cup \mathbb{S}_{\text{ele}}$, and $\hat{n}_{\partial\mathbb{V}}^{+} = -\hat{n}_{\mathbb{S}_{\text{in}}}$ on $\mathbb{S}_{\text{in}}$, and $\hat{n}_{\partial\mathbb{V}}^{+} = \hat{n}_{\mathbb{S}_{\text{out}}}$ on $\mathbb{S}_{\text{out}}$, thus the surface integral term in Eq. (2-2) can be decomposed as follows:

$$\oiint_{\partial\mathbb{V}} \left(\vec{\mathcal{E}} \times \vec{\mathcal{H}}\right) \cdot \hat{n}_{\partial\mathbb{V}}^{+} dS = -\iint_{\mathbb{S}_{\text{in}}} \left(\vec{\mathcal{E}} \times \vec{\mathcal{H}}\right) \cdot \hat{n}_{\mathbb{S}_{\text{in}}} dS + \iint_{\mathbb{S}_{\text{out}}} \left(\vec{\mathcal{E}} \times \vec{\mathcal{H}}\right) \cdot \hat{n}_{\mathbb{S}_{\text{out}}} dS$$
$$+ \iint_{\mathbb{S}_{\text{ele}}} \left(\vec{\mathcal{E}} \times \vec{\mathcal{H}}\right) \cdot \hat{n}_{\partial\mathbb{V}}^{+} dS$$
$$= -\iint_{\mathbb{S}_{\text{in}}} \left(\vec{\mathcal{E}} \times \vec{\mathcal{H}}\right) \cdot \hat{n}_{\mathbb{S}_{\text{in}}} dS + \iint_{\mathbb{S}_{\text{out}}} \left(\vec{\mathcal{E}} \times \vec{\mathcal{H}}\right) \cdot \hat{n}_{\mathbb{S}_{\text{out}}} dS \quad (2\text{-}3)$$

where the second equality is due to that $\iint_{\mathbb{S}_{\text{ele}}} (\vec{\mathcal{E}} \times \vec{\mathcal{H}}) \cdot \hat{n}_{\partial\mathbb{V}}^{+} dS = 0$ (since that $\vec{\mathcal{E}} \times \hat{n}_{\partial\mathbb{V}}^{+} = 0$ on whole electric wall $\mathbb{S}_{\text{ele}}$). Because of relations $\vec{J}^{\text{CV}} = \vec{\sigma} \cdot \vec{\mathcal{E}}$, $\vec{\mathcal{M}}^{\text{MV}} = \Delta\vec{\mu} \cdot \partial\vec{\mathcal{H}} / \partial t$, and $\vec{J}^{\text{PV}} = \Delta\vec{\varepsilon} \cdot \partial\vec{\mathcal{E}} / \partial t$, and that both $\Delta\vec{\mu}$ and $\Delta\vec{\varepsilon}$ are restricted to symmetrical dyads (for details please see the App. B1 in Ref. [13]), thus the field-source interaction terms in Eq. (2-2) can be alternatively written as follows:

$$\left\langle \vec{J}^{\text{CV}}, \vec{\mathcal{E}} \right\rangle_{\mathbb{V}} = \left\langle \vec{\sigma} \cdot \vec{\mathcal{E}}, \vec{\mathcal{E}} \right\rangle_{\mathbb{V}} \quad (2\text{-}4)$$

$$\left\langle \vec{\mathcal{H}}, \vec{\mathcal{M}}^{\text{MV}} \right\rangle_{\mathbb{V}} + \left\langle \vec{J}^{\text{PV}}, \vec{\mathcal{E}} \right\rangle_{\mathbb{V}} = \frac{d}{dt}\left[(1/2)\left\langle \vec{\mathcal{H}}, \Delta\vec{\mu} \cdot \vec{\mathcal{H}} \right\rangle_{\mathbb{V}} + (1/2)\left\langle \Delta\vec{\varepsilon} \cdot \vec{\mathcal{E}}, \vec{\mathcal{E}} \right\rangle_{\mathbb{V}}\right] \quad (2\text{-}5)$$

where we have used the condition that $\vec{\sigma}$, $\Delta\vec{\mu}$, and $\Delta\vec{\varepsilon}$ are independent of time variable $t$. Substituting the above Eqs. (2-3)~(2-5) into the previous Eq. (2-2), the Eq. (2-2) can be equivalently rewritten as the following one for any time variable $t$.

$$\iint_{\mathbb{S}_{\text{in}}} \left(\vec{\mathcal{E}} \times \vec{\mathcal{H}}\right) \cdot \hat{n}_{\mathbb{S}_{\text{in}}} dS = \iint_{\mathbb{S}_{\text{out}}} \left(\vec{\mathcal{E}} \times \vec{\mathcal{H}}\right) \cdot \hat{n}_{\mathbb{S}_{\text{out}}} dS + \overbrace{\left\langle \vec{\sigma} \cdot \vec{\mathcal{E}}, \vec{\mathcal{E}} \right\rangle_{\mathbb{V}}}^{\left\langle \vec{J}^{\text{CV}}, \vec{\mathcal{E}} \right\rangle_{\mathbb{V}}}$$
$$+ \frac{d}{dt}\left[(1/2)\left\langle \vec{\mathcal{H}}, \mu_0\vec{\mathcal{H}} \right\rangle_{\mathbb{V}} + (1/2)\left\langle \varepsilon_0\vec{\mathcal{E}}, \vec{\mathcal{E}} \right\rangle_{\mathbb{V}}\right]$$
$$\underbrace{+ \frac{d}{dt}\left[(1/2)\left\langle \vec{\mathcal{H}}, \Delta\vec{\mu} \cdot \vec{\mathcal{H}} \right\rangle_{\mathbb{V}} + (1/2)\left\langle \Delta\vec{\varepsilon} \cdot \vec{\mathcal{E}}, \vec{\mathcal{E}} \right\rangle_{\mathbb{V}}\right]}_{\left\langle \vec{\mathcal{H}}, \vec{\mathcal{M}}^{\text{MV}} \right\rangle_{\mathbb{V}} + \left\langle \vec{J}^{\text{PV}}, \vec{\mathcal{E}} \right\rangle_{\mathbb{V}}} \quad (2\text{-}6)$$

If we integrate the above Eq. (2-6) in a time interval $t_1 \sim t_2$ (where $t_1 < t_2$), we immediately have that

$$\begin{aligned}
&\int_{t_1}^{t_2} \left[ \iint_{\mathbb{S}_{\text{in}}} \left( \vec{\mathcal{E}} \times \vec{\mathcal{H}} \right) \cdot \hat{n}_{\mathbb{S}_{\text{in}}} \, dS \right] dt \\
= \; & \int_{t_1}^{t_2} \left[ \iint_{\mathbb{S}_{\text{out}}} \left( \vec{\mathcal{E}} \times \vec{\mathcal{H}} \right) \cdot \hat{n}_{\mathbb{S}_{\text{out}}} \, dS \right] dt + \overbrace{\int_{t_1}^{t_2} \left\langle \vec{\sigma} \cdot \vec{\mathcal{E}}, \vec{\mathcal{E}} \right\rangle_{\mathbb{V}} dt}^{\int_{t_1}^{t_2} \langle \vec{\mathcal{J}}^{\text{CV}}, \vec{\mathcal{E}} \rangle_{\mathbb{V}} dt} \\
& + \left[ \frac{1}{2} \left\langle \vec{\mathcal{H}}, \mu_0 \vec{\mathcal{H}} \right\rangle_{\mathbb{V}} + \frac{1}{2} \left\langle \varepsilon_0 \vec{\mathcal{E}}, \vec{\mathcal{E}} \right\rangle_{\mathbb{V}} \right]_{t=t_2} - \left[ \frac{1}{2} \left\langle \vec{\mathcal{H}}, \mu_0 \vec{\mathcal{H}} \right\rangle_{\mathbb{V}} + \frac{1}{2} \left\langle \varepsilon_0 \vec{\mathcal{E}}, \vec{\mathcal{E}} \right\rangle_{\mathbb{V}} \right]_{t=t_1} \\
& \underbrace{+ \left[ \frac{1}{2} \left\langle \vec{\mathcal{H}}, \Delta \vec{\mu} \cdot \vec{\mathcal{H}} \right\rangle_{\mathbb{V}} + \frac{1}{2} \left\langle \Delta \vec{\varepsilon} \cdot \vec{\mathcal{E}}, \vec{\mathcal{E}} \right\rangle_{\mathbb{V}} \right]_{t=t_2} - \left[ \frac{1}{2} \left\langle \vec{\mathcal{H}}, \Delta \vec{\mu} \cdot \vec{\mathcal{H}} \right\rangle_{\mathbb{V}} + \frac{1}{2} \left\langle \Delta \vec{\varepsilon} \cdot \vec{\mathcal{E}}, \vec{\mathcal{E}} \right\rangle_{\mathbb{V}} \right]_{t=t_1}}_{\int_{t_1}^{t_2} \left[ \langle \vec{\mathcal{H}}, \vec{\mathcal{M}}^{\text{MV}} \rangle_{\mathbb{V}} + \langle \vec{\mathcal{J}}^{\text{PV}}, \vec{\mathcal{E}} \rangle_{\mathbb{V}} \right] dt}
\end{aligned} \tag{2-7}$$

where $[\text{quantity}]_{t=t_1/t_2}$ represents the value of the quantity in time $t_1 / t_2$. Equation (2-7) has a very clear physical meaning — **in time interval $t_1 \sim t_2$, the net energy flowing into the input port of region $\mathbb{V}$ (i.e. the port $\mathbb{S}_{\text{in}}$) is transformed into four parts, in which**

**Part I** **flows out from the output port of region $\mathbb{V}$ (i.e. the port $\mathbb{S}_{\text{out}}$) to the exterior of $\mathbb{V}$, and**

**Part II** **is used to do the work on current $\vec{\mathcal{J}}^{\text{CV}}$ (which part is finally converted into *Joule heating energy*), and**

**Part III** **is used to increase the *magnetic field energy* and *electric field energy* stored in region $\mathbb{V}$, and**

**Part IV** **is used to do the work on currents $\vec{\mathcal{M}}^{\text{MV}}$ and $\vec{\mathcal{J}}^{\text{PV}}$ (which part is finally converted into the *magnetization energy* and *polarization energy* stored in the matter occupying region $\mathbb{V}$)**.

It is not difficult to find out that Eq. (2-7) is just a quantitative description for the *energy flowing process* during energy's passing though region $\mathbb{V}$, and also a quantitative description for the *work-energy transformation* during the interaction process between fields $\{\vec{\mathcal{E}}, \vec{\mathcal{H}}\}$ and currents $\{\vec{\mathcal{J}}^{\text{PV}} + \vec{\mathcal{J}}^{\text{CV}}, \vec{\mathcal{M}}^{\text{MV}}\}$. In fact, Eq. (2-7) is just the *Poynting's theorem* (*PtT*) and *work-energy principle* (*WEP*) corresponding to the two-port region. Thus, we call the power $\iint_{\mathbb{S}_{\text{in}}} (\vec{\mathcal{E}} \times \vec{\mathcal{H}}) \cdot \hat{n}_{\mathbb{S}_{\text{in}}} dS$ *input power*, and denote it as $\mathcal{P}^{\text{in}}$, i.e.,

$$\mathcal{P}^{\text{in}} = \iint_{\mathbb{S}_{\text{in}}} \left( \vec{\mathcal{E}} \times \vec{\mathcal{H}} \right) \cdot \hat{n}_{\mathbb{S}_{\text{in}}} dS \tag{2-8}$$

and call the corresponding operator *input power operator* (*IPO*). Similarly, we call the power $\iint_{\mathbb{S}_{\text{out}}} (\vec{\mathcal{E}} \times \vec{\mathcal{H}}) \cdot \hat{n}_{\mathbb{S}_{\text{out}}} dS$ *output power*, and denote it as $\mathcal{P}^{\text{out}}$, i.e.,

$$\mathcal{P}^{\text{out}} = \iint_{\mathbb{S}_{\text{out}}} \left(\vec{\mathcal{E}} \times \vec{\mathcal{H}}\right) \cdot \hat{n}_{\mathbb{S}_{\text{out}}} dS \tag{2-9}$$

and call the corresponding operator *output power operator* (*OPO*).

For time-harmonic EM problem, besides the above-mentioned time-domain versions of Eqs. (2-6) and (2-7) (where the former is for a time point and the latter is for a time interval), there also exists the following frequency-domain version

$$\begin{aligned}\overbrace{(1/2)\iint_{\mathbb{S}_{\text{in}}} \left(\vec{E} \times \vec{H}^{\dagger}\right) \cdot \hat{n}_{\mathbb{S}_{\text{in}}} dS}^{P^{\text{in}}} = & \overbrace{(1/2)\iint_{\mathbb{S}_{\text{out}}} \left(\vec{E} \times \vec{H}^{\dagger}\right) \cdot \hat{n}_{\mathbb{S}_{\text{out}}} dS}^{P^{\text{out}}} + (1/2)\left\langle \ddot{\sigma} \cdot \vec{E}, \vec{E} \right\rangle_{\mathbb{V}} \\ & + j\, 2\omega\left[(1/4)\left\langle \vec{H}, \ddot{\mu} \cdot \vec{H} \right\rangle_{\mathbb{V}} - (1/4)\left\langle \ddot{\varepsilon} \cdot \vec{E}, \vec{E} \right\rangle_{\mathbb{V}}\right] \end{aligned} \tag{2-10}$$

called *frequency-domain PtT* / *WEP*, and the $P^{\text{in}}$ and $P^{\text{out}}$ are the *frequency-domain input power* and *output power* respectively, where the coefficient 1/2 originates from the *time average* for a power-type quadratic quantity, and relations $\ddot{\mu} = \Delta\ddot{\mu} + \ddot{I}\mu_0$ and $\ddot{\varepsilon} = \Delta\ddot{\varepsilon} + \ddot{I}\varepsilon_0$ are employed. In the future Sec. 2.4, the frequency-domain PtT (2-10) will be utilized to derive the *power transport theorem* (*PTT*) for transceiving systems.

In addition, we also have the following relation between the *frequency-domain IPO* $P^{in}$ and time-domain IPO $\mathcal{P}^{\text{in}}$.

$$\text{Re}\left\{P^{\text{in}}\right\} = (1/T)\int_{t_0}^{t_0+T} \mathcal{P}^{\text{in}}\, dt \tag{2-11}$$

where $T$ is the *time period* of the time-harmonic EM field.

**The PtT/WEP discussed in this section quantitatively expresses the physical law governing the power exchange process between a region and its exterior.** In the following Sec. 2.3, we divide whole transceiving system into some regions. In the future Sec. 2.4, we apply the PtT/WEP discussed in this section to the regions obtained in Sec. 2.3, and provide the quantitative expression for the physical law governing the flowing process of the power passing through whole transceiving system.

## 2.3 Region Division and Region Integration for Transceiving System

A typical transceicing system is shown in Fig. 2-2. Clearly, whole transceiving system can be divided into two sub-systems *transmitting system* (simply called *transmitter*) and *receiving system* (simply called *receiver*).

In Sec. 1.4.1, we had qualitatively divided the sub-systems into some sub-structures: *generating structure* (simply called *generator*), *guiding structure in transmitter* (simply called *tra-guide*), *transmitting antenna structure* (simply called *tra-antenna*), *propagation medium* (simply called *medium*), *infinity*, *receiving antenna structure* (simply called *rec-antenna*), *guiding structure in receiver* (simply called *rec-guide*), and *absorbing structure* (simply called *absorber*). The structures are shown as below.

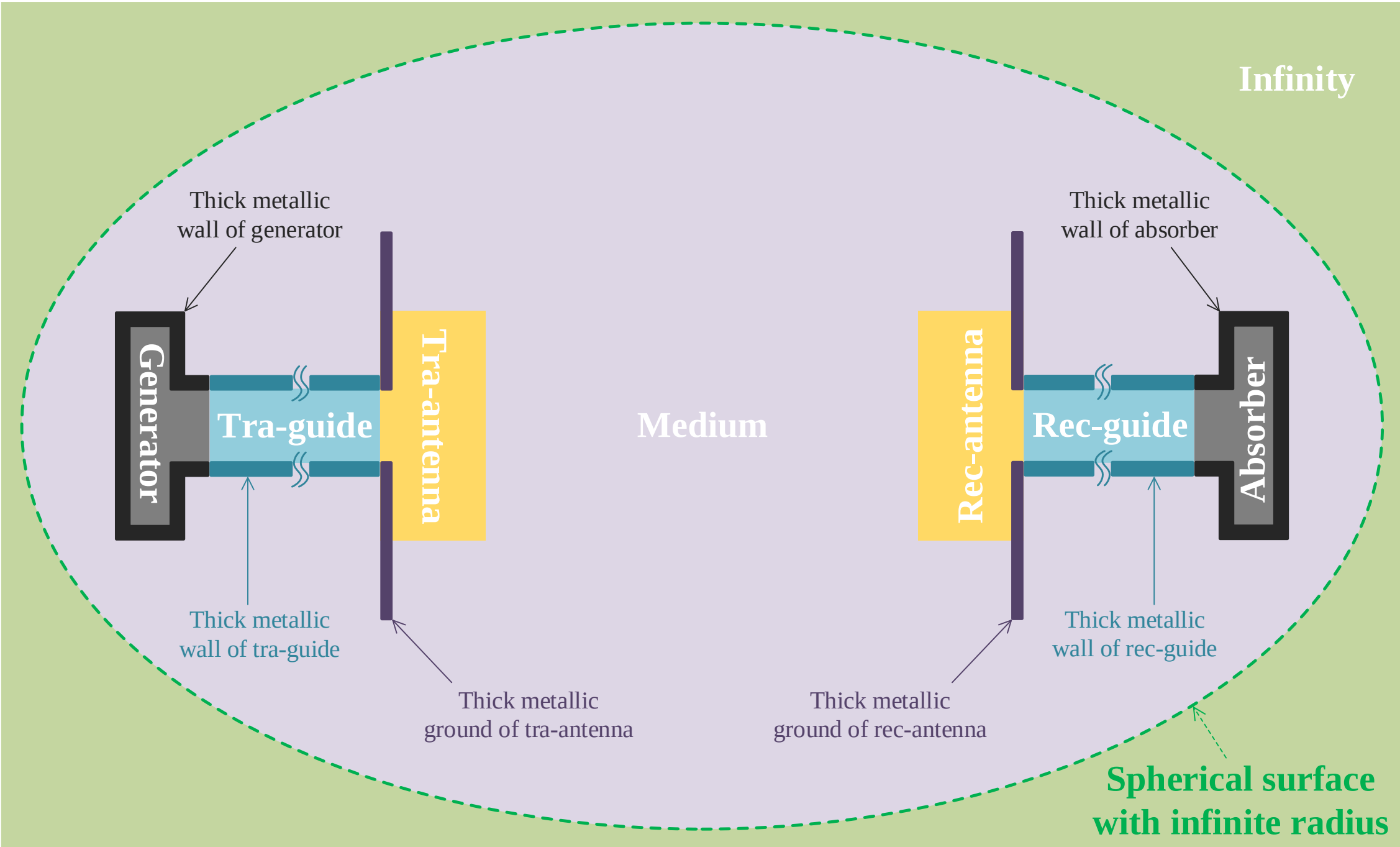

Figure 2-2 A typical transceiving system and its sub-structures

To mathematically apply the PtT/WEP to the regions occupied by the sub-structures obtained in Sec. 1.4.1, we need some rigorous & quantitative descriptions for the division given in Sec. 1.4.1. In the following parts of this section, we employ the language of *point set topology*[44] to achieve the rigorous & quantitative descriptions.

## 2.3.1 Region Division for Transmitting System

In this sub-section, we divide the regions occupied by transmitter and its sub-structures in a mathematically rigorous manner.

### 1. Dividing the Region Occupied by Generator

In Fig. 2-3, we show the above-mentioned generator separately from the other structures.

The inner cavity of the generator shown in Fig. 2-3 is denoted as $\mathbb{V}^{\mathrm{O}}$①, and its boundary $\partial\mathbb{V}^{\mathrm{O}}$ is not included in $\mathbb{V}^{\mathrm{O}}$②, and this implies that $\mathbb{V}^{\mathrm{O}}$ is an *open point set*,

① Here, superscript "O" originates from the letter "o" contained in word "generator".
② If boundary $\partial\mathbb{V}^{\mathrm{O}}$ is included in $\mathbb{V}^{\mathrm{O}}$, we will not have the differential form of Maxwell's equations in whole $\mathbb{V}^{\mathrm{O}}$, because the fields on $\partial\mathbb{V}^{\mathrm{O}}$ are not continuous and then not differentiable.

i.e., $\mathbb{V}^{\rm O} = {\rm int}\,\mathbb{V}^{\rm O}$ [44], where ${\rm int}\,\mathbb{V}^{\rm O}$ denotes the interior of $\mathbb{V}^{\rm O}$. In addition, it is assumed that: the generator has a *thick perfectly conducting electric wall* $\mathbb{W}^{\rm O\|M}$, which is utilized to separate $\mathbb{V}^{\rm O}$ from propagation medium, and the boundary of $\mathbb{W}^{\rm O\|M}$ is not included in $\mathbb{W}^{\rm O\|M}$, and this implies that the $\mathbb{W}^{\rm O\|M}$ is also an open point set, i.e., $\mathbb{W}^{\rm O\|M} = {\rm int}\,\mathbb{W}^{\rm O\|M}$. Above $\mathbb{V}^{\rm O}$ and $\mathbb{W}^{\rm O\|M}$ are shown in Fig. 2-3. Here, the letter "${\rm M}$" in the superscript of $\mathbb{W}^{\rm O\|M}$ is the acronym of "(propagation) medium", and the superscript "${\rm O} \| {\rm M}$" is to vividly express the fact that $\mathbb{W}^{\rm O\|M}$ separates generator and propagation medium.

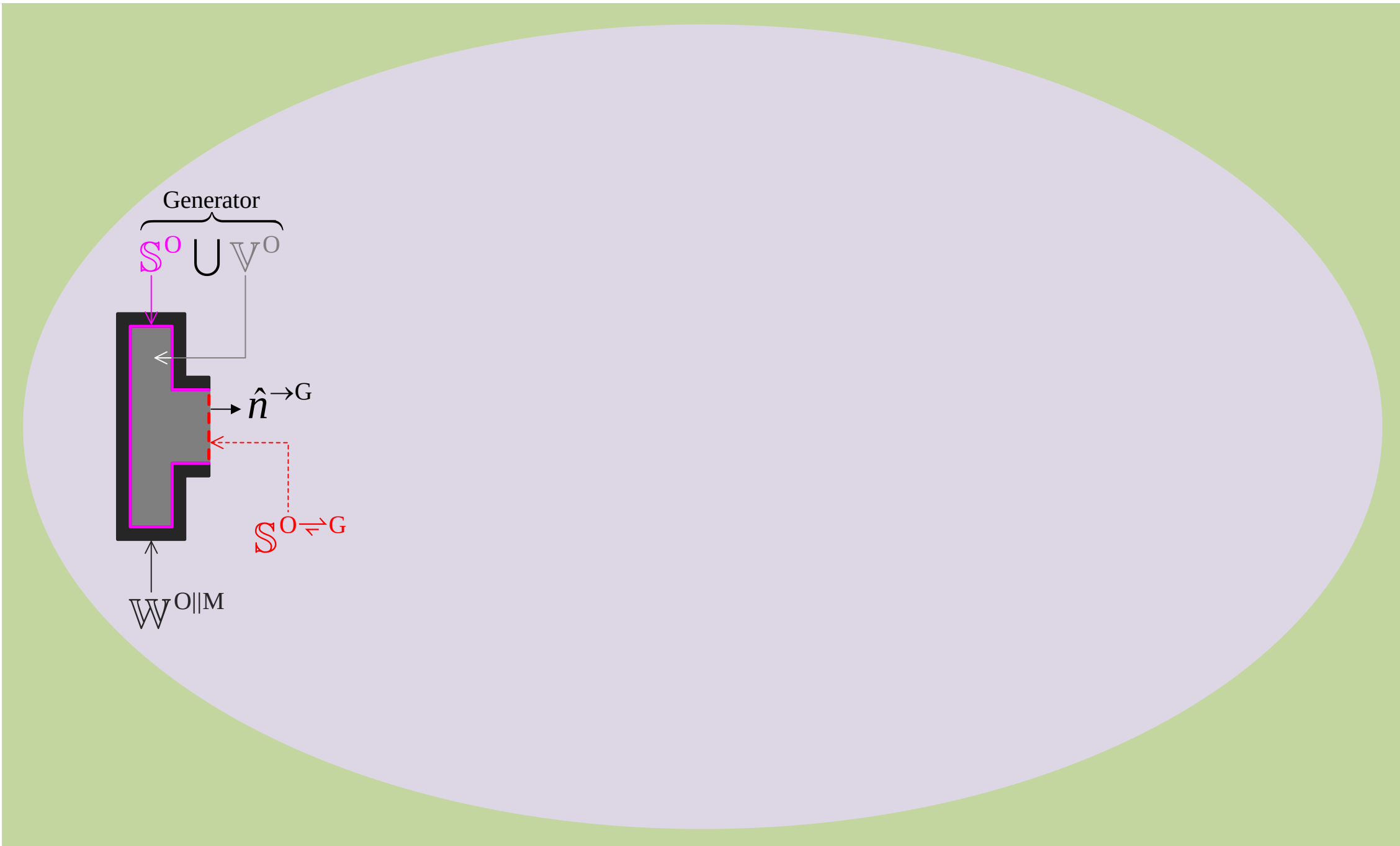

Figure 2-3 Region division for generator

In boundary $\partial \mathbb{V}^{\rm O}$, the electric wall part (i.e. the impenetrable part used to separate $\mathbb{V}^{\rm O}$ from $\mathbb{W}^{\rm O\|M}$) is denoted as $\mathbb{S}^{\rm O}$, and the part which is neither electric wall nor *magnetic wall* (i.e. the penetrable part used as the window for exchanging energy between $\mathbb{V}^{\rm O}$ and the exterior) is denoted as $\mathbb{S}^{\rm O \rightleftharpoons G}$, as shown in Fig. 2-3. For the generator shown in Fig. 2-3, it is obvious that: $\mathbb{S}^{\rm O}$ and $\mathbb{S}^{\rm O \rightleftharpoons G}$ constitute whole $\partial \mathbb{V}^{\rm O}$ (i.e., $\mathbb{S}^{\rm O} \cup \mathbb{S}^{\rm O \rightleftharpoons G} = \partial \mathbb{V}^{\rm O}$), and $\mathbb{S}^{\rm O}$ and $\mathbb{S}^{\rm O \rightleftharpoons G}$ are pairwise disjoint (i.e., $\mathbb{S}^{\rm O} \cap \mathbb{S}^{\rm O \rightleftharpoons G} = \varnothing$). The reason to use "$\rightleftharpoons$" in $\mathbb{S}^{\rm O \rightleftharpoons G}$ will be explained after finishing Secs. 2.4 and 2.5.

In this report, inner cavity $\mathbb{V}^{\rm O}$ and impenetrable sub-boundary $\mathbb{S}^{\rm O}$ are collectively treated as generator as shown in Fig. 2-3. It should be emphasized that thick electric wall $\mathbb{W}^{\rm O\|M}$ and penetrable sub-boundary $\mathbb{S}^{\rm O \rightleftharpoons G}$ are not included in the generator. In addition, cavity $\mathbb{V}^{\text{O}}$ can be lossy, inhomogeneous, and anisotropic, and its conductivity, permittivity, and permeability are denoted as $\ddot{\sigma}^{\text{O}}(\vec{r})$, $\ddot{\varepsilon}^{\text{O}}(\vec{r})$, and $\ddot{\mu}^{\text{O}}(\vec{r})$ respectively, or simply denoted as $\ddot{\sigma}^{\text{O}}$, $\ddot{\varepsilon}^{\text{O}}$, and $\ddot{\mu}^{\text{O}}$ respectively. Of cause, the $\mathbb{V}^{\text{O}}$ can also be homogeneous isotropic, and the results corresponding to the homogeneous isotropic version can be viewed as the special case derived from the inhomogeneous anisotropic version.

**2. Dividing the Region Occupied by Tra-guide**

In the following Fig. 2-4, we show the previously mentioned tra-guide separately from the other structures.

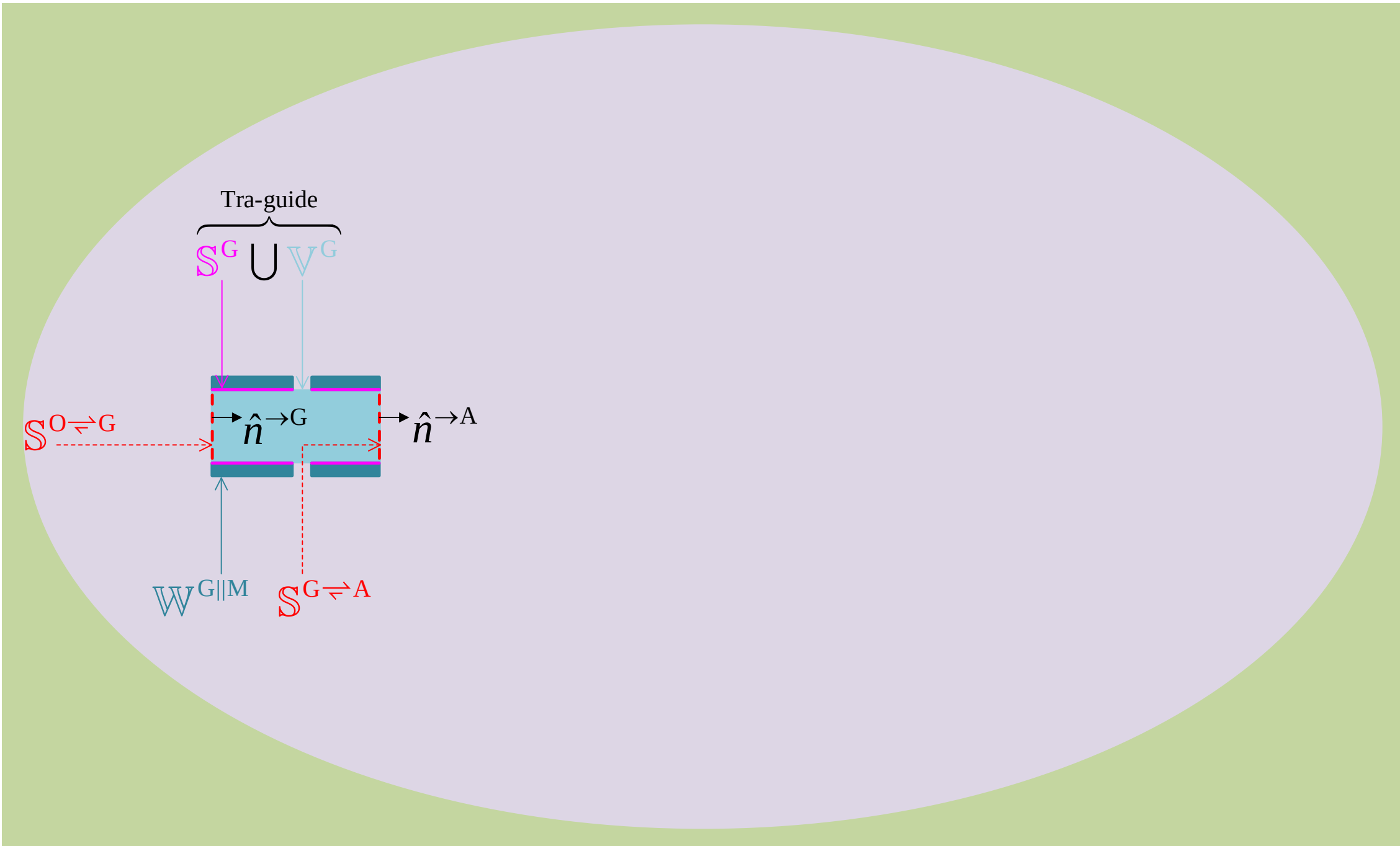

Figure 2-4 Region division for tra-guide

Similarly to denoting the topological structures related to the generator, the *inner cavity* and *thick electric wall* (which is utilized to separate the cavity from propagation medium) of the tra-guide are denoted as $\mathbb{V}^{\text{G}}$ and $\mathbb{W}^{\text{G}\|\text{M}}$ respectively, and it is restricted that $\mathbb{V}^{\text{G}} = \text{int}\,\mathbb{V}^{\text{G}}$ and $\mathbb{W}^{\text{G}\|\text{M}} = \text{int}\,\mathbb{W}^{\text{G}\|\text{M}}$, i.e., both $\mathbb{V}^{\text{G}}$ and $\mathbb{W}^{\text{G}\|\text{M}}$ don't include their boundaries for the same reasons explained previously. Above $\mathbb{V}^{\text{G}}$ and $\mathbb{W}^{\text{G}\|\text{M}}$ are shown in Fig. 2-4.

Similarly to the division for the boundary of generator cavity, the boundary of tra-guide cavity is divided as that $\partial\mathbb{V}^{\text{G}} = \mathbb{S}^{\text{O}\rightleftharpoons\text{G}} \cup \mathbb{S}^{\text{G}} \cup \mathbb{S}^{\text{G}\rightleftharpoons\text{A}}$. Here, sub-boundary $\mathbb{S}^{\text{O}\rightleftharpoons\text{G}}$ is the same as the one shown in the previous Fig. 2-3; for the tra-guide shown in Fig. 2-4, sub-boundary $\mathbb{S}^{G}$ is defined as the interface between $\mathbb{V}^{G}$ and $\mathbb{W}^{G\|M}$, i.e., $\mathbb{S}^{G} = \partial\mathbb{V}^{G} \cap \partial\mathbb{W}^{G\|M}$; for the tra-guide shown in Fig. 2-4, sub-boundary $\mathbb{S}^{G \rightleftharpoons A}$ is defined as the other part of $\partial\mathbb{V}^{G}$ except $\mathbb{S}^{O \rightleftharpoons G}$ and $\mathbb{S}^{G}$, i.e., $\mathbb{S}^{G \rightleftharpoons A} = \partial\mathbb{V}^{G} \setminus (\mathbb{S}^{O \rightleftharpoons G} \cup \mathbb{S}^{G})$. Obviously, sub-boundaries $\mathbb{S}^{O \rightleftharpoons G}$ and $\mathbb{S}^{G \rightleftharpoons A}$ are penetrable, and sub-boundary $\mathbb{S}^{G}$ is impenetrable, and the three sub-boundaries are pairwise disjoint, and they are shown in Fig. 2-4.

In this report, inner cavity $\mathbb{V}^{G}$ and impenetrable sub-boundary $\mathbb{S}^{G}$ are collectively treated as tra-guide, as shown in Fig. 2-4. It should be emphasized that the thick electric wall $\mathbb{W}^{G\|M}$ and the penetrable sub-boundaries $\mathbb{S}^{O \rightleftharpoons G}$ & $\mathbb{S}^{G \rightleftharpoons A}$ are not included in the tra-guide. In addition, cavity $\mathbb{V}^{G}$ can be lossy, inhomogeneous, and anisotropic, and its material parameters are denoted as $\vec{\vec{\sigma}}^{G}(\vec{r})$, $\vec{\vec{\varepsilon}}^{G}(\vec{r})$, and $\vec{\vec{\mu}}^{G}(\vec{r})$, or simply denoted as $\vec{\vec{\sigma}}^{G}$, $\vec{\vec{\varepsilon}}^{G}$, and $\vec{\vec{\mu}}^{G}$.

**3. Integrating the Regions Occupied by Generator and Tra-guide and the Port $\mathbb{S}^{O \rightleftharpoons G}$ — Feeding Structure (Feed)**

For the convenience of discussion, the generator, port $\mathbb{S}^{O \rightleftharpoons G}$, and tra-guide are collectively referred to as *feeding system* (simply called *feed*) as shown in Fig. 2-5.

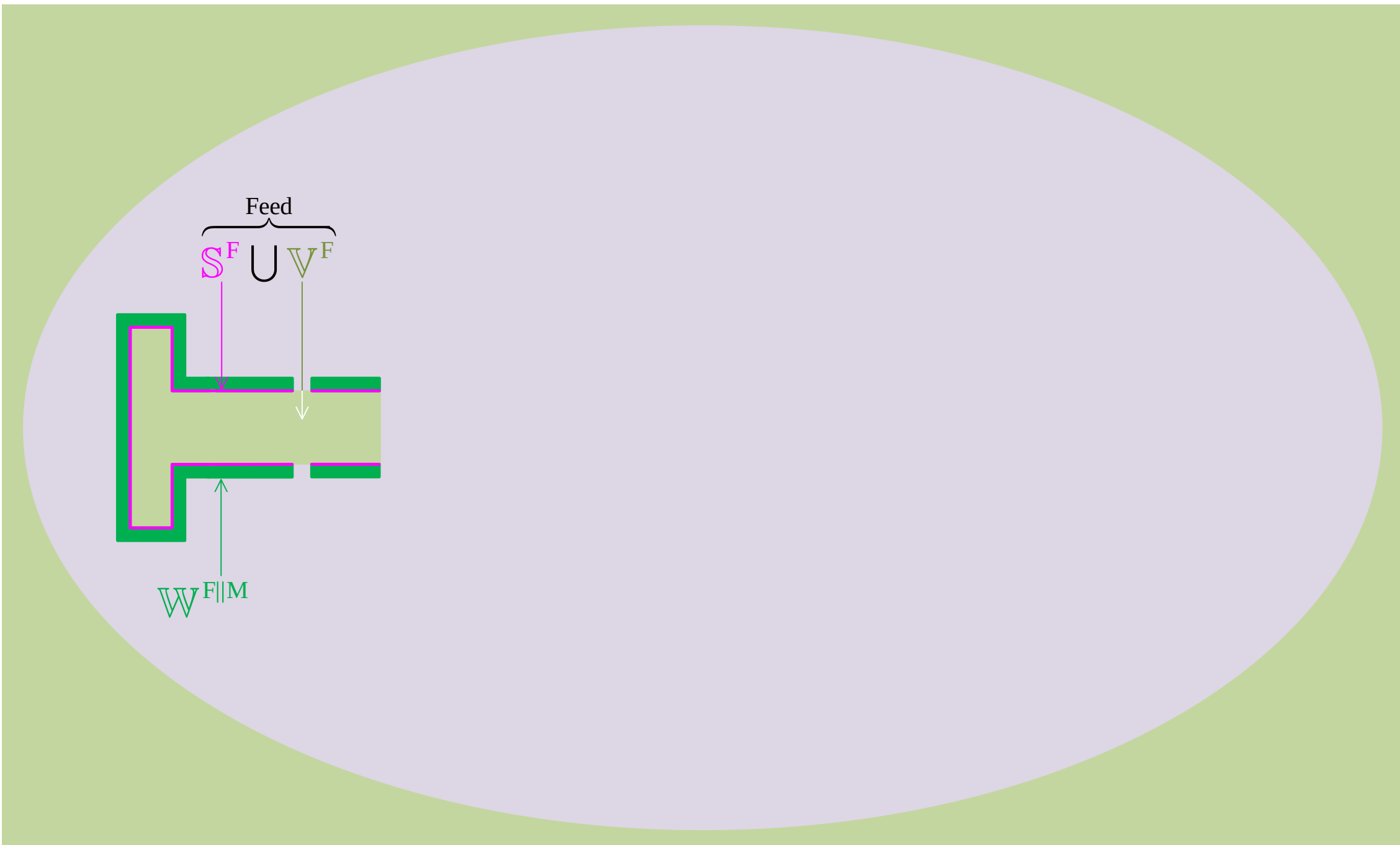

Figure 2-5 Region integration for generator, port $\mathbb{S}^{O \rightleftharpoons G}$, and tra-guide

**4. Dividing the Region Occupied by Tra-antenna**

In the following Fig. 2-6, we show the previously mentioned tra-antenna separately from the other structures.

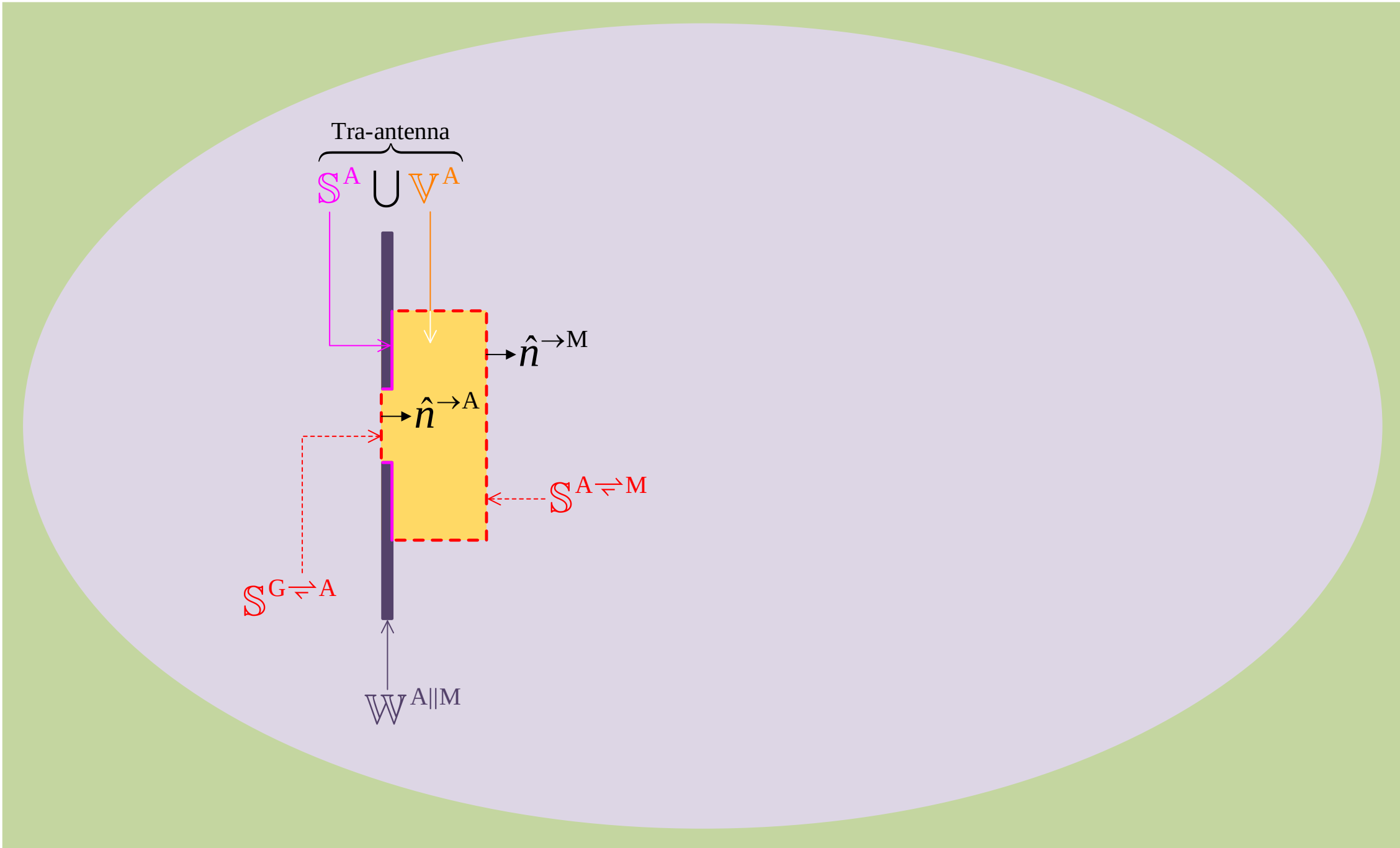

Figure 2-6 Region division for tra-antenna

As shown in the above Fig. 2-6, the tra-antenna considered here is a *material body* mounted on a *thick metallic ground plane*. In the tra-antenna, the region occupied by the material body is denoted as $\mathbb{V}^{A}$, and it is restricted that $\mathbb{V}^{A} = \operatorname{int} \mathbb{V}^{A}$ for guaranteeing the differential form of Maxwell's equations on whole $\mathbb{V}^{A}$. The ground plane utilized to partially separate tra-antenna from propagation medium is denoted as $\mathbb{W}^{A\|M}$, and it is restricted that $\mathbb{W}^{A\|M} = \operatorname{int} \mathbb{W}^{A\|M}$. The above-mentioned $\mathbb{V}^{A}$ and $\mathbb{W}^{A\|M}$ are shown in Fig. 2-6.

Following the previous conventions, the interface between $\mathbb{V}^{A}$ and $\mathbb{W}^{A\|M}$ (i.e. the impenetrable part of $\partial \mathbb{V}^{A}$) is denoted as $\mathbb{S}^{A}$, i.e., $\mathbb{S}^{A} = \partial \mathbb{V}^{A} \cap \partial \mathbb{W}^{A\|M}$; the interface between $\mathbb{V}^{A}$ and propagation medium is denoted as $\mathbb{S}^{A \rightleftharpoons M}$, and it is penetrable. In addition, it is obvious that the penetrable interface between $\mathbb{V}^{A}$ and $\mathbb{V}^{G}$ is just $\mathbb{S}^{G \rightleftharpoons A}$ (which had been shown in the previous Fig. 2-4), i.e., $\mathbb{S}^{G \rightleftharpoons A} = \partial \mathbb{V}^{G} \cap \partial \mathbb{V}^{A}$. The above-mentioned $\mathbb{S}^{G \rightleftharpoons A}$, $\mathbb{S}^{A}$, and $\mathbb{S}^{A \rightleftharpoons M}$ are shown in Fig. 2-6.

In this report, inner cavity $\mathbb{V}^{A}$ and impenetrable sub-boundary $\mathbb{S}^{A}$ are collectively treated as tra-antenna, as shown in Fig. 2-6. It should be emphasized that both of the penetrable sub-boundaries $\mathbb{S}^{G \rightleftharpoons A}$ and $\mathbb{S}^{A \rightleftharpoons M}$ are not included in tra-antenna. In addition, the cavity $\mathbb{V}^{A}$ can be lossy, inhomogeneous, and anisotropic, and its material parameters are denoted as $\vec{\sigma}^{A}(\vec{r})$, $\vec{\varepsilon}^{A}(\vec{r})$, and $\vec{\mu}^{A}(\vec{r})$, or simply as $\vec{\sigma}^{A}$, $\vec{\varepsilon}^{A}$, and $\vec{\mu}^{A}$.

**5. Integrating the Regions Occupied by Feed (Generator & Port $\mathbb{S}^{O \rightleftharpoons G}$ & Tra-guide) and Tra-antenna and the Port $\mathbb{S}^{G \rightleftharpoons A}$ — Transmitting System (Transmitter)**

Similarly to integrating generator, port $\mathbb{S}^{O \rightleftharpoons G}$, and tra-guide previously, the feeding structure, port $\mathbb{S}^{G \rightleftharpoons A}$, and tra-antenna are also treated as a whole, and it is just the *transmitting system* as shown in Fig. 2-7, and it is correspondingly denoted as $\mathbb{T}$.

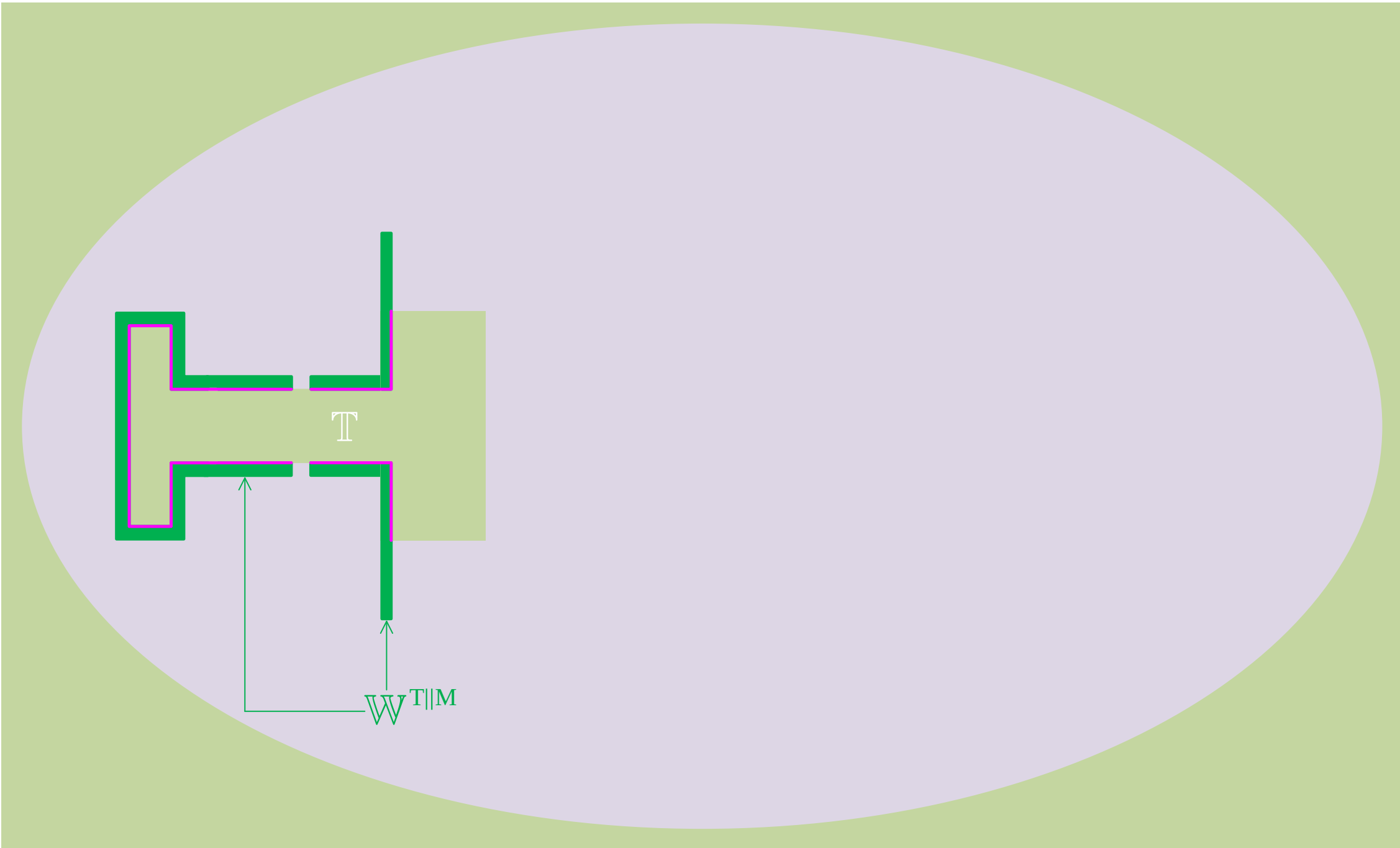

Figure 2-7 Region integration for feed, port $\mathbb{S}^{G \rightleftharpoons A}$, and tra-antenna

**6. Summary**

In summary, the region division for transmitting system (i.e. transmitter) $\mathbb{T}$ is given as follows:

$$\text{Transmitter } \mathbb{T} = \overbrace{\underbrace{\mathbb{S}^{O} \cup \mathbb{V}^{O}}_{\text{Generator}} \cup \underbrace{\mathbb{S}^{O \rightleftharpoons G}}_{\text{Output/Input Port of } \mathbb{V}^{O}/\mathbb{V}^{G}} \cup \overbrace{\mathbb{S}^{G} \cup \mathbb{V}^{G}}^{\text{Tra-guide}}}^{\text{Feed}} \cup \underbrace{\mathbb{S}^{G \rightleftharpoons A}}_{\text{Output/Input Port of } \mathbb{V}^{G}/\mathbb{V}^{A}} \cup \overbrace{\mathbb{S}^{A} \cup \mathbb{V}^{A}}^{\text{Tra-antenna}} \cup \underbrace{\mathbb{W}^{O\|M} \cup \mathbb{W}^{G\|M} \cup \mathbb{W}^{A\|M}}_{\text{Thick Electric Walls}} \tag{2-12}$$

and the various regions in the above division formulation (2-12) are shown in Fig. 2-8, where the ports marked by dotted lines are penetrable, and the thick electric walls $\mathbb{W}^{O\|M} \cup \mathbb{W}^{G\|M} \cup \mathbb{W}^{A\|M}$ are collectively denoted as $\mathbb{W}^{T\|M}$.

By employing the above region division formulation (2-12), the energy/power flow in transmitting system will be quantitatively discussed in the future Sec. 2.4.1.

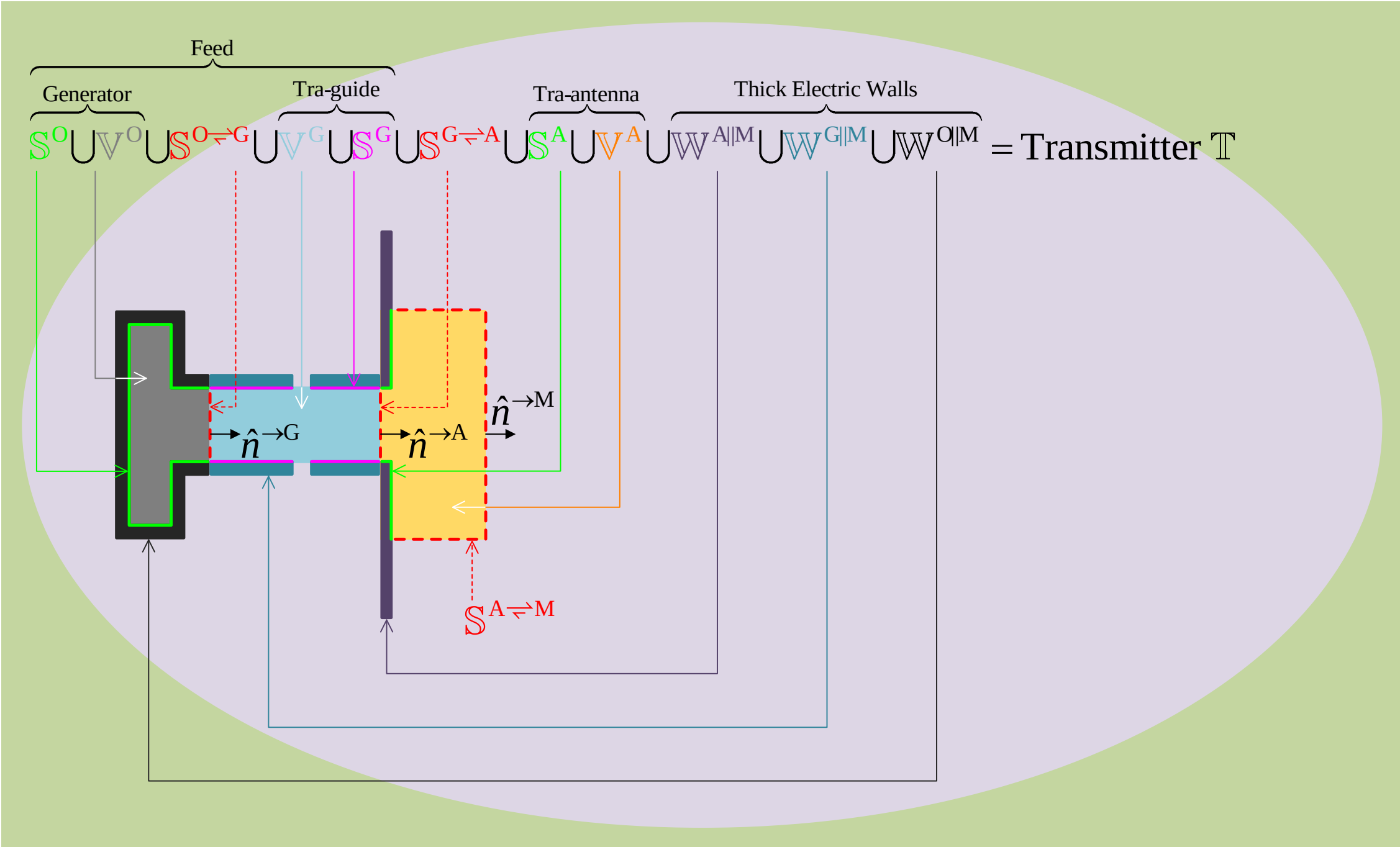

Figure 2-8 Region division for transmitting system

### 2.3.2 Region Division for Receiving System

In this sub-section, we divide the regions occupied by receiving system and its sub-structures in a mathematically rigorous manner. In Fig. 2-9, we show the previously mentioned receiving system separately from the transmitting system.

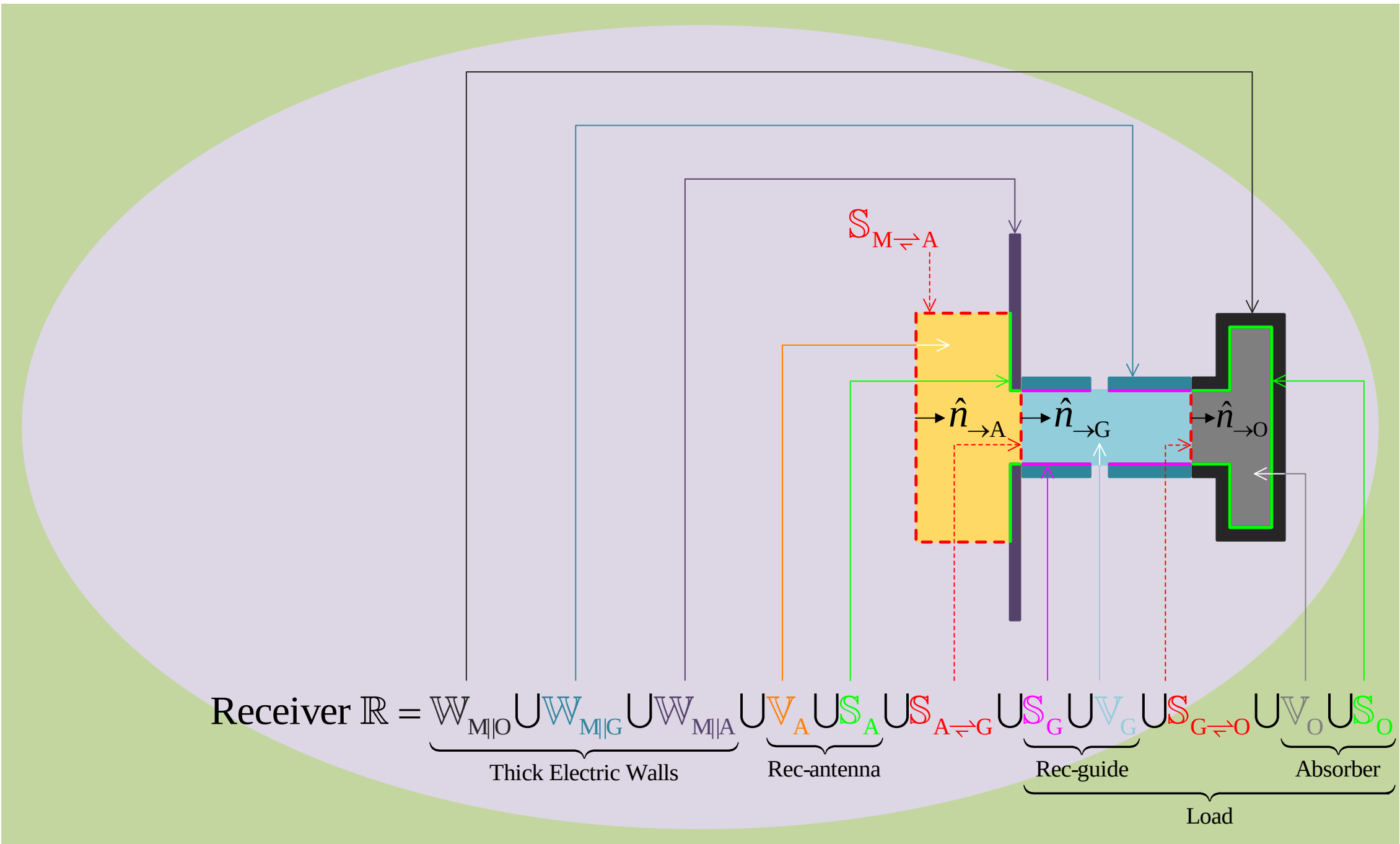

Figure 2-9 Region division for receiving system

Similarly to the previous region division for transmitting system, the whole receiving system (i.e. receiver) $\mathbb{R}$ can be divided as follows:

$$\text{Receiver } \mathbb{R} = \overbrace{\mathbb{W}_{\text{M}||\text{A}} \bigcup \mathbb{W}_{\text{M}||\text{G}} \bigcup \mathbb{W}_{\text{M}||\text{O}}}^{\text{Thick Electric Walls}} \bigcup \underbrace{\mathbb{V}_{\text{A}} \bigcup \mathbb{S}_{\text{A}}}_{\text{Rec-antenna}} \bigcup \overbrace{\mathbb{S}_{\text{A}\rightleftharpoons\text{G}}}^{\text{Output/Input Port of } \mathbb{V}_{\text{A}}/\mathbb{V}_{\text{G}}} \bigcup \underbrace{\underbrace{\mathbb{S}_{\text{G}} \bigcup \mathbb{V}_{\text{G}}}_{\text{Rec-guide}} \bigcup \overbrace{\mathbb{S}_{\text{G}\rightleftharpoons\text{O}}}^{\text{Output/Input Port of } \mathbb{V}_{\text{G}}/\mathbb{V}_{\text{O}}} \bigcup \underbrace{\mathbb{V}_{\text{O}} \bigcup \mathbb{S}_{\text{O}}}_{\text{Absorber}}}_{\text{Load}} \tag{2-13}$$

as shown in Fig. 2-9. To be distinguished from the region division for transmitting system, the regions, ports (penetrable interfaces), electric walls (impenetrable interfaces), and cavities corresponding to receiving system are only with subscripts but without any superscript.

Now, we explain the various symbols used in region division formulation (2-13) and region division figure Fig. 2-9 as follows:

- Receiving system $\mathbb{R}$ is constituted by rec-antenna, port $\mathbb{S}_{\text{A}\rightleftharpoons\text{G}}$ (the penetrable interface between rec-antenna and rec-guide), and *loading system* (simply called *load*). The load is constituted by rec-guide, port $\mathbb{S}_{\text{G}\rightleftharpoons\text{O}}$ (the penetrable interface between rec-guide and absorber), and absorber.
- Rec-antenna is constructed by electric wall $\mathbb{S}_{\text{A}}$ (an impenetrable interface) and a material body $\mathbb{V}_{\text{A}}$ placed on a thick metallic ground plane $\mathbb{W}_{\text{M}||\text{A}}$, where $\mathbb{V}_{\text{A}} = \text{int}\,\mathbb{V}_{\text{A}}$ and $\mathbb{W}_{\text{M}||\text{A}} = \text{int}\,\mathbb{W}_{\text{M}||\text{A}}$ and $\mathbb{S}_{\text{A}} = \partial\mathbb{V}_{\text{A}} \bigcap \partial\mathbb{W}_{\text{M}||\text{A}}$.
- Rec-guide is constituted by its inner cavity $\mathbb{V}_{\text{G}}$ and an impenetrable interface $\mathbb{S}_{\text{G}}$, where $\mathbb{V}_{\text{G}}$ doesn't include its boundary, i.e., $\mathbb{V}_{\text{G}} = \text{int}\,\mathbb{V}_{\text{G}}$. The thick metallic wall used to separate $\mathbb{V}_{\text{G}}$ from propagation medium is denoted as $\mathbb{W}_{\text{M}||\text{G}}$.
- Absorber is constructed by its inner cavity $\mathbb{V}_{\text{O}}$ and an impenetrable interface $\mathbb{S}_{\text{O}}$, where $\mathbb{V}_{\text{O}}$ doesn't include its boundary, i.e., $\mathbb{V}_{\text{O}} = \text{int}\,\mathbb{V}_{\text{O}}$. The thick metallic wall used to separate $\mathbb{V}_{\text{O}}$ from propagation medium is denoted as $\mathbb{W}_{\text{M}||\text{O}}$.

The thick metallic walls $\mathbb{W}_{\text{M}||\text{A}} \bigcup \mathbb{W}_{\text{M}||\text{G}} \bigcup \mathbb{W}_{\text{M}||\text{O}}$ are collectively denoted as $\mathbb{W}_{\text{M}||\text{R}}$. The region $\mathbb{V}_{\text{A/G/O}}$ can be lossy, inhomogeneous, and anisotropic, and its permeability, permittivity, and conductivity are denoted as $\ddot{\sigma}_{\text{A/G/O}}(\vec{r})$, $\ddot{\varepsilon}_{\text{A/G/O}}(\vec{r})$, and $\ddot{\mu}_{\text{A/G/O}}(\vec{r})$, or simply as $\ddot{\sigma}_{\text{A/G/O}}$, $\ddot{\varepsilon}_{\text{A/G/O}}$, and $\ddot{\mu}_{\text{A/G/O}}$. Of cause, $\mathbb{V}_{\text{A/G/O}}$ can also be homogeneous isotropic, and the results corresponding to the homogeneous isotropic version can be viewed as the special case of the inhomogeneous anisotropic version.

By employing the above region division formulation (2-13), the energy/power flow in receiving system will be quantitatively discussed in the future Sec. 2.4.3.

### 2.3.3 Region Division for Surrounding Environment

In this sub-section, we divide the region occupied by surrounding environment in a mathematically rigorous manner.

From Fig. 2-10, it is obvious that whole boundary of environment consists of two parts: a part locates at infinity, and it is called *the outer boundary of environment* or simply called *infinity*, and it is correspondingly denoted as $\mathbb{I}$; the other part doesn't locate at infinity, and it is called *the inner boundary of environment*. For the convenience of this report, the following surface $\mathbb{S}_{\mathrm{M}\to\mathrm{I}}^{\mathrm{M}\to\mathrm{I}}$ is defined.

$$\mathbb{S}_{\mathrm{M}\to\mathrm{I}}^{\mathrm{M}\to\mathrm{I}} = \{\vec{r} : \vec{r} \text{ belongs to the interior of } \mathbb{M}\text{, and tends to } \mathbb{I}.\} \tag{2-14}$$

and it can be considered to be a *spherical surface with infinite radius*. Here, we use both superscript and subscript "$\mathrm{M}\to\mathrm{I}$" on $\mathbb{S}_{\mathrm{M}\to\mathrm{I}}^{\mathrm{M}\to\mathrm{I}}$ is to emphasize that $\mathbb{S}_{\mathrm{M}\to\mathrm{I}}^{\mathrm{M}\to\mathrm{I}}$ will be employed in both transmitting problem and receiving problem. The reason we use "$\to$" in $\mathbb{S}_{\mathrm{M}\to\mathrm{I}}^{\mathrm{M}\to\mathrm{I}}$ instead of the "$\rightleftharpoons$" previously used in the other ports will be clear when the subsequent Secs. 2.4 and 2.5 are finished.

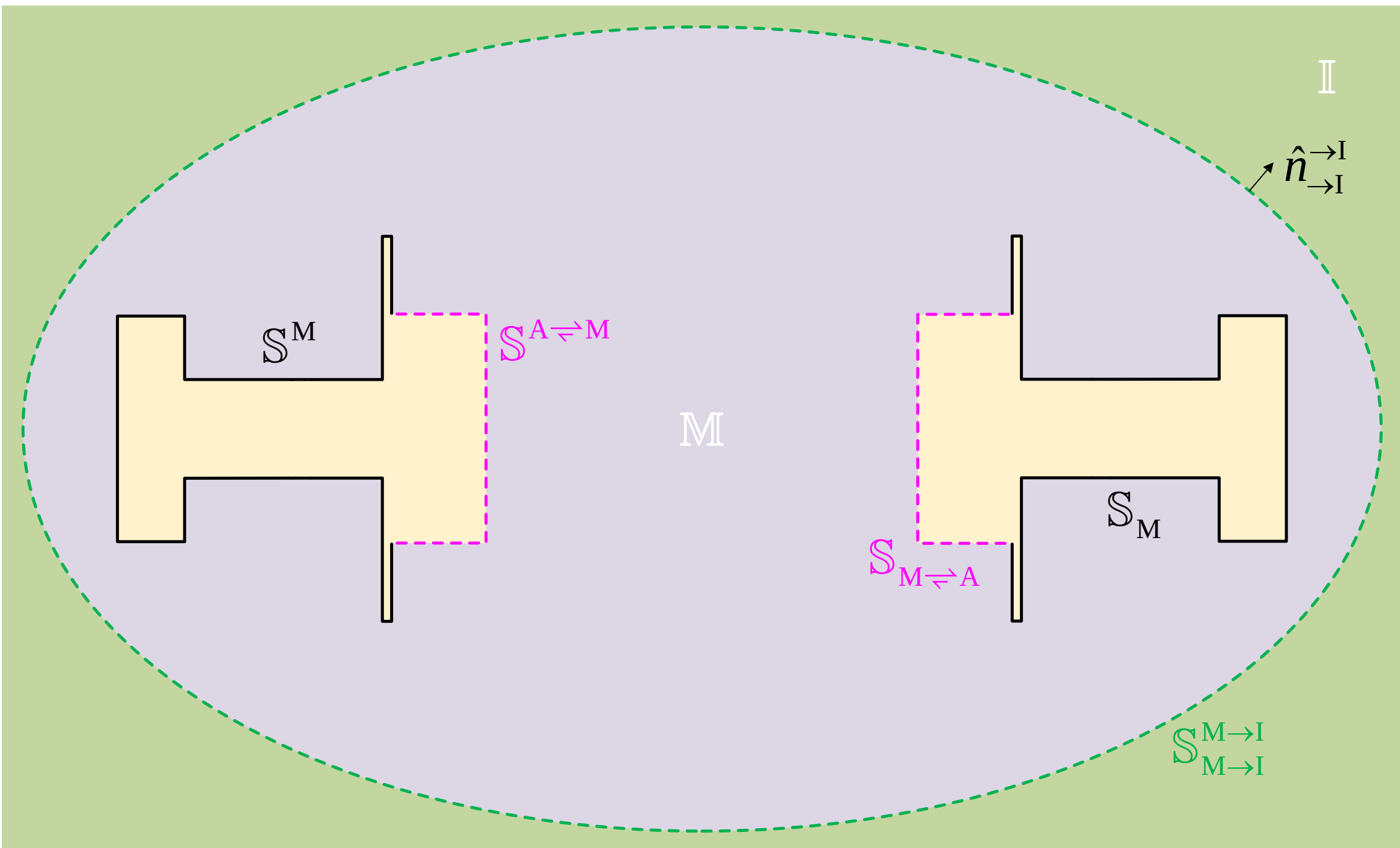

Figure 2-10 Region division for environment

Clearly, the $\mathbb{W}^{\mathrm{O}\|\mathrm{M}} \cup \mathbb{W}^{\mathrm{G}\|\mathrm{M}} \cup \mathbb{W}^{\mathrm{A}\|\mathrm{M}}$ shown in Figs. 2-8 is just the thick electric wall utilized to partially separate transmitter from propagation medium, so it is simply denoted as $\mathbb{W}^{\mathrm{T}\|\mathrm{M}}$. The impenetrable interface between $\mathbb{W}^{\mathrm{T}\|\mathrm{M}}$ and propagation medium is

denoted as $\mathbb{S}^{\rm M}$. Obviously, the $\mathbb{W}_{\rm M||A} \cup \mathbb{W}_{\rm M||G} \cup \mathbb{W}_{\rm M||O}$ shown in Figs. 2-9 is just the thick electric wall utilized to partially separate propagation medium from receiver, so it is simply denoted as $\mathbb{W}_{\rm M||R}$. The impenetrable interface between $\mathbb{W}_{\rm M||R}$ and propagation medium is denoted as $\mathbb{S}_{\rm M}$. In fact, $\mathbb{S}^{\rm M} \cup \mathbb{S}^{\rm A \rightleftharpoons M} \cup \mathbb{S}_{\rm M \rightleftharpoons A} \cup \mathbb{S}_{\rm M}$ is just the whole inner boundary of propagation medium, as shown in Fig. 2-10.①

As shown in Fig. 2-10, both $\mathbb{S}^{\rm M} \cup \mathbb{S}^{\rm A \rightleftharpoons M} \cup \mathbb{S}_{\rm M \rightleftharpoons A} \cup \mathbb{S}_{\rm M}$ and $\mathbb{S}_{\rm M \to I}^{\rm M \to I}$ are closed surfaces, and the open point set sandwiched between them is just the *propagation medium* (*medium*), so the sandwiched set is denoted as $\mathbb{M}$, and obviously $\partial \mathbb{M} = \mathbb{S}^{\rm M} \cup \mathbb{S}^{\rm A \rightleftharpoons M} \cup \mathbb{S}_{\rm M \to I}^{\rm M \to I} \cup \mathbb{S}_{\rm M \rightleftharpoons A} \cup \mathbb{S}_{\rm M}$[44]. Evidently, $\mathbb{S}_{\rm M \to I}^{\rm M \to I}$ is just the outer boundary of $\mathbb{M}$. Following the previous conventions, $\mathbb{S}^{\rm M} \cup \mathbb{M} \cup \mathbb{S}_{\rm M}$ is defined as the region occupied by propagation medium, and it should be emphasized that the penetrable surface $\mathbb{S}^{\rm A \rightleftharpoons M} \cup \mathbb{S}_{\rm M \to I}^{\rm M \to I} \cup \mathbb{S}_{\rm M \rightleftharpoons A}$ is not included in the propagation medium.

Based on the above these, we have the following region division formulation for environment $\mathbb{E}$.

$$\text{Environment } \mathbb{E} = \underbrace{\mathbb{S}^{\rm M} \cup \mathbb{M} \cup \mathbb{S}_{\rm M}}_{\text{Medium}} \cup \mathbb{S}_{\rm M \to I}^{\rm M \to I} \cup \underbrace{\mathbb{I}}_{\text{Infinity}} \tag{2-15}$$

In addition, the medium $\mathbb{M}$ can be lossy, inhomogeneous, and anisotropic, and its conductivity, permittivity, and permeability are denoted as $\vec{\sigma}_{\rm M}^{\rm M}(\vec{r})$, $\vec{\varepsilon}_{\rm M}^{\rm M}(\vec{r})$, and $\vec{\mu}_{\rm M}^{\rm M}(\vec{r})$ respectively, or simply denoted as $\vec{\sigma}_{\rm M}^{\rm M}$, $\vec{\varepsilon}_{\rm M}^{\rm M}$, and $\vec{\mu}_{\rm M}^{\rm M}$ respectively.

### 2.3.4 Region Integration for Transmitting Antenna and Grounding Structure — Augmented Transmitting Antenna

In the future Secs. 2.4.1 and 2.4.2, it will be exhibited that: **not only $\mathbb{S}^{\rm A}$ and $\mathbb{V}^{\rm A}$ have ability to modulate the energy to be released into medium, but also $\mathbb{S}^{\rm M}$ has the ability**. In addition, in the future Chaps. 4~6, it will be found out that: **to treat $\mathbb{S}^{\rm A} \cup \mathbb{V}^{\rm A}$ and $\mathbb{S}^{\rm M}$ as a whole will bring some conveniences to the *modal analysis* for transmitting problem**.

Based on these above, we introduce the following concept of *augmented transmitting antenna* (or simply denoted as *augmented tra-antenna*)

$$\text{Augmented Tra-antenna} = \mathbb{S}^{\rm M} \cup \overbrace{\mathbb{V}^{\rm A} \cup \mathbb{S}^{\rm A}}^{\text{Tra-antenna}} \tag{2-16}$$

which is the union of transmitting antenna and the *grounding structure* $\mathbb{S}^M$ *related to transmitting problem* (simply called *tra-ground*), as shown in Fig. 2-11.

---

① Strictly speaking, $\mathbb{W}^{\rm T||M} = {\rm int}({\rm cl}(\mathbb{W}^{\rm O||M} \cup \mathbb{W}^{\rm G||M} \cup \mathbb{W}^{\rm A||M}))$, and $\mathbb{W}_{\rm M||R} = {\rm int}({\rm cl}(\mathbb{W}_{\rm M||A} \cup \mathbb{W}_{\rm M||G} \cup \mathbb{W}_{\rm M||O}))$.

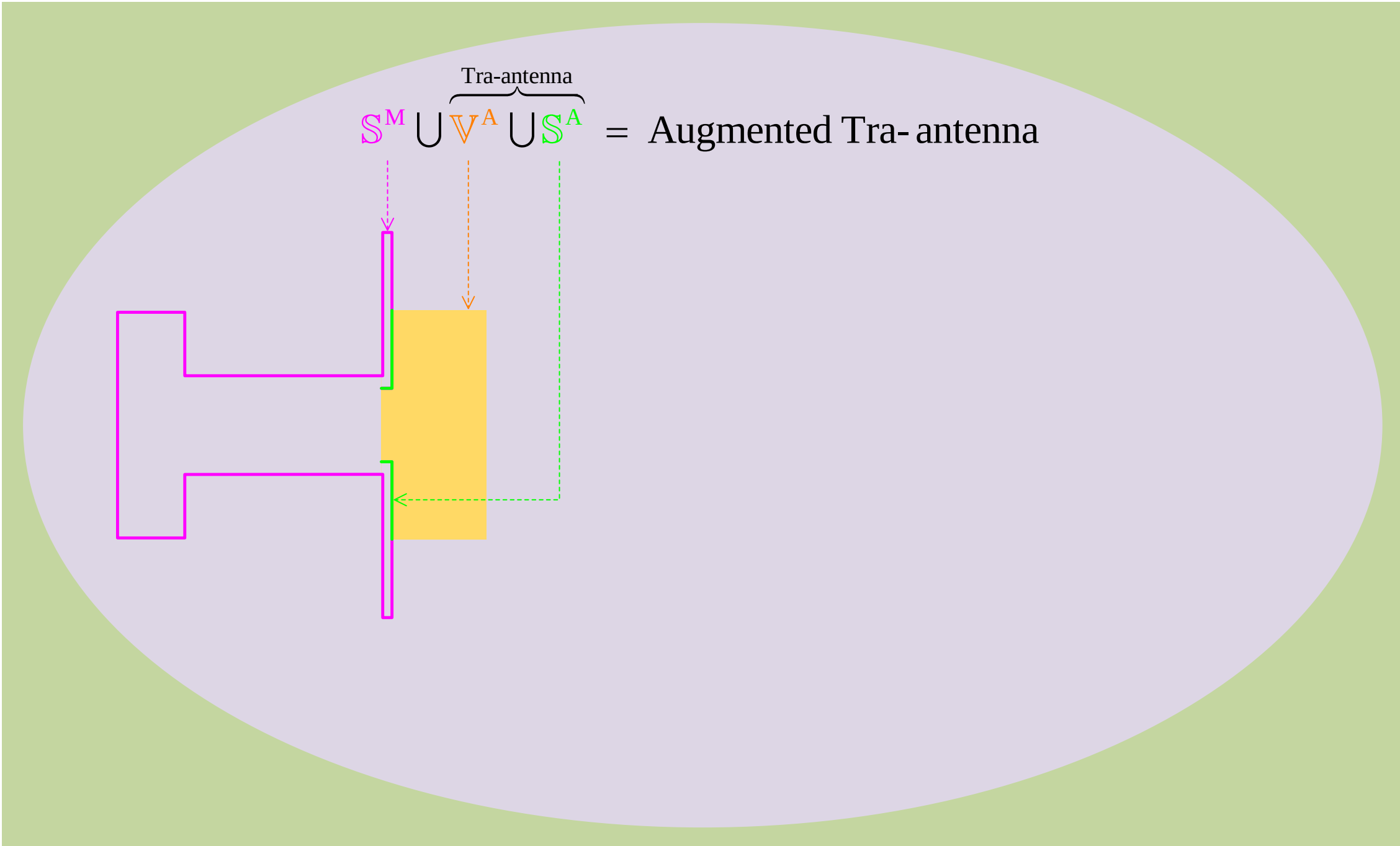

Figure 2-11 Augmented transmitting antenna

### 2.3.5 Region Integration for Receiving Antenna and Grounding Structure — Augmented Receiving Antenna

The rec-antenna $\mathbb{S}_{A} \cup \mathbb{V}_{A}$ and electric wall $\mathbb{S}_{M}$ mentioned in the previous Secs. 2.3.2 and 2.3.3 are shown in the following Fig. 2-12.

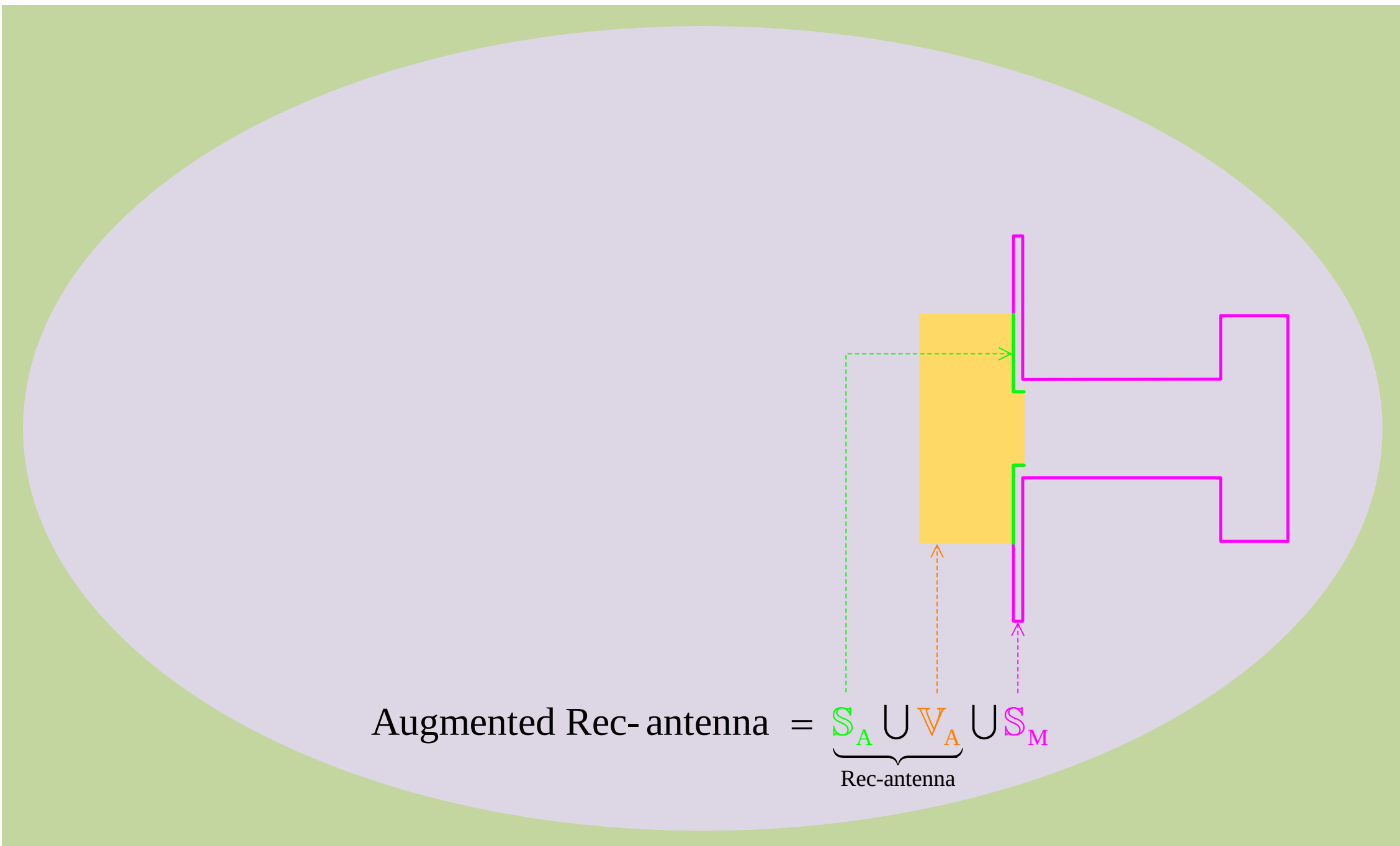

Figure 2-12 Augmented receiving antenna

In the future Secs. 2.4.1 and 2.4.2, it will also be exhibited that: **not only** $\mathbb{V}_{\rm A}$ **and** $\mathbb{S}_{\rm A}$ **have ability to modulate the energy to be inputted into receiving system, but also** $\mathbb{S}_{\rm M}$ **has the ability**. In addition, in the future Chaps. 4, 5 and 7, it will be found out that: **to treat** $\mathbb{S}_{\rm A} \bigcup \mathbb{V}_{\rm A}$ **and** $\mathbb{S}_{\rm M}$ **as a whole will bring some conveniences to the modal analysis for receiving problem**.

Based on these above, we introduce the following concept of *augmented receiving antenna* (or simply denoted as *augmented rec-antenna*)

$$\text{Augmented Rec-antenna} \;=\; \underbrace{\mathbb{S}_{\rm A} \bigcup \mathbb{V}_{\rm A}}_{\text{Rec-antenna}} \bigcup \mathbb{S}_{\rm M} \tag{2-17}$$

which is the union of the *grounding structure* $\mathbb{S}_{M}$ *related to receiving problem* (simply called *rec-ground*) and receiving antenna $\mathbb{S}_{\rm A} \bigcup \mathbb{V}_{\rm A}$, as shown in Fig. 2-12.

## 2.3.6 Section Summary

In summary, according to the division for transceiving system, whole *three-dimensional Euclidean space* can be correspondingly divided into some regions as follows:

$$\mathbb{S}^{\rm O} \bigcup \mathbb{V}^{\rm O} \bigcup \mathbb{S}^{\rm O \rightleftharpoons G} \bigcup \mathbb{S}^{\rm G} \bigcup \mathbb{V}^{\rm G} \bigcup \mathbb{S}^{\rm G \rightleftharpoons A} \bigcup \mathbb{S}^{\rm A} \bigcup \mathbb{V}^{\rm A} \bigcup \mathbb{W}^{\rm T \| M} \bigcup \mathbb{S}^{\rm A \rightleftharpoons M} \bigcup \mathbb{S}^{\rm M} \bigcup \mathbb{M} \bigcup \mathbb{S}_{\rm M} \bigcup \mathbb{S}_{\rm M \to I}^{\rm M \to I} \bigcup \mathbb{I} \bigcup \mathbb{S}_{\rm M \rightleftharpoons A} \bigcup \mathbb{W}_{\rm M \| R} \bigcup \mathbb{V}_{\rm A} \bigcup \mathbb{S}_{\rm A} \bigcup \mathbb{S}_{\rm A \rightleftharpoons G} \bigcup \mathbb{V}_{\rm G} \bigcup \mathbb{S}_{\rm G} \bigcup \mathbb{S}_{\rm G \rightleftharpoons O} \bigcup \mathbb{V}_{\rm O} \bigcup \mathbb{S}_{\rm O} \tag{2-18}$$

Transmitter $\mathbb{T}$
Feed
Generator
Tra-guide
Tra-antenna
O/I Port of $\mathbb{V}^{\rm O}/\mathbb{V}^{\rm G}$
O/I Port of $\mathbb{V}^{\rm G}/\mathbb{V}^{\rm A}$
O/I Port of $\mathbb{V}^{\rm A}/\mathbb{M}$
Environment $\mathbb{E}$
Medium
O/I Port of $\mathbb{M}/\mathbb{V}_{\rm A}$
Receiver $\mathbb{R}$
Load
Rec-antenna
Rec-guide
Absorber
O/I Port of $\mathbb{V}_{\rm A}/\mathbb{V}_{\rm G}$
O/I Port of $\mathbb{V}_{\rm G}/\mathbb{V}_{\rm O}$
Augmented Tra-antenna
Augmented Rec-antenna
Augmented Tra-guid-Tra-antenna Combined System (TGTA System)
Augmented Tra-antenna-Rec-antenna Combined System (TARA System)
Augmented Rec-antenna-Rec-guide Combined System (RARG System)

where "O/I Port" is the abbreviation for "Output/Input Port". In addition, for simplifying the symbolic system of the following discussions, we uniformly denote the material parameters of the various regions as follows:

$$\vec{\gamma}(\vec{r}) = \begin{cases} \vec{\gamma}^{\rm O}(\vec{r}) & , \quad \vec{r} \in \mathbb{V}^{\rm O} \\ \vec{\gamma}^{\rm G}(\vec{r}) & , \quad \vec{r} \in \mathbb{V}^{\rm G} \\ \vec{\gamma}^{\rm A}(\vec{r}) & , \quad \vec{r} \in \mathbb{V}^{\rm A} \\ \vec{\gamma}_{\rm M}^{\rm M}(\vec{r}) & , \quad \vec{r} \in \mathbb{M} \\ \vec{\gamma}_{\rm A}(\vec{r}) & , \quad \vec{r} \in \mathbb{V}_{\rm A} \\ \vec{\gamma}_{\rm G}(\vec{r}) & , \quad \vec{r} \in \mathbb{V}_{\rm G} \\ \vec{\gamma}_{\rm O}(\vec{r}) & , \quad \vec{r} \in \mathbb{V}_{\rm O} \end{cases} \tag{2-19}$$

where $\vec{\gamma} = \vec{\sigma} / \vec{\varepsilon} / \vec{\mu}$.

In the following Sec. 2.4, we will quantitatively discuss the flowing way of EM power in the above-mentioned various regions, by appling the PtT/WEP to the regions.

## 2.4 Power Flow in Transceiving System

In this section, by applying the PtT/WEP given in Sec. 2.2 to the regions obtained in Sec. 2.3, we provide a quantitative expression for the flowing way of the power/energy passing through the regions.

### 2.4.1 Power Flow in Transmitting System

This sub-section focuses on discussing the power/energy flow in the regions related to transmitting system.

**1. Power Generated by Generator**

The field $\vec{F}$ in region $\mathbb{V}^{\text{O}}$ satisfy the following inhomogeneous Maxwell's equations:

$$\nabla \times \vec{H} = j\omega\varepsilon_0\vec{E} + \vec{J}_{\text{imp}} \quad , \quad \vec{r} \in \mathbb{V}^{\text{O}} \tag{2-20a}$$

$$-\nabla \times \vec{E} = j\omega\mu_0\vec{H} \quad , \quad \vec{r} \in \mathbb{V}^{\text{O}} \tag{2-20b}$$

where $\vec{J}_{\text{imp}}$ is an *impressed source*, which is utilized to supply power/energy to transceiving system. Similar to the process for deriving Eq. (2-10) from Maxwell's equations (2-1), we can derive the following relation

$$P_{\text{imp}}^{\text{O}} = P_{\text{dis}}^{\text{O}} + P^{\text{O}\rightleftharpoons\text{G}} + j\, P_{\text{sto}}^{\text{O}} \tag{2-21}$$

from Maxwell's equations (2-20). In Eq. (2-21), $P_{\text{imp}}^{\text{O}}$ is the *power generated by the impressed source*, and $P_{\text{dis}}^{\text{O}}$ is the *power dissipated in generator* (which power is finally converted into *Joule heating*), and $P^{\text{O}\rightleftharpoons\text{G}}$ is the *power outputted from generator* (which power is finally inputted into tra-guide by passing through the port $\mathbb{S}^{\text{O}\rightleftharpoons\text{G}}$), and $P_{\text{sto}}^{\text{O}}$ is related to the *net energy stored in generator* (which power is finally used to contribute the EM energies stored in $\mathbb{V}^{\text{O}}$), and these powers are as follows:

$$P_{\text{imp}}^{\text{O}} = -(1/2)\left\langle \vec{J}_{\text{imp}}, \vec{E} \right\rangle_{\mathbb{V}^{\text{O}}} \tag{2-22a}$$

$$P_{\text{dis}}^{\text{O}} = (1/2)\left\langle \vec{\sigma}\cdot\vec{E}, \vec{E} \right\rangle_{\mathbb{V}^{\text{O}}} \tag{2-22b}$$

$$P^{\text{O}\rightleftharpoons\text{G}} = (1/2)\iint_{\mathbb{S}^{\text{O}\rightleftharpoons\text{G}}} \left(\vec{E}\times\vec{H}^{\dagger}\right)\cdot\hat{n}^{\rightarrow\text{G}} dS \tag{2-22c}$$

$$P_{\text{sto}}^{\text{O}} = 2\omega\left[(1/4)\left\langle \vec{H}, \vec{\mu}\cdot\vec{H} \right\rangle_{\mathbb{V}^{\text{O}}} - (1/4)\left\langle \vec{\varepsilon}\cdot\vec{E}, \vec{E} \right\rangle_{\mathbb{V}^{\text{O}}}\right] \tag{2-22d}$$

where unit vector $\hat{n}^{\rightarrow\text{G}}$ is the normal direction of port $\mathbb{S}^{\text{O}\rightleftharpoons\text{G}}$ and points to tra-guide as shown in Fig. 2-3, and $\mathbb{S}^{\text{O}\rightleftharpoons\text{G}}$ is called the *output port of generator* or the *input port of tra-guide*.

**2. Power Flow Passing Through Tra-guide**

Applying Eq. (2-10) to the region $\mathbb{V}^{\text{G}}$ occupied by tra-guide cavity, we can obtain the following relation

$$P^{\text{O}\rightleftharpoons\text{G}} = P_{\text{dis}}^{\text{G}} + P^{\text{G}\rightleftharpoons\text{A}} + j\,P_{\text{sto}}^{\text{G}} \tag{2-23}$$

In Eq. (2-23), $P^{\text{O}\rightleftharpoons\text{G}}$ (which was viewed as the *output power of generator* previously) can also be viewed as the *power inputted into tra-guide*, and $P_{\text{dis}}^{\text{G}}$ is the *power dissipated in tra-guide*, and $P^{\text{G}\rightleftharpoons\text{A}}$ is the *power outputted from tra-guide*, and $P_{\text{sto}}^{\text{G}}$ is related to the *net energy stored in tra-guide*, and these powers can be explicitly expressed as the following integrals

$$P_{\text{dis}}^{\text{G}} = (1/2)\left\langle \vec{\sigma}\cdot\vec{E},\vec{E}\right\rangle_{\mathbb{V}^{\text{G}}} \tag{2-24a}$$

$$P^{\text{G}\rightleftharpoons\text{A}} = (1/2)\iint_{\mathbb{S}^{\text{G}\rightleftharpoons\text{A}}}\left(\vec{E}\times\vec{H}^{\dagger}\right)\cdot\hat{n}^{\rightarrow\text{A}}dS \tag{2-24b}$$

$$P_{\text{sto}}^{\text{G}} = 2\omega\left[(1/4)\left\langle \vec{H},\vec{\mu}\cdot\vec{H}\right\rangle_{\mathbb{V}^{\text{G}}} - (1/4)\left\langle \vec{\varepsilon}\cdot\vec{E},\vec{E}\right\rangle_{\mathbb{V}^{\text{G}}}\right] \tag{2-24c}$$

where unit vector $\hat{n}^{\rightarrow\text{A}}$ is the normal direction of port $\mathbb{S}^{\text{G}\rightleftharpoons\text{A}}$ and points to tra-antenna as shown in Fig. 2-4, and $\mathbb{S}^{\text{G}\rightleftharpoons\text{A}}$ is called the *output port of tra-guide* or the *input port of tra-antenna*.

**3. Power Flow Passing Through Tra-antenna**

Applying Eq. (2-10) to the region $\mathbb{V}^{\text{A}}$ occupied by tra-antenna, we can obtain the following relation

$$P^{\text{G}\rightleftharpoons\text{A}} = P_{\text{dis}}^{\text{A}} + P^{\text{A}\rightleftharpoons\text{M}} + j\,P_{\text{sto}}^{\text{A}} \tag{2-25}$$

In Eq. (2-25), $P^{\text{G}\rightleftharpoons\text{A}}$ (which was viewed as the *output power of tra-guide* previously) can also be viewed as the *power inputted into tra-antenna*, and $P_{\text{dis}}^{\text{A}}$ is the *power dissipated in tra-antenna*, and $P^{\text{A}\rightleftharpoons\text{M}}$ is the *power outputted from tra-antenna*, and $P_{\text{sto}}^{\text{A}}$ is related to the *net energy stored in tra-antenna*, and these powers are as follows:

$$P_{\text{dis}}^{\text{A}} = (1/2)\left\langle \vec{\sigma}\cdot\vec{E},\vec{E}\right\rangle_{\mathbb{V}^{\text{A}}} \tag{2-26a}$$

$$P^{\text{A}\rightleftharpoons\text{M}} = (1/2)\iint_{\mathbb{S}^{\text{A}\rightleftharpoons\text{M}}}\left(\vec{E}\times\vec{H}^{\dagger}\right)\cdot\hat{n}^{\rightarrow\text{M}}dS \tag{2-26b}$$

$$P_{\text{sto}}^{\text{A}} = 2\omega\left[(1/4)\left\langle \vec{H},\vec{\mu}\cdot\vec{H}\right\rangle_{\mathbb{V}^{\text{A}}} - (1/4)\left\langle \vec{\varepsilon}\cdot\vec{E},\vec{E}\right\rangle_{\mathbb{V}^{\text{A}}}\right] \tag{2-26c}$$

where unit vector $\hat{n}^{\rightarrow\text{M}}$ is the normal direction of port $\mathbb{S}^{\text{A}\rightleftharpoons\text{M}}$ and points to propagation medium as shown in Fig. 2-6, and $\mathbb{S}^{\text{A}\rightleftharpoons\text{M}}$ is called the *output port of tra-antenna* or the *input port of medium*.

**4. Summary**

Now, we summarize the above-mentioned Eqs. (2-21), (2-23) and (2-25) related to transmitting system as follows:

$$P_{\text{imp}}^{\text{O}} = P_{\text{dis}}^{\text{O}} + \overbrace{P_{\text{dis}}^{\text{G}} + \overbrace{P_{\text{dis}}^{\text{A}} + P^{\text{A}\rightleftharpoons\text{M}} + j\,P_{\text{sto}}^{\text{A}}}^{P^{\text{G}\rightleftharpoons\text{A}}} + j\,P_{\text{sto}}^{\text{G}}}^{P^{\text{O}\rightleftharpoons\text{G}}} + j\,P_{\text{sto}}^{\text{O}} \tag{2-27}$$

and this is a quantitative expression for the flowing way of EM power/energy in whole transmitting system.

### 2.4.2 Power Flow in Surrounding Environment

Applying Eq. (2-10) to the region $\mathbb{M}$ occupied by propagation medium, we can obtain the following relation

$$P^{\text{A}\rightleftharpoons\text{M}} = P_{\text{Mdis}}^{\text{Mdis}} + \underbrace{P_{\text{M}\rightarrow\text{I}}^{\text{M}\rightarrow\text{I}}}_{P_{\text{sca}}^{\text{rad}}} + j\,P_{\text{Msto}}^{\text{Msto}} + P_{\text{M}\rightleftharpoons\text{A}} \tag{2-28}$$

In Eq. (2-28), $P^{\text{A}\rightleftharpoons\text{M}}$ (which was viewed as the *output power of tra-antenna* previously) can also be viewed as the *power inputted into propagation medium*, and $P_{\text{Mdis}}^{\text{Mdis}}$ is the *power dissipated in propagation medium*, and $P_{\text{M}\rightarrow\text{I}}^{\text{M}\rightarrow\text{I}}$ $(= P_{\text{sca}}^{\text{rad}})$ originates from both the *field radiated by transmitter* and the *field scattered by receiver*, and $P_{\text{Msto}}^{\text{Msto}}$ is related to the *net energy stored in propagation medium*, and $P_{\text{M}\rightleftharpoons\text{A}}$ is the *power inputted into rec-antenna*, and these powers are as follows:

$$P_{\text{Mdis}}^{\text{Mdis}} = (1/2)\left\langle \vec{\vec{\sigma}} \cdot \vec{E}, \vec{E} \right\rangle_{\mathbb{M}} \tag{2-29a}$$

$$P_{\text{M}\rightarrow\text{I}}^{\text{M}\rightarrow\text{I}} = (1/2)\oiint_{\mathbb{S}_{\text{M}\rightarrow\text{I}}^{\text{M}\rightarrow\text{I}}} \left(\vec{E} \times \vec{H}^{\dagger}\right) \cdot \hat{n}_{\rightarrow\text{I}}^{\rightarrow\text{I}}\, dS \tag{2-29b}$$

$$P_{\text{Msto}}^{\text{Msto}} = 2\omega\left[(1/4)\left\langle \vec{H}, \vec{\vec{\mu}} \cdot \vec{H} \right\rangle_{\mathbb{M}} - (1/4)\left\langle \vec{\vec{\varepsilon}} \cdot \vec{E}, \vec{E} \right\rangle_{\mathbb{M}}\right] \tag{2-29c}$$

$$P_{\text{M}\rightleftharpoons\text{A}} = (1/2)\iint_{\mathbb{S}_{\text{M}\rightleftharpoons\text{A}}} \left(\vec{E} \times \vec{H}^{\dagger}\right) \cdot \hat{n}_{\rightarrow\text{A}}\, dS \tag{2-29d}$$

where unit vector $\hat{n}_{\rightarrow\text{I}}^{\rightarrow\text{I}}$ is the normal direction vector of surface $\mathbb{S}_{\text{M}\rightarrow\text{I}}^{\text{M}\rightarrow\text{I}}$ and points to infinity as shown in Fig. 2-10, and $\hat{n}_{\rightarrow\text{A}}$ is the normal direction vector of port $\mathbb{S}_{\text{M}\rightleftharpoons\text{A}}$ and points to rec-antenna as shown in Fig. 2-9, and $\mathbb{S}_{\text{M}\rightleftharpoons\text{A}}$ is called the *input port of rec-antenna*.

### 2.4.3 Power Flow in Receiving System

This sub-section focuses on discussing the power/energy flow in the regions related to receiving system.

**1. Power Flow Passing Through Rec-antenna**

Applying Eq. (2-10) to the region $\mathbb{V}_{\text{A}}$ occupied by rec-antenna, we can obtain the following relation

$$P_{\text{M}\rightleftharpoons\text{A}} = P_{\text{A}}^{\text{dis}} + P_{\text{A}\rightleftharpoons\text{G}} + j\,P_{\text{A}}^{\text{sto}} \tag{2-30}$$

In Eq. (2-30), $P_{\text{M}\rightleftharpoons\text{A}}$ is the *power inputted into rec-antenna* as mentioned previously, and $P_{\text{A}}^{\text{dis}}$ is the *power dissipated in rec-antenna*, and $P_{\text{A}\rightleftharpoons\text{G}}$ is the *power outputted from rec-antenna*, and $P_{\text{A}}^{\text{sto}}$ is related to the *net energy stored in rec-antenna*, and these powers are as follows:

$$P_{\text{A}}^{\text{dis}} = (1/2)\left\langle \vec{\sigma}\cdot\vec{E},\vec{E}\right\rangle_{\mathbb{V}_{\text{A}}} \tag{2-31a}$$

$$P_{\text{A}\rightleftharpoons\text{G}} = (1/2)\iint_{\mathbb{S}_{\text{A}\rightleftharpoons\text{G}}}\left(\vec{E}\times\vec{H}^{\dagger}\right)\cdot\hat{n}_{\to\text{G}}\,dS \tag{2-31b}$$

$$P_{\text{A}}^{\text{sto}} = 2\omega\left[(1/4)\left\langle \vec{H},\vec{\mu}\cdot\vec{H}\right\rangle_{\mathbb{V}_{\text{A}}} - (1/4)\left\langle \vec{\varepsilon}\cdot\vec{E},\vec{E}\right\rangle_{\mathbb{V}_{\text{A}}}\right] \tag{2-31c}$$

where unit vector $\hat{n}_{\to\text{G}}$ is the normal direction of port $\mathbb{S}_{\text{A}\rightleftharpoons\text{G}}$ and points to rec-guide as shown in Fig. 2-9, and $\mathbb{S}_{\text{A}\rightleftharpoons\text{G}}$ is called the *output port of rec-antenna* or the *input port of rec-guide*.

**2. Power Flow Passing Through Rec-guide**

Applying Eq. (2-10) to the region $\mathbb{V}_{\text{G}}$ occupied by rec-guide cavity, we can obtain the following relation

$$P_{\text{A}\rightleftharpoons\text{G}} = P_{\text{G}}^{\text{dis}} + P_{\text{G}\rightleftharpoons\text{O}} + j\,P_{\text{G}}^{\text{sto}} \tag{2-32}$$

In Eq. (2-32), $P_{\text{A}\rightleftharpoons\text{G}}$ (which was viewed as the *output power of rec-antenna* previously) can also be viewed as the *power inputted into rec-guide*, and $P_{\text{G}}^{\text{dis}}$ is the *power dissipated in rec-guide*, and $P_{\text{G}\rightleftharpoons\text{O}}$ is the *power outputted from rec-guide*, and $P_{\text{G}}^{\text{sto}}$ is related to the *net energy stored in rec-guide*, and these powers can be mathematically expressed as follows:

$$P_{\text{G}}^{\text{dis}} = (1/2)\left\langle \vec{\sigma}\cdot\vec{E},\vec{E}\right\rangle_{\mathbb{V}_{\text{G}}} \tag{2-33a}$$

$$P_{\text{G}\rightleftharpoons\text{O}} = (1/2)\iint_{\mathbb{S}_{\text{G}\rightleftharpoons\text{O}}}\left(\vec{E}\times\vec{H}^{\dagger}\right)\cdot\hat{n}_{\to\text{O}}\,dS \tag{2-33b}$$

$$P_{\text{G}}^{\text{sto}} = 2\omega\left[(1/4)\left\langle \vec{H},\vec{\mu}\cdot\vec{H}\right\rangle_{\mathbb{V}_{\text{G}}} - (1/4)\left\langle \vec{\varepsilon}\cdot\vec{E},\vec{E}\right\rangle_{\mathbb{V}_{\text{G}}}\right] \tag{2-33c}$$

where unit vector $\hat{n}_{\to\text{O}}$ is the normal direction of port $\mathbb{S}_{\text{G}\rightleftharpoons\text{O}}$ and points to absorber as shown in Fig. 2-9, and $\mathbb{S}_{\text{G}\rightleftharpoons\text{O}}$ is called the *output port of rec-guide* or the *input port of absorber*.

**3. Power Flow Passing Into Absorber**

If the power absorbed in absorber is denoted as $P_{\mathrm{O}}^{\mathrm{abs}}$, the following relation can be derived from applying Eq. (2-10) to the region $\mathbb{V}_{\mathrm{O}}$ occupied by absorber cavity.

$$P_{\mathrm{G}\rightleftharpoons\mathrm{O}} = P_{\mathrm{O}}^{\mathrm{dis}} + P_{\mathrm{O}}^{\mathrm{abs}} + j\,P_{\mathrm{O}}^{\mathrm{sto}} \tag{2-34}$$

In the above Eq. (2-34), $P_{\mathrm{G}\rightleftharpoons\mathrm{O}}$ (which was viewed as the *output power of rec-guide* previously) can also be viewed as the *power inputted into absorber*, and $P_{\mathrm{O}}^{\mathrm{dis}}$ is the *power dissipated in absorber*, and $P_{\mathrm{O}}^{\mathrm{sto}}$ is related to the *net energy stored in absorber*, and these powers are as follows:

$$P_{\mathrm{O}}^{\mathrm{dis}} = (1/2)\left\langle \vec{\sigma}\cdot\vec{E},\vec{E}\right\rangle_{\mathbb{V}_{\mathrm{O}}} \tag{2-35a}$$

$$P_{\mathrm{O}}^{\mathrm{sto}} = 2\omega\left[(1/4)\left\langle \vec{H},\vec{\mu}\cdot\vec{H}\right\rangle_{\mathbb{V}_{\mathrm{O}}} - (1/4)\left\langle \vec{\varepsilon}\cdot\vec{E},\vec{E}\right\rangle_{\mathbb{V}_{\mathrm{O}}}\right] \tag{2-35b}$$

**4. Summary**

The above-mentioned Eqs. (2-30), (2-32) and (2-34) related to the receiving system are summarized as follows:

$$P_{\mathrm{M}\rightleftharpoons\mathrm{A}} = P_{\mathrm{A}}^{\mathrm{dis}} + \underbrace{P_{\mathrm{G}}^{\mathrm{dis}} + \underbrace{P_{\mathrm{O}}^{\mathrm{dis}} + P_{\mathrm{O}}^{\mathrm{abs}} + j\,P_{\mathrm{O}}^{\mathrm{sto}}}_{P_{\mathrm{G}\rightleftharpoons\mathrm{O}}} + j\,P_{\mathrm{G}}^{\mathrm{sto}}}_{P_{\mathrm{A}\rightleftharpoons\mathrm{G}}} + j\,P_{\mathrm{A}}^{\mathrm{sto}} \tag{2-36}$$

and this is a quantitative expression for the flowing way of power/energy in whole receiving system.

## 2.4.4 Section Summary

The relations (2-27), (2-28), and (2-36) can be uniformly written as the following *power transport theorem* (*PTT*)

$$P_{\mathrm{imp}}^{\mathrm{O}} = \left(P_{\mathrm{dis}}^{\mathrm{O}} + jP_{\mathrm{sto}}^{\mathrm{O}}\right) + \underbrace{\left(P_{\mathrm{dis}}^{\mathrm{G}} + jP_{\mathrm{sto}}^{\mathrm{G}}\right) + \underbrace{\left(P_{\mathrm{dis}}^{\mathrm{A}} + jP_{\mathrm{sto}}^{\mathrm{A}}\right) + \underbrace{\left(P_{\mathrm{Mdis}}^{\mathrm{Mdis}} + jP_{\mathrm{Msto}}^{\mathrm{Msto}}\right) + \overbrace{P_{\mathrm{M}\to\mathrm{I}}^{\mathrm{M}\to\mathrm{I}}}^{P_{\mathrm{sca}}^{\mathrm{rad}}} + \underbrace{\left(P_{\mathrm{A}}^{\mathrm{dis}} + jP_{\mathrm{A}}^{\mathrm{sto}}\right) + \underbrace{\left(P_{\mathrm{G}}^{\mathrm{dis}} + jP_{\mathrm{G}}^{\mathrm{sto}}\right) + \underbrace{\left(P_{\mathrm{O}}^{\mathrm{dis}} + jP_{\mathrm{O}}^{\mathrm{sto}}\right) + P_{\mathrm{O}}^{\mathrm{abs}}}_{P_{\mathrm{G}\rightleftharpoons\mathrm{O}}}}_{P_{\mathrm{A}\rightleftharpoons\mathrm{G}}}}_{P_{\mathrm{M}\rightleftharpoons\mathrm{A}}}}_{P^{\mathrm{A}\rightleftharpoons\mathrm{M}}}}_{P^{\mathrm{G}\rightleftharpoons\mathrm{A}}}}_{P^{\mathrm{O}\rightleftharpoons\mathrm{G}}} \tag{2-37}$$

and it has a clear physical meaning as below.

Along tra-guide, EM power/energy flows from generator to tra-antenna; with the modulations of {tra-antenna, tra-ground}, the power is released into medium; by passing through medium, the power finally reaches {receiver, infinity}; with the modulations of

{rec-ground, rec-antenna}, some powers/energies are inputted into receiver; along rec-guide, the powers flow from rec-antenna to absorber. At the same time, there also exist some EM powers/energies scattered by {receiver, medium}, and these scattered powers flow from {receiver, medium} to {transmitter, infinity} by passing through the propagation medium.

In the above processes, the carrier of EM power/energy is EM field. While carrying the power from generator to {absorber, infinity}, the field acts on {tra-guide, tra-antenna, tra-ground, medium, rec-ground, rec-antenna, rec-guide, absorber}, and the actions will lead to some induced currents, and the induced currents will correspondingly generate some EM fields. The EM fields generated by the induced currents will further react on generator; tra-guide, tra-antenna, tra-ground, medium, rec-ground, rec-antenna, rec-guide, and absorber also interact with each others by the EM fields generated by the induced currents on themselves.

Under time-harmonic excitation, the interactions among the EM fields and the structures will finally reach a *stationary state* (i.e. a *dynamic equilibrium*). At the stationary state, the EM powers/energies flowing in the structures are illustrated by the following *power flow graph*.

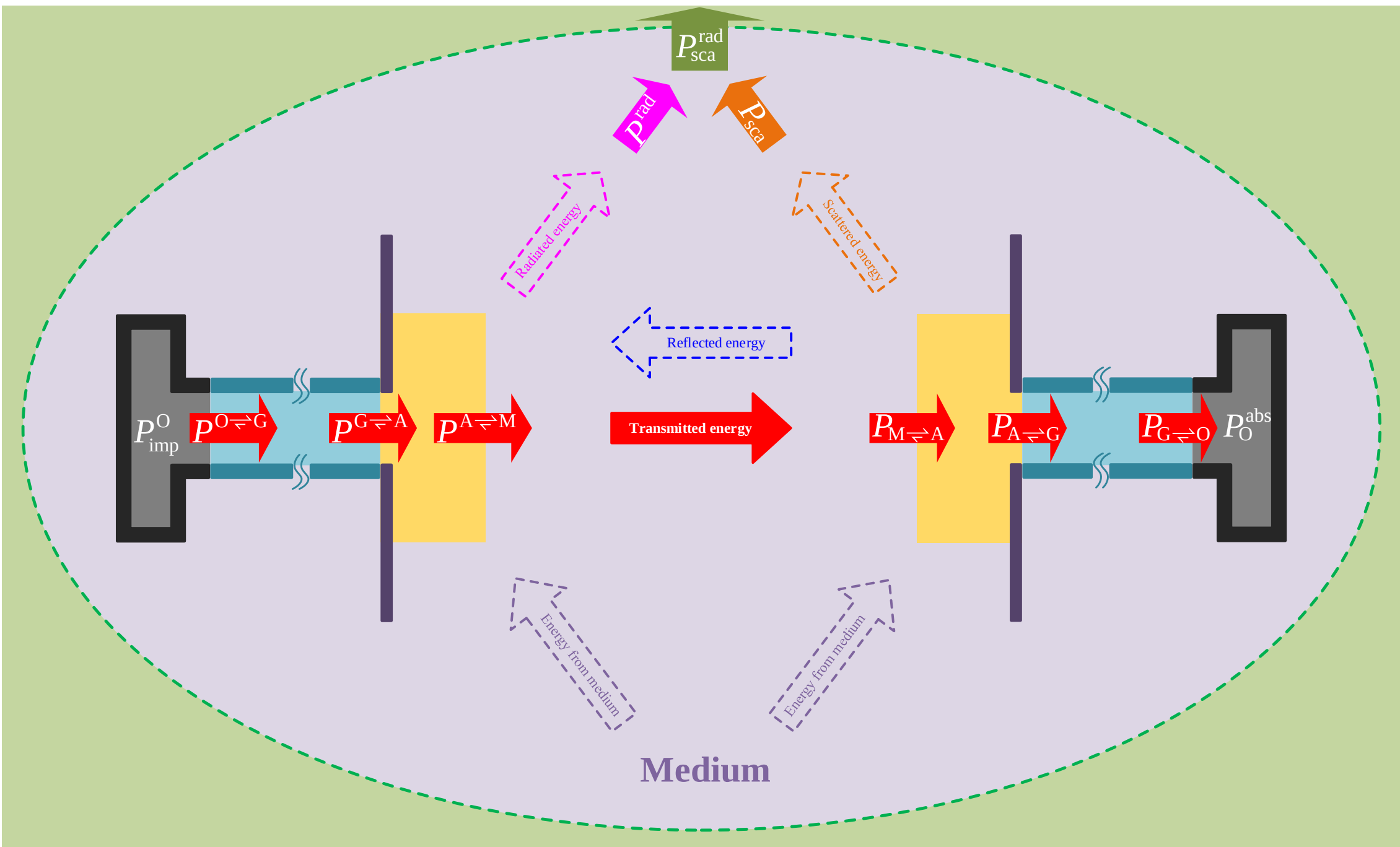

Figure 2-13 Power flow graph for the transceiving system working at transceiving state

From the above observations, it is not difficult to find out that: **on the penetrable**

**ports {$\mathbb{S}^{\text{O}\rightleftharpoons\text{G}}$, $\mathbb{S}^{\text{G}\rightleftharpoons\text{A}}$, $\mathbb{S}^{\text{A}\rightleftharpoons\text{M}}$, $\mathbb{S}_{\text{M}\rightleftharpoons\text{A}}$, $\mathbb{S}_{\text{A}\rightleftharpoons\text{G}}$, $\mathbb{S}_{\text{G}\rightleftharpoons\text{O}}$}, there also exist the powers flowing from the right to the left besides the powers flowing from the left to the right, and the powers {$P^{\text{O}\rightleftharpoons\text{G}}$, $P^{\text{G}\rightleftharpoons\text{A}}$, $P^{\text{A}\rightleftharpoons\text{M}}$, $P_{\text{M}\rightleftharpoons\text{A}}$, $P_{\text{A}\rightleftharpoons\text{G}}$, $P_{\text{G}\rightleftharpoons\text{O}}$} are the net power flows passing through the ports; there only exists the power passing through port $\mathbb{S}_{\text{M}\to\text{I}}^{\text{M}\to\text{I}}$ from medium to infinity, i.e., there doesn't exist any reflected power from infinity, because of the *Sommerfeld's radiation condition*.** In fact, these above are just the reason to utilize symbol "$\rightleftharpoons$" in the ports {$\mathbb{S}^{\text{O}\rightleftharpoons\text{G}}$, $\mathbb{S}^{\text{G}\rightleftharpoons\text{A}}$, $\mathbb{S}^{\text{A}\rightleftharpoons\text{M}}$, $\mathbb{S}_{\text{M}\rightleftharpoons\text{A}}$, $\mathbb{S}_{\text{A}\rightleftharpoons\text{G}}$, $\mathbb{S}_{\text{G}\rightleftharpoons\text{O}}$} and the powers {$P^{\text{O}\rightleftharpoons\text{G}}$, $P^{\text{G}\rightleftharpoons\text{A}}$, $P^{\text{A}\rightleftharpoons\text{M}}$, $P_{\text{M}\rightleftharpoons\text{A}}$, $P_{\text{A}\rightleftharpoons\text{G}}$, $P_{\text{G}\rightleftharpoons\text{O}}$}, and the reason to utilize symbol "$\rightarrow$" in the port $\mathbb{S}_{\text{M}\to\text{I}}^{\text{M}\to\text{I}}$ and the power $P_{\text{M}\to\text{I}}^{\text{M}\to\text{I}}$ $(= P_{\text{sca}}^{\text{rad}})$.

## 2.5 Classification for the Working States of Transmitter and Receiver

Just like dividing whole transceiving system into transmitting system and receiving system, whole *transceiving problem* can also be decomposed into two sub-problems *transmitting problem* and *receiving problem*.

The power flow process corresponding to the transmitting problem is qualitatively shown as the following power flow graph

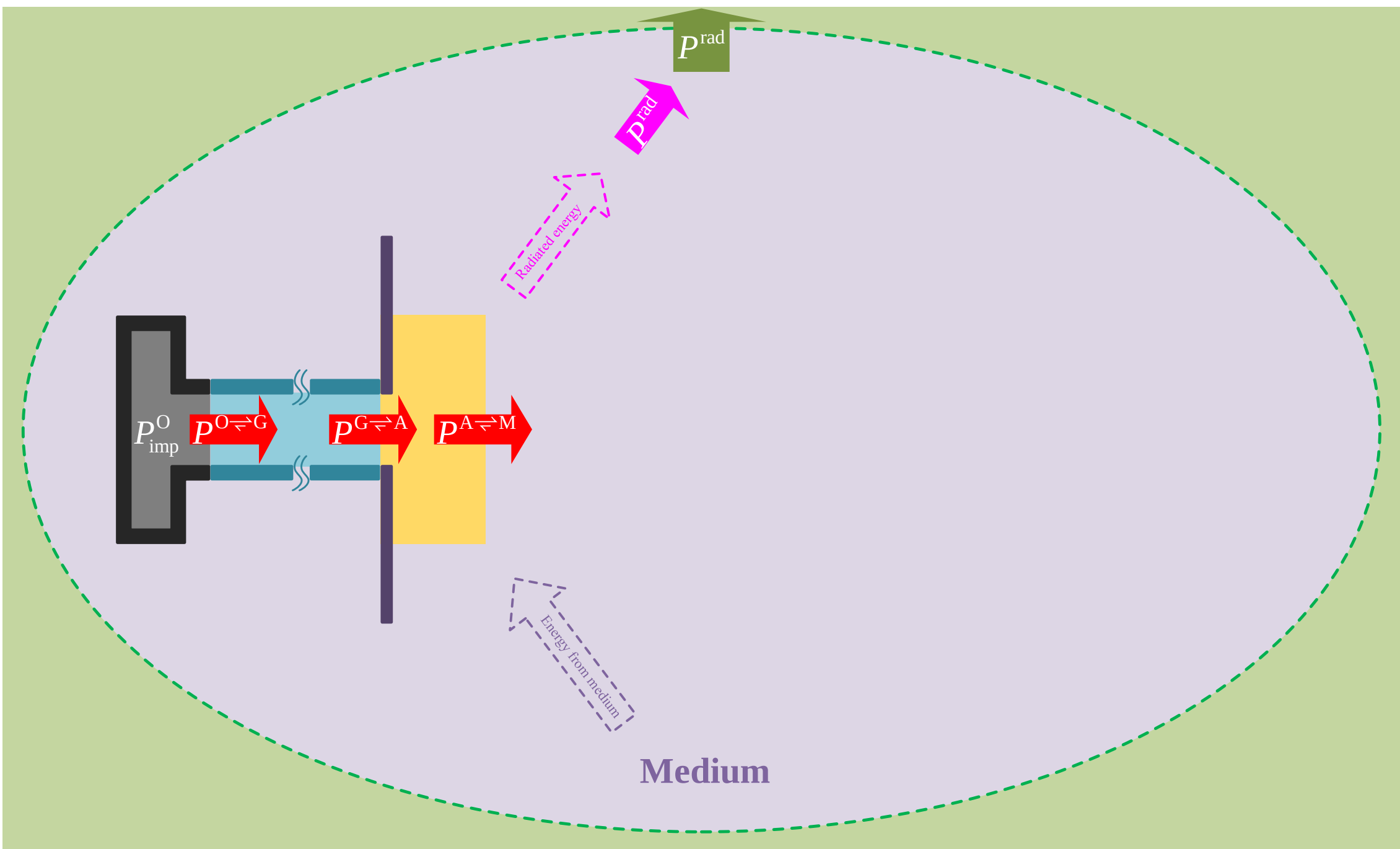

Figure 2-14 Power flow graph for the transmitting system placed in an arbitrary environment

and the power flow process corresponding to the receiving problem is qualitatively shown as the following power flow graph

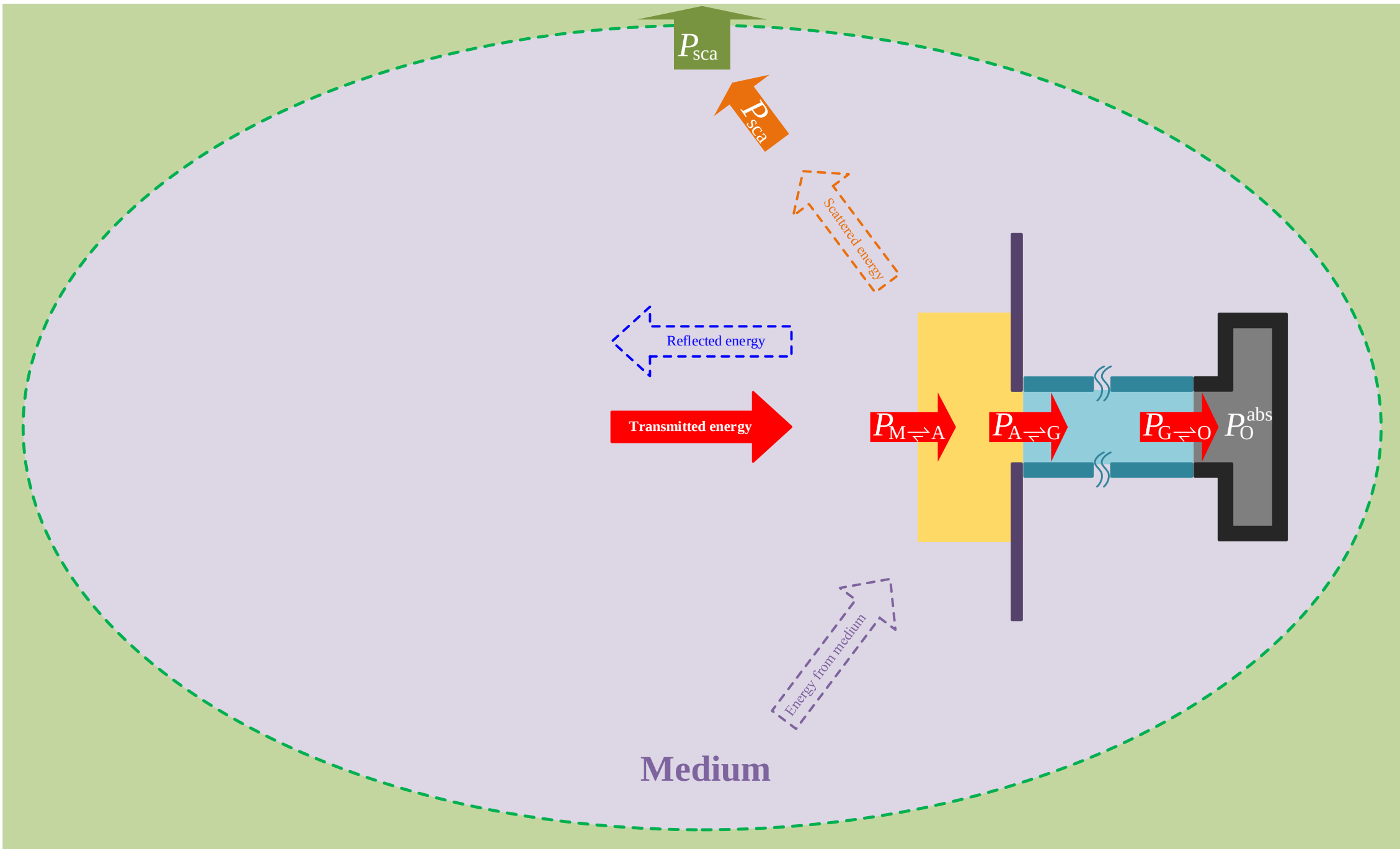

Figure 2-15 Power flow graph for the receiving system with an arbitrary load

In Fig. 2-14, if the propagation medium is *vacuum*, i.e., the transmitting system is placed in *free space*, then there will not exist any reflected power flowing from free space to transmitting system, and this conclusion can be simply proved as below. Suppose there exists a closed surface $\mathbb{S}$ enclosing whole transmitting system $\mathbb{T}$ as Fig. 2-16.

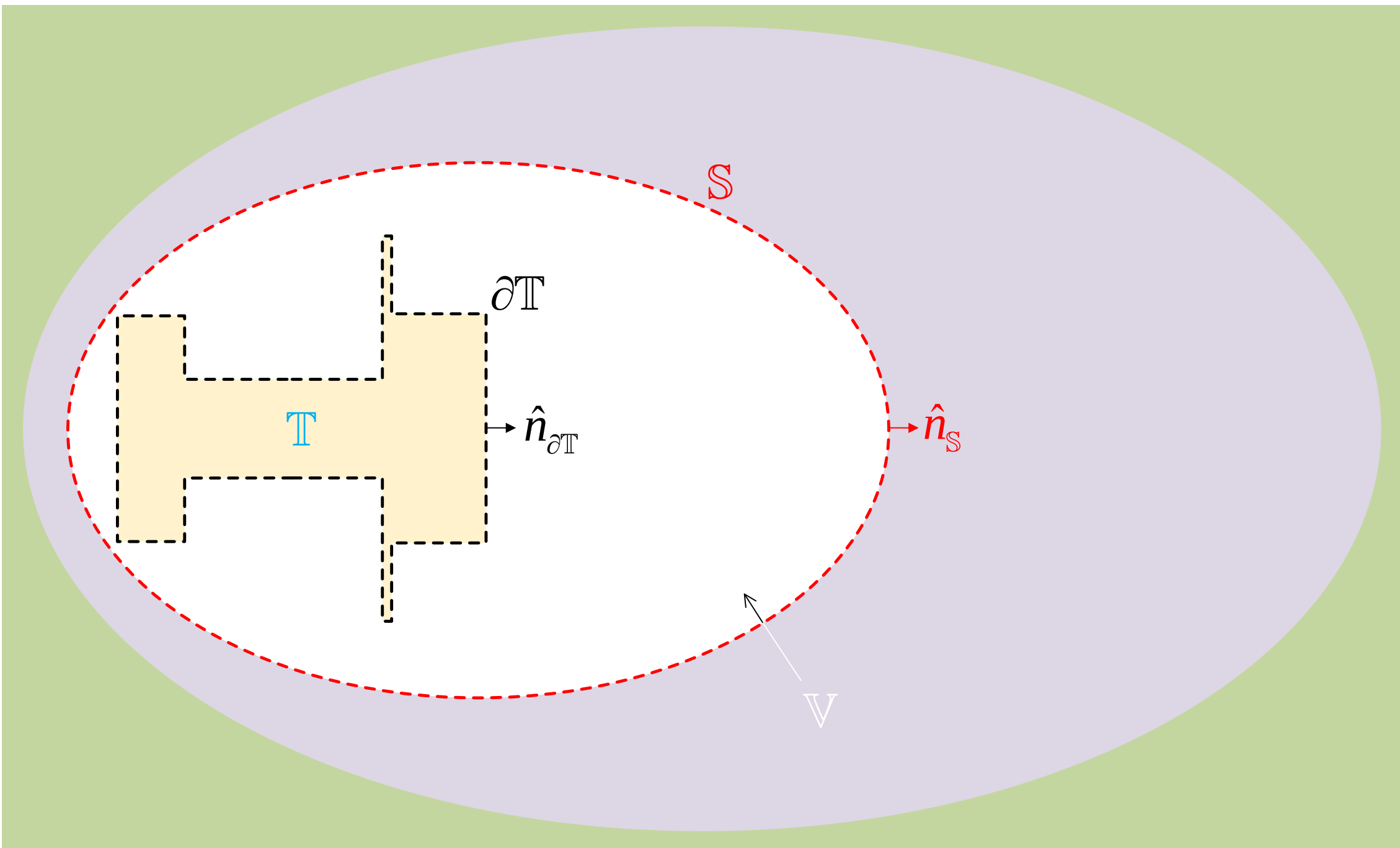

Figure 2-16 Region $\mathbb{V}$ sandwiched by two closed surfaces $\partial\mathbb{T}$ and $\mathbb{S}$

Here, $\partial\mathbb{T}$ is the boundary of $\mathbb{T}$, and $\mathbb{V}$ is the region sandwiched by $\mathbb{S}$ and $\partial\mathbb{T}$.

If the EM fields related to the transmitting problem shown in Fig. 2-16 are denoted as $\{\vec{E},\vec{H}\}$, then the fields satisfy the following homogeneous Maxwell's equations:

$$\nabla\times\vec{H} = j\omega\varepsilon_0\vec{E} \quad , \quad \vec{r}\in\mathbb{V} \tag{2-38a}$$

$$-\nabla\times\vec{E} = j\omega\mu_0\vec{H} \quad , \quad \vec{r}\in\mathbb{V} \tag{2-38b}$$

because there doesn't exist any source distributing on $\mathbb{V}$. Similar to deriving Eq. (2-10) from Maxwell's equations (2-1), the above Maxwell's equations (2-38) leads to

$$\frac{1}{2}\oiint_{\partial\mathbb{T}}\left(\vec{E}\times\vec{H}^\dagger\right)\cdot\hat{n}_{\partial\mathbb{T}}dS = \frac{1}{2}\oiint_{\mathbb{S}}\left(\vec{E}\times\vec{H}^\dagger\right)\cdot\hat{n}_{\mathbb{S}}dS + j2\omega\left[\frac{1}{4}\left\langle\vec{H},\mu_0\vec{H}\right\rangle_{\mathbb{V}} - \frac{1}{4}\left\langle\varepsilon_0\vec{E},\vec{E}\right\rangle_{\mathbb{V}}\right] \tag{2-39}$$

where $\hat{n}_{\partial\mathbb{T}}$ is the normal vector of $\partial\mathbb{T}$ and points to the interior of $\mathbb{V}$, and $\hat{n}_{\mathbb{S}}$ is the normal vector of $\mathbb{S}$ and points to the exterior of $\mathbb{V}$, as shown in Fig. 2-16. Because both $(1/4)<\vec{H},\mu_0\vec{H}>_{\mathbb{V}}$ and $(1/4)<\varepsilon_0\vec{E},\vec{E}>_{\mathbb{V}}$ are purely real, thus

$$\operatorname{Re}\left\{(1/2)\oiint_{\partial\mathbb{T}}\left(\vec{E}\times\vec{H}^\dagger\right)\cdot\hat{n}_{\partial\mathbb{T}}dS\right\} = \operatorname{Re}\left\{(1/2)\oiint_{\mathbb{S}}\left(\vec{E}\times\vec{H}^\dagger\right)\cdot\hat{n}_{\mathbb{S}}dS\right\} \tag{2-40}$$

and then

$$(1/T)\int_{t_0}^{t_0+T}\left\{\oiint_{\partial\mathbb{T}}\left(\vec{\mathcal{E}}\times\vec{\mathcal{H}}\right)\cdot\hat{n}_{\partial\mathbb{T}}dS\right\}dt = (1/T)\int_{t_0}^{t_0+T}\left\{\oiint_{\mathbb{S}}\left(\vec{\mathcal{E}}\times\vec{\mathcal{H}}\right)\cdot\hat{n}_{\mathbb{S}}dS\right\}dt \tag{2-41}$$

The Eq. (2-41) can be physically explained as that: **in any integral period there doesn't exist any net reflected power flowing from free space to transmitting system.** Thus, the power flow graph corresponding to the above special case can be illustrated as follows:

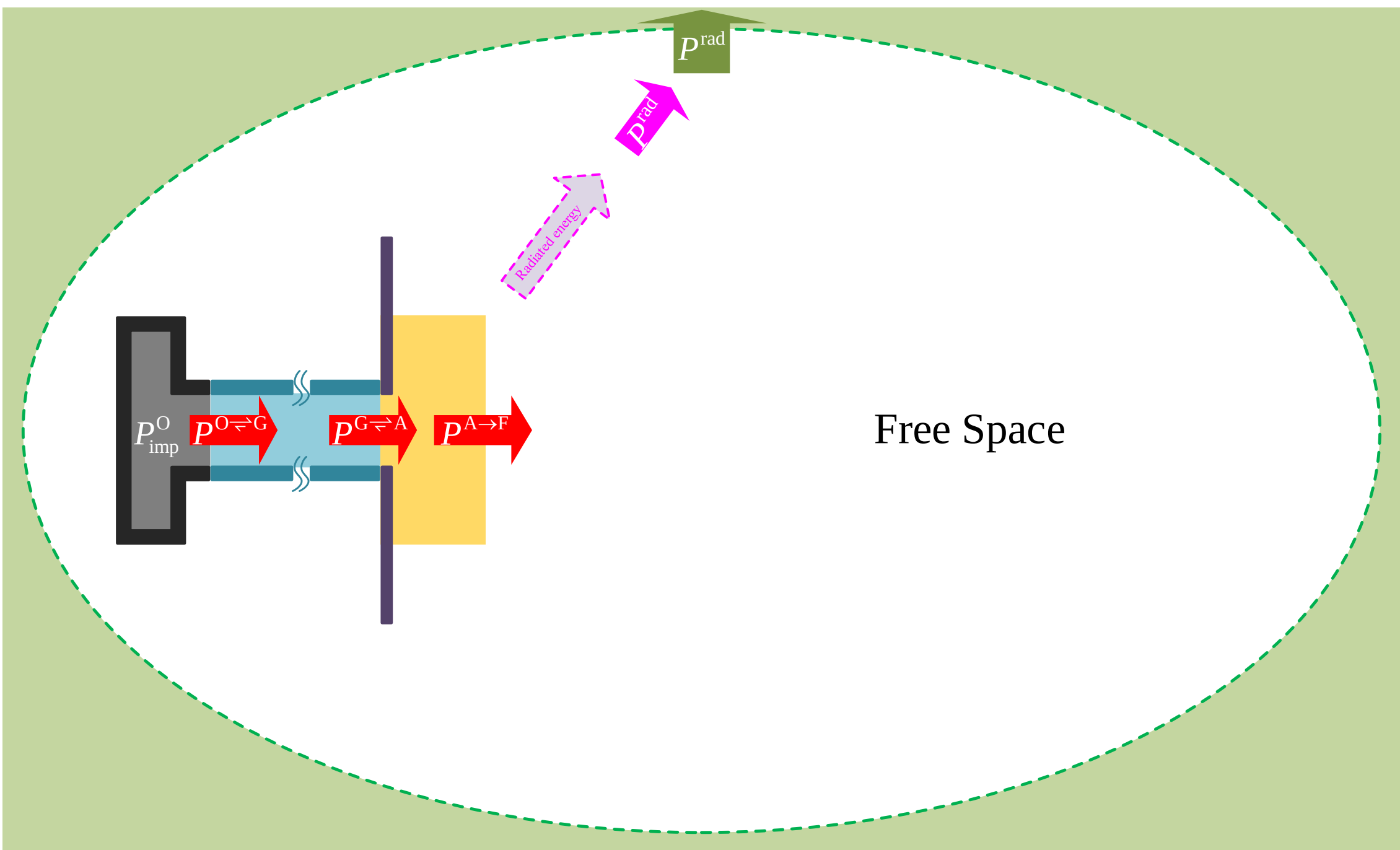

Figure 2-17 Power flow graph for the transmitting system placed in free space

where we use a white background to emphasize free space, and $P^{\mathrm{A}\rightarrow\mathrm{F}}$ is just the counterpart of the $P^{\mathrm{A}\rightleftharpoons\mathrm{M}}$ in the general environment case (where the superscript "$\mathrm{A}\rightarrow\mathrm{F}$" in $P^{\mathrm{A}\rightarrow\mathrm{F}}$ is to emphasize that there doesn't exist any power reflected from free space to tra-antenna).

Similarly, in the power flow graph shown as Fig. 2-15, if the loading system {rec-guide, port $\mathbb{S}_{\mathrm{G}\rightleftharpoons\mathrm{O}}$, absorber} works as a *perfectly matched load* (*PML*), then there will not exist any reflected power flowing from the PML to rec-antenna, and the power flow graph corresponding to this special case is illustrated as follows:

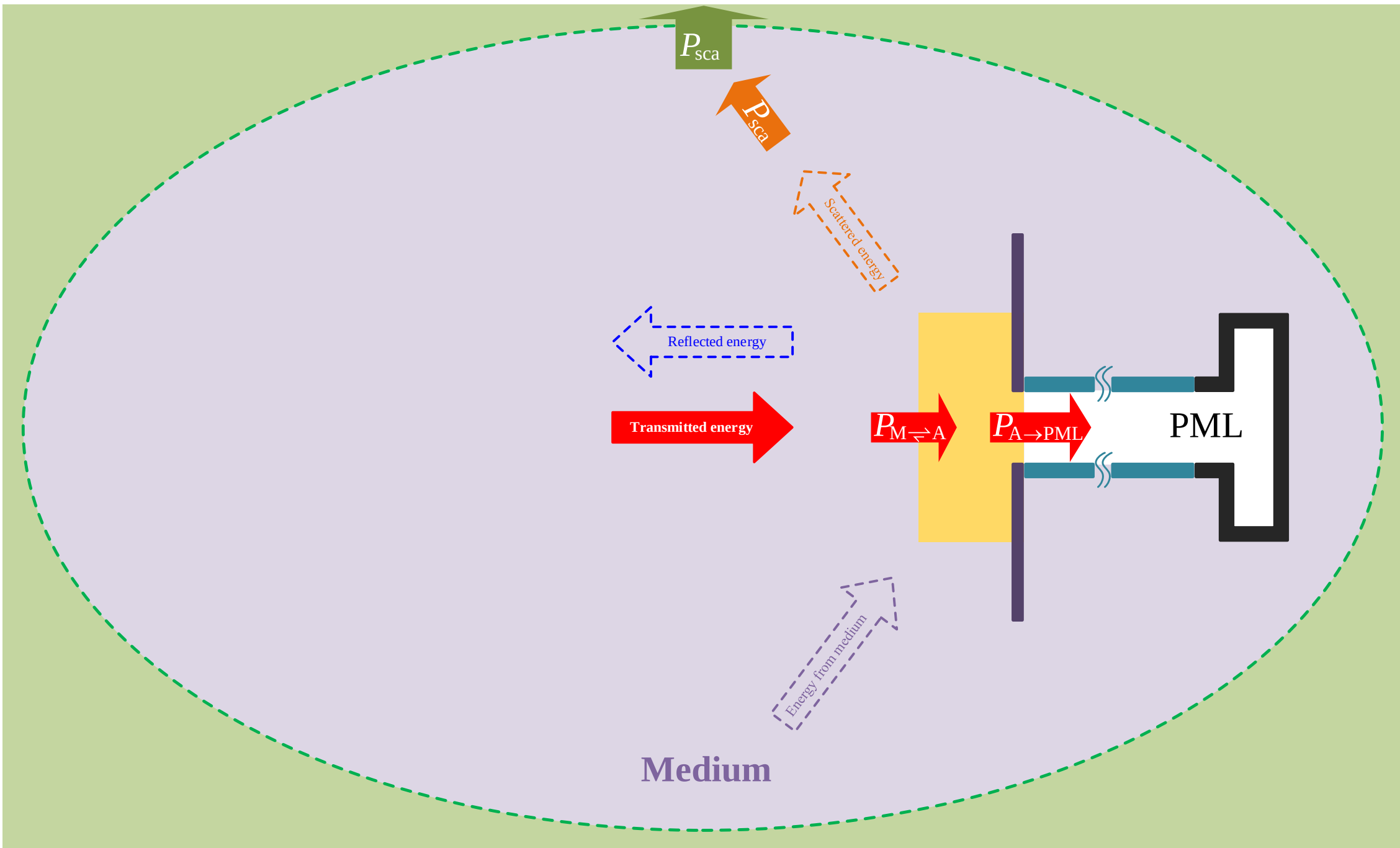

Figure 2-18 Power flow graph for the receiving system with perfectly matched load (PML)

where we use a white cavity to emphasize PML, and $P_{\mathrm{A}\rightarrow\mathrm{PML}}$ is just the counterpart of the $P_{\mathrm{A}\rightleftharpoons\mathrm{G}}$ in the general load case (where the subscript "$\mathrm{A}\rightarrow\mathrm{PML}$" in $P_{\mathrm{A}\rightarrow\mathrm{PML}}$ is to emphasize that there doesn't exist any power reflected from PML to rec-antenna).

Based on the above these as shown in the Figs. 2-14, 2-15, 2-17, and 2-18, we can observe that:

(a) the tra-antenna shown in power flow graph Fig. 2-14 works at a *compositely transmitting-receiving state*, because there simultaneously exist a power passing through it from tra-guide to medium and a power passing through it from medium to tra-guide;

(b) the rec-antenna shown in power flow graph Fig. 2-15 works at a *compositely*

*scattering-receiving-transmitting state*, because there simultaneously exist a power scattered from it to infinity, a power passing through it from medium to rec-guide, and a power passing through it from rec-guide to medium;

(c) the tra-antenna shown in power flow graph Fig. 2-17 works at a *quasi-purely transmitting state*, because there only exists a power propagating from transmitter to free space, but doesn't exist any power propagating from free space to transmitter;

(d) the rec-antenna shown in power flow graph Fig. 2-18 works at a *quasi-purely receiving state*, because there doesn't exist the power passing through it from PML to rec-antenna (though there simultaneously exist a power scattered from it to infinity and a power passing through it from medium to PML).

Due to the above observations, we call the tra-antenna working at quasi-purely transmitting state *quasi-purely transmitting antenna* or simply *quasi-pure tra-antenna*, and call the corresponding transmitting system and transmitting problem *quasi-purely transmitting system* and *quasi-purely transmitting problem* respectively; we call the rec-antenna working at quasi-purely receiving state *quasi-purely receiving antenna* or simply *quasi-pure rec-antenna*, and call the corresponding receiving system and receiving problem *quasi-purely receiving system* and *quasi-purely receiving problem* respectively.

Similarly, the guide carrying *purely traveling wave* (rather than *standing wave* or *traveling-standing wave*) is called *purely traveling-wave guide* or simply *traveling-wave guide*.

## 2.6 Chapter Summary

This chapter is the theoretical foundation of whole report. In this chapter, we finish the following works.

Firstly, we, from Maxwell's equations, derive the PtT/WEP corresponding to an arbitrary two-port region, and the PtT/WEP says that: **in any time interval, the net energy flowing into the input port of the region is transformed into four parts — a part flows out from the output port of the region, and a part is used to do the work on the conduction current in the region (this part is finally converted into Joule heating energy), and a part is used to increase the electric and magnetic field energies stored in the region, and a part is used to do the work on the polarization and magnetization currents in the region (this part is finally converted into the polarization and magnetization energies stored in the region)**. Based on the PtT/WEP, we introduce the concept of input power corresponding to the input port of the region.

Secondly, we divide whole three-dimensional Euclidean space into three regions — the region occupied by transmitter, the region occupied by environment, and the region occupied by receiver. At the same time, we divide the transmitter region into the sub-regions occupied by generator, tra-guide, tra-antenna, and tra-ground; we divide the environment region into the sub-regions occupied by medium and infinity; we divide the receiver region into the sub-regions occupied by rec-ground, rec-antenna, rec-guide, and absorber. By employing the region divisions, we also introduce the concepts of augmented tra-antenna (the union of tra-antenna and tra-ground) and augmented rec-antenna (the union of rec-ground and rec-antenna) etc., for the convenience of the following chapters.

Thirdly, by employing Poynting's theorem (PtT) / work-energy principle (WEP), we analyze the flowing processes of the EM powers in the above-mentioned regions, and derive the power transport theorem (PTT) governing the power flowing process in whole transceiving system. Then, we classify the working states of the various structures contained in transceiving system into some typical ones, such as quasi-purely transmitting state of tra-antenna and quasi-purely receiving state of rec-antenna etc.

In the following parts of this report, we will, under PTT framework, establish the *decoupling mode theory* (*DMT*) for the structures mentioned above, and, by orthogonalizing frequency-domain IPO, derive the *input-power-decoupled modes* (*IP-DMs*) distributing on the input ports of the structures. Specifically, in Chap. 3, we will construct the *traveling-wave-type IP-DMs* of various guiding structures by orthogonalizing the corresponding IPOs; in Chap. 4, we will construct the IP-DMs of tra-antenna, propagation medium, and rec-antenna by orthogonalizing the corresponding IPOs; in Chap. 5, we will discuss the *modal matching problem* on the interfaces among tra-guide, tra-antenna, medium, and rec-antenna, etc.; in Chap. 6, we will construct the IP-DMs of augmented tra-antenna by orthogonalizing the corresponding IPO; in Chap. 7, we will construct the IP-DMs of augmented rec-antenna by orthogonalizing the corresponding IPO; in Chap. 8, we will construct the IP-DMs of *augmented tra-guide-tra-antenna combined system* (*TGTA system*) and *augmented tra-antenna-rec-antenna combined system* (*TARA system*) by orthogonalizing the corresponding IPOs; in Chap. 9, the *modal decomposition* and a novel concept of *electric-magnetic decoupling factor* will be discussed by employing the obtained IP-DMs.

# Chapter 3 Input-Power-Decoupled Modes of Guiding Structure

**CHAPTER MOTIVATION:** This chapter is devoted to constructing the traveling-wave-type *input-power-decoupled modes* (*IP-DMs*) of various guiding structures by orthogonalizing the frequecny-domain *input power operator* (*IPO*) in *power transport theorem* (*PTT*) framework, and doing some necessary analysis and discussions for the related topics. The obtained IP-DMs don't have net energy exchange in integral period.

## 3.1 Chapter Introduction

As a type of effective structure used to guide the directional propagation of electromagnetic (EM) energy, *wave-guiding structure* (simply called *guiding structure* or *guide*) widely exists in the various EM systems, such as *transmitting system* and *receiving system* etc. The guiding structure can be classified into *metallic guiding structure* (such as a metallic tube), *material guiding structure* (such as *optical fiber*), and *metal-material composite guiding structure* (such as *microstrip transmission line*) [1~3,45].

Under classical *Sturm-Liouville theory* (*SLT*)[4] framework, the *eigen-modes* working on metallic guiding structures can be effectively constructed by using famous *eigen-mode theory* (*EMT*)[1~3], and the eigen-modes are usually energy-decoupled. The *SLT-based EMT* has been widely used to analyze and design various metallic guiding structures[1~3,45], but it cannot be easily generalized to antenna structures.

To resolve the above problem, this chapter, under *power transport theorem* (*PTT*) framework, establishes an alternative modal theory — *decoupling mode theory* (*DMT*) — for various guiding structures, and the theory will be further generalized to various antenna structures in future, such that the various sub-structures contained in *transceiving system* can be analyzed and designed by an uniform modal theory — *PTT-based DMT* (*PTT-DMT*). In addition, the fundamental modes constructed by PTT-DMT are always energy-decoupled, so the modes are called *energy-decoupled modes* (*DMs*) in this report.

## 3.2 IP-DMs of Metallic Guiding Structure

This section focuses on *uniform cylindrical metallic waveguides*. For the convenience of the following discussions, the terminology "uniform cylindrical metallic waveguide" is simplified to "*metallic guide*" in this section.

### 3.2.1 Preliminaries

In this subsection, some important conclusions related to metallic guides are discussed for the preparations of establishing the *PTT-based DMT for various metallic guides* (*PTT-MetGuid-DMT*). The geometry of a typical metallic guide is shown in the following Fig. 3-1.

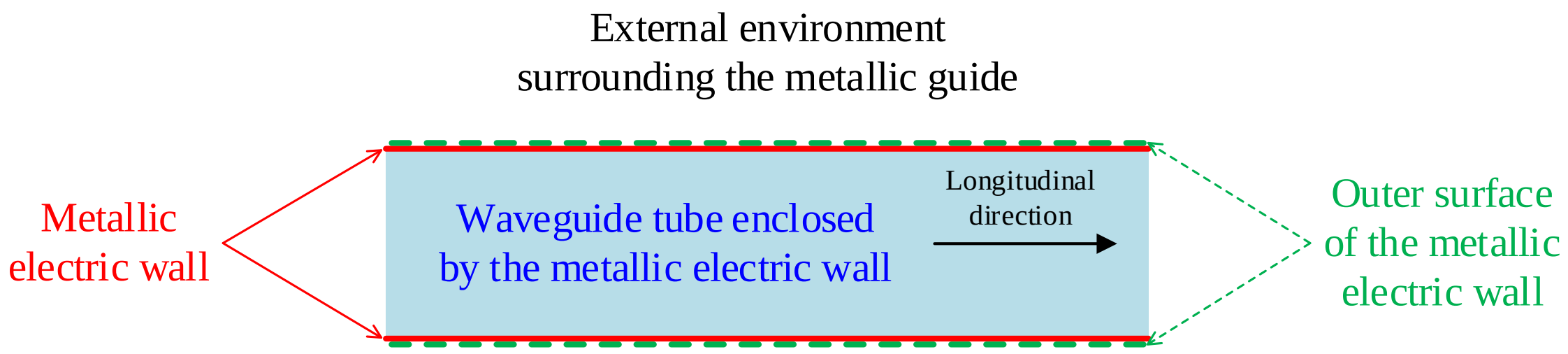

Figure 3-1 Geometry of a typical uniform cylindrical metallic waveguide

The topological structure of the metallic guide shown in Fig. 3-1 is illustrated in the following Fig. 3-2.

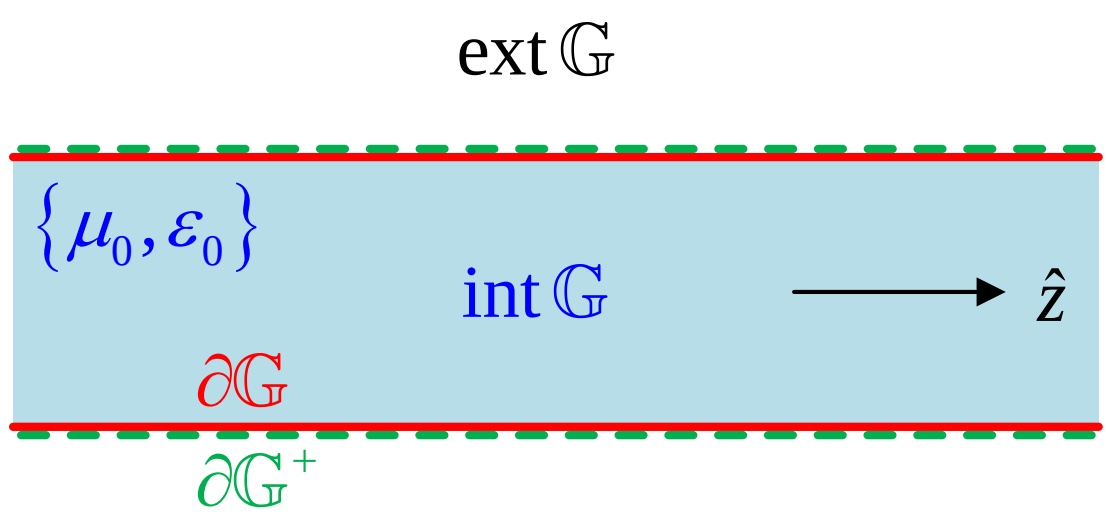

Figure 3-2 Topological structure of the metallic guide shown in Fig. 3-1

In the above Fig. 3-2, $\partial\mathbb{G}$ is the *metallic electric wall* used to separate the *guide tube* from *external environment*; $\partial\mathbb{G}^+$ is the outer surface of $\partial\mathbb{G}$; $\text{int}\,\mathbb{G}$ is the guide tube enclosed by $\partial\mathbb{G}$; $\text{ext}\,\mathbb{G}$ is the external environment surrounding the guide; the *longitudinal direction* of the guide is selected as Z-axis, and the direction vector of Z-axis is denoted as $\hat{z}$; parameters $\mu_0$ and $\varepsilon_0$ are respectively the *permeability* and *permittivity* of free space.

#### 3.2.1.1 Externally Shielding Effect of Metallic Guide

The metallic guide has *externally shielding effect* for the modes working in its tube, i.e., the modal fields $\vec{F}$ satisfy the following relation

$$\vec{F}\left(\vec{r}\right) = 0 \quad , \quad \vec{r} \in \text{ext}\,\mathbb{G} \tag{3-1}$$

where $\vec{F} = \vec{E} / \vec{H}$, and the region $\text{ext}\,\mathbb{G}$ is shown in Fig. 3-2.

In fact, this externally shielding effect originates from the following homogeneous tangential boundary conditions of the modal fields.

The tangential components of $\vec{E}$ are 0 on $\partial\mathbb{G}$. (3-2a)

The tangential components of $\vec{H}$ are 0 on $\partial\mathbb{G}^+$. (3-2b)

where the surfaces $\partial\mathbb{G}$ and $\partial\mathbb{G}^+$ are shown in Fig. 3-2.

#### 3.2.1.2 PTT Satisfied by the Modes Working in Metallic Guide

Now, we consider a section of the whole guide, and the topological structure of the section is shown in the following Fig. 3-3.

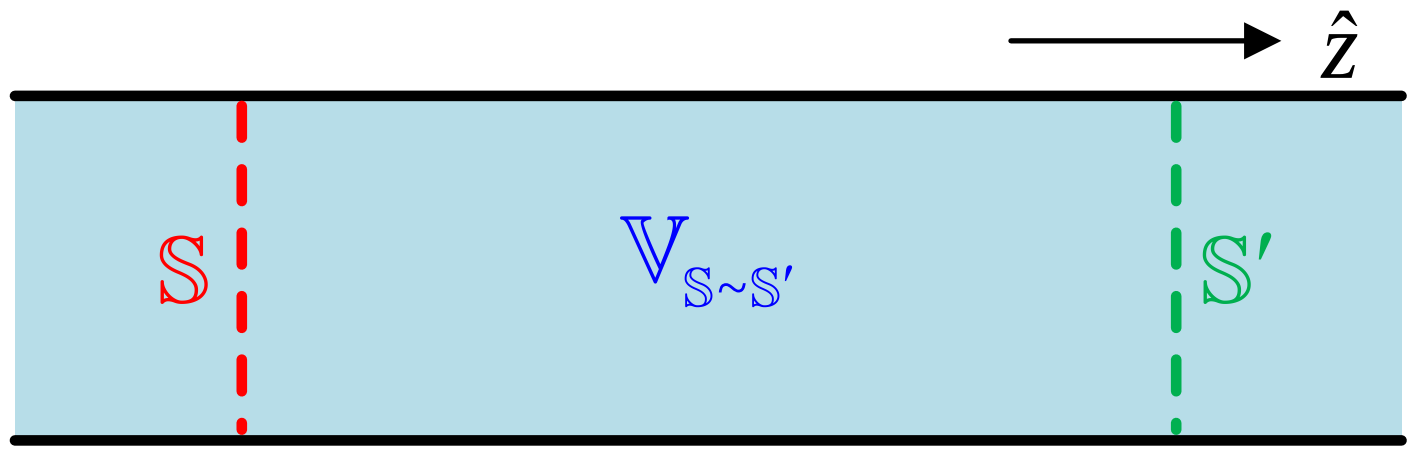

Figure 3-3 A section of metallic guide tube $\mathbb{V}_{\mathbb{S}\sim\mathbb{S}'}$ sandwiched between ports $\mathbb{S}$ and $\mathbb{S}'$

In Fig. 3-3, $\mathbb{S}$ and $\mathbb{S}'$ are respectively the *input port* and *output port* of the section, and $\mathbb{V}_{\mathbb{S}\sim\mathbb{S}'}$ is the region sandwiched between the ports.

For any mode working in the metallic guide tube, its modal fields $\{\vec{E},\vec{H}\}$ satisfy the following homogeneous *Maxwell's equations*:

$$\nabla\times\vec{H}(\vec{r}) = j\omega\varepsilon_0\vec{E}(\vec{r}) \quad , \quad \vec{r}\in\mathbb{V}_{\mathbb{S}\sim\mathbb{S}'} \tag{3-3a}$$

$$-\nabla\times\vec{E}(\vec{r}) = j\omega\mu_0\vec{H}(\vec{r}) \quad , \quad \vec{r}\in\mathbb{V}_{\mathbb{S}\sim\mathbb{S}'} \tag{3-3b}$$

Thus there exists the following *power transport theorem* (*PTT*):

$$\frac{1}{2}\iint_{\mathbb{S}}\left(\vec{E}\times\vec{H}^\dagger\right)\cdot\hat{z}dS = \frac{1}{2}\iint_{\mathbb{S}'}\left(\vec{E}\times\vec{H}^\dagger\right)\cdot\hat{z}dS + j2\omega\left[\frac{1}{4}\left\langle\vec{H},\mu_0\vec{H}\right\rangle_{\mathbb{V}_{\mathbb{S}\sim\mathbb{S}'}} - \frac{1}{4}\left\langle\varepsilon_0\vec{E},\vec{E}\right\rangle_{\mathbb{V}_{\mathbb{S}\sim\mathbb{S}'}}\right] \tag{3-4}$$

based on the results obtained in the previous Sec. 2.2, where the superscript "†" is the *complex conjugate operation* for a field or current, and the *inner product* is defined as the one used in the previous Chaps. 1 and 2.

Power transport theorem (3-4) implies that: **the modal energy flow passing through input port $\mathbb{S}$ is divided into two parts, and**

**Part I further passes through the output port $\mathbb{S}'$, and**

**Part II is used to contribute the increment of the EM field energies stored in the region $\mathbb{V}_{\mathbb{S}\sim\mathbb{S}'}$.**

Based on this observation and considering of the dimensions of terms $(1/2)\iint_{\mathbb{S}}(\vec{E}\times\vec{H}^{\dagger})\cdot\hat{z}dS$ and $(1/2)\iint_{\mathbb{S}'}(\vec{E}\times\vec{H}^{\dagger})\cdot\hat{z}dS$ , the first term is called the *input power inputted into input port* $\mathbb{S}$ , and is correspondingly denoted as $P_{\mathbb{S}}^{\text{in}}$ ; the second term is called the *output power outputted from output port* $\mathbb{S}'$ , and is correspondingly denoted as $P_{\mathbb{S}'}^{\text{out}}$ .

### 3.2.1.3 Traveling-wave Condition of the Modes Working in Metallic Guide

Now, we consider the metallic guide shown in the following Fig. 3-4. In the figure, $\mathbb{V}_{-\infty\sim z}$ and $\mathbb{V}_{z\sim+\infty}$ are respectively the left and right parts of the guide tube divided by a port $\mathbb{S}$ located at $z$ , and $\mathbb{S}$ is just the common boundary surface of regions $\mathbb{V}_{-\infty\sim z}$ and $\mathbb{V}_{z\sim+\infty}$ (i.e. the interface between $\mathbb{V}_{-\infty\sim z}$ and $\mathbb{V}_{z\sim+\infty}$ ); $\mathbb{S}_{-\infty\sim z}$ and $\mathbb{S}_{z\sim+\infty}$ are the metallic boundaries (i.e. the electric walls) of $\mathbb{V}_{-\infty\sim z}$ and $\mathbb{V}_{z\sim+\infty}$ respectively.

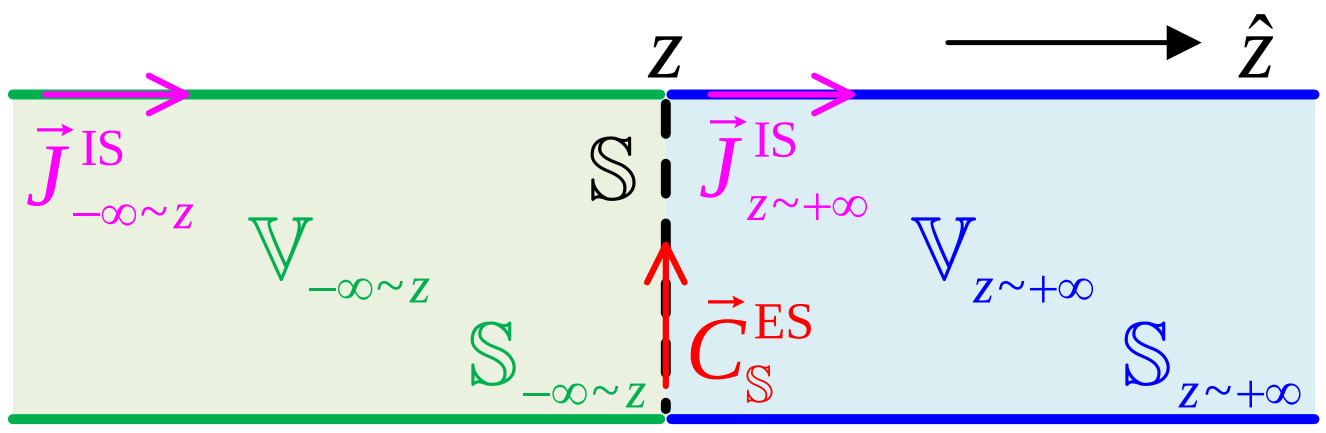

Figure 3-4 Equivalent current $\vec{C}_{\mathbb{S}}^{\text{ES}}$ distributing on a cross section of guide tube and induced currents $\vec{J}_{-\infty\sim z}^{\text{IS}}$ & $\vec{J}_{z\sim+\infty}^{\text{IS}}$ distributing on guide walls. In this figure, $\vec{C}=\vec{J}/\vec{M}$

In addition, $\vec{J}_{-\infty\sim z}^{\text{IS}}$ and $\vec{J}_{z\sim+\infty}^{\text{IS}}$ are the *induced surface electric currents* distributing on $\mathbb{S}_{-\infty\sim z}$ and $\mathbb{S}_{z\sim+\infty}$ respectively, where superscript "IS" is the acronym of "induced surface (electric current)"; $\{\vec{J}_{\mathbb{S}}^{\text{ES}},\vec{M}_{\mathbb{S}}^{\text{ES}}\}$ are the *equivalent surface currents* distributing on $\mathbb{S}$, where superscript "ES" is the acronym of "equivalent surface (current)". If the total modal electric and magnetic fields distributing on the guide tube are denoted as $\{\vec{E},\vec{H}\}$, then the currents $\{\vec{J}_{\mathbb{S}}^{\text{ES}},\vec{M}_{\mathbb{S}}^{\text{ES}}\}$ and fields $\{\vec{E},\vec{H}\}$ satisfy the following relations:

$$\vec{J}_{\mathbb{S}}^{\text{ES}}(\vec{r}) = \hat{z}\times\vec{H}(\vec{r}) \quad , \quad \vec{r}\in\mathbb{S} \tag{3-5a}$$

$$\vec{M}_{\mathbb{S}}^{\text{ES}}(\vec{r}) = \vec{E}(\vec{r})\times\hat{z} \quad , \quad \vec{r}\in\mathbb{S} \tag{3-5b}$$

based on the results given in the App. A of this report. In fact, these above relations are just the definitions for equivalent currents $\vec{J}_{\mathbb{S}}^{\text{ES}}$ and $\vec{M}_{\mathbb{S}}^{\text{ES}}$.

For any working mode (not restricted to the *traveling-wave mode*) in the metallic guide, if the fields generated by $\vec{J}_{-\infty}^{\text{imp}}+\vec{J}_{-\infty\sim z}^{\text{IS}}$ and $\vec{J}_{z\sim+\infty}^{\text{IS}}$ are denoted as $\vec{F}_{-\infty\sim z}$ and $\vec{F}_{z\sim+\infty}$ respectively (where $\vec{J}_{-\infty}^{\text{imp}}$ is the impressed source located as $z=-\infty$ ), then there exists the following *surface equivalence principle*:

$$\left.\begin{array}{l}\mathbb{V}_{-\infty\sim z}: -\vec{F}_{z\sim+\infty}\\ \mathbb{V}_{z\sim+\infty}: \quad \vec{F}_{-\infty\sim z}\end{array}\right\} = \iint_{\mathbb{S}} \vec{G}_0^{JF}(\vec{r},\vec{r}')\cdot\vec{J}_{\mathbb{S}}^{\mathrm{ES}}(\vec{r}')dS' + \iint_{\mathbb{S}} \vec{G}_0^{MF}(\vec{r},\vec{r}')\cdot\vec{M}_{\mathbb{S}}^{\mathrm{ES}}(\vec{r}')dS'$$
$$= \mathcal{F}_0\left(\vec{J}_{\mathbb{S}}^{\mathrm{ES}},\vec{M}_{\mathbb{S}}^{\mathrm{ES}}\right) \quad (3\text{-}6)$$

due to Eqs. (3-2)&(3-5) and the results given in the App. A of this report. In Eq. (3-6), $\vec{F}=\vec{E}/\vec{H}$, and correspondingly $\mathcal{F}_0=\mathcal{E}_0/\mathcal{H}_0$, and operators $\mathcal{E}_0$ and $\mathcal{H}_0$ are defined as

$$\mathcal{E}_0\left(\vec{J},\vec{M}\right) = -j\omega\mu_0\mathcal{L}_0\left(\vec{J}\right)-\mathcal{K}_0\left(\vec{M}\right) \quad (3\text{-}7a)$$

$$\mathcal{H}_0\left(\vec{J},\vec{M}\right) = -j\omega\varepsilon_0\mathcal{L}_0\left(\vec{M}\right)+\mathcal{K}_0\left(\vec{J}\right) \quad (3\text{-}7b)$$

in which

$$\mathcal{L}_0\left(\vec{X}\right) = \left(1+\frac{1}{k_0^2}\nabla\nabla\cdot\right)\int_{\Pi} G_0(\vec{r},\vec{r}')\vec{X}(\vec{r}')d\Pi' \quad (3\text{-}8a)$$

$$\mathcal{K}_0\left(\vec{X}\right) = \nabla\times\int_{\Pi} G_0(\vec{r},\vec{r}')\vec{X}(\vec{r}')d\Pi' \quad (3\text{-}8b)$$

where $G_0(\vec{r},\vec{r}')=e^{-jk_0|\vec{r}-\vec{r}'|}/4\pi|\vec{r}-\vec{r}'|$ is the *free-space scalar Green's function*, and $k_0=\omega\sqrt{\mu_0\varepsilon_0}$ is the *free-space wave number*. In summary,

$$\mathcal{F}_0\left(\vec{J}_{\mathbb{S}}^{\mathrm{ES}},\vec{M}_{\mathbb{S}}^{\mathrm{ES}}\right) = \begin{cases} -\mathcal{F}_0\left(\vec{J}_{z\sim+\infty}^{\mathrm{IS}},0\right) & = -\vec{F}_{z\sim+\infty}(\vec{r}) \quad , \quad \vec{r}\in\mathbb{V}_{-\infty\sim z} \\ \mathcal{F}_0\left(\vec{J}_{-\infty\sim z}^{\mathrm{IS}}+\vec{J}_{-\infty}^{\mathrm{imp}},0\right) & = \vec{F}_{-\infty\sim z}(\vec{r}) \quad , \quad \vec{r}\in\mathbb{V}_{z\sim+\infty} \end{cases} \quad (3\text{-}9)$$

In addition, it is obvious that the summation of $\vec{F}_{-\infty\sim z}$ and $\vec{F}_{z\sim+\infty}$ is juts the whole modal field $\vec{F}$, i.e., $\vec{F}=\vec{F}_{-\infty\sim z}+\vec{F}_{z\sim+\infty}$.

For a traveling-wave mode working in the guide, the induced current $\vec{J}_{z\sim+\infty}^{\mathrm{IS}}$ distributing on $\mathbb{S}_{z\sim+\infty}$ will not contribute to the field distributing in region $\mathbb{V}_{-\infty\sim z}$, i.e.,

$$\mathcal{F}_0\left(\vec{J}_{z\sim+\infty}^{\mathrm{IS}},0\right) = \vec{F}_{z\sim+\infty}(\vec{r}) = 0 \quad , \quad \vec{r}\in\mathbb{V}_{-\infty\sim z} \quad (3\text{-}10)$$

Thus, the traveling-wave modes in guide tube satisfy the following surface equivalence principle

$$\mathcal{F}_0\left(\vec{J}_{\mathbb{S}}^{\mathrm{ES}},\vec{M}_{\mathbb{S}}^{\mathrm{ES}}\right) = \begin{cases} -\mathcal{F}_0\left(\vec{J}_{z\sim+\infty}^{\mathrm{IS}},0\right) & = -\vec{F}_{z\sim+\infty}(\vec{r}) = 0 \quad , \quad \vec{r}\in\mathbb{V}_{-\infty\sim z} \\ \mathcal{F}_0\left(\vec{J}_{-\infty\sim z}^{\mathrm{IS}}+\vec{J}_{-\infty}^{\mathrm{imp}},0\right) & = \vec{F}_{-\infty\sim z}(\vec{r}) \quad , \quad \vec{r}\in\mathbb{V}_{z\sim+\infty} \end{cases} \quad (3\text{-}11)$$

and this formulation is particularly called the *traveling-wave condition* of the modes working in the guide tube. **In fact, the traveling-wave condition is just a counterpart of famous *Sommerfeld's radiation condition*.**

### 3.2.1.4 Some Corollaries of Traveling-wave Condition

In this sub-section, we will consider the metallic guide shown in the following Fig. 3-5, and focus on the traveling-wave modes (whose *modal waveguide wavelength along Z-direction* are denoted as $\lambda_z$).

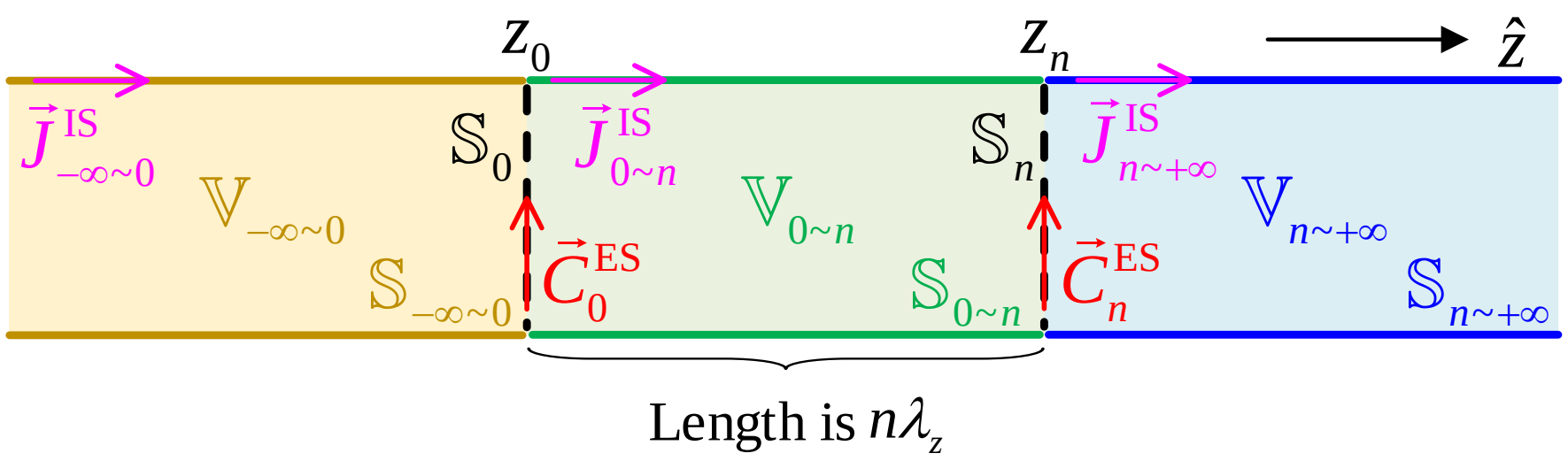

Figure 3-5 Equivalent currents distributing on two cross sections $\mathbb{S}_0 \,\&\, \mathbb{S}_n$ of guide tube and induced electric currents distributing on guide walls $\mathbb{S}_{-\infty\sim0} \cup \mathbb{S}_{0\sim n} \cup \mathbb{S}_{n\sim+\infty}$

In the figure, the $\mathbb{S}_0$ located at $z_0$ and the $\mathbb{S}_n$ located at $z_n$ are two pre-selected cross sections, such that $z_n - z_0 = n\lambda_z$ (where $n$ is a positive integer, i.e., $n \in \mathbb{N}$). The $\mathbb{S}_0$ and $\mathbb{S}_n$ divide whole guide tube into three parts $\mathbb{V}_{-\infty\sim0}$, $\mathbb{V}_{0\sim n}$, and $\mathbb{V}_{n\sim+\infty}$, and the electric walls of the three parts are denoted as $\mathbb{S}_{-\infty\sim0}$, $\mathbb{S}_{0\sim n}$, and $\mathbb{S}_{n\sim+\infty}$ respectively. The induced currents distributing on the electric walls are denoted as $\vec{J}^{\rm IS}_{-\infty\sim0}$, $\vec{J}^{\rm IS}_{0\sim n}$, and $\vec{J}^{\rm IS}_{n\sim+\infty}$ correspondingly, and the currents $\vec{J}^{\rm imp}_{-\infty} + \vec{J}^{\rm IS}_{-\infty\sim0}$, $\vec{J}^{\rm IS}_{0\sim n}$, and $\vec{J}^{\rm IS}_{n\sim+\infty}$ generate fields $\vec{F}_{-\infty\sim0}$, $\vec{F}_{0\sim n}$, and $\vec{F}_{n\sim+\infty}$ respectively (here $\vec{J}^{\rm imp}_{-\infty}$ is the impressed electric current at $z = -\infty$).

Below, we derive some important corollaries from traveling-wave condition (3-11), and they are the theoretical foundations for constructing traveling-wave-type IP-DMs.

**Corollary 1. Periodical Non-reflection Feature**

If the equivalent surface currents distributing on $\mathbb{S}_0$ and $\mathbb{S}_n$ are denoted as $\{\vec{J}^{\rm ES}_0, \vec{M}^{\rm ES}_0\}$ and $\{\vec{J}^{\rm ES}_n, \vec{M}^{\rm ES}_n\}$ respectively as shown in Fig. 3-5, then we rigorously have the following *source-field relationship*

$$\vec{F}(\vec{r}) = \mathcal{F}_0\left(\vec{J}^{\rm ES}_0 + \vec{J}^{\rm IS}_{0\sim n} - \vec{J}^{\rm ES}_n, \vec{M}^{\rm ES}_0 - \vec{M}^{\rm ES}_n\right) \quad , \quad \vec{r} \in \mathbb{V}_{0\sim n} \tag{3-12}$$

because surfaces $\mathbb{S}_0$, $\mathbb{S}_{0\sim n}$, and $\mathbb{S}_n$ constitute a closed surface enclosing region $\mathbb{V}_{0\sim n}$. The details of the proof for Eq. (3-12) can be found in Ref. [13] and the App. A of this report. Thus,

$$\vec{F}(\vec{r}) = \mathcal{F}_0\left(\vec{J}^{\rm ES}_0 + \vec{J}^{\rm IS}_{0\sim n} - \vec{J}^{\rm ES}_n, \vec{M}^{\rm ES}_0 - \vec{M}^{\rm ES}_n\right) \quad , \quad \vec{r} \in \mathbb{S}^-_n \tag{3-13}$$

because $\mathbb{S}^-_n \subset \mathbb{V}_{0\sim n}$, where $\mathbb{S}^-_n$ is the left-side surface of $\mathbb{S}_n$.

For traveling-wave modes, Eq. (3-11) implies that

$$\vec{F}_{n\sim+\infty}\left(\vec{r}\right) = 0 \quad , \quad \vec{r} \in \mathbb{S}_n^- \tag{3-14}$$

because of that $\vec{F}_{n\sim+\infty} = 0$ on $\mathbb{V}_{-\infty\sim0} \bigcup \mathbb{V}_{0\sim n}$ and that $\mathbb{S}_n^- \subset \mathbb{V}_{-\infty\sim0} \bigcup \mathbb{V}_{0\sim n}$, and then

$$\vec{F}_{n\sim+\infty}\left(\vec{r}\right) = 0 \quad , \quad \vec{r} \in \mathbb{S}_n^+ \tag{3-15}$$

due to the continuity of the $\vec{F}_{n\sim+\infty}$ on $\mathbb{S}_n$, where $\mathbb{S}_n^+$ is the right-side surface of $\mathbb{S}_n$, and thus

$$\vec{F}\left(\vec{r}\right) = \vec{F}_{-\infty\sim0}\left(\vec{r}\right)+\vec{F}_{0\sim n}\left(\vec{r}\right)+\vec{F}_{n\sim+\infty}\left(\vec{r}\right) = \overbrace{\vec{F}_{-\infty\sim0}\left(\vec{r}\right)+\vec{F}_{0\sim n}\left(\vec{r}\right)}^{\vec{F}_{-\infty\sim n}\left(\vec{r}\right)} \quad , \quad \vec{r} \in \mathbb{S}_n^+ \tag{3-16}$$

In addition, Eq. (3-11) also implies that

$$\mathcal{F}_0\left(\vec{J}_n^{\rm ES},\vec{M}_n^{\rm ES}\right) = \vec{F}_{-\infty\sim n} \quad , \quad \vec{r} \in \mathbb{S}_n^+ \tag{3-17}$$

because $\mathbb{S}_n^+ \subset \mathbb{V}_{n\sim+\infty}$, so

$$\vec{F}\left(\vec{r}\right) = \mathcal{F}_0\left(\vec{J}_n^{\rm ES},\vec{M}_n^{\rm ES}\right) \quad , \quad \vec{r} \in \mathbb{S}_n^+ \tag{3-18}$$

by employing Eqs. (3-16) and (3-17).

Based on Eqs. (3-13) and (3-18), we immediately have that

$$\left[\mathcal{F}_0\left(\vec{J}_0^{\rm ES}+\vec{J}_{0\sim n}^{\rm IS}-\vec{J}_n^{\rm ES},\vec{M}_0^{\rm ES}-\vec{M}_n^{\rm ES}\right)\right]_{\vec{r}_-\to\vec{r}}^{\tan} = \left[\mathcal{F}_0\left(\vec{J}_n^{\rm ES},\vec{M}_n^{\rm ES}\right)\right]_{\vec{r}_+\to\vec{r}}^{\tan} \quad , \quad \vec{r} \in \mathbb{S}_n \tag{3-19}$$

for traveling-wave modes, because of the tangential continuity of the $\vec{F}$ on $\mathbb{S}_n$, where $\vec{r}_- \in \mathbb{V}_{0\sim n}$ and $\vec{r}_+ \in \mathbb{V}_{n\sim+\infty}$ and the points approach the $\vec{r}$ on $\mathbb{S}_n$. Equation (3-19) reveals that: **for a traveling-wave mode, in region $\mathbb{V}_{0\sim n}$, there doesn't exist the "reflected" field generated by the induced currents $\vec{J}_{n\sim+\infty}^{\rm IS}$ on $\mathbb{S}_{n\sim+\infty}$.** This report calls it the *periodical non-reflection feature of traveling-wave modes*.

**Corollary 2. Periodical Invariance of Modal Input Power and Periodical Local Resonance**

Based on the *longitudinal periodicity of traveling-wave modes*, the input power $P_{\mathbb{S}_0}^{\rm in}$ of port $\mathbb{S}_0$ and the input power $P_{\mathbb{S}_n}^{\rm in}$ of port $\mathbb{S}_n$ satisfy the following relation

$$P_{\mathbb{S}_0}^{\rm in} = \left(1/2\right)\iint_{\mathbb{S}_0}\left(\vec{E}\times\vec{H}^\dagger\right)\cdot\hat{z}dS = \left(1/2\right)\iint_{\mathbb{S}_n}\left(\vec{E}\times\vec{H}^\dagger\right)\cdot\hat{z}dS = P_{\mathbb{S}_n}^{\rm in} \tag{3-20}$$

because the distance between $\mathbb{S}_0$ and $\mathbb{S}_n$ is $n\lambda_z$. Then, for **traveling-wave** modes, we have that

$$2\omega\left[\left(1/4\right)\left\langle\vec{H},\mu_0\vec{H}\right\rangle_{\mathbb{V}_{0\sim n}}-\left(1/4\right)\left\langle\varepsilon_0\vec{E},\vec{E}\right\rangle_{\mathbb{V}_{0\sim n}}\right] = 0 \tag{3-21}$$

by employing a similar relation to PTT (3-4).

Equation (3-20) is particularly called the *periodical invariance of input power for traveling-wave modes*, and Eq. (3-21) is particularly called the *periodical local resonance of traveling-wave modes*.

**Corollary 3. Modal Input Impedance, Modal Input Admittance, and Their Pure Reality**

Now, we define the following quantities

$$Z_{\mathbb{S}_0}^{\text{in}} = \frac{(1/2)\iint_{\mathbb{S}_0}\left(\vec{E}\times\vec{H}^\dagger\right)\cdot\hat{z}dS}{(1/2)\left\langle \vec{J}_0^{\text{ES}},\vec{J}_0^{\text{ES}}\right\rangle_{\mathbb{S}_0}} \tag{3-22a}$$

$$Y_{\mathbb{S}_0}^{\text{in}} = \frac{(1/2)\iint_{\mathbb{S}_0}\left(\vec{E}\times\vec{H}^\dagger\right)\cdot\hat{z}dS}{(1/2)\left\langle \vec{M}_0^{\text{ES}},\vec{M}_0^{\text{ES}}\right\rangle_{\mathbb{S}_0}} \tag{3-22b}$$

and they are respectively called *modal input impedance* and *modal input admittance* of port $\mathbb{S}_0$, because their dimensions are *ohmic* and *siemens* respectively. In addition, the $R_{\mathbb{S}_0}^{\text{in}} = \text{Re}\{Z_{\mathbb{S}_0}^{\text{in}}\}$ and $X_{\mathbb{S}_0}^{\text{in}} = \text{Im}\{Z_{\mathbb{S}_0}^{\text{in}}\}$ are just the corresponding *modal input resistance* and *modal input reactance* respectively, and the $G_{\mathbb{S}_0}^{\text{in}} = \text{Re}\{Y_{\mathbb{S}_0}^{\text{in}}\}$ and $B_{\mathbb{S}_0}^{\text{in}} = \text{Im}\{Y_{\mathbb{S}_0}^{\text{in}}\}$ are just the corresponding *modal input conductance* and *modal input susceptance* respectively.

In addition, for any traveling-wave mode we can further conclude here that

$$Z_{\mathbb{S}_0}^{\text{in}} = \overbrace{\text{Re}\left\{Z_{\mathbb{S}_0}^{\text{in}}\right\}}^{R_{\mathbb{S}_0}^{\text{in}}} + j\,\underbrace{\overbrace{\text{Im}\left\{Z_{\mathbb{S}_0}^{\text{in}}\right\}}^{X_{\mathbb{S}_0}^{\text{in}}}}_{0} = R_{\mathbb{S}_0}^{\text{in}} \tag{3-23a}$$

$$Y_{\mathbb{S}_0}^{\text{in}} = \underbrace{\text{Re}\left\{Y_{\mathbb{S}_0}^{\text{in}}\right\}}_{G_{\mathbb{S}_0}^{\text{in}}} + j\,\underbrace{\overbrace{\text{Im}\left\{Y_{\mathbb{S}_0}^{\text{in}}\right\}}^{0}}_{B_{\mathbb{S}_0}^{\text{in}}} = G_{\mathbb{S}_0}^{\text{in}} \tag{3-23b}$$

based on PTT (3-4), Eq. (3-21), and the fact that $(1/2)\iint_{\mathbb{S}_{+\infty}}\left(\vec{E}\times\vec{H}^\dagger\right)\cdot\hat{z}dS \in \mathbb{R}$ (here, $\mathbb{S}_{+\infty}$ is the cross section located at $z = +\infty$).

### 3.2.1.5 Dispersion Equation Satisfied by Traveling-wave Modes

As everyone knows, the *cut-off frequency* $f^c$ of a traveling-wave mode satisfies the following *dispersion equation*

$$\left(2\pi f^{\text{c}}\right)^2 \mu_0\varepsilon_0 + \left(\frac{2\pi}{\lambda_z}\right)^2 = (2\pi f)^2 \mu_0\varepsilon_0 \tag{3-24}$$

where $\lambda_z$ and $f$ are respectively the *waveguide wavelength along Z-axis* and *working frequency* of the traveling-wave mode.

### 3.2.2 Mathematical Description for Modal Space

In this sub-section, we consider a section of metalic guide as shown in the following Fig. 3-6.

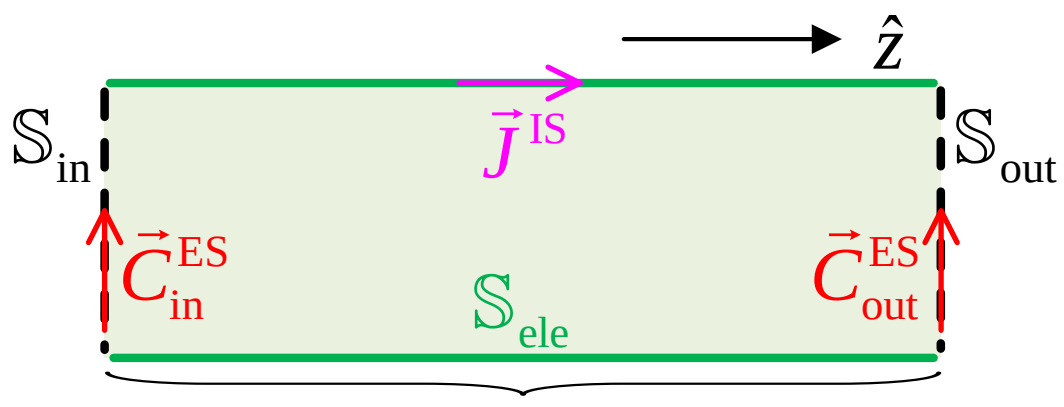

Figure 3-6 Geometry of a section of guide tube with practical longitudinal length $L$

In the above figure, $\mathbb{S}_{\text{in}}$ is the *input port* of the section, and $\mathbb{S}_{\text{out}}$ is the *output port* of the section, and $\mathbb{S}_{\text{ele}}$ is the *electric wall* of the section, and the currents $\vec{C}_0^{\text{ES}} \,\&\, \vec{C}_n^{\text{ES}}$ and $\vec{J}^{\text{IS}}$ are respectively the *equivalent surface currents* on $\mathbb{S}_{\text{in}} \,\&\, \mathbb{S}_{\text{out}}$ and the *induced surface current* on $\mathbb{S}_{\text{ele}}$, and the longitudinal length of the section is $L$[①]. In addition, the guide tube is with material parameters $\{\mu, \varepsilon\}$, where $\mu$ and $\varepsilon$ are constant scalars.

Below, we firstly establish the dependence relations between *independent variables* and *dependent variables*, and secondly, by employing the dependence relations, we restrict the discussed EM problem to the *modal space* constituted by the traveling-wave modes of the guide shown in Fig. 3-6, and thirdly we derive the traveling-wave-type IP-DMs in the modal space from orthogonalizing the frequency-domain *input power operator* (*IPO*) defined on the modal space.

#### 3.2.2.1 Integral Equations for Relating Independent Variables and Dependent Variables

Based on Eq. (3-12) and Eqs. (3-5a)&(3-5b), the following *integral equations* can be derived:

$$\left[\mathcal{H}\left(\vec{J}_{\text{in}}^{\text{ES}} + \vec{J}^{\text{IS}} - \vec{J}_{\text{out}}^{\text{ES}}, \vec{M}_{\text{in}}^{\text{ES}} - \vec{M}_{\text{out}}^{\text{ES}}\right)\right]_{\vec{r}' \to \vec{r}}^{\tan} = \vec{J}_{\text{in}}^{\text{ES}}(\vec{r}) \times \hat{z} \quad , \quad \vec{r} \in \mathbb{S}_{\text{in}} \tag{3-25a}$$

$$\left[\mathcal{E}\left(\vec{J}_{\text{in}}^{\text{ES}} + \vec{J}^{\text{IS}} - \vec{J}_{\text{out}}^{\text{ES}}, \vec{M}_{\text{in}}^{\text{ES}} - \vec{M}_{\text{out}}^{\text{ES}}\right)\right]_{\vec{r}' \to \vec{r}}^{\tan} = \hat{z} \times \vec{M}_{\text{in}}^{\text{ES}}(\vec{r}) \quad , \quad \vec{r} \in \mathbb{S}_{\text{in}} \tag{3-25b}$$

① For a traveling-wave mode, the longitudinal length of the section satisfies that $L = n\lambda_z$, where $\lambda_z$ is the waveguide wavelength of the traveling-wave mode.

Here, $\vec{r}'$ belongs to the guide tube shown in Fig. 3-6, and $\vec{r}'$ approaches the $\vec{r}$ on $\mathbb{S}_{\text{in}}$; the superscript "tan" represents the tangential components of the modal fields; the operators $\mathcal{E}$ and $\mathcal{H}$ are defined as follows:

$$\mathcal{E}\left(\vec{J},\vec{M}\right) = -j\omega\mu\mathcal{L}\left(\vec{J}\right)-\mathcal{K}\left(\vec{M}\right) \tag{3-26a}$$

$$\mathcal{H}\left(\vec{J},\vec{M}\right) = -j\omega\varepsilon\mathcal{L}\left(\vec{M}\right)+\mathcal{K}\left(\vec{J}\right) \tag{3-26b}$$

where operators $\mathcal{L}$ and $\mathcal{K}$ are defined similarly to the ones given in Eq. (3-8).

On $\mathbb{S}_{\text{ele}}$, the following *electric field integral equation* exists:

$$\left[\mathcal{E}\left(\vec{J}_{\text{in}}^{\text{ES}}+\vec{J}^{\text{IS}}-\vec{J}_{\text{out}}^{\text{ES}},\vec{M}_{\text{in}}^{\text{ES}}-\vec{M}_{\text{out}}^{\text{ES}}\right)\right]^{\tan} = \left[\vec{E}\left(\vec{r}\right)\right]^{\tan} = 0 \quad , \quad \vec{r}\in\mathbb{S}_{\text{ele}} \tag{3-27}$$

because of the homogeneous tangential electric field boundary condition on $\mathbb{S}_{\text{ele}}$.

Based on Eq. (3-19), the following integral equations can be derived:

$$\left[\mathcal{E}\left(\vec{J}_{\text{in}}^{\text{ES}}+\vec{J}^{\text{IS}}-\vec{J}_{\text{out}}^{\text{ES}},\vec{M}_{\text{in}}^{\text{ES}}-\vec{M}_{\text{out}}^{\text{ES}}\right)\right]_{\vec{r}_-\to\vec{r}}^{\tan} = \left[\mathcal{E}\left(\vec{J}_{\text{out}}^{\text{ES}},\vec{M}_{\text{out}}^{\text{ES}}\right)\right]_{\vec{r}_+\to\vec{r}}^{\tan} \quad , \quad \vec{r}\in\mathbb{S}_{\text{out}} \tag{3-28a}$$

$$\left[\mathcal{H}\left(\vec{J}_{\text{in}}^{\text{ES}}+\vec{J}^{\text{IS}}-\vec{J}_{\text{out}}^{\text{ES}},\vec{M}_{\text{in}}^{\text{ES}}-\vec{M}_{\text{out}}^{\text{ES}}\right)\right]_{\vec{r}_-\to\vec{r}}^{\tan} = \left[\mathcal{H}\left(\vec{J}_{\text{out}}^{\text{ES}},\vec{M}_{\text{out}}^{\text{ES}}\right)\right]_{\vec{r}_+\to\vec{r}}^{\tan} \quad , \quad \vec{r}\in\mathbb{S}_{\text{out}} \tag{3-28b}$$

where $\vec{r}_-$ and $\vec{r}_+$ locate on the left-side and right-side surfaces of $\mathbb{S}_n$, and the points approach the $\vec{r}$ on $\mathbb{S}_{\text{out}}$.

From the above Eqs. (3-25a)&(3-25b), (3-27), and (3-28a)&(3-28b), it is easy to find out that the currents $\{\vec{J}^{\text{IS}},\vec{J}_{\text{out}}^{\text{ES}},\vec{M}_{\text{in}}^{\text{ES}},\vec{M}_{\text{out}}^{\text{ES}}\}$ can be uniquely determined by current $\vec{J}_{\text{in}}^{\text{ES}}$, and the currents $\{\vec{J}_{\text{in}}^{\text{ES}},\vec{J}^{\text{IS}},\vec{J}_{\text{out}}^{\text{ES}},\vec{M}_{\text{out}}^{\text{ES}}\}$ can be uniquely determined by current $\vec{M}_{\text{in}}^{\text{ES}}$, because Eqs. (3-25a) and (3-25b) are theoretically equivalent to each other and Eqs. (3-25a)/(3-25b), (3-27), (3-28a), and (3-28b) are independent of each other.

#### 3.2.2.2 Matrix Equations for Relating Independent Variables and Dependent Variables

Equivalent currents $\vec{J}_{\text{in}}^{\text{ES}}$ & $\vec{M}_{\text{in}}^{\text{ES}}$, wall electric current $\vec{J}^{\text{IS}}$, and equivalent currents $\vec{J}_{\text{out}}^{\text{ES}}$ & $\vec{M}_{\text{out}}^{\text{ES}}$ can be respectively expanded in terms of some proper *basis functions* $\{\vec{b}_\xi^{\vec{J}_{\text{in}}^{\text{ES}}}\}$ & $\{\vec{b}_\xi^{\vec{M}_{\text{in}}^{\text{ES}}}\}$, $\{\vec{b}_\xi^{\vec{J}^{\text{IS}}}\}$, and $\{\vec{b}_\xi^{\vec{J}_{\text{out}}^{\text{ES}}}\}$ & $\{\vec{b}_\xi^{\vec{M}_{\text{out}}^{\text{ES}}}\}$ as follows:

$$\vec{C}_{\text{in}}^{\text{ES}}\left(\vec{r}\right) = \sum\nolimits_\xi a_\xi^{\vec{C}_{\text{in}}^{\text{ES}}}\vec{b}_\xi^{\vec{C}_{\text{in}}^{\text{ES}}} = \bar{\boldsymbol{B}}^{\vec{C}_{\text{in}}^{\text{ES}}}\cdot\bar{a}^{\vec{C}_{\text{in}}^{\text{ES}}} \quad , \quad \vec{r}\in\mathbb{S}_{\text{in}} \tag{3-29a}$$

$$\vec{J}^{\text{IS}}\left(\vec{r}\right) = \sum\nolimits_\xi a_\xi^{\vec{J}^{\text{IS}}}\vec{b}_\xi^{\vec{J}^{\text{IS}}} = \bar{\boldsymbol{B}}^{\vec{J}^{\text{IS}}}\cdot\bar{a}^{\vec{J}^{\text{IS}}} \quad , \quad \vec{r}\in\mathbb{S}_{\text{ele}} \tag{3-29b}$$

$$\vec{C}_{\text{out}}^{\text{ES}}\left(\vec{r}\right) = \sum\nolimits_\xi a_\xi^{\vec{C}_{\text{out}}^{\text{ES}}}\vec{b}_\xi^{\vec{C}_{\text{out}}^{\text{ES}}} = \bar{\boldsymbol{B}}^{\vec{C}_{\text{out}}^{\text{ES}}}\cdot\bar{a}^{\vec{C}_{\text{out}}^{\text{ES}}} \quad , \quad \vec{r}\in\mathbb{S}_{\text{out}} \tag{3-29c}$$

in which $\vec{C}=\vec{J}/\vec{M}$, and

$$\bar{\boldsymbol{B}}^{\vec{X}} = \begin{bmatrix} \vec{b}_1^{\vec{X}} & \vec{b}_2^{\vec{X}} & \cdots \end{bmatrix} \tag{3-30a}$$

$$\bar{a}^{\vec{X}} = \begin{bmatrix} a_1^{\vec{X}} & a_2^{\vec{X}} & \cdots \end{bmatrix}^T \tag{3-30b}$$

where $\vec{X} = \vec{J}_{\text{in}}^{\text{ES}} / \vec{M}_{\text{in}}^{\text{ES}} / \vec{J}^{\text{IS}} / \vec{J}_{\text{out}}^{\text{ES}} / \vec{M}_{\text{out}}^{\text{ES}}$, and the superscript "$T$" in Eq. (3-30b) denotes the *transpose operation* for a matrix or vector.

Substituting Eqs. (3-29a)~(3-29c) into Eqs. (3-25a) and (3-25b) and testing the Eqs. (3-25a) and (3-25b) with $\{\vec{b}_\xi^{\vec{M}_{\text{in}}^{\text{ES}}}\}$ and $\{\vec{b}_\xi^{\vec{J}_{\text{in}}^{\text{ES}}}\}$ respectively, the integral equations (3-25a) and (3-25b) are discretized into the following *matrix equations*:

$$\bar{\bar{Z}}^{\vec{M}_{\text{in}}^{\text{ES}}\vec{J}_{\text{in}}^{\text{ES}}} \cdot \bar{a}^{\vec{J}_{\text{in}}^{\text{ES}}} + \bar{\bar{Z}}^{\vec{M}_{\text{in}}^{\text{ES}}\vec{J}^{\text{IS}}} \cdot \bar{a}^{\vec{J}^{\text{IS}}} + \bar{\bar{Z}}^{\vec{M}_{\text{in}}^{\text{ES}}\vec{J}_{\text{out}}^{\text{ES}}} \cdot \bar{a}^{\vec{J}_{\text{out}}^{\text{ES}}} + \bar{\bar{Z}}^{\vec{M}_{\text{in}}^{\text{ES}}\vec{M}_{\text{in}}^{\text{ES}}} \cdot \bar{a}^{\vec{M}_{\text{in}}^{\text{ES}}} + \bar{\bar{Z}}^{\vec{M}_{\text{in}}^{\text{ES}}\vec{M}_{\text{out}}^{\text{ES}}} \cdot \bar{a}^{\vec{M}_{\text{out}}^{\text{ES}}} = 0 \tag{3-31a}$$

$$\bar{\bar{Z}}^{\vec{J}_{\text{in}}^{\text{ES}}\vec{J}_{\text{in}}^{\text{ES}}} \cdot \bar{a}^{\vec{J}_{\text{in}}^{\text{ES}}} + \bar{\bar{Z}}^{\vec{J}_{\text{in}}^{\text{ES}}\vec{J}^{\text{IS}}} \cdot \bar{a}^{\vec{J}^{\text{IS}}} + \bar{\bar{Z}}^{\vec{J}_{\text{in}}^{\text{ES}}\vec{J}_{\text{out}}^{\text{ES}}} \cdot \bar{a}^{\vec{J}_{\text{out}}^{\text{ES}}} + \bar{\bar{Z}}^{\vec{J}_{\text{in}}^{\text{ES}}\vec{M}_{\text{in}}^{\text{ES}}} \cdot \bar{a}^{\vec{M}_{\text{in}}^{\text{ES}}} + \bar{\bar{Z}}^{\vec{J}_{\text{in}}^{\text{ES}}\vec{M}_{\text{out}}^{\text{ES}}} \cdot \bar{a}^{\vec{M}_{\text{out}}^{\text{ES}}} = 0 \tag{3-31b}$$

In Eq. (3-31a), the elements of the various matrices are calculated as follows:

$$z_{\xi\zeta}^{\vec{M}_{\text{in}}^{\text{ES}}\vec{J}_{\text{in}}^{\text{ES}}} = \left\langle \vec{b}_\xi^{\vec{M}_{\text{in}}^{\text{ES}}}, \text{P.V.}\mathcal{K}\left(\vec{b}_\zeta^{\vec{J}_{\text{in}}^{\text{ES}}}\right) \right\rangle_{\mathbb{S}_{\text{in}}} - \left\langle \vec{b}_\xi^{\vec{M}_{\text{in}}^{\text{ES}}}, \frac{1}{2}\vec{b}_\zeta^{\vec{J}_{\text{in}}^{\text{ES}}} \times \hat{z} \right\rangle_{\mathbb{S}_{\text{in}}} \tag{3-32a}$$

$$z_{\xi\zeta}^{\vec{M}_{\text{in}}^{\text{ES}}\vec{J}^{\text{IS}}} = \left\langle \vec{b}_\xi^{\vec{M}_{\text{in}}^{\text{ES}}}, \mathcal{K}\left(\vec{b}_\zeta^{\vec{J}^{\text{IS}}}\right) \right\rangle_{\mathbb{S}_{\text{in}}} \tag{3-32b}$$

$$z_{\xi\zeta}^{\vec{M}_{\text{in}}^{\text{ES}}\vec{J}_{\text{out}}^{\text{ES}}} = \left\langle \vec{b}_\xi^{\vec{M}_{\text{in}}^{\text{ES}}}, \mathcal{K}\left(-\vec{b}_\zeta^{\vec{J}_{\text{out}}^{\text{ES}}}\right) \right\rangle_{\mathbb{S}_{\text{in}}} \tag{3-32c}$$

$$z_{\xi\zeta}^{\vec{M}_{\text{in}}^{\text{ES}}\vec{M}_{\text{in}}^{\text{ES}}} = \left\langle \vec{b}_\xi^{\vec{M}_{\text{in}}^{\text{ES}}}, -j\omega\varepsilon\mathcal{L}\left(\vec{b}_\zeta^{\vec{M}_{\text{in}}^{\text{ES}}}\right) \right\rangle_{\mathbb{S}_{\text{in}}} \tag{3-32d}$$

$$z_{\xi\zeta}^{\vec{M}_{\text{in}}^{\text{ES}}\vec{M}_{\text{out}}^{\text{ES}}} = \left\langle \vec{b}_\xi^{\vec{M}_{\text{in}}^{\text{ES}}}, -j\omega\varepsilon\mathcal{L}\left(-\vec{b}_\zeta^{\vec{M}_{\text{out}}^{\text{ES}}}\right) \right\rangle_{\mathbb{S}_{\text{in}}} \tag{3-32e}$$

In Eq. (3-31b), the elements of the various matrices are calculated as follows:

$$z_{\xi\zeta}^{\vec{J}_{\text{in}}^{\text{ES}}\vec{J}_{\text{in}}^{\text{ES}}} = \left\langle \vec{b}_\xi^{\vec{J}_{\text{in}}^{\text{ES}}}, -j\omega\mu\mathcal{L}\left(\vec{b}_\zeta^{\vec{J}_{\text{in}}^{\text{ES}}}\right) \right\rangle_{\mathbb{S}_{\text{in}}} \tag{3-33a}$$

$$z_{\xi\zeta}^{\vec{J}_{\text{in}}^{\text{ES}}\vec{J}^{\text{IS}}} = \left\langle \vec{b}_\xi^{\vec{J}_{\text{in}}^{\text{ES}}}, -j\omega\mu\mathcal{L}\left(\vec{b}_\zeta^{\vec{J}^{\text{IS}}}\right) \right\rangle_{\mathbb{S}_{\text{in}}} \tag{3-33b}$$

$$z_{\xi\zeta}^{\vec{J}_{\text{in}}^{\text{ES}}\vec{J}_{\text{out}}^{\text{ES}}} = \left\langle \vec{b}_\xi^{\vec{J}_{\text{in}}^{\text{ES}}}, -j\omega\mu\mathcal{L}\left(-\vec{b}_\zeta^{\vec{J}_{\text{out}}^{\text{ES}}}\right) \right\rangle_{\mathbb{S}_{\text{in}}} \tag{3-33c}$$

$$z_{\xi\zeta}^{\vec{J}_{\text{in}}^{\text{ES}}\vec{M}_{\text{in}}^{\text{ES}}} = \left\langle \vec{b}_\xi^{\vec{J}_{\text{in}}^{\text{ES}}}, -\text{P.V.}\mathcal{K}\left(\vec{b}_\zeta^{\vec{M}_{\text{in}}^{\text{ES}}}\right) \right\rangle_{\mathbb{S}_{\text{in}}} - \left\langle \vec{b}_\xi^{\vec{J}_{\text{in}}^{\text{ES}}}, \hat{z} \times \frac{1}{2}\vec{b}_\zeta^{\vec{M}_{\text{in}}^{\text{ES}}} \right\rangle_{\mathbb{S}_{\text{in}}} \tag{3-33d}$$

$$z_{\xi\zeta}^{\vec{J}_{\text{in}}^{\text{ES}}\vec{M}_{\text{out}}^{\text{ES}}} = \left\langle \vec{b}_\xi^{\vec{J}_{\text{in}}^{\text{ES}}}, -\mathcal{K}\left(-\vec{b}_\zeta^{\vec{M}_{\text{out}}^{\text{ES}}}\right) \right\rangle_{\mathbb{S}_{\text{in}}} \tag{3-33e}$$

In the above Eqs. (3-32a) and (3-33d), the symbol "$\text{P.V.}\mathcal{K}$" is the *principal value* of operator $\mathcal{K}$.

Substituting Eqs. (3-29a)~(3-29c) into Eq. (3-27) and testing the Eq. (3-27) with $\{\vec{b}_\xi^{\vec{J}^{\text{IS}}}\}$, the integral equation (3-27) is discretized into the following matrix equation:

$$\overline{\overline{Z}}^{\vec{J}^{\rm IS}\vec{J}_{\rm in}^{\rm ES}}\cdot\overline{a}^{\vec{J}_{\rm in}^{\rm ES}}+\overline{\overline{Z}}^{\vec{J}^{\rm IS}\vec{J}^{\rm IS}}\cdot\overline{a}^{\vec{J}^{\rm IS}}+\overline{\overline{Z}}^{\vec{J}^{\rm IS}\vec{J}_{\rm out}^{\rm ES}}\cdot\overline{a}^{\vec{J}_{\rm out}^{\rm ES}}+\overline{\overline{Z}}^{\vec{J}^{\rm IS}\vec{M}_{\rm in}^{\rm ES}}\cdot\overline{a}^{\vec{M}_{\rm in}^{\rm ES}}+\overline{\overline{Z}}^{\vec{J}^{\rm IS}\vec{M}_{\rm out}^{\rm ES}}\cdot\overline{a}^{\vec{M}_{\rm out}^{\rm ES}}=0 \quad (3\text{-}34)$$

where the elements of various matrices are calculated as follows:

$$z_{\xi\zeta}^{\vec{J}^{\rm IS}\vec{J}_{\rm in}^{\rm ES}} = \left\langle \vec{b}_{\xi}^{\vec{J}^{\rm IS}}, -j\omega\mu\mathcal{L}\left(\vec{b}_{\zeta}^{\vec{J}_{\rm in}^{\rm ES}}\right)\right\rangle_{\mathbb{S}_{\rm ele}} \quad (3\text{-}35a)$$

$$z_{\xi\zeta}^{\vec{J}^{\rm IS}\vec{J}^{\rm IS}} = \left\langle \vec{b}_{\xi}^{\vec{J}^{\rm IS}}, -j\omega\mu\mathcal{L}\left(\vec{b}_{\zeta}^{\vec{J}^{\rm IS}}\right)\right\rangle_{\mathbb{S}_{\rm ele}} \quad (3\text{-}35b)$$

$$z_{\xi\zeta}^{\vec{J}^{\rm IS}\vec{J}_{\rm out}^{\rm ES}} = \left\langle \vec{b}_{\xi}^{\vec{J}^{\rm IS}}, -j\omega\mu\mathcal{L}\left(-\vec{b}_{\zeta}^{\vec{J}_{\rm out}^{\rm ES}}\right)\right\rangle_{\mathbb{S}_{\rm ele}} \quad (3\text{-}35c)$$

$$z_{\xi\zeta}^{\vec{J}^{\rm IS}\vec{M}_{\rm in}^{\rm ES}} = \left\langle \vec{b}_{\xi}^{\vec{J}^{\rm IS}}, -\mathcal{K}\left(\vec{b}_{\zeta}^{\vec{M}_{\rm in}^{\rm ES}}\right)\right\rangle_{\mathbb{S}_{\rm ele}} \quad (3\text{-}35d)$$

$$z_{\xi\zeta}^{\vec{J}^{\rm IS}\vec{M}_{\rm out}^{\rm ES}} = \left\langle \vec{b}_{\xi}^{\vec{J}^{\rm IS}}, -\mathcal{K}\left(-\vec{b}_{\zeta}^{\vec{M}_{\rm out}^{\rm ES}}\right)\right\rangle_{\mathbb{S}_{\rm ele}} \quad (3\text{-}35e)$$

Substituting Eqs. (3-29a)~(3-29c) into Eqs. (3-28a) and (3-28b) and testing the Eqs. (3-28a) and (3-28b) with $\{\vec{b}_{\xi}^{\vec{J}_{\rm out}^{\rm ES}}\}$ and $\{\vec{b}_{\xi}^{\vec{M}_{\rm out}^{\rm ES}}\}$ respectively, the integral equations (3-28a) and (3-28b) are discretized into the following matrix equations:

$$\overline{\overline{Z}}^{\vec{J}_{\rm out}^{\rm ES}\vec{J}_{\rm in}^{\rm ES}}\cdot\overline{a}^{\vec{J}_{\rm in}^{\rm ES}}+\overline{\overline{Z}}^{\vec{J}_{\rm out}^{\rm ES}\vec{J}^{\rm IS}}\cdot\overline{a}^{\vec{J}^{\rm IS}}+\overline{\overline{Z}}^{\vec{J}_{\rm out}^{\rm ES}\vec{J}_{\rm out}^{\rm ES}}\cdot\overline{a}^{\vec{J}_{\rm out}^{\rm ES}}+\overline{\overline{Z}}^{\vec{J}_{\rm out}^{\rm ES}\vec{M}_{\rm in}^{\rm ES}}\cdot\overline{a}^{\vec{M}_{\rm in}^{\rm ES}}+\overline{\overline{Z}}^{\vec{J}_{\rm out}^{\rm ES}\vec{M}_{\rm out}^{\rm ES}}\cdot\overline{a}^{\vec{M}_{\rm out}^{\rm ES}}=0 \quad (3\text{-}36a)$$

$$\overline{\overline{Z}}^{\vec{M}_{\rm out}^{\rm ES}\vec{J}_{\rm in}^{\rm ES}}\cdot\overline{a}^{\vec{J}_{\rm in}^{\rm ES}}+\overline{\overline{Z}}^{\vec{M}_{\rm out}^{\rm ES}\vec{J}^{\rm IS}}\cdot\overline{a}^{\vec{J}^{\rm IS}}+\overline{\overline{Z}}^{\vec{M}_{\rm out}^{\rm ES}\vec{J}_{\rm out}^{\rm ES}}\cdot\overline{a}^{\vec{J}_{\rm out}^{\rm ES}}+\overline{\overline{Z}}^{\vec{M}_{\rm out}^{\rm ES}\vec{M}_{\rm in}^{\rm ES}}\cdot\overline{a}^{\vec{M}_{\rm in}^{\rm ES}}+\overline{\overline{Z}}^{\vec{M}_{\rm out}^{\rm ES}\vec{M}_{\rm out}^{\rm ES}}\cdot\overline{a}^{\vec{M}_{\rm out}^{\rm ES}}=0 \quad (3\text{-}36b)$$

In Eq. (3-36a), the elements of various matrices are calculated as follows:

$$z_{\xi\zeta}^{\vec{J}_{\rm out}^{\rm ES}\vec{J}_{\rm in}^{\rm ES}} = \left\langle \vec{b}_{\xi}^{\vec{J}_{\rm out}^{\rm ES}}, -j\omega\mu\mathcal{L}\left(\vec{b}_{\zeta}^{\vec{J}_{\rm in}^{\rm ES}}\right)\right\rangle_{\mathbb{S}_{\rm out}} \quad (3\text{-}37a)$$

$$z_{\xi\zeta}^{\vec{J}_{\rm out}^{\rm ES}\vec{J}^{\rm IS}} = \left\langle \vec{b}_{\xi}^{\vec{J}_{\rm out}^{\rm ES}}, -j\omega\mu\mathcal{L}\left(\vec{b}_{\zeta}^{\vec{J}^{\rm IS}}\right)\right\rangle_{\mathbb{S}_{\rm out}} \quad (3\text{-}37b)$$

$$z_{\xi\zeta}^{\vec{J}_{\rm out}^{\rm ES}\vec{J}_{\rm out}^{\rm ES}} = \left\langle \vec{b}_{\xi}^{\vec{J}_{\rm out}^{\rm ES}}, -j\omega\mu\mathcal{L}\left(-\vec{b}_{\zeta}^{\vec{J}_{\rm out}^{\rm ES}}\right)\right\rangle_{\mathbb{S}_{\rm out}} - \left\langle \vec{b}_{\xi}^{\vec{J}_{\rm out}^{\rm ES}}, -j\omega\mu\mathcal{L}\left(\vec{b}_{\zeta}^{\vec{J}_{\rm out}^{\rm ES}}\right)\right\rangle_{\mathbb{S}_{\rm out}} \quad (3\text{-}37c)$$

$$z_{\xi\zeta}^{\vec{J}_{\rm out}^{\rm ES}\vec{M}_{\rm in}^{\rm ES}} = \left\langle \vec{b}_{\xi}^{\vec{J}_{\rm out}^{\rm ES}}, -\mathcal{K}\left(\vec{b}_{\zeta}^{\vec{M}_{\rm in}^{\rm ES}}\right)\right\rangle_{\mathbb{S}_{\rm out}} \quad (3\text{-}37d)$$

$$z_{\xi\zeta}^{\vec{J}_{\rm out}^{\rm ES}\vec{M}_{\rm out}^{\rm ES}} = \left\langle \vec{b}_{\xi}^{\vec{J}_{\rm out}^{\rm ES}}, -\text{P.V.}\,\mathcal{K}\left(-\vec{b}_{\zeta}^{\vec{M}_{\rm out}^{\rm ES}}\right)\right\rangle_{\mathbb{S}_{\rm out}} - \left\langle \vec{b}_{\xi}^{\vec{J}_{\rm out}^{\rm ES}}, -\text{P.V.}\,\mathcal{K}\left(\vec{b}_{\zeta}^{\vec{M}_{\rm out}^{\rm ES}}\right)\right\rangle_{\mathbb{S}_{\rm out}} \quad (3\text{-}37e)$$

In Eq. (3-36b), the elements of the various matrices are calculated as follows:

$$z_{\xi\zeta}^{\vec{M}_{\rm out}^{\rm ES}\vec{J}_{\rm in}^{\rm ES}} = \left\langle \vec{b}_{\xi}^{\vec{M}_{\rm out}^{\rm ES}}, \mathcal{K}\left(\vec{b}_{\zeta}^{\vec{J}_{\rm in}^{\rm ES}}\right)\right\rangle_{\mathbb{S}_{\rm out}} \quad (3\text{-}38a)$$

$$z_{\xi\zeta}^{\vec{M}_{\rm out}^{\rm ES}\vec{J}^{\rm IS}} = \left\langle \vec{b}_{\xi}^{\vec{M}_{\rm out}^{\rm ES}}, \mathcal{K}\left(\vec{b}_{\zeta}^{\vec{J}^{\rm IS}}\right)\right\rangle_{\mathbb{S}_{\rm out}} \quad (3\text{-}38b)$$

$$z_{\xi\zeta}^{\vec{M}_{\rm out}^{\rm ES}\vec{J}_{\rm out}^{\rm ES}} = \left\langle \vec{b}_{\xi}^{\vec{M}_{\rm out}^{\rm ES}}, \text{P.V.}\,\mathcal{K}\left(-\vec{b}_{\zeta}^{\vec{J}_{\rm out}^{\rm ES}}\right)\right\rangle_{\mathbb{S}_{\rm out}} - \left\langle \vec{b}_{\xi}^{\vec{M}_{\rm out}^{\rm ES}}, \text{P.V.}\,\mathcal{K}\left(\vec{b}_{\zeta}^{\vec{J}_{\rm out}^{\rm ES}}\right)\right\rangle_{\mathbb{S}_{\rm out}} \quad (3\text{-}38c)$$

$$z_{\xi\zeta}^{\vec{M}_{\rm out}^{\rm ES}\vec{M}_{\rm in}^{\rm ES}} = \left\langle \vec{b}_{\xi}^{\vec{M}_{\rm out}^{\rm ES}}, -j\omega\varepsilon\mathcal{L}\left(\vec{b}_{\zeta}^{\vec{M}_{\rm in}^{\rm ES}}\right)\right\rangle_{\mathbb{S}_{\rm out}} \quad (3\text{-}38d)$$

$$z_{\xi\zeta}^{\vec{M}_{\rm out}^{\rm ES}\vec{M}_{\rm out}^{\rm ES}} = \left\langle \vec{b}_{\xi}^{\vec{M}_{\rm out}^{\rm ES}}, -j\omega\varepsilon\mathcal{L}\left(-\vec{b}_{\zeta}^{\vec{M}_{\rm out}^{\rm ES}}\right)\right\rangle_{\mathbb{S}_{\rm out}} - \left\langle \vec{b}_{\xi}^{\vec{M}_{\rm out}^{\rm ES}}, -j\omega\varepsilon\mathcal{L}\left(\vec{b}_{\zeta}^{\vec{M}_{\rm out}^{\rm ES}}\right)\right\rangle_{\mathbb{S}_{\rm out}} \quad (3\text{-}38e)$$

In the following subsections, by employing the above matrix equations we provide some rigorous mathematical descriptions for the modal space constituted by the travelling-wave modes satisfying Eqs. (3-31a)/(3-31b), (3-34), and (3-36a)&(3-36b), i.e., satisfying Eqs. (3-25a)/(3-25b), (3-27), and (3-28a)&(3-28b).

### 3.2.2.3 Mathematical Description for Modal Space — Scheme I: Dependent Variable Elimination (DVE)

Following the convention of Refs. [13,18,19], Eqs. (3-31a), (3-34), (3-36a), and (3-36b) are combined as follows:

$$\underbrace{\begin{bmatrix} \bar{\bar{I}}^{\vec{J}_{\text{in}}^{\text{ES}}} & 0 & 0 & 0 & 0 \\ 0 & \bar{\bar{Z}}^{\vec{M}_{\text{in}}^{\text{ES}}\vec{J}^{\text{IS}}} & \bar{\bar{Z}}^{\vec{M}_{\text{in}}^{\text{ES}}\vec{J}_{\text{out}}^{\text{ES}}} & \bar{\bar{Z}}^{\vec{M}_{\text{in}}^{\text{ES}}\vec{M}_{\text{in}}^{\text{ES}}} & \bar{\bar{Z}}^{\vec{M}_{\text{in}}^{\text{ES}}\vec{M}_{\text{out}}^{\text{ES}}} \\ 0 & \bar{\bar{Z}}^{\vec{J}^{\text{IS}}\vec{J}^{\text{IS}}} & \bar{\bar{Z}}^{\vec{J}^{\text{IS}}\vec{J}_{\text{out}}^{\text{ES}}} & \bar{\bar{Z}}^{\vec{J}^{\text{IS}}\vec{M}_{\text{in}}^{\text{ES}}} & \bar{\bar{Z}}^{\vec{J}^{\text{IS}}\vec{M}_{\text{out}}^{\text{ES}}} \\ 0 & \bar{\bar{Z}}^{\vec{J}_{\text{out}}^{\text{ES}}\vec{J}^{\text{IS}}} & \bar{\bar{Z}}^{\vec{J}_{\text{out}}^{\text{ES}}\vec{J}_{\text{out}}^{\text{ES}}} & \bar{\bar{Z}}^{\vec{J}_{\text{out}}^{\text{ES}}\vec{M}_{\text{in}}^{\text{ES}}} & \bar{\bar{Z}}^{\vec{J}_{\text{out}}^{\text{ES}}\vec{M}_{\text{out}}^{\text{ES}}} \\ 0 & \bar{\bar{Z}}^{\vec{M}_{\text{out}}^{\text{ES}}\vec{J}^{\text{IS}}} & \bar{\bar{Z}}^{\vec{M}_{\text{out}}^{\text{ES}}\vec{J}_{\text{out}}^{\text{ES}}} & \bar{\bar{Z}}^{\vec{M}_{\text{out}}^{\text{ES}}\vec{M}_{\text{in}}^{\text{ES}}} & \bar{\bar{Z}}^{\vec{M}_{\text{out}}^{\text{ES}}\vec{M}_{\text{out}}^{\text{ES}}} \end{bmatrix}}_{\bar{\bar{\Psi}}_1} \cdot \underbrace{\begin{bmatrix} \bar{a}^{\vec{J}_{\text{in}}^{\text{ES}}} \\ \bar{a}^{\vec{J}^{\text{IS}}} \\ \bar{a}^{\vec{J}_{\text{out}}^{\text{ES}}} \\ \bar{a}^{\vec{M}_{\text{in}}^{\text{ES}}} \\ \bar{a}^{\vec{M}_{\text{out}}^{\text{ES}}} \end{bmatrix}}_{\bar{a}^{\text{AV}}} = \underbrace{\begin{bmatrix} \bar{\bar{I}}^{\vec{J}_{\text{in}}^{\text{ES}}} \\ -\bar{\bar{Z}}^{\vec{M}_{\text{in}}^{\text{ES}}\vec{J}_{\text{in}}^{\text{ES}}} \\ -\bar{\bar{Z}}^{\vec{J}^{\text{IS}}\vec{J}_{\text{in}}^{\text{ES}}} \\ -\bar{\bar{Z}}^{\vec{J}_{\text{out}}^{\text{ES}}\vec{J}_{\text{in}}^{\text{ES}}} \\ -\bar{\bar{Z}}^{\vec{M}_{\text{out}}^{\text{ES}}\vec{J}_{\text{in}}^{\text{ES}}} \end{bmatrix}}_{\bar{\bar{\Psi}}_2} \cdot \bar{a}^{\vec{J}_{\text{in}}^{\text{ES}}} \tag{3-39}$$

and Eqs. (3-31b), (3-34), (3-36a), and (3-36b) are combined as follows:

$$\underbrace{\begin{bmatrix} 0 & 0 & 0 & \bar{\bar{I}}^{\vec{M}_{\text{in}}^{\text{ES}}} & 0 \\ \bar{\bar{Z}}^{\vec{J}_{\text{in}}^{\text{ES}}\vec{J}_{\text{in}}^{\text{ES}}} & \bar{\bar{Z}}^{\vec{J}_{\text{in}}^{\text{ES}}\vec{J}^{\text{IS}}} & \bar{\bar{Z}}^{\vec{J}_{\text{in}}^{\text{ES}}\vec{J}_{\text{out}}^{\text{ES}}} & 0 & \bar{\bar{Z}}^{\vec{J}_{\text{in}}^{\text{ES}}\vec{M}_{\text{out}}^{\text{ES}}} \\ \bar{\bar{Z}}^{\vec{J}^{\text{IS}}\vec{J}_{\text{in}}^{\text{ES}}} & \bar{\bar{Z}}^{\vec{J}^{\text{IS}}\vec{J}^{\text{IS}}} & \bar{\bar{Z}}^{\vec{J}^{\text{IS}}\vec{J}_{\text{out}}^{\text{ES}}} & 0 & \bar{\bar{Z}}^{\vec{J}^{\text{IS}}\vec{M}_{\text{out}}^{\text{ES}}} \\ \bar{\bar{Z}}^{\vec{J}_{\text{out}}^{\text{ES}}\vec{J}_{\text{in}}^{\text{ES}}} & \bar{\bar{Z}}^{\vec{J}_{\text{out}}^{\text{ES}}\vec{J}^{\text{IS}}} & \bar{\bar{Z}}^{\vec{J}_{\text{out}}^{\text{ES}}\vec{J}_{\text{out}}^{\text{ES}}} & 0 & \bar{\bar{Z}}^{\vec{J}_{\text{out}}^{\text{ES}}\vec{M}_{\text{out}}^{\text{ES}}} \\ \bar{\bar{Z}}^{\vec{M}_{\text{out}}^{\text{ES}}\vec{J}_{\text{in}}^{\text{ES}}} & \bar{\bar{Z}}^{\vec{M}_{\text{out}}^{\text{ES}}\vec{J}^{\text{IS}}} & \bar{\bar{Z}}^{\vec{M}_{\text{out}}^{\text{ES}}\vec{J}_{\text{out}}^{\text{ES}}} & 0 & \bar{\bar{Z}}^{\vec{M}_{\text{out}}^{\text{ES}}\vec{M}_{\text{out}}^{\text{ES}}} \end{bmatrix}}_{\bar{\bar{\Psi}}_3} \cdot \underbrace{\begin{bmatrix} \bar{a}^{\vec{J}_{\text{in}}^{\text{ES}}} \\ \bar{a}^{\vec{J}^{\text{IS}}} \\ \bar{a}^{\vec{J}_{\text{out}}^{\text{ES}}} \\ \bar{a}^{\vec{M}_{\text{in}}^{\text{ES}}} \\ \bar{a}^{\vec{M}_{\text{out}}^{\text{ES}}} \end{bmatrix}}_{\bar{a}^{\text{AV}}} = \underbrace{\begin{bmatrix} \bar{\bar{I}}^{\vec{J}_{\text{in}}^{\text{ES}}} \\ -\bar{\bar{Z}}^{\vec{J}_{\text{in}}^{\text{ES}}\vec{M}_{\text{in}}^{\text{ES}}} \\ -\bar{\bar{Z}}^{\vec{J}^{\text{IS}}\vec{M}_{\text{in}}^{\text{ES}}} \\ -\bar{\bar{Z}}^{\vec{J}_{\text{out}}^{\text{ES}}\vec{M}_{\text{in}}^{\text{ES}}} \\ -\bar{\bar{Z}}^{\vec{M}_{\text{out}}^{\text{ES}}\vec{M}_{\text{in}}^{\text{ES}}} \end{bmatrix}}_{\bar{\bar{\Psi}}_4} \cdot \bar{a}^{\vec{M}_{\text{in}}^{\text{ES}}} \tag{3-40}$$

where $\bar{\bar{I}}^{\vec{J}_{\text{in}}^{\text{ES}}}$ and $\bar{\bar{I}}^{\vec{M}_{\text{in}}^{\text{ES}}}$ are the *identity matrices* whose orders are the same as the row numbers of $\bar{a}^{\vec{J}_{\text{in}}^{\text{ES}}}$ and $\bar{a}^{\vec{M}_{\text{in}}^{\text{ES}}}$ respectively, and the superscript "AV" used in $\bar{a}^{\text{AV}}$ is to emphasize that the vector is constituted by the expansion coefficients of *all variables* (*AVs*) $\{\bar{a}^{\vec{J}_{\text{in}}^{\text{ES}}},\bar{a}^{\vec{J}^{\text{IS}}},\bar{a}^{\vec{J}_{\text{out}}^{\text{ES}}},\bar{a}^{\vec{M}_{\text{in}}^{\text{ES}}},\bar{a}^{\vec{M}_{\text{out}}^{\text{ES}}}\}$.

By solving the above Eqs. (3-39) and (3-40), the following transformations are obtained:

$$\bar{a}^{\text{AV}} = \overbrace{\left[\left(\bar{\bar{\Psi}}_1\right)^{-1} \cdot \bar{\bar{\Psi}}_2\right]}^{\bar{\bar{T}}^{\vec{J}_{\text{in}}^{\text{ES}} \to \text{AV}}} \cdot \bar{a}^{\vec{J}_{\text{in}}^{\text{ES}}} \tag{3-41}$$

$$\bar{a}^{\text{AV}} = \underbrace{\left[\left(\bar{\bar{\Psi}}_3\right)^{-1} \cdot \bar{\bar{\Psi}}_4\right]}_{\bar{\bar{T}}^{\vec{M}_{\text{in}}^{\text{ES}} \to \text{AV}}} \cdot \bar{a}^{\vec{M}_{\text{in}}^{\text{ES}}} \tag{3-42}$$

where the superscript “$-1$” is the *inversion operation* for a *non-singular matrix*. For simplifying the symbolic system of this report, the above two transformations are uniformly written as follows:

$$\overline{a}^{\rm AV} = \overline{\overline{T}}^{\rm BV\rightarrow AV} \cdot \overline{a}^{\rm BV} \tag{3-43}$$

where the superscript “BV” used in $\overline{a}^{\rm BV}$ is the acronym for the terminology “*basic variables*” (basic variables are the ones which are not only independent but also complete[13]).

In the future Sec. 3.2.3, the above transformation will be utilized to eliminate the dependent variables contained in the *generating operator of IP-DMs*, so this report calls the scheme *dependent variable elimination* (*DVE*).

#### 3.2.2.4 Mathematical Description for Modal Space — Scheme II: Solution/Definition Domain Compression (SDC/DDC)

In fact, the previous Eqs. (3-31a)/(3-31b), (3-34), (3-36a), and (3-36b) can be alternatively combined as follows:

$$\overbrace{\begin{bmatrix} \overline{\overline{Z}}^{\vec{M}_{\rm in}^{\rm ES}\vec{J}_{\rm in}^{\rm ES}} & \overline{\overline{Z}}^{\vec{M}_{\rm in}^{\rm ES}\vec{J}^{\rm IS}} & \overline{\overline{Z}}^{\vec{M}_{\rm in}^{\rm ES}\vec{J}_{\rm out}^{\rm ES}} & \overline{\overline{Z}}^{\vec{M}_{\rm in}^{\rm ES}\vec{M}_{\rm in}^{\rm ES}} & \overline{\overline{Z}}^{\vec{M}_{\rm in}^{\rm ES}\vec{M}_{\rm out}^{\rm ES}} \\ \overline{\overline{Z}}^{\vec{J}^{\rm IS}\vec{J}_{\rm in}^{\rm ES}} & \overline{\overline{Z}}^{\vec{J}^{\rm IS}\vec{J}^{\rm IS}} & \overline{\overline{Z}}^{\vec{J}^{\rm IS}\vec{J}_{\rm out}^{\rm ES}} & \overline{\overline{Z}}^{\vec{J}^{\rm IS}\vec{M}_{\rm in}^{\rm ES}} & \overline{\overline{Z}}^{\vec{J}^{\rm IS}\vec{M}_{\rm out}^{\rm ES}} \\ \overline{\overline{Z}}^{\vec{J}_{\rm out}^{\rm ES}\vec{J}_{\rm in}^{\rm ES}} & \overline{\overline{Z}}^{\vec{J}_{\rm out}^{\rm ES}\vec{J}^{\rm IS}} & \overline{\overline{Z}}^{\vec{J}_{\rm out}^{\rm ES}\vec{J}_{\rm out}^{\rm ES}} & \overline{\overline{Z}}^{\vec{J}_{\rm out}^{\rm ES}\vec{M}_{\rm in}^{\rm ES}} & \overline{\overline{Z}}^{\vec{J}_{\rm out}^{\rm ES}\vec{M}_{\rm out}^{\rm ES}} \\ \overline{\overline{Z}}^{\vec{M}_{\rm out}^{\rm ES}\vec{J}_{\rm in}^{\rm ES}} & \overline{\overline{Z}}^{\vec{M}_{\rm out}^{\rm ES}\vec{J}^{\rm IS}} & \overline{\overline{Z}}^{\vec{M}_{\rm out}^{\rm ES}\vec{J}_{\rm out}^{\rm ES}} & \overline{\overline{Z}}^{\vec{M}_{\rm out}^{\rm ES}\vec{M}_{\rm in}^{\rm ES}} & \overline{\overline{Z}}^{\vec{M}_{\rm out}^{\rm ES}\vec{M}_{\rm out}^{\rm ES}} \end{bmatrix}}^{\overline{\overline{\Psi}}_{\rm FCE}^{\rm DoJ}} \cdot \overline{a}^{\rm AV} = 0 \tag{3-44}$$

$$\underbrace{\begin{bmatrix} \overline{\overline{Z}}^{\vec{J}_{\rm in}^{\rm ES}\vec{J}_{\rm in}^{\rm ES}} & \overline{\overline{Z}}^{\vec{J}_{\rm in}^{\rm ES}\vec{J}^{\rm IS}} & \overline{\overline{Z}}^{\vec{J}_{\rm in}^{\rm ES}\vec{J}_{\rm out}^{\rm ES}} & \overline{\overline{Z}}^{\vec{J}_{\rm in}^{\rm ES}\vec{M}_{\rm in}^{\rm ES}} & \overline{\overline{Z}}^{\vec{J}_{\rm in}^{\rm ES}\vec{M}_{\rm out}^{\rm ES}} \\ \overline{\overline{Z}}^{\vec{J}^{\rm IS}\vec{J}_{\rm in}^{\rm ES}} & \overline{\overline{Z}}^{\vec{J}^{\rm IS}\vec{J}^{\rm IS}} & \overline{\overline{Z}}^{\vec{J}^{\rm IS}\vec{J}_{\rm out}^{\rm ES}} & \overline{\overline{Z}}^{\vec{J}^{\rm IS}\vec{M}_{\rm in}^{\rm ES}} & \overline{\overline{Z}}^{\vec{J}^{\rm IS}\vec{M}_{\rm out}^{\rm ES}} \\ \overline{\overline{Z}}^{\vec{J}_{\rm out}^{\rm ES}\vec{J}_{\rm in}^{\rm ES}} & \overline{\overline{Z}}^{\vec{J}_{\rm out}^{\rm ES}\vec{J}^{\rm IS}} & \overline{\overline{Z}}^{\vec{J}_{\rm out}^{\rm ES}\vec{J}_{\rm out}^{\rm ES}} & \overline{\overline{Z}}^{\vec{J}_{\rm out}^{\rm ES}\vec{M}_{\rm in}^{\rm ES}} & \overline{\overline{Z}}^{\vec{J}_{\rm out}^{\rm ES}\vec{M}_{\rm out}^{\rm ES}} \\ \overline{\overline{Z}}^{\vec{M}_{\rm out}^{\rm ES}\vec{J}_{\rm in}^{\rm ES}} & \overline{\overline{Z}}^{\vec{M}_{\rm out}^{\rm ES}\vec{J}^{\rm IS}} & \overline{\overline{Z}}^{\vec{M}_{\rm out}^{\rm ES}\vec{J}_{\rm out}^{\rm ES}} & \overline{\overline{Z}}^{\vec{M}_{\rm out}^{\rm ES}\vec{M}_{\rm in}^{\rm ES}} & \overline{\overline{Z}}^{\vec{M}_{\rm out}^{\rm ES}\vec{M}_{\rm out}^{\rm ES}} \end{bmatrix}}_{\overline{\overline{\Psi}}_{\rm FCE}^{\rm DoM}} \cdot \overline{a}^{\rm AV} = 0 \tag{3-45}$$

where the subscript “FCE” is the acronym of “field continuation equation”①, and the superscripts “DoJ” and “DoM” are the acronyms of “definition of $\vec{J}_{\rm in}^{\rm ES}$” and “definition of $\vec{M}_{\rm in}^{\rm ES}$” respectively②.

Equations (3-44) and (3-45) are theoretically equivalent to each other, and all their solutions constitute the modal space constituted by all the travelling-wave modes. If the

① because Eqs. (3-34), (3-36a), and (3-36b) are obtained from discretizing field continuation equations (3-27), (3-28a), and (3-28b).
② because Eqs. (3-31a) and (3-31b) are obtained from discretizing current definitions (3-25a) and (3-25b).

*basic solutions* (*BSs*) of the matrix equations are denoted as $\{\bar{s}_1^{\rm BS}, \bar{s}_2^{\rm BS}, \cdots\}$, then any travelling-wave mode in the modal space can be expanded in terms of the basic solutions as follows:

$$\bar{a}^{\rm AV} = \sum_i a_i^{\rm BS} \bar{s}_i^{\rm BS} = \underbrace{\left[ \bar{s}_1^{\rm BS}, \bar{s}_2^{\rm BS}, \cdots \right]}_{\bar{\bar{T}}^{\rm BS \to AV}} \cdot \underbrace{\begin{bmatrix} a_1^{\rm BS} \\ a_2^{\rm BS} \\ \vdots \end{bmatrix}}_{\bar{a}^{\rm BS}} \tag{3-46}$$

Obviously, we have the following relations:

$$\begin{aligned} \text{Modal Space} &= \text{nullspace}\left( \bar{\bar{\Psi}}_{\rm FCE}^{\rm DoJ/DoM} \right) \\ &= \text{span}\left\{ \bar{s}_1^{\rm BS}, \bar{s}_2^{\rm BS}, \cdots \right\} \\ &= \text{range}\left( \bar{\bar{T}}^{\rm BS \to AV} \right) \end{aligned} \tag{3-47}$$

where $\text{nullspace}(\bar{\bar{\Psi}}_{\rm FCE}^{\rm DoJ/DoM})$ denotes the *null space* of matrix $\bar{\bar{\Psi}}_{\rm FCE}^{\rm DoJ/DoM}$, and $\text{span}\{\bar{s}_1^{\rm BS}, \bar{s}_2^{\rm BS}, \cdots\}$ denotes the linear space spanned by $\{\bar{s}_1^{\rm BS}, \bar{s}_2^{\rm BS}, \cdots\}$, and $\text{range}(\bar{\bar{T}}^{\rm BS \to AV})$ denotes the *range* of matrix $\bar{\bar{T}}^{\rm BS \to AV}$.

In the future Sec. 3.2.3, the above Eq. (3-46) will be utilized to compress the *solution domain of the equation for calculating IP-DMs* (i.e. the *definition domain of the operator for generating IP-DMs*), so this report calls the scheme *solution/definition domain compression* (*SDC/DDC*).

## 3.2.3 Input Power Operator

This subsection discusses the *input power operator* (*IPO*), which is the *generating operator* of IP-DMs, and provides two different formulations for expressing the IPO.

### 3.2.3.1 Formulation I: Current Form

The IPO corresponding to port $\mathbb{S}_{\rm in}$ is as follows:

$$P^{\rm in} = (1/2) \iint_{\mathbb{S}_{\rm in}} \left( \vec{E} \times \vec{H}^\dagger \right) \cdot \hat{z} dS = (1/2) \left\langle \hat{z} \times \vec{J}_{\rm in}^{\rm ES}, \vec{M}_{\rm in}^{\rm ES} \right\rangle_{\mathbb{S}_{\rm in}} \tag{3-48}$$

where the second equality is based on Eqs. (3-5a) and (3-5b). Inserting Eq. (3-29a) into the above IPO (3-48), the operator is immediately discretized as follows:

$$P^{\rm in} = \left( \bar{a}^{\rm AV} \right)^\dagger \cdot \underbrace{\begin{bmatrix} 0 & 0 & 0 & \bar{\bar{C}}^{\vec{J}_{\rm in}^{\rm ES} \vec{M}_{\rm in}^{\rm ES}} & 0 \\ 0 & 0 & 0 & 0 & 0 \\ 0 & 0 & 0 & 0 & 0 \\ 0 & 0 & 0 & 0 & 0 \\ 0 & 0 & 0 & 0 & 0 \end{bmatrix}}_{\bar{\bar{P}}_{\rm curAV}^{\rm in}} \cdot \bar{a}^{\rm AV} \tag{3-49}$$

where the superscript "†" represents the *transpose conjugate operation* for a matrix or vector, and the elements of the sub-matrix $\overline{\overline{C}}^{\vec{J}_{\text{in}}^{\text{ES}}\vec{M}_{\text{in}}^{\text{ES}}}$ are calculated as that $c_{\xi\zeta}^{\vec{J}_{\text{in}}^{\text{ES}}\vec{M}_{\text{in}}^{\text{ES}}} = (1/2) < \hat{z}\times\vec{b}_{\xi}^{\vec{J}_{\text{in}}^{\text{ES}}}, \vec{b}_{\zeta}^{\vec{M}_{\text{in}}^{\text{ES}}} >_{\mathbb{S}_{\text{in}}}$.

Inserting Eqs. (3-43) and (3-46) into IPO (3-49), the IPO can be further written as follows:

$$P^{\text{in}} = \left(\bar{a}^{\text{BV}}\right)^{\dagger}\cdot\overbrace{\left[\left(\overline{\overline{T}}^{\text{BV}\rightarrow\text{AV}}\right)^{\dagger}\cdot\overline{\overline{P}}_{\text{curAV}}^{\text{in}}\cdot\overline{\overline{T}}^{\text{BV}\rightarrow\text{AV}}\right]}^{\overline{\overline{P}}_{\text{cur BV}}^{\text{in}}}\cdot\bar{a}^{\text{BV}} \tag{3-50}$$

$$P^{\text{in}} = \left(\bar{a}^{\text{BS}}\right)^{\dagger}\cdot\underbrace{\left[\left(\overline{\overline{T}}^{\text{BS}\rightarrow\text{AV}}\right)^{\dagger}\cdot\overline{\overline{P}}_{\text{cur AV}}^{\text{in}}\cdot\overline{\overline{T}}^{\text{BS}\rightarrow\text{AV}}\right]}_{\overline{\overline{P}}_{\text{curBS}}^{\text{in}}}\cdot\bar{a}^{\text{BS}} \tag{3-51}$$

or uniformly written as follows:

$$P^{\text{in}} = \bar{a}^{\dagger}\cdot\underbrace{\left(\overline{\overline{T}}^{\dagger}\cdot\overline{\overline{P}}_{\text{cur AV}}^{\text{in}}\cdot\overline{\overline{T}}\right)}_{\overline{\overline{P}}_{\text{cur}}^{\text{in}}}\cdot\bar{a} \tag{3-52}$$

where $\overline{\overline{T}} = \overline{\overline{T}}^{\text{BV}\rightarrow\text{AV}} / \overline{\overline{T}}^{\text{BS}\rightarrow\text{AV}}$ and $\overline{\overline{P}}_{\text{cur}}^{\text{in}} = \overline{\overline{P}}_{\text{cur BV}}^{\text{in}} / \overline{\overline{P}}_{\text{curBS}}^{\text{in}}$ and $\bar{a} = \bar{a}^{\text{BV}} / \bar{a}^{\text{BS}}$, and the subscript "cur" is to emphasize that $\overline{\overline{P}}_{\text{cur}}^{\text{in}}$ originates from discretizing the current form of IPO.

### 3.2.3.2 Formulation II: Field-Current Interaction Form

In fact, the IPO (3-48) also has the following equivalent expressions

$$\begin{aligned} P^{\text{in}} &= -(1/2)\left\langle \vec{J}_{\text{in}}^{\text{ES}}, \vec{E}\right\rangle_{\mathbb{S}_{\text{in}}} \\ &= -(1/2)\left\langle \vec{J}_{\text{in}}^{\text{ES}}, \mathcal{E}\left(\vec{J}_{\text{in}}^{\text{ES}} + \vec{J}^{\text{IS}} - \vec{J}_{\text{out}}^{\text{ES}}, \vec{M}_{\text{in}}^{\text{ES}} - \vec{M}_{\text{out}}^{\text{ES}}\right)\right\rangle_{\mathbb{S}_{\text{in}}^{+}} \end{aligned} \tag{3-53}$$

and

$$\begin{aligned} P^{\text{in}} &= -(1/2)\left\langle \vec{H}, \vec{M}_{\text{in}}^{\text{ES}}\right\rangle_{\mathbb{S}_{\text{in}}} \\ &= -(1/2)\left\langle \vec{M}_{\text{in}}^{\text{ES}}, \vec{H}\right\rangle_{\mathbb{S}_{\text{in}}}^{\dagger} \\ &= -(1/2)\left\langle \vec{M}_{\text{in}}^{\text{ES}}, \mathcal{H}\left(\vec{J}_{\text{in}}^{\text{ES}} + \vec{J}^{\text{IS}} - \vec{J}_{\text{out}}^{\text{ES}}, \vec{M}_{\text{in}}^{\text{ES}} - \vec{M}_{\text{out}}^{\text{ES}}\right)\right\rangle_{\mathbb{S}_{\text{in}}^{+}}^{\dagger} \end{aligned} \tag{3-54}$$

and they are particularly called the *field-current interaction forms of IPO* because they are expressed in terms of the interactions between fields and currents. In Eq. (3-53), the first equality is based on Eqs. (3-5a) and (3-5b); the second equality is based on Eq. (3-12). The equalities in Eq. (3-54) can be similarly explained.

Inserting Eqs. (3-29a)~(3-29c) into Eqs. (3-53) and (3-54), the interaction forms are immediately discretized as follows:

$$P^{\text{in}} = \left(\bar{a}^{\text{AV}}\right)^{\dagger} \cdot \overline{\overline{P}}{}^{\text{in}}_{\text{intAV}} \cdot \bar{a}^{\text{AV}} \tag{3-55}$$

in which

$$\overline{\overline{P}}{}^{\text{in}}_{\text{intAV}} = \begin{cases} \begin{bmatrix} \overline{\overline{P}}{}^{\vec{J}^{\text{ES}}_{\text{in}}\vec{J}^{\text{ES}}_{\text{in}}} & \overline{\overline{P}}{}^{\vec{J}^{\text{ES}}_{\text{in}}\vec{J}^{\text{IS}}} & \overline{\overline{P}}{}^{\vec{J}^{\text{ES}}_{\text{in}}\vec{J}^{\text{ES}}_{\text{out}}} & \overline{\overline{P}}{}^{\vec{J}^{\text{ES}}_{\text{in}}\vec{M}^{\text{ES}}_{\text{in}}} & \overline{\overline{P}}{}^{\vec{J}^{\text{ES}}_{\text{in}}\vec{M}^{\text{ES}}_{\text{out}}} \\ 0 & 0 & 0 & 0 & 0 \\ 0 & 0 & 0 & 0 & 0 \\ 0 & 0 & 0 & 0 & 0 \\ 0 & 0 & 0 & 0 & 0 \end{bmatrix} &, \text{ for Eq.}(3\text{-}53) \\ \begin{bmatrix} 0 & 0 & 0 & 0 & 0 \\ 0 & 0 & 0 & 0 & 0 \\ 0 & 0 & 0 & 0 & 0 \\ \overline{\overline{P}}{}^{\vec{M}^{\text{ES}}_{\text{in}}\vec{J}^{\text{ES}}_{\text{in}}} & \overline{\overline{P}}{}^{\vec{M}^{\text{ES}}_{\text{in}}\vec{J}^{\text{IS}}} & \overline{\overline{P}}{}^{\vec{M}^{\text{ES}}_{\text{in}}\vec{J}^{\text{ES}}_{\text{out}}} & \overline{\overline{P}}{}^{\vec{M}^{\text{ES}}_{\text{in}}\vec{M}^{\text{ES}}_{\text{in}}} & \overline{\overline{P}}{}^{\vec{M}^{\text{ES}}_{\text{in}}\vec{M}^{\text{ES}}_{\text{out}}} \\ 0 & 0 & 0 & 0 & 0 \end{bmatrix}^{\dagger} &, \text{ for Eq.}(3\text{-}54) \end{cases} \tag{3-56}$$

where the 0s are some *zero matrices* with proper row and column numbers, and the elements of the non-zero sub-matrices are calculated as follows:

$$p_{\xi\zeta}^{\vec{J}^{\text{ES}}_{\text{in}}\vec{J}^{\text{ES}}_{\text{in}}} = -(1/2)\left\langle \vec{b}_{\xi}^{\vec{J}^{\text{ES}}_{\text{in}}}, -j\omega\mu\mathcal{L}\left(\vec{b}_{\zeta}^{\vec{J}^{\text{ES}}_{\text{in}}}\right)\right\rangle_{\mathbb{S}_{\text{in}}} \tag{3-57a}$$

$$p_{\xi\zeta}^{\vec{J}^{\text{ES}}_{\text{in}}\vec{J}^{\text{IS}}} = -(1/2)\left\langle \vec{b}_{\xi}^{\vec{J}^{\text{ES}}_{\text{in}}}, -j\omega\mu\mathcal{L}\left(\vec{b}_{\zeta}^{\vec{J}^{\text{IS}}}\right)\right\rangle_{\mathbb{S}_{\text{in}}} \tag{3-57b}$$

$$p_{\xi\zeta}^{\vec{J}^{\text{ES}}_{\text{in}}\vec{J}^{\text{ES}}_{\text{out}}} = -(1/2)\left\langle \vec{b}_{\xi}^{\vec{J}^{\text{ES}}_{\text{in}}}, -j\omega\mu\mathcal{L}\left(-\vec{b}_{\zeta}^{\vec{J}^{\text{ES}}_{\text{out}}}\right)\right\rangle_{\mathbb{S}_{\text{in}}} \tag{3-57c}$$

$$p_{\xi\zeta}^{\vec{J}^{\text{ES}}_{\text{in}}\vec{M}^{\text{ES}}_{\text{in}}} = -(1/2)\left\langle \vec{b}_{\xi}^{\vec{J}^{\text{ES}}_{\text{in}}}, \hat{z}\times\frac{1}{2}\vec{b}_{\zeta}^{\vec{M}^{\text{ES}}_{\text{in}}} - \text{P.V.}\mathcal{K}\left(\vec{b}_{\zeta}^{\vec{M}^{\text{ES}}_{\text{in}}}\right)\right\rangle_{\mathbb{S}_{\text{in}}} \tag{3-57d}$$

$$p_{\xi\zeta}^{\vec{J}^{\text{ES}}_{\text{in}}\vec{M}^{\text{ES}}_{\text{out}}} = -(1/2)\left\langle \vec{b}_{\xi}^{\vec{J}^{\text{ES}}_{\text{in}}}, -\mathcal{K}\left(-\vec{b}_{\zeta}^{\vec{M}^{\text{ES}}_{\text{out}}}\right)\right\rangle_{\mathbb{S}_{\text{in}}} \tag{3-57e}$$

and

$$p_{\xi\zeta}^{\vec{M}^{\text{ES}}_{\text{in}}\vec{J}^{\text{ES}}_{\text{in}}} = -(1/2)\left\langle \vec{b}_{\xi}^{\vec{M}^{\text{ES}}_{\text{in}}}, \frac{1}{2}\vec{b}_{\zeta}^{\vec{J}^{\text{ES}}_{\text{in}}}\times\hat{z} + \text{P.V.}\mathcal{K}\left(\vec{b}_{\zeta}^{\vec{J}^{\text{ES}}_{\text{in}}}\right)\right\rangle_{\mathbb{S}_{\text{in}}} \tag{3-57f}$$

$$p_{\xi\zeta}^{\vec{M}^{\text{ES}}_{\text{in}}\vec{J}^{\text{IS}}} = -(1/2)\left\langle \vec{b}_{\xi}^{\vec{M}^{\text{ES}}_{\text{in}}}, \mathcal{K}\left(\vec{b}_{\zeta}^{\vec{J}^{\text{IS}}}\right)\right\rangle_{\mathbb{S}_{\text{in}}} \tag{3-57g}$$

$$p_{\xi\zeta}^{\vec{M}^{\text{ES}}_{\text{in}}\vec{J}^{\text{ES}}_{\text{out}}} = -(1/2)\left\langle \vec{b}_{\xi}^{\vec{M}^{\text{ES}}_{\text{in}}}, \mathcal{K}\left(-\vec{b}_{\zeta}^{\vec{J}^{\text{ES}}_{\text{out}}}\right)\right\rangle_{\mathbb{S}_{\text{in}}} \tag{3-57h}$$

$$p_{\xi\zeta}^{\vec{M}^{\text{ES}}_{\text{in}}\vec{M}^{\text{ES}}_{\text{in}}} = -(1/2)\left\langle \vec{b}_{\xi}^{\vec{M}^{\text{ES}}_{\text{in}}}, -j\omega\varepsilon\mathcal{L}\left(\vec{b}_{\zeta}^{\vec{M}^{\text{ES}}_{\text{in}}}\right)\right\rangle_{\mathbb{S}_{\text{in}}} \tag{3-57i}$$

$$p_{\xi\zeta}^{\vec{M}^{\text{ES}}_{\text{in}}\vec{M}^{\text{ES}}_{\text{out}}} = -(1/2)\left\langle \vec{b}_{\xi}^{\vec{M}^{\text{ES}}_{\text{in}}}, -j\omega\varepsilon\mathcal{L}\left(-\vec{b}_{\zeta}^{\vec{M}^{\text{ES}}_{\text{out}}}\right)\right\rangle_{\mathbb{S}_{\text{in}}} \tag{3-57j}$$

To obtain the IPO defined on modal space, we substitute Eqs. (3-43) and (3-46) into Eq. (3-55), and then we have that

$$P^{\text{in}} = \bar{a}^\dagger \cdot \underbrace{\left( \overline{\overline{T}}^\dagger \cdot \overline{\overline{P}}^{\text{in}}_{\text{intAV}} \cdot \overline{\overline{T}} \right)}_{\overline{\overline{P}}^{\text{in}}_{\text{int}}} \cdot \bar{a} \tag{3-58}$$

where $\overline{\overline{T}} = \overline{\overline{T}}^{\text{BV}\to\text{AV}} / \overline{\overline{T}}^{\text{BS}\to\text{AV}}$ and $\bar{a} = \bar{a}^{\text{BV}} / \bar{a}^{\text{BS}}$, and the subscript "int" is to emphasize that $\overline{\overline{P}}^{\text{in}}_{\text{int}}$ originates from discretizing the interaction form of IPO.

For the convenience of the following discussions, the IPOs (3-52) and (3-58) are uniformly written as follows:

$$P^{\text{in}} = \bar{a}^\dagger \cdot \overline{\overline{P}}^{\text{in}} \cdot \bar{a} \tag{3-59}$$

where $\overline{\overline{P}}^{\text{in}} = \overline{\overline{P}}^{\text{in}}_{\text{cur}} / \overline{\overline{P}}^{\text{in}}_{\text{int}}$ .

### 3.2.4 Input-Power-Decoupled Modes

This subsection focuses on constructing the *input-power-decoupled modes* (*IP-DMs*) in the modal space of the metallic guide shown in Fig. 3-6 (the figure is only a section of whole metallic guide), and discusses some related topics.

#### 3.2.4.1 Construction Method

Quadratic form matrix $\overline{\overline{P}}^{\text{in}}$ can be decomposed as follows:

$$\overline{\overline{P}}^{\text{in}} = \overline{\overline{P}}^{\text{in}}_{+} + j\, \overline{\overline{P}}^{\text{in}}_{-} \tag{3-60}$$

where

$$\overline{\overline{P}}^{\text{in}}_{+} = \frac{1}{2}\left[ \overline{\overline{P}}^{\text{in}} + \left( \overline{\overline{P}}^{\text{in}} \right)^\dagger \right] \tag{3-61a}$$

$$\overline{\overline{P}}^{\text{in}}_{-} = \frac{1}{2j}\left[ \overline{\overline{P}}^{\text{in}} - \left( \overline{\overline{P}}^{\text{in}} \right)^\dagger \right] \tag{3-61b}$$

When matrix $\overline{\overline{P}}^{\text{in}}_{+}$ is positive definite, there exists a non-singular matrix $\overline{\overline{A}}$ such that[46]:

$$\overline{\overline{A}}^\dagger \cdot \overline{\overline{P}}^{\text{in}} \cdot \overline{\overline{A}} = \text{diag}\left\{ P^{\text{in}}_1 , P^{\text{in}}_2 , \cdots \right\} \tag{3-62}$$

and the column vectors of $\overline{\overline{A}}$ can be obtained by solving the following *modal decoupling equation* (or simply called *decoupling equation*)

$$\overline{\overline{P}}^{\text{in}}_{-} \cdot \bar{\alpha}_\xi = \theta_\xi\, \overline{\overline{P}}^{\text{in}}_{+} \cdot \bar{\alpha}_\xi \tag{3-63}$$

Obviously, the structure of the above equation is similar to the one used in the traditional

CMT[8~12]. **The reason to call it modal decoupling equation is that: the modes calculated from the equation are energy-decoupled.** The reason to utilize symbol "$\theta_\xi$" instead of the traditional symbol "$\lambda_\xi$" will be clear in the Chap. 9 of this report.

If some derived modes $\{\bar{\alpha}_1, \bar{\alpha}_2, \cdots, \bar{\alpha}_d\}$ are *d*-order degenerate, then the following *Gram-Schmidt orthogonalization*[46] process is necessary.

$$\left.\begin{aligned} \bar{\alpha}_1 &= \bar{\alpha}_1{}' \\ \bar{\alpha}_2 - \chi_{12}\bar{\alpha}_1{}' &= \bar{\alpha}_2{}' \\ &\cdots \\ \bar{\alpha}_d - \cdots - \chi_{2d}\bar{\alpha}_2{}' - \chi_{1d}\bar{\alpha}_1{}' &= \bar{\alpha}_d{}' \end{aligned}\right\} \tag{3-64}$$

where the coefficients are calculated as follows:

$$\chi_{mn} = \left(\bar{\alpha}_m{}'\right)^\dagger \cdot \bar{\bar{P}}_+^{\text{in}} \cdot \bar{\alpha}_n \Big/ \left(\bar{\alpha}_m{}'\right)^\dagger \cdot \bar{\bar{P}}_+^{\text{in}} \cdot \bar{\alpha}_m{}' \tag{3-65}$$

The above-obtained new modes $\{\bar{\alpha}_1{}', \bar{\alpha}_2{}', \cdots, \bar{\alpha}_d{}'\}$ are input-power-decoupled.

The modal vectors are as follows:

$$\begin{bmatrix} \bar{\alpha}_\xi^{\vec{J}_{\text{in}}^{\text{ES}}} \\ \bar{\alpha}_\xi^{\vec{J}^{\text{IS}}} \\ \bar{\alpha}_\xi^{\vec{J}_{\text{out}}^{\text{ES}}} \\ \bar{\alpha}_\xi^{\vec{M}_{\text{in}}^{\text{ES}}} \\ \bar{\alpha}_\xi^{\vec{M}_{\text{out}}^{\text{ES}}} \end{bmatrix} = \bar{\alpha}_\xi^{\text{AV}} = \bar{\bar{T}} \cdot \bar{\alpha}_\xi \tag{3-66}$$

and then the corresponding modal currents are as follows:

$$\vec{C}_{\text{in};\xi}^{\text{ES}}\left(\vec{r}\right) = \bar{\boldsymbol{B}}^{\vec{C}_{\text{in}}^{\text{ES}}} \cdot \bar{\alpha}_\xi^{\vec{C}_{\text{in}}^{\text{ES}}} \quad , \quad \vec{r} \in \mathbb{S}_{\text{in}} \tag{3-67a}$$

$$\vec{J}_\xi^{\text{IS}}\left(\vec{r}\right) = \bar{\boldsymbol{B}}^{\vec{J}^{\text{IS}}} \cdot \bar{\alpha}_\xi^{\vec{J}^{\text{IS}}} \quad , \quad \vec{r} \in \mathbb{S}_{\text{ele}} \tag{3-67b}$$

$$\vec{C}_{\text{out};\xi}^{\text{ES}}\left(\vec{r}\right) = \bar{\boldsymbol{B}}^{\vec{C}_{\text{out}}^{\text{ES}}} \cdot \bar{\alpha}_\xi^{\vec{C}_{\text{out}}^{\text{ES}}} \quad , \quad \vec{r} \in \mathbb{S}_{\text{out}} \tag{3-67c}$$

and the corresponding modal fields can be expressed in terms of the modal currents as Eq. (3-12).

### 3.2.4.2 Modal Decoupling Relation and Modal Expansion

It is not difficult to prove that the modes derived above satisfy the following *modal decoupling relation*

$$\bar{\alpha}_\xi^\dagger \cdot \bar{\bar{P}}^{\text{in}} \cdot \bar{\alpha}_\zeta = \underbrace{\left[\text{Re}\left\{P_\xi^{\text{in}}\right\} + j\,\text{Im}\left\{P_\xi^{\text{in}}\right\}\right]}_{P_\xi^{\text{in}}} \delta_{\xi\zeta} \xrightarrow{\text{Normalizing Re}\{P_\xi^{\text{in}}\}\text{ to }1} \underbrace{\left(1 + j\,\theta_\xi\right)}_{\text{Normalized } P_\xi^{\text{in}}} \delta_{\xi\zeta} \tag{3-68}$$

in the form of matrix-vector multiplication, where $\delta_{\xi\zeta}$ is *Kronecker's symbol*. Then, we have the following field/current form of decoupling relation

$$\begin{aligned}\left(1+j\,\theta_{\xi}\right)\delta_{\xi\zeta} &= \left(1/2\right)\iint_{\mathbb{S}_{\text{in}}}\left(\vec{E}_{\zeta}\times\vec{H}_{\xi}^{\dagger}\right)\cdot\hat{z}dS \\ &= \left(1/2\right)\left\langle \hat{z}\times\vec{J}_{\text{in};\xi}^{\text{ES}},\vec{M}_{\text{in};\zeta}^{\text{ES}}\right\rangle_{\mathbb{S}_{\text{in}}}\end{aligned}\qquad(3\text{-}69)$$

The physical explanation why $\text{Re}\{P_{\xi}^{\text{in}}\}$ is normalized to 1 has been given in Ref. [18] and the previous Sec. 1.2.4.7. **Obviously, the above Eq. (3-69) is equivalent to the decoupling relation satisfied by classical *eigen-modes*[1,2].**

Because of the completeness of the IP-DMs, any traveling-wave mode in the guide tube can be expanded in terms of the IP-DMs as follows:

$$\bar{a} = \sum\nolimits_{\xi} c_{\xi}\bar{\alpha}_{\xi}\qquad(3\text{-}70)$$

$$\vec{C}^{\text{ES}} = \sum\nolimits_{\xi} c_{\xi}\vec{C}_{\xi}^{\text{ES}}\qquad(3\text{-}71)$$

$$\vec{F} = \sum\nolimits_{\xi} c_{\xi}\vec{F}_{\xi}\qquad(3\text{-}72)$$

Applying the Eqs. (3-68) and (3-69) to the above modal expasnions, we can obtain the explicit expression for the expansion coefficients as follows:

$$\begin{aligned}c_{\xi} &= \frac{\left(1/2\right)\iint_{\mathbb{S}_{\text{in}}}\left(\vec{E}\times\vec{H}_{\xi}^{\dagger}\right)\cdot\hat{z}dS}{1+j\,\theta_{\xi}} = \frac{\left(1/2\right)\iint_{\mathbb{S}_{\text{in}}}\left(\vec{E}_{\xi}\times\vec{H}^{\dagger}\right)\cdot\hat{z}dS}{1+j\,\theta_{\xi}} \\ &= \frac{-\left(1/2\right)\left\langle \vec{J}_{\text{in};\xi}^{\text{ES}},\vec{E}\right\rangle_{\mathbb{S}_{\text{in}}}}{1+j\,\theta_{\xi}} = \frac{-\left(1/2\right)\left\langle \vec{H},\vec{M}_{\text{in};\xi}^{\text{ES}}\right\rangle_{\mathbb{S}_{\text{in}}}}{1+j\,\theta_{\xi}}\end{aligned}\qquad(3\text{-}73)$$

In addition, Eq. (3-69) also implies the following *Parseval's identity*

$$\sum\nolimits_{\xi}\left|c_{\xi}\right|^{2} = \text{Re}\left\{\left(1/2\right)\iint_{\mathbb{S}_{\text{in}}}\left(\vec{E}\times\vec{H}^{\dagger}\right)\cdot\hat{z}dS\right\} = \frac{1}{T}\int_{t_0}^{t_0+T}\left[\iint_{\mathbb{S}_{\text{in}}}\left(\vec{\mathcal{E}}\times\vec{\mathcal{H}}\right)\cdot\hat{z}dS\right]dt\qquad(3\text{-}74)$$

where all the modal real powers $\text{Re}\{P_{\xi}^{\text{in}}\}$ have been normalized to 1.

### 3.2.4.3 Traveling-wave-type IP-DMs

We have *modal input impedance* $Z_{\xi}^{\text{in}}$ as follows:

$$Z_{\xi}^{\text{in}} = \frac{\left(1/2\right)\iint_{\mathbb{S}_{\text{in}}}\left(\vec{E}_{\xi}\times\vec{H}_{\xi}^{\dagger}\right)\cdot\hat{z}dS}{\left(1/2\right)\left\langle \vec{J}_{\text{in};\xi}^{\text{ES}},\vec{J}_{\text{in};\xi}^{\text{ES}}\right\rangle_{\mathbb{S}_{\text{in}}}} = \underbrace{\text{Re}\left\{Z_{\xi}^{\text{in}}\right\}}_{R_{\xi}^{\text{in}}} + j\,\underbrace{\text{Im}\left\{Z_{\xi}^{\text{in}}\right\}}_{X_{\xi}^{\text{in}}}\qquad(3\text{-}75)$$

where $R_{\xi}^{\text{in}}$ and $X_{\xi}^{\text{in}}$ are the *modal input resistance* and *modal input reactance* respectively.

Obviously, both $R_{\xi}^{\text{in}}$ and $G_{\xi}^{\text{in}}$ are real functions about *working frequency* $f$, so the *dispersion curves* of $R_{\xi}^{\text{in}}(f)$ and $G_{\xi}^{\text{in}}(f)$ can be easily obtained. If the condition "$f_{\xi}$ is the second point such that curve $R_{\xi}^{\text{in}}(f)$ or $G_{\xi}^{\text{in}}(f)$ reaches its local maximum" is satisfied, then it can be concluded here that frequency $f_{\xi}$ and the *cut-off frequency* $f_{\xi}^{c}$ *of the* $\xi$*-th traveling-wave IP-DM* satisfy the following *dispersion equation*

$$\left(2\pi f_{\xi}^{\text{c}}\right)^{2}\mu\varepsilon+\left(\frac{2\pi}{L}\right)^{2} = \left(2\pi f_{\xi}\right)^{2}\mu\varepsilon \tag{3-76}$$

based on the conclusions given in Sec. 3.2.1.5, where $L=\lambda_{z}$.

Hence, cut-off frequency $f_{\xi}^{\text{c}}$ can be calculated as follows:

$$f_{\xi}^{\text{c}} = \sqrt{\left(f_{\xi}\right)^{2}-\left(\frac{1}{L\sqrt{\mu\varepsilon}}\right)^{2}} \quad \text{with} \quad L=\lambda_{z} \tag{3-77}$$

Among all $\{f_{\xi}^{\text{c}}\}_{\xi=1}^{\infty}$, the smallest one corresponds to the *dominant traveling-wave mode* (or simply called *dominant mode*) of metallic guide, and it can be determined as follows:

$$f_{1}^{\text{c}} = \min\left\{f_{\xi}^{\text{c}}\right\} \tag{3-78}$$

The cut-off frequency $f_{\zeta}^{\text{c}}$ of *the* $\zeta$ *-th higher order traveling-wave mode* can be determined as the following recursive formulations

$$\left.\begin{aligned} f_{2}^{\text{c}} &= \min\left\{\left\{f_{\xi}^{\text{c}}\right\}\setminus\left\{f_{1}^{\text{c}}\right\}\right\} \\ &\cdots \\ f_{\zeta+1}^{\text{c}} &= \min\left\{\left\{f_{\xi}^{\text{c}}\right\}\setminus\left\{f_{1}^{\text{c}},f_{2}^{\text{c}},\cdots,f_{\zeta}^{\text{c}}\right\}\right\} \end{aligned}\right\} \tag{3-79}$$

In the following subsection, some typical numerical experiments are provided for verifying the results obtained above.

### 3.2.5 Numerical Examples Corresponding to Typical Structures

In this subsection, we consider two different kinds of typical metallic guides *circular metallic guide* and *metallic coaxial line,* and construct their travelling-wave-type IP-DMs by using the formulations given above.

#### 3.2.5.1 Traveling-wave-type IP-DMs of Circular Metallic Guide

A section of circular metallic guide is shown in Fig. 3-7, and its radius and length are 1cm and 5cm respectively. The guide tube is filled with homogeneous isotropic material whose *relative permeability* and *relative permittivity* are 1 and 5 respectively.

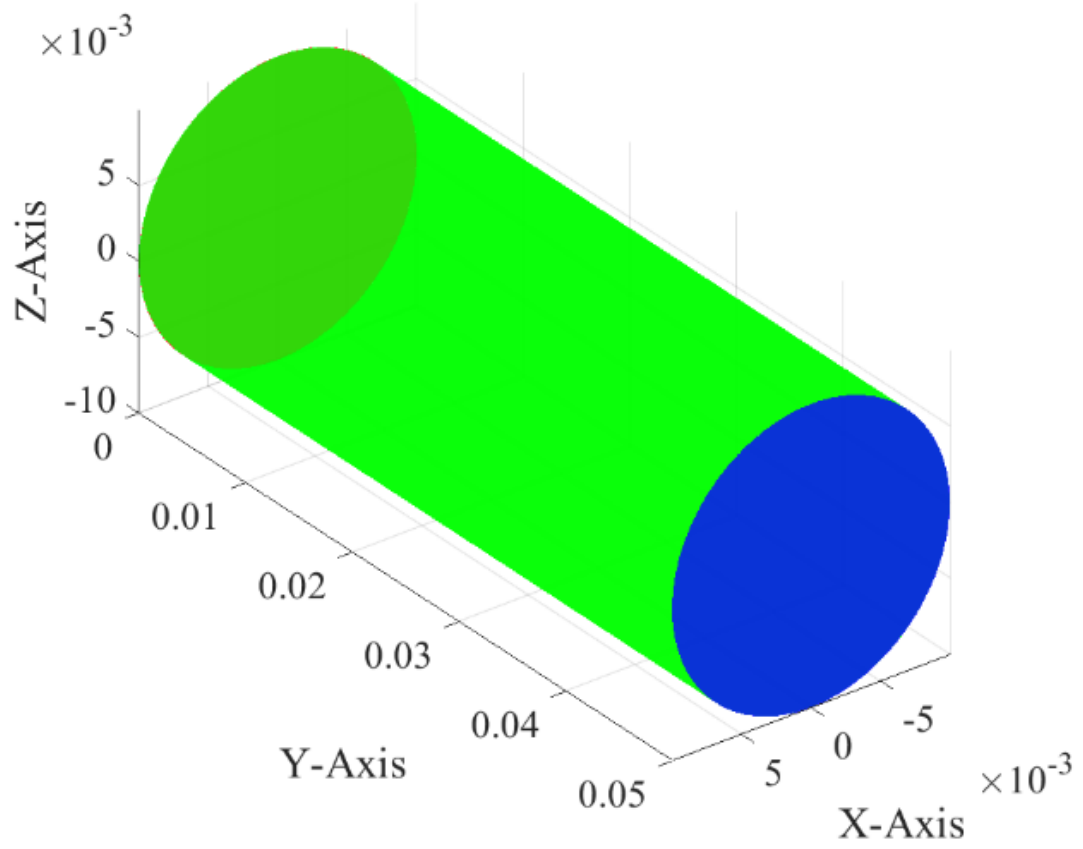

Figure 3-7 Geometry of a section of circular metallic guide

The boundary of the structure shown in Fig. 3-7 can be divided into three parts: *input port* $\mathbb{S}_{in}$, *output port* $\mathbb{S}_{out}$, and *guide electric wall* $\mathbb{S}_{ele}$, as shown in the following Fig. 3-8.

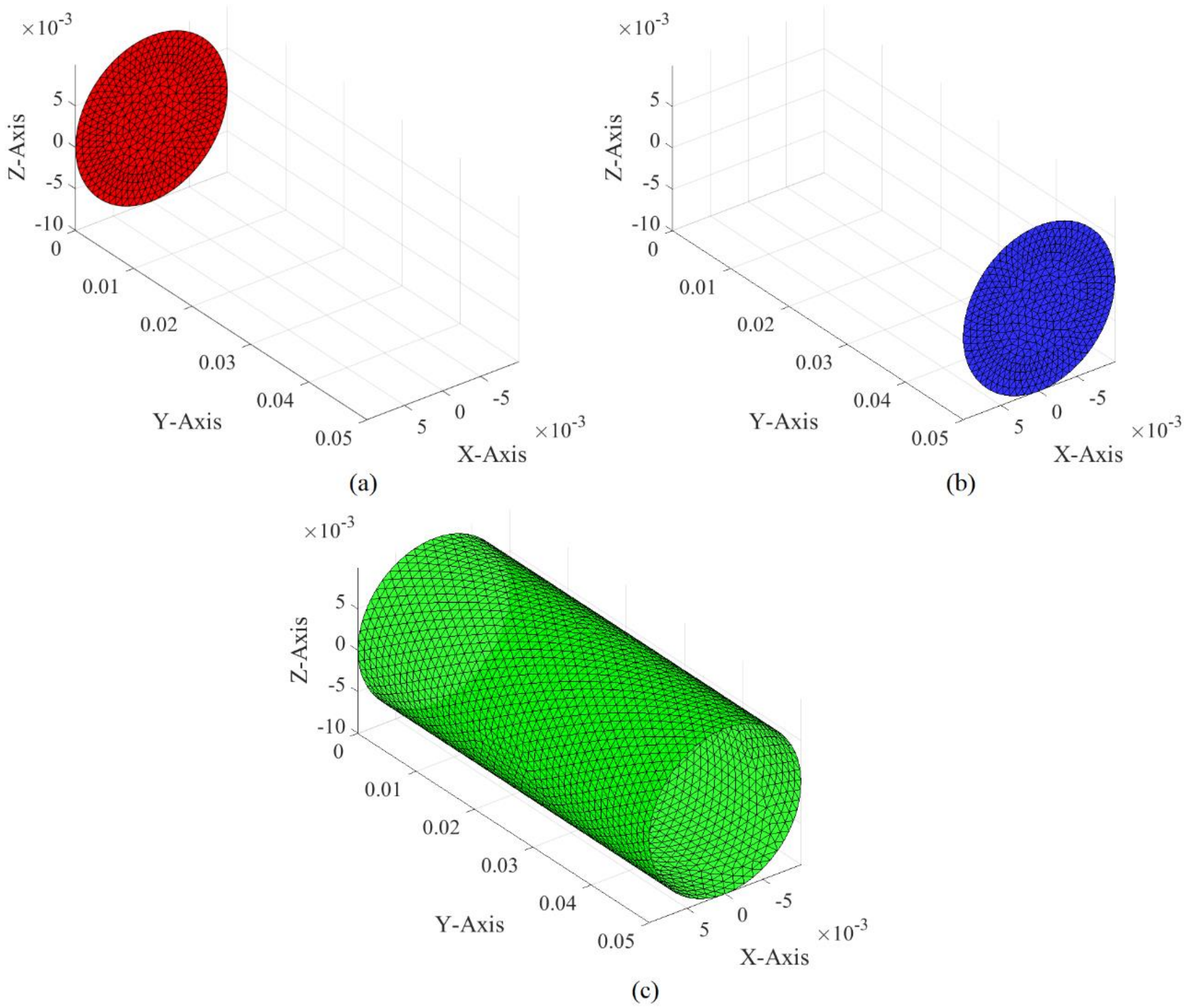

Figure 3-8 Topological structures and surface triangular meshes of the circular metallic guide shown in Fig. 3-7. (a) Mesh of input port $\mathbb{S}_{\text{in}}$; (b) mesh of output port $\mathbb{S}_{\text{out}}$; (c) mesh of guide electric wall $\mathbb{S}_{\text{ele}}$

By solving the Eq. (3-63) originating from Eqs. (3-42) and (3-54), the IP-DMs of the circular metallic guide are constructed, and the modal input resistance curves and modal input conductance curves of the first several typical IP-DMs are plotted in the following Fig. 3-9.

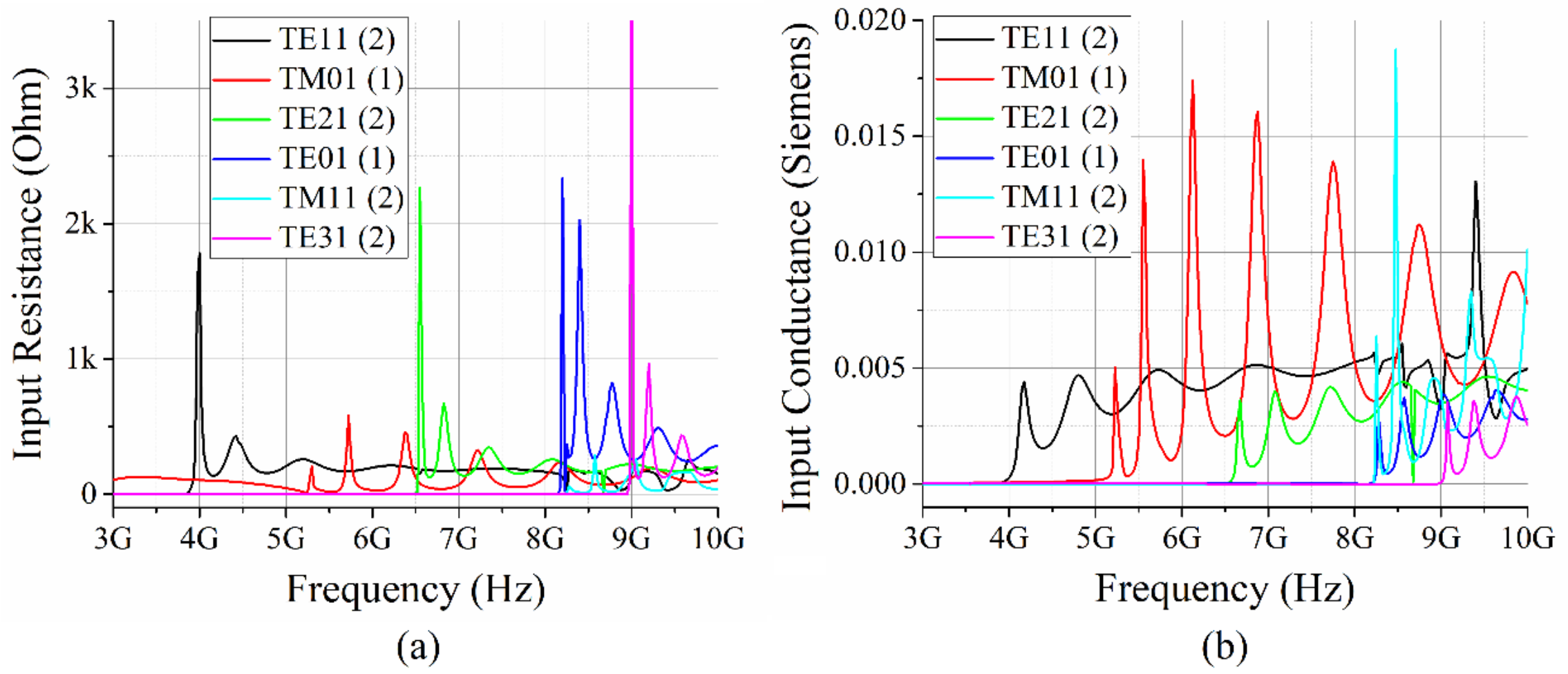

Figure 3-9 Modal input resistance and conductance curves of the first several typical IP-DMs. (a) Resistance curves; (b) conductance curves

Taking the first mode working at 4.175 GHz as an example (corresponding to the first local maximum of the conductance curve), we show its modal equivalent magnetic current $\vec{M}_{\text{in}}^{\text{ES}}$ distributing on input port $\mathbb{S}_{\text{in}}$ in the following Fig. 3-10

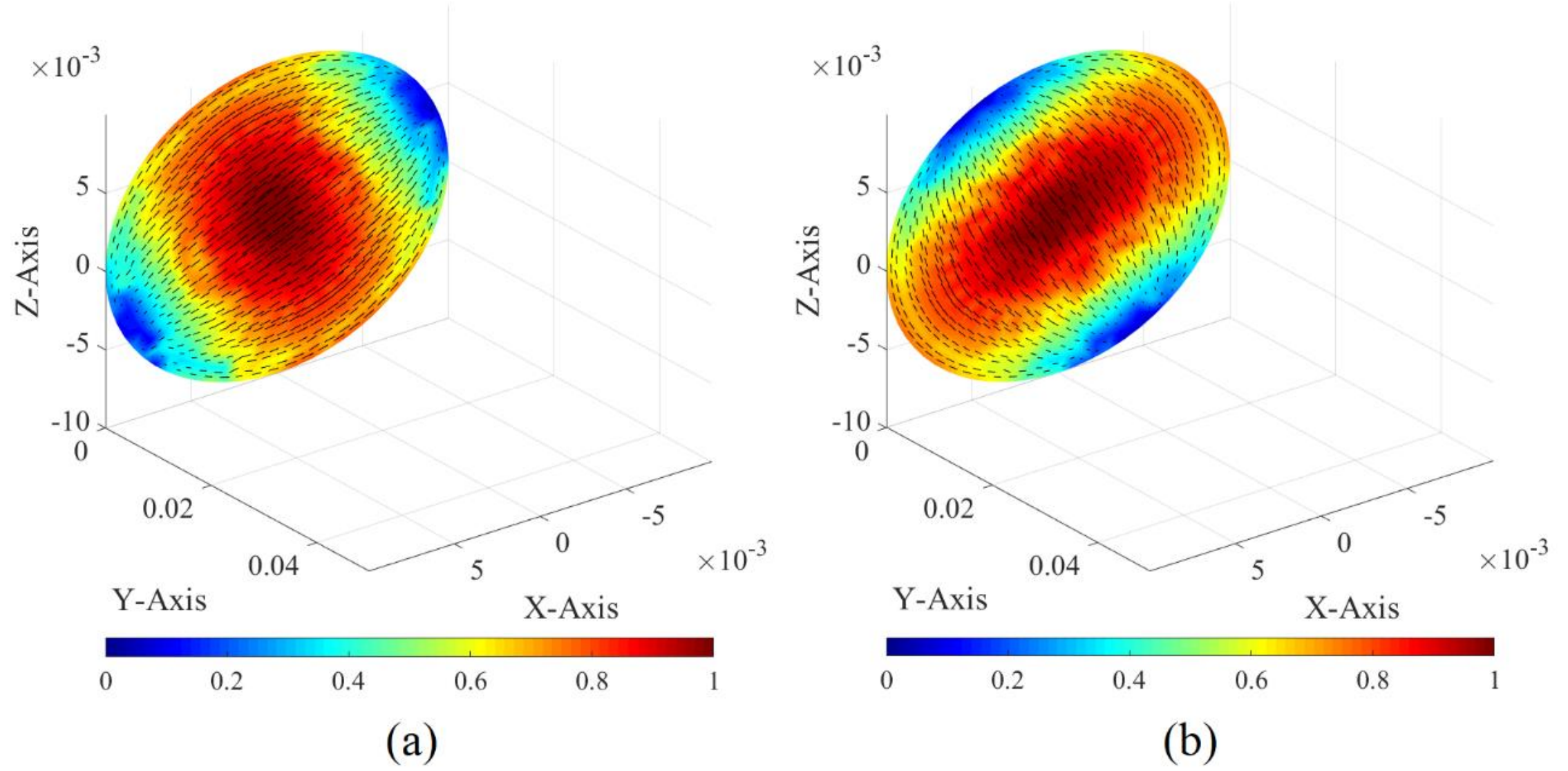

Figure 3-10 Modal equivalent magnetic current $\vec{M}_{\text{in}}^{\text{ES}}$ (of the first mode) distributing on input port $\mathbb{S}_{\text{in}}$. (a) The first degenerate state; (b) the second degenerate state

and the corresponding tangential modal electric field distributing on $\mathbb{S}_{\text{in}}$ can be determined as that $\vec{E}_{\text{tan}} = \hat{z} \times \vec{M}_{\text{in}}^{\text{ES}}$ on $\mathbb{S}_{\text{in}}$. From Fig. 3-10, it is easy to recognize that the mode is just the TE11 mode of circular metallic guide. Now, we separately plot the modal conductance curve of the TE11 mode in the following Fig. 3-11, and mark a series of critical points in the curve.

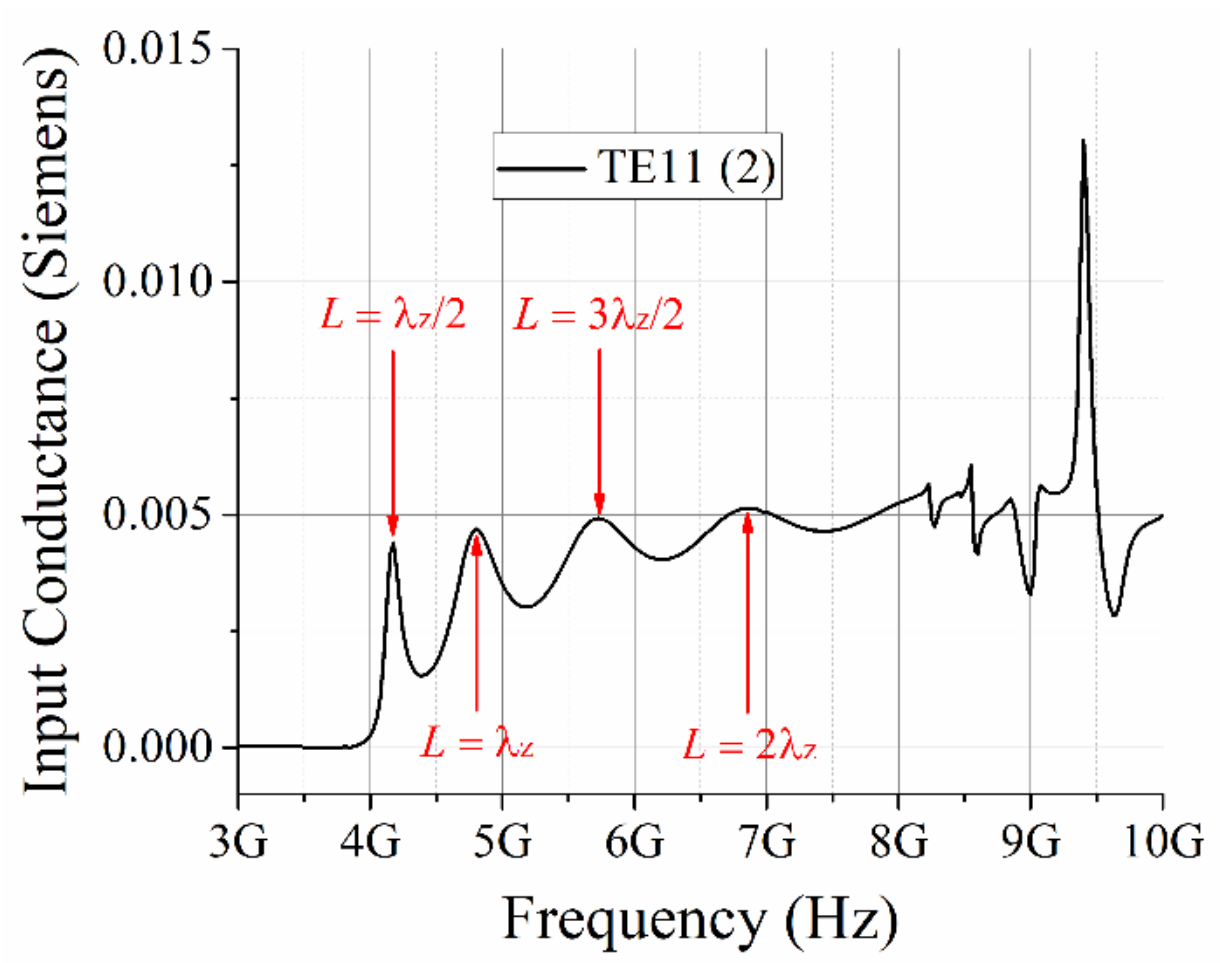

Figure 3-11 Some critical points in the modal input conductance curve of the TE11 mode shown in Figs. 3-9(b) and 3-10

The wall electric current and electric energy density distributions corresponding to the critical points are shown in the following Figs. 3-12 and 3-13 respectively, and the figures imply that: 5cm is equal to the $\lambda_z/2$, $\lambda_z$, $3\lambda_z/2$, and $2\lambda_z$ of the TE11 mode working at 4.175 GHz, 4.800 GHz, 5.725 GHz, and 6.875 GHz respectively.

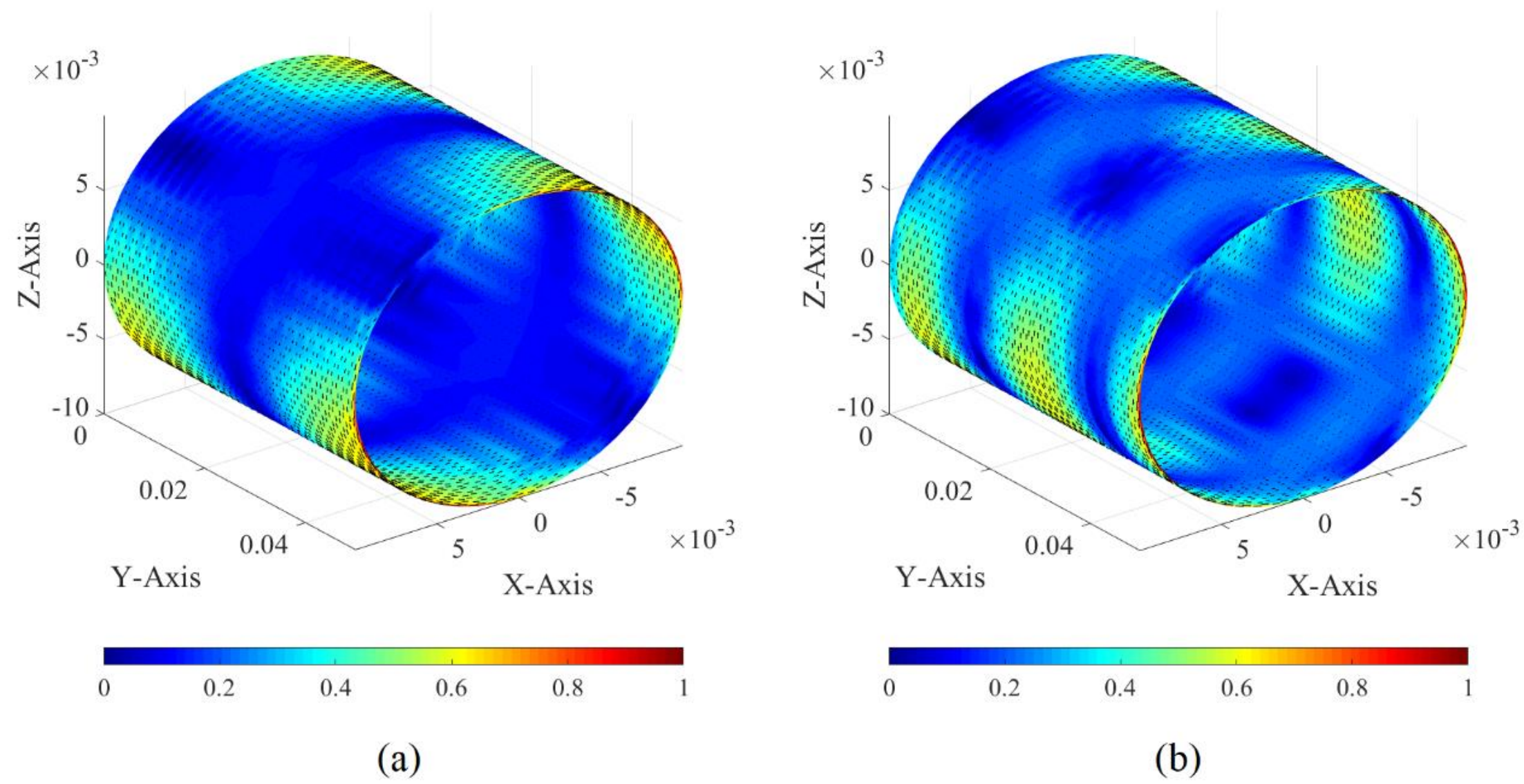

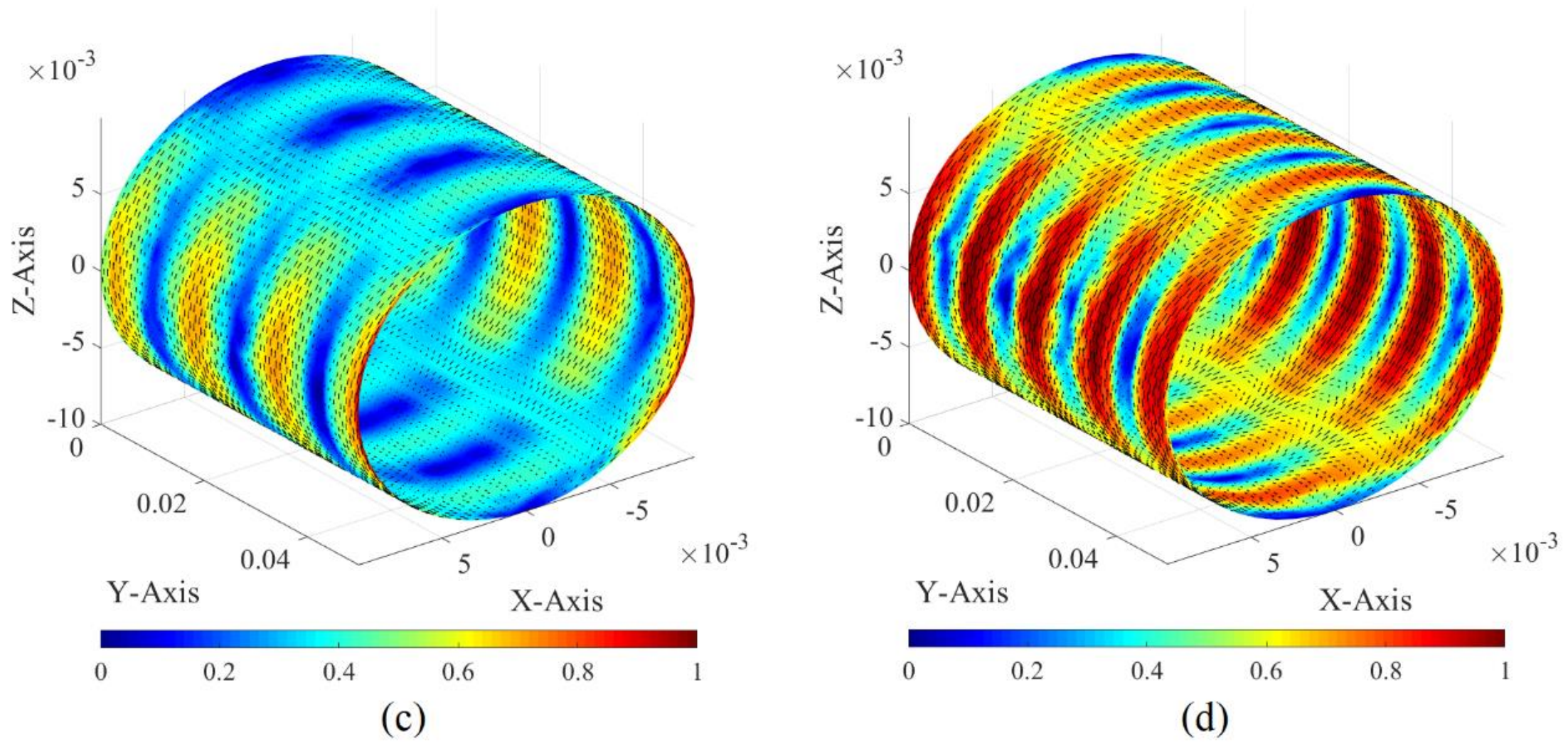

Figure 3-12 TE11 modal electric currents distributing on guide wall $\mathbb{S}_{ele}$. (a) 4.175 GHz; (b) 4.800 GHz; (c) 5.725 GHz; (d) 6.875 GHz

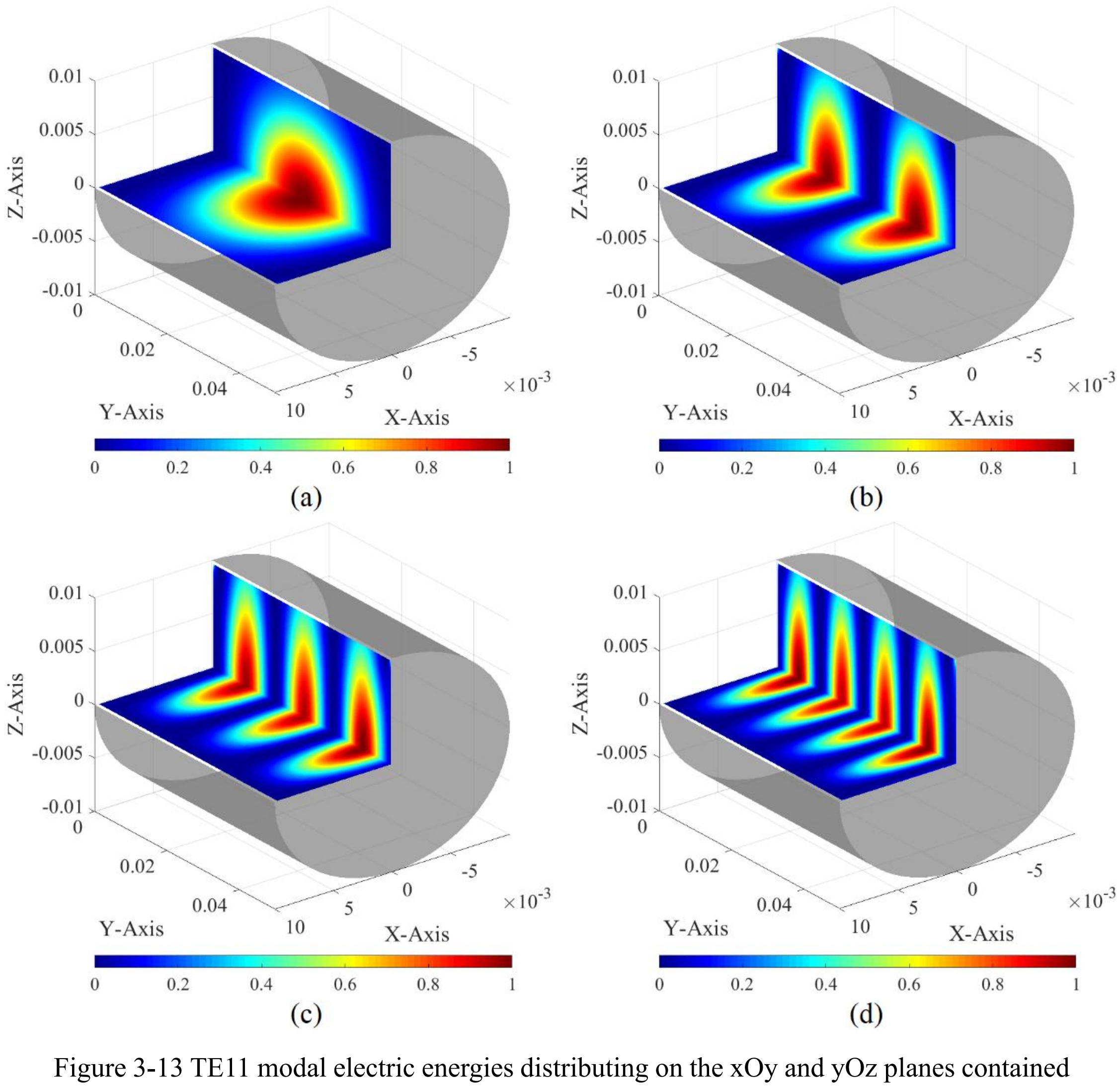

Figure 3-13 TE11 modal electric energies distributing on the xOy and yOz planes contained in guide tube. (a) 4.175 GHz; (b) 4.800 GHz; (c) 5.725 GHz; (d) 6.875 GHz

Based on the conclusions given in this section, it is easy to explain why the conductance curve achieves the local maximums at 4.800 GHz ( $\lambda_z = 5\text{cm}$ ) and 6.875 GHz ( $2\lambda_z = 5\text{cm}$ ). Now, we focus on explaining the reasons leading to the local maximums at 4.175 GHz ( $\lambda_z/2 = 5\text{cm}$ ) and 5.725 GHz ( $3\lambda_z/2 = 5\text{cm}$ ).

Because of the time-harmonic distributions of the modal fields along Z-axis, the modal fields must satisfy the relation that $\vec{F}(z) = -\vec{F}(z+\lambda_z/2) = \vec{F}(z+\lambda_z)$. Based on this observation, we can conclude that: the Eqs. (3-28a) and (3-28b) are also applicable to the case that the distance between $\mathbb{S}_{\text{in}}$ and $\mathbb{S}_{\text{out}}$ is $\lambda_z/2$. The case that the distance between $\mathbb{S}_{\text{in}}$ and $\mathbb{S}_{\text{out}}$ is $3\lambda_z/2$ can be similarly explained.

Because of these above, we can calculate the cutoff frequency of the TE11 mode from substituting $\{f_\xi = 4.175\text{GHz};\ \lambda_z = 10\text{cm};\ \mu = \mu_0,\ \varepsilon = 5\varepsilon_0\}$ into Eq. (3-77), and then the derived cutoff frequency is 3.9539 GHz. Similarly, we can also obtain the cutoff frequencies of the other modes shown in Fig. 3-9, and we list the obtained cutoff frequencies in the following Tab. 3-1.

Table 3-1 The cutoff frequencies (GHz) and the degeneracy degrees corresponding to the first several travelling-wave modes derived from PTT-MetGuid-DMT and the traditional SLT-based EMT for source-free regions (SLT-SFReg-EMT)[3]

| | **PTT-MetGuid-DMP** | | **SLT-SFReg-EMT[3]** |
|---|---|---|---|
| | **Recognized from the Local Maximum of $R_\xi$ Curve** | **Recognized from the Local Maximum of $G_\xi$ Curve** | |
| **TE11** | 3.7686 (2) | **3.9539 (2)** | 3.9288 (2) |
| **TM01** | **5.1276 (1)** | 5.0501 (1) | 5.1316 (1) |
| **TE21** | 6.4113 (2) | **6.5390 (2)** | 6.5171 (2) |
| **TE01** | 8.0897 (1) | **8.1657 (1)** | 8.1763 (1) |
| **TM11** | **8.1657 (2)** | 8.1403 (2) | 8.1763 (2) |
| **TE31** | 8.8996 (2) | **8.9754 (2)** | 8.9646 (2) |

By comparing the results derived from PTT-MetGuid-DMT and the traditional *Sturm-Liouville theory based eigen-mode theory for source-free regions* (*SLT-SFReg-EMT*), it is not difficult to find out that the results are agreed well with each other. At the same time, it is easy to observe that: for the TE modes, the results derived from the conductance curves are more desirable; for the TM modes, the results derived from the resistance curves are more desirable.

In addition, we also plot their modal equivalent electric or magnetic current distributions on input port $\mathbb{S}_{\text{in}}$ in the following Fig. 3-14 for readers' reference.

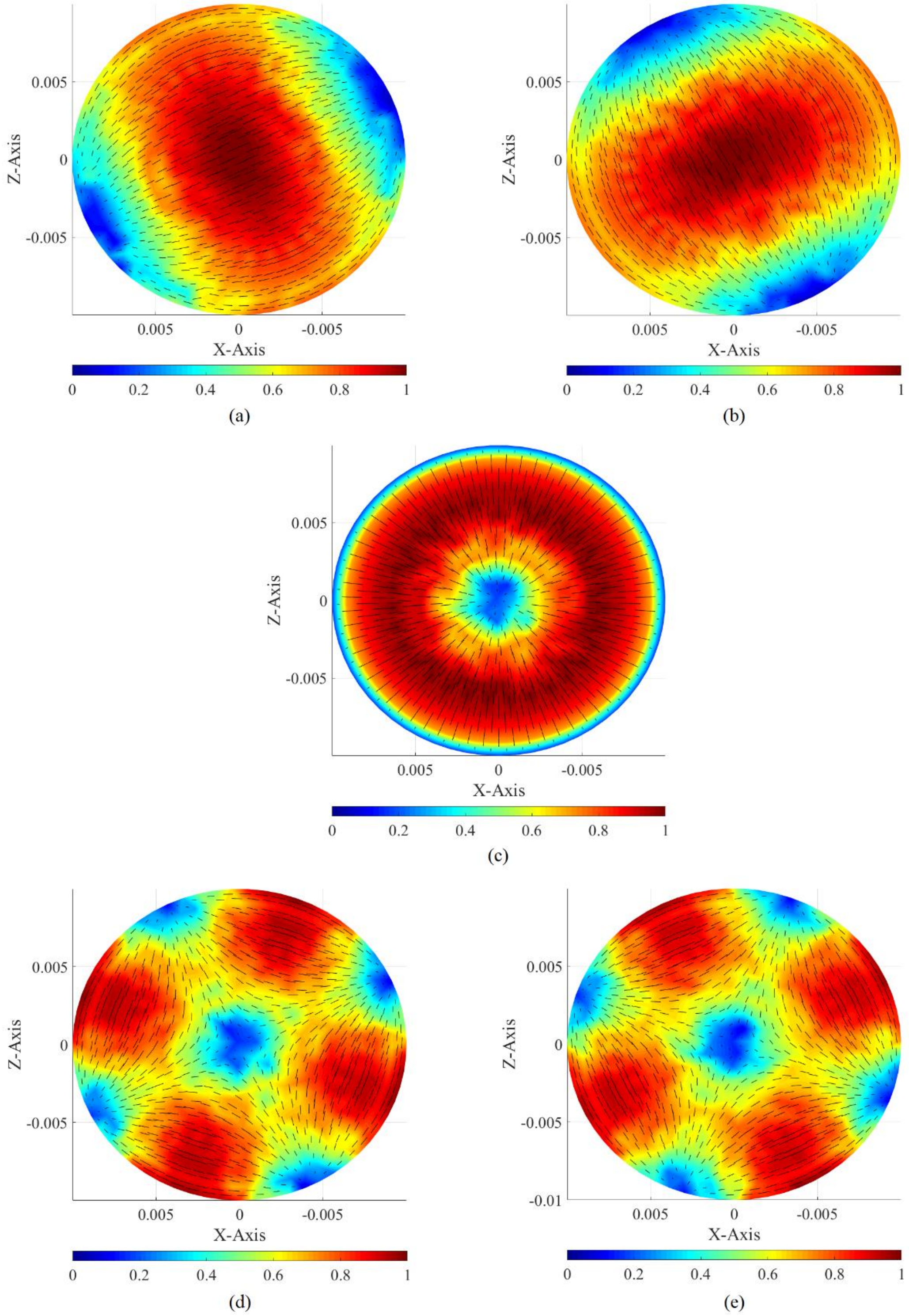

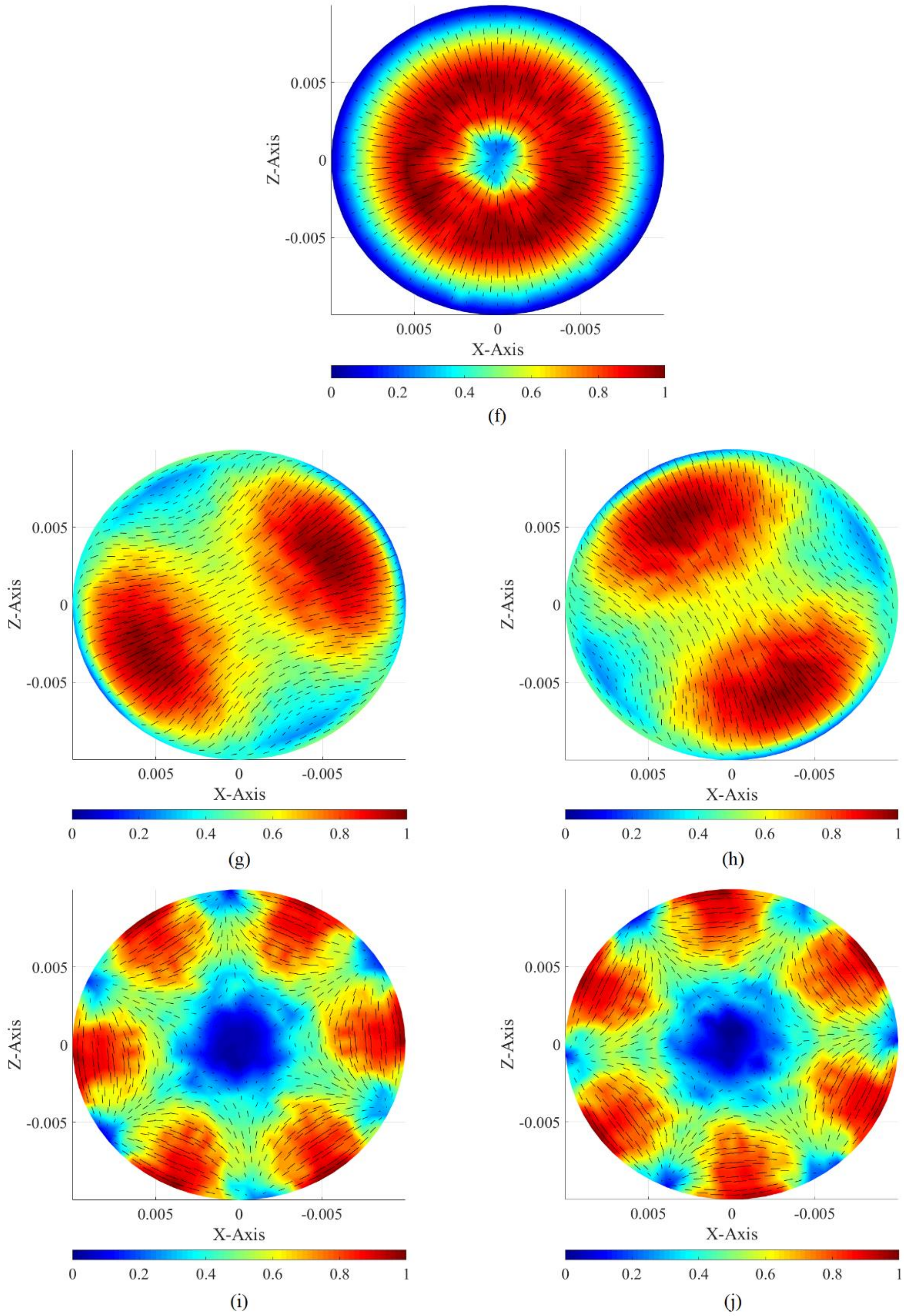

Figure 3-14 Modal equivalent currents on input port $\mathbb{S}_{\text{in}}$. (a) $\vec{M}_{\text{in}}^{\text{ES}}$ of the 1st degenerate state of TE11; (b) $\vec{M}_{\text{in}}^{\text{ES}}$ of the 2nd degenerate state of TE11; (c) $\vec{J}_{\text{in}}^{\text{ES}}$ of TM01; (d) $\vec{M}_{\text{in}}^{\text{ES}}$ of the 1st degenerate state of TE21; (e) $\vec{M}_{\text{in}}^{\text{ES}}$ of the 2nd degenerate state of TE21; (f) $\vec{M}_{\text{in}}^{\text{ES}}$ of TE01; (g) $\vec{J}_{\text{in}}^{\text{ES}}$ of the 1st degenerate state of TM11; (h) $\vec{J}_{\text{in}}^{\text{ES}}$ of the 2nd degenerate state of TM11; (i) $\vec{M}_{\text{in}}^{\text{ES}}$ of the 1st degenerate state of TE31; (j) $\vec{M}_{\text{in}}^{\text{ES}}$ of the 2nd degenerate state of TE31

Thus, we conclude here that the PTT-MetGuid-DMT indeed has ability to construct the travelling-wave modes of the metallic circular guide.

### 3.2.5.2 Traveling-wave-type IP-DMs of Metallic Coaxial Line

The geometry and topological structure of a section of metallic coaxial line are shown in the following Fig. 3-15 and 3-16 respectively.

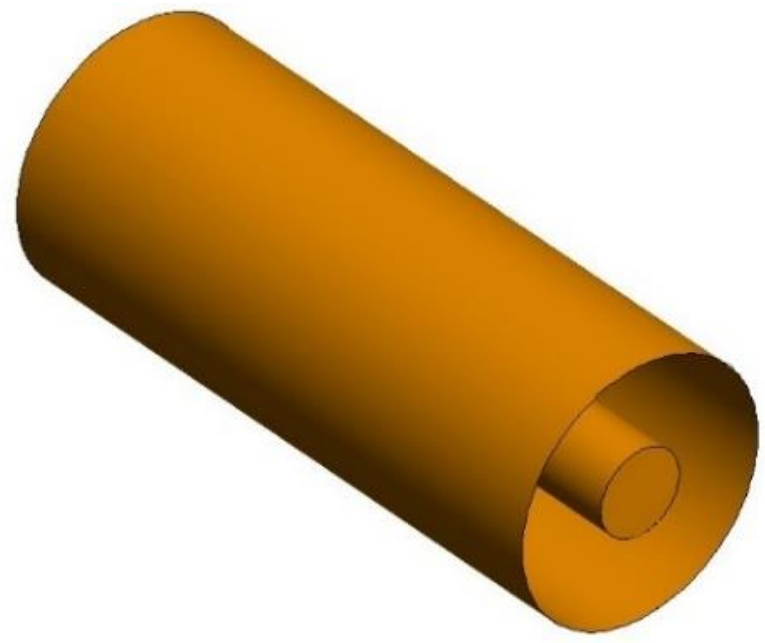

Figure 3-15 Geometry of metallic coaxial line

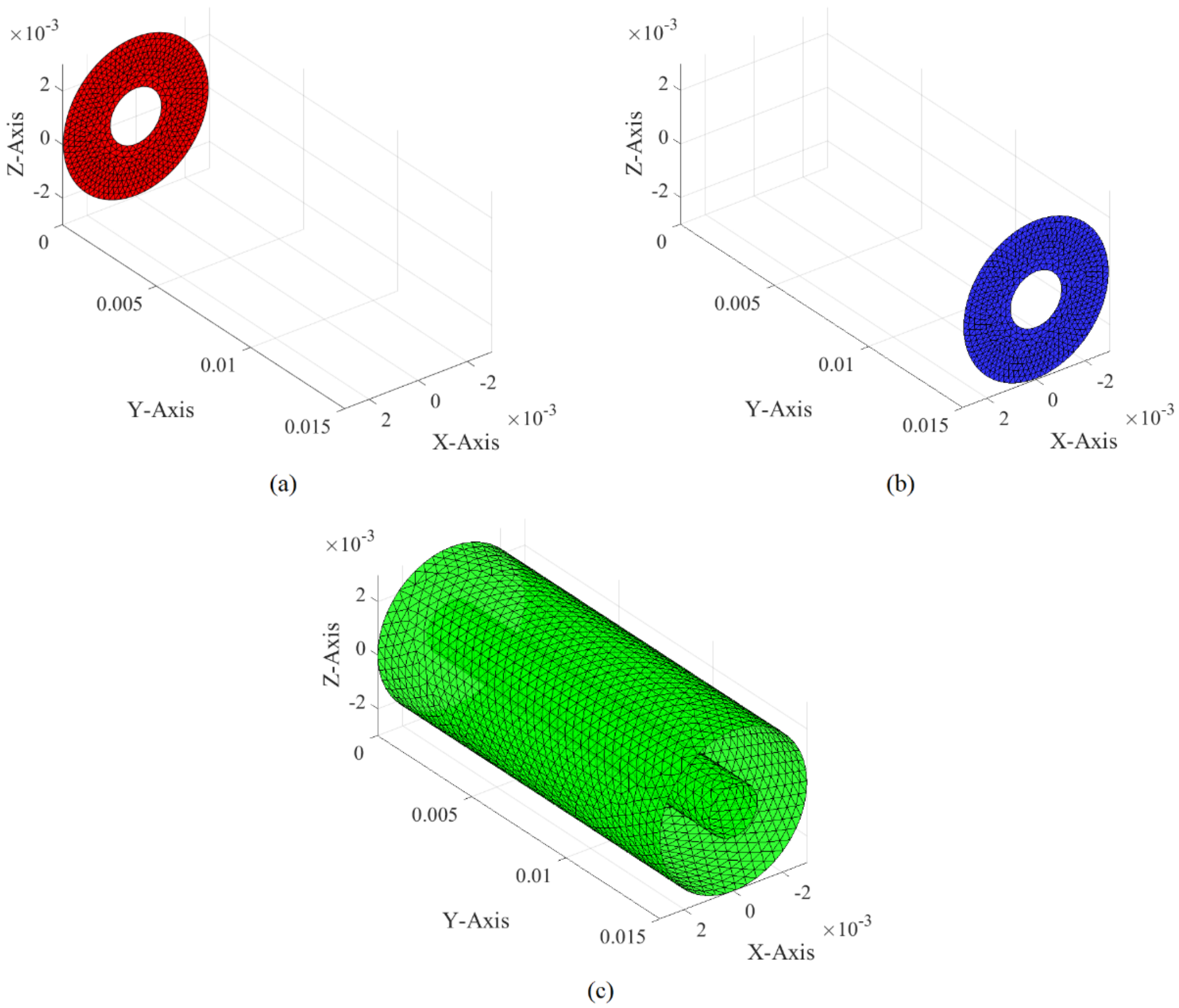

Figure 3-16 Topological structures and surface triangular meshes of the metallic coaxial line shown in Fig. 3-15. (a) Mesh of input port $\mathbb{S}_{in}$ ; (b) mesh of output port $\mathbb{S}_{out}$ ; (c) mesh of guide electric wall $\mathbb{S}_{ele}$

The inner and outer radiuses of the guide tube are $r=1\text{mm}$ and $R=3\text{mm}$ respectively, and the longitudinal length of the section is $L=15\text{mm}$. The guide tube is filled with lossless homogeneous isotropic material whose *relative permeability* and *relative permittivity* are $\mu_r=1$ and $\varepsilon_r=10$ respectively.

By solving the Eq. (3-63) originating from Eqs. (3-42) and (3-54), the IP-DMs of the coaxial line are constructed, and the modal input resistance curve of the dominant IP-DM is plotted in the following Fig. 3-17. Based on the previous discussions, we can conclude that: when the mode works at 3.05GHz, 6.10GHz, and 9.10GHz, there exist relations $\lambda_z/2=L=15\text{mm}$, $\lambda_z=L=15\text{mm}$, and $3\lambda_z/2=L=15\text{mm}$ respectively.

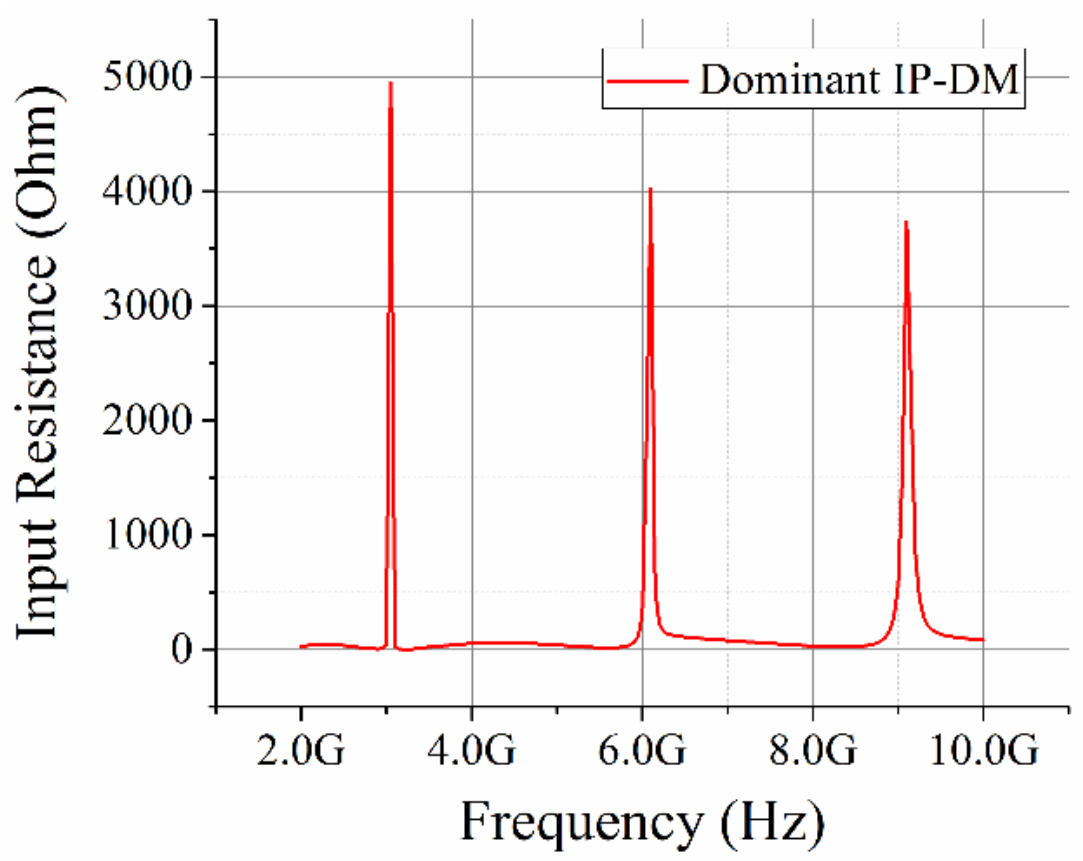

Figure 3-17 Modal input resistance curve of the dominant IP-DM

Taking the IP-DM mode working at 6.10GHz as an example, we plot the equivalent electric & magnetic currents and electric & magnetic fields on input port $\mathbb{S}_{\text{in}}$ in the following Fig. 3-18.

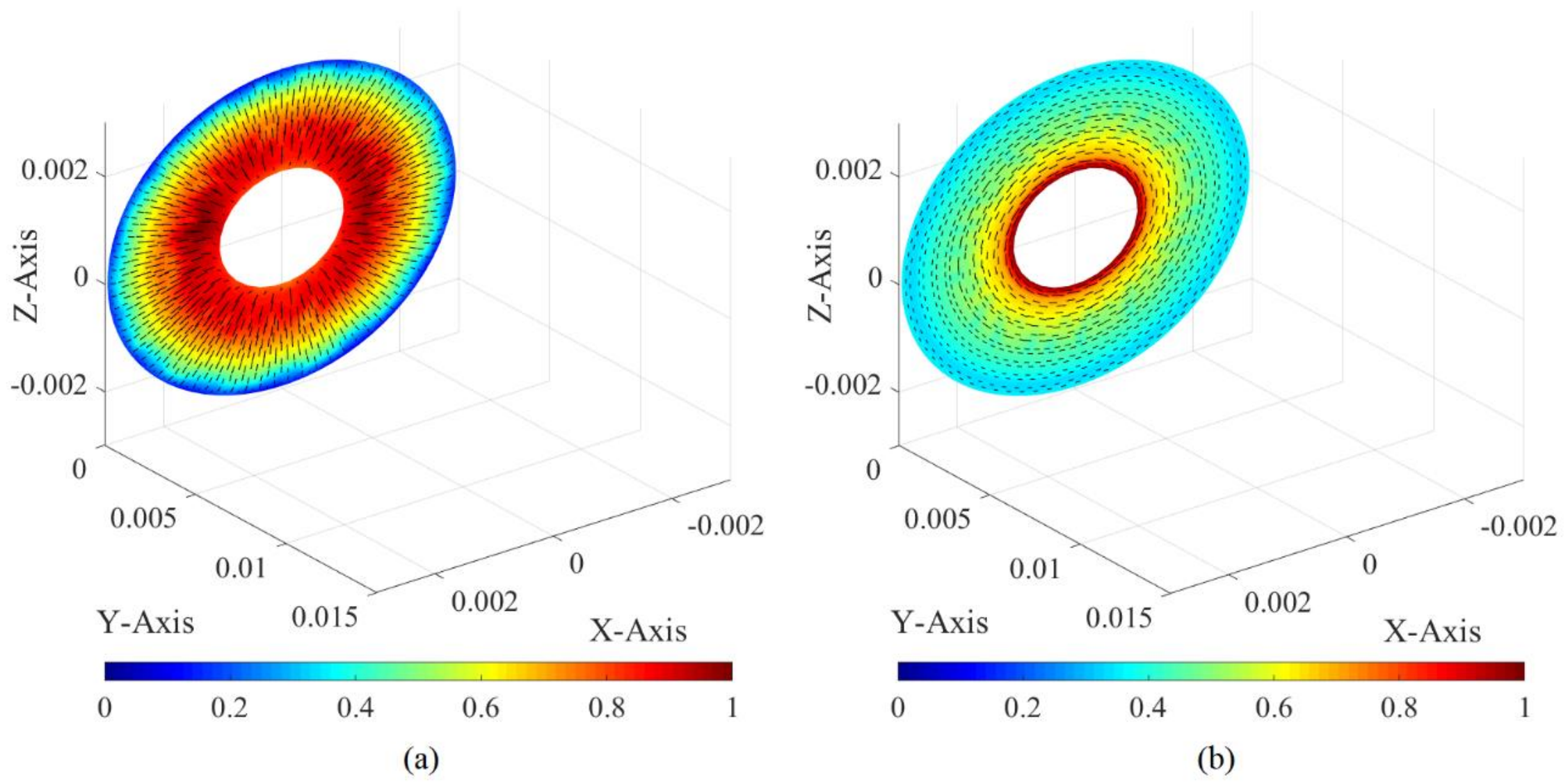

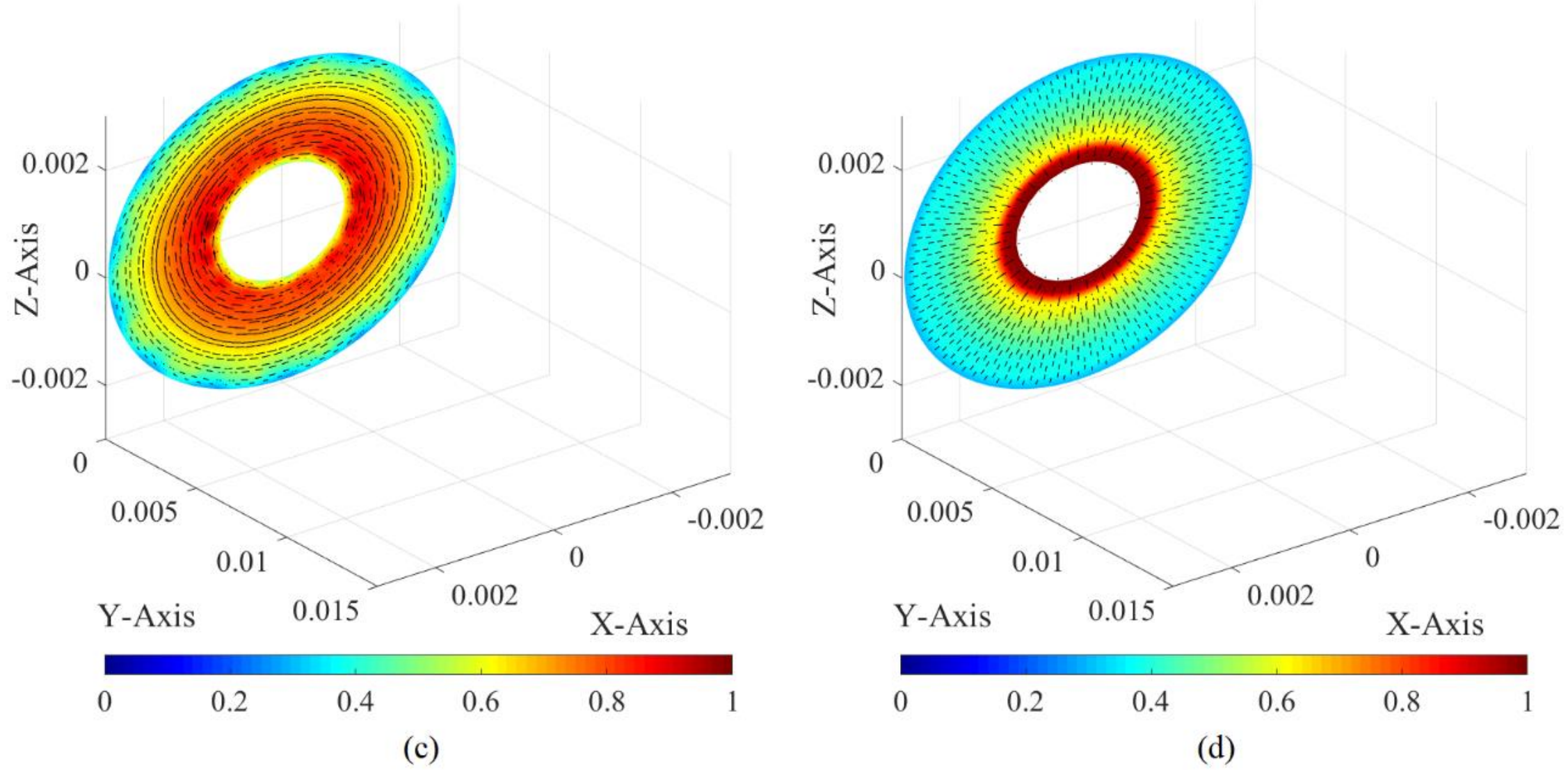

Figure 3-18 Modal equivalent currents and modal fields (of the mode shown in Fig. 3-17) distributing on input port $\mathbb{S}_{\text{in}}$ . (a) Modal equivalent electric current; (b) modal equivalent magnetic current; (c) modal magnetic field; (d) modal electric field

From Fig. 3-18, it is easy to recognize that the mode is just the TEM mode of metallic coaxial line. Now, we calculate the TEM working frequencies corresponding to $L = \lambda_z/2, \lambda_z, 3\lambda_z/2$ from PTT-MetGuid-DMP and analytical method, and list the results in Tab. 3-2, and also plot the current, field, and energy distributions in Fig. 3-19.

Table 3-2 Working frequencies (GHz) of the TEM mode calculated from PTT-MetGuid-DMT and analytical method

| | **PTT-MetGuid-DMP** | **Analytical Method** |
|---|---|---|
| **Working Frequency Corresponding to** $L = \lambda_z / 2$ | 3.05 | 3.16 |
| **Working Frequency Corresponding to** $L = \lambda_z$ | 6.10 | 6.32 |
| **Working Frequency Corresponding to** $L = 3\lambda_z / 2$ | 9.10 | 9.48 |

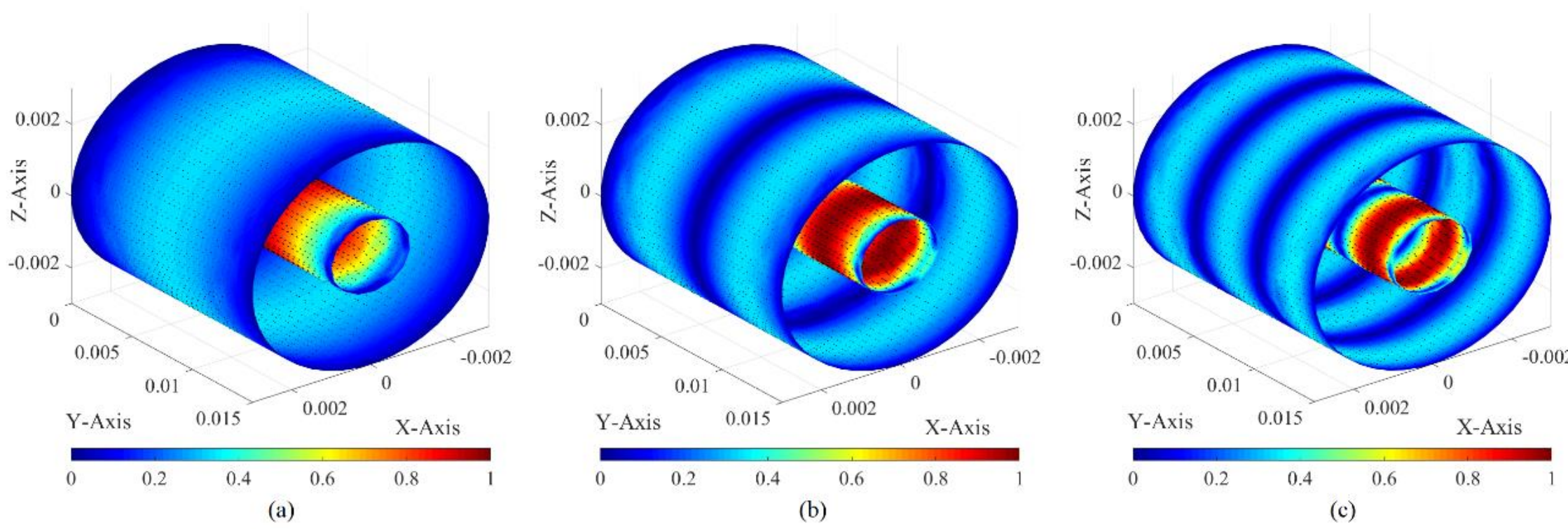

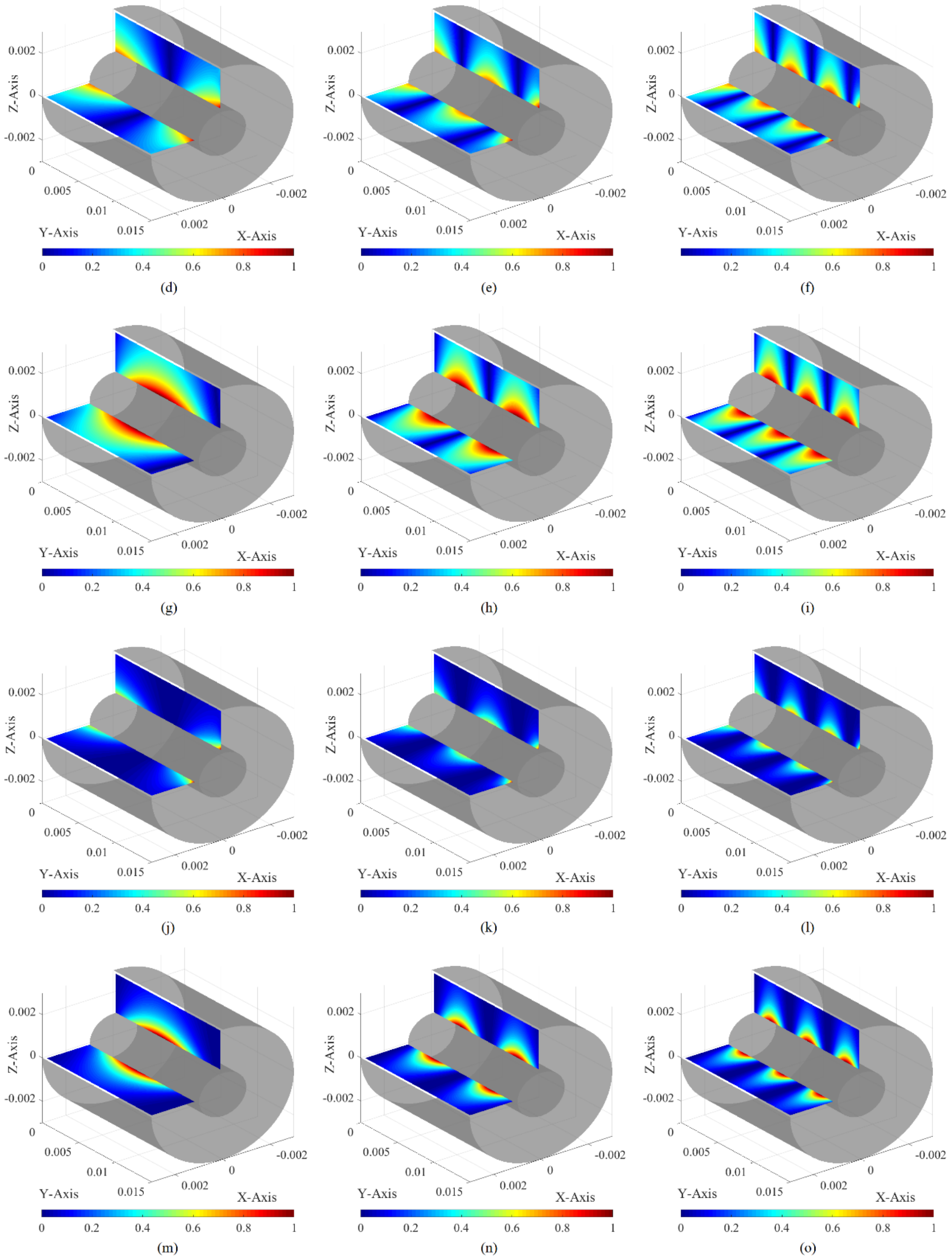

Figure 3-19 Distributions of the TEM modal wall electric current $\vec{J}^{\text{IS}}$, modal fields $\{\vec{E},\vec{H}\}$, and modal energy densities $\{w_{\text{ele}}, w_{\text{mag}}\}$. (a) $\vec{J}^{\text{IS}}$ at 3.05 GHz; (b) $\vec{J}^{\text{IS}}$ at 6.10 GHz; (c) $\vec{J}^{\text{IS}}$ at 9.10 GHz; (d) $|\vec{E}|$ at 3.05 GHz; (e) $|\vec{E}|$ at 6.10 GHz; (f) $|\vec{E}|$ at 9.10 GHz; (g) $|\vec{H}|$ at 3.05 GHz; (h) $|\vec{H}|$ at 6.10 GHz; (i) $|\vec{H}|$ at 9.10 GHz; (j) $w_{\text{ele}}$ at 3.05 GHz; (k) $w_{\text{ele}}$ at 6.10 GHz; (l) $w_{\text{ele}}$ at 9.10 GHz; (m) $w_{\text{mag}}$ at 3.05 GHz; (n) $w_{\text{mag}}$ at 6.10 GHz; (o) $w_{\text{mag}}$ at 9.10 GHz

Thus, we conclude here that the PTT-MetGuid-DMT indeed has ability to construct the travelling-wave modes of the metcallic coaxial line.

## 3.3 IP-DMs of Material Guiding Structure

Under PTT framework, we construct the IP-DMs of *material guiding structures* in this section. For simplifing terminology, the material guiding structures are simply called *material guides* in the following discussions. A typical material guide is shown in the following Fig. 3-20.

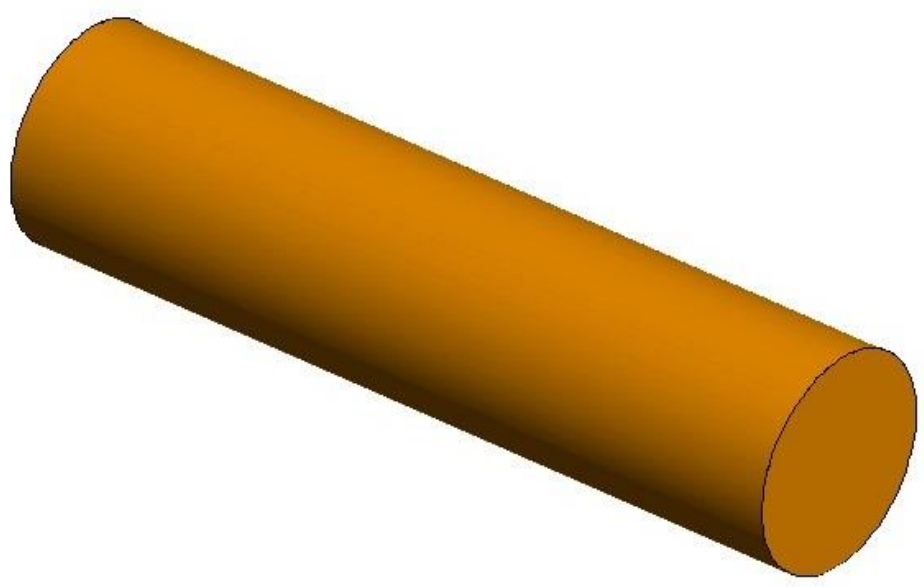

Figure 3-20 Geometry of a typical material guide

This section is organized as follows: in Sec. 3.3.1, the topological structure of the material guide is described in a rigorous mathematical language — *point set topology*[44], and the EM fields are expressed in terms of the currents distributing on the topological structure; in Sec. 3.3.2, the rigorous mathematical descriptions for the modal space constituted by the traveling-wave modes of the material guide are provided by employing the source-field relationships obtained in Sec. 3.3.1; in Sec. 3.3.3, the IPO defined on the modal space is derived; in Sec. 3.3.4, the traveling-wave IP-DMs in the modal space are construced by orthogonalizing the IPO, and some related topics, such as modal decoupling relation and Parseval's identity etc., are also simply discussed; in Sec. 3.3.5, some typical numerical examples are provided to verify the theory and formulations established in whole Sec. 3.3.

The above-mentioned general process for constructing traveling-wave IP-DMs is visually summarized in the following Fig. 3-21.

Mathematically Depicting Topological Structure

↓

Deriving Source- Field Relationships (SFRs)

↓

Mathematically Describing Modal Space by Employing the SFRs

↓

Deriving the Input Power Operator (IPO) Defined on the Modal Space

↓

Constructing Input-Power-Decoupled Modes (IP-DMs) by Orthogonalizing the IPO

Figure 3-21 A general process for constructing IP-DMs

### 3.3.1 Topological Structure and Source-Field Relationships

Now we consider a section of the material guide shown in the previous Fig. 3-20, and the longitudinal length of the section is $L$[①], and its topological structure is shown in the following Fig. 3-22.

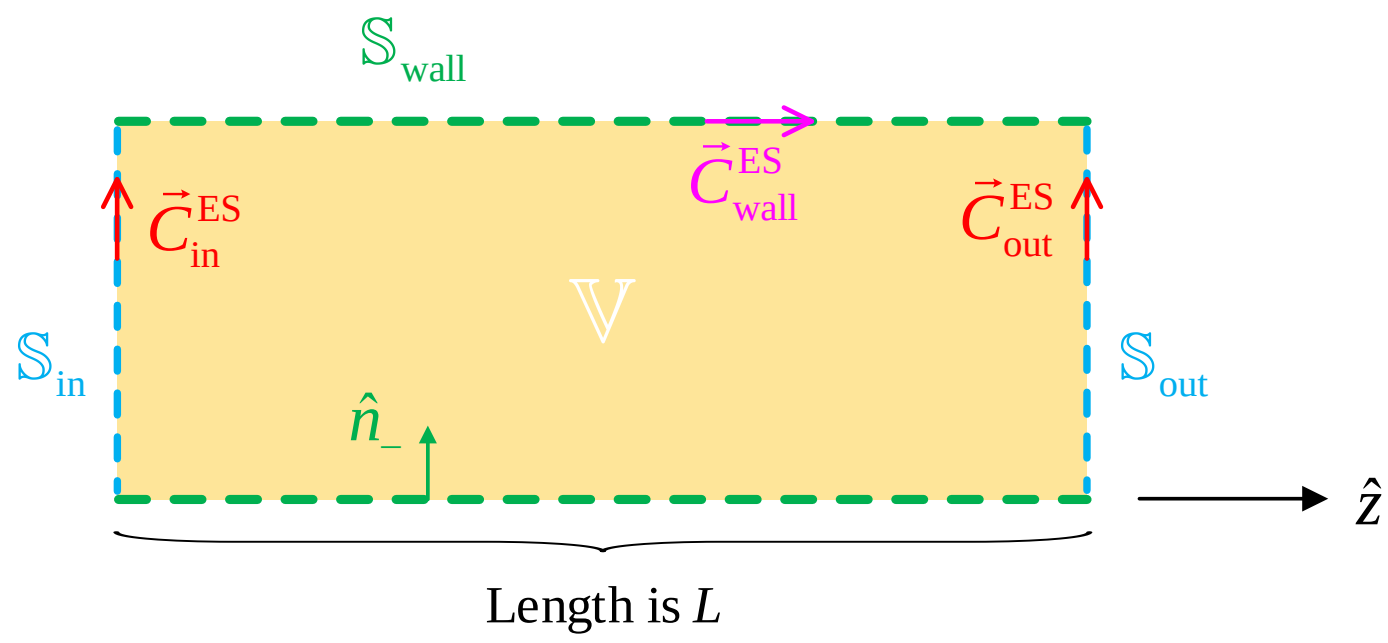

Figure 3-22 Topological structure of a section of the material guide shown in Fig. 3-20

In the above Fig. 3-22, the region occupied by the material guide tube is denoted as $\mathbb{V}$, where $\mathbb{V} = \text{int}\,\mathbb{V}$; the input port of $\mathbb{V}$ is denoted as $\mathbb{S}_{\text{in}}$; the output port of $\mathbb{V}$ is denoted as $\mathbb{S}_{\text{out}}$; the guide wall, i.e. the interface between $\mathbb{V}$ and environment, is denoted as $\mathbb{S}_{\text{wall}}$.

The equivalent surface currents distributing on $\mathbb{S}_{\text{in}}$ are denoted as $\{\vec{J}^{\text{ES}}_{\text{in}}, \vec{M}^{\text{ES}}_{\text{in}}\}$, or simply denoted as $\vec{C}^{\text{ES}}_{\text{in}}$ where $\vec{C} = \vec{J} / \vec{M}$. The equivalent surface currents distributing on $\mathbb{S}_{\text{out}}$ are denoted as $\{\vec{J}^{\text{ES}}_{\text{out}}, \vec{M}^{\text{ES}}_{\text{out}}\}$. The equivalent surface currents distributing on $\mathbb{S}_{\text{wall}}$ are denoted as $\{\vec{J}^{\text{ES}}_{\text{wall}}, \vec{M}^{\text{ES}}_{\text{wall}}\}$.

The above-mentined equivalent surface currents are defined in terms of the fields $\{\vec{E}, \vec{H}\}$ in the material guide as follows:

$$\vec{J}^{\text{ES}}_{\text{in/out}}(\vec{r}) = \hat{z} \times \vec{H}(\vec{r}) \quad , \quad \vec{r} \in \mathbb{S}_{\text{in/out}} \tag{3-80a}$$

$$\vec{M}^{\text{ES}}_{\text{in/out}}(\vec{r}) = \vec{E}(\vec{r}) \times \hat{z} \quad , \quad \vec{r} \in \mathbb{S}_{\text{in/out}} \tag{3-80b}$$

① For a traveling-wave mode, the longitudinal length of the selected section satisfies that $L = n\lambda_z$, where $\lambda_z$ is the waveguide wavelength of the traveling-wave mode and $n$ is a positive integer.

and

$$\vec{J}_{\text{wall}}^{\text{ES}}\left(\vec{r}\right) = \hat{n}_{-}\times\vec{H}\left(\vec{r}\right) \quad , \quad \vec{r}\in\mathbb{S}_{\text{wall}} \tag{3-81a}$$

$$\vec{M}_{\text{wall}}^{\text{ES}}\left(\vec{r}\right) = \vec{E}\left(\vec{r}\right)\times\hat{n}_{-} \quad , \quad \vec{r}\in\mathbb{S}_{\text{wall}} \tag{3-81b}$$

Here, $\hat{z}$ is the Z-directional vector, and $\hat{n}_{-}$ is the normal direction of $\mathbb{S}_{\text{wall}}$ and points to the interior of $\mathbb{V}$, as shown in Fig. 3-22.

Using the above currents, the field $\vec{F}$ in region $\mathbb{V}$ can be expressed as follows:

$$\vec{F}\left(\vec{r}\right) = \mathcal{F}\left(\vec{J}_{\text{in}}^{\text{ES}}+\vec{J}_{\text{wall}}^{\text{ES}}-\vec{J}_{\text{out}}^{\text{ES}},\vec{M}_{\text{in}}^{\text{ES}}+\vec{M}_{\text{wall}}^{\text{ES}}-\vec{M}_{\text{out}}^{\text{ES}}\right) \quad , \quad \vec{r}\in\mathbb{V} \tag{3-82}$$

in which $\vec{F}=\vec{E}/\vec{H}$ and correspondingly $\mathcal{F}=\mathcal{E}/\mathcal{H}$, and operators $\mathcal{E}$ and $\mathcal{H}$ have the same forms as the ones given in Eqs. (3-26a) and (3-26b). Based on the conclusions given in Sec. 3.2, the field $\vec{F}$ on $\mathbb{S}_{\text{out}}^{+}$ can be expressed as follows:

$$\vec{F}\left(\vec{r}\right) = \mathcal{F}\left(\vec{J}_{\text{out}}^{\text{ES}},\vec{M}_{\text{out}}^{\text{ES}}\right) \quad , \quad \vec{r}\in\mathbb{S}_{\text{out}}^{+} \tag{3-83}$$

where $\mathbb{S}_{\text{out}}^{+}$ is the right-side surface of $\mathbb{S}_{\text{out}}$. In addition, for the traveling-wave modes, the field $\vec{F}$ on $\mathbb{S}_{\text{wall}}^{+}$ can be approximately expressed as follows:

$$\vec{F}\left(\vec{r}\right) = \mathcal{F}_{0}\left(-\vec{J}_{\text{wall}}^{\text{ES}},-\vec{M}_{\text{wall}}^{\text{ES}}\right) \quad , \quad \vec{r}\in\mathbb{S}_{\text{wall}}^{+} \tag{3-84}$$

where $\mathbb{S}_{\text{wall}}^{+}$ is the outer boundary surface of the guide wall, and the operator $\mathcal{F}_{0}$ is the same as the one used in the previous chapters.

## 3.3.2 Mathematical Description for Modal Space

Substituting Eq. (3-82) into Eqs. (3-80a) and (3-80b), we obtain the following integral equations

$$\left[\mathcal{H}\left(\vec{J}_{\text{in}}^{\text{ES}}+\vec{J}_{\text{wall}}^{\text{ES}}-\vec{J}_{\text{out}}^{\text{ES}},\vec{M}_{\text{in}}^{\text{ES}}+\vec{M}_{\text{wall}}^{\text{ES}}-\vec{M}_{\text{out}}^{\text{ES}}\right)\right]_{\vec{r}_{-}\to\vec{r}}^{\tan} = \vec{J}_{\text{in}}^{\text{ES}}\left(\vec{r}\right)\times\hat{z} \quad , \quad \vec{r}\in\mathbb{S}_{\text{in}} \tag{3-85a}$$

$$\left[\mathcal{E}\left(\vec{J}_{\text{in}}^{\text{ES}}+\vec{J}_{\text{wall}}^{\text{ES}}-\vec{J}_{\text{out}}^{\text{ES}},\vec{M}_{\text{in}}^{\text{ES}}+\vec{M}_{\text{wall}}^{\text{ES}}-\vec{M}_{\text{out}}^{\text{ES}}\right)\right]_{\vec{r}_{-}\to\vec{r}}^{\tan} = \hat{z}\times\vec{M}_{\text{in}}^{\text{ES}}\left(\vec{r}\right) \quad , \quad \vec{r}\in\mathbb{S}_{\text{in}} \tag{3-85b}$$

about currents $\{\vec{J}_{\text{in}}^{\text{ES}},\vec{M}_{\text{in}}^{\text{ES}}\}$, $\{\vec{J}_{\text{wall}}^{\text{ES}},\vec{M}_{\text{wall}}^{\text{ES}}\}$, and $\{\vec{J}_{\text{out}}^{\text{ES}},\vec{M}_{\text{out}}^{\text{ES}}\}$, where $\vec{r}_{-}\in\text{int}\,\mathbb{V}$ and $\vec{r}_{-}$ approaches the point $\vec{r}$ on $\mathbb{S}_{\text{in}}$.

Utilizing Eqs. (3-82)&(3-84) and the tangential continuation condition satisfied by the field on $\mathbb{S}_{\text{wall}}$, we obtain the following integral equations

$$\begin{aligned}&\left[\mathcal{E}\left(\vec{J}_{\text{in}}^{\text{ES}}+\vec{J}_{\text{wall}}^{\text{ES}}-\vec{J}_{\text{out}}^{\text{ES}},\vec{M}_{\text{in}}^{\text{ES}}+\vec{M}_{\text{wall}}^{\text{ES}}-\vec{M}_{\text{out}}^{\text{ES}}\right)\right]_{\vec{r}_{-}\to\vec{r}}^{\tan}\\&=\left[\mathcal{E}_{0}\left(-\vec{J}_{\text{wall}}^{\text{ES}},-\vec{M}_{\text{wall}}^{\text{ES}}\right)\right]_{\vec{r}_{+}\to\vec{r}}^{\tan} \quad , \quad \vec{r}\in\mathbb{S}_{\text{wall}}\end{aligned} \tag{3-86a}$$

$$\left[\mathcal{H}\left(\vec{J}_{\text{in}}^{\text{ES}}+\vec{J}_{\text{wall}}^{\text{ES}}-\vec{J}_{\text{out}}^{\text{ES}},\vec{M}_{\text{in}}^{\text{ES}}+\vec{M}_{\text{wall}}^{\text{ES}}-\vec{M}_{\text{out}}^{\text{ES}}\right)\right]_{\vec{r}_-\to\vec{r}}^{\tan} = \left[\mathcal{H}_0\left(-\vec{J}_{\text{wall}}^{\text{ES}},-\vec{M}_{\text{wall}}^{\text{ES}}\right)\right]_{\vec{r}_+\to\vec{r}}^{\tan} \quad , \quad \vec{r}\in\mathbb{S}_{\text{wall}} \tag{3-86b}$$

about currents $\{\vec{J}_{\text{in}}^{\text{ES}},\vec{M}_{\text{in}}^{\text{ES}}\}$, $\{\vec{J}_{\text{wall}}^{\text{ES}},\vec{M}_{\text{wall}}^{\text{ES}}\}$, and $\{\vec{J}_{\text{out}}^{\text{ES}},\vec{M}_{\text{out}}^{\text{ES}}\}$, where $\vec{r}_-\in\text{int}\,\mathbb{V}$ and $\vec{r}_+\in\text{ext}\,\mathbb{V}$ and the points approach the point $\vec{r}$ on $\mathbb{S}_{\text{wall}}$.

Similar to deriving Eqs. (3-28a) and (3-28b) from Eq. (3-19), we can derive the following integral equations

$$\left[\mathcal{E}\left(\vec{J}_{\text{in}}^{\text{ES}}+\vec{J}_{\text{wall}}^{\text{ES}}-\vec{J}_{\text{out}}^{\text{ES}},\vec{M}_{\text{in}}^{\text{ES}}+\vec{M}_{\text{wall}}^{\text{ES}}-\vec{M}_{\text{out}}^{\text{ES}}\right)\right]_{\vec{r}_-\to\vec{r}}^{\tan} = \left[\mathcal{E}\left(\vec{J}_{\text{out}}^{\text{ES}},\vec{M}_{\text{out}}^{\text{ES}}\right)\right]_{\vec{r}_+\to\vec{r}}^{\tan} \quad , \quad \vec{r}\in\mathbb{S}_{\text{out}} \tag{3-87a}$$

$$\left[\mathcal{H}\left(\vec{J}_{\text{in}}^{\text{ES}}+\vec{J}_{\text{wall}}^{\text{ES}}-\vec{J}_{\text{out}}^{\text{ES}},\vec{M}_{\text{in}}^{\text{ES}}+\vec{M}_{\text{wall}}^{\text{ES}}-\vec{M}_{\text{out}}^{\text{ES}}\right)\right]_{\vec{r}_-\to\vec{r}}^{\tan} = \left[\mathcal{H}\left(\vec{J}_{\text{out}}^{\text{ES}},\vec{M}_{\text{out}}^{\text{ES}}\right)\right]_{\vec{r}_+\to\vec{r}}^{\tan} \quad , \quad \vec{r}\in\mathbb{S}_{\text{out}} \tag{3-87b}$$

about currents $\{\vec{J}_{\text{in}}^{\text{ES}},\vec{M}_{\text{in}}^{\text{ES}}\}$, $\{\vec{J}_{\text{wall}}^{\text{ES}},\vec{M}_{\text{wall}}^{\text{ES}}\}$, and $\{\vec{J}_{\text{out}}^{\text{ES}},\vec{M}_{\text{out}}^{\text{ES}}\}$, from Eqs. (3-82) and (3-83).

If the currents $\{\vec{J}_{\text{in}}^{\text{ES}},\vec{M}_{\text{in}}^{\text{ES}}\}$, $\{\vec{J}_{\text{wall}}^{\text{ES}},\vec{M}_{\text{wall}}^{\text{ES}}\}$, and $\{\vec{J}_{\text{out}}^{\text{ES}},\vec{M}_{\text{out}}^{\text{ES}}\}$ involved in Eqs. (3-85a)~(3-87b) are expanded in terms of some proper basis functions, and the equations are tested with the basis functions $\{\vec{b}_\xi^{\vec{M}_{\text{in}}^{\text{ES}}}\}$, $\{\vec{b}_\xi^{\vec{J}_{\text{in}}^{\text{ES}}}\}$, $\{\vec{b}_\xi^{\vec{J}_{\text{wall}}^{\text{ES}}}\}$, $\{\vec{b}_\xi^{\vec{M}_{\text{wall}}^{\text{ES}}}\}$, $\{\vec{b}_\xi^{\vec{J}_{\text{out}}^{\text{ES}}}\}$, and $\{\vec{b}_\xi^{\vec{M}_{\text{out}}^{\text{ES}}}\}$ respectively, then the integral equations are immediately discretized into the following matrix equations

$$\begin{aligned}0 \;=\;& \overline{\overline{Z}}^{\vec{M}_{\text{in}}^{\text{ES}}\vec{J}_{\text{in}}^{\text{ES}}}\cdot\overline{a}^{\vec{J}_{\text{in}}^{\text{ES}}}+\overline{\overline{Z}}^{\vec{M}_{\text{in}}^{\text{ES}}\vec{J}_{\text{wall}}^{\text{ES}}}\cdot\overline{a}^{\vec{J}_{\text{wall}}^{\text{ES}}}+\overline{\overline{Z}}^{\vec{M}_{\text{in}}^{\text{ES}}\vec{J}_{\text{out}}^{\text{ES}}}\cdot\overline{a}^{\vec{J}_{\text{out}}^{\text{ES}}}+\overline{\overline{Z}}^{\vec{M}_{\text{in}}^{\text{ES}}\vec{M}_{\text{in}}^{\text{ES}}}\cdot\overline{a}^{\vec{M}_{\text{in}}^{\text{ES}}}+\overline{\overline{Z}}^{\vec{M}_{\text{in}}^{\text{ES}}\vec{M}_{\text{wall}}^{\text{ES}}}\cdot\overline{a}^{\vec{M}_{\text{wall}}^{\text{ES}}}\\&+\overline{\overline{Z}}^{\vec{M}_{\text{in}}^{\text{ES}}\vec{M}_{\text{out}}^{\text{ES}}}\cdot\overline{a}^{\vec{M}_{\text{out}}^{\text{ES}}}\end{aligned} \tag{3-88a}$$

$$\begin{aligned}0 \;=\;& \overline{\overline{Z}}^{\vec{J}_{\text{in}}^{\text{ES}}\vec{J}_{\text{in}}^{\text{ES}}}\cdot\overline{a}^{\vec{J}_{\text{in}}^{\text{ES}}}+\overline{\overline{Z}}^{\vec{J}_{\text{in}}^{\text{ES}}\vec{J}_{\text{wall}}^{\text{ES}}}\cdot\overline{a}^{\vec{J}_{\text{wall}}^{\text{ES}}}+\overline{\overline{Z}}^{\vec{J}_{\text{in}}^{\text{ES}}\vec{J}_{\text{out}}^{\text{ES}}}\cdot\overline{a}^{\vec{J}_{\text{out}}^{\text{ES}}}+\overline{\overline{Z}}^{\vec{J}_{\text{in}}^{\text{ES}}\vec{M}_{\text{in}}^{\text{ES}}}\cdot\overline{a}^{\vec{M}_{\text{in}}^{\text{ES}}}+\overline{\overline{Z}}^{\vec{J}_{\text{in}}^{\text{ES}}\vec{M}_{\text{wall}}^{\text{ES}}}\cdot\overline{a}^{\vec{M}_{\text{wall}}^{\text{ES}}}\\&+\overline{\overline{Z}}^{\vec{J}_{\text{in}}^{\text{ES}}\vec{M}_{\text{out}}^{\text{ES}}}\cdot\overline{a}^{\vec{M}_{\text{out}}^{\text{ES}}}\end{aligned} \tag{3-88b}$$

and

$$\begin{aligned}0 \;=\;& \overline{\overline{Z}}^{\vec{J}_{\text{wall}}^{\text{ES}}\vec{J}_{\text{in}}^{\text{ES}}}\cdot\overline{a}^{\vec{J}_{\text{in}}^{\text{ES}}}+\overline{\overline{Z}}^{\vec{J}_{\text{wall}}^{\text{ES}}\vec{J}_{\text{wall}}^{\text{ES}}}\cdot\overline{a}^{\vec{J}_{\text{wall}}^{\text{ES}}}+\overline{\overline{Z}}^{\vec{J}_{\text{wall}}^{\text{ES}}\vec{J}_{\text{out}}^{\text{ES}}}\cdot\overline{a}^{\vec{J}_{\text{out}}^{\text{ES}}}+\overline{\overline{Z}}^{\vec{J}_{\text{wall}}^{\text{ES}}\vec{M}_{\text{in}}^{\text{ES}}}\cdot\overline{a}^{\vec{M}_{\text{in}}^{\text{ES}}}+\overline{\overline{Z}}^{\vec{J}_{\text{wall}}^{\text{ES}}\vec{M}_{\text{wall}}^{\text{ES}}}\cdot\overline{a}^{\vec{M}_{\text{wall}}^{\text{ES}}}\\&+\overline{\overline{Z}}^{\vec{J}_{\text{wall}}^{\text{ES}}\vec{M}_{\text{out}}^{\text{ES}}}\cdot\overline{a}^{\vec{M}_{\text{out}}^{\text{ES}}}\end{aligned} \tag{3-89a}$$

$$\begin{aligned}0 \;=\;& \overline{\overline{Z}}^{\vec{M}_{\text{wall}}^{\text{ES}}\vec{J}_{\text{in}}^{\text{ES}}}\cdot\overline{a}^{\vec{J}_{\text{in}}^{\text{ES}}}+\overline{\overline{Z}}^{\vec{M}_{\text{wall}}^{\text{ES}}\vec{J}_{\text{wall}}^{\text{ES}}}\cdot\overline{a}^{\vec{J}_{\text{wall}}^{\text{ES}}}+\overline{\overline{Z}}^{\vec{M}_{\text{wall}}^{\text{ES}}\vec{J}_{\text{out}}^{\text{ES}}}\cdot\overline{a}^{\vec{J}_{\text{out}}^{\text{ES}}}+\overline{\overline{Z}}^{\vec{M}_{\text{wall}}^{\text{ES}}\vec{M}_{\text{in}}^{\text{ES}}}\cdot\overline{a}^{\vec{M}_{\text{in}}^{\text{ES}}}+\overline{\overline{Z}}^{\vec{M}_{\text{wall}}^{\text{ES}}\vec{M}_{\text{wall}}^{\text{ES}}}\cdot\overline{a}^{\vec{M}_{\text{wall}}^{\text{ES}}}\\&+\overline{\overline{Z}}^{\vec{M}_{\text{wall}}^{\text{ES}}\vec{M}_{\text{out}}^{\text{ES}}}\cdot\overline{a}^{\vec{M}_{\text{out}}^{\text{ES}}}\end{aligned} \tag{3-89b}$$

and

$$\begin{aligned}0 \;=\;& \overline{\overline{Z}}^{\vec{J}_{\text{out}}^{\text{ES}}\vec{J}_{\text{in}}^{\text{ES}}}\cdot\overline{a}^{\vec{J}_{\text{in}}^{\text{ES}}}+\overline{\overline{Z}}^{\vec{J}_{\text{out}}^{\text{ES}}\vec{J}_{\text{wall}}^{\text{ES}}}\cdot\overline{a}^{\vec{J}_{\text{wall}}^{\text{ES}}}+\overline{\overline{Z}}^{\vec{J}_{\text{out}}^{\text{ES}}\vec{J}_{\text{out}}^{\text{ES}}}\cdot\overline{a}^{\vec{J}_{\text{out}}^{\text{ES}}}+\overline{\overline{Z}}^{\vec{J}_{\text{out}}^{\text{ES}}\vec{M}_{\text{in}}^{\text{ES}}}\cdot\overline{a}^{\vec{M}_{\text{in}}^{\text{ES}}}+\overline{\overline{Z}}^{\vec{J}_{\text{out}}^{\text{ES}}\vec{M}_{\text{wall}}^{\text{ES}}}\cdot\overline{a}^{\vec{M}_{\text{wall}}^{\text{ES}}}\\&+\overline{\overline{Z}}^{\vec{J}_{\text{out}}^{\text{ES}}\vec{M}_{\text{out}}^{\text{ES}}}\cdot\overline{a}^{\vec{M}_{\text{out}}^{\text{ES}}}\end{aligned} \tag{3-90a}$$

$$\begin{aligned}0 \;=\;& \overline{\overline{Z}}^{\vec{M}_{\text{out}}^{\text{ES}}\vec{J}_{\text{in}}^{\text{ES}}}\cdot\overline{a}^{\vec{J}_{\text{in}}^{\text{ES}}}+\overline{\overline{Z}}^{\vec{M}_{\text{out}}^{\text{ES}}\vec{J}_{\text{wall}}^{\text{ES}}}\cdot\overline{a}^{\vec{J}_{\text{wall}}^{\text{ES}}}+\overline{\overline{Z}}^{\vec{M}_{\text{out}}^{\text{ES}}\vec{J}_{\text{out}}^{\text{ES}}}\cdot\overline{a}^{\vec{J}_{\text{out}}^{\text{ES}}}+\overline{\overline{Z}}^{\vec{M}_{\text{out}}^{\text{ES}}\vec{M}_{\text{in}}^{\text{ES}}}\cdot\overline{a}^{\vec{M}_{\text{in}}^{\text{ES}}}+\overline{\overline{Z}}^{\vec{M}_{\text{out}}^{\text{ES}}\vec{M}_{\text{wall}}^{\text{ES}}}\cdot\overline{a}^{\vec{M}_{\text{wall}}^{\text{ES}}}\\&+\overline{\overline{Z}}^{\vec{M}_{\text{out}}^{\text{ES}}\vec{M}_{\text{out}}^{\text{ES}}}\cdot\overline{a}^{\vec{M}_{\text{out}}^{\text{ES}}}\end{aligned} \tag{3-90b}$$

In Eq. (3-88a), the matrix elements are calculated as follows:

$$z_{\xi\zeta}^{\vec{M}_{\rm in}^{\rm ES}\vec{J}_{\rm in}^{\rm ES}} = \left\langle \vec{b}_\xi^{\vec{M}_{\rm in}^{\rm ES}}, {\rm P.V.}\mathcal{K}\left(\vec{b}_\zeta^{\vec{J}_{\rm in}^{\rm ES}}\right)\right\rangle_{\mathbb{S}_{\rm in}} - (1/2)\left\langle \vec{b}_\xi^{\vec{M}_{\rm in}^{\rm ES}}, \vec{b}_\zeta^{\vec{J}_{\rm in}^{\rm ES}} \times \hat{z}\right\rangle_{\mathbb{S}_{\rm in}} \tag{3-91a}$$

$$z_{\xi\zeta}^{\vec{M}_{\rm in}^{\rm ES}\vec{J}_{\rm wall}^{\rm ES}} = \left\langle \vec{b}_\xi^{\vec{M}_{\rm in}^{\rm ES}}, \mathcal{K}\left(\vec{b}_\zeta^{\vec{J}_{\rm wall}^{\rm ES}}\right)\right\rangle_{\mathbb{S}_{\rm in}} \tag{3-91b}$$

$$z_{\xi\zeta}^{\vec{M}_{\rm in}^{\rm ES}\vec{J}_{\rm out}^{\rm ES}} = \left\langle \vec{b}_\xi^{\vec{M}_{\rm in}^{\rm ES}}, \mathcal{K}\left(-\vec{b}_\zeta^{\vec{J}_{\rm out}^{\rm ES}}\right)\right\rangle_{\mathbb{S}_{\rm in}} \tag{3-91c}$$

$$z_{\xi\zeta}^{\vec{M}_{\rm in}^{\rm ES}\vec{M}_{\rm in}^{\rm ES}} = \left\langle \vec{b}_\xi^{\vec{M}_{\rm in}^{\rm ES}}, -j\omega\varepsilon\mathcal{L}\left(\vec{b}_\zeta^{\vec{M}_{\rm in}^{\rm ES}}\right)\right\rangle_{\mathbb{S}_{\rm in}} \tag{3-91d}$$

$$z_{\xi\zeta}^{\vec{M}_{\rm in}^{\rm ES}\vec{M}_{\rm wall}^{\rm ES}} = \left\langle \vec{b}_\xi^{\vec{M}_{\rm in}^{\rm ES}}, -j\omega\varepsilon\mathcal{L}\left(\vec{b}_\zeta^{\vec{M}_{\rm wall}^{\rm ES}}\right)\right\rangle_{\mathbb{S}_{\rm in}} \tag{3-91e}$$

$$z_{\xi\zeta}^{\vec{M}_{\rm in}^{\rm ES}\vec{M}_{\rm out}^{\rm ES}} = \left\langle \vec{b}_\xi^{\vec{M}_{\rm in}^{\rm ES}}, -j\omega\varepsilon\mathcal{L}\left(-\vec{b}_\zeta^{\vec{M}_{\rm out}^{\rm ES}}\right)\right\rangle_{\mathbb{S}_{\rm in}} \tag{3-91f}$$

In Eq. (3-88b), the matrix elements are calculated as follows:

$$z_{\xi\zeta}^{\vec{J}_{\rm in}^{\rm ES}\vec{J}_{\rm in}^{\rm ES}} = \left\langle \vec{b}_\xi^{\vec{J}_{\rm in}^{\rm ES}}, -j\omega\mu\mathcal{L}\left(\vec{b}_\zeta^{\vec{J}_{\rm in}^{\rm ES}}\right)\right\rangle_{\mathbb{S}_{\rm in}} \tag{3-92a}$$

$$z_{\xi\zeta}^{\vec{J}_{\rm in}^{\rm ES}\vec{J}_{\rm wall}^{\rm ES}} = \left\langle \vec{b}_\xi^{\vec{J}_{\rm in}^{\rm ES}}, -j\omega\mu\mathcal{L}\left(\vec{b}_\zeta^{\vec{J}_{\rm wall}^{\rm ES}}\right)\right\rangle_{\mathbb{S}_{\rm in}} \tag{3-92b}$$

$$z_{\xi\zeta}^{\vec{J}_{\rm in}^{\rm ES}\vec{J}_{\rm out}^{\rm ES}} = \left\langle \vec{b}_\xi^{\vec{J}_{\rm in}^{\rm ES}}, -j\omega\mu\mathcal{L}\left(-\vec{b}_\zeta^{\vec{J}_{\rm out}^{\rm ES}}\right)\right\rangle_{\mathbb{S}_{\rm in}} \tag{3-92c}$$

$$z_{\xi\zeta}^{\vec{J}_{\rm in}^{\rm ES}\vec{M}_{\rm in}^{\rm ES}} = \left\langle \vec{b}_\xi^{\vec{J}_{\rm in}^{\rm ES}}, -{\rm P.V.}\mathcal{K}\left(\vec{b}_\zeta^{\vec{M}_{\rm in}^{\rm ES}}\right)\right\rangle_{\mathbb{S}_{\rm in}} - (1/2)\left\langle \vec{b}_\xi^{\vec{J}_{\rm in}^{\rm ES}}, \hat{z} \times \vec{b}_\zeta^{\vec{M}_{\rm in}^{\rm ES}}\right\rangle_{\mathbb{S}_{\rm in}} \tag{3-92d}$$

$$z_{\xi\zeta}^{\vec{J}_{\rm in}^{\rm ES}\vec{M}_{\rm wall}^{\rm ES}} = \left\langle \vec{b}_\xi^{\vec{J}_{\rm in}^{\rm ES}}, -\mathcal{K}\left(\vec{b}_\zeta^{\vec{M}_{\rm wall}^{\rm ES}}\right)\right\rangle_{\mathbb{S}_{\rm in}} \tag{3-92e}$$

$$z_{\xi\zeta}^{\vec{J}_{\rm in}^{\rm ES}\vec{M}_{\rm out}^{\rm ES}} = \left\langle \vec{b}_\xi^{\vec{J}_{\rm in}^{\rm ES}}, -\mathcal{K}\left(-\vec{b}_\zeta^{\vec{M}_{\rm out}^{\rm ES}}\right)\right\rangle_{\mathbb{S}_{\rm in}} \tag{3-92f}$$

In Eq. (3-89a), the matrix elements are calculated as follows:

$$z_{\xi\zeta}^{\vec{J}_{\rm wall}^{\rm ES}\vec{J}_{\rm in}^{\rm ES}} = \left\langle \vec{b}_\xi^{\vec{J}_{\rm wall}^{\rm ES}}, -j\omega\mu\left(\vec{b}_\zeta^{\vec{J}_{\rm in}^{\rm ES}}\right)\right\rangle_{\mathbb{S}_{\rm wall}} \tag{3-93a}$$

$$z_{\xi\zeta}^{\vec{J}_{\rm wall}^{\rm ES}\vec{J}_{\rm wall}^{\rm ES}} = \left\langle \vec{b}_\xi^{\vec{J}_{\rm wall}^{\rm ES}}, -j\omega\mu\left(\vec{b}_\zeta^{\vec{J}_{\rm wall}^{\rm ES}}\right)\right\rangle_{\mathbb{S}_{\rm wall}} - \left\langle \vec{b}_\xi^{\vec{J}_{\rm wall}^{\rm ES}}, -j\omega\mu_0\mathcal{L}_0\left(-\vec{b}_\zeta^{\vec{J}_{\rm wall}^{\rm ES}}\right)\right\rangle_{\mathbb{S}_{\rm wall}} \tag{3-93b}$$

$$z_{\xi\zeta}^{\vec{J}_{\rm wall}^{\rm ES}\vec{J}_{\rm out}^{\rm ES}} = \left\langle \vec{b}_\xi^{\vec{J}_{\rm wall}^{\rm ES}}, -j\omega\mu\left(-\vec{b}_\zeta^{\vec{J}_{\rm out}^{\rm ES}}\right)\right\rangle_{\mathbb{S}_{\rm wall}} \tag{3-93c}$$

$$z_{\xi\zeta}^{\vec{J}_{\rm wall}^{\rm ES}\vec{M}_{\rm in}^{\rm ES}} = \left\langle \vec{b}_\xi^{\vec{J}_{\rm wall}^{\rm ES}}, -\mathcal{K}\left(\vec{b}_\zeta^{\vec{M}_{\rm in}^{\rm ES}}\right)\right\rangle_{\mathbb{S}_{\rm wall}} \tag{3-93d}$$

$$z_{\xi\zeta}^{\vec{J}_{\rm wall}^{\rm ES}\vec{M}_{\rm wall}^{\rm ES}} = \left\langle \vec{b}_\xi^{\vec{J}_{\rm wall}^{\rm ES}}, -{\rm P.V.}\mathcal{K}\left(\vec{b}_\zeta^{\vec{M}_{\rm wall}^{\rm ES}}\right)\right\rangle_{\mathbb{S}_{\rm wall}} - \left\langle \vec{b}_\xi^{\vec{J}_{\rm wall}^{\rm ES}}, -{\rm P.V.}\mathcal{K}_0\left(-\vec{b}_\zeta^{\vec{M}_{\rm wall}^{\rm ES}}\right)\right\rangle_{\mathbb{S}_{\rm wall}} \tag{3-93e}$$

$$z_{\xi\zeta}^{\vec{J}_{\rm wall}^{\rm ES}\vec{M}_{\rm out}^{\rm ES}} = \left\langle \vec{b}_\xi^{\vec{J}_{\rm wall}^{\rm ES}}, -\mathcal{K}\left(-\vec{b}_\zeta^{\vec{M}_{\rm out}^{\rm ES}}\right)\right\rangle_{\mathbb{S}_{\rm wall}} \tag{3-93f}$$

In Eq. (3-89b), the matrix elements are calculated as follows:

$$z_{\xi\zeta}^{\vec{M}_{\text{wall}}^{\text{ES}}\vec{J}_{\text{in}}^{\text{ES}}} = \left\langle \vec{b}_{\xi}^{\vec{M}_{\text{wall}}^{\text{ES}}}, \mathcal{K}\left(\vec{b}_{\zeta}^{\vec{J}_{\text{in}}^{\text{ES}}}\right)\right\rangle_{\mathbb{S}_{\text{wall}}} \tag{3-94a}$$

$$z_{\xi\zeta}^{\vec{M}_{\text{wall}}^{\text{ES}}\vec{J}_{\text{wall}}^{\text{ES}}} = \left\langle \vec{b}_{\xi}^{\vec{M}_{\text{wall}}^{\text{ES}}}, \text{P.V.}\mathcal{K}\left(\vec{b}_{\zeta}^{\vec{J}_{\text{wall}}^{\text{ES}}}\right)\right\rangle_{\mathbb{S}_{\text{wall}}} - \left\langle \vec{b}_{\xi}^{\vec{M}_{\text{wall}}^{\text{ES}}}, \text{P.V.}\mathcal{K}_0\left(-\vec{b}_{\zeta}^{\vec{J}_{\text{wall}}^{\text{ES}}}\right)\right\rangle_{\mathbb{S}_{\text{wall}}} \tag{3-94b}$$

$$z_{\xi\zeta}^{\vec{M}_{\text{wall}}^{\text{ES}}\vec{J}_{\text{out}}^{\text{ES}}} = \left\langle \vec{b}_{\xi}^{\vec{M}_{\text{wall}}^{\text{ES}}}, \mathcal{K}\left(-\vec{b}_{\zeta}^{\vec{J}_{\text{out}}^{\text{ES}}}\right)\right\rangle_{\mathbb{S}_{\text{wall}}} \tag{3-94c}$$

$$z_{\xi\zeta}^{\vec{M}_{\text{wall}}^{\text{ES}}\vec{M}_{\text{in}}^{\text{ES}}} = \left\langle \vec{b}_{\xi}^{\vec{M}_{\text{wall}}^{\text{ES}}}, -j\omega\varepsilon\mathcal{L}\left(\vec{b}_{\zeta}^{\vec{M}_{\text{in}}^{\text{ES}}}\right)\right\rangle_{\mathbb{S}_{\text{wall}}} \tag{3-94d}$$

$$z_{\xi\zeta}^{\vec{M}_{\text{wall}}^{\text{ES}}\vec{M}_{\text{wall}}^{\text{ES}}} = \left\langle \vec{b}_{\xi}^{\vec{M}_{\text{wall}}^{\text{ES}}}, -j\omega\varepsilon\mathcal{L}\left(\vec{b}_{\zeta}^{\vec{M}_{\text{wall}}^{\text{ES}}}\right)\right\rangle_{\mathbb{S}_{\text{wall}}} - \left\langle \vec{b}_{\xi}^{\vec{M}_{\text{wall}}^{\text{ES}}}, -j\omega\varepsilon_0\mathcal{L}_0\left(-\vec{b}_{\zeta}^{\vec{M}_{\text{wall}}^{\text{ES}}}\right)\right\rangle_{\mathbb{S}_{\text{wall}}} \tag{3-94e}$$

$$z_{\xi\zeta}^{\vec{M}_{\text{wall}}^{\text{ES}}\vec{M}_{\text{out}}^{\text{ES}}} = \left\langle \vec{b}_{\xi}^{\vec{M}_{\text{wall}}^{\text{ES}}}, -j\omega\varepsilon\mathcal{L}\left(-\vec{b}_{\zeta}^{\vec{M}_{\text{out}}^{\text{ES}}}\right)\right\rangle_{\mathbb{S}_{\text{wall}}} \tag{3-94f}$$

In Eq. (3-90a), the matrix elements are calculated as follows:

$$z_{\xi\zeta}^{\vec{J}_{\text{out}}^{\text{ES}}\vec{J}_{\text{in}}^{\text{ES}}} = \left\langle \vec{b}_{\xi}^{\vec{J}_{\text{out}}^{\text{ES}}}, -j\omega\mu\mathcal{L}\left(\vec{b}_{\zeta}^{\vec{J}_{\text{in}}^{\text{ES}}}\right)\right\rangle_{\mathbb{S}_{\text{out}}} \tag{3-95a}$$

$$z_{\xi\zeta}^{\vec{J}_{\text{out}}^{\text{ES}}\vec{J}_{\text{wall}}^{\text{ES}}} = \left\langle \vec{b}_{\xi}^{\vec{J}_{\text{out}}^{\text{ES}}}, -j\omega\mu\mathcal{L}\left(\vec{b}_{\zeta}^{\vec{J}_{\text{wall}}^{\text{ES}}}\right)\right\rangle_{\mathbb{S}_{\text{out}}} \tag{3-95b}$$

$$z_{\xi\zeta}^{\vec{J}_{\text{out}}^{\text{ES}}\vec{J}_{\text{out}}^{\text{ES}}} = \left\langle \vec{b}_{\xi}^{\vec{J}_{\text{out}}^{\text{ES}}}, -j\omega\mu\mathcal{L}\left(-\vec{b}_{\zeta}^{\vec{J}_{\text{out}}^{\text{ES}}}\right)\right\rangle_{\mathbb{S}_{\text{out}}} - \left\langle \vec{b}_{\xi}^{\vec{J}_{\text{out}}^{\text{ES}}}, -j\omega\mu\mathcal{L}\left(\vec{b}_{\zeta}^{\vec{J}_{\text{out}}^{\text{ES}}}\right)\right\rangle_{\mathbb{S}_{\text{out}}} \tag{3-95c}$$

$$z_{\xi\zeta}^{\vec{J}_{\text{out}}^{\text{ES}}\vec{M}_{\text{in}}^{\text{ES}}} = \left\langle \vec{b}_{\xi}^{\vec{J}_{\text{out}}^{\text{ES}}}, -\mathcal{K}\left(\vec{b}_{\zeta}^{\vec{M}_{\text{in}}^{\text{ES}}}\right)\right\rangle_{\mathbb{S}_{\text{out}}} \tag{3-95d}$$

$$z_{\xi\zeta}^{\vec{J}_{\text{out}}^{\text{ES}}\vec{M}_{\text{wall}}^{\text{ES}}} = \left\langle \vec{b}_{\xi}^{\vec{J}_{\text{out}}^{\text{ES}}}, -\mathcal{K}\left(\vec{b}_{\zeta}^{\vec{M}_{\text{wall}}^{\text{ES}}}\right)\right\rangle_{\mathbb{S}_{\text{out}}} \tag{3-95e}$$

$$z_{\xi\zeta}^{\vec{J}_{\text{out}}^{\text{ES}}\vec{M}_{\text{out}}^{\text{ES}}} = \left\langle \vec{b}_{\xi}^{\vec{J}_{\text{out}}^{\text{ES}}}, -\text{P.V.}\mathcal{K}\left(-\vec{b}_{\zeta}^{\vec{M}_{\text{out}}^{\text{ES}}}\right)\right\rangle_{\mathbb{S}_{\text{out}}} - \left\langle \vec{b}_{\xi}^{\vec{J}_{\text{out}}^{\text{ES}}}, -\text{P.V.}\mathcal{K}\left(\vec{b}_{\zeta}^{\vec{M}_{\text{out}}^{\text{ES}}}\right)\right\rangle_{\mathbb{S}_{\text{out}}} \tag{3-95f}$$

In Eq. (3-90b), the matrix elements are calculated as follows:

$$z_{\xi\zeta}^{\vec{M}_{\text{out}}^{\text{ES}}\vec{J}_{\text{in}}^{\text{ES}}} = \left\langle \vec{b}_{\xi}^{\vec{M}_{\text{out}}^{\text{ES}}}, \mathcal{K}\left(\vec{b}_{\zeta}^{\vec{J}_{\text{in}}^{\text{ES}}}\right)\right\rangle_{\mathbb{S}_{\text{out}}} \tag{3-96a}$$

$$z_{\xi\zeta}^{\vec{M}_{\text{out}}^{\text{ES}}\vec{J}_{\text{wall}}^{\text{ES}}} = \left\langle \vec{b}_{\xi}^{\vec{M}_{\text{out}}^{\text{ES}}}, \mathcal{K}\left(\vec{b}_{\zeta}^{\vec{J}_{\text{wall}}^{\text{ES}}}\right)\right\rangle_{\mathbb{S}_{\text{out}}} \tag{3-96b}$$

$$z_{\xi\zeta}^{\vec{M}_{\text{out}}^{\text{ES}}\vec{J}_{\text{out}}^{\text{ES}}} = \left\langle \vec{b}_{\xi}^{\vec{M}_{\text{out}}^{\text{ES}}}, \text{P.V.}\mathcal{K}\left(-\vec{b}_{\zeta}^{\vec{J}_{\text{out}}^{\text{ES}}}\right)\right\rangle_{\mathbb{S}_{\text{out}}} - \left\langle \vec{b}_{\xi}^{\vec{M}_{\text{out}}^{\text{ES}}}, \text{P.V.}\mathcal{K}\left(\vec{b}_{\zeta}^{\vec{J}_{\text{out}}^{\text{ES}}}\right)\right\rangle_{\mathbb{S}_{\text{out}}} \tag{3-96c}$$

$$z_{\xi\zeta}^{\vec{M}_{\text{out}}^{\text{ES}}\vec{M}_{\text{in}}^{\text{ES}}} = \left\langle \vec{b}_{\xi}^{\vec{M}_{\text{out}}^{\text{ES}}}, -j\omega\varepsilon\mathcal{L}\left(\vec{b}_{\zeta}^{\vec{M}_{\text{in}}^{\text{ES}}}\right)\right\rangle_{\mathbb{S}_{\text{out}}} \tag{3-96d}$$

$$z_{\xi\zeta}^{\vec{M}_{\text{out}}^{\text{ES}}\vec{M}_{\text{wall}}^{\text{ES}}} = \left\langle \vec{b}_{\xi}^{\vec{M}_{\text{out}}^{\text{ES}}}, -j\omega\varepsilon\mathcal{L}\left(\vec{b}_{\zeta}^{\vec{M}_{\text{wall}}^{\text{ES}}}\right)\right\rangle_{\mathbb{S}_{\text{out}}} \tag{3-96e}$$

$$z_{\xi\zeta}^{\vec{M}_{\text{out}}^{\text{ES}}\vec{M}_{\text{out}}^{\text{ES}}} = \left\langle \vec{b}_{\xi}^{\vec{M}_{\text{out}}^{\text{ES}}}, -j\omega\varepsilon\mathcal{L}\left(-\vec{b}_{\zeta}^{\vec{M}_{\text{out}}^{\text{ES}}}\right)\right\rangle_{\mathbb{S}_{\text{out}}} - \left\langle \vec{b}_{\xi}^{\vec{M}_{\text{out}}^{\text{ES}}}, -j\omega\varepsilon\mathcal{L}\left(\vec{b}_{\zeta}^{\vec{M}_{\text{out}}^{\text{ES}}}\right)\right\rangle_{\mathbb{S}_{\text{out}}} \tag{3-96f}$$

Below, we propose two different schemes — *dependent variable elimination* (*DVE*) and *solution/definition domain compression* (*SDC/DDC*) — for mathematically describing the modal space corresponding to the material guide shown in Fig. 3-22.

**Scheme I: Dependent Variable Elimination (DVE)**

By employing the Eqs. (3-88a)~(3-90b), we can obtain the transformation from basic variables $\overline{a}^{\text{BV}}$ to all variables $\overline{a}^{\text{AV}}$ as follows:

$$\begin{bmatrix} \overline{a}^{\vec{J}_{\text{in}}^{\text{ES}}} \\ \overline{a}^{\vec{J}_{\text{wall}}^{\text{ES}}} \\ \overline{a}^{\vec{J}_{\text{out}}^{\text{ES}}} \\ \overline{a}^{\vec{M}_{\text{in}}^{\text{ES}}} \\ \overline{a}^{\vec{M}_{\text{wall}}^{\text{ES}}} \\ \overline{a}^{\vec{M}_{\text{out}}^{\text{ES}}} \end{bmatrix} = \overline{a}^{\text{AV}} = \overline{\overline{T}}^{\text{BV}\to\text{AV}} \cdot \overline{a}^{\text{BV}} = \begin{cases} \overline{\overline{T}}^{\vec{J}_{\text{in}}^{\text{ES}}\to\text{AV}} \cdot \overline{a}^{\vec{J}_{\text{in}}^{\text{ES}}} \\ \overline{\overline{T}}^{\vec{M}_{\text{in}}^{\text{ES}}\to\text{AV}} \cdot \overline{a}^{\vec{M}_{\text{in}}^{\text{ES}}} \end{cases} \tag{3-97}$$

in which $\overline{\overline{T}}^{\text{BV}\to\text{AV}} = \overline{\overline{T}}^{\vec{J}_{\text{in}}^{\text{ES}}\to\text{AV}} / \overline{\overline{T}}^{\vec{M}_{\text{in}}^{\text{ES}}\to\text{AV}}$, and correspondingly $\overline{a}^{\text{BV}} = \overline{a}^{\vec{J}_{\text{in}}^{\text{ES}}} / \overline{a}^{\vec{M}_{\text{in}}^{\text{ES}}}$. In addition,

$$\overline{\overline{T}}^{\vec{J}_{\text{in}}^{\text{ES}}\to\text{AV}} = \begin{bmatrix} \overline{\overline{I}}^{\vec{J}_{\text{in}}^{\text{ES}}} & 0 & 0 & 0 & 0 & 0 \\ 0 & \overline{\overline{Z}}^{\vec{M}_{\text{in}}^{\text{ES}}\vec{J}_{\text{wall}}^{\text{ES}}} & \overline{\overline{Z}}^{\vec{M}_{\text{in}}^{\text{ES}}\vec{J}_{\text{out}}^{\text{ES}}} & \overline{\overline{Z}}^{\vec{M}_{\text{in}}^{\text{ES}}\vec{M}_{\text{in}}^{\text{ES}}} & \overline{\overline{Z}}^{\vec{M}_{\text{in}}^{\text{ES}}\vec{M}_{\text{wall}}^{\text{ES}}} & \overline{\overline{Z}}^{\vec{M}_{\text{in}}^{\text{ES}}\vec{M}_{\text{out}}^{\text{ES}}} \\ 0 & \overline{\overline{Z}}^{\vec{J}_{\text{wall}}^{\text{ES}}\vec{J}_{\text{wall}}^{\text{ES}}} & \overline{\overline{Z}}^{\vec{J}_{\text{wall}}^{\text{ES}}\vec{J}_{\text{out}}^{\text{ES}}} & \overline{\overline{Z}}^{\vec{J}_{\text{wall}}^{\text{ES}}\vec{M}_{\text{in}}^{\text{ES}}} & \overline{\overline{Z}}^{\vec{J}_{\text{wall}}^{\text{ES}}\vec{M}_{\text{wall}}^{\text{ES}}} & \overline{\overline{Z}}^{\vec{J}_{\text{wall}}^{\text{ES}}\vec{M}_{\text{out}}^{\text{ES}}} \\ 0 & \overline{\overline{Z}}^{\vec{M}_{\text{wall}}^{\text{ES}}\vec{J}_{\text{wall}}^{\text{ES}}} & \overline{\overline{Z}}^{\vec{M}_{\text{wall}}^{\text{ES}}\vec{J}_{\text{out}}^{\text{ES}}} & \overline{\overline{Z}}^{\vec{M}_{\text{wall}}^{\text{ES}}\vec{M}_{\text{in}}^{\text{ES}}} & \overline{\overline{Z}}^{\vec{M}_{\text{wall}}^{\text{ES}}\vec{M}_{\text{wall}}^{\text{ES}}} & \overline{\overline{Z}}^{\vec{M}_{\text{wall}}^{\text{ES}}\vec{M}_{\text{out}}^{\text{ES}}} \\ 0 & \overline{\overline{Z}}^{\vec{J}_{\text{out}}^{\text{ES}}\vec{J}_{\text{wall}}^{\text{ES}}} & \overline{\overline{Z}}^{\vec{J}_{\text{out}}^{\text{ES}}\vec{J}_{\text{out}}^{\text{ES}}} & \overline{\overline{Z}}^{\vec{J}_{\text{out}}^{\text{ES}}\vec{M}_{\text{in}}^{\text{ES}}} & \overline{\overline{Z}}^{\vec{J}_{\text{out}}^{\text{ES}}\vec{M}_{\text{wall}}^{\text{ES}}} & \overline{\overline{Z}}^{\vec{J}_{\text{out}}^{\text{ES}}\vec{M}_{\text{out}}^{\text{ES}}} \\ 0 & \overline{\overline{Z}}^{\vec{M}_{\text{out}}^{\text{ES}}\vec{J}_{\text{wall}}^{\text{ES}}} & \overline{\overline{Z}}^{\vec{M}_{\text{out}}^{\text{ES}}\vec{J}_{\text{out}}^{\text{ES}}} & \overline{\overline{Z}}^{\vec{M}_{\text{out}}^{\text{ES}}\vec{M}_{\text{in}}^{\text{ES}}} & \overline{\overline{Z}}^{\vec{M}_{\text{out}}^{\text{ES}}\vec{M}_{\text{wall}}^{\text{ES}}} & \overline{\overline{Z}}^{\vec{M}_{\text{out}}^{\text{ES}}\vec{M}_{\text{out}}^{\text{ES}}} \end{bmatrix}^{-1} \cdot \begin{bmatrix} \overline{\overline{I}}^{\vec{J}_{\text{in}}^{\text{ES}}} \\ -\overline{\overline{Z}}^{\vec{M}_{\text{in}}^{\text{ES}}\vec{J}_{\text{in}}^{\text{ES}}} \\ -\overline{\overline{Z}}^{\vec{J}_{\text{wall}}^{\text{ES}}\vec{J}_{\text{in}}^{\text{ES}}} \\ -\overline{\overline{Z}}^{\vec{M}_{\text{wall}}^{\text{ES}}\vec{J}_{\text{in}}^{\text{ES}}} \\ -\overline{\overline{Z}}^{\vec{J}_{\text{out}}^{\text{ES}}\vec{J}_{\text{in}}^{\text{ES}}} \\ -\overline{\overline{Z}}^{\vec{M}_{\text{out}}^{\text{ES}}\vec{J}_{\text{in}}^{\text{ES}}} \end{bmatrix} \tag{3-98a}$$

$$\overline{\overline{T}}^{\vec{M}_{\text{in}}^{\text{ES}}\to\text{AV}} = \begin{bmatrix} 0 & 0 & 0 & \overline{\overline{I}}^{\vec{M}_{\text{in}}^{\text{ES}}} & 0 & 0 \\ \overline{\overline{Z}}^{\vec{J}_{\text{in}}^{\text{ES}}\vec{J}_{\text{in}}^{\text{ES}}} & \overline{\overline{Z}}^{\vec{J}_{\text{in}}^{\text{ES}}\vec{J}_{\text{wall}}^{\text{ES}}} & \overline{\overline{Z}}^{\vec{J}_{\text{in}}^{\text{ES}}\vec{J}_{\text{out}}^{\text{ES}}} & 0 & \overline{\overline{Z}}^{\vec{J}_{\text{in}}^{\text{ES}}\vec{M}_{\text{wall}}^{\text{ES}}} & \overline{\overline{Z}}^{\vec{J}_{\text{in}}^{\text{ES}}\vec{M}_{\text{out}}^{\text{ES}}} \\ \overline{\overline{Z}}^{\vec{J}_{\text{wall}}^{\text{ES}}\vec{J}_{\text{in}}^{\text{ES}}} & \overline{\overline{Z}}^{\vec{J}_{\text{wall}}^{\text{ES}}\vec{J}_{\text{wall}}^{\text{ES}}} & \overline{\overline{Z}}^{\vec{J}_{\text{wall}}^{\text{ES}}\vec{J}_{\text{out}}^{\text{ES}}} & 0 & \overline{\overline{Z}}^{\vec{J}_{\text{wall}}^{\text{ES}}\vec{M}_{\text{wall}}^{\text{ES}}} & \overline{\overline{Z}}^{\vec{J}_{\text{wall}}^{\text{ES}}\vec{M}_{\text{out}}^{\text{ES}}} \\ \overline{\overline{Z}}^{\vec{M}_{\text{wall}}^{\text{ES}}\vec{J}_{\text{in}}^{\text{ES}}} & \overline{\overline{Z}}^{\vec{M}_{\text{wall}}^{\text{ES}}\vec{J}_{\text{wall}}^{\text{ES}}} & \overline{\overline{Z}}^{\vec{M}_{\text{wall}}^{\text{ES}}\vec{J}_{\text{out}}^{\text{ES}}} & 0 & \overline{\overline{Z}}^{\vec{M}_{\text{wall}}^{\text{ES}}\vec{M}_{\text{wall}}^{\text{ES}}} & \overline{\overline{Z}}^{\vec{M}_{\text{wall}}^{\text{ES}}\vec{M}_{\text{out}}^{\text{ES}}} \\ \overline{\overline{Z}}^{\vec{J}_{\text{out}}^{\text{ES}}\vec{J}_{\text{in}}^{\text{ES}}} & \overline{\overline{Z}}^{\vec{J}_{\text{out}}^{\text{ES}}\vec{J}_{\text{wall}}^{\text{ES}}} & \overline{\overline{Z}}^{\vec{J}_{\text{out}}^{\text{ES}}\vec{J}_{\text{out}}^{\text{ES}}} & 0 & \overline{\overline{Z}}^{\vec{J}_{\text{out}}^{\text{ES}}\vec{M}_{\text{wall}}^{\text{ES}}} & \overline{\overline{Z}}^{\vec{J}_{\text{out}}^{\text{ES}}\vec{M}_{\text{out}}^{\text{ES}}} \\ \overline{\overline{Z}}^{\vec{M}_{\text{out}}^{\text{ES}}\vec{J}_{\text{in}}^{\text{ES}}} & \overline{\overline{Z}}^{\vec{M}_{\text{out}}^{\text{ES}}\vec{J}_{\text{wall}}^{\text{ES}}} & \overline{\overline{Z}}^{\vec{M}_{\text{out}}^{\text{ES}}\vec{J}_{\text{out}}^{\text{ES}}} & 0 & \overline{\overline{Z}}^{\vec{M}_{\text{out}}^{\text{ES}}\vec{M}_{\text{wall}}^{\text{ES}}} & \overline{\overline{Z}}^{\vec{M}_{\text{out}}^{\text{ES}}\vec{M}_{\text{out}}^{\text{ES}}} \end{bmatrix}^{-1} \cdot \begin{bmatrix} \overline{\overline{I}}^{\vec{M}_{\text{in}}^{\text{ES}}} \\ -\overline{\overline{Z}}^{\vec{J}_{\text{in}}^{\text{ES}}\vec{M}_{\text{in}}^{\text{ES}}} \\ -\overline{\overline{Z}}^{\vec{J}_{\text{wall}}^{\text{ES}}\vec{M}_{\text{in}}^{\text{ES}}} \\ -\overline{\overline{Z}}^{\vec{M}_{\text{wall}}^{\text{ES}}\vec{M}_{\text{in}}^{\text{ES}}} \\ -\overline{\overline{Z}}^{\vec{J}_{\text{out}}^{\text{ES}}\vec{M}_{\text{in}}^{\text{ES}}} \\ -\overline{\overline{Z}}^{\vec{M}_{\text{out}}^{\text{ES}}\vec{M}_{\text{in}}^{\text{ES}}} \end{bmatrix} \tag{3-98b}$$

In the above Eqs. (3-98a) and (3-98b), the $\overline{\overline{I}}^{\vec{J}_{\text{in}}^{\text{ES}}}$ and $\overline{\overline{I}}^{\vec{M}_{\text{in}}^{\text{ES}}}$ are two identity matrices with proper orders; the 0s are some zero matrices with proper row and column numbers.

**Scheme II: Solution/Definition Domain Compression (SDC/DDC)**

In addition, by employing the Eqs. (3-88a)~(3-90b), we can also obtain the following transformation

$$\overline{a}^{\text{AV}} = \overline{\overline{T}}^{\text{BS}\to\text{AV}} \cdot \overline{a}^{\text{BS}} \tag{3-99}$$

where $\overline{\overline{T}}^{\text{BS}\to\text{AV}} = \left[\overline{s}_1^{\text{BS}}, \overline{s}_2^{\text{BS}}, \cdots\right]$, and $\{\overline{s}_1^{\text{BS}}, \overline{s}_2^{\text{BS}}, \cdots\}$ are the basic solutions of the following two theoretically equivalent equations

$$\overline{\overline{\Psi}}_{\text{FCE}}^{\text{DoJ}} \cdot \bar{a}^{\text{AV}} = 0 \tag{3-100a}$$

$$\overline{\overline{\Psi}}_{\text{FCE}}^{\text{DoM}} \cdot \bar{a}^{\text{AV}} = 0 \tag{3-100b}$$

where

$$\overline{\overline{\Psi}}_{\text{FCE}}^{\text{DoJ}} = \begin{bmatrix} \overline{\overline{Z}}^{\vec{M}_{\text{in}}^{\text{ES}}\vec{J}_{\text{in}}^{\text{ES}}} & \overline{\overline{Z}}^{\vec{M}_{\text{in}}^{\text{ES}}\vec{J}_{\text{wall}}^{\text{ES}}} & \overline{\overline{Z}}^{\vec{M}_{\text{in}}^{\text{ES}}\vec{J}_{\text{out}}^{\text{ES}}} & \overline{\overline{Z}}^{\vec{M}_{\text{in}}^{\text{ES}}\vec{M}_{\text{in}}^{\text{ES}}} & \overline{\overline{Z}}^{\vec{M}_{\text{in}}^{\text{ES}}\vec{M}_{\text{wall}}^{\text{ES}}} & \overline{\overline{Z}}^{\vec{M}_{\text{in}}^{\text{ES}}\vec{M}_{\text{out}}^{\text{ES}}} \\ \overline{\overline{Z}}^{\vec{J}_{\text{wall}}^{\text{ES}}\vec{J}_{\text{in}}^{\text{ES}}} & \overline{\overline{Z}}^{\vec{J}_{\text{wall}}^{\text{ES}}\vec{J}_{\text{wall}}^{\text{ES}}} & \overline{\overline{Z}}^{\vec{J}_{\text{wall}}^{\text{ES}}\vec{J}_{\text{out}}^{\text{ES}}} & \overline{\overline{Z}}^{\vec{J}_{\text{wall}}^{\text{ES}}\vec{M}_{\text{in}}^{\text{ES}}} & \overline{\overline{Z}}^{\vec{J}_{\text{wall}}^{\text{ES}}\vec{M}_{\text{wall}}^{\text{ES}}} & \overline{\overline{Z}}^{\vec{J}_{\text{wall}}^{\text{ES}}\vec{M}_{\text{out}}^{\text{ES}}} \\ \overline{\overline{Z}}^{\vec{M}_{\text{wall}}^{\text{ES}}\vec{J}_{\text{in}}^{\text{ES}}} & \overline{\overline{Z}}^{\vec{M}_{\text{wall}}^{\text{ES}}\vec{J}_{\text{wall}}^{\text{ES}}} & \overline{\overline{Z}}^{\vec{M}_{\text{wall}}^{\text{ES}}\vec{J}_{\text{out}}^{\text{ES}}} & \overline{\overline{Z}}^{\vec{M}_{\text{wall}}^{\text{ES}}\vec{M}_{\text{in}}^{\text{ES}}} & \overline{\overline{Z}}^{\vec{M}_{\text{wall}}^{\text{ES}}\vec{M}_{\text{wall}}^{\text{ES}}} & \overline{\overline{Z}}^{\vec{M}_{\text{wall}}^{\text{ES}}\vec{M}_{\text{out}}^{\text{ES}}} \\ \overline{\overline{Z}}^{\vec{J}_{\text{out}}^{\text{ES}}\vec{J}_{\text{in}}^{\text{ES}}} & \overline{\overline{Z}}^{\vec{J}_{\text{out}}^{\text{ES}}\vec{J}_{\text{wall}}^{\text{ES}}} & \overline{\overline{Z}}^{\vec{J}_{\text{out}}^{\text{ES}}\vec{J}_{\text{out}}^{\text{ES}}} & \overline{\overline{Z}}^{\vec{J}_{\text{out}}^{\text{ES}}\vec{M}_{\text{in}}^{\text{ES}}} & \overline{\overline{Z}}^{\vec{J}_{\text{out}}^{\text{ES}}\vec{M}_{\text{wall}}^{\text{ES}}} & \overline{\overline{Z}}^{\vec{J}_{\text{out}}^{\text{ES}}\vec{M}_{\text{out}}^{\text{ES}}} \\ \overline{\overline{Z}}^{\vec{M}_{\text{out}}^{\text{ES}}\vec{J}_{\text{in}}^{\text{ES}}} & \overline{\overline{Z}}^{\vec{M}_{\text{out}}^{\text{ES}}\vec{J}_{\text{wall}}^{\text{ES}}} & \overline{\overline{Z}}^{\vec{M}_{\text{out}}^{\text{ES}}\vec{J}_{\text{out}}^{\text{ES}}} & \overline{\overline{Z}}^{\vec{M}_{\text{out}}^{\text{ES}}\vec{M}_{\text{in}}^{\text{ES}}} & \overline{\overline{Z}}^{\vec{M}_{\text{out}}^{\text{ES}}\vec{M}_{\text{wall}}^{\text{ES}}} & \overline{\overline{Z}}^{\vec{M}_{\text{out}}^{\text{ES}}\vec{M}_{\text{out}}^{\text{ES}}} \end{bmatrix} \tag{3-101a}$$

$$\overline{\overline{\Psi}}_{\text{FCE}}^{\text{DoM}} = \begin{bmatrix} \overline{\overline{Z}}^{\vec{J}_{\text{in}}^{\text{ES}}\vec{J}_{\text{in}}^{\text{ES}}} & \overline{\overline{Z}}^{\vec{J}_{\text{in}}^{\text{ES}}\vec{J}_{\text{wall}}^{\text{ES}}} & \overline{\overline{Z}}^{\vec{J}_{\text{in}}^{\text{ES}}\vec{J}_{\text{out}}^{\text{ES}}} & \overline{\overline{Z}}^{\vec{J}_{\text{in}}^{\text{ES}}\vec{M}_{\text{in}}^{\text{ES}}} & \overline{\overline{Z}}^{\vec{J}_{\text{in}}^{\text{ES}}\vec{M}_{\text{wall}}^{\text{ES}}} & \overline{\overline{Z}}^{\vec{J}_{\text{in}}^{\text{ES}}\vec{M}_{\text{out}}^{\text{ES}}} \\ \overline{\overline{Z}}^{\vec{J}_{\text{wall}}^{\text{ES}}\vec{J}_{\text{in}}^{\text{ES}}} & \overline{\overline{Z}}^{\vec{J}_{\text{wall}}^{\text{ES}}\vec{J}_{\text{wall}}^{\text{ES}}} & \overline{\overline{Z}}^{\vec{J}_{\text{wall}}^{\text{ES}}\vec{J}_{\text{out}}^{\text{ES}}} & \overline{\overline{Z}}^{\vec{J}_{\text{wall}}^{\text{ES}}\vec{M}_{\text{in}}^{\text{ES}}} & \overline{\overline{Z}}^{\vec{J}_{\text{wall}}^{\text{ES}}\vec{M}_{\text{wall}}^{\text{ES}}} & \overline{\overline{Z}}^{\vec{J}_{\text{wall}}^{\text{ES}}\vec{M}_{\text{out}}^{\text{ES}}} \\ \overline{\overline{Z}}^{\vec{M}_{\text{wall}}^{\text{ES}}\vec{J}_{\text{in}}^{\text{ES}}} & \overline{\overline{Z}}^{\vec{M}_{\text{wall}}^{\text{ES}}\vec{J}_{\text{wall}}^{\text{ES}}} & \overline{\overline{Z}}^{\vec{M}_{\text{wall}}^{\text{ES}}\vec{J}_{\text{out}}^{\text{ES}}} & \overline{\overline{Z}}^{\vec{M}_{\text{wall}}^{\text{ES}}\vec{M}_{\text{in}}^{\text{ES}}} & \overline{\overline{Z}}^{\vec{M}_{\text{wall}}^{\text{ES}}\vec{M}_{\text{wall}}^{\text{ES}}} & \overline{\overline{Z}}^{\vec{M}_{\text{wall}}^{\text{ES}}\vec{M}_{\text{out}}^{\text{ES}}} \\ \overline{\overline{Z}}^{\vec{J}_{\text{out}}^{\text{ES}}\vec{J}_{\text{in}}^{\text{ES}}} & \overline{\overline{Z}}^{\vec{J}_{\text{out}}^{\text{ES}}\vec{J}_{\text{wall}}^{\text{ES}}} & \overline{\overline{Z}}^{\vec{J}_{\text{out}}^{\text{ES}}\vec{J}_{\text{out}}^{\text{ES}}} & \overline{\overline{Z}}^{\vec{J}_{\text{out}}^{\text{ES}}\vec{M}_{\text{in}}^{\text{ES}}} & \overline{\overline{Z}}^{\vec{J}_{\text{out}}^{\text{ES}}\vec{M}_{\text{wall}}^{\text{ES}}} & \overline{\overline{Z}}^{\vec{J}_{\text{out}}^{\text{ES}}\vec{M}_{\text{out}}^{\text{ES}}} \\ \overline{\overline{Z}}^{\vec{M}_{\text{out}}^{\text{ES}}\vec{J}_{\text{in}}^{\text{ES}}} & \overline{\overline{Z}}^{\vec{M}_{\text{out}}^{\text{ES}}\vec{J}_{\text{wall}}^{\text{ES}}} & \overline{\overline{Z}}^{\vec{M}_{\text{out}}^{\text{ES}}\vec{J}_{\text{out}}^{\text{ES}}} & \overline{\overline{Z}}^{\vec{M}_{\text{out}}^{\text{ES}}\vec{M}_{\text{in}}^{\text{ES}}} & \overline{\overline{Z}}^{\vec{M}_{\text{out}}^{\text{ES}}\vec{M}_{\text{wall}}^{\text{ES}}} & \overline{\overline{Z}}^{\vec{M}_{\text{out}}^{\text{ES}}\vec{M}_{\text{out}}^{\text{ES}}} \end{bmatrix} \tag{3-101b}$$

where the superscripts and subscripts of $\overline{\overline{\Psi}}_{\text{FCE}}^{\text{DoJ}}$ and $\overline{\overline{\Psi}}_{\text{FCE}}^{\text{DoM}}$ have the same meanings as the ones used in Eqs. (3-44) and (3-45).

For the convenience of the following discussions, Eqs. (3-97) and (3-99) are uniformly written as follows:

$$\bar{a}^{\text{AV}} = \overline{\overline{T}} \cdot \bar{a} \tag{3-102}$$

where $\bar{a} = \bar{a}^{\text{BV}} / \bar{a}^{\text{BS}}$ and correspondingly $\overline{\overline{T}} = \overline{\overline{T}}^{\text{BV}\to\text{AV}} / \overline{\overline{T}}^{\text{BS}\to\text{AV}}$.

## 3.3.3 Input Power Operator

The IPO corresponding to the input port $\mathbb{S}_{\text{in}}$ shown in Fig. 3-22 is as follows:

$$P^{\text{in}} = (1/2)\iint_{\mathbb{S}_{\text{in}}} \left(\vec{E} \times \vec{H}^{\dagger}\right) \cdot \hat{z} dS \tag{3-103}$$

called the field form of the IPO. Below, we provide two alternative forms for the IPO.

**Formulation I: Current Form**

Based on Eqs. (3-80a) and (3-80b), we have that

$$P^{\text{in}} = (1/2)\left\langle \hat{z} \times \vec{J}_{\text{in}}^{\text{ES}}, \vec{M}_{\text{in}}^{\text{ES}} \right\rangle_{\mathbb{S}_{\text{in}}} \tag{3-104}$$

Substituting the expansion formulations of $\{\vec{J}_{\text{in}}^{\text{ES}}, \vec{M}_{\text{in}}^{\text{ES}}\}$ into the above IPO, the IPO is immediately discretized as follows:

$$P^{\text{in}} = \left(\bar{a}^{\text{AV}}\right)^{\dagger} \cdot \underbrace{\begin{bmatrix} 0 & 0 & 0 & \overline{\overline{C}}^{\vec{J}_{\text{in}}^{\text{ES}}\vec{M}_{\text{in}}^{\text{ES}}} & 0 & 0 \\ 0 & 0 & 0 & 0 & 0 & 0 \\ 0 & 0 & 0 & 0 & 0 & 0 \\ 0 & 0 & 0 & 0 & 0 & 0 \\ 0 & 0 & 0 & 0 & 0 & 0 \\ 0 & 0 & 0 & 0 & 0 & 0 \end{bmatrix}}_{\overline{\overline{P}}_{\text{curAV}}^{\text{in}}} \cdot \bar{a}^{\text{AV}} \tag{3-105}$$

where the elements of the sub-matrix $\overline{\overline{C}}^{\vec{J}_{\text{in}}^{\text{ES}}\vec{M}_{\text{in}}^{\text{ES}}}$ are calculated as that $c_{\xi\zeta}^{\vec{J}_{\text{in}}^{\text{ES}}\vec{M}_{\text{in}}^{\text{ES}}} = (1/2) < \hat{z} \times \vec{b}_{\xi}^{\vec{J}_{\text{in}}^{\text{ES}}}, \vec{b}_{\zeta}^{\vec{M}_{\text{in}}^{\text{ES}}} >_{\mathbb{S}_{\text{in}}}$. Substituting Eq. (3-102) into the above Eq. (3-105), we immediately have that

$$P^{\text{in}} = \bar{a}^{\dagger} \cdot \underbrace{\left(\overline{\overline{T}}^{\dagger} \cdot \overline{\overline{P}}_{\text{curAV}}^{\text{in}} \cdot \overline{\overline{T}}\right)}_{\overline{\overline{P}}_{\text{cur}}^{\text{in}}} \cdot \bar{a} \tag{3-106}$$

which is just the current form defined on modal space, and the subscript "cur" is to emphasize that the corresponding matrix is derived from discretizing the current form.

**Formulation II: Field-Current Interaction Form**

Similarly to the discussions for the previous metallic guide, we have the following alternative IPO expressions

$$P^{\text{in}} = -(1/2)\left\langle \vec{J}_{\text{in}}^{\text{ES}}, \mathcal{E}\left(\vec{J}_{\text{in}}^{\text{ES}} + \vec{J}_{\text{wall}}^{\text{ES}} - \vec{J}_{\text{out}}^{\text{ES}}, \vec{M}_{\text{in}}^{\text{ES}} + \vec{M}_{\text{wall}}^{\text{ES}} - \vec{M}_{\text{out}}^{\text{ES}}\right)\right\rangle_{\mathbb{S}_{\text{in}}^{+}} \tag{3-107a}$$

$$P^{\text{in}} = -(1/2)\left\langle \vec{M}_{\text{in}}^{\text{ES}}, \mathcal{H}\left(\vec{J}_{\text{in}}^{\text{ES}} + \vec{J}_{\text{wall}}^{\text{ES}} - \vec{J}_{\text{out}}^{\text{ES}}, \vec{M}_{\text{in}}^{\text{ES}} + \vec{M}_{\text{wall}}^{\text{ES}} - \vec{M}_{\text{out}}^{\text{ES}}\right)\right\rangle_{\mathbb{S}_{\text{in}}^{+}}^{\dagger} \tag{3-107b}$$

for the material guide, where integral domain $\mathbb{S}_{\text{in}}^{+}$ is the right-side surface of $\mathbb{S}_{\text{in}}$.

Substituting the expansion formulations of the currents into the above expressions, the expressions are immediately discretized as follows:

$$P^{\text{in}} = \left(\bar{a}^{\text{AV}}\right)^{\dagger} \cdot \overline{\overline{P}}_{\text{intAV}}^{\text{in}} \cdot \bar{a}^{\text{AV}} \tag{3-108}$$

in which

$$\overline{\overline{P}}_{\text{intAV}}^{\text{in}} = \begin{bmatrix} \overline{\overline{P}}^{\vec{J}_{\text{in}}^{\text{ES}}\vec{J}_{\text{in}}^{\text{ES}}} & \overline{\overline{P}}^{\vec{J}_{\text{in}}^{\text{ES}}\vec{J}_{\text{wall}}^{\text{ES}}} & \overline{\overline{P}}^{\vec{J}_{\text{in}}^{\text{ES}}\vec{J}_{\text{out}}^{\text{ES}}} & \overline{\overline{P}}^{\vec{J}_{\text{in}}^{\text{ES}}\vec{M}_{\text{in}}^{\text{ES}}} & \overline{\overline{P}}^{\vec{J}_{\text{in}}^{\text{ES}}\vec{M}_{\text{wall}}^{\text{ES}}} & \overline{\overline{P}}^{\vec{J}_{\text{in}}^{\text{ES}}\vec{M}_{\text{out}}^{\text{ES}}} \\ 0 & 0 & 0 & 0 & 0 & 0 \\ 0 & 0 & 0 & 0 & 0 & 0 \\ 0 & 0 & 0 & 0 & 0 & 0 \\ 0 & 0 & 0 & 0 & 0 & 0 \\ 0 & 0 & 0 & 0 & 0 & 0 \end{bmatrix} \tag{3-109a}$$

for Eq. (3-107a), or

$$\overline{\overline{P}}_{\text{intAV}}^{\text{in}} = \begin{bmatrix} 0 & 0 & 0 & 0 & 0 & 0 \\ 0 & 0 & 0 & 0 & 0 & 0 \\ 0 & 0 & 0 & 0 & 0 & 0 \\ \overline{\overline{P}}^{\vec{M}_{\text{in}}^{\text{ES}}\vec{J}_{\text{in}}^{\text{ES}}} & \overline{\overline{P}}^{\vec{M}_{\text{in}}^{\text{ES}}\vec{J}_{\text{wall}}^{\text{ES}}} & \overline{\overline{P}}^{\vec{M}_{\text{in}}^{\text{ES}}\vec{J}_{\text{out}}^{\text{ES}}} & \overline{\overline{P}}^{\vec{M}_{\text{in}}^{\text{ES}}\vec{M}_{\text{in}}^{\text{ES}}} & \overline{\overline{P}}^{\vec{M}_{\text{in}}^{\text{ES}}\vec{M}_{\text{wall}}^{\text{ES}}} & \overline{\overline{P}}^{\vec{M}_{\text{in}}^{\text{ES}}\vec{M}_{\text{out}}^{\text{ES}}} \\ 0 & 0 & 0 & 0 & 0 & 0 \\ 0 & 0 & 0 & 0 & 0 & 0 \end{bmatrix} \tag{3-109b}$$

for Eq. (3-107b), where the elements of the sub-matrices are as follows:

$$p_{\xi\zeta}^{\vec{J}_{\text{in}}^{\text{ES}}\vec{J}_{\text{in}}^{\text{ES}}} = -(1/2)\left\langle \vec{b}_{\xi}^{\vec{J}_{\text{in}}^{\text{ES}}}, -j\omega\mu\mathcal{L}\left(\vec{b}_{\zeta}^{\vec{J}_{\text{in}}^{\text{ES}}}\right)\right\rangle_{\mathbb{S}_{\text{in}}} \tag{3-110a}$$

$$p_{\xi\zeta}^{\vec{J}_{\text{in}}^{\text{ES}}\vec{J}_{\text{wall}}^{\text{ES}}} = -(1/2)\left\langle \vec{b}_{\xi}^{\vec{J}_{\text{in}}^{\text{ES}}}, -j\omega\mu\mathcal{L}\left(\vec{b}_{\zeta}^{\vec{J}_{\text{wall}}^{\text{ES}}}\right)\right\rangle_{\mathbb{S}_{\text{in}}} \tag{3-110b}$$

$$p_{\xi\zeta}^{\vec{J}_{\text{in}}^{\text{ES}}\vec{J}_{\text{out}}^{\text{ES}}} = -(1/2)\left\langle \vec{b}_{\xi}^{\vec{J}_{\text{in}}^{\text{ES}}}, -j\omega\mu\mathcal{L}\left(-\vec{b}_{\zeta}^{\vec{J}_{\text{out}}^{\text{ES}}}\right)\right\rangle_{\mathbb{S}_{\text{in}}} \tag{3-110c}$$

$$p_{\xi\zeta}^{\vec{J}_{\text{in}}^{\text{ES}}\vec{M}_{\text{in}}^{\text{ES}}} = -(1/2)\left\langle \vec{b}_{\xi}^{\vec{J}_{\text{in}}^{\text{ES}}}, \hat{z}\times\frac{1}{2}\vec{b}_{\zeta}^{\vec{M}_{\text{in}}^{\text{ES}}} - \text{P.V.}\,\mathcal{K}\left(\vec{b}_{\zeta}^{\vec{M}_{\text{in}}^{\text{ES}}}\right)\right\rangle_{\mathbb{S}_{\text{in}}} \tag{3-110d}$$

$$p_{\xi\zeta}^{\vec{J}_{\text{in}}^{\text{ES}}\vec{M}_{\text{wall}}^{\text{ES}}} = -(1/2)\left\langle \vec{b}_{\xi}^{\vec{J}_{\text{in}}^{\text{ES}}}, -\mathcal{K}\left(\vec{b}_{\zeta}^{\vec{M}_{\text{wall}}^{\text{ES}}}\right)\right\rangle_{\mathbb{S}_{\text{in}}} \tag{3-110e}$$

$$p_{\xi\zeta}^{\vec{J}_{\text{in}}^{\text{ES}}\vec{M}_{\text{out}}^{\text{ES}}} = -(1/2)\left\langle \vec{b}_{\xi}^{\vec{J}_{\text{in}}^{\text{ES}}}, -\mathcal{K}\left(-\vec{b}_{\zeta}^{\vec{M}_{\text{out}}^{\text{ES}}}\right)\right\rangle_{\mathbb{S}_{\text{in}}} \tag{3-110f}$$

and

$$p_{\xi\zeta}^{\vec{M}_{\text{in}}^{\text{ES}}\vec{J}_{\text{in}}^{\text{ES}}} = -(1/2)\left\langle \vec{b}_{\xi}^{\vec{M}_{\text{in}}^{\text{ES}}}, \frac{1}{2}\vec{b}_{\zeta}^{\vec{J}_{\text{in}}^{\text{ES}}}\times\hat{z} + \text{P.V.}\,\mathcal{K}\left(\vec{b}_{\zeta}^{\vec{J}_{\text{in}}^{\text{ES}}}\right)\right\rangle_{\mathbb{S}_{\text{in}}} \tag{3-110g}$$

$$p_{\xi\zeta}^{\vec{M}_{\text{in}}^{\text{ES}}\vec{J}_{\text{wall}}^{\text{ES}}} = -(1/2)\left\langle \vec{b}_{\xi}^{\vec{M}_{\text{in}}^{\text{ES}}}, \mathcal{K}\left(\vec{b}_{\zeta}^{\vec{J}_{\text{wall}}^{\text{ES}}}\right)\right\rangle_{\mathbb{S}_{\text{in}}} \tag{3-110h}$$

$$p_{\xi\zeta}^{\vec{M}_{\text{in}}^{\text{ES}}\vec{J}_{\text{out}}^{\text{ES}}} = -(1/2)\left\langle \vec{b}_{\xi}^{\vec{M}_{\text{in}}^{\text{ES}}}, \mathcal{K}\left(-\vec{b}_{\zeta}^{\vec{J}_{\text{out}}^{\text{ES}}}\right)\right\rangle_{\mathbb{S}_{\text{in}}} \tag{3-110i}$$

$$p_{\xi\zeta}^{\vec{M}_{\text{in}}^{\text{ES}}\vec{M}_{\text{in}}^{\text{ES}}} = -(1/2)\left\langle \vec{b}_{\xi}^{\vec{M}_{\text{in}}^{\text{ES}}}, -j\omega\varepsilon\mathcal{L}\left(\vec{b}_{\zeta}^{\vec{M}_{\text{in}}^{\text{ES}}}\right)\right\rangle_{\mathbb{S}_{\text{in}}} \tag{3-110j}$$

$$p_{\xi\zeta}^{\vec{M}_{\text{in}}^{\text{ES}}\vec{M}_{\text{wall}}^{\text{ES}}} = -(1/2)\left\langle \vec{b}_{\xi}^{\vec{M}_{\text{in}}^{\text{ES}}}, -j\omega\varepsilon\mathcal{L}\left(\vec{b}_{\zeta}^{\vec{M}_{\text{wall}}^{\text{ES}}}\right)\right\rangle_{\mathbb{S}_{\text{in}}} \tag{3-110k}$$

$$p_{\xi\zeta}^{\vec{M}_{\text{in}}^{\text{ES}}\vec{M}_{\text{out}}^{\text{ES}}} = -(1/2)\left\langle \vec{b}_{\xi}^{\vec{M}_{\text{in}}^{\text{ES}}}, -j\omega\varepsilon\mathcal{L}\left(-\vec{b}_{\zeta}^{\vec{M}_{\text{out}}^{\text{ES}}}\right)\right\rangle_{\mathbb{S}_{\text{in}}} \tag{3-110l}$$

To obtain the IPO defined on modal space, we substitute Eq. (3-102) into Eq. (3-108), and then we have that

$$P^{\text{in}} = \overline{a}^{\dagger}\cdot\underbrace{\left(\overline{\overline{T}}^{\dagger}\cdot\overline{\overline{P}}_{\text{intAV}}^{\text{in}}\cdot\overline{\overline{T}}\right)}_{\overline{\overline{P}}_{\text{int}}^{\text{in}}}\cdot\overline{a} \tag{3-111}$$

and the subscript "int" is to emphasize that the corresponding matrix is derived from discretizing the interaction form.

For the convenience of the following discussions, the IPOs (3-106) and (3-111) are uniformly written as follows:

$$P^{\text{in}} \;=\; \bar{a}^{\dagger} \cdot \overline{\overline{P}}^{\text{in}} \cdot \bar{a} \tag{3-112}$$

where $\overline{\overline{P}}^{\text{in}} = \overline{\overline{P}}^{\text{in}}_{\text{cur}} / \overline{\overline{P}}^{\text{in}}_{\text{int}}$ .

## 3.3.4 Input-Power-Decoupled Modes

Below, we construct the *input-power-decoupled modes* (*IP-DMs*) of the material guide shown in Fig. 3-22, by using the results obtained above.

**Construction Method**

The IP-DMs in modal space can be derived from solving the modal decoupling equation $\overline{\overline{P}}^{\text{in}}_{-} \cdot \bar{\alpha}_{\xi} = \theta_{\xi}\, \overline{\overline{P}}^{\text{in}}_{+} \cdot \bar{\alpha}_{\xi}$ defined on modal space, where $\overline{\overline{P}}^{\text{in}}_{+}$ and $\overline{\overline{P}}^{\text{in}}_{-}$ are the positive and negative Hermitian parts of matrix $\overline{\overline{P}}^{\text{in}}$ .

If some derived modes $\{\bar{\alpha}_1, \bar{\alpha}_2, \cdots, \bar{\alpha}_d\}$ are *d*-order degenerate, then the Gram-Schmidt orthogonalization process given in the previous Sec. 3.2.4.1 is necessary, and it is not repeated here.

**Modal Decoupling Relation and Parseval's Identity**

The IP-DMs constructed above satisfy the following decoupling relation

$$(1/2)\iint_{\mathbb{S}_{\text{in}}} \left(\vec{E}_{\zeta} \times \vec{H}^{\dagger}_{\xi}\right) \cdot \hat{z} dS \;=\; \left(1 + j\,\theta_{\xi}\right) \delta_{\xi\zeta} \tag{3-113}$$

and the relation implies that the IP-DMs don't have net energy coupling in integral period.

By employing the decoupling relation, we have the following Parseval's identity

$$\sum\nolimits_{\xi} \left|c_{\xi}\right|^2 \;=\; \left(1/T\right) \int_{t_0}^{t_0+T} \left[ \iint_{\mathbb{S}_{\text{in}}} \left(\vec{\mathcal{E}} \times \vec{\mathcal{H}}\right) \cdot \hat{z} dS \right] dt \tag{3-114}$$

with modal expansion coefficient $c_{\xi}$ as follows:

$$c_{\xi} \;=\; \frac{-(1/2)\left\langle \vec{J}^{\text{ES}}_{\text{in};\xi}, \vec{E} \right\rangle_{\mathbb{S}_{\text{in}}}}{1 + j\,\theta_{\xi}} \;=\; \frac{-(1/2)\left\langle \vec{H}, \vec{M}^{\text{ES}}_{\text{in};\xi} \right\rangle_{\mathbb{S}_{\text{in}}}}{1 + j\,\theta_{\xi}} \tag{3-115}$$

where $\{\vec{E}, \vec{H}\}$ are previously known.

**Traveling-wave-type IP-DMs**

Similarly to the previous Sec. 3.2.4.3, the following modal input impedance and admittance are defined

$$Z_\xi^{\rm in} = \frac{(1/2)\iint_{\mathbb{S}_{\rm in}} \left(\vec{E}_\xi \times \vec{H}_\xi^\dagger\right)\cdot \hat{z} dS}{(1/2)\left\langle \vec{J}_{{\rm in};\xi}^{\rm ES}, \vec{J}_{{\rm in};\xi}^{\rm ES}\right\rangle_{\mathbb{S}_{\rm in}}} = \overbrace{{\rm Re}\left\{Z_\xi^{\rm in}\right\}}^{R_\xi^{\rm in}} + j\, \overbrace{{\rm Im}\left\{Z_\xi^{\rm in}\right\}}^{X_\xi^{\rm in}} \tag{3-116a}$$

$$Y_\xi^{\rm in} = \frac{(1/2)\iint_{\mathbb{S}_{\rm in}} \left(\vec{E}_\xi \times \vec{H}_\xi^\dagger\right)\cdot \hat{z} dS}{(1/2)\left\langle \vec{M}_{{\rm in};\xi}^{\rm ES}, \vec{M}_{{\rm in};\xi}^{\rm ES}\right\rangle_{\mathbb{S}_{\rm in}}} = \underbrace{{\rm Re}\left\{Y_\xi^{\rm in}\right\}}_{G_\xi^{\rm in}} + j\, \underbrace{{\rm Im}\left\{Y_\xi^{\rm in}\right\}}_{B_\xi^{\rm in}} \tag{3-116b}$$

Employing the above $Z_\xi^{\rm in}$ and $Y_\xi^{\rm in}$, the traveling-wave-type IP-DMs can be effectively recognized as doing in Secs. 3.2.4 and 3.2.5, and the corresponding cut-off frequencies can be easily calculated like Eq. (3-77).

### 3.3.5 Numerical Examples Corresponding to Typical Structures

In this subsection, we consider a typical elliptical material guide, and construct its travelling-wave-type IP-DMs by using the formulations given above. The cross section of the guide (with $\mu_{\rm r} = 1$ and $\varepsilon_{\rm r} = 20$) is a $4\,{\rm mm} \times 6\,{\rm mm}$ ellipse. The geometry of a section of the guide (whose longitudinal length is $L = 24\,{\rm mm}$) is shown in Fig. 3-23

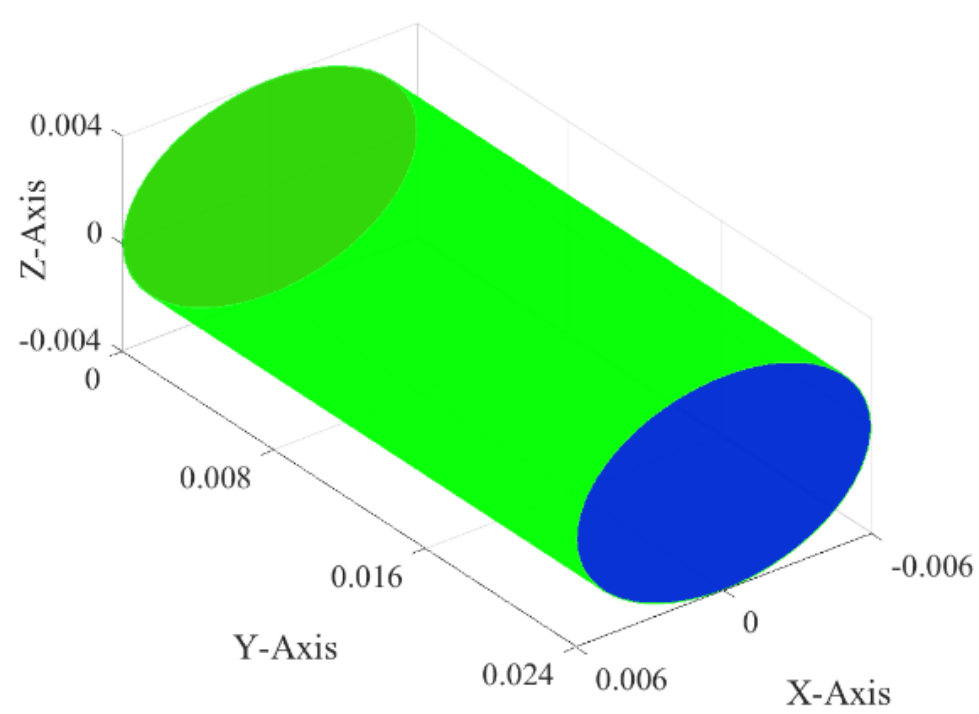

Figure 3-23 Geometry of a section of elliptical material guide

and the corresponding topological structures are shown in the following Fig. 3-24

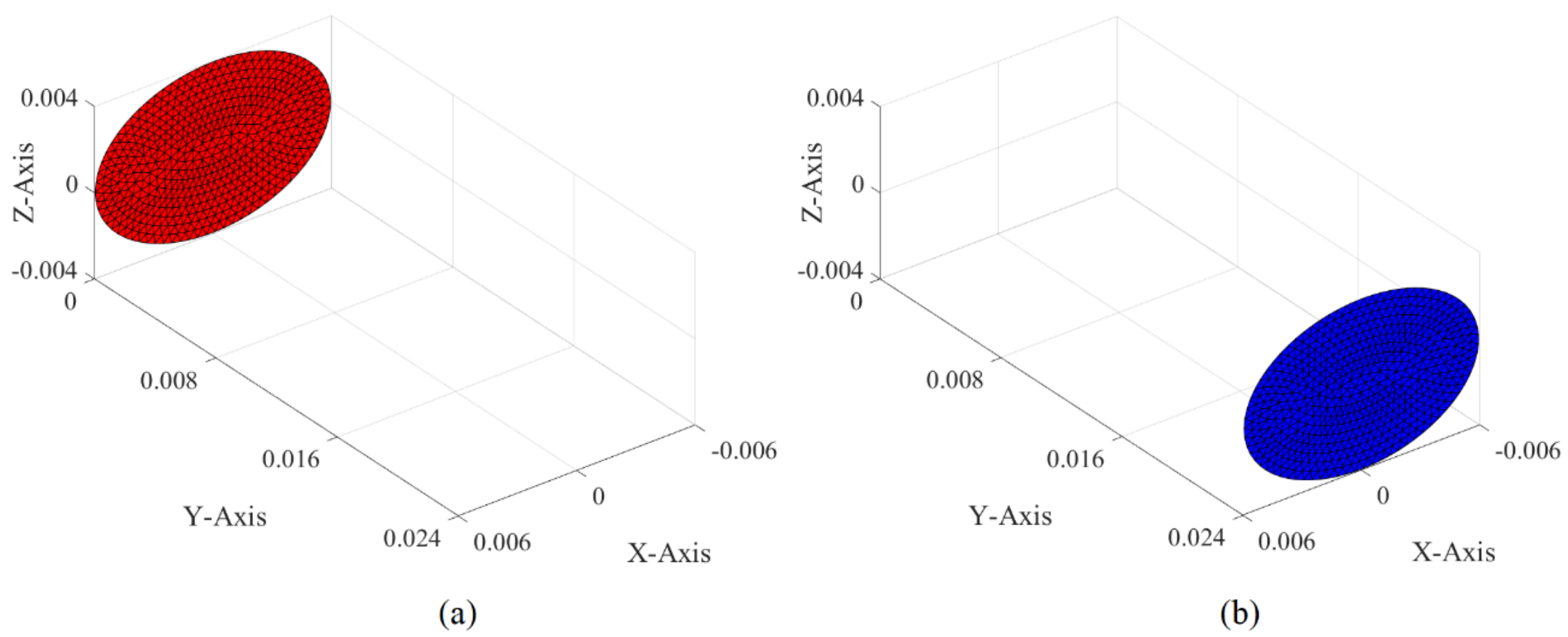

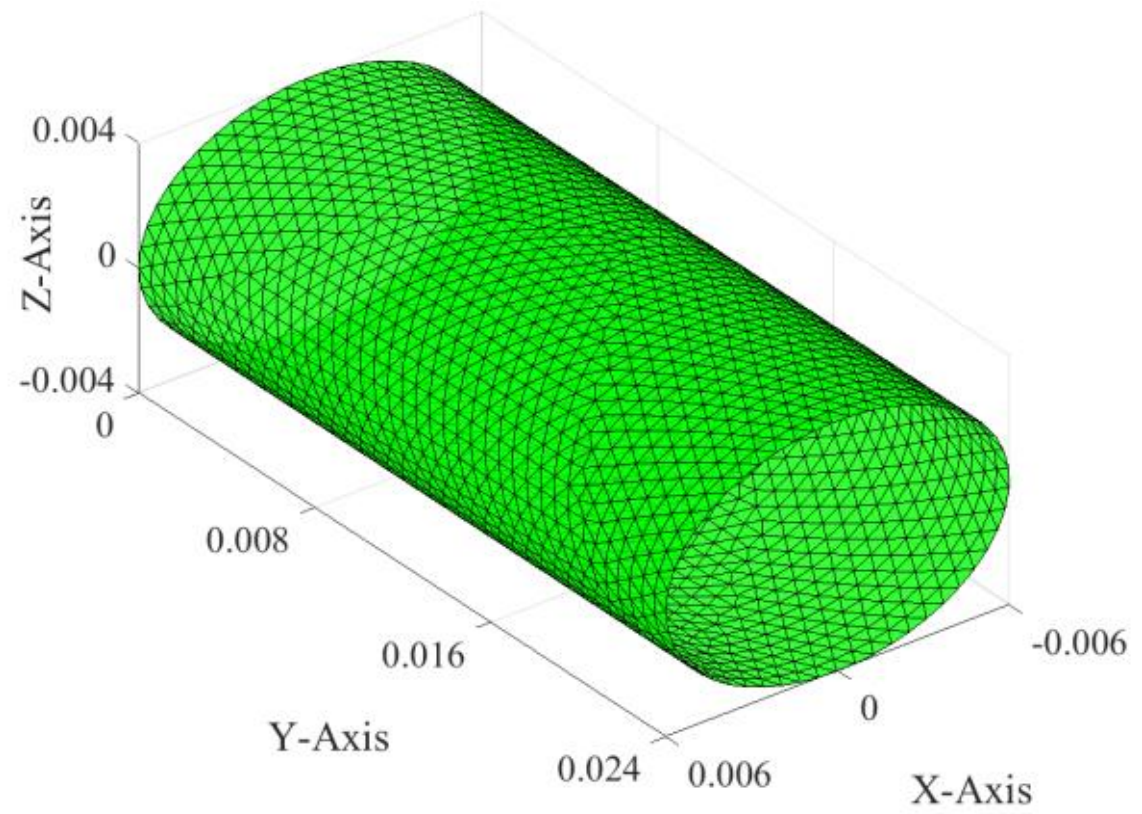

(c)

Figure 3-24 Topological structures and surface triangular meshes of the elliptical material guide shown in Fig. 3-23. (a) Mesh of input port $\mathbb{S}_{in}$ ; (b) mesh of output port $\mathbb{S}_{out}$ ; (c) mesh of guide wall $\mathbb{S}_{wall}$

By orthogonalizing the HM-DoM-based IPO (3-112), the IP-DMs of the material guide are constructed, and the modal input resistance curves of the first several typical IP-DMs are plotted in the following Fig. 3-25.

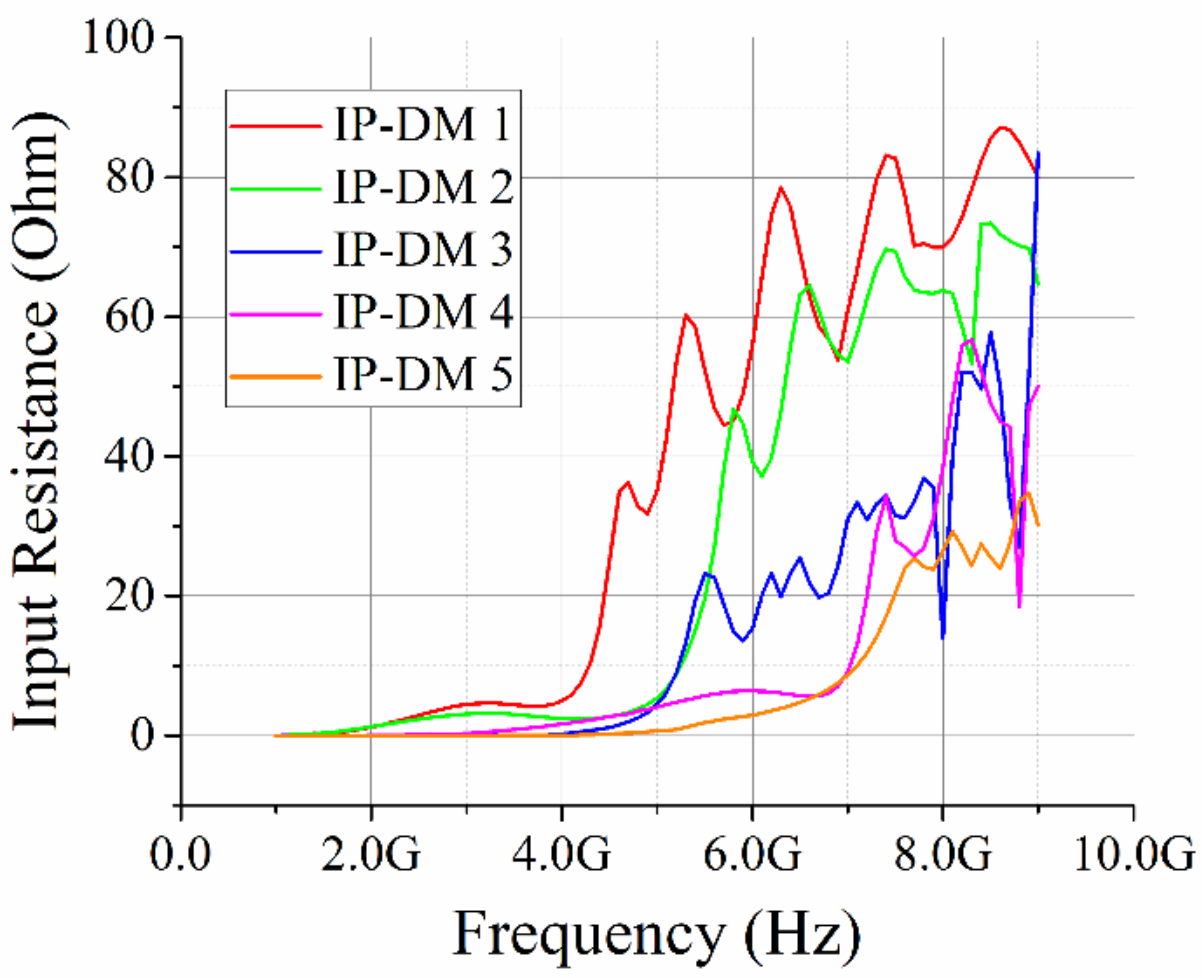

Figure 3-25 Modal input resistance curve of the first several typical IP-DMs

For visually exhibiting the IP-DM 1 (working at 4.7GHz), we plot its modal electric energy density distributing on the material guide tube in the following Fig. 3-26 (b); we plot its modal equivalent electric current distributing on the input port in the following Fig. 3-26 (c); we plot its modal equivalent magnetic current distributing on the input port in the following Fig. 3-26 (d); we plot its modal equivalent electric current distributing on the guide wall in the following Fig. 3-26 (e); we plot its modal equivalent magnetic current distributing on the guide wall in the following Fig. 3-26 (f).

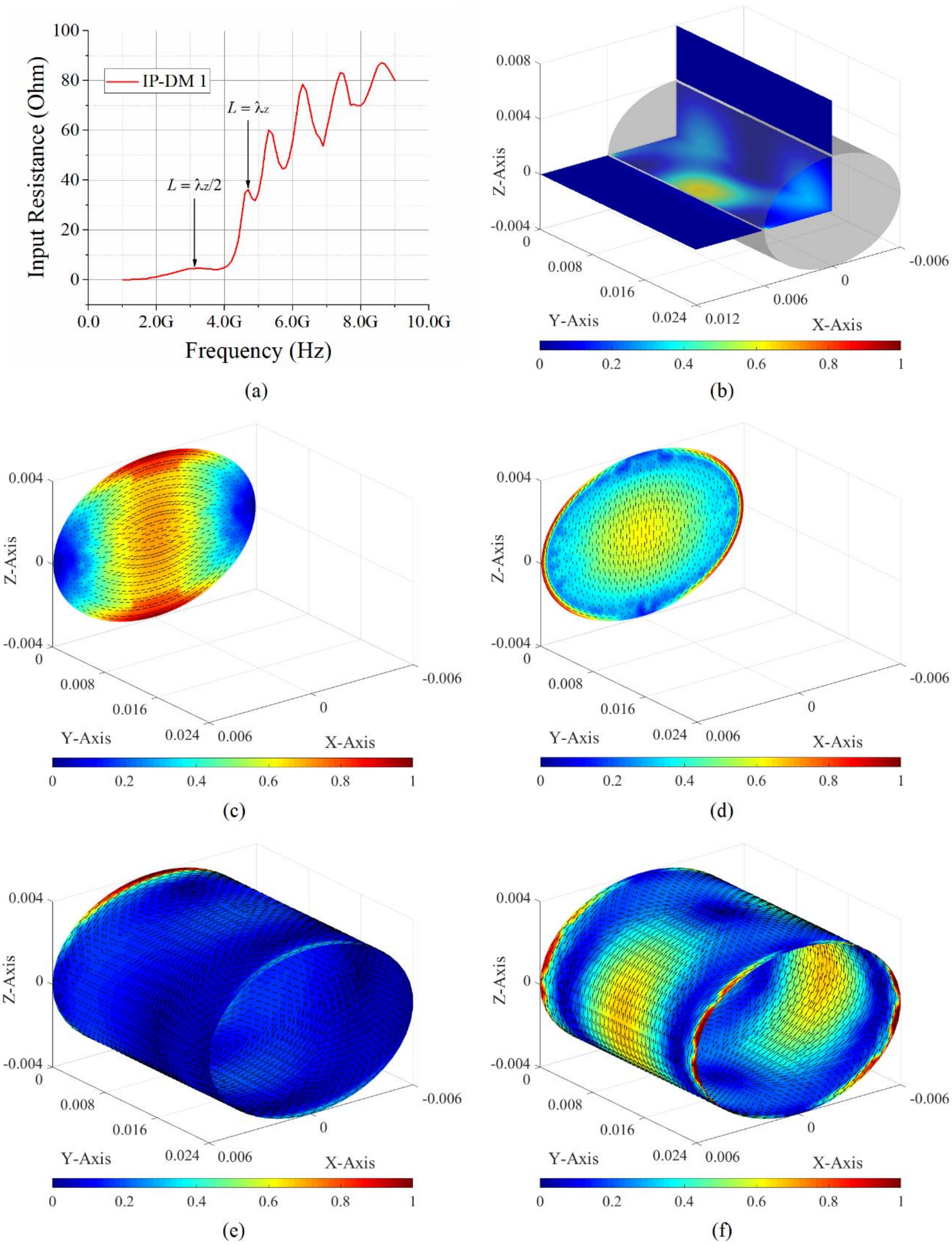

Figure 3-26 Modal quantities of the IP-DM 1 working at 4.7GHz. (a) Resistance curve; (b) electric energy density distribution; (c) equivalent electric current on input port; (d) equivalent magnetic current on input port; (e) equivalent electric current on guide wall; (f) equivalent magnetic current on guide wall

For the IP-DM 2 (working at 5.8GHz) shown in Fig. 3-25, its some typical modal quantities are shown in the following Fig. 3-27.

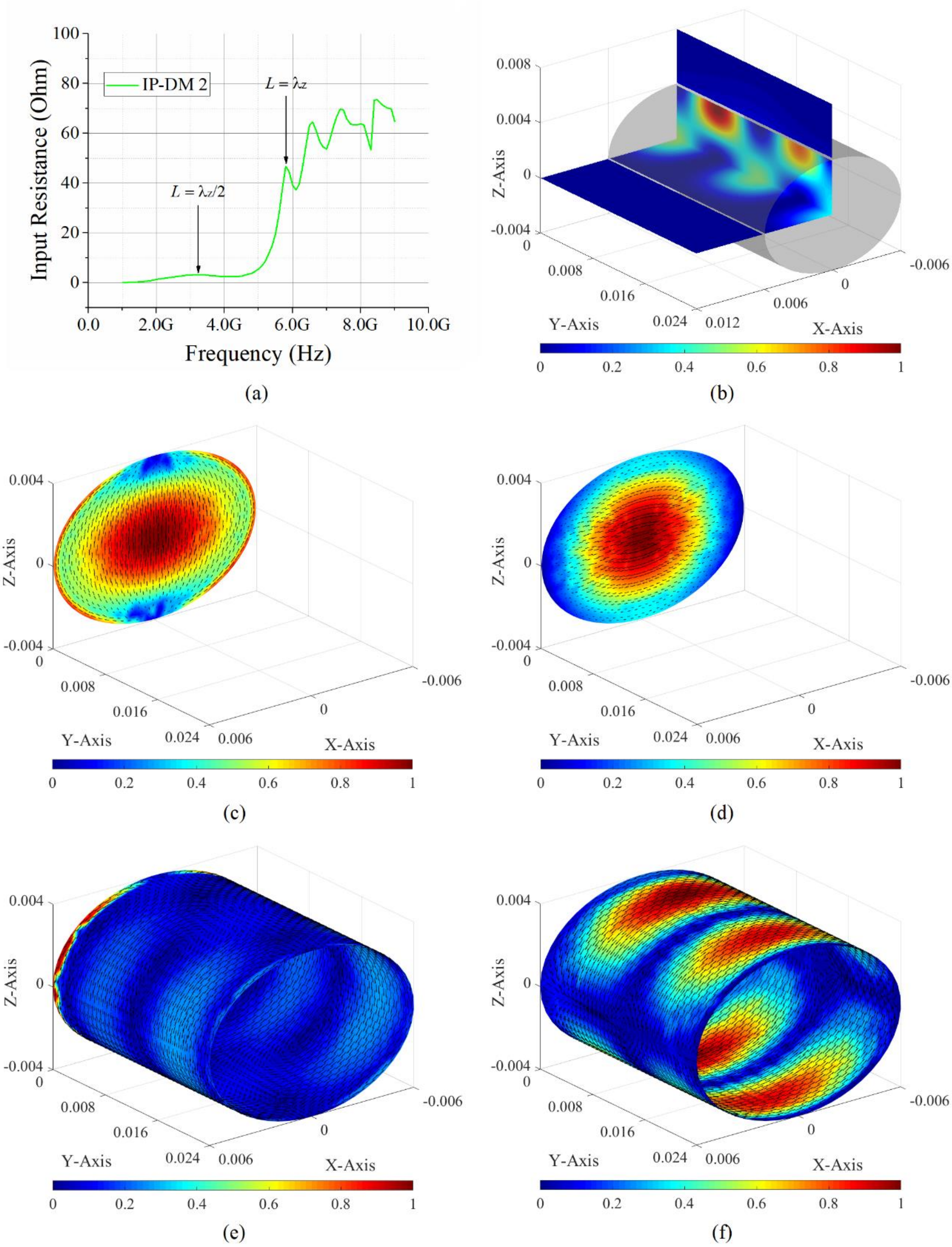

Figure 3-27 Modal quantities of IP-DM 2 (working at 5.8GHz). (a) Resistance curve; (b) electric energy density distribution; (c) equivalent electric current on input port; (d) equivalent magnetic current on input port; (e) equivalent electric current on guide wall; (f) equivalent magnetic current on guide wall

For the IP-DM 3 (working at 6.2GHz) shown in Fig. 3-25, its some typical modal quantities are shown in the following Fig. 3-28.

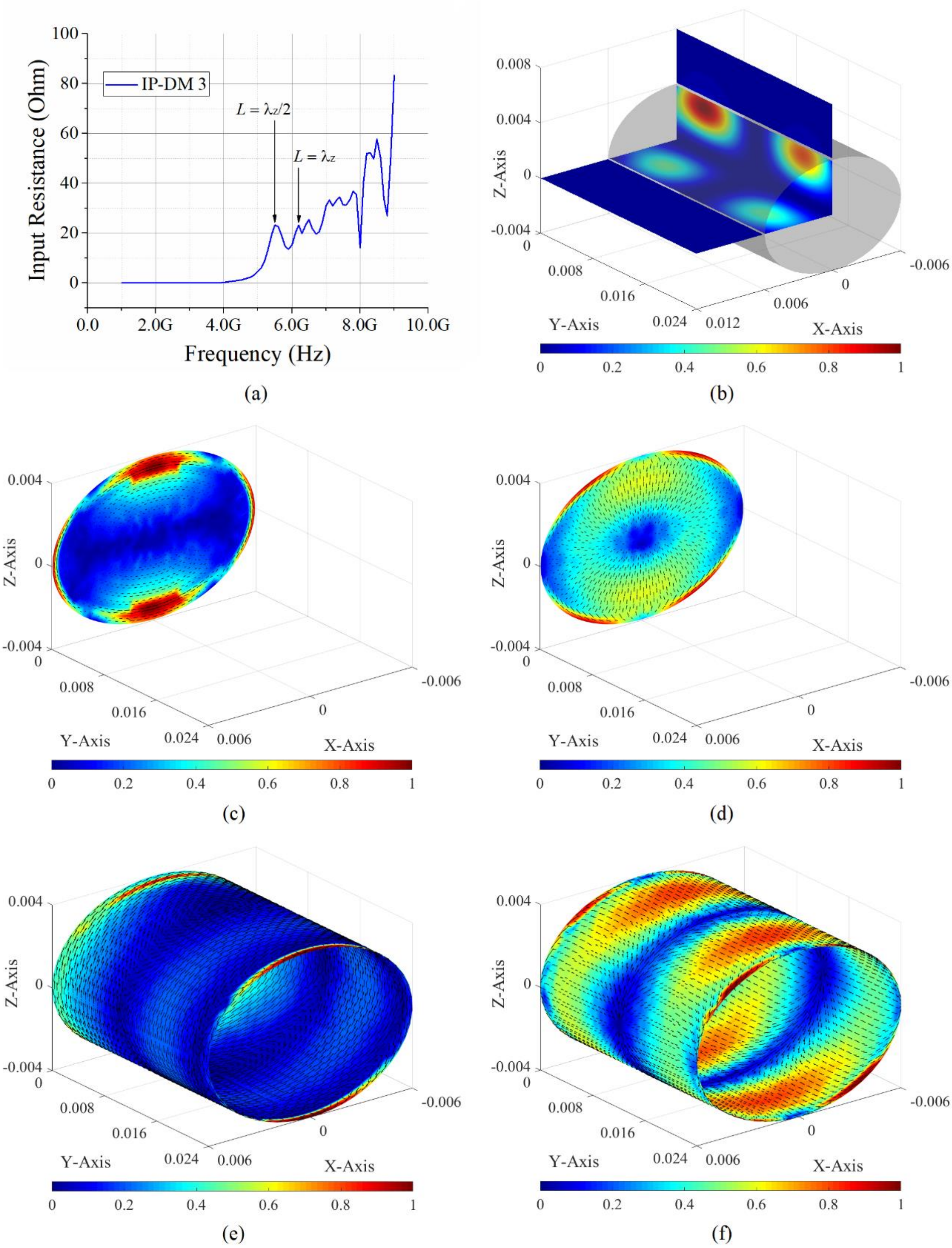

Figure 3-28 Modal quantities of IP-DM 3 (working at 6.2GHz). (a) Resistance curve; (b) electric energy density distribution; (c) equivalent electric current on input port; (d) equivalent magnetic current on input port; (e) equivalent electric current on guide wall; (f) equivalent magnetic current on guide wall

For the IP-DM 4 (working at 7.4GHz) shown in Fig. 3-25, its some typical modal quantities are shown in the following Fig. 3-29.

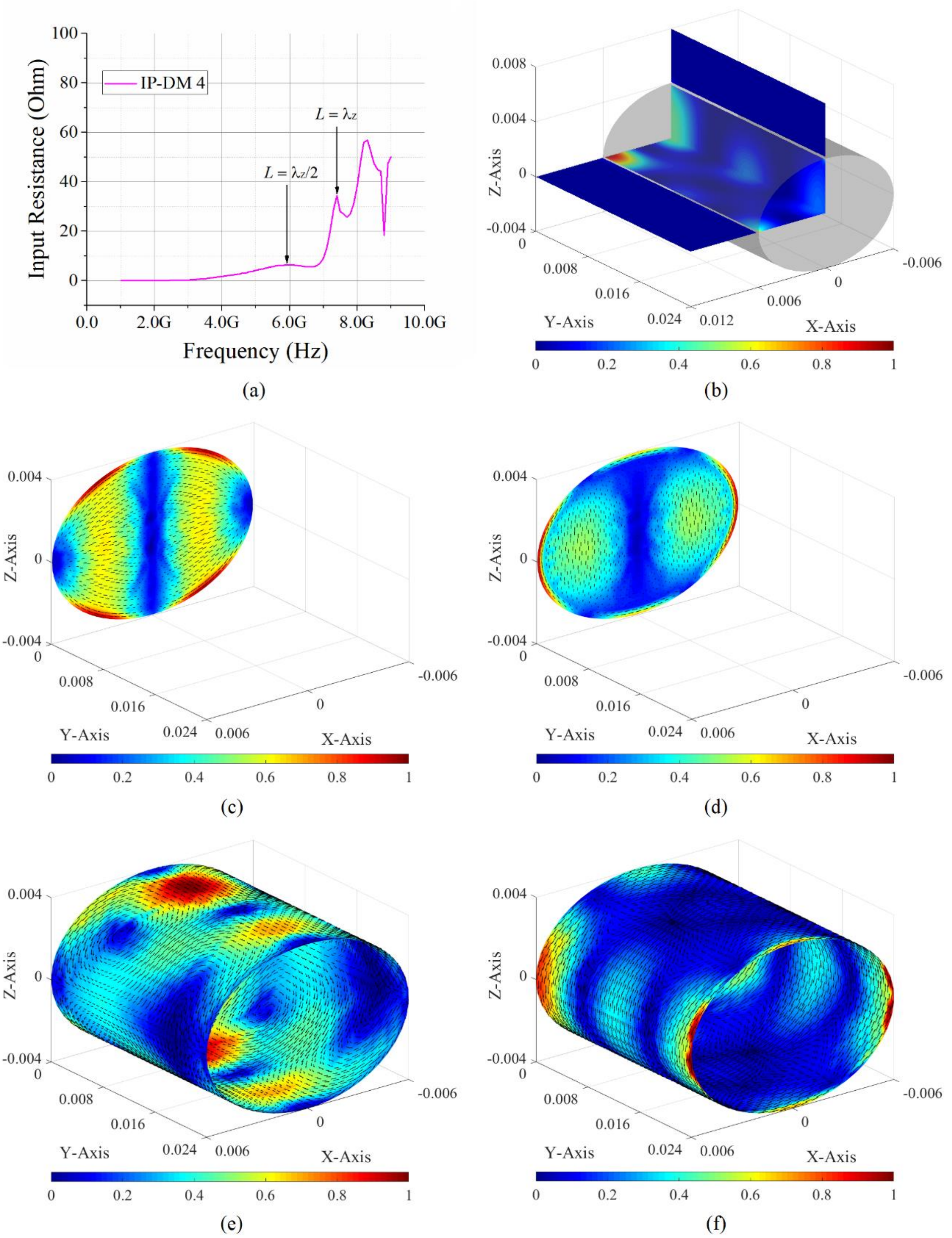

Figure 3-29 Modal quantities of IP-DM 4 (working at 7.4GHz). (a) Resistance curve; (b) electric energy density distribution; (c) equivalent electric current on input port; (d) equivalent magnetic current on input port; (e) equivalent electric current on guide wall; (f) equivalent magnetic current on guide wall

For the IP-DM 5 (working at 8.1GHz) shown in Fig. 3-25, its some typical modal quantities are shown in the following Fig. 3-30.

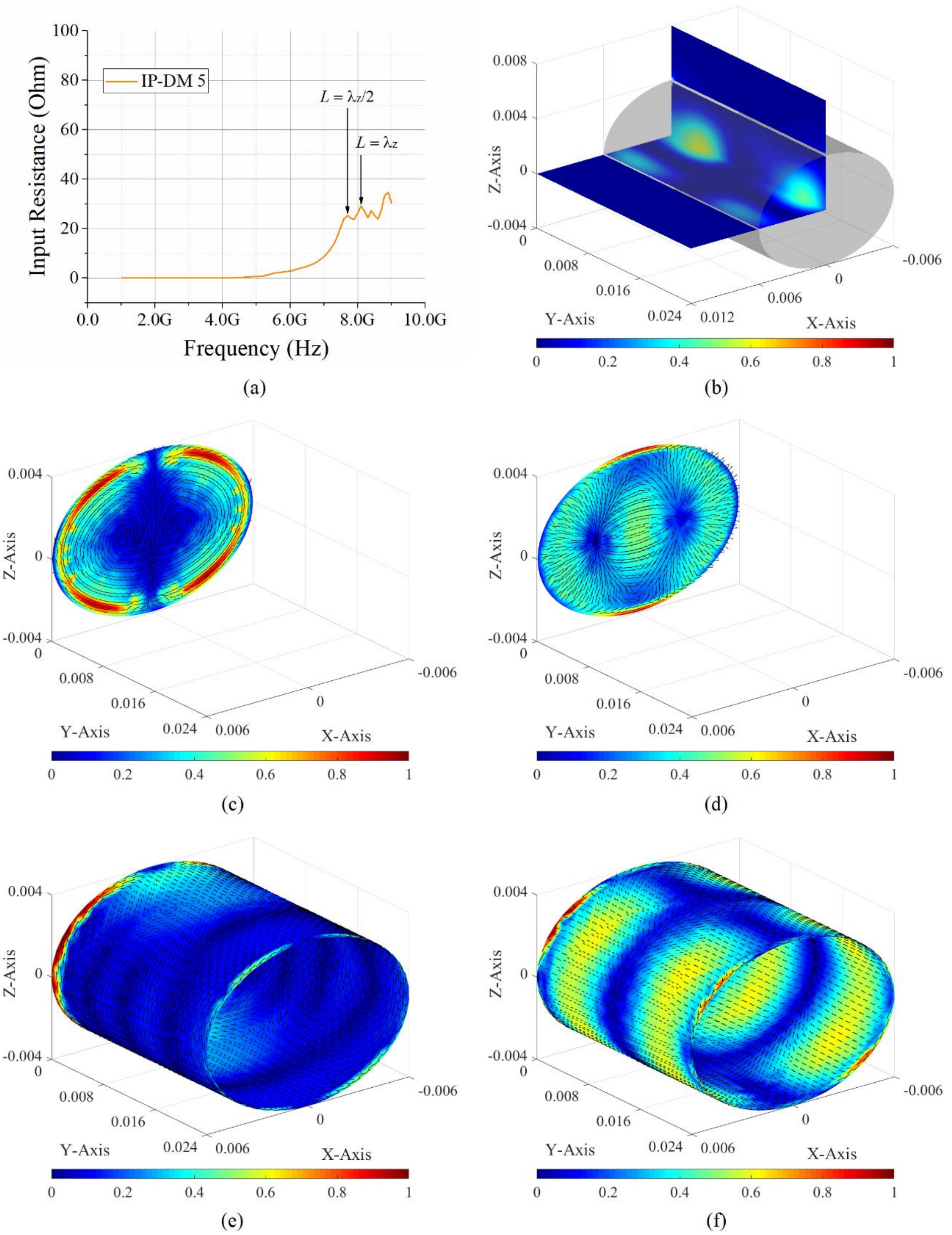

Figure 3-30 Modal quantities of IP-DM 5 (working at 8.1GHz). (a) Resistance curve; (b) electric energy density distribution; (c) equivalent electric current on input port; (d) equivalent magnetic current on input port; (e) equivalent electric current on guide wall; (f) equivalent magnetic current on guide wall

Thus, we conclude here that the PTT-MatGuid-DMT indeed has ability to construct the travelling-wave modes of the material guide.

## 3.4 IP-DMs of Metal-Material Composite Guiding Structure

Under PTT framework, we further generalize the above-obtained results for metallic and material guiding structures to metal-material composite guiding structures. In this section, we focus on constructing the IP-DMs of *microstrip transmission lines*. A typical microstrip line is shown in the following Fig. 3-31.

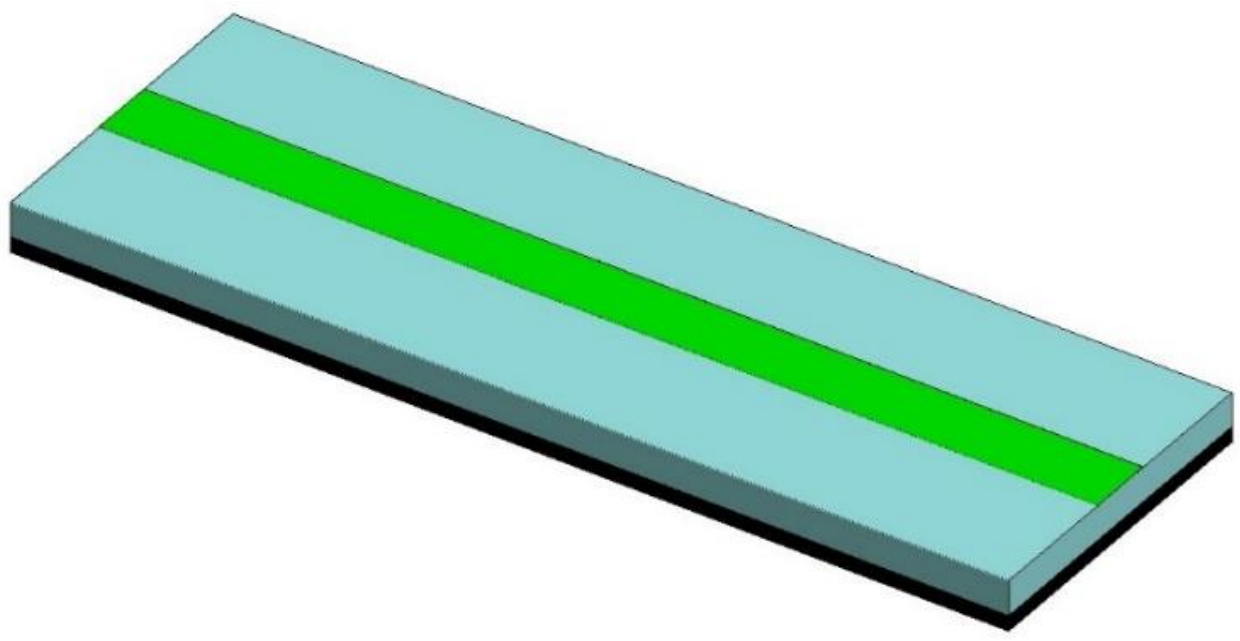

Figure 3-31 Geometry of a typical microstrip transmission line

The sub-sections contained in this Sec. 3.4 are organized similarly to the sub-sections of the previous Secs. 3.2 and 3.3.

### 3.4.1 Topological Structure and Source-Field Relationships

Now we consider a section of the microstrip line shown in Fig. 3-31, and its longitudinal length is $L$ [①], and its *longitudinal section view* and *transverse section view* are shown in the following Fig. 3-32.

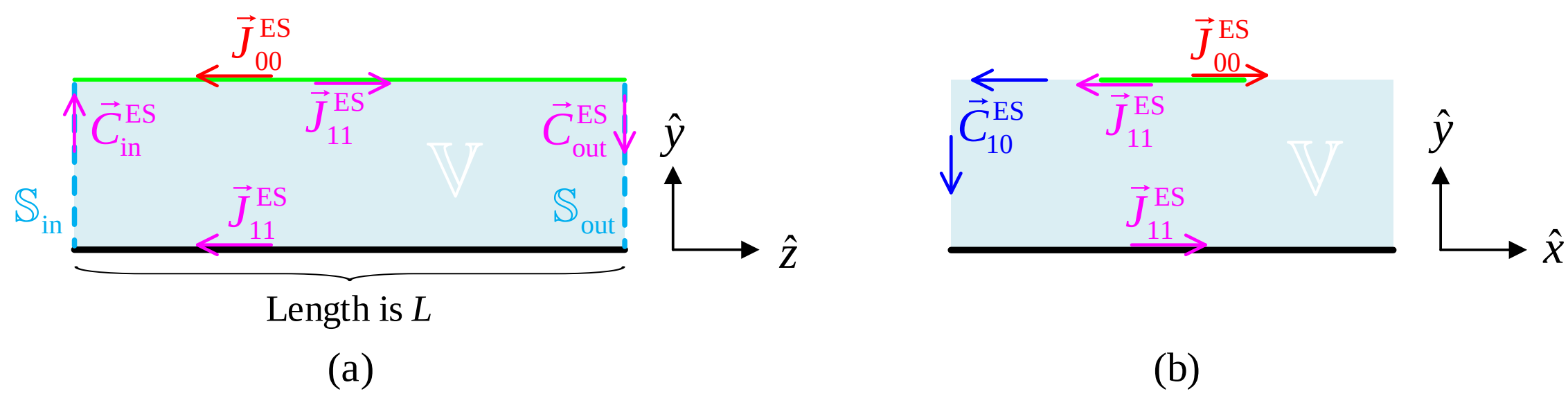

Figure 3-32 Section views of the microstrip line shown in Fig. 3-31. (a) Longitudinal section view and (b) transverse section view

In the Fig. 3-32, the region occupied by the *material substrate* is denoted as $\mathbb{V}$; the input and output ports of $\mathbb{V}$ are denoted as $\mathbb{S}_{\rm in}$ and $\mathbb{S}_{\rm out}$ respectively; the interface between $\mathbb{V}$ and *metallic ground plane* and the interface between $\mathbb{V}$ and *metallic strip* are

① For a traveling-wave mode, the longitudinal length of the section satisfies that $L = n\lambda_z$, where $\lambda_z$ is the waveguide wavelength of the traveling-wave mode, and $n$ is a positive integer.

collectively denoted as $\mathbb{S}_{11}$; the interface between $\mathbb{V}$ and environment is denoted as $\mathbb{S}_{10}$; the interface between metallic strip and environment is denoted as $\mathbb{S}_{00}$.

The equivalent surface currents distributing on $\mathbb{S}_{\text{in}}$ are denoted as $\{\vec{J}_{\text{in}}^{\text{ES}}, \vec{M}_{\text{in}}^{\text{ES}}\}$, or simply denoted as $\vec{C}_{\text{in}}^{\text{ES}}$ where $\vec{C} = \vec{J} / \vec{M}$; the equivalent surface currents distributing on $\mathbb{S}_{\text{out}}$ are denoted as $\{\vec{J}_{\text{out}}^{\text{ES}}, \vec{M}_{\text{out}}^{\text{ES}}\}$; the equivalent surface magnetic currents distributing on $\mathbb{S}_{11}$ and $\mathbb{S}_{00}$ are zero due to the homogeneous tangential electric field boundary condition, and the equivalent surface electric currents distributing on $\mathbb{S}_{11}$ and $\mathbb{S}_{00}$ are denoted as $\vec{J}_{11}^{\text{ES}}$ and $\vec{J}_{00}^{\text{ES}}$ respectively; the equivalent surface currents distributing on $\mathbb{S}_{10}$ are denoted as $\{\vec{J}_{10}^{\text{ES}}, \vec{M}_{10}^{\text{ES}}\}$. The above-mentined $\{\vec{J}_{\text{in}}^{\text{ES}}, \vec{M}_{\text{in}}^{\text{ES}}\}$ are defined in terms of the total modal fields $\{\vec{E}, \vec{H}\}$ as follows:

$$\vec{J}_{\text{in}}^{\text{ES}}(\vec{r}) = \hat{z} \times \vec{H}(\vec{r}) \quad , \quad \vec{r} \in \mathbb{S}_{\text{in}} \tag{3-117a}$$

$$\vec{M}_{\text{in}}^{\text{ES}}(\vec{r}) = \vec{E}(\vec{r}) \times \hat{z} \quad , \quad \vec{r} \in \mathbb{S}_{\text{in}} \tag{3-117b}$$

where $\hat{z}$ is the Z-directional vector, and the other currents are defined similarly.

Using the above currents, the field $\vec{F}$ in region $\mathbb{V}$ can be expressed as the following operator form

$$\vec{F}(\vec{r}) = \mathcal{F}\left(\vec{J}_{\text{in}}^{\text{ES}} + \vec{J}_{11}^{\text{ES}} + \vec{J}_{10}^{\text{ES}} - \vec{J}_{\text{out}}^{\text{ES}}, \vec{M}_{\text{in}}^{\text{ES}} + \vec{M}_{10}^{\text{ES}} - \vec{M}_{\text{out}}^{\text{ES}}\right) \quad , \quad \vec{r} \in \mathbb{V} \tag{3-118}$$

in which $\vec{F} = \vec{E} / \vec{H}$ and correspondingly $\mathcal{F} = \mathcal{E} / \mathcal{H}$, and operators $\mathcal{E}$ and $\mathcal{H}$ have the same forms as the ones given in Eqs. (3-26a) and (3-26b). Based on the conclusions given in Sec. 3.2, the field $\vec{F}$ on $\mathbb{S}_{\text{out}}^{+}$ can be expressed as follows:

$$\vec{F}(\vec{r}) = \mathcal{F}\left(\vec{J}_{\text{out}}^{\text{ES}}, \vec{M}_{\text{out}}^{\text{ES}}\right) \quad , \quad \vec{r} \in \mathbb{S}_{\text{out}}^{+} \tag{3-119}$$

where $\mathbb{S}_{\text{out}}^{+}$ is the right-side surface of $\mathbb{S}_{\text{out}}$. In addition, for the traveling-wave modes, the field $\vec{F}$ on $\mathbb{S}_{10}^{+} \bigcup \mathbb{S}_{00}^{+}$ can be approximately expressed as follows:

$$\vec{F}(\vec{r}) = \mathcal{F}_0\left(-\vec{J}_{10}^{\text{ES}} + \vec{J}_{00}^{\text{ES}}, -\vec{M}_{10}^{\text{ES}}\right) \quad , \quad \vec{r} \in \mathbb{S}_{10}^{+} \bigcup \mathbb{S}_{00}^{+} \tag{3-120}$$

where $\mathbb{S}_{10}^{+} \bigcup \mathbb{S}_{00}^{+}$ is the outer boundary surface of $\mathbb{S}_{10} \bigcup \mathbb{S}_{00}$ locating in environment, and the operator $\mathcal{F}_0$ is the same as the one used in the previous chapters.

### 3.4.2 Mathematical Description for Modal Space

Substituting Eq. (3-118) into Eqs. (3-117a) and (3-117b), we obtain the following integral equations

$$\left[\mathcal{H}\left(\vec{J}_{\text{in}}^{\text{ES}} + \vec{J}_{11}^{\text{ES}} + \vec{J}_{10}^{\text{ES}} - \vec{J}_{\text{out}}^{\text{ES}}, \vec{M}_{\text{in}}^{\text{ES}} + \vec{M}_{10}^{\text{ES}} - \vec{M}_{\text{out}}^{\text{ES}}\right)\right]_{\vec{r}_{-} \to \vec{r}}^{\tan} = \vec{J}_{\text{in}}^{\text{ES}}(\vec{r}) \times \hat{z} \quad , \quad \vec{r} \in \mathbb{S}_{\text{in}} \tag{3-121a}$$

$$\Big[\mathcal{E}\Big(\vec{J}_{\text{in}}^{\text{ES}}+\vec{J}_{11}^{\text{ES}}+\vec{J}_{10}^{\text{ES}}-\vec{J}_{\text{out}}^{\text{ES}},\vec{M}_{\text{in}}^{\text{ES}}+\vec{M}_{10}^{\text{ES}}-\vec{M}_{\text{out}}^{\text{ES}}\Big)\Big]_{\vec{r}_-\to\vec{r}}^{\tan} = \hat{z}\times\vec{M}_{\text{in}}^{\text{ES}}(\vec{r}) \ , \quad \vec{r}\in\mathbb{S}_{\text{in}} \quad (3\text{-}121\text{b})$$

satisfied by the currents, where $\vec{r}_-\in\text{int}\,\mathbb{V}$ and $\vec{r}_-$ approaches the point $\vec{r}$ on $\mathbb{S}_{\text{in}}$.

Utilizing Eqs. (3-118) and (3-120) and the homogeneous tangential electric field boundary conditions on $\mathbb{S}_{11}$ and $\mathbb{S}_{00}$, we obtain the following integral equations

$$\Big[\mathcal{E}\Big(\vec{J}_{\text{in}}^{\text{ES}}+\vec{J}_{11}^{\text{ES}}+\vec{J}_{10}^{\text{ES}}-\vec{J}_{\text{out}}^{\text{ES}},\vec{M}_{\text{in}}^{\text{ES}}+\vec{M}_{10}^{\text{ES}}-\vec{M}_{\text{out}}^{\text{ES}}\Big)\Big]_{\vec{r}_-\to\vec{r}}^{\tan} = 0 \ , \quad \vec{r}\in\mathbb{S}_{11} \quad (3\text{-}122)$$

$$\Big[\mathcal{E}_0\Big(-\vec{J}_{10}^{\text{ES}}+\vec{J}_{00}^{\text{ES}},-\vec{M}_{10}^{\text{ES}}\Big)\Big]_{\vec{r}_+\to\vec{r}}^{\tan} = 0 \ , \quad \vec{r}\in\mathbb{S}_{00} \quad (3\text{-}123)$$

satisfied by the currents, where $\vec{r}_-$ locates in $\mathbb{V}$ and approaches the point $\vec{r}$ on surface $\mathbb{S}_{11}$, and $\vec{r}_+$ locates in environment and approaches the point $\vec{r}$ on $\mathbb{S}_{00}$.

Utilizing Eqs. (3-118)&(3-120) and the tangential continuation conditions satisfied by the fields on $\mathbb{S}_{10}$, we obtain the following integral equations

$$\begin{aligned}&\Big[\mathcal{E}\Big(\vec{J}_{\text{in}}^{\text{ES}}+\vec{J}_{11}^{\text{ES}}+\vec{J}_{10}^{\text{ES}}-\vec{J}_{\text{out}}^{\text{ES}},\vec{M}_{\text{in}}^{\text{ES}}+\vec{M}_{10}^{\text{ES}}-\vec{M}_{\text{out}}^{\text{ES}}\Big)\Big]_{\vec{r}_-\to\vec{r}}^{\tan}\\ =&\Big[\mathcal{E}_0\Big(-\vec{J}_{10}^{\text{ES}}+\vec{J}_{00}^{\text{ES}},-\vec{M}_{10}^{\text{ES}}\Big)\Big]_{\vec{r}_+\to\vec{r}}^{\tan} \ , \quad \vec{r}\in\mathbb{S}_{10} \quad (3\text{-}124\text{a})\end{aligned}$$

$$\begin{aligned}&\Big[\mathcal{H}\Big(\vec{J}_{\text{in}}^{\text{ES}}+\vec{J}_{11}^{\text{ES}}+\vec{J}_{10}^{\text{ES}}-\vec{J}_{\text{out}}^{\text{ES}},\vec{M}_{\text{in}}^{\text{ES}}+\vec{M}_{10}^{\text{ES}}-\vec{M}_{\text{out}}^{\text{ES}}\Big)\Big]_{\vec{r}_-\to\vec{r}}^{\tan}\\ =&\Big[\mathcal{H}_0\Big(-\vec{J}_{10}^{\text{ES}}+\vec{J}_{00}^{\text{ES}},-\vec{M}_{10}^{\text{ES}}\Big)\Big]_{\vec{r}_+\to\vec{r}}^{\tan} \ , \quad \vec{r}\in\mathbb{S}_{10} \quad (3\text{-}124\text{b})\end{aligned}$$

satisfied by the currents, where the points $\vec{r}_-$ and $\vec{r}_+$ can be explained similarly to the ones appearing in Eqs. (3-121)~(3-123).

Similar to obtaining Eqs. (3-28a) and (3-28b), we can obtain the following integral equations

$$\begin{aligned}&\Big[\mathcal{E}\Big(\vec{J}_{\text{in}}^{\text{ES}}+\vec{J}_{11}^{\text{ES}}+\vec{J}_{10}^{\text{ES}}-\vec{J}_{\text{out}}^{\text{ES}},\vec{M}_{\text{in}}^{\text{ES}}+\vec{M}_{10}^{\text{ES}}-\vec{M}_{\text{out}}^{\text{ES}}\Big)\Big]_{\vec{r}_-\to\vec{r}}^{\tan}\\ =&\Big[\mathcal{E}\Big(\vec{J}_{\text{out}}^{\text{ES}},\vec{M}_{\text{out}}^{\text{ES}}\Big)\Big]_{\vec{r}_+\to\vec{r}}^{\tan} \ , \quad \vec{r}\in\mathbb{S}_{\text{out}} \quad (3\text{-}125\text{a})\end{aligned}$$

$$\begin{aligned}&\Big[\mathcal{H}\Big(\vec{J}_{\text{in}}^{\text{ES}}+\vec{J}_{11}^{\text{ES}}+\vec{J}_{10}^{\text{ES}}-\vec{J}_{\text{out}}^{\text{ES}},\vec{M}_{\text{in}}^{\text{ES}}+\vec{M}_{10}^{\text{ES}}-\vec{M}_{\text{out}}^{\text{ES}}\Big)\Big]_{\vec{r}_-\to\vec{r}}^{\tan}\\ =&\Big[\mathcal{H}\Big(\vec{J}_{\text{out}}^{\text{ES}},\vec{M}_{\text{out}}^{\text{ES}}\Big)\Big]_{\vec{r}_+\to\vec{r}}^{\tan} \ , \quad \vec{r}\in\mathbb{S}_{\text{out}} \quad (3\text{-}125\text{b})\end{aligned}$$

satisfied by the currents, where points $\vec{r}_-$ and $\vec{r}_+$ approach $\mathbb{S}_{\text{out}}$ from the left and right sides of $\mathbb{S}_{\text{out}}$ respectively.

If the currents involved in Eqs. (3-121a)~(3-125b) are expanded in terms of some proper basis functions, and the equations are tested with $\{\vec{b}_\xi^{\vec{M}_{\text{in}}^{\text{ES}}}\}$, $\{\vec{b}_\xi^{\vec{J}_{\text{in}}^{\text{ES}}}\}$, $\{\vec{b}_\xi^{\vec{J}_{11}^{\text{ES}}}\}$,

$\{\vec{b}_\xi^{\vec{J}_{00}^{\rm ES}}\}$, $\{\vec{b}_\xi^{\vec{J}_{10}^{\rm ES}}\}$, $\{\vec{b}_\xi^{\vec{M}_{10}^{\rm ES}}\}$, $\{\vec{b}_\xi^{\vec{J}_{\rm out}^{\rm ES}}\}$, and $\{\vec{b}_\xi^{\vec{M}_{\rm out}^{\rm ES}}\}$ respectively, then the integral equations are immediately discretized into the following matrix equations

$$\begin{aligned} 0 \;=\;& \overline{\overline{Z}}^{\vec{M}_{\rm in}^{\rm ES}\vec{J}_{\rm in}^{\rm ES}}\cdot\overline{a}^{\vec{J}_{\rm in}^{\rm ES}} + \overline{\overline{Z}}^{\vec{M}_{\rm in}^{\rm ES}\vec{J}_{11}^{\rm ES}}\cdot\overline{a}^{\vec{J}_{11}^{\rm ES}} + \overline{\overline{Z}}^{\vec{M}_{\rm in}^{\rm ES}\vec{J}_{10}^{\rm ES}}\cdot\overline{a}^{\vec{J}_{10}^{\rm ES}} + \overline{\overline{Z}}^{\vec{M}_{\rm in}^{\rm ES}\vec{J}_{\rm out}^{\rm ES}}\cdot\overline{a}^{\vec{J}_{\rm out}^{\rm ES}} + \overline{\overline{Z}}^{\vec{M}_{\rm in}^{\rm ES}\vec{M}_{\rm in}^{\rm ES}}\cdot\overline{a}^{\vec{M}_{\rm in}^{\rm ES}} \\ &+ \overline{\overline{Z}}^{\vec{M}_{\rm in}^{\rm ES}\vec{M}_{10}^{\rm ES}}\cdot\overline{a}^{\vec{M}_{10}^{\rm ES}} + \overline{\overline{Z}}^{\vec{M}_{\rm in}^{\rm ES}\vec{M}_{\rm out}^{\rm ES}}\cdot\overline{a}^{\vec{M}_{\rm out}^{\rm ES}} \end{aligned} \tag{3-126a}$$

$$\begin{aligned} 0 \;=\;& \overline{\overline{Z}}^{\vec{J}_{\rm in}^{\rm ES}\vec{J}_{\rm in}^{\rm ES}}\cdot\overline{a}^{\vec{J}_{\rm in}^{\rm ES}} + \overline{\overline{Z}}^{\vec{J}_{\rm in}^{\rm ES}\vec{J}_{11}^{\rm ES}}\cdot\overline{a}^{\vec{J}_{11}^{\rm ES}} + \overline{\overline{Z}}^{\vec{J}_{\rm in}^{\rm ES}\vec{J}_{10}^{\rm ES}}\cdot\overline{a}^{\vec{J}_{10}^{\rm ES}} + \overline{\overline{Z}}^{\vec{J}_{\rm in}^{\rm ES}\vec{J}_{\rm out}^{\rm ES}}\cdot\overline{a}^{\vec{J}_{\rm out}^{\rm ES}} + \overline{\overline{Z}}^{\vec{J}_{\rm in}^{\rm ES}\vec{M}_{\rm in}^{\rm ES}}\cdot\overline{a}^{\vec{M}_{\rm in}^{\rm ES}} \\ &+ \overline{\overline{Z}}^{\vec{J}_{\rm in}^{\rm ES}\vec{M}_{10}^{\rm ES}}\cdot\overline{a}^{\vec{M}_{10}^{\rm ES}} + \overline{\overline{Z}}^{\vec{J}_{\rm in}^{\rm ES}\vec{M}_{\rm out}^{\rm ES}}\cdot\overline{a}^{\vec{M}_{\rm out}^{\rm ES}} \end{aligned} \tag{3-126b}$$

and

$$\begin{aligned} 0 \;=\;& \overline{\overline{Z}}^{\vec{J}_{11}^{\rm ES}\vec{J}_{\rm in}^{\rm ES}}\cdot\overline{a}^{\vec{J}_{\rm in}^{\rm ES}} + \overline{\overline{Z}}^{\vec{J}_{11}^{\rm ES}\vec{J}_{11}^{\rm ES}}\cdot\overline{a}^{\vec{J}_{11}^{\rm ES}} + \overline{\overline{Z}}^{\vec{J}_{11}^{\rm ES}\vec{J}_{10}^{\rm ES}}\cdot\overline{a}^{\vec{J}_{10}^{\rm ES}} + \overline{\overline{Z}}^{\vec{J}_{11}^{\rm ES}\vec{J}_{\rm out}^{\rm ES}}\cdot\overline{a}^{\vec{J}_{\rm out}^{\rm ES}} + \overline{\overline{Z}}^{\vec{J}_{11}^{\rm ES}\vec{M}_{\rm in}^{\rm ES}}\cdot\overline{a}^{\vec{M}_{\rm in}^{\rm ES}} \\ &+ \overline{\overline{Z}}^{\vec{J}_{11}^{\rm ES}\vec{M}_{10}^{\rm ES}}\cdot\overline{a}^{\vec{M}_{10}^{\rm ES}} + \overline{\overline{Z}}^{\vec{J}_{11}^{\rm ES}\vec{M}_{\rm out}^{\rm ES}}\cdot\overline{a}^{\vec{M}_{\rm out}^{\rm ES}} \end{aligned} \tag{3-127}$$

$$0 \;=\; \overline{\overline{Z}}^{\vec{J}_{00}^{\rm ES}\vec{J}_{10}^{\rm ES}}\cdot\overline{a}^{\vec{J}_{10}^{\rm ES}} + \overline{\overline{Z}}^{\vec{J}_{00}^{\rm ES}\vec{J}_{00}^{\rm ES}}\cdot\overline{a}^{\vec{J}_{00}^{\rm ES}} + \overline{\overline{Z}}^{\vec{J}_{00}^{\rm ES}\vec{M}_{10}^{\rm ES}}\cdot\overline{a}^{\vec{M}_{10}^{\rm ES}} \tag{3-128}$$

and

$$\begin{aligned} 0 \;=\;& \overline{\overline{Z}}^{\vec{J}_{10}^{\rm ES}\vec{J}_{\rm in}^{\rm ES}}\cdot\overline{a}^{\vec{J}_{\rm in}^{\rm ES}} + \overline{\overline{Z}}^{\vec{J}_{10}^{\rm ES}\vec{J}_{11}^{\rm ES}}\cdot\overline{a}^{\vec{J}_{11}^{\rm ES}} + \overline{\overline{Z}}^{\vec{J}_{10}^{\rm ES}\vec{J}_{10}^{\rm ES}}\cdot\overline{a}^{\vec{J}_{10}^{\rm ES}} + \overline{\overline{Z}}^{\vec{J}_{10}^{\rm ES}\vec{J}_{00}^{\rm ES}}\cdot\overline{a}^{\vec{J}_{00}^{\rm ES}} + \overline{\overline{Z}}^{\vec{J}_{10}^{\rm ES}\vec{J}_{\rm out}^{\rm ES}}\cdot\overline{a}^{\vec{J}_{\rm out}^{\rm ES}} \\ &+ \overline{\overline{Z}}^{\vec{J}_{10}^{\rm ES}\vec{M}_{\rm in}^{\rm ES}}\cdot\overline{a}^{\vec{M}_{\rm in}^{\rm ES}} + \overline{\overline{Z}}^{\vec{J}_{10}^{\rm ES}\vec{M}_{10}^{\rm ES}}\cdot\overline{a}^{\vec{M}_{10}^{\rm ES}} + \overline{\overline{Z}}^{\vec{J}_{10}^{\rm ES}\vec{M}_{\rm out}^{\rm ES}}\cdot\overline{a}^{\vec{M}_{\rm out}^{\rm ES}} \end{aligned} \tag{3-129a}$$

$$\begin{aligned} 0 \;=\;& \overline{\overline{Z}}^{\vec{M}_{10}^{\rm ES}\vec{J}_{\rm in}^{\rm ES}}\cdot\overline{a}^{\vec{J}_{\rm in}^{\rm ES}} + \overline{\overline{Z}}^{\vec{M}_{10}^{\rm ES}\vec{J}_{11}^{\rm ES}}\cdot\overline{a}^{\vec{J}_{11}^{\rm ES}} + \overline{\overline{Z}}^{\vec{M}_{10}^{\rm ES}\vec{J}_{10}^{\rm ES}}\cdot\overline{a}^{\vec{J}_{10}^{\rm ES}} + \overline{\overline{Z}}^{\vec{M}_{10}^{\rm ES}\vec{J}_{00}^{\rm ES}}\cdot\overline{a}^{\vec{J}_{00}^{\rm ES}} + \overline{\overline{Z}}^{\vec{M}_{10}^{\rm ES}\vec{J}_{\rm out}^{\rm ES}}\cdot\overline{a}^{\vec{J}_{\rm out}^{\rm ES}} \\ &+ \overline{\overline{Z}}^{\vec{M}_{10}^{\rm ES}\vec{M}_{\rm in}^{\rm ES}}\cdot\overline{a}^{\vec{M}_{\rm in}^{\rm ES}} + \overline{\overline{Z}}^{\vec{M}_{10}^{\rm ES}\vec{M}_{10}^{\rm ES}}\cdot\overline{a}^{\vec{M}_{10}^{\rm ES}} + \overline{\overline{Z}}^{\vec{M}_{10}^{\rm ES}\vec{M}_{\rm out}^{\rm ES}}\cdot\overline{a}^{\vec{M}_{\rm out}^{\rm ES}} \end{aligned} \tag{3-129b}$$

and

$$\begin{aligned} 0 \;=\;& \overline{\overline{Z}}^{\vec{J}_{\rm out}^{\rm ES}\vec{J}_{\rm in}^{\rm ES}}\cdot\overline{a}^{\vec{J}_{\rm in}^{\rm ES}} + \overline{\overline{Z}}^{\vec{J}_{\rm out}^{\rm ES}\vec{J}_{11}^{\rm ES}}\cdot\overline{a}^{\vec{J}_{11}^{\rm ES}} + \overline{\overline{Z}}^{\vec{J}_{\rm out}^{\rm ES}\vec{J}_{10}^{\rm ES}}\cdot\overline{a}^{\vec{J}_{10}^{\rm ES}} + \overline{\overline{Z}}^{\vec{J}_{\rm out}^{\rm ES}\vec{J}_{\rm out}^{\rm ES}}\cdot\overline{a}^{\vec{J}_{\rm out}^{\rm ES}} + \overline{\overline{Z}}^{\vec{J}_{\rm out}^{\rm ES}\vec{M}_{\rm in}^{\rm ES}}\cdot\overline{a}^{\vec{M}_{\rm in}^{\rm ES}} \\ &+ \overline{\overline{Z}}^{\vec{J}_{\rm out}^{\rm ES}\vec{M}_{10}^{\rm ES}}\cdot\overline{a}^{\vec{M}_{10}^{\rm ES}} + \overline{\overline{Z}}^{\vec{J}_{\rm out}^{\rm ES}\vec{M}_{\rm out}^{\rm ES}}\cdot\overline{a}^{\vec{M}_{\rm out}^{\rm ES}} \end{aligned} \tag{3-130a}$$

$$\begin{aligned} 0 \;=\;& \overline{\overline{Z}}^{\vec{M}_{\rm out}^{\rm ES}\vec{J}_{\rm in}^{\rm ES}}\cdot\overline{a}^{\vec{J}_{\rm in}^{\rm ES}} + \overline{\overline{Z}}^{\vec{M}_{\rm out}^{\rm ES}\vec{J}_{11}^{\rm ES}}\cdot\overline{a}^{\vec{J}_{11}^{\rm ES}} + \overline{\overline{Z}}^{\vec{M}_{\rm out}^{\rm ES}\vec{J}_{10}^{\rm ES}}\cdot\overline{a}^{\vec{J}_{10}^{\rm ES}} + \overline{\overline{Z}}^{\vec{M}_{\rm out}^{\rm ES}\vec{J}_{\rm out}^{\rm ES}}\cdot\overline{a}^{\vec{J}_{\rm out}^{\rm ES}} + \overline{\overline{Z}}^{\vec{M}_{\rm out}^{\rm ES}\vec{M}_{\rm in}^{\rm ES}}\cdot\overline{a}^{\vec{M}_{\rm in}^{\rm ES}} \\ &+ \overline{\overline{Z}}^{\vec{M}_{\rm out}^{\rm ES}\vec{M}_{10}^{\rm ES}}\cdot\overline{a}^{\vec{M}_{10}^{\rm ES}} + \overline{\overline{Z}}^{\vec{M}_{\rm out}^{\rm ES}\vec{M}_{\rm out}^{\rm ES}}\cdot\overline{a}^{\vec{M}_{\rm out}^{\rm ES}} \end{aligned} \tag{3-130b}$$

The formulations used to calculate the elements of the matrices in the above matrix equations are similar to the formulations given in Eqs. (3-91a)~(3-96f), and they are explicitly given in the App. D1 of this report.

Employing the above Eqs. (3-126a)~(3-130b), we can obtain the transformation

$$\begin{bmatrix} \overline{a}^{\vec{J}_{\rm in}^{\rm ES}} \\ \overline{a}^{\vec{J}_{11}^{\rm ES}} \\ \overline{a}^{\vec{J}_{10}^{\rm ES}} \\ \overline{a}^{\vec{J}_{00}^{\rm ES}} \\ \overline{a}^{\vec{J}_{\rm out}^{\rm ES}} \\ \overline{a}^{\vec{M}_{\rm in}^{\rm ES}} \\ \overline{a}^{\vec{M}_{10}^{\rm ES}} \\ \overline{a}^{\vec{M}_{\rm out}^{\rm ES}} \end{bmatrix} = \overline{a}^{\rm AV} = \overline{\overline{T}}\cdot\overline{a} \tag{3-131}$$

where the calculation formulation for transformation matrix $\overline{\overline{T}}$ is given in the App. D1 of this report.

### 3.4.3 Input Power Operator

The IPO corresponding to the input port $\mathbb{S}_{\text{in}}$ shown in Fig. 3-32 is as follows:

$$\begin{aligned} P^{\text{in}} &= (1/2)\iint_{\mathbb{S}_{\text{in}}} \left(\vec{E}\times\vec{H}^{\dagger}\right)\cdot\hat{z}dS \\ &= (1/2)\left\langle \hat{z}\times\vec{J}_{\text{in}}^{\text{ES}},\vec{M}_{\text{in}}^{\text{ES}}\right\rangle_{\mathbb{S}_{\text{in}}} \\ &= -(1/2)\left\langle \vec{J}_{\text{in}}^{\text{ES}},\mathcal{E}\left(\vec{J}_{\text{in}}^{\text{ES}}+\vec{J}_{11}^{\text{ES}}+\vec{J}_{10}^{\text{ES}}-\vec{J}_{\text{out}}^{\text{ES}},\vec{M}_{\text{in}}^{\text{ES}}+\vec{M}_{10}^{\text{ES}}-\vec{M}_{\text{out}}^{\text{ES}}\right)\right\rangle_{\mathbb{S}_{\text{in}}^{-}} \\ &= -(1/2)\left\langle \vec{M}_{\text{in}}^{\text{ES}},\mathcal{H}\left(\vec{J}_{\text{in}}^{\text{ES}}+\vec{J}_{11}^{\text{ES}}+\vec{J}_{10}^{\text{ES}}-\vec{J}_{\text{out}}^{\text{ES}},\vec{M}_{\text{in}}^{\text{ES}}+\vec{M}_{10}^{\text{ES}}-\vec{M}_{\text{out}}^{\text{ES}}\right)\right\rangle_{\mathbb{S}_{\text{in}}^{-}}^{\dagger} \end{aligned} \tag{3-132}$$

Here, the right-hand side of the first equality is the field form of IPO, and the right-hand side of the second equality is the current form of IPO, and the right-hand sides of the third and fourth equalities are the current-field interaction forms of IPO.

By discretizing IPO (3-132) and utilizing transformation (3-131), we derive the following matrix form of the IPO

$$P^{\text{in}} = \bar{a}^{\dagger}\cdot\overline{\overline{P}}^{\text{in}}\cdot\bar{a} \tag{3-133}$$

and the formulation for calculating the quadratic matrix $\overline{\overline{P}}^{\text{in}}$ is given in the App. D1 of this report.

### 3.4.4 Input-Power-Decoupled Modes

The IP-DMs in modal space can be derived from solving the modal decoupling equation $\overline{\overline{P}}_{-}^{\text{in}}\cdot\bar{\alpha}_{\xi} = \theta_{\xi}\,\overline{\overline{P}}_{+}^{\text{in}}\cdot\bar{\alpha}_{\xi}$ defined on modal space, where $\overline{\overline{P}}_{+}^{\text{in}}$ and $\overline{\overline{P}}_{-}^{\text{in}}$ are the positive and negative Hermitian parts of the matrix $\overline{\overline{P}}^{\text{in}}$ given in Eq. (3-133). If some derived modes $\{\bar{\alpha}_1,\bar{\alpha}_2,\cdots,\bar{\alpha}_d\}$ are $d$-order degenerate, then the Gram-Schmidt orthogonalization process given in the previous Sec. 3.2.4.1 is necessary, and it is not repeated here.

The IP-DMs constructed above satisfy the following decoupling relation

$$\left(1+j\,\theta_{\xi}\right)\delta_{\xi\zeta} = (1/2)\iint_{\mathbb{S}_{\text{in}}}\left(\vec{E}_{\zeta}\times\vec{H}_{\xi}^{\dagger}\right)\cdot\hat{z}dS = (1/2)\left\langle \hat{z}\times\vec{J}_{\text{in};\xi}^{\text{ES}},\vec{M}_{\text{in};\zeta}^{\text{ES}}\right\rangle_{\mathbb{S}_{\text{in}}} \tag{3-134}$$

and the relation implies that the IP-DMs don't have net energy coupling in integral period. By employing the decoupling relation, we have the following Parseval's identity

$$\sum_{\xi}\left|c_{\xi}\right|^2 = (1/T)\int_{t_0}^{t_0+T}\left[\iint_{\mathbb{S}_{\text{in}}}\left(\vec{\mathcal{E}}\times\vec{\mathcal{H}}\right)\cdot\hat{z}dS\right]dt \tag{3-135}$$

in which $\{\vec{\mathcal{E}},\vec{\mathcal{H}}\}$ are the time-domain fields, and the expansion coefficients $c_\xi$ have expression $c_\xi = -(1/2) < \vec{J}_{\text{in};\xi}^{\text{ES}}, \vec{E} >_{\mathbb{S}_{\text{in}}} \big/ (1+j\theta_\xi) = -(1/2) < \vec{H}, \vec{M}_{\text{in};\xi}^{\text{ES}} >_{\mathbb{S}_{\text{in}}} \big/ (1+j\theta_\xi)$, and $\{\vec{E},\vec{H}\}$ are some previously known fields distributing on input port $\mathbb{S}_{\text{in}}$.

Similarly to the previous sections, the modal input impedance and admittance can be defined. Employing the modal input impedance and admittance, the traveling-wave-type IP-DMs can be effectively recognized as doing in Secs. 3.2.4 and 3.2.5, and the corresponding cut-off frequencies can be easily calculated like Eq. (3-77).

### 3.4.5 Numerical Examples Corresponding to Typical Structures

In this subsection, we consider the microstrip line shown in Fig. 3-31, and construct its travelling-wave IP-DMs by using the formulations given above. The size of the microstrip line (with $\mu_r = 1$ and $\varepsilon_r = 20$) is shown in the following Fig. 3-33.

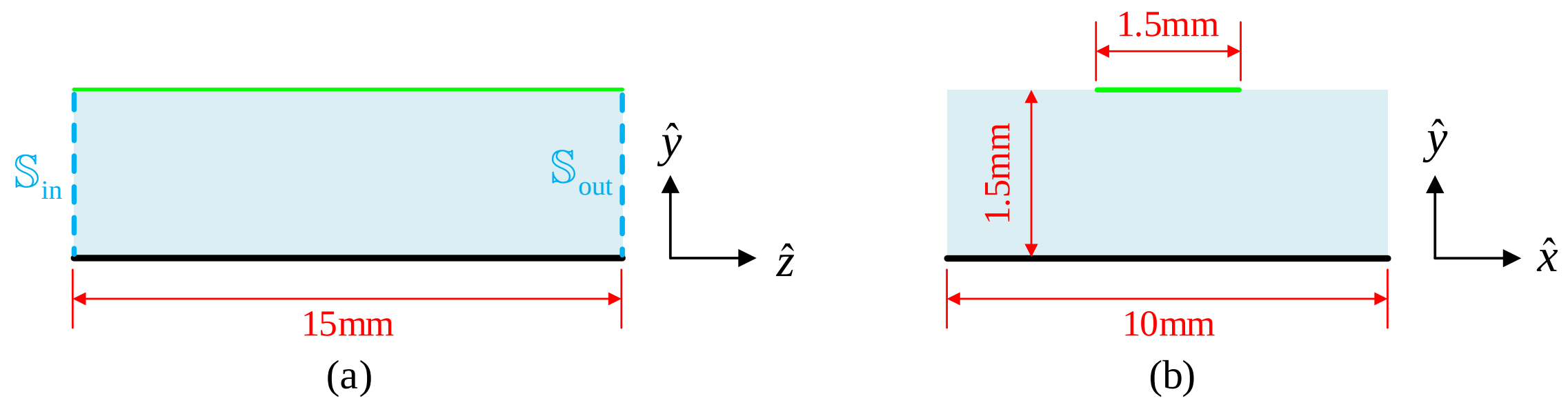

Figure 3-33 Size of the microstrip line considered in this subsection

By orthogonalizing the JE-DoJ-based IPO (3-133), the IP-DMs of the microstrip line are constructed, and the modal input resistance curve of the dominant IP-DM is plotted in the following Fig. 3-34.

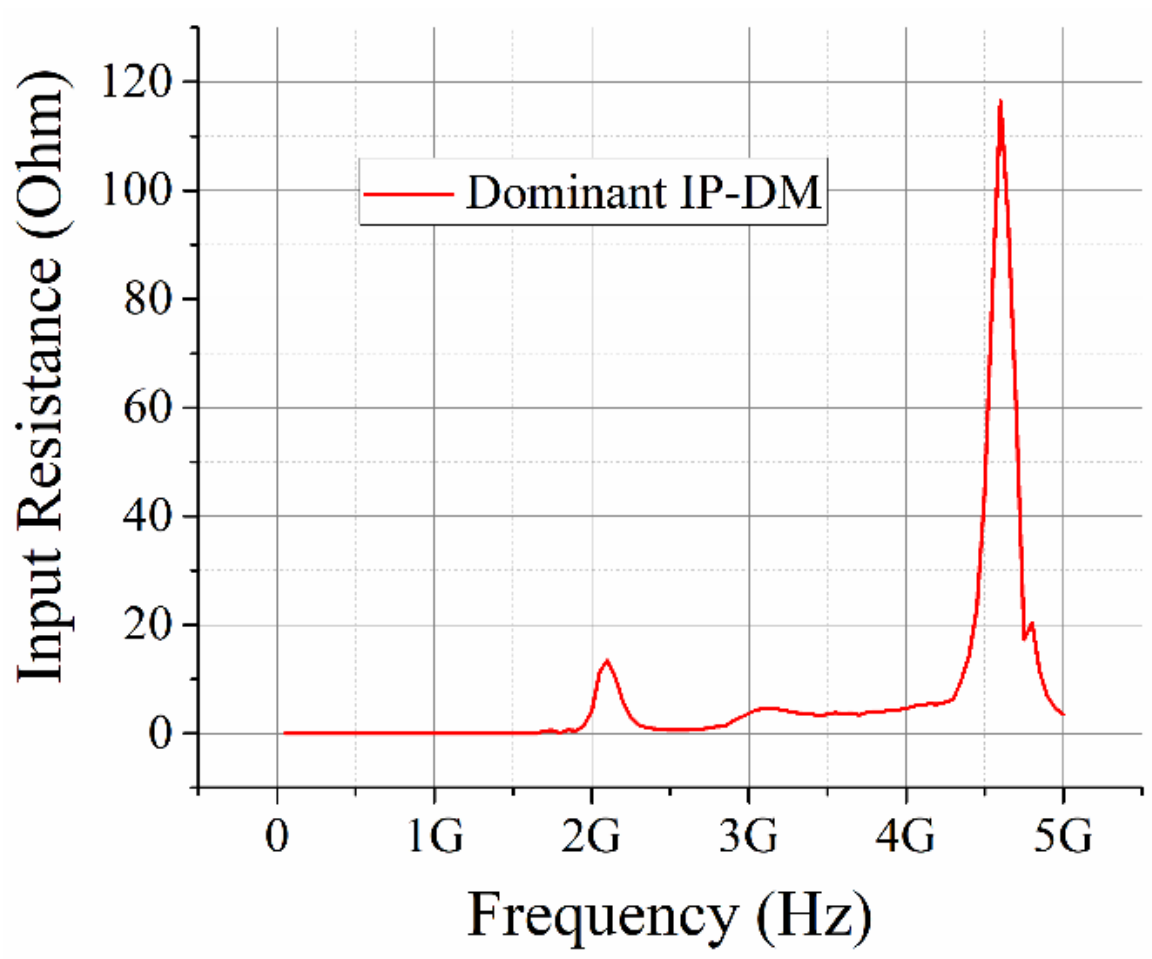

Figure 3-34 Modal input resistance curve of the dominant IP-DM

Taking the dominant IP-DM working at 4.6GHz as an example, we plot the field distributions and energy density distributions in the following Fig. 3-35.

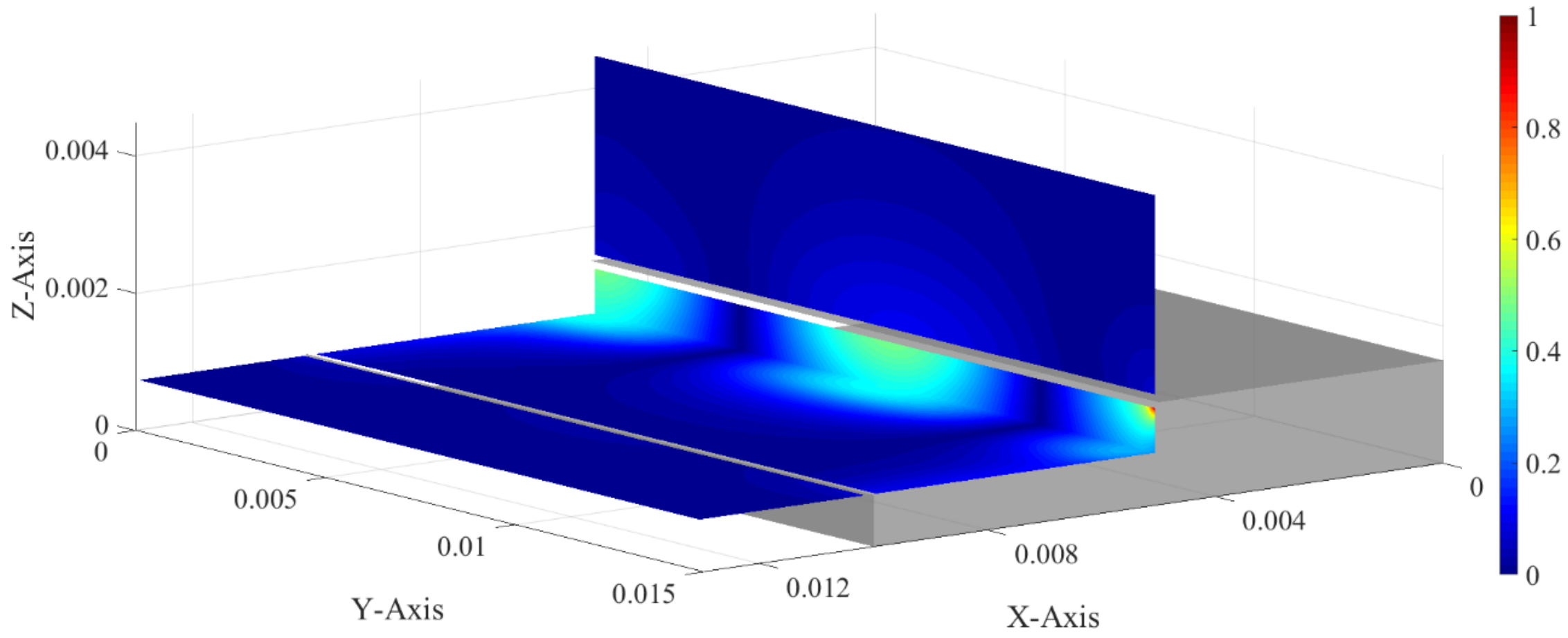

(a)

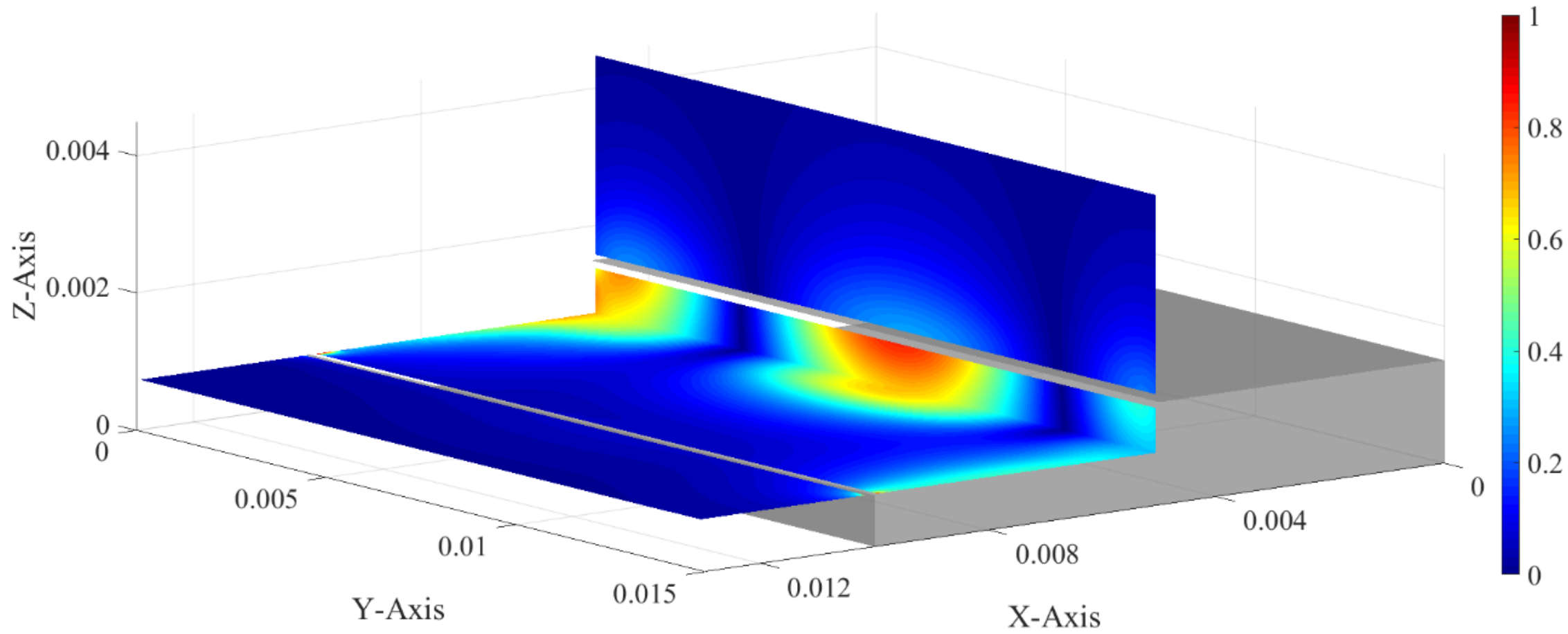

(b)

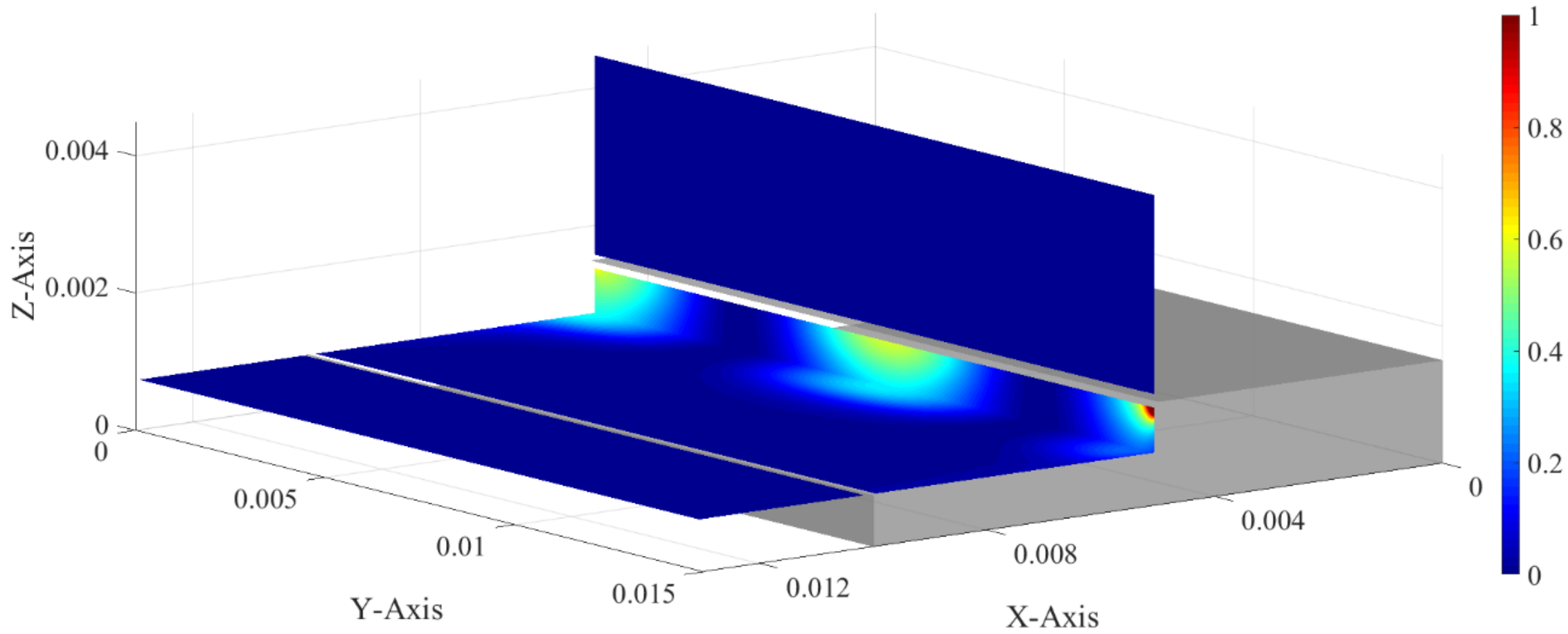

(c)

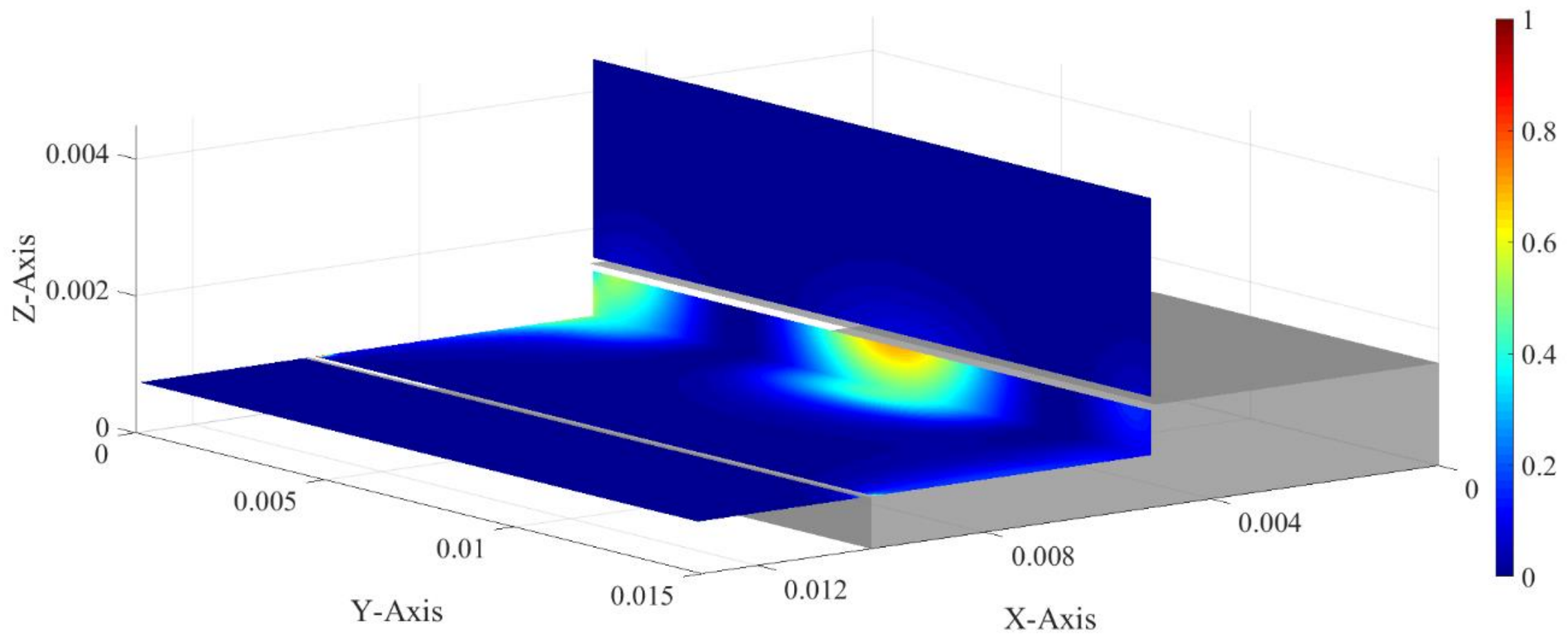

(d)

Figure 3-35 Modal field and energy density distributions of the IP-DM (working at 4.6GHz) shown in Fig. 3-34. (a) electric field distribution; (b) magnetic field distribution; (c) electric energy density distribution; (d) magnetic energy density distribution

Thus, we conclude here that the PTT-ComGuid-DMT indeed has ability to construct the travelling-wave modes of the microstrip line.

## 3.5 IP-DMs of the Guiding Structure With Finite Length and Arbitrary Cross Section

Under PTT framework, we further generalize the above-obtained results for the *cylindrical guiding structures with infinite lengthes and homogeneous cross sections* to the *guiding structures with finite lengthes and arbitrary cross sections*. The geometry of a typical guide with finite length and arbitrary cross section is shown in Fig. 3-36, and we focus on constructing the IP-DMs of the guide in this section.

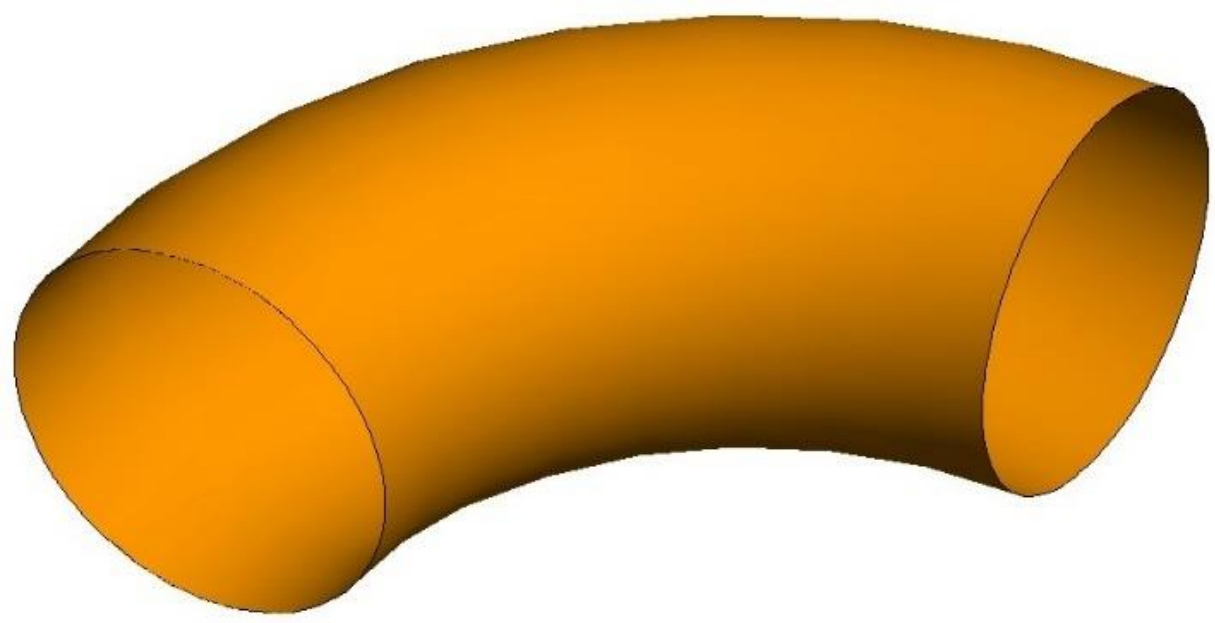

Figure 3-36 Geometry of a typical guide with finite length and arbitrary cross section

The sub-sections contained in this Sec. 3.5 are organized similarly to the sub-sections of the previous Secs. 3.2~3.4.

### 3.5.1 Topological Structure and Source-Field Relationships

For the guide shown in Fig. 3-36, the region occupied by the guide tube is denoted as $\mathbb{V}$; the input and output ports of $\mathbb{V}$ are denoted as $\mathbb{S}_{\text{in}}$ and $\mathbb{S}_{\text{out}}$ respectively; the guide electric wall of $\mathbb{V}$ is denoted as $\mathbb{S}_{\text{ele}}$. The material parameters of $\mathbb{V}$ are $\{\mu,\varepsilon\}$, where $\mu$ and $\varepsilon$ are scalar constants.

The equivalent surface currents distributing on $\mathbb{S}_{\text{in}}$ are denoted as $\{\vec{J}_{\text{in}}^{\text{ES}},\vec{M}_{\text{in}}^{\text{ES}}\}$; the equivalent surface currents distributing on $\mathbb{S}_{\text{out}}$ are denoted as $\{\vec{J}_{\text{out}}^{\text{ES}},\vec{M}_{\text{out}}^{\text{ES}}\}$; the equivalent surface magnetic current distributing on $\mathbb{S}_{\text{ele}}$ is zero due to the homogeneous tangential electric field boundary condition, and the equivalent surface electric current distributing on $\mathbb{S}_{\text{ele}}$ is denoted as $\vec{J}^{\text{IS}}$.

The above-mentined $\{\vec{J}_{\text{in}}^{\text{ES}},\vec{M}_{\text{in}}^{\text{ES}}\}$ are defined in terms of the total modal fields $\{\vec{E},\vec{H}\}$ as follows:

$$\vec{J}_{\text{in}}^{\text{ES}}(\vec{r}) = \hat{n}_{-}\times\vec{H}(\vec{r}) \quad , \quad \vec{r}\in\mathbb{S}_{\text{in}} \tag{3-136a}$$

$$\vec{M}_{\text{in}}^{\text{ES}}(\vec{r}) = \vec{E}(\vec{r})\times\hat{n}_{-} \quad , \quad \vec{r}\in\mathbb{S}_{\text{in}} \tag{3-136b}$$

where $\hat{n}_{-}$ is the normal direction of $\mathbb{S}_{\text{in}}$ and points to the interior of $\mathbb{V}$. The above-mentined $\{\vec{J}_{\text{out}}^{\text{ES}},\vec{M}_{\text{out}}^{\text{ES}}\}$ are defined in terms of the total modal fields $\{\vec{E},\vec{H}\}$ as follows:

$$\vec{J}_{\text{out}}^{\text{ES}}(\vec{r}) = \hat{n}_{+}\times\vec{H}(\vec{r}) \quad , \quad \vec{r}\in\mathbb{S}_{\text{out}} \tag{3-137a}$$

$$\vec{M}_{\text{out}}^{\text{ES}}(\vec{r}) = \vec{E}(\vec{r})\times\hat{n}_{+} \quad , \quad \vec{r}\in\mathbb{S}_{\text{out}} \tag{3-137b}$$

where $\hat{n}_{+}$ is the normal direction of $\mathbb{S}_{\text{out}}$ and points to the exterior of $\mathbb{V}$.

Using the above currents, the field $\vec{F}$ in region $\mathbb{V}$ can be expressed as follows:

$$\vec{F}(\vec{r}) = \mathcal{F}\left(\vec{J}_{\text{in}}^{\text{ES}}+\vec{J}^{\text{IS}}-\vec{J}_{\text{out}}^{\text{ES}},\vec{M}_{\text{in}}^{\text{ES}}-\vec{M}_{\text{out}}^{\text{ES}}\right) \quad , \quad \vec{r}\in\mathbb{V} \tag{3-138}$$

in which $\vec{F}=\vec{E}/\vec{H}$ and correspondingly $\mathcal{F}=\mathcal{E}/\mathcal{H}$, and operators $\mathcal{E}$ and $\mathcal{H}$ are the same as the ones used in the previous sections. Based on the conclusions given in Sec. 3.2, the field $\vec{F}$ on $\mathbb{S}_{\text{out}}^{+}$ can be approximately expressed as follows:

$$\vec{F}(\vec{r}) = \mathcal{F}\left(\vec{J}_{\text{out}}^{\text{ES}},\vec{M}_{\text{out}}^{\text{ES}}\right) \quad , \quad \vec{r}\in\mathbb{S}_{\text{out}}^{+} \tag{3-139}$$

where $\mathbb{S}_{\text{out}}^{+}$ is the outer surface of $\mathbb{S}_{\text{out}}$.

### 3.5.2 Mathematical Description for Modal Space

Substituting Eq. (3-138) into Eqs. (3-136a) and (3-136b), we obtain the following integral equations

$$\vec{J}_{\text{in}}^{\text{ES}}\left(\vec{r}\right)\times\hat{n}_{-} \;=\left[\mathcal{H}\left(\vec{J}_{\text{in}}^{\text{ES}}+\vec{J}^{\text{IS}}-\vec{J}_{\text{out}}^{\text{ES}},\vec{M}_{\text{in}}^{\text{ES}}-\vec{M}_{\text{out}}^{\text{ES}}\right)\right]_{\vec{r}'\to\vec{r}}^{\tan} \quad , \quad \vec{r}\in\mathbb{S}_{\text{in}} \tag{3-140a}$$

$$\hat{n}_{-}\times\vec{M}_{\text{in}}^{\text{ES}}\left(\vec{r}\right)=\left[\mathcal{E}\left(\vec{J}_{\text{in}}^{\text{ES}}+\vec{J}^{\text{IS}}-\vec{J}_{\text{out}}^{\text{ES}},\vec{M}_{\text{in}}^{\text{ES}}-\vec{M}_{\text{out}}^{\text{ES}}\right)\right]_{\vec{r}'\to\vec{r}}^{\tan} \quad , \quad \vec{r}\in\mathbb{S}_{\text{in}} \tag{3-140b}$$

satisfied by the currents, where $\vec{r}'\in\text{int}\,\mathbb{V}$ and $\vec{r}'$ approaches the point $\vec{r}$ on surface $\mathbb{S}_{\text{in}}$.

Utilizing Eq. (3-138) and the homogeneous tangential electric field boundary condition on $\mathbb{S}_{\text{ele}}$, we obtain the following electric field integral equation

$$\left[\mathcal{E}\left(\vec{J}_{\text{in}}^{\text{ES}}+\vec{J}^{\text{IS}}-\vec{J}_{\text{out}}^{\text{ES}},\vec{M}_{\text{in}}^{\text{ES}}-\vec{M}_{\text{out}}^{\text{ES}}\right)\right]^{\tan}=0 \quad , \quad \vec{r}\in\mathbb{S}_{\text{ele}} \tag{3-141}$$

satisfied by the currents.

Utilizing Eqs. (3-138)&(3-139) and the tangential continuation condition satisfied by the field on $\mathbb{S}_{\text{out}}$, we obtain the following integral equations

$$\left[\mathcal{E}\left(\vec{J}_{\text{in}}^{\text{ES}}+\vec{J}^{\text{IS}}-\vec{J}_{\text{out}}^{\text{ES}},\vec{M}_{\text{in}}^{\text{ES}}-\vec{M}_{\text{out}}^{\text{ES}}\right)\right]_{\vec{r}_{-}\to\vec{r}}^{\tan}=\left[\mathcal{E}\left(\vec{J}_{\text{out}}^{\text{ES}},\vec{M}_{\text{out}}^{\text{ES}}\right)\right]_{\vec{r}_{+}\to\vec{r}}^{\tan} \quad , \quad \vec{r}\in\mathbb{S}_{\text{out}} \tag{3-142a}$$

$$\left[\mathcal{H}\left(\vec{J}_{\text{in}}^{\text{ES}}+\vec{J}^{\text{IS}}-\vec{J}_{\text{out}}^{\text{ES}},\vec{M}_{\text{in}}^{\text{ES}}-\vec{M}_{\text{out}}^{\text{ES}}\right)\right]_{\vec{r}_{-}\to\vec{r}}^{\tan}=\left[\mathcal{H}\left(\vec{J}_{\text{out}}^{\text{ES}},\vec{M}_{\text{out}}^{\text{ES}}\right)\right]_{\vec{r}_{+}\to\vec{r}}^{\tan} \quad , \quad \vec{r}\in\mathbb{S}_{\text{out}} \tag{3-142b}$$

satisfied by the currents, where $\vec{r}_{-}$ locates in the interior of $\mathbb{V}$ and approaches the point $\vec{r}$ on $\mathbb{S}_{\text{out}}$, and $\vec{r}_{+}$ locates in the exterior of $\mathbb{V}$ and approaches the point $\vec{r}$ on $\mathbb{S}_{\text{out}}$.

If the currents involved in Eqs. (3-140a)~(3-142b) are expanded in terms of some proper basis functions, and the equations are tested with $\{\vec{b}_{\xi}^{\vec{M}_{\text{in}}^{\text{ES}}}\}$, $\{\vec{b}_{\xi}^{\vec{J}_{\text{in}}^{\text{ES}}}\}$, $\{\vec{b}_{\xi}^{\vec{J}^{\text{IS}}}\}$, $\{\vec{b}_{\xi}^{\vec{J}_{\text{out}}^{\text{ES}}}\}$, and $\{\vec{b}_{\xi}^{\vec{M}_{\text{out}}^{\text{ES}}}\}$ respectively, then the integral equations are immediately discretized into the following matrix equations

$$\bar{\bar{Z}}^{\vec{M}_{\text{in}}^{\text{ES}}\vec{J}_{\text{in}}^{\text{ES}}}\cdot\bar{a}^{\vec{J}_{\text{in}}^{\text{ES}}}+\bar{\bar{Z}}^{\vec{M}_{\text{in}}^{\text{ES}}\vec{J}^{\text{IS}}}\cdot\bar{a}^{\vec{J}^{\text{IS}}}+\bar{\bar{Z}}^{\vec{M}_{\text{in}}^{\text{ES}}\vec{J}_{\text{out}}^{\text{ES}}}\cdot\bar{a}^{\vec{J}_{\text{out}}^{\text{ES}}}+\bar{\bar{Z}}^{\vec{M}_{\text{in}}^{\text{ES}}\vec{M}_{\text{in}}^{\text{ES}}}\cdot\bar{a}^{\vec{M}_{\text{in}}^{\text{ES}}}+\bar{\bar{Z}}^{\vec{M}_{\text{in}}^{\text{ES}}\vec{M}_{\text{out}}^{\text{ES}}}\cdot\bar{a}^{\vec{M}_{\text{out}}^{\text{ES}}}=0 \tag{3-143a}$$

$$\bar{\bar{Z}}^{\vec{J}_{\text{in}}^{\text{ES}}\vec{J}_{\text{in}}^{\text{ES}}}\cdot\bar{a}^{\vec{J}_{\text{in}}^{\text{ES}}}+\bar{\bar{Z}}^{\vec{J}_{\text{in}}^{\text{ES}}\vec{J}^{\text{IS}}}\cdot\bar{a}^{\vec{J}^{\text{IS}}}+\bar{\bar{Z}}^{\vec{J}_{\text{in}}^{\text{ES}}\vec{J}_{\text{out}}^{\text{ES}}}\cdot\bar{a}^{\vec{J}_{\text{out}}^{\text{ES}}}+\bar{\bar{Z}}^{\vec{J}_{\text{in}}^{\text{ES}}\vec{M}_{\text{in}}^{\text{ES}}}\cdot\bar{a}^{\vec{M}_{\text{in}}^{\text{ES}}}+\bar{\bar{Z}}^{\vec{J}_{\text{in}}^{\text{ES}}\vec{M}_{\text{out}}^{\text{ES}}}\cdot\bar{a}^{\vec{M}_{\text{out}}^{\text{ES}}}=0 \tag{3-143b}$$

and

$$\bar{\bar{Z}}^{\vec{J}^{\text{IS}}\vec{J}_{\text{in}}^{\text{ES}}}\cdot\bar{a}^{\vec{J}_{\text{in}}^{\text{ES}}}+\bar{\bar{Z}}^{\vec{J}^{\text{IS}}\vec{J}^{\text{IS}}}\cdot\bar{a}^{\vec{J}^{\text{IS}}}+\bar{\bar{Z}}^{\vec{J}^{\text{IS}}\vec{J}_{\text{out}}^{\text{ES}}}\cdot\bar{a}^{\vec{J}_{\text{out}}^{\text{ES}}}+\bar{\bar{Z}}^{\vec{J}^{\text{IS}}\vec{M}_{\text{in}}^{\text{ES}}}\cdot\bar{a}^{\vec{M}_{\text{in}}^{\text{ES}}}+\bar{\bar{Z}}^{\vec{J}^{\text{IS}}\vec{M}_{\text{out}}^{\text{ES}}}\cdot\bar{a}^{\vec{M}_{\text{out}}^{\text{ES}}}=0 \tag{3-144}$$

and

$$\bar{\bar{Z}}^{\vec{J}_{\text{out}}^{\text{ES}}\vec{J}_{\text{in}}^{\text{ES}}}\cdot\bar{a}^{\vec{J}_{\text{in}}^{\text{ES}}}+\bar{\bar{Z}}^{\vec{J}_{\text{out}}^{\text{ES}}\vec{J}^{\text{IS}}}\cdot\bar{a}^{\vec{J}^{\text{IS}}}+\bar{\bar{Z}}^{\vec{J}_{\text{out}}^{\text{ES}}\vec{J}_{\text{out}}^{\text{ES}}}\cdot\bar{a}^{\vec{J}_{\text{out}}^{\text{ES}}}+\bar{\bar{Z}}^{\vec{J}_{\text{out}}^{\text{ES}}\vec{M}_{\text{in}}^{\text{ES}}}\cdot\bar{a}^{\vec{M}_{\text{in}}^{\text{ES}}}+\bar{\bar{Z}}^{\vec{J}_{\text{out}}^{\text{ES}}\vec{M}_{\text{out}}^{\text{ES}}}\cdot\bar{a}^{\vec{M}_{\text{out}}^{\text{ES}}}=0 \tag{3-145a}$$

$$\bar{\bar{Z}}^{\vec{M}_{\text{out}}^{\text{ES}}\vec{J}_{\text{in}}^{\text{ES}}}\cdot\bar{a}^{\vec{J}_{\text{in}}^{\text{ES}}}+\bar{\bar{Z}}^{\vec{M}_{\text{out}}^{\text{ES}}\vec{J}^{\text{IS}}}\cdot\bar{a}^{\vec{J}^{\text{IS}}}+\bar{\bar{Z}}^{\vec{M}_{\text{out}}^{\text{ES}}\vec{J}_{\text{out}}^{\text{ES}}}\cdot\bar{a}^{\vec{J}_{\text{out}}^{\text{ES}}}+\bar{\bar{Z}}^{\vec{M}_{\text{out}}^{\text{ES}}\vec{M}_{\text{in}}^{\text{ES}}}\cdot\bar{a}^{\vec{M}_{\text{in}}^{\text{ES}}}+\bar{\bar{Z}}^{\vec{M}_{\text{out}}^{\text{ES}}\vec{M}_{\text{out}}^{\text{ES}}}\cdot\bar{a}^{\vec{M}_{\text{out}}^{\text{ES}}}=0 \tag{3-145b}$$

The formulations used to calculate the elements of the matrices in the above matrix equations are similar to the formulations given in Eqs. (3-91a)~(3-96f), and they are explicitly given in the App. D2 of this report.

Employing the above Eqs. (3-143a)~(3-145b), we can obtain the following transformation

$$\begin{bmatrix} \bar{a}^{\vec{J}_{\text{in}}^{\text{ES}}} \\ \bar{a}^{\vec{J}^{\text{IS}}} \\ \bar{a}^{\vec{J}_{\text{out}}^{\text{ES}}} \\ \bar{a}^{\vec{M}_{\text{in}}^{\text{ES}}} \\ \bar{a}^{\vec{M}_{\text{out}}^{\text{ES}}} \end{bmatrix} = \bar{a}^{\text{AV}} = \bar{\bar{T}} \cdot \bar{a} \tag{3-146}$$

where the calculation formulation for transformation matrix $\bar{\bar{T}}$ is given in the App. D2 of this report.

## 3.5.3 Input Power Operator

The IPO corresponding to the input port $\mathbb{S}_{\text{in}}$ shown in Fig. 3-36 is as follows:

$$\begin{aligned} P^{\text{in}} &= (1/2)\iint_{\mathbb{S}_{\text{in}}} \left(\vec{E}\times\vec{H}^{\dagger}\right)\cdot\hat{z}dS \\ &= (1/2)\left\langle \hat{z}\times\vec{J}_{\text{in}}^{\text{ES}},\vec{M}_{\text{in}}^{\text{ES}}\right\rangle_{\mathbb{S}_{\text{in}}} \\ &= -(1/2)\left\langle \vec{J}_{\text{in}}^{\text{ES}},\mathcal{E}\left(\vec{J}_{\text{in}}^{\text{ES}}+\vec{J}^{\text{IS}}-\vec{J}_{\text{out}}^{\text{ES}},\vec{M}_{\text{in}}^{\text{ES}}-\vec{M}_{\text{out}}^{\text{ES}}\right)\right\rangle_{\mathbb{S}_{\text{in}}^{-}} \\ &= -(1/2)\left\langle \vec{M}_{\text{in}}^{\text{ES}},\mathcal{H}\left(\vec{J}_{\text{in}}^{\text{ES}}+\vec{J}^{\text{IS}}-\vec{J}_{\text{out}}^{\text{ES}},\vec{M}_{\text{in}}^{\text{ES}}-\vec{M}_{\text{out}}^{\text{ES}}\right)\right\rangle_{\mathbb{S}_{\text{in}}^{-}}^{\dagger} \end{aligned} \tag{3-147}$$

Here, the right-hand side of the first equality is the field form of IPO, and the right-hand side of the second equality is the current form of IPO, and the right-hand sides of the third and fourth equalities are the field-current interaction forms of IPO.

By discretizing IPO (3-147) and utilizing transformation (3-146), we derive the following matrix form of the IPO

$$P^{\text{in}} = \bar{a}^{\dagger}\cdot\bar{\bar{P}}^{\text{in}}\cdot\bar{a} \tag{3-148}$$

and the formulation for calculating the quadratic matrix $\bar{\bar{P}}^{\text{in}}$ is given in the App. D2 of this report.

## 3.5.4 Input-Power-Decoupled Modes

The IP-DMs in modal space can be derived from solving the modal decoupling equation $\bar{\bar{P}}_{-}^{\text{in}}\cdot\bar{\alpha}_{\xi} = \theta_{\xi}\,\bar{\bar{P}}_{+}^{\text{in}}\cdot\bar{\alpha}_{\xi}$ defined on modal space, where $\bar{\bar{P}}_{+}^{\text{in}}$ and $\bar{\bar{P}}_{-}^{\text{in}}$ are the positive and negative Hermitian parts of the matrix $\bar{\bar{P}}^{\text{in}}$ given in Eq. (3-148). If some derived modes $\{\bar{\alpha}_1,\bar{\alpha}_2,\cdots,\bar{\alpha}_d\}$ are *d*-order degenerate, then the Gram-Schmidt orthogonalization process given in previous Sec. 3.2.4.1 is necessary, and it is not repeated here.

The IP-DMs constructed above satisfy the following decoupling relation

$$\left(1+ j\,\theta_{\xi}\right)\delta_{\xi\zeta} \;=\; \left(1/2\right)\iint_{\mathbb{S}_{\text{in}}}\left(\vec{E}_{\zeta}\times\vec{H}_{\xi}^{\dagger}\right)\cdot\hat{z}dS \;=\; \left(1/2\right)\left\langle \hat{z}\times\vec{J}_{\text{in};\xi}^{\text{ES}},\vec{M}_{\text{in};\zeta}^{\text{ES}}\right\rangle_{\mathbb{S}_{\text{in}}} \tag{3-149}$$

and the relation implies that the IP-DMs don't have net energy coupling in integral period. By employing the decoupling relation, we have the following Parseval's identity

$$\sum_{\xi}\left|c_{\xi}\right|^{2} \;=\; \left(1/T\right)\int_{t_0}^{t_0+T}\left[\iint_{\mathbb{S}_{\text{in}}}\left(\vec{\mathcal{E}}\times\vec{\mathcal{H}}\right)\cdot\hat{z}dS\right]dt \tag{3-150}$$

in which $\{\vec{\mathcal{E}},\vec{\mathcal{H}}\}$ are the time-domain fields, and the expansion coefficients have expression $c_{\xi}=-(1/2)<\vec{J}_{\text{in};\xi}^{\text{ES}},\vec{E}>_{\mathbb{S}_{\text{in}}}\big/(1+j\theta_{\xi})=-(1/2)<\vec{H},\vec{M}_{\text{in};\xi}^{\text{ES}}>_{\mathbb{S}_{\text{in}}}\big/(1+j\theta_{\xi})$, and $\{\vec{E},\vec{H}\}$ are some previously known fields distributing on input port $\mathbb{S}_{\text{in}}$.

Similarly to the previous sections, the modal input impedance and admittance can be defined and employed to recognize travelling-wave-type IP-DMs, and they will not be repeated here.

## 3.5.5 Numerical Examples Corresponding to Typical Structures

In this subsection, we consider two kinds of non-standard guides, and construct their travelling-wave-type IP-DMs by using the formulations given above.

### 3.5.5.1 Non-standard Guide I

The geometry of non-standard guide I is shown in Fig. 3-37. The cross section of the guide tube is circular and has radius 1cm. As shown in Fig. 3-37, the input port and output port of the guide locate in $x=3\text{cm}$ plane and $y=3\text{cm}$ plane respectively.

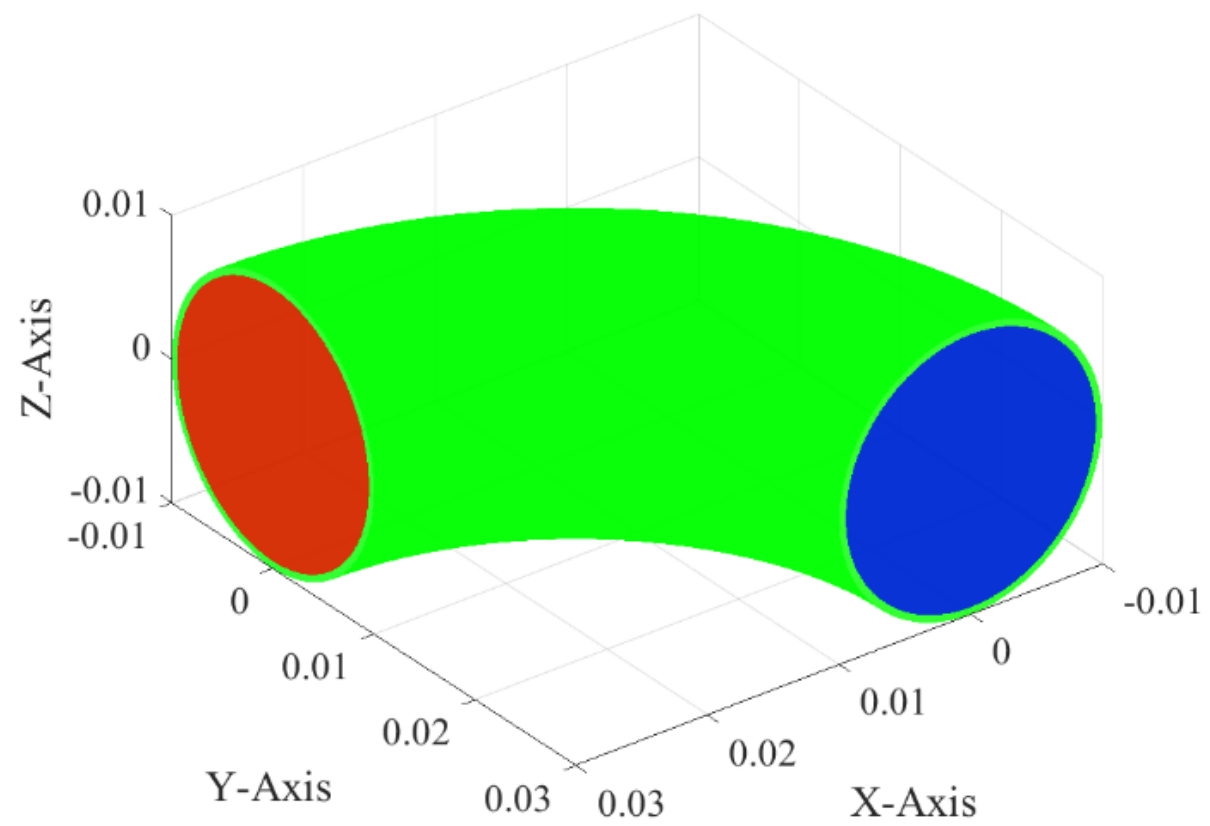

Figure 3-37 Geometry of non-standard guide I

The guide tube is filled by the material with $\mu_{\text{r}}=1$ and $\varepsilon_{\text{r}}=5$. The topological structures of the guide shown in Fig. 3-37 are illustrated in the following Fig. 3-38.

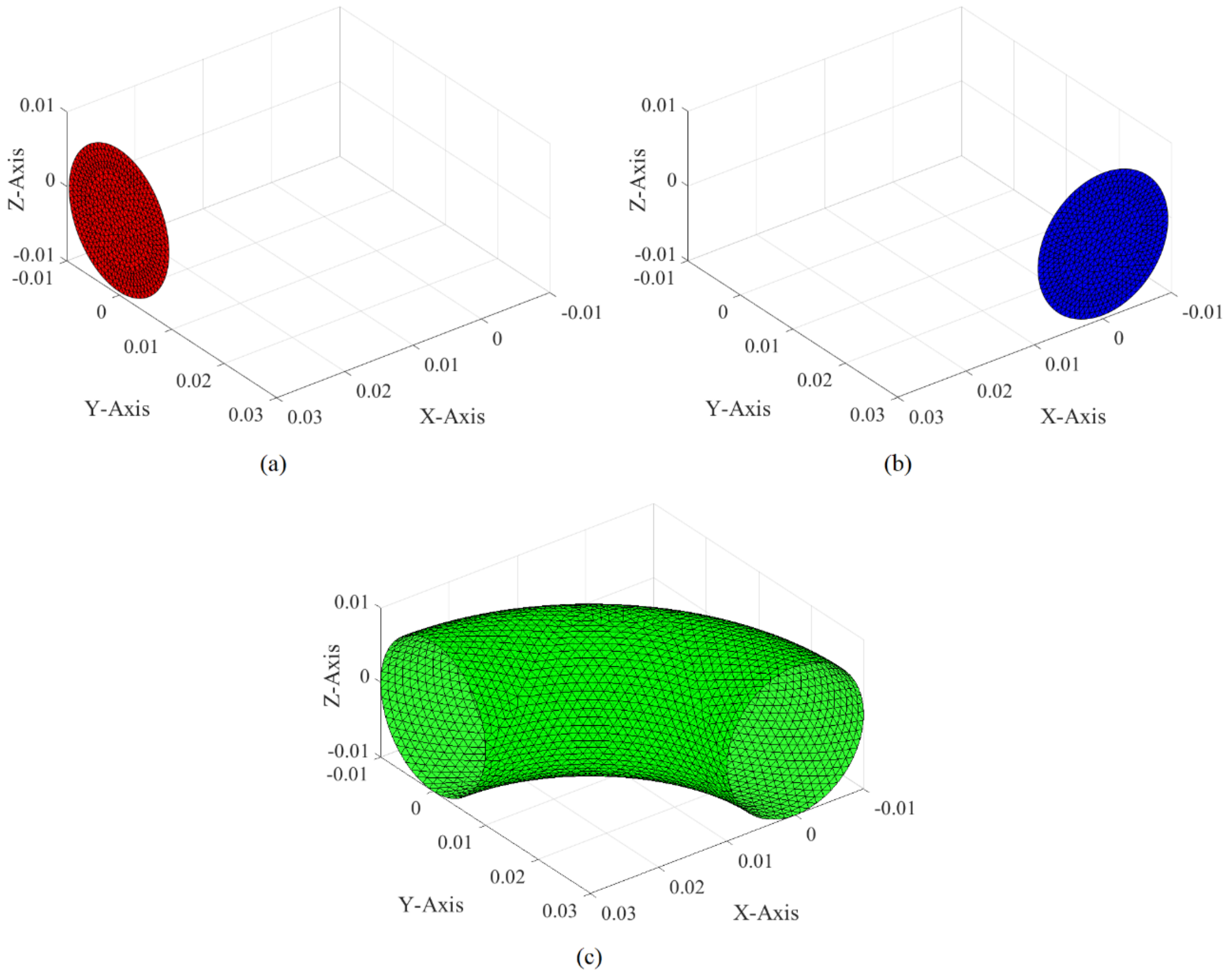

Figure 3-38 Topological structures and surface triangular meshes of the guide shown in Fig. 3-37. (a) Mesh of input port $\mathbb{S}_{in}$ ; (b) mesh of output port $\mathbb{S}_{out}$ ; (c) mesh of guide electric wall $\mathbb{S}_{ele}$

By orthogonalizing the HM-DoM-based IPO (3-148), the IP-DMs of the guide are constructed, and the modal input conductance curves of the first several typical IP-DMs are plotted in the following Fig. 3-39.

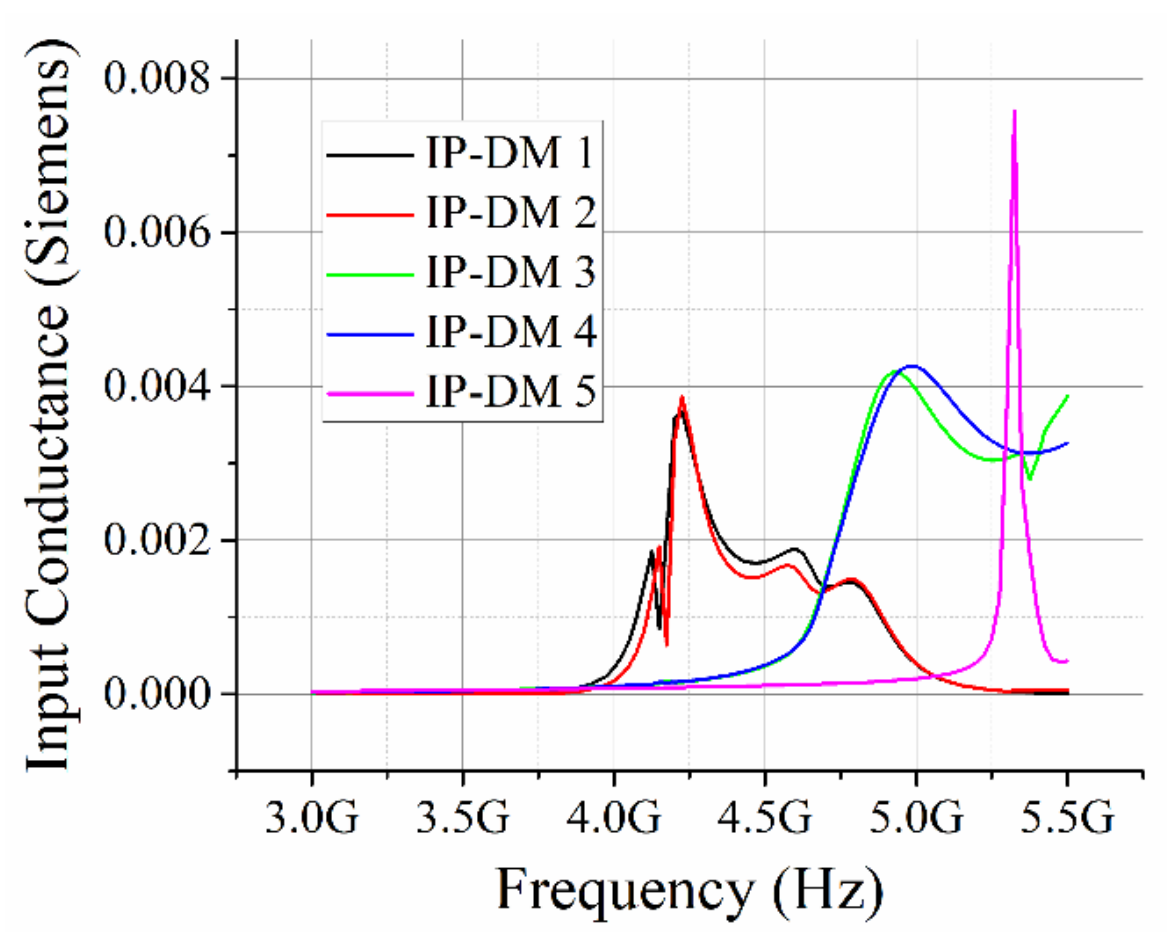

Figure 3-39 Modal input conductance curves of some typical IP-DMs

For the IP-DM 1 (at 4.225GHz), its modal quantities are shown as follows:

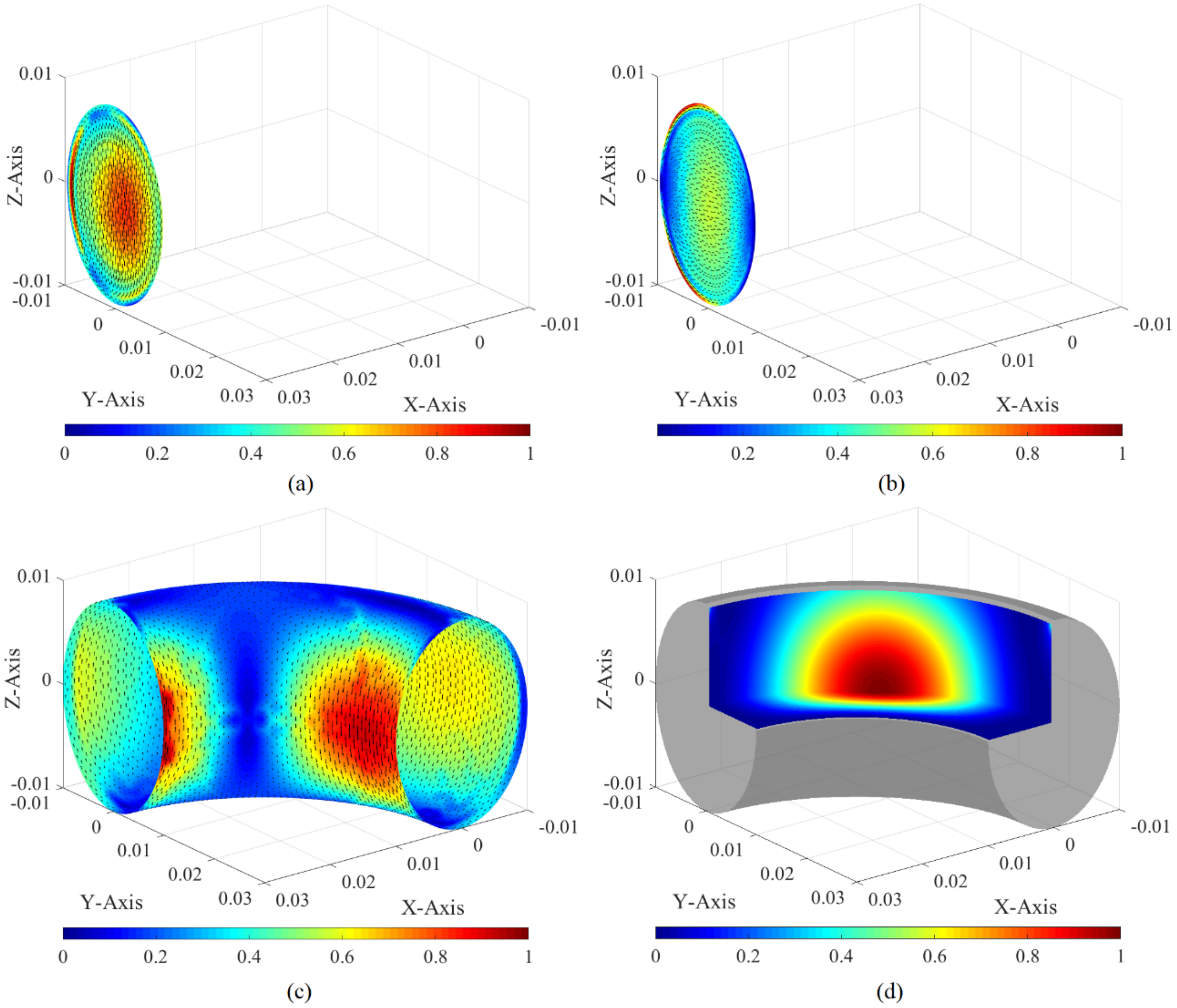

Figure 3-40 Modal quantity distributions of the IP-DM 1 (working at 4.225GHz). (a) Equivalent electric current on $\mathbb{S}_{in}$; (b) equivalent magnetic current on $\mathbb{S}_{in}$; (c) equivalent electric current on $\mathbb{S}_{ele}$; (d) electric energy density in guide tube

For the IP-DM 2 (at 4.225GHz), its modal quantities are shown as follows:

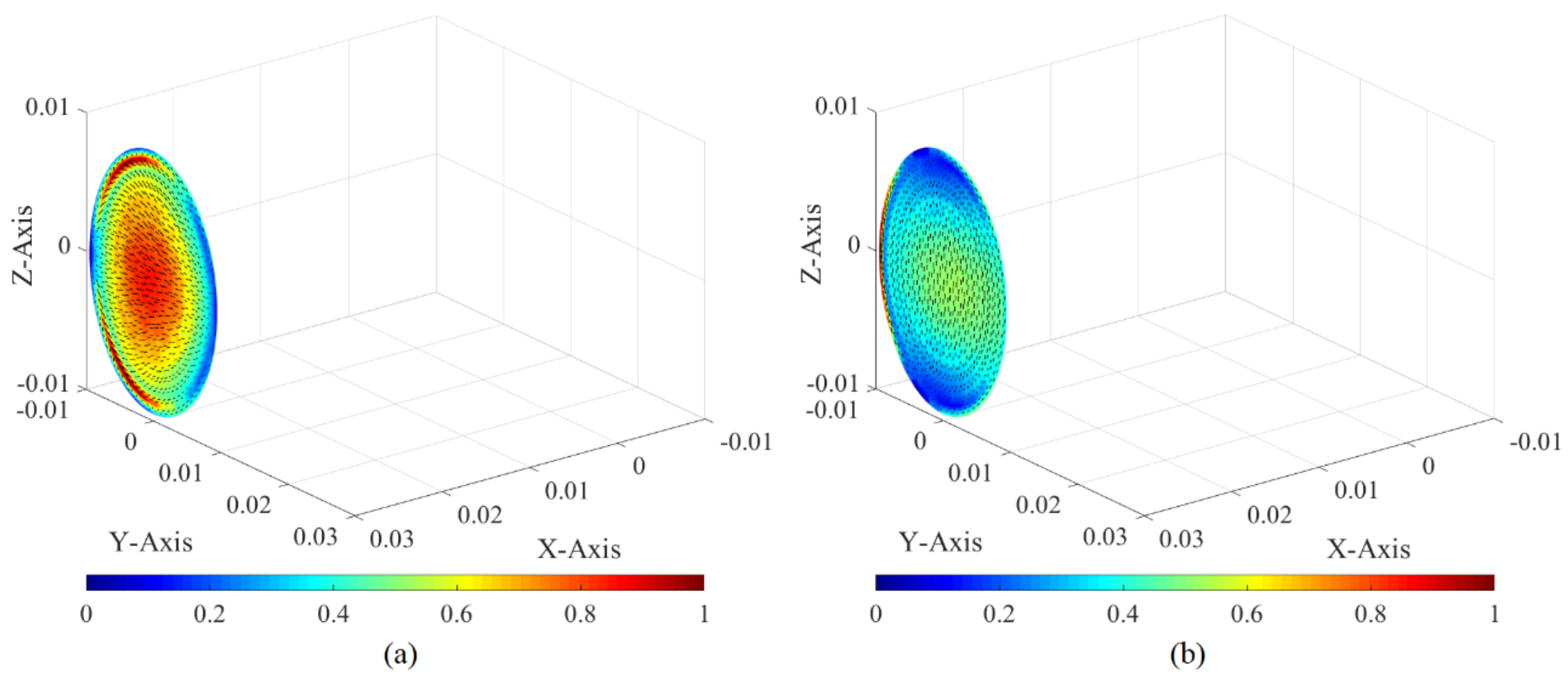

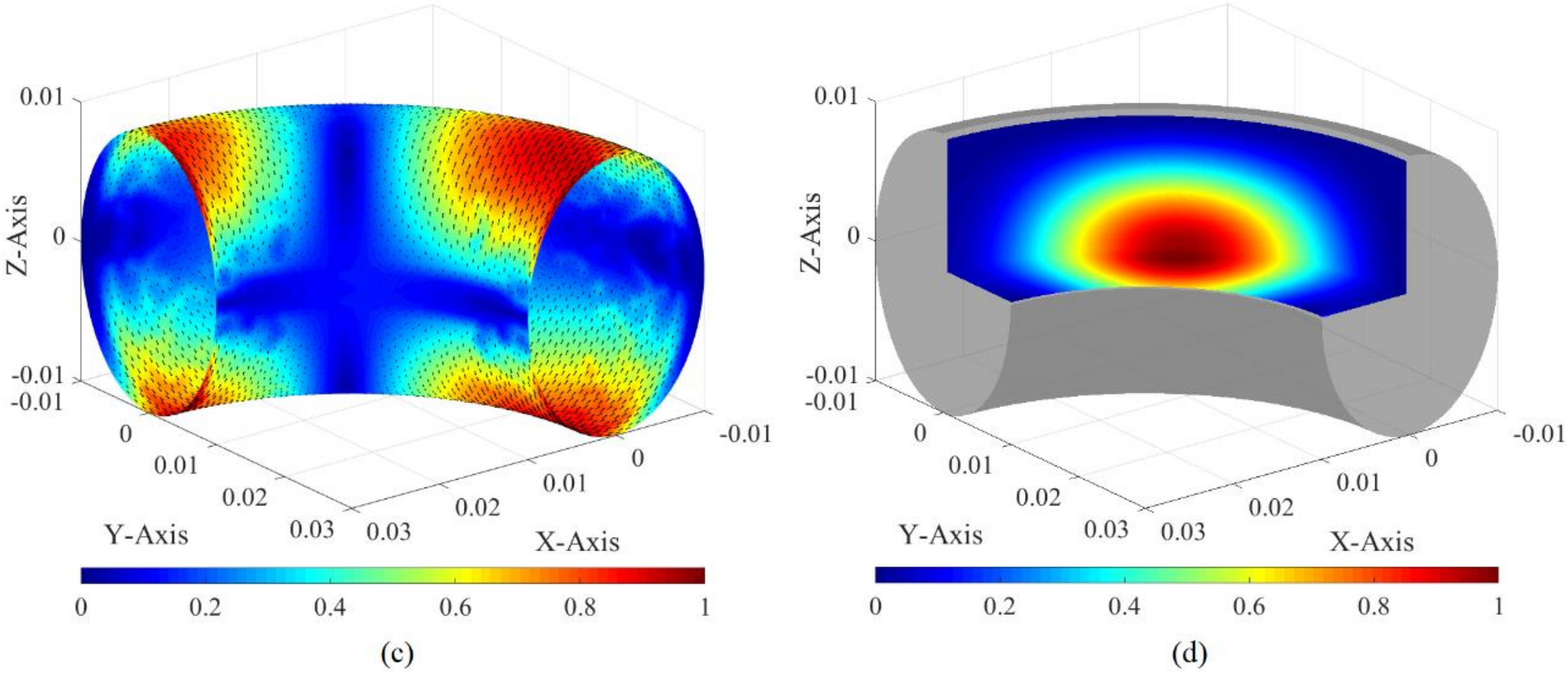

Figure 3-41 Modal quantity distributions of the IP-DM 2 (working at 4.225GHz). (a) Equivalent electric current on $\mathbb{S}_{\text{in}}$; (b) equivalent magnetic current on $\mathbb{S}_{\text{in}}$; (c) equivalent electric current on $\mathbb{S}_{\text{ele}}$; (d) electric energy density in guide tube

For the IP-DM 3 (at 4.925GHz), its modal quantities are shown as follows:

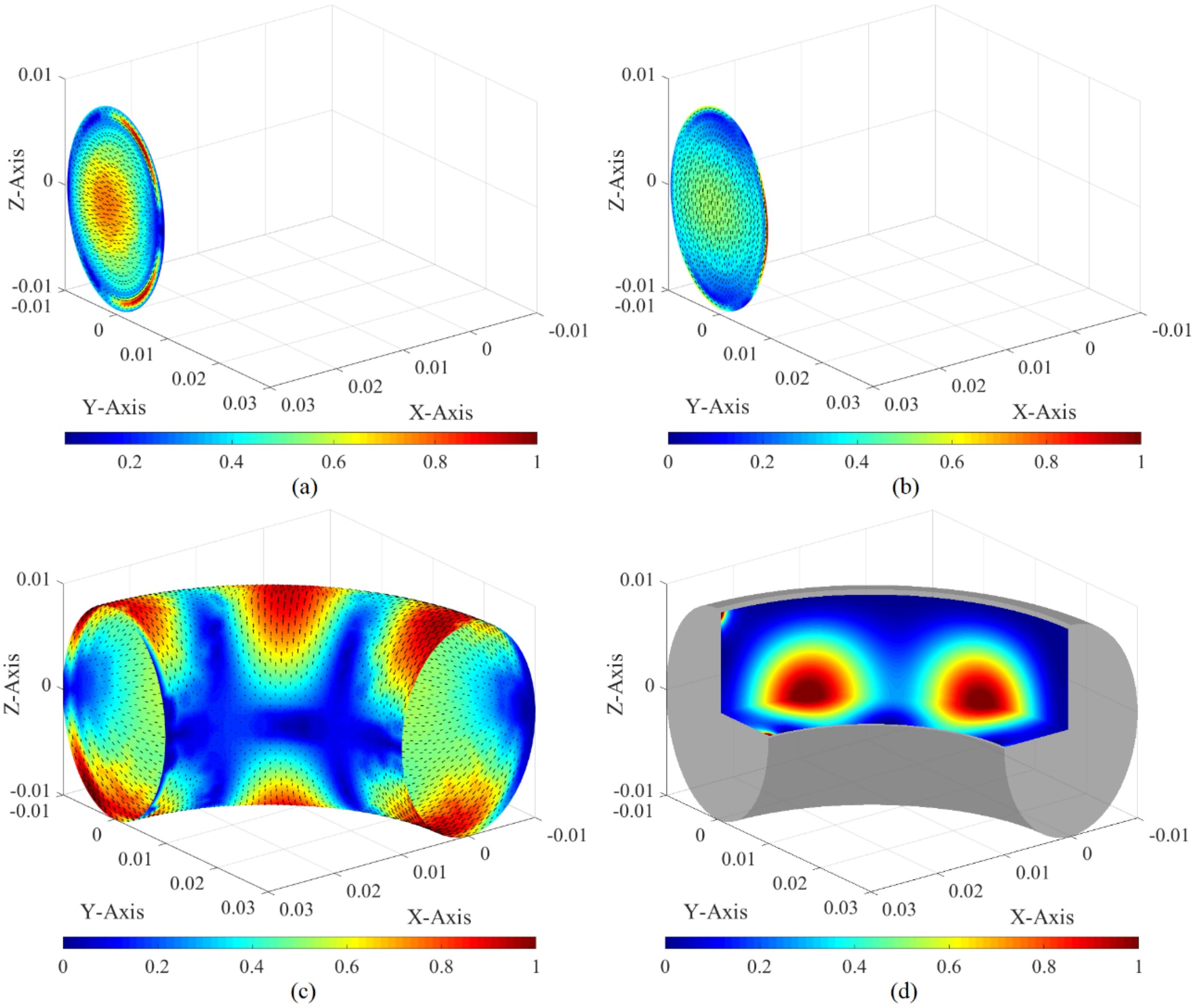

Figure 3-42 Modal quantity distributions of the IP-DM 3 (working at 4.925GHz). (a) Equivalent electric current on $\mathbb{S}_{\text{in}}$; (b) equivalent magnetic current on $\mathbb{S}_{\text{in}}$; (c) equivalent electric current on $\mathbb{S}_{\text{ele}}$; (d) electric energy density in guide tube

For the IP-DM 4 (at 4.975GHz), its modal quantities are shown as follows:

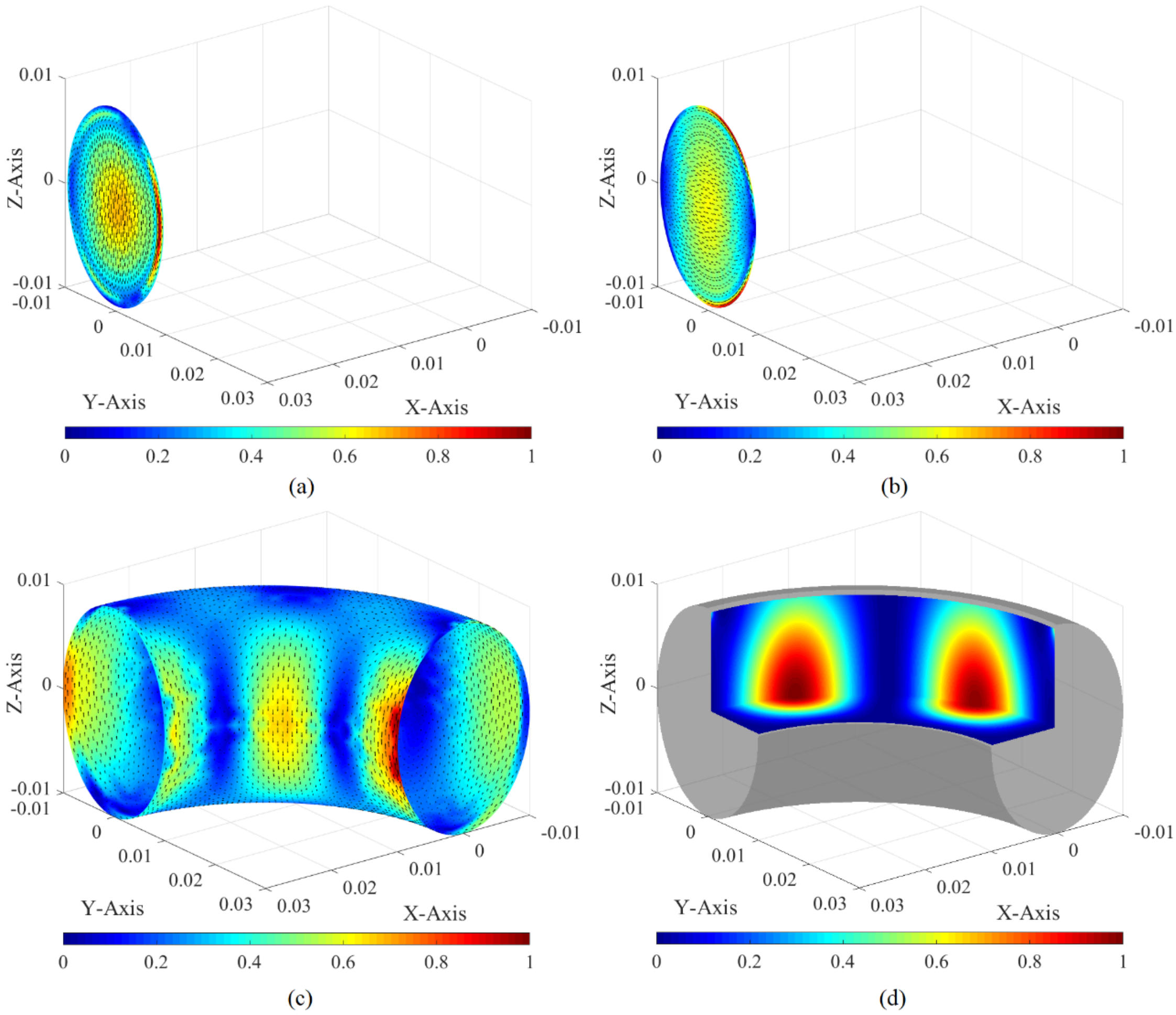

Figure 3-43 Modal quantity distributions of the IP-DM 4 (working at 4.975GHz). (a) Equivalent electric current on $\mathbb{S}_{\text{in}}$; (b) equivalent magnetic current on $\mathbb{S}_{\text{in}}$; (c) equivalent electric current on $\mathbb{S}_{\text{ele}}$; (d) electric energy density in guide tube

For the IP-DM 4 (at 5.325GHz), its modal quantities are shown as follows:

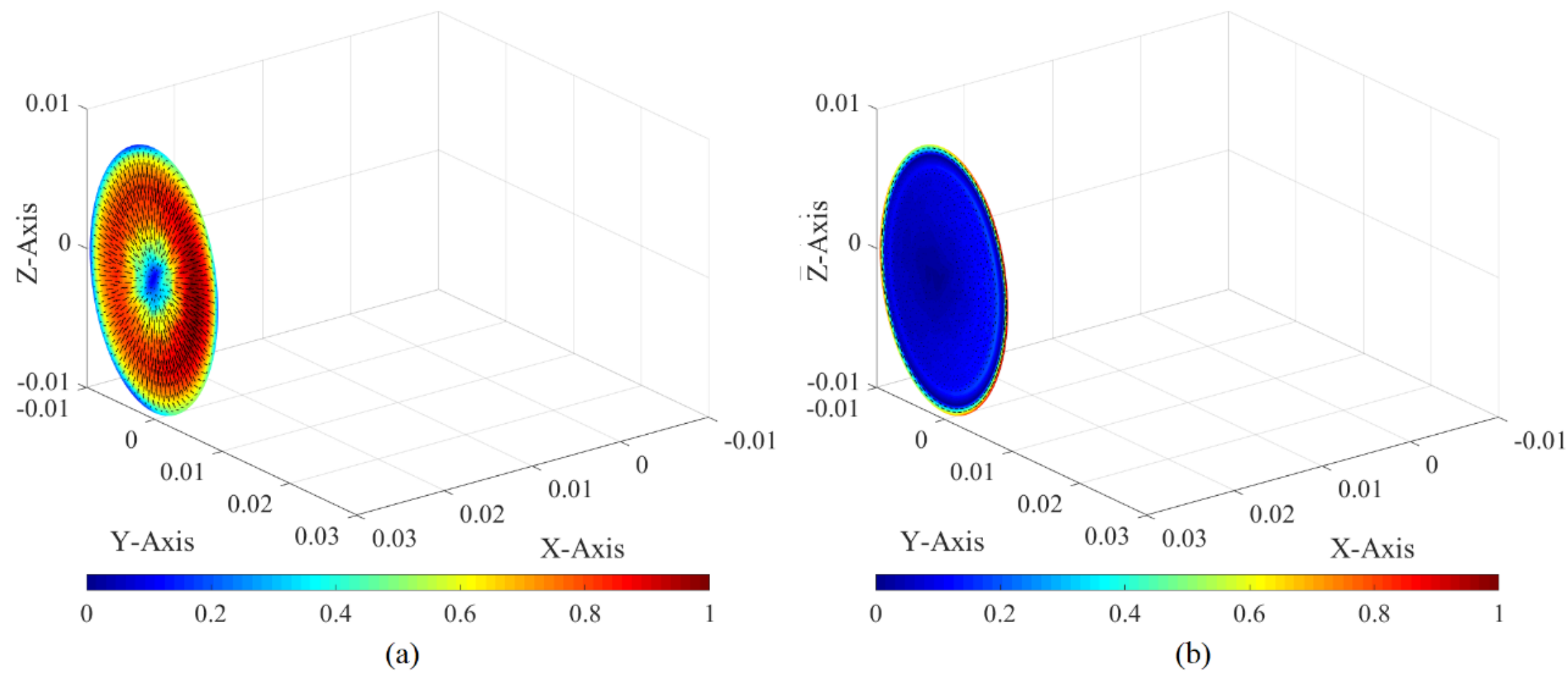

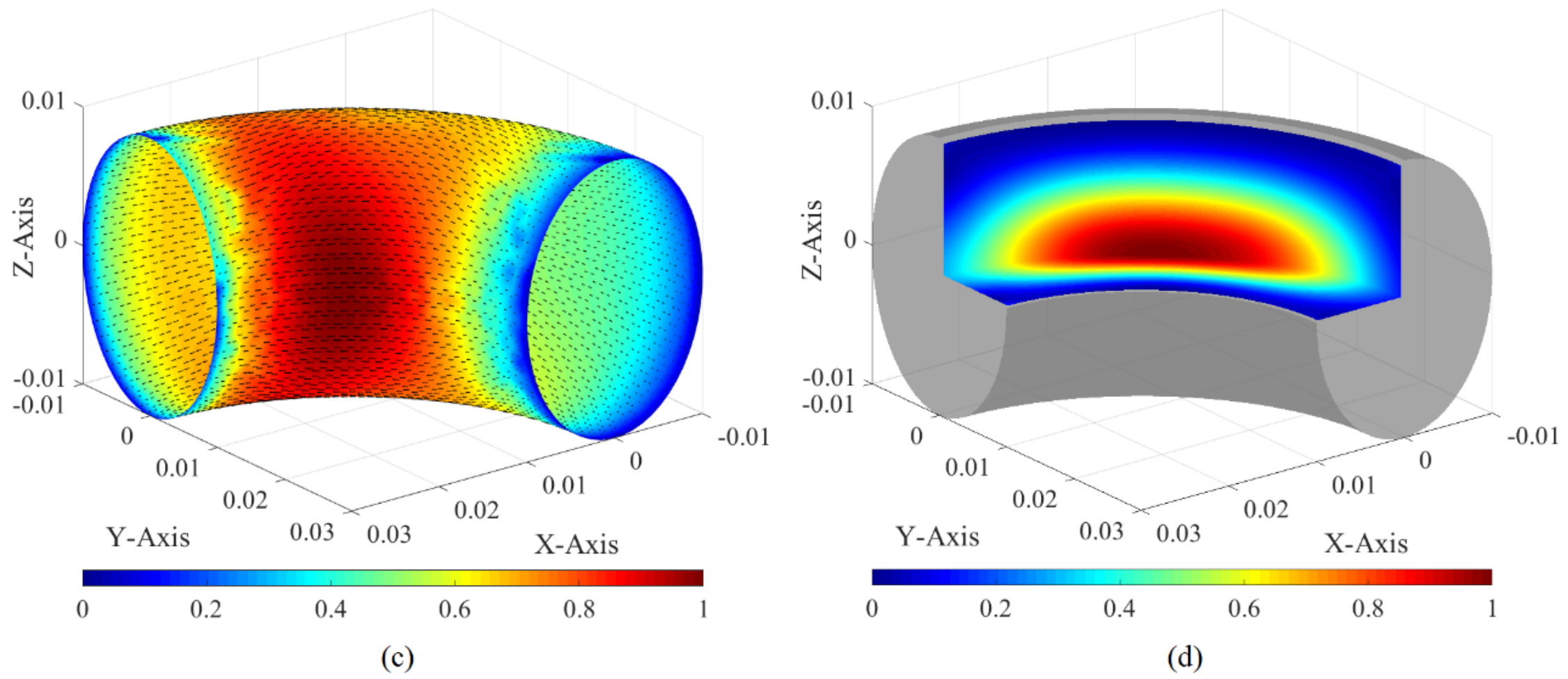

Figure 3-44 Modal quantity distributions of the IP-DM 5 (working at 5.325GHz). (a) Equivalent electric current on $\mathbb{S}_{\rm in}$; (b) equivalent magnetic current on $\mathbb{S}_{\rm in}$; (c) equivalent electric current on $\mathbb{S}_{\rm ele}$; (d) electric energy density in guide tube

In the following subsection, we construct the IP-DMs of another typical non-standard guide.

### 3.5.5.2 Non-standard Guide II

The geometry of non-standard guide II is shown in the following Fig. 3-45. The guide electric wall is the surface constituted by that curve $z = 0.01 + 0.003\sin(3\pi y / 0.05)$ spins around Y-axis. As shown in Fig. 3-45, the input port locates in zOx plane, and the output port locates in $y = 5\,\text{cm}$ plane.

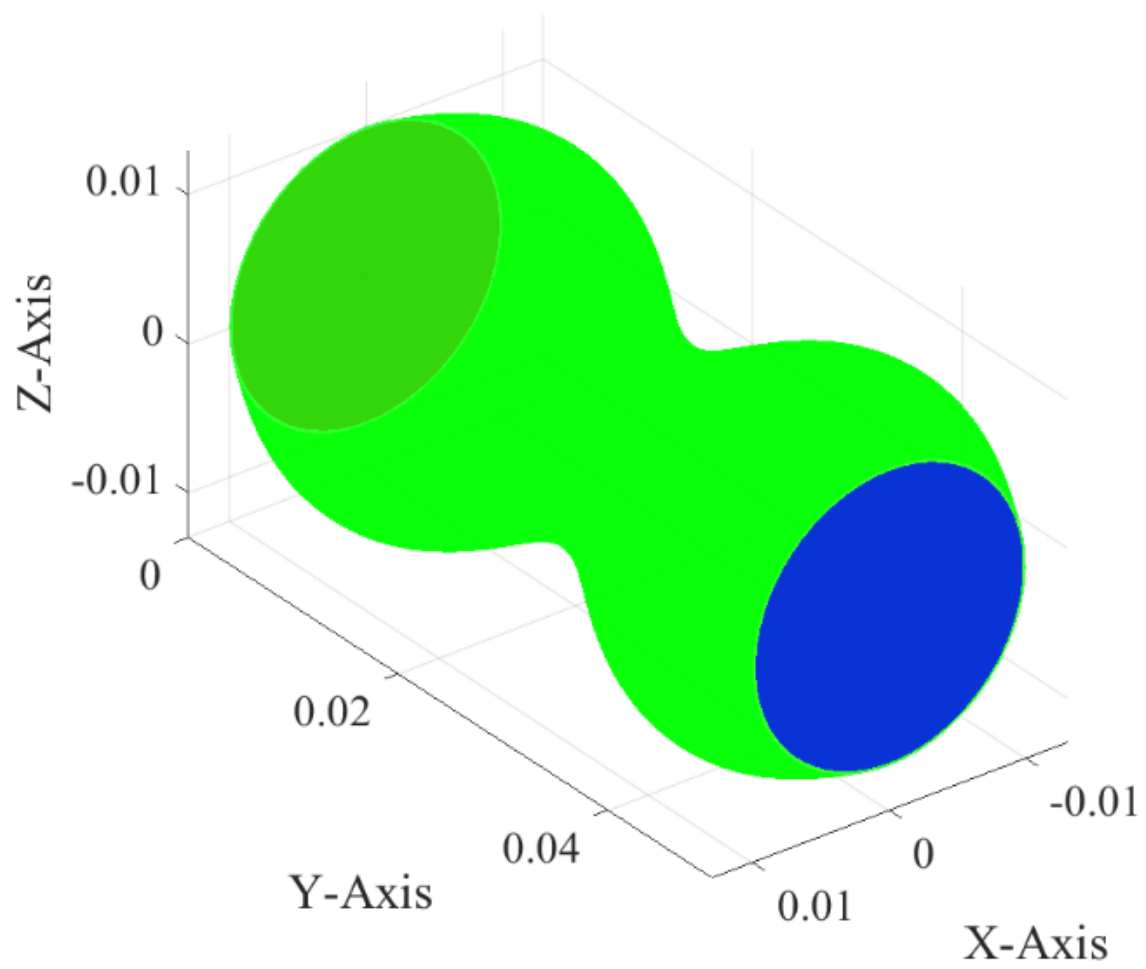

Figure 3-45 Geometry of non-standard guide II

The guide tube is filled by the material with $\mu_{\rm r} = 1$ and $\varepsilon_{\rm r} = 5$. The topological structure of the guide shown in Fig. 3-45 is shown in the following Fig. 3-46.

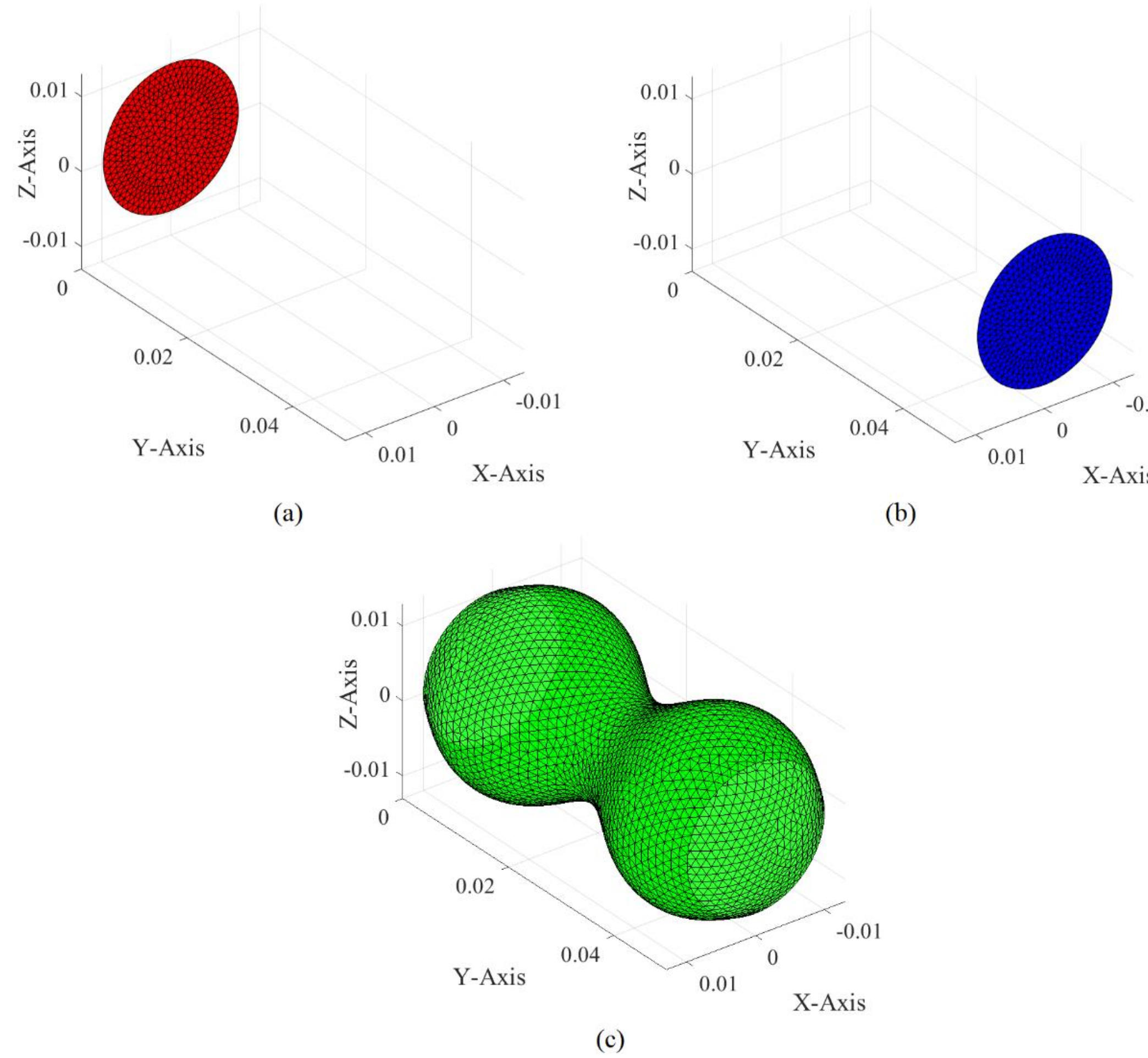

Figure 3-46 Topological structures and surface triangular meshes of the guide shown in Fig. 3-45. (a) Mesh of input port $\mathbb{S}_{\text{in}}$ ; (b) mesh of output port $\mathbb{S}_{\text{out}}$ ; (c) mesh of guide electric wall $\mathbb{S}_{\text{ele}}$

By orthogonalizing the JE-DoJ-based IPO (3-148), the modal input resistance curves of some typical IP-DMs are obtained as shown in the following Fig. 3-47.

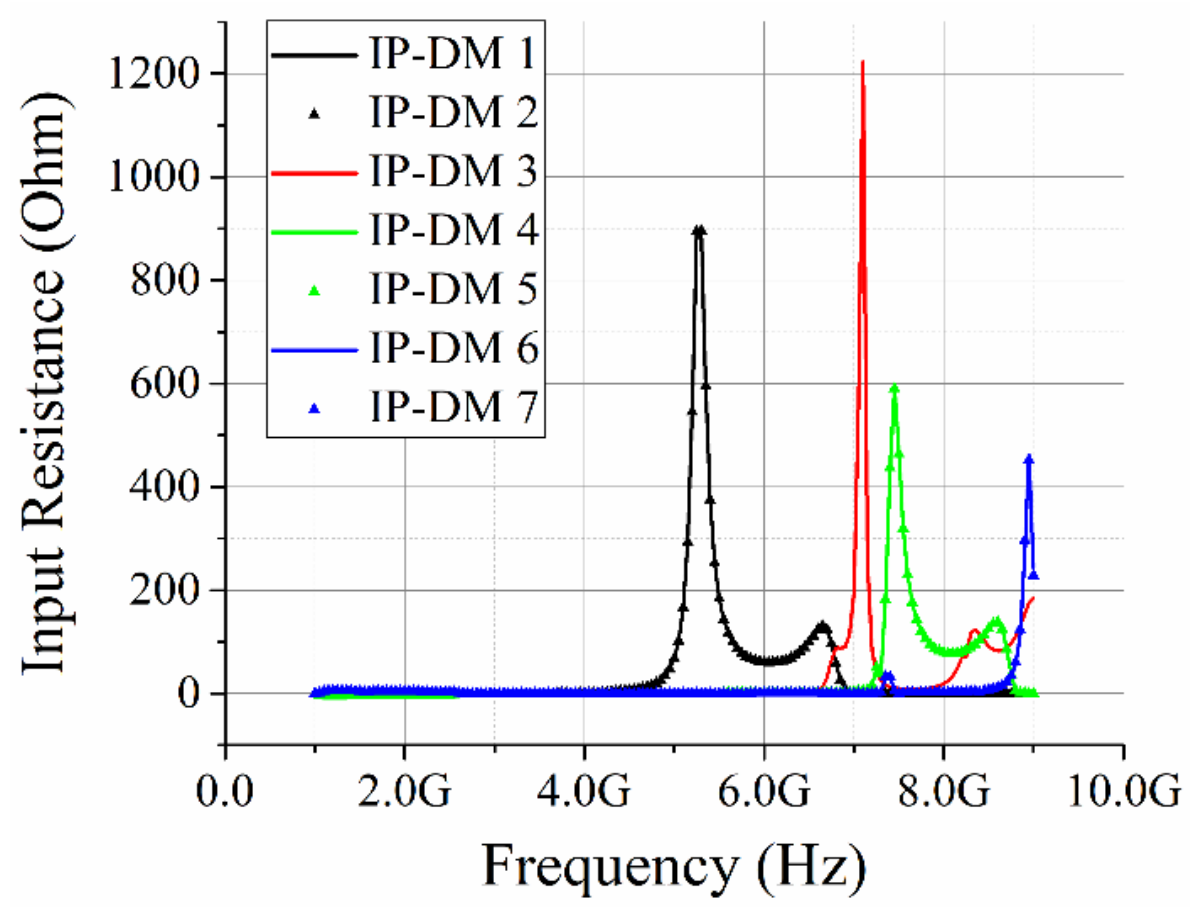

Figure 3-47 Modal input resistance curves of some typical IP-DMs

For the IP-DM 1 (at 5.30GHz), its modal quantities are shown in Fig. 3-48.

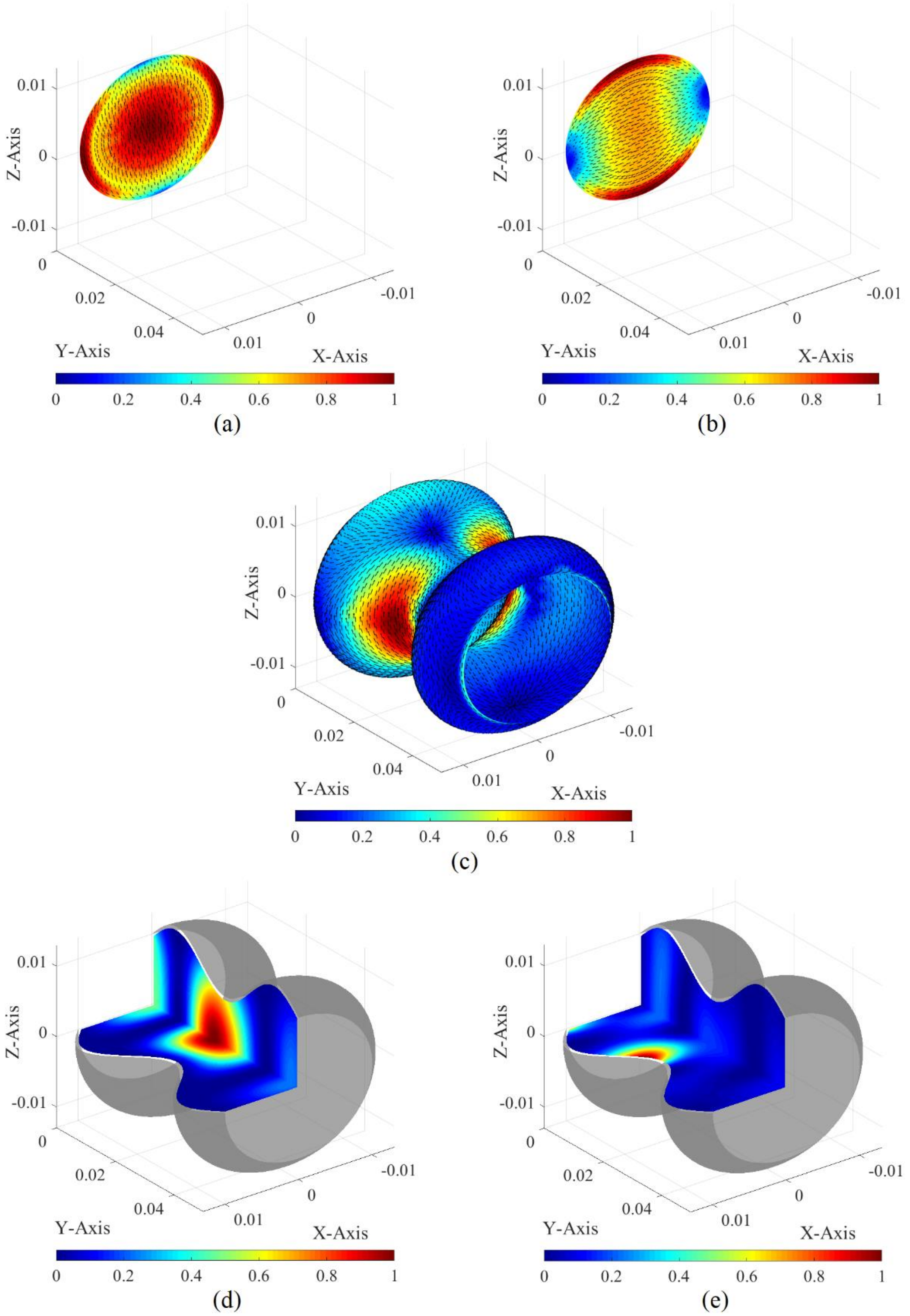

Figure 3-48 Modal quantity distributions of the IP-DM 1 (working at 5.30GHz). (a) Equivalent electric current on $\mathbb{S}_{in}$; (b) equivalent magnetic current on $\mathbb{S}_{in}$; (c) equivalent electric current on $\mathbb{S}_{ele}$; (d) electric energy density in guide tube; (e) magnetic energy density in guide tube

For the IP-DM 2 (at 5.30GHz), its modal quantities are shown in Fig. 3-49.

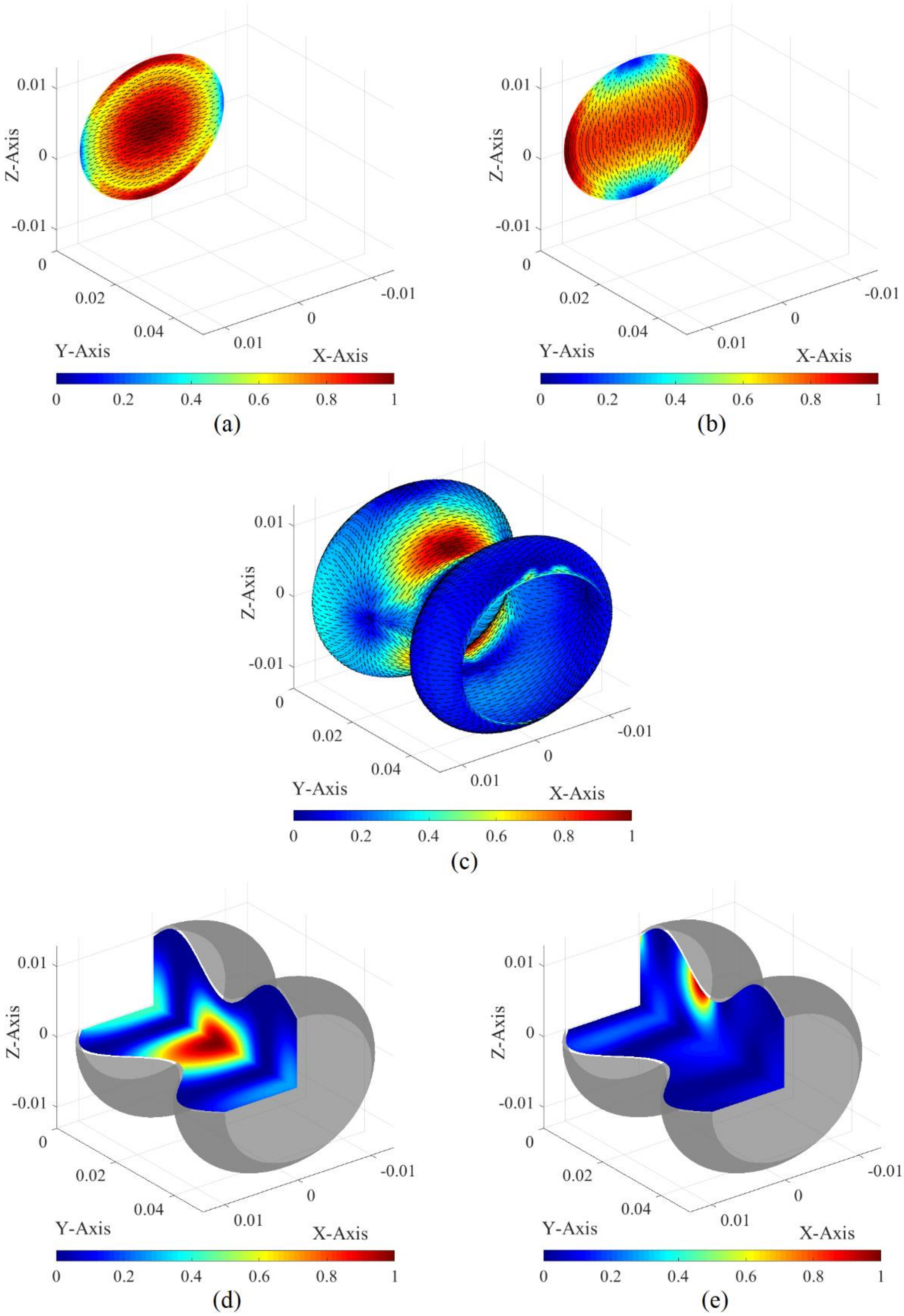

Figure 3-49 Modal quantity distributions of the IP-DM 2 (working at 5.30GHz). (a) Equivalent electric current on $\mathbb{S}_{\text{in}}$ ; (b) equivalent magnetic current on $\mathbb{S}_{\text{in}}$ ; (c) equivalent electric current on $\mathbb{S}_{\text{ele}}$ ; (d) electric energy density in guide tube; (e) magnetic energy density in guide tube

For the IP-DM 3 (at 7.10GHz), its modal quantities are shown in Fig. 3-50.

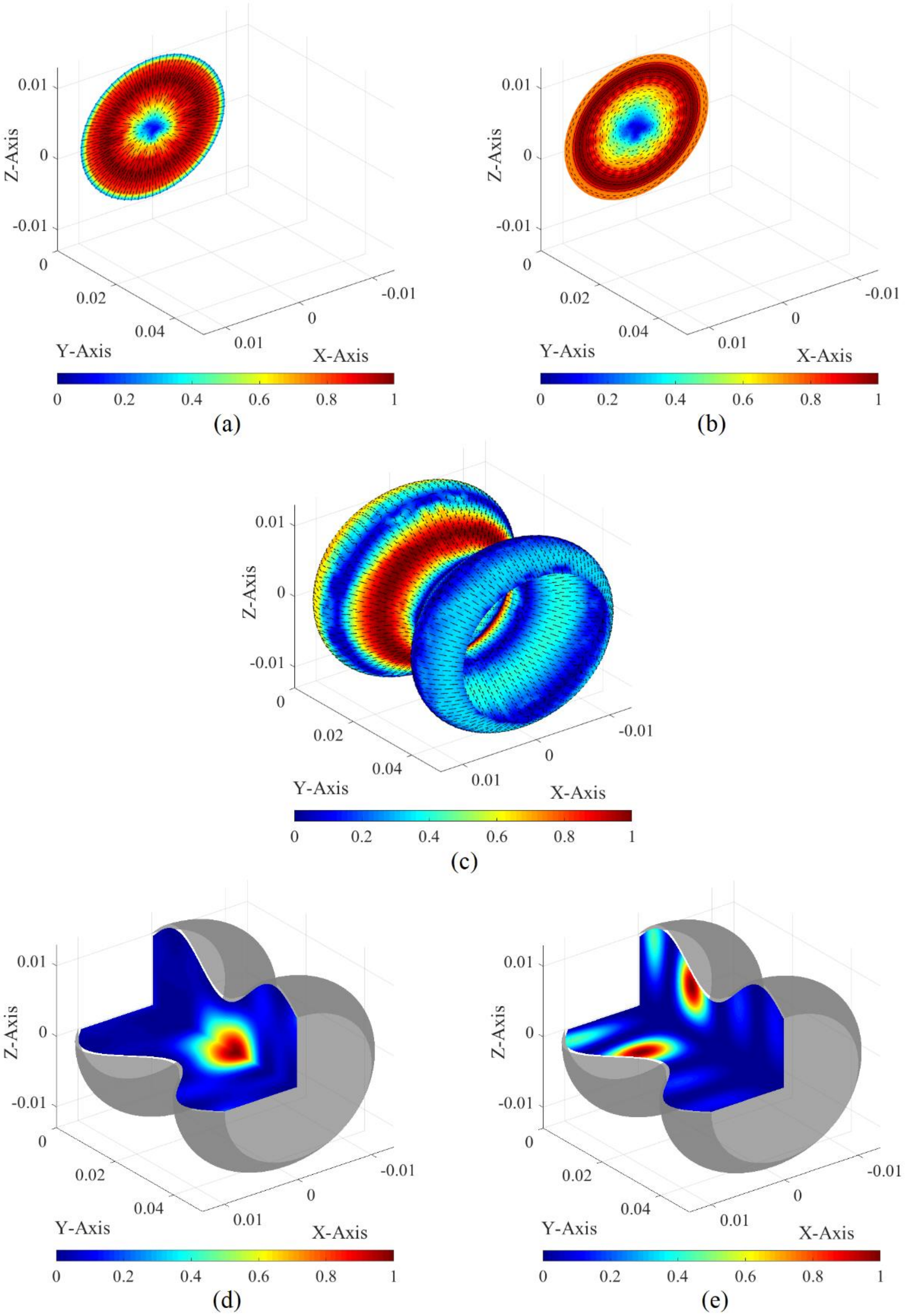

Figure 3-50 Modal quantity distributions of the IP-DM 3 (working at 7.10GHz). (a) Equivalent electric current on $\mathbb{S}_{in}$ ; (b) equivalent magnetic current on $\mathbb{S}_{in}$ ; (c) equivalent electric current on $\mathbb{S}_{ele}$ ; (d) electric energy density in guide tube; (e) magnetic energy density in guide tube

For the IP-DM 4 (at 7.45GHz), its modal quantities are shown in Fig. 3-51.

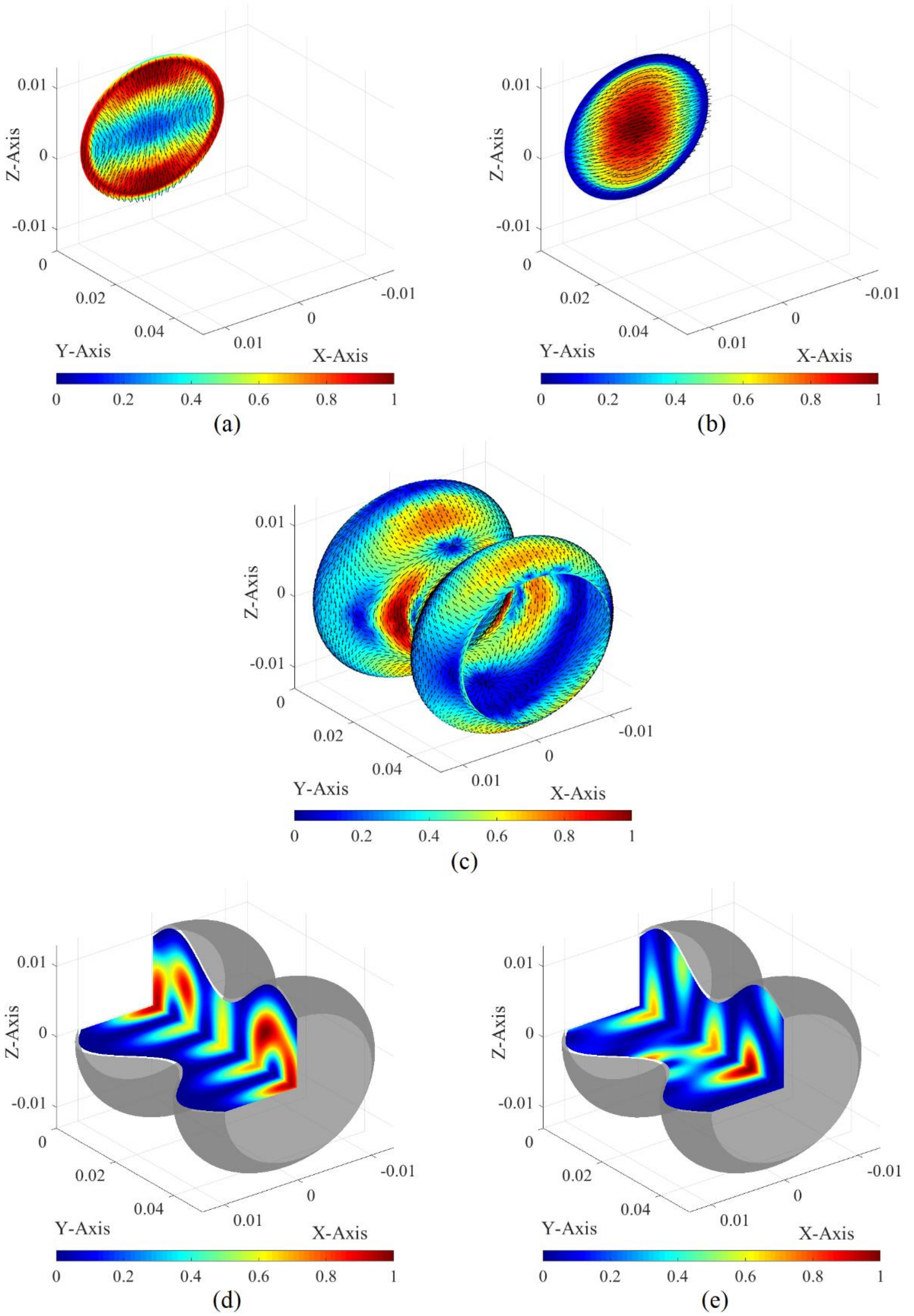

Figure 3-51 Modal quantity distributions of the IP-DM 4 (working at 7.45GHz). (a) Equivalent electric current on $\mathbb{S}_{in}$; (b) equivalent magnetic current on $\mathbb{S}_{in}$; (c) equivalent electric current on $\mathbb{S}_{ele}$; (d) electric energy density in guide tube; (e) magnetic energy density in guide tube

For the IP-DM 5 (at 7.45GHz), its modal quantities are shown in Fig. 3-52.

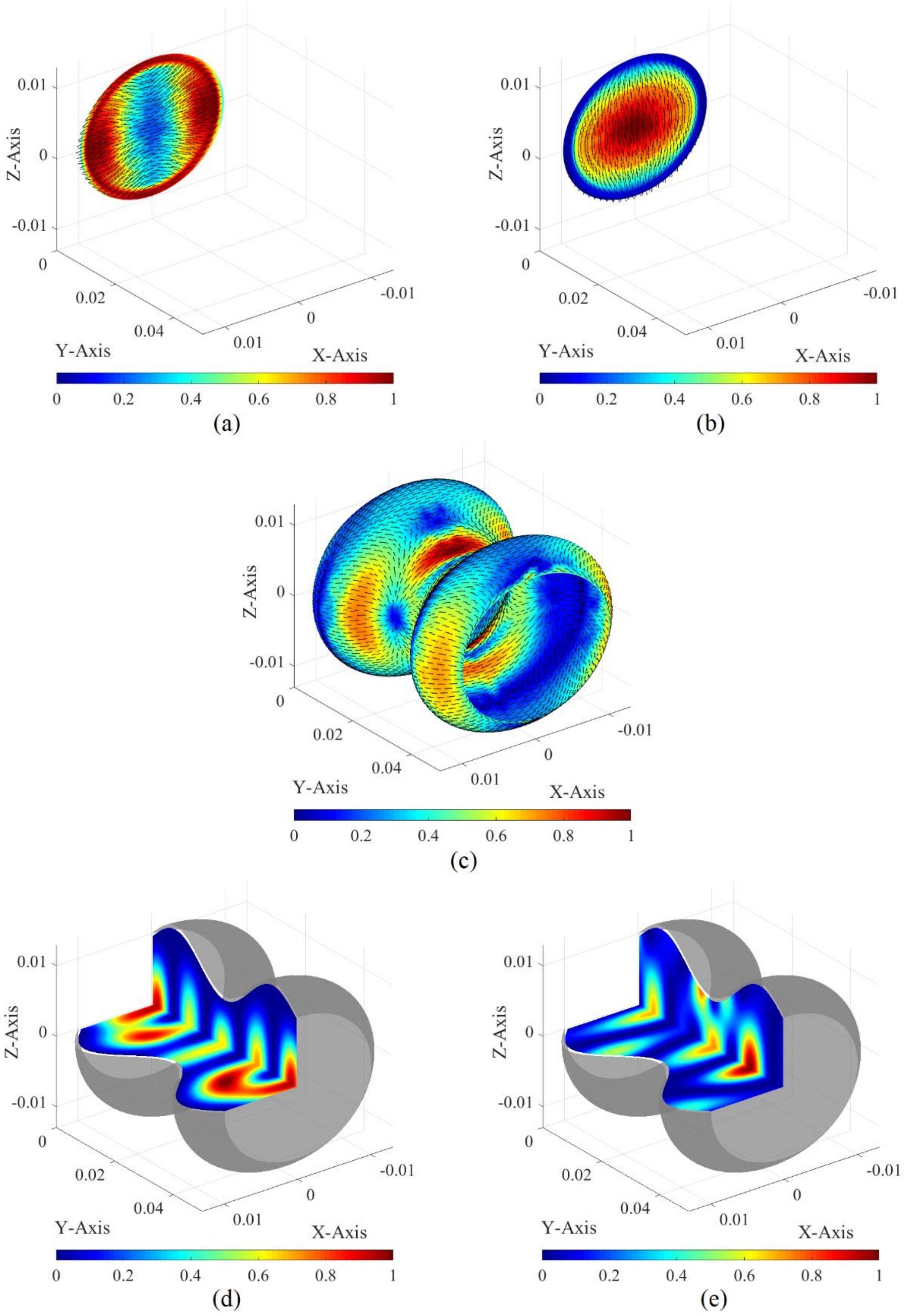

Figure 3-52 Modal quantity distributions of the IP-DM 5 (working at 7.45GHz). (a) Equivalent electric current on $\mathbb{S}_{\text{in}}$; (b) equivalent magnetic current on $\mathbb{S}_{\text{in}}$; (c) equivalent electric current on $\mathbb{S}_{\text{ele}}$; (d) electric energy density in guide tube; (e) magnetic energy density in guide tube

For the IP-DM 6 (at 8.95GHz), its modal quantities are shown in Fig. 3-53.

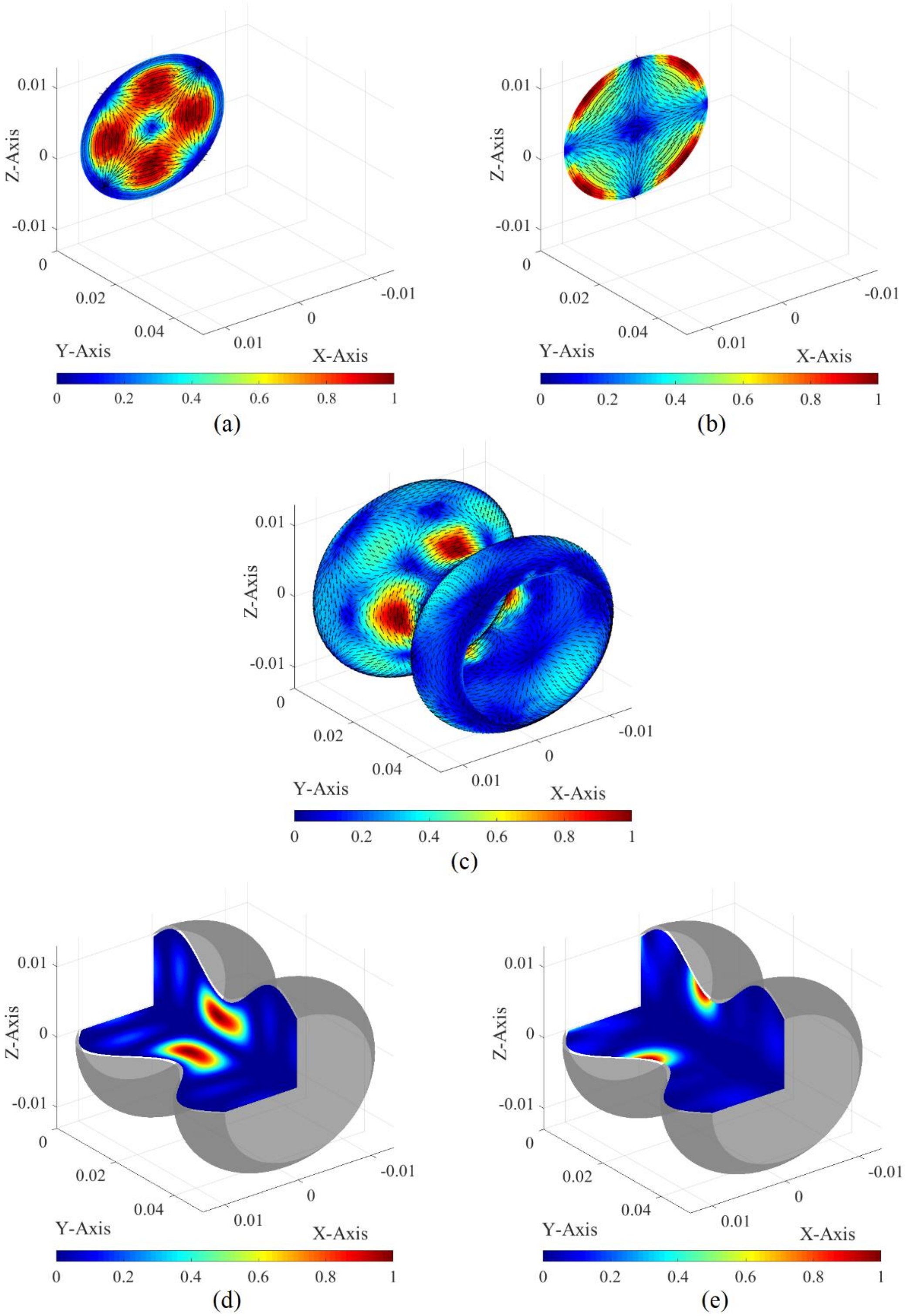

Figure 3-53 Modal quantity distributions of the IP-DM 6 (working at 8.95GHz). (a) Equivalent electric current on $\mathbb{S}_{in}$; (b) equivalent magnetic current on $\mathbb{S}_{in}$; (c) equivalent electric current on $\mathbb{S}_{ele}$; (d) electric energy density in guide tube; (e) magnetic energy density in guide tube

For the IP-DM 7 (at 8.95GHz), its modal quantities are shown in Fig. 3-54.

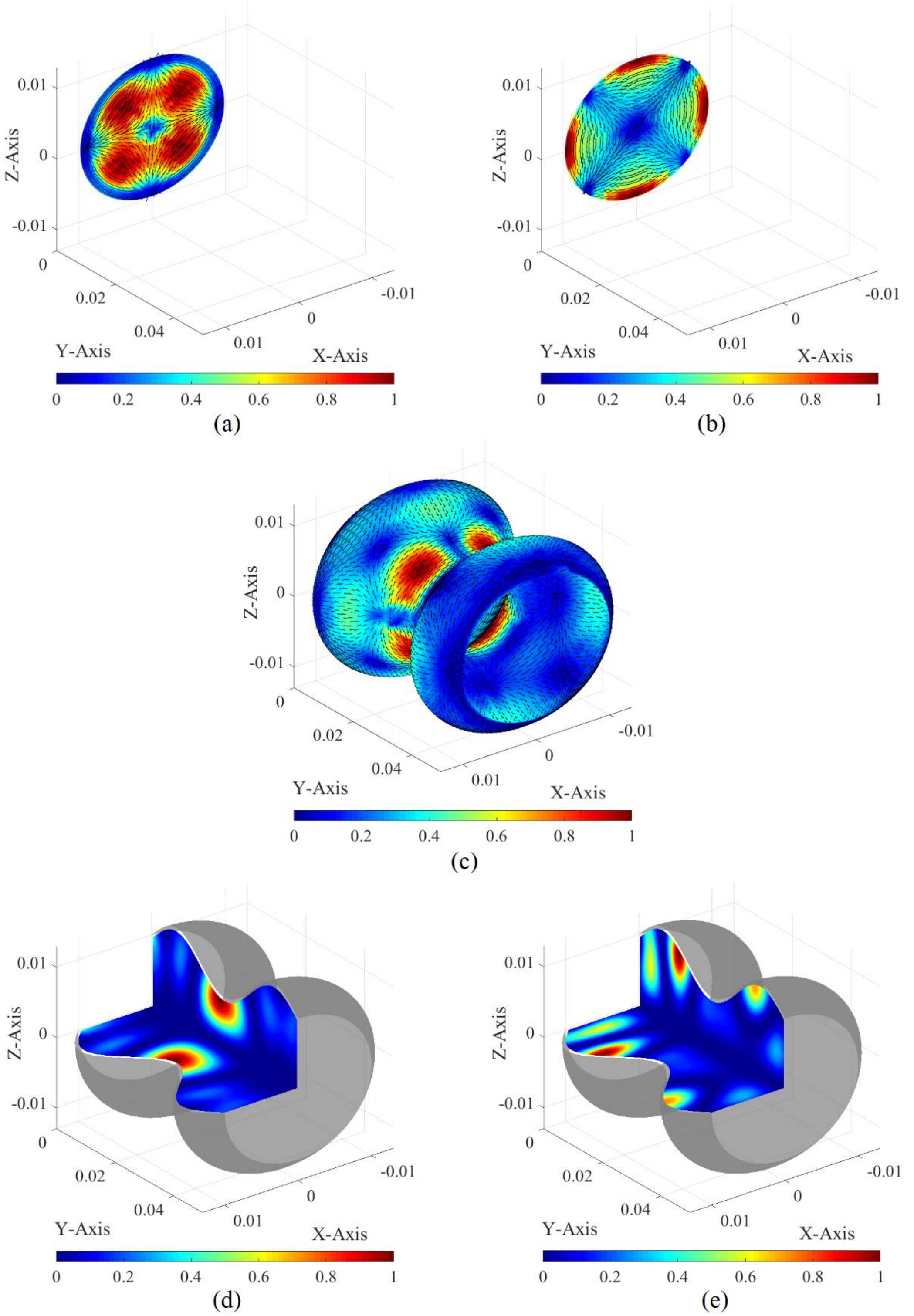

Figure 3-54 Modal quantity distributions of the IP-DM 7 (working at 8.95GHz). (a) Equivalent electric current on $\mathbb{S}_{\text{in}}$ ; (b) equivalent magnetic current on $\mathbb{S}_{\text{in}}$ ; (c) equivalent electric current on $\mathbb{S}_{\text{ele}}$ ; (d) electric energy density in guide tube; (e) magnetic energy density in guide tube

## 3.6 IP-DMs of Free Space

Now we consider some currents $\{\vec{J},\vec{M}\}$ distributing in *source region*, and the source region is surrounded by free space, as shown in Fig. 3-55. The currents $\{\vec{J},\vec{M}\}$ generate some fields $\{\vec{E},\vec{H}\}$ in whole *three-dimensional Euclidean space* $\mathbb{E}^3$.

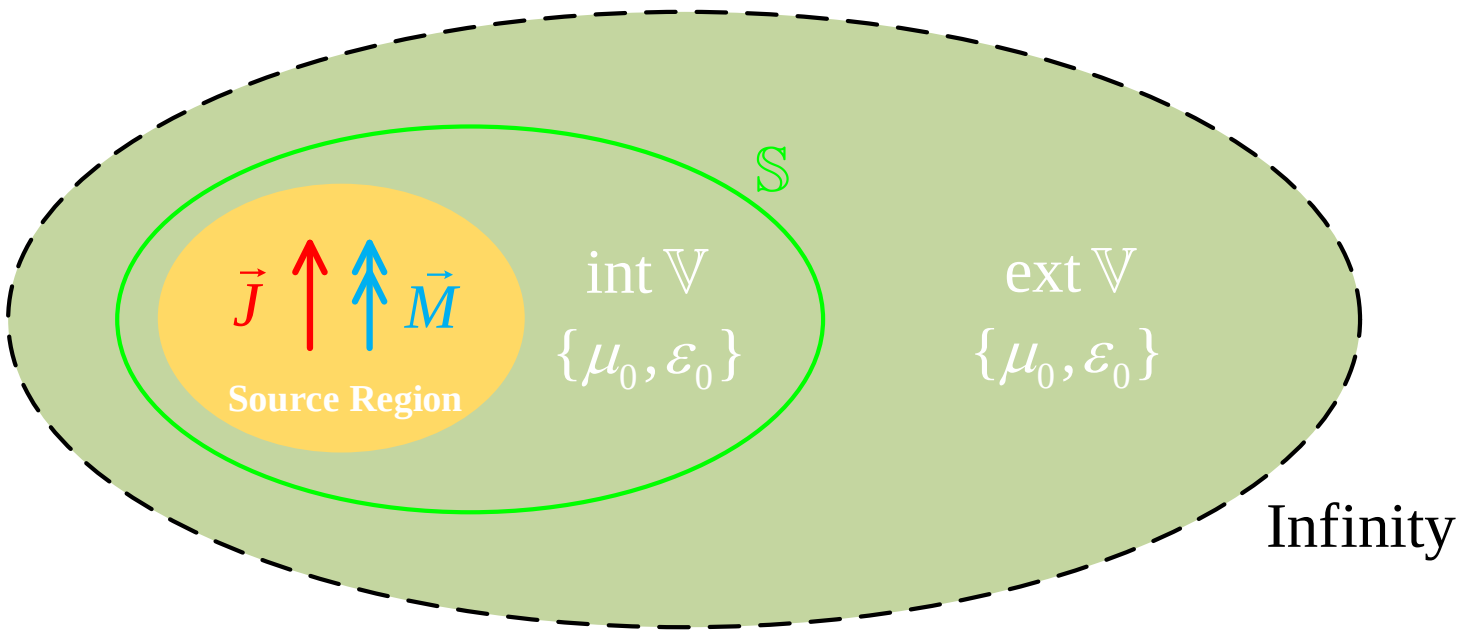

Figure 3-55 Surface $\mathbb{S}$ divides whole $\mathbb{E}^3$ into two parts $\text{int}\,\mathbb{V}$ and $\text{ext}\,\mathbb{V}$. Whole source region belongs to $\text{int}\,\mathbb{V}$. $\text{ext}\,\mathbb{V}$ and $\text{int}\,\mathbb{V}\setminus\{\text{source region}\}$ are with free-space material parameters $\{\mu_0,\varepsilon_0\}$

If $\mathbb{S}$ is a closed surface enclosing the whole source region, and the region enclosed by $\mathbb{S}$ is denoted as $\mathbb{V}$ (where the source region belongs to $\mathbb{V}$, but it is not necessarily identical to $\mathbb{V}$), then there exists the following *Huygens-Fresnel principle*:

$$\vec{G}_0^{JF} * \left(\hat{n}\times\vec{H}\right)+\vec{G}_0^{MF} * \left(\vec{E}\times\hat{n}\right) = \begin{cases} \vec{F}\left(\vec{r}\right) & , \quad \vec{r}\in \text{ext}\,\mathbb{V} \\ 0 & , \quad \vec{r}\in \text{int}\,\mathbb{V} \end{cases} \tag{3-151}$$

In Eq. (3-151), $\text{int}\,\mathbb{V}$ and $\text{ext}\,\mathbb{V}$ are the interior and exterior of $\mathbb{V}$ respectively, and $\text{int}\,\mathbb{V}\bigcup\mathbb{S}\bigcup\text{ext}\,\mathbb{V}=\mathbb{E}^3$; $\hat{n}$ is the normal direction of $\mathbb{S}$, and it points to $\text{ext}\,\mathbb{V}$.

The above Huygens-Fresnel principle implies that: in free space, the fields $\{\vec{E},\vec{H}\}$ propagate from the source region to infinity, and there doesn't exist any reflection. This feature is very similar to the one satisfied by the travelling-wave modes working in the various guiding structures discussed in the previous Secs. 3.2~3.4. Thus, **this report treats the free space as a guiding structure used to guide EM energy from the source region to infinity**.

This section focuses on constructing the IP-DMs of the free space, and has a similar organization as the previous sections.

### 3.6.1 Topological Structure and Source-Field Relationships

The region occupied by free space is denoted as $\mathbb{F}$. The boundary of $\mathbb{F}$ is constituted by two closed surfaces $\mathbb{S}$ and $\mathbb{S}_\infty$, and $\mathbb{S}$ encloses whole source region, and $\mathbb{S}_\infty$ is a spherical surface with infinite radius. The penetrable part of $\mathbb{S}$ is just the input port of $\mathbb{F}$, and it is denoted as $\mathbb{S}_{\text{in}}$ correspondingly; the impenetrable part of $\mathbb{S}$ is usually electric wall, and it is denoted as $\mathbb{S}_{\text{ele}}$ correspondingly; surface $\mathbb{S}_{\infty}$ is just the output port of $\mathbb{F}$, and the fields automatically satisfy Sommerfeld's radiation condition on the surface.

The equivalent surface currents distributing on $\mathbb{S}_{\text{in}}$ are denoted as $\{\vec{J}_{\text{in}}^{\text{ES}}, \vec{M}_{\text{in}}^{\text{ES}}\}$, and they are defined in terms of the total modal fields $\{\vec{E}, \vec{H}\}$ as follows:

$$\vec{J}_{\text{in}}^{\text{ES}}(\vec{r}) = \hat{n} \times \vec{H}(\vec{r}) \quad , \quad \vec{r} \in \mathbb{S}_{\text{in}} \tag{3-152a}$$

$$\vec{M}_{\text{in}}^{\text{ES}}(\vec{r}) = \vec{E}(\vec{r}) \times \hat{n} \quad , \quad \vec{r} \in \mathbb{S}_{\text{in}} \tag{3-152b}$$

where $\hat{n}$ is the normal direction of $\mathbb{S}_{\text{in}}$ and points to the interior of $\mathbb{F}$. The equivalent surface magnetic current distributing on $\mathbb{S}_{\text{ele}}$ is zero due to the homogeneous tangential electric field boundary condition, and the equivalent surface electric current distributing on $\mathbb{S}_{\text{ele}}$ is denoted as $\vec{J}^{\text{IS}}$. The equivalent surface currents distributing on $\mathbb{S}_{\infty}$ will not contribute to the fields in $\mathbb{F}$ due to Sommerfeld's radiation condition, so there is no need to introduce them.

Using the above currents, the field $\vec{F}$ in regions $\mathbb{F}$ can be expressed as the following surface equivalence principle

$$\vec{F}(\vec{r}) = \mathcal{F}_0\left(\vec{J}_{\text{in}}^{\text{ES}} + \vec{J}^{\text{IS}}, \vec{M}_{\text{in}}^{\text{ES}}\right) \quad , \quad \vec{r} \in \mathbb{F} \tag{3-153}$$

in which $\vec{F} = \vec{E} / \vec{H}$ and correspondingly $\mathcal{F} = \mathcal{E}_0 / \mathcal{H}_0$, and operators $\mathcal{E}_0$ and $\mathcal{H}_0$ are the same as the ones used in the previous sections.

### 3.6.2 Mathematical Description for Modal Space

Inserting surface equivalence principle (3-153) into current definitions (3-152a) and (3-152b), we derive the following integral equations

$$\vec{J}_{\text{in}}^{\text{ES}}(\vec{r}) \times \hat{n} = \left[\mathcal{H}_0\left(\vec{J}_{\text{in}}^{\text{ES}} + \vec{J}^{\text{IS}}, \vec{M}_{\text{in}}^{\text{ES}}\right)\right]_{\vec{r}' \to \vec{r}}^{\tan} \quad , \quad \vec{r} \in \mathbb{S}_{\text{in}} \tag{3-154a}$$

$$\hat{n} \times \vec{M}_{\text{in}}^{\text{ES}}(\vec{r}) = \left[\mathcal{E}_0\left(\vec{J}_{\text{in}}^{\text{ES}} + \vec{J}^{\text{IS}}, \vec{M}_{\text{in}}^{\text{ES}}\right)\right]_{\vec{r}' \to \vec{r}}^{\tan} \quad , \quad \vec{r} \in \mathbb{S}_{\text{in}} \tag{3-154b}$$

satisfied by the currents, where $\vec{r}' \in \text{int}\,\mathbb{F}$ and $\vec{r}'$ approaches the point $\vec{r}$ on $\mathbb{S}_{\text{in}}$.

Based on the homogeneous tangential electric field boundary condition on $\mathbb{S}_{\text{ele}}$ and surface equivalence principle (3-153), we derive the following electric field integral equation

$$0 \ = \ \left[\mathcal{E}_0\left(\vec{J}_{\text{in}}^{\text{ES}} + \vec{J}^{\text{IS}}, \vec{M}_{\text{in}}^{\text{ES}}\right)\right]^{\tan} \quad , \quad \vec{r} \in \mathbb{S}_{\text{ele}} \tag{3-155}$$

satisfied by the currents.

If the currents involved in Eqs. (3-154a)~(3-155) are expanded in terms of some proper basis functions, and the equations are tested with $\{\vec{b}_\xi^{\vec{M}_{\text{in}}^{\text{ES}}}\}$, $\{\vec{b}_\xi^{\vec{J}_{\text{in}}^{\text{ES}}}\}$, and $\{\vec{b}_\xi^{\vec{J}^{\text{IS}}}\}$ respectively, then the integral equations are immediately discretized into the following matrix equations

$$\overline{\overline{Z}}^{\vec{M}_{\text{in}}^{\text{ES}}\vec{J}_{\text{in}}^{\text{ES}}} \cdot \overline{a}^{\vec{J}_{\text{in}}^{\text{ES}}} + \overline{\overline{Z}}^{\vec{M}_{\text{in}}^{\text{ES}}\vec{J}^{\text{IS}}} \cdot \overline{a}^{\vec{J}^{\text{IS}}} + \overline{\overline{Z}}^{\vec{M}_{\text{in}}^{\text{ES}}\vec{M}_{\text{in}}^{\text{ES}}} \cdot \overline{a}^{\vec{M}_{\text{in}}^{\text{ES}}} \ = \ 0 \tag{3-156a}$$

$$\overline{\overline{Z}}^{\vec{J}_{\text{in}}^{\text{ES}}\vec{J}_{\text{in}}^{\text{ES}}} \cdot \overline{a}^{\vec{J}_{\text{in}}^{\text{ES}}} + \overline{\overline{Z}}^{\vec{J}_{\text{in}}^{\text{ES}}\vec{J}^{\text{IS}}} \cdot \overline{a}^{\vec{J}^{\text{IS}}} + \overline{\overline{Z}}^{\vec{J}_{\text{in}}^{\text{ES}}\vec{M}_{\text{in}}^{\text{ES}}} \cdot \overline{a}^{\vec{M}_{\text{in}}^{\text{ES}}} \ = \ 0 \tag{3-156b}$$

and

$$\overline{\overline{Z}}^{\vec{J}^{\text{IS}}\vec{J}_{\text{in}}^{\text{ES}}} \cdot \overline{a}^{\vec{J}_{\text{in}}^{\text{ES}}} + \overline{\overline{Z}}^{\vec{J}^{\text{IS}}\vec{J}^{\text{IS}}} \cdot \overline{a}^{\vec{J}^{\text{IS}}} + \overline{\overline{Z}}^{\vec{J}^{\text{IS}}\vec{M}_{\text{in}}^{\text{ES}}} \cdot \overline{a}^{\vec{M}_{\text{in}}^{\text{ES}}} \ = \ 0 \tag{3-157}$$

The formulations used to calculate the elements of the matrices in the above matrix equations are explicitly given in the App. D3 of this report.

Employing the above Eqs. (3-156a)~(3-157), we can obtain the following transformation

$$\begin{bmatrix} \overline{a}^{\vec{J}_{\text{in}}^{\text{ES}}} \\ \overline{a}^{\vec{J}^{\text{IS}}} \\ \overline{a}^{\vec{M}_{\text{in}}^{\text{ES}}} \end{bmatrix} \ = \ \overline{a}^{\text{AV}} \ = \ \overline{\overline{T}} \cdot \overline{a} \tag{3-158}$$

where the calculation formulation for transformation matrix $\overline{\overline{T}}$ is given in the App. D3 of this report.

## 3.6.3 Input Power Operator

The IPO corresponding to the input port $\mathbb{S}_{\text{in}}$ is as follows:

$$\begin{aligned} P^{\text{in}} \ &= \ (1/2)\iint_{\mathbb{S}_{\text{in}}} \left(\vec{E} \times \vec{H}^\dagger\right) \cdot \hat{z}dS \\ &= \ (1/2)\left\langle \hat{z} \times \vec{J}_{\text{in}}^{\text{ES}}, \vec{M}_{\text{in}}^{\text{ES}} \right\rangle_{\mathbb{S}_{\text{in}}} \\ &= -(1/2)\left\langle \vec{J}_{\text{in}}^{\text{ES}}, \mathcal{E}_0\left(\vec{J}_{\text{in}}^{\text{ES}} + \vec{J}^{\text{IS}}, \vec{M}_{\text{in}}^{\text{ES}}\right)\right\rangle_{\mathbb{S}_{\text{in}}^-} \\ &= -(1/2)\left\langle \vec{M}_{\text{in}}^{\text{ES}}, \mathcal{H}_0\left(\vec{J}_{\text{in}}^{\text{ES}} + \vec{J}^{\text{IS}}, \vec{M}_{\text{in}}^{\text{ES}}\right)\right\rangle_{\mathbb{S}_{\text{in}}^-}^\dagger \end{aligned} \tag{3-159}$$

where the integral surface $\mathbb{S}_{\text{in}}^-$ is the side of $\mathbb{S}_{\text{in}}$ locating in $\text{int}\,\mathbb{F}$.

By discretizing IPO (3-159) and utilizing transformation (3-158), we derive the following matrix form of the IPO

$$P^{\text{in}} = \bar{a}^{\dagger} \cdot \bar{\bar{P}}^{\text{in}} \cdot \bar{a} \tag{3-160}$$

and the formulation for calculating the quadratic matrix $\bar{\bar{P}}^{\text{in}}$ is given in the App. D3 of this report.

### 3.6.4 Input-Power-Decoupled Modes

The IP-DMs in modal space can be derived from solving the modal decoupling equation $\bar{\bar{P}}_{-}^{\text{in}} \cdot \bar{\alpha}_{\xi} = \theta_{\xi} \, \bar{\bar{P}}_{+}^{\text{in}} \cdot \bar{\alpha}_{\xi}$ defined on modal space, where $\bar{\bar{P}}_{+}^{\text{in}}$ and $\bar{\bar{P}}_{-}^{\text{in}}$ are the positive and negative Hermitian parts of the matrix $\bar{\bar{P}}^{\text{in}}$ given in Eq. (3-160). If some derived modes $\{\bar{\alpha}_1, \bar{\alpha}_2, \cdots, \bar{\alpha}_d\}$ are $d$-order degenerate, then the Gram-Schmidt orthogonalization process given in previous Sec. 3.2.4.1 is necessary, and it is not repeated here.

The IP-DMs constructed above satisfy the following decoupling relation

$$\left(1 + j\,\theta_{\xi}\right)\delta_{\xi\zeta} = \left(1/2\right)\iint_{\mathbb{S}_{\text{in}}}\left(\vec{E}_{\zeta} \times \vec{H}_{\xi}^{\dagger}\right) \cdot \hat{z}\,dS = \left(1/2\right)\left\langle \hat{z} \times \vec{J}_{\text{in};\xi}^{\text{ES}}, \vec{M}_{\text{in};\zeta}^{\text{ES}} \right\rangle_{\mathbb{S}_{\text{in}}} \tag{3-161}$$

and the relation implies that the IP-DMs don't have net energy coupling in integral period. By employing the decoupling relation, we have the following Parseval's identity

$$\sum_{\xi}\left|c_{\xi}\right|^2 = \left(1/T\right)\int_{t_0}^{t_0+T}\left[\iint_{\mathbb{S}_{\text{in}}}\left(\vec{\mathcal{E}} \times \vec{\mathcal{H}}\right) \cdot \hat{z}\,dS\right]dt \tag{3-162}$$

in which $\{\vec{\mathcal{E}}, \vec{\mathcal{H}}\}$ are the time-domain fields, and the expansion coefficients have expression $c_{\xi} = -(1/2) < \vec{J}_{\text{in};\xi}^{\text{ES}}, \vec{E} >_{\mathbb{S}_{\text{in}}} \big/ (1 + j\theta_{\xi}) = -(1/2) < \vec{H}, \vec{M}_{\text{in};\xi}^{\text{ES}} >_{\mathbb{S}_{\text{in}}} \big/ (1 + j\theta_{\xi})$ , and $\{\vec{E}, \vec{H}\}$ are some previously known fields distributing on input port $\mathbb{S}_{\text{in}}$ .

Similarly to the previous sections, the modal input impedance and admittance can be defined, and it will not be repeated here.

### 3.6.5 Numerical Examples Corresponding to Typical Structures

Here, we select the input port of free space as a spherical surface with radius 3cm.

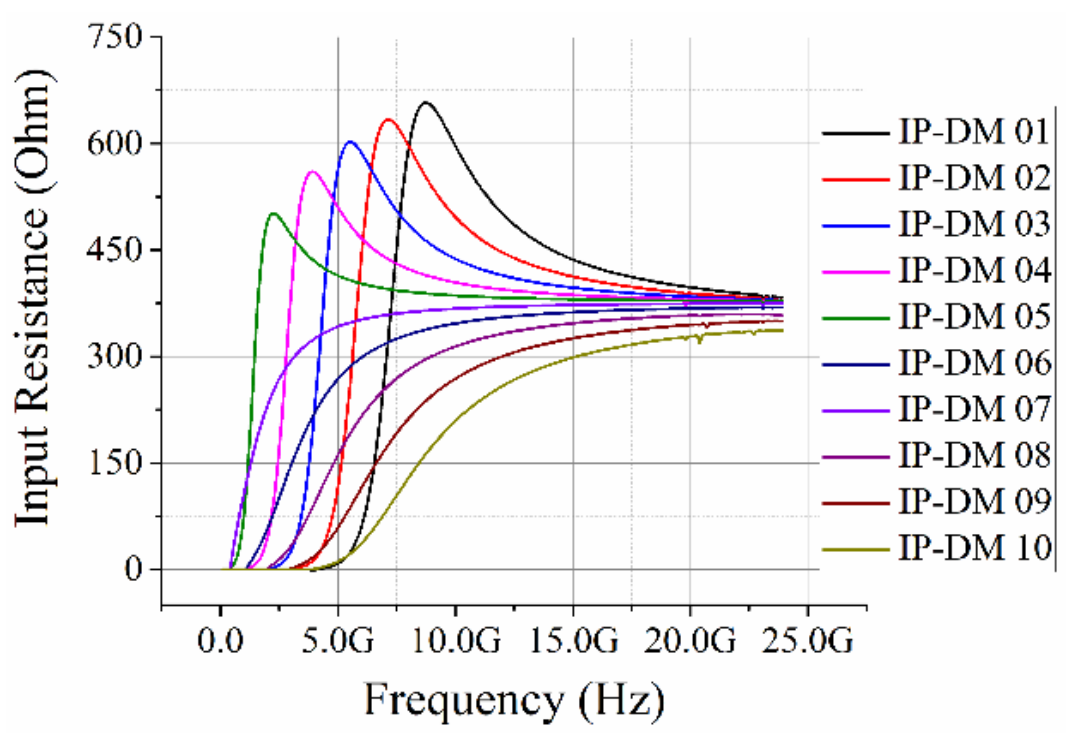

Figure 3-56 Modal input resistance curves of some typical IP-DMs

By orthogonalizing the JE-DoJ-based IPO (3-160), we obtain the IP-DMs of free space. The input resistance curves of some typical IP-DMs are shown in Fig. 3-56.

The equivalent electric currents and radiation patterns of 10 typical IP-DMs are shown in the following Fig. 3-57.

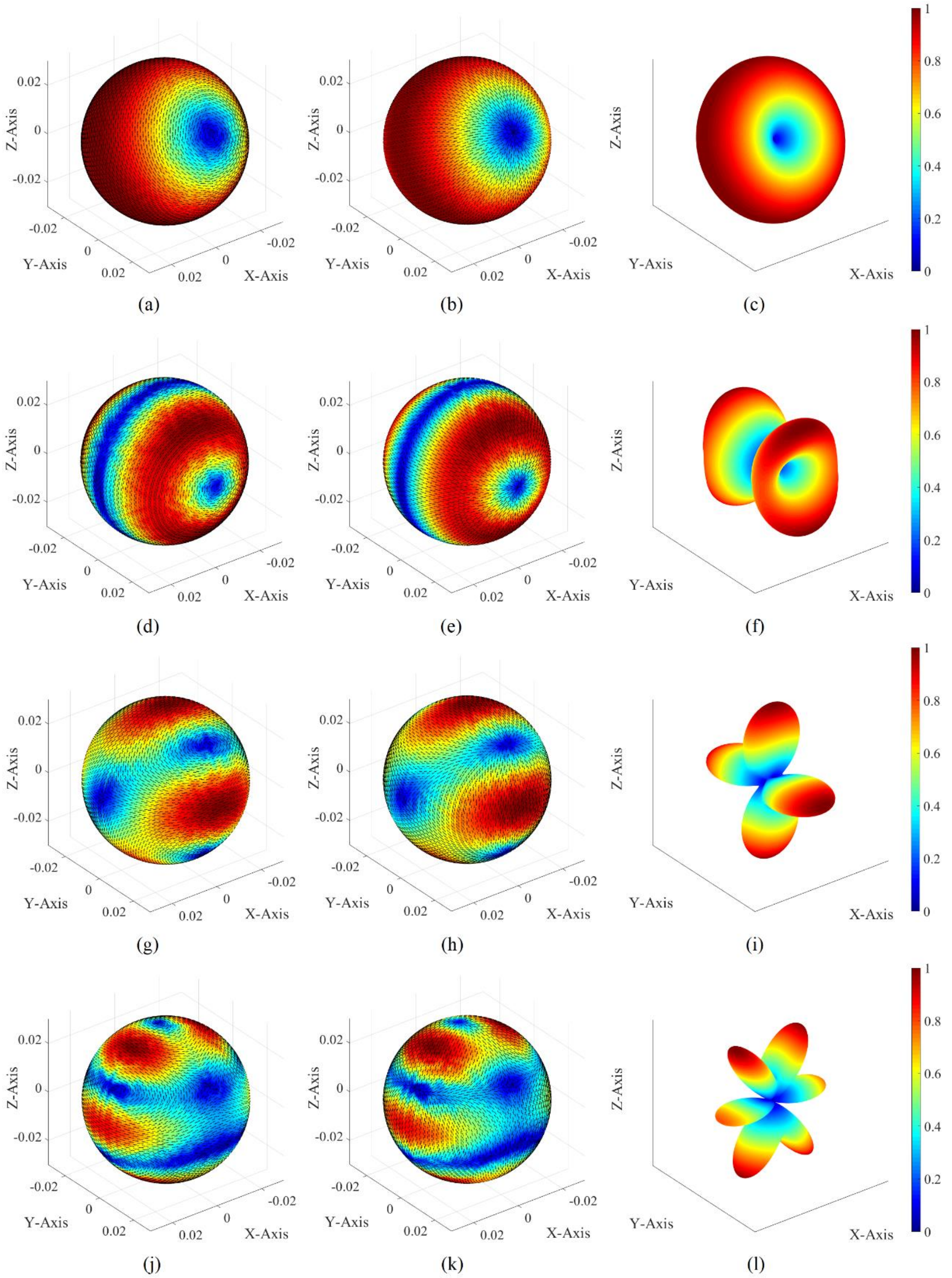

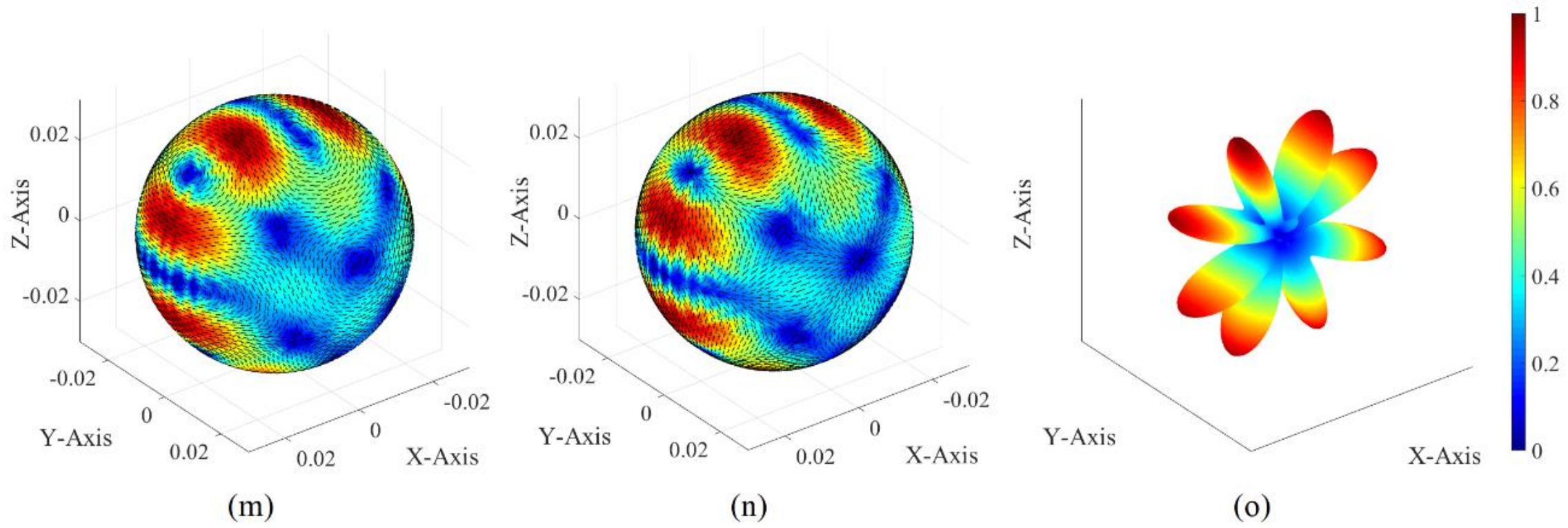

Figure 3-57 Modal equivalent electric currents and radiation patterns of 10 typical IP-DMs

## 3.7 Chapter Summary

This chapter theoretically discusses the *travelling-wave condition* and its some important corollaries first, and then establishes the *power transport theorem* (*PTT*) based *decoupling mode theory* (*DMT*) for guiding structures — *PTT-Guid-DMT* — by employing the condition and corollaries.

By orthogonalizing the frequency-domain *input power operator* (*IPO*) (i.e. solving the *modal decoupling equation*), Sec. 3.2 constructs the *input-power-decoupled modes* (*IP-DMs*) of metallic guiding structures, and the IP-DMs of some more complicated guiding structures, such as material and metal-material composite guiding structures etc., are similarly constructed in Secs. 3.3~3.5. In addition, the IP-DMs of free space are also similarly constructed in Sec. 3.6.

The IP-DMs satisfy a similar *energy decoupling relation* to the classical *eigen-modes* of guiding structures. By employing the energy decoupling relation, it is found out that the IP-DMs don't have net energy exchange in any integral period and the IP-DMs satisfy the famous *Parseval's identity*.

An effective method for recognizing the traveling-wave modes in whole IP-DM set is developed, and the formulation for calculating the cut-off frequencies of the traveling-wave-type IP-DMs is also derived.

# *Chapter 4 Input-Power-Decoupled Modes of Transmitting Antenna, Propagation Medium, and Receiving Antenna

**CHAPTER MOTIVATION:** This chapter focuses on constructing the *input-power-decoupled modes* of some sub-structures decomposed from transceiving system by orthogonalizing the *input power operator* in *power transport theorem* framework, and doing some necessary analysis and discussions for the related topics. This chapter is only a transitional chapter for introducing the future Chaps. 6~8.

## 4.1 Chapter Introduction

This chapter is devoted to constructing the *input-power-decoupled modes* (*IP-DMs*) of *transmitting antenna* (simply called *tra-antenna*), *propagation medium* (simply called *medium*), and *receiving antenna* (simply called *rec-antenna*) separately.

The separate treatment for tra-antenna, medium, and rec-antenna ignores the interactions among the structures, and then leads to an inevitable utilization of the *modal matching process* among the structures (for the details of the process, please see the subsequent Chap. 5). However, the modal matching process is very tedious as exhibited in Chap. 5.

To both consider the interactions and avoid the modal matching process, we will propose an alternative scheme for constructing the IP-DMs of the tra-antenna, medium, and rec-antenna with interactions in the future chapters, and the core idea of the scheme is to combine the structures into some systems (such as tra-antenna-medium combined system simply called *augmented tra-antenna* and medium-rec-antenna combined system simply called *augmented rec-antenna* etc., for details please see Secs. 2.3.4 and 2.3.5) and to directly construct the IP-DMs of the systems (for details please see Chaps. 6~8).

In fact, both this chapter and the following Chap. 5 are only the transitional chapters for introducing the future Chaps. 6~8, so we will not exhibit any numerical experiment in this chapter. Some typical numerical experiments related to augmented tra-antenna, augmented rec-antenna, and other combined systems will be exhibited in the future Chaps. 6~8.

As a typical example, the transmitting-receiving problem shown in the following Fig. 4-1 is considered in this chapter.

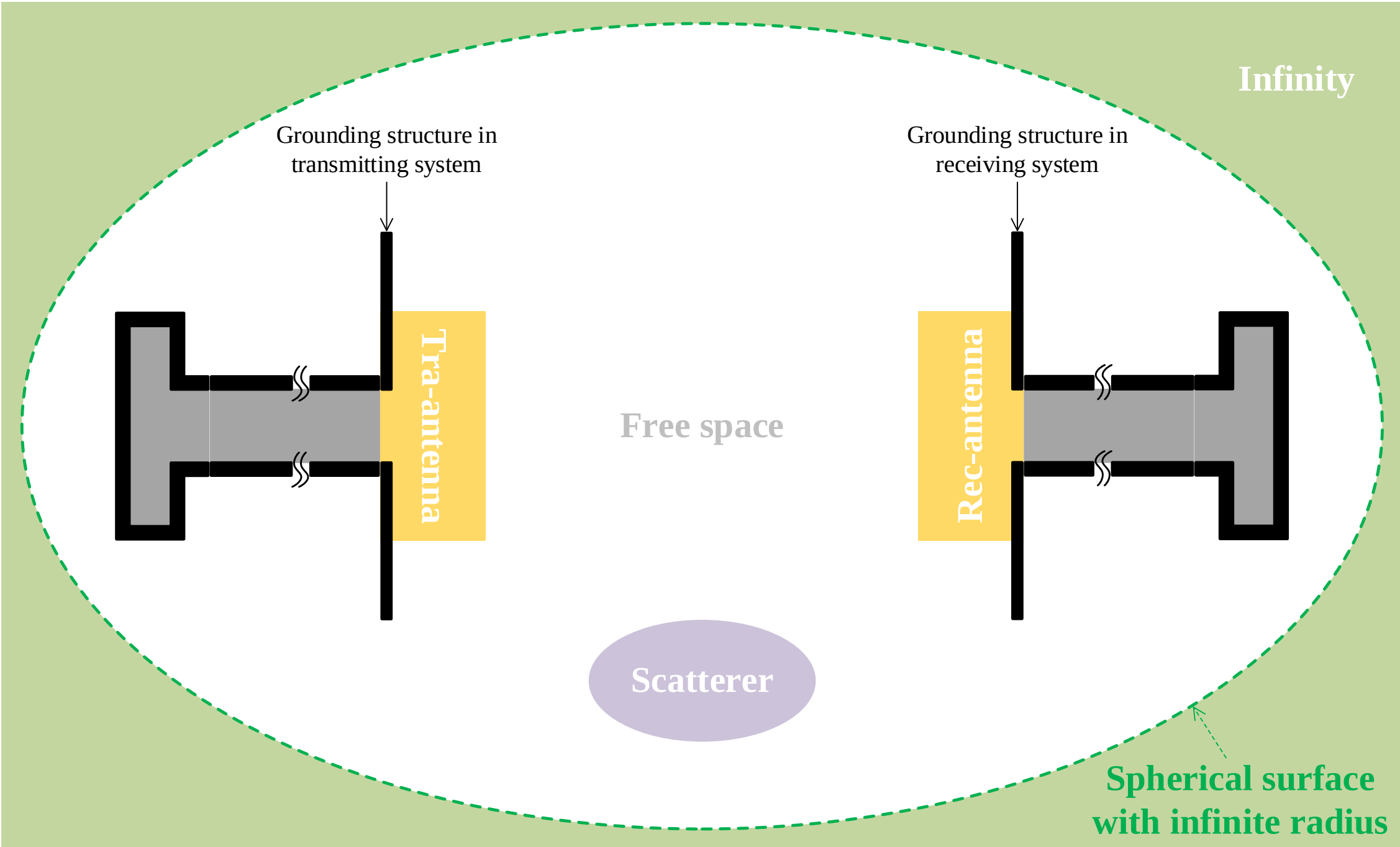

Figure 4-1 A typical transmitting-receiving problem considered in this chapter

Here, the tra-antenna is a *waveguide-fed DRA*, and the propagation medium is the *free space* with a *material scatterer*, and the rec-antenna is a *waveguide-loaded DRA*.

## 4.2 IP-DMs of Transmitting Antenna

Seperately from the other structures, the tra-antenna is shown as follows:

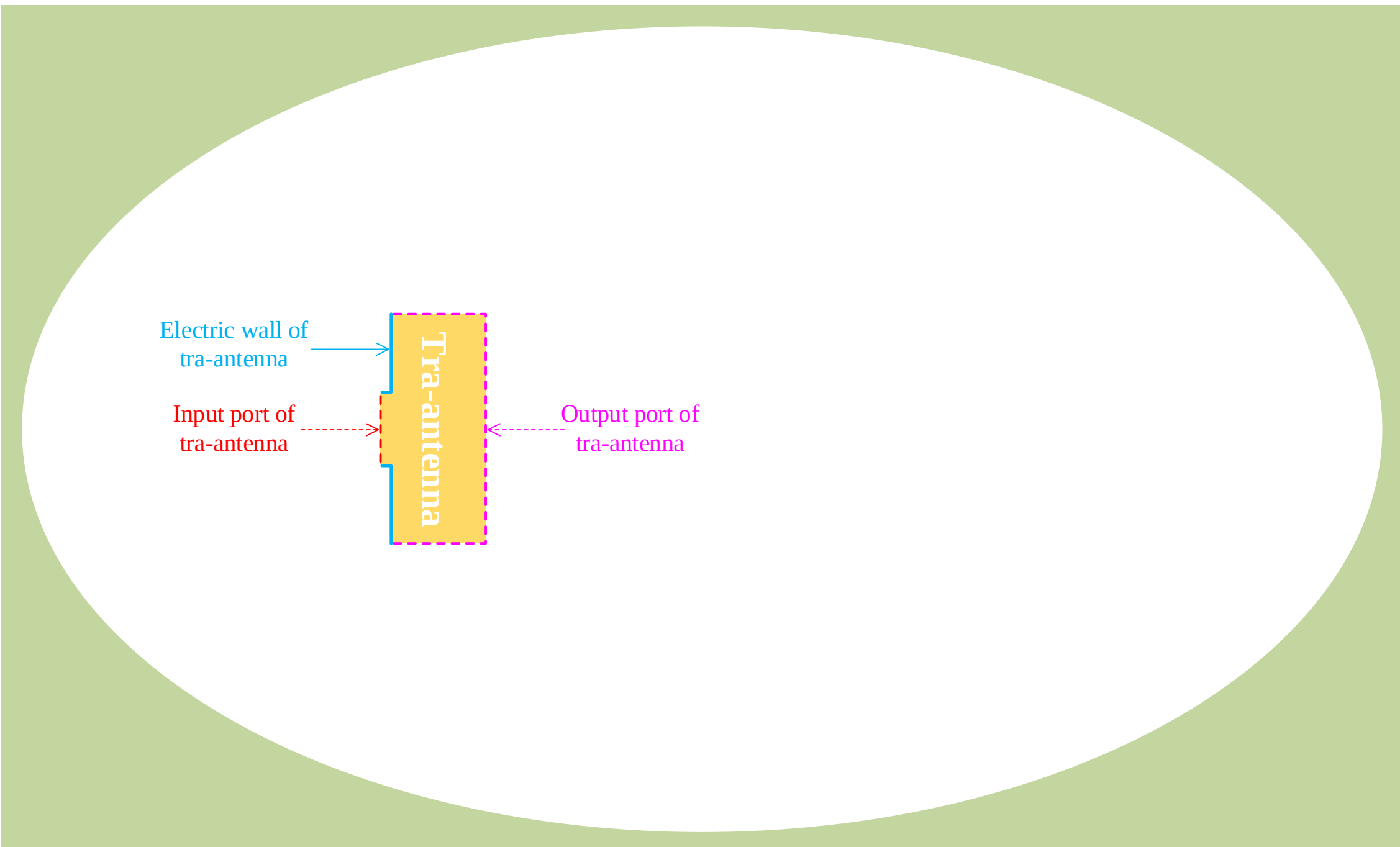

Figure 4-2 Geometry of the tra-antenna shown in Fig. 4-1

and this section focuses on discussing the tra-antenna.

The general process for constructing the IP-DMs of the tra-antenna is similar to the one used in the previous Chap. 3, and it is visually summarized in the following Fig. 4-3.

Mathematically Depicting Topological Structure

↓

Deriving Source- Field Relationships (SFRs)

↓

Mathematically Describing Modal Space by Employing the SFRs

↓

Deriving the Input Power Operator (IPO) Defined on the Modal Space

↓

Constructing Input- Power- Decoupled Modes (IP- DMs) by Orthogonalizing the IPO

Figure 4-3 A general process for constructing IP-DMs, which is the same as the one shown in Fig. 3-21

## 4.2.1 Topological Structure and Source-Field Relationships

The topological structure of the tra-antenna shown in Fig. 4-2 is detailedly exhibited in the following Fig. 4-4.

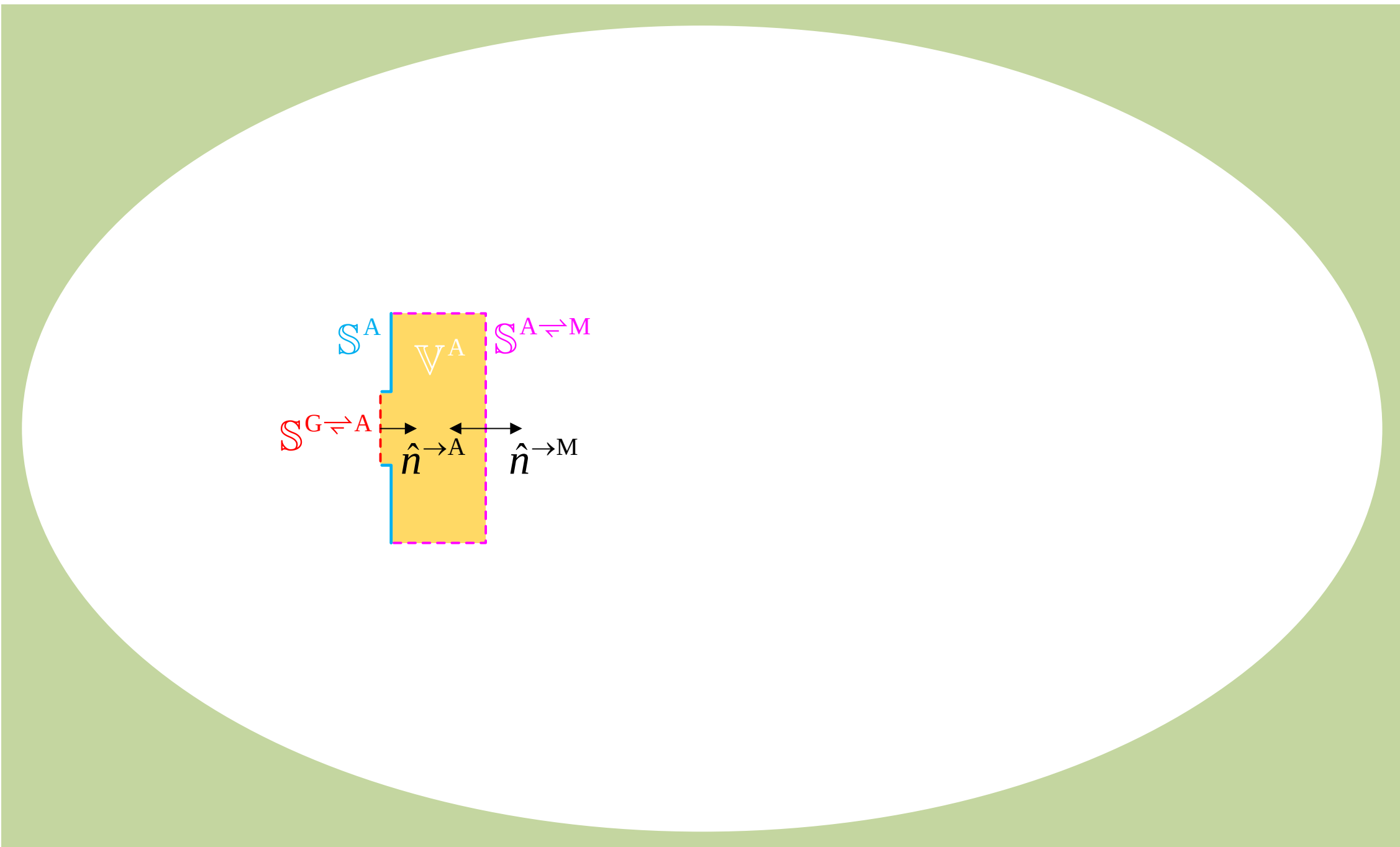

Figure 4-4 Detailed topological structure of the tra-antenna shown in Fig.4-2 and the normal directions of the input and output ports

In the above Fig. 4-4, surface $\mathbb{S}^{\text{G}\rightleftharpoons\text{A}}$ denotes the *input port* of the tra-antenna. The region occupied by the material body, i.e. the DRA, is denoted as $\mathbb{V}^{\text{A}}$. The interface between $\mathbb{V}^{\text{A}}$ and the *thick metallic ground plane* is denoted as $\mathbb{S}^{\text{A}}$; the interface between $\mathbb{V}^{\text{A}}$ and propagation medium is denoted as $\mathbb{S}^{\text{A}\rightleftharpoons\text{M}}$.

Clearly, $\mathbb{S}^{\text{G}\rightleftharpoons\text{A}}$, $\mathbb{S}^{\text{A}}$, and $\mathbb{S}^{\text{A}\rightleftharpoons\text{M}}$ constitute a closed surface, and the closed surface is just the boundary of $\mathbb{V}^{\text{A}}$, i.e., $\partial\mathbb{V}^{\text{A}} = \mathbb{S}^{\text{G}\rightleftharpoons\text{A}} \bigcup \mathbb{S}^{\text{A}} \bigcup \mathbb{S}^{\text{A}\rightleftharpoons\text{M}}$. In addition, the magnetic permeability, dielectric permeativity, and electric conductivity of $\mathbb{V}^{\text{A}}$ are denoted as $\ddot{\mu}$, $\ddot{\varepsilon}$, and $\ddot{\sigma}$ respectively.

If the equivalent surface currents distributing on $\mathbb{S}^{\text{G}\rightleftharpoons\text{A}}$ are denoted as $\{\vec{J}^{\text{G}\rightleftharpoons\text{A}},\vec{M}^{\text{G}\rightleftharpoons\text{A}}\}$, and the equivalent surface electric current distributing on $\mathbb{S}^{\text{A}}$ is denoted as $\vec{J}^{\text{A}}$ ①, and the equivalent surface currents distributing on $\mathbb{S}^{\text{A}\rightleftharpoons\text{M}}$ are denoted as $\{\vec{J}^{\text{A}\rightleftharpoons\text{M}},\vec{M}^{\text{A}\rightleftharpoons\text{M}}\}$, then the field distributing on $\mathbb{V}^{\text{A}}$ can be expressed as the following *surface equivalence principle*

$$\vec{F}(\vec{r}) = \mathcal{F}\left(\vec{J}^{\text{G}\rightleftharpoons\text{A}} + \vec{J}^{\text{A}\rightleftharpoons\text{M}} + \vec{J}^{\text{A}}, \vec{M}^{\text{G}\rightleftharpoons\text{A}} + \vec{M}^{\text{A}\rightleftharpoons\text{M}}\right) \quad , \quad \vec{r} \in \mathbb{V}^{\text{A}} \tag{4-1}$$

where $\vec{F} = \vec{E}/\vec{H}$, and correspondingly $\mathcal{F} = \mathcal{E}/\mathcal{H}$, and the operator is defined as that $\mathcal{F}(\vec{J},\vec{M}) = \ddot{G}^{JF} * \vec{J} + \ddot{G}^{MF} * \vec{M}$ (here, $\ddot{G}^{JF}$ and $\ddot{G}^{MF}$ are the *dyadic Green's functions* corresponding to the region $\mathbb{V}^{\text{A}}$ with material parameters $\{\ddot{\mu},\ddot{\varepsilon},\ddot{\sigma}\}$).

The currents $\{\vec{J}^{\text{G}\rightleftharpoons\text{A}},\vec{M}^{\text{G}\rightleftharpoons\text{A}}\}$ and fields $\{\vec{E},\vec{H}\}$ involved in Eq. (4-1) satisfy the following relations

$$\hat{n}^{\rightarrow\text{A}} \times \left[\vec{H}\left(\vec{r}^{\,\text{A}}\right)\right]_{\vec{r}^{\text{A}}\to\vec{r}} = \vec{J}^{\text{G}\rightleftharpoons\text{A}}(\vec{r}) \quad , \quad \vec{r} \in \mathbb{S}^{\text{G}\rightleftharpoons\text{A}} \tag{4-2a}$$

$$\left[\vec{E}\left(\vec{r}^{\,\text{A}}\right)\right]_{\vec{r}^{\text{A}}\to\vec{r}} \times \hat{n}^{\rightarrow\text{A}} = \vec{M}^{\text{G}\rightleftharpoons\text{A}}(\vec{r}) \quad , \quad \vec{r} \in \mathbb{S}^{\text{G}\rightleftharpoons\text{A}} \tag{4-2b}$$

and the currents $\{\vec{J}^{\text{A}\rightleftharpoons\text{M}},\vec{M}^{\text{A}\rightleftharpoons\text{M}}\}$ and fields $\{\vec{E},\vec{H}\}$ involved in Eq. (4-1) satisfy the following relations

$$\hat{n}^{\rightarrow\text{A}} \times \left[\vec{H}\left(\vec{r}^{\,\text{A}}\right)\right]_{\vec{r}^{\text{A}}\to\vec{r}} = \vec{J}^{\text{A}\rightleftharpoons\text{M}}(\vec{r}) \quad , \quad \vec{r} \in \mathbb{S}^{\text{A}\rightleftharpoons\text{M}} \tag{4-3a}$$

$$\left[\vec{E}\left(\vec{r}^{\,\text{A}}\right)\right]_{\vec{r}^{\text{A}}\to\vec{r}} \times \hat{n}^{\rightarrow\text{A}} = \vec{M}^{\text{A}\rightleftharpoons\text{M}}(\vec{r}) \quad , \quad \vec{r} \in \mathbb{S}^{\text{A}\rightleftharpoons\text{M}} \tag{4-3b}$$

In the above Eqs. (4-2a)&(4-2b) and (4-3a)&(4-3b), point $\vec{r}^{\,\text{A}}$ belongs to $\mathbb{V}^{\text{A}}$ and approaches the point $\vec{r}$ on $\mathbb{S}^{\text{G}\rightleftharpoons\text{A}} \bigcup \mathbb{S}^{\text{A}\rightleftharpoons\text{M}}$, and $\hat{n}^{\rightarrow\text{A}}$ is the normal direction of $\partial\mathbb{V}^{\text{A}}$ ($= \mathbb{S}^{\text{G}\rightleftharpoons\text{A}} \bigcup \mathbb{S}^{\text{A}} \bigcup \mathbb{S}^{\text{A}\rightleftharpoons\text{M}}$) and points to the interior of $\mathbb{V}^{\text{A}}$ as shown in Fig. 4-4.

① The equivalent surface electric current distributing on $\mathbb{S}^{\text{A}}$ is equal to the induced surface electric current distributing on $\mathbb{S}^{\text{A}}$ [13]. The equivalent surface magnetic current distributing on $\mathbb{S}^{\text{A}}$ is zero, because of the homogeneous tangential electric field boundary condition on $\mathbb{S}^{\text{A}}$ [13].

### 4.2.2 Mathematical Description for Modal Space

Combining the Eq. (4-1) with Eqs. (4-2a)&(4-2b), we can obtain the following *integral equations*

$$\left[\mathcal{H}\left(\vec{J}^{\mathrm{G}\rightleftharpoons\mathrm{A}}+\vec{J}^{\mathrm{A}\rightleftharpoons\mathrm{M}}+\vec{J}^{\mathrm{A}},\vec{M}^{\mathrm{G}\rightleftharpoons\mathrm{A}}+\vec{M}^{\mathrm{A}\rightleftharpoons\mathrm{M}}\right)\right]_{\vec{r}^{\mathrm{A}}\to\vec{r}}^{\tan} = \vec{J}^{\mathrm{G}\rightleftharpoons\mathrm{A}}\left(\vec{r}\right)\times\hat{n}^{\to\mathrm{A}} \quad , \quad \vec{r}\in\mathbb{S}^{\mathrm{G}\rightleftharpoons\mathrm{A}} \tag{4-4a}$$

$$\left[\mathcal{E}\left(\vec{J}^{\mathrm{G}\rightleftharpoons\mathrm{A}}+\vec{J}^{\mathrm{A}\rightleftharpoons\mathrm{M}}+\vec{J}^{\mathrm{A}},\vec{M}^{\mathrm{G}\rightleftharpoons\mathrm{A}}+\vec{M}^{\mathrm{A}\rightleftharpoons\mathrm{M}}\right)\right]_{\vec{r}^{\mathrm{A}}\to\vec{r}}^{\tan} = \hat{n}^{\to\mathrm{A}}\times\vec{M}^{\mathrm{G}\rightleftharpoons\mathrm{A}}\left(\vec{r}\right) \quad , \quad \vec{r}\in\mathbb{S}^{\mathrm{G}\rightleftharpoons\mathrm{A}} \tag{4-4b}$$

about currents $\{\vec{J}^{\mathrm{G}\rightleftharpoons\mathrm{A}},\vec{M}^{\mathrm{G}\rightleftharpoons\mathrm{A}}\}$, $\{\vec{J}^{\mathrm{A}\rightleftharpoons\mathrm{M}},\vec{M}^{\mathrm{A}\rightleftharpoons\mathrm{M}}\}$, and $\vec{J}^{\mathrm{A}}$, where the superscript "tan" represents the tangential component of the fields. Combining the Eq. (4-1) with Eqs. (4-3a)&(4-3b), we can obtain the following integral equations

$$\left[\mathcal{H}\left(\vec{J}^{\mathrm{G}\rightleftharpoons\mathrm{A}}+\vec{J}^{\mathrm{A}\rightleftharpoons\mathrm{M}}+\vec{J}^{\mathrm{A}},\vec{M}^{\mathrm{G}\rightleftharpoons\mathrm{A}}+\vec{M}^{\mathrm{A}\rightleftharpoons\mathrm{M}}\right)\right]_{\vec{r}^{\mathrm{A}}\to\vec{r}}^{\tan} = \vec{J}^{\mathrm{A}\rightleftharpoons\mathrm{M}}\left(\vec{r}\right)\times\hat{n}^{\to\mathrm{A}} \quad , \quad \vec{r}\in\mathbb{S}^{\mathrm{A}\rightleftharpoons\mathrm{M}} \tag{4-5a}$$

$$\left[\mathcal{E}\left(\vec{J}^{\mathrm{G}\rightleftharpoons\mathrm{A}}+\vec{J}^{\mathrm{A}\rightleftharpoons\mathrm{M}}+\vec{J}^{\mathrm{A}},\vec{M}^{\mathrm{G}\rightleftharpoons\mathrm{A}}+\vec{M}^{\mathrm{A}\rightleftharpoons\mathrm{M}}\right)\right]_{\vec{r}^{\mathrm{A}}\to\vec{r}}^{\tan} = \hat{n}^{\to\mathrm{A}}\times\vec{M}^{\mathrm{A}\rightleftharpoons\mathrm{M}}\left(\vec{r}\right) \quad , \quad \vec{r}\in\mathbb{S}^{\mathrm{A}\rightleftharpoons\mathrm{M}} \tag{4-5b}$$

about currents $\{\vec{J}^{\mathrm{G}\rightleftharpoons\mathrm{A}},\vec{M}^{\mathrm{G}\rightleftharpoons\mathrm{A}}\}$, $\{\vec{J}^{\mathrm{A}\rightleftharpoons\mathrm{M}},\vec{M}^{\mathrm{A}\rightleftharpoons\mathrm{M}}\}$, and $\vec{J}^{\mathrm{A}}$.

Based on Eq. (4-1) and the homogeneous tangential electric field boundary condition on $\mathbb{S}^{\mathrm{A}}$, we have the following *electric field integral equation*

$$\left[\mathcal{E}\left(\vec{J}^{\mathrm{G}\rightleftharpoons\mathrm{A}}+\vec{J}^{\mathrm{A}\rightleftharpoons\mathrm{M}}+\vec{J}^{\mathrm{A}},\vec{M}^{\mathrm{G}\rightleftharpoons\mathrm{A}}+\vec{M}^{\mathrm{A}\rightleftharpoons\mathrm{M}}\right)\right]_{\vec{r}^{\mathrm{A}}\to\vec{r}}^{\tan} = 0 \quad , \quad \vec{r}\in\mathbb{S}^{\mathrm{A}} \tag{4-6}$$

about currents $\{\vec{J}^{\mathrm{G}\rightleftharpoons\mathrm{A}},\vec{M}^{\mathrm{G}\rightleftharpoons\mathrm{A}}\}$, $\{\vec{J}^{\mathrm{A}\rightleftharpoons\mathrm{M}},\vec{M}^{\mathrm{A}\rightleftharpoons\mathrm{M}}\}$, and $\vec{J}^{\mathrm{A}}$.

The above Eqs. (4-4a)~(4-6) are a complete mathematical description for the *modal space* of the tra-antenna shown in Figs. 4-2 and 4-4. If the currents $\{\vec{J}^{\mathrm{G}\rightleftharpoons\mathrm{A}},\vec{M}^{\mathrm{G}\rightleftharpoons\mathrm{A}}\}$, $\{\vec{J}^{\mathrm{A}\rightleftharpoons\mathrm{M}},\vec{M}^{\mathrm{A}\rightleftharpoons\mathrm{M}}\}$, and $\vec{J}^{\mathrm{A}}$ are expanded in terms of some *basis functions* as follows:

$$\vec{C}^{x}\left(\vec{r}\right) = \sum\nolimits_{\xi} a_{\xi}^{\vec{C}^{x}}\vec{b}_{\xi}^{\vec{C}^{x}} = \underbrace{\begin{bmatrix}\vec{b}_{1}^{\vec{C}^{x}} & \vec{b}_{2}^{\vec{C}^{x}} & \cdots\end{bmatrix}}_{\vec{\mathbf{B}}^{\vec{C}^{x}}}\cdot\underbrace{\begin{bmatrix}a_{1}^{\vec{C}^{x}}\\ a_{2}^{\vec{C}^{x}}\\ \vdots\end{bmatrix}}_{\bar{a}^{\vec{C}^{x}}} \quad , \quad \vec{r}\in\mathbb{S}^{x} \tag{4-7}$$

and Eqs. (4-4a), (4-4b), (4-5a), (4-5b), and (4-6) are tested with $\{\vec{b}_{\xi}^{\vec{M}^{\mathrm{G}\rightleftharpoons\mathrm{A}}}\}$, $\{\vec{b}_{\xi}^{\vec{J}^{\mathrm{G}\rightleftharpoons\mathrm{A}}}\}$, $\{\vec{b}_{\xi}^{\vec{M}^{\mathrm{A}\rightleftharpoons\mathrm{M}}}\}$, $\{\vec{b}_{\xi}^{\vec{J}^{\mathrm{A}\rightleftharpoons\mathrm{M}}}\}$, and $\{\vec{b}_{\xi}^{\vec{J}^{\mathrm{A}}}\}$ respectively, then the integral equations are immediately discretized into the following *matrix equations*

$$\begin{aligned}0 = {} & \bar{\bar{Z}}^{\vec{M}^{\mathrm{G}\rightleftharpoons\mathrm{A}}\vec{J}^{\mathrm{G}\rightleftharpoons\mathrm{A}}}\cdot\bar{a}^{\vec{J}^{\mathrm{G}\rightleftharpoons\mathrm{A}}}+\bar{\bar{Z}}^{\vec{M}^{\mathrm{G}\rightleftharpoons\mathrm{A}}\vec{J}^{\mathrm{A}\rightleftharpoons\mathrm{M}}}\cdot\bar{a}^{\vec{J}^{\mathrm{A}\rightleftharpoons\mathrm{M}}}+\bar{\bar{Z}}^{\vec{M}^{\mathrm{G}\rightleftharpoons\mathrm{A}}\vec{J}^{\mathrm{A}}}\cdot\bar{a}^{\vec{J}^{\mathrm{A}}}+\bar{\bar{Z}}^{\vec{M}^{\mathrm{G}\rightleftharpoons\mathrm{A}}\vec{M}^{\mathrm{G}\rightleftharpoons\mathrm{A}}}\cdot\bar{a}^{\vec{M}^{\mathrm{G}\rightleftharpoons\mathrm{A}}}\\ & +\bar{\bar{Z}}^{\vec{M}^{\mathrm{G}\rightleftharpoons\mathrm{A}}\vec{M}^{\mathrm{A}\rightleftharpoons\mathrm{M}}}\cdot\bar{a}^{\vec{M}^{\mathrm{A}\rightleftharpoons\mathrm{M}}}\end{aligned} \tag{4-8a}$$

$$\begin{aligned}0 = {} & \bar{\bar{Z}}^{\vec{J}^{\mathrm{G}\rightleftharpoons\mathrm{A}}\vec{J}^{\mathrm{G}\rightleftharpoons\mathrm{A}}}\cdot\bar{a}^{\vec{J}^{\mathrm{G}\rightleftharpoons\mathrm{A}}}+\bar{\bar{Z}}^{\vec{J}^{\mathrm{G}\rightleftharpoons\mathrm{A}}\vec{J}^{\mathrm{A}\rightleftharpoons\mathrm{M}}}\cdot\bar{a}^{\vec{J}^{\mathrm{A}\rightleftharpoons\mathrm{M}}}+\bar{\bar{Z}}^{\vec{J}^{\mathrm{G}\rightleftharpoons\mathrm{A}}\vec{J}^{\mathrm{A}}}\cdot\bar{a}^{\vec{J}^{\mathrm{A}}}+\bar{\bar{Z}}^{\vec{J}^{\mathrm{G}\rightleftharpoons\mathrm{A}}\vec{M}^{\mathrm{G}\rightleftharpoons\mathrm{A}}}\cdot\bar{a}^{\vec{M}^{\mathrm{G}\rightleftharpoons\mathrm{A}}}\\ & +\bar{\bar{Z}}^{\vec{J}^{\mathrm{G}\rightleftharpoons\mathrm{A}}\vec{M}^{\mathrm{A}\rightleftharpoons\mathrm{M}}}\cdot\bar{a}^{\vec{M}^{\mathrm{A}\rightleftharpoons\mathrm{M}}}\end{aligned} \tag{4-8b}$$

and

$$\begin{aligned} 0 = \; & \overline{\overline{Z}}^{\vec{M}^{\mathrm{A}\rightleftharpoons\mathrm{M}}\vec{J}^{\mathrm{G}\rightleftharpoons\mathrm{A}}} \cdot \overline{a}^{\vec{J}^{\mathrm{G}\rightleftharpoons\mathrm{A}}} + \overline{\overline{Z}}^{\vec{M}^{\mathrm{A}\rightleftharpoons\mathrm{M}}\vec{J}^{\mathrm{A}\rightleftharpoons\mathrm{M}}} \cdot \overline{a}^{\vec{J}^{\mathrm{A}\rightleftharpoons\mathrm{M}}} + \overline{\overline{Z}}^{\vec{M}^{\mathrm{A}\rightleftharpoons\mathrm{M}}\vec{J}^{\mathrm{A}}} \cdot \overline{a}^{\vec{J}^{\mathrm{A}}} + \overline{\overline{Z}}^{\vec{M}^{\mathrm{A}\rightleftharpoons\mathrm{M}}\vec{M}^{\mathrm{G}\rightleftharpoons\mathrm{A}}} \cdot \overline{a}^{\vec{M}^{\mathrm{G}\rightleftharpoons\mathrm{A}}} \\ & + \overline{\overline{Z}}^{\vec{M}^{\mathrm{A}\rightleftharpoons\mathrm{M}}\vec{M}^{\mathrm{A}\rightleftharpoons\mathrm{M}}} \cdot \overline{a}^{\vec{M}^{\mathrm{A}\rightleftharpoons\mathrm{M}}} \end{aligned} \tag{4-9a}$$

$$\begin{aligned} 0 = \; & \overline{\overline{Z}}^{\vec{J}^{\mathrm{A}\rightleftharpoons\mathrm{M}}\vec{J}^{\mathrm{G}\rightleftharpoons\mathrm{A}}} \cdot \overline{a}^{\vec{J}^{\mathrm{G}\rightleftharpoons\mathrm{A}}} + \overline{\overline{Z}}^{\vec{J}^{\mathrm{A}\rightleftharpoons\mathrm{M}}\vec{J}^{\mathrm{A}\rightleftharpoons\mathrm{M}}} \cdot \overline{a}^{\vec{J}^{\mathrm{A}\rightleftharpoons\mathrm{M}}} + \overline{\overline{Z}}^{\vec{J}^{\mathrm{A}\rightleftharpoons\mathrm{M}}\vec{J}^{\mathrm{A}}} \cdot \overline{a}^{\vec{J}^{\mathrm{A}}} + \overline{\overline{Z}}^{\vec{J}^{\mathrm{A}\rightleftharpoons\mathrm{M}}\vec{M}^{\mathrm{G}\rightleftharpoons\mathrm{A}}} \cdot \overline{a}^{\vec{M}^{\mathrm{G}\rightleftharpoons\mathrm{A}}} \\ & + \overline{\overline{Z}}^{\vec{J}^{\mathrm{A}\rightleftharpoons\mathrm{M}}\vec{M}^{\mathrm{A}\rightleftharpoons\mathrm{M}}} \cdot \overline{a}^{\vec{M}^{\mathrm{A}\rightleftharpoons\mathrm{M}}} \end{aligned} \tag{4-9b}$$

and

$$\begin{aligned} 0 = \; & \overline{\overline{Z}}^{\vec{J}^{\mathrm{A}}\vec{J}^{\mathrm{G}\rightleftharpoons\mathrm{A}}} \cdot \overline{a}^{\vec{J}^{\mathrm{G}\rightleftharpoons\mathrm{A}}} + \overline{\overline{Z}}^{\vec{J}^{\mathrm{A}}\vec{J}^{\mathrm{A}\rightleftharpoons\mathrm{M}}} \cdot \overline{a}^{\vec{J}^{\mathrm{A}\rightleftharpoons\mathrm{M}}} + \overline{\overline{Z}}^{\vec{J}^{\mathrm{A}}\vec{J}^{\mathrm{A}}} \cdot \overline{a}^{\vec{J}^{\mathrm{A}}} + \overline{\overline{Z}}^{\vec{J}^{\mathrm{A}}\vec{M}^{\mathrm{G}\rightleftharpoons\mathrm{A}}} \cdot \overline{a}^{\vec{M}^{\mathrm{G}\rightleftharpoons\mathrm{A}}} \\ & + \overline{\overline{Z}}^{\vec{J}^{\mathrm{A}}\vec{M}^{\mathrm{A}\rightleftharpoons\mathrm{M}}} \cdot \overline{a}^{\vec{M}^{\mathrm{A}\rightleftharpoons\mathrm{M}}} \end{aligned} \tag{4-10}$$

The formulations used to calculate the elements of the matrices in Eq. (4-8a) are as follows:

$$z_{\xi\zeta}^{\vec{M}^{\mathrm{G}\rightleftharpoons\mathrm{A}}\vec{J}^{\mathrm{G}\rightleftharpoons\mathrm{A}}} = \left\langle \vec{b}_{\xi}^{\vec{M}^{\mathrm{G}\rightleftharpoons\mathrm{A}}}, \mathcal{H}\left(\vec{b}_{\zeta}^{\vec{J}^{\mathrm{G}\rightleftharpoons\mathrm{A}}}, 0\right) \right\rangle_{\underset{\sim}{\mathbb{S}}^{\mathrm{G}\rightleftharpoons\mathrm{A}}} - \left\langle \vec{b}_{\xi}^{\vec{M}^{\mathrm{G}\rightleftharpoons\mathrm{A}}}, \vec{b}_{\zeta}^{\vec{J}^{\mathrm{G}\rightleftharpoons\mathrm{A}}} \times \hat{n}^{\rightarrow\mathrm{A}} \right\rangle_{\underset{\sim}{\mathbb{S}}^{\mathrm{G}\rightleftharpoons\mathrm{A}}} \tag{4-11a}$$

$$z_{\xi\zeta}^{\vec{M}^{\mathrm{G}\rightleftharpoons\mathrm{A}}\vec{J}^{\mathrm{A}\rightleftharpoons\mathrm{M}}} = \left\langle \vec{b}_{\xi}^{\vec{M}^{\mathrm{G}\rightleftharpoons\mathrm{A}}}, \mathcal{H}\left(\vec{b}_{\zeta}^{\vec{J}^{\mathrm{A}\rightleftharpoons\mathrm{M}}}, 0\right) \right\rangle_{\underset{\sim}{\mathbb{S}}^{\mathrm{G}\rightleftharpoons\mathrm{A}}} \tag{4-11b}$$

$$z_{\xi\zeta}^{\vec{M}^{\mathrm{G}\rightleftharpoons\mathrm{A}}\vec{J}^{\mathrm{A}}} = \left\langle \vec{b}_{\xi}^{\vec{M}^{\mathrm{G}\rightleftharpoons\mathrm{A}}}, \mathcal{H}\left(\vec{b}_{\zeta}^{\vec{J}^{\mathrm{A}}}, 0\right) \right\rangle_{\underset{\sim}{\mathbb{S}}^{\mathrm{G}\rightleftharpoons\mathrm{A}}} \tag{4-11c}$$

$$z_{\xi\zeta}^{\vec{M}^{\mathrm{G}\rightleftharpoons\mathrm{A}}\vec{M}^{\mathrm{G}\rightleftharpoons\mathrm{A}}} = \left\langle \vec{b}_{\xi}^{\vec{M}^{\mathrm{G}\rightleftharpoons\mathrm{A}}}, \mathcal{H}\left(0, \vec{b}_{\zeta}^{\vec{M}^{\mathrm{G}\rightleftharpoons\mathrm{A}}}\right) \right\rangle_{\underset{\sim}{\mathbb{S}}^{\mathrm{G}\rightleftharpoons\mathrm{A}}} \tag{4-11d}$$

$$z_{\xi\zeta}^{\vec{M}^{\mathrm{G}\rightleftharpoons\mathrm{A}}\vec{M}^{\mathrm{A}\rightleftharpoons\mathrm{M}}} = \left\langle \vec{b}_{\xi}^{\vec{M}^{\mathrm{G}\rightleftharpoons\mathrm{A}}}, \mathcal{H}\left(0, \vec{b}_{\zeta}^{\vec{M}^{\mathrm{A}\rightleftharpoons\mathrm{M}}}\right) \right\rangle_{\underset{\sim}{\mathbb{S}}^{\mathrm{G}\rightleftharpoons\mathrm{A}}} \tag{4-11e}$$

where integral surface $\underset{\sim}{\mathbb{S}}^{\mathrm{G}\rightleftharpoons\mathrm{A}}$ is an inner sub-boundary of $\mathbb{V}^{\mathrm{A}}$ as shown in the following Fig. 4-5.

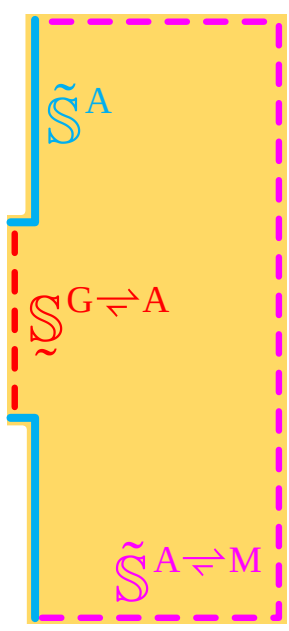

Figure 4-5 Integral surfaces used in Eqs. (4-11)~(4-15)

The formulations for calculating the elements of the matrices in Eq. (4-8b) are as follows:

$$z_{\xi\zeta}^{\vec{J}^{\mathrm{G}\rightleftharpoons\mathrm{A}}\vec{J}^{\mathrm{G}\rightleftharpoons\mathrm{A}}} = \left\langle \vec{b}_{\xi}^{\vec{J}^{\mathrm{G}\rightleftharpoons\mathrm{A}}}, \mathcal{E}\left(\vec{b}_{\zeta}^{\vec{J}^{\mathrm{G}\rightleftharpoons\mathrm{A}}}, 0\right) \right\rangle_{\underset{\sim}{\mathbb{S}}^{\mathrm{G}\rightleftharpoons\mathrm{A}}} \tag{4-12a}$$

$$z_{\xi\zeta}^{\vec{J}^{\mathrm{G}\rightleftharpoons\mathrm{A}}\vec{J}^{\mathrm{A}\rightleftharpoons\mathrm{M}}} = \left\langle \vec{b}_{\xi}^{\vec{J}^{\mathrm{G}\rightleftharpoons\mathrm{A}}}, \mathcal{E}\left(\vec{b}_{\zeta}^{\vec{J}^{\mathrm{A}\rightleftharpoons\mathrm{M}}}, 0\right) \right\rangle_{\underset{\sim}{\mathbb{S}}^{\mathrm{G}\rightleftharpoons\mathrm{A}}} \tag{4-12b}$$

$$z_{\xi\zeta}^{\vec{J}^{\mathrm{G}\rightleftharpoons\mathrm{A}}\vec{J}^{\mathrm{A}}} = \left\langle \vec{b}_{\xi}^{\vec{J}^{\mathrm{G}\rightleftharpoons\mathrm{A}}}, \mathcal{E}\left(\vec{b}_{\zeta}^{\vec{J}^{\mathrm{A}}}, 0\right)\right\rangle_{\tilde{\mathbb{S}}^{\mathrm{G}\rightleftharpoons\mathrm{A}}} \tag{4-12c}$$

$$z_{\xi\zeta}^{\vec{J}^{\mathrm{G}\rightleftharpoons\mathrm{A}}\vec{M}^{\mathrm{G}\rightleftharpoons\mathrm{A}}} = \left\langle \vec{b}_{\xi}^{\vec{J}^{\mathrm{G}\rightleftharpoons\mathrm{A}}}, \mathcal{E}\left(0, \vec{b}_{\zeta}^{\vec{M}^{\mathrm{G}\rightleftharpoons\mathrm{A}}}\right)\right\rangle_{\tilde{\mathbb{S}}^{\mathrm{G}\rightleftharpoons\mathrm{A}}} - \left\langle \vec{b}_{\xi}^{\vec{J}^{\mathrm{G}\rightleftharpoons\mathrm{A}}}, \hat{n}^{\rightarrow\mathrm{A}} \times \vec{b}_{\zeta}^{\vec{M}^{\mathrm{G}\rightleftharpoons\mathrm{A}}}\right\rangle_{\mathbb{S}^{\mathrm{G}\rightleftharpoons\mathrm{A}}} \tag{4-12d}$$

$$z_{\xi\zeta}^{\vec{J}^{\mathrm{G}\rightleftharpoons\mathrm{A}}\vec{M}^{\mathrm{A}\rightleftharpoons\mathrm{M}}} = \left\langle \vec{b}_{\xi}^{\vec{J}^{\mathrm{G}\rightleftharpoons\mathrm{A}}}, \mathcal{E}\left(0, \vec{b}_{\zeta}^{\vec{M}^{\mathrm{A}\rightleftharpoons\mathrm{M}}}\right)\right\rangle_{\tilde{\mathbb{S}}^{\mathrm{G}\rightleftharpoons\mathrm{A}}} \tag{4-12e}$$

The formulations used to calculate the elements of the matrices in Eq. (4-9a) are as follows:

$$z_{\xi\zeta}^{\vec{M}^{\mathrm{A}\rightleftharpoons\mathrm{M}}\vec{J}^{\mathrm{G}\rightleftharpoons\mathrm{A}}} = \left\langle \vec{b}_{\xi}^{\vec{M}^{\mathrm{A}\rightleftharpoons\mathrm{M}}}, \mathcal{H}\left(\vec{b}_{\zeta}^{\vec{J}^{\mathrm{G}\rightleftharpoons\mathrm{A}}}, 0\right)\right\rangle_{\tilde{\mathbb{S}}^{\mathrm{A}\rightleftharpoons\mathrm{M}}} \tag{4-13a}$$

$$z_{\xi\zeta}^{\vec{M}^{\mathrm{A}\rightleftharpoons\mathrm{M}}\vec{J}^{\mathrm{A}\rightleftharpoons\mathrm{M}}} = \left\langle \vec{b}_{\xi}^{\vec{M}^{\mathrm{A}\rightleftharpoons\mathrm{M}}}, \mathcal{H}\left(\vec{b}_{\zeta}^{\vec{J}^{\mathrm{A}\rightleftharpoons\mathrm{M}}}, 0\right)\right\rangle_{\tilde{\mathbb{S}}^{\mathrm{A}\rightleftharpoons\mathrm{M}}} - \left\langle \vec{b}_{\xi}^{\vec{M}^{\mathrm{A}\rightleftharpoons\mathrm{M}}}, \vec{b}_{\zeta}^{\vec{J}^{\mathrm{A}\rightleftharpoons\mathrm{M}}} \times \hat{n}^{\rightarrow\mathrm{A}}\right\rangle_{\mathbb{S}^{\mathrm{A}\rightleftharpoons\mathrm{M}}} \tag{4-13b}$$

$$z_{\xi\zeta}^{\vec{M}^{\mathrm{A}\rightleftharpoons\mathrm{M}}\vec{J}^{\mathrm{A}}} = \left\langle \vec{b}_{\xi}^{\vec{M}^{\mathrm{A}\rightleftharpoons\mathrm{M}}}, \mathcal{H}\left(\vec{b}_{\zeta}^{\vec{J}^{\mathrm{A}}}, 0\right)\right\rangle_{\tilde{\mathbb{S}}^{\mathrm{A}\rightleftharpoons\mathrm{M}}} \tag{4-13c}$$

$$z_{\xi\zeta}^{\vec{M}^{\mathrm{A}\rightleftharpoons\mathrm{M}}\vec{M}^{\mathrm{G}\rightleftharpoons\mathrm{A}}} = \left\langle \vec{b}_{\xi}^{\vec{M}^{\mathrm{A}\rightleftharpoons\mathrm{M}}}, \mathcal{H}\left(0, \vec{b}_{\zeta}^{\vec{M}^{\mathrm{G}\rightleftharpoons\mathrm{A}}}\right)\right\rangle_{\tilde{\mathbb{S}}^{\mathrm{A}\rightleftharpoons\mathrm{M}}} \tag{4-13d}$$

$$z_{\xi\zeta}^{\vec{M}^{\mathrm{A}\rightleftharpoons\mathrm{M}}\vec{M}^{\mathrm{A}\rightleftharpoons\mathrm{M}}} = \left\langle \vec{b}_{\xi}^{\vec{M}^{\mathrm{A}\rightleftharpoons\mathrm{M}}}, \mathcal{H}\left(0, \vec{b}_{\zeta}^{\vec{M}^{\mathrm{A}\rightleftharpoons\mathrm{M}}}\right)\right\rangle_{\tilde{\mathbb{S}}^{\mathrm{A}\rightleftharpoons\mathrm{M}}} \tag{4-13e}$$

where integral surface $\tilde{\mathbb{S}}^{\mathrm{A}\rightleftharpoons\mathrm{M}}$ is an inner sub-boundary of $\mathbb{V}^{\mathrm{A}}$ as shown in Fig. 4-5. The formulations used to calculate the elements of the matrices in Eq. (4-9b) are as follows:

$$z_{\xi\zeta}^{\vec{J}^{\mathrm{A}\rightleftharpoons\mathrm{M}}\vec{J}^{\mathrm{G}\rightleftharpoons\mathrm{A}}} = \left\langle \vec{b}_{\xi}^{\vec{J}^{\mathrm{A}\rightleftharpoons\mathrm{M}}}, \mathcal{E}\left(\vec{b}_{\zeta}^{\vec{J}^{\mathrm{G}\rightleftharpoons\mathrm{A}}}, 0\right)\right\rangle_{\tilde{\mathbb{S}}^{\mathrm{A}\rightleftharpoons\mathrm{M}}} \tag{4-14a}$$

$$z_{\xi\zeta}^{\vec{J}^{\mathrm{A}\rightleftharpoons\mathrm{M}}\vec{J}^{\mathrm{A}\rightleftharpoons\mathrm{M}}} = \left\langle \vec{b}_{\xi}^{\vec{J}^{\mathrm{A}\rightleftharpoons\mathrm{M}}}, \mathcal{E}\left(\vec{b}_{\zeta}^{\vec{J}^{\mathrm{A}\rightleftharpoons\mathrm{M}}}, 0\right)\right\rangle_{\tilde{\mathbb{S}}^{\mathrm{A}\rightleftharpoons\mathrm{M}}} \tag{4-14b}$$

$$z_{\xi\zeta}^{\vec{J}^{\mathrm{A}\rightleftharpoons\mathrm{M}}\vec{J}^{\mathrm{A}}} = \left\langle \vec{b}_{\xi}^{\vec{J}^{\mathrm{A}\rightleftharpoons\mathrm{M}}}, \mathcal{E}\left(\vec{b}_{\zeta}^{\vec{J}^{\mathrm{A}}}, 0\right)\right\rangle_{\tilde{\mathbb{S}}^{\mathrm{A}\rightleftharpoons\mathrm{M}}} \tag{4-14c}$$

$$z_{\xi\zeta}^{\vec{J}^{\mathrm{A}\rightleftharpoons\mathrm{M}}\vec{M}^{\mathrm{G}\rightleftharpoons\mathrm{A}}} = \left\langle \vec{b}_{\xi}^{\vec{J}^{\mathrm{A}\rightleftharpoons\mathrm{M}}}, \mathcal{E}\left(0, \vec{b}_{\zeta}^{\vec{M}^{\mathrm{G}\rightleftharpoons\mathrm{A}}}\right)\right\rangle_{\tilde{\mathbb{S}}^{\mathrm{A}\rightleftharpoons\mathrm{M}}} \tag{4-14d}$$

$$z_{\xi\zeta}^{\vec{J}^{\mathrm{A}\rightleftharpoons\mathrm{M}}\vec{M}^{\mathrm{A}\rightleftharpoons\mathrm{M}}} = \left\langle \vec{b}_{\xi}^{\vec{J}^{\mathrm{A}\rightleftharpoons\mathrm{M}}}, \mathcal{E}\left(0, \vec{b}_{\zeta}^{\vec{M}^{\mathrm{A}\rightleftharpoons\mathrm{M}}}\right)\right\rangle_{\tilde{\mathbb{S}}^{\mathrm{A}\rightleftharpoons\mathrm{M}}} - \left\langle \vec{b}_{\xi}^{\vec{J}^{\mathrm{A}\rightleftharpoons\mathrm{M}}}, \hat{n}^{\rightarrow\mathrm{A}} \times \vec{b}_{\zeta}^{\vec{M}^{\mathrm{A}\rightleftharpoons\mathrm{M}}}\right\rangle_{\mathbb{S}^{\mathrm{A}\rightleftharpoons\mathrm{M}}} \tag{4-14e}$$

The formulations used to calculate the elements of the matrices in Eq. (4-10) are as follows:

$$z_{\xi\zeta}^{\vec{J}^{\mathrm{A}}\vec{J}^{\mathrm{G}\rightleftharpoons\mathrm{A}}} = \left\langle \vec{b}_{\xi}^{\vec{J}^{\mathrm{A}}}, \mathcal{E}\left(\vec{b}_{\zeta}^{\vec{J}^{\mathrm{G}\rightleftharpoons\mathrm{A}}}, 0\right)\right\rangle_{\tilde{\mathbb{S}}^{\mathrm{A}}} \tag{4-15a}$$

$$z_{\xi\zeta}^{\vec{J}^{\mathrm{A}}\vec{J}^{\mathrm{A}\rightleftharpoons\mathrm{M}}} = \left\langle \vec{b}_{\xi}^{\vec{J}^{\mathrm{A}}}, \mathcal{E}\left(\vec{b}_{\zeta}^{\vec{J}^{\mathrm{A}\rightleftharpoons\mathrm{M}}}, 0\right)\right\rangle_{\tilde{\mathbb{S}}^{\mathrm{A}}} \tag{4-15b}$$

$$z_{\xi\zeta}^{\vec{J}^{\mathrm{A}}\vec{J}^{\mathrm{A}}} = \left\langle \vec{b}_{\xi}^{\vec{J}^{\mathrm{A}}}, \mathcal{E}\left(\vec{b}_{\zeta}^{\vec{J}^{\mathrm{A}}}, 0\right)\right\rangle_{\tilde{\mathbb{S}}^{\mathrm{A}}} \tag{4-15c}$$

$$z_{\xi\zeta}^{\vec{J}^{\mathrm{A}}\vec{M}^{\mathrm{G}\rightleftharpoons\mathrm{A}}} = \left\langle \vec{b}_{\xi}^{\vec{J}^{\mathrm{A}}}, \mathcal{E}\left(0, \vec{b}_{\zeta}^{\vec{M}^{\mathrm{G}\rightleftharpoons\mathrm{A}}}\right)\right\rangle_{\tilde{\mathbb{S}}^{\mathrm{A}}} \tag{4-15d}$$

$$z_{\xi\zeta}^{\vec{J}^{\mathrm{A}}\vec{M}^{\mathrm{A}\rightleftharpoons\mathrm{M}}} = \left\langle \vec{b}_{\xi}^{\vec{J}^{\mathrm{A}}}, \mathcal{E}\left(0, \vec{b}_{\zeta}^{\vec{M}^{\mathrm{A}\rightleftharpoons\mathrm{M}}}\right)\right\rangle_{\tilde{\mathbb{S}}^{\mathrm{A}}} \tag{4-15e}$$

where integral surface $\tilde{\mathbb{S}}^{\mathrm{A}}$ is an inner sub-boundary of $\mathbb{V}^{\mathrm{A}}$ as shown in Fig. 4-5.

Below, we propose two somewhat different schemes for mathematically describing the modal space of the tra-antenna shown in Figs. 4-2 and 4-4.

#### 4.2.2.1 Scheme I: Dependent Variable Elimination (DVE)

By properly assembling the above-mentioned matrix equations (4-8a), (4-8b), (4-9a), (4-9b), and (4-10), we have the following two theoretically equivalent *augmented matrix equations*

$$\overline{\overline{\Psi}}_1 \cdot \bar{a}^{\mathrm{AV}} = \overline{\overline{\Psi}}_2 \cdot \bar{a}^{\vec{J}^{\mathrm{G}\rightleftharpoons\mathrm{A}}} \tag{4-16}$$

$$\overline{\overline{\Psi}}_3 \cdot \bar{a}^{\mathrm{AV}} = \overline{\overline{\Psi}}_4 \cdot \bar{a}^{\vec{M}^{\mathrm{G}\rightleftharpoons\mathrm{A}}} \tag{4-17}$$

in which

$$\overline{\overline{\Psi}}_1 = \begin{bmatrix} \overline{\overline{I}}^{\vec{J}^{\mathrm{G}\rightleftharpoons\mathrm{A}}} & 0 & 0 & 0 & 0 \\ 0 & \overline{\overline{Z}}^{\vec{M}^{\mathrm{G}\rightleftharpoons\mathrm{A}}\vec{J}^{\mathrm{A}\rightleftharpoons\mathrm{M}}} & \overline{\overline{Z}}^{\vec{M}^{\mathrm{G}\rightleftharpoons\mathrm{A}}\vec{J}^{\mathrm{A}}} & \overline{\overline{Z}}^{\vec{M}^{\mathrm{G}\rightleftharpoons\mathrm{A}}\vec{M}^{\mathrm{G}\rightleftharpoons\mathrm{A}}} & \overline{\overline{Z}}^{\vec{M}^{\mathrm{G}\rightleftharpoons\mathrm{A}}\vec{M}^{\mathrm{A}\rightleftharpoons\mathrm{M}}} \\ 0 & \overline{\overline{Z}}^{\vec{M}^{\mathrm{A}\rightleftharpoons\mathrm{M}}\vec{J}^{\mathrm{A}\rightleftharpoons\mathrm{M}}} & \overline{\overline{Z}}^{\vec{M}^{\mathrm{A}\rightleftharpoons\mathrm{M}}\vec{J}^{\mathrm{A}}} & \overline{\overline{Z}}^{\vec{M}^{\mathrm{A}\rightleftharpoons\mathrm{M}}\vec{M}^{\mathrm{G}\rightleftharpoons\mathrm{A}}} & \overline{\overline{Z}}^{\vec{M}^{\mathrm{A}\rightleftharpoons\mathrm{M}}\vec{M}^{\mathrm{A}\rightleftharpoons\mathrm{M}}} \\ 0 & \overline{\overline{Z}}^{\vec{J}^{\mathrm{A}\rightleftharpoons\mathrm{M}}\vec{J}^{\mathrm{A}\rightleftharpoons\mathrm{M}}} & \overline{\overline{Z}}^{\vec{J}^{\mathrm{A}\rightleftharpoons\mathrm{M}}\vec{J}^{\mathrm{A}}} & \overline{\overline{Z}}^{\vec{J}^{\mathrm{A}\rightleftharpoons\mathrm{M}}\vec{M}^{\mathrm{G}\rightleftharpoons\mathrm{A}}} & \overline{\overline{Z}}^{\vec{J}^{\mathrm{A}\rightleftharpoons\mathrm{M}}\vec{M}^{\mathrm{A}\rightleftharpoons\mathrm{M}}} \\ 0 & \overline{\overline{Z}}^{\vec{J}^{\mathrm{A}}\vec{J}^{\mathrm{A}\rightleftharpoons\mathrm{M}}} & \overline{\overline{Z}}^{\vec{J}^{\mathrm{A}}\vec{J}^{\mathrm{A}}} & \overline{\overline{Z}}^{\vec{J}^{\mathrm{A}}\vec{M}^{\mathrm{G}\rightleftharpoons\mathrm{A}}} & \overline{\overline{Z}}^{\vec{J}^{\mathrm{A}}\vec{M}^{\mathrm{A}\rightleftharpoons\mathrm{M}}} \end{bmatrix} \tag{4-18a}$$

$$\overline{\overline{\Psi}}_2 = \begin{bmatrix} \overline{\overline{I}}^{\vec{J}^{\mathrm{G}\rightleftharpoons\mathrm{A}}} \\ -\overline{\overline{Z}}^{\vec{M}^{\mathrm{G}\rightleftharpoons\mathrm{A}}\vec{J}^{\mathrm{G}\rightleftharpoons\mathrm{A}}} \\ -\overline{\overline{Z}}^{\vec{M}^{\mathrm{A}\rightleftharpoons\mathrm{M}}\vec{J}^{\mathrm{G}\rightleftharpoons\mathrm{A}}} \\ -\overline{\overline{Z}}^{\vec{J}^{\mathrm{A}\rightleftharpoons\mathrm{M}}\vec{J}^{\mathrm{G}\rightleftharpoons\mathrm{A}}} \\ -\overline{\overline{Z}}^{\vec{J}^{\mathrm{A}}\vec{J}^{\mathrm{G}\rightleftharpoons\mathrm{A}}} \end{bmatrix} \tag{4-18b}$$

$$\bar{a}^{\mathrm{AV}} = \begin{bmatrix} \bar{a}^{\vec{J}^{\mathrm{G}\rightleftharpoons\mathrm{A}}} \\ \bar{a}^{\vec{J}^{\mathrm{A}\rightleftharpoons\mathrm{M}}} \\ \bar{a}^{\vec{J}^{\mathrm{A}}} \\ \bar{a}^{\vec{M}^{\mathrm{G}\rightleftharpoons\mathrm{A}}} \\ \bar{a}^{\vec{M}^{\mathrm{A}\rightleftharpoons\mathrm{M}}} \end{bmatrix} \tag{4-18c}$$

and

$$\overline{\overline{\Psi}}_3 = \begin{bmatrix} 0 & 0 & 0 & \overline{\overline{I}}^{\vec{M}^{\mathrm{G}\rightleftharpoons\mathrm{A}}} & 0 \\ \overline{\overline{Z}}^{\vec{J}^{\mathrm{G}\rightleftharpoons\mathrm{A}}\vec{J}^{\mathrm{G}\rightleftharpoons\mathrm{A}}} & \overline{\overline{Z}}^{\vec{J}^{\mathrm{G}\rightleftharpoons\mathrm{A}}\vec{J}^{\mathrm{A}\rightleftharpoons\mathrm{M}}} & \overline{\overline{Z}}^{\vec{J}^{\mathrm{G}\rightleftharpoons\mathrm{A}}\vec{J}^{\mathrm{A}}} & 0 & \overline{\overline{Z}}^{\vec{J}^{\mathrm{G}\rightleftharpoons\mathrm{A}}\vec{M}^{\mathrm{A}\rightleftharpoons\mathrm{M}}} \\ \overline{\overline{Z}}^{\vec{M}^{\mathrm{A}\rightleftharpoons\mathrm{M}}\vec{J}^{\mathrm{G}\rightleftharpoons\mathrm{A}}} & \overline{\overline{Z}}^{\vec{M}^{\mathrm{A}\rightleftharpoons\mathrm{M}}\vec{J}^{\mathrm{A}\rightleftharpoons\mathrm{M}}} & \overline{\overline{Z}}^{\vec{M}^{\mathrm{A}\rightleftharpoons\mathrm{M}}\vec{J}^{\mathrm{A}}} & 0 & \overline{\overline{Z}}^{\vec{M}^{\mathrm{A}\rightleftharpoons\mathrm{M}}\vec{M}^{\mathrm{A}\rightleftharpoons\mathrm{M}}} \\ \overline{\overline{Z}}^{\vec{J}^{\mathrm{A}\rightleftharpoons\mathrm{M}}\vec{J}^{\mathrm{G}\rightleftharpoons\mathrm{A}}} & \overline{\overline{Z}}^{\vec{J}^{\mathrm{A}\rightleftharpoons\mathrm{M}}\vec{J}^{\mathrm{A}\rightleftharpoons\mathrm{M}}} & \overline{\overline{Z}}^{\vec{J}^{\mathrm{A}\rightleftharpoons\mathrm{M}}\vec{J}^{\mathrm{A}}} & 0 & \overline{\overline{Z}}^{\vec{J}^{\mathrm{A}\rightleftharpoons\mathrm{M}}\vec{M}^{\mathrm{A}\rightleftharpoons\mathrm{M}}} \\ \overline{\overline{Z}}^{\vec{J}^{\mathrm{A}}\vec{J}^{\mathrm{G}\rightleftharpoons\mathrm{A}}} & \overline{\overline{Z}}^{\vec{J}^{\mathrm{A}}\vec{J}^{\mathrm{A}\rightleftharpoons\mathrm{M}}} & \overline{\overline{Z}}^{\vec{J}^{\mathrm{A}}\vec{J}^{\mathrm{A}}} & 0 & \overline{\overline{Z}}^{\vec{J}^{\mathrm{A}}\vec{M}^{\mathrm{A}\rightleftharpoons\mathrm{M}}} \end{bmatrix} \tag{4-19a}$$

$$\overline{\overline{\Psi}}_4 = \begin{bmatrix} \overline{\overline{I}}^{\vec{M}^{\mathrm{G}\rightleftharpoons\mathrm{A}}} \\ -\overline{\overline{Z}}^{\vec{J}^{\mathrm{G}\rightleftharpoons\mathrm{A}}\vec{M}^{\mathrm{G}\rightleftharpoons\mathrm{A}}} \\ -\overline{\overline{Z}}^{\vec{M}^{\mathrm{A}\rightleftharpoons\mathrm{M}}\vec{M}^{\mathrm{G}\rightleftharpoons\mathrm{A}}} \\ -\overline{\overline{Z}}^{\vec{J}^{\mathrm{A}\rightleftharpoons\mathrm{M}}\vec{M}^{\mathrm{G}\rightleftharpoons\mathrm{A}}} \\ -\overline{\overline{Z}}^{\vec{J}^{\mathrm{A}}\vec{M}^{\mathrm{G}\rightleftharpoons\mathrm{A}}} \end{bmatrix} \tag{4-19b}$$

where $\overline{\overline{I}}^{\vec{J}^{\mathrm{G}\rightleftharpoons\mathrm{A}}}$ and $\overline{\overline{I}}^{\vec{M}^{\mathrm{G}\rightleftharpoons\mathrm{A}}}$ are the *identity matrices* whose orders are the same as the dimensions of $\overline{a}^{\vec{J}^{\mathrm{G}\rightleftharpoons\mathrm{A}}}$ and $\overline{a}^{\vec{M}^{\mathrm{G}\rightleftharpoons\mathrm{A}}}$, the 0s are some *zero matrices* with proper row numbers and column numbers.

By solving the above augmented matrix equations, there exist the following transformations from $\overline{a}^{\vec{J}^{\mathrm{G}\rightleftharpoons\mathrm{A}}}$ to $\overline{a}^{\mathrm{AV}}$ and from $\overline{a}^{\vec{M}^{\mathrm{G}\rightleftharpoons\mathrm{A}}}$ to $\overline{a}^{\mathrm{AV}}$.

$$\overline{a}^{\mathrm{AV}} = \overbrace{\left(\overline{\overline{\Psi}}_1\right)^{-1}\cdot\overline{\overline{\Psi}}_2}^{\overline{\overline{T}}^{\vec{J}^{\mathrm{G}\rightleftharpoons\mathrm{A}}\rightarrow\mathrm{AV}}}\cdot\overline{a}^{\vec{J}^{\mathrm{G}\rightleftharpoons\mathrm{A}}} \tag{4-20}$$

$$\overline{a}^{\mathrm{AV}} = \underbrace{\left(\overline{\overline{\Psi}}_3\right)^{-1}\cdot\overline{\overline{\Psi}}_4}_{\overline{\overline{T}}^{\vec{M}^{\mathrm{G}\rightleftharpoons\mathrm{A}}\rightarrow\mathrm{AV}}}\cdot\overline{a}^{\vec{M}^{\mathrm{G}\rightleftharpoons\mathrm{A}}} \tag{4-21}$$

and they can be uniformly written as that $\overline{a}^{\mathrm{AV}} = \overline{\overline{T}}^{\mathrm{BV}\rightarrow\mathrm{AV}}\cdot\overline{a}^{\mathrm{BV}}$ for simplifying the symbolic system of the following discussions, where $\overline{\overline{T}}^{\mathrm{BV}\rightarrow\mathrm{AV}} = \overline{\overline{T}}^{\vec{J}^{\mathrm{G}\rightleftharpoons\mathrm{A}}\rightarrow\mathrm{AV}} / \overline{\overline{T}}^{\vec{M}^{\mathrm{G}\rightleftharpoons\mathrm{A}}\rightarrow\mathrm{AV}}$ and $\overline{a}^{\mathrm{BV}} = \overline{a}^{\vec{J}^{\mathrm{G}\rightleftharpoons\mathrm{A}}} / \overline{a}^{\vec{M}^{\mathrm{G}\rightleftharpoons\mathrm{A}}}$ correspondingly.

#### 4.2.2.2 Scheme II: Solution/Definition Domain Compression (SDC/DDC)

In fact, the Eqs. (4-8a), (4-8b), (4-9a), (4-9b), and (4-10) can also be alternatively combined as follows:

$$\overline{\overline{\Psi}}_{\mathrm{FCE}}^{\mathrm{DoJ}}\cdot\overline{a}^{\mathrm{AV}} = 0 \tag{4-22}$$

$$\overline{\overline{\Psi}}_{\mathrm{FCE}}^{\mathrm{DoM}}\cdot\overline{a}^{\mathrm{AV}} = 0 \tag{4-23}$$

in which

$$\overline{\overline{\Psi}}_{\mathrm{FCE}}^{\mathrm{DoJ}} = \begin{bmatrix} \overline{\overline{Z}}^{\vec{M}^{\mathrm{G}\rightleftharpoons\mathrm{A}}\vec{J}^{\mathrm{G}\rightleftharpoons\mathrm{A}}} & \overline{\overline{Z}}^{\vec{M}^{\mathrm{G}\rightleftharpoons\mathrm{A}}\vec{J}^{\mathrm{A}\rightleftharpoons\mathrm{M}}} & \overline{\overline{Z}}^{\vec{M}^{\mathrm{G}\rightleftharpoons\mathrm{A}}\vec{J}^{\mathrm{A}}} & \overline{\overline{Z}}^{\vec{M}^{\mathrm{G}\rightleftharpoons\mathrm{A}}\vec{M}^{\mathrm{G}\rightleftharpoons\mathrm{A}}} & \overline{\overline{Z}}^{\vec{M}^{\mathrm{G}\rightleftharpoons\mathrm{A}}\vec{M}^{\mathrm{A}\rightleftharpoons\mathrm{M}}} \\ \overline{\overline{Z}}^{\vec{M}^{\mathrm{A}\rightleftharpoons\mathrm{M}}\vec{J}^{\mathrm{G}\rightleftharpoons\mathrm{A}}} & \overline{\overline{Z}}^{\vec{M}^{\mathrm{A}\rightleftharpoons\mathrm{M}}\vec{J}^{\mathrm{A}\rightleftharpoons\mathrm{M}}} & \overline{\overline{Z}}^{\vec{M}^{\mathrm{A}\rightleftharpoons\mathrm{M}}\vec{J}^{\mathrm{A}}} & \overline{\overline{Z}}^{\vec{M}^{\mathrm{A}\rightleftharpoons\mathrm{M}}\vec{M}^{\mathrm{G}\rightleftharpoons\mathrm{A}}} & \overline{\overline{Z}}^{\vec{M}^{\mathrm{A}\rightleftharpoons\mathrm{M}}\vec{M}^{\mathrm{A}\rightleftharpoons\mathrm{M}}} \\ \overline{\overline{Z}}^{\vec{J}^{\mathrm{A}\rightleftharpoons\mathrm{M}}\vec{J}^{\mathrm{G}\rightleftharpoons\mathrm{A}}} & \overline{\overline{Z}}^{\vec{J}^{\mathrm{A}\rightleftharpoons\mathrm{M}}\vec{J}^{\mathrm{A}\rightleftharpoons\mathrm{M}}} & \overline{\overline{Z}}^{\vec{J}^{\mathrm{A}\rightleftharpoons\mathrm{M}}\vec{J}^{\mathrm{A}}} & \overline{\overline{Z}}^{\vec{J}^{\mathrm{A}\rightleftharpoons\mathrm{M}}\vec{M}^{\mathrm{G}\rightleftharpoons\mathrm{A}}} & \overline{\overline{Z}}^{\vec{J}^{\mathrm{A}\rightleftharpoons\mathrm{M}}\vec{M}^{\mathrm{A}\rightleftharpoons\mathrm{M}}} \\ \overline{\overline{Z}}^{\vec{J}^{\mathrm{A}}\vec{J}^{\mathrm{G}\rightleftharpoons\mathrm{A}}} & \overline{\overline{Z}}^{\vec{J}^{\mathrm{A}}\vec{J}^{\mathrm{A}\rightleftharpoons\mathrm{M}}} & \overline{\overline{Z}}^{\vec{J}^{\mathrm{A}}\vec{J}^{\mathrm{A}}} & \overline{\overline{Z}}^{\vec{J}^{\mathrm{A}}\vec{M}^{\mathrm{G}\rightleftharpoons\mathrm{A}}} & \overline{\overline{Z}}^{\vec{J}^{\mathrm{A}}\vec{M}^{\mathrm{A}\rightleftharpoons\mathrm{M}}} \end{bmatrix} \tag{4-24}$$

$$\overline{\overline{\Psi}}_{\mathrm{FCE}}^{\mathrm{DoM}} = \begin{bmatrix} \overline{\overline{Z}}^{\vec{J}^{\mathrm{G}\rightleftharpoons\mathrm{A}}\vec{J}^{\mathrm{G}\rightleftharpoons\mathrm{A}}} & \overline{\overline{Z}}^{\vec{J}^{\mathrm{G}\rightleftharpoons\mathrm{A}}\vec{J}^{\mathrm{A}\rightleftharpoons\mathrm{M}}} & \overline{\overline{Z}}^{\vec{J}^{\mathrm{G}\rightleftharpoons\mathrm{A}}\vec{J}^{\mathrm{A}}} & \overline{\overline{Z}}^{\vec{J}^{\mathrm{G}\rightleftharpoons\mathrm{A}}\vec{M}^{\mathrm{G}\rightleftharpoons\mathrm{A}}} & \overline{\overline{Z}}^{\vec{J}^{\mathrm{G}\rightleftharpoons\mathrm{A}}\vec{M}^{\mathrm{A}\rightleftharpoons\mathrm{M}}} \\ \overline{\overline{Z}}^{\vec{M}^{\mathrm{A}\rightleftharpoons\mathrm{M}}\vec{J}^{\mathrm{G}\rightleftharpoons\mathrm{A}}} & \overline{\overline{Z}}^{\vec{M}^{\mathrm{A}\rightleftharpoons\mathrm{M}}\vec{J}^{\mathrm{A}\rightleftharpoons\mathrm{M}}} & \overline{\overline{Z}}^{\vec{M}^{\mathrm{A}\rightleftharpoons\mathrm{M}}\vec{J}^{\mathrm{A}}} & \overline{\overline{Z}}^{\vec{M}^{\mathrm{A}\rightleftharpoons\mathrm{M}}\vec{M}^{\mathrm{G}\rightleftharpoons\mathrm{A}}} & \overline{\overline{Z}}^{\vec{M}^{\mathrm{A}\rightleftharpoons\mathrm{M}}\vec{M}^{\mathrm{A}\rightleftharpoons\mathrm{M}}} \\ \overline{\overline{Z}}^{\vec{J}^{\mathrm{A}\rightleftharpoons\mathrm{M}}\vec{J}^{\mathrm{G}\rightleftharpoons\mathrm{A}}} & \overline{\overline{Z}}^{\vec{J}^{\mathrm{A}\rightleftharpoons\mathrm{M}}\vec{J}^{\mathrm{A}\rightleftharpoons\mathrm{M}}} & \overline{\overline{Z}}^{\vec{J}^{\mathrm{A}\rightleftharpoons\mathrm{M}}\vec{J}^{\mathrm{A}}} & \overline{\overline{Z}}^{\vec{J}^{\mathrm{A}\rightleftharpoons\mathrm{M}}\vec{M}^{\mathrm{G}\rightleftharpoons\mathrm{A}}} & \overline{\overline{Z}}^{\vec{J}^{\mathrm{A}\rightleftharpoons\mathrm{M}}\vec{M}^{\mathrm{A}\rightleftharpoons\mathrm{M}}} \\ \overline{\overline{Z}}^{\vec{J}^{\mathrm{A}}\vec{J}^{\mathrm{G}\rightleftharpoons\mathrm{A}}} & \overline{\overline{Z}}^{\vec{J}^{\mathrm{A}}\vec{J}^{\mathrm{A}\rightleftharpoons\mathrm{M}}} & \overline{\overline{Z}}^{\vec{J}^{\mathrm{A}}\vec{J}^{\mathrm{A}}} & \overline{\overline{Z}}^{\vec{J}^{\mathrm{A}}\vec{M}^{\mathrm{G}\rightleftharpoons\mathrm{A}}} & \overline{\overline{Z}}^{\vec{J}^{\mathrm{A}}\vec{M}^{\mathrm{A}\rightleftharpoons\mathrm{M}}} \end{bmatrix} \tag{4-25}$$

where the subscript "FCE" is the acronym for "*field continuation equation*", and the superscripts "DoJ" and "DoM" are the acronyms for "*definition of* $\vec{J}^{G \rightleftharpoons A}$" and "*definition of* $\vec{M}^{G \rightleftharpoons A}$".

Theoretically, the Eqs. (4-22) and (4-23) are equivalent to each other, and they have the same *solution space*. If the *basic solutions* (*BS*) used to span the solution space are denoted as $\{\bar{s}_1^{\text{BS}}, \bar{s}_2^{\text{BS}}, \cdots\}$, then any mode contained in the solution space can be expanded as follows:

$$\bar{a}^{\text{AV}} = \sum_i a_i^{\text{BS}} \bar{s}_i^{\text{BS}} = \underbrace{\left[ \bar{s}_1^{\text{BS}}, \bar{s}_2^{\text{BS}}, \cdots \right]}_{\bar{\bar{T}}^{\text{BS}\to\text{AV}}} \cdot \underbrace{\begin{bmatrix} a_1^{\text{BS}} \\ a_2^{\text{BS}} \\ \vdots \end{bmatrix}}_{\bar{a}^{\text{BS}}} \tag{4-26}$$

where the solution space is just the modal space of the tra-antenna shown in the previous Figs. 4-2 and 4-4.

For the convenience of the following discussions, Eqs. (4-20), (4-21), and (4-26) are uniformly written as follows:

$$\bar{a}^{\text{AV}} = \bar{\bar{T}} \cdot \bar{a} \tag{4-27}$$

where $\bar{a} = \bar{a}^{\text{BV}} / \bar{a}^{\text{BS}}$ and correspondingly $\bar{\bar{T}} = \bar{\bar{T}}^{\text{BV}\to\text{AV}} / \bar{\bar{T}}^{\text{BS}\to\text{AV}}$.

### 4.2.3 Power Transport Theorem and Input Power Operator

In this subsection, we provide the *power transport theorem* (*PTT*) and *input power operator* (*IPO*) of the tra-antenna shown in Figs. 4-2 and 4-4.

#### 4.2.3.1 Power Transport Theorem

Applying the results obtained in Chap. 2 to the tra-antenna shown in Figs. 4-2 and 4-4, we immediately have the following PTT for the tra-antenna

$$P^{\text{G} \rightleftharpoons \text{A}} = P_{\text{dis}}^{\text{A}} + P^{\text{A} \rightleftharpoons \text{M}} + j\, P_{\text{sto}}^{\text{A}} \tag{4-28}$$

where $P^{\text{G} \rightleftharpoons \text{A}}$ is the *input power* inputted into the tra-antenna, and $P_{\text{dis}}^{\text{A}}$ is the power dissipated in the tra-antenna, and $P^{\text{A} \rightleftharpoons \text{M}}$ is the *output power* outputted from the tra-antenna, and $P_{\text{sto}}^{\text{A}}$ is the power corresponding to the stored energy in the tra-antenna.

The above-mentioned various powers are as follows:

$$P^{\text{G} \rightleftharpoons \text{A}} = (1/2) \iint_{\mathbb{S}^{\text{G} \rightleftharpoons \text{A}}} \left( \vec{E} \times \vec{H}^{\dagger} \right) \cdot \hat{n}^{\to \text{A}} dS \tag{4-29a}$$

$$P_{\text{dis}}^{\text{A}} = (1/2) \left\langle \vec{\sigma} \cdot \vec{E}, \vec{E} \right\rangle_{\mathbb{V}^{\text{A}}} \tag{4-29b}$$

$$P^{\text{A}\rightleftharpoons\text{M}} = (1/2)\iint_{\mathbb{S}^{\text{A}\rightleftharpoons\text{M}}} \left(\vec{E}\times\vec{H}^\dagger\right)\cdot\hat{n}^{\to\text{M}} dS \tag{4-29c}$$

$$P_{\text{sto}}^{\text{A}} = 2\omega\left[(1/4)\left\langle\vec{H},\vec{\vec{\mu}}\cdot\vec{H}\right\rangle_{\mathbb{V}^\text{A}} - (1/4)\left\langle\vec{\vec{\varepsilon}}\cdot\vec{E},\vec{E}\right\rangle_{\mathbb{V}^\text{A}}\right] \tag{4-29d}$$

where $\hat{n}^{\to\text{M}}$ is the normal direction of $\mathbb{S}^{\text{A}\rightleftharpoons\text{M}}$ and points to free space as shown in Fig. 4-4 (obviously, $\hat{n}^{\to\text{M}} = -\hat{n}^{\to\text{A}}$ on $\mathbb{S}^{\text{A}\rightleftharpoons\text{M}}$).

#### 4.2.3.2 Input Power Operator — Formulation I: Current Form

Based on Eqs. (4-2a)&(4-2b) and the tangential continuity of the $\{\vec{E},\vec{H}\}$ on $\mathbb{S}^{\text{G}\rightleftharpoons\text{A}}$, the IPO $P^{\text{G}\rightleftharpoons\text{A}}$ given in Eq. (4-29a) can be alternatively written as follows:

$$P^{\text{G}\rightleftharpoons\text{A}} = (1/2)\left\langle\hat{n}^{\to\text{A}}\times\vec{J}^{\text{G}\rightleftharpoons\text{A}},\vec{M}^{\text{G}\rightleftharpoons\text{A}}\right\rangle_{\mathbb{S}^{\text{G}\rightleftharpoons\text{A}}} \tag{4-30}$$

and it is just the *current form of IPO*.

Inserting Eq. (4-7) into the above current form, the current form is immediately discretized as follows:

$$P^{\text{G}\rightleftharpoons\text{A}} = \left(\bar{a}^{\text{AV}}\right)^\dagger\cdot\underbrace{\begin{bmatrix} 0 & 0 & 0 & \overline{\overline{C}}^{\vec{J}^{\text{G}\rightleftharpoons\text{A}}\vec{M}^{\text{G}\rightleftharpoons\text{A}}} & 0 \\ 0 & 0 & 0 & 0 & 0 \\ 0 & 0 & 0 & 0 & 0 \\ 0 & 0 & 0 & 0 & 0 \\ 0 & 0 & 0 & 0 & 0 \end{bmatrix}}_{\overline{\overline{P}}_{\text{curAV}}^{\text{G}\rightleftharpoons\text{A}}}\cdot\bar{a}^{\text{AV}} \tag{4-31}$$

where the elements of sub-matrix $\overline{\overline{C}}^{\vec{J}^{\text{G}\rightleftharpoons\text{A}}\vec{M}^{\text{G}\rightleftharpoons\text{A}}}$ are calculated as that $c_{\xi\zeta}^{\vec{J}^{\text{G}\rightleftharpoons\text{A}}\vec{M}^{\text{G}\rightleftharpoons\text{A}}} = (1/2) < \hat{n}^{\to\text{A}}\times\vec{J}_\xi^{\text{G}\rightleftharpoons\text{A}},\vec{M}_\zeta^{\text{G}\rightleftharpoons\text{A}} >_{\mathbb{S}^{\text{G}\rightleftharpoons\text{A}}}$. To obtain the IPO defined on modal space, we substitute Eq. (4-27) into the above Eq. (4-31), and then we have that

$$P^{\text{G}\rightleftharpoons\text{A}} = \bar{a}^\dagger\cdot\underbrace{\left(\overline{\overline{T}}^\dagger\cdot\overline{\overline{P}}_{\text{curAV}}^{\text{G}\rightleftharpoons\text{A}}\cdot\overline{\overline{T}}\right)}_{\overline{\overline{P}}_{\text{cur}}^{\text{G}\rightleftharpoons\text{A}}}\cdot\bar{a} \tag{4-32}$$

where subscript "cur" is to emphasize that $\overline{\overline{P}}_{\text{cur}}^{\text{G}\rightleftharpoons\text{A}}$ originates from discretizing the current form of IPO.

#### 4.2.3.3 Input Power Operator — Formulation II: Field-Current Interaction Forms

Similarly to the results obtained in the previous Chap. 3, the IPO $P^{\text{G}\rightleftharpoons\text{A}}$ given in Eq. (4-29a) also has the following equivalent expressions

$$\begin{aligned} P^{\text{G}\rightleftharpoons\text{A}} &= -(1/2)\left\langle\vec{J}^{\text{G}\rightleftharpoons\text{A}},\vec{E}\right\rangle_{\mathbb{S}^{\text{G}\rightleftharpoons\text{A}}} \\ &= -(1/2)\left\langle\vec{J}^{\text{G}\rightleftharpoons\text{A}},\mathcal{E}\left(\vec{J}^{\text{G}\rightleftharpoons\text{A}}+\vec{J}^{\text{A}\rightleftharpoons\text{M}}+\vec{J}^{\text{A}},\vec{M}^{\text{G}\rightleftharpoons\text{A}}+\vec{M}^{\text{A}\rightleftharpoons\text{M}}\right)\right\rangle_{\tilde{\mathbb{S}}^{\text{G}\rightleftharpoons\text{A}}} \end{aligned} \tag{4-33a}$$

$$
\begin{aligned}
P^{\mathrm{G}\rightleftharpoons\mathrm{A}} &= -(1/2)\left\langle \vec{M}^{\mathrm{G}\rightleftharpoons\mathrm{A}}, \vec{H} \right\rangle_{\mathbb{S}^{\mathrm{G}\rightleftharpoons\mathrm{A}}}^{\dagger} \\
&= -(1/2)\left\langle \vec{M}^{\mathrm{G}\rightleftharpoons\mathrm{A}}, \mathcal{H}\left( \vec{J}^{\mathrm{G}\rightleftharpoons\mathrm{A}} + \vec{J}^{\mathrm{A}\rightleftharpoons\mathrm{M}} + \vec{J}^{\mathrm{A}}, \vec{M}^{\mathrm{G}\rightleftharpoons\mathrm{A}} + \vec{M}^{\mathrm{A}\rightleftharpoons\mathrm{M}} \right) \right\rangle_{\mathbb{S}^{\mathrm{G}\rightleftharpoons\mathrm{A}}}^{\dagger} \quad (4\text{-}33\text{b})
\end{aligned}
$$

and they are just the *field-current interaction forms of IPO*.

Substituting Eq. (4-7) into the above interaction forms, they are immediately discretized as follows:

$$
P^{\mathrm{G}\rightleftharpoons\mathrm{A}} = \left(\bar{a}^{\mathrm{AV}}\right)^{\dagger} \cdot \bar{\bar{P}}_{\mathrm{intAV}}^{\mathrm{G}\rightleftharpoons\mathrm{A}} \cdot \bar{a}^{\mathrm{AV}} \tag{4-34}
$$

in which

$$
\bar{\bar{P}}_{\mathrm{intAV}}^{\mathrm{G}\rightleftharpoons\mathrm{A}} = \begin{cases}
\begin{bmatrix}
\bar{\bar{P}}^{\vec{J}^{\mathrm{G}\rightleftharpoons\mathrm{A}}\vec{J}^{\mathrm{G}\rightleftharpoons\mathrm{A}}} & \bar{\bar{P}}^{\vec{J}^{\mathrm{G}\rightleftharpoons\mathrm{A}}\vec{J}^{\mathrm{A}\rightleftharpoons\mathrm{M}}} & \bar{\bar{P}}^{\vec{J}^{\mathrm{G}\rightleftharpoons\mathrm{A}}\vec{J}^{\mathrm{A}}} & \bar{\bar{P}}^{\vec{J}^{\mathrm{G}\rightleftharpoons\mathrm{A}}\vec{M}^{\mathrm{G}\rightleftharpoons\mathrm{A}}} & \bar{\bar{P}}^{\vec{J}^{\mathrm{G}\rightleftharpoons\mathrm{A}}\vec{M}^{\mathrm{A}\rightleftharpoons\mathrm{M}}} \\
0 & 0 & 0 & 0 & 0 \\
0 & 0 & 0 & 0 & 0 \\
0 & 0 & 0 & 0 & 0 \\
0 & 0 & 0 & 0 & 0
\end{bmatrix} & \text{for Eq.}\,(4\text{-}33\text{a}) \\
\begin{bmatrix}
0 & 0 & 0 & 0 & 0 \\
0 & 0 & 0 & 0 & 0 \\
0 & 0 & 0 & 0 & 0 \\
\bar{\bar{P}}^{\vec{M}^{\mathrm{G}\rightleftharpoons\mathrm{A}}\vec{J}^{\mathrm{G}\rightleftharpoons\mathrm{A}}} & \bar{\bar{P}}^{\vec{M}^{\mathrm{G}\rightleftharpoons\mathrm{A}}\vec{J}^{\mathrm{A}\rightleftharpoons\mathrm{M}}} & \bar{\bar{P}}^{\vec{M}^{\mathrm{G}\rightleftharpoons\mathrm{A}}\vec{J}^{\mathrm{A}}} & \bar{\bar{P}}^{\vec{M}^{\mathrm{G}\rightleftharpoons\mathrm{A}}\vec{M}^{\mathrm{G}\rightleftharpoons\mathrm{A}}} & \bar{\bar{P}}^{\vec{M}^{\mathrm{G}\rightleftharpoons\mathrm{A}}\vec{M}^{\mathrm{A}\rightleftharpoons\mathrm{M}}} \\
0 & 0 & 0 & 0 & 0
\end{bmatrix}^{\dagger} & \text{for Eq.}\,(4\text{-}33\text{b})
\end{cases} \tag{4-35}
$$

where the elements of the sub-matrices are calculated as follows:

$$
p_{\xi\zeta}^{\vec{J}^{\mathrm{G}\rightleftharpoons\mathrm{A}}\vec{J}^{\mathrm{G}\rightleftharpoons\mathrm{A}}} = -(1/2)\left\langle \vec{b}_{\xi}^{\vec{J}^{\mathrm{G}\rightleftharpoons\mathrm{A}}}, \mathcal{E}\left(\vec{b}_{\zeta}^{\vec{J}^{\mathrm{G}\rightleftharpoons\mathrm{A}}}, 0\right) \right\rangle_{\mathbb{S}^{\mathrm{G}\rightleftharpoons\mathrm{A}}} \tag{4-36a}
$$

$$
p_{\xi\zeta}^{\vec{J}^{\mathrm{G}\rightleftharpoons\mathrm{A}}\vec{J}^{\mathrm{A}\rightleftharpoons\mathrm{M}}} = -(1/2)\left\langle \vec{b}_{\xi}^{\vec{J}^{\mathrm{G}\rightleftharpoons\mathrm{A}}}, \mathcal{E}\left(\vec{b}_{\zeta}^{\vec{J}^{\mathrm{A}\rightleftharpoons\mathrm{M}}}, 0\right) \right\rangle_{\mathbb{S}^{\mathrm{G}\rightleftharpoons\mathrm{A}}} \tag{4-36b}
$$

$$
p_{\xi\zeta}^{\vec{J}^{\mathrm{G}\rightleftharpoons\mathrm{A}}\vec{J}^{\mathrm{A}}} = -(1/2)\left\langle \vec{b}_{\xi}^{\vec{J}^{\mathrm{G}\rightleftharpoons\mathrm{A}}}, \mathcal{E}\left(\vec{b}_{\zeta}^{\vec{J}^{\mathrm{A}}}, 0\right) \right\rangle_{\mathbb{S}^{\mathrm{G}\rightleftharpoons\mathrm{A}}} \tag{4-36c}
$$

$$
p_{\xi\zeta}^{\vec{J}^{\mathrm{G}\rightleftharpoons\mathrm{A}}\vec{M}^{\mathrm{G}\rightleftharpoons\mathrm{A}}} = -(1/2)\left\langle \vec{b}_{\xi}^{\vec{J}^{\mathrm{G}\rightleftharpoons\mathrm{A}}}, \mathcal{E}\left(0, \vec{b}_{\zeta}^{\vec{M}^{\mathrm{G}\rightleftharpoons\mathrm{A}}}\right) \right\rangle_{\mathbb{S}^{\mathrm{G}\rightleftharpoons\mathrm{A}}} \tag{4-36d}
$$

$$
p_{\xi\zeta}^{\vec{J}^{\mathrm{G}\rightleftharpoons\mathrm{A}}\vec{M}^{\mathrm{A}\rightleftharpoons\mathrm{M}}} = -(1/2)\left\langle \vec{b}_{\xi}^{\vec{J}^{\mathrm{G}\rightleftharpoons\mathrm{A}}}, \mathcal{E}\left(0, \vec{b}_{\zeta}^{\vec{M}^{\mathrm{A}\rightleftharpoons\mathrm{M}}}\right) \right\rangle_{\mathbb{S}^{\mathrm{G}\rightleftharpoons\mathrm{A}}} \tag{4-36e}
$$

and

$$
p_{\xi\zeta}^{\vec{M}^{\mathrm{G}\rightleftharpoons\mathrm{A}}\vec{J}^{\mathrm{G}\rightleftharpoons\mathrm{A}}} = -(1/2)\left\langle \vec{b}_{\xi}^{\vec{M}^{\mathrm{G}\rightleftharpoons\mathrm{A}}}, \mathcal{H}\left(\vec{b}_{\zeta}^{\vec{J}^{\mathrm{G}\rightleftharpoons\mathrm{A}}}, 0\right) \right\rangle_{\mathbb{S}^{\mathrm{G}\rightleftharpoons\mathrm{A}}} \tag{4-36f}
$$

$$
p_{\xi\zeta}^{\vec{M}^{\mathrm{G}\rightleftharpoons\mathrm{A}}\vec{J}^{\mathrm{A}\rightleftharpoons\mathrm{M}}} = -(1/2)\left\langle \vec{b}_{\xi}^{\vec{M}^{\mathrm{G}\rightleftharpoons\mathrm{A}}}, \mathcal{H}\left(\vec{b}_{\zeta}^{\vec{J}^{\mathrm{A}\rightleftharpoons\mathrm{M}}}, 0\right) \right\rangle_{\mathbb{S}^{\mathrm{G}\rightleftharpoons\mathrm{A}}} \tag{4-36g}
$$

$$
p_{\xi\zeta}^{\vec{M}^{\mathrm{G}\rightleftharpoons\mathrm{A}}\vec{J}^{\mathrm{A}}} = -(1/2)\left\langle \vec{b}_{\xi}^{\vec{M}^{\mathrm{G}\rightleftharpoons\mathrm{A}}}, \mathcal{H}\left(\vec{b}_{\zeta}^{\vec{J}^{\mathrm{A}}}, 0\right) \right\rangle_{\mathbb{S}^{\mathrm{G}\rightleftharpoons\mathrm{A}}} \tag{4-36h}
$$

$$p_{\xi\zeta}^{\vec{M}^{\text{G}\rightleftharpoons\text{A}}\vec{M}^{\text{G}\rightleftharpoons\text{A}}} = -(1/2)\Big\langle \vec{b}_{\xi}^{\vec{M}^{\text{G}\rightleftharpoons\text{A}}}, \mathcal{H}\Big(0, \vec{b}_{\zeta}^{\vec{M}^{\text{G}\rightleftharpoons\text{A}}}\Big)\Big\rangle_{\underset{\sim}{\mathbb{S}}^{\text{G}\rightleftharpoons\text{A}}} \tag{4-36i}$$

$$p_{\xi\zeta}^{\vec{M}^{\text{G}\rightleftharpoons\text{A}}\vec{M}^{\text{A}\rightleftharpoons\text{M}}} = -(1/2)\Big\langle \vec{b}_{\xi}^{\vec{M}^{\text{G}\rightleftharpoons\text{A}}}, \mathcal{H}\Big(0, \vec{b}_{\zeta}^{\vec{M}^{\text{A}\rightleftharpoons\text{M}}}\Big)\Big\rangle_{\underset{\sim}{\mathbb{S}}^{\text{G}\rightleftharpoons\text{A}}} \tag{4-36j}$$

where the integral surface $\underset{\sim}{\mathbb{S}}^{\text{G}\rightleftharpoons\text{A}}$ is shown in Fig. 4-5.

To obtain the IPO defined on modal space, we substitute the previous Eq. (4-27) into the above Eq. (4-34), and then we have that

$$P^{\text{G}\rightleftharpoons\text{A}} = \bar{a}^{\dagger} \cdot \underbrace{\Big(\overline{\overline{T}}^{\dagger} \cdot \overline{\overline{P}}_{\text{intAV}}^{\text{G}\rightleftharpoons\text{A}} \cdot \overline{\overline{T}}\Big)}_{\overline{\overline{P}}_{\text{int}}^{\text{G}\rightleftharpoons\text{A}}} \cdot \bar{a} \tag{4-37}$$

where the subscript "int" is to emphasize that $\overline{\overline{P}}_{\text{int}}^{\text{G}\rightleftharpoons\text{A}}$ originates from discretizing the interaction form of IPO.

For the convenience of the following discussions, the Eqs. (4-32) and (4-37) are uniformly written as follows:

$$P^{\text{G}\rightleftharpoons\text{A}} = \bar{a}^{\dagger} \cdot \overline{\overline{P}}^{\text{G}\rightleftharpoons\text{A}} \cdot \bar{a} \tag{4-38}$$

where $\overline{\overline{P}}^{\text{G}\rightleftharpoons\text{A}} = \overline{\overline{P}}_{\text{cur}}^{\text{G}\rightleftharpoons\text{A}} / \overline{\overline{P}}_{\text{int}}^{\text{G}\rightleftharpoons\text{A}}$.

## 4.2.4 Input-Power-Decoupled Modes

Below, we construct the *input-power-decoupled modes* (*IP-DMs*) of the tra-antenna shown in Figs. 4-2 and 4-4, by using the results obtained above.

### 4.2.4.1 Construction Method

The IP-DMs in the modal space can be derived from solving the following *modal decoupling equation* (or simply called *decoupling equation*)

$$\overline{\overline{P}}_{-}^{\text{G}\rightleftharpoons\text{A}} \cdot \bar{\alpha}_{\xi} = \theta_{\xi}\, \overline{\overline{P}}_{+}^{\text{G}\rightleftharpoons\text{A}} \cdot \bar{\alpha}_{\xi} \tag{4-39}$$

defined on modal space, where $\overline{\overline{P}}_{+}^{\text{G}\rightleftharpoons\text{A}}$ and $\overline{\overline{P}}_{-}^{\text{G}\rightleftharpoons\text{A}}$ are the *positive and negative Hermitian parts* obtained from the *Toeplitz's decomposition* for the $\overline{\overline{P}}^{\text{G}\rightleftharpoons\text{A}}$ in Eq. (4-38). The reason to use symbol "$\theta_{\xi}$" instead of symbol "$\lambda_{\xi}$" will be given in Chap. 9.

If some derived modes $\{\bar{\alpha}_1, \bar{\alpha}_2, \cdots, \bar{\alpha}_d\}$ are $d$-order degenerate, the following *Gram-Schmidt orthogonalization*[46] process is necessary.

$$\left.\begin{aligned} \bar{\alpha}_1 &= \bar{\alpha}_1{}' \\ \bar{\alpha}_2 - \chi_{12}\bar{\alpha}_1{}' &= \bar{\alpha}_2{}' \\ &\cdots \\ \bar{\alpha}_d - \cdots - \chi_{2d}\bar{\alpha}_2{}' - \chi_{1d}\bar{\alpha}_1{}' &= \bar{\alpha}_d{}' \end{aligned}\right\} \tag{4-40}$$

where the coefficients are calculated as follows:

$$\chi_{mn} = \frac{\left(\bar{\alpha}_m{}'\right)^{\dagger} \cdot \overline{\overline{P}}_{+}^{\mathrm{G}\rightleftharpoons\mathrm{A}} \cdot \bar{\alpha}_n}{\left(\bar{\alpha}_m{}'\right)^{\dagger} \cdot \overline{\overline{P}}_{+}^{\mathrm{G}\rightleftharpoons\mathrm{A}} \cdot \bar{\alpha}_m{}'} \tag{4-41}$$

The obtained new modes $\{\bar{\alpha}_1{}', \bar{\alpha}_2{}', \cdots, \bar{\alpha}_d{}'\}$ are input-power-decoupled with each other.

### 4.2.4.2 Modal Decoupling Relation and Parseval's Identity

The modal vectors constructed in the above subsection satisfy the following *modal decoupling relation*

$$\begin{aligned}&\bar{\alpha}_\xi^{\dagger} \cdot \overline{\overline{P}}^{\mathrm{G}\rightleftharpoons\mathrm{A}} \cdot \bar{\alpha}_\zeta \\ &= \underbrace{\left[\mathrm{Re}\left\{P_\xi^{\mathrm{G}\rightleftharpoons\mathrm{A}}\right\} + j\,\mathrm{Im}\left\{P_\xi^{\mathrm{G}\rightleftharpoons\mathrm{A}}\right\}\right]}_{P_\xi^{\mathrm{G}\rightleftharpoons\mathrm{A}}} \delta_{\xi\zeta} \xrightarrow{\text{Normalizing } \mathrm{Re}\{P_\xi^{\mathrm{G}\rightleftharpoons\mathrm{A}}\} \text{ to } 1} \underbrace{\left(1 + j\,\theta_\xi\right)}_{\text{Normalized } P_\xi^{\mathrm{G}\rightleftharpoons\mathrm{A}}} \delta_{\xi\zeta}\end{aligned} \tag{4-42}$$

where $P_\xi^{\mathrm{G}\rightleftharpoons\mathrm{A}}$ is so-called *modal input power* corresponding to the $\xi$-th IP-DM. The physical explanation why $\mathrm{Re}\{P_\xi^{\mathrm{G}\rightleftharpoons\mathrm{A}}\}$ is normalized to 1 has been given in Ref. [18]. The above matrix-vector multiplication decoupling relation can be alternatively written as the following more physical form

$$(1/2)\iint_{\mathbb{S}^{\mathrm{G}\rightleftharpoons\mathrm{A}}} \left(\vec{E}_\zeta \times \vec{H}_\xi^{\dagger}\right) \cdot \hat{n}^{\rightarrow\mathrm{A}} dS = \left(1 + j\,\theta_\xi\right)\delta_{\xi\zeta} \tag{4-43}$$

and then

$$(1/T)\int_{t_0}^{t_0+T}\left[\iint_{\mathbb{S}^{\mathrm{G}\rightleftharpoons\mathrm{A}}} \left(\vec{\mathcal{E}}_\zeta \times \vec{\mathcal{H}}_\xi\right) \cdot \hat{n}^{\rightarrow\mathrm{A}} dS\right] dt = \delta_{\xi\zeta} \tag{4-44}$$

and Eq. (4-44) has a very clear physical meaning — **the modes obtained above are energy-decoupled in any integral period**.

By employing the above decoupling relation, we have the following *Parseval's identity*

$$\sum_\xi \left|c_\xi\right|^2 = (1/T)\int_{t_0}^{t_0+T}\left[\iint_{\mathbb{S}^{\mathrm{G}\rightleftharpoons\mathrm{A}}} \left(\vec{\mathcal{E}} \times \vec{\mathcal{H}}\right) \cdot \hat{n}^{\rightarrow\mathrm{A}} dS\right] dt \tag{4-45}$$

where $c_\xi$ is the *modal expansion coefficient* used in *modal expansion formulation* and can be explicitly calculated as follows:

$$c_\xi = \frac{-(1/2)\left\langle \vec{J}_\xi^{\mathrm{G}\rightleftharpoons\mathrm{A}}, \vec{E}\right\rangle_{\mathbb{S}^{\mathrm{G}\rightleftharpoons\mathrm{A}}}}{1 + j\,\theta_\xi} = \frac{-(1/2)\left\langle \vec{H}, \vec{M}_\xi^{\mathrm{G}\rightleftharpoons\mathrm{A}}\right\rangle_{\mathbb{S}^{\mathrm{G}\rightleftharpoons\mathrm{A}}}}{1 + j\,\theta_\xi} \tag{4-46}$$

where $\{\vec{E}, \vec{H}\}$ are some previously known fields distributing on input port $\mathbb{S}^{\mathrm{G}\rightleftharpoons\mathrm{A}}$.

#### 4.2.4.3 Modal Quantities

For quantitatively describing the modal features, we usually employ

$$\mathrm{MS}_\xi = \frac{1}{|1 + j\,\theta_\xi|} \tag{4-47}$$

called *modal significance* (*MS*), and

$$Z_\xi^{\mathrm{G}\rightleftharpoons\mathrm{A}} = \frac{P_\xi^{\mathrm{G}\rightleftharpoons\mathrm{A}}}{(1/2)\left\langle \vec{J}_\xi^{\mathrm{G}\rightleftharpoons\mathrm{A}}, \vec{J}_\xi^{\mathrm{G}\rightleftharpoons\mathrm{A}} \right\rangle_{\mathbb{S}^{\mathrm{G}\rightleftharpoons\mathrm{A}}}} = \overbrace{\mathrm{Re}\left\{Z_\xi^{\mathrm{G}\rightleftharpoons\mathrm{A}}\right\}}^{R_\xi^{\mathrm{G}\rightleftharpoons\mathrm{A}}} + j\,\overbrace{\mathrm{Im}\left\{Z_\xi^{\mathrm{G}\rightleftharpoons\mathrm{A}}\right\}}^{X_\xi^{\mathrm{G}\rightleftharpoons\mathrm{A}}} \tag{4-48a}$$

$$Y_\xi^{\mathrm{G}\rightleftharpoons\mathrm{A}} = \frac{P_\xi^{\mathrm{G}\rightleftharpoons\mathrm{A}}}{(1/2)\left\langle \vec{M}_\xi^{\mathrm{G}\rightleftharpoons\mathrm{A}}, \vec{M}_\xi^{\mathrm{G}\rightleftharpoons\mathrm{A}} \right\rangle_{\mathbb{S}^{\mathrm{G}\rightleftharpoons\mathrm{A}}}} = \underbrace{\mathrm{Re}\left\{Y_\xi^{\mathrm{G}\rightleftharpoons\mathrm{A}}\right\}}_{G_\xi^{\mathrm{G}\rightleftharpoons\mathrm{A}}} + j\,\underbrace{\mathrm{Im}\left\{Y_\xi^{\mathrm{G}\rightleftharpoons\mathrm{A}}\right\}}_{B_\xi^{\mathrm{G}\rightleftharpoons\mathrm{A}}} \tag{4-48b}$$

called *modal input impedance* (*MII*) and *modal input admittance* (*MIA*) respectively.

The MS and {MII, MIA} quantitatively depict the modal weight in whole modal expansion formulation and the allocation way for the energy carried by the mode.

### 4.3 IP-DMs of Propagation Medium

In this section, the propagation medium, i.e. the free space with a material scatterer, shown in Fig. 4-1 is focused on seperately from the other structures, as shown in the following Fig. 4-6.

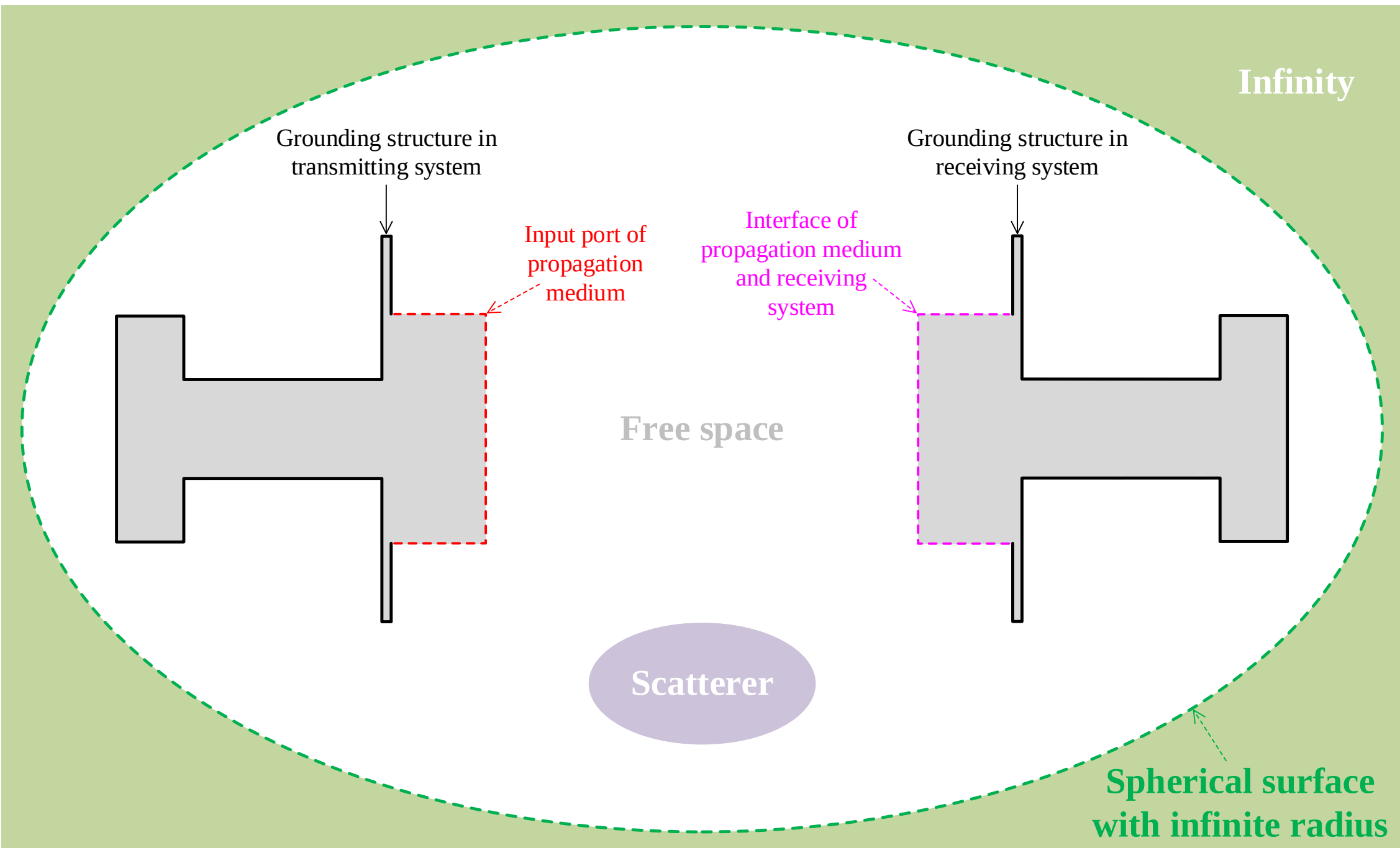

Figure 4-6 Geometry of the propagation medium shown in Fig. 4-1

The general process for constructing the corresponding IP-DMs is shown in Fig. 4-3.

### 4.3.1 Topological Structure and Source-Field Relationships

The topological structure of the propagation medium shown in Fig. 4-6 is exhibited in the following Fig. 4-7.

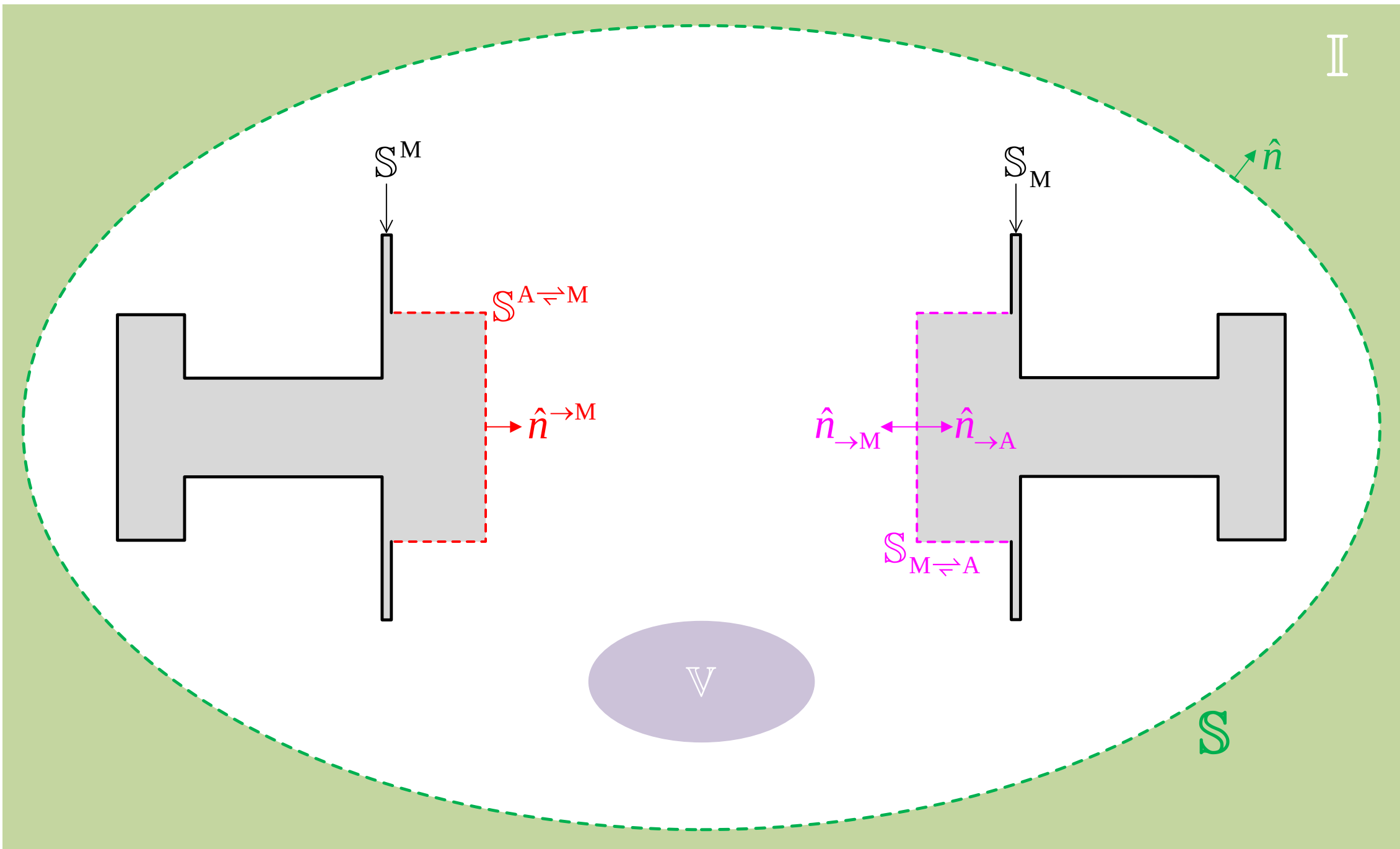

Figure 4-7 Topological structure of the propagation medium shown in Fig. 4-6

In the above Fig. 4-7, surface $\mathbb{S}^{\text{A}\rightleftharpoons\text{M}}$ denotes the input port of the propagation medium. The region occupied by propagation medium is denoted as $\mathbb{M}$, and $\mathbb{M}$ is constituted by two parts — free space $\mathbb{F}$ and material scatterer $\mathbb{V}$ — as shown in Fig. 4-7. The interface between $\mathbb{M}$ and the *grounding structure of transmitting system* is denoted as $\mathbb{S}^{\text{M}}$, and the interface between $\mathbb{M}$ and the *grounding structure of receiving system* is denoted as $\mathbb{S}_{\text{M}}$; the interface between $\mathbb{M}$ and the rec-antenna is denoted as $\mathbb{S}_{\text{M}\rightleftharpoons\text{A}}$. $\mathbb{S}$ is a spherical surface with infinite radius.

Clearly, $\mathbb{S}^{\text{A}\rightleftharpoons\text{M}}$, $\mathbb{S}^{\text{M}}$, $\mathbb{S}$, $\mathbb{S}_{\text{M}}$, and $\mathbb{S}_{\text{A}\rightleftharpoons\text{M}}$ constitute the whole boundary of propagation medium, i.e., $\partial\mathbb{M}=\mathbb{S}^{\text{A}\rightleftharpoons\text{M}}\bigcup\mathbb{S}^{\text{M}}\bigcup\mathbb{S}\bigcup\mathbb{S}_{\text{M}}\bigcup\mathbb{S}_{\text{M}\rightleftharpoons\text{A}}$. In addition, the permeability, permeativity, and conductivity of $\mathbb{V}$ are denoted as $\ddot{\mu}$, $\ddot{\varepsilon}$, and $\ddot{\sigma}$ respectively.

If the equivalent surface currents distributing on $\mathbb{S}^{\text{A}\rightleftharpoons\text{M}}$ are denoted as $\{\vec{J}^{\text{A}\rightleftharpoons\text{M}},\vec{M}^{\text{A}\rightleftharpoons\text{M}}\}$, and the equivalent surface electric current distributing on $\mathbb{S}^{\text{M}}/\mathbb{S}_{\text{M}}$ is denoted as $\vec{J}^{\rm M} / \vec{J}_{\rm M}$ ①, and the equivalent volume currents distributing on $\mathbb{V}$ are denoted as $\{\vec{J},\vec{M}\}$, then the field distributing on $\mathbb{M}$ can be expressed as follows:

$$\vec{F}(\vec{r}) = \mathcal{F}_0\left(\vec{J}^{\rm A\rightleftharpoons M} + \vec{J}^{\rm M} + \vec{J} + \vec{J}_{\rm M} + \vec{J}_{\rm M\rightleftharpoons A}, \vec{M}^{\rm A\rightleftharpoons M} + \vec{M} + \vec{M}_{\rm M\rightleftharpoons A}\right) \quad , \quad \vec{r} \in \mathbb{M} \quad \text{(4-49)}$$

where $\vec{F} = \vec{E} / \vec{H}$, and correspondingly $\mathcal{F}_0 = \mathcal{E}_0 / \mathcal{H}_0$, and the operators are the same as the ones used in the previous Chaps. 2 and 3. The currents $\{\vec{J}^{\rm A\rightleftharpoons M},\vec{M}^{\rm A\rightleftharpoons M}\}$ and the fields $\{\vec{E},\vec{H}\}$ in Eq. (4-49) satisfy the following relations

$$\hat{n}^{\rightarrow \rm M} \times \left[\vec{H}(\vec{r}^{\rm M})\right]_{\vec{r}^{\rm M}\to\vec{r}} = \vec{J}^{\rm A\rightleftharpoons M}(\vec{r}) \quad , \quad \vec{r} \in \mathbb{S}^{\rm A\rightleftharpoons M} \quad \text{(4-50a)}$$

$$\left[\vec{E}(\vec{r}^{\rm M})\right]_{\vec{r}^{\rm M}\to\vec{r}} \times \hat{n}^{\rightarrow \rm M} = \vec{M}^{\rm A\rightleftharpoons M}(\vec{r}) \quad , \quad \vec{r} \in \mathbb{S}^{\rm A\rightleftharpoons M} \quad \text{(4-50b)}$$

and the currents $\{\vec{J},\vec{M}\}$ and fields $\{\vec{E},\vec{H}\}$ in Eq. (4-49) satisfy the following relations

$$j\omega\Delta\vec{\vec{\varepsilon}}_{\rm c} \cdot \vec{E}(\vec{r}) = \vec{J}(\vec{r}) \quad , \quad \vec{r} \in \mathbb{V} \quad \text{(4-51a)}$$

$$j\omega\Delta\vec{\vec{\mu}} \cdot \vec{H}(\vec{r}) = \vec{M}(\vec{r}) \quad , \quad \vec{r} \in \mathbb{V} \quad \text{(4-51b)}$$

and the currents $\{\vec{J}_{\rm M\rightleftharpoons A},\vec{M}_{\rm M\rightleftharpoons A}\}$ and fields $\{\vec{E},\vec{H}\}$ in Eq. (4-49) satisfy the following relations

$$\hat{n}_{\rightarrow \rm M} \times \left[\vec{H}(\vec{r}_{\rm M})\right]_{\vec{r}_{\rm M}\to\vec{r}} = \vec{J}_{\rm M\rightleftharpoons A}(\vec{r}) \quad , \quad \vec{r} \in \mathbb{S}_{\rm M\rightleftharpoons A} \quad \text{(4-52a)}$$

$$\left[\vec{E}(\vec{r}_{\rm M})\right]_{\vec{r}_{\rm M}\to\vec{r}} \times \hat{n}_{\rightarrow \rm M} = \vec{M}_{\rm M\rightleftharpoons A}(\vec{r}) \quad , \quad \vec{r} \in \mathbb{S}_{\rm M\rightleftharpoons A} \quad \text{(4-52b)}$$

In the above Eqs. (4-50)~(4-52), point $\vec{r}^{\rm M} / \vec{r}_{\rm M}$ belongs to $\mathbb{M}$ and $\vec{r}^{\rm M} / \vec{r}_{\rm M}$ approaches the point $\vec{r}$ on $\mathbb{S}^{\rm A\rightleftharpoons M} / \mathbb{S}_{\rm M\rightleftharpoons A}$; and $\hat{n}^{\rightarrow \rm M} / \hat{n}_{\rightarrow \rm M}$ is the normal direction of $\mathbb{S}^{\rm A\rightleftharpoons M}$ / $\mathbb{S}_{\rm M\rightleftharpoons A}$ and points to the interior of $\mathbb{M}$ as shown in Fig. 4-7; $\Delta\vec{\vec{\varepsilon}}_{\rm c} = (\vec{\vec{\varepsilon}} - j\vec{\vec{\sigma}} / \omega) - \vec{\vec{I}}\varepsilon_0$ and $\Delta\vec{\vec{\mu}} = \vec{\vec{\mu}} - \vec{\vec{I}}\mu_0$.

## 4.3.2 Mathematical Description for Modal Space

Combining the Eq. (4-49) with Eqs. (4-50a)&(4-50b), we can obtain the following integral equations

$$\left[\mathcal{H}_0\left(\vec{J}^{\rm A\rightleftharpoons M} + \vec{J}^{\rm M} + \vec{J} + \vec{J}_{\rm M} + \vec{J}_{\rm M\rightleftharpoons A}, \vec{M}^{\rm A\rightleftharpoons M} + \vec{M} + \vec{M}_{\rm M\rightleftharpoons A}\right)\right]^{\tan}_{\vec{r}^{\rm M}\to\vec{r}} = \vec{J}^{\rm A\rightleftharpoons M}(\vec{r}) \times \hat{n}^{\rightarrow \rm M} \quad , \quad \vec{r} \in \mathbb{S}^{\rm A\rightleftharpoons M} \quad \text{(4-53a)}$$

$$\left[\mathcal{E}_0\left(\vec{J}^{\rm A\rightleftharpoons M} + \vec{J}^{\rm M} + \vec{J} + \vec{J}_{\rm M} + \vec{J}_{\rm M\rightleftharpoons A}, \vec{M}^{\rm A\rightleftharpoons M} + \vec{M} + \vec{M}_{\rm M\rightleftharpoons A}\right)\right]^{\tan}_{\vec{r}^{\rm M}\to\vec{r}} = \hat{n}^{\rightarrow \rm M} \times \vec{M}^{\rm A\rightleftharpoons M}(\vec{r}) \quad , \quad \vec{r} \in \mathbb{S}^{\rm A\rightleftharpoons M} \quad \text{(4-53b)}$$

---

① The equivalent surface electric current distributing on $\mathbb{S}^{\rm M} / \mathbb{S}_{\rm M}$ is equal to the induced surface electric current distributing on $\mathbb{S}^{\rm M} / \mathbb{S}_{\rm M}$ [13]. The equivalent surface magnetic current distributing on $\mathbb{S}^{\rm M} / \mathbb{S}_{\rm M}$ is zero, because of the homogeneous tangential electric field boundary condition on $\mathbb{S}^{\rm M} / \mathbb{S}_{\rm M}$ [13].

about currents $\{\vec{J}^{\mathrm{A}\rightleftharpoons\mathrm{M}},\vec{M}^{\mathrm{A}\rightleftharpoons\mathrm{M}}\}$, $\vec{J}^{\mathrm{M}}$, $\{\vec{J},\vec{M}\}$, $\vec{J}_{\mathrm{M}}$, and $\{\vec{J}_{\mathrm{M}\rightleftharpoons\mathrm{A}},\vec{M}_{\mathrm{M}\rightleftharpoons\mathrm{A}}\}$, where the superscript "tan" represents that only the tangential components of the fields satisfy the equations.

Combining the Eq. (4-49) with Eqs. (4-51a)&(4-51b), we can obtain the following integral equations

$$\begin{aligned}&\mathcal{E}_0\left(\vec{J}^{\mathrm{A}\rightleftharpoons\mathrm{M}}+\vec{J}^{\mathrm{M}}+\vec{J}+\vec{J}_{\mathrm{M}}+\vec{J}_{\mathrm{M}\rightleftharpoons\mathrm{A}},\vec{M}^{\mathrm{A}\rightleftharpoons\mathrm{M}}+\vec{M}+\vec{M}_{\mathrm{M}\rightleftharpoons\mathrm{A}}\right)\\&=\left(j\omega\Delta\vec{\vec{\varepsilon}}_{\mathrm{c}}\right)^{-1}\cdot\vec{J}\left(\vec{r}\right)\qquad,\qquad\vec{r}\in\mathbb{V}\qquad(4\text{-}54\mathrm{a})\end{aligned}$$

$$\begin{aligned}&\mathcal{H}_0\left(\vec{J}^{\mathrm{A}\rightleftharpoons\mathrm{M}}+\vec{J}^{\mathrm{M}}+\vec{J}+\vec{J}_{\mathrm{M}}+\vec{J}_{\mathrm{M}\rightleftharpoons\mathrm{A}},\vec{M}^{\mathrm{A}\rightleftharpoons\mathrm{M}}+\vec{M}+\vec{M}_{\mathrm{M}\rightleftharpoons\mathrm{A}}\right)\\&=\left(j\omega\Delta\vec{\vec{\mu}}\right)^{-1}\cdot\vec{M}\left(\vec{r}\right)\qquad,\qquad\vec{r}\in\mathbb{V}\qquad(4\text{-}54\mathrm{b})\end{aligned}$$

about currents $\{\vec{J}^{\mathrm{A}\rightleftharpoons\mathrm{M}},\vec{M}^{\mathrm{A}\rightleftharpoons\mathrm{M}}\}$, $\vec{J}^{\mathrm{M}}$, $\{\vec{J},\vec{M}\}$, $\vec{J}_{\mathrm{M}}$, and $\{\vec{J}_{\mathrm{M}\rightleftharpoons\mathrm{A}},\vec{M}_{\mathrm{M}\rightleftharpoons\mathrm{A}}\}$.

Based on Eq. (4-49) and the homogeneous tangential electric field boundary conditions on surfaces $\mathbb{S}^{\mathrm{M}}$ and $\mathbb{S}_{\mathrm{M}}$, we can obtain the following electric field integral equation

$$\left[\mathcal{E}_0\left(\vec{J}^{\mathrm{A}\rightleftharpoons\mathrm{M}}+\vec{J}^{\mathrm{M}}+\vec{J}+\vec{J}_{\mathrm{M}}+\vec{J}_{\mathrm{M}\rightleftharpoons\mathrm{A}},\vec{M}^{\mathrm{A}\rightleftharpoons\mathrm{M}}+\vec{M}+\vec{M}_{\mathrm{M}\rightleftharpoons\mathrm{A}}\right)\right]_{\vec{r}^{\mathrm{M}}\to\vec{r}}^{\tan}=0\quad,\quad\vec{r}\in\mathbb{S}^{\mathrm{M}}\quad(4\text{-}55)$$

$$\left[\mathcal{E}_0\left(\vec{J}^{\mathrm{A}\rightleftharpoons\mathrm{M}}+\vec{J}^{\mathrm{M}}+\vec{J}+\vec{J}_{\mathrm{M}}+\vec{J}_{\mathrm{M}\rightleftharpoons\mathrm{A}},\vec{M}^{\mathrm{A}\rightleftharpoons\mathrm{M}}+\vec{M}+\vec{M}_{\mathrm{M}\rightleftharpoons\mathrm{A}}\right)\right]_{\vec{r}_{\mathrm{M}}\to\vec{r}}^{\tan}=0\quad,\quad\vec{r}\in\mathbb{S}_{\mathrm{M}}\quad(4\text{-}56)$$

about currents $\{\vec{J}^{\mathrm{A}\rightleftharpoons\mathrm{M}},\vec{M}^{\mathrm{A}\rightleftharpoons\mathrm{M}}\}$, $\vec{J}^{\mathrm{M}}$, $\{\vec{J},\vec{M}\}$, $\vec{J}_{\mathrm{M}}$, and $\{\vec{J}_{\mathrm{M}\rightleftharpoons\mathrm{A}},\vec{M}_{\mathrm{M}\rightleftharpoons\mathrm{A}}\}$.

Combining the Eq. (4-49) with Eqs. (4-52a)&(4-52b), we can obtain the following integral equations

$$\begin{aligned}&\left[\mathcal{H}_0\left(\vec{J}^{\mathrm{A}\rightleftharpoons\mathrm{M}}+\vec{J}^{\mathrm{M}}+\vec{J}+\vec{J}_{\mathrm{M}}+\vec{J}_{\mathrm{M}\rightleftharpoons\mathrm{A}},\vec{M}^{\mathrm{A}\rightleftharpoons\mathrm{M}}+\vec{M}+\vec{M}_{\mathrm{M}\rightleftharpoons\mathrm{A}}\right)\right]_{\vec{r}_{\mathrm{M}}\to\vec{r}}^{\tan}\\&=\vec{J}_{\mathrm{M}\rightleftharpoons\mathrm{A}}\left(\vec{r}\right)\times\hat{n}_{\to\mathrm{M}}\qquad,\qquad\vec{r}\in\mathbb{S}_{\mathrm{M}\rightleftharpoons\mathrm{A}}\qquad(4\text{-}57\mathrm{a})\end{aligned}$$

$$\begin{aligned}&\left[\mathcal{E}_0\left(\vec{J}^{\mathrm{A}\rightleftharpoons\mathrm{M}}+\vec{J}^{\mathrm{M}}+\vec{J}+\vec{J}_{\mathrm{M}}+\vec{J}_{\mathrm{M}\rightleftharpoons\mathrm{A}},\vec{M}^{\mathrm{A}\rightleftharpoons\mathrm{M}}+\vec{M}+\vec{M}_{\mathrm{M}\rightleftharpoons\mathrm{A}}\right)\right]_{\vec{r}_{\mathrm{M}}\to\vec{r}}^{\tan}\\&=\hat{n}_{\to\mathrm{M}}\times\vec{M}_{\mathrm{M}\rightleftharpoons\mathrm{A}}\left(\vec{r}\right)\qquad,\qquad\vec{r}\in\mathbb{S}_{\mathrm{M}\rightleftharpoons\mathrm{A}}\qquad(4\text{-}57\mathrm{b})\end{aligned}$$

about currents $\{\vec{J}^{\mathrm{A}\rightleftharpoons\mathrm{M}},\vec{M}^{\mathrm{A}\rightleftharpoons\mathrm{M}}\}$, $\vec{J}^{\mathrm{M}}$, $\{\vec{J},\vec{M}\}$, $\vec{J}_{\mathrm{M}}$, and $\{\vec{J}_{\mathrm{M}\rightleftharpoons\mathrm{A}},\vec{M}_{\mathrm{M}\rightleftharpoons\mathrm{A}}\}$.

The above Eqs. (4-53a)~(4-57b) are a complete mathematical description for the modal space of the propagation medium shown in Figs. 4-6 and 4-7. If the currents $\{\vec{J}^{\mathrm{A}\rightleftharpoons\mathrm{M}},\vec{M}^{\mathrm{A}\rightleftharpoons\mathrm{M}}\}$, $\vec{J}^{\mathrm{M}}$, $\{\vec{J},\vec{M}\}$, $\vec{J}_{\mathrm{M}}$, and $\{\vec{J}_{\mathrm{M}\rightleftharpoons\mathrm{A}},\vec{M}_{\mathrm{M}\rightleftharpoons\mathrm{A}}\}$ are expanded in terms of some proper basis functions, and Eqs. (4-53a), (4-53b), (4-54a), (4-54b), (4-55), (4-56), (4-57a), and (4-57b) are tested with $\{\vec{b}_\xi^{\vec{M}^{\mathrm{A}\rightleftharpoons\mathrm{M}}}\}$, $\{\vec{b}_\xi^{\vec{J}^{\mathrm{A}\rightleftharpoons\mathrm{M}}}\}$, $\{\vec{b}_\xi^{\vec{J}}\}$, $\{\vec{b}_\xi^{\vec{M}}\}$, $\{\vec{b}_\xi^{\vec{J}^{\mathrm{M}}}\}$, $\{\vec{b}_\xi^{\vec{J}_{\mathrm{M}}}\}$, $\{\vec{b}_\xi^{\vec{M}_{\mathrm{M}\rightleftharpoons\mathrm{A}}}\}$, and $\{\vec{b}_\xi^{\vec{J}_{\mathrm{M}\rightleftharpoons\mathrm{A}}}\}$ respectively, then the integral equations are immediately discretized into the following matrix equations

$$\begin{aligned}0 = \ & \overline{\overline{Z}}^{\vec{M}^{\mathrm{A}\rightleftharpoons\mathrm{M}}\vec{J}^{\mathrm{A}\rightleftharpoons\mathrm{M}}} \cdot \overline{a}^{\vec{J}^{\mathrm{A}\rightleftharpoons\mathrm{M}}} + \overline{\overline{Z}}^{\vec{M}^{\mathrm{A}\rightleftharpoons\mathrm{M}}\vec{J}^{\mathrm{M}}} \cdot \overline{a}^{\vec{J}^{\mathrm{M}}} + \overline{\overline{Z}}^{\vec{M}^{\mathrm{A}\rightleftharpoons\mathrm{M}}\vec{J}} \cdot \overline{a}^{\vec{J}} + \overline{\overline{Z}}^{\vec{M}^{\mathrm{A}\rightleftharpoons\mathrm{M}}\vec{J}_{\mathrm{M}}} \cdot \overline{a}^{\vec{J}_{\mathrm{M}}} + \overline{\overline{Z}}^{\vec{M}^{\mathrm{A}\rightleftharpoons\mathrm{M}}\vec{J}_{\mathrm{M}\rightleftharpoons\mathrm{A}}} \cdot \overline{a}^{\vec{J}_{\mathrm{M}\rightleftharpoons\mathrm{A}}} \\ & + \overline{\overline{Z}}^{\vec{M}^{\mathrm{A}\rightleftharpoons\mathrm{M}}\vec{M}^{\mathrm{A}\rightleftharpoons\mathrm{M}}} \cdot \overline{a}^{\vec{M}^{\mathrm{A}\rightleftharpoons\mathrm{M}}} + \overline{\overline{Z}}^{\vec{M}^{\mathrm{A}\rightleftharpoons\mathrm{M}}\vec{M}} \cdot \overline{a}^{\vec{M}} + \overline{\overline{Z}}^{\vec{M}^{\mathrm{A}\rightleftharpoons\mathrm{M}}\vec{M}_{\mathrm{M}\rightleftharpoons\mathrm{A}}} \cdot \overline{a}^{\vec{M}_{\mathrm{M}\rightleftharpoons\mathrm{A}}} \end{aligned} \tag{4-58a}$$

$$\begin{aligned}0 = \ & \overline{\overline{Z}}^{\vec{J}^{\mathrm{A}\rightleftharpoons\mathrm{M}}\vec{J}^{\mathrm{A}\rightleftharpoons\mathrm{M}}} \cdot \overline{a}^{\vec{J}^{\mathrm{A}\rightleftharpoons\mathrm{M}}} + \overline{\overline{Z}}^{\vec{J}^{\mathrm{A}\rightleftharpoons\mathrm{M}}\vec{J}^{\mathrm{M}}} \cdot \overline{a}^{\vec{J}^{\mathrm{M}}} + \overline{\overline{Z}}^{\vec{J}^{\mathrm{A}\rightleftharpoons\mathrm{M}}\vec{J}} \cdot \overline{a}^{\vec{J}} + \overline{\overline{Z}}^{\vec{J}^{\mathrm{A}\rightleftharpoons\mathrm{M}}\vec{J}_{\mathrm{M}}} \cdot \overline{a}^{\vec{J}_{\mathrm{M}}} + \overline{\overline{Z}}^{\vec{J}^{\mathrm{A}\rightleftharpoons\mathrm{M}}\vec{J}_{\mathrm{M}\rightleftharpoons\mathrm{A}}} \cdot \overline{a}^{\vec{J}_{\mathrm{M}\rightleftharpoons\mathrm{A}}} \\ & + \overline{\overline{Z}}^{\vec{J}^{\mathrm{A}\rightleftharpoons\mathrm{M}}\vec{M}^{\mathrm{A}\rightleftharpoons\mathrm{M}}} \cdot \overline{a}^{\vec{M}^{\mathrm{A}\rightleftharpoons\mathrm{M}}} + \overline{\overline{Z}}^{\vec{J}^{\mathrm{A}\rightleftharpoons\mathrm{M}}\vec{M}} \cdot \overline{a}^{\vec{M}} + \overline{\overline{Z}}^{\vec{J}^{\mathrm{A}\rightleftharpoons\mathrm{M}}\vec{M}_{\mathrm{M}\rightleftharpoons\mathrm{A}}} \cdot \overline{a}^{\vec{M}_{\mathrm{M}\rightleftharpoons\mathrm{A}}} \end{aligned} \tag{4-58b}$$

and

$$\begin{aligned}0 = \ & \overline{\overline{Z}}^{\vec{J}\vec{J}^{\mathrm{A}\rightleftharpoons\mathrm{M}}} \cdot \overline{a}^{\vec{J}^{\mathrm{A}\rightleftharpoons\mathrm{M}}} + \overline{\overline{Z}}^{\vec{J}\vec{J}^{\mathrm{M}}} \cdot \overline{a}^{\vec{J}^{\mathrm{M}}} + \overline{\overline{Z}}^{\vec{J}\vec{J}} \cdot \overline{a}^{\vec{J}} + \overline{\overline{Z}}^{\vec{J}\vec{J}_{\mathrm{M}}} \cdot \overline{a}^{\vec{J}_{\mathrm{M}}} + \overline{\overline{Z}}^{\vec{J}\vec{J}_{\mathrm{M}\rightleftharpoons\mathrm{A}}} \cdot \overline{a}^{\vec{J}_{\mathrm{M}\rightleftharpoons\mathrm{A}}} + \overline{\overline{Z}}^{\vec{J}\vec{M}^{\mathrm{A}\rightleftharpoons\mathrm{M}}} \cdot \overline{a}^{\vec{M}^{\mathrm{A}\rightleftharpoons\mathrm{M}}} \\ & + \overline{\overline{Z}}^{\vec{J}\vec{M}} \cdot \overline{a}^{\vec{M}} + \overline{\overline{Z}}^{\vec{J}\vec{M}_{\mathrm{M}\rightleftharpoons\mathrm{A}}} \cdot \overline{a}^{\vec{M}_{\mathrm{M}\rightleftharpoons\mathrm{A}}} \end{aligned} \tag{4-59a}$$

$$\begin{aligned}0 = \ & \overline{\overline{Z}}^{\vec{M}\vec{J}^{\mathrm{A}\rightleftharpoons\mathrm{M}}} \cdot \overline{a}^{\vec{J}^{\mathrm{A}\rightleftharpoons\mathrm{M}}} + \overline{\overline{Z}}^{\vec{M}\vec{J}^{\mathrm{M}}} \cdot \overline{a}^{\vec{J}^{\mathrm{M}}} + \overline{\overline{Z}}^{\vec{M}\vec{J}} \cdot \overline{a}^{\vec{J}} + \overline{\overline{Z}}^{\vec{M}\vec{J}_{\mathrm{M}}} \cdot \overline{a}^{\vec{J}_{\mathrm{M}}} + \overline{\overline{Z}}^{\vec{M}\vec{J}_{\mathrm{M}\rightleftharpoons\mathrm{A}}} \cdot \overline{a}^{\vec{J}_{\mathrm{M}\rightleftharpoons\mathrm{A}}} + \overline{\overline{Z}}^{\vec{M}\vec{M}^{\mathrm{A}\rightleftharpoons\mathrm{M}}} \cdot \overline{a}^{\vec{M}^{\mathrm{A}\rightleftharpoons\mathrm{M}}} \\ & + \overline{\overline{Z}}^{\vec{M}\vec{M}} \cdot \overline{a}^{\vec{M}} + \overline{\overline{Z}}^{\vec{M}\vec{M}_{\mathrm{M}\rightleftharpoons\mathrm{A}}} \cdot \overline{a}^{\vec{M}_{\mathrm{M}\rightleftharpoons\mathrm{A}}} \end{aligned} \tag{4-59b}$$

and

$$\begin{aligned}0 = \ & \overline{\overline{Z}}^{\vec{J}^{\mathrm{M}}\vec{J}^{\mathrm{A}\rightleftharpoons\mathrm{M}}} \cdot \overline{a}^{\vec{J}^{\mathrm{A}\rightleftharpoons\mathrm{M}}} + \overline{\overline{Z}}^{\vec{J}^{\mathrm{M}}\vec{J}^{\mathrm{M}}} \cdot \overline{a}^{\vec{J}^{\mathrm{M}}} + \overline{\overline{Z}}^{\vec{J}^{\mathrm{M}}\vec{J}} \cdot \overline{a}^{\vec{J}} + \overline{\overline{Z}}^{\vec{J}^{\mathrm{M}}\vec{J}_{\mathrm{M}}} \cdot \overline{a}^{\vec{J}_{\mathrm{M}}} + \overline{\overline{Z}}^{\vec{J}^{\mathrm{M}}\vec{J}_{\mathrm{M}\rightleftharpoons\mathrm{A}}} \cdot \overline{a}^{\vec{J}_{\mathrm{M}\rightleftharpoons\mathrm{A}}} + \overline{\overline{Z}}^{\vec{J}^{\mathrm{M}}\vec{M}^{\mathrm{A}\rightleftharpoons\mathrm{M}}} \cdot \overline{a}^{\vec{M}^{\mathrm{A}\rightleftharpoons\mathrm{M}}} \\ & + \overline{\overline{Z}}^{\vec{J}^{\mathrm{M}}\vec{M}} \cdot \overline{a}^{\vec{M}} + \overline{\overline{Z}}^{\vec{J}^{\mathrm{M}}\vec{M}_{\mathrm{M}\rightleftharpoons\mathrm{A}}} \cdot \overline{a}^{\vec{M}_{\mathrm{M}\rightleftharpoons\mathrm{A}}} \end{aligned} \tag{4-60}$$

and

$$\begin{aligned}0 = \ & \overline{\overline{Z}}^{\vec{J}_{\mathrm{M}}\vec{J}^{\mathrm{A}\rightleftharpoons\mathrm{M}}} \cdot \overline{a}^{\vec{J}^{\mathrm{A}\rightleftharpoons\mathrm{M}}} + \overline{\overline{Z}}^{\vec{J}_{\mathrm{M}}\vec{J}^{\mathrm{M}}} \cdot \overline{a}^{\vec{J}^{\mathrm{M}}} + \overline{\overline{Z}}^{\vec{J}_{\mathrm{M}}\vec{J}} \cdot \overline{a}^{\vec{J}} + \overline{\overline{Z}}^{\vec{J}_{\mathrm{M}}\vec{J}_{\mathrm{M}}} \cdot \overline{a}^{\vec{J}_{\mathrm{M}}} + \overline{\overline{Z}}^{\vec{J}_{\mathrm{M}}\vec{J}_{\mathrm{M}\rightleftharpoons\mathrm{A}}} \cdot \overline{a}^{\vec{J}_{\mathrm{M}\rightleftharpoons\mathrm{A}}} + \overline{\overline{Z}}^{\vec{J}_{\mathrm{M}}\vec{M}^{\mathrm{A}\rightleftharpoons\mathrm{M}}} \cdot \overline{a}^{\vec{M}^{\mathrm{A}\rightleftharpoons\mathrm{M}}} \\ & + \overline{\overline{Z}}^{\vec{J}_{\mathrm{M}}\vec{M}} \cdot \overline{a}^{\vec{M}} + \overline{\overline{Z}}^{\vec{J}_{\mathrm{M}}\vec{M}_{\mathrm{M}\rightleftharpoons\mathrm{A}}} \cdot \overline{a}^{\vec{M}_{\mathrm{M}\rightleftharpoons\mathrm{A}}} \end{aligned} \tag{4-61}$$

and

$$\begin{aligned}0 = \ & \overline{\overline{Z}}^{\vec{M}_{\mathrm{M}\rightleftharpoons\mathrm{A}}\vec{J}^{\mathrm{A}\rightleftharpoons\mathrm{M}}} \cdot \overline{a}^{\vec{J}^{\mathrm{A}\rightleftharpoons\mathrm{M}}} + \overline{\overline{Z}}^{\vec{M}_{\mathrm{M}\rightleftharpoons\mathrm{A}}\vec{J}^{\mathrm{M}}} \cdot \overline{a}^{\vec{J}^{\mathrm{M}}} + \overline{\overline{Z}}^{\vec{M}_{\mathrm{M}\rightleftharpoons\mathrm{A}}\vec{J}} \cdot \overline{a}^{\vec{J}} + \overline{\overline{Z}}^{\vec{M}_{\mathrm{M}\rightleftharpoons\mathrm{A}}\vec{J}_{\mathrm{M}}} \cdot \overline{a}^{\vec{J}_{\mathrm{M}}} + \overline{\overline{Z}}^{\vec{M}_{\mathrm{M}\rightleftharpoons\mathrm{A}}\vec{J}_{\mathrm{M}\rightleftharpoons\mathrm{A}}} \cdot \overline{a}^{\vec{J}_{\mathrm{M}\rightleftharpoons\mathrm{A}}} \\ & + \overline{\overline{Z}}^{\vec{M}_{\mathrm{M}\rightleftharpoons\mathrm{A}}\vec{M}^{\mathrm{A}\rightleftharpoons\mathrm{M}}} \cdot \overline{a}^{\vec{M}^{\mathrm{A}\rightleftharpoons\mathrm{M}}} + \overline{\overline{Z}}^{\vec{M}_{\mathrm{M}\rightleftharpoons\mathrm{A}}\vec{M}} \cdot \overline{a}^{\vec{M}} + \overline{\overline{Z}}^{\vec{M}_{\mathrm{M}\rightleftharpoons\mathrm{A}}\vec{M}_{\mathrm{M}\rightleftharpoons\mathrm{A}}} \cdot \overline{a}^{\vec{M}_{\mathrm{M}\rightleftharpoons\mathrm{A}}} \end{aligned} \tag{4-62a}$$

$$\begin{aligned}0 = \ & \overline{\overline{Z}}^{\vec{J}_{\mathrm{M}\rightleftharpoons\mathrm{A}}\vec{J}^{\mathrm{A}\rightleftharpoons\mathrm{M}}} \cdot \overline{a}^{\vec{J}^{\mathrm{A}\rightleftharpoons\mathrm{M}}} + \overline{\overline{Z}}^{\vec{J}_{\mathrm{M}\rightleftharpoons\mathrm{A}}\vec{J}^{\mathrm{M}}} \cdot \overline{a}^{\vec{J}^{\mathrm{M}}} + \overline{\overline{Z}}^{\vec{J}_{\mathrm{M}\rightleftharpoons\mathrm{A}}\vec{J}} \cdot \overline{a}^{\vec{J}} + \overline{\overline{Z}}^{\vec{J}_{\mathrm{M}\rightleftharpoons\mathrm{A}}\vec{J}_{\mathrm{M}}} \cdot \overline{a}^{\vec{J}_{\mathrm{M}}} + \overline{\overline{Z}}^{\vec{J}_{\mathrm{M}\rightleftharpoons\mathrm{A}}\vec{J}_{\mathrm{M}\rightleftharpoons\mathrm{A}}} \cdot \overline{a}^{\vec{J}_{\mathrm{M}\rightleftharpoons\mathrm{A}}} \\ & + \overline{\overline{Z}}^{\vec{J}_{\mathrm{M}\rightleftharpoons\mathrm{A}}\vec{M}^{\mathrm{A}\rightleftharpoons\mathrm{M}}} \cdot \overline{a}^{\vec{M}^{\mathrm{A}\rightleftharpoons\mathrm{M}}} + \overline{\overline{Z}}^{\vec{J}_{\mathrm{M}\rightleftharpoons\mathrm{A}}\vec{M}} \cdot \overline{a}^{\vec{M}} + \overline{\overline{Z}}^{\vec{J}_{\mathrm{M}\rightleftharpoons\mathrm{A}}\vec{M}_{\mathrm{M}\rightleftharpoons\mathrm{A}}} \cdot \overline{a}^{\vec{M}_{\mathrm{M}\rightleftharpoons\mathrm{A}}} \end{aligned} \tag{4-62b}$$

The formulations used to calculate the elements of the matrices in Eq. (4-58a) are as

$$z_{\xi\zeta}^{\vec{M}^{\mathrm{A}\rightleftharpoons\mathrm{M}}\vec{J}^{\mathrm{A}\rightleftharpoons\mathrm{M}}} = \left\langle \vec{b}_{\xi}^{\vec{M}^{\mathrm{A}\rightleftharpoons\mathrm{M}}}, \mathrm{P.V.}\,\mathcal{K}_0\left(\vec{b}_{\zeta}^{\vec{J}^{\mathrm{A}\rightleftharpoons\mathrm{M}}}\right)\right\rangle_{\mathbb{S}^{\mathrm{A}\rightleftharpoons\mathrm{M}}} - \left\langle \vec{b}_{\xi}^{\vec{M}^{\mathrm{A}\rightleftharpoons\mathrm{M}}}, \frac{1}{2}\vec{b}_{\zeta}^{\vec{J}^{\mathrm{A}\rightleftharpoons\mathrm{M}}} \times \hat{n}^{\rightarrow\mathrm{M}}\right\rangle_{\mathbb{S}^{\mathrm{A}\rightleftharpoons\mathrm{M}}} \tag{4-63a}$$

$$z_{\xi\zeta}^{\vec{M}^{\mathrm{A}\rightleftharpoons\mathrm{M}}\vec{J}^{\mathrm{M}}} = \left\langle \vec{b}_{\xi}^{\vec{M}^{\mathrm{A}\rightleftharpoons\mathrm{M}}}, \mathcal{K}_0\left(\vec{b}_{\zeta}^{\vec{J}^{\mathrm{M}}}\right)\right\rangle_{\mathbb{S}^{\mathrm{A}\rightleftharpoons\mathrm{M}}} \tag{4-63b}$$

$$z_{\xi\zeta}^{\vec{M}^{\mathrm{A}\rightleftharpoons\mathrm{M}}\vec{J}} = \left\langle \vec{b}_{\xi}^{\vec{M}^{\mathrm{A}\rightleftharpoons\mathrm{M}}}, \mathcal{K}_0\left(\vec{b}_{\zeta}^{\vec{J}}\right)\right\rangle_{\mathbb{S}^{\mathrm{A}\rightleftharpoons\mathrm{M}}} \tag{4-63c}$$

$$z_{\xi\zeta}^{\vec{M}^{\mathrm{A}\rightleftharpoons\mathrm{M}}\vec{J}_{\mathrm{M}}} = \left\langle \vec{b}_{\xi}^{\vec{M}^{\mathrm{A}\rightleftharpoons\mathrm{M}}}, \mathcal{K}_0\left(\vec{b}_{\zeta}^{\vec{J}_{\mathrm{M}}}\right)\right\rangle_{\mathbb{S}^{\mathrm{A}\rightleftharpoons\mathrm{M}}} \tag{4-63d}$$

$$z_{\xi\zeta}^{\vec{M}^{\mathrm{A}\rightleftharpoons\mathrm{M}}\vec{J}_{\mathrm{M}\rightleftharpoons\mathrm{A}}} = \left\langle \vec{b}_{\xi}^{\vec{M}^{\mathrm{A}\rightleftharpoons\mathrm{M}}}, \mathcal{K}_0\left(\vec{b}_{\zeta}^{\vec{J}_{\mathrm{M}\rightleftharpoons\mathrm{A}}}\right)\right\rangle_{\mathbb{S}^{\mathrm{A}\rightleftharpoons\mathrm{M}}} \tag{4-63e}$$

$$z_{\xi\zeta}^{\vec{M}^{\text{A}\rightleftharpoons\text{M}}\vec{M}^{\text{A}\rightleftharpoons\text{M}}} = \left\langle \vec{b}_{\xi}^{\vec{M}^{\text{A}\rightleftharpoons\text{M}}}, -j\omega\varepsilon_0\mathcal{L}_0\left(\vec{b}_{\zeta}^{\vec{M}^{\text{A}\rightleftharpoons\text{M}}}\right)\right\rangle_{\mathbb{S}^{\text{A}\rightleftharpoons\text{M}}} \tag{4-63f}$$

$$z_{\xi\zeta}^{\vec{M}^{\text{A}\rightleftharpoons\text{M}}\vec{M}} = \left\langle \vec{b}_{\xi}^{\vec{M}^{\text{A}\rightleftharpoons\text{M}}}, -j\omega\varepsilon_0\mathcal{L}_0\left(\vec{b}_{\zeta}^{\vec{M}}\right)\right\rangle_{\mathbb{S}^{\text{A}\rightleftharpoons\text{M}}} \tag{4-63g}$$

$$z_{\xi\zeta}^{\vec{M}^{\text{A}\rightleftharpoons\text{M}}\vec{M}_{\text{M}\rightleftharpoons\text{A}}} = \left\langle \vec{b}_{\xi}^{\vec{M}^{\text{A}\rightleftharpoons\text{M}}}, -j\omega\varepsilon_0\mathcal{L}_0\left(\vec{b}_{\zeta}^{\vec{M}_{\text{M}\rightleftharpoons\text{A}}}\right)\right\rangle_{\mathbb{S}^{\text{A}\rightleftharpoons\text{M}}} \tag{4-63h}$$

The formulations used to calculate the elements of the matrices in Eq. (4-58b) are as follows:

$$z_{\xi\zeta}^{\vec{J}^{\text{A}\rightleftharpoons\text{M}}\vec{J}^{\text{A}\rightleftharpoons\text{M}}} = \left\langle \vec{b}_{\xi}^{\vec{J}^{\text{A}\rightleftharpoons\text{M}}}, -j\omega\mu_0\mathcal{L}_0\left(\vec{b}_{\zeta}^{\vec{J}^{\text{A}\rightleftharpoons\text{M}}}\right)\right\rangle_{\mathbb{S}^{\text{A}\rightleftharpoons\text{M}}} \tag{4-64a}$$

$$z_{\xi\zeta}^{\vec{J}^{\text{A}\rightleftharpoons\text{M}}\vec{J}^{\text{M}}} = \left\langle \vec{b}_{\xi}^{\vec{J}^{\text{A}\rightleftharpoons\text{M}}}, -j\omega\mu_0\mathcal{L}_0\left(\vec{b}_{\zeta}^{\vec{J}^{\text{M}}}\right)\right\rangle_{\mathbb{S}^{\text{A}\rightleftharpoons\text{M}}} \tag{4-64b}$$

$$z_{\xi\zeta}^{\vec{J}^{\text{A}\rightleftharpoons\text{M}}\vec{J}} = \left\langle \vec{b}_{\xi}^{\vec{J}^{\text{A}\rightleftharpoons\text{M}}}, -j\omega\mu_0\mathcal{L}_0\left(\vec{b}_{\zeta}^{\vec{J}}\right)\right\rangle_{\mathbb{S}^{\text{A}\rightleftharpoons\text{M}}} \tag{4-64c}$$

$$z_{\xi\zeta}^{\vec{J}^{\text{A}\rightleftharpoons\text{M}}\vec{J}_{\text{M}}} = \left\langle \vec{b}_{\xi}^{\vec{J}^{\text{A}\rightleftharpoons\text{M}}}, -j\omega\mu_0\mathcal{L}_0\left(\vec{b}_{\zeta}^{\vec{J}_{\text{M}}}\right)\right\rangle_{\mathbb{S}^{\text{A}\rightleftharpoons\text{M}}} \tag{4-64d}$$

$$z_{\xi\zeta}^{\vec{J}^{\text{A}\rightleftharpoons\text{M}}\vec{J}_{\text{M}\rightleftharpoons\text{A}}} = \left\langle \vec{b}_{\xi}^{\vec{J}^{\text{A}\rightleftharpoons\text{M}}}, -j\omega\mu_0\mathcal{L}_0\left(\vec{b}_{\zeta}^{\vec{J}_{\text{M}\rightleftharpoons\text{A}}}\right)\right\rangle_{\mathbb{S}^{\text{A}\rightleftharpoons\text{M}}} \tag{4-64e}$$

$$z_{\xi\zeta}^{\vec{J}^{\text{A}\rightleftharpoons\text{M}}\vec{M}^{\text{A}\rightleftharpoons\text{M}}} = \left\langle \vec{b}_{\xi}^{\vec{J}^{\text{A}\rightleftharpoons\text{M}}}, -\text{P.V.}\,\mathcal{K}_0\left(\vec{b}_{\zeta}^{\vec{M}^{\text{A}\rightleftharpoons\text{M}}}\right)\right\rangle_{\mathbb{S}^{\text{A}\rightleftharpoons\text{M}}} - \left\langle \vec{b}_{\xi}^{\vec{J}^{\text{A}\rightleftharpoons\text{M}}}, \hat{n}^{\rightarrow\text{M}} \times \frac{1}{2}\vec{b}_{\zeta}^{\vec{M}^{\text{A}\rightleftharpoons\text{M}}}\right\rangle_{\mathbb{S}^{\text{A}\rightleftharpoons\text{M}}} \tag{4-64f}$$

$$z_{\xi\zeta}^{\vec{J}^{\text{A}\rightleftharpoons\text{M}}\vec{M}} = \left\langle \vec{b}_{\xi}^{\vec{J}^{\text{A}\rightleftharpoons\text{M}}}, -\mathcal{K}_0\left(\vec{b}_{\zeta}^{\vec{M}}\right)\right\rangle_{\mathbb{S}^{\text{A}\rightleftharpoons\text{M}}} \tag{4-64g}$$

$$z_{\xi\zeta}^{\vec{J}^{\text{A}\rightleftharpoons\text{M}}\vec{M}_{\text{M}\rightleftharpoons\text{A}}} = \left\langle \vec{b}_{\xi}^{\vec{J}^{\text{A}\rightleftharpoons\text{M}}}, -\mathcal{K}_0\left(\vec{b}_{\zeta}^{\vec{M}_{\text{M}\rightleftharpoons\text{A}}}\right)\right\rangle_{\mathbb{S}^{\text{A}\rightleftharpoons\text{M}}} \tag{4-64h}$$

The formulations used to calculate the elements of the matrices in Eq. (4-59a) are as follows:

$$z_{\xi\zeta}^{\vec{J}\vec{J}^{\text{A}\rightleftharpoons\text{M}}} = \left\langle \vec{b}_{\xi}^{\vec{J}}, -j\omega\mu_0\mathcal{L}_0\left(\vec{b}_{\zeta}^{\vec{J}^{\text{A}\rightleftharpoons\text{M}}}\right)\right\rangle_{\mathbb{V}} \tag{4-65a}$$

$$z_{\xi\zeta}^{\vec{J}\vec{J}^{\text{M}}} = \left\langle \vec{b}_{\xi}^{\vec{J}}, -j\omega\mu_0\mathcal{L}_0\left(\vec{b}_{\zeta}^{\vec{J}^{\text{M}}}\right)\right\rangle_{\mathbb{V}} \tag{4-65b}$$

$$z_{\xi\zeta}^{\vec{J}\vec{J}} = \left\langle \vec{b}_{\xi}^{\vec{J}}, -j\omega\mu_0\mathcal{L}_0\left(\vec{b}_{\zeta}^{\vec{J}}\right)\right\rangle_{\mathbb{V}} - \left\langle \vec{b}_{\xi}^{\vec{J}}, \left(j\omega\Delta\vec{\vec{\varepsilon}}_{\text{c}}\right)^{-1}\cdot\vec{b}_{\zeta}^{\vec{J}}\right\rangle_{\mathbb{V}} \tag{4-65c}$$

$$z_{\xi\zeta}^{\vec{J}\vec{J}_{\text{M}}} = \left\langle \vec{b}_{\xi}^{\vec{J}}, -j\omega\mu_0\mathcal{L}_0\left(\vec{b}_{\zeta}^{\vec{J}_{\text{M}}}\right)\right\rangle_{\mathbb{V}} \tag{4-65d}$$

$$z_{\xi\zeta}^{\vec{J}\vec{J}_{\text{M}\rightleftharpoons\text{A}}} = \left\langle \vec{b}_{\xi}^{\vec{J}}, -j\omega\mu_0\mathcal{L}_0\left(\vec{b}_{\zeta}^{\vec{J}_{\text{M}\rightleftharpoons\text{A}}}\right)\right\rangle_{\mathbb{V}} \tag{4-65e}$$

$$z_{\xi\zeta}^{\vec{J}\vec{M}^{\text{A}\rightleftharpoons\text{M}}} = \left\langle \vec{b}_{\xi}^{\vec{J}}, -\mathcal{K}_0\left(\vec{b}_{\zeta}^{\vec{M}^{\text{A}\rightleftharpoons\text{M}}}\right)\right\rangle_{\mathbb{V}} \tag{4-65f}$$

$$z_{\xi\zeta}^{\vec{J}\vec{M}} = \left\langle \vec{b}_{\xi}^{\vec{J}}, -\mathcal{K}_0\left(\vec{b}_{\zeta}^{\vec{M}}\right)\right\rangle_{\mathbb{V}} \tag{4-65g}$$

$$z_{\xi\zeta}^{\vec{J}\vec{M}_{\text{M}\rightleftharpoons\text{A}}} = \left\langle \vec{b}_{\xi}^{\vec{J}}, -\mathcal{K}_0\left(\vec{b}_{\zeta}^{\vec{M}_{\text{M}\rightleftharpoons\text{A}}}\right)\right\rangle_{\mathbb{V}} \tag{4-65h}$$

The formulations used to calculate the elements of the matrices in Eq. (4-59b) are as

follows:

$$z_{\xi\zeta}^{\vec{M}\vec{J}^{\mathrm{A\rightleftharpoons M}}} = \left\langle \vec{b}_{\xi}^{\vec{M}}, \mathcal{K}_0\left(\vec{b}_{\zeta}^{\vec{J}^{\mathrm{A\rightleftharpoons M}}}\right)\right\rangle_{\mathbb{V}} \tag{4-66a}$$

$$z_{\xi\zeta}^{\vec{M}\vec{J}^{\mathrm{M}}} = \left\langle \vec{b}_{\xi}^{\vec{M}}, \mathcal{K}_0\left(\vec{b}_{\zeta}^{\vec{J}^{\mathrm{M}}}\right)\right\rangle_{\mathbb{V}} \tag{4-66b}$$

$$z_{\xi\zeta}^{\vec{M}\vec{J}} = \left\langle \vec{b}_{\xi}^{\vec{M}}, \mathcal{K}_0\left(\vec{b}_{\zeta}^{\vec{J}}\right)\right\rangle_{\mathbb{V}} \tag{4-66c}$$

$$z_{\xi\zeta}^{\vec{M}\vec{J}_{\mathrm{M}}} = \left\langle \vec{b}_{\xi}^{\vec{M}}, \mathcal{K}_0\left(\vec{b}_{\zeta}^{\vec{J}_{\mathrm{M}}}\right)\right\rangle_{\mathbb{V}} \tag{4-66d}$$

$$z_{\xi\zeta}^{\vec{M}\vec{J}_{\mathrm{M\rightleftharpoons A}}} = \left\langle \vec{b}_{\xi}^{\vec{M}}, \mathcal{K}_0\left(\vec{b}_{\zeta}^{\vec{J}_{\mathrm{M\rightleftharpoons A}}}\right)\right\rangle_{\mathbb{V}} \tag{4-66e}$$

$$z_{\xi\zeta}^{\vec{M}\vec{M}^{\mathrm{A\rightleftharpoons M}}} = \left\langle \vec{b}_{\xi}^{\vec{M}}, -j\omega\varepsilon_0\mathcal{L}_0\left(\vec{b}_{\zeta}^{\vec{M}^{\mathrm{A\rightleftharpoons M}}}\right)\right\rangle_{\mathbb{V}} \tag{4-66f}$$

$$z_{\xi\zeta}^{\vec{M}\vec{M}} = \left\langle \vec{b}_{\xi}^{\vec{M}}, -j\omega\varepsilon_0\mathcal{L}_0\left(\vec{b}_{\zeta}^{\vec{M}}\right)\right\rangle_{\mathbb{V}} - \left\langle \vec{b}_{\xi}^{\vec{M}}, \left(j\omega\Delta\vec{\vec{\mu}}\right)^{-1}\cdot\vec{b}_{\zeta}^{\vec{M}}\right\rangle_{\mathbb{V}} \tag{4-66g}$$

$$z_{\xi\zeta}^{\vec{M}\vec{M}_{\mathrm{M\rightleftharpoons A}}} = \left\langle \vec{b}_{\xi}^{\vec{M}}, -j\omega\varepsilon_0\mathcal{L}_0\left(\vec{b}_{\zeta}^{\vec{M}_{\mathrm{M\rightleftharpoons A}}}\right)\right\rangle_{\mathbb{V}} \tag{4-66h}$$

The formulations used to calculate the elements of the matrices in Eq. (4-60) are as follows:

$$z_{\xi\zeta}^{\vec{J}^{\mathrm{M}}\vec{J}^{\mathrm{A\rightleftharpoons M}}} = \left\langle \vec{b}_{\xi}^{\vec{J}^{\mathrm{M}}}, -j\omega\mu_0\mathcal{L}_0\left(\vec{b}_{\zeta}^{\vec{J}^{\mathrm{A\rightleftharpoons M}}}\right)\right\rangle_{\mathbb{S}^{\mathrm{M}}} \tag{4-67a}$$

$$z_{\xi\zeta}^{\vec{J}^{\mathrm{M}}\vec{J}^{\mathrm{M}}} = \left\langle \vec{b}_{\xi}^{\vec{J}^{\mathrm{M}}}, -j\omega\mu_0\mathcal{L}_0\left(\vec{b}_{\zeta}^{\vec{J}^{\mathrm{M}}}\right)\right\rangle_{\mathbb{S}^{\mathrm{M}}} \tag{4-67b}$$

$$z_{\xi\zeta}^{\vec{J}^{\mathrm{M}}\vec{J}} = \left\langle \vec{b}_{\xi}^{\vec{J}^{\mathrm{M}}}, -j\omega\mu_0\mathcal{L}_0\left(\vec{b}_{\zeta}^{\vec{J}}\right)\right\rangle_{\mathbb{S}^{\mathrm{M}}} \tag{4-67c}$$

$$z_{\xi\zeta}^{\vec{J}^{\mathrm{M}}\vec{J}_{\mathrm{M}}} = \left\langle \vec{b}_{\xi}^{\vec{J}^{\mathrm{M}}}, -j\omega\mu_0\mathcal{L}_0\left(\vec{b}_{\zeta}^{\vec{J}_{\mathrm{M}}}\right)\right\rangle_{\mathbb{S}^{\mathrm{M}}} \tag{4-67d}$$

$$z_{\xi\zeta}^{\vec{J}^{\mathrm{M}}\vec{J}_{\mathrm{M\rightleftharpoons A}}} = \left\langle \vec{b}_{\xi}^{\vec{J}^{\mathrm{M}}}, -j\omega\mu_0\mathcal{L}_0\left(\vec{b}_{\zeta}^{\vec{J}_{\mathrm{M\rightleftharpoons A}}}\right)\right\rangle_{\mathbb{S}^{\mathrm{M}}} \tag{4-67e}$$

$$z_{\xi\zeta}^{\vec{J}^{\mathrm{M}}\vec{M}^{\mathrm{A\rightleftharpoons M}}} = \left\langle \vec{b}_{\xi}^{\vec{J}^{\mathrm{M}}}, -\mathcal{K}_0\left(\vec{b}_{\zeta}^{\vec{M}^{\mathrm{A\rightleftharpoons M}}}\right)\right\rangle_{\mathbb{S}^{\mathrm{M}}} \tag{4-67f}$$

$$z_{\xi\zeta}^{\vec{J}^{\mathrm{M}}\vec{M}} = \left\langle \vec{b}_{\xi}^{\vec{J}^{\mathrm{M}}}, -\mathcal{K}_0\left(\vec{b}_{\zeta}^{\vec{M}}\right)\right\rangle_{\mathbb{S}^{\mathrm{M}}} \tag{4-67g}$$

$$z_{\xi\zeta}^{\vec{J}^{\mathrm{M}}\vec{M}_{\mathrm{M\rightleftharpoons A}}} = \left\langle \vec{b}_{\xi}^{\vec{J}^{\mathrm{M}}}, -\mathcal{K}_0\left(\vec{b}_{\zeta}^{\vec{M}_{\mathrm{M\rightleftharpoons A}}}\right)\right\rangle_{\mathbb{S}^{\mathrm{M}}} \tag{4-67h}$$

The formulations used to calculate the elements of the matrices in Eq. (4-61) are as follows:

$$z_{\xi\zeta}^{\vec{J}_{\mathrm{M}}\vec{J}^{\mathrm{A\rightleftharpoons M}}} = \left\langle \vec{b}_{\xi}^{\vec{J}_{\mathrm{M}}}, -j\omega\mu_0\mathcal{L}_0\left(\vec{b}_{\zeta}^{\vec{J}^{\mathrm{A\rightleftharpoons M}}}\right)\right\rangle_{\mathbb{S}_{\mathrm{M}}} \tag{4-68a}$$

$$z_{\xi\zeta}^{\vec{J}_{\mathrm{M}}\vec{J}^{\mathrm{M}}} = \left\langle \vec{b}_{\xi}^{\vec{J}_{\mathrm{M}}}, -j\omega\mu_0\mathcal{L}_0\left(\vec{b}_{\zeta}^{\vec{J}^{\mathrm{M}}}\right)\right\rangle_{\mathbb{S}_{\mathrm{M}}} \tag{4-68b}$$

$$z_{\xi\zeta}^{\vec{J}_{\mathrm{M}}\vec{J}} = \left\langle \vec{b}_{\xi}^{\vec{J}_{\mathrm{M}}}, -j\omega\mu_0\mathcal{L}_0\left(\vec{b}_{\zeta}^{\vec{J}}\right)\right\rangle_{\mathbb{S}_{\mathrm{M}}} \tag{4-68c}$$

$$z_{\xi\zeta}^{\vec{J}_{\mathrm{M}}\vec{J}_{\mathrm{M}}} = \left\langle \vec{b}_{\xi}^{\vec{J}_{\mathrm{M}}}, -j\omega\mu_0\mathcal{L}_0\left(\vec{b}_{\zeta}^{\vec{J}_{\mathrm{M}}}\right)\right\rangle_{\mathbb{S}_{\mathrm{M}}} \tag{4-68d}$$

$$z_{\xi\zeta}^{\vec{J}_{\mathrm{M}}\vec{J}_{\mathrm{M}\rightleftharpoons\mathrm{A}}} = \left\langle \vec{b}_{\xi}^{\vec{J}_{\mathrm{M}}}, -j\omega\mu_0\mathcal{L}_0\left(\vec{b}_{\zeta}^{\vec{J}_{\mathrm{M}}}\right)\right\rangle_{\mathbb{S}_{\mathrm{M}}} \tag{4-68e}$$

$$z_{\xi\zeta}^{\vec{J}_{\mathrm{M}}\vec{M}^{\mathrm{A}\rightleftharpoons\mathrm{M}}} = \left\langle \vec{b}_{\xi}^{\vec{J}_{\mathrm{M}}}, -\mathcal{K}_0\left(\vec{b}_{\zeta}^{\vec{M}^{\mathrm{A}\rightleftharpoons\mathrm{M}}}\right)\right\rangle_{\mathbb{S}_{\mathrm{M}}} \tag{4-68f}$$

$$z_{\xi\zeta}^{\vec{J}_{\mathrm{M}}\vec{M}} = \left\langle \vec{b}_{\xi}^{\vec{J}_{\mathrm{M}}}, -\mathcal{K}_0\left(\vec{b}_{\zeta}^{\vec{M}}\right)\right\rangle_{\mathbb{S}_{\mathrm{M}}} \tag{4-68g}$$

$$z_{\xi\zeta}^{\vec{J}_{\mathrm{M}}\vec{M}_{\mathrm{M}\rightleftharpoons\mathrm{A}}} = \left\langle \vec{b}_{\xi}^{\vec{J}_{\mathrm{M}}}, -\mathcal{K}_0\left(\vec{b}_{\zeta}^{\vec{M}_{\mathrm{M}\rightleftharpoons\mathrm{A}}}\right)\right\rangle_{\mathbb{S}_{\mathrm{M}}} \tag{4-68h}$$

The formulations used to calculate the elements of the matrices in Eq. (4-62a) are as follows:

$$z_{\xi\zeta}^{\vec{M}_{\mathrm{M}\rightleftharpoons\mathrm{A}}\vec{J}^{\mathrm{A}\rightleftharpoons\mathrm{M}}} = \left\langle \vec{b}_{\xi}^{\vec{M}_{\mathrm{M}\rightleftharpoons\mathrm{A}}}, \mathcal{K}_0\left(\vec{b}_{\zeta}^{\vec{J}^{\mathrm{A}\rightleftharpoons\mathrm{M}}}\right)\right\rangle_{\mathbb{S}_{\mathrm{M}\rightleftharpoons\mathrm{A}}} \tag{4-69a}$$

$$z_{\xi\zeta}^{\vec{M}_{\mathrm{M}\rightleftharpoons\mathrm{A}}\vec{J}^{\mathrm{M}}} = \left\langle \vec{b}_{\xi}^{\vec{M}_{\mathrm{M}\rightleftharpoons\mathrm{A}}}, \mathcal{K}_0\left(\vec{b}_{\zeta}^{\vec{J}^{\mathrm{M}}}\right)\right\rangle_{\mathbb{S}_{\mathrm{M}\rightleftharpoons\mathrm{A}}} \tag{4-69b}$$

$$z_{\xi\zeta}^{\vec{M}_{\mathrm{M}\rightleftharpoons\mathrm{A}}\vec{J}} = \left\langle \vec{b}_{\xi}^{\vec{M}_{\mathrm{M}\rightleftharpoons\mathrm{A}}}, \mathcal{K}_0\left(\vec{b}_{\zeta}^{\vec{J}}\right)\right\rangle_{\mathbb{S}_{\mathrm{M}\rightleftharpoons\mathrm{A}}} \tag{4-69c}$$

$$z_{\xi\zeta}^{\vec{M}_{\mathrm{M}\rightleftharpoons\mathrm{A}}\vec{J}_{\mathrm{M}}} = \left\langle \vec{b}_{\xi}^{\vec{M}_{\mathrm{M}\rightleftharpoons\mathrm{A}}}, \mathcal{K}_0\left(\vec{b}_{\zeta}^{\vec{J}_{\mathrm{M}}}\right)\right\rangle_{\mathbb{S}_{\mathrm{M}\rightleftharpoons\mathrm{A}}} \tag{4-69d}$$

$$z_{\xi\zeta}^{\vec{M}_{\mathrm{M}\rightleftharpoons\mathrm{A}}\vec{J}_{\mathrm{M}\rightleftharpoons\mathrm{A}}} = \left\langle \vec{b}_{\xi}^{\vec{M}_{\mathrm{M}\rightleftharpoons\mathrm{A}}}, \mathrm{P.V.}\,\mathcal{K}_0\left(\vec{b}_{\zeta}^{\vec{J}_{\mathrm{M}\rightleftharpoons\mathrm{A}}}\right)\right\rangle_{\mathbb{S}_{\mathrm{M}\rightleftharpoons\mathrm{A}}} - \left\langle \vec{b}_{\xi}^{\vec{M}_{\mathrm{M}\rightleftharpoons\mathrm{A}}}, \frac{1}{2}\vec{b}_{\zeta}^{\vec{J}_{\mathrm{M}\rightleftharpoons\mathrm{A}}} \times \hat{n}_{\rightarrow\mathrm{M}}\right\rangle_{\mathbb{S}_{\mathrm{M}\rightleftharpoons\mathrm{A}}} \tag{4-69e}$$

$$z_{\xi\zeta}^{\vec{M}_{\mathrm{M}\rightleftharpoons\mathrm{A}}\vec{M}^{\mathrm{A}\rightleftharpoons\mathrm{M}}} = \left\langle \vec{b}_{\xi}^{\vec{M}_{\mathrm{M}\rightleftharpoons\mathrm{A}}}, -j\omega\varepsilon_0\mathcal{L}_0\left(\vec{b}_{\zeta}^{\vec{M}^{\mathrm{A}\rightleftharpoons\mathrm{M}}}\right)\right\rangle_{\mathbb{S}_{\mathrm{M}\rightleftharpoons\mathrm{A}}} \tag{4-69f}$$

$$z_{\xi\zeta}^{\vec{M}_{\mathrm{M}\rightleftharpoons\mathrm{A}}\vec{M}} = \left\langle \vec{b}_{\xi}^{\vec{M}_{\mathrm{M}\rightleftharpoons\mathrm{A}}}, -j\omega\varepsilon_0\mathcal{L}_0\left(\vec{b}_{\zeta}^{\vec{M}}\right)\right\rangle_{\mathbb{S}_{\mathrm{M}\rightleftharpoons\mathrm{A}}} \tag{4-69g}$$

$$z_{\xi\zeta}^{\vec{M}_{\mathrm{M}\rightleftharpoons\mathrm{A}}\vec{M}_{\mathrm{M}\rightleftharpoons\mathrm{A}}} = \left\langle \vec{b}_{\xi}^{\vec{M}_{\mathrm{M}\rightleftharpoons\mathrm{A}}}, -j\omega\varepsilon_0\mathcal{L}_0\left(\vec{b}_{\zeta}^{\vec{M}_{\mathrm{M}\rightleftharpoons\mathrm{A}}}\right)\right\rangle_{\mathbb{S}_{\mathrm{M}\rightleftharpoons\mathrm{A}}} \tag{4-69h}$$

The formulations used to calculate the elements of the matrices in Eq. (4-62b) are as follows:

$$z_{\xi\zeta}^{\vec{J}_{\mathrm{M}\rightleftharpoons\mathrm{A}}\vec{J}^{\mathrm{A}\rightleftharpoons\mathrm{M}}} = \left\langle \vec{b}_{\xi}^{\vec{J}_{\mathrm{M}\rightleftharpoons\mathrm{A}}}, -j\omega\mu_0\mathcal{L}_0\left(\vec{b}_{\zeta}^{\vec{J}^{\mathrm{A}\rightleftharpoons\mathrm{M}}}\right)\right\rangle_{\mathbb{S}_{\mathrm{M}\rightleftharpoons\mathrm{A}}} \tag{4-70a}$$

$$z_{\xi\zeta}^{\vec{J}_{\mathrm{M}\rightleftharpoons\mathrm{A}}\vec{J}^{\mathrm{M}}} = \left\langle \vec{b}_{\xi}^{\vec{J}_{\mathrm{M}\rightleftharpoons\mathrm{A}}}, -j\omega\mu_0\mathcal{L}_0\left(\vec{b}_{\zeta}^{\vec{J}^{\mathrm{M}}}\right)\right\rangle_{\mathbb{S}_{\mathrm{M}\rightleftharpoons\mathrm{A}}} \tag{4-70b}$$

$$z_{\xi\zeta}^{\vec{J}_{\mathrm{M}\rightleftharpoons\mathrm{A}}\vec{J}} = \left\langle \vec{b}_{\xi}^{\vec{J}_{\mathrm{M}\rightleftharpoons\mathrm{A}}}, -j\omega\mu_0\mathcal{L}_0\left(\vec{b}_{\zeta}^{\vec{J}}\right)\right\rangle_{\mathbb{S}_{\mathrm{M}\rightleftharpoons\mathrm{A}}} \tag{4-70c}$$

$$z_{\xi\zeta}^{\vec{J}_{\mathrm{M}\rightleftharpoons\mathrm{A}}\vec{J}_{\mathrm{M}}} = \left\langle \vec{b}_{\xi}^{\vec{J}_{\mathrm{M}\rightleftharpoons\mathrm{A}}}, -j\omega\mu_0\mathcal{L}_0\left(\vec{b}_{\zeta}^{\vec{J}_{\mathrm{M}}}\right)\right\rangle_{\mathbb{S}_{\mathrm{M}\rightleftharpoons\mathrm{A}}} \tag{4-70d}$$

$$z_{\xi\zeta}^{\vec{J}_{\mathrm{M}\rightleftharpoons\mathrm{A}}\vec{J}_{\mathrm{M}\rightleftharpoons\mathrm{A}}} = \left\langle \vec{b}_{\xi}^{\vec{J}_{\mathrm{M}\rightleftharpoons\mathrm{A}}}, -j\omega\mu_0\mathcal{L}_0\left(\vec{b}_{\zeta}^{\vec{J}_{\mathrm{M}\rightleftharpoons\mathrm{A}}}\right)\right\rangle_{\mathbb{S}_{\mathrm{M}\rightleftharpoons\mathrm{A}}} \tag{4-70e}$$

$$z_{\xi\zeta}^{\vec{J}_{\mathrm{M}\rightleftharpoons\mathrm{A}}\vec{M}^{\mathrm{A}\rightleftharpoons\mathrm{M}}} = \left\langle \vec{b}_{\xi}^{\vec{J}_{\mathrm{M}\rightleftharpoons\mathrm{A}}}, -\mathcal{K}_0\left(\vec{b}_{\zeta}^{\vec{M}^{\mathrm{A}\rightleftharpoons\mathrm{M}}}\right)\right\rangle_{\mathbb{S}_{\mathrm{M}\rightleftharpoons\mathrm{A}}} \tag{4-70f}$$

$$z_{\xi\zeta}^{\vec{J}_{\mathrm{M}\rightleftharpoons\mathrm{A}}\vec{M}} \;=\; \left\langle \vec{b}_{\xi}^{\vec{J}_{\mathrm{M}\rightleftharpoons\mathrm{A}}}, -\mathcal{K}_0\left(\vec{b}_{\zeta}^{\vec{M}}\right)\right\rangle_{\mathbb{S}_{\mathrm{M}\rightleftharpoons\mathrm{A}}} \tag{4-70g}$$

$$z_{\xi\zeta}^{\vec{J}_{\mathrm{M}\rightleftharpoons\mathrm{A}}\vec{M}_{\mathrm{M}\rightleftharpoons\mathrm{A}}} \;=\; \left\langle \vec{b}_{\xi}^{\vec{J}_{\mathrm{M}\rightleftharpoons\mathrm{A}}}, -\mathrm{P.V.}\,\mathcal{K}_0\left(\vec{b}_{\zeta}^{\vec{M}_{\mathrm{M}\rightleftharpoons\mathrm{A}}}\right)\right\rangle_{\mathbb{S}_{\mathrm{M}\rightleftharpoons\mathrm{A}}} - \left\langle \vec{b}_{\xi}^{\vec{J}_{\mathrm{M}\rightleftharpoons\mathrm{A}}}, \hat{n}_{\rightarrow\mathrm{M}}\times\frac{1}{2}\vec{b}_{\zeta}^{\vec{M}_{\mathrm{M}\rightleftharpoons\mathrm{A}}}\right\rangle_{\mathbb{S}_{\mathrm{M}\rightleftharpoons\mathrm{A}}} \tag{4-70h}$$

Below, we propose two different schemes for mathematically describing modal space by employing the above matrix equations.

**Scheme I: Dependent Variable Elimination (DVE)**

Using Eqs. (4-58a)~(4-62b), we can obtain the following transformation from *basic variables* (*BVs*) $\bar{a}^{BV}$ to *all variables* (*AVs*) $\bar{a}^{AV}$.

$$\begin{bmatrix} \bar{a}^{\vec{J}^{\mathrm{A}\rightleftharpoons\mathrm{M}}} \\ \bar{a}^{\vec{J}^{\mathrm{M}}} \\ \bar{a}^{\vec{J}} \\ \bar{a}^{\vec{J}_{\mathrm{M}}} \\ \bar{a}^{\vec{J}_{\mathrm{M}\rightleftharpoons\mathrm{A}}} \\ \bar{a}^{\vec{M}^{\mathrm{A}\rightleftharpoons\mathrm{M}}} \\ \bar{a}^{\vec{M}} \\ \bar{a}^{\vec{M}_{\mathrm{M}\rightleftharpoons\mathrm{A}}} \end{bmatrix} = \bar{a}^{\mathrm{AV}} = \overline{\overline{T}}^{\mathrm{BV}\rightarrow\mathrm{AV}}\cdot\bar{a}^{\mathrm{BV}} \tag{4-71}$$

where $\bar{a}^{\mathrm{BV}} = \bar{a}^{\vec{J}^{\mathrm{A}\rightleftharpoons\mathrm{M}}} / \bar{a}^{\vec{M}^{\mathrm{A}\rightleftharpoons\mathrm{M}}}$. Specifically, when $\bar{a}^{\mathrm{BV}} = \bar{a}^{\vec{J}^{\mathrm{A}\rightleftharpoons\mathrm{M}}}$, we have that

$$\begin{aligned}
\overline{\overline{T}}^{\mathrm{BV}\rightarrow\mathrm{AV}} &= \overline{\overline{T}}^{\vec{J}^{\mathrm{A}\rightleftharpoons\mathrm{M}}\rightarrow\mathrm{AV}} \\
&= \begin{bmatrix}
\overline{\overline{I}}^{\vec{J}^{\mathrm{A}\rightleftharpoons\mathrm{M}}} & 0 & 0 & 0 & 0 & 0 & 0 & 0 \\
0 & \overline{\overline{Z}}^{\vec{M}^{\mathrm{A}\rightleftharpoons\mathrm{M}}\vec{J}^{\mathrm{M}}} & \overline{\overline{Z}}^{\vec{M}^{\mathrm{A}\rightleftharpoons\mathrm{M}}\vec{J}} & \overline{\overline{Z}}^{\vec{M}^{\mathrm{A}\rightleftharpoons\mathrm{M}}\vec{J}_{\mathrm{M}}} & \overline{\overline{Z}}^{\vec{M}^{\mathrm{A}\rightleftharpoons\mathrm{M}}\vec{J}_{\mathrm{M}\rightleftharpoons\mathrm{A}}} & \overline{\overline{Z}}^{\vec{M}^{\mathrm{A}\rightleftharpoons\mathrm{M}}\vec{M}^{\mathrm{A}\rightleftharpoons\mathrm{M}}} & \overline{\overline{Z}}^{\vec{M}^{\mathrm{A}\rightleftharpoons\mathrm{M}}\vec{M}} & \overline{\overline{Z}}^{\vec{M}^{\mathrm{A}\rightleftharpoons\mathrm{M}}\vec{M}_{\mathrm{M}\rightleftharpoons\mathrm{A}}} \\
0 & \overline{\overline{Z}}^{\vec{J}^{\mathrm{M}}\vec{J}^{\mathrm{M}}} & \overline{\overline{Z}}^{\vec{J}^{\mathrm{M}}\vec{J}} & \overline{\overline{Z}}^{\vec{J}^{\mathrm{M}}\vec{J}_{\mathrm{M}}} & \overline{\overline{Z}}^{\vec{J}^{\mathrm{M}}\vec{J}_{\mathrm{M}\rightleftharpoons\mathrm{A}}} & \overline{\overline{Z}}^{\vec{J}^{\mathrm{M}}\vec{M}^{\mathrm{A}\rightleftharpoons\mathrm{M}}} & \overline{\overline{Z}}^{\vec{J}^{\mathrm{M}}\vec{M}} & \overline{\overline{Z}}^{\vec{J}^{\mathrm{M}}\vec{M}_{\mathrm{M}\rightleftharpoons\mathrm{A}}} \\
0 & \overline{\overline{Z}}^{\vec{J}\vec{J}^{\mathrm{M}}} & \overline{\overline{Z}}^{\vec{J}\vec{J}} & \overline{\overline{Z}}^{\vec{J}\vec{J}_{\mathrm{M}}} & \overline{\overline{Z}}^{\vec{J}\vec{J}_{\mathrm{M}\rightleftharpoons\mathrm{A}}} & \overline{\overline{Z}}^{\vec{J}\vec{M}^{\mathrm{A}\rightleftharpoons\mathrm{M}}} & \overline{\overline{Z}}^{\vec{J}\vec{M}} & \overline{\overline{Z}}^{\vec{J}\vec{M}_{\mathrm{M}\rightleftharpoons\mathrm{A}}} \\
0 & \overline{\overline{Z}}^{\vec{J}_{\mathrm{M}}\vec{J}^{\mathrm{M}}} & \overline{\overline{Z}}^{\vec{J}_{\mathrm{M}}\vec{J}} & \overline{\overline{Z}}^{\vec{J}_{\mathrm{M}}\vec{J}_{\mathrm{M}}} & \overline{\overline{Z}}^{\vec{J}_{\mathrm{M}}\vec{J}_{\mathrm{M}\rightleftharpoons\mathrm{A}}} & \overline{\overline{Z}}^{\vec{J}_{\mathrm{M}}\vec{M}^{\mathrm{A}\rightleftharpoons\mathrm{M}}} & \overline{\overline{Z}}^{\vec{J}_{\mathrm{M}}\vec{M}} & \overline{\overline{Z}}^{\vec{J}_{\mathrm{M}}\vec{M}_{\mathrm{M}\rightleftharpoons\mathrm{A}}} \\
0 & \overline{\overline{Z}}^{\vec{M}\vec{J}^{\mathrm{M}}} & \overline{\overline{Z}}^{\vec{M}\vec{J}} & \overline{\overline{Z}}^{\vec{M}\vec{J}_{\mathrm{M}}} & \overline{\overline{Z}}^{\vec{M}\vec{J}_{\mathrm{M}\rightleftharpoons\mathrm{A}}} & \overline{\overline{Z}}^{\vec{M}\vec{M}^{\mathrm{A}\rightleftharpoons\mathrm{M}}} & \overline{\overline{Z}}^{\vec{M}\vec{M}} & \overline{\overline{Z}}^{\vec{M}\vec{M}_{\mathrm{M}\rightleftharpoons\mathrm{A}}} \\
0 & \overline{\overline{Z}}^{\vec{M}_{\mathrm{M}\rightleftharpoons\mathrm{A}}\vec{J}^{\mathrm{M}}} & \overline{\overline{Z}}^{\vec{M}_{\mathrm{M}\rightleftharpoons\mathrm{A}}\vec{J}} & \overline{\overline{Z}}^{\vec{M}_{\mathrm{M}\rightleftharpoons\mathrm{A}}\vec{J}_{\mathrm{M}}} & \overline{\overline{Z}}^{\vec{M}_{\mathrm{M}\rightleftharpoons\mathrm{A}}\vec{J}_{\mathrm{M}\rightleftharpoons\mathrm{A}}} & \overline{\overline{Z}}^{\vec{M}_{\mathrm{M}\rightleftharpoons\mathrm{A}}\vec{M}^{\mathrm{A}\rightleftharpoons\mathrm{M}}} & \overline{\overline{Z}}^{\vec{M}_{\mathrm{M}\rightleftharpoons\mathrm{A}}\vec{M}} & \overline{\overline{Z}}^{\vec{M}_{\mathrm{M}\rightleftharpoons\mathrm{A}}\vec{M}_{\mathrm{M}\rightleftharpoons\mathrm{A}}} \\
0 & \overline{\overline{Z}}^{\vec{J}_{\mathrm{M}\rightleftharpoons\mathrm{A}}\vec{J}^{\mathrm{M}}} & \overline{\overline{Z}}^{\vec{J}_{\mathrm{M}\rightleftharpoons\mathrm{A}}\vec{J}} & \overline{\overline{Z}}^{\vec{J}_{\mathrm{M}\rightleftharpoons\mathrm{A}}\vec{J}_{\mathrm{M}}} & \overline{\overline{Z}}^{\vec{J}_{\mathrm{M}\rightleftharpoons\mathrm{A}}\vec{J}_{\mathrm{M}\rightleftharpoons\mathrm{A}}} & \overline{\overline{Z}}^{\vec{J}_{\mathrm{M}\rightleftharpoons\mathrm{A}}\vec{M}^{\mathrm{A}\rightleftharpoons\mathrm{M}}} & \overline{\overline{Z}}^{\vec{J}_{\mathrm{M}\rightleftharpoons\mathrm{A}}\vec{M}} & \overline{\overline{Z}}^{\vec{J}_{\mathrm{M}\rightleftharpoons\mathrm{A}}\vec{M}_{\mathrm{M}\rightleftharpoons\mathrm{A}}}
\end{bmatrix}^{-1}
\cdot
\begin{bmatrix}
\overline{\overline{I}}^{\vec{J}^{\mathrm{A}\rightleftharpoons\mathrm{M}}} \\
-\overline{\overline{Z}}^{\vec{M}^{\mathrm{A}\rightleftharpoons\mathrm{M}}\vec{J}^{\mathrm{A}\rightleftharpoons\mathrm{M}}} \\
-\overline{\overline{Z}}^{\vec{J}^{\mathrm{M}}\vec{J}^{\mathrm{A}\rightleftharpoons\mathrm{M}}} \\
-\overline{\overline{Z}}^{\vec{J}\vec{J}^{\mathrm{A}\rightleftharpoons\mathrm{M}}} \\
-\overline{\overline{Z}}^{\vec{J}_{\mathrm{M}}\vec{J}^{\mathrm{A}\rightleftharpoons\mathrm{M}}} \\
-\overline{\overline{Z}}^{\vec{M}\vec{J}^{\mathrm{A}\rightleftharpoons\mathrm{M}}} \\
-\overline{\overline{Z}}^{\vec{M}_{\mathrm{M}\rightleftharpoons\mathrm{A}}\vec{J}^{\mathrm{A}\rightleftharpoons\mathrm{M}}} \\
-\overline{\overline{Z}}^{\vec{J}_{\mathrm{M}\rightleftharpoons\mathrm{A}}\vec{J}^{\mathrm{A}\rightleftharpoons\mathrm{M}}}
\end{bmatrix}
\end{aligned} \tag{4-72}$$

When $\bar{a}^{\mathrm{BV}} = \bar{a}^{\vec{M}^{\mathrm{A}\rightleftharpoons\mathrm{F}}}$, we have that

$$\begin{aligned}
\overline{\overline{T}}^{\mathrm{BV}\rightarrow\mathrm{AV}} &= \overline{\overline{T}}^{\vec{M}^{\mathrm{A}\rightleftharpoons\mathrm{M}}\rightarrow\mathrm{AV}} \\
&= \begin{bmatrix}
0 & 0 & 0 & 0 & 0 & \overline{\overline{I}}^{\vec{M}^{\mathrm{A}\rightleftharpoons\mathrm{M}}} & 0 & 0 \\
\overline{\overline{Z}}^{\vec{J}^{\mathrm{A}\rightleftharpoons\mathrm{M}}\vec{J}^{\mathrm{A}\rightleftharpoons\mathrm{M}}} & \overline{\overline{Z}}^{\vec{J}^{\mathrm{A}\rightleftharpoons\mathrm{M}}\vec{J}^{\mathrm{M}}} & \overline{\overline{Z}}^{\vec{J}^{\mathrm{A}\rightleftharpoons\mathrm{M}}\vec{J}} & \overline{\overline{Z}}^{\vec{J}^{\mathrm{A}\rightleftharpoons\mathrm{M}}\vec{J}_{\mathrm{M}}} & \overline{\overline{Z}}^{\vec{J}^{\mathrm{A}\rightleftharpoons\mathrm{M}}\vec{J}_{\mathrm{M}\rightleftharpoons\mathrm{A}}} & 0 & \overline{\overline{Z}}^{\vec{J}^{\mathrm{A}\rightleftharpoons\mathrm{M}}\vec{M}} & \overline{\overline{Z}}^{\vec{J}^{\mathrm{A}\rightleftharpoons\mathrm{M}}\vec{M}_{\mathrm{M}\rightleftharpoons\mathrm{A}}} \\
\overline{\overline{Z}}^{\vec{J}^{\mathrm{M}}\vec{J}^{\mathrm{A}\rightleftharpoons\mathrm{M}}} & \overline{\overline{Z}}^{\vec{J}^{\mathrm{M}}\vec{J}^{\mathrm{M}}} & \overline{\overline{Z}}^{\vec{J}^{\mathrm{M}}\vec{J}} & \overline{\overline{Z}}^{\vec{J}^{\mathrm{M}}\vec{J}_{\mathrm{M}}} & \overline{\overline{Z}}^{\vec{J}^{\mathrm{M}}\vec{J}_{\mathrm{M}\rightleftharpoons\mathrm{A}}} & 0 & \overline{\overline{Z}}^{\vec{J}^{\mathrm{M}}\vec{M}} & \overline{\overline{Z}}^{\vec{J}^{\mathrm{M}}\vec{M}_{\mathrm{M}\rightleftharpoons\mathrm{A}}} \\
\overline{\overline{Z}}^{\vec{J}\vec{J}^{\mathrm{A}\rightleftharpoons\mathrm{M}}} & \overline{\overline{Z}}^{\vec{J}\vec{J}^{\mathrm{M}}} & \overline{\overline{Z}}^{\vec{J}\vec{J}} & \overline{\overline{Z}}^{\vec{J}\vec{J}_{\mathrm{M}}} & \overline{\overline{Z}}^{\vec{J}\vec{J}_{\mathrm{M}\rightleftharpoons\mathrm{A}}} & 0 & \overline{\overline{Z}}^{\vec{J}\vec{M}} & \overline{\overline{Z}}^{\vec{J}\vec{M}_{\mathrm{M}\rightleftharpoons\mathrm{A}}} \\
\overline{\overline{Z}}^{\vec{J}_{\mathrm{M}}\vec{J}^{\mathrm{A}\rightleftharpoons\mathrm{M}}} & \overline{\overline{Z}}^{\vec{J}_{\mathrm{M}}\vec{J}^{\mathrm{M}}} & \overline{\overline{Z}}^{\vec{J}_{\mathrm{M}}\vec{J}} & \overline{\overline{Z}}^{\vec{J}_{\mathrm{M}}\vec{J}_{\mathrm{M}}} & \overline{\overline{Z}}^{\vec{J}_{\mathrm{M}}\vec{J}_{\mathrm{M}\rightleftharpoons\mathrm{A}}} & 0 & \overline{\overline{Z}}^{\vec{J}_{\mathrm{M}}\vec{M}} & \overline{\overline{Z}}^{\vec{J}_{\mathrm{M}}\vec{M}_{\mathrm{M}\rightleftharpoons\mathrm{A}}} \\
\overline{\overline{Z}}^{\vec{M}\vec{J}^{\mathrm{A}\rightleftharpoons\mathrm{M}}} & \overline{\overline{Z}}^{\vec{M}\vec{J}^{\mathrm{M}}} & \overline{\overline{Z}}^{\vec{M}\vec{J}} & \overline{\overline{Z}}^{\vec{M}\vec{J}_{\mathrm{M}}} & \overline{\overline{Z}}^{\vec{M}\vec{J}_{\mathrm{M}\rightleftharpoons\mathrm{A}}} & 0 & \overline{\overline{Z}}^{\vec{M}\vec{M}} & \overline{\overline{Z}}^{\vec{M}\vec{M}_{\mathrm{M}\rightleftharpoons\mathrm{A}}} \\
\overline{\overline{Z}}^{\vec{M}_{\mathrm{M}\rightleftharpoons\mathrm{A}}\vec{J}^{\mathrm{A}\rightleftharpoons\mathrm{M}}} & \overline{\overline{Z}}^{\vec{M}_{\mathrm{M}\rightleftharpoons\mathrm{A}}\vec{J}^{\mathrm{M}}} & \overline{\overline{Z}}^{\vec{M}_{\mathrm{M}\rightleftharpoons\mathrm{A}}\vec{J}} & \overline{\overline{Z}}^{\vec{M}_{\mathrm{M}\rightleftharpoons\mathrm{A}}\vec{J}_{\mathrm{M}}} & \overline{\overline{Z}}^{\vec{M}_{\mathrm{M}\rightleftharpoons\mathrm{A}}\vec{J}_{\mathrm{M}\rightleftharpoons\mathrm{A}}} & 0 & \overline{\overline{Z}}^{\vec{M}_{\mathrm{M}\rightleftharpoons\mathrm{A}}\vec{M}} & \overline{\overline{Z}}^{\vec{M}_{\mathrm{M}\rightleftharpoons\mathrm{A}}\vec{M}_{\mathrm{M}\rightleftharpoons\mathrm{A}}} \\
\overline{\overline{Z}}^{\vec{J}_{\mathrm{M}\rightleftharpoons\mathrm{A}}\vec{J}^{\mathrm{A}\rightleftharpoons\mathrm{M}}} & \overline{\overline{Z}}^{\vec{J}_{\mathrm{M}\rightleftharpoons\mathrm{A}}\vec{J}^{\mathrm{M}}} & \overline{\overline{Z}}^{\vec{J}_{\mathrm{M}\rightleftharpoons\mathrm{A}}\vec{J}} & \overline{\overline{Z}}^{\vec{J}_{\mathrm{M}\rightleftharpoons\mathrm{A}}\vec{J}_{\mathrm{M}}} & \overline{\overline{Z}}^{\vec{J}_{\mathrm{M}\rightleftharpoons\mathrm{A}}\vec{J}_{\mathrm{M}\rightleftharpoons\mathrm{A}}} & 0 & \overline{\overline{Z}}^{\vec{J}_{\mathrm{M}\rightleftharpoons\mathrm{A}}\vec{M}} & \overline{\overline{Z}}^{\vec{J}_{\mathrm{M}\rightleftharpoons\mathrm{A}}\vec{M}_{\mathrm{M}\rightleftharpoons\mathrm{A}}}
\end{bmatrix}^{-1}
\cdot
\begin{bmatrix}
\overline{\overline{I}}^{\vec{M}^{\mathrm{A}\rightleftharpoons\mathrm{M}}} \\
-\overline{\overline{Z}}^{\vec{J}^{\mathrm{A}\rightleftharpoons\mathrm{M}}\vec{M}^{\mathrm{A}\rightleftharpoons\mathrm{M}}} \\
-\overline{\overline{Z}}^{\vec{J}^{\mathrm{M}}\vec{M}^{\mathrm{A}\rightleftharpoons\mathrm{M}}} \\
-\overline{\overline{Z}}^{\vec{J}\vec{M}^{\mathrm{A}\rightleftharpoons\mathrm{M}}} \\
-\overline{\overline{Z}}^{\vec{J}_{\mathrm{M}}\vec{M}^{\mathrm{A}\rightleftharpoons\mathrm{M}}} \\
-\overline{\overline{Z}}^{\vec{M}\vec{M}^{\mathrm{A}\rightleftharpoons\mathrm{M}}} \\
-\overline{\overline{Z}}^{\vec{M}_{\mathrm{M}\rightleftharpoons\mathrm{A}}\vec{M}^{\mathrm{A}\rightleftharpoons\mathrm{M}}} \\
-\overline{\overline{Z}}^{\vec{J}_{\mathrm{M}\rightleftharpoons\mathrm{A}}\vec{M}^{\mathrm{A}\rightleftharpoons\mathrm{M}}}
\end{bmatrix}
\end{aligned} \tag{4-73}$$

**Scheme II: Solution/Definition Domain Compression (SDC/DDC)**

Using Eqs. (4-58a)~(4-62b), we can also obtain the following transformation

$$\bar{a}^{\mathrm{AV}} = \underbrace{\left[\bar{s}_1^{\mathrm{BS}}, \bar{s}_2^{\mathrm{BS}}, \cdots\right]}_{\overline{\overline{T}}^{\mathrm{BS}\to\mathrm{AV}}} \cdot \underbrace{\begin{bmatrix} a_1^{\mathrm{BS}} \\ a_2^{\mathrm{BS}} \\ \vdots \end{bmatrix}}_{\bar{a}^{\mathrm{BS}}} \tag{4-74}$$

in which $\{\bar{s}_1^{\mathrm{BS}}, \bar{s}_2^{\mathrm{BS}}, \cdots\}$ are the *basic solutions* (*BSs*) of the following equation

$$\overline{\overline{\Psi}}_{\mathrm{FCE}}^{\mathrm{DoJ}} \cdot \bar{a}^{\mathrm{AV}} = 0 \quad \text{or} \quad \overline{\overline{\Psi}}_{\mathrm{FCE}}^{\mathrm{DoM}} \cdot \bar{a}^{\mathrm{AV}} = 0 \tag{4-75}$$

where

$$\overline{\overline{\Psi}}_{\mathrm{FCE}}^{\mathrm{DoJ}} = \begin{bmatrix}
\overline{\overline{Z}}^{\vec{M}^{\mathrm{A\rightleftharpoons M}}\vec{J}^{\mathrm{A\rightleftharpoons M}}} & \overline{\overline{Z}}^{\vec{M}^{\mathrm{A\rightleftharpoons M}}\vec{J}^{\mathrm{M}}} & \overline{\overline{Z}}^{\vec{M}^{\mathrm{A\rightleftharpoons M}}\vec{J}} & \overline{\overline{Z}}^{\vec{M}^{\mathrm{A\rightleftharpoons M}}\vec{J}_{\mathrm{M}}} & \overline{\overline{Z}}^{\vec{M}^{\mathrm{A\rightleftharpoons M}}\vec{J}_{\mathrm{M\rightleftharpoons A}}} & \overline{\overline{Z}}^{\vec{M}^{\mathrm{A\rightleftharpoons M}}\vec{M}^{\mathrm{A\rightleftharpoons M}}} & \overline{\overline{Z}}^{\vec{M}^{\mathrm{A\rightleftharpoons M}}\vec{M}} & \overline{\overline{Z}}^{\vec{M}^{\mathrm{A\rightleftharpoons M}}\vec{M}_{\mathrm{M\rightleftharpoons A}}} \\
\overline{\overline{Z}}^{\vec{J}^{\mathrm{M}}\vec{J}^{\mathrm{A\rightleftharpoons M}}} & \overline{\overline{Z}}^{\vec{J}^{\mathrm{M}}\vec{J}^{\mathrm{M}}} & \overline{\overline{Z}}^{\vec{J}^{\mathrm{M}}\vec{J}} & \overline{\overline{Z}}^{\vec{J}^{\mathrm{M}}\vec{J}_{\mathrm{M}}} & \overline{\overline{Z}}^{\vec{J}^{\mathrm{M}}\vec{J}_{\mathrm{M\rightleftharpoons A}}} & \overline{\overline{Z}}^{\vec{J}^{\mathrm{M}}\vec{M}^{\mathrm{A\rightleftharpoons M}}} & \overline{\overline{Z}}^{\vec{J}^{\mathrm{M}}\vec{M}} & \overline{\overline{Z}}^{\vec{J}^{\mathrm{M}}\vec{M}_{\mathrm{M\rightleftharpoons A}}} \\
\overline{\overline{Z}}^{\vec{J}\vec{J}^{\mathrm{A\rightleftharpoons M}}} & \overline{\overline{Z}}^{\vec{J}\vec{J}^{\mathrm{M}}} & \overline{\overline{Z}}^{\vec{J}\vec{J}} & \overline{\overline{Z}}^{\vec{J}\vec{J}_{\mathrm{M}}} & \overline{\overline{Z}}^{\vec{J}\vec{J}_{\mathrm{M\rightleftharpoons A}}} & \overline{\overline{Z}}^{\vec{J}\vec{M}^{\mathrm{A\rightleftharpoons M}}} & \overline{\overline{Z}}^{\vec{J}\vec{M}} & \overline{\overline{Z}}^{\vec{J}\vec{M}_{\mathrm{M\rightleftharpoons A}}} \\
\overline{\overline{Z}}^{\vec{J}_{\mathrm{M}}\vec{J}^{\mathrm{A\rightleftharpoons M}}} & \overline{\overline{Z}}^{\vec{J}_{\mathrm{M}}\vec{J}^{\mathrm{M}}} & \overline{\overline{Z}}^{\vec{J}_{\mathrm{M}}\vec{J}} & \overline{\overline{Z}}^{\vec{J}_{\mathrm{M}}\vec{J}_{\mathrm{M}}} & \overline{\overline{Z}}^{\vec{J}_{\mathrm{M}}\vec{J}_{\mathrm{M\rightleftharpoons A}}} & \overline{\overline{Z}}^{\vec{J}_{\mathrm{M}}\vec{M}^{\mathrm{A\rightleftharpoons M}}} & \overline{\overline{Z}}^{\vec{J}_{\mathrm{M}}\vec{M}} & \overline{\overline{Z}}^{\vec{J}_{\mathrm{M}}\vec{M}_{\mathrm{M\rightleftharpoons A}}} \\
\overline{\overline{Z}}^{\vec{M}\vec{J}^{\mathrm{A\rightleftharpoons M}}} & \overline{\overline{Z}}^{\vec{M}\vec{J}^{\mathrm{M}}} & \overline{\overline{Z}}^{\vec{M}\vec{J}} & \overline{\overline{Z}}^{\vec{M}\vec{J}_{\mathrm{M}}} & \overline{\overline{Z}}^{\vec{M}\vec{J}_{\mathrm{M\rightleftharpoons A}}} & \overline{\overline{Z}}^{\vec{M}\vec{M}^{\mathrm{A\rightleftharpoons M}}} & \overline{\overline{Z}}^{\vec{M}\vec{M}} & \overline{\overline{Z}}^{\vec{M}\vec{M}_{\mathrm{M\rightleftharpoons A}}} \\
\overline{\overline{Z}}^{\vec{M}_{\mathrm{M\rightleftharpoons A}}\vec{J}^{\mathrm{A\rightleftharpoons M}}} & \overline{\overline{Z}}^{\vec{M}_{\mathrm{M\rightleftharpoons A}}\vec{J}^{\mathrm{M}}} & \overline{\overline{Z}}^{\vec{M}_{\mathrm{M\rightleftharpoons A}}\vec{J}} & \overline{\overline{Z}}^{\vec{M}_{\mathrm{M\rightleftharpoons A}}\vec{J}_{\mathrm{M}}} & \overline{\overline{Z}}^{\vec{M}_{\mathrm{M\rightleftharpoons A}}\vec{J}_{\mathrm{M\rightleftharpoons A}}} & \overline{\overline{Z}}^{\vec{M}_{\mathrm{M\rightleftharpoons A}}\vec{M}^{\mathrm{A\rightleftharpoons M}}} & \overline{\overline{Z}}^{\vec{M}_{\mathrm{M\rightleftharpoons A}}\vec{M}} & \overline{\overline{Z}}^{\vec{M}_{\mathrm{M\rightleftharpoons A}}\vec{M}_{\mathrm{M\rightleftharpoons A}}} \\
\overline{\overline{Z}}^{\vec{J}_{\mathrm{M\rightleftharpoons A}}\vec{J}^{\mathrm{A\rightleftharpoons M}}} & \overline{\overline{Z}}^{\vec{J}_{\mathrm{M\rightleftharpoons A}}\vec{J}^{\mathrm{M}}} & \overline{\overline{Z}}^{\vec{J}_{\mathrm{M\rightleftharpoons A}}\vec{J}} & \overline{\overline{Z}}^{\vec{J}_{\mathrm{M\rightleftharpoons A}}\vec{J}_{\mathrm{M}}} & \overline{\overline{Z}}^{\vec{J}_{\mathrm{M\rightleftharpoons A}}\vec{J}_{\mathrm{M\rightleftharpoons A}}} & \overline{\overline{Z}}^{\vec{J}_{\mathrm{M\rightleftharpoons A}}\vec{M}^{\mathrm{A\rightleftharpoons M}}} & \overline{\overline{Z}}^{\vec{J}_{\mathrm{M\rightleftharpoons A}}\vec{M}} & \overline{\overline{Z}}^{\vec{J}_{\mathrm{M\rightleftharpoons A}}\vec{M}_{\mathrm{M\rightleftharpoons A}}}
\end{bmatrix} \tag{4-76a}$$

$$\overline{\overline{\Psi}}_{\mathrm{FCE}}^{\mathrm{DoM}} = \begin{bmatrix}
\overline{\overline{Z}}^{\vec{J}^{\mathrm{A\rightleftharpoons M}}\vec{J}^{\mathrm{A\rightleftharpoons M}}} & \overline{\overline{Z}}^{\vec{J}^{\mathrm{A\rightleftharpoons M}}\vec{J}^{\mathrm{M}}} & \overline{\overline{Z}}^{\vec{J}^{\mathrm{A\rightleftharpoons M}}\vec{J}} & \overline{\overline{Z}}^{\vec{J}^{\mathrm{A\rightleftharpoons M}}\vec{J}_{\mathrm{M}}} & \overline{\overline{Z}}^{\vec{J}^{\mathrm{A\rightleftharpoons M}}\vec{J}_{\mathrm{M\rightleftharpoons A}}} & \overline{\overline{Z}}^{\vec{J}^{\mathrm{A\rightleftharpoons M}}\vec{M}^{\mathrm{A\rightleftharpoons M}}} & \overline{\overline{Z}}^{\vec{J}^{\mathrm{A\rightleftharpoons M}}\vec{M}} & \overline{\overline{Z}}^{\vec{J}^{\mathrm{A\rightleftharpoons M}}\vec{M}_{\mathrm{M\rightleftharpoons A}}} \\
\overline{\overline{Z}}^{\vec{J}^{\mathrm{M}}\vec{J}^{\mathrm{A\rightleftharpoons M}}} & \overline{\overline{Z}}^{\vec{J}^{\mathrm{M}}\vec{J}^{\mathrm{M}}} & \overline{\overline{Z}}^{\vec{J}^{\mathrm{M}}\vec{J}} & \overline{\overline{Z}}^{\vec{J}^{\mathrm{M}}\vec{J}_{\mathrm{M}}} & \overline{\overline{Z}}^{\vec{J}^{\mathrm{M}}\vec{J}_{\mathrm{M\rightleftharpoons A}}} & \overline{\overline{Z}}^{\vec{J}^{\mathrm{M}}\vec{M}^{\mathrm{A\rightleftharpoons M}}} & \overline{\overline{Z}}^{\vec{J}^{\mathrm{M}}\vec{M}} & \overline{\overline{Z}}^{\vec{J}^{\mathrm{M}}\vec{M}_{\mathrm{M\rightleftharpoons A}}} \\
\overline{\overline{Z}}^{\vec{J}\vec{J}^{\mathrm{A\rightleftharpoons M}}} & \overline{\overline{Z}}^{\vec{J}\vec{J}^{\mathrm{M}}} & \overline{\overline{Z}}^{\vec{J}\vec{J}} & \overline{\overline{Z}}^{\vec{J}\vec{J}_{\mathrm{M}}} & \overline{\overline{Z}}^{\vec{J}\vec{J}_{\mathrm{M\rightleftharpoons A}}} & \overline{\overline{Z}}^{\vec{J}\vec{M}^{\mathrm{A\rightleftharpoons M}}} & \overline{\overline{Z}}^{\vec{J}\vec{M}} & \overline{\overline{Z}}^{\vec{J}\vec{M}_{\mathrm{M\rightleftharpoons A}}} \\
\overline{\overline{Z}}^{\vec{J}_{\mathrm{M}}\vec{J}^{\mathrm{A\rightleftharpoons M}}} & \overline{\overline{Z}}^{\vec{J}_{\mathrm{M}}\vec{J}^{\mathrm{M}}} & \overline{\overline{Z}}^{\vec{J}_{\mathrm{M}}\vec{J}} & \overline{\overline{Z}}^{\vec{J}_{\mathrm{M}}\vec{J}_{\mathrm{M}}} & \overline{\overline{Z}}^{\vec{J}_{\mathrm{M}}\vec{J}_{\mathrm{M\rightleftharpoons A}}} & \overline{\overline{Z}}^{\vec{J}_{\mathrm{M}}\vec{M}^{\mathrm{A\rightleftharpoons M}}} & \overline{\overline{Z}}^{\vec{J}_{\mathrm{M}}\vec{M}} & \overline{\overline{Z}}^{\vec{J}_{\mathrm{M}}\vec{M}_{\mathrm{M\rightleftharpoons A}}} \\
\overline{\overline{Z}}^{\vec{M}\vec{J}^{\mathrm{A\rightleftharpoons M}}} & \overline{\overline{Z}}^{\vec{M}\vec{J}^{\mathrm{M}}} & \overline{\overline{Z}}^{\vec{M}\vec{J}} & \overline{\overline{Z}}^{\vec{M}\vec{J}_{\mathrm{M}}} & \overline{\overline{Z}}^{\vec{M}\vec{J}_{\mathrm{M\rightleftharpoons A}}} & \overline{\overline{Z}}^{\vec{M}\vec{M}^{\mathrm{A\rightleftharpoons M}}} & \overline{\overline{Z}}^{\vec{M}\vec{M}} & \overline{\overline{Z}}^{\vec{M}\vec{M}_{\mathrm{M\rightleftharpoons A}}} \\
\overline{\overline{Z}}^{\vec{M}_{\mathrm{M\rightleftharpoons A}}\vec{J}^{\mathrm{A\rightleftharpoons M}}} & \overline{\overline{Z}}^{\vec{M}_{\mathrm{M\rightleftharpoons A}}\vec{J}^{\mathrm{M}}} & \overline{\overline{Z}}^{\vec{M}_{\mathrm{M\rightleftharpoons A}}\vec{J}} & \overline{\overline{Z}}^{\vec{M}_{\mathrm{M\rightleftharpoons A}}\vec{J}_{\mathrm{M}}} & \overline{\overline{Z}}^{\vec{M}_{\mathrm{M\rightleftharpoons A}}\vec{J}_{\mathrm{M\rightleftharpoons A}}} & \overline{\overline{Z}}^{\vec{M}_{\mathrm{M\rightleftharpoons A}}\vec{M}^{\mathrm{A\rightleftharpoons M}}} & \overline{\overline{Z}}^{\vec{M}_{\mathrm{M\rightleftharpoons A}}\vec{M}} & \overline{\overline{Z}}^{\vec{M}_{\mathrm{M\rightleftharpoons A}}\vec{M}_{\mathrm{M\rightleftharpoons A}}} \\
\overline{\overline{Z}}^{\vec{J}_{\mathrm{M\rightleftharpoons A}}\vec{J}^{\mathrm{A\rightleftharpoons M}}} & \overline{\overline{Z}}^{\vec{J}_{\mathrm{M\rightleftharpoons A}}\vec{J}^{\mathrm{M}}} & \overline{\overline{Z}}^{\vec{J}_{\mathrm{M\rightleftharpoons A}}\vec{J}} & \overline{\overline{Z}}^{\vec{J}_{\mathrm{M\rightleftharpoons A}}\vec{J}_{\mathrm{M}}} & \overline{\overline{Z}}^{\vec{J}_{\mathrm{M\rightleftharpoons A}}\vec{J}_{\mathrm{M\rightleftharpoons A}}} & \overline{\overline{Z}}^{\vec{J}_{\mathrm{M\rightleftharpoons A}}\vec{M}^{\mathrm{A\rightleftharpoons M}}} & \overline{\overline{Z}}^{\vec{J}_{\mathrm{M\rightleftharpoons A}}\vec{M}} & \overline{\overline{Z}}^{\vec{J}_{\mathrm{M\rightleftharpoons A}}\vec{M}_{\mathrm{M\rightleftharpoons A}}}
\end{bmatrix} \tag{4-76b}$$

Here, the subscript "FCE" and superscripts "DoJ"&"DoM" have the same meanings as the ones used in Eqs. (4-24) and (4-25).

For the convenience of the following discussions, Eqs. (4-71) and (4-74) are uniformly written as follows:

$$\bar{a}^{\mathrm{AV}} = \overline{\overline{T}} \cdot \bar{a} \tag{4-77}$$

where $\bar{a} = \bar{a}^{\mathrm{BV}} / \bar{a}^{\mathrm{BS}}$ and correspondingly $\overline{\overline{T}} = \overline{\overline{T}}^{\mathrm{BV}\to\mathrm{AV}} / \overline{\overline{T}}^{\mathrm{BS}\to\mathrm{AV}}$.

### 4.3.3 Power Transport Theorem and Input Power Operator

Applying the results obtained in previous Chap. 2 to the propagation medium shown in Figs. 4-6 and 4-7, we immediately have the power transport theorem (PTT) for the propagation medium as follows:

$$P^{\mathrm{A}\rightleftharpoons\mathrm{M}} = P_{\mathrm{dis}}^{\mathrm{M}} + P_{\mathrm{rad}}^{\mathrm{I}} + j\,P_{\mathrm{sto}}^{\mathrm{M}} + P_{\mathrm{M}\rightleftharpoons\mathrm{A}} \tag{4-78}$$

where $P^{\mathrm{A}\rightleftharpoons\mathrm{M}}$ is the input power inputted into the propagation medium $\mathbb{M}$, and $P_{\mathrm{dis}}^{\mathrm{M}}$ is the power dissipated in $\mathbb{M}$, and $P_{\mathrm{rad}}^{\mathrm{I}}$ is the power radiated to infinity, and $P_{\mathrm{sto}}^{\mathrm{M}}$ is the power corresponding to the stored energy in $\mathbb{M}$, and $P_{\mathrm{M}\rightleftharpoons\mathrm{A}}$ is the power inputted into receiving system.

The above-mentioned various powers are as follows:

$$P^{\mathrm{A}\rightleftharpoons\mathrm{M}} = (1/2)\iint_{\mathbb{S}^{\mathrm{A}\rightleftharpoons\mathrm{M}}}\left(\vec{E}\times\vec{H}^{\dagger}\right)\cdot\hat{n}^{\to\mathrm{M}}dS \tag{4-79a}$$

$$P_{\mathrm{dis}}^{\mathrm{M}} = (1/2)\left\langle\vec{\vec{\sigma}}\cdot\vec{E},\vec{E}\right\rangle_{\mathbb{V}} \tag{4-79b}$$

$$P_{\mathrm{rad}}^{\mathrm{I}} = (1/2)\iint_{\mathbb{S}}\left(\vec{E}\times\vec{H}^{\dagger}\right)\cdot\hat{n}dS \tag{4-79c}$$

$$P_{\mathrm{sto}}^{\mathrm{M}} = 2\omega\left\{\left[\frac{1}{4}\left\langle\vec{H},\mu_0\vec{H}\right\rangle_{\mathbb{F}} - \frac{1}{4}\left\langle\varepsilon_0\vec{E},\vec{E}\right\rangle_{\mathbb{F}}\right] + \left[\frac{1}{4}\left\langle\vec{H},\vec{\vec{\mu}}\cdot\vec{H}\right\rangle_{\mathbb{V}} - \frac{1}{4}\left\langle\vec{\vec{\varepsilon}}\cdot\vec{E},\vec{E}\right\rangle_{\mathbb{V}}\right]\right\} \tag{4-79d}$$

$$P_{\mathrm{M}\rightleftharpoons\mathrm{A}} = (1/2)\iint_{\mathbb{S}_{\mathrm{M}\rightleftharpoons\mathrm{A}}}\left(\vec{E}\times\vec{H}^{\dagger}\right)\cdot\hat{n}_{\to\mathrm{A}}dS \tag{4-79e}$$

where $\hat{n}^{\to\mathrm{M}}$ is the normal direction of $\mathbb{S}^{\mathrm{A}\rightleftharpoons\mathrm{M}}$ and points to $\mathbb{M}$, and $\hat{n}$ is the normal direction of $\mathbb{S}$ and points to infinity, and $\hat{n}_{\to\mathrm{A}}$ is the normal direction of $\mathbb{S}_{\mathrm{M}\rightleftharpoons\mathrm{A}}$ and points to rec-antenna, as shown in Figs. 4-6 and 4-7.

Based on Eqs. (4-50a)&(4-50b) and the tangential continuity of the $\{\vec{E},\vec{H}\}$ on $\mathbb{S}^{\mathrm{A}\rightleftharpoons\mathrm{M}}$, the IPO $P^{\mathrm{A}\rightleftharpoons\mathrm{M}}$ given in Eq. (4-79a) can be alternatively written as the following ones

$$\begin{aligned}
P^{\mathrm{A}\rightleftharpoons\mathrm{M}} &= (1/2)\left\langle\hat{n}^{\to\mathrm{M}}\times\vec{J}^{\mathrm{A}\rightleftharpoons\mathrm{M}},\vec{M}^{\mathrm{A}\rightleftharpoons\mathrm{M}}\right\rangle_{\mathbb{S}^{\mathrm{A}\rightleftharpoons\mathrm{M}}}\\
&= -\frac{1}{2}\left\langle\vec{J}^{\mathrm{A}\rightleftharpoons\mathrm{M}},\mathcal{E}_0\left(\vec{J}^{\mathrm{A}\rightleftharpoons\mathrm{M}}+\vec{J}^{\mathrm{M}}+\vec{J}+\vec{J}_{\mathrm{M}}+\vec{J}_{\mathrm{M}\rightleftharpoons\mathrm{A}},\vec{M}^{\mathrm{A}\rightleftharpoons\mathrm{M}}+\vec{M}+\vec{M}_{\mathrm{M}\rightleftharpoons\mathrm{A}}\right)\right\rangle_{\underset{\sim}{\mathbb{S}}^{\mathrm{A}\rightleftharpoons\mathrm{M}}}\\
&= -\frac{1}{2}\left\langle\vec{M}^{\mathrm{A}\rightleftharpoons\mathrm{M}},\mathcal{H}_0\left(\vec{J}^{\mathrm{A}\rightleftharpoons\mathrm{M}}+\vec{J}^{\mathrm{M}}+\vec{J}+\vec{J}_{\mathrm{M}}+\vec{J}_{\mathrm{M}\rightleftharpoons\mathrm{A}},\vec{M}^{\mathrm{A}\rightleftharpoons\mathrm{M}}+\vec{M}+\vec{M}_{\mathrm{M}\rightleftharpoons\mathrm{A}}\right)\right\rangle_{\underset{\sim}{\mathbb{S}}^{\mathrm{A}\rightleftharpoons\mathrm{M}}}^{\dagger}
\end{aligned} \tag{4-80}$$

Here, the right-hand side of the first equality is the current form of IPO, and the right-hand sides of the second and third equalities are respectively the JE and HM versions of the interaction form for IPO. The integral surface $\underset{\sim}{\mathbb{S}}^{\mathrm{A}\rightleftharpoons\mathrm{M}}$ is an inner sub-boundary of propagation medium.

By discretizing IPO (4-80) and utilizing transformation (4-77), we derive the following matrix form of the IPO

$$P^{\mathrm{A}\rightleftharpoons\mathrm{M}} = \bar{a}^{\dagger} \cdot \bar{\bar{P}}^{\mathrm{A}\rightleftharpoons\mathrm{M}} \cdot \bar{a} \tag{4-81}$$

and the formulation for calculating the quadratic matrix $\bar{\bar{P}}^{\mathrm{A}\rightleftharpoons\mathrm{M}}$ is given in the App. D4 of this report.

### 4.3.4 Input-Power-Decoupled Modes

The IP-DMs in modal space can be derived from solving the modal decoupling equation $\bar{\bar{P}}_{-}^{\mathrm{A}\rightleftharpoons\mathrm{M}} \cdot \bar{\alpha}_{\xi} = \theta_{\xi} \bar{\bar{P}}_{+}^{\mathrm{A}\rightleftharpoons\mathrm{M}} \cdot \bar{\alpha}_{\xi}$ defined on modal space, where $\bar{\bar{P}}_{+}^{\mathrm{A}\rightleftharpoons\mathrm{M}}$ and $\bar{\bar{P}}_{-}^{\mathrm{A}\rightleftharpoons\mathrm{M}}$ are the positive and negative Hermitian parts of matrix $\bar{\bar{P}}^{\mathrm{A}\rightleftharpoons\mathrm{M}}$. If some derived modes $\{\bar{\alpha}_1, \bar{\alpha}_2, \cdots, \bar{\alpha}_d\}$ are *d*-order degenerate, then the Gram-Schmidt orthogonalization process given in the previous Sec. 4.2.4.1 is necessary, and it is not repeated here.

Similarly to the previous Sec. 4.2, the modal fields constructed above satisfy the following decoupling relation

$$(1/2) \iint_{\mathbb{S}^{\mathrm{A}\rightleftharpoons\mathrm{M}}} \left( \vec{E}_{\zeta} \times \vec{H}_{\xi}^{\dagger} \right) \cdot \hat{n}^{\rightarrow\mathrm{M}} dS = \left( 1 + j\,\theta_{\xi} \right) \delta_{\xi\zeta} \tag{4-82}$$

and the relation implies that **the IP-DMs don't have net energy coupling in integral period**. By employing the decoupling relation, we have the following Parseval's identity

$$\sum_{\xi} \left| c_{\xi} \right|^2 = (1/T) \int_{t_0}^{t_0+T} \left[ \iint_{\mathbb{S}^{\mathrm{A}\rightleftharpoons\mathrm{M}}} \left( \vec{\mathcal{E}} \times \vec{\mathcal{H}} \right) \cdot \hat{n}^{\rightarrow\mathrm{M}} dS \right] dt \tag{4-83}$$

in which $\{\vec{\mathcal{E}}, \vec{\mathcal{H}}\}$ are the time-domain fields, and the expansion coefficients $c_{\xi}$ have expression $c_{\xi} = -(1/2) < \vec{J}_{\xi}^{\mathrm{A}\rightleftharpoons\mathrm{M}}, \vec{E} >_{\mathbb{S}^{\mathrm{A}\rightleftharpoons\mathrm{M}}} \big/ (1 + j\,\theta_{\xi}) = -(1/2) < \vec{H}, \vec{M}_{\xi}^{\mathrm{A}\rightleftharpoons\mathrm{M}} >_{\mathbb{S}^{\mathrm{A}\rightleftharpoons\mathrm{M}}} \big/ (1 + j\,\theta_{\xi})$, where $\{\vec{E}, \vec{H}\}$ is a previously known field distributing on input port $\mathbb{S}^{\mathrm{A}\rightleftharpoons\mathrm{M}}$.

Just like the results obtained in the previous Sec. 4.2.4.3, we can also define the modal significance $\mathrm{MS}_{\xi} = 1/|1 + j\,\theta_{\xi}|$, modal input impedance $Z_{\xi}^{\mathrm{A}\rightleftharpoons\mathrm{M}} = P_{\xi}^{\mathrm{A}\rightleftharpoons\mathrm{M}} \big/ \left[ (1/2) < \vec{J}_{\xi}^{\mathrm{A}\rightleftharpoons\mathrm{M}}, \vec{J}_{\xi}^{\mathrm{A}\rightleftharpoons\mathrm{M}} >_{\mathbb{S}^{\mathrm{A}\rightleftharpoons\mathrm{M}}} \right]$, and modal input admittance $Y_{\xi}^{\mathrm{A}\rightleftharpoons\mathrm{M}} = P_{\xi}^{\mathrm{A}\rightleftharpoons\mathrm{M}} \big/ \left[ (1/2) < \vec{M}_{\xi}^{\mathrm{A}\rightleftharpoons\mathrm{M}}, \vec{M}_{\xi}^{\mathrm{A}\rightleftharpoons\mathrm{M}} >_{\mathbb{S}^{\mathrm{A}\rightleftharpoons\mathrm{M}}} \right]$, to quantitatively describe the modal features of the propagation medium shown in Figs. 4-6 and 4-7.

## 4.4 IP-DMs of Receiving Antenna

In this section, the rec-antenna, i.e. the waveguide-loaded *dielectric resonator antenna* (*DRA*), shown in Fig. 4-1 is focused on seperately from the other structures, as shown in the following Fig. 4-8.

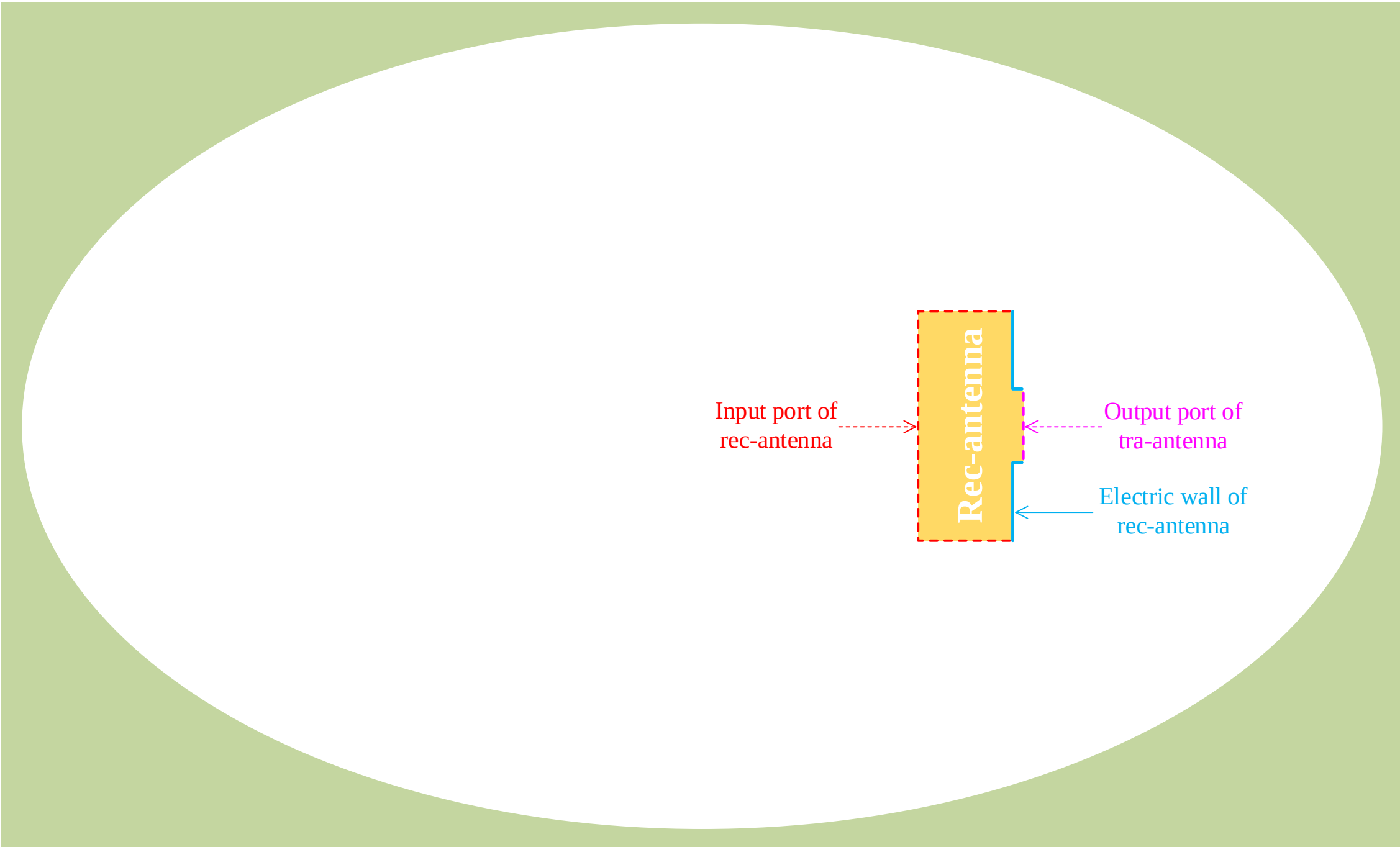

Figure 4-8 Geometry of the rec-antenna shown in Fig. 4-1

The general process for constructing the IP-DMs of the rec-antenna is similar to the one used in the previous Sec. 4.2.

### 4.4.1 Topological Structure and Source-Field Relationships

The topological structure of the rec-antenna is detailedly exhibited in Fig. 4-9.

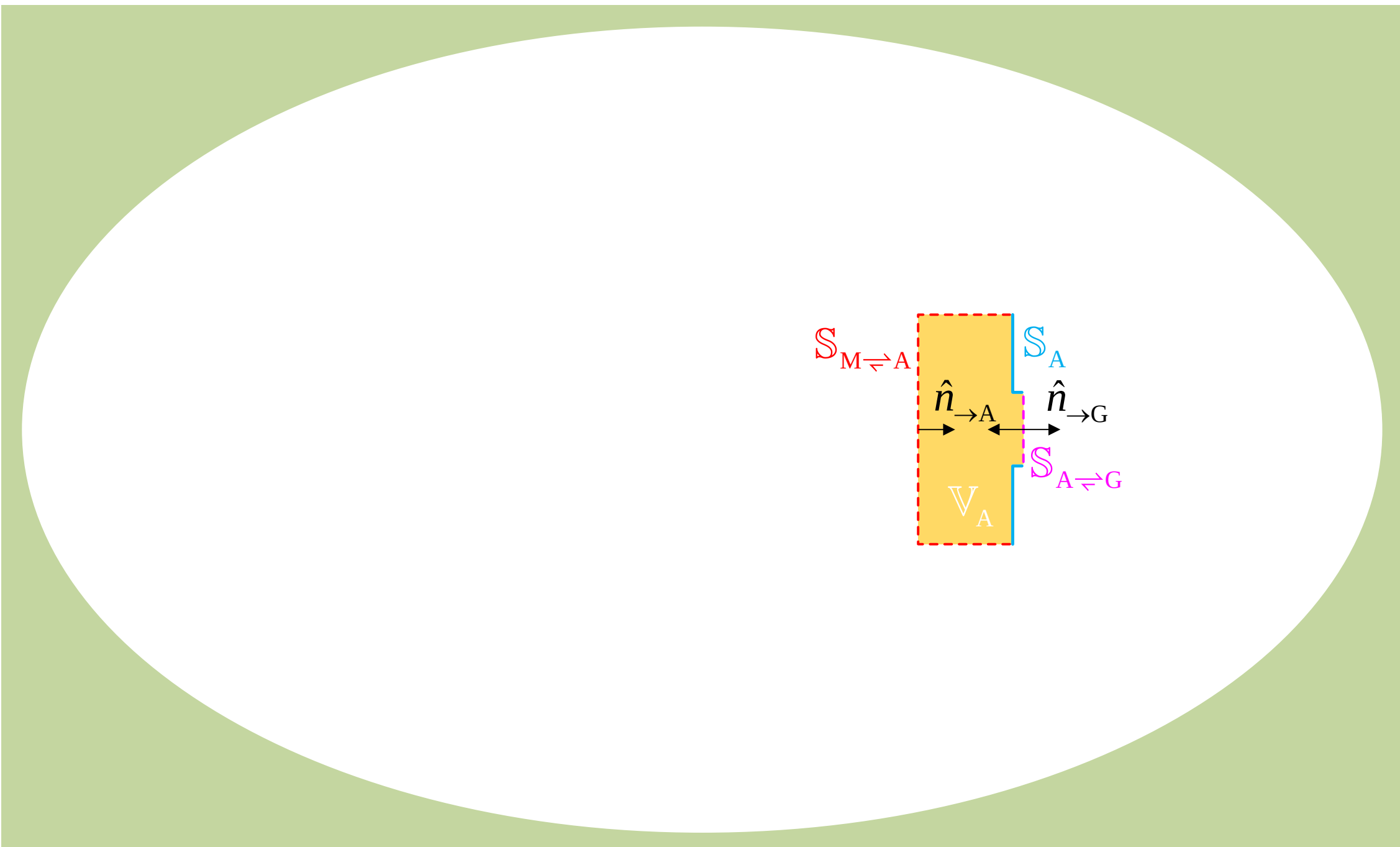

Figure 4-9 Detailed topological structure of the rec-antenna shown in Fig.4-8

In the above Fig. 4-9, surface $\mathbb{S}_{\text{M}\rightleftharpoons\text{A}}$ denotes the input port of the rec-antenna. The region occupied by the material body, i.e. the DRA, is denoted as $\mathbb{V}_{\text{A}}$. The interface between $\mathbb{V}_{\text{A}}$ and the *thick metallic ground plane* is denoted as $\mathbb{S}_{\text{A}}$; the interface between $\mathbb{V}_{\text{A}}$ and the loading waveguide is denoted as $\mathbb{S}_{\text{A}\rightleftharpoons\text{G}}$.

Clearly, surfaces $\mathbb{S}_{\text{M}\rightleftharpoons\text{A}}$, $\mathbb{S}_{\text{A}}$, and $\mathbb{S}_{\text{A}\rightleftharpoons\text{G}}$ constitute a closed surface, and the closed surface is just the whole boundary of $\mathbb{V}_{\text{A}}$, i.e., $\partial\mathbb{V}_{\text{A}} = \mathbb{S}_{\text{M}\rightleftharpoons\text{A}} \bigcup \mathbb{S}_{\text{A}} \bigcup \mathbb{S}_{\text{A}\rightleftharpoons\text{G}}$. In addition, the permeability, permeativity, and conductivity of $\mathbb{V}_{\text{A}}$ are denoted as $\ddot{\mu}$, $\ddot{\varepsilon}$, and $\ddot{\sigma}$ respectively.

If the equivalent surface currents distributing on $\mathbb{S}_{\text{M}\rightleftharpoons\text{A}}$ are denoted as $\{\vec{J}_{\text{M}\rightleftharpoons\text{A}}, \vec{M}_{\text{M}\rightleftharpoons\text{A}}\}$, and the equivalent surface electric current distributing on $\mathbb{S}_{\text{A}}$ is denoted as $\vec{J}_{\text{A}}$ ①, and the equivalent surface currents distributing on $\mathbb{S}_{\text{A}\rightleftharpoons\text{G}}$ are denoted as $\{\vec{J}_{\text{A}\rightleftharpoons\text{G}}, \vec{M}_{\text{A}\rightleftharpoons\text{G}}\}$, then the field distributing on $\mathbb{V}_{\text{A}}$ can be expressed as the following operator form

$$\vec{F}(\vec{r}) = \mathcal{F}\left(\vec{J}_{\text{M}\rightleftharpoons\text{A}} + \vec{J}_{\text{A}\rightleftharpoons\text{G}} + \vec{J}_{\text{A}}, \vec{M}_{\text{M}\rightleftharpoons\text{A}} + \vec{M}_{\text{A}\rightleftharpoons\text{G}}\right) \quad , \quad \vec{r} \in \mathbb{V}_{\text{A}} \tag{4-84}$$

where $\vec{F} = \vec{E} / \vec{H}$, and correspondingly $\mathcal{F} = \mathcal{E} / \mathcal{H}$, and the operator is defined as that $\mathcal{F}(\vec{J}, \vec{M}) = \ddot{G}^{JF} * \vec{J} + \ddot{G}^{MF} * \vec{M}$ (here, $\ddot{G}^{JF}$ and $\ddot{G}^{MF}$ are the dyadic Green's functions corresponding to the region $\mathbb{V}_{\text{A}}$ with material parameters $\{\ddot{\mu}, \ddot{\varepsilon}, \ddot{\sigma}\}$).

The currents $\{\vec{J}_{\text{M}\rightleftharpoons\text{A}}, \vec{M}_{\text{M}\rightleftharpoons\text{A}}\}$ and fields $\{\vec{E}, \vec{H}\}$ involved in Eq. (4-84) satisfy the following relations

$$\hat{n}_{\to\text{A}} \times \left[\vec{H}(\vec{r}_{\text{A}})\right]_{\vec{r}_{\text{A}} \to \vec{r}} = \vec{J}_{\text{M}\rightleftharpoons\text{A}}(\vec{r}) \quad , \quad \vec{r} \in \mathbb{S}_{\text{M}\rightleftharpoons\text{A}} \tag{4-85a}$$

$$\left[\vec{E}(\vec{r}_{\text{A}})\right]_{\vec{r}_{\text{A}} \to \vec{r}} \times \hat{n}_{\to\text{A}} = \vec{M}_{\text{M}\rightleftharpoons\text{A}}(\vec{r}) \quad , \quad \vec{r} \in \mathbb{S}_{\text{M}\rightleftharpoons\text{A}} \tag{4-85b}$$

and the currents $\{\vec{J}_{\text{A}\rightleftharpoons\text{G}}, \vec{M}_{\text{A}\rightleftharpoons\text{G}}\}$ and fields $\{\vec{E}, \vec{H}\}$ involved in Eq. (4-84) satisfy the following relations

$$\hat{n}_{\to\text{A}} \times \left[\vec{H}(\vec{r}_{\text{A}})\right]_{\vec{r}_{\text{A}} \to \vec{r}} = \vec{J}_{\text{A}\rightleftharpoons\text{G}}(\vec{r}) \quad , \quad \vec{r} \in \mathbb{S}_{\text{A}\rightleftharpoons\text{G}} \tag{4-86a}$$

$$\left[\vec{E}(\vec{r}_{\text{A}})\right]_{\vec{r}_{\text{A}} \to \vec{r}} \times \hat{n}_{\to\text{A}} = \vec{M}_{\text{A}\rightleftharpoons\text{G}}(\vec{r}) \quad , \quad \vec{r} \in \mathbb{S}_{\text{A}\rightleftharpoons\text{G}} \tag{4-86b}$$

In the above Eqs. (4-85a)&(4-85b) and (4-86a)&(4-86b), point $\vec{r}_{\text{A}}$ belongs to $\mathbb{V}_{\text{A}}$ and $\vec{r}_{\text{A}}$ approaches the point $\vec{r}$ on $\mathbb{S}_{\text{M}\rightleftharpoons\text{A}} \bigcup \mathbb{S}_{\text{A}\rightleftharpoons\text{G}}$, and $\hat{n}_{\to\text{A}}$ is the normal direction of $\partial\mathbb{V}_{\text{A}}$ ($= \mathbb{S}_{\text{M}\rightleftharpoons\text{A}} \bigcup \mathbb{S}_{\text{A}} \bigcup \mathbb{S}_{\text{A}\rightleftharpoons\text{G}}$) and points to the interior of $\mathbb{V}_{\text{A}}$ as shown in Fig. 4-9.

---

① The equivalent surface electric current distributing on $\mathbb{S}_{\text{A}}$ is equal to the induced surface electric current distributing on $\mathbb{S}_{\text{A}}$ [13]. The equivalent surface magnetic current distributing on $\mathbb{S}_{\text{A}}$ is zero, because of the homogeneous tangential electric field boundary condition on $\mathbb{S}_{\text{A}}$ [13].

### 4.4.2 Mathematical Description for Modal Space

Combining the Eq. (4-84) with Eqs. (4-85a)&(4-85b), we can obtain the following integral equations

$$\left[\mathcal{H}\left(\vec{J}_{\mathrm{M}\rightleftharpoons\mathrm{A}}+\vec{J}_{\mathrm{A}\rightleftharpoons\mathrm{G}}+\vec{J}_{\mathrm{A}},\vec{M}_{\mathrm{M}\rightleftharpoons\mathrm{A}}+\vec{M}_{\mathrm{A}\rightleftharpoons\mathrm{G}}\right)\right]_{\vec{r}_{\mathrm{A}}\to\vec{r}}^{\tan} = \vec{J}_{\mathrm{M}\rightleftharpoons\mathrm{A}}\left(\vec{r}\right)\times\hat{n}_{\to\mathrm{A}} \quad , \quad \vec{r}\in\mathbb{S}_{\mathrm{M}\rightleftharpoons\mathrm{A}} \quad (4\text{-}87\mathrm{a})$$

$$\left[\mathcal{E}\left(\vec{J}_{\mathrm{M}\rightleftharpoons\mathrm{A}}+\vec{J}_{\mathrm{A}\rightleftharpoons\mathrm{G}}+\vec{J}_{\mathrm{A}},\vec{M}_{\mathrm{M}\rightleftharpoons\mathrm{A}}+\vec{M}_{\mathrm{A}\rightleftharpoons\mathrm{G}}\right)\right]_{\vec{r}_{\mathrm{A}}\to\vec{r}}^{\tan} = \hat{n}_{\to\mathrm{A}}\times\vec{M}_{\mathrm{M}\rightleftharpoons\mathrm{A}}\left(\vec{r}\right) \quad , \quad \vec{r}\in\mathbb{S}_{\mathrm{M}\rightleftharpoons\mathrm{A}} \quad (4\text{-}87\mathrm{b})$$

about currents $\{\vec{J}_{\mathrm{M}\rightleftharpoons\mathrm{A}},\vec{M}_{\mathrm{M}\rightleftharpoons\mathrm{A}}\}$, $\{\vec{J}_{\mathrm{A}\rightleftharpoons\mathrm{G}},\vec{M}_{\mathrm{A}\rightleftharpoons\mathrm{G}}\}$, and $\vec{J}_{\mathrm{A}}$, where the superscript "tan" represents the tangential components of the fields. Combining the Eq. (4-84) with Eqs. (4-86a)&(4-86b), we can obtain the following integral equations

$$\left[\mathcal{H}\left(\vec{J}_{\mathrm{M}\rightleftharpoons\mathrm{A}}+\vec{J}_{\mathrm{A}\rightleftharpoons\mathrm{G}}+\vec{J}_{\mathrm{A}},\vec{M}_{\mathrm{M}\rightleftharpoons\mathrm{A}}+\vec{M}_{\mathrm{A}\rightleftharpoons\mathrm{G}}\right)\right]_{\vec{r}_{\mathrm{A}}\to\vec{r}}^{\tan} = \vec{J}_{\mathrm{A}\rightleftharpoons\mathrm{G}}\left(\vec{r}\right)\times\hat{n}_{\to\mathrm{A}} \quad , \quad \vec{r}\in\mathbb{S}_{\mathrm{A}\rightleftharpoons\mathrm{G}} \quad (4\text{-}88\mathrm{a})$$

$$\left[\mathcal{E}\left(\vec{J}_{\mathrm{M}\rightleftharpoons\mathrm{A}}+\vec{J}_{\mathrm{A}\rightleftharpoons\mathrm{G}}+\vec{J}_{\mathrm{A}},\vec{M}_{\mathrm{M}\rightleftharpoons\mathrm{A}}+\vec{M}_{\mathrm{A}\rightleftharpoons\mathrm{G}}\right)\right]_{\vec{r}_{\mathrm{A}}\to\vec{r}}^{\tan} = \hat{n}_{\to\mathrm{A}}\times\vec{M}_{\mathrm{A}\rightleftharpoons\mathrm{G}}\left(\vec{r}\right) \quad , \quad \vec{r}\in\mathbb{S}_{\mathrm{A}\rightleftharpoons\mathrm{G}} \quad (4\text{-}88\mathrm{b})$$

about currents $\{\vec{J}_{\mathrm{M}\rightleftharpoons\mathrm{A}},\vec{M}_{\mathrm{M}\rightleftharpoons\mathrm{A}}\}$, $\{\vec{J}_{\mathrm{A}\rightleftharpoons\mathrm{G}},\vec{M}_{\mathrm{A}\rightleftharpoons\mathrm{G}}\}$, and $\vec{J}_{\mathrm{A}}$.

Based on Eq. (4-84) and the homogeneous tangential electric field boundary condition on $\mathbb{S}_{\mathrm{A}}$, we have the following electric field integral equation

$$\left[\mathcal{E}\left(\vec{J}_{\mathrm{M}\rightleftharpoons\mathrm{A}}+\vec{J}_{\mathrm{A}\rightleftharpoons\mathrm{G}}+\vec{J}_{\mathrm{A}},\vec{M}_{\mathrm{M}\rightleftharpoons\mathrm{A}}+\vec{M}_{\mathrm{A}\rightleftharpoons\mathrm{G}}\right)\right]_{\vec{r}_{\mathrm{A}}\to\vec{r}}^{\tan} = 0 \quad , \quad \vec{r}\in\mathbb{S}_{\mathrm{A}} \quad (4\text{-}89)$$

about currents $\{\vec{J}_{\mathrm{M}\rightleftharpoons\mathrm{A}},\vec{M}_{\mathrm{M}\rightleftharpoons\mathrm{A}}\}$, $\{\vec{J}_{\mathrm{A}\rightleftharpoons\mathrm{G}},\vec{M}_{\mathrm{A}\rightleftharpoons\mathrm{G}}\}$, and $\vec{J}_{\mathrm{A}}$.

The above Eqs. (4-87a)~(4-89) are a complete mathematical description for the modal space of the rec-antenna shown in Figs. 4-8 and 4-9. If the currents $\{\vec{J}_{\mathrm{M}\rightleftharpoons\mathrm{A}},\vec{M}_{\mathrm{M}\rightleftharpoons\mathrm{A}}\}$, $\{\vec{J}_{\mathrm{A}\rightleftharpoons\mathrm{G}},\vec{M}_{\mathrm{A}\rightleftharpoons\mathrm{G}}\}$, and $\vec{J}_{\mathrm{A}}$ are expanded in terms of some proper basis functions, and Eqs. (4-87a), (4-87b), (4-88a), (4-88b), and (4-89) are tested with $\{\vec{b}_{\xi}^{\vec{M}_{\mathrm{M}\rightleftharpoons\mathrm{A}}}\}$, $\{\vec{b}_{\xi}^{\vec{J}_{\mathrm{M}\rightleftharpoons\mathrm{A}}}\}$, $\{\vec{b}_{\xi}^{\vec{M}_{\mathrm{A}\rightleftharpoons\mathrm{G}}}\}$, $\{\vec{b}_{\xi}^{\vec{J}_{\mathrm{A}\rightleftharpoons\mathrm{G}}}\}$, and $\{\vec{b}_{\xi}^{\vec{J}_{\mathrm{A}}}\}$ respectively, then the integral equations are immediately discretized into the following matrix equations

$$\begin{aligned}0 = {} & \overline{\overline{Z}}^{\vec{M}_{\mathrm{M}\rightleftharpoons\mathrm{A}}\vec{J}_{\mathrm{M}\rightleftharpoons\mathrm{A}}}\cdot\overline{a}^{\vec{J}_{\mathrm{M}\rightleftharpoons\mathrm{A}}}+\overline{\overline{Z}}^{\vec{M}_{\mathrm{M}\rightleftharpoons\mathrm{A}}\vec{J}_{\mathrm{A}\rightleftharpoons\mathrm{G}}}\cdot\overline{a}^{\vec{J}_{\mathrm{A}\rightleftharpoons\mathrm{G}}}+\overline{\overline{Z}}^{\vec{M}_{\mathrm{M}\rightleftharpoons\mathrm{A}}\vec{J}_{\mathrm{A}}}\cdot\overline{a}^{\vec{J}_{\mathrm{A}}}+\overline{\overline{Z}}^{\vec{M}_{\mathrm{M}\rightleftharpoons\mathrm{A}}\vec{M}_{\mathrm{M}\rightleftharpoons\mathrm{A}}}\cdot\overline{a}^{\vec{M}_{\mathrm{M}\rightleftharpoons\mathrm{A}}}\\&+\overline{\overline{Z}}^{\vec{M}_{\mathrm{M}\rightleftharpoons\mathrm{A}}\vec{M}_{\mathrm{A}\rightleftharpoons\mathrm{G}}}\cdot\overline{a}^{\vec{M}_{\mathrm{A}\rightleftharpoons\mathrm{G}}}\end{aligned} \quad (4\text{-}90\mathrm{a})$$

$$\begin{aligned}0 = {} & \overline{\overline{Z}}^{\vec{J}_{\mathrm{M}\rightleftharpoons\mathrm{A}}\vec{J}_{\mathrm{M}\rightleftharpoons\mathrm{A}}}\cdot\overline{a}^{\vec{J}_{\mathrm{M}\rightleftharpoons\mathrm{A}}}+\overline{\overline{Z}}^{\vec{J}_{\mathrm{M}\rightleftharpoons\mathrm{A}}\vec{J}_{\mathrm{A}\rightleftharpoons\mathrm{G}}}\cdot\overline{a}^{\vec{J}_{\mathrm{A}\rightleftharpoons\mathrm{G}}}+\overline{\overline{Z}}^{\vec{J}_{\mathrm{M}\rightleftharpoons\mathrm{A}}\vec{J}_{\mathrm{A}}}\cdot\overline{a}^{\vec{J}_{\mathrm{A}}}+\overline{\overline{Z}}^{\vec{J}_{\mathrm{M}\rightleftharpoons\mathrm{A}}\vec{M}_{\mathrm{M}\rightleftharpoons\mathrm{A}}}\cdot\overline{a}^{\vec{M}_{\mathrm{M}\rightleftharpoons\mathrm{A}}}\\&+\overline{\overline{Z}}^{\vec{J}_{\mathrm{M}\rightleftharpoons\mathrm{A}}\vec{M}_{\mathrm{A}\rightleftharpoons\mathrm{G}}}\cdot\overline{a}^{\vec{M}_{\mathrm{A}\rightleftharpoons\mathrm{G}}}\end{aligned} \quad (4\text{-}90\mathrm{b})$$

and

$$\begin{aligned}0 = {} & \overline{\overline{Z}}^{\vec{M}_{\mathrm{A}\rightleftharpoons\mathrm{G}}\vec{J}_{\mathrm{M}\rightleftharpoons\mathrm{A}}}\cdot\overline{a}^{\vec{J}_{\mathrm{M}\rightleftharpoons\mathrm{A}}}+\overline{\overline{Z}}^{\vec{M}_{\mathrm{A}\rightleftharpoons\mathrm{G}}\vec{J}_{\mathrm{A}\rightleftharpoons\mathrm{G}}}\cdot\overline{a}^{\vec{J}_{\mathrm{A}\rightleftharpoons\mathrm{G}}}+\overline{\overline{Z}}^{\vec{M}_{\mathrm{A}\rightleftharpoons\mathrm{G}}\vec{J}_{\mathrm{A}}}\cdot\overline{a}^{\vec{J}_{\mathrm{A}}}+\overline{\overline{Z}}^{\vec{M}_{\mathrm{A}\rightleftharpoons\mathrm{G}}\vec{M}_{\mathrm{M}\rightleftharpoons\mathrm{A}}}\cdot\overline{a}^{\vec{M}_{\mathrm{M}\rightleftharpoons\mathrm{A}}}\\&+\overline{\overline{Z}}^{\vec{M}_{\mathrm{A}\rightleftharpoons\mathrm{G}}\vec{M}_{\mathrm{A}\rightleftharpoons\mathrm{G}}}\cdot\overline{a}^{\vec{M}_{\mathrm{A}\rightleftharpoons\mathrm{G}}}\end{aligned} \quad (4\text{-}91\mathrm{a})$$

$$0 \;=\; \bar{\bar{Z}}^{\vec{J}_{\mathrm{A}\rightleftharpoons\mathrm{G}}\vec{J}_{\mathrm{M}\rightleftharpoons\mathrm{A}}} \cdot \bar{a}^{\vec{J}_{\mathrm{M}\rightleftharpoons\mathrm{A}}} + \bar{\bar{Z}}^{\vec{J}_{\mathrm{A}\rightleftharpoons\mathrm{G}}\vec{J}_{\mathrm{A}\rightleftharpoons\mathrm{G}}} \cdot \bar{a}^{\vec{J}_{\mathrm{A}\rightleftharpoons\mathrm{G}}} + \bar{\bar{Z}}^{\vec{J}_{\mathrm{A}\rightleftharpoons\mathrm{G}}\vec{J}_{\mathrm{A}}} \cdot \bar{a}^{\vec{J}_{\mathrm{A}}} + \bar{\bar{Z}}^{\vec{J}_{\mathrm{A}\rightleftharpoons\mathrm{G}}\vec{M}_{\mathrm{M}\rightleftharpoons\mathrm{A}}} \cdot \bar{a}^{\vec{M}_{\mathrm{M}\rightleftharpoons\mathrm{A}}} + \bar{\bar{Z}}^{\vec{J}_{\mathrm{A}\rightleftharpoons\mathrm{G}}\vec{M}_{\mathrm{A}\rightleftharpoons\mathrm{G}}} \cdot \bar{a}^{\vec{M}_{\mathrm{A}\rightleftharpoons\mathrm{G}}} \tag{4-91b}$$

and

$$0 \;=\; \bar{\bar{Z}}^{\vec{J}_{\mathrm{A}}\vec{J}_{\mathrm{M}\rightleftharpoons\mathrm{A}}} \cdot \bar{a}^{\vec{J}_{\mathrm{M}\rightleftharpoons\mathrm{A}}} + \bar{\bar{Z}}^{\vec{J}_{\mathrm{A}}\vec{J}_{\mathrm{A}\rightleftharpoons\mathrm{G}}} \cdot \bar{a}^{\vec{J}_{\mathrm{A}\rightleftharpoons\mathrm{G}}} + \bar{\bar{Z}}^{\vec{J}_{\mathrm{A}}\vec{J}_{\mathrm{A}}} \cdot \bar{a}^{\vec{J}_{\mathrm{A}}} + \bar{\bar{Z}}^{\vec{J}_{\mathrm{A}}\vec{M}_{\mathrm{M}\rightleftharpoons\mathrm{A}}} \cdot \bar{a}^{\vec{M}_{\mathrm{M}\rightleftharpoons\mathrm{A}}} + \bar{\bar{Z}}^{\vec{J}_{\mathrm{A}}\vec{M}_{\mathrm{A}\rightleftharpoons\mathrm{G}}} \cdot \bar{a}^{\vec{M}_{\mathrm{A}\rightleftharpoons\mathrm{G}}} \tag{4-92}$$

The formulations used to calculate the elements of the matrices in the above matrix equations are similar to the formulations given in Eqs. (4-11)~(4-15), and they are explicitly given in the App. D5 of this report.

By employing the above matrix equations (4-90a)~(4-92), we can obtain the following transformation

$$\begin{bmatrix} \bar{a}^{\vec{J}_{\mathrm{M}\rightleftharpoons\mathrm{A}}} \\ \bar{a}^{\vec{J}_{\mathrm{A}\rightleftharpoons\mathrm{G}}} \\ \bar{a}^{\vec{J}_{\mathrm{A}}} \\ \bar{a}^{\vec{M}_{\mathrm{M}\rightleftharpoons\mathrm{A}}} \\ \bar{a}^{\vec{M}_{\mathrm{A}\rightleftharpoons\mathrm{G}}} \end{bmatrix} = \bar{a}^{\mathrm{AV}} \;=\; \bar{\bar{T}} \cdot \bar{a} \tag{4-93}$$

and the calculation formulation for transformation matrix $\bar{\bar{T}}$ is given in the App. D5 of this report.

### 4.4.3 Power Transport Theorem and Input Power Operator

Applying the results obtained in Chap. 2 to the rec-antenna shown in Figs. 4-8 and 4-9, we immediately have the following PTT for the rec-antenna

$$P_{\mathrm{M}\rightleftharpoons\mathrm{A}} = P_{\mathrm{A}}^{\mathrm{dis}} + P_{\mathrm{A}\rightleftharpoons\mathrm{G}} + j\, P_{\mathrm{A}}^{\mathrm{sto}} \tag{4-94}$$

where $P_{\mathrm{M}\rightleftharpoons\mathrm{A}}$ is the input power inputted into the rec-antenna, and $P_{\mathrm{A}}^{\mathrm{dis}}$ is the power dissipated in the rec-antenna, and $P_{\mathrm{A}\rightleftharpoons\mathrm{G}}$ is the power outputted from the rec-antenna, and $P_{\mathrm{A}}^{\mathrm{sto}}$ is the power corresponding to the stored energy in the rec-antenna.

The above-mentioned various powers are expressed as the following integral expressions:

$$P_{\mathrm{M}\rightleftharpoons\mathrm{A}} \;=\; (1/2)\iint_{\mathbb{S}_{\mathrm{M}\rightleftharpoons\mathrm{A}}} \left(\vec{E}\times\vec{H}^{\dagger}\right)\cdot\hat{n}_{\rightarrow\mathrm{A}}\,dS \tag{4-95a}$$

$$P_{\mathrm{A}}^{\mathrm{dis}} \;=\; (1/2)\left\langle \ddot{\sigma}\cdot\vec{E},\vec{E}\right\rangle_{\mathbb{V}_{\mathrm{A}}} \tag{4-95b}$$

$$P_{\mathrm{A}\rightleftharpoons\mathrm{G}} \;=\; (1/2)\iint_{\mathbb{S}_{\mathrm{A}\rightleftharpoons\mathrm{G}}} \left(\vec{E}\times\vec{H}^{\dagger}\right)\cdot\hat{n}_{\rightarrow\mathrm{G}}\,dS \tag{4-95c}$$

$$P_{\rm A}^{\rm sto} \quad = \quad 2\omega\Big[\big(1/4\big)\big\langle \vec{H},\vec{\vec{\mu}}\cdot\vec{H}\big\rangle_{\mathbb{V}_{\rm A}} - \big(1/4\big)\big\langle \vec{\vec{\varepsilon}}\cdot\vec{E},\vec{E}\big\rangle_{\mathbb{V}_{\rm A}}\Big] \tag{4-95d}$$

where $\hat{n}_{\to{\rm A}}$ is the normal direction of $\mathbb{S}_{\rm M\rightleftharpoons A}$ and points to $\mathbb{V}_{\rm A}$, and $\hat{n}_{\to{\rm G}}$ is the normal direction of $\mathbb{S}_{\rm A\rightleftharpoons G}$ and points to the loading waveguide, as shown in Fig. 4-9.

Based on Eqs. (4-85a)&(4-85b) and the tangential continuity of the $\{\vec{E},\vec{H}\}$ on $\mathbb{S}_{\rm M\rightleftharpoons A}$, the IPO $P_{\rm M\rightleftharpoons A}$ given in Eq. (4-95a) can be alternatively written as follows:

$$\begin{aligned} P_{\rm M\rightleftharpoons A} &= \big(1/2\big)\big\langle \hat{n}_{\to{\rm A}}\times\vec{J}_{\rm M\rightleftharpoons A},\vec{M}_{\rm M\rightleftharpoons A}\big\rangle_{\mathbb{S}_{\rm M\rightleftharpoons A}} \\ &= -\big(1/2\big)\Big\langle \vec{J}_{\rm M\rightleftharpoons A},\mathcal{E}\big(\vec{J}_{\rm M\rightleftharpoons A}+\vec{J}_{\rm A\rightleftharpoons G}+\vec{J}_{\rm A},\vec{M}_{\rm M\rightleftharpoons A}+\vec{M}_{\rm A\rightleftharpoons G}\big)\Big\rangle_{\underset{\sim}{\mathbb{S}}_{\rm M\rightleftharpoons A}} \\ &= -\big(1/2\big)\Big\langle \vec{M}_{\rm M\rightleftharpoons A},\mathcal{H}\big(\vec{J}_{\rm M\rightleftharpoons A}+\vec{J}_{\rm A\rightleftharpoons G}+\vec{J}_{\rm A},\vec{M}_{\rm M\rightleftharpoons A}+\vec{M}_{\rm A\rightleftharpoons G}\big)\Big\rangle^{\dagger}_{\underset{\sim}{\mathbb{S}}_{\rm M\rightleftharpoons A}} \end{aligned} \tag{4-96}$$

Here, the right-hand side of the first equality is the current form of IPO, and the right-hand sides of the second and third equalities are the interaction forms of IPO. The integral surface $\underset{\sim}{\mathbb{S}}_{\rm M\rightleftharpoons A}$ is an inner sub-boundary of rec-antenna.

By discretizing IPO (4-96) and utilizing transformation (4-93), we derive the following matrix form of the IPO

$$P_{\rm M\rightleftharpoons A} = \bar{a}^{\dagger}\cdot\bar{\bar{P}}_{\rm M\rightleftharpoons A}\cdot\bar{a} \tag{4-97}$$

and the detailed formulation for calculating the quadratic matrix $\bar{\bar{P}}_{\rm M\rightleftharpoons A}$ is given in the App. D5 of this report.

### 4.4.4 Input-Power-Decoupled Modes

The IP-DMs in modal space can be derived from solving the modal decoupling equation $\bar{\bar{P}}_{\rm M\rightleftharpoons A}^{-}\cdot\bar{\alpha}_{\xi} = \theta_{\xi}\bar{\bar{P}}_{\rm M\rightleftharpoons A}^{+}\cdot\bar{\alpha}_{\xi}$ defined on modal space, where $\bar{\bar{P}}_{\rm M\rightleftharpoons A}^{+}$ and $\bar{\bar{P}}_{\rm M\rightleftharpoons A}^{-}$ are the positive and negative Hermitian parts of matrix $\bar{\bar{P}}_{\rm M\rightleftharpoons A}$. If some derived modes $\{\bar{\alpha}_1,\bar{\alpha}_2,\cdots,\bar{\alpha}_d\}$ are $d$-order degenerate, then the Gram-Schmidt orthogonalization process given in the previous Sec. 4.2.4.1 is necessary, and it is not repeated here.

The modal fields constructed above satisfy the following decoupling relation

$$\big(1/2\big)\iint_{\mathbb{S}_{\rm M\rightleftharpoons A}}\big(\vec{E}_{\zeta}\times\vec{H}_{\xi}^{\dagger}\big)\cdot\hat{n}_{\to{\rm A}}dS \;=\; \big(1+j\,\theta_{\xi}\big)\delta_{\xi\zeta} \tag{4-98}$$

and the relation implies that **the IP-DMs don't have net energy coupling in integral period**. By employing the decoupling relation, we have the following Parseval's identity

$$\sum\nolimits_{\xi}\big|c_{\xi}\big|^2 \;=\; \big(1/T\big)\int_{t_0}^{t_0+T}\Big[\iint_{\mathbb{S}_{\rm M\rightleftharpoons A}}\big(\vec{\mathcal{E}}\times\vec{\mathcal{H}}\big)\cdot\hat{n}_{\to{\rm A}}dS\Big]dt \tag{4-99}$$

in which $\{\vec{\mathcal{E}},\vec{\mathcal{H}}\}$ are the time-domain fields, and the expansion coefficients $c_\xi$ have expression $c_\xi = -(1/2) < \vec{J}^{\xi}_{\mathrm{M}\rightleftharpoons\mathrm{A}}, \vec{E} >_{\mathbb{S}_{\mathrm{M}\rightleftharpoons\mathrm{A}}} \big/ (1 + j\,\theta_\xi) = -(1/2) < \vec{H}, \vec{M}^{\xi}_{\mathrm{M}\rightleftharpoons\mathrm{A}} >_{\mathbb{S}_{\mathrm{M}\rightleftharpoons\mathrm{A}}} \big/ (1 + j\,\theta_\xi)$, where $\{\vec{E},\vec{H}\}$ is a previously known field distributing on input port $\mathbb{S}_{\mathrm{M}\rightleftharpoons\mathrm{A}}$.

Just like the tra-antenna discussed in the previous Sec. 4.2.4.3, we can also define the modal significance $\mathrm{MS}_\xi = 1\big/|1 + j\,\theta_\xi|$, modal input impedance $Z^{\xi}_{\mathrm{M}\rightleftharpoons\mathrm{A}} = P^{\xi}_{\mathrm{M}\rightleftharpoons\mathrm{A}} \Big/ \Big[(1/2) < \vec{J}^{\xi}_{\mathrm{M}\rightleftharpoons\mathrm{A}}, \vec{J}^{\xi}_{\mathrm{M}\rightleftharpoons\mathrm{A}} >_{\mathbb{S}_{\mathrm{M}\rightleftharpoons\mathrm{A}}}\Big]$, and modal input admittance $Y^{\xi}_{\mathrm{M}\rightleftharpoons\mathrm{A}} = P^{\xi}_{\mathrm{M}\rightleftharpoons\mathrm{A}} \Big/ \Big[(1/2) < \vec{M}^{\xi}_{\mathrm{M}\rightleftharpoons\mathrm{A}}, \vec{M}^{\xi}_{\mathrm{M}\rightleftharpoons\mathrm{A}} >_{\mathbb{S}_{\mathrm{M}\rightleftharpoons\mathrm{A}}}\Big]$, to quantitatively describe the modal features of the rec-antenna shown in Figs. 4-8 and 4-9.

## 4.5 Chapter Summary

In this chapter, the *tra-antenna*, *propagation-medium*, and *rec-antenna* in transceiving system are separately treated as some *two-port regions*. All the two-port regions have a penetrable *input port* and a penetrable *output port* separated by an impenetrable *electric wall*.

The power inputted into the input port is called *input power*, and the corresponding *input power operator* (*IPO*) is derived under *power transport theorem* (*PTT*) framework. Employing the IPO as the *generating operator* of the fundamental modes of *modal space*, the *input-power-decoupled modes* (*IP-DMs*) of the tra-antenna, propagation medium, and rec-antenna are constructed by using *orthogonalizing IPO method*.

Because the tra-antenna, propagation medium, and rec-antenna are separately treated in this chapter, thus the interactions among them and the interactions among their IP-DMs are not considered in this chapter. In the following Chap. 5, the interactions are quantitatively discussed by employing *modal matching technique*. As the transitional chapters, this chapter and the following Chap. 5 will be utilized as the introduction for the future Chap. 6 ~ Chap. 8.

# *Chapter 5 Input-Power-Decoupled Mode Based Modal Matching

**CHAPTER MOTIVATION:** This Chap. 5 and the previous Chap. 4 collectively constitute the explanatory chapters used to explain why we introduce the concepts of *augmented tra-antenna* and *augmented rec-antenna* in the previous Chap. 2, and also constitute the transitional chapters used to introduce the subsequent Chaps. 6~8.

## 5.1 Chapter Introduction

In the previous Chaps. 3 and 4, the *input-power-decoupled modes* (*IP-DMs*) of *guiding structure* (simply called *guide*), *transmitting antenna* (simply called *tra-antenna*), *propagation medium* (simply called *medium*), and *receiving antenna* (simply called *rec-antenna*) had been constructed separately. But the interactions among the IP-DMs of guide, the IP-DMs of tra-antenna, the IP-DMs of propagation medium, and the the IP-DMs of rec-antenna have not been considered. This chapter focuses on studying the interactions among the IP-DMs by employing *modal matching* technique.

A systematical and deep discussion for modal matching technique can be found in Refs. [47~70]. The general steps to do modal matching are as follows: separately construct the fundamental modes of the different regions → expand the fields in the regions in terms of the corresponding fundamental modes → derive the equations satisfied by the modal expansion coefficients by matching the field continuation conditions on the interfaces among the regions → determine the modal expansion coefficients by solving the equations.

The first step had been done in the previous Chaps. 3 and 4. This Chp. 5 focuses on discussing the other steps.

## 5.2 A Single Two-port Region — Working State Decomposition

Now, we consider the two-port region $\mathbb{V}$ shown in the following Fig. 5-1

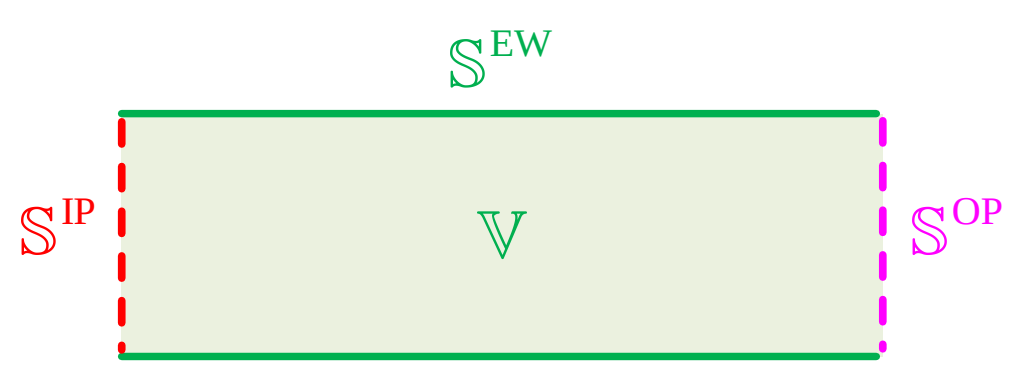

Figure 5-1 Geometry of a two-port region $\mathbb{V}$ with input port $\mathbb{S}^{IP}$ and output port $\mathbb{S}^{OP}$

and the region is with *permeability* $\vec{\vec{\mu}}$, *permittivity* $\vec{\vec{\varepsilon}}$, and *conductivity* $\vec{\vec{\sigma}}$. In the figure, $\mathbb{S}^{\text{IP}}$ is the *input port* of the region, and $\mathbb{S}^{\text{OP}}$ is the *output port* of the region, and $\mathbb{S}^{\text{EW}}$ is the *electric wall* used to separate the region from *surrounding environment*.

Under the excitations of the *primary source* located at the left-hand side of $\mathbb{S}^{\text{IP}}$ and the *primary source* located at the right-hand side of $\mathbb{S}^{\text{OP}}$, some induced surface currents $\vec{J}_{\text{IS}}$ and some induced volume currents $\{\vec{J}_{\text{IV}},\vec{M}_{\text{IV}}\}$ will be induced on $\mathbb{S}^{\text{EW}}$ and $\mathbb{V}$ respectively. For the convenience of the following discussions, the $\vec{J}_{\text{IS}}$ and $\vec{J}_{\text{IV}}$ are collectively denoted as $\vec{J}_{\text{ind}}$ (i.e., $\vec{J}_{\text{ind}}=\vec{J}_{\text{IS}}+\vec{J}_{\text{IV}}$), and the $\vec{M}_{\text{IV}}$ is similarly denoted as $\vec{M}_{\text{ind}}$ (i.e., $\vec{M}_{\text{ind}}=\vec{M}_{\text{IV}}$). In addition, if the *equivalent surface currents* on $\mathbb{S}^{\text{IP}}$ and $\mathbb{S}^{\text{OP}}$ are denoted as $\{\vec{J}^{\text{IP}},\vec{M}^{\text{IP}}\}$ and $\{\vec{J}^{\text{OP}},\vec{M}^{\text{OP}}\}$ respectively, then the field in $\mathbb{V}$ can be expressed as follows:

$$\vec{F}(\vec{r}) = \mathcal{F}_0\left(\vec{J}^{\text{IP}}+\vec{J}_{\text{ind}}+\vec{J}^{\text{OP}},\vec{M}^{\text{IP}}+\vec{M}_{\text{ind}}+\vec{M}^{\text{OP}}\right) \quad , \quad \vec{r}\in\mathbb{V} \tag{5-1}$$

where $\vec{F}=\vec{E}/\vec{H}$, and correspondingly $\mathcal{F}_0=\mathcal{E}_0/\mathcal{H}_0$, and the operators are the same as the ones used in the previous chapters. In addition, the currents $\{\vec{J}^{\text{IP}},\vec{M}^{\text{IP}}\}$ and $\{\vec{J}^{\text{OP}},\vec{M}^{\text{OP}}\}$ are defined as that $\vec{J}^{\text{IP/OP}}=\hat{n}^{\rightarrow\text{V}}\times\vec{H}$ on $\mathbb{S}^{\text{IP/OP}}$, and $\vec{M}^{\text{IP/OP}}=\vec{E}\times\hat{n}^{\rightarrow\text{V}}$ on $\mathbb{S}^{\text{IP/OP}}$, where $\hat{n}^{\rightarrow\text{V}}$ is the normal direction of $\mathbb{S}^{\text{IP/OP}}$ and points to the interior of $\mathbb{V}$.

If the currents $\{\vec{J}^{\text{IP}},\vec{M}^{\text{IP}}\}$, $\{\vec{J}_{\text{ind}},\vec{M}_{\text{ind}}\}$, and $\{\vec{J}^{\text{OP}},\vec{M}^{\text{OP}}\}$ are decomposed as follows:

$$\vec{C}^{\text{IP}} = \vec{C}^{\text{IP}}_{\text{pri}} + \vec{C}^{\text{IP}}_{\text{sec}} \tag{5-2a}$$
$$\vec{C}_{\text{ind}} = \vec{C}^{\text{IP}}_{\text{ind}} + \vec{C}^{\text{OP}}_{\text{ind}} \tag{5-2b}$$
$$\vec{C}^{\text{OP}} = \vec{C}^{\text{OP}}_{\text{sec}} + \vec{C}^{\text{OP}}_{\text{pri}} \tag{5-2c}$$

where $\vec{C}=\vec{J}/\vec{M}$, such that

$$\vec{J}^{\text{IP}}_{\text{ind}} = j\omega\Delta\vec{\vec{\varepsilon}}_{\text{c}}\cdot\mathcal{E}_0\left(\vec{J}^{\text{IP}}_{\text{pri}}+\vec{J}^{\text{IP}}_{\text{ind}}+\vec{J}^{\text{OP}}_{\text{sec}},\vec{M}^{\text{IP}}_{\text{pri}}+\vec{M}^{\text{IP}}_{\text{ind}}+\vec{M}^{\text{OP}}_{\text{sec}}\right) \quad , \quad \vec{r}\in\mathbb{V} \tag{5-3a}$$
$$\vec{M}^{\text{IP}}_{\text{ind}} = j\omega\Delta\vec{\vec{\mu}}\cdot\mathcal{H}_0\left(\vec{J}^{\text{IP}}_{\text{pri}}+\vec{J}^{\text{IP}}_{\text{ind}}+\vec{J}^{\text{OP}}_{\text{sec}},\vec{M}^{\text{IP}}_{\text{pri}}+\vec{M}^{\text{IP}}_{\text{ind}}+\vec{M}^{\text{OP}}_{\text{sec}}\right) \quad , \quad \vec{r}\in\mathbb{V} \tag{5-3b}$$
$$0 = \left[\mathcal{E}_0\left(\vec{J}^{\text{IP}}_{\text{pri}}+\vec{J}^{\text{IP}}_{\text{ind}}+\vec{J}^{\text{OP}}_{\text{sec}},\vec{M}^{\text{IP}}_{\text{pri}}+\vec{M}^{\text{IP}}_{\text{ind}}+\vec{M}^{\text{OP}}_{\text{sec}}\right)\right]^{\tan} \quad , \quad \vec{r}\in\mathbb{S}^{\text{EW}} \tag{5-3c}$$

and then it automatically has that

$$\vec{J}^{\text{OP}}_{\text{ind}} = j\omega\Delta\vec{\vec{\varepsilon}}_{\text{c}}\cdot\mathcal{E}_0\left(\vec{J}^{\text{IP}}_{\text{sec}}+\vec{J}^{\text{OP}}_{\text{ind}}+\vec{J}^{\text{OP}}_{\text{pri}},\vec{M}^{\text{IP}}_{\text{sec}}+\vec{M}^{\text{OP}}_{\text{ind}}+\vec{M}^{\text{OP}}_{\text{pri}}\right) \quad , \quad \vec{r}\in\mathbb{V} \tag{5-4a}$$
$$\vec{M}^{\text{OP}}_{\text{ind}} = j\omega\Delta\vec{\vec{\mu}}\cdot\mathcal{H}_0\left(\vec{J}^{\text{IP}}_{\text{sec}}+\vec{J}^{\text{OP}}_{\text{ind}}+\vec{J}^{\text{OP}}_{\text{pri}},\vec{M}^{\text{IP}}_{\text{sec}}+\vec{M}^{\text{OP}}_{\text{ind}}+\vec{M}^{\text{OP}}_{\text{pri}}\right) \quad , \quad \vec{r}\in\mathbb{V} \tag{5-4b}$$
$$0 = \left[\mathcal{E}_0\left(\vec{J}^{\text{IP}}_{\text{sec}}+\vec{J}^{\text{OP}}_{\text{ind}}+\vec{J}^{\text{OP}}_{\text{pri}},\vec{M}^{\text{IP}}_{\text{sec}}+\vec{M}^{\text{OP}}_{\text{ind}}+\vec{M}^{\text{OP}}_{\text{pri}}\right)\right]^{\tan} \quad , \quad \vec{r}\in\mathbb{S}^{\text{EW}} \tag{5-4c}$$

Thus, the sub-currents $\{\vec{J}_{\text{ind}}^{\text{IP}},\vec{M}_{\text{ind}}^{\text{IP}};\vec{J}_{\text{sec}}^{\text{OP}},\vec{M}_{\text{sec}}^{\text{OP}}\}/\{\vec{J}_{\text{sec}}^{\text{IP}},\vec{M}_{\text{sec}}^{\text{IP}};\vec{J}_{\text{ind}}^{\text{OP}},\vec{M}_{\text{ind}}^{\text{OP}}\}$ can be viewed as the currents resulted from the action of the field generated by $\{\vec{J}_{\text{pri}}^{\text{IP}},\vec{M}_{\text{pri}}^{\text{IP}}\}/\{\vec{J}_{\text{pri}}^{\text{OP}},\vec{M}_{\text{pri}}^{\text{OP}}\}$.

Thus, the field $\vec{F}$ in $\mathbb{V}$ can be decomposed as follows:

$$\vec{F}(\vec{r}) = \vec{F}^{\text{IP}}(\vec{r})+\vec{F}^{\text{OP}}(\vec{r}) \quad , \quad \vec{r}\in\mathbb{V} \tag{5-5}$$

where

$$\vec{F}^{\text{IP}}(\vec{r}) = \mathcal{F}_0\left(\vec{J}_{\text{pri}}^{\text{IP}}+\vec{J}_{\text{ind}}^{\text{IP}}+\vec{J}_{\text{sec}}^{\text{OP}},\vec{M}_{\text{pri}}^{\text{IP}}+\vec{M}_{\text{ind}}^{\text{IP}}+\vec{M}_{\text{sec}}^{\text{OP}}\right) \quad , \quad \vec{r}\in\mathbb{V} \tag{5-6a}$$

$$\vec{F}^{\text{OP}}(\vec{r}) = \mathcal{F}_0\left(\vec{J}_{\text{sec}}^{\text{IP}}+\vec{J}_{\text{ind}}^{\text{OP}}+\vec{J}_{\text{pri}}^{\text{OP}},\vec{M}_{\text{sec}}^{\text{IP}}+\vec{M}_{\text{ind}}^{\text{OP}}+\vec{M}_{\text{pri}}^{\text{OP}}\right) \quad , \quad \vec{r}\in\mathbb{V} \tag{5-6b}$$

Here, the $\vec{F}^{\text{IP}}$ is originated from the primary source located at the left-hand side of port $\mathbb{S}^{\text{IP}}$, and the $\vec{F}^{\text{OP}}$ is originated from the primary source located at the right-hand side of port $\mathbb{S}^{\text{OP}}$.

## 5.3 Two Cascaded Two-port Regions — Local Reflection and Transmission Operators

Now, we consider two cascaded two-port regions $\mathbb{V}_n$ and $\mathbb{V}_{n+1}$ as shown in the following Fig. 5-2.

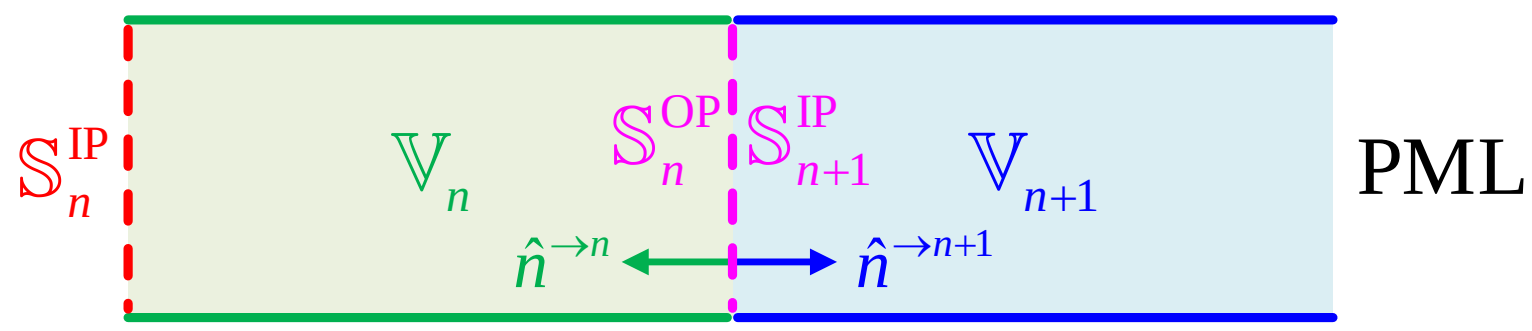

Figure 5-2 Geometry of two cascaded two-port regions $\mathbb{V}_n$ and $\mathbb{V}_{n+1}$

In the figure, the region $\mathbb{V}_n$ is loaded by the region $\mathbb{V}_{n+1}$, i.e., the region $\mathbb{V}_{n+1}$ is driven by the region $\mathbb{V}_n$; the region $\mathbb{V}_{n+1}$ is loaded by a *perfectly matched load* (*PML*). In addition, $\mathbb{S}_n^{\text{IP}}$ and $\mathbb{S}_n^{\text{OP}}$ are respectively the input and output ports of $\mathbb{V}_n$, and $\mathbb{S}_{n+1}^{\text{IP}}$ is the input port of $\mathbb{V}_{n+1}$. It is evident that $\mathbb{S}_n^{\text{OP}}=\mathbb{S}_{n+1}^{\text{IP}}$.

By a similar analysis given in Refs. [47~70], it can be concluded that: the field in $\mathbb{V}_n$ is constituted by two parts — the field $\vec{F}_n^{\text{IP}}$ "propagating from $\mathbb{S}_n^{\text{IP}}$ to $\mathbb{S}_n^{\text{OP}}$" and the field $\vec{F}_n^{\text{OP}}$ "propagating from $\mathbb{S}_n^{\text{OP}}$ to $\mathbb{S}_n^{\text{IP}}$"; the field in $\mathbb{V}_{n+1}$ is constituted by only one part, i.e. the field $\vec{F}_{n+1}^{\text{IP}}$ "propagating from $\mathbb{S}_{n+1}^{\text{IP}}$ to the PML". Thus, we have the following relations

$$\vec{F}(\vec{r}) = \vec{F}_n^{\text{IP}}(\vec{r})+\vec{F}_n^{\text{OP}}(\vec{r}) \quad , \quad \vec{r}\in\mathbb{V}_n \tag{5-7}$$

$$\vec{F}(\vec{r}) = \vec{F}_{n+1}^{\text{IP}}(\vec{r}) \quad , \quad \vec{r}\in\mathbb{V}_{n+1} \tag{5-8}$$

On the interface $\mathbb{S}_n^{\rm OP}$ $(=\mathbb{S}_{n+1}^{\rm IP})$ between $\mathbb{V}_n$ and $\mathbb{V}_{n+1}$, the field satisfies the following tangential continuation condition

$$\left[\vec{F}_n^{\rm IP}(\vec{r}_n)+\vec{F}_n^{\rm OP}(\vec{r}_n)\right]_{\vec{r}_n\to\vec{r}}^{\tan} = \left[\vec{F}_{n+1}^{\rm IP}(\vec{r}_{n+1})\right]_{\vec{r}_{n+1}\to\vec{r}}^{\tan} \quad , \quad \vec{r}\in\mathbb{S}_n^{\rm OP}=\mathbb{S}_{n+1}^{\rm IP} \tag{5-9}$$

where $\vec{r}_n\in\mathbb{V}_n$ and $\vec{r}_{n+1}\in\mathbb{V}_{n+1}$, and both $\vec{r}_n$ and $\vec{r}_{n+1}$ approach the point $\vec{r}$ on $\mathbb{S}_n^{\rm OP}$ $(=\mathbb{S}_{n+1}^{\rm IP})$. For the convenience of the following discussions, we rewrite Eq. (5-9) as follows:

$$\begin{aligned}\hat{n}^{\to n}\times\left[\vec{E}_n^{\rm IP}(\vec{r}_n)+\vec{E}_n^{\rm OP}(\vec{r}_n)\right]_{\vec{r}_n\to\vec{r}} &= \hat{n}^{\to n}\times\left[\vec{E}_{n+1}^{\rm IP}(\vec{r}_{n+1})\right]_{\vec{r}_{n+1}\to\vec{r}}\\ &= -\,\hat{n}^{\to n+1}\times\left[\vec{E}_{n+1}^{\rm IP}(\vec{r}_{n+1})\right]_{\vec{r}_{n+1}\to\vec{r}} \quad , \quad \vec{r}\in\mathbb{S}_n^{\rm OP}=\mathbb{S}_{n+1}^{\rm IP}\end{aligned} \tag{5-10a}$$

$$\begin{aligned}\left[\vec{H}_n^{\rm IP}(\vec{r}_n)+\vec{H}_n^{\rm OP}(\vec{r}_n)\right]_{\vec{r}_n\to\vec{r}}\times\hat{n}^{\to n} &= \left[\vec{H}_{n+1}^{\rm IP}(\vec{r}_{n+1})\right]_{\vec{r}_{n+1}\to\vec{r}}\times\hat{n}^{\to n}\\ &= -\left[\vec{H}_{n+1}^{\rm IP}(\vec{r}_{n+1})\right]_{\vec{r}_{n+1}\to\vec{r}}\times\hat{n}^{\to n+1} \quad , \quad \vec{r}\in\mathbb{S}_n^{\rm OP}=\mathbb{S}_{n+1}^{\rm IP}\end{aligned} \tag{5-10b}$$

in which $\hat{n}^{\to n}$ and $\hat{n}^{\to n+1}$ are the normal directions of $\mathbb{S}_n^{\rm OP}$ $(=\mathbb{S}_{n+1}^{\rm IP})$ and points to the interiors of $\mathbb{V}_n$ and $\mathbb{V}_{n+1}$ respectively, and it is obvious that $\hat{n}^{\to n}=-\hat{n}^{\to n+1}$ as shown in Fig. 5-2.

The fields in Eqs. (5-10a) and (5-10b) can be expanded in terms of the IP-DMs in the corresponding regions as follows:

$$\vec{F}_n^{\rm IP}(\vec{r}) = \sum_{\xi=1}^{\Xi_n^{\rm IP}} c_{n;\xi}^{\rm IP}\vec{E}_{n;\xi}^{\rm IP}(\vec{r}) = \bar{\boldsymbol{F}}_n^{\rm IP}\cdot\bar{c}_n^{\rm IP} \quad , \quad \vec{r}\in\mathbb{V}_n \tag{5-11a}$$

$$\vec{F}_n^{\rm OP}(\vec{r}) = \sum_{\xi=1}^{\Xi_n^{\rm OP}} c_{n;\xi}^{\rm OP}\vec{E}_{n;\xi}^{\rm OP}(\vec{r}) = \bar{\boldsymbol{F}}_n^{\rm OP}\cdot\bar{c}_n^{\rm OP} \quad , \quad \vec{r}\in\mathbb{V}_n \tag{5-11b}$$

$$\vec{F}_{n+1}^{\rm IP}(\vec{r}) = \sum_{\xi=1}^{\Xi_{n+1}^{\rm IP}} c_{n+1;\xi}^{\rm IP}\vec{E}_{n+1;\xi}^{\rm IP}(\vec{r}) = \bar{\boldsymbol{F}}_{n+1}^{\rm IP}\cdot\bar{c}_{n+1}^{\rm IP} \quad , \quad \vec{r}\in\mathbb{V}_{n+1} \tag{5-11c}$$

where $\vec{F}=\vec{E}/\vec{H}$ and correspondingly $\bar{\boldsymbol{F}}=\bar{\boldsymbol{E}}/\bar{\boldsymbol{H}}$, and

$$\bar{\boldsymbol{F}}_n^{\rm IP} = \begin{bmatrix}\vec{F}_{n;1}^{\rm IP} & \cdots & \vec{F}_{n;\Xi_n^{\rm IP}}^{\rm IP}\end{bmatrix} \quad , \quad \bar{c}_n^{\rm IP} = \begin{bmatrix}c_{n;1}^{\rm IP} & \cdots & c_{n;\Xi_n^{\rm IP}}^{\rm IP}\end{bmatrix}^T \tag{5-12a}$$

$$\bar{\boldsymbol{F}}_n^{\rm OP} = \begin{bmatrix}\vec{F}_{n;1}^{\rm OP} & \cdots & \vec{F}_{n;\Xi_n^{\rm OP}}^{\rm OP}\end{bmatrix} \quad , \quad \bar{c}_n^{\rm OP} = \begin{bmatrix}c_{n;1}^{\rm OP} & \cdots & c_{n;\Xi_n^{\rm OP}}^{\rm OP}\end{bmatrix}^T \tag{5-12b}$$

$$\bar{\boldsymbol{F}}_{n+1}^{\rm IP} = \begin{bmatrix}\vec{F}_{n+1;1}^{\rm IP} & \cdots & \vec{F}_{n+1;\Xi_{n+1}^{\rm IP}}^{\rm IP}\end{bmatrix} \quad , \quad \bar{c}_{n+1}^{\rm IP} = \begin{bmatrix}c_{n+1;1}^{\rm IP} & \cdots & c_{n+1;\Xi_{n+1}^{\rm IP}}^{\rm IP}\end{bmatrix}^T \tag{5-12c}$$

where the superscript "$T$" represents the *transpose operation* for a matrix or vector.

Inserting the above Eqs. (5-11a)~(5-11c) into the previous Eqs. (5-10a) and (5-10b), we immediately have that

$$0 = \left(\hat{n}^{\rightarrow n} \times \bar{\boldsymbol{E}}_n^{\text{IP}}\right) \cdot \bar{c}_n^{\text{IP}} + \left(\hat{n}^{\rightarrow n} \times \bar{\boldsymbol{E}}_n^{\text{OP}}\right) \cdot \bar{c}_n^{\text{OP}} + \left(\hat{n}^{\rightarrow n+1} \times \bar{\boldsymbol{E}}_{n+1}^{\text{IP}}\right) \cdot \bar{c}_{n+1}^{\text{IP}} \tag{5-13}$$

$$0 = \left(\bar{\boldsymbol{H}}_n^{\text{IP}} \times \hat{n}^{\rightarrow n}\right) \cdot \bar{c}_n^{\text{IP}} + \left(\bar{\boldsymbol{H}}_n^{\text{OP}} \times \hat{n}^{\rightarrow n}\right) \cdot \bar{c}_n^{\text{OP}} + \left(\bar{\boldsymbol{H}}_{n+1}^{\text{IP}} \times \hat{n}^{\rightarrow n+1}\right) \cdot \bar{c}_{n+1}^{\text{IP}} \tag{5-14}$$

on $\mathbb{S}_n^{\text{OP}}$ $(=\mathbb{S}_{n+1}^{\text{IP}})$, were the symbol "$\hat{n} \times \bar{\boldsymbol{F}}$" is defined as follows:

$$\hat{n} \times \bar{\boldsymbol{F}} = \left[\hat{n} \times \vec{F}_1 \quad \hat{n} \times \vec{F}_2 \quad \cdots \quad \hat{n} \times \vec{F}_{\Xi}\right] \tag{5-15}$$

Left multiplying Eq. (5-13) with $(\bar{\boldsymbol{H}}_n^{\text{OP}})^\dagger$ and left multiplying Eq. (5-14) with $(\bar{\boldsymbol{E}}_{n+1}^{\text{IP}})^\dagger$ and integrating on whole $\mathbb{S}_n^{\text{OP}}$ $(=\mathbb{S}_{n+1}^{\text{IP}})$, we have that

$$0 = \overbrace{\left\langle \bar{\boldsymbol{H}}_n^{\text{OP}}, \hat{n}^{\rightarrow n} \times \bar{\boldsymbol{E}}_n^{\text{IP}} \right\rangle_{\mathbb{S}_n^{\text{OP}}}}^{\bar{\bar{Z}}_{n,n}^{\text{OP,IP}}} \cdot \bar{c}_n^{\text{IP}} + \overbrace{\left\langle \bar{\boldsymbol{H}}_n^{\text{OP}}, \hat{n}^{\rightarrow n} \times \bar{\boldsymbol{E}}_n^{\text{OP}} \right\rangle_{\mathbb{S}_n^{\text{OP}}}}^{\bar{\bar{\Lambda}}_{n,n}^{\text{OP,OP}}} \cdot \bar{c}_n^{\text{OP}} + \overbrace{\left\langle \bar{\boldsymbol{H}}_n^{\text{OP}}, \hat{n}^{\rightarrow n+1} \times \bar{\boldsymbol{E}}_{n+1}^{\text{IP}} \right\rangle_{\mathbb{S}_n^{\text{OP}}}}^{\bar{\bar{Z}}_{n,n+1}^{\text{OP,IP}}} \cdot \bar{c}_{n+1}^{\text{IP}} \tag{5-16}$$

$$0 = \underbrace{\left\langle \bar{\boldsymbol{E}}_{n+1}^{\text{IP}}, \bar{\boldsymbol{H}}_n^{\text{IP}} \times \hat{n}^{\rightarrow n} \right\rangle_{\mathbb{S}_n^{\text{OP}}}}_{\bar{\bar{Z}}_{n+1,n}^{\text{IP,IP}}} \cdot \bar{c}_n^{\text{IP}} + \underbrace{\left\langle \bar{\boldsymbol{E}}_{n+1}^{\text{IP}}, \bar{\boldsymbol{H}}_n^{\text{OP}} \times \hat{n}^{\rightarrow n} \right\rangle_{\mathbb{S}_n^{\text{OP}}}}_{\bar{\bar{Z}}_{n+1,n}^{\text{IP,OP}}} \cdot \bar{c}_n^{\text{OP}} + \underbrace{\left\langle \bar{\boldsymbol{E}}_{n+1}^{\text{IP}}, \bar{\boldsymbol{H}}_{n+1}^{\text{IP}} \times \hat{n}^{\rightarrow n+1} \right\rangle_{\mathbb{S}_n^{\text{OP}}}}_{\bar{\bar{\Lambda}}_{n+1,n+1}^{\text{IP,IP}}} \cdot \bar{c}_{n+1}^{\text{IP}} \tag{5-17}$$

where the elements of matrices $\bar{\bar{Z}}_{n,n}^{\text{OP,IP}}$, $\bar{\bar{\Lambda}}_{n,n}^{\text{OP,OP}}$, $\bar{\bar{Z}}_{n,n+1}^{\text{OP,IP}}$, $\bar{\bar{Z}}_{n+1,n}^{\text{IP,IP}}$, $\bar{\bar{Z}}_{n+1,n}^{\text{IP,OP}}$, and $\bar{\bar{\Lambda}}_{n+1,n+1}^{\text{IP,IP}}$ are as follows:

$$\left[\bar{\bar{Z}}_{n,n}^{\text{OP,IP}}\right]_{\xi\zeta} = \left\langle \vec{H}_{n;\xi}^{\text{OP}}, \hat{n}^{\rightarrow n} \times \vec{E}_{n;\zeta}^{\text{IP}} \right\rangle_{\mathbb{S}_n^{\text{OP}}=\mathbb{S}_{n+1}^{\text{IP}}} \tag{5-18a}$$

$$\left[\bar{\bar{\Lambda}}_{n,n}^{\text{OP,OP}}\right]_{\xi\zeta} = \left\langle \vec{H}_{n;\xi}^{\text{OP}}, \hat{n}^{\rightarrow n} \times \vec{E}_{n;\zeta}^{\text{OP}} \right\rangle_{\mathbb{S}_n^{\text{OP}}=\mathbb{S}_{n+1}^{\text{IP}}} \tag{5-18b}$$

$$\left[\bar{\bar{Z}}_{n,n+1}^{\text{OP,IP}}\right]_{\xi\zeta} = \left\langle \vec{H}_{n;\xi}^{\text{OP}}, \hat{n}^{\rightarrow n+1} \times \vec{E}_{n+1;\zeta}^{\text{IP}} \right\rangle_{\mathbb{S}_n^{\text{OP}}=\mathbb{S}_{n+1}^{\text{IP}}} \tag{5-18c}$$

and

$$\left[\bar{\bar{Z}}_{n+1,n}^{\text{IP,IP}}\right]_{\xi\zeta} = \left\langle \vec{E}_{n+1;\xi}^{\text{IP}}, \vec{H}_{n;\zeta}^{\text{IP}} \times \hat{n}^{\rightarrow n} \right\rangle_{\mathbb{S}_n^{\text{OP}}=\mathbb{S}_{n+1}^{\text{IP}}} = \left\langle \vec{H}_{n;\zeta}^{\text{IP}}, \hat{n}^{\rightarrow n} \times \vec{E}_{n+1;\xi}^{\text{IP}} \right\rangle_{\mathbb{S}_n^{\text{OP}}=\mathbb{S}_{n+1}^{\text{IP}}}^\dagger \tag{5-19a}$$

$$\left[\bar{\bar{Z}}_{n+1,n}^{\text{IP,OP}}\right]_{\xi\zeta} = \left\langle \vec{E}_{n+1;\xi}^{\text{IP}}, \vec{H}_{n;\zeta}^{\text{OP}} \times \hat{n}^{\rightarrow n} \right\rangle_{\mathbb{S}_n^{\text{OP}}=\mathbb{S}_{n+1}^{\text{IP}}} = \left\langle \vec{H}_{n;\zeta}^{\text{OP}}, \hat{n}^{\rightarrow n} \times \vec{E}_{n+1;\xi}^{\text{IP}} \right\rangle_{\mathbb{S}_n^{\text{OP}}=\mathbb{S}_{n+1}^{\text{IP}}}^\dagger \tag{5-19b}$$

$$\left[\bar{\bar{\Lambda}}_{n+1,n+1}^{\text{IP,IP}}\right]_{\xi\zeta} = \left\langle \vec{E}_{n+1;\xi}^{\text{IP}}, \vec{H}_{n+1;\zeta}^{\text{IP}} \times \hat{n}^{\rightarrow n+1} \right\rangle_{\mathbb{S}_n^{\text{OP}}=\mathbb{S}_{n+1}^{\text{IP}}} = \left\langle \vec{H}_{n+1;\zeta}^{\text{IP}}, \hat{n}^{\rightarrow n+1} \times \vec{E}_{n+1;\xi}^{\text{IP}} \right\rangle_{\mathbb{S}_n^{\text{OP}}=\mathbb{S}_{n+1}^{\text{IP}}}^\dagger \tag{5-19c}$$

Obviously, the matrices $\bar{\bar{\Lambda}}_{n,n}^{\text{OP,OP}}$ and $\bar{\bar{\Lambda}}_{n+1,n+1}^{\text{IP,IP}}$ are diagonal matrices because of the decoupling relations satisfied by the IP-DMs.

Assembling Eqs. (5-16) and (5-17), we immediately have the following augmented matrix equation

$$\begin{bmatrix} \bar{\bar{\Lambda}}_{n,n}^{\text{OP,OP}} & \bar{\bar{Z}}_{n,n+1}^{\text{OP,IP}} \\ \bar{\bar{Z}}_{n+1,n}^{\text{IP,OP}} & \bar{\bar{\Lambda}}_{n+1,n+1}^{\text{IP,IP}} \end{bmatrix} \cdot \begin{bmatrix} \bar{c}_n^{\text{OP}} \\ \bar{c}_{n+1}^{\text{IP}} \end{bmatrix} = -\begin{bmatrix} \bar{\bar{Z}}_{n,n}^{\text{OP,IP}} \\ \bar{\bar{Z}}_{n+1,n}^{\text{IP,IP}} \end{bmatrix} \cdot \bar{c}_n^{\text{IP}} \tag{5-20}$$

Solving the above matrix equation, we have the following transformation from $\overline{c}_n^{\text{IP}}$ to $\overline{c}_n^{\text{OP}}$ and $\overline{c}_{n+1}^{\text{IP}}$.

$$\begin{bmatrix} \overline{c}_n^{\text{OP}} \\ \overline{c}_{n+1}^{\text{IP}} \end{bmatrix} = \underbrace{-\begin{bmatrix} \overline{\overline{\Lambda}}_{n,n}^{\text{OP,OP}} & \overline{\overline{Z}}_{n,n+1}^{\text{OP,IP}} \\ \overline{\overline{Z}}_{n+1,n}^{\text{IP,OP}} & \overline{\overline{\Lambda}}_{n+1,n+1}^{\text{IP,IP}} \end{bmatrix}^{-1} \cdot \begin{bmatrix} \overline{\overline{Z}}_{n,n}^{\text{OP,IP}} \\ \overline{\overline{Z}}_{n+1,n}^{\text{IP,IP}} \end{bmatrix}}_{\begin{bmatrix} \overline{\overline{r}}_{n,n+1} \\ \overline{\overline{t}}_{n,n+1} \end{bmatrix}} \cdot \overline{c}_n^{\text{IP}} \tag{5-21}$$

where $\overline{\overline{r}}_{n,n+1}$ and $\overline{\overline{t}}_{n,n+1}$ have the same row numbers as $\overline{c}_n^{\text{OP}}$ and $\overline{c}_{n+1}^{\text{IP}}$ respectively, and then

$$\overline{c}_n^{\text{OP}} = \overline{\overline{r}}_{n,n+1} \cdot \overline{c}_n^{\text{IP}} \tag{5-22}$$

$$\overline{c}_{n+1}^{\text{IP}} = \overline{\overline{t}}_{n,n+1} \cdot \overline{c}_n^{\text{IP}} \tag{5-23}$$

Following the convention used in Refs. [47~70], the above transformation matrices $\overline{\overline{r}}_{n,n+1}$ and $\overline{\overline{t}}_{n,n+1}$ are called the *local reflection matrix from* $\mathbb{V}_{n+1}$ *to* $\mathbb{V}_n$ and the *local transmission matrix from* $\mathbb{V}_n$ *to* $\mathbb{V}_{n+1}$ respectively.

## 5.4 Multiple Cascaded Two-port Regions — Global Reflection and Transmission Operators

Now, we further consider three cascaded regions $\mathbb{V}_{n-1}$, $\mathbb{V}_n$, and $\mathbb{V}_{n+1}$ as shown in the following Fig. 5-3.

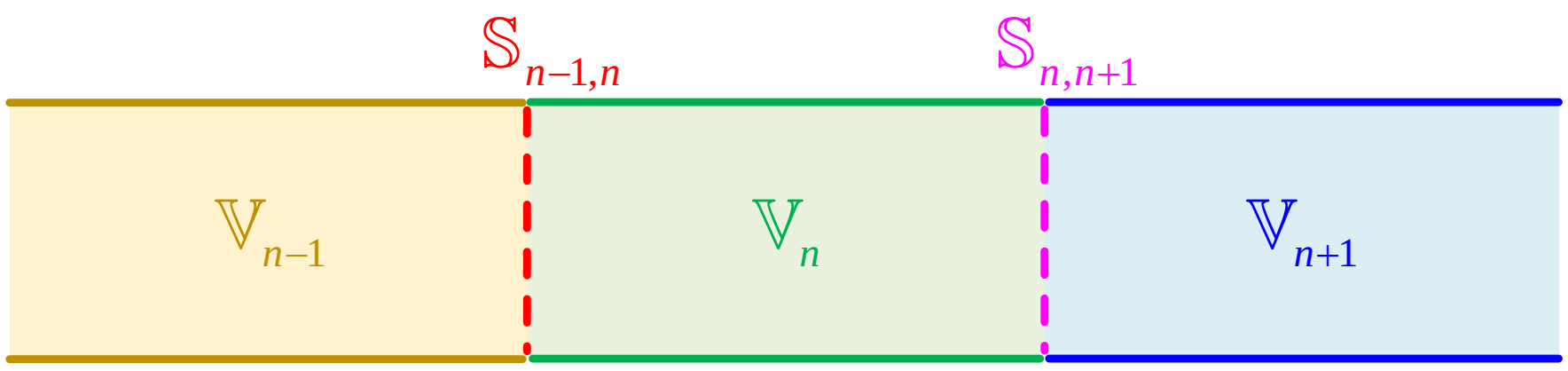

Figure 5-3 Geometry of multiplely cascaded two-port regions $\mathbb{V}_{n-1}$, $\mathbb{V}_n$, and $\mathbb{V}_{n+1}$

In the above figure, the region $\mathbb{V}_{n-1}$ is loaded by the region $\mathbb{V}_n$, i.e., the region $\mathbb{V}_n$ is driven by the region $\mathbb{V}_{n-1}$; the region $\mathbb{V}_n$ is loaded by the region $\mathbb{V}_{n+1}$, i.e., the region $\mathbb{V}_{n+1}$ is driven by the region $\mathbb{V}_n$. For the convenience of the following discussions, the interface between $\mathbb{V}_{n-1}$ and $\mathbb{V}_n$ is denoted as $\mathbb{S}_{n-1,n}$, and the interface between $\mathbb{V}_n$ and $\mathbb{V}_{n+1}$ is denoted as $\mathbb{S}_{n,n+1}$.

Similarly to the analysis given in Refs. [47~70], it can be concluded that: both the reflection of $\overline{c}_n^{\text{IP}}$ from $\mathbb{S}_{n,n+1}$ and the transmission of $\overline{c}_{n+1}^{\text{OP}}$ on $\mathbb{S}_{n,n+1}$ contribute to the $\overline{c}_n^{\text{OP}}$, and then

$$\overline{c}_n^{\rm OP} = \overline{\overline{r}}_{n,n+1} \cdot \overline{c}_n^{\rm IP} + \overline{\overline{t}}_{n+1,n} \cdot \overline{c}_{n+1}^{\rm OP} \tag{5-24}$$

Similarly, both the reflection of $\overline{c}_n^{\rm OP}$ from $\mathbb{S}_{n,n-1}$ and the transmission of $\overline{c}_{n-1}^{\rm IP}$ on $\mathbb{S}_{n-1,n}$ contribute to the $\overline{c}_n^{\rm IP}$, and then

$$\overline{c}_n^{\rm IP} = \overline{\overline{r}}_{n,n-1} \cdot \overline{c}_n^{\rm OP} + \overline{\overline{t}}_{n-1,n} \cdot \overline{c}_{n-1}^{\rm IP} \tag{5-25}$$

The *causality* and *linearity* imply the following linear transformations from $\overline{c}_n^{\rm IP}$ to $\overline{c}_n^{\rm OP}$ and from $\overline{c}_{n+1}^{\rm IP}$ to $\overline{c}_{n+1}^{\rm OP}$

$$\overline{c}_n^{\rm OP} = \overline{\overline{R}}_{n,n+1} \cdot \overline{c}_n^{\rm IP} \tag{5-26}$$

$$\overline{c}_{n+1}^{\rm OP} = \overline{\overline{R}}_{n+1,n+2} \cdot \overline{c}_{n+1}^{\rm IP} \tag{5-27}$$

Substituting the above Eqs. (5-26) and (5-27) into the previous Eqs. (5-24) and (5-25), we immediately have that

$$\overbrace{\overline{\overline{R}}_{n,n+1} \cdot \overline{c}_n^{\rm IP}}^{\overline{c}_n^{\rm OP}} = \overline{\overline{r}}_{n,n+1} \cdot \overline{c}_n^{\rm IP} + \overline{\overline{t}}_{n+1,n} \cdot \overbrace{\overline{\overline{R}}_{n+1,n+2} \cdot \overline{c}_{n+1}^{\rm IP}}^{\overline{c}_{n+1}^{\rm OP}} \tag{5-28}$$

$$\overline{c}_n^{\rm IP} = \overline{\overline{r}}_{n,n-1} \cdot \underbrace{\overline{\overline{R}}_{n,n+1} \cdot \overline{c}_n^{\rm IP}}_{\overline{c}_n^{\rm OP}} + \overline{\overline{t}}_{n-1,n} \cdot \overline{c}_{n-1}^{\rm IP} \tag{5-29}$$

Similar to the above Eq. (5-29), we also have that

$$\overline{c}_{n+1}^{\rm IP} = \overline{\overline{r}}_{n+1,n} \cdot \underbrace{\overline{\overline{R}}_{n+1,n+2} \cdot \overline{c}_{n+1}^{\rm IP}}_{\overline{c}_{n+1}^{\rm OP}} + \overline{\overline{t}}_{n,n+1} \cdot \overline{c}_n^{\rm IP} \tag{5-30}$$

Solving the above Eq. (5-30), it is easy to obtain that

$$\overline{c}_{n+1}^{\rm IP} = \underbrace{\left(\overline{\overline{I}} - \overline{\overline{r}}_{n+1,n} \cdot \overline{\overline{R}}_{n+1,n+2}\right)^{-1} \cdot \overline{\overline{t}}_{n,n+1}}_{\overline{\overline{T}}_{n,n+1}} \cdot \overline{c}_n^{\rm IP} \tag{5-31}$$

and then we have the following *iteration relation*

$$\overline{\overline{T}}_{n,n+1} = \left(\overline{\overline{I}} - \overline{\overline{r}}_{n+1,n} \cdot \overline{\overline{R}}_{n+1,n+2}\right)^{-1} \cdot \overline{\overline{t}}_{n,n+1} \tag{5-32}$$

for matrix $\overline{\overline{T}}$. Inserting the above Eq. (5-31) into the previous Eq. (5-28), it is easy to derive that

$$\overline{\overline{R}}_{n,n+1} \cdot \overline{c}_n^{\rm IP} = \overline{\overline{r}}_{n,n+1} \cdot \overline{c}_n^{\rm IP} + \overline{\overline{t}}_{n+1,n} \cdot \overline{\overline{R}}_{n+1,n+2} \cdot \underbrace{\left(\overline{\overline{I}} - \overline{\overline{r}}_{n+1,n} \cdot \overline{\overline{R}}_{n+1,n+2}\right)^{-1} \cdot \overline{\overline{t}}_{n,n+1}}_{\overline{\overline{T}}_{n,n+1}} \cdot \overline{c}_n^{\rm IP} \tag{5-33}$$

Because of the arbitrariness of $\overline{c}_n^{\rm IP}$, the above Eq. (5-33) implies the following iteration relation

$$\overline{\overline{R}}_{n,n+1} = \overline{\overline{r}}_{n,n+1} + \overline{\overline{t}}_{n+1,n} \cdot \overline{\overline{R}}_{n+1,n+2} \cdot \underbrace{\left( \overline{\overline{I}} - \overline{\overline{r}}_{n+1,n} \cdot \overline{\overline{R}}_{n+1,n+2} \right)^{-1} \cdot \overline{\overline{t}}_{n,n+1}}_{\overline{\overline{T}}_{n,n+1}} \tag{5-34}$$

for matrix $\overline{\overline{R}}$.

Now, we properly rewrite the above-obtained iteration relations (5-32) and (5-34) as follows:

$$\overline{\overline{T}}_{n-2,n-1} = \left( \overline{\overline{I}} - \overline{\overline{r}}_{n-1,n-2} \cdot \overline{\overline{R}}_{n-1,n} \right)^{-1} \cdot \overline{\overline{t}}_{n-2,n-1} \tag{5-35}$$

$$\overline{\overline{R}}_{n-2,n-1} = \overline{\overline{r}}_{n-2,n-1} + \overline{\overline{t}}_{n-1,n-2} \cdot \overline{\overline{R}}_{n-1,n} \cdot \overline{\overline{T}}_{n-2,n-1} \tag{5-36}$$

with the following *initial condition*

$$\overline{\overline{R}}_{N-1,N} = \overline{\overline{r}}_{N-1,N} \tag{5-37}$$

where $N$ is the *index number* of the last region in the cascaded network shown in the Fig. 5-3. Following the convention used in Refs. [47~70], the above transformation matrices $\overline{\overline{R}}$ and $\overline{\overline{T}}$ are called *global reflection matrix* and *global transmission matrix* respectively.

## 5.5 Expose on Multi-region Matching Problem

Now supposing that only one $\mathbb{S}_n^{\text{IP}}$-based mode and only one $\mathbb{S}_n^{\text{OP}}$-based mode work in region $\mathbb{V}_n$, and also supposing that only one $\mathbb{S}_{n+1}^{\text{IP}}$-based mode works in region $\mathbb{V}_{n+1}$, then the matrices in Sec. 5.3 are reduced to the following simple forms

$$Z_{n,n}^{\text{OP,IP}} = \left\langle \vec{H}_n^{\text{OP}}, \hat{n}^{\to n} \times \vec{E}_n^{\text{IP}} \right\rangle_{\mathbb{S}_{n,n+1}} \tag{5-38a}$$

$$\Lambda_{n,n}^{\text{OP,OP}} = \left\langle \vec{H}_n^{\text{OP}}, \hat{n}^{\to n} \times \vec{E}_n^{\text{OP}} \right\rangle_{\mathbb{S}_{n,n+1}} \tag{5-38b}$$

$$Z_{n,n+1}^{\text{OP,IP}} = \left\langle \vec{H}_n^{\text{OP}}, \hat{n}^{\to n+1} \times \vec{E}_{n+1}^{\text{IP}} \right\rangle_{\mathbb{S}_{n,n+1}} \tag{5-38c}$$

and

$$Z_{n+1,n}^{\text{IP,IP}} = \left\langle \vec{E}_{n+1}^{\text{IP}}, \vec{H}_n^{\text{IP}} \times \hat{n}^{\to n} \right\rangle_{\mathbb{S}_{n,n+1}} \tag{5-39a}$$

$$Z_{n+1,n}^{\text{IP,OP}} = \left\langle \vec{E}_{n+1}^{\text{IP}}, \vec{H}_n^{\text{OP}} \times \hat{n}^{\to n} \right\rangle_{\mathbb{S}_{n,n+1}} \tag{5-39b}$$

$$\Lambda_{n+1,n+1}^{\text{IP,IP}} = \left\langle \vec{E}_{n+1}^{\text{IP}}, \vec{H}_{n+1}^{\text{IP}} \times \hat{n}^{\to n+1} \right\rangle_{\mathbb{S}_{n,n+1}} \tag{5-39c}$$

Thus, the reflection matrix $\overline{\overline{r}}_{n,n+1}$ and transmission matrix $\overline{\overline{t}}_{n,n+1}$ are reduced to *reflection coefficient* $r_{n,n+1}$ and *transmission coefficient* $t_{n,n+1}$ respectively, and the coefficients have the following explicit expressions

$$r_{n,n+1} = \frac{Z_{n,n+1}^{\text{OP,IP}} Z_{n+1,n}^{\text{IP,IP}} - \Lambda_{n+1,n+1}^{\text{IP,IP}} Z_{n,n}^{\text{OP,IP}}}{\begin{vmatrix} \Lambda_{n,n}^{\text{OP,OP}} & Z_{n,n+1}^{\text{OP,IP}} \\ Z_{n+1,n}^{\text{IP,OP}} & \Lambda_{n+1,n+1}^{\text{IP,IP}} \end{vmatrix}} \tag{5-40}$$

$$t_{n,n+1} = \frac{Z_{n+1,n}^{\text{IP,OP}} Z_{n,n}^{\text{OP,IP}} - \Lambda_{n,n}^{\text{OP,OP}} Z_{n+1,n}^{\text{IP,IP}}}{\begin{vmatrix} \Lambda_{n,n}^{\text{OP,OP}} & Z_{n,n+1}^{\text{OP,IP}} \\ Z_{n+1,n}^{\text{IP,OP}} & \Lambda_{n+1,n+1}^{\text{IP,IP}} \end{vmatrix}} \tag{5-41}$$

where the denominator is the *determinant* of the corresponding matrix. It is obvious to exist the relations that

$$\left.\begin{aligned} \vec{H}_{n+1}^{\text{IP}} &= \vec{H}_{n}^{\text{IP}} \\ \vec{E}_{n}^{\text{IP}} &= \vec{E}_{n+1}^{\text{IP}} \end{aligned}\right\} \text{on } \mathbb{S}_{n,n+1} \quad \Rightarrow \quad \Lambda_{n+1,n+1}^{\text{IP,IP}} Z_{n,n}^{\text{OP,IP}} = Z_{n,n+1}^{\text{OP,IP}} Z_{n+1,n}^{\text{IP,IP}} \quad \Rightarrow \quad r_{n,n+1} = 0 \tag{5-42}$$

This implies that: **when the $\mathbb{S}_n^{\text{IP}}$ -based mode and the $\mathbb{S}_{n+1}^{\text{IP}}$ -based mode have the same tangential component on interface $\mathbb{S}_{n,n+1}$, there will not exist any $\mathbb{S}_n^{\text{OP}}$ -based mode reflected from $\mathbb{S}_{n,n+1}$.**

However, the modes obtained in the previous Chaps. 3 and 4 were constructed by seperately treating guide, tra-anetnna, propagation medium, and rec-antenna respectively (as shown in Fig. 5-4), and the separate treatment can not guarantee the automatic matching for the tangential components on the interfaces.

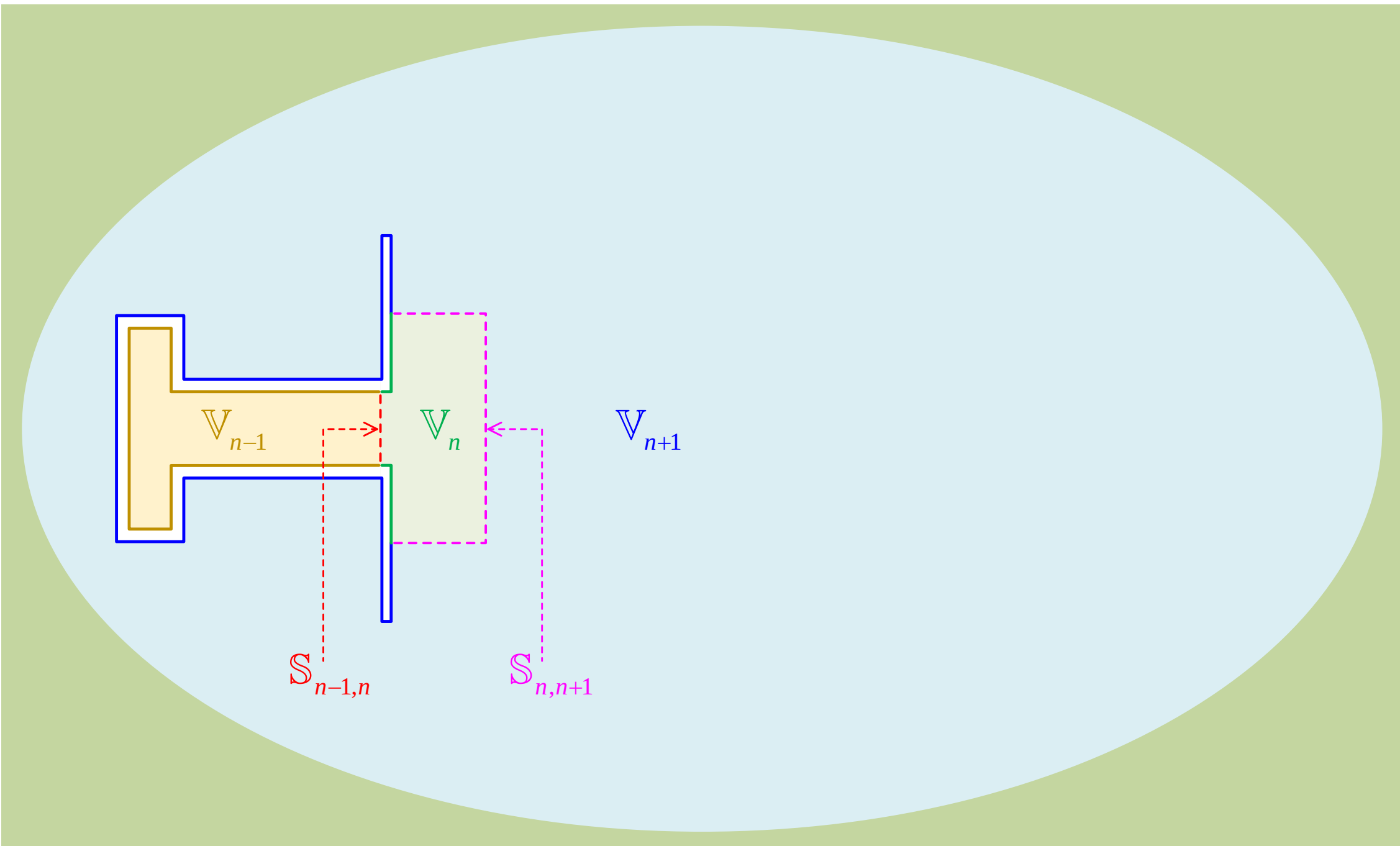

Figure 5-4 Modal matching among tra-guide $\mathbb{V}_{n-1}$, tra-antenna $\mathbb{V}_n$, and propagation medium $\mathbb{V}_{n+1}$

Thus, the modal matching process proposed in the previous Secs. 5.3 and 5.4 is necessary for the modal analysis of guide-antenna system, antenna-medium system, and medium-antenna system, etc. But unfortunately, the modal matching process is relatively troublesome as exhibited above.

By the above analysis, it is not difficult to observe that: **the reason leading to the indispensability for modal matching process originates from seperately treating the regions occupied by guide, tra-antenna, propagation medium, and rec-antenna.** Based on this observation, we propose some schemes for avoiding the modal matching process in the subsequent Chaps. 6, 7, and 8.

## 5.6 Chapter Summary

In this chapter, the interactions among two-port regions are discussed quantitatively by employing the *input-power-decoupled modes* (*IP-DMs*) and using *modal matching technique*.

When some incident IP-DMs incident on the interface between two regions, some reflected IP-DMs and some transmitted IP-DMs will be resulted. The reflected IP-DMs and incident IP-DMs are quantitatively related to each other by *reflection matrix operator*, and the transmitted IP-DMs and incident IP-DMs are quantitatively related to each other by *transmission matrix operator*.

If the IP-DMs of tra-guide, tra-antenna, propagation medium, rec-antenna, and rec-guide are constructed seperately, then the modal matching process and the reflection&transmission matrix operators are indispensable for modal analysis. Unfortunately, the modal matching process, i.e. the process for deriving the reflection&transmission matrix operators, is cumbersome.

To effectively avoid the modal matching process, we had introduced the concepts of *augmented tra-antenna* and *augmented rec-antenna* in the previous Secs. 2.3.4 and 2.3.5, and will construct the IP-DMs of the augmented antennas in the subsequent Chaps. 6~8.

# Chapter 6 Input-Power-Decoupled Modes of Augmented Transmitting Antenna (Including Grounding Structure)

**CHAPTER MOTIVATION:** This chapter focuses on constructing the *input-power-decoupled modes* (*IP-DMs*) of *augmented transmitting antenna*① by orthogonalizing the frequecny-domain *input power operator* (*IPO*) obtained in *power transport theorem* (*PTT*) framework, and does some necessary analysis and discussions for the related topics. The obtained IP-DMs, which are inputted into the augmented transmitting antenna, don't have net energy exchange in any integral period.

## 6.1 Chapter Introduction

*Transmitting system* (simply called *transmitter*) is a device used for generating and radiating electromagnetic (EM) signals or powers. *Transmitting antenna* (simply called *tra-antenna*) is an important component in transmitter, and is used for modulating/ controling the powers to be released from transmitter to *free space* (the simplest *propagation medium*). The *working states* of tra-antenna are usually called *working modes* or *modes* simply, and all possible modes constitute a linear space — *modal space*.

Besides the tra-antenna itself, the modes also depend on the interaction between tra-antenna and free space. In fact, the interaction can be physically explained as follows: besides the tra-antenna, the *grounding structure* (used for partially separating transmitter from free space, for the rigorous mathematical definition please see Sec. 2.3.4) also has ability to modulate the powers to be released to free space. As exhibited in the previous Chaps. 4 and 5, if the modal analysis for tra-antenna and free space are done separately, then a *modal matching process* is inevitable, and the process is relatively cumbersome.

A very natural idea for resolving this problem is to treat tra-antenna and free space as a whole and to do the modal analysis for the whole directly. Employing this idea and considering the following facts,

Fact 1. the outer boundary of free space locates at *infinity*, and the *Green's functions* on it satisfy *Sommerfeld's radiation condition* automatically;

Fact 2. the penetrable part of the inner boundary of free space is just the interface

① Augmented transmitting antenna is the union of transmitting antenna and grounding structure, as mathematically defined in Sec. 2.3.4.

between free space and tra-antenna (i.e. the *output port* of tra-antenna);

Fact 3. the impenetrable part of the inner boundary of free space is just the grounding structure;

this chapter treats the tra-antenna and grounding structure as a whole — *augmented transmitting antenna* (simply called *augmented tra-antenna*), and a mathematically rigorous discussion for this treatment can be found in Sec. 2.3.4.

This chapter is organized as follows: Sec. 6.2 focuses on establishing the *power transport theorem* (*PTT*) based *decoupling mode theory* (*DMT*) for *augmented metallic tra-antennas*, and constructing the corresponding *input-power-decoupled modes* (*IP-DMs*); using a similar method, Sec. 6.3 establishes the PTT-based DMT for *augmented material tra-antennas*, and constructs the corresponding IP-DMs; Secs. 6.4~6.6 and 6.7 further generalize the results obtained in Secs. 6.2 and 6.3 to *augmented metal-material composite tra-antennas* and *tra-antenna array*; Sec. 6.8 concludes this chapter.

All the tra-antennas discussed in this chapter are augmented. To simplify the terminology of this chapter, the determiner "augmented" is usually omitted in this chapter.

## 6.2 IP-DMs of Augmented Metallic Transmitting Antenna

In the realm of EM engineering, there exist many different kinds of metallic tra-antennas[71~78], such as *metallic monopole antenna*[79,80], *metallic dipole antenna*[80,81], *metallic curl antenna*[82,83], *metallic conical antenna*[84~86], *metallic patch antenna*[87~91], *metallic planar inverted-F antenna* (*PIFA*)[92,93], *metallic cavity antenna*[94~97], *metallic horn antenna*[98~100], and *metallic reflector antenna*[101~104], etc.

As a kind of typical metallic tra-antenna, *metallic horn fed metallic reflector antennas* have been widely applied to point-to-point communication and deep-space communication[105~109], because of its high gain and narrow beam features[71~78,98~104]. In this section, taking the *metallic corrugated horn fed metallic reflector antenna* shown in Fig. 6-1 as a typical example, we will discuss the fundamental principle of the *PTT-based DMT for metallic tra-antennas* (*PTT-MetTraAnt-DMT*), and provide the detailed process for constructing the IP-DMs of the metallic tra-antenna.

The fundamental principle established for the tra-antenna in Fig. 6-1 is directly applicable to all kinds of the metallic tra-antennas mentioned above, and the specific process focusing on the tra-antenna in Fig. 6-1 can also be easily generalized to the other types of the metallic tra-antennas above mentioned.

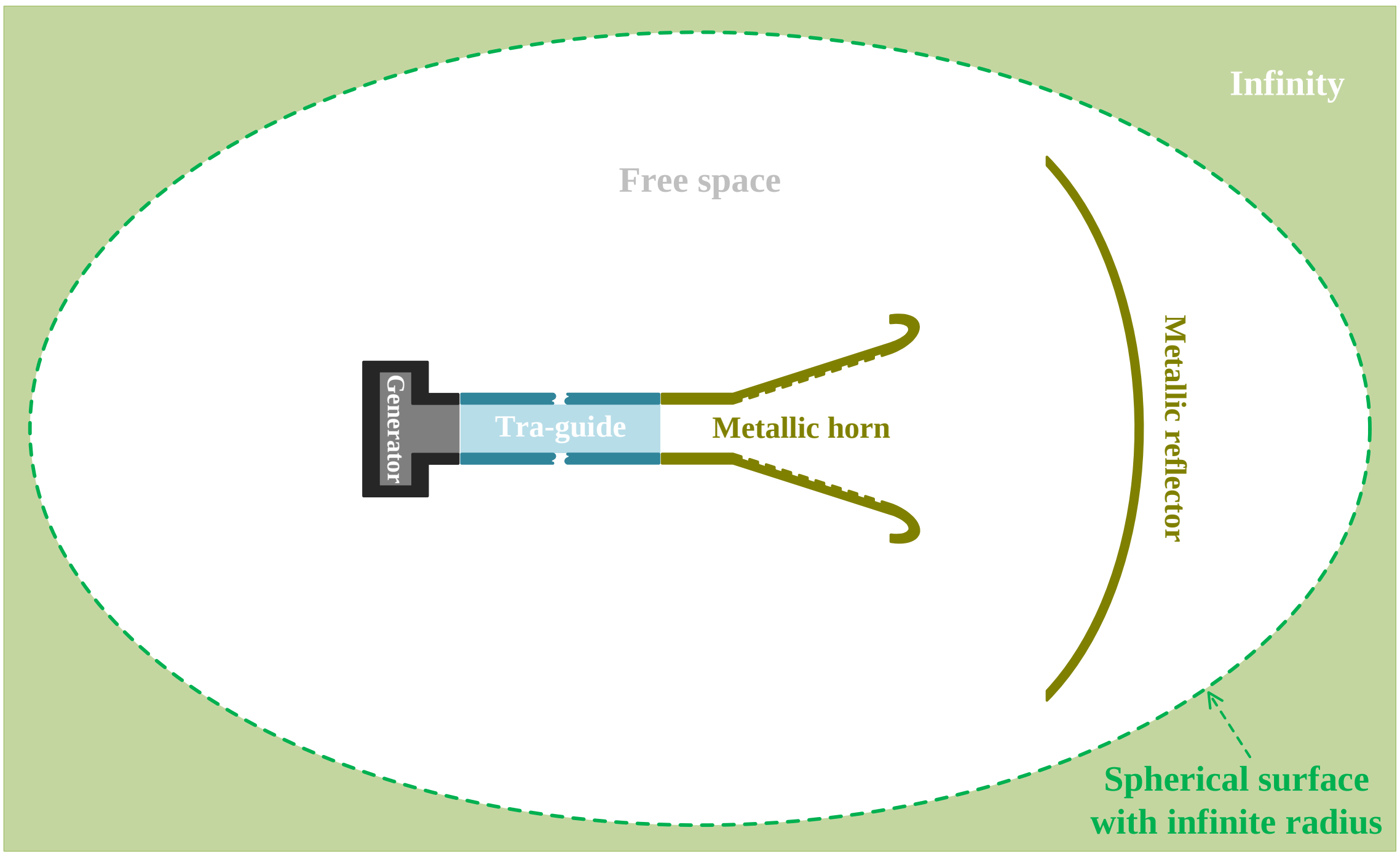

Figure 6-1 Geometry of a typical metallic corrugated horn fed metallic reflector antenna, which is placed in free space

This section is organized as follows:

in Sec. 6.2.1, the *topological structure* of the metallic tra-antenna shown in Fig. 6-1 is described in a rigorous mathematical language — *point set topology*[44], and the EM fields distributing in *three-dimensional Euclidean space* are expressed in terms of the currents distributing on the topological structure;

in Sec. 6.2.2, the rigorous mathematical descriptions for the *modal space* of the metallic tra-antenna are provided by employing the *source-field relationships* obtained in Sec. 6.2.1 and some necessary boundary conditions;

in Sec. 6.2.3, PTT and the *input power operator* (*IPO*) defined on the modal space are derived;

in Sec. 6.2.4, the IP-DMs used to span the modal space are construced by orthogonalizing the IPO obtained in Sec. 6.2.3, and some related topics, such as *modal decoupling relation* and *Parseval's identity* etc., are also discussed;

in Sec. 6.2.5, some typical numerical examples are provided to verify the theory and formulations established in Secs. 6.2.1~6.2.4.

## 6.2.1 Topological Structure and Source-Field Relationships

The topological structure of the metallic tra-antenna in Fig. 6-1 is exhibited in Fig. 6-2.

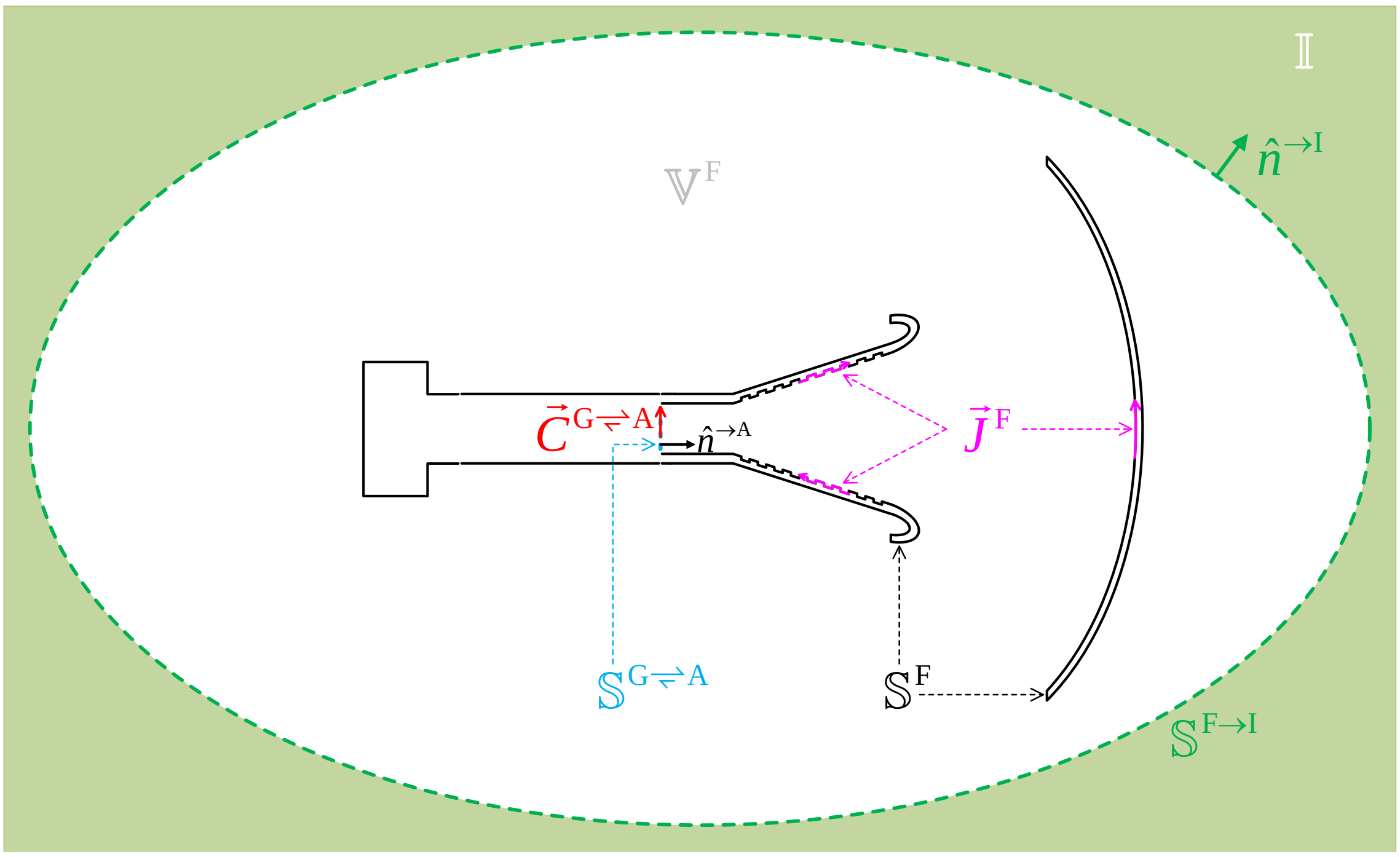

Figure 6-2 Topological structure of the metallic tra-antenna shown in Fig. 6-1

In the above figure, surface $\mathbb{S}^{\mathrm{G}\rightleftharpoons\mathrm{A}}$ denotes the *input port* of the tra-antenna, and $\mathbb{S}^{\mathrm{F}}$ is the interface between the augmented tra-antenna and free space, and $\mathbb{S}^{\mathrm{F}\to\mathrm{I}}$ is a *spherical surface with infinite radius*, and $\mathbb{V}^{\mathrm{F}}$ is the region occupied by whole free space. The explanations for the superscripts of $\mathbb{S}^{\mathrm{G}\rightleftharpoons\mathrm{A}}$, $\mathbb{S}^{\mathrm{F}}$, $\mathbb{S}^{\mathrm{F}\to\mathrm{I}}$, and $\mathbb{V}^{\mathrm{F}}$ can be found in Sec. 2.3. Clearly, $\mathbb{S}^{\mathrm{G}\rightleftharpoons\mathrm{A}}$ and $\mathbb{S}^{\mathrm{F}}$ constitute a closed surface, and the closed surface $\mathbb{S}^{\mathrm{G}\rightleftharpoons\mathrm{A}}\cup\mathbb{S}^{\mathrm{F}}$ and closed surface $\mathbb{S}^{\mathrm{F}\to\mathrm{I}}$ sandwich whole $\mathbb{V}^{\mathrm{F}}$. Thus, closed surfaces $\mathbb{S}^{\mathrm{G}\rightleftharpoons\mathrm{A}}\cup\mathbb{S}^{\mathrm{F}}$ and $\mathbb{S}^{\mathrm{F}\to\mathrm{I}}$ constitute the whole boundary of $\mathbb{V}^{\mathrm{F}}$, i.e., $\partial\mathbb{V}^{\mathrm{F}}=\mathbb{S}^{\mathrm{G}\rightleftharpoons\mathrm{A}}\cup\mathbb{S}^{\mathrm{F}}\cup\mathbb{S}^{\mathrm{F}\to\mathrm{I}}$.

If the *equivalent surface currents* distributing on $\mathbb{S}^{\mathrm{G}\rightleftharpoons\mathrm{A}}$ are denoted as $\{\vec{J}^{\mathrm{G}\rightleftharpoons\mathrm{A}},\vec{M}^{\mathrm{G}\rightleftharpoons\mathrm{A}}\}$, and the *equivalent surface electric current* distributing on $\mathbb{S}^{\mathrm{F}}$ is denoted as $\vec{J}^{\mathrm{F}}$ ①, then the field distributing on $\mathbb{V}^{\mathrm{F}}$ can be expressed as follows:

$$\vec{F}(\vec{r}) = \mathcal{F}_0\left(\vec{J}^{\mathrm{G}\rightleftharpoons\mathrm{A}}+\vec{J}^{\mathrm{F}},\vec{M}^{\mathrm{G}\rightleftharpoons\mathrm{A}}\right) \quad , \quad \vec{r}\in\mathbb{V}^{\mathrm{F}} \tag{6-1}$$

where $\vec{F}=\vec{E}/\vec{H}$, and correspondingly $\mathcal{F}_0=\mathcal{E}_0/\mathcal{H}_0$, and operator $\mathcal{F}_0$ is the same as the one used in the previous chapters. The currents $\{\vec{J}^{\mathrm{G}\rightleftharpoons\mathrm{A}},\vec{M}^{\mathrm{G}\rightleftharpoons\mathrm{A}}\}$ and fields $\{\vec{E},\vec{H}\}$ in Eq. (6-1) satisfy the following relations

$$\vec{J}^{\mathrm{G}\rightleftharpoons\mathrm{A}}(\vec{r}) = \hat{n}^{\to\mathrm{A}}\times\vec{H}(\vec{r}) \quad , \quad \vec{r}\in\mathbb{S}^{\mathrm{G}\rightleftharpoons\mathrm{A}} \tag{6-2a}$$

① The equivalent surface electric current distributing on $\mathbb{S}^{\mathrm{F}}$ is equal to the *induced surface electric current* distributing on $\mathbb{S}^{\mathrm{F}}$ [13]. The *equivalent surface magnetic current* distributing on $\mathbb{S}^{\mathrm{F}}$ is zero, because of the homogeneous tangential electric field boundary condition on $\mathbb{S}^{\mathrm{F}}$ [13].

$$\vec{M}^{\mathrm{G}\rightleftharpoons\mathrm{A}}(\vec{r}) = \vec{E}(\vec{r})\times\hat{n}^{\to\mathrm{A}} \quad , \quad \vec{r}\in\mathbb{S}^{\mathrm{G}\rightleftharpoons\mathrm{A}} \tag{6-2b}$$

where $\hat{n}^{\to\mathrm{A}}$ is the *normal direction* of $\mathbb{S}^{\mathrm{G}\rightleftharpoons\mathrm{A}}$ and points to the tra-antenna side (as shown in Fig. 6-2) rather than the tra-guide side. In fact, Eqs. (6-2a) and (6-2b) are just the definitions for currents $\vec{J}^{\mathrm{G}\rightleftharpoons\mathrm{A}}$ and $\vec{M}^{\mathrm{G}\rightleftharpoons\mathrm{A}}$ respectively.

### 6.2.2 Mathematical Description for Modal Space

Substituting Eq. (6-1) into Eqs. (6-2a) and (6-2b), we obtain the following *integral equations*

$$\vec{J}^{\mathrm{G}\rightleftharpoons\mathrm{A}}(\vec{r})\times\hat{n}^{\to\mathrm{A}} = \left[\mathcal{H}_0\left(\vec{J}^{\mathrm{G}\rightleftharpoons\mathrm{A}}+\vec{J}^{\mathrm{F}},\vec{M}^{\mathrm{G}\rightleftharpoons\mathrm{A}}\right)\right]_{\vec{r}^{\mathrm{F}}\to\vec{r}}^{\tan} \quad , \quad \vec{r}\in\mathbb{S}^{\mathrm{G}\rightleftharpoons\mathrm{A}} \tag{6-3a}$$

$$\hat{n}^{\to\mathrm{A}}\times\vec{M}^{\mathrm{G}\rightleftharpoons\mathrm{A}}(\vec{r}) = \left[\mathcal{E}_0\left(\vec{J}^{\mathrm{G}\rightleftharpoons\mathrm{A}}+\vec{J}^{\mathrm{F}},\vec{M}^{\mathrm{G}\rightleftharpoons\mathrm{A}}\right)\right]_{\vec{r}^{\mathrm{F}}\to\vec{r}}^{\tan} \quad , \quad \vec{r}\in\mathbb{S}^{\mathrm{G}\rightleftharpoons\mathrm{A}} \tag{6-3b}$$

about currents $\{\vec{J}^{\mathrm{G}\rightleftharpoons\mathrm{A}},\vec{M}^{\mathrm{G}\rightleftharpoons\mathrm{A}}\}$ and $\vec{J}^{\mathrm{F}}$. Based on Eq. (6-1) and the *homogeneous tangential electric field boundary condition* on $\mathbb{S}^{\mathrm{F}}$, we have the following *electric field integral equation*

$$0 = \left[\mathcal{E}_0\left(\vec{J}^{\mathrm{G}\rightleftharpoons\mathrm{A}}+\vec{J}^{\mathrm{F}},\vec{M}^{\mathrm{G}\rightleftharpoons\mathrm{A}}\right)\right]_{\vec{r}^{\mathrm{F}}\to\vec{r}}^{\tan} \quad , \quad \vec{r}\in\mathbb{S}^{\mathrm{F}} \tag{6-4}$$

about currents $\{\vec{J}^{\mathrm{G}\rightleftharpoons\mathrm{A}},\vec{M}^{\mathrm{G}\rightleftharpoons\mathrm{A}}\}$ and $\vec{J}^{\mathrm{F}}$. In Eqs. (6-3a), (6-3b), and (6-4), $\vec{r}^{\mathrm{F}}$ is the point in $\mathbb{V}^{\mathrm{F}}$, and $\vec{r}^{\mathrm{F}}$ approaches the point $\vec{r}$ on $\mathbb{S}^{\mathrm{G}\rightleftharpoons\mathrm{A}}\bigcup\mathbb{S}^{\mathrm{F}}$.

If the currents $\{\vec{J}^{\mathrm{G}\rightleftharpoons\mathrm{A}},\vec{M}^{\mathrm{G}\rightleftharpoons\mathrm{A}}\}$ and $\vec{J}^{\mathrm{F}}$ contained in Eqs. (6-3a), (6-3b), and (6-4) are expanded in terms of some proper *basis functions* as follows:

$$\vec{X}(\vec{r}) = \sum_\xi a_\xi^{\vec{X}}\vec{b}_\xi^{\vec{X}} = \underbrace{\begin{bmatrix}\vec{b}_1^{\vec{X}} & \vec{b}_2^{\vec{X}} & \cdots\end{bmatrix}}_{\bar{\vec{B}}^{\vec{X}}}\cdot\underbrace{\begin{bmatrix}a_1^{\vec{X}}\\ a_2^{\vec{X}}\\ \vdots\end{bmatrix}}_{\bar{a}^{\vec{X}}} \quad , \quad \vec{r}\in\mathbb{X} \tag{6-5}$$

(where $\vec{X}=\vec{J}^{\mathrm{G}\rightleftharpoons\mathrm{A}}/\vec{M}^{\mathrm{G}\rightleftharpoons\mathrm{A}}/\vec{J}^{\mathrm{F}}$ and correspondingly $\mathbb{X}=\mathbb{S}^{\mathrm{G}\rightleftharpoons\mathrm{A}}/\mathbb{S}^{\mathrm{G}\rightleftharpoons\mathrm{A}}/\mathbb{S}^{\mathrm{F}}$) and Eqs. (6-3a), (6-3b), and (6-4) are tested with $\{\vec{b}_\xi^{\vec{M}^{\mathrm{G}\rightleftharpoons\mathrm{A}}}\}$, $\{\vec{b}_\xi^{\vec{J}^{\mathrm{G}\rightleftharpoons\mathrm{A}}}\}$, and $\{\vec{b}_\xi^{\vec{J}^{\mathrm{F}}}\}$ respectively, then the integral equations are immediately discretized into the following *matrix equations*

$$\bar{\bar{Z}}^{\vec{M}^{\mathrm{G}\rightleftharpoons\mathrm{A}}\vec{J}^{\mathrm{G}\rightleftharpoons\mathrm{A}}}\cdot\bar{a}^{\vec{J}^{\mathrm{G}\rightleftharpoons\mathrm{A}}}+\bar{\bar{Z}}^{\vec{M}^{\mathrm{G}\rightleftharpoons\mathrm{A}}\vec{J}^{\mathrm{F}}}\cdot\bar{a}^{\vec{J}^{\mathrm{F}}}+\bar{\bar{Z}}^{\vec{M}^{\mathrm{G}\rightleftharpoons\mathrm{A}}\vec{M}^{\mathrm{G}\rightleftharpoons\mathrm{A}}}\cdot\bar{a}^{\vec{M}^{\mathrm{G}\rightleftharpoons\mathrm{A}}} = 0 \tag{6-6a}$$

$$\bar{\bar{Z}}^{\vec{J}^{\mathrm{G}\rightleftharpoons\mathrm{A}}\vec{J}^{\mathrm{G}\rightleftharpoons\mathrm{A}}}\cdot\bar{a}^{\vec{J}^{\mathrm{G}\rightleftharpoons\mathrm{A}}}+\bar{\bar{Z}}^{\vec{J}^{\mathrm{G}\rightleftharpoons\mathrm{A}}\vec{J}^{\mathrm{F}}}\cdot\bar{a}^{\vec{J}^{\mathrm{F}}}+\bar{\bar{Z}}^{\vec{J}^{\mathrm{G}\rightleftharpoons\mathrm{A}}\vec{M}^{\mathrm{G}\rightleftharpoons\mathrm{A}}}\cdot\bar{a}^{\vec{M}^{\mathrm{G}\rightleftharpoons\mathrm{A}}} = 0 \tag{6-6b}$$

and

$$\bar{\bar{Z}}^{\vec{J}^{\mathrm{F}}\vec{J}^{\mathrm{G}\rightleftharpoons\mathrm{A}}}\cdot\bar{a}^{\vec{J}^{\mathrm{G}\rightleftharpoons\mathrm{A}}}+\bar{\bar{Z}}^{\vec{J}^{\mathrm{F}}\vec{J}^{\mathrm{F}}}\cdot\bar{a}^{\vec{J}^{\mathrm{F}}}+\bar{\bar{Z}}^{\vec{J}^{\mathrm{F}}\vec{M}^{\mathrm{G}\rightleftharpoons\mathrm{A}}}\cdot\bar{a}^{\vec{M}^{\mathrm{G}\rightleftharpoons\mathrm{A}}} = 0 \tag{6-7}$$

The formulations used to calculate the elements of the matrices in Eq. (6-6a) are as follows:

$$z_{\xi\zeta}^{\vec{M}^{\mathrm{G}\rightleftharpoons\mathrm{A}}\vec{J}^{\mathrm{G}\rightleftharpoons\mathrm{A}}} = \left\langle \vec{b}_{\xi}^{\vec{M}^{\mathrm{G}\rightleftharpoons\mathrm{A}}}, \hat{n}^{\rightarrow\mathrm{A}} \times \frac{1}{2}\vec{b}_{\zeta}^{\vec{J}^{\mathrm{G}\rightleftharpoons\mathrm{A}}} + \mathrm{P.V.}\mathcal{K}_0\left(\vec{b}_{\zeta}^{\vec{J}^{\mathrm{G}\rightleftharpoons\mathrm{A}}}\right)\right\rangle_{\mathbb{S}^{\mathrm{G}\rightleftharpoons\mathrm{A}}} \tag{6-8a}$$

$$z_{\xi\zeta}^{\vec{M}^{\mathrm{G}\rightleftharpoons\mathrm{A}}\vec{J}^{\mathrm{F}}} = \left\langle \vec{b}_{\xi}^{\vec{M}^{\mathrm{G}\rightleftharpoons\mathrm{A}}}, \mathcal{K}_0\left(\vec{b}_{\zeta}^{\vec{J}^{\mathrm{F}}}\right)\right\rangle_{\mathbb{S}^{\mathrm{G}\rightleftharpoons\mathrm{A}}} \tag{6-8b}$$

$$z_{\xi\zeta}^{\vec{M}^{\mathrm{G}\rightleftharpoons\mathrm{A}}\vec{M}^{\mathrm{G}\rightleftharpoons\mathrm{A}}} = \left\langle \vec{b}_{\xi}^{\vec{M}^{\mathrm{G}\rightleftharpoons\mathrm{A}}}, -j\omega\varepsilon_0\mathcal{L}_0\left(\vec{b}_{\zeta}^{\vec{M}^{\mathrm{G}\rightleftharpoons\mathrm{A}}}\right)\right\rangle_{\mathbb{S}^{\mathrm{G}\rightleftharpoons\mathrm{A}}} \tag{6-8c}$$

and the formulations used to calculate the elements of the matrices in Eq. (6-6b) are as follows:

$$z_{\xi\zeta}^{\vec{J}^{\mathrm{G}\rightleftharpoons\mathrm{A}}\vec{J}^{\mathrm{G}\rightleftharpoons\mathrm{A}}} = \left\langle \vec{b}_{\xi}^{\vec{J}^{\mathrm{G}\rightleftharpoons\mathrm{A}}}, -j\omega\mu_0\mathcal{L}_0\left(\vec{b}_{\zeta}^{\vec{J}^{\mathrm{G}\rightleftharpoons\mathrm{A}}}\right)\right\rangle_{\mathbb{S}^{\mathrm{G}\rightleftharpoons\mathrm{A}}} \tag{6-9a}$$

$$z_{\xi\zeta}^{\vec{J}^{\mathrm{G}\rightleftharpoons\mathrm{A}}\vec{J}^{\mathrm{F}}} = \left\langle \vec{b}_{\xi}^{\vec{J}^{\mathrm{G}\rightleftharpoons\mathrm{A}}}, -j\omega\mu_0\mathcal{L}_0\left(\vec{b}_{\zeta}^{\vec{J}^{\mathrm{F}}}\right)\right\rangle_{\mathbb{S}^{\mathrm{G}\rightleftharpoons\mathrm{A}}} \tag{6-9b}$$

$$z_{\xi\zeta}^{\vec{J}^{\mathrm{G}\rightleftharpoons\mathrm{A}}\vec{M}^{\mathrm{G}\rightleftharpoons\mathrm{A}}} = \left\langle \vec{b}_{\xi}^{\vec{J}^{\mathrm{G}\rightleftharpoons\mathrm{A}}}, \frac{1}{2}\vec{b}_{\zeta}^{\vec{M}^{\mathrm{G}\rightleftharpoons\mathrm{A}}} \times \hat{n}^{\rightarrow\mathrm{A}} - \mathrm{P.V.}\mathcal{K}_0\left(\vec{b}_{\zeta}^{\vec{M}^{\mathrm{G}\rightleftharpoons\mathrm{A}}}\right)\right\rangle_{\mathbb{S}^{\mathrm{G}\rightleftharpoons\mathrm{A}}} \tag{6-9c}$$

and the formulations used to calculate the elements of the matrices in Eq. (6-7) are as follows:

$$z_{\xi\zeta}^{\vec{J}^{\mathrm{F}}\vec{J}^{\mathrm{G}\rightleftharpoons\mathrm{A}}} = \left\langle \vec{b}_{\xi}^{\vec{J}^{\mathrm{F}}}, -j\omega\mu_0\mathcal{L}_0\left(\vec{b}_{\zeta}^{\vec{J}^{\mathrm{G}\rightleftharpoons\mathrm{A}}}\right)\right\rangle_{\mathbb{S}^{\mathrm{F}}} \tag{6-10a}$$

$$z_{\xi\zeta}^{\vec{J}^{\mathrm{F}}\vec{J}^{\mathrm{F}}} = \left\langle \vec{b}_{\xi}^{\vec{J}^{\mathrm{F}}}, -j\omega\mu_0\mathcal{L}_0\left(\vec{b}_{\zeta}^{\vec{J}^{\mathrm{F}}}\right)\right\rangle_{\mathbb{S}^{\mathrm{F}}} \tag{6-10b}$$

$$z_{\xi\zeta}^{\vec{J}^{\mathrm{F}}\vec{M}^{\mathrm{G}\rightleftharpoons\mathrm{A}}} = \left\langle \vec{b}_{\xi}^{\vec{J}^{\mathrm{F}}}, -\mathcal{K}_0\left(\vec{b}_{\zeta}^{\vec{M}^{\mathrm{G}\rightleftharpoons\mathrm{A}}}\right)\right\rangle_{\mathbb{S}^{\mathrm{F}}} \tag{6-10c}$$

In the above Eqs. (6-8a) and (6-9c), symbol "$\mathrm{P.V.}\mathcal{K}_0$" represents the *principal value* of operator $\mathcal{K}_0$.

In the following parts of this subsection, by employing the above-obtained matrix equations, we give a rigorous mathematical description to the modal space of the tra-antenna shown in Figs. 6-1 and 6-2.

### 6.2.2.1 Scheme I: Dependent Variable Elimination (DVE)

By properly assembling the Eqs. (6-6a), (6-6b), and (6-7), we have the following two theoretically equivalent *augmented matrix equations*

$$\underbrace{\begin{bmatrix} \bar{\bar{I}}^{\vec{J}^{\mathrm{G}\rightleftharpoons\mathrm{A}}} & 0 & 0 \\ 0 & \bar{\bar{Z}}^{\vec{M}^{\mathrm{G}\rightleftharpoons\mathrm{A}}\vec{J}^{\mathrm{F}}} & \bar{\bar{Z}}^{\vec{M}^{\mathrm{G}\rightleftharpoons\mathrm{A}}\vec{M}^{\mathrm{G}\rightleftharpoons\mathrm{A}}} \\ 0 & \bar{\bar{Z}}^{\vec{J}^{\mathrm{F}}\vec{J}^{\mathrm{F}}} & \bar{\bar{Z}}^{\vec{J}^{\mathrm{F}}\vec{M}^{\mathrm{G}\rightleftharpoons\mathrm{A}}} \end{bmatrix}}_{\bar{\bar{\Psi}}_1} \cdot \underbrace{\begin{bmatrix} \bar{a}^{\vec{J}^{\mathrm{G}\rightleftharpoons\mathrm{A}}} \\ \bar{a}^{\vec{J}^{\mathrm{F}}} \\ \bar{a}^{\vec{M}^{\mathrm{G}\rightleftharpoons\mathrm{A}}} \end{bmatrix}}_{\bar{a}^{\mathrm{AV}}} = \underbrace{\begin{bmatrix} \bar{\bar{I}}^{\vec{J}^{\mathrm{G}\rightleftharpoons\mathrm{A}}} \\ -\bar{\bar{Z}}^{\vec{M}^{\mathrm{G}\rightleftharpoons\mathrm{A}}\vec{J}^{\mathrm{G}\rightleftharpoons\mathrm{A}}} \\ -\bar{\bar{Z}}^{\vec{J}^{\mathrm{F}}\vec{J}^{\mathrm{G}\rightleftharpoons\mathrm{A}}} \end{bmatrix}}_{\bar{\bar{\Psi}}_2} \cdot \bar{a}^{\vec{J}^{\mathrm{G}\rightleftharpoons\mathrm{A}}} \tag{6-11}$$

$$\underbrace{\begin{bmatrix} \overline{\overline{Z}}^{\vec{J}^{\text{G}\rightleftharpoons\text{A}}\vec{J}^{\text{G}\rightleftharpoons\text{A}}} & \overline{\overline{Z}}^{\vec{J}^{\text{G}\rightleftharpoons\text{A}}\vec{J}^{\text{F}}} & 0 \\ \overline{\overline{Z}}^{\vec{J}^{\text{F}}\vec{J}^{\text{G}\rightleftharpoons\text{A}}} & \overline{\overline{Z}}^{\vec{J}^{\text{F}}\vec{J}^{\text{F}}} & 0 \\ 0 & 0 & \overline{\overline{I}}^{\vec{M}^{\text{G}\rightleftharpoons\text{A}}} \end{bmatrix}}_{\overline{\overline{\Psi}}_3} \cdot \underbrace{\begin{bmatrix} \overline{a}^{\vec{J}^{\text{G}\rightleftharpoons\text{A}}} \\ \overline{a}^{\vec{J}^{\text{F}}} \\ \overline{a}^{\vec{M}^{\text{G}\rightleftharpoons\text{A}}} \end{bmatrix}}_{\overline{a}^{\text{AV}}} = \underbrace{\begin{bmatrix} -\overline{\overline{Z}}^{\vec{J}^{\text{G}\rightleftharpoons\text{A}}\vec{M}^{\text{G}\rightleftharpoons\text{A}}} \\ -\overline{\overline{Z}}^{\vec{J}^{\text{F}}\vec{M}^{\text{G}\rightleftharpoons\text{A}}} \\ \overline{\overline{I}}^{\vec{M}^{\text{G}\rightleftharpoons\text{A}}} \end{bmatrix}}_{\overline{\overline{\Psi}}_4} \cdot \overline{a}^{\vec{M}^{\text{G}\rightleftharpoons\text{A}}} \tag{6-12}$$

where $\overline{\overline{I}}^{\vec{J}^{\text{G}\rightleftharpoons\text{A}}}$ and $\overline{\overline{I}}^{\vec{M}^{\text{G}\rightleftharpoons\text{A}}}$ are two *identity matrices* with the same orders as the dimensions of $\overline{a}^{\vec{J}^{\text{G}\rightleftharpoons\text{A}}}$ and $\overline{a}^{\vec{M}^{\text{G}\rightleftharpoons\text{A}}}$ respectively, and the superscript "AV" is the acronym of "*all variables* (related to the tra-antenna)".

By solving the above matrix equations, there exist the following transformations from $\overline{a}^{\vec{J}^{\text{G}\rightleftharpoons\text{A}}}$ to $\overline{a}^{\text{AV}}$ and from $\overline{a}^{\vec{M}^{\text{G}\rightleftharpoons\text{A}}}$ to $\overline{a}^{\text{AV}}$

$$\overline{a}^{\text{AV}} = \overbrace{\left(\overline{\overline{\Psi}}_1\right)^{-1} \cdot \overline{\overline{\Psi}}_2}^{\overline{\overline{T}}^{\vec{J}^{\text{G}\rightleftharpoons\text{A}}\rightarrow\text{AV}}} \cdot \overline{a}^{\vec{J}^{\text{G}\rightleftharpoons\text{A}}} \tag{6-13}$$

$$\overline{a}^{\text{AV}} = \underbrace{\left(\overline{\overline{\Psi}}_3\right)^{-1} \cdot \overline{\overline{\Psi}}_4}_{\overline{\overline{T}}^{\vec{M}^{\text{G}\rightleftharpoons\text{A}}\rightarrow\text{AV}}} \cdot \overline{a}^{\vec{M}^{\text{G}\rightleftharpoons\text{A}}} \tag{6-14}$$

and the transformations can be uniformly written as follows:

$$\overline{\overline{a}}^{\text{AV}} = \overline{\overline{T}}^{\text{BV}\rightarrow\text{AV}} \cdot \overline{a}^{\text{BV}} \tag{6-15}$$

for simplifying the symbolic system of the following discussions, where $\overline{a}^{\text{BV}} = \overline{a}^{\vec{J}^{\text{G}\rightleftharpoons\text{A}}} / \overline{a}^{\vec{M}^{\text{G}\rightleftharpoons\text{A}}}$ and correspondingly $\overline{\overline{T}}^{\text{BV}\rightarrow\text{AV}} = \overline{\overline{T}}^{\vec{J}^{\text{G}\rightleftharpoons\text{A}}\rightarrow\text{AV}} / \overline{\overline{T}}^{\vec{M}^{\text{G}\rightleftharpoons\text{A}}\rightarrow\text{AV}}$, and the superscript "BV" is the acronym of "*basic variables* (which are both complete and independent for describing the working state of the augmented tra-antenna)".

#### 6.2.2.2 Scheme II: Solution/Definition Domain Compression (SDC/DDC)

In fact, the Eqs. (6-6a), (6-6b), and (6-7) can also be alternatively assembled as the following augmented matrix equations

$$\overbrace{\begin{bmatrix} \overline{\overline{Z}}^{\vec{M}^{\text{G}\rightleftharpoons\text{A}}\vec{J}^{\text{G}\rightleftharpoons\text{A}}} & \overline{\overline{Z}}^{\vec{M}^{\text{G}\rightleftharpoons\text{A}}\vec{J}^{\text{F}}} & \overline{\overline{Z}}^{\vec{M}^{\text{G}\rightleftharpoons\text{A}}\vec{M}^{\text{G}\rightleftharpoons\text{A}}} \\ \overline{\overline{Z}}^{\vec{J}^{\text{F}}\vec{J}^{\text{G}\rightleftharpoons\text{A}}} & \overline{\overline{Z}}^{\vec{J}^{\text{F}}\vec{J}^{\text{F}}} & \overline{\overline{Z}}^{\vec{J}^{\text{F}}\vec{M}^{\text{G}\rightleftharpoons\text{A}}} \end{bmatrix}}^{\overline{\overline{\Psi}}_{\text{FCE}}^{\text{DoJ}}} \cdot \overline{\overline{a}}^{\text{AV}} = 0 \tag{6-16}$$

$$\underbrace{\begin{bmatrix} \overline{\overline{Z}}^{\vec{J}^{\text{G}\rightleftharpoons\text{A}}\vec{J}^{\text{G}\rightleftharpoons\text{A}}} & \overline{\overline{Z}}^{\vec{J}^{\text{G}\rightleftharpoons\text{A}}\vec{J}^{\text{F}}} & \overline{\overline{Z}}^{\vec{J}^{\text{G}\rightleftharpoons\text{A}}\vec{M}^{\text{G}\rightleftharpoons\text{A}}} \\ \overline{\overline{Z}}^{\vec{J}^{\text{F}}\vec{J}^{\text{G}\rightleftharpoons\text{A}}} & \overline{\overline{Z}}^{\vec{J}^{\text{F}}\vec{J}^{\text{F}}} & \overline{\overline{Z}}^{\vec{J}^{\text{F}}\vec{M}^{\text{G}\rightleftharpoons\text{A}}} \end{bmatrix}}_{\overline{\overline{\Psi}}_{\text{FCE}}^{\text{DoM}}} \cdot \overline{\overline{a}}^{\text{AV}} = 0 \tag{6-17}$$

where the superscript "DoJ/DoM" is the acronym of "*definition of* $\vec{J}^{G\rightleftharpoons A} / \vec{M}^{G\rightleftharpoons A}$", and the subscript "FCE" is the acronym of "*field continuation equation*".

Theoretically, the Eqs. (6-16) and (6-17) are equivalent to each other, and they have the same *solution space*, and the solution space is just the modal space of the tra-antenna shown in Figs. 6-1 and 6-2. If the *basic solutions* (*BSs*) used to span the solution space / modal space are denoted as $\{\bar{s}_1^{\rm BS},\bar{s}_2^{\rm BS},\cdots\}$, then any mode contained in the space can be expanded as follows:

$$\bar{a}^{\rm AV} = \sum_i a_i^{\rm BS}\bar{s}_i^{\rm BS} = \underbrace{\left[\bar{s}_1^{\rm BS},\bar{s}_2^{\rm BS},\cdots\right]}_{\bar{\bar{T}}^{\rm BS\to AV}}\cdot\underbrace{\begin{bmatrix} a_1^{\rm BS} \\ a_2^{\rm BS} \\ \vdots \end{bmatrix}}_{\bar{a}^{\rm BS}} \tag{6-18}$$

as explained in Ref. [18].

For the convenience of the following discussions, Eqs. (6-15) and (6-18) are uniformly written as follows:

$$\bar{a}^{\rm AV} = \bar{\bar{T}}\cdot\bar{a} \tag{6-19}$$

where $\bar{a} = \bar{a}^{\rm BV} / \bar{a}^{\rm BS}$ and correspondingly $\bar{\bar{T}} = \bar{\bar{T}}^{\rm BV\to AV} / \bar{\bar{T}}^{\rm BS\to AV}$.

### 6.2.3 Power Transport Theorem and Input Power Operator

In this subsection, we provide the *power transport theorem* (*PTT*) used to quantitatively describe the flowing way of the power passing through the tra-antenna shown in Figs. 6-1 and 6-2, and then obtain the *input power operator* (*IPO*) defined on modal space.

#### 6.2.3.1 Power Transport Theorem

Applying the results obtained in Chap. 2 to the tra-antenna shown in Figs. 6-1 and 6-2, we immediately have the following PTT for the tra-antenna.

$$P^{\rm G\rightleftharpoons A} = P_{\rm rad}^{\rm I} + j\,P_{\rm sto}^{\rm F} = P^{\rm A\rightleftharpoons F} \tag{6-20}$$

where $P^{\rm G\rightleftharpoons A}$ is the *input power* inputted into the tra-antenna, and $P_{\rm rad}^{\rm I}$ is the *radiated power* arriving at infinity (here, the superscript “I” is just the acronym of “infinity”), and $P_{\rm sto}^{\rm F}$ is the power corresponding to the energy stored in free space.

The various powers contained in PTT (6-20) are as follows:

$$P^{\rm G\rightleftharpoons A} = (1/2)\iint_{\mathbb{S}^{\rm G\rightleftharpoons A}}\left(\vec{E}\times\vec{H}^{\dagger}\right)\cdot\hat{n}^{\rightarrow\rm A}dS \tag{6-21a}$$

$$P_{\rm rad}^{\rm I} = (1/2)\oiint_{\mathbb{S}^{\rm F\to I}}\left(\vec{E}\times\vec{H}^{\dagger}\right)\cdot\hat{n}^{\rightarrow\rm I}dS \tag{6-21b}$$

$$P_{\rm sto}^{\rm F} = 2\omega\left[(1/4)\left\langle\vec{H},\mu_0\vec{H}\right\rangle_{\mathbb{V}^{\rm F}} - (1/4)\left\langle\varepsilon_0\vec{E},\vec{E}\right\rangle_{\mathbb{V}^{\rm F}}\right] \tag{6-21c}$$

where $\hat{n}^{\rightarrow \mathrm{I}}$ is the *normal direction vector* of $\mathbb{S}^{\mathrm{F} \rightarrow \mathrm{I}}$ and points to infinity as shown in Fig. 6-2.

#### 6.2.3.2 Input Power Operator — Formulation I: Current Form

Based on Eqs. (6-2a)&(6-2b) and the tangential continuity of the $\{\vec{E}, \vec{H}\}$ on $\mathbb{S}^{\mathrm{G} \rightleftharpoons \mathrm{A}}$, the IPO $P^{\mathrm{G} \rightleftharpoons \mathrm{A}}$ given in Eq. (6-21a) can be alternatively written as follows:

$$P^{\mathrm{G} \rightleftharpoons \mathrm{A}} = (1/2)\left\langle \hat{n}^{\rightarrow \mathrm{A}} \times \vec{J}^{\mathrm{G} \rightleftharpoons \mathrm{A}}, \vec{M}^{\mathrm{G} \rightleftharpoons \mathrm{A}} \right\rangle_{\mathbb{S}^{\mathrm{G} \rightleftharpoons \mathrm{A}}} \tag{6-22}$$

and it is particularly called the *current form of IPO* (because the expression contains currents only).

Inserting Eq. (6-5) into the above current form, the current form is immediately discretized as follows:

$$P^{\mathrm{G} \rightleftharpoons \mathrm{A}} = \left(\bar{a}^{\mathrm{AV}}\right)^{\dagger} \cdot \underbrace{\begin{bmatrix} 0 & 0 & \overline{\overline{C}}^{\vec{J}^{\mathrm{G} \rightleftharpoons \mathrm{A}} \vec{M}^{\mathrm{G} \rightleftharpoons \mathrm{A}}} \\ 0 & 0 & 0 \\ 0 & 0 & 0 \end{bmatrix}}_{\overline{\overline{P}}_{\mathrm{curAV}}^{\mathrm{G} \rightleftharpoons \mathrm{A}}} \cdot \bar{a}^{\mathrm{AV}} \tag{6-23}$$

where the elements of sub-matrix $\overline{\overline{C}}^{\vec{J}^{\mathrm{G} \rightleftharpoons \mathrm{A}} \vec{M}^{\mathrm{G} \rightleftharpoons \mathrm{A}}}$ are calculated as that $c_{\xi\zeta}^{\vec{J}^{\mathrm{G} \rightleftharpoons \mathrm{A}} \vec{M}^{\mathrm{G} \rightleftharpoons \mathrm{A}}} = (1/2) < \hat{n}^{\rightarrow \mathrm{A}} \times \vec{b}_{\xi}^{\vec{J}^{\mathrm{G} \rightleftharpoons \mathrm{A}}}, \vec{b}_{\zeta}^{\vec{M}^{\mathrm{G} \rightleftharpoons \mathrm{A}}} >_{\mathbb{S}^{\mathrm{G} \rightleftharpoons \mathrm{A}}}$. To obtain the IPO defined on modal space, we substitute the previously obtained transformation (6-19) into the above matrix form (6-23), and then we have that

$$P^{\mathrm{G} \rightleftharpoons \mathrm{A}} = \bar{a}^{\dagger} \cdot \underbrace{\left( \overline{\overline{T}}^{\dagger} \cdot \overline{\overline{P}}_{\mathrm{curAV}}^{\mathrm{G} \rightleftharpoons \mathrm{A}} \cdot \overline{\overline{T}} \right)}_{\overline{\overline{P}}_{\mathrm{cur}}^{\mathrm{G} \rightleftharpoons \mathrm{A}}} \cdot \bar{a} \tag{6-24}$$

where subscript "cur" is to emphasize that $\overline{\overline{P}}_{\mathrm{cur}}^{\mathrm{G} \rightleftharpoons \mathrm{A}}$ originates from discretizing the current form of IPO.

#### 6.2.3.3 Input Power Operator — Formulation II: Field-Current Interaction Forms

In fact, the IPO $P^{\mathrm{G} \rightleftharpoons \mathrm{A}}$ given in Eqs. (6-21a) and (6-22) also has the following equivalent expressions

$$\begin{aligned} P^{\mathrm{G} \rightleftharpoons \mathrm{A}} &= -(1/2)\left\langle \vec{J}^{\mathrm{G} \rightleftharpoons \mathrm{A}}, \vec{E} \right\rangle_{\mathbb{S}^{\mathrm{G} \rightleftharpoons \mathrm{A}}} \\ &= -(1/2)\left\langle \vec{J}^{\mathrm{G} \rightleftharpoons \mathrm{A}}, \mathcal{E}_0\left( \vec{J}^{\mathrm{G} \rightleftharpoons \mathrm{A}} + \vec{J}^{\mathrm{F}}, \vec{M}^{\mathrm{G} \rightleftharpoons \mathrm{A}} \right) \right\rangle_{\mathbb{S}^{\mathrm{G} \rightleftharpoons \mathrm{A}}} \end{aligned} \tag{6-25}$$

and

$$\begin{aligned} P^{\mathrm{G}\rightleftharpoons\mathrm{A}} &= -(1/2)\left\langle \vec{M}^{\mathrm{G}\rightleftharpoons\mathrm{A}}, \vec{H} \right\rangle_{\underset{\sim}{\mathbb{S}}^{\mathrm{G}\rightleftharpoons\mathrm{A}}}^{\dagger} \\ &= -(1/2)\left\langle \vec{M}^{\mathrm{G}\rightleftharpoons\mathrm{A}}, \mathcal{H}_0\left(\vec{J}^{\mathrm{G}\rightleftharpoons\mathrm{A}} + \vec{J}^{\mathrm{F}}, \vec{M}^{\mathrm{G}\rightleftharpoons\mathrm{A}}\right) \right\rangle_{\underset{\sim}{\mathbb{S}}^{\mathrm{G}\rightleftharpoons\mathrm{A}}}^{\dagger} \end{aligned} \tag{6-26}$$

and they are particularly called the *field-current interaction forms of IPO* because they are expressed in terms of the interactions between fields and currents. In the interaction forms, integral surface $\underset{\sim}{\mathbb{S}}^{\mathrm{G}\rightleftharpoons\mathrm{A}}$ is the one locating on the tra-antenna side of $\mathbb{S}^{\mathrm{G}\rightleftharpoons\mathrm{A}}$ rather than the tra-guide side of $\mathbb{S}^{\mathrm{G}\rightleftharpoons\mathrm{A}}$, as shown in Fig. 6-3.

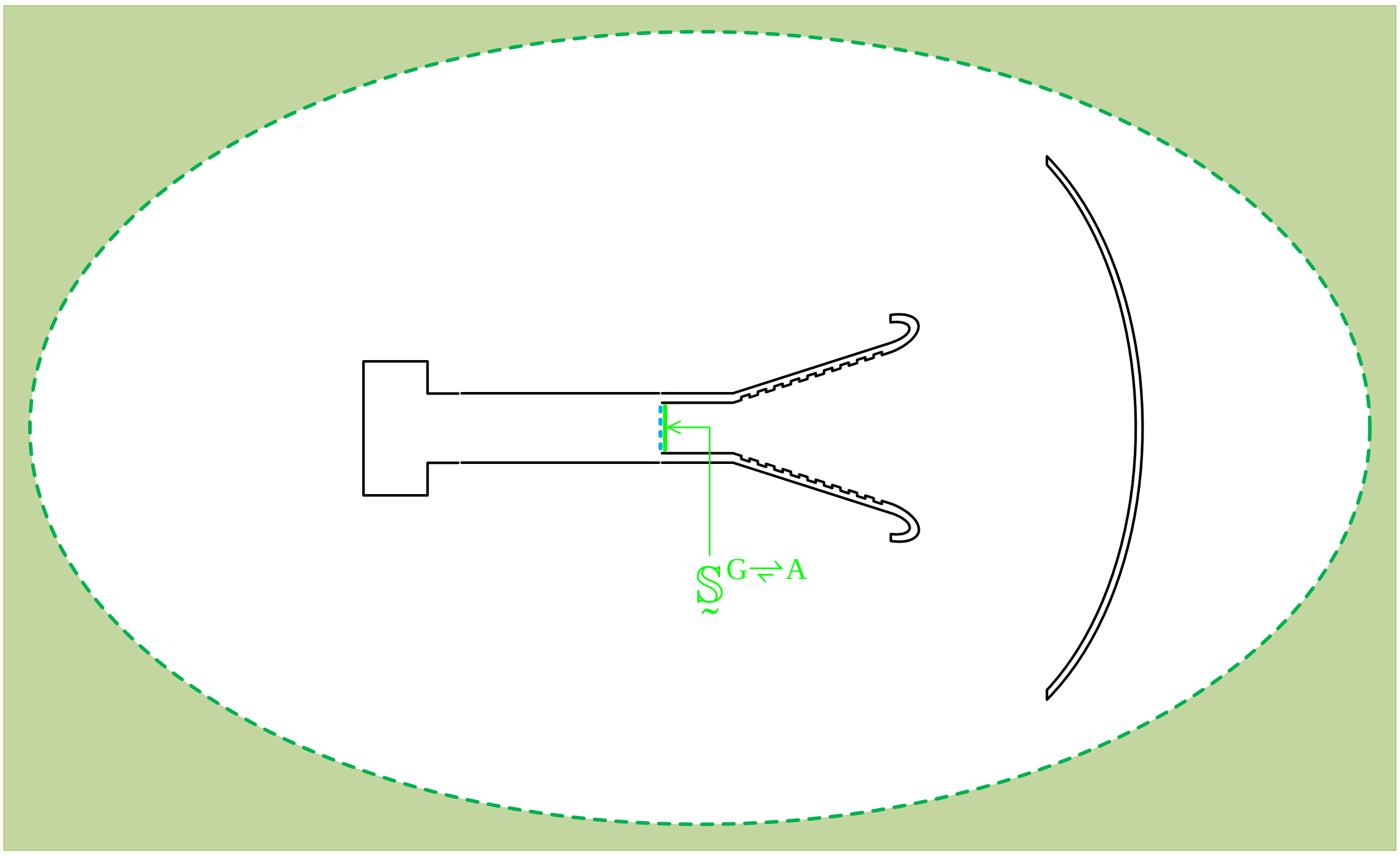

Figure 6-3 Integral surface $\underset{\sim}{\mathbb{S}}^{\mathrm{G}\rightleftharpoons\mathrm{A}}$ used in Eqs. (6-25) and (6-26)

Inserting Eq. (6-5) into Eqs. (6-25) and (6-26), the interaction forms are immediately discretized as follows:

$$P^{\mathrm{G}\rightleftharpoons\mathrm{A}} = \left(\bar{a}^{\mathrm{AV}}\right)^{\dagger} \cdot \overline{\overline{P}}_{\mathrm{intAV}}^{\mathrm{G}\rightleftharpoons\mathrm{A}} \cdot \bar{a}^{\mathrm{AV}} \tag{6-27}$$

in which

$$\overline{\overline{P}}_{\mathrm{intAV}}^{\mathrm{G}\rightleftharpoons\mathrm{A}} = \begin{cases} \begin{bmatrix} \overline{\overline{P}}^{\vec{J}^{\mathrm{G}\rightleftharpoons\mathrm{A}}\vec{J}^{\mathrm{G}\rightleftharpoons\mathrm{A}}} & \overline{\overline{P}}^{\vec{J}^{\mathrm{G}\rightleftharpoons\mathrm{A}}\vec{J}^{\mathrm{F}}} & \overline{\overline{P}}^{\vec{J}^{\mathrm{G}\rightleftharpoons\mathrm{A}}\vec{M}^{\mathrm{G}\rightleftharpoons\mathrm{A}}} \\ 0 & 0 & 0 \\ 0 & 0 & 0 \end{bmatrix} & \text{for Eq.(6-25)} \\ \begin{bmatrix} 0 & 0 & 0 \\ 0 & 0 & 0 \\ \overline{\overline{P}}^{\vec{M}^{\mathrm{G}\rightleftharpoons\mathrm{A}}\vec{J}^{\mathrm{G}\rightleftharpoons\mathrm{A}}} & \overline{\overline{P}}^{\vec{M}^{\mathrm{G}\rightleftharpoons\mathrm{A}}\vec{J}^{\mathrm{F}}} & \overline{\overline{P}}^{\vec{M}^{\mathrm{G}\rightleftharpoons\mathrm{A}}\vec{M}^{\mathrm{G}\rightleftharpoons\mathrm{A}}} \end{bmatrix}^{\dagger} & \text{for Eq.(6-26)} \end{cases} \tag{6-28}$$

where the elements of the sub-matrices can be calculated by using the following formulations

$$p_{\xi\zeta}^{\vec{J}^{\mathrm{G}\rightleftharpoons\mathrm{A}}\vec{J}^{\mathrm{G}\rightleftharpoons\mathrm{A}}} = -(1/2)\left\langle \vec{b}_{\xi}^{\vec{J}^{\mathrm{G}\rightleftharpoons\mathrm{A}}}, -j\omega\mu_0\mathcal{L}_0\left(\vec{b}_{\zeta}^{\vec{J}^{\mathrm{G}\rightleftharpoons\mathrm{A}}}\right)\right\rangle_{\mathbb{S}^{\mathrm{G}\rightleftharpoons\mathrm{A}}} \tag{6-29a}$$

$$p_{\xi\zeta}^{\vec{J}^{\mathrm{G}\rightleftharpoons\mathrm{A}}\vec{J}^{\mathrm{F}}} = -(1/2)\left\langle \vec{b}_{\xi}^{\vec{J}^{\mathrm{G}\rightleftharpoons\mathrm{A}}}, -j\omega\mu_0\mathcal{L}_0\left(\vec{b}_{\zeta}^{\vec{J}^{\mathrm{F}}}\right)\right\rangle_{\mathbb{S}^{\mathrm{G}\rightleftharpoons\mathrm{A}}} \tag{6-29b}$$

$$p_{\xi\zeta}^{\vec{J}^{\mathrm{G}\rightleftharpoons\mathrm{A}}\vec{M}^{\mathrm{G}\rightleftharpoons\mathrm{A}}} = -(1/2)\left\langle \vec{b}_{\xi}^{\vec{J}^{\mathrm{G}\rightleftharpoons\mathrm{A}}}, \hat{n}^{\rightarrow\mathrm{A}}\times\frac{1}{2}\vec{b}_{\zeta}^{\vec{M}^{\mathrm{G}\rightleftharpoons\mathrm{A}}} - \mathrm{P.V.}\,\mathcal{K}_0\left(\vec{b}_{\zeta}^{\vec{M}^{\mathrm{G}\rightleftharpoons\mathrm{A}}}\right)\right\rangle_{\mathbb{S}^{\mathrm{G}\rightleftharpoons\mathrm{A}}} \tag{6-29c}$$

and

$$p_{\xi\zeta}^{\vec{M}^{\mathrm{G}\rightleftharpoons\mathrm{A}}\vec{J}^{\mathrm{G}\rightleftharpoons\mathrm{A}}} = -(1/2)\left\langle \vec{b}_{\xi}^{\vec{M}^{\mathrm{G}\rightleftharpoons\mathrm{A}}}, \frac{1}{2}\vec{b}_{\zeta}^{\vec{J}^{\mathrm{G}\rightleftharpoons\mathrm{A}}}\times\hat{n}^{\rightarrow\mathrm{A}} + \mathrm{P.V.}\,\mathcal{K}_0\left(\vec{b}_{\zeta}^{\vec{J}^{\mathrm{G}\rightleftharpoons\mathrm{A}}}\right)\right\rangle_{\mathbb{S}^{\mathrm{G}\rightleftharpoons\mathrm{A}}} \tag{6-29d}$$

$$p_{\xi\zeta}^{\vec{M}^{\mathrm{G}\rightleftharpoons\mathrm{A}}\vec{J}^{\mathrm{F}}} = -(1/2)\left\langle \vec{b}_{\xi}^{\vec{M}^{\mathrm{G}\rightleftharpoons\mathrm{A}}}, \mathcal{K}_0\left(\vec{b}_{\zeta}^{\vec{J}^{\mathrm{F}}}\right)\right\rangle_{\mathbb{S}^{\mathrm{G}\rightleftharpoons\mathrm{A}}} \tag{6-29e}$$

$$p_{\xi\zeta}^{\vec{M}^{\mathrm{G}\rightleftharpoons\mathrm{A}}\vec{M}^{\mathrm{G}\rightleftharpoons\mathrm{A}}} = -(1/2)\left\langle \vec{b}_{\xi}^{\vec{M}^{\mathrm{G}\rightleftharpoons\mathrm{A}}}, -j\omega\varepsilon_0\mathcal{L}_0\left(\vec{b}_{\zeta}^{\vec{M}^{\mathrm{G}\rightleftharpoons\mathrm{A}}}\right)\right\rangle_{\mathbb{S}^{\mathrm{G}\rightleftharpoons\mathrm{A}}} \tag{6-29f}$$

To obtain the IPO defined on modal space, we substitute Eq. (6-19) into Eq. (6-27), and then we have that

$$P^{\mathrm{G}\rightleftharpoons\mathrm{A}} = \bar{a}^{\dagger}\cdot\underbrace{\left(\bar{\bar{T}}^{\dagger}\cdot\bar{\bar{P}}_{\mathrm{intAV}}^{\mathrm{G}\rightleftharpoons\mathrm{A}}\cdot\bar{\bar{T}}\right)}_{\bar{\bar{P}}_{\mathrm{int}}^{\mathrm{G}\rightleftharpoons\mathrm{A}}}\cdot\bar{a} \tag{6-30}$$

where subscript "int" is to emphasize that $\bar{\bar{P}}_{\mathrm{int}}^{\mathrm{G}\rightleftharpoons\mathrm{A}}$ originates from discretizing the interaction form of IPO.

For the convenience of the following discussions, the IPOs (6-24) and (6-30) are uniformly written as follows:

$$P^{\mathrm{G}\rightleftharpoons\mathrm{A}} = \bar{a}^{\dagger}\cdot\bar{\bar{P}}^{\mathrm{G}\rightleftharpoons\mathrm{A}}\cdot\bar{a} \tag{6-31}$$

where $\bar{\bar{P}}^{\mathrm{G}\rightleftharpoons\mathrm{A}} = \bar{\bar{P}}_{\mathrm{cur}}^{\mathrm{G}\rightleftharpoons\mathrm{A}} / \bar{\bar{P}}_{\mathrm{int}}^{\mathrm{G}\rightleftharpoons\mathrm{A}}$.

### 6.2.4 Input-Power-Decoupled Modes

This subsection focuses on constructing the *input-power-decoupled modes* (*IP-DMs*) in the modal space of the tra-antenna shown in Figs. 6-1 ~ 6-3, and discusses some related topics.

#### 6.2.4.1 Construction Method

The IP-DMs in the modal space can be derived from solving the following *modal decoupling equation* (or simply called *decoupling equation*)

$$\overline{\overline{P}}_{-}^{\mathrm{G} \rightleftharpoons \mathrm{A}} \cdot \overline{\alpha}_{\xi} \ = \ \theta_{\xi} \ \overline{\overline{P}}_{+}^{\mathrm{G} \rightleftharpoons \mathrm{A}} \cdot \overline{\alpha}_{\xi} \tag{6-32}$$

defined on modal space, where $\overline{\overline{P}}_{+}^{\mathrm{G} \rightleftharpoons \mathrm{A}}$ and $\overline{\overline{P}}_{-}^{\mathrm{G} \rightleftharpoons \mathrm{A}}$ are the *positive and negative Hermitian parts* obtained from the *Toeplitz's decomposition* for the IPO matrix $\overline{\overline{P}}^{\mathrm{G} \rightleftharpoons \mathrm{A}}$ given in Eq. (6-31).

If some derived modes $\{\overline{\alpha}_1, \overline{\alpha}_2, \cdots, \overline{\alpha}_d\}$ are *d*-order degenerate, then the following *Gram-Schmidt orthogonalization process*[46] is necessary.

$$\left.\begin{aligned} \overline{\alpha}_1 &= \overline{\alpha}_1{}' \\ \overline{\alpha}_2 - \chi_{12}\overline{\alpha}_1{}' &= \overline{\alpha}_2{}' \\ &\cdots \\ \overline{\alpha}_d - \ \cdots \ - \chi_{2d}\overline{\alpha}_2{}' - \chi_{1d}\overline{\alpha}_1{}' &= \overline{\alpha}_d{}' \end{aligned}\right\} \tag{6-33}$$

where the coefficients are calculated as follows:

$$\chi_{mn} \ = \ \frac{\left(\overline{\alpha}_m{}'\right)^{\dagger} \cdot \overline{\overline{P}}_{+}^{\mathrm{G} \rightleftharpoons \mathrm{A}} \cdot \overline{\alpha}_n}{\left(\overline{\alpha}_m{}'\right)^{\dagger} \cdot \overline{\overline{P}}_{+}^{\mathrm{G} \rightleftharpoons \mathrm{A}} \cdot \overline{\alpha}_m{}'} \tag{6-34}$$

The above-obtained new modes $\{\overline{\alpha}_1{}', \overline{\alpha}_2{}', \cdots, \overline{\alpha}_d{}'\}$ are input-power-decoupled with each other.

#### 6.2.4.2 Modal Decoupling Relation and Parseval's Identity

The modal vectors constructed in the above subsection satisfy the following *modal decoupling relation*

$$\begin{aligned} & \overline{\alpha}_{\xi}^{\dagger} \cdot \overline{\overline{P}}^{\mathrm{G} \rightleftharpoons \mathrm{A}} \cdot \overline{\alpha}_{\zeta} \\ &= \underbrace{\left[ \mathrm{Re}\left\{ P_{\xi}^{\mathrm{G} \rightleftharpoons \mathrm{A}} \right\} + j \, \mathrm{Im}\left\{ P_{\xi}^{\mathrm{G} \rightleftharpoons \mathrm{A}} \right\} \right]}_{P_{\xi}^{\mathrm{G} \rightleftharpoons \mathrm{A}}} \delta_{\xi\zeta} \xrightarrow{\text{Normalizing } \mathrm{Re}\{P_{\xi}^{\mathrm{G} \rightleftharpoons \mathrm{A}}\} \text{ to } 1} \underbrace{\left(1 + j \, \theta_{\xi}\right)}_{\text{Normalized } P_{\xi}^{\mathrm{G} \rightleftharpoons \mathrm{A}}} \ \delta_{\xi\zeta} \end{aligned} \tag{6-35}$$

where $P_{\xi}^{\mathrm{G} \rightleftharpoons \mathrm{A}}$ is the *modal input power* corresponding to the $\xi$-th IP-DM. The physical explanation why $\mathrm{Re}\{P_{\xi}^{\mathrm{G} \rightleftharpoons \mathrm{A}}\}$ is normalized to 1 has been given in Ref. [18]. In fact, the above matrix-vector multiplication decoupling relation can also be alternatively written as the following more physical form

$$(1/2) \iint_{\mathbb{S}^{\mathrm{G} \rightleftharpoons \mathrm{A}}} \left( \vec{E}_{\zeta} \times \vec{H}_{\xi}^{\dagger} \right) \cdot \hat{n}^{\rightarrow \mathrm{A}} dS \ = \ \left(1 + j \, \theta_{\xi}\right) \delta_{\xi\zeta} \tag{6-36}$$

and Eq. (6-36) has a very clear physical meaning — **the modes obtained above don't have net energy exchange in any integral period**.

By employing the above decoupling relation, we have the following famous *Parseval's identity*

$$\sum_{\xi}\left|c_{\xi}\right|^{2} = \left(1/T\right)\int_{t_0}^{t_0+T}\left[\iint_{\mathbb{S}^{\mathrm{G}\rightleftharpoons\mathrm{A}}}\left(\vec{\mathcal{E}}\times\vec{\mathcal{H}}\right)\cdot\hat{n}^{\rightarrow\mathrm{A}}dS\right]dt \tag{6-37}$$

where $c_{\xi}$ is the modal expansion coefficient used in *modal expansion formulation* and can be explicitly calculated as follows:

$$c_{\xi} = \frac{-\left(1/2\right)\left\langle \vec{J}_{\xi}^{\mathrm{G}\rightleftharpoons\mathrm{A}},\vec{E}\right\rangle_{\mathbb{S}^{\mathrm{G}\rightleftharpoons\mathrm{A}}}}{1+j\,\theta_{\xi}} = \frac{-\left(1/2\right)\left\langle \vec{H},\vec{M}_{\xi}^{\mathrm{G}\rightleftharpoons\mathrm{A}}\right\rangle_{\mathbb{S}^{\mathrm{G}\rightleftharpoons\mathrm{A}}}}{1+j\,\theta_{\xi}} \tag{6-38}$$

where $\{\vec{E},\vec{H}\}$ is a previously known field distributing on input port $\mathbb{S}^{\mathrm{G}\rightleftharpoons\mathrm{A}}$.

### 6.2.4.3 Modal Quantities

For quantitatively describing the modal features, the following modal quantities are usually used

$$\mathrm{MS}_{\xi} = \frac{1}{\left|1+j\,\theta_{\xi}\right|} \tag{6-39}$$

called *modal significance* (*MS*), and

$$Z_{\xi}^{\mathrm{G}\rightleftharpoons\mathrm{A}} = \frac{P_{\xi}^{\mathrm{G}\rightleftharpoons\mathrm{A}}}{\left(1/2\right)\left\langle \vec{J}_{\xi}^{\mathrm{G}\rightleftharpoons\mathrm{A}},\vec{J}_{\xi}^{\mathrm{G}\rightleftharpoons\mathrm{A}}\right\rangle_{\mathbb{S}^{\mathrm{G}\rightleftharpoons\mathrm{A}}}} = \overbrace{\mathrm{Re}\left\{Z_{\xi}^{\mathrm{G}\rightleftharpoons\mathrm{A}}\right\}}^{R_{\xi}^{\mathrm{G}\rightleftharpoons\mathrm{A}}} + j\,\overbrace{\mathrm{Im}\left\{Z_{\xi}^{\mathrm{G}\rightleftharpoons\mathrm{A}}\right\}}^{X_{\xi}^{\mathrm{G}\rightleftharpoons\mathrm{A}}} \tag{6-40a}$$

called *modal input impedance* (*MII*), and

$$Y_{\xi}^{\mathrm{G}\rightleftharpoons\mathrm{A}} = \frac{P_{\xi}^{\mathrm{G}\rightleftharpoons\mathrm{A}}}{\left(1/2\right)\left\langle \vec{M}_{\xi}^{\mathrm{G}\rightleftharpoons\mathrm{A}},\vec{M}_{\xi}^{\mathrm{G}\rightleftharpoons\mathrm{A}}\right\rangle_{\mathbb{S}^{\mathrm{G}\rightleftharpoons\mathrm{A}}}} = \underbrace{\mathrm{Re}\left\{Y_{\xi}^{\mathrm{G}\rightleftharpoons\mathrm{A}}\right\}}_{G_{\xi}^{\mathrm{G}\rightleftharpoons\mathrm{A}}} + j\,\underbrace{\mathrm{Im}\left\{Y_{\xi}^{\mathrm{G}\rightleftharpoons\mathrm{A}}\right\}}_{B_{\xi}^{\mathrm{G}\rightleftharpoons\mathrm{A}}} \tag{6-40b}$$

called *modal input admittance* (*MIA*).

The above MS quantitatively depicts the *modal weight in whole modal expansion formulation*. The above MII and MIA quantitatively depict the *allocation way for the energy carried by the mode*.

## 6.2.5 Numerical Examples Corresponding to Typical Structures

In this subsection, we calculate the IP-DMs of three typical metallic tra-antennas — *metallic horn antenna* (Sce. 6.2.5.1), *metallic reflector antenna* (Sce. 6.2.5.2) and *metallic horn fed metallic reflector antenna* (Sce. 6.2.5.3) — by using the above formulations.

### 6.2.5.1 Metallic Horn Antenna

Here, we consider a metallic horn tra-antenna, and show its geometrical size in the following Fig. 6-4.

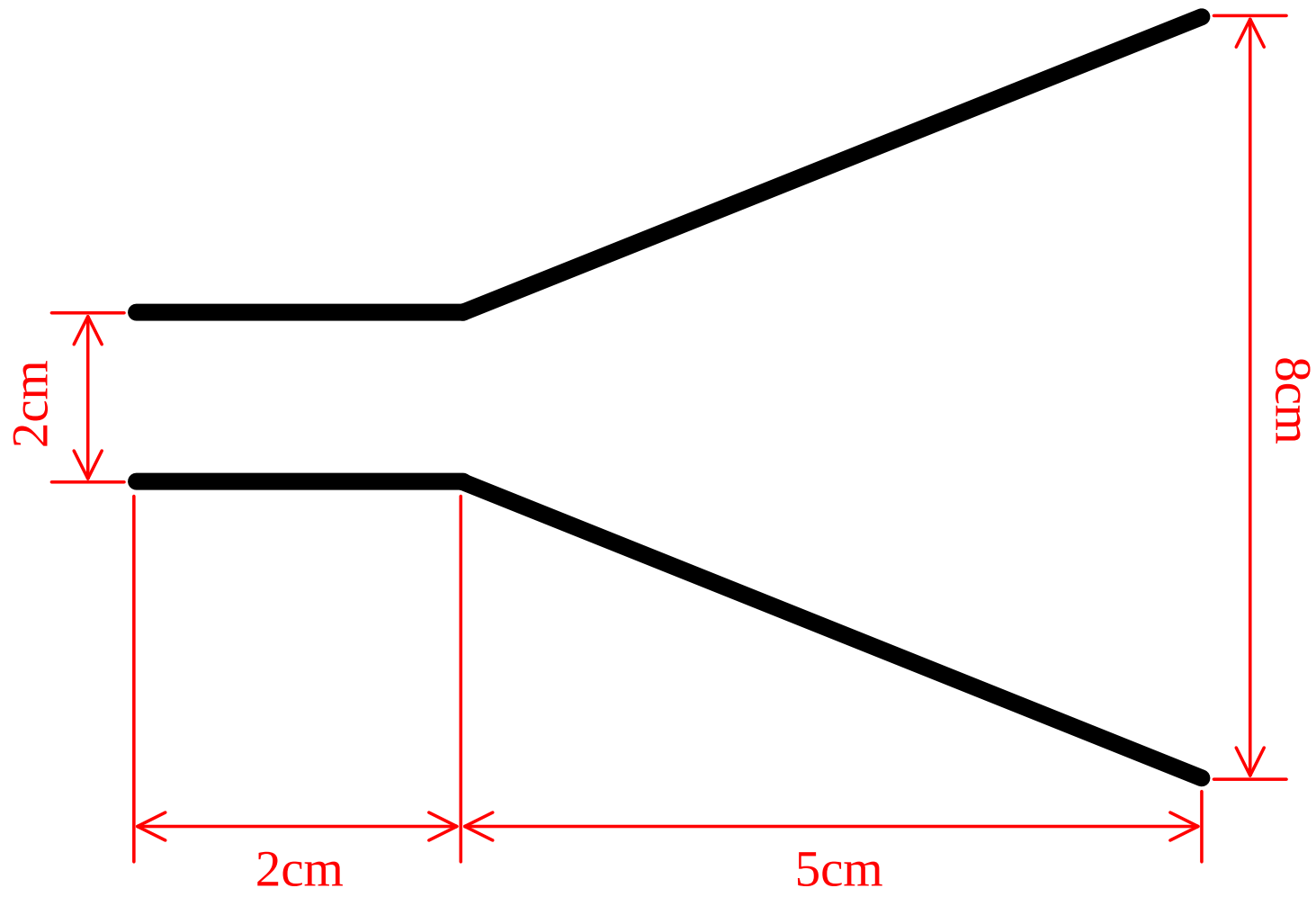

Figure 6-4 Geometrical size of a metallic horn tra-antenna

The topological structures and surface triangular meshes of the tra-antenna are shown in the following Fig. 6-5.

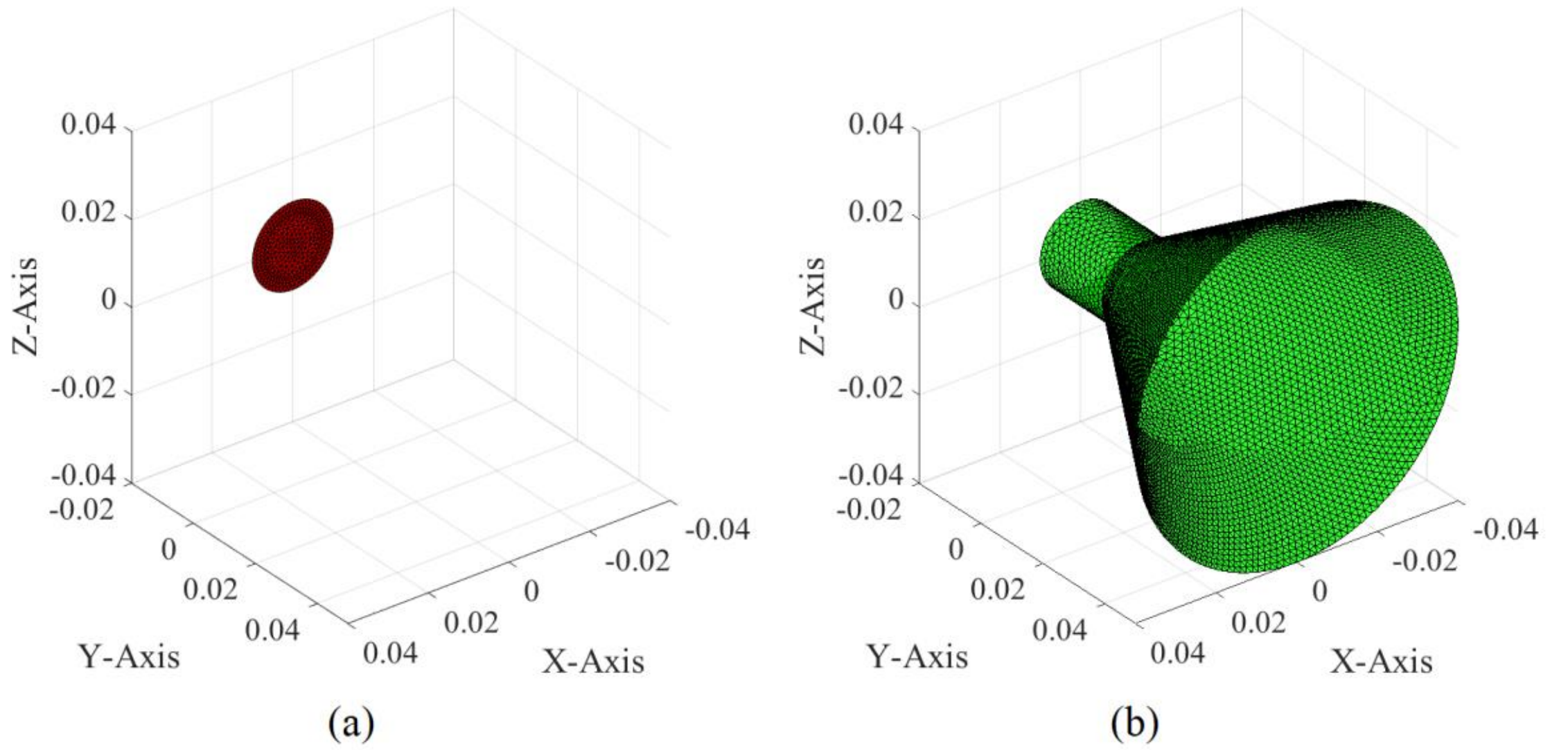

Figure 6-5 Topological structures and surface triangular meshes of (a) input port and (b) metallic horn

By orthogonalizing the JE interaction form of IPO with DoJ-based DVE scheme, we obtain the IP-DMs of the tra-antenna. The modal input resistances and MSs of the first several typical modes are shown in the following Fig. 6-6.

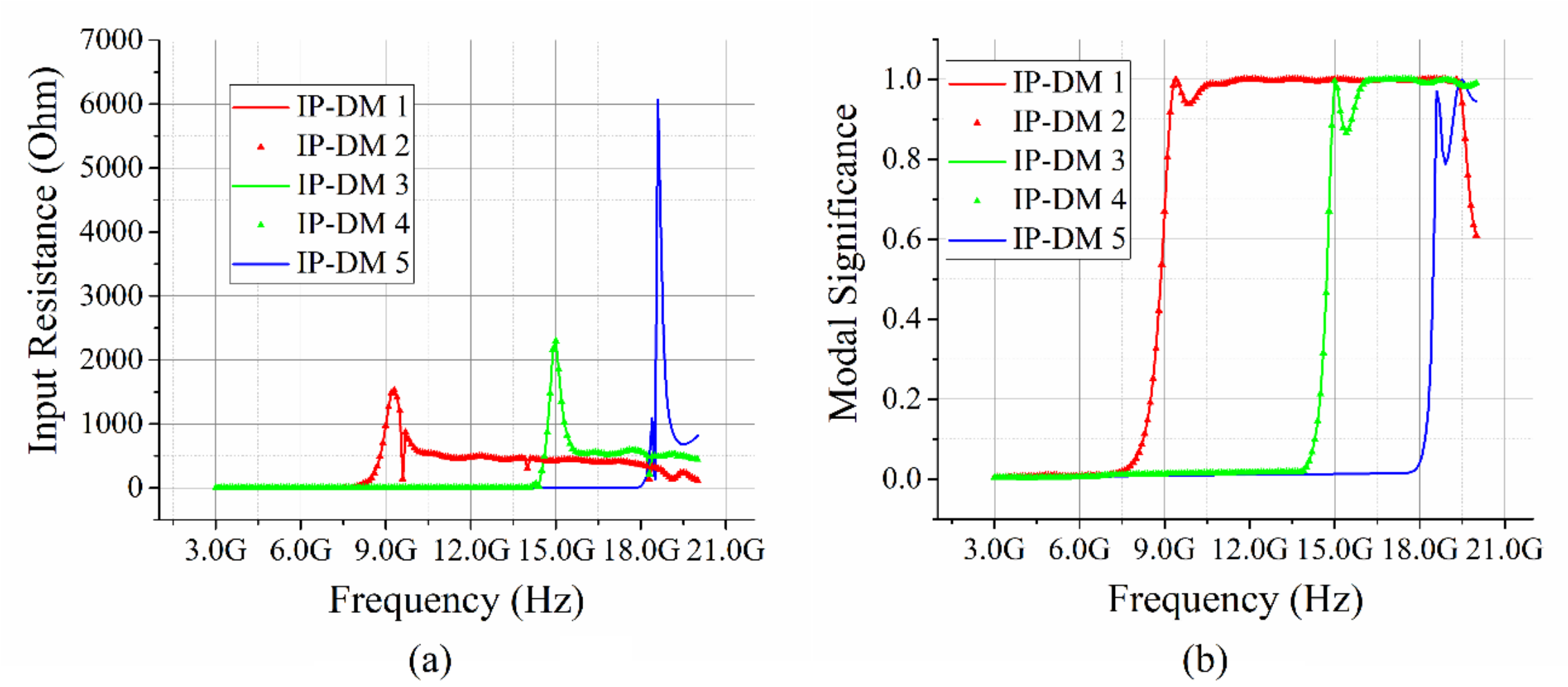

Figure 6-6 (a) Modal input resistances and (b) MSs of the first several typical IP-DMs

The port equivalent currents of IP-DM 1 and IP-DM 2 are shown as follows:

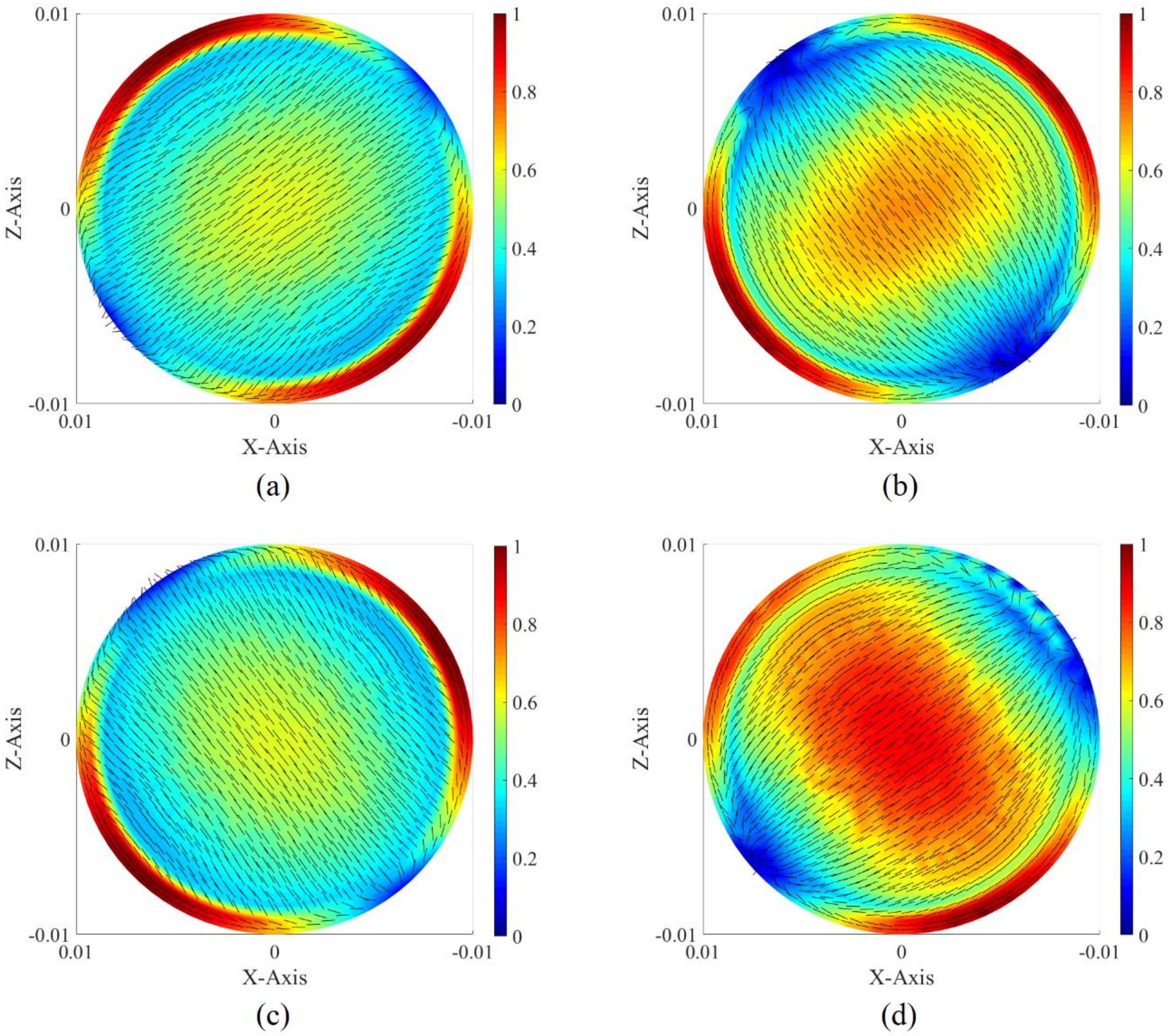

Figure 6-7 Equivalent currents on input port. (a) $\vec{J}^{G\rightleftharpoons A}$ of IP-DM 1; (b) $\vec{M}^{G\rightleftharpoons A}$ of IP-DM 1; (c) $\vec{J}^{G\rightleftharpoons A}$ of IP-DM 2; (d) $\vec{M}^{G\rightleftharpoons A}$ of IP-DM 2

Obviously, the IP-DM 1 and IP-DM 2 are a pair of *degenerate states*. For the first degenerate state (working at 9.3 GHz), its $\vec{J}^{\mathrm{F}}$ distribution, radiation pattern, electric field distribution, and magnetic field distribution are shown in the following Fig. 6-8.

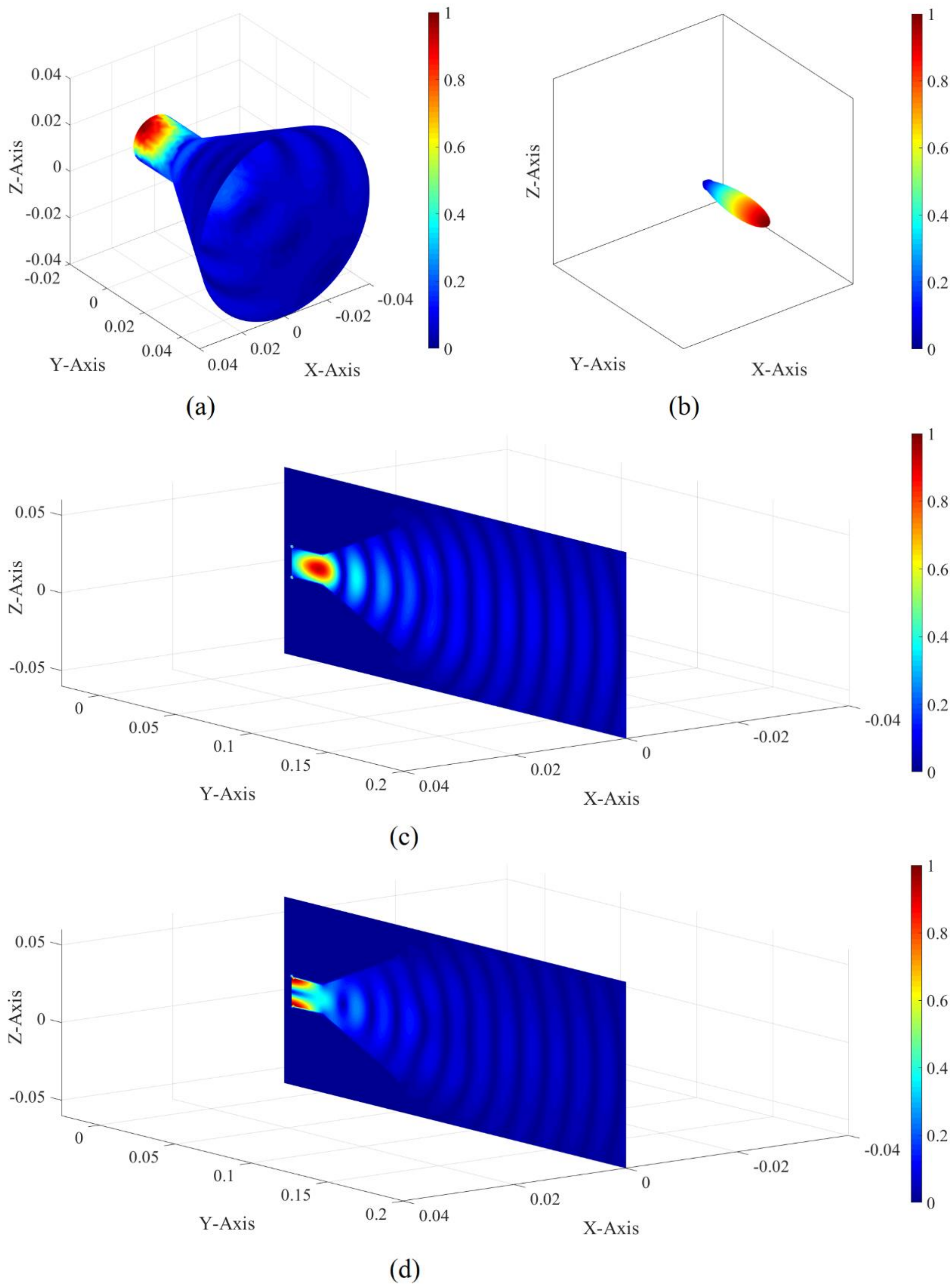

Figure 6-8 (a) Wall electric current distribution, (b) radiation pattern, (c) electric field distribution, and (d) magnetic field distribution of IP-DM 1 (at 9.3 GHz)

Because the IP-DM 1 and IP-DM 2 are spatially degenerate as illustrated by their modal port equivalent currents shown in Fig. 6-7, then the modal wall electric current distribution, modal radiation pattern, modal electric field distribution, and modal magnetic field distribution corresponding to the second degenerate state are completely similar to the figures shown in the above Fig. 6-8(a) ~ Fig. 6-8(d) (except a spatial rotation around Y-axis). Thus, we don't explicitly provide the distributions here for shortening the length of this report.

The port equivalent electric and magnetic currents of IP-DM 3 and IP-DM 4 are shown in the following Fig. 6-9.

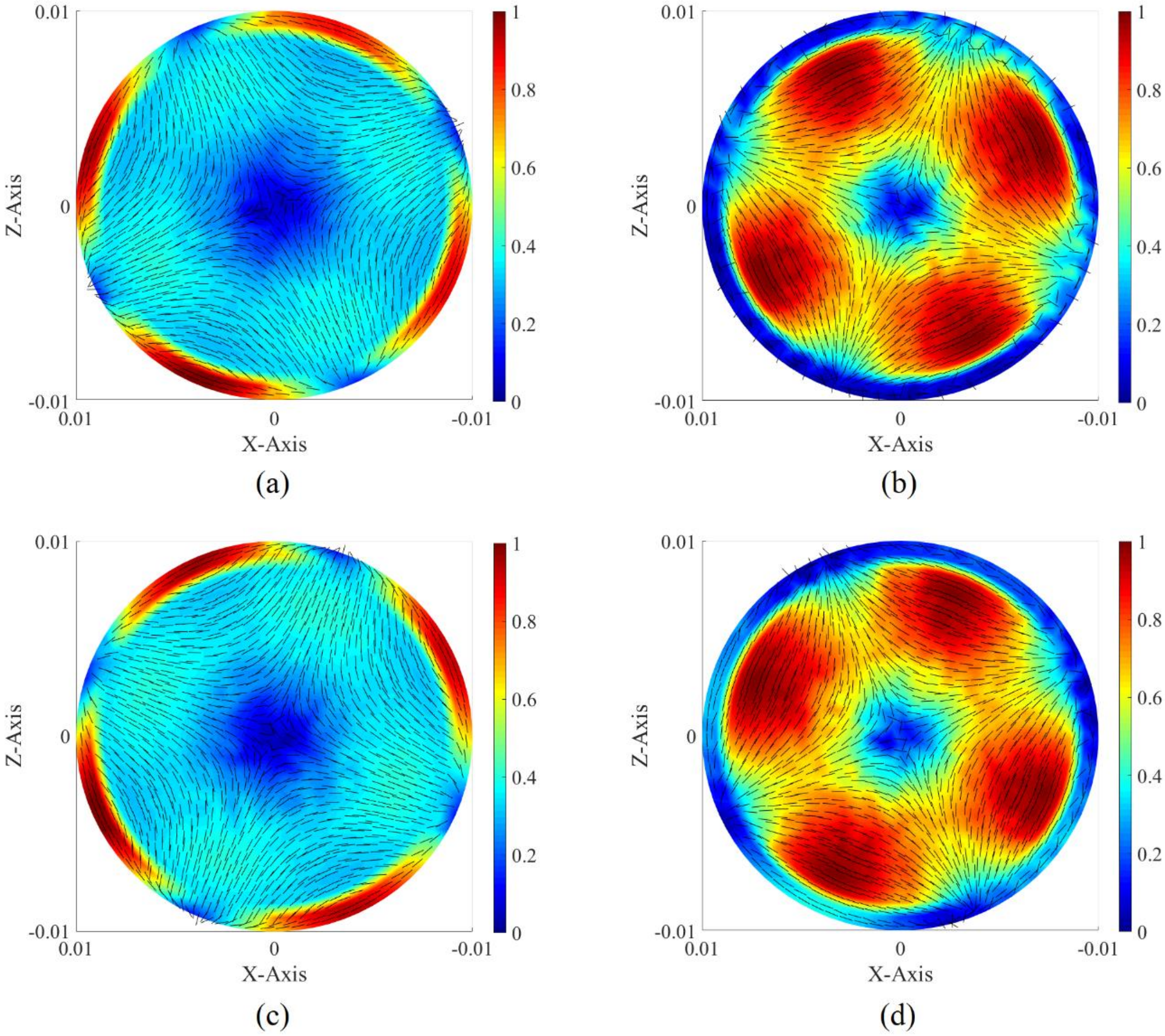

Figure 6-9 Equivalent currents on input port. (a) $\vec{J}^{\mathrm{G}\rightleftharpoons\mathrm{A}}$ of IP-DM 3; (b) $\vec{M}^{\mathrm{G}\rightleftharpoons\mathrm{A}}$ of IP-DM 3; (c) $\vec{J}^{\mathrm{G}\rightleftharpoons\mathrm{A}}$ of IP-DM 4; (d) $\vec{M}^{\mathrm{G}\rightleftharpoons\mathrm{A}}$ of IP-DM 4

Obviously, the IP-DM 3 and IP-DM 4 are also a pair of degenerate states. For the first degenerate state (working at 14.9 GHz), its wall electric current distribution, radiation pattern, electric field distribution, and magnetic field distribution are shown in the following Fig. 6-10.

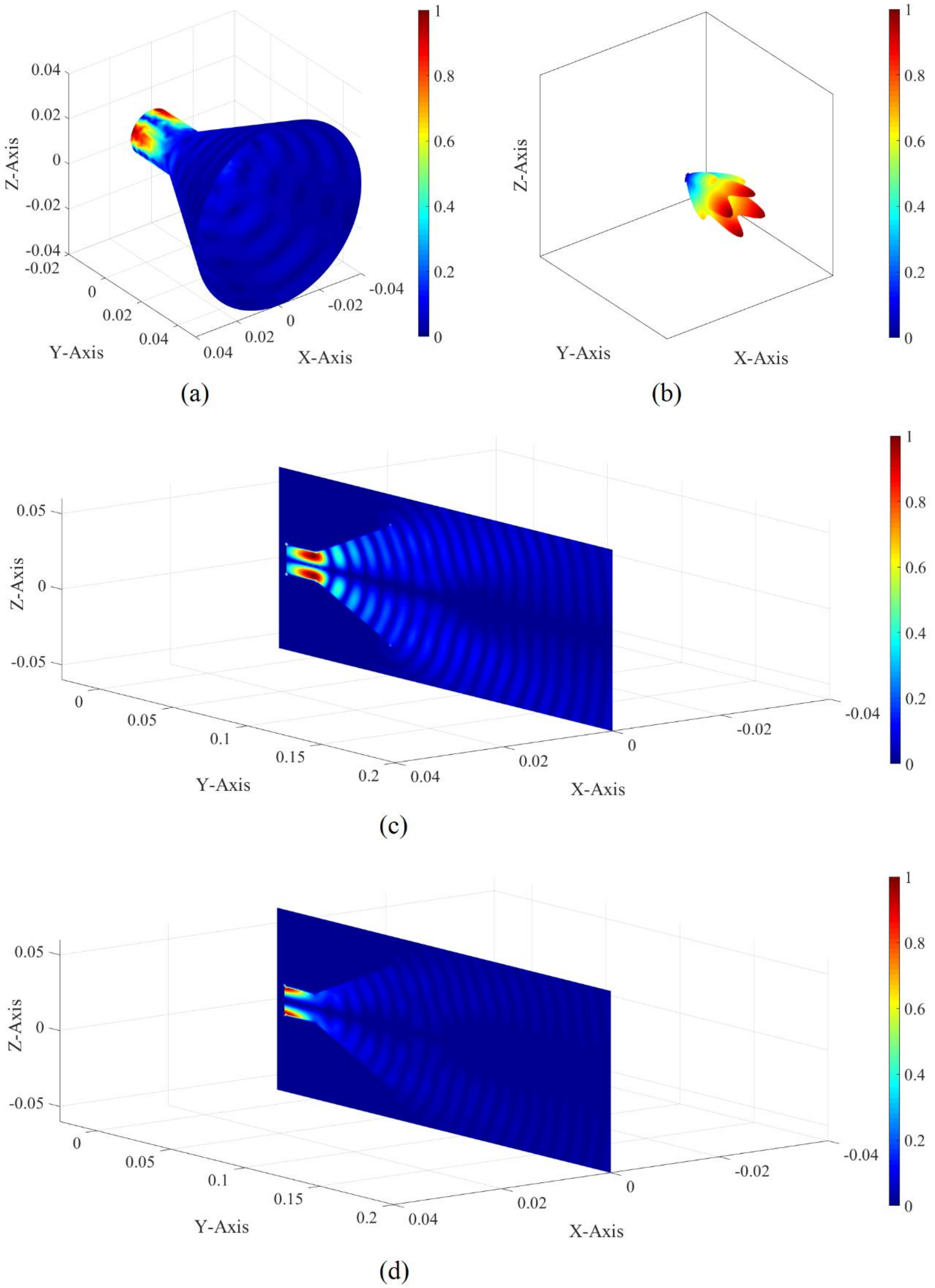

Figure 6-10 (a) Wall electric current distribution, (b) radiation pattern, (c) electric field distribution, and (d) magnetic field distribution of IP-DM 3 (at 14.9 GHz)

From the modal resistance curves and MS curves shown in the previous Fig. 6-6, it is not difficult to find out that: there doesn't exist any degenerate state for IP-DM 5. The port equivalent electric and magnetic currents of IP-DM 5 are shown in the following Fig. 6-11.

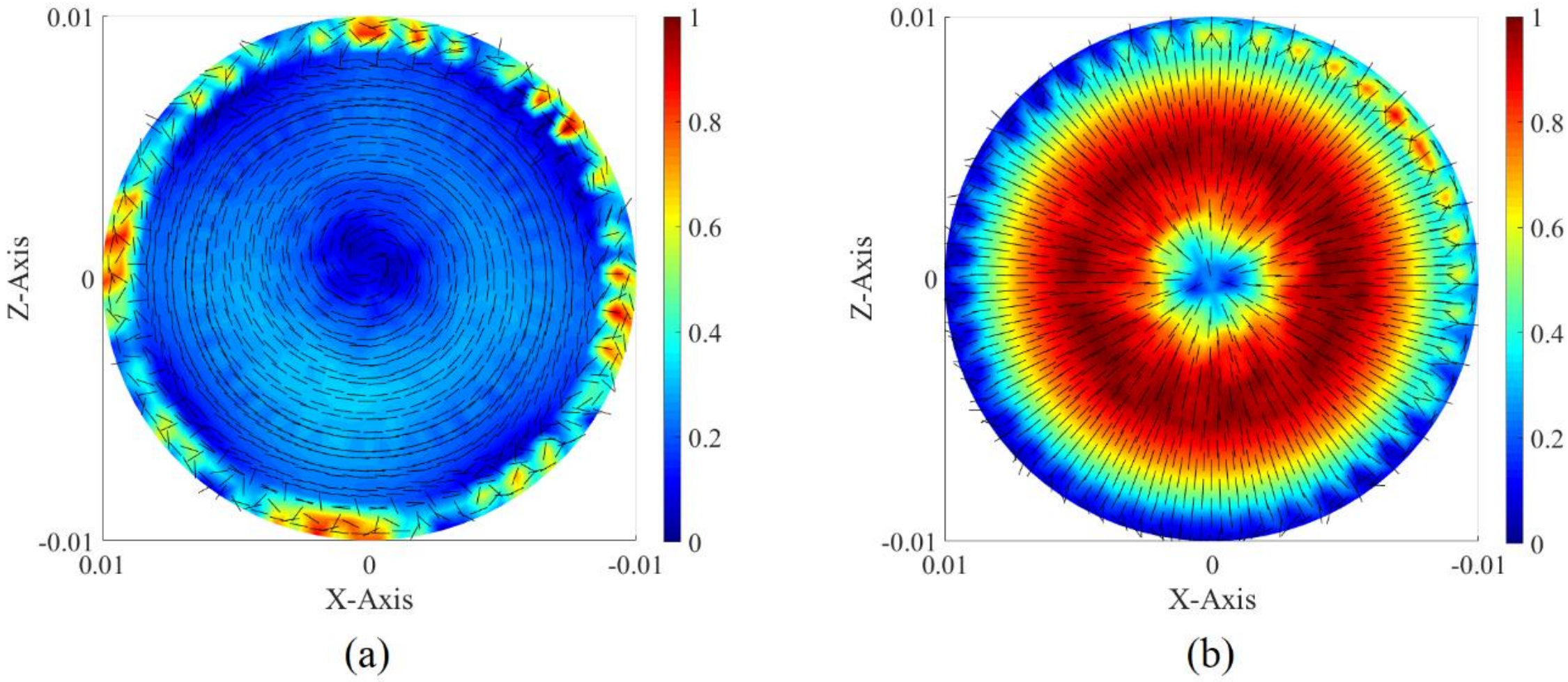

Figure 6-11 Equivalent currents on input port. (a) $\vec{J}^{\text{G}\rightleftharpoons\text{A}}$ of IP-DM 5; (b) $\vec{M}^{\text{G}\rightleftharpoons\text{A}}$ of IP-DM 5

Obviously, the modal port currents are rotationally symmetrical about Y-axis, and this is just the reason why this IP-DM 5 doesn't have any other degenerate state. For the IP-DM 5 (working at 18.6 GHz), its modal electric current distributing on the metallic wall of the horn tra-antenna, its modal radiation pattern at infinity, its modal electric field distributing on yOz plane, and its modal magnetic field distributing on yOz plane are shown in the following Fig. 6-12.

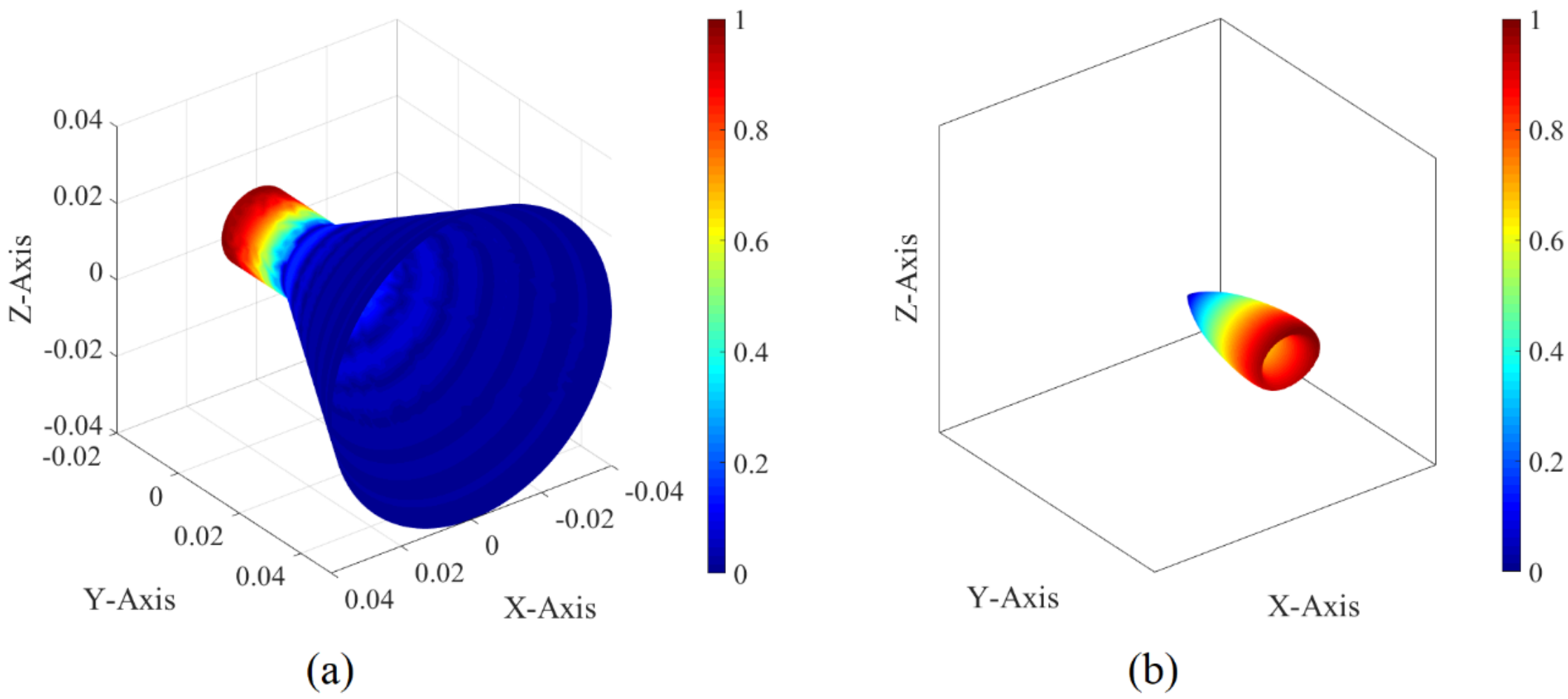

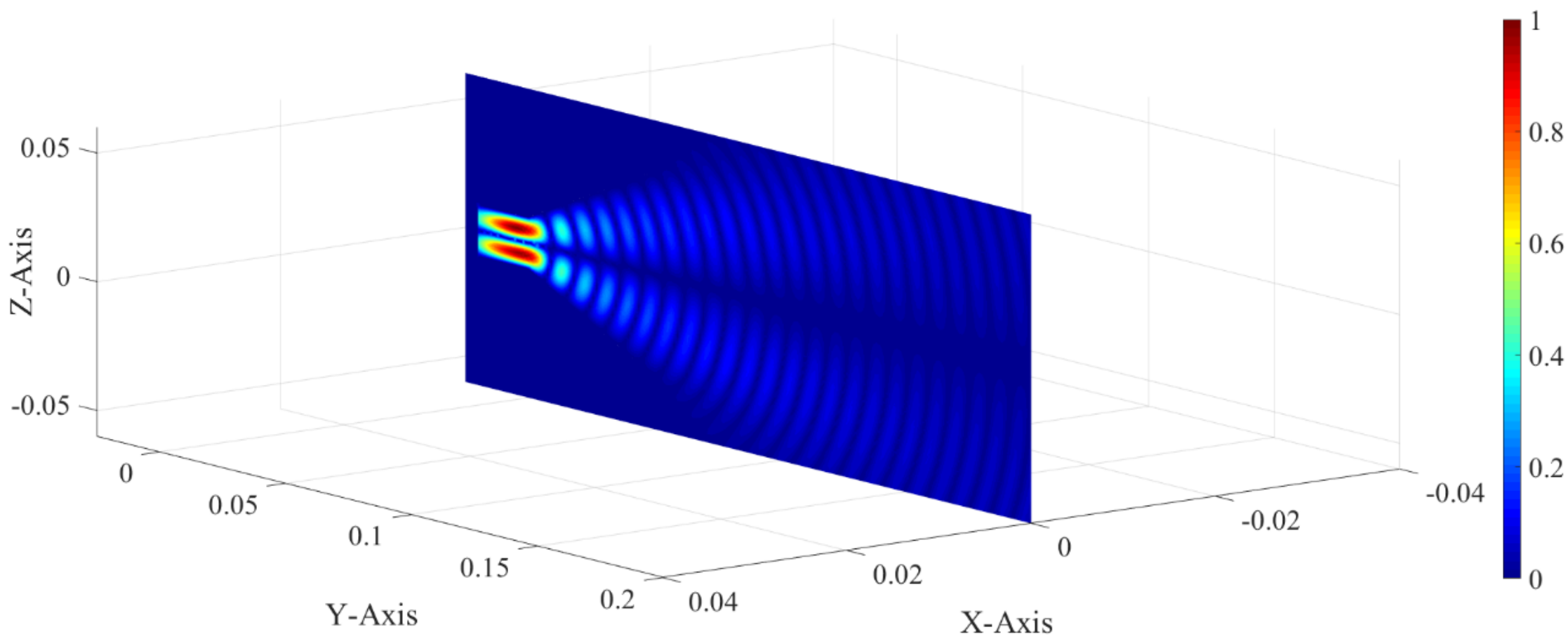

(c)

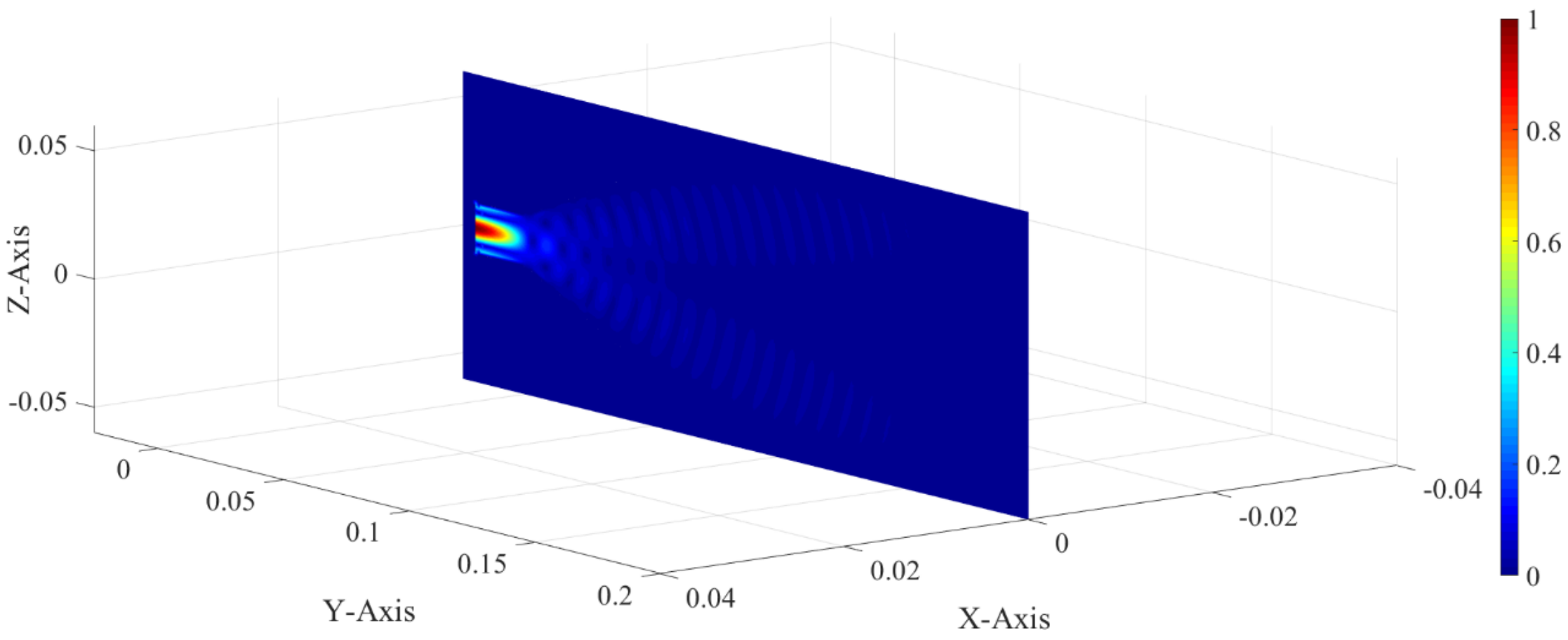

(d)

Figure 6-12 (a) Wall electric current distribution, (b) radiation pattern, (c) electric field distribution, and (d) magnetic field distribution of IP-DM 5 (at 18.6 GHz)

### 6.2.5.2 Metallic Reflector Antenna

Here, we consider a *metallic paraboloidal reflector tra-antenna* shown in Fig. 6-13

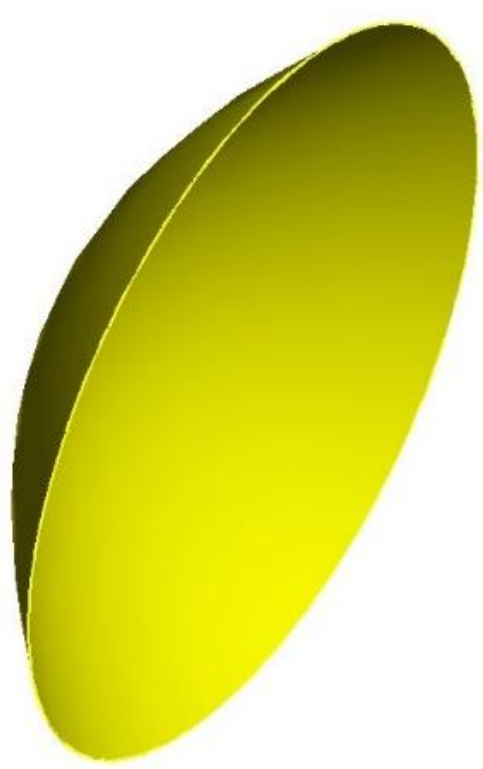

Figure 6-13 Geometry of a paraboloidal reflector with radius 10cm and focal depth 5cm

Its radius and focal depth are 10cm and 5cm respectively. The tra-antenna is fed by a circular input port locating at the focal point, and the radius of the port is 1 cm. The topological structures and surface triangular meshes of the port and reflector are shown in the following Fig. 6-14.

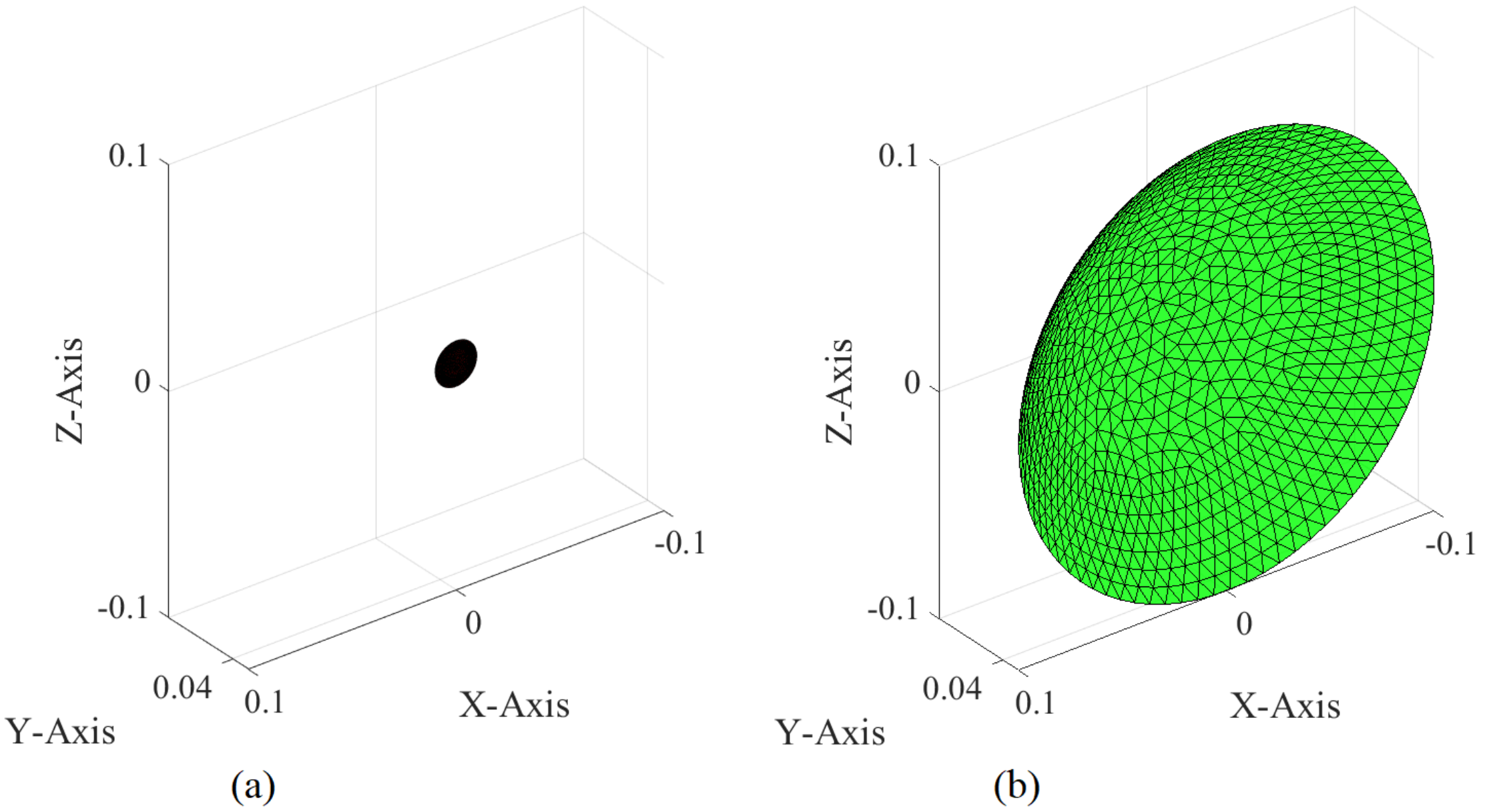

Figure 6-14 Topological structures and surface triangular meshes of (a) input port and (b) metallic reflector

By orthogonalizing the JE-DoJ-based formulation of IPO, we obtain the IP-DMs of the tra-antenna. The modal input resistances of the first several typical modes are shown in the following Fig. 6-15.

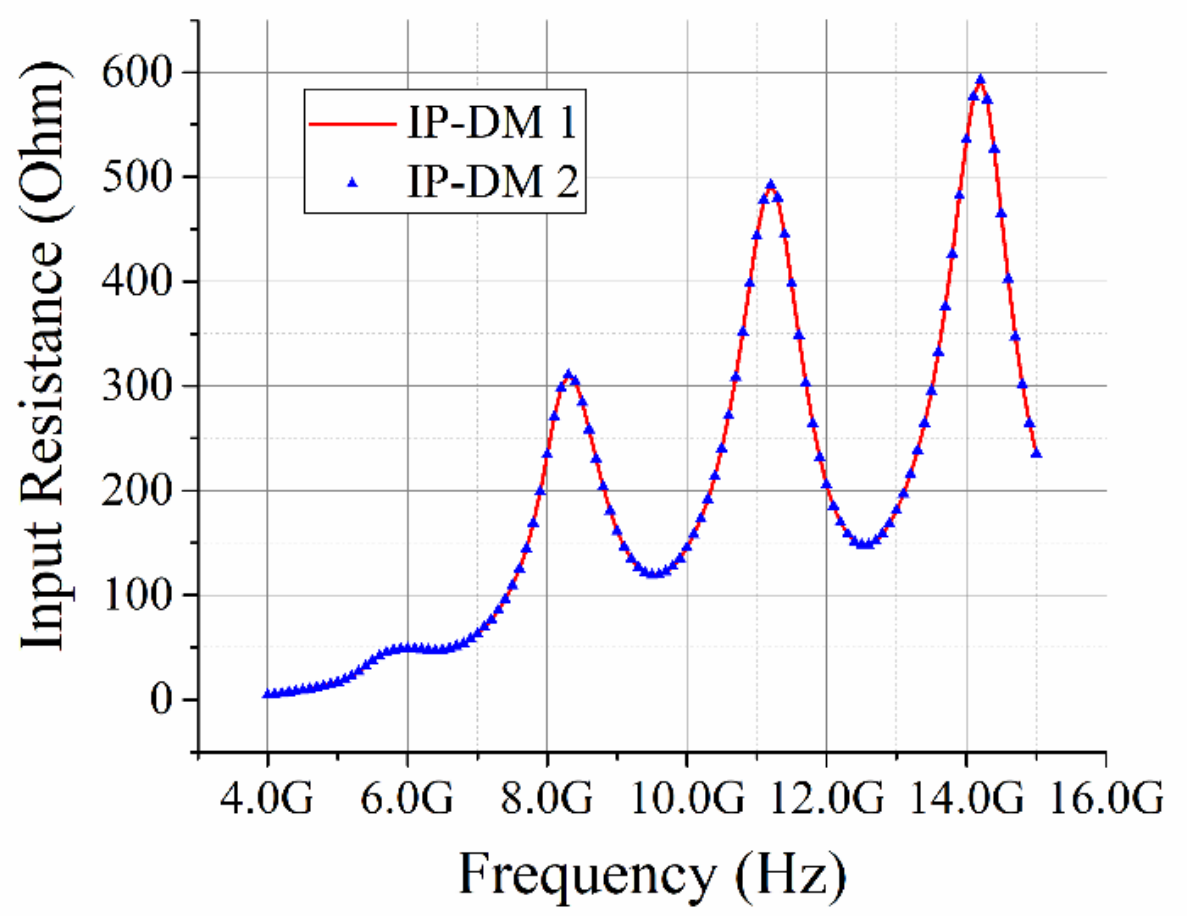

Figure 6-15 Modal input resistances of the first several typical IP-DMs

The port equivalent electric and magnetic currents of IP-DM 1 and IP-DM 2 are shown in the following Fig. 6-16.

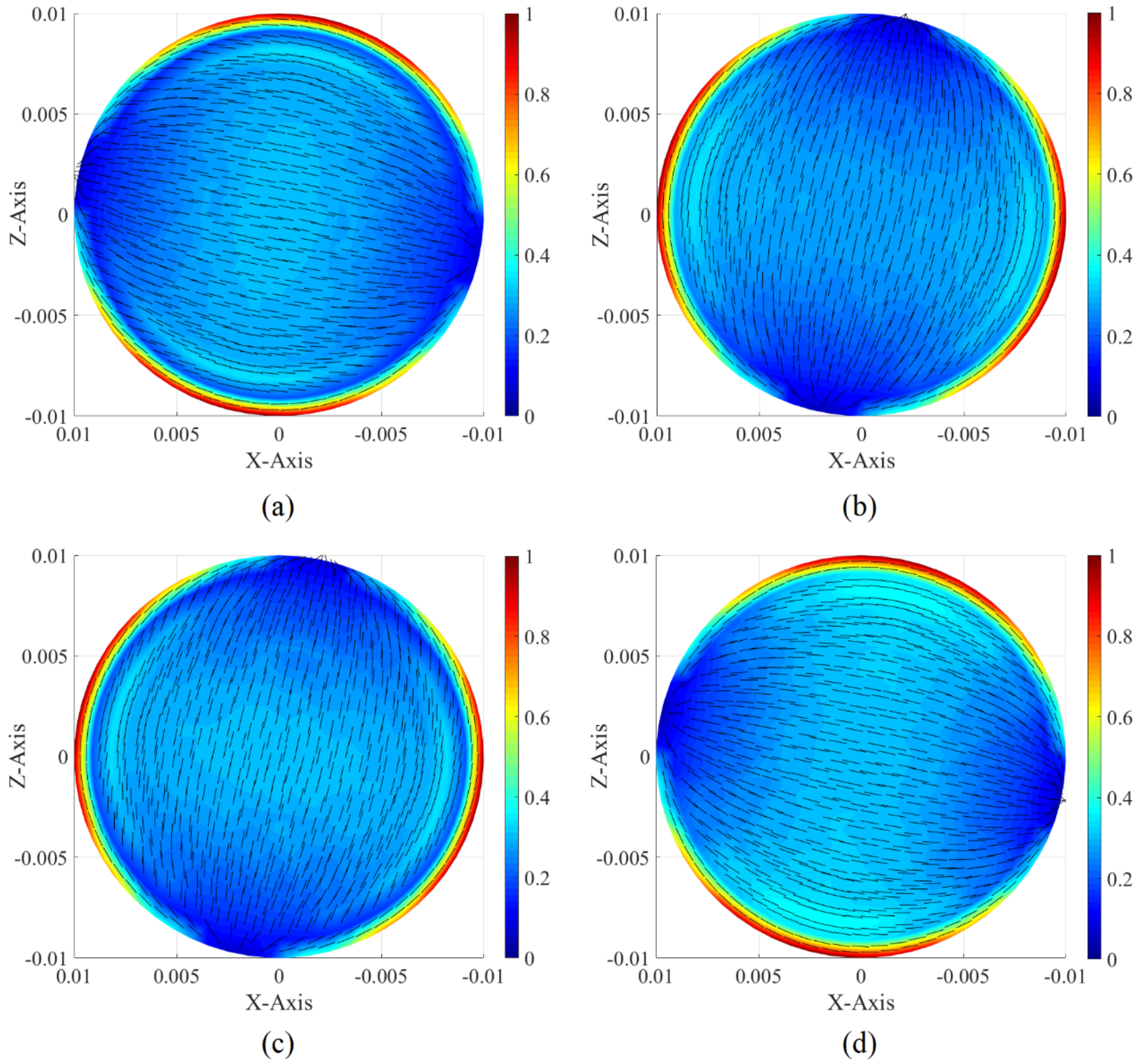

Figure 6-16 Equivalent currents on input port. (a) $\vec{J}^{G\rightleftharpoons A}$ of IP-DM 1; (b) $\vec{M}^{G\rightleftharpoons A}$ of IP-DM 1; (c) $\vec{J}^{G\rightleftharpoons A}$ of IP-DM 2; (d) $\vec{M}^{G\rightleftharpoons A}$ of IP-DM 2

Obviously, the IP-DM 1 and IP-DM 2 are a pair of degenerate states. For the first degenerate state (working at 8.3 GHz), its modal $\vec{J}^{F}$ distributing on the metallic reflector, modal radiation pattern at infinity, and modal electric field distributing on yOz plane are shown in the following Fig. 6-17. Because the IP-DM 1 and IP-DM 2 are spatially degenerate as illustrated by their modal port equivalent currents shown in the above Fig. 6-16, then the modal $\vec{J}^{F}$ distribution, modal radiation pattern, and modal electric field distribution corresponding to the second degenerate state are completely similar to the figures shown in Fig. 6-17(a) ~ Fig. 6-17(c) (except a spatial rotation around Y-axis). Thus, we don't explicitly provide the distributions here for shortening the length of this report.

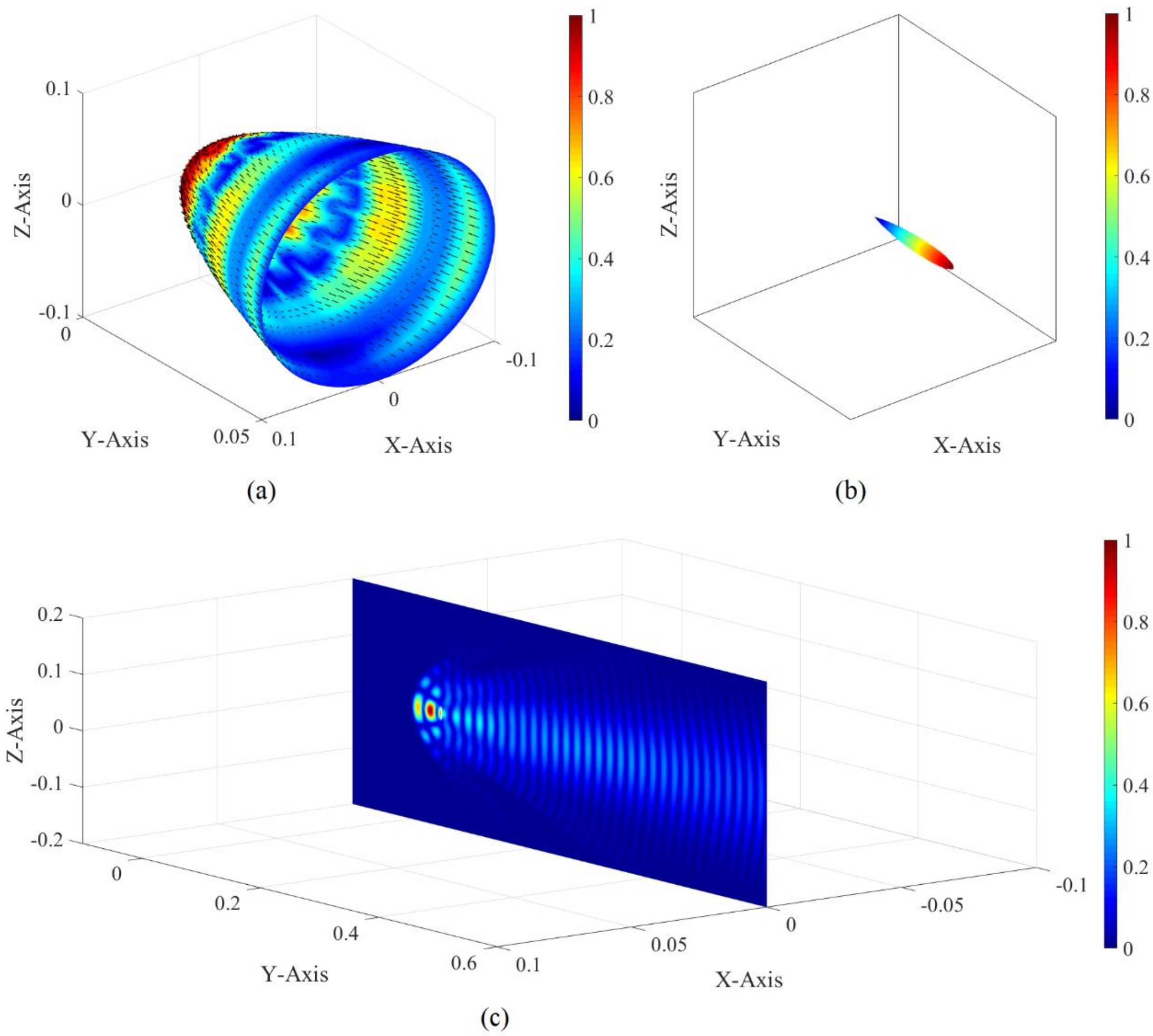

Figure 6-17 (a) Wall electric current distribution, (b) radiation pattern, and (c) electric field distribution of IP-DM 1 (at 8.3 GHz)

For the IP-DM 1 working at 11.2 GHz, its far-field radiation patter is similar to Fig. 6-17(b), and its electric field distribution is shown in the following Fig. 6-18.

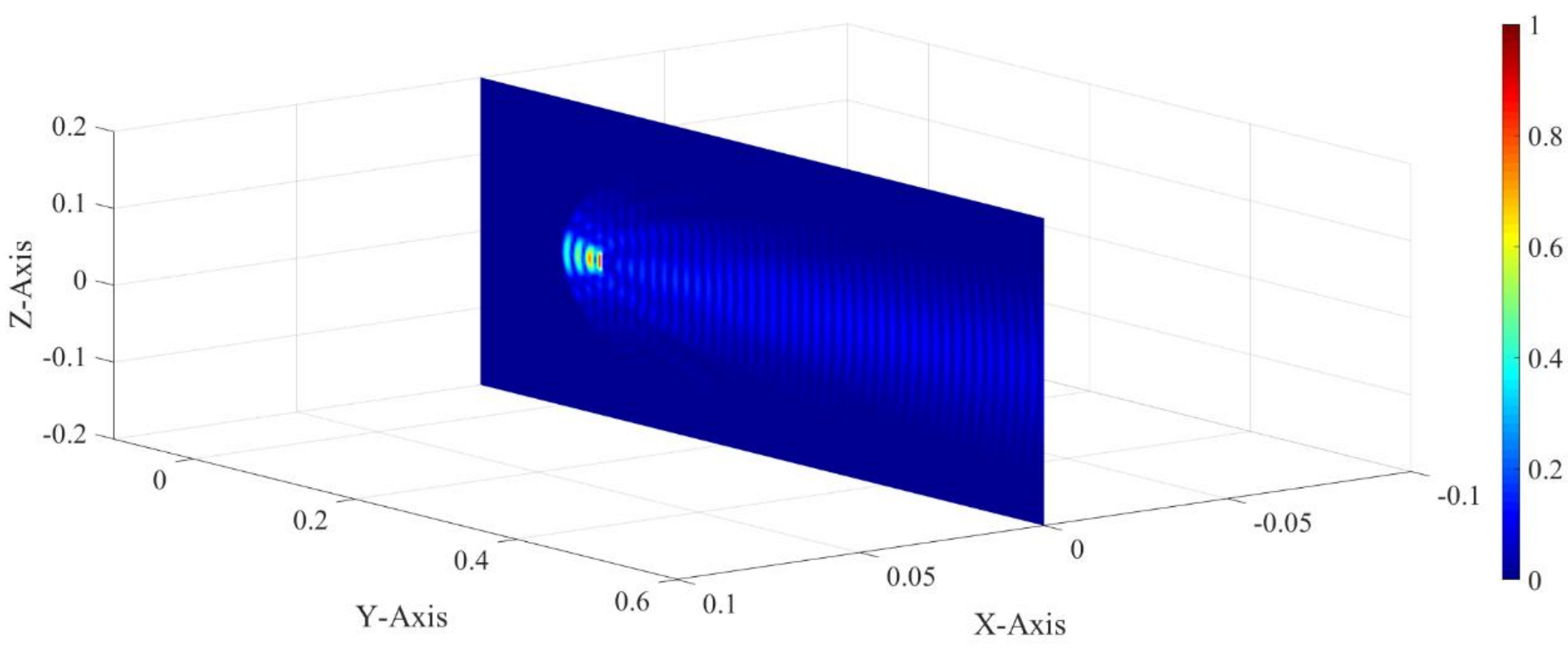

Figure 6-18 Electric field distribution of IP-DM 1 (at 11.2 GHz)

For the IP-DM 1 working at 14.2 GHz, its far-field radiation patter is also similar to the one shown in the previous Fig. 6-17(b), and its electric field distribution is shown in the following Fig. 6-19.

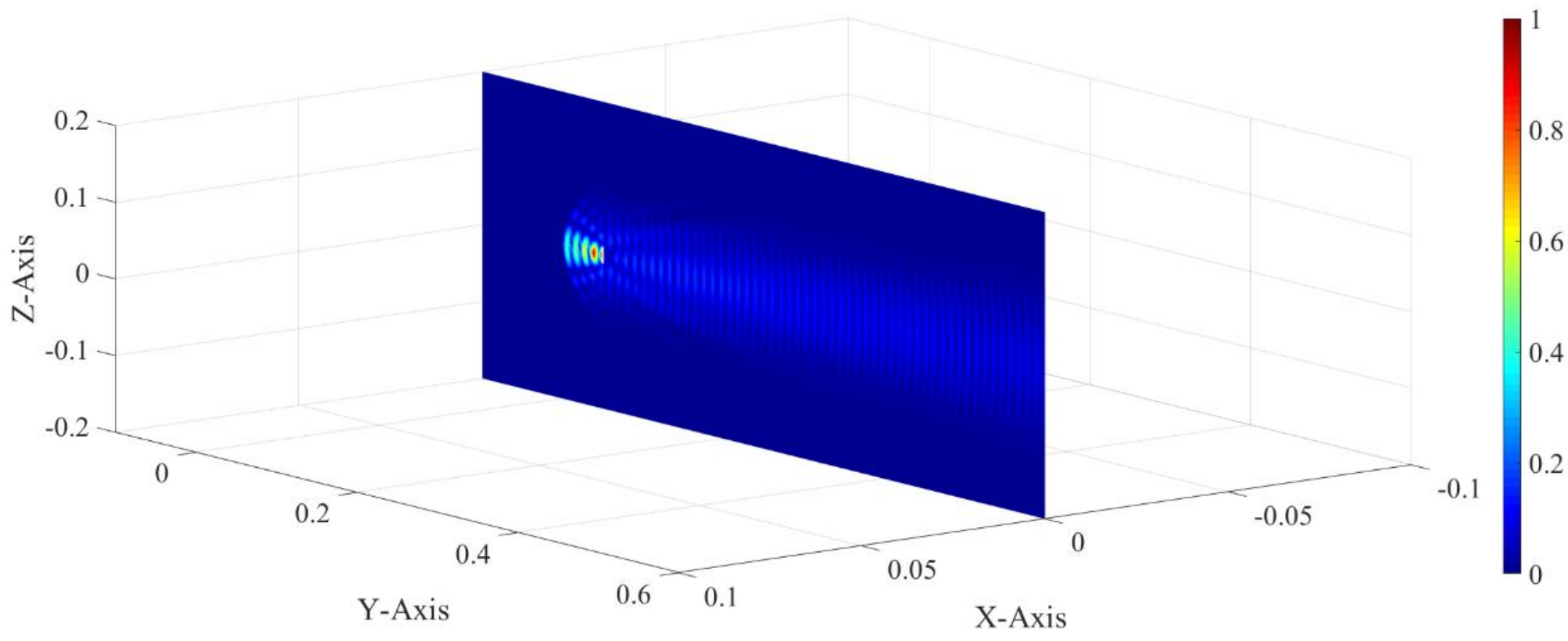

Figure 6-19 Electric field distribution of IP-DM 1 (at 14.2 GHz)

### 6.2.5.3 Metallic Horn Fed Metallic Reflector Antenna

Now we consider a *metallic horn fed metallic parabolic reflector tra-antenna* as shown in the following Fig. 6-20.

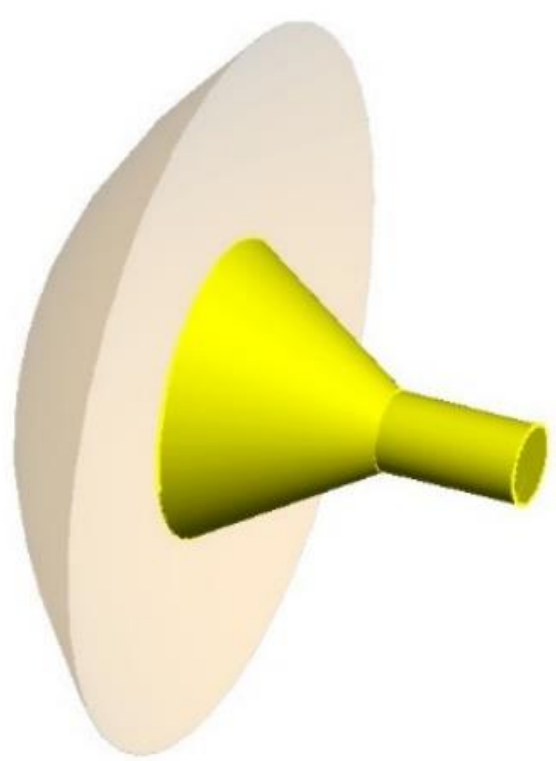

Figure 6-20 Geometry of a metallic horn fed metallic parabolic reflector tra-antenna

The metallic horn has the same size as the one considered in Sec. 6.2.5.1, except its thick electric wall. The metallic reflector is completely the same as the one considered in the previous Sec. 6.2.5.2. The metallic horn is placed at the focal point of the metallic reflector.

The topological structures and surface triangular meshes of the tra-antenna are shown in the following Fig. 6-21.

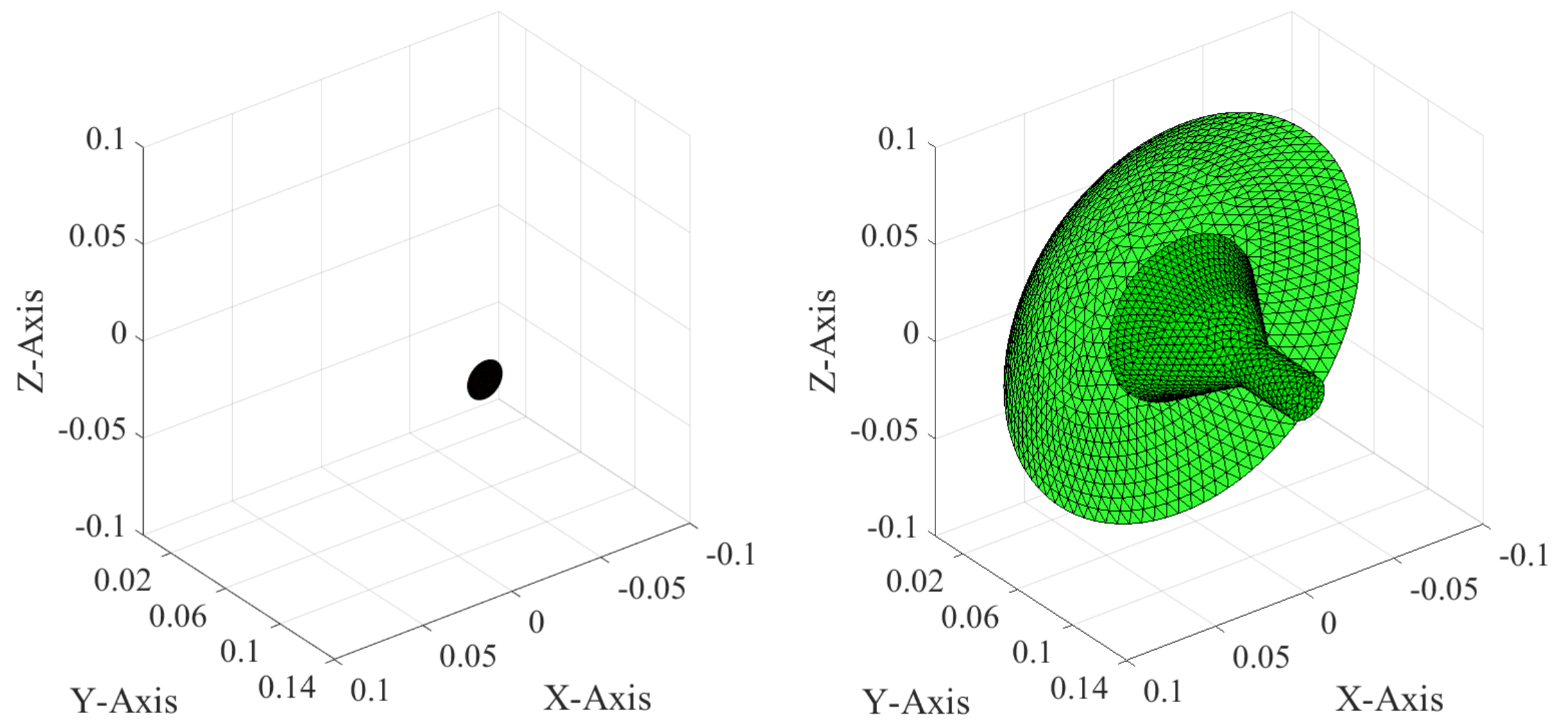

(a) (b)

Figure 6-21 Topological structures and surface triangular meshes of (a) input port and (b) tra-antenna

By orthogonalizing the JE-DoJ-based and HM-DoM-based formulations of IPO given in the previous Sec. 6.2.3, we obtain the IP-DMs of the tra-antenna. The modal input resistance curves of the first several typical modes are shown in the following Fig. 6-22.

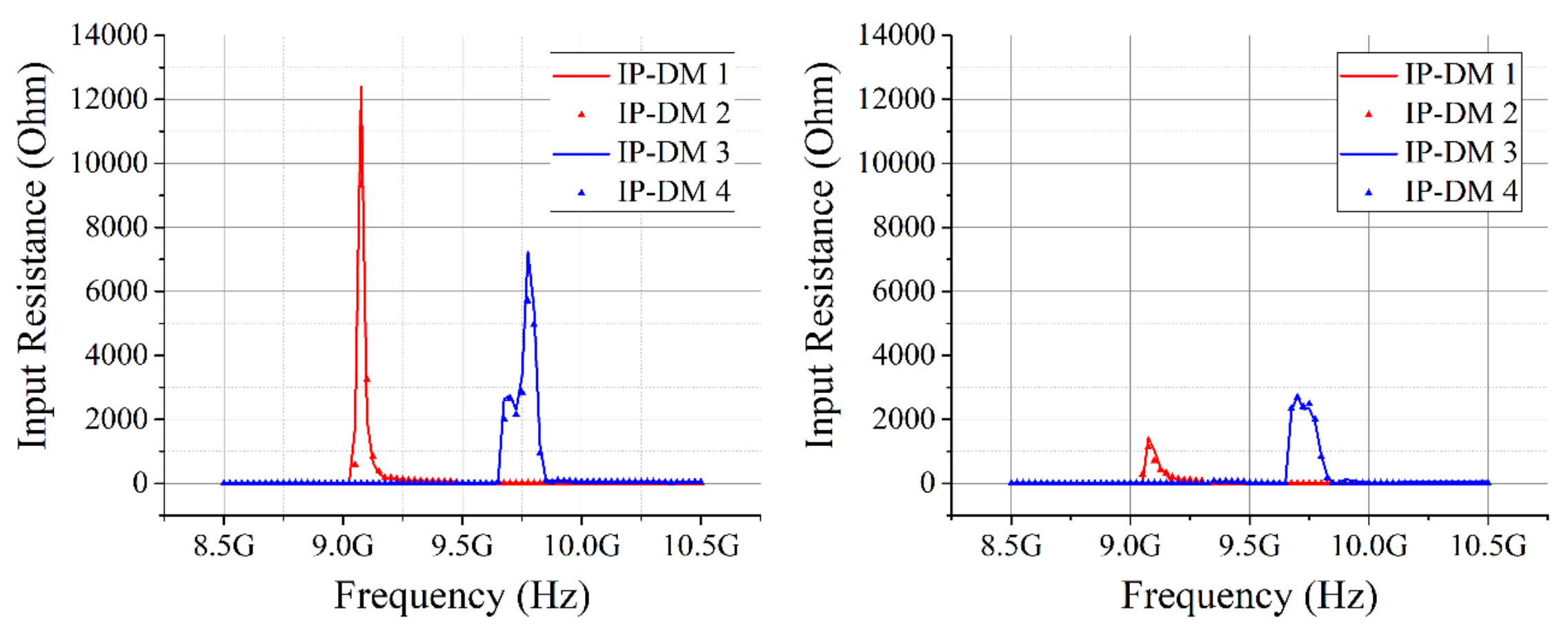

(a) (b)

Figure 6-22 Modal resistance curves of the first several typical IP-DMs. (a) JE-DoJ-based results; (b) HM-DoM-based results

Taking the JE-DoJ-based IP-DM 1 and IP-DM 2 working at 9.1 GHz as typical examples, we plot their equivalent electric current and equivalent magnetic current distributing on input port in the following Fig. 6-23. The port current distributions of the HM-DoM-based IP-DM 1 and IP-DM 2 working at 9.1 GHz are similar to the ones illustrated in the figure.

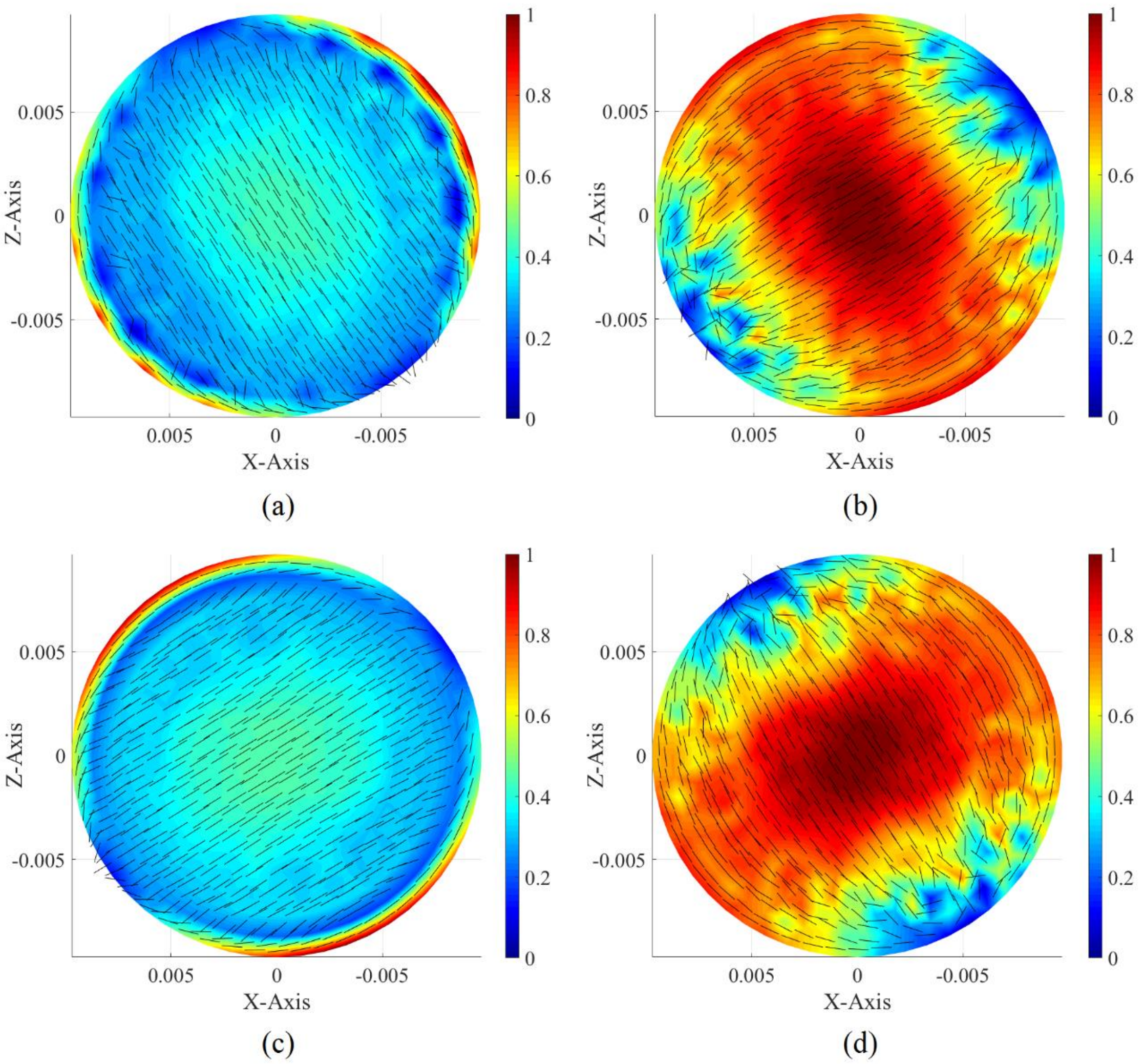

Figure 6-23 Equivalent currents on input port. (a) $\vec{J}^{\mathrm{G}\rightleftharpoons\mathrm{A}}$ of IP-DM 1; (b) $\vec{M}^{\mathrm{G}\rightleftharpoons\mathrm{A}}$ of IP-DM 1; (c) $\vec{J}^{\mathrm{G}\rightleftharpoons\mathrm{A}}$ of IP-DM 2; (d) $\vec{M}^{\mathrm{G}\rightleftharpoons\mathrm{A}}$ of IP-DM 2 (at 9.1 GHz)

By comparing the above Figs. 6-23(a)&(b) and 6-23(c)&(d), it is not difficult to find out that the IP-DM 1 and IP-DM 2 are a pair of degenerate modes, and the degeneracy phenomenon originates from the fact that the tra-anetnna shown in Fig. 6-20 satisfies a spatial rotation symmetry about Y-axis.

Taking the above IP-DM 1 working at 9.1 GHz as an example, we also plot its modal induced electric current distributing on the metallic horn and reflector, modal far-field radiation pattern, and modal electric field distribution, in the following Fig. 6-24.

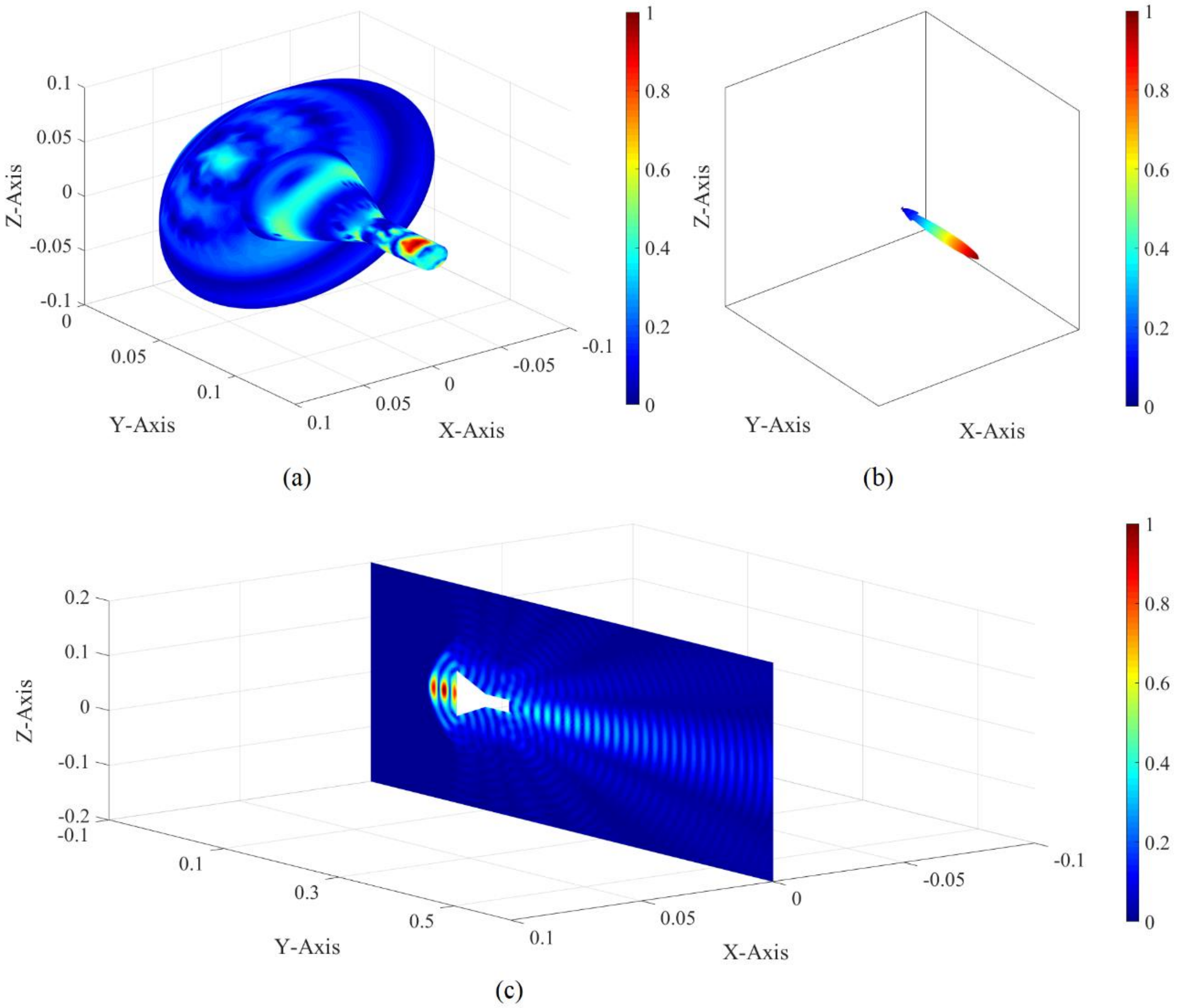

Figure 6-24 (a) Wall electric current distribution, (b) radiation pattern, and (c) electric field distribution of IP-DM 1 (at 9.1 GHz)

The above numerical examples illustrate the effectiveness of the theory and formulations established in this section. In the subsequent sections, we will further generalize the theory and formulations to some other more complicated tra-antennas.

## 6.3 IP-DMs of Augmented Material Transmitting Antenna

*Material tra-antennas*, such as *material rod antenna*[110~119] and *material horn antenna*[120~128] etc., are usually treated as a kind of *travelling-wave antennas*. They have many important engineering applications due to their attractive features, such as broad bandwidth, high gain, high polarization purity, low sidelobe level, low loss, low radar-cross section, and ease of fabrication, etc.

Taking the material horn shown in the following Fig. 6-25 as a typical example, this section, under PTT framework, discusses the processes to construct IP-DMs.

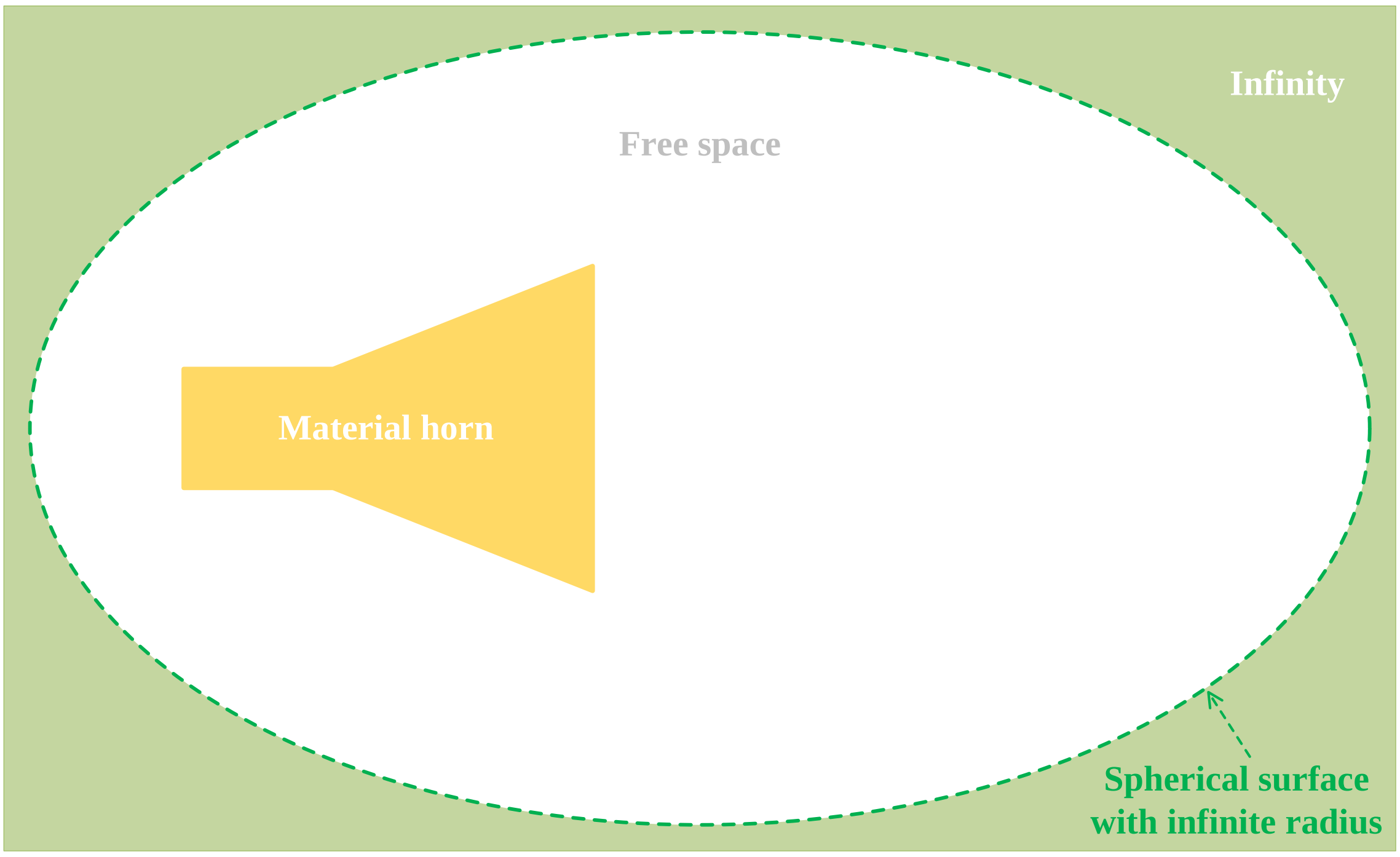

Figure 6-25 Geometry of a typical material horn antenna

The principle and process for constructing the IP-DMs of the other kinds of material tra-antennas are similar.

The structure of this section is organized being completely similar to the previous Sec. 6.2, and the organization is visually summarized in the following Fig. 6-26.

Mathematically Depicting Topological Structure

↓

Deriving Source- Field Relationships (SFRs)

↓

Mathematically Describing Modal Space by Employing the SFRs

↓

Deriving the Input Power Operator (IPO) Defined on the Modal Space

↓

Constructing Input- Power- Decoupled Modes (IP- DMs) by Orthogonalizing the IPO

Figure 6-26 Organization for this section

### 6.3.1 Topological Structure and Source-Field Relationships

The topological structure of the tra-antenna shown in the previous Fig. 6-25 is illustrated in the following Fig. 6-27

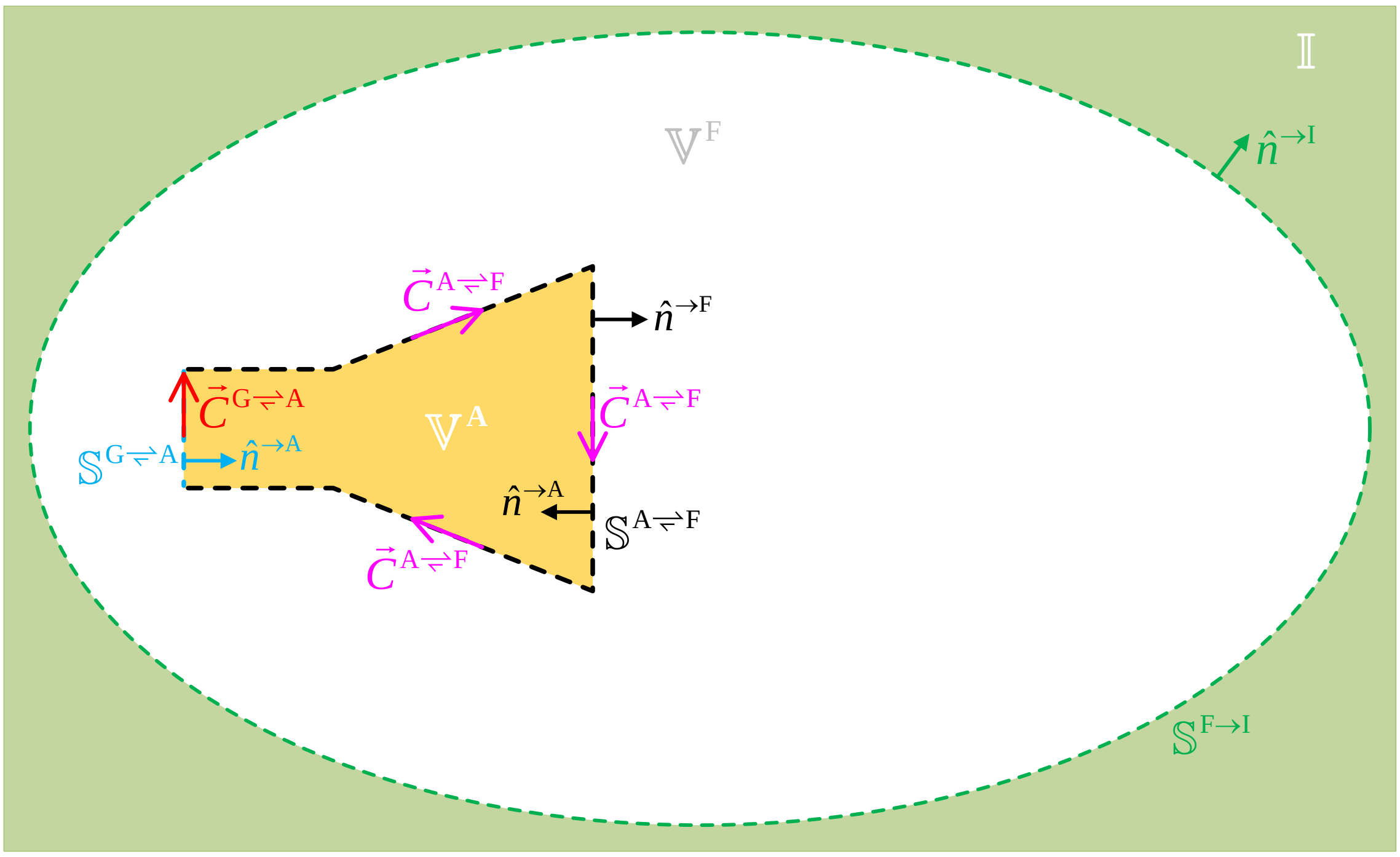

Figure 6-27 Topological structure of the tra-antenna shown in Fig. 6-25

In the figure, surface $\mathbb{S}^{\text{G}\rightleftharpoons\text{A}}$ denotes the input port of the tra-antenna, and $\mathbb{V}^{\text{A}}$ is the region occupied by the tra-antenna, and $\mathbb{V}^{\text{F}}$ is the region occupied by free space, and $\mathbb{S}^{\text{A}\rightleftharpoons\text{F}}$ is the interface between $\mathbb{V}^{\text{A}}$ and $\mathbb{V}^{\text{F}}$, i.e., the *output port* of the tra-antenna, and $\mathbb{S}^{\text{F}\to\text{I}}$ is a spherical surface with infinite radius. Clearly, $\mathbb{S}^{\text{G}\rightleftharpoons\text{A}}$ and $\mathbb{S}^{\text{A}\rightleftharpoons\text{F}}$ constitute a closed surface, and the closed surface is just the boundary of $\mathbb{V}^{\text{A}}$, i.e., $\partial\mathbb{V}^{\text{A}} = \mathbb{S}^{\text{G}\rightleftharpoons\text{A}} \bigcup \mathbb{S}^{\text{A}\rightleftharpoons\text{F}}$. In addition, the permeability, permeativity, and conductivity of $\mathbb{V}^{\text{A}}$ are denoted as $\ddot{\mu}$, $\ddot{\varepsilon}$, and $\ddot{\sigma}$ respectively.

If the equivalent surface currents distributing on $\mathbb{S}^{\text{G}\rightleftharpoons\text{A}}$ are denoted as $\{\vec{J}^{\text{G}\rightleftharpoons\text{A}}, \vec{M}^{\text{G}\rightleftharpoons\text{A}}\}$, and the equivalent surface currents distributing on $\mathbb{S}^{\text{A}\rightleftharpoons\text{F}}$ are denoted as $\{\vec{J}^{\text{A}\rightleftharpoons\text{F}}, \vec{M}^{\text{A}\rightleftharpoons\text{F}}\}$, then the field distributing on $\mathbb{V}^{\text{A}}$ can be expressed as follows:

$$\vec{F}(\vec{r}) = \mathcal{F}\left(\vec{J}^{\text{G}\rightleftharpoons\text{A}} + \vec{J}^{\text{A}\rightleftharpoons\text{F}}, \vec{M}^{\text{G}\rightleftharpoons\text{A}} + \vec{M}^{\text{A}\rightleftharpoons\text{F}}\right) \quad , \quad \vec{r} \in \mathbb{V}^{\text{A}} \tag{6-41}$$

where $\vec{F} = \vec{E} / \vec{H}$, and correspondingly $\mathcal{F} = \mathcal{E} / \mathcal{H}$, and the operator is defined as that $\mathcal{F}(\vec{J}, \vec{M}) = \ddot{G}^{JF} * \vec{J} + \ddot{G}^{MF} * \vec{M}$ (here, $\ddot{G}^{JF}$ and $\ddot{G}^{MF}$ are the *dyadic Green's functions* corresponding to the region $\mathbb{V}^{\text{A}}$ with material parameters $\{\ddot{\mu}, \ddot{\varepsilon}, \ddot{\sigma}\}$).

The currents $\{\vec{J}^{\text{G}\rightleftharpoons\text{A}}, \vec{M}^{\text{G}\rightleftharpoons\text{A}}\}$ and fields $\{\vec{E}, \vec{H}\}$ in Eq. (6-41) satisfy the following relations

$$\hat{n}^{\to\text{A}} \times \left[\vec{H}(\vec{r}^{\text{A}})\right]_{\vec{r}^{\text{A}} \to \vec{r}} = \vec{J}^{\text{G}\rightleftharpoons\text{A}}(\vec{r}) \quad , \quad \vec{r} \in \mathbb{S}^{\text{G}\rightleftharpoons\text{A}} \tag{6-42a}$$

$$\left[\vec{E}\left(\vec{r}^{\rm A}\right)\right]_{\vec{r}^{\rm A}\to\vec{r}}\times\hat{n}^{\to\rm A} = \vec{M}^{\rm G\rightleftharpoons A}\left(\vec{r}\right) \quad , \quad \vec{r}\in\mathbb{S}^{\rm G\rightleftharpoons A} \tag{6-42b}$$

and the currents $\{\vec{J}^{\rm A\rightleftharpoons F},\vec{M}^{\rm A\rightleftharpoons F}\}$ and fields $\{\vec{E},\vec{H}\}$ in Eq. (6-41) satisfy the following relations

$$\hat{n}^{\to\rm A}\times\left[\vec{H}\left(\vec{r}^{\rm A}\right)\right]_{\vec{r}^{\rm A}\to\vec{r}} = \vec{J}^{\rm A\rightleftharpoons F}\left(\vec{r}\right) \quad , \quad \vec{r}\in\mathbb{S}^{\rm A\rightleftharpoons F} \tag{6-43a}$$

$$\left[\vec{E}\left(\vec{r}^{\rm A}\right)\right]_{\vec{r}^{\rm A}\to\vec{r}}\times\hat{n}^{\to\rm A} = \vec{M}^{\rm A\rightleftharpoons F}\left(\vec{r}\right) \quad , \quad \vec{r}\in\mathbb{S}^{\rm A\rightleftharpoons F} \tag{6-43b}$$

In the above Eqs. (6-42a)&(6-42b) and (6-43a)&(6-43b), point $\vec{r}^{\rm A}$ belongs to $\mathbb{V}^{\rm A}$ and approaches the point $\vec{r}$ on $\mathbb{S}^{\rm G\rightleftharpoons A}\bigcup\mathbb{S}^{\rm A\rightleftharpoons F}$, and $\hat{n}^{\to\rm A}$ is the normal direction of $\partial\mathbb{V}^{\rm A}$ ($=\mathbb{S}^{\rm G\rightleftharpoons A}\bigcup\mathbb{S}^{\rm A\rightleftharpoons F}$) and points to the interior of $\mathbb{V}^{\rm A}$.

In addition, the field distributing on $\mathbb{V}^{\rm F}$ can be approximately expressed as follows:

$$\vec{F}\left(\vec{r}\right) = \mathcal{F}_0\left(-\vec{J}^{\rm A\rightleftharpoons F},-\vec{M}^{\rm A\rightleftharpoons F}\right) \quad , \quad \vec{r}\in\mathbb{V}^{\rm F} \tag{6-44}$$

where $\vec{F}=\vec{E}/\vec{H}$, and correspondingly $\mathcal{F}_0=\mathcal{E}_0/\mathcal{H}_0$, and the operator is the same as the one used in the previous chapters.

### 6.3.2 Mathematical Description for Modal Space

Combining Eq. (6-41) with Eq. (6-42), we obtain the following integral equations

$$\left[\mathcal{H}\left(\vec{J}^{\rm G\rightleftharpoons A}+\vec{J}^{\rm A\rightleftharpoons F},\vec{M}^{\rm G\rightleftharpoons A}+\vec{M}^{\rm A\rightleftharpoons F}\right)\right]_{\vec{r}^{\rm A}\to\vec{r}}^{\tan} = \vec{J}^{\rm G\rightleftharpoons A}\left(\vec{r}\right)\times\hat{n}^{\to\rm A} \quad , \quad \vec{r}\in\mathbb{S}^{\rm G\rightleftharpoons A} \tag{6-45a}$$

$$\left[\mathcal{E}\left(\vec{J}^{\rm G\rightleftharpoons A}+\vec{J}^{\rm A\rightleftharpoons F},\vec{M}^{\rm G\rightleftharpoons A}+\vec{M}^{\rm A\rightleftharpoons F}\right)\right]_{\vec{r}^{\rm A}\to\vec{r}}^{\tan} = \hat{n}^{\to\rm A}\times\vec{M}^{\rm G\rightleftharpoons A}\left(\vec{r}\right) \quad , \quad \vec{r}\in\mathbb{S}^{\rm G\rightleftharpoons A} \tag{6-45b}$$

about currents $\{\vec{J}^{\rm G\rightleftharpoons A},\vec{M}^{\rm G\rightleftharpoons A}\}$ and $\{\vec{J}^{\rm A\rightleftharpoons F},\vec{M}^{\rm A\rightleftharpoons F}\}$.

Using Eqs. (6-41) and (6-44) and the tangential field continuation condition on $\mathbb{S}^{\rm A\rightleftharpoons F}$, we have the following integral equations

$$\left[\mathcal{E}\left(\vec{J}^{\rm G\rightleftharpoons A}+\vec{J}^{\rm A\rightleftharpoons F},\vec{M}^{\rm G\rightleftharpoons A}+\vec{M}^{\rm A\rightleftharpoons F}\right)\right]_{\vec{r}^{\rm A}\to\vec{r}}^{\tan} = \left[\mathcal{E}_0\left(-\vec{J}^{\rm A\rightleftharpoons F},-\vec{M}^{\rm A\rightleftharpoons F}\right)\right]_{\vec{r}^{\rm F}\to\vec{r}}^{\tan} \quad , \quad \vec{r}\in\mathbb{S}^{\rm A\rightleftharpoons F} \tag{6-46a}$$

$$\left[\mathcal{H}\left(\vec{J}^{\rm G\rightleftharpoons A}+\vec{J}^{\rm A\rightleftharpoons F},\vec{M}^{\rm G\rightleftharpoons A}+\vec{M}^{\rm A\rightleftharpoons F}\right)\right]_{\vec{r}^{\rm A}\to\vec{r}}^{\tan} = \left[\mathcal{H}_0\left(-\vec{J}^{\rm A\rightleftharpoons F},-\vec{M}^{\rm A\rightleftharpoons F}\right)\right]_{\vec{r}^{\rm F}\to\vec{r}}^{\tan} \quad , \quad \vec{r}\in\mathbb{S}^{\rm A\rightleftharpoons F} \tag{6-46b}$$

about currents $\{\vec{J}^{\rm G\rightleftharpoons A},\vec{M}^{\rm G\rightleftharpoons A}\}$ and $\{\vec{J}^{\rm A\rightleftharpoons F},\vec{M}^{\rm A\rightleftharpoons F}\}$, where point $\vec{r}^{\rm F}$ belongs to $\mathbb{V}^{\rm F}$ and approaches the point $\vec{r}$ on $\mathbb{S}^{\rm A\rightleftharpoons F}$.

If the currents contained in Eqs. (6-45a)~(6-46b) are expanded in terms of some proper basis functions, and Eqs. (6-45a), (6-45b), (6-46a), and (6-46b) are tested with $\{\vec{b}_\xi^{\vec{M}^{\rm G\rightleftharpoons A}}\}$, $\{\vec{b}_\xi^{\vec{J}^{\rm G\rightleftharpoons A}}\}$, $\{\vec{b}_\xi^{\vec{J}^{\rm A\rightleftharpoons F}}\}$, and $\{\vec{b}_\xi^{\vec{M}^{\rm A\rightleftharpoons F}}\}$ respectively, then the integral equations are immediately discretized into the following matrix equations

$$\overline{\overline{Z}}^{\vec{M}^{G\rightleftharpoons A}\vec{J}^{G\rightleftharpoons A}}\cdot\overline{a}^{\vec{J}^{G\rightleftharpoons A}}+\overline{\overline{Z}}^{\vec{M}^{G\rightleftharpoons A}\vec{J}^{A\rightleftharpoons F}}\cdot\overline{a}^{\vec{J}^{A\rightleftharpoons F}}+\overline{\overline{Z}}^{\vec{M}^{G\rightleftharpoons A}\vec{M}^{G\rightleftharpoons A}}\cdot\overline{a}^{\vec{M}^{G\rightleftharpoons A}}+\overline{\overline{Z}}^{\vec{M}^{G\rightleftharpoons A}\vec{M}^{A\rightleftharpoons F}}\cdot\overline{a}^{\vec{M}^{A\rightleftharpoons F}}=0 \quad (6\text{-}47a)$$

$$\overline{\overline{Z}}^{\vec{J}^{G\rightleftharpoons A}\vec{J}^{G\rightleftharpoons A}}\cdot\overline{a}^{\vec{J}^{G\rightleftharpoons A}}+\overline{\overline{Z}}^{\vec{J}^{G\rightleftharpoons A}\vec{J}^{A\rightleftharpoons F}}\cdot\overline{a}^{\vec{J}^{A\rightleftharpoons F}}+\overline{\overline{Z}}^{\vec{J}^{G\rightleftharpoons A}\vec{M}^{G\rightleftharpoons A}}\cdot\overline{a}^{\vec{M}^{G\rightleftharpoons A}}+\overline{\overline{Z}}^{\vec{J}^{G\rightleftharpoons A}\vec{M}^{A\rightleftharpoons F}}\cdot\overline{a}^{\vec{M}^{A\rightleftharpoons F}}=0 \quad (6\text{-}47b)$$

and

$$\overline{\overline{Z}}^{\vec{J}^{A\rightleftharpoons F}\vec{J}^{G\rightleftharpoons A}}\cdot\overline{a}^{\vec{J}^{G\rightleftharpoons A}}+\overline{\overline{Z}}^{\vec{J}^{A\rightleftharpoons F}\vec{J}^{A\rightleftharpoons F}}\cdot\overline{a}^{\vec{J}^{A\rightleftharpoons F}}+\overline{\overline{Z}}^{\vec{J}^{A\rightleftharpoons F}\vec{M}^{G\rightleftharpoons A}}\cdot\overline{a}^{\vec{M}^{G\rightleftharpoons A}}+\overline{\overline{Z}}^{\vec{J}^{A\rightleftharpoons F}\vec{M}^{A\rightleftharpoons F}}\cdot\overline{a}^{\vec{M}^{A\rightleftharpoons F}}=0 \quad (6\text{-}48a)$$

$$\overline{\overline{Z}}^{\vec{M}^{A\rightleftharpoons F}\vec{J}^{G\rightleftharpoons A}}\cdot\overline{a}^{\vec{J}^{G\rightleftharpoons A}}+\overline{\overline{Z}}^{\vec{M}^{A\rightleftharpoons F}\vec{J}^{A\rightleftharpoons F}}\cdot\overline{a}^{\vec{J}^{A\rightleftharpoons F}}+\overline{\overline{Z}}^{\vec{M}^{A\rightleftharpoons F}\vec{M}^{G\rightleftharpoons A}}\cdot\overline{a}^{\vec{M}^{G\rightleftharpoons A}}+\overline{\overline{Z}}^{\vec{M}^{A\rightleftharpoons F}\vec{M}^{A\rightleftharpoons F}}\cdot\overline{a}^{\vec{M}^{A\rightleftharpoons F}}=0 \quad (6\text{-}48b)$$

The formulations used to calculate the elements of the matrices in Eq. (6-47a) are as follows:

$$z_{\xi\zeta}^{\vec{M}^{G\rightleftharpoons A}\vec{J}^{G\rightleftharpoons A}}=\left\langle\vec{b}_{\xi}^{\vec{M}^{G\rightleftharpoons A}},\mathcal{H}\left(\vec{b}_{\zeta}^{\vec{J}^{G\rightleftharpoons A}}\right)\right\rangle_{\underset{\sim}{\mathbb{S}}^{G\rightleftharpoons A}}-\left\langle\vec{b}_{\xi}^{\vec{M}^{G\rightleftharpoons A}},\vec{b}_{\zeta}^{\vec{J}^{G\rightleftharpoons A}}\times\hat{n}^{\rightarrow A}\right\rangle_{\underset{\sim}{\mathbb{S}}^{G\rightleftharpoons A}} \quad (6\text{-}49a)$$

$$z_{\xi\zeta}^{\vec{M}^{G\rightleftharpoons A}\vec{J}^{A\rightleftharpoons F}}=\left\langle\vec{b}_{\xi}^{\vec{M}^{G\rightleftharpoons A}},\mathcal{H}\left(\vec{b}_{\zeta}^{\vec{J}^{A\rightleftharpoons F}}\right)\right\rangle_{\underset{\sim}{\mathbb{S}}^{G\rightleftharpoons A}} \quad (6\text{-}49b)$$

$$z_{\xi\zeta}^{\vec{M}^{G\rightleftharpoons A}\vec{M}^{G\rightleftharpoons A}}=\left\langle\vec{b}_{\xi}^{\vec{M}^{G\rightleftharpoons A}},\mathcal{H}\left(\vec{b}_{\zeta}^{\vec{M}^{G\rightleftharpoons A}}\right)\right\rangle_{\underset{\sim}{\mathbb{S}}^{G\rightleftharpoons A}} \quad (6\text{-}49c)$$

$$z_{\xi\zeta}^{\vec{M}^{G\rightleftharpoons A}\vec{M}^{A\rightleftharpoons F}}=\left\langle\vec{b}_{\xi}^{\vec{M}^{G\rightleftharpoons A}},\mathcal{H}\left(\vec{b}_{\zeta}^{\vec{M}^{A\rightleftharpoons F}}\right)\right\rangle_{\underset{\sim}{\mathbb{S}}^{G\rightleftharpoons A}} \quad (6\text{-}49d)$$

where integral surface $\underset{\sim}{\mathbb{S}}^{G\rightleftharpoons A}$ is an inner sub-surface of the tra-antenna as shown in the following Fig. 6-28.

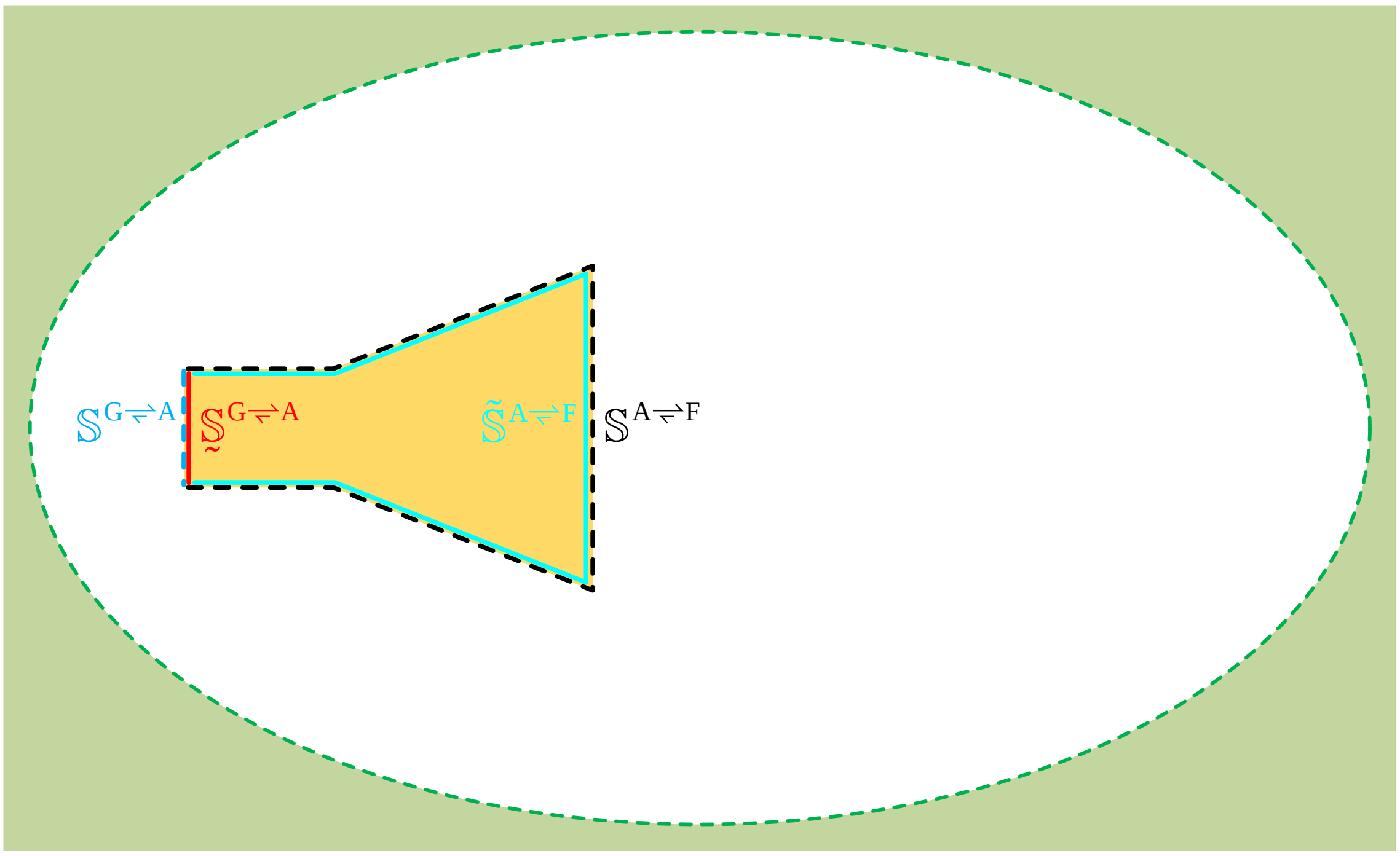

Figure 6-28 Integral surfaces used in Eqs. (6-49)~(6-52)

The formulations used to calculate the elements of the matrices in Eq. (6-47b) are as follows:

$$z_{\xi\zeta}^{\vec{J}^{\text{G}\rightleftharpoons\text{A}}\vec{J}^{\text{G}\rightleftharpoons\text{A}}} = \left\langle \vec{b}_{\xi}^{\vec{J}^{\text{G}\rightleftharpoons\text{A}}}, \mathcal{E}\left(\vec{b}_{\zeta}^{\vec{J}^{\text{G}\rightleftharpoons\text{A}}}\right)\right\rangle_{\mathbb{S}^{\text{G}\rightleftharpoons\text{A}}} \tag{6-50a}$$

$$z_{\xi\zeta}^{\vec{J}^{\text{G}\rightleftharpoons\text{A}}\vec{J}^{\text{A}\rightleftharpoons\text{F}}} = \left\langle \vec{b}_{\xi}^{\vec{J}^{\text{G}\rightleftharpoons\text{A}}}, \mathcal{E}\left(\vec{b}_{\zeta}^{\vec{J}^{\text{A}\rightleftharpoons\text{F}}}\right)\right\rangle_{\mathbb{S}^{\text{G}\rightleftharpoons\text{A}}} \tag{6-50b}$$

$$z_{\xi\zeta}^{\vec{J}^{\text{G}\rightleftharpoons\text{A}}\vec{M}^{\text{G}\rightleftharpoons\text{A}}} = \left\langle \vec{b}_{\xi}^{\vec{J}^{\text{G}\rightleftharpoons\text{A}}}, \mathcal{E}\left(\vec{b}_{\zeta}^{\vec{M}^{\text{G}\rightleftharpoons\text{A}}}\right)\right\rangle_{\mathbb{S}^{\text{G}\rightleftharpoons\text{A}}} - \left\langle \vec{b}_{\xi}^{\vec{J}^{\text{G}\rightleftharpoons\text{A}}}, \hat{n}^{\rightarrow\text{A}} \times \vec{b}_{\zeta}^{\vec{M}^{\text{G}\rightleftharpoons\text{A}}}\right\rangle_{\mathbb{S}^{\text{G}\rightleftharpoons\text{A}}} \tag{6-50c}$$

$$z_{\xi\zeta}^{\vec{J}^{\text{G}\rightleftharpoons\text{A}}\vec{M}^{\text{A}\rightleftharpoons\text{F}}} = \left\langle \vec{b}_{\xi}^{\vec{J}^{\text{G}\rightleftharpoons\text{A}}}, \mathcal{E}\left(\vec{b}_{\zeta}^{\vec{M}^{\text{A}\rightleftharpoons\text{F}}}\right)\right\rangle_{\mathbb{S}^{\text{G}\rightleftharpoons\text{A}}} \tag{6-50d}$$

In the above Eqs. (6-49a)~(6-50d), we have abbreviated the symbols $\mathcal{F}(\vec{J},0)$ and $\mathcal{F}(0,\vec{M})$ to $\mathcal{F}(\vec{J})$ and $\mathcal{F}(\vec{M})$ for simplifying the symbolic system of this report. The formulations used to calculate the elements of the matrices in Eq. (6-48a) are as follows:

$$z_{\xi\zeta}^{\vec{J}^{\text{A}\rightleftharpoons\text{F}}\vec{J}^{\text{G}\rightleftharpoons\text{A}}} = \left\langle \vec{b}_{\xi}^{\vec{J}^{\text{A}\rightleftharpoons\text{F}}}, \mathcal{E}\left(\vec{b}_{\zeta}^{\vec{J}^{\text{G}\rightleftharpoons\text{A}}}\right)\right\rangle_{\tilde{\mathbb{S}}^{\text{A}\rightleftharpoons\text{F}}} \tag{6-51a}$$

$$z_{\xi\zeta}^{\vec{J}^{\text{A}\rightleftharpoons\text{F}}\vec{J}^{\text{A}\rightleftharpoons\text{F}}} = \left\langle \vec{b}_{\xi}^{\vec{J}^{\text{A}\rightleftharpoons\text{F}}}, \mathcal{E}\left(\vec{b}_{\zeta}^{\vec{J}^{\text{A}\rightleftharpoons\text{F}}}\right)\right\rangle_{\tilde{\mathbb{S}}^{\text{A}\rightleftharpoons\text{F}}} - \left\langle \vec{b}_{\xi}^{\vec{J}^{\text{A}\rightleftharpoons\text{F}}}, -j\omega\mu_0\mathcal{L}_0\left(-\vec{b}_{\zeta}^{\vec{J}^{\text{A}\rightleftharpoons\text{F}}}\right)\right\rangle_{\mathbb{S}^{\text{A}\rightleftharpoons\text{F}}} \tag{6-51b}$$

$$z_{\xi\zeta}^{\vec{J}^{\text{A}\rightleftharpoons\text{F}}\vec{M}^{\text{G}\rightleftharpoons\text{A}}} = \left\langle \vec{b}_{\xi}^{\vec{J}^{\text{A}\rightleftharpoons\text{F}}}, \mathcal{E}\left(\vec{b}_{\zeta}^{\vec{M}^{\text{G}\rightleftharpoons\text{A}}}\right)\right\rangle_{\tilde{\mathbb{S}}^{\text{A}\rightleftharpoons\text{F}}} \tag{6-51c}$$

$$\begin{aligned} z_{\xi\zeta}^{\vec{J}^{\text{A}\rightleftharpoons\text{F}}\vec{M}^{\text{A}\rightleftharpoons\text{F}}} &= \left\langle \vec{b}_{\xi}^{\vec{J}^{\text{A}\rightleftharpoons\text{F}}}, \mathcal{E}\left(\vec{b}_{\zeta}^{\vec{M}^{\text{A}\rightleftharpoons\text{F}}}\right)\right\rangle_{\tilde{\mathbb{S}}^{\text{A}\rightleftharpoons\text{F}}} \\ &\quad - \left\langle \vec{b}_{\xi}^{\vec{J}^{\text{A}\rightleftharpoons\text{F}}}, \hat{n}^{\rightarrow\text{A}} \times \frac{1}{2}\vec{b}_{\zeta}^{\vec{M}^{\text{A}\rightleftharpoons\text{F}}} - \text{P.V.}\,\mathcal{K}_0\left(-\vec{b}_{\zeta}^{\vec{M}^{\text{A}\rightleftharpoons\text{F}}}\right)\right\rangle_{\mathbb{S}^{\text{A}\rightleftharpoons\text{F}}} \end{aligned} \tag{6-51d}$$

where integral surface $\tilde{\mathbb{S}}^{\text{A}\rightleftharpoons\text{F}}$ is an inner sub-surface of the tra-antenna as shown in Fig. 6-28. The formulations used to calculate the elements of the matrices in Eq. (6-48b) are as follows:

$$z_{\xi\zeta}^{\vec{M}^{\text{A}\rightleftharpoons\text{F}}\vec{J}^{\text{G}\rightleftharpoons\text{A}}} = \left\langle \vec{b}_{\xi}^{\vec{M}^{\text{A}\rightleftharpoons\text{F}}}, \mathcal{H}\left(\vec{b}_{\zeta}^{\vec{J}^{\text{G}\rightleftharpoons\text{A}}}\right)\right\rangle_{\tilde{\mathbb{S}}^{\text{A}\rightleftharpoons\text{F}}} \tag{6-52a}$$

$$\begin{aligned} z_{\xi\zeta}^{\vec{M}^{\text{A}\rightleftharpoons\text{F}}\vec{J}^{\text{A}\rightleftharpoons\text{F}}} &= \left\langle \vec{b}_{\xi}^{\vec{M}^{\text{A}\rightleftharpoons\text{F}}}, \mathcal{H}\left(\vec{b}_{\zeta}^{\vec{J}^{\text{A}\rightleftharpoons\text{F}}}\right)\right\rangle_{\tilde{\mathbb{S}}^{\text{A}\rightleftharpoons\text{F}}} \\ &\quad - \left\langle \vec{b}_{\xi}^{\vec{M}^{\text{A}\rightleftharpoons\text{F}}}, \frac{1}{2}\vec{b}_{\zeta}^{\vec{J}^{\text{A}\rightleftharpoons\text{F}}} \times \hat{n}^{\rightarrow\text{A}} + \text{P.V.}\,\mathcal{K}_0\left(-\vec{b}_{\zeta}^{\vec{J}^{\text{A}\rightleftharpoons\text{F}}}\right)\right\rangle_{\mathbb{S}^{\text{A}\rightleftharpoons\text{F}}} \end{aligned} \tag{6-52b}$$

$$z_{\xi\zeta}^{\vec{M}^{\text{A}\rightleftharpoons\text{F}}\vec{M}^{\text{G}\rightleftharpoons\text{A}}} = \left\langle \vec{b}_{\xi}^{\vec{M}^{\text{A}\rightleftharpoons\text{F}}}, \mathcal{H}\left(\vec{b}_{\zeta}^{\vec{M}^{\text{G}\rightleftharpoons\text{A}}}\right)\right\rangle_{\tilde{\mathbb{S}}^{\text{A}\rightleftharpoons\text{F}}} \tag{6-52c}$$

$$z_{\xi\zeta}^{\vec{M}^{\text{A}\rightleftharpoons\text{F}}\vec{M}^{\text{A}\rightleftharpoons\text{F}}} = \left\langle \vec{b}_{\xi}^{\vec{M}^{\text{A}\rightleftharpoons\text{F}}}, \mathcal{H}\left(\vec{b}_{\zeta}^{\vec{M}^{\text{A}\rightleftharpoons\text{F}}}\right)\right\rangle_{\tilde{\mathbb{S}}^{\text{A}\rightleftharpoons\text{F}}} - \left\langle \vec{b}_{\xi}^{\vec{M}^{\text{A}\rightleftharpoons\text{F}}}, -j\omega\varepsilon_0\mathcal{L}_0\left(-\vec{b}_{\zeta}^{\vec{M}^{\text{A}\rightleftharpoons\text{F}}}\right)\right\rangle_{\mathbb{S}^{\text{A}\rightleftharpoons\text{F}}} \tag{6-52d}$$

Below, we propose two different schemes for mathematically describing the modal space corresponding to the tra-antenna shown in Figs. 6-25 and 6-27.

**Scheme I: Dependent Variable Elimination (DVE)**

By employing the Eqs. (6-47a), (6-47b), (6-48a), and (6-48b), we can obtain the transformation from basic variables $\bar{a}^{\text{BV}}$ to all variables $\bar{a}^{\text{AV}}$ as follows:

$$\overline{a}^{\mathrm{AV}} = \overline{\overline{T}}^{\mathrm{BV}\to\mathrm{AV}} \cdot \overline{a}^{\mathrm{BV}} \tag{6-53}$$

in which

$$\overline{a}^{\mathrm{AV}} = \begin{bmatrix} \overline{a}^{\vec{J}^{\mathrm{G}\rightleftharpoons\mathrm{A}}} \\ \overline{a}^{\vec{J}^{\mathrm{A}\rightleftharpoons\mathrm{F}}} \\ \overline{a}^{\vec{M}^{\mathrm{G}\rightleftharpoons\mathrm{A}}} \\ \overline{a}^{\vec{M}^{\mathrm{A}\rightleftharpoons\mathrm{F}}} \end{bmatrix} \tag{6-54}$$

and $\overline{a}^{\mathrm{BV}} = \overline{a}^{\vec{J}^{\mathrm{G}\rightleftharpoons\mathrm{A}}} / \overline{a}^{\vec{M}^{\mathrm{G}\rightleftharpoons\mathrm{A}}}$, and correspondingly $\overline{\overline{T}}^{\mathrm{BV}\to\mathrm{AV}} = \overline{\overline{T}}^{\vec{J}^{\mathrm{G}\rightleftharpoons\mathrm{A}}\to\mathrm{AV}} / \overline{\overline{T}}^{\vec{M}^{\mathrm{G}\rightleftharpoons\mathrm{A}}\to\mathrm{AV}}$ where

$$\overline{\overline{T}}^{\vec{J}^{\mathrm{G}\rightleftharpoons\mathrm{A}}\to\mathrm{AV}} = \begin{bmatrix} \overline{\overline{I}}^{\vec{J}^{\mathrm{G}\rightleftharpoons\mathrm{A}}} & 0 & 0 & 0 \\ 0 & \overline{\overline{Z}}^{\vec{M}^{\mathrm{G}\rightleftharpoons\mathrm{A}}\vec{J}^{\mathrm{A}\rightleftharpoons\mathrm{F}}} & \overline{\overline{Z}}^{\vec{M}^{\mathrm{G}\rightleftharpoons\mathrm{A}}\vec{M}^{\mathrm{G}\rightleftharpoons\mathrm{A}}} & \overline{\overline{Z}}^{\vec{M}^{\mathrm{G}\rightleftharpoons\mathrm{A}}\vec{M}^{\mathrm{A}\rightleftharpoons\mathrm{F}}} \\ 0 & \overline{\overline{Z}}^{\vec{J}^{\mathrm{A}\rightleftharpoons\mathrm{F}}\vec{J}^{\mathrm{A}\rightleftharpoons\mathrm{F}}} & \overline{\overline{Z}}^{\vec{J}^{\mathrm{A}\rightleftharpoons\mathrm{F}}\vec{M}^{\mathrm{G}\rightleftharpoons\mathrm{A}}} & \overline{\overline{Z}}^{\vec{J}^{\mathrm{A}\rightleftharpoons\mathrm{F}}\vec{M}^{\mathrm{A}\rightleftharpoons\mathrm{F}}} \\ 0 & \overline{\overline{Z}}^{\vec{M}^{\mathrm{A}\rightleftharpoons\mathrm{F}}\vec{J}^{\mathrm{A}\rightleftharpoons\mathrm{F}}} & \overline{\overline{Z}}^{\vec{M}^{\mathrm{A}\rightleftharpoons\mathrm{F}}\vec{M}^{\mathrm{G}\rightleftharpoons\mathrm{A}}} & \overline{\overline{Z}}^{\vec{M}^{\mathrm{A}\rightleftharpoons\mathrm{F}}\vec{M}^{\mathrm{A}\rightleftharpoons\mathrm{F}}} \end{bmatrix}^{-1} \cdot \begin{bmatrix} \overline{\overline{I}}^{\vec{J}^{\mathrm{G}\rightleftharpoons\mathrm{A}}} \\ -\overline{\overline{Z}}^{\vec{M}^{\mathrm{G}\rightleftharpoons\mathrm{A}}\vec{J}^{\mathrm{G}\rightleftharpoons\mathrm{A}}} \\ -\overline{\overline{Z}}^{\vec{J}^{\mathrm{A}\rightleftharpoons\mathrm{F}}\vec{J}^{\mathrm{G}\rightleftharpoons\mathrm{A}}} \\ -\overline{\overline{Z}}^{\vec{M}^{\mathrm{A}\rightleftharpoons\mathrm{F}}\vec{J}^{\mathrm{G}\rightleftharpoons\mathrm{A}}} \end{bmatrix} \tag{6-55a}$$

$$\overline{\overline{T}}^{\vec{M}^{\mathrm{G}\rightleftharpoons\mathrm{A}}\to\mathrm{AV}} = \begin{bmatrix} \overline{\overline{Z}}^{\vec{J}^{\mathrm{G}\rightleftharpoons\mathrm{A}}\vec{J}^{\mathrm{G}\rightleftharpoons\mathrm{A}}} & \overline{\overline{Z}}^{\vec{J}^{\mathrm{G}\rightleftharpoons\mathrm{A}}\vec{J}^{\mathrm{A}\rightleftharpoons\mathrm{F}}} & 0 & \overline{\overline{Z}}^{\vec{J}^{\mathrm{G}\rightleftharpoons\mathrm{A}}\vec{M}^{\mathrm{A}\rightleftharpoons\mathrm{F}}} \\ \overline{\overline{Z}}^{\vec{J}^{\mathrm{A}\rightleftharpoons\mathrm{F}}\vec{J}^{\mathrm{G}\rightleftharpoons\mathrm{A}}} & \overline{\overline{Z}}^{\vec{J}^{\mathrm{A}\rightleftharpoons\mathrm{F}}\vec{J}^{\mathrm{A}\rightleftharpoons\mathrm{F}}} & 0 & \overline{\overline{Z}}^{\vec{J}^{\mathrm{A}\rightleftharpoons\mathrm{F}}\vec{M}^{\mathrm{A}\rightleftharpoons\mathrm{F}}} \\ 0 & 0 & \overline{\overline{I}}^{\vec{M}^{\mathrm{G}\rightleftharpoons\mathrm{A}}} & 0 \\ \overline{\overline{Z}}^{\vec{M}^{\mathrm{A}\rightleftharpoons\mathrm{F}}\vec{J}^{\mathrm{G}\rightleftharpoons\mathrm{A}}} & \overline{\overline{Z}}^{\vec{M}^{\mathrm{A}\rightleftharpoons\mathrm{F}}\vec{J}^{\mathrm{A}\rightleftharpoons\mathrm{F}}} & 0 & \overline{\overline{Z}}^{\vec{M}^{\mathrm{A}\rightleftharpoons\mathrm{F}}\vec{M}^{\mathrm{A}\rightleftharpoons\mathrm{FS}}} \end{bmatrix}^{-1} \cdot \begin{bmatrix} -\overline{\overline{Z}}^{\vec{J}^{\mathrm{G}\rightleftharpoons\mathrm{A}}\vec{M}^{\mathrm{G}\rightleftharpoons\mathrm{A}}} \\ -\overline{\overline{Z}}^{\vec{J}^{\mathrm{A}\rightleftharpoons\mathrm{F}}\vec{M}^{\mathrm{G}\rightleftharpoons\mathrm{A}}} \\ \overline{\overline{I}}^{\vec{M}^{\mathrm{G}\rightleftharpoons\mathrm{A}}} \\ -\overline{\overline{Z}}^{\vec{M}^{\mathrm{A}\rightleftharpoons\mathrm{F}}\vec{M}^{\mathrm{G}\rightleftharpoons\mathrm{A}}} \end{bmatrix} \tag{6-55b}$$

The 0s in the above Eqs. (6-55a) and (6-55b) are some zero matrices with proper row numbers and column numbers.

**Scheme II: Solution/Definition Domain Compression (SDC/DDC)**

In addition, by employing the Eqs. (6-47a), (6-47b), (6-48a), and (6-48b), we can also obtain the following transformation

$$\overline{a}^{\mathrm{AV}} = \overline{\overline{T}}^{\mathrm{BS}\to\mathrm{AV}} \cdot \overline{a}^{\mathrm{BS}} \tag{6-56}$$

where $\overline{\overline{T}}^{\mathrm{BS}\to\mathrm{AV}} = [\overline{s}_1^{\mathrm{BS}}, \overline{s}_2^{\mathrm{BS}}, \cdots]$ and $\{\overline{s}_1^{\mathrm{BS}}, \overline{s}_2^{\mathrm{BS}}, \cdots\}$ are the basic solutions of the following two theoretically equivalent equations

$$\overline{\overline{\Psi}}_{\mathrm{FCE}}^{\mathrm{DoJ}} \cdot \overline{a}^{\mathrm{AV}} = 0 \tag{6-57a}$$

$$\overline{\overline{\Psi}}_{\mathrm{FCE}}^{\mathrm{DoM}} \cdot \overline{a}^{\mathrm{AV}} = 0 \tag{6-57b}$$

in which

$$\overline{\overline{\Psi}}_{\mathrm{FCE}}^{\mathrm{DoJ}} = \begin{bmatrix} \overline{\overline{Z}}^{\vec{M}^{\mathrm{G}\rightleftharpoons\mathrm{A}}\vec{J}^{\mathrm{G}\rightleftharpoons\mathrm{A}}} & \overline{\overline{Z}}^{\vec{M}^{\mathrm{G}\rightleftharpoons\mathrm{A}}\vec{J}^{\mathrm{A}\rightleftharpoons\mathrm{F}}} & \overline{\overline{Z}}^{\vec{M}^{\mathrm{G}\rightleftharpoons\mathrm{A}}\vec{M}^{\mathrm{G}\rightleftharpoons\mathrm{A}}} & \overline{\overline{Z}}^{\vec{M}^{\mathrm{G}\rightleftharpoons\mathrm{A}}\vec{M}^{\mathrm{A}\rightleftharpoons\mathrm{F}}} \\ \overline{\overline{Z}}^{\vec{J}^{\mathrm{A}\rightleftharpoons\mathrm{F}}\vec{J}^{\mathrm{G}\rightleftharpoons\mathrm{A}}} & \overline{\overline{Z}}^{\vec{J}^{\mathrm{A}\rightleftharpoons\mathrm{F}}\vec{J}^{\mathrm{A}\rightleftharpoons\mathrm{F}}} & \overline{\overline{Z}}^{\vec{J}^{\mathrm{A}\rightleftharpoons\mathrm{F}}\vec{M}^{\mathrm{G}\rightleftharpoons\mathrm{A}}} & \overline{\overline{Z}}^{\vec{J}^{\mathrm{A}\rightleftharpoons\mathrm{F}}\vec{M}^{\mathrm{A}\rightleftharpoons\mathrm{F}}} \\ \overline{\overline{Z}}^{\vec{M}^{\mathrm{A}\rightleftharpoons\mathrm{F}}\vec{J}^{\mathrm{G}\rightleftharpoons\mathrm{A}}} & \overline{\overline{Z}}^{\vec{M}^{\mathrm{A}\rightleftharpoons\mathrm{F}}\vec{J}^{\mathrm{A}\rightleftharpoons\mathrm{F}}} & \overline{\overline{Z}}^{\vec{M}^{\mathrm{A}\rightleftharpoons\mathrm{F}}\vec{M}^{\mathrm{G}\rightleftharpoons\mathrm{A}}} & \overline{\overline{Z}}^{\vec{M}^{\mathrm{A}\rightleftharpoons\mathrm{F}}\vec{M}^{\mathrm{A}\rightleftharpoons\mathrm{F}}} \end{bmatrix} \tag{6-58a}$$

$$\bar{\bar{\Psi}}_{\text{FCE}}^{\text{DoM}} = \begin{bmatrix} \bar{\bar{Z}}^{\vec{J}^{\text{G}\rightleftharpoons\text{A}}\vec{J}^{\text{G}\rightleftharpoons\text{A}}} & \bar{\bar{Z}}^{\vec{J}^{\text{G}\rightleftharpoons\text{A}}\vec{J}^{\text{A}\rightleftharpoons\text{F}}} & \bar{\bar{Z}}^{\vec{J}^{\text{G}\rightleftharpoons\text{A}}\vec{M}^{\text{G}\rightleftharpoons\text{A}}} & \bar{\bar{Z}}^{\vec{J}^{\text{G}\rightleftharpoons\text{A}}\vec{M}^{\text{A}\rightleftharpoons\text{F}}} \\ \bar{\bar{Z}}^{\vec{J}^{\text{A}\rightleftharpoons\text{F}}\vec{J}^{\text{G}\rightleftharpoons\text{A}}} & \bar{\bar{Z}}^{\vec{J}^{\text{A}\rightleftharpoons\text{F}}\vec{J}^{\text{A}\rightleftharpoons\text{F}}} & \bar{\bar{Z}}^{\vec{J}^{\text{A}\rightleftharpoons\text{F}}\vec{M}^{\text{G}\rightleftharpoons\text{A}}} & \bar{\bar{Z}}^{\vec{J}^{\text{A}\rightleftharpoons\text{F}}\vec{M}^{\text{A}\rightleftharpoons\text{F}}} \\ \bar{\bar{Z}}^{\vec{M}^{\text{A}\rightleftharpoons\text{F}}\vec{J}^{\text{G}\rightleftharpoons\text{A}}} & \bar{\bar{Z}}^{\vec{M}^{\text{A}\rightleftharpoons\text{F}}\vec{J}^{\text{A}\rightleftharpoons\text{F}}} & \bar{\bar{Z}}^{\vec{M}^{\text{A}\rightleftharpoons\text{F}}\vec{M}^{\text{G}\rightleftharpoons\text{A}}} & \bar{\bar{Z}}^{\vec{M}^{\text{A}\rightleftharpoons\text{F}}\vec{M}^{\text{A}\rightleftharpoons\text{FS}}} \end{bmatrix} \tag{6-58b}$$

where the superscript and subscript of matrices $\bar{\bar{\Psi}}_{\text{FCE}}^{\text{DoJ}}$ and $\bar{\bar{\Psi}}_{\text{FCE}}^{\text{DoM}}$ have the same meanings as the ones used in Eqs. (6-16) and (6-17).

For the convenience of the following discussions, Eqs. (6-53) and (6-56) are uniformly written as follows:

$$\bar{a}^{\text{AV}} = \bar{\bar{T}} \cdot \bar{a} \tag{6-59}$$

where $\bar{a} = \bar{a}^{\text{BV}} / \bar{a}^{\text{BS}}$ and correspondingly $\bar{\bar{T}} = \bar{\bar{T}}^{\text{BV}\to\text{AV}} / \bar{\bar{T}}^{\text{BS}\to\text{AV}}$.

## 6.3.3 Power Transport Theorem and Input Power Operator

In this subsection, we provide the *power transport theorem* (*PTT*) and *input power operator* (*IPO*) of the tra-antenna shown in Figs. 6-25 and 6-27.

**Power Transport Theorem (PTT)**

Applying the results obtained in Chap. 2 to the tra-antenna shown in Figs. 6-25 and 6-27, we immediately have the following PTT for the tra-antenna

$$P^{\text{G}\rightleftharpoons\text{A}} = P_{\text{dis}}^{\text{A}} + \underbrace{P_{\text{rad}}^{\text{I}} + j\, P_{\text{sto}}^{\text{F}}}_{P^{\text{A}\rightleftharpoons\text{F}}} + j\, P_{\text{sto}}^{\text{A}} \tag{6-60}$$

where $P^{\text{G}\rightleftharpoons\text{A}}$ and $P^{\text{A}\rightleftharpoons\text{F}}$ are respectively the powers inputted into and outputted from the tra-antenna ($P^{\text{A}\rightleftharpoons\text{F}}$ is also the power inputted into free space), and $P_{\text{dis}}^{\text{A}}$ is the power dissipated in the tra-antenna, and $P_{\text{rad}}^{\text{I}}$ is the radiated power arriving at infinity, and $P_{\text{sto}}^{\text{F}}$ is the power corresponding to the stored energy in free space, and $P_{\text{sto}}^{\text{A}}$ is the power corresponding to the stored energy in the tra-antenna.

The above-mentioned various powers are as follows:

$$P^{\text{G}\rightleftharpoons\text{A}} = (1/2)\iint_{\mathbb{S}^{\text{G}\rightleftharpoons\text{A}}} \left(\vec{E}\times\vec{H}^{\dagger}\right)\cdot\hat{n}^{\to\text{A}} dS \tag{6-61a}$$

$$P^{\text{A}\rightleftharpoons\text{F}} = (1/2)\iint_{\mathbb{S}^{\text{A}\rightleftharpoons\text{F}}} \left(\vec{E}\times\vec{H}^{\dagger}\right)\cdot\hat{n}^{\to\text{F}} dS \tag{6-61b}$$

$$P_{\text{dis}}^{\text{A}} = (1/2)\left\langle \vec{\vec{\sigma}}\cdot\vec{E},\vec{E}\right\rangle_{\mathbb{V}^{\text{A}}} \tag{6-61c}$$

$$P_{\text{rad}}^{\text{I}} = (1/2)\oiint_{\mathbb{S}^{\text{F}\to\text{I}}} \left(\vec{E}\times\vec{H}^{\dagger}\right)\cdot\hat{n}^{\to\text{I}} dS \tag{6-61d}$$

$$P_{\text{sto}}^{\text{F}} = 2\omega\left[(1/4)\left\langle \vec{H},\mu_0\vec{H}\right\rangle_{\mathbb{V}^{\text{F}}} - (1/4)\left\langle \varepsilon_0\vec{E},\vec{E}\right\rangle_{\mathbb{V}^{\text{F}}}\right] \tag{6-61e}$$

$$P_{\text{sto}}^{\text{A}} = 2\omega\left[(1/4)\left\langle \vec{H},\vec{\vec{\mu}}\cdot\vec{H}\right\rangle_{\mathbb{V}^{\text{A}}} - (1/4)\left\langle \vec{\vec{\varepsilon}}\cdot\vec{E},\vec{E}\right\rangle_{\mathbb{V}^{\text{A}}}\right] \tag{6-61f}$$

where $\hat{n}^{\rightarrow \mathrm{F}}$ is the normal direction of $\mathbb{S}^{\mathrm{A} \rightleftharpoons \mathrm{F}}$ and points to $\mathbb{V}^{\mathrm{F}}$, and $\hat{n}^{\rightarrow \mathrm{I}}$ is the normal direction of $\mathbb{S}^{\mathrm{I}}$ and points to infinity, as shown in Fig. 6-27.

**Input Power Operator — Formulation I: Current Form**

Based on Eqs. (6-42a)&(6-42b) and the tangential continuity of the $\{\vec{E},\vec{H}\}$ on $\mathbb{S}^{\mathrm{G} \rightleftharpoons \mathrm{A}}$, the IPO $P^{\mathrm{G} \rightleftharpoons \mathrm{A}}$ given in Eq. (6-61a) can be alternatively written as follows:

$$P^{\mathrm{G} \rightleftharpoons \mathrm{A}} = (1/2)\left\langle \hat{n}^{\rightarrow \mathrm{A}} \times \vec{J}^{\mathrm{G} \rightleftharpoons \mathrm{A}}, \vec{M}^{\mathrm{G} \rightleftharpoons \mathrm{A}} \right\rangle_{\mathbb{S}^{\mathrm{G} \rightleftharpoons \mathrm{A}}} \tag{6-62}$$

called the current form of IPO, and it can be discretized as follows:

$$P^{\mathrm{G} \rightleftharpoons \mathrm{A}} = \left(\bar{a}^{\mathrm{AV}}\right)^{\dagger} \cdot \underbrace{\begin{bmatrix} 0 & 0 & \bar{\bar{C}}^{\vec{J}^{\mathrm{G} \rightleftharpoons \mathrm{A}} \vec{M}^{\mathrm{G} \rightleftharpoons \mathrm{A}}} & 0 \\ 0 & 0 & 0 & 0 \\ 0 & 0 & 0 & 0 \\ 0 & 0 & 0 & 0 \end{bmatrix}}_{\bar{\bar{P}}_{\mathrm{curAV}}^{\mathrm{G} \rightleftharpoons \mathrm{A}}} \cdot \bar{a}^{\mathrm{AV}} = \bar{a}^{\dagger} \cdot \underbrace{\left(\bar{\bar{T}}^{\dagger} \cdot \bar{\bar{P}}_{\mathrm{curAV}}^{\mathrm{G} \rightleftharpoons \mathrm{A}} \cdot \bar{\bar{T}}\right)}_{\bar{\bar{P}}_{\mathrm{cur}}^{\mathrm{G} \rightleftharpoons \mathrm{A}}} \cdot \bar{a} \tag{6-63}$$

where the elements of the sub-matrix $\bar{\bar{C}}^{\vec{J}^{\mathrm{G} \rightleftharpoons \mathrm{A}} \vec{M}^{\mathrm{G} \rightleftharpoons \mathrm{A}}}$ are calculated as that $c_{\xi\zeta}^{\vec{J}^{\mathrm{G} \rightleftharpoons \mathrm{A}} \vec{M}^{\mathrm{G} \rightleftharpoons \mathrm{A}}} = (1/2) < \hat{n}^{\rightarrow \mathrm{A}} \times \vec{b}_{\xi}^{\vec{J}^{\mathrm{G} \rightleftharpoons \mathrm{A}}}, \vec{b}_{\zeta}^{\vec{M}^{\mathrm{G} \rightleftharpoons \mathrm{A}}} >_{\mathbb{S}^{\mathrm{G} \rightleftharpoons \mathrm{A}}}$, and the second equality employs transformation (6-59), and the subscript "cur" had been explained in the previous sections.

**Input Power Operator — Formulation II: Field-Current Interaction Form**

Actually, IPO $P^{\mathrm{G} \rightleftharpoons \mathrm{A}}$ also has the following theoretically equivalent expressions

$$\begin{aligned} P^{\mathrm{G} \rightleftharpoons \mathrm{A}} &= -(1/2)\left\langle \vec{J}^{\mathrm{G} \rightleftharpoons \mathrm{A}}, \vec{E} \right\rangle_{\mathbb{S}^{\mathrm{G} \rightleftharpoons \mathrm{A}}} \\ &= -(1/2)\left\langle \vec{J}^{\mathrm{G} \rightleftharpoons \mathrm{A}}, \mathcal{E}\left(\vec{J}^{\mathrm{G} \rightleftharpoons \mathrm{A}} + \vec{J}^{\mathrm{A} \rightleftharpoons \mathrm{F}}, \vec{M}^{\mathrm{G} \rightleftharpoons \mathrm{A}} + \vec{M}^{\mathrm{A} \rightleftharpoons \mathrm{F}}\right)\right\rangle_{\tilde{\mathbb{S}}^{\mathrm{G} \rightleftharpoons \mathrm{A}}} \end{aligned} \tag{6-64a}$$

$$\begin{aligned} P^{\mathrm{G} \rightleftharpoons \mathrm{A}} &= -(1/2)\left\langle \vec{M}^{\mathrm{G} \rightleftharpoons \mathrm{A}}, \vec{H} \right\rangle_{\mathbb{S}^{\mathrm{G} \rightleftharpoons \mathrm{A}}}^{\dagger} \\ &= -(1/2)\left\langle \vec{M}^{\mathrm{G} \rightleftharpoons \mathrm{A}}, \mathcal{H}\left(\vec{J}^{\mathrm{G} \rightleftharpoons \mathrm{A}} + \vec{J}^{\mathrm{A} \rightleftharpoons \mathrm{F}}, \vec{M}^{\mathrm{G} \rightleftharpoons \mathrm{A}} + \vec{M}^{\mathrm{A} \rightleftharpoons \mathrm{F}}\right)\right\rangle_{\tilde{\mathbb{S}}^{\mathrm{G} \rightleftharpoons \mathrm{A}}}^{\dagger} \end{aligned} \tag{6-64b}$$

called the field-current interaction forms of IPO, and it can be discretized as follows:

$$P^{\mathrm{G} \rightleftharpoons \mathrm{A}} = \left(\bar{a}^{\mathrm{AV}}\right)^{\dagger} \cdot \bar{\bar{P}}_{\mathrm{intAV}}^{\mathrm{G} \rightleftharpoons \mathrm{A}} \cdot \bar{a}^{\mathrm{AV}} = \bar{a}^{\dagger} \cdot \underbrace{\left(\bar{\bar{T}}^{\dagger} \cdot \bar{\bar{P}}_{\mathrm{intAV}}^{\mathrm{G} \rightleftharpoons \mathrm{A}} \cdot \bar{\bar{T}}\right)}_{\bar{\bar{P}}_{\mathrm{int}}^{\mathrm{G} \rightleftharpoons \mathrm{A}}} \cdot \bar{a} \tag{6-65}$$

where the second equality is based on transformation (6-59), and

$$\bar{\bar{P}}_{\mathrm{intAV}}^{\mathrm{G} \rightleftharpoons \mathrm{A}} = \begin{bmatrix} \bar{\bar{P}}^{\vec{J}^{\mathrm{G} \rightleftharpoons \mathrm{A}} \vec{J}^{\mathrm{G} \rightleftharpoons \mathrm{A}}} & \bar{\bar{P}}^{\vec{J}^{\mathrm{G} \rightleftharpoons \mathrm{A}} \vec{J}^{\mathrm{A} \rightleftharpoons \mathrm{F}}} & \bar{\bar{P}}^{\vec{J}^{\mathrm{G} \rightleftharpoons \mathrm{A}} \vec{M}^{\mathrm{G} \rightleftharpoons \mathrm{A}}} & \bar{\bar{P}}^{\vec{J}^{\mathrm{G} \rightleftharpoons \mathrm{A}} \vec{M}^{\mathrm{A} \rightleftharpoons \mathrm{F}}} \\ 0 & 0 & 0 & 0 \\ 0 & 0 & 0 & 0 \\ 0 & 0 & 0 & 0 \end{bmatrix} \tag{6-66a}$$

for Eq. (6-64a), or

$$\overline{\overline{P}}{}^{\text{G}\rightleftharpoons\text{A}}_{\text{intAV}} = \begin{bmatrix} 0 & 0 & 0 & 0 \\ 0 & 0 & 0 & 0 \\ \overline{\overline{P}}{}^{\vec{M}^{\text{G}\rightleftharpoons\text{A}}\vec{J}^{\text{G}\rightleftharpoons\text{A}}} & \overline{\overline{P}}{}^{\vec{M}^{\text{G}\rightleftharpoons\text{A}}\vec{J}^{\text{A}\rightleftharpoons\text{F}}} & \overline{\overline{P}}{}^{\vec{M}^{\text{G}\rightleftharpoons\text{A}}\vec{M}^{\text{G}\rightleftharpoons\text{A}}} & \overline{\overline{P}}{}^{\vec{M}^{\text{G}\rightleftharpoons\text{A}}\vec{M}^{\text{A}\rightleftharpoons\text{F}}} \\ 0 & 0 & 0 & 0 \end{bmatrix}^{\dagger} \tag{6-66b}$$

for Eq. (6-64b), with matrix elements

$$p_{\xi\zeta}^{\vec{J}^{\text{G}\rightleftharpoons\text{A}}\vec{J}^{\text{G}\rightleftharpoons\text{A}}} = -(1/2)\left\langle \vec{b}_{\xi}^{\vec{J}^{\text{G}\rightleftharpoons\text{A}}}, \mathcal{E}\left(\vec{b}_{\zeta}^{\vec{J}^{\text{G}\rightleftharpoons\text{A}}}\right)\right\rangle_{\underset{\sim}{\mathbb{S}}^{\text{G}\rightleftharpoons\text{A}}} \tag{6-67a}$$

$$p_{\xi\zeta}^{\vec{J}^{\text{G}\rightleftharpoons\text{A}}\vec{J}^{\text{A}\rightleftharpoons\text{F}}} = -(1/2)\left\langle \vec{b}_{\xi}^{\vec{J}^{\text{G}\rightleftharpoons\text{A}}}, \mathcal{E}\left(\vec{b}_{\zeta}^{\vec{J}^{\text{A}\rightleftharpoons\text{F}}}\right)\right\rangle_{\underset{\sim}{\mathbb{S}}^{\text{G}\rightleftharpoons\text{A}}} \tag{6-67b}$$

$$p_{\xi\zeta}^{\vec{J}^{\text{G}\rightleftharpoons\text{A}}\vec{M}^{\text{G}\rightleftharpoons\text{A}}} = -(1/2)\left\langle \vec{b}_{\xi}^{\vec{J}^{\text{G}\rightleftharpoons\text{A}}}, \mathcal{E}\left(\vec{b}_{\zeta}^{\vec{M}^{\text{G}\rightleftharpoons\text{A}}}\right)\right\rangle_{\underset{\sim}{\mathbb{S}}^{\text{G}\rightleftharpoons\text{A}}} \tag{6-67c}$$

$$p_{\xi\zeta}^{\vec{J}^{\text{G}\rightleftharpoons\text{A}}\vec{M}^{\text{A}\rightleftharpoons\text{F}}} = -(1/2)\left\langle \vec{b}_{\xi}^{\vec{J}^{\text{G}\rightleftharpoons\text{A}}}, \mathcal{E}\left(\vec{b}_{\zeta}^{\vec{M}^{\text{A}\rightleftharpoons\text{F}}}\right)\right\rangle_{\underset{\sim}{\mathbb{S}}^{\text{G}\rightleftharpoons\text{A}}} \tag{6-67d}$$

and

$$p_{\xi\zeta}^{\vec{M}^{\text{G}\rightleftharpoons\text{A}}\vec{J}^{\text{G}\rightleftharpoons\text{A}}} = -(1/2)\left\langle \vec{b}_{\xi}^{\vec{M}^{\text{G}\rightleftharpoons\text{A}}}, \mathcal{H}\left(\vec{b}_{\zeta}^{\vec{J}^{\text{G}\rightleftharpoons\text{A}}}\right)\right\rangle_{\underset{\sim}{\mathbb{S}}^{\text{G}\rightleftharpoons\text{A}}} \tag{6-67e}$$

$$p_{\xi\zeta}^{\vec{M}^{\text{G}\rightleftharpoons\text{A}}\vec{J}^{\text{A}\rightleftharpoons\text{F}}} = -(1/2)\left\langle \vec{b}_{\xi}^{\vec{M}^{\text{G}\rightleftharpoons\text{A}}}, \mathcal{H}\left(\vec{b}_{\zeta}^{\vec{J}^{\text{A}\rightleftharpoons\text{F}}}\right)\right\rangle_{\underset{\sim}{\mathbb{S}}^{\text{G}\rightleftharpoons\text{A}}} \tag{6-67f}$$

$$p_{\xi\zeta}^{\vec{M}^{\text{G}\rightleftharpoons\text{A}}\vec{M}^{\text{G}\rightleftharpoons\text{A}}} = -(1/2)\left\langle \vec{b}_{\xi}^{\vec{M}^{\text{G}\rightleftharpoons\text{A}}}, \mathcal{H}\left(\vec{b}_{\zeta}^{\vec{M}^{\text{G}\rightleftharpoons\text{A}}}\right)\right\rangle_{\underset{\sim}{\mathbb{S}}^{\text{G}\rightleftharpoons\text{A}}} \tag{6-67g}$$

$$p_{\xi\zeta}^{\vec{M}^{\text{G}\rightleftharpoons\text{A}}\vec{M}^{\text{A}\rightleftharpoons\text{F}}} = -(1/2)\left\langle \vec{b}_{\xi}^{\vec{M}^{\text{G}\rightleftharpoons\text{A}}}, \mathcal{H}\left(\vec{b}_{\zeta}^{\vec{M}^{\text{A}\rightleftharpoons\text{F}}}\right)\right\rangle_{\underset{\sim}{\mathbb{S}}^{\text{G}\rightleftharpoons\text{A}}} \tag{6-67h}$$

where the subscript "int" had been explained in the previous sections.

For the convenience of the following discussions, the Eqs. (6-63) and (6-65) are uniformly written as follows:

$$P^{\text{G}\rightleftharpoons\text{A}} = \bar{a}^{\dagger} \cdot \overline{\overline{P}}{}^{\text{G}\rightleftharpoons\text{A}} \cdot \bar{a} \tag{6-68}$$

where $\overline{\overline{P}}{}^{\text{G}\rightleftharpoons\text{A}} = \overline{\overline{P}}{}^{\text{G}\rightleftharpoons\text{A}}_{\text{cur}} \,/\, \overline{\overline{P}}{}^{\text{G}\rightleftharpoons\text{A}}_{\text{int}}$ .

## 6.3.4 Input-Power-Decoupled Modes

Below, we construct the *input-power-decoupled modes* (*IP-DMs*) of the tra-antenna shown in Figs. 6-25 and 6-27, by using the results obtained above.

**Construction Method**

The IP-DMs in modal space can be derived from solving the modal decoupling equation $\overline{\overline{P}}{}^{\text{G}\rightleftharpoons\text{A}}_{-} \cdot \bar{\alpha}_{\xi} = \theta_{\xi} \overline{\overline{P}}{}^{\text{G}\rightleftharpoons\text{A}}_{+} \cdot \bar{\alpha}_{\xi}$ defined on modal space, where $\overline{\overline{P}}{}^{\text{G}\rightleftharpoons\text{A}}_{+}$ and $\overline{\overline{P}}{}^{\text{G}\rightleftharpoons\text{A}}_{-}$

are the positive and negative Hermitian parts of matrix $\overline{\overline{P}}^{\mathrm{G}\rightleftharpoons\mathrm{A}}$.

If some derived modes $\{\bar{\alpha}_1, \bar{\alpha}_2, \cdots, \bar{\alpha}_d\}$ are $d$-order degenerate, then the Gram-Schmidt orthogonalization process given in the previous Sec. 6.2.4.1 is necessary, and it is not repeated here.

**Modal Decoupling Relation and Parseval's Identity**

The modal fields constructed above satisfy the following decoupling relation

$$(1/2)\iint_{\mathbb{S}^{\mathrm{G}\rightleftharpoons\mathrm{A}}}\left(\vec{E}_\zeta \times \vec{H}_\xi^\dagger\right)\cdot \hat{n}^{\rightarrow\mathrm{A}} dS = \left(1+j\,\theta_\xi\right)\delta_{\xi\zeta} \tag{6-69}$$

and the relation implies that **the IP-DMs don't have net energy coupling in one period**.

By employing the above decoupling relation, we have the following Parseval's identity

$$\sum_\xi \left|c_\xi\right|^2 = \left(1/T\right)\int_{t_0}^{t_0+T}\left[\iint_{\mathbb{S}^{\mathrm{G}\rightleftharpoons\mathrm{A}}}\left(\vec{\mathcal{E}}\times\vec{\mathcal{H}}\right)\cdot\hat{n}^{\rightarrow\mathrm{A}}dS\right]dt \tag{6-70}$$

with modal expansion coefficient $c_\xi$ as that $c_\xi = -(1/2) < \vec{J}_\xi^{\mathrm{G}\rightleftharpoons\mathrm{A}}, \vec{E} >_{\mathbb{S}^{\mathrm{G}\rightleftharpoons\mathrm{A}}} \Big/ (1+j\,\theta_\xi)$ $= -(1/2) < \vec{H}, \vec{M}_\xi^{\mathrm{G}\rightleftharpoons\mathrm{A}} >_{\mathbb{S}^{\mathrm{G}\rightleftharpoons\mathrm{A}}} \Big/ (1+j\,\theta_\xi)$, where $\{\vec{E}, \vec{H}\}$ are some previously known fields distributing on input port $\mathbb{S}^{\mathrm{G}\rightleftharpoons\mathrm{A}}$.

**Modal Quantities**

Just like the metallic tra-antenna discussed in Sec. 6.2, we can also define the modal significance $\mathrm{MS}_\xi = 1\big/|1+j\,\theta_\xi|$ and modal input impedance $Z_\xi^{\mathrm{G}\rightleftharpoons\mathrm{A}} = P_\xi^{\mathrm{G}\rightleftharpoons\mathrm{A}} \Big/ \left[(1/2) < \vec{J}_\xi^{\mathrm{G}\rightleftharpoons\mathrm{A}}, \vec{J}_\xi^{\mathrm{G}\rightleftharpoons\mathrm{A}} >_{\mathbb{S}^{\mathrm{G}\rightleftharpoons\mathrm{A}}}\right]$ and modal input admittance $Y_\xi^{\mathrm{G}\rightleftharpoons\mathrm{A}} = P_\xi^{\mathrm{G}\rightleftharpoons\mathrm{A}} \Big/ \left[(1/2) < \vec{M}_\xi^{\mathrm{G}\rightleftharpoons\mathrm{A}}, \vec{M}_\xi^{\mathrm{G}\rightleftharpoons\mathrm{A}} >_{\mathbb{S}^{\mathrm{G}\rightleftharpoons\mathrm{A}}}\right]$ for the tra-antenna shown in Figs. 6-25 and 6-27 to quantitatively describe the modal features of the tra-antenna.

## 6.3.5 Numerical Examples Corresponding to Typical Structures

In this section, we consider the lossless material rod tra-antenna shown in the following Fig. 6-29. Its radius and length are 5mm and 20mm respectively, and its relative permeability and relative permittivity are 1 and 10 respectively.

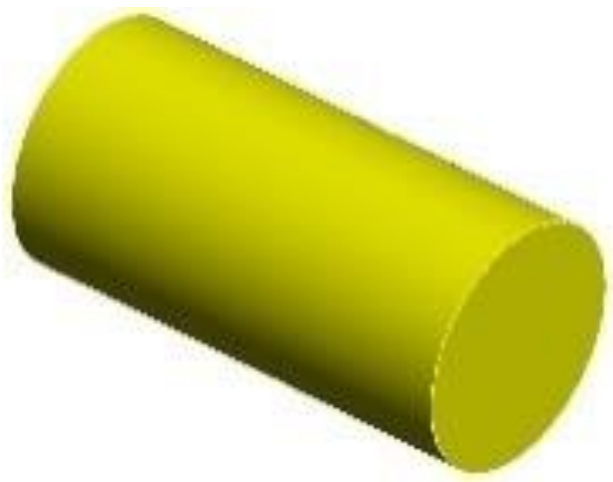

Figure 6-29 Geometry of a material tra-antenna

The topological structures and surface triangular meshes of the input and output ports of the tra-antenna are shown in the following Fig. 6-30.

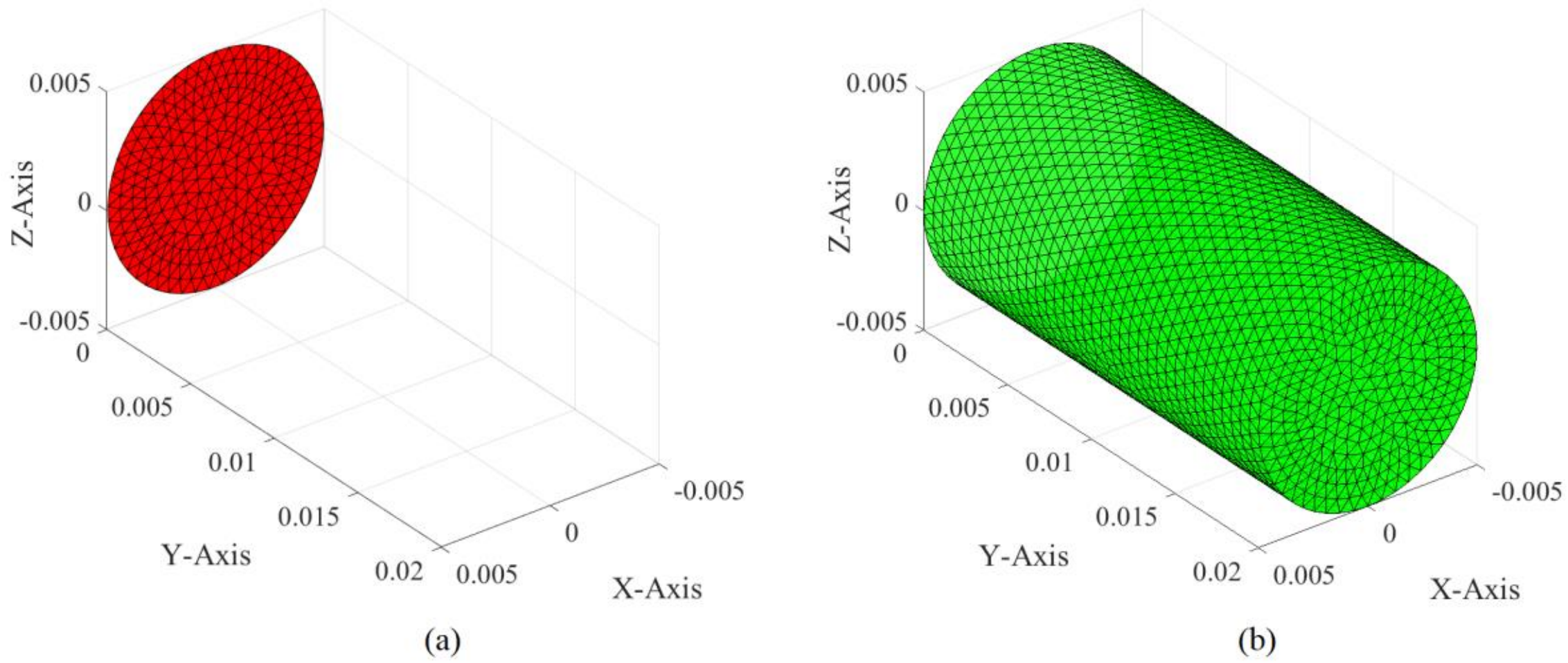

Figure 6-30 Topological structures and surface triangular meshes of (a) inport port and (b) output port of the material tra-antenna shown in Fig. 6-29

By employing the JE-DoJ-based and HM-DoM-based formulations of IPO given in the previous Eq. (6-68), we calculate the IP-DMs of the material tra-antenna. The modal input resistance curves of the first several typical modes are shown in the following Fig. 6-31.

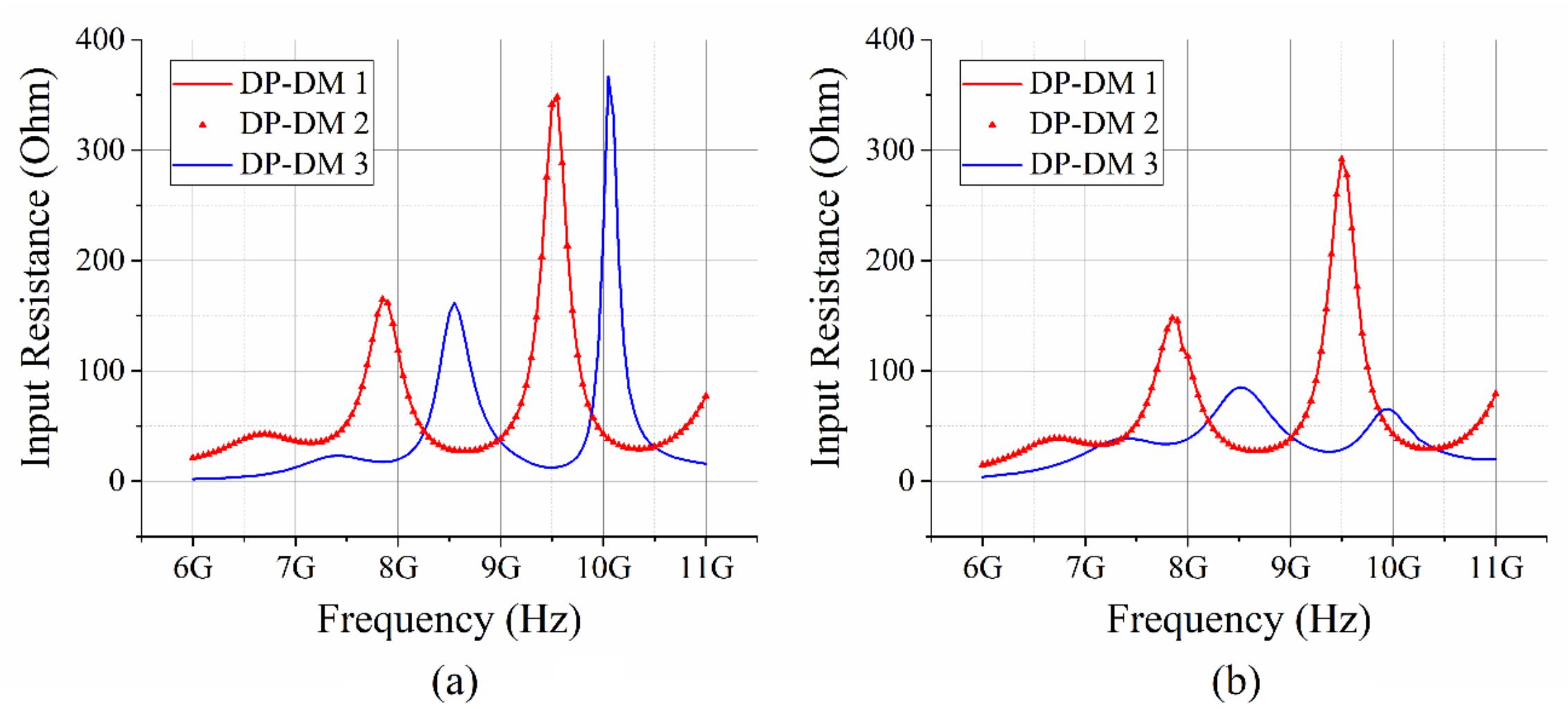

Figure 6-31 Modal input resistance curves of the first several typical IP-DMs. (a) JE-DoJ-based results; (b) HM-DoM-based results

The input port equivalent magnetic currents of the JE-DoJ-based IP-DMs 1 and 2 are shown in the following Fig. 6-32.

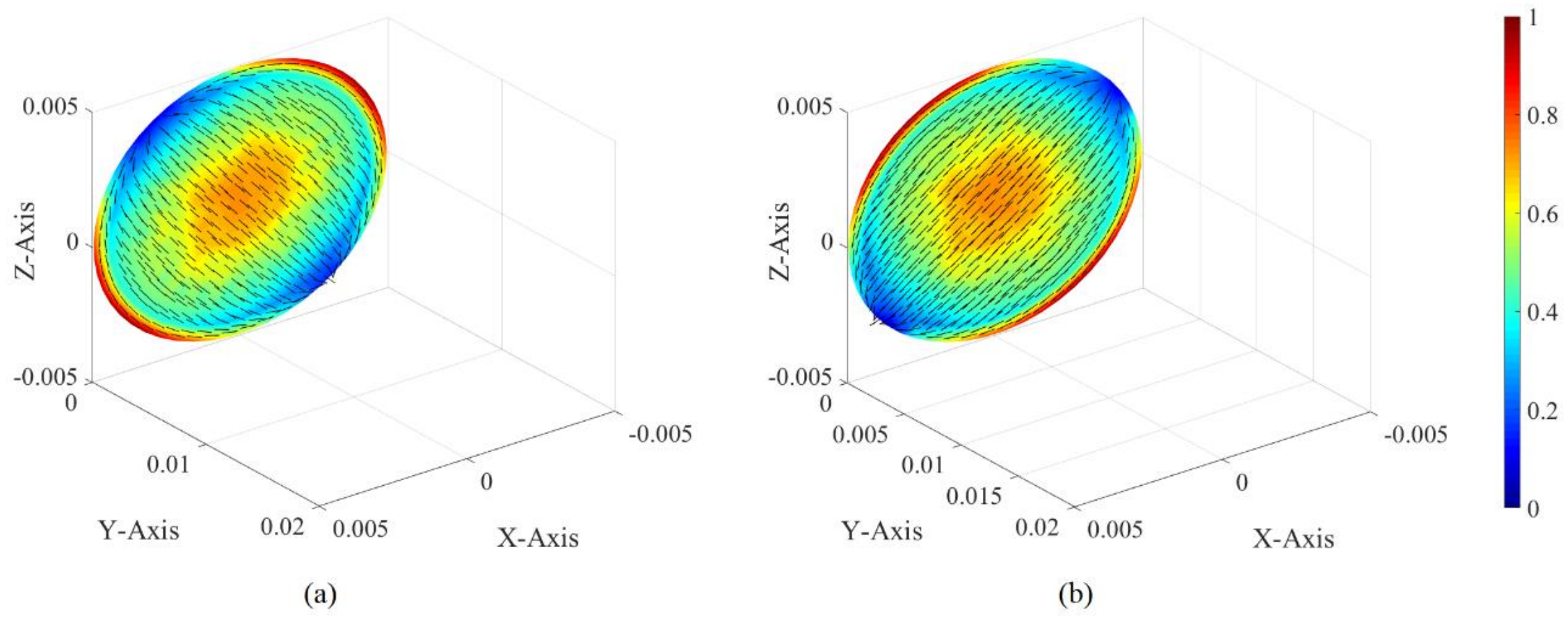

Figure 6-32 Input port equivalent magnetic currents of the JE-DoJ-based (a) IP-DM 1 and (b) IP-DM 2

Obviously, the JE-DoJ-based IP-DMs 1 and 2 are a pair of degenerate modes. Taking the frist generate state working at 6.70 GHz as an example, its modal induced volume electric current, radiation pattern, and electric field are shown in the following Fig. 6-33.

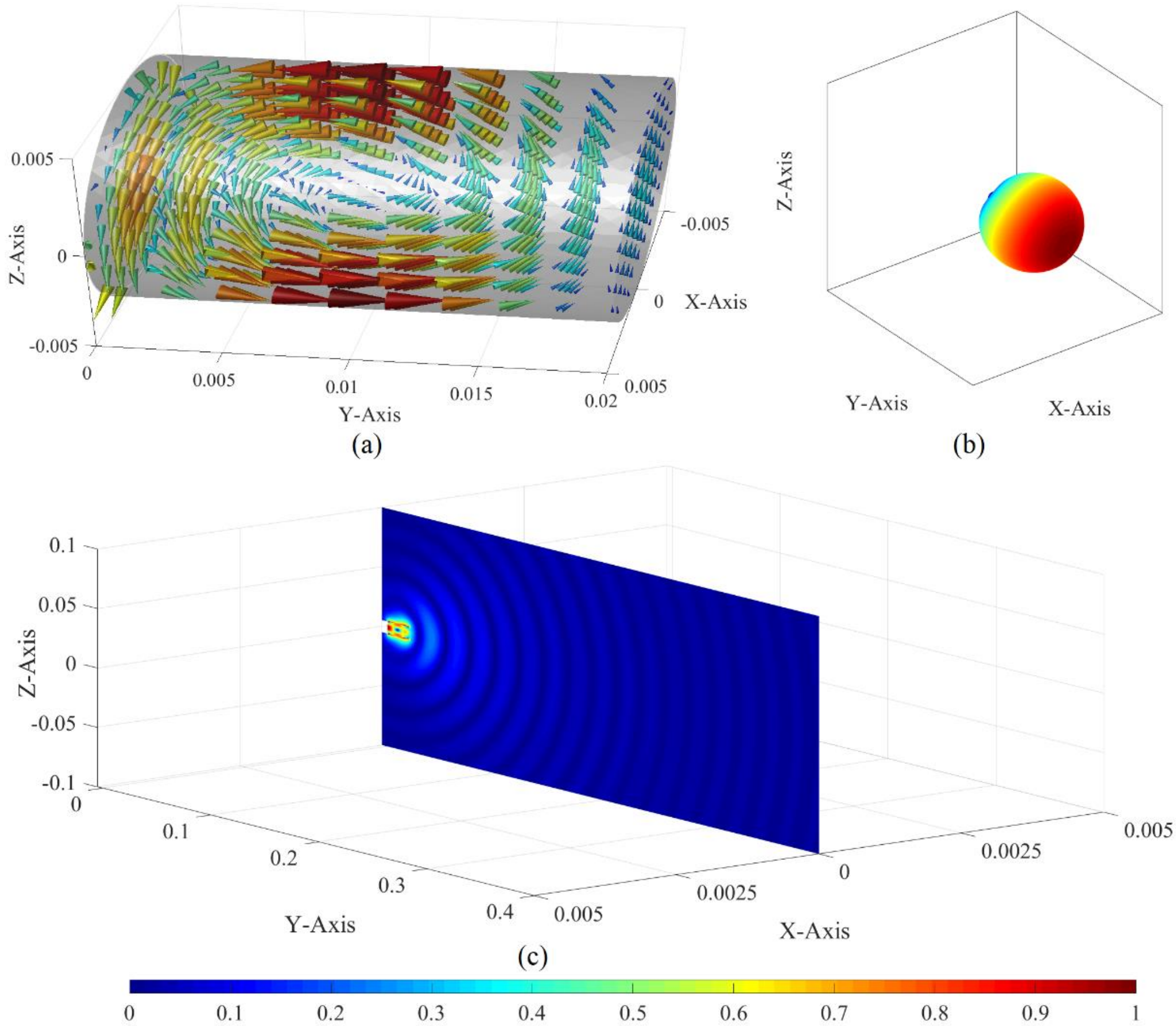

Figure 6-33 Modal (a) induced volume electric current, (b) radiation pattern, and (c) electric field of the first degenerate state of JE-DoJ-based IP-DM 1 (at 6.70 GHz)

For the frist generate state working at 7.85 GHz, its modal induced volume electric current, modal radiation pattern, and modal electric field are shown in the following Fig. 6-34.

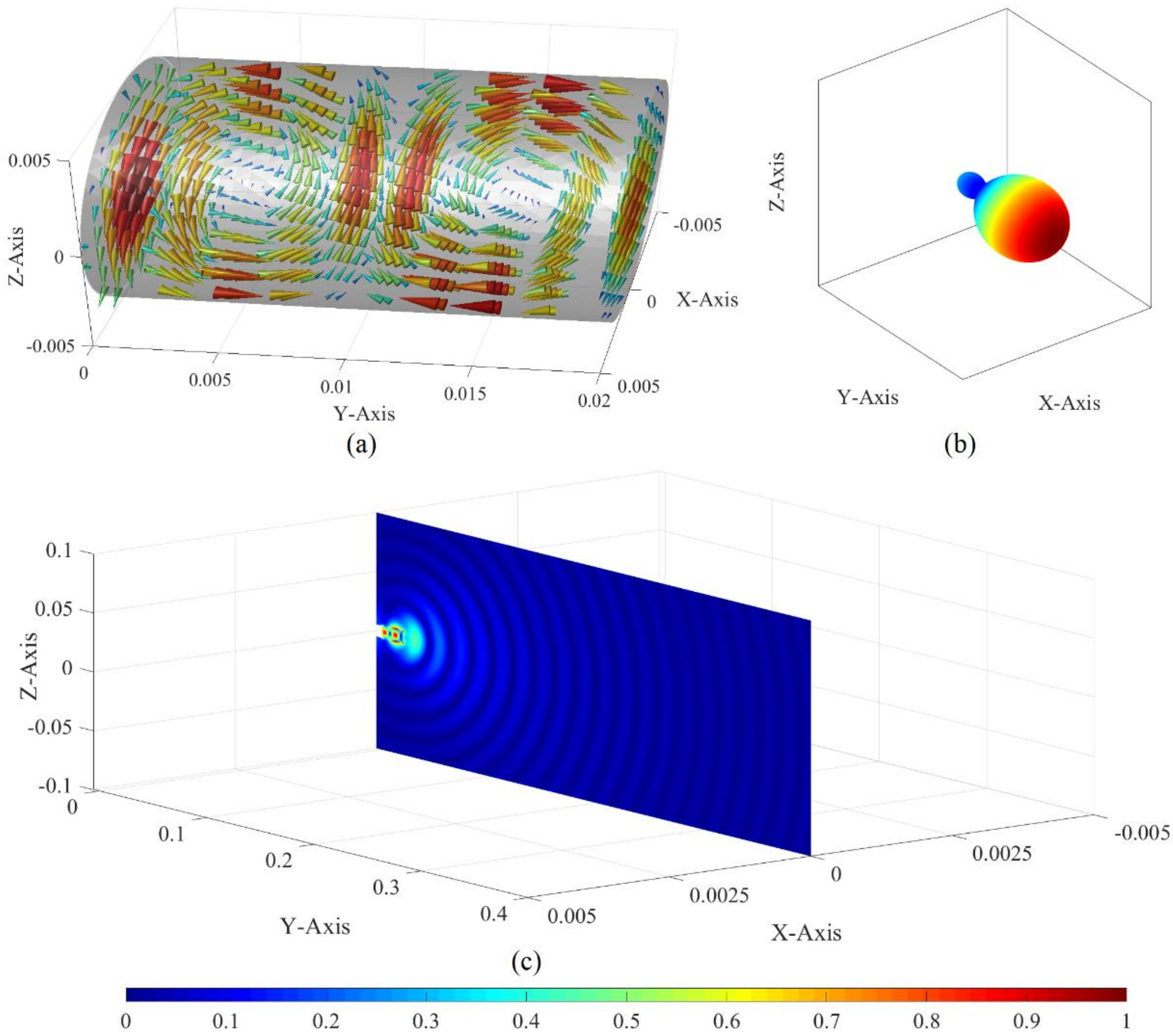

Figure 6-34 Modal (a) induced volume electric current, (b) radiation pattern, and (c) electric field of the first degenerate state of JE-DoJ-based IP-DM 1 (at 7.85 GHz)

For the frist generate state working at 9.55 GHz, its modal induced volume electric current, modal radiation pattern, and modal electric field are shown in the following Fig. 6-35.

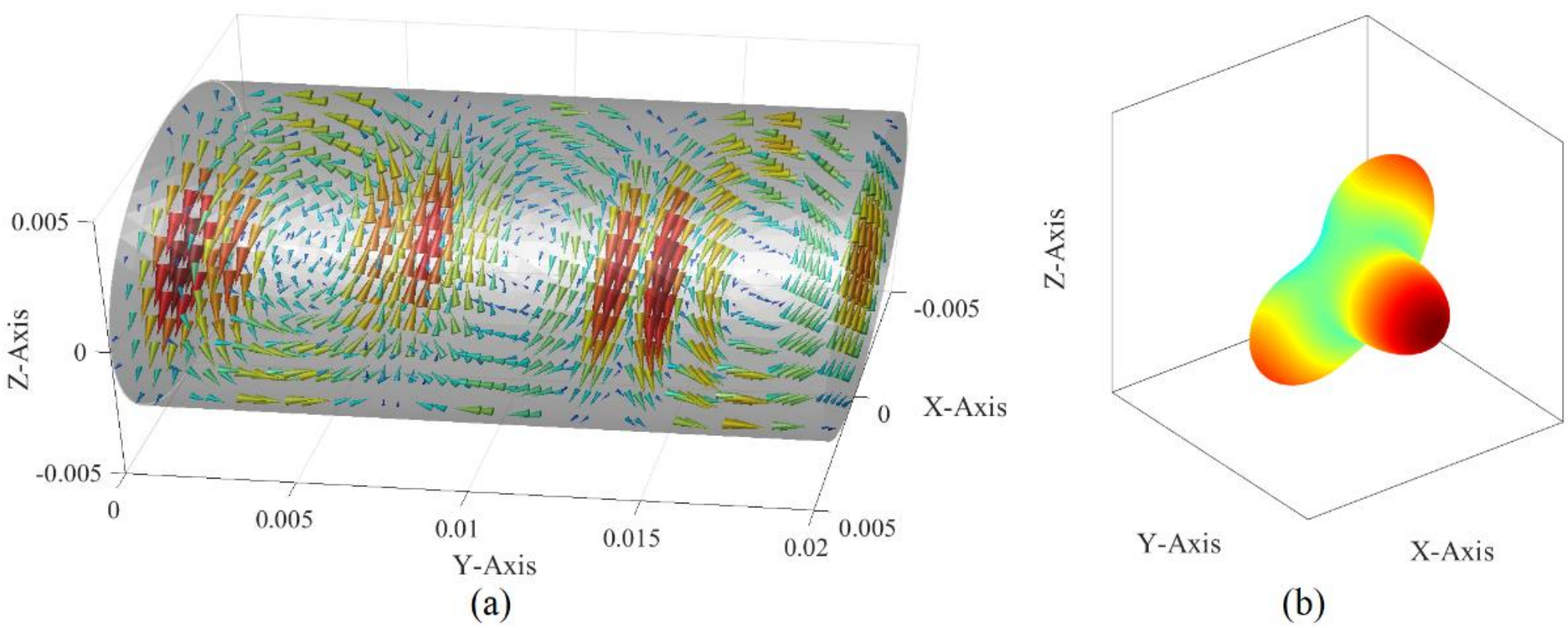

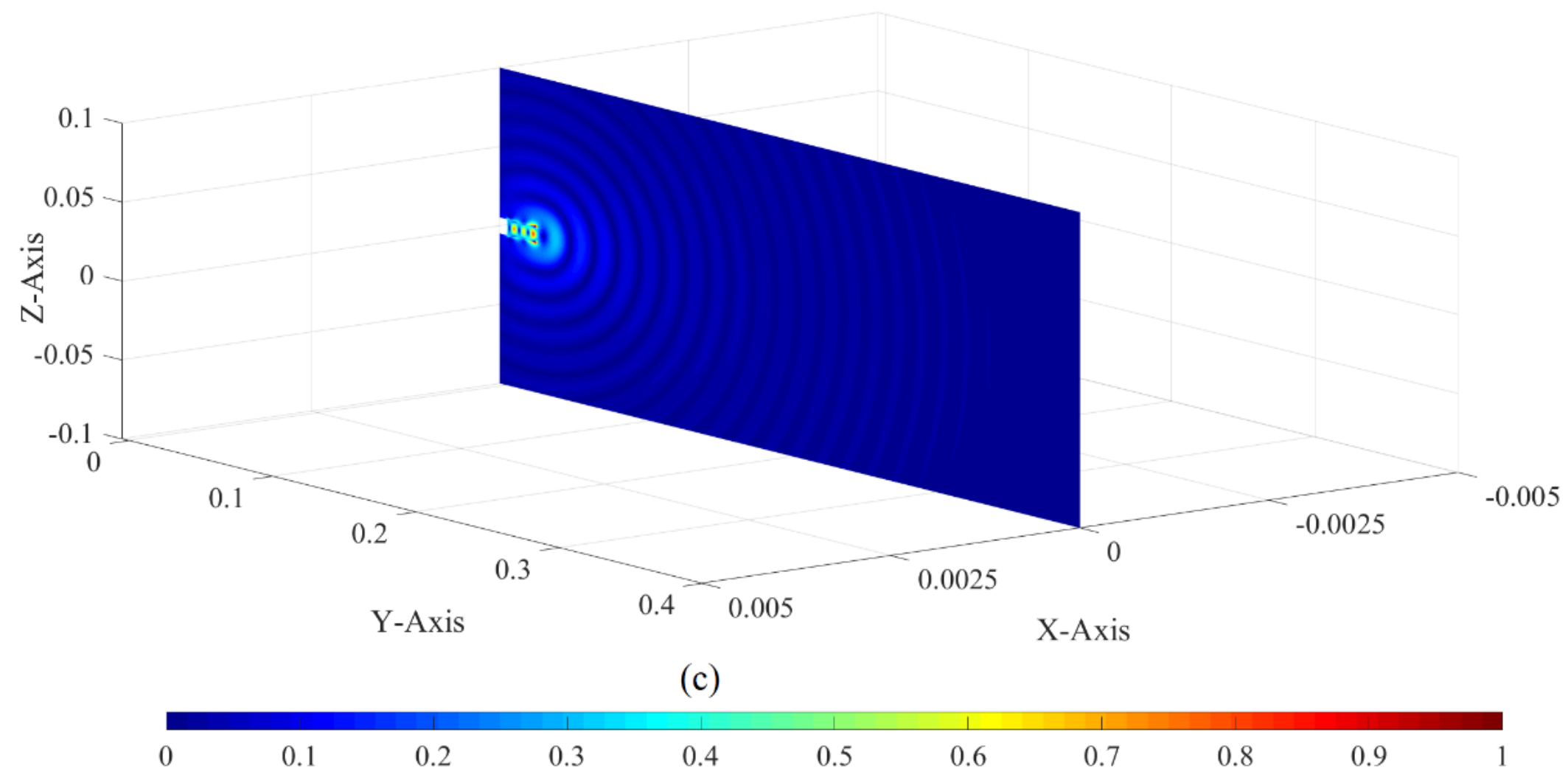

Figure 6-35 Modal (a) induced volume electric current, (b) radiation pattern, and (c) electric field of the first degenerate state of JE-DoJ-based IP-DM 1 (at 9.55 GHz)

The input port equivalent magnetic current of the JE-DoJ-based IP-DM 3 is shown in the following Fig. 6-36.

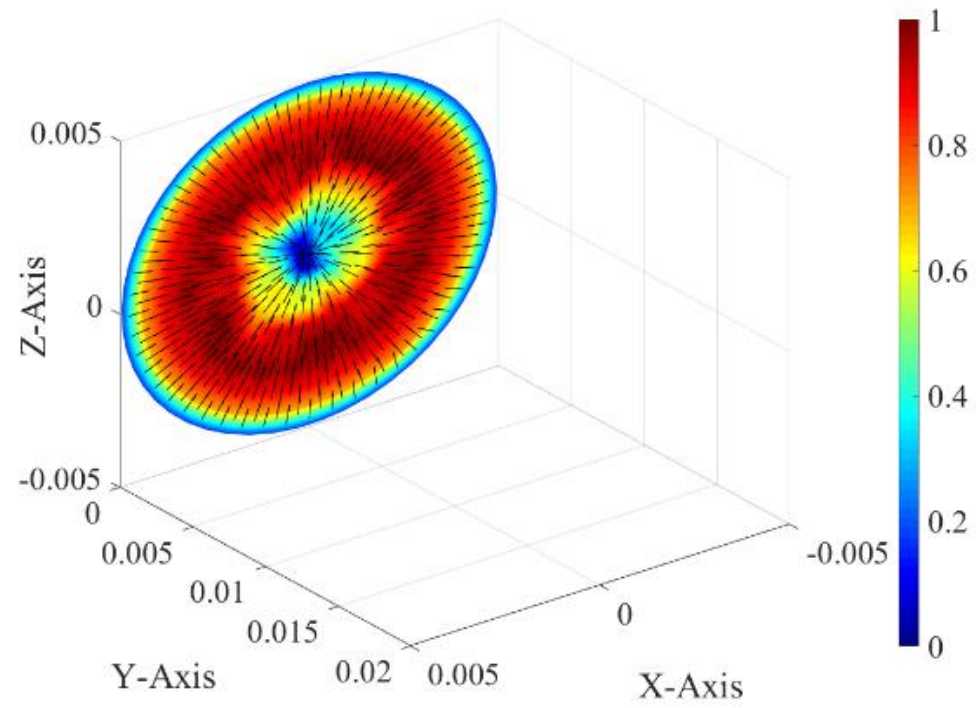

Figure 6-36 Input port equivalent magnetic current of the JE-DoJ-based IP-DM 3

For the IP-DM 3 working at 7.40 GHz, its modal induced volume electric current, modal radiation pattern, and modal electric field are shown in the following Fig. 6-37.

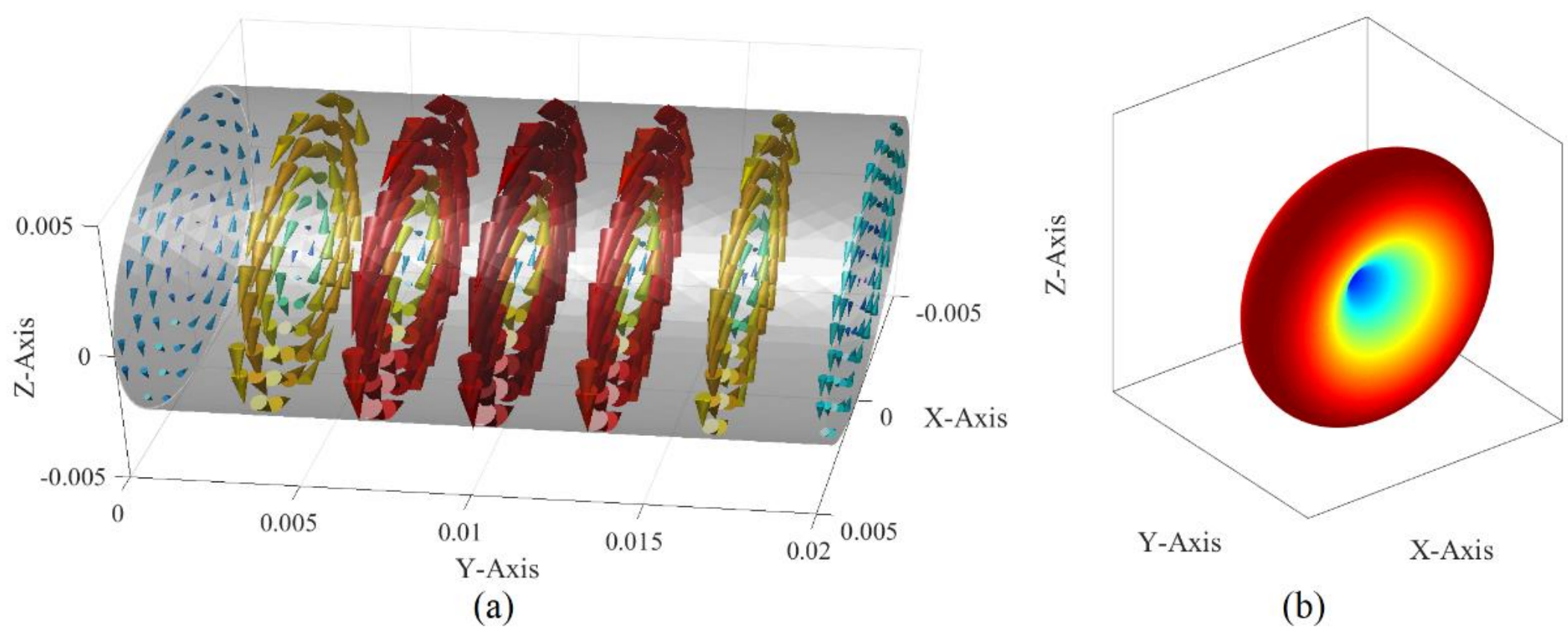

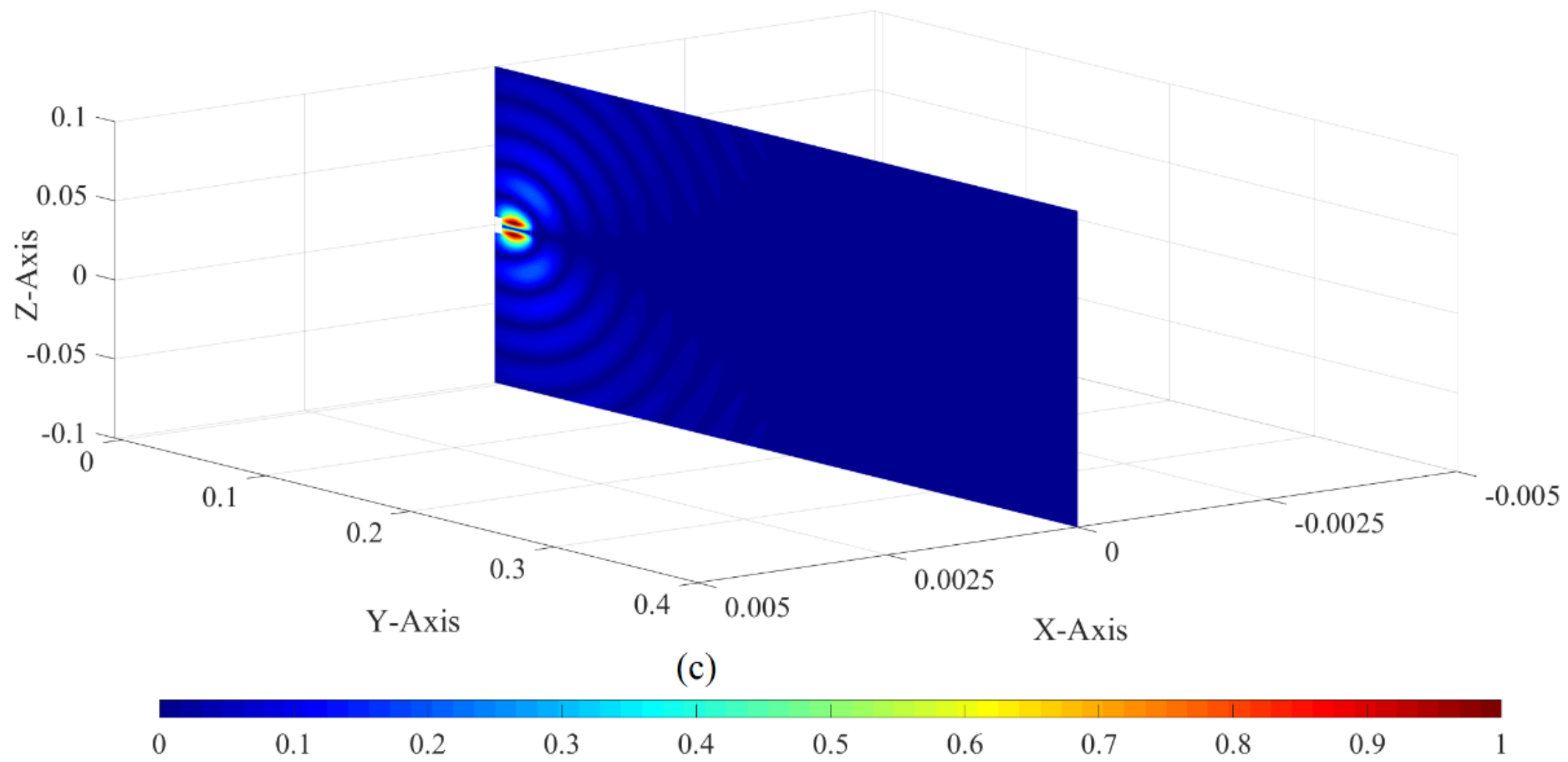

Figure 6-37 Modal (a) induced volume electric current, (b) radiation pattern, and (c) electric field of the first degenerate state of JE-DoJ-based IP-DM 3 (at 7.40 GHz)

For the IP-DM 3 working at 8.55 GHz, its modal induced volume electric current, modal radiation pattern, and modal electric field are shown in the following Fig. 6-38.

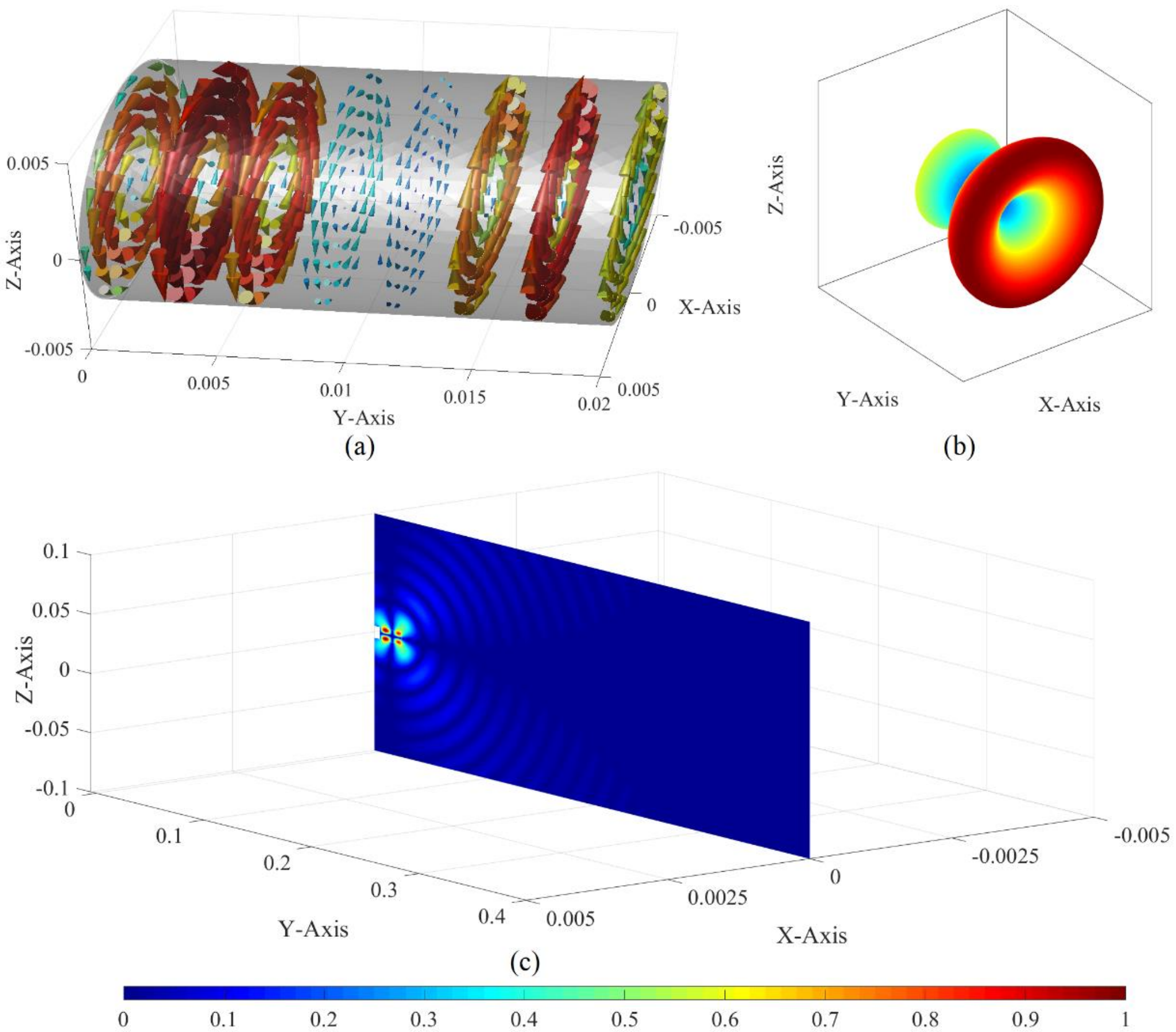

Figure 6-38 Modal (a) induced volume electric current, (b) radiation pattern, and (c) electric field of the first degenerate state of JE-DoJ-based IP-DM 3 (at 8.55 GHz)

For the IP-DM 3 working at 10.05 GHz, its modal induced volume electric current, modal radiation pattern, and modal electric field are shown in the following Fig. 6-39.

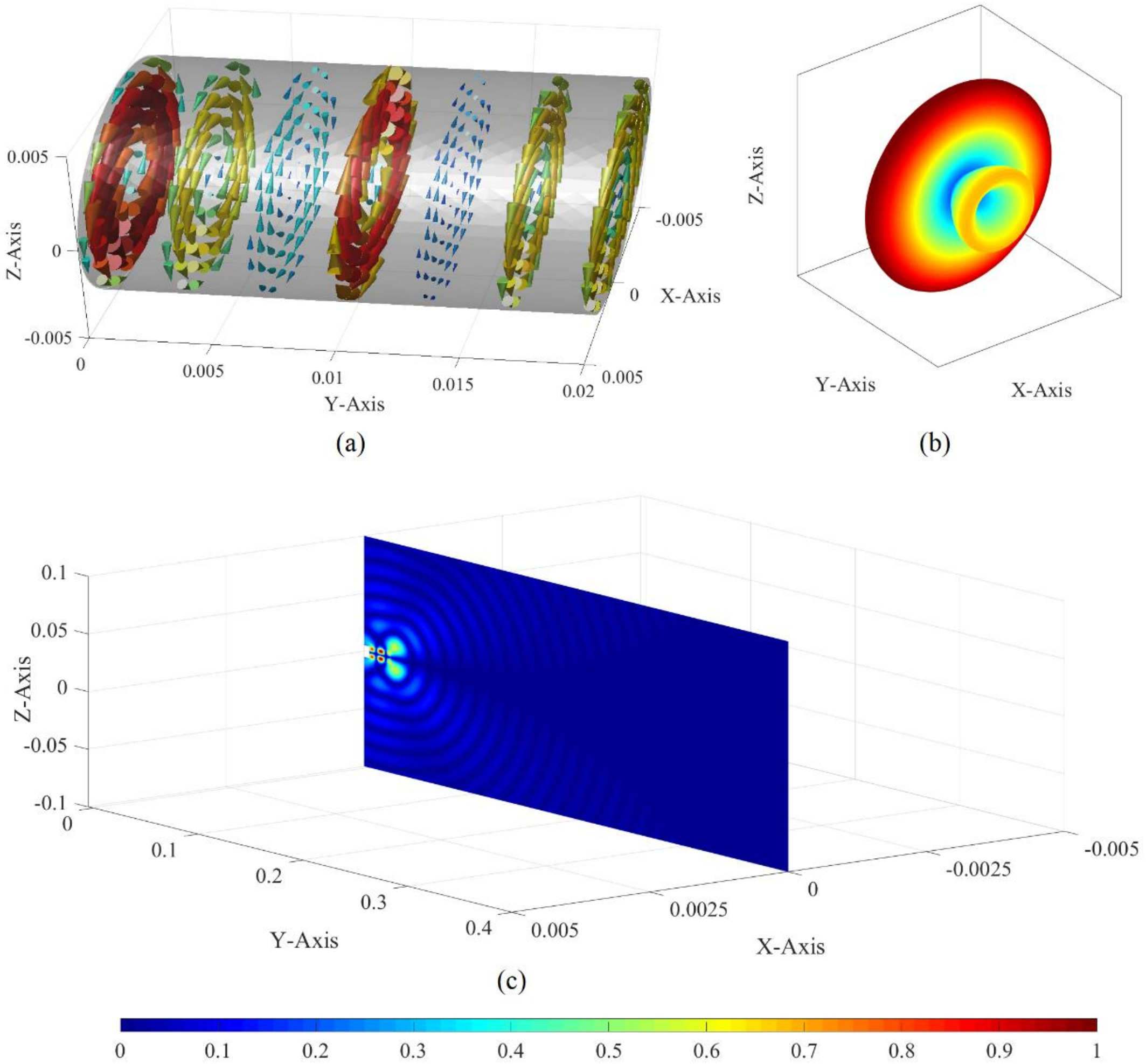

Figure 6-39 Modal (a) induced volume electric current, (b) radiation pattern, and (c) electric field of the first degenerate state of JE-DoJ-based IP-DM 3 (at 10.05 GHz)

## 6.4 IP-DMs of Augmented Metal-Material Composite Transmitting Antenna — Type I: Coaxial Metallic Probe Fed Stacked Dielectric Resonator Antenna on Metallic Ground Plane

*Dielectric resonator antennas* (*DRAs*) have been widely used in antenna engineering society, because of their low loss, high radiation efficiency, broad bandwidth, ease of excitation, compact size, light weight, low cost, and freedom of shape, etc. features[129~131]. According to the difference in the aspect of topological structure, DRA can be categorized into single-block DRA[132~164] and multiple-block DRA[165,166]. According to the difference in the aspect of geometrical shape, the single-block DRA can be categorized into semi-spherical DRA[132~142], circular-cylindrical DRA[143~152], rectangular DRA[153~159], circular-conical DRA[160], and some other more complicated ones[161~164]. The usually used feeding methods for DRA include coaxial-probe feeding[132,133,143,144,153,161,162,165], waveguide-probe feeding[145,146], and aperture/slot feeding[134~142,147~152,154~160,163,164,166], etc., and some reviews for the various feeding methods can be found in Refs. [167,168].

Taking the *coaxial metallic probe fed stacked rectangular DRA* shown in Fig. 6-40 as a typical example, this section, under PTT framework, discusses the processes to construct the IP-DMs of *metal-material composite augmented tra-antenna*. The other kinds of DRAs can be similarly discussed.

As shown in Fig. 6-40, the tra-antenna is a two-layered material body mounted on a *thick metallic ground plane*, and the tra-antenna is excited by a *thick metallic probe* driven by *coaxial line*. In addition, the environment surrounding the tra-antenna is free space.

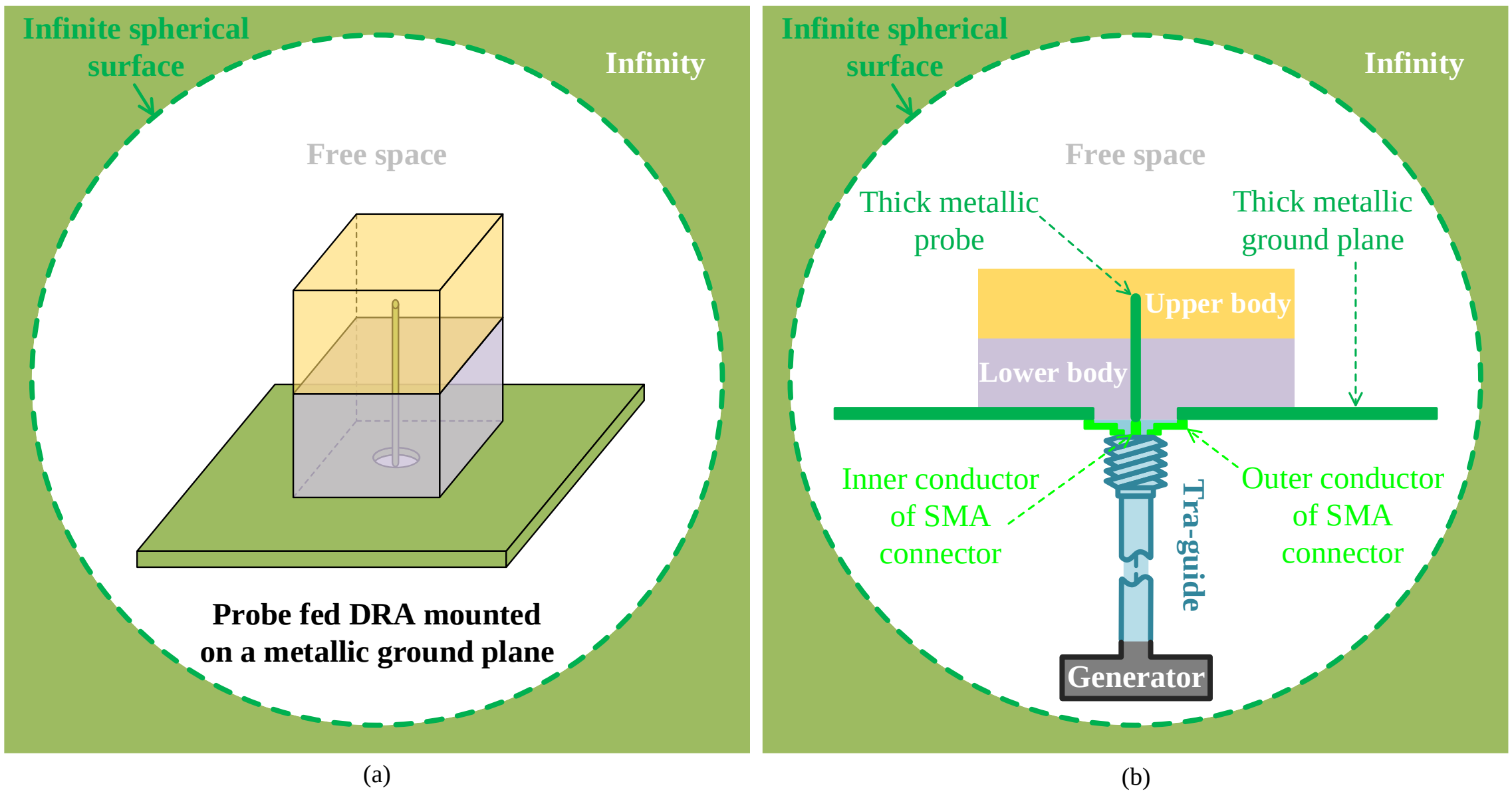

Figure 6-40 Geometry of a typical coaxial metallic probe fed stacked dielectric resonator antenna mounted on a thick metallic ground plane

The organization of this section is completely similar to the previous Secs. 6.2 and 6.3, i.e., “topological structure → source-field relationships → modal space → power transport theorem → input power operator → input-power-decoupled modes”.

## 6.4.1 Topological Structure

The topological structure of the tra-antenna shown in the previous Fig. 6-40 is illustrated in the following Fig. 6-41.

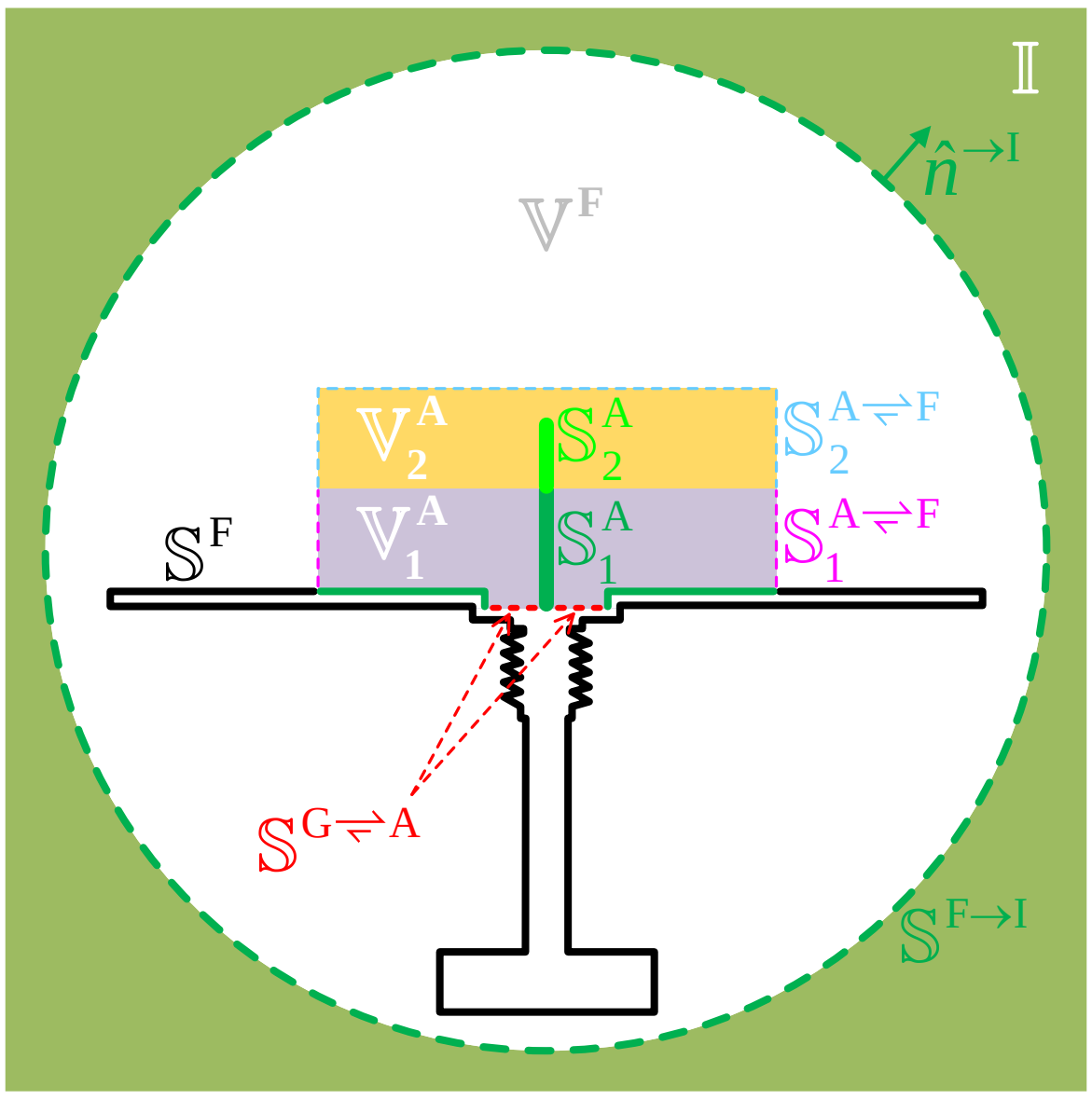

Figure 6-41 Topological structure of the tra-antenna shown in Fig. 6-40

In the above figure, surface $\mathbb{S}^{\mathrm{G}\rightleftharpoons\mathrm{A}}$ denotes the input port of the tra-antenna. The regions occupied by the lower and upper material bodies are denoted as $\mathbb{V}_1^{\mathrm{A}}$ and $\mathbb{V}_2^{\mathrm{A}}$ respectively; the region occupied by free space is denoted as $\mathbb{V}^{\mathrm{F}}$.

The interface between $\mathbb{V}_1^{\mathrm{A}}$ and ground plane and the interface between $\mathbb{V}_1^{\mathrm{A}}$ and metallic probe are collectively denoted as $\mathbb{S}_1^{\mathrm{A}}$; the interface between $\mathbb{V}_2^{\mathrm{A}}$ and metallic probe is denoted as $\mathbb{S}_2^{\mathrm{A}}$; the interface between $\mathbb{V}_1^{\mathrm{A}} / \mathbb{V}_2^{\mathrm{A}}$ and $\mathbb{V}^{\mathrm{F}}$ is denoted as $\mathbb{S}_1^{\mathrm{A}\rightleftharpoons\mathrm{F}} / \mathbb{S}_2^{\mathrm{A}\rightleftharpoons\mathrm{F}}$; the interface between $\mathbb{V}_1^{\mathrm{A}}$ and $\mathbb{V}_2^{\mathrm{A}}$ is denoted as $\mathbb{S}_{12}^{\mathrm{A}\rightleftharpoons\mathrm{A}}$; the interface between $\mathbb{V}^{\mathrm{F}}$ and the metallic wall of whole transmitting system is just the *grounding structure*, and it is denoted as $\mathbb{S}^{\mathrm{F}}$. Here, $\mathbb{S}^{\mathrm{F}\rightarrow\mathrm{I}}$ is a spherical surface with infinite radius.

Clearly, $\mathbb{S}^{\mathrm{G}\rightleftharpoons\mathrm{A}}$, $\mathbb{S}_1^{\mathrm{A}\rightleftharpoons\mathrm{F}}$, $\mathbb{S}_{12}^{\mathrm{A}\rightleftharpoons\mathrm{A}}$, and $\mathbb{S}_1^{\mathrm{A}}$ constitute a closed surface, and the closed surface is just the whole boundary of $\mathbb{V}_1^{\mathrm{A}}$, i.e., $\partial\mathbb{V}_1^{\mathrm{A}} = \mathbb{S}^{\mathrm{G}\rightleftharpoons\mathrm{A}} \cup \mathbb{S}_1^{\mathrm{A}\rightleftharpoons\mathrm{F}} \cup \mathbb{S}_{12}^{\mathrm{A}\rightleftharpoons\mathrm{A}} \cup \mathbb{S}_1^{\mathrm{A}}$; $\mathbb{S}_{12}^{\mathrm{A}\rightleftharpoons\mathrm{A}}$, $\mathbb{S}_2^{\mathrm{A}\rightleftharpoons\mathrm{F}}$, and $\mathbb{S}_2^{\mathrm{A}}$ also constitute a closed surface, and the closed surface is just the whole boundary of $\mathbb{V}_2^{\mathrm{A}}$, i.e., $\partial\mathbb{V}_2^{\mathrm{A}} = \mathbb{S}_{12}^{\mathrm{A}\rightleftharpoons\mathrm{A}} \cup \mathbb{S}_2^{\mathrm{A}\rightleftharpoons\mathrm{F}} \cup \mathbb{S}_2^{\mathrm{A}}$. Similarly, $\mathbb{S}_1^{\mathrm{A}\rightleftharpoons\mathrm{F}}$, $\mathbb{S}_2^{\mathrm{A}\rightleftharpoons\mathrm{F}}$, $\mathbb{S}^{\mathrm{F}}$, and $\mathbb{S}^{\mathrm{F}\rightarrow\mathrm{I}}$ constitute the whole boundary of $\mathbb{V}^{\mathrm{F}}$, i.e., $\partial\mathbb{V}^{\mathrm{F}} = \mathbb{S}_1^{\mathrm{A}\rightleftharpoons\mathrm{F}} \cup \mathbb{S}_2^{\mathrm{A}\rightleftharpoons\mathrm{F}} \cup \mathbb{S}^{\mathrm{F}} \cup \mathbb{S}^{\mathrm{F}\rightarrow\mathrm{I}}$.

In addition, the permeability, permeativity, and conductivity of $\mathbb{V}_1^{\mathrm{A}}$ are denoted as $\ddot{\mu}_1$, $\ddot{\varepsilon}_1$, and $\ddot{\sigma}_1$ respectively, and the permeability, permeativity, and conductivity of $\mathbb{V}_2^{\mathrm{A}}$ are denoted as $\ddot{\mu}_2$, $\ddot{\varepsilon}_2$, and $\ddot{\sigma}_2$ respectively.

### 6.4.2 Source-Field Relationships

If the equivalent surface currents distributing on $\mathbb{S}^{\mathrm{G}\rightleftharpoons\mathrm{A}}$ are denoted as $\{\vec{J}^{\mathrm{G}\rightleftharpoons\mathrm{A}}, \vec{M}^{\mathrm{G}\rightleftharpoons\mathrm{A}}\}$, and the equivalent surface currents distributing on $\mathbb{S}_1^{\mathrm{A}\rightleftharpoons\mathrm{F}}$ are denoted as

$\{\vec{J}_1^{\mathrm{A}\rightleftharpoons\mathrm{F}},\vec{M}_1^{\mathrm{A}\rightleftharpoons\mathrm{F}}\}$, and the equivalent surface currents distributing on $\mathbb{S}_{12}^{\mathrm{A}\rightleftharpoons\mathrm{A}}$ are denoted as $\{\vec{J}_{12}^{\mathrm{A}\rightleftharpoons\mathrm{A}},\vec{M}_{12}^{\mathrm{A}\rightleftharpoons\mathrm{A}}\}$, and the equivalent surface electric current distributing on $\mathbb{S}_1^{\mathrm{A}}$ is denoted as $\vec{J}_1^{\mathrm{A}}$ ①, then the field distributing on $\mathbb{V}_1^{\mathrm{A}}$ can be expressed as follows:

$$\vec{F}(\vec{r})=\mathcal{F}_1\left(\vec{J}^{\mathrm{G}\rightleftharpoons\mathrm{A}}+\vec{J}_1^{\mathrm{A}\rightleftharpoons\mathrm{F}}+\vec{J}_{12}^{\mathrm{A}\rightleftharpoons\mathrm{A}}+\vec{J}_1^{\mathrm{A}},\vec{M}^{\mathrm{G}\rightleftharpoons\mathrm{A}}+\vec{M}_1^{\mathrm{A}\rightleftharpoons\mathrm{F}}+\vec{M}_{12}^{\mathrm{A}\rightleftharpoons\mathrm{A}}\right)\ ,\ \vec{r}\in\mathbb{V}_1^{\mathrm{A}} \quad (6\text{-}71)$$

where $\vec{F}=\vec{E}/\vec{H}$, and correspondingly $\mathcal{F}_1=\mathcal{E}_1/\mathcal{H}_1$, and the operator is defined as that $\mathcal{F}_1(\vec{J},\vec{M})=\overset{\leftrightarrow}{G}{}_1^{JF}*\vec{J}+\overset{\leftrightarrow}{G}{}_1^{MF}*\vec{M}$ (here, $\overset{\leftrightarrow}{G}{}_1^{JF}$ and $\overset{\leftrightarrow}{G}{}_1^{MF}$ are the dyadic Green's functions corresponding to the region $\mathbb{V}_1^{\mathrm{A}}$ with material parameters $\{\overset{\leftrightarrow}{\mu}_1,\overset{\leftrightarrow}{\varepsilon}_1,\overset{\leftrightarrow}{\sigma}_1\}$). The currents $\{\vec{J}^{\mathrm{G}\rightleftharpoons\mathrm{A}},\vec{M}^{\mathrm{G}\rightleftharpoons\mathrm{A}}\}$ & $\{\vec{J}_1^{\mathrm{A}\rightleftharpoons\mathrm{F}},\vec{M}_1^{\mathrm{A}\rightleftharpoons\mathrm{F}}\}$ & $\{\vec{J}_{12}^{\mathrm{A}\rightleftharpoons\mathrm{A}},\vec{M}_{12}^{\mathrm{A}\rightleftharpoons\mathrm{A}}\}$ and fields $\{\vec{E},\vec{H}\}$ in Eq. (6-71) satisfy the following relations

$$\hat{n}_1^{\to\mathrm{A}}\times\left[\vec{H}\left(\vec{r}_1^{\,\mathrm{A}}\right)\right]_{\vec{r}_1^{\,\mathrm{A}}\to\vec{r}}=\vec{J}(\vec{r})\quad,\quad\vec{r}\in\mathbb{X} \quad (6\text{-}72\text{a})$$

$$\left[\vec{E}\left(\vec{r}_1^{\,\mathrm{A}}\right)\right]_{\vec{r}_1^{\,\mathrm{A}}\to\vec{r}}\times\hat{n}_1^{\to\mathrm{A}}=\vec{M}(\vec{r})\quad,\quad\vec{r}\in\mathbb{X} \quad (6\text{-}72\text{b})$$

In the above Eq. (6-72), $\vec{J}=\vec{J}^{\mathrm{G}\rightleftharpoons\mathrm{A}}/\vec{J}_1^{\mathrm{A}\rightleftharpoons\mathrm{F}}/\vec{J}_{12}^{\mathrm{A}\rightleftharpoons\mathrm{A}}$ and $\vec{M}=\vec{M}^{\mathrm{G}\rightleftharpoons\mathrm{A}}/\vec{M}_1^{\mathrm{A}\rightleftharpoons\mathrm{F}}/\vec{M}_{12}^{\mathrm{A}\rightleftharpoons\mathrm{A}}$, and correspondingly $\mathbb{X}=\mathbb{S}^{\mathrm{G}\rightleftharpoons\mathrm{A}}/\mathbb{S}_1^{\mathrm{A}\rightleftharpoons\mathrm{F}}/\mathbb{S}_{12}^{\mathrm{A}\rightleftharpoons\mathrm{A}}$; point $\vec{r}_1^{\,\mathrm{A}}$ belongs to $\mathbb{V}_1^{\mathrm{A}}$, and approaches the point $\vec{r}$ on $\mathbb{S}^{\mathrm{G}\rightleftharpoons\mathrm{A}}\bigcup\mathbb{S}_1^{\mathrm{A}\rightleftharpoons\mathrm{F}}\bigcup\mathbb{S}_{12}^{\mathrm{A}\rightleftharpoons\mathrm{A}}$; $\hat{n}_1^{\to\mathrm{A}}$ is the normal direction of $\mathbb{S}^{\mathrm{G}\rightleftharpoons\mathrm{A}}\bigcup\mathbb{S}_1^{\mathrm{A}\rightleftharpoons\mathrm{F}}\bigcup\mathbb{S}_{12}^{\mathrm{A}\rightleftharpoons\mathrm{A}}$, and points to the interior of $\mathbb{V}_1^{\mathrm{A}}$.

If the equivalent surface currents distributing on $\mathbb{S}_2^{\mathrm{A}\rightleftharpoons\mathrm{F}}$ are denoted as $\{\vec{J}_2^{\mathrm{A}\rightleftharpoons\mathrm{F}},\vec{M}_2^{\mathrm{A}\rightleftharpoons\mathrm{F}}\}$, and the equivalent surface electric current distributing on $\mathbb{S}_2^{\mathrm{A}}$ is denoted as $\vec{J}_2^{\mathrm{A}}$ ②, then the field distributing on $\mathbb{V}_2^{\mathrm{A}}$ can be expressed as follows:

$$\vec{F}(\vec{r})=\mathcal{F}_2\left(-\vec{J}_{12}^{\mathrm{A}\rightleftharpoons\mathrm{A}}+\vec{J}_2^{\mathrm{A}\rightleftharpoons\mathrm{F}}+\vec{J}_2^{\mathrm{A}},-\vec{M}_{12}^{\mathrm{A}\rightleftharpoons\mathrm{A}}+\vec{M}_2^{\mathrm{A}\rightleftharpoons\mathrm{F}}\right)\quad,\quad\vec{r}\in\mathbb{V}_2^{\mathrm{A}} \quad (6\text{-}73)$$

where $\vec{F}=\vec{E}/\vec{H}$, and correspondingly $\mathcal{F}_2=\mathcal{E}_2/\mathcal{H}_2$, and the operator is defined as that $\mathcal{F}_2(\vec{J},\vec{M})=\overset{\leftrightarrow}{G}{}_2^{JF}*\vec{J}+\overset{\leftrightarrow}{G}{}_2^{MF}*\vec{M}$ (here, $\overset{\leftrightarrow}{G}{}_2^{JF}$ and $\overset{\leftrightarrow}{G}{}_2^{MF}$ are the dyadic Green's functions corresponding to the region $\mathbb{V}_2^{\mathrm{A}}$ with material parameters $\{\overset{\leftrightarrow}{\mu}_2,\overset{\leftrightarrow}{\varepsilon}_2,\overset{\leftrightarrow}{\sigma}_2\}$). The currents $\{\vec{J}_2^{\mathrm{A}\rightleftharpoons\mathrm{F}},\vec{M}_2^{\mathrm{A}\rightleftharpoons\mathrm{F}}\}$ and fields $\{\vec{E},\vec{H}\}$ in Eq. (6-73) satisfy the following relations

$$\hat{n}_2^{\to\mathrm{A}}\times\left[\vec{H}\left(\vec{r}_2^{\,\mathrm{A}}\right)\right]_{\vec{r}_2^{\,\mathrm{A}}\to\vec{r}}=\vec{J}_2^{\mathrm{A}\rightleftharpoons\mathrm{F}}(\vec{r})\quad,\quad\vec{r}\in\mathbb{S}_2^{\mathrm{A}\rightleftharpoons\mathrm{F}} \quad (6\text{-}74\text{a})$$

$$\left[\vec{E}\left(\vec{r}_2^{\,\mathrm{A}}\right)\right]_{\vec{r}_2^{\,\mathrm{A}}\to\vec{r}}\times\hat{n}_2^{\to\mathrm{A}}=\vec{M}_2^{\mathrm{A}\rightleftharpoons\mathrm{F}}(\vec{r})\quad,\quad\vec{r}\in\mathbb{S}_2^{\mathrm{A}\rightleftharpoons\mathrm{F}} \quad (6\text{-}74\text{b})$$

---

① The equivalent surface electric current distributing on $\mathbb{S}_1^{\mathrm{A}}$ is equal to the induced surface electric current distributing on $\mathbb{S}_1^{\mathrm{A}}$ [13]. The equivalent surface magnetic current distributing on $\mathbb{S}_1^{\mathrm{A}}$ is zero, because of the homogeneous tangential electric field boundary condition on $\mathbb{S}_1^{\mathrm{A}}$ [13].

② The equivalent surface electric current distributing on $\mathbb{S}_2^{\mathrm{A}}$ is equal to the induced surface electric current distributing on $\mathbb{S}_2^{\mathrm{A}}$ [13]. The equivalent surface magnetic current distributing on $\mathbb{S}_2^{\mathrm{A}}$ is zero, because of the homogeneous tangential electric field boundary condition on $\mathbb{S}_2^{\mathrm{A}}$ [13].

In the above Eq. (6-74), point $\vec{r}_2^{\,\text{A}}$ belongs to region $\mathbb{V}_2^{\text{A}}$, and approaches the point $\vec{r}$ on surface $\mathbb{S}_2^{\text{A}\rightleftharpoons\text{F}}$; $\hat{n}_2^{\rightarrow\text{A}}$ is the normal direction of surface $\mathbb{S}_2^{\text{A}\rightleftharpoons\text{F}}$, and points to the interior of region $\mathbb{V}_2^{\text{A}}$.

If the equivalent surface electric current distributing on $\mathbb{S}^{\text{F}}$ is denoted as $\vec{J}^{\,\text{F}}$ ①, then the field distributing on $\mathbb{V}^{\text{F}}$ can be expressed as follows:

$$\vec{F}\left(\vec{r}\right) = \mathcal{F}_0\left(-\vec{J}_1^{\,\text{A}\rightleftharpoons\text{F}} - \vec{J}_2^{\,\text{A}\rightleftharpoons\text{F}} + \vec{J}^{\,\text{F}}, -\vec{M}_1^{\text{A}\rightleftharpoons\text{F}} - \vec{M}_2^{\text{A}\rightleftharpoons\text{F}}\right) \quad , \quad \vec{r} \in \mathbb{V}^{\text{F}} \tag{6-75}$$

where $\vec{F} = \vec{E} / \vec{H}$, and correspondingly $\mathcal{F}_0 = \mathcal{E}_0 / \mathcal{H}_0$, and the operators are the same as the ones used in the previous Sec. 6.2.

### 6.4.3 Mathematical Description for Modal Space

Combining the Eq. (6-71) with Eq. (6-72), we immediately obtain the following integral equations

$$\begin{aligned}&\left[\mathcal{H}_1\left(\vec{J}^{\,\text{G}\rightleftharpoons\text{A}} + \vec{J}_1^{\,\text{A}\rightleftharpoons\text{F}} + \vec{J}_{12}^{\,\text{A}\rightleftharpoons\text{A}} + \vec{J}_1^{\,\text{A}}, \vec{M}^{\text{G}\rightleftharpoons\text{A}} + \vec{M}_1^{\text{A}\rightleftharpoons\text{F}} + \vec{M}_{12}^{\text{A}\rightleftharpoons\text{A}}\right)\right]_{\vec{r}_1^{\,\text{A}}\to\vec{r}}^{\tan} \\ &= \vec{J}^{\,\text{G}\rightleftharpoons\text{A}}\left(\vec{r}\right)\times\hat{n}_1^{\rightarrow\text{A}} \quad , \quad \vec{r}\in\mathbb{S}^{\text{G}\rightleftharpoons\text{A}}\end{aligned} \tag{6-76a}$$

$$\begin{aligned}&\left[\mathcal{E}_1\left(\vec{J}^{\,\text{G}\rightleftharpoons\text{A}} + \vec{J}_1^{\,\text{A}\rightleftharpoons\text{F}} + \vec{J}_{12}^{\,\text{A}\rightleftharpoons\text{A}} + \vec{J}_1^{\,\text{A}}, \vec{M}^{\text{G}\rightleftharpoons\text{A}} + \vec{M}_1^{\text{A}\rightleftharpoons\text{F}} + \vec{M}_{12}^{\text{A}\rightleftharpoons\text{A}}\right)\right]_{\vec{r}_1^{\,\text{A}}\to\vec{r}}^{\tan} \\ &= \hat{n}_1^{\rightarrow\text{A}}\times\vec{M}^{\text{G}\rightleftharpoons\text{A}}\left(\vec{r}\right) \quad , \quad \vec{r}\in\mathbb{S}^{\text{G}\rightleftharpoons\text{A}}\end{aligned} \tag{6-76b}$$

about currents $\{\vec{J}^{\,\text{G}\rightleftharpoons\text{A}}, \vec{M}^{\text{G}\rightleftharpoons\text{A}}\}$, $\{\vec{J}_1^{\,\text{A}\rightleftharpoons\text{F}}, \vec{M}_1^{\text{A}\rightleftharpoons\text{F}}\}$, $\{\vec{J}_{12}^{\,\text{A}\rightleftharpoons\text{A}}, \vec{M}_{12}^{\text{A}\rightleftharpoons\text{A}}\}$, and $\vec{J}_1^{\,\text{A}}$, where the superscript "tan" represents the tangential component of the field.

Using Eqs. (6-71)&(6-75) and employing the tangential field continuation condition on $\mathbb{S}_1^{\text{A}\rightleftharpoons\text{F}}$, there exist the following integral equations

$$\begin{aligned}&\left[\mathcal{E}_1\left(\vec{J}^{\,\text{G}\rightleftharpoons\text{A}} + \vec{J}_1^{\,\text{A}\rightleftharpoons\text{F}} + \vec{J}_{12}^{\,\text{A}\rightleftharpoons\text{A}} + \vec{J}_1^{\,\text{A}}, \vec{M}^{\text{G}\rightleftharpoons\text{A}} + \vec{M}_1^{\text{A}\rightleftharpoons\text{F}} + \vec{M}_{12}^{\text{A}\rightleftharpoons\text{A}}\right)\right]_{\vec{r}_1^{\,\text{A}}\to\vec{r}}^{\tan} \\ &= \left[\mathcal{E}_0\left(-\vec{J}_1^{\,\text{A}\rightleftharpoons\text{F}} - \vec{J}_2^{\,\text{A}\rightleftharpoons\text{F}} + \vec{J}^{\,\text{F}}, -\vec{M}_1^{\text{A}\rightleftharpoons\text{F}} - \vec{M}_2^{\text{A}\rightleftharpoons\text{F}}\right)\right]_{\vec{r}^{\,\text{F}}\to\vec{r}}^{\tan} \quad , \quad \vec{r}\in\mathbb{S}_1^{\text{A}\rightleftharpoons\text{F}}\end{aligned} \tag{6-77a}$$

$$\begin{aligned}&\left[\mathcal{H}_1\left(\vec{J}^{\,\text{G}\rightleftharpoons\text{A}} + \vec{J}_1^{\,\text{A}\rightleftharpoons\text{F}} + \vec{J}_{12}^{\,\text{A}\rightleftharpoons\text{A}} + \vec{J}_1^{\,\text{A}}, \vec{M}^{\text{G}\rightleftharpoons\text{A}} + \vec{M}_1^{\text{A}\rightleftharpoons\text{F}} + \vec{M}_{12}^{\text{A}\rightleftharpoons\text{A}}\right)\right]_{\vec{r}_1^{\,\text{A}}\to\vec{r}}^{\tan} \\ &= \left[\mathcal{H}_0\left(-\vec{J}_1^{\,\text{A}\rightleftharpoons\text{F}} - \vec{J}_2^{\,\text{A}\rightleftharpoons\text{F}} + \vec{J}^{\,\text{F}}, -\vec{M}_1^{\text{A}\rightleftharpoons\text{F}} - \vec{M}_2^{\text{A}\rightleftharpoons\text{F}}\right)\right]_{\vec{r}^{\,\text{F}}\to\vec{r}}^{\tan} \quad , \quad \vec{r}\in\mathbb{S}_1^{\text{A}\rightleftharpoons\text{F}}\end{aligned} \tag{6-77b}$$

about currents $\{\vec{J}^{\,\text{G}\rightleftharpoons\text{A}}, \vec{M}^{\text{G}\rightleftharpoons\text{A}}\}$, $\{\vec{J}_1^{\,\text{A}\rightleftharpoons\text{F}}, \vec{M}_1^{\text{A}\rightleftharpoons\text{F}}\}$, $\{\vec{J}_{12}^{\,\text{A}\rightleftharpoons\text{A}}, \vec{M}_{12}^{\text{A}\rightleftharpoons\text{A}}\}$, $\{\vec{J}_2^{\,\text{A}\rightleftharpoons\text{F}}, \vec{M}_2^{\text{A}\rightleftharpoons\text{F}}\}$, $\vec{J}_1^{\,\text{A}}$, and $\vec{J}^{\,\text{F}}$, where point $\vec{r}^{\,\text{F}}$ belongs to $\mathbb{V}^{\text{F}}$ and approaches the point $\vec{r}$ on $\mathbb{S}_1^{\text{A}\rightleftharpoons\text{F}}$.

① The equivalent surface electric current distributing on $\mathbb{S}^{\text{F}}$ is equal to the induced surface electric current distributing on $\mathbb{S}^{\text{F}}$ [13]. The equivalent surface magnetic current distributing on $\mathbb{S}^{\text{F}}$ is zero, because of the homogeneous tangential electric field boundary condition on $\mathbb{S}^{\text{F}}$ [13].

Using Eqs. (6-71)&(6-73) and employing the tangential field continuation condition on $\mathbb{S}_{12}^{\text{A}\rightleftharpoons\text{A}}$, there exist the following integral equations

$$\begin{aligned}
&\left[\mathcal{E}_1\left(\vec{J}^{\text{G}\rightleftharpoons\text{A}}+\vec{J}_1^{\text{A}\rightleftharpoons\text{F}}+\vec{J}_{12}^{\text{A}\rightleftharpoons\text{A}}+\vec{J}_1^{\text{A}},\vec{M}^{\text{G}\rightleftharpoons\text{A}}+\vec{M}_1^{\text{A}\rightleftharpoons\text{F}}+\vec{M}_{12}^{\text{A}\rightleftharpoons\text{A}}\right)\right]_{\vec{r}_1^{\text{A}}\to\vec{r}}^{\tan}\\
&=\left[\mathcal{E}_2\left(-\vec{J}_{12}^{\text{A}\rightleftharpoons\text{A}}+\vec{J}_2^{\text{A}\rightleftharpoons\text{F}}+\vec{J}_2^{\text{A}},-\vec{M}_{12}^{\text{A}\rightleftharpoons\text{A}}+\vec{M}_2^{\text{A}\rightleftharpoons\text{F}}\right)\right]_{\vec{r}_2^{\text{A}}\to\vec{r}}^{\tan} \quad , \quad \vec{r}\in\mathbb{S}_{12}^{\text{A}\rightleftharpoons\text{A}} \quad (6\text{-}78\text{a})\\
&\left[\mathcal{H}_1\left(\vec{J}^{\text{G}\rightleftharpoons\text{A}}+\vec{J}_1^{\text{A}\rightleftharpoons\text{F}}+\vec{J}_{12}^{\text{A}\rightleftharpoons\text{A}}+\vec{J}_1^{\text{A}},\vec{M}^{\text{G}\rightleftharpoons\text{A}}+\vec{M}_1^{\text{A}\rightleftharpoons\text{F}}+\vec{M}_{12}^{\text{A}\rightleftharpoons\text{A}}\right)\right]_{\vec{r}_1^{\text{A}}\to\vec{r}}^{\tan}\\
&=\left[\mathcal{H}_2\left(-\vec{J}_{12}^{\text{A}\rightleftharpoons\text{A}}+\vec{J}_2^{\text{A}\rightleftharpoons\text{F}}+\vec{J}_2^{\text{A}},-\vec{M}_{12}^{\text{A}\rightleftharpoons\text{A}}+\vec{M}_2^{\text{A}\rightleftharpoons\text{F}}\right)\right]_{\vec{r}_2^{\text{A}}\to\vec{r}}^{\tan} \quad , \quad \vec{r}\in\mathbb{S}_{12}^{\text{A}\rightleftharpoons\text{A}} \quad (6\text{-}78\text{b})
\end{aligned}$$

about currents $\{\vec{J}^{\text{G}\rightleftharpoons\text{A}},\vec{M}^{\text{G}\rightleftharpoons\text{A}}\}$, $\{\vec{J}_1^{\text{A}\rightleftharpoons\text{F}},\vec{M}_1^{\text{A}\rightleftharpoons\text{F}}\}$, $\{\vec{J}_{12}^{\text{A}\rightleftharpoons\text{A}},\vec{M}_{12}^{\text{A}\rightleftharpoons\text{A}}\}$, $\{\vec{J}_2^{\text{A}\rightleftharpoons\text{F}},\vec{M}_2^{\text{A}\rightleftharpoons\text{F}}\}$, $\vec{J}_1^{\text{A}}$, and $\vec{J}_2^{\text{A}}$, where point $\vec{r}_2^{\text{A}}$ belongs to $\mathbb{V}_2^{\text{A}}$ and approaches the point $\vec{r}$ on $\mathbb{S}_{12}^{\text{A}\rightleftharpoons\text{A}}$. Using Eqs. (6-73)&(6-75) and employing the tangential field continuation condition on $\mathbb{S}_2^{\text{A}\rightleftharpoons\text{F}}$, there exist the following integral equations

$$\begin{aligned}
&\left[\mathcal{E}_2\left(-\vec{J}_{12}^{\text{A}\rightleftharpoons\text{A}}+\vec{J}_2^{\text{A}\rightleftharpoons\text{F}}+\vec{J}_2^{\text{A}},-\vec{M}_{12}^{\text{A}\rightleftharpoons\text{A}}+\vec{M}_2^{\text{A}\rightleftharpoons\text{F}}\right)\right]_{\vec{r}_2^{\text{A}}\to\vec{r}}^{\tan}\\
&=\left[\mathcal{E}_0\left(-\vec{J}_1^{\text{A}\rightleftharpoons\text{F}}-\vec{J}_2^{\text{A}\rightleftharpoons\text{F}}+\vec{J}^{\text{F}},-\vec{M}_1^{\text{A}\rightleftharpoons\text{F}}-\vec{M}_2^{\text{A}\rightleftharpoons\text{F}}\right)\right]_{\vec{r}^{\text{F}}\to\vec{r}}^{\tan} \quad , \quad \vec{r}\in\mathbb{S}_2^{\text{A}\rightleftharpoons\text{F}} \quad (6\text{-}79\text{a})\\
&\left[\mathcal{H}_2\left(-\vec{J}_{12}^{\text{A}\rightleftharpoons\text{A}}+\vec{J}_2^{\text{A}\rightleftharpoons\text{F}}+\vec{J}_2^{\text{A}},-\vec{M}_{12}^{\text{A}\rightleftharpoons\text{A}}+\vec{M}_2^{\text{A}\rightleftharpoons\text{F}}\right)\right]_{\vec{r}_2^{\text{A}}\to\vec{r}}^{\tan}\\
&=\left[\mathcal{H}_0\left(-\vec{J}_1^{\text{A}\rightleftharpoons\text{F}}-\vec{J}_2^{\text{A}\rightleftharpoons\text{F}}+\vec{J}^{\text{F}},-\vec{M}_1^{\text{A}\rightleftharpoons\text{F}}-\vec{M}_2^{\text{A}\rightleftharpoons\text{F}}\right)\right]_{\vec{r}^{\text{F}}\to\vec{r}}^{\tan} \quad , \quad \vec{r}\in\mathbb{S}_2^{\text{A}\rightleftharpoons\text{F}} \quad (6\text{-}79\text{b})
\end{aligned}$$

about currents $\{\vec{J}_1^{\text{A}\rightleftharpoons\text{F}},\vec{M}_1^{\text{A}\rightleftharpoons\text{F}}\}$, $\{\vec{J}_{12}^{\text{A}\rightleftharpoons\text{A}},\vec{M}_{12}^{\text{A}\rightleftharpoons\text{A}}\}$, $\{\vec{J}_2^{\text{A}\rightleftharpoons\text{F}},\vec{M}_2^{\text{A}\rightleftharpoons\text{F}}\}$, $\vec{J}_2^{\text{A}}$, and $\vec{J}^{\text{F}}$.

Based on Eq. (6-71) and the homogeneous tangential electric field boundary condition on $\mathbb{S}_1^{\text{A}}$, we have the following electric field integral equation

$$\begin{aligned}
&\left[\mathcal{E}_1\left(\vec{J}^{\text{G}\rightleftharpoons\text{A}}+\vec{J}_1^{\text{A}\rightleftharpoons\text{F}}+\vec{J}_{12}^{\text{A}\rightleftharpoons\text{A}}+\vec{J}_1^{\text{A}},\vec{M}^{\text{G}\rightleftharpoons\text{A}}+\vec{M}_1^{\text{A}\rightleftharpoons\text{F}}+\vec{M}_{12}^{\text{A}\rightleftharpoons\text{A}}\right)\right]_{\vec{r}_1^{\text{A}}\to\vec{r}}^{\tan}\\
&=0 \quad , \quad \vec{r}\in\mathbb{S}_1^{\text{A}} \quad (6\text{-}80)
\end{aligned}$$

about currents $\{\vec{J}^{\text{G}\rightleftharpoons\text{A}},\vec{M}^{\text{G}\rightleftharpoons\text{A}}\}$, $\{\vec{J}_1^{\text{A}\rightleftharpoons\text{F}},\vec{M}_1^{\text{A}\rightleftharpoons\text{F}}\}$, $\{\vec{J}_{12}^{\text{A}\rightleftharpoons\text{A}},\vec{M}_{12}^{\text{A}\rightleftharpoons\text{A}}\}$, and $\vec{J}_1^{\text{A}}$. Based on Eq. (6-73) and the homogeneous tangential electric field boundary condition on $\mathbb{S}_2^{\text{A}}$, we have the following electric field integral equation

$$\left[\mathcal{E}_2\left(-\vec{J}_{12}^{\text{A}\rightleftharpoons\text{A}}+\vec{J}_2^{\text{A}\rightleftharpoons\text{F}}+\vec{J}_2^{\text{A}},-\vec{M}_{12}^{\text{A}\rightleftharpoons\text{A}}+\vec{M}_2^{\text{A}\rightleftharpoons\text{F}}\right)\right]_{\vec{r}_2^{\text{A}}\to\vec{r}}^{\tan}=0 \quad , \quad \vec{r}\in\mathbb{S}_2^{\text{A}} \quad (6\text{-}81)$$

about currents $\{\vec{J}_{12}^{\text{A}\rightleftharpoons\text{A}},\vec{M}_{12}^{\text{A}\rightleftharpoons\text{A}}\}$, $\{\vec{J}_2^{\text{A}\rightleftharpoons\text{F}},\vec{M}_2^{\text{A}\rightleftharpoons\text{F}}\}$, and $\vec{J}_2^{\text{A}}$. Based on Eq. (6-75) and the homogeneous tangential electric field boundary condition on $\mathbb{S}^{\text{F}}$, we have the following electric field integral equation

$$\left[\mathcal{E}_0\left(-\vec{J}_1^{\mathrm{A}\rightleftharpoons\mathrm{F}}-\vec{J}_2^{\mathrm{A}\rightleftharpoons\mathrm{F}}+\vec{J}^{\mathrm{F}},-\vec{M}_1^{\mathrm{A}\rightleftharpoons\mathrm{F}}-\vec{M}_2^{\mathrm{A}\rightleftharpoons\mathrm{F}}\right)\right]_{\vec{r}^{\mathrm{F}}\to\vec{r}}^{\tan}=0\quad,\quad\vec{r}\in\mathbb{S}^{\mathrm{F}}\qquad(6\text{-}82)$$

about currents $\{\vec{J}_1^{\mathrm{A}\rightleftharpoons\mathrm{F}},\vec{M}_1^{\mathrm{A}\rightleftharpoons\mathrm{F}}\}$, $\{\vec{J}_2^{\mathrm{A}\rightleftharpoons\mathrm{F}},\vec{M}_2^{\mathrm{A}\rightleftharpoons\mathrm{F}}\}$, and $\vec{J}^{\mathrm{F}}$.

The above Eqs. (6-76a)~(6-82) are a complete mathematical description for the modal space of the tra-antenna shown in Fig. 6-40. If the currents contained in the above integral equations are expanded in terms of some proper basis functions, and the Eqs. (6-76a), (6-76b), (6-77a), (6-77b), (6-78a), (6-78b), (6-79a), (6-79b), (6-80), (6-81), and (6-82) are tested with $\{\vec{b}_\xi^{\vec{M}^{\mathrm{G}\rightleftharpoons\mathrm{A}}}\}$, $\{\vec{b}_\xi^{\vec{J}^{\mathrm{G}\rightleftharpoons\mathrm{A}}}\}$, $\{\vec{b}_\xi^{\vec{J}_1^{\mathrm{A}\rightleftharpoons\mathrm{F}}}\}$, $\{\vec{b}_\xi^{\vec{M}_1^{\mathrm{A}\rightleftharpoons\mathrm{F}}}\}$, $\{\vec{b}_\xi^{\vec{J}_{12}^{\mathrm{A}\rightleftharpoons\mathrm{A}}}\}$, $\{\vec{b}_\xi^{\vec{M}_{12}^{\mathrm{A}\rightleftharpoons\mathrm{A}}}\}$, $\{\vec{b}_\xi^{\vec{J}_2^{\mathrm{A}\rightleftharpoons\mathrm{F}}}\}$, $\{\vec{b}_\xi^{\vec{M}_2^{\mathrm{A}\rightleftharpoons\mathrm{F}}}\}$, $\{\vec{b}_\xi^{\vec{J}_1^{\mathrm{A}}}\}$, $\{\vec{b}_\xi^{\vec{J}_2^{\mathrm{A}}}\}$, and $\{\vec{b}_\xi^{\vec{J}^{\mathrm{F}}}\}$ respectively, then the integral equations are immediately discretized into the following matrix equations

$$\begin{aligned}&\overline{\overline{Z}}^{\vec{M}^{\mathrm{G}\rightleftharpoons\mathrm{A}}\vec{J}^{\mathrm{G}\rightleftharpoons\mathrm{A}}}\cdot\overline{a}^{\vec{J}^{\mathrm{G}\rightleftharpoons\mathrm{A}}}+\overline{\overline{Z}}^{\vec{M}^{\mathrm{G}\rightleftharpoons\mathrm{A}}\vec{J}_1^{\mathrm{A}\rightleftharpoons\mathrm{F}}}\cdot\overline{a}^{\vec{J}_1^{\mathrm{A}\rightleftharpoons\mathrm{F}}}+\overline{\overline{Z}}^{\vec{M}^{\mathrm{G}\rightleftharpoons\mathrm{A}}\vec{J}_{12}^{\mathrm{A}\rightleftharpoons\mathrm{A}}}\cdot\overline{a}^{\vec{J}_{12}^{\mathrm{A}\rightleftharpoons\mathrm{A}}}+\overline{\overline{Z}}^{\vec{M}^{\mathrm{G}\rightleftharpoons\mathrm{A}}\vec{J}_1^{\mathrm{A}}}\cdot\overline{a}^{\vec{J}_1^{\mathrm{A}}}\\&+\overline{\overline{Z}}^{\vec{M}^{\mathrm{G}\rightleftharpoons\mathrm{A}}\vec{M}^{\mathrm{G}\rightleftharpoons\mathrm{A}}}\cdot\overline{a}^{\vec{M}^{\mathrm{G}\rightleftharpoons\mathrm{A}}}+\overline{\overline{Z}}^{\vec{M}^{\mathrm{G}\rightleftharpoons\mathrm{A}}\vec{M}_1^{\mathrm{A}\rightleftharpoons\mathrm{F}}}\cdot\overline{a}^{\vec{M}_1^{\mathrm{A}\rightleftharpoons\mathrm{F}}}+\overline{\overline{Z}}^{\vec{M}^{\mathrm{G}\rightleftharpoons\mathrm{A}}\vec{M}_{12}^{\mathrm{A}\rightleftharpoons\mathrm{A}}}\cdot\overline{a}^{\vec{M}_{12}^{\mathrm{A}\rightleftharpoons\mathrm{A}}}=0\qquad(6\text{-}83\mathrm{a})\end{aligned}$$

$$\begin{aligned}&\overline{\overline{Z}}^{\vec{J}^{\mathrm{G}\rightleftharpoons\mathrm{A}}\vec{J}^{\mathrm{G}\rightleftharpoons\mathrm{A}}}\cdot\overline{a}^{\vec{J}^{\mathrm{G}\rightleftharpoons\mathrm{A}}}+\overline{\overline{Z}}^{\vec{J}^{\mathrm{G}\rightleftharpoons\mathrm{A}}\vec{J}_1^{\mathrm{A}\rightleftharpoons\mathrm{F}}}\cdot\overline{a}^{\vec{J}_1^{\mathrm{A}\rightleftharpoons\mathrm{F}}}+\overline{\overline{Z}}^{\vec{J}^{\mathrm{G}\rightleftharpoons\mathrm{A}}\vec{J}_{12}^{\mathrm{A}\rightleftharpoons\mathrm{A}}}\cdot\overline{a}^{\vec{J}_{12}^{\mathrm{A}\rightleftharpoons\mathrm{A}}}+\overline{\overline{Z}}^{\vec{J}^{\mathrm{G}\rightleftharpoons\mathrm{A}}\vec{J}_1^{\mathrm{A}}}\cdot\overline{a}^{\vec{J}_1^{\mathrm{A}}}\\&+\overline{\overline{Z}}^{\vec{J}^{\mathrm{G}\rightleftharpoons\mathrm{A}}\vec{M}^{\mathrm{G}\rightleftharpoons\mathrm{A}}}\cdot\overline{a}^{\vec{M}^{\mathrm{G}\rightleftharpoons\mathrm{A}}}+\overline{\overline{Z}}^{\vec{J}^{\mathrm{G}\rightleftharpoons\mathrm{A}}\vec{M}_1^{\mathrm{A}\rightleftharpoons\mathrm{F}}}\cdot\overline{a}^{\vec{M}_1^{\mathrm{A}\rightleftharpoons\mathrm{F}}}+\overline{\overline{Z}}^{\vec{J}^{\mathrm{G}\rightleftharpoons\mathrm{A}}\vec{M}_{12}^{\mathrm{A}\rightleftharpoons\mathrm{A}}}\cdot\overline{a}^{\vec{M}_{12}^{\mathrm{A}\rightleftharpoons\mathrm{A}}}=0\qquad(6\text{-}83\mathrm{b})\end{aligned}$$

and

$$\begin{aligned}&\overline{\overline{Z}}^{\vec{J}_1^{\mathrm{A}\rightleftharpoons\mathrm{F}}\vec{J}^{\mathrm{G}\rightleftharpoons\mathrm{A}}}\cdot\overline{a}^{\vec{J}^{\mathrm{G}\rightleftharpoons\mathrm{A}}}+\overline{\overline{Z}}^{\vec{J}_1^{\mathrm{A}\rightleftharpoons\mathrm{F}}\vec{J}_1^{\mathrm{A}\rightleftharpoons\mathrm{F}}}\cdot\overline{a}^{\vec{J}_1^{\mathrm{A}\rightleftharpoons\mathrm{F}}}+\overline{\overline{Z}}^{\vec{J}_1^{\mathrm{A}\rightleftharpoons\mathrm{F}}\vec{J}_{12}^{\mathrm{A}\rightleftharpoons\mathrm{A}}}\cdot\overline{a}^{\vec{J}_{12}^{\mathrm{A}\rightleftharpoons\mathrm{A}}}+\overline{\overline{Z}}^{\vec{J}_1^{\mathrm{A}\rightleftharpoons\mathrm{F}}\vec{J}_2^{\mathrm{A}\rightleftharpoons\mathrm{F}}}\cdot\overline{a}^{\vec{J}_2^{\mathrm{A}\rightleftharpoons\mathrm{F}}}\\&+\overline{\overline{Z}}^{\vec{J}_1^{\mathrm{A}\rightleftharpoons\mathrm{F}}\vec{J}_1^{\mathrm{A}}}\cdot\overline{a}^{\vec{J}_1^{\mathrm{A}}}+\overline{\overline{Z}}^{\vec{J}_1^{\mathrm{A}\rightleftharpoons\mathrm{F}}\vec{J}^{\mathrm{F}}}\cdot\overline{a}^{\vec{J}^{\mathrm{F}}}+\overline{\overline{Z}}^{\vec{J}_1^{\mathrm{A}\rightleftharpoons\mathrm{F}}\vec{M}^{\mathrm{G}\rightleftharpoons\mathrm{A}}}\cdot\overline{a}^{\vec{M}^{\mathrm{G}\rightleftharpoons\mathrm{A}}}+\overline{\overline{Z}}^{\vec{J}_1^{\mathrm{A}\rightleftharpoons\mathrm{F}}\vec{M}_1^{\mathrm{A}\rightleftharpoons\mathrm{F}}}\cdot\overline{a}^{\vec{M}_1^{\mathrm{A}\rightleftharpoons\mathrm{F}}}\\&+\overline{\overline{Z}}^{\vec{J}_1^{\mathrm{A}\rightleftharpoons\mathrm{F}}\vec{M}_{12}^{\mathrm{A}\rightleftharpoons\mathrm{A}}}\cdot\overline{a}^{\vec{M}_{12}^{\mathrm{A}\rightleftharpoons\mathrm{A}}}+\overline{\overline{Z}}^{\vec{J}_1^{\mathrm{A}\rightleftharpoons\mathrm{F}}\vec{M}_2^{\mathrm{A}\rightleftharpoons\mathrm{F}}}\cdot\overline{a}^{\vec{M}_2^{\mathrm{A}\rightleftharpoons\mathrm{F}}}=0\qquad(6\text{-}84\mathrm{a})\end{aligned}$$

$$\begin{aligned}&\overline{\overline{Z}}^{\vec{M}_1^{\mathrm{A}\rightleftharpoons\mathrm{F}}\vec{J}^{\mathrm{G}\rightleftharpoons\mathrm{A}}}\cdot\overline{a}^{\vec{J}^{\mathrm{G}\rightleftharpoons\mathrm{A}}}+\overline{\overline{Z}}^{\vec{M}_1^{\mathrm{A}\rightleftharpoons\mathrm{F}}\vec{J}_1^{\mathrm{A}\rightleftharpoons\mathrm{F}}}\cdot\overline{a}^{\vec{J}_1^{\mathrm{A}\rightleftharpoons\mathrm{F}}}+\overline{\overline{Z}}^{\vec{M}_1^{\mathrm{A}\rightleftharpoons\mathrm{F}}\vec{J}_{12}^{\mathrm{A}\rightleftharpoons\mathrm{A}}}\cdot\overline{a}^{\vec{J}_{12}^{\mathrm{A}\rightleftharpoons\mathrm{A}}}+\overline{\overline{Z}}^{\vec{M}_1^{\mathrm{A}\rightleftharpoons\mathrm{F}}\vec{J}_2^{\mathrm{A}\rightleftharpoons\mathrm{F}}}\cdot\overline{a}^{\vec{J}_2^{\mathrm{A}\rightleftharpoons\mathrm{F}}}\\&+\overline{\overline{Z}}^{\vec{M}_1^{\mathrm{A}\rightleftharpoons\mathrm{F}}\vec{J}_1^{\mathrm{A}}}\cdot\overline{a}^{\vec{J}_1^{\mathrm{A}}}+\overline{\overline{Z}}^{\vec{M}_1^{\mathrm{A}\rightleftharpoons\mathrm{F}}\vec{J}^{\mathrm{F}}}\cdot\overline{a}^{\vec{J}^{\mathrm{F}}}+\overline{\overline{Z}}^{\vec{M}_1^{\mathrm{A}\rightleftharpoons\mathrm{F}}\vec{M}^{\mathrm{G}\rightleftharpoons\mathrm{A}}}\cdot\overline{a}^{\vec{M}^{\mathrm{G}\rightleftharpoons\mathrm{A}}}+\overline{\overline{Z}}^{\vec{M}_1^{\mathrm{A}\rightleftharpoons\mathrm{F}}\vec{M}_1^{\mathrm{A}\rightleftharpoons\mathrm{F}}}\cdot\overline{a}^{\vec{M}_1^{\mathrm{A}\rightleftharpoons\mathrm{F}}}\\&+\overline{\overline{Z}}^{\vec{M}_1^{\mathrm{A}\rightleftharpoons\mathrm{F}}\vec{M}_{12}^{\mathrm{A}\rightleftharpoons\mathrm{A}}}\cdot\overline{a}^{\vec{M}_{12}^{\mathrm{A}\rightleftharpoons\mathrm{A}}}+\overline{\overline{Z}}^{\vec{M}_1^{\mathrm{A}\rightleftharpoons\mathrm{F}}\vec{M}_2^{\mathrm{A}\rightleftharpoons\mathrm{F}}}\cdot\overline{a}^{\vec{M}_2^{\mathrm{A}\rightleftharpoons\mathrm{F}}}=0\qquad(6\text{-}84\mathrm{b})\end{aligned}$$

and

$$\begin{aligned}&\overline{\overline{Z}}^{\vec{J}_{12}^{\mathrm{A}\rightleftharpoons\mathrm{A}}\vec{J}^{\mathrm{G}\rightleftharpoons\mathrm{A}}}\cdot\overline{a}^{\vec{J}^{\mathrm{G}\rightleftharpoons\mathrm{A}}}+\overline{\overline{Z}}^{\vec{J}_{12}^{\mathrm{A}\rightleftharpoons\mathrm{A}}\vec{J}_1^{\mathrm{A}\rightleftharpoons\mathrm{F}}}\cdot\overline{a}^{\vec{J}_1^{\mathrm{A}\rightleftharpoons\mathrm{F}}}+\overline{\overline{Z}}^{\vec{J}_{12}^{\mathrm{A}\rightleftharpoons\mathrm{A}}\vec{J}_{12}^{\mathrm{A}\rightleftharpoons\mathrm{A}}}\cdot\overline{a}^{\vec{J}_{12}^{\mathrm{A}\rightleftharpoons\mathrm{A}}}+\overline{\overline{Z}}^{\vec{J}_{12}^{\mathrm{A}\rightleftharpoons\mathrm{A}}\vec{J}_2^{\mathrm{A}\rightleftharpoons\mathrm{F}}}\cdot\overline{a}^{\vec{J}_2^{\mathrm{A}\rightleftharpoons\mathrm{F}}}\\&+\overline{\overline{Z}}^{\vec{J}_{12}^{\mathrm{A}\rightleftharpoons\mathrm{A}}\vec{J}_1^{\mathrm{A}}}\cdot\overline{a}^{\vec{J}_1^{\mathrm{A}}}+\overline{\overline{Z}}^{\vec{J}_{12}^{\mathrm{A}\rightleftharpoons\mathrm{A}}\vec{J}_2^{\mathrm{A}}}\cdot\overline{a}^{\vec{J}_2^{\mathrm{A}}}+\overline{\overline{Z}}^{\vec{J}_{12}^{\mathrm{A}\rightleftharpoons\mathrm{A}}\vec{M}^{\mathrm{G}\rightleftharpoons\mathrm{A}}}\cdot\overline{a}^{\vec{M}^{\mathrm{G}\rightleftharpoons\mathrm{A}}}+\overline{\overline{Z}}^{\vec{J}_{12}^{\mathrm{A}\rightleftharpoons\mathrm{A}}\vec{M}_1^{\mathrm{A}\rightleftharpoons\mathrm{F}}}\cdot\overline{a}^{\vec{M}_1^{\mathrm{A}\rightleftharpoons\mathrm{F}}}\\&+\overline{\overline{Z}}^{\vec{J}_{12}^{\mathrm{A}\rightleftharpoons\mathrm{A}}\vec{M}_{12}^{\mathrm{A}\rightleftharpoons\mathrm{A}}}\cdot\overline{a}^{\vec{M}_{12}^{\mathrm{A}\rightleftharpoons\mathrm{A}}}+\overline{\overline{Z}}^{\vec{J}_{12}^{\mathrm{A}\rightleftharpoons\mathrm{A}}\vec{M}_2^{\mathrm{A}\rightleftharpoons\mathrm{F}}}\cdot\overline{a}^{\vec{M}_2^{\mathrm{A}\rightleftharpoons\mathrm{F}}}=0\qquad(6\text{-}85\mathrm{a})\end{aligned}$$

$$\begin{aligned}&\overline{\overline{Z}}^{\vec{M}_{12}^{\mathrm{A}\rightleftharpoons\mathrm{A}}\vec{J}^{\mathrm{G}\rightleftharpoons\mathrm{A}}}\cdot\overline{a}^{\vec{J}^{\mathrm{G}\rightleftharpoons\mathrm{A}}}+\overline{\overline{Z}}^{\vec{M}_{12}^{\mathrm{A}\rightleftharpoons\mathrm{A}}\vec{J}_1^{\mathrm{A}\rightleftharpoons\mathrm{F}}}\cdot\overline{a}^{\vec{J}_1^{\mathrm{A}\rightleftharpoons\mathrm{F}}}+\overline{\overline{Z}}^{\vec{M}_{12}^{\mathrm{A}\rightleftharpoons\mathrm{A}}\vec{J}_{12}^{\mathrm{A}\rightleftharpoons\mathrm{A}}}\cdot\overline{a}^{\vec{J}_{12}^{\mathrm{A}\rightleftharpoons\mathrm{A}}}+\overline{\overline{Z}}^{\vec{M}_{12}^{\mathrm{A}\rightleftharpoons\mathrm{A}}\vec{J}_2^{\mathrm{A}\rightleftharpoons\mathrm{F}}}\cdot\overline{a}^{\vec{J}_2^{\mathrm{A}\rightleftharpoons\mathrm{F}}}\\&+\overline{\overline{Z}}^{\vec{M}_{12}^{\mathrm{A}\rightleftharpoons\mathrm{A}}\vec{J}_1^{\mathrm{A}}}\cdot\overline{a}^{\vec{J}_1^{\mathrm{A}}}+\overline{\overline{Z}}^{\vec{M}_{12}^{\mathrm{A}\rightleftharpoons\mathrm{A}}\vec{J}_2^{\mathrm{A}}}\cdot\overline{a}^{\vec{J}_2^{\mathrm{A}}}+\overline{\overline{Z}}^{\vec{M}_{12}^{\mathrm{A}\rightleftharpoons\mathrm{A}}\vec{M}^{\mathrm{G}\rightleftharpoons\mathrm{A}}}\cdot\overline{a}^{\vec{M}^{\mathrm{G}\rightleftharpoons\mathrm{A}}}+\overline{\overline{Z}}^{\vec{M}_{12}^{\mathrm{A}\rightleftharpoons\mathrm{A}}\vec{M}_1^{\mathrm{A}\rightleftharpoons\mathrm{F}}}\cdot\overline{a}^{\vec{M}_1^{\mathrm{A}\rightleftharpoons\mathrm{F}}}\\&+\overline{\overline{Z}}^{\vec{M}_{12}^{\mathrm{A}\rightleftharpoons\mathrm{A}}\vec{M}_{12}^{\mathrm{A}\rightleftharpoons\mathrm{A}}}\cdot\overline{a}^{\vec{M}_{12}^{\mathrm{A}\rightleftharpoons\mathrm{A}}}+\overline{\overline{Z}}^{\vec{M}_{12}^{\mathrm{A}\rightleftharpoons\mathrm{A}}\vec{M}_2^{\mathrm{A}\rightleftharpoons\mathrm{F}}}\cdot\overline{a}^{\vec{M}_2^{\mathrm{A}\rightleftharpoons\mathrm{F}}}=0\qquad(6\text{-}85\mathrm{b})\end{aligned}$$

and

$$\begin{aligned}&\overline{\overline{Z}}^{\vec{J}_2^{\mathrm{A}\rightleftharpoons\mathrm{F}}\vec{J}_1^{\mathrm{A}\rightleftharpoons\mathrm{F}}}\cdot\overline{a}^{\vec{J}_1^{\mathrm{A}\rightleftharpoons\mathrm{F}}}+\overline{\overline{Z}}^{\vec{J}_2^{\mathrm{A}\rightleftharpoons\mathrm{F}}\vec{J}_{12}^{\mathrm{A}\rightleftharpoons\mathrm{A}}}\cdot\overline{a}^{\vec{J}_{12}^{\mathrm{A}\rightleftharpoons\mathrm{A}}}+\overline{\overline{Z}}^{\vec{J}_2^{\mathrm{A}\rightleftharpoons\mathrm{F}}\vec{J}_2^{\mathrm{A}\rightleftharpoons\mathrm{F}}}\cdot\overline{a}^{\vec{J}_2^{\mathrm{A}\rightleftharpoons\mathrm{F}}}+\overline{\overline{Z}}^{\vec{J}_2^{\mathrm{A}\rightleftharpoons\mathrm{F}}\vec{J}_2^{\mathrm{A}}}\cdot\overline{a}^{\vec{J}_2^{\mathrm{A}}}+\overline{\overline{Z}}^{\vec{J}_2^{\mathrm{A}\rightleftharpoons\mathrm{F}}\vec{J}^{\mathrm{F}}}\cdot\overline{a}^{\vec{J}^{\mathrm{F}}}\\&+\overline{\overline{Z}}^{\vec{J}_2^{\mathrm{A}\rightleftharpoons\mathrm{F}}\vec{M}_1^{\mathrm{A}\rightleftharpoons\mathrm{F}}}\cdot\overline{a}^{\vec{M}_1^{\mathrm{A}\rightleftharpoons\mathrm{F}}}+\overline{\overline{Z}}^{\vec{J}_2^{\mathrm{A}\rightleftharpoons\mathrm{F}}\vec{M}_{12}^{\mathrm{A}\rightleftharpoons\mathrm{A}}}\cdot\overline{a}^{\vec{M}_{12}^{\mathrm{A}\rightleftharpoons\mathrm{A}}}+\overline{\overline{Z}}^{\vec{J}_2^{\mathrm{A}\rightleftharpoons\mathrm{F}}\vec{M}_2^{\mathrm{A}\rightleftharpoons\mathrm{F}}}\cdot\overline{a}^{\vec{M}_2^{\mathrm{A}\rightleftharpoons\mathrm{F}}}=0\qquad(6\text{-}86\mathrm{a})\end{aligned}$$

$$\overline{\overline{Z}}^{\vec{M}_2^{\mathrm{A}\rightleftharpoons\mathrm{F}}\vec{J}_1^{\mathrm{A}\rightleftharpoons\mathrm{F}}}\cdot\overline{a}^{\vec{J}_1^{\mathrm{A}\rightleftharpoons\mathrm{F}}}+\overline{\overline{Z}}^{\vec{M}_2^{\mathrm{A}\rightleftharpoons\mathrm{F}}\vec{J}_{12}^{\mathrm{A}\rightleftharpoons\mathrm{A}}}\cdot\overline{a}^{\vec{J}_{12}^{\mathrm{A}\rightleftharpoons\mathrm{A}}}+\overline{\overline{Z}}^{\vec{M}_2^{\mathrm{A}\rightleftharpoons\mathrm{F}}\vec{J}_2^{\mathrm{A}\rightleftharpoons\mathrm{F}}}\cdot\overline{a}^{\vec{J}_2^{\mathrm{A}\rightleftharpoons\mathrm{F}}}+\overline{\overline{Z}}^{\vec{M}_2^{\mathrm{A}\rightleftharpoons\mathrm{F}}\vec{J}_2^{\mathrm{A}}}\cdot\overline{a}^{\vec{J}_2^{\mathrm{A}}}+\overline{\overline{Z}}^{\vec{M}_2^{\mathrm{A}\rightleftharpoons\mathrm{F}}\vec{J}^{\mathrm{F}}}\cdot\overline{a}^{\vec{J}^{\mathrm{F}}}$$
$$+\overline{\overline{Z}}^{\vec{M}_2^{\mathrm{A}\rightleftharpoons\mathrm{F}}\vec{M}_1^{\mathrm{A}\rightleftharpoons\mathrm{F}}}\cdot\overline{a}^{\vec{M}_1^{\mathrm{A}\rightleftharpoons\mathrm{F}}}+\overline{\overline{Z}}^{\vec{M}_2^{\mathrm{A}\rightleftharpoons\mathrm{F}}\vec{M}_{12}^{\mathrm{A}\rightleftharpoons\mathrm{A}}}\cdot\overline{a}^{\vec{M}_{12}^{\mathrm{A}\rightleftharpoons\mathrm{A}}}+\overline{\overline{Z}}^{\vec{M}_2^{\mathrm{A}\rightleftharpoons\mathrm{F}}\vec{M}_2^{\mathrm{A}\rightleftharpoons\mathrm{F}}}\cdot\overline{a}^{\vec{M}_2^{\mathrm{A}\rightleftharpoons\mathrm{F}}}=0 \qquad (6\text{-}86\text{b})$$

and

$$\overline{\overline{Z}}^{\vec{J}_1^{\mathrm{A}}\vec{J}^{\mathrm{G}\rightleftharpoons\mathrm{A}}}\cdot\overline{a}^{\vec{J}^{\mathrm{G}\rightleftharpoons\mathrm{A}}}+\overline{\overline{Z}}^{\vec{J}_1^{\mathrm{A}}\vec{J}_1^{\mathrm{A}\rightleftharpoons\mathrm{F}}}\cdot\overline{a}^{\vec{J}_1^{\mathrm{A}\rightleftharpoons\mathrm{F}}}+\overline{\overline{Z}}^{\vec{J}_1^{\mathrm{A}}\vec{J}_{12}^{\mathrm{A}\rightleftharpoons\mathrm{A}}}\cdot\overline{a}^{\vec{J}_{12}^{\mathrm{A}\rightleftharpoons\mathrm{A}}}+\overline{\overline{Z}}^{\vec{J}_1^{\mathrm{A}}\vec{J}_1^{\mathrm{A}}}\cdot\overline{a}^{\vec{J}_1^{\mathrm{A}}}+\overline{\overline{Z}}^{\vec{J}_1^{\mathrm{A}}\vec{M}^{\mathrm{G}\rightleftharpoons\mathrm{A}}}\cdot\overline{a}^{\vec{M}^{\mathrm{G}\rightleftharpoons\mathrm{A}}}$$
$$+\overline{\overline{Z}}^{\vec{J}_1^{\mathrm{A}}\vec{M}_1^{\mathrm{A}\rightleftharpoons\mathrm{F}}}\cdot\overline{a}^{\vec{M}_1^{\mathrm{A}\rightleftharpoons\mathrm{F}}}+\overline{\overline{Z}}^{\vec{J}_1^{\mathrm{A}}\vec{M}_{12}^{\mathrm{A}\rightleftharpoons\mathrm{A}}}\cdot\overline{a}^{\vec{M}_{12}^{\mathrm{A}\rightleftharpoons\mathrm{A}}}=0 \qquad (6\text{-}87)$$

and

$$\overline{\overline{Z}}^{\vec{J}_2^{\mathrm{A}}\vec{J}_{12}^{\mathrm{A}\rightleftharpoons\mathrm{A}}}\cdot\overline{a}^{\vec{J}_{12}^{\mathrm{A}\rightleftharpoons\mathrm{A}}}+\overline{\overline{Z}}^{\vec{J}_2^{\mathrm{A}}\vec{J}_2^{\mathrm{A}\rightleftharpoons\mathrm{F}}}\cdot\overline{a}^{\vec{J}_2^{\mathrm{A}\rightleftharpoons\mathrm{F}}}+\overline{\overline{Z}}^{\vec{J}_2^{\mathrm{A}}\vec{J}_2^{\mathrm{A}}}\cdot\overline{a}^{\vec{J}_2^{\mathrm{A}}}+\overline{\overline{Z}}^{\vec{J}_2^{\mathrm{A}}\vec{M}_{12}^{\mathrm{A}\rightleftharpoons\mathrm{A}}}\cdot\overline{a}^{\vec{M}_{12}^{\mathrm{A}\rightleftharpoons\mathrm{A}}}+\overline{\overline{Z}}^{\vec{J}_2^{\mathrm{A}}\vec{M}_2^{\mathrm{A}\rightleftharpoons\mathrm{F}}}\cdot\overline{a}^{\vec{M}_2^{\mathrm{A}\rightleftharpoons\mathrm{F}}}$$
$$=0 \qquad (6\text{-}88)$$

and

$$\overline{\overline{Z}}^{\vec{J}^{\mathrm{F}}\vec{J}_1^{\mathrm{A}\rightleftharpoons\mathrm{F}}}\cdot\overline{a}^{\vec{J}_1^{\mathrm{A}\rightleftharpoons\mathrm{F}}}+\overline{\overline{Z}}^{\vec{J}^{\mathrm{F}}\vec{J}_2^{\mathrm{A}\rightleftharpoons\mathrm{F}}}\cdot\overline{a}^{\vec{J}_2^{\mathrm{A}\rightleftharpoons\mathrm{F}}}+\overline{\overline{Z}}^{\vec{J}^{\mathrm{F}}\vec{J}^{\mathrm{F}}}\cdot\overline{a}^{\vec{J}^{\mathrm{F}}}+\overline{\overline{Z}}^{\vec{J}^{\mathrm{F}}\vec{M}_1^{\mathrm{A}\rightleftharpoons\mathrm{F}}}\cdot\overline{a}^{\vec{M}_1^{\mathrm{A}\rightleftharpoons\mathrm{F}}}+\overline{\overline{Z}}^{\vec{J}^{\mathrm{F}}\vec{M}_2^{\mathrm{A}\rightleftharpoons\mathrm{F}}}\cdot\overline{a}^{\vec{M}_2^{\mathrm{A}\rightleftharpoons\mathrm{F}}}$$
$$=0 \qquad (6\text{-}89)$$

The formulations used to calculate the elements of the matrices in the above matrix equations are similar to the ones given in Eqs. (6-8a)~(6-10c) and Eqs. (6-49a)~(6-52d), and they are explicitly given in the App. D6 of this report. By employing the above matrix equations, we propose two schemes for mathematically describing modal space as below.

Employing the above Eqs. (6-83a)~(6-89), we can obtain the following transformation

$$\overline{a}^{\mathrm{AV}}=\overline{\overline{T}}\cdot\overline{a} \qquad (6\text{-}90)$$

where

$$\overline{a}^{\mathrm{AV}}=\begin{bmatrix}\overline{a}^{\vec{J}^{\mathrm{G}\rightleftharpoons\mathrm{A}}}\\ \overline{a}^{\vec{J}_1^{\mathrm{A}\rightleftharpoons\mathrm{F}}}\\ \overline{a}^{\vec{J}_{12}^{\mathrm{A}\rightleftharpoons\mathrm{A}}}\\ \overline{a}^{\vec{J}_2^{\mathrm{A}\rightleftharpoons\mathrm{F}}}\\ \overline{a}^{\vec{J}_1^{\mathrm{A}}}\\ \overline{a}^{\vec{J}_2^{\mathrm{A}}}\\ \overline{a}^{\vec{J}^{\mathrm{F}}}\\ \overline{a}^{\vec{M}^{\mathrm{G}\rightleftharpoons\mathrm{A}}}\\ \overline{a}^{\vec{M}_1^{\mathrm{A}\rightleftharpoons\mathrm{F}}}\\ \overline{a}^{\vec{M}_{12}^{\mathrm{A}\rightleftharpoons\mathrm{A}}}\\ \overline{a}^{\vec{M}_2^{\mathrm{A}\rightleftharpoons\mathrm{F}}}\end{bmatrix} \qquad (6\text{-}91)$$

and the calculation formulation for the transformation matrix $\overline{\overline{T}}$ is given in the App. D6 of this report.

### 6.4.4 Power Transport Theorem and Input Power Operator

Applying the results obtained in Chap. 2 to the tra-antenna shown in Fig. 6-40, we immediately have the following PTT for the tra-antenna

$$P^{\mathrm{G}\rightleftharpoons\mathrm{A}} = P_{\mathrm{dis}}^{\mathrm{A}} + \underbrace{P_{\mathrm{rad}}^{\mathrm{I}} + j\,P_{\mathrm{sto}}^{\mathrm{F}}}_{P^{\mathrm{A}\rightleftharpoons\mathrm{F}}} + j\,P_{\mathrm{sto}}^{\mathrm{A}} \tag{6-92}$$

where $P^{\mathrm{G}\rightleftharpoons\mathrm{A}}$ is the net power inputted into the tra-antenna, and $P_{\mathrm{dis}}^{\mathrm{A}}$ is the power dissipated in the tra-antenna, and $P_{\mathrm{rad}}^{\mathrm{I}}$ is the radiated power arriving at infinity, and $P_{\mathrm{sto}}^{\mathrm{F}}$ is the power corresponding to the stored energy in free space, and $P_{\mathrm{sto}}^{\mathrm{A}}$ is the power corresponding to the stored energy in the tra-antenna, and $P^{\mathrm{A}\rightleftharpoons\mathrm{F}}$ is the net power outputted from the tra-antenna (also the net power inputted into free space).

The above-mentioned various powers are as follows:

$$P^{\mathrm{G}\rightleftharpoons\mathrm{A}} = (1/2)\iint_{\mathbb{S}^{\mathrm{G}\rightleftharpoons\mathrm{A}}} \left(\vec{E}\times\vec{H}^{\dagger}\right)\cdot\hat{n}^{\rightarrow\mathrm{A}} dS \tag{6-93a}$$

$$P^{\mathrm{A}\rightleftharpoons\mathrm{F}} = (1/2)\iint_{\mathbb{S}_1^{\mathrm{A}\rightleftharpoons\mathrm{F}}} \left(\vec{E}\times\vec{H}^{\dagger}\right)\cdot\hat{n}^{\rightarrow\mathrm{F}} dS + (1/2)\iint_{\mathbb{S}_2^{\mathrm{A}\rightleftharpoons\mathrm{F}}} \left(\vec{E}\times\vec{H}^{\dagger}\right)\cdot\hat{n}^{\rightarrow\mathrm{F}} dS \tag{6-93b}$$

$$P_{\mathrm{dis}}^{\mathrm{A}} = (1/2)\left\langle \vec{\vec{\sigma}}_1\cdot\vec{E},\vec{E}\right\rangle_{\mathbb{V}_1^{\mathrm{A}}} + (1/2)\left\langle \vec{\vec{\sigma}}_2\cdot\vec{E},\vec{E}\right\rangle_{\mathbb{V}_2^{\mathrm{A}}} \tag{6-93c}$$

$$P_{\mathrm{rad}}^{\mathrm{I}} = (1/2)\oiint_{\mathbb{S}^{\mathrm{FS}\rightarrow\mathrm{I}}} \left(\vec{E}\times\vec{H}^{\dagger}\right)\cdot\hat{n}^{\rightarrow\mathrm{I}} dS \tag{6-93d}$$

$$P_{\mathrm{sto}}^{\mathrm{F}} = 2\omega\left[(1/4)\left\langle \vec{H},\mu_0\vec{H}\right\rangle_{\mathbb{V}^{\mathrm{F}}} - (1/4)\left\langle \varepsilon_0\vec{E},\vec{E}\right\rangle_{\mathbb{V}^{\mathrm{F}}}\right] \tag{6-93e}$$

$$\begin{aligned} P_{\mathrm{sto}}^{\mathrm{A}} = & \ 2\omega\left[(1/4)\left\langle \vec{H},\vec{\vec{\mu}}_1\cdot\vec{H}\right\rangle_{\mathbb{V}_1^{\mathrm{A}}} - (1/4)\left\langle \vec{\vec{\varepsilon}}_1\cdot\vec{E},\vec{E}\right\rangle_{\mathbb{V}_1^{\mathrm{A}}}\right] \\ & + 2\omega\left[(1/4)\left\langle \vec{H},\vec{\vec{\mu}}_2\cdot\vec{H}\right\rangle_{\mathbb{V}_2^{\mathrm{A}}} - (1/4)\left\langle \vec{\vec{\varepsilon}}_2\cdot\vec{E},\vec{E}\right\rangle_{\mathbb{V}_2^{\mathrm{A}}}\right] \end{aligned} \tag{6-93f}$$

where $\hat{n}^{\rightarrow\mathrm{A}}$ is the normal direction of $\mathbb{S}^{\mathrm{G}\rightleftharpoons\mathrm{A}}$ and points to $\mathbb{V}_1^{\mathrm{A}}$, and $\hat{n}^{\rightarrow\mathrm{F}}$ is the normal direction of $\mathbb{S}_1^{\mathrm{A}\rightleftharpoons\mathrm{F}}\bigcup\mathbb{S}_2^{\mathrm{A}\rightleftharpoons\mathrm{F}}$ and points to $\mathbb{V}^{\mathrm{F}}$, and $\hat{n}^{\rightarrow\mathrm{I}}$ is the normal direction of $\mathbb{S}^{\mathrm{F}\rightarrow\mathrm{I}}$ and points to infinity.

Based on Eqs. (6-72a)&(6-72b) and the tangential continuity of the $\{\vec{E},\vec{H}\}$ on $\mathbb{S}^{\mathrm{G}\rightleftharpoons\mathrm{A}}$, the IPO $P^{\mathrm{G}\rightleftharpoons\mathrm{A}}$ given in Eq. (6-93a) can be alternatively written as follows:

$$\begin{aligned} P^{\mathrm{G}\rightleftharpoons\mathrm{A}} = & \ (1/2)\left\langle \hat{n}^{\rightarrow\mathrm{A}}\times\vec{J}^{\mathrm{G}\rightleftharpoons\mathrm{A}},\vec{M}^{\mathrm{G}\rightleftharpoons\mathrm{A}}\right\rangle_{\mathbb{S}^{\mathrm{G}\rightleftharpoons\mathrm{A}}} \\ = & -\frac{1}{2}\left\langle \vec{J}^{\mathrm{G}\rightleftharpoons\mathrm{A}},\mathcal{E}_1\left(\vec{J}^{\mathrm{G}\rightleftharpoons\mathrm{A}}+\vec{J}_1^{\mathrm{A}\rightleftharpoons\mathrm{F}}+\vec{J}_{12}^{\mathrm{A}\rightleftharpoons\mathrm{A}}+\vec{J}_1^{\mathrm{A}},\vec{M}^{\mathrm{G}\rightleftharpoons\mathrm{A}}+\vec{M}_1^{\mathrm{A}\rightleftharpoons\mathrm{F}}+\vec{M}_{12}^{\mathrm{A}\rightleftharpoons\mathrm{A}}\right)\right\rangle_{\mathbb{S}_-^{\mathrm{G}\rightleftharpoons\mathrm{A}}} \\ = & -\frac{1}{2}\left\langle \vec{M}^{\mathrm{G}\rightleftharpoons\mathrm{A}},\mathcal{H}_1\left(\vec{J}^{\mathrm{G}\rightleftharpoons\mathrm{A}}+\vec{J}_1^{\mathrm{A}\rightleftharpoons\mathrm{F}}+\vec{J}_{12}^{\mathrm{A}\rightleftharpoons\mathrm{A}}+\vec{J}_1^{\mathrm{A}},\vec{M}^{\mathrm{G}\rightleftharpoons\mathrm{A}}+\vec{M}_1^{\mathrm{A}\rightleftharpoons\mathrm{F}}+\vec{M}_{12}^{\mathrm{A}\rightleftharpoons\mathrm{A}}\right)\right\rangle_{\mathbb{S}_-^{\mathrm{G}\rightleftharpoons\mathrm{A}}}^{\dagger} \end{aligned} \tag{6-94}$$

Here, the right-hand side of the first equality is the current form of IPO, and the right-hand sides of the second and third equalities are the interaction forms of IPO.

By discretizing IPO (6-94) and utilizing transformation (6-90), we derive the following matrix form of the IPO

$$P^{\mathrm{G}\rightleftharpoons\mathrm{A}} \ = \ \bar{a}^{\dagger} \cdot \bar{\bar{P}}^{\mathrm{G}\rightleftharpoons\mathrm{A}} \cdot \bar{a} \tag{6-95}$$

and the formulations for calculating the quadratic matrix $\bar{\bar{P}}^{\mathrm{G}\rightleftharpoons\mathrm{A}}$ are similar to the ones used in Eqs. (6-23)&(6-29a)~(6-29f) and (6-63)&(6-67a)~(6-67h), and are explicitly given in the App. D6 of this report.

## 6.4.5 Input-Power-Decoupled Modes

The IP-DMs in modal space can be derived from solving the modal decoupling equation $\bar{\bar{P}}_{-}^{\mathrm{G}\rightleftharpoons\mathrm{A}} \cdot \bar{\alpha}_{\xi} = \theta_{\xi} \bar{\bar{P}}_{+}^{\mathrm{G}\rightleftharpoons\mathrm{A}} \cdot \bar{\alpha}_{\xi}$ defined on modal space, where $\bar{\bar{P}}_{+}^{\mathrm{G}\rightleftharpoons\mathrm{A}}$ and $\bar{\bar{P}}_{-}^{\mathrm{G}\rightleftharpoons\mathrm{A}}$ are the positive and negative Hermitian parts of matrix $\bar{\bar{P}}^{\mathrm{G}\rightleftharpoons\mathrm{A}}$. If some derived modes $\{\bar{\alpha}_1, \bar{\alpha}_2, \cdots, \bar{\alpha}_d\}$ are *d*-order degenerate, then the Gram-Schmidt orthogonalization process given in previous Sec. 4.2.4.1 is necessary, and it is not repeated here.

The modal fields constructed above satisfy the following frequency-domain power-decoupling relation

$$(1/2)\iint_{\mathbb{S}^{\mathrm{G}\rightleftharpoons\mathrm{A}}} \left(\vec{E}_{\zeta} \times \vec{H}_{\xi}^{\dagger}\right) \cdot \hat{n}^{\rightarrow\mathrm{A}} dS \ = \ \left(1 + j\,\theta_{\xi}\right)\delta_{\xi\zeta} \tag{6-96}$$

and the relation implies that **the IP-DMs don't have net energy coupling in one period**. By employing the decoupling relation, we have the following Parseval's identity

$$\sum_{\xi} \left|c_{\xi}\right|^{2} \ = \ \left(1/T\right)\int_{t_0}^{t_0+T}\left[\iint_{\mathbb{S}^{\mathrm{G}\rightleftharpoons\mathrm{A}}} \left(\vec{\mathcal{E}} \times \vec{\mathcal{H}}\right) \cdot \hat{n}^{\rightarrow\mathrm{A}} dS\right] dt \tag{6-97}$$

in which $\{\vec{\mathcal{E}}, \vec{\mathcal{H}}\}$ are the time-domain fields, and the expansion coefficients $c_{\xi}$ have expression $c_{\xi} = -(1/2) < \vec{J}_{\xi}^{\mathrm{G}\rightleftharpoons\mathrm{A}}, \vec{E} >_{\mathbb{S}^{\mathrm{G}\rightleftharpoons\mathrm{A}}} \big/ (1 + j\,\theta_{\xi}) = -(1/2) < \vec{H}, \vec{M}_{\xi}^{\mathrm{G}\rightleftharpoons\mathrm{A}} >_{\mathbb{S}^{\mathrm{G}\rightleftharpoons\mathrm{A}}} \big/ (1 + j\,\theta_{\xi})$, where $\{\vec{E}, \vec{H}\}$ is a previously known field distributing on input port $\mathbb{S}^{\mathrm{G}\rightleftharpoons\mathrm{A}}$.

Just like the metallic tra-antenna discussed in Sec. 6.2.4.3, we can also define the modal significance $\mathrm{MS}_{\xi} = 1 \big/ |1 + j\,\theta_{\xi}|$, modal input impedance $Z_{\xi}^{\mathrm{G}\rightleftharpoons\mathrm{A}} = P_{\xi}^{\mathrm{G}\rightleftharpoons\mathrm{A}} \big/ \left[(1/2) < \vec{J}_{\xi}^{\mathrm{G}\rightleftharpoons\mathrm{A}}, \vec{J}_{\xi}^{\mathrm{G}\rightleftharpoons\mathrm{A}} >_{\mathbb{S}^{\mathrm{G}\rightleftharpoons\mathrm{A}}}\right]$, and modal input admittance $Y_{\xi}^{\mathrm{G}\rightleftharpoons\mathrm{A}} = P_{\xi}^{\mathrm{G}\rightleftharpoons\mathrm{A}} \big/ \left[(1/2) < \vec{M}_{\xi}^{\mathrm{G}\rightleftharpoons\mathrm{A}}, \vec{M}_{\xi}^{\mathrm{G}\rightleftharpoons\mathrm{A}} >_{\mathbb{S}^{\mathrm{G}\rightleftharpoons\mathrm{A}}}\right]$ to quantitatively describe the modal features of the tra-antenna shown in Fig. 6-40.

## 6.4.6 Numerical Examples Corresponding to Typical Structures

In this subsection, we consider a circularly cylindrical DRA mounted on a circular ground plane, and the DRA is fed by a coaxial probe, as shown in Fig. 6-42.

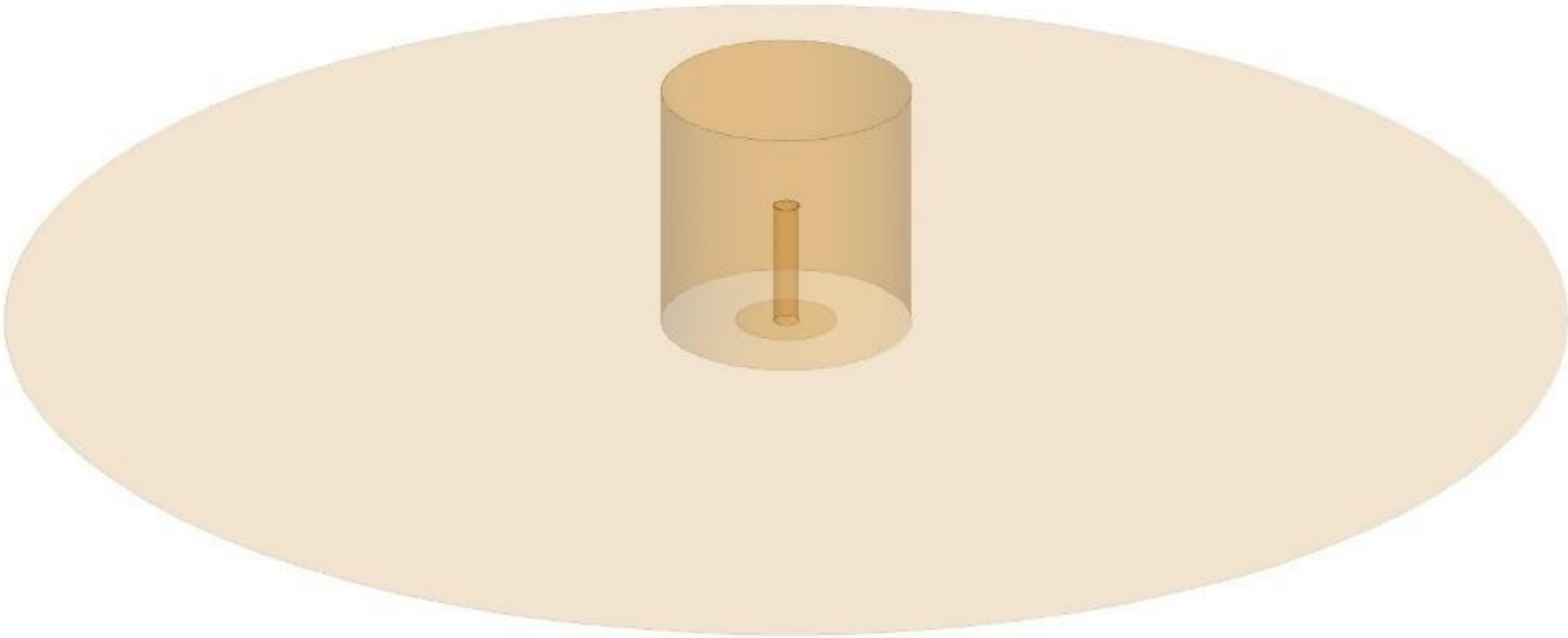

Figure 6-42 Geometry of a coaxial probe fed circularly cylindrical DRA mounted on a circular ground plane

The relative permeability, relative permittivity, and conductivity of the DRA are 1, 10, and 0 respectively. The size of the tra-antenna is shown in the following Fig. 6-43.

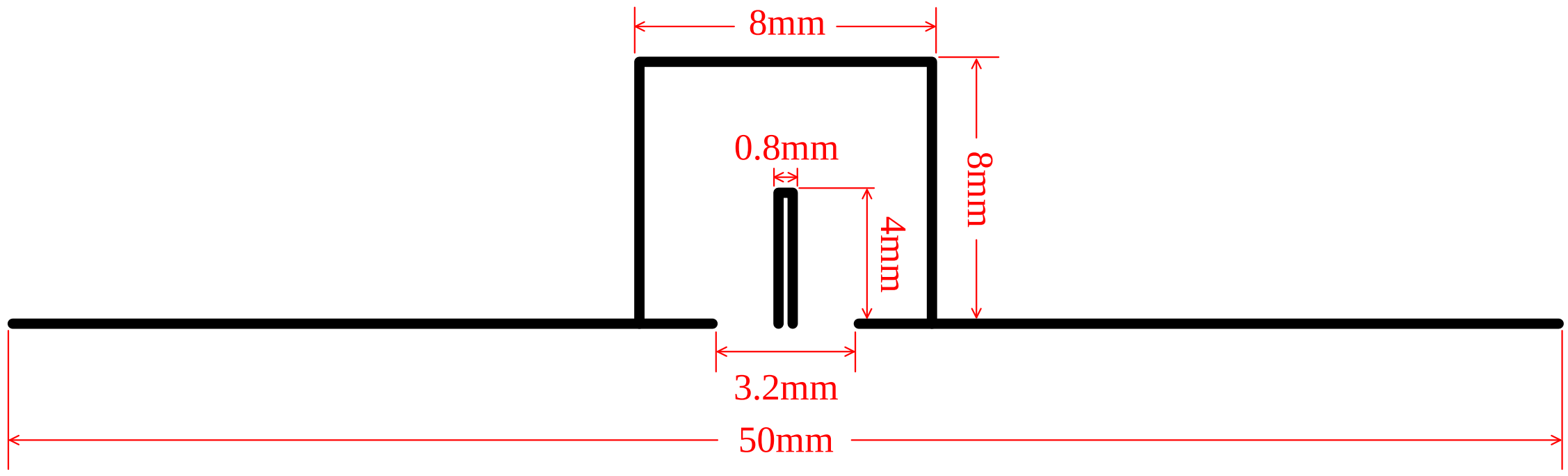

Figure 6-43 Size of the tra-antenna shown in Fig. 6-42

The topological structure and surface triangular meshes of the augmented tra-antenna, i.e. the union of the DRA and ground, are shown in the following Fig. 6-44.

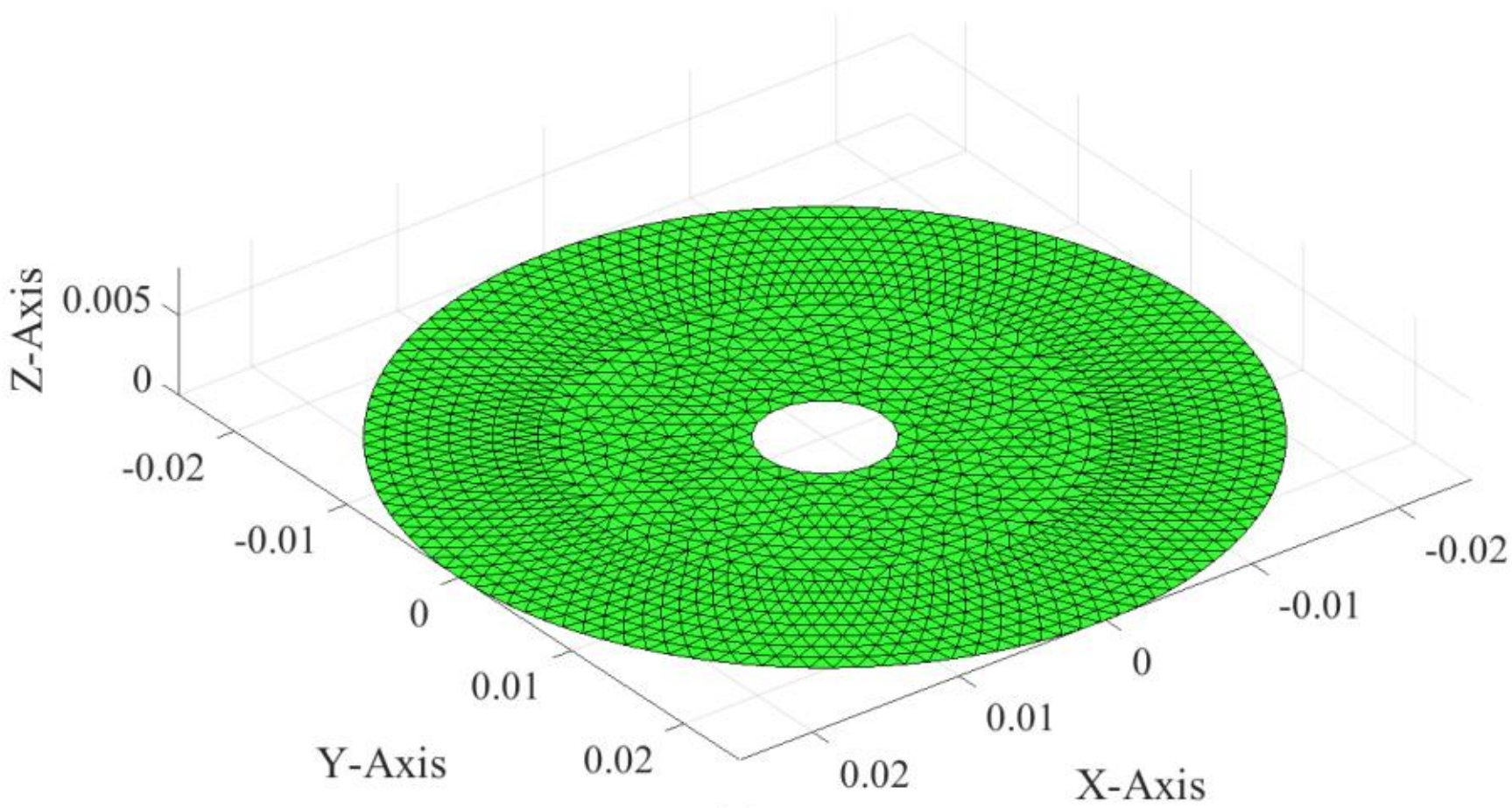

(a)

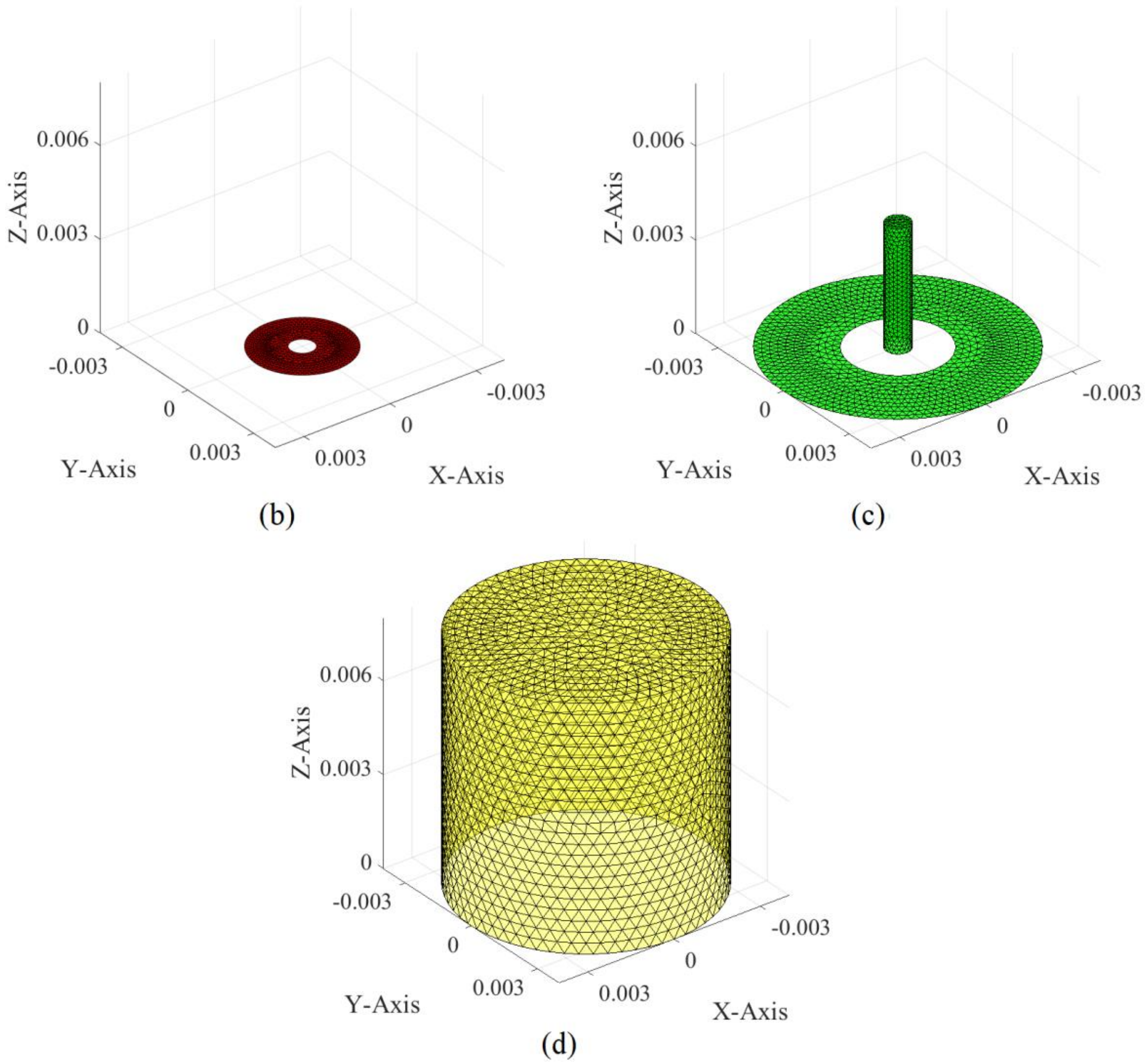

Figure 6-44 Topological structures and surface triangular meshes of the augmented tra-antenna shown in Fig. 6-42. (a) $\mathbb{S}^{F}$; (b) $\mathbb{S}^{G\rightleftharpoons A}$; (c) $\mathbb{S}^{A}$; (d) $\mathbb{S}^{A\rightleftharpoons F}$

Employing the JE-DoJ-based formulation of IPO, we calculate the IP-DMs of the tra-antenna, and plot some typical modal input resistance curves in following Fig. 6-45.

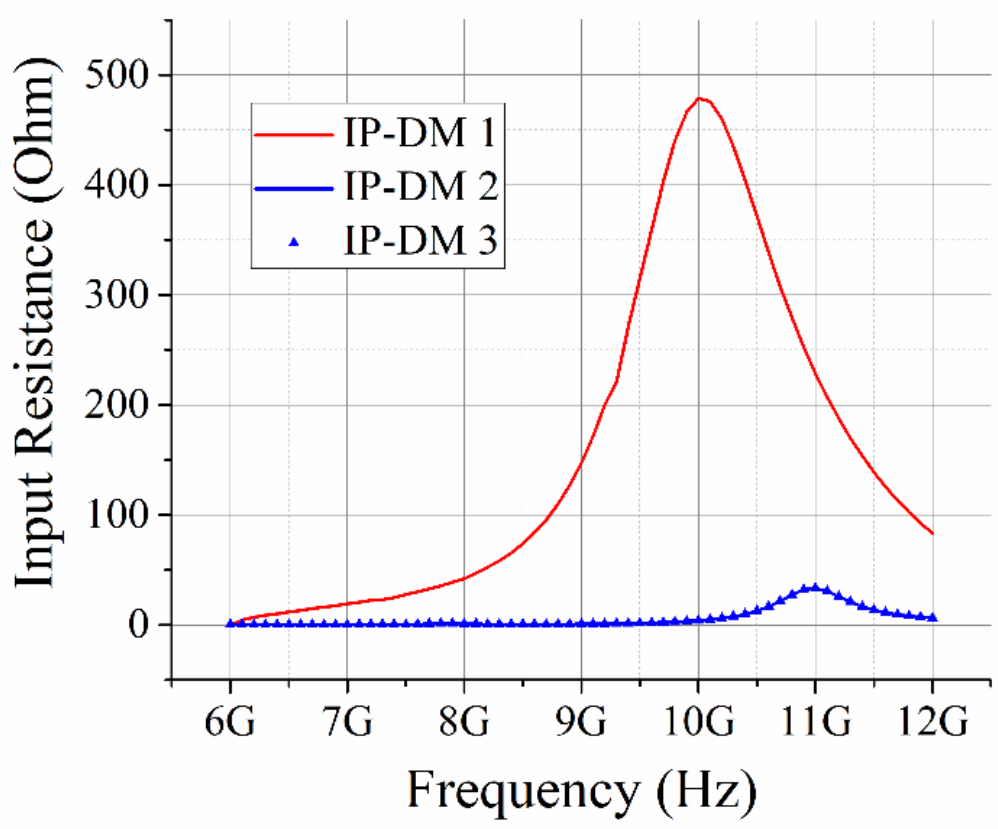

Figure 6-45 Modal resistance curves of the first several typical IP-DMs

For the IP-DM 1 working at 10 GHz, its modal $\vec{J}^{\text{G}\rightleftharpoons\text{A}}$ and $\vec{M}^{\text{G}\rightleftharpoons\text{A}}$ are shown in the following Fig. 6-46

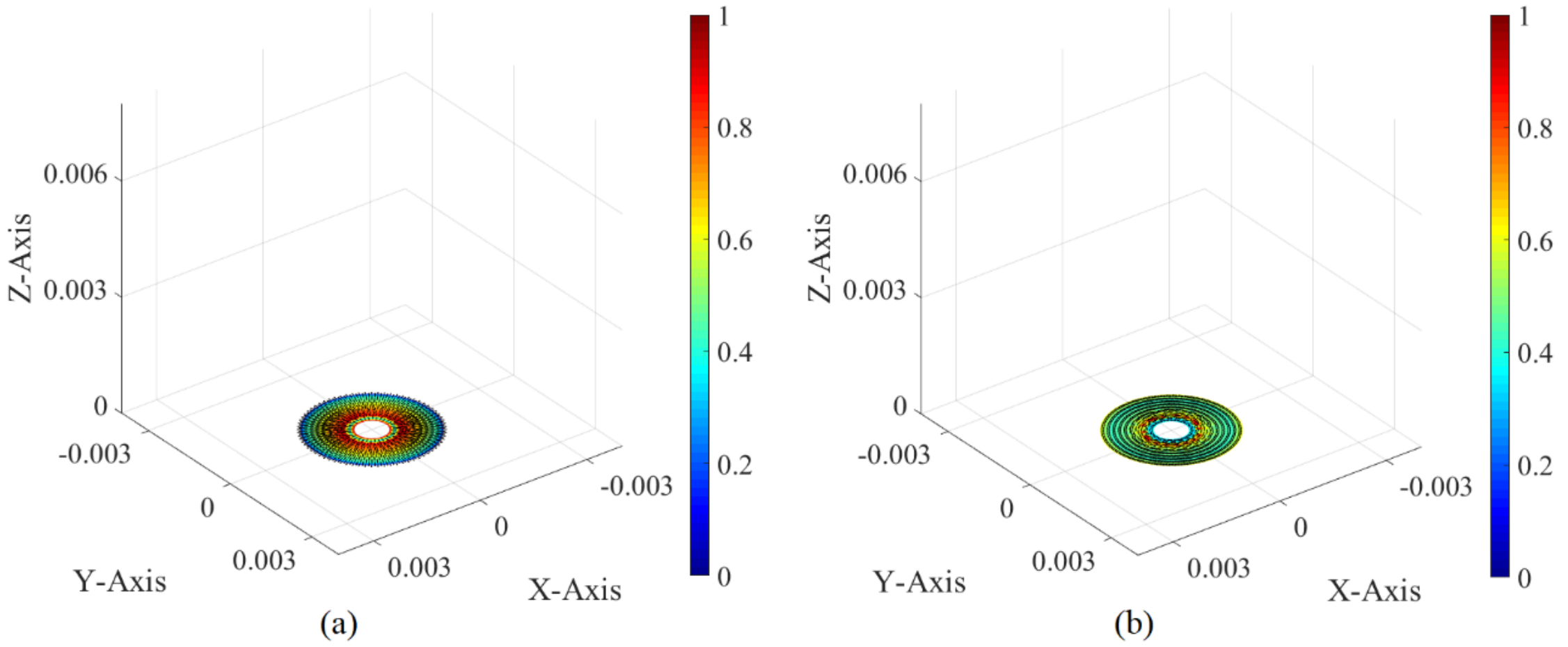

Figure 6-46 (a) Modal $\vec{J}^{\text{G}\rightleftharpoons\text{A}}$ and (b) modal $\vec{M}^{\text{G}\rightleftharpoons\text{A}}$ of the IP-DM 1 at 10 GHz

and its modal $\vec{J}^{\text{A}}$ is shown in the following Fig. 6-47

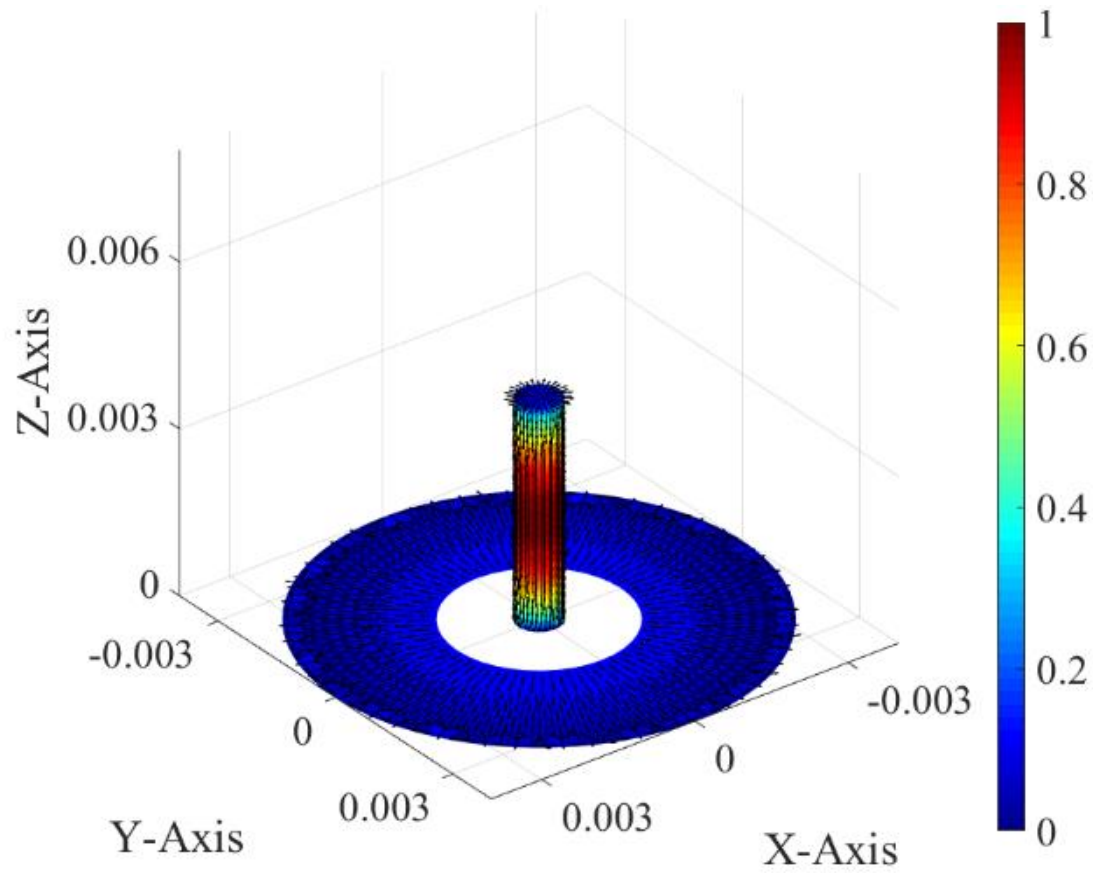

Figure 6-47 Modal $\vec{J}^{\text{A}}$ of the IP-DM 1 at 10 GHz

and its modal $\vec{J}^{\text{A}\rightleftharpoons\text{F}}$ and $\vec{M}^{\text{A}\rightleftharpoons\text{F}}$ are shown in the following Fig. 6-48

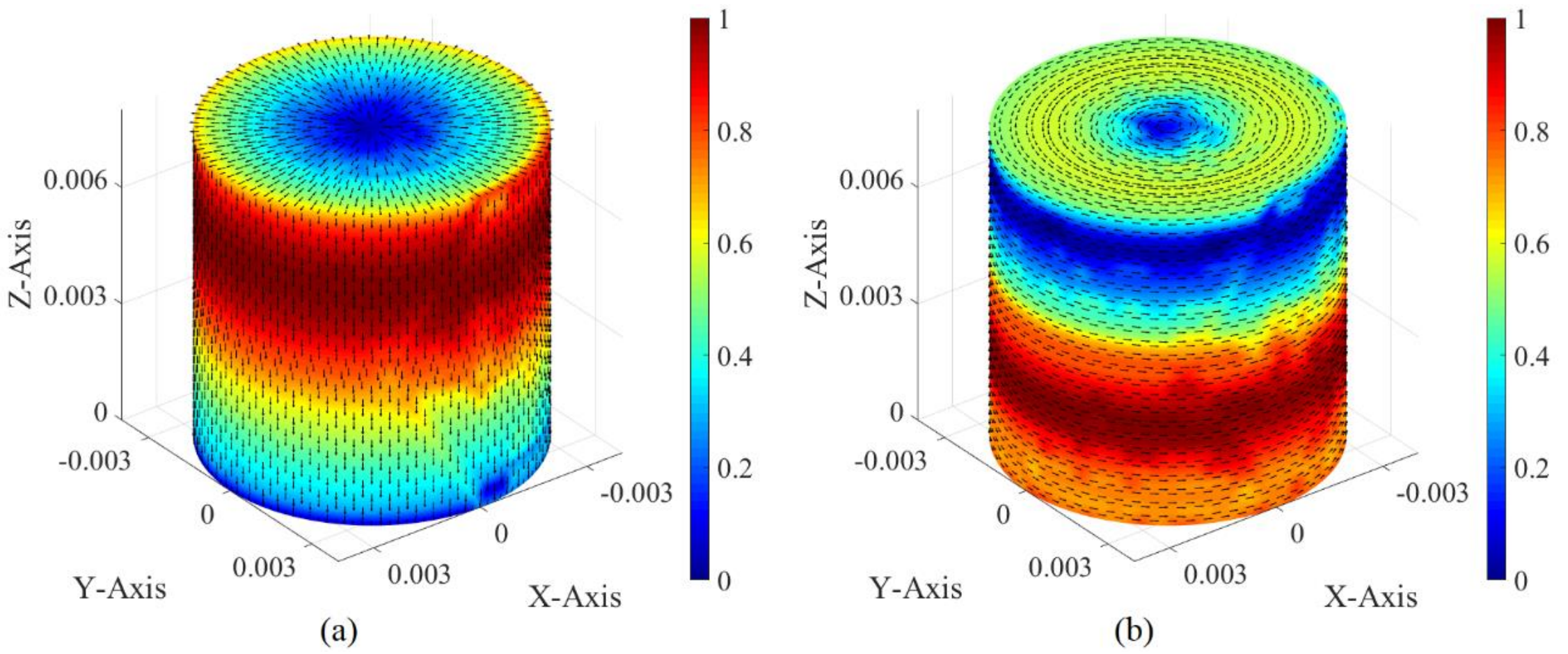

Figure 6-48 (a) Modal $\vec{J}^{\text{A}\rightleftharpoons\text{F}}$ and (b) modal $\vec{M}^{\text{A}\rightleftharpoons\text{F}}$ of the IP-DM 1 at 10 GHz

and its modal $\vec{J}^{\mathrm{F}}$ is shown in the following Fig. 6-49

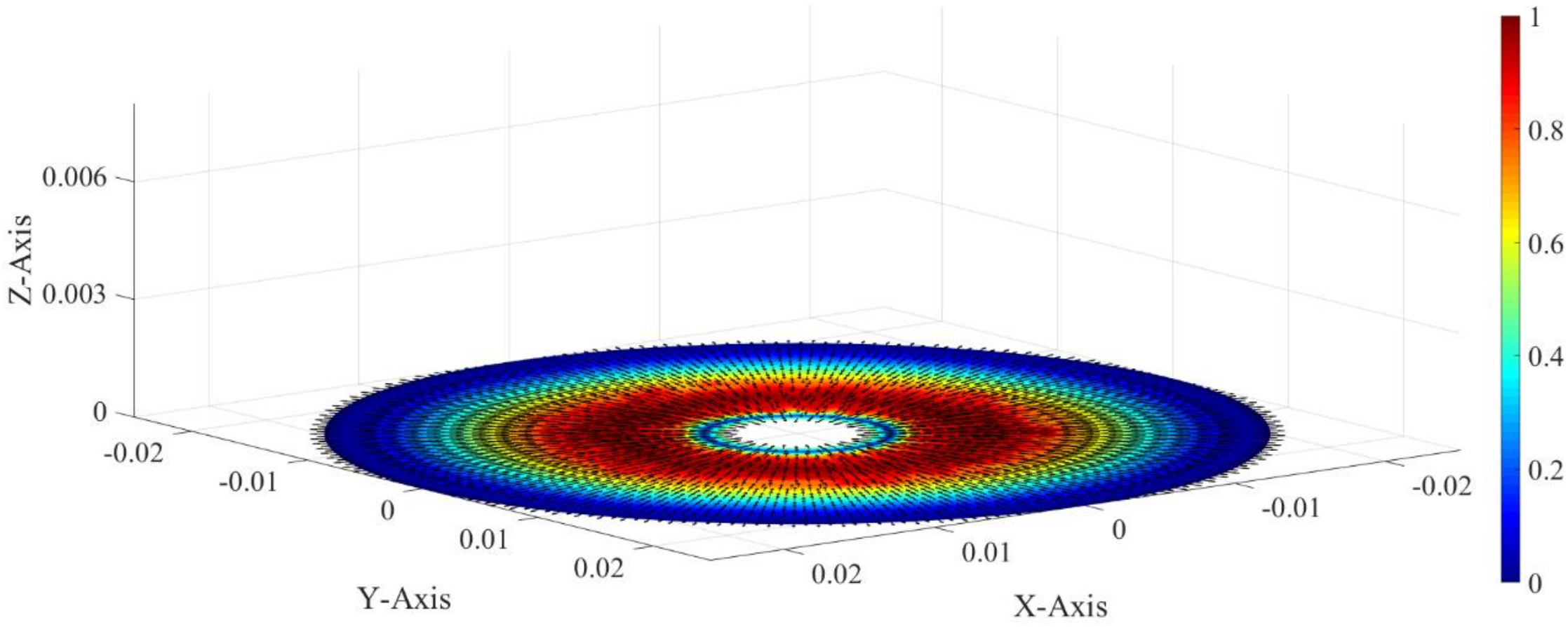

Figure 6-49 Modal $\vec{J}^{\mathrm{F}}$ of the IP-DM 1 at 10 GHz

and its modal induced volume electric current on DRA is shown in following Fig. 6-50

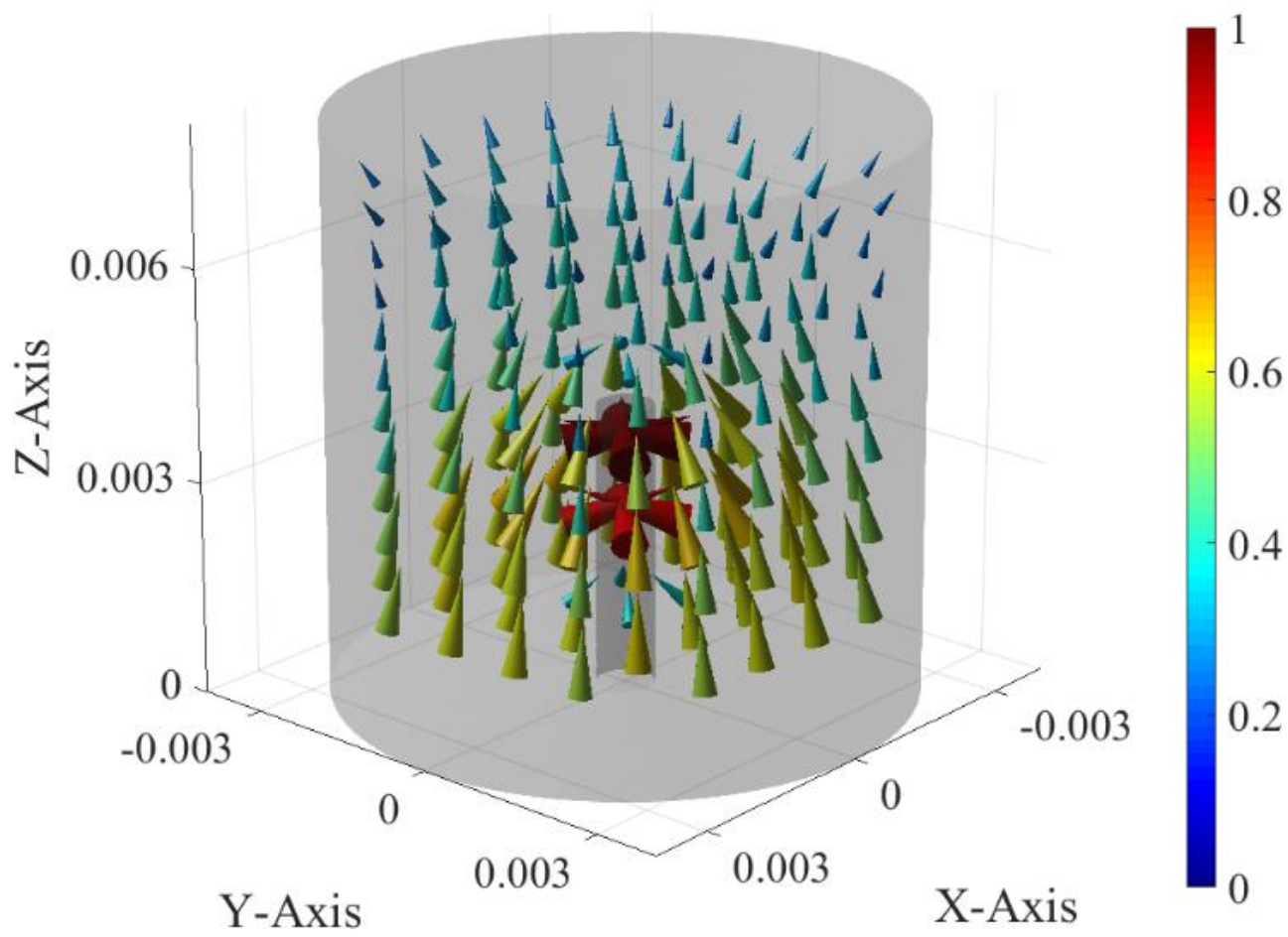

Figure 6-50 Modal induced volume electric current of the IP-DM 1 at 10 GHz

and its modal electric field distribution on yOz plane is shown in the following Fig. 6-51

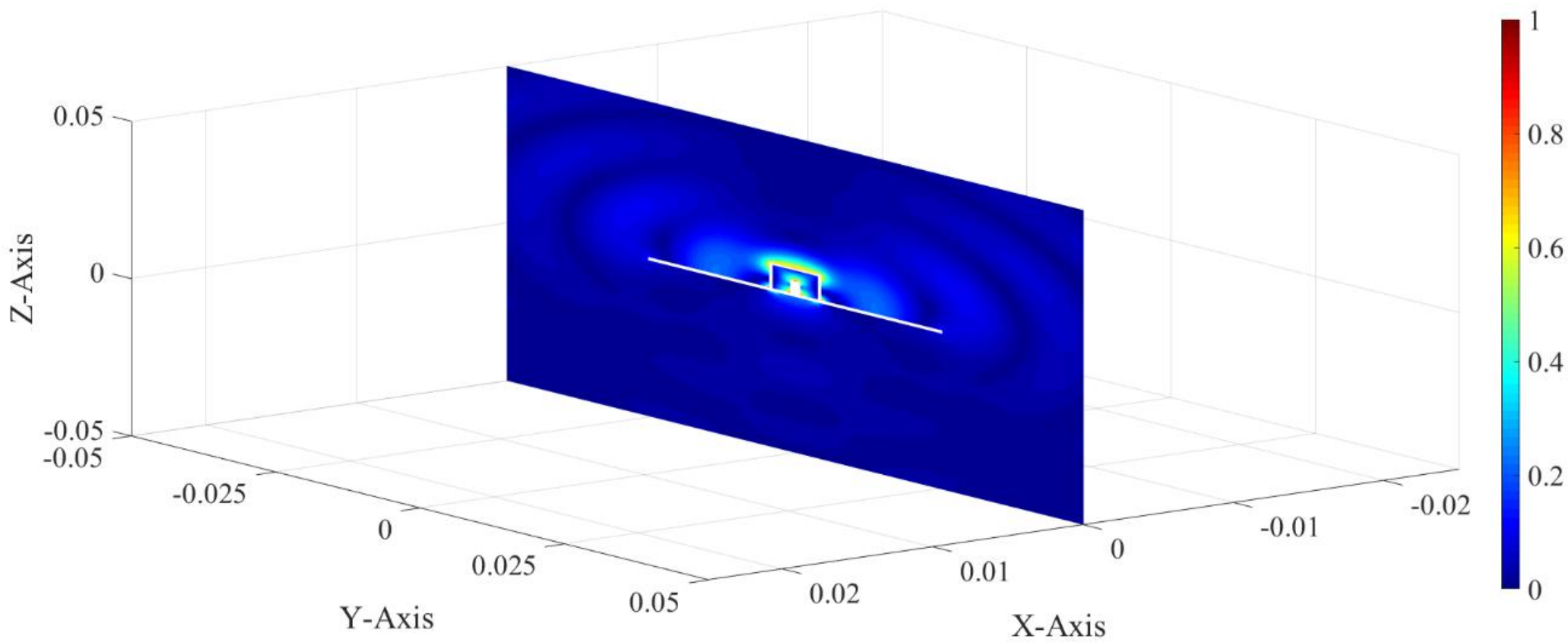

Figure 6-51 Modal electric field distribution of the IP-DM 1 at 10 GHz

and its modal radiation pattern is shown in the following Fig. 6-52

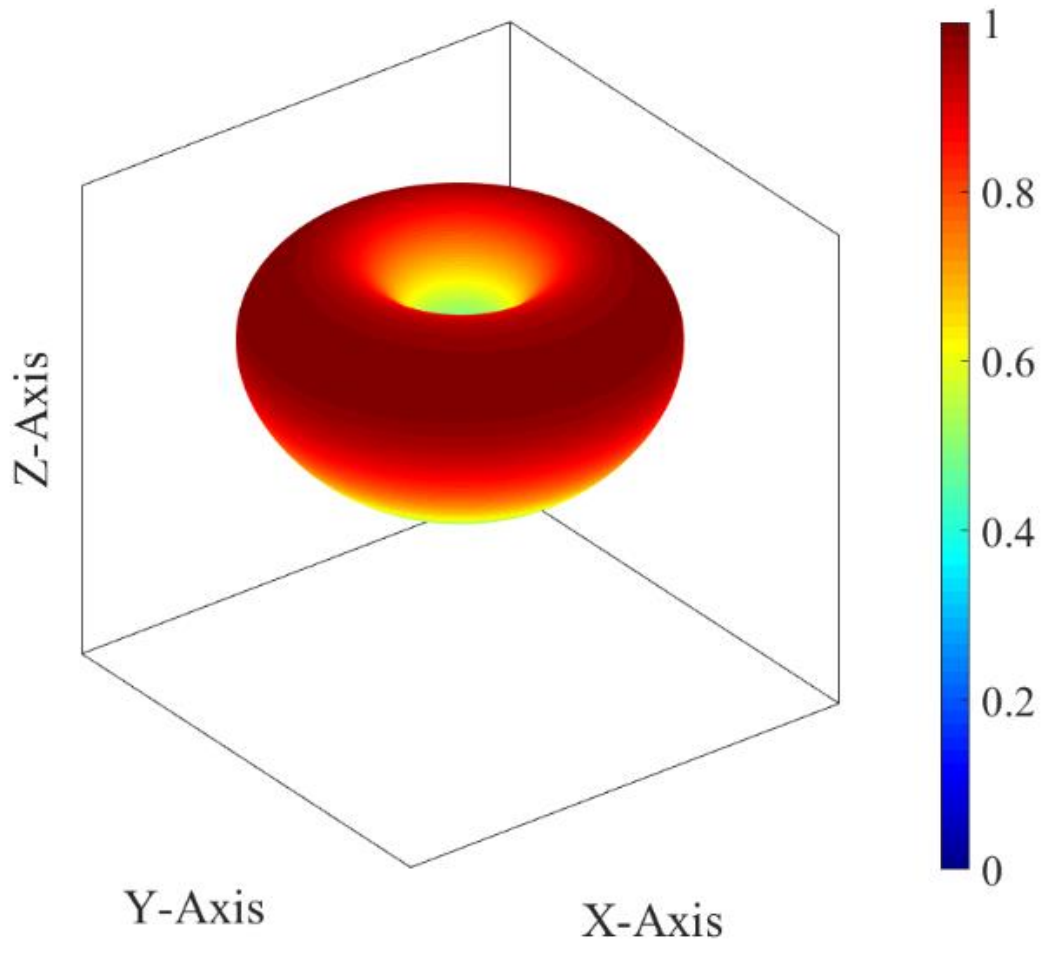

Figure 6-52 Modal radiation pattern of the IP-DM 1 at 10 GHz

## 6.5 IP-DMs of Augmented Metal-Material Composite Transmitting Antenna — Type II: Aperture Fed Stacked Printed Antenna

Printed tra-antenna has been widely used in high-speed automobiles, high-performance aircrafts and missiles, because of its low profile, simple&inexpensive to manufacture and mechanical robustness, etc. features[169~178]. The usually used feeding methods for printed tra-antenna include coaxial-probe feeding[179~188] and aperture feeding[189~198] etc.

As a typical example, the tra-antenna shown in Fig. 6-53 is focused on in this section, and it is an *aperture fed stacked printed antenna mounted on a thick metallic ground plane*.

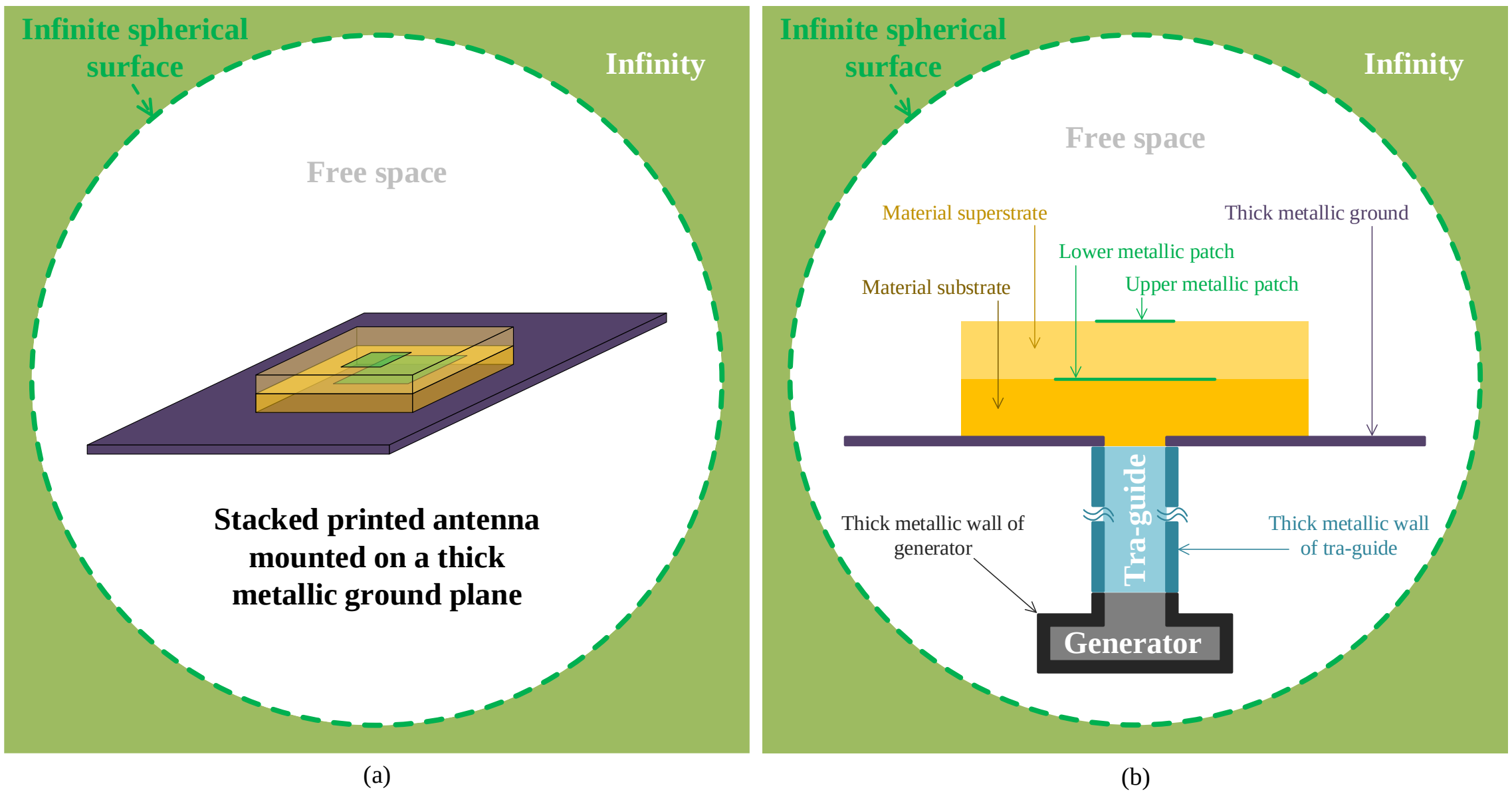

Figure 6-53 Geometry of a typical aperture fed stacked printed antenna

The fundamental principle and general process for constructing the IP-DMs of the above printed tra-antenna are completely similar to the ones used in the previous Secs. 6.2, 6.3, and 6.4.

### 6.5.1 Topological Structure

The topological structure of the tra-antenna shown in Fig. 6-53 is exhibited in the following Fig. 6-54.

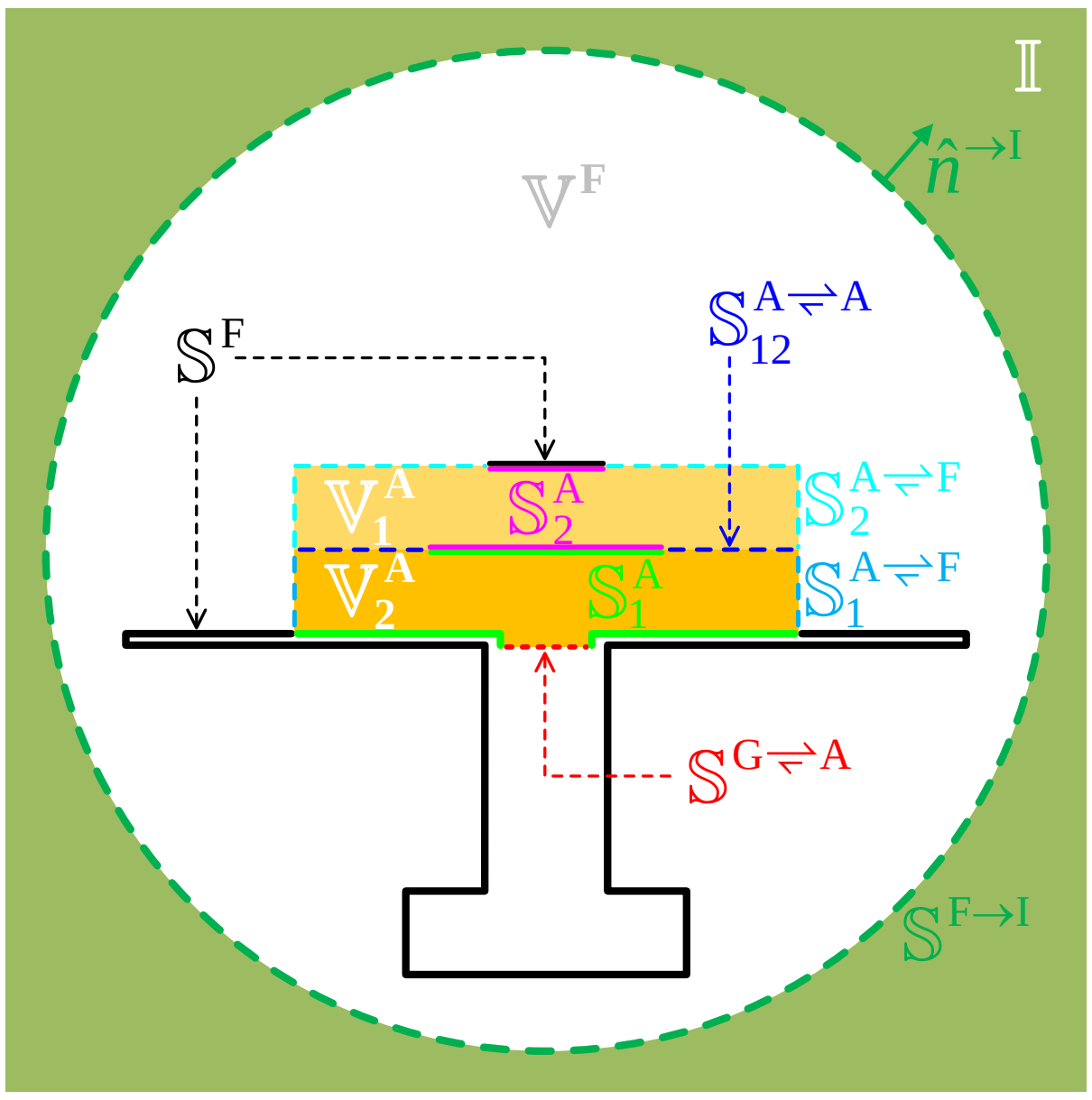

Figure 6-54 Topological structure of the tra-antenna shown in Fig. 6-53

In the above figure, $\mathbb{S}^{G \rightleftharpoons A}$ denotes the input port of the tra-antenna. The region occupied by the lower material body is denoted as $\mathbb{V}_1^{A}$, and the region occupied by the upper material body is denoted as $\mathbb{V}_2^{A}$; the region occupied by free space is denoted as $\mathbb{V}^{F}$.

The interface between $\mathbb{V}_1^{A}$ and ground plane and the interface between $\mathbb{V}_1^{A}$ and the lower metallic patch are collectively denoted as $\mathbb{S}_1^{A}$; the penetrable interface between $\mathbb{V}_1^{A}$ and $\mathbb{V}_2^{A}$ is denoted as $\mathbb{S}_{12}^{A \rightleftharpoons A}$; the interface between $\mathbb{V}_2^{A}$ and the lower metallic patch and the interface between $\mathbb{V}_2^{A}$ and the upper metallic patch are collectively denoted as $\mathbb{S}_2^{A}$; the interface between $\mathbb{V}_1^{A}$ and $\mathbb{V}^{F}$ is denoted as $\mathbb{S}_1^{A \rightleftharpoons F}$, and the interface between $\mathbb{V}_2^{A}$ and $\mathbb{V}^{F}$ is denoted as $\mathbb{S}_2^{A \rightleftharpoons F}$; the interface between $\mathbb{V}^{F}$ and ground plane and the interface between $\mathbb{V}^{F}$ and the upper metallic patch are collectively denoted as $\mathbb{S}^{F}$. The closed surface $\mathbb{S}^{S \rightarrow I}$ is a spherical surface with infinite radius.

Clearly, surfaces $\mathbb{S}^{G \rightleftharpoons A}$, $\mathbb{S}_1^{A \rightleftharpoons F}$, $\mathbb{S}_{12}^{A \rightleftharpoons A}$, and $\mathbb{S}_1^{A}$ constitute a closed surface, and the closed surface is just the boundary of $\mathbb{V}_1^{A}$, i.e., $\partial \mathbb{V}_1^{A} = \mathbb{S}^{G \rightleftharpoons A} \bigcup \mathbb{S}_1^{A \rightleftharpoons F} \bigcup \mathbb{S}_{12}^{A \rightleftharpoons A} \bigcup \mathbb{S}_1^{A}$;

surfaces $\mathbb{S}_{12}^{\mathrm{A} \rightleftharpoons \mathrm{A}}$, $\mathbb{S}_{2}^{\mathrm{A} \rightleftharpoons \mathrm{F}}$, and $\mathbb{S}_{2}^{\mathrm{A}}$ constitute a closed surface, and the closed surface is just the boundary of $\mathbb{V}_{2}^{\mathrm{A}}$, i.e., $\partial \mathbb{V}_{2}^{\mathrm{A}} = \mathbb{S}_{12}^{\mathrm{A} \rightleftharpoons \mathrm{A}} \bigcup \mathbb{S}_{2}^{\mathrm{A} \rightleftharpoons \mathrm{F}} \bigcup \mathbb{S}_{2}^{\mathrm{A}}$; surfaces $\mathbb{S}_{1}^{\mathrm{A} \rightleftharpoons \mathrm{F}}$, $\mathbb{S}_{2}^{\mathrm{A} \rightleftharpoons \mathrm{F}}$, $\mathbb{S}^{\mathrm{F}}$, and $\mathbb{S}^{\mathrm{F} \rightarrow \mathrm{I}}$ constitute the whole boundary of $\mathbb{V}^{\mathrm{F}}$, i.e., $\partial \mathbb{V}^{\mathrm{F}} = \mathbb{S}_{1}^{\mathrm{A} \rightleftharpoons \mathrm{F}} \bigcup \mathbb{S}_{2}^{\mathrm{A} \rightleftharpoons \mathrm{F}} \bigcup \mathbb{S}^{\mathrm{F}} \bigcup \mathbb{S}^{\mathrm{F} \rightarrow \mathrm{I}}$.

In addition, the permeability, permeativity, and conductivity of $\mathbb{V}_{1}^{\mathrm{A}}$ are denoted as $\ddot{\mu}_1$, $\ddot{\varepsilon}_1$, and $\ddot{\sigma}_1$ respectively; the permeability, permeativity, and conductivity of $\mathbb{V}_{2}^{\mathrm{A}}$ are denoted as $\ddot{\mu}_2$, $\ddot{\varepsilon}_2$, and $\ddot{\sigma}_2$ respectively; the permeability, permeativity, and conductivity of $\mathbb{V}^{\mathrm{F}}$ are $\mu_0$, $\varepsilon_0$, and 0 respectively.

It is not difficult to observe that: **the topological structure of the above tra-antenna shown in Figs. 6-53 and 6-54 has a one-to-one correspondence with the topological structure of the tra-antenna discussed in the previous Sec. 6.4.** In fact, this observation implies that: **the processes and formulations used to construct the IP-DMs of the above tra-antenna shown in Figs. 6-53 and 6-54 are the same as the ones given in the previous Sec. 6.4.** Thus, the processes and formulations are not repeated here for reducing the length of this report. But, we will provide some typical examples related to the printed tra-antenna in the following subsection.

### 6.5.2 Numerical Examples Corresponding to Typical Structures

In this subsection, we consider a rectangular microstrip antenna (with a circular metallic patch) mounted on a rectangular metallic ground plane, and the microstrip antenna is fed by a circular aperture, as shown in the following Fig. 6-55.

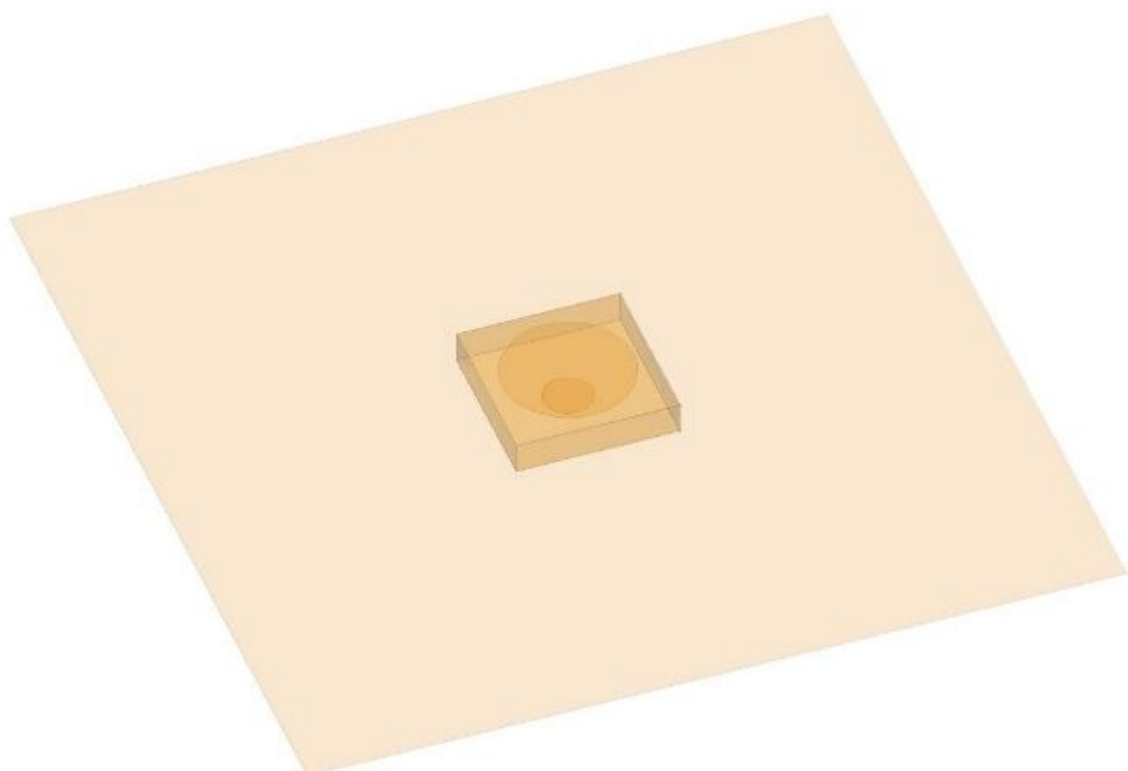

Figure 6-55 Geometry of a circular aperture fed rectangular microstrip antenna (with a circular metallic patch) mounted on a rectangular metallic ground plane

The relative permeability, relative permittivity, and conductivity of the material substrate are 1, 5, and 0 respectively. The size of the tra-antenna is shown in the following Fig. 6-56.

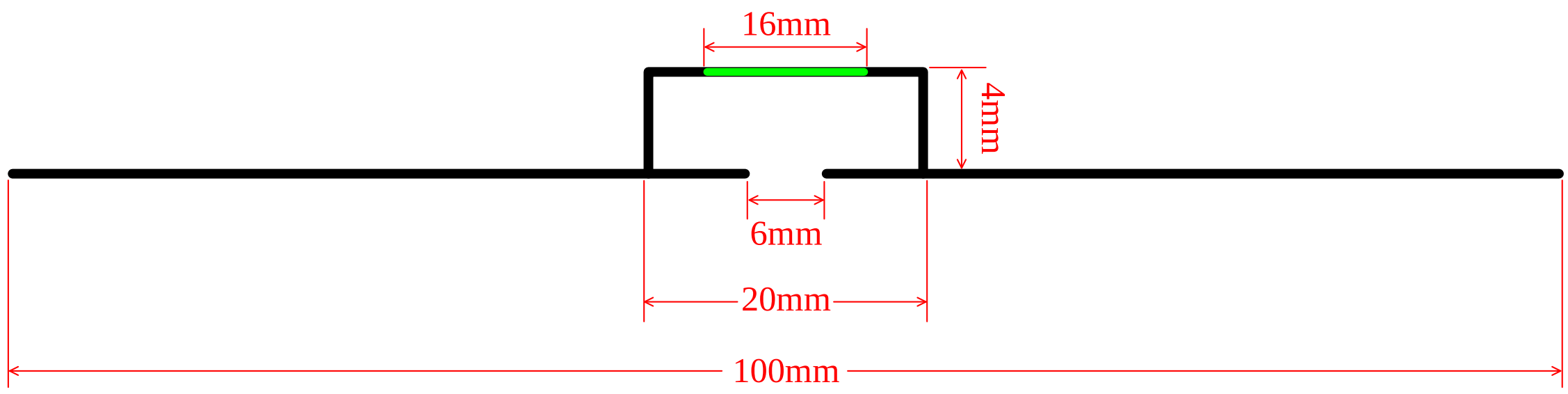

Figure 6-56 Size of tra-antenna shown in Fig. 6-55

The topological structure and surface triangular meshes of the augmented tra-antenna, i.e. the union of the microstrip antenna and ground, are shown in the following Fig. 6-57.

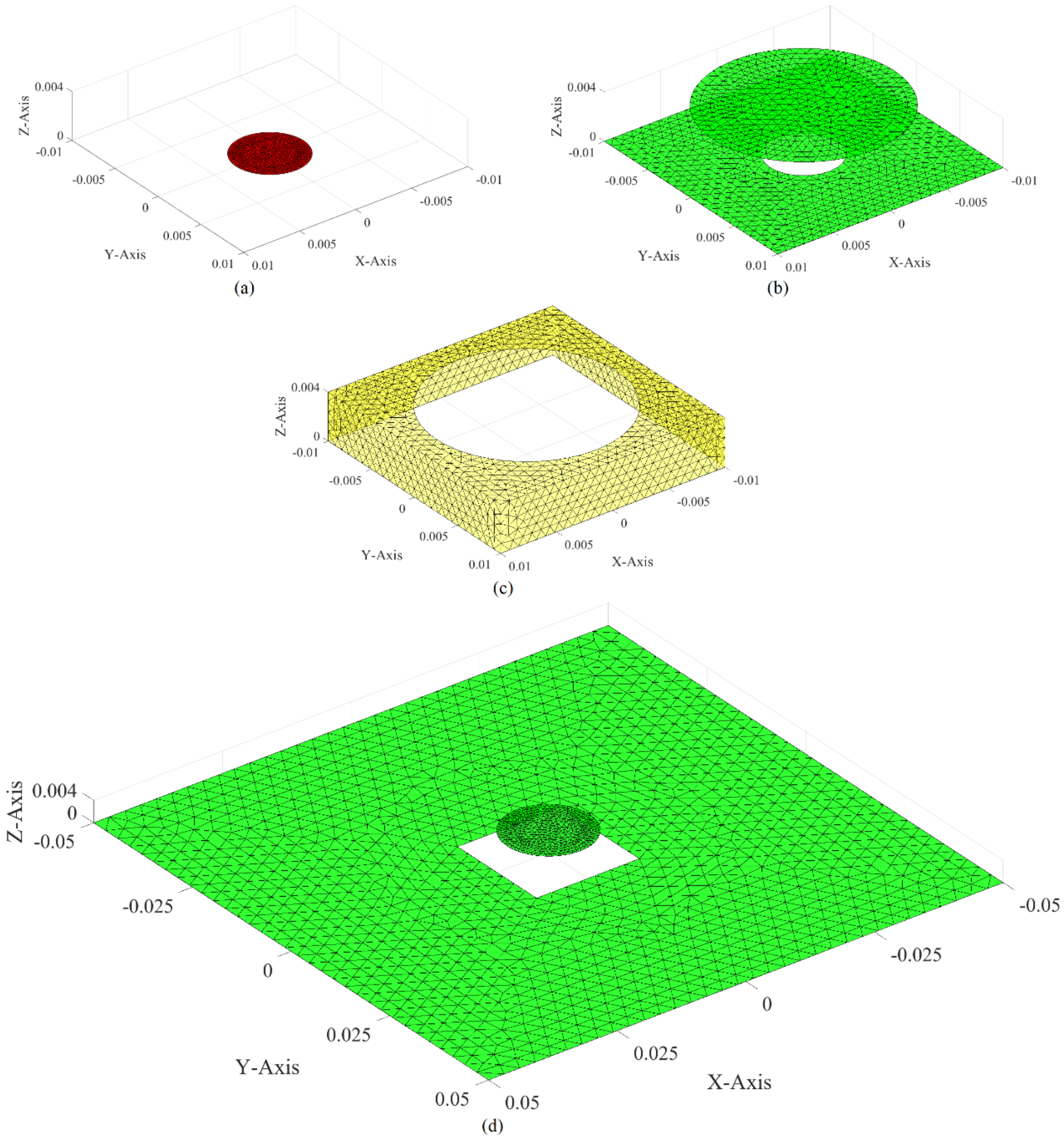

Figure 6-57 Topological structures and surface triangular meshes of the augmented tra-antenna shown in Fig. 6-55. (a) $\mathbb{S}^{G\rightleftharpoons A}$ ; (b) $\mathbb{S}^{A}$ ; (c) $\mathbb{S}^{A\rightleftharpoons F}$ ; (d) $\mathbb{S}^{F}$

Completely the same as the previous Sec. 6.4.6, we calculate the IP-DMs of the augmented tra-antenna by employing the JE-DoJ-based formulation of IPO. The modal input resistance curves of the first several typical modes are shown in the following Fig. 6-58.

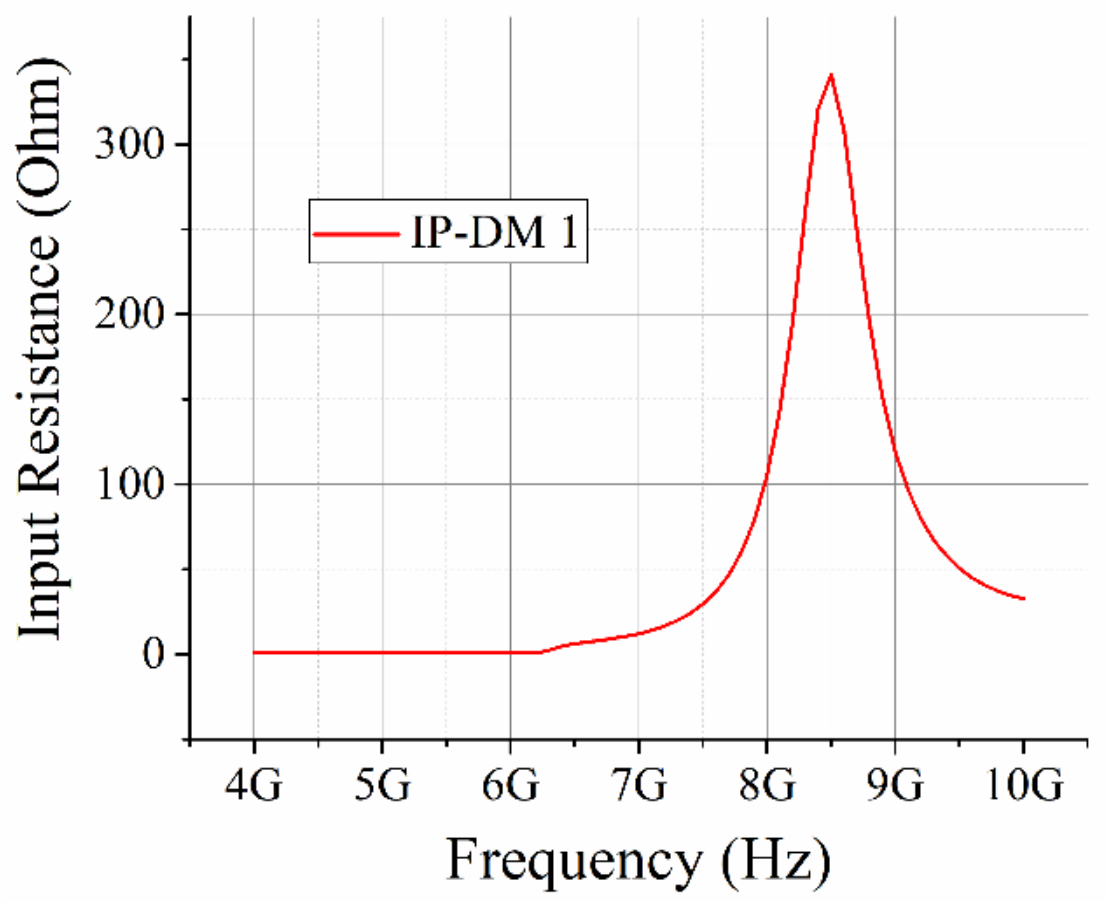

Figure 6-58 Modal resistance curves of the first several typical IP-DMs

For the IP-DM 1 working at 8.5 GHz, its modal $\vec{J}^{\mathrm{G}\rightleftharpoons\mathrm{A}}$ and $\vec{M}^{\mathrm{G}\rightleftharpoons\mathrm{A}}$ are shown in the following Fig. 6-59

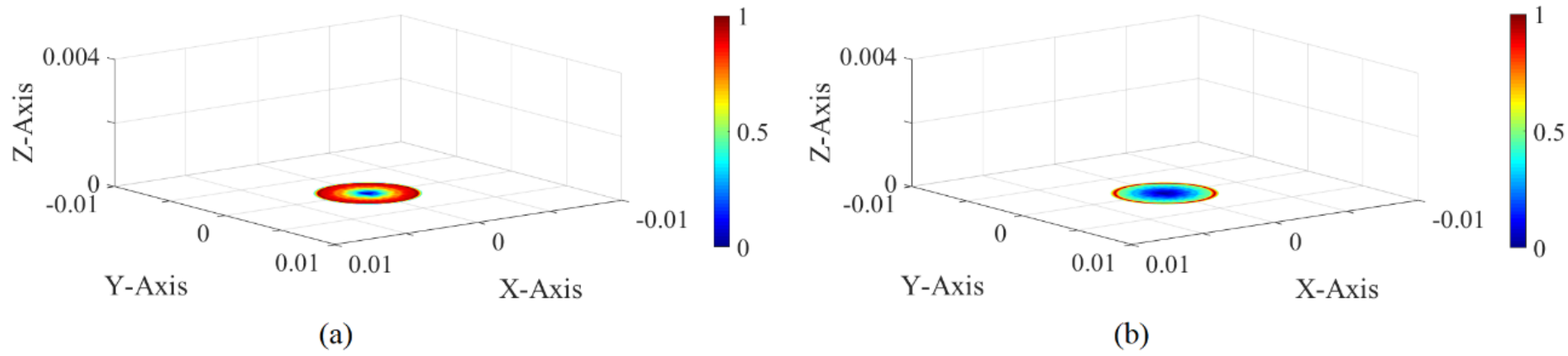

Figure 6-59 (a) Modal $\vec{J}^{\mathrm{G}\rightleftharpoons\mathrm{A}}$ and (b) modal $\vec{M}^{\mathrm{G}\rightleftharpoons\mathrm{A}}$ of the IP-DM 1 at 8.5 GHz

and its modal $\vec{J}^{\mathrm{A}}$ is shown in the following Fig. 6-60

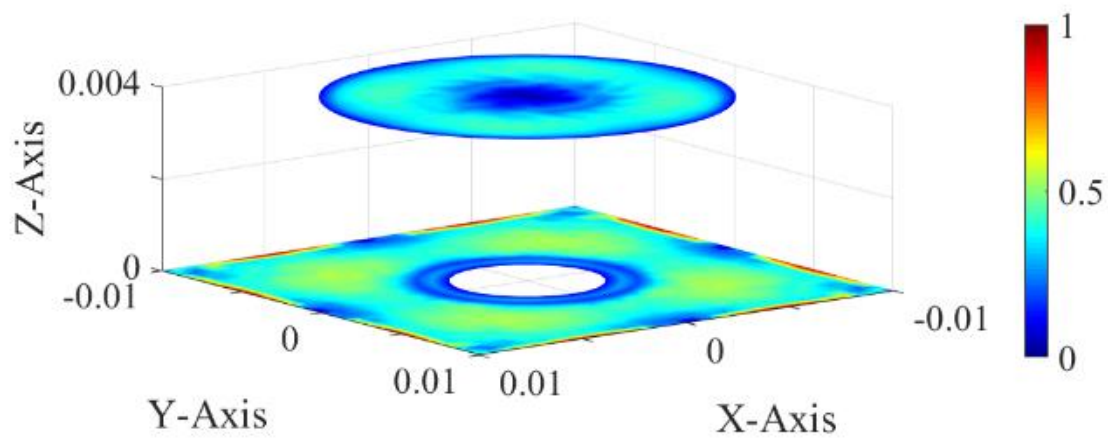

Figure 6-60 Modal $\vec{J}^{\mathrm{A}}$ of the IP-DM 1 at 8.5 GHz

and its modal $\vec{J}^{\mathrm{A}\rightleftharpoons\mathrm{F}}$ and $\vec{M}^{\mathrm{A}\rightleftharpoons\mathrm{F}}$ are shown in the following Fig. 6-61

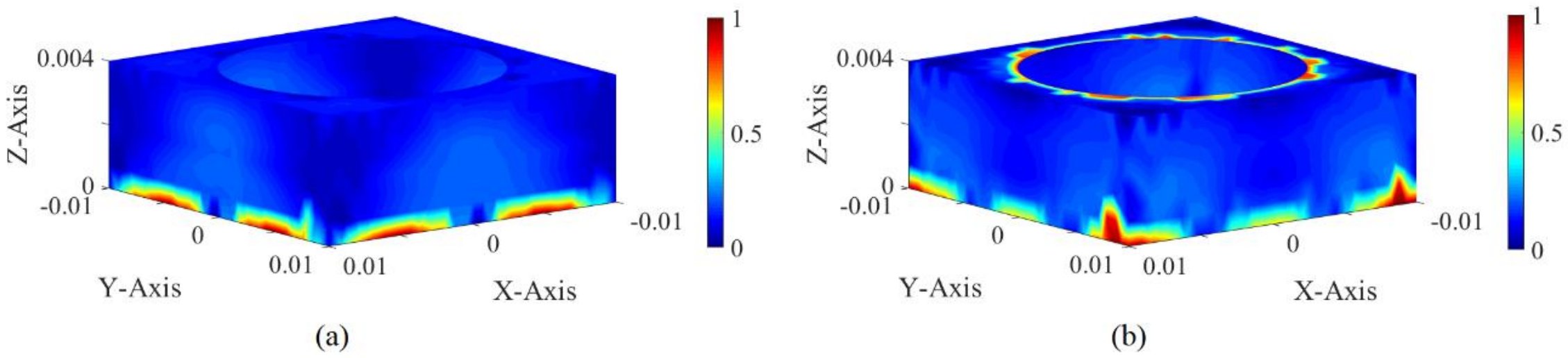

Figure 6-61 (a) Modal $\vec{J}^{\text{A}\rightleftharpoons\text{F}}$ and (b) modal $\vec{M}^{\text{A}\rightleftharpoons\text{F}}$ of the IP-DM 1 at 8.5 GHz

and its modal $\vec{J}^{\text{F}}$ is shown in the following Fig. 6-62

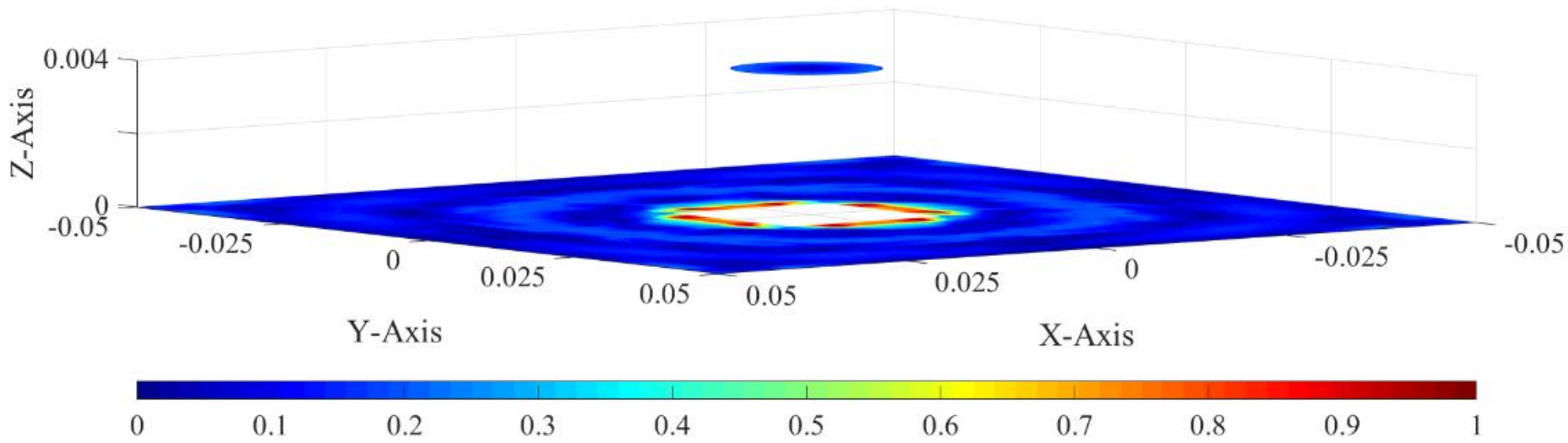

Figure 6-62 Modal $\vec{J}^{\text{F}}$ of the IP-DM 1 at 8.5 GHz

In addition, we can also obtain the modal induced volume electric current on the substrate by using the above-mentioned various equivalent surface electric and magnetic currents and by employing the surface equivalence principle, and we visually exhibit the modal induced volume electric current corresponding to the IP-DM 1 working at 8.5 GHz in the following Fig. 6-63

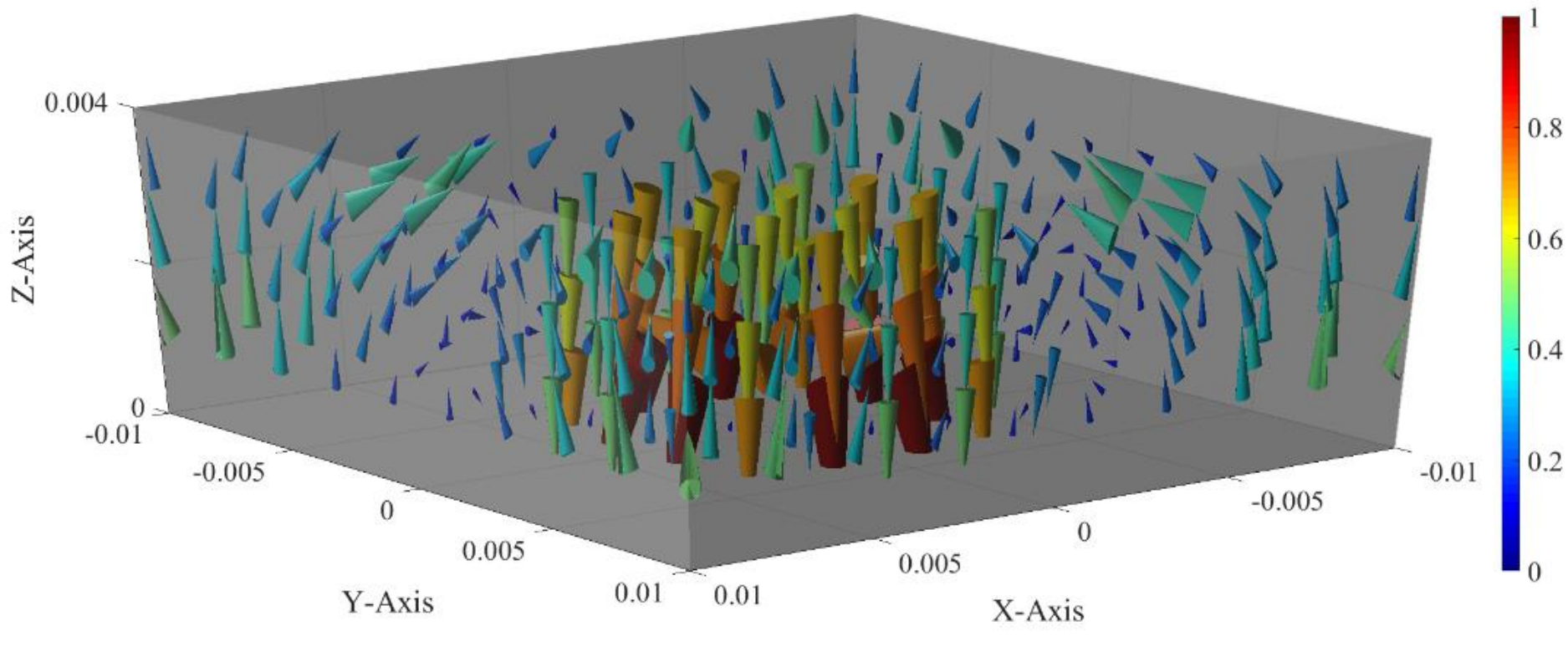

Figure 6-63 Modal induced volume electric current of the IP-DM 1 at 8.5 GHz

Similarly, we can also calculate the modal electric field distribution based on the currents and surface equivalence principle, and we plot the modal electric field on yOz plane in the following Fig. 6-64

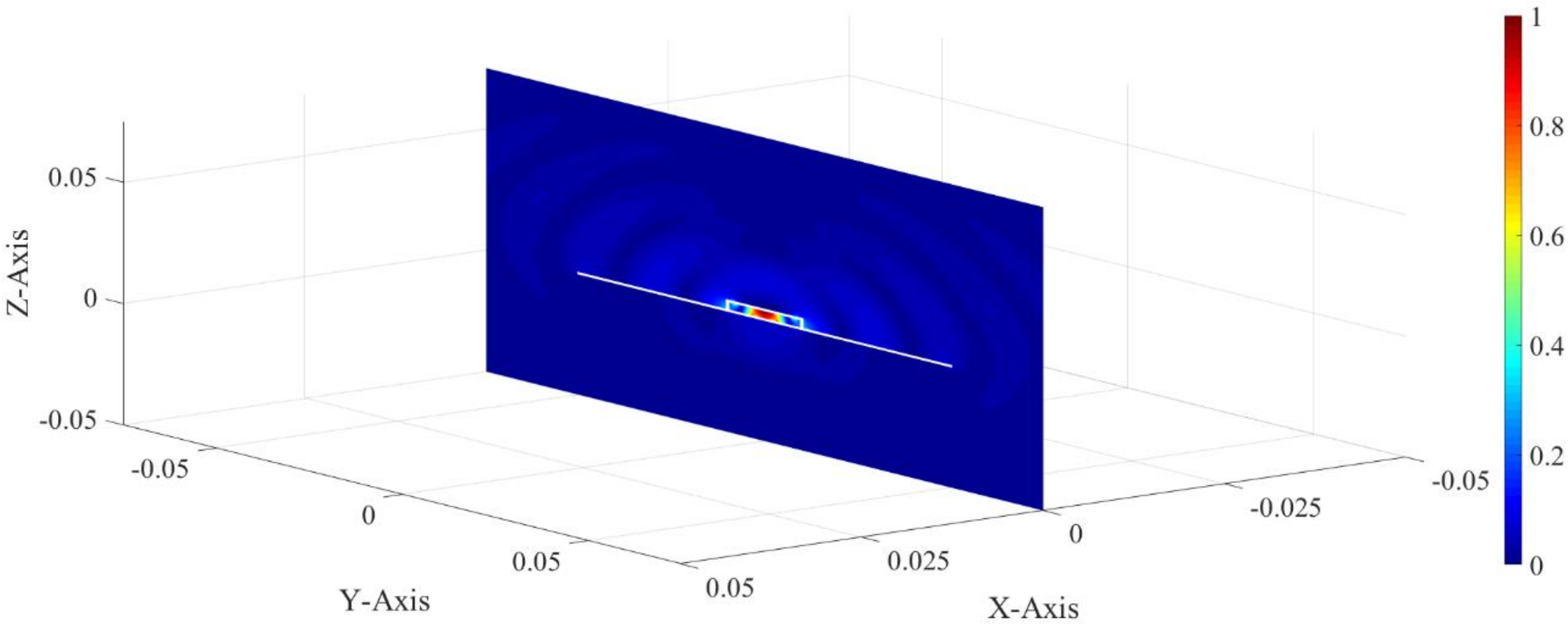

Figure 6-64 Modal electric field distribution of the IP-DM 1 at 8.5 GHz

The modal radiation pattern of the mode is shown in the following Fig. 6-65

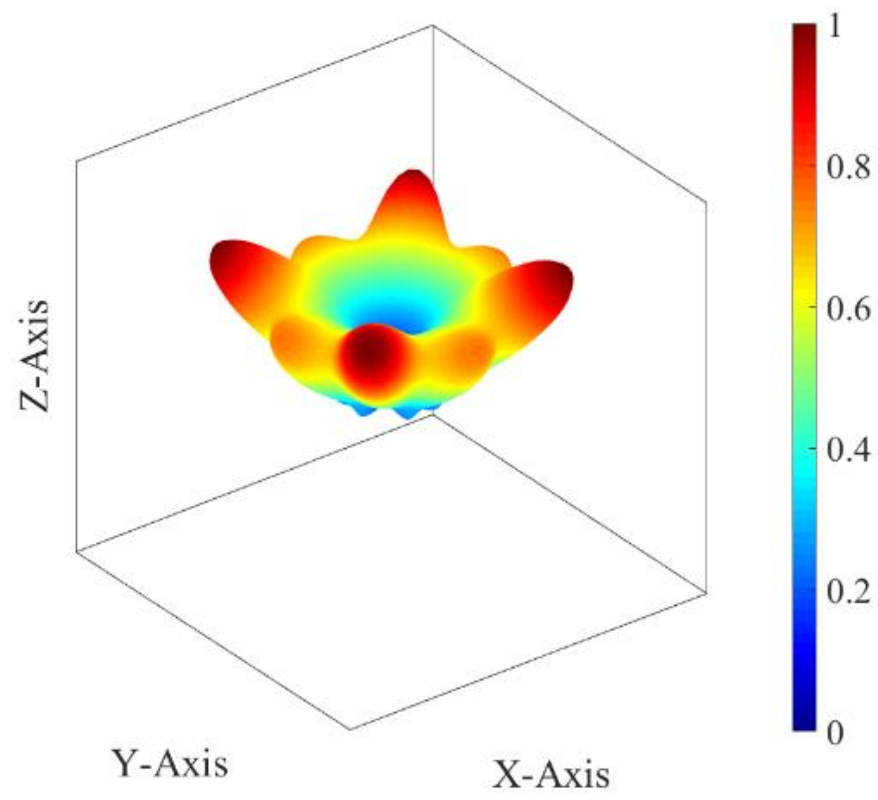

Figure 6-65 Modal radiation pattern of the IP-DM 1 at 8.5 GHz

## 6.6 IP-DMs of Augmented Metal-Material Composite Transmitting Antenna — Type III: Meta-horn Fed Layered Material Lens Antenna

In many applications (such as deep-space communication and point-to-point communication), the high gain and narrow beam antenna can significantly improve the EM performance of whole EM system[71~78,98~100,199~201]. Reflector antenna, which was discussed in the previous Sec. 6.2, is a typical reference in terms of high gain and narrow beam applications, but it usually needs a relatively large diameter[71~78,101~104]. In high gain and narrow beam applications, a better alternative to reflector antenna is *lens antenna*[71~78,199~201].

According to the differences of the materials of lenses, the lens antennas can be categorized to material lens antennas (homogeneous material lens antennas[202~205], inhomogeneous material lens antennas[206~208], and layered material lens antennas[209~212], etc.), metallic lens antennas[213~215], and metamaterial lens antennas[216~223], etc.

In this section, a *meta-horn fed layered material lens antenna* (which is placed in free space) is considered, and it is shown in the following Fig. 6-66

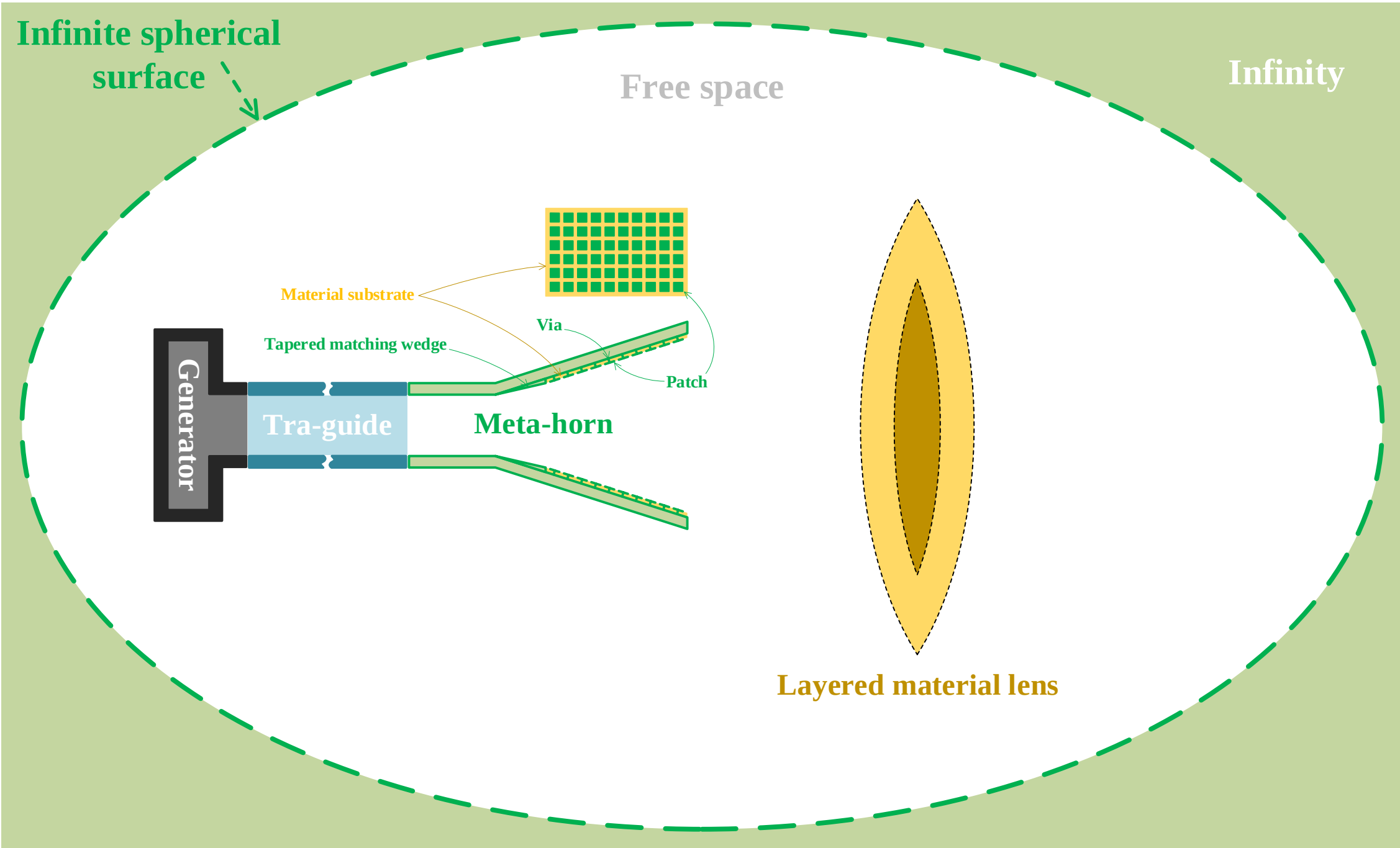

Figure 6-66 Geometry of a typical meta-horn fed layered material lens antenna

**In fact, there exist two different ways to treat the transmitting system shown in Fig. 6-66. The first way, which I prefer, is to treat the whole of {horn, lens} as tra-antenna; the second way is to treat the horn as tra-antenna and to treat the lens as a part of environment. Based on the conclusions given in Chap. 2, the first way corresponds to a *quasi-purely transmitting problem*, but the second way corresponds to a *compositely transmitting-receiving problem*. In this section, we will, under PTT framework, focus on constructing the IP-DMs corresponding to the tra-antenna working at the *quasi-purely transmitting state* (i.e. corresponding to the first way).**

The subsections contained in this section are organized similar to the subsections of Secs. 6.2~6.4, i.e., "topological structure → source-field relationships → modal space → power transport theorem → input power operator → input-power-decoupled modes".

## 6.6.1 Topological Structure

The topological structure of the tra-antenna shown in Fig. 6-66 is exhibited in Fig. 6-67.

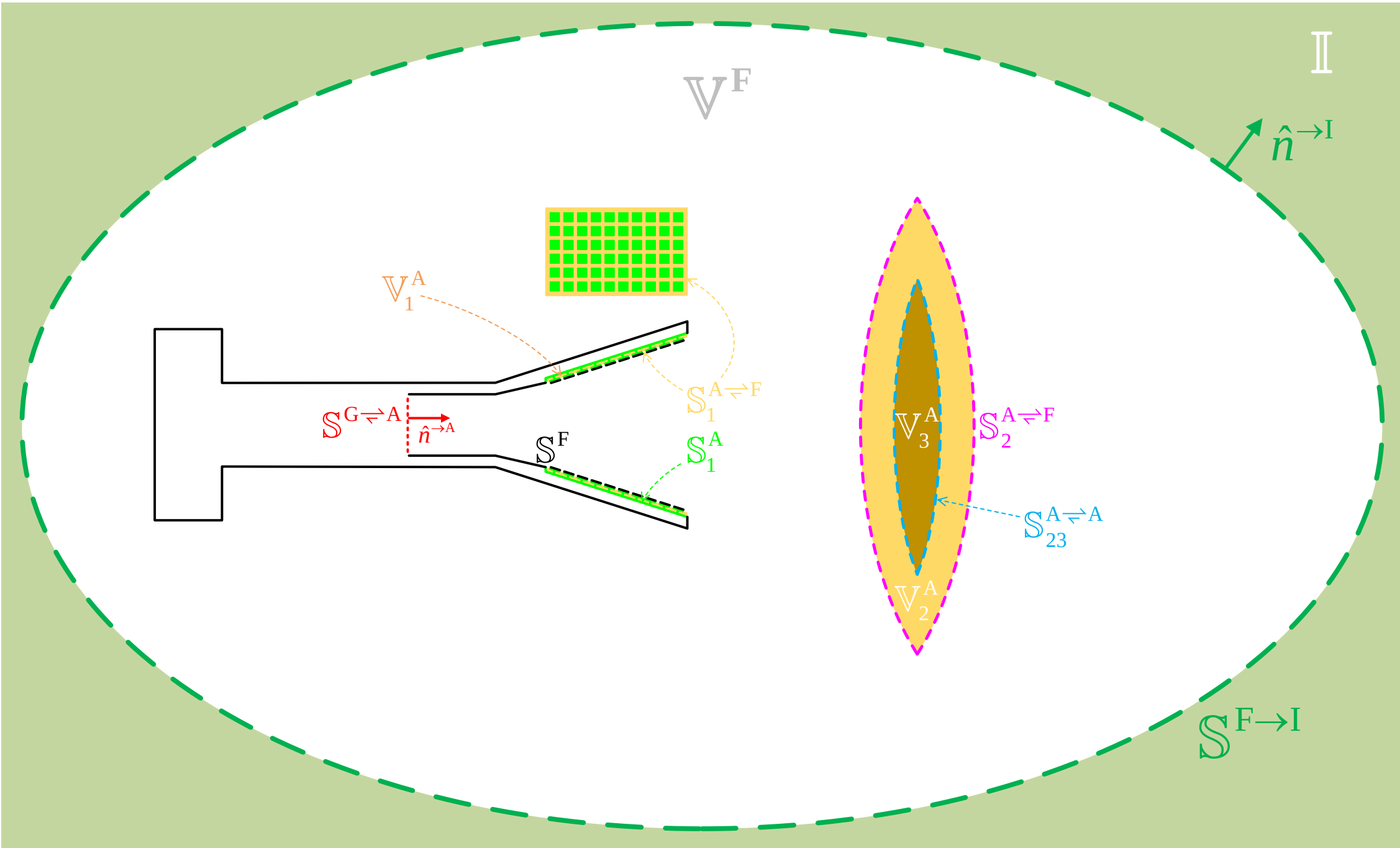

Figure 6-67 Topological structure of the tra-antenna shown in Fig. 6-66

In the above Fig. 6-67, surface $\mathbb{S}^{G\rightleftharpoons A}$ denotes the input port of the tra-antenna. The region occupied by the *material substrate* in the meta-horn is denoted as $\mathbb{V}_1^{A}$, and the region occupied by the *outer layer* of the two-layered material lens is denoted as $\mathbb{V}_2^{A}$, and the region occupied by the *inner layer* of the two-layered material lens is denoted as $\mathbb{V}_3^{A}$. The region occupied by free space is denoted as $\mathbb{V}^{F}$.

The interfaces between $\mathbb{V}_1^{A}$ and {metallic horn, metallic vias, metallic patches} are collectively denoted as $\mathbb{S}_1^{A}$; the interface between $\mathbb{V}_1^{A}$ and $\mathbb{V}^{F}$ is denoted as $\mathbb{S}_1^{A\rightleftharpoons F}$; the interface between $\mathbb{V}_2^{A}$ and $\mathbb{V}^{F}$ is denoted as $\mathbb{S}_2^{A\rightleftharpoons F}$; the interface between $\mathbb{V}_2^{A}$ and $\mathbb{V}_3^{A}$ is denoted as $\mathbb{S}_{23}^{A\rightleftharpoons A}$; the interface between $\mathbb{V}^{F}$ and {metallic horn, metallic patches, thick metallic wall} are collectively denoted as $\mathbb{S}^{F}$ (i.e. the grounding structure). The closed surface $\mathbb{S}^{F\rightarrow I}$ is a spherical surface with infinite radius.

Clearly, surfaces $\mathbb{S}_1^{A}$ and $\mathbb{S}_1^{A\rightleftharpoons F}$ constitute the boundary of $\mathbb{V}_1^{A}$, i.e., $\partial\mathbb{V}_1^{A}=\mathbb{S}_1^{A}\cup\mathbb{S}_1^{A\rightleftharpoons F}$; surfaces $\mathbb{S}_2^{A\rightleftharpoons F}$ and $\mathbb{S}_{23}^{A\rightleftharpoons A}$ constitute the boundary of $\mathbb{V}_2^{A}$, i.e., $\partial\mathbb{V}_2^{A}=\mathbb{S}_2^{A\rightleftharpoons F}\cup\mathbb{S}_{23}^{A\rightleftharpoons A}$; surface $\mathbb{S}_{23}^{A\rightleftharpoons A}$ is just the whole boundary of $\mathbb{V}_3^{A}$, i.e., $\partial\mathbb{V}_3^{A}=\mathbb{S}_{23}^{A\rightleftharpoons A}$; surfaces $\mathbb{S}^{G\rightleftharpoons A}$, $\mathbb{S}_1^{A\rightleftharpoons F}$, $\mathbb{S}_2^{A\rightleftharpoons F}$, $\mathbb{S}^{F}$, and $\mathbb{S}^{F\rightarrow I}$ constitute the boundary of $\mathbb{V}^{F}$, i.e., $\partial\mathbb{V}^{F}=\mathbb{S}^{G\rightleftharpoons A}\cup\mathbb{S}_1^{A\rightleftharpoons F}\cup\mathbb{S}_2^{A\rightleftharpoons F}\cup\mathbb{S}^{F}\cup\mathbb{S}^{F\rightarrow I}$.

In addition, the permeability, permeativity, and conductivity of $\mathbb{V}_1^{A}$ are denoted as $\overleftrightarrow{\mu}_1$, $\overleftrightarrow{\varepsilon}_1$, and $\overleftrightarrow{\sigma}_1$ respectively; the permeability, permeativity, and conductivity of $\mathbb{V}_2^{A}$ are denoted as $\overleftrightarrow{\mu}_2$, $\overleftrightarrow{\varepsilon}_2$, and $\overleftrightarrow{\sigma}_2$ respectively; the permeability, permeativity, and conductivity of $\mathbb{V}_3^{\text{A}}$ are denoted as $\ddot{\mu}_3$, $\ddot{\varepsilon}_3$, and $\ddot{\sigma}_3$ respectively; the permeability, permeativity, and conductivity of $\mathbb{V}^{\text{F}}$ are $\mu_0$, $\varepsilon_0$, and 0 respectively.

## 6.6.2 Source-Field Relationships

If the equivalent surface currents distributing on $\mathbb{S}_1^{\text{A}\rightleftharpoons\text{F}}$ are denoted as $\{\vec{J}_1^{\text{A}\rightleftharpoons\text{F}},\vec{M}_1^{\text{A}\rightleftharpoons\text{F}}\}$, and the equivalent electric current distributing on $\mathbb{S}_1^{\text{A}}$ is denoted as $\vec{J}_1^{\text{A}}$ ①, then the field distributing on $\mathbb{V}_1^{\text{A}}$ can be expressed as follows:

$$\vec{F}(\vec{r}) = \mathcal{F}_1\left(\vec{J}_1^{\text{A}\rightleftharpoons\text{F}}+\vec{J}_1^{\text{A}},\vec{M}_1^{\text{A}\rightleftharpoons\text{F}}\right) \quad , \quad \vec{r}\in\mathbb{V}_1^{\text{A}} \tag{6-98}$$

where $\vec{F}=\vec{E}/\vec{H}$, and correspondingly $\mathcal{F}_1=\mathcal{E}_1/\mathcal{H}_1$, and the operator is defined as that $\mathcal{F}_1(\vec{J},\vec{M})=\ddot{G}_1^{JF}*\vec{J}+\ddot{G}_1^{MF}*\vec{M}$ (here, $\ddot{G}_1^{JF}$ and $\ddot{G}_1^{MF}$ are the Green's functions of the region $\mathbb{V}_1^{\text{A}}$ with material parameters $\{\ddot{\mu}_1,\ddot{\varepsilon}_1,\ddot{\sigma}_1\}$). The currents $\{\vec{J}_1^{\text{A}\rightleftharpoons\text{F}},\vec{M}_1^{\text{A}\rightleftharpoons\text{F}}\}$ and $\vec{J}_1^{\text{A}}$ and fields $\{\vec{E},\vec{H}\}$ in Eq. (6-98) satisfy the following relations

$$\hat{n}_1^{\rightarrow\text{A}}\times\left[\vec{H}\left(\vec{r}_1^{\,\text{A}}\right)\right]_{\vec{r}_1^{\,\text{A}}\to\vec{r}} = \vec{J}^{x}(\vec{r}) \quad , \quad \vec{r}\in\mathbb{S}^{x} \tag{6-99a}$$

$$\left[\vec{E}\left(\vec{r}_1^{\,\text{A}}\right)\right]_{\vec{r}_1^{\,\text{A}}\to\vec{r}}\times\hat{n}_1^{\rightarrow\text{A}} = \vec{M}_1^{\text{A}\rightleftharpoons\text{F}}(\vec{r}) \quad , \quad \vec{r}\in\mathbb{S}_1^{\text{A}\rightleftharpoons\text{F}} \tag{6-99b}$$

In Eq. (6-99a), $\vec{J}^{x}=\vec{J}_1^{\text{A}\rightleftharpoons\text{F}}/\vec{J}_1^{\text{A}}$, and $\mathbb{S}^{x}=\mathbb{S}_1^{\text{A}\rightleftharpoons\text{F}}/\mathbb{S}_1^{\text{A}}$ correspondingly; point $\vec{r}_1^{\,\text{A}}$ belongs to $\mathbb{V}_1^{\text{A}}$, and approaches the point $\vec{r}$ on $\mathbb{S}^{x}$; $\hat{n}_1^{\rightarrow\text{A}}$ is the normal direction of $\partial\mathbb{V}_1^{\text{A}}$ $(=\mathbb{S}_1^{\text{A}\rightleftharpoons\text{F}}\bigcup\mathbb{S}_1^{\text{A}})$, and points to the interior of $\mathbb{V}_1^{\text{A}}$.

If the equivalent surface currents distributing on $\mathbb{S}_2^{\text{A}\rightleftharpoons\text{F}}$ are denoted as $\{\vec{J}_2^{\text{A}\rightleftharpoons\text{F}},\vec{M}_2^{\text{A}\rightleftharpoons\text{F}}\}$, and the equivalent surface currents distributing on $\mathbb{S}_{23}^{\text{A}\rightleftharpoons\text{A}}$ are denoted as $\{\vec{J}_{23}^{\text{A}\rightleftharpoons\text{A}},\vec{M}_{23}^{\text{A}\rightleftharpoons\text{A}}\}$, then the field distributing on $\mathbb{V}_2^{\text{A}}$ can be expressed as follows:

$$\vec{F}(\vec{r}) = \mathcal{F}_2\left(\vec{J}_2^{\text{A}\rightleftharpoons\text{F}}+\vec{J}_{23}^{\text{A}\rightleftharpoons\text{A}},\vec{M}_2^{\text{A}\rightleftharpoons\text{F}}+\vec{M}_{23}^{\text{A}\rightleftharpoons\text{A}}\right) \quad , \quad \vec{r}\in\mathbb{V}_2^{\text{A}} \tag{6-100}$$

where $\vec{F}=\vec{E}/\vec{H}$, and correspondingly $\mathcal{F}_2=\mathcal{E}_2/\mathcal{H}_2$, and the operator is defined as that $\mathcal{F}_2(\vec{J},\vec{M})=\ddot{G}_2^{JF}*\vec{J}+\ddot{G}_2^{MF}*\vec{M}$ (here, $\ddot{G}_2^{JF}$ and $\ddot{G}_2^{MF}$ are the dyadic Green's functions corresponding to the region $\mathbb{V}_2^{\text{A}}$ with material parameters $\{\ddot{\mu}_2,\ddot{\varepsilon}_2,\ddot{\sigma}_2\}$). The currents $\{\vec{J}_2^{\text{A}\rightleftharpoons\text{F}},\vec{M}_2^{\text{A}\rightleftharpoons\text{F}}\}$ & $\{\vec{J}_{23}^{\text{A}\rightleftharpoons\text{A}},\vec{M}_{23}^{\text{A}\rightleftharpoons\text{A}}\}$ and fields $\{\vec{E},\vec{H}\}$ in Eq. (6-100) satisfy the following relations

$$\hat{n}_2^{\rightarrow\text{A}}\times\left[\vec{H}\left(\vec{r}_2^{\,\text{A}}\right)\right]_{\vec{r}_2^{\,\text{A}}\to\vec{r}} = \vec{J}^{x}(\vec{r}) \quad , \quad \vec{r}\in\mathbb{S}^{x} \tag{6-101a}$$

$$\left[\vec{E}\left(\vec{r}_2^{\,\text{A}}\right)\right]_{\vec{r}_2^{\,\text{A}}\to\vec{r}}\times\hat{n}_2^{\rightarrow\text{A}} = \vec{M}^{x}(\vec{r}) \quad , \quad \vec{r}\in\mathbb{S}^{x} \tag{6-101b}$$

① The equivalent magnetic current distributing on $\mathbb{S}_1^{\text{A}}$ is zero, because of the homogeneous tangential electric field boundary condition on $\mathbb{S}_1^{\text{A}}$ [13].

In Eqs. (6-101a) and (6-101b), $\vec{J}^{x} = \vec{J}_2^{A\rightleftharpoons F} / \vec{J}_{23}^{A\rightleftharpoons A}$, and $\vec{M}^{x} = \vec{M}_2^{A\rightleftharpoons F} / \vec{M}_{23}^{A\rightleftharpoons A}$, and $\mathbb{S}^{x} = \mathbb{S}_2^{A\rightleftharpoons F} / \mathbb{S}_{23}^{A\rightleftharpoons A}$ correspondingly; point $\vec{r}_2^{A}$ belongs to $\mathbb{V}_2^{A}$, and approaches the point $\vec{r}$ on $\mathbb{S}^{x}$; $\hat{n}_2^{\rightarrow A}$ is the normal direction of $\partial\mathbb{V}_2^{A}$ ($= \mathbb{S}_2^{A\rightleftharpoons F} \bigcup \mathbb{S}_{23}^{A\rightleftharpoons A}$), and points to the interior of $\mathbb{V}_2^{A}$.

The field distributing on $\mathbb{V}_3^{A}$ can be expressed as follows:

$$\vec{F}\left(\vec{r}\right) = \mathcal{F}_3\left(-\vec{J}_{23}^{A\rightleftharpoons A}, -\vec{M}_{23}^{A\rightleftharpoons A}\right) \quad , \quad \vec{r} \in \mathbb{V}_3^{A} \tag{6-102}$$

where $\vec{F} = \vec{E} / \vec{H}$, and correspondingly $\mathcal{F}_3 = \mathcal{E}_3 / \mathcal{H}_3$, and the operator is defined as that $\mathcal{F}_3(\vec{J}, \vec{M}) = \vec{\vec{G}}_3^{JF} * \vec{J} + \vec{\vec{G}}_3^{MF} * \vec{M}$ (here, $\vec{\vec{G}}_3^{JF}$ and $\vec{\vec{G}}_3^{MF}$ are the dyadic Green's functions corresponding to the region $\mathbb{V}_3^{A}$ with material parameters $\{\vec{\vec{\mu}}_3, \vec{\vec{\varepsilon}}_3, \vec{\vec{\sigma}}_3\}$).

If the equivalent surface currents distributing on $\mathbb{S}^{G\rightleftharpoons A}$ are denoted as $\{\vec{J}^{G\rightleftharpoons A}, \vec{M}^{G\rightleftharpoons A}\}$, and the equivalent surface electric current distributing on $\mathbb{S}^{F}$ is denoted as $\vec{J}^{F}$ ①, then the field distributing on $\mathbb{V}^{F}$ can be expressed as follows:

$$\vec{F}\left(\vec{r}\right) = \mathcal{F}_0\left(\vec{J}^{G\rightleftharpoons A} - \vec{J}_1^{A\rightleftharpoons F} - \vec{J}_2^{A\rightleftharpoons F} + \vec{J}^{F}, \vec{M}^{G\rightleftharpoons A} - \vec{M}_1^{A\rightleftharpoons F} - \vec{M}_2^{A\rightleftharpoons F}\right) \quad , \quad \vec{r} \in \mathbb{V}^{F} \tag{6-103}$$

where $\vec{F} = \vec{E} / \vec{H}$, and correspondingly $\mathcal{F}_0 = \mathcal{E}_0 / \mathcal{H}_0$, and the operators are the same as the ones used in the previous chapters. The currents $\{\vec{J}^{G\rightleftharpoons A}, \vec{M}^{G\rightleftharpoons A}\}$ and fields $\{\vec{E}, \vec{H}\}$ in Eq. (6-103) satisfy the following relations

$$\hat{n}^{\rightarrow A} \times \left[\vec{H}\left(\vec{r}^{F}\right)\right]_{\vec{r}^{F} \to \vec{r}} = \vec{J}^{G\rightleftharpoons A}\left(\vec{r}\right) \quad , \quad \vec{r} \in \mathbb{S}^{G\rightleftharpoons A} \tag{6-104a}$$

$$\left[\vec{E}\left(\vec{r}^{F}\right)\right]_{\vec{r}^{F} \to \vec{r}} \times \hat{n}^{\rightarrow A} = \vec{M}^{G\rightleftharpoons A}\left(\vec{r}\right) \quad , \quad \vec{r} \in \mathbb{S}^{G\rightleftharpoons A} \tag{6-104b}$$

where point $\vec{r}^{F}$ belongs to $\mathbb{V}^{F}$ and approaches the point $\vec{r}$ on $\mathbb{S}^{G\rightleftharpoons A}$, and $\hat{n}^{\rightarrow A}$ is the normal direction of $\mathbb{S}^{G\rightleftharpoons A}$ and points to tra-antenna.

### 6.6.3 Mathematical Description for Modal Space

Combining Eq. (6-103) with Eq. (6-104), we immediately obtain the following integral equations

$$\begin{aligned}&\left[\mathcal{H}_0\left(\vec{J}^{G\rightleftharpoons A} - \vec{J}_1^{A\rightleftharpoons F} - \vec{J}_2^{A\rightleftharpoons F} + \vec{J}^{F}, \vec{M}^{G\rightleftharpoons A} - \vec{M}_1^{A\rightleftharpoons F} - \vec{M}_2^{A\rightleftharpoons F}\right)\right]_{\vec{r}^{F} \to \vec{r}}^{\tan} \\ &= \vec{J}^{G\rightleftharpoons A}\left(\vec{r}\right) \times \hat{n}^{\rightarrow A} \quad , \quad \vec{r} \in \mathbb{S}^{G\rightleftharpoons A}\end{aligned} \tag{6-105a}$$

$$\begin{aligned}&\left[\mathcal{E}_0\left(\vec{J}^{G\rightleftharpoons A} - \vec{J}_1^{A\rightleftharpoons F} - \vec{J}_2^{A\rightleftharpoons F} + \vec{J}^{F}, \vec{M}^{G\rightleftharpoons A} - \vec{M}_1^{A\rightleftharpoons F} - \vec{M}_2^{A\rightleftharpoons F}\right)\right]_{\vec{r}^{F} \to \vec{r}}^{\tan} \\ &= \hat{n}^{\rightarrow A} \times \vec{M}^{G\rightleftharpoons A}\left(\vec{r}\right) \quad , \quad \vec{r} \in \mathbb{S}^{G\rightleftharpoons A}\end{aligned} \tag{6-105b}$$

① The equivalent surface magnetic current distributing on $\mathbb{S}^{F}$ is zero, because of the homogeneous tangential electric field boundary condition on $\mathbb{S}^{F}$ [13].

about currents $\{\vec{J}^{\mathrm{G}\rightleftharpoons\mathrm{A}},\vec{M}^{\mathrm{G}\rightleftharpoons\mathrm{A}}\}$, $\{\vec{J}_1^{\mathrm{A}\rightleftharpoons\mathrm{F}},\vec{M}_1^{\mathrm{A}\rightleftharpoons\mathrm{F}}\}$, $\{\vec{J}_2^{\mathrm{A}\rightleftharpoons\mathrm{F}},\vec{M}_2^{\mathrm{A}\rightleftharpoons\mathrm{F}}\}$, and $\vec{J}^{\mathrm{F}}$, where the superscript "tan" represents the tangential component of the field.

Using Eqs. (6-98)&(6-103) and employing the tangential field continuation condition on $\mathbb{S}_1^{\mathrm{A}\rightleftharpoons\mathrm{F}}$, there exist the following integral equations

$$\begin{aligned}&\Big[\mathcal{E}_0\Big(\vec{J}^{\mathrm{G}\rightleftharpoons\mathrm{A}}-\vec{J}_1^{\mathrm{A}\rightleftharpoons\mathrm{F}}-\vec{J}_2^{\mathrm{A}\rightleftharpoons\mathrm{F}}+\vec{J}^{\mathrm{F}},\vec{M}^{\mathrm{G}\rightleftharpoons\mathrm{A}}-\vec{M}_1^{\mathrm{A}\rightleftharpoons\mathrm{F}}-\vec{M}_2^{\mathrm{A}\rightleftharpoons\mathrm{F}}\Big)\Big]_{\vec{r}^{\mathrm{F}}\to\vec{r}}^{\tan}\\&=\Big[\mathcal{E}_1\Big(\vec{J}_1^{\mathrm{A}\rightleftharpoons\mathrm{F}}+\vec{J}_1^{\mathrm{A}},\vec{M}_1^{\mathrm{A}\rightleftharpoons\mathrm{F}}\Big)\Big]_{\vec{r}_1^{\mathrm{A}}\to\vec{r}}^{\tan}\quad,\quad\vec{r}\in\mathbb{S}_1^{\mathrm{A}\rightleftharpoons\mathrm{F}}\qquad(6\text{-}106\text{a})\end{aligned}$$

$$\begin{aligned}&\Big[\mathcal{H}_0\Big(\vec{J}^{\mathrm{G}\rightleftharpoons\mathrm{A}}-\vec{J}_1^{\mathrm{A}\rightleftharpoons\mathrm{F}}-\vec{J}_2^{\mathrm{A}\rightleftharpoons\mathrm{F}}+\vec{J}^{\mathrm{F}},\vec{M}^{\mathrm{G}\rightleftharpoons\mathrm{A}}-\vec{M}_1^{\mathrm{A}\rightleftharpoons\mathrm{F}}-\vec{M}_2^{\mathrm{A}\rightleftharpoons\mathrm{F}}\Big)\Big]_{\vec{r}^{\mathrm{F}}\to\vec{r}}^{\tan}\\&=\Big[\mathcal{H}_1\Big(\vec{J}_1^{\mathrm{A}\rightleftharpoons\mathrm{F}}+\vec{J}_1^{\mathrm{A}},\vec{M}_1^{\mathrm{A}\rightleftharpoons\mathrm{F}}\Big)\Big]_{\vec{r}_1^{\mathrm{A}}\to\vec{r}}^{\tan}\quad,\quad\vec{r}\in\mathbb{S}_1^{\mathrm{A}\rightleftharpoons\mathrm{F}}\qquad(6\text{-}106\text{b})\end{aligned}$$

about currents $\{\vec{J}^{\mathrm{G}\rightleftharpoons\mathrm{A}},\vec{M}^{\mathrm{G}\rightleftharpoons\mathrm{A}}\}$, $\{\vec{J}_1^{\mathrm{A}\rightleftharpoons\mathrm{F}},\vec{M}_1^{\mathrm{A}\rightleftharpoons\mathrm{F}}\}$, $\{\vec{J}_2^{\mathrm{A}\rightleftharpoons\mathrm{F}},\vec{M}_2^{\mathrm{A}\rightleftharpoons\mathrm{F}}\}$, $\vec{J}_1^{\mathrm{A}}$, and $\vec{J}^{\mathrm{F}}$. Using Eqs. (6-100)&(6-103) and employing the tangential field continuation condition on $\mathbb{S}_2^{\mathrm{A}\rightleftharpoons\mathrm{F}}$, there exist the following integral equations

$$\begin{aligned}&\Big[\mathcal{E}_0\Big(\vec{J}^{\mathrm{G}\rightleftharpoons\mathrm{A}}-\vec{J}_1^{\mathrm{A}\rightleftharpoons\mathrm{F}}-\vec{J}_2^{\mathrm{A}\rightleftharpoons\mathrm{F}}+\vec{J}^{\mathrm{F}},\vec{M}^{\mathrm{G}\rightleftharpoons\mathrm{A}}-\vec{M}_1^{\mathrm{A}\rightleftharpoons\mathrm{F}}-\vec{M}_2^{\mathrm{A}\rightleftharpoons\mathrm{F}}\Big)\Big]_{\vec{r}^{\mathrm{F}}\to\vec{r}}^{\tan}\\&=\Big[\mathcal{E}_2\Big(\vec{J}_2^{\mathrm{A}\rightleftharpoons\mathrm{F}}+\vec{J}_{23}^{\mathrm{A}\rightleftharpoons\mathrm{A}},\vec{M}_2^{\mathrm{A}\rightleftharpoons\mathrm{F}}+\vec{M}_{23}^{\mathrm{A}\rightleftharpoons\mathrm{A}}\Big)\Big]_{\vec{r}_2^{\mathrm{A}}\to\vec{r}}^{\tan}\quad,\quad\vec{r}\in\mathbb{S}_2^{\mathrm{A}\rightleftharpoons\mathrm{F}}\qquad(6\text{-}107\text{a})\end{aligned}$$

$$\begin{aligned}&\Big[\mathcal{H}_0\Big(\vec{J}^{\mathrm{G}\rightleftharpoons\mathrm{A}}-\vec{J}_1^{\mathrm{A}\rightleftharpoons\mathrm{F}}-\vec{J}_2^{\mathrm{A}\rightleftharpoons\mathrm{F}}+\vec{J}^{\mathrm{F}},\vec{M}^{\mathrm{G}\rightleftharpoons\mathrm{A}}-\vec{M}_1^{\mathrm{A}\rightleftharpoons\mathrm{F}}-\vec{M}_2^{\mathrm{A}\rightleftharpoons\mathrm{F}}\Big)\Big]_{\vec{r}^{\mathrm{F}}\to\vec{r}}^{\tan}\\&=\Big[\mathcal{H}_2\Big(\vec{J}_2^{\mathrm{A}\rightleftharpoons\mathrm{F}}+\vec{J}_{23}^{\mathrm{A}\rightleftharpoons\mathrm{A}},\vec{M}_2^{\mathrm{A}\rightleftharpoons\mathrm{F}}+\vec{M}_{23}^{\mathrm{A}\rightleftharpoons\mathrm{A}}\Big)\Big]_{\vec{r}_2^{\mathrm{A}}\to\vec{r}}^{\tan}\quad,\quad\vec{r}\in\mathbb{S}_2^{\mathrm{A}\rightleftharpoons\mathrm{F}}\qquad(6\text{-}107\text{b})\end{aligned}$$

about currents $\{\vec{J}^{\mathrm{G}\rightleftharpoons\mathrm{A}},\vec{M}^{\mathrm{G}\rightleftharpoons\mathrm{A}}\}$, $\{\vec{J}_1^{\mathrm{A}\rightleftharpoons\mathrm{F}},\vec{M}_1^{\mathrm{A}\rightleftharpoons\mathrm{F}}\}$, $\{\vec{J}_2^{\mathrm{A}\rightleftharpoons\mathrm{F}},\vec{M}_2^{\mathrm{A}\rightleftharpoons\mathrm{F}}\}$, $\{\vec{J}_{23}^{\mathrm{A}\rightleftharpoons\mathrm{A}},\vec{M}_{23}^{\mathrm{A}\rightleftharpoons\mathrm{A}}\}$, and $\vec{J}^{\mathrm{F}}$. Using Eqs. (6-100)&(6-102) and employing the tangential field continuation condition on $\mathbb{S}_{23}^{\mathrm{A}\rightleftharpoons\mathrm{A}}$, there exist the following integral equations

$$\Big[\mathcal{E}_2\Big(\vec{J}_2^{\mathrm{A}\rightleftharpoons\mathrm{F}}+\vec{J}_{23}^{\mathrm{A}\rightleftharpoons\mathrm{A}},\vec{M}_2^{\mathrm{A}\rightleftharpoons\mathrm{F}}+\vec{M}_{23}^{\mathrm{A}\rightleftharpoons\mathrm{A}}\Big)\Big]_{\vec{r}_2^{\mathrm{A}}\to\vec{r}}^{\tan}=\Big[\mathcal{E}_3\Big(-\vec{J}_{23}^{\mathrm{A}\rightleftharpoons\mathrm{A}},-\vec{M}_{23}^{\mathrm{A}\rightleftharpoons\mathrm{A}}\Big)\Big]_{\vec{r}_3^{\mathrm{A}}\to\vec{r}}^{\tan},\ \vec{r}\in\mathbb{S}_{23}^{\mathrm{A}\rightleftharpoons\mathrm{A}}\ (6\text{-}108\text{a})$$

$$\Big[\mathcal{H}_2\Big(\vec{J}_2^{\mathrm{A}\rightleftharpoons\mathrm{F}}+\vec{J}_{23}^{\mathrm{A}\rightleftharpoons\mathrm{A}},\vec{M}_2^{\mathrm{A}\rightleftharpoons\mathrm{F}}+\vec{M}_{23}^{\mathrm{A}\rightleftharpoons\mathrm{A}}\Big)\Big]_{\vec{r}_2^{\mathrm{A}}\to\vec{r}}^{\tan}=\Big[\mathcal{H}_3\Big(-\vec{J}_{23}^{\mathrm{A}\rightleftharpoons\mathrm{A}},-\vec{M}_{23}^{\mathrm{A}\rightleftharpoons\mathrm{A}}\Big)\Big]_{\vec{r}_3^{\mathrm{A}}\to\vec{r}}^{\tan},\ \vec{r}\in\mathbb{S}_{23}^{\mathrm{A}\rightleftharpoons\mathrm{A}}\ (6\text{-}108\text{b})$$

about currents $\{\vec{J}_2^{\mathrm{A}\rightleftharpoons\mathrm{F}},\vec{M}_2^{\mathrm{A}\rightleftharpoons\mathrm{F}}\}$ and $\{\vec{J}_{23}^{\mathrm{A}\rightleftharpoons\mathrm{A}},\vec{M}_{23}^{\mathrm{A}\rightleftharpoons\mathrm{A}}\}$, where point $\vec{r}_3^{\mathrm{A}}$ belongs to $\mathbb{V}_3^{\mathrm{A}}$ and approaches the point $\vec{r}$ on $\mathbb{S}_{23}^{\mathrm{A}\rightleftharpoons\mathrm{A}}$.

Based on Eq. (6-98) and the homogeneous tangential electric field boundary condition on $\mathbb{S}_1^{\mathrm{A}}$, we have the following electric field integral equation

$$\Big[\mathcal{E}_1\Big(\vec{J}_1^{\mathrm{A}\rightleftharpoons\mathrm{F}}+\vec{J}_1^{\mathrm{A}},\vec{M}_1^{\mathrm{A}\rightleftharpoons\mathrm{F}}\Big)\Big]_{\vec{r}_1^{\mathrm{A}}\to\vec{r}}^{\tan}=0\quad,\quad\vec{r}\in\mathbb{S}_1^{\mathrm{A}}\qquad(6\text{-}109)$$

about currents $\{\vec{J}_1^{\mathrm{A}\rightleftharpoons\mathrm{F}},\vec{M}_1^{\mathrm{A}\rightleftharpoons\mathrm{F}}\}$ and $\vec{J}_1^{\mathrm{A}}$. Based on Eq. (6-103) and the homogeneous tangential electric field boundary condition on $\mathbb{S}^{\rm F}$, we have the following electric field integral equation

$$\left[\mathcal{E}_0\left(\vec{J}^{\,{\rm G}\rightleftharpoons{\rm A}}-\vec{J}_1^{\,{\rm A}\rightleftharpoons{\rm F}}-\vec{J}_2^{\,{\rm A}\rightleftharpoons{\rm F}}+\vec{J}^{\,{\rm F}},\vec{M}^{\,{\rm G}\rightleftharpoons{\rm A}}-\vec{M}_1^{\,{\rm A}\rightleftharpoons{\rm F}}-\vec{M}_2^{\,{\rm A}\rightleftharpoons{\rm F}}\right)\right]_{\vec{r}^{\,\rm F}\to\vec{r}}^{\rm tan}=0\ ,\ \vec{r}\in\mathbb{S}^{\rm F}\quad(6\text{-}110)$$

about currents $\{\vec{J}^{\,{\rm G}\rightleftharpoons{\rm A}},\vec{M}^{\,{\rm G}\rightleftharpoons{\rm A}}\}$, $\{\vec{J}_1^{\,{\rm A}\rightleftharpoons{\rm F}},\vec{M}_1^{\,{\rm A}\rightleftharpoons{\rm F}}\}$, $\{\vec{J}_2^{\,{\rm A}\rightleftharpoons{\rm F}},\vec{M}_2^{\,{\rm A}\rightleftharpoons{\rm F}}\}$, and $\vec{J}^{\,\rm F}$.

If the above currents $\{\vec{J}^{\,{\rm G}\rightleftharpoons{\rm A}},\vec{M}^{\,{\rm G}\rightleftharpoons{\rm A}}\}$, $\{\vec{J}_1^{\,{\rm A}\rightleftharpoons{\rm F}},\vec{M}_1^{\,{\rm A}\rightleftharpoons{\rm F}}\}$, $\{\vec{J}_2^{\,{\rm A}\rightleftharpoons{\rm F}},\vec{M}_2^{\,{\rm A}\rightleftharpoons{\rm F}}\}$, $\{\vec{J}_{23}^{\,{\rm A}\rightleftharpoons{\rm A}},\vec{M}_{23}^{\,{\rm A}\rightleftharpoons{\rm A}}\}$, $\vec{J}_1^{\,\rm A}$, and $\vec{J}^{\,\rm F}$ are expanded in terms of some proper basis functions, and the Eqs. (6-105a), (6-105b), (6-106a), (6-106b), (6-107a), (6-107b), (6-108a), (6-108b), (6-109), and (6-110) are tested with $\{\vec{b}_\xi^{\vec{M}^{{\rm G}\rightleftharpoons{\rm A}}}\}$, $\{\vec{b}_\xi^{\vec{J}^{{\rm G}\rightleftharpoons{\rm A}}}\}$, $\{\vec{b}_\xi^{\vec{J}_1^{{\rm A}\rightleftharpoons{\rm F}}}\}$, $\{\vec{b}_\xi^{\vec{M}_1^{{\rm A}\rightleftharpoons{\rm F}}}\}$, $\{\vec{b}_\xi^{\vec{J}_2^{{\rm A}\rightleftharpoons{\rm F}}}\}$, $\{\vec{b}_\xi^{\vec{M}_2^{{\rm A}\rightleftharpoons{\rm F}}}\}$, $\{\vec{b}_\xi^{\vec{J}_{23}^{{\rm A}\rightleftharpoons{\rm A}}}\}$, $\{\vec{b}_\xi^{\vec{M}_{23}^{{\rm A}\rightleftharpoons{\rm A}}}\}$, $\{\vec{b}_\xi^{\vec{J}_1^{\rm A}}\}$, and $\{\vec{b}_\xi^{\vec{J}^{\rm F}}\}$ respectively, then the integral equations are immediately discretized into the following matrix equations

$$\begin{aligned}&\overline{\overline{Z}}^{\vec{M}^{{\rm G}\rightleftharpoons{\rm A}}\vec{J}^{{\rm G}\rightleftharpoons{\rm A}}}\cdot\overline{a}^{\vec{J}^{{\rm G}\rightleftharpoons{\rm A}}}+\overline{\overline{Z}}^{\vec{M}^{{\rm G}\rightleftharpoons{\rm A}}\vec{J}_1^{{\rm A}\rightleftharpoons{\rm F}}}\cdot\overline{a}^{\vec{J}_1^{{\rm A}\rightleftharpoons{\rm F}}}+\overline{\overline{Z}}^{\vec{M}^{{\rm G}\rightleftharpoons{\rm A}}\vec{J}_2^{{\rm A}\rightleftharpoons{\rm F}}}\cdot\overline{a}^{\vec{J}_2^{{\rm A}\rightleftharpoons{\rm F}}}+\overline{\overline{Z}}^{\vec{M}^{{\rm G}\rightleftharpoons{\rm A}}\vec{J}^{\rm F}}\cdot\overline{a}^{\vec{J}^{\rm F}}\\&+\overline{\overline{Z}}^{\vec{M}^{{\rm G}\rightleftharpoons{\rm A}}\vec{M}^{{\rm G}\rightleftharpoons{\rm A}}}\cdot\overline{a}^{\vec{M}^{{\rm G}\rightleftharpoons{\rm A}}}+\overline{\overline{Z}}^{\vec{M}^{{\rm G}\rightleftharpoons{\rm A}}\vec{M}_1^{{\rm A}\rightleftharpoons{\rm F}}}\cdot\overline{a}^{\vec{M}_1^{{\rm A}\rightleftharpoons{\rm F}}}+\overline{\overline{Z}}^{\vec{M}^{{\rm G}\rightleftharpoons{\rm A}}\vec{M}_2^{{\rm A}\rightleftharpoons{\rm F}}}\cdot\overline{a}^{\vec{M}_2^{{\rm A}\rightleftharpoons{\rm F}}}=0\qquad(6\text{-}111{\rm a})\\&\overline{\overline{Z}}^{\vec{J}^{{\rm G}\rightleftharpoons{\rm A}}\vec{J}^{{\rm G}\rightleftharpoons{\rm A}}}\cdot\overline{a}^{\vec{J}^{{\rm G}\rightleftharpoons{\rm A}}}+\overline{\overline{Z}}^{\vec{J}^{{\rm G}\rightleftharpoons{\rm A}}\vec{J}_1^{{\rm A}\rightleftharpoons{\rm F}}}\cdot\overline{a}^{\vec{J}_1^{{\rm A}\rightleftharpoons{\rm F}}}+\overline{\overline{Z}}^{\vec{J}^{{\rm G}\rightleftharpoons{\rm A}}\vec{J}_2^{{\rm A}\rightleftharpoons{\rm F}}}\cdot\overline{a}^{\vec{J}_2^{{\rm A}\rightleftharpoons{\rm F}}}+\overline{\overline{Z}}^{\vec{J}^{{\rm G}\rightleftharpoons{\rm A}}\vec{J}^{\rm F}}\cdot\overline{a}^{\vec{J}^{\rm F}}\\&+\overline{\overline{Z}}^{\vec{J}^{{\rm G}\rightleftharpoons{\rm A}}\vec{M}^{{\rm G}\rightleftharpoons{\rm A}}}\cdot\overline{a}^{\vec{M}^{{\rm G}\rightleftharpoons{\rm A}}}+\overline{\overline{Z}}^{\vec{J}^{{\rm G}\rightleftharpoons{\rm A}}\vec{M}_1^{{\rm A}\rightleftharpoons{\rm F}}}\cdot\overline{a}^{\vec{M}_1^{{\rm A}\rightleftharpoons{\rm F}}}+\overline{\overline{Z}}^{\vec{J}^{{\rm G}\rightleftharpoons{\rm A}}\vec{M}_2^{{\rm A}\rightleftharpoons{\rm F}}}\cdot\overline{a}^{\vec{M}_2^{{\rm A}\rightleftharpoons{\rm F}}}=0\qquad(6\text{-}111{\rm b})\end{aligned}$$

and

$$\begin{aligned}&\overline{\overline{Z}}^{\vec{J}_1^{{\rm A}\rightleftharpoons{\rm F}}\vec{J}^{{\rm G}\rightleftharpoons{\rm A}}}\cdot\overline{a}^{\vec{J}^{{\rm G}\rightleftharpoons{\rm A}}}+\overline{\overline{Z}}^{\vec{J}_1^{{\rm A}\rightleftharpoons{\rm F}}\vec{J}_1^{{\rm A}\rightleftharpoons{\rm F}}}\cdot\overline{a}^{\vec{J}_1^{{\rm A}\rightleftharpoons{\rm F}}}+\overline{\overline{Z}}^{\vec{J}_1^{{\rm A}\rightleftharpoons{\rm F}}\vec{J}_2^{{\rm A}\rightleftharpoons{\rm F}}}\cdot\overline{a}^{\vec{J}_2^{{\rm A}\rightleftharpoons{\rm F}}}+\overline{\overline{Z}}^{\vec{J}_1^{{\rm A}\rightleftharpoons{\rm F}}\vec{J}_1^{\rm A}}\cdot\overline{a}^{\vec{J}_1^{\rm A}}+\overline{\overline{Z}}^{\vec{J}_1^{{\rm A}\rightleftharpoons{\rm F}}\vec{J}^{\rm F}}\cdot\overline{a}^{\vec{J}^{\rm F}}\\&+\overline{\overline{Z}}^{\vec{J}_1^{{\rm A}\rightleftharpoons{\rm F}}\vec{M}^{{\rm G}\rightleftharpoons{\rm A}}}\cdot\overline{a}^{\vec{M}^{{\rm G}\rightleftharpoons{\rm A}}}+\overline{\overline{Z}}^{\vec{J}_1^{{\rm A}\rightleftharpoons{\rm F}}\vec{M}_1^{{\rm A}\rightleftharpoons{\rm F}}}\cdot\overline{a}^{\vec{M}_1^{{\rm A}\rightleftharpoons{\rm F}}}+\overline{\overline{Z}}^{\vec{J}_1^{{\rm A}\rightleftharpoons{\rm F}}\vec{M}_2^{{\rm A}\rightleftharpoons{\rm F}}}\cdot\overline{a}^{\vec{M}_2^{{\rm A}\rightleftharpoons{\rm F}}}=0\qquad(6\text{-}112{\rm a})\\&\overline{\overline{Z}}^{\vec{M}_1^{{\rm A}\rightleftharpoons{\rm F}}\vec{J}^{{\rm G}\rightleftharpoons{\rm A}}}\cdot\overline{a}^{\vec{J}^{{\rm G}\rightleftharpoons{\rm A}}}+\overline{\overline{Z}}^{\vec{M}_1^{{\rm A}\rightleftharpoons{\rm F}}\vec{J}_1^{{\rm A}\rightleftharpoons{\rm F}}}\cdot\overline{a}^{\vec{J}_1^{{\rm A}\rightleftharpoons{\rm F}}}+\overline{\overline{Z}}^{\vec{M}_1^{{\rm A}\rightleftharpoons{\rm F}}\vec{J}_2^{{\rm A}\rightleftharpoons{\rm F}}}\cdot\overline{a}^{\vec{J}_2^{{\rm A}\rightleftharpoons{\rm F}}}+\overline{\overline{Z}}^{\vec{M}_1^{{\rm A}\rightleftharpoons{\rm F}}\vec{J}_1^{\rm A}}\cdot\overline{a}^{\vec{J}_1^{\rm A}}+\overline{\overline{Z}}^{\vec{M}_1^{{\rm A}\rightleftharpoons{\rm F}}\vec{J}^{\rm F}}\cdot\overline{a}^{\vec{J}^{\rm F}}\\&+\overline{\overline{Z}}^{\vec{M}_1^{{\rm A}\rightleftharpoons{\rm F}}\vec{M}^{{\rm G}\rightleftharpoons{\rm A}}}\cdot\overline{a}^{\vec{M}^{{\rm G}\rightleftharpoons{\rm A}}}+\overline{\overline{Z}}^{\vec{M}_1^{{\rm A}\rightleftharpoons{\rm F}}\vec{M}_1^{{\rm A}\rightleftharpoons{\rm F}}}\cdot\overline{a}^{\vec{M}_1^{{\rm A}\rightleftharpoons{\rm F}}}+\overline{\overline{Z}}^{\vec{M}_1^{{\rm A}\rightleftharpoons{\rm F}}\vec{M}_2^{{\rm A}\rightleftharpoons{\rm F}}}\cdot\overline{a}^{\vec{M}_2^{{\rm A}\rightleftharpoons{\rm F}}}=0\qquad(6\text{-}112{\rm b})\end{aligned}$$

and

$$\begin{aligned}&\overline{\overline{Z}}^{\vec{J}_2^{{\rm A}\rightleftharpoons{\rm F}}\vec{J}^{{\rm G}\rightleftharpoons{\rm A}}}\cdot\overline{a}^{\vec{J}^{{\rm G}\rightleftharpoons{\rm A}}}+\overline{\overline{Z}}^{\vec{J}_2^{{\rm A}\rightleftharpoons{\rm F}}\vec{J}_1^{{\rm A}\rightleftharpoons{\rm F}}}\cdot\overline{a}^{\vec{J}_1^{{\rm A}\rightleftharpoons{\rm F}}}+\overline{\overline{Z}}^{\vec{J}_2^{{\rm A}\rightleftharpoons{\rm F}}\vec{J}_2^{{\rm A}\rightleftharpoons{\rm F}}}\cdot\overline{a}^{\vec{J}_2^{{\rm A}\rightleftharpoons{\rm F}}}+\overline{\overline{Z}}^{\vec{J}_2^{{\rm A}\rightleftharpoons{\rm F}}\vec{J}_{23}^{{\rm A}\rightleftharpoons{\rm A}}}\cdot\overline{a}^{\vec{J}_{23}^{{\rm A}\rightleftharpoons{\rm A}}}\\&+\overline{\overline{Z}}^{\vec{J}_2^{{\rm A}\rightleftharpoons{\rm F}}\vec{J}^{\rm F}}\cdot\overline{a}^{\vec{J}^{\rm F}}+\overline{\overline{Z}}^{\vec{J}_2^{{\rm A}\rightleftharpoons{\rm F}}\vec{M}^{{\rm G}\rightleftharpoons{\rm A}}}\cdot\overline{a}^{\vec{M}^{{\rm G}\rightleftharpoons{\rm A}}}+\overline{\overline{Z}}^{\vec{J}_2^{{\rm A}\rightleftharpoons{\rm F}}\vec{M}_1^{{\rm A}\rightleftharpoons{\rm F}}}\cdot\overline{a}^{\vec{M}_1^{{\rm A}\rightleftharpoons{\rm F}}}+\overline{\overline{Z}}^{\vec{J}_2^{{\rm A}\rightleftharpoons{\rm F}}\vec{M}_2^{{\rm A}\rightleftharpoons{\rm F}}}\cdot\overline{a}^{\vec{M}_2^{{\rm A}\rightleftharpoons{\rm F}}}\\&+\overline{\overline{Z}}^{\vec{J}_2^{{\rm A}\rightleftharpoons{\rm F}}\vec{M}_{23}^{{\rm A}\rightleftharpoons{\rm A}}}\cdot\overline{a}^{\vec{M}_{23}^{{\rm A}\rightleftharpoons{\rm A}}}=0\qquad(6\text{-}113{\rm a})\\&\overline{\overline{Z}}^{\vec{M}_2^{{\rm A}\rightleftharpoons{\rm F}}\vec{J}^{{\rm G}\rightleftharpoons{\rm A}}}\cdot\overline{a}^{\vec{J}^{{\rm G}\rightleftharpoons{\rm A}}}+\overline{\overline{Z}}^{\vec{M}_2^{{\rm A}\rightleftharpoons{\rm F}}\vec{J}_1^{{\rm A}\rightleftharpoons{\rm F}}}\cdot\overline{a}^{\vec{J}_1^{{\rm A}\rightleftharpoons{\rm F}}}+\overline{\overline{Z}}^{\vec{M}_2^{{\rm A}\rightleftharpoons{\rm F}}\vec{J}_2^{{\rm A}\rightleftharpoons{\rm F}}}\cdot\overline{a}^{\vec{J}_2^{{\rm A}\rightleftharpoons{\rm F}}}+\overline{\overline{Z}}^{\vec{M}_2^{{\rm A}\rightleftharpoons{\rm F}}\vec{J}_{23}^{{\rm A}\rightleftharpoons{\rm A}}}\cdot\overline{a}^{\vec{J}_{23}^{{\rm A}\rightleftharpoons{\rm A}}}\\&+\overline{\overline{Z}}^{\vec{M}_2^{{\rm A}\rightleftharpoons{\rm F}}\vec{J}^{\rm F}}\cdot\overline{a}^{\vec{J}^{\rm F}}+\overline{\overline{Z}}^{\vec{M}_2^{{\rm A}\rightleftharpoons{\rm F}}\vec{M}^{{\rm G}\rightleftharpoons{\rm A}}}\cdot\overline{a}^{\vec{M}^{{\rm G}\rightleftharpoons{\rm A}}}+\overline{\overline{Z}}^{\vec{M}_2^{{\rm A}\rightleftharpoons{\rm F}}\vec{M}_1^{{\rm A}\rightleftharpoons{\rm F}}}\cdot\overline{a}^{\vec{M}_1^{{\rm A}\rightleftharpoons{\rm F}}}+\overline{\overline{Z}}^{\vec{M}_2^{{\rm A}\rightleftharpoons{\rm F}}\vec{M}_2^{{\rm A}\rightleftharpoons{\rm F}}}\cdot\overline{a}^{\vec{M}_2^{{\rm A}\rightleftharpoons{\rm F}}}\\&+\overline{\overline{Z}}^{\vec{M}_2^{{\rm A}\rightleftharpoons{\rm F}}\vec{M}_{23}^{{\rm A}\rightleftharpoons{\rm A}}}\cdot\overline{a}^{\vec{M}_{23}^{{\rm A}\rightleftharpoons{\rm A}}}=0\qquad(6\text{-}113{\rm b})\end{aligned}$$

and

$$\begin{aligned}&\overline{\overline{Z}}^{\vec{J}_{23}^{{\rm A}\rightleftharpoons{\rm A}}\vec{J}_2^{{\rm A}\rightleftharpoons{\rm F}}}\cdot\overline{a}^{\vec{J}_2^{{\rm A}\rightleftharpoons{\rm F}}}+\overline{\overline{Z}}^{\vec{J}_{23}^{{\rm A}\rightleftharpoons{\rm A}}\vec{J}_{23}^{{\rm A}\rightleftharpoons{\rm A}}}\cdot\overline{a}^{\vec{J}_{23}^{{\rm A}\rightleftharpoons{\rm A}}}+\overline{\overline{Z}}^{\vec{J}_{23}^{{\rm A}\rightleftharpoons{\rm A}}\vec{M}_2^{{\rm A}\rightleftharpoons{\rm F}}}\cdot\overline{a}^{\vec{M}_2^{{\rm A}\rightleftharpoons{\rm F}}}+\overline{\overline{Z}}^{\vec{J}_{23}^{{\rm A}\rightleftharpoons{\rm A}}\vec{M}_{23}^{{\rm A}\rightleftharpoons{\rm A}}}\cdot\overline{a}^{\vec{M}_{23}^{{\rm A}\rightleftharpoons{\rm A}}}=0\quad(6\text{-}114{\rm a})\\&\overline{\overline{Z}}^{\vec{M}_{23}^{{\rm A}\rightleftharpoons{\rm A}}\vec{J}_2^{{\rm A}\rightleftharpoons{\rm F}}}\cdot\overline{a}^{\vec{J}_2^{{\rm A}\rightleftharpoons{\rm F}}}+\overline{\overline{Z}}^{\vec{M}_{23}^{{\rm A}\rightleftharpoons{\rm A}}\vec{J}_{23}^{{\rm A}\rightleftharpoons{\rm A}}}\cdot\overline{a}^{\vec{J}_{23}^{{\rm A}\rightleftharpoons{\rm A}}}+\overline{\overline{Z}}^{\vec{M}_{23}^{{\rm A}\rightleftharpoons{\rm A}}\vec{M}_2^{{\rm A}\rightleftharpoons{\rm F}}}\cdot\overline{a}^{\vec{M}_2^{{\rm A}\rightleftharpoons{\rm F}}}+\overline{\overline{Z}}^{\vec{M}_{23}^{{\rm A}\rightleftharpoons{\rm A}}\vec{M}_{23}^{{\rm A}\rightleftharpoons{\rm A}}}\cdot\overline{a}^{\vec{M}_{23}^{{\rm A}\rightleftharpoons{\rm A}}}=0\quad(6\text{-}114{\rm b})\end{aligned}$$

and

$$\overline{\overline{Z}}^{\vec{J}_1^{\rm A}\vec{J}_1^{\rm A \rightleftharpoons F}} \cdot \overline{a}^{\vec{J}_1^{\rm A \rightleftharpoons F}} + \overline{\overline{Z}}^{\vec{J}_1^{\rm A}\vec{J}_1^{\rm A}} \cdot \overline{a}^{\vec{J}_1^{\rm A}} + \overline{\overline{Z}}^{\vec{J}_1^{\rm A}\vec{M}_1^{\rm A \rightleftharpoons F}} \cdot \overline{a}^{\vec{M}_1^{\rm A \rightleftharpoons F}} = 0 \tag{6-115}$$

and

$$\overline{\overline{Z}}^{\vec{J}^{\rm F}\vec{J}^{\rm G \rightleftharpoons A}} \cdot \overline{a}^{\vec{J}^{\rm G \rightleftharpoons A}} + \overline{\overline{Z}}^{\vec{J}^{\rm F}\vec{J}_1^{\rm A \rightleftharpoons F}} \cdot \overline{a}^{\vec{J}_1^{\rm A \rightleftharpoons F}} + \overline{\overline{Z}}^{\vec{J}^{\rm F}\vec{J}_2^{\rm A \rightleftharpoons F}} \cdot \overline{a}^{\vec{J}_2^{\rm A \rightleftharpoons F}} + \overline{\overline{Z}}^{\vec{J}^{\rm F}\vec{J}^{\rm F}} \cdot \overline{a}^{\vec{J}^{\rm F}} + \overline{\overline{Z}}^{\vec{J}^{\rm F}\vec{M}^{\rm G \rightleftharpoons A}} \cdot \overline{a}^{\vec{M}^{\rm G \rightleftharpoons A}}$$
$$+ \overline{\overline{Z}}^{\vec{J}^{\rm F}\vec{M}_1^{\rm A \rightleftharpoons F}} \cdot \overline{a}^{\vec{M}_1^{\rm A \rightleftharpoons F}} + \overline{\overline{Z}}^{\vec{J}^{\rm F}\vec{M}_2^{\rm A \rightleftharpoons F}} \cdot \overline{a}^{\vec{M}_2^{\rm A \rightleftharpoons F}} = 0 \tag{6-116}$$

The formulations used to calculate the elements of the matrices in the above matrix equations are similar to the ones given in Eqs. (6-8a)~(6-10c) and Eqs. (6-49a)~(6-52d), and they are explicitly given in the App. D7 of this report. By employing the above matrix equations, we propose two schemes for mathematically describing modal space as below.

Employing the above Eqs. (6-111a)~(6-116), we can obtain the following transformation

$$\overline{a}^{\rm AV} = \overline{\overline{T}} \cdot \overline{a} \tag{6-117}$$

where

$$\overline{a}^{\rm AV} = \begin{bmatrix} \overline{a}^{\vec{J}^{\rm G \rightleftharpoons A}} \\ \overline{a}^{\vec{J}_1^{\rm A \rightleftharpoons F}} \\ \overline{a}^{\vec{J}_2^{\rm A \rightleftharpoons F}} \\ \overline{a}^{\vec{J}_{23}^{\rm A \rightleftharpoons A}} \\ \overline{a}^{\vec{J}_1^{\rm A}} \\ \overline{a}^{\vec{J}^{\rm F}} \\ \overline{a}^{\vec{M}^{\rm G \rightleftharpoons A}} \\ \overline{a}^{\vec{M}_1^{\rm A \rightleftharpoons F}} \\ \overline{a}^{\vec{M}_2^{\rm A \rightleftharpoons F}} \\ \overline{a}^{\vec{M}_{23}^{\rm A \rightleftharpoons A}} \end{bmatrix} \tag{6-118}$$

and the calculation formulation for transformation matrix $\overline{\overline{T}}$ is given in the App. D7 of this report.

### 6.6.4 Power Transport Theorem and Input Power Operator

Applying the results obtained in Chap. 2 to the tra-antenna shown in Fig. 6-66, we immediately have the following PTT for the tra-antenna

$$P^{\rm G \rightleftharpoons A} = P_{\rm rad}^{\rm I} + \underbrace{P_{\rm dis}^{\rm A} + j\, P_{\rm sto}^{\rm A}}_{-P^{\rm A \rightleftharpoons F}} + j\, P_{\rm sto}^{\rm F} \tag{6-119}$$

where $P^{\rm G \rightleftharpoons A}$ is the input power inputted into the tra-antenna, and $P_{\rm dis}^{\rm A}$ is the power dissipated in the tra-antenna, and $P_{\rm rad}^{\rm I}$ is the radiated power arriving at infinity, and $P_{\rm sto}^{\rm F}$ is the power corresponding to the stored energy in free space, and $P_{\rm sto}^{\rm A}$ is the power corresponding to the stored energy in the tra-antenna.

The above-mentioned various powers are as follows:

$$P^{\rm G\rightleftharpoons A} = (1/2)\iint_{\mathbb{S}^{\rm G\rightleftharpoons A}} \left(\vec{E}\times\vec{H}^{\dagger}\right)\cdot\hat{n}^{\rightarrow \rm A} dS \tag{6-120a}$$

$$P^{\rm A\rightleftharpoons F} = (1/2)\iint_{\mathbb{S}_1^{\rm A\rightleftharpoons F}} \left(\vec{E}\times\vec{H}^{\dagger}\right)\cdot\hat{n}^{\rightarrow \rm F} dS + (1/2)\iint_{\mathbb{S}_2^{\rm A\rightleftharpoons F}} \left(\vec{E}\times\vec{H}^{\dagger}\right)\cdot\hat{n}^{\rightarrow \rm F} dS \tag{6-120b}$$

$$P_{\rm dis}^{\rm A} = (1/2)\left\langle\vec{\sigma}_1\cdot\vec{E},\vec{E}\right\rangle_{\mathbb{V}_1^{\rm A}} + (1/2)\left\langle\vec{\sigma}_2\cdot\vec{E},\vec{E}\right\rangle_{\mathbb{V}_2^{\rm A}} + (1/2)\left\langle\vec{\sigma}_3\cdot\vec{E},\vec{E}\right\rangle_{\mathbb{V}_3^{\rm A}} \tag{6-120c}$$

$$P_{\rm rad}^{\rm I} = (1/2)\oiint_{\mathbb{S}^{\rm F\to I}} \left(\vec{E}\times\vec{H}^{\dagger}\right)\cdot\hat{n}^{\rightarrow \rm I} dS \tag{6-120d}$$

$$P_{\rm sto}^{\rm F} = 2\omega\left[(1/4)\left\langle\vec{H},\mu_0\vec{H}\right\rangle_{\mathbb{V}^{\rm F}} - (1/4)\left\langle\varepsilon_0\vec{E},\vec{E}\right\rangle_{\mathbb{V}^{\rm F}}\right] \tag{6-120e}$$

$$\begin{aligned} P_{\rm sto}^{\rm A} = & 2\omega\left[(1/4)\left\langle\vec{H},\vec{\mu}_1\cdot\vec{H}\right\rangle_{\mathbb{V}_1^{\rm A}} - (1/4)\left\langle\vec{\varepsilon}_1\cdot\vec{E},\vec{E}\right\rangle_{\mathbb{V}_1^{\rm A}}\right] \\ & + 2\omega\left[(1/4)\left\langle\vec{H},\vec{\mu}_2\cdot\vec{H}\right\rangle_{\mathbb{V}_2^{\rm A}} - (1/4)\left\langle\vec{\varepsilon}_2\cdot\vec{E},\vec{E}\right\rangle_{\mathbb{V}_2^{\rm A}}\right] \\ & + 2\omega\left[(1/4)\left\langle\vec{H},\vec{\mu}_3\cdot\vec{H}\right\rangle_{\mathbb{V}_3^{\rm A}} - (1/4)\left\langle\vec{\varepsilon}_3\cdot\vec{E},\vec{E}\right\rangle_{\mathbb{V}_3^{\rm A}}\right] \end{aligned} \tag{6-120f}$$

where $\hat{n}^{\rightarrow \rm F}$ is the normal direction of $\mathbb{S}_1^{\rm A\rightleftharpoons F}\bigcup\mathbb{S}_2^{\rm A\rightleftharpoons F}$ and points to $\mathbb{V}^{\rm F}$, and $\hat{n}^{\rightarrow \rm I}$ is the normal direction of $\mathbb{S}^{\rm F\to I}$ and points to infinity.

Based on Eqs. (6-104a)&(6-104b) and the tangential continuity of the $\{\vec{E},\vec{H}\}$ on $\mathbb{S}^{\rm G\rightleftharpoons A}$, the IPO $P^{\rm G\rightleftharpoons A}$ given in Eq. (6-120a) can be alternatively written as follows:

$$\begin{aligned} P^{\rm G\rightleftharpoons A} = & (1/2)\left\langle\hat{n}^{\rightarrow \rm A}\times\vec{J}^{\rm G\rightleftharpoons A},\vec{M}^{\rm G\rightleftharpoons A}\right\rangle_{\mathbb{S}^{\rm G\rightleftharpoons A}} \\ = & -\frac{1}{2}\left\langle\vec{J}^{\rm G\rightleftharpoons A},\mathcal{E}_0\left(\vec{J}^{\rm G\rightleftharpoons A}-\vec{J}_1^{\rm A\rightleftharpoons F}-\vec{J}_2^{\rm A\rightleftharpoons F}+\vec{J}^{\rm F},\vec{M}^{\rm G\rightleftharpoons A}-\vec{M}_1^{\rm A\rightleftharpoons F}-\vec{M}_2^{\rm A\rightleftharpoons F}\right)\right\rangle_{\mathbb{S}^{\rm G\rightleftharpoons A}} \\ = & -\frac{1}{2}\left\langle\vec{M}^{\rm G\rightleftharpoons A},\mathcal{H}_0\left(\vec{J}^{\rm G\rightleftharpoons A}-\vec{J}_1^{\rm A\rightleftharpoons F}-\vec{J}_2^{\rm A\rightleftharpoons F}+\vec{J}^{\rm F},\vec{M}^{\rm G\rightleftharpoons A}-\vec{M}_1^{\rm A\rightleftharpoons F}-\vec{M}_2^{\rm A\rightleftharpoons F}\right)\right\rangle_{\mathbb{S}^{\rm G\rightleftharpoons A}}^{\dagger} \end{aligned} \tag{6-121}$$

Here, the right-hand side of the first equality is the current form of IPO, and the right-hand sides of the second and third equalities are the interaction forms of IPO.

By discretizing IPO (6-121) and utilizing transformation (6-117), we derive the following matrix form of the IPO

$$P^{\rm G\rightleftharpoons A} = \bar{a}^{\dagger}\cdot\overline{\overline{P}}^{\rm G\rightleftharpoons A}\cdot\bar{a} \tag{6-122}$$

and the formulation for calculating matrix $\overline{\overline{P}}^{\rm G\rightleftharpoons A}$ is given in the App. D7 of this report.

### 6.6.5 Input-Power-Decoupled Modes

The IP-DMs in modal space can be derived from solving the modal decoupling equation

$\overline{\overline{P}}_{-}^{\mathrm{G}\rightleftharpoons\mathrm{A}} \cdot \bar{\alpha}_{\xi} = \theta_{\xi} \overline{\overline{P}}_{+}^{\mathrm{G}\rightleftharpoons\mathrm{A}} \cdot \bar{\alpha}_{\xi}$ defined on modal space, where $\overline{\overline{P}}_{+}^{\mathrm{G}\rightleftharpoons\mathrm{A}}$ and $\overline{\overline{P}}_{-}^{\mathrm{G}\rightleftharpoons\mathrm{A}}$ are the positive and negative Hermitian parts of matrix $\overline{\overline{P}}^{\mathrm{G}\rightleftharpoons\mathrm{A}}$. If some derived modes $\{\bar{\alpha}_1, \bar{\alpha}_2, \cdots, \bar{\alpha}_d\}$ are $d$-order degenerate, then the Gram-Schmidt orthogonalization process given in previous Sec. 4.2.4.1 is necessary, and it is not repeated here.

The modal fields constructed above satisfy the following decoupling relation

$$(1/2)\iint_{\mathbb{S}^{\mathrm{G}\rightleftharpoons\mathrm{A}}} \left(\vec{E}_{\zeta} \times \vec{H}_{\xi}^{\dagger}\right) \cdot \hat{n}^{\rightarrow\mathrm{A}} dS \;=\; \left(1 + j\,\theta_{\xi}\right) \delta_{\xi\zeta} \tag{6-123}$$

and the relation implies that **the IP-DMs don't have net energy coupling in one period**. By employing the decoupling relation, we have the following Parseval's identity

$$\sum_{\xi} \left|c_{\xi}\right|^2 \;=\; \left(1/T\right) \int_{t_0}^{t_0+T} \left[\iint_{\mathbb{S}^{\mathrm{G}\rightleftharpoons\mathrm{A}}} \left(\vec{\mathcal{E}} \times \vec{\mathcal{H}}\right) \cdot \hat{n}^{\rightarrow\mathrm{A}} dS\right] dt \tag{6-124}$$

in which $\{\vec{\mathcal{E}}, \vec{\mathcal{H}}\}$ are the time-domain fields, and the expansion coefficients $c_{\xi}$ have expression $c_{\xi} = -(1/2) < \vec{J}_{\xi}^{\mathrm{G}\rightleftharpoons\mathrm{A}}, \vec{E} >_{\mathbb{S}^{\mathrm{G}\rightleftharpoons\mathrm{A}}} \big/ (1 + j\,\theta_{\xi}) = -(1/2) < \vec{H}, \vec{M}_{\xi}^{\mathrm{G}\rightleftharpoons\mathrm{A}} >_{\mathbb{S}^{\mathrm{G}\rightleftharpoons\mathrm{A}}} \big/ (1 + j\,\theta_{\xi})$, where $\{\vec{E}, \vec{H}\}$ is a previously known field distributing on input port $\mathbb{S}^{\mathrm{G}\rightleftharpoons\mathrm{A}}$.

Just like the metallic tra-antenna discussed in Sec. 6.2.4.3, we can also define the modal significance $\mathrm{MS}_{\xi} = 1/|1 + j\,\theta_{\xi}|$, modal input impedance $Z_{\xi}^{\mathrm{G}\rightleftharpoons\mathrm{A}} = P_{\xi}^{\mathrm{G}\rightleftharpoons\mathrm{A}} \big/ \left[(1/2) < \vec{J}_{\xi}^{\mathrm{G}\rightleftharpoons\mathrm{A}}, \vec{J}_{\xi}^{\mathrm{G}\rightleftharpoons\mathrm{A}} >_{\mathbb{S}^{\mathrm{G}\rightleftharpoons\mathrm{A}}}\right]$, and modal input admittance $Y_{\xi}^{\mathrm{G}\rightleftharpoons\mathrm{A}} = P_{\xi}^{\mathrm{G}\rightleftharpoons\mathrm{A}} \big/ \left[(1/2) < \vec{M}_{\xi}^{\mathrm{G}\rightleftharpoons\mathrm{A}}, \vec{M}_{\xi}^{\mathrm{G}\rightleftharpoons\mathrm{A}} >_{\mathbb{S}^{\mathrm{G}\rightleftharpoons\mathrm{A}}}\right]$, to quantitatively describe the modal features of the tra-antenna shown in Fig. 6-66.

## 6.6.6 Numerical Examples Corresponding to Typical Structures

In this section, we consider a metallic circular horn fed material lens tra-antenna as shown in the following Fig. 6-68.

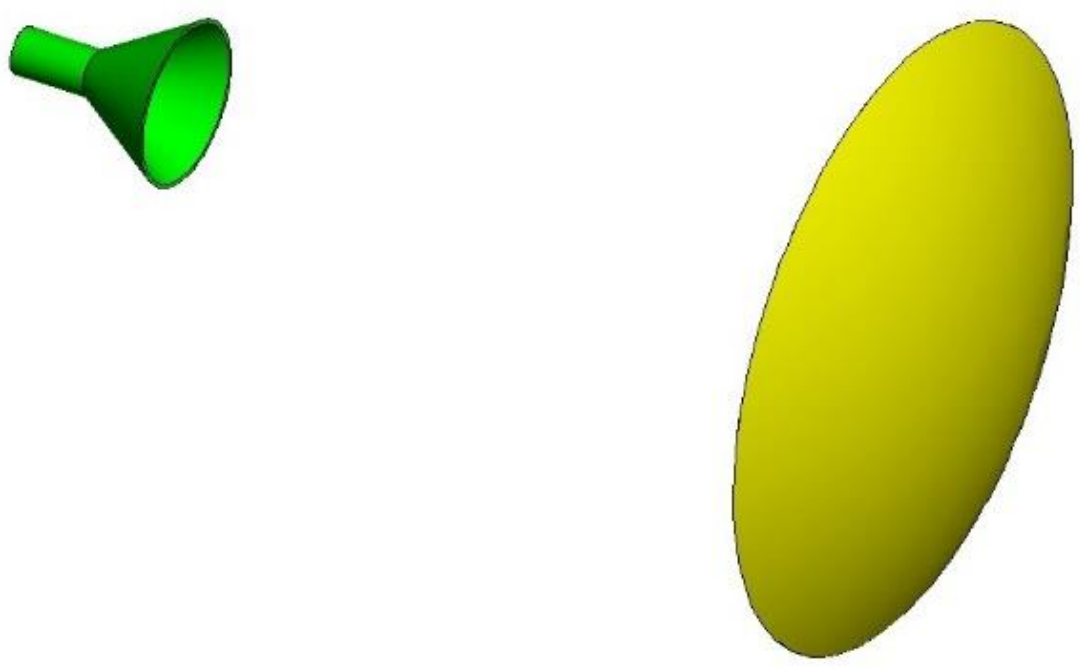

Figure 6-68 Geometry of a metallic horn fed material lens tra-antenna

The feeding horn is placed at the focal point of the lens. The size of the tra-antenna is illustrated in the following Fig. 6-69.

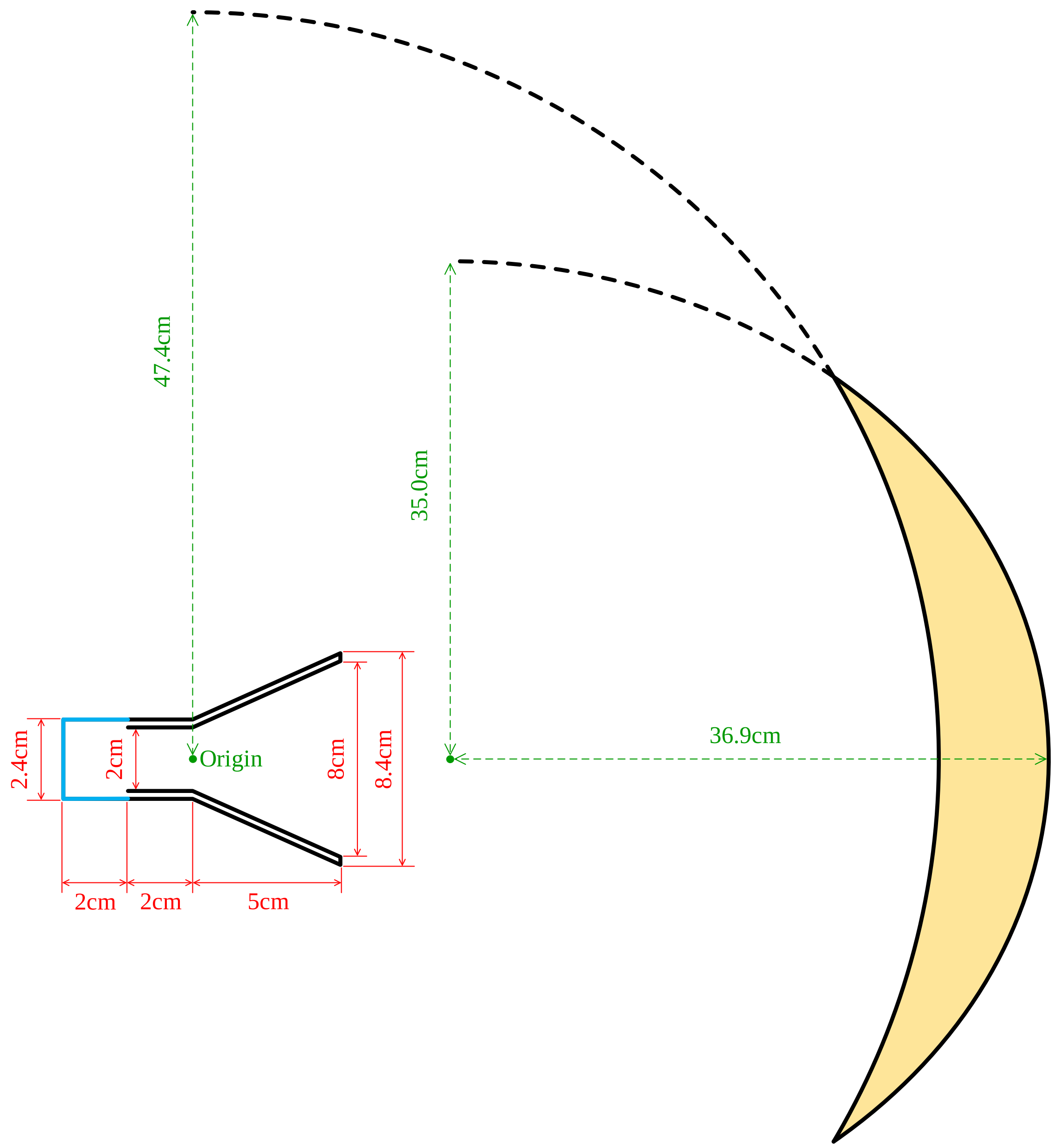

Figure 6-69 Size of the tra-antenna shown in Fig. 6-68

The metallic horn is a thick horn as shown in the above Fig. 6-69. The blue surface is just the interface between feeding structure and free space, and it is a metallic wall. The “Origin” in Fig. 6-69 is just the coordinate origin O with coordinates $\{x=0, y=0, z=0\}$. The lens is characterized by a sphere with radius 47.4cm and a ellipsoid with X-, Y- and Z- radiuses 35.0cm, 36.9cm and 35.0cm. The relative permeability, relative permittivity, and conductivity of the material lens are 1, 10, and 0 respectively. The topological structures and surface triangular meshes of the tra-antenna are shown in Fig. 6-70.

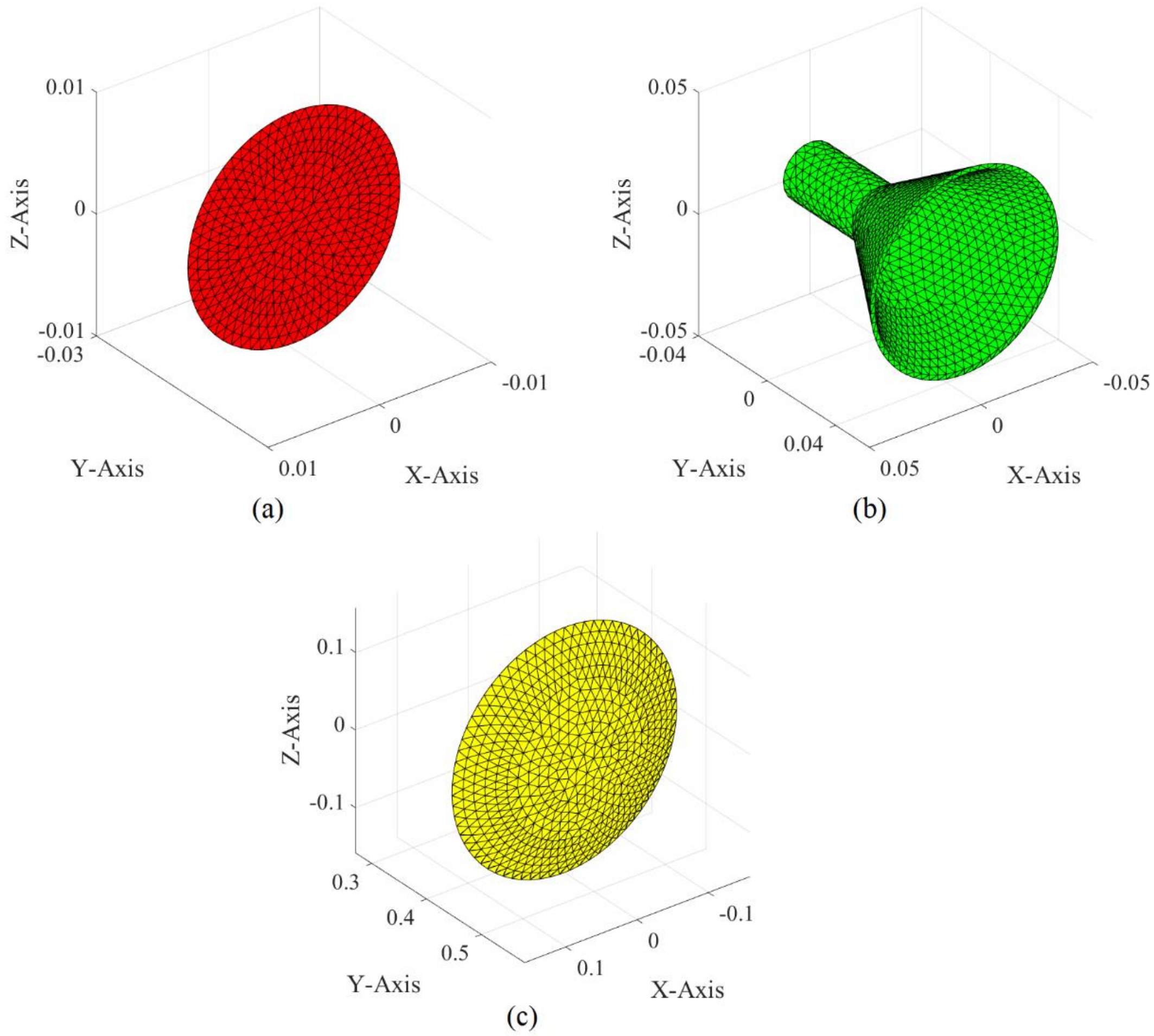

Figure 6-70 Topological structures and surface triangular meshes of (a) input port $\mathbb{S}^{G \rightleftharpoons A}$, (b) interface $\mathbb{S}^{F}$ between ground structure and free space, and (c) interface $\mathbb{S}^{A \rightleftharpoons F}$ between material lens and free space

Using the JE-DoJ-based and HM-DoM-based formulations of IPO, we calculate the IP-DMs of the tra-antenna, and plot the corresponding resistance curves in Fig. 6-71.

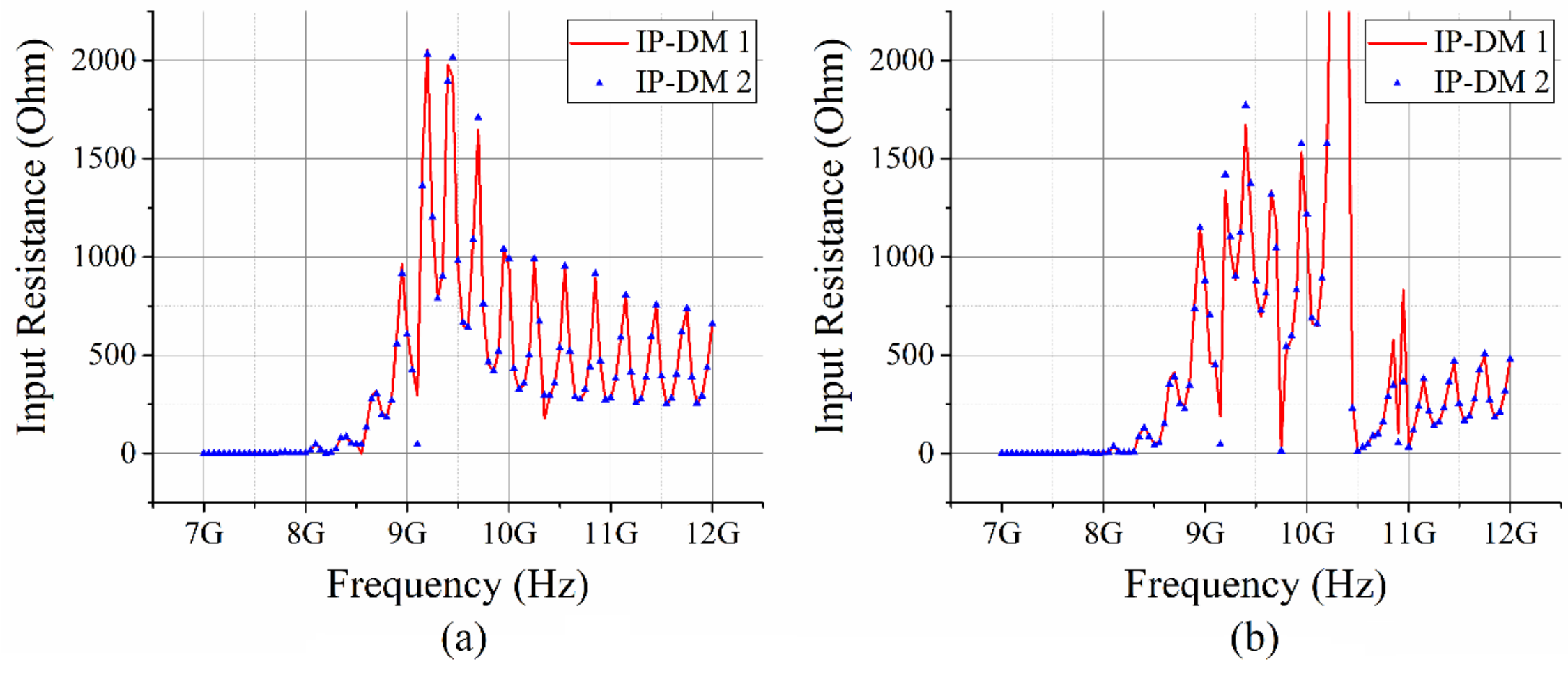

Figure 6-71 Modal input resistance curves of some typical IP-DMs. (a) JE-DoJ-based results; (b) HM-DoM-based results

The modal equivalent electric and magnetic currents distributing on the input port are shown in the following Fig. 6-72.

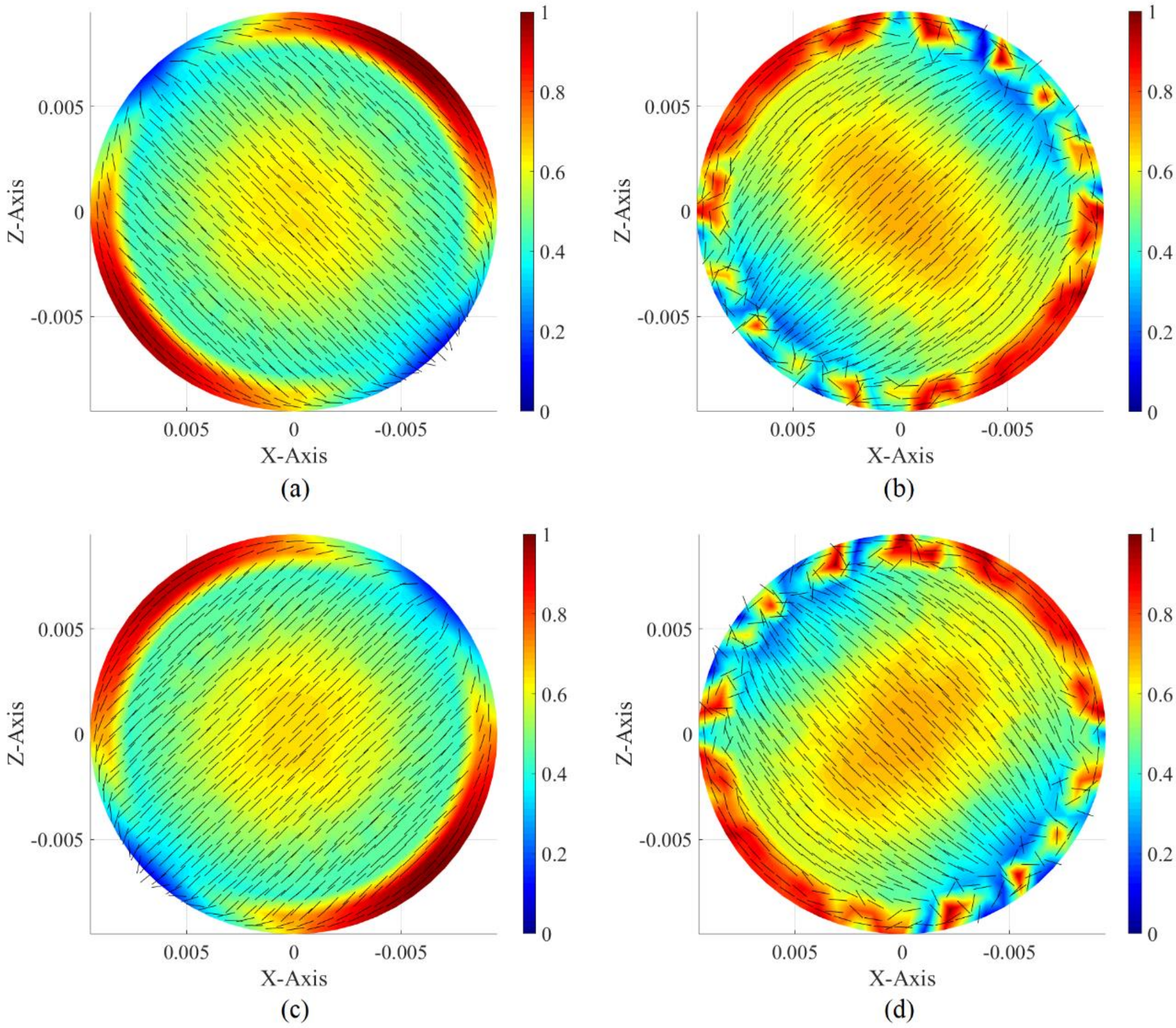

Figure 6-72 Modal port currents of the IP-DMs shown in Fig. 6-71. (a) Electric current of IP-DM 1; (b) magnetic current of IP-DM 1; (c) electric current of IP-DM 2; (d) magnetic current of IP-DM 2

By comparing the above Figs. 6-72(a)&(b) and 6-72(c)&(d), it is not difficult to find out that the port currents of the two modes are rotationally symmetrical around Y-axis, i.e., the IP-DM 1 and IP-DM 2 are a pair of degenerate states. The reason leading to the degeneracy phenomenon is that the tra-antenna is rotationally symmetrical about Y-axis as shown in Figs. 6-68 ~ 6-70.

Taking the IP-DM 1 working at 9.2 GHz as a typical example, we plot its modal electric field distributing on yOz coordinate plane and its modal far-field radiation pattern in the following Fig. 6-73.

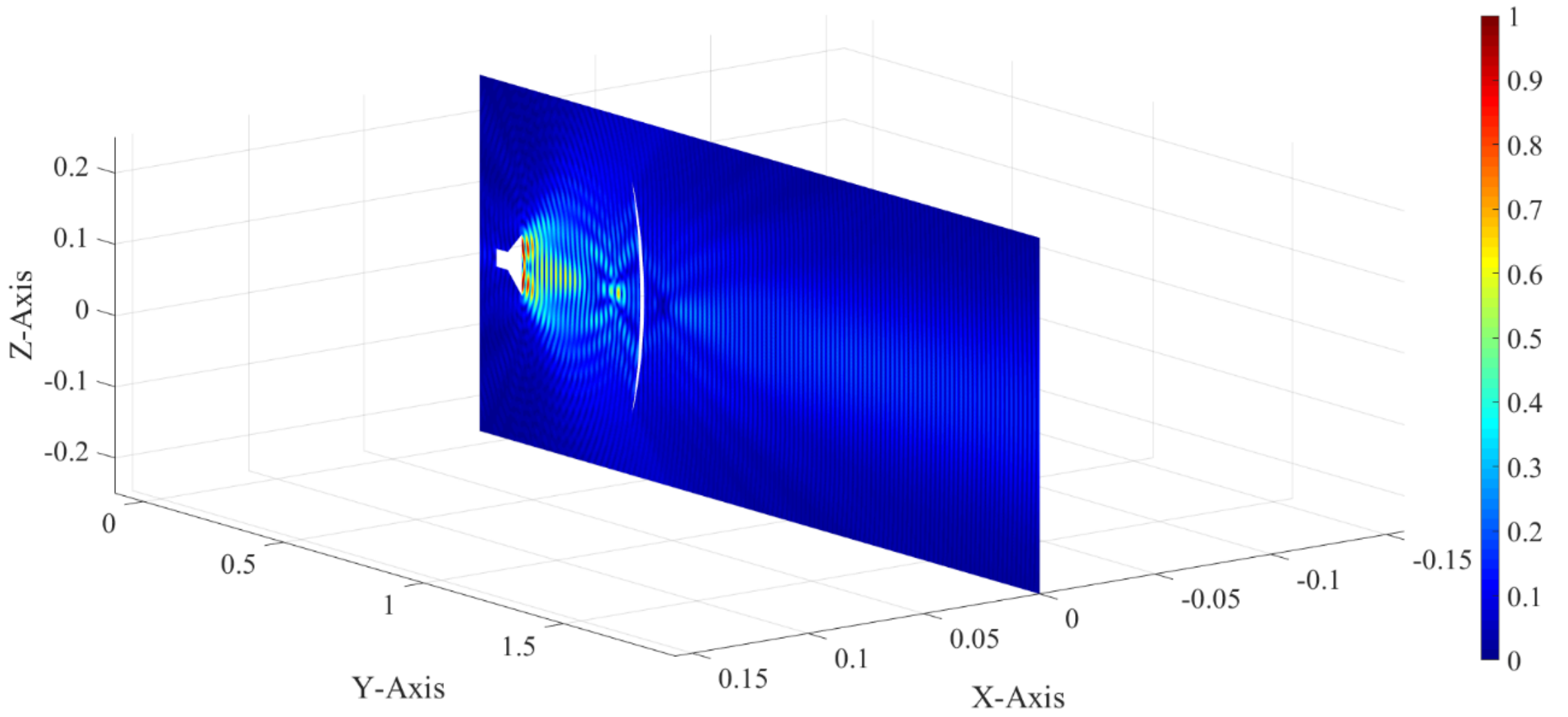

(a)

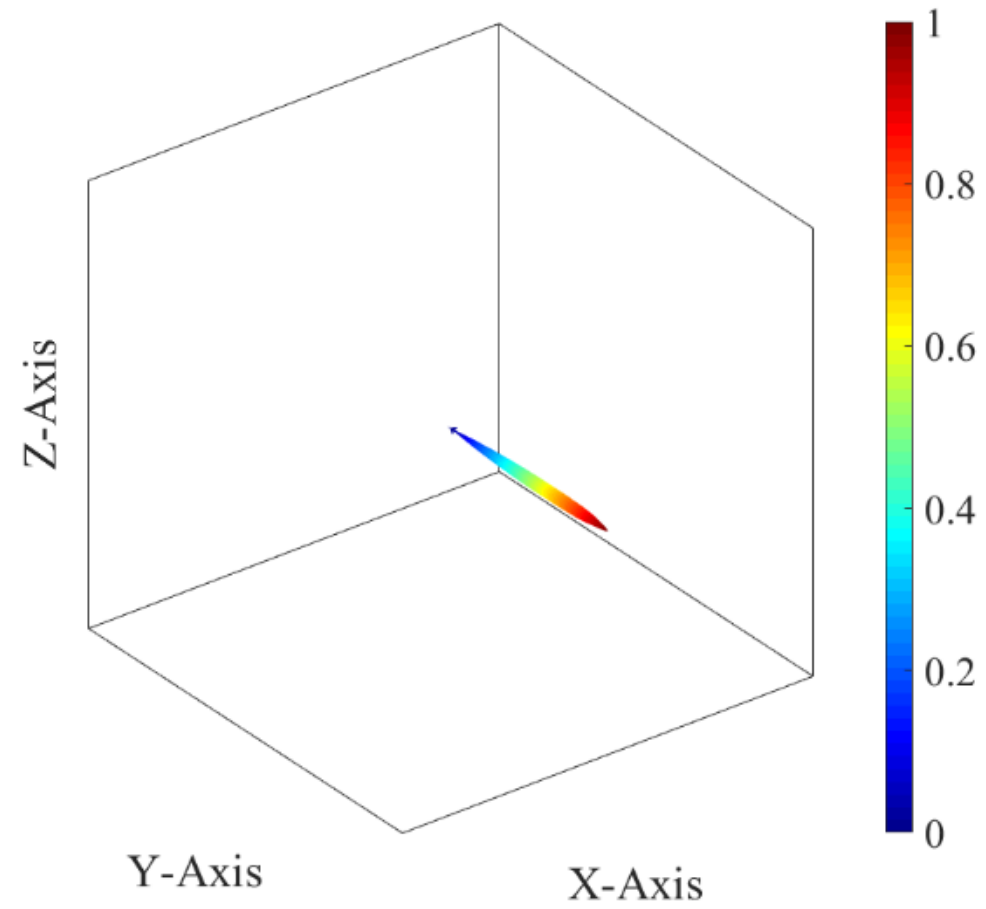

(b)

Figure 6-73 Modal (a) electric field distribution and (b) far-field radiation pattern of the IP-DM 1 working at 9.2 GHz

## 6.7 IP-DMs of Augmented Transmitting Antenna Array

In this section, we further generalize the theory and formulations established in the previous Sec. 6.2 (for a single metallic tra-antenna), Sec. 6.3 (for a single material tra-antenna), and Secs. 6.4~6.6 (for a single metal-material composite tra-antenna) to *tra-antenna array*.

Here, we only focus on the metallic tra-antenna array case, and the material tra-antenna array and composite tra-antenna array cases can be similarly discussed. As a typical example, we consider the *two-element metallic horn array* shown in the following Fig. 6-74, and the array is placed in free space.

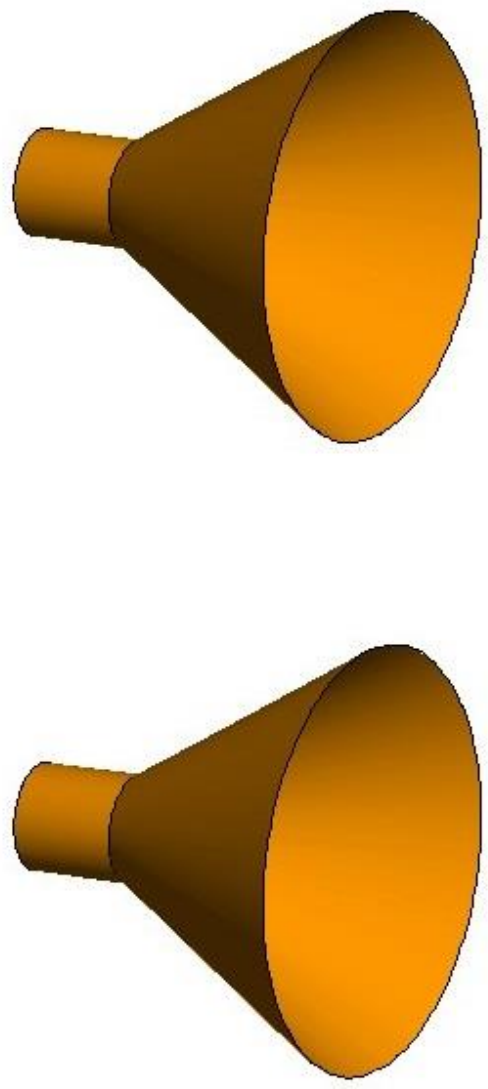

Figure 6-74 Geometry of a two-element metallic horn array

The topological structure of the two-element metallic horn array is illustrated in the following Fig. 6-75.

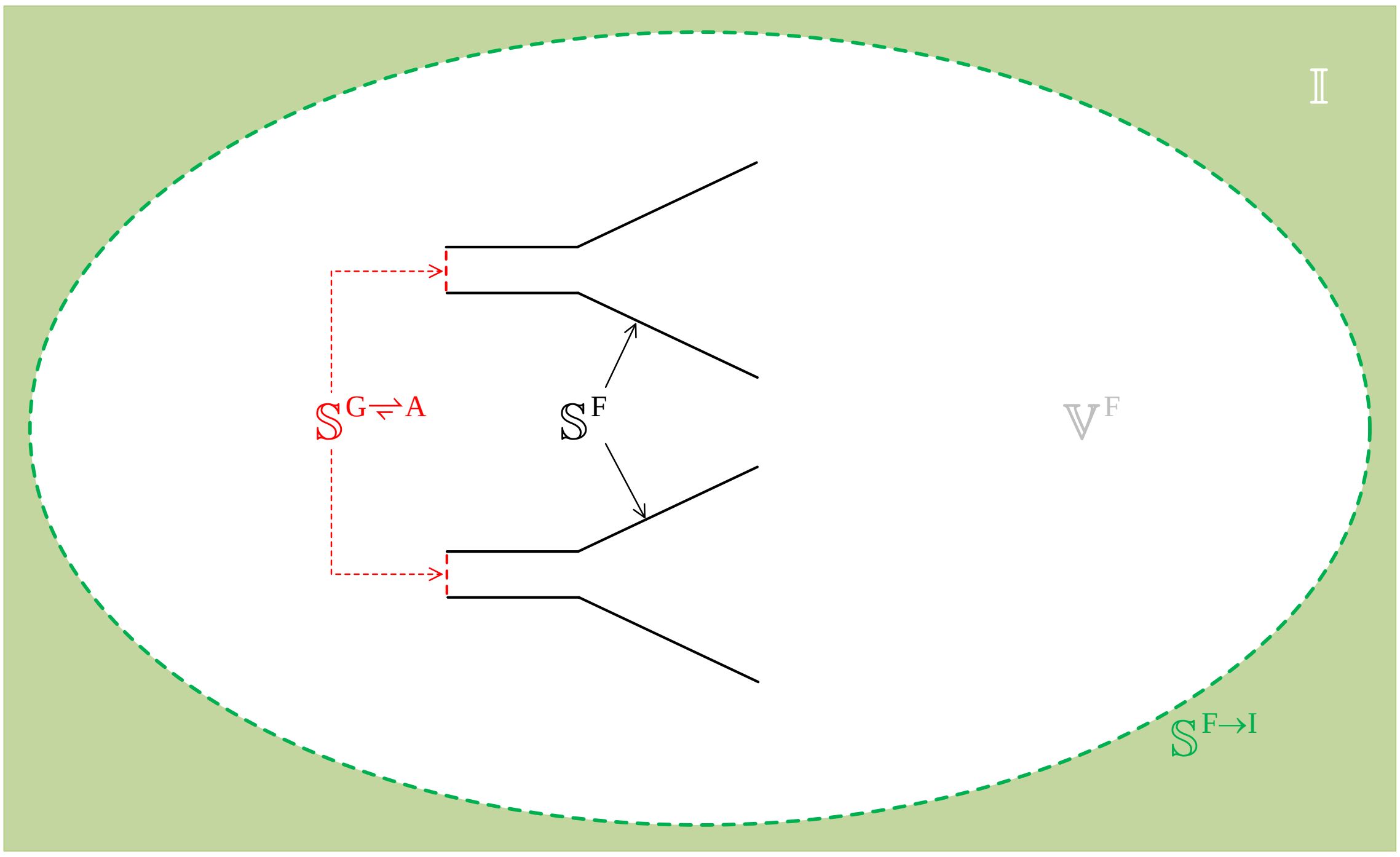

Figure 6-75 Topological structure of the metallic horn array shown in Fig. 6-74

In the above Fig. 6-75, surface $\mathbb{S}^{G \rightleftharpoons A}$ is the input port of the array, and $\mathbb{S}^{F}$ is the interface between the array and free space, and $\mathbb{S}^{F \to I}$ is a spherical surface with infinite radius, and $\mathbb{V}^{F}$ is the region occupied by free space. Obviously, the metallic tra-antenna array has a completely identical topological structure to the single metallic tra-antenna shown in the previous Fig. 6-2. Thus the corresponding formulations are also identical to each other, and they will not be repeated here.

Now, we provide a typical numerical example as below. The surface triangular meshes of the example is shown in the following Fig. 6-76.

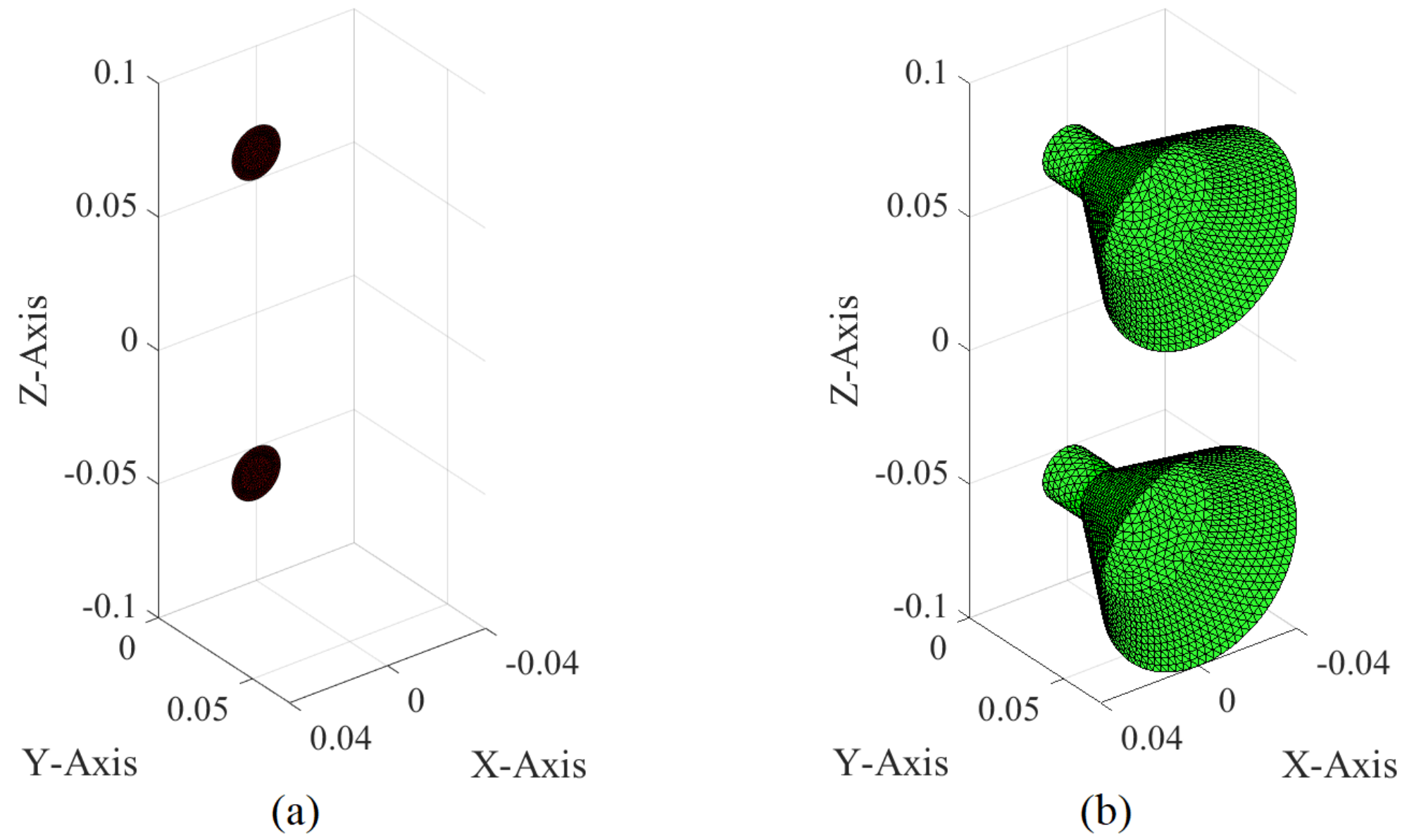

Figure 6-76 Surface triangular meshes of a two-element metallic horn array. (a) Meshes of input port; (b) meshes of metallic horns

The two metallic horns have the completely same size as the one considered in the previous Sec. 6.2.5.1. The distance between the axises of the metallic horns is 12cm.

The resistance curves of some typical IP-DMs of the array are shown in the following Fig. 6-77.

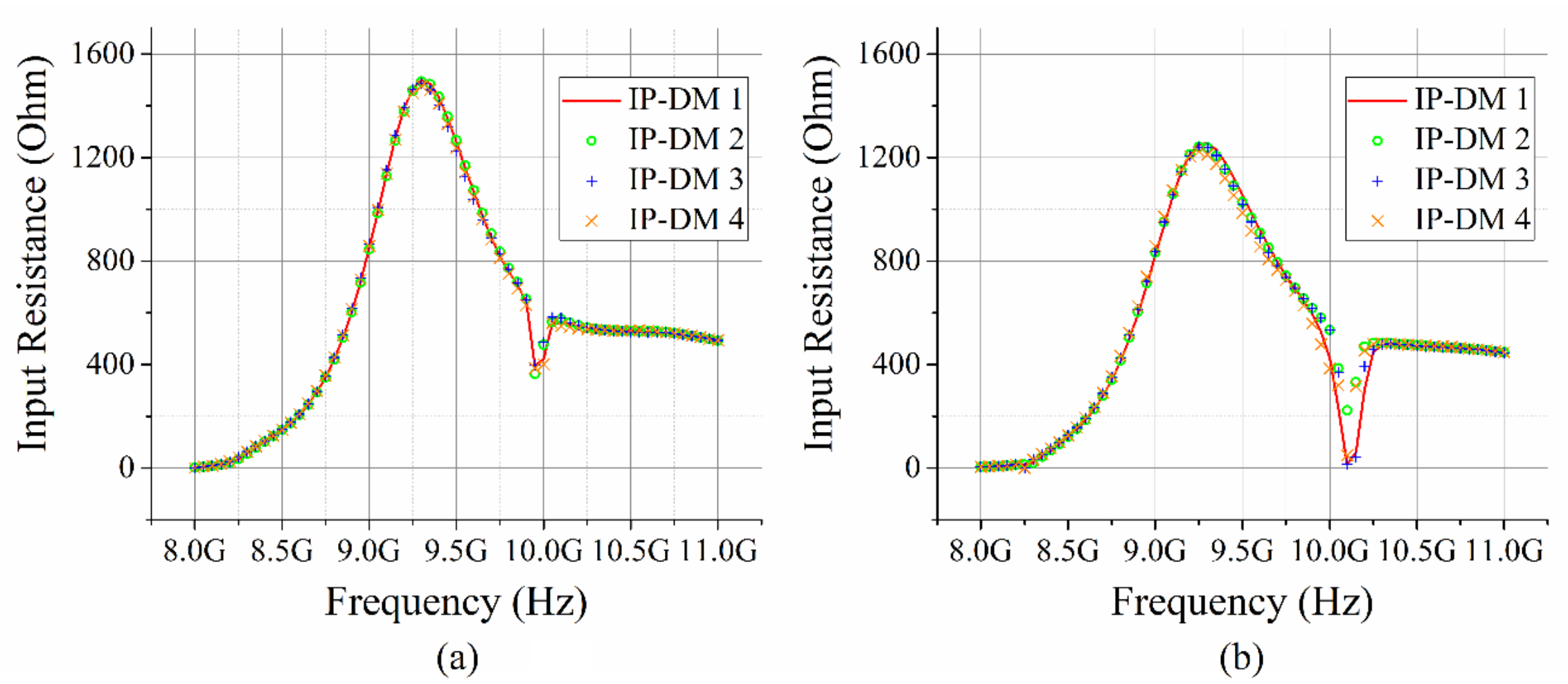

Figure 6-77 Modal resistance curves of the array. (a) JE-DoJ-based results; (b) HM-DoM-based results

For the IP-DM 1 at 9.3 GHz, its some modal quantities are plotted in Fig. 6-78.

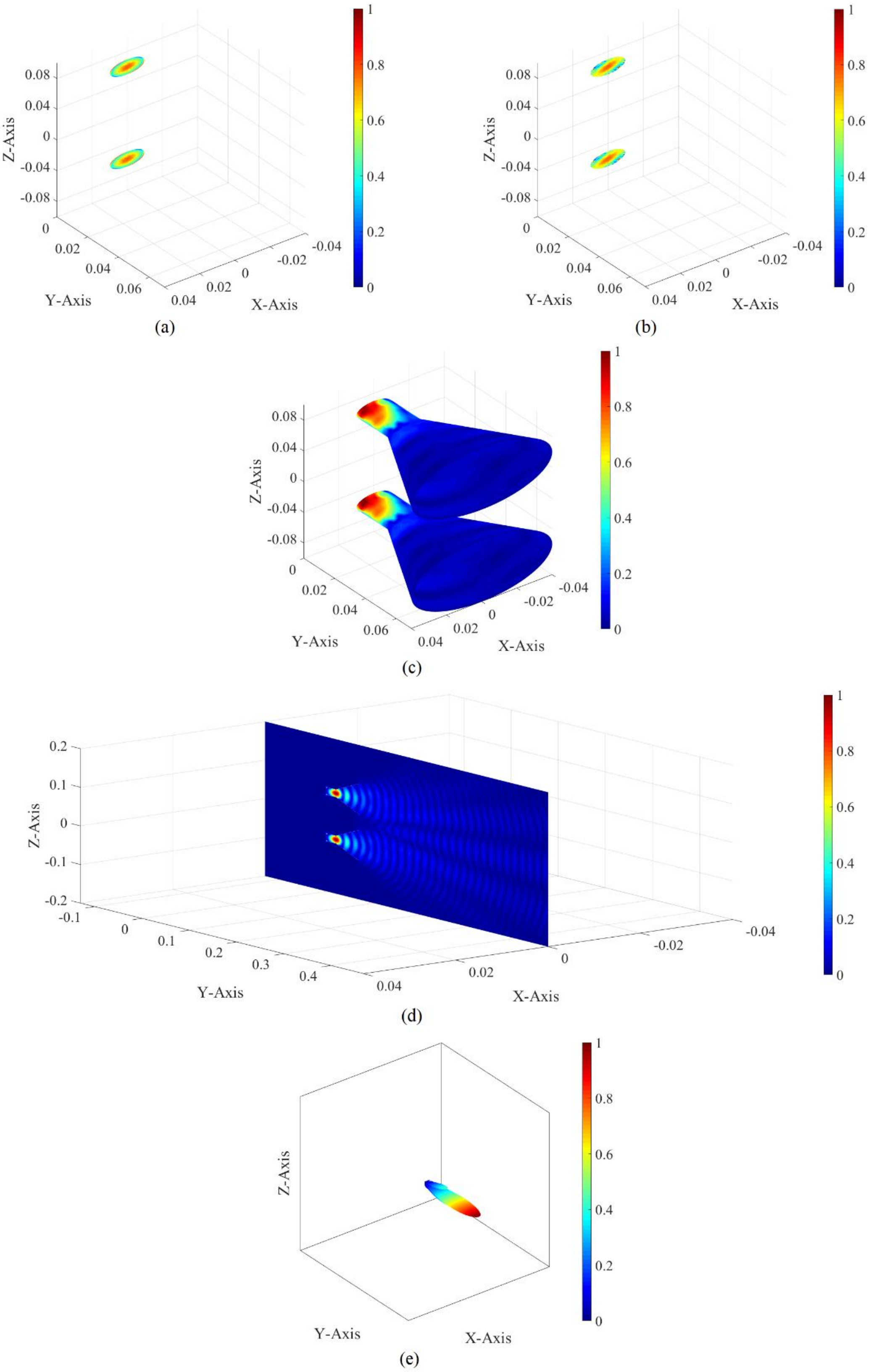

Figure 6-78 Modal (a) port electric current, (b) port magnetic current, (c) wall electric current, (d) electric field, and (e) radiation pattern of the IP-DM 1 at 9.3 GHz

For the IP-DM 2 at 9.3 GHz, its some modal quantities are plotted in Fig. 6-79.

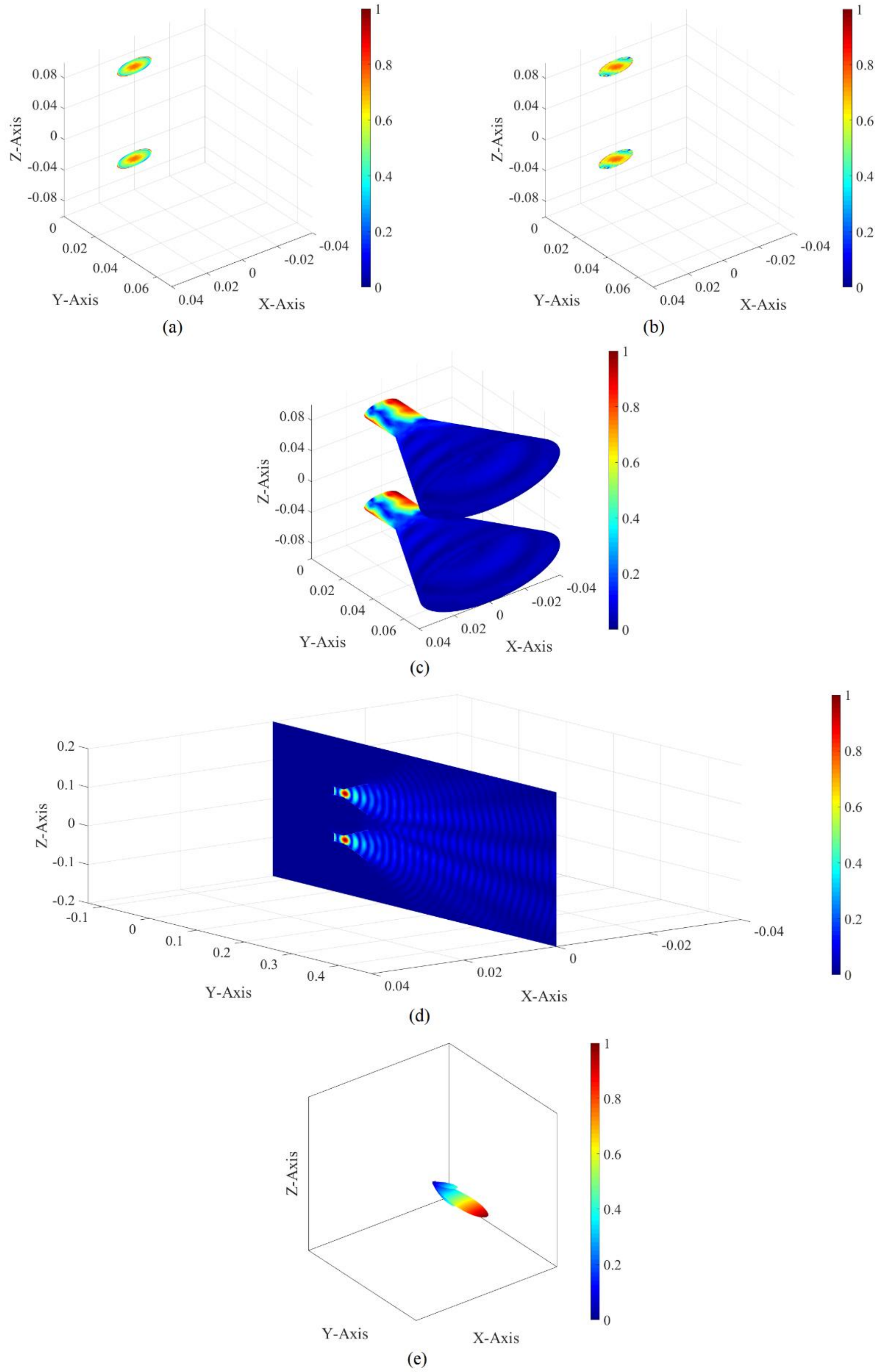

Figure 6-79 Modal (a) port electric current, (b) port magnetic current, (c) wall electric current, (d) electric field, and (e) radiation pattern of the IP-DM 2 at 9.3 GHz

For the IP-DM 3 at 9.3 GHz, its some modal quantities are plotted in Fig. 6-80.

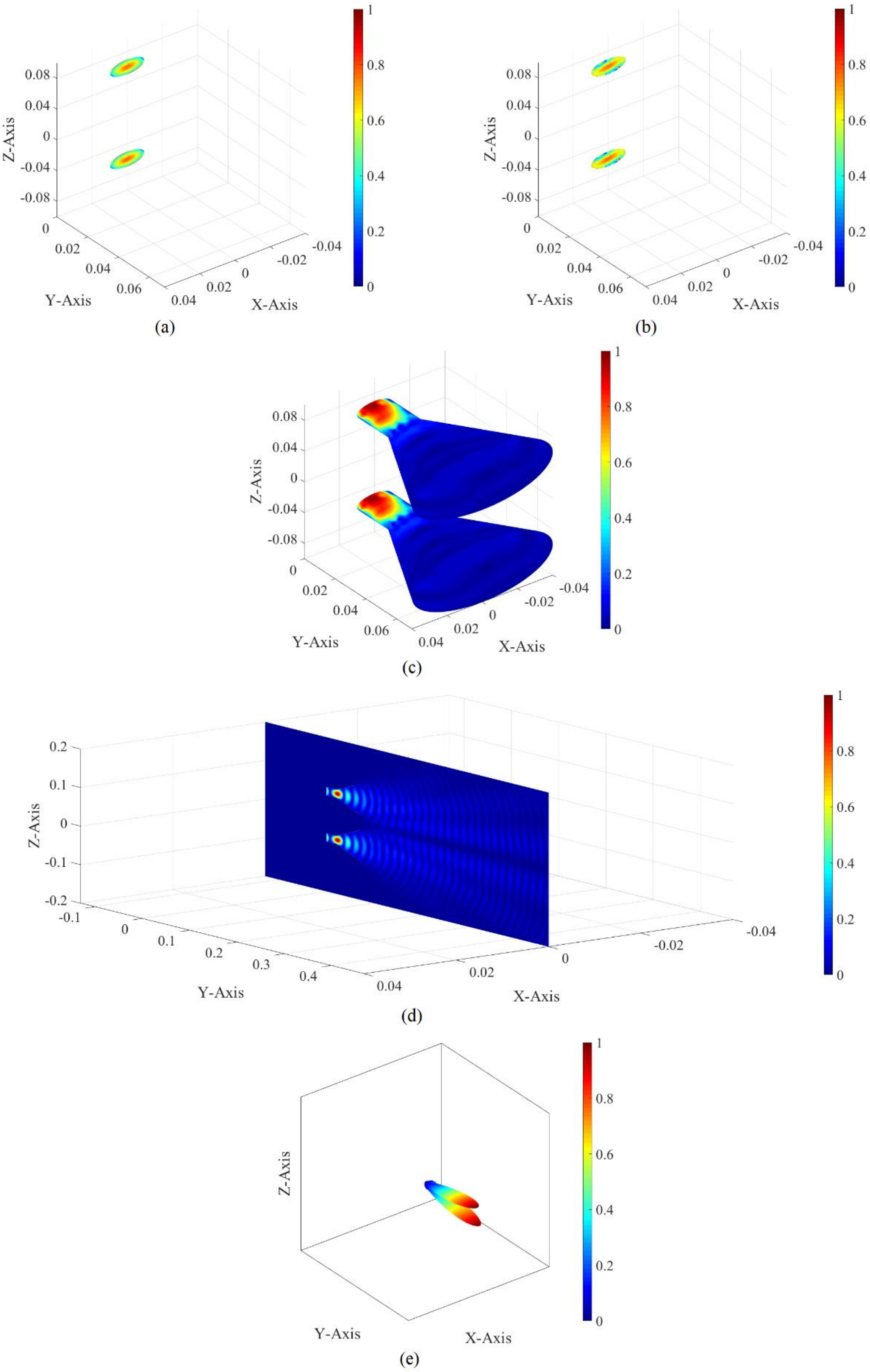

Figure 6-80 Modal (a) port electric current, (b) port magnetic current, (c) wall electric current, (d) electric field, and (e) radiation pattern of the IP-DM 3 at 9.3 GHz

For the IP-DM 4 at 9.3 GHz, its some modal quantities are plotted in Fig. 6-81.

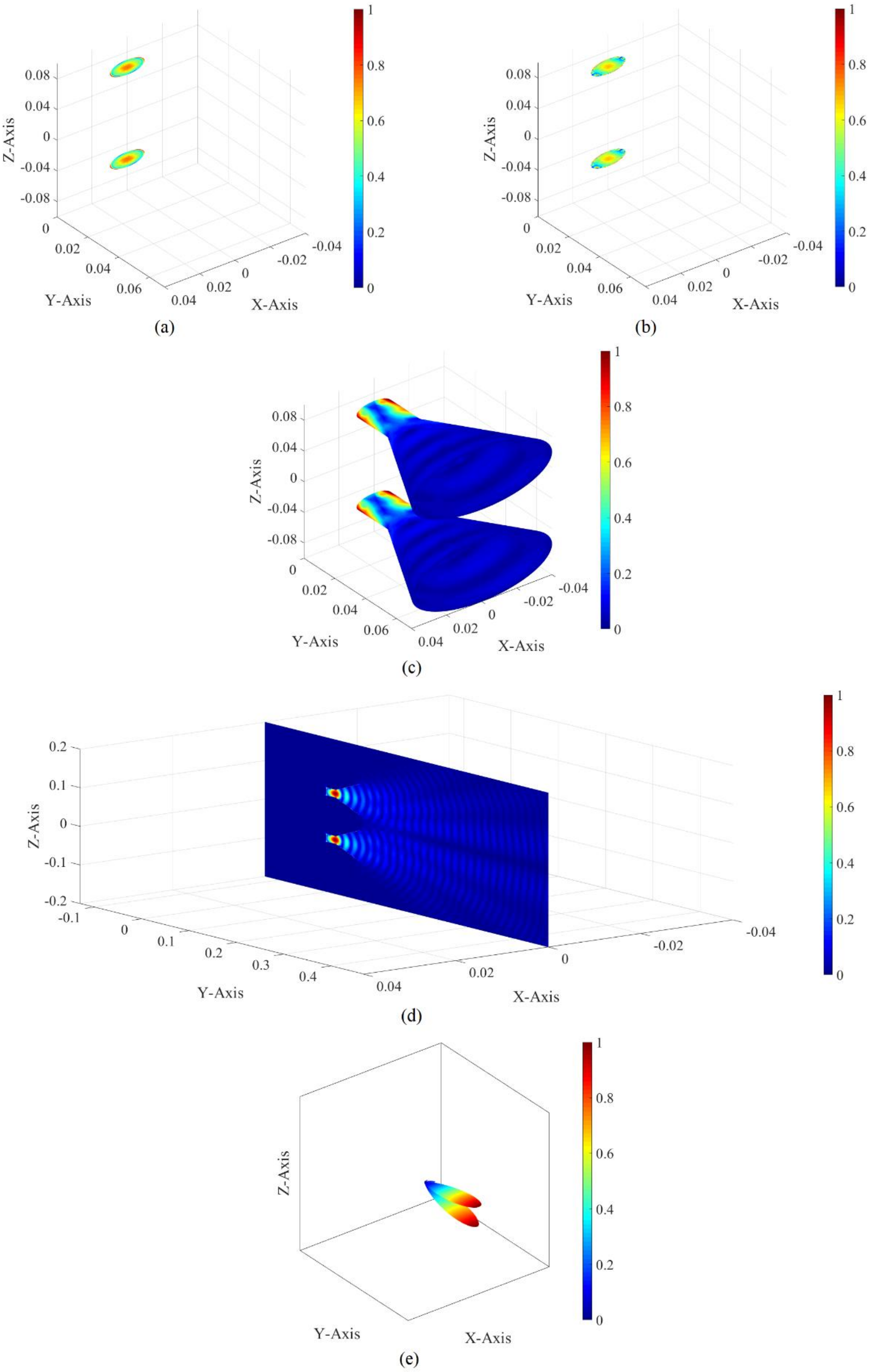

Figure 6-81 Modal (a) port electric current, (b) port magnetic current, (c) wall electric current, (d) electric field, and (e) radiation pattern of the IP-DM 4 at 9.3 GHz

## 6.8 Chapter Summary

In this chapter, the *power transport theorem* (*PTT*) based *decoupling mode theory* (*DMT*) for *augmented transmitting antennas* (*augmented tra-antennas*) — *PTT-TraAnt-DMT* — is established.

By orthogonalizing frequency-domain *input power operator* (*IPO*), Sec. 6.2 constructs the *input-power-decoupled modes* (*IP-DMs*) of metallic augmented tra-antennas, and the IP-DMs of material and composite augmented tra-antennas are similarly constructed in Sec. 6.3 and Secs. 6.4~6.6 respectively. In addition, the IP-DMs of augmented tra-antenna arrays are also similarly constructed in Sec. 6.7.

The IP-DMs of augmented tra-antennas satisfy a similar *energy-decoupling relation* to the IP-DMs of the *wave-guiding structures* discussed in Chap. 3. By employing the energy-decoupling relation, it is found out that the IP-DMs don't have net energy exchange in any integral period and the IP-DMs satisfy the famous *Parseval's identity*.

# Chapter 7 Input-Power-Decoupled Modes of Augmented Receiving Antenna (Including Grounding Structure)

**CHAPTER MOTIVATION:** The main aim of this chapter is to construct the *input-power-decoupled modes* (*IP-DMs*) of *augmented receiving antenna* (*augmented rec-antenna*) by orthogonalizing the frequecny-domain *input power operator* (*IPO*) in *power transport theorem* (*PTT*) framework. The obtained IP-DMs, which are inputted into the augmented rec-antenna, don't have net energy exchange in any integral period.

## 7.1 Chapter Introduction

*Receiving system* is a device for collecting radio-frequency power/energy from its surrounding environment and converting the collected power/energy into perceptible signals[37]. *Receiving antenna* (simply denoted as *rec-antenna*) is an important part in whole receiving system, and it is used to modulate the power to be inputted into the receiving system. Thus the modal analysis for rec-antenna is valuable for designing not only rec-antenna itself but also whole receiving system.

In receiving system, the rec-antenna is usually mounted on a *metallic grounding structure* (simply denoted as *rec-ground*), and the rec-ground can mechanically support the rec-antenna. In fact, besides the rec-antenna, the rec-ground also has ability to modulate the *input power* of receiving system.

As exhibited in the previous Chaps. 4 and 5, a relatively complicated *modal matching* process will be inevitable, if the modal analysis for rec-antenna and rec-ground are done separately. Following the idea of the previous Chap. 6, this chapter treats the rec-antenna and rec-ground as a whole — *augmented rec-antenna*, and a rigorous mathematical description for the augmented rec-antenna can be found in Sec. 2.3.5.

Under *power transport theorem* (*PTT*) framework, this chapter establishes the *decoupling mode theory* (*DMT*) for the augmented rec-antenna. The *PTT-based DMT for augmented rec-antenna* (*PTT-RecAnt-DMT*) has ability to construct a set of *input-power-decoupled modes* (*IP-DMs*), and the IP-DMs don't have net energy exchange in any integral period.

This chapter is organized as follows: Sec. 7.2 focuses on the metallic augmented rec-antennas driven by a definite transmitting system; Sec. 7.3 focuses on the metallic augmented rec-antennas driven by an arbitrary transmitting system; Sec. 7.4 focuses on the the metal-material composite augmented rec-antennas driven by an arbitrary transmitting system.

All the rec-antennas discussed in this chapter are augmented, so the determiner "augmented" is usually omitted in the following parts of this chapter to simplify the symbolic system.

## 7.2 IP-DMs of Metallic Receiving Antenna (Driven by a Definite Transmitting System)

As a typical example, the receiving problem shown in Fig. 7-1 is focused on in this section, and the rec-antenna is a *waveguide-loaded metallic horn antenna*.

In this section, we consider the case that the rec-antenna is driven by a *metallic horn transmitting antenna* (which is fed by a *metallic waveguide*) as shown in the following Fig. 7-1. In the subsequent Sec. 7.3, we will generalize the results obtained in this section to a more general case that the rec-antenna is driven by an arbitrary transmitter.

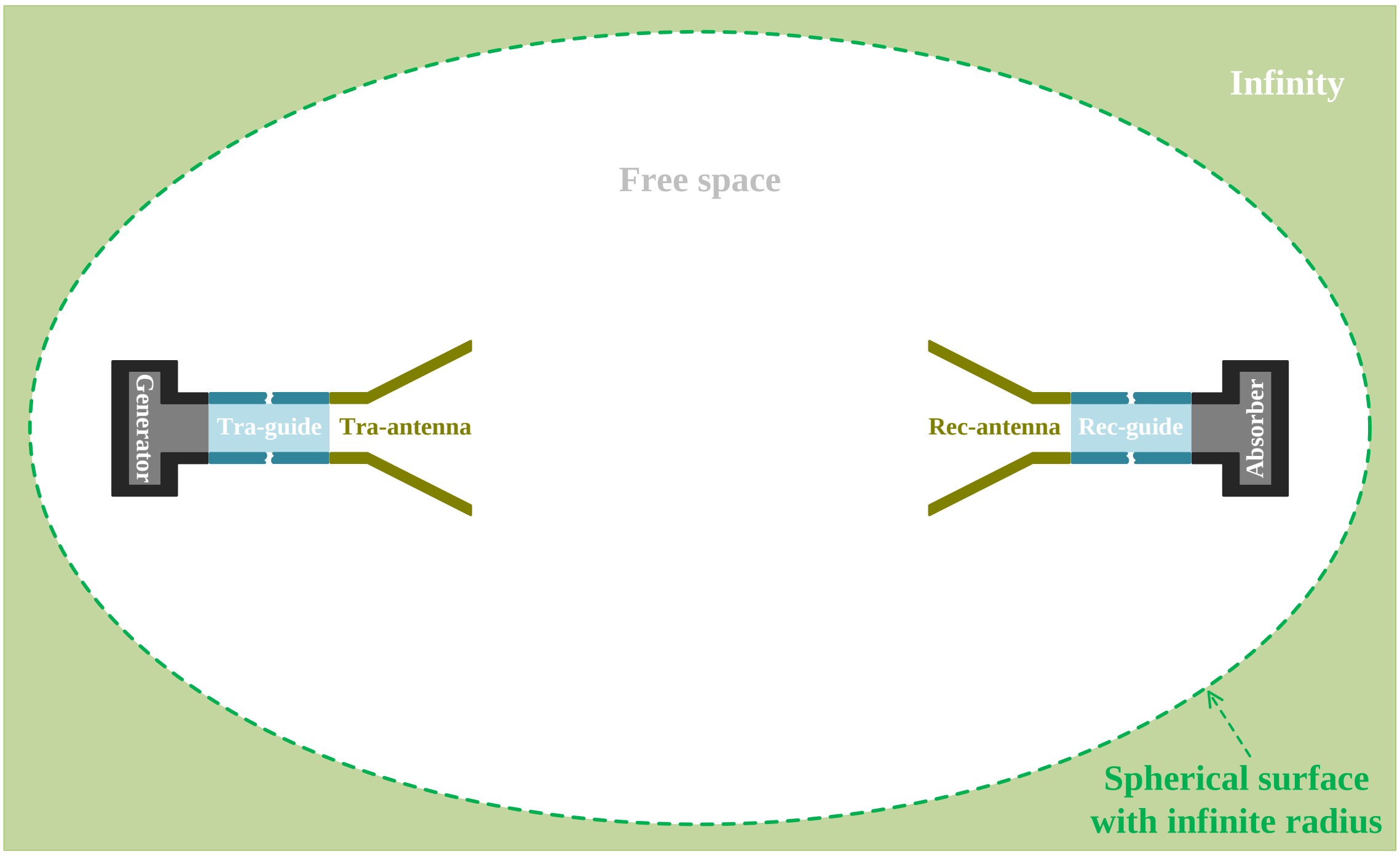

Figure 7-1 Geometries of the augmented rec-antenna and its driver considered in this section

### 7.2.1 Topological Structure and Source-Field Relationships

The topological structures of the rec-antenna and *transmitting antenna* (simply called *tra-antenna*) shown in above Fig. 7-1 are specifically exhibited in the following Fig. 7-2.

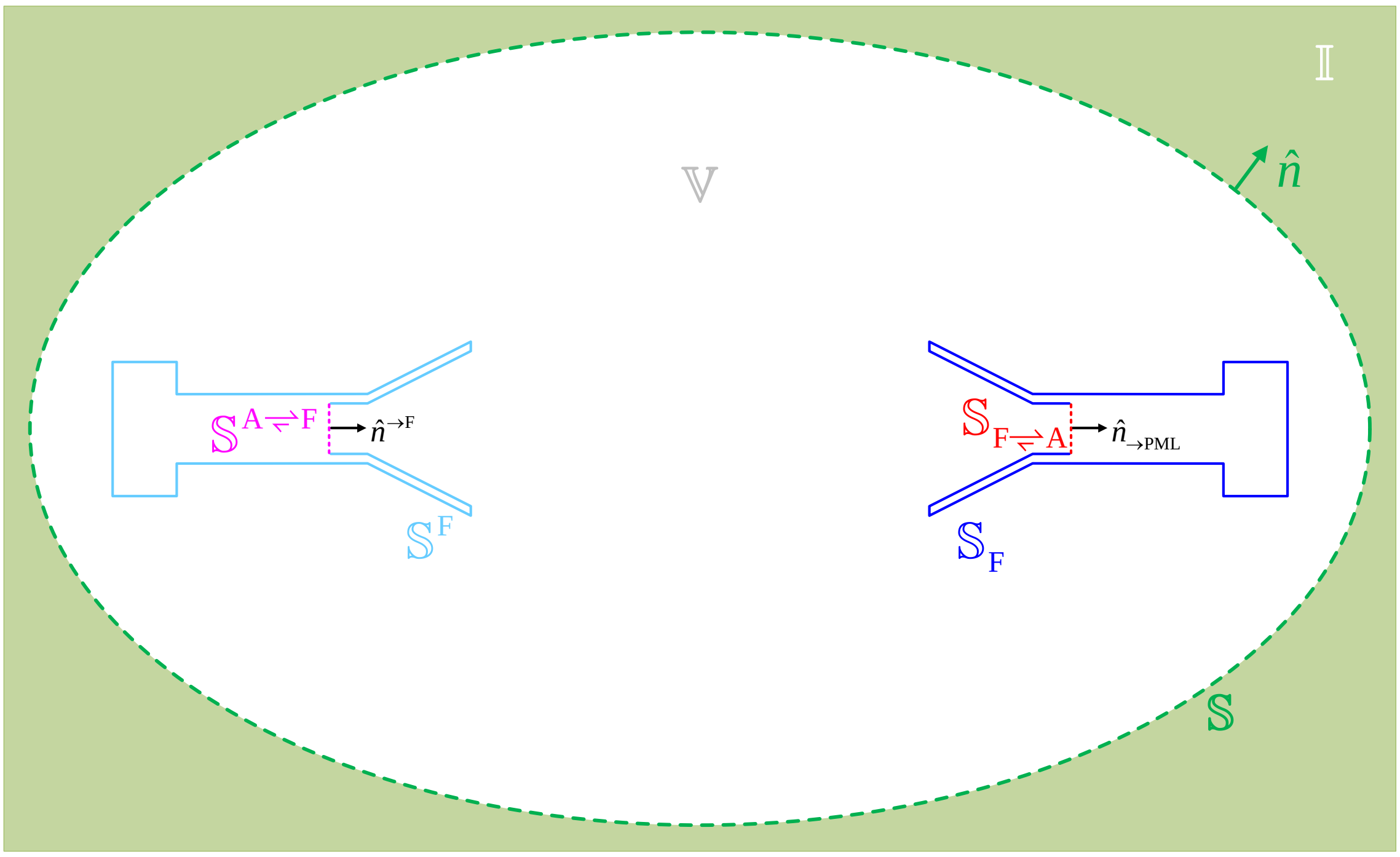

Figure 7-2 Topological structures of the rec-antenna and tra-antenna shown in Fig. 7-1

In the above Fig. 7-2, $\mathbb{S}^{A \rightleftharpoons F}$ is the *output port* of the transmitting system, and $\mathbb{S}^{F}$ is the impenetrable interface between *transmitting system* and *free space*. The free space is denoted as $\mathbb{V}$, and its outer boundary is denoted as $\mathbb{S}$, which is a spherical surface with infinite radius. Surface $\mathbb{S}_{F}$ is the impenetrable interface between *receiving system* and free space, and the *input port* of the receiving system is denoted as $\mathbb{S}_{F \rightleftharpoons A}$. Clearly, surfaces $\mathbb{S}^{A \rightleftharpoons F}$, $\mathbb{S}^{F}$, $\mathbb{S}$, $\mathbb{S}_{F}$, and $\mathbb{S}_{F \rightleftharpoons A}$ constitute the whole boundary of $\mathbb{V}$, i.e., $\partial \mathbb{V} = \mathbb{S}^{A \rightleftharpoons F} \cup \mathbb{S}^{F} \cup \mathbb{S} \cup \mathbb{S}_{F} \cup \mathbb{S}_{F \rightleftharpoons A}$.

The $\hat{n}^{\to F}$ shown in Fig. 7-2 is the *normal direction* of $\mathbb{S}^{A \rightleftharpoons F}$, and it points to the interior of $\mathbb{V}$; the $\hat{n}_{\to PML}$ shown in Fig. 7-2 is the normal direction of $\mathbb{S}_{F \rightleftharpoons A}$, and it points to the interior of loading structure, where the loading structure is treated as *perfectly matched load* (*PML*). In addition, the *magnetic permeability*, *dielectric permittivity*, and *electric conductivity* of free space are $\mu_0$, $\varepsilon_0$, and 0 respectively.

If the *equivalent surface currents* distributing on $\mathbb{S}^{A \rightleftharpoons F}$ are denoted as $\{\vec{J}^{A \rightleftharpoons F}, \vec{M}^{A \rightleftharpoons F}\}$, and the *equivalent surface electric currents* distributing on $\mathbb{S}^{F}$ and $\mathbb{S}_{F}$ are denoted as $\vec{J}^{F}$ and $\vec{J}_{F}$ respectively ①, and the equivalent surface currents distributing on $\mathbb{S}_{F \rightleftharpoons A}$ are denoted as $\{\vec{J}_{F \rightleftharpoons A}, \vec{M}_{F \rightleftharpoons A}\}$, then the field distributing on $\mathbb{V}$ can be expressed as follows:

① The equivalent surface magnetic currents distributing on $\mathbb{S}^{F}$ and $\mathbb{S}_{F}$ are zero, because of the homogeneous tangential electric field boundary conditions on $\mathbb{S}^{F}$ and $\mathbb{S}_{F}$ [13].

$$\vec{F}(\vec{r}) = \mathcal{F}_0\left(\vec{J}^{\mathrm{A}\rightleftharpoons\mathrm{F}} + \vec{J}^{\mathrm{F}} + \vec{J}_{\mathrm{F}} - \vec{J}_{\mathrm{F}\rightleftharpoons\mathrm{A}}, \vec{M}^{\mathrm{A}\rightleftharpoons\mathrm{F}} - \vec{M}_{\mathrm{F}\rightleftharpoons\mathrm{A}}\right) \quad , \quad \vec{r} \in \mathbb{V} \tag{7-1}$$

where $\vec{F} = \vec{E} / \vec{H}$, and correspondingly $\mathcal{F}_0 = \mathcal{E}_0 / \mathcal{H}_0$, and the operators are the same as the ones used in the previous chapters.

The currents $\{\vec{J}^{\mathrm{A}\rightleftharpoons\mathrm{F}}, \vec{M}^{\mathrm{A}\rightleftharpoons\mathrm{F}}\}$ and fields $\{\vec{E}, \vec{H}\}$ in Eq. (7-1) satisfy the following relations

$$\hat{n}^{\to\mathrm{F}} \times \left[\vec{H}(\vec{r}^{\mathrm{F}})\right]_{\vec{r}^{\mathrm{F}}\to\vec{r}} = \vec{J}^{\mathrm{A}\rightleftharpoons\mathrm{F}}(\vec{r}) \quad , \quad \vec{r} \in \mathbb{S}^{\mathrm{A}\rightleftharpoons\mathrm{F}} \tag{7-2a}$$

$$\left[\vec{E}(\vec{r}^{\mathrm{F}})\right]_{\vec{r}^{\mathrm{F}}\to\vec{r}} \times \hat{n}^{\to\mathrm{F}} = \vec{M}^{\mathrm{A}\rightleftharpoons\mathrm{F}}(\vec{r}) \quad , \quad \vec{r} \in \mathbb{S}^{\mathrm{A}\rightleftharpoons\mathrm{F}} \tag{7-2b}$$

where point $\vec{r}^{\mathrm{F}}$ belongs to $\mathbb{V}$ and approaches the point $\vec{r}$ on $\mathbb{S}^{\mathrm{A}\rightleftharpoons\mathrm{F}}$. The currents $\{\vec{J}_{\mathrm{F}\rightleftharpoons\mathrm{A}}, \vec{M}_{\mathrm{F}\rightleftharpoons\mathrm{A}}\}$ and fields $\{\vec{E}, \vec{H}\}$ in Eq. (7-1) satisfy the following relations

$$\hat{n}_{\to\mathrm{PML}} \times \left[\vec{H}(\vec{r}_{\mathrm{F}})\right]_{\vec{r}_{\mathrm{F}}\to\vec{r}} = \vec{J}_{\mathrm{F}\rightleftharpoons\mathrm{A}}(\vec{r}) \quad , \quad \vec{r} \in \mathbb{S}_{\mathrm{F}\rightleftharpoons\mathrm{A}} \tag{7-3a}$$

$$\left[\vec{E}(\vec{r}_{\mathrm{F}})\right]_{\vec{r}_{\mathrm{F}}\to\vec{r}} \times \hat{n}_{\to\mathrm{PML}} = \vec{M}_{\mathrm{F}\rightleftharpoons\mathrm{A}}(\vec{r}) \quad , \quad \vec{r} \in \mathbb{S}_{\mathrm{F}\rightleftharpoons\mathrm{A}} \tag{7-3b}$$

where point $\vec{r}_{\mathrm{F}}$ belongs to $\mathbb{V}$ and approaches the point $\vec{r}$ on $\mathbb{S}_{\mathrm{F}\rightleftharpoons\mathrm{A}}$.

### 7.2.2 Mathematical Description for Modal Space

Substituting Eq. (7-1) into Eqs. (7-2a) and (7-2b), we immediately obtain the following *integral equations*

$$\left[\mathcal{H}_0\left(\vec{J}^{\mathrm{A}\rightleftharpoons\mathrm{F}} + \vec{J}^{\mathrm{F}} + \vec{J}_{\mathrm{F}} - \vec{J}_{\mathrm{F}\rightleftharpoons\mathrm{A}}, \vec{M}^{\mathrm{A}\rightleftharpoons\mathrm{F}} - \vec{M}_{\mathrm{F}\rightleftharpoons\mathrm{A}}\right)\right]^{\tan}_{\vec{r}^{\mathrm{F}}\to\vec{r}} = \vec{J}^{\mathrm{A}\rightleftharpoons\mathrm{F}}(\vec{r}) \times \hat{n}^{\to\mathrm{F}} \quad , \quad \vec{r} \in \mathbb{S}^{\mathrm{A}\rightleftharpoons\mathrm{F}} \tag{7-4a}$$

$$\left[\mathcal{E}_0\left(\vec{J}^{\mathrm{A}\rightleftharpoons\mathrm{F}} + \vec{J}^{\mathrm{F}} + \vec{J}_{\mathrm{F}} - \vec{J}_{\mathrm{F}\rightleftharpoons\mathrm{A}}, \vec{M}^{\mathrm{A}\rightleftharpoons\mathrm{F}} - \vec{M}_{\mathrm{F}\rightleftharpoons\mathrm{A}}\right)\right]^{\tan}_{\vec{r}^{\mathrm{F}}\to\vec{r}} = \hat{n}^{\to\mathrm{F}} \times \vec{M}^{\mathrm{A}\rightleftharpoons\mathrm{F}}(\vec{r}) \ , \quad \vec{r} \in \mathbb{S}^{\mathrm{A}\rightleftharpoons\mathrm{F}} \tag{7-4b}$$

about currents $\{\vec{J}^{\mathrm{A}\rightleftharpoons\mathrm{F}}, \vec{M}^{\mathrm{A}\rightleftharpoons\mathrm{F}}\}$, $\vec{J}^{\mathrm{F}}$, $\vec{J}_{\mathrm{F}}$, and $\{\vec{J}_{\mathrm{F}\rightleftharpoons\mathrm{A}}, \vec{M}_{\mathrm{F}\rightleftharpoons\mathrm{A}}\}$, where the superscript "tan" represents the *tangential component* of the field.

Based on Eq. (7-1) and the *homogeneous tangential electric field boundary condition* on electric walls $\mathbb{S}^{\mathrm{F}}$ and $\mathbb{S}_{\mathrm{F}}$, we have the following *electric field integral equations*

$$\left[\mathcal{E}_0\left(\vec{J}^{\mathrm{A}\rightleftharpoons\mathrm{F}} + \vec{J}^{\mathrm{F}} + \vec{J}_{\mathrm{F}} - \vec{J}_{\mathrm{F}\rightleftharpoons\mathrm{A}}, \vec{M}^{\mathrm{A}\rightleftharpoons\mathrm{F}} - \vec{M}_{\mathrm{F}\rightleftharpoons\mathrm{A}}\right)\right]^{\tan} = 0 \quad , \quad \vec{r} \in \mathbb{S}^{\mathrm{F}} \tag{7-5}$$

$$\left[\mathcal{E}_0\left(\vec{J}^{\mathrm{A}\rightleftharpoons\mathrm{F}} + \vec{J}^{\mathrm{F}} + \vec{J}_{\mathrm{F}} - \vec{J}_{\mathrm{F}\rightleftharpoons\mathrm{A}}, \vec{M}^{\mathrm{A}\rightleftharpoons\mathrm{F}} - \vec{M}_{\mathrm{F}\rightleftharpoons\mathrm{A}}\right)\right]^{\tan} = 0 \quad , \quad \vec{r} \in \mathbb{S}_{\mathrm{F}} \tag{7-6}$$

about currents $\{\vec{J}^{\mathrm{A}\rightleftharpoons\mathrm{F}}, \vec{M}^{\mathrm{A}\rightleftharpoons\mathrm{F}}\}$, $\vec{J}^{\mathrm{F}}$, $\vec{J}_{\mathrm{F}}$, and $\{\vec{J}_{\mathrm{F}\rightleftharpoons\mathrm{A}}, \vec{M}_{\mathrm{F}\rightleftharpoons\mathrm{A}}\}$.

Based on Eq. (7-1) and a similar consideration as the previous Chap. 3, we immediately obtain the following integral equations

$$\begin{aligned}&\left[\mathcal{E}_0\left(\vec{J}^{\mathrm{A}\rightleftharpoons\mathrm{F}}+\vec{J}^{\mathrm{F}}+\vec{J}_{\mathrm{F}}-\vec{J}_{\mathrm{F}\rightleftharpoons\mathrm{A}},\vec{M}^{\mathrm{A}\rightleftharpoons\mathrm{F}}-\vec{M}_{\mathrm{F}\rightleftharpoons\mathrm{A}}\right)\right]^{\tan}_{\vec{r}_{\mathrm{F}}\to\vec{r}}\\ &=\left[\mathcal{E}_0\left(\vec{J}_{\mathrm{F}\rightleftharpoons\mathrm{A}},\vec{M}_{\mathrm{F}\rightleftharpoons\mathrm{A}}\right)\right]^{\tan}_{\vec{r}_{\mathrm{PML}}\to\vec{r}}\quad,\quad\vec{r}\in\mathbb{S}_{\mathrm{F}\rightleftharpoons\mathrm{A}}\end{aligned}\tag{7-7a}$$

$$\begin{aligned}&\left[\mathcal{H}_0\left(\vec{J}^{\mathrm{A}\rightleftharpoons\mathrm{F}}+\vec{J}^{\mathrm{F}}+\vec{J}_{\mathrm{F}}-\vec{J}_{\mathrm{F}\rightleftharpoons\mathrm{A}},\vec{M}^{\mathrm{A}\rightleftharpoons\mathrm{F}}-\vec{M}_{\mathrm{F}\rightleftharpoons\mathrm{A}}\right)\right]^{\tan}_{\vec{r}_{\mathrm{F}}\to\vec{r}}\\ &=\left[\mathcal{H}_0\left(\vec{J}_{\mathrm{F}\rightleftharpoons\mathrm{A}},\vec{M}_{\mathrm{F}\rightleftharpoons\mathrm{A}}\right)\right]^{\tan}_{\vec{r}_{\mathrm{PML}}\to\vec{r}}\quad,\quad\vec{r}\in\mathbb{S}_{\mathrm{F}\rightleftharpoons\mathrm{A}}\end{aligned}\tag{7-7b}$$

about currents $\{\vec{J}^{\mathrm{A}\rightleftharpoons\mathrm{F}},\vec{M}^{\mathrm{A}\rightleftharpoons\mathrm{F}}\}$, $\vec{J}^{\mathrm{F}}$, $\vec{J}_{\mathrm{F}}$, and $\{\vec{J}_{\mathrm{F}\rightleftharpoons\mathrm{A}},\vec{M}_{\mathrm{F}\rightleftharpoons\mathrm{A}}\}$, where point $\vec{r}_{\mathrm{PML}}$ belongs to the region of PML and approaches the point $\vec{r}$ on $\mathbb{S}_{\mathrm{F}\rightleftharpoons\mathrm{A}}$.

If the currents $\{\vec{J}^{\mathrm{A}\rightleftharpoons\mathrm{F}},\vec{M}^{\mathrm{A}\rightleftharpoons\mathrm{F}}\}$, $\vec{J}^{\mathrm{F}}$, $\vec{J}_{\mathrm{F}}$, and $\{\vec{J}_{\mathrm{F}\rightleftharpoons\mathrm{A}},\vec{M}_{\mathrm{F}\rightleftharpoons\mathrm{A}}\}$ are expanded in terms of some proper basis functions as follows:

$$\vec{X}\left(\vec{r}\right)=\sum\nolimits_{\xi}a_{\xi}^{\vec{X}}\vec{b}_{\xi}^{\vec{X}}=\underbrace{\begin{bmatrix}\vec{b}_1^{\vec{X}} & \vec{b}_2^{\vec{X}} & \cdots\end{bmatrix}}_{\bar{\boldsymbol{B}}^{\vec{X}}}\cdot\underbrace{\begin{bmatrix}a_1^{\vec{X}}\\ a_2^{\vec{X}}\\ \vdots\end{bmatrix}}_{\bar{a}^{\vec{X}}}\quad,\quad\vec{r}\in\mathbb{X}\tag{7-8}$$

and Eqs. (7-4a), (7-4b), (7-5), (7-6), (7-7a), and (7-7b) are tested with $\{\vec{b}_{\xi}^{\vec{M}^{\mathrm{A}\rightleftharpoons\mathrm{F}}}\}$, $\{\vec{b}_{\xi}^{\vec{J}^{\mathrm{A}\rightleftharpoons\mathrm{F}}}\}$, $\{\vec{b}_{\xi}^{\vec{J}^{\mathrm{F}}}\}$, $\{\vec{b}_{\xi}^{\vec{J}_{\mathrm{F}}}\}$, $\{\vec{b}_{\xi}^{\vec{J}_{\mathrm{F}\rightleftharpoons\mathrm{A}}}\}$, and $\{\vec{b}_{\xi}^{\vec{M}_{\mathrm{F}\rightleftharpoons\mathrm{A}}}\}$ respectively, then the integral equations are immediately discretized into the following *matrix equations*

$$\begin{aligned}&\overline{\overline{Z}}^{\vec{M}^{\mathrm{A}\rightleftharpoons\mathrm{F}}\vec{J}^{\mathrm{A}\rightleftharpoons\mathrm{F}}}\cdot\bar{a}^{\vec{J}^{\mathrm{A}\rightleftharpoons\mathrm{F}}}+\overline{\overline{Z}}^{\vec{M}^{\mathrm{A}\rightleftharpoons\mathrm{F}}\vec{J}^{\mathrm{F}}}\cdot\bar{a}^{\vec{J}^{\mathrm{F}}}+\overline{\overline{Z}}^{\vec{M}^{\mathrm{A}\rightleftharpoons\mathrm{F}}\vec{J}_{\mathrm{F}}}\cdot\bar{a}^{\vec{J}_{\mathrm{F}}}+\overline{\overline{Z}}^{\vec{M}^{\mathrm{A}\rightleftharpoons\mathrm{F}}\vec{J}_{\mathrm{F}\rightleftharpoons\mathrm{A}}}\cdot\bar{a}^{\vec{J}_{\mathrm{F}\rightleftharpoons\mathrm{A}}}+\overline{\overline{Z}}^{\vec{M}^{\mathrm{A}\rightleftharpoons\mathrm{F}}\vec{M}^{\mathrm{A}\rightleftharpoons\mathrm{F}}}\cdot\bar{a}^{\vec{M}^{\mathrm{A}\rightleftharpoons\mathrm{F}}}\\ &+\overline{\overline{Z}}^{\vec{M}^{\mathrm{A}\rightleftharpoons\mathrm{F}}\vec{M}_{\mathrm{F}\rightleftharpoons\mathrm{A}}}\cdot\bar{a}^{\vec{M}_{\mathrm{F}\rightleftharpoons\mathrm{A}}}=0\end{aligned}\tag{7-9a}$$

$$\begin{aligned}&\overline{\overline{Z}}^{\vec{J}^{\mathrm{A}\rightleftharpoons\mathrm{F}}\vec{J}^{\mathrm{A}\rightleftharpoons\mathrm{F}}}\cdot\bar{a}^{\vec{J}^{\mathrm{A}\rightleftharpoons\mathrm{F}}}+\overline{\overline{Z}}^{\vec{J}^{\mathrm{A}\rightleftharpoons\mathrm{F}}\vec{J}^{\mathrm{F}}}\cdot\bar{a}^{\vec{J}^{\mathrm{F}}}+\overline{\overline{Z}}^{\vec{J}^{\mathrm{A}\rightleftharpoons\mathrm{F}}\vec{J}_{\mathrm{F}}}\cdot\bar{a}^{\vec{J}_{\mathrm{F}}}+\overline{\overline{Z}}^{\vec{J}^{\mathrm{A}\rightleftharpoons\mathrm{F}}\vec{J}_{\mathrm{F}\rightleftharpoons\mathrm{A}}}\cdot\bar{a}^{\vec{J}_{\mathrm{F}\rightleftharpoons\mathrm{A}}}+\overline{\overline{Z}}^{\vec{J}^{\mathrm{A}\rightleftharpoons\mathrm{F}}\vec{M}^{\mathrm{A}\rightleftharpoons\mathrm{F}}}\cdot\bar{a}^{\vec{M}^{\mathrm{A}\rightleftharpoons\mathrm{F}}}\\ &+\overline{\overline{Z}}^{\vec{J}^{\mathrm{A}\rightleftharpoons\mathrm{F}}\vec{M}_{\mathrm{F}\rightleftharpoons\mathrm{A}}}\cdot\bar{a}^{\vec{M}_{\mathrm{F}\rightleftharpoons\mathrm{A}}}=0\end{aligned}\tag{7-9b}$$

and

$$\begin{aligned}&\overline{\overline{Z}}^{\vec{J}^{\mathrm{F}}\vec{J}^{\mathrm{A}\rightleftharpoons\mathrm{F}}}\cdot\bar{a}^{\vec{J}^{\mathrm{A}\rightleftharpoons\mathrm{F}}}+\overline{\overline{Z}}^{\vec{J}^{\mathrm{F}}\vec{J}^{\mathrm{F}}}\cdot\bar{a}^{\vec{J}^{\mathrm{F}}}+\overline{\overline{Z}}^{\vec{J}^{\mathrm{F}}\vec{J}_{\mathrm{F}}}\cdot\bar{a}^{\vec{J}_{\mathrm{F}}}+\overline{\overline{Z}}^{\vec{J}^{\mathrm{F}}\vec{J}_{\mathrm{F}\rightleftharpoons\mathrm{A}}}\cdot\bar{a}^{\vec{J}_{\mathrm{F}\rightleftharpoons\mathrm{A}}}+\overline{\overline{Z}}^{\vec{J}^{\mathrm{F}}\vec{M}^{\mathrm{A}\rightleftharpoons\mathrm{F}}}\cdot\bar{a}^{\vec{M}^{\mathrm{A}\rightleftharpoons\mathrm{F}}}\\ &+\overline{\overline{Z}}^{\vec{J}^{\mathrm{F}}\vec{M}_{\mathrm{F}\rightleftharpoons\mathrm{A}}}\cdot\bar{a}^{\vec{M}_{\mathrm{F}\rightleftharpoons\mathrm{A}}}=0\end{aligned}\tag{7-10}$$

$$\begin{aligned}&\overline{\overline{Z}}^{\vec{J}_{\mathrm{F}}\vec{J}^{\mathrm{A}\rightleftharpoons\mathrm{F}}}\cdot\bar{a}^{\vec{J}^{\mathrm{A}\rightleftharpoons\mathrm{F}}}+\overline{\overline{Z}}^{\vec{J}_{\mathrm{F}}\vec{J}^{\mathrm{F}}}\cdot\bar{a}^{\vec{J}^{\mathrm{F}}}+\overline{\overline{Z}}^{\vec{J}_{\mathrm{F}}\vec{J}_{\mathrm{F}}}\cdot\bar{a}^{\vec{J}_{\mathrm{F}}}+\overline{\overline{Z}}^{\vec{J}_{\mathrm{F}}\vec{J}_{\mathrm{F}\rightleftharpoons\mathrm{A}}}\cdot\bar{a}^{\vec{J}_{\mathrm{F}\rightleftharpoons\mathrm{A}}}+\overline{\overline{Z}}^{\vec{J}_{\mathrm{F}}\vec{M}^{\mathrm{A}\rightleftharpoons\mathrm{F}}}\cdot\bar{a}^{\vec{M}^{\mathrm{A}\rightleftharpoons\mathrm{F}}}\\ &+\overline{\overline{Z}}^{\vec{J}_{\mathrm{F}}\vec{M}_{\mathrm{F}\rightleftharpoons\mathrm{A}}}\cdot\bar{a}^{\vec{M}_{\mathrm{F}\rightleftharpoons\mathrm{A}}}=0\end{aligned}\tag{7-11}$$

and

$$\begin{aligned}&\overline{\overline{Z}}^{\vec{J}_{\mathrm{F}\rightleftharpoons\mathrm{A}}\vec{J}^{\mathrm{A}\rightleftharpoons\mathrm{F}}}\cdot\bar{a}^{\vec{J}^{\mathrm{A}\rightleftharpoons\mathrm{F}}}+\overline{\overline{Z}}^{\vec{J}_{\mathrm{F}\rightleftharpoons\mathrm{A}}\vec{J}^{\mathrm{F}}}\cdot\bar{a}^{\vec{J}^{\mathrm{F}}}+\overline{\overline{Z}}^{\vec{J}_{\mathrm{F}\rightleftharpoons\mathrm{A}}\vec{J}_{\mathrm{F}}}\cdot\bar{a}^{\vec{J}_{\mathrm{F}}}+\overline{\overline{Z}}^{\vec{J}_{\mathrm{F}\rightleftharpoons\mathrm{A}}\vec{J}_{\mathrm{F}\rightleftharpoons\mathrm{A}}}\cdot\bar{a}^{\vec{J}_{\mathrm{F}\rightleftharpoons\mathrm{A}}}+\overline{\overline{Z}}^{\vec{J}_{\mathrm{F}\rightleftharpoons\mathrm{A}}\vec{M}^{\mathrm{A}\rightleftharpoons\mathrm{F}}}\cdot\bar{a}^{\vec{M}^{\mathrm{A}\rightleftharpoons\mathrm{F}}}\\ &+\overline{\overline{Z}}^{\vec{J}_{\mathrm{F}\rightleftharpoons\mathrm{A}}\vec{M}_{\mathrm{F}\rightleftharpoons\mathrm{A}}}\cdot\bar{a}^{\vec{M}_{\mathrm{F}\rightleftharpoons\mathrm{A}}}=0\end{aligned}\tag{7-12a}$$

$$\begin{aligned}&\overline{\overline{Z}}^{\vec{M}_{\mathrm{F}\rightleftharpoons\mathrm{A}}\vec{J}^{\mathrm{A}\rightleftharpoons\mathrm{F}}}\cdot\bar{a}^{\vec{J}^{\mathrm{A}\rightleftharpoons\mathrm{F}}}+\overline{\overline{Z}}^{\vec{M}_{\mathrm{F}\rightleftharpoons\mathrm{A}}\vec{J}^{\mathrm{F}}}\cdot\bar{a}^{\vec{J}^{\mathrm{F}}}+\overline{\overline{Z}}^{\vec{M}_{\mathrm{F}\rightleftharpoons\mathrm{A}}\vec{J}_{\mathrm{F}}}\cdot\bar{a}^{\vec{J}_{\mathrm{F}}}+\overline{\overline{Z}}^{\vec{M}_{\mathrm{F}\rightleftharpoons\mathrm{A}}\vec{J}_{\mathrm{F}\rightleftharpoons\mathrm{A}}}\cdot\bar{a}^{\vec{J}_{\mathrm{F}\rightleftharpoons\mathrm{A}}}+\overline{\overline{Z}}^{\vec{M}_{\mathrm{F}\rightleftharpoons\mathrm{A}}\vec{M}^{\mathrm{A}\rightleftharpoons\mathrm{F}}}\cdot\bar{a}^{\vec{M}^{\mathrm{A}\rightleftharpoons\mathrm{F}}}\\ &+\overline{\overline{Z}}^{\vec{M}_{\mathrm{F}\rightleftharpoons\mathrm{A}}\vec{M}_{\mathrm{F}\rightleftharpoons\mathrm{A}}}\cdot\bar{a}^{\vec{M}_{\mathrm{F}\rightleftharpoons\mathrm{A}}}=0\end{aligned}\tag{7-12b}$$

The formulations used to calculate the elements of the matrices in Eq. (7-9a) are as

$$z_{\xi\zeta}^{\vec{M}^{\mathrm{A}\rightleftharpoons\mathrm{F}}\vec{J}^{\mathrm{A}\rightleftharpoons\mathrm{F}}} = \left\langle \vec{b}_{\xi}^{\vec{M}^{\mathrm{A}\rightleftharpoons\mathrm{F}}}, \hat{n}^{\rightarrow\mathrm{F}} \times \frac{1}{2}\vec{b}_{\zeta}^{\vec{J}^{\mathrm{A}\rightleftharpoons\mathrm{F}}} + \mathrm{P.V.}\mathcal{K}_0\left(\vec{b}_{\zeta}^{\vec{J}^{\mathrm{A}\rightleftharpoons\mathrm{F}}}\right)\right\rangle_{\mathbb{S}^{\mathrm{A}\rightleftharpoons\mathrm{F}}} \tag{7-13a}$$

$$z_{\xi\zeta}^{\vec{M}^{\mathrm{A}\rightleftharpoons\mathrm{F}}\vec{J}^{\mathrm{F}}} = \left\langle \vec{b}_{\xi}^{\vec{M}^{\mathrm{A}\rightleftharpoons\mathrm{F}}}, \mathcal{K}_0\left(\vec{b}_{\zeta}^{\vec{J}^{\mathrm{F}}}\right)\right\rangle_{\mathbb{S}^{\mathrm{A}\rightleftharpoons\mathrm{F}}} \tag{7-13b}$$

$$z_{\xi\zeta}^{\vec{M}^{\mathrm{A}\rightleftharpoons\mathrm{F}}\vec{J}_{\mathrm{F}}} = \left\langle \vec{b}_{\xi}^{\vec{M}^{\mathrm{A}\rightleftharpoons\mathrm{F}}}, \mathcal{K}_0\left(\vec{b}_{\zeta}^{\vec{J}_{\mathrm{F}}}\right)\right\rangle_{\mathbb{S}^{\mathrm{A}\rightleftharpoons\mathrm{F}}} \tag{7-13c}$$

$$z_{\xi\zeta}^{\vec{M}^{\mathrm{A}\rightleftharpoons\mathrm{F}}\vec{J}_{\mathrm{F}\rightleftharpoons\mathrm{A}}} = \left\langle \vec{b}_{\xi}^{\vec{M}^{\mathrm{A}\rightleftharpoons\mathrm{F}}}, \mathcal{K}_0\left(-\vec{b}_{\zeta}^{\vec{J}_{\mathrm{F}\rightleftharpoons\mathrm{A}}}\right)\right\rangle_{\mathbb{S}^{\mathrm{A}\rightleftharpoons\mathrm{F}}} \tag{7-13d}$$

$$z_{\xi\zeta}^{\vec{M}^{\mathrm{A}\rightleftharpoons\mathrm{F}}\vec{M}^{\mathrm{A}\rightleftharpoons\mathrm{F}}} = \left\langle \vec{b}_{\xi}^{\vec{M}^{\mathrm{A}\rightleftharpoons\mathrm{F}}}, -j\omega\varepsilon_0\mathcal{L}_0\left(\vec{b}_{\zeta}^{\vec{M}^{\mathrm{A}\rightleftharpoons\mathrm{F}}}\right)\right\rangle_{\mathbb{S}^{\mathrm{A}\rightleftharpoons\mathrm{F}}} \tag{7-13e}$$

$$z_{\xi\zeta}^{\vec{M}^{\mathrm{A}\rightleftharpoons\mathrm{F}}\vec{M}_{\mathrm{F}\rightleftharpoons\mathrm{A}}} = \left\langle \vec{b}_{\xi}^{\vec{M}^{\mathrm{A}\rightleftharpoons\mathrm{F}}}, -j\omega\varepsilon_0\mathcal{L}_0\left(-\vec{b}_{\zeta}^{\vec{M}_{\mathrm{F}\rightleftharpoons\mathrm{A}}}\right)\right\rangle_{\mathbb{S}^{\mathrm{A}\rightleftharpoons\mathrm{F}}} \tag{7-13f}$$

The formulations used to calculate the elements of the matrices in Eq. (7-9b) are as

$$z_{\xi\zeta}^{\vec{J}^{\mathrm{A}\rightleftharpoons\mathrm{F}}\vec{J}^{\mathrm{A}\rightleftharpoons\mathrm{F}}} = \left\langle \vec{b}_{\xi}^{\vec{J}^{\mathrm{A}\rightleftharpoons\mathrm{F}}}, -j\omega\mu_0\mathcal{L}_0\left(\vec{b}_{\zeta}^{\vec{J}^{\mathrm{A}\rightleftharpoons\mathrm{F}}}\right)\right\rangle_{\mathbb{S}^{\mathrm{A}\rightleftharpoons\mathrm{F}}} \tag{7-14a}$$

$$z_{\xi\zeta}^{\vec{J}^{\mathrm{A}\rightleftharpoons\mathrm{F}}\vec{J}^{\mathrm{F}}} = \left\langle \vec{b}_{\xi}^{\vec{J}^{\mathrm{A}\rightleftharpoons\mathrm{F}}}, -j\omega\mu_0\mathcal{L}_0\left(\vec{b}_{\zeta}^{\vec{J}^{\mathrm{F}}}\right)\right\rangle_{\mathbb{S}^{\mathrm{A}\rightleftharpoons\mathrm{F}}} \tag{7-14b}$$

$$z_{\xi\zeta}^{\vec{J}^{\mathrm{A}\rightleftharpoons\mathrm{F}}\vec{J}_{\mathrm{F}}} = \left\langle \vec{b}_{\xi}^{\vec{J}^{\mathrm{A}\rightleftharpoons\mathrm{F}}}, -j\omega\mu_0\mathcal{L}_0\left(\vec{b}_{\zeta}^{\vec{J}_{\mathrm{F}}}\right)\right\rangle_{\mathbb{S}^{\mathrm{A}\rightleftharpoons\mathrm{F}}} \tag{7-14c}$$

$$z_{\xi\zeta}^{\vec{J}^{\mathrm{A}\rightleftharpoons\mathrm{F}}\vec{J}_{\mathrm{F}\rightleftharpoons\mathrm{A}}} = \left\langle \vec{b}_{\xi}^{\vec{J}^{\mathrm{A}\rightleftharpoons\mathrm{F}}}, -j\omega\mu_0\mathcal{L}_0\left(-\vec{b}_{\zeta}^{\vec{J}_{\mathrm{F}\rightleftharpoons\mathrm{A}}}\right)\right\rangle_{\mathbb{S}^{\mathrm{A}\rightleftharpoons\mathrm{F}}} \tag{7-14d}$$

$$z_{\xi\zeta}^{\vec{J}^{\mathrm{A}\rightleftharpoons\mathrm{F}}\vec{M}^{\mathrm{A}\rightleftharpoons\mathrm{F}}} = \left\langle \vec{b}_{\xi}^{\vec{J}^{\mathrm{A}\rightleftharpoons\mathrm{F}}}, \frac{1}{2}\vec{b}_{\zeta}^{\vec{M}^{\mathrm{A}\rightleftharpoons\mathrm{F}}} \times \hat{n}^{\rightarrow\mathrm{F}} - \mathrm{P.V.}\mathcal{K}_0\left(\vec{b}_{\zeta}^{\vec{M}^{\mathrm{A}\rightleftharpoons\mathrm{F}}}\right)\right\rangle_{\mathbb{S}^{\mathrm{A}\rightleftharpoons\mathrm{F}}} \tag{7-14e}$$

$$z_{\xi\zeta}^{\vec{J}^{\mathrm{A}\rightleftharpoons\mathrm{F}}\vec{M}_{\mathrm{F}\rightleftharpoons\mathrm{A}}} = \left\langle \vec{b}_{\xi}^{\vec{J}^{\mathrm{A}\rightleftharpoons\mathrm{F}}}, -\mathcal{K}_0\left(-\vec{b}_{\zeta}^{\vec{M}_{\mathrm{F}\rightleftharpoons\mathrm{A}}}\right)\right\rangle_{\mathbb{S}^{\mathrm{A}\rightleftharpoons\mathrm{F}}} \tag{7-14f}$$

The formulations used to calculate the elements of the matrices in Eq. (7-10) are as

$$z_{\xi\zeta}^{\vec{J}^{\mathrm{F}}\vec{J}^{\mathrm{A}\rightleftharpoons\mathrm{F}}} = \left\langle \vec{b}_{\xi}^{\vec{J}^{\mathrm{F}}}, -j\omega\mu_0\mathcal{L}_0\left(\vec{b}_{\zeta}^{\vec{J}^{\mathrm{A}\rightleftharpoons\mathrm{F}}}\right)\right\rangle_{\mathbb{S}^{\mathrm{F}}} \tag{7-15a}$$

$$z_{\xi\zeta}^{\vec{J}^{\mathrm{F}}\vec{J}^{\mathrm{F}}} = \left\langle \vec{b}_{\xi}^{\vec{J}^{\mathrm{F}}}, -j\omega\mu_0\mathcal{L}_0\left(\vec{b}_{\zeta}^{\vec{J}^{\mathrm{F}}}\right)\right\rangle_{\mathbb{S}^{\mathrm{F}}} \tag{7-15b}$$

$$z_{\xi\zeta}^{\vec{J}^{\mathrm{F}}\vec{J}_{\mathrm{F}}} = \left\langle \vec{b}_{\xi}^{\vec{J}^{\mathrm{F}}}, -j\omega\mu_0\mathcal{L}_0\left(\vec{b}_{\zeta}^{\vec{J}_{\mathrm{F}}}\right)\right\rangle_{\mathbb{S}^{\mathrm{F}}} \tag{7-15c}$$

$$z_{\xi\zeta}^{\vec{J}^{\mathrm{F}}\vec{J}_{\mathrm{F}\rightleftharpoons\mathrm{A}}} = \left\langle \vec{b}_{\xi}^{\vec{J}^{\mathrm{F}}}, -j\omega\mu_0\mathcal{L}_0\left(-\vec{b}_{\zeta}^{\vec{J}_{\mathrm{F}\rightleftharpoons\mathrm{A}}}\right)\right\rangle_{\mathbb{S}^{\mathrm{F}}} \tag{7-15d}$$

$$z_{\xi\zeta}^{\vec{J}^{\mathrm{F}}\vec{M}^{\mathrm{A}\rightleftharpoons\mathrm{F}}} = \left\langle \vec{b}_{\xi}^{\vec{J}^{\mathrm{F}}}, -\mathcal{K}_0\left(\vec{b}_{\zeta}^{\vec{M}^{\mathrm{A}\rightleftharpoons\mathrm{F}}}\right)\right\rangle_{\mathbb{S}^{\mathrm{F}}} \tag{7-15e}$$

$$z_{\xi\zeta}^{\vec{J}^{\mathrm{F}}\vec{M}_{\mathrm{F}\rightleftharpoons\mathrm{A}}} = \left\langle \vec{b}_{\xi}^{\vec{J}^{\mathrm{F}}}, -\mathcal{K}_0\left(-\vec{b}_{\zeta}^{\vec{M}_{\mathrm{F}\rightleftharpoons\mathrm{A}}}\right)\right\rangle_{\mathbb{S}^{\mathrm{F}}} \tag{7-15f}$$

The formulations used to calculate the elements of the matrices in Eq. (7-11) are as

$$z_{\xi\zeta}^{\vec{J}_{\mathrm{F}}\vec{J}^{\mathrm{A}\rightleftharpoons\mathrm{F}}} = \left\langle \vec{b}_{\xi}^{\vec{J}_{\mathrm{F}}}, -j\omega\mu_0\mathcal{L}_0\left(\vec{b}_{\zeta}^{\vec{J}^{\mathrm{A}\rightleftharpoons\mathrm{F}}}\right)\right\rangle_{\mathbb{S}_{\mathrm{F}}} \tag{7-16a}$$

$$z_{\xi\zeta}^{\vec{J}_{\rm F}\vec{J}^{\rm F}} = \left\langle \vec{b}_{\xi}^{\vec{J}_{\rm F}}, -j\omega\mu_0\mathcal{L}_0\left(\vec{b}_{\zeta}^{\vec{J}^{\rm F}}\right)\right\rangle_{\mathbb{S}_{\rm F}} \tag{7-16b}$$

$$z_{\xi\zeta}^{\vec{J}_{\rm F}\vec{J}_{\rm F}} = \left\langle \vec{b}_{\xi}^{\vec{J}_{\rm F}}, -j\omega\mu_0\mathcal{L}_0\left(\vec{b}_{\zeta}^{\vec{J}_{\rm F}}\right)\right\rangle_{\mathbb{S}_{\rm F}} \tag{7-16c}$$

$$z_{\xi\zeta}^{\vec{J}_{\rm F}\vec{J}_{\rm F\rightleftharpoons A}} = \left\langle \vec{b}_{\xi}^{\vec{J}_{\rm F}}, -j\omega\mu_0\mathcal{L}_0\left(-\vec{b}_{\zeta}^{\vec{J}_{\rm F\rightleftharpoons A}}\right)\right\rangle_{\mathbb{S}_{\rm F}} \tag{7-16d}$$

$$z_{\xi\zeta}^{\vec{J}_{\rm F}\vec{M}^{\rm A\rightleftharpoons F}} = \left\langle \vec{b}_{\xi}^{\vec{J}_{\rm F}}, -\mathcal{K}_0\left(\vec{b}_{\zeta}^{\vec{M}^{\rm A\rightleftharpoons F}}\right)\right\rangle_{\mathbb{S}_{\rm F}} \tag{7-16e}$$

$$z_{\xi\zeta}^{\vec{J}_{\rm F}\vec{M}_{\rm F\rightleftharpoons A}} = \left\langle \vec{b}_{\xi}^{\vec{J}_{\rm F}}, -\mathcal{K}_0\left(-\vec{b}_{\zeta}^{\vec{M}_{\rm F\rightleftharpoons A}}\right)\right\rangle_{\mathbb{S}_{\rm F}} \tag{7-16f}$$

The formulations used to calculate the elements of the matrices in Eq. (7-12a) are as follows:

$$z_{\xi\zeta}^{\vec{J}_{\rm F\rightleftharpoons A}\vec{J}^{\rm A\rightleftharpoons F}} = \left\langle \vec{b}_{\xi}^{\vec{J}_{\rm F\rightleftharpoons A}}, -j\omega\mu_0\mathcal{L}_0\left(\vec{b}_{\zeta}^{\vec{J}^{\rm A\rightleftharpoons F}}\right)\right\rangle_{\mathbb{S}_{\rm F\rightleftharpoons A}} \tag{7-17a}$$

$$z_{\xi\zeta}^{\vec{J}_{\rm F\rightleftharpoons A}\vec{J}^{\rm F}} = \left\langle \vec{b}_{\xi}^{\vec{J}_{\rm F\rightleftharpoons A}}, -j\omega\mu_0\mathcal{L}_0\left(\vec{b}_{\zeta}^{\vec{J}^{\rm F}}\right)\right\rangle_{\mathbb{S}_{\rm F\rightleftharpoons A}} \tag{7-17b}$$

$$z_{\xi\zeta}^{\vec{J}_{\rm F\rightleftharpoons A}\vec{J}_{\rm F}} = \left\langle \vec{b}_{\xi}^{\vec{J}_{\rm F\rightleftharpoons A}}, -j\omega\mu_0\mathcal{L}_0\left(\vec{b}_{\zeta}^{\vec{J}_{\rm F}}\right)\right\rangle_{\mathbb{S}_{\rm F\rightleftharpoons A}} \tag{7-17c}$$

$$z_{\xi\zeta}^{\vec{J}_{\rm F\rightleftharpoons A}\vec{J}_{\rm F\rightleftharpoons A}} = \left\langle \vec{b}_{\xi}^{\vec{J}_{\rm F\rightleftharpoons A}}, -j\omega\mu_0\mathcal{L}_0\left(-\vec{b}_{\zeta}^{\vec{J}_{\rm F\rightleftharpoons A}}\right)\right\rangle_{\mathbb{S}_{\rm F\rightleftharpoons A}} - \left\langle \vec{b}_{\xi}^{\vec{J}_{\rm F\rightleftharpoons A}}, -j\omega\mu_0\mathcal{L}_0\left(\vec{b}_{\zeta}^{\vec{J}_{\rm F\rightleftharpoons A}}\right)\right\rangle_{\mathbb{S}_{\rm F\rightleftharpoons A}} \tag{7-17d}$$

$$z_{\xi\zeta}^{\vec{J}_{\rm F\rightleftharpoons A}\vec{M}^{\rm A\rightleftharpoons F}} = \left\langle \vec{b}_{\xi}^{\vec{J}_{\rm F\rightleftharpoons A}}, -\mathcal{K}_0\left(\vec{b}_{\zeta}^{\vec{M}^{\rm A\rightleftharpoons F}}\right)\right\rangle_{\mathbb{S}_{\rm F\rightleftharpoons A}} \tag{7-17e}$$

$$z_{\xi\zeta}^{\vec{J}_{\rm F\rightleftharpoons A}\vec{M}_{\rm F\rightleftharpoons A}} = \left\langle \vec{b}_{\xi}^{\vec{J}_{\rm F\rightleftharpoons A}}, -{\rm P.V.}\,\mathcal{K}_0\left(-\vec{b}_{\zeta}^{\vec{M}_{\rm F\rightleftharpoons A}}\right)\right\rangle_{\mathbb{S}_{\rm F\rightleftharpoons A}} - \left\langle \vec{b}_{\xi}^{\vec{J}_{\rm F\rightleftharpoons A}}, -{\rm P.V.}\,\mathcal{K}_0\left(\vec{b}_{\zeta}^{\vec{M}_{\rm F\rightleftharpoons A}}\right)\right\rangle_{\mathbb{S}_{\rm F\rightleftharpoons A}} \tag{7-17f}$$

The formulations used to calculate the elements of the matrices in Eq. (7-12b) are as follows:

$$z_{\xi\zeta}^{\vec{M}_{\rm F\rightleftharpoons A}\vec{J}^{\rm A\rightleftharpoons F}} = \left\langle \vec{b}_{\xi}^{\vec{M}_{\rm F\rightleftharpoons A}}, \mathcal{K}_0\left(\vec{b}_{\zeta}^{\vec{J}^{\rm A\rightleftharpoons F}}\right)\right\rangle_{\mathbb{S}_{\rm F\rightleftharpoons A}} \tag{7-18a}$$

$$z_{\xi\zeta}^{\vec{M}_{\rm F\rightleftharpoons A}\vec{J}^{\rm F}} = \left\langle \vec{b}_{\xi}^{\vec{M}_{\rm F\rightleftharpoons A}}, \mathcal{K}_0\left(\vec{b}_{\zeta}^{\vec{J}^{\rm F}}\right)\right\rangle_{\mathbb{S}_{\rm F\rightleftharpoons A}} \tag{7-18b}$$

$$z_{\xi\zeta}^{\vec{M}_{\rm F\rightleftharpoons A}\vec{J}_{\rm F}} = \left\langle \vec{b}_{\xi}^{\vec{M}_{\rm F\rightleftharpoons A}}, \mathcal{K}_0\left(\vec{b}_{\zeta}^{\vec{J}_{\rm F}}\right)\right\rangle_{\mathbb{S}_{\rm F\rightleftharpoons A}} \tag{7-18c}$$

$$z_{\xi\zeta}^{\vec{M}_{\rm F\rightleftharpoons A}\vec{J}_{\rm F\rightleftharpoons A}} = \left\langle \vec{b}_{\xi}^{\vec{M}_{\rm F\rightleftharpoons A}}, {\rm P.V.}\,\mathcal{K}_0\left(-\vec{b}_{\zeta}^{\vec{J}_{\rm F\rightleftharpoons A}}\right)\right\rangle_{\mathbb{S}_{\rm F\rightleftharpoons A}} - \left\langle \vec{b}_{\xi}^{\vec{M}_{\rm F\rightleftharpoons A}}, {\rm P.V.}\,\mathcal{K}_0\left(\vec{b}_{\zeta}^{\vec{J}_{\rm F\rightleftharpoons A}}\right)\right\rangle_{\mathbb{S}_{\rm F\rightleftharpoons A}} \tag{7-18d}$$

$$z_{\xi\zeta}^{\vec{M}_{\rm F\rightleftharpoons A}\vec{M}^{\rm A\rightleftharpoons F}} = \left\langle \vec{b}_{\xi}^{\vec{M}_{\rm F\rightleftharpoons A}}, -j\omega\varepsilon_0\mathcal{L}_0\left(\vec{b}_{\zeta}^{\vec{M}^{\rm A\rightleftharpoons F}}\right)\right\rangle_{\mathbb{S}_{\rm F\rightleftharpoons A}} \tag{7-18e}$$

$$z_{\xi\zeta}^{\vec{M}_{\rm F\rightleftharpoons A}\vec{M}_{\rm F\rightleftharpoons A}} = \left\langle \vec{b}_{\xi}^{\vec{M}_{\rm F\rightleftharpoons A}}, -j\omega\varepsilon_0\mathcal{L}_0\left(-\vec{b}_{\zeta}^{\vec{M}_{\rm F\rightleftharpoons A}}\right)\right\rangle_{\mathbb{S}_{\rm F\rightleftharpoons A}} - \left\langle \vec{b}_{\xi}^{\vec{M}_{\rm F\rightleftharpoons A}}, -j\omega\varepsilon_0\mathcal{L}_0\left(\vec{b}_{\zeta}^{\vec{M}_{\rm F\rightleftharpoons A}}\right)\right\rangle_{\mathbb{S}_{\rm F\rightleftharpoons A}} \tag{7-18f}$$

Below, we propose two different schemes for mathematically describing the modal space of the rec-antenna shown in Figs. 7-1 and 7-2.

#### 7.2.2.1 Scheme I: Dependent Variable Elimination (DVE)

By properly assembling the Eqs. (7-9a)~(7-12b), we have the following two theoretically equivalent *augmented matrix equations*

$$\bar{\bar{\Psi}}_1 \cdot \bar{a}^{\mathrm{AV}} = \bar{\bar{\Psi}}_2 \cdot \bar{a}^{\vec{J}^{\mathrm{A}\rightleftharpoons\mathrm{F}}} \tag{7-19}$$

$$\bar{\bar{\Psi}}_3 \cdot \bar{a}^{\mathrm{AV}} = \bar{\bar{\Psi}}_4 \cdot \bar{a}^{\vec{M}^{\mathrm{A}\rightleftharpoons\mathrm{F}}} \tag{7-20}$$

in which

$$\bar{\bar{\Psi}}_1 = \begin{bmatrix} \bar{\bar{I}}^{\vec{J}^{\mathrm{A}\rightleftharpoons\mathrm{F}}} & 0 & 0 & 0 & 0 & 0 \\ 0 & \bar{\bar{Z}}^{\vec{M}^{\mathrm{A}\rightleftharpoons\mathrm{F}}\vec{J}^{\mathrm{F}}} & \bar{\bar{Z}}^{\vec{M}^{\mathrm{A}\rightleftharpoons\mathrm{F}}\vec{J}_{\mathrm{F}}} & \bar{\bar{Z}}^{\vec{M}^{\mathrm{A}\rightleftharpoons\mathrm{F}}\vec{J}_{\mathrm{F}\rightleftharpoons\mathrm{A}}} & \bar{\bar{Z}}^{\vec{M}^{\mathrm{A}\rightleftharpoons\mathrm{F}}\vec{M}^{\mathrm{A}\rightleftharpoons\mathrm{F}}} & \bar{\bar{Z}}^{\vec{M}^{\mathrm{A}\rightleftharpoons\mathrm{F}}\vec{M}_{\mathrm{F}\rightleftharpoons\mathrm{A}}} \\ 0 & \bar{\bar{Z}}^{\vec{J}^{\mathrm{F}}\vec{J}^{\mathrm{F}}} & \bar{\bar{Z}}^{\vec{J}^{\mathrm{F}}\vec{J}_{\mathrm{F}}} & \bar{\bar{Z}}^{\vec{J}^{\mathrm{F}}\vec{J}_{\mathrm{F}\rightleftharpoons\mathrm{A}}} & \bar{\bar{Z}}^{\vec{J}^{\mathrm{F}}\vec{M}^{\mathrm{A}\rightleftharpoons\mathrm{F}}} & \bar{\bar{Z}}^{\vec{J}^{\mathrm{F}}\vec{M}_{\mathrm{F}\rightleftharpoons\mathrm{A}}} \\ 0 & \bar{\bar{Z}}^{\vec{J}_{\mathrm{F}}\vec{J}^{\mathrm{F}}} & \bar{\bar{Z}}^{\vec{J}_{\mathrm{F}}\vec{J}_{\mathrm{F}}} & \bar{\bar{Z}}^{\vec{J}_{\mathrm{F}}\vec{J}_{\mathrm{F}\rightleftharpoons\mathrm{A}}} & \bar{\bar{Z}}^{\vec{J}_{\mathrm{F}}\vec{M}^{\mathrm{A}\rightleftharpoons\mathrm{F}}} & \bar{\bar{Z}}^{\vec{J}_{\mathrm{F}}\vec{M}_{\mathrm{F}\rightleftharpoons\mathrm{A}}} \\ 0 & \bar{\bar{Z}}^{\vec{J}_{\mathrm{F}\rightleftharpoons\mathrm{A}}\vec{J}^{\mathrm{F}}} & \bar{\bar{Z}}^{\vec{J}_{\mathrm{F}\rightleftharpoons\mathrm{A}}\vec{J}_{\mathrm{F}}} & \bar{\bar{Z}}^{\vec{J}_{\mathrm{F}\rightleftharpoons\mathrm{A}}\vec{J}_{\mathrm{F}\rightleftharpoons\mathrm{A}}} & \bar{\bar{Z}}^{\vec{J}_{\mathrm{F}\rightleftharpoons\mathrm{A}}\vec{M}^{\mathrm{A}\rightleftharpoons\mathrm{F}}} & \bar{\bar{Z}}^{\vec{J}_{\mathrm{F}\rightleftharpoons\mathrm{A}}\vec{M}_{\mathrm{F}\rightleftharpoons\mathrm{A}}} \\ 0 & \bar{\bar{Z}}^{\vec{M}_{\mathrm{F}\rightleftharpoons\mathrm{A}}\vec{J}^{\mathrm{F}}} & \bar{\bar{Z}}^{\vec{M}_{\mathrm{F}\rightleftharpoons\mathrm{A}}\vec{J}_{\mathrm{F}}} & \bar{\bar{Z}}^{\vec{M}_{\mathrm{F}\rightleftharpoons\mathrm{A}}\vec{J}_{\mathrm{F}\rightleftharpoons\mathrm{A}}} & \bar{\bar{Z}}^{\vec{M}_{\mathrm{F}\rightleftharpoons\mathrm{A}}\vec{M}^{\mathrm{A}\rightleftharpoons\mathrm{F}}} & \bar{\bar{Z}}^{\vec{M}_{\mathrm{F}\rightleftharpoons\mathrm{A}}\vec{M}_{\mathrm{F}\rightleftharpoons\mathrm{A}}} \end{bmatrix} \tag{7-21a}$$

$$\bar{\bar{\Psi}}_2 = \begin{bmatrix} \bar{\bar{I}}^{\vec{J}^{\mathrm{A}\rightleftharpoons\mathrm{F}}} \\ -\bar{\bar{Z}}^{\vec{M}^{\mathrm{A}\rightleftharpoons\mathrm{F}}\vec{J}^{\mathrm{A}\rightleftharpoons\mathrm{F}}} \\ -\bar{\bar{Z}}^{\vec{J}^{\mathrm{F}}\vec{J}^{\mathrm{A}\rightleftharpoons\mathrm{F}}} \\ -\bar{\bar{Z}}^{\vec{J}_{\mathrm{F}}\vec{J}^{\mathrm{A}\rightleftharpoons\mathrm{F}}} \\ -\bar{\bar{Z}}^{\vec{J}_{\mathrm{F}\rightleftharpoons\mathrm{A}}\vec{J}^{\mathrm{A}\rightleftharpoons\mathrm{F}}} \\ -\bar{\bar{Z}}^{\vec{M}_{\mathrm{F}\rightleftharpoons\mathrm{A}}\vec{J}^{\mathrm{A}\rightleftharpoons\mathrm{F}}} \end{bmatrix} \tag{7-21b}$$

$$\bar{a}^{\mathrm{AV}} = \begin{bmatrix} \bar{a}^{\vec{J}^{\mathrm{A}\rightleftharpoons\mathrm{F}}} \\ \bar{a}^{\vec{J}^{\mathrm{F}}} \\ \bar{a}^{\vec{J}_{\mathrm{F}}} \\ \bar{a}^{\vec{J}_{\mathrm{F}\rightleftharpoons\mathrm{A}}} \\ \bar{a}^{\vec{M}^{\mathrm{A}\rightleftharpoons\mathrm{F}}} \\ \bar{a}^{\vec{M}_{\mathrm{F}\rightleftharpoons\mathrm{A}}} \end{bmatrix} \tag{7-22}$$

and

$$\bar{\bar{\Psi}}_3 = \begin{bmatrix} 0 & 0 & 0 & 0 & \bar{\bar{I}}^{\vec{M}^{\mathrm{A}\rightleftharpoons\mathrm{F}}} & 0 \\ \bar{\bar{Z}}^{\vec{J}^{\mathrm{A}\rightleftharpoons\mathrm{F}}\vec{J}^{\mathrm{A}\rightleftharpoons\mathrm{F}}} & \bar{\bar{Z}}^{\vec{J}^{\mathrm{A}\rightleftharpoons\mathrm{F}}\vec{J}^{\mathrm{F}}} & \bar{\bar{Z}}^{\vec{J}^{\mathrm{A}\rightleftharpoons\mathrm{F}}\vec{J}_{\mathrm{F}}} & \bar{\bar{Z}}^{\vec{J}^{\mathrm{A}\rightleftharpoons\mathrm{F}}\vec{J}_{\mathrm{F}\rightleftharpoons\mathrm{A}}} & 0 & \bar{\bar{Z}}^{\vec{J}^{\mathrm{A}\rightleftharpoons\mathrm{F}}\vec{M}_{\mathrm{F}\rightleftharpoons\mathrm{A}}} \\ \bar{\bar{Z}}^{\vec{J}^{\mathrm{F}}\vec{J}^{\mathrm{A}\rightleftharpoons\mathrm{F}}} & \bar{\bar{Z}}^{\vec{J}^{\mathrm{F}}\vec{J}^{\mathrm{F}}} & \bar{\bar{Z}}^{\vec{J}^{\mathrm{F}}\vec{J}_{\mathrm{F}}} & \bar{\bar{Z}}^{\vec{J}^{\mathrm{F}}\vec{J}_{\mathrm{F}\rightleftharpoons\mathrm{A}}} & 0 & \bar{\bar{Z}}^{\vec{J}^{\mathrm{F}}\vec{M}_{\mathrm{F}\rightleftharpoons\mathrm{A}}} \\ \bar{\bar{Z}}^{\vec{J}_{\mathrm{F}}\vec{J}^{\mathrm{A}\rightleftharpoons\mathrm{F}}} & \bar{\bar{Z}}^{\vec{J}_{\mathrm{F}}\vec{J}^{\mathrm{F}}} & \bar{\bar{Z}}^{\vec{J}_{\mathrm{F}}\vec{J}_{\mathrm{F}}} & \bar{\bar{Z}}^{\vec{J}_{\mathrm{F}}\vec{J}_{\mathrm{F}\rightleftharpoons\mathrm{A}}} & 0 & \bar{\bar{Z}}^{\vec{J}_{\mathrm{F}}\vec{M}_{\mathrm{F}\rightleftharpoons\mathrm{A}}} \\ \bar{\bar{Z}}^{\vec{J}_{\mathrm{F}\rightleftharpoons\mathrm{A}}\vec{J}^{\mathrm{A}\rightleftharpoons\mathrm{F}}} & \bar{\bar{Z}}^{\vec{J}_{\mathrm{F}\rightleftharpoons\mathrm{A}}\vec{J}^{\mathrm{F}}} & \bar{\bar{Z}}^{\vec{J}_{\mathrm{F}\rightleftharpoons\mathrm{A}}\vec{J}_{\mathrm{F}}} & \bar{\bar{Z}}^{\vec{J}_{\mathrm{F}\rightleftharpoons\mathrm{A}}\vec{J}_{\mathrm{F}\rightleftharpoons\mathrm{A}}} & 0 & \bar{\bar{Z}}^{\vec{J}_{\mathrm{F}\rightleftharpoons\mathrm{A}}\vec{M}_{\mathrm{F}\rightleftharpoons\mathrm{A}}} \\ \bar{\bar{Z}}^{\vec{M}_{\mathrm{F}\rightleftharpoons\mathrm{A}}\vec{J}^{\mathrm{A}\rightleftharpoons\mathrm{F}}} & \bar{\bar{Z}}^{\vec{M}_{\mathrm{F}\rightleftharpoons\mathrm{A}}\vec{J}^{\mathrm{F}}} & \bar{\bar{Z}}^{\vec{M}_{\mathrm{F}\rightleftharpoons\mathrm{A}}\vec{J}_{\mathrm{F}}} & \bar{\bar{Z}}^{\vec{M}_{\mathrm{F}\rightleftharpoons\mathrm{A}}\vec{J}_{\mathrm{F}\rightleftharpoons\mathrm{A}}} & 0 & \bar{\bar{Z}}^{\vec{M}_{\mathrm{F}\rightleftharpoons\mathrm{A}}\vec{M}_{\mathrm{F}\rightleftharpoons\mathrm{A}}} \end{bmatrix} \tag{7-23a}$$

$$\overline{\overline{\Psi}}_4 = \begin{bmatrix} \overline{\overline{I}}^{\vec{M}^{\mathrm{A}\rightleftharpoons\mathrm{F}}} \\ -\overline{\overline{Z}}^{\vec{J}^{\mathrm{A}\rightleftharpoons\mathrm{F}}\vec{M}^{\mathrm{A}\rightleftharpoons\mathrm{F}}} \\ -\overline{\overline{Z}}^{\vec{J}^{\mathrm{F}}\vec{M}^{\mathrm{A}\rightleftharpoons\mathrm{F}}} \\ -\overline{\overline{Z}}^{\vec{J}_{\mathrm{F}}\vec{M}^{\mathrm{A}\rightleftharpoons\mathrm{F}}} \\ -\overline{\overline{Z}}^{\vec{J}_{\mathrm{F}\rightleftharpoons\mathrm{A}}\vec{M}^{\mathrm{A}\rightleftharpoons\mathrm{F}}} \\ -\overline{\overline{Z}}^{\vec{M}_{\mathrm{F}\rightleftharpoons\mathrm{A}}\vec{M}^{\mathrm{A}\rightleftharpoons\mathrm{F}}} \end{bmatrix} \tag{7-23b}$$

By solving the above matrix equations, there exist the following transformations from $\overline{a}^{\vec{J}^{\mathrm{A}\rightleftharpoons\mathrm{F}}}$ to $\overline{a}^{\mathrm{AV}}$ and from $\overline{a}^{\vec{M}^{\mathrm{A}\rightleftharpoons\mathrm{F}}}$ to $\overline{a}^{\mathrm{AV}}$

$$\overline{a}^{\mathrm{AV}} = \overbrace{\left(\overline{\overline{\Psi}}_1\right)^{-1}\cdot\overline{\overline{\Psi}}_2}^{\overline{\overline{T}}^{\vec{J}^{\mathrm{A}\rightleftharpoons\mathrm{F}}\rightarrow\mathrm{AV}}}\cdot\overline{a}^{\vec{J}^{\mathrm{A}\rightleftharpoons\mathrm{F}}} \tag{7-24}$$

$$\overline{a}^{\mathrm{AV}} = \underbrace{\left(\overline{\overline{\Psi}}_3\right)^{-1}\cdot\overline{\overline{\Psi}}_4}_{\overline{\overline{T}}^{\vec{M}^{\mathrm{A}\rightleftharpoons\mathrm{F}}\rightarrow\mathrm{AV}}}\cdot\overline{a}^{\vec{M}^{\mathrm{A}\rightleftharpoons\mathrm{F}}} \tag{7-25}$$

and they can be uniformly written as that $\overline{a}^{\mathrm{AV}} = \overline{\overline{T}}^{\mathrm{BV}\rightarrow\mathrm{AV}}\cdot\overline{a}^{\mathrm{BV}}$, for simplifying the symbolic system of the following discussions.

### 7.2.2.2 Scheme II: Solution/Definition Domain Compression (SDC/DDC)

In fact, the Eqs. (7-9a)~(7-12b) can also be alternatively combined as follows:

$$\overline{\overline{\Psi}}_{\mathrm{FCE}}^{\mathrm{DoJ}}\cdot\overline{a}^{\mathrm{AV}} = 0 \tag{7-26}$$

$$\overline{\overline{\Psi}}_{\mathrm{FCE}}^{\mathrm{DoM}}\cdot\overline{a}^{\mathrm{AV}} = 0 \tag{7-27}$$

in which

$$\overline{\overline{\Psi}}_{\mathrm{FCE}}^{\mathrm{DoJ}} = \begin{bmatrix}
\overline{\overline{Z}}^{\vec{M}^{\mathrm{A}\rightleftharpoons\mathrm{F}}\vec{J}^{\mathrm{A}\rightleftharpoons\mathrm{F}}} & \overline{\overline{Z}}^{\vec{M}^{\mathrm{A}\rightleftharpoons\mathrm{F}}\vec{J}^{\mathrm{F}}} & \overline{\overline{Z}}^{\vec{M}^{\mathrm{A}\rightleftharpoons\mathrm{F}}\vec{J}_{\mathrm{F}}} & \overline{\overline{Z}}^{\vec{M}^{\mathrm{A}\rightleftharpoons\mathrm{F}}\vec{J}_{\mathrm{F}\rightleftharpoons\mathrm{A}}} & \overline{\overline{Z}}^{\vec{M}^{\mathrm{A}\rightleftharpoons\mathrm{F}}\vec{M}^{\mathrm{A}\rightleftharpoons\mathrm{F}}} & \overline{\overline{Z}}^{\vec{M}^{\mathrm{A}\rightleftharpoons\mathrm{F}}\vec{M}_{\mathrm{F}\rightleftharpoons\mathrm{A}}} \\
\overline{\overline{Z}}^{\vec{J}^{\mathrm{F}}\vec{J}^{\mathrm{A}\rightleftharpoons\mathrm{F}}} & \overline{\overline{Z}}^{\vec{J}^{\mathrm{F}}\vec{J}^{\mathrm{F}}} & \overline{\overline{Z}}^{\vec{J}^{\mathrm{F}}\vec{J}_{\mathrm{F}}} & \overline{\overline{Z}}^{\vec{J}^{\mathrm{F}}\vec{J}_{\mathrm{F}\rightleftharpoons\mathrm{A}}} & \overline{\overline{Z}}^{\vec{J}^{\mathrm{F}}\vec{M}^{\mathrm{A}\rightleftharpoons\mathrm{F}}} & \overline{\overline{Z}}^{\vec{J}^{\mathrm{F}}\vec{M}_{\mathrm{F}\rightleftharpoons\mathrm{A}}} \\
\overline{\overline{Z}}^{\vec{J}_{\mathrm{F}}\vec{J}^{\mathrm{A}\rightleftharpoons\mathrm{F}}} & \overline{\overline{Z}}^{\vec{J}_{\mathrm{F}}\vec{J}^{\mathrm{F}}} & \overline{\overline{Z}}^{\vec{J}_{\mathrm{F}}\vec{J}_{\mathrm{F}}} & \overline{\overline{Z}}^{\vec{J}_{\mathrm{F}}\vec{J}_{\mathrm{F}\rightleftharpoons\mathrm{A}}} & \overline{\overline{Z}}^{\vec{J}_{\mathrm{F}}\vec{M}^{\mathrm{A}\rightleftharpoons\mathrm{F}}} & \overline{\overline{Z}}^{\vec{J}_{\mathrm{F}}\vec{M}_{\mathrm{F}\rightleftharpoons\mathrm{A}}} \\
\overline{\overline{Z}}^{\vec{J}_{\mathrm{F}\rightleftharpoons\mathrm{A}}\vec{J}^{\mathrm{A}\rightleftharpoons\mathrm{F}}} & \overline{\overline{Z}}^{\vec{J}_{\mathrm{F}\rightleftharpoons\mathrm{A}}\vec{J}^{\mathrm{F}}} & \overline{\overline{Z}}^{\vec{J}_{\mathrm{F}\rightleftharpoons\mathrm{A}}\vec{J}_{\mathrm{F}}} & \overline{\overline{Z}}^{\vec{J}_{\mathrm{F}\rightleftharpoons\mathrm{A}}\vec{J}_{\mathrm{F}\rightleftharpoons\mathrm{A}}} & \overline{\overline{Z}}^{\vec{J}_{\mathrm{F}\rightleftharpoons\mathrm{A}}\vec{M}^{\mathrm{A}\rightleftharpoons\mathrm{F}}} & \overline{\overline{Z}}^{\vec{J}_{\mathrm{F}\rightleftharpoons\mathrm{A}}\vec{M}_{\mathrm{F}\rightleftharpoons\mathrm{A}}} \\
\overline{\overline{Z}}^{\vec{M}_{\mathrm{F}\rightleftharpoons\mathrm{A}}\vec{J}^{\mathrm{A}\rightleftharpoons\mathrm{F}}} & \overline{\overline{Z}}^{\vec{M}_{\mathrm{F}\rightleftharpoons\mathrm{A}}\vec{J}^{\mathrm{F}}} & \overline{\overline{Z}}^{\vec{M}_{\mathrm{F}\rightleftharpoons\mathrm{A}}\vec{J}_{\mathrm{F}}} & \overline{\overline{Z}}^{\vec{M}_{\mathrm{F}\rightleftharpoons\mathrm{A}}\vec{J}_{\mathrm{F}\rightleftharpoons\mathrm{A}}} & \overline{\overline{Z}}^{\vec{M}_{\mathrm{F}\rightleftharpoons\mathrm{A}}\vec{M}^{\mathrm{A}\rightleftharpoons\mathrm{F}}} & \overline{\overline{Z}}^{\vec{M}_{\mathrm{F}\rightleftharpoons\mathrm{A}}\vec{M}_{\mathrm{F}\rightleftharpoons\mathrm{A}}}
\end{bmatrix} \tag{7-28}$$

$$\overline{\overline{\Psi}}_{\mathrm{FCE}}^{\mathrm{DoM}} = \begin{bmatrix}
\overline{\overline{Z}}^{\vec{J}^{\mathrm{A}\rightleftharpoons\mathrm{F}}\vec{J}^{\mathrm{A}\rightleftharpoons\mathrm{F}}} & \overline{\overline{Z}}^{\vec{J}^{\mathrm{A}\rightleftharpoons\mathrm{F}}\vec{J}^{\mathrm{F}}} & \overline{\overline{Z}}^{\vec{J}^{\mathrm{A}\rightleftharpoons\mathrm{F}}\vec{J}_{\mathrm{F}}} & \overline{\overline{Z}}^{\vec{J}^{\mathrm{A}\rightleftharpoons\mathrm{F}}\vec{J}_{\mathrm{F}\rightleftharpoons\mathrm{A}}} & \overline{\overline{Z}}^{\vec{J}^{\mathrm{A}\rightleftharpoons\mathrm{F}}\vec{M}^{\mathrm{A}\rightleftharpoons\mathrm{F}}} & \overline{\overline{Z}}^{\vec{J}^{\mathrm{A}\rightleftharpoons\mathrm{F}}\vec{M}_{\mathrm{F}\rightleftharpoons\mathrm{A}}} \\
\overline{\overline{Z}}^{\vec{J}^{\mathrm{F}}\vec{J}^{\mathrm{A}\rightleftharpoons\mathrm{F}}} & \overline{\overline{Z}}^{\vec{J}^{\mathrm{F}}\vec{J}^{\mathrm{F}}} & \overline{\overline{Z}}^{\vec{J}^{\mathrm{F}}\vec{J}_{\mathrm{F}}} & \overline{\overline{Z}}^{\vec{J}^{\mathrm{F}}\vec{J}_{\mathrm{F}\rightleftharpoons\mathrm{A}}} & \overline{\overline{Z}}^{\vec{J}^{\mathrm{F}}\vec{M}^{\mathrm{A}\rightleftharpoons\mathrm{F}}} & \overline{\overline{Z}}^{\vec{J}^{\mathrm{F}}\vec{M}_{\mathrm{F}\rightleftharpoons\mathrm{A}}} \\
\overline{\overline{Z}}^{\vec{J}_{\mathrm{F}}\vec{J}^{\mathrm{A}\rightleftharpoons\mathrm{F}}} & \overline{\overline{Z}}^{\vec{J}_{\mathrm{F}}\vec{J}^{\mathrm{F}}} & \overline{\overline{Z}}^{\vec{J}_{\mathrm{F}}\vec{J}_{\mathrm{F}}} & \overline{\overline{Z}}^{\vec{J}_{\mathrm{F}}\vec{J}_{\mathrm{F}\rightleftharpoons\mathrm{A}}} & \overline{\overline{Z}}^{\vec{J}_{\mathrm{F}}\vec{M}^{\mathrm{A}\rightleftharpoons\mathrm{F}}} & \overline{\overline{Z}}^{\vec{J}_{\mathrm{F}}\vec{M}_{\mathrm{F}\rightleftharpoons\mathrm{A}}} \\
\overline{\overline{Z}}^{\vec{J}_{\mathrm{F}\rightleftharpoons\mathrm{A}}\vec{J}^{\mathrm{A}\rightleftharpoons\mathrm{F}}} & \overline{\overline{Z}}^{\vec{J}_{\mathrm{F}\rightleftharpoons\mathrm{A}}\vec{J}^{\mathrm{F}}} & \overline{\overline{Z}}^{\vec{J}_{\mathrm{F}\rightleftharpoons\mathrm{A}}\vec{J}_{\mathrm{F}}} & \overline{\overline{Z}}^{\vec{J}_{\mathrm{F}\rightleftharpoons\mathrm{A}}\vec{J}_{\mathrm{F}\rightleftharpoons\mathrm{A}}} & \overline{\overline{Z}}^{\vec{J}_{\mathrm{F}\rightleftharpoons\mathrm{A}}\vec{M}^{\mathrm{A}\rightleftharpoons\mathrm{F}}} & \overline{\overline{Z}}^{\vec{J}_{\mathrm{F}\rightleftharpoons\mathrm{A}}\vec{M}_{\mathrm{F}\rightleftharpoons\mathrm{A}}} \\
\overline{\overline{Z}}^{\vec{M}_{\mathrm{F}\rightleftharpoons\mathrm{A}}\vec{J}^{\mathrm{A}\rightleftharpoons\mathrm{F}}} & \overline{\overline{Z}}^{\vec{M}_{\mathrm{F}\rightleftharpoons\mathrm{A}}\vec{J}^{\mathrm{F}}} & \overline{\overline{Z}}^{\vec{M}_{\mathrm{F}\rightleftharpoons\mathrm{A}}\vec{J}_{\mathrm{F}}} & \overline{\overline{Z}}^{\vec{M}_{\mathrm{F}\rightleftharpoons\mathrm{A}}\vec{J}_{\mathrm{F}\rightleftharpoons\mathrm{A}}} & \overline{\overline{Z}}^{\vec{M}_{\mathrm{F}\rightleftharpoons\mathrm{A}}\vec{M}^{\mathrm{A}\rightleftharpoons\mathrm{F}}} & \overline{\overline{Z}}^{\vec{M}_{\mathrm{F}\rightleftharpoons\mathrm{A}}\vec{M}_{\mathrm{F}\rightleftharpoons\mathrm{A}}}
\end{bmatrix} \tag{7-29}$$

Theoretically, the Eqs. (7-26) and (7-27) are equivalent to each other, and they have the same *solution space*. If the *basic solutions* used to span the space are denoted as $\{\bar{s}_1^{\text{BS}}, \bar{s}_2^{\text{BS}}, \cdots\}$, then any mode contained in the space can be expanded as follows:

$$\bar{a}^{\text{AV}} = \sum_i a_i^{\text{BS}} \bar{s}_i^{\text{BS}} = \underbrace{\left[\bar{s}_1^{\text{BS}}, \bar{s}_2^{\text{BS}}, \cdots\right]}_{\bar{\bar{T}}^{\text{BS}\to\text{AV}}} \cdot \underbrace{\begin{bmatrix} a_1^{\text{BS}} \\ a_2^{\text{BS}} \\ \vdots \end{bmatrix}}_{\bar{a}^{\text{BS}}} \tag{7-30}$$

where the solution space is just the modal space of the rec-antenna shown in Fig. 7-1.

For the convenience of the following discussions, Eqs. (7-24), (7-25), and (7-30) are uniformly written as follows:

$$\bar{a}^{\text{AV}} = \bar{\bar{T}} \cdot \bar{a} \tag{7-31}$$

where $\bar{a} = \bar{a}^{\text{BV}} / \bar{a}^{\text{BS}}$ and correspondingly $\bar{\bar{T}} = \bar{\bar{T}}^{\text{BV}\to\text{AV}} / \bar{\bar{T}}^{\text{BS}\to\text{AV}}$.

## 7.2.3 Power Transport Theorem and Input Power Operator

In this subsection, we provide the *power transport theorem* (*PTT*) and *input power operator* (*IPO*) corresponding to the rec-antenna shown in Figs. 7-1 and 7-2.

### 7.2.3.1 Power Transport Theorem

Applying the results obtained in Chap. 2 to the rec-antenna shown in Figs. 7-1 and 7-2, we immediately have the following PTT for the rec-antenna

$$P^{\text{A}\rightleftharpoons\text{F}} = P_{\text{rad}}^{\text{I}} + j\, P_{\text{sto}}^{\text{F}} + P_{\text{F}\rightleftharpoons\text{A}} \tag{7-32}$$

Here $P_{\text{F}\rightleftharpoons\text{A}}$ is the power inputted into rec-antenna and also the power inputted into PML.

The above-mentioned power $P_{\text{F}\rightleftharpoons\text{A}}$ is as follows:

$$P_{\text{F}\rightleftharpoons\text{A}} = (1/2) \iint_{\mathbb{S}_{\text{F}\rightleftharpoons\text{A}}} \left(\vec{E} \times \vec{H}^{\dagger}\right) \cdot \hat{n}_{\to\text{PML}} dS \tag{7-33}$$

where $\hat{n}_{\to\text{PML}}$ is the normal direction of $\mathbb{S}_{\text{F}\rightleftharpoons\text{A}}$ and points to PML as shown in Fig. 7-2.

### 7.2.3.2 Input Power Operator — Formulation I: Current Form

Based on Eqs. (7-3a)&(7-3b) and the tangential continuity of the $\{\vec{E}, \vec{H}\}$ on $\mathbb{S}_{\text{F}\rightleftharpoons\text{A}}$, the IPO $P_{\text{F}\rightleftharpoons\text{A}}$ given in Eq. (7-33) can be alternatively written as follows:

$$P_{\text{F}\rightleftharpoons\text{A}} = (1/2) \left\langle \hat{n}_{\to\text{PML}} \times \vec{J}_{\text{F}\rightleftharpoons\text{A}}, \vec{M}_{\text{F}\rightleftharpoons\text{A}} \right\rangle_{\mathbb{S}_{\text{F}\rightleftharpoons\text{A}}} \tag{7-34}$$

and it is just the *current form of IPO*.

Inserting Eq. (7-8) into the above current form, the current form is immediately discretized as follows:

$$P_{\mathrm{F}\rightleftharpoons\mathrm{A}} = \left(\bar{a}^{\mathrm{AV}}\right)^{\dagger} \cdot \underbrace{\begin{bmatrix} 0 & 0 & 0 & 0 & 0 & 0 \\ 0 & 0 & 0 & 0 & 0 & 0 \\ 0 & 0 & 0 & 0 & 0 & 0 \\ 0 & 0 & 0 & 0 & 0 & \overline{\overline{C}}^{\vec{J}_{\mathrm{F}\rightleftharpoons\mathrm{A}}\vec{M}_{\mathrm{F}\rightleftharpoons\mathrm{A}}} \\ 0 & 0 & 0 & 0 & 0 & 0 \\ 0 & 0 & 0 & 0 & 0 & 0 \end{bmatrix}}_{\overline{\overline{P}}_{\mathrm{F}\rightleftharpoons\mathrm{A}}^{\mathrm{curAV}}} \cdot \bar{a}^{\mathrm{AV}} \tag{7-35}$$

where the elements of sub-matrix $\overline{\overline{C}}^{\vec{J}_{\mathrm{F}\rightleftharpoons\mathrm{A}}\vec{M}_{\mathrm{F}\rightleftharpoons\mathrm{A}}}$ are calculated as that $c_{\xi\zeta}^{\vec{J}_{\mathrm{F}\rightleftharpoons\mathrm{A}}\vec{M}_{\mathrm{F}\rightleftharpoons\mathrm{A}}} = (1/2) < \hat{n}_{\to\mathrm{PML}} \times \vec{b}_{\xi}^{\vec{J}_{\mathrm{F}\rightleftharpoons\mathrm{A}}}, \vec{b}_{\zeta}^{\vec{M}_{\mathrm{F}\rightleftharpoons\mathrm{A}}} >_{\mathbb{S}_{\mathrm{F}\rightleftharpoons\mathrm{A}}}$. To obtain the IPO defined on modal space, we substitute Eq. (7-31) into the above Eq. (7-35), and then we have that

$$P_{\mathrm{F}\rightleftharpoons\mathrm{A}} = \bar{a}^{\dagger} \cdot \underbrace{\left(\overline{\overline{T}}^{\dagger} \cdot \overline{\overline{P}}_{\mathrm{F}\rightleftharpoons\mathrm{A}}^{\mathrm{curAV}} \cdot \overline{\overline{T}}\right)}_{\overline{\overline{P}}_{\mathrm{F}\rightleftharpoons\mathrm{A}}^{\mathrm{cur}}} \cdot \bar{a} \tag{7-36}$$

where the superscript "cur" is to emphasize that the matrix originates from discretizing the current form of IPO.

#### 7.2.3.3 Input Power Operator — Formulation II: Field-Current Interaction Forms

Actually, IPO $P_{\mathrm{F}\rightleftharpoons\mathrm{A}}$ also has the following equivalent expression

$$\begin{aligned} P_{\mathrm{F}\rightleftharpoons\mathrm{A}} &= -(1/2)\left\langle \vec{J}_{\mathrm{F}\rightleftharpoons\mathrm{A}}, \vec{E} \right\rangle_{\mathbb{S}_{\mathrm{F}\rightleftharpoons\mathrm{A}}} \\ &= -(1/2)\left\langle \vec{M}_{\mathrm{F}\rightleftharpoons\mathrm{A}}, \vec{H} \right\rangle_{\mathbb{S}_{\mathrm{F}\rightleftharpoons\mathrm{A}}}^{\dagger} \\ &= -(1/2)\left\langle \vec{J}_{\mathrm{F}\rightleftharpoons\mathrm{A}}, \mathcal{E}_0\left(\vec{J}^{\mathrm{A}\rightleftharpoons\mathrm{F}} + \vec{J}^{\mathrm{F}} + \vec{J}_{\mathrm{F}} - \vec{J}_{\mathrm{F}\rightleftharpoons\mathrm{A}}, \vec{M}^{\mathrm{A}\rightleftharpoons\mathrm{F}} - \vec{M}_{\mathrm{F}\rightleftharpoons\mathrm{A}}\right)\right\rangle_{\tilde{\mathbb{S}}_{\mathrm{F}\rightleftharpoons\mathrm{A}}} \\ &= -(1/2)\left\langle \vec{M}_{\mathrm{F}\rightleftharpoons\mathrm{A}}, \mathcal{H}_0\left(\vec{J}^{\mathrm{A}\rightleftharpoons\mathrm{F}} + \vec{J}^{\mathrm{F}} + \vec{J}_{\mathrm{F}} - \vec{J}_{\mathrm{F}\rightleftharpoons\mathrm{A}}, \vec{M}^{\mathrm{A}\rightleftharpoons\mathrm{F}} - \vec{M}_{\mathrm{F}\rightleftharpoons\mathrm{A}}\right)\right\rangle_{\tilde{\mathbb{S}}_{\mathrm{F}\rightleftharpoons\mathrm{A}}}^{\dagger} \end{aligned} \tag{7-37}$$

and they are just the *field-current interaction forms of IPO*.

Substituting Eq. (7-8) into Eq. (7-37), the interaction form is immediately discretized as follows:

$$P_{\mathrm{F}\rightleftharpoons\mathrm{A}} = \left(\bar{a}^{\mathrm{AV}}\right)^{\dagger} \cdot \overline{\overline{P}}_{\mathrm{F}\rightleftharpoons\mathrm{A}}^{\mathrm{intAV}} \cdot \bar{a}^{\mathrm{AV}} \tag{7-38}$$

in which

$$\overline{\overline{P}}_{\mathrm{F}\rightleftharpoons\mathrm{A}}^{\mathrm{intAV}} = \begin{bmatrix} 0 & 0 & 0 & 0 & 0 & 0 \\ 0 & 0 & 0 & 0 & 0 & 0 \\ 0 & 0 & 0 & 0 & 0 & 0 \\ \overline{\overline{P}}^{\vec{J}_{\mathrm{F}\rightleftharpoons\mathrm{A}}\vec{J}^{\mathrm{A}\rightleftharpoons\mathrm{F}}} & \overline{\overline{P}}^{\vec{J}_{\mathrm{F}\rightleftharpoons\mathrm{A}}\vec{J}^{\mathrm{F}}} & \overline{\overline{P}}^{\vec{J}_{\mathrm{F}\rightleftharpoons\mathrm{A}}\vec{J}_{\mathrm{F}}} & \overline{\overline{P}}^{\vec{J}_{\mathrm{F}\rightleftharpoons\mathrm{A}}\vec{J}_{\mathrm{F}\rightleftharpoons\mathrm{A}}} & \overline{\overline{P}}^{\vec{J}_{\mathrm{F}\rightleftharpoons\mathrm{A}}\vec{M}^{\mathrm{A}\rightleftharpoons\mathrm{F}}} & \overline{\overline{P}}^{\vec{J}_{\mathrm{F}\rightleftharpoons\mathrm{A}}\vec{M}_{\mathrm{F}\rightleftharpoons\mathrm{A}}} \\ 0 & 0 & 0 & 0 & 0 & 0 \\ 0 & 0 & 0 & 0 & 0 & 0 \end{bmatrix} \tag{7-39a}$$

for the JE interaction version, and

$$\overline{\overline{P}}_{\text{F}\rightleftharpoons\text{A}}^{\text{intAV}} = \begin{bmatrix} 0 & 0 & 0 & 0 & 0 & 0 \\ 0 & 0 & 0 & 0 & 0 & 0 \\ 0 & 0 & 0 & 0 & 0 & 0 \\ 0 & 0 & 0 & 0 & 0 & 0 \\ 0 & 0 & 0 & 0 & 0 & 0 \\ \overline{\overline{P}}^{\vec{M}_{\text{F}\rightleftharpoons\text{A}}\vec{J}^{\text{A}\rightleftharpoons\text{F}}} & \overline{\overline{P}}^{\vec{M}_{\text{F}\rightleftharpoons\text{A}}\vec{J}^{\text{F}}} & \overline{\overline{P}}^{\vec{M}_{\text{F}\rightleftharpoons\text{A}}\vec{J}_{\text{F}}} & \overline{\overline{P}}^{\vec{M}_{\text{F}\rightleftharpoons\text{A}}\vec{J}_{\text{F}\rightleftharpoons\text{A}}} & \overline{\overline{P}}^{\vec{M}_{\text{F}\rightleftharpoons\text{A}}\vec{M}^{\text{A}\rightleftharpoons\text{F}}} & \overline{\overline{P}}^{\vec{M}_{\text{F}\rightleftharpoons\text{A}}\vec{M}_{\text{F}\rightleftharpoons\text{A}}} \end{bmatrix}^{\dagger} \tag{7-39b}$$

for the HM interaction version,where the elements of the sub-matrices are calculated as

$$p_{\xi\zeta}^{\vec{J}_{\text{F}\rightleftharpoons\text{A}}\vec{J}^{\text{A}\rightleftharpoons\text{F}}} = -(1/2)\left\langle \vec{b}_{\xi}^{\vec{J}_{\text{F}\rightleftharpoons\text{A}}}, -j\omega\mu_0\mathcal{L}_0\left(\vec{b}_{\zeta}^{\vec{J}^{\text{A}\rightleftharpoons\text{F}}}\right)\right\rangle_{\mathbb{S}_{\text{F}\rightleftharpoons\text{A}}} \tag{7-40a}$$

$$p_{\xi\zeta}^{\vec{J}_{\text{F}\rightleftharpoons\text{A}}\vec{J}^{\text{F}}} = -(1/2)\left\langle \vec{b}_{\xi}^{\vec{J}_{\text{F}\rightleftharpoons\text{A}}}, -j\omega\mu_0\mathcal{L}_0\left(\vec{b}_{\zeta}^{\vec{J}^{\text{F}}}\right)\right\rangle_{\mathbb{S}_{\text{F}\rightleftharpoons\text{A}}} \tag{7-40b}$$

$$p_{\xi\zeta}^{\vec{J}_{\text{F}\rightleftharpoons\text{A}}\vec{J}_{\text{F}}} = -(1/2)\left\langle \vec{b}_{\xi}^{\vec{J}_{\text{F}\rightleftharpoons\text{A}}}, -j\omega\mu_0\mathcal{L}_0\left(\vec{b}_{\zeta}^{\vec{J}_{\text{F}}}\right)\right\rangle_{\mathbb{S}_{\text{F}\rightleftharpoons\text{A}}} \tag{7-40c}$$

$$p_{\xi\zeta}^{\vec{J}_{\text{F}\rightleftharpoons\text{A}}\vec{J}_{\text{F}\rightleftharpoons\text{A}}} = -(1/2)\left\langle \vec{b}_{\xi}^{\vec{J}_{\text{F}\rightleftharpoons\text{A}}}, -j\omega\mu_0\mathcal{L}_0\left(-\vec{b}_{\zeta}^{\vec{J}_{\text{F}\rightleftharpoons\text{A}}}\right)\right\rangle_{\mathbb{S}_{\text{F}\rightleftharpoons\text{A}}} \tag{7-40d}$$

$$p_{\xi\zeta}^{\vec{J}_{\text{F}\rightleftharpoons\text{A}}\vec{M}^{\text{A}\rightleftharpoons\text{F}}} = -(1/2)\left\langle \vec{b}_{\xi}^{\vec{J}_{\text{F}\rightleftharpoons\text{A}}}, -\mathcal{K}_0\left(\vec{b}_{\zeta}^{\vec{M}^{\text{A}\rightleftharpoons\text{F}}}\right)\right\rangle_{\mathbb{S}_{\text{F}\rightleftharpoons\text{A}}} \tag{7-40e}$$

$$p_{\xi\zeta}^{\vec{J}_{\text{F}\rightleftharpoons\text{A}}\vec{M}_{\text{F}\rightleftharpoons\text{A}}} = -(1/2)\left\langle \vec{b}_{\xi}^{\vec{J}_{\text{F}\rightleftharpoons\text{A}}}, \hat{n}_{\rightarrow\text{PML}} \times \frac{1}{2}\vec{b}_{\zeta}^{\vec{M}_{\text{F}\rightleftharpoons\text{A}}} - \text{P.V.}\mathcal{K}_0\left(-\vec{b}_{\zeta}^{\vec{M}_{\text{F}\rightleftharpoons\text{A}}}\right)\right\rangle_{\mathbb{S}_{\text{F}\rightleftharpoons\text{A}}} \tag{7-40f}$$

and

$$p_{\xi\zeta}^{\vec{M}_{\text{F}\rightleftharpoons\text{A}}\vec{J}^{\text{A}\rightleftharpoons\text{F}}} = -(1/2)\left\langle \vec{b}_{\xi}^{\vec{M}_{\text{F}\rightleftharpoons\text{A}}}, \mathcal{K}_0\left(\vec{b}_{\zeta}^{\vec{J}^{\text{A}\rightleftharpoons\text{F}}}\right)\right\rangle_{\mathbb{S}_{\text{F}\rightleftharpoons\text{A}}} \tag{7-40g}$$

$$p_{\xi\zeta}^{\vec{M}_{\text{F}\rightleftharpoons\text{A}}\vec{J}^{\text{F}}} = -(1/2)\left\langle \vec{b}_{\xi}^{\vec{M}_{\text{F}\rightleftharpoons\text{A}}}, \mathcal{K}_0\left(\vec{b}_{\zeta}^{\vec{J}^{\text{F}}}\right)\right\rangle_{\mathbb{S}_{\text{F}\rightleftharpoons\text{A}}} \tag{7-40h}$$

$$p_{\xi\zeta}^{\vec{M}_{\text{F}\rightleftharpoons\text{A}}\vec{J}_{\text{F}}} = -(1/2)\left\langle \vec{b}_{\xi}^{\vec{M}_{\text{F}\rightleftharpoons\text{A}}}, \mathcal{K}_0\left(\vec{b}_{\zeta}^{\vec{J}_{\text{F}}}\right)\right\rangle_{\mathbb{S}_{\text{F}\rightleftharpoons\text{A}}} \tag{7-40i}$$

$$p_{\xi\zeta}^{\vec{M}_{\text{F}\rightleftharpoons\text{A}}\vec{J}_{\text{F}\rightleftharpoons\text{A}}} = -(1/2)\left\langle \vec{b}_{\xi}^{\vec{M}_{\text{F}\rightleftharpoons\text{A}}}, \frac{1}{2}\vec{b}_{\zeta}^{\vec{J}_{\text{F}\rightleftharpoons\text{A}}} \times \hat{n}_{\rightarrow\text{PML}} + \text{P.V.}\mathcal{K}_0\left(-\vec{b}_{\zeta}^{\vec{J}_{\text{F}\rightleftharpoons\text{A}}}\right)\right\rangle_{\mathbb{S}_{\text{F}\rightleftharpoons\text{A}}} \tag{7-40j}$$

$$p_{\xi\zeta}^{\vec{M}_{\text{F}\rightleftharpoons\text{A}}\vec{M}^{\text{A}\rightleftharpoons\text{F}}} = -(1/2)\left\langle \vec{b}_{\xi}^{\vec{M}_{\text{F}\rightleftharpoons\text{A}}}, -j\omega\varepsilon_0\mathcal{L}_0\left(\vec{b}_{\zeta}^{\vec{M}^{\text{A}\rightleftharpoons\text{F}}}\right)\right\rangle_{\mathbb{S}_{\text{F}\rightleftharpoons\text{A}}} \tag{7-40k}$$

$$p_{\xi\zeta}^{\vec{M}_{\text{F}\rightleftharpoons\text{A}}\vec{M}_{\text{F}\rightleftharpoons\text{A}}} = -(1/2)\left\langle \vec{b}_{\xi}^{\vec{M}_{\text{F}\rightleftharpoons\text{A}}}, -j\omega\varepsilon_0\mathcal{L}_0\left(-\vec{b}_{\zeta}^{\vec{M}_{\text{F}\rightleftharpoons\text{A}}}\right)\right\rangle_{\mathbb{S}_{\text{F}\rightleftharpoons\text{A}}} \tag{7-40l}$$

To obtain the IPO defined on modal space, we substitute Eq. (7-31) into the above Eq. (7-38), and then have that

$$P_{\text{F}\rightleftharpoons\text{A}} = \bar{a}^{\dagger} \cdot \underbrace{\left(\overline{\overline{T}}^{\dagger} \cdot \overline{\overline{P}}_{\text{F}\rightleftharpoons\text{A}}^{\text{intAV}} \cdot \overline{\overline{T}}\right)}_{\overline{\overline{P}}_{\text{F}\rightleftharpoons\text{A}}^{\text{int}}} \cdot \bar{a} \tag{7-41}$$

where the superscript "int" is to emphasize that the matrix originates from discretizing the interaction form of IPO.

For the convenience of the following discussions, the Eqs. (7-36) and (7-41) are uniformly written as follows:

$$P_{\mathrm{F}\rightleftharpoons\mathrm{A}} \ = \ \bar{a}^{\dagger} \cdot \overline{\overline{P}}_{\mathrm{F}\rightleftharpoons\mathrm{A}} \cdot \bar{a} \tag{7-42}$$

where $\overline{\overline{P}}_{\mathrm{F}\rightleftharpoons\mathrm{A}} = \overline{\overline{P}}{}^{\mathrm{cur}}_{\mathrm{F}\rightleftharpoons\mathrm{A}} \,/\, \overline{\overline{P}}{}^{\mathrm{int}}_{\mathrm{F}\rightleftharpoons\mathrm{A}}$ .

## 7.3.4 Input-Power-Decoupled Modes

This subsection focuses on constructing the *input-power-decoupled modes* (*IP-DMs*) in the modal space of the rec-antenna shown in Figs. 7-1 and 7-2, and discusses some related topics.

### 7.2.4.1 Construction Method

The IP-DMs contained in modal space can be derived from solving the following *modal decoupling equation* (or simply called *decoupling equation*)

$$\overline{\overline{P}}{}^{-}_{\mathrm{F}\rightleftharpoons\mathrm{A}} \cdot \bar{\alpha}_{\xi} \ = \ \theta_{\xi} \, \overline{\overline{P}}{}^{+}_{\mathrm{F}\rightleftharpoons\mathrm{A}} \cdot \bar{\alpha}_{\xi} \tag{7-43}$$

defined on modal space. Here $\overline{\overline{P}}{}^{+}_{\mathrm{F}\rightleftharpoons\mathrm{A}}$ and $\overline{\overline{P}}{}^{-}_{\mathrm{F}\rightleftharpoons\mathrm{A}}$ are the *positive and negative Hermitian parts* obtained from the *Toeplitz's decomposition* for the $\overline{\overline{P}}_{\mathrm{F}\rightleftharpoons\mathrm{A}}$ given in Eq. (7-42).

If some derived modes $\{\bar{\alpha}_1, \bar{\alpha}_2, \cdots, \bar{\alpha}_d\}$ are *d*-order degenerate, then the following *Gram-Schmidt orthogonalization* process[46] is necessary.

$$\left.\begin{aligned} \bar{\alpha}_1 &= \bar{\alpha}_1{}' \\ \bar{\alpha}_2 - \chi_{12}\bar{\alpha}_1{}' &= \bar{\alpha}_2{}' \\ &\cdots \\ \bar{\alpha}_d - \cdots - \chi_{2d}\bar{\alpha}_2{}' - \chi_{1d}\bar{\alpha}_1{}' &= \bar{\alpha}_d{}' \end{aligned}\right\} \tag{7-44}$$

where the coefficients are calculated as follows:

$$\chi_{mn} \ = \ \frac{\left(\bar{\alpha}_m{}'\right)^{\dagger} \cdot \overline{\overline{P}}{}^{+}_{\mathrm{F}\rightleftharpoons\mathrm{A}} \cdot \bar{\alpha}_n}{\left(\bar{\alpha}_m{}'\right)^{\dagger} \cdot \overline{\overline{P}}{}^{+}_{\mathrm{F}\rightleftharpoons\mathrm{A}} \cdot \bar{\alpha}_m{}'} \tag{7-45}$$

The above-obtained new modes $\{\bar{\alpha}_1{}', \bar{\alpha}_2{}', \cdots, \bar{\alpha}_d{}'\}$ are input-power-decoupled with each other.

### 7.2.4.2 Modal Decoupling Relation and Parseval's Identity

The modal vectors constructed in the above subsection satisfy the following *modal decoupling relation*

$$\bar{\alpha}_{\xi}^{\dagger}\cdot\bar{\bar{P}}_{\text{F}\rightleftharpoons\text{A}}\cdot\bar{\alpha}_{\zeta} = \underbrace{\left[\text{Re}\left\{P_{\text{F}\rightleftharpoons\text{A}}^{\xi}\right\}+j\,\text{Im}\left\{P_{\text{F}\rightleftharpoons\text{A}}^{\xi}\right\}\right]}_{P_{\text{F}\rightleftharpoons\text{A}}^{\xi}}\delta_{\xi\zeta} \xrightarrow{\text{Re}\{P_{\text{F}\rightleftharpoons\text{A}}^{\xi}\}=1} \left(1+j\,\theta_{\xi}\right)\delta_{\xi\zeta} \quad (7\text{-}46)$$

where $P_{\text{F}\rightleftharpoons\text{A}}^{\xi}$ is *modal input power*. At the same time, we also have that

$$(1/2)\iint_{\mathbb{S}_{\text{F}\rightleftharpoons\text{A}}}\left(\vec{E}_{\zeta}\times\vec{H}_{\xi}^{\dagger}\right)\cdot\hat{n}_{\rightarrow\text{PML}}dS = \left(1+j\,\theta_{\xi}\right)\delta_{\xi\zeta} \quad (7\text{-}47)$$

and this implies that the IP-DMs don't have energy exchange in any integral period.

By employing the above decoupling relations, we have the following *Parseval's identity*

$$\sum_{\xi}\left|c_{\xi}\right|^{2} = (1/T)\int_{t_0}^{t_0+T}\left[\iint_{\mathbb{S}_{\text{F}\rightleftharpoons\text{A}}}\left(\vec{\mathcal{E}}\times\vec{\mathcal{H}}\right)\cdot\hat{n}_{\rightarrow\text{PML}}dS\right]dt \quad (7\text{-}48)$$

where $c_{\xi}$ is the *modal expansion coefficient* used in modal expansion formulation and can be explicitly calculated as follows:

$$c_{\xi} = \frac{-(1/2)\left\langle\vec{J}_{\text{F}\rightleftharpoons\text{A}}^{\xi},\vec{E}\right\rangle_{\mathbb{S}_{\text{F}\rightleftharpoons\text{A}}}}{1+j\,\theta_{\xi}} = \frac{-(1/2)\left\langle\vec{H},\vec{M}_{\text{F}\rightleftharpoons\text{A}}^{\xi}\right\rangle_{\mathbb{S}_{\text{F}\rightleftharpoons\text{A}}}}{1+j\,\theta_{\xi}} \quad (7\text{-}49)$$

where $\{\vec{E},\vec{H}\}$ are some previously known EM fields distributing on the port $\mathbb{S}_{\text{F}\rightleftharpoons\text{A}}$.

### 7.2.4.3 Modal Quantities

For quantitatively describing the modal features, the following modal quantities are usually used

$$\text{MS}_{\xi} = \frac{1}{\left|1+j\,\theta_{\xi}\right|} \quad (7\text{-}50)$$

which is called *modal significance* (*MS*), and

$$Z_{\text{F}\rightleftharpoons\text{A}}^{\xi} = \frac{P_{\text{F}\rightleftharpoons\text{A}}^{\xi}}{(1/2)\left\langle\vec{J}_{\text{F}\rightleftharpoons\text{A}}^{\xi},\vec{J}_{\text{F}\rightleftharpoons\text{A}}^{\xi}\right\rangle_{\mathbb{S}_{\text{F}\rightleftharpoons\text{A}}}} = \underbrace{\text{Re}\left\{Z_{\text{F}\rightleftharpoons\text{A}}^{\xi}\right\}}_{R_{\text{F}\rightleftharpoons\text{A}}^{\xi}}+j\,\underbrace{\text{Im}\left\{Z_{\text{F}\rightleftharpoons\text{A}}^{\xi}\right\}}_{X_{\text{F}\rightleftharpoons\text{A}}^{\xi}} \quad (7\text{-}51\text{a})$$

which is called *modal input impedance* (*MII*) because its dimension is Ohm, and

$$Y_{\text{F}\rightleftharpoons\text{A}}^{\xi} = \frac{P_{\text{F}\rightleftharpoons\text{A}}^{\xi}}{(1/2)\left\langle\vec{M}_{\text{F}\rightleftharpoons\text{A}}^{\xi},\vec{M}_{\text{F}\rightleftharpoons\text{A}}^{\xi}\right\rangle_{\mathbb{S}_{\text{F}\rightleftharpoons\text{A}}}} = \underbrace{\text{Re}\left\{Y_{\text{F}\rightleftharpoons\text{A}}^{\xi}\right\}}_{G_{\text{F}\rightleftharpoons\text{A}}^{\xi}}+j\,\underbrace{\text{Im}\left\{Y_{\text{F}\rightleftharpoons\text{A}}^{\xi}\right\}}_{B_{\text{F}\rightleftharpoons\text{A}}^{\xi}} \quad (7\text{-}51\text{b})$$

which is called *modal input admittance* (*MIA*) because its dimension is Siemens.

The above MS quantitatively depicts the *modal weight* in whole modal expansion formulation. The above MII and MIA quantitatively depict the allocation way for the energy carried by the mode.

## 7.3 IP-DMs of Metallic Receiving Antenna (Driven by an Arbitrary Transmitting System)

As a typical example, the receiving problem shown in the following Fig. 7-3 is focused on in this section, and the rec-antenna is a *metallic-waveguide-loaded metallic horn antenna*.

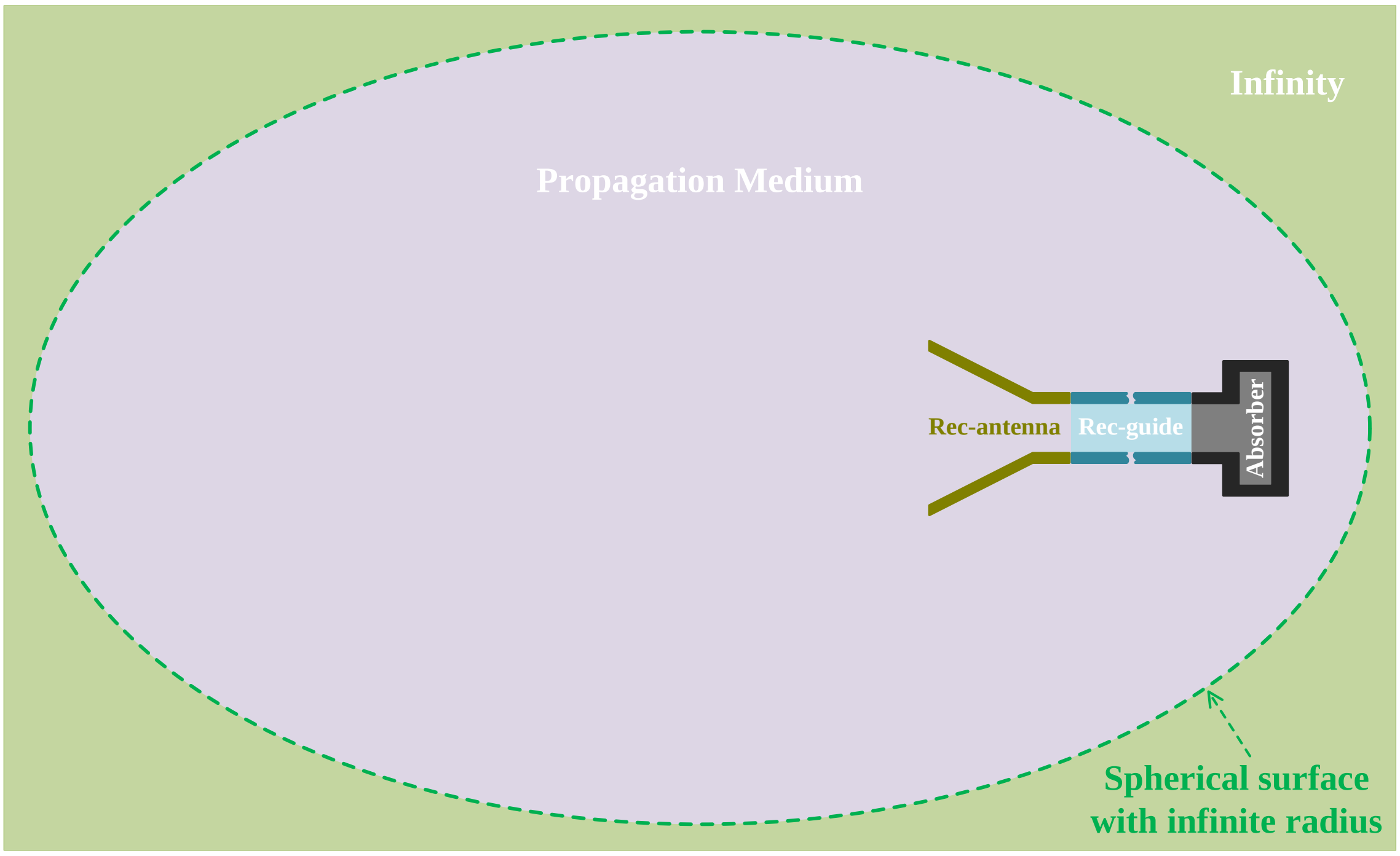

Figure 7-3 Geometry of the augmented rec-antenna considered in this section

The rec-antenna considered in this section is the same as the one discussed in the previous section. However, the rec-antenna considered in this section is driven by an arbitrary transmitting system, rather than the previous one driven by a specific transmitting system. Thus the results obtained in this section are more general than the ones obtained in the previous section.

The organization of this section is completely similar to the previous Sec. 7.2, i.e., "topological structure (Sec. 7.3.1) → source-field relationships (Sec. 7.3.1) → modal space (Sec. 7.3.2) → power transport theorem (Sec. 7.3.3) → input power operator (Sec. 7.3.3) → input-power-decoupled modes (Sec. 7.3.4)".

### 7.3.1 Topological Structure and Source-Field Relationships

The topological structure corresponding to the rec-antenna shown in the previous Fig. 7-3 is specifically illustrated in the following Fig. 7-4.

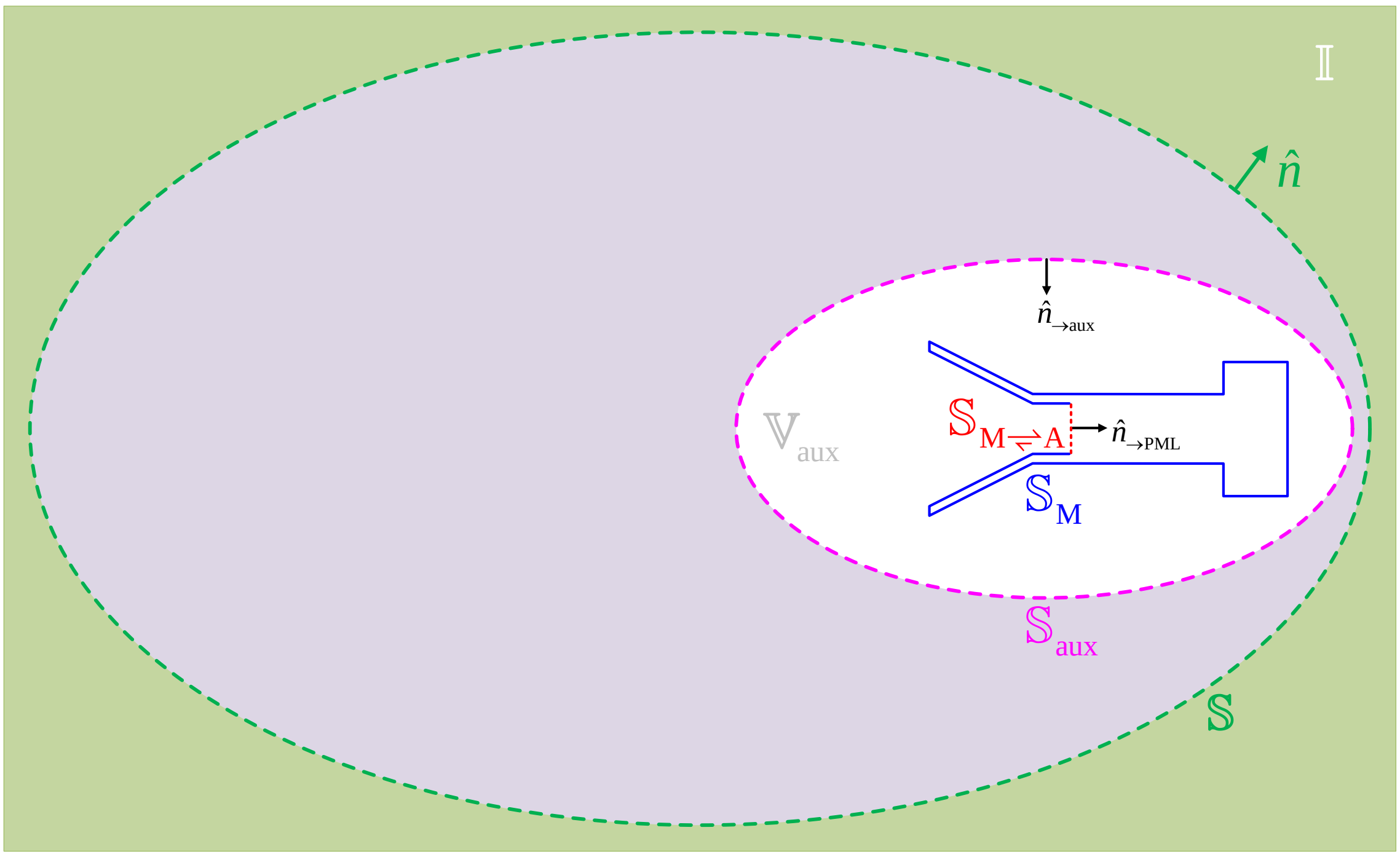

Figure 7-4 Topological structure of the rec-antenna shown in Fig. 7-3

In the above Fig. 7-4, $\mathbb{S}_{\text{aux}}$ is a finite *auxiliary surface* enclosing whole receiving system, and it is a penetrable surface. The region sandwiched by the auxiliary surface and receiving system is denoted as $\mathbb{V}_{\text{aux}}$, and there doesn't exist any source distributing in $\mathbb{V}_{\text{aux}}$ (i.e., the permeability and permittivity of $\mathbb{V}_{\text{aux}}$ are $\mu_0$ and $\varepsilon_0$ respectively). The impenetrable interface between $\mathbb{V}_{\text{aux}}$ and receiving system is just the impenetrable interface between propagation medium and receiving system, so it is denoted as $\mathbb{S}_{\text{M}}$.

The penetrable interface between $\mathbb{V}_{\text{aux}}$ and rec-antenna is just the input port of rec-antenna, and it is denoted as $\mathbb{S}_{\text{M}\rightleftharpoons\text{A}}$. In addition, for the receiving problem shown in Fig. 7-4, surface $\mathbb{S}_{\text{M}\rightleftharpoons\text{A}}$ is also the input port of PML.

Clearly, surfaces $\mathbb{S}_{\text{aux}}$, $\mathbb{S}_{\text{M}}$, and $\mathbb{S}_{\text{M}\rightleftharpoons\text{A}}$ constitute the whole boundary of $\mathbb{V}_{\text{aux}}$, i.e., $\partial\mathbb{V}_{\text{aux}} = \mathbb{S}_{\text{aux}} \cup \mathbb{S}_{\text{M}} \cup \mathbb{S}_{\text{M}\rightleftharpoons\text{A}}$.

If the equivalent surface currents distributing on $\mathbb{S}_{\text{aux}}$ are denoted as $\{\vec{J}_{\text{aux}}, \vec{M}_{\text{aux}}\}$, and the equivalent surface electric current distributing on $\mathbb{S}_{\text{M}}$ is denoted as $\vec{J}_{\text{M}}$ ①, and the equivalent surface currents distributing on $\mathbb{S}_{\text{M}\rightleftharpoons\text{A}}$ are denoted as $\{\vec{J}_{\text{M}\rightleftharpoons\text{A}}, \vec{M}_{\text{M}\rightleftharpoons\text{A}}\}$, then the field distributing on $\mathbb{V}_{\text{aux}}$ can be expressed as follows:

$$\vec{F}(\vec{r}) = \mathcal{F}_0\left(\vec{J}_{\text{aux}} + \vec{J}_{\text{M}} - \vec{J}_{\text{M}\rightleftharpoons\text{A}}, \vec{M}_{\text{aux}} - \vec{M}_{\text{M}\rightleftharpoons\text{A}}\right) \quad , \quad \vec{r} \in \mathbb{V}_{\text{aux}} \tag{7-52}$$

① The equivalent surface magnetic current distributing on $\mathbb{S}_{\text{M}}$ is zero, because of the homogeneous tangential electric field boundary condition on $\mathbb{S}_{\text{M}}$ [13].

where $\vec{F}=\vec{E}/\vec{H}$, and correspondingly $\mathcal{F}_0=\mathcal{E}_0/\mathcal{H}_0$, and the operators are the same as the ones used in the previous Sec. 7.2. The currents $\{\vec{J}_{\text{aux}},\vec{M}_{\text{aux}}\}$ and fields $\{\vec{E},\vec{H}\}$ in Eq. (7-52) satisfy the following relations

$$\hat{n}_{\to\text{aux}}\times\left[\vec{H}\left(\vec{r}_{\text{aux}}\right)\right]_{\vec{r}_{\text{aux}}\to\vec{r}} = \vec{J}_{\text{aux}}\left(\vec{r}\right) \quad , \quad \vec{r}\in\mathbb{S}_{\text{aux}} \tag{7-53a}$$

$$\left[\vec{E}\left(\vec{r}_{\text{aux}}\right)\right]_{\vec{r}_{\text{aux}}\to\vec{r}}\times\hat{n}_{\to\text{aux}} = \vec{M}_{\text{aux}}\left(\vec{r}\right) \quad , \quad \vec{r}\in\mathbb{S}_{\text{aux}} \tag{7-53b}$$

where point $\vec{r}_{\text{aux}}$ belongs to $\mathbb{V}_{\text{aux}}$ and approaches the point $\vec{r}$ on $\mathbb{S}_{\text{aux}}$, and $\hat{n}_{\to\text{aux}}$ is the normal direction of $\mathbb{S}_{\text{aux}}$ and points to the interior of $\mathbb{V}_{\text{aux}}$ as shown in Fig. 7-4. The $\{\vec{J}_{\text{M}\rightleftharpoons\text{A}},\vec{M}_{\text{M}\rightleftharpoons\text{A}}\}$ and $\{\vec{E},\vec{H}\}$ in Eq. (7-52) satisfy the following relations

$$\hat{n}_{\to\text{PML}}\times\left[\vec{H}\left(\vec{r}_{\text{aux}}\right)\right]_{\vec{r}_{\text{aux}}\to\vec{r}} = \vec{J}_{\text{M}\rightleftharpoons\text{A}}\left(\vec{r}\right) \quad , \quad \vec{r}\in\mathbb{S}_{\text{M}\rightleftharpoons\text{A}} \tag{7-54a}$$

$$\left[\vec{E}\left(\vec{r}_{\text{aux}}\right)\right]_{\vec{r}_{\text{aux}}\to\vec{r}}\times\hat{n}_{\to\text{PML}} = \vec{M}_{\text{M}\rightleftharpoons\text{A}}\left(\vec{r}\right) \quad , \quad \vec{r}\in\mathbb{S}_{\text{M}\rightleftharpoons\text{A}} \tag{7-54b}$$

where point $\vec{r}_{\text{aux}}$ belongs to $\mathbb{V}_{\text{aux}}$ and approaches the point $\vec{r}$ on $\mathbb{S}_{\text{M}\rightleftharpoons\text{A}}$, and $\hat{n}_{\to\text{PML}}$ is the normal direction of $\mathbb{S}_{\text{M}\rightleftharpoons\text{A}}$ and points to the interior of loading structure as shown in Fig. 7-4 (here, the loading structure is a PML).

### 7.3.2 Mathematical Description for Modal Space

Substituting Eq. (7-52) into Eqs. (7-53a) and (7-53b), we immediately obtain the following integral equations

$$\left[\mathcal{H}_0\left(\vec{J}_{\text{aux}}+\vec{J}_{\text{M}}-\vec{J}_{\text{M}\rightleftharpoons\text{A}},\vec{M}_{\text{aux}}-\vec{M}_{\text{M}\rightleftharpoons\text{A}}\right)\right]_{\vec{r}_{\text{aux}}\to\vec{r}}^{\tan} = \vec{J}_{\text{aux}}\left(\vec{r}\right)\times\hat{n}_{\to\text{aux}} \quad , \ \vec{r}\in\mathbb{S}_{\text{aux}} \tag{7-55a}$$

$$\left[\mathcal{E}_0\left(\vec{J}_{\text{aux}}+\vec{J}_{\text{M}}-\vec{J}_{\text{M}\rightleftharpoons\text{A}},\vec{M}_{\text{aux}}-\vec{M}_{\text{M}\rightleftharpoons\text{A}}\right)\right]_{\vec{r}_{\text{aux}}\to\vec{r}}^{\tan} = \hat{n}_{\to\text{aux}}\times\vec{M}_{\text{aux}}\left(\vec{r}\right) , \ \vec{r}\in\mathbb{S}_{\text{aux}} \tag{7-55b}$$

about currents $\{\vec{J}_{\text{aux}},\vec{M}_{\text{aux}}\}$, $\{\vec{J}_{\text{M}\rightleftharpoons\text{A}},\vec{M}_{\text{M}\rightleftharpoons\text{A}}\}$, and $\vec{J}_{\text{M}}$, where the superscript "tan" represents the tangential component of the field.

Based on Eq. (7-52) and the homogeneous tangential electric field boundary condition on $\mathbb{S}_{\text{M}}$, we have the following electric field integral equation

$$\left[\mathcal{E}_0\left(\vec{J}_{\text{aux}}+\vec{J}_{\text{M}}-\vec{J}_{\text{M}\rightleftharpoons\text{A}},\vec{M}_{\text{aux}}-\vec{M}_{\text{M}\rightleftharpoons\text{A}}\right)\right]_{\vec{r}_{\text{aux}}\to\vec{r}}^{\tan} = 0 \quad , \quad \vec{r}\in\mathbb{S}_{\text{M}} \tag{7-56}$$

about currents $\{\vec{J}_{\text{aux}},\vec{M}_{\text{aux}}\}$, $\{\vec{J}_{\text{M}\rightleftharpoons\text{A}},\vec{M}_{\text{M}\rightleftharpoons\text{A}}\}$, and $\vec{J}_{\text{M}}$.

Similarly to the previous Sec. 7.2, we have the following integral equations

$$\begin{aligned}&\left[\mathcal{E}_0\left(\vec{J}_{\text{aux}}+\vec{J}_{\text{M}}-\vec{J}_{\text{M}\rightleftharpoons\text{A}},\vec{M}_{\text{aux}}-\vec{M}_{\text{M}\rightleftharpoons\text{A}}\right)\right]_{\vec{r}_{\text{aux}}\to\vec{r}}^{\tan}\\ &=\left[\mathcal{E}_0\left(\vec{J}_{\text{M}\rightleftharpoons\text{A}},\vec{M}_{\text{M}\rightleftharpoons\text{A}}\right)\right]_{\vec{r}_{\text{PML}}\to\vec{r}}^{\tan} \quad , \quad \vec{r}\in\mathbb{S}_{\text{M}\rightleftharpoons\text{A}}\end{aligned} \tag{7-57a}$$

$$\left[\mathcal{H}_0\left(\vec{J}_{\text{aux}}+\vec{J}_{\text{M}}-\vec{J}_{\text{M}\rightleftharpoons\text{A}},\vec{M}_{\text{aux}}-\vec{M}_{\text{M}\rightleftharpoons\text{A}}\right)\right]_{\vec{r}_{\text{aux}}\to\vec{r}}^{\tan}$$
$$=\left[\mathcal{H}_0\left(\vec{J}_{\text{M}\rightleftharpoons\text{A}},\vec{M}_{\text{M}\rightleftharpoons\text{A}}\right)\right]_{\vec{r}_{\text{PML}}\to\vec{r}}^{\tan}\quad,\qquad \vec{r}\in\mathbb{S}_{\text{M}\rightleftharpoons\text{A}}\qquad(7\text{-}57\text{b})$$

about currents $\{\vec{J}_{\text{aux}},\vec{M}_{\text{aux}}\}$, $\{\vec{J}_{\text{M}\rightleftharpoons\text{A}},\vec{M}_{\text{M}\rightleftharpoons\text{A}}\}$, and $\vec{J}_{\text{M}}$, where $\vec{r}_{\text{PML}}$ belongs to the interior of PML and approaches the $\vec{r}$ on $\mathbb{S}_{\text{M}\rightleftharpoons\text{A}}$.

The above Eqs. (6-55a)~(6-57b) are a complete mathematical description for the modal space of the rec-antenna shown in Fig. 7-4. If the currents $\{\vec{J}_{\text{aux}},\vec{M}_{\text{aux}}\}$, $\vec{J}_{\text{M}}$, and $\{\vec{J}_{\text{M}\rightleftharpoons\text{A}},\vec{M}_{\text{M}\rightleftharpoons\text{A}}\}$ are expanded in terms of some proper basis functions and (6-55a)~(6-57b) are tested with $\{\vec{b}_\xi^{\vec{M}_{\text{aux}}}\}$, $\{\vec{b}_\xi^{\vec{J}_{\text{aux}}}\}$, $\{\vec{b}_\xi^{\vec{J}_{\text{M}}}\}$, $\{\vec{b}_\xi^{\vec{J}_{\text{M}\rightleftharpoons\text{A}}}\}$, and $\{\vec{b}_\xi^{\vec{M}_{\text{M}\rightleftharpoons\text{A}}}\}$ respectively, then the integral equations are immediately discretized into the following matrix equations

$$\overline{\overline{Z}}^{\vec{M}_{\text{aux}}\vec{J}_{\text{aux}}}\cdot\overline{a}^{\vec{J}_{\text{aux}}}+\overline{\overline{Z}}^{\vec{M}_{\text{aux}}\vec{J}_{\text{M}}}\cdot\overline{a}^{\vec{J}_{\text{M}}}+\overline{\overline{Z}}^{\vec{M}_{\text{aux}}\vec{J}_{\text{M}\rightleftharpoons\text{A}}}\cdot\overline{a}^{\vec{J}_{\text{M}\rightleftharpoons\text{A}}}+\overline{\overline{Z}}^{\vec{M}_{\text{aux}}\vec{M}_{\text{aux}}}\cdot\overline{a}^{\vec{M}_{\text{aux}}}+\overline{\overline{Z}}^{\vec{M}_{\text{aux}}\vec{M}_{\text{M}\rightleftharpoons\text{A}}}\cdot\overline{a}^{\vec{M}_{\text{M}\rightleftharpoons\text{A}}}$$
$$=0\qquad(7\text{-}58\text{a})$$
$$\overline{\overline{Z}}^{\vec{J}_{\text{aux}}\vec{J}_{\text{aux}}}\cdot\overline{a}^{\vec{J}_{\text{aux}}}+\overline{\overline{Z}}^{\vec{J}_{\text{aux}}\vec{J}_{\text{M}}}\cdot\overline{a}^{\vec{J}_{\text{M}}}+\overline{\overline{Z}}^{\vec{J}_{\text{aux}}\vec{J}_{\text{M}\rightleftharpoons\text{A}}}\cdot\overline{a}^{\vec{J}_{\text{M}\rightleftharpoons\text{A}}}+\overline{\overline{Z}}^{\vec{J}_{\text{aux}}\vec{M}_{\text{aux}}}\cdot\overline{a}^{\vec{M}_{\text{aux}}}+\overline{\overline{Z}}^{\vec{J}_{\text{aux}}\vec{M}_{\text{M}\rightleftharpoons\text{A}}}\cdot\overline{a}^{\vec{M}_{\text{M}\rightleftharpoons\text{A}}}$$
$$=0\qquad(7\text{-}58\text{b})$$

and

$$\overline{\overline{Z}}^{\vec{J}_{\text{M}}\vec{J}_{\text{aux}}}\cdot\overline{a}^{\vec{J}_{\text{aux}}}+\overline{\overline{Z}}^{\vec{J}_{\text{M}}\vec{J}_{\text{M}}}\cdot\overline{a}^{\vec{J}_{\text{M}}}+\overline{\overline{Z}}^{\vec{J}_{\text{M}}\vec{J}_{\text{M}\rightleftharpoons\text{A}}}\cdot\overline{a}^{\vec{J}_{\text{M}\rightleftharpoons\text{A}}}+\overline{\overline{Z}}^{\vec{J}_{\text{M}}\vec{M}_{\text{aux}}}\cdot\overline{a}^{\vec{M}_{\text{aux}}}+\overline{\overline{Z}}^{\vec{J}_{\text{M}}\vec{M}_{\text{M}\rightleftharpoons\text{A}}}\cdot\overline{a}^{\vec{M}_{\text{M}\rightleftharpoons\text{A}}}=0\qquad(7\text{-}59)$$

and

$$\overline{\overline{Z}}^{\vec{J}_{\text{M}\rightleftharpoons\text{A}}\vec{J}_{\text{aux}}}\cdot\overline{a}^{\vec{J}_{\text{aux}}}+\overline{\overline{Z}}^{\vec{J}_{\text{M}\rightleftharpoons\text{A}}\vec{J}_{\text{M}}}\cdot\overline{a}^{\vec{J}_{\text{M}}}+\overline{\overline{Z}}^{\vec{J}_{\text{M}\rightleftharpoons\text{A}}\vec{J}_{\text{M}\rightleftharpoons\text{A}}}\cdot\overline{a}^{\vec{J}_{\text{M}\rightleftharpoons\text{A}}}+\overline{\overline{Z}}^{\vec{J}_{\text{M}\rightleftharpoons\text{A}}\vec{M}_{\text{aux}}}\cdot\overline{a}^{\vec{M}_{\text{aux}}}+\overline{\overline{Z}}^{\vec{J}_{\text{M}\rightleftharpoons\text{A}}\vec{M}_{\text{M}\rightleftharpoons\text{A}}}\cdot\overline{a}^{\vec{M}_{\text{M}\rightleftharpoons\text{A}}}$$
$$=0\qquad(7\text{-}60\text{a})$$
$$\overline{\overline{Z}}^{\vec{M}_{\text{M}\rightleftharpoons\text{A}}\vec{J}_{\text{aux}}}\cdot\overline{a}^{\vec{J}_{\text{aux}}}+\overline{\overline{Z}}^{\vec{M}_{\text{M}\rightleftharpoons\text{A}}\vec{J}_{\text{M}}}\cdot\overline{a}^{\vec{J}_{\text{M}}}+\overline{\overline{Z}}^{\vec{M}_{\text{M}\rightleftharpoons\text{A}}\vec{J}_{\text{M}\rightleftharpoons\text{A}}}\cdot\overline{a}^{\vec{J}_{\text{M}\rightleftharpoons\text{A}}}+\overline{\overline{Z}}^{\vec{M}_{\text{M}\rightleftharpoons\text{A}}\vec{M}_{\text{aux}}}\cdot\overline{a}^{\vec{M}_{\text{aux}}}+\overline{\overline{Z}}^{\vec{M}_{\text{M}\rightleftharpoons\text{A}}\vec{M}_{\text{M}\rightleftharpoons\text{A}}}\cdot\overline{a}^{\vec{M}_{\text{M}\rightleftharpoons\text{A}}}$$
$$=0\qquad(7\text{-}60\text{b})$$

The formulations used to calculate the elements of the matrices in the above matrix equations are similar to the formulations given in Eqs. (7-13a)~(7-18f), and they are explicitly listed in the App. D8 of this report. By employing the above matrix equations, we propose two different schemes for mathematically depicting modal space as below.

Employing the above Eqs. (7-58a)~(7-60b), we can obtain the following transformation

$$\begin{bmatrix}\overline{a}^{\vec{J}_{\text{aux}}}\\\overline{a}^{\vec{J}_{\text{M}}}\\\overline{a}^{\vec{J}_{\text{M}\rightleftharpoons\text{A}}}\\\overline{a}^{\vec{M}_{\text{aux}}}\\\overline{a}^{\vec{M}_{\text{M}\rightleftharpoons\text{A}}}\end{bmatrix}=\overline{a}^{\text{AV}}=\overline{\overline{T}}\cdot\overline{a}\qquad(7\text{-}61)$$

and the calculation formulation for transformation matrix $\overline{\overline{T}}$ is given in the App. D8 of this report.

### 7.3.3 Power Transport Theorem and Input Power Operator

Applying the results obtained in Chap. 2 to the rec-antenna shown in Fig. 7-4, we immediately have the following PTT for the rec-antenna

$$P_{\text{aux}\rightleftharpoons\text{V}} = P_{\text{M}\rightleftharpoons\text{A}} + j\, P_{\text{V}}^{\text{sto}} \tag{7-62}$$

where

$$P_{\text{M}\rightleftharpoons\text{A}} \ = \ (1/2)\iint_{\mathbb{S}_{\text{M}\rightleftharpoons\text{A}}} \left(\vec{E}\times\vec{H}^{\dagger}\right)\cdot\hat{n}_{\rightarrow\text{PML}}dS \tag{7-63}$$

called the input power inputted into rec-antenna, where $\hat{n}_{\rightarrow\text{PML}}$ is the normal direction of $\mathbb{S}_{\text{M}\rightleftharpoons\text{A}}$ and points to the interior of PML as shown in Fig. 7-4.

Based on Eqs. (7-54a)&(7-54b) and the tangential continuity of the $\{\vec{E},\vec{H}\}$ on $\mathbb{S}_{\text{M}\rightleftharpoons\text{A}}$, the IPO $P_{\text{M}\rightleftharpoons\text{A}}$ given in Eq. (7-63) can be alternatively written as follows:

$$P_{\text{M}\rightleftharpoons\text{A}} \ = \ (1/2)\left\langle \hat{n}_{\rightarrow\text{PML}}\times\vec{J}_{\text{M}\rightleftharpoons\text{A}},\vec{M}_{\text{M}\rightleftharpoons\text{A}}\right\rangle_{\mathbb{S}_{\text{M}\rightleftharpoons\text{A}}} \tag{7-64}$$

or alternatively as follows:

$$\begin{aligned} P_{\text{M}\rightleftharpoons\text{A}} &= -(1/2)\left\langle \vec{J}_{\text{M}\rightleftharpoons\text{A}},\vec{E}\right\rangle_{\mathbb{S}_{\text{M}\rightleftharpoons\text{A}}} \\ &= -(1/2)\left\langle \vec{J}_{\text{M}\rightleftharpoons\text{A}},\mathcal{E}_0\left(\vec{J}_{\text{aux}}+\vec{J}_{\text{M}}-\vec{J}_{\text{M}\rightleftharpoons\text{A}},\vec{M}_{\text{aux}}-\vec{M}_{\text{M}\rightleftharpoons\text{A}}\right)\right\rangle_{\tilde{\mathbb{S}}_{\text{M}\rightleftharpoons\text{A}}} \end{aligned} \tag{7-65a}$$

$$\begin{aligned} P_{\text{M}\rightleftharpoons\text{A}} &= -(1/2)\left\langle \vec{M}_{\text{M}\rightleftharpoons\text{A}},\vec{H}\right\rangle^{\dagger}_{\mathbb{S}_{\text{M}\rightleftharpoons\text{A}}} \\ &= -(1/2)\left\langle \vec{M}_{\text{M}\rightleftharpoons\text{A}},\mathcal{H}_0\left(\vec{J}_{\text{aux}}+\vec{J}_{\text{M}}-\vec{J}_{\text{M}\rightleftharpoons\text{A}},\vec{M}_{\text{aux}}-\vec{M}_{\text{M}\rightleftharpoons\text{A}}\right)\right\rangle^{\dagger}_{\tilde{\mathbb{S}}_{\text{M}\rightleftharpoons\text{A}}} \end{aligned} \tag{7-65b}$$

Here, the Eq. (7-64) is the current form of IPO, and the Eq. (7-65a) & (7-65b) are the JE & HM interaction forms of IPO.

By discretizing IPOs (7-64)&(7-65) and utilizing transformation (7-61), we derive the following matrix form of the IPO

$$P_{\text{M}\rightleftharpoons\text{A}} \ = \ \bar{a}^{\dagger}\cdot\overline{\overline{P}}_{\text{M}\rightleftharpoons\text{A}}\cdot\bar{a} \tag{7-66}$$

and the formulation for calculating the quadratic matrix $P_{\text{M}\rightleftharpoons\text{A}}$ is given in the App. D8 of this report.

### 7.3.4 Input-Power-Decoupled Modes

The IP-DMs in modal space can be derived from solving the modal decoupling equation $\overline{\overline{P}}_{\text{M}\rightleftharpoons\text{A}}^{-} \cdot \bar{\alpha}_{\xi} = \theta_{\xi} \overline{\overline{P}}_{\text{M}\rightleftharpoons\text{A}}^{+} \cdot \bar{\alpha}_{\xi}$ defined on modal space, where $\overline{\overline{P}}_{\text{M}\rightleftharpoons\text{A}}^{+}$ and $\overline{\overline{P}}_{\text{M}\rightleftharpoons\text{A}}^{-}$ are the positive and negative Hermitian parts of matrix $\overline{\overline{P}}_{\text{M}\rightleftharpoons\text{A}}$. If some derived modes $\{\bar{\alpha}_1, \bar{\alpha}_2, \cdots, \bar{\alpha}_d\}$ are $d$-order degenerate, then the Gram-Schmidt orthogonalization process given in previous Sec. 7.2.4.1 is necessary, and it is not repeated here.

The modal vectors constructed above satisfy the following decoupling relation

$$(1/2)\iint_{\mathbb{S}_{\text{M}\rightleftharpoons\text{A}}} \left(\vec{E}_{\zeta} \times \vec{H}_{\xi}^{\dagger}\right) \cdot \hat{n}_{\rightarrow\text{PML}} dS \; = \; \left(1 + j\,\theta_{\xi}\right)\delta_{\xi\zeta} \tag{7-67}$$

and the relation implies that **the IP-DMs don't have net energy coupling in one period**. By employing the decoupling relation, we have the following Parseval's identity

$$\sum_{\xi} \left|c_{\xi}\right|^2 \; = \; \left(1/T\right)\int_{t_0}^{t_0+T} \left[\iint_{\mathbb{S}_{\text{M}\rightleftharpoons\text{A}}} \left(\vec{\mathcal{E}} \times \vec{\mathcal{H}}\right) \cdot \hat{n}_{\rightarrow\text{PML}} dS\right] dt \tag{7-68}$$

in which $\{\vec{\mathcal{E}}, \vec{\mathcal{H}}\}$ are the time-domain fields, and the expansion coefficients $c_{\xi}$ have expression $c_{\xi} = -(1/2) < \vec{J}_{\text{M}\rightleftharpoons\text{A}}^{\xi}, \vec{E} >_{\mathbb{S}_{\text{M}\rightleftharpoons\text{A}}} \big/ (1 + j\,\theta_{\xi}) = -(1/2) < \vec{H}, \vec{M}_{\text{M}\rightleftharpoons\text{A}}^{\xi} >_{\mathbb{S}_{\text{M}\rightleftharpoons\text{A}}} \big/ (1 + j\,\theta_{\xi})$, where $\{\vec{E}, \vec{H}\}$ is a previously known field distributing on input port $\mathbb{S}_{\text{M}\rightleftharpoons\text{A}}$.

Just like the metallic rec-antenna discussed in Sec. 7.2.4.3, we can also define the modal significance $\text{MS}_{\xi} = 1 \big/ |1 + j\,\theta_{\xi}|$, modal input impedance $Z_{\text{M}\rightleftharpoons\text{A}}^{\xi} = P_{\text{M}\rightleftharpoons\text{A}}^{\xi} \big/ (1/2) < \vec{J}_{\text{M}\rightleftharpoons\text{A}}^{\xi}, \vec{J}_{\text{M}\rightleftharpoons\text{A}}^{\xi} >_{\mathbb{S}_{\text{M}\rightleftharpoons\text{A}}}$, and modal input admittance $Y_{\text{M}\rightleftharpoons\text{A}}^{\xi} = P_{\text{M}\rightleftharpoons\text{A}}^{\xi} \big/ (1/2) < \vec{M}_{\text{M}\rightleftharpoons\text{A}}^{\xi}, \vec{M}_{\text{M}\rightleftharpoons\text{A}}^{\xi} >_{\mathbb{S}_{\text{M}\rightleftharpoons\text{A}}}$ to quantitatively describe the modal features of the rec-antenna shown in Fig. 7-4.

### 7.3.5 Numerical Examples Corresponding to Typical Structures

Theoretically, the generating operator of IP-DMs should be selected as $P_{\text{M}\rightleftharpoons\text{A}}$ just like the previous Sec. 7.3.4 does. However, we find out that: the numerical performance of IPO $P_{\text{M}\rightleftharpoons\text{A}}$ is not desired in the aspect of generating IP-DMs, because the corresponding matrix $\overline{\overline{P}}_{\text{M}\rightleftharpoons\text{A}}^{+}$ are not positive definite. As a compromise scheme, we alternatively select the operator $P_{\text{aux}\rightleftharpoons\text{V}}$ appearing in Eq. (7-62) as the generaing operator for the IP-DMs of rec-antenna.

In this subsection, we consider a metallic horn rec-antenna driven by an auxiliary spherical surface. The geometries of the metallic receiving horn and auxiliary driving surface are shown in the following Fig. 7-5. The metallic receiving horn is the one with 1/4 size of the metallic feeding horn considered in the previous Sec. 6.2.5.3. The auxiliary driving surface is with radius 2cm.

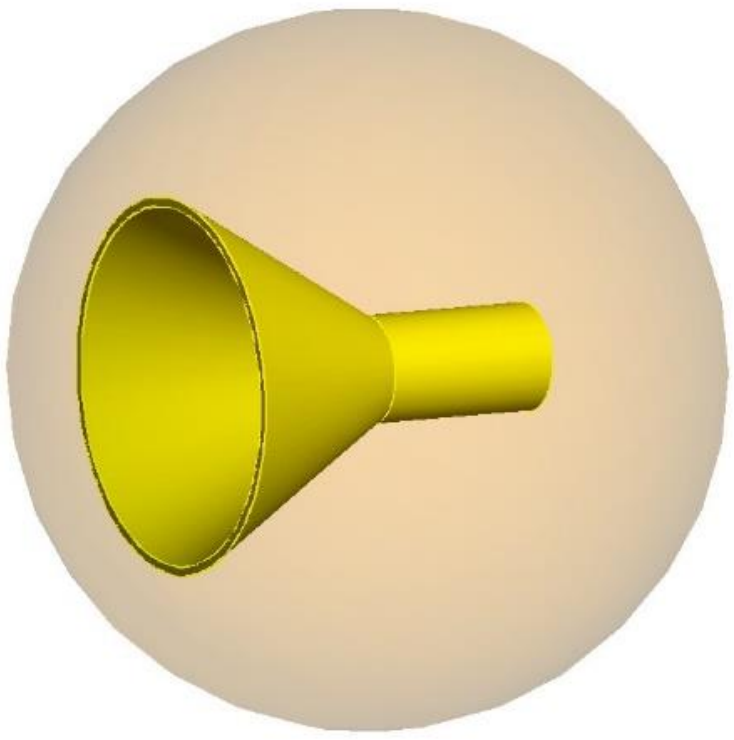

Figure 7-5 Geometry of a metallic receiving horn driven by a spherical surface

The topological structure and surface triangular meshes of the input port, receiving horn, and output port are shown in the following Fig. 7-6.

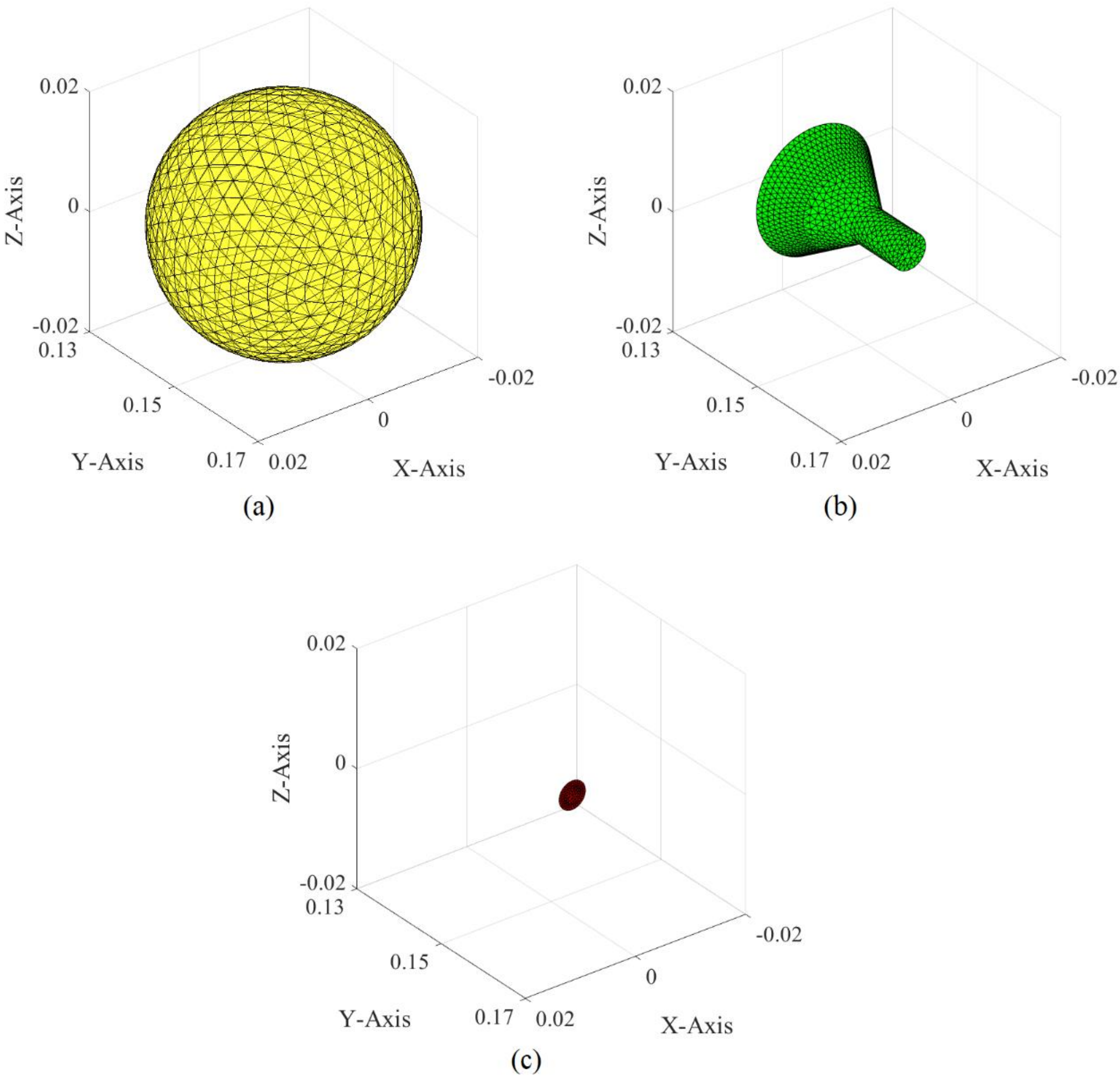

Figure 7-6 Topological structure and surface triangular meshes of (a) input port, (b) receiving horn, and (c) output port

Using the JE-DoJ-based and HM-DoM-based formulations of IPO, we calculate the IP-DMs of the metallic receiving horn. The input resistance curves of the first several typical IP-DMs are plotted in the following Fig. 7-7.

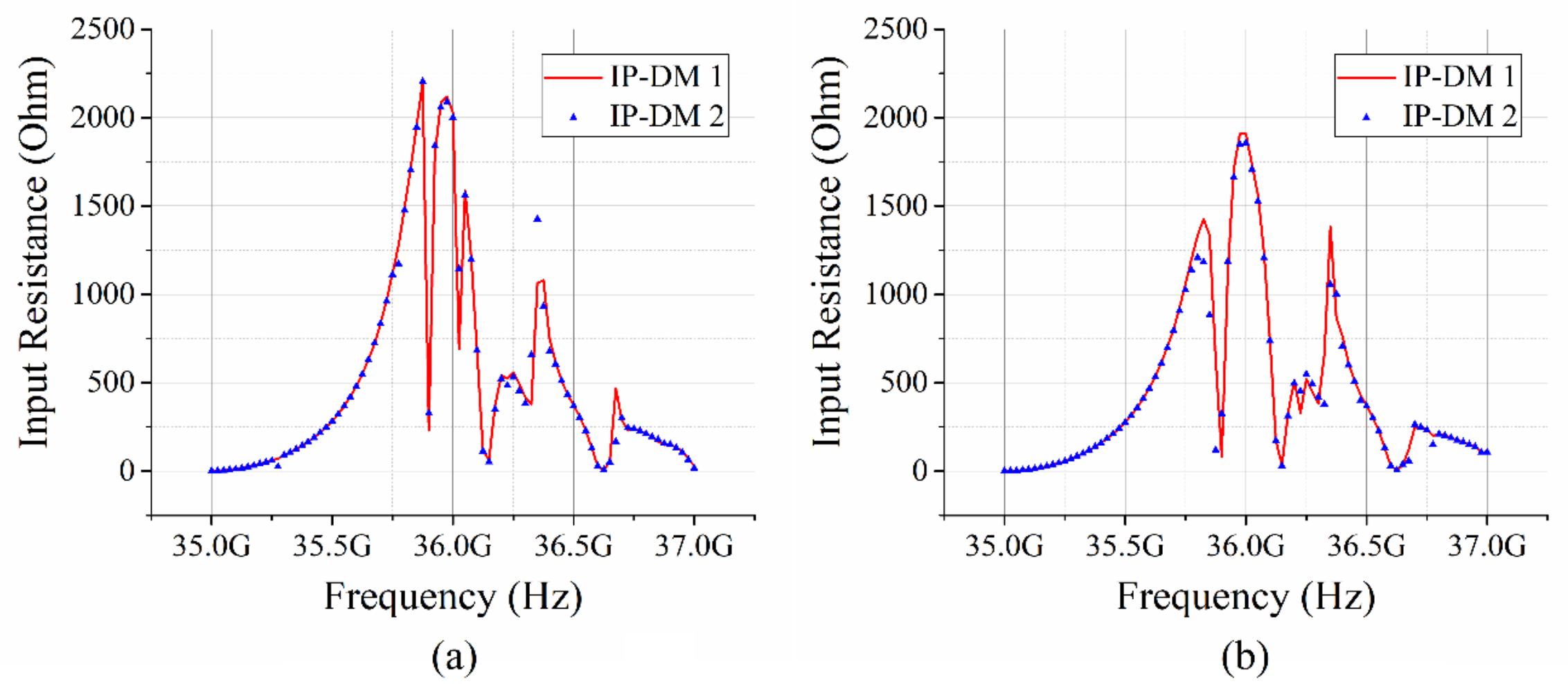

Figure 7-7 Some typical modal input resistance curves calculated from the (a) JE-DoJ-based and (b) HM-DoM-based formulations of IPO

Taking the IP-DM 1 shown in Fig. 7-7(a) as an example, its modal electric and magnetic currents distributing on input port are illustrated in the following Fig. 7-8.

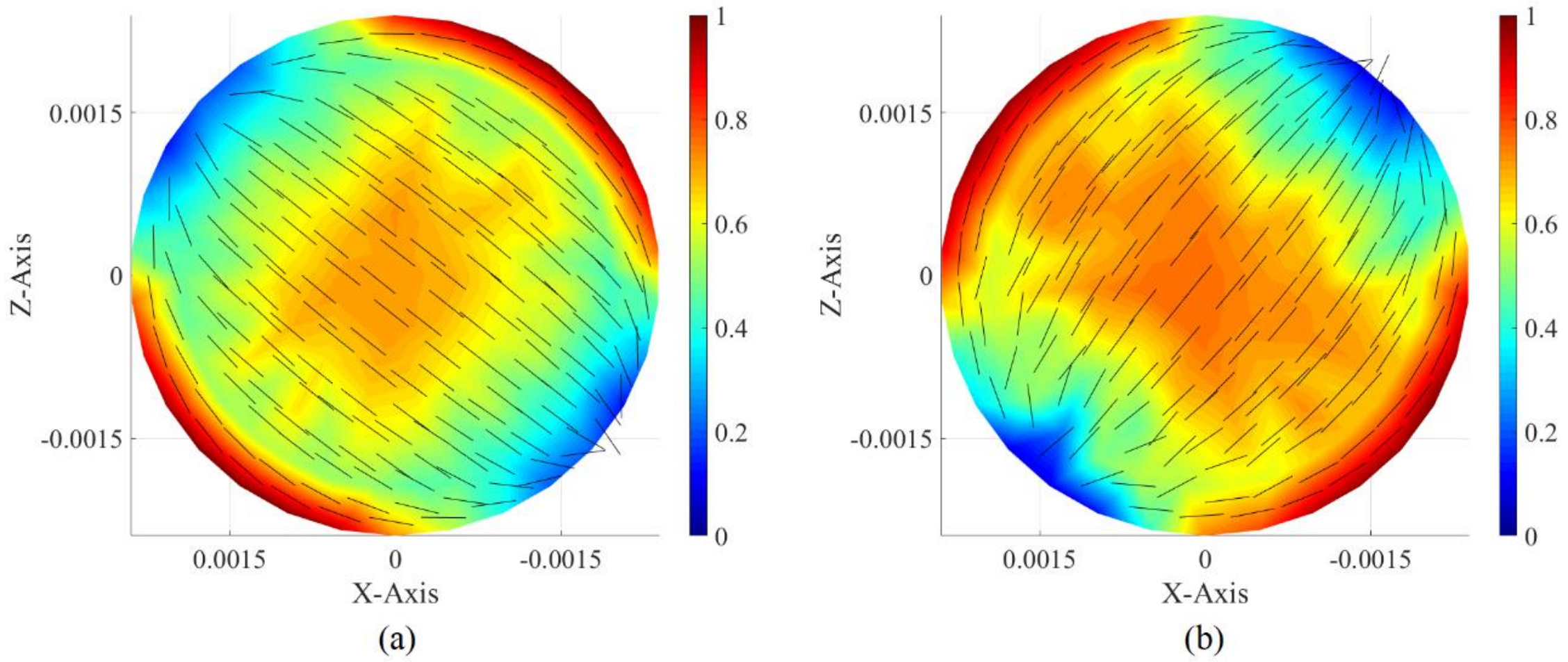

Figure 7-8 Modal (a) electric and (b) magnetic currents distributing on input port

For the above-mentioned mode working at 35.875 GHz and 35.975 GHz, we plot their modal electric fields with a series of time points $t = \{ 0.10T, 0.20T, 0.30T, 0.40T, 0.50T \}$ in the following Fig. 7-9, where $T$ is the time period of the time-harmonic field.

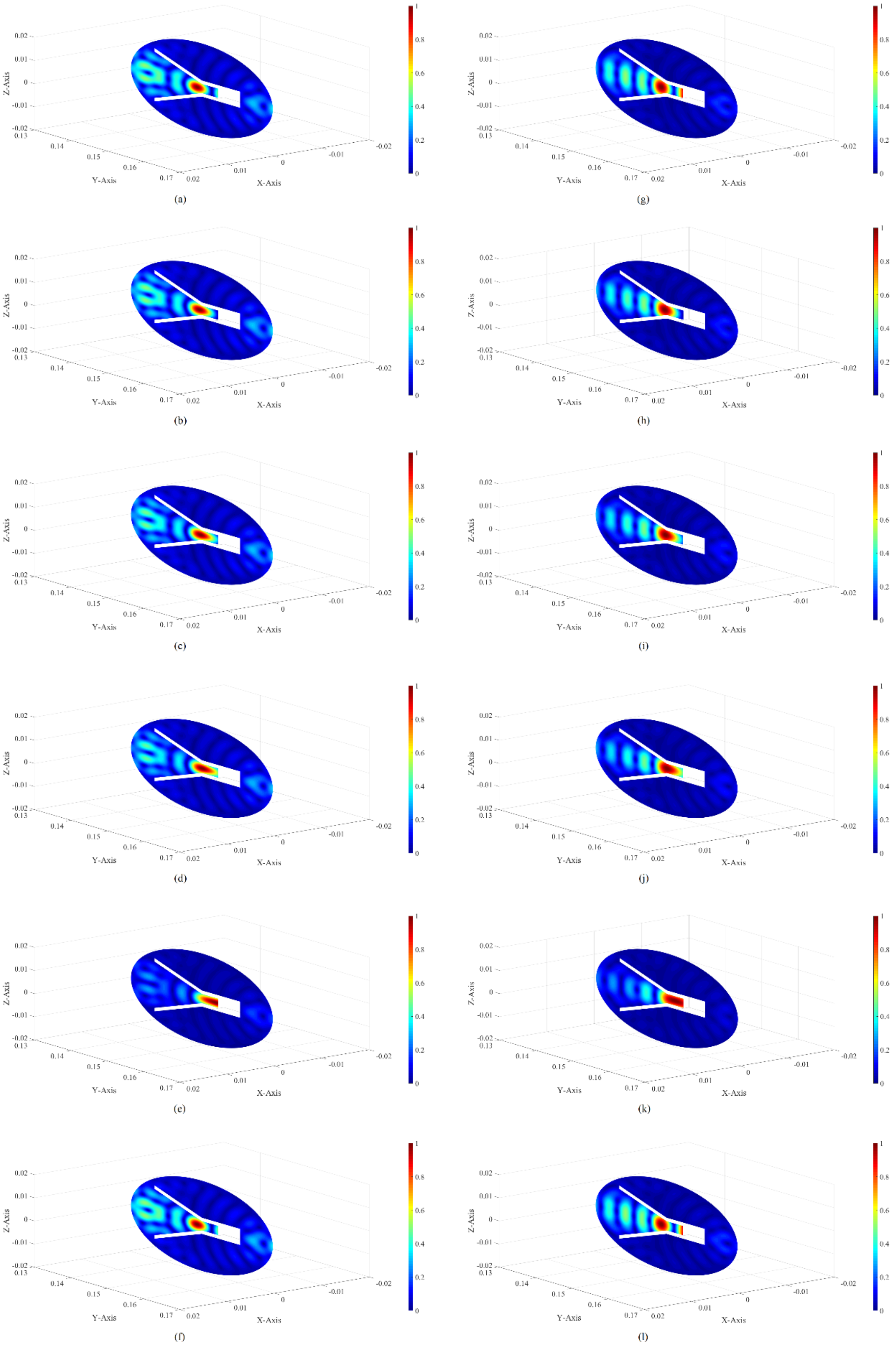

Figure 7-9 Modal electric field working at 35.875 GHz with time points (a) 0.0*T*, (b) 0.1*T*, (c) 0.2*T*, (d) 0.3*T*, (e) 0.4*T*, and (f) 0.5*T*. Modal electric field working at 35.975 GHz with time points (g) 0.0*T*, (h) 0.1*T*, (i) 0.2*T*, (j) 0.3*T*, (k) 0.4*T*, and (l) 0.5*T*

## 7.4 IP-DMs of Metal-Material Composite Receiving Antenna (Driven by an Arbitrary Transmitting System)

As a typical example, the receiving problem shown in Fig. 7-10 is focused on in this section, and the rec-antenna is a *coaxial-loaded metamaterial-inspired stacked printed antenna*.

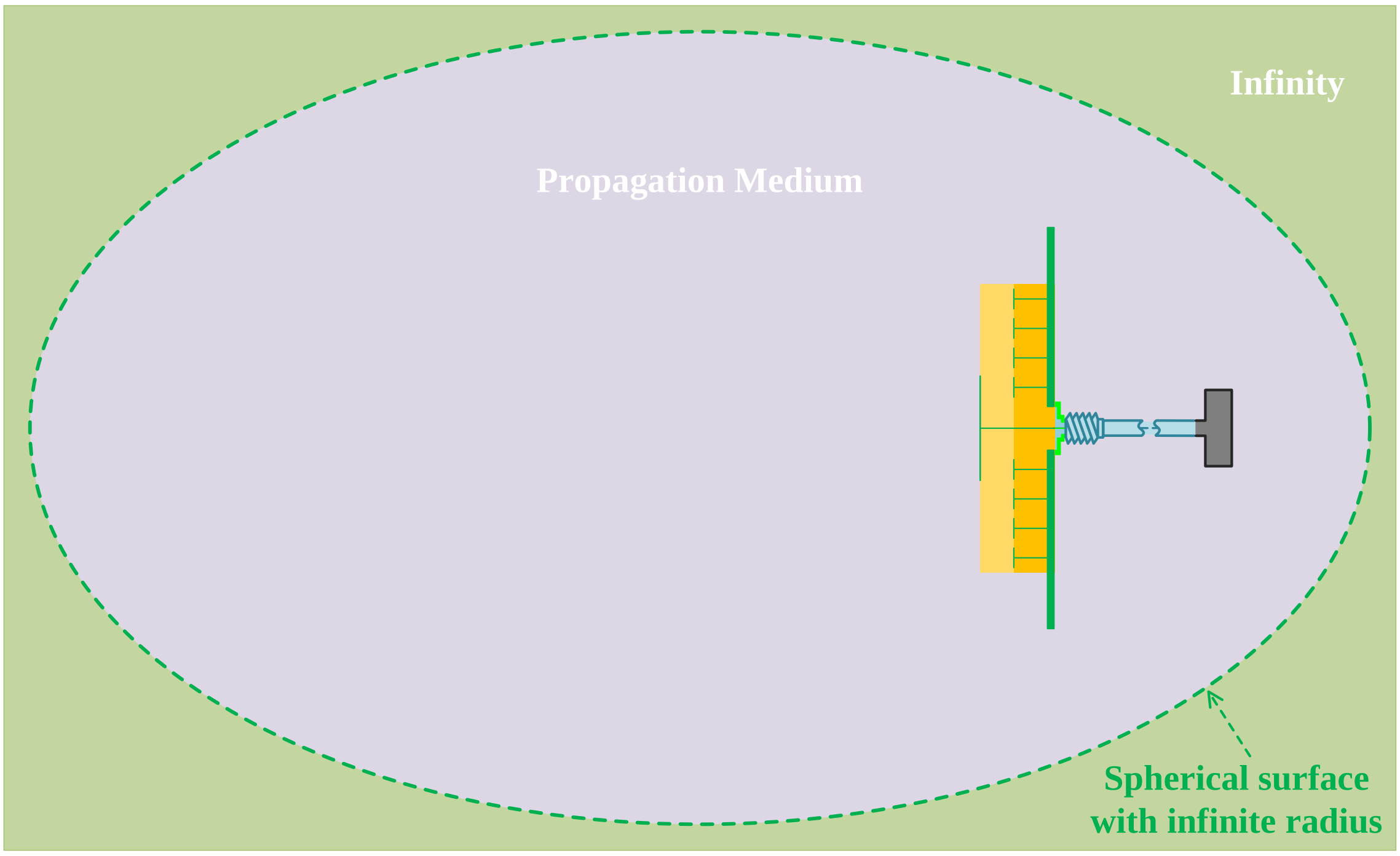

Figure 7-10 Receiving problem considered in this section

The geometry of the rec-antenna shown in the above Fig. 7-10 are illustrated in the following Fig. 7-11, where Fig. 7-11(a) shows the top view of the rec-antenna, and Fig. 7-11(b) shows the lateral view of the rec-antenna, and Fig. 7-11(c) shows the cross section view of the rec-antenna.

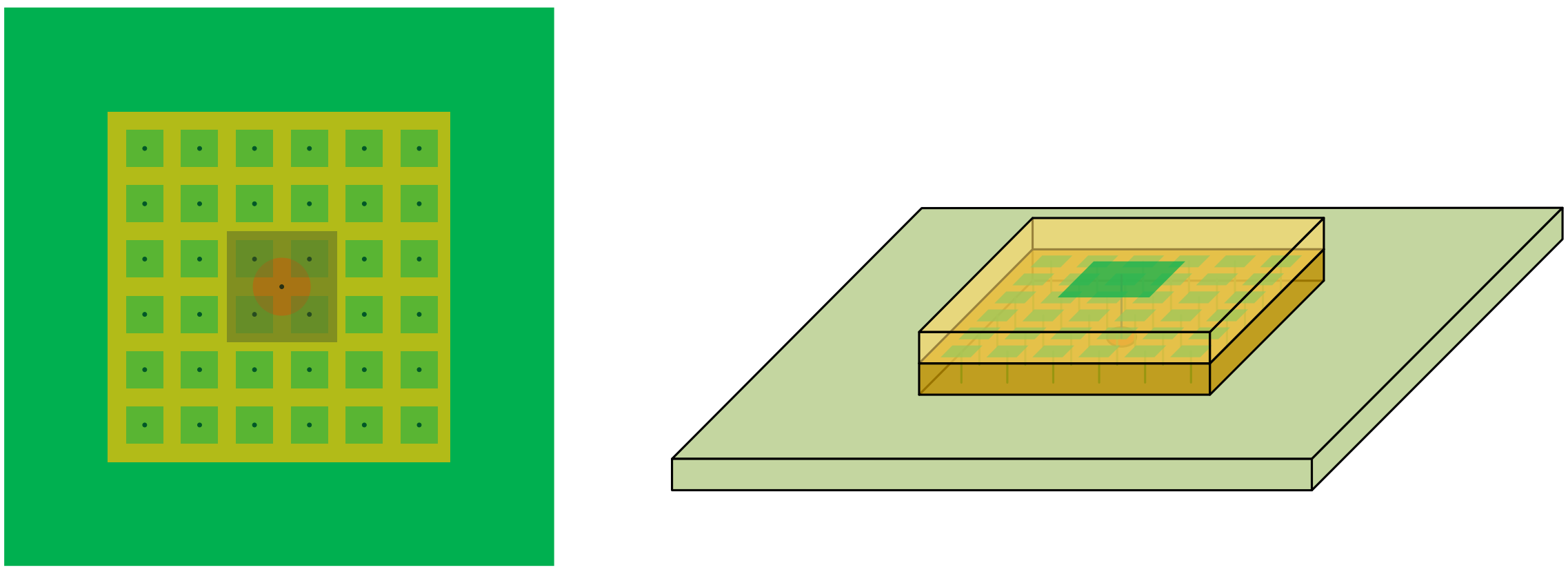

(a) (b)

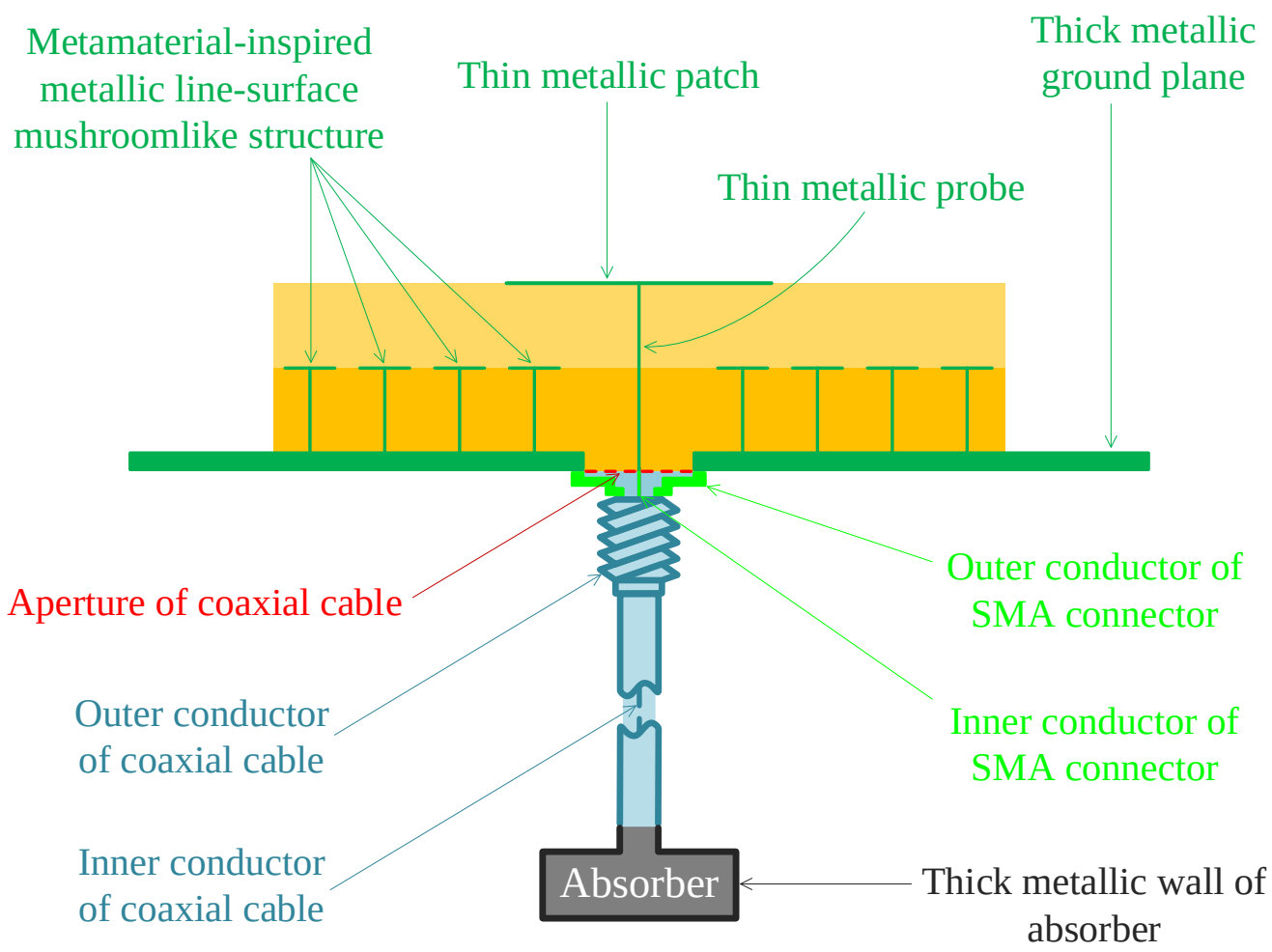

(c)

Figure 7-11 Geometry of the rec-antenna. (a) Top view; (b) lateral view; (c) cross section

For the generality, we consider the case that the rec-antenna is driven by an arbitrary transmitter (as shown in Fig. 7-12) in this section just like the previous Sec. 7.3 did.

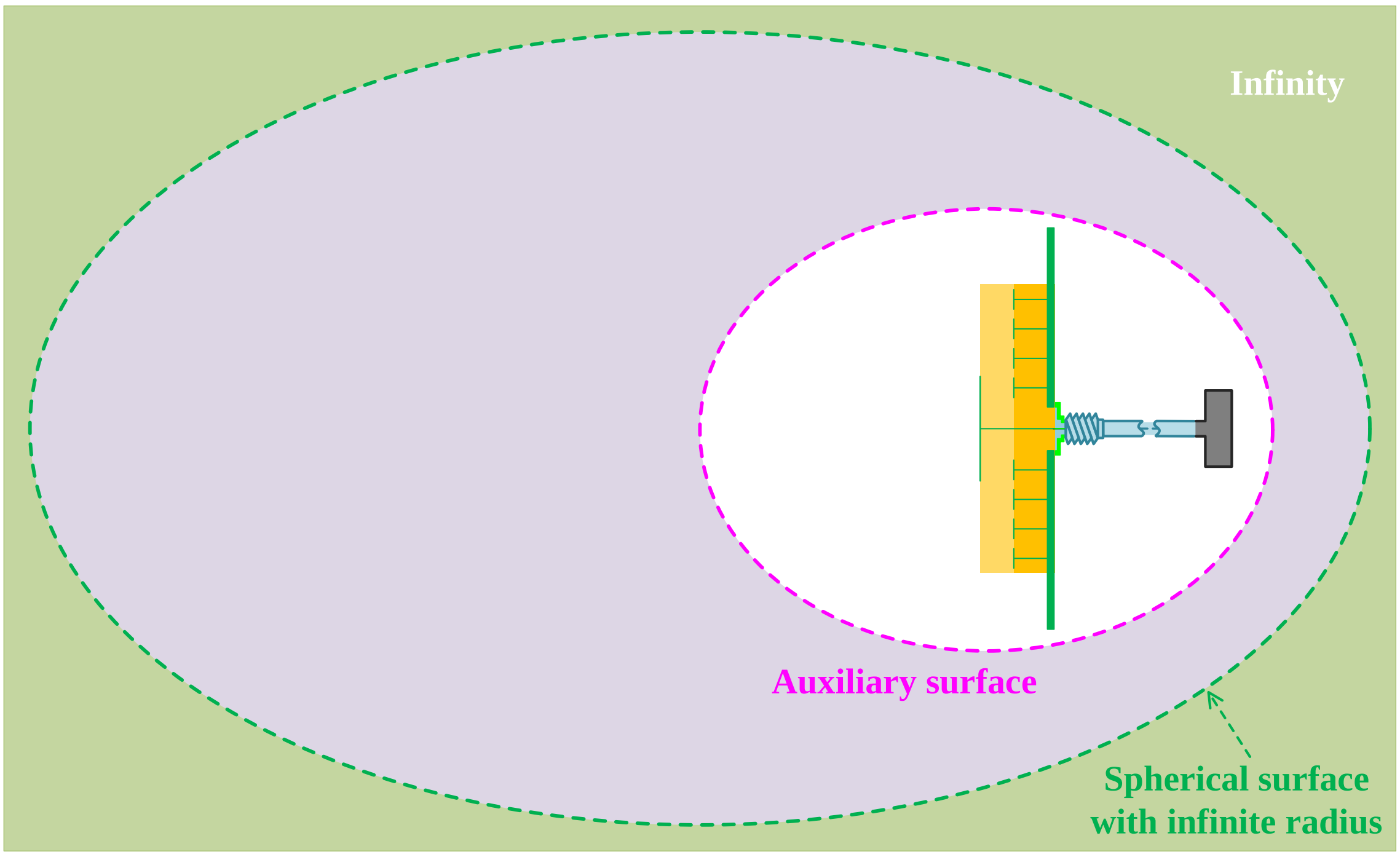

Figure 7-12 The rec-antenna driven by an arbitrary transmitter

The organization of this section is completely similar to the previous Secs. 7.2 and 7.3.

## 7.4.1 Topological Structure

The topological structure of the rec-antenna shown in Fig. 7-12 is illustrated in the following Fig. 7-13.

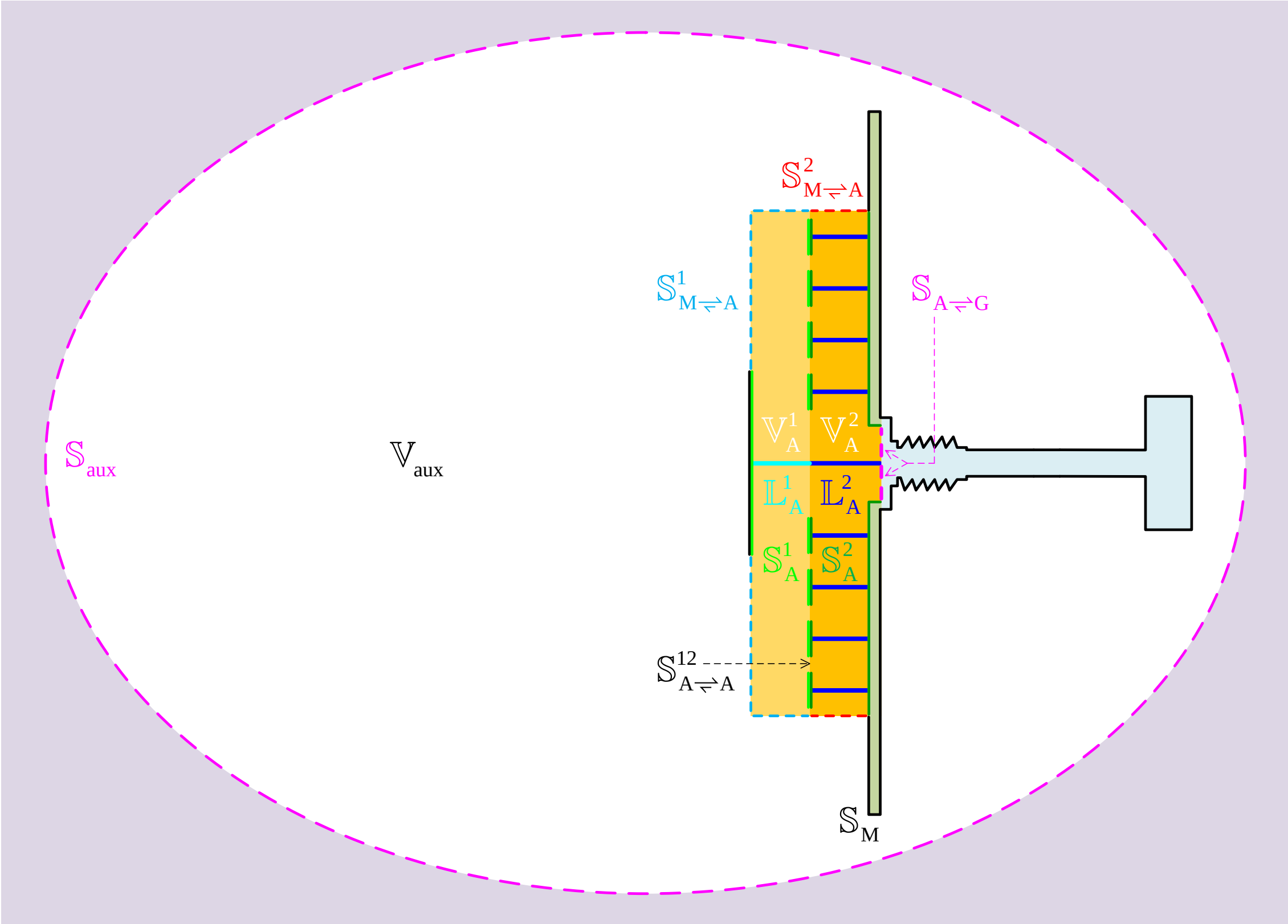

Figure 7-13 Topological structure of the rec-antenna shown in Fig. 7-12

In the above Fig. 7-13, $\mathbb{S}_{aux}$ is a finite *auxiliary surface* enclosing whole receiving system, and it is a penetrable surface. The region sandwiched between the auxiliary surface and receiving system is denoted as $\mathbb{V}_{aux}$, and there doesn't exist any source distributing in $\mathbb{V}_{aux}$. The impenetrable interface between $\mathbb{V}_{aux}$ and receiving system is just the impenetrable interface between propagation medium and receiving system, so it is denoted as $\mathbb{S}_M$.

The region occupied by the upper material superstrate is denoted as $\mathbb{V}^1_A$, and the region occupied by the lower material substrate is denoted as $\mathbb{V}^2_A$.

The penetrable interface between $\mathbb{V}_{aux}$ and $\mathbb{V}^1_A$ is just the penetrable interface between propagation medium and $\mathbb{V}^1_A$, so it is denoted as $\mathbb{S}^1_{M\rightleftharpoons A}$. The interface between $\mathbb{V}^1_A$ and the upper metallic patch and the interface between $\mathbb{V}^1_A$ and the metallic patches in mushroomlike structure are collectively denoted as $\mathbb{S}^1_A$; the boundary between $\mathbb{V}^1_A$ and the thin metallic probe is denoted as $\mathbb{L}^1_A$.

The penetrable interface between $\mathbb{V}^1_A$ and $\mathbb{V}^2_A$ is denoted as $\mathbb{S}^{12}_{A\rightleftharpoons A}$.

The penetrable interface between $\mathbb{V}_{aux}$ and $\mathbb{V}^2_A$ is just the penetrable interface between propagation medium and $\mathbb{V}^2_A$, so it is denoted as $\mathbb{S}^2_{M\rightleftharpoons A}$. The interface between $\mathbb{V}_{\rm A}^2$ and the metallic patches in mushroomlike structure and the interface between $\mathbb{V}_{\rm A}^2$ and the thick metallic ground plane are collectively denoted as $\mathbb{S}_{\rm A}^2$; the boundary between $\mathbb{V}_{\rm A}^2$ and thin metallic probe and the boundary between $\mathbb{V}_{\rm A}^2$ and the metallic lines in mushroomlike structure are collectively denoted as $\mathbb{L}_{\rm A}^2$. The penetrable interface between $\mathbb{V}_{\rm A}^2$ and the PML is just the output port of the rec-antenna, and it is denoted as $\mathbb{S}_{\rm A \rightleftharpoons G}$.

Clearly, $\mathbb{S}_{\rm aux}$, $\mathbb{S}_{\rm M}$, $\mathbb{S}_{\rm M \rightleftharpoons A}^1$, and $\mathbb{S}_{\rm M \rightleftharpoons A}^2$ constitute the boundary of $\mathbb{V}_{\rm aux}$, i.e., $\partial \mathbb{V}_{\rm aux} = \mathbb{S}_{\rm aux} \bigcup \mathbb{S}_{\rm M} \bigcup \mathbb{S}_{\rm M \rightleftharpoons A}^1 \bigcup \mathbb{S}_{\rm M \rightleftharpoons A}^2$, and $\mathbb{S}_{\rm M \rightleftharpoons A}^1 \bigcup \mathbb{S}_{\rm M \rightleftharpoons A}^2$ is just the input port of the rec-antenna. Surfaces $\mathbb{S}_{\rm M \rightleftharpoons A}^1 \,\&\, \mathbb{S}_{\rm A}^1 \,\&\, \mathbb{S}_{\rm A \rightleftharpoons A}^{12}$ and line $\mathbb{L}_{\rm A}^1$ constitute the whole boundary of $\mathbb{V}_{\rm A}^1$, i.e., $\partial \mathbb{V}_{\rm A}^1 = \mathbb{S}_{\rm M \rightleftharpoons A}^1 \bigcup \mathbb{S}_{\rm A}^1 \bigcup \mathbb{S}_{\rm A \rightleftharpoons A}^{12} \bigcup \mathbb{L}_{\rm A}^1$. Surfaces $\mathbb{S}_{\rm M \rightleftharpoons A}^2 \,\&\, \mathbb{S}_{\rm A \rightleftharpoons A}^{12} \,\&\, \mathbb{S}_{\rm A}^2 \,\&\, \mathbb{S}_{\rm A \rightleftharpoons G}$ and line $\mathbb{L}_{\rm A}^2$ constitute the whole boundary of $\mathbb{V}_{\rm A}^2$, i.e., $\partial \mathbb{V}_{\rm A}^2 = \mathbb{S}_{\rm M \rightleftharpoons A}^2 \bigcup \mathbb{S}_{\rm A \rightleftharpoons A}^{12} \bigcup \mathbb{S}_{\rm A}^2 \bigcup \mathbb{S}_{\rm A \rightleftharpoons G} \bigcup \mathbb{L}_{\rm A}^2$.

In addition, the permeability, permeativity, and conductivity of $\mathbb{V}_{\rm A}^1$ are denoted as $\ddot{\mu}_1$, $\ddot{\varepsilon}_1$, and $\ddot{\sigma}_1$ respectively; the permeability, permeativity, and conductivity of $\mathbb{V}_{\rm A}^2$ are denoted as $\ddot{\mu}_2$, $\ddot{\varepsilon}_2$, and $\ddot{\sigma}_2$ respectively; the permeability and permeativity of $\mathbb{V}_{\rm aux}$ are just the ones corresponding to free space, so they are denoted as $\mu_0$ and $\varepsilon_0$ respectively.

## 7.4.2 Source-Field Relationships

If the equivalent surface currents distributing on $\mathbb{S}_{\rm aux}$ are denoted as $\{\vec{J}_{\rm aux}, \vec{M}_{\rm aux}\}$, and the equivalent surface currents distributing on $\mathbb{S}_{\rm M \rightleftharpoons A}^1$ and $\mathbb{S}_{\rm M \rightleftharpoons A}^2$ are denoted as $\{\vec{J}_{\rm M \rightleftharpoons A}^1, \vec{M}_{\rm M \rightleftharpoons A}^1\}$ and $\{\vec{J}_{\rm M \rightleftharpoons A}^2, \vec{M}_{\rm M \rightleftharpoons A}^2\}$ respectively, and the equivalent surface electric current distributing on $\mathbb{S}_{\rm M}$ is denoted as $\vec{J}_{\rm M}$ ①, then the field distributing on $\mathbb{V}_{\rm aux}$ can be expressed as follows:

$$\vec{F}(\vec{r}) \;=\; \mathcal{F}_0\left(\vec{J}_{\rm aux} - \vec{J}_{\rm M \rightleftharpoons A}^1 - \vec{J}_{\rm M \rightleftharpoons A}^2 + \vec{J}_{\rm M}, \vec{M}_{\rm aux} - \vec{M}_{\rm M \rightleftharpoons A}^1 - \vec{M}_{\rm M \rightleftharpoons A}^2\right) \;,\; \vec{r} \in \mathbb{V}_{\rm aux} \quad \text{(7-69)}$$

where $\vec{F} = \vec{E} / \vec{H}$, and correspondingly $\mathcal{F}_0 = \mathcal{E}_0 / \mathcal{H}_0$, and the operators are the same as the ones used in the previous chapters. The currents $\{\vec{J}_{\rm aux}, \vec{M}_{\rm aux}\}$ and fields $\{\vec{E}, \vec{H}\}$ in Eq. (7-69) satisfy the following relations

$$\hat{n}_{\rightarrow \rm aux} \times \left[\vec{H}\left(\vec{r}_{\rm aux}\right)\right]_{\vec{r}_{\rm aux} \to \vec{r}} \;=\; \vec{J}_{\rm aux}\left(\vec{r}\right) \quad , \quad \vec{r} \in \mathbb{S}_{\rm aux} \quad \text{(7-70a)}$$

$$\left[\vec{E}\left(\vec{r}_{\rm aux}\right)\right]_{\vec{r}_{\rm aux} \to \vec{r}} \times \hat{n}_{\rightarrow \rm aux} \;=\; \vec{M}_{\rm aux}\left(\vec{r}\right) \quad , \quad \vec{r} \in \mathbb{S}_{\rm aux} \quad \text{(7-70b)}$$

① The equivalent surface magnetic current distributing on $\mathbb{S}_{\rm M}$ is zero, because of the homogeneous tangential electric field boundary condition on $\mathbb{S}_{\rm M}$ [13].

where point $\vec{r}_{\text{aux}}$ belongs to $\mathbb{V}_{\text{aux}}$ and approaches the point $\vec{r}$ on $\mathbb{S}_{\text{aux}}$, and $\hat{n}_{\rightarrow\text{aux}}$ is the normal direction of $\mathbb{S}_{\text{aux}}$ and points to the interior of $\mathbb{V}_{\text{aux}}$. The currents $\{\vec{J}^{1}_{\text{M}\rightleftharpoons\text{A}},\vec{M}^{1}_{\text{M}\rightleftharpoons\text{A}}\}$ & $\{\vec{J}^{2}_{\text{M}\rightleftharpoons\text{A}},\vec{M}^{2}_{\text{M}\rightleftharpoons\text{A}}\}$ and fields $\{\vec{E},\vec{H}\}$ in Eq. (7-69) satisfy the following relations

$$\hat{n}^{1}_{\rightarrow\text{A}}\times\left[\vec{H}\left(\vec{r}_{\text{aux}}\right)\right]_{\vec{r}_{\text{aux}}\to\vec{r}} = \vec{J}^{1}_{\text{M}\rightleftharpoons\text{A}}\left(\vec{r}\right) \quad , \quad \vec{r}\in\mathbb{S}^{1}_{\text{M}\rightleftharpoons\text{A}} \tag{7-71a}$$

$$\left[\vec{E}\left(\vec{r}_{\text{aux}}\right)\right]_{\vec{r}_{\text{aux}}\to\vec{r}}\times\hat{n}^{1}_{\rightarrow\text{A}} = \vec{M}^{1}_{\text{M}\rightleftharpoons\text{A}}\left(\vec{r}\right) \quad , \quad \vec{r}\in\mathbb{S}^{1}_{\text{M}\rightleftharpoons\text{A}} \tag{7-71b}$$

and

$$\hat{n}^{2}_{\rightarrow\text{A}}\times\left[\vec{H}\left(\vec{r}_{\text{aux}}\right)\right]_{\vec{r}_{\text{aux}}\to\vec{r}} = \vec{J}^{2}_{\text{M}\rightleftharpoons\text{A}}\left(\vec{r}\right) \quad , \quad \vec{r}\in\mathbb{S}^{2}_{\text{M}\rightleftharpoons\text{A}} \tag{7-72a}$$

$$\left[\vec{E}\left(\vec{r}_{\text{aux}}\right)\right]_{\vec{r}_{\text{aux}}\to\vec{r}}\times\hat{n}^{2}_{\rightarrow\text{A}} = \vec{M}^{2}_{\text{M}\rightleftharpoons\text{A}}\left(\vec{r}\right) \quad , \quad \vec{r}\in\mathbb{S}^{2}_{\text{M}\rightleftharpoons\text{A}} \tag{7-72b}$$

where point $\vec{r}_{\text{aux}}$ belongs to $\mathbb{V}_{\text{aux}}$ and approaches the point $\vec{r}$ on $\mathbb{S}^{1}_{\text{M}\rightleftharpoons\text{A}}\bigcup\mathbb{S}^{2}_{\text{M}\rightleftharpoons\text{A}}$, and $\hat{n}^{1}_{\rightarrow\text{A}}$ and $\hat{n}^{2}_{\rightarrow\text{A}}$ are the normal directions of $\mathbb{S}^{1}_{\text{M}\rightleftharpoons\text{A}}$ and $\mathbb{S}^{2}_{\text{M}\rightleftharpoons\text{A}}$ and point to the interiors of $\mathbb{V}^{1}_{\text{A}}$ and $\mathbb{V}^{2}_{\text{A}}$ respectively.

If the equivalent surface currents distributing on $\mathbb{S}^{12}_{\text{A}\rightleftharpoons\text{A}}$ are denoted as $\{\vec{J}^{12}_{\text{A}\rightleftharpoons\text{A}},\vec{M}^{12}_{\text{A}\rightleftharpoons\text{A}}\}$, and the equivalent electric currents distributing on $\mathbb{S}^{1}_{\text{A}}$ and $\mathbb{L}^{1}_{\text{A}}$ are collectively denoted as $\vec{J}^{1}_{\text{A}}$ ①, then the field distributing on $\mathbb{V}^{1}_{\text{A}}$ can be expressed as follows:

$$\vec{F}\left(\vec{r}\right) = \mathcal{F}_{1}\left(\vec{J}^{1}_{\text{M}\rightleftharpoons\text{A}}+\vec{J}^{12}_{\text{A}\rightleftharpoons\text{A}}+\vec{J}^{1}_{\text{A}},\vec{M}^{1}_{\text{M}\rightleftharpoons\text{A}}+\vec{M}^{12}_{\text{A}\rightleftharpoons\text{A}}\right) \quad , \quad \vec{r}\in\mathbb{V}^{1}_{\text{A}} \tag{7-73}$$

where $\vec{F}=\vec{E}\,/\,\vec{H}$, and correspondingly $\mathcal{F}_{1}=\mathcal{E}_{1}\,/\,\mathcal{H}_{1}$, and the operator is defined as that $\mathcal{F}_{1}(\vec{J},\vec{M})=\overset{\leftrightarrow}{G}{}^{JF}_{1}*\vec{J}+\overset{\leftrightarrow}{G}{}^{MF}_{1}*\vec{M}$ (here, $\overset{\leftrightarrow}{G}{}^{JF}_{1}$ and $\overset{\leftrightarrow}{G}{}^{MF}_{1}$ are the dyadic Green's functions corresponding to the region $\mathbb{V}^{1}_{\text{A}}$ with material parameters $\{\overset{\leftrightarrow}{\mu}_{1},\overset{\leftrightarrow}{\varepsilon}_{1},\overset{\leftrightarrow}{\sigma}_{1}\}$). The currents $\{\vec{J}^{12}_{\text{A}\rightleftharpoons\text{A}},\vec{M}^{12}_{\text{A}\rightleftharpoons\text{A}}\}$ and fields $\{\vec{E},\vec{H}\}$ in Eq. (7-73) satisfy the following relations

$$\hat{n}^{1}_{\rightarrow\text{A}}\times\left[\vec{H}\left(\vec{r}^{1}_{\text{A}}\right)\right]_{\vec{r}^{1}_{\text{A}}\to\vec{r}} = \vec{J}^{12}_{\text{A}\rightleftharpoons\text{A}}\left(\vec{r}\right) \quad , \quad \vec{r}\in\mathbb{S}^{12}_{\text{A}\rightleftharpoons\text{A}} \tag{7-74a}$$

$$\left[\vec{E}\left(\vec{r}^{1}_{\text{A}}\right)\right]_{\vec{r}^{1}_{\text{A}}\to\vec{r}}\times\hat{n}^{1}_{\rightarrow\text{A}} = \vec{M}^{12}_{\text{A}\rightleftharpoons\text{A}}\left(\vec{r}\right) \quad , \quad \vec{r}\in\mathbb{S}^{12}_{\text{A}\rightleftharpoons\text{A}} \tag{7-74b}$$

where point $\vec{r}^{1}_{\text{A}}$ belongs to $\mathbb{V}^{1}_{\text{A}}$ and approaches the point $\vec{r}$ on $\mathbb{S}^{12}_{\text{A}\rightleftharpoons\text{A}}$, and $\hat{n}^{1}_{\rightarrow\text{A}}$ is the normal direction of $\mathbb{S}^{12}_{\text{A}\rightleftharpoons\text{A}}$ and points to the interior of $\mathbb{V}^{1}_{\text{A}}$.

If the equivalent surface currents distributing on $\mathbb{S}_{\text{A}\rightleftharpoons\text{G}}$ are denoted as $\{\vec{J}_{\text{A}\rightleftharpoons\text{G}},\vec{M}_{\text{A}\rightleftharpoons\text{G}}\}$, and the equivalent electric currents distributing on $\mathbb{S}^{2}_{\text{A}}$ and $\mathbb{L}^{2}_{\text{A}}$ are

① The equivalent magnetic current distributing on $\mathbb{S}^{1}_{\text{A}}\bigcup\mathbb{L}^{1}_{\text{A}}$ is zero, because of the homogeneous tangential electric field boundary condition on $\mathbb{S}^{1}_{\text{A}}\bigcup\mathbb{L}^{1}_{\text{A}}$ [13].

collectively denoted as $\vec{J}_{\rm A}^{2}$ ①, then the field distributing on $\mathbb{V}_{\rm A}^{2}$ can be expressed as

$$\vec{F}\left(\vec{r}\right) = \mathcal{F}_2\left(\vec{J}_{\rm M\rightleftharpoons A}^{2} - \vec{J}_{\rm A\rightleftharpoons A}^{12} + \vec{J}_{\rm A\rightleftharpoons G} + \vec{J}_{\rm A}^{2}, \vec{M}_{\rm M\rightleftharpoons A}^{2} - \vec{M}_{\rm A\rightleftharpoons A}^{12} + \vec{M}_{\rm A\rightleftharpoons G}\right) \quad , \quad \vec{r} \in \mathbb{V}_{\rm A}^{2} \quad (7\text{-}75)$$

where $\vec{F} = \vec{E} / \vec{H}$, and correspondingly $\mathcal{F}_2 = \mathcal{E}_2 / \mathcal{H}_2$, and the operator is defined as that $\mathcal{F}_2(\vec{J},\vec{M}) = \vec{G}_2^{JF} * \vec{J} + \vec{G}_2^{MF} * \vec{M}$ (here, $\vec{G}_2^{JF}$ and $\vec{G}_2^{MF}$ are the dyadic Green's functions corresponding to the region $\mathbb{V}_{\rm A}^{2}$ with material parameters $\{\vec{\mu}_2, \vec{\varepsilon}_2, \vec{\sigma}_2\}$). The currents $\{\vec{J}_{\rm A\rightleftharpoons G}, \vec{M}_{\rm A\rightleftharpoons G}\}$ and fields $\{\vec{E}, \vec{H}\}$ in Eq. (7-75) satisfy the following relations

$$\hat{n}_{\rightarrow \rm A}^{2} \times \left[\vec{H}\left(\vec{r}_{\rm A}^{\,2}\right)\right]_{\vec{r}_{\rm A}^{\,2} \to \vec{r}} = \vec{J}_{\rm A\rightleftharpoons G}\left(\vec{r}\right) \quad , \quad \vec{r} \in \mathbb{S}_{\rm A\rightleftharpoons G} \quad (7\text{-}76a)$$

$$\left[\vec{E}\left(\vec{r}_{\rm A}^{\,2}\right)\right]_{\vec{r}_{\rm A}^{\,2} \to \vec{r}} \times \hat{n}_{\rightarrow \rm A}^{2} = \vec{M}_{\rm A\rightleftharpoons G}\left(\vec{r}\right) \quad , \quad \vec{r} \in \mathbb{S}_{\rm A\rightleftharpoons G} \quad (7\text{-}76b)$$

where point $\vec{r}_{\rm A}^{\,2}$ belongs to $\mathbb{V}_{\rm A}^{2}$ and approaches the point $\vec{r}$ on $\mathbb{S}_{\rm A\rightleftharpoons G}$, and $\hat{n}_{\rightarrow \rm A}^{2}$ is the normal direction of $\mathbb{S}_{\rm A\rightleftharpoons G}$ and points to the interior of $\mathbb{V}_{\rm A}^{2}$.

### 7.4.3 Mathematical Description for Modal Space

Substituting Eq. (7-69) into Eqs. (7-70a) and (7-70b), we immediately obtain the following integral equations

$$\begin{aligned}
&\left[\mathcal{H}_0\left(\vec{J}_{\rm aux} - \vec{J}_{\rm M\rightleftharpoons A}^{1} - \vec{J}_{\rm M\rightleftharpoons A}^{2} + \vec{J}_{\rm M}, \vec{M}_{\rm aux} - \vec{M}_{\rm M\rightleftharpoons A}^{1} - \vec{M}_{\rm M\rightleftharpoons A}^{2}\right)\right]_{\vec{r}_{\rm aux} \to \vec{r}}^{\rm tan} \\
&= \vec{J}_{\rm aux}\left(\vec{r}\right) \times \hat{n}_{\rightarrow \rm aux} \quad , \quad \vec{r} \in \mathbb{S}_{\rm aux} \quad (7\text{-}77a)
\end{aligned}$$

$$\begin{aligned}
&\left[\mathcal{E}_0\left(\vec{J}_{\rm aux} - \vec{J}_{\rm M\rightleftharpoons A}^{1} - \vec{J}_{\rm M\rightleftharpoons A}^{2} + \vec{J}_{\rm M}, \vec{M}_{\rm aux} - \vec{M}_{\rm M\rightleftharpoons A}^{1} - \vec{M}_{\rm M\rightleftharpoons A}^{2}\right)\right]_{\vec{r}_{\rm aux} \to \vec{r}}^{\rm tan} \\
&= \hat{n}_{\rightarrow \rm aux} \times \vec{M}_{\rm aux}\left(\vec{r}\right) \quad , \quad \vec{r} \in \mathbb{S}_{\rm aux} \quad (7\text{-}77b)
\end{aligned}$$

about currents $\{\vec{J}_{\rm aux}, \vec{M}_{\rm aux}\}$, $\{\vec{J}_{\rm M\rightleftharpoons A}^{1}, \vec{M}_{\rm M\rightleftharpoons A}^{1}\}$, $\{\vec{J}_{\rm M\rightleftharpoons A}^{2}, \vec{M}_{\rm M\rightleftharpoons A}^{2}\}$, and $\vec{J}_{\rm M}$, where the superscript "tan" represents the tangential component of the field.

Using Eqs. (7-69)&(7-73) and employing the tangential field continuation condition on $\mathbb{S}_{\rm M\rightleftharpoons A}^{1}$, there exist the following integral equations

$$\begin{aligned}
&\left[\mathcal{E}_0\left(\vec{J}_{\rm aux} - \vec{J}_{\rm M\rightleftharpoons A}^{1} - \vec{J}_{\rm M\rightleftharpoons A}^{2} + \vec{J}_{\rm M}, \vec{M}_{\rm aux} - \vec{M}_{\rm M\rightleftharpoons A}^{1} - \vec{M}_{\rm M\rightleftharpoons A}^{2}\right)\right]_{\vec{r}_{\rm aux} \to \vec{r}}^{\rm tan} \\
&= \left[\mathcal{E}_1\left(\vec{J}_{\rm M\rightleftharpoons A}^{1} + \vec{J}_{\rm A\rightleftharpoons A}^{12} + \vec{J}_{\rm A}^{1}, \vec{M}_{\rm M\rightleftharpoons A}^{1} + \vec{M}_{\rm A\rightleftharpoons A}^{12}\right)\right]_{\vec{r}_{\rm A}^{\,1} \to \vec{r}}^{\rm tan} \quad , \quad \vec{r} \in \mathbb{S}_{\rm M\rightleftharpoons A}^{1} \quad (7\text{-}78a)
\end{aligned}$$

$$\begin{aligned}
&\left[\mathcal{H}_0\left(\vec{J}_{\rm aux} - \vec{J}_{\rm M\rightleftharpoons A}^{1} - \vec{J}_{\rm M\rightleftharpoons A}^{2} + \vec{J}_{\rm M}, \vec{M}_{\rm aux} - \vec{M}_{\rm M\rightleftharpoons A}^{1} - \vec{M}_{\rm M\rightleftharpoons A}^{2}\right)\right]_{\vec{r}_{\rm aux} \to \vec{r}}^{\rm tan} \\
&= \left[\mathcal{H}_1\left(\vec{J}_{\rm M\rightleftharpoons A}^{1} + \vec{J}_{\rm A\rightleftharpoons A}^{12} + \vec{J}_{\rm A}^{1}, \vec{M}_{\rm M\rightleftharpoons A}^{1} + \vec{M}_{\rm A\rightleftharpoons A}^{12}\right)\right]_{\vec{r}_{\rm A}^{\,1} \to \vec{r}}^{\rm tan} \quad , \quad \vec{r} \in \mathbb{S}_{\rm M\rightleftharpoons A}^{1} \quad (7\text{-}78b)
\end{aligned}$$

① The equivalent magnetic current distributing on $\mathbb{S}_{\rm A}^{2} \cup \mathbb{L}_{\rm A}^{2}$ is zero, because of the homogeneous tangential electric field boundary condition on $\mathbb{S}_{\rm A}^{2} \cup \mathbb{L}_{\rm A}^{2}$ [13].

about currents $\{\vec{J}_{\text{aux}},\vec{M}_{\text{aux}}\}$, $\{\vec{J}^{1}_{\text{M}\rightleftharpoons\text{A}},\vec{M}^{1}_{\text{M}\rightleftharpoons\text{A}}\}$, $\{\vec{J}^{2}_{\text{M}\rightleftharpoons\text{A}},\vec{M}^{2}_{\text{M}\rightleftharpoons\text{A}}\}$, $\{\vec{J}^{12}_{\text{A}\rightleftharpoons\text{A}},\vec{M}^{12}_{\text{A}\rightleftharpoons\text{A}}\}$, $\vec{J}_{\text{M}}$, and $\vec{J}^{1}_{\text{A}}$. Using Eqs. (7-69)&(7-75) and employing the tangential field continuation condition on $\mathbb{S}^{2}_{\text{M}\rightleftharpoons\text{A}}$, there exist the following integral equations

$$\left[\mathcal{E}_0\left(\vec{J}_{\text{aux}}-\vec{J}^{1}_{\text{M}\rightleftharpoons\text{A}}-\vec{J}^{2}_{\text{M}\rightleftharpoons\text{A}}+\vec{J}_{\text{M}},\vec{M}_{\text{aux}}-\vec{M}^{1}_{\text{M}\rightleftharpoons\text{A}}-\vec{M}^{2}_{\text{M}\rightleftharpoons\text{A}}\right)\right]^{\tan}_{\vec{r}_{\text{aux}}\to\vec{r}}$$
$$=\left[\mathcal{E}_2\left(\vec{J}^{2}_{\text{M}\rightleftharpoons\text{A}}-\vec{J}^{12}_{\text{A}\rightleftharpoons\text{A}}+\vec{J}_{\text{A}\rightleftharpoons\text{G}}+\vec{J}^{2}_{\text{A}},\vec{M}^{2}_{\text{M}\rightleftharpoons\text{A}}-\vec{M}^{12}_{\text{A}\rightleftharpoons\text{A}}+\vec{M}_{\text{A}\rightleftharpoons\text{G}}\right)\right]^{\tan}_{\vec{r}^{2}_{\text{A}}\to\vec{r}} \quad,\ \vec{r}\in\mathbb{S}^{2}_{\text{M}\rightleftharpoons\text{A}} \quad (7\text{-}79\text{a})$$

$$\left[\mathcal{H}_0\left(\vec{J}_{\text{aux}}-\vec{J}^{1}_{\text{M}\rightleftharpoons\text{A}}-\vec{J}^{2}_{\text{M}\rightleftharpoons\text{A}}+\vec{J}_{\text{M}},\vec{M}_{\text{aux}}-\vec{M}^{1}_{\text{M}\rightleftharpoons\text{A}}-\vec{M}^{2}_{\text{M}\rightleftharpoons\text{A}}\right)\right]^{\tan}_{\vec{r}_{\text{aux}}\to\vec{r}}$$
$$=\left[\mathcal{H}_2\left(\vec{J}^{2}_{\text{M}\rightleftharpoons\text{A}}-\vec{J}^{12}_{\text{A}\rightleftharpoons\text{A}}+\vec{J}_{\text{A}\rightleftharpoons\text{G}}+\vec{J}^{2}_{\text{A}},\vec{M}^{2}_{\text{M}\rightleftharpoons\text{A}}-\vec{M}^{12}_{\text{A}\rightleftharpoons\text{A}}+\vec{M}_{\text{A}\rightleftharpoons\text{G}}\right)\right]^{\tan}_{\vec{r}^{2}_{\text{A}}\to\vec{r}} ,\ \vec{r}\in\mathbb{S}^{2}_{\text{M}\rightleftharpoons\text{A}} \quad (7\text{-}79\text{b})$$

about currents $\{\vec{J}_{\text{aux}},\vec{M}_{\text{aux}}\}$, $\{\vec{J}^{1}_{\text{M}\rightleftharpoons\text{A}},\vec{M}^{1}_{\text{M}\rightleftharpoons\text{A}}\}$, $\{\vec{J}^{2}_{\text{M}\rightleftharpoons\text{A}},\vec{M}^{2}_{\text{M}\rightleftharpoons\text{A}}\}$, $\{\vec{J}^{12}_{\text{A}\rightleftharpoons\text{A}},\vec{M}^{12}_{\text{A}\rightleftharpoons\text{A}}\}$, $\{\vec{J}_{\text{A}\rightleftharpoons\text{G}},\vec{M}_{\text{A}\rightleftharpoons\text{G}}\}$, $\vec{J}_{\text{M}}$, and $\vec{J}^{2}_{\text{A}}$. Using Eqs. (7-73)&(7-75) and employing the tangential field continuation condition on $\mathbb{S}^{12}_{\text{A}\rightleftharpoons\text{A}}$, there exist the following integral equations

$$\left[\mathcal{E}_1\left(\vec{J}^{1}_{\text{M}\rightleftharpoons\text{A}}+\vec{J}^{12}_{\text{A}\rightleftharpoons\text{A}}+\vec{J}^{1}_{\text{A}},\vec{M}^{1}_{\text{M}\rightleftharpoons\text{A}}+\vec{M}^{12}_{\text{A}\rightleftharpoons\text{A}}\right)\right]^{\tan}_{\vec{r}^{1}_{\text{A}}\to\vec{r}}$$
$$=\left[\mathcal{E}_2\left(\vec{J}^{2}_{\text{M}\rightleftharpoons\text{A}}-\vec{J}^{12}_{\text{A}\rightleftharpoons\text{A}}+\vec{J}_{\text{A}\rightleftharpoons\text{G}}+\vec{J}^{2}_{\text{A}},\vec{M}^{2}_{\text{M}\rightleftharpoons\text{A}}-\vec{M}^{12}_{\text{A}\rightleftharpoons\text{A}}+\vec{M}_{\text{A}\rightleftharpoons\text{G}}\right)\right]^{\tan}_{\vec{r}^{2}_{\text{A}}\to\vec{r}} \quad,\ \vec{r}\in\mathbb{S}^{12}_{\text{A}\rightleftharpoons\text{A}} \quad (7\text{-}80\text{a})$$

$$\left[\mathcal{H}_1\left(\vec{J}^{1}_{\text{M}\rightleftharpoons\text{A}}+\vec{J}^{12}_{\text{A}\rightleftharpoons\text{A}}+\vec{J}^{1}_{\text{A}},\vec{M}^{1}_{\text{M}\rightleftharpoons\text{A}}+\vec{M}^{12}_{\text{A}\rightleftharpoons\text{A}}\right)\right]^{\tan}_{\vec{r}^{1}_{\text{A}}\to\vec{r}}$$
$$=\left[\mathcal{H}_2\left(\vec{J}^{2}_{\text{M}\rightleftharpoons\text{A}}-\vec{J}^{12}_{\text{A}\rightleftharpoons\text{A}}+\vec{J}_{\text{A}\rightleftharpoons\text{G}}+\vec{J}^{2}_{\text{A}},\vec{M}^{2}_{\text{M}\rightleftharpoons\text{A}}-\vec{M}^{12}_{\text{A}\rightleftharpoons\text{A}}+\vec{M}_{\text{A}\rightleftharpoons\text{G}}\right)\right]^{\tan}_{\vec{r}^{2}_{\text{A}}\to\vec{r}} ,\ \vec{r}\in\mathbb{S}^{12}_{\text{A}\rightleftharpoons\text{A}} \quad (7\text{-}80\text{b})$$

about currents $\{\vec{J}^{1}_{\text{M}\rightleftharpoons\text{A}},\vec{M}^{1}_{\text{M}\rightleftharpoons\text{A}}\}$, $\{\vec{J}^{2}_{\text{M}\rightleftharpoons\text{A}},\vec{M}^{2}_{\text{M}\rightleftharpoons\text{A}}\}$, $\{\vec{J}^{12}_{\text{A}\rightleftharpoons\text{A}},\vec{M}^{12}_{\text{A}\rightleftharpoons\text{A}}\}$, $\{\vec{J}_{\text{A}\rightleftharpoons\text{G}},\vec{M}_{\text{A}\rightleftharpoons\text{G}}\}$, $\vec{J}^{1}_{\text{A}}$, and $\vec{J}^{2}_{\text{A}}$.

Similarly to the previous Sec. 7.2, we have the following integral equations

$$\left[\mathcal{E}_2\left(\vec{J}^{2}_{\text{M}\rightleftharpoons\text{A}}-\vec{J}^{12}_{\text{A}\rightleftharpoons\text{A}}+\vec{J}_{\text{A}\rightleftharpoons\text{G}}+\vec{J}^{2}_{\text{A}},\vec{M}^{2}_{\text{M}\rightleftharpoons\text{A}}-\vec{M}^{12}_{\text{A}\rightleftharpoons\text{A}}+\vec{M}_{\text{A}\rightleftharpoons\text{G}}\right)\right]^{\tan}_{\vec{r}^{2}_{\text{A}}\to\vec{r}}$$
$$=\left[\mathcal{E}_2\left(-\vec{J}_{\text{A}\rightleftharpoons\text{G}},-\vec{M}_{\text{A}\rightleftharpoons\text{G}}\right)\right]^{\tan}_{\vec{r}_{\text{PML}}\to\vec{r}} \quad,\quad \vec{r}\in\mathbb{S}_{\text{A}\rightleftharpoons\text{G}} \quad (7\text{-}81\text{a})$$

$$\left[\mathcal{H}_2\left(\vec{J}^{2}_{\text{M}\rightleftharpoons\text{A}}-\vec{J}^{12}_{\text{A}\rightleftharpoons\text{A}}+\vec{J}_{\text{A}\rightleftharpoons\text{G}}+\vec{J}^{2}_{\text{A}},\vec{M}^{2}_{\text{M}\rightleftharpoons\text{A}}-\vec{M}^{12}_{\text{A}\rightleftharpoons\text{A}}+\vec{M}_{\text{A}\rightleftharpoons\text{G}}\right)\right]^{\tan}_{\vec{r}^{2}_{\text{A}}\to\vec{r}}$$
$$=\left[\mathcal{H}_2\left(-\vec{J}_{\text{A}\rightleftharpoons\text{G}},-\vec{M}_{\text{A}\rightleftharpoons\text{G}}\right)\right]^{\tan}_{\vec{r}_{\text{PML}}\to\vec{r}} \quad,\quad \vec{r}\in\mathbb{S}_{\text{A}\rightleftharpoons\text{G}} \quad (7\text{-}81\text{b})$$

about currents $\{\vec{J}^{2}_{\text{M}\rightleftharpoons\text{A}},\vec{M}^{2}_{\text{M}\rightleftharpoons\text{A}}\}$, $\{\vec{J}^{12}_{\text{A}\rightleftharpoons\text{A}},\vec{M}^{12}_{\text{A}\rightleftharpoons\text{A}}\}$, $\{\vec{J}_{\text{A}\rightleftharpoons\text{G}},\vec{M}_{\text{A}\rightleftharpoons\text{G}}\}$, and $\vec{J}^{2}_{\text{A}}$, where $\vec{r}_{\text{PML}}$ belongs to PML and approaches the point $\vec{r}$ on $\mathbb{S}_{\text{A}\rightleftharpoons\text{G}}$.

Based on Eq. (7-69) and the homogeneous tangential electric field boundary condition on $\mathbb{S}_{\text{M}}$, we have the following electric field integral equation

$$\Big[\mathcal{E}_0\Big(\vec{J}_{\text{aux}}-\vec{J}^1_{\text{M}\rightleftharpoons\text{A}}-\vec{J}^2_{\text{M}\rightleftharpoons\text{A}}+\vec{J}_{\text{M}},\vec{M}_{\text{aux}}-\vec{M}^1_{\text{M}\rightleftharpoons\text{A}}-\vec{M}^2_{\text{M}\rightleftharpoons\text{A}}\Big)\Big]^{\tan}_{\vec{r}_{\text{aux}}\to\vec{r}}=0\quad,\quad\vec{r}\in\mathbb{S}_{\text{M}}\quad(7\text{-}82)$$

about currents $\{\vec{J}_{\text{aux}},\vec{M}_{\text{aux}}\}$, $\{\vec{J}^1_{\text{M}\rightleftharpoons\text{A}},\vec{M}^1_{\text{M}\rightleftharpoons\text{A}}\}$, $\{\vec{J}^2_{\text{M}\rightleftharpoons\text{A}},\vec{M}^2_{\text{M}\rightleftharpoons\text{A}}\}$, and $\vec{J}_{\text{M}}$. Based on Eq. (7-73) and the homogeneous tangential electric field boundary condition on $\mathbb{S}^1_{\text{A}}$, we have the following electric field integral equation

$$\Big[\mathcal{E}_1\Big(\vec{J}^1_{\text{M}\rightleftharpoons\text{A}}+\vec{J}^{12}_{\text{A}\rightleftharpoons\text{A}}+\vec{J}^1_{\text{A}},\vec{M}^1_{\text{M}\rightleftharpoons\text{A}}+\vec{M}^{12}_{\text{A}\rightleftharpoons\text{A}}\Big)\Big]^{\tan}_{\vec{r}^1_{\text{A}}\to\vec{r}}=0\quad,\quad\vec{r}\in\mathbb{S}^1_{\text{A}}\qquad(7\text{-}83)$$

about currents $\{\vec{J}^1_{\text{M}\rightleftharpoons\text{A}},\vec{M}^1_{\text{M}\rightleftharpoons\text{A}}\}$, $\{\vec{J}^{12}_{\text{A}\rightleftharpoons\text{A}},\vec{M}^{12}_{\text{A}\rightleftharpoons\text{A}}\}$, and $\vec{J}^1_{\text{A}}$. Based on Eq. (7-75) and the homogeneous tangential electric field boundary condition on $\mathbb{S}^2_{\text{A}}$, we have the following electric field integral equation

$$\Big[\mathcal{E}_2\Big(\vec{J}^2_{\text{M}\rightleftharpoons\text{A}}-\vec{J}^{12}_{\text{A}\rightleftharpoons\text{A}}+\vec{J}_{\text{A}\rightleftharpoons\text{G}}+\vec{J}^2_{\text{A}},\vec{M}^2_{\text{M}\rightleftharpoons\text{A}}-\vec{M}^{12}_{\text{A}\rightleftharpoons\text{A}}+\vec{M}_{\text{A}\rightleftharpoons\text{G}}\Big)\Big]^{\tan}_{\vec{r}^2_{\text{A}}\to\vec{r}}=0\quad,\quad\vec{r}\in\mathbb{S}^2_{\text{A}}\quad(7\text{-}84)$$

about currents $\{\vec{J}^2_{\text{M}\rightleftharpoons\text{A}},\vec{M}^2_{\text{M}\rightleftharpoons\text{A}}\}$, $\{\vec{J}^{12}_{\text{A}\rightleftharpoons\text{A}},\vec{M}^{12}_{\text{A}\rightleftharpoons\text{A}}\}$, $\{\vec{J}_{\text{A}\rightleftharpoons\text{G}},\vec{M}_{\text{A}\rightleftharpoons\text{G}}\}$, and $\vec{J}^2_{\text{A}}$.

The above Eqs. (6-77a)~(6-84) constitute a complete mathematical description for the modal space of the rec-antenna shown in Fig. 7-13. If the currents $\{\vec{J}_{\text{aux}},\vec{M}_{\text{aux}}\}$, $\{\vec{J}^1_{\text{M}\rightleftharpoons\text{A}},\vec{M}^1_{\text{M}\rightleftharpoons\text{A}}\}$, $\{\vec{J}^2_{\text{M}\rightleftharpoons\text{A}},\vec{M}^2_{\text{M}\rightleftharpoons\text{A}}\}$, $\{\vec{J}^{12}_{\text{A}\rightleftharpoons\text{A}},\vec{M}^{12}_{\text{A}\rightleftharpoons\text{A}}\}$, $\{\vec{J}_{\text{A}\rightleftharpoons\text{G}},\vec{M}_{\text{A}\rightleftharpoons\text{G}}\}$, $\vec{J}_{\text{M}}$, $\vec{J}^1_{\text{A}}$, and $\vec{J}^2_{\text{A}}$ are expanded in terms of some proper basis functions, and Eqs. (7-77a), (7-77b), (7-78a), (7-78b), (7-79a), (7-79b), (7-80a), (7-80b), (7-81a), (7-81b), (7-82), (7-83), and (7-84) are tested with $\{\vec{b}_\xi^{\vec{M}_{\text{aux}}}\}$, $\{\vec{b}_\xi^{\vec{J}_{\text{aux}}}\}$, $\{\vec{b}_\xi^{\vec{J}^1_{\text{M}\rightleftharpoons\text{A}}}\}$, $\{\vec{b}_\xi^{\vec{M}^1_{\text{M}\rightleftharpoons\text{A}}}\}$, $\{\vec{b}_\xi^{\vec{J}^2_{\text{M}\rightleftharpoons\text{A}}}\}$, $\{\vec{b}_\xi^{\vec{M}^2_{\text{M}\rightleftharpoons\text{A}}}\}$, $\{\vec{b}_\xi^{\vec{J}^{12}_{\text{A}\rightleftharpoons\text{A}}}\}$, $\{\vec{b}_\xi^{\vec{M}^{12}_{\text{A}\rightleftharpoons\text{A}}}\}$, $\{\vec{b}_\xi^{\vec{J}_{\text{A}\rightleftharpoons\text{G}}}\}$, $\{\vec{b}_\xi^{\vec{M}_{\text{A}\rightleftharpoons\text{G}}}\}$, $\{\vec{b}_\xi^{\vec{J}_{\text{M}}}\}$, $\{\vec{b}_\xi^{\vec{J}^1_{\text{A}}}\}$, and $\{\vec{b}_\xi^{\vec{J}^2_{\text{A}}}\}$ respectively, then the integral equations are immediately discretized into the following matrix equations

$$\begin{aligned}&\overline{\overline{Z}}^{\vec{M}_{\text{aux}}\vec{J}_{\text{aux}}}\cdot\overline{a}^{\vec{J}_{\text{aux}}}+\overline{\overline{Z}}^{\vec{M}_{\text{aux}}\vec{J}^1_{\text{M}\rightleftharpoons\text{A}}}\cdot\overline{a}^{\vec{J}^1_{\text{M}\rightleftharpoons\text{A}}}+\overline{\overline{Z}}^{\vec{M}_{\text{aux}}\vec{J}^2_{\text{M}\rightleftharpoons\text{A}}}\cdot\overline{a}^{\vec{J}^2_{\text{M}\rightleftharpoons\text{A}}}+\overline{\overline{Z}}^{\vec{M}_{\text{aux}}\vec{J}_{\text{M}}}\cdot\overline{a}^{\vec{J}_{\text{M}}}\\&+\overline{\overline{Z}}^{\vec{M}_{\text{aux}}\vec{M}_{\text{aux}}}\cdot\overline{a}^{\vec{M}_{\text{aux}}}+\overline{\overline{Z}}^{\vec{M}_{\text{aux}}\vec{M}^1_{\text{M}\rightleftharpoons\text{A}}}\cdot\overline{a}^{\vec{M}^1_{\text{M}\rightleftharpoons\text{A}}}+\overline{\overline{Z}}^{\vec{M}_{\text{aux}}\vec{M}^2_{\text{M}\rightleftharpoons\text{A}}}\cdot\overline{a}^{\vec{M}^2_{\text{M}\rightleftharpoons\text{A}}}=0\end{aligned}\qquad(7\text{-}85\text{a})$$

$$\begin{aligned}&\overline{\overline{Z}}^{\vec{J}_{\text{aux}}\vec{J}_{\text{aux}}}\cdot\overline{a}^{\vec{J}_{\text{aux}}}+\overline{\overline{Z}}^{\vec{J}_{\text{aux}}\vec{J}^1_{\text{M}\rightleftharpoons\text{A}}}\cdot\overline{a}^{\vec{J}^1_{\text{M}\rightleftharpoons\text{A}}}+\overline{\overline{Z}}^{\vec{J}_{\text{aux}}\vec{J}^2_{\text{M}\rightleftharpoons\text{A}}}\cdot\overline{a}^{\vec{J}^2_{\text{M}\rightleftharpoons\text{A}}}+\overline{\overline{Z}}^{\vec{J}_{\text{aux}}\vec{J}_{\text{M}}}\cdot\overline{a}^{\vec{J}_{\text{M}}}\\&+\overline{\overline{Z}}^{\vec{J}_{\text{aux}}\vec{M}_{\text{aux}}}\cdot\overline{a}^{\vec{M}_{\text{aux}}}+\overline{\overline{Z}}^{\vec{J}_{\text{aux}}\vec{M}^1_{\text{M}\rightleftharpoons\text{A}}}\cdot\overline{a}^{\vec{M}^1_{\text{M}\rightleftharpoons\text{A}}}+\overline{\overline{Z}}^{\vec{J}_{\text{aux}}\vec{M}^2_{\text{M}\rightleftharpoons\text{A}}}\cdot\overline{a}^{\vec{M}^2_{\text{M}\rightleftharpoons\text{A}}}=0\end{aligned}\qquad(7\text{-}85\text{b})$$

and

$$\begin{aligned}&\overline{\overline{Z}}^{\vec{J}^1_{\text{M}\rightleftharpoons\text{A}}\vec{J}_{\text{aux}}}\cdot\overline{a}^{\vec{J}_{\text{aux}}}+\overline{\overline{Z}}^{\vec{J}^1_{\text{M}\rightleftharpoons\text{A}}\vec{J}^1_{\text{M}\rightleftharpoons\text{A}}}\cdot\overline{a}^{\vec{J}^1_{\text{M}\rightleftharpoons\text{A}}}+\overline{\overline{Z}}^{\vec{J}^1_{\text{M}\rightleftharpoons\text{A}}\vec{J}^2_{\text{M}\rightleftharpoons\text{A}}}\cdot\overline{a}^{\vec{J}^2_{\text{M}\rightleftharpoons\text{A}}}+\overline{\overline{Z}}^{\vec{J}^1_{\text{M}\rightleftharpoons\text{A}}\vec{J}^{12}_{\text{A}\rightleftharpoons\text{A}}}\cdot\overline{a}^{\vec{J}^{12}_{\text{A}\rightleftharpoons\text{A}}}\\&+\overline{\overline{Z}}^{\vec{J}^1_{\text{M}\rightleftharpoons\text{A}}\vec{J}_{\text{M}}}\cdot\overline{a}^{\vec{J}_{\text{M}}}+\overline{\overline{Z}}^{\vec{J}^1_{\text{M}\rightleftharpoons\text{A}}\vec{J}^1_{\text{A}}}\cdot\overline{a}^{\vec{J}^1_{\text{A}}}+\overline{\overline{Z}}^{\vec{J}^1_{\text{M}\rightleftharpoons\text{A}}\vec{M}_{\text{aux}}}\cdot\overline{a}^{\vec{M}_{\text{aux}}}+\overline{\overline{Z}}^{\vec{J}^1_{\text{M}\rightleftharpoons\text{A}}\vec{M}^1_{\text{M}\rightleftharpoons\text{A}}}\cdot\overline{a}^{\vec{M}^1_{\text{M}\rightleftharpoons\text{A}}}\\&+\overline{\overline{Z}}^{\vec{J}^1_{\text{M}\rightleftharpoons\text{A}}\vec{M}^2_{\text{M}\rightleftharpoons\text{A}}}\cdot\overline{a}^{\vec{M}^2_{\text{M}\rightleftharpoons\text{A}}}+\overline{\overline{Z}}^{\vec{J}^1_{\text{M}\rightleftharpoons\text{A}}\vec{M}^{12}_{\text{A}\rightleftharpoons\text{A}}}\cdot\overline{a}^{\vec{M}^{12}_{\text{A}\rightleftharpoons\text{A}}}=0\end{aligned}\qquad(7\text{-}86\text{a})$$

$$\begin{aligned}&\overline{\overline{Z}}^{\vec{M}^1_{\text{M}\rightleftharpoons\text{A}}\vec{J}_{\text{aux}}}\cdot\overline{a}^{\vec{J}_{\text{aux}}}+\overline{\overline{Z}}^{\vec{M}^1_{\text{M}\rightleftharpoons\text{A}}\vec{J}^1_{\text{M}\rightleftharpoons\text{A}}}\cdot\overline{a}^{\vec{J}^1_{\text{M}\rightleftharpoons\text{A}}}+\overline{\overline{Z}}^{\vec{M}^1_{\text{M}\rightleftharpoons\text{A}}\vec{J}^2_{\text{M}\rightleftharpoons\text{A}}}\cdot\overline{a}^{\vec{J}^2_{\text{M}\rightleftharpoons\text{A}}}+\overline{\overline{Z}}^{\vec{M}^1_{\text{M}\rightleftharpoons\text{A}}\vec{J}^{12}_{\text{A}\rightleftharpoons\text{A}}}\cdot\overline{a}^{\vec{J}^{12}_{\text{A}\rightleftharpoons\text{A}}}\\&+\overline{\overline{Z}}^{\vec{M}^1_{\text{M}\rightleftharpoons\text{A}}\vec{J}_{\text{M}}}\cdot\overline{a}^{\vec{J}_{\text{M}}}+\overline{\overline{Z}}^{\vec{M}^1_{\text{M}\rightleftharpoons\text{A}}\vec{J}^1_{\text{A}}}\cdot\overline{a}^{\vec{J}^1_{\text{A}}}+\overline{\overline{Z}}^{\vec{M}^1_{\text{M}\rightleftharpoons\text{A}}\vec{M}_{\text{aux}}}\cdot\overline{a}^{\vec{M}_{\text{aux}}}+\overline{\overline{Z}}^{\vec{M}^1_{\text{M}\rightleftharpoons\text{A}}\vec{M}^1_{\text{M}\rightleftharpoons\text{A}}}\cdot\overline{a}^{\vec{M}^1_{\text{M}\rightleftharpoons\text{A}}}\\&+\overline{\overline{Z}}^{\vec{M}^1_{\text{M}\rightleftharpoons\text{A}}\vec{M}^2_{\text{M}\rightleftharpoons\text{A}}}\cdot\overline{a}^{\vec{M}^2_{\text{M}\rightleftharpoons\text{A}}}+\overline{\overline{Z}}^{\vec{M}^1_{\text{M}\rightleftharpoons\text{A}}\vec{M}^{12}_{\text{A}\rightleftharpoons\text{A}}}\cdot\overline{a}^{\vec{M}^{12}_{\text{A}\rightleftharpoons\text{A}}}=0\end{aligned}\qquad(7\text{-}86\text{b})$$

and

$$\begin{aligned}
&\overline{\overline{Z}}^{\vec{J}^2_{\mathrm{M\rightleftharpoons A}}\vec{J}_{\mathrm{aux}}}\cdot\overline{a}^{\vec{J}_{\mathrm{aux}}}+\overline{\overline{Z}}^{\vec{J}^2_{\mathrm{M\rightleftharpoons A}}\vec{J}^1_{\mathrm{M\rightleftharpoons A}}}\cdot\overline{a}^{\vec{J}^1_{\mathrm{M\rightleftharpoons A}}}+\overline{\overline{Z}}^{\vec{J}^2_{\mathrm{M\rightleftharpoons A}}\vec{J}^2_{\mathrm{M\rightleftharpoons A}}}\cdot\overline{a}^{\vec{J}^2_{\mathrm{M\rightleftharpoons A}}}+\overline{\overline{Z}}^{\vec{J}^2_{\mathrm{M\rightleftharpoons A}}\vec{J}^{12}_{\mathrm{A\rightleftharpoons A}}}\cdot\overline{a}^{\vec{J}^{12}_{\mathrm{A\rightleftharpoons A}}}\\
&+\overline{\overline{Z}}^{\vec{J}^2_{\mathrm{M\rightleftharpoons A}}\vec{J}_{\mathrm{A\rightleftharpoons G}}}\cdot\overline{a}^{\vec{J}_{\mathrm{A\rightleftharpoons G}}}+\overline{\overline{Z}}^{\vec{J}^2_{\mathrm{M\rightleftharpoons A}}\vec{J}_{\mathrm{M}}}\cdot\overline{a}^{\vec{J}_{\mathrm{M}}}+\overline{\overline{Z}}^{\vec{J}^2_{\mathrm{M\rightleftharpoons A}}\vec{J}^2_{\mathrm{A}}}\cdot\overline{a}^{\vec{J}^2_{\mathrm{A}}}+\overline{\overline{Z}}^{\vec{J}^2_{\mathrm{M\rightleftharpoons A}}\vec{M}_{\mathrm{aux}}}\cdot\overline{a}^{\vec{M}_{\mathrm{aux}}}+\overline{\overline{Z}}^{\vec{J}^2_{\mathrm{M\rightleftharpoons A}}\vec{M}^1_{\mathrm{M\rightleftharpoons A}}}\cdot\overline{a}^{\vec{M}^1_{\mathrm{M\rightleftharpoons A}}}\\
&+\overline{\overline{Z}}^{\vec{J}^2_{\mathrm{M\rightleftharpoons A}}\vec{M}^2_{\mathrm{M\rightleftharpoons A}}}\cdot\overline{a}^{\vec{M}^2_{\mathrm{M\rightleftharpoons A}}}+\overline{\overline{Z}}^{\vec{J}^2_{\mathrm{M\rightleftharpoons A}}\vec{M}^{12}_{\mathrm{A\rightleftharpoons A}}}\cdot\overline{a}^{\vec{M}^{12}_{\mathrm{A\rightleftharpoons A}}}+\overline{\overline{Z}}^{\vec{J}^2_{\mathrm{M\rightleftharpoons A}}\vec{M}_{\mathrm{A\rightleftharpoons G}}}\cdot\overline{a}^{\vec{M}_{\mathrm{A\rightleftharpoons G}}}=0 \qquad (7\text{-}87\mathrm{a})
\end{aligned}$$

$$\begin{aligned}
&\overline{\overline{Z}}^{\vec{M}^2_{\mathrm{M\rightleftharpoons A}}\vec{J}_{\mathrm{aux}}}\cdot\overline{a}^{\vec{J}_{\mathrm{aux}}}+\overline{\overline{Z}}^{\vec{M}^2_{\mathrm{M\rightleftharpoons A}}\vec{J}^1_{\mathrm{M\rightleftharpoons A}}}\cdot\overline{a}^{\vec{J}^1_{\mathrm{M\rightleftharpoons A}}}+\overline{\overline{Z}}^{\vec{M}^2_{\mathrm{M\rightleftharpoons A}}\vec{J}^2_{\mathrm{M\rightleftharpoons A}}}\cdot\overline{a}^{\vec{J}^2_{\mathrm{M\rightleftharpoons A}}}+\overline{\overline{Z}}^{\vec{M}^2_{\mathrm{M\rightleftharpoons A}}\vec{J}^{12}_{\mathrm{A\rightleftharpoons A}}}\cdot\overline{a}^{\vec{J}^{12}_{\mathrm{A\rightleftharpoons A}}}\\
&+\overline{\overline{Z}}^{\vec{M}^2_{\mathrm{M\rightleftharpoons A}}\vec{J}_{\mathrm{A\rightleftharpoons G}}}\cdot\overline{a}^{\vec{J}_{\mathrm{A\rightleftharpoons G}}}+\overline{\overline{Z}}^{\vec{M}^2_{\mathrm{M\rightleftharpoons A}}\vec{J}_{\mathrm{M}}}\cdot\overline{a}^{\vec{J}_{\mathrm{M}}}+\overline{\overline{Z}}^{\vec{M}^2_{\mathrm{M\rightleftharpoons A}}\vec{J}^2_{\mathrm{A}}}\cdot\overline{a}^{\vec{J}^2_{\mathrm{A}}}+\overline{\overline{Z}}^{\vec{M}^2_{\mathrm{M\rightleftharpoons A}}\vec{M}_{\mathrm{aux}}}\cdot\overline{a}^{\vec{M}_{\mathrm{aux}}}\\
&+\overline{\overline{Z}}^{\vec{M}^2_{\mathrm{M\rightleftharpoons A}}\vec{M}^1_{\mathrm{M\rightleftharpoons A}}}\cdot\overline{a}^{\vec{M}^1_{\mathrm{M\rightleftharpoons A}}}+\overline{\overline{Z}}^{\vec{M}^2_{\mathrm{M\rightleftharpoons A}}\vec{M}^2_{\mathrm{M\rightleftharpoons A}}}\cdot\overline{a}^{\vec{M}^2_{\mathrm{M\rightleftharpoons A}}}+\overline{\overline{Z}}^{\vec{M}^2_{\mathrm{M\rightleftharpoons A}}\vec{M}^{12}_{\mathrm{A\rightleftharpoons A}}}\cdot\overline{a}^{\vec{M}^{12}_{\mathrm{A\rightleftharpoons A}}}\\
&+\overline{\overline{Z}}^{\vec{M}^2_{\mathrm{M\rightleftharpoons A}}\vec{M}_{\mathrm{A\rightleftharpoons G}}}\cdot\overline{a}^{\vec{M}_{\mathrm{A\rightleftharpoons G}}}=0 \qquad (7\text{-}87\mathrm{b})
\end{aligned}$$

and

$$\begin{aligned}
&\overline{\overline{Z}}^{\vec{J}^{12}_{\mathrm{A\rightleftharpoons A}}\vec{J}^1_{\mathrm{M\rightleftharpoons A}}}\cdot\overline{a}^{\vec{J}^1_{\mathrm{M\rightleftharpoons A}}}+\overline{\overline{Z}}^{\vec{J}^{12}_{\mathrm{A\rightleftharpoons A}}\vec{J}^2_{\mathrm{M\rightleftharpoons A}}}\cdot\overline{a}^{\vec{J}^2_{\mathrm{M\rightleftharpoons A}}}+\overline{\overline{Z}}^{\vec{J}^{12}_{\mathrm{A\rightleftharpoons A}}\vec{J}^{12}_{\mathrm{A\rightleftharpoons A}}}\cdot\overline{a}^{\vec{J}^{12}_{\mathrm{A\rightleftharpoons A}}}+\overline{\overline{Z}}^{\vec{J}^{12}_{\mathrm{A\rightleftharpoons A}}\vec{J}_{\mathrm{A\rightleftharpoons G}}}\cdot\overline{a}^{\vec{J}_{\mathrm{A\rightleftharpoons G}}}\\
&+\overline{\overline{Z}}^{\vec{J}^{12}_{\mathrm{A\rightleftharpoons A}}\vec{J}^1_{\mathrm{A}}}\cdot\overline{a}^{\vec{J}^1_{\mathrm{A}}}+\overline{\overline{Z}}^{\vec{J}^{12}_{\mathrm{A\rightleftharpoons A}}\vec{J}^2_{\mathrm{A}}}\cdot\overline{a}^{\vec{J}^2_{\mathrm{A}}}+\overline{\overline{Z}}^{\vec{J}^{12}_{\mathrm{A\rightleftharpoons A}}\vec{M}^1_{\mathrm{M\rightleftharpoons A}}}\cdot\overline{a}^{\vec{M}^1_{\mathrm{M\rightleftharpoons A}}}+\overline{\overline{Z}}^{\vec{J}^{12}_{\mathrm{A\rightleftharpoons A}}\vec{M}^2_{\mathrm{M\rightleftharpoons A}}}\cdot\overline{a}^{\vec{M}^2_{\mathrm{M\rightleftharpoons A}}}\\
&+\overline{\overline{Z}}^{\vec{J}^{12}_{\mathrm{A\rightleftharpoons A}}\vec{M}^{12}_{\mathrm{A\rightleftharpoons A}}}\cdot\overline{a}^{\vec{M}^{12}_{\mathrm{A\rightleftharpoons A}}}+\overline{\overline{Z}}^{\vec{J}^{12}_{\mathrm{A\rightleftharpoons A}}\vec{M}_{\mathrm{A\rightleftharpoons G}}}\cdot\overline{a}^{\vec{M}_{\mathrm{A\rightleftharpoons G}}}=0 \qquad (7\text{-}88\mathrm{a})
\end{aligned}$$

$$\begin{aligned}
&\overline{\overline{Z}}^{\vec{M}^{12}_{\mathrm{A\rightleftharpoons A}}\vec{J}^1_{\mathrm{M\rightleftharpoons A}}}\cdot\overline{a}^{\vec{J}^1_{\mathrm{M\rightleftharpoons A}}}+\overline{\overline{Z}}^{\vec{M}^{12}_{\mathrm{A\rightleftharpoons A}}\vec{J}^2_{\mathrm{M\rightleftharpoons A}}}\cdot\overline{a}^{\vec{J}^2_{\mathrm{M\rightleftharpoons A}}}+\overline{\overline{Z}}^{\vec{M}^{12}_{\mathrm{A\rightleftharpoons A}}\vec{J}^{12}_{\mathrm{A\rightleftharpoons A}}}\cdot\overline{a}^{\vec{J}^{12}_{\mathrm{A\rightleftharpoons A}}}+\overline{\overline{Z}}^{\vec{M}^{12}_{\mathrm{A\rightleftharpoons A}}\vec{J}_{\mathrm{A\rightleftharpoons G}}}\cdot\overline{a}^{\vec{J}_{\mathrm{A\rightleftharpoons G}}}\\
&+\overline{\overline{Z}}^{\vec{M}^{12}_{\mathrm{A\rightleftharpoons A}}\vec{J}^1_{\mathrm{A}}}\cdot\overline{a}^{\vec{J}^1_{\mathrm{A}}}+\overline{\overline{Z}}^{\vec{M}^{12}_{\mathrm{A\rightleftharpoons A}}\vec{J}^2_{\mathrm{A}}}\cdot\overline{a}^{\vec{J}^2_{\mathrm{A}}}+\overline{\overline{Z}}^{\vec{M}^{12}_{\mathrm{A\rightleftharpoons A}}\vec{M}^1_{\mathrm{M\rightleftharpoons A}}}\cdot\overline{a}^{\vec{M}^1_{\mathrm{M\rightleftharpoons A}}}+\overline{\overline{Z}}^{\vec{M}^{12}_{\mathrm{A\rightleftharpoons A}}\vec{M}^2_{\mathrm{M\rightleftharpoons A}}}\cdot\overline{a}^{\vec{M}^2_{\mathrm{M\rightleftharpoons A}}}\\
&+\overline{\overline{Z}}^{\vec{M}^{12}_{\mathrm{A\rightleftharpoons A}}\vec{M}^{12}_{\mathrm{A\rightleftharpoons A}}}\cdot\overline{a}^{\vec{M}^{12}_{\mathrm{A\rightleftharpoons A}}}+\overline{\overline{Z}}^{\vec{M}^{12}_{\mathrm{A\rightleftharpoons A}}\vec{M}_{\mathrm{A\rightleftharpoons G}}}\cdot\overline{a}^{\vec{M}_{\mathrm{A\rightleftharpoons G}}}=0 \qquad (7\text{-}88\mathrm{b})
\end{aligned}$$

and

$$\begin{aligned}
&\overline{\overline{Z}}^{\vec{J}_{\mathrm{A\rightleftharpoons G}}\vec{J}^2_{\mathrm{M\rightleftharpoons A}}}\cdot\overline{a}^{\vec{J}^2_{\mathrm{M\rightleftharpoons A}}}+\overline{\overline{Z}}^{\vec{J}_{\mathrm{A\rightleftharpoons G}}\vec{J}^{12}_{\mathrm{A\rightleftharpoons A}}}\cdot\overline{a}^{\vec{J}^{12}_{\mathrm{A\rightleftharpoons A}}}+\overline{\overline{Z}}^{\vec{J}_{\mathrm{A\rightleftharpoons G}}\vec{J}_{\mathrm{A\rightleftharpoons G}}}\cdot\overline{a}^{\vec{J}_{\mathrm{A\rightleftharpoons G}}}+\overline{\overline{Z}}^{\vec{J}_{\mathrm{A\rightleftharpoons G}}\vec{J}^2_{\mathrm{A}}}\cdot\overline{a}^{\vec{J}^2_{\mathrm{A}}}\\
&+\overline{\overline{Z}}^{\vec{J}_{\mathrm{A\rightleftharpoons G}}\vec{M}^2_{\mathrm{M\rightleftharpoons A}}}\cdot\overline{a}^{\vec{M}^2_{\mathrm{M\rightleftharpoons A}}}+\overline{\overline{Z}}^{\vec{J}_{\mathrm{A\rightleftharpoons G}}\vec{M}^{12}_{\mathrm{A\rightleftharpoons A}}}\cdot\overline{a}^{\vec{M}^{12}_{\mathrm{A\rightleftharpoons A}}}+\overline{\overline{Z}}^{\vec{J}_{\mathrm{A\rightleftharpoons G}}\vec{M}_{\mathrm{A\rightleftharpoons G}}}\cdot\overline{a}^{\vec{M}_{\mathrm{A\rightleftharpoons G}}}=0 \qquad (7\text{-}89\mathrm{a})
\end{aligned}$$

$$\begin{aligned}
&\overline{\overline{Z}}^{\vec{M}_{\mathrm{A\rightleftharpoons G}}\vec{J}^2_{\mathrm{M\rightleftharpoons A}}}\cdot\overline{a}^{\vec{J}^2_{\mathrm{M\rightleftharpoons A}}}+\overline{\overline{Z}}^{\vec{M}_{\mathrm{A\rightleftharpoons G}}\vec{J}^{12}_{\mathrm{A\rightleftharpoons A}}}\cdot\overline{a}^{\vec{J}^{12}_{\mathrm{A\rightleftharpoons A}}}+\overline{\overline{Z}}^{\vec{M}_{\mathrm{A\rightleftharpoons G}}\vec{J}_{\mathrm{A\rightleftharpoons G}}}\cdot\overline{a}^{\vec{J}_{\mathrm{A\rightleftharpoons G}}}+\overline{\overline{Z}}^{\vec{M}_{\mathrm{A\rightleftharpoons G}}\vec{J}^2_{\mathrm{A}}}\cdot\overline{a}^{\vec{J}^2_{\mathrm{A}}}\\
&+\overline{\overline{Z}}^{\vec{M}_{\mathrm{A\rightleftharpoons G}}\vec{M}^2_{\mathrm{M\rightleftharpoons A}}}\cdot\overline{a}^{\vec{M}^2_{\mathrm{M\rightleftharpoons A}}}+\overline{\overline{Z}}^{\vec{M}_{\mathrm{A\rightleftharpoons G}}\vec{M}^{12}_{\mathrm{A\rightleftharpoons A}}}\cdot\overline{a}^{\vec{M}^{12}_{\mathrm{A\rightleftharpoons A}}}+\overline{\overline{Z}}^{\vec{M}_{\mathrm{A\rightleftharpoons G}}\vec{M}_{\mathrm{A\rightleftharpoons G}}}\cdot\overline{a}^{\vec{M}_{\mathrm{A\rightleftharpoons G}}}=0 \qquad (7\text{-}89\mathrm{b})
\end{aligned}$$

and

$$\begin{aligned}
&\overline{\overline{Z}}^{\vec{J}_{\mathrm{M}}\vec{J}_{\mathrm{aux}}}\cdot\overline{a}^{\vec{J}_{\mathrm{aux}}}+\overline{\overline{Z}}^{\vec{J}_{\mathrm{M}}\vec{J}^1_{\mathrm{M\rightleftharpoons A}}}\cdot\overline{a}^{\vec{J}^1_{\mathrm{M\rightleftharpoons A}}}+\overline{\overline{Z}}^{\vec{J}_{\mathrm{M}}\vec{J}^2_{\mathrm{M\rightleftharpoons A}}}\cdot\overline{a}^{\vec{J}^2_{\mathrm{M\rightleftharpoons A}}}+\overline{\overline{Z}}^{\vec{J}_{\mathrm{M}}\vec{J}_{\mathrm{M}}}\cdot\overline{a}^{\vec{J}_{\mathrm{M}}}+\overline{\overline{Z}}^{\vec{J}_{\mathrm{M}}\vec{M}_{\mathrm{aux}}}\cdot\overline{a}^{\vec{M}_{\mathrm{aux}}}\\
&+\overline{\overline{Z}}^{\vec{J}_{\mathrm{M}}\vec{M}^1_{\mathrm{M\rightleftharpoons A}}}\cdot\overline{a}^{\vec{M}^1_{\mathrm{M\rightleftharpoons A}}}+\overline{\overline{Z}}^{\vec{J}_{\mathrm{M}}\vec{M}^2_{\mathrm{M\rightleftharpoons A}}}\cdot\overline{a}^{\vec{M}^2_{\mathrm{M\rightleftharpoons A}}}=0 \qquad (7\text{-}90)
\end{aligned}$$

and

$$\begin{aligned}
&\overline{\overline{Z}}^{\vec{J}^1_{\mathrm{A}}\vec{J}^1_{\mathrm{M\rightleftharpoons A}}}\cdot\overline{a}^{\vec{J}^1_{\mathrm{M\rightleftharpoons A}}}+\overline{\overline{Z}}^{\vec{J}^1_{\mathrm{A}}\vec{J}^{12}_{\mathrm{A\rightleftharpoons A}}}\cdot\overline{a}^{\vec{J}^{12}_{\mathrm{A\rightleftharpoons A}}}+\overline{\overline{Z}}^{\vec{J}^1_{\mathrm{A}}\vec{J}^1_{\mathrm{A}}}\cdot\overline{a}^{\vec{J}^1_{\mathrm{A}}}+\overline{\overline{Z}}^{\vec{J}^1_{\mathrm{A}}\vec{M}^1_{\mathrm{M\rightleftharpoons A}}}\cdot\overline{a}^{\vec{M}^1_{\mathrm{M\rightleftharpoons A}}}\\
&+\overline{\overline{Z}}^{\vec{J}^1_{\mathrm{A}}\vec{M}^{12}_{\mathrm{A\rightleftharpoons A}}}\cdot\overline{a}^{\vec{M}^{12}_{\mathrm{A\rightleftharpoons A}}}=0 \qquad (7\text{-}91)
\end{aligned}$$

and

$$\begin{aligned}
&\overline{\overline{Z}}^{\vec{J}^2_{\mathrm{A}}\vec{J}^2_{\mathrm{M\rightleftharpoons A}}}\cdot\overline{a}^{\vec{J}^2_{\mathrm{M\rightleftharpoons A}}}+\overline{\overline{Z}}^{\vec{J}^2_{\mathrm{A}}\vec{J}^{12}_{\mathrm{A\rightleftharpoons A}}}\cdot\overline{a}^{\vec{J}^{12}_{\mathrm{A\rightleftharpoons A}}}+\overline{\overline{Z}}^{\vec{J}^2_{\mathrm{A}}\vec{J}_{\mathrm{A\rightleftharpoons G}}}\cdot\overline{a}^{\vec{J}_{\mathrm{A\rightleftharpoons G}}}+\overline{\overline{Z}}^{\vec{J}^2_{\mathrm{A}}\vec{J}^2_{\mathrm{A}}}\cdot\overline{a}^{\vec{J}^2_{\mathrm{A}}}+\overline{\overline{Z}}^{\vec{J}^2_{\mathrm{A}}\vec{M}^2_{\mathrm{M\rightleftharpoons A}}}\cdot\overline{a}^{\vec{M}^2_{\mathrm{M\rightleftharpoons A}}}\\
&+\overline{\overline{Z}}^{\vec{J}^2_{\mathrm{A}}\vec{M}^{12}_{\mathrm{A\rightleftharpoons A}}}\cdot\overline{a}^{\vec{M}^{12}_{\mathrm{A\rightleftharpoons A}}}+\overline{\overline{Z}}^{\vec{J}^2_{\mathrm{A}}\vec{M}_{\mathrm{A\rightleftharpoons G}}}\cdot\overline{a}^{\vec{M}_{\mathrm{A\rightleftharpoons G}}}=0 \qquad (7\text{-}92)
\end{aligned}$$

The formulations used to calculate the elements of the matrices in the above matrix equations are similar to the ones given in Eqs. (7-13a)~(7-18f), and they are explicitly given in the App. D9 of this report. By employing the above matrix equations, we propose two different schemes for mathematically describing modal space as below.

Employing the above Eqs. (7-85a)~(7-92), we can obtain the following transformation

$$\begin{bmatrix} \bar{a}^{\vec{J}_{\text{aux}}} \\ \bar{a}^{\vec{J}^{1}_{\text{M}\rightleftharpoons\text{A}}} \\ \bar{a}^{\vec{J}^{2}_{\text{M}\rightleftharpoons\text{A}}} \\ \bar{a}^{\vec{J}^{12}_{\text{A}\rightleftharpoons\text{A}}} \\ \bar{a}^{\vec{J}_{\text{A}\rightleftharpoons\text{G}}} \\ \bar{a}^{\vec{J}_{\text{M}}} \\ \bar{a}^{\vec{J}^{1}_{\text{A}}} \\ \bar{a}^{\vec{J}^{2}_{\text{A}}} \\ \bar{a}^{\vec{M}_{\text{aux}}} \\ \bar{a}^{\vec{M}^{1}_{\text{M}\rightleftharpoons\text{A}}} \\ \bar{a}^{\vec{M}^{2}_{\text{M}\rightleftharpoons\text{A}}} \\ \bar{a}^{\vec{M}^{12}_{\text{A}\rightleftharpoons\text{A}}} \\ \bar{a}^{\vec{M}_{\text{A}\rightleftharpoons\text{G}}} \end{bmatrix} = \bar{a}^{\text{AV}} = \bar{\bar{T}} \cdot \bar{a} \tag{6-93}$$

and the calculation formulation for transformation matrix $\bar{\bar{T}}$ is given in the App. D9 of this report.

### 7.4.4 Power Transport Theorem and Input Power Operator

Applying the results obtained in Chap. 2 to the rec-antenna shown in Fig. 7-13, we immediately have the following PTT for the rec-antenna

$$P_{\text{aux}\rightleftharpoons\text{V}} = \underbrace{\overbrace{\underbrace{P_{\text{A}}^{1\text{dis}} + P_{\text{A}}^{2\text{dis}}}_{P_{\text{A}}^{\text{dis}}} + P_{\text{A}\rightleftharpoons\text{G}} + j\,\underbrace{P_{\text{A}}^{2\text{sto}} + P_{\text{A}}^{1\text{sto}}}_{P_{\text{A}}^{\text{sto}}}}^{P^{1}_{\text{M}\rightleftharpoons\text{A}} + P^{2}_{\text{M}\rightleftharpoons\text{A}}}}_{P_{\text{M}\rightleftharpoons\text{A}}} + j\,P_{\text{V}}^{\text{sto}} \tag{7-94}$$

where $P^{1}_{\text{M}\rightleftharpoons\text{A}}$ and $P^{2}_{\text{M}\rightleftharpoons\text{A}}$ are the input powers inputted into $\mathbb{V}_{\text{A}}^{1}$ and $\mathbb{V}_{\text{A}}^{2}$ from medium, and $P_{\text{A}}^{1\text{dis}}$ and $P_{\text{A}}^{2\text{dis}}$ are the powers dissipated in $\mathbb{V}_{\text{A}}^{1}$ and $\mathbb{V}_{\text{A}}^{2}$, and $P_{\text{A}}^{2\text{sto}}$ and $P_{\text{A}}^{1\text{sto}}$ are the powers corresponding to the stored energies in $\mathbb{V}_{\text{A}}^{1}$ and $\mathbb{V}_{\text{A}}^{2}$, and $P_{\text{A}\rightleftharpoons\text{G}}$ is the output power outputted from rec-antenna to loading structure.

The above-mentioned various powers are as follows:

$$P_{\text{aux}\rightleftharpoons\text{V}} = (1/2)\iint_{\mathbb{S}_{\text{aux}}} \left(\vec{E} \times \vec{H}^{\dagger}\right) \cdot \hat{n}_{\rightarrow\text{V}}\, dS \tag{7-95a}$$

$$P_{\text{M}\rightleftharpoons\text{A}} = \underbrace{(1/2)\iint_{\mathbb{S}^1_{\text{M}\rightleftharpoons\text{A}}}\left(\vec{E}\times\vec{H}^\dagger\right)\cdot\hat{n}^1_{\rightarrow\text{A}}dS}_{P^1_{\text{M}\rightleftharpoons\text{A}}} + \underbrace{(1/2)\iint_{\mathbb{S}^2_{\text{M}\rightleftharpoons\text{A}}}\left(\vec{E}\times\vec{H}^\dagger\right)\cdot\hat{n}^2_{\rightarrow\text{A}}dS}_{P^2_{\text{M}\rightleftharpoons\text{A}}} \tag{7-95b}$$

$$P^{\text{dis}}_{\text{A}} = \underbrace{(1/2)\left\langle\vec{\vec{\sigma}}_1\cdot\vec{E},\vec{E}\right\rangle_{\mathbb{V}^1_{\text{A}}}}_{P^{1\,\text{dis}}_{\text{A}}} + \underbrace{(1/2)\left\langle\vec{\vec{\sigma}}_2\cdot\vec{E},\vec{E}\right\rangle_{\mathbb{V}^2_{\text{A}}}}_{P^{2\,\text{dis}}_{\text{A}}} \tag{7-95c}$$

$$P_{\text{A}\rightleftharpoons\text{G}} = (1/2)\iint_{\mathbb{S}_{\text{A}\rightleftharpoons\text{G}}}\left(\vec{E}\times\vec{H}^\dagger\right)\cdot\hat{n}_{\rightarrow\text{PML}}dS \tag{7-95d}$$

$$\begin{aligned} P^{\text{sto}}_{\text{A}} = & \overbrace{2\omega\left[(1/4)\left\langle\vec{H},\vec{\vec{\mu}}_1\cdot\vec{H}\right\rangle_{\mathbb{V}^1_{\text{A}}} - (1/4)\left\langle\vec{\vec{\varepsilon}}_1\cdot\vec{E},\vec{E}\right\rangle_{\mathbb{V}^1_{\text{A}}}\right]}^{P^{1\,\text{sto}}_{\text{A}}} \\ & + \underbrace{2\omega\left[(1/4)\left\langle\vec{H},\vec{\vec{\mu}}_2\cdot\vec{H}\right\rangle_{\mathbb{V}^2_{\text{A}}} - (1/4)\left\langle\vec{\vec{\varepsilon}}_2\cdot\vec{E},\vec{E}\right\rangle_{\mathbb{V}^2_{\text{A}}}\right]}_{P^{2\,\text{sto}}_{\text{A}}} \end{aligned} \tag{7-95e}$$

where $\hat{n}_{\rightarrow\text{V}}$ is normal direction of $\mathbb{S}_{\text{aux}}$ and points to $\mathbb{V}$, and $\hat{n}^{1/2}_{\rightarrow\text{A}}$ is the normal direction of $\mathbb{S}^{1/2}_{\text{M}\rightleftharpoons\text{A}}$ and points to $\mathbb{V}^{1/2}_{\text{A}}$, and $\hat{n}_{\rightarrow\text{PML}}$ is the normal direction of $\mathbb{S}_{\text{A}\rightleftharpoons\text{G}}$ and points to PML.

Based on Eqs. (7-71a)~(7-72b) and the tangential continuity of the $\{\vec{E},\vec{H}\}$ on $\mathbb{S}_{\text{M}\rightleftharpoons\text{A}}$, the IPO $P_{\text{M}\rightleftharpoons\text{A}}$ given in Eq. (7-95b) can be alternatively written as follows:

$$\begin{aligned} P_{\text{M}\rightleftharpoons\text{A}} = & (1/2)\left\langle\hat{n}^1_{\rightarrow\text{A}}\times\vec{J}^1_{\text{M}\rightleftharpoons\text{A}},\vec{M}^1_{\text{M}\rightleftharpoons\text{A}}\right\rangle_{\mathbb{S}^1_{\text{M}\rightleftharpoons\text{A}}} + (1/2)\left\langle\hat{n}^2_{\rightarrow\text{A}}\times\vec{J}^2_{\text{M}\rightleftharpoons\text{A}},\vec{M}^2_{\text{M}\rightleftharpoons\text{A}}\right\rangle_{\mathbb{S}^2_{\text{M}\rightleftharpoons\text{A}}} \\ = & -\frac{1}{2}\left\langle\vec{J}^1_{\text{M}\rightleftharpoons\text{A}},\mathcal{E}_0\left(\vec{J}_{\text{aux}}-\vec{J}^1_{\text{M}\rightleftharpoons\text{A}}-\vec{J}^2_{\text{M}\rightleftharpoons\text{A}}+\vec{J}_{\text{M}},\vec{M}_{\text{aux}}-\vec{M}^1_{\text{M}\rightleftharpoons\text{A}}-\vec{M}^2_{\text{M}\rightleftharpoons\text{A}}\right)\right\rangle_{\tilde{\mathbb{S}}^1_{\text{M}\rightleftharpoons\text{A}}} \\ & -\frac{1}{2}\left\langle\vec{J}^2_{\text{M}\rightleftharpoons\text{A}},\mathcal{E}_0\left(\vec{J}_{\text{aux}}-\vec{J}^1_{\text{M}\rightleftharpoons\text{A}}-\vec{J}^2_{\text{M}\rightleftharpoons\text{A}}+\vec{J}_{\text{M}},\vec{M}_{\text{aux}}-\vec{M}^1_{\text{M}\rightleftharpoons\text{A}}-\vec{M}^2_{\text{M}\rightleftharpoons\text{A}}\right)\right\rangle_{\tilde{\mathbb{S}}^2_{\text{M}\rightleftharpoons\text{A}}} \\ = & -\frac{1}{2}\left\langle\vec{M}^1_{\text{M}\rightleftharpoons\text{A}},\mathcal{H}_0\left(\vec{J}_{\text{aux}}-\vec{J}^1_{\text{M}\rightleftharpoons\text{A}}-\vec{J}^2_{\text{M}\rightleftharpoons\text{A}}+\vec{J}_{\text{M}},\vec{M}_{\text{aux}}-\vec{M}^1_{\text{M}\rightleftharpoons\text{A}}-\vec{M}^2_{\text{M}\rightleftharpoons\text{A}}\right)\right\rangle^\dagger_{\tilde{\mathbb{S}}^1_{\text{M}\rightleftharpoons\text{A}}} \\ & -\frac{1}{2}\left\langle\vec{M}^2_{\text{M}\rightleftharpoons\text{A}},\mathcal{H}_0\left(\vec{J}_{\text{aux}}-\vec{J}^1_{\text{M}\rightleftharpoons\text{A}}-\vec{J}^2_{\text{M}\rightleftharpoons\text{A}}+\vec{J}_{\text{M}},\vec{M}_{\text{aux}}-\vec{M}^1_{\text{M}\rightleftharpoons\text{A}}-\vec{M}^2_{\text{M}\rightleftharpoons\text{A}}\right)\right\rangle^\dagger_{\tilde{\mathbb{S}}^2_{\text{M}\rightleftharpoons\text{A}}} \end{aligned} \tag{7-96}$$

Here, the right-hand side of the first equality is the current form of IPO, and the right-hand sides of the second and third equalities are the interaction forms of IPO.

By discretizing IPO (7-96) and utilizing transformation (7-93), we derive the following matrix form of the IPO

$$P_{\text{M}\rightleftharpoons\text{A}} = \bar{a}^\dagger\cdot\underbrace{\left(\bar{\bar{T}}^\dagger\cdot\bar{\bar{P}}^{\text{AV}}_{\text{M}\rightleftharpoons\text{A}}\cdot\bar{\bar{T}}\right)}_{\bar{\bar{P}}_{\text{M}\rightleftharpoons\text{A}}}\cdot\bar{a} \tag{7-97}$$

and the formulation for calculating the quadratic matrix $\bar{\bar{P}}^{\text{AV}}_{\text{M}\rightleftharpoons\text{A}}$ is given in the App. D9 of this report.

### 7.4.5 Input-Power-Decoupled Modes

The IP-DMs in modal space can be derived from solving the modal decoupling equation $\overline{\overline{P}}{}_{\mathrm{M}\rightleftharpoons\mathrm{A}}^{-} \cdot \bar{\alpha}_{\xi} = \theta_{\xi} \overline{\overline{P}}{}_{\mathrm{M}\rightleftharpoons\mathrm{A}}^{+} \cdot \bar{\alpha}_{\xi}$ defined on modal space, where $\overline{\overline{P}}{}_{\mathrm{M}\rightleftharpoons\mathrm{A}}^{+}$ and $\overline{\overline{P}}{}_{\mathrm{M}\rightleftharpoons\mathrm{A}}^{-}$ are the positive and negative Hermitian parts of matrix $\overline{\overline{P}}_{\mathrm{M}\rightleftharpoons\mathrm{A}}$. If some derived modes $\{\bar{\alpha}_1, \bar{\alpha}_2, \cdots, \bar{\alpha}_d\}$ are *d*-order degenerate, then the Gram-Schmidt orthogonalization process given in previous Sec. 7.2.4.1 is necessary, and it is not repeated here.

The modal vectors constructed above satisfy the following decoupling relation

$$\frac{1}{2}\iint_{\mathbb{S}_{\mathrm{M}\rightleftharpoons\mathrm{A}}^{1}} \left(\vec{E}_{\zeta} \times \vec{H}_{\xi}^{\dagger}\right) \cdot \hat{n}_{\to\mathrm{A}}^{1} dS + \frac{1}{2}\iint_{\mathbb{S}_{\mathrm{M}\rightleftharpoons\mathrm{A}}^{2}} \left(\vec{E}_{\zeta} \times \vec{H}_{\xi}^{\dagger}\right) \cdot \hat{n}_{\to\mathrm{A}}^{2} dS = \left(1 + j\,\theta_{\xi}\right) \delta_{\xi\zeta} \quad (7\text{-}98)$$

and the relation implies that **the IP-DMs don't have net energy coupling in one period**. By employing the decoupling relation, we have the following Parseval's identity

$$\sum_{\xi} \left|c_{\xi}\right|^{2} = \frac{1}{T}\int_{t_0}^{t_0+T} \left[\iint_{\mathbb{S}_{\mathrm{M}\rightleftharpoons\mathrm{A}}^{1}} \left(\vec{\mathcal{E}} \times \vec{\mathcal{H}}\right) \cdot \hat{n}_{\to\mathrm{A}}^{1} dS + \iint_{\mathbb{S}_{\mathrm{M}\rightleftharpoons\mathrm{A}}^{2}} \left(\vec{\mathcal{E}} \times \vec{\mathcal{H}}\right) \cdot \hat{n}_{\to\mathrm{A}}^{2} dS\right] dt \quad (7\text{-}99)$$

in which $\{\vec{\mathcal{E}}, \vec{\mathcal{H}}\}$ are the time-domain fields, and the expansion coefficients $c_{\xi}$ have the explicit expression that $c_{\xi} = -(1/2) < \vec{J}_{\mathrm{M}\rightleftharpoons\mathrm{A}}^{1;\xi} + \vec{J}_{\mathrm{M}\rightleftharpoons\mathrm{A}}^{2;\xi}, \vec{E} >_{\mathbb{S}_{\mathrm{M}\rightleftharpoons\mathrm{A}}^{1} \cup \mathbb{S}_{\mathrm{M}\rightleftharpoons\mathrm{A}}^{2}} \Big/ (1 + j\,\theta_{\xi})$ $= -(1/2) < \vec{H}, \vec{M}_{\mathrm{M}\rightleftharpoons\mathrm{A}}^{1;\xi} + \vec{M}_{\mathrm{M}\rightleftharpoons\mathrm{A}}^{2;\xi} >_{\mathbb{S}_{\mathrm{M}\rightleftharpoons\mathrm{A}}^{1} \cup \mathbb{S}_{\mathrm{M}\rightleftharpoons\mathrm{A}}^{2}} \Big/ (1 + j\,\theta_{\xi})$, where $\{\vec{E}, \vec{H}\}$ is a previously known field distributing on input port $\mathbb{S}_{\mathrm{M}\rightleftharpoons\mathrm{A}}^{1} \cup \mathbb{S}_{\mathrm{M}\rightleftharpoons\mathrm{A}}^{2}$.

Just like the metallic rec-antenna discussed in Sec. 7.2.4.3, we can also define the modal significance $\mathrm{MS}_{\xi} = 1 \big/ |1 + j\,\theta_{\xi}|$, and modal input impedance

$$\begin{aligned} Z_{\mathrm{M}\rightleftharpoons\mathrm{A}}^{\xi} &= \frac{P_{\mathrm{M}\rightleftharpoons\mathrm{A}}^{\xi}}{(1/2)\left\langle \vec{J}_{\mathrm{M}\rightleftharpoons\mathrm{A}}^{1;\xi} + \vec{J}_{\mathrm{M}\rightleftharpoons\mathrm{A}}^{2;\xi}, \vec{J}_{\mathrm{M}\rightleftharpoons\mathrm{A}}^{1;\xi} + \vec{J}_{\mathrm{M}\rightleftharpoons\mathrm{A}}^{2;\xi} \right\rangle_{\mathbb{S}_{\mathrm{M}\rightleftharpoons\mathrm{A}}^{1} \cup \mathbb{S}_{\mathrm{M}\rightleftharpoons\mathrm{A}}^{2}}} \\ &= \underbrace{\mathrm{Re}\left\{Z_{\mathrm{M}\rightleftharpoons\mathrm{A}}^{\xi}\right\}}_{R_{\mathrm{M}\rightleftharpoons\mathrm{A}}^{\xi}} + j\, \underbrace{\mathrm{Im}\left\{Z_{\mathrm{M}\rightleftharpoons\mathrm{A}}^{\xi}\right\}}_{X_{\mathrm{M}\rightleftharpoons\mathrm{A}}^{\xi}} \end{aligned} \quad (7\text{-}100\mathrm{a})$$

and modal input admittance

$$\begin{aligned} Y_{\mathrm{M}\rightleftharpoons\mathrm{A}}^{\xi} &= \frac{P_{\mathrm{M}\rightleftharpoons\mathrm{A}}^{\xi}}{(1/2)\left\langle \vec{M}_{\mathrm{M}\rightleftharpoons\mathrm{A}}^{1;\xi} + \vec{M}_{\mathrm{M}\rightleftharpoons\mathrm{A}}^{2;\xi}, \vec{M}_{\mathrm{M}\rightleftharpoons\mathrm{A}}^{1;\xi} + \vec{M}_{\mathrm{M}\rightleftharpoons\mathrm{A}}^{2;\xi} \right\rangle_{\mathbb{S}_{\mathrm{M}\rightleftharpoons\mathrm{A}}^{1} \cup \mathbb{S}_{\mathrm{M}\rightleftharpoons\mathrm{A}}^{2}}} \\ &= \underbrace{\mathrm{Re}\left\{Y_{\mathrm{M}\rightleftharpoons\mathrm{A}}^{\xi}\right\}}_{G_{\mathrm{M}\rightleftharpoons\mathrm{A}}^{\xi}} + j\, \underbrace{\mathrm{Im}\left\{Y_{\mathrm{M}\rightleftharpoons\mathrm{A}}^{\xi}\right\}}_{B_{\mathrm{M}\rightleftharpoons\mathrm{A}}^{\xi}} \end{aligned} \quad (7\text{-}100\mathrm{b})$$

to quantitatively describe the modal features of the rec-antenna shown in Fig. 7-13.

### 7.4.6 Numerical Examples Corresponding to Typical Structures

Similarly to the previous Sec. 7.3.5, this Sec. 7.4.6 also uses the $P_{\mathrm{aux}\rightleftharpoons\mathrm{V}}$ as the generating operator of IP-DMs.

Here, we consider a metallic waveguide loaded cylindrical *dielectric resonator antenna* (*DRA*) mounted on a circular ground plane, and the rec-antenna is excited by a elliptical auxiliary surface, as shown in the following Fig. 7-14.

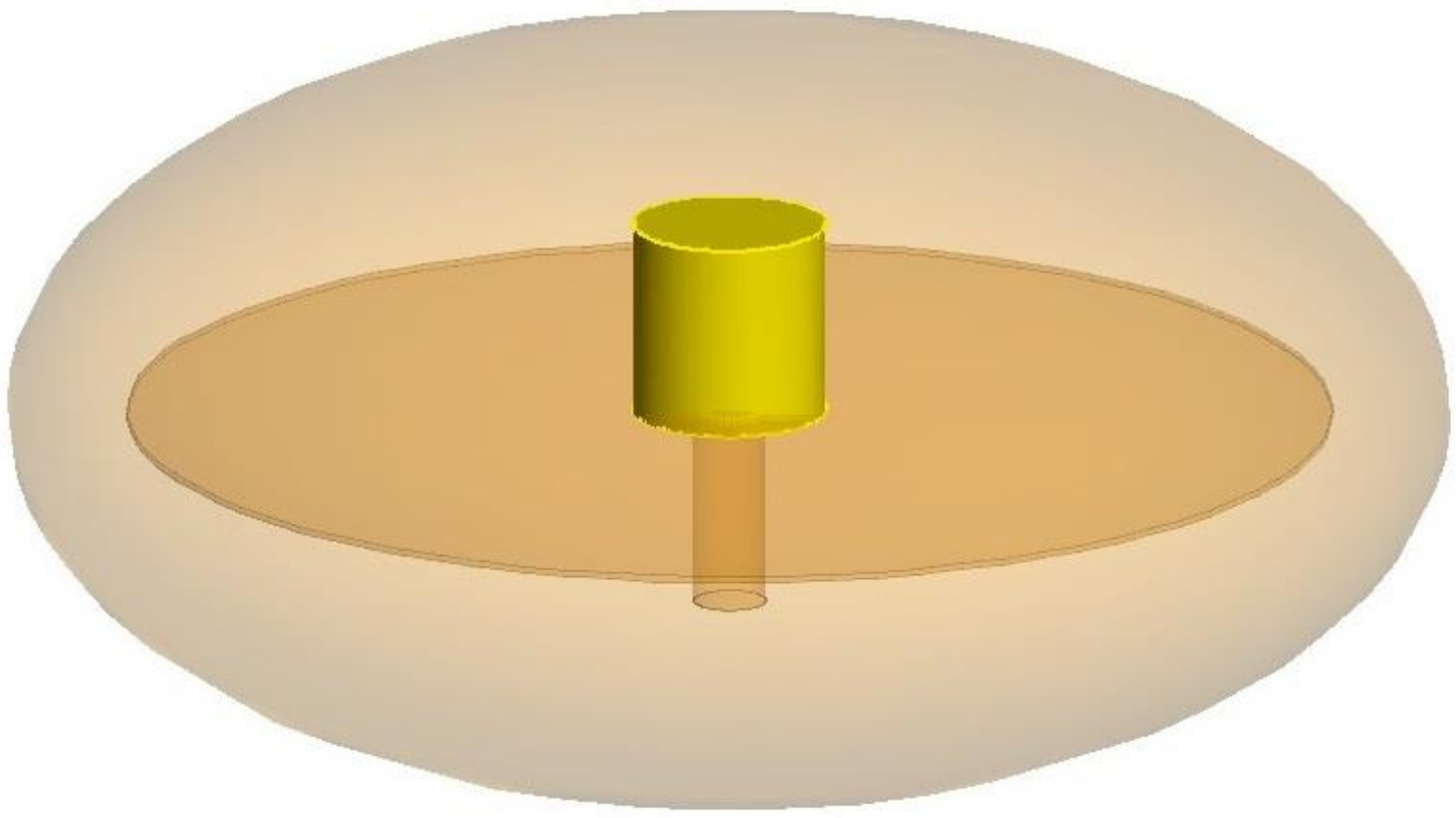

Figure 7-14 Geometry of a metallic waveguide loaded cylindrical DRA mounted on a circular ground plane

The geometrical sizes of the rec-antenna and auxiliary surface are shown in the following Fig. 7-15.

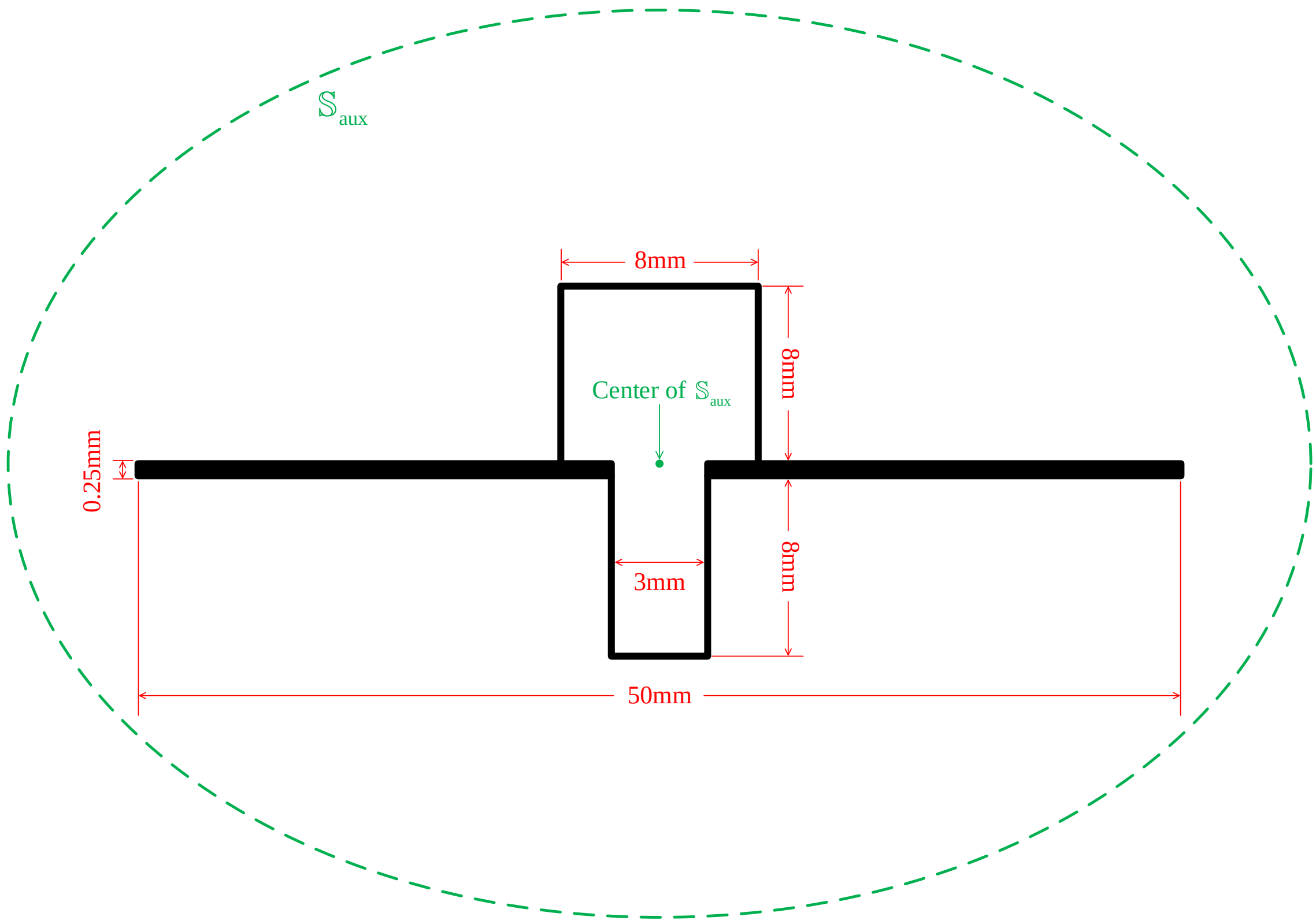

Figure 7-15 Geometrical sizes of the rec-antenna and auxiliary surface $\mathbb{S}_{aux}$. Here, $\mathbb{S}_{aux}$ is elliptical sphere with two 60mm major axises and a 30mm minor axis

Using the JE-DoJ-based IPO, we calculate the IP-DMs of the receiving DRA. The input resistance curves of the first several typical IP-DMs are plotted in Fig. 7-16.

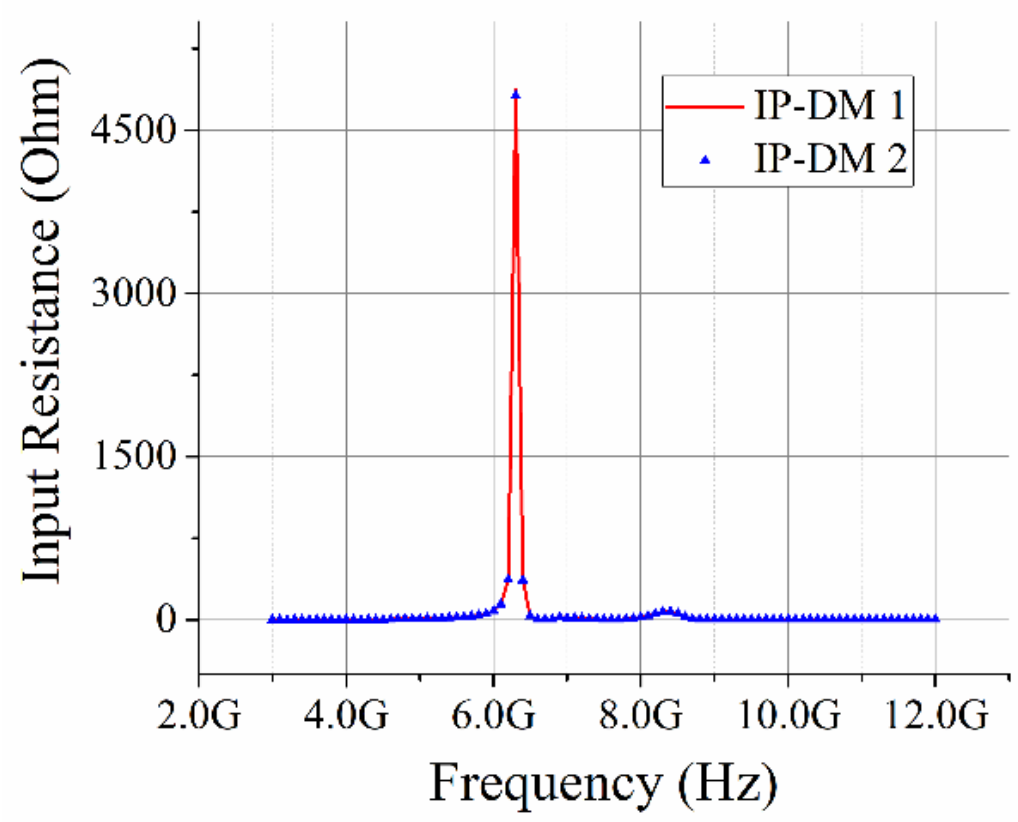

Figure 7-16 Some typical modal input resistance curves

Taking the IP-DM 1 shown in Fig. 7-16 as an example, its modal electric and magnetic currents distributing on $\mathbb{S}_{aux}$ are illustrated in the following Fig. 7-17

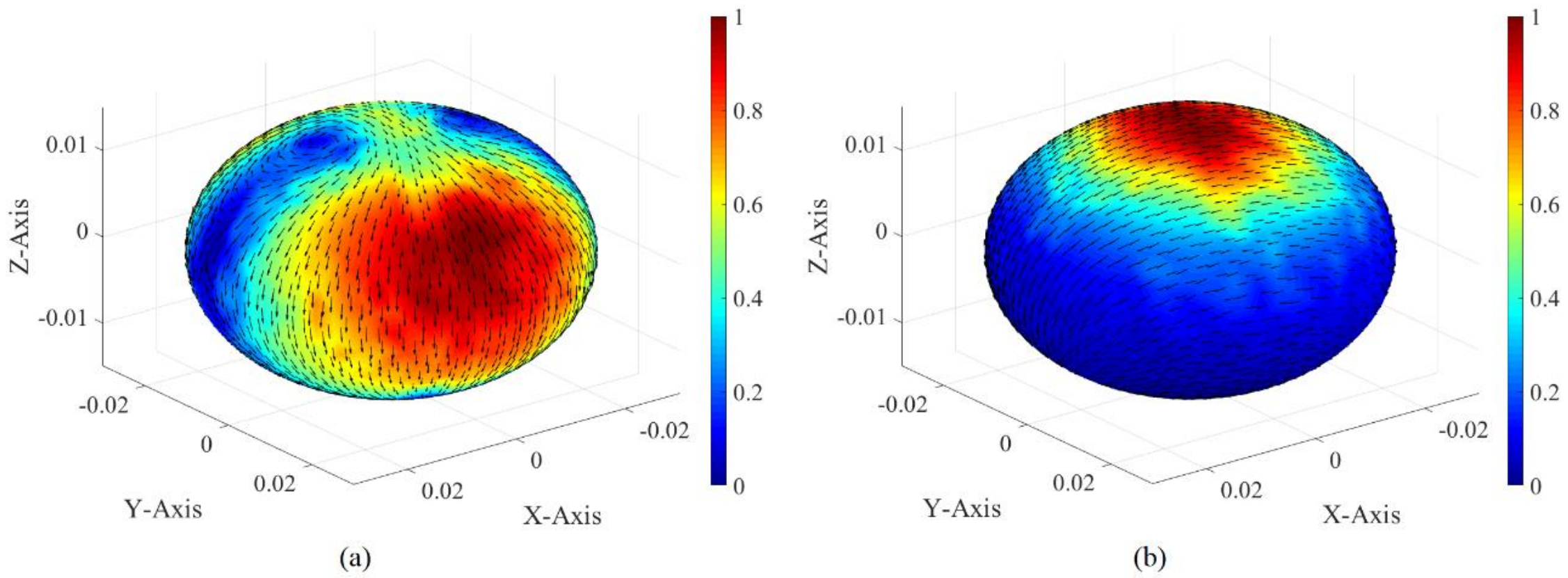

Figure 7-17 Modal (a) electric and (b) magnetic currents distributing on $\mathbb{S}_{aux}$

and its modal electric and magnetic currents on $\mathbb{S}_{M\rightleftharpoons A}$ are illustrated in Fig. 7-18

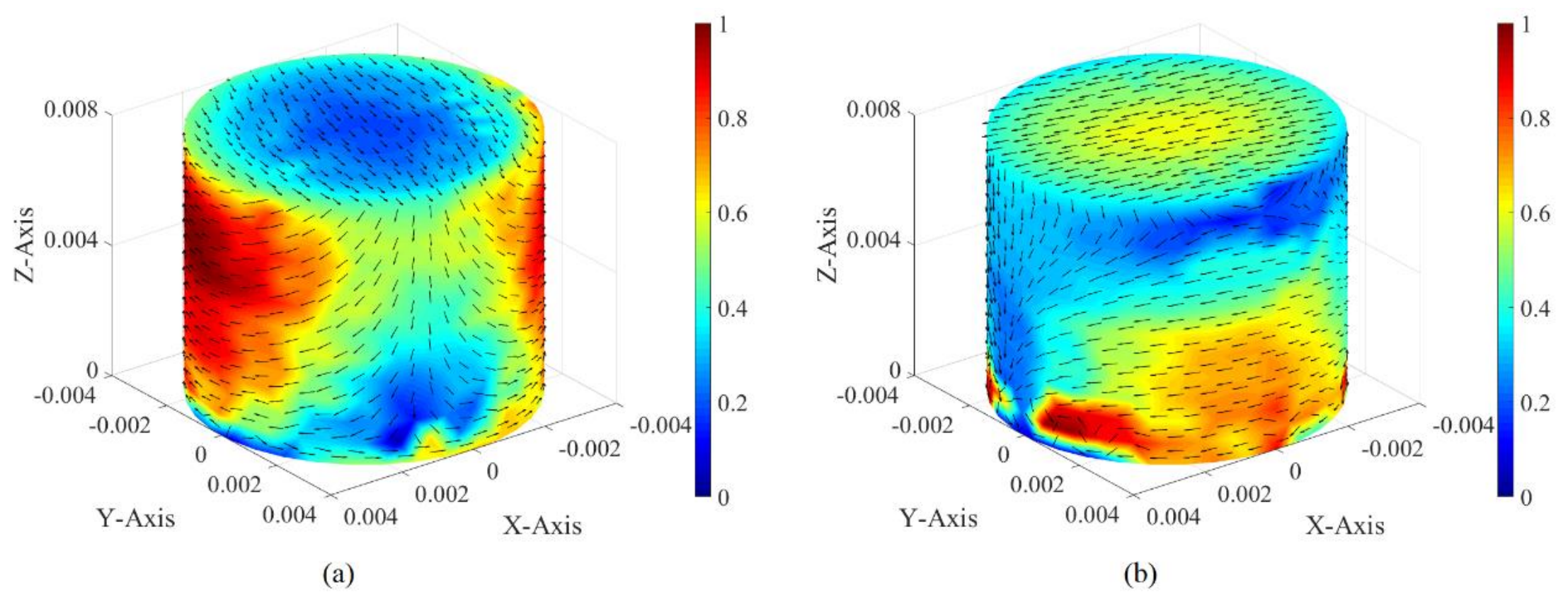

Figure 7-18 Modal (a) electric and (b) magnetic currents distributing on $\mathbb{S}_{M\rightleftharpoons A}$

## 7.5 Chapter Summary

In this chapter, the *power transport theorem* (*PTT*) based *decoupling mode theory* (*DMT*) for *augmented receiving antennas* (*augmented rec-antennas*) — *PTT-RecAnt-DMT* — is established.

By orthogonalizing frequency-domain *input power operator* (*IPO*), Sec. 7.3 constructs the *input-power-decoupled modes* (*IP-DMs*) of metallic augmented rec-antennas, and the IP-DMs of composite augmented rec-antennas are similarly constructed in Sec. 7.4.

The IP-DMs of augmented rec-antennas satisfy a similar *energy decoupling relation* to the IP-DMs of the *guiding structures* discussed in Chap. 3 and the augmented transmitting antennas discussed in Chap. 6. By employing the energy decoupling relation, it is found out that the IP-DMs don't have net energy exchange in any integrated period and the IP-DMs satisfy the famous *Parseval's identity*.

# Chapter 8 Input-Power-Decoupled Modes of Some Combined Systems

**CHAPTER MOTIVATION:** This chapter is devoted to constructing the *input-power-decoupled modes* (*IP-DMs*) of some combined systems, such as "the *one constituted by guiding structure & augmented transmitting antenna*" and "the *one constituted by augmented transmitting antenna & augmented receiving antenna*", by orthogonalizing frequency-domain *input power operator* (*IPO*) under *power transport theorem* (*PTT*) framework. The obtained IP-DMs, which are inputted into the combined systems, don't have net energy exchange in any integral period.

## 8.1 Chapter Introduction

As exhibited in the previous Chap. 4, the IP-DMs of *transmitting antenna* (simply called *tra-antenna*), *propagation medium*, and *receiving antenna* (simply called *rece-antenna*) can be separately constructed without considering their interactions. However, in practical engineering, the interactions are usually strong, and then the interactions cannot be completely ignored.

There are two ways to consider the interactions. The first way is to employ *modal matching* technique. The second way, which we prefer, is to treat tra-antenna and propagation medium as a whole — *augmented tra-antenna*, and to treat propagation medium and rec-antenna as a whole — *augmented rec-antenna*.

The detailed processes for executing the modal matching were exhibited in Chap. 5. The rigorous mathematical descriptions for the concepts of augmented tra-antenna and augmented rec-antenna were given in Sec. 2.3, and the detailed processes for constructing the IP-DMs of the augmented antennas were provided in Chaps. 6 and 7. From Chaps. 4~7, it is easy to find out that: the second way not only has a simpler mathematical treatment, but also has a clearer *physical picture* for revealing the power transport process.

In fact, the *guiding structure in transmitting system* (simply called *tra-guide*), the augmented tra-antenna, the augmented rec-antenna, and the *guiding structure in receiving system* (simply called *rec-guide*) also interact with each other. This chapter focuses on the interactions among these structures, by further generalizing the ideas proposed in the previous Sec. 2.3 and Chaps. 6&7.

This chapter is organized as follows: Sec. 8.2 treats the tra-guide and augmented tra-antenna as a whole — *augmented tra-guide-tra-antenna combined system* (simply called *augmented TGTA-combined system* or more simply as *TGTA system*), and constructs the IP-DMs of the system; Sec. 8.3 treats the augmented tra-antenna and augmented rec-antenna as a whole — *augmented tra-antenna-rec-antenna combined system* (simply called *augmented TARA-combined system* or more simply as *TARA system*), and constructs the IP-DMs of the system.

The interaction between augmented rec-antenna and rec-guide can be similarly discussed just like discussing TGTA system. The interaction between *transmitting system* and *receiving system* can be similarly discussed just like discussing TARA system.

## 8.2 IP-DMs of Augmented Tra-guide-Tra-antenna Combined System

In this section, we consider the *microstrip line fed microstrip patch antenna mounted on thick metallic ground plane* as shown in the following Fig 8-1.

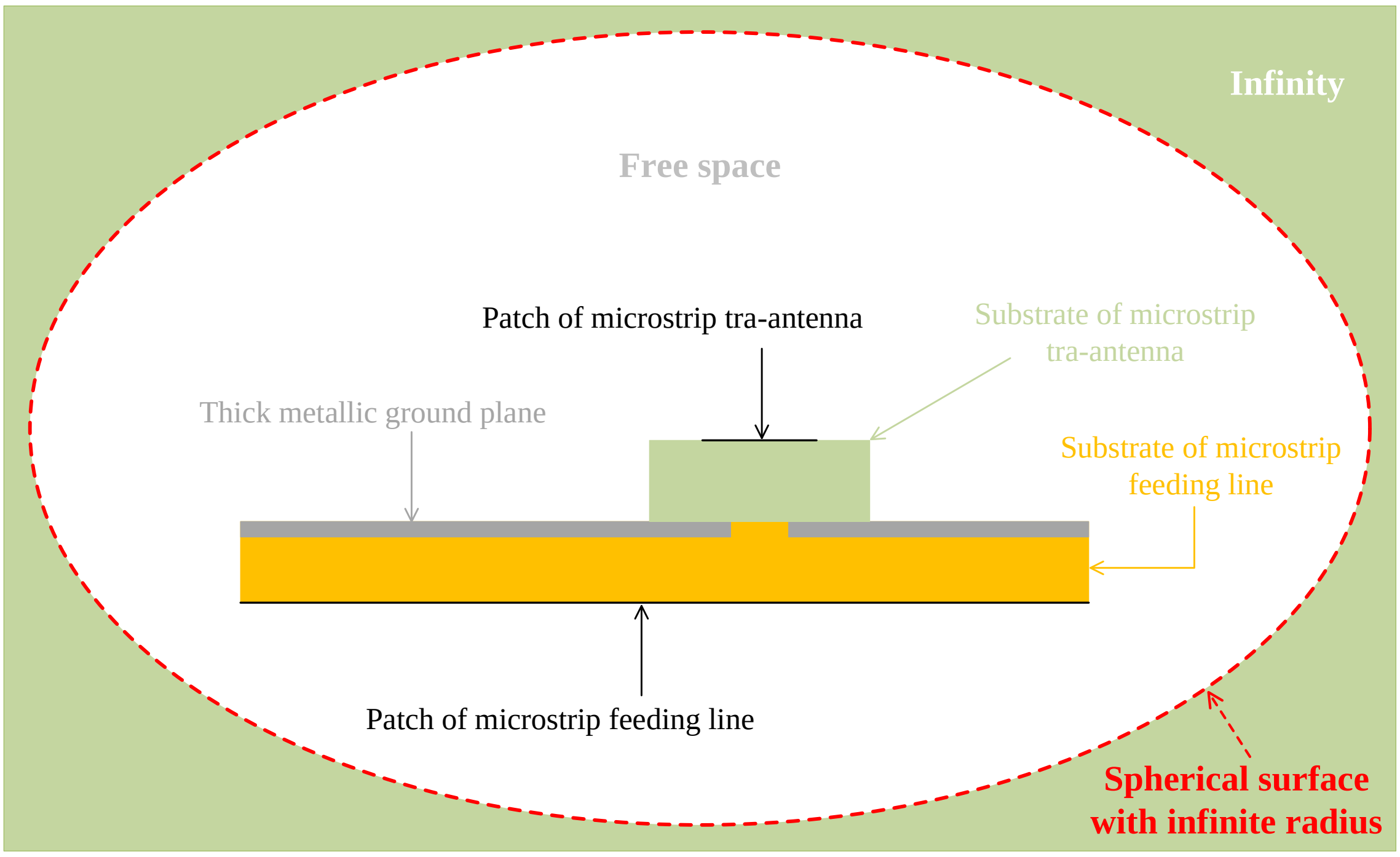

Figure 8-1 Geometry of the TGTA system considered in this section

Here, we treat {microstrip line, ground plane, microstrip antenna} as a whole — augmented tra-guide-tra-antenna combined system (TGTA system). The TGTA system is placed in *free space*.

Similarly to the previous sections, this section is organized as the following Fig 8-2

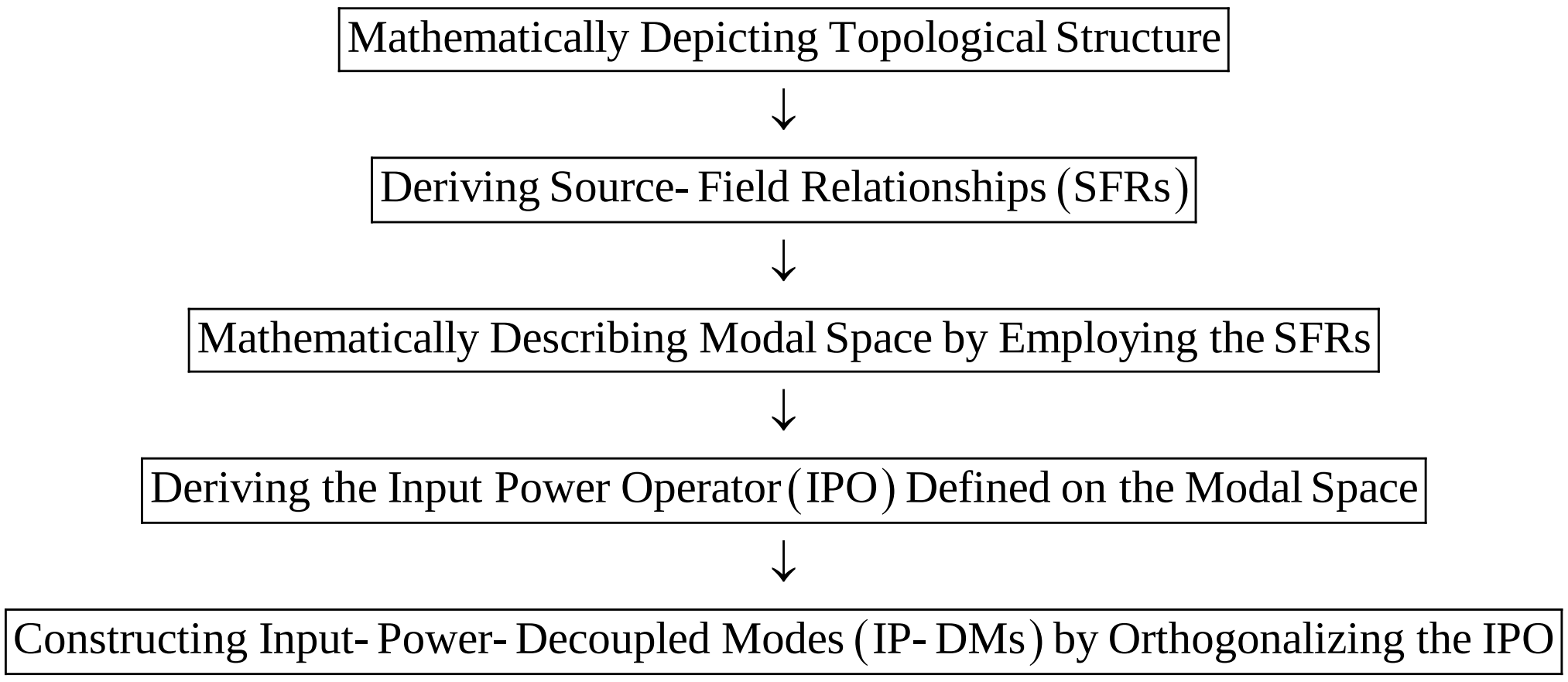

Figure 8-2 Organization for this section

### 8.2.1 Topological Structure

The topological structure of the TGTA system shown in Fig. 8-1 is illustrated in the following Fig. 8-3.

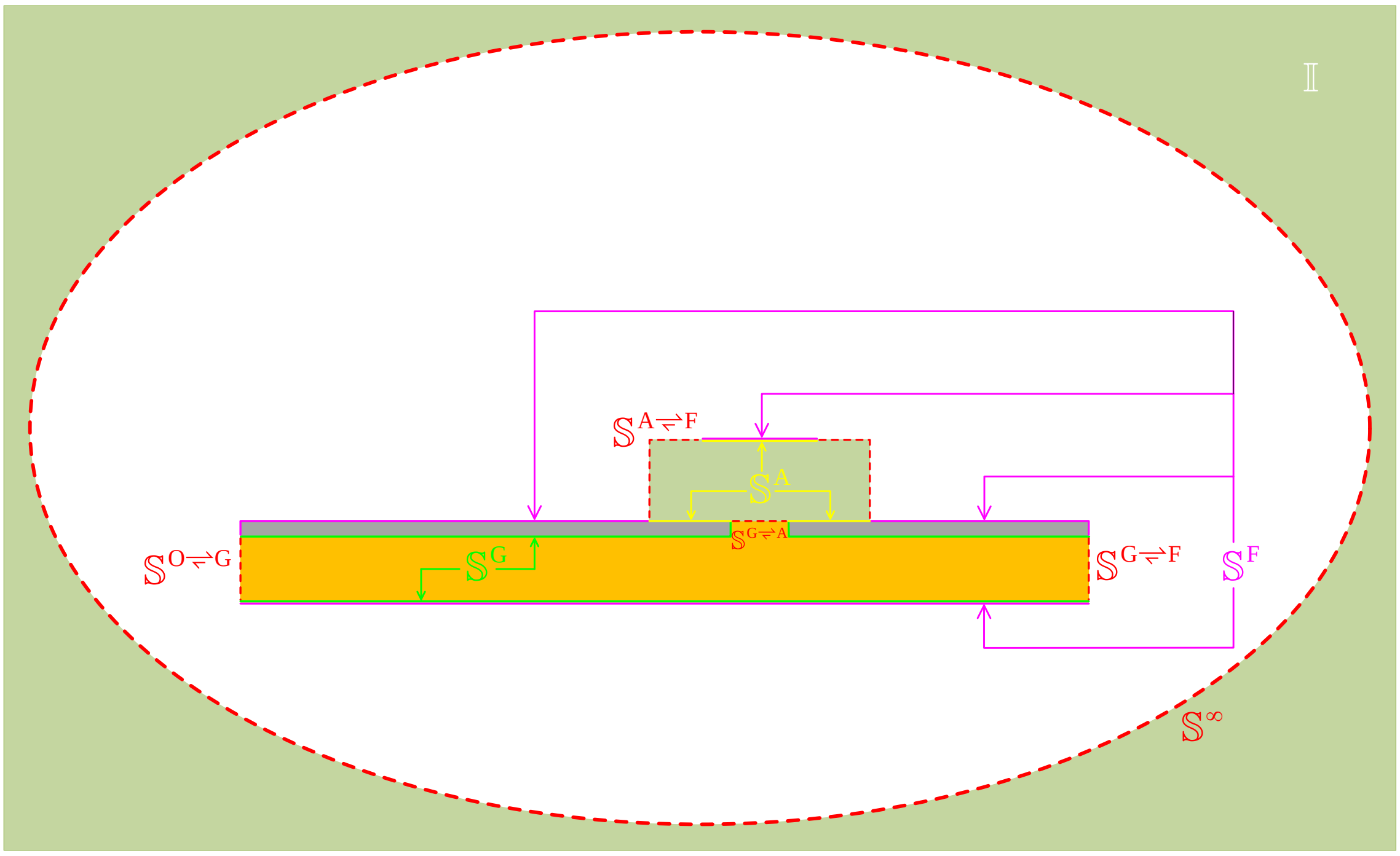

Figure 8-3 Topological structure of the TGTA system shown in Fig. 8-1

In the above Fig. 8-3, the region occupied by the substrate of microstrip line is denoted as $\mathbb{V}^{G}$; the region occupied by the substrate of microstrip antenna is denoted as $\mathbb{V}^{A}$; the region occupied by free space is denoted as $\mathbb{V}^{F}$.

The *input port* of $\mathbb{V}^{G}$, i.e. the interface between $\mathbb{V}^{G}$ and *generator*, is denoted as $\mathbb{S}^{O\rightleftharpoons G}$. The impenetrable *electric wall* of $\mathbb{V}^{G}$, i.e. the interface between $\mathbb{V}^{G}$ and

{metallic strip of microstrip line, metallic ground plane}, is denoted as $\mathbb{S}^{\rm G}$. The penetrable interface between $\mathbb{V}^{\rm G}$ and $\mathbb{V}^{\rm A}$ is denoted as $\mathbb{S}^{\rm G \rightleftharpoons A}$. The penetrable interface between $\mathbb{V}^{\rm G}$ and $\mathbb{V}^{\rm F}$ is denoted as $\mathbb{S}^{\rm G \rightleftharpoons F}$. The whole boundary of $\mathbb{V}^{\rm G}$ is denoted as $\partial\mathbb{V}^{\rm G}$, and it is obvious that $\partial\mathbb{V}^{\rm G} = \mathbb{S}^{\rm O \rightleftharpoons G} \bigcup \mathbb{S}^{\rm G} \bigcup \mathbb{S}^{\rm G \rightleftharpoons A} \bigcup \mathbb{S}^{\rm G \rightleftharpoons F}$. In addition, the *inner normal direction* of $\partial\mathbb{V}^{\rm G}$ is denoted as $\hat{n}^{\rightarrow \rm G}$.

The input port of $\mathbb{V}^{\rm A}$ is just the $\mathbb{S}^{\rm G \rightleftharpoons A}$. The impenetrable electric wall of $\mathbb{V}^{\rm A}$, i.e. the inferface between $\mathbb{V}^{\rm A}$ and metallic ground plane, is denoted as $\mathbb{S}^{\rm A}$. The penetrable interface between $\mathbb{V}^{\rm A}$ and $\mathbb{V}^{\rm F}$ is denoted as $\mathbb{S}^{\rm A \rightleftharpoons F}$. The whole boundary of $\mathbb{V}^{\rm A}$ is denoted as $\partial\mathbb{V}^{\rm A}$, and it is obvious that $\partial\mathbb{V}^{\rm A} = \mathbb{S}^{\rm G \rightleftharpoons A} \bigcup \mathbb{S}^{\rm A} \bigcup \mathbb{S}^{\rm A \rightleftharpoons F}$. In addition, the inner normal direction of $\partial\mathbb{V}^{\rm A}$ is denoted as $\hat{n}^{\rightarrow \rm A}$.

The input port of $\mathbb{V}^{\rm F}$ is just the union of $\mathbb{S}^{\rm A \rightleftharpoons F}$ and $\mathbb{S}^{\rm G \rightleftharpoons F}$, i.e. $\mathbb{S}^{\rm A \rightleftharpoons F} \bigcup \mathbb{S}^{\rm G \rightleftharpoons F}$. The impenetrable electric wall of $\mathbb{V}^{\rm F}$, i.e. the inferface between $\mathbb{V}^{\rm F}$ and {metallic ground plane, metallic strip of microstrip line, metallic patch of microstrip antenna} is denoted as $\mathbb{S}^{\rm F}$. The outport port of $\mathbb{V}^{\rm F}$, i.e. the interface between $\mathbb{V}^{\rm F}$ and *infinity*, is denoted as $\mathbb{S}^{\infty}$. The whole boundary of $\mathbb{V}^{\rm F}$ is denoted as $\partial\mathbb{V}^{\rm F}$, and it is obvious that $\partial\mathbb{V}^{\rm F} = \mathbb{S}^{\rm A \rightleftharpoons F} \bigcup \mathbb{S}^{\rm F} \bigcup \mathbb{S}^{\rm G \rightleftharpoons F} \bigcup \mathbb{S}^{\infty}$. In addition, the inner normal direction of $\partial\mathbb{V}^{\rm F}$ is denoted as $\hat{n}^{\rightarrow \rm F}$, and the *outer normal direction* of $\mathbb{S}^{\infty}$ is denoted as $\hat{n}^{\rightarrow \infty}$.

The *magnetic permeability*, *dielectric permittivity*, and *electric conductivity* of $\mathbb{V}^{\rm G}$ are denoted as $\overleftrightarrow{\mu}^{\rm G}(\vec{r})$, $\overleftrightarrow{\varepsilon}^{\rm G}(\vec{r})$, and $\overleftrightarrow{\sigma}^{\rm G}(\vec{r})$ respectively. The permeability, permittivity, and conductivity of $\mathbb{V}^{\rm A}$ are denoted as $\overleftrightarrow{\mu}^{\rm A}(\vec{r})$, $\overleftrightarrow{\varepsilon}^{\rm A}(\vec{r})$, and $\overleftrightarrow{\sigma}^{\rm A}(\vec{r})$ respectively. The permeability, permittivity, and conductivity of $\mathbb{V}^{\rm F}$ are $\mu_0$, $\varepsilon_0$, and 0 respectively. To simplify the symbolic system of the following discussions, the above-mentioned material parameters are uniformly denoted as $\overleftrightarrow{\gamma}(\vec{r})$, as follows:

$$\overleftrightarrow{\gamma}(\vec{r}) = \begin{cases} \overleftrightarrow{\gamma}^{\rm G}(\vec{r}) & , \quad \vec{r} \in \mathbb{V}^{\rm G} \\ \overleftrightarrow{\gamma}^{\rm A}(\vec{r}) & , \quad \vec{r} \in \mathbb{V}^{\rm A} \end{cases} \tag{8-1}$$

where $\overleftrightarrow{\gamma} = \overleftrightarrow{\mu} / \overleftrightarrow{\varepsilon} / \overleftrightarrow{\sigma}$.

## 8.2.2 Source-Field Relationships

If the *equivalent surface currents* distributing on $\mathbb{S}^{\rm O \rightleftharpoons G}$ are denoted as $\{\vec{J}^{\rm O \rightleftharpoons G}, \vec{M}^{\rm O \rightleftharpoons G}\}$, and the *equivalent surface electric current* distributing on $\mathbb{S}^{\rm G}$ is denoted as $\vec{J}^{\rm G}$, and the equivalent surface currents distributing on $\mathbb{S}^{\rm G \rightleftharpoons A}$ are denoted as $\{\vec{J}^{\rm G \rightleftharpoons A}, \vec{M}^{\rm G \rightleftharpoons A}\}$, and the equivalent surface currents distributing on $\mathbb{S}^{\rm G \rightleftharpoons F}$ are denoted as $\{\vec{J}^{\rm G \rightleftharpoons F}, \vec{M}^{\rm G \rightleftharpoons F}\}$, then the field distributing on $\mathbb{V}^{\rm G}$ can be expressed as follows:

$$\vec{F}(\vec{r}) = \mathcal{F}_{\rm G}\left(\vec{J}^{\rm O\rightleftharpoons G} + \vec{J}^{\rm G} - \vec{J}^{\rm G\rightleftharpoons A} + \vec{J}^{\rm G\rightleftharpoons F}, \vec{M}^{\rm O\rightleftharpoons G} - \vec{M}^{\rm G\rightleftharpoons A} + \vec{M}^{\rm G\rightleftharpoons F}\right), \quad \vec{r} \in \mathbb{V}^{\rm G} \quad (8\text{-}2)$$

where $\vec{F} = \vec{E} / \vec{H}$, and correspondingly $\mathcal{F}_{\rm G} = \mathcal{E}_{\rm G} / \mathcal{H}_{\rm G}$, and the operator is defined as that $\mathcal{F}_{\rm G}(\vec{J},\vec{M}) = \vec{\vec{G}}_{\rm G}^{JF} * \vec{J} + \vec{\vec{G}}_{\rm G}^{MF} * \vec{M}$ (here, $\vec{\vec{G}}_{\rm G}^{JF}$ and $\vec{\vec{G}}_{\rm G}^{MF}$ are the *dyadic Green's functions* corresponding to the region $\mathbb{V}^{\rm G}$ with parameters $\{\vec{\vec{\mu}}^{\rm G}, \vec{\vec{\varepsilon}}^{\rm G}, \vec{\vec{\sigma}}^{\rm G}\}$ ). The currents $\{\vec{J}^{\rm O\rightleftharpoons G}, \vec{M}^{\rm O\rightleftharpoons G}\}$ and fields $\{\vec{E},\vec{H}\}$ in Eq. (8-2) satisfy the following relations

$$\hat{n}^{\rightarrow \rm G} \times \left[\vec{H}\left(\vec{r}^{\rm G}\right)\right]_{\vec{r}^{\rm G} \to \vec{r}} = \vec{J}^{\rm O\rightleftharpoons G}(\vec{r}) \quad , \quad \vec{r} \in \mathbb{S}^{\rm O\rightleftharpoons G} \quad (8\text{-}3a)$$

$$\left[\vec{E}\left(\vec{r}^{\rm G}\right)\right]_{\vec{r}^{\rm G} \to \vec{r}} \times \hat{n}^{\rightarrow \rm G} = \vec{M}^{\rm O\rightleftharpoons G}(\vec{r}) \quad , \quad \vec{r} \in \mathbb{S}^{\rm O\rightleftharpoons G} \quad (8\text{-}3b)$$

and the currents $\{\vec{J}^{\rm G\rightleftharpoons A}, \vec{M}^{\rm G\rightleftharpoons A}\}$ and fields $\{\vec{E},\vec{H}\}$ in Eq. (8-2) satisfy the following relations

$$\hat{n}^{\rightarrow \rm A} \times \left[\vec{H}\left(\vec{r}^{\rm G}\right)\right]_{\vec{r}^{\rm G} \to \vec{r}} = \vec{J}^{\rm G\rightleftharpoons A}(\vec{r}) \quad , \quad \vec{r} \in \mathbb{S}^{\rm G\rightleftharpoons A} \quad (8\text{-}4a)$$

$$\left[\vec{E}\left(\vec{r}^{\rm G}\right)\right]_{\vec{r}^{\rm G} \to \vec{r}} \times \hat{n}^{\rightarrow \rm A} = \vec{M}^{\rm G\rightleftharpoons A}(\vec{r}) \quad , \quad \vec{r} \in \mathbb{S}^{\rm G\rightleftharpoons A} \quad (8\text{-}4b)$$

and the currents $\{\vec{J}^{\rm G\rightleftharpoons F}, \vec{M}^{\rm G\rightleftharpoons F}\}$ and fields $\{\vec{E},\vec{H}\}$ in Eq. (8-2) satisfy the following relations

$$\hat{n}^{\rightarrow \rm G} \times \left[\vec{H}\left(\vec{r}^{\rm G}\right)\right]_{\vec{r}^{\rm G} \to \vec{r}} = \vec{J}^{\rm G\rightleftharpoons F}(\vec{r}) \quad , \quad \vec{r} \in \mathbb{S}^{\rm G\rightleftharpoons F} \quad (8\text{-}5a)$$

$$\left[\vec{E}\left(\vec{r}^{\rm G}\right)\right]_{\vec{r}^{\rm G} \to \vec{r}} \times \hat{n}^{\rightarrow \rm G} = \vec{M}^{\rm G\rightleftharpoons F}(\vec{r}) \quad , \quad \vec{r} \in \mathbb{S}^{\rm G\rightleftharpoons F} \quad (8\text{-}5b)$$

In the above Eqs. (8-3)~(8-5), point $\vec{r}^{\rm G}$ belongs to region $\mathbb{V}^{\rm G}$, and approaches the point $\vec{r}$ on $\mathbb{S}^{\rm O\rightleftharpoons G} \bigcup \mathbb{S}^{\rm G\rightleftharpoons A} \bigcup \mathbb{S}^{\rm G\rightleftharpoons F}$.

If the equivalent surface electric current distributing on $\mathbb{S}^{\rm A}$ is denoted as $\vec{J}^{\rm A}$, and the equivalent surface currents distributing on $\mathbb{S}^{\rm A\rightleftharpoons F}$ are denoted as $\{\vec{J}^{\rm A\rightleftharpoons F}, \vec{M}^{\rm A\rightleftharpoons F}\}$, then the field distributing on $\mathbb{V}^{\rm A}$ can be expressed as follows:

$$\vec{F}(\vec{r}) = \mathcal{F}_{\rm A}\left(\vec{J}^{\rm G\rightleftharpoons A} + \vec{J}^{\rm A} - \vec{J}^{\rm A\rightleftharpoons F}, \vec{M}^{\rm G\rightleftharpoons A} - \vec{M}^{\rm A\rightleftharpoons F}\right) \quad , \quad \vec{r} \in \mathbb{V}^{\rm A} \quad (8\text{-}6)$$

where $\vec{F} = \vec{E} / \vec{H}$, and correspondingly $\mathcal{F}_{\rm A} = \mathcal{E}_{\rm A} / \mathcal{H}_{\rm A}$, and the operator is defined as that $\mathcal{F}_{\rm A}(\vec{J},\vec{M}) = \vec{\vec{G}}_{\rm A}^{JF} * \vec{J} + \vec{\vec{G}}_{\rm A}^{MF} * \vec{M}$ (here, $\vec{\vec{G}}_{\rm A}^{JF}$ and $\vec{\vec{G}}_{\rm A}^{MF}$ are the dyadic Green's functions corresponding to the region $\mathbb{V}^{\rm A}$ with parameters $\{\vec{\vec{\mu}}^{\rm A}, \vec{\vec{\varepsilon}}^{\rm A}, \vec{\vec{\sigma}}^{\rm A}\}$ ). The currents $\{\vec{J}^{\rm A\rightleftharpoons F}, \vec{M}^{\rm A\rightleftharpoons F}\}$ and fields $\{\vec{E},\vec{H}\}$ in Eq. (8-6) satisfy the following relations

$$\hat{n}^{\rightarrow \rm F} \times \left[\vec{H}\left(\vec{r}^{\rm A}\right)\right]_{\vec{r}^{\rm A} \to \vec{r}} = \vec{J}^{\rm A\rightleftharpoons F}(\vec{r}) \quad , \quad \vec{r} \in \mathbb{S}^{\rm A\rightleftharpoons F} \quad (8\text{-}7a)$$

$$\left[\vec{E}\left(\vec{r}^{\rm A}\right)\right]_{\vec{r}^{\rm A} \to \vec{r}} \times \hat{n}^{\rightarrow \rm F} = \vec{M}^{\rm A\rightleftharpoons F}(\vec{r}) \quad , \quad \vec{r} \in \mathbb{S}^{\rm A\rightleftharpoons F} \quad (8\text{-}7b)$$

where point $\vec{r}^{\rm A}$ belongs to region $\mathbb{V}^{\rm A}$, and approaches the point $\vec{r}$ on $\mathbb{S}^{\rm A\rightleftharpoons F}$.

If the equivalent surface electric current distributing on $\mathbb{S}^{\rm F}$ is denoted as $\vec{J}^{\rm F}$, then the field distributing on $\mathbb{V}^{\rm F}$ can be expressed as follows:

$$\vec{F}\left(\vec{r}\right)=\mathcal{F}_0\left(\vec{J}^{\rm A\rightleftharpoons F}+\vec{J}^{\rm F}-\vec{J}^{\rm G\rightleftharpoons F},\vec{M}^{\rm A\rightleftharpoons F}-\vec{M}^{\rm G\rightleftharpoons F}\right) \quad , \quad \vec{r}\in\mathbb{V}^{\rm F} \tag{8-8}$$

where $\vec{F}=\vec{E}/\vec{H}$, and correspondingly $\mathcal{F}_0=\mathcal{E}_0/\mathcal{H}_0$, and the operator is the same as the one used in the previous chapters.

## 8.2.3 Mathematical Description for Modal Space

In this subsection, we establish some rigorous mathematical descriptions for the *modal space* of the TGTA system shown in Fig. 8-1, by employing the *source-field relationships* given in the above Sec. 8.2.2.

### 8.2.3.1 Integral Equations

Substituting Eq. (8-2) into Eq. (8-3), we have the following *integral equations*

$$\begin{aligned}&\left[\mathcal{H}_{\rm G}\left(\vec{J}^{\rm O\rightleftharpoons G}+\vec{J}^{\rm G}-\vec{J}^{\rm G\rightleftharpoons A}+\vec{J}^{\rm G\rightleftharpoons F},\vec{M}^{\rm O\rightleftharpoons G}-\vec{M}^{\rm G\rightleftharpoons A}+\vec{M}^{\rm G\rightleftharpoons F}\right)\right]_{\vec{r}^{\rm G}\to\vec{r}}^{\rm tan}\\&=\vec{J}^{\rm O\rightleftharpoons G}\left(\vec{r}\right)\times\hat{n}^{\to\rm G} \qquad , \qquad \vec{r}\in\mathbb{S}^{\rm O\rightleftharpoons G}\end{aligned} \tag{8-9a}$$

$$\begin{aligned}&\left[\mathcal{E}_{\rm G}\left(\vec{J}^{\rm O\rightleftharpoons G}+\vec{J}^{\rm G}-\vec{J}^{\rm G\rightleftharpoons A}+\vec{J}^{\rm G\rightleftharpoons F},\vec{M}^{\rm O\rightleftharpoons G}-\vec{M}^{\rm G\rightleftharpoons A}+\vec{M}^{\rm G\rightleftharpoons F}\right)\right]_{\vec{r}^{\rm G}\to\vec{r}}^{\rm tan}\\&=\hat{n}^{\to\rm G}\times\vec{M}^{\rm O\rightleftharpoons G}\left(\vec{r}\right) \qquad , \qquad \vec{r}\in\mathbb{S}^{\rm O\rightleftharpoons G}\end{aligned} \tag{8-9b}$$

about currents $\{\vec{J}^{\rm O\rightleftharpoons G},\vec{M}^{\rm O\rightleftharpoons G}\}$, $\vec{J}^{\rm G}$, $\{\vec{J}^{\rm G\rightleftharpoons A},\vec{M}^{\rm G\rightleftharpoons A}\}$, and $\{\vec{J}^{\rm G\rightleftharpoons F},\vec{M}^{\rm G\rightleftharpoons F}\}$.

Based on Eq. (8-2) and the *homogeneous tangential electric field boundary condition* on $\mathbb{S}^{\rm G}$, we have the following *electric field integral equations*

$$\left[\mathcal{E}_{\rm G}\left(\vec{J}^{\rm O\rightleftharpoons G}+\vec{J}^{\rm G}-\vec{J}^{\rm G\rightleftharpoons A}+\vec{J}^{\rm G\rightleftharpoons F},\vec{M}^{\rm O\rightleftharpoons G}-\vec{M}^{\rm G\rightleftharpoons A}+\vec{M}^{\rm G\rightleftharpoons F}\right)\right]_{\vec{r}^{\rm G}\to\vec{r}}^{\rm tan}=0 \quad , \quad \vec{r}\in\mathbb{S}^{\rm G} \tag{8-10}$$

about currents $\{\vec{J}^{\rm O\rightleftharpoons G},\vec{M}^{\rm O\rightleftharpoons G}\}$, $\vec{J}^{\rm G}$, $\{\vec{J}^{\rm G\rightleftharpoons A},\vec{M}^{\rm G\rightleftharpoons A}\}$, and $\{\vec{J}^{\rm G\rightleftharpoons F},\vec{M}^{\rm G\rightleftharpoons F}\}$.

Based on Eqs. (8-2)&(8-6) and the *tangential field continuation condition* on $\mathbb{S}^{\rm G\rightleftharpoons A}$, we have the following integral equations

$$\begin{aligned}&\left[\mathcal{E}_{\rm G}\left(\vec{J}^{\rm O\rightleftharpoons G}+\vec{J}^{\rm G}-\vec{J}^{\rm G\rightleftharpoons A}+\vec{J}^{\rm G\rightleftharpoons F},\vec{M}^{\rm O\rightleftharpoons G}-\vec{M}^{\rm G\rightleftharpoons A}+\vec{M}^{\rm G\rightleftharpoons F}\right)\right]_{\vec{r}^{\rm G}\to\vec{r}}^{\rm tan}\\&=\left[\mathcal{E}_{\rm A}\left(\vec{J}^{\rm G\rightleftharpoons A}+\vec{J}^{\rm A}-\vec{J}^{\rm A\rightleftharpoons F},\vec{M}^{\rm G\rightleftharpoons A}-\vec{M}^{\rm A\rightleftharpoons F}\right)\right]_{\vec{r}^{\rm A}\to\vec{r}}^{\rm tan} \qquad , \qquad \vec{r}\in\mathbb{S}^{\rm G\rightleftharpoons A}\end{aligned} \tag{8-11a}$$

$$\begin{aligned}&\left[\mathcal{H}_{\rm G}\left(\vec{J}^{\rm O\rightleftharpoons G}+\vec{J}^{\rm G}-\vec{J}^{\rm G\rightleftharpoons A}+\vec{J}^{\rm G\rightleftharpoons F},\vec{M}^{\rm O\rightleftharpoons G}-\vec{M}^{\rm G\rightleftharpoons A}+\vec{M}^{\rm G\rightleftharpoons F}\right)\right]_{\vec{r}^{\rm G}\to\vec{r}}^{\rm tan}\\&=\left[\mathcal{H}_{\rm A}\left(\vec{J}^{\rm G\rightleftharpoons A}+\vec{J}^{\rm A}-\vec{J}^{\rm A\rightleftharpoons F},\vec{M}^{\rm G\rightleftharpoons A}-\vec{M}^{\rm A\rightleftharpoons F}\right)\right]_{\vec{r}^{\rm A}\to\vec{r}}^{\rm tan} \qquad , \qquad \vec{r}\in\mathbb{S}^{\rm G\rightleftharpoons A}\end{aligned} \tag{8-11b}$$

about currents $\{\vec{J}^{\mathrm{O}\rightleftharpoons\mathrm{G}},\vec{M}^{\mathrm{O}\rightleftharpoons\mathrm{G}}\}$, $\vec{J}^{\mathrm{G}}$, $\{\vec{J}^{\mathrm{G}\rightleftharpoons\mathrm{A}},\vec{M}^{\mathrm{G}\rightleftharpoons\mathrm{A}}\}$, $\vec{J}^{\mathrm{A}}$, $\{\vec{J}^{\mathrm{A}\rightleftharpoons\mathrm{F}},\vec{M}^{\mathrm{A}\rightleftharpoons\mathrm{F}}\}$, and $\{\vec{J}^{\mathrm{G}\rightleftharpoons\mathrm{F}},\vec{M}^{\mathrm{G}\rightleftharpoons\mathrm{F}}\}$.

Based on Eq. (8-6) and the homogeneous tangential electric field boundary condition on $\mathbb{S}^{\mathrm{A}}$, we have the following electric field integral equations

$$\left[\mathcal{E}_{\mathrm{A}}\left(\vec{J}^{\mathrm{G}\rightleftharpoons\mathrm{A}}+\vec{J}^{\mathrm{A}}-\vec{J}^{\mathrm{A}\rightleftharpoons\mathrm{F}},\vec{M}^{\mathrm{G}\rightleftharpoons\mathrm{A}}-\vec{M}^{\mathrm{A}\rightleftharpoons\mathrm{F}}\right)\right]_{\vec{r}^{\mathrm{A}}\to\vec{r}}^{\tan}=0 \quad , \quad \vec{r}\in\mathbb{S}^{\mathrm{A}} \tag{8-12}$$

about currents $\{\vec{J}^{\mathrm{G}\rightleftharpoons\mathrm{A}},\vec{M}^{\mathrm{G}\rightleftharpoons\mathrm{A}}\}$, $\vec{J}^{\mathrm{A}}$, and $\{\vec{J}^{\mathrm{A}\rightleftharpoons\mathrm{F}},\vec{M}^{\mathrm{A}\rightleftharpoons\mathrm{F}}\}$.

Based on Eqs. (8-6)&(8-8) and the tangential field continuation condition on $\mathbb{S}^{\mathrm{A}\rightleftharpoons\mathrm{F}}$, we have the following integral equations

$$\begin{aligned}&\left[\mathcal{E}_{\mathrm{A}}\left(\vec{J}^{\mathrm{G}\rightleftharpoons\mathrm{A}}+\vec{J}^{\mathrm{A}}-\vec{J}^{\mathrm{A}\rightleftharpoons\mathrm{F}},\vec{M}^{\mathrm{G}\rightleftharpoons\mathrm{A}}-\vec{M}^{\mathrm{A}\rightleftharpoons\mathrm{F}}\right)\right]_{\vec{r}^{\mathrm{A}}\to\vec{r}}^{\tan}\\&=\left[\mathcal{E}_{0}\left(\vec{J}^{\mathrm{A}\rightleftharpoons\mathrm{F}}+\vec{J}^{\mathrm{F}}-\vec{J}^{\mathrm{G}\rightleftharpoons\mathrm{F}},\vec{M}^{\mathrm{A}\rightleftharpoons\mathrm{F}}-\vec{M}^{\mathrm{G}\rightleftharpoons\mathrm{F}}\right)\right]_{\vec{r}^{\mathrm{F}}\to\vec{r}}^{\tan} \quad , \quad \vec{r}\in\mathbb{S}^{\mathrm{A}\rightleftharpoons\mathrm{F}}\end{aligned} \tag{8-13a}$$

$$\begin{aligned}&\left[\mathcal{H}_{\mathrm{A}}\left(\vec{J}^{\mathrm{G}\rightleftharpoons\mathrm{A}}+\vec{J}^{\mathrm{A}}-\vec{J}^{\mathrm{A}\rightleftharpoons\mathrm{F}},\vec{M}^{\mathrm{G}\rightleftharpoons\mathrm{A}}-\vec{M}^{\mathrm{A}\rightleftharpoons\mathrm{F}}\right)\right]_{\vec{r}^{\mathrm{A}}\to\vec{r}}^{\tan}\\&=\left[\mathcal{H}_{0}\left(\vec{J}^{\mathrm{A}\rightleftharpoons\mathrm{F}}+\vec{J}^{\mathrm{F}}-\vec{J}^{\mathrm{G}\rightleftharpoons\mathrm{F}},\vec{M}^{\mathrm{A}\rightleftharpoons\mathrm{F}}-\vec{M}^{\mathrm{G}\rightleftharpoons\mathrm{F}}\right)\right]_{\vec{r}^{\mathrm{F}}\to\vec{r}}^{\tan} \quad , \quad \vec{r}\in\mathbb{S}^{\mathrm{A}\rightleftharpoons\mathrm{F}}\end{aligned} \tag{8-13b}$$

about currents $\{\vec{J}^{\mathrm{G}\rightleftharpoons\mathrm{A}},\vec{M}^{\mathrm{G}\rightleftharpoons\mathrm{A}}\}$, $\vec{J}^{\mathrm{A}}$, $\{\vec{J}^{\mathrm{A}\rightleftharpoons\mathrm{F}},\vec{M}^{\mathrm{A}\rightleftharpoons\mathrm{F}}\}$, $\vec{J}^{\mathrm{F}}$, and $\{\vec{J}^{\mathrm{G}\rightleftharpoons\mathrm{F}},\vec{M}^{\mathrm{G}\rightleftharpoons\mathrm{F}}\}$.

Based on Eq. (8-8) and the homogeneous tangential electric field boundary condition on $\mathbb{S}^{\mathrm{F}}$, we have the following electric field integral equations

$$\left[\mathcal{E}_{0}\left(\vec{J}^{\mathrm{A}\rightleftharpoons\mathrm{F}}+\vec{J}^{\mathrm{F}}-\vec{J}^{\mathrm{G}\rightleftharpoons\mathrm{F}},\vec{M}^{\mathrm{A}\rightleftharpoons\mathrm{F}}-\vec{M}^{\mathrm{G}\rightleftharpoons\mathrm{F}}\right)\right]_{\vec{r}^{\mathrm{F}}\to\vec{r}}^{\tan}=0 \quad , \quad \vec{r}\in\mathbb{S}^{\mathrm{F}} \tag{8-14}$$

about currents $\{\vec{J}^{\mathrm{A}\rightleftharpoons\mathrm{F}},\vec{M}^{\mathrm{A}\rightleftharpoons\mathrm{F}}\}$, $\vec{J}^{\mathrm{F}}$, and $\{\vec{J}^{\mathrm{G}\rightleftharpoons\mathrm{F}},\vec{M}^{\mathrm{G}\rightleftharpoons\mathrm{F}}\}$.

Based on Eqs. (8-2)&(8-8) and the tangential field continuation condition on $\mathbb{S}^{\mathrm{G}\rightleftharpoons\mathrm{F}}$, we have the following integral equations

$$\begin{aligned}&\left[\mathcal{E}_{\mathrm{G}}\left(\vec{J}^{\mathrm{O}\rightleftharpoons\mathrm{G}}+\vec{J}^{\mathrm{G}}-\vec{J}^{\mathrm{G}\rightleftharpoons\mathrm{A}}+\vec{J}^{\mathrm{G}\rightleftharpoons\mathrm{F}},\vec{M}^{\mathrm{O}\rightleftharpoons\mathrm{G}}-\vec{M}^{\mathrm{G}\rightleftharpoons\mathrm{A}}+\vec{M}^{\mathrm{G}\rightleftharpoons\mathrm{F}}\right)\right]_{\vec{r}^{\mathrm{G}}\to\vec{r}}^{\tan}\\&=\left[\mathcal{E}_{0}\left(\vec{J}^{\mathrm{A}\rightleftharpoons\mathrm{F}}+\vec{J}^{\mathrm{F}}-\vec{J}^{\mathrm{G}\rightleftharpoons\mathrm{F}},\vec{M}^{\mathrm{A}\rightleftharpoons\mathrm{F}}-\vec{M}^{\mathrm{G}\rightleftharpoons\mathrm{F}}\right)\right]_{\vec{r}^{\mathrm{F}}\to\vec{r}}^{\tan} \quad , \quad \vec{r}\in\mathbb{S}^{\mathrm{G}\rightleftharpoons\mathrm{F}}\end{aligned} \tag{8-15a}$$

$$\begin{aligned}&\left[\mathcal{H}_{\mathrm{G}}\left(\vec{J}^{\mathrm{O}\rightleftharpoons\mathrm{G}}+\vec{J}^{\mathrm{G}}-\vec{J}^{\mathrm{G}\rightleftharpoons\mathrm{A}}+\vec{J}^{\mathrm{G}\rightleftharpoons\mathrm{F}},\vec{M}^{\mathrm{O}\rightleftharpoons\mathrm{G}}-\vec{M}^{\mathrm{G}\rightleftharpoons\mathrm{A}}+\vec{M}^{\mathrm{G}\rightleftharpoons\mathrm{F}}\right)\right]_{\vec{r}^{\mathrm{G}}\to\vec{r}}^{\tan}\\&=\left[\mathcal{H}_{0}\left(\vec{J}^{\mathrm{A}\rightleftharpoons\mathrm{F}}+\vec{J}^{\mathrm{F}}-\vec{J}^{\mathrm{G}\rightleftharpoons\mathrm{F}},\vec{M}^{\mathrm{A}\rightleftharpoons\mathrm{F}}-\vec{M}^{\mathrm{G}\rightleftharpoons\mathrm{F}}\right)\right]_{\vec{r}^{\mathrm{F}}\to\vec{r}}^{\tan} \quad , \quad \vec{r}\in\mathbb{S}^{\mathrm{G}\rightleftharpoons\mathrm{F}}\end{aligned} \tag{8-15b}$$

about currents $\{\vec{J}^{\mathrm{O}\rightleftharpoons\mathrm{G}},\vec{M}^{\mathrm{O}\rightleftharpoons\mathrm{G}}\}$, $\vec{J}^{\mathrm{G}}$, $\{\vec{J}^{\mathrm{G}\rightleftharpoons\mathrm{A}},\vec{M}^{\mathrm{G}\rightleftharpoons\mathrm{A}}\}$, $\{\vec{J}^{\mathrm{A}\rightleftharpoons\mathrm{F}},\vec{M}^{\mathrm{A}\rightleftharpoons\mathrm{F}}\}$, $\vec{J}^{\mathrm{F}}$, and $\{\vec{J}^{\mathrm{G}\rightleftharpoons\mathrm{F}},\vec{M}^{\mathrm{G}\rightleftharpoons\mathrm{F}}\}$.

The above Eqs. (8-9a)~(8-15b) constitute a complete mathematical description for the modal space of the TGTA system shown in Fig. 8-1.

### 8.2.3.2 Matrix Equations

If the above-mentioned currents are expanded in terms of some proper *current basis functions* as follows:

$$\vec{C}^x\left(\vec{r}\right)=\sum\nolimits_{\xi}a_{\xi}^{\vec{C}^x}\vec{b}_{\xi}^{\vec{C}^x}=\underbrace{\begin{bmatrix}\vec{b}_1^{\vec{C}^x} & \vec{b}_2^{\vec{C}^x} & \cdots\end{bmatrix}}_{\bar{B}^{\vec{C}^x}}\cdot\underbrace{\begin{bmatrix}a_1^{\vec{C}^x}\\ a_2^{\vec{C}^x}\\ \vdots\end{bmatrix}}_{\bar{a}^{\vec{C}^x}}\quad,\quad \vec{r}\in\mathbb{S}^x \tag{8-16}$$

where $\vec{C}^x$ denotes the various currents mentioned above, and the Eqs. (8-9a)~(8-15b) are tested by the basis functions $\{\vec{b}_{\xi}^{\vec{M}^{\text{O}\rightleftharpoons\text{G}}}\}$, $\{\vec{b}_{\xi}^{\vec{J}^{\text{O}\rightleftharpoons\text{G}}}\}$, $\{\vec{b}_{\xi}^{\vec{J}^{\text{G}}}\}$, $\{\vec{b}_{\xi}^{\vec{J}^{\text{G}\rightleftharpoons\text{A}}}\}$, $\{\vec{b}_{\xi}^{\vec{M}^{\text{G}\rightleftharpoons\text{A}}}\}$, $\{\vec{b}_{\xi}^{\vec{J}^{\text{A}}}\}$, $\{\vec{b}_{\xi}^{\vec{J}^{\text{A}\rightleftharpoons\text{F}}}\}$, $\{\vec{b}_{\xi}^{\vec{M}^{\text{A}\rightleftharpoons\text{F}}}\}$, $\{\vec{b}_{\xi}^{\vec{J}^{\text{F}}}\}$, $\{\vec{b}_{\xi}^{\vec{J}^{\text{G}\rightleftharpoons\text{F}}}\}$, and $\{\vec{b}_{\xi}^{\vec{M}^{\text{G}\rightleftharpoons\text{F}}}\}$ respectively with the *inner product* used in the previous chapters, then the integral equations are immediately discretized into the following *matrix equations*

$$\begin{aligned}&\bar{\bar{Z}}^{\vec{M}^{\text{O}\rightleftharpoons\text{G}}\vec{J}^{\text{O}\rightleftharpoons\text{G}}}\cdot\bar{a}^{\vec{J}^{\text{O}\rightleftharpoons\text{G}}}+\bar{\bar{Z}}^{\vec{M}^{\text{O}\rightleftharpoons\text{G}}\vec{J}^{\text{G}}}\cdot\bar{a}^{\vec{J}^{\text{G}}}+\bar{\bar{Z}}^{\vec{M}^{\text{O}\rightleftharpoons\text{G}}\vec{J}^{\text{G}\rightleftharpoons\text{A}}}\cdot\bar{a}^{\vec{J}^{\text{G}\rightleftharpoons\text{A}}}+\bar{\bar{Z}}^{\vec{M}^{\text{O}\rightleftharpoons\text{G}}\vec{J}^{\text{G}\rightleftharpoons\text{F}}}\cdot\bar{a}^{\vec{J}^{\text{G}\rightleftharpoons\text{F}}}\\&+\bar{\bar{Z}}^{\vec{M}^{\text{O}\rightleftharpoons\text{G}}\vec{M}^{\text{O}\rightleftharpoons\text{G}}}\cdot\bar{a}^{\vec{M}^{\text{O}\rightleftharpoons\text{G}}}+\bar{\bar{Z}}^{\vec{M}^{\text{O}\rightleftharpoons\text{G}}\vec{M}^{\text{G}\rightleftharpoons\text{A}}}\cdot\bar{a}^{\vec{M}^{\text{G}\rightleftharpoons\text{A}}}+\bar{\bar{Z}}^{\vec{M}^{\text{O}\rightleftharpoons\text{G}}\vec{M}^{\text{G}\rightleftharpoons\text{F}}}\cdot\bar{a}^{\vec{M}^{\text{G}\rightleftharpoons\text{F}}}=0\end{aligned}\tag{8-17a}$$

$$\begin{aligned}&\bar{\bar{Z}}^{\vec{J}^{\text{O}\rightleftharpoons\text{G}}\vec{J}^{\text{O}\rightleftharpoons\text{G}}}\cdot\bar{a}^{\vec{J}^{\text{O}\rightleftharpoons\text{G}}}+\bar{\bar{Z}}^{\vec{J}^{\text{O}\rightleftharpoons\text{G}}\vec{J}^{\text{G}}}\cdot\bar{a}^{\vec{J}^{\text{G}}}+\bar{\bar{Z}}^{\vec{J}^{\text{O}\rightleftharpoons\text{G}}\vec{J}^{\text{G}\rightleftharpoons\text{A}}}\cdot\bar{a}^{\vec{J}^{\text{G}\rightleftharpoons\text{A}}}+\bar{\bar{Z}}^{\vec{J}^{\text{O}\rightleftharpoons\text{G}}\vec{J}^{\text{G}\rightleftharpoons\text{F}}}\cdot\bar{a}^{\vec{J}^{\text{G}\rightleftharpoons\text{F}}}\\&+\bar{\bar{Z}}^{\vec{J}^{\text{O}\rightleftharpoons\text{G}}\vec{M}^{\text{O}\rightleftharpoons\text{G}}}\cdot\bar{a}^{\vec{M}^{\text{O}\rightleftharpoons\text{G}}}+\bar{\bar{Z}}^{\vec{J}^{\text{O}\rightleftharpoons\text{G}}\vec{M}^{\text{G}\rightleftharpoons\text{A}}}\cdot\bar{a}^{\vec{M}^{\text{G}\rightleftharpoons\text{A}}}+\bar{\bar{Z}}^{\vec{J}^{\text{O}\rightleftharpoons\text{G}}\vec{M}^{\text{G}\rightleftharpoons\text{F}}}\cdot\bar{a}^{\vec{M}^{\text{G}\rightleftharpoons\text{F}}}=0\end{aligned}\tag{8-17b}$$

and

$$\begin{aligned}&\bar{\bar{Z}}^{\vec{J}^{\text{G}}\vec{J}^{\text{O}\rightleftharpoons\text{G}}}\cdot\bar{a}^{\vec{J}^{\text{O}\rightleftharpoons\text{G}}}+\bar{\bar{Z}}^{\vec{J}^{\text{G}}\vec{J}^{\text{G}}}\cdot\bar{a}^{\vec{J}^{\text{G}}}+\bar{\bar{Z}}^{\vec{J}^{\text{G}}\vec{J}^{\text{G}\rightleftharpoons\text{A}}}\cdot\bar{a}^{\vec{J}^{\text{G}\rightleftharpoons\text{A}}}+\bar{\bar{Z}}^{\vec{J}^{\text{G}}\vec{J}^{\text{G}\rightleftharpoons\text{F}}}\cdot\bar{a}^{\vec{J}^{\text{G}\rightleftharpoons\text{F}}}+\bar{\bar{Z}}^{\vec{J}^{\text{G}}\vec{M}^{\text{O}\rightleftharpoons\text{G}}}\cdot\bar{a}^{\vec{M}^{\text{O}\rightleftharpoons\text{G}}}\\&+\bar{\bar{Z}}^{\vec{J}^{\text{G}}\vec{M}^{\text{G}\rightleftharpoons\text{A}}}\cdot\bar{a}^{\vec{M}^{\text{G}\rightleftharpoons\text{A}}}+\bar{\bar{Z}}^{\vec{J}^{\text{G}}\vec{M}^{\text{G}\rightleftharpoons\text{F}}}\cdot\bar{a}^{\vec{M}^{\text{G}\rightleftharpoons\text{F}}}=0\end{aligned}\tag{8-18}$$

and

$$\begin{aligned}&\bar{\bar{Z}}^{\vec{J}^{\text{G}\rightleftharpoons\text{A}}\vec{J}^{\text{O}\rightleftharpoons\text{G}}}\cdot\bar{a}^{\vec{J}^{\text{O}\rightleftharpoons\text{G}}}+\bar{\bar{Z}}^{\vec{J}^{\text{G}\rightleftharpoons\text{A}}\vec{J}^{\text{G}}}\cdot\bar{a}^{\vec{J}^{\text{G}}}+\bar{\bar{Z}}^{\vec{J}^{\text{G}\rightleftharpoons\text{A}}\vec{J}^{\text{G}\rightleftharpoons\text{A}}}\cdot\bar{a}^{\vec{J}^{\text{G}\rightleftharpoons\text{A}}}+\bar{\bar{Z}}^{\vec{J}^{\text{G}\rightleftharpoons\text{A}}\vec{J}^{\text{A}}}\cdot\bar{a}^{\vec{J}^{\text{A}}}+\bar{\bar{Z}}^{\vec{J}^{\text{G}\rightleftharpoons\text{A}}\vec{J}^{\text{A}\rightleftharpoons\text{F}}}\cdot\bar{a}^{\vec{J}^{\text{A}\rightleftharpoons\text{F}}}\\&+\bar{\bar{Z}}^{\vec{J}^{\text{G}\rightleftharpoons\text{A}}\vec{J}^{\text{G}\rightleftharpoons\text{F}}}\cdot\bar{a}^{\vec{J}^{\text{G}\rightleftharpoons\text{F}}}+\bar{\bar{Z}}^{\vec{J}^{\text{G}\rightleftharpoons\text{A}}\vec{M}^{\text{O}\rightleftharpoons\text{G}}}\cdot\bar{a}^{\vec{M}^{\text{O}\rightleftharpoons\text{G}}}+\bar{\bar{Z}}^{\vec{J}^{\text{G}\rightleftharpoons\text{A}}\vec{M}^{\text{G}\rightleftharpoons\text{A}}}\cdot\bar{a}^{\vec{M}^{\text{G}\rightleftharpoons\text{A}}}+\bar{\bar{Z}}^{\vec{J}^{\text{G}\rightleftharpoons\text{A}}\vec{M}^{\text{A}\rightleftharpoons\text{F}}}\cdot\bar{a}^{\vec{M}^{\text{A}\rightleftharpoons\text{F}}}\\&+\bar{\bar{Z}}^{\vec{J}^{\text{G}\rightleftharpoons\text{A}}\vec{M}^{\text{G}\rightleftharpoons\text{F}}}\cdot\bar{a}^{\vec{M}^{\text{G}\rightleftharpoons\text{F}}}=0\end{aligned}\tag{8-19a}$$

$$\begin{aligned}&\bar{\bar{Z}}^{\vec{M}^{\text{G}\rightleftharpoons\text{A}}\vec{J}^{\text{O}\rightleftharpoons\text{G}}}\cdot\bar{a}^{\vec{J}^{\text{O}\rightleftharpoons\text{G}}}+\bar{\bar{Z}}^{\vec{M}^{\text{G}\rightleftharpoons\text{A}}\vec{J}^{\text{G}}}\cdot\bar{a}^{\vec{J}^{\text{G}}}+\bar{\bar{Z}}^{\vec{M}^{\text{G}\rightleftharpoons\text{A}}\vec{J}^{\text{G}\rightleftharpoons\text{A}}}\cdot\bar{a}^{\vec{J}^{\text{G}\rightleftharpoons\text{A}}}+\bar{\bar{Z}}^{\vec{M}^{\text{G}\rightleftharpoons\text{A}}\vec{J}^{\text{A}}}\cdot\bar{a}^{\vec{J}^{\text{A}}}+\bar{\bar{Z}}^{\vec{M}^{\text{G}\rightleftharpoons\text{A}}\vec{J}^{\text{A}\rightleftharpoons\text{F}}}\cdot\bar{a}^{\vec{J}^{\text{A}\rightleftharpoons\text{F}}}\\&+\bar{\bar{Z}}^{\vec{M}^{\text{G}\rightleftharpoons\text{A}}\vec{J}^{\text{G}\rightleftharpoons\text{F}}}\cdot\bar{a}^{\vec{J}^{\text{G}\rightleftharpoons\text{F}}}+\bar{\bar{Z}}^{\vec{M}^{\text{G}\rightleftharpoons\text{A}}\vec{M}^{\text{O}\rightleftharpoons\text{G}}}\cdot\bar{a}^{\vec{M}^{\text{O}\rightleftharpoons\text{G}}}+\bar{\bar{Z}}^{\vec{M}^{\text{G}\rightleftharpoons\text{A}}\vec{M}^{\text{G}\rightleftharpoons\text{A}}}\cdot\bar{a}^{\vec{M}^{\text{G}\rightleftharpoons\text{A}}}+\bar{\bar{Z}}^{\vec{M}^{\text{G}\rightleftharpoons\text{A}}\vec{M}^{\text{A}\rightleftharpoons\text{F}}}\cdot\bar{a}^{\vec{M}^{\text{A}\rightleftharpoons\text{F}}}\\&+\bar{\bar{Z}}^{\vec{M}^{\text{G}\rightleftharpoons\text{A}}\vec{M}^{\text{G}\rightleftharpoons\text{F}}}\cdot\bar{a}^{\vec{M}^{\text{G}\rightleftharpoons\text{F}}}=0\end{aligned}\tag{8-19b}$$

and

$$\begin{aligned}&\bar{\bar{Z}}^{\vec{J}^{\text{A}}\vec{J}^{\text{G}\rightleftharpoons\text{A}}}\cdot\bar{a}^{\vec{J}^{\text{G}\rightleftharpoons\text{A}}}+\bar{\bar{Z}}^{\vec{J}^{\text{A}}\vec{J}^{\text{A}}}\cdot\bar{a}^{\vec{J}^{\text{A}}}+\bar{\bar{Z}}^{\vec{J}^{\text{A}}\vec{J}^{\text{A}\rightleftharpoons\text{F}}}\cdot\bar{a}^{\vec{J}^{\text{A}\rightleftharpoons\text{F}}}+\bar{\bar{Z}}^{\vec{J}^{\text{A}}\vec{M}^{\text{G}\rightleftharpoons\text{A}}}\cdot\bar{a}^{\vec{M}^{\text{G}\rightleftharpoons\text{A}}}+\bar{\bar{Z}}^{\vec{J}^{\text{A}}\vec{M}^{\text{A}\rightleftharpoons\text{F}}}\cdot\bar{a}^{\vec{M}^{\text{A}\rightleftharpoons\text{F}}}\\&=0\end{aligned}\tag{8-20}$$

and

$$\overline{\overline{Z}}^{\vec{J}^{\mathrm{A}\rightleftharpoons\mathrm{F}}\vec{J}^{\mathrm{G}\rightleftharpoons\mathrm{A}}}\cdot\bar{a}^{\vec{J}^{\mathrm{G}\rightleftharpoons\mathrm{A}}}+\overline{\overline{Z}}^{\vec{J}^{\mathrm{A}\rightleftharpoons\mathrm{F}}\vec{J}^{\mathrm{A}}}\cdot\bar{a}^{\vec{J}^{\mathrm{A}}}+\overline{\overline{Z}}^{\vec{J}^{\mathrm{A}\rightleftharpoons\mathrm{F}}\vec{J}^{\mathrm{A}\rightleftharpoons\mathrm{F}}}\cdot\bar{a}^{\vec{J}^{\mathrm{A}\rightleftharpoons\mathrm{F}}}+\overline{\overline{Z}}^{\vec{J}^{\mathrm{A}\rightleftharpoons\mathrm{F}}\vec{J}^{\mathrm{F}}}\cdot\bar{a}^{\vec{J}^{\mathrm{F}}}+\overline{\overline{Z}}^{\vec{J}^{\mathrm{A}\rightleftharpoons\mathrm{F}}\vec{J}^{\mathrm{G}\rightleftharpoons\mathrm{F}}}\cdot\bar{a}^{\vec{J}^{\mathrm{G}\rightleftharpoons\mathrm{F}}}$$
$$+\overline{\overline{Z}}^{\vec{J}^{\mathrm{A}\rightleftharpoons\mathrm{F}}\vec{M}^{\mathrm{G}\rightleftharpoons\mathrm{A}}}\cdot\bar{a}^{\vec{M}^{\mathrm{G}\rightleftharpoons\mathrm{A}}}+\overline{\overline{Z}}^{\vec{J}^{\mathrm{A}\rightleftharpoons\mathrm{F}}\vec{M}^{\mathrm{A}\rightleftharpoons\mathrm{F}}}\cdot\bar{a}^{\vec{M}^{\mathrm{A}\rightleftharpoons\mathrm{F}}}+\overline{\overline{Z}}^{\vec{J}^{\mathrm{A}\rightleftharpoons\mathrm{F}}\vec{M}^{\mathrm{G}\rightleftharpoons\mathrm{F}}}\cdot\bar{a}^{\vec{M}^{\mathrm{G}\rightleftharpoons\mathrm{F}}}=0 \tag{8-21a}$$

$$\overline{\overline{Z}}^{\vec{M}^{\mathrm{A}\rightleftharpoons\mathrm{F}}\vec{J}^{\mathrm{G}\rightleftharpoons\mathrm{A}}}\cdot\bar{a}^{\vec{J}^{\mathrm{G}\rightleftharpoons\mathrm{A}}}+\overline{\overline{Z}}^{\vec{M}^{\mathrm{A}\rightleftharpoons\mathrm{F}}\vec{J}^{\mathrm{A}}}\cdot\bar{a}^{\vec{J}^{\mathrm{A}}}+\overline{\overline{Z}}^{\vec{M}^{\mathrm{A}\rightleftharpoons\mathrm{F}}\vec{J}^{\mathrm{A}\rightleftharpoons\mathrm{F}}}\cdot\bar{a}^{\vec{J}^{\mathrm{A}\rightleftharpoons\mathrm{F}}}+\overline{\overline{Z}}^{\vec{M}^{\mathrm{A}\rightleftharpoons\mathrm{F}}\vec{J}^{\mathrm{F}}}\cdot\bar{a}^{\vec{J}^{\mathrm{F}}}+\overline{\overline{Z}}^{\vec{M}^{\mathrm{A}\rightleftharpoons\mathrm{F}}\vec{J}^{\mathrm{G}\rightleftharpoons\mathrm{F}}}\cdot\bar{a}^{\vec{J}^{\mathrm{G}\rightleftharpoons\mathrm{F}}}$$
$$+\overline{\overline{Z}}^{\vec{M}^{\mathrm{A}\rightleftharpoons\mathrm{F}}\vec{M}^{\mathrm{G}\rightleftharpoons\mathrm{A}}}\cdot\bar{a}^{\vec{M}^{\mathrm{G}\rightleftharpoons\mathrm{A}}}+\overline{\overline{Z}}^{\vec{M}^{\mathrm{A}\rightleftharpoons\mathrm{F}}\vec{M}^{\mathrm{A}\rightleftharpoons\mathrm{F}}}\cdot\bar{a}^{\vec{M}^{\mathrm{A}\rightleftharpoons\mathrm{F}}}+\overline{\overline{Z}}^{\vec{M}^{\mathrm{A}\rightleftharpoons\mathrm{F}}\vec{M}^{\mathrm{G}\rightleftharpoons\mathrm{F}}}\cdot\bar{a}^{\vec{M}^{\mathrm{G}\rightleftharpoons\mathrm{F}}}=0 \tag{8-21b}$$

and

$$\overline{\overline{Z}}^{\vec{J}^{\mathrm{F}}\vec{J}^{\mathrm{A}\rightleftharpoons\mathrm{F}}}\cdot\bar{a}^{\vec{J}^{\mathrm{A}\rightleftharpoons\mathrm{F}}}+\overline{\overline{Z}}^{\vec{J}^{\mathrm{F}}\vec{J}^{\mathrm{F}}}\cdot\bar{a}^{\vec{J}^{\mathrm{F}}}+\overline{\overline{Z}}^{\vec{J}^{\mathrm{F}}\vec{J}^{\mathrm{G}\rightleftharpoons\mathrm{F}}}\cdot\bar{a}^{\vec{J}^{\mathrm{G}\rightleftharpoons\mathrm{F}}}+\overline{\overline{Z}}^{\vec{J}^{\mathrm{F}}\vec{M}^{\mathrm{A}\rightleftharpoons\mathrm{F}}}\cdot\bar{a}^{\vec{M}^{\mathrm{A}\rightleftharpoons\mathrm{F}}}+\overline{\overline{Z}}^{\vec{J}^{\mathrm{F}}\vec{M}^{\mathrm{G}\rightleftharpoons\mathrm{F}}}\cdot\bar{a}^{\vec{M}^{\mathrm{G}\rightleftharpoons\mathrm{F}}}$$
$$=0 \tag{8-22}$$

and

$$\overline{\overline{Z}}^{\vec{J}^{\mathrm{G}\rightleftharpoons\mathrm{F}}\vec{J}^{\mathrm{O}\rightleftharpoons\mathrm{G}}}\cdot\bar{a}^{\vec{J}^{\mathrm{O}\rightleftharpoons\mathrm{G}}}+\overline{\overline{Z}}^{\vec{J}^{\mathrm{G}\rightleftharpoons\mathrm{F}}\vec{J}^{\mathrm{G}}}\cdot\bar{a}^{\vec{J}^{\mathrm{G}}}+\overline{\overline{Z}}^{\vec{J}^{\mathrm{G}\rightleftharpoons\mathrm{F}}\vec{J}^{\mathrm{G}\rightleftharpoons\mathrm{A}}}\cdot\bar{a}^{\vec{J}^{\mathrm{G}\rightleftharpoons\mathrm{A}}}+\overline{\overline{Z}}^{\vec{J}^{\mathrm{G}\rightleftharpoons\mathrm{F}}\vec{J}^{\mathrm{A}\rightleftharpoons\mathrm{F}}}\cdot\bar{a}^{\vec{J}^{\mathrm{A}\rightleftharpoons\mathrm{F}}}+\overline{\overline{Z}}^{\vec{J}^{\mathrm{G}\rightleftharpoons\mathrm{F}}\vec{J}^{\mathrm{F}}}\cdot\bar{a}^{\vec{J}^{\mathrm{F}}}$$
$$+\overline{\overline{Z}}^{\vec{J}^{\mathrm{G}\rightleftharpoons\mathrm{F}}\vec{J}^{\mathrm{G}\rightleftharpoons\mathrm{F}}}\cdot\bar{a}^{\vec{J}^{\mathrm{G}\rightleftharpoons\mathrm{F}}}+\overline{\overline{Z}}^{\vec{J}^{\mathrm{G}\rightleftharpoons\mathrm{F}}\vec{M}^{\mathrm{O}\rightleftharpoons\mathrm{G}}}\cdot\bar{a}^{\vec{M}^{\mathrm{O}\rightleftharpoons\mathrm{G}}}+\overline{\overline{Z}}^{\vec{J}^{\mathrm{G}\rightleftharpoons\mathrm{F}}\vec{M}^{\mathrm{G}\rightleftharpoons\mathrm{A}}}\cdot\bar{a}^{\vec{M}^{\mathrm{G}\rightleftharpoons\mathrm{A}}}+\overline{\overline{Z}}^{\vec{J}^{\mathrm{G}\rightleftharpoons\mathrm{F}}\vec{M}^{\mathrm{A}\rightleftharpoons\mathrm{F}}}\cdot\bar{a}^{\vec{M}^{\mathrm{A}\rightleftharpoons\mathrm{F}}}$$
$$+\overline{\overline{Z}}^{\vec{J}^{\mathrm{G}\rightleftharpoons\mathrm{F}}\vec{M}^{\mathrm{G}\rightleftharpoons\mathrm{F}}}\cdot\bar{a}^{\vec{M}^{\mathrm{G}\rightleftharpoons\mathrm{F}}}=0 \tag{8-23a}$$

$$\overline{\overline{Z}}^{\vec{M}^{\mathrm{G}\rightleftharpoons\mathrm{F}}\vec{J}^{\mathrm{O}\rightleftharpoons\mathrm{G}}}\cdot\bar{a}^{\vec{J}^{\mathrm{O}\rightleftharpoons\mathrm{G}}}+\overline{\overline{Z}}^{\vec{M}^{\mathrm{G}\rightleftharpoons\mathrm{F}}\vec{J}^{\mathrm{G}}}\cdot\bar{a}^{\vec{J}^{\mathrm{G}}}+\overline{\overline{Z}}^{\vec{M}^{\mathrm{G}\rightleftharpoons\mathrm{F}}\vec{J}^{\mathrm{G}\rightleftharpoons\mathrm{A}}}\cdot\bar{a}^{\vec{J}^{\mathrm{G}\rightleftharpoons\mathrm{A}}}+\overline{\overline{Z}}^{\vec{M}^{\mathrm{G}\rightleftharpoons\mathrm{F}}\vec{J}^{\mathrm{A}\rightleftharpoons\mathrm{F}}}\cdot\bar{a}^{\vec{J}^{\mathrm{A}\rightleftharpoons\mathrm{F}}}+\overline{\overline{Z}}^{\vec{M}^{\mathrm{G}\rightleftharpoons\mathrm{F}}\vec{J}^{\mathrm{F}}}\cdot\bar{a}^{\vec{J}^{\mathrm{F}}}$$
$$+\overline{\overline{Z}}^{\vec{M}^{\mathrm{G}\rightleftharpoons\mathrm{F}}\vec{J}^{\mathrm{G}\rightleftharpoons\mathrm{F}}}\cdot\bar{a}^{\vec{J}^{\mathrm{G}\rightleftharpoons\mathrm{F}}}+\overline{\overline{Z}}^{\vec{M}^{\mathrm{G}\rightleftharpoons\mathrm{F}}\vec{M}^{\mathrm{O}\rightleftharpoons\mathrm{G}}}\cdot\bar{a}^{\vec{M}^{\mathrm{O}\rightleftharpoons\mathrm{G}}}+\overline{\overline{Z}}^{\vec{M}^{\mathrm{G}\rightleftharpoons\mathrm{F}}\vec{M}^{\mathrm{G}\rightleftharpoons\mathrm{A}}}\cdot\bar{a}^{\vec{M}^{\mathrm{G}\rightleftharpoons\mathrm{A}}}+\overline{\overline{Z}}^{\vec{M}^{\mathrm{G}\rightleftharpoons\mathrm{F}}\vec{M}^{\mathrm{A}\rightleftharpoons\mathrm{F}}}\cdot\bar{a}^{\vec{M}^{\mathrm{A}\rightleftharpoons\mathrm{F}}}$$
$$+\overline{\overline{Z}}^{\vec{M}^{\mathrm{G}\rightleftharpoons\mathrm{F}}\vec{M}^{\mathrm{G}\rightleftharpoons\mathrm{F}}}\cdot\bar{a}^{\vec{M}^{\mathrm{G}\rightleftharpoons\mathrm{F}}}=0 \tag{8-23b}$$

The formulations used to calculate the elements of the matrices in Eq. (8-17a) are as follows:

$$z_{\xi\zeta}^{\vec{M}^{\mathrm{O}\rightleftharpoons\mathrm{G}}\vec{J}^{\mathrm{O}\rightleftharpoons\mathrm{G}}} = \left\langle \vec{b}_{\xi}^{\vec{M}^{\mathrm{O}\rightleftharpoons\mathrm{G}}}, \mathcal{H}_{\mathrm{G}}\left(\vec{b}_{\zeta}^{\vec{J}^{\mathrm{O}\rightleftharpoons\mathrm{G}}}\right)\right\rangle_{\underset{\sim}{\mathbb{S}}^{\mathrm{O}\rightleftharpoons\mathrm{G}}} + \left\langle \vec{b}_{\xi}^{\vec{M}^{\mathrm{O}\rightleftharpoons\mathrm{G}}}, \hat{n}^{\rightarrow\mathrm{G}}\times\vec{b}_{\zeta}^{\vec{J}^{\mathrm{O}\rightleftharpoons\mathrm{G}}}\right\rangle_{\mathbb{S}^{\mathrm{O}\rightleftharpoons\mathrm{G}}} \tag{8-24a}$$

$$z_{\xi\zeta}^{\vec{M}^{\mathrm{O}\rightleftharpoons\mathrm{G}}\vec{J}^{\mathrm{G}}} = \left\langle \vec{b}_{\xi}^{\vec{M}^{\mathrm{O}\rightleftharpoons\mathrm{G}}}, \mathcal{H}_{\mathrm{G}}\left(\vec{b}_{\zeta}^{\vec{J}^{\mathrm{G}}}\right)\right\rangle_{\underset{\sim}{\mathbb{S}}^{\mathrm{O}\rightleftharpoons\mathrm{G}}} \tag{8-24b}$$

$$z_{\xi\zeta}^{\vec{M}^{\mathrm{O}\rightleftharpoons\mathrm{G}}\vec{J}^{\mathrm{G}\rightleftharpoons\mathrm{A}}} = \left\langle \vec{b}_{\xi}^{\vec{M}^{\mathrm{O}\rightleftharpoons\mathrm{G}}}, \mathcal{H}_{\mathrm{G}}\left(-\vec{b}_{\zeta}^{\vec{J}^{\mathrm{G}\rightleftharpoons\mathrm{A}}}\right)\right\rangle_{\underset{\sim}{\mathbb{S}}^{\mathrm{O}\rightleftharpoons\mathrm{G}}} \tag{8-24c}$$

$$z_{\xi\zeta}^{\vec{M}^{\mathrm{O}\rightleftharpoons\mathrm{G}}\vec{J}^{\mathrm{G}\rightleftharpoons\mathrm{F}}} = \left\langle \vec{b}_{\xi}^{\vec{M}^{\mathrm{O}\rightleftharpoons\mathrm{G}}}, \mathcal{H}_{\mathrm{G}}\left(\vec{b}_{\zeta}^{\vec{J}^{\mathrm{G}\rightleftharpoons\mathrm{F}}}\right)\right\rangle_{\underset{\sim}{\mathbb{S}}^{\mathrm{O}\rightleftharpoons\mathrm{G}}} \tag{8-24d}$$

$$z_{\xi\zeta}^{\vec{M}^{\mathrm{O}\rightleftharpoons\mathrm{G}}\vec{M}^{\mathrm{O}\rightleftharpoons\mathrm{G}}} = \left\langle \vec{b}_{\xi}^{\vec{M}^{\mathrm{O}\rightleftharpoons\mathrm{G}}}, \mathcal{H}_{\mathrm{G}}\left(\vec{b}_{\zeta}^{\vec{M}^{\mathrm{O}\rightleftharpoons\mathrm{G}}}\right)\right\rangle_{\underset{\sim}{\mathbb{S}}^{\mathrm{O}\rightleftharpoons\mathrm{G}}} \tag{8-24e}$$

$$z_{\xi\zeta}^{\vec{M}^{\mathrm{O}\rightleftharpoons\mathrm{G}}\vec{M}^{\mathrm{G}\rightleftharpoons\mathrm{A}}} = \left\langle \vec{b}_{\xi}^{\vec{M}^{\mathrm{O}\rightleftharpoons\mathrm{G}}}, \mathcal{H}_{\mathrm{G}}\left(-\vec{b}_{\zeta}^{\vec{M}^{\mathrm{G}\rightleftharpoons\mathrm{A}}}\right)\right\rangle_{\underset{\sim}{\mathbb{S}}^{\mathrm{O}\rightleftharpoons\mathrm{G}}} \tag{8-24f}$$

$$z_{\xi\zeta}^{\vec{M}^{\mathrm{O}\rightleftharpoons\mathrm{G}}\vec{M}^{\mathrm{G}\rightleftharpoons\mathrm{F}}} = \left\langle \vec{b}_{\xi}^{\vec{M}^{\mathrm{O}\rightleftharpoons\mathrm{G}}}, \mathcal{H}_{\mathrm{G}}\left(\vec{b}_{\zeta}^{\vec{M}^{\mathrm{G}\rightleftharpoons\mathrm{F}}}\right)\right\rangle_{\underset{\sim}{\mathbb{S}}^{\mathrm{O}\rightleftharpoons\mathrm{G}}} \tag{8-24g}$$

where the integral surface $\underset{\sim}{\mathbb{S}}^{\mathrm{O}\rightleftharpoons\mathrm{G}}$ is a surface approaching $\mathbb{S}^{\mathrm{O}\rightleftharpoons\mathrm{G}}$ from the side of $\mathbb{V}^{\mathrm{G}}$ and it is shown in the following Fig. 8-4

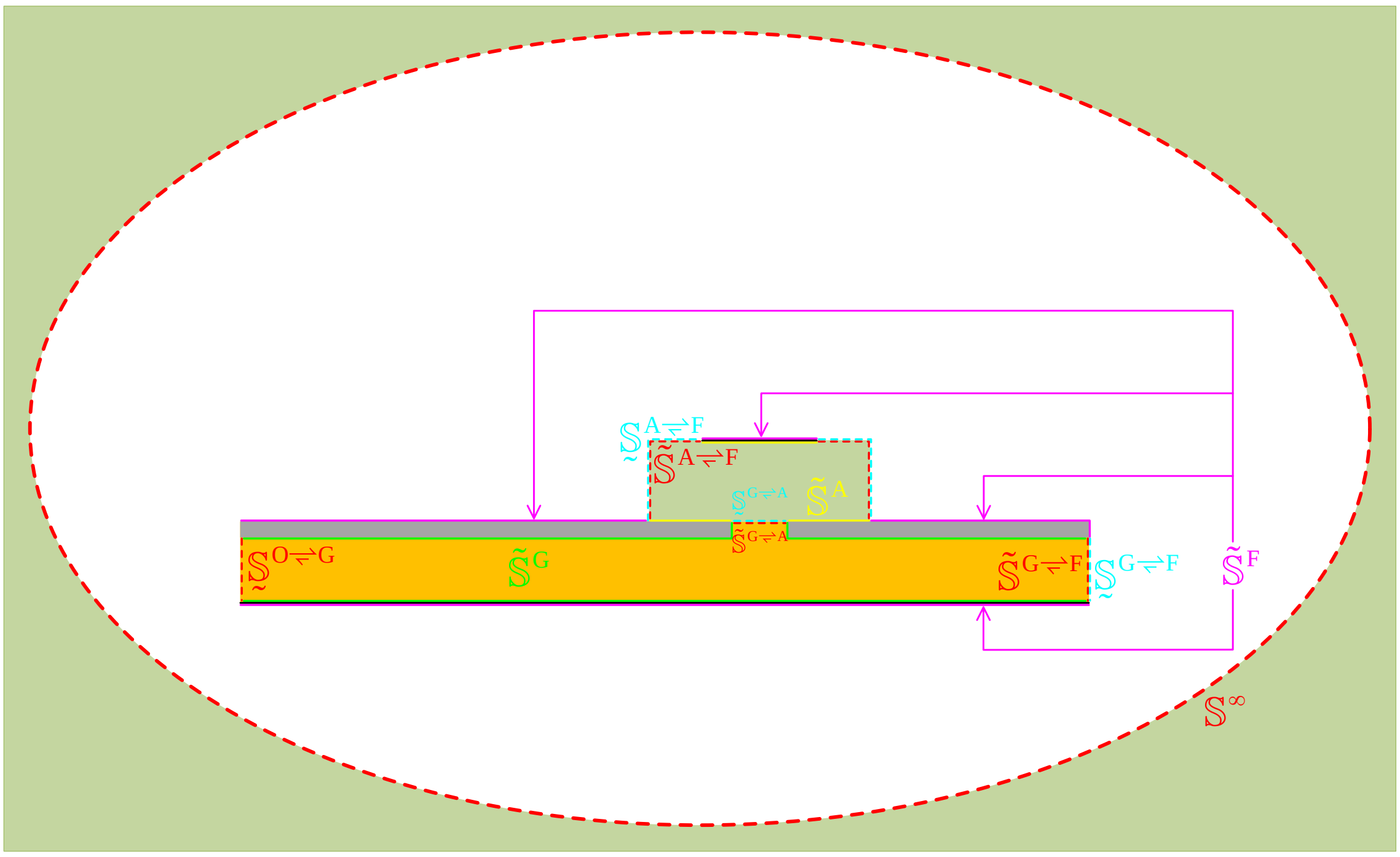

Figure 8-4 Integral surfaces used in Eqs. (8-24)~(8-34)

The formulations used to calculate the elements of the matrices in Eq. (8-17b) are as follows:

$$z_{\xi\zeta}^{\vec{J}^{O\rightleftharpoons G}\vec{J}^{O\rightleftharpoons G}} = \left\langle \vec{b}_{\xi}^{\vec{J}^{O\rightleftharpoons G}}, \mathcal{E}_{G}\left(\vec{b}_{\zeta}^{\vec{J}^{O\rightleftharpoons G}}\right)\right\rangle_{\underset{\sim}{\mathbb{S}}^{O\rightleftharpoons G}} \tag{8-25a}$$

$$z_{\xi\zeta}^{\vec{J}^{O\rightleftharpoons G}\vec{J}^{G}} = \left\langle \vec{b}_{\xi}^{\vec{J}^{O\rightleftharpoons G}}, \mathcal{E}_{G}\left(\vec{b}_{\zeta}^{\vec{J}^{G}}\right)\right\rangle_{\underset{\sim}{\mathbb{S}}^{O\rightleftharpoons G}} \tag{8-25b}$$

$$z_{\xi\zeta}^{\vec{J}^{O\rightleftharpoons G}\vec{J}^{G\rightleftharpoons A}} = \left\langle \vec{b}_{\xi}^{\vec{J}^{O\rightleftharpoons G}}, \mathcal{E}_{G}\left(-\vec{b}_{\zeta}^{\vec{J}^{G\rightleftharpoons A}}\right)\right\rangle_{\underset{\sim}{\mathbb{S}}^{O\rightleftharpoons G}} \tag{8-25c}$$

$$z_{\xi\zeta}^{\vec{J}^{O\rightleftharpoons G}\vec{J}^{G\rightleftharpoons F}} = \left\langle \vec{b}_{\xi}^{\vec{J}^{O\rightleftharpoons G}}, \mathcal{E}_{G}\left(\vec{b}_{\zeta}^{\vec{J}^{G\rightleftharpoons F}}\right)\right\rangle_{\underset{\sim}{\mathbb{S}}^{O\rightleftharpoons G}} \tag{8-25d}$$

$$z_{\xi\zeta}^{\vec{J}^{O\rightleftharpoons G}\vec{M}^{O\rightleftharpoons G}} = \left\langle \vec{b}_{\xi}^{\vec{J}^{O\rightleftharpoons G}}, \mathcal{E}_{G}\left(\vec{b}_{\zeta}^{\vec{M}^{O\rightleftharpoons G}}\right)\right\rangle_{\underset{\sim}{\mathbb{S}}^{O\rightleftharpoons G}} + \left\langle \vec{b}_{\xi}^{\vec{J}^{O\rightleftharpoons G}}, \vec{b}_{\zeta}^{\vec{M}^{O\rightleftharpoons G}} \times \hat{n}^{\rightarrow G}\right\rangle_{\underset{\sim}{\mathbb{S}}^{O\rightleftharpoons G}} \tag{8-25e}$$

$$z_{\xi\zeta}^{\vec{J}^{O\rightleftharpoons G}\vec{M}^{G\rightleftharpoons A}} = \left\langle \vec{b}_{\xi}^{\vec{J}^{O\rightleftharpoons G}}, \mathcal{E}_{G}\left(-\vec{b}_{\zeta}^{\vec{M}^{G\rightleftharpoons A}}\right)\right\rangle_{\underset{\sim}{\mathbb{S}}^{O\rightleftharpoons G}} \tag{8-25f}$$

$$z_{\xi\zeta}^{\vec{J}^{O\rightleftharpoons G}\vec{M}^{G\rightleftharpoons F}} = \left\langle \vec{b}_{\xi}^{\vec{J}^{O\rightleftharpoons G}}, \mathcal{E}_{G}\left(\vec{b}_{\zeta}^{\vec{M}^{G\rightleftharpoons F}}\right)\right\rangle_{\underset{\sim}{\mathbb{S}}^{O\rightleftharpoons G}} \tag{8-25g}$$

The formulations used to calculate the elements of the matrices in Eq. (8-18) are as follows:

$$z_{\xi\zeta}^{\vec{J}^{G}\vec{J}^{O\rightleftharpoons G}} = \left\langle \vec{b}_{\xi}^{\vec{J}^{G}}, \mathcal{E}_{G}\left(\vec{b}_{\zeta}^{\vec{J}^{O\rightleftharpoons G}}\right)\right\rangle_{\tilde{\mathbb{S}}^{G}} \tag{8-26a}$$

$$z_{\xi\zeta}^{\vec{J}^{G}\vec{J}^{G}} = \left\langle \vec{b}_{\xi}^{\vec{J}^{G}}, \mathcal{E}_{G}\left(\vec{b}_{\zeta}^{\vec{J}^{G}}\right)\right\rangle_{\tilde{\mathbb{S}}^{G}} \tag{8-26b}$$

$$z_{\xi\zeta}^{\vec{J}^{G}\vec{J}^{G\rightleftharpoons A}} = \left\langle \vec{b}_{\xi}^{\vec{J}^{G}}, \mathcal{E}_{G}\left(-\vec{b}_{\zeta}^{\vec{J}^{G\rightleftharpoons A}}\right)\right\rangle_{\tilde{\mathbb{S}}^{G}} \tag{8-26c}$$

$$z_{\xi\zeta}^{\vec{J}^{\mathrm{G}}\vec{J}^{\mathrm{G}\rightleftharpoons\mathrm{F}}} = \left\langle \vec{b}_{\xi}^{\vec{J}^{\mathrm{G}}}, \mathcal{E}_{\mathrm{G}}\left(\vec{b}_{\zeta}^{\vec{J}^{\mathrm{G}\rightleftharpoons\mathrm{F}}}\right)\right\rangle_{\tilde{\mathbb{S}}^{\mathrm{G}}} \tag{8-26d}$$

$$z_{\xi\zeta}^{\vec{J}^{\mathrm{G}}\vec{M}^{\mathrm{O}\rightleftharpoons\mathrm{G}}} = \left\langle \vec{b}_{\xi}^{\vec{J}^{\mathrm{G}}}, \mathcal{E}_{\mathrm{G}}\left(\vec{b}_{\zeta}^{\vec{M}^{\mathrm{O}\rightleftharpoons\mathrm{G}}}\right)\right\rangle_{\tilde{\mathbb{S}}^{\mathrm{G}}} \tag{8-26e}$$

$$z_{\xi\zeta}^{\vec{J}^{\mathrm{G}}\vec{M}^{\mathrm{G}\rightleftharpoons\mathrm{A}}} = \left\langle \vec{b}_{\xi}^{\vec{J}^{\mathrm{G}}}, \mathcal{E}_{\mathrm{G}}\left(-\vec{b}_{\zeta}^{\vec{M}^{\mathrm{G}\rightleftharpoons\mathrm{A}}}\right)\right\rangle_{\tilde{\mathbb{S}}^{\mathrm{G}}} \tag{8-26f}$$

$$z_{\xi\zeta}^{\vec{J}^{\mathrm{G}}\vec{M}^{\mathrm{G}\rightleftharpoons\mathrm{F}}} = \left\langle \vec{b}_{\xi}^{\vec{J}^{\mathrm{G}}}, \mathcal{E}_{\mathrm{G}}\left(\vec{b}_{\zeta}^{\vec{M}^{\mathrm{G}\rightleftharpoons\mathrm{F}}}\right)\right\rangle_{\tilde{\mathbb{S}}^{\mathrm{G}}} \tag{8-26g}$$

where the integral surface $\tilde{\mathbb{S}}^{\mathrm{G}}$ is shown in Fig. 8-4. The formulations used to calculate the elements of the matrices in Eq. (8-19a) are as follows:

$$z_{\xi\zeta}^{\vec{J}^{\mathrm{G}\rightleftharpoons\mathrm{A}}\vec{J}^{\mathrm{O}\rightleftharpoons\mathrm{G}}} = \left\langle \vec{b}_{\xi}^{\vec{J}^{\mathrm{G}\rightleftharpoons\mathrm{A}}}, \mathcal{E}_{\mathrm{G}}\left(\vec{b}_{\zeta}^{\vec{J}^{\mathrm{O}\rightleftharpoons\mathrm{G}}}\right)\right\rangle_{\tilde{\mathbb{S}}^{\mathrm{G}\rightleftharpoons\mathrm{A}}} \tag{8-27a}$$

$$z_{\xi\zeta}^{\vec{J}^{\mathrm{G}\rightleftharpoons\mathrm{A}}\vec{J}^{\mathrm{G}}} = \left\langle \vec{b}_{\xi}^{\vec{J}^{\mathrm{G}\rightleftharpoons\mathrm{A}}}, \mathcal{E}_{\mathrm{G}}\left(\vec{b}_{\zeta}^{\vec{J}^{\mathrm{G}}}\right)\right\rangle_{\tilde{\mathbb{S}}^{\mathrm{G}\rightleftharpoons\mathrm{A}}} \tag{8-27b}$$

$$z_{\xi\zeta}^{\vec{J}^{\mathrm{G}\rightleftharpoons\mathrm{A}}\vec{J}^{\mathrm{G}\rightleftharpoons\mathrm{A}}} = \left\langle \vec{b}_{\xi}^{\vec{J}^{\mathrm{G}\rightleftharpoons\mathrm{A}}}, \mathcal{E}_{\mathrm{G}}\left(-\vec{b}_{\zeta}^{\vec{J}^{\mathrm{G}\rightleftharpoons\mathrm{A}}}\right)\right\rangle_{\tilde{\mathbb{S}}^{\mathrm{G}\rightleftharpoons\mathrm{A}}} + \left\langle \vec{b}_{\xi}^{\vec{J}^{\mathrm{G}\rightleftharpoons\mathrm{A}}}, -\mathcal{E}_{\mathrm{A}}\left(\vec{b}_{\zeta}^{\vec{J}^{\mathrm{G}\rightleftharpoons\mathrm{A}}}\right)\right\rangle_{\underset{\sim}{\mathbb{S}}^{\mathrm{G}\rightleftharpoons\mathrm{A}}} \tag{8-27c}$$

$$z_{\xi\zeta}^{\vec{J}^{\mathrm{G}\rightleftharpoons\mathrm{A}}\vec{J}^{\mathrm{A}}} = \left\langle \vec{b}_{\xi}^{\vec{J}^{\mathrm{G}\rightleftharpoons\mathrm{A}}}, -\mathcal{E}_{\mathrm{A}}\left(\vec{b}_{\zeta}^{\vec{J}^{\mathrm{A}}}\right)\right\rangle_{\underset{\sim}{\mathbb{S}}^{\mathrm{G}\rightleftharpoons\mathrm{A}}} \tag{8-27d}$$

$$z_{\xi\zeta}^{\vec{J}^{\mathrm{G}\rightleftharpoons\mathrm{A}}\vec{J}^{\mathrm{A}\rightleftharpoons\mathrm{F}}} = \left\langle \vec{b}_{\xi}^{\vec{J}^{\mathrm{G}\rightleftharpoons\mathrm{A}}}, -\mathcal{E}_{\mathrm{A}}\left(-\vec{b}_{\zeta}^{\vec{J}^{\mathrm{A}\rightleftharpoons\mathrm{F}}}\right)\right\rangle_{\underset{\sim}{\mathbb{S}}^{\mathrm{G}\rightleftharpoons\mathrm{A}}} \tag{8-27e}$$

$$z_{\xi\zeta}^{\vec{J}^{\mathrm{G}\rightleftharpoons\mathrm{A}}\vec{J}^{\mathrm{G}\rightleftharpoons\mathrm{F}}} = \left\langle \vec{b}_{\xi}^{\vec{J}^{\mathrm{G}\rightleftharpoons\mathrm{A}}}, \mathcal{E}_{\mathrm{G}}\left(\vec{b}_{\zeta}^{\vec{J}^{\mathrm{G}\rightleftharpoons\mathrm{F}}}\right)\right\rangle_{\tilde{\mathbb{S}}^{\mathrm{G}\rightleftharpoons\mathrm{A}}} \tag{8-27f}$$

$$z_{\xi\zeta}^{\vec{J}^{\mathrm{G}\rightleftharpoons\mathrm{A}}\vec{M}^{\mathrm{O}\rightleftharpoons\mathrm{G}}} = \left\langle \vec{b}_{\xi}^{\vec{J}^{\mathrm{G}\rightleftharpoons\mathrm{A}}}, \mathcal{E}_{\mathrm{G}}\left(\vec{b}_{\zeta}^{\vec{M}^{\mathrm{O}\rightleftharpoons\mathrm{G}}}\right)\right\rangle_{\tilde{\mathbb{S}}^{\mathrm{G}\rightleftharpoons\mathrm{A}}} \tag{8-27g}$$

$$z_{\xi\zeta}^{\vec{J}^{\mathrm{G}\rightleftharpoons\mathrm{A}}\vec{M}^{\mathrm{G}\rightleftharpoons\mathrm{A}}} = \left\langle \vec{b}_{\xi}^{\vec{J}^{\mathrm{G}\rightleftharpoons\mathrm{A}}}, \mathcal{E}_{\mathrm{G}}\left(-\vec{b}_{\zeta}^{\vec{M}^{\mathrm{G}\rightleftharpoons\mathrm{A}}}\right)\right\rangle_{\tilde{\mathbb{S}}^{\mathrm{G}\rightleftharpoons\mathrm{A}}} + \left\langle \vec{b}_{\xi}^{\vec{J}^{\mathrm{G}\rightleftharpoons\mathrm{A}}}, -\mathcal{E}_{\mathrm{A}}\left(\vec{b}_{\zeta}^{\vec{M}^{\mathrm{G}\rightleftharpoons\mathrm{A}}}\right)\right\rangle_{\underset{\sim}{\mathbb{S}}^{\mathrm{G}\rightleftharpoons\mathrm{A}}} \tag{8-27h}$$

$$z_{\xi\zeta}^{\vec{J}^{\mathrm{G}\rightleftharpoons\mathrm{A}}\vec{M}^{\mathrm{A}\rightleftharpoons\mathrm{F}}} = \left\langle \vec{b}_{\xi}^{\vec{J}^{\mathrm{G}\rightleftharpoons\mathrm{A}}}, -\mathcal{E}_{\mathrm{A}}\left(-\vec{b}_{\zeta}^{\vec{M}^{\mathrm{A}\rightleftharpoons\mathrm{F}}}\right)\right\rangle_{\underset{\sim}{\mathbb{S}}^{\mathrm{G}\rightleftharpoons\mathrm{A}}} \tag{8-27i}$$

$$z_{\xi\zeta}^{\vec{J}^{\mathrm{G}\rightleftharpoons\mathrm{A}}\vec{M}^{\mathrm{G}\rightleftharpoons\mathrm{F}}} = \left\langle \vec{b}_{\xi}^{\vec{J}^{\mathrm{G}\rightleftharpoons\mathrm{A}}}, \mathcal{E}_{\mathrm{G}}\left(\vec{b}_{\zeta}^{\vec{M}^{\mathrm{G}\rightleftharpoons\mathrm{F}}}\right)\right\rangle_{\tilde{\mathbb{S}}^{\mathrm{G}\rightleftharpoons\mathrm{A}}} \tag{8-27j}$$

where the integral surface $\tilde{\mathbb{S}}^{\mathrm{G}\rightleftharpoons\mathrm{A}}$ is shown in Fig. 8-4. The formulations used to calculate the elements of the matrices in Eq. (8-19b) are as follows:

$$z_{\xi\zeta}^{\vec{M}^{\mathrm{G}\rightleftharpoons\mathrm{A}}\vec{J}^{\mathrm{O}\rightleftharpoons\mathrm{G}}} = \left\langle \vec{b}_{\xi}^{\vec{M}^{\mathrm{G}\rightleftharpoons\mathrm{A}}}, \mathcal{H}_{\mathrm{G}}\left(\vec{b}_{\zeta}^{\vec{J}^{\mathrm{O}\rightleftharpoons\mathrm{G}}}\right)\right\rangle_{\tilde{\mathbb{S}}^{\mathrm{G}\rightleftharpoons\mathrm{A}}} \tag{8-28a}$$

$$z_{\xi\zeta}^{\vec{M}^{\mathrm{G}\rightleftharpoons\mathrm{A}}\vec{J}^{\mathrm{G}}} = \left\langle \vec{b}_{\xi}^{\vec{M}^{\mathrm{G}\rightleftharpoons\mathrm{A}}}, \mathcal{H}_{\mathrm{G}}\left(\vec{b}_{\zeta}^{\vec{J}^{\mathrm{G}}}\right)\right\rangle_{\tilde{\mathbb{S}}^{\mathrm{G}\rightleftharpoons\mathrm{A}}} \tag{8-28b}$$

$$z_{\xi\zeta}^{\vec{M}^{\mathrm{G}\rightleftharpoons\mathrm{A}}\vec{J}^{\mathrm{G}\rightleftharpoons\mathrm{A}}} = \left\langle \vec{b}_{\xi}^{\vec{M}^{\mathrm{G}\rightleftharpoons\mathrm{A}}}, \mathcal{H}_{\mathrm{G}}\left(-\vec{b}_{\zeta}^{\vec{J}^{\mathrm{G}\rightleftharpoons\mathrm{A}}}\right)\right\rangle_{\tilde{\mathbb{S}}^{\mathrm{G}\rightleftharpoons\mathrm{A}}} + \left\langle \vec{b}_{\xi}^{\vec{M}^{\mathrm{G}\rightleftharpoons\mathrm{A}}}, -\mathcal{H}_{\mathrm{A}}\left(\vec{b}_{\zeta}^{\vec{J}^{\mathrm{G}\rightleftharpoons\mathrm{A}}}\right)\right\rangle_{\underset{\sim}{\mathbb{S}}^{\mathrm{G}\rightleftharpoons\mathrm{A}}} \tag{8-28c}$$

$$z_{\xi\zeta}^{\vec{M}^{\mathrm{G}\rightleftharpoons\mathrm{A}}\vec{J}^{\mathrm{A}}} = \left\langle \vec{b}_{\xi}^{\vec{M}^{\mathrm{G}\rightleftharpoons\mathrm{A}}}, -\mathcal{H}_{\mathrm{A}}\left(\vec{b}_{\zeta}^{\vec{J}^{\mathrm{A}}}\right)\right\rangle_{\underset{\sim}{\mathbb{S}}^{\mathrm{G}\rightleftharpoons\mathrm{A}}} \tag{8-28d}$$

$$z_{\xi\zeta}^{\vec{M}^{\mathrm{G}\rightleftharpoons\mathrm{A}}\vec{J}^{\mathrm{A}\rightleftharpoons\mathrm{F}}} = \left\langle \vec{b}_{\xi}^{\vec{M}^{\mathrm{G}\rightleftharpoons\mathrm{A}}}, -\mathcal{H}_{\mathrm{A}}\left(-\vec{b}_{\zeta}^{\vec{J}^{\mathrm{A}\rightleftharpoons\mathrm{F}}}\right)\right\rangle_{\underset{\sim}{\mathbb{S}}^{\mathrm{G}\rightleftharpoons\mathrm{A}}} \tag{8-28e}$$

$$z_{\xi\zeta}^{\vec{M}^{\mathrm{G}\rightleftharpoons\mathrm{A}}\vec{J}^{\mathrm{G}\rightleftharpoons\mathrm{F}}} = \left\langle \vec{b}_{\xi}^{\vec{M}^{\mathrm{G}\rightleftharpoons\mathrm{A}}}, \mathcal{H}_{\mathrm{G}}\left(\vec{b}_{\zeta}^{\vec{J}^{\mathrm{G}\rightleftharpoons\mathrm{F}}}\right)\right\rangle_{\tilde{\mathbb{S}}^{\mathrm{G}\rightleftharpoons\mathrm{A}}} \tag{8-28f}$$

$$z_{\xi\zeta}^{\vec{M}^{\mathrm{G}\rightleftharpoons\mathrm{A}}\vec{M}^{\mathrm{O}\rightleftharpoons\mathrm{G}}} = \left\langle \vec{b}_{\xi}^{\vec{M}^{\mathrm{G}\rightleftharpoons\mathrm{A}}}, \mathcal{H}_{\mathrm{G}}\left(\vec{b}_{\zeta}^{\vec{M}^{\mathrm{O}\rightleftharpoons\mathrm{G}}}\right)\right\rangle_{\tilde{\mathbb{S}}^{\mathrm{G}\rightleftharpoons\mathrm{A}}} \tag{8-28g}$$

$$z_{\xi\zeta}^{\vec{M}^{\mathrm{G}\rightleftharpoons\mathrm{A}}\vec{M}^{\mathrm{G}\rightleftharpoons\mathrm{A}}} = \left\langle \vec{b}_{\xi}^{\vec{M}^{\mathrm{G}\rightleftharpoons\mathrm{A}}}, \mathcal{H}_{\mathrm{G}}\left(-\vec{b}_{\zeta}^{\vec{M}^{\mathrm{G}\rightleftharpoons\mathrm{A}}}\right)\right\rangle_{\tilde{\mathbb{S}}^{\mathrm{G}\rightleftharpoons\mathrm{A}}} + \left\langle \vec{b}_{\xi}^{\vec{M}^{\mathrm{G}\rightleftharpoons\mathrm{A}}}, -\mathcal{H}_{\mathrm{A}}\left(\vec{b}_{\zeta}^{\vec{M}^{\mathrm{G}\rightleftharpoons\mathrm{A}}}\right)\right\rangle_{\underset{\sim}{\mathbb{S}}^{\mathrm{G}\rightleftharpoons\mathrm{A}}} \tag{8-28h}$$

$$z_{\xi\zeta}^{\vec{M}^{\mathrm{G}\rightleftharpoons\mathrm{A}}\vec{M}^{\mathrm{A}\rightleftharpoons\mathrm{F}}} = \left\langle \vec{b}_{\xi}^{\vec{M}^{\mathrm{G}\rightleftharpoons\mathrm{A}}}, -\mathcal{H}_{\mathrm{A}}\left(-\vec{b}_{\zeta}^{\vec{M}^{\mathrm{A}\rightleftharpoons\mathrm{F}}}\right)\right\rangle_{\underset{\sim}{\mathbb{S}}^{\mathrm{G}\rightleftharpoons\mathrm{A}}} \tag{8-28i}$$

$$z_{\xi\zeta}^{\vec{M}^{\mathrm{G}\rightleftharpoons\mathrm{A}}\vec{M}^{\mathrm{G}\rightleftharpoons\mathrm{F}}} = \left\langle \vec{b}_{\xi}^{\vec{M}^{\mathrm{G}\rightleftharpoons\mathrm{A}}}, \mathcal{H}_{\mathrm{G}}\left(\vec{b}_{\zeta}^{\vec{M}^{\mathrm{G}\rightleftharpoons\mathrm{F}}}\right)\right\rangle_{\tilde{\mathbb{S}}^{\mathrm{G}\rightleftharpoons\mathrm{A}}} \tag{8-28j}$$

The formulations used to calculate the elements of the matrices in Eq. (8-20) are as follows:

$$z_{\xi\zeta}^{\vec{J}^{\mathrm{A}}\vec{J}^{\mathrm{G}\rightleftharpoons\mathrm{A}}} = \left\langle \vec{b}_{\xi}^{\vec{J}^{\mathrm{A}}}, \mathcal{E}_{\mathrm{A}}\left(\vec{b}_{\zeta}^{\vec{J}^{\mathrm{G}\rightleftharpoons\mathrm{A}}}\right)\right\rangle_{\tilde{\mathbb{S}}^{\mathrm{A}}} \tag{8-29a}$$

$$z_{\xi\zeta}^{\vec{J}^{\mathrm{A}}\vec{J}^{\mathrm{A}}} = \left\langle \vec{b}_{\xi}^{\vec{J}^{\mathrm{A}}}, \mathcal{E}_{\mathrm{A}}\left(\vec{b}_{\zeta}^{\vec{J}^{\mathrm{A}}}\right)\right\rangle_{\tilde{\mathbb{S}}^{\mathrm{A}}} \tag{8-29b}$$

$$z_{\xi\zeta}^{\vec{J}^{\mathrm{A}}\vec{J}^{\mathrm{A}\rightleftharpoons\mathrm{F}}} = \left\langle \vec{b}_{\xi}^{\vec{J}^{\mathrm{A}}}, \mathcal{E}_{\mathrm{A}}\left(-\vec{b}_{\zeta}^{\vec{J}^{\mathrm{A}\rightleftharpoons\mathrm{F}}}\right)\right\rangle_{\tilde{\mathbb{S}}^{\mathrm{A}}} \tag{8-29c}$$

$$z_{\xi\zeta}^{\vec{J}^{\mathrm{A}}\vec{M}^{\mathrm{G}\rightleftharpoons\mathrm{A}}} = \left\langle \vec{b}_{\xi}^{\vec{J}^{\mathrm{A}}}, \mathcal{E}_{\mathrm{A}}\left(\vec{b}_{\zeta}^{\vec{M}^{\mathrm{G}\rightleftharpoons\mathrm{A}}}\right)\right\rangle_{\tilde{\mathbb{S}}^{\mathrm{A}}} \tag{8-29d}$$

$$z_{\xi\zeta}^{\vec{J}^{\mathrm{A}}\vec{M}^{\mathrm{A}\rightleftharpoons\mathrm{F}}} = \left\langle \vec{b}_{\xi}^{\vec{J}^{\mathrm{A}}}, \mathcal{E}_{\mathrm{A}}\left(-\vec{b}_{\zeta}^{\vec{M}^{\mathrm{A}\rightleftharpoons\mathrm{F}}}\right)\right\rangle_{\tilde{\mathbb{S}}^{\mathrm{A}}} \tag{8-29e}$$

where the integral surface $\tilde{\mathbb{S}}^{\mathrm{A}}$ is shown in Fig. 8-4. The formulations used to calculate the elements of the matrices in Eq. (8-21a) are as follows:

$$z_{\xi\zeta}^{\vec{J}^{\mathrm{A}\rightleftharpoons\mathrm{F}}\vec{J}^{\mathrm{G}\rightleftharpoons\mathrm{A}}} = \left\langle \vec{b}_{\xi}^{\vec{J}^{\mathrm{A}\rightleftharpoons\mathrm{F}}}, \mathcal{E}_{\mathrm{A}}\left(\vec{b}_{\zeta}^{\vec{J}^{\mathrm{G}\rightleftharpoons\mathrm{A}}}\right)\right\rangle_{\tilde{\mathbb{S}}^{\mathrm{A}\rightleftharpoons\mathrm{F}}} \tag{8-30a}$$

$$z_{\xi\zeta}^{\vec{J}^{\mathrm{A}\rightleftharpoons\mathrm{F}}\vec{J}^{\mathrm{A}}} = \left\langle \vec{b}_{\xi}^{\vec{J}^{\mathrm{A}\rightleftharpoons\mathrm{F}}}, \mathcal{E}_{\mathrm{A}}\left(\vec{b}_{\zeta}^{\vec{J}^{\mathrm{A}}}\right)\right\rangle_{\tilde{\mathbb{S}}^{\mathrm{A}\rightleftharpoons\mathrm{F}}} \tag{8-30b}$$

$$z_{\xi\zeta}^{\vec{J}^{\mathrm{A}\rightleftharpoons\mathrm{F}}\vec{J}^{\mathrm{A}\rightleftharpoons\mathrm{F}}} = \left\langle \vec{b}_{\xi}^{\vec{J}^{\mathrm{A}\rightleftharpoons\mathrm{F}}}, \mathcal{E}_{\mathrm{A}}\left(-\vec{b}_{\zeta}^{\vec{J}^{\mathrm{A}\rightleftharpoons\mathrm{F}}}\right)\right\rangle_{\tilde{\mathbb{S}}^{\mathrm{A}\rightleftharpoons\mathrm{F}}} + \left\langle \vec{b}_{\xi}^{\vec{J}^{\mathrm{A}\rightleftharpoons\mathrm{F}}}, -\mathcal{E}_{0}\left(\vec{b}_{\zeta}^{\vec{J}^{\mathrm{A}\rightleftharpoons\mathrm{F}}}\right)\right\rangle_{\underset{\sim}{\mathbb{S}}^{\mathrm{A}\rightleftharpoons\mathrm{F}}} \tag{8-30c}$$

$$z_{\xi\zeta}^{\vec{J}^{\mathrm{A}\rightleftharpoons\mathrm{F}}\vec{J}^{\mathrm{F}}} = \left\langle \vec{b}_{\xi}^{\vec{J}^{\mathrm{A}\rightleftharpoons\mathrm{F}}}, -\mathcal{E}_{0}\left(\vec{b}_{\zeta}^{\vec{J}^{\mathrm{F}}}\right)\right\rangle_{\underset{\sim}{\mathbb{S}}^{\mathrm{A}\rightleftharpoons\mathrm{F}}} \tag{8-30d}$$

$$z_{\xi\zeta}^{\vec{J}^{\mathrm{A}\rightleftharpoons\mathrm{F}}\vec{J}^{\mathrm{G}\rightleftharpoons\mathrm{F}}} = \left\langle \vec{b}_{\xi}^{\vec{J}^{\mathrm{A}\rightleftharpoons\mathrm{F}}}, -\mathcal{E}_{0}\left(-\vec{b}_{\zeta}^{\vec{J}^{\mathrm{G}\rightleftharpoons\mathrm{F}}}\right)\right\rangle_{\underset{\sim}{\mathbb{S}}^{\mathrm{A}\rightleftharpoons\mathrm{F}}} \tag{8-30e}$$

$$z_{\xi\zeta}^{\vec{J}^{\mathrm{A}\rightleftharpoons\mathrm{F}}\vec{M}^{\mathrm{G}\rightleftharpoons\mathrm{A}}} = \left\langle \vec{b}_{\xi}^{\vec{J}^{\mathrm{A}\rightleftharpoons\mathrm{F}}}, \mathcal{E}_{\mathrm{A}}\left(\vec{b}_{\zeta}^{\vec{M}^{\mathrm{G}\rightleftharpoons\mathrm{A}}}\right)\right\rangle_{\tilde{\mathbb{S}}^{\mathrm{A}\rightleftharpoons\mathrm{F}}} \tag{8-30f}$$

$$z_{\xi\zeta}^{\vec{J}^{\mathrm{A}\rightleftharpoons\mathrm{F}}\vec{M}^{\mathrm{A}\rightleftharpoons\mathrm{F}}} = \left\langle \vec{b}_{\xi}^{\vec{J}^{\mathrm{A}\rightleftharpoons\mathrm{F}}}, \mathcal{E}_{\mathrm{A}}\left(-\vec{b}_{\zeta}^{\vec{M}^{\mathrm{A}\rightleftharpoons\mathrm{F}}}\right)\right\rangle_{\tilde{\mathbb{S}}^{\mathrm{A}\rightleftharpoons\mathrm{F}}} + \left\langle \vec{b}_{\xi}^{\vec{J}^{\mathrm{A}\rightleftharpoons\mathrm{F}}}, -\mathcal{E}_{0}\left(\vec{b}_{\zeta}^{\vec{M}^{\mathrm{A}\rightleftharpoons\mathrm{F}}}\right)\right\rangle_{\underset{\sim}{\mathbb{S}}^{\mathrm{A}\rightleftharpoons\mathrm{F}}} \tag{8-30g}$$

$$z_{\xi\zeta}^{\vec{J}^{\mathrm{A}\rightleftharpoons\mathrm{F}}\vec{M}^{\mathrm{G}\rightleftharpoons\mathrm{F}}} = \left\langle \vec{b}_{\xi}^{\vec{J}^{\mathrm{A}\rightleftharpoons\mathrm{F}}}, -\mathcal{E}_{0}\left(-\vec{b}_{\zeta}^{\vec{M}^{\mathrm{G}\rightleftharpoons\mathrm{F}}}\right)\right\rangle_{\underset{\sim}{\mathbb{S}}^{\mathrm{A}\rightleftharpoons\mathrm{F}}} \tag{8-30h}$$

where the integral surfaces $\tilde{\mathbb{S}}^{\mathrm{A}\rightleftharpoons\mathrm{F}}$ and $\underset{\sim}{\mathbb{S}}^{\mathrm{A}\rightleftharpoons\mathrm{F}}$ are shown in Fig. 8-4. The formulations used to calculate the elements of the matrices in Eq. (8-21b) are as follows:

$$z_{\xi\zeta}^{\vec{M}^{\mathrm{A}\rightleftharpoons\mathrm{F}}\vec{J}^{\mathrm{G}\rightleftharpoons\mathrm{A}}} = \left\langle \vec{b}_{\xi}^{\vec{M}^{\mathrm{A}\rightleftharpoons\mathrm{F}}}, \mathcal{H}_{\mathrm{A}}\left(\vec{b}_{\zeta}^{\vec{J}^{\mathrm{G}\rightleftharpoons\mathrm{A}}}\right)\right\rangle_{\tilde{\mathbb{S}}^{\mathrm{A}\rightleftharpoons\mathrm{F}}} \tag{8-31a}$$

$$z_{\xi\zeta}^{\vec{M}^{\mathrm{A}\rightleftharpoons\mathrm{F}}\vec{J}^{\mathrm{A}}} = \left\langle \vec{b}_{\xi}^{\vec{M}^{\mathrm{A}\rightleftharpoons\mathrm{F}}}, \mathcal{H}_{\mathrm{A}}\left(\vec{b}_{\zeta}^{\vec{J}^{\mathrm{A}}}\right)\right\rangle_{\tilde{\mathbb{S}}^{\mathrm{A}\rightleftharpoons\mathrm{F}}} \tag{8-31b}$$

$$z_{\xi\zeta}^{\vec{M}^{\mathrm{A}\rightleftharpoons\mathrm{F}}\vec{J}^{\mathrm{A}\rightleftharpoons\mathrm{F}}} = \left\langle \vec{b}_{\xi}^{\vec{M}^{\mathrm{A}\rightleftharpoons\mathrm{F}}}, \mathcal{H}_{\mathrm{A}}\left(-\vec{b}_{\zeta}^{\vec{J}^{\mathrm{A}\rightleftharpoons\mathrm{F}}}\right)\right\rangle_{\tilde{\mathbb{S}}^{\mathrm{A}\rightleftharpoons\mathrm{F}}} + \left\langle \vec{b}_{\xi}^{\vec{J}^{\mathrm{A}\rightleftharpoons\mathrm{F}}}, -\mathcal{H}_{0}\left(\vec{b}_{\zeta}^{\vec{J}^{\mathrm{A}\rightleftharpoons\mathrm{F}}}\right)\right\rangle_{\utilde{\mathbb{S}}^{\mathrm{A}\rightleftharpoons\mathrm{F}}} \tag{8-31c}$$

$$z_{\xi\zeta}^{\vec{M}^{\mathrm{A}\rightleftharpoons\mathrm{F}}\vec{J}^{\mathrm{F}}} = \left\langle \vec{b}_{\xi}^{\vec{M}^{\mathrm{A}\rightleftharpoons\mathrm{F}}}, -\mathcal{H}_{0}\left(\vec{b}_{\zeta}^{\vec{J}^{\mathrm{F}}}\right)\right\rangle_{\utilde{\mathbb{S}}^{\mathrm{A}\rightleftharpoons\mathrm{F}}} \tag{8-31d}$$

$$z_{\xi\zeta}^{\vec{M}^{\mathrm{A}\rightleftharpoons\mathrm{F}}\vec{J}^{\mathrm{G}\rightleftharpoons\mathrm{F}}} = \left\langle \vec{b}_{\xi}^{\vec{M}^{\mathrm{A}\rightleftharpoons\mathrm{F}}}, -\mathcal{H}_{0}\left(-\vec{b}_{\zeta}^{\vec{J}^{\mathrm{G}\rightleftharpoons\mathrm{F}}}\right)\right\rangle_{\utilde{\mathbb{S}}^{\mathrm{A}\rightleftharpoons\mathrm{F}}} \tag{8-31e}$$

$$z_{\xi\zeta}^{\vec{M}^{\mathrm{A}\rightleftharpoons\mathrm{F}}\vec{M}^{\mathrm{G}\rightleftharpoons\mathrm{A}}} = \left\langle \vec{b}_{\xi}^{\vec{M}^{\mathrm{A}\rightleftharpoons\mathrm{F}}}, \mathcal{H}_{\mathrm{A}}\left(\vec{b}_{\zeta}^{\vec{M}^{\mathrm{G}\rightleftharpoons\mathrm{A}}}\right)\right\rangle_{\tilde{\mathbb{S}}^{\mathrm{A}\rightleftharpoons\mathrm{F}}} \tag{8-31f}$$

$$z_{\xi\zeta}^{\vec{M}^{\mathrm{A}\rightleftharpoons\mathrm{F}}\vec{M}^{\mathrm{A}\rightleftharpoons\mathrm{F}}} = \left\langle \vec{b}_{\xi}^{\vec{M}^{\mathrm{A}\rightleftharpoons\mathrm{F}}}, \mathcal{H}_{\mathrm{A}}\left(-\vec{b}_{\zeta}^{\vec{M}^{\mathrm{A}\rightleftharpoons\mathrm{F}}}\right)\right\rangle_{\tilde{\mathbb{S}}^{\mathrm{A}\rightleftharpoons\mathrm{F}}} + \left\langle \vec{b}_{\xi}^{\vec{J}^{\mathrm{A}\rightleftharpoons\mathrm{F}}}, -\mathcal{H}_{0}\left(\vec{b}_{\zeta}^{\vec{M}^{\mathrm{A}\rightleftharpoons\mathrm{F}}}\right)\right\rangle_{\utilde{\mathbb{S}}^{\mathrm{A}\rightleftharpoons\mathrm{F}}} \tag{8-31g}$$

$$z_{\xi\zeta}^{\vec{M}^{\mathrm{A}\rightleftharpoons\mathrm{F}}\vec{M}^{\mathrm{G}\rightleftharpoons\mathrm{F}}} = \left\langle \vec{b}_{\xi}^{\vec{M}^{\mathrm{A}\rightleftharpoons\mathrm{F}}}, -\mathcal{H}_{0}\left(-\vec{b}_{\zeta}^{\vec{M}^{\mathrm{G}\rightleftharpoons\mathrm{F}}}\right)\right\rangle_{\utilde{\mathbb{S}}^{\mathrm{A}\rightleftharpoons\mathrm{F}}} \tag{8-31h}$$

The formulations used to calculate the elements of the matrices in Eq. (8-22) are as follows:

$$z_{\xi\zeta}^{\vec{J}^{\mathrm{F}}\vec{J}^{\mathrm{A}\rightleftharpoons\mathrm{F}}} = \left\langle \vec{b}_{\xi}^{\vec{J}^{\mathrm{F}}}, \mathcal{E}_{0}\left(\vec{b}_{\zeta}^{\vec{J}^{\mathrm{A}\rightleftharpoons\mathrm{F}}}\right)\right\rangle_{\tilde{\mathbb{S}}^{\mathrm{F}}} \tag{8-32a}$$

$$z_{\xi\zeta}^{\vec{J}^{\mathrm{F}}\vec{J}^{\mathrm{F}}} = \left\langle \vec{b}_{\xi}^{\vec{J}^{\mathrm{F}}}, \mathcal{E}_{0}\left(\vec{b}_{\zeta}^{\vec{J}^{\mathrm{F}}}\right)\right\rangle_{\tilde{\mathbb{S}}^{\mathrm{F}}} \tag{8-32b}$$

$$z_{\xi\zeta}^{\vec{J}^{\mathrm{F}}\vec{J}^{\mathrm{G}\rightleftharpoons\mathrm{F}}} = \left\langle \vec{b}_{\xi}^{\vec{J}^{\mathrm{F}}}, \mathcal{E}_{0}\left(-\vec{b}_{\zeta}^{\vec{J}^{\mathrm{G}\rightleftharpoons\mathrm{F}}}\right)\right\rangle_{\tilde{\mathbb{S}}^{\mathrm{F}}} \tag{8-32c}$$

$$z_{\xi\zeta}^{\vec{J}^{\mathrm{F}}\vec{M}^{\mathrm{A}\rightleftharpoons\mathrm{F}}} = \left\langle \vec{b}_{\xi}^{\vec{J}^{\mathrm{F}}}, \mathcal{E}_{0}\left(\vec{b}_{\zeta}^{\vec{M}^{\mathrm{A}\rightleftharpoons\mathrm{F}}}\right)\right\rangle_{\tilde{\mathbb{S}}^{\mathrm{F}}} \tag{8-32d}$$

$$z_{\xi\zeta}^{\vec{J}^{\mathrm{F}}\vec{M}^{\mathrm{G}\rightleftharpoons\mathrm{F}}} = \left\langle \vec{b}_{\xi}^{\vec{J}^{\mathrm{F}}}, \mathcal{E}_{0}\left(-\vec{b}_{\zeta}^{\vec{M}^{\mathrm{G}\rightleftharpoons\mathrm{F}}}\right)\right\rangle_{\tilde{\mathbb{S}}^{\mathrm{F}}} \tag{8-32e}$$

where the integral surface $\tilde{\mathbb{S}}^{\mathrm{F}}$ is shown in Fig. 8-4. The formulations used to calculate the elements of the matrices in Eq. (8-23a) are as follows:

$$z_{\xi\zeta}^{\vec{J}^{\mathrm{G}\rightleftharpoons\mathrm{F}}\vec{J}^{\mathrm{O}\rightleftharpoons\mathrm{G}}} = \left\langle \vec{b}_{\xi}^{\vec{J}^{\mathrm{G}\rightleftharpoons\mathrm{F}}}, \mathcal{E}_{\mathrm{G}}\left(\vec{b}_{\zeta}^{\vec{J}^{\mathrm{O}\rightleftharpoons\mathrm{G}}}\right)\right\rangle_{\tilde{\mathbb{S}}^{\mathrm{G}\rightleftharpoons\mathrm{F}}} \tag{8-33a}$$

$$z_{\xi\zeta}^{\vec{J}^{\mathrm{G}\rightleftharpoons\mathrm{F}}\vec{J}^{\mathrm{G}}} = \left\langle \vec{b}_{\xi}^{\vec{J}^{\mathrm{G}\rightleftharpoons\mathrm{F}}}, \mathcal{E}_{\mathrm{G}}\left(\vec{b}_{\zeta}^{\vec{J}^{\mathrm{G}}}\right)\right\rangle_{\tilde{\mathbb{S}}^{\mathrm{G}\rightleftharpoons\mathrm{F}}} \tag{8-33b}$$

$$z_{\xi\zeta}^{\vec{J}^{\mathrm{G}\rightleftharpoons\mathrm{F}}\vec{J}^{\mathrm{G}\rightleftharpoons\mathrm{A}}} = \left\langle \vec{b}_{\xi}^{\vec{J}^{\mathrm{G}\rightleftharpoons\mathrm{F}}}, \mathcal{E}_{\mathrm{G}}\left(-\vec{b}_{\zeta}^{\vec{J}^{\mathrm{G}\rightleftharpoons\mathrm{A}}}\right)\right\rangle_{\tilde{\mathbb{S}}^{\mathrm{G}\rightleftharpoons\mathrm{F}}} \tag{8-33c}$$

$$z_{\xi\zeta}^{\vec{J}^{\mathrm{G}\rightleftharpoons\mathrm{F}}\vec{J}^{\mathrm{A}\rightleftharpoons\mathrm{F}}} = \left\langle \vec{b}_{\xi}^{\vec{J}^{\mathrm{G}\rightleftharpoons\mathrm{F}}}, -\mathcal{E}_{0}\left(\vec{b}_{\zeta}^{\vec{J}^{\mathrm{A}\rightleftharpoons\mathrm{F}}}\right)\right\rangle_{\utilde{\mathbb{S}}^{\mathrm{G}\rightleftharpoons\mathrm{F}}} \tag{8-33d}$$

$$z_{\xi\zeta}^{\vec{J}^{\mathrm{G}\rightleftharpoons\mathrm{F}}\vec{J}^{\mathrm{F}}} = \left\langle \vec{b}_{\xi}^{\vec{J}^{\mathrm{G}\rightleftharpoons\mathrm{F}}}, -\mathcal{E}_{0}\left(\vec{b}_{\zeta}^{\vec{J}^{\mathrm{F}}}\right)\right\rangle_{\utilde{\mathbb{S}}^{\mathrm{G}\rightleftharpoons\mathrm{F}}} \tag{8-33e}$$

$$z_{\xi\zeta}^{\vec{J}^{\mathrm{G}\rightleftharpoons\mathrm{F}}\vec{J}^{\mathrm{G}\rightleftharpoons\mathrm{F}}} = \left\langle \vec{b}_{\xi}^{\vec{J}^{\mathrm{G}\rightleftharpoons\mathrm{F}}}, \mathcal{E}_{\mathrm{G}}\left(\vec{b}_{\zeta}^{\vec{J}^{\mathrm{G}\rightleftharpoons\mathrm{F}}}\right)\right\rangle_{\tilde{\mathbb{S}}^{\mathrm{G}\rightleftharpoons\mathrm{F}}} + \left\langle \vec{b}_{\xi}^{\vec{J}^{\mathrm{G}\rightleftharpoons\mathrm{F}}}, -\mathcal{E}_{0}\left(-\vec{b}_{\zeta}^{\vec{J}^{\mathrm{G}\rightleftharpoons\mathrm{F}}}\right)\right\rangle_{\utilde{\mathbb{S}}^{\mathrm{G}\rightleftharpoons\mathrm{F}}} \tag{8-33f}$$

$$z_{\xi\zeta}^{\vec{J}^{\mathrm{G}\rightleftharpoons\mathrm{F}}\vec{M}^{\mathrm{O}\rightleftharpoons\mathrm{G}}} = \left\langle \vec{b}_{\xi}^{\vec{J}^{\mathrm{G}\rightleftharpoons\mathrm{F}}}, \mathcal{E}_{\mathrm{G}}\left(\vec{b}_{\zeta}^{\vec{M}^{\mathrm{O}\rightleftharpoons\mathrm{G}}}\right)\right\rangle_{\tilde{\mathbb{S}}^{\mathrm{G}\rightleftharpoons\mathrm{F}}} \tag{8-33g}$$

$$z_{\xi\zeta}^{\vec{J}^{\mathrm{G}\rightleftharpoons\mathrm{F}}\vec{M}^{\mathrm{G}\rightleftharpoons\mathrm{A}}} = \left\langle \vec{b}_{\xi}^{\vec{J}^{\mathrm{G}\rightleftharpoons\mathrm{F}}}, \mathcal{E}_{\mathrm{G}}\left(-\vec{b}_{\zeta}^{\vec{M}^{\mathrm{G}\rightleftharpoons\mathrm{A}}}\right)\right\rangle_{\tilde{\mathbb{S}}^{\mathrm{G}\rightleftharpoons\mathrm{F}}} \tag{8-33h}$$

$$z_{\xi\zeta}^{\vec{J}^{\mathrm{G}\rightleftharpoons\mathrm{F}}\vec{M}^{\mathrm{A}\rightleftharpoons\mathrm{F}}} = \left\langle \vec{b}_{\xi}^{\vec{J}^{\mathrm{G}\rightleftharpoons\mathrm{F}}}, -\mathcal{E}_0\left(\vec{b}_{\zeta}^{\vec{M}^{\mathrm{A}\rightleftharpoons\mathrm{F}}}\right)\right\rangle_{\underset{\sim}{\mathbb{S}}^{\mathrm{G}\rightleftharpoons\mathrm{F}}} \tag{8-33i}$$

$$z_{\xi\zeta}^{\vec{J}^{\mathrm{G}\rightleftharpoons\mathrm{F}}\vec{M}^{\mathrm{G}\rightleftharpoons\mathrm{F}}} = \left\langle \vec{b}_{\xi}^{\vec{J}^{\mathrm{G}\rightleftharpoons\mathrm{F}}}, \mathcal{E}_{\mathrm{G}}\left(\vec{b}_{\zeta}^{\vec{M}^{\mathrm{G}\rightleftharpoons\mathrm{F}}}\right)\right\rangle_{\tilde{\mathbb{S}}^{\mathrm{G}\rightleftharpoons\mathrm{F}}} + \left\langle \vec{b}_{\xi}^{\vec{J}^{\mathrm{G}\rightleftharpoons\mathrm{F}}}, -\mathcal{E}_0\left(-\vec{b}_{\zeta}^{\vec{M}^{\mathrm{G}\rightleftharpoons\mathrm{F}}}\right)\right\rangle_{\underset{\sim}{\mathbb{S}}^{\mathrm{G}\rightleftharpoons\mathrm{F}}} \tag{8-33j}$$

where the integral surfaces $\tilde{\mathbb{S}}^{\mathrm{G}\rightleftharpoons\mathrm{F}}$ and $\underset{\sim}{\mathbb{S}}^{\mathrm{G}\rightleftharpoons\mathrm{F}}$ are as the ones shown in previous Fig. 8-4. The formulations used to calculate the elements of the matrices in Eq. (8-23b) are as follows:

$$z_{\xi\zeta}^{\vec{M}^{\mathrm{G}\rightleftharpoons\mathrm{F}}\vec{J}^{\mathrm{O}\rightleftharpoons\mathrm{G}}} = \left\langle \vec{b}_{\xi}^{\vec{M}^{\mathrm{G}\rightleftharpoons\mathrm{F}}}, \mathcal{H}_{\mathrm{G}}\left(\vec{b}_{\zeta}^{\vec{J}^{\mathrm{O}\rightleftharpoons\mathrm{G}}}\right)\right\rangle_{\tilde{\mathbb{S}}^{\mathrm{G}\rightleftharpoons\mathrm{F}}} \tag{8-34a}$$

$$z_{\xi\zeta}^{\vec{M}^{\mathrm{G}\rightleftharpoons\mathrm{F}}\vec{J}^{\mathrm{G}}} = \left\langle \vec{b}_{\xi}^{\vec{M}^{\mathrm{G}\rightleftharpoons\mathrm{F}}}, \mathcal{H}_{\mathrm{G}}\left(\vec{b}_{\zeta}^{\vec{J}^{\mathrm{G}}}\right)\right\rangle_{\tilde{\mathbb{S}}^{\mathrm{G}\rightleftharpoons\mathrm{F}}} \tag{8-34b}$$

$$z_{\xi\zeta}^{\vec{M}^{\mathrm{G}\rightleftharpoons\mathrm{F}}\vec{J}^{\mathrm{G}\rightleftharpoons\mathrm{A}}} = \left\langle \vec{b}_{\xi}^{\vec{M}^{\mathrm{G}\rightleftharpoons\mathrm{F}}}, \mathcal{H}_{\mathrm{G}}\left(-\vec{b}_{\zeta}^{\vec{J}^{\mathrm{G}\rightleftharpoons\mathrm{A}}}\right)\right\rangle_{\tilde{\mathbb{S}}^{\mathrm{G}\rightleftharpoons\mathrm{F}}} \tag{8-34c}$$

$$z_{\xi\zeta}^{\vec{M}^{\mathrm{G}\rightleftharpoons\mathrm{F}}\vec{J}^{\mathrm{A}\rightleftharpoons\mathrm{F}}} = \left\langle \vec{b}_{\xi}^{\vec{M}^{\mathrm{G}\rightleftharpoons\mathrm{F}}}, -\mathcal{H}_0\left(\vec{b}_{\zeta}^{\vec{J}^{\mathrm{A}\rightleftharpoons\mathrm{F}}}\right)\right\rangle_{\underset{\sim}{\mathbb{S}}^{\mathrm{G}\rightleftharpoons\mathrm{F}}} \tag{8-34d}$$

$$z_{\xi\zeta}^{\vec{M}^{\mathrm{G}\rightleftharpoons\mathrm{F}}\vec{J}^{\mathrm{F}}} = \left\langle \vec{b}_{\xi}^{\vec{M}^{\mathrm{G}\rightleftharpoons\mathrm{F}}}, -\mathcal{H}_0\left(\vec{b}_{\zeta}^{\vec{J}^{\mathrm{F}}}\right)\right\rangle_{\underset{\sim}{\mathbb{S}}^{\mathrm{G}\rightleftharpoons\mathrm{F}}} \tag{8-34e}$$

$$z_{\xi\zeta}^{\vec{M}^{\mathrm{G}\rightleftharpoons\mathrm{F}}\vec{J}^{\mathrm{G}\rightleftharpoons\mathrm{F}}} = \left\langle \vec{b}_{\xi}^{\vec{M}^{\mathrm{G}\rightleftharpoons\mathrm{F}}}, \mathcal{H}_{\mathrm{G}}\left(\vec{b}_{\zeta}^{\vec{J}^{\mathrm{G}\rightleftharpoons\mathrm{F}}}\right)\right\rangle_{\tilde{\mathbb{S}}^{\mathrm{G}\rightleftharpoons\mathrm{F}}} + \left\langle \vec{b}_{\xi}^{\vec{M}^{\mathrm{G}\rightleftharpoons\mathrm{F}}}, -\mathcal{H}_0\left(-\vec{b}_{\zeta}^{\vec{J}^{\mathrm{G}\rightleftharpoons\mathrm{F}}}\right)\right\rangle_{\underset{\sim}{\mathbb{S}}^{\mathrm{G}\rightleftharpoons\mathrm{F}}} \tag{8-34f}$$

$$z_{\xi\zeta}^{\vec{M}^{\mathrm{G}\rightleftharpoons\mathrm{F}}\vec{M}^{\mathrm{O}\rightleftharpoons\mathrm{G}}} = \left\langle \vec{b}_{\xi}^{\vec{M}^{\mathrm{G}\rightleftharpoons\mathrm{F}}}, \mathcal{H}_{\mathrm{G}}\left(\vec{b}_{\zeta}^{\vec{M}^{\mathrm{O}\rightleftharpoons\mathrm{G}}}\right)\right\rangle_{\tilde{\mathbb{S}}^{\mathrm{G}\rightleftharpoons\mathrm{F}}} \tag{8-34g}$$

$$z_{\xi\zeta}^{\vec{M}^{\mathrm{G}\rightleftharpoons\mathrm{F}}\vec{M}^{\mathrm{G}\rightleftharpoons\mathrm{A}}} = \left\langle \vec{b}_{\xi}^{\vec{M}^{\mathrm{G}\rightleftharpoons\mathrm{F}}}, \mathcal{H}_{\mathrm{G}}\left(-\vec{b}_{\zeta}^{\vec{M}^{\mathrm{G}\rightleftharpoons\mathrm{A}}}\right)\right\rangle_{\tilde{\mathbb{S}}^{\mathrm{G}\rightleftharpoons\mathrm{F}}} \tag{8-34h}$$

$$z_{\xi\zeta}^{\vec{M}^{\mathrm{G}\rightleftharpoons\mathrm{F}}\vec{M}^{\mathrm{A}\rightleftharpoons\mathrm{F}}} = \left\langle \vec{b}_{\xi}^{\vec{M}^{\mathrm{G}\rightleftharpoons\mathrm{F}}}, -\mathcal{H}_0\left(\vec{b}_{\zeta}^{\vec{M}^{\mathrm{A}\rightleftharpoons\mathrm{F}}}\right)\right\rangle_{\underset{\sim}{\mathbb{S}}^{\mathrm{G}\rightleftharpoons\mathrm{F}}} \tag{8-34i}$$

$$z_{\xi\zeta}^{\vec{M}^{\mathrm{G}\rightleftharpoons\mathrm{F}}\vec{M}^{\mathrm{G}\rightleftharpoons\mathrm{F}}} = \left\langle \vec{b}_{\xi}^{\vec{M}^{\mathrm{G}\rightleftharpoons\mathrm{F}}}, \mathcal{H}_{\mathrm{G}}\left(\vec{b}_{\zeta}^{\vec{M}^{\mathrm{G}\rightleftharpoons\mathrm{F}}}\right)\right\rangle_{\tilde{\mathbb{S}}^{\mathrm{G}\rightleftharpoons\mathrm{F}}} + \left\langle \vec{b}_{\xi}^{\vec{M}^{\mathrm{G}\rightleftharpoons\mathrm{F}}}, -\mathcal{H}_0\left(-\vec{b}_{\zeta}^{\vec{M}^{\mathrm{G}\rightleftharpoons\mathrm{F}}}\right)\right\rangle_{\underset{\sim}{\mathbb{S}}^{\mathrm{G}\rightleftharpoons\mathrm{F}}} \tag{8-34j}$$

Below, we will propose two somewhat different schemes — *dependent variable elimination* (*DVE*) and *solution*/*definition domain compression* (*SDC*/*DDC*) — for mathematically describing the *modal space* of the TGTA system shown in previous Fig. 8-1.

### 8.2.3.3 Scheme I: Dependent Variable Elimination (DVE)

By properly assembling the matrix equation (8-17a) and matrix equations (8-18)~(8-23b), we have the following *augmented matrix equation*

$$\overline{\overline{\Psi}}_1 \cdot \bar{a}^{\mathrm{AV}} = \overline{\overline{\Psi}}_2 \cdot \bar{a}^{\vec{J}^{\mathrm{G}\rightleftharpoons\mathrm{A}}} \tag{8-35}$$

in which the augmented matrices $\overline{\overline{\Psi}}_1 \,\&\, \overline{\overline{\Psi}}_2$ and augmented vector $\bar{a}^{\mathrm{AV}}$ are given as follows:

$$
\overline{\overline{\Psi}}_1 = \begin{bmatrix}
\overline{\overline{I}}^{\vec{J}^{\mathrm{O}\rightleftharpoons\mathrm{G}}} & 0 & 0 & 0 & 0 & 0 & 0 & 0 & 0 & 0 & 0 \\
0 & \overline{\overline{Z}}^{\vec{M}^{\mathrm{O}\rightleftharpoons\mathrm{G}}\vec{J}^{\mathrm{G}}} & \overline{\overline{Z}}^{\vec{M}^{\mathrm{O}\rightleftharpoons\mathrm{G}}\vec{J}^{\mathrm{G}\rightleftharpoons\mathrm{A}}} & 0 & 0 & 0 & \overline{\overline{Z}}^{\vec{M}^{\mathrm{O}\rightleftharpoons\mathrm{G}}\vec{J}^{\mathrm{G}\rightleftharpoons\mathrm{F}}} & \overline{\overline{Z}}^{\vec{M}^{\mathrm{O}\rightleftharpoons\mathrm{G}}\vec{M}^{\mathrm{O}\rightleftharpoons\mathrm{G}}} & \overline{\overline{Z}}^{\vec{M}^{\mathrm{O}\rightleftharpoons\mathrm{G}}\vec{M}^{\mathrm{G}\rightleftharpoons\mathrm{A}}} & 0 & \overline{\overline{Z}}^{\vec{M}^{\mathrm{O}\rightleftharpoons\mathrm{G}}\vec{M}^{\mathrm{G}\rightleftharpoons\mathrm{F}}} \\
0 & \overline{\overline{Z}}^{\vec{J}^{\mathrm{G}}\vec{J}^{\mathrm{G}}} & \overline{\overline{Z}}^{\vec{J}^{\mathrm{G}}\vec{J}^{\mathrm{G}\rightleftharpoons\mathrm{A}}} & 0 & 0 & 0 & \overline{\overline{Z}}^{\vec{J}^{\mathrm{G}}\vec{J}^{\mathrm{G}\rightleftharpoons\mathrm{F}}} & \overline{\overline{Z}}^{\vec{J}^{\mathrm{G}}\vec{M}^{\mathrm{O}\rightleftharpoons\mathrm{G}}} & \overline{\overline{Z}}^{\vec{J}^{\mathrm{G}}\vec{M}^{\mathrm{G}\rightleftharpoons\mathrm{A}}} & 0 & \overline{\overline{Z}}^{\vec{J}^{\mathrm{G}}\vec{M}^{\mathrm{G}\rightleftharpoons\mathrm{F}}} \\
0 & \overline{\overline{Z}}^{\vec{J}^{\mathrm{G}\rightleftharpoons\mathrm{A}}\vec{J}^{\mathrm{G}}} & \overline{\overline{Z}}^{\vec{J}^{\mathrm{G}\rightleftharpoons\mathrm{A}}\vec{J}^{\mathrm{G}\rightleftharpoons\mathrm{A}}} & \overline{\overline{Z}}^{\vec{J}^{\mathrm{G}\rightleftharpoons\mathrm{A}}\vec{J}^{\mathrm{A}}} & \overline{\overline{Z}}^{\vec{J}^{\mathrm{G}\rightleftharpoons\mathrm{A}}\vec{J}^{\mathrm{A}\rightleftharpoons\mathrm{F}}} & 0 & \overline{\overline{Z}}^{\vec{J}^{\mathrm{G}\rightleftharpoons\mathrm{A}}\vec{J}^{\mathrm{G}\rightleftharpoons\mathrm{F}}} & \overline{\overline{Z}}^{\vec{J}^{\mathrm{G}\rightleftharpoons\mathrm{A}}\vec{M}^{\mathrm{O}\rightleftharpoons\mathrm{G}}} & \overline{\overline{Z}}^{\vec{J}^{\mathrm{G}\rightleftharpoons\mathrm{A}}\vec{M}^{\mathrm{G}\rightleftharpoons\mathrm{A}}} & \overline{\overline{Z}}^{\vec{J}^{\mathrm{G}\rightleftharpoons\mathrm{A}}\vec{M}^{\mathrm{A}\rightleftharpoons\mathrm{F}}} & \overline{\overline{Z}}^{\vec{J}^{\mathrm{G}\rightleftharpoons\mathrm{A}}\vec{M}^{\mathrm{G}\rightleftharpoons\mathrm{F}}} \\
0 & \overline{\overline{Z}}^{\vec{M}^{\mathrm{G}\rightleftharpoons\mathrm{A}}\vec{J}^{\mathrm{G}}} & \overline{\overline{Z}}^{\vec{M}^{\mathrm{G}\rightleftharpoons\mathrm{A}}\vec{J}^{\mathrm{G}\rightleftharpoons\mathrm{A}}} & \overline{\overline{Z}}^{\vec{M}^{\mathrm{G}\rightleftharpoons\mathrm{A}}\vec{J}^{\mathrm{A}}} & \overline{\overline{Z}}^{\vec{M}^{\mathrm{G}\rightleftharpoons\mathrm{A}}\vec{J}^{\mathrm{A}\rightleftharpoons\mathrm{F}}} & 0 & \overline{\overline{Z}}^{\vec{M}^{\mathrm{G}\rightleftharpoons\mathrm{A}}\vec{J}^{\mathrm{G}\rightleftharpoons\mathrm{F}}} & \overline{\overline{Z}}^{\vec{M}^{\mathrm{G}\rightleftharpoons\mathrm{A}}\vec{M}^{\mathrm{O}\rightleftharpoons\mathrm{G}}} & \overline{\overline{Z}}^{\vec{M}^{\mathrm{G}\rightleftharpoons\mathrm{A}}\vec{M}^{\mathrm{G}\rightleftharpoons\mathrm{A}}} & \overline{\overline{Z}}^{\vec{M}^{\mathrm{G}\rightleftharpoons\mathrm{A}}\vec{M}^{\mathrm{A}\rightleftharpoons\mathrm{F}}} & \overline{\overline{Z}}^{\vec{M}^{\mathrm{G}\rightleftharpoons\mathrm{A}}\vec{M}^{\mathrm{G}\rightleftharpoons\mathrm{F}}} \\
0 & 0 & \overline{\overline{Z}}^{\vec{J}^{\mathrm{A}}\vec{J}^{\mathrm{G}\rightleftharpoons\mathrm{A}}} & \overline{\overline{Z}}^{\vec{J}^{\mathrm{A}}\vec{J}^{\mathrm{A}}} & \overline{\overline{Z}}^{\vec{J}^{\mathrm{A}}\vec{J}^{\mathrm{A}\rightleftharpoons\mathrm{F}}} & 0 & 0 & 0 & \overline{\overline{Z}}^{\vec{J}^{\mathrm{A}}\vec{M}^{\mathrm{G}\rightleftharpoons\mathrm{A}}} & \overline{\overline{Z}}^{\vec{J}^{\mathrm{A}}\vec{M}^{\mathrm{A}\rightleftharpoons\mathrm{F}}} & 0 \\
0 & 0 & \overline{\overline{Z}}^{\vec{J}^{\mathrm{A}\rightleftharpoons\mathrm{F}}\vec{J}^{\mathrm{G}\rightleftharpoons\mathrm{A}}} & \overline{\overline{Z}}^{\vec{J}^{\mathrm{A}\rightleftharpoons\mathrm{F}}\vec{J}^{\mathrm{A}}} & \overline{\overline{Z}}^{\vec{J}^{\mathrm{A}\rightleftharpoons\mathrm{F}}\vec{J}^{\mathrm{A}\rightleftharpoons\mathrm{F}}} & \overline{\overline{Z}}^{\vec{J}^{\mathrm{A}\rightleftharpoons\mathrm{F}}\vec{J}^{\mathrm{F}}} & \overline{\overline{Z}}^{\vec{J}^{\mathrm{A}\rightleftharpoons\mathrm{F}}\vec{J}^{\mathrm{G}\rightleftharpoons\mathrm{F}}} & 0 & \overline{\overline{Z}}^{\vec{J}^{\mathrm{A}\rightleftharpoons\mathrm{F}}\vec{M}^{\mathrm{G}\rightleftharpoons\mathrm{A}}} & \overline{\overline{Z}}^{\vec{J}^{\mathrm{A}\rightleftharpoons\mathrm{F}}\vec{M}^{\mathrm{A}\rightleftharpoons\mathrm{F}}} & \overline{\overline{Z}}^{\vec{J}^{\mathrm{A}\rightleftharpoons\mathrm{F}}\vec{M}^{\mathrm{G}\rightleftharpoons\mathrm{F}}} \\
0 & 0 & \overline{\overline{Z}}^{\vec{M}^{\mathrm{A}\rightleftharpoons\mathrm{F}}\vec{J}^{\mathrm{G}\rightleftharpoons\mathrm{A}}} & \overline{\overline{Z}}^{\vec{M}^{\mathrm{A}\rightleftharpoons\mathrm{F}}\vec{J}^{\mathrm{A}}} & \overline{\overline{Z}}^{\vec{M}^{\mathrm{A}\rightleftharpoons\mathrm{F}}\vec{J}^{\mathrm{A}\rightleftharpoons\mathrm{F}}} & \overline{\overline{Z}}^{\vec{M}^{\mathrm{A}\rightleftharpoons\mathrm{F}}\vec{J}^{\mathrm{F}}} & \overline{\overline{Z}}^{\vec{M}^{\mathrm{A}\rightleftharpoons\mathrm{F}}\vec{J}^{\mathrm{G}\rightleftharpoons\mathrm{F}}} & 0 & \overline{\overline{Z}}^{\vec{M}^{\mathrm{A}\rightleftharpoons\mathrm{F}}\vec{M}^{\mathrm{G}\rightleftharpoons\mathrm{A}}} & \overline{\overline{Z}}^{\vec{M}^{\mathrm{A}\rightleftharpoons\mathrm{F}}\vec{M}^{\mathrm{A}\rightleftharpoons\mathrm{F}}} & \overline{\overline{Z}}^{\vec{M}^{\mathrm{A}\rightleftharpoons\mathrm{F}}\vec{M}^{\mathrm{G}\rightleftharpoons\mathrm{F}}} \\
0 & 0 & 0 & 0 & \overline{\overline{Z}}^{\vec{J}^{\mathrm{F}}\vec{J}^{\mathrm{A}\rightleftharpoons\mathrm{F}}} & \overline{\overline{Z}}^{\vec{J}^{\mathrm{F}}\vec{J}^{\mathrm{F}}} & \overline{\overline{Z}}^{\vec{J}^{\mathrm{F}}\vec{J}^{\mathrm{G}\rightleftharpoons\mathrm{F}}} & 0 & 0 & \overline{\overline{Z}}^{\vec{J}^{\mathrm{F}}\vec{M}^{\mathrm{A}\rightleftharpoons\mathrm{F}}} & \overline{\overline{Z}}^{\vec{J}^{\mathrm{F}}\vec{M}^{\mathrm{G}\rightleftharpoons\mathrm{F}}} \\
0 & \overline{\overline{Z}}^{\vec{J}^{\mathrm{G}\rightleftharpoons\mathrm{F}}\vec{J}^{\mathrm{G}}} & \overline{\overline{Z}}^{\vec{J}^{\mathrm{G}\rightleftharpoons\mathrm{F}}\vec{J}^{\mathrm{G}\rightleftharpoons\mathrm{A}}} & 0 & \overline{\overline{Z}}^{\vec{J}^{\mathrm{G}\rightleftharpoons\mathrm{F}}\vec{J}^{\mathrm{A}\rightleftharpoons\mathrm{F}}} & \overline{\overline{Z}}^{\vec{J}^{\mathrm{G}\rightleftharpoons\mathrm{F}}\vec{J}^{\mathrm{F}}} & \overline{\overline{Z}}^{\vec{J}^{\mathrm{G}\rightleftharpoons\mathrm{F}}\vec{J}^{\mathrm{G}\rightleftharpoons\mathrm{F}}} & \overline{\overline{Z}}^{\vec{J}^{\mathrm{G}\rightleftharpoons\mathrm{F}}\vec{M}^{\mathrm{O}\rightleftharpoons\mathrm{G}}} & \overline{\overline{Z}}^{\vec{J}^{\mathrm{G}\rightleftharpoons\mathrm{F}}\vec{M}^{\mathrm{G}\rightleftharpoons\mathrm{A}}} & \overline{\overline{Z}}^{\vec{J}^{\mathrm{G}\rightleftharpoons\mathrm{F}}\vec{M}^{\mathrm{A}\rightleftharpoons\mathrm{F}}} & \overline{\overline{Z}}^{\vec{J}^{\mathrm{G}\rightleftharpoons\mathrm{F}}\vec{M}^{\mathrm{G}\rightleftharpoons\mathrm{F}}} \\
0 & \overline{\overline{Z}}^{\vec{M}^{\mathrm{G}\rightleftharpoons\mathrm{F}}\vec{J}^{\mathrm{G}}} & \overline{\overline{Z}}^{\vec{M}^{\mathrm{G}\rightleftharpoons\mathrm{F}}\vec{J}^{\mathrm{G}\rightleftharpoons\mathrm{A}}} & 0 & \overline{\overline{Z}}^{\vec{M}^{\mathrm{G}\rightleftharpoons\mathrm{F}}\vec{J}^{\mathrm{A}\rightleftharpoons\mathrm{F}}} & \overline{\overline{Z}}^{\vec{M}^{\mathrm{G}\rightleftharpoons\mathrm{F}}\vec{J}^{\mathrm{F}}} & \overline{\overline{Z}}^{\vec{M}^{\mathrm{G}\rightleftharpoons\mathrm{F}}\vec{J}^{\mathrm{G}\rightleftharpoons\mathrm{F}}} & \overline{\overline{Z}}^{\vec{M}^{\mathrm{G}\rightleftharpoons\mathrm{F}}\vec{M}^{\mathrm{O}\rightleftharpoons\mathrm{G}}} & \overline{\overline{Z}}^{\vec{M}^{\mathrm{G}\rightleftharpoons\mathrm{F}}\vec{M}^{\mathrm{G}\rightleftharpoons\mathrm{A}}} & \overline{\overline{Z}}^{\vec{M}^{\mathrm{G}\rightleftharpoons\mathrm{F}}\vec{M}^{\mathrm{A}\rightleftharpoons\mathrm{F}}} & \overline{\overline{Z}}^{\vec{M}^{\mathrm{G}\rightleftharpoons\mathrm{F}}\vec{M}^{\mathrm{G}\rightleftharpoons\mathrm{F}}}
\end{bmatrix} \tag{8-36a}
$$

$$
\overline{\overline{\Psi}}_2 = \begin{bmatrix}
\overline{\overline{I}}^{\vec{J}^{\mathrm{O}\rightleftharpoons\mathrm{G}}} \\
-\overline{\overline{Z}}^{\vec{M}^{\mathrm{O}\rightleftharpoons\mathrm{G}}\vec{J}^{\mathrm{O}\rightleftharpoons\mathrm{G}}} \\
-\overline{\overline{Z}}^{\vec{J}^{\mathrm{G}}\vec{J}^{\mathrm{O}\rightleftharpoons\mathrm{G}}} \\
-\overline{\overline{Z}}^{\vec{J}^{\mathrm{G}\rightleftharpoons\mathrm{A}}\vec{J}^{\mathrm{O}\rightleftharpoons\mathrm{G}}} \\
-\overline{\overline{Z}}^{\vec{M}^{\mathrm{G}\rightleftharpoons\mathrm{A}}\vec{J}^{\mathrm{O}\rightleftharpoons\mathrm{G}}} \\
0 \\
0 \\
0 \\
0 \\
-\overline{\overline{Z}}^{\vec{J}^{\mathrm{G}\rightleftharpoons\mathrm{F}}\vec{J}^{\mathrm{O}\rightleftharpoons\mathrm{G}}} \\
-\overline{\overline{Z}}^{\vec{M}^{\mathrm{G}\rightleftharpoons\mathrm{F}}\vec{J}^{\mathrm{O}\rightleftharpoons\mathrm{G}}}
\end{bmatrix} \tag{8-36b}
$$

and

$$
\bar{a}^{\mathrm{AV}} = \begin{bmatrix}
\bar{a}^{\vec{J}^{\mathrm{O}\rightleftharpoons\mathrm{G}}} \\
\bar{a}^{\vec{J}^{\mathrm{G}}} \\
\bar{a}^{\vec{J}^{\mathrm{G}\rightleftharpoons\mathrm{A}}} \\
\bar{a}^{\vec{J}^{\mathrm{A}}} \\
\bar{a}^{\vec{J}^{\mathrm{A}\rightleftharpoons\mathrm{F}}} \\
\bar{a}^{\vec{J}^{\mathrm{F}}} \\
\bar{a}^{\vec{J}^{\mathrm{G}\rightleftharpoons\mathrm{F}}} \\
\bar{a}^{\vec{M}^{\mathrm{O}\rightleftharpoons\mathrm{G}}} \\
\bar{a}^{\vec{M}^{\mathrm{G}\rightleftharpoons\mathrm{A}}} \\
\bar{a}^{\vec{M}^{\mathrm{A}\rightleftharpoons\mathrm{F}}} \\
\bar{a}^{\vec{M}^{\mathrm{G}\rightleftharpoons\mathrm{F}}}
\end{bmatrix} \tag{8-37}
$$

By properly assembling the matrix equations (8-17b) and (8-18)~(8-23b), we have the following another augmented matrix equation

$$
\overline{\overline{\Psi}}_3 \cdot \bar{a}^{\mathrm{AV}} = \overline{\overline{\Psi}}_4 \cdot \bar{a}^{\vec{M}^{\mathrm{G}\rightleftharpoons\mathrm{A}}} \tag{8-38}
$$

in which

$$\overline{\overline{\Psi}}_3=\begin{bmatrix}
0 & 0 & 0 & 0 & 0 & 0 & 0 & \overline{\overline{I}}^{\vec{M}^{O\rightleftharpoons G}} & 0 & 0 & 0\\
\overline{\overline{Z}}^{\vec{J}^{O\rightleftharpoons G}\vec{J}^{O\rightleftharpoons G}} & \overline{\overline{Z}}^{\vec{J}^{O\rightleftharpoons G}\vec{J}^{G}} & \overline{\overline{Z}}^{\vec{J}^{O\rightleftharpoons G}\vec{J}^{G\rightleftharpoons A}} & 0 & 0 & 0 & \overline{\overline{Z}}^{\vec{J}^{O\rightleftharpoons G}\vec{J}^{G\rightleftharpoons F}} & 0 & \overline{\overline{Z}}^{\vec{J}^{O\rightleftharpoons G}\vec{M}^{G\rightleftharpoons A}} & 0 & \overline{\overline{Z}}^{\vec{J}^{O\rightleftharpoons G}\vec{M}^{G\rightleftharpoons F}}\\
\overline{\overline{Z}}^{\vec{J}^{G}\vec{J}^{O\rightleftharpoons G}} & \overline{\overline{Z}}^{\vec{J}^{G}\vec{J}^{G}} & \overline{\overline{Z}}^{\vec{J}^{G}\vec{J}^{G\rightleftharpoons A}} & 0 & 0 & 0 & \overline{\overline{Z}}^{\vec{J}^{G}\vec{J}^{G\rightleftharpoons F}} & 0 & \overline{\overline{Z}}^{\vec{J}^{G}\vec{M}^{G\rightleftharpoons A}} & 0 & \overline{\overline{Z}}^{\vec{J}^{G}\vec{M}^{G\rightleftharpoons F}}\\
\overline{\overline{Z}}^{\vec{J}^{G\rightleftharpoons A}\vec{J}^{O\rightleftharpoons G}} & \overline{\overline{Z}}^{\vec{J}^{G\rightleftharpoons A}\vec{J}^{G}} & \overline{\overline{Z}}^{\vec{J}^{G\rightleftharpoons A}\vec{J}^{G\rightleftharpoons A}} & \overline{\overline{Z}}^{\vec{J}^{G\rightleftharpoons A}\vec{J}^{A}} & \overline{\overline{Z}}^{\vec{J}^{G\rightleftharpoons A}\vec{J}^{A\rightleftharpoons F}} & 0 & \overline{\overline{Z}}^{\vec{J}^{G\rightleftharpoons A}\vec{J}^{G\rightleftharpoons F}} & 0 & \overline{\overline{Z}}^{\vec{J}^{G\rightleftharpoons A}\vec{M}^{G\rightleftharpoons A}} & \overline{\overline{Z}}^{\vec{J}^{G\rightleftharpoons A}\vec{M}^{A\rightleftharpoons F}} & \overline{\overline{Z}}^{\vec{J}^{G\rightleftharpoons A}\vec{M}^{G\rightleftharpoons F}}\\
\overline{\overline{Z}}^{\vec{M}^{G\rightleftharpoons A}\vec{J}^{O\rightleftharpoons G}} & \overline{\overline{Z}}^{\vec{M}^{G\rightleftharpoons A}\vec{J}^{G}} & \overline{\overline{Z}}^{\vec{M}^{G\rightleftharpoons A}\vec{J}^{G\rightleftharpoons A}} & \overline{\overline{Z}}^{\vec{M}^{G\rightleftharpoons A}\vec{J}^{A}} & \overline{\overline{Z}}^{\vec{M}^{G\rightleftharpoons A}\vec{J}^{A\rightleftharpoons F}} & 0 & \overline{\overline{Z}}^{\vec{M}^{G\rightleftharpoons A}\vec{J}^{G\rightleftharpoons F}} & 0 & \overline{\overline{Z}}^{\vec{M}^{G\rightleftharpoons A}\vec{M}^{G\rightleftharpoons A}} & \overline{\overline{Z}}^{\vec{M}^{G\rightleftharpoons A}\vec{M}^{A\rightleftharpoons F}} & \overline{\overline{Z}}^{\vec{M}^{G\rightleftharpoons A}\vec{M}^{G\rightleftharpoons F}}\\
0 & 0 & \overline{\overline{Z}}^{\vec{J}^{A}\vec{J}^{G\rightleftharpoons A}} & \overline{\overline{Z}}^{\vec{J}^{A}\vec{J}^{A}} & \overline{\overline{Z}}^{\vec{J}^{A}\vec{J}^{A\rightleftharpoons F}} & 0 & 0 & 0 & \overline{\overline{Z}}^{\vec{J}^{A}\vec{M}^{G\rightleftharpoons A}} & \overline{\overline{Z}}^{\vec{J}^{A}\vec{M}^{A\rightleftharpoons F}} & 0\\
0 & 0 & \overline{\overline{Z}}^{\vec{J}^{A\rightleftharpoons F}\vec{J}^{G\rightleftharpoons A}} & \overline{\overline{Z}}^{\vec{J}^{A\rightleftharpoons F}\vec{J}^{A}} & \overline{\overline{Z}}^{\vec{J}^{A\rightleftharpoons F}\vec{J}^{A\rightleftharpoons F}} & \overline{\overline{Z}}^{\vec{J}^{A\rightleftharpoons F}\vec{J}^{F}} & \overline{\overline{Z}}^{\vec{J}^{A\rightleftharpoons F}\vec{J}^{G\rightleftharpoons F}} & 0 & \overline{\overline{Z}}^{\vec{J}^{A\rightleftharpoons F}\vec{M}^{G\rightleftharpoons A}} & \overline{\overline{Z}}^{\vec{J}^{A\rightleftharpoons F}\vec{M}^{A\rightleftharpoons F}} & \overline{\overline{Z}}^{\vec{J}^{A\rightleftharpoons F}\vec{M}^{G\rightleftharpoons F}}\\
0 & 0 & \overline{\overline{Z}}^{\vec{M}^{A\rightleftharpoons F}\vec{J}^{G\rightleftharpoons A}} & \overline{\overline{Z}}^{\vec{M}^{A\rightleftharpoons F}\vec{J}^{A}} & \overline{\overline{Z}}^{\vec{M}^{A\rightleftharpoons F}\vec{J}^{A\rightleftharpoons F}} & \overline{\overline{Z}}^{\vec{M}^{A\rightleftharpoons F}\vec{J}^{F}} & \overline{\overline{Z}}^{\vec{M}^{A\rightleftharpoons F}\vec{J}^{G\rightleftharpoons F}} & 0 & \overline{\overline{Z}}^{\vec{M}^{A\rightleftharpoons F}\vec{M}^{G\rightleftharpoons A}} & \overline{\overline{Z}}^{\vec{M}^{A\rightleftharpoons F}\vec{M}^{A\rightleftharpoons F}} & \overline{\overline{Z}}^{\vec{M}^{A\rightleftharpoons F}\vec{M}^{G\rightleftharpoons F}}\\
0 & 0 & 0 & 0 & \overline{\overline{Z}}^{\vec{J}^{F}\vec{J}^{A\rightleftharpoons F}} & \overline{\overline{Z}}^{\vec{J}^{F}\vec{J}^{F}} & \overline{\overline{Z}}^{\vec{J}^{F}\vec{J}^{G\rightleftharpoons F}} & 0 & 0 & \overline{\overline{Z}}^{\vec{J}^{F}\vec{M}^{A\rightleftharpoons F}} & \overline{\overline{Z}}^{\vec{J}^{F}\vec{M}^{G\rightleftharpoons F}}\\
\overline{\overline{Z}}^{\vec{J}^{G\rightleftharpoons F}\vec{J}^{O\rightleftharpoons G}} & \overline{\overline{Z}}^{\vec{J}^{G\rightleftharpoons F}\vec{J}^{G}} & \overline{\overline{Z}}^{\vec{J}^{G\rightleftharpoons F}\vec{J}^{G\rightleftharpoons A}} & 0 & \overline{\overline{Z}}^{\vec{J}^{G\rightleftharpoons F}\vec{J}^{A\rightleftharpoons F}} & \overline{\overline{Z}}^{\vec{J}^{G\rightleftharpoons F}\vec{J}^{F}} & \overline{\overline{Z}}^{\vec{J}^{G\rightleftharpoons F}\vec{J}^{G\rightleftharpoons F}} & 0 & \overline{\overline{Z}}^{\vec{J}^{G\rightleftharpoons F}\vec{M}^{G\rightleftharpoons A}} & \overline{\overline{Z}}^{\vec{J}^{G\rightleftharpoons F}\vec{M}^{A\rightleftharpoons F}} & \overline{\overline{Z}}^{\vec{J}^{G\rightleftharpoons F}\vec{M}^{G\rightleftharpoons F}}\\
\overline{\overline{Z}}^{\vec{M}^{G\rightleftharpoons F}\vec{J}^{O\rightleftharpoons G}} & \overline{\overline{Z}}^{\vec{M}^{G\rightleftharpoons F}\vec{J}^{G}} & \overline{\overline{Z}}^{\vec{M}^{G\rightleftharpoons F}\vec{J}^{G\rightleftharpoons A}} & 0 & \overline{\overline{Z}}^{\vec{M}^{G\rightleftharpoons F}\vec{J}^{A\rightleftharpoons F}} & \overline{\overline{Z}}^{\vec{M}^{G\rightleftharpoons F}\vec{J}^{F}} & \overline{\overline{Z}}^{\vec{M}^{G\rightleftharpoons F}\vec{J}^{G\rightleftharpoons F}} & 0 & \overline{\overline{Z}}^{\vec{M}^{G\rightleftharpoons F}\vec{M}^{G\rightleftharpoons A}} & \overline{\overline{Z}}^{\vec{M}^{G\rightleftharpoons F}\vec{M}^{A\rightleftharpoons F}} & \overline{\overline{Z}}^{\vec{M}^{G\rightleftharpoons F}\vec{M}^{G\rightleftharpoons F}}
\end{bmatrix} \tag{8-39a}$$

$$\overline{\overline{\Psi}}_4=\begin{bmatrix}
\overline{\overline{I}}^{\vec{M}^{O\rightleftharpoons G}}\\
-\overline{\overline{Z}}^{\vec{J}^{O\rightleftharpoons G}\vec{M}^{O\rightleftharpoons G}}\\
-\overline{\overline{Z}}^{\vec{J}^{G}\vec{M}^{O\rightleftharpoons G}}\\
-\overline{\overline{Z}}^{\vec{J}^{G\rightleftharpoons A}\vec{M}^{O\rightleftharpoons G}}\\
-\overline{\overline{Z}}^{\vec{M}^{G\rightleftharpoons A}\vec{M}^{O\rightleftharpoons G}}\\
0\\
0\\
0\\
0\\
-\overline{\overline{Z}}^{\vec{J}^{G\rightleftharpoons F}\vec{M}^{O\rightleftharpoons G}}\\
-\overline{\overline{Z}}^{\vec{M}^{G\rightleftharpoons F}\vec{M}^{O\rightleftharpoons G}}
\end{bmatrix} \tag{8-39b}$$

where the 0s are some zero matrices with proper row numbers and column numbers.

By solving the above augmented matrix equations, there exist the following transformations from $\bar{a}^{\vec{J}^{G\rightleftharpoons A}}$ to $\bar{a}^{AV}$ and from $\bar{a}^{\vec{M}^{G\rightleftharpoons A}}$ to $\bar{a}^{AV}$.

$$\bar{a}^{AV} = \overbrace{\left(\overline{\overline{\Psi}}_1\right)^{-1}\cdot\overline{\overline{\Psi}}_2}^{\overline{\overline{T}}^{\vec{J}^{G\rightleftharpoons A}\to AV}}\cdot\bar{a}^{\vec{J}^{G\rightleftharpoons A}} \tag{8-40}$$

$$\bar{a}^{AV} = \underbrace{\left(\overline{\overline{\Psi}}_3\right)^{-1}\cdot\overline{\overline{\Psi}}_4}_{\overline{\overline{T}}^{\vec{M}^{G\rightleftharpoons A}\to AV}}\cdot\bar{a}^{\vec{M}^{G\rightleftharpoons A}} \tag{8-41}$$

and they can be uniformly written as follows:

$$\bar{a}^{AV} = \overline{\overline{T}}^{BV\to AV}\cdot\bar{a}^{BV} \tag{8-42}$$

where $\overline{\overline{T}}^{BV\to AV} = \overline{\overline{T}}^{\vec{J}^{G\rightleftharpoons A}\to AV} / \overline{\overline{T}}^{\vec{M}^{G\rightleftharpoons A}\to AV}$, and correspondingly $\bar{a}^{BV} = \bar{a}^{\vec{J}^{G\rightleftharpoons A}} / \bar{a}^{\vec{M}^{G\rightleftharpoons A}}$.

### 8.2.3.4 Scheme II: Solution/Definition Domain Compression (SDC/DDC)

In fact, the matrix equation (8-17a) and matrix equations (8-18)~(8-23b) can be alternatively combined as follows:

$$\overline{\overline{\Psi}}{}_{\rm FCE}^{\rm DoJ} \cdot \overline{a}^{\rm AV} = 0 \tag{8-43}$$

where

$$\overline{\overline{\Psi}}{}_{\rm FCE}^{\rm DoJ} = \begin{bmatrix}
\overline{\overline{Z}}^{\vec{M}^{O\rightleftharpoons G}\vec{J}^{O\rightleftharpoons G}} & \overline{\overline{Z}}^{\vec{M}^{O\rightleftharpoons G}\vec{J}^{G}} & \overline{\overline{Z}}^{\vec{M}^{O\rightleftharpoons G}\vec{J}^{G\rightleftharpoons A}} & 0 & 0 & 0 & \overline{\overline{Z}}^{\vec{M}^{O\rightleftharpoons G}\vec{J}^{G\rightleftharpoons F}} & \overline{\overline{Z}}^{\vec{M}^{O\rightleftharpoons G}\vec{M}^{O\rightleftharpoons G}} & \overline{\overline{Z}}^{\vec{M}^{O\rightleftharpoons G}\vec{M}^{G\rightleftharpoons A}} & 0 & \overline{\overline{Z}}^{\vec{M}^{O\rightleftharpoons G}\vec{M}^{G\rightleftharpoons F}} \\
\overline{\overline{Z}}^{\vec{J}^{G}\vec{J}^{O\rightleftharpoons G}} & \overline{\overline{Z}}^{\vec{J}^{G}\vec{J}^{G}} & \overline{\overline{Z}}^{\vec{J}^{G}\vec{J}^{G\rightleftharpoons A}} & 0 & 0 & 0 & \overline{\overline{Z}}^{\vec{J}^{G}\vec{J}^{G\rightleftharpoons F}} & \overline{\overline{Z}}^{\vec{J}^{G}\vec{M}^{O\rightleftharpoons G}} & \overline{\overline{Z}}^{\vec{J}^{G}\vec{M}^{G\rightleftharpoons A}} & 0 & \overline{\overline{Z}}^{\vec{J}^{G}\vec{M}^{G\rightleftharpoons F}} \\
\overline{\overline{Z}}^{\vec{J}^{G\rightleftharpoons A}\vec{J}^{O\rightleftharpoons G}} & \overline{\overline{Z}}^{\vec{J}^{G\rightleftharpoons A}\vec{J}^{G}} & \overline{\overline{Z}}^{\vec{J}^{G\rightleftharpoons A}\vec{J}^{G\rightleftharpoons A}} & \overline{\overline{Z}}^{\vec{J}^{G\rightleftharpoons A}\vec{J}^{A}} & \overline{\overline{Z}}^{\vec{J}^{G\rightleftharpoons A}\vec{J}^{A\rightleftharpoons F}} & 0 & \overline{\overline{Z}}^{\vec{J}^{G\rightleftharpoons A}\vec{J}^{G\rightleftharpoons F}} & \overline{\overline{Z}}^{\vec{J}^{G\rightleftharpoons A}\vec{M}^{O\rightleftharpoons G}} & \overline{\overline{Z}}^{\vec{J}^{G\rightleftharpoons A}\vec{M}^{G\rightleftharpoons A}} & \overline{\overline{Z}}^{\vec{J}^{G\rightleftharpoons A}\vec{M}^{A\rightleftharpoons F}} & \overline{\overline{Z}}^{\vec{J}^{G\rightleftharpoons A}\vec{M}^{G\rightleftharpoons F}} \\
\overline{\overline{Z}}^{\vec{M}^{G\rightleftharpoons A}\vec{J}^{O\rightleftharpoons G}} & \overline{\overline{Z}}^{\vec{M}^{G\rightleftharpoons A}\vec{J}^{G}} & \overline{\overline{Z}}^{\vec{M}^{G\rightleftharpoons A}\vec{J}^{G\rightleftharpoons A}} & \overline{\overline{Z}}^{\vec{M}^{G\rightleftharpoons A}\vec{J}^{A}} & \overline{\overline{Z}}^{\vec{M}^{G\rightleftharpoons A}\vec{J}^{A\rightleftharpoons F}} & 0 & \overline{\overline{Z}}^{\vec{M}^{G\rightleftharpoons A}\vec{J}^{G\rightleftharpoons F}} & \overline{\overline{Z}}^{\vec{M}^{G\rightleftharpoons A}\vec{M}^{O\rightleftharpoons G}} & \overline{\overline{Z}}^{\vec{M}^{G\rightleftharpoons A}\vec{M}^{G\rightleftharpoons A}} & \overline{\overline{Z}}^{\vec{M}^{G\rightleftharpoons A}\vec{M}^{A\rightleftharpoons F}} & \overline{\overline{Z}}^{\vec{M}^{G\rightleftharpoons A}\vec{M}^{G\rightleftharpoons F}} \\
0 & 0 & \overline{\overline{Z}}^{\vec{J}^{A}\vec{J}^{G\rightleftharpoons A}} & \overline{\overline{Z}}^{\vec{J}^{A}\vec{J}^{A}} & \overline{\overline{Z}}^{\vec{J}^{A}\vec{J}^{A\rightleftharpoons F}} & 0 & 0 & 0 & \overline{\overline{Z}}^{\vec{J}^{A}\vec{M}^{G\rightleftharpoons A}} & \overline{\overline{Z}}^{\vec{J}^{A}\vec{M}^{A\rightleftharpoons F}} & 0 \\
0 & 0 & \overline{\overline{Z}}^{\vec{J}^{A\rightleftharpoons F}\vec{J}^{G\rightleftharpoons A}} & \overline{\overline{Z}}^{\vec{J}^{A\rightleftharpoons F}\vec{J}^{A}} & \overline{\overline{Z}}^{\vec{J}^{A\rightleftharpoons F}\vec{J}^{A\rightleftharpoons F}} & \overline{\overline{Z}}^{\vec{J}^{A\rightleftharpoons F}\vec{J}^{F}} & \overline{\overline{Z}}^{\vec{J}^{A\rightleftharpoons F}\vec{J}^{G\rightleftharpoons F}} & 0 & \overline{\overline{Z}}^{\vec{J}^{A\rightleftharpoons F}\vec{M}^{G\rightleftharpoons A}} & \overline{\overline{Z}}^{\vec{J}^{A\rightleftharpoons F}\vec{M}^{A\rightleftharpoons F}} & \overline{\overline{Z}}^{\vec{J}^{A\rightleftharpoons F}\vec{M}^{G\rightleftharpoons F}} \\
0 & 0 & \overline{\overline{Z}}^{\vec{M}^{A\rightleftharpoons F}\vec{J}^{G\rightleftharpoons A}} & \overline{\overline{Z}}^{\vec{M}^{A\rightleftharpoons F}\vec{J}^{A}} & \overline{\overline{Z}}^{\vec{M}^{A\rightleftharpoons F}\vec{J}^{A\rightleftharpoons F}} & \overline{\overline{Z}}^{\vec{M}^{A\rightleftharpoons F}\vec{J}^{F}} & \overline{\overline{Z}}^{\vec{M}^{A\rightleftharpoons F}\vec{J}^{G\rightleftharpoons F}} & 0 & \overline{\overline{Z}}^{\vec{M}^{A\rightleftharpoons F}\vec{M}^{G\rightleftharpoons A}} & \overline{\overline{Z}}^{\vec{M}^{A\rightleftharpoons F}\vec{M}^{A\rightleftharpoons F}} & \overline{\overline{Z}}^{\vec{M}^{A\rightleftharpoons F}\vec{M}^{G\rightleftharpoons F}} \\
0 & 0 & 0 & 0 & \overline{\overline{Z}}^{\vec{J}^{F}\vec{J}^{A\rightleftharpoons F}} & \overline{\overline{Z}}^{\vec{J}^{F}\vec{J}^{F}} & \overline{\overline{Z}}^{\vec{J}^{F}\vec{J}^{G\rightleftharpoons F}} & 0 & 0 & \overline{\overline{Z}}^{\vec{J}^{F}\vec{M}^{A\rightleftharpoons F}} & \overline{\overline{Z}}^{\vec{J}^{F}\vec{M}^{G\rightleftharpoons F}} \\
\overline{\overline{Z}}^{\vec{J}^{G\rightleftharpoons F}\vec{J}^{O\rightleftharpoons G}} & \overline{\overline{Z}}^{\vec{J}^{G\rightleftharpoons F}\vec{J}^{G}} & \overline{\overline{Z}}^{\vec{J}^{G\rightleftharpoons F}\vec{J}^{G\rightleftharpoons A}} & 0 & \overline{\overline{Z}}^{\vec{J}^{G\rightleftharpoons F}\vec{J}^{A\rightleftharpoons F}} & \overline{\overline{Z}}^{\vec{J}^{G\rightleftharpoons F}\vec{J}^{F}} & \overline{\overline{Z}}^{\vec{J}^{G\rightleftharpoons F}\vec{J}^{G\rightleftharpoons F}} & \overline{\overline{Z}}^{\vec{J}^{G\rightleftharpoons F}\vec{M}^{O\rightleftharpoons G}} & \overline{\overline{Z}}^{\vec{J}^{G\rightleftharpoons F}\vec{M}^{G\rightleftharpoons A}} & \overline{\overline{Z}}^{\vec{J}^{G\rightleftharpoons F}\vec{M}^{A\rightleftharpoons F}} & \overline{\overline{Z}}^{\vec{J}^{G\rightleftharpoons F}\vec{M}^{G\rightleftharpoons F}} \\
\overline{\overline{Z}}^{\vec{M}^{G\rightleftharpoons F}\vec{J}^{O\rightleftharpoons G}} & \overline{\overline{Z}}^{\vec{M}^{G\rightleftharpoons F}\vec{J}^{G}} & \overline{\overline{Z}}^{\vec{M}^{G\rightleftharpoons F}\vec{J}^{G\rightleftharpoons A}} & 0 & \overline{\overline{Z}}^{\vec{M}^{G\rightleftharpoons F}\vec{J}^{A\rightleftharpoons F}} & \overline{\overline{Z}}^{\vec{M}^{G\rightleftharpoons F}\vec{J}^{F}} & \overline{\overline{Z}}^{\vec{M}^{G\rightleftharpoons F}\vec{J}^{G\rightleftharpoons F}} & \overline{\overline{Z}}^{\vec{M}^{G\rightleftharpoons F}\vec{M}^{O\rightleftharpoons G}} & \overline{\overline{Z}}^{\vec{M}^{G\rightleftharpoons F}\vec{M}^{G\rightleftharpoons A}} & \overline{\overline{Z}}^{\vec{M}^{G\rightleftharpoons F}\vec{M}^{A\rightleftharpoons F}} & \overline{\overline{Z}}^{\vec{M}^{G\rightleftharpoons F}\vec{M}^{G\rightleftharpoons F}}
\end{bmatrix} \tag{8-44}$$

Similarly, the matrix equation (8-17b) and matrix equations (8-18)~(8-23b) can also be alternatively combined as follows:

$$\overline{\overline{\Psi}}{}_{\rm FCE}^{\rm DoM} \cdot \overline{a}^{\rm AV} = 0 \tag{8-45}$$

where

$$\overline{\overline{\Psi}}{}_{\rm FCE}^{\rm DoM} = \begin{bmatrix}
\overline{\overline{Z}}^{\vec{J}^{O\rightleftharpoons G}\vec{J}^{O\rightleftharpoons G}} & \overline{\overline{Z}}^{\vec{J}^{O\rightleftharpoons G}\vec{J}^{G}} & \overline{\overline{Z}}^{\vec{J}^{O\rightleftharpoons G}\vec{J}^{G\rightleftharpoons A}} & 0 & 0 & 0 & \overline{\overline{Z}}^{\vec{J}^{O\rightleftharpoons G}\vec{J}^{G\rightleftharpoons F}} & \overline{\overline{Z}}^{\vec{J}^{O\rightleftharpoons G}\vec{M}^{O\rightleftharpoons G}} & \overline{\overline{Z}}^{\vec{J}^{O\rightleftharpoons G}\vec{M}^{G\rightleftharpoons A}} & 0 & \overline{\overline{Z}}^{\vec{J}^{O\rightleftharpoons G}\vec{M}^{G\rightleftharpoons F}} \\
\overline{\overline{Z}}^{\vec{J}^{G}\vec{J}^{O\rightleftharpoons G}} & \overline{\overline{Z}}^{\vec{J}^{G}\vec{J}^{G}} & \overline{\overline{Z}}^{\vec{J}^{G}\vec{J}^{G\rightleftharpoons A}} & 0 & 0 & 0 & \overline{\overline{Z}}^{\vec{J}^{G}\vec{J}^{G\rightleftharpoons F}} & \overline{\overline{Z}}^{\vec{J}^{G}\vec{M}^{O\rightleftharpoons G}} & \overline{\overline{Z}}^{\vec{J}^{G}\vec{M}^{G\rightleftharpoons A}} & 0 & \overline{\overline{Z}}^{\vec{J}^{G}\vec{M}^{G\rightleftharpoons F}} \\
\overline{\overline{Z}}^{\vec{J}^{G\rightleftharpoons A}\vec{J}^{O\rightleftharpoons G}} & \overline{\overline{Z}}^{\vec{J}^{G\rightleftharpoons A}\vec{J}^{G}} & \overline{\overline{Z}}^{\vec{J}^{G\rightleftharpoons A}\vec{J}^{G\rightleftharpoons A}} & \overline{\overline{Z}}^{\vec{J}^{G\rightleftharpoons A}\vec{J}^{A}} & \overline{\overline{Z}}^{\vec{J}^{G\rightleftharpoons A}\vec{J}^{A\rightleftharpoons F}} & 0 & \overline{\overline{Z}}^{\vec{J}^{G\rightleftharpoons A}\vec{J}^{G\rightleftharpoons F}} & \overline{\overline{Z}}^{\vec{J}^{G\rightleftharpoons A}\vec{M}^{O\rightleftharpoons G}} & \overline{\overline{Z}}^{\vec{J}^{G\rightleftharpoons A}\vec{M}^{G\rightleftharpoons A}} & \overline{\overline{Z}}^{\vec{J}^{G\rightleftharpoons A}\vec{M}^{A\rightleftharpoons F}} & \overline{\overline{Z}}^{\vec{J}^{G\rightleftharpoons A}\vec{M}^{G\rightleftharpoons F}} \\
\overline{\overline{Z}}^{\vec{M}^{G\rightleftharpoons A}\vec{J}^{O\rightleftharpoons G}} & \overline{\overline{Z}}^{\vec{M}^{G\rightleftharpoons A}\vec{J}^{G}} & \overline{\overline{Z}}^{\vec{M}^{G\rightleftharpoons A}\vec{J}^{G\rightleftharpoons A}} & \overline{\overline{Z}}^{\vec{M}^{G\rightleftharpoons A}\vec{J}^{A}} & \overline{\overline{Z}}^{\vec{M}^{G\rightleftharpoons A}\vec{J}^{A\rightleftharpoons F}} & 0 & \overline{\overline{Z}}^{\vec{M}^{G\rightleftharpoons A}\vec{J}^{G\rightleftharpoons F}} & \overline{\overline{Z}}^{\vec{M}^{G\rightleftharpoons A}\vec{M}^{O\rightleftharpoons G}} & \overline{\overline{Z}}^{\vec{M}^{G\rightleftharpoons A}\vec{M}^{G\rightleftharpoons A}} & \overline{\overline{Z}}^{\vec{M}^{G\rightleftharpoons A}\vec{M}^{A\rightleftharpoons F}} & \overline{\overline{Z}}^{\vec{M}^{G\rightleftharpoons A}\vec{M}^{G\rightleftharpoons F}} \\
0 & 0 & \overline{\overline{Z}}^{\vec{J}^{A}\vec{J}^{G\rightleftharpoons A}} & \overline{\overline{Z}}^{\vec{J}^{A}\vec{J}^{A}} & \overline{\overline{Z}}^{\vec{J}^{A}\vec{J}^{A\rightleftharpoons F}} & 0 & 0 & 0 & \overline{\overline{Z}}^{\vec{J}^{A}\vec{M}^{G\rightleftharpoons A}} & \overline{\overline{Z}}^{\vec{J}^{A}\vec{M}^{A\rightleftharpoons F}} & 0 \\
0 & 0 & \overline{\overline{Z}}^{\vec{J}^{A\rightleftharpoons F}\vec{J}^{G\rightleftharpoons A}} & \overline{\overline{Z}}^{\vec{J}^{A\rightleftharpoons F}\vec{J}^{A}} & \overline{\overline{Z}}^{\vec{J}^{A\rightleftharpoons F}\vec{J}^{A\rightleftharpoons F}} & \overline{\overline{Z}}^{\vec{J}^{A\rightleftharpoons F}\vec{J}^{F}} & \overline{\overline{Z}}^{\vec{J}^{A\rightleftharpoons F}\vec{J}^{G\rightleftharpoons F}} & 0 & \overline{\overline{Z}}^{\vec{J}^{A\rightleftharpoons F}\vec{M}^{G\rightleftharpoons A}} & \overline{\overline{Z}}^{\vec{J}^{A\rightleftharpoons F}\vec{M}^{A\rightleftharpoons F}} & \overline{\overline{Z}}^{\vec{J}^{A\rightleftharpoons F}\vec{M}^{G\rightleftharpoons F}} \\
0 & 0 & \overline{\overline{Z}}^{\vec{M}^{A\rightleftharpoons F}\vec{J}^{G\rightleftharpoons A}} & \overline{\overline{Z}}^{\vec{M}^{A\rightleftharpoons F}\vec{J}^{A}} & \overline{\overline{Z}}^{\vec{M}^{A\rightleftharpoons F}\vec{J}^{A\rightleftharpoons F}} & \overline{\overline{Z}}^{\vec{M}^{A\rightleftharpoons F}\vec{J}^{F}} & \overline{\overline{Z}}^{\vec{M}^{A\rightleftharpoons F}\vec{J}^{G\rightleftharpoons F}} & 0 & \overline{\overline{Z}}^{\vec{M}^{A\rightleftharpoons F}\vec{M}^{G\rightleftharpoons A}} & \overline{\overline{Z}}^{\vec{M}^{A\rightleftharpoons F}\vec{M}^{A\rightleftharpoons F}} & \overline{\overline{Z}}^{\vec{M}^{A\rightleftharpoons F}\vec{M}^{G\rightleftharpoons F}} \\
0 & 0 & 0 & 0 & \overline{\overline{Z}}^{\vec{J}^{F}\vec{J}^{A\rightleftharpoons F}} & \overline{\overline{Z}}^{\vec{J}^{F}\vec{J}^{F}} & \overline{\overline{Z}}^{\vec{J}^{F}\vec{J}^{G\rightleftharpoons F}} & 0 & 0 & \overline{\overline{Z}}^{\vec{J}^{F}\vec{M}^{A\rightleftharpoons F}} & \overline{\overline{Z}}^{\vec{J}^{F}\vec{M}^{G\rightleftharpoons F}} \\
\overline{\overline{Z}}^{\vec{J}^{G\rightleftharpoons F}\vec{J}^{O\rightleftharpoons G}} & \overline{\overline{Z}}^{\vec{J}^{G\rightleftharpoons F}\vec{J}^{G}} & \overline{\overline{Z}}^{\vec{J}^{G\rightleftharpoons F}\vec{J}^{G\rightleftharpoons A}} & 0 & \overline{\overline{Z}}^{\vec{J}^{G\rightleftharpoons F}\vec{J}^{A\rightleftharpoons F}} & \overline{\overline{Z}}^{\vec{J}^{G\rightleftharpoons F}\vec{J}^{F}} & \overline{\overline{Z}}^{\vec{J}^{G\rightleftharpoons F}\vec{J}^{G\rightleftharpoons F}} & \overline{\overline{Z}}^{\vec{J}^{G\rightleftharpoons F}\vec{M}^{O\rightleftharpoons G}} & \overline{\overline{Z}}^{\vec{J}^{G\rightleftharpoons F}\vec{M}^{G\rightleftharpoons A}} & \overline{\overline{Z}}^{\vec{J}^{G\rightleftharpoons F}\vec{M}^{A\rightleftharpoons F}} & \overline{\overline{Z}}^{\vec{J}^{G\rightleftharpoons F}\vec{M}^{G\rightleftharpoons F}} \\
\overline{\overline{Z}}^{\vec{M}^{G\rightleftharpoons F}\vec{J}^{O\rightleftharpoons G}} & \overline{\overline{Z}}^{\vec{M}^{G\rightleftharpoons F}\vec{J}^{G}} & \overline{\overline{Z}}^{\vec{M}^{G\rightleftharpoons F}\vec{J}^{G\rightleftharpoons A}} & 0 & \overline{\overline{Z}}^{\vec{M}^{G\rightleftharpoons F}\vec{J}^{A\rightleftharpoons F}} & \overline{\overline{Z}}^{\vec{M}^{G\rightleftharpoons F}\vec{J}^{F}} & \overline{\overline{Z}}^{\vec{M}^{G\rightleftharpoons F}\vec{J}^{G\rightleftharpoons F}} & \overline{\overline{Z}}^{\vec{M}^{G\rightleftharpoons F}\vec{M}^{O\rightleftharpoons G}} & \overline{\overline{Z}}^{\vec{M}^{G\rightleftharpoons F}\vec{M}^{G\rightleftharpoons A}} & \overline{\overline{Z}}^{\vec{M}^{G\rightleftharpoons F}\vec{M}^{A\rightleftharpoons F}} & \overline{\overline{Z}}^{\vec{M}^{G\rightleftharpoons F}\vec{M}^{G\rightleftharpoons F}}
\end{bmatrix} \tag{8-46}$$

In the above equations, the subscript "FCE" is the acronym for "*field continuation condition*", and the superscripts "DoJ" and "DoM" are the acronyms for "*definition of* $\vec{J}^{G\rightleftharpoons A}$" and "*definition of* $\vec{M}^{G\rightleftharpoons A}$" respectively.

Theoretically, the Eqs. (8-43) and (8-45) are equivalent to each other, and they have the same *solution space*. If the *basic solutions* used to span the solution space are denoted as $\{\overline{s}_1^{\rm BS}, \overline{s}_2^{\rm BS}, \cdots\}$, then any mode contained in the solution space can be expanded as follows:

$$\bar{a}^{\text{AV}} = \sum_i a_i^{\text{BS}} \bar{s}_i^{\text{BS}} = \underbrace{\left[ \bar{s}_1^{\text{BS}}, \bar{s}_2^{\text{BS}}, \cdots \right]}_{\bar{\bar{T}}^{\text{BS}\to\text{AV}}} \cdot \underbrace{\begin{bmatrix} a_1^{\text{BS}} \\ a_2^{\text{BS}} \\ \vdots \end{bmatrix}}_{\bar{a}^{\text{BS}}} \tag{8-47}$$

where the solution space is just the modal space of the TGTA system shown in the previous Fig. 8-1.

For the convenience of the following discussions, Eqs. (8-42) and (8-47) are uniformly written as follows:

$$\bar{a}^{\text{AV}} = \bar{\bar{T}} \cdot \bar{a} \tag{8-48}$$

where $\bar{a} = \bar{a}^{\text{BV}} / \bar{a}^{\text{BS}}$ and correspondingly $\bar{\bar{T}} = \bar{\bar{T}}^{\text{BV}\to\text{AV}} / \bar{\bar{T}}^{\text{BS}\to\text{AV}}$ .

## 8.2.4 Power Transport Theorem and Input Power Operator

In this section, we provide the *power transport theorem* (*PTT*) and *input power operator* (*IPO*) of the TGTA system shown in Fig. 8-1.

### 8.2.4.1 Power Transport Theorem

Applying the results obtained in Chap. 2 to the TGTA system shown in Fig. 8-1, we immediately have the following PTT for the TGTA system

$$P^{\text{O}\rightleftharpoons\text{G}} = P_{\text{dis}}^{\text{G}} + P_{\text{dis}}^{\text{A}} + \overbrace{\underbrace{P_{\text{rad}}^{\infty} + j\, P_{\text{sto}}^{\text{F}}}_{P^{\text{TGTA}\rightleftharpoons\text{F}}}}^{P^{\text{A}\rightleftharpoons\text{F}} + P^{\text{G}\rightleftharpoons\text{F}}} + j\, P_{\text{sto}}^{\text{A}} + j\, P_{\text{sto}}^{\text{G}} \tag{8-49}$$

In the above PTT (8-49), $P^{\text{O}\rightleftharpoons\text{G}}$ is the power inputted into the tra-guide; $P^{\text{A}\rightleftharpoons\text{F}}$ is the power passing through $\mathbb{S}^{\text{A}\rightleftharpoons\text{F}}$; $P^{\text{G}\rightleftharpoons\text{F}}$ is the power passing through $\mathbb{S}^{\text{G}\rightleftharpoons\text{F}}$; $P^{\text{TGTA}\rightleftharpoons\text{F}}$ is the power outputted from the whole TGTA system; $P_{\text{dis}}^{\text{G}}$ and $P_{\text{dis}}^{\text{A}}$ are the powers dissipated in tra-guide and tra-antenna respectively; $P_{\text{rad}}^{\infty}$ is the power radiated by whole TGTA system; $P_{\text{sto}}^{\text{F}}$, $P_{\text{sto}}^{\text{A}}$, and $P_{\text{sto}}^{\text{G}}$ are the powers corresponding to the energies stored in free space, tra-antenna, and tra-guide respectively.

The above-mentioned various powers are as follows:

$$P^{\text{O}\rightleftharpoons\text{G}} = (1/2) \iint_{\mathbb{S}^{\text{O}\rightleftharpoons\text{G}}} \left( \vec{E} \times \vec{H}^{\dagger} \right) \cdot \hat{n}^{\to\text{G}} dS \tag{8-50a}$$

$$P^{\text{A}\rightleftharpoons\text{F}} = (1/2) \iint_{\mathbb{S}^{\text{A}\rightleftharpoons\text{F}}} \left( \vec{E} \times \vec{H}^{\dagger} \right) \cdot \hat{n}^{\to\text{F}} dS \tag{8-50b}$$

$$P^{\text{G}\rightleftharpoons\text{F}} = (1/2) \iint_{\mathbb{S}^{\text{G}\rightleftharpoons\text{F}}} \left( \vec{E} \times \vec{H}^{\dagger} \right) \cdot \hat{n}^{\to\text{F}} dS \tag{8-50c}$$

$$P^{\text{TGTA}\rightleftharpoons\text{F}} = (1/2) \iint_{\mathbb{S}^{\text{G}\rightleftharpoons\text{F}} \cup \mathbb{S}^{\text{A}\rightleftharpoons\text{F}}} \left( \vec{E} \times \vec{H}^{\dagger} \right) \cdot \hat{n}^{\to\text{F}} dS \tag{8-50d}$$

$$
\begin{aligned}
P_{\text{rad}}^{\infty} &= (1/2) \oiint_{\mathbb{S}^{\infty}} \left(\vec{E} \times \vec{H}^{\dagger}\right) \cdot \hat{n}^{\rightarrow\infty} dS && \text{(8-50e)} \\
P_{\text{dis}}^{\text{G}} &= (1/2) \left\langle \vec{\vec{\sigma}} \cdot \vec{E}, \vec{E} \right\rangle_{\mathbb{V}^{\text{G}}} && \text{(8-50f)} \\
P_{\text{dis}}^{\text{A}} &= (1/2) \left\langle \vec{\vec{\sigma}} \cdot \vec{E}, \vec{E} \right\rangle_{\mathbb{V}^{\text{A}}} && \text{(8-50g)} \\
P_{\text{sto}}^{\text{F}} &= 2\omega \left[ (1/4) \left\langle \vec{H}, \mu_0 \vec{H} \right\rangle_{\mathbb{V}^{\text{F}}} - (1/4) \left\langle \varepsilon_0 \vec{E}, \vec{E} \right\rangle_{\mathbb{V}^{\text{F}}} \right] && \text{(8-50h)} \\
P_{\text{sto}}^{\text{A}} &= 2\omega \left[ (1/4) \left\langle \vec{H}, \vec{\vec{\mu}} \cdot \vec{H} \right\rangle_{\mathbb{V}^{\text{A}}} - (1/4) \left\langle \vec{\vec{\varepsilon}} \cdot \vec{E}, \vec{E} \right\rangle_{\mathbb{V}^{\text{A}}} \right] && \text{(8-50i)} \\
P_{\text{sto}}^{\text{G}} &= 2\omega \left[ (1/4) \left\langle \vec{H}, \vec{\vec{\mu}} \cdot \vec{H} \right\rangle_{\mathbb{V}^{\text{G}}} - (1/4) \left\langle \vec{\vec{\varepsilon}} \cdot \vec{E}, \vec{E} \right\rangle_{\mathbb{V}^{\text{G}}} \right] && \text{(8-50j)}
\end{aligned}
$$

where the various unit vectors were defined in Sec. 8.2.1.

#### 8.2.4.2 Input Power Operator — Formulation I: Current Form

Based on the Eqs. (8-3a)&(8-3b) and the tangential continuity of the fields $\{\vec{E}, \vec{H}\}$ on input port $\mathbb{S}^{\text{O}\rightleftharpoons\text{G}}$, the IPO $P^{\text{O}\rightleftharpoons\text{G}}$ given in Eq. (8-50a) can be alternatively written as follows:

$$
P^{\text{O}\rightleftharpoons\text{G}} = (1/2) \left\langle \hat{n}^{\rightarrow\text{G}} \times \vec{J}^{\text{O}\rightleftharpoons\text{G}}, \vec{M}^{\text{O}\rightleftharpoons\text{G}} \right\rangle_{\mathbb{S}^{\text{O}\rightleftharpoons\text{G}}} \tag{8-51}
$$

and it is just the *current form of IPO* (or alternatively called *JM form*).

Inserting Eq. (8-16) into the above current form, the current form is immediately discretized as follows:

$$
P^{\text{O}\rightleftharpoons\text{G}} = \left(\bar{a}^{\text{AV}}\right)^{\dagger} \cdot \underbrace{\begin{bmatrix}
0 & 0 & 0 & 0 & 0 & 0 & 0 & \overline{\overline{C}}^{\vec{J}^{\text{O}\rightleftharpoons\text{G}}\vec{M}^{\text{O}\rightleftharpoons\text{G}}} & 0 & 0 & 0 \\
0 & 0 & 0 & 0 & 0 & 0 & 0 & 0 & 0 & 0 & 0 \\
0 & 0 & 0 & 0 & 0 & 0 & 0 & 0 & 0 & 0 & 0 \\
0 & 0 & 0 & 0 & 0 & 0 & 0 & 0 & 0 & 0 & 0 \\
0 & 0 & 0 & 0 & 0 & 0 & 0 & 0 & 0 & 0 & 0 \\
0 & 0 & 0 & 0 & 0 & 0 & 0 & 0 & 0 & 0 & 0 \\
0 & 0 & 0 & 0 & 0 & 0 & 0 & 0 & 0 & 0 & 0 \\
0 & 0 & 0 & 0 & 0 & 0 & 0 & 0 & 0 & 0 & 0 \\
0 & 0 & 0 & 0 & 0 & 0 & 0 & 0 & 0 & 0 & 0 \\
0 & 0 & 0 & 0 & 0 & 0 & 0 & 0 & 0 & 0 & 0 \\
0 & 0 & 0 & 0 & 0 & 0 & 0 & 0 & 0 & 0 & 0
\end{bmatrix}}_{\overline{\overline{P}}_{\text{curAV}}^{\text{O}\rightleftharpoons\text{G}}} \cdot \bar{a}^{\text{AV}} \tag{8-52}
$$

where the elements of sub-matrix $\overline{\overline{C}}^{\vec{J}^{\text{O}\rightleftharpoons\text{G}}\vec{M}^{\text{O}\rightleftharpoons\text{G}}}$ are calculated as follows:

$$
c_{\xi\zeta}^{\text{O}\rightleftharpoons\text{G}} = (1/2) \left\langle \hat{n}^{\rightarrow\text{G}} \times \vec{b}_{\xi}^{\vec{J}^{\text{O}\rightleftharpoons\text{G}}}, \vec{b}_{\zeta}^{\vec{M}^{\text{O}\rightleftharpoons\text{G}}} \right\rangle_{\mathbb{S}^{\text{O}\rightleftharpoons\text{G}}} \tag{8-53}
$$

To obtain the IPO defined on modal space, we substitute Eq. (8-48) into the above Eq. (8-

52), and then we have that

$$P^{\mathrm{O}\rightleftharpoons\mathrm{G}} = \bar{a}^{\dagger} \cdot \underbrace{\left( \bar{\bar{T}}^{\dagger} \cdot \bar{\bar{P}}_{\mathrm{curAV}}^{\mathrm{O}\rightleftharpoons\mathrm{G}} \cdot \bar{\bar{T}} \right)}_{\bar{\bar{P}}_{\mathrm{cur}}^{\mathrm{O}\rightleftharpoons\mathrm{G}}} \cdot \bar{a} \tag{8-54}$$

where subscript "cur" is to emphasize that $\bar{\bar{P}}_{\mathrm{cur}}^{\mathrm{O}\rightleftharpoons\mathrm{G}}$ originates from discretizing the current form of IPO.

#### 8.2.4.3 Input Power Operator — Formulation II: Field-Current Interaction Form

Using source-field relation (8-2), the IPO $P^{\mathrm{O}\rightleftharpoons\mathrm{G}}$ given in Eq. (8-50a) can be further rewritten as follows:

$$\begin{aligned} P^{\mathrm{O}\rightleftharpoons\mathrm{G}} &= -(1/2)\left\langle \vec{J}^{\mathrm{O}\rightleftharpoons\mathrm{G}}, \mathcal{E}_{\mathrm{G}}\left( \vec{J}^{\mathrm{O}\rightleftharpoons\mathrm{G}} + \vec{J}^{\mathrm{G}} - \vec{J}^{\mathrm{G}\rightleftharpoons\mathrm{A}} + \vec{J}^{\mathrm{G}\rightleftharpoons\mathrm{F}}, \vec{M}^{\mathrm{O}\rightleftharpoons\mathrm{G}} - \vec{M}^{\mathrm{G}\rightleftharpoons\mathrm{A}} + \vec{M}^{\mathrm{G}\rightleftharpoons\mathrm{F}} \right) \right\rangle_{\mathbb{S}^{\mathrm{O}\rightleftharpoons\mathrm{G}}} \\ &= -(1/2)\left\langle \vec{M}^{\mathrm{O}\rightleftharpoons\mathrm{G}}, \mathcal{H}_{\mathrm{G}}\left( \vec{J}^{\mathrm{O}\rightleftharpoons\mathrm{G}} + \vec{J}^{\mathrm{G}} - \vec{J}^{\mathrm{G}\rightleftharpoons\mathrm{A}} + \vec{J}^{\mathrm{G}\rightleftharpoons\mathrm{F}}, \vec{M}^{\mathrm{O}\rightleftharpoons\mathrm{G}} - \vec{M}^{\mathrm{G}\rightleftharpoons\mathrm{A}} + \vec{M}^{\mathrm{G}\rightleftharpoons\mathrm{F}} \right) \right\rangle_{\mathbb{S}^{\mathrm{O}\rightleftharpoons\mathrm{G}}}^{\dagger} \end{aligned} \tag{8-55}$$

and it is just the *field-current interaction forms of IPO* (or alternatively called *JE form* and *HM form*).

Inserting Eq. (8-16) into the above interaction form, the interaction form is immediately discretized as follows:

$$P^{\mathrm{O}\rightleftharpoons\mathrm{G}} = \left( \bar{a}^{\mathrm{AV}} \right)^{\dagger} \cdot \bar{\bar{P}}_{\mathrm{intAV}}^{\mathrm{O}\rightleftharpoons\mathrm{G}} \cdot \bar{a}^{\mathrm{AV}} \tag{8-56}$$

in which

$$\bar{\bar{P}}_{\mathrm{intAV}}^{\mathrm{O}\rightleftharpoons\mathrm{G}} = \begin{bmatrix} \bar{\bar{P}}^{\vec{J}^{\mathrm{O}\rightleftharpoons\mathrm{G}}\vec{J}^{\mathrm{O}\rightleftharpoons\mathrm{G}}} & \bar{\bar{P}}^{\vec{J}^{\mathrm{O}\rightleftharpoons\mathrm{G}}\vec{J}^{\mathrm{G}}} & \bar{\bar{P}}^{\vec{J}^{\mathrm{O}\rightleftharpoons\mathrm{G}}\vec{J}^{\mathrm{G}\rightleftharpoons\mathrm{A}}} & 0 & 0 & 0 & \bar{\bar{P}}^{\vec{J}^{\mathrm{O}\rightleftharpoons\mathrm{G}}\vec{J}^{\mathrm{G}\rightleftharpoons\mathrm{F}}} & \bar{\bar{P}}^{\vec{J}^{\mathrm{O}\rightleftharpoons\mathrm{G}}\vec{M}^{\mathrm{O}\rightleftharpoons\mathrm{G}}} & \bar{\bar{P}}^{\vec{J}^{\mathrm{O}\rightleftharpoons\mathrm{G}}\vec{M}^{\mathrm{G}\rightleftharpoons\mathrm{A}}} & 0 & \bar{\bar{P}}^{\vec{J}^{\mathrm{O}\rightleftharpoons\mathrm{G}}\vec{M}^{\mathrm{G}\rightleftharpoons\mathrm{F}}} \\ 0 & 0 & 0 & 0 & 0 & 0 & 0 & 0 & 0 & 0 & 0 \\ 0 & 0 & 0 & 0 & 0 & 0 & 0 & 0 & 0 & 0 & 0 \\ 0 & 0 & 0 & 0 & 0 & 0 & 0 & 0 & 0 & 0 & 0 \\ 0 & 0 & 0 & 0 & 0 & 0 & 0 & 0 & 0 & 0 & 0 \\ 0 & 0 & 0 & 0 & 0 & 0 & 0 & 0 & 0 & 0 & 0 \\ 0 & 0 & 0 & 0 & 0 & 0 & 0 & 0 & 0 & 0 & 0 \\ 0 & 0 & 0 & 0 & 0 & 0 & 0 & 0 & 0 & 0 & 0 \\ 0 & 0 & 0 & 0 & 0 & 0 & 0 & 0 & 0 & 0 & 0 \\ 0 & 0 & 0 & 0 & 0 & 0 & 0 & 0 & 0 & 0 & 0 \\ 0 & 0 & 0 & 0 & 0 & 0 & 0 & 0 & 0 & 0 & 0 \end{bmatrix} \tag{8-57a}$$

for the *JE interaction version*, or

$$\overline{\overline{P}}_{\text{intAV}}^{\text{O}\rightleftharpoons\text{G}}=\begin{bmatrix} 0 & 0 & 0 & 0\;0\;0 & 0 & 0 & 0 & 0 & 0 \\ 0 & 0 & 0 & 0\;0\;0 & 0 & 0 & 0 & 0 & 0 \\ 0 & 0 & 0 & 0\;0\;0 & 0 & 0 & 0 & 0 & 0 \\ 0 & 0 & 0 & 0\;0\;0 & 0 & 0 & 0 & 0 & 0 \\ 0 & 0 & 0 & 0\;0\;0 & 0 & 0 & 0 & 0 & 0 \\ 0 & 0 & 0 & 0\;0\;0 & 0 & 0 & 0 & 0 & 0 \\ 0 & 0 & 0 & 0\;0\;0 & 0 & 0 & 0 & 0 & 0 \\ \overline{\overline{P}}^{\vec{M}^{\text{O}\rightleftharpoons\text{G}}\vec{J}^{\text{O}\rightleftharpoons\text{G}}} & \overline{\overline{P}}^{\vec{M}^{\text{O}\rightleftharpoons\text{G}}\vec{J}^{\text{G}}} & \overline{\overline{P}}^{\vec{M}^{\text{O}\rightleftharpoons\text{G}}\vec{J}^{\text{G}\rightleftharpoons\text{A}}} & 0\;0\;0 & \overline{\overline{P}}^{\vec{M}^{\text{O}\rightleftharpoons\text{G}}\vec{J}^{\text{G}\rightleftharpoons\text{F}}} & \overline{\overline{P}}^{\vec{M}^{\text{O}\rightleftharpoons\text{G}}\vec{M}^{\text{O}\rightleftharpoons\text{G}}} & \overline{\overline{P}}^{\vec{M}^{\text{O}\rightleftharpoons\text{G}}\vec{M}^{\text{G}\rightleftharpoons\text{A}}} & 0 & \overline{\overline{P}}^{\vec{M}^{\text{O}\rightleftharpoons\text{G}}\vec{M}^{\text{G}\rightleftharpoons\text{F}}} \\ 0 & 0 & 0 & 0\;0\;0 & 0 & 0 & 0 & 0 & 0 \\ 0 & 0 & 0 & 0\;0\;0 & 0 & 0 & 0 & 0 & 0 \\ 0 & 0 & 0 & 0\;0\;0 & 0 & 0 & 0 & 0 & 0 \end{bmatrix}^{\dagger} \tag{8-57b}$$

for the *HM interaction version*. The elements of the sub-matrices are calculated as follows:

$$p_{\xi\zeta}^{\vec{J}^{\text{O}\rightleftharpoons\text{G}}\vec{J}^{\text{O}\rightleftharpoons\text{G}}} = -(1/2)\left\langle \vec{b}_{\xi}^{\vec{J}^{\text{O}\rightleftharpoons\text{G}}}, \mathcal{E}_{\text{G}}\left(\vec{b}_{\zeta}^{\vec{J}^{\text{O}\rightleftharpoons\text{G}}}\right)\right\rangle_{\mathbb{S}^{\text{O}\rightleftharpoons\text{G}}} \tag{8-58a}$$

$$p_{\xi\zeta}^{\vec{J}^{\text{O}\rightleftharpoons\text{G}}\vec{J}^{\text{G}}} = -(1/2)\left\langle \vec{b}_{\xi}^{\vec{J}^{\text{O}\rightleftharpoons\text{G}}}, \mathcal{E}_{\text{G}}\left(\vec{b}_{\zeta}^{\vec{J}^{\text{G}}}\right)\right\rangle_{\mathbb{S}^{\text{O}\rightleftharpoons\text{G}}} \tag{8-58b}$$

$$p_{\xi\zeta}^{\vec{J}^{\text{O}\rightleftharpoons\text{G}}\vec{J}^{\text{G}\rightleftharpoons\text{A}}} = -(1/2)\left\langle \vec{b}_{\xi}^{\vec{J}^{\text{O}\rightleftharpoons\text{G}}}, \mathcal{E}_{\text{G}}\left(-\vec{b}_{\zeta}^{\vec{J}^{\text{G}\rightleftharpoons\text{A}}}\right)\right\rangle_{\mathbb{S}^{\text{O}\rightleftharpoons\text{G}}} \tag{8-58c}$$

$$p_{\xi\zeta}^{\vec{J}^{\text{O}\rightleftharpoons\text{G}}\vec{J}^{\text{G}\rightleftharpoons\text{F}}} = -(1/2)\left\langle \vec{b}_{\xi}^{\vec{J}^{\text{O}\rightleftharpoons\text{G}}}, \mathcal{E}_{\text{G}}\left(\vec{b}_{\zeta}^{\vec{J}^{\text{G}\rightleftharpoons\text{F}}}\right)\right\rangle_{\mathbb{S}^{\text{O}\rightleftharpoons\text{G}}} \tag{8-58d}$$

$$p_{\xi\zeta}^{\vec{J}^{\text{O}\rightleftharpoons\text{G}}\vec{M}^{\text{O}\rightleftharpoons\text{G}}} = -(1/2)\left\langle \vec{b}_{\xi}^{\vec{J}^{\text{O}\rightleftharpoons\text{G}}}, \mathcal{E}_{\text{G}}\left(\vec{b}_{\zeta}^{\vec{M}^{\text{O}\rightleftharpoons\text{G}}}\right)\right\rangle_{\mathbb{S}^{\text{O}\rightleftharpoons\text{G}}} \tag{8-58e}$$

$$p_{\xi\zeta}^{\vec{J}^{\text{O}\rightleftharpoons\text{G}}\vec{M}^{\text{G}\rightleftharpoons\text{A}}} = -(1/2)\left\langle \vec{b}_{\xi}^{\vec{J}^{\text{O}\rightleftharpoons\text{G}}}, \mathcal{E}_{\text{G}}\left(-\vec{b}_{\zeta}^{\vec{M}^{\text{G}\rightleftharpoons\text{A}}}\right)\right\rangle_{\mathbb{S}^{\text{O}\rightleftharpoons\text{G}}} \tag{8-58f}$$

$$p_{\xi\zeta}^{\vec{J}^{\text{O}\rightleftharpoons\text{G}}\vec{M}^{\text{G}\rightleftharpoons\text{F}}} = -(1/2)\left\langle \vec{b}_{\xi}^{\vec{J}^{\text{O}\rightleftharpoons\text{G}}}, \mathcal{E}_{\text{G}}\left(\vec{b}_{\zeta}^{\vec{M}^{\text{G}\rightleftharpoons\text{F}}}\right)\right\rangle_{\mathbb{S}^{\text{O}\rightleftharpoons\text{G}}} \tag{8-58g}$$

and

$$p_{\xi\zeta}^{\vec{M}^{\text{O}\rightleftharpoons\text{G}}\vec{J}^{\text{O}\rightleftharpoons\text{G}}} = -(1/2)\left\langle \vec{b}_{\xi}^{\vec{M}^{\text{O}\rightleftharpoons\text{G}}}, \mathcal{H}_{\text{G}}\left(\vec{b}_{\zeta}^{\vec{J}^{\text{O}\rightleftharpoons\text{G}}}\right)\right\rangle_{\mathbb{S}^{\text{O}\rightleftharpoons\text{G}}} \tag{8-59a}$$

$$p_{\xi\zeta}^{\vec{M}^{\text{O}\rightleftharpoons\text{G}}\vec{J}^{\text{G}}} = -(1/2)\left\langle \vec{b}_{\xi}^{\vec{M}^{\text{O}\rightleftharpoons\text{G}}}, \mathcal{H}_{\text{G}}\left(\vec{b}_{\zeta}^{\vec{J}^{\text{G}}}\right)\right\rangle_{\mathbb{S}^{\text{O}\rightleftharpoons\text{G}}} \tag{8-59b}$$

$$p_{\xi\zeta}^{\vec{M}^{\text{O}\rightleftharpoons\text{G}}\vec{J}^{\text{G}\rightleftharpoons\text{A}}} = -(1/2)\left\langle \vec{b}_{\xi}^{\vec{M}^{\text{O}\rightleftharpoons\text{G}}}, \mathcal{H}_{\text{G}}\left(-\vec{b}_{\zeta}^{\vec{J}^{\text{G}\rightleftharpoons\text{A}}}\right)\right\rangle_{\mathbb{S}^{\text{O}\rightleftharpoons\text{G}}} \tag{8-59c}$$

$$p_{\xi\zeta}^{\vec{M}^{\text{O}\rightleftharpoons\text{G}}\vec{J}^{\text{G}\rightleftharpoons\text{F}}} = -(1/2)\left\langle \vec{b}_{\xi}^{\vec{M}^{\text{O}\rightleftharpoons\text{G}}}, \mathcal{H}_{\text{G}}\left(\vec{b}_{\zeta}^{\vec{J}^{\text{G}\rightleftharpoons\text{F}}}\right)\right\rangle_{\mathbb{S}^{\text{O}\rightleftharpoons\text{G}}} \tag{8-59d}$$

$$p_{\xi\zeta}^{\vec{M}^{\text{O}\rightleftharpoons\text{G}}\vec{M}^{\text{O}\rightleftharpoons\text{G}}} = -(1/2)\left\langle \vec{b}_{\xi}^{\vec{M}^{\text{O}\rightleftharpoons\text{G}}}, \mathcal{H}_{\text{G}}\left(\vec{b}_{\zeta}^{\vec{M}^{\text{O}\rightleftharpoons\text{G}}}\right)\right\rangle_{\mathbb{S}^{\text{O}\rightleftharpoons\text{G}}} \tag{8-59e}$$

$$p_{\xi\zeta}^{\vec{M}^{\text{O}\rightleftharpoons\text{G}}\vec{M}^{\text{G}\rightleftharpoons\text{A}}} = -(1/2)\left\langle \vec{b}_{\xi}^{\vec{M}^{\text{O}\rightleftharpoons\text{G}}}, \mathcal{H}_{\text{G}}\left(-\vec{b}_{\zeta}^{\vec{M}^{\text{G}\rightleftharpoons\text{A}}}\right)\right\rangle_{\mathbb{S}^{\text{O}\rightleftharpoons\text{G}}} \tag{8-59f}$$

$$p_{\xi\zeta}^{\vec{M}^{\text{O}\rightleftharpoons\text{G}}\vec{M}^{\text{G}\rightleftharpoons\text{F}}} = -(1/2)\left\langle \vec{b}_{\xi}^{\vec{M}^{\text{O}\rightleftharpoons\text{G}}}, \mathcal{H}_{\text{G}}\left(\vec{b}_{\zeta}^{\vec{M}^{\text{G}\rightleftharpoons\text{F}}}\right)\right\rangle_{\mathbb{S}^{\text{O}\rightleftharpoons\text{G}}} \tag{8-59g}$$

where the integral surface $\underset{\sim}{\mathbb{S}}^{\text{O}\rightleftharpoons\text{G}}$ is shown in Fig. 8-4.

To obtain the IPO defined on modal space, we substitute Eq. (8-48) into the above Eq. (8-56), and then we have that

$$P^{\text{O}\rightleftharpoons\text{G}} = \bar{a}^{\dagger} \cdot \underbrace{\left(\bar{\bar{T}}^{\dagger} \cdot \bar{\bar{P}}_{\text{intAV}}^{\text{O}\rightleftharpoons\text{G}} \cdot \bar{\bar{T}}\right)}_{\bar{\bar{P}}_{\text{int}}^{\text{O}\rightleftharpoons\text{G}}} \cdot \bar{a} \tag{8-60}$$

where subscript "int" is to emphasize that $\bar{\bar{P}}_{\text{int}}^{\text{O}\rightleftharpoons\text{G}}$ originates from discretizing the interaction form of IPO.

For the convenience of the following discussions, the Eqs. (8-54) and (8-60) are uniformly written as follows:

$$P^{\text{O}\rightleftharpoons\text{G}} = \bar{a}^{\dagger} \cdot \bar{\bar{P}}^{\text{O}\rightleftharpoons\text{G}} \cdot \bar{a} \tag{8-61}$$

where $\bar{\bar{P}}^{\text{O}\rightleftharpoons\text{G}} = \bar{\bar{P}}_{\text{cur}}^{\text{O}\rightleftharpoons\text{G}} / \bar{\bar{P}}_{\text{int}}^{\text{O}\rightleftharpoons\text{G}}$.

## 8.2.5 Input-Power-Decoupled Modes

Below, we construct the *input-power-decoupled modes* (*IP-DMs*) of the TGTA system shown in Fig. 8-1, by using the results obtained above.

### 8.2.5.1 Construction Method

The IP-DMs in the modal space can be derived from solving the following *modal decoupling equation* (or simply called *decoupling equation*)

$$\bar{\bar{P}}_{-}^{\text{O}\rightleftharpoons\text{G}} \cdot \bar{\alpha}_{\xi} = \theta_{\xi}\, \bar{\bar{P}}_{+}^{\text{O}\rightleftharpoons\text{G}} \cdot \bar{\alpha}_{\xi} \tag{8-62}$$

defined on modal space, where $\bar{\bar{P}}_{+}^{\text{O}\rightleftharpoons\text{G}}$ and $\bar{\bar{P}}_{-}^{\text{O}\rightleftharpoons\text{G}}$ are the *positive and negative Hermitian parts* obtained from the *Toeplitz's decomposition* for the $\bar{\bar{P}}^{\text{O}\rightleftharpoons\text{G}}$ given in Eq. (8-61). The reason to use symbol "$\theta_{\xi}$" instead of the traditional symbol "$\lambda_{\xi}$" will be given in the subsequent Chap. 9.

If some derived modes $\{\bar{\alpha}_1, \bar{\alpha}_2, \cdots, \bar{\alpha}_d\}$ are *d*-order degenerate, then the following *Gram-Schmidt orthogonalization* process is necessary.

$$\left.\begin{aligned} \bar{\alpha}_1 &= \bar{\alpha}_1{}' \\ \bar{\alpha}_2 - \chi_{12}\bar{\alpha}_1{}' &= \bar{\alpha}_2{}' \\ &\cdots \\ \bar{\alpha}_d - \cdots - \chi_{2d}\bar{\alpha}_2{}' - \chi_{1d}\bar{\alpha}_1{}' &= \bar{\alpha}_d{}' \end{aligned}\right\} \tag{8-63}$$

where the coefficients are calculated as that $\chi_{mn} = \left(\bar{\alpha}_m{}'\right)^{\dagger} \cdot \bar{\bar{P}}_{+}^{\text{O}\rightleftharpoons\text{G}} \cdot \bar{\alpha}_n \Big/ \left(\bar{\alpha}_m{}'\right)^{\dagger} \cdot \bar{\bar{P}}_{+}^{\text{O}\rightleftharpoons\text{G}} \cdot \bar{\alpha}_m{}'$.

The above-obtained new modes $\{\bar{\alpha}_1{}', \bar{\alpha}_2{}', \cdots, \bar{\alpha}_d{}'\}$ are input-power-decoupled with each other.

### 8.2.5.2 Modal Decoupling Relation and Parseval's Identity

The modal vectors constructed in the above subsection satisfy the following *modal decoupling relation*

$$\begin{aligned} & \bar{\alpha}_\xi^\dagger \cdot \overline{\overline{P}}^{\mathrm{O} \rightleftharpoons \mathrm{G}} \cdot \bar{\alpha}_\zeta \\ = & \underbrace{\left[ \mathrm{Re}\left\{ P_\xi^{\mathrm{O} \rightleftharpoons \mathrm{G}} \right\} + j \, \mathrm{Im}\left\{ P_\xi^{\mathrm{O} \rightleftharpoons \mathrm{G}} \right\} \right]}_{P_\xi^{\mathrm{O} \rightleftharpoons \mathrm{G}}} \delta_{\xi\zeta} \xrightarrow{\text{Normalizing } \mathrm{Re}\{P_\xi^{\mathrm{O} \rightleftharpoons \mathrm{G}}\} \text{ to } 1} \underbrace{\left( 1 + j \, \theta_\xi \right)}_{\text{Normalized } P_\xi^{\mathrm{O} \rightleftharpoons \mathrm{G}}} \delta_{\xi\zeta} \end{aligned} \tag{8-64}$$

where $P_\xi^{\mathrm{O} \rightleftharpoons \mathrm{G}}$ is *modal input power* corresponding to the $\xi$-th IP-DM. The physical explanation why $\mathrm{Re}\{P_\xi^{\mathrm{O} \rightleftharpoons \mathrm{G}}\}$ is normalized to 1 has been given in Ref. [18]. The above matrix-vector multiplication decoupling relation can be alternatively written as the following more physical form

$$(1/2) \iint_{\mathbb{S}^{\mathrm{O} \rightleftharpoons \mathrm{G}}} \left( \vec{E}_\zeta \times \vec{H}_\xi^\dagger \right) \cdot \hat{n}^{\rightarrow \mathrm{G}} dS = \left( 1 + j \, \theta_\xi \right) \delta_{\xi\zeta} \tag{8-65}$$

and then

$$(1/T) \int_{t_0}^{t_0 + T} \left[ \iint_{\mathbb{S}^{\mathrm{O} \rightleftharpoons \mathrm{G}}} \left( \vec{\mathcal{E}}_\zeta \times \vec{\mathcal{H}}_\xi \right) \cdot \hat{n}^{\rightarrow \mathrm{G}} dS \right] dt = \delta_{\xi\zeta} \tag{8-66}$$

and Eq. (8-66) has a very clear physical meaning — **the modes obtained above are energy-decoupled in any integral period**.

By employing the above decoupling relation, we have the following *Parseval's identity*

$$\sum_\xi \left| c_\xi \right|^2 = (1/T) \int_{t_0}^{t_0 + T} \left[ \iint_{\mathbb{S}^{\mathrm{O} \rightleftharpoons \mathrm{G}}} \left( \vec{\mathcal{E}} \times \vec{\mathcal{H}} \right) \cdot \hat{n}^{\rightarrow \mathrm{G}} dS \right] dt \tag{8-67}$$

where $c_\xi$ is the *modal expansion coefficient* used in modal expansion formulation and can be explicitly calculated as follows:

$$c_\xi = \frac{-(1/2) \left\langle \vec{J}_\xi^{\mathrm{O} \rightleftharpoons \mathrm{G}}, \vec{E} \right\rangle_{\mathbb{S}^{\mathrm{O} \rightleftharpoons \mathrm{G}}}}{1 + j \, \theta_\xi} = \frac{-(1/2) \left\langle \vec{H}, \vec{M}_\xi^{\mathrm{O} \rightleftharpoons \mathrm{G}} \right\rangle_{\mathbb{S}^{\mathrm{O} \rightleftharpoons \mathrm{G}}}}{1 + j \, \theta_\xi} \tag{8-68}$$

where $\{\vec{E}, \vec{H}\}$ are some previously known fields distributing on input port $\mathbb{S}^{\mathrm{O} \rightleftharpoons \mathrm{G}}$.

### 8.2.5.3 Modal Quantities

For quantitatively describing the modal features, the following modal quantities are usually used

$$\text{MS}_{\xi} = \frac{1}{\left|1 + j\,\theta_{\xi}\right|} \tag{8-69}$$

called *modal significance* (*MS*), and

$$Z_{\xi}^{\text{O}\rightleftharpoons\text{G}} = \frac{P_{\xi}^{\text{O}\rightleftharpoons\text{G}}}{(1/2)\left\langle \vec{J}_{\xi}^{\text{O}\rightleftharpoons\text{G}}, \vec{J}_{\xi}^{\text{O}\rightleftharpoons\text{G}} \right\rangle_{\mathbb{S}^{\text{O}\rightleftharpoons\text{G}}}} = \underbrace{\text{Re}\left\{Z_{\xi}^{\text{O}\rightleftharpoons\text{G}}\right\}}_{R_{\xi}^{\text{O}\rightleftharpoons\text{G}}} + j\,\underbrace{\text{Im}\left\{Z_{\xi}^{\text{O}\rightleftharpoons\text{G}}\right\}}_{X_{\xi}^{\text{O}\rightleftharpoons\text{G}}} \tag{8-70a}$$

called *modal input impedance* (*MII*), and

$$Y_{\xi}^{\text{O}\rightleftharpoons\text{G}} = \frac{P_{\xi}^{\text{O}\rightleftharpoons\text{G}}}{(1/2)\left\langle \vec{M}_{\xi}^{\text{O}\rightleftharpoons\text{G}}, \vec{M}_{\xi}^{\text{O}\rightleftharpoons\text{G}} \right\rangle_{\mathbb{S}^{\text{O}\rightleftharpoons\text{G}}}} = \underbrace{\text{Re}\left\{Y_{\xi}^{\text{O}\rightleftharpoons\text{G}}\right\}}_{G_{\xi}^{\text{O}\rightleftharpoons\text{G}}} + j\,\underbrace{\text{Im}\left\{Y_{\xi}^{\text{O}\rightleftharpoons\text{G}}\right\}}_{B_{\xi}^{\text{O}\rightleftharpoons\text{G}}} \tag{8-70b}$$

called *modal input admittance* (*MIA*).

The above MS quantitatively depict the modal weight in whole modal expansion formulation. The above MII and MIA quantitatively depict the allocation way for the energy carried by the mode.

### 8.2.6 Numerical Examples Corresponding to Typical Structures

To verify the theory and formulations established above, we provide several typical numerical examples as below.

#### 8.2.6.1 Waveguide-fed Single Metallic Horn

In this subsection, we consider the TGTA system shown in the following Fig. 8-5, which is constituted by a *curved metallic tra-guide* and a *metallic horn tra-antenna*. The tra-antenna is fed by the tra-guide.

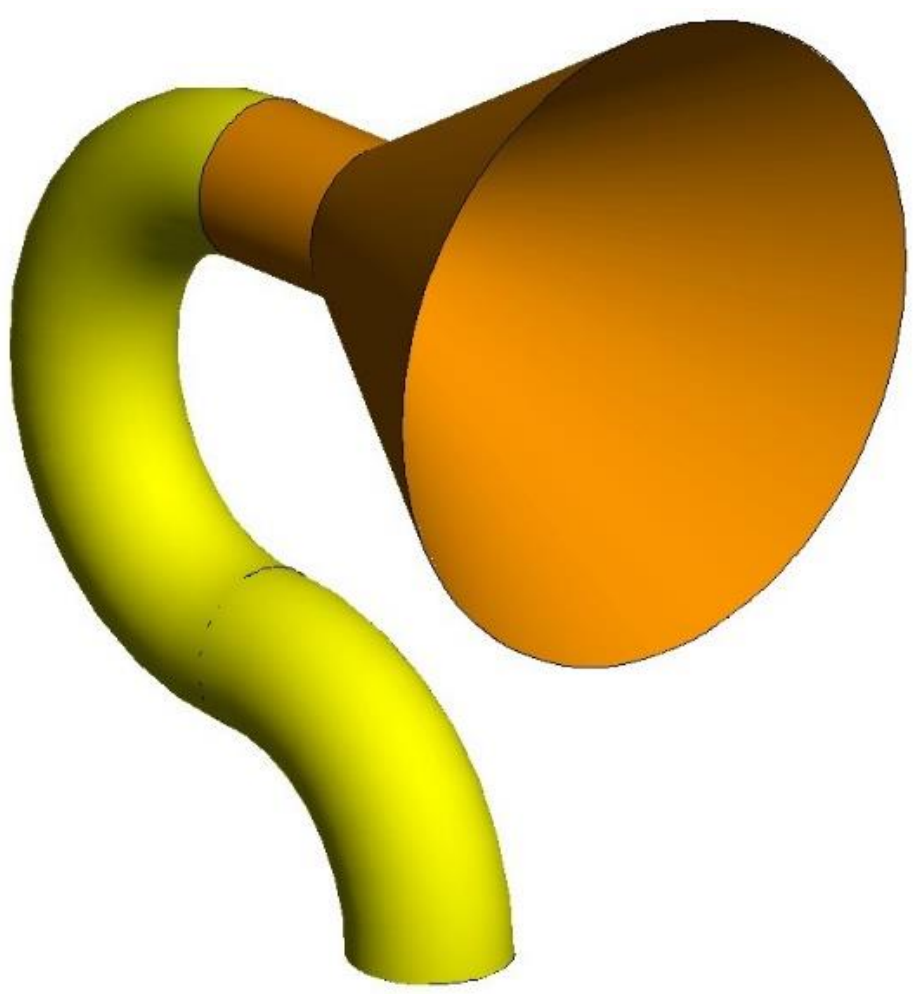

Figure 8-5 Geometry of a TGTA system constituted by a curved metallic tra-guide and a metallic horn tra-antenna

The geometrical size of the TGTA system is shown in the following Fig. 8-6.

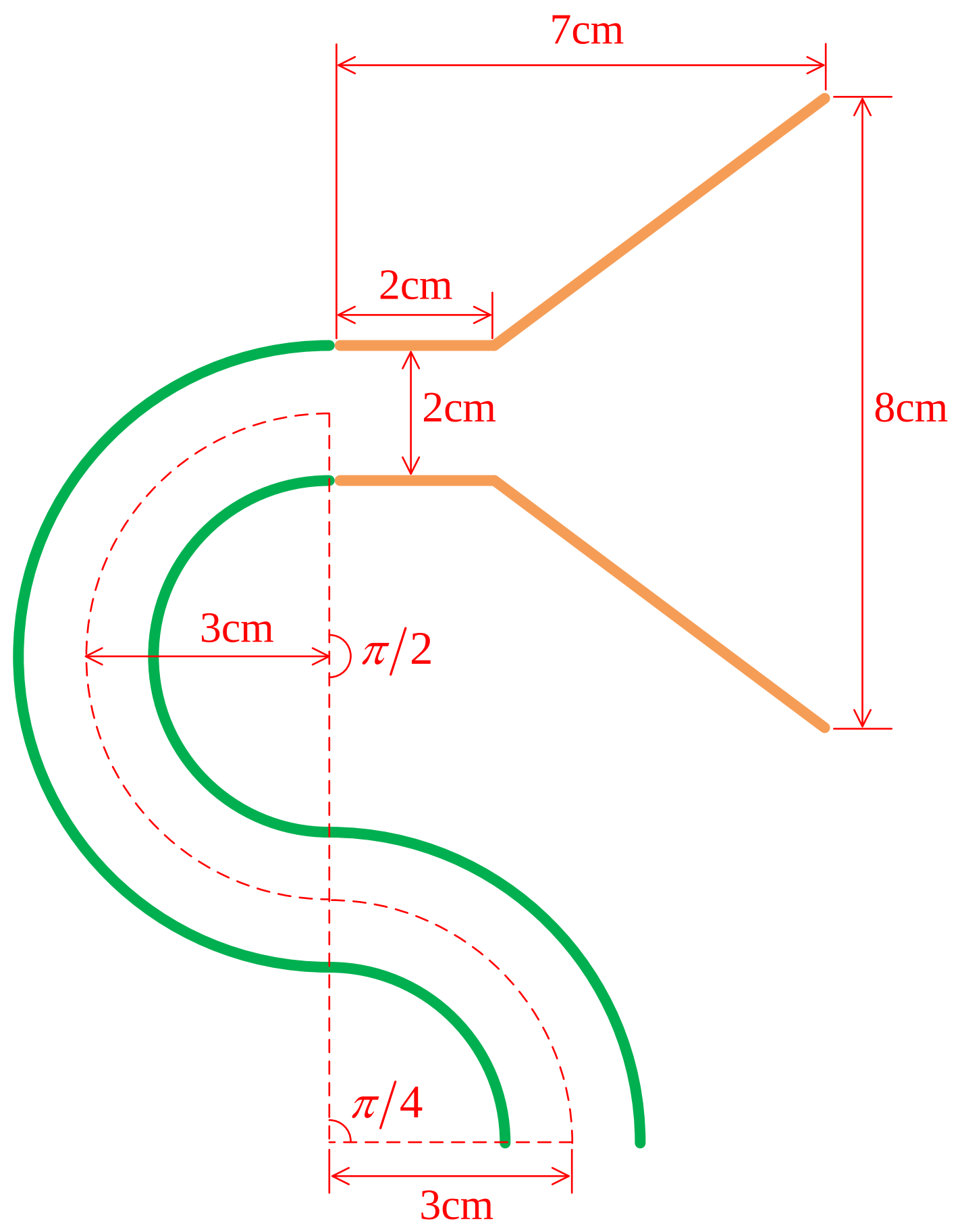

Figure 8-6 Geometrical size of the TGTA system shown in Fig. 8-5

The topological structures and surface triangular meshes of the TGTA system are shown in the following Fig. 8-7.

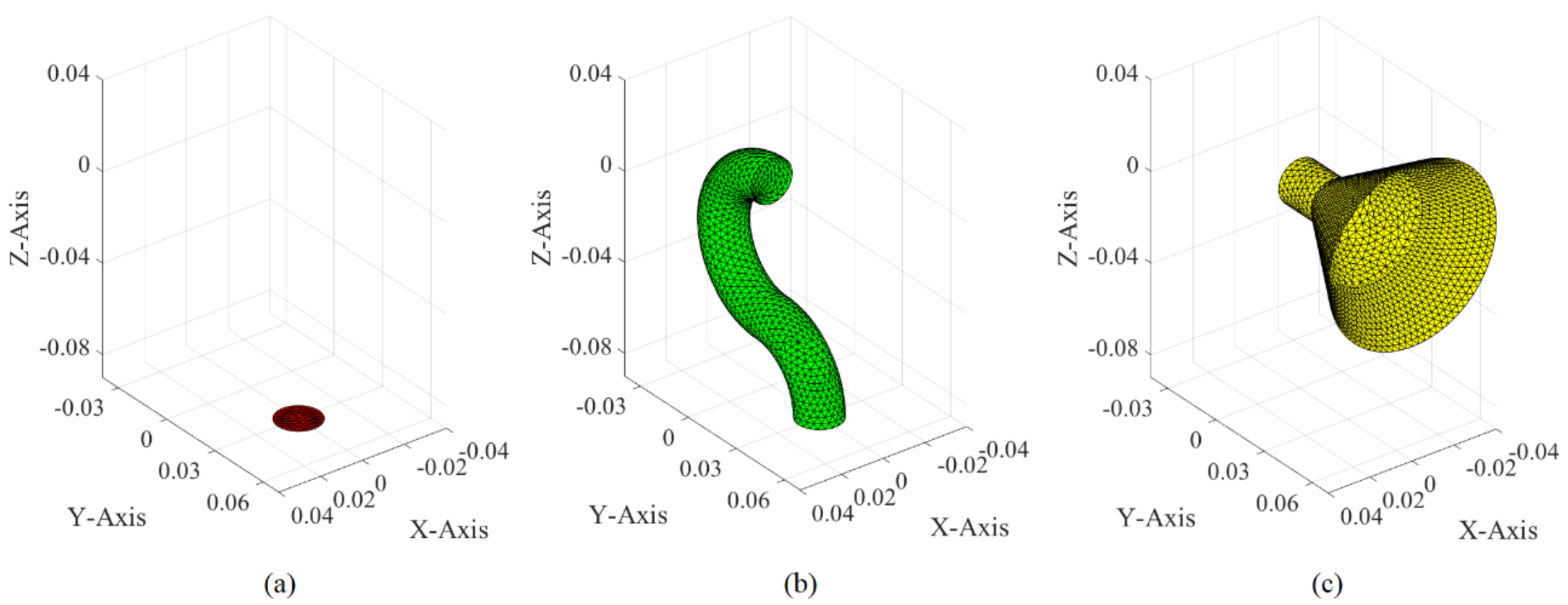

Figure 8-7 Topological structures and surface triangular meshes of the TGTA system shown in Fig. 8-5. (a) Input port; (b) curved metallic guide; (c) metallic horn

Using the JE-DoJ-based and HM-DoM-based formulations of IPO, we calculate the IP-DMs of the TGTA system, and we plot the corresponding modal input resistance curves in the following Fig. 8-8.

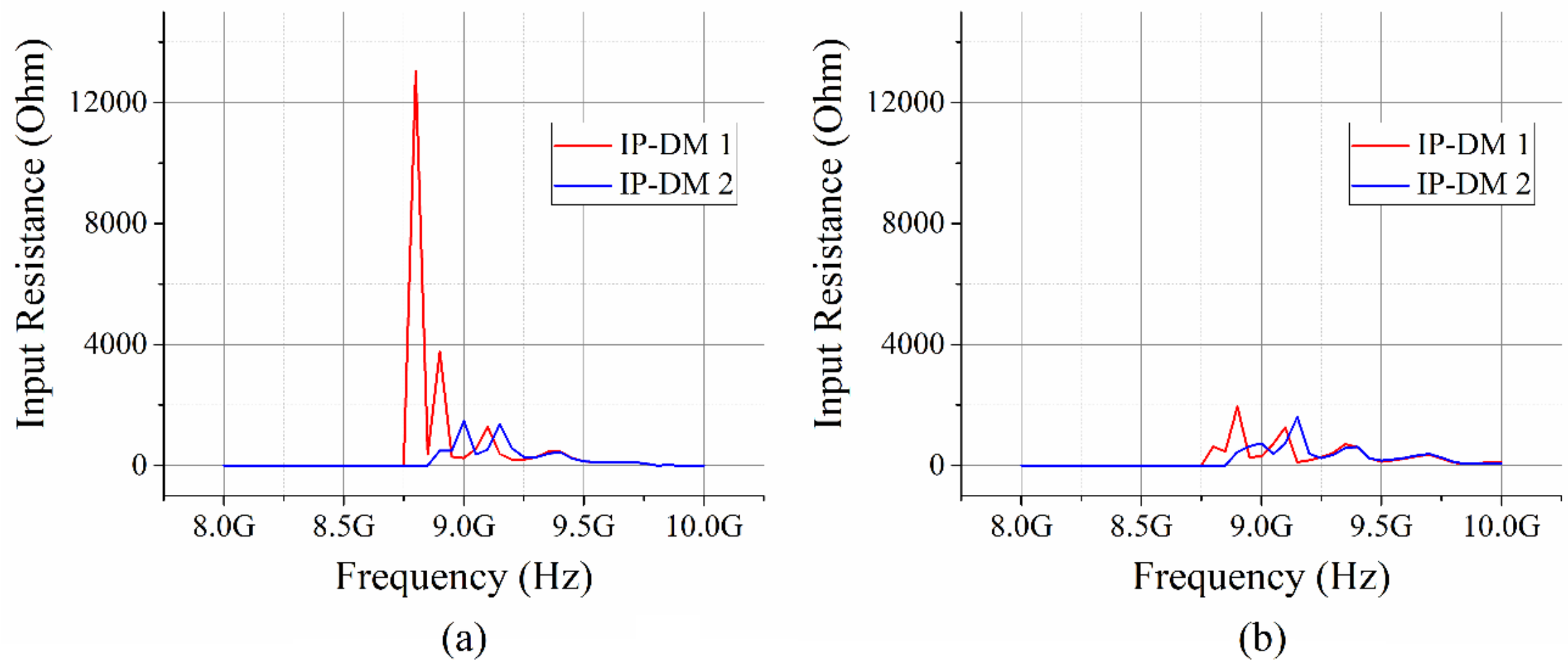

Figure 8-8 Modal input resistance curves of IP-DMs. (a) JE-DoJ-based results; (b) HM-DoM-based results

Taking the JE-DoJ-based IP-DM 1 shown in Fig. 8-8(a) as an example, its equivalent surface electric and magnetic currents distributing on input port are shown in Fig. 8-9.

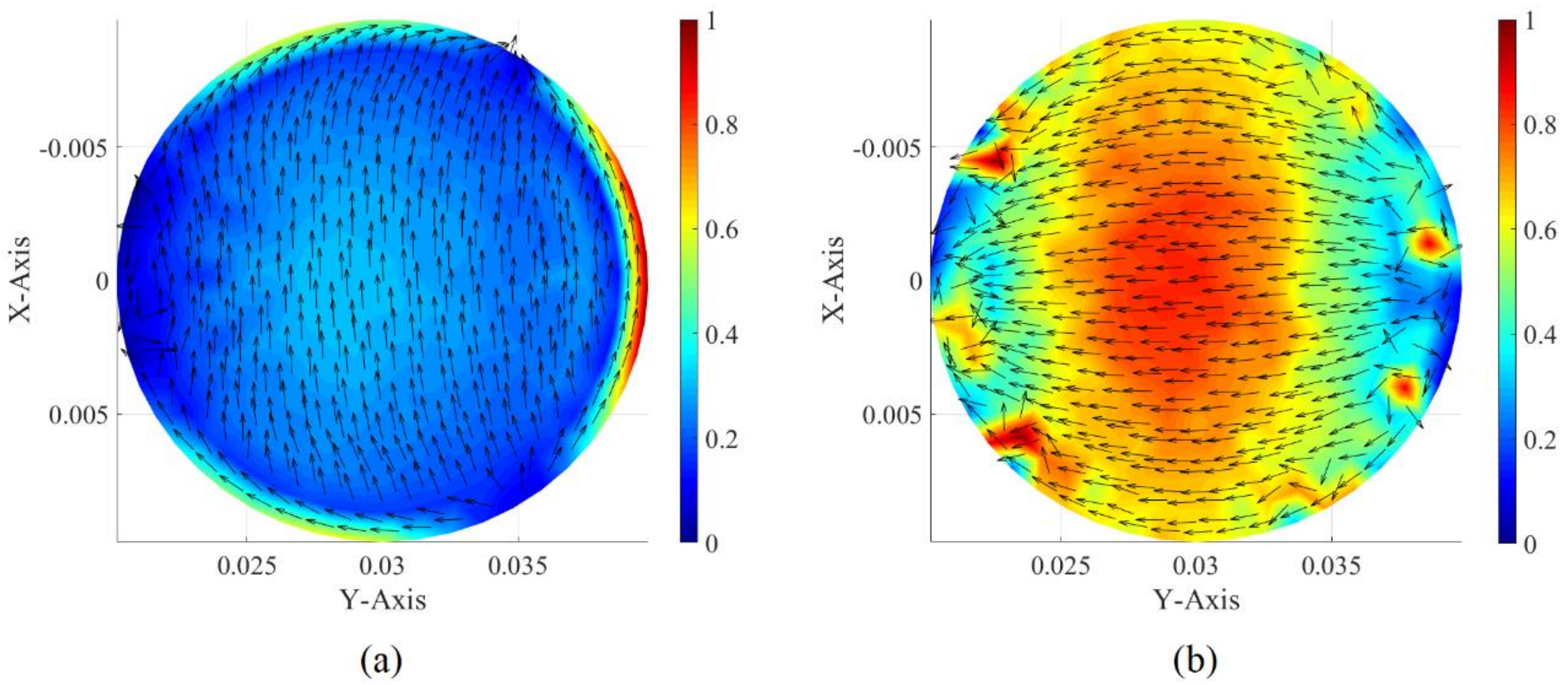

Figure 8-9 Modal equivalent (a) electric and (b) magnetic currents of the JE-DoJ-based IP-DM 1 shown in Fig. 8-8

From the Fig. 8-8(a), it is easy to find out that the IP-DM 1 curve reaches the local peaks at 8.8 GHz, 8.9 GHz, 9.1 GHz, and 9.4 GHz. The modal induced electric currents (on the metallic electric wall) corresponding to the four frequencies are shown in Fig. 8-10.

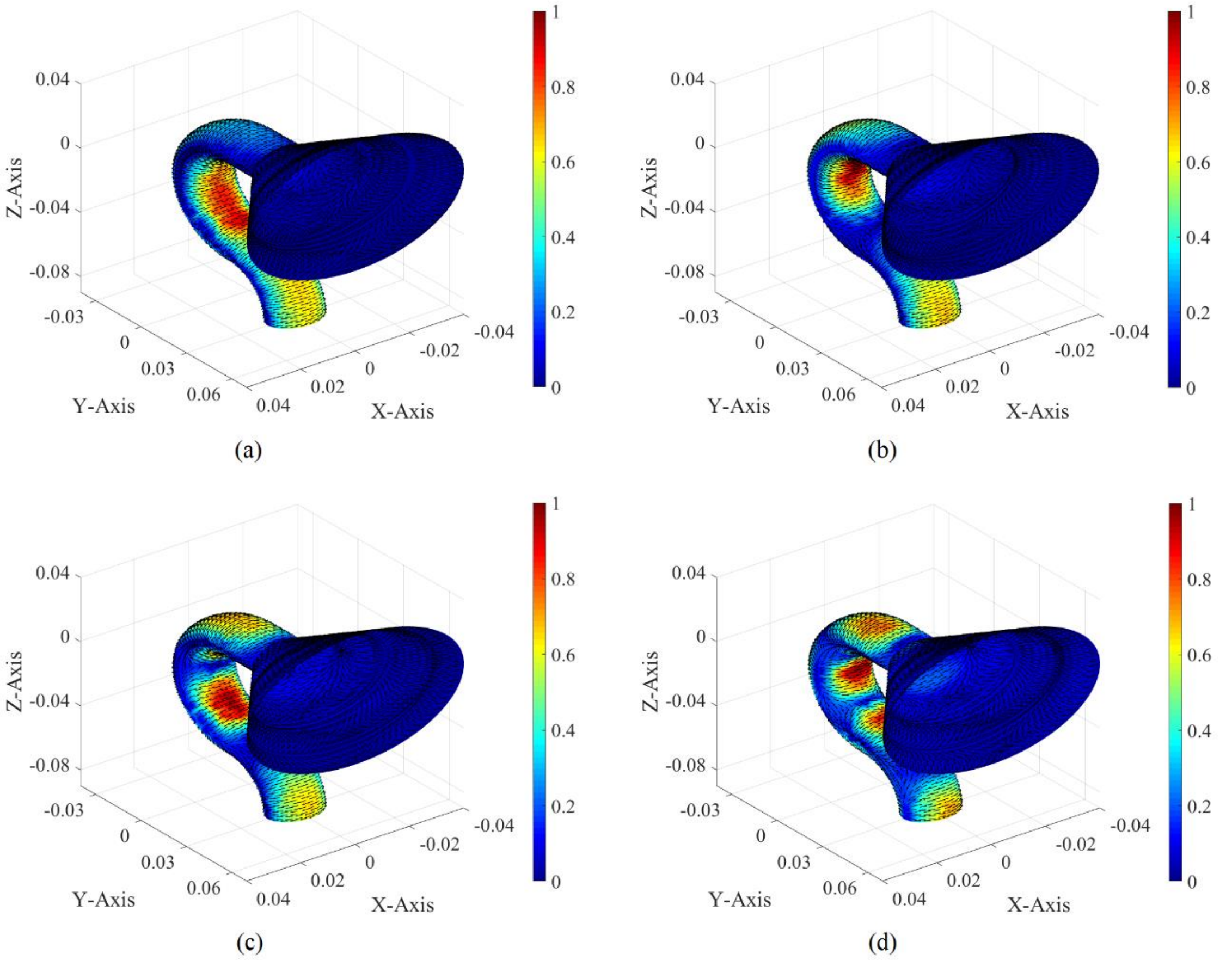

Figure 8-10 Modal induced electric current (on metallic electric wall) of the JE-DoJ-based IP-DM 1 shown in Fig. 8-8. (a) 8.8 GHz; (b) 8.9 GHz; (c) 9.1 GHz; (d) 9.4 GHz

In addition, we also plot the magnitude distributions of the modal electric field (on yOz plane) of the JE-DoJ-based IP-DM 1 (shown in the previous Fig. 8-8(a)) working at the four peak frequencies 8.8 GHz, 8.9 GHz, 9.1 GHz, and 9.4 GHz in the following Fig. 8-11.

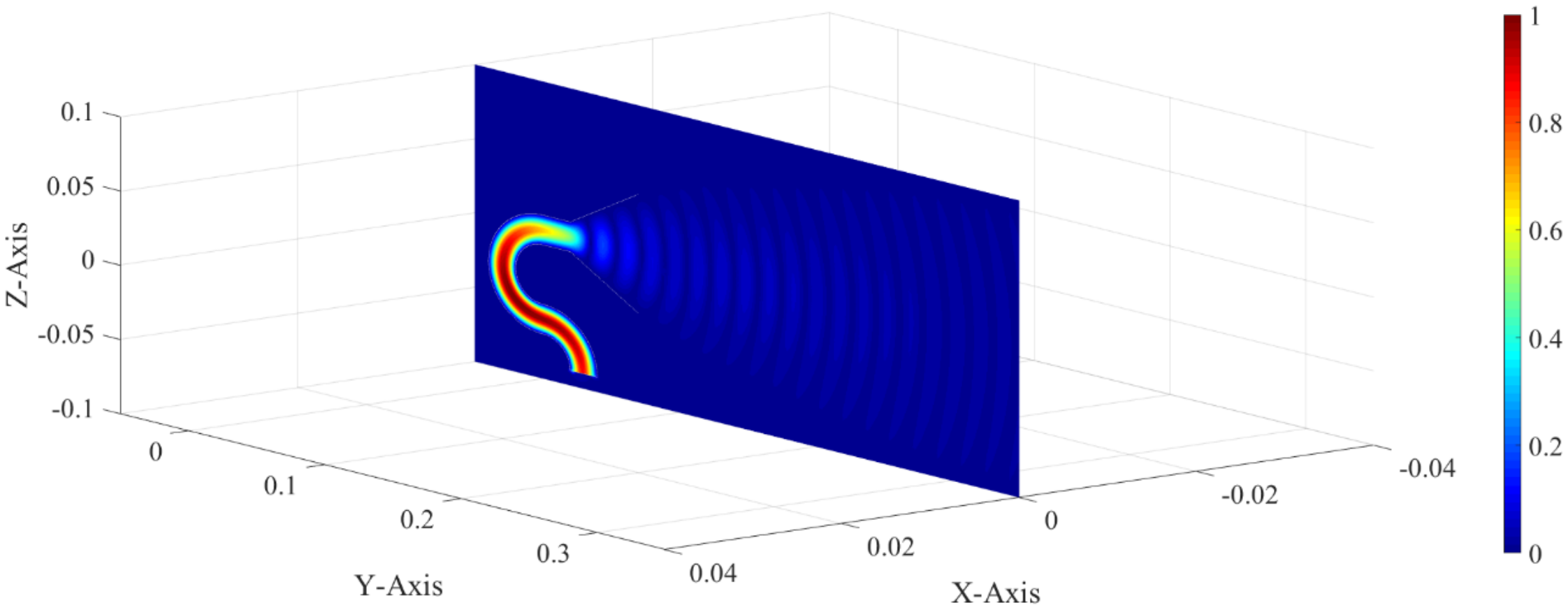

(a)

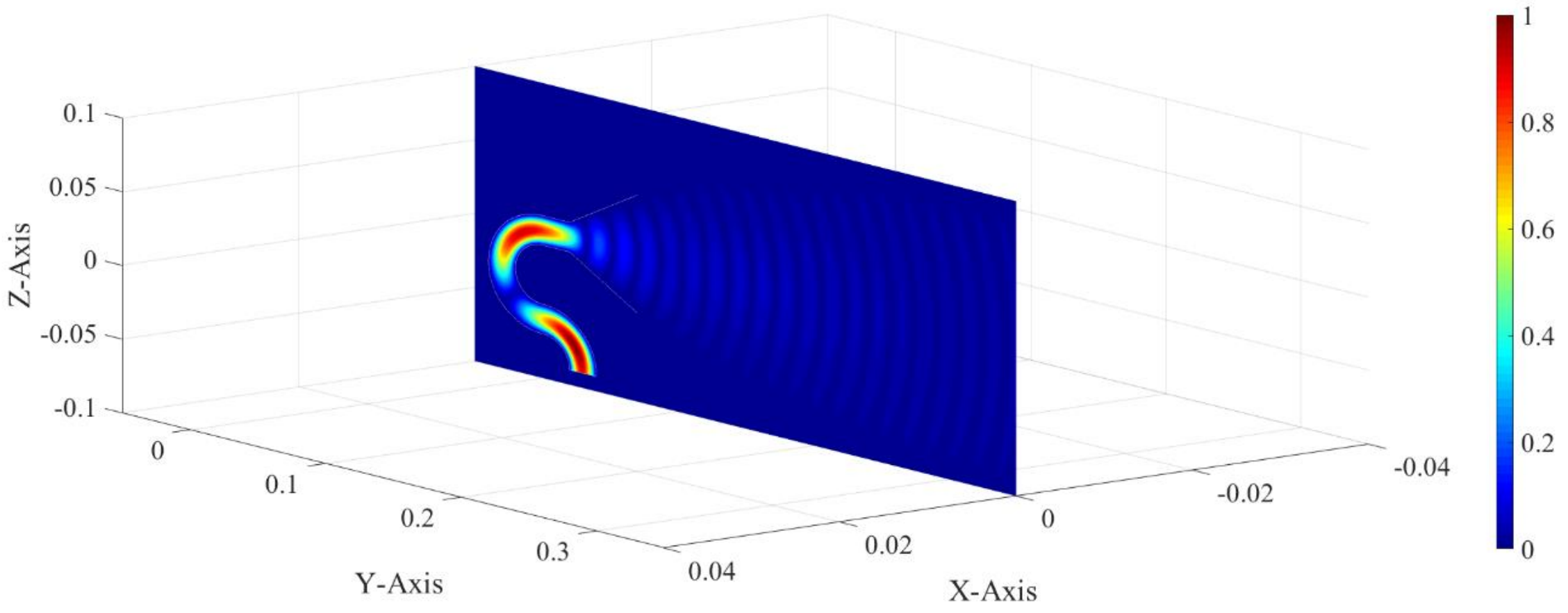

(b)

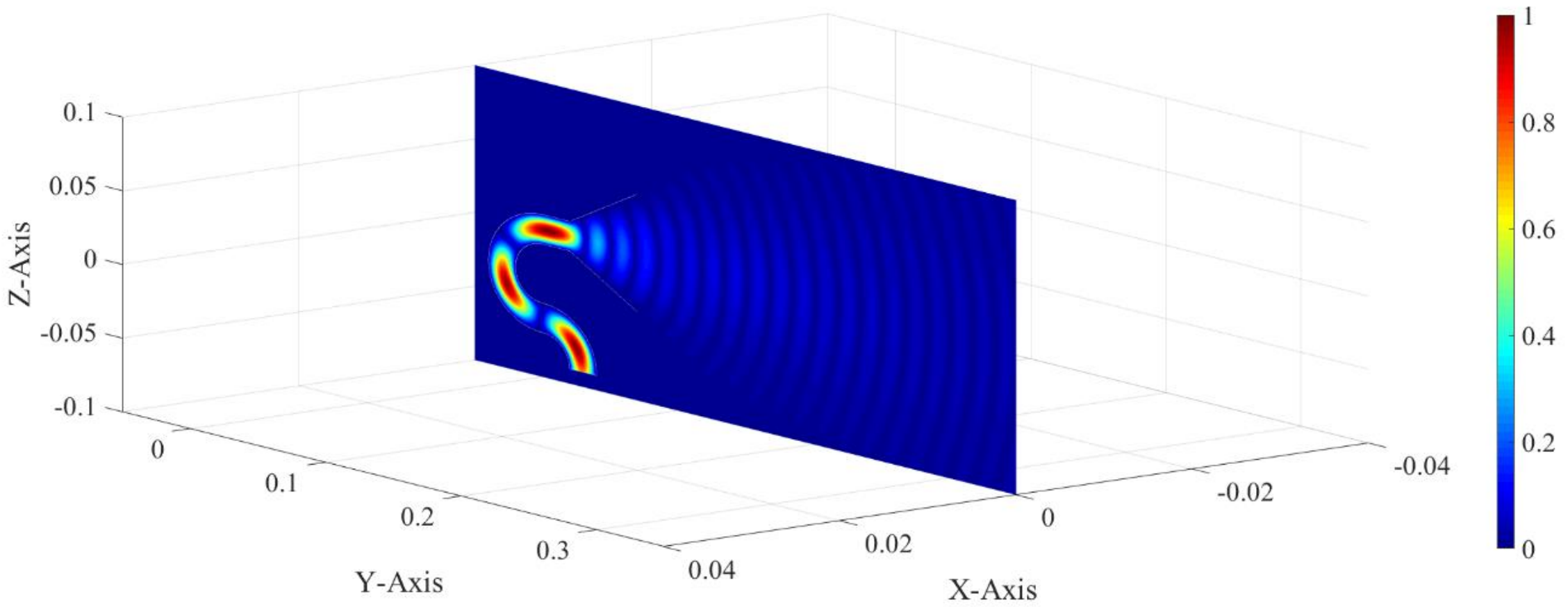

(c)

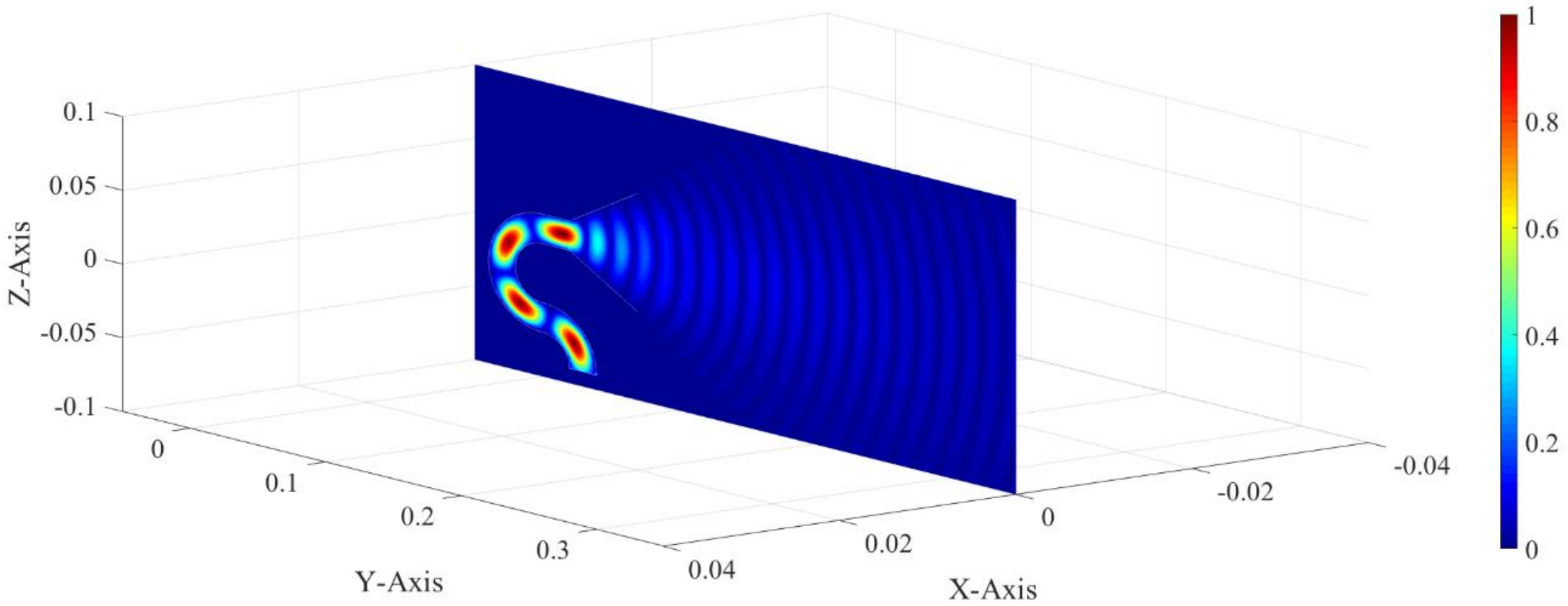

(d)

Figure 8-11 Modal electric field distribution (on yOz plane) of the JE-DoJ-based IP-DM 1 shown in Fig. 8-8. (a) 8.8 GHz; (b) 8.9 GHz; (c) 9.1 GHz; (d) 9.4 GHz

At last, we visually plot the modal far-field radiation pattern of the JE-DoJ-based IP-DM 1 in the following Fig. 8-12.

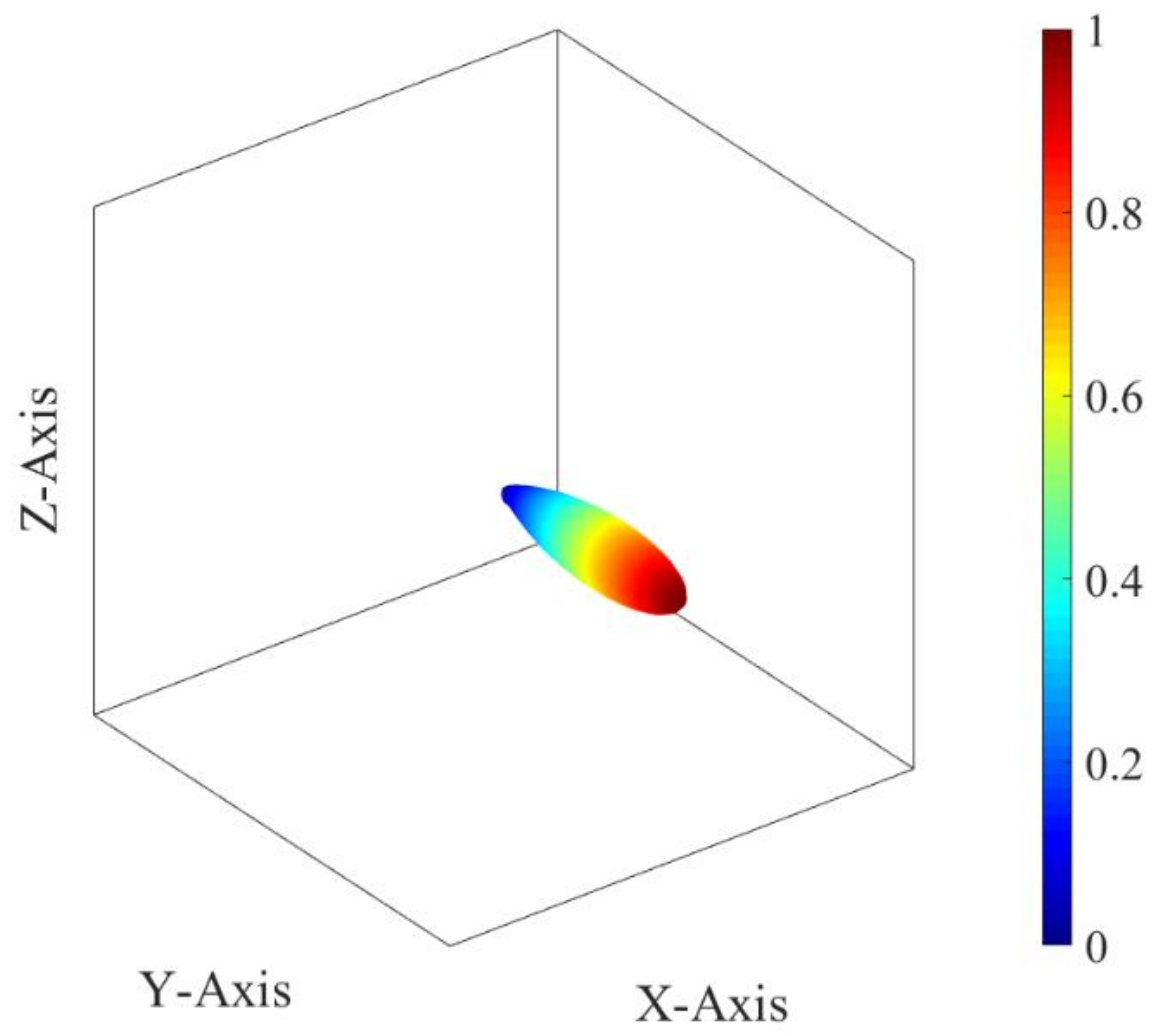

Figure 8-12 Modal radiation pattern of the JE-DoJ-based IP-DM 1 shown in Fig. 8-8

#### 8.2.6.2 Waveguide-fed Two-element Metallic Horn Array

In this subsection, we consider the TGTA system shown in the following Fig. 8-13, which is constituted by a *curved metallic guide* (the green part in the Fig. 8-13) and a *two-element metallic horn array* (the orange part in the Fig. 8-13). The horn array is fed by the guide.

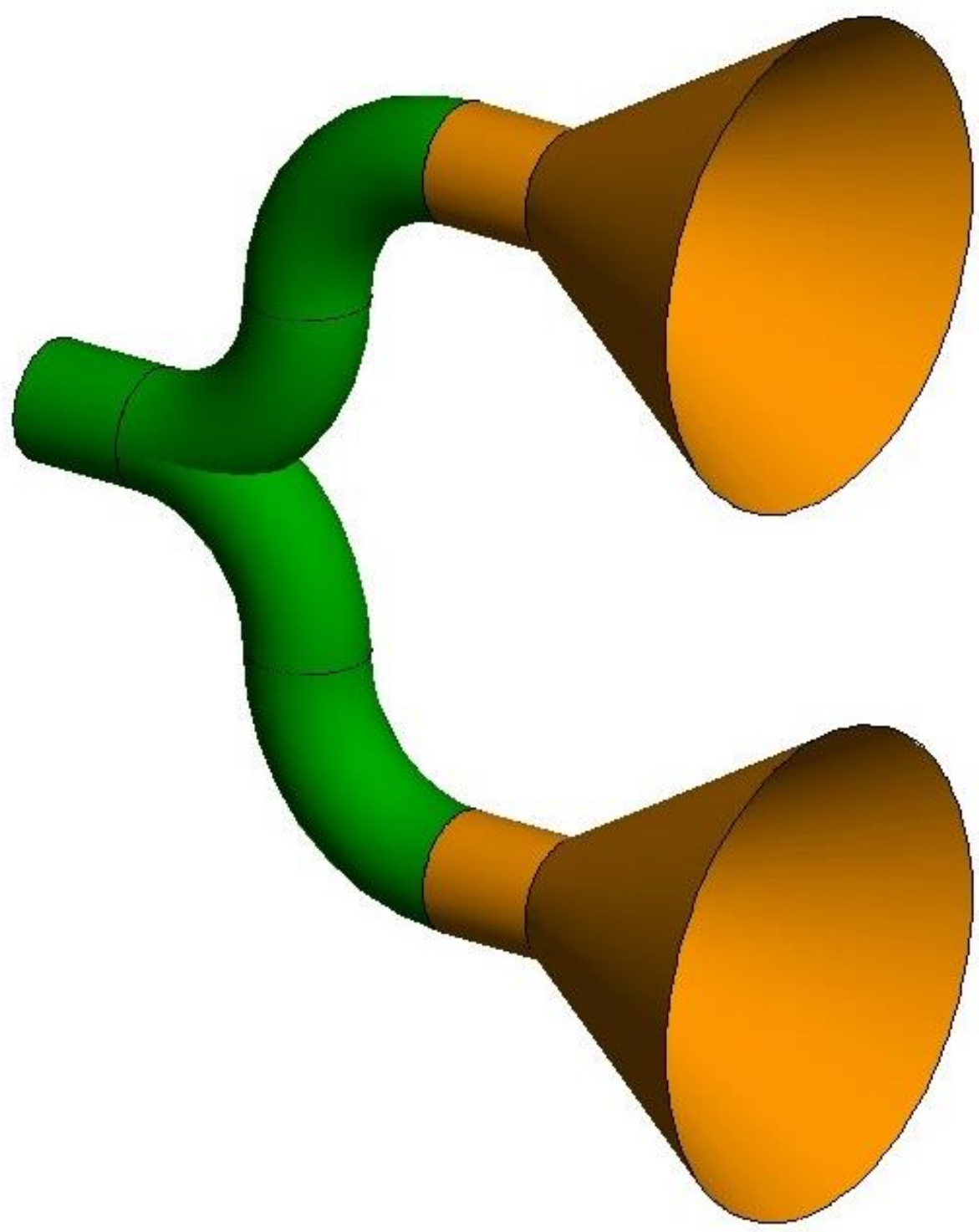

Figure 8-13 Geometry of a TGTA system constituted by a curved metallic guide and a two-element metallic horn array

The geometrical size of the TGTA system is shown in the following Fig. 8-14.

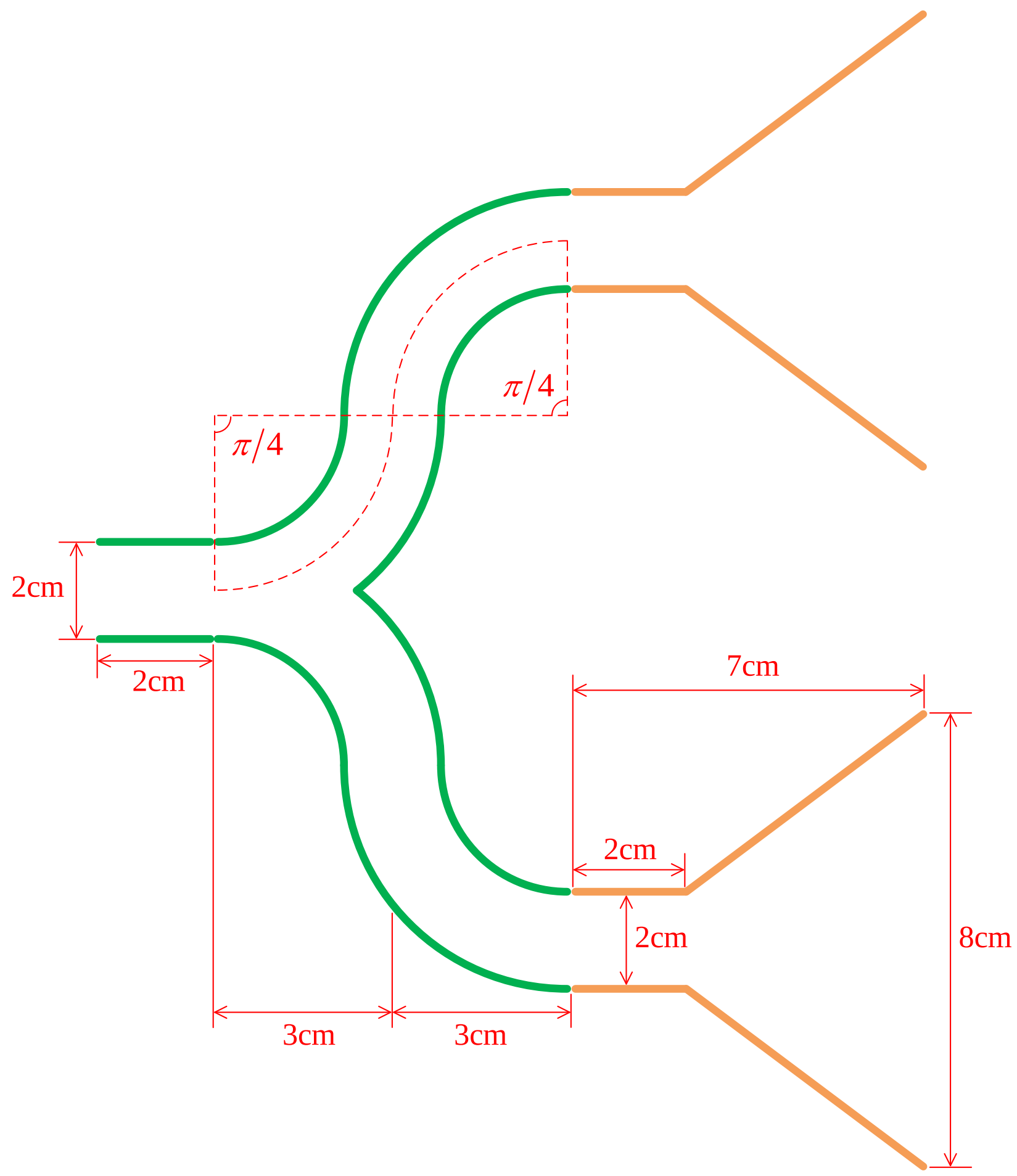

Figure 8-14 Geometrical size of the TGTA system shown in Fig. 8-13

The topological structures and surface triangular meshes of the TGTA system are shown in the following Fig. 8-15.

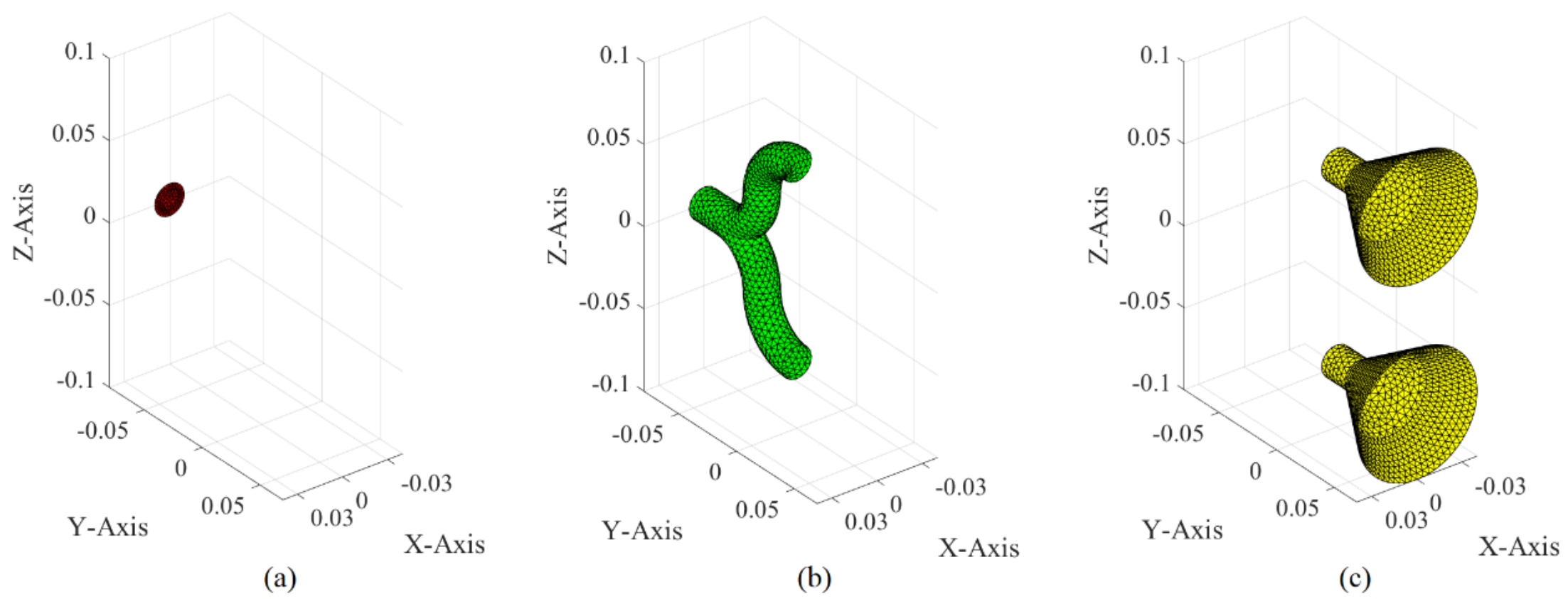

Figure 8-15 Topological structures and surface triangular meshes of the TGTA system shown in Fig. 8-13. (a) Input port; (b) curved metallic guide; (c) metallic horn array

Using the JE-DoJ-based and HM-DoM-based formulations of IPO, we calculate the IP-DMs of the TGTA system, and we plot the corresponding modal input resistance curves in the following Fig. 8-16.

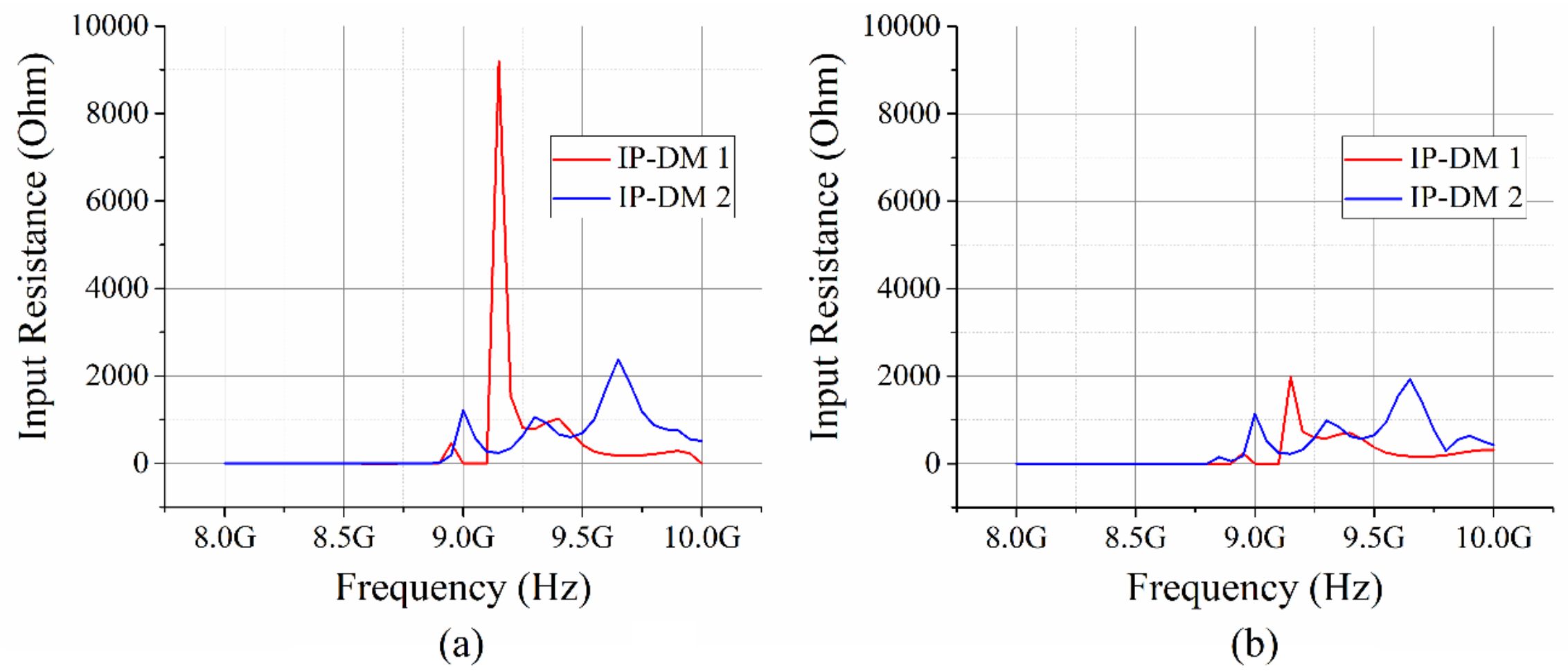

Figure 8-16 Modal input resistance curves of IP-DMs. (a) JE-DoJ-based results; (b) HM-DoM-based results

Taking the JE-DoJ-based IP-DM 1 shown in Fig. 8-16(a) as an example, its equivalent surface electric and magnetic currents distributing on input port are shown in Fig. 8-17.

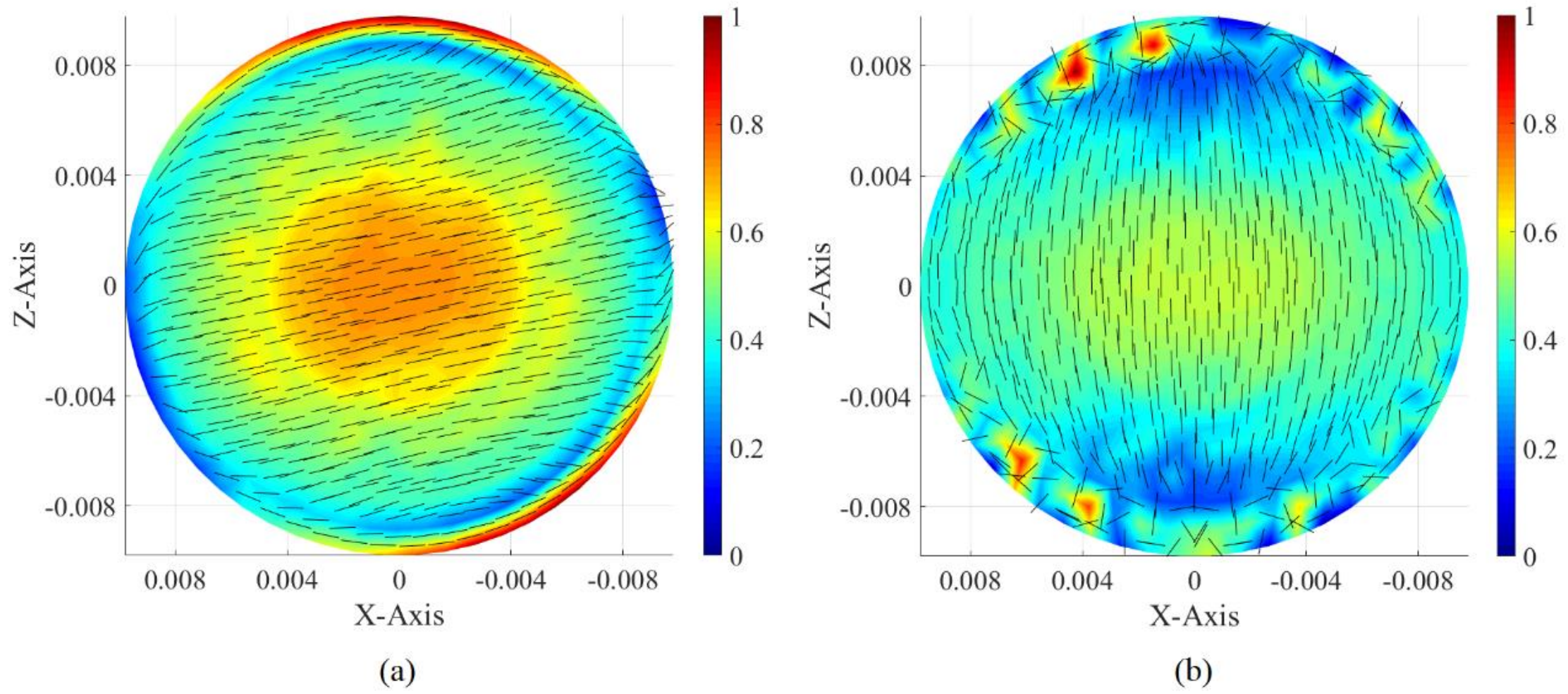

Figure 8-17 Modal equivalent (a) electric and (b) magnetic currents of the JE-DoJ-based IP-DM 1 shown in Fig. 8-16

From Fig. 8-16(a), it is easy to find out that the IP-DM 1 curve reaches the local peaks at 8.95 GHz, 9.15 GHz, 9.40 GHz, and 9.90 GHz. The modal induced electric currents (on the metallic electric wall) corresponding to the four frequencies are shown in Fig. 8-18.

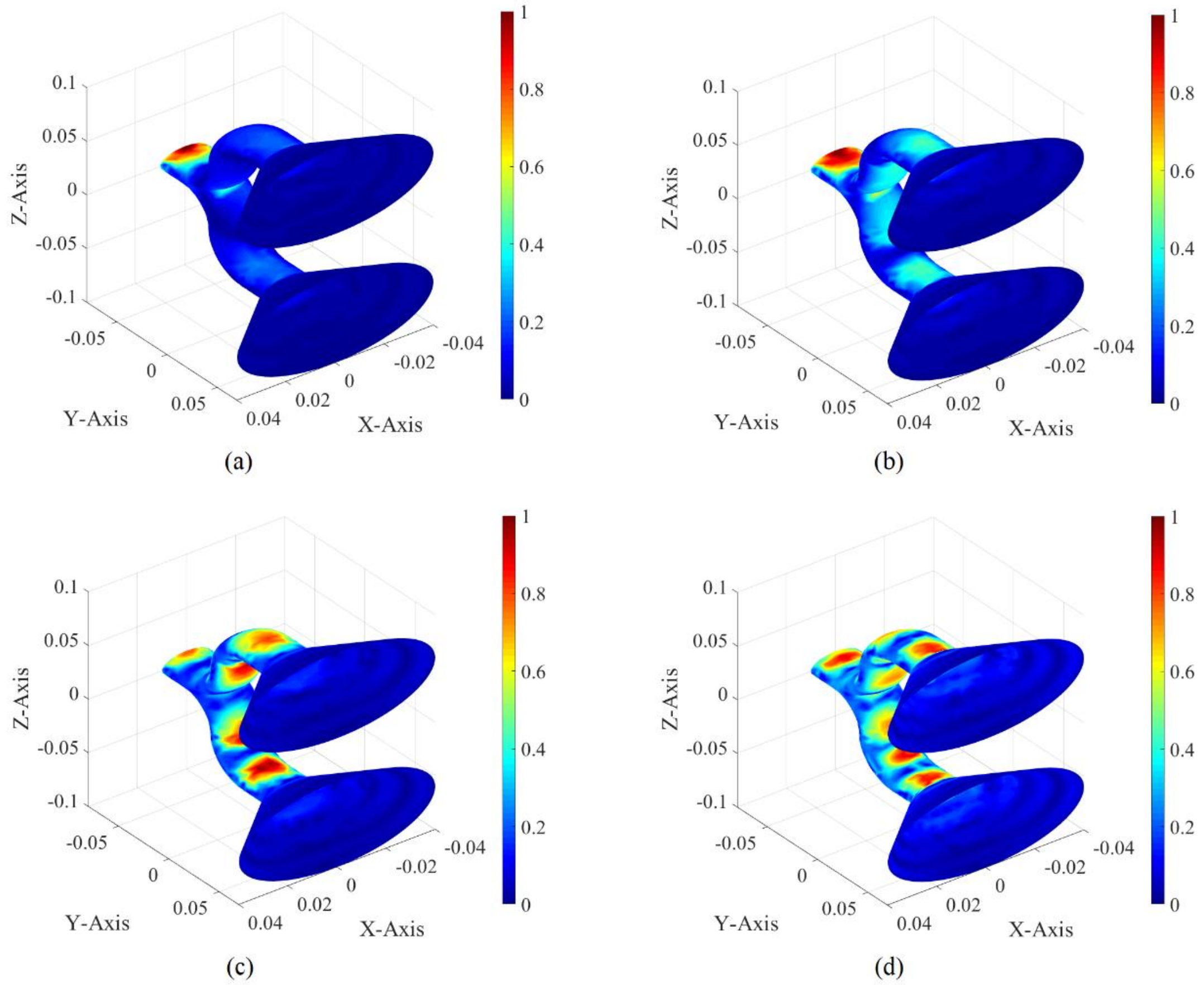

Figure 8-18 Modal induced electric current (on metallic electric wall) of the JE-DoJ-based IP-DM 1 in Fig. 8-16. (a) 8.95 GHz; (b) 9.15 GHz; (c) 9.40 GHz; (d) 9.90 GHz

In addition, we also plot the magnitude distributions of the modal electric field (on yOz plane) of the JE-DoJ-based IP-DM 1 (shown in the previous Fig. 8-16 (a)) working at the four peak frequencies 8.95 GHz, 9.15 GHz, 9.40 GHz, and 9.90 GHz in the following Fig. 8-19.

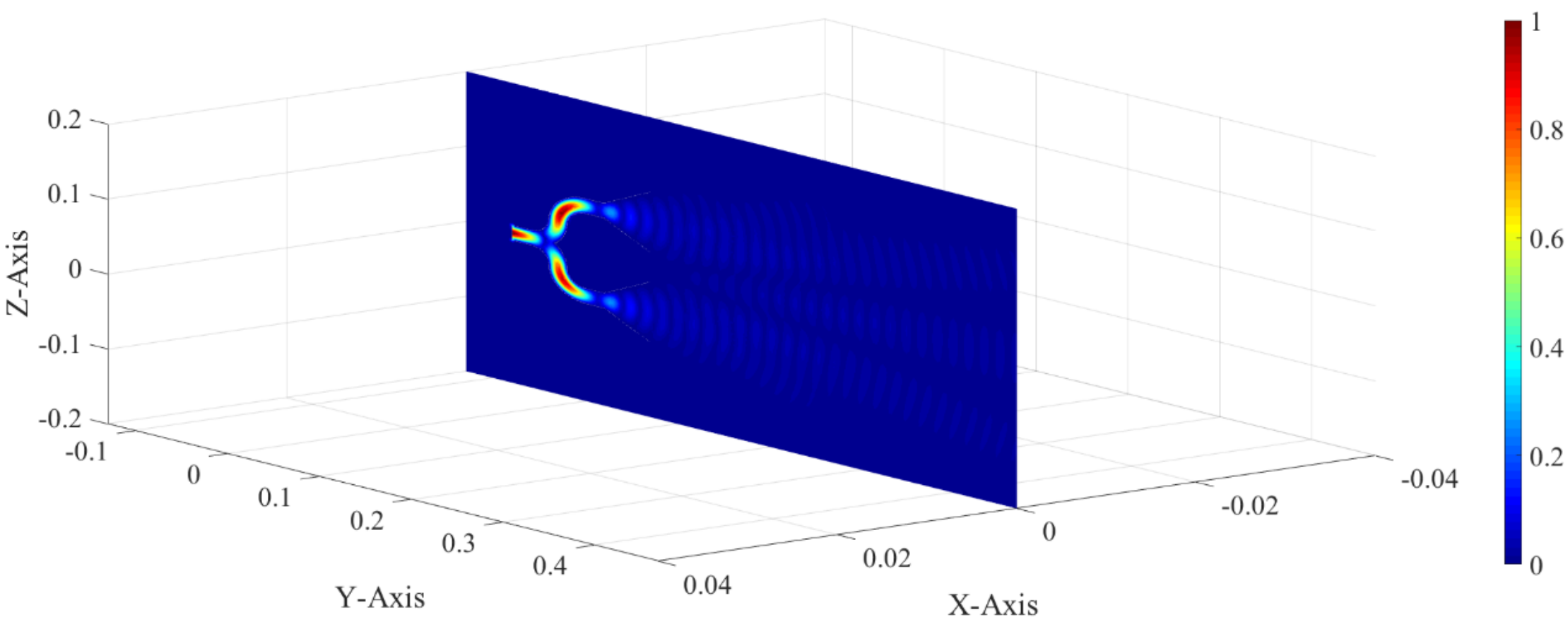

(a)

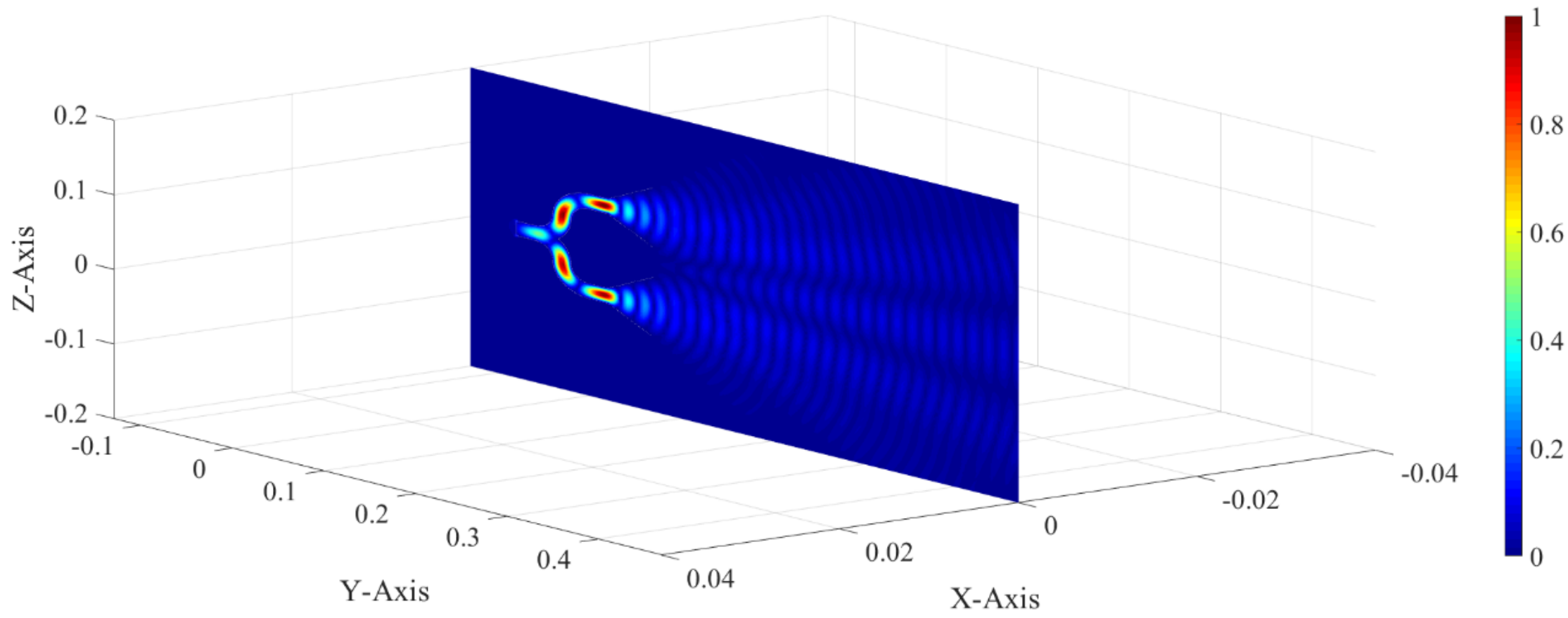

(b)

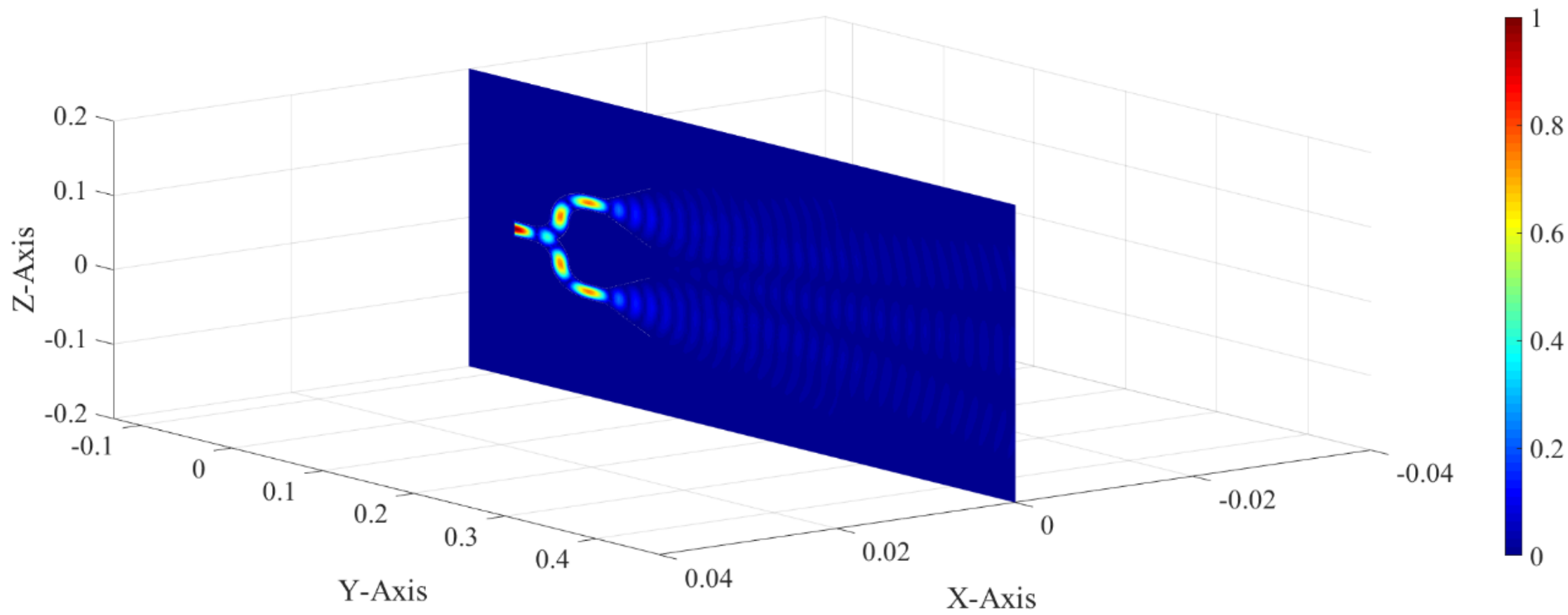

(c)

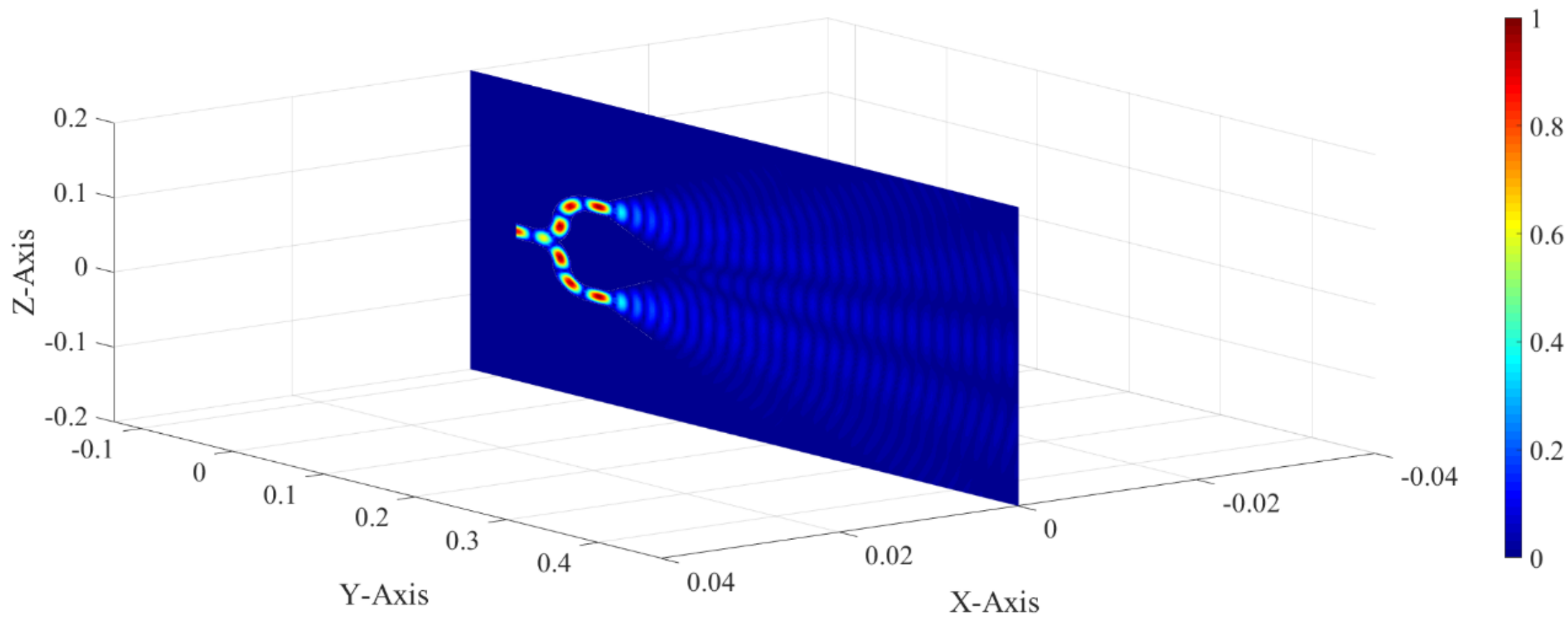

(d)

Figure 8-19 Modal electric field distribution (on yOz plane) of the JE-DoJ-based IP-DM 1 shown in Fig. 8-16. (a) 8.95 GHz; (b) 9.15 GHz; (c) 9.40 GHz; (d) 9.90 GHz

At last, we visually plot the modal far-field radiation pattern of the JE-DoJ-based IP-DM 1 in the following Fig. 8-20.

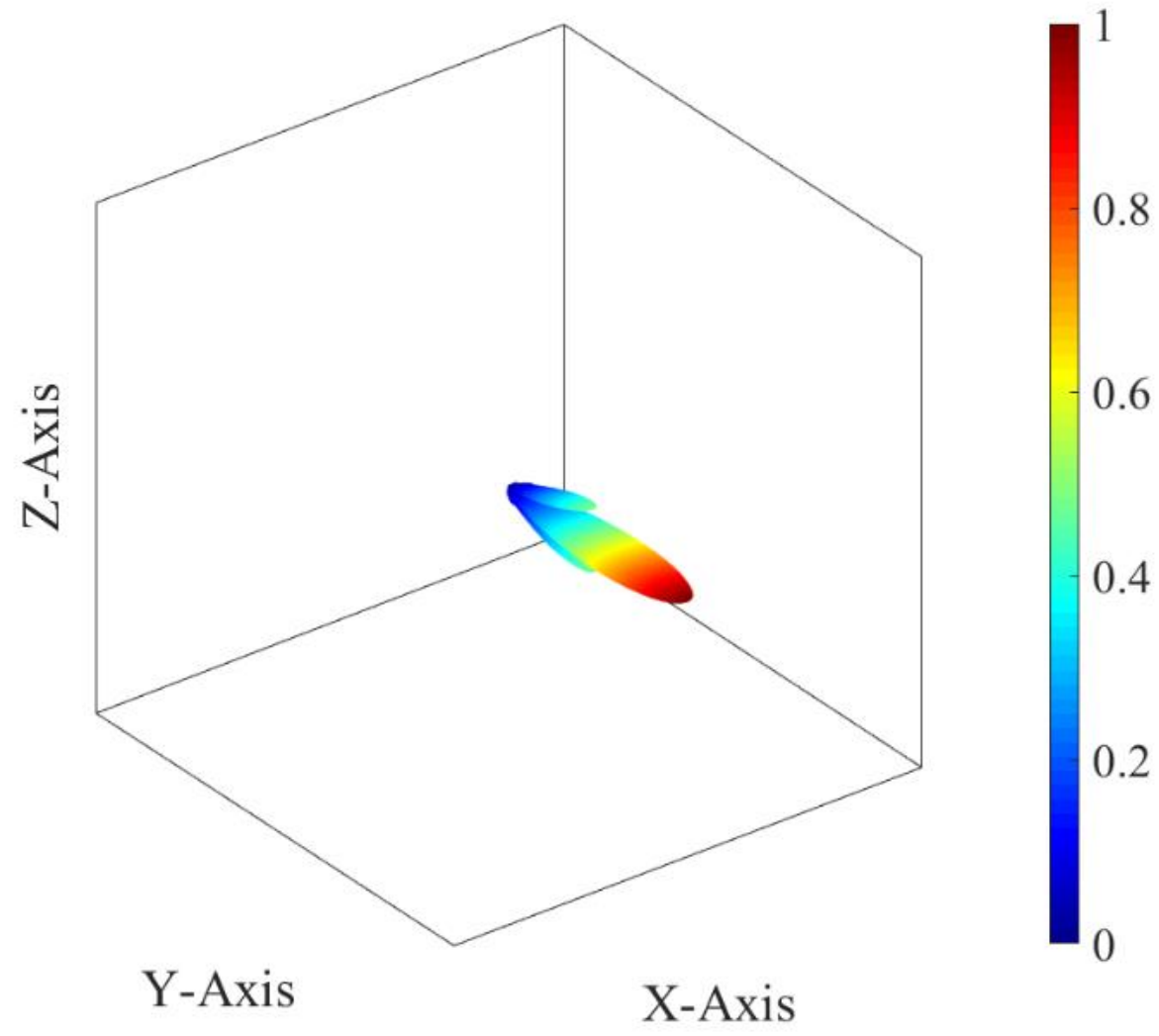

Figure 8-20 Modal radiation pattern of the JE-DoJ-based IP-DM 1 shown in Fig. 8-16

## 8.3 IP-DMs of Tra-antenna-Rec-antenna Combined System

In this section, we consider the *transmitting-receiving problem* shown in the following Fig. 8-21.

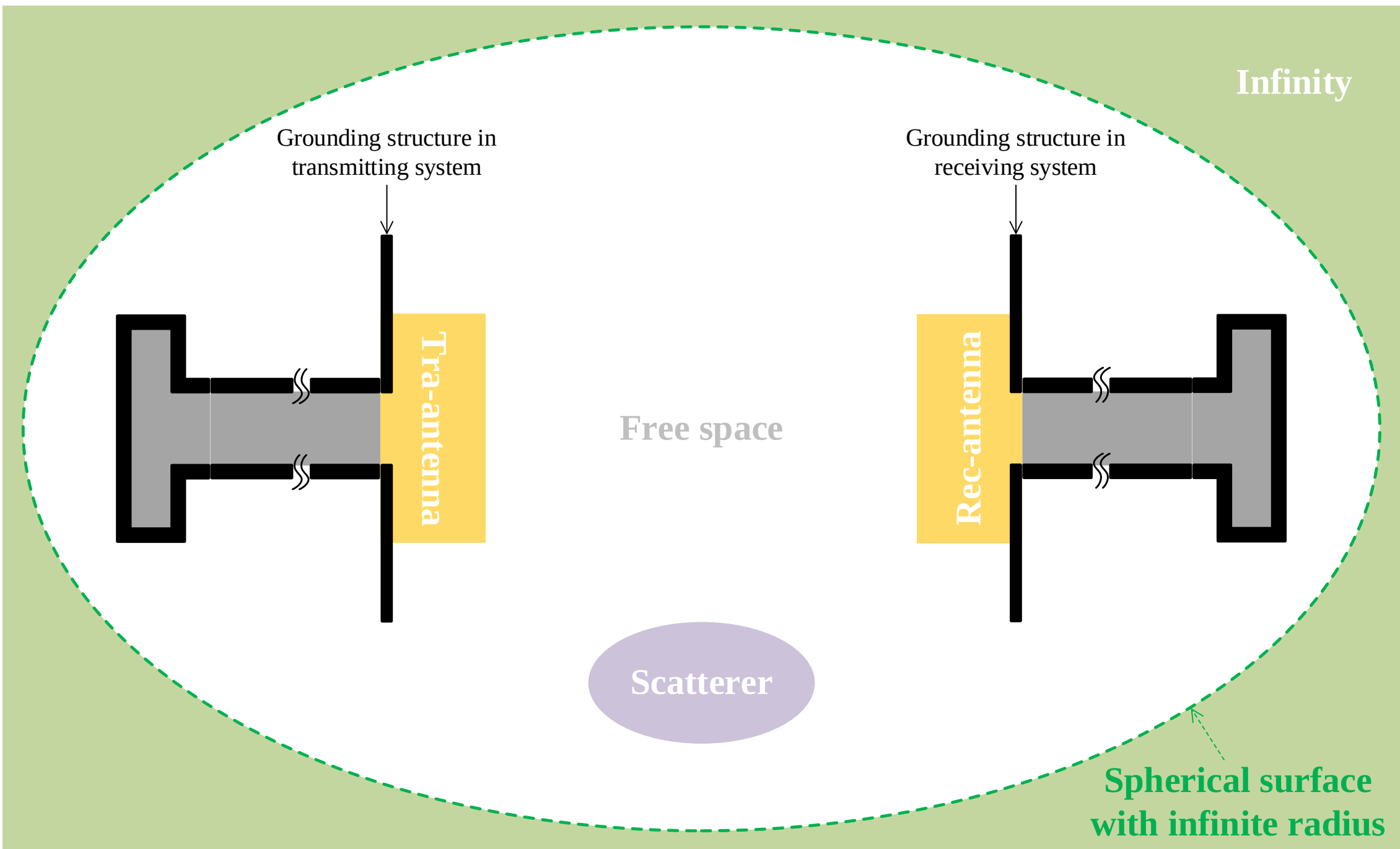

Figure 8-21 Geometry of the TARA system considered in this section

The tra-antenna is a waveguide-fed DRA, and the rec-antenna is a waveguide-loaded DRA, and both the DRAs are mounted on thick metallic ground planes. The transceiving system is placed in a propagation medium containing a material scatterer. Here, we treat the {tra-antenna, propagation medium, rec-antenna, ground planes} as a whole — *augmented tra-antenna-rec-antenna combined system* (simply called *augmented TARA-combined system* or more simply as *TARA system*).

Similarly to the previous Sec. 8.2, this section is organized as that "topological structure → source-field relationships → modal space → power transport theorem → input power operator → input-power-decoupled modes".

## 8.3.1 Topological Structure

The topological structure of the TARA system shown in Fig. 8-21 is illustrated in the following Fig. 8-22.

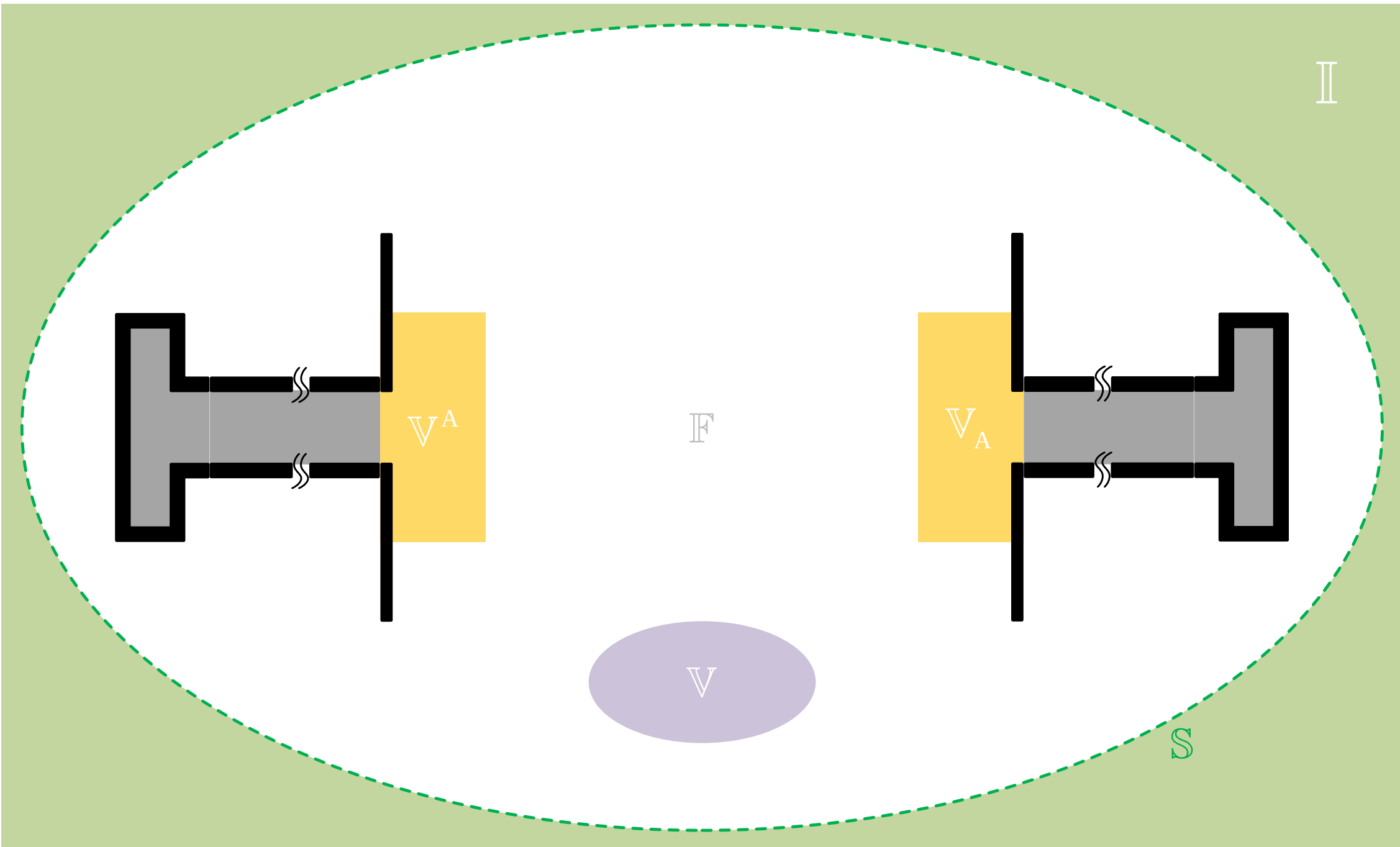

Figure 8-22 Topological structure of the TARA system shown in Fig. 8-21

In the above Fig. 8-22, the region occupied by the transmitting DRA is denoted as $\mathbb{V}^{A}$, and the region occupied by the receiving DRA is denoted as $\mathbb{V}_{A}$; the region occupied by whole propagation medium is denoted as $\mathbb{M}$; the material scatterer located in $\mathbb{M}$ is denoted as $\mathbb{V}$; the region occupied by free space is denoted as $\mathbb{F}$. Obviously, $\mathbb{M}=\mathbb{F}\bigcup\mathbb{V}$ and $\mathbb{F}\bigcap\mathbb{V}=\varnothing$.

The input port of $\mathbb{V}^{A}$, i.e. the interface between $\mathbb{V}^{A}$ and tra-guide, is denoted as $\mathbb{S}^{G\rightleftharpoons A}$. The electric wall of $\mathbb{V}^{A}$, i.e. the interface between $\mathbb{V}^{A}$ and ground plane, is denoted as $\mathbb{S}^{A}$. The output port of $\mathbb{V}^{A}$, i.e. the interface between $\mathbb{V}^{A}$ and propagation medium, is denoted as $\mathbb{S}^{\text{A}\rightleftharpoons\text{M}}$. The inner normal direction of $\partial\mathbb{V}^{\text{A}}$ is denoted as $\hat{n}^{\to\text{A}}$, and it points to the interior of $\mathbb{V}^{\text{A}}$.

The input port of $\mathbb{M}$ is just the $\mathbb{S}^{\text{A}\rightleftharpoons\text{M}}$. The electric wall of $\mathbb{M}$ contains two parts; the first part is the interface between $\mathbb{M}$ and the grounding structure of transmitting system, and it is denoted as $\mathbb{S}^{\text{M}}$; the second part is the interface between $\mathbb{M}$ and the grounding structure of receiving system, and it is denoted as $\mathbb{S}_{\text{M}}$. The output port of $\mathbb{M}$ also contains two parts; the first part, which is the interface between $\mathbb{M}$ and infinity, is denoted as $\mathbb{S}$; the second part, which is the interface between $\mathbb{M}$ and receiving DRA, is denoted as $\mathbb{S}_{\text{M}\rightleftharpoons\text{A}}$. The outer normal direction of $\mathbb{S}$ is denoted as $\hat{n}$, and it points to infinity.

The input port of $\mathbb{V}_{\text{A}}$ is just the $\mathbb{S}_{\text{M}\rightleftharpoons\text{A}}$. The electric wall of $\mathbb{V}_{\text{A}}$, i.e. the interface between $\mathbb{V}_{\text{A}}$ and ground plane, is denoted as $\mathbb{S}_{\text{A}}$. The output port of $\mathbb{V}_{\text{A}}$, i.e. the interface between $\mathbb{V}_{\text{A}}$ and rec-guide, is denoted as $\mathbb{S}_{\text{A}\rightleftharpoons\text{G}}$. The inner normal direction of $\partial\mathbb{V}_{\text{A}}$ is denoted as $\hat{n}_{\to\text{A}}$, and it points to the interior of $\mathbb{V}_{\text{A}}$.

The permeability, permittivity, and conductivity of the transmitting DRA are denoted as $\ddot{\mu}_{\text{T}}(\vec{r})$, $\ddot{\varepsilon}_{\text{T}}(\vec{r})$, and $\ddot{\sigma}_{\text{T}}(\vec{r})$ respectively; the permeability, permittivity, and conductivity of the scatterer are denoted as $\ddot{\mu}_{\text{S}}(\vec{r})$, $\ddot{\varepsilon}_{\text{S}}(\vec{r})$, and $\ddot{\sigma}_{\text{S}}(\vec{r})$ respectively; the permeability, permittivity, and conductivity of the receiving DRA are denoted as $\ddot{\mu}_{\text{R}}(\vec{r})$, $\ddot{\varepsilon}_{\text{R}}(\vec{r})$, and $\ddot{\sigma}_{\text{R}}(\vec{r})$ respectively; the permeability and permittivity of free space are denoted as $\mu_0$ and $\varepsilon_0$ respectively. To simplify the symbolic system of the following discussions, the above material parameters are uniformly denoted as $\ddot{\gamma}(\vec{r})$, i.e.,

$$\ddot{\gamma}(\vec{r}) = \begin{cases} \ddot{\gamma}_{\text{T}}(\vec{r}) & , \quad \vec{r} \in \mathbb{V}^{\text{A}} \\ \ddot{\gamma}_{\text{S}}(\vec{r}) & , \quad \vec{r} \in \mathbb{V} \\ \ddot{\gamma}_{\text{R}}(\vec{r}) & , \quad \vec{r} \in \mathbb{V}_{\text{A}} \end{cases} \tag{8-71}$$

where $\ddot{\gamma} = \ddot{\mu} / \ddot{\varepsilon} / \ddot{\sigma}$.

## 8.3.2 Source-Field Relationships

If the equivalent surface currents distributing on $\mathbb{S}^{\text{G}\rightleftharpoons\text{A}}$ are denoted as $\{\vec{J}^{\text{G}\rightleftharpoons\text{A}}, \vec{M}^{\text{G}\rightleftharpoons\text{A}}\}$, and the equivalent surface electric current distributing on $\mathbb{S}^{\text{A}}$ is denoted as $\vec{J}^{\text{A}}$, and the equivalent surface currents distributing on $\mathbb{S}^{\text{A}\rightleftharpoons\text{M}}$ are denoted as $\{\vec{J}^{\text{A}\rightleftharpoons\text{M}}, \vec{M}^{\text{A}\rightleftharpoons\text{M}}\}$, then the field distributing on $\mathbb{V}^{\text{A}}$ can be expressed as follows:

$$\vec{F}(\vec{r}) = \mathcal{F}_{\text{T}}\left(\vec{J}^{\text{G}\rightleftharpoons\text{A}} + \vec{J}^{\text{A}} + \vec{J}^{\text{A}\rightleftharpoons\text{M}}, \vec{M}^{\text{G}\rightleftharpoons\text{A}} + \vec{M}^{\text{A}\rightleftharpoons\text{M}}\right) \quad , \quad \vec{r} \in \mathbb{V}^{\text{A}} \tag{8-72}$$

where $\vec{F} = \vec{E} / \vec{H}$, and correspondingly $\mathcal{F}_{\text{T}} = \mathcal{E}_{\text{T}} / \mathcal{H}_{\text{T}}$, and the operator is defined as that $\mathcal{F}_{\text{T}}(\vec{J},\vec{M}) = \vec{G}_{\text{T}}^{JF} * \vec{J} + \vec{G}_{\text{T}}^{MF} * \vec{M}$ (here, $\vec{G}_{\text{T}}^{JF}$ and $\vec{G}_{\text{T}}^{MF}$ are the dyadic Green's functions of the region $\mathbb{V}^{\text{A}}$ with parameters $\{\vec{\mu}_{\text{T}}, \vec{\varepsilon}_{\text{T}}, \vec{\sigma}_{\text{T}}\}$). The currents $\{\vec{J}^{\text{G}\rightleftharpoons\text{A}}, \vec{M}^{\text{G}\rightleftharpoons\text{A}}\}$ and fields $\{\vec{E},\vec{H}\}$ in Eq. (8-72) satisfy the following relations

$$\hat{n}^{\to\text{A}} \times \left[ \vec{H}\left(\vec{r}^{\,\text{A}}\right) \right]_{\vec{r}^{\,\text{A}} \to \vec{r}} = \vec{J}^{\text{G}\rightleftharpoons\text{A}}\left(\vec{r}\right) \quad , \quad \vec{r} \in \mathbb{S}^{\text{G}\rightleftharpoons\text{A}} \tag{8-73a}$$

$$\left[ \vec{E}\left(\vec{r}^{\,\text{A}}\right) \right]_{\vec{r}^{\,\text{A}} \to \vec{r}} \times \hat{n}^{\to\text{A}} = \vec{M}^{\text{G}\rightleftharpoons\text{A}}\left(\vec{r}\right) \quad , \quad \vec{r} \in \mathbb{S}^{\text{G}\rightleftharpoons\text{A}} \tag{8-73b}$$

and the currents $\{\vec{J}^{\text{A}\rightleftharpoons\text{M}}, \vec{M}^{\text{A}\rightleftharpoons\text{M}}\}$ and fields $\{\vec{E},\vec{H}\}$ in Eq. (8-72) satisfy the following relations

$$\hat{n}^{\to\text{A}} \times \left[ \vec{H}\left(\vec{r}^{\,\text{A}}\right) \right]_{\vec{r}^{\,\text{A}} \to \vec{r}} = \vec{J}^{\text{A}\rightleftharpoons\text{M}}\left(\vec{r}\right) \quad , \quad \vec{r} \in \mathbb{S}^{\text{A}\rightleftharpoons\text{M}} \tag{8-74a}$$

$$\left[ \vec{E}\left(\vec{r}^{\,\text{A}}\right) \right]_{\vec{r}^{\,\text{A}} \to \vec{r}} \times \hat{n}^{\to\text{A}} = \vec{M}^{\text{A}\rightleftharpoons\text{M}}\left(\vec{r}\right) \quad , \quad \vec{r} \in \mathbb{S}^{\text{A}\rightleftharpoons\text{M}} \tag{8-74b}$$

In the above Eqs. (8-73) and (8-74), point $\vec{r}^{\,\text{A}}$ belongs to region $\mathbb{V}^{\text{A}}$, and approaches the point $\vec{r}$ on $\mathbb{S}^{\text{G}\rightleftharpoons\text{A}} \bigcup \mathbb{S}^{\text{A}\rightleftharpoons\text{M}}$.

If the equivalent surface electric current distributing on $\mathbb{S}^{\text{M}}$ is denoted as $\vec{J}^{\text{M}}$, and the equivalent volume currents distributing on $\mathbb{V}$ are denoted as $\{\vec{J},\vec{M}\}$, and the equivalent surface electric current distributing on $\mathbb{S}_{\text{M}}$ is denoted as $\vec{J}_{\text{M}}$, and the equivalent surface currents distributing on $\mathbb{S}_{\text{M}\rightleftharpoons\text{A}}$ are denoted as $\{\vec{J}_{\text{M}\rightleftharpoons\text{A}}, \vec{M}_{\text{M}\rightleftharpoons\text{A}}\}$, then the field distributing on $\mathbb{M}$ can be expressed as follows:

$$\vec{F}\left(\vec{r}\right) = \mathcal{F}_0\left( -\vec{J}^{\text{A}\rightleftharpoons\text{M}} + \vec{J}^{\text{M}} + \vec{J} + \vec{J}_{\text{M}} - \vec{J}_{\text{M}\rightleftharpoons\text{A}}, -\vec{M}^{\text{A}\rightleftharpoons\text{M}} + \vec{M} - \vec{M}_{\text{M}\rightleftharpoons\text{A}} \right) \ , \ \vec{r} \in \mathbb{M} \tag{8-75}$$

where $\vec{F} = \vec{E} / \vec{H}$, and correspondingly $\mathcal{F}_0 = \mathcal{E}_0 / \mathcal{H}_0$, and the operator is the same as the one used previously. The currents $\{\vec{J},\vec{M}\}$ and fields $\{\vec{E},\vec{H}\}$ in Eq. (8-75) satisfy the following relations

$$j\omega\Delta\vec{\varepsilon}_{\text{c}} \cdot \vec{E}\left(\vec{r}\right) = \vec{J}\left(\vec{r}\right) \quad , \quad \vec{r} \in \mathbb{V} \tag{8-76a}$$

$$j\omega\Delta\vec{\mu} \cdot \vec{H}\left(\vec{r}\right) = \vec{M}\left(\vec{r}\right) \quad , \quad \vec{r} \in \mathbb{V} \tag{8-76b}$$

and the currents $\{\vec{J}_{\text{M}\rightleftharpoons\text{A}}, \vec{M}_{\text{M}\rightleftharpoons\text{A}}\}$ and fields $\{\vec{E},\vec{H}\}$ in Eq. (8-75) satisfy the following relations

$$\hat{n}_{\to\text{A}} \times \left[ \vec{H}\left(\vec{r}_{\text{M}}\right) \right]_{\vec{r}_{\text{M}} \to \vec{r}} = \vec{J}_{\text{M}\rightleftharpoons\text{A}}\left(\vec{r}\right) \quad , \quad \vec{r} \in \mathbb{S}_{\text{M}\rightleftharpoons\text{A}} \tag{8-77a}$$

$$\left[ \vec{E}\left(\vec{r}_{\text{M}}\right) \right]_{\vec{r}_{\text{M}} \to \vec{r}} \times \hat{n}_{\to\text{A}} = \vec{M}_{\text{M}\rightleftharpoons\text{A}}\left(\vec{r}\right) \quad , \quad \vec{r} \in \mathbb{S}_{\text{M}\rightleftharpoons\text{A}} \tag{8-77b}$$

In the above Eq. (8-77), point $\vec{r}_{\text{M}}$ belongs to region $\mathbb{M}$, and approaches the point $\vec{r}$ on $\mathbb{S}_{\text{M}\rightleftharpoons\text{A}}$.

If the equivalent surface electric current distributing on $\mathbb{S}_{\text{A}}$ is denoted as $\vec{J}_{\text{A}}$, and the equivalent surface currents distributing on $\mathbb{S}_{\text{A}\rightleftharpoons\text{G}}$ are denoted as $\{\vec{J}_{\text{A}\rightleftharpoons\text{G}},\vec{M}_{\text{A}\rightleftharpoons\text{G}}\}$, then the field distributing on $\mathbb{V}_{\text{A}}$ can be expressed as follows:

$$\vec{F}\left(\vec{r}\right)=\mathcal{F}_{\text{R}}\left(\vec{J}_{\text{M}\rightleftharpoons\text{A}}+\vec{J}_{\text{A}}+\vec{J}_{\text{A}\rightleftharpoons\text{G}},\vec{M}_{\text{M}\rightleftharpoons\text{A}}+\vec{M}_{\text{A}\rightleftharpoons\text{G}}\right) \quad , \quad \vec{r}\in\mathbb{V}_{\text{A}} \tag{8-78}$$

where $\vec{F}=\vec{E}\,/\,\vec{H}$, and correspondingly $\mathcal{F}_{\text{R}}=\mathcal{E}_{\text{R}}\,/\,\mathcal{H}_{\text{R}}$, and the operator is defined as that $\mathcal{F}_{\text{R}}(\vec{J},\vec{M})=\overset{\leftrightarrow}{G}{}_{\text{R}}^{JF}*\vec{J}+\overset{\leftrightarrow}{G}{}_{\text{R}}^{MF}*\vec{M}$ (here, $\overset{\leftrightarrow}{G}{}_{\text{R}}^{JF}$ and $\overset{\leftrightarrow}{G}{}_{\text{R}}^{MF}$ are the dyadic Green's functions of the region $\mathbb{V}_{\text{A}}$ with parameters $\{\overset{\leftrightarrow}{\mu}_{\text{R}},\overset{\leftrightarrow}{\varepsilon}_{\text{R}},\overset{\leftrightarrow}{\sigma}_{\text{R}}\}$). The currents $\{\vec{J}_{\text{A}\rightleftharpoons\text{G}},\vec{M}_{\text{A}\rightleftharpoons\text{G}}\}$ and fields $\{\vec{E},\vec{H}\}$ in Eq. (8-78) satisfy the following relations

$$\hat{n}_{\rightarrow\text{A}}\times\left[\vec{H}\left(\vec{r}_{\text{A}}\right)\right]_{\vec{r}_{\text{A}}\to\vec{r}}=\vec{J}_{\text{A}\rightleftharpoons\text{G}}\left(\vec{r}\right) \quad , \quad \vec{r}\in\mathbb{S}_{\text{A}\rightleftharpoons\text{G}} \tag{8-79a}$$

$$\left[\vec{E}\left(\vec{r}_{\text{A}}\right)\right]_{\vec{r}_{\text{A}}\to\vec{r}}\times\hat{n}_{\rightarrow\text{A}}=\vec{M}_{\text{A}\rightleftharpoons\text{G}}\left(\vec{r}\right) \quad , \quad \vec{r}\in\mathbb{S}_{\text{A}\rightleftharpoons\text{G}} \tag{8-79b}$$

In the above Eq. (8-79), point $\vec{r}_{\text{A}}$ belongs to region $\mathbb{V}_{\text{A}}$, and approaches the point $\vec{r}$ on $\mathbb{S}_{\text{A}\rightleftharpoons\text{G}}$.

### 8.3.3 Mathematical Description for Modal Space

In this subsection, we establish some rigorous mathematical descriptions for the modal space of the TARA system shown in Fig. 8-21, by employing the source-field relationships given in the above Sec. 8.3.2.

#### 8.3.3.1 Integral Equations

Substituting Eq. (8-72) into Eq. (8-73), we have the following integral equations

$$\left[\mathcal{H}_{\text{T}}\left(\vec{J}^{\text{G}\rightleftharpoons\text{A}}+\vec{J}^{\text{A}}+\vec{J}^{\text{A}\rightleftharpoons\text{M}},\vec{M}^{\text{G}\rightleftharpoons\text{A}}+\vec{M}^{\text{A}\rightleftharpoons\text{M}}\right)\right]_{\vec{r}^{\text{A}}\to\vec{r}}^{\tan}=\vec{J}^{\text{G}\rightleftharpoons\text{A}}\left(\vec{r}\right)\times\hat{n}^{\rightarrow\text{A}} \quad , \quad \vec{r}\in\mathbb{S}^{\text{G}\rightleftharpoons\text{A}} \tag{8-80a}$$

$$\left[\mathcal{E}_{\text{T}}\left(\vec{J}^{\text{G}\rightleftharpoons\text{A}}+\vec{J}^{\text{A}}+\vec{J}^{\text{A}\rightleftharpoons\text{M}},\vec{M}^{\text{G}\rightleftharpoons\text{A}}+\vec{M}^{\text{A}\rightleftharpoons\text{M}}\right)\right]_{\vec{r}^{\text{A}}\to\vec{r}}^{\tan}=\hat{n}^{\rightarrow\text{A}}\times\vec{M}^{\text{G}\rightleftharpoons\text{A}}\left(\vec{r}\right), \quad \vec{r}\in\mathbb{S}^{\text{G}\rightleftharpoons\text{A}} \tag{8-80b}$$

about currents $\{\vec{J}^{\text{G}\rightleftharpoons\text{A}},\vec{M}^{\text{G}\rightleftharpoons\text{A}}\}$, $\vec{J}^{\text{A}}$, and $\{\vec{J}^{\text{A}\rightleftharpoons\text{M}},\vec{M}^{\text{A}\rightleftharpoons\text{M}}\}$.

Based on Eq. (8-72) and the homogeneous tangential electric field boundary condition on $\mathbb{S}^{\text{A}}$, we have the following electric field integral equations

$$\left[\mathcal{E}_{\text{T}}\left(\vec{J}^{\text{G}\rightleftharpoons\text{A}}+\vec{J}^{\text{A}}+\vec{J}^{\text{A}\rightleftharpoons\text{M}},\vec{M}^{\text{G}\rightleftharpoons\text{A}}+\vec{M}^{\text{A}\rightleftharpoons\text{M}}\right)\right]_{\vec{r}^{\text{A}}\to\vec{r}}^{\tan}=0 \quad , \quad \vec{r}\in\mathbb{S}^{\text{A}} \tag{8-81}$$

about currents $\{\vec{J}^{\text{G}\rightleftharpoons\text{A}},\vec{M}^{\text{G}\rightleftharpoons\text{A}}\}$, $\vec{J}^{\text{A}}$, and $\{\vec{J}^{\text{A}\rightleftharpoons\text{M}},\vec{M}^{\text{A}\rightleftharpoons\text{M}}\}$.

Based on Eqs. (8-72)&(8-75) and the tangential field continuation condition on $\mathbb{S}^{\text{A}\rightleftharpoons\text{M}}$, we have the following integral equations

$$\Big[\mathcal{E}_{\rm T}\Big(\vec{J}^{\rm G\rightleftharpoons A}+\vec{J}^{\rm A}+\vec{J}^{\rm A\rightleftharpoons M},\vec{M}^{\rm G\rightleftharpoons A}+\vec{M}^{\rm A\rightleftharpoons M}\Big)\Big]_{\vec{r}^{\rm A}\to\vec{r}}^{\tan}$$
$$=\Big[\mathcal{E}_0\Big(-\vec{J}^{\rm A\rightleftharpoons M}+\vec{J}^{\rm M}+\vec{J}+\vec{J}_{\rm M}-\vec{J}_{\rm M\rightleftharpoons A},-\vec{M}^{\rm A\rightleftharpoons M}+\vec{M}-\vec{M}_{\rm M\rightleftharpoons A}\Big)\Big]_{\vec{r}^{\rm M}\to\vec{r}}^{\tan}\ ,\ \vec{r}\in\mathbb{S}^{\rm A\rightleftharpoons M}\quad(8\text{-}82\text{a})$$
$$\Big[\mathcal{H}_{\rm T}\Big(\vec{J}^{\rm G\rightleftharpoons A}+\vec{J}^{\rm A}+\vec{J}^{\rm A\rightleftharpoons M},\vec{M}^{\rm G\rightleftharpoons A}+\vec{M}^{\rm A\rightleftharpoons M}\Big)\Big]_{\vec{r}^{\rm A}\to\vec{r}}^{\tan}$$
$$=\Big[\mathcal{H}_0\Big(-\vec{J}^{\rm A\rightleftharpoons M}+\vec{J}^{\rm M}+\vec{J}+\vec{J}_{\rm M}-\vec{J}_{\rm M\rightleftharpoons A},-\vec{M}^{\rm A\rightleftharpoons M}+\vec{M}-\vec{M}_{\rm M\rightleftharpoons A}\Big)\Big]_{\vec{r}^{\rm M}\to\vec{r}}^{\tan}\ ,\ \vec{r}\in\mathbb{S}^{\rm A\rightleftharpoons M}\quad(8\text{-}82\text{b})$$

about currents $\{\vec{J}^{\rm G\rightleftharpoons A},\vec{M}^{\rm G\rightleftharpoons A}\}$, $\vec{J}^{\rm A}$, $\{\vec{J}^{\rm A\rightleftharpoons M},\vec{M}^{\rm A\rightleftharpoons M}\}$, $\vec{J}^{\rm M}$, $\{\vec{J},\vec{M}\}$, $\vec{J}_{\rm M}$, and $\{\vec{J}_{\rm M\rightleftharpoons A},\vec{M}_{\rm M\rightleftharpoons A}\}$.

Based on Eq. (8-75) and the homogeneous tangential electric field boundary condition on $\mathbb{S}^{\rm M}$, we have the following electric field integral equation

$$\Big[\mathcal{E}_0\Big(-\vec{J}^{\rm A\rightleftharpoons M}+\vec{J}^{\rm M}+\vec{J}+\vec{J}_{\rm M}-\vec{J}_{\rm M\rightleftharpoons A},-\vec{M}^{\rm A\rightleftharpoons M}+\vec{M}-\vec{M}_{\rm M\rightleftharpoons A}\Big)\Big]_{\vec{r}^{\rm M}\to\vec{r}}^{\tan}=0\ ,\quad\vec{r}\in\mathbb{S}^{\rm M}\ (8\text{-}83)$$

about currents $\{\vec{J}^{\rm A\rightleftharpoons M},\vec{M}^{\rm A\rightleftharpoons M}\}$, $\vec{J}^{\rm M}$, $\{\vec{J},\vec{M}\}$, $\vec{J}_{\rm M}$, and $\{\vec{J}_{\rm M\rightleftharpoons A},\vec{M}_{\rm M\rightleftharpoons A}\}$.

Substituting Eq. (8-75) into Eq. (8-76), we have the following integral equations

$$\mathcal{E}_0\Big(-\vec{J}^{\rm A\rightleftharpoons M}+\vec{J}^{\rm M}+\vec{J}+\vec{J}_{\rm M}-\vec{J}_{\rm M\rightleftharpoons A},-\vec{M}^{\rm A\rightleftharpoons M}+\vec{M}-\vec{M}_{\rm M\rightleftharpoons A}\Big)$$
$$=\big(j\omega\Delta\vec{\vec{\varepsilon}}_{\rm c}\big)^{-1}\cdot\vec{J}(\vec{r})\qquad,\qquad\vec{r}\in\mathbb{V}\quad(8\text{-}84\text{a})$$
$$\mathcal{H}_0\Big(-\vec{J}^{\rm A\rightleftharpoons M}+\vec{J}^{\rm M}+\vec{J}+\vec{J}_{\rm M}-\vec{J}_{\rm M\rightleftharpoons A},-\vec{M}^{\rm A\rightleftharpoons M}+\vec{M}-\vec{M}_{\rm M\rightleftharpoons A}\Big)$$
$$=\big(j\omega\Delta\vec{\vec{\mu}}\big)^{-1}\cdot\vec{M}(\vec{r})\qquad,\qquad\vec{r}\in\mathbb{V}\quad(8\text{-}84\text{b})$$

about currents $\{\vec{J}^{\rm A\rightleftharpoons M},\vec{M}^{\rm A\rightleftharpoons M}\}$, $\vec{J}^{\rm M}$, $\{\vec{J},\vec{M}\}$, $\vec{J}_{\rm M}$, and $\{\vec{J}_{\rm M\rightleftharpoons A},\vec{M}_{\rm M\rightleftharpoons A}\}$.

Based on Eq. (8-75) and the homogeneous tangential electric field boundary condition on $\mathbb{S}_{\rm M}$, we have the following electric field integral equation

$$\Big[\mathcal{E}_0\Big(-\vec{J}^{\rm A\rightleftharpoons M}+\vec{J}^{\rm M}+\vec{J}+\vec{J}_{\rm M}-\vec{J}_{\rm M\rightleftharpoons A},-\vec{M}^{\rm A\rightleftharpoons M}+\vec{M}-\vec{M}_{\rm M\rightleftharpoons A}\Big)\Big]_{\vec{r}_{\rm M}\to\vec{r}}^{\tan}=0\ ,\quad\vec{r}\in\mathbb{S}_{\rm M}\ (8\text{-}85)$$

about currents $\{\vec{J}^{\rm A\rightleftharpoons M},\vec{M}^{\rm A\rightleftharpoons M}\}$, $\vec{J}^{\rm M}$, $\{\vec{J},\vec{M}\}$, $\vec{J}_{\rm M}$, and $\{\vec{J}_{\rm M\rightleftharpoons A},\vec{M}_{\rm M\rightleftharpoons A}\}$.

Based on Eqs. (8-75)&(8-78) and the tangential field continuation condition on $\mathbb{S}_{\rm M\rightleftharpoons A}$, we have the following integral equations

$$\Big[\mathcal{E}_0\Big(-\vec{J}^{\rm A\rightleftharpoons M}+\vec{J}^{\rm M}+\vec{J}+\vec{J}_{\rm M}-\vec{J}_{\rm M\rightleftharpoons A},-\vec{M}^{\rm A\rightleftharpoons M}+\vec{M}-\vec{M}_{\rm M\rightleftharpoons A}\Big)\Big]_{\vec{r}_{\rm M}\to\vec{r}}^{\tan}$$
$$=\Big[\mathcal{E}_{\rm R}\Big(\vec{J}_{\rm M\rightleftharpoons A}+\vec{J}_{\rm A}+\vec{J}_{\rm A\rightleftharpoons G},\vec{M}_{\rm M\rightleftharpoons A}+\vec{M}_{\rm A\rightleftharpoons G}\Big)\Big]_{\vec{r}_{\rm A}\to\vec{r}}^{\tan}\qquad,\qquad\vec{r}\in\mathbb{S}_{\rm M\rightleftharpoons A}\quad(8\text{-}86\text{a})$$
$$\Big[\mathcal{H}_0\Big(-\vec{J}^{\rm A\rightleftharpoons M}+\vec{J}^{\rm M}+\vec{J}+\vec{J}_{\rm M}-\vec{J}_{\rm M\rightleftharpoons A},-\vec{M}^{\rm A\rightleftharpoons M}+\vec{M}-\vec{M}_{\rm M\rightleftharpoons A}\Big)\Big]_{\vec{r}_{\rm M}\to\vec{r}}^{\tan}$$
$$=\Big[\mathcal{H}_{\rm R}\Big(\vec{J}_{\rm M\rightleftharpoons A}+\vec{J}_{\rm A}+\vec{J}_{\rm A\rightleftharpoons G},\vec{M}_{\rm M\rightleftharpoons A}+\vec{M}_{\rm A\rightleftharpoons G}\Big)\Big]_{\vec{r}_{\rm A}\to\vec{r}}^{\tan}\qquad,\qquad\vec{r}\in\mathbb{S}_{\rm M\rightleftharpoons A}\quad(8\text{-}86\text{b})$$

about currents $\{\vec{J}^{\mathrm{A}\rightleftharpoons\mathrm{M}},\vec{M}^{\mathrm{A}\rightleftharpoons\mathrm{M}}\}$, $\vec{J}^{\mathrm{M}}$, $\{\vec{J},\vec{M}\}$, $\vec{J}_{\mathrm{M}}$, $\{\vec{J}_{\mathrm{M}\rightleftharpoons\mathrm{A}},\vec{M}_{\mathrm{M}\rightleftharpoons\mathrm{A}}\}$, $\vec{J}_{\mathrm{A}}$, and $\{\vec{J}_{\mathrm{A}\rightleftharpoons\mathrm{G}},\vec{M}_{\mathrm{A}\rightleftharpoons\mathrm{G}}\}$.

Based on Eq. (8-78) and the homogeneous tangential electric field boundary condition on $\mathbb{S}_{\mathrm{A}}$, we have the following electric field integral equation

$$\left[\mathcal{E}_{\mathrm{R}}\left(\vec{J}_{\mathrm{M}\rightleftharpoons\mathrm{A}}+\vec{J}_{\mathrm{A}}+\vec{J}_{\mathrm{A}\rightleftharpoons\mathrm{G}},\vec{M}_{\mathrm{M}\rightleftharpoons\mathrm{A}}+\vec{M}_{\mathrm{A}\rightleftharpoons\mathrm{G}}\right)\right]_{\vec{r}_{\mathrm{A}}\to\vec{r}}^{\tan}=0 \quad , \quad \vec{r}\in\mathbb{S}_{\mathrm{A}} \tag{8-87}$$

about currents $\{\vec{J}_{\mathrm{M}\rightleftharpoons\mathrm{A}},\vec{M}_{\mathrm{M}\rightleftharpoons\mathrm{A}}\}$, $\vec{J}_{\mathrm{A}}$, and $\{\vec{J}_{\mathrm{A}\rightleftharpoons\mathrm{G}},\vec{M}_{\mathrm{A}\rightleftharpoons\mathrm{G}}\}$.

Substituting Eq. (8-78) into Eq. (8-79), we have the following integral equations

$$\left[\mathcal{H}_{\mathrm{R}}\left(\vec{J}_{\mathrm{M}\rightleftharpoons\mathrm{A}}+\vec{J}_{\mathrm{A}}+\vec{J}_{\mathrm{A}\rightleftharpoons\mathrm{G}},\vec{M}_{\mathrm{M}\rightleftharpoons\mathrm{A}}+\vec{M}_{\mathrm{A}\rightleftharpoons\mathrm{G}}\right)\right]_{\vec{r}_{\mathrm{A}}\to\vec{r}}=\vec{J}_{\mathrm{A}\rightleftharpoons\mathrm{G}}\left(\vec{r}\right)\times\hat{n}_{\to\mathrm{A}} \quad , \quad \vec{r}\in\mathbb{S}_{\mathrm{A}\rightleftharpoons\mathrm{G}} \tag{8-88a}$$

$$\left[\mathcal{E}_{\mathrm{R}}\left(\vec{J}_{\mathrm{M}\rightleftharpoons\mathrm{A}}+\vec{J}_{\mathrm{A}}+\vec{J}_{\mathrm{A}\rightleftharpoons\mathrm{G}},\vec{M}_{\mathrm{M}\rightleftharpoons\mathrm{A}}+\vec{M}_{\mathrm{A}\rightleftharpoons\mathrm{G}}\right)\right]_{\vec{r}_{\mathrm{A}}\to\vec{r}}=\hat{n}_{\to\mathrm{A}}\times\vec{M}_{\mathrm{A}\rightleftharpoons\mathrm{G}}\left(\vec{r}\right) \quad , \quad \vec{r}\in\mathbb{S}_{\mathrm{A}\rightleftharpoons\mathrm{G}} \tag{8-88b}$$

about currents $\{\vec{J}_{\mathrm{M}\rightleftharpoons\mathrm{A}},\vec{M}_{\mathrm{M}\rightleftharpoons\mathrm{A}}\}$, $\vec{J}_{\mathrm{A}}$, and $\{\vec{J}_{\mathrm{A}\rightleftharpoons\mathrm{G}},\vec{M}_{\mathrm{A}\rightleftharpoons\mathrm{G}}\}$.

The above Eqs. (8-80a)~(8-88b) constitute a complete mathematical description for the *modal space* of the TARA system shown in Figs. 8-21 and 8-22.

### 8.3.3.2 Matrix Equations

If the above-mentioned currents are expanded in terms of some proper basis functions as follows:

$$\vec{C}^{x}\left(\vec{r}\right)=\sum_{\xi}a_{\xi}^{\vec{C}^{x}}\vec{b}_{\xi}^{\vec{C}^{x}}=\underbrace{\begin{bmatrix}\vec{b}_{1}^{\vec{C}^{x}} & \vec{b}_{2}^{\vec{C}^{x}} & \cdots\end{bmatrix}}_{\bar{\boldsymbol{B}}^{\vec{C}^{x}}}\cdot\underbrace{\begin{bmatrix}a_{1}^{\vec{C}^{x}}\\ a_{2}^{\vec{C}^{x}}\\ \vdots\end{bmatrix}}_{\bar{a}^{\vec{C}^{x}}} \quad , \quad \vec{r}\in\mathbb{S}^{x} \tag{8-89}$$

and Eqs. (4-80a)~(4-88b) are tested with $\{\vec{b}_{\xi}^{\vec{M}^{\mathrm{G}\rightleftharpoons\mathrm{A}}}\}$, $\{\vec{b}_{\xi}^{\vec{J}^{\mathrm{G}\rightleftharpoons\mathrm{A}}}\}$, $\{\vec{b}_{\xi}^{\vec{J}^{\mathrm{A}}}\}$, $\{\vec{b}_{\xi}^{\vec{J}^{\mathrm{A}\rightleftharpoons\mathrm{M}}}\}$, $\{\vec{b}_{\xi}^{\vec{M}^{\mathrm{A}\rightleftharpoons\mathrm{M}}}\}$, $\{\vec{b}_{\xi}^{\vec{J}^{\mathrm{M}}}\}$, $\{\vec{b}_{\xi}^{\vec{J}}\}$, $\{\vec{b}_{\xi}^{\vec{M}}\}$, $\{\vec{b}_{\xi}^{\vec{J}_{\mathrm{M}}}\}$, $\{\vec{b}_{\xi}^{\vec{J}_{\mathrm{M}\rightleftharpoons\mathrm{A}}}\}$, $\{\vec{b}_{\xi}^{\vec{M}_{\mathrm{M}\rightleftharpoons\mathrm{A}}}\}$, $\{\vec{b}_{\xi}^{\vec{J}_{\mathrm{A}}}\}$, $\{\vec{b}_{\xi}^{\vec{M}_{\mathrm{A}\rightleftharpoons\mathrm{G}}}\}$, and $\{\vec{b}_{\xi}^{\vec{J}_{\mathrm{A}\rightleftharpoons\mathrm{G}}}\}$ respectively, then the integral equations are immediately discretized into the following *matrix equations*

$$\begin{aligned}&\bar{\bar{Z}}^{\vec{M}^{\mathrm{G}\rightleftharpoons\mathrm{A}}\vec{J}^{\mathrm{G}\rightleftharpoons\mathrm{A}}}\cdot\bar{a}^{\vec{J}^{\mathrm{G}\rightleftharpoons\mathrm{A}}}+\bar{\bar{Z}}^{\vec{M}^{\mathrm{G}\rightleftharpoons\mathrm{A}}\vec{J}^{\mathrm{A}}}\cdot\bar{a}^{\vec{J}^{\mathrm{A}}}+\bar{\bar{Z}}^{\vec{M}^{\mathrm{G}\rightleftharpoons\mathrm{A}}\vec{J}^{\mathrm{A}\rightleftharpoons\mathrm{M}}}\cdot\bar{a}^{\vec{J}^{\mathrm{A}\rightleftharpoons\mathrm{M}}}+\bar{\bar{Z}}^{\vec{M}^{\mathrm{G}\rightleftharpoons\mathrm{A}}\vec{M}^{\mathrm{G}\rightleftharpoons\mathrm{A}}}\cdot\bar{a}^{\vec{M}^{\mathrm{G}\rightleftharpoons\mathrm{A}}}\\&+\bar{\bar{Z}}^{\vec{M}^{\mathrm{G}\rightleftharpoons\mathrm{A}}\vec{M}^{\mathrm{A}\rightleftharpoons\mathrm{M}}}\cdot\bar{a}^{\vec{M}^{\mathrm{A}\rightleftharpoons\mathrm{M}}}=0\end{aligned} \tag{8-90a}$$

$$\begin{aligned}&\bar{\bar{Z}}^{\vec{J}^{\mathrm{G}\rightleftharpoons\mathrm{A}}\vec{J}^{\mathrm{G}\rightleftharpoons\mathrm{A}}}\cdot\bar{a}^{\vec{J}^{\mathrm{G}\rightleftharpoons\mathrm{A}}}+\bar{\bar{Z}}^{\vec{J}^{\mathrm{G}\rightleftharpoons\mathrm{A}}\vec{J}^{\mathrm{A}}}\cdot\bar{a}^{\vec{J}^{\mathrm{A}}}+\bar{\bar{Z}}^{\vec{J}^{\mathrm{G}\rightleftharpoons\mathrm{A}}\vec{J}^{\mathrm{A}\rightleftharpoons\mathrm{M}}}\cdot\bar{a}^{\vec{J}^{\mathrm{A}\rightleftharpoons\mathrm{M}}}+\bar{\bar{Z}}^{\vec{J}^{\mathrm{G}\rightleftharpoons\mathrm{A}}\vec{M}^{\mathrm{G}\rightleftharpoons\mathrm{A}}}\cdot\bar{a}^{\vec{M}^{\mathrm{G}\rightleftharpoons\mathrm{A}}}\\&+\bar{\bar{Z}}^{\vec{J}^{\mathrm{G}\rightleftharpoons\mathrm{A}}\vec{M}^{\mathrm{A}\rightleftharpoons\mathrm{M}}}\cdot\bar{a}^{\vec{M}^{\mathrm{A}\rightleftharpoons\mathrm{M}}}=0\end{aligned} \tag{8-90b}$$

and

$$\overline{\overline{Z}}^{\vec{J}^{A}\vec{J}^{G\rightleftharpoons A}}\cdot\overline{a}^{\vec{J}^{G\rightleftharpoons A}}+\overline{\overline{Z}}^{\vec{J}^{A}\vec{J}^{A}}\cdot\overline{a}^{\vec{J}^{A}}+\overline{\overline{Z}}^{\vec{J}^{A}\vec{J}^{A\rightleftharpoons M}}\cdot\overline{a}^{\vec{J}^{A\rightleftharpoons M}}+\overline{\overline{Z}}^{\vec{J}^{A}\vec{M}^{G\rightleftharpoons A}}\cdot\overline{a}^{\vec{M}^{G\rightleftharpoons A}}+\overline{\overline{Z}}^{\vec{J}^{A}\vec{M}^{A\rightleftharpoons M}}\cdot\overline{a}^{\vec{M}^{A\rightleftharpoons M}}$$
$$=0 \qquad (8\text{-}91)$$

and

$$\overline{\overline{Z}}^{\vec{J}^{A\rightleftharpoons M}\vec{J}^{G\rightleftharpoons A}}\cdot\overline{a}^{\vec{J}^{G\rightleftharpoons A}}+\overline{\overline{Z}}^{\vec{J}^{A\rightleftharpoons M}\vec{J}^{A}}\cdot\overline{a}^{\vec{J}^{A}}+\overline{\overline{Z}}^{\vec{J}^{A\rightleftharpoons M}\vec{J}^{A\rightleftharpoons M}}\cdot\overline{a}^{\vec{J}^{A\rightleftharpoons M}}+\overline{\overline{Z}}^{\vec{J}^{A\rightleftharpoons M}\vec{J}^{M}}\cdot\overline{a}^{\vec{J}^{M}}+\overline{\overline{Z}}^{\vec{J}^{A\rightleftharpoons M}\vec{J}}\cdot\overline{a}^{\vec{J}}$$
$$+\overline{\overline{Z}}^{\vec{J}^{A\rightleftharpoons M}\vec{J}_{M}}\cdot\overline{a}^{\vec{J}_{M}}+\overline{\overline{Z}}^{\vec{J}^{A\rightleftharpoons M}\vec{J}_{M\rightleftharpoons A}}\cdot\overline{a}^{\vec{J}_{M\rightleftharpoons A}}+\overline{\overline{Z}}^{\vec{J}^{A\rightleftharpoons M}\vec{M}^{G\rightleftharpoons A}}\cdot\overline{a}^{\vec{M}^{G\rightleftharpoons A}}+\overline{\overline{Z}}^{\vec{J}^{A\rightleftharpoons M}\vec{M}^{A\rightleftharpoons M}}\cdot\overline{a}^{\vec{M}^{A\rightleftharpoons M}}$$
$$+\overline{\overline{Z}}^{\vec{J}^{A\rightleftharpoons M}\vec{M}}\cdot\overline{a}^{\vec{M}}+\overline{\overline{Z}}^{\vec{J}^{A\rightleftharpoons M}\vec{M}_{M\rightleftharpoons A}}\cdot\overline{a}^{\vec{M}_{M\rightleftharpoons A}}=0 \qquad (8\text{-}92a)$$

$$\overline{\overline{Z}}^{\vec{M}^{A\rightleftharpoons M}\vec{J}^{G\rightleftharpoons A}}\cdot\overline{a}^{\vec{J}^{G\rightleftharpoons A}}+\overline{\overline{Z}}^{\vec{M}^{A\rightleftharpoons M}\vec{J}^{A}}\cdot\overline{a}^{\vec{J}^{A}}+\overline{\overline{Z}}^{\vec{M}^{A\rightleftharpoons M}\vec{J}^{A\rightleftharpoons M}}\cdot\overline{a}^{\vec{J}^{A\rightleftharpoons M}}+\overline{\overline{Z}}^{\vec{M}^{A\rightleftharpoons M}\vec{J}^{M}}\cdot\overline{a}^{\vec{J}^{M}}+\overline{\overline{Z}}^{\vec{M}^{A\rightleftharpoons M}\vec{J}}\cdot\overline{a}^{\vec{J}}$$
$$+\overline{\overline{Z}}^{\vec{M}^{A\rightleftharpoons M}\vec{J}_{M}}\cdot\overline{a}^{\vec{J}_{M}}+\overline{\overline{Z}}^{\vec{M}^{A\rightleftharpoons M}\vec{J}_{M\rightleftharpoons A}}\cdot\overline{a}^{\vec{J}_{M\rightleftharpoons A}}+\overline{\overline{Z}}^{\vec{M}^{A\rightleftharpoons M}\vec{M}^{G\rightleftharpoons A}}\cdot\overline{a}^{\vec{M}^{G\rightleftharpoons A}}+\overline{\overline{Z}}^{\vec{M}^{A\rightleftharpoons M}\vec{M}^{A\rightleftharpoons M}}\cdot\overline{a}^{\vec{M}^{A\rightleftharpoons M}}$$
$$+\overline{\overline{Z}}^{\vec{M}^{A\rightleftharpoons M}\vec{M}}\cdot\overline{a}^{\vec{M}}+\overline{\overline{Z}}^{\vec{M}^{A\rightleftharpoons M}\vec{M}_{M\rightleftharpoons A}}\cdot\overline{a}^{\vec{M}_{M\rightleftharpoons A}}=0 \qquad (8\text{-}92b)$$

and

$$\overline{\overline{Z}}^{\vec{J}^{M}\vec{J}^{A\rightleftharpoons M}}\cdot\overline{a}^{\vec{J}^{A\rightleftharpoons M}}+\overline{\overline{Z}}^{\vec{J}^{M}\vec{J}^{M}}\cdot\overline{a}^{\vec{J}^{M}}+\overline{\overline{Z}}^{\vec{J}^{M}\vec{J}}\cdot\overline{a}^{\vec{J}}+\overline{\overline{Z}}^{\vec{J}^{M}\vec{J}_{M}}\cdot\overline{a}^{\vec{J}_{M}}+\overline{\overline{Z}}^{\vec{J}^{M}\vec{J}_{M\rightleftharpoons A}}\cdot\overline{a}^{\vec{J}_{M\rightleftharpoons A}}$$
$$+\overline{\overline{Z}}^{\vec{J}^{M}\vec{M}^{A\rightleftharpoons M}}\cdot\overline{a}^{\vec{M}^{A\rightleftharpoons M}}+\overline{\overline{Z}}^{\vec{J}^{M}\vec{M}}\cdot\overline{a}^{\vec{M}}+\overline{\overline{Z}}^{\vec{J}^{M}\vec{M}_{M\rightleftharpoons A}}\cdot\overline{a}^{\vec{M}_{M\rightleftharpoons A}}=0 \qquad (8\text{-}93)$$

and

$$\overline{\overline{Z}}^{\vec{J}\vec{J}^{A\rightleftharpoons M}}\cdot\overline{a}^{\vec{J}^{A\rightleftharpoons M}}+\overline{\overline{Z}}^{\vec{J}\vec{J}^{M}}\cdot\overline{a}^{\vec{J}^{M}}+\overline{\overline{Z}}^{\vec{J}\vec{J}}\cdot\overline{a}^{\vec{J}}+\overline{\overline{Z}}^{\vec{J}\vec{J}_{M}}\cdot\overline{a}^{\vec{J}_{M}}+\overline{\overline{Z}}^{\vec{J}\vec{J}_{M\rightleftharpoons A}}\cdot\overline{a}^{\vec{J}_{M\rightleftharpoons A}}+\overline{\overline{Z}}^{\vec{J}\vec{M}^{A\rightleftharpoons M}}\cdot\overline{a}^{\vec{M}^{A\rightleftharpoons M}}$$
$$+\overline{\overline{Z}}^{\vec{J}\vec{M}}\cdot\overline{a}^{\vec{M}}+\overline{\overline{Z}}^{\vec{J}\vec{M}_{M\rightleftharpoons A}}\cdot\overline{a}^{\vec{M}_{M\rightleftharpoons A}}=0 \qquad (8\text{-}94a)$$

$$\overline{\overline{Z}}^{\vec{M}\vec{J}^{A\rightleftharpoons M}}\cdot\overline{a}^{\vec{J}^{A\rightleftharpoons M}}+\overline{\overline{Z}}^{\vec{M}\vec{J}^{M}}\cdot\overline{a}^{\vec{J}^{M}}+\overline{\overline{Z}}^{\vec{M}\vec{J}}\cdot\overline{a}^{\vec{J}}+\overline{\overline{Z}}^{\vec{M}\vec{J}_{M}}\cdot\overline{a}^{\vec{J}_{M}}+\overline{\overline{Z}}^{\vec{M}\vec{J}_{M\rightleftharpoons A}}\cdot\overline{a}^{\vec{J}_{M\rightleftharpoons A}}+\overline{\overline{Z}}^{\vec{M}\vec{M}^{A\rightleftharpoons M}}\cdot\overline{a}^{\vec{M}^{A\rightleftharpoons M}}$$
$$+\overline{\overline{Z}}^{\vec{M}\vec{M}}\cdot\overline{a}^{\vec{M}}+\overline{\overline{Z}}^{\vec{M}\vec{M}_{M\rightleftharpoons A}}\cdot\overline{a}^{\vec{M}_{M\rightleftharpoons A}}=0 \qquad (8\text{-}94b)$$

and

$$\overline{\overline{Z}}^{\vec{J}_{M}\vec{J}^{A\rightleftharpoons M}}\cdot\overline{a}^{\vec{J}^{A\rightleftharpoons M}}+\overline{\overline{Z}}^{\vec{J}_{M}\vec{J}^{M}}\cdot\overline{a}^{\vec{J}^{M}}+\overline{\overline{Z}}^{\vec{J}_{M}\vec{J}}\cdot\overline{a}^{\vec{J}}+\overline{\overline{Z}}^{\vec{J}_{M}\vec{J}_{M}}\cdot\overline{a}^{\vec{J}_{M}}+\overline{\overline{Z}}^{\vec{J}_{M}\vec{J}_{M\rightleftharpoons A}}\cdot\overline{a}^{\vec{J}_{M\rightleftharpoons A}}+\overline{\overline{Z}}^{\vec{J}_{M}\vec{M}^{A\rightleftharpoons M}}\cdot\overline{a}^{\vec{M}^{A\rightleftharpoons M}}$$
$$+\overline{\overline{Z}}^{\vec{J}_{M}\vec{M}}\cdot\overline{a}^{\vec{M}}+\overline{\overline{Z}}^{\vec{J}_{M}\vec{M}_{M\rightleftharpoons A}}\cdot\overline{a}^{\vec{M}_{M\rightleftharpoons A}}=0 \qquad (8\text{-}95)$$

and

$$\overline{\overline{Z}}^{\vec{J}_{M\rightleftharpoons A}\vec{J}^{A\rightleftharpoons M}}\cdot\overline{a}^{\vec{J}^{A\rightleftharpoons M}}+\overline{\overline{Z}}^{\vec{J}_{M\rightleftharpoons A}\vec{J}^{M}}\cdot\overline{a}^{\vec{J}^{M}}+\overline{\overline{Z}}^{\vec{J}_{M\rightleftharpoons A}\vec{J}}\cdot\overline{a}^{\vec{J}}+\overline{\overline{Z}}^{\vec{J}_{M\rightleftharpoons A}\vec{J}_{M}}\cdot\overline{a}^{\vec{J}_{M}}+\overline{\overline{Z}}^{\vec{J}_{M\rightleftharpoons A}\vec{J}_{M\rightleftharpoons A}}\cdot\overline{a}^{\vec{J}_{M\rightleftharpoons A}}$$
$$+\overline{\overline{Z}}^{\vec{J}_{M\rightleftharpoons A}\vec{J}_{A}}\cdot\overline{a}^{\vec{J}_{A}}+\overline{\overline{Z}}^{\vec{J}_{M\rightleftharpoons A}\vec{J}_{A\rightleftharpoons G}}\cdot\overline{a}^{\vec{J}_{A\rightleftharpoons G}}+\overline{\overline{Z}}^{\vec{J}_{M\rightleftharpoons A}\vec{M}^{A\rightleftharpoons M}}\cdot\overline{a}^{\vec{M}^{A\rightleftharpoons M}}+\overline{\overline{Z}}^{\vec{J}_{M\rightleftharpoons A}\vec{M}}\cdot\overline{a}^{\vec{M}}$$
$$+\overline{\overline{Z}}^{\vec{J}_{M\rightleftharpoons A}\vec{M}_{M\rightleftharpoons A}}\cdot\overline{a}^{\vec{M}_{M\rightleftharpoons A}}+\overline{\overline{Z}}^{\vec{J}_{M\rightleftharpoons A}\vec{M}_{A\rightleftharpoons G}}\cdot\overline{a}^{\vec{M}_{A\rightleftharpoons G}}=0 \qquad (8\text{-}96a)$$

$$\overline{\overline{Z}}^{\vec{M}_{M\rightleftharpoons A}\vec{J}^{A\rightleftharpoons M}}\cdot\overline{a}^{\vec{J}^{A\rightleftharpoons M}}+\overline{\overline{Z}}^{\vec{M}_{M\rightleftharpoons A}\vec{J}^{M}}\cdot\overline{a}^{\vec{J}^{M}}+\overline{\overline{Z}}^{\vec{M}_{M\rightleftharpoons A}\vec{J}}\cdot\overline{a}^{\vec{J}}+\overline{\overline{Z}}^{\vec{M}_{M\rightleftharpoons A}\vec{J}_{M}}\cdot\overline{a}^{\vec{J}_{M}}+\overline{\overline{Z}}^{\vec{M}_{M\rightleftharpoons A}\vec{J}_{M\rightleftharpoons A}}\cdot\overline{a}^{\vec{J}_{M\rightleftharpoons A}}$$
$$+\overline{\overline{Z}}^{\vec{M}_{M\rightleftharpoons A}\vec{J}_{A}}\cdot\overline{a}^{\vec{J}_{A}}+\overline{\overline{Z}}^{\vec{M}_{M\rightleftharpoons A}\vec{J}_{A\rightleftharpoons G}}\cdot\overline{a}^{\vec{J}_{A\rightleftharpoons G}}+\overline{\overline{Z}}^{\vec{M}_{M\rightleftharpoons A}\vec{M}^{A\rightleftharpoons M}}\cdot\overline{a}^{\vec{M}^{A\rightleftharpoons M}}+\overline{\overline{Z}}^{\vec{M}_{M\rightleftharpoons A}\vec{M}}\cdot\overline{a}^{\vec{M}}$$
$$+\overline{\overline{Z}}^{\vec{M}_{M\rightleftharpoons A}\vec{M}_{M\rightleftharpoons A}}\cdot\overline{a}^{\vec{M}_{M\rightleftharpoons A}}+\overline{\overline{Z}}^{\vec{M}_{M\rightleftharpoons A}\vec{M}_{A\rightleftharpoons G}}\cdot\overline{a}^{\vec{M}_{A\rightleftharpoons G}}=0 \qquad (8\text{-}96b)$$

and

$$\overline{\overline{Z}}^{\vec{J}_{A}\vec{J}_{M\rightleftharpoons A}}\cdot\overline{a}^{\vec{J}_{M\rightleftharpoons A}}+\overline{\overline{Z}}^{\vec{J}_{A}\vec{J}_{A}}\cdot\overline{a}^{\vec{J}_{A}}+\overline{\overline{Z}}^{\vec{J}_{A}\vec{J}_{A\rightleftharpoons G}}\cdot\overline{a}^{\vec{J}_{A\rightleftharpoons G}}+\overline{\overline{Z}}^{\vec{J}_{A}\vec{M}_{M\rightleftharpoons A}}\cdot\overline{a}^{\vec{M}_{M\rightleftharpoons A}}+\overline{\overline{Z}}^{\vec{J}_{A}\vec{M}_{A\rightleftharpoons G}}\cdot\overline{a}^{\vec{M}_{A\rightleftharpoons G}}=0 \quad (8\text{-}97)$$

and

$$\overline{\overline{Z}}^{\vec{M}_{\mathrm{A}\rightleftharpoons\mathrm{G}}\vec{J}_{\mathrm{M}\rightleftharpoons\mathrm{A}}}\cdot\overline{a}^{\vec{J}_{\mathrm{M}\rightleftharpoons\mathrm{A}}}+\overline{\overline{Z}}^{\vec{M}_{\mathrm{A}\rightleftharpoons\mathrm{G}}\vec{J}_{\mathrm{A}}}\cdot\overline{a}^{\vec{J}_{\mathrm{A}}}+\overline{\overline{Z}}^{\vec{M}_{\mathrm{A}\rightleftharpoons\mathrm{G}}\vec{J}_{\mathrm{A}\rightleftharpoons\mathrm{G}}}\cdot\overline{a}^{\vec{J}_{\mathrm{A}\rightleftharpoons\mathrm{G}}}+\overline{\overline{Z}}^{\vec{M}_{\mathrm{A}\rightleftharpoons\mathrm{G}}\vec{M}_{\mathrm{M}\rightleftharpoons\mathrm{A}}}\cdot\overline{a}^{\vec{M}_{\mathrm{M}\rightleftharpoons\mathrm{A}}}$$
$$+\overline{\overline{Z}}^{\vec{M}_{\mathrm{A}\rightleftharpoons\mathrm{G}}\vec{M}_{\mathrm{A}\rightleftharpoons\mathrm{G}}}\cdot\overline{a}^{\vec{M}_{\mathrm{A}\rightleftharpoons\mathrm{G}}}=0 \tag{8-98a}$$

$$\overline{\overline{Z}}^{\vec{J}_{\mathrm{A}\rightleftharpoons\mathrm{G}}\vec{J}_{\mathrm{M}\rightleftharpoons\mathrm{A}}}\cdot\overline{a}^{\vec{J}_{\mathrm{M}\rightleftharpoons\mathrm{A}}}+\overline{\overline{Z}}^{\vec{J}_{\mathrm{A}\rightleftharpoons\mathrm{G}}\vec{J}_{\mathrm{A}}}\cdot\overline{a}^{\vec{J}_{\mathrm{A}}}+\overline{\overline{Z}}^{\vec{J}_{\mathrm{A}\rightleftharpoons\mathrm{G}}\vec{J}_{\mathrm{A}\rightleftharpoons\mathrm{G}}}\cdot\overline{a}^{\vec{J}_{\mathrm{A}\rightleftharpoons\mathrm{G}}}+\overline{\overline{Z}}^{\vec{J}_{\mathrm{A}\rightleftharpoons\mathrm{G}}\vec{M}_{\mathrm{M}\rightleftharpoons\mathrm{A}}}\cdot\overline{a}^{\vec{M}_{\mathrm{M}\rightleftharpoons\mathrm{A}}}$$
$$+\overline{\overline{Z}}^{\vec{J}_{\mathrm{A}\rightleftharpoons\mathrm{G}}\vec{M}_{\mathrm{A}\rightleftharpoons\mathrm{G}}}\cdot\overline{a}^{\vec{M}_{\mathrm{A}\rightleftharpoons\mathrm{G}}}=0 \tag{8-98b}$$

The formulations used to calculate the elements of the matrices in Eq. (8-90a) are as follows:

$$z_{\xi\zeta}^{\vec{M}^{\mathrm{G}\rightleftharpoons\mathrm{A}}\vec{J}^{\mathrm{G}\rightleftharpoons\mathrm{A}}} = \left\langle \vec{b}_{\xi}^{\vec{M}^{\mathrm{G}\rightleftharpoons\mathrm{A}}}, \mathcal{H}_{\mathrm{T}}\left(\vec{b}_{\zeta}^{\vec{J}^{\mathrm{G}\rightleftharpoons\mathrm{A}}}\right)\right\rangle_{\mathbb{S}^{\mathrm{G}\rightleftharpoons\mathrm{A}}} + \left\langle \vec{b}_{\xi}^{\vec{M}^{\mathrm{G}\rightleftharpoons\mathrm{A}}}, \hat{n}^{\rightarrow\mathrm{A}} \times \vec{b}_{\zeta}^{\vec{J}^{\mathrm{G}\rightleftharpoons\mathrm{A}}}\right\rangle_{\mathbb{S}^{\mathrm{G}\rightleftharpoons\mathrm{A}}} \tag{8-99a}$$

$$z_{\xi\zeta}^{\vec{M}^{\mathrm{G}\rightleftharpoons\mathrm{A}}\vec{J}^{\mathrm{A}}} = \left\langle \vec{b}_{\xi}^{\vec{M}^{\mathrm{G}\rightleftharpoons\mathrm{A}}}, \mathcal{H}_{\mathrm{T}}\left(\vec{b}_{\zeta}^{\vec{J}^{\mathrm{A}}}\right)\right\rangle_{\mathbb{S}^{\mathrm{G}\rightleftharpoons\mathrm{A}}} \tag{8-99b}$$

$$z_{\xi\zeta}^{\vec{M}^{\mathrm{G}\rightleftharpoons\mathrm{A}}\vec{J}^{\mathrm{A}\rightleftharpoons\mathrm{M}}} = \left\langle \vec{b}_{\xi}^{\vec{M}^{\mathrm{G}\rightleftharpoons\mathrm{A}}}, \mathcal{H}_{\mathrm{T}}\left(\vec{b}_{\zeta}^{\vec{J}^{\mathrm{A}\rightleftharpoons\mathrm{M}}}\right)\right\rangle_{\mathbb{S}^{\mathrm{G}\rightleftharpoons\mathrm{A}}} \tag{8-99c}$$

$$z_{\xi\zeta}^{\vec{M}^{\mathrm{G}\rightleftharpoons\mathrm{A}}\vec{M}^{\mathrm{G}\rightleftharpoons\mathrm{A}}} = \left\langle \vec{b}_{\xi}^{\vec{M}^{\mathrm{G}\rightleftharpoons\mathrm{A}}}, \mathcal{H}_{\mathrm{T}}\left(\vec{b}_{\zeta}^{\vec{M}^{\mathrm{G}\rightleftharpoons\mathrm{A}}}\right)\right\rangle_{\mathbb{S}^{\mathrm{G}\rightleftharpoons\mathrm{A}}} \tag{8-99d}$$

$$z_{\xi\zeta}^{\vec{M}^{\mathrm{G}\rightleftharpoons\mathrm{A}}\vec{M}^{\mathrm{A}\rightleftharpoons\mathrm{M}}} = \left\langle \vec{b}_{\xi}^{\vec{M}^{\mathrm{G}\rightleftharpoons\mathrm{A}}}, \mathcal{H}_{\mathrm{T}}\left(\vec{b}_{\zeta}^{\vec{M}^{\mathrm{A}\rightleftharpoons\mathrm{M}}}\right)\right\rangle_{\mathbb{S}^{\mathrm{G}\rightleftharpoons\mathrm{A}}} \tag{8-99e}$$

where the integral surface $\mathbb{S}^{\mathrm{G}\rightleftharpoons\mathrm{A}}$ is shown in the following Fig. 8-23

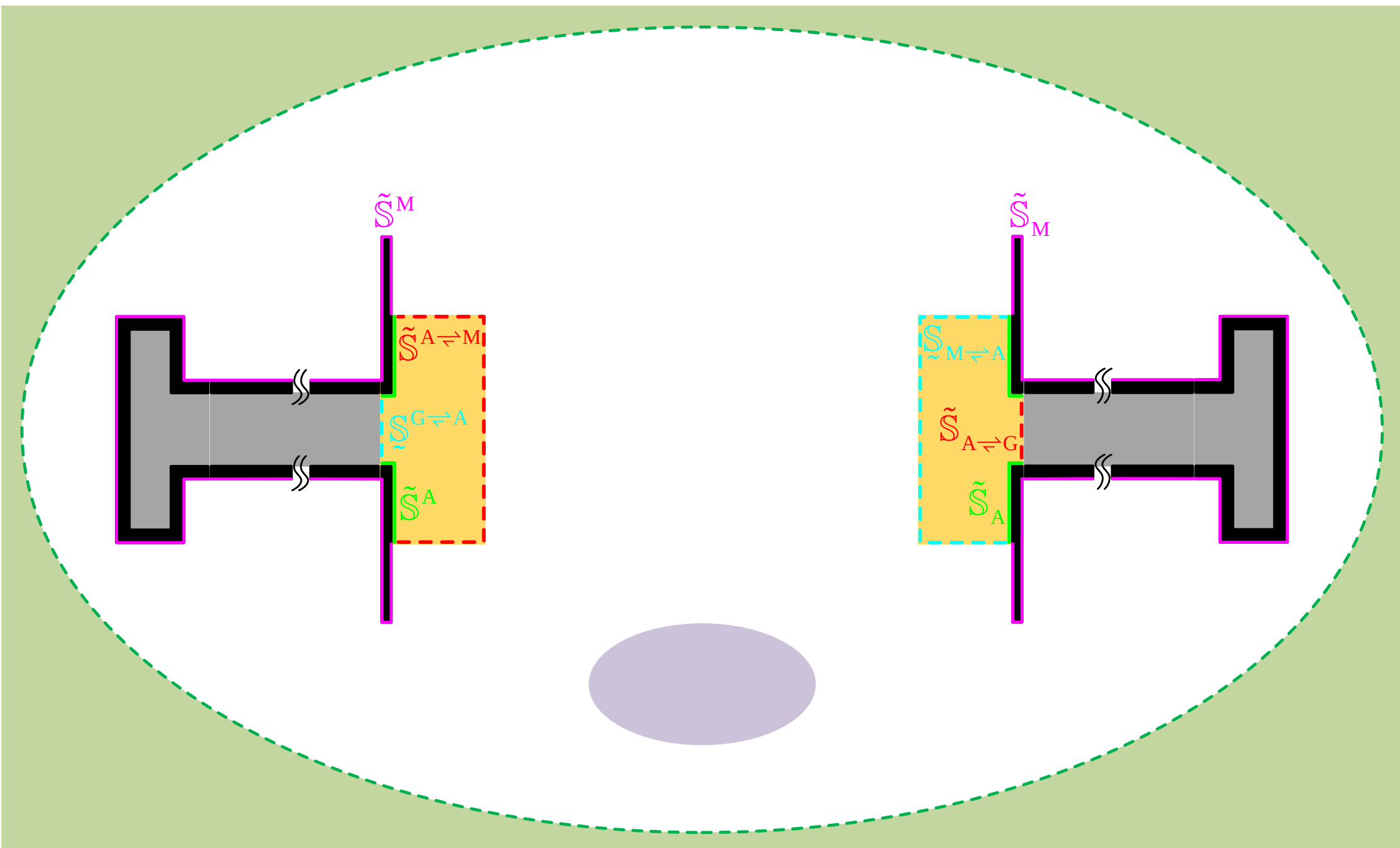

Figure 8-23 Integral surfaces used in Eqs. (8-99)~(8-112)

The formulations for calculating the elements of the matrices in Eq. (8-90b) are as follows:

$$z_{\xi\zeta}^{\vec{J}^{\text{G}\rightleftharpoons\text{A}}\vec{J}^{\text{G}\rightleftharpoons\text{A}}} = \left\langle \vec{b}_{\xi}^{\vec{J}^{\text{G}\rightleftharpoons\text{A}}}, \mathcal{E}_{\text{T}}\left(\vec{b}_{\zeta}^{\vec{J}^{\text{G}\rightleftharpoons\text{A}}}\right)\right\rangle_{\mathbb{S}^{\text{G}\rightleftharpoons\text{A}}} \tag{8-100a}$$

$$z_{\xi\zeta}^{\vec{J}^{\text{G}\rightleftharpoons\text{A}}\vec{J}^{\text{A}}} = \left\langle \vec{b}_{\xi}^{\vec{J}^{\text{G}\rightleftharpoons\text{A}}}, \mathcal{E}_{\text{T}}\left(\vec{b}_{\zeta}^{\vec{J}^{\text{A}}}\right)\right\rangle_{\mathbb{S}^{\text{G}\rightleftharpoons\text{A}}} \tag{8-100b}$$

$$z_{\xi\zeta}^{\vec{J}^{\text{G}\rightleftharpoons\text{A}}\vec{J}^{\text{A}\rightleftharpoons\text{M}}} = \left\langle \vec{b}_{\xi}^{\vec{J}^{\text{G}\rightleftharpoons\text{A}}}, \mathcal{E}_{\text{T}}\left(\vec{b}_{\zeta}^{\vec{J}^{\text{A}\rightleftharpoons\text{M}}}\right)\right\rangle_{\mathbb{S}^{\text{G}\rightleftharpoons\text{A}}} \tag{8-100c}$$

$$z_{\xi\zeta}^{\vec{J}^{\text{G}\rightleftharpoons\text{A}}\vec{M}^{\text{G}\rightleftharpoons\text{A}}} = \left\langle \vec{b}_{\xi}^{\vec{J}^{\text{G}\rightleftharpoons\text{A}}}, \mathcal{E}_{\text{T}}\left(\vec{b}_{\zeta}^{\vec{M}^{\text{G}\rightleftharpoons\text{A}}}\right)\right\rangle_{\mathbb{S}^{\text{G}\rightleftharpoons\text{A}}} + \left\langle \vec{b}_{\xi}^{\vec{J}^{\text{G}\rightleftharpoons\text{A}}}, \vec{b}_{\zeta}^{\vec{M}^{\text{G}\rightleftharpoons\text{A}}} \times \hat{n}^{\rightarrow\text{A}}\right\rangle_{\mathbb{S}^{\text{G}\rightleftharpoons\text{A}}} \tag{8-100d}$$

$$z_{\xi\zeta}^{\vec{J}^{\text{G}\rightleftharpoons\text{A}}\vec{M}^{\text{A}\rightleftharpoons\text{M}}} = \left\langle \vec{b}_{\xi}^{\vec{J}^{\text{G}\rightleftharpoons\text{A}}}, \mathcal{E}_{\text{T}}\left(\vec{b}_{\zeta}^{\vec{M}^{\text{A}\rightleftharpoons\text{M}}}\right)\right\rangle_{\mathbb{S}^{\text{G}\rightleftharpoons\text{A}}} \tag{8-100e}$$

The formulations used to calculate the elements of the matrices in Eq. (8-91) are as follows:

$$z_{\xi\zeta}^{\vec{J}^{\text{A}}\vec{J}^{\text{G}\rightleftharpoons\text{A}}} = \left\langle \vec{b}_{\xi}^{\vec{J}^{\text{A}}}, \mathcal{E}_{\text{T}}\left(\vec{b}_{\zeta}^{\vec{J}^{\text{G}\rightleftharpoons\text{A}}}\right)\right\rangle_{\tilde{\mathbb{S}}^{\text{A}}} \tag{8-101a}$$

$$z_{\xi\zeta}^{\vec{J}^{\text{A}}\vec{J}^{\text{A}}} = \left\langle \vec{b}_{\xi}^{\vec{J}^{\text{A}}}, \mathcal{E}_{\text{T}}\left(\vec{b}_{\zeta}^{\vec{J}^{\text{A}}}\right)\right\rangle_{\tilde{\mathbb{S}}^{\text{A}}} \tag{8-101b}$$

$$z_{\xi\zeta}^{\vec{J}^{\text{A}}\vec{J}^{\text{A}\rightleftharpoons\text{M}}} = \left\langle \vec{b}_{\xi}^{\vec{J}^{\text{A}}}, \mathcal{E}_{\text{T}}\left(\vec{b}_{\zeta}^{\vec{J}^{\text{A}\rightleftharpoons\text{M}}}\right)\right\rangle_{\tilde{\mathbb{S}}^{\text{A}}} \tag{8-101c}$$

$$z_{\xi\zeta}^{\vec{J}^{\text{A}}\vec{M}^{\text{G}\rightleftharpoons\text{A}}} = \left\langle \vec{b}_{\xi}^{\vec{J}^{\text{A}}}, \mathcal{E}_{\text{T}}\left(\vec{b}_{\zeta}^{\vec{M}^{\text{G}\rightleftharpoons\text{A}}}\right)\right\rangle_{\tilde{\mathbb{S}}^{\text{A}}} \tag{8-101d}$$

$$z_{\xi\zeta}^{\vec{J}^{\text{A}}\vec{M}^{\text{A}\rightleftharpoons\text{M}}} = \left\langle \vec{b}_{\xi}^{\vec{J}^{\text{A}}}, \mathcal{E}_{\text{T}}\left(\vec{b}_{\zeta}^{\vec{M}^{\text{A}\rightleftharpoons\text{M}}}\right)\right\rangle_{\tilde{\mathbb{S}}^{\text{A}}} \tag{8-101e}$$

where the integral surface $\tilde{\mathbb{S}}^{\text{A}}$ is shown in Fig. 8-23. The formulations used to calculate the elements of the matrices in Eq. (8-92a) are as follows:

$$z_{\xi\zeta}^{\vec{J}^{\text{A}\rightleftharpoons\text{M}}\vec{J}^{\text{G}\rightleftharpoons\text{A}}} = \left\langle \vec{b}_{\xi}^{\vec{J}^{\text{A}\rightleftharpoons\text{M}}}, \mathcal{E}_{\text{T}}\left(\vec{b}_{\zeta}^{\vec{J}^{\text{G}\rightleftharpoons\text{A}}}\right)\right\rangle_{\tilde{\mathbb{S}}^{\text{A}\rightleftharpoons\text{M}}} \tag{8-102a}$$

$$z_{\xi\zeta}^{\vec{J}^{\text{A}\rightleftharpoons\text{M}}\vec{J}^{\text{A}}} = \left\langle \vec{b}_{\xi}^{\vec{J}^{\text{A}\rightleftharpoons\text{M}}}, \mathcal{E}_{\text{T}}\left(\vec{b}_{\zeta}^{\vec{J}^{\text{A}}}\right)\right\rangle_{\tilde{\mathbb{S}}^{\text{A}\rightleftharpoons\text{M}}} \tag{8-102b}$$

$$z_{\xi\zeta}^{\vec{J}^{\text{A}\rightleftharpoons\text{M}}\vec{J}^{\text{A}\rightleftharpoons\text{M}}} = \left\langle \vec{b}_{\xi}^{\vec{J}^{\text{A}\rightleftharpoons\text{M}}}, \mathcal{E}_{\text{T}}\left(\vec{b}_{\zeta}^{\vec{J}^{\text{A}\rightleftharpoons\text{M}}}\right)\right\rangle_{\tilde{\mathbb{S}}^{\text{A}\rightleftharpoons\text{M}}} - \left\langle \vec{b}_{\xi}^{\vec{J}^{\text{A}\rightleftharpoons\text{M}}}, -j\omega\mu_0\mathcal{L}_0\left(-\vec{b}_{\zeta}^{\vec{J}^{\text{A}\rightleftharpoons\text{M}}}\right)\right\rangle_{\mathbb{S}^{\text{A}\rightleftharpoons\text{M}}} \tag{8-102c}$$

$$z_{\xi\zeta}^{\vec{J}^{\text{A}\rightleftharpoons\text{M}}\vec{J}^{\text{M}}} = -\left\langle \vec{b}_{\xi}^{\vec{J}^{\text{A}\rightleftharpoons\text{M}}}, -j\omega\mu_0\mathcal{L}_0\left(\vec{b}_{\zeta}^{\vec{J}^{\text{M}}}\right)\right\rangle_{\mathbb{S}^{\text{A}\rightleftharpoons\text{M}}} \tag{8-102d}$$

$$z_{\xi\zeta}^{\vec{J}^{\text{A}\rightleftharpoons\text{M}}\vec{J}} = -\left\langle \vec{b}_{\xi}^{\vec{J}^{\text{A}\rightleftharpoons\text{M}}}, -j\omega\mu_0\mathcal{L}_0\left(\vec{b}_{\zeta}^{\vec{J}}\right)\right\rangle_{\mathbb{S}^{\text{A}\rightleftharpoons\text{M}}} \tag{8-102e}$$

$$z_{\xi\zeta}^{\vec{J}^{\text{A}\rightleftharpoons\text{M}}\vec{J}_{\text{M}}} = -\left\langle \vec{b}_{\xi}^{\vec{J}^{\text{A}\rightleftharpoons\text{M}}}, -j\omega\mu_0\mathcal{L}_0\left(\vec{b}_{\zeta}^{\vec{J}_{\text{M}}}\right)\right\rangle_{\mathbb{S}^{\text{A}\rightleftharpoons\text{M}}} \tag{8-102f}$$

$$z_{\xi\zeta}^{\vec{J}^{\text{A}\rightleftharpoons\text{M}}\vec{J}_{\text{M}\rightleftharpoons\text{A}}} = -\left\langle \vec{b}_{\xi}^{\vec{J}^{\text{A}\rightleftharpoons\text{M}}}, -j\omega\mu_0\mathcal{L}_0\left(-\vec{b}_{\zeta}^{\vec{J}_{\text{M}\rightleftharpoons\text{A}}}\right)\right\rangle_{\mathbb{S}^{\text{A}\rightleftharpoons\text{M}}} \tag{8-102g}$$

$$z_{\xi\zeta}^{\vec{J}^{\text{A}\rightleftharpoons\text{M}}\vec{M}^{\text{G}\rightleftharpoons\text{A}}} = \left\langle \vec{b}_{\xi}^{\vec{J}^{\text{A}\rightleftharpoons\text{M}}}, \mathcal{E}_{\text{T}}\left(\vec{b}_{\zeta}^{\vec{M}^{\text{G}\rightleftharpoons\text{A}}}\right)\right\rangle_{\tilde{\mathbb{S}}^{\text{A}\rightleftharpoons\text{M}}} \tag{8-102h}$$

$$\begin{aligned} z_{\xi\zeta}^{\vec{J}^{\text{A}\rightleftharpoons\text{M}}\vec{M}^{\text{A}\rightleftharpoons\text{M}}} = & \left\langle \vec{b}_{\xi}^{\vec{J}^{\text{A}\rightleftharpoons\text{M}}}, \mathcal{E}_{\text{T}}\left(\vec{b}_{\zeta}^{\vec{M}^{\text{A}\rightleftharpoons\text{M}}}\right)\right\rangle_{\tilde{\mathbb{S}}^{\text{A}\rightleftharpoons\text{M}}} \\ & - \left\langle \vec{b}_{\xi}^{\vec{J}^{\text{A}\rightleftharpoons\text{M}}}, \hat{n}^{\rightarrow\text{A}} \times \frac{1}{2}\vec{b}_{\zeta}^{\vec{M}^{\text{A}\rightleftharpoons\text{M}}} - \text{P.V.}\,\mathcal{K}_0\left(-\vec{b}_{\zeta}^{\vec{M}^{\text{A}\rightleftharpoons\text{M}}}\right)\right\rangle_{\mathbb{S}^{\text{A}\rightleftharpoons\text{M}}} \end{aligned} \tag{8-102i}$$

$$z_{\xi\zeta}^{\vec{J}^{\mathrm{A}\rightleftharpoons\mathrm{M}}\vec{M}} = -\left\langle \vec{b}_{\xi}^{\vec{J}^{\mathrm{A}\rightleftharpoons\mathrm{M}}}, -\mathcal{K}_0\left(\vec{b}_{\zeta}^{\vec{M}}\right)\right\rangle_{\mathbb{S}^{\mathrm{A}\rightleftharpoons\mathrm{M}}} \tag{8-102j}$$

$$z_{\xi\zeta}^{\vec{J}^{\mathrm{A}\rightleftharpoons\mathrm{M}}\vec{M}_{\mathrm{M}\rightleftharpoons\mathrm{A}}} = -\left\langle \vec{b}_{\xi}^{\vec{J}^{\mathrm{A}\rightleftharpoons\mathrm{M}}}, -\mathcal{K}_0\left(-\vec{b}_{\zeta}^{\vec{M}_{\mathrm{M}\rightleftharpoons\mathrm{A}}}\right)\right\rangle_{\mathbb{S}^{\mathrm{A}\rightleftharpoons\mathrm{M}}} \tag{8-102k}$$

where the integral surface $\tilde{\mathbb{S}}^{\mathrm{A}\rightleftharpoons\mathrm{M}}$ is shown in Fig. 8-23. The formulations used to calculate the elements of the matrices in Eq. (8-92b) are as follows:

$$z_{\xi\zeta}^{\vec{M}^{\mathrm{A}\rightleftharpoons\mathrm{M}}\vec{J}^{\mathrm{G}\rightleftharpoons\mathrm{A}}} = \left\langle \vec{b}_{\xi}^{\vec{M}^{\mathrm{A}\rightleftharpoons\mathrm{M}}}, \mathcal{H}_{\mathrm{T}}\left(\vec{b}_{\zeta}^{\vec{J}^{\mathrm{G}\rightleftharpoons\mathrm{A}}}\right)\right\rangle_{\tilde{\mathbb{S}}^{\mathrm{A}\rightleftharpoons\mathrm{M}}} \tag{8-103a}$$

$$z_{\xi\zeta}^{\vec{M}^{\mathrm{A}\rightleftharpoons\mathrm{M}}\vec{J}^{\mathrm{A}}} = \left\langle \vec{b}_{\xi}^{\vec{M}^{\mathrm{A}\rightleftharpoons\mathrm{M}}}, \mathcal{H}_{\mathrm{T}}\left(\vec{b}_{\zeta}^{\vec{J}^{\mathrm{A}}}\right)\right\rangle_{\tilde{\mathbb{S}}^{\mathrm{A}\rightleftharpoons\mathrm{M}}} \tag{8-103b}$$

$$\begin{aligned} z_{\xi\zeta}^{\vec{M}^{\mathrm{A}\rightleftharpoons\mathrm{M}}\vec{J}^{\mathrm{A}\rightleftharpoons\mathrm{M}}} = & \left\langle \vec{b}_{\xi}^{\vec{M}^{\mathrm{A}\rightleftharpoons\mathrm{M}}}, \mathcal{H}_{\mathrm{T}}\left(\vec{b}_{\zeta}^{\vec{J}^{\mathrm{A}\rightleftharpoons\mathrm{M}}}\right)\right\rangle_{\tilde{\mathbb{S}}^{\mathrm{A}\rightleftharpoons\mathrm{M}}} \\ & - \left\langle \vec{b}_{\xi}^{\vec{M}^{\mathrm{A}\rightleftharpoons\mathrm{M}}}, \frac{1}{2}\vec{b}_{\zeta}^{\vec{J}^{\mathrm{A}\rightleftharpoons\mathrm{M}}} \times \hat{n}^{\rightarrow\mathrm{A}} + \mathrm{P.V.}\,\mathcal{K}_0\left(-\vec{b}_{\zeta}^{\vec{J}^{\mathrm{A}\rightleftharpoons\mathrm{M}}}\right)\right\rangle_{\mathbb{S}^{\mathrm{A}\rightleftharpoons\mathrm{M}}} \end{aligned} \tag{8-103c}$$

$$z_{\xi\zeta}^{\vec{M}^{\mathrm{A}\rightleftharpoons\mathrm{M}}\vec{J}^{\mathrm{M}}} = -\left\langle \vec{b}_{\xi}^{\vec{M}^{\mathrm{A}\rightleftharpoons\mathrm{M}}}, \mathcal{K}_0\left(\vec{b}_{\zeta}^{\vec{J}^{\mathrm{M}}}\right)\right\rangle_{\mathbb{S}^{\mathrm{A}\rightleftharpoons\mathrm{M}}} \tag{8-103d}$$

$$z_{\xi\zeta}^{\vec{M}^{\mathrm{A}\rightleftharpoons\mathrm{M}}\vec{J}} = -\left\langle \vec{b}_{\xi}^{\vec{M}^{\mathrm{A}\rightleftharpoons\mathrm{M}}}, \mathcal{K}_0\left(\vec{b}_{\zeta}^{\vec{J}}\right)\right\rangle_{\mathbb{S}^{\mathrm{A}\rightleftharpoons\mathrm{M}}} \tag{8-103e}$$

$$z_{\xi\zeta}^{\vec{M}^{\mathrm{A}\rightleftharpoons\mathrm{M}}\vec{J}_{\mathrm{M}}} = -\left\langle \vec{b}_{\xi}^{\vec{M}^{\mathrm{A}\rightleftharpoons\mathrm{M}}}, \mathcal{K}_0\left(\vec{b}_{\zeta}^{\vec{J}_{\mathrm{M}}}\right)\right\rangle_{\mathbb{S}^{\mathrm{A}\rightleftharpoons\mathrm{M}}} \tag{8-103f}$$

$$z_{\xi\zeta}^{\vec{M}^{\mathrm{A}\rightleftharpoons\mathrm{M}}\vec{J}_{\mathrm{M}\rightleftharpoons\mathrm{A}}} = -\left\langle \vec{b}_{\xi}^{\vec{M}^{\mathrm{A}\rightleftharpoons\mathrm{M}}}, \mathcal{K}_0\left(-\vec{b}_{\zeta}^{\vec{J}_{\mathrm{M}\rightleftharpoons\mathrm{A}}}\right)\right\rangle_{\mathbb{S}^{\mathrm{A}\rightleftharpoons\mathrm{M}}} \tag{8-103g}$$

$$z_{\xi\zeta}^{\vec{M}^{\mathrm{A}\rightleftharpoons\mathrm{M}}\vec{M}^{\mathrm{G}\rightleftharpoons\mathrm{A}}} = \left\langle \vec{b}_{\xi}^{\vec{M}^{\mathrm{A}\rightleftharpoons\mathrm{M}}}, \mathcal{H}_{\mathrm{T}}\left(\vec{b}_{\zeta}^{\vec{M}^{\mathrm{G}\rightleftharpoons\mathrm{A}}}\right)\right\rangle_{\tilde{\mathbb{S}}^{\mathrm{A}\rightleftharpoons\mathrm{M}}} \tag{8-103h}$$

$$z_{\xi\zeta}^{\vec{M}^{\mathrm{A}\rightleftharpoons\mathrm{M}}\vec{M}^{\mathrm{A}\rightleftharpoons\mathrm{M}}} = \left\langle \vec{b}_{\xi}^{\vec{M}^{\mathrm{A}\rightleftharpoons\mathrm{M}}}, \mathcal{H}_{\mathrm{T}}\left(\vec{b}_{\zeta}^{\vec{M}^{\mathrm{A}\rightleftharpoons\mathrm{M}}}\right)\right\rangle_{\tilde{\mathbb{S}}^{\mathrm{A}\rightleftharpoons\mathrm{M}}} - \left\langle \vec{b}_{\xi}^{\vec{M}^{\mathrm{A}\rightleftharpoons\mathrm{M}}}, -j\omega\varepsilon_0\mathcal{L}_0\left(-\vec{b}_{\zeta}^{\vec{M}^{\mathrm{A}\rightleftharpoons\mathrm{M}}}\right)\right\rangle_{\mathbb{S}^{\mathrm{A}\rightleftharpoons\mathrm{M}}} \tag{8-103i}$$

$$z_{\xi\zeta}^{\vec{M}^{\mathrm{A}\rightleftharpoons\mathrm{M}}\vec{M}} = -\left\langle \vec{b}_{\xi}^{\vec{M}^{\mathrm{A}\rightleftharpoons\mathrm{M}}}, -j\omega\varepsilon_0\mathcal{L}_0\left(\vec{b}_{\zeta}^{\vec{M}}\right)\right\rangle_{\mathbb{S}^{\mathrm{A}\rightleftharpoons\mathrm{M}}} \tag{8-103j}$$

$$z_{\xi\zeta}^{\vec{M}^{\mathrm{A}\rightleftharpoons\mathrm{M}}\vec{M}_{\mathrm{M}\rightleftharpoons\mathrm{A}}} = -\left\langle \vec{b}_{\xi}^{\vec{M}^{\mathrm{A}\rightleftharpoons\mathrm{M}}}, -j\omega\varepsilon_0\mathcal{L}_0\left(-\vec{b}_{\zeta}^{\vec{M}_{\mathrm{M}\rightleftharpoons\mathrm{A}}}\right)\right\rangle_{\mathbb{S}^{\mathrm{A}\rightleftharpoons\mathrm{M}}} \tag{8-103k}$$

The formulations used to calculate the elements of the matrices in Eq. (8-93) are as follows:

$$z_{\xi\zeta}^{\vec{J}^{\mathrm{M}}\vec{J}^{\mathrm{A}\rightleftharpoons\mathrm{M}}} = \left\langle \vec{b}_{\xi}^{\vec{J}^{\mathrm{M}}}, -j\omega\mu_0\mathcal{L}_0\left(-\vec{b}_{\zeta}^{\vec{J}^{\mathrm{A}\rightleftharpoons\mathrm{M}}}\right)\right\rangle_{\mathbb{S}^{\mathrm{M}}} \tag{8-104a}$$

$$z_{\xi\zeta}^{\vec{J}^{\mathrm{M}}\vec{J}^{\mathrm{M}}} = \left\langle \vec{b}_{\xi}^{\vec{J}^{\mathrm{M}}}, -j\omega\mu_0\mathcal{L}_0\left(\vec{b}_{\zeta}^{\vec{J}^{\mathrm{M}}}\right)\right\rangle_{\mathbb{S}^{\mathrm{M}}} \tag{8-104b}$$

$$z_{\xi\zeta}^{\vec{J}^{\mathrm{M}}\vec{J}} = \left\langle \vec{b}_{\xi}^{\vec{J}^{\mathrm{M}}}, -j\omega\mu_0\mathcal{L}_0\left(\vec{b}_{\zeta}^{\vec{J}}\right)\right\rangle_{\mathbb{S}^{\mathrm{M}}} \tag{8-104c}$$

$$z_{\xi\zeta}^{\vec{J}^{\mathrm{M}}\vec{J}_{\mathrm{M}}} = \left\langle \vec{b}_{\xi}^{\vec{J}^{\mathrm{M}}}, -j\omega\mu_0\mathcal{L}_0\left(\vec{b}_{\zeta}^{\vec{J}_{\mathrm{M}}}\right)\right\rangle_{\mathbb{S}^{\mathrm{M}}} \tag{8-104d}$$

$$z_{\xi\zeta}^{\vec{J}^{\mathrm{M}}\vec{J}_{\mathrm{M}\rightleftharpoons\mathrm{A}}} = \left\langle \vec{b}_{\xi}^{\vec{J}^{\mathrm{M}}}, -j\omega\mu_0\mathcal{L}_0\left(-\vec{b}_{\zeta}^{\vec{J}_{\mathrm{M}\rightleftharpoons\mathrm{A}}}\right)\right\rangle_{\mathbb{S}^{\mathrm{M}}} \tag{8-104e}$$

$$z_{\xi\zeta}^{\vec{J}^{\mathrm{M}}\vec{M}^{\mathrm{A}\rightleftharpoons\mathrm{M}}} = \left\langle \vec{b}_{\xi}^{\vec{J}^{\mathrm{M}}}, -\mathcal{K}_0\left(-\vec{b}_{\zeta}^{\vec{M}^{\mathrm{A}\rightleftharpoons\mathrm{M}}}\right)\right\rangle_{\mathbb{S}^{\mathrm{M}}} \tag{8-104f}$$

$$z_{\xi\zeta}^{\vec{J}^{\mathrm{M}}\vec{M}} = \left\langle \vec{b}_{\xi}^{\vec{J}^{\mathrm{M}}}, -\mathcal{K}_0\left(\vec{b}_{\zeta}^{\vec{M}}\right)\right\rangle_{\mathbb{S}^{\mathrm{M}}} \tag{8-104g}$$

$$z_{\xi\zeta}^{\vec{J}^{\mathrm{M}}\vec{M}_{\mathrm{M}\rightleftharpoons\mathrm{A}}} = \left\langle \vec{b}_{\xi}^{\vec{J}^{\mathrm{M}}}, -\mathcal{K}_0\left(-\vec{b}_{\zeta}^{\vec{M}_{\mathrm{M}\rightleftharpoons\mathrm{A}}}\right)\right\rangle_{\mathbb{S}^{\mathrm{M}}} \tag{8-104h}$$

The formulations used to calculate the elements of the matrices in Eq. (8-94a) are as follows:

$$z_{\xi\zeta}^{\vec{J}\vec{J}^{\mathrm{A}\rightleftharpoons\mathrm{M}}} = \left\langle \vec{b}_{\xi}^{\vec{J}}, -j\omega\mu_0\mathcal{L}_0\left(-\vec{b}_{\zeta}^{\vec{J}^{\mathrm{A}\rightleftharpoons\mathrm{M}}}\right)\right\rangle_{\mathbb{V}} \tag{8-105a}$$

$$z_{\xi\zeta}^{\vec{J}\vec{J}^{\mathrm{M}}} = \left\langle \vec{b}_{\xi}^{\vec{J}}, -j\omega\mu_0\mathcal{L}_0\left(\vec{b}_{\zeta}^{\vec{J}^{\mathrm{M}}}\right)\right\rangle_{\mathbb{V}} \tag{8-105b}$$

$$z_{\xi\zeta}^{\vec{J}\vec{J}} = \left\langle \vec{b}_{\xi}^{\vec{J}}, -j\omega\mu_0\mathcal{L}_0\left(\vec{b}_{\zeta}^{\vec{J}}\right)\right\rangle_{\mathbb{V}} - \left\langle \vec{b}_{\xi}^{\vec{J}}, \left(j\omega\Delta\vec{\vec{\varepsilon}}_{\mathrm{c}}\right)^{-1}\cdot\vec{b}_{\zeta}^{\vec{J}}\right\rangle_{\mathbb{V}} \tag{8-105c}$$

$$z_{\xi\zeta}^{\vec{J}\vec{J}_{\mathrm{M}}} = \left\langle \vec{b}_{\xi}^{\vec{J}}, -j\omega\mu_0\mathcal{L}_0\left(\vec{b}_{\zeta}^{\vec{J}_{\mathrm{M}}}\right)\right\rangle_{\mathbb{V}} \tag{8-105d}$$

$$z_{\xi\zeta}^{\vec{J}\vec{J}_{\mathrm{M}\rightleftharpoons\mathrm{A}}} = \left\langle \vec{b}_{\xi}^{\vec{J}}, -j\omega\mu_0\mathcal{L}_0\left(-\vec{b}_{\zeta}^{\vec{J}_{\mathrm{M}\rightleftharpoons\mathrm{A}}}\right)\right\rangle_{\mathbb{V}} \tag{8-105e}$$

$$z_{\xi\zeta}^{\vec{J}\vec{M}^{\mathrm{A}\rightleftharpoons\mathrm{M}}} = \left\langle \vec{b}_{\xi}^{\vec{J}}, -\mathcal{K}_0\left(-\vec{b}_{\zeta}^{\vec{M}^{\mathrm{A}\rightleftharpoons\mathrm{M}}}\right)\right\rangle_{\mathbb{V}} \tag{8-105f}$$

$$z_{\xi\zeta}^{\vec{J}\vec{M}} = \left\langle \vec{b}_{\xi}^{\vec{J}}, -\mathcal{K}_0\left(\vec{b}_{\zeta}^{\vec{M}}\right)\right\rangle_{\mathbb{V}} \tag{8-105g}$$

$$z_{\xi\zeta}^{\vec{J}\vec{M}_{\mathrm{M}\rightleftharpoons\mathrm{A}}} = \left\langle \vec{b}_{\xi}^{\vec{J}}, -\mathcal{K}_0\left(-\vec{b}_{\zeta}^{\vec{M}_{\mathrm{M}\rightleftharpoons\mathrm{A}}}\right)\right\rangle_{\mathbb{V}} \tag{8-105h}$$

The formulations used to calculate the elements of the matrices in Eq. (8-94b) are as follows:

$$z_{\xi\zeta}^{\vec{M}\vec{J}^{\mathrm{A}\rightleftharpoons\mathrm{M}}} = \left\langle \vec{b}_{\xi}^{\vec{M}}, \mathcal{K}_0\left(-\vec{b}_{\zeta}^{\vec{J}^{\mathrm{A}\rightleftharpoons\mathrm{M}}}\right)\right\rangle_{\mathbb{V}} \tag{8-106a}$$

$$z_{\xi\zeta}^{\vec{M}\vec{J}^{\mathrm{M}}} = \left\langle \vec{b}_{\xi}^{\vec{M}}, \mathcal{K}_0\left(\vec{b}_{\zeta}^{\vec{J}^{\mathrm{M}}}\right)\right\rangle_{\mathbb{V}} \tag{8-106b}$$

$$z_{\xi\zeta}^{\vec{M}\vec{J}} = \left\langle \vec{b}_{\xi}^{\vec{M}}, \mathcal{K}_0\left(\vec{b}_{\zeta}^{\vec{J}}\right)\right\rangle_{\mathbb{V}} \tag{8-106c}$$

$$z_{\xi\zeta}^{\vec{M}\vec{J}_{\mathrm{M}}} = \left\langle \vec{b}_{\xi}^{\vec{M}}, \mathcal{K}_0\left(\vec{b}_{\zeta}^{\vec{J}_{\mathrm{M}}}\right)\right\rangle_{\mathbb{V}} \tag{8-106d}$$

$$z_{\xi\zeta}^{\vec{M}\vec{J}_{\mathrm{M}\rightleftharpoons\mathrm{A}}} = \left\langle \vec{b}_{\xi}^{\vec{M}}, \mathcal{K}_0\left(-\vec{b}_{\zeta}^{\vec{J}_{\mathrm{M}\rightleftharpoons\mathrm{A}}}\right)\right\rangle_{\mathbb{V}} \tag{8-106e}$$

$$z_{\xi\zeta}^{\vec{M}\vec{M}^{\mathrm{A}\rightleftharpoons\mathrm{M}}} = \left\langle \vec{b}_{\xi}^{\vec{M}}, -j\omega\varepsilon_0\mathcal{L}_0\left(-\vec{b}_{\zeta}^{\vec{M}^{\mathrm{A}\rightleftharpoons\mathrm{M}}}\right)\right\rangle_{\mathbb{V}} \tag{8-106f}$$

$$z_{\xi\zeta}^{\vec{M}\vec{M}} = \left\langle \vec{b}_{\xi}^{\vec{M}}, -j\omega\varepsilon_0\mathcal{L}_0\left(\vec{b}_{\zeta}^{\vec{M}}\right)\right\rangle_{\mathbb{V}} - \left\langle \vec{b}_{\xi}^{\vec{M}}, \left(j\omega\Delta\vec{\vec{\mu}}\right)^{-1}\cdot\vec{b}_{\zeta}^{\vec{M}}\right\rangle_{\mathbb{V}} \tag{8-106g}$$

$$z_{\xi\zeta}^{\vec{M}\vec{M}_{\mathrm{M}\rightleftharpoons\mathrm{A}}} = \left\langle \vec{b}_{\xi}^{\vec{M}}, -j\omega\varepsilon_0\mathcal{L}_0\left(-\vec{b}_{\zeta}^{\vec{M}_{\mathrm{M}\rightleftharpoons\mathrm{A}}}\right)\right\rangle_{\mathbb{V}} \tag{8-106h}$$

The formulations for calculating the elements of the matrices in Eq. (8-95) are as follows:

$$z_{\xi\zeta}^{\vec{J}_{\mathrm{M}}\vec{J}^{\mathrm{A}\rightleftharpoons\mathrm{M}}} = \left\langle \vec{b}_{\xi}^{\vec{J}_{\mathrm{M}}}, -j\omega\mu_0\mathcal{L}_0\left(-\vec{b}_{\zeta}^{\vec{J}^{\mathrm{A}\rightleftharpoons\mathrm{M}}}\right)\right\rangle_{\mathbb{S}_{\mathrm{M}}} \tag{8-107a}$$

$$z_{\xi\zeta}^{\vec{J}_{\rm M}\vec{J}^{\rm M}} = \left\langle \vec{b}_{\xi}^{\vec{J}_{\rm M}}, -j\omega\mu_0\mathcal{L}_0\left(\vec{b}_{\zeta}^{\vec{J}^{\rm M}}\right)\right\rangle_{\mathbb{S}_{\rm M}} \tag{8-107b}$$

$$z_{\xi\zeta}^{\vec{J}_{\rm M}\vec{J}} = \left\langle \vec{b}_{\xi}^{\vec{J}_{\rm M}}, -j\omega\mu_0\mathcal{L}_0\left(\vec{b}_{\zeta}^{\vec{J}}\right)\right\rangle_{\mathbb{S}_{\rm M}} \tag{8-107c}$$

$$z_{\xi\zeta}^{\vec{J}_{\rm M}\vec{J}_{\rm M}} = \left\langle \vec{b}_{\xi}^{\vec{J}_{\rm M}}, -j\omega\mu_0\mathcal{L}_0\left(\vec{b}_{\zeta}^{\vec{J}_{\rm M}}\right)\right\rangle_{\mathbb{S}_{\rm M}} \tag{8-107d}$$

$$z_{\xi\zeta}^{\vec{J}_{\rm M}\vec{J}_{\rm M\rightleftharpoons A}} = \left\langle \vec{b}_{\xi}^{\vec{J}_{\rm M}}, -j\omega\mu_0\mathcal{L}_0\left(-\vec{b}_{\zeta}^{\vec{J}_{\rm M\rightleftharpoons A}}\right)\right\rangle_{\mathbb{S}_{\rm M}} \tag{8-107e}$$

$$z_{\xi\zeta}^{\vec{J}_{\rm M}\vec{M}^{\rm A\rightleftharpoons M}} = \left\langle \vec{b}_{\xi}^{\vec{J}_{\rm M}}, -\mathcal{K}_0\left(-\vec{b}_{\zeta}^{\vec{M}^{\rm A\rightleftharpoons M}}\right)\right\rangle_{\mathbb{S}_{\rm M}} \tag{8-107f}$$

$$z_{\xi\zeta}^{\vec{J}_{\rm M}\vec{M}} = \left\langle \vec{b}_{\xi}^{\vec{J}_{\rm M}}, -\mathcal{K}_0\left(\vec{b}_{\zeta}^{\vec{M}}\right)\right\rangle_{\mathbb{S}_{\rm M}} \tag{8-107g}$$

$$z_{\xi\zeta}^{\vec{J}_{\rm M}\vec{M}_{\rm M\rightleftharpoons A}} = \left\langle \vec{b}_{\xi}^{\vec{J}_{\rm M}}, -\mathcal{K}_0\left(-\vec{b}_{\zeta}^{\vec{M}_{\rm M\rightleftharpoons A}}\right)\right\rangle_{\mathbb{S}_{\rm M}} \tag{8-107h}$$

The formulations used to calculate the elements of the matrices in Eq. (8-96a) are as follows:

$$z_{\xi\zeta}^{\vec{J}_{\rm M\rightleftharpoons A}\vec{J}^{\rm A\rightleftharpoons M}} = \left\langle \vec{b}_{\xi}^{\vec{J}_{\rm M\rightleftharpoons A}}, -j\omega\mu_0\mathcal{L}_0\left(-\vec{b}_{\zeta}^{\vec{J}^{\rm A\rightleftharpoons M}}\right)\right\rangle_{\mathbb{S}_{\rm M\rightleftharpoons A}} \tag{8-108a}$$

$$z_{\xi\zeta}^{\vec{J}_{\rm M\rightleftharpoons A}\vec{J}^{\rm M}} = \left\langle \vec{b}_{\xi}^{\vec{J}_{\rm M\rightleftharpoons A}}, -j\omega\mu_0\mathcal{L}_0\left(\vec{b}_{\zeta}^{\vec{J}^{\rm M}}\right)\right\rangle_{\mathbb{S}_{\rm M\rightleftharpoons A}} \tag{8-108b}$$

$$z_{\xi\zeta}^{\vec{J}_{\rm M\rightleftharpoons A}\vec{J}} = \left\langle \vec{b}_{\xi}^{\vec{J}_{\rm M\rightleftharpoons A}}, -j\omega\mu_0\mathcal{L}_0\left(\vec{b}_{\zeta}^{\vec{J}}\right)\right\rangle_{\mathbb{S}_{\rm M\rightleftharpoons A}} \tag{8-108c}$$

$$z_{\xi\zeta}^{\vec{J}_{\rm M\rightleftharpoons A}\vec{J}_{\rm M}} = \left\langle \vec{b}_{\xi}^{\vec{J}_{\rm M\rightleftharpoons A}}, -j\omega\mu_0\mathcal{L}_0\left(\vec{b}_{\zeta}^{\vec{J}_{\rm M}}\right)\right\rangle_{\mathbb{S}_{\rm M\rightleftharpoons A}} \tag{8-108d}$$

$$z_{\xi\zeta}^{\vec{J}_{\rm M\rightleftharpoons A}\vec{J}_{\rm M\rightleftharpoons A}} = \left\langle \vec{b}_{\xi}^{\vec{J}_{\rm M\rightleftharpoons A}}, -j\omega\mu_0\mathcal{L}_0\left(-\vec{b}_{\zeta}^{\vec{J}_{\rm M\rightleftharpoons A}}\right)\right\rangle_{\mathbb{S}_{\rm M\rightleftharpoons A}} - \left\langle \vec{b}_{\xi}^{\vec{J}_{\rm M\rightleftharpoons A}}, \mathcal{E}_{\rm R}\left(\vec{b}_{\zeta}^{\vec{J}_{\rm M\rightleftharpoons A}}\right)\right\rangle_{\underset{\sim}{\mathbb{S}}_{\rm M\rightleftharpoons A}} \tag{8-108e}$$

$$z_{\xi\zeta}^{\vec{J}_{\rm M\rightleftharpoons A}\vec{J}_{\rm A}} = -\left\langle \vec{b}_{\xi}^{\vec{J}_{\rm M\rightleftharpoons A}}, \mathcal{E}_{\rm R}\left(\vec{b}_{\zeta}^{\vec{J}_{\rm A}}\right)\right\rangle_{\underset{\sim}{\mathbb{S}}_{\rm M\rightleftharpoons A}} \tag{8-108f}$$

$$z_{\xi\zeta}^{\vec{J}_{\rm M\rightleftharpoons A}\vec{J}_{\rm A\rightleftharpoons G}} = -\left\langle \vec{b}_{\xi}^{\vec{J}_{\rm M\rightleftharpoons A}}, \mathcal{E}_{\rm R}\left(\vec{b}_{\zeta}^{\vec{J}_{\rm A\rightleftharpoons G}}\right)\right\rangle_{\underset{\sim}{\mathbb{S}}_{\rm M\rightleftharpoons A}} \tag{8-108g}$$

$$z_{\xi\zeta}^{\vec{J}_{\rm M\rightleftharpoons A}\vec{M}^{\rm A\rightleftharpoons M}} = \left\langle \vec{b}_{\xi}^{\vec{J}_{\rm M\rightleftharpoons A}}, -\mathcal{K}_0\left(-\vec{b}_{\zeta}^{\vec{M}^{\rm A\rightleftharpoons M}}\right)\right\rangle_{\mathbb{S}_{\rm M\rightleftharpoons A}} \tag{8-108h}$$

$$z_{\xi\zeta}^{\vec{J}_{\rm M\rightleftharpoons A}\vec{M}} = \left\langle \vec{b}_{\xi}^{\vec{J}_{\rm M\rightleftharpoons A}}, -\mathcal{K}_0\left(\vec{b}_{\zeta}^{\vec{M}}\right)\right\rangle_{\mathbb{S}_{\rm M\rightleftharpoons A}} \tag{8-108i}$$

$$\begin{aligned} z_{\xi\zeta}^{\vec{J}_{\rm M\rightleftharpoons A}\vec{M}_{\rm M\rightleftharpoons A}} = & \left\langle \vec{b}_{\xi}^{\vec{J}_{\rm M\rightleftharpoons A}}, \hat{n}_{\rightarrow \rm A} \times \frac{1}{2}\vec{b}_{\zeta}^{\vec{M}_{\rm M\rightleftharpoons A}} - {\rm P.V.}\,\mathcal{K}_0\left(-\vec{b}_{\zeta}^{\vec{M}_{\rm M\rightleftharpoons A}}\right)\right\rangle_{\tilde{\mathbb{S}}_{\rm M\rightleftharpoons A}} \\ & - \left\langle \vec{b}_{\xi}^{\vec{J}_{\rm M\rightleftharpoons A}}, \mathcal{E}_{\rm R}\left(\vec{b}_{\zeta}^{\vec{M}_{\rm M\rightleftharpoons A}}\right)\right\rangle_{\underset{\sim}{\mathbb{S}}_{\rm M\rightleftharpoons A}} \end{aligned} \tag{8-108 j}$$

$$z_{\xi\zeta}^{\vec{J}_{\rm M\rightleftharpoons A}\vec{M}_{\rm A\rightleftharpoons G}} = -\left\langle \vec{b}_{\xi}^{\vec{J}_{\rm M\rightleftharpoons A}}, \mathcal{E}_{\rm R}\left(\vec{b}_{\zeta}^{\vec{M}_{\rm A\rightleftharpoons G}}\right)\right\rangle_{\underset{\sim}{\mathbb{S}}_{\rm M\rightleftharpoons A}} \tag{8-108k}$$

where the integral surface $\underset{\sim}{\mathbb{S}}_{\rm M\rightleftharpoons A}$ is shown in Fig. 8-23. The formulations used to calculate the elements of the matrices in Eq. (8-96b) are as follows:

$$z_{\xi\zeta}^{\vec{M}_{\mathrm{M}\rightleftharpoons\mathrm{A}}\vec{J}^{\mathrm{A}\rightleftharpoons\mathrm{M}}} = \left\langle \vec{b}_{\xi}^{\vec{M}_{\mathrm{M}\rightleftharpoons\mathrm{A}}}, \mathcal{K}_0\left(-\vec{b}_{\zeta}^{\vec{J}^{\mathrm{A}\rightleftharpoons\mathrm{M}}}\right)\right\rangle_{\mathbb{S}_{\mathrm{M}\rightleftharpoons\mathrm{A}}} \tag{8-109a}$$

$$z_{\xi\zeta}^{\vec{M}_{\mathrm{M}\rightleftharpoons\mathrm{A}}\vec{J}^{\mathrm{M}}} = \left\langle \vec{b}_{\xi}^{\vec{M}_{\mathrm{M}\rightleftharpoons\mathrm{A}}}, \mathcal{K}_0\left(\vec{b}_{\zeta}^{\vec{J}^{\mathrm{M}}}\right)\right\rangle_{\mathbb{S}_{\mathrm{M}\rightleftharpoons\mathrm{A}}} \tag{8-109b}$$

$$z_{\xi\zeta}^{\vec{M}_{\mathrm{M}\rightleftharpoons\mathrm{A}}\vec{J}} = \left\langle \vec{b}_{\xi}^{\vec{M}_{\mathrm{M}\rightleftharpoons\mathrm{A}}}, \mathcal{K}_0\left(\vec{b}_{\zeta}^{\vec{J}}\right)\right\rangle_{\mathbb{S}_{\mathrm{M}\rightleftharpoons\mathrm{A}}} \tag{8-109c}$$

$$z_{\xi\zeta}^{\vec{M}_{\mathrm{M}\rightleftharpoons\mathrm{A}}\vec{J}_{\mathrm{M}}} = \left\langle \vec{b}_{\xi}^{\vec{M}_{\mathrm{M}\rightleftharpoons\mathrm{A}}}, \mathcal{K}_0\left(\vec{b}_{\zeta}^{\vec{J}_{\mathrm{M}}}\right)\right\rangle_{\mathbb{S}_{\mathrm{M}\rightleftharpoons\mathrm{A}}} \tag{8-109d}$$

$$\begin{aligned} z_{\xi\zeta}^{\vec{M}_{\mathrm{M}\rightleftharpoons\mathrm{A}}\vec{J}_{\mathrm{M}\rightleftharpoons\mathrm{A}}} = & \left\langle \vec{b}_{\xi}^{\vec{M}_{\mathrm{M}\rightleftharpoons\mathrm{A}}}, (1/2)\vec{b}_{\zeta}^{\vec{J}_{\mathrm{M}\rightleftharpoons\mathrm{A}}}\times\hat{n}_{\rightarrow\mathrm{A}} + \mathrm{P.V.}\,\mathcal{K}_0\left(-\vec{b}_{\zeta}^{\vec{J}_{\mathrm{M}\rightleftharpoons\mathrm{A}}}\right)\right\rangle_{\mathbb{S}_{\mathrm{M}\rightleftharpoons\mathrm{A}}} \\ & -\left\langle \vec{b}_{\xi}^{\vec{M}_{\mathrm{M}\rightleftharpoons\mathrm{A}}}, \mathcal{H}_{\mathrm{R}}\left(\vec{b}_{\zeta}^{\vec{J}_{\mathrm{M}\rightleftharpoons\mathrm{A}}}\right)\right\rangle_{\underset{\sim}{\mathbb{S}}_{\mathrm{M}\rightleftharpoons\mathrm{A}}} \end{aligned} \tag{8-109e}$$

$$z_{\xi\zeta}^{\vec{M}_{\mathrm{M}\rightleftharpoons\mathrm{A}}\vec{J}_{\mathrm{A}}} = -\left\langle \vec{b}_{\xi}^{\vec{M}_{\mathrm{M}\rightleftharpoons\mathrm{A}}}, \mathcal{H}_{\mathrm{R}}\left(\vec{b}_{\zeta}^{\vec{J}_{\mathrm{A}}}\right)\right\rangle_{\underset{\sim}{\mathbb{S}}_{\mathrm{M}\rightleftharpoons\mathrm{A}}} \tag{8-109f}$$

$$z_{\xi\zeta}^{\vec{M}_{\mathrm{M}\rightleftharpoons\mathrm{A}}\vec{J}_{\mathrm{A}\rightleftharpoons\mathrm{G}}} = -\left\langle \vec{b}_{\xi}^{\vec{M}_{\mathrm{M}\rightleftharpoons\mathrm{A}}}, \mathcal{H}_{\mathrm{R}}\left(\vec{b}_{\zeta}^{\vec{J}_{\mathrm{A}\rightleftharpoons\mathrm{G}}}\right)\right\rangle_{\underset{\sim}{\mathbb{S}}_{\mathrm{M}\rightleftharpoons\mathrm{A}}} \tag{8-109g}$$

$$z_{\xi\zeta}^{\vec{M}_{\mathrm{M}\rightleftharpoons\mathrm{A}}\vec{M}^{\mathrm{A}\rightleftharpoons\mathrm{M}}} = \left\langle \vec{b}_{\xi}^{\vec{M}_{\mathrm{M}\rightleftharpoons\mathrm{A}}}, -j\omega\varepsilon_0\mathcal{L}_0\left(-\vec{b}_{\zeta}^{\vec{M}^{\mathrm{A}\rightleftharpoons\mathrm{M}}}\right)\right\rangle_{\mathbb{S}_{\mathrm{M}\rightleftharpoons\mathrm{A}}} \tag{8-109h}$$

$$z_{\xi\zeta}^{\vec{M}_{\mathrm{M}\rightleftharpoons\mathrm{A}}\vec{M}} = \left\langle \vec{b}_{\xi}^{\vec{M}_{\mathrm{M}\rightleftharpoons\mathrm{A}}}, -j\omega\varepsilon_0\mathcal{L}_0\left(\vec{b}_{\zeta}^{\vec{M}}\right)\right\rangle_{\mathbb{S}_{\mathrm{M}\rightleftharpoons\mathrm{A}}} \tag{8-109i}$$

$$z_{\xi\zeta}^{\vec{M}_{\mathrm{M}\rightleftharpoons\mathrm{A}}\vec{M}_{\mathrm{M}\rightleftharpoons\mathrm{A}}} = \left\langle \vec{b}_{\xi}^{\vec{M}_{\mathrm{M}\rightleftharpoons\mathrm{A}}}, -j\omega\varepsilon_0\mathcal{L}_0\left(-\vec{b}_{\zeta}^{\vec{M}_{\mathrm{M}\rightleftharpoons\mathrm{A}}}\right)\right\rangle_{\mathbb{S}_{\mathrm{M}\rightleftharpoons\mathrm{A}}} - \left\langle \vec{b}_{\xi}^{\vec{M}_{\mathrm{M}\rightleftharpoons\mathrm{A}}}, \mathcal{H}_{\mathrm{R}}\left(\vec{b}_{\zeta}^{\vec{M}_{\mathrm{M}\rightleftharpoons\mathrm{A}}}\right)\right\rangle_{\underset{\sim}{\mathbb{S}}_{\mathrm{M}\rightleftharpoons\mathrm{A}}} \tag{8-109j}$$

$$z_{\xi\zeta}^{\vec{M}_{\mathrm{M}\rightleftharpoons\mathrm{A}}\vec{M}_{\mathrm{A}\rightleftharpoons\mathrm{G}}} = -\left\langle \vec{b}_{\xi}^{\vec{M}_{\mathrm{M}\rightleftharpoons\mathrm{A}}}, \mathcal{H}_{\mathrm{R}}\left(\vec{b}_{\zeta}^{\vec{M}_{\mathrm{A}\rightleftharpoons\mathrm{G}}}\right)\right\rangle_{\underset{\sim}{\mathbb{S}}_{\mathrm{M}\rightleftharpoons\mathrm{A}}} \tag{8-109k}$$

The formulations used to calculate the elements of the matrices in Eq. (8-97) are as follows:

$$z_{\xi\zeta}^{\vec{J}_{\mathrm{A}}\vec{J}_{\mathrm{M}\rightleftharpoons\mathrm{A}}} = \left\langle \vec{b}_{\xi}^{\vec{J}_{\mathrm{A}}}, \mathcal{E}_{\mathrm{R}}\left(\vec{b}_{\zeta}^{\vec{J}_{\mathrm{M}\rightleftharpoons\mathrm{A}}}\right)\right\rangle_{\tilde{\mathbb{S}}_{\mathrm{A}}} \tag{8-110a}$$

$$z_{\xi\zeta}^{\vec{J}_{\mathrm{A}}\vec{J}_{\mathrm{A}}} = \left\langle \vec{b}_{\xi}^{\vec{J}_{\mathrm{A}}}, \mathcal{E}_{\mathrm{R}}\left(\vec{b}_{\zeta}^{\vec{J}_{\mathrm{A}}}\right)\right\rangle_{\tilde{\mathbb{S}}_{\mathrm{A}}} \tag{8-110b}$$

$$z_{\xi\zeta}^{\vec{J}_{\mathrm{A}}\vec{J}_{\mathrm{A}\rightleftharpoons\mathrm{G}}} = \left\langle \vec{b}_{\xi}^{\vec{J}_{\mathrm{A}}}, \mathcal{E}_{\mathrm{R}}\left(\vec{b}_{\zeta}^{\vec{J}_{\mathrm{A}\rightleftharpoons\mathrm{G}}}\right)\right\rangle_{\tilde{\mathbb{S}}_{\mathrm{A}}} \tag{8-110c}$$

$$z_{\xi\zeta}^{\vec{J}_{\mathrm{A}}\vec{M}_{\mathrm{M}\rightleftharpoons\mathrm{A}}} = \left\langle \vec{b}_{\xi}^{\vec{J}_{\mathrm{A}}}, \mathcal{E}_{\mathrm{R}}\left(\vec{b}_{\zeta}^{\vec{M}_{\mathrm{M}\rightleftharpoons\mathrm{A}}}\right)\right\rangle_{\tilde{\mathbb{S}}_{\mathrm{A}}} \tag{8-110d}$$

$$z_{\xi\zeta}^{\vec{J}_{\mathrm{A}}\vec{M}_{\mathrm{A}\rightleftharpoons\mathrm{G}}} = \left\langle \vec{b}_{\xi}^{\vec{J}_{\mathrm{A}}}, \mathcal{E}_{\mathrm{R}}\left(\vec{b}_{\zeta}^{\vec{M}_{\mathrm{A}\rightleftharpoons\mathrm{G}}}\right)\right\rangle_{\tilde{\mathbb{S}}_{\mathrm{A}}} \tag{8-110e}$$

where the integral surface $\tilde{\mathbb{S}}_{\mathrm{A}}$ is shown in Fig. 8-23. The formulations used to calculate the elements of the matrices in Eq. (8-98a) are as follows:

$$z_{\xi\zeta}^{\vec{M}_{\mathrm{A}\rightleftharpoons\mathrm{G}}\vec{J}_{\mathrm{M}\rightleftharpoons\mathrm{A}}} = \left\langle \vec{b}_{\xi}^{\vec{M}_{\mathrm{A}\rightleftharpoons\mathrm{G}}}, \mathcal{H}_{\mathrm{R}}\left(\vec{b}_{\zeta}^{\vec{J}_{\mathrm{M}\rightleftharpoons\mathrm{A}}}\right)\right\rangle_{\tilde{\mathbb{S}}_{\mathrm{A}\rightleftharpoons\mathrm{G}}} \tag{8-111a}$$

$$z_{\xi\zeta}^{\vec{M}_{\mathrm{A}\rightleftharpoons\mathrm{G}}\vec{J}_{\mathrm{A}}} = \left\langle \vec{b}_{\xi}^{\vec{M}_{\mathrm{A}\rightleftharpoons\mathrm{G}}}, \mathcal{H}_{\mathrm{R}}\left(\vec{b}_{\zeta}^{\vec{J}_{\mathrm{A}}}\right)\right\rangle_{\tilde{\mathbb{S}}_{\mathrm{A}\rightleftharpoons\mathrm{G}}} \tag{8-111b}$$

$$z_{\xi\zeta}^{\vec{M}_{\rm A\rightleftharpoons G}\vec{J}_{\rm A\rightleftharpoons G}} = \left\langle \vec{b}_{\xi}^{\vec{M}_{\rm A\rightleftharpoons G}}, \mathcal{H}_{\rm R}\left(\vec{b}_{\zeta}^{\vec{J}_{\rm A\rightleftharpoons G}}\right)\right\rangle_{\tilde{\mathbb{S}}_{\rm A\rightleftharpoons G}} - \left\langle \vec{b}_{\xi}^{\vec{M}_{\rm A\rightleftharpoons G}}, \vec{b}_{\zeta}^{\vec{J}_{\rm A\rightleftharpoons G}} \times \hat{n}_{\rightarrow \rm A}\right\rangle_{\mathbb{S}_{\rm A\rightleftharpoons G}} \tag{8-111c}$$

$$z_{\xi\zeta}^{\vec{M}_{\rm A\rightleftharpoons G}\vec{M}_{\rm M\rightleftharpoons A}} = \left\langle \vec{b}_{\xi}^{\vec{M}_{\rm A\rightleftharpoons G}}, \mathcal{H}_{\rm R}\left(\vec{b}_{\zeta}^{\vec{M}_{\rm M\rightleftharpoons A}}\right)\right\rangle_{\tilde{\mathbb{S}}_{\rm A\rightleftharpoons G}} \tag{8-111d}$$

$$z_{\xi\zeta}^{\vec{M}_{\rm A\rightleftharpoons G}\vec{M}_{\rm A\rightleftharpoons G}} = \left\langle \vec{b}_{\xi}^{\vec{M}_{\rm A\rightleftharpoons G}}, \mathcal{H}_{\rm R}\left(\vec{b}_{\zeta}^{\vec{M}_{\rm A\rightleftharpoons G}}\right)\right\rangle_{\tilde{\mathbb{S}}_{\rm A\rightleftharpoons G}} \tag{8-111e}$$

where the integral surface $\tilde{\mathbb{S}}_{\rm A\rightleftharpoons G}$ is shown in Fig. 8-23. The formulations used to calculate the elements of the matrices in Eq. (8-98b) are as follows:

$$z_{\xi\zeta}^{\vec{J}_{\rm A\rightleftharpoons G}\vec{J}_{\rm M\rightleftharpoons A}} = \left\langle \vec{b}_{\xi}^{\vec{J}_{\rm A\rightleftharpoons G}}, \mathcal{E}_{\rm R}\left(\vec{b}_{\zeta}^{\vec{J}_{\rm M\rightleftharpoons A}}\right)\right\rangle_{\tilde{\mathbb{S}}_{\rm A\rightleftharpoons G}} \tag{8-112a}$$

$$z_{\xi\zeta}^{\vec{J}_{\rm A\rightleftharpoons G}\vec{J}_{\rm A}} = \left\langle \vec{b}_{\xi}^{\vec{J}_{\rm A\rightleftharpoons G}}, \mathcal{E}_{\rm R}\left(\vec{b}_{\zeta}^{\vec{J}_{\rm A}}\right)\right\rangle_{\tilde{\mathbb{S}}_{\rm A\rightleftharpoons G}} \tag{8-112b}$$

$$z_{\xi\zeta}^{\vec{J}_{\rm A\rightleftharpoons G}\vec{J}_{\rm A\rightleftharpoons G}} = \left\langle \vec{b}_{\xi}^{\vec{J}_{\rm A\rightleftharpoons G}}, \mathcal{E}_{\rm R}\left(\vec{b}_{\zeta}^{\vec{J}_{\rm A\rightleftharpoons G}}\right)\right\rangle_{\tilde{\mathbb{S}}_{\rm A\rightleftharpoons G}} \tag{8-112c}$$

$$z_{\xi\zeta}^{\vec{J}_{\rm A\rightleftharpoons G}\vec{M}_{\rm M\rightleftharpoons A}} = \left\langle \vec{b}_{\xi}^{\vec{J}_{\rm A\rightleftharpoons G}}, \mathcal{E}_{\rm R}\left(\vec{b}_{\zeta}^{\vec{M}_{\rm M\rightleftharpoons A}}\right)\right\rangle_{\tilde{\mathbb{S}}_{\rm A\rightleftharpoons G}} \tag{8-112d}$$

$$z_{\xi\zeta}^{\vec{J}_{\rm A\rightleftharpoons G}\vec{M}_{\rm A\rightleftharpoons G}} = \left\langle \vec{b}_{\xi}^{\vec{J}_{\rm A\rightleftharpoons G}}, \mathcal{E}_{\rm R}\left(\vec{b}_{\zeta}^{\vec{M}_{\rm A\rightleftharpoons G}}\right)\right\rangle_{\tilde{\mathbb{S}}_{\rm A\rightleftharpoons G}} - \left\langle \vec{b}_{\xi}^{\vec{J}_{\rm A\rightleftharpoons G}}, \hat{n}_{\rightarrow \rm A} \times \vec{b}_{\zeta}^{\vec{M}_{\rm A\rightleftharpoons G}}\right\rangle_{\mathbb{S}_{\rm A\rightleftharpoons G}} \tag{8-112e}$$

Below, we propose two schemes for mathematically describing the modal space of the TRTA system shown in Figs. 8-21 and 8-22.

### 8.3.3.3 Scheme I: Dependent Variable Elimination (DVE)

By properly assembling the matrix equations (8-90a) and (8-91)~(8-98b), we have the following augmented matrix equation

$$\bar{\bar{\Psi}}_1 \cdot \bar{a}^{\rm AV} = \bar{\bar{\Psi}}_2 \cdot \bar{a}^{\vec{J}^{\rm G\rightleftharpoons A}} \tag{8-113}$$

in which

$$\bar{\bar{\Psi}}_1 = \begin{bmatrix}
\bar{\bar{I}}^{\vec{J}^{\rm G\rightleftharpoons A}} & 0 & 0 & 0 & 0 & 0 & 0 & 0 & 0 & 0 & 0 & 0 & 0 & 0 \\
0 & \bar{\bar{Z}}^{\vec{M}^{\rm G\rightleftharpoons A}\vec{J}^{\rm A}} & \bar{\bar{Z}}^{\vec{M}^{\rm G\rightleftharpoons A}\vec{J}^{\rm A\rightleftharpoons M}} & 0 & 0 & 0 & 0 & 0 & 0 & \bar{\bar{Z}}^{\vec{M}^{\rm G\rightleftharpoons A}\vec{M}^{\rm G\rightleftharpoons A}} & \bar{\bar{Z}}^{\vec{M}^{\rm G\rightleftharpoons A}\vec{M}^{\rm A\rightleftharpoons M}} & 0 & 0 & 0 \\
0 & \bar{\bar{Z}}^{\vec{J}^{\rm A}\vec{J}^{\rm A}} & \bar{\bar{Z}}^{\vec{J}^{\rm A}\vec{J}^{\rm A\rightleftharpoons M}} & 0 & 0 & 0 & 0 & 0 & 0 & \bar{\bar{Z}}^{\vec{J}^{\rm A}\vec{M}^{\rm G\rightleftharpoons A}} & \bar{\bar{Z}}^{\vec{J}^{\rm A}\vec{M}^{\rm A\rightleftharpoons M}} & 0 & 0 & 0 \\
0 & \bar{\bar{Z}}^{\vec{J}^{\rm A\rightleftharpoons M}\vec{J}^{\rm A}} & \bar{\bar{Z}}^{\vec{J}^{\rm A\rightleftharpoons M}\vec{J}^{\rm A\rightleftharpoons M}} & \bar{\bar{Z}}^{\vec{J}^{\rm A\rightleftharpoons M}\vec{J}^{\rm M}} & \bar{\bar{Z}}^{\vec{J}^{\rm A\rightleftharpoons M}\vec{J}} & \bar{\bar{Z}}^{\vec{J}^{\rm A\rightleftharpoons M}\vec{J}_{\rm M}} & \bar{\bar{Z}}^{\vec{J}^{\rm A\rightleftharpoons M}\vec{J}_{\rm M\rightleftharpoons A}} & 0 & 0 & \bar{\bar{Z}}^{\vec{J}^{\rm A\rightleftharpoons M}\vec{M}^{\rm G\rightleftharpoons A}} & \bar{\bar{Z}}^{\vec{J}^{\rm A\rightleftharpoons M}\vec{M}^{\rm A\rightleftharpoons M}} & \bar{\bar{Z}}^{\vec{J}^{\rm A\rightleftharpoons M}\vec{M}} & \bar{\bar{Z}}^{\vec{J}^{\rm A\rightleftharpoons M}\vec{M}_{\rm M\rightleftharpoons A}} & 0 \\
0 & \bar{\bar{Z}}^{\vec{M}^{\rm A\rightleftharpoons M}\vec{J}^{\rm A}} & \bar{\bar{Z}}^{\vec{M}^{\rm A\rightleftharpoons M}\vec{J}^{\rm A\rightleftharpoons M}} & \bar{\bar{Z}}^{\vec{M}^{\rm A\rightleftharpoons M}\vec{J}^{\rm M}} & \bar{\bar{Z}}^{\vec{M}^{\rm A\rightleftharpoons M}\vec{J}} & \bar{\bar{Z}}^{\vec{M}^{\rm A\rightleftharpoons M}\vec{J}_{\rm M}} & \bar{\bar{Z}}^{\vec{M}^{\rm A\rightleftharpoons M}\vec{J}_{\rm M\rightleftharpoons A}} & 0 & 0 & \bar{\bar{Z}}^{\vec{M}^{\rm A\rightleftharpoons M}\vec{M}^{\rm G\rightleftharpoons A}} & \bar{\bar{Z}}^{\vec{M}^{\rm A\rightleftharpoons M}\vec{M}^{\rm A\rightleftharpoons M}} & \bar{\bar{Z}}^{\vec{M}^{\rm A\rightleftharpoons M}\vec{M}} & \bar{\bar{Z}}^{\vec{M}^{\rm A\rightleftharpoons M}\vec{M}_{\rm M\rightleftharpoons A}} & 0 \\
0 & 0 & \bar{\bar{Z}}^{\vec{J}^{\rm M}\vec{J}^{\rm A\rightleftharpoons M}} & \bar{\bar{Z}}^{\vec{J}^{\rm M}\vec{J}^{\rm M}} & \bar{\bar{Z}}^{\vec{J}^{\rm M}\vec{J}} & \bar{\bar{Z}}^{\vec{J}^{\rm M}\vec{J}_{\rm M}} & \bar{\bar{Z}}^{\vec{J}^{\rm M}\vec{J}_{\rm M\rightleftharpoons A}} & 0 & 0 & 0 & \bar{\bar{Z}}^{\vec{J}^{\rm M}\vec{M}^{\rm A\rightleftharpoons M}} & \bar{\bar{Z}}^{\vec{J}^{\rm M}\vec{M}} & \bar{\bar{Z}}^{\vec{J}^{\rm M}\vec{M}_{\rm M\rightleftharpoons A}} & 0 \\
0 & 0 & \bar{\bar{Z}}^{\vec{J}\vec{J}^{\rm A\rightleftharpoons M}} & \bar{\bar{Z}}^{\vec{J}\vec{J}^{\rm M}} & \bar{\bar{Z}}^{\vec{J}\vec{J}} & \bar{\bar{Z}}^{\vec{J}\vec{J}_{\rm M}} & \bar{\bar{Z}}^{\vec{J}\vec{J}_{\rm M\rightleftharpoons A}} & 0 & 0 & 0 & \bar{\bar{Z}}^{\vec{J}\vec{M}^{\rm A\rightleftharpoons M}} & \bar{\bar{Z}}^{\vec{J}\vec{M}} & \bar{\bar{Z}}^{\vec{J}\vec{M}_{\rm M\rightleftharpoons A}} & 0 \\
0 & 0 & \bar{\bar{Z}}^{\vec{M}\vec{J}^{\rm A\rightleftharpoons M}} & \bar{\bar{Z}}^{\vec{M}\vec{J}^{\rm M}} & \bar{\bar{Z}}^{\vec{M}\vec{J}} & \bar{\bar{Z}}^{\vec{M}\vec{J}_{\rm M}} & \bar{\bar{Z}}^{\vec{M}\vec{J}_{\rm M\rightleftharpoons A}} & 0 & 0 & 0 & \bar{\bar{Z}}^{\vec{M}\vec{M}^{\rm A\rightleftharpoons M}} & \bar{\bar{Z}}^{\vec{M}\vec{M}} & \bar{\bar{Z}}^{\vec{M}\vec{M}_{\rm M\rightleftharpoons A}} & 0 \\
0 & 0 & \bar{\bar{Z}}^{\vec{J}_{\rm M}\vec{J}^{\rm A\rightleftharpoons M}} & \bar{\bar{Z}}^{\vec{J}_{\rm M}\vec{J}^{\rm M}} & \bar{\bar{Z}}^{\vec{J}_{\rm M}\vec{J}} & \bar{\bar{Z}}^{\vec{J}_{\rm M}\vec{J}_{\rm M}} & \bar{\bar{Z}}^{\vec{J}_{\rm M}\vec{J}_{\rm M\rightleftharpoons A}} & 0 & 0 & 0 & \bar{\bar{Z}}^{\vec{J}_{\rm M}\vec{M}^{\rm A\rightleftharpoons M}} & \bar{\bar{Z}}^{\vec{J}_{\rm M}\vec{M}} & \bar{\bar{Z}}^{\vec{J}_{\rm M}\vec{M}_{\rm M\rightleftharpoons A}} & 0 \\
0 & 0 & \bar{\bar{Z}}^{\vec{J}_{\rm M\rightleftharpoons A}\vec{J}^{\rm A\rightleftharpoons M}} & \bar{\bar{Z}}^{\vec{J}_{\rm M\rightleftharpoons A}\vec{J}^{\rm M}} & \bar{\bar{Z}}^{\vec{J}_{\rm M\rightleftharpoons A}\vec{J}} & \bar{\bar{Z}}^{\vec{J}_{\rm M\rightleftharpoons A}\vec{J}_{\rm M}} & \bar{\bar{Z}}^{\vec{J}_{\rm M\rightleftharpoons A}\vec{J}_{\rm M\rightleftharpoons A}} & \bar{\bar{Z}}^{\vec{J}_{\rm M\rightleftharpoons A}\vec{J}_{\rm A}} & \bar{\bar{Z}}^{\vec{J}_{\rm M\rightleftharpoons A}\vec{J}_{\rm A\rightleftharpoons G}} & 0 & \bar{\bar{Z}}^{\vec{J}_{\rm M\rightleftharpoons A}\vec{M}^{\rm A\rightleftharpoons M}} & \bar{\bar{Z}}^{\vec{J}_{\rm M\rightleftharpoons A}\vec{M}} & \bar{\bar{Z}}^{\vec{J}_{\rm M\rightleftharpoons A}\vec{M}_{\rm M\rightleftharpoons A}} & \bar{\bar{Z}}^{\vec{J}_{\rm M\rightleftharpoons A}\vec{M}_{\rm A\rightleftharpoons G}} \\
0 & 0 & \bar{\bar{Z}}^{\vec{M}_{\rm M\rightleftharpoons A}\vec{J}^{\rm A\rightleftharpoons M}} & \bar{\bar{Z}}^{\vec{M}_{\rm M\rightleftharpoons A}\vec{J}^{\rm M}} & \bar{\bar{Z}}^{\vec{M}_{\rm M\rightleftharpoons A}\vec{J}} & \bar{\bar{Z}}^{\vec{M}_{\rm M\rightleftharpoons A}\vec{J}_{\rm M}} & \bar{\bar{Z}}^{\vec{M}_{\rm M\rightleftharpoons A}\vec{J}_{\rm M\rightleftharpoons A}} & \bar{\bar{Z}}^{\vec{M}_{\rm M\rightleftharpoons A}\vec{J}_{\rm A}} & \bar{\bar{Z}}^{\vec{M}_{\rm M\rightleftharpoons A}\vec{J}_{\rm A\rightleftharpoons G}} & 0 & \bar{\bar{Z}}^{\vec{M}_{\rm M\rightleftharpoons A}\vec{M}^{\rm A\rightleftharpoons M}} & \bar{\bar{Z}}^{\vec{M}_{\rm M\rightleftharpoons A}\vec{M}} & \bar{\bar{Z}}^{\vec{M}_{\rm M\rightleftharpoons A}\vec{M}_{\rm M\rightleftharpoons A}} & \bar{\bar{Z}}^{\vec{M}_{\rm M\rightleftharpoons A}\vec{M}_{\rm A\rightleftharpoons G}} \\
0 & 0 & 0 & 0 & 0 & 0 & \bar{\bar{Z}}^{\vec{J}_{\rm A}\vec{J}_{\rm M\rightleftharpoons A}} & \bar{\bar{Z}}^{\vec{J}_{\rm A}\vec{J}_{\rm A}} & \bar{\bar{Z}}^{\vec{J}_{\rm A}\vec{J}_{\rm A\rightleftharpoons G}} & 0 & 0 & 0 & \bar{\bar{Z}}^{\vec{J}_{\rm A}\vec{M}_{\rm M\rightleftharpoons A}} & \bar{\bar{Z}}^{\vec{J}_{\rm A}\vec{M}_{\rm A\rightleftharpoons G}} \\
0 & 0 & 0 & 0 & 0 & 0 & \bar{\bar{Z}}^{\vec{M}_{\rm A\rightleftharpoons G}\vec{J}_{\rm M\rightleftharpoons A}} & \bar{\bar{Z}}^{\vec{M}_{\rm A\rightleftharpoons G}\vec{J}_{\rm A}} & \bar{\bar{Z}}^{\vec{M}_{\rm A\rightleftharpoons G}\vec{J}_{\rm A\rightleftharpoons G}} & 0 & 0 & 0 & \bar{\bar{Z}}^{\vec{M}_{\rm A\rightleftharpoons G}\vec{M}_{\rm M\rightleftharpoons A}} & \bar{\bar{Z}}^{\vec{M}_{\rm A\rightleftharpoons G}\vec{M}_{\rm A\rightleftharpoons G}} \\
0 & 0 & 0 & 0 & 0 & 0 & \bar{\bar{Z}}^{\vec{J}_{\rm A\rightleftharpoons G}\vec{J}_{\rm M\rightleftharpoons A}} & \bar{\bar{Z}}^{\vec{J}_{\rm A\rightleftharpoons G}\vec{J}_{\rm A}} & \bar{\bar{Z}}^{\vec{J}_{\rm A\rightleftharpoons G}\vec{J}_{\rm A\rightleftharpoons G}} & 0 & 0 & 0 & \bar{\bar{Z}}^{\vec{J}_{\rm A\rightleftharpoons G}\vec{M}_{\rm M\rightleftharpoons A}} & \bar{\bar{Z}}^{\vec{J}_{\rm A\rightleftharpoons G}\vec{M}_{\rm A\rightleftharpoons G}}
\end{bmatrix} \tag{8-114a}$$

$$\bar{\bar{\Psi}}_2=\begin{bmatrix} \bar{\bar{I}}^{\vec{J}^{\mathrm{G\rightleftharpoons A}}} \\ -\bar{\bar{Z}}^{\vec{M}^{\mathrm{G\rightleftharpoons A}}\vec{J}^{\mathrm{G\rightleftharpoons A}}} \\ -\bar{\bar{Z}}^{\vec{J}^{\mathrm{A}}\vec{J}^{\mathrm{G\rightleftharpoons A}}} \\ -\bar{\bar{Z}}^{\vec{J}^{\mathrm{A\rightleftharpoons M}}\vec{J}^{\mathrm{G\rightleftharpoons A}}} \\ -\bar{\bar{Z}}^{\vec{M}^{\mathrm{A\rightleftharpoons M}}\vec{J}^{\mathrm{G\rightleftharpoons A}}} \\ 0 \\ 0 \\ 0 \\ 0 \\ 0 \\ 0 \\ 0 \\ 0 \\ 0 \end{bmatrix} \tag{8-114b}$$

and

$$\bar{a}^{\mathrm{AV}}=\begin{bmatrix} \bar{a}^{\vec{J}^{\mathrm{G\rightleftharpoons A}}} \\ \bar{a}^{\vec{J}^{\mathrm{A}}} \\ \bar{a}^{\vec{J}^{\mathrm{A\rightleftharpoons M}}} \\ \bar{a}^{\vec{J}^{\mathrm{M}}} \\ \bar{a}^{\vec{J}} \\ \bar{a}^{\vec{J}_{\mathrm{M}}} \\ \bar{a}^{\vec{J}_{\mathrm{M\rightleftharpoons A}}} \\ \bar{a}^{\vec{J}_{\mathrm{A}}} \\ \bar{a}^{\vec{J}_{\mathrm{A\rightleftharpoons G}}} \\ \bar{a}^{\vec{M}^{\mathrm{G\rightleftharpoons A}}} \\ \bar{a}^{\vec{M}^{\mathrm{A\rightleftharpoons M}}} \\ \bar{a}^{\vec{M}} \\ \bar{a}^{\vec{M}_{\mathrm{M\rightleftharpoons A}}} \\ \bar{a}^{\vec{M}_{\mathrm{A\rightleftharpoons G}}} \end{bmatrix} \tag{8-115}$$

By properly assembling the matrix equations (8-90b) and (8-91)~(8-98b), we have the following another augmented matrix equation

$$\bar{\bar{\Psi}}_3\cdot\bar{a}^{\mathrm{AV}}=\bar{\bar{\Psi}}_4\cdot\bar{a}^{\vec{M}^{\mathrm{G\rightleftharpoons A}}} \tag{8-116}$$

in which

$$\bar{\bar{\Psi}}_3=\begin{bmatrix}
0 & 0 & 0 & 0 & 0 & 0 & 0 & 0 & 0 & \bar{\bar{I}}^{\vec{M}^{\mathrm{G\rightleftharpoons A}}} & 0 & 0 & 0 & 0 \\
\bar{\bar{Z}}^{\vec{J}^{\mathrm{G\rightleftharpoons A}}\vec{J}^{\mathrm{G\rightleftharpoons A}}} & \bar{\bar{Z}}^{\vec{J}^{\mathrm{G\rightleftharpoons A}}\vec{J}^{\mathrm{A}}} & \bar{\bar{Z}}^{\vec{J}^{\mathrm{G\rightleftharpoons A}}\vec{J}^{\mathrm{A\rightleftharpoons M}}} & 0 & 0 & 0 & 0 & 0 & 0 & 0 & \bar{\bar{Z}}^{\vec{J}^{\mathrm{G\rightleftharpoons A}}\vec{M}^{\mathrm{A\rightleftharpoons M}}} & 0 & 0 & 0 \\
\bar{\bar{Z}}^{\vec{J}^{\mathrm{A}}\vec{J}^{\mathrm{G\rightleftharpoons A}}} & \bar{\bar{Z}}^{\vec{J}^{\mathrm{A}}\vec{J}^{\mathrm{A}}} & \bar{\bar{Z}}^{\vec{J}^{\mathrm{A}}\vec{J}^{\mathrm{A\rightleftharpoons M}}} & 0 & 0 & 0 & 0 & 0 & 0 & 0 & \bar{\bar{Z}}^{\vec{J}^{\mathrm{A}}\vec{M}^{\mathrm{A\rightleftharpoons M}}} & 0 & 0 & 0 \\
\bar{\bar{Z}}^{\vec{J}^{\mathrm{A\rightleftharpoons M}}\vec{J}^{\mathrm{G\rightleftharpoons A}}} & \bar{\bar{Z}}^{\vec{J}^{\mathrm{A\rightleftharpoons M}}\vec{J}^{\mathrm{A}}} & \bar{\bar{Z}}^{\vec{J}^{\mathrm{A\rightleftharpoons M}}\vec{J}^{\mathrm{A\rightleftharpoons M}}} & \bar{\bar{Z}}^{\vec{J}^{\mathrm{A\rightleftharpoons M}}\vec{J}^{\mathrm{M}}} & \bar{\bar{Z}}^{\vec{J}^{\mathrm{A\rightleftharpoons M}}\vec{J}} & \bar{\bar{Z}}^{\vec{J}^{\mathrm{A\rightleftharpoons M}}\vec{J}_{\mathrm{M}}} & \bar{\bar{Z}}^{\vec{J}^{\mathrm{A\rightleftharpoons M}}\vec{J}_{\mathrm{M\rightleftharpoons A}}} & 0 & 0 & 0 & \bar{\bar{Z}}^{\vec{J}^{\mathrm{A\rightleftharpoons M}}\vec{M}^{\mathrm{A\rightleftharpoons M}}} & \bar{\bar{Z}}^{\vec{J}^{\mathrm{A\rightleftharpoons M}}\vec{M}} & \bar{\bar{Z}}^{\vec{J}^{\mathrm{A\rightleftharpoons M}}\vec{M}_{\mathrm{M\rightleftharpoons A}}} & 0 \\
\bar{\bar{Z}}^{\vec{M}^{\mathrm{A\rightleftharpoons M}}\vec{J}^{\mathrm{G\rightleftharpoons A}}} & \bar{\bar{Z}}^{\vec{M}^{\mathrm{A\rightleftharpoons M}}\vec{J}^{\mathrm{A}}} & \bar{\bar{Z}}^{\vec{M}^{\mathrm{A\rightleftharpoons M}}\vec{J}^{\mathrm{A\rightleftharpoons M}}} & \bar{\bar{Z}}^{\vec{M}^{\mathrm{A\rightleftharpoons M}}\vec{J}^{\mathrm{M}}} & \bar{\bar{Z}}^{\vec{M}^{\mathrm{A\rightleftharpoons M}}\vec{J}} & \bar{\bar{Z}}^{\vec{M}^{\mathrm{A\rightleftharpoons M}}\vec{J}_{\mathrm{M}}} & \bar{\bar{Z}}^{\vec{M}^{\mathrm{A\rightleftharpoons M}}\vec{J}_{\mathrm{M\rightleftharpoons A}}} & 0 & 0 & 0 & \bar{\bar{Z}}^{\vec{M}^{\mathrm{A\rightleftharpoons M}}\vec{M}^{\mathrm{A\rightleftharpoons M}}} & \bar{\bar{Z}}^{\vec{M}^{\mathrm{A\rightleftharpoons M}}\vec{M}} & \bar{\bar{Z}}^{\vec{M}^{\mathrm{A\rightleftharpoons M}}\vec{M}_{\mathrm{M\rightleftharpoons A}}} & 0 \\
0 & 0 & \bar{\bar{Z}}^{\vec{J}^{\mathrm{M}}\vec{J}^{\mathrm{A\rightleftharpoons M}}} & \bar{\bar{Z}}^{\vec{J}^{\mathrm{M}}\vec{J}^{\mathrm{M}}} & \bar{\bar{Z}}^{\vec{J}^{\mathrm{M}}\vec{J}} & \bar{\bar{Z}}^{\vec{J}^{\mathrm{M}}\vec{J}_{\mathrm{M}}} & \bar{\bar{Z}}^{\vec{J}^{\mathrm{M}}\vec{J}_{\mathrm{M\rightleftharpoons A}}} & 0 & 0 & 0 & \bar{\bar{Z}}^{\vec{J}^{\mathrm{M}}\vec{M}^{\mathrm{A\rightleftharpoons M}}} & \bar{\bar{Z}}^{\vec{J}^{\mathrm{M}}\vec{M}} & \bar{\bar{Z}}^{\vec{J}^{\mathrm{M}}\vec{M}_{\mathrm{M\rightleftharpoons A}}} & 0 \\
0 & 0 & \bar{\bar{Z}}^{\vec{J}\vec{J}^{\mathrm{A\rightleftharpoons M}}} & \bar{\bar{Z}}^{\vec{J}\vec{J}^{\mathrm{M}}} & \bar{\bar{Z}}^{\vec{J}\vec{J}} & \bar{\bar{Z}}^{\vec{J}\vec{J}_{\mathrm{M}}} & \bar{\bar{Z}}^{\vec{J}\vec{J}_{\mathrm{M\rightleftharpoons A}}} & 0 & 0 & 0 & \bar{\bar{Z}}^{\vec{J}\vec{M}^{\mathrm{A\rightleftharpoons M}}} & \bar{\bar{Z}}^{\vec{J}\vec{M}} & \bar{\bar{Z}}^{\vec{J}\vec{M}_{\mathrm{M\rightleftharpoons A}}} & 0 \\
0 & 0 & \bar{\bar{Z}}^{\vec{M}\vec{J}^{\mathrm{A\rightleftharpoons M}}} & \bar{\bar{Z}}^{\vec{M}\vec{J}^{\mathrm{M}}} & \bar{\bar{Z}}^{\vec{M}\vec{J}} & \bar{\bar{Z}}^{\vec{M}\vec{J}_{\mathrm{M}}} & \bar{\bar{Z}}^{\vec{M}\vec{J}_{\mathrm{M\rightleftharpoons A}}} & 0 & 0 & 0 & \bar{\bar{Z}}^{\vec{M}\vec{M}^{\mathrm{A\rightleftharpoons M}}} & \bar{\bar{Z}}^{\vec{M}\vec{M}} & \bar{\bar{Z}}^{\vec{M}\vec{M}_{\mathrm{M\rightleftharpoons A}}} & 0 \\
0 & 0 & \bar{\bar{Z}}^{\vec{J}_{\mathrm{M}}\vec{J}^{\mathrm{A\rightleftharpoons M}}} & \bar{\bar{Z}}^{\vec{J}_{\mathrm{M}}\vec{J}^{\mathrm{M}}} & \bar{\bar{Z}}^{\vec{J}_{\mathrm{M}}\vec{J}} & \bar{\bar{Z}}^{\vec{J}_{\mathrm{M}}\vec{J}_{\mathrm{M}}} & \bar{\bar{Z}}^{\vec{J}_{\mathrm{M}}\vec{J}_{\mathrm{M\rightleftharpoons A}}} & 0 & 0 & 0 & \bar{\bar{Z}}^{\vec{J}_{\mathrm{M}}\vec{M}^{\mathrm{A\rightleftharpoons M}}} & \bar{\bar{Z}}^{\vec{J}_{\mathrm{M}}\vec{M}} & \bar{\bar{Z}}^{\vec{J}_{\mathrm{M}}\vec{M}_{\mathrm{M\rightleftharpoons A}}} & 0 \\
0 & 0 & \bar{\bar{Z}}^{\vec{J}_{\mathrm{M\rightleftharpoons A}}\vec{J}^{\mathrm{A\rightleftharpoons M}}} & \bar{\bar{Z}}^{\vec{J}_{\mathrm{M\rightleftharpoons A}}\vec{J}^{\mathrm{M}}} & \bar{\bar{Z}}^{\vec{J}_{\mathrm{M\rightleftharpoons A}}\vec{J}} & \bar{\bar{Z}}^{\vec{J}_{\mathrm{M\rightleftharpoons A}}\vec{J}_{\mathrm{M}}} & \bar{\bar{Z}}^{\vec{J}_{\mathrm{M\rightleftharpoons A}}\vec{J}_{\mathrm{M\rightleftharpoons A}}} & \bar{\bar{Z}}^{\vec{J}_{\mathrm{M\rightleftharpoons A}}\vec{J}_{\mathrm{A}}} & \bar{\bar{Z}}^{\vec{J}_{\mathrm{M\rightleftharpoons A}}\vec{J}_{\mathrm{A\rightleftharpoons G}}} & 0 & \bar{\bar{Z}}^{\vec{J}_{\mathrm{M\rightleftharpoons A}}\vec{M}^{\mathrm{A\rightleftharpoons M}}} & \bar{\bar{Z}}^{\vec{J}_{\mathrm{M\rightleftharpoons A}}\vec{M}} & \bar{\bar{Z}}^{\vec{J}_{\mathrm{M\rightleftharpoons A}}\vec{M}_{\mathrm{M\rightleftharpoons A}}} & \bar{\bar{Z}}^{\vec{J}_{\mathrm{M\rightleftharpoons A}}\vec{M}_{\mathrm{A\rightleftharpoons G}}} \\
0 & 0 & \bar{\bar{Z}}^{\vec{M}_{\mathrm{M\rightleftharpoons A}}\vec{J}^{\mathrm{A\rightleftharpoons M}}} & \bar{\bar{Z}}^{\vec{M}_{\mathrm{M\rightleftharpoons A}}\vec{J}^{\mathrm{M}}} & \bar{\bar{Z}}^{\vec{M}_{\mathrm{M\rightleftharpoons A}}\vec{J}} & \bar{\bar{Z}}^{\vec{M}_{\mathrm{M\rightleftharpoons A}}\vec{J}_{\mathrm{M}}} & \bar{\bar{Z}}^{\vec{M}_{\mathrm{M\rightleftharpoons A}}\vec{J}_{\mathrm{M\rightleftharpoons A}}} & \bar{\bar{Z}}^{\vec{M}_{\mathrm{M\rightleftharpoons A}}\vec{J}_{\mathrm{A}}} & \bar{\bar{Z}}^{\vec{M}_{\mathrm{M\rightleftharpoons A}}\vec{J}_{\mathrm{A\rightleftharpoons G}}} & 0 & \bar{\bar{Z}}^{\vec{M}_{\mathrm{M\rightleftharpoons A}}\vec{M}^{\mathrm{A\rightleftharpoons M}}} & \bar{\bar{Z}}^{\vec{M}_{\mathrm{M\rightleftharpoons A}}\vec{M}} & \bar{\bar{Z}}^{\vec{M}_{\mathrm{M\rightleftharpoons A}}\vec{M}_{\mathrm{M\rightleftharpoons A}}} & \bar{\bar{Z}}^{\vec{M}_{\mathrm{M\rightleftharpoons A}}\vec{M}_{\mathrm{A\rightleftharpoons G}}} \\
0 & 0 & 0 & 0 & 0 & 0 & \bar{\bar{Z}}^{\vec{J}_{\mathrm{A}}\vec{J}_{\mathrm{M\rightleftharpoons A}}} & \bar{\bar{Z}}^{\vec{J}_{\mathrm{A}}\vec{J}_{\mathrm{A}}} & \bar{\bar{Z}}^{\vec{J}_{\mathrm{A}}\vec{J}_{\mathrm{A\rightleftharpoons G}}} & 0 & 0 & 0 & \bar{\bar{Z}}^{\vec{J}_{\mathrm{A}}\vec{M}_{\mathrm{M\rightleftharpoons A}}} & \bar{\bar{Z}}^{\vec{J}_{\mathrm{A}}\vec{M}_{\mathrm{A\rightleftharpoons G}}} \\
0 & 0 & 0 & 0 & 0 & 0 & \bar{\bar{Z}}^{\vec{M}_{\mathrm{A\rightleftharpoons G}}\vec{J}_{\mathrm{M\rightleftharpoons A}}} & \bar{\bar{Z}}^{\vec{M}_{\mathrm{A\rightleftharpoons G}}\vec{J}_{\mathrm{A}}} & \bar{\bar{Z}}^{\vec{M}_{\mathrm{A\rightleftharpoons G}}\vec{J}_{\mathrm{A\rightleftharpoons G}}} & 0 & 0 & 0 & \bar{\bar{Z}}^{\vec{M}_{\mathrm{A\rightleftharpoons G}}\vec{M}_{\mathrm{M\rightleftharpoons A}}} & \bar{\bar{Z}}^{\vec{M}_{\mathrm{A\rightleftharpoons G}}\vec{M}_{\mathrm{A\rightleftharpoons G}}} \\
0 & 0 & 0 & 0 & 0 & 0 & \bar{\bar{Z}}^{\vec{J}_{\mathrm{A\rightleftharpoons G}}\vec{J}_{\mathrm{M\rightleftharpoons A}}} & \bar{\bar{Z}}^{\vec{J}_{\mathrm{A\rightleftharpoons G}}\vec{J}_{\mathrm{A}}} & \bar{\bar{Z}}^{\vec{J}_{\mathrm{A\rightleftharpoons G}}\vec{J}_{\mathrm{A\rightleftharpoons G}}} & 0 & 0 & 0 & \bar{\bar{Z}}^{\vec{J}_{\mathrm{A\rightleftharpoons G}}\vec{M}_{\mathrm{M\rightleftharpoons A}}} & \bar{\bar{Z}}^{\vec{J}_{\mathrm{A\rightleftharpoons G}}\vec{M}_{\mathrm{A\rightleftharpoons G}}}
\end{bmatrix} \tag{8-117a}$$

$$\overline{\overline{\Psi}}_4 = \begin{bmatrix} \overline{\overline{I}}^{\vec{M}^{\mathrm{G}\rightleftharpoons\mathrm{A}}} \\ -\overline{\overline{Z}}^{\vec{J}^{\mathrm{G}\rightleftharpoons\mathrm{A}}\vec{M}^{\mathrm{G}\rightleftharpoons\mathrm{A}}} \\ -\overline{\overline{Z}}^{\vec{J}^{\mathrm{A}}\vec{M}^{\mathrm{G}\rightleftharpoons\mathrm{A}}} \\ -\overline{\overline{Z}}^{\vec{J}^{\mathrm{A}\rightleftharpoons\mathrm{M}}\vec{M}^{\mathrm{G}\rightleftharpoons\mathrm{A}}} \\ -\overline{\overline{Z}}^{\vec{M}^{\mathrm{A}\rightleftharpoons\mathrm{M}}\vec{M}^{\mathrm{G}\rightleftharpoons\mathrm{A}}} \\ 0 \\ 0 \\ 0 \\ 0 \\ 0 \\ 0 \\ 0 \\ 0 \\ 0 \end{bmatrix} \tag{8-117b}$$

where the 0s are some zero matrices with proper row numbers and column numbers.

By solving the above augmented matrix equations, there exist the following transformations from $\overline{a}^{\vec{J}^{\mathrm{G}\rightleftharpoons\mathrm{A}}}$ to $\overline{a}^{\mathrm{AV}}$ and from $\overline{a}^{\vec{M}^{\mathrm{G}\rightleftharpoons\mathrm{A}}}$ to $\overline{a}^{\mathrm{AV}}$.

$$\overline{a}^{\mathrm{AV}} = \overbrace{\left(\overline{\overline{\Psi}}_1\right)^{-1} \cdot \overline{\overline{\Psi}}_2}^{\overline{\overline{T}}^{\vec{J}^{\mathrm{G}\rightleftharpoons\mathrm{A}}\rightarrow\mathrm{AV}}} \cdot \overline{a}^{\vec{J}^{\mathrm{G}\rightleftharpoons\mathrm{A}}} \tag{8-118}$$

$$\overline{a}^{\mathrm{AV}} = \underbrace{\left(\overline{\overline{\Psi}}_3\right)^{-1} \cdot \overline{\overline{\Psi}}_4}_{\overline{\overline{T}}^{\vec{M}^{\mathrm{G}\rightleftharpoons\mathrm{A}}\rightarrow\mathrm{AV}}} \cdot \overline{a}^{\vec{M}^{\mathrm{G}\rightleftharpoons\mathrm{A}}} \tag{8-119}$$

and they can be uniformly written as follows:

$$\overline{a}^{\mathrm{AV}} = \overline{\overline{T}}^{\mathrm{BV}\rightarrow\mathrm{AV}} \cdot \overline{a}^{\mathrm{BV}} \tag{8-120}$$

where $\overline{\overline{T}}^{\mathrm{BV}\rightarrow\mathrm{AV}} = \overline{\overline{T}}^{\vec{J}^{\mathrm{G}\rightleftharpoons\mathrm{A}}\rightarrow\mathrm{AV}} / \overline{\overline{T}}^{\vec{M}^{\mathrm{G}\rightleftharpoons\mathrm{A}}\rightarrow\mathrm{AV}}$, and correspondingly $\overline{a}^{\mathrm{BV}} = \overline{a}^{\vec{J}^{\mathrm{G}\rightleftharpoons\mathrm{A}}} / \overline{a}^{\vec{M}^{\mathrm{G}\rightleftharpoons\mathrm{A}}}$.

### 8.3.3.4 Scheme II: Solution/Definition Domain Compression (SDC/DDC)

In fact, the Eqs. (8-90a) and (8-91)~(8-98b) can be alternatively combined as follows:

$$\overline{\overline{\Psi}}_{\mathrm{FCE}}^{\mathrm{DoJ}} \cdot \overline{a}^{\mathrm{AV}} = 0 \tag{8-121}$$

where

$$\overline{\overline{\Psi}}_{\mathrm{FCE}}^{\mathrm{DoJ}} = \begin{bmatrix}
\overline{\overline{Z}}^{\vec{M}^{\mathrm{G}\rightleftharpoons\mathrm{A}}\vec{J}^{\mathrm{G}\rightleftharpoons\mathrm{A}}} & \overline{\overline{Z}}^{\vec{M}^{\mathrm{G}\rightleftharpoons\mathrm{A}}\vec{J}^{\mathrm{A}}} & \overline{\overline{Z}}^{\vec{M}^{\mathrm{G}\rightleftharpoons\mathrm{A}}\vec{J}^{\mathrm{A}\rightleftharpoons\mathrm{M}}} & 0 & 0 & 0 & 0 & 0 & 0 & \overline{\overline{Z}}^{\vec{M}^{\mathrm{G}\rightleftharpoons\mathrm{A}}\vec{M}^{\mathrm{G}\rightleftharpoons\mathrm{A}}} & \overline{\overline{Z}}^{\vec{M}^{\mathrm{G}\rightleftharpoons\mathrm{A}}\vec{M}^{\mathrm{A}\rightleftharpoons\mathrm{M}}} & 0 & 0 & 0 \\
\overline{\overline{Z}}^{\vec{J}^{\mathrm{A}}\vec{J}^{\mathrm{G}\rightleftharpoons\mathrm{A}}} & \overline{\overline{Z}}^{\vec{J}^{\mathrm{A}}\vec{J}^{\mathrm{A}}} & \overline{\overline{Z}}^{\vec{J}^{\mathrm{A}}\vec{J}^{\mathrm{A}\rightleftharpoons\mathrm{M}}} & 0 & 0 & 0 & 0 & 0 & 0 & \overline{\overline{Z}}^{\vec{J}^{\mathrm{A}}\vec{M}^{\mathrm{G}\rightleftharpoons\mathrm{A}}} & \overline{\overline{Z}}^{\vec{J}^{\mathrm{A}}\vec{M}^{\mathrm{A}\rightleftharpoons\mathrm{M}}} & 0 & 0 & 0 \\
\overline{\overline{Z}}^{\vec{J}^{\mathrm{A}\rightleftharpoons\mathrm{M}}\vec{J}^{\mathrm{G}\rightleftharpoons\mathrm{A}}} & \overline{\overline{Z}}^{\vec{J}^{\mathrm{A}\rightleftharpoons\mathrm{M}}\vec{J}^{\mathrm{A}}} & \overline{\overline{Z}}^{\vec{J}^{\mathrm{A}\rightleftharpoons\mathrm{M}}\vec{J}^{\mathrm{A}\rightleftharpoons\mathrm{M}}} & \overline{\overline{Z}}^{\vec{J}^{\mathrm{A}\rightleftharpoons\mathrm{M}}\vec{J}^{\mathrm{M}}} & \overline{\overline{Z}}^{\vec{J}^{\mathrm{A}\rightleftharpoons\mathrm{M}}\vec{J}} & \overline{\overline{Z}}^{\vec{J}^{\mathrm{A}\rightleftharpoons\mathrm{M}}\vec{J}_{\mathrm{M}}} & \overline{\overline{Z}}^{\vec{J}^{\mathrm{A}\rightleftharpoons\mathrm{M}}\vec{J}_{\mathrm{M}\rightleftharpoons\mathrm{A}}} & 0 & 0 & \overline{\overline{Z}}^{\vec{J}^{\mathrm{A}\rightleftharpoons\mathrm{M}}\vec{M}^{\mathrm{G}\rightleftharpoons\mathrm{A}}} & \overline{\overline{Z}}^{\vec{J}^{\mathrm{A}\rightleftharpoons\mathrm{M}}\vec{M}^{\mathrm{A}\rightleftharpoons\mathrm{M}}} & \overline{\overline{Z}}^{\vec{J}^{\mathrm{A}\rightleftharpoons\mathrm{M}}\vec{M}} & \overline{\overline{Z}}^{\vec{J}^{\mathrm{A}\rightleftharpoons\mathrm{M}}\vec{M}_{\mathrm{M}\rightleftharpoons\mathrm{A}}} & 0 \\
\overline{\overline{Z}}^{\vec{M}^{\mathrm{A}\rightleftharpoons\mathrm{M}}\vec{J}^{\mathrm{G}\rightleftharpoons\mathrm{A}}} & \overline{\overline{Z}}^{\vec{M}^{\mathrm{A}\rightleftharpoons\mathrm{M}}\vec{J}^{\mathrm{A}}} & \overline{\overline{Z}}^{\vec{M}^{\mathrm{A}\rightleftharpoons\mathrm{M}}\vec{J}^{\mathrm{A}\rightleftharpoons\mathrm{M}}} & \overline{\overline{Z}}^{\vec{M}^{\mathrm{A}\rightleftharpoons\mathrm{M}}\vec{J}^{\mathrm{M}}} & \overline{\overline{Z}}^{\vec{M}^{\mathrm{A}\rightleftharpoons\mathrm{M}}\vec{J}} & \overline{\overline{Z}}^{\vec{M}^{\mathrm{A}\rightleftharpoons\mathrm{M}}\vec{J}_{\mathrm{M}}} & \overline{\overline{Z}}^{\vec{M}^{\mathrm{A}\rightleftharpoons\mathrm{M}}\vec{J}_{\mathrm{M}\rightleftharpoons\mathrm{A}}} & 0 & 0 & \overline{\overline{Z}}^{\vec{M}^{\mathrm{A}\rightleftharpoons\mathrm{M}}\vec{M}^{\mathrm{G}\rightleftharpoons\mathrm{A}}} & \overline{\overline{Z}}^{\vec{M}^{\mathrm{A}\rightleftharpoons\mathrm{M}}\vec{M}^{\mathrm{A}\rightleftharpoons\mathrm{M}}} & \overline{\overline{Z}}^{\vec{M}^{\mathrm{A}\rightleftharpoons\mathrm{M}}\vec{M}} & \overline{\overline{Z}}^{\vec{M}^{\mathrm{A}\rightleftharpoons\mathrm{M}}\vec{M}_{\mathrm{M}\rightleftharpoons\mathrm{A}}} & 0 \\
0 & 0 & \overline{\overline{Z}}^{\vec{J}^{\mathrm{M}}\vec{J}^{\mathrm{A}\rightleftharpoons\mathrm{M}}} & \overline{\overline{Z}}^{\vec{J}^{\mathrm{M}}\vec{J}^{\mathrm{M}}} & \overline{\overline{Z}}^{\vec{J}^{\mathrm{M}}\vec{J}} & \overline{\overline{Z}}^{\vec{J}^{\mathrm{M}}\vec{J}_{\mathrm{M}}} & \overline{\overline{Z}}^{\vec{J}^{\mathrm{M}}\vec{J}_{\mathrm{M}\rightleftharpoons\mathrm{A}}} & 0 & 0 & 0 & \overline{\overline{Z}}^{\vec{J}^{\mathrm{M}}\vec{M}^{\mathrm{A}\rightleftharpoons\mathrm{M}}} & \overline{\overline{Z}}^{\vec{J}^{\mathrm{M}}\vec{M}} & \overline{\overline{Z}}^{\vec{J}^{\mathrm{M}}\vec{M}_{\mathrm{M}\rightleftharpoons\mathrm{A}}} & 0 \\
0 & 0 & \overline{\overline{Z}}^{\vec{J}\vec{J}^{\mathrm{A}\rightleftharpoons\mathrm{M}}} & \overline{\overline{Z}}^{\vec{J}\vec{J}^{\mathrm{M}}} & \overline{\overline{Z}}^{\vec{J}\vec{J}} & \overline{\overline{Z}}^{\vec{J}\vec{J}_{\mathrm{M}}} & \overline{\overline{Z}}^{\vec{J}\vec{J}_{\mathrm{M}\rightleftharpoons\mathrm{A}}} & 0 & 0 & 0 & \overline{\overline{Z}}^{\vec{J}\vec{M}^{\mathrm{A}\rightleftharpoons\mathrm{M}}} & \overline{\overline{Z}}^{\vec{J}\vec{M}} & \overline{\overline{Z}}^{\vec{J}\vec{M}_{\mathrm{M}\rightleftharpoons\mathrm{A}}} & 0 \\
0 & 0 & \overline{\overline{Z}}^{\vec{M}\vec{J}^{\mathrm{A}\rightleftharpoons\mathrm{M}}} & \overline{\overline{Z}}^{\vec{M}\vec{J}^{\mathrm{M}}} & \overline{\overline{Z}}^{\vec{M}\vec{J}} & \overline{\overline{Z}}^{\vec{M}\vec{J}_{\mathrm{M}}} & \overline{\overline{Z}}^{\vec{M}\vec{J}_{\mathrm{M}\rightleftharpoons\mathrm{A}}} & 0 & 0 & 0 & \overline{\overline{Z}}^{\vec{M}\vec{M}^{\mathrm{A}\rightleftharpoons\mathrm{M}}} & \overline{\overline{Z}}^{\vec{M}\vec{M}} & \overline{\overline{Z}}^{\vec{M}\vec{M}_{\mathrm{M}\rightleftharpoons\mathrm{A}}} & 0 \\
0 & 0 & \overline{\overline{Z}}^{\vec{J}_{\mathrm{M}}\vec{J}^{\mathrm{A}\rightleftharpoons\mathrm{M}}} & \overline{\overline{Z}}^{\vec{J}_{\mathrm{M}}\vec{J}^{\mathrm{M}}} & \overline{\overline{Z}}^{\vec{J}_{\mathrm{M}}\vec{J}} & \overline{\overline{Z}}^{\vec{J}_{\mathrm{M}}\vec{J}_{\mathrm{M}}} & \overline{\overline{Z}}^{\vec{J}_{\mathrm{M}}\vec{J}_{\mathrm{M}\rightleftharpoons\mathrm{A}}} & 0 & 0 & 0 & \overline{\overline{Z}}^{\vec{J}_{\mathrm{M}}\vec{M}^{\mathrm{A}\rightleftharpoons\mathrm{M}}} & \overline{\overline{Z}}^{\vec{J}_{\mathrm{M}}\vec{M}} & \overline{\overline{Z}}^{\vec{J}_{\mathrm{M}}\vec{M}_{\mathrm{M}\rightleftharpoons\mathrm{A}}} & 0 \\
0 & 0 & \overline{\overline{Z}}^{\vec{J}_{\mathrm{M}\rightleftharpoons\mathrm{A}}\vec{J}^{\mathrm{A}\rightleftharpoons\mathrm{M}}} & \overline{\overline{Z}}^{\vec{J}_{\mathrm{M}\rightleftharpoons\mathrm{A}}\vec{J}^{\mathrm{M}}} & \overline{\overline{Z}}^{\vec{J}_{\mathrm{M}\rightleftharpoons\mathrm{A}}\vec{J}} & \overline{\overline{Z}}^{\vec{J}_{\mathrm{M}\rightleftharpoons\mathrm{A}}\vec{J}_{\mathrm{M}}} & \overline{\overline{Z}}^{\vec{J}_{\mathrm{M}\rightleftharpoons\mathrm{A}}\vec{J}_{\mathrm{M}\rightleftharpoons\mathrm{A}}} & \overline{\overline{Z}}^{\vec{J}_{\mathrm{M}\rightleftharpoons\mathrm{A}}\vec{J}_{\mathrm{A}}} & \overline{\overline{Z}}^{\vec{J}_{\mathrm{M}\rightleftharpoons\mathrm{A}}\vec{J}_{\mathrm{A}\rightleftharpoons\mathrm{G}}} & 0 & \overline{\overline{Z}}^{\vec{J}_{\mathrm{M}\rightleftharpoons\mathrm{A}}\vec{M}^{\mathrm{A}\rightleftharpoons\mathrm{M}}} & \overline{\overline{Z}}^{\vec{J}_{\mathrm{M}\rightleftharpoons\mathrm{A}}\vec{M}} & \overline{\overline{Z}}^{\vec{J}_{\mathrm{M}\rightleftharpoons\mathrm{A}}\vec{M}_{\mathrm{M}\rightleftharpoons\mathrm{A}}} & \overline{\overline{Z}}^{\vec{J}_{\mathrm{M}\rightleftharpoons\mathrm{A}}\vec{M}_{\mathrm{A}\rightleftharpoons\mathrm{G}}} \\
0 & 0 & \overline{\overline{Z}}^{\vec{M}_{\mathrm{M}\rightleftharpoons\mathrm{A}}\vec{J}^{\mathrm{A}\rightleftharpoons\mathrm{M}}} & \overline{\overline{Z}}^{\vec{M}_{\mathrm{M}\rightleftharpoons\mathrm{A}}\vec{J}^{\mathrm{M}}} & \overline{\overline{Z}}^{\vec{M}_{\mathrm{M}\rightleftharpoons\mathrm{A}}\vec{J}} & \overline{\overline{Z}}^{\vec{M}_{\mathrm{M}\rightleftharpoons\mathrm{A}}\vec{J}_{\mathrm{M}}} & \overline{\overline{Z}}^{\vec{M}_{\mathrm{M}\rightleftharpoons\mathrm{A}}\vec{J}_{\mathrm{M}\rightleftharpoons\mathrm{A}}} & \overline{\overline{Z}}^{\vec{M}_{\mathrm{M}\rightleftharpoons\mathrm{A}}\vec{J}_{\mathrm{A}}} & \overline{\overline{Z}}^{\vec{M}_{\mathrm{M}\rightleftharpoons\mathrm{A}}\vec{J}_{\mathrm{A}\rightleftharpoons\mathrm{G}}} & 0 & \overline{\overline{Z}}^{\vec{M}_{\mathrm{M}\rightleftharpoons\mathrm{A}}\vec{M}^{\mathrm{A}\rightleftharpoons\mathrm{M}}} & \overline{\overline{Z}}^{\vec{M}_{\mathrm{M}\rightleftharpoons\mathrm{A}}\vec{M}} & \overline{\overline{Z}}^{\vec{M}_{\mathrm{M}\rightleftharpoons\mathrm{A}}\vec{M}_{\mathrm{M}\rightleftharpoons\mathrm{A}}} & \overline{\overline{Z}}^{\vec{M}_{\mathrm{M}\rightleftharpoons\mathrm{A}}\vec{M}_{\mathrm{A}\rightleftharpoons\mathrm{G}}} \\
0 & 0 & 0 & 0 & 0 & 0 & \overline{\overline{Z}}^{\vec{J}_{\mathrm{A}}\vec{J}_{\mathrm{M}\rightleftharpoons\mathrm{A}}} & \overline{\overline{Z}}^{\vec{J}_{\mathrm{A}}\vec{J}_{\mathrm{A}}} & \overline{\overline{Z}}^{\vec{J}_{\mathrm{A}}\vec{J}_{\mathrm{A}\rightleftharpoons\mathrm{G}}} & 0 & 0 & 0 & \overline{\overline{Z}}^{\vec{J}_{\mathrm{A}}\vec{M}_{\mathrm{M}\rightleftharpoons\mathrm{A}}} & \overline{\overline{Z}}^{\vec{J}_{\mathrm{A}}\vec{M}_{\mathrm{A}\rightleftharpoons\mathrm{G}}} \\
0 & 0 & 0 & 0 & 0 & 0 & \overline{\overline{Z}}^{\vec{M}_{\mathrm{A}\rightleftharpoons\mathrm{G}}\vec{J}_{\mathrm{M}\rightleftharpoons\mathrm{A}}} & \overline{\overline{Z}}^{\vec{M}_{\mathrm{A}\rightleftharpoons\mathrm{G}}\vec{J}_{\mathrm{A}}} & \overline{\overline{Z}}^{\vec{M}_{\mathrm{A}\rightleftharpoons\mathrm{G}}\vec{J}_{\mathrm{A}\rightleftharpoons\mathrm{G}}} & 0 & 0 & 0 & \overline{\overline{Z}}^{\vec{M}_{\mathrm{A}\rightleftharpoons\mathrm{G}}\vec{M}_{\mathrm{M}\rightleftharpoons\mathrm{A}}} & \overline{\overline{Z}}^{\vec{M}_{\mathrm{A}\rightleftharpoons\mathrm{G}}\vec{M}_{\mathrm{A}\rightleftharpoons\mathrm{G}}} \\
0 & 0 & 0 & 0 & 0 & 0 & \overline{\overline{Z}}^{\vec{J}_{\mathrm{A}\rightleftharpoons\mathrm{G}}\vec{J}_{\mathrm{M}\rightleftharpoons\mathrm{A}}} & \overline{\overline{Z}}^{\vec{J}_{\mathrm{A}\rightleftharpoons\mathrm{G}}\vec{J}_{\mathrm{A}}} & \overline{\overline{Z}}^{\vec{J}_{\mathrm{A}\rightleftharpoons\mathrm{G}}\vec{J}_{\mathrm{A}\rightleftharpoons\mathrm{G}}} & 0 & 0 & 0 & \overline{\overline{Z}}^{\vec{J}_{\mathrm{A}\rightleftharpoons\mathrm{G}}\vec{M}_{\mathrm{M}\rightleftharpoons\mathrm{A}}} & \overline{\overline{Z}}^{\vec{J}_{\mathrm{A}\rightleftharpoons\mathrm{G}}\vec{M}_{\mathrm{A}\rightleftharpoons\mathrm{G}}}
\end{bmatrix} \tag{8-122}$$

Similarly, the Eqs. (8-90b) and (8-91)~(8-98b) can be alternatively combined as follows:

$$\overline{\overline{\Psi}}_{\mathrm{FCE}}^{\mathrm{DoM}} \cdot \bar{a}^{\mathrm{AV}} = 0 \tag{8-123}$$

where

$$\overline{\overline{\Psi}}_{\mathrm{FCE}}^{\mathrm{DoM}} = \begin{bmatrix}
\overline{\overline{Z}}^{\vec{J}^{\mathrm{G\rightleftharpoons A}}\vec{J}^{\mathrm{G\rightleftharpoons A}}} & \overline{\overline{Z}}^{\vec{J}^{\mathrm{G\rightleftharpoons A}}\vec{J}^{\mathrm{A}}} & \overline{\overline{Z}}^{\vec{J}^{\mathrm{G\rightleftharpoons A}}\vec{J}^{\mathrm{A\rightleftharpoons M}}} & 0 & 0 & 0 & 0 & 0 & 0 & \overline{\overline{Z}}^{\vec{J}^{\mathrm{G\rightleftharpoons A}}\vec{M}^{\mathrm{G\rightleftharpoons A}}} & \overline{\overline{Z}}^{\vec{J}^{\mathrm{G\rightleftharpoons A}}\vec{M}^{\mathrm{A\rightleftharpoons M}}} & 0 & 0 & 0 \\
\overline{\overline{Z}}^{\vec{J}^{\mathrm{A}}\vec{J}^{\mathrm{G\rightleftharpoons A}}} & \overline{\overline{Z}}^{\vec{J}^{\mathrm{A}}\vec{J}^{\mathrm{A}}} & \overline{\overline{Z}}^{\vec{J}^{\mathrm{A}}\vec{J}^{\mathrm{A\rightleftharpoons M}}} & 0 & 0 & 0 & 0 & 0 & 0 & \overline{\overline{Z}}^{\vec{J}^{\mathrm{A}}\vec{M}^{\mathrm{G\rightleftharpoons A}}} & \overline{\overline{Z}}^{\vec{J}^{\mathrm{A}}\vec{M}^{\mathrm{A\rightleftharpoons M}}} & 0 & 0 & 0 \\
\overline{\overline{Z}}^{\vec{J}^{\mathrm{A\rightleftharpoons M}}\vec{J}^{\mathrm{G\rightleftharpoons A}}} & \overline{\overline{Z}}^{\vec{J}^{\mathrm{A\rightleftharpoons M}}\vec{J}^{\mathrm{A}}} & \overline{\overline{Z}}^{\vec{J}^{\mathrm{A\rightleftharpoons M}}\vec{J}^{\mathrm{A\rightleftharpoons M}}} & \overline{\overline{Z}}^{\vec{J}^{\mathrm{A\rightleftharpoons M}}\vec{J}^{\mathrm{M}}} & \overline{\overline{Z}}^{\vec{J}^{\mathrm{A\rightleftharpoons M}}\vec{J}} & \overline{\overline{Z}}^{\vec{J}^{\mathrm{A\rightleftharpoons M}}\vec{J}_{\mathrm{M}}} & \overline{\overline{Z}}^{\vec{J}^{\mathrm{A\rightleftharpoons M}}\vec{J}_{\mathrm{M\rightleftharpoons A}}} & 0 & 0 & \overline{\overline{Z}}^{\vec{J}^{\mathrm{A\rightleftharpoons M}}\vec{M}^{\mathrm{G\rightleftharpoons A}}} & \overline{\overline{Z}}^{\vec{J}^{\mathrm{A\rightleftharpoons M}}\vec{M}^{\mathrm{A\rightleftharpoons M}}} & \overline{\overline{Z}}^{\vec{J}^{\mathrm{A\rightleftharpoons M}}\vec{M}} & \overline{\overline{Z}}^{\vec{J}^{\mathrm{A\rightleftharpoons M}}\vec{M}_{\mathrm{M\rightleftharpoons A}}} & 0 \\
\overline{\overline{Z}}^{\vec{M}^{\mathrm{A\rightleftharpoons M}}\vec{J}^{\mathrm{G\rightleftharpoons A}}} & \overline{\overline{Z}}^{\vec{M}^{\mathrm{A\rightleftharpoons M}}\vec{J}^{\mathrm{A}}} & \overline{\overline{Z}}^{\vec{M}^{\mathrm{A\rightleftharpoons M}}\vec{J}^{\mathrm{A\rightleftharpoons M}}} & \overline{\overline{Z}}^{\vec{M}^{\mathrm{A\rightleftharpoons M}}\vec{J}^{\mathrm{M}}} & \overline{\overline{Z}}^{\vec{M}^{\mathrm{A\rightleftharpoons M}}\vec{J}} & \overline{\overline{Z}}^{\vec{M}^{\mathrm{A\rightleftharpoons M}}\vec{J}_{\mathrm{M}}} & \overline{\overline{Z}}^{\vec{M}^{\mathrm{A\rightleftharpoons M}}\vec{J}_{\mathrm{M\rightleftharpoons A}}} & 0 & 0 & \overline{\overline{Z}}^{\vec{M}^{\mathrm{A\rightleftharpoons M}}\vec{M}^{\mathrm{G\rightleftharpoons A}}} & \overline{\overline{Z}}^{\vec{M}^{\mathrm{A\rightleftharpoons M}}\vec{M}^{\mathrm{A\rightleftharpoons M}}} & \overline{\overline{Z}}^{\vec{M}^{\mathrm{A\rightleftharpoons M}}\vec{M}} & \overline{\overline{Z}}^{\vec{M}^{\mathrm{A\rightleftharpoons M}}\vec{M}_{\mathrm{M\rightleftharpoons A}}} & 0 \\
0 & 0 & \overline{\overline{Z}}^{\vec{J}^{\mathrm{M}}\vec{J}^{\mathrm{A\rightleftharpoons M}}} & \overline{\overline{Z}}^{\vec{J}^{\mathrm{M}}\vec{J}^{\mathrm{M}}} & \overline{\overline{Z}}^{\vec{J}^{\mathrm{M}}\vec{J}} & \overline{\overline{Z}}^{\vec{J}^{\mathrm{M}}\vec{J}_{\mathrm{M}}} & \overline{\overline{Z}}^{\vec{J}^{\mathrm{M}}\vec{J}_{\mathrm{M\rightleftharpoons A}}} & 0 & 0 & 0 & \overline{\overline{Z}}^{\vec{J}^{\mathrm{M}}\vec{M}^{\mathrm{A\rightleftharpoons M}}} & \overline{\overline{Z}}^{\vec{J}^{\mathrm{M}}\vec{M}} & \overline{\overline{Z}}^{\vec{J}^{\mathrm{M}}\vec{M}_{\mathrm{M\rightleftharpoons A}}} & 0 \\
0 & 0 & \overline{\overline{Z}}^{\vec{J}\vec{J}^{\mathrm{A\rightleftharpoons M}}} & \overline{\overline{Z}}^{\vec{J}\vec{J}^{\mathrm{M}}} & \overline{\overline{Z}}^{\vec{J}\vec{J}} & \overline{\overline{Z}}^{\vec{J}\vec{J}_{\mathrm{M}}} & \overline{\overline{Z}}^{\vec{J}\vec{J}_{\mathrm{M\rightleftharpoons A}}} & 0 & 0 & 0 & \overline{\overline{Z}}^{\vec{J}\vec{M}^{\mathrm{A\rightleftharpoons M}}} & \overline{\overline{Z}}^{\vec{J}\vec{M}} & \overline{\overline{Z}}^{\vec{J}\vec{M}_{\mathrm{M\rightleftharpoons A}}} & 0 \\
0 & 0 & \overline{\overline{Z}}^{\vec{M}\vec{J}^{\mathrm{A\rightleftharpoons M}}} & \overline{\overline{Z}}^{\vec{M}\vec{J}^{\mathrm{M}}} & \overline{\overline{Z}}^{\vec{M}\vec{J}} & \overline{\overline{Z}}^{\vec{M}\vec{J}_{\mathrm{M}}} & \overline{\overline{Z}}^{\vec{M}\vec{J}_{\mathrm{M\rightleftharpoons A}}} & 0 & 0 & 0 & \overline{\overline{Z}}^{\vec{M}\vec{M}^{\mathrm{A\rightleftharpoons M}}} & \overline{\overline{Z}}^{\vec{M}\vec{M}} & \overline{\overline{Z}}^{\vec{M}\vec{M}_{\mathrm{M\rightleftharpoons A}}} & 0 \\
0 & 0 & \overline{\overline{Z}}^{\vec{J}_{\mathrm{M}}\vec{J}^{\mathrm{A\rightleftharpoons M}}} & \overline{\overline{Z}}^{\vec{J}_{\mathrm{M}}\vec{J}^{\mathrm{M}}} & \overline{\overline{Z}}^{\vec{J}_{\mathrm{M}}\vec{J}} & \overline{\overline{Z}}^{\vec{J}_{\mathrm{M}}\vec{J}_{\mathrm{M}}} & \overline{\overline{Z}}^{\vec{J}_{\mathrm{M}}\vec{J}_{\mathrm{M\rightleftharpoons A}}} & 0 & 0 & 0 & \overline{\overline{Z}}^{\vec{J}_{\mathrm{M}}\vec{M}^{\mathrm{A\rightleftharpoons M}}} & \overline{\overline{Z}}^{\vec{J}_{\mathrm{M}}\vec{M}} & \overline{\overline{Z}}^{\vec{J}_{\mathrm{M}}\vec{M}_{\mathrm{M\rightleftharpoons A}}} & 0 \\
0 & 0 & \overline{\overline{Z}}^{\vec{J}_{\mathrm{M\rightleftharpoons A}}\vec{J}^{\mathrm{A\rightleftharpoons M}}} & \overline{\overline{Z}}^{\vec{J}_{\mathrm{M\rightleftharpoons A}}\vec{J}^{\mathrm{M}}} & \overline{\overline{Z}}^{\vec{J}_{\mathrm{M\rightleftharpoons A}}\vec{J}} & \overline{\overline{Z}}^{\vec{J}_{\mathrm{M\rightleftharpoons A}}\vec{J}_{\mathrm{M}}} & \overline{\overline{Z}}^{\vec{J}_{\mathrm{M\rightleftharpoons A}}\vec{J}_{\mathrm{M\rightleftharpoons A}}} & \overline{\overline{Z}}^{\vec{J}_{\mathrm{M\rightleftharpoons A}}\vec{J}_{\mathrm{A}}} & \overline{\overline{Z}}^{\vec{J}_{\mathrm{M\rightleftharpoons A}}\vec{J}_{\mathrm{A\rightleftharpoons G}}} & 0 & \overline{\overline{Z}}^{\vec{J}_{\mathrm{M\rightleftharpoons A}}\vec{M}^{\mathrm{A\rightleftharpoons M}}} & \overline{\overline{Z}}^{\vec{J}_{\mathrm{M\rightleftharpoons A}}\vec{M}} & \overline{\overline{Z}}^{\vec{J}_{\mathrm{M\rightleftharpoons A}}\vec{M}_{\mathrm{M\rightleftharpoons A}}} & \overline{\overline{Z}}^{\vec{J}_{\mathrm{M\rightleftharpoons A}}\vec{M}_{\mathrm{A\rightleftharpoons G}}} \\
0 & 0 & \overline{\overline{Z}}^{\vec{M}_{\mathrm{M\rightleftharpoons A}}\vec{J}^{\mathrm{A\rightleftharpoons M}}} & \overline{\overline{Z}}^{\vec{M}_{\mathrm{M\rightleftharpoons A}}\vec{J}^{\mathrm{M}}} & \overline{\overline{Z}}^{\vec{M}_{\mathrm{M\rightleftharpoons A}}\vec{J}} & \overline{\overline{Z}}^{\vec{M}_{\mathrm{M\rightleftharpoons A}}\vec{J}_{\mathrm{M}}} & \overline{\overline{Z}}^{\vec{M}_{\mathrm{M\rightleftharpoons A}}\vec{J}_{\mathrm{M\rightleftharpoons A}}} & \overline{\overline{Z}}^{\vec{M}_{\mathrm{M\rightleftharpoons A}}\vec{J}_{\mathrm{A}}} & \overline{\overline{Z}}^{\vec{M}_{\mathrm{M\rightleftharpoons A}}\vec{J}_{\mathrm{A\rightleftharpoons G}}} & 0 & \overline{\overline{Z}}^{\vec{M}_{\mathrm{M\rightleftharpoons A}}\vec{M}^{\mathrm{A\rightleftharpoons M}}} & \overline{\overline{Z}}^{\vec{M}_{\mathrm{M\rightleftharpoons A}}\vec{M}} & \overline{\overline{Z}}^{\vec{M}_{\mathrm{M\rightleftharpoons A}}\vec{M}_{\mathrm{M\rightleftharpoons A}}} & \overline{\overline{Z}}^{\vec{M}_{\mathrm{M\rightleftharpoons A}}\vec{M}_{\mathrm{A\rightleftharpoons G}}} \\
0 & 0 & 0 & 0 & 0 & 0 & \overline{\overline{Z}}^{\vec{J}_{\mathrm{A}}\vec{J}_{\mathrm{M\rightleftharpoons A}}} & \overline{\overline{Z}}^{\vec{J}_{\mathrm{A}}\vec{J}_{\mathrm{A}}} & \overline{\overline{Z}}^{\vec{J}_{\mathrm{A}}\vec{J}_{\mathrm{A\rightleftharpoons G}}} & 0 & 0 & 0 & \overline{\overline{Z}}^{\vec{J}_{\mathrm{A}}\vec{M}_{\mathrm{M\rightleftharpoons A}}} & \overline{\overline{Z}}^{\vec{J}_{\mathrm{A}}\vec{M}_{\mathrm{A\rightleftharpoons G}}} \\
0 & 0 & 0 & 0 & 0 & 0 & \overline{\overline{Z}}^{\vec{M}_{\mathrm{A\rightleftharpoons G}}\vec{J}_{\mathrm{M\rightleftharpoons A}}} & \overline{\overline{Z}}^{\vec{M}_{\mathrm{A\rightleftharpoons G}}\vec{J}_{\mathrm{A}}} & \overline{\overline{Z}}^{\vec{M}_{\mathrm{A\rightleftharpoons G}}\vec{J}_{\mathrm{A\rightleftharpoons G}}} & 0 & 0 & 0 & \overline{\overline{Z}}^{\vec{M}_{\mathrm{A\rightleftharpoons G}}\vec{M}_{\mathrm{M\rightleftharpoons A}}} & \overline{\overline{Z}}^{\vec{M}_{\mathrm{A\rightleftharpoons G}}\vec{M}_{\mathrm{A\rightleftharpoons G}}} \\
0 & 0 & 0 & 0 & 0 & 0 & \overline{\overline{Z}}^{\vec{J}_{\mathrm{A\rightleftharpoons G}}\vec{J}_{\mathrm{M\rightleftharpoons A}}} & \overline{\overline{Z}}^{\vec{J}_{\mathrm{A\rightleftharpoons G}}\vec{J}_{\mathrm{A}}} & \overline{\overline{Z}}^{\vec{J}_{\mathrm{A\rightleftharpoons G}}\vec{J}_{\mathrm{A\rightleftharpoons G}}} & 0 & 0 & 0 & \overline{\overline{Z}}^{\vec{J}_{\mathrm{A\rightleftharpoons G}}\vec{M}_{\mathrm{M\rightleftharpoons A}}} & \overline{\overline{Z}}^{\vec{J}_{\mathrm{A\rightleftharpoons G}}\vec{M}_{\mathrm{A\rightleftharpoons G}}}
\end{bmatrix} \tag{8-124}$$

In the above equations, the subscript "FCE" is the acronym for "*field continuation equation*", and the superscripts "DoJ" and "DoM" are the acronyms for "*definition of* $\vec{J}^{G\rightleftharpoons A}$" and "*definition of* $\vec{M}^{G\rightleftharpoons A}$".

Theoretically, the Eqs. (8-121) and (8-123) are equivalent to each other, and they have the same *solution space*. If the *basic solutions* used to span the solution space are denoted as $\{\bar{s}_1^{\mathrm{BS}}, \bar{s}_2^{\mathrm{BS}}, \cdots\}$, then any mode contained in the solution space can be expanded as follows:

$$\bar{a}^{\mathrm{AV}} = \sum_i a_i^{\mathrm{BS}} \bar{s}_i^{\mathrm{BS}} = \underbrace{\begin{bmatrix} \bar{s}_1^{\mathrm{BS}}, \bar{s}_2^{\mathrm{BS}}, \cdots \end{bmatrix}}_{\overline{\overline{T}}^{\mathrm{BS}\to\mathrm{AV}}} \cdot \underbrace{\begin{bmatrix} a_1^{\mathrm{BS}} \\ a_2^{\mathrm{BS}} \\ \vdots \end{bmatrix}}_{\bar{a}^{\mathrm{BS}}} \tag{8-125}$$

where the solution space is just the modal space of the TARA system shown in Fig. 8-21.

For the convenience of the following discussions, Eqs. (8-120) and (8-125) are uniformly written as follows:

$$\bar{a}^{\mathrm{AV}} = \overline{\overline{T}} \cdot \bar{a} \tag{8-126}$$

where $\bar{a} = \bar{a}^{\mathrm{BV}} / \bar{a}^{\mathrm{BS}}$ and correspondingly $\overline{\overline{T}} = \overline{\overline{T}}^{\mathrm{BV}\to\mathrm{AV}} / \overline{\overline{T}}^{\mathrm{BS}\to\mathrm{AV}}$.

## 8.3.4 Power Transport Theorem and Input Power Operator

In this section, we provide the *power transport theorem* (*PTT*) and *input power operator* (*IPO*) of the TARA system shown in Figs. 8-21 and 8-22.

#### 8.3.4.1 Power Transport Theorem

Applying the results obtained in Chap. 2 to the TARA system shown in Figs. 8-21 and 8-22, we immediately have the following PTT for the TARA system

$$P^{\mathrm{G}\rightleftharpoons\mathrm{A}} = P_{\mathrm{dis}}^{\mathrm{A}} + \underbrace{P_{\mathrm{M\,dis}}^{\mathrm{M\,dis}} + P_{\mathrm{sca}}^{\mathrm{rad}} + \underbrace{P_{\mathrm{A}}^{\mathrm{dis}} + P_{\mathrm{A}\rightleftharpoons\mathrm{G}} + j\,P_{\mathrm{A}}^{\mathrm{sto}}}_{P_{\mathrm{M}\rightleftharpoons\mathrm{A}}} + j\,P_{\mathrm{Msto}}^{\mathrm{Msto}}}_{P^{\mathrm{A}\rightleftharpoons\mathrm{M}}} + j\,P_{\mathrm{sto}}^{\mathrm{A}} \tag{8-127}$$

In the above PTT (8-127), $P^{\mathrm{G}\rightleftharpoons\mathrm{A}}$ is the power inputted into the transmitting DRA; $P^{\mathrm{A}\rightleftharpoons\mathrm{M}}$ is power outputted from the transmitting DRA, and also the power inputted into propagation medium; $P_{\mathrm{M}\rightleftharpoons\mathrm{A}}$ is a part of the power outputted from propagation medium, and also the power inputted into the receiving DRA; $P_{\mathrm{A}\rightleftharpoons\mathrm{G}}$ is the power outputted from the receiving DRA, and also the power inputted into rec-guide; $P_{\mathrm{sca}}^{\mathrm{rad}}$ is the other part of the output power from medium, and also the summation of the powers radiated by transmitting system and scattered by receiving system. In addition, $P_{\mathrm{dis}}^{\mathrm{A}}$, $P_{\mathrm{M\,dis}}^{\mathrm{M\,dis}}$, and $P_{\mathrm{A}}^{\mathrm{dis}}$ are the powers dissipated in transmitting DRA, propagation medium, and receiving DRA respectively; $P_{\mathrm{sto}}^{\mathrm{A}}$, $P_{\mathrm{Msto}}^{\mathrm{Msto}}$, and $P_{\mathrm{A}}^{\mathrm{sto}}$ are the powers corresponding to the energies stored in transmitting DRA, propagation medium, and receiving DRA respectively.

The above-mentioned various powers are as follows:

$$P^{\mathrm{G}\rightleftharpoons\mathrm{A}} = (1/2)\iint_{\mathbb{S}^{\mathrm{G}\rightleftharpoons\mathrm{A}}} \left(\vec{E}\times\vec{H}^{\dagger}\right)\cdot\hat{n}^{\to\mathrm{A}}dS \tag{8-128a}$$

$$P^{\mathrm{A}\rightleftharpoons\mathrm{M}} = (1/2)\iint_{\mathbb{S}^{\mathrm{A}\rightleftharpoons\mathrm{M}}} \left(\vec{E}\times\vec{H}^{\dagger}\right)\cdot\hat{n}^{\to\mathrm{M}}dS \tag{8-128b}$$

$$P_{\mathrm{M}\rightleftharpoons\mathrm{A}} = (1/2)\iint_{\mathbb{S}_{\mathrm{M}\rightleftharpoons\mathrm{A}}} \left(\vec{E}\times\vec{H}^{\dagger}\right)\cdot\hat{n}_{\to\mathrm{A}}dS \tag{8-128c}$$

$$P_{\mathrm{A}\rightleftharpoons\mathrm{G}} = (1/2)\iint_{\mathbb{S}_{\mathrm{A}\rightleftharpoons\mathrm{G}}} \left(\vec{E}\times\vec{H}^{\dagger}\right)\cdot\hat{n}_{\to\mathrm{G}}dS \tag{8-128d}$$

$$P_{\mathrm{sca}}^{\mathrm{rad}} = (1/2)\oiint_{\mathbb{S}} \left(\vec{E}\times\vec{H}^{\dagger}\right)\cdot\hat{n}dS \tag{8-128e}$$

$$P_{\mathrm{dis}}^{\mathrm{A}} = (1/2)\left\langle\vec{\sigma}\cdot\vec{E},\vec{E}\right\rangle_{\mathbb{V}^{\mathrm{A}}} \tag{8-128f}$$

$$P_{\mathrm{M\,dis}}^{\mathrm{M\,dis}} = (1/2)\left\langle\vec{\sigma}\cdot\vec{E},\vec{E}\right\rangle_{\mathbb{V}} \tag{8-128g}$$

$$P_{\mathrm{A}}^{\mathrm{dis}} = (1/2)\left\langle\vec{\sigma}\cdot\vec{E},\vec{E}\right\rangle_{\mathbb{V}_{\mathrm{A}}} \tag{8-128h}$$

$$P_{\mathrm{sto}}^{\mathrm{A}} = 2\omega\left[(1/4)\left\langle\vec{H},\vec{\mu}\cdot\vec{H}\right\rangle_{\mathbb{V}^{\mathrm{A}}} - (1/4)\left\langle\vec{\varepsilon}\cdot\vec{E},\vec{E}\right\rangle_{\mathbb{V}^{\mathrm{A}}}\right] \tag{8-128i}$$

$$\begin{aligned} P_{\mathrm{Msto}}^{\mathrm{Msto}} = 2\omega\Big\{&\left[(1/4)\left\langle\vec{H},\mu_0\vec{H}\right\rangle_{\mathbb{F}} - (1/4)\left\langle\varepsilon_0\vec{E},\vec{E}\right\rangle_{\mathbb{F}}\right] \\ &+\left[(1/4)\left\langle\vec{H},\vec{\mu}\cdot\vec{H}\right\rangle_{\mathbb{V}} - (1/4)\left\langle\vec{\varepsilon}\cdot\vec{E},\vec{E}\right\rangle_{\mathbb{V}}\right]\Big\} \end{aligned} \tag{8-128j}$$

$$P_{\mathrm{A}}^{\mathrm{sto}} = 2\omega\left[(1/4)\left\langle\vec{H},\vec{\mu}\cdot\vec{H}\right\rangle_{\mathbb{V}_{\mathrm{A}}} - (1/4)\left\langle\vec{\varepsilon}\cdot\vec{E},\vec{E}\right\rangle_{\mathbb{V}_{\mathrm{A}}}\right] \tag{8-128k}$$

where $\hat{n}^{\to\mathrm{M}}$ is the normal direction of $\mathbb{S}^{\mathrm{A}\rightleftharpoons\mathrm{M}}$ and points to propagation medium, and $\hat{n}_{\rightarrow \mathrm{G}}$ is the normal direction of $\mathbb{S}_{\mathrm{A} \rightleftharpoons \mathrm{G}}$ and points to rec-guide, and the integral domain $\mathbb{F}$ is the region occupied by free space.

### 8.3.4.2 Input Power Operator — Formulation I: Current Form

Based on Eqs. (8-73a)&(8-73b) and the tangential continuity of the $\{\vec{E},\vec{H}\}$ on $\mathbb{S}^{\mathrm{G} \rightleftharpoons \mathrm{A}}$, the IPO $P^{\mathrm{G} \rightleftharpoons \mathrm{A}}$ given in Eq. (8-128a) can be alternatively written as follows:

$$P^{\mathrm{G} \rightleftharpoons \mathrm{A}} = (1/2)\left\langle \hat{n}^{\rightarrow \mathrm{A}} \times \vec{J}^{\mathrm{G} \rightleftharpoons \mathrm{A}}, \vec{M}^{\mathrm{G} \rightleftharpoons \mathrm{A}} \right\rangle_{\mathbb{S}^{\mathrm{G} \rightleftharpoons \mathrm{A}}} \tag{8-129}$$

and it is just the *current form of IPO*.

Inserting Eq. (8-89) into the above current form, the current form is immediately discretized as follows:

$$P^{\mathrm{G} \rightleftharpoons \mathrm{A}} = \left(\bar{a}^{\mathrm{AV}}\right)^{\dagger} \cdot \underbrace{\begin{bmatrix} 0 & 0 & 0 & 0 & 0 & 0 & 0 & 0 & 0 & \overline{\overline{C}}^{\vec{J}^{\mathrm{G} \rightleftharpoons \mathrm{A}} \vec{M}^{\mathrm{G} \rightleftharpoons \mathrm{A}}} & 0 & 0 & 0 & 0 \\ 0 & 0 & 0 & 0 & 0 & 0 & 0 & 0 & 0 & 0 & 0 & 0 & 0 & 0 \\ 0 & 0 & 0 & 0 & 0 & 0 & 0 & 0 & 0 & 0 & 0 & 0 & 0 & 0 \\ 0 & 0 & 0 & 0 & 0 & 0 & 0 & 0 & 0 & 0 & 0 & 0 & 0 & 0 \\ 0 & 0 & 0 & 0 & 0 & 0 & 0 & 0 & 0 & 0 & 0 & 0 & 0 & 0 \\ 0 & 0 & 0 & 0 & 0 & 0 & 0 & 0 & 0 & 0 & 0 & 0 & 0 & 0 \\ 0 & 0 & 0 & 0 & 0 & 0 & 0 & 0 & 0 & 0 & 0 & 0 & 0 & 0 \\ 0 & 0 & 0 & 0 & 0 & 0 & 0 & 0 & 0 & 0 & 0 & 0 & 0 & 0 \\ 0 & 0 & 0 & 0 & 0 & 0 & 0 & 0 & 0 & 0 & 0 & 0 & 0 & 0 \\ 0 & 0 & 0 & 0 & 0 & 0 & 0 & 0 & 0 & 0 & 0 & 0 & 0 & 0 \\ 0 & 0 & 0 & 0 & 0 & 0 & 0 & 0 & 0 & 0 & 0 & 0 & 0 & 0 \\ 0 & 0 & 0 & 0 & 0 & 0 & 0 & 0 & 0 & 0 & 0 & 0 & 0 & 0 \\ 0 & 0 & 0 & 0 & 0 & 0 & 0 & 0 & 0 & 0 & 0 & 0 & 0 & 0 \\ 0 & 0 & 0 & 0 & 0 & 0 & 0 & 0 & 0 & 0 & 0 & 0 & 0 & 0 \end{bmatrix}}_{\overline{\overline{P}}_{\mathrm{curAV}}^{\mathrm{G} \rightleftharpoons \mathrm{A}}} \cdot \bar{a}^{\mathrm{AV}} \tag{8-130}$$

where the elements of sub-matrix $\overline{\overline{C}}^{\vec{J}^{\mathrm{G} \rightleftharpoons \mathrm{A}} \vec{M}^{\mathrm{G} \rightleftharpoons \mathrm{A}}}$ are calculated as follows:

$$c_{\xi\zeta}^{\mathrm{G} \rightleftharpoons \mathrm{A}} = (1/2)\left\langle \hat{n}^{\rightarrow \mathrm{A}} \times \vec{b}_{\xi}^{\vec{J}^{\mathrm{G} \rightleftharpoons \mathrm{A}}}, \vec{b}_{\zeta}^{\vec{M}^{\mathrm{G} \rightleftharpoons \mathrm{A}}} \right\rangle_{\mathbb{S}^{\mathrm{G} \rightleftharpoons \mathrm{A}}} \tag{8-131}$$

To obtain the IPO defined on modal space, we substitute Eq. (8-126) into the above Eq. (8-130), and then we have that

$$P^{\mathrm{G} \rightleftharpoons \mathrm{A}} = \bar{a}^{\dagger} \cdot \underbrace{\left(\overline{\overline{T}}^{\dagger} \cdot \overline{\overline{P}}_{\mathrm{curAV}}^{\mathrm{G} \rightleftharpoons \mathrm{A}} \cdot \overline{\overline{T}}\right)}_{\overline{\overline{P}}_{\mathrm{cur}}^{\mathrm{G} \rightleftharpoons \mathrm{A}}} \cdot \bar{a} \tag{8-132}$$

where subscript "cur" is to emphasize that $\overline{\overline{P}}_{\mathrm{cur}}^{\mathrm{G} \rightleftharpoons \mathrm{A}}$ originates from discretizing the current form of IPO.

#### 8.3.4.3 Input Power Operator — Formulation II: Field-Current Interaction Form

Using source-field relation (8-72), the IPO $P^{\text{G}\rightleftharpoons\text{A}}$ given in Eq. (8-128a) can be further rewritten as follows:

$$\begin{aligned} P^{\text{G}\rightleftharpoons\text{A}} &= -(1/2)\left\langle \vec{J}^{\text{G}\rightleftharpoons\text{A}}, \mathcal{E}_{\text{T}}\left(\vec{J}^{\text{G}\rightleftharpoons\text{A}} + \vec{J}^{\text{A}} + \vec{J}^{\text{A}\rightleftharpoons\text{M}}, \vec{M}^{\text{G}\rightleftharpoons\text{A}} + \vec{M}^{\text{A}\rightleftharpoons\text{M}}\right)\right\rangle_{\mathbb{S}^{\text{G}\rightleftharpoons\text{A}}} \\ &= -(1/2)\left\langle \vec{M}^{\text{G}\rightleftharpoons\text{A}}, \mathcal{H}_{\text{T}}\left(\vec{J}^{\text{G}\rightleftharpoons\text{A}} + \vec{J}^{\text{A}} + \vec{J}^{\text{A}\rightleftharpoons\text{M}}, \vec{M}^{\text{G}\rightleftharpoons\text{A}} + \vec{M}^{\text{A}\rightleftharpoons\text{M}}\right)\right\rangle^{\dagger}_{\mathbb{S}^{\text{G}\rightleftharpoons\text{A}}} \end{aligned} \tag{8-133}$$

and it is just the *field-current interaction form of IPO*.

Inserting Eq. (8-89) into the above interaction form, the interaction form is immediately discretized as follows:

$$P^{\text{G}\rightleftharpoons\text{A}} = \left(\bar{a}^{\text{AV}}\right)^{\dagger} \cdot \bar{\bar{P}}^{\text{G}\rightleftharpoons\text{A}}_{\text{intAV}} \cdot \bar{a}^{\text{AV}} \tag{8-134}$$

in which

$$\bar{\bar{P}}^{\text{G}\rightleftharpoons\text{A}}_{\text{intAV}} = \begin{bmatrix} \bar{\bar{P}}^{\vec{J}^{\text{G}\rightleftharpoons\text{A}}\vec{J}^{\text{G}\rightleftharpoons\text{A}}} & \bar{\bar{P}}^{\vec{J}^{\text{G}\rightleftharpoons\text{A}}\vec{J}^{\text{A}}} & \bar{\bar{P}}^{\vec{J}^{\text{G}\rightleftharpoons\text{A}}\vec{J}^{\text{A}\rightleftharpoons\text{M}}} & 0 & 0 & 0 & 0 & 0 & 0 & \bar{\bar{P}}^{\vec{J}^{\text{G}\rightleftharpoons\text{A}}\vec{M}^{\text{G}\rightleftharpoons\text{A}}} & \bar{\bar{P}}^{\vec{J}^{\text{G}\rightleftharpoons\text{A}}\vec{M}^{\text{A}\rightleftharpoons\text{M}}} & 0 & 0 & 0 \\ 0 & 0 & 0 & 0 & 0 & 0 & 0 & 0 & 0 & 0 & 0 & 0 & 0 & 0 \\ 0 & 0 & 0 & 0 & 0 & 0 & 0 & 0 & 0 & 0 & 0 & 0 & 0 & 0 \\ 0 & 0 & 0 & 0 & 0 & 0 & 0 & 0 & 0 & 0 & 0 & 0 & 0 & 0 \\ 0 & 0 & 0 & 0 & 0 & 0 & 0 & 0 & 0 & 0 & 0 & 0 & 0 & 0 \\ 0 & 0 & 0 & 0 & 0 & 0 & 0 & 0 & 0 & 0 & 0 & 0 & 0 & 0 \\ 0 & 0 & 0 & 0 & 0 & 0 & 0 & 0 & 0 & 0 & 0 & 0 & 0 & 0 \\ 0 & 0 & 0 & 0 & 0 & 0 & 0 & 0 & 0 & 0 & 0 & 0 & 0 & 0 \\ 0 & 0 & 0 & 0 & 0 & 0 & 0 & 0 & 0 & 0 & 0 & 0 & 0 & 0 \\ 0 & 0 & 0 & 0 & 0 & 0 & 0 & 0 & 0 & 0 & 0 & 0 & 0 & 0 \\ 0 & 0 & 0 & 0 & 0 & 0 & 0 & 0 & 0 & 0 & 0 & 0 & 0 & 0 \\ 0 & 0 & 0 & 0 & 0 & 0 & 0 & 0 & 0 & 0 & 0 & 0 & 0 & 0 \\ 0 & 0 & 0 & 0 & 0 & 0 & 0 & 0 & 0 & 0 & 0 & 0 & 0 & 0 \\ 0 & 0 & 0 & 0 & 0 & 0 & 0 & 0 & 0 & 0 & 0 & 0 & 0 & 0 \end{bmatrix} \tag{8-135a}$$

for the JE version, and

$$\bar{\bar{P}}^{\text{G}\rightleftharpoons\text{A}}_{\text{intAV}} = \begin{bmatrix} 0 & 0 & 0 & 0 & 0 & 0 & 0 & 0 & 0 & 0 & 0 & 0 & 0 & 0 \\ 0 & 0 & 0 & 0 & 0 & 0 & 0 & 0 & 0 & 0 & 0 & 0 & 0 & 0 \\ 0 & 0 & 0 & 0 & 0 & 0 & 0 & 0 & 0 & 0 & 0 & 0 & 0 & 0 \\ 0 & 0 & 0 & 0 & 0 & 0 & 0 & 0 & 0 & 0 & 0 & 0 & 0 & 0 \\ 0 & 0 & 0 & 0 & 0 & 0 & 0 & 0 & 0 & 0 & 0 & 0 & 0 & 0 \\ 0 & 0 & 0 & 0 & 0 & 0 & 0 & 0 & 0 & 0 & 0 & 0 & 0 & 0 \\ 0 & 0 & 0 & 0 & 0 & 0 & 0 & 0 & 0 & 0 & 0 & 0 & 0 & 0 \\ 0 & 0 & 0 & 0 & 0 & 0 & 0 & 0 & 0 & 0 & 0 & 0 & 0 & 0 \\ 0 & 0 & 0 & 0 & 0 & 0 & 0 & 0 & 0 & 0 & 0 & 0 & 0 & 0 \\ \bar{\bar{P}}^{\vec{M}^{\text{G}\rightleftharpoons\text{A}}\vec{J}^{\text{G}\rightleftharpoons\text{A}}} & \bar{\bar{P}}^{\vec{M}^{\text{G}\rightleftharpoons\text{A}}\vec{J}^{\text{A}}} & \bar{\bar{P}}^{\vec{M}^{\text{G}\rightleftharpoons\text{A}}\vec{J}^{\text{A}\rightleftharpoons\text{M}}} & 0 & 0 & 0 & 0 & 0 & 0 & \bar{\bar{P}}^{\vec{M}^{\text{G}\rightleftharpoons\text{A}}\vec{M}^{\text{G}\rightleftharpoons\text{A}}} & \bar{\bar{P}}^{\vec{M}^{\text{G}\rightleftharpoons\text{A}}\vec{M}^{\text{A}\rightleftharpoons\text{M}}} & 0 & 0 & 0 \\ 0 & 0 & 0 & 0 & 0 & 0 & 0 & 0 & 0 & 0 & 0 & 0 & 0 & 0 \\ 0 & 0 & 0 & 0 & 0 & 0 & 0 & 0 & 0 & 0 & 0 & 0 & 0 & 0 \\ 0 & 0 & 0 & 0 & 0 & 0 & 0 & 0 & 0 & 0 & 0 & 0 & 0 & 0 \\ 0 & 0 & 0 & 0 & 0 & 0 & 0 & 0 & 0 & 0 & 0 & 0 & 0 & 0 \end{bmatrix}^{\dagger} \tag{8-135b}$$

for the HM version, where the elements of the sub-matrices are calculated as follows:

$$p_{\xi\zeta}^{\vec{J}^{G\rightleftharpoons A}\vec{J}^{G\rightleftharpoons A}} = -(1/2)\left\langle \vec{b}_{\xi}^{\vec{J}^{G\rightleftharpoons A}}, \mathcal{E}_{T}\left(\vec{b}_{\zeta}^{\vec{J}^{G\rightleftharpoons A}}\right)\right\rangle_{\underset{\sim}{\mathbb{S}}^{G\rightleftharpoons A}} \tag{8-136a}$$

$$p_{\xi\zeta}^{\vec{J}^{G\rightleftharpoons A}\vec{J}^{A}} = -(1/2)\left\langle \vec{b}_{\xi}^{\vec{J}^{G\rightleftharpoons A}}, \mathcal{E}_{T}\left(\vec{b}_{\zeta}^{\vec{J}^{A}}\right)\right\rangle_{\underset{\sim}{\mathbb{S}}^{G\rightleftharpoons A}} \tag{8-136b}$$

$$p_{\xi\zeta}^{\vec{J}^{G\rightleftharpoons A}\vec{J}^{A\rightleftharpoons M}} = -(1/2)\left\langle \vec{b}_{\xi}^{\vec{J}^{G\rightleftharpoons A}}, \mathcal{E}_{T}\left(\vec{b}_{\zeta}^{\vec{J}^{A\rightleftharpoons M}}\right)\right\rangle_{\underset{\sim}{\mathbb{S}}^{G\rightleftharpoons A}} \tag{8-136c}$$

$$p_{\xi\zeta}^{\vec{J}^{G\rightleftharpoons A}\vec{M}^{G\rightleftharpoons A}} = -(1/2)\left\langle \vec{b}_{\xi}^{\vec{J}^{G\rightleftharpoons A}}, \mathcal{E}_{T}\left(\vec{b}_{\zeta}^{\vec{M}^{G\rightleftharpoons A}}\right)\right\rangle_{\underset{\sim}{\mathbb{S}}^{G\rightleftharpoons A}} \tag{8-136d}$$

$$p_{\xi\zeta}^{\vec{J}^{G\rightleftharpoons A}\vec{M}^{A\rightleftharpoons M}} = -(1/2)\left\langle \vec{b}_{\xi}^{\vec{J}^{G\rightleftharpoons A}}, \mathcal{E}_{T}\left(\vec{b}_{\zeta}^{\vec{M}^{A\rightleftharpoons M}}\right)\right\rangle_{\underset{\sim}{\mathbb{S}}^{G\rightleftharpoons A}} \tag{8-136e}$$

and

$$p_{\xi\zeta}^{\vec{M}^{G\rightleftharpoons A}\vec{J}^{G\rightleftharpoons A}} = -(1/2)\left\langle \vec{b}_{\xi}^{\vec{M}^{G\rightleftharpoons A}}, \mathcal{H}_{T}\left(\vec{b}_{\zeta}^{\vec{J}^{G\rightleftharpoons A}}\right)\right\rangle_{\underset{\sim}{\mathbb{S}}^{G\rightleftharpoons A}} \tag{8-136f}$$

$$p_{\xi\zeta}^{\vec{M}^{G\rightleftharpoons A}\vec{J}^{A}} = -(1/2)\left\langle \vec{b}_{\xi}^{\vec{M}^{G\rightleftharpoons A}}, \mathcal{H}_{T}\left(\vec{b}_{\zeta}^{\vec{J}^{A}}\right)\right\rangle_{\underset{\sim}{\mathbb{S}}^{G\rightleftharpoons A}} \tag{8-136g}$$

$$p_{\xi\zeta}^{\vec{M}^{G\rightleftharpoons A}\vec{J}^{A\rightleftharpoons M}} = -(1/2)\left\langle \vec{b}_{\xi}^{\vec{M}^{G\rightleftharpoons A}}, \mathcal{H}_{T}\left(\vec{b}_{\zeta}^{\vec{J}^{A\rightleftharpoons M}}\right)\right\rangle_{\underset{\sim}{\mathbb{S}}^{G\rightleftharpoons A}} \tag{8-136h}$$

$$p_{\xi\zeta}^{\vec{M}^{G\rightleftharpoons A}\vec{M}^{G\rightleftharpoons A}} = -(1/2)\left\langle \vec{b}_{\xi}^{\vec{M}^{G\rightleftharpoons A}}, \mathcal{H}_{T}\left(\vec{b}_{\zeta}^{\vec{M}^{G\rightleftharpoons A}}\right)\right\rangle_{\underset{\sim}{\mathbb{S}}^{G\rightleftharpoons A}} \tag{8-136i}$$

$$p_{\xi\zeta}^{\vec{M}^{G\rightleftharpoons A}\vec{M}^{A\rightleftharpoons M}} = -(1/2)\left\langle \vec{b}_{\xi}^{\vec{M}^{G\rightleftharpoons A}}, \mathcal{H}_{T}\left(\vec{b}_{\zeta}^{\vec{M}^{A\rightleftharpoons M}}\right)\right\rangle_{\underset{\sim}{\mathbb{S}}^{G\rightleftharpoons A}} \tag{8-136j}$$

where the integral surface $\underset{\sim}{\mathbb{S}}^{G\rightleftharpoons A}$ is shown in Fig. 8-23.

To obtain the IPO defined on modal space, we substitute Eq. (8-126) into the above Eq. (8-134), and then we have that

$$P^{G\rightleftharpoons A} = \bar{a}^{\dagger} \cdot \underbrace{\left(\overline{\overline{T}}^{\dagger} \cdot \overline{\overline{P}}_{\text{intAV}}^{G\rightleftharpoons A} \cdot \overline{\overline{T}}\right)}_{\overline{\overline{P}}_{\text{int}}^{G\rightleftharpoons A}} \cdot \bar{a} \tag{8-137}$$

where subscript "int" is to emphasize that $\overline{\overline{P}}_{\text{int}}^{G\rightleftharpoons A}$ originates from discretizing the interaction form of IPO.

For the convenience of the following discussions, the Eqs. (8-132) and (8-137) are uniformly written as follows:

$$P^{G\rightleftharpoons A} = \bar{a}^{\dagger} \cdot \overline{\overline{P}}^{G\rightleftharpoons A} \cdot \bar{a} \tag{8-138}$$

where $\overline{\overline{P}}^{G\rightleftharpoons A} = \overline{\overline{P}}_{\text{cur}}^{G\rightleftharpoons A} / \overline{\overline{P}}_{\text{int}}^{G\rightleftharpoons A}$.

### 8.3.5 Input-Power-Decoupled Modes

Below, we construct the *input-power-decoupled modes* (*IP-DMs*) of the TARA system shown in Figs. 8-21 and 8-22, by using the results obtained above.

### 8.3.5.1 Construction Method

The IP-DMs contained in the modal space can be derived from solving the following modal decoupling equation

$$\overline{\overline{P}}_{-}^{\mathrm{G}\rightleftharpoons\mathrm{A}} \cdot \bar{\alpha}_{\xi} = \theta_{\xi}\, \overline{\overline{P}}_{+}^{\mathrm{G}\rightleftharpoons\mathrm{A}} \cdot \bar{\alpha}_{\xi} \tag{8-139}$$

defined on modal space, where $\overline{\overline{P}}_{+}^{\mathrm{G}\rightleftharpoons\mathrm{A}}$ and $\overline{\overline{P}}_{-}^{\mathrm{G}\rightleftharpoons\mathrm{A}}$ are the *positive and negative Hermitian parts* obtained from the *Toeplitz's decomposition* for the $\overline{\overline{P}}^{\mathrm{G}\rightleftharpoons\mathrm{A}}$ given in Eq. (8-138). The reason to use symbol "$\theta_{\xi}$" instead of the conventional symbol "$\lambda_{\xi}$" will be given in the subsequent Chap. 9.

If some derived modes $\{\bar{\alpha}_1, \bar{\alpha}_2, \cdots, \bar{\alpha}_d\}$ are *d*-order degenerate, then the following *Gram-Schmidt orthogonalization* process is necessary.

$$\left.\begin{aligned} \bar{\alpha}_1 &= \bar{\alpha}_1{}' \\ \bar{\alpha}_2 - \chi_{12}\bar{\alpha}_1{}' &= \bar{\alpha}_2{}' \\ &\cdots \\ \bar{\alpha}_d - \cdots - \chi_{2d}\bar{\alpha}_2{}' - \chi_{1d}\bar{\alpha}_1{}' &= \bar{\alpha}_d{}' \end{aligned}\right\} \tag{8-140}$$

where the coefficients are calculated as follows:

$$\chi_{mn} = \frac{\left(\bar{\alpha}_m{}'\right)^{\dagger} \cdot \overline{\overline{P}}_{+}^{\mathrm{G}\rightleftharpoons\mathrm{A}} \cdot \bar{\alpha}_n}{\left(\bar{\alpha}_m{}'\right)^{\dagger} \cdot \overline{\overline{P}}_{+}^{\mathrm{G}\rightleftharpoons\mathrm{A}} \cdot \bar{\alpha}_m{}'} \tag{8-141}$$

The above-obtained new modes $\{\bar{\alpha}_1{}', \bar{\alpha}_2{}', \cdots, \bar{\alpha}_d{}'\}$ are input-power-decoupled with each other.

### 8.3.5.2 Modal Decoupling Relation and Parseval's Identity

The modal vectors constructed in the above subsection satisfy the following *modal decoupling relation*

$$\begin{aligned} &\bar{\alpha}_{\xi}^{\dagger} \cdot \overline{\overline{P}}^{\mathrm{G}\rightleftharpoons\mathrm{A}} \cdot \bar{\alpha}_{\zeta} \\ =\ &\underbrace{\left[\mathrm{Re}\left\{P_{\xi}^{\mathrm{G}\rightleftharpoons\mathrm{A}}\right\} + j\,\mathrm{Im}\left\{P_{\xi}^{\mathrm{G}\rightleftharpoons\mathrm{A}}\right\}\right]}_{P_{\xi}^{\mathrm{G}\rightleftharpoons\mathrm{A}}} \delta_{\xi\zeta} \xrightarrow{\text{Normalizing } \mathrm{Re}\{P_{\xi}^{\mathrm{G}\rightleftharpoons\mathrm{A}}\} \text{ to } 1} \underbrace{\left(1 + j\,\theta_{\xi}\right)}_{\text{Normalized } P_{\xi}^{\mathrm{G}\rightleftharpoons\mathrm{A}}} \delta_{\xi\zeta} \end{aligned} \tag{8-142}$$

where $P_{\xi}^{\mathrm{G}\rightleftharpoons\mathrm{A}}$ is the *modal input power* corresponding to the $\xi$-th IP-DM. The physical explanation why $\mathrm{Re}\{P_{\xi}^{\mathrm{G}\rightleftharpoons\mathrm{A}}\}$ is normalized to 1 has been given in Ref. [18]. The above matrix-vector multiplication decoupling relation can be alternatively written as the following more physical form

$$(1/2)\iint_{\mathbb{S}^{\mathrm{G}\rightleftharpoons\mathrm{A}}}\left(\vec{E}_{\zeta}\times\vec{H}_{\xi}^{\dagger}\right)\cdot\hat{n}^{\rightarrow\mathrm{A}}dS \;=\; \left(1+j\,\theta_{\xi}\right)\delta_{\xi\zeta} \tag{8-143}$$

and then

$$(1/T)\int_{t_0}^{t_0+T}\left[\iint_{\mathbb{S}^{\mathrm{G}\rightleftharpoons\mathrm{A}}}\left(\vec{\mathcal{E}}_{\zeta}\times\vec{\mathcal{H}}_{\xi}\right)\cdot\hat{n}^{\rightarrow\mathrm{A}}dS\right]dt \;=\; \delta_{\xi\zeta} \tag{8-144}$$

and Eq. (8-144) has a very clear physical meaning — **the modes obtained above are energy-decoupled in any integral period**.

By employing the above decoupling relation, we have the following *Parseval's identity*

$$\sum_{\xi}\left|c_{\xi}\right|^{2} \;=\; (1/T)\int_{t_0}^{t_0+T}\left[\iint_{\mathbb{S}^{\mathrm{G}\rightleftharpoons\mathrm{A}}}\left(\vec{\mathcal{E}}\times\vec{\mathcal{H}}\right)\cdot\hat{n}^{\rightarrow\mathrm{A}}dS\right]dt \tag{8-145}$$

where $c_{\xi}$ is the modal expansion coefficient used in modal expansion formulation and can be explicitly calculated as follows:

$$c_{\xi} = \frac{-(1/2)\left\langle \vec{J}_{\xi}^{\mathrm{G}\rightleftharpoons\mathrm{A}},\vec{E}\right\rangle_{\mathbb{S}^{\mathrm{G}\rightleftharpoons\mathrm{A}}}}{1+j\,\theta_{\xi}} = \frac{-(1/2)\left\langle \vec{H},\vec{M}_{\xi}^{\mathrm{G}\rightleftharpoons\mathrm{A}}\right\rangle_{\mathbb{S}^{\mathrm{G}\rightleftharpoons\mathrm{A}}}}{1+j\,\theta_{\xi}} \tag{8-146}$$

where $\{\vec{E},\vec{H}\}$ is a previously known field distributing on input port $\mathbb{S}^{\mathrm{G}\rightleftharpoons\mathrm{A}}$.

### 8.3.5.3 Modal Quantities

For quantitatively describing the modal features, the following modal quantities are usually used

$$\mathrm{MS}_{\xi} = \frac{1}{\left|1+j\,\theta_{\xi}\right|} \tag{8-147}$$

called modal significance (MS), and

$$Z_{\xi}^{\mathrm{G}\rightleftharpoons\mathrm{A}} \;=\; \frac{P_{\xi}^{\mathrm{G}\rightleftharpoons\mathrm{A}}}{(1/2)\left\langle \vec{J}_{\xi}^{\mathrm{G}\rightleftharpoons\mathrm{A}},\vec{J}_{\xi}^{\mathrm{G}\rightleftharpoons\mathrm{A}}\right\rangle_{\mathbb{S}^{\mathrm{G}\rightleftharpoons\mathrm{A}}}} \;=\; \underbrace{\mathrm{Re}\left\{Z_{\xi}^{\mathrm{G}\rightleftharpoons\mathrm{A}}\right\}}_{R_{\xi}^{\mathrm{G}\rightleftharpoons\mathrm{A}}}+j\,\underbrace{\mathrm{Im}\left\{Z_{\xi}^{\mathrm{G}\rightleftharpoons\mathrm{A}}\right\}}_{X_{\xi}^{\mathrm{G}\rightleftharpoons\mathrm{A}}} \tag{8-148a}$$

called modal input impedance (MII), and

$$Y_{\xi}^{\mathrm{G}\rightleftharpoons\mathrm{A}} \;=\; \frac{P_{\xi}^{\mathrm{G}\rightleftharpoons\mathrm{A}}}{(1/2)\left\langle \vec{M}_{\xi}^{\mathrm{G}\rightleftharpoons\mathrm{A}},\vec{M}_{\xi}^{\mathrm{G}\rightleftharpoons\mathrm{A}}\right\rangle_{\mathbb{S}^{\mathrm{G}\rightleftharpoons\mathrm{A}}}} \;=\; \underbrace{\mathrm{Re}\left\{Y_{\xi}^{\mathrm{G}\rightleftharpoons\mathrm{A}}\right\}}_{G_{\xi}^{\mathrm{G}\rightleftharpoons\mathrm{A}}}+j\,\underbrace{\mathrm{Im}\left\{Y_{\xi}^{\mathrm{G}\rightleftharpoons\mathrm{A}}\right\}}_{B_{\xi}^{\mathrm{G}\rightleftharpoons\mathrm{A}}} \tag{8-148b}$$

called modal input admittance (MIA).

The above MS quantitatively depict the modal weight in whole modal expansion formulation. The above MII and MIA quantitatively depict the allocation way for the energy carried by the mode.

### 8.3.6 Numerical Examples Corresponding to Typical Structures

To verify the theory and formulations established above, we provide a typical numerical example as below.

Here, we consider the TARA system shown in the following Fig. 8-24, which is constituted by a *metallic guide fed metallic transmitting horn* and a *metallic guide loaded metallic receiving horn*.

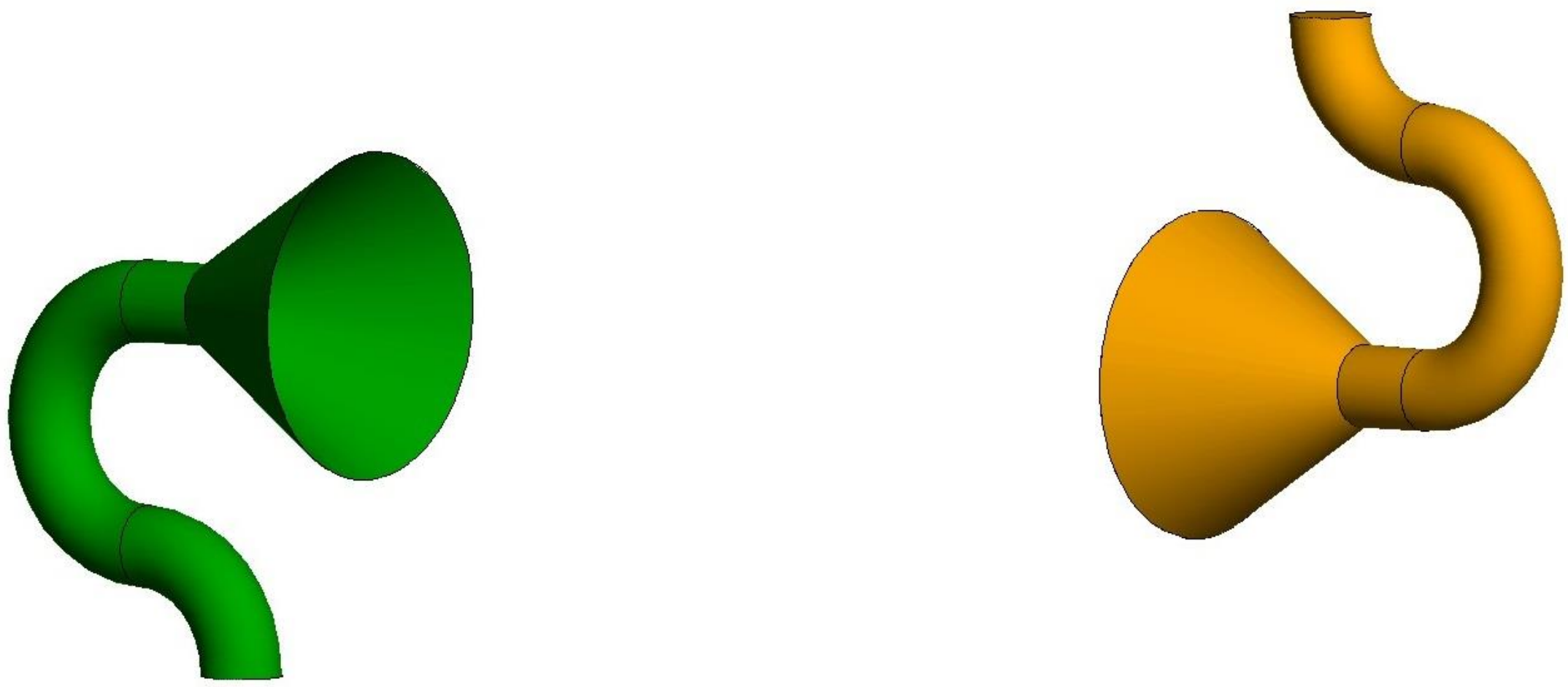

Figure 8-24 Geometry of a TARA system constituted by a metallic guide fed metallic transmitting horn and a metallic guide loaded metallic receiving horn

The geometrical size of the TARA system is shown in the following Fig. 8-25.

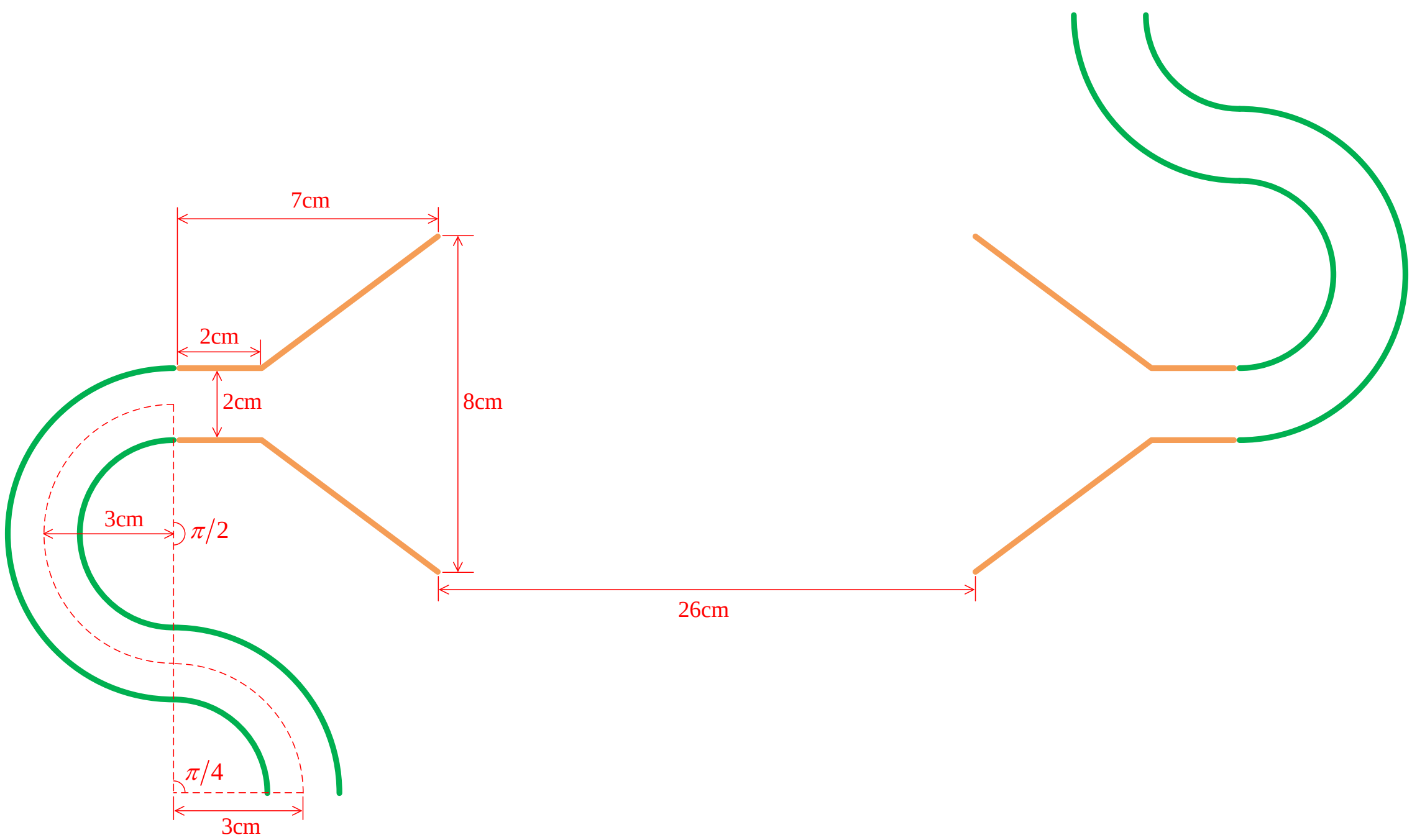

Figure 8-25 Geometrical size of the TARA system shown in Fig. 8-24

The topological structures and surface triangular meshes of the TARA system are shown in the following Fig. 8-26.

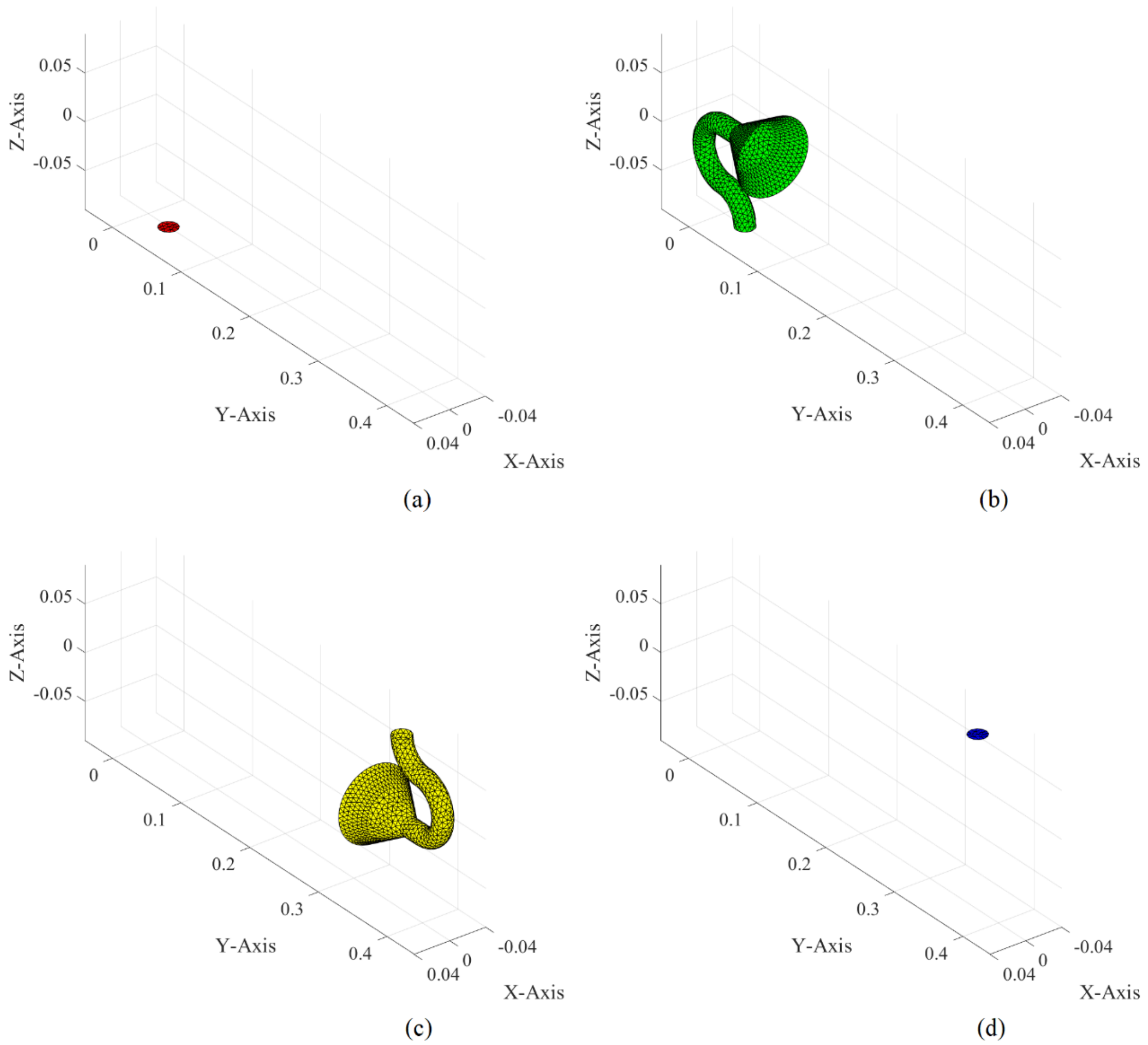

Figure 8-26 Topological structures and surface triangular meshes of the TARA system shown in Fig. 8-24. (a) Input port; (b) augmented tra-antenna; (c) augmented rec-antenna; (d) output port

By orthogonalizing the previously given JE-DoJ-based and HM-DoM-based formulations of IPO, we calculate the IP-DMs of the TARA system, and we plot the modal input resistance curves corresponding to the first several lower order modes in the following Fig. 8-27. From the curves shown in the following Fig. 8-27(a) and Fig. 8-27(b), it is not difficult to observe the fact that: the frequencies corresponding to the peaks of the JE-DoJ-based curves and HM-DoM-based curves are basically consistent with each other, though the peak values of the two kinds of modal input resistance curves have a little difference.

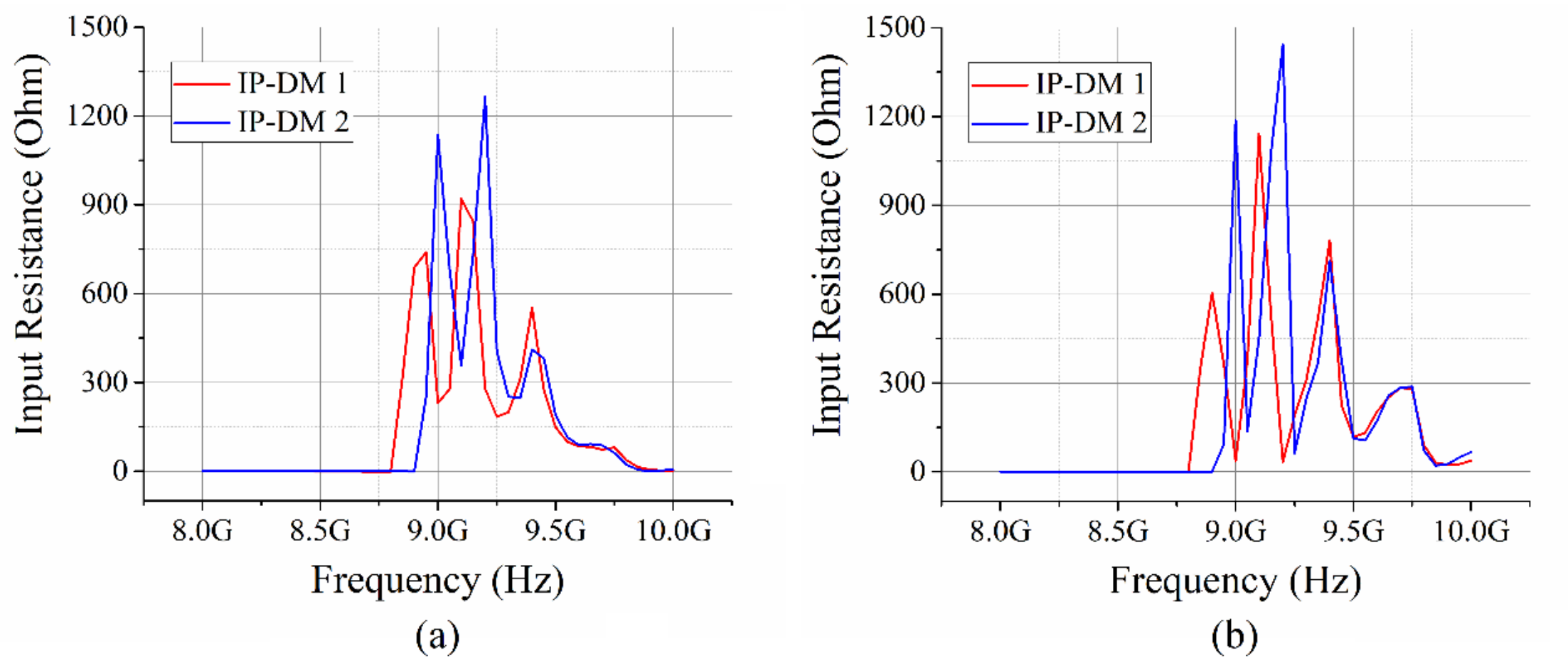

Figure 8-27 Modal input resistance curves of IP-DMs. (a) JE-DoJ-based results; (b) HM-DoM-based results

For the IP-DM 1 in Fig. 8-27(a), it is easy to find out that the curve reaches the local peaks at 8.95 GHz, 9.10 GHz, and 9.40 GHz. The modal electric field distributions (on yOz plane) corresponding to the three frequencies are shown in Fig. 8-28.

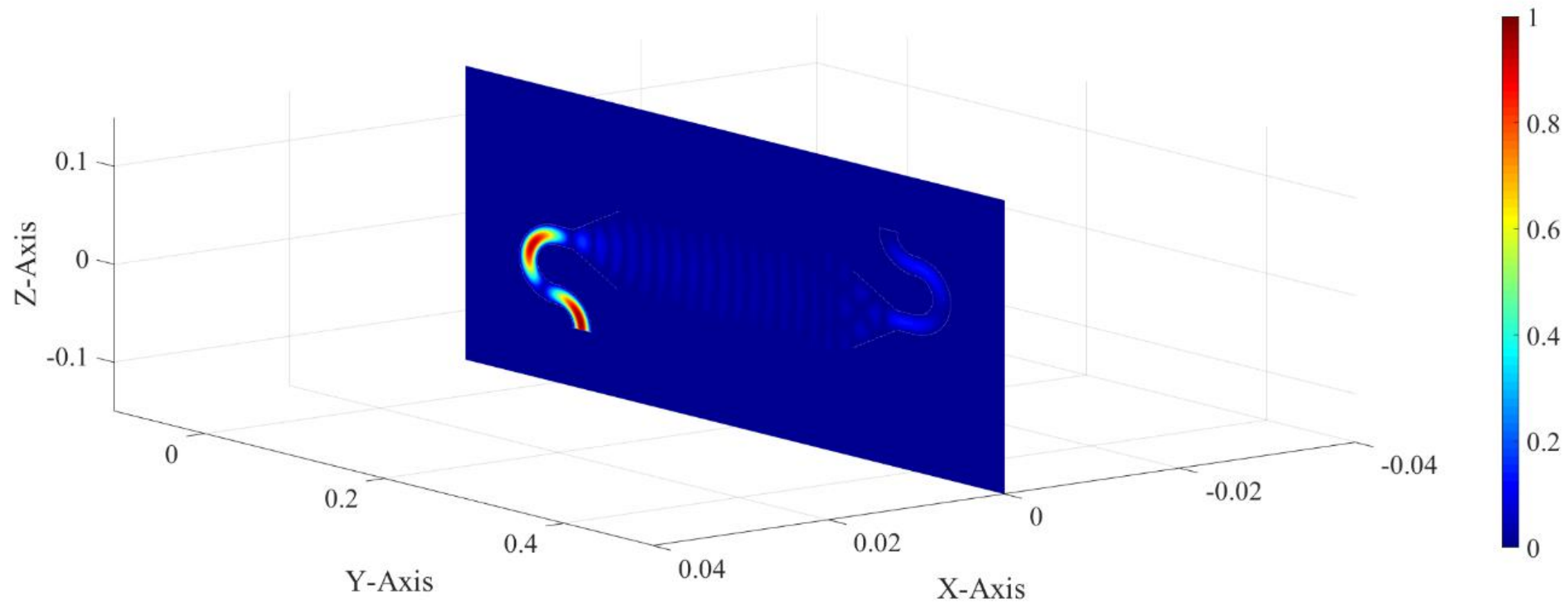

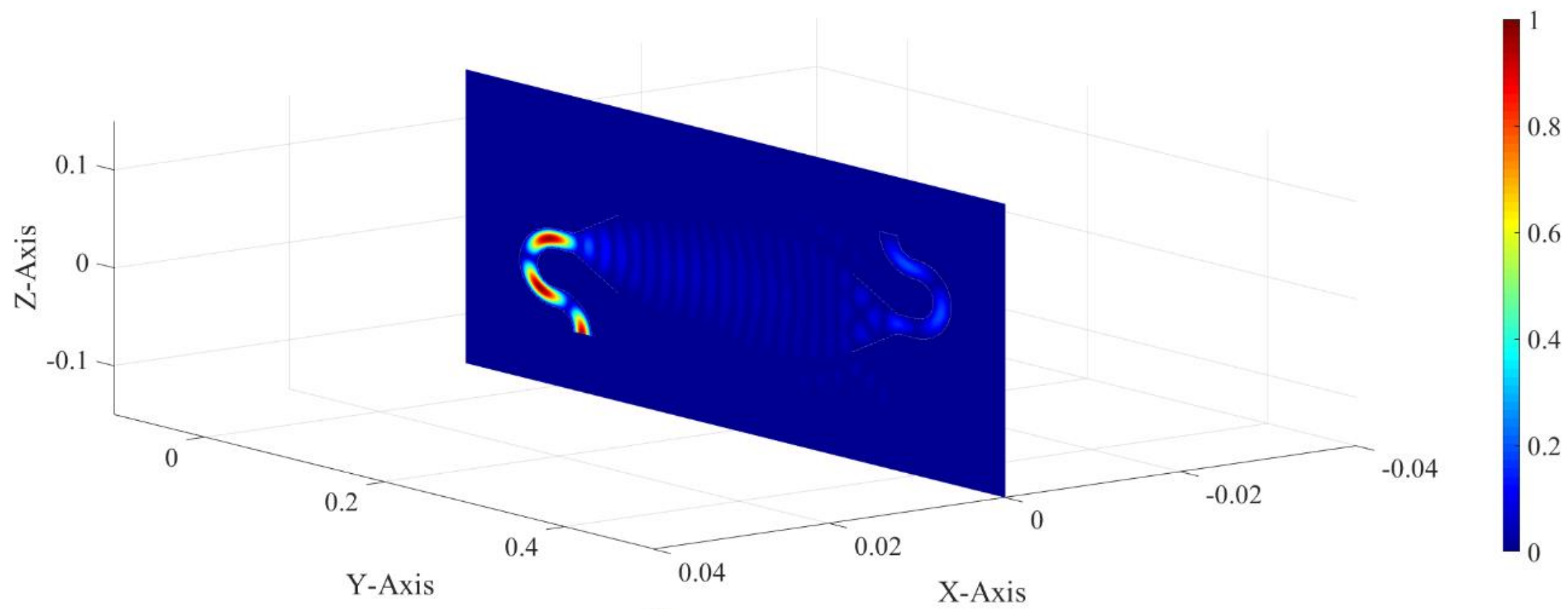

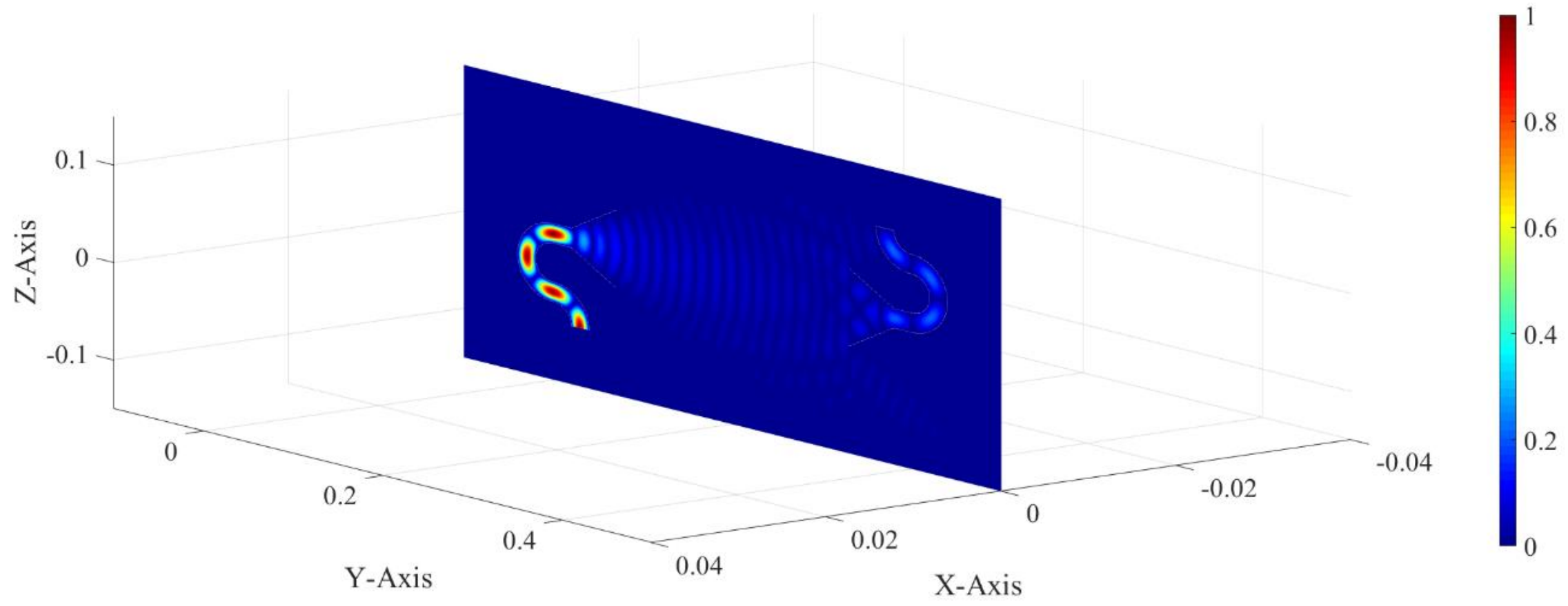

(c)

Figure 8-28 Modal electric field distribution (on yOz plane) of the JE-DoJ-based IP-DM 1 shown in Fig. 8-27(a). (a) 8.95 GHz; (b) 9.10 GHz; (c) 9.40 GHz

## 8.4 Chapter Summary

In this chapter, the *power transport theorem* (*PTT*) based *decoupling mode theories* (*DMTs*) for *augmented tra-guide-tra-antenna combined system* (*TGTA system*) — *PTT-TGTA-DMT* — and for *augmented tra-antenna-rec-antenna combined system* (*TARA system*) — *PTT-TARA-DMT* — are established.

By orthogonalizing frequency-domain *input power operator* (*IPO*), Sec. 8.2 constructs the *input-power-decoupled modes* (*IP-DMs*) of TGTA system, and Sec. 8.3 constructs the IP-DMs of TARA system. The IP-DMs of the other kinds of combined systems can be similarly constructed.

The IP-DMs of the combined systems satisfy a similar *energy decoupling relation* to the IP-DMs of the *guiding structures* discussed in Chap. 3, the *augmented tra-antennas* discussed in Chap. 6, and the *augmented rec-antennas* discussed in Chap. 7. By employing the energy decoupling relation, it is found out that the IP-DMs don't have net energy exchange in any integral period and the IP-DMs satisfy the famous *Parseval's identity*.

# Chapter 9 Modal Decomposition and Electric-Magnetic Decoupling Factor

**CHAPTER MOTIVATION:** For any *working mode* of a *transceiving system* (or its any sub-structure, such as *tra-/rec-guide*, *augmented tra-/rec-antenna*, *TGTA/RARG system*, and *TARA system*, etc.), it is decomposed into three energy-decoupled components — *purely capacitive*, *resonant*, and *inductive components* — in this chapter. Based on the *modal decomposition*, this chapter generalizes the conventional concept of *quantity factor* (i.e. *Q-factor*) to a generalized version — *electric-magnetic decoupling factor* (denoted as *Θ-factor*). **The Θ-factor is a quantitative description for the *coupling degree* between the electric energy and magnetic energy carried by the working mode.**

## 9.1 Chapter Introduction

Traditionally, *quantity factor*, i.e. *Q-factor*, is defined as the ratio of $2\omega(W^{\rm mag}+W^{\rm ele})$ to $P^{\rm rad}+P^{\rm dis}$ [224~236]. Here, $W^{\rm mag}$ and $W^{\rm ele}$ are the *magnetic and electric energies* carried by the working mode; $P^{\rm rad}+P^{\rm dis}$ is the *active power* of the working mode; $\omega$ is the *angular frequency* of the working mode. The Q-fator has had several engineering applications on the analysis and designs for antennas and cavities[237~251].

A main difficulty for evaluating the Q-factor of a mode originates from the term $W^{\rm mag}+W^{\rm ele}$, because of the facts that:

Fact 1. The energies $W^{\rm mag}$ and $W^{\rm ele}$ involve the integral in whole *three-dimensional Euclidean space*, and the integral domain is an infinite one.

Fact 2. The famous *complex Poynting's theorem* cannot be directly employed during calculating Q-factor, because the imaginary part of complex Poynting's flux is propotional to $W^{\rm mag}-W^{\rm ele}$ rather than $W^{\rm mag}+W^{\rm ele}$.

To resolve the above problem, scholars have done some studies[230,236]. However, there has not been a widely recognized calculation method for Q-factor.

In fact, to introduce the concept of Q-factor into electromagnetism society is to quantitatively depict the *coupling degree* between the electric and magnetic energies carried by the mode. This chapter is devoted to providing an alternative way to depict the coupling degree, by introducing a novel concept — *electric-magnetic decoupling factor*, which is simply denoted as *Θ-factor* to be distinguished from the traditional Q-factor.

## 9.2 Modal Decomposition

For an *input port* $\mathbb{S}$, the corresponding *input power* $P$ has the following *integral expression*

$$P = (1/2)\iint_{\mathbb{S}} \left(\vec{E}\times\vec{H}^{\dagger}\right)\cdot\hat{n}dS \tag{9-1}$$

where $\vec{E}$ and $\vec{H}$ are the electric and magnetic fields distributing on $\mathbb{S}$, and $\hat{n}$ is the *unit normal vector* of $\mathbb{S}$. Similar to the previous chapters, the $P$ also has the following *matrix expression*

$$P = \bar{a}^{\dagger}\cdot\bar{\bar{P}}\cdot\bar{a} \tag{9-2}$$

where $\bar{\bar{P}}$ is a square matrix called *input power matrix*, and $\bar{a}$ is a column vector called *modal vector*.

Because of the completeness of *input-power-decoupled modes* (*IP-DMs*) $\{\bar{\alpha}_m\}$, the modal vector $\bar{a}$ can be expanded in terms of the superposition of IP-DMs $\{\bar{\alpha}_m\}$ as

$$\bar{a} = \sum c_m\bar{\alpha}_m \tag{9-3}$$

where expansion coefficient $c_m$ can be explicitly calculated as follows:

$$c_m = \frac{-(1/2)\left\langle\hat{n}\times\vec{H}_m,\vec{E}\right\rangle_{\mathbb{S}}}{P_m} = \frac{-(1/2)\left\langle\vec{H},\vec{E}_m\times\hat{n}\right\rangle_{\mathbb{S}}}{P_m} \tag{9-4}$$

due to the *energy-decoupling feature* of the IP-DMs. In Eq. (9-4), $\vec{E}_m$ and $\vec{H}_m$ are the *m*-th modal electric and magnetic fields, and $P_m$ is the corresponding modal input power, i.e., $P_m = (1/2)\iint_{\mathbb{S}}(\vec{E}_m\times\vec{H}_m^{\dagger})\cdot\hat{n}dS = \bar{\alpha}_m^{\dagger}\cdot\bar{\bar{P}}\cdot\bar{\alpha}_m$.

Following the ideas proposed in Refs. [13] and [252], the whole IP-DM set $\{\bar{\alpha}_m\}$ can be decomposed into three parts — *purely inductive IP-DM set* $\{\bar{\alpha}_\zeta^{ind}\}$ (which is constituted by all inductive IP-DMs), *purely resonant IP-DM set* $\{\bar{\alpha}_\varsigma^{res}\}$ (which is constituted by all resonant IP-DMs), and *purely capacitive IP-DM set* $\{\bar{\alpha}_\xi^{cap}\}$ (which is constituted by all capacitive IP-DMs). Some careful explanations for using modifier "purely" can be found in Refs. [13] and [252], and they are not repeated here.

Thus the modal expansion formulation (9-3) can be alternatively written as the following more illuminating form

$$\bar{a} = \sum c_\zeta^{\rm ind}\bar{\alpha}_\zeta^{\rm ind} + \sum c_\varsigma^{\rm res}\bar{\alpha}_\varsigma^{\rm res} + \sum c_\xi^{\rm cap}\bar{\alpha}_\xi^{\rm cap} \tag{9-5}$$

Following the convention used in Refs. [13] and [252], the three building block terms $\sum c_\zeta^{\rm ind}\bar{\alpha}_\zeta^{\rm ind}$, $\sum c_\varsigma^{\rm res}\bar{\alpha}_\varsigma^{\rm res}$, and $\sum c_\xi^{\rm cap}\bar{\alpha}_\xi^{\rm cap}$ contained in the right-hand side of Eq. (9-5) are denoted as $\bar{\beta}^{\text{ind}}$, $\bar{\beta}^{\text{res}}$, and $\bar{\beta}^{\text{cap}}$ respectively, and then

$$\bar{a} \;=\; \bar{\beta}^{\text{ind}} + \bar{\beta}^{\text{res}} + \bar{\beta}^{\text{cap}} \tag{9-6}$$

where $\bar{\beta}^{\text{ind}}$, $\bar{\beta}^{\text{res}}$, and $\bar{\beta}^{\text{cap}}$ are called *purely inductive term*, *purely resonant term*, and *purely capacitive term* respectively. Clearly, the three components are energy-decoupled.

## 9.3 Electric-Magnetic Decoupling Factor

Because of the *energy-decoupling feature* satisfied by the IP-DMs, the following relations are evident.

$$\text{Re}\left\{\left(\bar{\beta}^{\text{ind}}\right)^{\dagger} \cdot \bar{\bar{P}} \cdot \bar{\beta}^{\text{ind}}\right\} \;=\; \sum \left|c_{\zeta}^{\text{ind}}\right|^2 \text{Re}\left\{P_{\zeta}^{\text{ind}}\right\} \;>\; 0 \tag{9-7a}$$

$$\text{Re}\left\{\left(\bar{\beta}^{\text{res}}\right)^{\dagger} \cdot \bar{\bar{P}} \cdot \bar{\beta}^{\text{res}}\right\} \;=\; \sum \left|c_{\varsigma}^{\text{res}}\right|^2 \text{Re}\left\{P_{\varsigma}^{\text{res}}\right\} \;>\; 0 \tag{9-7b}$$

$$\text{Re}\left\{\left(\bar{\beta}^{\text{cap}}\right)^{\dagger} \cdot \bar{\bar{P}} \cdot \bar{\beta}^{\text{cap}}\right\} \;=\; \sum \left|c_{\xi}^{\text{cap}}\right|^2 \text{Re}\left\{P_{\xi}^{\text{cap}}\right\} \;>\; 0 \tag{9-7c}$$

and

$$\text{Im}\left\{\left(\bar{\beta}^{\text{ind}}\right)^{\dagger} \cdot \bar{\bar{P}} \cdot \bar{\beta}^{\text{ind}}\right\} \;=\; \sum \left|c_{\zeta}^{\text{ind}}\right|^2 \text{Im}\left\{P_{\zeta}^{\text{ind}}\right\} \;>\; 0 \tag{9-8a}$$

$$\text{Im}\left\{\left(\bar{\beta}^{\text{res}}\right)^{\dagger} \cdot \bar{\bar{P}} \cdot \bar{\beta}^{\text{res}}\right\} \;=\; \sum \left|c_{\varsigma}^{\text{res}}\right|^2 \text{Im}\left\{P_{\varsigma}^{\text{res}}\right\} \;=\; 0 \tag{9-8b}$$

$$\text{Im}\left\{\left(\bar{\beta}^{\text{cap}}\right)^{\dagger} \cdot \bar{\bar{P}} \cdot \bar{\beta}^{\text{cap}}\right\} \;=\; \sum \left|c_{\xi}^{\text{cap}}\right|^2 \text{Im}\left\{P_{\xi}^{\text{cap}}\right\} \;<\; 0 \tag{9-8c}$$

where $P_{\zeta}^{\text{ind}}$, $P_{\varsigma}^{\text{res}}$, and $P_{\xi}^{\text{cap}}$ denote the modal input powers of inductive IP-DM $\bar{\alpha}_{\zeta}^{\text{ind}}$, resonant IP-DM $\bar{\alpha}_{\varsigma}^{\text{res}}$, and capacitive IP-DM $\bar{\alpha}_{\xi}^{\text{cap}}$ respectively.

Based on the above observations, we introduce a novel concept of *electric-magnetic decoupling factor* (simply denoted as *Θ-factor*) for any working mode $\bar{a}$ as follows:

$$\begin{aligned}
\Theta(\bar{a}) &= \frac{\left|\text{Im}\left\{\left(\bar{\beta}^{\text{ind}}\right)^{\dagger} \cdot \bar{\bar{P}} \cdot \bar{\beta}^{\text{ind}}\right\}\right| + \left|\text{Im}\left\{\left(\bar{\beta}^{\text{res}}\right)^{\dagger} \cdot \bar{\bar{P}} \cdot \bar{\beta}^{\text{res}}\right\}\right| + \left|\text{Im}\left\{\left(\bar{\beta}^{\text{cap}}\right)^{\dagger} \cdot \bar{\bar{P}} \cdot \bar{\beta}^{\text{cap}}\right\}\right|}{\left|\text{Re}\left\{\left(\bar{\beta}^{\text{ind}}\right)^{\dagger} \cdot \bar{\bar{P}} \cdot \bar{\beta}^{\text{ind}}\right\}\right| + \left|\text{Re}\left\{\left(\bar{\beta}^{\text{res}}\right)^{\dagger} \cdot \bar{\bar{P}} \cdot \bar{\beta}^{\text{res}}\right\}\right| + \left|\text{Re}\left\{\left(\bar{\beta}^{\text{cap}}\right)^{\dagger} \cdot \bar{\bar{P}} \cdot \bar{\beta}^{\text{cap}}\right\}\right|} \\
&= \frac{\text{Im}\left\{\left(\bar{\beta}^{\text{ind}}\right)^{\dagger} \cdot \bar{\bar{P}} \cdot \bar{\beta}^{\text{ind}}\right\} + \text{Im}\left\{\left(\bar{\beta}^{\text{res}}\right)^{\dagger} \cdot \bar{\bar{P}} \cdot \bar{\beta}^{\text{res}}\right\} - \text{Im}\left\{\left(\bar{\beta}^{\text{cap}}\right)^{\dagger} \cdot \bar{\bar{P}} \cdot \bar{\beta}^{\text{cap}}\right\}}{\text{Re}\left\{\left(\bar{\beta}^{\text{ind}}\right)^{\dagger} \cdot \bar{\bar{P}} \cdot \bar{\beta}^{\text{ind}}\right\} + \text{Re}\left\{\left(\bar{\beta}^{\text{res}}\right)^{\dagger} \cdot \bar{\bar{P}} \cdot \bar{\beta}^{\text{res}}\right\} + \text{Re}\left\{\left(\bar{\beta}^{\text{cap}}\right)^{\dagger} \cdot \bar{\bar{P}} \cdot \bar{\beta}^{\text{cap}}\right\}} \\
&= \frac{\text{Im}\left\{\left(\bar{\beta}^{\text{ind}} + \bar{\beta}^{\text{res}} \pm \bar{\beta}^{\text{cap}}\right)^{\dagger} \cdot \bar{\bar{P}} \cdot \left(\bar{\beta}^{\text{ind}} + \bar{\beta}^{\text{res}} \mp \bar{\beta}^{\text{cap}}\right)\right\}}{\text{Re}\left\{\left(\bar{\beta}^{\text{ind}} + \bar{\beta}^{\text{res}} + \bar{\beta}^{\text{cap}}\right)^{\dagger} \cdot \bar{\bar{P}} \cdot \left(\bar{\beta}^{\text{ind}} + \bar{\beta}^{\text{res}} + \bar{\beta}^{\text{cap}}\right)\right\}} \\
&= \frac{\text{Im}\left\{\left(\bar{\beta}^{\text{ind}} \pm \bar{\beta}^{\text{cap}}\right)^{\dagger} \cdot \bar{\bar{P}} \cdot \left(\bar{\beta}^{\text{ind}} \mp \bar{\beta}^{\text{cap}}\right)\right\}}{\text{Re}\left\{\bar{a}^{\dagger} \cdot \bar{\bar{P}} \cdot \bar{a}\right\}}
\end{aligned} \tag{9-9}$$

where the last equality is due to Eq. (9-6). Clearly, the Θ-factor quantitatively describes the *coupling degree* between the electric energy and magnetic energy carried by the working mode, i.e., **the smaller Θ-factor is, the stronger electric energy and magnetic energy couple with each other**. In fact, the Θ-factor can also be viewed as a generalization for the traditional Q-factor, and this is just the reason to select symbol "Θ".

In particular, for the modes which don't include the term $\overline{\beta}^{\text{cap}}$ and then can be expressed as form $\overline{\beta}^{\text{ind}} + \overline{\beta}^{\text{res}}$, their Θ-factors can be expressed as follows:

$$\Theta\left(\overline{\beta}^{\text{ind}} + \overline{\beta}^{\text{res}}\right) = \frac{\text{Im}\{P\}}{\text{Re}\{P\}} \tag{9-10a}$$

for the purely resonant modes, which only includes the term $\overline{\beta}^{\text{res}}$, their Θ-factors can be expressed as follows:

$$\Theta\left(\overline{\beta}^{\text{res}}\right) = 0 \tag{9-10b}$$

for the modes which don't include the term $\overline{\beta}^{\text{ind}}$ and then can be expressed as form $\overline{\beta}^{\text{res}} + \overline{\beta}^{\text{cap}}$, their Θ-factors can be expressed as follows:

$$\Theta\left(\overline{\beta}^{\text{res}} + \overline{\beta}^{\text{cap}}\right) = -\frac{\text{Im}\{P\}}{\text{Re}\{P\}} \tag{9-10c}$$

More particularly, the Θ-factors of the various IP-DMs themselves can be expressed as the following uniform formulation

$$\Theta\left(\overline{\alpha}_m\right) = \left|\theta_m\right| \tag{9-11}$$

where $\theta_m$ is just the one involved in the *modal decoupling equation* $\overline{\overline{P}}_- \cdot \overline{\alpha}_m = \theta_m \overline{\overline{P}}_+ \cdot \overline{\alpha}_m$ used to derive IP-DMs. In fact, this is just the reason why we utilize symbol $\theta_m$ in the equation rather than the traditional symbol $\lambda_m$. Above these reveal the fact that: **unlike the *characteristic value* $\lambda_m$ involved in the *characteristic equation* for *scattering systems*, the *decoupling factor* $\theta_m$ involved in the *decoupling equation* for *transceiving systems* has a clear physical meaning**.

## 9.4 On Modal Significance

Substituting the above relation (9-11) into the definition of *modal significance* (*MS*), we immediately derive the following relation between MS and Θ-factor.

$$\text{MS}_m = \frac{1}{\left|1 + j\,\theta_m\right|} = \frac{1}{\left|1 + j\left|\theta_m\right|\right|} = \frac{1}{\left|1 + j\,\Theta\left(\overline{\alpha}_m\right)\right|} \tag{9-12}$$

where we have used the fact that $\theta_m$ is always real. This implies that MS is a monotonically decreasing function of Θ-factor. Thus, the larger MS is, the stronger modal electric-magnetic energy-coupling is. In fact, this is just the third physical explanation of modal quantity MS. **Now, we list all three known physical explanations of MS as follows:**

**Physical explanation 1. MS is a quantitative description for the weight of a certain IP-DM contained in whole IP-DM-based modal expansion formulation;**

**Physical explanation 2. MS is a quantitative description for the weight of the modal active power contained in the whole modal power of the IP-DM[18];**

**Physical explanation 3. MS is a quantitative description for the coupling degree of the electric energy and magnetic energy carried by the IP-DM.**

"The three somewhat different physical meanings are collected in a single body" reflects the fact that: the MS is a very important physical quantity in decoupling mode theory[18].

## 9.5 Chapter Summary

This chapter can be viewed as a theoretical application for the IP-DMs constructed in the previous chapters.

Based on the completeness of the IP-DMs, any working mode is expressed in terms of the superposition of the IP-DMs. Employing the input-power-decoupling feature satisfied by the IP-DMs, any working mode is decomposed into three energy-decoupled components — purely inductive component, purely resonant component, and purely capacitive component.

Utilizing the modal decomposition, a relatively novel concept — electric-magnetic decoupling factor (Θ-factor) — is introduced. The Θ-factor satisfies the feature that: the smaller Θ-factor is, the stronger electric energy and magnetic energy couple with. Thus, the Θ-factor has ability to quantitatively describe the coupling degree between the electric energy and magnetic energy carried by any working mode. The Θ-factor can also be viewed as a generalization for the traditional Q-factor.

In fact, it is also exhibited that the scalar quantity $\theta$ involved in modal decoupling equation $\overline{\overline{P}}_- \cdot \bar{\alpha} = \theta \overline{\overline{P}}_+ \cdot \bar{\alpha}$ (which was used to calculate IP-DMs in the previous chapters) is just the Θ-factor of the IP-DMs themselves (except sometimes necessary scalar multiplier “$-1$”). This is just the reason to select symbol $\theta$ in the modal decoupling equation, rather than the traditional symbol $\lambda$.

In addition, the three different physical meanings of modal significance (MS) are summarized. “The three somewhat different physical meanings are collected in a single body” reflects the fact that: the MS is a very important physical quantity in decoupling mode theory.

# Chapter 10 Conclusions

**CHAPTER MOTIVATION:** This Postdoctoral Research Report "Research on the Power Transport Theorem Based Decoupling Mode Theory for Transceiving Systems" is a companion volume of the author's Doctoral Dissertation "Research on the Work-Energy Principle Based Characteristic Mode Theory for Scattering Systems". The central purpose of this chapter is to summarize the works done in this research report.

Electromagnetic (EM) transmitting-receiving system (or simply called transceiving system) is a device that transmits EM energy to and/or receives EM energy from the surrounding environment. A typical transceiving system placed in surrounding environment is shown in the following Fig. 10-1.

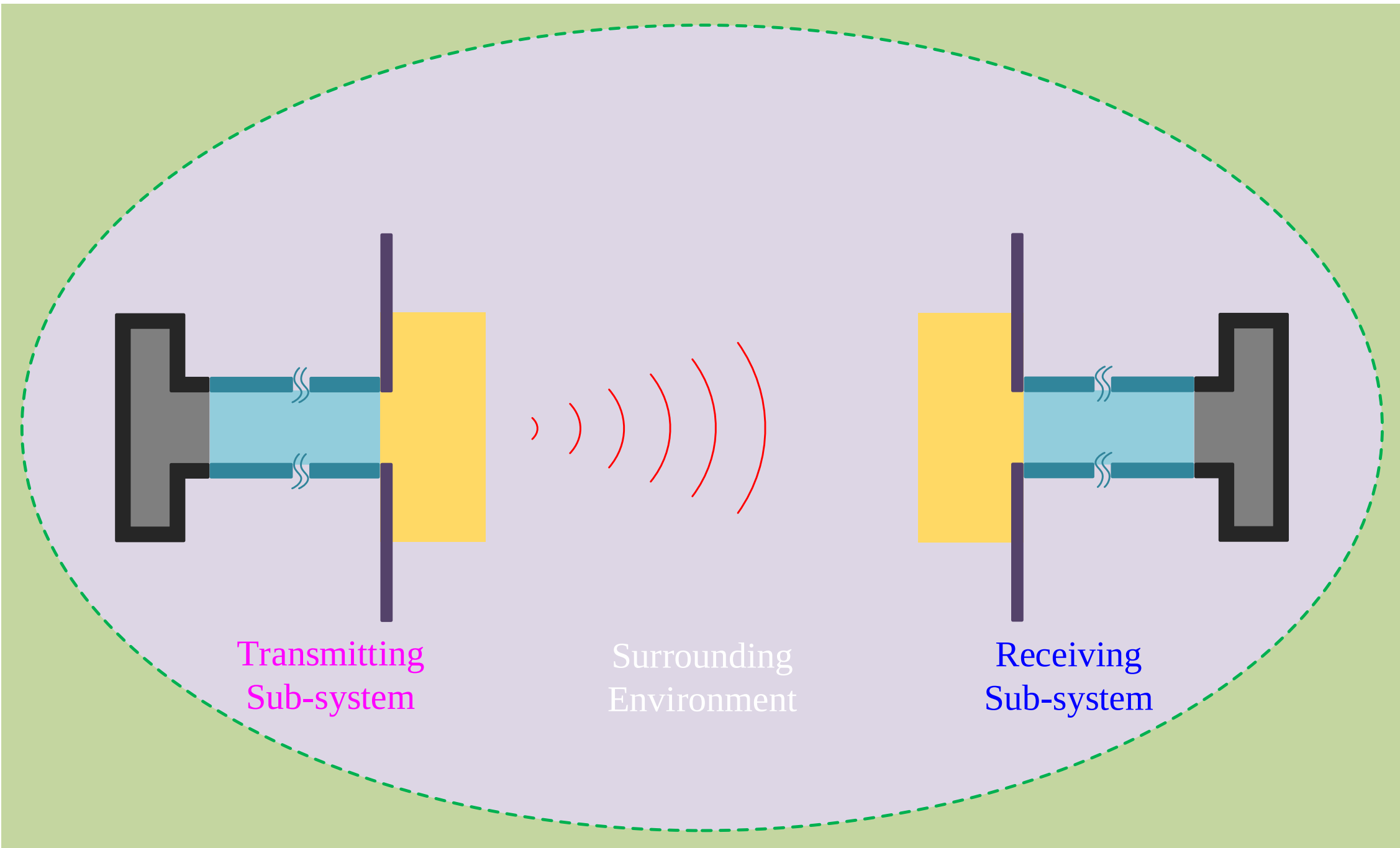

Figure 10-1 A typical transmitting-receiving system, which is placed in a surrounding environment

Transceiving systems have many important applications in EM engineering domain. To effectively study the working states (or alternatively called working modes, or more simply called modes) of a objective transceiving system is very valuable for both the theoretical analysis and engineering design for the system.

Eigen-mode theory (EMT) and characteristic mode theory (CMT) are usually employed to do the modal analysis for transceiving systems. However, their physical pictures clearly reveal the fact that: the EMT and CMT are the modal theories for wave-guiding systems and scattering systems, rather than for transceiving systems. Thus, the EMT-based and CMT-based modal analysis for transceiving systems is only approximate, but by no means rigorous. “How to effectively establish a rigorous modal theory / modal analysis for transceiving systems?” is one of the most important challenges existing in the realms of EM theory and engineering.

Focusing on the challenge, this report does some novel studies in the following 8 main aspects.

**Aspect I. Divisions for Transceiving System and Surrounding Environment**

Whole transceiving system, which is placed in a surrounding environment (simply called environment), is divided into two sub-systems — transmitting system and receiving system. The transmitting system, which is simply called transmitter in this report, is a device or circuit that generates radio-frequency EM energy, controled or modulated, which can be radiated by a transmitting antenna and converted into freely propagating signals. The receiving system, which is simply called receiver in this report, is a device or circuit that collects radio-frequency EM energy, distributing in environment, which can be catched by a receiving antenna and converted into perceptible signals.

The transmitter is further divided into three sub-structures — generating structure, feeding waveguide structure, and transmitting antenna structure. The generating structure, which is simply called generator in this report, is a device used to generate EM energy. The transmitting antenna structure, which is simply called tra-antenna in this report, is a device used to control or modulate the energy to be released into environment. The feeding waveguide structure, which is simply called tra-guide in this report, is a device used to guide the energy from generator to tra-antenna and then to environment.

Similarly, the receiver is also divided into three sub-structures — receiving antenna structure, loading waveguide structure, and absorbing structure. The receiving antenna structure, which is simply called rec-antenna in this report, is a device used to collect or catch the EM energy distributing in environment. The absorbing structure, which is simply called absorber in this report, is a device used to convert the energy into perceptible signal. The loading waveguide structure, which is simply called rec-guide in this report, is a device used to guide the energy from rec-antenna to absorber.

In addition, the environment is divided into two parts — propagation medium and infinity. The infinity is used to receive some EM energies radiated by transmitter and scattered by receiver. The propagation medium, which is simply called medium in this report, is used as the area for propagating the radiated energy from transmitter to {receiver, infinity} and the scattered energy from receiver to {transmitter, infinity}.

Taking the transceiving system shown in Fig. 10-1 as an example, its various sub-structures are illustrated in the following Fig. 10-2.

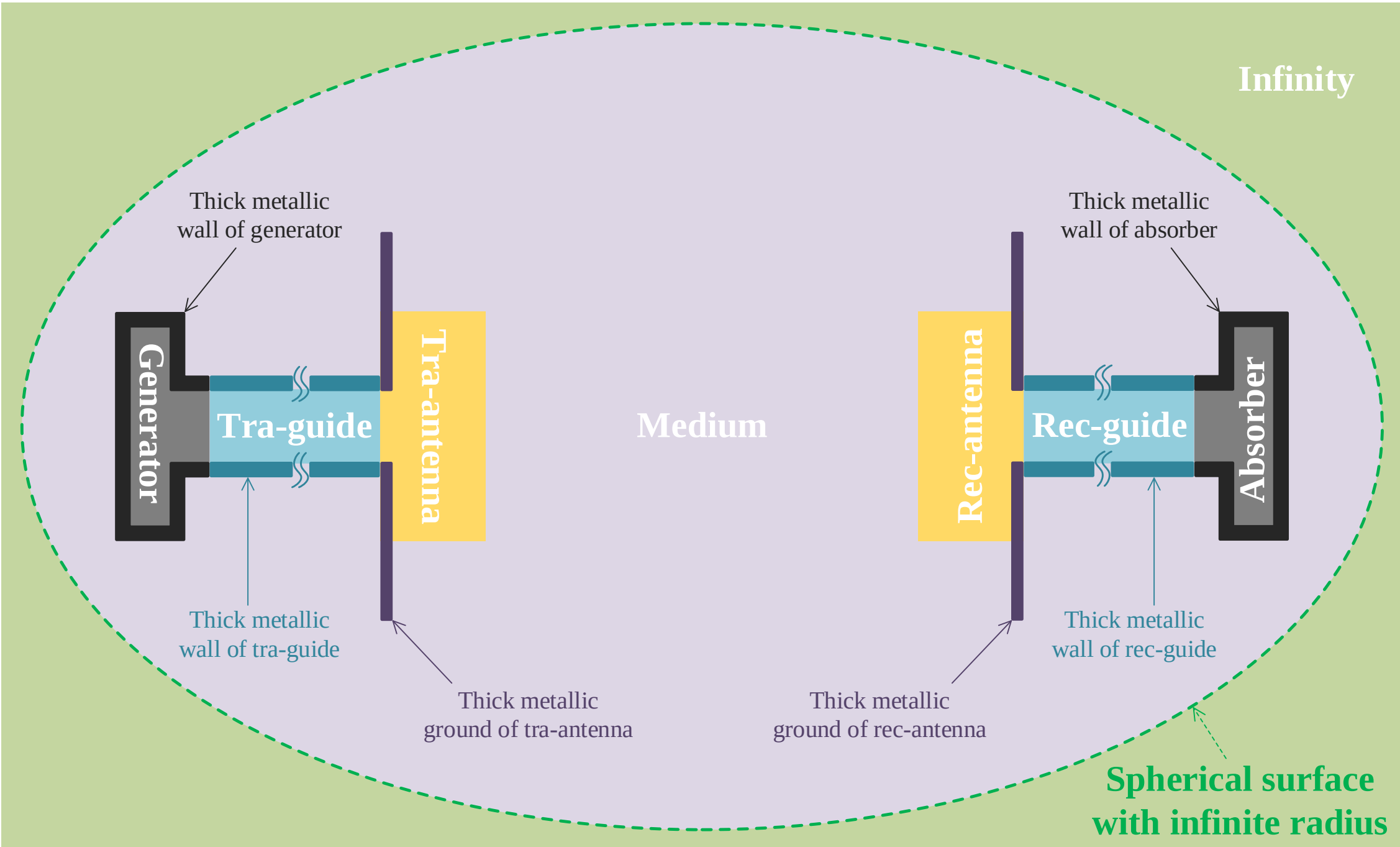

Figure 10-2 Division for the transceiving system shown in Fig. 10-1

In the above figure, there exist two different kinds of structures — two-port structure and one-port structure. The two-port structure consists of an input port, which is used to input energy from its preceding structure to it, and an output port, which is used to output energy from it to its following structure. However, the one-port structure has only one port. For the transceiving system shown in the above Fig. 10-2, it is obvious that the tra-guide, tra-antenna, medium, rec-antenna, and rec-guide are two-port structures, and the generator, infinity, and absorber are one-port structures.

**Aspect II. Division for Three-dimensional Euclidean Space**

In the language of point set topology, the above-mentioned two-port structures and one-port structures can be one-to-one corresponded to some two-port regions and one-port regions in three-dimensional Euclidean space $\mathbb{E}^3$. Thus the above divisions for transceiving system and surrounding environment lead to the following division for $\mathbb{E}^3$.

$$\overbrace{\underbrace{\overbrace{\overbrace{\mathbb{S}^{O}\cup\mathbb{V}^{O}}^{\text{Generator}}\cup\mathbb{S}^{O\rightleftharpoons G}\cup\overbrace{\mathbb{S}^{G}\cup\mathbb{V}^{G}}^{\text{Tra-guide}}}^{\text{Feed}}\cup\mathbb{S}^{G\rightleftharpoons A}\cup\overbrace{\mathbb{S}^{A}\cup\mathbb{V}^{A}}^{\text{Tra-antenna}}\cup\mathbb{W}^{T\|M}}_{\text{Transmitter } \mathbb{T}}\cup\mathbb{S}^{A\rightleftharpoons M}\cup\overbrace{\overbrace{\mathbb{S}^{M}\cup\mathbb{M}\cup\mathbb{S}_{M}}^{\text{Medium}}\cup\mathbb{S}^{M\to I}_{M\to I}\cup\mathbb{I}}^{\text{Environment } \mathbb{E}}\cup\mathbb{S}_{M\rightleftharpoons A}\cup\underbrace{\mathbb{W}_{M\|R}\cup\overbrace{\mathbb{V}_{A}\cup\mathbb{S}_{A}}^{\text{Rec-antenna}}\cup\mathbb{S}_{A\rightleftharpoons G}\cup\overbrace{\overbrace{\mathbb{V}_{G}\cup\mathbb{S}_{G}}^{\text{Rec-guide}}\cup\mathbb{S}_{G\rightleftharpoons O}\cup\overbrace{\mathbb{V}_{O}\cup\mathbb{S}_{O}}^{\text{Absorber}}}^{\text{Load}}}_{\text{Receiver } \mathbb{R}}}^{\text{Three-dimensional Euclidean Space } \mathbb{E}^3} \tag{10-1}$$

O/I Port of $\mathbb{V}^{O}/\mathbb{V}^{G}$; O/I Port of $\mathbb{V}^{G}/\mathbb{V}^{A}$; O/I Port of $\mathbb{V}^{A}/\mathbb{M}$; O/I Port of $\mathbb{M}/\mathbb{V}_{A}$; O/I Port of $\mathbb{V}_{A}/\mathbb{V}_{G}$; O/I Port of $\mathbb{V}_{G}/\mathbb{V}_{O}$

Augmented Tra-antenna; Augmented Rec-antenna

Augmented Tra-guid-Tra-antenna Combined System (TGTA System)

Augmented Tra-antenna-Rec-antenna Combined System (TARA System)

Augmented Rec-antenna-Rec-guide Combined System (RARG System)

which is visually shown as follows:

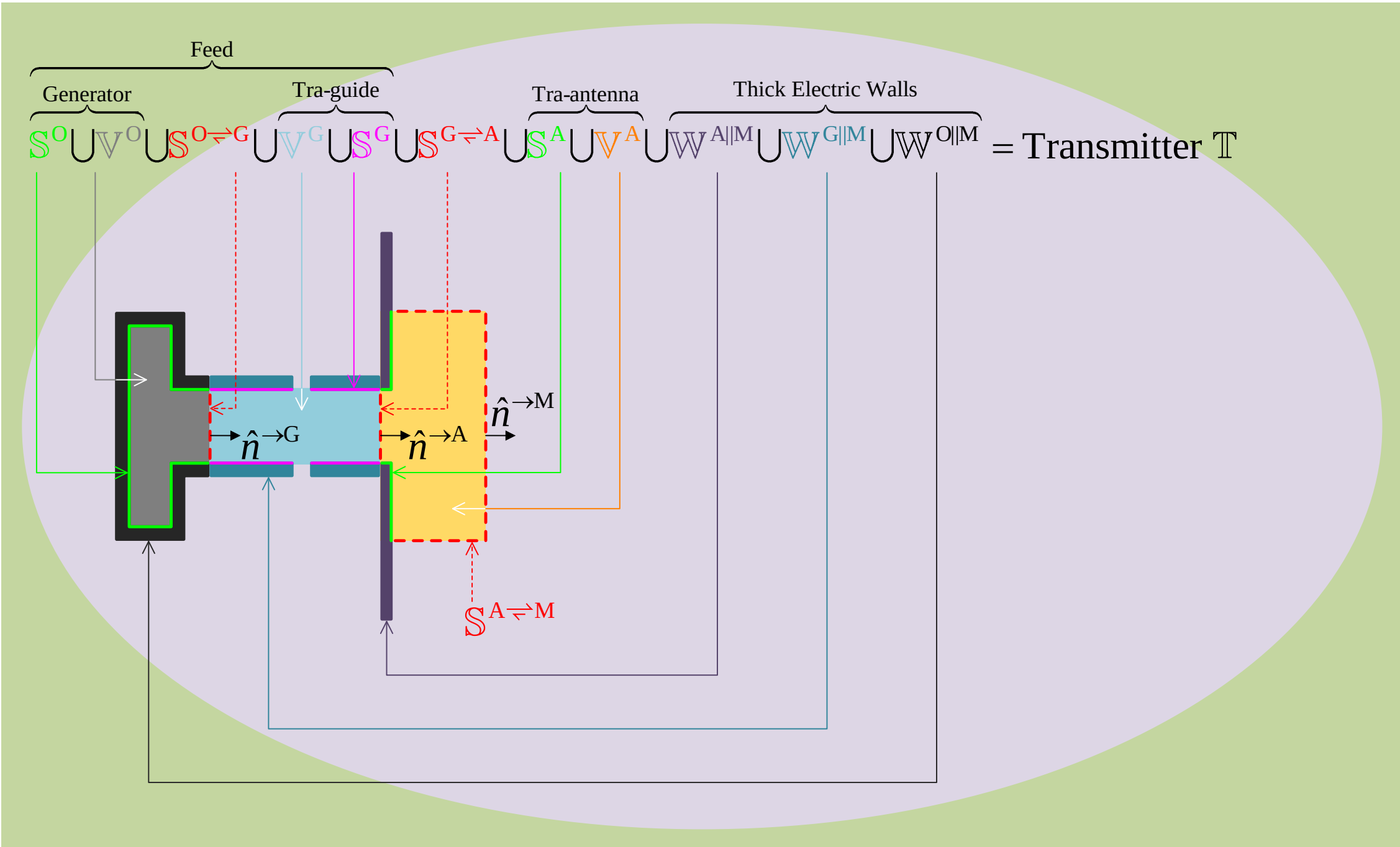

(a)

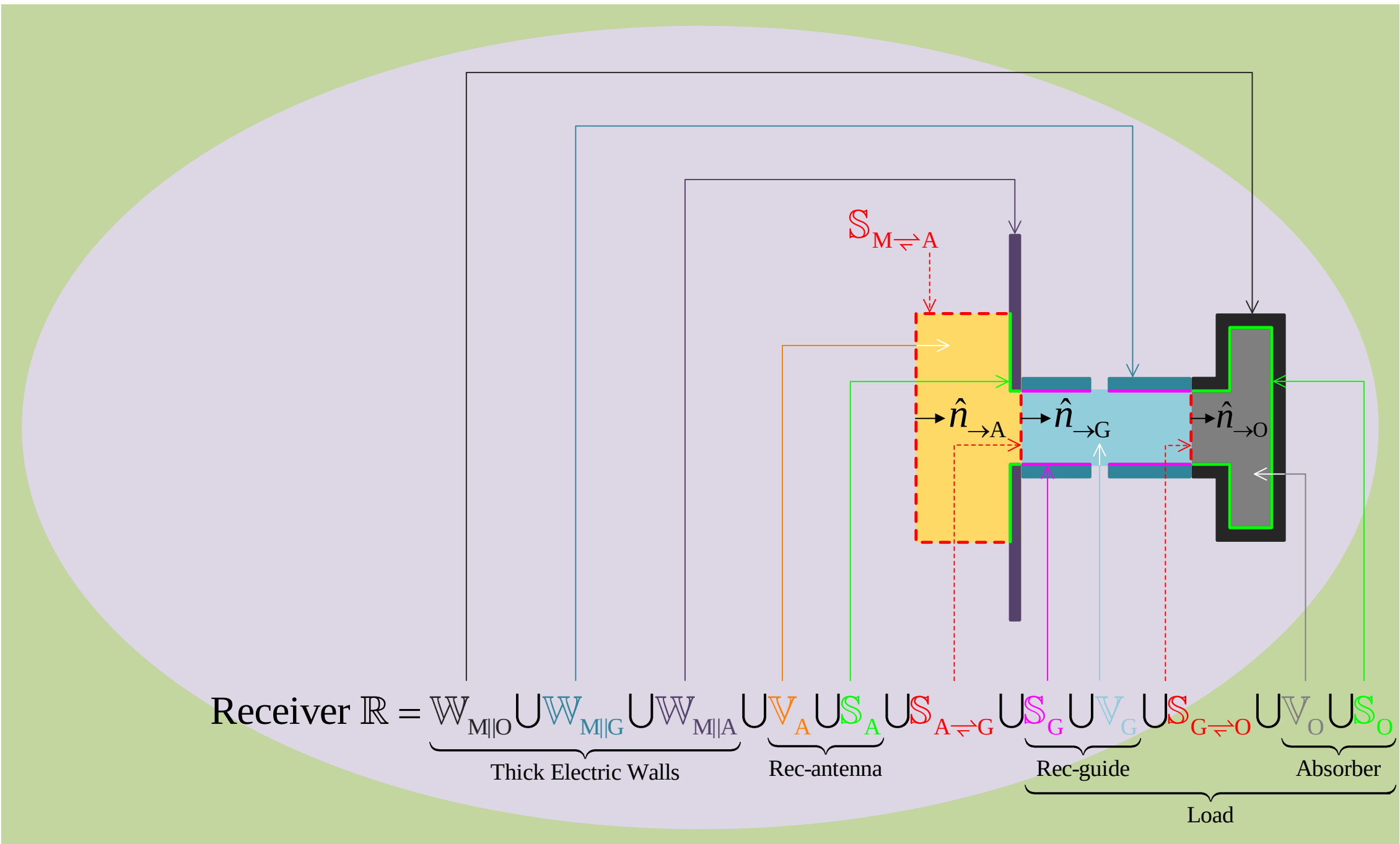

(b)

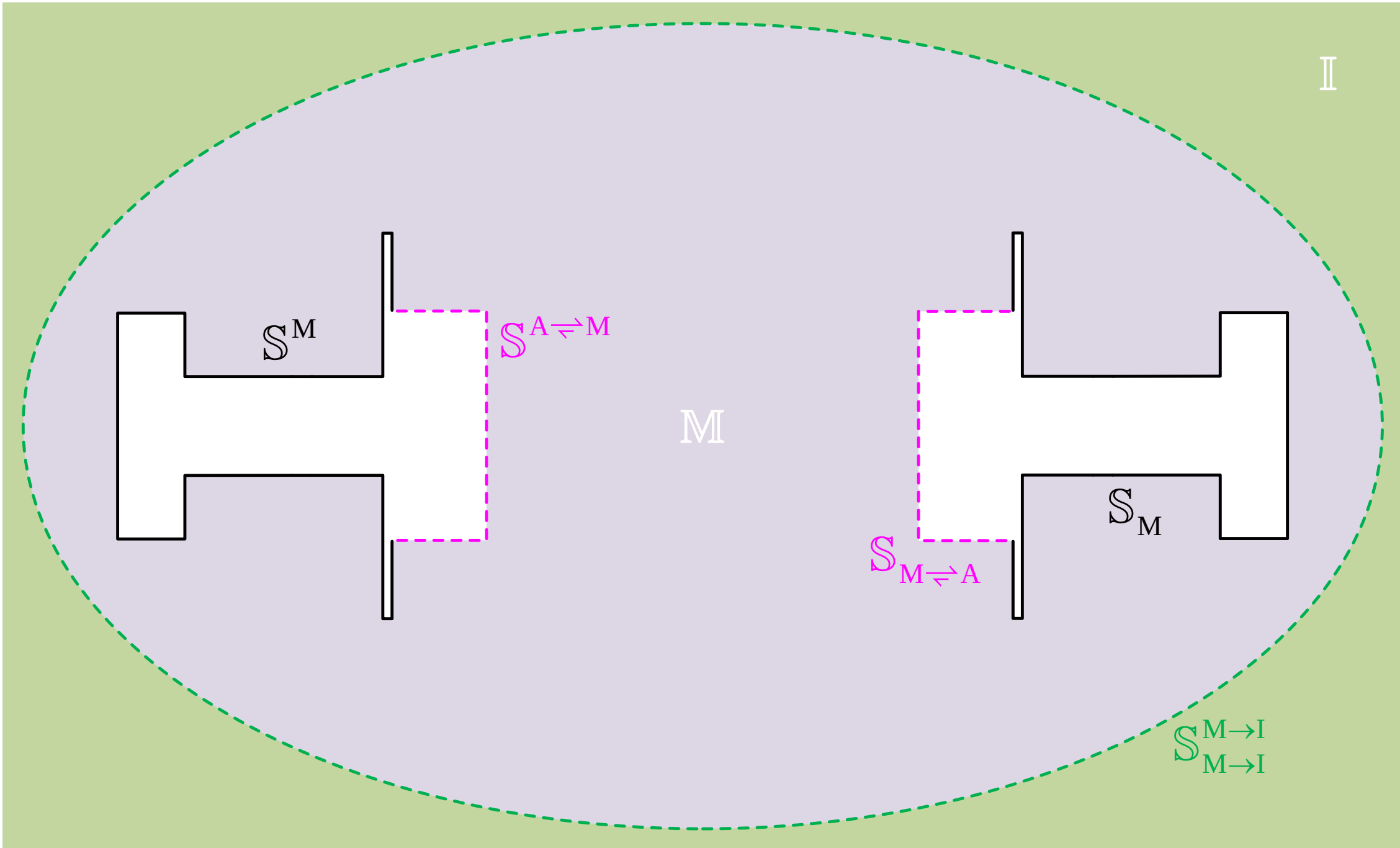

(c)

Figure 10-3 Detailed topological structure of three-dimensional Euclidean space. (a) Topology of the region occupied by transmitter; (b) topology of the region occupied by receiver; (c) topology of the region occupied by environment

The rigorous mathematical definitions for the various symbols in the above Eq. (10-1) and Fig. 10-3 had been given in Sec. 2.3 in detail, and they are simply summarized as follows:

$\mathbb{S}^{O}$ is the thin electric wall used to separate generator from medium;

$\mathbb{V}^{O}$ is the cavity occupied by generator;

$\mathbb{S}^{O \rightleftharpoons G}$ is the output port of generator, and also the input port of tra-guide;

$\mathbb{S}^{G}$ is the thin electric wall used to separate tra-guide from medium;

$\mathbb{V}^{G}$ is the cavity occupied by tra-guide;

$\mathbb{S}^{G \rightleftharpoons A}$ is the output port of tra-guide, and also the input port of tra-antenna;

$\mathbb{S}^{A}$ is the interface between tra-antenna and ground plane, and is a thin electric wall;

$\mathbb{V}^{A}$ is the region occupied by the material part of tra-antenna;

$\mathbb{W}^{T \| M}$ is the thick electric wall used to separate transmitter from medium;

$\mathbb{S}^{A \rightleftharpoons M}$ is the output port of tra-antenna, and also the input port of medium;

$\mathbb{S}^{M}$ is the thin electric wall used to separate medium from transmitter, i.e. the grounding structure related to transmitter, and it is simply denoted as tra-ground;

$\mathbb{M}$ is the region occupied by medium;

$\mathbb{S}_{\text{M}}$ is the thin electric wall used to separate medium from receiver, i.e. the grounding structure related to receiver, and it is simply denoted as rec-ground;

$\mathbb{S}_{\text{M}\to\text{I}}^{\text{M}\to\text{I}}$ is a spherical surface with infinite radius, and is used to collect the EM energies which are radiated by transmitter and/or scattered by receiver and finally reach infinity;

$\mathbb{S}_{\text{M}\rightleftharpoons\text{A}}$ is the input port of rec-antenna, and also a part of the output port of medium;

$\mathbb{W}_{\text{M}||\text{R}}$ is the thick electric wall used to separate receiver from medium;

$\mathbb{V}_{\text{A}}$ is the region occupied by the material part of rec-antenna;

$\mathbb{S}_{\text{A}}$ is the interface between rec-antenna and ground plane, and is a thin electric wall;

$\mathbb{S}_{\text{A}\rightleftharpoons\text{G}}$ is the output port of rec-antenna, and also the input port of rec-guide;

$\mathbb{V}_{\text{G}}$ is the cavity occupied by rec-guide;

$\mathbb{S}_{\text{G}}$ is the thin electric wall used to separate rec-guide from medium;

$\mathbb{S}_{\text{G}\rightleftharpoons\text{O}}$ is the output port of tra-guide, and also the input port of absorber;

$\mathbb{V}_{\text{O}}$ is the cavity occupied by absorber;

$\mathbb{S}_{\text{O}}$ is the thin electric wall used to separate absorber from medium.

Here, to use symbol "$\rightleftharpoons$" in the various ports is to emphasize that the energies passing through the ports are bi-directional. The reason to use symbol "$\rightarrow$" in $\mathbb{S}_{\text{M}\to\text{I}}^{\text{M}\to\text{I}}$ rather than "$\rightleftharpoons$" is that Sommerfeld's radiation condition on $\mathbb{S}_{\text{M}\to\text{I}}^{\text{M}\to\text{I}}$ guarantees the uni-directional propagation of the energy passing through $\mathbb{S}_{\text{M}\to\text{I}}^{\text{M}\to\text{I}}$.

**Aspect III. Power Transport Theorem (PTT) for Transceiving System**

For any two-port region $\mathbb{V}$ shown in Fig. 10-4, whose input port $\mathbb{S}_{\text{in}}$ and output port $\mathbb{S}_{\text{out}}$ are seperated by an impenetrable electric wall $\mathbb{S}_{\text{ele}}$,

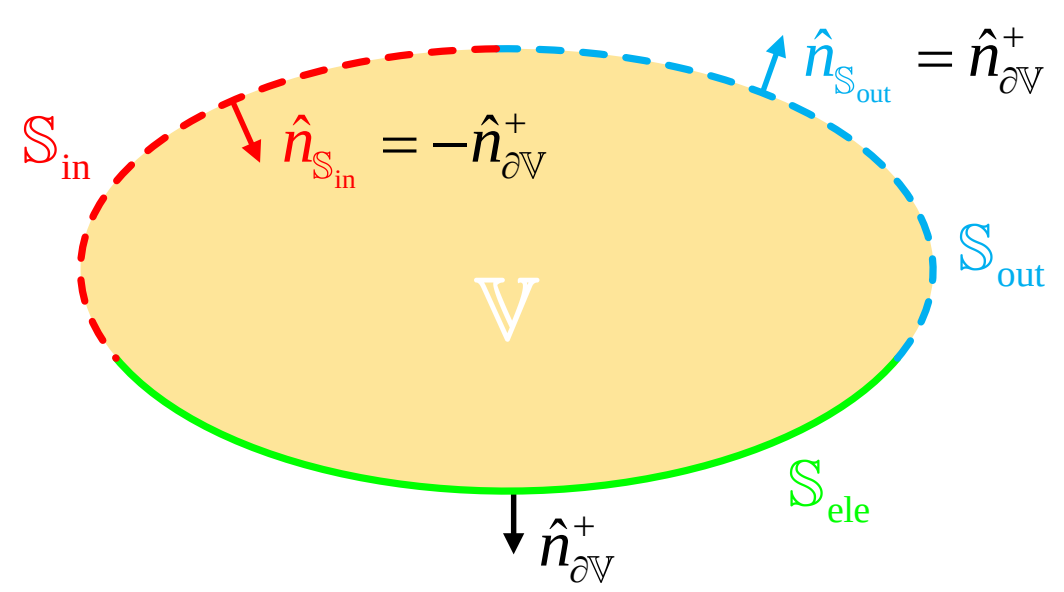

Figure 10-4 Topological structure of a typical two-port region

where $\hat{n}_{\mathbb{S}_{\text{in}}}$, $\hat{n}_{\mathbb{S}_{\text{out}}}$, and $\hat{n}_{\partial\mathbb{V}}^{+}$ are the normal directions of $\mathbb{S}_{\text{in}}$, $\mathbb{S}_{\text{out}}$, and $\partial\mathbb{V}$ respectively, the Maxwell's equations on $\mathbb{V}$ implies the following time-domain Poynting's theorem (PtT)

$$\int_{t_1}^{t_2}\left[\iint_{\mathbb{S}_{\text{in}}}\left(\vec{\mathcal{E}}\times\vec{\mathcal{H}}\right)\cdot\hat{n}_{\mathbb{S}_{\text{in}}}dS\right]dt$$

$$= \int_{t_1}^{t_2}\left[\iint_{\mathbb{S}_{\text{out}}}\left(\vec{\mathcal{E}}\times\vec{\mathcal{H}}\right)\cdot\hat{n}_{\mathbb{S}_{\text{out}}}dS\right]dt+\int_{t_1}^{t_2}\left\langle\vec{\sigma}\cdot\vec{\mathcal{E}},\vec{\mathcal{E}}\right\rangle_{\mathbb{V}}dt$$

$$+\left[\frac{1}{2}\left\langle\vec{\mathcal{H}},\vec{\mu}\cdot\vec{\mathcal{H}}\right\rangle_{\mathbb{V}}+\frac{1}{2}\left\langle\vec{\varepsilon}\cdot\vec{\mathcal{E}},\vec{\mathcal{E}}\right\rangle_{\mathbb{V}}\right]_{t=t_2}-\left[\frac{1}{2}\left\langle\vec{\mathcal{H}},\vec{\mu}\cdot\vec{\mathcal{H}}\right\rangle_{\mathbb{V}}+\frac{1}{2}\left\langle\vec{\varepsilon}\cdot\vec{\mathcal{E}},\vec{\mathcal{E}}\right\rangle_{\mathbb{V}}\right]_{t=t_1} \quad (10\text{-}2\text{a})$$

and the following frequency-domain PtT

$$\overbrace{\frac{1}{2}\iint_{\mathbb{S}_{\text{in}}}\left(\vec{E}\times\vec{H}^{\dagger}\right)\cdot\hat{n}_{\mathbb{S}_{\text{in}}}dS}^{P_{\text{in}}}$$

$$=\underbrace{\frac{1}{2}\iint_{\mathbb{S}_{\text{out}}}\left(\vec{E}\times\vec{H}^{\dagger}\right)\cdot\hat{n}_{\mathbb{S}_{\text{out}}}dS}_{P_{\text{out}}}+\underbrace{\frac{1}{2}\left\langle\vec{\sigma}\cdot\vec{E},\vec{E}\right\rangle_{\mathbb{V}}}_{P_{\text{dis}}}+j\underbrace{2\omega\left[\frac{1}{4}\left\langle\vec{H},\vec{\mu}\cdot\vec{H}\right\rangle_{\mathbb{V}}-\frac{1}{4}\left\langle\vec{\varepsilon}\cdot\vec{E},\vec{E}\right\rangle_{\mathbb{V}}\right]}_{P_{\text{sto}}} \quad (10\text{-}2\text{b})$$

where $P_{\text{in}}$ and $P_{\text{out}}$ are called input power and output power respectively. In fact, the above time-domain PtT can be alternatively expressed as follows:

$$\int_{t_1}^{t_2}\left[\iint_{\mathbb{S}_{\text{in}}}\left(\vec{\mathcal{E}}\times\vec{\mathcal{H}}\right)\cdot\hat{n}_{\mathbb{S}_{\text{in}}}dS\right]dt$$

$$=\int_{t_1}^{t_2}\left[\iint_{\mathbb{S}_{\text{out}}}\left(\vec{\mathcal{E}}\times\vec{\mathcal{H}}\right)\cdot\hat{n}_{\mathbb{S}_{\text{out}}}dS\right]dt+\overbrace{\int_{t_1}^{t_2}\left\langle\vec{\sigma}\cdot\vec{\mathcal{E}},\vec{\mathcal{E}}\right\rangle_{\mathbb{V}}dt}^{\int_{t_1}^{t_2}\left\langle\vec{\mathcal{J}}^{\text{CV}},\vec{\mathcal{E}}\right\rangle_{\mathbb{V}}dt}$$

$$+\left[\frac{1}{2}\left\langle\vec{\mathcal{H}},\mu_0\vec{\mathcal{H}}\right\rangle_{\mathbb{V}}+\frac{1}{2}\left\langle\varepsilon_0\vec{\mathcal{E}},\vec{\mathcal{E}}\right\rangle_{\mathbb{V}}\right]_{t=t_2}-\left[\frac{1}{2}\left\langle\vec{\mathcal{H}},\mu_0\vec{\mathcal{H}}\right\rangle_{\mathbb{V}}+\frac{1}{2}\left\langle\varepsilon_0\vec{\mathcal{E}},\vec{\mathcal{E}}\right\rangle_{\mathbb{V}}\right]_{t=t_1}$$

$$+\underbrace{\left[\frac{1}{2}\left\langle\vec{\mathcal{H}},\Delta\vec{\mu}\cdot\vec{\mathcal{H}}\right\rangle_{\mathbb{V}}+\frac{1}{2}\left\langle\Delta\vec{\varepsilon}\cdot\vec{\mathcal{E}},\vec{\mathcal{E}}\right\rangle_{\mathbb{V}}\right]_{t=t_2}-\left[\frac{1}{2}\left\langle\vec{\mathcal{H}},\Delta\vec{\mu}\cdot\vec{\mathcal{H}}\right\rangle_{\mathbb{V}}+\frac{1}{2}\left\langle\Delta\vec{\varepsilon}\cdot\vec{\mathcal{E}},\vec{\mathcal{E}}\right\rangle_{\mathbb{V}}\right]_{t=t_1}}_{\int_{t_1}^{t_2}\left[\left\langle\vec{\mathcal{H}},\vec{\mathcal{M}}^{\text{MV}}\right\rangle_{\mathbb{V}}+\left\langle\vec{\mathcal{J}}^{\text{PV}},\vec{\mathcal{E}}\right\rangle_{\mathbb{V}}\right]dt} \quad (10\text{-}3)$$

and this alternative expression has a very clear physical meaning: in whole time interval $t_1 \sim t_2$, the net energy flowing into input port $\mathbb{S}_{\text{in}}$ is transformed into four parts, in which

Part 1 flows out from output port $\mathbb{S}_{\text{out}}$, and

Part 2 is used to do the work on conduction volume electric current $\vec{\mathcal{J}}^{\text{CV}}$ (which part is finally converted into Joule heating energy), and

Part 3 is used to increase the electric field energy and magnetic field energy stored in region $\mathbb{V}$, and

Part 4 is used to do the work on polarization volume electric current $\vec{\mathcal{J}}^{\text{PV}}$ and magnetization volume magnetic current $\vec{\mathcal{M}}^{\text{MV}}$ (which part is finally converted into the polarization and magnetization energies stored in the matter occupying region $\mathbb{V}$).

Evidently, Eq. (10-3) is just the work-energy principle (WEP) used to quantitatively depict the work-energy transformation phenomenon occuring in any two-port region. Thus, just like the scatterer case discussed in Sec. 1.2.4.4, the PtT and WEP are also equivalent to each other in the two-port region case.

The above frequency-domain PtT (10-2b) for a single two-port structure can be easily generalized to various cascaded combined systems, such as the following

$$P^{\mathrm{G}\rightleftharpoons\mathrm{A}} = P_{\mathrm{dis}}^{\mathrm{A}} + \underbrace{P_{\mathrm{dis}}^{\mathrm{M}} + P^{\mathrm{M}\to\mathrm{I}} + j\,P_{\mathrm{sto}}^{\mathrm{M}}}_{P^{\mathrm{A}\rightleftharpoons\mathrm{M}}} + j\,P_{\mathrm{sto}}^{\mathrm{A}} \tag{10-4}$$

for tra-antenna-medium combined system (alternatively called augmented tra-antenna system in this report), the following

$$P^{\mathrm{O}\rightleftharpoons\mathrm{G}} = P_{\mathrm{dis}}^{\mathrm{G}} + \underbrace{P_{\mathrm{dis}}^{\mathrm{A}} + \underbrace{P_{\mathrm{dis}}^{\mathrm{M}} + P^{\mathrm{M}\to\mathrm{I}} + j\,P_{\mathrm{sto}}^{\mathrm{M}}}_{P^{\mathrm{A}\rightleftharpoons\mathrm{M}}} + j\,P_{\mathrm{sto}}^{\mathrm{A}}}_{P^{\mathrm{G}\rightleftharpoons\mathrm{A}}} + j\,P_{\mathrm{sto}}^{\mathrm{G}} \tag{10-5}$$

for tra-guide-tra-antenna-medium combined system (simply called augmented TGTA system in this report), and the following

$$P^{\mathrm{G}\rightleftharpoons\mathrm{A}} = P_{\mathrm{dis}}^{\mathrm{A}} + \underbrace{P_{\mathrm{Mdis}}^{\mathrm{Mdis}} + P_{\mathrm{M}\to\mathrm{I}}^{\mathrm{M}\to\mathrm{I}} + \underbrace{P_{\mathrm{A}}^{\mathrm{dis}} + P_{\mathrm{A}\rightleftharpoons\mathrm{G}} + j\,P_{\mathrm{A}}^{\mathrm{sto}}}_{P_{\mathrm{M}\rightleftharpoons\mathrm{A}}} + j\,P_{\mathrm{Msto}}^{\mathrm{Msto}}}_{P^{\mathrm{A}\rightleftharpoons\mathrm{M}}} + j\,P_{\mathrm{sto}}^{\mathrm{A}} \tag{10-6}$$

for tra-antenna-medium-rec-antenna combined system (simply called augmented TARA system in this report), etc.

In fact, this report has further derived the following more general power flow equation

$$P_{\mathrm{imp}}^{\mathrm{O}} = \left(P_{\mathrm{dis}}^{\mathrm{O}} + jP_{\mathrm{sto}}^{\mathrm{O}}\right) + \underbrace{\left(P_{\mathrm{dis}}^{\mathrm{G}} + jP_{\mathrm{sto}}^{\mathrm{G}}\right) + \underbrace{\left(P_{\mathrm{dis}}^{\mathrm{A}} + jP_{\mathrm{sto}}^{\mathrm{A}}\right) + \underbrace{\left(P_{\mathrm{Mdis}}^{\mathrm{Mdis}} + jP_{\mathrm{Msto}}^{\mathrm{Msto}}\right) + \overbrace{P_{\mathrm{M}\to\mathrm{I}}^{\mathrm{M}\to\mathrm{I}}}^{P_{\mathrm{sca}}^{\mathrm{rad}}} + \underbrace{\left(P_{\mathrm{A}}^{\mathrm{dis}} + jP_{\mathrm{A}}^{\mathrm{sto}}\right) + \underbrace{\left(P_{\mathrm{G}}^{\mathrm{dis}} + jP_{\mathrm{G}}^{\mathrm{sto}}\right) + \underbrace{\left(P_{\mathrm{O}}^{\mathrm{dis}} + jP_{\mathrm{O}}^{\mathrm{sto}}\right) + P_{\mathrm{O}}^{\mathrm{abs}}}_{P_{\mathrm{G}\rightleftharpoons\mathrm{O}}}}_{P_{\mathrm{A}\rightleftharpoons\mathrm{G}}}}_{P_{\mathrm{M}\rightleftharpoons\mathrm{A}}}}_{P^{\mathrm{A}\rightleftharpoons\mathrm{M}}}}_{P^{\mathrm{G}\rightleftharpoons\mathrm{A}}}}_{P^{\mathrm{O}\rightleftharpoons\mathrm{G}}} \tag{10-7}$$

for whole transceiving system, where

$P_{\mathrm{imp}}^{\mathrm{O}}$ is the power generated by the impressed source in generator, and it is transformed into three parts — a part $P_{\mathrm{dis}}^{\mathrm{O}}$ dissipated in generator, a part $P_{\mathrm{sto}}^{\mathrm{O}}$ contributing to the energy stored in generator, and a part $P^{\mathrm{O}\rightleftharpoons\mathrm{G}}$ transported from generator to tra-guide by passing through port $\mathbb{S}^{\mathrm{O}\rightleftharpoons\mathrm{G}}$;

$P^{\mathrm{O}\rightleftharpoons\mathrm{G}}$ is the net power inputted into tra-guide, and it is transformed into three parts — a part $P_{\mathrm{dis}}^{\mathrm{G}}$ dissipated in tra-guide, a part $P_{\mathrm{sto}}^{\mathrm{G}}$ contributing to the energy stored in tra-guide, and a part $P^{\mathrm{G}\rightleftharpoons\mathrm{A}}$ transported from tra-guide to tra-antenna by passing through port $\mathbb{S}^{\mathrm{G}\rightleftharpoons\mathrm{A}}$;

$P^{\mathrm{G}\rightleftharpoons\mathrm{A}}$ is the net power inputted into tra-antenna, and it is transformed into three parts — a part $P_{\mathrm{dis}}^{\mathrm{A}}$ dissipated in tra-antenna, a part $P_{\mathrm{sto}}^{\mathrm{A}}$ contributing to the energy stored in tra-antenna, and a part $P^{\mathrm{A}\rightleftharpoons\mathrm{M}}$ transported from tra-antenna to medium by passing through port $\mathbb{S}^{\mathrm{A}\rightleftharpoons\mathrm{M}}$;

$P^{\mathrm{A}\rightleftharpoons\mathrm{M}}$ is the net power inputted into medium, and it is transformed into four parts — a part $P_{\mathrm{Mdis}}^{\mathrm{Mdis}}$ dissipated in medium, a part $P_{\mathrm{Msto}}^{\mathrm{Msto}}$ contributing to the energy stored in medium, a part $P_{\mathrm{M}\to\mathrm{I}}^{\mathrm{M}\to\mathrm{I}}$ passing through port $\mathbb{S}_{\mathrm{M}\to\mathrm{I}}^{\mathrm{M}\to\mathrm{I}}$, and a part $P_{\mathrm{M}\rightleftharpoons\mathrm{A}}$ transported from medium to rec-antenna by passing through port $\mathbb{S}_{\mathrm{M}\rightleftharpoons\mathrm{A}}$;

$P_{\mathrm{M}\to\mathrm{I}}^{\mathrm{M}\to\mathrm{I}}$ originates from four parts — a part radiated by transmitter, a part scattered by receiver, a part scattered by medium, and a part scattered by transmitter (because there also exists the reaction from receiver&medium to transmitter, i.e., the EM fields generated by receiver&medium will also be scattered by transmitter);

$P_{\mathrm{M}\rightleftharpoons\mathrm{A}}$ is the net power inputted into rec-antenna, and it is transformed into three parts — a part $P_{\mathrm{A}}^{\mathrm{dis}}$ dissipated in rec-antenna, a part $P_{\mathrm{A}}^{\mathrm{sto}}$ contributing to the energy stored in rec-antenna, and a part $P_{\mathrm{A}\rightleftharpoons\mathrm{G}}$ transported from rec-antenna to rec-guide by passing through port $\mathbb{S}_{\mathrm{A}\rightleftharpoons\mathrm{G}}$;

$P_{\mathrm{A}\rightleftharpoons\mathrm{G}}$ is the net power inputted into rec-guide, and it is transformed into three parts — a part $P_{\mathrm{G}}^{\mathrm{dis}}$ dissipated in rec-guide, a part $P_{\mathrm{G}}^{\mathrm{sto}}$ contributing to the energy stored in rec-guide, and a part $P_{\mathrm{G}\rightleftharpoons\mathrm{O}}$ transported from rec-guide to absorber by passing through port $\mathbb{S}_{\mathrm{G}\rightleftharpoons\mathrm{O}}$;

$P_{\mathrm{G}\rightleftharpoons\mathrm{O}}$ is the net power inputted into absorber, and it is transformed into three parts — a part $P_{\mathrm{O}}^{\mathrm{dis}}$ dissipated in absorber, a part $P_{\mathrm{O}}^{\mathrm{sto}}$ contributing to the energy stored in absorber, and a part $P_{\mathrm{O}}^{\mathrm{abs}}$ converted into some kinds of energies by absorber.

Here, the superscripts and subscripts used in the various input and output powers are the same as the ones used in the corresponding input and output ports.

Clearly, the above Eqs. (10-4)~(10-6) and (10-7) rigorously govern the transporting process of the power flowing in the various combined systems and whole {transceiving system, surrounding environment}. The power transporting process is visually illustrated in the following power flow graph.

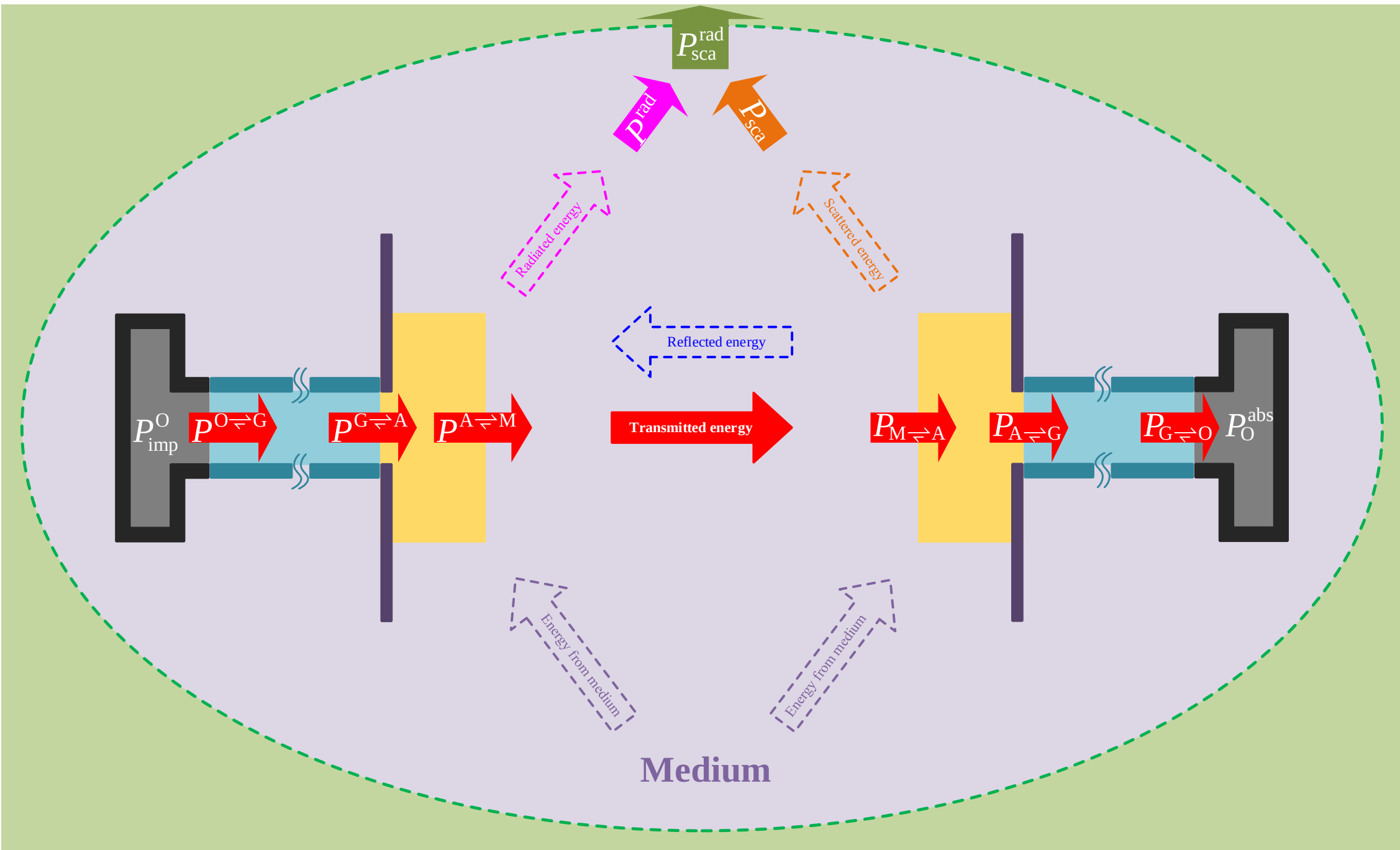

Figure 10-5 Power flow graph for the transceiving system shown in Fig. 10-1

Thus in this report the above-mentioned power equations are collectively referred to as power transport theorem (PTT) for transceiving systems. Here, the acronyms of Poynting's theorem and power transport theorem are selected as PtT and PTT respectively is to emphasize that the latter can be viewed as a generalization for the former.

**Aspect IV. Power Transport Theorem Based Decoupling Mode Theory (PTT-DMT) for Transceiving Systems**

In the PTT for any objective transceiving system, the source term, which is called input power by this report, is the one used to sustain a steady power transport in the system. Under the PTT framework, this report establishes a novel modal theory — decoupling mode theory (DMT) — for transceiving systems. The PTT-based DMT for transceiving systems (PTT-TRSys-DMT) is different from the classical Sturm-Liouville theory based eigen-mode theory for wave-guiding systems (SLT-WGSys-EMT) and the conventional work-energy principle based characteristic mode theory for scattering systems (WEP-ScaSys-CMT).

The PTT-TRSys-DMT is devoted to constructing a set of decoupled modes by orthogonalizing input power operator (IPO), and the modes satisfy the following frequency-domain input-power-decoupling relation

$$(1/2) \iint_{\text{Input Port}} \left(\vec{E}_m \times \vec{H}_n^\dagger\right) \cdot d\vec{S} = \left(1 + j\,\theta_m\right)\delta_{mn} \tag{10-8}$$

so the modes are correspondingly called input-power-decoupled modes (IP-DMs). Obviously, the modes also satisfy the following time-domain energy-decoupling relation

$$(1/T)\int_{t_0}^{t_0+T}\left[\iint_{\text{Input Port}}\left(\vec{\mathcal{E}}_m\times\vec{\mathcal{H}}_n\right)\cdot d\vec{S}\right]dt \ = \ \delta_{mn} \tag{10-9}$$

and thus the modes are energy-decoupled in any integral period.

The energy-decoupling feature leads to the following Parseval's identity immediately.

$$(1/T)\int_{t_0}^{t_0+T}\left[\iint_{\text{Input Port}}\left(\vec{\mathcal{E}}\times\vec{\mathcal{H}}\right)\cdot d\vec{S}\right]dt \ = \ \sum_m |c_m|^2 \tag{10-10}$$

for any working mode with time-domain EM fields $\{\vec{\mathcal{E}},\vec{\mathcal{H}}\}$, where $c_m$ is the modal expansion coefficient used to linearly expand the working mode in terms of the superposition of IP-DMs, and

$$c_m \ = \ \frac{(1/2)\iint_{\text{Input Port}}\left(\vec{E}\times\vec{H}_m^{\dagger}\right)\cdot d\vec{S}}{1+j\,\theta_m} \ = \ \frac{(1/2)\iint_{\text{Input Port}}\left(\vec{E}_m\times\vec{H}^{\dagger}\right)\cdot d\vec{S}}{1+j\,\theta_m} \tag{10-11}$$

where fields $\{\vec{E},\vec{H}\}$ are previously known.

Now, the fundamental process for constructing the IP-DMs of an objective transceiving system is summarized in the following Fig. 10-6.

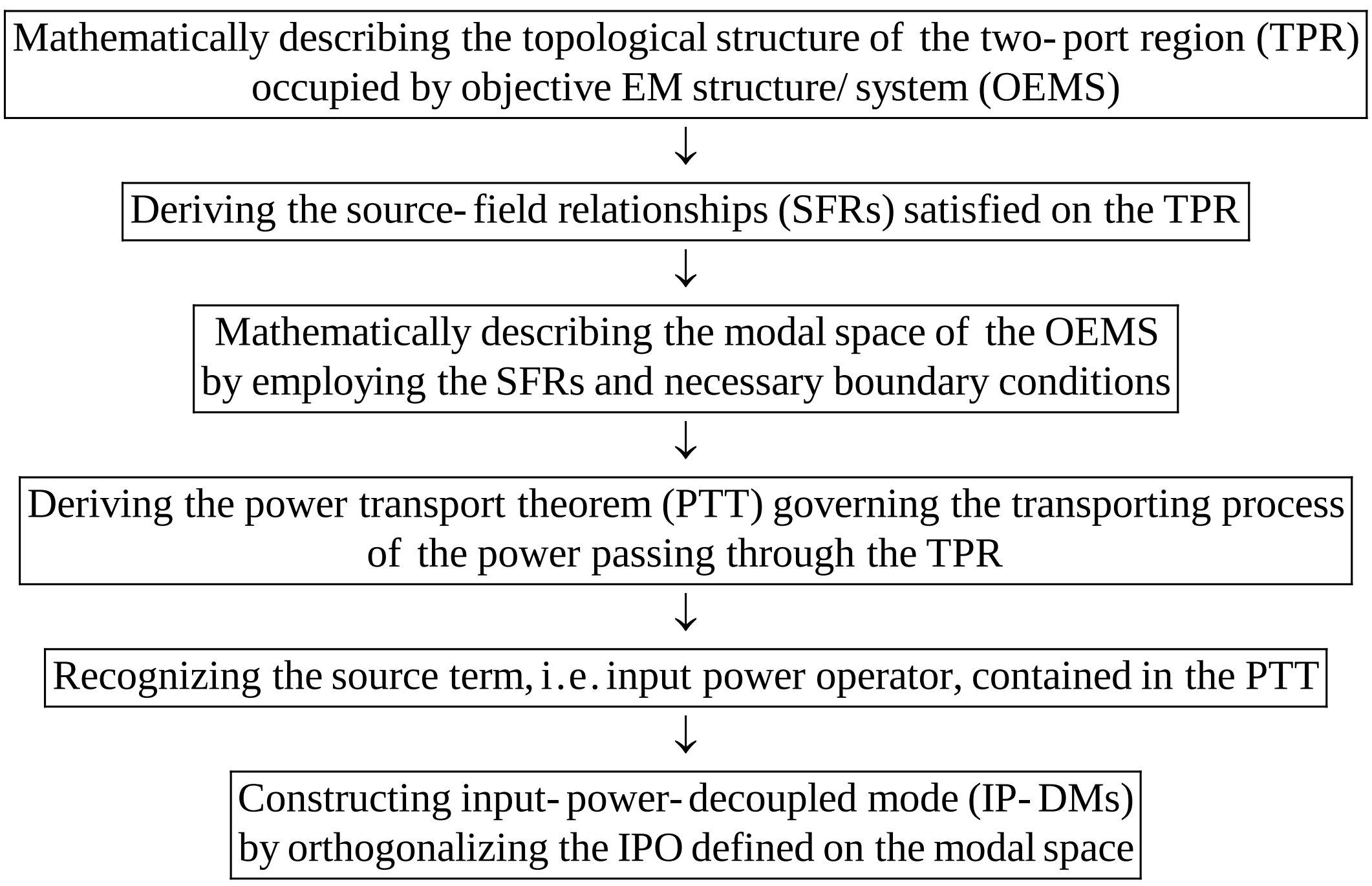

Figure 10-6 General process for constructing the IP-DMs of any objective system

**Aspect V. PTT-DMT for a Single Two-port Region/Structure and Modal Matching for Some Cascaded Two-port Regions/Structures**

Focusing on some typical two-port structures (such as tra-guide/rec-guide, tra-antenna, medium, and rec-antenna), this report constructs their IP-DMs separately. During the process for constructing IP-DMs, the interaction between the objective structure and the other structures is ignored temporarily. After obtaining the IP-DMs of the various structures, the interactions among the IP-DMs of different structures must be considered.

The classical modal matching technique is employed in this report to consider the interactions among the IP-DMs of different structures. The general steps to do modal matching are shown in the following Fig. 10-7.

Expanding the electromagnteic fields in two cascaded structures
in terms of the input-power-decoupled modes (IP-DMs) of the structures

↓

Establishing the integral equations (IEs) of the expansion coefficients
by matching the boundary conditions on the interface between the structures

↓

Discretizing the IEs into some matrix equations
by using method of moments and employing the IP-DMs as testing functions

↓

Deriving the local reflection and transmission matrices
by solving the above matrix equations

↓

Establishing the matrix equations of global reflection and transmission matrices
by employing energy conservation law and local reflection and transmission matrices

↓

Deriving the iteration relations for global reflection and transmission matrices
by solving the above matrix equations

Figure 10-7 General steps to do modal matching

**Aspect VI. PTT-DMT for the Combined Systems Constituted by Some Cascaded Two-port Regions/Structures**

In fact, there are two different ways to consider the interactions among the various structures, as shown in the following Fig. 10-8.

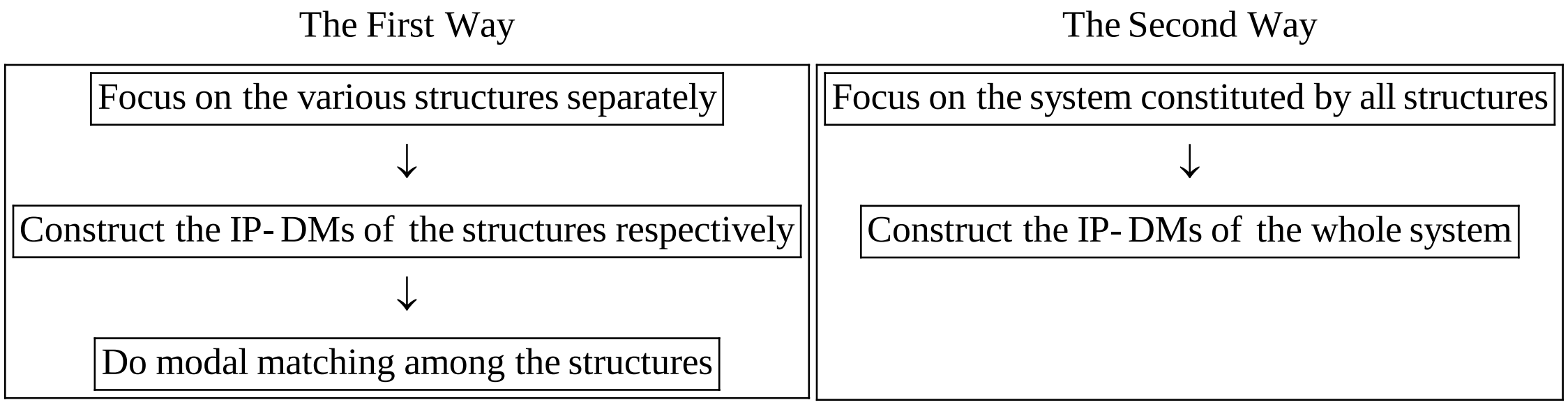

Figure 10-8 Two differential schemes to consider the interactions among the structures

The first way, which is relatively tedious as exhibited in Fig.10-7, is to separately construct the IP-DMs of various structures first, and then to do the modal matching among the different structures. The second way, which this report prefers, is to integrate all related structures into a combined system first, and then to construct the IP-DMs of the system directly.

Specifically, the second way treats tra-antenna and medium as a combined system — augmented tra-antenna, and constructs the IP-DMs of the system at once; the second way treats medium and rec-antenna as a combined system — augmented rec-antenna, and constructs the IP-DMs of the system at once; the second way treats tra-guide and augmented tra-antenna as a combined system — augmented tra-guide-tra-antenna combined system (simply called TGTA system in this report), and constructs the IP-DMs of the system at once; the second way treats augmented tra-antenna and augmented rec-antenna as a combined system — augmented tra-antenna-rec-antenna combined system (simply called TARA system in this report), and constructs the IP-DMs of the system at once. The treatments for other combined systems, such as the system constituted by {medium, rec-antenna, rec-guide} and the system constituted by {tra-guide, tra-antenna, medium, rec-antenna, rec-guide}, are similar.

**Aspect VII. Modal Decomposition for any Working Mode of a Transceiving System**

For a certain objective transceiving system, its any working mode $\bar{a}$ can be linearly expanded in terms of the superposition of its IP-DMs as follows:

$$\bar{a} = \sum c_m \bar{\alpha}_m \tag{10-12}$$

with the expansion coefficients given in Eq. (10-11).

In addition, for a certain objective transceiving system, its all IP-DMs $\{\bar{\alpha}_m\}$ are classified into three categories — purely inductive IP-DMs $\{\bar{\alpha}_\zeta^{\text{ind}}\}$ (whose imaginary powers are positive), purely resonant IP-DMs $\{\overline{\alpha}_{\varsigma}^{\text{res}}\}$ (whose imaginary powers are zero), and purely capacitive IP-DMs $\{\overline{\alpha}_{\xi}^{\text{cap}}\}$ (whose imaginary powers are negative) — according to their imaginary powers.

Based on above these, any working mode of a certain objective transceiving system is decomposed into the superposition of three energy-decoupled components as follows:

$$\overline{a} = \underbrace{\sum c_{\zeta}^{\text{ind}} \overline{\alpha}_{\zeta}^{\text{ind}}}_{\overline{\beta}^{\text{ind}}} + \underbrace{\sum c_{\varsigma}^{\text{res}} \overline{\alpha}_{\varsigma}^{\text{res}}}_{\overline{\beta}^{\text{res}}} + \underbrace{\sum c_{\xi}^{\text{cap}} \overline{\alpha}_{\xi}^{\text{cap}}}_{\overline{\beta}^{\text{cap}}} \tag{10-13}$$

where $\overline{\beta}^{\text{ind}}$ is called purely inductive component (which is constituted by all purely inductive IP-DMs), and $\overline{\beta}^{\text{res}}$ is called purely resonant component (which is constituted by all purely resonant IP-DMs), and $\overline{\beta}^{\text{cap}}$ is called purely capacitive component (which is constituted by all purely capacitive IP-DMs).

**Aspect VIII. Electric-Magnetic Decoupling Factor for any Working Mode of a Transceiving System**

By Employing the above modal decomposition, this report introduces a novel concept of electric-magnetic decoupling factor (simply denoted as Θ-factor) for any working mode as follows:

$$\Theta\left(\overline{a}\right) = \frac{\operatorname{Im}\left\{\left(\overline{\beta}^{\text{ind}} \pm \overline{\beta}^{\text{cap}}\right)^{\dagger} \cdot \overline{\overline{P}}_{\text{in}} \cdot \left(\overline{\beta}^{\text{ind}} \mp \overline{\beta}^{\text{cap}}\right)\right\}}{\operatorname{Re}\left\{\overline{a} \cdot \overline{\overline{P}}_{\text{in}} \cdot \overline{a}\right\}} \tag{10-14}$$

where $\overline{\overline{P}}_{\text{in}}$ is the quadratic matrix of the input power. This report finds out that the Θ-factor is a quantitative depiction for the coupling degree between the electric energy and magnetic energy carried by the mode, and the relation between the Θ-factor and electric-magnetic coupling degree is monotonically decreasing, i.e., the smaller the Θ-factor is the stronger the electric-magnetic coupling is.

In particular, the Θ-factors of the $\overline{\beta}^{\text{ind}} + \overline{\beta}^{\text{res}}$, $\overline{\beta}^{\text{res}}$, and $\overline{\beta}^{\text{res}} + \overline{\beta}^{\text{cap}}$ components are as follows:

$$\Theta\left(\overline{\beta}^{\text{ind}} + \overline{\beta}^{\text{res}}\right) = +\frac{\operatorname{Im}\{P_{\text{in}}\}}{\operatorname{Re}\{P_{\text{in}}\}} \tag{10-15a}$$

$$\Theta\left(\overline{\beta}^{\text{res}}\right) = 0 \tag{10-15b}$$

$$\Theta\left(\overline{\beta}^{\text{res}} + \overline{\beta}^{\text{cap}}\right) = -\frac{\operatorname{Im}\{P_{\text{in}}\}}{\operatorname{Re}\{P_{\text{in}}\}} \tag{10-15c}$$

More particularly, the Θ-factors of the purely inductive, resonant, and capacitive IP-DMs can be uniformally expressed as follows:

$$\Theta\left(\bar{\alpha}_m\right) = \left|\theta_m\right| \tag{10-16}$$

In fact, the above relations clearly reveal the physical meaning of the $\theta_m$ calculated from the following modal decoupling equation

Input-power-decoupled mode

Θ-factor

$$\frac{1}{2j}\left(\bar{\bar{P}}_{\rm in}-\bar{\bar{P}}_{\rm in}^{\dagger}\right)\cdot\bar{\alpha}_m = \theta_m\,\frac{1}{2}\left(\bar{\bar{P}}_{\rm in}+\bar{\bar{P}}_{\rm in}^{\dagger}\right)\cdot\bar{\alpha}_m \tag{10-17}$$

Hermitian parts of input power operator

i.e., $|\theta_m|$ is just the Θ-factor of the IP-DM itself.

In addition, the $|\theta_m|$ and modal significance (MS) exists a one-to-one correspondence, because $\mathrm{MS}_m = 1/|1+j\theta_m|$. Thus the above physical interpretation for decoupling factor $\theta_\xi$ also provides a third physical interpretation to MS, i.e. the quantitative depiction for the electric-magnetic coupling degree of a IP-DM. Now, we summarize all three known physical explanations of MS as follows:

Physical Explanation 1. MS is a quantitative depiction for the weight of a certain IP-DM in whole IP-DM-based modal expansion formulation;

Physical Explanation 2. MS is a quantitative depiction for the weight of the modal active power contained in the whole modal power of the IP-DM;

Physical Explanation 3. MS is a quantitative depiction for the coupling degree of the electric energy and magnetic energy carried by the IP-DM.

"The three somewhat different physical meanings are collected in a single body" reflects the fact that: the MS is a very important physical quantity in DMT.

**SUMMARY.** At the last of this report, we use a single sentence to summarize the physical purpose of decoupling mode theory — to construct a set of energy-decoupled modes for any objective transceiving system by orthogonalizing the frequency-domain input power operator in power transport theorem framework.

The decoupling mode theory (DMT) has both similarity and differences with conventional eigen-mode theory (EMT) and characteristic mode theory (CMT) as below.

Similarity. DMT, EMT, and CMT are the modal theories focusing on constructing energy-decoupled modes.

Difference 1. DMT, EMT, and CMT focus on different objective EM systems. Specifically, DMT focuses on transceiving systems; EMT focuses on wave-guiding systems; CMT focuses on scattering systems.

Difference 2. DMT, EMT, and CMT are supported by different frameworks. Specifically, DMT is supported by power transport theorem framework; EMT is supported by Sturm-Liouville theory framework; CMT is supported by work-energy principle framework.

Difference 3. DMT, EMT, and CMT utilize different modal generating operators. Specifically, DMT utilizes input power operator; EMT utilizes Sturm-Liouville operator / Helmholtz's operator; CMT utilizes driving power operator.

# Appendices

**MOTIVATION:** In this part, some valuable conclusions and detailed formulations related to this report are provided for the convenience of readers's reference.

## Appendix A Source-Field Relationships

The electromagnetic (EM) fields distributing in a pre-selected region can be expressed in terms of some currents, and the expression way is usually not single. In this App. A, four different expression ways are provided. Specifically, the fields are expressed in terms of the related *induced currents* in App. A1; the fields are expressed in terms of the related *Huygens' second currents* in App. A2; the fields are expressed in terms of some proper *equivalent currents* in App. A3; the fields are expressed in terms of the induced and equivalent currents in App. A4.

## A1 Induction Principle

Under the excitation of the EM fields acting on an EM object (such as a *material body*, *metallic body*, *metallic surface*, or *metallic wire*), some currents will be induced on the object, and the induced currents will generate some EM fields. The fields are expressed as the functions of the induced currents in this App. A1.

### A1.1 Material Body Case

Now, suppose there is a material body $\mathbb{V}$ under the action of EM fields $\{\vec{E},\vec{H}\}$, where $\{\vec{E},\vec{H}\}$ are the **total** fields. The action will induce a *volume conduction electric current* $\vec{J}_{\mathbb{V}}^{con}$, a *volume polarization electric current* $\vec{J}_{\mathbb{V}}^{pol}$, and a *volume magnetization magnetic current* $\vec{M}_{\mathbb{V}}^{mag}$ on the body $\mathbb{V}$, as shown in the following Fig. A-1.

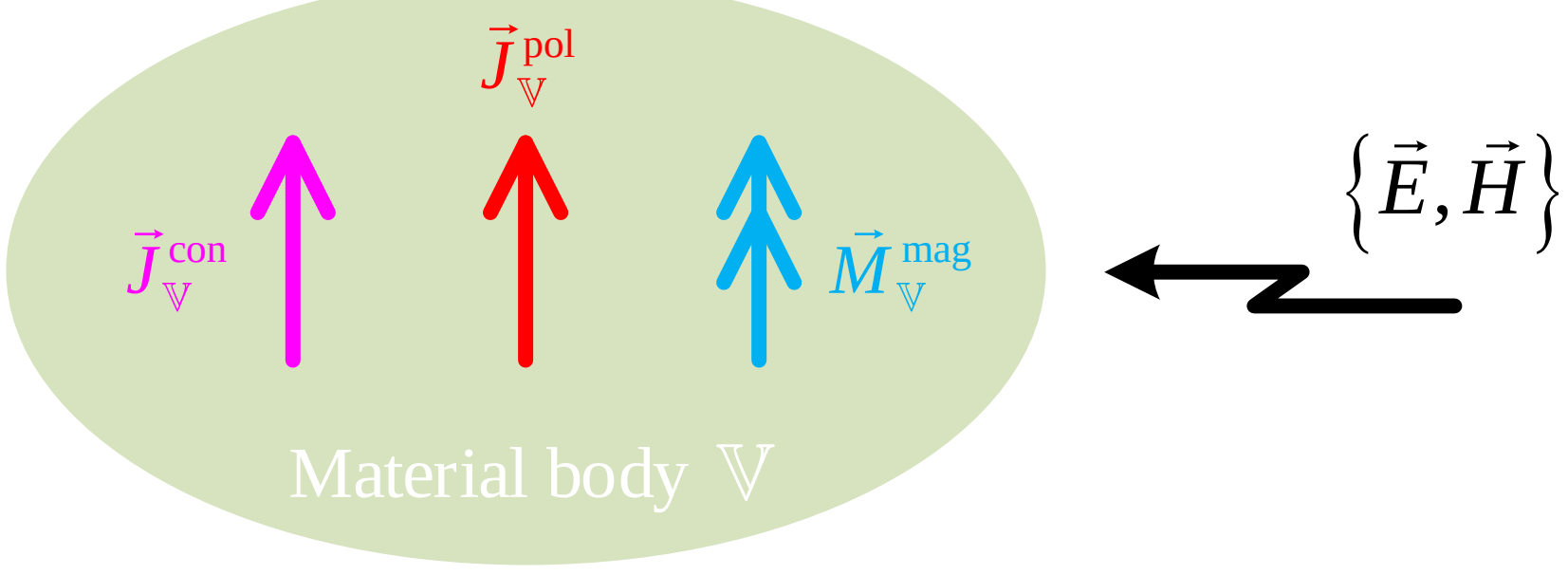

Figure A-1 A material body $\mathbb{V}$ under the action of EM fields $\{\vec{E},\vec{H}\}$

If the *electric conductivity*, *dielectric permitivity*, and *magnetic permeability* of the body $\mathbb{V}$ are $\vec{\vec{\sigma}}_{\mathbb{V}}$, $\vec{\vec{\varepsilon}}_{\mathbb{V}}$, and $\vec{\vec{\mu}}_{\mathbb{V}}$ respectively, then the currents $\{\vec{J}_{\mathbb{V}}^{\rm con},\vec{J}_{\mathbb{V}}^{\rm pol},\vec{M}_{\mathbb{V}}^{\rm mag}\}$ and the fields $\{\vec{E},\vec{H}\}$ satisfy the following relations

$$\vec{J}_{\mathbb{V}}^{\rm con}(\vec{r}) = \vec{\vec{\sigma}}_{\mathbb{V}}\cdot\vec{E}(\vec{r}) \quad , \quad \vec{r}\in\mathbb{V} \tag{A-1a}$$

$$\vec{J}_{\mathbb{V}}^{\rm pol}(\vec{r}) = j\omega\Delta\vec{\vec{\varepsilon}}_{\mathbb{V}}\cdot\vec{E}(\vec{r}) \quad , \quad \vec{r}\in\mathbb{V} \tag{A-1b}$$

$$\vec{M}_{\mathbb{V}}^{\rm mag}(\vec{r}) = j\omega\Delta\vec{\vec{\mu}}_{\mathbb{V}}\cdot\vec{H}(\vec{r}) \quad , \quad \vec{r}\in\mathbb{V} \tag{A-1c}$$

in which $\Delta\vec{\vec{\varepsilon}}_{\mathbb{V}}=\vec{\vec{\varepsilon}}_{\mathbb{V}}-\vec{\vec{I}}\varepsilon_0$ and $\Delta\vec{\vec{\mu}}_{\mathbb{V}}=\vec{\vec{\mu}}_{\mathbb{V}}-\vec{\vec{I}}\mu_0$ (where $\vec{\vec{I}}=\hat{x}\hat{x}+\hat{y}\hat{y}+\hat{z}\hat{z}$ is *unit dyad*).

The currents $\{\vec{J}_{\mathbb{V}}^{\rm con},\vec{J}_{\mathbb{V}}^{\rm pol},\vec{M}_{\mathbb{V}}^{\rm mag}\}$ will generate a field $\vec{F}_{\mathbb{V}}$ on whole *three-dimensional Euclidean sace* $\mathbb{E}^3$, and $\vec{F}_{\mathbb{V}}$ can be expressed in terms of the currents as

$$\begin{aligned}\vec{F}_{\mathbb{V}}(\vec{r}) &= \iiint_{\mathbb{V}}\vec{\vec{G}}_0^{JF}(\vec{r},\vec{r}\,')\cdot\left[\vec{J}_{\mathbb{V}}^{\rm con}(\vec{r}\,')+\vec{J}_{\mathbb{V}}^{\rm pol}(\vec{r}\,')\right]dV'+\iiint_{\mathbb{V}}\vec{\vec{G}}_0^{MF}(\vec{r},\vec{r}\,')\cdot\vec{M}_{\mathbb{V}}^{\rm mag}(\vec{r}\,')dV' \\ &=\left[\vec{\vec{G}}_0^{JF}*\left(\vec{J}_{\mathbb{V}}^{\rm con}+\vec{J}_{\mathbb{V}}^{\rm pol}\right)+\vec{\vec{G}}_0^{MF}*\vec{M}_{\mathbb{V}}^{\rm mag}\right]_{\mathbb{V}} \quad , \quad \vec{r}\in\mathbb{E}^3\end{aligned} \tag{A-2}$$

in which $\{\vec{\vec{G}}_0^{JF},\vec{\vec{G}}_0^{MF}\}$ are the *free-space dyadic Green's functions*, and symbol "$*$" is the *convolution integral operation*.

## A1.2 Metallic Body Case

Now, suppose there is a metallic body $\mathbb{V}$ under the action of EM fields $\{\vec{E},\vec{H}\}$, where $\{\vec{E},\vec{H}\}$ are the **total** fields. The action will induce a surface electric current $\vec{J}_{\partial\mathbb{V}}$ on the boundary $\partial\mathbb{V}$ of $\mathbb{V}$, as shown in the following Fig. A-2.

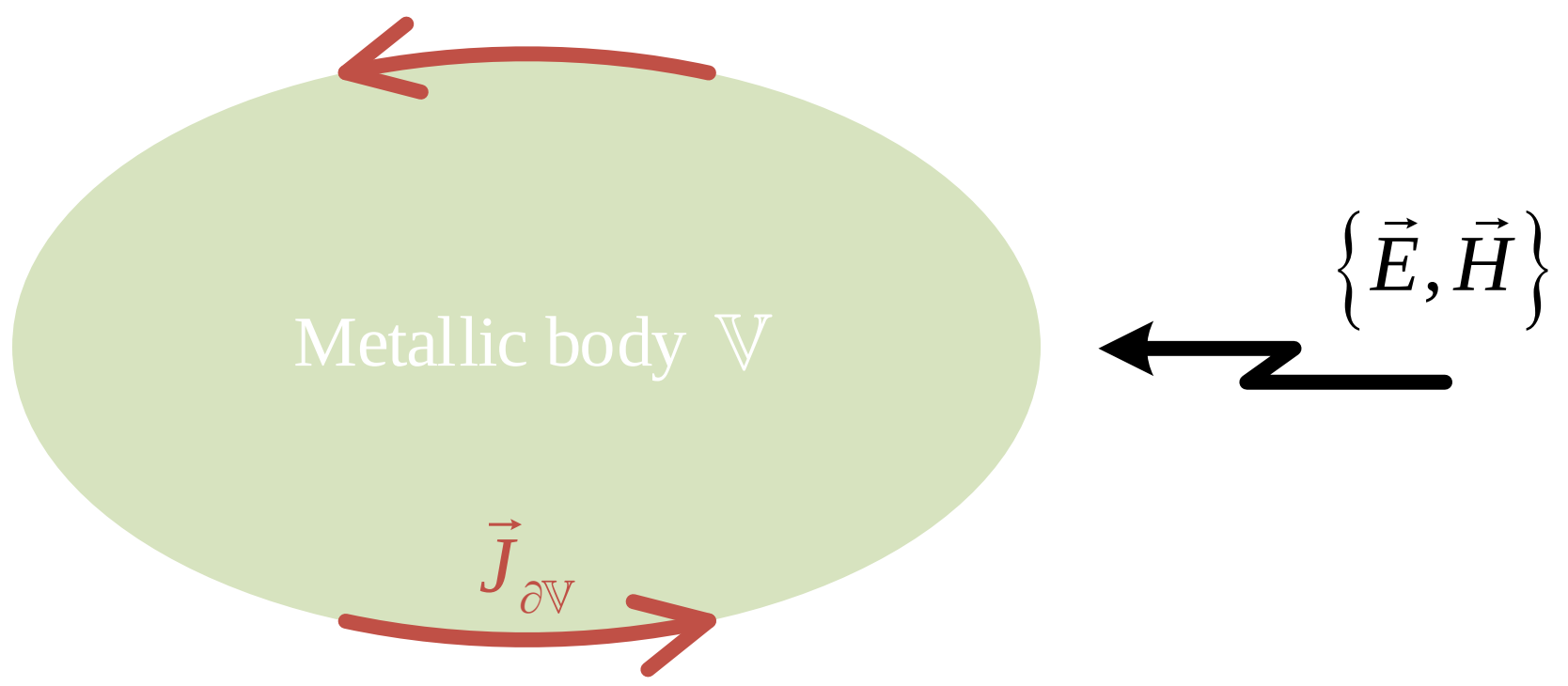

Figure A-2 A metallic body $\mathbb{V}$ under the action of EM fields $\{\vec{E},\vec{H}\}$

The above-mentioned electric current $\vec{J}_{\partial\mathbb{V}}$ and magnetic field $\vec{H}$ satisfy the following relation

$$\vec{J}_{\partial\mathbb{V}}(\vec{r}) = \hat{n}_{\partial\mathbb{V}}^{+}\times\left[\vec{H}(\vec{r}_{\mathbb{V}}^{\,+})\right]_{\vec{r}_{\mathbb{V}}^{\,+}\to\vec{r}} \quad , \quad \vec{r}\in\partial\mathbb{V} \tag{A-3}$$

In the above Eq. (A-3), $\vec{r}_{\mathbb{V}}^{\,+}$ is a point in the exterior of $\mathbb{V}$, and approaches the point $\vec{r}$

on $\partial\mathbb{V}$; $\hat{n}^+_{\partial\mathbb{V}}$ is the *normal direction* of $\partial\mathbb{V}$, and points to the exterior of $\mathbb{V}$.

The current $\vec{J}_{\partial\mathbb{V}}$ will generate a field $\vec{F}_{\partial\mathbb{V}}$ on $\mathbb{E}^3$, and the field can be expressed in terms of the current $\vec{J}_{\partial\mathbb{V}}$ as follows:

$$\vec{F}_{\partial\mathbb{V}}(\vec{r}) = \overset{\leftrightarrow}{G}{}^{JF}_0 * \vec{J}_{\partial\mathbb{V}} \quad , \quad \vec{r} \in \mathbb{E}^3 \setminus \partial\mathbb{V} \tag{A-4}$$

in which $\mathbb{E}^3 \setminus \partial\mathbb{V}$ denotes the *complementary set* of the $\partial\mathbb{V}$ on $\mathbb{E}^3$.

## A1.3 Metallic Surface Case

Now, suppose there is a metallic surface $\mathbb{S}$ under the action of EM fields $\{\vec{E},\vec{H}\}$, where $\{\vec{E},\vec{H}\}$ are the **total** fields. The action will induce a surface electric current $\vec{J}_{\mathbb{S}}$ on the surface $\mathbb{S}$, as shown in the following Fig. A-3.

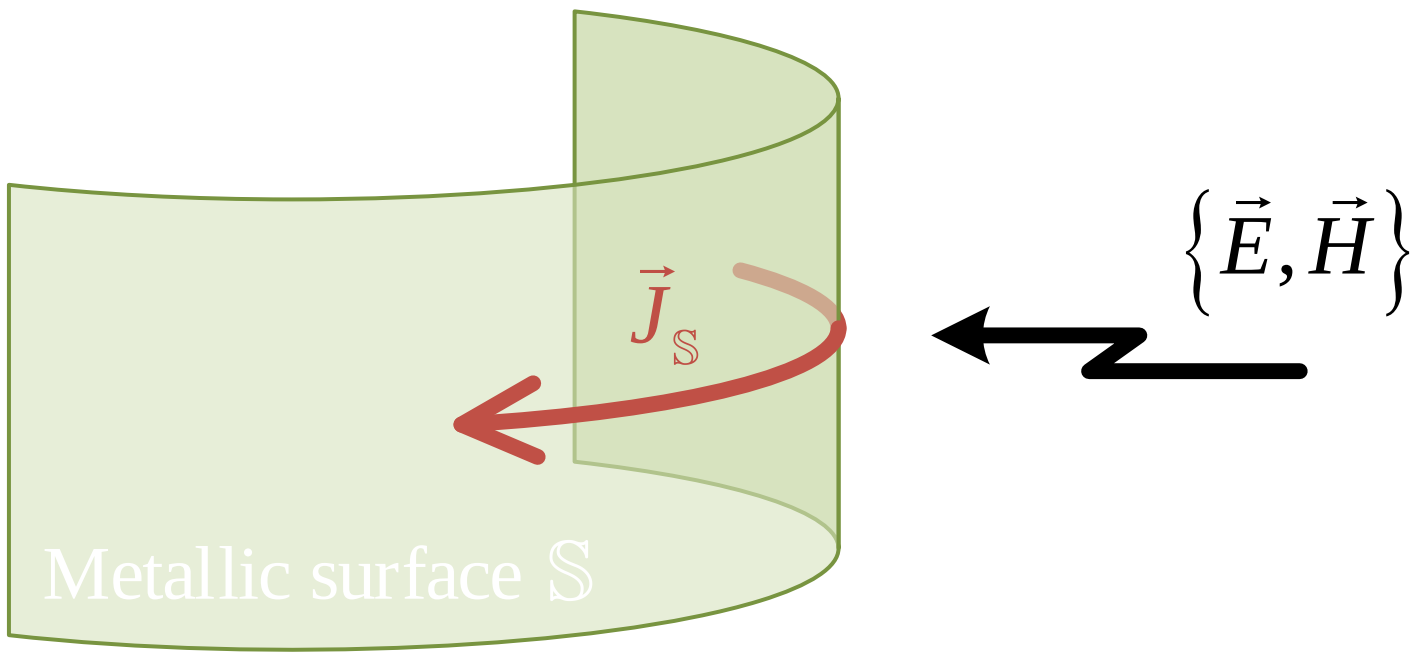

Figure A-3 A metallic surface $\mathbb{S}$ under the action of EM fields $\{\vec{E},\vec{H}\}$

The electric current $\vec{J}_{\mathbb{S}}$ and magnetic field $\vec{H}$ satisfy the following relation

$$\vec{J}_{\mathbb{S}}(\vec{r}) = \hat{n}^+_{\mathbb{S}} \times \left[\vec{H}(\vec{r}^{\,+}_{\mathbb{S}}) - \vec{H}(\vec{r}^{\,-}_{\mathbb{S}})\right]_{\vec{r}^{\,\pm}_{\mathbb{S}} \to \vec{r}} \quad , \quad \vec{r} \in \mathbb{S} \tag{A-5}$$

In the above Eq. (A-5), $\vec{r}^{\,+}_{\mathbb{S}}$ is a point in the plus side of $\mathbb{S}$, and $\vec{r}^{\,-}_{\mathbb{S}}$ is a point in the minus side of $\mathbb{S}$, and the points approach the point $\vec{r}$ on $\mathbb{S}$; $\hat{n}^+_{\mathbb{S}}$ is the normal direction of $\mathbb{S}$, and points to the plus side of $\mathbb{S}$.

The current $\vec{J}_{\mathbb{S}}$ will generate a field $\vec{F}_{\mathbb{S}}$ on $\mathbb{E}^3$, and the field can be expressed in terms of the current $\vec{J}_{\mathbb{S}}$ as follows:

$$\vec{F}_{\mathbb{S}}(\vec{r}) = \overset{\leftrightarrow}{G}{}^{JF}_0 * \vec{J}_{\mathbb{S}} \quad , \quad \vec{r} \in \mathbb{E}^3 \setminus \mathbb{S} \tag{A-6}$$

in which $\mathbb{E}^3 \setminus \mathbb{S}$ denotes the complementary set of the $\mathbb{S}$ on $\mathbb{E}^3$.

## A1.4 Metallic Wire Case

Now, suppose there is a metallic wire $\mathbb{L}$ under the action of EM fields $\{\vec{E},\vec{H}\}$, where $\{\vec{E},\vec{H}\}$ are the **total** fields. The action will induce a line electric current $\vec{J}_{\mathbb{L}}$ on $\mathbb{L}$, as shown in the following Fig. A-4.

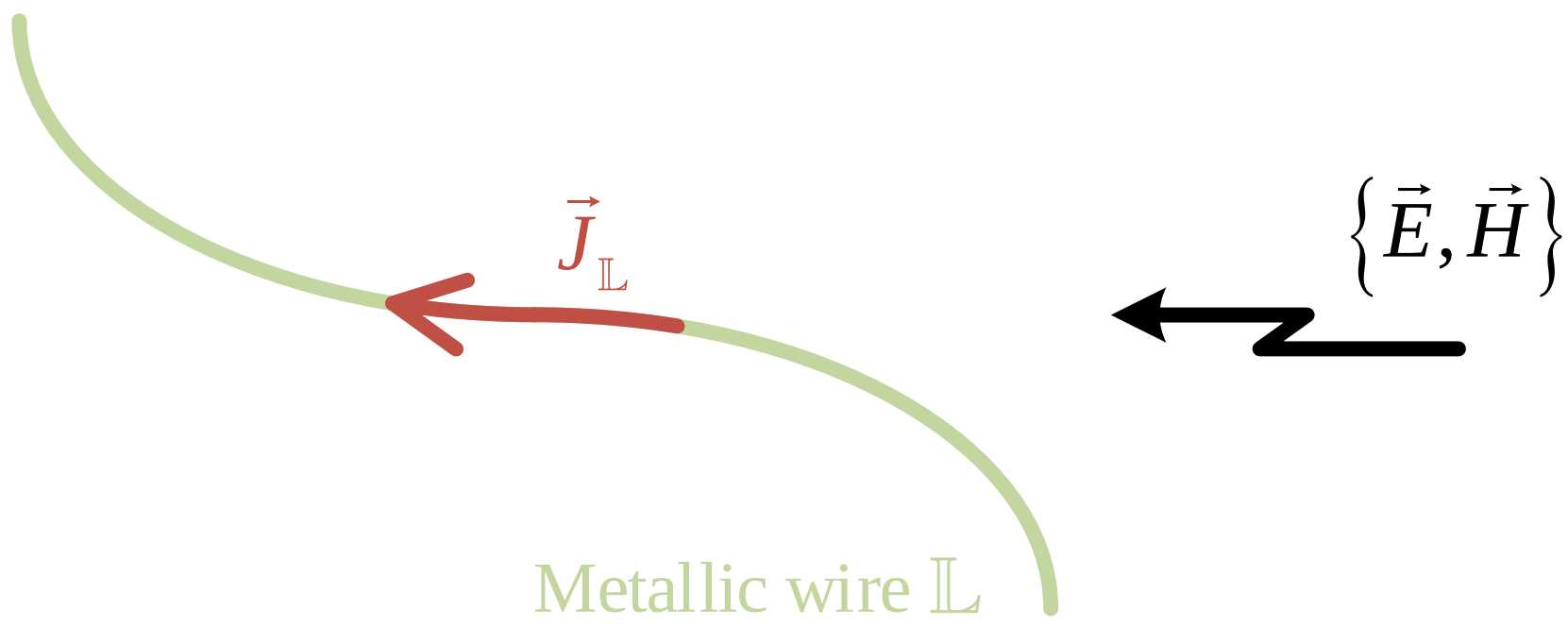

Figure A-4 A metallic wire $\mathbb{L}$ under the action of EM fields $\{\vec{E},\vec{H}\}$

The electric current $\vec{J}_{\mathbb{L}}$ and magnetic field $\vec{H}$ satisfy the following relation

$$\vec{J}_{\mathbb{L}}(\vec{r}) = \hat{e}_{\mathbb{L}} \lim_{\vec{r}'\to\vec{r}} \oint_{\mathbb{C}(\vec{r}')} \vec{H}(\vec{r}')\cdot\hat{e}_{\mathbb{C}} dl' \quad , \quad \vec{r}\in\mathbb{L} \tag{A-7}$$

In the above Eq. (A-7), the integral domain $\mathbb{C}$ is a circle surrounding $\mathbb{L}$; point $\vec{r}'$ is on $\mathbb{C}$, and approaches the point $\vec{r}$ on $\mathbb{L}$; $\hat{e}_{\mathbb{C}}$ is the *tangential reference direction* of $\mathbb{C}$, and $\hat{e}_{\mathbb{L}}$ is the tangential reference direction of $\mathbb{L}$, and $\hat{e}_{\mathbb{C}}$ and $\hat{e}_{\mathbb{L}}$ satisfy *right-hand rule*.

The current $\vec{J}_{\mathbb{L}}$ will generate a field $\vec{F}_{\mathbb{L}}$ on $\mathbb{E}^3$, and the field can be expressed in terms of the current $\vec{J}_{\mathbb{L}}$ as follows:

$$\vec{F}_{\mathbb{L}}(\vec{r}) = \overleftrightarrow{G}_0^{JF} * \vec{J}_{\mathbb{L}} \quad , \quad \vec{r} = \mathbb{E}^3 \setminus \mathbb{L} \tag{A-8}$$

in which $\mathbb{E}^3 \setminus \mathbb{L}$ denotes the complementary set of the $\mathbb{L}$ on $\mathbb{E}^3$.

## A2 Huygens-Fresnel Principle

Now, suppose there is a closed surface $\mathbb{S}$ dividing whole $\mathbb{E}^3$ into two regions $\mathbb{V}_1$ and $\mathbb{V}_2$, as shown in the following Fig. A-5.

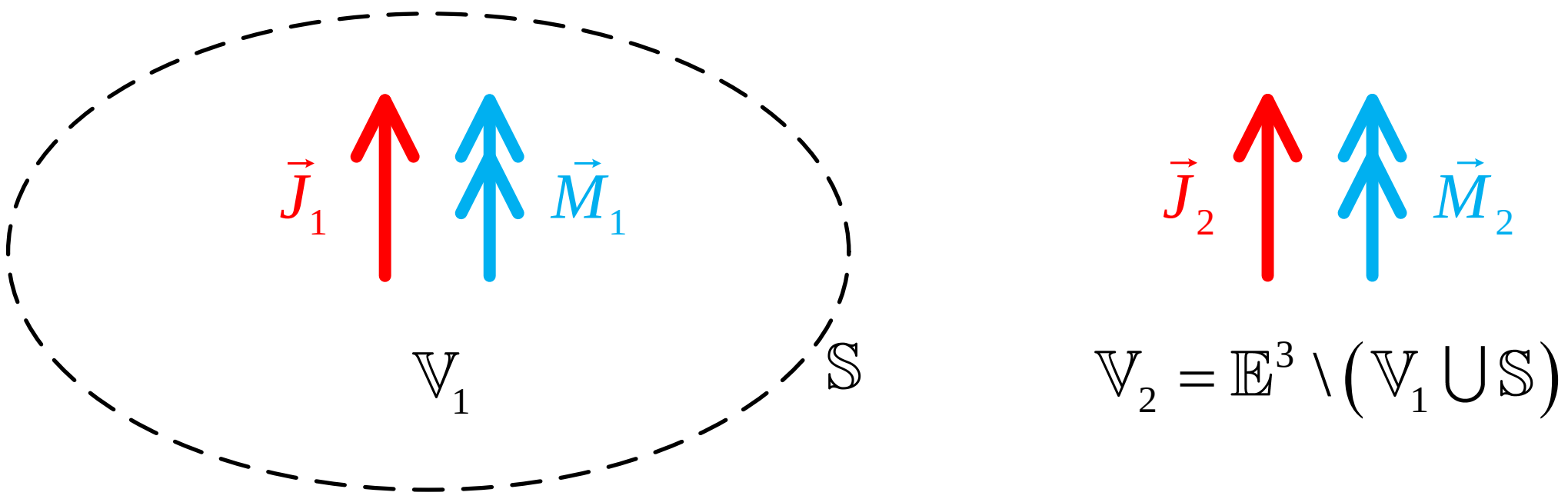

Figure A-5 Two regions $\mathbb{V}_1$ & $\mathbb{V}_2$ determined by a closed surface $\mathbb{S}$

The currents distributing on $\mathbb{V}_1$ and $\mathbb{V}_2$ are denoted as $\{\vec{J}_1,\vec{M}_1\}$ and $\{\vec{J}_2,\vec{M}_2\}$ respectively. The fields generated by $\{\vec{J}_1,\vec{M}_1\}$ and $\{\vec{J}_2,\vec{M}_2\}$ are denoted as $\vec{F}_1$ and

$\vec{F}_2$ respectively. *Huygens-Fresnel principle* gives the following *source-field relationships*

$$\left.\begin{matrix} \mathbb{V}_1 : & 0 \\ \mathbb{V}_2 : & \vec{F}_1(\vec{r}) \end{matrix}\right\} = \left[ \vec{G}_0^{JF} * \left( \hat{n}_{\mathbb{S}}^{\rightarrow 2} \times \vec{H}_1 \right) + \vec{G}_0^{MF} * \left( \vec{E}_1 \times \hat{n}_{\mathbb{S}}^{\rightarrow 2} \right) \right]_{\mathbb{S}} \tag{A-9}$$

$$\left.\begin{matrix} \mathbb{V}_1 : & \vec{F}_2(\vec{r}) \\ \mathbb{V}_2 : & 0 \end{matrix}\right\} = \left[ \vec{G}_0^{JF} * \left( \hat{n}_{\mathbb{S}}^{\rightarrow 1} \times \vec{H}_2 \right) + \vec{G}_0^{MF} * \left( \vec{E}_2 \times \hat{n}_{\mathbb{S}}^{\rightarrow 1} \right) \right]_{\mathbb{S}} \tag{A-10}$$

In the above Eqs. (A-9) and (A-10), $\hat{n}_{\mathbb{S}}^{\rightarrow 2}$ is the normal direction of $\mathbb{S}$, and points to the interior of region $\mathbb{V}_2$; $\hat{n}_{\mathbb{S}}^{\rightarrow 1}$ is the normal direction of $\mathbb{S}$, and points to the interior of region $\mathbb{V}_1$. It is obvious that $\hat{n}_{\mathbb{S}}^{\rightarrow 2} = -\hat{n}_{\mathbb{S}}^{\rightarrow 1}$ on whole $\mathbb{S}$.

In addition, the $\{\hat{n}_{\mathbb{S}}^{\rightarrow 2} \times \vec{H}_1, \vec{E}_1 \times \hat{n}_{\mathbb{S}}^{\rightarrow 2}\}$ and $\{\hat{n}_{\mathbb{S}}^{\rightarrow 1} \times \vec{H}_2, \vec{E}_2 \times \hat{n}_{\mathbb{S}}^{\rightarrow 1}\}$ are usually called the *Huygens' second sources* corresponding to the fields $\{\vec{E}_1, \vec{H}_1\}$ and $\{\vec{E}_2, \vec{H}_2\}$ respectively.

## A3 Equivalence Principles

The source-field relationships (*induction principle* and *Huygens-Fresnel principle*) given in the above App. A1 and App. A2 have very clear *physical pictures*. In this App. A3, we provide some alternative source-field relationships (*surface equivalence principle* and *line-surface equivalence principle*), which are more mathematical and more directly applicable to the contents discussed in this report.

### A3.1 Surface Equivalence Principle

Here, we still consider the EM problem discussed in App. A2. Because of that $\hat{n}_{\mathbb{S}}^{\rightarrow 2} = -\hat{n}_{\mathbb{S}}^{\rightarrow 1}$, Huygens-Fresnel principle (A-9) can be alternatively written as the following equivalent form

$$\left.\begin{matrix} \mathbb{V}_1 : & 0 \\ \mathbb{V}_2 : & -\vec{F}_1(\vec{r}) \end{matrix}\right\} = \left[ \vec{G}_0^{JF} * \left( \hat{n}_{\mathbb{S}}^{\rightarrow 1} \times \vec{H}_1 \right) + \vec{G}_0^{MF} * \left( \vec{E}_1 \times \hat{n}_{\mathbb{S}}^{\rightarrow 1} \right) \right]_{\mathbb{S}} \tag{A-11}$$

The summation of Huygens-Fresnel principles (A-10) and (A-11) gives the *surface equivalence principle* that

$$\begin{aligned} \left.\begin{matrix} \mathbb{V}_1 : & \vec{F}_2(\vec{r}) \\ \mathbb{V}_2 : & -\vec{F}_1(\vec{r}) \end{matrix}\right\} &= \left\{ \vec{G}_0^{JF} * \left[ \hat{n}_{\mathbb{S}}^{\rightarrow 1} \times \left( \vec{H}_1 + \vec{H}_2 \right) \right] + \vec{G}_0^{MF} * \left[ \left( \vec{E}_1 + \vec{E}_2 \right) \times \hat{n}_{\mathbb{S}}^{\rightarrow 1} \right] \right\}_{\mathbb{S}} \\ &= \left[ \vec{G}_0^{JF} * \vec{J}_{\mathbb{S}}^{\text{ES}} + \vec{G}_0^{MF} * \vec{M}_{\mathbb{S}}^{\text{ES}} \right]_{\mathbb{S}} \end{aligned} \tag{A-12}$$

in which *equivalent surface currents* $\{ \vec{J}_{\mathbb{S}}^{ES}, \vec{M}_{\mathbb{S}}^{ES} \}$ are defined as that

$$\vec{J}_{\mathbb{S}}^{\text{ES}}(\vec{r}) = \hat{n}_{\mathbb{S}}^{\rightarrow 1} \times \overbrace{\left(\vec{H}_1 + \vec{H}_2\right)}^{\vec{H}} \tag{A-13a}$$

$$\vec{M}_{\mathbb{S}}^{\text{ES}}(\vec{r}) = \underbrace{\left(\vec{E}_1 + \vec{E}_2\right)}_{\vec{E}} \times \hat{n}_{\mathbb{S}}^{\rightarrow 1} \tag{A-13b}$$

where $\{\vec{E},\vec{H}\}$ are the **total** fields distributing on $\mathbb{E}^3$.

Similarly, there also exists the following surface equivalence principle formulating the total fields $\{\vec{E},\vec{H}\}$ on $\mathbb{V}_1$ in terms of the equivalent surface currents $\{\vec{J}_{\mathbb{S}}^{\text{ES}},\vec{M}_{\mathbb{S}}^{\text{ES}}\}$.

$$\vec{F}(\vec{r}) = \left[\overleftrightarrow{G}_1^{JF} * \vec{J}_{\mathbb{S}}^{\text{ES}} + \overleftrightarrow{G}_1^{MF} * \vec{M}_{\mathbb{S}}^{\text{ES}}\right]_{\mathbb{S}} \quad , \quad \vec{r} \in \mathbb{V}_1 \tag{A-14}$$

if the $\{\vec{J}_1,\vec{M}_1\}$ shown in Fig. A-5 are some induced volume currents instead of some induced surface/line currents or *impressed currents*. Here, $\{\overleftrightarrow{G}_1^{JF},\overleftrightarrow{G}_1^{MF}\}$ are the *dyadic Green's functions corresponding to the region* $V_1$ *with parameters* $\{\mu_1,\varepsilon_1,\sigma_1\}$.

## A3.2 Line-Surface Equivalence Principle

When some metallic wires are completely or partially submerged in some material bodies, there will exist some *equivalent line electric currents* distributing on the EM objectice besides the equivalent surface electric currents. Thus the corresponding equivalence principle is called *line-surface equivalence principle*.

A complete and detailed discussion for line-surface equivalence principle can be found in the Secs. 5.2~5.4 of Ref. [13], and it will not be repeated here.

# A4 Extinction Theorem

Here, we still consider the EM problem discussed in App. A2. Based on the results given in App. A1, there exists the following *induction principle*

$$\left.\begin{aligned} \mathbb{V}_1 &: \vec{F}_1(\vec{r}) \\ \mathbb{V}_2 &: \vec{F}_1(\vec{r}) \end{aligned}\right\} = \left[\overleftrightarrow{G}_0^{JF} * \vec{J}_1 + \overleftrightarrow{G}_0^{MF} * \vec{M}_1\right]_{\mathbb{V}_1} \tag{A-15}$$

The summation of the previous surface equivalence principle (A-12) and the above induction principle (A-15) gives the following source-field relationship

$$\left.\begin{aligned} \mathbb{V}_1 &: \vec{F}(\vec{r}) \\ \mathbb{V}_2 &: \ 0 \end{aligned}\right\} = \left[\overleftrightarrow{G}_0^{JF} * \vec{J}_{\mathbb{S}}^{\text{ES}} + \overleftrightarrow{G}_0^{MF} * \vec{M}_{\mathbb{S}}^{\text{ES}}\right]_{\mathbb{S}} + \left[\overleftrightarrow{G}_0^{JF} * \vec{J}_1 + \overleftrightarrow{G}_0^{MF} * \vec{M}_1\right]_{\mathbb{V}_1} \tag{A-16}$$

where $\vec{F} = \vec{F}_1 + \vec{F}_2$ is the **total** field generated by $\{\vec{J}_1,\vec{M}_1\}$ and $\{\vec{J}_2,\vec{M}_2\}$ together, and the relationship is usually called *extinction theorem* because it results null field in region $\mathbb{V}_2$.

## Appendix B Numerical Analysis

In this report, the *decoupling mode theory* (*DMT*) for any objective *transceiving system* is established under *power transport theorem* (*PTT*) framework, and it can effectively construct a set of *fundamental modes* used to span the *modal space* of the objective transceiving system.

The PTT is a quantitative mathematical expression used to govern the transporting process of the power flowing in transceiving system. In the PTT, there exists a source term — *input power operator* (*IPO*) — used to sustain a steady power transport process. DMT constructs the fundamental modes by orthogonalizing the frequency-domain IPO, so the fundamental modes are called *input-power-decoupled modes* (*IP-DMs*), and the IPO is viewed as the generating operator of IP-DMs. In addition, the *definition domain* of IPO must be properly restricted to the modal space before calculating IP-DMs, or some *spurious modes* will be resulted.

This report provides three expressions to IPO — JM, JE, and HM expressions, and provides four schemes used to mathematically restrict definition domain — DVE-DoJ, DVE-DoM, SDC-DoJ, and SDC-DoM schemes, and they are listed in Tab. B-1.

Table B-1 Various ways for expressing the generating operator defined on modal space

| | **DVE-DoJ** | **DVE-DoM** | **SDC-DoJ** | **SDC-DoM** |
|---|---|---|---|---|
| **JM** | | | | |
| **JE** | Desired | | Desired | |
| **HM** | | Desired | | Desired |

Theoretically all the ways mentioned above are correct, but this is not the case numerical. Based on some numerical experiments, we find out that: the combination ways marked as "Desired" have more acceptable numerical performances than the other ways.

Besides the *modal decoupling equation* $\overline{\overline{P}}_{-} \cdot \bar{\alpha}_{\xi} \;=\; \theta_{\xi}\, \overline{\overline{P}}_{+} \cdot \bar{\alpha}_{\xi}$ given in the main body of this report, we also propose an alternative modal decoupling equation as follows:

$$\left(\overline{\overline{P}}_{-} + \ell \cdot \overline{\overline{\Psi}}^{\dagger} \cdot \overline{\overline{\Psi}}\right) \cdot \bar{\alpha}_{\xi} \;=\; \theta_{\xi}\, \overline{\overline{P}}_{+} \cdot \bar{\alpha}_{\xi} \tag{B-1}$$

for calculating IP-DMs, where $\overline{\overline{\Psi}} = \overline{\overline{\Psi}}_{\rm FCE}^{\rm DoJ} / \overline{\overline{\Psi}}_{\rm FCE}^{\rm DoM}$ (for details of $\overline{\overline{\Psi}}_{\rm FCE}^{\rm DoJ/DoM}$ please see the main body of this report) and $\ell$ is a relatively large real number (for example $\ell = 10^{10}$). Clearly, the first several (B-1)-based modes whose $\theta_{\xi}$ are relatively small are physical.

# Appendix C Surface-Volume Formulations of the PTT-Based DMT for Transceiving Systems

In the main body of this report, all the formulations are established by employing the surface equivalence principle or line-surface equivalence principle, so the resulted formulations involve the surface currents or {line currents, surface currents} only. Based on this reason, the formulations exhibited in the main body of this report can be collectively referred to as the *surface formulations* or *line-surface formulations* of the PTT-based DMT for transceiving systems.

Alternatively, this App. C provides some other formulations expressed in terms of induced surface electric currents and {induced volume electric currents, induced volume magnetic currents}. The alternative formulations are correspondingly called *surface-volume formulations*.

## C1 Surface-Volume Formulation of the PTT-Based DMT for the Material Guiding Structure Discussed in Sec. 3.3

By establishing the surface-volume formulation of the PTT-based DMT for material guiding structure, the isotropic material case can be generalized to anisotropic material case.

The process to establish the surface-volume formulation is similar to the process given in the following App. C2, and it will not be explicitly provided here for condensing the length of this report.

## C2 Surface-Volume Formulation of the PTT-Based DMT for the Metal-Material Composite Guiding Structure Discussed in Sec. 3.4

In this section, we provide the surface-volume formulation of the PTT-based DMT for the composite guiding structure discussed in Sec. 3.4, and generalize the material part from isotropic to anisotropic. The geometry of the guide is shown in the following Fig. C-1.

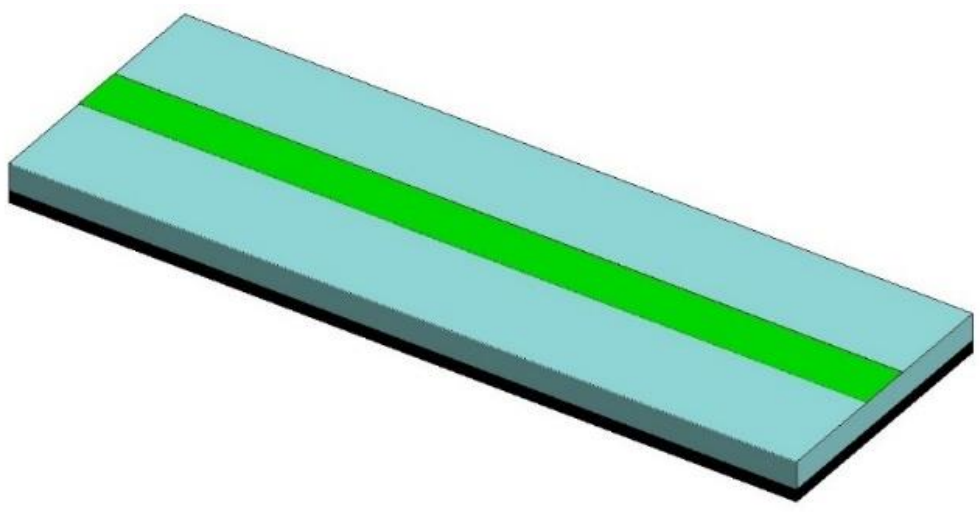

Figure C-1 Geometry of a microstrip transmission line with anisotropic material substrate

and the topological structure of a section of the guide is illustrated in the following Fig. C-2.

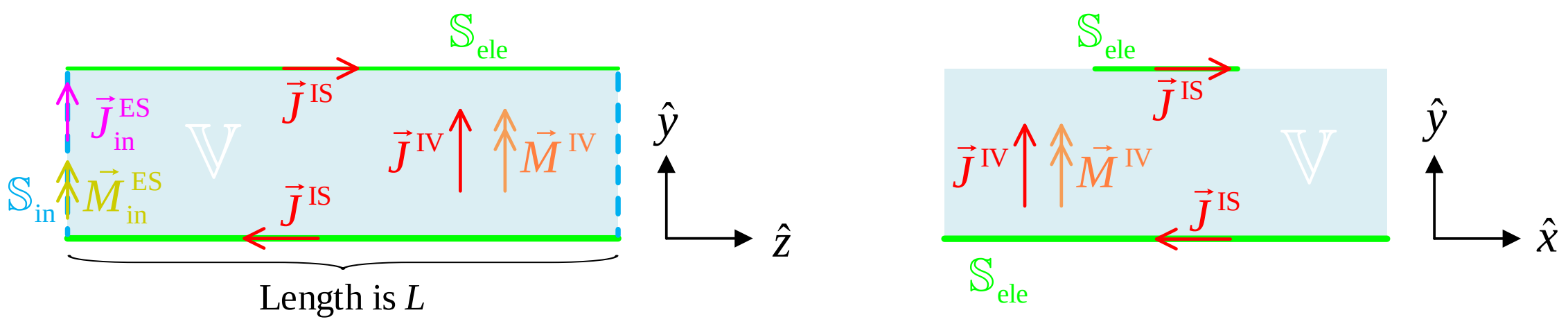

(a) (b)

Figure C-2 Topological structure of a section of the guide shown in Fig. C-1. (a) Longitudinal section view; (b) transverse section view

In the above Fig. C-2, the region occupied by the anisotropic material is denoted as $\mathbb{V}$; the *input port* of $\mathbb{V}$ is denoted as $\mathbb{S}_{\text{in}}$; the metallic boundary surface is denoted as $\mathbb{S}_{\text{ele}}$. The permeability and permittivity of $\mathbb{V}$ are $\vec{\vec{\mu}}$ and $\vec{\vec{\varepsilon}}$ respectively.

The equivalent surface currents distributing on $\mathbb{S}_{\text{in}}$ are denoted as $\{\vec{J}_{\text{in}}^{\text{ES}},\vec{M}_{\text{in}}^{\text{ES}}\}$; the induced volume currents distributing on $\mathbb{V}$ are denoted as $\{\vec{J}^{\,\text{IV}},\vec{M}^{\,\text{IV}}\}$; the induced surface electric current distributing on $\mathbb{S}_{\text{ele}}$ is denoted as $\vec{J}^{\,\text{IS}}$. The $\{\vec{J}_{\text{in}}^{\text{ES}},\vec{M}_{\text{in}}^{\text{ES}}\}$ are defined in terms of the modal total fields $\{\vec{E},\vec{H}\}$ as follows:

$$\vec{J}_{\text{in}}^{\text{ES}}(\vec{r}) = \hat{z}\times\vec{H}(\vec{r}) \quad , \quad \vec{r}\in\mathbb{S}_{\text{in}} \tag{C-1a}$$
$$\vec{M}_{\text{in}}^{\text{ES}}(\vec{r}) = \vec{E}(\vec{r})\times\hat{z} \quad , \quad \vec{r}\in\mathbb{S}_{\text{in}} \tag{C-1b}$$

where $\hat{z}$ is the Z-directional vector. The $\{\vec{J}^{\,\text{IV}},\vec{M}^{\,\text{IV}}\}$ and the modal total fields $\{\vec{E},\vec{H}\}$ satisfy the following relations

$$\vec{J}^{\,\text{IV}}(\vec{r}) = j\omega\Delta\vec{\vec{\varepsilon}}\cdot\vec{E}(\vec{r}) \quad , \quad \vec{r}\in\mathbb{V} \tag{C-2a}$$
$$\vec{M}^{\,\text{IV}}(\vec{r}) = j\omega\Delta\vec{\vec{\mu}}\cdot\vec{H}(\vec{r}) \quad , \quad \vec{r}\in\mathbb{V} \tag{C-2b}$$

based on the conclusions given in App. A1.1.

Using the above currents, the field $\vec{F}$ in region $\mathbb{V}$ can be expressed as the following operator form

$$\vec{F}(\vec{r}) = \mathcal{F}_0\left(\vec{J}_{\text{in}}^{\text{ES}}+\vec{J}^{\,\text{IS}}+\vec{J}^{\,\text{IV}},\vec{M}_{\text{in}}^{\text{ES}}+\vec{M}^{\,\text{IV}}\right) \quad , \quad \vec{r}\in\mathbb{V} \tag{C-3}$$

in which $\vec{F}=\vec{E}/\vec{H}$ and correspondingly $\mathcal{F}_0=\mathcal{E}_0/\mathcal{H}_0$, and operators $\mathcal{E}_0$ and $\mathcal{H}_0$ are the same as the ones used in the main body of this report.

Substituting Eq. (C-3) into Eqs. (C-1a) and (3-1b), we immediately obtain the following *integral equations*

$$\left[\mathcal{H}_0\left(\vec{J}_{\text{in}}^{\text{ES}}+\vec{J}^{\text{IS}}+\vec{J}^{\text{IV}},\vec{M}_{\text{in}}^{\text{ES}}+\vec{M}^{\text{IV}}\right)\right]_{\vec{r}'\to\vec{r}}^{\tan} = \vec{J}_{\text{in}}^{\text{ES}}\left(\vec{r}\right)\times\hat{z} \quad , \quad \vec{r}\in\mathbb{S}_{\text{in}} \tag{C-4a}$$

$$\left[\mathcal{E}_0\left(\vec{J}_{\text{in}}^{\text{ES}}+\vec{J}^{\text{IS}}+\vec{J}^{\text{IV}},\vec{M}_{\text{in}}^{\text{ES}}+\vec{M}^{\text{IV}}\right)\right]_{\vec{r}'\to\vec{r}}^{\tan} = \hat{z}\times\vec{M}_{\text{in}}^{\text{ES}}\left(\vec{r}\right) \quad , \quad \vec{r}\in\mathbb{S}_{\text{in}} \tag{C-4b}$$

satisfied by the related currents, where $\vec{r}'\in\mathbb{V}$ and approaches the point $\vec{r}$ on surface $\mathbb{S}_{\text{in}}$.

Utilizing Eq. (C-3) and the *homogeneous tangential electric field boundary condition* on $\mathbb{S}_{\text{ele}}$, we obtain the following *electric field integral equation*

$$\left[\mathcal{E}_0\left(\vec{J}_{\text{in}}^{\text{ES}}+\vec{J}^{\text{IS}}+\vec{J}^{\text{IV}},\vec{M}_{\text{in}}^{\text{ES}}+\vec{M}^{\text{IV}}\right)\right]^{\tan} = 0 \quad , \quad \vec{r}\in\mathbb{S}_{\text{ele}} \tag{C-5}$$

satisfied by the currents.

Substituting Eq. (C-3) into Eqs. (C-2a) and (C-2b), we obtain the following integral equations

$$\mathcal{E}_0\left(\vec{J}_{\text{in}}^{\text{ES}}+\vec{J}^{\text{IS}}+\vec{J}^{\text{IV}},\vec{M}_{\text{in}}^{\text{ES}}+\vec{M}^{\text{IV}}\right) = \left(j\omega\Delta\ddot{\varepsilon}\right)^{-1}\cdot\vec{J}^{\text{IV}}\left(\vec{r}\right) \quad , \quad \vec{r}\in\mathbb{V} \tag{C-6a}$$

$$\mathcal{H}_0\left(\vec{J}_{\text{in}}^{\text{ES}}+\vec{J}^{\text{IS}}+\vec{J}^{\text{IV}},\vec{M}_{\text{in}}^{\text{ES}}+\vec{M}^{\text{IV}}\right) = \left(j\omega\Delta\ddot{\mu}\right)^{-1}\cdot\vec{M}^{\text{IV}}\left(\vec{r}\right) \quad , \quad \vec{r}\in\mathbb{V} \tag{C-6b}$$

satisfied by the currents.

If the currents contained in Eqs. (C-4a)~(C-6b) are expanded in terms of some proper *basis functions*, and the equations are tested with $\{\vec{b}_{\xi}^{\vec{M}_{\text{in}}^{\text{ES}}}\}$, $\{\vec{b}_{\xi}^{\vec{J}_{\text{in}}^{\text{ES}}}\}$, $\{\vec{b}_{\xi}^{\vec{J}^{\text{IS}}}\}$, $\{\vec{b}_{\xi}^{\vec{J}^{\text{IV}}}\}$, and $\{\vec{b}_{\xi}^{\vec{M}^{\text{IV}}}\}$ respectively, then the integral equations are immediately discretized into the following *matrix equations*

$$\overline{\overline{Z}}^{\vec{M}_{\text{in}}^{\text{ES}}\vec{J}_{\text{in}}^{\text{ES}}}\cdot\overline{a}^{\vec{J}_{\text{in}}^{\text{ES}}}+\overline{\overline{Z}}^{\vec{M}_{\text{in}}^{\text{ES}}\vec{J}^{\text{IS}}}\cdot\overline{a}^{\vec{J}^{\text{IS}}}+\overline{\overline{Z}}^{\vec{M}_{\text{in}}^{\text{ES}}\vec{J}^{\text{IV}}}\cdot\overline{a}^{\vec{J}^{\text{IV}}}+\overline{\overline{Z}}^{\vec{M}_{\text{in}}^{\text{ES}}\vec{M}_{\text{in}}^{\text{ES}}}\cdot\overline{a}^{\vec{M}_{\text{in}}^{\text{ES}}}+\overline{\overline{Z}}^{\vec{M}_{\text{in}}^{\text{ES}}\vec{J}^{\text{IV}}}\cdot\overline{a}^{\vec{J}^{\text{IV}}} = 0 \tag{C-7a}$$

$$\overline{\overline{Z}}^{\vec{J}_{\text{in}}^{\text{ES}}\vec{J}_{\text{in}}^{\text{ES}}}\cdot\overline{a}^{\vec{J}_{\text{in}}^{\text{ES}}}+\overline{\overline{Z}}^{\vec{J}_{\text{in}}^{\text{ES}}\vec{J}^{\text{IS}}}\cdot\overline{a}^{\vec{J}^{\text{IS}}}+\overline{\overline{Z}}^{\vec{J}_{\text{in}}^{\text{ES}}\vec{J}^{\text{IV}}}\cdot\overline{a}^{\vec{J}^{\text{IV}}}+\overline{\overline{Z}}^{\vec{J}_{\text{in}}^{\text{ES}}\vec{M}_{\text{in}}^{\text{ES}}}\cdot\overline{a}^{\vec{M}_{\text{in}}^{\text{ES}}}+\overline{\overline{Z}}^{\vec{J}_{\text{in}}^{\text{ES}}\vec{M}^{\text{IV}}}\cdot\overline{a}^{\vec{M}^{\text{IV}}} = 0 \tag{C-7b}$$

and

$$\overline{\overline{Z}}^{\vec{J}^{\text{IS}}\vec{J}_{\text{in}}^{\text{ES}}}\cdot\overline{a}^{\vec{J}_{\text{in}}^{\text{ES}}}+\overline{\overline{Z}}^{\vec{J}^{\text{IS}}\vec{J}^{\text{IS}}}\cdot\overline{a}^{\vec{J}^{\text{IS}}}+\overline{\overline{Z}}^{\vec{J}^{\text{IS}}\vec{J}^{\text{IV}}}\cdot\overline{a}^{\vec{J}^{\text{IV}}}+\overline{\overline{Z}}^{\vec{J}^{\text{IS}}\vec{M}_{\text{in}}^{\text{ES}}}\cdot\overline{a}^{\vec{M}_{\text{in}}^{\text{ES}}}+\overline{\overline{Z}}^{\vec{J}^{\text{IS}}\vec{M}^{\text{IV}}}\cdot\overline{a}^{\vec{M}^{\text{IV}}} = 0 \tag{C-8}$$

and

$$\overline{\overline{Z}}^{\vec{J}^{\text{IV}}\vec{J}_{\text{in}}^{\text{ES}}}\cdot\overline{a}^{\vec{J}_{\text{in}}^{\text{ES}}}+\overline{\overline{Z}}^{\vec{J}^{\text{IV}}\vec{J}^{\text{IS}}}\cdot\overline{a}^{\vec{J}^{\text{IS}}}+\overline{\overline{Z}}^{\vec{J}^{\text{IV}}\vec{J}^{\text{IV}}}\cdot\overline{a}^{\vec{J}^{\text{IV}}}+\overline{\overline{Z}}^{\vec{J}^{\text{IV}}\vec{M}_{\text{in}}^{\text{ES}}}\cdot\overline{a}^{\vec{M}_{\text{in}}^{\text{ES}}}+\overline{\overline{Z}}^{\vec{J}^{\text{IV}}\vec{M}^{\text{IV}}}\cdot\overline{a}^{\vec{M}^{\text{IV}}} = 0 \tag{C-9a}$$

$$\overline{\overline{Z}}^{\vec{M}^{\text{IV}}\vec{J}_{\text{in}}^{\text{ES}}}\cdot\overline{a}^{\vec{J}_{\text{in}}^{\text{ES}}}+\overline{\overline{Z}}^{\vec{M}^{\text{IV}}\vec{J}^{\text{IS}}}\cdot\overline{a}^{\vec{J}^{\text{IS}}}+\overline{\overline{Z}}^{\vec{M}^{\text{IV}}\vec{J}^{\text{IV}}}\cdot\overline{a}^{\vec{J}^{\text{IV}}}+\overline{\overline{Z}}^{\vec{M}^{\text{IV}}\vec{M}_{\text{in}}^{\text{ES}}}\cdot\overline{a}^{\vec{M}_{\text{in}}^{\text{ES}}}+\overline{\overline{Z}}^{\vec{M}^{\text{IV}}\vec{M}^{\text{IV}}}\cdot\overline{a}^{\vec{M}^{\text{IV}}} = 0 \tag{C-9b}$$

The formulations used to calculate the elements of the matrices in the above matrix equations are as follows:

$$z_{\xi\zeta}^{\vec{M}_{\text{in}}^{\text{ES}}\vec{J}_{\text{in}}^{\text{ES}}} = \left\langle\vec{b}_{\xi}^{\vec{M}_{\text{in}}^{\text{ES}}},\text{P.V.}\mathcal{K}_0\left(\vec{b}_{\zeta}^{\vec{J}_{\text{in}}^{\text{ES}}}\right)\right\rangle_{\mathbb{S}_{\text{in}}}-\left(1/2\right)\left\langle\vec{b}_{\xi}^{\vec{M}_{\text{in}}^{\text{ES}}},\vec{b}_{\zeta}^{\vec{J}_{\text{in}}^{\text{ES}}}\times\hat{z}\right\rangle_{\mathbb{S}_{\text{in}}} \tag{C-10a}$$

$$z_{\xi\zeta}^{\vec{M}_{\text{in}}^{\text{ES}}\vec{J}^{\text{IS}}} = \left\langle\vec{b}_{\xi}^{\vec{M}_{\text{in}}^{\text{ES}}},\mathcal{K}_0\left(\vec{b}_{\zeta}^{\vec{J}^{\text{IS}}}\right)\right\rangle_{\mathbb{S}_{\text{in}}} \tag{C-10b}$$

$$z_{\xi\zeta}^{\vec{M}_{\rm in}^{\rm ES}\vec{J}^{\rm IV}} = \left\langle \vec{b}_{\xi}^{\vec{M}_{\rm in}^{\rm ES}}, \mathcal{K}_0\left(\vec{b}_{\zeta}^{\vec{J}^{\rm IV}}\right)\right\rangle_{\mathbb{S}_{\rm in}} \tag{C-10c}$$

$$z_{\xi\zeta}^{\vec{M}_{\rm in}^{\rm ES}\vec{M}_{\rm in}^{\rm ES}} = \left\langle \vec{b}_{\xi}^{\vec{M}_{\rm in}^{\rm ES}}, -j\omega\varepsilon_0\mathcal{L}_0\left(\vec{b}_{\zeta}^{\vec{M}_{\rm in}^{\rm ES}}\right)\right\rangle_{\mathbb{S}_{\rm in}} \tag{C-10d}$$

$$z_{\xi\zeta}^{\vec{M}_{\rm in}^{\rm ES}\vec{J}^{\rm IV}} = \left\langle \vec{b}_{\xi}^{\vec{M}_{\rm in}^{\rm ES}}, -j\omega\varepsilon_0\mathcal{L}_0\left(\vec{b}_{\zeta}^{\vec{M}^{\rm IV}}\right)\right\rangle_{\mathbb{S}_{\rm in}} \tag{C-10e}$$

and

$$z_{\xi\zeta}^{\vec{J}_{\rm in}^{\rm ES}\vec{J}_{\rm in}^{\rm ES}} = \left\langle \vec{b}_{\xi}^{\vec{J}_{\rm in}^{\rm ES}}, -j\omega\mu_0\mathcal{L}_0\left(\vec{b}_{\zeta}^{\vec{J}_{\rm in}^{\rm ES}}\right)\right\rangle_{\mathbb{S}_{\rm in}} \tag{C-11a}$$

$$z_{\xi\zeta}^{\vec{J}_{\rm in}^{\rm ES}\vec{J}^{\rm IS}} = \left\langle \vec{b}_{\xi}^{\vec{J}_{\rm in}^{\rm ES}}, -j\omega\mu_0\mathcal{L}_0\left(\vec{b}_{\zeta}^{\vec{J}^{\rm IS}}\right)\right\rangle_{\mathbb{S}_{\rm in}} \tag{C-11b}$$

$$z_{\xi\zeta}^{\vec{J}_{\rm in}^{\rm ES}\vec{J}^{\rm IV}} = \left\langle \vec{b}_{\xi}^{\vec{J}_{\rm in}^{\rm ES}}, -j\omega\mu_0\mathcal{L}_0\left(\vec{b}_{\zeta}^{\vec{J}^{\rm IV}}\right)\right\rangle_{\mathbb{S}_{\rm in}} \tag{C-11c}$$

$$z_{\xi\zeta}^{\vec{J}_{\rm in}^{\rm ES}\vec{M}_{\rm in}^{\rm ES}} = \left\langle \vec{b}_{\xi}^{\vec{J}_{\rm in}^{\rm ES}}, -\text{P.V.}\,\mathcal{K}_0\left(\vec{b}_{\zeta}^{\vec{M}_{\rm in}^{\rm ES}}\right)\right\rangle_{\mathbb{S}_{\rm in}} - (1/2)\left\langle \vec{b}_{\xi}^{\vec{J}_{\rm in}^{\rm ES}}, \hat{z}\times\vec{b}_{\zeta}^{\vec{M}_{\rm in}^{\rm ES}}\right\rangle_{\mathbb{S}_{\rm in}} \tag{C-11d}$$

$$z_{\xi\zeta}^{\vec{J}_{\rm in}^{\rm ES}\vec{M}^{\rm IV}} = \left\langle \vec{b}_{\xi}^{\vec{J}_{\rm in}^{\rm ES}}, -\mathcal{K}_0\left(\vec{b}_{\zeta}^{\vec{M}^{\rm IV}}\right)\right\rangle_{\mathbb{S}_{\rm in}} \tag{C-11e}$$

and

$$z_{\xi\zeta}^{\vec{J}^{\rm IS}\vec{J}_{\rm in}^{\rm ES}} = \left\langle \vec{b}_{\xi}^{\vec{J}^{\rm IS}}, -j\omega\mu_0\mathcal{L}_0\left(\vec{b}_{\zeta}^{\vec{J}_{\rm in}^{\rm ES}}\right)\right\rangle_{\mathbb{S}_{\rm ele}} \tag{C-12a}$$

$$z_{\xi\zeta}^{\vec{J}^{\rm IS}\vec{J}^{\rm IS}} = \left\langle \vec{b}_{\xi}^{\vec{J}^{\rm IS}}, -j\omega\mu_0\mathcal{L}_0\left(\vec{b}_{\zeta}^{\vec{J}^{\rm IS}}\right)\right\rangle_{\mathbb{S}_{\rm ele}} \tag{C-12b}$$

$$z_{\xi\zeta}^{\vec{J}^{\rm IS}\vec{J}^{\rm IV}} = \left\langle \vec{b}_{\xi}^{\vec{J}^{\rm IS}}, -j\omega\mu_0\mathcal{L}_0\left(\vec{b}_{\zeta}^{\vec{J}^{\rm IV}}\right)\right\rangle_{\mathbb{S}_{\rm ele}} \tag{C-12c}$$

$$z_{\xi\zeta}^{\vec{J}^{\rm IS}\vec{M}_{\rm in}^{\rm ES}} = \left\langle \vec{b}_{\xi}^{\vec{J}^{\rm IS}}, -\mathcal{K}_0\left(\vec{b}_{\zeta}^{\vec{M}_{\rm in}^{\rm ES}}\right)\right\rangle_{\mathbb{S}_{\rm ele}} \tag{C-12d}$$

$$z_{\xi\zeta}^{\vec{J}^{\rm IS}\vec{M}^{\rm IV}} = \left\langle \vec{b}_{\xi}^{\vec{J}^{\rm IS}}, -\mathcal{K}_0\left(\vec{b}_{\zeta}^{\vec{M}^{\rm IV}}\right)\right\rangle_{\mathbb{S}_{\rm ele}} \tag{C-12e}$$

and

$$z_{\xi\zeta}^{\vec{J}^{\rm IV}\vec{J}_{\rm in}^{\rm ES}} = \left\langle \vec{b}_{\xi}^{\vec{J}^{\rm IV}}, -j\omega\mu_0\mathcal{L}_0\left(\vec{b}_{\zeta}^{\vec{J}_{\rm in}^{\rm ES}}\right)\right\rangle_{\mathbb{V}} \tag{C-13a}$$

$$z_{\xi\zeta}^{\vec{J}^{\rm IV}\vec{J}^{\rm IS}} = \left\langle \vec{b}_{\xi}^{\vec{J}^{\rm IV}}, -j\omega\mu_0\mathcal{L}_0\left(\vec{b}_{\zeta}^{\vec{J}^{\rm IS}}\right)\right\rangle_{\mathbb{V}} \tag{C-13b}$$

$$z_{\xi\zeta}^{\vec{J}^{\rm IV}\vec{J}^{\rm IV}} = \left\langle \vec{b}_{\xi}^{\vec{J}^{\rm IV}}, -j\omega\mu_0\mathcal{L}_0\left(\vec{b}_{\zeta}^{\vec{J}^{\rm IV}}\right)\right\rangle_{\mathbb{V}} - \left\langle \vec{b}_{\xi}^{\vec{J}^{\rm IV}}, \left(j\omega\Delta\overset{\leftrightarrow}{\varepsilon}\right)^{-1}\cdot\vec{b}_{\zeta}^{\vec{J}^{\rm IV}}\right\rangle_{\mathbb{V}} \tag{C-13c}$$

$$z_{\xi\zeta}^{\vec{J}^{\rm IV}\vec{M}_{\rm in}^{\rm ES}} = \left\langle \vec{b}_{\xi}^{\vec{J}^{\rm IV}}, -\mathcal{K}_0\left(\vec{b}_{\zeta}^{\vec{M}_{\rm in}^{\rm ES}}\right)\right\rangle_{\mathbb{V}} \tag{C-13d}$$

$$z_{\xi\zeta}^{\vec{J}^{\rm IV}\vec{M}^{\rm IV}} = \left\langle \vec{b}_{\xi}^{\vec{J}^{\rm IV}}, -\mathcal{K}_0\left(\vec{b}_{\zeta}^{\vec{M}^{\rm IV}}\right)\right\rangle_{\mathbb{V}} \tag{C-13e}$$

and

$$z_{\xi\zeta}^{\vec{M}^{\rm IV}\vec{J}_{\rm in}^{\rm ES}} = \left\langle \vec{b}_{\xi}^{\vec{M}^{\rm IV}}, \mathcal{K}_0\left(\vec{b}_{\zeta}^{\vec{J}_{\rm in}^{\rm ES}}\right)\right\rangle_{\mathbb{V}} \tag{C-14a}$$

$$z_{\xi\zeta}^{\vec{M}^{\text{IV}}\vec{J}^{\text{IS}}} = \left\langle \vec{b}_{\xi}^{\vec{M}^{\text{IV}}}, \mathcal{K}_0\left(\vec{b}_{\zeta}^{\vec{J}^{\text{IS}}}\right)\right\rangle_{\mathbb{V}} \tag{C-14b}$$

$$z_{\xi\zeta}^{\vec{M}^{\text{IV}}\vec{J}^{\text{IV}}} = \left\langle \vec{b}_{\xi}^{\vec{M}^{\text{IV}}}, \mathcal{K}_0\left(\vec{b}_{\zeta}^{\vec{J}^{\text{IV}}}\right)\right\rangle_{\mathbb{V}} \tag{C-14c}$$

$$z_{\xi\zeta}^{\vec{M}^{\text{IV}}\vec{M}_{\text{in}}^{\text{ES}}} = \left\langle \vec{b}_{\xi}^{\vec{M}^{\text{IV}}}, -j\omega\varepsilon_0\mathcal{L}_0\left(\vec{b}_{\zeta}^{\vec{M}_{\text{in}}^{\text{ES}}}\right)\right\rangle_{\mathbb{V}} \tag{C-14d}$$

$$z_{\xi\zeta}^{\vec{M}^{\text{IV}}\vec{M}^{\text{IV}}} = \left\langle \vec{b}_{\xi}^{\vec{M}^{\text{IV}}}, -j\omega\varepsilon_0\mathcal{L}_0\left(\vec{b}_{\zeta}^{\vec{M}^{\text{IV}}}\right)\right\rangle_{\mathbb{V}} - \left\langle \vec{b}_{\xi}^{\vec{M}^{\text{IV}}}, \left(j\omega\Delta\vec{\vec{\mu}}\right)^{-1}\cdot\vec{b}_{\zeta}^{\vec{M}^{\text{IV}}}\right\rangle_{\mathbb{V}} \tag{C-14e}$$

By employing the above matrix equations, we propose the scheme for rigorously describing the *modal space* as below.

Employing the above Eqs. (C-7a)~(C-9b), we can obtain the following transformation

$$\begin{bmatrix} \bar{a}^{\vec{J}_{\text{in}}^{\text{ES}}} \\ \bar{a}^{\vec{J}^{\text{IS}}} \\ \bar{a}^{\vec{J}^{\text{IV}}} \\ \bar{a}^{\vec{M}_{\text{in}}^{\text{ES}}} \\ \bar{a}^{\vec{J}^{\text{IV}}} \end{bmatrix} = \bar{a}^{\text{AV}} = \bar{\bar{T}}\cdot\bar{a} \tag{C-15}$$

where the transformation matrix $\bar{\bar{T}}$ is $\bar{\bar{T}}^{\vec{J}_{\text{in}}^{\text{ES}}\rightarrow\text{AV}}/\bar{\bar{T}}^{\vec{M}_{\text{in}}^{\text{ES}}\rightarrow\text{AV}}/\bar{\bar{T}}^{\text{BS}\rightarrow\text{AV}}$ and the vector $\bar{a}$ is $\bar{a}^{\vec{J}_{\text{in}}^{\text{ES}}}/\bar{a}^{\vec{M}_{\text{in}}^{\text{ES}}}/\bar{a}^{\text{BS}}$ correspondingly. Here, the matrices $\bar{\bar{T}}^{\vec{J}_{\text{in}}^{\text{ES}}\rightarrow\text{AV}}$, $\bar{\bar{T}}^{\vec{M}_{\text{in}}^{\text{ES}}\rightarrow\text{AV}}$, and $\bar{\bar{T}}^{\text{BS}\rightarrow\text{AV}}$ are as follows:

$$\bar{\bar{T}}^{\vec{J}_{\text{in}}^{\text{ES}}\rightarrow\text{AV}} = \left(\bar{\bar{\Psi}}_1\right)^{-1}\cdot\bar{\bar{\Psi}}_2 \tag{C-16a}$$

$$\bar{\bar{T}}^{\vec{M}_{\text{in}}^{\text{ES}}\rightarrow\text{AV}} = \left(\bar{\bar{\Psi}}_3\right)^{-1}\cdot\bar{\bar{\Psi}}_4 \tag{C-16b}$$

and

$$\bar{\bar{T}}^{\text{BS}\rightarrow\text{AV}} = \text{nullspace}\left(\bar{\bar{\Psi}}^{\text{DoJ/DoM}}\right) \tag{C-17}$$

in which

$$\bar{\bar{\Psi}}_1 = \begin{bmatrix} \bar{\bar{I}}^{\vec{J}_{\text{in}}^{\text{ES}}} & 0 & 0 & 0 & 0 \\ 0 & \bar{\bar{Z}}^{\vec{M}_{\text{in}}^{\text{ES}}\vec{J}^{\text{IS}}} & \bar{\bar{Z}}^{\vec{M}_{\text{in}}^{\text{ES}}\vec{J}^{\text{IV}}} & \bar{\bar{Z}}^{\vec{M}_{\text{in}}^{\text{ES}}\vec{M}_{\text{in}}^{\text{ES}}} & \bar{\bar{Z}}^{\vec{M}_{\text{in}}^{\text{ES}}\vec{J}^{\text{IV}}} \\ 0 & \bar{\bar{Z}}^{\vec{J}^{\text{IS}}\vec{J}^{\text{IS}}} & \bar{\bar{Z}}^{\vec{J}^{\text{IS}}\vec{J}^{\text{IV}}} & \bar{\bar{Z}}^{\vec{J}^{\text{IS}}\vec{M}_{\text{in}}^{\text{ES}}} & \bar{\bar{Z}}^{\vec{J}^{\text{IS}}\vec{M}^{\text{IV}}} \\ 0 & \bar{\bar{Z}}^{\vec{J}^{\text{IV}}\vec{J}^{\text{IS}}} & \bar{\bar{Z}}^{\vec{J}^{\text{IV}}\vec{J}^{\text{IV}}} & \bar{\bar{Z}}^{\vec{J}^{\text{IV}}\vec{M}_{\text{in}}^{\text{ES}}} & \bar{\bar{Z}}^{\vec{J}^{\text{IV}}\vec{M}^{\text{IV}}} \\ 0 & \bar{\bar{Z}}^{\vec{M}^{\text{IV}}\vec{J}^{\text{IS}}} & \bar{\bar{Z}}^{\vec{M}^{\text{IV}}\vec{J}^{\text{IV}}} & \bar{\bar{Z}}^{\vec{M}^{\text{IV}}\vec{M}_{\text{in}}^{\text{ES}}} & \bar{\bar{Z}}^{\vec{M}^{\text{IV}}\vec{M}^{\text{IV}}} \end{bmatrix} \tag{C-18a}$$

$$\bar{\bar{\Psi}}_2 = \begin{bmatrix} \bar{\bar{I}}^{\vec{J}_{\text{in}}^{\text{ES}}} \\ -\bar{\bar{Z}}^{\vec{M}_{\text{in}}^{\text{ES}}\vec{J}_{\text{in}}^{\text{ES}}} \\ -\bar{\bar{Z}}^{\vec{J}^{\text{IS}}\vec{J}_{\text{in}}^{\text{ES}}} \\ -\bar{\bar{Z}}^{\vec{J}^{\text{IV}}\vec{J}_{\text{in}}^{\text{ES}}} \\ -\bar{\bar{Z}}^{\vec{M}^{\text{IV}}\vec{J}_{\text{in}}^{\text{ES}}} \end{bmatrix} \tag{C-18b}$$

and

$$\overline{\overline{\Psi}}_3 = \begin{bmatrix} 0 & 0 & 0 & \overline{\overline{I}}^{\vec{M}_{\rm in}^{\rm ES}} & 0 \\ \overline{\overline{Z}}^{\vec{J}_{\rm in}^{\rm ES}\vec{J}_{\rm in}^{\rm ES}} & \overline{\overline{Z}}^{\vec{J}_{\rm in}^{\rm ES}\vec{J}^{\rm IS}} & \overline{\overline{Z}}^{\vec{J}_{\rm in}^{\rm ES}\vec{J}^{\rm IV}} & 0 & \overline{\overline{Z}}^{\vec{J}_{\rm in}^{\rm ES}\vec{M}^{\rm IV}} \\ \overline{\overline{Z}}^{\vec{J}^{\rm IS}\vec{J}_{\rm in}^{\rm ES}} & \overline{\overline{Z}}^{\vec{J}^{\rm IS}\vec{J}^{\rm IS}} & \overline{\overline{Z}}^{\vec{J}^{\rm IS}\vec{J}^{\rm IV}} & 0 & \overline{\overline{Z}}^{\vec{J}^{\rm IS}\vec{M}^{\rm IV}} \\ \overline{\overline{Z}}^{\vec{J}^{\rm IV}\vec{J}_{\rm in}^{\rm ES}} & \overline{\overline{Z}}^{\vec{J}^{\rm IV}\vec{J}^{\rm IS}} & \overline{\overline{Z}}^{\vec{J}^{\rm IV}\vec{J}^{\rm IV}} & 0 & \overline{\overline{Z}}^{\vec{J}^{\rm IV}\vec{M}^{\rm IV}} \\ \overline{\overline{Z}}^{\vec{M}^{\rm IV}\vec{J}_{\rm in}^{\rm ES}} & \overline{\overline{Z}}^{\vec{M}^{\rm IV}\vec{J}^{\rm IS}} & \overline{\overline{Z}}^{\vec{M}^{\rm IV}\vec{J}^{\rm IV}} & 0 & \overline{\overline{Z}}^{\vec{M}^{\rm IV}\vec{M}^{\rm IV}} \end{bmatrix} \tag{C-19a}$$

$$\overline{\overline{\Psi}}_4 = \begin{bmatrix} \overline{\overline{I}}^{\vec{M}_{\rm in}^{\rm ES}} \\ -\overline{\overline{Z}}^{\vec{J}_{\rm in}^{\rm ES}\vec{M}_{\rm in}^{\rm ES}} \\ -\overline{\overline{Z}}^{\vec{J}^{\rm IS}\vec{M}_{\rm in}^{\rm ES}} \\ -\overline{\overline{Z}}^{\vec{J}^{\rm IV}\vec{M}_{\rm in}^{\rm ES}} \\ -\overline{\overline{Z}}^{\vec{M}^{\rm IV}\vec{M}_{\rm in}^{\rm ES}} \end{bmatrix} \tag{C-19b}$$

and

$$\overline{\overline{\Psi}}^{\rm DoJ} = \begin{bmatrix} \overline{\overline{Z}}^{\vec{M}_{\rm in}^{\rm ES}\vec{J}_{\rm in}^{\rm ES}} & \overline{\overline{Z}}^{\vec{M}_{\rm in}^{\rm ES}\vec{J}^{\rm IS}} & \overline{\overline{Z}}^{\vec{M}_{\rm in}^{\rm ES}\vec{J}^{\rm IV}} & \overline{\overline{Z}}^{\vec{M}_{\rm in}^{\rm ES}\vec{M}_{\rm in}^{\rm ES}} & \overline{\overline{Z}}^{\vec{M}_{\rm in}^{\rm ES}\vec{J}^{\rm IV}} \\ \overline{\overline{Z}}^{\vec{J}^{\rm IS}\vec{J}_{\rm in}^{\rm ES}} & \overline{\overline{Z}}^{\vec{J}^{\rm IS}\vec{J}^{\rm IS}} & \overline{\overline{Z}}^{\vec{J}^{\rm IS}\vec{J}^{\rm IV}} & \overline{\overline{Z}}^{\vec{J}^{\rm IS}\vec{M}_{\rm in}^{\rm ES}} & \overline{\overline{Z}}^{\vec{J}^{\rm IS}\vec{M}^{\rm IV}} \\ \overline{\overline{Z}}^{\vec{J}^{\rm IV}\vec{J}_{\rm in}^{\rm ES}} & \overline{\overline{Z}}^{\vec{J}^{\rm IV}\vec{J}^{\rm IS}} & \overline{\overline{Z}}^{\vec{J}^{\rm IV}\vec{J}^{\rm IV}} & \overline{\overline{Z}}^{\vec{J}^{\rm IV}\vec{M}_{\rm in}^{\rm ES}} & \overline{\overline{Z}}^{\vec{J}^{\rm IV}\vec{M}^{\rm IV}} \\ \overline{\overline{Z}}^{\vec{M}^{\rm IV}\vec{J}_{\rm in}^{\rm ES}} & \overline{\overline{Z}}^{\vec{M}^{\rm IV}\vec{J}^{\rm IS}} & \overline{\overline{Z}}^{\vec{M}^{\rm IV}\vec{J}^{\rm IV}} & \overline{\overline{Z}}^{\vec{M}^{\rm IV}\vec{M}_{\rm in}^{\rm ES}} & \overline{\overline{Z}}^{\vec{M}^{\rm IV}\vec{M}^{\rm IV}} \end{bmatrix} \tag{C-20a}$$

$$\overline{\overline{\Psi}}^{\rm DoM} = \begin{bmatrix} \overline{\overline{Z}}^{\vec{J}_{\rm in}^{\rm ES}\vec{J}_{\rm in}^{\rm ES}} & \overline{\overline{Z}}^{\vec{J}_{\rm in}^{\rm ES}\vec{J}^{\rm IS}} & \overline{\overline{Z}}^{\vec{J}_{\rm in}^{\rm ES}\vec{J}^{\rm IV}} & \overline{\overline{Z}}^{\vec{J}_{\rm in}^{\rm ES}\vec{M}_{\rm in}^{\rm ES}} & \overline{\overline{Z}}^{\vec{J}_{\rm in}^{\rm ES}\vec{M}^{\rm IV}} \\ \overline{\overline{Z}}^{\vec{J}^{\rm IS}\vec{J}_{\rm in}^{\rm ES}} & \overline{\overline{Z}}^{\vec{J}^{\rm IS}\vec{J}^{\rm IS}} & \overline{\overline{Z}}^{\vec{J}^{\rm IS}\vec{J}^{\rm IV}} & \overline{\overline{Z}}^{\vec{J}^{\rm IS}\vec{M}_{\rm in}^{\rm ES}} & \overline{\overline{Z}}^{\vec{J}^{\rm IS}\vec{M}^{\rm IV}} \\ \overline{\overline{Z}}^{\vec{J}^{\rm IV}\vec{J}_{\rm in}^{\rm ES}} & \overline{\overline{Z}}^{\vec{J}^{\rm IV}\vec{J}^{\rm IS}} & \overline{\overline{Z}}^{\vec{J}^{\rm IV}\vec{J}^{\rm IV}} & \overline{\overline{Z}}^{\vec{J}^{\rm IV}\vec{M}_{\rm in}^{\rm ES}} & \overline{\overline{Z}}^{\vec{J}^{\rm IV}\vec{M}^{\rm IV}} \\ \overline{\overline{Z}}^{\vec{M}^{\rm IV}\vec{J}_{\rm in}^{\rm ES}} & \overline{\overline{Z}}^{\vec{M}^{\rm IV}\vec{J}^{\rm IS}} & \overline{\overline{Z}}^{\vec{M}^{\rm IV}\vec{J}^{\rm IV}} & \overline{\overline{Z}}^{\vec{M}^{\rm IV}\vec{M}_{\rm in}^{\rm ES}} & \overline{\overline{Z}}^{\vec{M}^{\rm IV}\vec{M}^{\rm IV}} \end{bmatrix} \tag{C-20b}$$

The IPO corresponding to the input port $\mathbb{S}_{\rm in}$ shown in Fig. C-2 can be written as follows:

$$\begin{aligned} P^{\rm in} &= (1/2)\iint_{\mathbb{S}_{\rm in}} \left(\vec{E}\times\vec{H}^{\dagger}\right)\cdot\hat{z}dS \\ &= (1/2)\left\langle \hat{z}\times\vec{J}_{\rm in}^{\rm ES},\vec{M}_{\rm in}^{\rm ES}\right\rangle_{\mathbb{S}_{\rm in}} \\ &= -(1/2)\left\langle \vec{J}_{\rm in}^{\rm ES},\mathcal{E}_0\left(\vec{J}_{\rm in}^{\rm ES}+\vec{J}^{\rm IS}+\vec{J}^{\rm IV},\vec{M}_{\rm in}^{\rm ES}+\vec{M}^{\rm IV}\right)\right\rangle_{\mathbb{S}_{\rm in}^{-}} \\ &= -(1/2)\left\langle \vec{M}_{\rm in}^{\rm ES},\mathcal{H}_0\left(\vec{J}_{\rm in}^{\rm ES}+\vec{J}^{\rm IS}+\vec{J}^{\rm IV},\vec{M}_{\rm in}^{\rm ES}+\vec{M}^{\rm IV}\right)\right\rangle_{\mathbb{S}_{\rm in}^{-}}^{\dagger} \end{aligned} \tag{C- 21}$$

Here, the right-hand side of the first equality is the *field form of IPO*, and the right-hand side of the second equality is the *current form of IPO*, and the right-hand sides of the third and fourth equalities are the *field-current interaction forms of IPO*.

By discretizing IPO (C-21) and utilizing transformation (C-15), we derive the following matrix form of the IPO

$$P^{\rm in} = \bar{a}^{\dagger}\cdot\underbrace{\overline{\overline{T}}^{\dagger}\cdot\overline{\overline{P}}_{\rm AV}^{\rm in}\cdot\overline{\overline{T}}}_{\overline{\overline{P}}^{\rm in}}\cdot\bar{a} \tag{C-22}$$

and the formulation for calculating the quadratic matrix $\overline{\overline{P}}{}_{\text{AV}}^{\text{in}}$ is as follows:

$$\overline{\overline{P}}{}_{\text{AV}}^{\text{in}} = \begin{bmatrix} 0 & 0 & 0 & \overline{\overline{C}}{}^{\vec{J}_{\text{in}}^{\text{ES}}\vec{M}_{\text{in}}^{\text{ES}}} & 0 \\ 0 & 0 & 0 & 0 & 0 \\ 0 & 0 & 0 & 0 & 0 \\ 0 & 0 & 0 & 0 & 0 \\ 0 & 0 & 0 & 0 & 0 \end{bmatrix} \tag{C-23}$$

corresponding to the second equality in IPO (C-21), and

$$\overline{\overline{P}}{}_{\text{AV}}^{\text{in}} = \begin{bmatrix} \overline{\overline{P}}{}^{\vec{J}_{\text{in}}^{\text{ES}}\vec{J}_{\text{in}}^{\text{ES}}} & \overline{\overline{P}}{}^{\vec{J}_{\text{in}}^{\text{ES}}\vec{J}^{\text{IS}}} & \overline{\overline{P}}{}^{\vec{J}_{\text{in}}^{\text{ES}}\vec{J}^{\text{IV}}} & \overline{\overline{P}}{}^{\vec{J}_{\text{in}}^{\text{ES}}\vec{M}_{\text{in}}^{\text{ES}}} & \overline{\overline{P}}{}^{\vec{J}_{\text{in}}^{\text{ES}}\vec{M}^{\text{IV}}} \\ 0 & 0 & 0 & 0 & 0 \\ 0 & 0 & 0 & 0 & 0 \\ 0 & 0 & 0 & 0 & 0 \\ 0 & 0 & 0 & 0 & 0 \end{bmatrix} \tag{C-24a}$$

corresponding to the third equality in IPO (C-21), and

$$\overline{\overline{P}}{}_{\text{AV}}^{\text{in}} = \begin{bmatrix} 0 & 0 & 0 & 0 & 0 \\ 0 & 0 & 0 & 0 & 0 \\ 0 & 0 & 0 & 0 & 0 \\ \overline{\overline{P}}{}^{\vec{M}_{\text{in}}^{\text{ES}}\vec{J}_{\text{in}}^{\text{ES}}} & \overline{\overline{P}}{}^{\vec{M}_{\text{in}}^{\text{ES}}\vec{J}^{\text{IS}}} & \overline{\overline{P}}{}^{\vec{M}_{\text{in}}^{\text{ES}}\vec{J}^{\text{IV}}} & \overline{\overline{P}}{}^{\vec{M}_{\text{in}}^{\text{ES}}\vec{M}_{\text{in}}^{\text{ES}}} & \overline{\overline{P}}{}^{\vec{M}_{\text{in}}^{\text{ES}}\vec{M}^{\text{IV}}} \\ 0 & 0 & 0 & 0 & 0 \end{bmatrix}^{\dagger} \tag{C-24b}$$

corresponding to the fourth equality in IPO (C-21). The elements of the above various sub-matrices are as follows:

$$c_{\xi\zeta}^{\vec{J}_{\text{in}}^{\text{ES}}\vec{M}_{\text{in}}^{\text{ES}}} = (1/2)\left\langle \hat{z}\times\vec{b}_{\xi}^{\vec{J}_{\text{in}}^{\text{ES}}}, \vec{b}_{\zeta}^{\vec{M}_{\text{in}}^{\text{ES}}} \right\rangle_{\mathbb{S}_{\text{in}}} \tag{C-25}$$

and

$$p_{\xi\zeta}^{\vec{J}_{\text{in}}^{\text{ES}}\vec{J}_{\text{in}}^{\text{ES}}} = -(1/2)\left\langle \vec{b}_{\xi}^{\vec{J}_{\text{in}}^{\text{ES}}}, -j\omega\mu_0\mathcal{L}_0\left(\vec{b}_{\zeta}^{\vec{J}_{\text{in}}^{\text{ES}}}\right) \right\rangle_{\mathbb{S}_{\text{in}}} \tag{C-26a}$$

$$p_{\xi\zeta}^{\vec{J}_{\text{in}}^{\text{ES}}\vec{J}^{\text{IS}}} = -(1/2)\left\langle \vec{b}_{\xi}^{\vec{J}_{\text{in}}^{\text{ES}}}, -j\omega\mu_0\mathcal{L}_0\left(\vec{b}_{\zeta}^{\vec{J}^{\text{IS}}}\right) \right\rangle_{\mathbb{S}_{\text{in}}} \tag{C-26b}$$

$$p_{\xi\zeta}^{\vec{J}_{\text{in}}^{\text{ES}}\vec{J}^{\text{IV}}} = -(1/2)\left\langle \vec{b}_{\xi}^{\vec{J}_{\text{in}}^{\text{ES}}}, -j\omega\mu_0\mathcal{L}_0\left(\vec{b}_{\zeta}^{\vec{J}^{\text{IV}}}\right) \right\rangle_{\mathbb{S}_{\text{in}}} \tag{C-26c}$$

$$p_{\xi\zeta}^{\vec{J}_{\text{in}}^{\text{ES}}\vec{M}_{\text{in}}^{\text{ES}}} = -(1/2)\left\langle \vec{b}_{\xi}^{\vec{J}_{\text{in}}^{\text{ES}}}, \hat{z}\times\frac{1}{2}\vec{b}_{\zeta}^{\vec{M}_{\text{in}}^{\text{ES}}} - \text{P.V.}\,\mathcal{K}_0\left(\vec{b}_{\zeta}^{\vec{M}_{\text{in}}^{\text{ES}}}\right) \right\rangle_{\mathbb{S}_{\text{in}}} \tag{C-26d}$$

$$p_{\xi\zeta}^{\vec{J}_{\text{in}}^{\text{ES}}\vec{M}^{\text{IV}}} = -(1/2)\left\langle \vec{b}_{\xi}^{\vec{J}_{\text{in}}^{\text{ES}}}, -\mathcal{K}_0\left(\vec{b}_{\zeta}^{\vec{M}^{\text{IV}}}\right) \right\rangle_{\mathbb{S}_{\text{in}}} \tag{C-26e}$$

and

$$p_{\xi\zeta}^{\vec{M}_{\text{in}}^{\text{ES}}\vec{J}_{\text{in}}^{\text{ES}}} = -(1/2)\left\langle \vec{b}_{\xi}^{\vec{M}_{\text{in}}^{\text{ES}}}, \frac{1}{2}\vec{b}_{\zeta}^{\vec{J}_{\text{in}}^{\text{ES}}}\times\hat{z} + \text{P.V.}\,\mathcal{K}_0\left(\vec{b}_{\zeta}^{\vec{J}_{\text{in}}^{\text{ES}}}\right) \right\rangle_{\mathbb{S}_{\text{in}}} \tag{C-26f}$$

$$p_{\xi\zeta}^{\vec{M}_{\text{in}}^{\text{ES}}\vec{J}^{\text{IS}}} = -(1/2)\left\langle \vec{b}_{\xi}^{\vec{M}_{\text{in}}^{\text{ES}}}, \mathcal{K}_0\left(\vec{b}_{\zeta}^{\vec{J}^{\text{IS}}}\right) \right\rangle_{\mathbb{S}_{\text{in}}} \tag{C-26g}$$

$$p_{\xi\zeta}^{\vec{M}_{\text{in}}^{\text{ES}}\vec{J}^{\text{IV}}} = -\left(1/2\right)\left\langle \vec{b}_{\xi}^{\vec{M}_{\text{in}}^{\text{ES}}}, \mathcal{K}_0\left(\vec{b}_{\zeta}^{\vec{J}^{\text{IV}}}\right)\right\rangle_{\mathbb{S}_{\text{in}}} \tag{C-26h}$$

$$p_{\xi\zeta}^{\vec{M}_{\text{in}}^{\text{ES}}\vec{M}_{\text{in}}^{\text{ES}}} = -\left(1/2\right)\left\langle \vec{b}_{\xi}^{\vec{M}_{\text{in}}^{\text{ES}}}, -j\omega\varepsilon_0\mathcal{L}_0\left(\vec{b}_{\zeta}^{\vec{M}_{\text{in}}^{\text{ES}}}\right)\right\rangle_{\mathbb{S}_{\text{in}}} \tag{C-26i}$$

$$p_{\xi\zeta}^{\vec{M}_{\text{in}}^{\text{ES}}\vec{M}^{\text{IV}}} = -\left(1/2\right)\left\langle \vec{b}_{\xi}^{\vec{M}_{\text{in}}^{\text{ES}}}, -j\omega\varepsilon_0\mathcal{L}_0\left(\vec{b}_{\zeta}^{\vec{M}^{\text{IV}}}\right)\right\rangle_{\mathbb{S}_{\text{in}}} \tag{C-26j}$$

where symbol "$\text{P.V.}\mathcal{K}_0$" represents the *principal value* of operator $\mathcal{K}_0$.

The *input-power-decoupled modes* (*IP-DMs*) in modal space can be derived from solving the modal decoupling equation $\overline{\overline{P}}_{-}^{\text{in}} \cdot \bar{\alpha}_{\xi} = \theta_{\xi}\, \overline{\overline{P}}_{+}^{\text{in}} \cdot \bar{\alpha}_{\xi}$ defined on modal space, where $\overline{\overline{P}}_{+}^{\text{in}}$ and $\overline{\overline{P}}_{-}^{\text{in}}$ are the *positive and negative Hermitian parts* of the matrix $\overline{\overline{P}}^{\text{in}}$ given in Eq. (C-22). If some derived modes $\{\bar{\alpha}_1, \bar{\alpha}_2, \cdots, \bar{\alpha}_d\}$ are *d*-order degenerate, then the *Gram-Schmidt orthogonalization process* given in previous Sec. 3.2.4.1 is necessary, and it is not repeated here.

The IP-DMs constructed above satisfy the following decoupling relation

$$\left(1 + j\,\theta_{\xi}\right)\delta_{\xi\zeta} = \left(1/2\right)\iint_{\mathbb{S}_{\text{in}}}\left(\vec{E}_{\zeta} \times \vec{H}_{\xi}^{\dagger}\right)\cdot\hat{z}dS = \left(1/2\right)\left\langle \hat{z} \times \vec{J}_{\text{in};\xi}^{\text{ES}}, \vec{M}_{\text{in};\zeta}^{\text{ES}}\right\rangle_{\mathbb{S}_{\text{in}}} \tag{C-27}$$

and the relation implies that the IP-DMs don't have net energy coupling in one period. By employing the decoupling relation, we have the following *Parseval's identity*

$$\sum_{\xi}\left|c_{\xi}\right|^2 = \left(1/T\right)\int_{t_0}^{t_0+T}\left[\iint_{\mathbb{S}_{\text{in}}}\left(\vec{\mathcal{E}} \times \vec{\mathcal{H}}\right)\cdot\hat{z}dS\right]dt \tag{C-28}$$

in which $\{\vec{\mathcal{E}}, \vec{\mathcal{H}}\}$ are the time-domain fields, and the expansion coefficients have expression $c_{\xi} = -(1/2) < \vec{J}_{\text{in};\xi}^{\text{ES}}, \vec{E} >_{\mathbb{S}_{\text{in}}} \big/ (1 + j\theta_{\xi}) = -(1/2) < \vec{H}, \vec{M}_{\text{in};\xi}^{\text{ES}} >_{\mathbb{S}_{\text{in}}} \big/ (1 + j\theta_{\xi})$, and $\{\vec{E}, \vec{H}\}$ are some previously known fields distributing on input port $\mathbb{S}_{\text{in}}$.

Similarly to Secs. 3.2~3.6, the *modal input impedance* and *modal input admittance* can be defined. Employing the impedance and admittance, the *traveling-wave-type IP-DMs* can be effectively recognized as doing in Sec. 3.2.4.3, and the corresponding *cut-off frequencies* can be easily calculated like Eq. (3-77).

## C3 Surface-Volume Formulation of the PTT-Based DMT for the Augmented Tra-antenna Discussed in Sec. 6.3

The process to establish the surface-volume formulation of the PTT-based DMT for the material augmented tra-antenna discussed in Sec. 6.3 is similar to the process given in the following App. C4, and it will not be explicitly provided here for condensing the length of this report.

## C4 Surface-Volume Formulation of the PTT-Based DMT for the Augmented Tra-antennas Discussed in Secs. 6.4 and 6.5

The topological structure of the *augmented tra-antenna* discussed in this App. C4 is the same as the one shown in Fig. 6-41 / 6-54.

If the equivalent surface currents distributing on $\mathbb{S}^{\mathrm{G}\rightleftharpoons\mathrm{A}}$ are denoted as $\{\vec{J}^{\mathrm{G}\rightleftharpoons\mathrm{A}},\vec{M}^{\mathrm{G}\rightleftharpoons\mathrm{A}}\}$, and the induced volume currents distributing on $\mathbb{V}_1^{\mathrm{A}}$ and $\mathbb{V}_2^{\mathrm{A}}$ are denoted as $\{\vec{J}_1^{\mathrm{IV}},\vec{M}_1^{\mathrm{IV}}\}$ and $\{\vec{J}_2^{\mathrm{IV}},\vec{M}_2^{\mathrm{IV}}\}$ respectively, and the induced surface electric current distributing on $\mathbb{S}^{\mathrm{ele}}$ is denoted as $\vec{J}^{\mathrm{IS}}$, then the field distributing on $\mathbb{V}_1^{\mathrm{A}}\bigcup\mathbb{V}_2^{\mathrm{A}}\bigcup\mathbb{V}^{\mathrm{F}}$ can be expressed as follows:

$$\vec{F}\left(\vec{r}\right)=\mathcal{F}_0\left(\vec{J}^{\mathrm{G}\rightleftharpoons\mathrm{A}}+\vec{J}_1^{\mathrm{IV}}+\vec{J}_2^{\mathrm{IV}}+\vec{J}^{\mathrm{IS}},\vec{M}^{\mathrm{G}\rightleftharpoons\mathrm{A}}+\vec{M}_1^{\mathrm{IV}}+\vec{M}_2^{\mathrm{IV}}\right)\ ,\ \vec{r}\in\mathbb{V}_1^{\mathrm{A}}\bigcup\mathbb{V}_2^{\mathrm{A}}\bigcup\mathbb{V}^{\mathrm{F}} \quad \text{(C-29)}$$

where $\vec{F}=\vec{E}\,/\,\vec{H}$, and correspondingly $\mathcal{F}_0=\mathcal{E}_0\,/\,\mathcal{H}_0$, and the operator is the same as the one used in Sec. 6.4. The currents $\{\vec{J}^{\mathrm{G}\rightleftharpoons\mathrm{A}},\vec{M}^{\mathrm{G}\rightleftharpoons\mathrm{A}}\}$ and fields $\{\vec{E},\vec{H}\}$ in Eq. (C-29) satisfy the following relations

$$\hat{n}^{\rightarrow\mathrm{A}}\times\left[\vec{H}\left(\vec{r}^{\,\mathrm{A}}\right)\right]_{\vec{r}^{\mathrm{A}}\to\vec{r}}=\vec{J}^{\mathrm{G}\rightleftharpoons\mathrm{A}}\left(\vec{r}\right)\quad,\quad\vec{r}\in\mathbb{S}^{\mathrm{G}\rightleftharpoons\mathrm{A}} \quad \text{(C-30a)}$$

$$\left[\vec{E}\left(\vec{r}^{\,\mathrm{A}}\right)\right]_{\vec{r}^{\mathrm{A}}\to\vec{r}}\times\hat{n}^{\rightarrow\mathrm{A}}=\vec{M}^{\mathrm{G}\rightleftharpoons\mathrm{A}}\left(\vec{r}\right)\quad,\quad\vec{r}\in\mathbb{S}^{\mathrm{G}\rightleftharpoons\mathrm{A}} \quad \text{(C-30b)}$$

In the above Eq. (C-30), point $\vec{r}^{\,\mathrm{A}}$ belongs to the region occupied by the tra-antenna, and $\vec{r}^{\,\mathrm{A}}$ approaches the point $\vec{r}$ on $\mathbb{S}^{\mathrm{G}\rightleftharpoons\mathrm{A}}$; $\hat{n}^{\rightarrow\mathrm{A}}$ is the normal direction of $\mathbb{S}^{\mathrm{G}\rightleftharpoons\mathrm{A}}$, and $\hat{n}^{\rightarrow\mathrm{A}}$ points to the tra-antenna. The currents $\{\vec{J}_1^{\mathrm{IV}},\vec{M}_1^{\mathrm{IV}}\}\,\&\,\{\vec{J}_2^{\mathrm{IV}},\vec{M}_2^{\mathrm{IV}}\}$ and fields $\{\vec{E},\vec{H}\}$ in Eq. (C-29) satisfy the following relations

$$\vec{E}\left(\vec{r}\right)=\left(j\omega\Delta\vec{\vec{\varepsilon}}_1^{\,\mathrm{c}}\right)^{-1}\cdot\vec{J}_1^{\mathrm{IV}}\quad,\quad\vec{r}\in\mathbb{V}_1^{\mathrm{A}} \quad \text{(C-31a)}$$

$$\vec{H}\left(\vec{r}\right)=\left(j\omega\Delta\vec{\vec{\mu}}_1\right)^{-1}\cdot\vec{M}_1^{\mathrm{IV}}\quad,\quad\vec{r}\in\mathbb{V}_1^{\mathrm{A}} \quad \text{(C-31b)}$$

and

$$\vec{E}\left(\vec{r}\right)=\left(j\omega\Delta\vec{\vec{\varepsilon}}_2^{\,\mathrm{c}}\right)^{-1}\cdot\vec{J}_2^{\mathrm{IV}}\quad,\quad\vec{r}\in\mathbb{V}_2^{\mathrm{A}} \quad \text{(C-32a)}$$

$$\vec{H}\left(\vec{r}\right)=\left(j\omega\Delta\vec{\vec{\mu}}_2\right)^{-1}\cdot\vec{M}_2^{\mathrm{IV}}\quad,\quad\vec{r}\in\mathbb{V}_2^{\mathrm{A}} \quad \text{(C-32b)}$$

In the above Eqs. (C-31) and (C-32), $\Delta\vec{\vec{\varepsilon}}_{1/2}^{\,\mathrm{c}}=(\vec{\vec{\varepsilon}}_{1/2}-j\,\vec{\vec{\sigma}}_{1/2}/\omega)-\vec{\vec{I}}\,\varepsilon_0$, and $\Delta\vec{\vec{\mu}}_{1/2}=\vec{\vec{\mu}}_{1/2}-\vec{\vec{I}}\,\mu_0$.

Substituting Eq. (C-29) into Eqs. (C-30a) and (C-30b), we obtain the following integral equations

$$\left[\mathcal{H}_0\left(\vec{J}^{\mathrm{G}\rightleftharpoons\mathrm{A}}+\vec{J}_1^{\mathrm{IV}}+\vec{J}_2^{\mathrm{IV}}+\vec{J}^{\mathrm{IS}},\vec{M}^{\mathrm{G}\rightleftharpoons\mathrm{A}}+\vec{M}_1^{\mathrm{IV}}+\vec{M}_2^{\mathrm{IV}}\right)\right]_{\vec{r}^{\mathrm{A}}\to\vec{r}}^{\tan}=\vec{J}^{\mathrm{G}\rightleftharpoons\mathrm{A}}\times\hat{n}^{\rightarrow\mathrm{A}}\quad,\quad\vec{r}\in\mathbb{S}^{\mathrm{G}\rightleftharpoons\mathrm{A}} \quad \text{(C-33a)}$$

$$\left[\mathcal{E}_0\left(\vec{J}^{\,\mathrm{G}\rightleftharpoons\mathrm{A}}+\vec{J}_1^{\,\mathrm{IV}}+\vec{J}_2^{\,\mathrm{IV}}+\vec{J}^{\,\mathrm{IS}},\vec{M}^{\,\mathrm{G}\rightleftharpoons\mathrm{A}}+\vec{M}_1^{\,\mathrm{IV}}+\vec{M}_2^{\,\mathrm{IV}}\right)\right]_{\vec{r}^{\,\mathrm{A}}\to\vec{r}}^{\tan} = \hat{n}^{\to\mathrm{A}}\times\vec{M}^{\,\mathrm{G}\rightleftharpoons\mathrm{A}} , \quad \vec{r}\in\mathbb{S}^{\mathrm{G}\rightleftharpoons\mathrm{A}} \quad \text{(C-33b)}$$

about currents $\{\vec{J}^{\,\mathrm{G}\rightleftharpoons\mathrm{A}},\vec{M}^{\,\mathrm{G}\rightleftharpoons\mathrm{A}}\}$, $\{\vec{J}_1^{\,\mathrm{IV}},\vec{M}_1^{\,\mathrm{IV}}\}$, $\{\vec{J}_2^{\,\mathrm{IV}},\vec{M}_2^{\,\mathrm{IV}}\}$, and $\vec{J}^{\,\mathrm{IS}}$, where the superscript "tan" represents the tangential component of the field. Substituting Eq. (C-29) into Eqs. (C-31a)&(C-31b) and (C-32a)&(C-32b), we obtain the following integral equations

$$\mathcal{E}_0\left(\vec{J}^{\,\mathrm{G}\rightleftharpoons\mathrm{A}}+\vec{J}_1^{\,\mathrm{IV}}+\vec{J}_2^{\,\mathrm{IV}}+\vec{J}^{\,\mathrm{IS}},\vec{M}^{\,\mathrm{G}\rightleftharpoons\mathrm{A}}+\vec{M}_1^{\,\mathrm{IV}}+\vec{M}_2^{\,\mathrm{IV}}\right) = \left(j\omega\Delta\ddot{\varepsilon}_1^{\mathrm{c}}\right)^{-1}\cdot\vec{J}_1^{\,\mathrm{IV}} , \quad \vec{r}\in\mathbb{V}_1^{\mathrm{A}} \quad \text{(C-34a)}$$

$$\mathcal{H}_0\left(\vec{J}^{\,\mathrm{G}\rightleftharpoons\mathrm{A}}+\vec{J}_1^{\,\mathrm{IV}}+\vec{J}_2^{\,\mathrm{IV}}+\vec{J}^{\,\mathrm{IS}},\vec{M}^{\,\mathrm{G}\rightleftharpoons\mathrm{A}}+\vec{M}_1^{\,\mathrm{IV}}+\vec{M}_2^{\,\mathrm{IV}}\right) = \left(j\omega\Delta\ddot{\mu}_1\right)^{-1}\cdot\vec{M}_1^{\,\mathrm{IV}} , \quad \vec{r}\in\mathbb{V}_1^{\mathrm{A}} \quad \text{(C-34b)}$$

and

$$\mathcal{E}_0\left(\vec{J}^{\,\mathrm{G}\rightleftharpoons\mathrm{A}}+\vec{J}_1^{\,\mathrm{IV}}+\vec{J}_2^{\,\mathrm{IV}}+\vec{J}^{\,\mathrm{IS}},\vec{M}^{\,\mathrm{G}\rightleftharpoons\mathrm{A}}+\vec{M}_1^{\,\mathrm{IV}}+\vec{M}_2^{\,\mathrm{IV}}\right) = \left(j\omega\Delta\ddot{\varepsilon}_2^{\mathrm{c}}\right)^{-1}\cdot\vec{J}_2^{\,\mathrm{IV}} , \quad \vec{r}\in\mathbb{V}_2^{\mathrm{A}} \quad \text{(C-35a)}$$

$$\mathcal{H}_0\left(\vec{J}^{\,\mathrm{G}\rightleftharpoons\mathrm{A}}+\vec{J}_1^{\,\mathrm{IV}}+\vec{J}_2^{\,\mathrm{IV}}+\vec{J}^{\,\mathrm{IS}},\vec{M}^{\,\mathrm{G}\rightleftharpoons\mathrm{A}}+\vec{M}_1^{\,\mathrm{IV}}+\vec{M}_2^{\,\mathrm{IV}}\right) = \left(j\omega\Delta\ddot{\mu}_2\right)^{-1}\cdot\vec{M}_2^{\,\mathrm{IV}} , \quad \vec{r}\in\mathbb{V}_2^{\mathrm{A}} \quad \text{(C-35b)}$$

about currents $\{\vec{J}^{\,\mathrm{G}\rightleftharpoons\mathrm{A}},\vec{M}^{\,\mathrm{G}\rightleftharpoons\mathrm{A}}\}$, $\{\vec{J}_1^{\,\mathrm{IV}},\vec{M}_1^{\,\mathrm{IV}}\}$, $\{\vec{J}_2^{\,\mathrm{IV}},\vec{M}_2^{\,\mathrm{IV}}\}$, and $\vec{J}^{\,\mathrm{IS}}$. Based on Eq. (C-29) and the homogeneous tangential electric field boundary condition on $\mathbb{S}^{\mathrm{ele}}$, we have the following electric field integral equation

$$\left[\mathcal{E}_0\left(\vec{J}^{\,\mathrm{G}\rightleftharpoons\mathrm{A}}+\vec{J}_1^{\,\mathrm{IV}}+\vec{J}_2^{\,\mathrm{IV}}+\vec{J}^{\,\mathrm{IS}},\vec{M}^{\,\mathrm{G}\rightleftharpoons\mathrm{A}}+\vec{M}_1^{\,\mathrm{IV}}+\vec{M}_2^{\,\mathrm{IV}}\right)\right]^{\tan} = 0 , \quad \vec{r}\in\mathbb{S}^{\mathrm{ele}} \quad \text{(C-36)}$$

about currents $\{\vec{J}^{\,\mathrm{G}\rightleftharpoons\mathrm{A}},\vec{M}^{\,\mathrm{G}\rightleftharpoons\mathrm{A}}\}$, $\{\vec{J}_1^{\,\mathrm{IV}},\vec{M}_1^{\,\mathrm{IV}}\}$, $\{\vec{J}_2^{\,\mathrm{IV}},\vec{M}_2^{\,\mathrm{IV}}\}$, and $\vec{J}^{\,\mathrm{IS}}$.

If the above-mentioned currents are expanded in terms of some proper basis functions, and the Eqs. (C-33a)~(C-36) are tested with $\{\vec{b}_\xi^{\vec{M}^{\mathrm{G}\rightleftharpoons\mathrm{A}}}\}$, $\{\vec{b}_\xi^{\vec{J}^{\mathrm{G}\rightleftharpoons\mathrm{A}}}\}$, $\{\vec{b}_\xi^{\vec{J}_1^{\mathrm{IV}}}\}$, $\{\vec{b}_\xi^{\vec{M}_1^{\mathrm{IV}}}\}$, $\{\vec{b}_\xi^{\vec{J}_2^{\mathrm{IV}}}\}$, $\{\vec{b}_\xi^{\vec{M}_2^{\mathrm{IV}}}\}$, and $\{\vec{b}_\xi^{\vec{J}^{\mathrm{IS}}}\}$ respectively, then the equations are immediately discretized into the following *matrix equations*

$$\begin{aligned} 0 \;=\;& \overline{\overline{Z}}^{\vec{M}^{\mathrm{G}\rightleftharpoons\mathrm{A}}\vec{J}^{\mathrm{G}\rightleftharpoons\mathrm{A}}}\cdot\overline{a}^{\vec{J}^{\mathrm{G}\rightleftharpoons\mathrm{A}}} + \overline{\overline{Z}}^{\vec{M}^{\mathrm{G}\rightleftharpoons\mathrm{A}}\vec{J}_1^{\mathrm{IV}}}\cdot\overline{a}^{\vec{J}_1^{\mathrm{IV}}} + \overline{\overline{Z}}^{\vec{M}^{\mathrm{G}\rightleftharpoons\mathrm{A}}\vec{J}_2^{\mathrm{IV}}}\cdot\overline{a}^{\vec{J}_2^{\mathrm{IV}}} + \overline{\overline{Z}}^{\vec{M}^{\mathrm{G}\rightleftharpoons\mathrm{A}}\vec{J}^{\mathrm{IS}}}\cdot\overline{a}^{\vec{J}^{\mathrm{IS}}} \\ &+ \overline{\overline{Z}}^{\vec{M}^{\mathrm{G}\rightleftharpoons\mathrm{A}}\vec{M}^{\mathrm{G}\rightleftharpoons\mathrm{A}}}\cdot\overline{a}^{\vec{M}^{\mathrm{G}\rightleftharpoons\mathrm{A}}} + \overline{\overline{Z}}^{\vec{M}^{\mathrm{G}\rightleftharpoons\mathrm{A}}\vec{M}_1^{\mathrm{IV}}}\cdot\overline{a}^{\vec{M}_1^{\mathrm{IV}}} + \overline{\overline{Z}}^{\vec{M}^{\mathrm{G}\rightleftharpoons\mathrm{A}}\vec{M}_2^{\mathrm{IV}}}\cdot\overline{a}^{\vec{M}_2^{\mathrm{IV}}} \qquad \text{(6-37a)} \end{aligned}$$

$$\begin{aligned} 0 \;=\;& \overline{\overline{Z}}^{\vec{J}^{\mathrm{G}\rightleftharpoons\mathrm{A}}\vec{J}^{\mathrm{G}\rightleftharpoons\mathrm{A}}}\cdot\overline{a}^{\vec{J}^{\mathrm{G}\rightleftharpoons\mathrm{A}}} + \overline{\overline{Z}}^{\vec{J}^{\mathrm{G}\rightleftharpoons\mathrm{A}}\vec{J}_1^{\mathrm{IV}}}\cdot\overline{a}^{\vec{J}_1^{\mathrm{IV}}} + \overline{\overline{Z}}^{\vec{J}^{\mathrm{G}\rightleftharpoons\mathrm{A}}\vec{J}_2^{\mathrm{IV}}}\cdot\overline{a}^{\vec{J}_2^{\mathrm{IV}}} + \overline{\overline{Z}}^{\vec{J}^{\mathrm{G}\rightleftharpoons\mathrm{A}}\vec{J}^{\mathrm{IS}}}\cdot\overline{a}^{\vec{J}^{\mathrm{IS}}} \\ &+ \overline{\overline{Z}}^{\vec{J}^{\mathrm{G}\rightleftharpoons\mathrm{A}}\vec{M}^{\mathrm{G}\rightleftharpoons\mathrm{A}}}\cdot\overline{a}^{\vec{M}^{\mathrm{G}\rightleftharpoons\mathrm{A}}} + \overline{\overline{Z}}^{\vec{J}^{\mathrm{G}\rightleftharpoons\mathrm{A}}\vec{M}_1^{\mathrm{IV}}}\cdot\overline{a}^{\vec{M}_1^{\mathrm{IV}}} + \overline{\overline{Z}}^{\vec{J}^{\mathrm{G}\rightleftharpoons\mathrm{A}}\vec{M}_2^{\mathrm{IV}}}\cdot\overline{a}^{\vec{M}_2^{\mathrm{IV}}} \qquad \text{(6-37b)} \end{aligned}$$

and

$$\begin{aligned} 0 \;=\;& \overline{\overline{Z}}^{\vec{J}_1^{\mathrm{IV}}\vec{J}^{\mathrm{G}\rightleftharpoons\mathrm{A}}}\cdot\overline{a}^{\vec{J}^{\mathrm{G}\rightleftharpoons\mathrm{A}}} + \overline{\overline{Z}}^{\vec{J}_1^{\mathrm{IV}}\vec{J}_1^{\mathrm{IV}}}\cdot\overline{a}^{\vec{J}_1^{\mathrm{IV}}} + \overline{\overline{Z}}^{\vec{J}_1^{\mathrm{IV}}\vec{J}_2^{\mathrm{IV}}}\cdot\overline{a}^{\vec{J}_2^{\mathrm{IV}}} + \overline{\overline{Z}}^{\vec{J}_1^{\mathrm{IV}}\vec{J}^{\mathrm{IS}}}\cdot\overline{a}^{\vec{J}^{\mathrm{IS}}} + \overline{\overline{Z}}^{\vec{J}_1^{\mathrm{IV}}\vec{M}^{\mathrm{G}\rightleftharpoons\mathrm{A}}}\cdot\overline{a}^{\vec{M}^{\mathrm{G}\rightleftharpoons\mathrm{A}}} \\ &+ \overline{\overline{Z}}^{\vec{J}_1^{\mathrm{IV}}\vec{M}_1^{\mathrm{IV}}}\cdot\overline{a}^{\vec{M}_1^{\mathrm{IV}}} + \overline{\overline{Z}}^{\vec{J}_1^{\mathrm{IV}}\vec{M}_2^{\mathrm{IV}}}\cdot\overline{a}^{\vec{M}_2^{\mathrm{IV}}} \qquad \text{(6-38a)} \end{aligned}$$

$$\begin{aligned} 0 \;=\;& \overline{\overline{Z}}^{\vec{M}_1^{\mathrm{IV}}\vec{J}^{\mathrm{G}\rightleftharpoons\mathrm{A}}}\cdot\overline{a}^{\vec{J}^{\mathrm{G}\rightleftharpoons\mathrm{A}}} + \overline{\overline{Z}}^{\vec{M}_1^{\mathrm{IV}}\vec{J}_1^{\mathrm{IV}}}\cdot\overline{a}^{\vec{J}_1^{\mathrm{IV}}} + \overline{\overline{Z}}^{\vec{M}_1^{\mathrm{IV}}\vec{J}_2^{\mathrm{IV}}}\cdot\overline{a}^{\vec{J}_2^{\mathrm{IV}}} + \overline{\overline{Z}}^{\vec{M}_1^{\mathrm{IV}}\vec{J}^{\mathrm{IS}}}\cdot\overline{a}^{\vec{J}^{\mathrm{IS}}} + \overline{\overline{Z}}^{\vec{M}_1^{\mathrm{IV}}\vec{M}^{\mathrm{G}\rightleftharpoons\mathrm{A}}}\cdot\overline{a}^{\vec{M}^{\mathrm{G}\rightleftharpoons\mathrm{A}}} \\ &+ \overline{\overline{Z}}^{\vec{M}_1^{\mathrm{IV}}\vec{M}_1^{\mathrm{IV}}}\cdot\overline{a}^{\vec{M}_1^{\mathrm{IV}}} + \overline{\overline{Z}}^{\vec{M}_1^{\mathrm{IV}}\vec{M}_2^{\mathrm{IV}}}\cdot\overline{a}^{\vec{M}_2^{\mathrm{IV}}} \qquad \text{(6-38b)} \end{aligned}$$

and

$$0 = \overline{\overline{Z}}^{\vec{J}_2^{\text{IV}}\vec{J}^{\text{G}\rightleftharpoons\text{A}}} \cdot \overline{a}^{\vec{J}^{\text{G}\rightleftharpoons\text{A}}} + \overline{\overline{Z}}^{\vec{J}_2^{\text{IV}}\vec{J}_1^{\text{IV}}} \cdot \overline{a}^{\vec{J}_1^{\text{IV}}} + \overline{\overline{Z}}^{\vec{J}_2^{\text{IV}}\vec{J}_2^{\text{IV}}} \cdot \overline{a}^{\vec{J}_2^{\text{IV}}} + \overline{\overline{Z}}^{\vec{J}_2^{\text{IV}}\vec{J}^{\text{IS}}} \cdot \overline{a}^{\vec{J}^{\text{IS}}} + \overline{\overline{Z}}^{\vec{J}_2^{\text{IV}}\vec{M}^{\text{G}\rightleftharpoons\text{A}}} \cdot \overline{a}^{\vec{M}^{\text{G}\rightleftharpoons\text{A}}} + \overline{\overline{Z}}^{\vec{J}_2^{\text{IV}}\vec{M}_1^{\text{IV}}} \cdot \overline{a}^{\vec{M}_1^{\text{IV}}} + \overline{\overline{Z}}^{\vec{J}_2^{\text{IV}}\vec{M}_2^{\text{IV}}} \cdot \overline{a}^{\vec{M}_2^{\text{IV}}} \quad (6\text{-}39\text{a})$$

$$0 = \overline{\overline{Z}}^{\vec{M}_2^{\text{IV}}\vec{J}^{\text{G}\rightleftharpoons\text{A}}} \cdot \overline{a}^{\vec{J}^{\text{G}\rightleftharpoons\text{A}}} + \overline{\overline{Z}}^{\vec{M}_2^{\text{IV}}\vec{J}_1^{\text{IV}}} \cdot \overline{a}^{\vec{J}_1^{\text{IV}}} + \overline{\overline{Z}}^{\vec{M}_2^{\text{IV}}\vec{J}_2^{\text{IV}}} \cdot \overline{a}^{\vec{J}_2^{\text{IV}}} + \overline{\overline{Z}}^{\vec{M}_2^{\text{IV}}\vec{J}^{\text{IS}}} \cdot \overline{a}^{\vec{J}^{\text{IS}}} + \overline{\overline{Z}}^{\vec{M}_2^{\text{IV}}\vec{M}^{\text{G}\rightleftharpoons\text{A}}} \cdot \overline{a}^{\vec{M}^{\text{G}\rightleftharpoons\text{A}}} + \overline{\overline{Z}}^{\vec{M}_2^{\text{IV}}\vec{M}_1^{\text{IV}}} \cdot \overline{a}^{\vec{M}_1^{\text{IV}}} + \overline{\overline{Z}}^{\vec{M}_2^{\text{IV}}\vec{M}_2^{\text{IV}}} \cdot \overline{a}^{\vec{M}_2^{\text{IV}}} \quad (6\text{-}39\text{b})$$

and

$$0 = \overline{\overline{Z}}^{\vec{J}^{\text{IS}}\vec{J}^{\text{G}\rightleftharpoons\text{A}}} \cdot \overline{a}^{\vec{J}^{\text{G}\rightleftharpoons\text{A}}} + \overline{\overline{Z}}^{\vec{J}^{\text{IS}}\vec{J}_1^{\text{IV}}} \cdot \overline{a}^{\vec{J}_1^{\text{IV}}} + \overline{\overline{Z}}^{\vec{J}^{\text{IS}}\vec{J}_2^{\text{IV}}} \cdot \overline{a}^{\vec{J}_2^{\text{IV}}} + \overline{\overline{Z}}^{\vec{J}^{\text{IS}}\vec{J}^{\text{IS}}} \cdot \overline{a}^{\vec{J}^{\text{IS}}} + \overline{\overline{Z}}^{\vec{J}^{\text{IS}}\vec{M}^{\text{G}\rightleftharpoons\text{A}}} \cdot \overline{a}^{\vec{M}^{\text{G}\rightleftharpoons\text{A}}} + \overline{\overline{Z}}^{\vec{J}^{\text{IS}}\vec{M}_1^{\text{IV}}} \cdot \overline{a}^{\vec{M}_1^{\text{IV}}} + \overline{\overline{Z}}^{\vec{J}^{\text{IS}}\vec{M}_2^{\text{IV}}} \cdot \overline{a}^{\vec{M}_2^{\text{IV}}} \quad (6\text{-}40)$$

about vectors $\{\overline{a}^{\vec{J}^{\text{G}\rightleftharpoons\text{A}}}, \overline{a}^{\vec{M}^{\text{G}\rightleftharpoons\text{A}}}\}$, $\{\overline{a}^{\vec{J}_1^{\text{IV}}}, \overline{a}^{\vec{M}_1^{\text{IV}}}\}$, $\{\overline{a}^{\vec{J}_2^{\text{IV}}}, \overline{a}^{\vec{M}_2^{\text{IV}}}\}$, and $\overline{a}^{\vec{J}^{\text{IS}}}$.

The formulations used to calculate the elements of the matrices in Eq. (C-37a) are as follows:

$$z_{\xi\zeta}^{\vec{M}^{\text{G}\rightleftharpoons\text{A}}\vec{J}^{\text{G}\rightleftharpoons\text{A}}} = \left\langle \vec{b}_{\xi}^{\vec{M}^{\text{G}\rightleftharpoons\text{A}}}, \hat{n}^{\rightarrow\text{A}} \times \frac{1}{2}\vec{b}_{\zeta}^{\vec{J}^{\text{G}\rightleftharpoons\text{A}}} + \text{P.V.}\mathcal{K}_0\left(\vec{b}_{\zeta}^{\vec{J}^{\text{G}\rightleftharpoons\text{A}}}\right) \right\rangle_{\mathbb{S}^{\text{G}\rightleftharpoons\text{A}}} \quad \text{(C-41a)}$$

$$z_{\xi\zeta}^{\vec{M}^{\text{G}\rightleftharpoons\text{A}}\vec{J}_1^{\text{IV}}} = \left\langle \vec{b}_{\xi}^{\vec{M}^{\text{G}\rightleftharpoons\text{A}}}, \mathcal{K}_0\left(\vec{b}_{\zeta}^{\vec{J}_1^{\text{IV}}}\right) \right\rangle_{\mathbb{S}^{\text{G}\rightleftharpoons\text{A}}} \quad \text{(C-41b)}$$

$$z_{\xi\zeta}^{\vec{M}^{\text{G}\rightleftharpoons\text{A}}\vec{J}_2^{\text{IV}}} = \left\langle \vec{b}_{\xi}^{\vec{M}^{\text{G}\rightleftharpoons\text{A}}}, \mathcal{K}_0\left(\vec{b}_{\zeta}^{\vec{J}_2^{\text{IV}}}\right) \right\rangle_{\mathbb{S}^{\text{G}\rightleftharpoons\text{A}}} \quad \text{(C-41c)}$$

$$z_{\xi\zeta}^{\vec{M}^{\text{G}\rightleftharpoons\text{A}}\vec{J}^{\text{IS}}} = \left\langle \vec{b}_{\xi}^{\vec{M}^{\text{G}\rightleftharpoons\text{A}}}, \mathcal{K}_0\left(\vec{b}_{\zeta}^{\vec{J}^{\text{IS}}}\right) \right\rangle_{\mathbb{S}^{\text{G}\rightleftharpoons\text{A}}} \quad \text{(C-41d)}$$

$$z_{\xi\zeta}^{\vec{M}^{\text{G}\rightleftharpoons\text{A}}\vec{M}^{\text{G}\rightleftharpoons\text{A}}} = \left\langle \vec{b}_{\xi}^{\vec{M}^{\text{G}\rightleftharpoons\text{A}}}, -j\omega\varepsilon_0\mathcal{L}_0\left(\vec{b}_{\zeta}^{\vec{M}^{\text{G}\rightleftharpoons\text{A}}}\right) \right\rangle_{\mathbb{S}^{\text{G}\rightleftharpoons\text{A}}} \quad \text{(C-41e)}$$

$$z_{\xi\zeta}^{\vec{M}^{\text{G}\rightleftharpoons\text{A}}\vec{M}_1^{\text{IV}}} = \left\langle \vec{b}_{\xi}^{\vec{M}^{\text{G}\rightleftharpoons\text{A}}}, -j\omega\varepsilon_0\mathcal{L}_0\left(\vec{b}_{\zeta}^{\vec{M}_1^{\text{IV}}}\right) \right\rangle_{\mathbb{S}^{\text{G}\rightleftharpoons\text{A}}} \quad \text{(C-41f)}$$

$$z_{\xi\zeta}^{\vec{M}^{\text{G}\rightleftharpoons\text{A}}\vec{M}_2^{\text{IV}}} = \left\langle \vec{b}_{\xi}^{\vec{M}^{\text{G}\rightleftharpoons\text{A}}}, -j\omega\varepsilon_0\mathcal{L}_0\left(\vec{b}_{\zeta}^{\vec{M}_2^{\text{IV}}}\right) \right\rangle_{\mathbb{S}^{\text{G}\rightleftharpoons\text{A}}} \quad \text{(C-41g)}$$

The formulations used to calculate the elements of the matrices in Eq. (C-37b) are as follows:

$$z_{\xi\zeta}^{\vec{J}^{\text{G}\rightleftharpoons\text{A}}\vec{J}^{\text{G}\rightleftharpoons\text{A}}} = \left\langle \vec{b}_{\xi}^{\vec{J}^{\text{G}\rightleftharpoons\text{A}}}, -j\omega\mu_0\mathcal{L}_0\left(\vec{b}_{\zeta}^{\vec{J}^{\text{G}\rightleftharpoons\text{A}}}\right) \right\rangle_{\mathbb{S}^{\text{G}\rightleftharpoons\text{A}}} \quad \text{(C-42a)}$$

$$z_{\xi\zeta}^{\vec{J}^{\text{G}\rightleftharpoons\text{A}}\vec{J}_1^{\text{IV}}} = \left\langle \vec{b}_{\xi}^{\vec{J}^{\text{G}\rightleftharpoons\text{A}}}, -j\omega\mu_0\mathcal{L}_0\left(\vec{b}_{\zeta}^{\vec{J}_1^{\text{IV}}}\right) \right\rangle_{\mathbb{S}^{\text{G}\rightleftharpoons\text{A}}} \quad \text{(C-42b)}$$

$$z_{\xi\zeta}^{\vec{J}^{\text{G}\rightleftharpoons\text{A}}\vec{J}_2^{\text{IV}}} = \left\langle \vec{b}_{\xi}^{\vec{J}^{\text{G}\rightleftharpoons\text{A}}}, -j\omega\mu_0\mathcal{L}_0\left(\vec{b}_{\zeta}^{\vec{J}_2^{\text{IV}}}\right) \right\rangle_{\mathbb{S}^{\text{G}\rightleftharpoons\text{A}}} \quad \text{(C-42c)}$$

$$z_{\xi\zeta}^{\vec{J}^{\text{G}\rightleftharpoons\text{A}}\vec{J}^{\text{IS}}} = \left\langle \vec{b}_{\xi}^{\vec{J}^{\text{G}\rightleftharpoons\text{A}}}, -j\omega\mu_0\mathcal{L}_0\left(\vec{b}_{\zeta}^{\vec{J}^{\text{IS}}}\right) \right\rangle_{\mathbb{S}^{\text{G}\rightleftharpoons\text{A}}} \quad \text{(C-42d)}$$

$$z_{\xi\zeta}^{\vec{J}^{\text{G}\rightleftharpoons\text{A}}\vec{M}^{\text{G}\rightleftharpoons\text{A}}} = \left\langle \vec{b}_{\xi}^{\vec{J}^{\text{G}\rightleftharpoons\text{A}}}, \frac{1}{2}\vec{b}_{\zeta}^{\vec{M}^{\text{G}\rightleftharpoons\text{A}}} \times \hat{n}^{\rightarrow\text{A}} - \text{P.V.}\mathcal{K}_0\left(\vec{b}_{\zeta}^{\vec{M}^{\text{G}\rightleftharpoons\text{A}}}\right) \right\rangle_{\mathbb{S}^{\text{G}\rightleftharpoons\text{A}}} \quad \text{(C-42e)}$$

$$z_{\xi\zeta}^{\vec{J}^{\text{G}\rightleftharpoons\text{A}}\vec{M}_1^{\text{IV}}} = \left\langle \vec{b}_{\xi}^{\vec{J}^{\text{G}\rightleftharpoons\text{A}}}, -\mathcal{K}_0\left(\vec{b}_{\zeta}^{\vec{M}_1^{\text{IV}}}\right) \right\rangle_{\mathbb{S}^{\text{G}\rightleftharpoons\text{A}}} \quad \text{(C-42f)}$$

$$z_{\xi\zeta}^{\vec{J}^{\mathrm{G}\rightleftharpoons\mathrm{A}}\vec{M}_2^{\mathrm{IV}}} = \left\langle \vec{b}_\xi^{\vec{J}^{\mathrm{G}\rightleftharpoons\mathrm{A}}}, -\mathcal{K}_0\left(\vec{b}_\zeta^{\vec{M}_2^{\mathrm{IV}}}\right)\right\rangle_{\mathbb{S}^{\mathrm{G}\rightleftharpoons\mathrm{A}}} \tag{C-42g}$$

The formulations used to calculate the elements of the matrices in Eq. (C-38a) are as follows:

$$z_{\xi\zeta}^{\vec{J}_1^{\mathrm{IV}}\vec{J}^{\mathrm{G}\rightleftharpoons\mathrm{A}}} = \left\langle \vec{b}_\xi^{\vec{J}_1^{\mathrm{IV}}}, -j\omega\mu_0\mathcal{L}_0\left(\vec{b}_\zeta^{\vec{J}^{\mathrm{G}\rightleftharpoons\mathrm{A}}}\right)\right\rangle_{\mathbb{V}_1^{\mathrm{A}}} \tag{C-43a}$$

$$z_{\xi\zeta}^{\vec{J}_1^{\mathrm{IV}}\vec{J}_1^{\mathrm{IV}}} = \left\langle \vec{b}_\xi^{\vec{J}_1^{\mathrm{IV}}}, -j\omega\mu_0\mathcal{L}_0\left(\vec{b}_\zeta^{\vec{J}_1^{\mathrm{IV}}}\right) - \left(j\omega\Delta\ddot{\varepsilon}_1^{\mathrm{c}}\right)^{-1}\cdot\vec{b}_\zeta^{\vec{J}_1^{\mathrm{IV}}}\right\rangle_{\mathbb{V}_1^{\mathrm{A}}} \tag{C-43b}$$

$$z_{\xi\zeta}^{\vec{J}_1^{\mathrm{IV}}\vec{J}_2^{\mathrm{IV}}} = \left\langle \vec{b}_\xi^{\vec{J}_1^{\mathrm{IV}}}, -j\omega\mu_0\mathcal{L}_0\left(\vec{b}_\zeta^{\vec{J}_2^{\mathrm{IV}}}\right)\right\rangle_{\mathbb{V}_1^{\mathrm{A}}} \tag{C-43c}$$

$$z_{\xi\zeta}^{\vec{J}_1^{\mathrm{IV}}\vec{J}^{\mathrm{IS}}} = \left\langle \vec{b}_\xi^{\vec{J}_1^{\mathrm{IV}}}, -j\omega\mu_0\mathcal{L}_0\left(\vec{b}_\zeta^{\vec{J}^{\mathrm{IS}}}\right)\right\rangle_{\mathbb{V}_1^{\mathrm{A}}} \tag{C-43d}$$

$$z_{\xi\zeta}^{\vec{J}_1^{\mathrm{IV}}\vec{M}^{\mathrm{G}\rightleftharpoons\mathrm{A}}} = \left\langle \vec{b}_\xi^{\vec{J}_1^{\mathrm{IV}}}, -\mathcal{K}_0\left(\vec{b}_\zeta^{\vec{M}^{\mathrm{G}\rightleftharpoons\mathrm{A}}}\right)\right\rangle_{\mathbb{V}_1^{\mathrm{A}}} \tag{C-43e}$$

$$z_{\xi\zeta}^{\vec{J}_1^{\mathrm{IV}}\vec{M}_1^{\mathrm{IV}}} = \left\langle \vec{b}_\xi^{\vec{J}_1^{\mathrm{IV}}}, -\mathcal{K}_0\left(\vec{b}_\zeta^{\vec{M}_1^{\mathrm{IV}}}\right)\right\rangle_{\mathbb{V}_1^{\mathrm{A}}} \tag{C-43f}$$

$$z_{\xi\zeta}^{\vec{J}_1^{\mathrm{IV}}\vec{M}_2^{\mathrm{IV}}} = \left\langle \vec{b}_\xi^{\vec{J}_1^{\mathrm{IV}}}, -\mathcal{K}_0\left(\vec{b}_\zeta^{\vec{M}_2^{\mathrm{IV}}}\right)\right\rangle_{\mathbb{V}_1^{\mathrm{A}}} \tag{C-43g}$$

The formulations used to calculate the elements of the matrices in Eq. (C-38b) are as follows:

$$z_{\xi\zeta}^{\vec{M}_1^{\mathrm{IV}}\vec{J}^{\mathrm{G}\rightleftharpoons\mathrm{A}}} = \left\langle \vec{b}_\xi^{\vec{M}_1^{\mathrm{IV}}}, \mathcal{K}_0\left(\vec{b}_\zeta^{\vec{J}^{\mathrm{G}\rightleftharpoons\mathrm{A}}}\right)\right\rangle_{\mathbb{V}_1^{\mathrm{A}}} \tag{C-44a}$$

$$z_{\xi\zeta}^{\vec{M}_1^{\mathrm{IV}}\vec{J}_1^{\mathrm{IV}}} = \left\langle \vec{b}_\xi^{\vec{M}_1^{\mathrm{IV}}}, \mathcal{K}_0\left(\vec{b}_\zeta^{\vec{J}_1^{\mathrm{IV}}}\right)\right\rangle_{\mathbb{V}_1^{\mathrm{A}}} \tag{C-44b}$$

$$z_{\xi\zeta}^{\vec{M}_1^{\mathrm{IV}}\vec{J}_2^{\mathrm{IV}}} = \left\langle \vec{b}_\xi^{\vec{M}_1^{\mathrm{IV}}}, \mathcal{K}_0\left(\vec{b}_\zeta^{\vec{J}_2^{\mathrm{IV}}}\right)\right\rangle_{\mathbb{V}_1^{\mathrm{A}}} \tag{C-44c}$$

$$z_{\xi\zeta}^{\vec{M}_1^{\mathrm{IV}}\vec{J}^{\mathrm{IS}}} = \left\langle \vec{b}_\xi^{\vec{M}_1^{\mathrm{IV}}}, \mathcal{K}_0\left(\vec{b}_\zeta^{\vec{J}^{\mathrm{IS}}}\right)\right\rangle_{\mathbb{V}_1^{\mathrm{A}}} \tag{C-44d}$$

$$z_{\xi\zeta}^{\vec{M}_1^{\mathrm{IV}}\vec{M}^{\mathrm{G}\rightleftharpoons\mathrm{A}}} = \left\langle \vec{b}_\xi^{\vec{M}_1^{\mathrm{IV}}}, -j\omega\varepsilon_0\mathcal{L}_0\left(\vec{b}_\zeta^{\vec{M}^{\mathrm{G}\rightleftharpoons\mathrm{A}}}\right)\right\rangle_{\mathbb{V}_1^{\mathrm{A}}} \tag{C-44e}$$

$$z_{\xi\zeta}^{\vec{M}_1^{\mathrm{IV}}\vec{M}_1^{\mathrm{IV}}} = \left\langle \vec{b}_\xi^{\vec{M}_1^{\mathrm{IV}}}, -j\omega\varepsilon_0\mathcal{L}_0\left(\vec{b}_\zeta^{\vec{M}_1^{\mathrm{IV}}}\right) - \left(j\omega\Delta\ddot{\mu}_1\right)^{-1}\cdot\vec{b}_\zeta^{\vec{M}_1^{\mathrm{IV}}}\right\rangle_{\mathbb{V}_1^{\mathrm{A}}} \tag{C-44f}$$

$$z_{\xi\zeta}^{\vec{M}_1^{\mathrm{IV}}\vec{M}_2^{\mathrm{IV}}} = \left\langle \vec{b}_\xi^{\vec{M}_1^{\mathrm{IV}}}, -j\omega\varepsilon_0\mathcal{L}_0\left(\vec{b}_\zeta^{\vec{M}_2^{\mathrm{IV}}}\right)\right\rangle_{\mathbb{V}_1^{\mathrm{A}}} \tag{C-44g}$$

The formulations used to calculate the elements of the matrices in Eq. (C-39a) are as follows:

$$z_{\xi\zeta}^{\vec{J}_2^{\mathrm{IV}}\vec{J}^{\mathrm{G}\rightleftharpoons\mathrm{A}}} = \left\langle \vec{b}_\xi^{\vec{J}_2^{\mathrm{IV}}}, -j\omega\mu_0\mathcal{L}_0\left(\vec{b}_\zeta^{\vec{J}^{\mathrm{G}\rightleftharpoons\mathrm{A}}}\right)\right\rangle_{\mathbb{V}_2^{\mathrm{A}}} \tag{C-45a}$$

$$z_{\xi\zeta}^{\vec{J}_2^{\mathrm{IV}}\vec{J}_1^{\mathrm{IV}}} = \left\langle \vec{b}_\xi^{\vec{J}_2^{\mathrm{IV}}}, -j\omega\mu_0\mathcal{L}_0\left(\vec{b}_\zeta^{\vec{J}_1^{\mathrm{IV}}}\right)\right\rangle_{\mathbb{V}_2^{\mathrm{A}}} \tag{C-45b}$$

$$z_{\xi\zeta}^{\vec{J}_2^{\rm IV}\vec{J}_2^{\rm IV}} = \left\langle \vec{b}_\xi^{\vec{J}_2^{\rm IV}}, -j\omega\mu_0\mathcal{L}_0\left(\vec{b}_\zeta^{\vec{J}_2^{\rm IV}}\right) - \left(j\omega\Delta\vec{\vec{\varepsilon}}_2^{\rm c}\right)^{-1}\cdot\vec{b}_\zeta^{\vec{J}_2^{\rm IV}} \right\rangle_{\mathbb{V}_2^{\rm A}} \tag{C-45c}$$

$$z_{\xi\zeta}^{\vec{J}_2^{\rm IV}\vec{J}^{\rm IS}} = \left\langle \vec{b}_\xi^{\vec{J}_2^{\rm IV}}, -j\omega\mu_0\mathcal{L}_0\left(\vec{b}_\zeta^{\vec{J}^{\rm IS}}\right)\right\rangle_{\mathbb{V}_2^{\rm A}} \tag{C-45d}$$

$$z_{\xi\zeta}^{\vec{J}_2^{\rm IV}\vec{M}^{\rm G\rightleftharpoons A}} = \left\langle \vec{b}_\xi^{\vec{J}_2^{\rm IV}}, -\mathcal{K}_0\left(\vec{b}_\zeta^{\vec{M}^{\rm G\rightleftharpoons A}}\right)\right\rangle_{\mathbb{V}_2^{\rm A}} \tag{C-45e}$$

$$z_{\xi\zeta}^{\vec{J}_2^{\rm IV}\vec{M}_1^{\rm IV}} = \left\langle \vec{b}_\xi^{\vec{J}_2^{\rm IV}}, -\mathcal{K}_0\left(\vec{b}_\zeta^{\vec{M}_1^{\rm IV}}\right)\right\rangle_{\mathbb{V}_2^{\rm A}} \tag{C-45f}$$

$$z_{\xi\zeta}^{\vec{J}_2^{\rm IV}\vec{M}_2^{\rm IV}} = \left\langle \vec{b}_\xi^{\vec{J}_2^{\rm IV}}, -\mathcal{K}_0\left(\vec{b}_\zeta^{\vec{M}_2^{\rm IV}}\right)\right\rangle_{\mathbb{V}_2^{\rm A}} \tag{C-45g}$$

The formulations used to calculate the elements of the matrices in Eq. (C-39b) are as follows:

$$z_{\xi\zeta}^{\vec{M}_2^{\rm IV}\vec{J}^{\rm G\rightleftharpoons A}} = \left\langle \vec{b}_\xi^{\vec{M}_2^{\rm IV}}, \mathcal{K}_0\left(\vec{b}_\zeta^{\vec{J}^{\rm G\rightleftharpoons A}}\right)\right\rangle_{\mathbb{V}_2^{\rm A}} \tag{C-46a}$$

$$z_{\xi\zeta}^{\vec{M}_2^{\rm IV}\vec{J}_1^{\rm IV}} = \left\langle \vec{b}_\xi^{\vec{M}_2^{\rm IV}}, \mathcal{K}_0\left(\vec{b}_\zeta^{\vec{J}_1^{\rm IV}}\right)\right\rangle_{\mathbb{V}_2^{\rm A}} \tag{C-46b}$$

$$z_{\xi\zeta}^{\vec{M}_2^{\rm IV}\vec{J}_2^{\rm IV}} = \left\langle \vec{b}_\xi^{\vec{M}_2^{\rm IV}}, \mathcal{K}_0\left(\vec{b}_\zeta^{\vec{J}_2^{\rm IV}}\right)\right\rangle_{\mathbb{V}_2^{\rm A}} \tag{C-46c}$$

$$z_{\xi\zeta}^{\vec{M}_2^{\rm IV}\vec{J}^{\rm IS}} = \left\langle \vec{b}_\xi^{\vec{M}_2^{\rm IV}}, \mathcal{K}_0\left(\vec{b}_\zeta^{\vec{J}^{\rm IS}}\right)\right\rangle_{\mathbb{V}_2^{\rm A}} \tag{C-46d}$$

$$z_{\xi\zeta}^{\vec{M}_2^{\rm IV}\vec{M}^{\rm G\rightleftharpoons A}} = \left\langle \vec{b}_\xi^{\vec{M}_2^{\rm IV}}, -j\omega\varepsilon_0\mathcal{L}_0\left(\vec{b}_\zeta^{\vec{M}^{\rm G\rightleftharpoons A}}\right)\right\rangle_{\mathbb{V}_2^{\rm A}} \tag{C-46e}$$

$$z_{\xi\zeta}^{\vec{M}_2^{\rm IV}\vec{M}_1^{\rm IV}} = \left\langle \vec{b}_\xi^{\vec{M}_2^{\rm IV}}, -j\omega\varepsilon_0\mathcal{L}_0\left(\vec{b}_\zeta^{\vec{M}_1^{\rm IV}}\right)\right\rangle_{\mathbb{V}_2^{\rm A}} \tag{C-46f}$$

$$z_{\xi\zeta}^{\vec{M}_2^{\rm IV}\vec{M}_2^{\rm IV}} = \left\langle \vec{b}_\xi^{\vec{M}_2^{\rm IV}}, -j\omega\varepsilon_0\mathcal{L}_0\left(\vec{b}_\zeta^{\vec{M}_2^{\rm IV}}\right) - \left(j\omega\Delta\vec{\vec{\mu}}_2\right)^{-1}\cdot\vec{b}_\zeta^{\vec{M}_2^{\rm IV}} \right\rangle_{\mathbb{V}_2^{\rm A}} \tag{C-46g}$$

The formulations used to calculate the elements of the matrices in Eq. (C-40) are as follows:

$$z_{\xi\zeta}^{\vec{J}^{\rm IS}\vec{J}^{\rm G\rightleftharpoons A}} = \left\langle \vec{b}_\xi^{\vec{J}^{\rm IS}}, -j\omega\mu_0\mathcal{L}_0\left(\vec{b}_\zeta^{\vec{J}^{\rm G\rightleftharpoons A}}\right)\right\rangle_{\mathbb{S}^{\rm ele}} \tag{C-47a}$$

$$z_{\xi\zeta}^{\vec{J}^{\rm IS}\vec{J}_1^{\rm IV}} = \left\langle \vec{b}_\xi^{\vec{J}^{\rm IS}}, -j\omega\mu_0\mathcal{L}_0\left(\vec{b}_\zeta^{\vec{J}_1^{\rm IV}}\right)\right\rangle_{\mathbb{S}^{\rm ele}} \tag{C-47b}$$

$$z_{\xi\zeta}^{\vec{J}^{\rm IS}\vec{J}_2^{\rm IV}} = \left\langle \vec{b}_\xi^{\vec{J}^{\rm IS}}, -j\omega\mu_0\mathcal{L}_0\left(\vec{b}_\zeta^{\vec{J}_2^{\rm IV}}\right)\right\rangle_{\mathbb{S}^{\rm ele}} \tag{C-47c}$$

$$z_{\xi\zeta}^{\vec{J}^{\rm IS}\vec{J}^{\rm IS}} = \left\langle \vec{b}_\xi^{\vec{J}^{\rm IS}}, -j\omega\mu_0\mathcal{L}_0\left(\vec{b}_\zeta^{\vec{J}^{\rm IS}}\right)\right\rangle_{\mathbb{S}^{\rm ele}} \tag{C-47d}$$

$$z_{\xi\zeta}^{\vec{J}^{\rm IS}\vec{M}^{\rm G\rightleftharpoons A}} = \left\langle \vec{b}_\xi^{\vec{J}^{\rm IS}}, -\mathcal{K}_0\left(\vec{b}_\zeta^{\vec{M}^{\rm G\rightleftharpoons A}}\right)\right\rangle_{\mathbb{S}^{\rm ele}} \tag{C-47e}$$

$$z_{\xi\zeta}^{\vec{J}^{\rm IS}\vec{M}_1^{\rm IV}} = \left\langle \vec{b}_\xi^{\vec{J}^{\rm IS}}, -\mathcal{K}_0\left(\vec{b}_\zeta^{\vec{M}_1^{\rm IV}}\right)\right\rangle_{\mathbb{S}^{\rm ele}} \tag{C-47f}$$

$$z_{\xi\zeta}^{\vec{J}^{\rm IS}\vec{M}_2^{\rm IV}} = \left\langle \vec{b}_\xi^{\vec{J}^{\rm IS}}, -\mathcal{K}_0\left(\vec{b}_\zeta^{\vec{M}_2^{\rm IV}}\right)\right\rangle_{\mathbb{S}^{\rm ele}} \tag{C-47g}$$

By properly assembling the Eqs. (C-37a)~(C-40), we have the following two theoretically equivalent augmented matrix equations

$$\overline{\overline{\Psi}}_1 \cdot \bar{a}^{\mathrm{AV}} = \overline{\overline{\Psi}}_2 \cdot \bar{a}^{\vec{J}^{\mathrm{G}\rightleftharpoons\mathrm{A}}} \qquad \text{(C-48)}$$

$$\overline{\overline{\Psi}}_3 \cdot \bar{a}^{\mathrm{AV}} = \overline{\overline{\Psi}}_4 \cdot \bar{a}^{\vec{M}^{\mathrm{G}\rightleftharpoons\mathrm{A}}} \qquad \text{(C-49)}$$

in which

$$\overline{\overline{\Psi}}_1 = \begin{bmatrix} \overline{\overline{I}}^{\vec{J}^{\mathrm{G}\rightleftharpoons\mathrm{A}}} & 0 & 0 & 0 & 0 & 0 & 0 \\ 0 & \overline{\overline{Z}}^{\vec{M}^{\mathrm{G}\rightleftharpoons\mathrm{A}}\vec{J}_1^{\mathrm{IV}}} & \overline{\overline{Z}}^{\vec{M}^{\mathrm{G}\rightleftharpoons\mathrm{A}}\vec{J}_2^{\mathrm{IV}}} & \overline{\overline{Z}}^{\vec{M}^{\mathrm{G}\rightleftharpoons\mathrm{A}}\vec{J}^{\mathrm{IS}}} & \overline{\overline{Z}}^{\vec{M}^{\mathrm{G}\rightleftharpoons\mathrm{A}}\vec{M}^{\mathrm{G}\rightleftharpoons\mathrm{A}}} & \overline{\overline{Z}}^{\vec{M}^{\mathrm{G}\rightleftharpoons\mathrm{A}}\vec{M}_1^{\mathrm{IV}}} & \overline{\overline{Z}}^{\vec{M}^{\mathrm{G}\rightleftharpoons\mathrm{A}}\vec{M}_2^{\mathrm{IV}}} \\ 0 & \overline{\overline{Z}}^{\vec{J}_1^{\mathrm{IV}}\vec{J}_1^{\mathrm{IV}}} & \overline{\overline{Z}}^{\vec{J}_1^{\mathrm{IV}}\vec{J}_2^{\mathrm{IV}}} & \overline{\overline{Z}}^{\vec{J}_1^{\mathrm{IV}}\vec{J}^{\mathrm{IS}}} & \overline{\overline{Z}}^{\vec{J}_1^{\mathrm{IV}}\vec{M}^{\mathrm{G}\rightleftharpoons\mathrm{A}}} & \overline{\overline{Z}}^{\vec{J}_1^{\mathrm{IV}}\vec{M}_1^{\mathrm{IV}}} & \overline{\overline{Z}}^{\vec{J}_1^{\mathrm{IV}}\vec{M}_2^{\mathrm{IV}}} \\ 0 & \overline{\overline{Z}}^{\vec{M}_1^{\mathrm{IV}}\vec{J}_1^{\mathrm{IV}}} & \overline{\overline{Z}}^{\vec{M}_1^{\mathrm{IV}}\vec{J}_2^{\mathrm{IV}}} & \overline{\overline{Z}}^{\vec{M}_1^{\mathrm{IV}}\vec{J}^{\mathrm{IS}}} & \overline{\overline{Z}}^{\vec{M}_1^{\mathrm{IV}}\vec{M}^{\mathrm{G}\rightleftharpoons\mathrm{A}}} & \overline{\overline{Z}}^{\vec{M}_1^{\mathrm{IV}}\vec{M}_1^{\mathrm{IV}}} & \overline{\overline{Z}}^{\vec{M}_1^{\mathrm{IV}}\vec{M}_2^{\mathrm{IV}}} \\ 0 & \overline{\overline{Z}}^{\vec{J}_2^{\mathrm{IV}}\vec{J}_1^{\mathrm{IV}}} & \overline{\overline{Z}}^{\vec{J}_2^{\mathrm{IV}}\vec{J}_2^{\mathrm{IV}}} & \overline{\overline{Z}}^{\vec{J}_2^{\mathrm{IV}}\vec{J}^{\mathrm{IS}}} & \overline{\overline{Z}}^{\vec{J}_2^{\mathrm{IV}}\vec{M}^{\mathrm{G}\rightleftharpoons\mathrm{A}}} & \overline{\overline{Z}}^{\vec{J}_2^{\mathrm{IV}}\vec{M}_1^{\mathrm{IV}}} & \overline{\overline{Z}}^{\vec{J}_2^{\mathrm{IV}}\vec{M}_2^{\mathrm{IV}}} \\ 0 & \overline{\overline{Z}}^{\vec{M}_2^{\mathrm{IV}}\vec{J}_1^{\mathrm{IV}}} & \overline{\overline{Z}}^{\vec{M}_2^{\mathrm{IV}}\vec{J}_2^{\mathrm{IV}}} & \overline{\overline{Z}}^{\vec{M}_2^{\mathrm{IV}}\vec{J}^{\mathrm{IS}}} & \overline{\overline{Z}}^{\vec{M}_2^{\mathrm{IV}}\vec{M}^{\mathrm{G}\rightleftharpoons\mathrm{A}}} & \overline{\overline{Z}}^{\vec{M}_2^{\mathrm{IV}}\vec{M}_1^{\mathrm{IV}}} & \overline{\overline{Z}}^{\vec{M}_2^{\mathrm{IV}}\vec{M}_2^{\mathrm{IV}}} \\ 0 & \overline{\overline{Z}}^{\vec{J}^{\mathrm{IS}}\vec{J}_1^{\mathrm{IV}}} & \overline{\overline{Z}}^{\vec{J}^{\mathrm{IS}}\vec{J}_2^{\mathrm{IV}}} & \overline{\overline{Z}}^{\vec{J}^{\mathrm{IS}}\vec{J}^{\mathrm{IS}}} & \overline{\overline{Z}}^{\vec{J}^{\mathrm{IS}}\vec{M}^{\mathrm{G}\rightleftharpoons\mathrm{A}}} & \overline{\overline{Z}}^{\vec{J}^{\mathrm{IS}}\vec{M}_1^{\mathrm{IV}}} & \overline{\overline{Z}}^{\vec{J}^{\mathrm{IS}}\vec{M}_2^{\mathrm{IV}}} \end{bmatrix} \qquad \text{(C-50a)}$$

$$\overline{\overline{\Psi}}_2 = \begin{bmatrix} \overline{\overline{I}}^{\vec{J}^{\mathrm{G}\rightleftharpoons\mathrm{A}}} \\ -\overline{\overline{Z}}^{\vec{M}^{\mathrm{G}\rightleftharpoons\mathrm{A}}\vec{J}^{\mathrm{G}\rightleftharpoons\mathrm{A}}} \\ -\overline{\overline{Z}}^{\vec{J}_1^{\mathrm{IV}}\vec{J}^{\mathrm{G}\rightleftharpoons\mathrm{A}}} \\ -\overline{\overline{Z}}^{\vec{M}_1^{\mathrm{IV}}\vec{J}^{\mathrm{G}\rightleftharpoons\mathrm{A}}} \\ -\overline{\overline{Z}}^{\vec{J}_2^{\mathrm{IV}}\vec{J}^{\mathrm{G}\rightleftharpoons\mathrm{A}}} \\ -\overline{\overline{Z}}^{\vec{M}_2^{\mathrm{IV}}\vec{J}^{\mathrm{G}\rightleftharpoons\mathrm{A}}} \\ -\overline{\overline{Z}}^{\vec{J}^{\mathrm{IS}}\vec{J}^{\mathrm{G}\rightleftharpoons\mathrm{A}}} \end{bmatrix} \qquad \text{(C-50b)}$$

$$\bar{a}^{\mathrm{AV}} = \begin{bmatrix} \bar{a}^{\vec{J}^{\mathrm{G}\rightleftharpoons\mathrm{A}}} \\ \bar{a}^{\vec{J}_1^{\mathrm{IV}}} \\ \bar{a}^{\vec{J}_2^{\mathrm{IV}}} \\ \bar{a}^{\vec{J}^{\mathrm{IS}}} \\ \bar{a}^{\vec{M}^{\mathrm{G}\rightleftharpoons\mathrm{A}}} \\ \bar{a}^{\vec{M}_1^{\mathrm{IV}}} \\ \bar{a}^{\vec{M}_2^{\mathrm{IV}}} \end{bmatrix} \qquad \text{(C-50c)}$$

and

$$\overline{\overline{\Psi}}_3 = \begin{bmatrix} 0 & 0 & 0 & 0 & \overline{\overline{I}}^{\vec{M}^{\mathrm{G}\rightleftharpoons\mathrm{A}}} & 0 & 0 \\ \overline{\overline{Z}}^{\vec{J}^{\mathrm{G}\rightleftharpoons\mathrm{A}}\vec{J}^{\mathrm{G}\rightleftharpoons\mathrm{A}}} & \overline{\overline{Z}}^{\vec{J}^{\mathrm{G}\rightleftharpoons\mathrm{A}}\vec{J}_1^{\mathrm{IV}}} & \overline{\overline{Z}}^{\vec{J}^{\mathrm{G}\rightleftharpoons\mathrm{A}}\vec{J}_2^{\mathrm{IV}}} & \overline{\overline{Z}}^{\vec{J}^{\mathrm{G}\rightleftharpoons\mathrm{A}}\vec{J}^{\mathrm{IS}}} & 0 & \overline{\overline{Z}}^{\vec{J}^{\mathrm{G}\rightleftharpoons\mathrm{A}}\vec{M}_1^{\mathrm{IV}}} & \overline{\overline{Z}}^{\vec{J}^{\mathrm{G}\rightleftharpoons\mathrm{A}}\vec{M}_2^{\mathrm{IV}}} \\ \overline{\overline{Z}}^{\vec{J}_1^{\mathrm{IV}}\vec{J}^{\mathrm{G}\rightleftharpoons\mathrm{A}}} & \overline{\overline{Z}}^{\vec{J}_1^{\mathrm{IV}}\vec{J}_1^{\mathrm{IV}}} & \overline{\overline{Z}}^{\vec{J}_1^{\mathrm{IV}}\vec{J}_2^{\mathrm{IV}}} & \overline{\overline{Z}}^{\vec{J}_1^{\mathrm{IV}}\vec{J}^{\mathrm{IS}}} & 0 & \overline{\overline{Z}}^{\vec{J}_1^{\mathrm{IV}}\vec{M}_1^{\mathrm{IV}}} & \overline{\overline{Z}}^{\vec{J}_1^{\mathrm{IV}}\vec{M}_2^{\mathrm{IV}}} \\ \overline{\overline{Z}}^{\vec{M}_1^{\mathrm{IV}}\vec{J}^{\mathrm{G}\rightleftharpoons\mathrm{A}}} & \overline{\overline{Z}}^{\vec{M}_1^{\mathrm{IV}}\vec{J}_1^{\mathrm{IV}}} & \overline{\overline{Z}}^{\vec{M}_1^{\mathrm{IV}}\vec{J}_2^{\mathrm{IV}}} & \overline{\overline{Z}}^{\vec{M}_1^{\mathrm{IV}}\vec{J}^{\mathrm{IS}}} & 0 & \overline{\overline{Z}}^{\vec{M}_1^{\mathrm{IV}}\vec{M}_1^{\mathrm{IV}}} & \overline{\overline{Z}}^{\vec{M}_1^{\mathrm{IV}}\vec{M}_2^{\mathrm{IV}}} \\ \overline{\overline{Z}}^{\vec{J}_2^{\mathrm{IV}}\vec{J}^{\mathrm{G}\rightleftharpoons\mathrm{A}}} & \overline{\overline{Z}}^{\vec{J}_2^{\mathrm{IV}}\vec{J}_1^{\mathrm{IV}}} & \overline{\overline{Z}}^{\vec{J}_2^{\mathrm{IV}}\vec{J}_2^{\mathrm{IV}}} & \overline{\overline{Z}}^{\vec{J}_2^{\mathrm{IV}}\vec{J}^{\mathrm{IS}}} & 0 & \overline{\overline{Z}}^{\vec{J}_2^{\mathrm{IV}}\vec{M}_1^{\mathrm{IV}}} & \overline{\overline{Z}}^{\vec{J}_2^{\mathrm{IV}}\vec{M}_2^{\mathrm{IV}}} \\ \overline{\overline{Z}}^{\vec{M}_2^{\mathrm{IV}}\vec{J}^{\mathrm{G}\rightleftharpoons\mathrm{A}}} & \overline{\overline{Z}}^{\vec{M}_2^{\mathrm{IV}}\vec{J}_1^{\mathrm{IV}}} & \overline{\overline{Z}}^{\vec{M}_2^{\mathrm{IV}}\vec{J}_2^{\mathrm{IV}}} & \overline{\overline{Z}}^{\vec{M}_2^{\mathrm{IV}}\vec{J}^{\mathrm{IS}}} & 0 & \overline{\overline{Z}}^{\vec{M}_2^{\mathrm{IV}}\vec{M}_1^{\mathrm{IV}}} & \overline{\overline{Z}}^{\vec{M}_2^{\mathrm{IV}}\vec{M}_2^{\mathrm{IV}}} \\ \overline{\overline{Z}}^{\vec{J}^{\mathrm{IS}}\vec{J}^{\mathrm{G}\rightleftharpoons\mathrm{A}}} & \overline{\overline{Z}}^{\vec{J}^{\mathrm{IS}}\vec{J}_1^{\mathrm{IV}}} & \overline{\overline{Z}}^{\vec{J}^{\mathrm{IS}}\vec{J}_2^{\mathrm{IV}}} & \overline{\overline{Z}}^{\vec{J}^{\mathrm{IS}}\vec{J}^{\mathrm{IS}}} & 0 & \overline{\overline{Z}}^{\vec{J}^{\mathrm{IS}}\vec{M}_1^{\mathrm{IV}}} & \overline{\overline{Z}}^{\vec{J}^{\mathrm{IS}}\vec{M}_2^{\mathrm{IV}}} \end{bmatrix} \qquad \text{(C-51a)}$$

$$\overline{\overline{\Psi}}_4 = \begin{bmatrix} \overline{\overline{I}}^{\vec{M}^{\mathrm{G}\rightleftharpoons\mathrm{A}}} \\ -\overline{\overline{Z}}^{\vec{J}^{\mathrm{G}\rightleftharpoons\mathrm{A}}\vec{M}^{\mathrm{G}\rightleftharpoons\mathrm{A}}} \\ -\overline{\overline{Z}}^{\vec{J}_1^{\mathrm{IV}}\vec{M}^{\mathrm{G}\rightleftharpoons\mathrm{A}}} \\ -\overline{\overline{Z}}^{\vec{M}_1^{\mathrm{IV}}\vec{M}^{\mathrm{G}\rightleftharpoons\mathrm{A}}} \\ -\overline{\overline{Z}}^{\vec{J}_2^{\mathrm{IV}}\vec{M}^{\mathrm{G}\rightleftharpoons\mathrm{A}}} \\ -\overline{\overline{Z}}^{\vec{M}_2^{\mathrm{IV}}\vec{M}^{\mathrm{G}\rightleftharpoons\mathrm{A}}} \\ -\overline{\overline{Z}}^{\vec{J}^{\mathrm{IS}}\vec{M}^{\mathrm{G}\rightleftharpoons\mathrm{A}}} \end{bmatrix} \tag{C-51b}$$

where the superscript "AV" is the acronym of "*all variables* (of the related augmented tra-antenna)". By solving the above matrix equations, there exist the following transformations from $\overline{a}^{\vec{J}^{\mathrm{G}\rightleftharpoons\mathrm{A}}}$ to $\overline{a}^{\mathrm{AV}}$ and from $\overline{a}^{\vec{M}^{\mathrm{G}\rightleftharpoons\mathrm{A}}}$ to $\overline{a}^{\mathrm{AV}}$

$$\overline{a}^{\mathrm{AV}} = \overbrace{\left(\overline{\overline{\Psi}}_1\right)^{-1} \cdot \overline{\overline{\Psi}}_2}^{\overline{\overline{T}}^{\vec{J}^{\mathrm{G}\rightleftharpoons\mathrm{A}}\rightarrow\mathrm{AV}}} \cdot \overline{a}^{\vec{J}^{\mathrm{G}\rightleftharpoons\mathrm{A}}} \tag{C-52}$$

$$\overline{a}^{\mathrm{AV}} = \underbrace{\left(\overline{\overline{\Psi}}_3\right)^{-1} \cdot \overline{\overline{\Psi}}_4}_{\overline{\overline{T}}^{\vec{M}^{\mathrm{G}\rightleftharpoons\mathrm{A}}\rightarrow\mathrm{AV}}} \cdot \overline{a}^{\vec{M}^{\mathrm{G}\rightleftharpoons\mathrm{A}}} \tag{C-53}$$

where the superscript "$-1$" represents the inversion of the related matrix.

In fact, the Eqs. (C-37a)~(C-40) can also be alternatively assembled as follows:

$$\overline{\overline{\Psi}}_{\mathrm{FCE}}^{\mathrm{DoJ}} \cdot \overline{a}^{\mathrm{AV}} = 0 \tag{C-54}$$

$$\overline{\overline{\Psi}}_{\mathrm{FCE}}^{\mathrm{DoM}} \cdot \overline{a}^{\mathrm{AV}} = 0 \tag{C-55}$$

where

$$\overline{\overline{\Psi}}_{\mathrm{FCE}}^{\mathrm{DoJ}} = \begin{bmatrix}
\overline{\overline{Z}}^{\vec{M}^{\mathrm{G}\rightleftharpoons\mathrm{A}}\vec{J}^{\mathrm{G}\rightleftharpoons\mathrm{A}}} & \overline{\overline{Z}}^{\vec{M}^{\mathrm{G}\rightleftharpoons\mathrm{A}}\vec{J}_1^{\mathrm{IV}}} & \overline{\overline{Z}}^{\vec{M}^{\mathrm{G}\rightleftharpoons\mathrm{A}}\vec{J}_2^{\mathrm{IV}}} & \overline{\overline{Z}}^{\vec{M}^{\mathrm{G}\rightleftharpoons\mathrm{A}}\vec{J}^{\mathrm{IS}}} & \overline{\overline{Z}}^{\vec{M}^{\mathrm{G}\rightleftharpoons\mathrm{A}}\vec{M}^{\mathrm{G}\rightleftharpoons\mathrm{A}}} & \overline{\overline{Z}}^{\vec{M}^{\mathrm{G}\rightleftharpoons\mathrm{A}}\vec{M}_1^{\mathrm{IV}}} & \overline{\overline{Z}}^{\vec{M}^{\mathrm{G}\rightleftharpoons\mathrm{A}}\vec{M}_2^{\mathrm{IV}}} \\
\overline{\overline{Z}}^{\vec{J}_1^{\mathrm{IV}}\vec{J}^{\mathrm{G}\rightleftharpoons\mathrm{A}}} & \overline{\overline{Z}}^{\vec{J}_1^{\mathrm{IV}}\vec{J}_1^{\mathrm{IV}}} & \overline{\overline{Z}}^{\vec{J}_1^{\mathrm{IV}}\vec{J}_2^{\mathrm{IV}}} & \overline{\overline{Z}}^{\vec{J}_1^{\mathrm{IV}}\vec{J}^{\mathrm{IS}}} & \overline{\overline{Z}}^{\vec{J}_1^{\mathrm{IV}}\vec{M}^{\mathrm{G}\rightleftharpoons\mathrm{A}}} & \overline{\overline{Z}}^{\vec{J}_1^{\mathrm{IV}}\vec{M}_1^{\mathrm{IV}}} & \overline{\overline{Z}}^{\vec{J}_1^{\mathrm{IV}}\vec{M}_2^{\mathrm{IV}}} \\
\overline{\overline{Z}}^{\vec{M}_1^{\mathrm{IV}}\vec{J}^{\mathrm{G}\rightleftharpoons\mathrm{A}}} & \overline{\overline{Z}}^{\vec{M}_1^{\mathrm{IV}}\vec{J}_1^{\mathrm{IV}}} & \overline{\overline{Z}}^{\vec{M}_1^{\mathrm{IV}}\vec{J}_2^{\mathrm{IV}}} & \overline{\overline{Z}}^{\vec{M}_1^{\mathrm{IV}}\vec{J}^{\mathrm{IS}}} & \overline{\overline{Z}}^{\vec{M}_1^{\mathrm{IV}}\vec{M}^{\mathrm{G}\rightleftharpoons\mathrm{A}}} & \overline{\overline{Z}}^{\vec{M}_1^{\mathrm{IV}}\vec{M}_1^{\mathrm{IV}}} & \overline{\overline{Z}}^{\vec{M}_1^{\mathrm{IV}}\vec{M}_2^{\mathrm{IV}}} \\
\overline{\overline{Z}}^{\vec{J}_2^{\mathrm{IV}}\vec{J}^{\mathrm{G}\rightleftharpoons\mathrm{A}}} & \overline{\overline{Z}}^{\vec{J}_2^{\mathrm{IV}}\vec{J}_1^{\mathrm{IV}}} & \overline{\overline{Z}}^{\vec{J}_2^{\mathrm{IV}}\vec{J}_2^{\mathrm{IV}}} & \overline{\overline{Z}}^{\vec{J}_2^{\mathrm{IV}}\vec{J}^{\mathrm{IS}}} & \overline{\overline{Z}}^{\vec{J}_2^{\mathrm{IV}}\vec{M}^{\mathrm{G}\rightleftharpoons\mathrm{A}}} & \overline{\overline{Z}}^{\vec{J}_2^{\mathrm{IV}}\vec{M}_1^{\mathrm{IV}}} & \overline{\overline{Z}}^{\vec{J}_2^{\mathrm{IV}}\vec{M}_2^{\mathrm{IV}}} \\
\overline{\overline{Z}}^{\vec{M}_2^{\mathrm{IV}}\vec{J}^{\mathrm{G}\rightleftharpoons\mathrm{A}}} & \overline{\overline{Z}}^{\vec{M}_2^{\mathrm{IV}}\vec{J}_1^{\mathrm{IV}}} & \overline{\overline{Z}}^{\vec{M}_2^{\mathrm{IV}}\vec{J}_2^{\mathrm{IV}}} & \overline{\overline{Z}}^{\vec{M}_2^{\mathrm{IV}}\vec{J}^{\mathrm{IS}}} & \overline{\overline{Z}}^{\vec{M}_2^{\mathrm{IV}}\vec{M}^{\mathrm{G}\rightleftharpoons\mathrm{A}}} & \overline{\overline{Z}}^{\vec{M}_2^{\mathrm{IV}}\vec{M}_1^{\mathrm{IV}}} & \overline{\overline{Z}}^{\vec{M}_2^{\mathrm{IV}}\vec{M}_2^{\mathrm{IV}}} \\
\overline{\overline{Z}}^{\vec{J}^{\mathrm{IS}}\vec{J}^{\mathrm{G}\rightleftharpoons\mathrm{A}}} & \overline{\overline{Z}}^{\vec{J}^{\mathrm{IS}}\vec{J}_1^{\mathrm{IV}}} & \overline{\overline{Z}}^{\vec{J}^{\mathrm{IS}}\vec{J}_2^{\mathrm{IV}}} & \overline{\overline{Z}}^{\vec{J}^{\mathrm{IS}}\vec{J}^{\mathrm{IS}}} & \overline{\overline{Z}}^{\vec{J}^{\mathrm{IS}}\vec{M}^{\mathrm{G}\rightleftharpoons\mathrm{A}}} & \overline{\overline{Z}}^{\vec{J}^{\mathrm{IS}}\vec{M}_1^{\mathrm{IV}}} & \overline{\overline{Z}}^{\vec{J}^{\mathrm{IS}}\vec{M}_2^{\mathrm{IV}}}
\end{bmatrix} \tag{C-56}$$

$$\overline{\overline{\Psi}}_{\mathrm{FCE}}^{\mathrm{DoM}} = \begin{bmatrix}
\overline{\overline{Z}}^{\vec{J}^{\mathrm{G}\rightleftharpoons\mathrm{A}}\vec{J}^{\mathrm{G}\rightleftharpoons\mathrm{A}}} & \overline{\overline{Z}}^{\vec{J}^{\mathrm{G}\rightleftharpoons\mathrm{A}}\vec{J}_1^{\mathrm{IV}}} & \overline{\overline{Z}}^{\vec{J}^{\mathrm{G}\rightleftharpoons\mathrm{A}}\vec{J}_2^{\mathrm{IV}}} & \overline{\overline{Z}}^{\vec{J}^{\mathrm{G}\rightleftharpoons\mathrm{A}}\vec{J}^{\mathrm{IS}}} & \overline{\overline{Z}}^{\vec{J}^{\mathrm{G}\rightleftharpoons\mathrm{A}}\vec{M}^{\mathrm{G}\rightleftharpoons\mathrm{A}}} & \overline{\overline{Z}}^{\vec{J}^{\mathrm{G}\rightleftharpoons\mathrm{A}}\vec{M}_1^{\mathrm{IV}}} & \overline{\overline{Z}}^{\vec{J}^{\mathrm{G}\rightleftharpoons\mathrm{A}}\vec{M}_2^{\mathrm{IV}}} \\
\overline{\overline{Z}}^{\vec{J}_1^{\mathrm{IV}}\vec{J}^{\mathrm{G}\rightleftharpoons\mathrm{A}}} & \overline{\overline{Z}}^{\vec{J}_1^{\mathrm{IV}}\vec{J}_1^{\mathrm{IV}}} & \overline{\overline{Z}}^{\vec{J}_1^{\mathrm{IV}}\vec{J}_2^{\mathrm{IV}}} & \overline{\overline{Z}}^{\vec{J}_1^{\mathrm{IV}}\vec{J}^{\mathrm{IS}}} & \overline{\overline{Z}}^{\vec{J}_1^{\mathrm{IV}}\vec{M}^{\mathrm{G}\rightleftharpoons\mathrm{A}}} & \overline{\overline{Z}}^{\vec{J}_1^{\mathrm{IV}}\vec{M}_1^{\mathrm{IV}}} & \overline{\overline{Z}}^{\vec{J}_1^{\mathrm{IV}}\vec{M}_2^{\mathrm{IV}}} \\
\overline{\overline{Z}}^{\vec{M}_1^{\mathrm{IV}}\vec{J}^{\mathrm{G}\rightleftharpoons\mathrm{A}}} & \overline{\overline{Z}}^{\vec{M}_1^{\mathrm{IV}}\vec{J}_1^{\mathrm{IV}}} & \overline{\overline{Z}}^{\vec{M}_1^{\mathrm{IV}}\vec{J}_2^{\mathrm{IV}}} & \overline{\overline{Z}}^{\vec{M}_1^{\mathrm{IV}}\vec{J}^{\mathrm{IS}}} & \overline{\overline{Z}}^{\vec{M}_1^{\mathrm{IV}}\vec{M}^{\mathrm{G}\rightleftharpoons\mathrm{A}}} & \overline{\overline{Z}}^{\vec{M}_1^{\mathrm{IV}}\vec{M}_1^{\mathrm{IV}}} & \overline{\overline{Z}}^{\vec{M}_1^{\mathrm{IV}}\vec{M}_2^{\mathrm{IV}}} \\
\overline{\overline{Z}}^{\vec{J}_2^{\mathrm{IV}}\vec{J}^{\mathrm{G}\rightleftharpoons\mathrm{A}}} & \overline{\overline{Z}}^{\vec{J}_2^{\mathrm{IV}}\vec{J}_1^{\mathrm{IV}}} & \overline{\overline{Z}}^{\vec{J}_2^{\mathrm{IV}}\vec{J}_2^{\mathrm{IV}}} & \overline{\overline{Z}}^{\vec{J}_2^{\mathrm{IV}}\vec{J}^{\mathrm{IS}}} & \overline{\overline{Z}}^{\vec{J}_2^{\mathrm{IV}}\vec{M}^{\mathrm{G}\rightleftharpoons\mathrm{A}}} & \overline{\overline{Z}}^{\vec{J}_2^{\mathrm{IV}}\vec{M}_1^{\mathrm{IV}}} & \overline{\overline{Z}}^{\vec{J}_2^{\mathrm{IV}}\vec{M}_2^{\mathrm{IV}}} \\
\overline{\overline{Z}}^{\vec{M}_2^{\mathrm{IV}}\vec{J}^{\mathrm{G}\rightleftharpoons\mathrm{A}}} & \overline{\overline{Z}}^{\vec{M}_2^{\mathrm{IV}}\vec{J}_1^{\mathrm{IV}}} & \overline{\overline{Z}}^{\vec{M}_2^{\mathrm{IV}}\vec{J}_2^{\mathrm{IV}}} & \overline{\overline{Z}}^{\vec{M}_2^{\mathrm{IV}}\vec{J}^{\mathrm{IS}}} & \overline{\overline{Z}}^{\vec{M}_2^{\mathrm{IV}}\vec{M}^{\mathrm{G}\rightleftharpoons\mathrm{A}}} & \overline{\overline{Z}}^{\vec{M}_2^{\mathrm{IV}}\vec{M}_1^{\mathrm{IV}}} & \overline{\overline{Z}}^{\vec{M}_2^{\mathrm{IV}}\vec{M}_2^{\mathrm{IV}}} \\
\overline{\overline{Z}}^{\vec{J}^{\mathrm{IS}}\vec{J}^{\mathrm{G}\rightleftharpoons\mathrm{A}}} & \overline{\overline{Z}}^{\vec{J}^{\mathrm{IS}}\vec{J}_1^{\mathrm{IV}}} & \overline{\overline{Z}}^{\vec{J}^{\mathrm{IS}}\vec{J}_2^{\mathrm{IV}}} & \overline{\overline{Z}}^{\vec{J}^{\mathrm{IS}}\vec{J}^{\mathrm{IS}}} & \overline{\overline{Z}}^{\vec{J}^{\mathrm{IS}}\vec{M}^{\mathrm{G}\rightleftharpoons\mathrm{A}}} & \overline{\overline{Z}}^{\vec{J}^{\mathrm{IS}}\vec{M}_1^{\mathrm{IV}}} & \overline{\overline{Z}}^{\vec{J}^{\mathrm{IS}}\vec{M}_2^{\mathrm{IV}}}
\end{bmatrix} \tag{C-57}$$

where the superscripts "DoJ" and "DoM" are the acronyms for "definition of $\vec{J}^{\,\mathrm{G}\rightleftharpoons\mathrm{A}}$" and "definition of $\vec{M}^{\,\mathrm{G}\rightleftharpoons\mathrm{A}}$" respectively, and the subscript "FCE" is the acronym of "field continuation equation". Theoretically, the Eqs. (C-54) and (C-55) are equivalent to each other, and they have the same *solution space*, and the solution space is just the modal space of the tra-antenna shown in Fig. 6-41 / 6-54. If the *basic solutions* (*BSs*) used to span the solution space / modal space are denoted as $\{\bar{s}_1^{\mathrm{BS}}, \bar{s}_2^{\mathrm{BS}}, \cdots\}$, then any mode contained in the space can be expanded as follows:

$$\bar{a}^{\mathrm{AV}} = \sum_i a_i^{\mathrm{BS}} \bar{s}_i^{\mathrm{BS}} = \underbrace{\left[\bar{s}_1^{\mathrm{BS}}, \bar{s}_2^{\mathrm{BS}}, \cdots\right]}_{\bar{\bar{T}}^{\mathrm{BS}\to\mathrm{AV}}} \cdot \underbrace{\begin{bmatrix} a_1^{\mathrm{BS}} \\ a_2^{\mathrm{BS}} \\ \vdots \end{bmatrix}}_{\bar{a}^{\mathrm{BS}}} \tag{C-58}$$

For the convenience of the following discussions, Eqs. (C-52)&(C-53) and (C-58) are uniformly written as follows:

$$\bar{a}^{\mathrm{AV}} = \bar{\bar{T}} \cdot \bar{a} \tag{C-59}$$

where $\bar{a} = \bar{a}^{\vec{J}^{\mathrm{G}\rightleftharpoons\mathrm{A}}} / \bar{a}^{\vec{M}^{\mathrm{G}\rightleftharpoons\mathrm{A}}} / \bar{a}^{\mathrm{BS}}$ and correspondingly $\bar{\bar{T}} = \bar{\bar{T}}^{\vec{J}^{\mathrm{G}\rightleftharpoons\mathrm{A}}\to\mathrm{AV}} / \bar{\bar{T}}^{\vec{M}^{\mathrm{G}\rightleftharpoons\mathrm{A}}\to\mathrm{AV}} / \bar{\bar{T}}^{\mathrm{BS}\to\mathrm{AV}}$.

Based on Eqs. (C-29) and (C-30), the IPO $P^{\mathrm{G}\rightleftharpoons\mathrm{A}} = (1/2)\iint_{\mathbb{S}^{\mathrm{G}\rightleftharpoons\mathrm{A}}} (\vec{E} \times \vec{H}^\dagger) \cdot \hat{n}^{\to\mathrm{A}} dS$ can be rewritten as follows:

$$\begin{aligned} P^{\mathrm{G}\rightleftharpoons\mathrm{A}} = \ & (1/2)\left\langle \hat{n}^{\to\mathrm{A}} \times \vec{J}^{\,\mathrm{G}\rightleftharpoons\mathrm{A}}, \vec{M}^{\,\mathrm{G}\rightleftharpoons\mathrm{A}} \right\rangle_{\mathbb{S}^{\mathrm{G}\rightleftharpoons\mathrm{A}}} \\ & = -(1/2)\left\langle \vec{J}^{\,\mathrm{G}\rightleftharpoons\mathrm{A}}, \mathcal{E}_0\left( \vec{J}^{\,\mathrm{G}\rightleftharpoons\mathrm{A}} + \vec{J}_1^{\,\mathrm{IV}} + \vec{J}_2^{\,\mathrm{IV}} + \vec{J}^{\,\mathrm{IS}}, \vec{M}^{\,\mathrm{G}\rightleftharpoons\mathrm{A}} + \vec{M}_1^{\,\mathrm{IV}} + \vec{M}_2^{\,\mathrm{IV}} \right) \right\rangle_{\mathbb{S}^{\mathrm{G}\rightleftharpoons\mathrm{A}}} \\ & = -(1/2)\left\langle \vec{M}^{\,\mathrm{G}\rightleftharpoons\mathrm{A}}, \mathcal{H}_0\left( \vec{J}^{\,\mathrm{G}\rightleftharpoons\mathrm{A}} + \vec{J}_1^{\,\mathrm{IV}} + \vec{J}_2^{\,\mathrm{IV}} + \vec{J}^{\,\mathrm{IS}}, \vec{M}^{\,\mathrm{G}\rightleftharpoons\mathrm{A}} + \vec{M}_1^{\,\mathrm{IV}} + \vec{M}_2^{\,\mathrm{IV}} \right) \right\rangle_{\mathbb{S}^{\mathrm{G}\rightleftharpoons\mathrm{A}}}^{\dagger} \end{aligned} \tag{C-60}$$

where integral surface $\mathbb{S}^{\mathrm{G}\rightleftharpoons\mathrm{A}}$ is an inner sub-boundary of $\mathbb{V}_1^{\mathrm{A}}$. Expanding the currents involved in Eq. (C-60), the IPO is discretized as follows:

$$P^{\mathrm{G}\rightleftharpoons\mathrm{A}} = \left(\bar{a}^{\mathrm{AV}}\right)^\dagger \cdot \bar{\bar{P}}_{\mathrm{curAV}}^{\mathrm{G}\rightleftharpoons\mathrm{A}} \cdot \bar{a}^{\mathrm{AV}} = \left(\bar{a}^{\mathrm{AV}}\right)^\dagger \cdot \bar{\bar{P}}_{\mathrm{intAV}}^{\mathrm{G}\rightleftharpoons\mathrm{A}} \cdot \bar{a}^{\mathrm{AV}} \tag{C-61}$$

in which

$$\bar{\bar{P}}_{\mathrm{curAV}}^{\mathrm{G}\rightleftharpoons\mathrm{A}} = \begin{bmatrix} 0 & 0 & 0 & 0 & \bar{\bar{C}}^{\vec{J}^{\mathrm{G}\rightleftharpoons\mathrm{A}}\vec{M}^{\mathrm{G}\rightleftharpoons\mathrm{A}}} & 0 & 0 \\ 0 & 0 & 0 & 0 & 0 & 0 & 0 \\ 0 & 0 & 0 & 0 & 0 & 0 & 0 \\ 0 & 0 & 0 & 0 & 0 & 0 & 0 \\ 0 & 0 & 0 & 0 & 0 & 0 & 0 \\ 0 & 0 & 0 & 0 & 0 & 0 & 0 \\ 0 & 0 & 0 & 0 & 0 & 0 & 0 \end{bmatrix} \tag{C-62}$$

corresponding to the first equality in Eq. (C-61), and

$$\overline{\overline{P}}_{\mathrm{intAV}}^{\mathrm{G}\rightleftharpoons\mathrm{A}}=\begin{bmatrix}\overline{\overline{P}}^{\vec{J}^{\mathrm{G}\rightleftharpoons\mathrm{A}}\vec{J}^{\mathrm{G}\rightleftharpoons\mathrm{A}}} & \overline{\overline{P}}^{\vec{J}^{\mathrm{G}\rightleftharpoons\mathrm{A}}\vec{J}_1^{\mathrm{IV}}} & \overline{\overline{P}}^{\vec{J}^{\mathrm{G}\rightleftharpoons\mathrm{A}}\vec{J}_2^{\mathrm{IV}}} & \overline{\overline{P}}^{\vec{J}^{\mathrm{G}\rightleftharpoons\mathrm{A}}\vec{J}^{\mathrm{IS}}} & \overline{\overline{P}}^{\vec{J}^{\mathrm{G}\rightleftharpoons\mathrm{A}}\vec{M}^{\mathrm{G}\rightleftharpoons\mathrm{A}}} & \overline{\overline{P}}^{\vec{J}^{\mathrm{G}\rightleftharpoons\mathrm{A}}\vec{M}_1^{\mathrm{IV}}} & \overline{\overline{P}}^{\vec{J}^{\mathrm{G}\rightleftharpoons\mathrm{A}}\vec{M}_2^{\mathrm{IV}}}\\ 0&0&0&0&0&0&0\\ 0&0&0&0&0&0&0\\ 0&0&0&0&0&0&0\\ 0&0&0&0&0&0&0\\ 0&0&0&0&0&0&0\\ 0&0&0&0&0&0&0\end{bmatrix} \tag{C-63a}$$

corresponding to the second equality in Eq. (C-61), and

$$\overline{\overline{P}}_{\mathrm{intAV}}^{\mathrm{G}\rightleftharpoons\mathrm{A}}=\begin{bmatrix}0&0&0&0&0&0&0\\ 0&0&0&0&0&0&0\\ 0&0&0&0&0&0&0\\ 0&0&0&0&0&0&0\\ \overline{\overline{P}}^{\vec{M}^{\mathrm{G}\rightleftharpoons\mathrm{A}}\vec{J}^{\mathrm{G}\rightleftharpoons\mathrm{A}}} & \overline{\overline{P}}^{\vec{M}^{\mathrm{G}\rightleftharpoons\mathrm{A}}\vec{J}_1^{\mathrm{IV}}} & \overline{\overline{P}}^{\vec{M}^{\mathrm{G}\rightleftharpoons\mathrm{A}}\vec{J}_2^{\mathrm{IV}}} & \overline{\overline{P}}^{\vec{M}^{\mathrm{G}\rightleftharpoons\mathrm{A}}\vec{J}^{\mathrm{IS}}} & \overline{\overline{P}}^{\vec{M}^{\mathrm{G}\rightleftharpoons\mathrm{A}}\vec{M}^{\mathrm{G}\rightleftharpoons\mathrm{A}}} & \overline{\overline{P}}^{\vec{M}^{\mathrm{G}\rightleftharpoons\mathrm{A}}\vec{M}_1^{\mathrm{IV}}} & \overline{\overline{P}}^{\vec{M}^{\mathrm{G}\rightleftharpoons\mathrm{A}}\vec{M}_2^{\mathrm{IV}}}\\ 0&0&0&0&0&0&0\\ 0&0&0&0&0&0&0\end{bmatrix}^{\dagger} \tag{C-63b}$$

corresponding to the third equality in Eq. (C-61), where the elements of the submatrices are as follows:

$$c_{\xi\zeta}^{\vec{J}^{\mathrm{G}\rightleftharpoons\mathrm{A}}\vec{M}^{\mathrm{G}\rightleftharpoons\mathrm{A}}} = (1/2)\left\langle \hat{n}^{\rightarrow\mathrm{A}}\times\vec{b}_{\xi}^{\vec{J}^{\mathrm{G}\rightleftharpoons\mathrm{A}}},\vec{b}_{\zeta}^{\vec{M}^{\mathrm{G}\rightleftharpoons\mathrm{A}}}\right\rangle_{\mathbb{S}^{\mathrm{G}\rightleftharpoons\mathrm{A}}} \tag{C-64}$$

and

$$p_{\xi\zeta}^{\vec{J}^{\mathrm{G}\rightleftharpoons\mathrm{A}}\vec{J}^{\mathrm{G}\rightleftharpoons\mathrm{A}}} = -(1/2)\left\langle \vec{b}_{\xi}^{\vec{J}^{\mathrm{G}\rightleftharpoons\mathrm{A}}},-j\omega\mu_0\mathcal{L}_0\left(\vec{b}_{\zeta}^{\vec{J}^{\mathrm{G}\rightleftharpoons\mathrm{A}}}\right)\right\rangle_{\mathbb{S}^{\mathrm{G}\rightleftharpoons\mathrm{A}}} \tag{C-65a}$$

$$p_{\xi\zeta}^{\vec{J}^{\mathrm{G}\rightleftharpoons\mathrm{A}}\vec{J}_1^{\mathrm{IV}}} = -(1/2)\left\langle \vec{b}_{\xi}^{\vec{J}^{\mathrm{G}\rightleftharpoons\mathrm{A}}},-j\omega\mu_0\mathcal{L}_0\left(\vec{b}_{\zeta}^{\vec{J}_1^{\mathrm{IV}}}\right)\right\rangle_{\mathbb{S}^{\mathrm{G}\rightleftharpoons\mathrm{A}}} \tag{C-65b}$$

$$p_{\xi\zeta}^{\vec{J}^{\mathrm{G}\rightleftharpoons\mathrm{A}}\vec{J}_2^{\mathrm{IV}}} = -(1/2)\left\langle \vec{b}_{\xi}^{\vec{J}^{\mathrm{G}\rightleftharpoons\mathrm{A}}},-j\omega\mu_0\mathcal{L}_0\left(\vec{b}_{\zeta}^{\vec{J}_2^{\mathrm{IV}}}\right)\right\rangle_{\mathbb{S}^{\mathrm{G}\rightleftharpoons\mathrm{A}}} \tag{C-65c}$$

$$p_{\xi\zeta}^{\vec{J}^{\mathrm{G}\rightleftharpoons\mathrm{A}}\vec{J}^{\mathrm{IS}}} = -(1/2)\left\langle \vec{b}_{\xi}^{\vec{J}^{\mathrm{G}\rightleftharpoons\mathrm{A}}},-j\omega\mu_0\mathcal{L}_0\left(\vec{b}_{\zeta}^{\vec{J}^{\mathrm{IS}}}\right)\right\rangle_{\mathbb{S}^{\mathrm{G}\rightleftharpoons\mathrm{A}}} \tag{C-65d}$$

$$p_{\xi\zeta}^{\vec{J}^{\mathrm{G}\rightleftharpoons\mathrm{A}}\vec{M}^{\mathrm{G}\rightleftharpoons\mathrm{A}}} = -(1/2)\left\langle \vec{b}_{\xi}^{\vec{J}^{\mathrm{G}\rightleftharpoons\mathrm{A}}},\hat{n}^{\rightarrow\mathrm{A}}\times\frac{1}{2}\vec{b}_{\zeta}^{\vec{M}^{\mathrm{G}\rightleftharpoons\mathrm{A}}}-\mathrm{P.V.}\,\mathcal{K}_0\left(\vec{b}_{\zeta}^{\vec{M}^{\mathrm{G}\rightleftharpoons\mathrm{A}}}\right)\right\rangle_{\mathbb{S}^{\mathrm{G}\rightleftharpoons\mathrm{A}}} \tag{C-65e}$$

$$p_{\xi\zeta}^{\vec{J}^{\mathrm{G}\rightleftharpoons\mathrm{A}}\vec{M}_1^{\mathrm{IV}}} = -(1/2)\left\langle \vec{b}_{\xi}^{\vec{J}^{\mathrm{G}\rightleftharpoons\mathrm{A}}},-\mathcal{K}_0\left(\vec{b}_{\zeta}^{\vec{M}_1^{\mathrm{IV}}}\right)\right\rangle_{\mathbb{S}^{\mathrm{G}\rightleftharpoons\mathrm{A}}} \tag{C-65f}$$

$$p_{\xi\zeta}^{\vec{J}^{\mathrm{G}\rightleftharpoons\mathrm{A}}\vec{M}_2^{\mathrm{IV}}} = -(1/2)\left\langle \vec{b}_{\xi}^{\vec{J}^{\mathrm{G}\rightleftharpoons\mathrm{A}}},-\mathcal{K}_0\left(\vec{b}_{\zeta}^{\vec{M}_2^{\mathrm{IV}}}\right)\right\rangle_{\mathbb{S}^{\mathrm{G}\rightleftharpoons\mathrm{A}}} \tag{C-65g}$$

and

$$p_{\xi\zeta}^{\vec{M}^{\mathrm{G}\rightleftharpoons\mathrm{A}}\vec{J}^{\mathrm{G}\rightleftharpoons\mathrm{A}}} = -(1/2)\left\langle \vec{b}_{\xi}^{\vec{M}^{\mathrm{G}\rightleftharpoons\mathrm{A}}},\frac{1}{2}\vec{b}_{\zeta}^{\vec{J}^{\mathrm{G}\rightleftharpoons\mathrm{A}}}\times\hat{n}^{\rightarrow\mathrm{A}}+\mathrm{P.V.}\,\mathcal{K}_0\left(\vec{b}_{\zeta}^{\vec{J}^{\mathrm{G}\rightleftharpoons\mathrm{A}}}\right)\right\rangle_{\mathbb{S}^{\mathrm{G}\rightleftharpoons\mathrm{A}}} \tag{C-65h}$$

$$p_{\xi\zeta}^{\vec{M}^{\mathrm{G}\rightleftharpoons\mathrm{A}}\vec{J}_1^{\mathrm{IV}}} = -(1/2)\left\langle \vec{b}_{\xi}^{\vec{M}^{\mathrm{G}\rightleftharpoons\mathrm{A}}},\mathcal{K}_0\left(\vec{b}_{\zeta}^{\vec{J}_1^{\mathrm{IV}}}\right)\right\rangle_{\mathbb{S}^{\mathrm{G}\rightleftharpoons\mathrm{A}}} \tag{C-65i}$$

$$p_{\xi\zeta}^{\vec{M}^{\text{G}\rightleftharpoons\text{A}}\vec{J}_2^{\text{IV}}} = -(1/2)\left\langle \vec{b}_\xi^{\vec{M}^{\text{G}\rightleftharpoons\text{A}}}, \mathcal{K}_0\left(\vec{b}_\zeta^{\vec{J}_2^{\text{IV}}}\right)\right\rangle_{\mathbb{S}^{\text{G}\rightleftharpoons\text{A}}} \tag{C-65j}$$

$$p_{\xi\zeta}^{\vec{M}^{\text{G}\rightleftharpoons\text{A}}\vec{J}^{\text{IS}}} = -(1/2)\left\langle \vec{b}_\xi^{\vec{M}^{\text{G}\rightleftharpoons\text{A}}}, \mathcal{K}_0\left(\vec{b}_\zeta^{\vec{J}^{\text{IS}}}\right)\right\rangle_{\mathbb{S}^{\text{G}\rightleftharpoons\text{A}}} \tag{C-65k}$$

$$p_{\xi\zeta}^{\vec{M}^{\text{G}\rightleftharpoons\text{A}}\vec{M}^{\text{G}\rightleftharpoons\text{A}}} = -(1/2)\left\langle \vec{b}_\xi^{\vec{M}^{\text{G}\rightleftharpoons\text{A}}}, -j\omega\varepsilon_0\mathcal{L}_0\left(\vec{b}_\zeta^{\vec{M}^{\text{G}\rightleftharpoons\text{A}}}\right)\right\rangle_{\mathbb{S}^{\text{G}\rightleftharpoons\text{A}}} \tag{C-65l}$$

$$p_{\xi\zeta}^{\vec{M}^{\text{G}\rightleftharpoons\text{A}}\vec{M}_1^{\text{IV}}} = -(1/2)\left\langle \vec{b}_\xi^{\vec{M}^{\text{G}\rightleftharpoons\text{A}}}, -j\omega\varepsilon_0\mathcal{L}_0\left(\vec{b}_\zeta^{\vec{M}_1^{\text{IV}}}\right)\right\rangle_{\mathbb{S}^{\text{G}\rightleftharpoons\text{A}}} \tag{C-65m}$$

$$p_{\xi\zeta}^{\vec{M}^{\text{G}\rightleftharpoons\text{A}}\vec{M}_2^{\text{IV}}} = -(1/2)\left\langle \vec{b}_\xi^{\vec{M}^{\text{G}\rightleftharpoons\text{A}}}, -j\omega\varepsilon_0\mathcal{L}_0\left(\vec{b}_\zeta^{\vec{M}_2^{\text{IV}}}\right)\right\rangle_{\mathbb{S}^{\text{G}\rightleftharpoons\text{A}}} \tag{C-65n}$$

To obtain the IPO defined on modal space, we substitute Eq. (C-59) into the Eq. (C-61), and then we have that

$$P^{\text{G}\rightleftharpoons\text{A}} = \bar{a}^\dagger \cdot \underbrace{\left(\bar{\bar{T}}^\dagger \cdot \bar{\bar{P}}_{\text{curAV}}^{\text{G}\rightleftharpoons\text{A}} \cdot \bar{\bar{T}}\right)}_{\bar{\bar{P}}_{\text{cur}}^{\text{G}\rightleftharpoons\text{A}}} \cdot \bar{a} = \bar{a}^\dagger \cdot \underbrace{\left(\bar{\bar{T}}^\dagger \cdot \bar{\bar{P}}_{\text{intAV}}^{\text{G}\rightleftharpoons\text{A}} \cdot \bar{\bar{T}}\right)}_{\bar{\bar{P}}_{\text{int}}^{\text{G}\rightleftharpoons\text{A}}} \cdot \bar{a} \tag{C-66}$$

where subscripts "cur" and "int" are to emphasize that $\bar{\bar{P}}_{\text{cur}}^{\text{G}\rightleftharpoons\text{A}}$ and $\bar{\bar{P}}_{\text{int}}^{\text{G}\rightleftharpoons\text{A}}$ respectively originate from discretizing the current and interaction forms of IPO.

The process of constructing IP-DMs from orthogonalizing $\bar{\bar{P}}_{\text{cur}}^{\text{G}\rightleftharpoons\text{A}}$ and $\bar{\bar{P}}_{\text{int}}^{\text{G}\rightleftharpoons\text{A}}$ is completely the same as the one used in Sec. 6.2.4.

## C5 Surface-Volume Formulation of the PTT-Based DMT for the Augmented Tra-antenna Discussed in Sec. 6.6

The topological structure of the *augmented tra-antenna* discussed in this App. C5 is the same as the one shown in Fig. 6-67.

Because of the action of the field on $\mathbb{V}_1^{\text{A}}$, $\mathbb{V}_2^{\text{A}}$, and $\mathbb{V}_3^{\text{A}}$, some volume currents will be induced on the material bodies, and the currents are denoted as $\{\vec{J}_1^{\text{IV}}, \vec{M}_1^{\text{IV}}\}$, $\{\vec{J}_2^{\text{IV}}, \vec{M}_2^{\text{IV}}\}$, and $\{\vec{J}_3^{\text{IV}}, \vec{M}_3^{\text{IV}}\}$ respectively. For simplifying the symbolic system of the following discussions, we define region $\mathbb{V}^{\text{A}}$, material parameter $\vec{\gamma}$, and current $\vec{C}^{\text{IV}}$ as follows:

$$\mathbb{V}^{\text{A}} = \mathbb{V}_1^{\text{A}} \cup \mathbb{V}_2^{\text{A}} \cup \mathbb{V}_3^{\text{A}} \tag{C-67}$$

and

$$\vec{\gamma}(\vec{r}) = \begin{cases} \vec{\gamma}_1(\vec{r}) & , \quad \vec{r} \in \mathbb{V}_1^{\text{A}} \\ \vec{\gamma}_2(\vec{r}) & , \quad \vec{r} \in \mathbb{V}_2^{\text{A}} \\ \vec{\gamma}_3(\vec{r}) & , \quad \vec{r} \in \mathbb{V}_3^{\text{A}} \end{cases} \tag{C-68}$$

and

$$\vec{C}^{\text{IV}}(\vec{r}) = \begin{cases} \vec{C}_1^{\text{IV}}(\vec{r}) & , \quad \vec{r} \in \mathbb{V}_1^{\text{A}} \\ \vec{C}_2^{\text{IV}}(\vec{r}) & , \quad \vec{r} \in \mathbb{V}_2^{\text{A}} \\ \vec{C}_3^{\text{IV}}(\vec{r}) & , \quad \vec{r} \in \mathbb{V}_3^{\text{A}} \end{cases} \tag{C-69}$$

If the equivalent surface currents distributing on $\mathbb{S}^{\text{G}\rightleftharpoons\text{A}}$ are denoted as $\{\vec{J}^{\text{G}\rightleftharpoons\text{A}}, \vec{M}^{\text{G}\rightleftharpoons\text{A}}\}$, and the induced surface electric current distributing on $\mathbb{S}^{\text{ele}}$ is denoted as $\vec{J}^{\text{IS}}$, and the induced line electric current distributing on $\mathbb{L}^{\text{ele}}$ is denoted as $\vec{J}^{\text{IL}}$, then the field distributing on $\mathbb{V}^{\text{A}} \bigcup \mathbb{V}^{\text{F}}$ can be expressed as follows:

$$\vec{F}(\vec{r}) = \mathcal{F}_0\left(\vec{J}^{\text{G}\rightleftharpoons\text{A}} + \vec{J}^{\text{IV}} + \vec{J}^{\text{IS}} + \vec{J}^{\text{IL}}, \vec{M}^{\text{G}\rightleftharpoons\text{A}} + \vec{M}^{\text{IV}}\right) \quad , \quad \vec{r} \in \mathbb{V}^{\text{A}} \bigcup \mathbb{V}^{\text{F}} \tag{C-70}$$

where $\vec{F} = \vec{E} / \vec{H}$, and correspondingly $\mathcal{F}_0 = \mathcal{E}_0 / \mathcal{H}_0$, and the operator is the same as the one used in Sec. 6.6. The currents $\{\vec{J}^{\text{G}\rightleftharpoons\text{A}}, \vec{M}^{\text{G}\rightleftharpoons\text{A}}\}$ and the fields $\{\vec{E}, \vec{H}\}$ in Eq. (C-70) satisfy the following relations

$$\hat{n}^{\rightarrow\text{A}} \times \left[\vec{H}(\vec{r}^{\text{F}})\right]_{\vec{r}^{\text{F}} \to \vec{r}} = \vec{J}^{\text{G}\rightleftharpoons\text{A}}(\vec{r}) \quad , \quad \vec{r} \in \mathbb{S}^{\text{G}\rightleftharpoons\text{A}} \tag{C-71a}$$

$$\left[\vec{E}(\vec{r}^{\text{F}})\right]_{\vec{r}^{\text{F}} \to \vec{r}} \times \hat{n}^{\rightarrow\text{A}} = \vec{M}^{\text{G}\rightleftharpoons\text{A}}(\vec{r}) \quad , \quad \vec{r} \in \mathbb{S}^{\text{G}\rightleftharpoons\text{A}} \tag{C-71b}$$

In the above Eq. (C-71), point $\vec{r}^{\text{F}}$ belongs to the region occupied by *free space*, and $\vec{r}^{\text{F}}$ approaches the point $\vec{r}$ on $\mathbb{S}^{\text{G}\rightleftharpoons\text{A}}$; $\hat{n}^{\rightarrow\text{A}}$ is the normal direction of $\mathbb{S}^{\text{G}\rightleftharpoons\text{A}}$, and $\hat{n}^{\rightarrow\text{A}}$ points to the tra-antenna, as shown in Fig. 6-67. The currents $\{\vec{J}^{\text{IV}}, \vec{M}^{\text{IV}}\}$ and the fields $\{\vec{E}, \vec{H}\}$ in (C-70) satisfy the following relations

$$\vec{E}(\vec{r}) = \left(j\omega\Delta\vec{\vec{\varepsilon}}^{\text{c}}\right)^{-1} \cdot \vec{J}^{\text{IV}} \quad , \quad \vec{r} \in \mathbb{V}^{\text{A}} \tag{C-72a}$$

$$\vec{H}(\vec{r}) = \left(j\omega\Delta\vec{\vec{\mu}}\right)^{-1} \cdot \vec{M}^{\text{IV}} \quad , \quad \vec{r} \in \mathbb{V}^{\text{A}} \tag{C-72b}$$

In the above Eq. (C-72), $\Delta\vec{\vec{\varepsilon}}^{\text{c}} = (\vec{\vec{\varepsilon}} - j\,\vec{\vec{\sigma}}/\omega) - \vec{\vec{I}}\,\varepsilon_0$, and $\Delta\vec{\vec{\mu}} = \vec{\vec{\mu}} - \vec{\vec{I}}\,\mu_0$.

Substituting Eq. (C-70) into Eqs. (C-71a) and (C-71b), we obtain the following integral equations

$$\left[\mathcal{H}_0\left(\vec{J}^{\text{G}\rightleftharpoons\text{A}} + \vec{J}^{\text{IV}} + \vec{J}^{\text{IS}} + \vec{J}^{\text{IL}}, \vec{M}^{\text{G}\rightleftharpoons\text{A}} + \vec{M}^{\text{IV}}\right)\right]_{\vec{r}^{\text{F}} \to \vec{r}}^{\text{tan}} = \vec{J}^{\text{G}\rightleftharpoons\text{A}} \times \hat{n}^{\rightarrow\text{A}} \quad , \ \vec{r} \in \mathbb{S}^{\text{G}\rightleftharpoons\text{A}} \tag{C-73a}$$

$$\left[\mathcal{E}_0\left(\vec{J}^{\text{G}\rightleftharpoons\text{A}} + \vec{J}^{\text{IV}} + \vec{J}^{\text{IS}} + \vec{J}^{\text{IL}}, \vec{M}^{\text{G}\rightleftharpoons\text{A}} + \vec{M}^{\text{IV}}\right)\right]_{\vec{r}^{\text{F}} \to \vec{r}}^{\text{tan}} = \hat{n}^{\rightarrow\text{A}} \times \vec{M}^{\text{G}\rightleftharpoons\text{A}} \ , \ \vec{r} \in \mathbb{S}^{\text{G}\rightleftharpoons\text{A}} \tag{C-73b}$$

about currents $\{\vec{J}^{\text{G}\rightleftharpoons\text{A}}, \vec{M}^{\text{G}\rightleftharpoons\text{A}}\}$, $\{\vec{J}^{\text{IV}}, \vec{M}^{\text{IV}}\}$, $\vec{J}^{\text{IS}}$, and $\vec{J}^{\text{IL}}$, where the superscript "tan" represents the tangential component of the field. Substituting Eq. (C-70) into Eqs. (C-72a) and (C-72b), we obtain the following integral equations

$$\mathcal{E}_0\left(\vec{J}^{\text{G}\rightleftharpoons\text{A}} + \vec{J}^{\text{IV}} + \vec{J}^{\text{IS}} + \vec{J}^{\text{IL}}, \vec{M}^{\text{G}\rightleftharpoons\text{A}} + \vec{M}^{\text{IV}}\right) = \left(j\omega\Delta\vec{\vec{\varepsilon}}^{\text{c}}\right)^{-1} \cdot \vec{J}^{\text{IV}} \quad , \ \vec{r} \in \mathbb{V}^{\text{A}} \tag{C-74a}$$

$$\mathcal{H}_0\left(\vec{J}^{\text{G}\rightleftharpoons\text{A}}+\vec{J}^{\text{IV}}+\vec{J}^{\text{IS}}+\vec{J}^{\text{IL}},\vec{M}^{\text{G}\rightleftharpoons\text{A}}+\vec{M}^{\text{IV}}\right)=\left(j\omega\Delta\vec{\mu}\right)^{-1}\cdot\vec{M}^{\text{IV}} \quad , \quad \vec{r}\in\mathbb{V}^{\text{A}} \quad \text{(C-74b)}$$

about currents $\{\vec{J}^{\text{G}\rightleftharpoons\text{A}},\vec{M}^{\text{G}\rightleftharpoons\text{A}}\}$, $\{\vec{J}^{\text{IV}},\vec{M}^{\text{IV}}\}$, $\vec{J}^{\text{IS}}$, and $\vec{J}^{\text{IL}}$. Based on Eq. (C-70) and the homogeneous tangential electric field boundary conditions on $\mathbb{S}^{\text{ele}}$ and $\mathbb{L}^{\text{ele}}$, we have the following electric field integral equations

$$\left[\mathcal{E}_0\left(\vec{J}^{\text{G}\rightleftharpoons\text{A}}+\vec{J}^{\text{IV}}+\vec{J}^{\text{IS}}+\vec{J}^{\text{IL}},\vec{M}^{\text{G}\rightleftharpoons\text{A}}+\vec{M}^{\text{IV}}\right)\right]^{\text{tan}}=0 \quad , \quad \vec{r}\in\mathbb{S}^{\text{ele}} \quad \text{(C-75)}$$

$$\left[\mathcal{E}_0\left(\vec{J}^{\text{G}\rightleftharpoons\text{A}}+\vec{J}^{\text{IV}}+\vec{J}^{\text{IS}}+\vec{J}^{\text{IL}},\vec{M}^{\text{G}\rightleftharpoons\text{A}}+\vec{M}^{\text{IV}}\right)\right]^{\text{tan}}=0 \quad , \quad \vec{r}\in\mathbb{L}^{\text{ele}} \quad \text{(C-76)}$$

about currents $\{\vec{J}^{\text{G}\rightleftharpoons\text{A}},\vec{M}^{\text{G}\rightleftharpoons\text{A}}\}$, $\{\vec{J}^{\text{IV}},\vec{M}^{\text{IV}}\}$, $\vec{J}^{\text{IS}}$, and $\vec{J}^{\text{IL}}$.

If the above-mentioned currents are expanded in terms of some proper basis functions, and the Eqs. (C-73a)~(C-76) are tested with $\{\vec{b}_\xi^{\vec{M}^{\text{G}\rightleftharpoons\text{A}}}\}$, $\{\vec{b}_\xi^{\vec{J}^{\text{G}\rightleftharpoons\text{A}}}\}$, $\{\vec{b}_\xi^{\vec{J}^{\text{IV}}}\}$, $\{\vec{b}_\xi^{\vec{M}^{\text{IV}}}\}$, $\{\vec{b}_\xi^{\vec{J}^{\text{IS}}}\}$, and $\{\vec{b}_\xi^{\vec{J}^{\text{IL}}}\}$ respectively, then the equations are immediately discretized into the following matrix equations

$$\begin{aligned}0 \;=\;& \overline{\overline{Z}}^{\vec{M}^{\text{G}\rightleftharpoons\text{A}}\vec{J}^{\text{G}\rightleftharpoons\text{A}}}\cdot\overline{a}^{\vec{J}^{\text{G}\rightleftharpoons\text{A}}}+\overline{\overline{Z}}^{\vec{M}^{\text{G}\rightleftharpoons\text{A}}\vec{J}^{\text{IV}}}\cdot\overline{a}^{\vec{J}^{\text{IV}}}+\overline{\overline{Z}}^{\vec{M}^{\text{G}\rightleftharpoons\text{A}}\vec{J}^{\text{IS}}}\cdot\overline{a}^{\vec{J}^{\text{IS}}}+\overline{\overline{Z}}^{\vec{M}^{\text{G}\rightleftharpoons\text{A}}\vec{J}^{\text{IL}}}\cdot\overline{a}^{\vec{J}^{\text{IL}}}\\&+\overline{\overline{Z}}^{\vec{M}^{\text{G}\rightleftharpoons\text{A}}\vec{M}^{\text{G}\rightleftharpoons\text{A}}}\cdot\overline{a}^{\vec{M}^{\text{G}\rightleftharpoons\text{A}}}+\overline{\overline{Z}}^{\vec{M}^{\text{G}\rightleftharpoons\text{A}}\vec{M}^{\text{IV}}}\cdot\overline{a}^{\vec{M}^{\text{IV}}} \qquad \text{(C-77a)}\\0 \;=\;& \overline{\overline{Z}}^{\vec{J}^{\text{G}\rightleftharpoons\text{A}}\vec{J}^{\text{G}\rightleftharpoons\text{A}}}\cdot\overline{a}^{\vec{J}^{\text{G}\rightleftharpoons\text{A}}}+\overline{\overline{Z}}^{\vec{J}^{\text{G}\rightleftharpoons\text{A}}\vec{J}^{\text{IV}}}\cdot\overline{a}^{\vec{J}^{\text{IV}}}+\overline{\overline{Z}}^{\vec{J}^{\text{G}\rightleftharpoons\text{A}}\vec{J}^{\text{IS}}}\cdot\overline{a}^{\vec{J}^{\text{IS}}}+\overline{\overline{Z}}^{\vec{J}^{\text{G}\rightleftharpoons\text{A}}\vec{J}^{\text{IL}}}\cdot\overline{a}^{\vec{J}^{\text{IL}}}\\&+\overline{\overline{Z}}^{\vec{J}^{\text{G}\rightleftharpoons\text{A}}\vec{M}^{\text{G}\rightleftharpoons\text{A}}}\cdot\overline{a}^{\vec{M}^{\text{G}\rightleftharpoons\text{A}}}+\overline{\overline{Z}}^{\vec{J}^{\text{G}\rightleftharpoons\text{A}}\vec{M}^{\text{IV}}}\cdot\overline{a}^{\vec{M}^{\text{IV}}} \qquad \text{(C-77b)}\end{aligned}$$

and

$$\begin{aligned}0 \;=\;& \overline{\overline{Z}}^{\vec{J}^{\text{IV}}\vec{J}^{\text{G}\rightleftharpoons\text{A}}}\cdot\overline{a}^{\vec{J}^{\text{G}\rightleftharpoons\text{A}}}+\overline{\overline{Z}}^{\vec{J}^{\text{IV}}\vec{J}^{\text{IV}}}\cdot\overline{a}^{\vec{J}^{\text{IV}}}+\overline{\overline{Z}}^{\vec{J}^{\text{IV}}\vec{J}^{\text{IS}}}\cdot\overline{a}^{\vec{J}^{\text{IS}}}+\overline{\overline{Z}}^{\vec{J}^{\text{IV}}\vec{J}^{\text{IL}}}\cdot\overline{a}^{\vec{J}^{\text{IL}}}\\&+\overline{\overline{Z}}^{\vec{J}^{\text{IV}}\vec{M}^{\text{G}\rightleftharpoons\text{A}}}\cdot\overline{a}^{\vec{M}^{\text{G}\rightleftharpoons\text{A}}}+\overline{\overline{Z}}^{\vec{J}^{\text{IV}}\vec{M}^{\text{IV}}}\cdot\overline{a}^{\vec{M}^{\text{IV}}} \qquad \text{(C-78a)}\\0 \;=\;& \overline{\overline{Z}}^{\vec{M}^{\text{IV}}\vec{J}^{\text{G}\rightleftharpoons\text{A}}}\cdot\overline{a}^{\vec{J}^{\text{G}\rightleftharpoons\text{A}}}+\overline{\overline{Z}}^{\vec{M}^{\text{IV}}\vec{J}^{\text{IV}}}\cdot\overline{a}^{\vec{J}^{\text{IV}}}+\overline{\overline{Z}}^{\vec{M}^{\text{IV}}\vec{J}^{\text{IS}}}\cdot\overline{a}^{\vec{J}^{\text{IS}}}+\overline{\overline{Z}}^{\vec{M}^{\text{IV}}\vec{J}^{\text{IL}}}\cdot\overline{a}^{\vec{J}^{\text{IL}}}\\&+\overline{\overline{Z}}^{\vec{M}^{\text{IV}}\vec{M}^{\text{G}\rightleftharpoons\text{A}}}\cdot\overline{a}^{\vec{M}^{\text{G}\rightleftharpoons\text{A}}}+\overline{\overline{Z}}^{\vec{M}^{\text{IV}}\vec{M}^{\text{IV}}}\cdot\overline{a}^{\vec{M}^{\text{IV}}} \qquad \text{(C-78b)}\end{aligned}$$

and

$$\begin{aligned}0 \;=\;& \overline{\overline{Z}}^{\vec{J}^{\text{IS}}\vec{J}^{\text{G}\rightleftharpoons\text{A}}}\cdot\overline{a}^{\vec{J}^{\text{G}\rightleftharpoons\text{A}}}+\overline{\overline{Z}}^{\vec{J}^{\text{IS}}\vec{J}^{\text{IV}}}\cdot\overline{a}^{\vec{J}^{\text{IV}}}+\overline{\overline{Z}}^{\vec{J}^{\text{IS}}\vec{J}^{\text{IS}}}\cdot\overline{a}^{\vec{J}^{\text{IS}}}+\overline{\overline{Z}}^{\vec{J}^{\text{IS}}\vec{J}^{\text{IL}}}\cdot\overline{a}^{\vec{J}^{\text{IL}}}\\&+\overline{\overline{Z}}^{\vec{J}^{\text{IS}}\vec{M}^{\text{G}\rightleftharpoons\text{A}}}\cdot\overline{a}^{\vec{M}^{\text{G}\rightleftharpoons\text{A}}}+\overline{\overline{Z}}^{\vec{J}^{\text{IS}}\vec{M}^{\text{IV}}}\cdot\overline{a}^{\vec{M}^{\text{IV}}} \qquad \text{(C-79)}\\0 \;=\;& \overline{\overline{Z}}^{\vec{J}^{\text{IL}}\vec{J}^{\text{G}\rightleftharpoons\text{A}}}\cdot\overline{a}^{\vec{J}^{\text{G}\rightleftharpoons\text{A}}}+\overline{\overline{Z}}^{\vec{J}^{\text{IL}}\vec{J}^{\text{IV}}}\cdot\overline{a}^{\vec{J}^{\text{IV}}}+\overline{\overline{Z}}^{\vec{J}^{\text{IL}}\vec{J}^{\text{IS}}}\cdot\overline{a}^{\vec{J}^{\text{IS}}}+\overline{\overline{Z}}^{\vec{J}^{\text{IL}}\vec{J}^{\text{IL}}}\cdot\overline{a}^{\vec{J}^{\text{IL}}}\\&+\overline{\overline{Z}}^{\vec{J}^{\text{IL}}\vec{M}^{\text{G}\rightleftharpoons\text{A}}}\cdot\overline{a}^{\vec{M}^{\text{G}\rightleftharpoons\text{A}}}+\overline{\overline{Z}}^{\vec{J}^{\text{IL}}\vec{M}^{\text{IV}}}\cdot\overline{a}^{\vec{M}^{\text{IV}}} \qquad \text{(C-80)}\end{aligned}$$

about the corresponding current expansion vectors $\{\overline{a}^{\vec{J}^{\text{G}\rightleftharpoons\text{A}}},\overline{a}^{\vec{M}^{\text{G}\rightleftharpoons\text{A}}}\}$, $\{\overline{a}^{\vec{J}^{\text{IV}}},\overline{a}^{\vec{M}^{\text{IV}}}\}$, $\overline{a}^{\vec{J}^{\text{IS}}}$, and $\overline{a}^{\vec{J}^{\text{IL}}}$.

The formulations used to calculate the elements of the matrices in Eq. (C-77a) are as follows:

$$z_{\xi\zeta}^{\vec{M}^{\mathrm{G\rightleftharpoons A}}\vec{J}^{\mathrm{G\rightleftharpoons A}}} = \left\langle \vec{b}_{\xi}^{\vec{M}^{\mathrm{G\rightleftharpoons A}}}, \hat{n}^{\to\mathrm{A}} \times \frac{1}{2}\vec{b}_{\zeta}^{\vec{J}^{\mathrm{G\rightleftharpoons A}}} + \mathrm{P.V.}\mathcal{K}_0\left(\vec{b}_{\zeta}^{\vec{J}^{\mathrm{G\rightleftharpoons A}}}\right)\right\rangle_{\mathbb{S}^{\mathrm{G\rightleftharpoons A}}} \tag{C-81a}$$

$$z_{\xi\zeta}^{\vec{M}^{\mathrm{G\rightleftharpoons A}}\vec{J}^{\mathrm{IV}}} = \left\langle \vec{b}_{\xi}^{\vec{M}^{\mathrm{G\rightleftharpoons A}}}, \mathcal{K}_0\left(\vec{b}_{\zeta}^{\vec{J}^{\mathrm{IV}}}\right)\right\rangle_{\mathbb{S}^{\mathrm{G\rightleftharpoons A}}} \tag{C-81b}$$

$$z_{\xi\zeta}^{\vec{M}^{\mathrm{G\rightleftharpoons A}}\vec{J}^{\mathrm{IS}}} = \left\langle \vec{b}_{\xi}^{\vec{M}^{\mathrm{G\rightleftharpoons A}}}, \mathcal{K}_0\left(\vec{b}_{\zeta}^{\vec{J}^{\mathrm{IS}}}\right)\right\rangle_{\mathbb{S}^{\mathrm{G\rightleftharpoons A}}} \tag{C-81c}$$

$$z_{\xi\zeta}^{\vec{M}^{\mathrm{G\rightleftharpoons A}}\vec{J}^{\mathrm{IL}}} = \left\langle \vec{b}_{\xi}^{\vec{M}^{\mathrm{G\rightleftharpoons A}}}, \mathcal{K}_0\left(\vec{b}_{\zeta}^{\vec{J}^{\mathrm{IL}}}\right)\right\rangle_{\mathbb{S}^{\mathrm{G\rightleftharpoons A}}} \tag{C-81d}$$

$$z_{\xi\zeta}^{\vec{M}^{\mathrm{G\rightleftharpoons A}}\vec{M}^{\mathrm{G\rightleftharpoons A}}} = \left\langle \vec{b}_{\xi}^{\vec{M}^{\mathrm{G\rightleftharpoons A}}}, -j\omega\varepsilon_0\mathcal{L}_0\left(\vec{b}_{\zeta}^{\vec{M}^{\mathrm{G\rightleftharpoons A}}}\right)\right\rangle_{\mathbb{S}^{\mathrm{G\rightleftharpoons A}}} \tag{C-81e}$$

$$z_{\xi\zeta}^{\vec{M}^{\mathrm{G\rightleftharpoons A}}\vec{M}^{\mathrm{IV}}} = \left\langle \vec{b}_{\xi}^{\vec{M}^{\mathrm{G\rightleftharpoons A}}}, -j\omega\varepsilon_0\mathcal{L}_0\left(\vec{b}_{\zeta}^{\vec{M}^{\mathrm{IV}}}\right)\right\rangle_{\mathbb{S}^{\mathrm{G\rightleftharpoons A}}} \tag{C-81f}$$

The formulations used to calculate the elements of the matrices in Eq. (C-77b) are as follows:

$$z_{\xi\zeta}^{\vec{J}^{\mathrm{G\rightleftharpoons A}}\vec{J}^{\mathrm{G\rightleftharpoons A}}} = \left\langle \vec{b}_{\xi}^{\vec{J}^{\mathrm{G\rightleftharpoons A}}}, -j\omega\mu_0\mathcal{L}_0\left(\vec{b}_{\zeta}^{\vec{J}^{\mathrm{G\rightleftharpoons A}}}\right)\right\rangle_{\mathbb{S}^{\mathrm{G\rightleftharpoons A}}} \tag{C-82a}$$

$$z_{\xi\zeta}^{\vec{J}^{\mathrm{G\rightleftharpoons A}}\vec{J}^{\mathrm{IV}}} = \left\langle \vec{b}_{\xi}^{\vec{J}^{\mathrm{G\rightleftharpoons A}}}, -j\omega\mu_0\mathcal{L}_0\left(\vec{b}_{\zeta}^{\vec{J}^{\mathrm{IV}}}\right)\right\rangle_{\mathbb{S}^{\mathrm{G\rightleftharpoons A}}} \tag{C-82b}$$

$$z_{\xi\zeta}^{\vec{J}^{\mathrm{G\rightleftharpoons A}}\vec{J}^{\mathrm{IS}}} = \left\langle \vec{b}_{\xi}^{\vec{J}^{\mathrm{G\rightleftharpoons A}}}, -j\omega\mu_0\mathcal{L}_0\left(\vec{b}_{\zeta}^{\vec{J}^{\mathrm{IS}}}\right)\right\rangle_{\mathbb{S}^{\mathrm{G\rightleftharpoons A}}} \tag{C-82c}$$

$$z_{\xi\zeta}^{\vec{J}^{\mathrm{G\rightleftharpoons A}}\vec{J}^{\mathrm{IL}}} = \left\langle \vec{b}_{\xi}^{\vec{J}^{\mathrm{G\rightleftharpoons A}}}, -j\omega\mu_0\mathcal{L}_0\left(\vec{b}_{\zeta}^{\vec{J}^{\mathrm{IL}}}\right)\right\rangle_{\mathbb{S}^{\mathrm{G\rightleftharpoons A}}} \tag{C-82d}$$

$$z_{\xi\zeta}^{\vec{J}^{\mathrm{G\rightleftharpoons A}}\vec{M}^{\mathrm{G\rightleftharpoons A}}} = \left\langle \vec{b}_{\xi}^{\vec{J}^{\mathrm{G\rightleftharpoons A}}}, \frac{1}{2}\vec{b}_{\zeta}^{\vec{M}^{\mathrm{G\rightleftharpoons A}}} \times \hat{n}^{\to\mathrm{A}} - \mathrm{P.V.}\mathcal{K}_0\left(\vec{b}_{\zeta}^{\vec{M}^{\mathrm{G\rightleftharpoons A}}}\right)\right\rangle_{\mathbb{S}^{\mathrm{G\rightleftharpoons A}}} \tag{C-82e}$$

$$z_{\xi\zeta}^{\vec{J}^{\mathrm{G\rightleftharpoons A}}\vec{M}^{\mathrm{IV}}} = \left\langle \vec{b}_{\xi}^{\vec{J}^{\mathrm{G\rightleftharpoons A}}}, -\mathcal{K}_0\left(\vec{b}_{\zeta}^{\vec{M}^{\mathrm{IV}}}\right)\right\rangle_{\mathbb{S}^{\mathrm{G\rightleftharpoons A}}} \tag{C-82f}$$

The formulations used to calculate the elements of the matrices in Eq. (C-78a) are as follows:

$$z_{\xi\zeta}^{\vec{J}^{\mathrm{IV}}\vec{J}^{\mathrm{G\rightleftharpoons A}}} = \left\langle \vec{b}_{\xi}^{\vec{J}^{\mathrm{IV}}}, -j\omega\mu_0\mathcal{L}_0\left(\vec{b}_{\zeta}^{\vec{J}^{\mathrm{G\rightleftharpoons A}}}\right)\right\rangle_{\mathbb{V}^{\mathrm{A}}} \tag{C-83a}$$

$$z_{\xi\zeta}^{\vec{J}^{\mathrm{IV}}\vec{J}^{\mathrm{IV}}} = \left\langle \vec{b}_{\xi}^{\vec{J}^{\mathrm{IV}}}, -j\omega\mu_0\mathcal{L}_0\left(\vec{b}_{\zeta}^{\vec{J}^{\mathrm{IV}}}\right) - \left(j\omega\Delta\vec{\vec{\varepsilon}}^{\,\mathrm{c}}\right)^{-1}\cdot\vec{b}_{\zeta}^{\vec{J}^{\mathrm{IV}}}\right\rangle_{\mathbb{V}^{\mathrm{A}}} \tag{C-83b}$$

$$z_{\xi\zeta}^{\vec{J}^{\mathrm{IV}}\vec{J}^{\mathrm{IS}}} = \left\langle \vec{b}_{\xi}^{\vec{J}^{\mathrm{IV}}}, -j\omega\mu_0\mathcal{L}_0\left(\vec{b}_{\zeta}^{\vec{J}^{\mathrm{IS}}}\right)\right\rangle_{\mathbb{V}^{\mathrm{A}}} \tag{C-83c}$$

$$z_{\xi\zeta}^{\vec{J}^{\mathrm{IV}}\vec{J}^{\mathrm{IL}}} = \left\langle \vec{b}_{\xi}^{\vec{J}^{\mathrm{IV}}}, -j\omega\mu_0\mathcal{L}_0\left(\vec{b}_{\zeta}^{\vec{J}^{\mathrm{IL}}}\right)\right\rangle_{\mathbb{V}^{\mathrm{A}}} \tag{C-83d}$$

$$z_{\xi\zeta}^{\vec{J}^{\mathrm{IV}}\vec{M}^{\mathrm{G\rightleftharpoons A}}} = \left\langle \vec{b}_{\xi}^{\vec{J}^{\mathrm{IV}}}, -\mathcal{K}_0\left(\vec{b}_{\zeta}^{\vec{M}^{\mathrm{G\rightleftharpoons A}}}\right)\right\rangle_{\mathbb{V}^{\mathrm{A}}} \tag{C-83e}$$

$$z_{\xi\zeta}^{\vec{J}^{\mathrm{IV}}\vec{M}^{\mathrm{IV}}} = \left\langle \vec{b}_{\xi}^{\vec{J}^{\mathrm{IV}}}, -\mathcal{K}_0\left(\vec{b}_{\zeta}^{\vec{M}^{\mathrm{IV}}}\right)\right\rangle_{\mathbb{V}^{\mathrm{A}}} \tag{C-83f}$$

The formulations used to calculate the elements of the matrices in Eq. (C-78b) are as follows:

$$z_{\xi\zeta}^{\vec{M}^{\text{IV}}\vec{J}^{\text{G}\rightleftharpoons\text{A}}} = \left\langle \vec{b}_{\xi}^{\vec{M}^{\text{IV}}}, \mathcal{K}_0\left(\vec{b}_{\zeta}^{\vec{J}^{\text{G}\rightleftharpoons\text{A}}}\right)\right\rangle_{\mathbb{V}^{\text{A}}} \tag{C-84a}$$

$$z_{\xi\zeta}^{\vec{M}^{\text{IV}}\vec{J}^{\text{IV}}} = \left\langle \vec{b}_{\xi}^{\vec{M}^{\text{IV}}}, \mathcal{K}_0\left(\vec{b}_{\zeta}^{\vec{J}^{\text{IV}}}\right)\right\rangle_{\mathbb{V}^{\text{A}}} \tag{C-84b}$$

$$z_{\xi\zeta}^{\vec{M}^{\text{IV}}\vec{J}^{\text{IS}}} = \left\langle \vec{b}_{\xi}^{\vec{M}^{\text{IV}}}, \mathcal{K}_0\left(\vec{b}_{\zeta}^{\vec{J}^{\text{IS}}}\right)\right\rangle_{\mathbb{V}^{\text{A}}} \tag{C-84c}$$

$$z_{\xi\zeta}^{\vec{M}^{\text{IV}}\vec{J}^{\text{IL}}} = \left\langle \vec{b}_{\xi}^{\vec{M}^{\text{IV}}}, \mathcal{K}_0\left(\vec{b}_{\zeta}^{\vec{J}^{\text{IL}}}\right)\right\rangle_{\mathbb{V}^{\text{A}}} \tag{C-84d}$$

$$z_{\xi\zeta}^{\vec{M}^{\text{IV}}\vec{M}^{\text{G}\rightleftharpoons\text{A}}} = \left\langle \vec{b}_{\xi}^{\vec{M}^{\text{IV}}}, -j\omega\varepsilon_0\mathcal{L}_0\left(\vec{b}_{\zeta}^{\vec{M}^{\text{G}\rightleftharpoons\text{A}}}\right)\right\rangle_{\mathbb{V}^{\text{A}}} \tag{C-84e}$$

$$z_{\xi\zeta}^{\vec{M}^{\text{IV}}\vec{M}^{\text{IV}}} = \left\langle \vec{b}_{\xi}^{\vec{M}^{\text{IV}}}, -j\omega\varepsilon_0\mathcal{L}_0\left(\vec{b}_{\zeta}^{\vec{M}^{\text{IV}}}\right) - \left(j\omega\Delta\vec{\vec{\mu}}\right)^{-1}\cdot\vec{b}_{\zeta}^{\vec{M}^{\text{IV}}}\right\rangle_{\mathbb{V}^{\text{A}}} \tag{C-84f}$$

The formulations used to calculate the elements of the matrices in Eq. (C-79) are as follows:

$$z_{\xi\zeta}^{\vec{J}^{\text{IS}}\vec{J}^{\text{G}\rightleftharpoons\text{A}}} = \left\langle \vec{b}_{\xi}^{\vec{J}^{\text{IS}}}, -j\omega\mu_0\mathcal{L}_0\left(\vec{b}_{\zeta}^{\vec{J}^{\text{G}\rightleftharpoons\text{A}}}\right)\right\rangle_{\mathbb{S}^{\text{ele}}} \tag{C-85a}$$

$$z_{\xi\zeta}^{\vec{J}^{\text{IS}}\vec{J}^{\text{IV}}} = \left\langle \vec{b}_{\xi}^{\vec{J}^{\text{IS}}}, -j\omega\mu_0\mathcal{L}_0\left(\vec{b}_{\zeta}^{\vec{J}^{\text{IV}}}\right)\right\rangle_{\mathbb{S}^{\text{ele}}} \tag{C-85b}$$

$$z_{\xi\zeta}^{\vec{J}^{\text{IS}}\vec{J}^{\text{IS}}} = \left\langle \vec{b}_{\xi}^{\vec{J}^{\text{IS}}}, -j\omega\mu_0\mathcal{L}_0\left(\vec{b}_{\zeta}^{\vec{J}^{\text{IS}}}\right)\right\rangle_{\mathbb{S}^{\text{ele}}} \tag{C-85c}$$

$$z_{\xi\zeta}^{\vec{J}^{\text{IS}}\vec{J}^{\text{IL}}} = \left\langle \vec{b}_{\xi}^{\vec{J}^{\text{IS}}}, -j\omega\mu_0\mathcal{L}_0\left(\vec{b}_{\zeta}^{\vec{J}^{\text{IL}}}\right)\right\rangle_{\mathbb{S}^{\text{ele}}} \tag{C-85d}$$

$$z_{\xi\zeta}^{\vec{J}^{\text{IS}}\vec{M}^{\text{G}\rightleftharpoons\text{A}}} = \left\langle \vec{b}_{\xi}^{\vec{J}^{\text{IS}}}, -\mathcal{K}_0\left(\vec{b}_{\zeta}^{\vec{M}^{\text{G}\rightleftharpoons\text{A}}}\right)\right\rangle_{\mathbb{S}^{\text{ele}}} \tag{C-85e}$$

$$z_{\xi\zeta}^{\vec{J}^{\text{IS}}\vec{M}^{\text{IV}}} = \left\langle \vec{b}_{\xi}^{\vec{J}^{\text{IS}}}, -\mathcal{K}_0\left(\vec{b}_{\zeta}^{\vec{M}^{\text{IV}}}\right)\right\rangle_{\mathbb{S}^{\text{ele}}} \tag{C-85f}$$

The formulations used to calculate the elements of the matrices in Eq. (C-80) are as follows:

$$z_{\xi\zeta}^{\vec{J}^{\text{IL}}\vec{J}^{\text{G}\rightleftharpoons\text{A}}} = \left\langle \vec{b}_{\xi}^{\vec{J}^{\text{IL}}}, -j\omega\mu_0\mathcal{L}_0\left(\vec{b}_{\zeta}^{\vec{J}^{\text{G}\rightleftharpoons\text{A}}}\right)\right\rangle_{\mathbb{L}^{\text{ele}}} \tag{C-86a}$$

$$z_{\xi\zeta}^{\vec{J}^{\text{IL}}\vec{J}^{\text{IV}}} = \left\langle \vec{b}_{\xi}^{\vec{J}^{\text{IL}}}, -j\omega\mu_0\mathcal{L}_0\left(\vec{b}_{\zeta}^{\vec{J}^{\text{IV}}}\right)\right\rangle_{\mathbb{L}^{\text{ele}}} \tag{C-86b}$$

$$z_{\xi\zeta}^{\vec{J}^{\text{IL}}\vec{J}^{\text{IS}}} = \left\langle \vec{b}_{\xi}^{\vec{J}^{\text{IL}}}, -j\omega\mu_0\mathcal{L}_0\left(\vec{b}_{\zeta}^{\vec{J}^{\text{IS}}}\right)\right\rangle_{\mathbb{L}^{\text{ele}}} \tag{C-86c}$$

$$z_{\xi\zeta}^{\vec{J}^{\text{IL}}\vec{J}^{\text{IL}}} = \left\langle \vec{b}_{\xi}^{\vec{J}^{\text{IL}}}, -j\omega\mu_0\mathcal{L}_0\left(\vec{b}_{\zeta}^{\vec{J}^{\text{IL}}}\right)\right\rangle_{\mathbb{L}^{\text{ele}}} \tag{C-86d}$$

$$z_{\xi\zeta}^{\vec{J}^{\text{IL}}\vec{M}^{\text{G}\rightleftharpoons\text{A}}} = \left\langle \vec{b}_{\xi}^{\vec{J}^{\text{IL}}}, -\mathcal{K}_0\left(\vec{b}_{\zeta}^{\vec{M}^{\text{G}\rightleftharpoons\text{A}}}\right)\right\rangle_{\mathbb{L}^{\text{ele}}} \tag{C-86e}$$

$$z_{\xi\zeta}^{\vec{J}^{\text{IL}}\vec{M}^{\text{IV}}} = \left\langle \vec{b}_{\xi}^{\vec{J}^{\text{IL}}}, -\mathcal{K}_0\left(\vec{b}_{\zeta}^{\vec{M}^{\text{IV}}}\right)\right\rangle_{\mathbb{L}^{\text{ele}}} \tag{C-86f}$$

Employing the Eqs. (C-77a)~(C-80), we have the following transformation from $\bar{a}$ to $\bar{a}^{\text{AV}}$

$$\begin{bmatrix} \bar{a}^{\vec{J}^{\mathrm{G}\rightleftharpoons\mathrm{A}}} \\ \bar{a}^{\vec{J}^{\mathrm{IV}}} \\ \bar{a}^{\vec{J}^{\mathrm{IS}}} \\ \bar{a}^{\vec{J}^{\mathrm{IL}}} \\ \bar{a}^{\vec{M}^{\mathrm{G}\rightleftharpoons\mathrm{A}}} \\ \bar{a}^{\vec{M}^{\mathrm{IV}}} \end{bmatrix} = \bar{a}^{\mathrm{AV}} = \bar{\bar{T}} \cdot \bar{a} \tag{C-87}$$

In the above Eq. (C-87), $\bar{\bar{T}} = \bar{\bar{T}}^{\vec{J}^{\mathrm{G}\rightleftharpoons\mathrm{A}}\to\mathrm{AV}} / \bar{\bar{T}}^{\vec{M}^{\mathrm{G}\rightleftharpoons\mathrm{A}}\to\mathrm{AV}} / \bar{\bar{T}}^{\mathrm{BS}\to\mathrm{AV}}$ and correspondingly $\bar{a} = \bar{a}^{\vec{J}^{\mathrm{G}\rightleftharpoons\mathrm{A}}} / \bar{a}^{\vec{M}^{\mathrm{G}\rightleftharpoons\mathrm{A}}} / \bar{a}^{\mathrm{BS}}$, and

$$\bar{\bar{T}}^{\vec{J}^{\mathrm{G}\rightleftharpoons\mathrm{A}}\to\mathrm{AV}} = \left(\bar{\bar{\Psi}}_1\right)^{-1} \cdot \bar{\bar{\Psi}}_2 \tag{C-88a}$$

$$\bar{\bar{T}}^{\vec{M}^{\mathrm{G}\rightleftharpoons\mathrm{A}}\to\mathrm{AV}} = \left(\bar{\bar{\Psi}}_3\right)^{-1} \cdot \bar{\bar{\Psi}}_4 \tag{C-88b}$$

and

$$\bar{\bar{T}}^{\mathrm{BS}\to\mathrm{AV}} = \mathrm{nullspace}\left(\bar{\bar{\Psi}}^{\mathrm{DoJ/DoM}}_{\mathrm{FCE}}\right) \tag{C-89}$$

where

$$\bar{\bar{\Psi}}_1 = \begin{bmatrix} \bar{\bar{I}}^{\vec{J}^{\mathrm{G}\rightleftharpoons\mathrm{A}}} & 0 & 0 & 0 & 0 & 0 \\ 0 & \bar{\bar{Z}}^{\vec{M}^{\mathrm{G}\rightleftharpoons\mathrm{A}}\vec{J}^{\mathrm{IV}}} & \bar{\bar{Z}}^{\vec{M}^{\mathrm{G}\rightleftharpoons\mathrm{A}}\vec{J}^{\mathrm{IS}}} & \bar{\bar{Z}}^{\vec{M}^{\mathrm{G}\rightleftharpoons\mathrm{A}}\vec{J}^{\mathrm{IL}}} & \bar{\bar{Z}}^{\vec{M}^{\mathrm{G}\rightleftharpoons\mathrm{A}}\vec{M}^{\mathrm{G}\rightleftharpoons\mathrm{A}}} & \bar{\bar{Z}}^{\vec{M}^{\mathrm{G}\rightleftharpoons\mathrm{A}}\vec{M}^{\mathrm{IV}}} \\ 0 & \bar{\bar{Z}}^{\vec{J}^{\mathrm{IV}}\vec{J}^{\mathrm{IV}}} & \bar{\bar{Z}}^{\vec{J}^{\mathrm{IV}}\vec{J}^{\mathrm{IS}}} & \bar{\bar{Z}}^{\vec{J}^{\mathrm{IV}}\vec{J}^{\mathrm{IL}}} & \bar{\bar{Z}}^{\vec{J}^{\mathrm{IV}}\vec{M}^{\mathrm{G}\rightleftharpoons\mathrm{A}}} & \bar{\bar{Z}}^{\vec{J}^{\mathrm{IV}}\vec{M}^{\mathrm{IV}}} \\ 0 & \bar{\bar{Z}}^{\vec{M}^{\mathrm{IV}}\vec{J}^{\mathrm{IV}}} & \bar{\bar{Z}}^{\vec{M}^{\mathrm{IV}}\vec{J}^{\mathrm{IS}}} & \bar{\bar{Z}}^{\vec{M}^{\mathrm{IV}}\vec{J}^{\mathrm{IL}}} & \bar{\bar{Z}}^{\vec{M}^{\mathrm{IV}}\vec{M}^{\mathrm{G}\rightleftharpoons\mathrm{A}}} & \bar{\bar{Z}}^{\vec{M}^{\mathrm{IV}}\vec{M}^{\mathrm{IV}}} \\ 0 & \bar{\bar{Z}}^{\vec{J}^{\mathrm{IS}}\vec{J}^{\mathrm{IV}}} & \bar{\bar{Z}}^{\vec{J}^{\mathrm{IS}}\vec{J}^{\mathrm{IS}}} & \bar{\bar{Z}}^{\vec{J}^{\mathrm{IS}}\vec{J}^{\mathrm{IL}}} & \bar{\bar{Z}}^{\vec{J}^{\mathrm{IS}}\vec{M}^{\mathrm{G}\rightleftharpoons\mathrm{A}}} & \bar{\bar{Z}}^{\vec{J}^{\mathrm{IS}}\vec{M}^{\mathrm{IV}}} \\ 0 & \bar{\bar{Z}}^{\vec{J}^{\mathrm{IL}}\vec{J}^{\mathrm{IV}}} & \bar{\bar{Z}}^{\vec{J}^{\mathrm{IL}}\vec{J}^{\mathrm{IS}}} & \bar{\bar{Z}}^{\vec{J}^{\mathrm{IL}}\vec{J}^{\mathrm{IL}}} & \bar{\bar{Z}}^{\vec{J}^{\mathrm{IL}}\vec{M}^{\mathrm{G}\rightleftharpoons\mathrm{A}}} & \bar{\bar{Z}}^{\vec{J}^{\mathrm{IL}}\vec{M}^{\mathrm{IV}}} \end{bmatrix} \tag{C-90a}$$

$$\bar{\bar{\Psi}}_2 = \begin{bmatrix} \bar{\bar{I}}^{\vec{J}^{\mathrm{G}\rightleftharpoons\mathrm{A}}} \\ -\bar{\bar{Z}}^{\vec{M}^{\mathrm{G}\rightleftharpoons\mathrm{A}}\vec{J}^{\mathrm{G}\rightleftharpoons\mathrm{A}}} \\ -\bar{\bar{Z}}^{\vec{J}^{\mathrm{IV}}\vec{J}^{\mathrm{G}\rightleftharpoons\mathrm{A}}} \\ -\bar{\bar{Z}}^{\vec{M}^{\mathrm{IV}}\vec{J}^{\mathrm{G}\rightleftharpoons\mathrm{A}}} \\ -\bar{\bar{Z}}^{\vec{J}^{\mathrm{IS}}\vec{J}^{\mathrm{G}\rightleftharpoons\mathrm{A}}} \\ -\bar{\bar{Z}}^{\vec{J}^{\mathrm{IL}}\vec{J}^{\mathrm{G}\rightleftharpoons\mathrm{A}}} \end{bmatrix} \tag{C-90b}$$

and

$$\bar{\bar{\Psi}}_3 = \begin{bmatrix} 0 & 0 & 0 & 0 & \bar{\bar{I}}^{\vec{M}^{\mathrm{G}\rightleftharpoons\mathrm{A}}} & 0 \\ \bar{\bar{Z}}^{\vec{J}^{\mathrm{G}\rightleftharpoons\mathrm{A}}\vec{J}^{\mathrm{G}\rightleftharpoons\mathrm{A}}} & \bar{\bar{Z}}^{\vec{J}^{\mathrm{G}\rightleftharpoons\mathrm{A}}\vec{J}^{\mathrm{IV}}} & \bar{\bar{Z}}^{\vec{J}^{\mathrm{G}\rightleftharpoons\mathrm{A}}\vec{J}^{\mathrm{IS}}} & \bar{\bar{Z}}^{\vec{J}^{\mathrm{G}\rightleftharpoons\mathrm{A}}\vec{J}^{\mathrm{IL}}} & 0 & \bar{\bar{Z}}^{\vec{J}^{\mathrm{G}\rightleftharpoons\mathrm{A}}\vec{M}^{\mathrm{IV}}} \\ \bar{\bar{Z}}^{\vec{J}^{\mathrm{IV}}\vec{J}^{\mathrm{G}\rightleftharpoons\mathrm{A}}} & \bar{\bar{Z}}^{\vec{J}^{\mathrm{IV}}\vec{J}^{\mathrm{IV}}} & \bar{\bar{Z}}^{\vec{J}^{\mathrm{IV}}\vec{J}^{\mathrm{IS}}} & \bar{\bar{Z}}^{\vec{J}^{\mathrm{IV}}\vec{J}^{\mathrm{IL}}} & 0 & \bar{\bar{Z}}^{\vec{J}^{\mathrm{IV}}\vec{M}^{\mathrm{IV}}} \\ \bar{\bar{Z}}^{\vec{M}^{\mathrm{IV}}\vec{J}^{\mathrm{G}\rightleftharpoons\mathrm{A}}} & \bar{\bar{Z}}^{\vec{M}^{\mathrm{IV}}\vec{J}^{\mathrm{IV}}} & \bar{\bar{Z}}^{\vec{M}^{\mathrm{IV}}\vec{J}^{\mathrm{IS}}} & \bar{\bar{Z}}^{\vec{M}^{\mathrm{IV}}\vec{J}^{\mathrm{IL}}} & 0 & \bar{\bar{Z}}^{\vec{M}^{\mathrm{IV}}\vec{M}^{\mathrm{IV}}} \\ \bar{\bar{Z}}^{\vec{J}^{\mathrm{IS}}\vec{J}^{\mathrm{G}\rightleftharpoons\mathrm{A}}} & \bar{\bar{Z}}^{\vec{J}^{\mathrm{IS}}\vec{J}^{\mathrm{IV}}} & \bar{\bar{Z}}^{\vec{J}^{\mathrm{IS}}\vec{J}^{\mathrm{IS}}} & \bar{\bar{Z}}^{\vec{J}^{\mathrm{IS}}\vec{J}^{\mathrm{IL}}} & 0 & \bar{\bar{Z}}^{\vec{J}^{\mathrm{IS}}\vec{M}^{\mathrm{IV}}} \\ \bar{\bar{Z}}^{\vec{J}^{\mathrm{IL}}\vec{J}^{\mathrm{G}\rightleftharpoons\mathrm{A}}} & \bar{\bar{Z}}^{\vec{J}^{\mathrm{IL}}\vec{J}^{\mathrm{IV}}} & \bar{\bar{Z}}^{\vec{J}^{\mathrm{IL}}\vec{J}^{\mathrm{IS}}} & \bar{\bar{Z}}^{\vec{J}^{\mathrm{IL}}\vec{J}^{\mathrm{IL}}} & 0 & \bar{\bar{Z}}^{\vec{J}^{\mathrm{IL}}\vec{M}^{\mathrm{IV}}} \end{bmatrix} \tag{C-91a}$$

$$\overline{\overline{\Psi}}_4=\begin{bmatrix}\overline{\overline{I}}^{\vec{M}^{\mathrm{G}\rightleftharpoons\mathrm{A}}}\\-\overline{\overline{Z}}^{\vec{J}^{\mathrm{G}\rightleftharpoons\mathrm{A}}\vec{M}^{\mathrm{G}\rightleftharpoons\mathrm{A}}}\\-\overline{\overline{Z}}^{\vec{J}^{\mathrm{IV}}\vec{M}^{\mathrm{G}\rightleftharpoons\mathrm{A}}}\\-\overline{\overline{Z}}^{\vec{M}^{\mathrm{IV}}\vec{M}^{\mathrm{G}\rightleftharpoons\mathrm{A}}}\\-\overline{\overline{Z}}^{\vec{J}^{\mathrm{IS}}\vec{M}^{\mathrm{G}\rightleftharpoons\mathrm{A}}}\\-\overline{\overline{Z}}^{\vec{J}^{\mathrm{IL}}\vec{M}^{\mathrm{G}\rightleftharpoons\mathrm{A}}}\end{bmatrix}\tag{C-91b}$$

and

$$\overline{\overline{\Psi}}_{\mathrm{FCE}}^{\mathrm{DoJ}}=\begin{bmatrix}\overline{\overline{Z}}^{\vec{M}^{\mathrm{G}\rightleftharpoons\mathrm{A}}\vec{J}^{\mathrm{G}\rightleftharpoons\mathrm{A}}} & \overline{\overline{Z}}^{\vec{M}^{\mathrm{G}\rightleftharpoons\mathrm{A}}\vec{J}^{\mathrm{IV}}} & \overline{\overline{Z}}^{\vec{M}^{\mathrm{G}\rightleftharpoons\mathrm{A}}\vec{J}^{\mathrm{IS}}} & \overline{\overline{Z}}^{\vec{M}^{\mathrm{G}\rightleftharpoons\mathrm{A}}\vec{J}^{\mathrm{IL}}} & \overline{\overline{Z}}^{\vec{M}^{\mathrm{G}\rightleftharpoons\mathrm{A}}\vec{M}^{\mathrm{G}\rightleftharpoons\mathrm{A}}} & \overline{\overline{Z}}^{\vec{M}^{\mathrm{G}\rightleftharpoons\mathrm{A}}\vec{M}^{\mathrm{IV}}}\\ \overline{\overline{Z}}^{\vec{J}^{\mathrm{IV}}\vec{J}^{\mathrm{G}\rightleftharpoons\mathrm{A}}} & \overline{\overline{Z}}^{\vec{J}^{\mathrm{IV}}\vec{J}^{\mathrm{IV}}} & \overline{\overline{Z}}^{\vec{J}^{\mathrm{IV}}\vec{J}^{\mathrm{IS}}} & \overline{\overline{Z}}^{\vec{J}^{\mathrm{IV}}\vec{J}^{\mathrm{IL}}} & \overline{\overline{Z}}^{\vec{J}^{\mathrm{IV}}\vec{M}^{\mathrm{G}\rightleftharpoons\mathrm{A}}} & \overline{\overline{Z}}^{\vec{J}^{\mathrm{IV}}\vec{M}^{\mathrm{IV}}}\\ \overline{\overline{Z}}^{\vec{M}^{\mathrm{IV}}\vec{J}^{\mathrm{G}\rightleftharpoons\mathrm{A}}} & \overline{\overline{Z}}^{\vec{M}^{\mathrm{IV}}\vec{J}^{\mathrm{IV}}} & \overline{\overline{Z}}^{\vec{M}^{\mathrm{IV}}\vec{J}^{\mathrm{IS}}} & \overline{\overline{Z}}^{\vec{M}^{\mathrm{IV}}\vec{J}^{\mathrm{IL}}} & \overline{\overline{Z}}^{\vec{M}^{\mathrm{IV}}\vec{M}^{\mathrm{G}\rightleftharpoons\mathrm{A}}} & \overline{\overline{Z}}^{\vec{M}^{\mathrm{IV}}\vec{M}^{\mathrm{IV}}}\\ \overline{\overline{Z}}^{\vec{J}^{\mathrm{IS}}\vec{J}^{\mathrm{G}\rightleftharpoons\mathrm{A}}} & \overline{\overline{Z}}^{\vec{J}^{\mathrm{IS}}\vec{J}^{\mathrm{IV}}} & \overline{\overline{Z}}^{\vec{J}^{\mathrm{IS}}\vec{J}^{\mathrm{IS}}} & \overline{\overline{Z}}^{\vec{J}^{\mathrm{IS}}\vec{J}^{\mathrm{IL}}} & \overline{\overline{Z}}^{\vec{J}^{\mathrm{IS}}\vec{M}^{\mathrm{G}\rightleftharpoons\mathrm{A}}} & \overline{\overline{Z}}^{\vec{J}^{\mathrm{IS}}\vec{M}^{\mathrm{IV}}}\\ \overline{\overline{Z}}^{\vec{J}^{\mathrm{IL}}\vec{J}^{\mathrm{G}\rightleftharpoons\mathrm{A}}} & \overline{\overline{Z}}^{\vec{J}^{\mathrm{IL}}\vec{J}^{\mathrm{IV}}} & \overline{\overline{Z}}^{\vec{J}^{\mathrm{IL}}\vec{J}^{\mathrm{IS}}} & \overline{\overline{Z}}^{\vec{J}^{\mathrm{IL}}\vec{J}^{\mathrm{IL}}} & \overline{\overline{Z}}^{\vec{J}^{\mathrm{IL}}\vec{M}^{\mathrm{G}\rightleftharpoons\mathrm{A}}} & \overline{\overline{Z}}^{\vec{J}^{\mathrm{IL}}\vec{M}^{\mathrm{IV}}}\end{bmatrix}\tag{C-92a}$$

$$\overline{\overline{\Psi}}_{\mathrm{FCE}}^{\mathrm{DoM}}=\begin{bmatrix}\overline{\overline{Z}}^{\vec{J}^{\mathrm{G}\rightleftharpoons\mathrm{A}}\vec{J}^{\mathrm{G}\rightleftharpoons\mathrm{A}}} & \overline{\overline{Z}}^{\vec{J}^{\mathrm{G}\rightleftharpoons\mathrm{A}}\vec{J}^{\mathrm{IV}}} & \overline{\overline{Z}}^{\vec{J}^{\mathrm{G}\rightleftharpoons\mathrm{A}}\vec{J}^{\mathrm{IS}}} & \overline{\overline{Z}}^{\vec{J}^{\mathrm{G}\rightleftharpoons\mathrm{A}}\vec{J}^{\mathrm{IL}}} & \overline{\overline{Z}}^{\vec{J}^{\mathrm{G}\rightleftharpoons\mathrm{A}}\vec{M}^{\mathrm{G}\rightleftharpoons\mathrm{A}}} & \overline{\overline{Z}}^{\vec{J}^{\mathrm{G}\rightleftharpoons\mathrm{A}}\vec{M}^{\mathrm{IV}}}\\ \overline{\overline{Z}}^{\vec{J}^{\mathrm{IV}}\vec{J}^{\mathrm{G}\rightleftharpoons\mathrm{A}}} & \overline{\overline{Z}}^{\vec{J}^{\mathrm{IV}}\vec{J}^{\mathrm{IV}}} & \overline{\overline{Z}}^{\vec{J}^{\mathrm{IV}}\vec{J}^{\mathrm{IS}}} & \overline{\overline{Z}}^{\vec{J}^{\mathrm{IV}}\vec{J}^{\mathrm{IL}}} & \overline{\overline{Z}}^{\vec{J}^{\mathrm{IV}}\vec{M}^{\mathrm{G}\rightleftharpoons\mathrm{A}}} & \overline{\overline{Z}}^{\vec{J}^{\mathrm{IV}}\vec{M}^{\mathrm{IV}}}\\ \overline{\overline{Z}}^{\vec{M}^{\mathrm{IV}}\vec{J}^{\mathrm{G}\rightleftharpoons\mathrm{A}}} & \overline{\overline{Z}}^{\vec{M}^{\mathrm{IV}}\vec{J}^{\mathrm{IV}}} & \overline{\overline{Z}}^{\vec{M}^{\mathrm{IV}}\vec{J}^{\mathrm{IS}}} & \overline{\overline{Z}}^{\vec{M}^{\mathrm{IV}}\vec{J}^{\mathrm{IL}}} & \overline{\overline{Z}}^{\vec{M}^{\mathrm{IV}}\vec{M}^{\mathrm{G}\rightleftharpoons\mathrm{A}}} & \overline{\overline{Z}}^{\vec{M}^{\mathrm{IV}}\vec{M}^{\mathrm{IV}}}\\ \overline{\overline{Z}}^{\vec{J}^{\mathrm{IS}}\vec{J}^{\mathrm{G}\rightleftharpoons\mathrm{A}}} & \overline{\overline{Z}}^{\vec{J}^{\mathrm{IS}}\vec{J}^{\mathrm{IV}}} & \overline{\overline{Z}}^{\vec{J}^{\mathrm{IS}}\vec{J}^{\mathrm{IS}}} & \overline{\overline{Z}}^{\vec{J}^{\mathrm{IS}}\vec{J}^{\mathrm{IL}}} & \overline{\overline{Z}}^{\vec{J}^{\mathrm{IS}}\vec{M}^{\mathrm{G}\rightleftharpoons\mathrm{A}}} & \overline{\overline{Z}}^{\vec{J}^{\mathrm{IS}}\vec{M}^{\mathrm{IV}}}\\ \overline{\overline{Z}}^{\vec{J}^{\mathrm{IL}}\vec{J}^{\mathrm{G}\rightleftharpoons\mathrm{A}}} & \overline{\overline{Z}}^{\vec{J}^{\mathrm{IL}}\vec{J}^{\mathrm{IV}}} & \overline{\overline{Z}}^{\vec{J}^{\mathrm{IL}}\vec{J}^{\mathrm{IS}}} & \overline{\overline{Z}}^{\vec{J}^{\mathrm{IL}}\vec{J}^{\mathrm{IL}}} & \overline{\overline{Z}}^{\vec{J}^{\mathrm{IL}}\vec{M}^{\mathrm{G}\rightleftharpoons\mathrm{A}}} & \overline{\overline{Z}}^{\vec{J}^{\mathrm{IL}}\vec{M}^{\mathrm{IV}}}\end{bmatrix}\tag{C-92b}$$

Based on Eqs. (C-70) and (C-71), the IPO $P^{\mathrm{G}\rightleftharpoons\mathrm{A}}=(1/2)\iint_{\mathbb{S}^{\mathrm{G}\rightleftharpoons\mathrm{A}}}(\vec{E}\times\vec{H}^{\dagger})\cdot\hat{n}^{\rightarrow\mathrm{A}}dS$ can be rewritten as follows:

$$\begin{aligned}P^{\mathrm{G}\rightleftharpoons\mathrm{A}}=&\ \left(1/2\right)\left\langle\hat{n}^{\rightarrow\mathrm{A}}\times\vec{J}^{\mathrm{G}\rightleftharpoons\mathrm{A}},\vec{M}^{\mathrm{G}\rightleftharpoons\mathrm{A}}\right\rangle_{\mathbb{S}^{\mathrm{G}\rightleftharpoons\mathrm{A}}}\\=&-\left(1/2\right)\left\langle\vec{J}^{\mathrm{G}\rightleftharpoons\mathrm{A}},\mathcal{E}_0\left(\vec{J}^{\mathrm{G}\rightleftharpoons\mathrm{A}}+\vec{J}^{\mathrm{IV}}+\vec{J}^{\mathrm{IS}}+\vec{J}^{\mathrm{IL}},\vec{M}^{\mathrm{G}\rightleftharpoons\mathrm{A}}+\vec{M}^{\mathrm{IV}}\right)\right\rangle_{\underset{\sim}{\mathbb{S}}^{\mathrm{G}\rightleftharpoons\mathrm{A}}}\\=&-\left(1/2\right)\left\langle\vec{M}^{\mathrm{G}\rightleftharpoons\mathrm{A}},\mathcal{H}_0\left(\vec{J}^{\mathrm{G}\rightleftharpoons\mathrm{A}}+\vec{J}^{\mathrm{IV}}+\vec{J}^{\mathrm{IS}}+\vec{J}^{\mathrm{IL}},\vec{M}^{\mathrm{G}\rightleftharpoons\mathrm{A}}+\vec{M}^{\mathrm{IV}}\right)\right\rangle_{\underset{\sim}{\mathbb{S}}^{\mathrm{G}\rightleftharpoons\mathrm{A}}}^{\dagger}\end{aligned}\tag{C-93}$$

where integral surface $\underset{\sim}{\mathbb{S}}^{\mathrm{G}\rightleftharpoons\mathrm{A}}$ locates at the tra-antenna side of $\mathbb{S}^{\mathrm{G}\rightleftharpoons\mathrm{A}}$. The IPO can be discretized as follows:

$$P^{\mathrm{G}\rightleftharpoons\mathrm{A}}=\left(\bar{a}^{\mathrm{AV}}\right)^{\dagger}\cdot\overline{\overline{P}}_{\mathrm{curAV}}^{\mathrm{G}\rightleftharpoons\mathrm{A}}\cdot\bar{a}^{\mathrm{AV}}=\left(\bar{a}^{\mathrm{AV}}\right)^{\dagger}\cdot\overline{\overline{P}}_{\mathrm{intAV}}^{\mathrm{G}\rightleftharpoons\mathrm{A}}\cdot\bar{a}^{\mathrm{AV}}\tag{C-94}$$

in which

$$\overline{\overline{P}}_{\mathrm{curAV}}^{\mathrm{G}\rightleftharpoons\mathrm{A}}=\begin{bmatrix}0&0&0&0&\overline{\overline{C}}^{\vec{J}^{\mathrm{G}\rightleftharpoons\mathrm{A}}\vec{M}^{\mathrm{G}\rightleftharpoons\mathrm{A}}}&0\\0&0&0&0&0&0\\0&0&0&0&0&0\\0&0&0&0&0&0\\0&0&0&0&0&0\\0&0&0&0&0&0\end{bmatrix}\tag{C-95}$$

corresponding to the first equality in Eq. (C-93), and

$$\overline{\overline{P}}{}_{\text{intAV}}^{\text{G}\rightleftharpoons\text{A}} = \begin{bmatrix} \overline{\overline{P}}{}^{\vec{J}^{\text{G}\rightleftharpoons\text{A}}\vec{J}^{\text{G}\rightleftharpoons\text{A}}} & \overline{\overline{P}}{}^{\vec{J}^{\text{G}\rightleftharpoons\text{A}}\vec{J}^{\text{IV}}} & \overline{\overline{P}}{}^{\vec{J}^{\text{G}\rightleftharpoons\text{A}}\vec{J}^{\text{IS}}} & \overline{\overline{P}}{}^{\vec{J}^{\text{G}\rightleftharpoons\text{A}}\vec{J}^{\text{IL}}} & \overline{\overline{P}}{}^{\vec{J}^{\text{G}\rightleftharpoons\text{A}}\vec{M}^{\text{G}\rightleftharpoons\text{A}}} & \overline{\overline{P}}{}^{\vec{J}^{\text{G}\rightleftharpoons\text{A}}\vec{M}^{\text{IV}}} \\ 0 & 0 & 0 & 0 & 0 & 0 \\ 0 & 0 & 0 & 0 & 0 & 0 \\ 0 & 0 & 0 & 0 & 0 & 0 \\ 0 & 0 & 0 & 0 & 0 & 0 \\ 0 & 0 & 0 & 0 & 0 & 0 \end{bmatrix} \tag{C-96a}$$

corresponding to the second equality in Eq. (C-93), and

$$\overline{\overline{P}}{}_{\text{intAV}}^{\text{G}\rightleftharpoons\text{A}} = \begin{bmatrix} 0 & 0 & 0 & 0 & 0 & 0 \\ 0 & 0 & 0 & 0 & 0 & 0 \\ 0 & 0 & 0 & 0 & 0 & 0 \\ 0 & 0 & 0 & 0 & 0 & 0 \\ \overline{\overline{P}}{}^{\vec{M}^{\text{G}\rightleftharpoons\text{A}}\vec{J}^{\text{G}\rightleftharpoons\text{A}}} & \overline{\overline{P}}{}^{\vec{M}^{\text{G}\rightleftharpoons\text{A}}\vec{J}^{\text{IV}}} & \overline{\overline{P}}{}^{\vec{M}^{\text{G}\rightleftharpoons\text{A}}\vec{J}^{\text{IS}}} & \overline{\overline{P}}{}^{\vec{M}^{\text{G}\rightleftharpoons\text{A}}\vec{J}^{\text{IL}}} & \overline{\overline{P}}{}^{\vec{M}^{\text{G}\rightleftharpoons\text{A}}\vec{M}^{\text{G}\rightleftharpoons\text{A}}} & \overline{\overline{P}}{}^{\vec{M}^{\text{G}\rightleftharpoons\text{A}}\vec{M}^{\text{IV}}} \\ 0 & 0 & 0 & 0 & 0 & 0 \end{bmatrix}^{\dagger} \tag{C-96b}$$

corresponding to the third equality in Eq. (C-93), where the elements of the submatrices are as follows:

$$c_{\xi\zeta}^{\vec{J}^{\text{G}\rightleftharpoons\text{A}}\vec{M}^{\text{G}\rightleftharpoons\text{A}}} = (1/2)\left\langle \hat{n}^{\rightarrow\text{A}} \times \vec{b}_{\xi}^{\vec{J}^{\text{G}\rightleftharpoons\text{A}}}, \vec{b}_{\zeta}^{\vec{M}^{\text{G}\rightleftharpoons\text{A}}} \right\rangle_{\mathbb{S}^{\text{G}\rightleftharpoons\text{A}}} \tag{C-97}$$

and

$$p_{\xi\zeta}^{\vec{J}^{\text{G}\rightleftharpoons\text{A}}\vec{J}^{\text{G}\rightleftharpoons\text{A}}} = -(1/2)\left\langle \vec{b}_{\xi}^{\vec{J}^{\text{G}\rightleftharpoons\text{A}}}, -j\omega\mu_0\mathcal{L}_0\left(\vec{b}_{\zeta}^{\vec{J}^{\text{G}\rightleftharpoons\text{A}}}\right)\right\rangle_{\mathbb{S}^{\text{G}\rightleftharpoons\text{A}}} \tag{C-98a}$$

$$p_{\xi\zeta}^{\vec{J}^{\text{G}\rightleftharpoons\text{A}}\vec{J}^{\text{IV}}} = -(1/2)\left\langle \vec{b}_{\xi}^{\vec{J}^{\text{G}\rightleftharpoons\text{A}}}, -j\omega\mu_0\mathcal{L}_0\left(\vec{b}_{\zeta}^{\vec{J}^{\text{IV}}}\right)\right\rangle_{\mathbb{S}^{\text{G}\rightleftharpoons\text{A}}} \tag{C-98b}$$

$$p_{\xi\zeta}^{\vec{J}^{\text{G}\rightleftharpoons\text{A}}\vec{J}^{\text{IS}}} = -(1/2)\left\langle \vec{b}_{\xi}^{\vec{J}^{\text{G}\rightleftharpoons\text{A}}}, -j\omega\mu_0\mathcal{L}_0\left(\vec{b}_{\zeta}^{\vec{J}^{\text{IS}}}\right)\right\rangle_{\mathbb{S}^{\text{G}\rightleftharpoons\text{A}}} \tag{C-98c}$$

$$p_{\xi\zeta}^{\vec{J}^{\text{G}\rightleftharpoons\text{A}}\vec{J}^{\text{IL}}} = -(1/2)\left\langle \vec{b}_{\xi}^{\vec{J}^{\text{G}\rightleftharpoons\text{A}}}, -j\omega\mu_0\mathcal{L}_0\left(\vec{b}_{\zeta}^{\vec{J}^{\text{IL}}}\right)\right\rangle_{\mathbb{S}^{\text{G}\rightleftharpoons\text{A}}} \tag{C-98d}$$

$$p_{\xi\zeta}^{\vec{J}^{\text{G}\rightleftharpoons\text{A}}\vec{M}^{\text{G}\rightleftharpoons\text{A}}} = -(1/2)\left\langle \vec{b}_{\xi}^{\vec{J}^{\text{G}\rightleftharpoons\text{A}}}, \hat{n}^{\rightarrow\text{A}} \times \frac{1}{2}\vec{b}_{\zeta}^{\vec{M}^{\text{G}\rightleftharpoons\text{A}}} - \text{P.V.}\,\mathcal{K}_0\left(\vec{b}_{\zeta}^{\vec{M}^{\text{G}\rightleftharpoons\text{A}}}\right)\right\rangle_{\mathbb{S}^{\text{G}\rightleftharpoons\text{A}}} \tag{C-98e}$$

$$p_{\xi\zeta}^{\vec{J}^{\text{G}\rightleftharpoons\text{A}}\vec{M}^{\text{IV}}} = -(1/2)\left\langle \vec{b}_{\xi}^{\vec{J}^{\text{G}\rightleftharpoons\text{A}}}, -\mathcal{K}_0\left(\vec{b}_{\zeta}^{\vec{M}^{\text{IV}}}\right)\right\rangle_{\mathbb{S}^{\text{G}\rightleftharpoons\text{A}}} \tag{C-98f}$$

and

$$p_{\xi\zeta}^{\vec{M}^{\text{G}\rightleftharpoons\text{A}}\vec{J}^{\text{G}\rightleftharpoons\text{A}}} = -(1/2)\left\langle \vec{b}_{\xi}^{\vec{M}^{\text{G}\rightleftharpoons\text{A}}}, \frac{1}{2}\vec{b}_{\zeta}^{\vec{J}^{\text{G}\rightleftharpoons\text{A}}} \times \hat{n}^{\rightarrow\text{A}} + \text{P.V.}\,\mathcal{K}_0\left(\vec{b}_{\zeta}^{\vec{J}^{\text{G}\rightleftharpoons\text{A}}}\right)\right\rangle_{\mathbb{S}^{\text{G}\rightleftharpoons\text{A}}} \tag{C-98g}$$

$$p_{\xi\zeta}^{\vec{M}^{\text{G}\rightleftharpoons\text{A}}\vec{J}^{\text{IV}}} = -(1/2)\left\langle \vec{b}_{\xi}^{\vec{M}^{\text{G}\rightleftharpoons\text{A}}}, \mathcal{K}_0\left(\vec{b}_{\zeta}^{\vec{J}^{\text{IV}}}\right)\right\rangle_{\mathbb{S}^{\text{G}\rightleftharpoons\text{A}}} \tag{C-98h}$$

$$p_{\xi\zeta}^{\vec{M}^{\text{G}\rightleftharpoons\text{A}}\vec{J}^{\text{IS}}} = -(1/2)\left\langle \vec{b}_{\xi}^{\vec{M}^{\text{G}\rightleftharpoons\text{A}}}, \mathcal{K}_0\left(\vec{b}_{\zeta}^{\vec{J}^{\text{IS}}}\right)\right\rangle_{\mathbb{S}^{\text{G}\rightleftharpoons\text{A}}} \tag{C-98i}$$

$$p_{\xi\zeta}^{\vec{M}^{\text{G}\rightleftharpoons\text{A}}\vec{J}^{\text{IL}}} = -(1/2)\left\langle \vec{b}_{\xi}^{\vec{M}^{\text{G}\rightleftharpoons\text{A}}}, \mathcal{K}_0\left(\vec{b}_{\zeta}^{\vec{J}^{\text{IL}}}\right)\right\rangle_{\mathbb{S}^{\text{G}\rightleftharpoons\text{A}}} \tag{C-98j}$$

$$p_{\xi\zeta}^{\vec{M}^{\text{G}\rightleftharpoons\text{A}}\vec{M}^{\text{G}\rightleftharpoons\text{A}}} = -(1/2)\left\langle \vec{b}_{\xi}^{\vec{M}^{\text{G}\rightleftharpoons\text{A}}}, -j\omega\varepsilon_0\mathcal{L}_0\left(\vec{b}_{\zeta}^{\vec{M}^{\text{G}\rightleftharpoons\text{A}}}\right)\right\rangle_{\mathbb{S}^{\text{G}\rightleftharpoons\text{A}}} \tag{C-98k}$$

$$p_{\xi\zeta}^{\vec{M}^{\text{G}\rightleftharpoons\text{A}}\vec{M}^{\text{IV}}} = -(1/2)\left\langle \vec{b}_{\xi}^{\vec{M}^{\text{G}\rightleftharpoons\text{A}}}, -j\omega\varepsilon_0\mathcal{L}_0\left(\vec{b}_{\zeta}^{\vec{M}^{\text{IV}}}\right)\right\rangle_{\mathbb{S}^{\text{G}\rightleftharpoons\text{A}}} \tag{C-98l}$$

To obtain the IPO defined on modal space, we substitute Eq. (C-87) into the Eq. (C-94), and then we have that

$$P^{\text{G}\rightleftharpoons\text{A}} = \bar{a}^{\dagger}\cdot\underbrace{\left(\bar{\bar{T}}^{\dagger}\cdot\bar{\bar{P}}_{\text{curAV}}^{\text{G}\rightleftharpoons\text{A}}\cdot\bar{\bar{T}}\right)}_{\bar{\bar{P}}_{\text{cur}}^{\text{G}\rightleftharpoons\text{A}}}\cdot\bar{a} = \bar{a}^{\dagger}\cdot\underbrace{\left(\bar{\bar{T}}^{\dagger}\cdot\bar{\bar{P}}_{\text{intAV}}^{\text{G}\rightleftharpoons\text{A}}\cdot\bar{\bar{T}}\right)}_{\bar{\bar{P}}_{\text{int}}^{\text{G}\rightleftharpoons\text{A}}}\cdot\bar{a} \tag{C-99}$$

where subscripts "cur" and "int" are to emphasize that $\bar{\bar{P}}_{\text{cur}}^{\text{G}\rightleftharpoons\text{A}}$ and $\bar{\bar{P}}_{\text{int}}^{\text{G}\rightleftharpoons\text{A}}$ respectively originate from discretizing the current and interaction forms of IPO.

The process of constructing IP-DMs from orthogonalizing $\bar{\bar{P}}_{\text{cur}}^{\text{G}\rightleftharpoons\text{A}}$ and $\bar{\bar{P}}_{\text{int}}^{\text{G}\rightleftharpoons\text{A}}$ is completely the same as the one used in Sec. 6.6.

## C6 Surface-Volume Formulation of the PTT-Based DMT for the Augmented Rec-antenna Discussed in Sec. 7.4

The topological structure of the *augmented rec-antenna* discussed in this App. C6 is the same as the one shown in Fig. 7-13.

Because of the action of the field on $\mathbb{V}_{\text{A}}^1$, $\mathbb{V}_{\text{A}}^2$, and $\mathbb{V}_{\text{A}}^3$, some volume currents will be induced on the material bodies, and the currents are denoted as $\{\vec{J}_{\text{IV}}^1, \vec{M}_{\text{IV}}^1\}$, $\{\vec{J}_{\text{IV}}^2, \vec{M}_{\text{IV}}^2\}$, and $\{\vec{J}_{\text{IV}}^3, \vec{M}_{\text{IV}}^3\}$ respectively. For simplifying the symbolic system of the following discussions, we define region $\mathbb{V}_{\text{A}}$, material parameter $\vec{\gamma}$, and current $\vec{C}_{\text{IV}}$ as

$$\mathbb{V}_{\text{A}} = \mathbb{V}_{\text{A}}^1 \cup \mathbb{V}_{\text{A}}^2 \cup \mathbb{V}_{\text{A}}^3 \tag{C-100}$$

and

$$\vec{\gamma}(\vec{r}) = \begin{cases} \vec{\gamma}_1(\vec{r}) & , \quad \vec{r}\in\mathbb{V}_{\text{A}}^1 \\ \vec{\gamma}_2(\vec{r}) & , \quad \vec{r}\in\mathbb{V}_{\text{A}}^2 \\ \vec{\gamma}_3(\vec{r}) & , \quad \vec{r}\in\mathbb{V}_{\text{A}}^3 \end{cases} \tag{C-101}$$

and

$$\vec{C}_{\text{IV}}(\vec{r}) = \begin{cases} \vec{C}_{\text{IV}}^1(\vec{r}) & , \quad \vec{r}\in\mathbb{V}_{\text{A}}^1 \\ \vec{C}_{\text{IV}}^2(\vec{r}) & , \quad \vec{r}\in\mathbb{V}_{\text{A}}^2 \\ \vec{C}_{\text{IV}}^3(\vec{r}) & , \quad \vec{r}\in\mathbb{V}_{\text{A}}^3 \end{cases} \tag{C-102}$$

If the equivalent surface currents distributing on $\mathbb{S}_{\text{aux}}$ are denoted as $\{\vec{J}_{\text{aux}}, \vec{M}_{\text{aux}}\}$, and the induced surface electric current distributing on electric wall $\mathbb{S}_{\text{ele}}$ is denoted as $\vec{J}_{\text{IS}}$, and the induced line electric current distributing on electric wall $\mathbb{L}_{\text{ele}}$ is denoted as $\vec{J}_{\text{IL}}$, then the field distributing on $\mathbb{V}_{\text{A}} \bigcup \mathbb{V}_{\text{aux}}$ can be expressed as follows:

$$\vec{F}\left(\vec{r}\right) = \mathcal{F}_0\left(\vec{J}_{\text{aux}} + \vec{J}_{\text{IV}} + \vec{J}_{\text{IS}} + \vec{J}_{\text{IL}}, \vec{M}_{\text{aux}} + \vec{M}_{\text{IV}}\right) \quad , \quad \vec{r} \in \mathbb{V}_{\text{A}} \bigcup \mathbb{V}_{\text{aux}} \tag{C-103}$$

where $\vec{F} = \vec{E} / \vec{H}$, and correspondingly $\mathcal{F}_0 = \mathcal{E}_0 / \mathcal{H}_0$, and the operator is the same as the one used in Sec. 7.4. The currents $\{\vec{J}^{\text{G} \rightleftharpoons \text{A}}, \vec{M}^{\text{G} \rightleftharpoons \text{A}}\}$ and the fields $\{\vec{E}, \vec{H}\}$ in Eq. (C-103) satisfy the following relations

$$\hat{n}_{\rightarrow \text{aux}} \times \left[\vec{H}\left(\vec{r}_{\text{aux}}\right)\right]_{\vec{r}_{\text{aux}} \to \vec{r}} = \vec{J}_{\text{aux}}\left(\vec{r}\right) \quad , \quad \vec{r} \in \mathbb{S}_{\text{aux}} \tag{C-104a}$$

$$\left[\vec{E}\left(\vec{r}_{\text{aux}}\right)\right]_{\vec{r}_{\text{aux}} \to \vec{r}} \times \hat{n}_{\rightarrow \text{aux}} = \vec{M}_{\text{aux}}\left(\vec{r}\right) \quad , \quad \vec{r} \in \mathbb{S}_{\text{aux}} \tag{C-104b}$$

In the above Eq. (C-104), point $\vec{r}_{\text{aux}}$ belongs to region $\mathbb{V}_{\text{aux}}$, and $\vec{r}_{\text{aux}}$ approaches the point $\vec{r}$ on $\mathbb{S}_{\text{aux}}$; $\hat{n}_{\rightarrow \text{aux}}$ is the normal direction of $\mathbb{S}_{\text{aux}}$, and $\hat{n}_{\rightarrow \text{aux}}$ points to the interior of $\mathbb{V}_{\text{aux}}$. The currents $\{\vec{J}_{\text{IV}}, \vec{M}_{\text{IV}}\}$ and the fields $\{\vec{E}, \vec{H}\}$ in (C-103) satisfy the following relations

$$\vec{E}\left(\vec{r}\right) = \left(j\omega\Delta\vec{\vec{\varepsilon}}^{\text{c}}\right)^{-1} \cdot \vec{J}_{\text{IV}} \quad , \quad \vec{r} \in \mathbb{V}_{\text{A}} \tag{C-105a}$$

$$\vec{H}\left(\vec{r}\right) = \left(j\omega\Delta\vec{\vec{\mu}}\right)^{-1} \cdot \vec{M}_{\text{IV}} \quad , \quad \vec{r} \in \mathbb{V}_{\text{A}} \tag{C-105b}$$

In the above Eq. (C-105), $\Delta\vec{\vec{\varepsilon}}^{\text{c}} = (\vec{\vec{\varepsilon}} - j\,\vec{\vec{\sigma}}/\omega) - \vec{\vec{I}}\,\varepsilon_0$, and $\Delta\vec{\vec{\mu}} = \vec{\vec{\mu}} - \vec{\vec{I}}\,\mu_0$.

Substituting Eq. (C-103) into Eqs. (C-104a) and (C-104b), we obtain the following integral equations

$$\left[\mathcal{H}_0\left(\vec{J}_{\text{aux}} + \vec{J}_{\text{IV}} + \vec{J}_{\text{IS}} + \vec{J}_{\text{IL}}, \vec{M}_{\text{aux}} + \vec{M}_{\text{IV}}\right)\right]^{\tan}_{\vec{r}_{\text{aux}} \to \vec{r}} = \vec{J}_{\text{aux}} \times \hat{n}_{\rightarrow \text{aux}} \quad , \quad \vec{r} \in \mathbb{S}_{\text{aux}} \tag{C-106a}$$

$$\left[\mathcal{E}_0\left(\vec{J}_{\text{aux}} + \vec{J}_{\text{IV}} + \vec{J}_{\text{IS}} + \vec{J}_{\text{IL}}, \vec{M}_{\text{aux}} + \vec{M}_{\text{IV}}\right)\right]^{\tan}_{\vec{r}_{\text{aux}} \to \vec{r}} = \hat{n}_{\rightarrow \text{aux}} \times \vec{M}_{\text{aux}} \quad , \quad \vec{r} \in \mathbb{S}_{\text{aux}} \tag{C-106b}$$

about currents $\{\vec{J}_{\text{aux}}, \vec{M}_{\text{aux}}\}$, $\{\vec{J}_{\text{IV}}, \vec{M}_{\text{IV}}\}$, $\vec{J}_{\text{IS}}$, and $\vec{J}_{\text{IL}}$, where the superscript "tan" represents the tangential component of the field. Substituting Eq. (C-103) into Eqs. (C-105a) and (C-105b), we obtain the following integral equations

$$\mathcal{E}_0\left(\vec{J}_{\text{aux}} + \vec{J}_{\text{IV}} + \vec{J}_{\text{IS}} + \vec{J}_{\text{IL}}, \vec{M}_{\text{aux}} + \vec{M}_{\text{IV}}\right) = \left(j\omega\Delta\vec{\vec{\varepsilon}}^{\text{c}}\right)^{-1} \cdot \vec{J}_{\text{IV}} \quad , \quad \vec{r} \in \mathbb{V}_{\text{A}} \tag{C-107a}$$

$$\mathcal{H}_0\left(\vec{J}_{\text{aux}} + \vec{J}_{\text{IV}} + \vec{J}_{\text{IS}} + \vec{J}_{\text{IL}}, \vec{M}_{\text{aux}} + \vec{M}_{\text{IV}}\right) = \left(j\omega\Delta\vec{\vec{\mu}}\right)^{-1} \cdot \vec{M}_{\text{IV}} \quad , \quad \vec{r} \in \mathbb{V}_{\text{A}} \tag{C-107b}$$

about currents $\{\vec{J}_{\text{aux}}, \vec{M}_{\text{aux}}\}$, $\{\vec{J}_{\text{IV}}, \vec{M}_{\text{IV}}\}$, $\vec{J}_{\text{IS}}$, and $\vec{J}_{\text{IL}}$. Based on Eq. (C-103) and the homogeneous tangential electric field boundary conditions on $\mathbb{S}_{\text{ele}}$ and $\mathbb{L}_{\text{ele}}$, we have the following electric field integral equations

$$\left[\mathcal{E}_0\left(\vec{J}_{\text{aux}}+\vec{J}_{\text{IV}}+\vec{J}_{\text{IS}}+\vec{J}_{\text{IL}},\vec{M}_{\text{aux}}+\vec{M}_{\text{IV}}\right)\right]^{\tan}=0 \quad , \quad \vec{r}\in\mathbb{S}_{\text{ele}} \tag{C-108}$$

$$\left[\mathcal{E}_0\left(\vec{J}_{\text{aux}}+\vec{J}_{\text{IV}}+\vec{J}_{\text{IS}}+\vec{J}_{\text{IL}},\vec{M}_{\text{aux}}+\vec{M}_{\text{IV}}\right)\right]^{\tan}=0 \quad , \quad \vec{r}\in\mathbb{L}_{\text{ele}} \tag{C-109}$$

about currents $\{\vec{J}_{\text{aux}},\vec{M}_{\text{aux}}\}$, $\{\vec{J}_{\text{IV}},\vec{M}_{\text{IV}}\}$, $\vec{J}_{\text{IS}}$, and $\vec{J}_{\text{IL}}$.

If the above-mentioned currents are expanded in terms of some proper basis functions, and the Eqs. (C-106a)~(C-109) are tested with $\{\vec{b}_\xi^{\vec{M}_{\text{aux}}}\}$, $\{\vec{b}_\xi^{\vec{J}_{\text{aux}}}\}$, $\{\vec{b}_\xi^{\vec{J}_{\text{IV}}}\}$, $\{\vec{b}_\xi^{\vec{M}_{\text{IV}}}\}$, $\{\vec{b}_\xi^{\vec{J}_{\text{IS}}}\}$, and $\{\vec{b}_\xi^{\vec{J}_{\text{IL}}}\}$ respectively, then the equations are immediately discretized into the following matrix equations

$$\bar{\bar{Z}}^{\vec{M}_{\text{aux}}\vec{J}_{\text{aux}}}\cdot\bar{a}^{\vec{J}_{\text{aux}}}+\bar{\bar{Z}}^{\vec{M}_{\text{aux}}\vec{J}_{\text{IV}}}\cdot\bar{a}^{\vec{J}_{\text{IV}}}+\bar{\bar{Z}}^{\vec{M}_{\text{aux}}\vec{J}_{\text{IS}}}\cdot\bar{a}^{\vec{J}_{\text{IS}}}+\bar{\bar{Z}}^{\vec{M}_{\text{aux}}\vec{J}_{\text{IL}}}\cdot\bar{a}^{\vec{J}_{\text{IL}}}+\bar{\bar{Z}}^{\vec{M}_{\text{aux}}\vec{M}_{\text{aux}}}\cdot\bar{a}^{\vec{M}_{\text{aux}}}+\bar{\bar{Z}}^{\vec{M}_{\text{aux}}\vec{M}_{\text{IV}}}\cdot\bar{a}^{\vec{M}_{\text{IV}}}=0 \tag{C-110a}$$

$$\bar{\bar{Z}}^{\vec{J}_{\text{aux}}\vec{J}_{\text{aux}}}\cdot\bar{a}^{\vec{J}_{\text{aux}}}+\bar{\bar{Z}}^{\vec{J}_{\text{aux}}\vec{J}_{\text{IV}}}\cdot\bar{a}^{\vec{J}_{\text{IV}}}+\bar{\bar{Z}}^{\vec{J}_{\text{aux}}\vec{J}_{\text{IS}}}\cdot\bar{a}^{\vec{J}_{\text{IS}}}+\bar{\bar{Z}}^{\vec{J}_{\text{aux}}\vec{J}_{\text{IL}}}\cdot\bar{a}^{\vec{J}_{\text{IL}}}+\bar{\bar{Z}}^{\vec{J}_{\text{aux}}\vec{M}_{\text{aux}}}\cdot\bar{a}^{\vec{M}_{\text{aux}}}+\bar{\bar{Z}}^{\vec{J}_{\text{aux}}\vec{M}_{\text{IV}}}\cdot\bar{a}^{\vec{M}_{\text{IV}}}=0 \tag{C-110b}$$

and

$$\bar{\bar{Z}}^{\vec{J}_{\text{IV}}\vec{J}_{\text{aux}}}\cdot\bar{a}^{\vec{J}_{\text{aux}}}+\bar{\bar{Z}}^{\vec{J}_{\text{IV}}\vec{J}_{\text{IV}}}\cdot\bar{a}^{\vec{J}_{\text{IV}}}+\bar{\bar{Z}}^{\vec{J}_{\text{IV}}\vec{J}_{\text{IS}}}\cdot\bar{a}^{\vec{J}_{\text{IS}}}+\bar{\bar{Z}}^{\vec{J}_{\text{IV}}\vec{J}_{\text{IL}}}\cdot\bar{a}^{\vec{J}_{\text{IL}}}+\bar{\bar{Z}}^{\vec{J}_{\text{IV}}\vec{M}_{\text{aux}}}\cdot\bar{a}^{\vec{M}_{\text{aux}}}+\bar{\bar{Z}}^{\vec{J}_{\text{IV}}\vec{M}_{\text{IV}}}\cdot\bar{a}^{\vec{M}_{\text{IV}}}=0 \tag{C-111a}$$

$$\bar{\bar{Z}}^{\vec{M}_{\text{IV}}\vec{J}_{\text{aux}}}\cdot\bar{a}^{\vec{J}_{\text{aux}}}+\bar{\bar{Z}}^{\vec{M}_{\text{IV}}\vec{J}_{\text{IV}}}\cdot\bar{a}^{\vec{J}_{\text{IV}}}+\bar{\bar{Z}}^{\vec{M}_{\text{IV}}\vec{J}_{\text{IS}}}\cdot\bar{a}^{\vec{J}_{\text{IS}}}+\bar{\bar{Z}}^{\vec{M}_{\text{IV}}\vec{J}_{\text{IL}}}\cdot\bar{a}^{\vec{J}_{\text{IL}}}+\bar{\bar{Z}}^{\vec{M}_{\text{IV}}\vec{M}_{\text{aux}}}\cdot\bar{a}^{\vec{M}_{\text{aux}}}+\bar{\bar{Z}}^{\vec{M}_{\text{IV}}\vec{M}_{\text{IV}}}\cdot\bar{a}^{\vec{M}_{\text{IV}}}=0 \tag{C-111b}$$

and

$$\bar{\bar{Z}}^{\vec{J}_{\text{IS}}\vec{J}_{\text{aux}}}\cdot\bar{a}^{\vec{J}_{\text{aux}}}+\bar{\bar{Z}}^{\vec{J}_{\text{IS}}\vec{J}_{\text{IV}}}\cdot\bar{a}^{\vec{J}_{\text{IV}}}+\bar{\bar{Z}}^{\vec{J}_{\text{IS}}\vec{J}_{\text{IS}}}\cdot\bar{a}^{\vec{J}_{\text{IS}}}+\bar{\bar{Z}}^{\vec{J}_{\text{IS}}\vec{J}_{\text{IL}}}\cdot\bar{a}^{\vec{J}_{\text{IL}}}+\bar{\bar{Z}}^{\vec{J}_{\text{IS}}\vec{M}_{\text{aux}}}\cdot\bar{a}^{\vec{M}_{\text{aux}}}+\bar{\bar{Z}}^{\vec{J}_{\text{IS}}\vec{M}_{\text{IV}}}\cdot\bar{a}^{\vec{M}_{\text{IV}}}=0 \tag{C-112}$$

$$\bar{\bar{Z}}^{\vec{J}_{\text{IL}}\vec{J}_{\text{aux}}}\cdot\bar{a}^{\vec{J}_{\text{aux}}}+\bar{\bar{Z}}^{\vec{J}_{\text{IL}}\vec{J}_{\text{IV}}}\cdot\bar{a}^{\vec{J}_{\text{IV}}}+\bar{\bar{Z}}^{\vec{J}_{\text{IL}}\vec{J}_{\text{IS}}}\cdot\bar{a}^{\vec{J}_{\text{IS}}}+\bar{\bar{Z}}^{\vec{J}_{\text{IL}}\vec{J}_{\text{IL}}}\cdot\bar{a}^{\vec{J}_{\text{IL}}}+\bar{\bar{Z}}^{\vec{J}_{\text{IL}}\vec{M}_{\text{aux}}}\cdot\bar{a}^{\vec{M}_{\text{aux}}}+\bar{\bar{Z}}^{\vec{J}_{\text{IL}}\vec{M}_{\text{IV}}}\cdot\bar{a}^{\vec{M}_{\text{IV}}}=0 \tag{C-113}$$

about vectors $\{\bar{a}^{\vec{J}_{\text{aux}}},\bar{a}^{\vec{M}_{\text{aux}}}\}$, $\{\bar{a}^{\vec{J}_{\text{IV}}},\bar{a}^{\vec{M}_{\text{IV}}}\}$, $\bar{a}^{\vec{J}_{\text{IS}}}$, and $\bar{a}^{\vec{J}_{\text{IL}}}$.

The formulations used to calculate the elements of the matrices in Eq. (C-110a) are as follows:

$$z_{\xi\zeta}^{\vec{M}_{\text{aux}}\vec{J}_{\text{aux}}}=\left\langle\vec{b}_\xi^{\vec{M}_{\text{aux}}},\hat{n}_{\to\text{aux}}\times\frac{1}{2}\vec{b}_\zeta^{\vec{J}_{\text{aux}}}+\text{P.V.}\,\mathcal{K}_0\left(\vec{b}_\zeta^{\vec{J}_{\text{aux}}}\right)\right\rangle_{\mathbb{S}_{\text{aux}}} \tag{C-114a}$$

$$z_{\xi\zeta}^{\vec{M}_{\text{aux}}\vec{J}_{\text{IV}}}=\left\langle\vec{b}_\xi^{\vec{M}_{\text{aux}}},\mathcal{K}_0\left(\vec{b}_\zeta^{\vec{J}_{\text{IV}}}\right)\right\rangle_{\mathbb{S}_{\text{aux}}} \tag{C-114b}$$

$$z_{\xi\zeta}^{\vec{M}_{\text{aux}}\vec{J}_{\text{IS}}}=\left\langle\vec{b}_\xi^{\vec{M}_{\text{aux}}},\mathcal{K}_0\left(\vec{b}_\zeta^{\vec{J}_{\text{IS}}}\right)\right\rangle_{\mathbb{S}_{\text{aux}}} \tag{C-114c}$$

$$z_{\xi\zeta}^{\vec{M}_{\text{aux}}\vec{J}_{\text{IL}}}=\left\langle\vec{b}_\xi^{\vec{M}_{\text{aux}}},\mathcal{K}_0\left(\vec{b}_\zeta^{\vec{J}_{\text{IL}}}\right)\right\rangle_{\mathbb{S}_{\text{aux}}} \tag{C-114d}$$

$$z_{\xi\zeta}^{\vec{M}_{\text{aux}}\vec{M}_{\text{aux}}}=\left\langle\vec{b}_\xi^{\vec{M}_{\text{aux}}},-j\omega\varepsilon_0\mathcal{L}_0\left(\vec{b}_\zeta^{\vec{M}_{\text{aux}}}\right)\right\rangle_{\mathbb{S}_{\text{aux}}} \tag{C-114e}$$

$$z_{\xi\zeta}^{\vec{M}_{\text{aux}}\vec{M}_{\text{IV}}}=\left\langle\vec{b}_\xi^{\vec{M}_{\text{aux}}},-j\omega\varepsilon_0\mathcal{L}_0\left(\vec{b}_\zeta^{\vec{M}_{\text{IV}}}\right)\right\rangle_{\mathbb{S}_{\text{aux}}} \tag{C-114f}$$

The formulations used to calculate the elements of the matrices in Eq. (C-110b) are as follows:

$$z_{\xi\zeta}^{\vec{J}_{\rm aux}\vec{J}_{\rm aux}} = \left\langle \vec{b}_{\xi}^{\vec{J}_{\rm aux}}, -j\omega\mu_0\mathcal{L}_0\left(\vec{b}_{\zeta}^{\vec{J}_{\rm aux}}\right)\right\rangle_{\mathbb{S}_{\rm aux}} \tag{C-115a}$$

$$z_{\xi\zeta}^{\vec{J}_{\rm aux}\vec{J}_{\rm IV}} = \left\langle \vec{b}_{\xi}^{\vec{J}_{\rm aux}}, -j\omega\mu_0\mathcal{L}_0\left(\vec{b}_{\zeta}^{\vec{J}_{\rm IV}}\right)\right\rangle_{\mathbb{S}_{\rm aux}} \tag{C-115b}$$

$$z_{\xi\zeta}^{\vec{J}_{\rm aux}\vec{J}_{\rm IS}} = \left\langle \vec{b}_{\xi}^{\vec{J}_{\rm aux}}, -j\omega\mu_0\mathcal{L}_0\left(\vec{b}_{\zeta}^{\vec{J}_{\rm IS}}\right)\right\rangle_{\mathbb{S}_{\rm aux}} \tag{C-115c}$$

$$z_{\xi\zeta}^{\vec{J}_{\rm aux}\vec{J}_{\rm IL}} = \left\langle \vec{b}_{\xi}^{\vec{J}_{\rm aux}}, -j\omega\mu_0\mathcal{L}_0\left(\vec{b}_{\zeta}^{\vec{J}_{\rm IL}}\right)\right\rangle_{\mathbb{S}_{\rm aux}} \tag{C-115d}$$

$$z_{\xi\zeta}^{\vec{J}_{\rm aux}\vec{M}_{\rm aux}} = \left\langle \vec{b}_{\xi}^{\vec{J}_{\rm aux}}, \frac{1}{2}\vec{b}_{\zeta}^{\vec{M}_{\rm aux}}\times\hat{n}_{\rightarrow\rm aux} - {\rm P.V.}\,\mathcal{K}_0\left(\vec{b}_{\zeta}^{\vec{M}_{\rm aux}}\right)\right\rangle_{\mathbb{S}_{\rm aux}} \tag{C-115e}$$

$$z_{\xi\zeta}^{\vec{J}_{\rm aux}\vec{M}_{\rm IV}} = \left\langle \vec{b}_{\xi}^{\vec{J}_{\rm aux}}, -\mathcal{K}_0\left(\vec{b}_{\zeta}^{\vec{M}_{\rm IV}}\right)\right\rangle_{\mathbb{S}_{\rm aux}} \tag{C-115f}$$

The formulations used to calculate the elements of the matrices in Eq. (C-111a) are as follows:

$$z_{\xi\zeta}^{\vec{J}_{\rm IV}\vec{J}_{\rm aux}} = \left\langle \vec{b}_{\xi}^{\vec{J}_{\rm IV}}, -j\omega\mu_0\mathcal{L}_0\left(\vec{b}_{\zeta}^{\vec{J}_{\rm aux}}\right)\right\rangle_{\mathbb{V}_{\rm A}} \tag{C-116a}$$

$$z_{\xi\zeta}^{\vec{J}_{\rm IV}\vec{J}_{\rm IV}} = \left\langle \vec{b}_{\xi}^{\vec{J}_{\rm IV}}, -j\omega\mu_0\mathcal{L}_0\left(\vec{b}_{\zeta}^{\vec{J}_{\rm IV}}\right) - \left(j\omega\Delta\vec{\vec{\varepsilon}}^{\rm c}\right)^{-1}\cdot\vec{b}_{\zeta}^{\vec{J}_{\rm IV}}\right\rangle_{\mathbb{V}_{\rm A}} \tag{C-116b}$$

$$z_{\xi\zeta}^{\vec{J}_{\rm IV}\vec{J}_{\rm IS}} = \left\langle \vec{b}_{\xi}^{\vec{J}_{\rm IV}}, -j\omega\mu_0\mathcal{L}_0\left(\vec{b}_{\zeta}^{\vec{J}_{\rm IS}}\right)\right\rangle_{\mathbb{V}_{\rm A}} \tag{C-116c}$$

$$z_{\xi\zeta}^{\vec{J}_{\rm IV}\vec{J}_{\rm IL}} = \left\langle \vec{b}_{\xi}^{\vec{J}_{\rm IV}}, -j\omega\mu_0\mathcal{L}_0\left(\vec{b}_{\zeta}^{\vec{J}_{\rm IL}}\right)\right\rangle_{\mathbb{V}_{\rm A}} \tag{C-116d}$$

$$z_{\xi\zeta}^{\vec{J}_{\rm IV}\vec{M}_{\rm aux}} = \left\langle \vec{b}_{\xi}^{\vec{J}_{\rm IV}}, -\mathcal{K}_0\left(\vec{b}_{\zeta}^{\vec{M}_{\rm aux}}\right)\right\rangle_{\mathbb{V}_{\rm A}} \tag{C-116e}$$

$$z_{\xi\zeta}^{\vec{J}_{\rm IV}\vec{M}_{\rm IV}} = \left\langle \vec{b}_{\xi}^{\vec{J}_{\rm IV}}, -\mathcal{K}_0\left(\vec{b}_{\zeta}^{\vec{M}_{\rm IV}}\right)\right\rangle_{\mathbb{V}_{\rm A}} \tag{C-116f}$$

The formulations used to calculate the elements of the matrices in Eq. (C-111b) are as follows:

$$z_{\xi\zeta}^{\vec{M}_{\rm IV}\vec{J}_{\rm aux}} = \left\langle \vec{b}_{\xi}^{\vec{M}_{\rm IV}}, \mathcal{K}_0\left(\vec{b}_{\zeta}^{\vec{J}_{\rm aux}}\right)\right\rangle_{\mathbb{V}_{\rm A}} \tag{C-117a}$$

$$z_{\xi\zeta}^{\vec{M}_{\rm IV}\vec{J}_{\rm IV}} = \left\langle \vec{b}_{\xi}^{\vec{M}_{\rm IV}}, \mathcal{K}_0\left(\vec{b}_{\zeta}^{\vec{J}_{\rm IV}}\right)\right\rangle_{\mathbb{V}_{\rm A}} \tag{C-117b}$$

$$z_{\xi\zeta}^{\vec{M}_{\rm IV}\vec{J}_{\rm IS}} = \left\langle \vec{b}_{\xi}^{\vec{M}_{\rm IV}}, \mathcal{K}_0\left(\vec{b}_{\zeta}^{\vec{J}_{\rm IS}}\right)\right\rangle_{\mathbb{V}_{\rm A}} \tag{C-117c}$$

$$z_{\xi\zeta}^{\vec{M}_{\rm IV}\vec{J}_{\rm IL}} = \left\langle \vec{b}_{\xi}^{\vec{M}_{\rm IV}}, \mathcal{K}_0\left(\vec{b}_{\zeta}^{\vec{J}_{\rm IL}}\right)\right\rangle_{\mathbb{V}_{\rm A}} \tag{C-117d}$$

$$z_{\xi\zeta}^{\vec{M}_{\rm IV}\vec{M}_{\rm aux}} = \left\langle \vec{b}_{\xi}^{\vec{M}_{\rm IV}}, -j\omega\varepsilon_0\mathcal{L}_0\left(\vec{b}_{\zeta}^{\vec{M}_{\rm aux}}\right)\right\rangle_{\mathbb{V}_{\rm A}} \tag{C-117e}$$

$$z_{\xi\zeta}^{\vec{M}_{\rm IV}\vec{M}_{\rm IV}} = \left\langle \vec{b}_{\xi}^{\vec{M}_{\rm IV}}, -j\omega\varepsilon_0\mathcal{L}_0\left(\vec{b}_{\zeta}^{\vec{M}_{\rm IV}}\right) - \left(j\omega\Delta\vec{\vec{\mu}}\right)^{-1}\cdot\vec{b}_{\zeta}^{\vec{M}_{\rm IV}}\right\rangle_{\mathbb{V}_{\rm A}} \tag{C-117f}$$

The formulas used to calculate the elements of the matrices in Eq. (C-112) are as follows:

$$z_{\xi\zeta}^{\vec{J}_{\rm IS}\vec{J}_{\rm aux}} = \left\langle \vec{b}_{\xi}^{\vec{J}_{\rm IS}}, -j\omega\mu_0\mathcal{L}_0\left(\vec{b}_{\zeta}^{\vec{J}_{\rm aux}}\right)\right\rangle_{\mathbb{S}_{\rm ele}} \tag{C-118a}$$

$$z_{\xi\zeta}^{\vec{J}_{\rm IS}\vec{J}_{\rm IV}} = \left\langle \vec{b}_{\xi}^{\vec{J}_{\rm IS}}, -j\omega\mu_0\mathcal{L}_0\left(\vec{b}_{\zeta}^{\vec{J}_{\rm IV}}\right)\right\rangle_{\mathbb{S}_{\rm ele}} \tag{C-118b}$$

$$z_{\xi\zeta}^{\vec{J}_{\rm IS}\vec{J}_{\rm IS}} = \left\langle \vec{b}_{\xi}^{\vec{J}_{\rm IS}}, -j\omega\mu_0\mathcal{L}_0\left(\vec{b}_{\zeta}^{\vec{J}_{\rm IS}}\right)\right\rangle_{\mathbb{S}_{\rm ele}} \tag{C-118c}$$

$$z_{\xi\zeta}^{\vec{J}_{\rm IS}\vec{J}_{\rm IL}} = \left\langle \vec{b}_{\xi}^{\vec{J}_{\rm IS}}, -j\omega\mu_0\mathcal{L}_0\left(\vec{b}_{\zeta}^{\vec{J}_{\rm IL}}\right)\right\rangle_{\mathbb{S}_{\rm ele}} \tag{C-118d}$$

$$z_{\xi\zeta}^{\vec{J}_{\rm IS}\vec{M}_{\rm aux}} = \left\langle \vec{b}_{\xi}^{\vec{J}_{\rm IS}}, -\mathcal{K}_0\left(\vec{b}_{\zeta}^{\vec{M}_{\rm aux}}\right)\right\rangle_{\mathbb{S}_{\rm ele}} \tag{C-118e}$$

$$z_{\xi\zeta}^{\vec{J}_{\rm IS}\vec{M}_{\rm IV}} = \left\langle \vec{b}_{\xi}^{\vec{J}_{\rm IS}}, -\mathcal{K}_0\left(\vec{b}_{\zeta}^{\vec{M}_{\rm IV}}\right)\right\rangle_{\mathbb{S}_{\rm ele}} \tag{C-118f}$$

The formulations used to calculate the elements of the matrices in Eq. (C-113) are as follows:

$$z_{\xi\zeta}^{\vec{J}_{\rm IL}\vec{J}_{\rm aux}} = \left\langle \vec{b}_{\xi}^{\vec{J}_{\rm IL}}, -j\omega\mu_0\mathcal{L}_0\left(\vec{b}_{\zeta}^{\vec{J}_{\rm aux}}\right)\right\rangle_{\mathbb{L}_{\rm ele}} \tag{C-119a}$$

$$z_{\xi\zeta}^{\vec{J}_{\rm IL}\vec{J}_{\rm IV}} = \left\langle \vec{b}_{\xi}^{\vec{J}_{\rm IL}}, -j\omega\mu_0\mathcal{L}_0\left(\vec{b}_{\zeta}^{\vec{J}_{\rm IV}}\right)\right\rangle_{\mathbb{L}_{\rm ele}} \tag{C-119b}$$

$$z_{\xi\zeta}^{\vec{J}_{\rm IL}\vec{J}_{\rm IS}} = \left\langle \vec{b}_{\xi}^{\vec{J}_{\rm IL}}, -j\omega\mu_0\mathcal{L}_0\left(\vec{b}_{\zeta}^{\vec{J}_{\rm IS}}\right)\right\rangle_{\mathbb{L}_{\rm ele}} \tag{C-119c}$$

$$z_{\xi\zeta}^{\vec{J}_{\rm IL}\vec{J}_{\rm IL}} = \left\langle \vec{b}_{\xi}^{\vec{J}_{\rm IL}}, -j\omega\mu_0\mathcal{L}_0\left(\vec{b}_{\zeta}^{\vec{J}_{\rm IL}}\right)\right\rangle_{\mathbb{L}_{\rm ele}} \tag{C-119d}$$

$$z_{\xi\zeta}^{\vec{J}_{\rm IL}\vec{M}_{\rm aux}} = \left\langle \vec{b}_{\xi}^{\vec{J}_{\rm IL}}, -\mathcal{K}_0\left(\vec{b}_{\zeta}^{\vec{M}_{\rm aux}}\right)\right\rangle_{\mathbb{L}_{\rm ele}} \tag{C-119e}$$

$$z_{\xi\zeta}^{\vec{J}_{\rm IL}\vec{M}_{\rm IV}} = \left\langle \vec{b}_{\xi}^{\vec{J}_{\rm IL}}, -\mathcal{K}_0\left(\vec{b}_{\zeta}^{\vec{M}_{\rm IV}}\right)\right\rangle_{\mathbb{L}_{\rm ele}} \tag{C-119f}$$

Employing the Eqs. (C-110a)~(C-113), we have the following transformation from $\bar{a}$ to $\bar{a}^{\rm AV}$

$$\begin{bmatrix} \bar{a}^{\vec{J}_{\rm aux}} \\ \bar{a}^{\vec{J}_{\rm IV}} \\ \bar{a}^{\vec{J}_{\rm IS}} \\ \bar{a}^{\vec{J}_{\rm IL}} \\ \bar{a}^{\vec{M}_{\rm aux}} \\ \bar{a}^{\vec{M}_{\rm IV}} \end{bmatrix} = \bar{a}^{\rm AV} = \bar{\bar{T}} \cdot \bar{a} \tag{C-120}$$

In the above Eq. (C-120), $\bar{\bar{T}} = \bar{\bar{T}}^{\vec{J}_{\rm aux}\to{\rm AV}} / \bar{\bar{T}}^{\vec{M}_{\rm aux}\to{\rm AV}} / \bar{\bar{T}}^{\rm BS\to AV}$ and correspondingly $\bar{a} = \bar{a}^{\vec{J}_{\rm aux}} / \bar{a}^{\vec{M}_{\rm aux}} / \bar{a}^{\rm BS}$, and

$$\bar{\bar{T}}^{\vec{J}_{\rm aux}\to{\rm AV}} = \left(\bar{\bar{\Psi}}_1\right)^{-1} \cdot \bar{\bar{\Psi}}_2 \tag{C-121a}$$

$$\bar{\bar{T}}^{\vec{M}_{\rm aux}\to{\rm AV}} = \left(\bar{\bar{\Psi}}_3\right)^{-1} \cdot \bar{\bar{\Psi}}_4 \tag{C-121b}$$

and

$$\overline{\overline{T}}^{\text{BS}\to\text{AV}} = \text{nullspace}\left(\overline{\overline{\Psi}}_{\text{FCE}}^{\text{DoJ/DoM}}\right) \tag{C-122}$$

where

$$\overline{\overline{\Psi}}_1 = \begin{bmatrix} \overline{\overline{I}}^{\vec{J}_{\text{aux}}} & 0 & 0 & 0 & 0 & 0 \\ 0 & \overline{\overline{Z}}^{\vec{M}_{\text{aux}}\vec{J}_{\text{IV}}} & \overline{\overline{Z}}^{\vec{M}_{\text{aux}}\vec{J}_{\text{IS}}} & \overline{\overline{Z}}^{\vec{M}_{\text{aux}}\vec{J}_{\text{IL}}} & \overline{\overline{Z}}^{\vec{M}_{\text{aux}}\vec{M}_{\text{aux}}} & \overline{\overline{Z}}^{\vec{M}_{\text{aux}}\vec{M}_{\text{IV}}} \\ 0 & \overline{\overline{Z}}^{\vec{J}_{\text{IV}}\vec{J}_{\text{IV}}} & \overline{\overline{Z}}^{\vec{J}_{\text{IV}}\vec{J}_{\text{IS}}} & \overline{\overline{Z}}^{\vec{J}_{\text{IV}}\vec{J}_{\text{IL}}} & \overline{\overline{Z}}^{\vec{J}_{\text{IV}}\vec{M}_{\text{aux}}} & \overline{\overline{Z}}^{\vec{J}_{\text{IV}}\vec{M}_{\text{IV}}} \\ 0 & \overline{\overline{Z}}^{\vec{M}_{\text{IV}}\vec{J}_{\text{IV}}} & \overline{\overline{Z}}^{\vec{M}_{\text{IV}}\vec{J}_{\text{IS}}} & \overline{\overline{Z}}^{\vec{M}_{\text{IV}}\vec{J}_{\text{IL}}} & \overline{\overline{Z}}^{\vec{M}_{\text{IV}}\vec{M}_{\text{aux}}} & \overline{\overline{Z}}^{\vec{M}_{\text{IV}}\vec{M}_{\text{IV}}} \\ 0 & \overline{\overline{Z}}^{\vec{J}_{\text{IS}}\vec{J}_{\text{IV}}} & \overline{\overline{Z}}^{\vec{J}_{\text{IS}}\vec{J}_{\text{IS}}} & \overline{\overline{Z}}^{\vec{J}_{\text{IS}}\vec{J}_{\text{IL}}} & \overline{\overline{Z}}^{\vec{J}_{\text{IS}}\vec{M}_{\text{aux}}} & \overline{\overline{Z}}^{\vec{J}_{\text{IS}}\vec{M}_{\text{IV}}} \\ 0 & \overline{\overline{Z}}^{\vec{J}_{\text{IL}}\vec{J}_{\text{IV}}} & \overline{\overline{Z}}^{\vec{J}_{\text{IL}}\vec{J}_{\text{IS}}} & \overline{\overline{Z}}^{\vec{J}_{\text{IL}}\vec{J}_{\text{IL}}} & \overline{\overline{Z}}^{\vec{J}_{\text{IL}}\vec{M}_{\text{aux}}} & \overline{\overline{Z}}^{\vec{J}_{\text{IL}}\vec{M}_{\text{IV}}} \end{bmatrix} \tag{C-123a}$$

$$\overline{\overline{\Psi}}_2 = \begin{bmatrix} \overline{\overline{I}}^{\vec{J}_{\text{aux}}} \\ -\overline{\overline{Z}}^{\vec{M}_{\text{aux}}\vec{J}_{\text{aux}}} \\ -\overline{\overline{Z}}^{\vec{J}_{\text{IV}}\vec{J}_{\text{aux}}} \\ -\overline{\overline{Z}}^{\vec{M}_{\text{IV}}\vec{J}_{\text{aux}}} \\ -\overline{\overline{Z}}^{\vec{J}_{\text{IS}}\vec{J}_{\text{aux}}} \\ -\overline{\overline{Z}}^{\vec{J}_{\text{IL}}\vec{J}_{\text{aux}}} \end{bmatrix} \tag{C-123b}$$

and

$$\overline{\overline{\Psi}}_3 = \begin{bmatrix} 0 & 0 & 0 & 0 & \overline{\overline{I}}^{\vec{M}_{\text{aux}}} & 0 \\ \overline{\overline{Z}}^{\vec{J}_{\text{aux}}\vec{J}_{\text{aux}}} & \overline{\overline{Z}}^{\vec{J}_{\text{aux}}\vec{J}_{\text{IV}}} & \overline{\overline{Z}}^{\vec{J}_{\text{aux}}\vec{J}_{\text{IS}}} & \overline{\overline{Z}}^{\vec{J}_{\text{aux}}\vec{J}_{\text{IL}}} & 0 & \overline{\overline{Z}}^{\vec{J}_{\text{aux}}\vec{M}_{\text{IV}}} \\ \overline{\overline{Z}}^{\vec{J}_{\text{IV}}\vec{J}_{\text{aux}}} & \overline{\overline{Z}}^{\vec{J}_{\text{IV}}\vec{J}_{\text{IV}}} & \overline{\overline{Z}}^{\vec{J}_{\text{IV}}\vec{J}_{\text{IS}}} & \overline{\overline{Z}}^{\vec{J}_{\text{IV}}\vec{J}_{\text{IL}}} & 0 & \overline{\overline{Z}}^{\vec{J}_{\text{IV}}\vec{M}_{\text{IV}}} \\ \overline{\overline{Z}}^{\vec{M}_{\text{IV}}\vec{J}_{\text{aux}}} & \overline{\overline{Z}}^{\vec{M}_{\text{IV}}\vec{J}_{\text{IV}}} & \overline{\overline{Z}}^{\vec{M}_{\text{IV}}\vec{J}_{\text{IS}}} & \overline{\overline{Z}}^{\vec{M}_{\text{IV}}\vec{J}_{\text{IL}}} & 0 & \overline{\overline{Z}}^{\vec{M}_{\text{IV}}\vec{M}_{\text{IV}}} \\ \overline{\overline{Z}}^{\vec{J}_{\text{IS}}\vec{J}_{\text{aux}}} & \overline{\overline{Z}}^{\vec{J}_{\text{IS}}\vec{J}_{\text{IV}}} & \overline{\overline{Z}}^{\vec{J}_{\text{IS}}\vec{J}_{\text{IS}}} & \overline{\overline{Z}}^{\vec{J}_{\text{IS}}\vec{J}_{\text{IL}}} & 0 & \overline{\overline{Z}}^{\vec{J}_{\text{IS}}\vec{M}_{\text{IV}}} \\ \overline{\overline{Z}}^{\vec{J}_{\text{IL}}\vec{J}_{\text{aux}}} & \overline{\overline{Z}}^{\vec{J}_{\text{IL}}\vec{J}_{\text{IV}}} & \overline{\overline{Z}}^{\vec{J}_{\text{IL}}\vec{J}_{\text{IS}}} & \overline{\overline{Z}}^{\vec{J}_{\text{IL}}\vec{J}_{\text{IL}}} & 0 & \overline{\overline{Z}}^{\vec{J}_{\text{IL}}\vec{M}_{\text{IV}}} \end{bmatrix} \tag{C-124a}$$

$$\overline{\overline{\Psi}}_4 = \begin{bmatrix} \overline{\overline{I}}^{\vec{M}_{\text{aux}}} \\ -\overline{\overline{Z}}^{\vec{J}_{\text{aux}}\vec{M}_{\text{aux}}} \\ -\overline{\overline{Z}}^{\vec{J}_{\text{IV}}\vec{M}_{\text{aux}}} \\ -\overline{\overline{Z}}^{\vec{M}_{\text{IV}}\vec{M}_{\text{aux}}} \\ -\overline{\overline{Z}}^{\vec{J}_{\text{IS}}\vec{M}_{\text{aux}}} \\ -\overline{\overline{Z}}^{\vec{J}_{\text{IL}}\vec{M}_{\text{aux}}} \end{bmatrix} \tag{C-124b}$$

and

$$\overline{\overline{\Psi}}_{\text{FCE}}^{\text{DoJ}} = \begin{bmatrix} \overline{\overline{Z}}^{\vec{M}_{\text{aux}}\vec{J}_{\text{aux}}} & \overline{\overline{Z}}^{\vec{M}_{\text{aux}}\vec{J}_{\text{IV}}} & \overline{\overline{Z}}^{\vec{M}_{\text{aux}}\vec{J}_{\text{IS}}} & \overline{\overline{Z}}^{\vec{M}_{\text{aux}}\vec{J}_{\text{IL}}} & \overline{\overline{Z}}^{\vec{M}_{\text{aux}}\vec{M}_{\text{aux}}} & \overline{\overline{Z}}^{\vec{M}_{\text{aux}}\vec{M}_{\text{IV}}} \\ \overline{\overline{Z}}^{\vec{J}_{\text{IV}}\vec{J}_{\text{aux}}} & \overline{\overline{Z}}^{\vec{J}_{\text{IV}}\vec{J}_{\text{IV}}} & \overline{\overline{Z}}^{\vec{J}_{\text{IV}}\vec{J}_{\text{IS}}} & \overline{\overline{Z}}^{\vec{J}_{\text{IV}}\vec{J}_{\text{IL}}} & \overline{\overline{Z}}^{\vec{J}_{\text{IV}}\vec{M}_{\text{aux}}} & \overline{\overline{Z}}^{\vec{J}_{\text{IV}}\vec{M}_{\text{IV}}} \\ \overline{\overline{Z}}^{\vec{M}_{\text{IV}}\vec{J}_{\text{aux}}} & \overline{\overline{Z}}^{\vec{M}_{\text{IV}}\vec{J}_{\text{IV}}} & \overline{\overline{Z}}^{\vec{M}_{\text{IV}}\vec{J}_{\text{IS}}} & \overline{\overline{Z}}^{\vec{M}_{\text{IV}}\vec{J}_{\text{IL}}} & \overline{\overline{Z}}^{\vec{M}_{\text{IV}}\vec{M}_{\text{aux}}} & \overline{\overline{Z}}^{\vec{M}_{\text{IV}}\vec{M}_{\text{IV}}} \\ \overline{\overline{Z}}^{\vec{J}_{\text{IS}}\vec{J}_{\text{aux}}} & \overline{\overline{Z}}^{\vec{J}_{\text{IS}}\vec{J}_{\text{IV}}} & \overline{\overline{Z}}^{\vec{J}_{\text{IS}}\vec{J}_{\text{IS}}} & \overline{\overline{Z}}^{\vec{J}_{\text{IS}}\vec{J}_{\text{IL}}} & \overline{\overline{Z}}^{\vec{J}_{\text{IS}}\vec{M}_{\text{aux}}} & \overline{\overline{Z}}^{\vec{J}_{\text{IS}}\vec{M}_{\text{IV}}} \\ \overline{\overline{Z}}^{\vec{J}_{\text{IL}}\vec{J}_{\text{aux}}} & \overline{\overline{Z}}^{\vec{J}_{\text{IL}}\vec{J}_{\text{IV}}} & \overline{\overline{Z}}^{\vec{J}_{\text{IL}}\vec{J}_{\text{IS}}} & \overline{\overline{Z}}^{\vec{J}_{\text{IL}}\vec{J}_{\text{IL}}} & \overline{\overline{Z}}^{\vec{J}_{\text{IL}}\vec{M}_{\text{aux}}} & \overline{\overline{Z}}^{\vec{J}_{\text{IL}}\vec{M}_{\text{IV}}} \end{bmatrix} \tag{C-125a}$$

$$\overline{\overline{\Psi}}_{\text{FCE}}^{\text{DoM}}=\begin{bmatrix} \overline{\overline{Z}}^{\vec{J}_{\text{aux}}\vec{J}_{\text{aux}}} & \overline{\overline{Z}}^{\vec{J}_{\text{aux}}\vec{J}_{\text{IV}}} & \overline{\overline{Z}}^{\vec{J}_{\text{aux}}\vec{J}_{\text{IS}}} & \overline{\overline{Z}}^{\vec{J}_{\text{aux}}\vec{J}_{\text{IL}}} & \overline{\overline{Z}}^{\vec{J}_{\text{aux}}\vec{M}_{\text{aux}}} & \overline{\overline{Z}}^{\vec{J}_{\text{aux}}\vec{M}_{\text{IV}}} \\ \overline{\overline{Z}}^{\vec{J}_{\text{IV}}\vec{J}_{\text{aux}}} & \overline{\overline{Z}}^{\vec{J}_{\text{IV}}\vec{J}_{\text{IV}}} & \overline{\overline{Z}}^{\vec{J}_{\text{IV}}\vec{J}_{\text{IS}}} & \overline{\overline{Z}}^{\vec{J}_{\text{IV}}\vec{J}_{\text{IL}}} & \overline{\overline{Z}}^{\vec{J}_{\text{IV}}\vec{M}_{\text{aux}}} & \overline{\overline{Z}}^{\vec{J}_{\text{IV}}\vec{M}_{\text{IV}}} \\ \overline{\overline{Z}}^{\vec{M}_{\text{IV}}\vec{J}_{\text{aux}}} & \overline{\overline{Z}}^{\vec{M}_{\text{IV}}\vec{J}_{\text{IV}}} & \overline{\overline{Z}}^{\vec{M}_{\text{IV}}\vec{J}_{\text{IS}}} & \overline{\overline{Z}}^{\vec{M}_{\text{IV}}\vec{J}_{\text{IL}}} & \overline{\overline{Z}}^{\vec{M}_{\text{IV}}\vec{M}_{\text{aux}}} & \overline{\overline{Z}}^{\vec{M}_{\text{IV}}\vec{M}_{\text{IV}}} \\ \overline{\overline{Z}}^{\vec{J}_{\text{IS}}\vec{J}_{\text{aux}}} & \overline{\overline{Z}}^{\vec{J}_{\text{IS}}\vec{J}_{\text{IV}}} & \overline{\overline{Z}}^{\vec{J}_{\text{IS}}\vec{J}_{\text{IS}}} & \overline{\overline{Z}}^{\vec{J}_{\text{IS}}\vec{J}_{\text{IL}}} & \overline{\overline{Z}}^{\vec{J}_{\text{IS}}\vec{M}_{\text{aux}}} & \overline{\overline{Z}}^{\vec{J}_{\text{IS}}\vec{M}_{\text{IV}}} \\ \overline{\overline{Z}}^{\vec{J}_{\text{IL}}\vec{J}_{\text{aux}}} & \overline{\overline{Z}}^{\vec{J}_{\text{IL}}\vec{J}_{\text{IV}}} & \overline{\overline{Z}}^{\vec{J}_{\text{IL}}\vec{J}_{\text{IS}}} & \overline{\overline{Z}}^{\vec{J}_{\text{IL}}\vec{J}_{\text{IL}}} & \overline{\overline{Z}}^{\vec{J}_{\text{IL}}\vec{M}_{\text{aux}}} & \overline{\overline{Z}}^{\vec{J}_{\text{IL}}\vec{M}_{\text{IV}}} \end{bmatrix} \tag{C-125b}$$

Based on Eqs. (C-103) and (C-104), the IPO $P_{\text{aux}}^{\text{in}}=(1/2)\iint_{\mathbb{S}_{\text{aux}}}(\vec{E}\times\vec{H}^{\dagger})\cdot\hat{n}_{\rightarrow\text{aux}}dS$ can be rewritten as follows:

$$\begin{aligned} P_{\text{aux}}^{\text{in}} &= (1/2)\left\langle \hat{n}_{\rightarrow\text{aux}}\times\vec{J}_{\text{aux}},\vec{M}_{\text{aux}}\right\rangle_{\mathbb{S}_{\text{aux}}} \\ &= -(1/2)\left\langle \vec{J}_{\text{aux}},\mathcal{E}_0\left(\vec{J}_{\text{aux}}+\vec{J}_{\text{IV}}+\vec{J}_{\text{IS}}+\vec{J}_{\text{IL}},\vec{M}_{\text{aux}}+\vec{M}_{\text{IV}}\right)\right\rangle_{\tilde{\mathbb{S}}_{\text{aux}}} \\ &= -(1/2)\left\langle \vec{M}_{\text{aux}},\mathcal{H}_0\left(\vec{J}_{\text{aux}}+\vec{J}_{\text{IV}}+\vec{J}_{\text{IS}}+\vec{J}_{\text{IL}},\vec{M}_{\text{aux}}+\vec{M}_{\text{IV}}\right)\right\rangle_{\tilde{\mathbb{S}}_{\text{aux}}}^{\dagger} \end{aligned} \tag{C-126}$$

where integral surface $\tilde{\mathbb{S}}_{\text{aux}}$ locates at the $\mathbb{V}_{\text{aux}}$ side of $\mathbb{S}_{\text{aux}}$. The IPO can be discretized as follows:

$$P_{\text{aux}}^{\text{in}} = \left(\bar{a}^{\text{AV}}\right)^{\dagger}\cdot\overline{\overline{P}}_{\text{aux}}^{\text{curAV}}\cdot\bar{a}^{\text{AV}} = \left(\bar{a}^{\text{AV}}\right)^{\dagger}\cdot\overline{\overline{P}}_{\text{aux}}^{\text{intAV}}\cdot\bar{a}^{\text{AV}} \tag{C-127}$$

in which

$$\overline{\overline{P}}_{\text{aux}}^{\text{curAV}}=\begin{bmatrix} 0 & 0 & 0 & 0 & \overline{\overline{C}}^{\vec{J}_{\text{aux}}\vec{M}_{\text{aux}}} & 0 \\ 0 & 0 & 0 & 0 & 0 & 0 \\ 0 & 0 & 0 & 0 & 0 & 0 \\ 0 & 0 & 0 & 0 & 0 & 0 \\ 0 & 0 & 0 & 0 & 0 & 0 \\ 0 & 0 & 0 & 0 & 0 & 0 \end{bmatrix} \tag{C-128}$$

corresponding to the first equality in Eq. (C-127), and

$$\overline{\overline{P}}_{\text{aux}}^{\text{intAV}}=\begin{bmatrix} \overline{\overline{P}}^{\vec{J}_{\text{aux}}\vec{J}_{\text{aux}}} & \overline{\overline{P}}^{\vec{J}_{\text{aux}}\vec{J}_{\text{IV}}} & \overline{\overline{P}}^{\vec{J}_{\text{aux}}\vec{J}_{\text{IS}}} & \overline{\overline{P}}^{\vec{J}_{\text{aux}}\vec{J}_{\text{IL}}} & \overline{\overline{P}}^{\vec{J}_{\text{aux}}\vec{M}_{\text{aux}}} & \overline{\overline{P}}^{\vec{J}_{\text{aux}}\vec{M}_{\text{IV}}} \\ 0 & 0 & 0 & 0 & 0 & 0 \\ 0 & 0 & 0 & 0 & 0 & 0 \\ 0 & 0 & 0 & 0 & 0 & 0 \\ 0 & 0 & 0 & 0 & 0 & 0 \\ 0 & 0 & 0 & 0 & 0 & 0 \end{bmatrix} \tag{C-129a}$$

corresponding to the second equality in Eq. (C-127), and

$$\overline{\overline{P}}_{\text{aux}}^{\text{intAV}}=\begin{bmatrix} 0 & 0 & 0 & 0 & 0 & 0 \\ 0 & 0 & 0 & 0 & 0 & 0 \\ 0 & 0 & 0 & 0 & 0 & 0 \\ 0 & 0 & 0 & 0 & 0 & 0 \\ \overline{\overline{P}}^{\vec{M}_{\text{aux}}\vec{J}_{\text{aux}}} & \overline{\overline{P}}^{\vec{M}_{\text{aux}}\vec{J}_{\text{IV}}} & \overline{\overline{P}}^{\vec{M}_{\text{aux}}\vec{J}_{\text{IS}}} & \overline{\overline{P}}^{\vec{M}_{\text{aux}}\vec{J}_{\text{IL}}} & \overline{\overline{P}}^{\vec{M}_{\text{aux}}\vec{M}_{\text{aux}}} & \overline{\overline{P}}^{\vec{M}_{\text{aux}}\vec{M}_{\text{IV}}} \\ 0 & 0 & 0 & 0 & 0 & 0 \end{bmatrix}^{\dagger} \tag{C-129b}$$

corresponding to the third equality in Eq. (C-127), where the elements of the submatrices are as follows:

$$c_{\xi\zeta}^{\vec{J}_{\text{aux}}\vec{M}_{\text{aux}}} = (1/2)\left\langle \hat{n}_{\to\text{aux}} \times \vec{b}_{\xi}^{\vec{J}_{\text{aux}}}, \vec{b}_{\zeta}^{\vec{M}_{\text{aux}}} \right\rangle_{\mathbb{S}_{\text{aux}}} \tag{C-130}$$

and

$$p_{\xi\zeta}^{\vec{J}_{\text{aux}}\vec{J}_{\text{aux}}} = -(1/2)\left\langle \vec{b}_{\xi}^{\vec{J}_{\text{aux}}}, -j\omega\mu_0\mathcal{L}_0\left(\vec{b}_{\zeta}^{\vec{J}_{\text{aux}}}\right)\right\rangle_{\mathbb{S}_{\text{aux}}} \tag{C-131a}$$

$$p_{\xi\zeta}^{\vec{J}_{\text{aux}}\vec{J}_{\text{IV}}} = -(1/2)\left\langle \vec{b}_{\xi}^{\vec{J}_{\text{aux}}}, -j\omega\mu_0\mathcal{L}_0\left(\vec{b}_{\zeta}^{\vec{J}_{\text{IV}}}\right)\right\rangle_{\mathbb{S}_{\text{aux}}} \tag{C-131b}$$

$$p_{\xi\zeta}^{\vec{J}_{\text{aux}}\vec{J}_{\text{IS}}} = -(1/2)\left\langle \vec{b}_{\xi}^{\vec{J}_{\text{aux}}}, -j\omega\mu_0\mathcal{L}_0\left(\vec{b}_{\zeta}^{\vec{J}_{\text{IS}}}\right)\right\rangle_{\mathbb{S}_{\text{aux}}} \tag{C-131c}$$

$$p_{\xi\zeta}^{\vec{J}_{\text{aux}}\vec{J}_{\text{IL}}} = -(1/2)\left\langle \vec{b}_{\xi}^{\vec{J}_{\text{aux}}}, -j\omega\mu_0\mathcal{L}_0\left(\vec{b}_{\zeta}^{\vec{J}_{\text{IL}}}\right)\right\rangle_{\mathbb{S}_{\text{aux}}} \tag{C-131d}$$

$$p_{\xi\zeta}^{\vec{J}_{\text{aux}}\vec{M}_{\text{aux}}} = -(1/2)\left\langle \vec{b}_{\xi}^{\vec{J}_{\text{aux}}}, \hat{n}_{\to\text{aux}} \times \frac{1}{2}\vec{b}_{\zeta}^{\vec{M}_{\text{aux}}} - \text{P.V.}\,\mathcal{K}_0\left(\vec{b}_{\zeta}^{\vec{M}_{\text{aux}}}\right)\right\rangle_{\mathbb{S}_{\text{aux}}} \tag{C-131e}$$

$$p_{\xi\zeta}^{\vec{J}_{\text{aux}}\vec{M}_{\text{IV}}} = -(1/2)\left\langle \vec{b}_{\xi}^{\vec{J}_{\text{aux}}}, -\mathcal{K}_0\left(\vec{b}_{\zeta}^{\vec{M}_{\text{IV}}}\right)\right\rangle_{\mathbb{S}_{\text{aux}}} \tag{C-131f}$$

and

$$p_{\xi\zeta}^{\vec{M}_{\text{aux}}\vec{J}_{\text{aux}}} = -(1/2)\left\langle \vec{b}_{\xi}^{\vec{M}_{\text{aux}}}, \frac{1}{2}\vec{b}_{\zeta}^{\vec{J}_{\text{aux}}} \times \hat{n}_{\to\text{aux}} + \text{P.V.}\,\mathcal{K}_0\left(\vec{b}_{\zeta}^{\vec{J}_{\text{aux}}}\right)\right\rangle_{\mathbb{S}_{\text{aux}}} \tag{C-131g}$$

$$p_{\xi\zeta}^{\vec{M}_{\text{aux}}\vec{J}_{\text{IV}}} = -(1/2)\left\langle \vec{b}_{\xi}^{\vec{M}_{\text{aux}}}, \mathcal{K}_0\left(\vec{b}_{\zeta}^{\vec{J}_{\text{IV}}}\right)\right\rangle_{\mathbb{S}_{\text{aux}}} \tag{C-131h}$$

$$p_{\xi\zeta}^{\vec{M}_{\text{aux}}\vec{J}_{\text{IS}}} = -(1/2)\left\langle \vec{b}_{\xi}^{\vec{M}_{\text{aux}}}, \mathcal{K}_0\left(\vec{b}_{\zeta}^{\vec{J}_{\text{IS}}}\right)\right\rangle_{\mathbb{S}_{\text{aux}}} \tag{C-131i}$$

$$p_{\xi\zeta}^{\vec{M}_{\text{aux}}\vec{J}_{\text{IL}}} = -(1/2)\left\langle \vec{b}_{\xi}^{\vec{M}_{\text{aux}}}, \mathcal{K}_0\left(\vec{b}_{\zeta}^{\vec{J}_{\text{IL}}}\right)\right\rangle_{\mathbb{S}_{\text{aux}}} \tag{C-131j}$$

$$p_{\xi\zeta}^{\vec{M}_{\text{aux}}\vec{M}_{\text{aux}}} = -(1/2)\left\langle \vec{b}_{\xi}^{\vec{M}_{\text{aux}}}, -j\omega\varepsilon_0\mathcal{L}_0\left(\vec{b}_{\zeta}^{\vec{M}_{\text{aux}}}\right)\right\rangle_{\mathbb{S}_{\text{aux}}} \tag{C-131k}$$

$$p_{\xi\zeta}^{\vec{M}_{\text{aux}}\vec{M}_{\text{IV}}} = -(1/2)\left\langle \vec{b}_{\xi}^{\vec{M}_{\text{aux}}}, -j\omega\varepsilon_0\mathcal{L}_0\left(\vec{b}_{\zeta}^{\vec{M}_{\text{IV}}}\right)\right\rangle_{\mathbb{S}_{\text{aux}}} \tag{C-131l}$$

To obtain the IPO defined on modal space, we substitute Eq. (C-120) into the Eq. (C-127), and then we have that

$$P_{\text{aux}}^{\text{in}} = \bar{a}^{\dagger} \cdot \underbrace{\left(\bar{\bar{T}}^{\dagger} \cdot \bar{\bar{P}}_{\text{aux}}^{\text{curAV}} \cdot \bar{\bar{T}}\right)}_{\bar{\bar{P}}_{\text{aux}}^{\text{cur}}} \cdot \bar{a} = \bar{a}^{\dagger} \cdot \underbrace{\left(\bar{\bar{T}}^{\dagger} \cdot \bar{\bar{P}}_{\text{aux}}^{\text{intAV}} \cdot \bar{\bar{T}}\right)}_{\bar{\bar{P}}_{\text{aux}}^{\text{int}}} \cdot \bar{a} \tag{C-132}$$

where subscripts "cur" and "int" are to emphasize that $\bar{\bar{P}}_{\text{aux}}^{\text{cur}}$ and $\bar{\bar{P}}_{\text{aux}}^{\text{int}}$ respectively originate from discretizing the current and interaction forms of IPO.

The process of constructing IP-DMs from orthogonalizing $\bar{\bar{P}}_{\text{aux}}^{\text{cur}}$ and $\bar{\bar{P}}_{\text{aux}}^{\text{int}}$ is completely the same as the one used in Sec. 7.4.

## C7 Surface-Volume Formulation of the PTT-Based DMT for the Augmented Tra-guide-Tra-antenna System Discussed in Sec. 8.2

The topological structure of the *tra-guide-tra-antenna system* (*TGTA system*) discussed in this App. C7 is the same as the one shown in Fig. 8-3.

Because of the action of the field on $\mathbb{V}^{\text{G}}$ and $\mathbb{V}^{\text{A}}$, some volume currents will be induced on $\mathbb{V}^{\text{G}}$ and $\mathbb{V}^{\text{A}}$, and the currents are denoted as $\{\vec{J}^{\text{IV}},\vec{M}^{\text{IV}}\}$. For simplifying the symbolic system of the following discussions, we define region $\mathbb{V}$ and material parameter $\ddot{\gamma}$ as follows:

$$\mathbb{V} \ = \ \mathbb{V}^{\text{G}} \bigcup \mathbb{V}^{\text{A}} \tag{C-133}$$

and

$$\ddot{\gamma}(\vec{r}) \ = \ \begin{cases} \ddot{\gamma}^{\text{G}}(\vec{r}) & , \quad \vec{r} \in \mathbb{V}^{\text{G}} \\ \ddot{\gamma}^{\text{A}}(\vec{r}) & , \quad \vec{r} \in \mathbb{V}^{\text{A}} \end{cases} \tag{C-134}$$

If the equivalent surface currents distributing on $\mathbb{S}^{\text{O}\rightleftharpoons\text{G}}$ are denoted as $\{\vec{J}^{\text{O}\rightleftharpoons\text{G}},\vec{M}^{\text{O}\rightleftharpoons\text{G}}\}$, and the induced surface electric current distributing on electric wall $\mathbb{S}^{\text{ele}}$ is denoted as $\vec{J}^{\text{IS}}$, then the field distributing on $\mathbb{V}\bigcup\mathbb{V}^{\text{F}}$ can be expressed as follows:

$$\vec{F}(\vec{r}) \ = \ \mathcal{F}_0\left(\vec{J}^{\text{O}\rightleftharpoons\text{G}} + \vec{J}^{\text{IV}} + \vec{J}^{\text{IS}}, \vec{M}^{\text{O}\rightleftharpoons\text{G}} + \vec{M}^{\text{IV}}\right) \quad , \quad \vec{r} \in \mathbb{V}\bigcup\mathbb{V}^{\text{F}} \tag{C-135}$$

where $\vec{F} = \vec{E} / \vec{H}$, and correspondingly $\mathcal{F}_0 = \mathcal{E}_0 / \mathcal{H}_0$, and the operator is the same as the one used in Sec. 8.2. The currents $\{\vec{J}^{\text{O}\rightleftharpoons\text{G}},\vec{M}^{\text{O}\rightleftharpoons\text{G}}\}$ and the fields $\{\vec{E},\vec{H}\}$ in Eq. (C-135) satisfy the following relations

$$\hat{n}^{\rightarrow\text{G}} \times \left[\vec{H}\left(\vec{r}^{\,\text{V}}\right)\right]_{\vec{r}^{\,\text{V}}\rightarrow\vec{r}} \ = \ \vec{J}^{\text{O}\rightleftharpoons\text{G}}(\vec{r}) \quad , \quad \vec{r} \in \mathbb{S}^{\text{O}\rightleftharpoons\text{G}} \tag{C-136a}$$

$$\left[\vec{E}\left(\vec{r}^{\,\text{V}}\right)\right]_{\vec{r}^{\,\text{V}}\rightarrow\vec{r}} \times \hat{n}^{\rightarrow\text{G}} \ = \ \vec{M}^{\text{O}\rightleftharpoons\text{G}}(\vec{r}) \quad , \quad \vec{r} \in \mathbb{S}^{\text{O}\rightleftharpoons\text{G}} \tag{C-136b}$$

In the above Eq. (C-136), point $\vec{r}^{\,\text{V}}$ belongs to region $\mathbb{V}$, and $\vec{r}^{\,\text{V}}$ approaches the point $\vec{r}$ on $\mathbb{S}^{\text{O}\rightleftharpoons\text{G}}$; $\hat{n}^{\rightarrow\text{G}}$ is the normal direction of $\mathbb{S}^{\text{O}\rightleftharpoons\text{G}}$, and $\hat{n}^{\rightarrow\text{G}}$ points to the interior of $\mathbb{V}$. The currents $\{\vec{J}^{\text{IV}},\vec{M}^{\text{IV}}\}$ and the fields $\{\vec{E},\vec{H}\}$ in (C-135) satisfy the following relations

$$\vec{E}(\vec{r}) = \left(j\omega\Delta\ddot{\varepsilon}^{\text{c}}\right)^{-1}\cdot\vec{J}^{\text{IV}} \quad , \quad \vec{r} \in \mathbb{V} \tag{C-137a}$$

$$\vec{H}(\vec{r}) = \left(j\omega\Delta\ddot{\mu}\right)^{-1}\cdot\vec{M}^{\text{IV}} \quad , \quad \vec{r} \in \mathbb{V} \tag{C-137b}$$

In the above Eq. (C-137), $\Delta\ddot{\varepsilon}^{\text{c}} = (\ddot{\varepsilon} - j\,\ddot{\sigma}/\omega) - \ddot{I}\,\varepsilon_0$, and $\Delta\ddot{\mu} = \ddot{\mu} - \ddot{I}\,\mu_0$.

Substituting Eq. (C-135) into Eqs. (C-136a) and (C-136b), we obtain the following integral equations

$$\left[\mathcal{H}_0\left(\vec{J}^{\mathrm{O}\rightleftharpoons\mathrm{G}}+\vec{J}^{\mathrm{IV}}+\vec{J}^{\mathrm{IS}},\vec{M}^{\mathrm{O}\rightleftharpoons\mathrm{G}}+\vec{M}^{\mathrm{IV}}\right)\right]_{\vec{r}^{\mathrm{V}}\to\vec{r}}^{\tan} = \vec{J}^{\mathrm{O}\rightleftharpoons\mathrm{G}}\times\hat{n}^{\to\mathrm{G}} \quad , \quad \vec{r}\in\mathbb{S}^{\mathrm{O}\rightleftharpoons\mathrm{G}} \quad \text{(C-138a)}$$

$$\left[\mathcal{E}_0\left(\vec{J}^{\mathrm{O}\rightleftharpoons\mathrm{G}}+\vec{J}^{\mathrm{IV}}+\vec{J}^{\mathrm{IS}},\vec{M}^{\mathrm{O}\rightleftharpoons\mathrm{G}}+\vec{M}^{\mathrm{IV}}\right)\right]_{\vec{r}^{\mathrm{V}}\to\vec{r}}^{\tan} = \hat{n}^{\to\mathrm{G}}\times\vec{M}^{\mathrm{O}\rightleftharpoons\mathrm{G}} \quad , \quad \vec{r}\in\mathbb{S}^{\mathrm{O}\rightleftharpoons\mathrm{G}} \quad \text{(C-138b)}$$

about currents $\{\vec{J}^{\mathrm{O}\rightleftharpoons\mathrm{G}},\vec{M}^{\mathrm{O}\rightleftharpoons\mathrm{G}}\}$, $\{\vec{J}^{\mathrm{IV}},\vec{M}^{\mathrm{IV}}\}$, and $\vec{J}^{\mathrm{IS}}$, where the superscript "tan" represents the tangential component of the field. Substituting Eq. (C-135) into Eqs. (C-137a) and (C-137b), we obtain the following integral equations

$$\mathcal{E}_0\left(\vec{J}^{\mathrm{O}\rightleftharpoons\mathrm{G}}+\vec{J}^{\mathrm{IV}}+\vec{J}^{\mathrm{IS}},\vec{M}^{\mathrm{O}\rightleftharpoons\mathrm{G}}+\vec{M}^{\mathrm{IV}}\right) = \left(j\omega\Delta\vec{\vec{\varepsilon}}^{\,\mathrm{c}}\right)^{-1}\cdot\vec{J}^{\mathrm{IV}} \quad , \quad \vec{r}\in\mathbb{V} \quad \text{(C-139a)}$$

$$\mathcal{H}_0\left(\vec{J}^{\mathrm{O}\rightleftharpoons\mathrm{G}}+\vec{J}^{\mathrm{IV}}+\vec{J}^{\mathrm{IS}},\vec{M}^{\mathrm{O}\rightleftharpoons\mathrm{G}}+\vec{M}^{\mathrm{IV}}\right) = \left(j\omega\Delta\vec{\vec{\mu}}\right)^{-1}\cdot\vec{M}^{\mathrm{IV}} \quad , \quad \vec{r}\in\mathbb{V} \quad \text{(C-139b)}$$

about currents $\{\vec{J}^{\mathrm{O}\rightleftharpoons\mathrm{G}},\vec{M}^{\mathrm{O}\rightleftharpoons\mathrm{G}}\}$, $\{\vec{J}^{\mathrm{IV}},\vec{M}^{\mathrm{IV}}\}$, and $\vec{J}^{\mathrm{IS}}$. Based on Eq. (C-135) and the homogeneous tangential electric field boundary conditions on $\mathbb{S}^{\mathrm{ele}}$, we have the following electric field integral equations

$$\left[\mathcal{E}_0\left(\vec{J}^{\mathrm{O}\rightleftharpoons\mathrm{G}}+\vec{J}^{\mathrm{IV}}+\vec{J}^{\mathrm{IS}},\vec{M}^{\mathrm{O}\rightleftharpoons\mathrm{G}}+\vec{M}^{\mathrm{IV}}\right)\right]^{\tan} = 0 \quad , \quad \vec{r}\in\mathbb{S}^{\mathrm{ele}} \quad \text{(C-140)}$$

about currents $\{\vec{J}^{\mathrm{O}\rightleftharpoons\mathrm{G}},\vec{M}^{\mathrm{O}\rightleftharpoons\mathrm{G}}\}$, $\{\vec{J}^{\mathrm{IV}},\vec{M}^{\mathrm{IV}}\}$, and $\vec{J}^{\mathrm{IS}}$.

If the above-mentioned currents are expanded in terms of some proper basis functions, and the Eqs. (C-138a)~(C-140) are tested with $\{\vec{b}_\xi^{\vec{M}^{\mathrm{O}\rightleftharpoons\mathrm{G}}}\}$, $\{\vec{b}_\xi^{\vec{J}^{\mathrm{O}\rightleftharpoons\mathrm{G}}}\}$, $\{\vec{b}_\xi^{\vec{J}^{\mathrm{IV}}}\}$, $\{\vec{b}_\xi^{\vec{M}^{\mathrm{IV}}}\}$, and $\{\vec{b}_\xi^{\vec{J}^{\mathrm{IS}}}\}$ respectively, then the equations are immediately discretized into the following matrix equations

$$0=\overline{\overline{Z}}^{\vec{M}^{\mathrm{O}\rightleftharpoons\mathrm{G}}\vec{J}^{\mathrm{O}\rightleftharpoons\mathrm{G}}}\cdot\overline{a}^{\vec{J}^{\mathrm{O}\rightleftharpoons\mathrm{G}}}+\overline{\overline{Z}}^{\vec{M}^{\mathrm{O}\rightleftharpoons\mathrm{G}}\vec{J}^{\mathrm{IV}}}\cdot\overline{a}^{\vec{J}^{\mathrm{IV}}}+\overline{\overline{Z}}^{\vec{M}^{\mathrm{O}\rightleftharpoons\mathrm{G}}\vec{J}^{\mathrm{IS}}}\cdot\overline{a}^{\vec{J}^{\mathrm{IS}}}+\overline{\overline{Z}}^{\vec{M}^{\mathrm{O}\rightleftharpoons\mathrm{G}}\vec{M}^{\mathrm{O}\rightleftharpoons\mathrm{G}}}\cdot\overline{a}^{\vec{M}^{\mathrm{O}\rightleftharpoons\mathrm{G}}}+\overline{\overline{Z}}^{\vec{M}^{\mathrm{O}\rightleftharpoons\mathrm{G}}\vec{M}^{\mathrm{IV}}}\cdot\overline{a}^{\vec{M}^{\mathrm{IV}}} \quad \text{(C-141a)}$$

$$0=\overline{\overline{Z}}^{\vec{J}^{\mathrm{O}\rightleftharpoons\mathrm{G}}\vec{J}^{\mathrm{O}\rightleftharpoons\mathrm{G}}}\cdot\overline{a}^{\vec{J}^{\mathrm{O}\rightleftharpoons\mathrm{G}}}+\overline{\overline{Z}}^{\vec{J}^{\mathrm{O}\rightleftharpoons\mathrm{G}}\vec{J}^{\mathrm{IV}}}\cdot\overline{a}^{\vec{J}^{\mathrm{IV}}}+\overline{\overline{Z}}^{\vec{J}^{\mathrm{O}\rightleftharpoons\mathrm{G}}\vec{J}^{\mathrm{IS}}}\cdot\overline{a}^{\vec{J}^{\mathrm{IS}}}+\overline{\overline{Z}}^{\vec{J}^{\mathrm{O}\rightleftharpoons\mathrm{G}}\vec{M}^{\mathrm{O}\rightleftharpoons\mathrm{G}}}\cdot\overline{a}^{\vec{M}^{\mathrm{O}\rightleftharpoons\mathrm{G}}}+\overline{\overline{Z}}^{\vec{J}^{\mathrm{O}\rightleftharpoons\mathrm{G}}\vec{M}^{\mathrm{IV}}}\cdot\overline{a}^{\vec{M}^{\mathrm{IV}}} \quad \text{(C-141b)}$$

and

$$0=\overline{\overline{Z}}^{\vec{J}^{\mathrm{IV}}\vec{J}^{\mathrm{O}\rightleftharpoons\mathrm{G}}}\cdot\overline{a}^{\vec{J}^{\mathrm{O}\rightleftharpoons\mathrm{G}}}+\overline{\overline{Z}}^{\vec{J}^{\mathrm{IV}}\vec{J}^{\mathrm{IV}}}\cdot\overline{a}^{\vec{J}^{\mathrm{IV}}}+\overline{\overline{Z}}^{\vec{J}^{\mathrm{IV}}\vec{J}^{\mathrm{IS}}}\cdot\overline{a}^{\vec{J}^{\mathrm{IS}}}+\overline{\overline{Z}}^{\vec{J}^{\mathrm{IV}}\vec{M}^{\mathrm{O}\rightleftharpoons\mathrm{G}}}\cdot\overline{a}^{\vec{M}^{\mathrm{O}\rightleftharpoons\mathrm{G}}}+\overline{\overline{Z}}^{\vec{J}^{\mathrm{IV}}\vec{M}^{\mathrm{IV}}}\cdot\overline{a}^{\vec{M}^{\mathrm{IV}}} \quad \text{(C-142a)}$$

$$0=\overline{\overline{Z}}^{\vec{M}^{\mathrm{IV}}\vec{J}^{\mathrm{O}\rightleftharpoons\mathrm{G}}}\cdot\overline{a}^{\vec{J}^{\mathrm{O}\rightleftharpoons\mathrm{G}}}+\overline{\overline{Z}}^{\vec{M}^{\mathrm{IV}}\vec{J}^{\mathrm{IV}}}\cdot\overline{a}^{\vec{J}^{\mathrm{IV}}}+\overline{\overline{Z}}^{\vec{M}^{\mathrm{IV}}\vec{J}^{\mathrm{IS}}}\cdot\overline{a}^{\vec{J}^{\mathrm{IS}}}+\overline{\overline{Z}}^{\vec{M}^{\mathrm{IV}}\vec{M}^{\mathrm{O}\rightleftharpoons\mathrm{G}}}\cdot\overline{a}^{\vec{M}^{\mathrm{O}\rightleftharpoons\mathrm{G}}}+\overline{\overline{Z}}^{\vec{M}^{\mathrm{IV}}\vec{M}^{\mathrm{IV}}}\cdot\overline{a}^{\vec{M}^{\mathrm{IV}}} \quad \text{(C-142b)}$$

and

$$0=\overline{\overline{Z}}^{\vec{J}^{\mathrm{IS}}\vec{J}^{\mathrm{O}\rightleftharpoons\mathrm{G}}}\cdot\overline{a}^{\vec{J}^{\mathrm{O}\rightleftharpoons\mathrm{G}}}+\overline{\overline{Z}}^{\vec{J}^{\mathrm{IS}}\vec{J}^{\mathrm{IV}}}\cdot\overline{a}^{\vec{J}^{\mathrm{IV}}}+\overline{\overline{Z}}^{\vec{J}^{\mathrm{IS}}\vec{J}^{\mathrm{IS}}}\cdot\overline{a}^{\vec{J}^{\mathrm{IS}}}+\overline{\overline{Z}}^{\vec{J}^{\mathrm{IS}}\vec{M}^{\mathrm{O}\rightleftharpoons\mathrm{G}}}\cdot\overline{a}^{\vec{M}^{\mathrm{O}\rightleftharpoons\mathrm{G}}}+\overline{\overline{Z}}^{\vec{J}^{\mathrm{IS}}\vec{M}^{\mathrm{IV}}}\cdot\overline{a}^{\vec{M}^{\mathrm{IV}}} \quad \text{(C-143)}$$

about vectors $\{\overline{a}^{\vec{J}^{\mathrm{O}\rightleftharpoons\mathrm{G}}},\overline{a}^{\vec{M}^{\mathrm{O}\rightleftharpoons\mathrm{G}}}\}$, $\{\overline{a}^{\vec{J}^{\mathrm{IV}}},\overline{a}^{\vec{M}^{\mathrm{IV}}}\}$, and $\overline{a}^{\vec{J}^{\mathrm{IS}}}$.

The formulations used to calculate the elements of the matrices in Eq. (C-141a) are as follows:

$$z_{\xi\zeta}^{\vec{M}^{\mathrm{O}\rightleftharpoons\mathrm{G}}\vec{J}^{\mathrm{O}\rightleftharpoons\mathrm{G}}} = \left\langle \vec{b}_{\xi}^{\vec{M}^{\mathrm{O}\rightleftharpoons\mathrm{G}}}, \hat{n}^{\rightarrow\mathrm{G}} \times \frac{1}{2}\vec{b}_{\zeta}^{\vec{J}^{\mathrm{O}\rightleftharpoons\mathrm{G}}} + \mathrm{P.V.}\,\mathcal{K}_0\left(\vec{b}_{\zeta}^{\vec{J}^{\mathrm{O}\rightleftharpoons\mathrm{G}}}\right)\right\rangle_{\mathbb{S}^{\mathrm{O}\rightleftharpoons\mathrm{G}}} \tag{C-144a}$$

$$z_{\xi\zeta}^{\vec{M}^{\mathrm{O}\rightleftharpoons\mathrm{G}}\vec{J}^{\mathrm{IV}}} = \left\langle \vec{b}_{\xi}^{\vec{M}^{\mathrm{O}\rightleftharpoons\mathrm{G}}}, \mathcal{K}_0\left(\vec{b}_{\zeta}^{\vec{J}^{\mathrm{IV}}}\right)\right\rangle_{\mathbb{S}^{\mathrm{O}\rightleftharpoons\mathrm{G}}} \tag{C-144b}$$

$$z_{\xi\zeta}^{\vec{M}^{\mathrm{O}\rightleftharpoons\mathrm{G}}\vec{J}^{\mathrm{IS}}} = \left\langle \vec{b}_{\xi}^{\vec{M}^{\mathrm{O}\rightleftharpoons\mathrm{G}}}, \mathcal{K}_0\left(\vec{b}_{\zeta}^{\vec{J}^{\mathrm{IS}}}\right)\right\rangle_{\mathbb{S}^{\mathrm{O}\rightleftharpoons\mathrm{G}}} \tag{C-144c}$$

$$z_{\xi\zeta}^{\vec{M}^{\mathrm{O}\rightleftharpoons\mathrm{G}}\vec{M}^{\mathrm{O}\rightleftharpoons\mathrm{G}}} = \left\langle \vec{b}_{\xi}^{\vec{M}^{\mathrm{O}\rightleftharpoons\mathrm{G}}}, -j\omega\varepsilon_0\mathcal{L}_0\left(\vec{b}_{\zeta}^{\vec{M}^{\mathrm{O}\rightleftharpoons\mathrm{G}}}\right)\right\rangle_{\mathbb{S}^{\mathrm{O}\rightleftharpoons\mathrm{G}}} \tag{C-144d}$$

$$z_{\xi\zeta}^{\vec{M}^{\mathrm{O}\rightleftharpoons\mathrm{G}}\vec{M}^{\mathrm{IV}}} = \left\langle \vec{b}_{\xi}^{\vec{M}^{\mathrm{O}\rightleftharpoons\mathrm{G}}}, -j\omega\varepsilon_0\mathcal{L}_0\left(\vec{b}_{\zeta}^{\vec{M}^{\mathrm{IV}}}\right)\right\rangle_{\mathbb{S}^{\mathrm{O}\rightleftharpoons\mathrm{G}}} \tag{C-144e}$$

The formulations used to calculate the elements of the matrices in Eq. (C-141b) are as follows:

$$z_{\xi\zeta}^{\vec{J}^{\mathrm{O}\rightleftharpoons\mathrm{G}}\vec{J}^{\mathrm{O}\rightleftharpoons\mathrm{G}}} = \left\langle \vec{b}_{\xi}^{\vec{J}^{\mathrm{O}\rightleftharpoons\mathrm{G}}}, -j\omega\mu_0\mathcal{L}_0\left(\vec{b}_{\zeta}^{\vec{J}^{\mathrm{O}\rightleftharpoons\mathrm{G}}}\right)\right\rangle_{\mathbb{S}^{\mathrm{O}\rightleftharpoons\mathrm{G}}} \tag{C-145a}$$

$$z_{\xi\zeta}^{\vec{J}^{\mathrm{O}\rightleftharpoons\mathrm{G}}\vec{J}^{\mathrm{IV}}} = \left\langle \vec{b}_{\xi}^{\vec{J}^{\mathrm{O}\rightleftharpoons\mathrm{G}}}, -j\omega\mu_0\mathcal{L}_0\left(\vec{b}_{\zeta}^{\vec{J}^{\mathrm{IV}}}\right)\right\rangle_{\mathbb{S}^{\mathrm{O}\rightleftharpoons\mathrm{G}}} \tag{C-145b}$$

$$z_{\xi\zeta}^{\vec{J}^{\mathrm{O}\rightleftharpoons\mathrm{G}}\vec{J}^{\mathrm{IS}}} = \left\langle \vec{b}_{\xi}^{\vec{J}^{\mathrm{O}\rightleftharpoons\mathrm{G}}}, -j\omega\mu_0\mathcal{L}_0\left(\vec{b}_{\zeta}^{\vec{J}^{\mathrm{IS}}}\right)\right\rangle_{\mathbb{S}^{\mathrm{O}\rightleftharpoons\mathrm{G}}} \tag{C-145c}$$

$$z_{\xi\zeta}^{\vec{J}^{\mathrm{O}\rightleftharpoons\mathrm{G}}\vec{M}^{\mathrm{O}\rightleftharpoons\mathrm{G}}} = \left\langle \vec{b}_{\xi}^{\vec{J}^{\mathrm{O}\rightleftharpoons\mathrm{G}}}, \frac{1}{2}\vec{b}_{\zeta}^{\vec{M}^{\mathrm{O}\rightleftharpoons\mathrm{G}}} \times \hat{n}^{\rightarrow\mathrm{G}} - \mathrm{P.V.}\,\mathcal{K}_0\left(\vec{b}_{\zeta}^{\vec{M}^{\mathrm{O}\rightleftharpoons\mathrm{G}}}\right)\right\rangle_{\mathbb{S}^{\mathrm{O}\rightleftharpoons\mathrm{G}}} \tag{C-145d}$$

$$z_{\xi\zeta}^{\vec{J}^{\mathrm{O}\rightleftharpoons\mathrm{G}}\vec{M}^{\mathrm{IV}}} = \left\langle \vec{b}_{\xi}^{\vec{J}^{\mathrm{O}\rightleftharpoons\mathrm{G}}}, -\mathcal{K}_0\left(\vec{b}_{\zeta}^{\vec{M}^{\mathrm{IV}}}\right)\right\rangle_{\mathbb{S}^{\mathrm{O}\rightleftharpoons\mathrm{G}}} \tag{C-145e}$$

The formulations used to calculate the elements of the matrices in Eq. (C-142a) are as follows:

$$z_{\xi\zeta}^{\vec{J}^{\mathrm{IV}}\vec{J}^{\mathrm{O}\rightleftharpoons\mathrm{G}}} = \left\langle \vec{b}_{\xi}^{\vec{J}^{\mathrm{IV}}}, -j\omega\mu_0\mathcal{L}_0\left(\vec{b}_{\zeta}^{\vec{J}^{\mathrm{O}\rightleftharpoons\mathrm{G}}}\right)\right\rangle_{\mathbb{V}} \tag{C-146a}$$

$$z_{\xi\zeta}^{\vec{J}^{\mathrm{IV}}\vec{J}^{\mathrm{IV}}} = \left\langle \vec{b}_{\xi}^{\vec{J}^{\mathrm{IV}}}, -j\omega\mu_0\mathcal{L}_0\left(\vec{b}_{\zeta}^{\vec{J}^{\mathrm{IV}}}\right) - \left(j\omega\Delta\vec{\vec{\varepsilon}}^{\,\mathrm{c}}\right)^{-1}\cdot\vec{b}_{\zeta}^{\vec{J}^{\mathrm{IV}}}\right\rangle_{\mathbb{V}} \tag{C-146b}$$

$$z_{\xi\zeta}^{\vec{J}^{\mathrm{IV}}\vec{J}^{\mathrm{IS}}} = \left\langle \vec{b}_{\xi}^{\vec{J}^{\mathrm{IV}}}, -j\omega\mu_0\mathcal{L}_0\left(\vec{b}_{\zeta}^{\vec{J}^{\mathrm{IS}}}\right)\right\rangle_{\mathbb{V}} \tag{C-146c}$$

$$z_{\xi\zeta}^{\vec{J}^{\mathrm{IV}}\vec{M}^{\mathrm{O}\rightleftharpoons\mathrm{G}}} = \left\langle \vec{b}_{\xi}^{\vec{J}^{\mathrm{IV}}}, -\mathcal{K}_0\left(\vec{b}_{\zeta}^{\vec{M}^{\mathrm{O}\rightleftharpoons\mathrm{G}}}\right)\right\rangle_{\mathbb{V}} \tag{C-146d}$$

$$z_{\xi\zeta}^{\vec{J}^{\mathrm{IV}}\vec{M}^{\mathrm{IV}}} = \left\langle \vec{b}_{\xi}^{\vec{J}^{\mathrm{IV}}}, -\mathcal{K}_0\left(\vec{b}_{\zeta}^{\vec{M}^{\mathrm{IV}}}\right)\right\rangle_{\mathbb{V}} \tag{C-146e}$$

The formulations used to calculate the elements of the matrices in Eq. (C-142b) are as follows:

$$z_{\xi\zeta}^{\vec{M}^{\mathrm{IV}}\vec{J}^{\mathrm{O}\rightleftharpoons\mathrm{G}}} = \left\langle \vec{b}_{\xi}^{\vec{M}^{\mathrm{IV}}}, \mathcal{K}_0\left(\vec{b}_{\zeta}^{\vec{J}^{\mathrm{O}\rightleftharpoons\mathrm{G}}}\right)\right\rangle_{\mathbb{V}} \tag{C-147a}$$

$$z_{\xi\zeta}^{\vec{M}^{\mathrm{IV}}\vec{J}^{\mathrm{IV}}} = \left\langle \vec{b}_{\xi}^{\vec{M}^{\mathrm{IV}}}, \mathcal{K}_0\left(\vec{b}_{\zeta}^{\vec{J}^{\mathrm{IV}}}\right)\right\rangle_{\mathbb{V}} \tag{C-147b}$$

$$z_{\xi\zeta}^{\vec{M}^{\mathrm{IV}}\vec{J}^{\mathrm{IS}}} = \left\langle \vec{b}_{\xi}^{\vec{M}^{\mathrm{IV}}}, \mathcal{K}_0\left(\vec{b}_{\zeta}^{\vec{J}^{\mathrm{IS}}}\right)\right\rangle_{\mathbb{V}} \tag{C-147c}$$

$$z_{\xi\zeta}^{\vec{M}^{\rm IV}\vec{M}^{\rm O\rightleftharpoons G}} \;=\; \left\langle \vec{b}_{\xi}^{\vec{M}^{\rm IV}}, -j\omega\varepsilon_0\mathcal{L}_0\left(\vec{b}_{\zeta}^{\vec{M}^{\rm O\rightleftharpoons G}}\right)\right\rangle_{\mathbb{V}} \tag{C-147d}$$

$$z_{\xi\zeta}^{\vec{M}^{\rm IV}\vec{M}^{\rm IV}} \;=\; \left\langle \vec{b}_{\xi}^{\vec{M}^{\rm IV}}, -j\omega\varepsilon_0\mathcal{L}_0\left(\vec{b}_{\zeta}^{\vec{M}^{\rm IV}}\right) - \left(j\omega\Delta\ddot{\mu}\right)^{-1}\cdot\vec{b}_{\zeta}^{\vec{M}^{\rm IV}}\right\rangle_{\mathbb{V}} \tag{C-147e}$$

The formulations used to calculate the elements of the matrices in Eq. (C-143) are as follows:

$$z_{\xi\zeta}^{\vec{J}^{\rm IS}\vec{J}^{\rm O\rightleftharpoons G}} \;=\; \left\langle \vec{b}_{\xi}^{\vec{J}^{\rm IS}}, -j\omega\mu_0\mathcal{L}_0\left(\vec{b}_{\zeta}^{\vec{J}^{\rm O\rightleftharpoons G}}\right)\right\rangle_{\mathbb{S}^{\rm ele}} \tag{C-148a}$$

$$z_{\xi\zeta}^{\vec{J}^{\rm IS}\vec{J}^{\rm IV}} \;=\; \left\langle \vec{b}_{\xi}^{\vec{J}^{\rm IS}}, -j\omega\mu_0\mathcal{L}_0\left(\vec{b}_{\zeta}^{\vec{J}^{\rm IV}}\right)\right\rangle_{\mathbb{S}^{\rm ele}} \tag{C-148b}$$

$$z_{\xi\zeta}^{\vec{J}^{\rm IS}\vec{J}^{\rm IS}} \;=\; \left\langle \vec{b}_{\xi}^{\vec{J}^{\rm IS}}, -j\omega\mu_0\mathcal{L}_0\left(\vec{b}_{\zeta}^{\vec{J}^{\rm IS}}\right)\right\rangle_{\mathbb{S}^{\rm ele}} \tag{C-148c}$$

$$z_{\xi\zeta}^{\vec{J}^{\rm IS}\vec{M}^{\rm O\rightleftharpoons G}} \;=\; \left\langle \vec{b}_{\xi}^{\vec{J}^{\rm IS}}, -\mathcal{K}_0\left(\vec{b}_{\zeta}^{\vec{M}^{\rm O\rightleftharpoons G}}\right)\right\rangle_{\mathbb{S}^{\rm ele}} \tag{C-148d}$$

$$z_{\xi\zeta}^{\vec{J}^{\rm IS}\vec{M}^{\rm IV}} \;=\; \left\langle \vec{b}_{\xi}^{\vec{J}^{\rm IS}}, -\mathcal{K}_0\left(\vec{b}_{\zeta}^{\vec{M}^{\rm IV}}\right)\right\rangle_{\mathbb{S}^{\rm ele}} \tag{C-148e}$$

Using the Eqs. (C-141a)~(C-143), we have the following transformation

$$\begin{bmatrix} \bar{a}^{\vec{J}^{\rm O\rightleftharpoons G}} \\ \bar{a}^{\vec{J}^{\rm IV}} \\ \bar{a}^{\vec{J}^{\rm IS}} \\ \bar{a}^{\vec{M}^{\rm O\rightleftharpoons G}} \\ \bar{a}^{\vec{M}^{\rm IV}} \end{bmatrix} \;=\; \bar{a}^{\rm AV} \;=\; \overline{\overline{T}}\cdot\bar{a} \tag{C-149}$$

In the above Eq. (C-149), $\overline{\overline{T}} = \overline{\overline{T}}^{\vec{J}^{\rm O\rightleftharpoons G}\rightarrow{\rm AV}} / \overline{\overline{T}}^{\vec{M}^{\rm O\rightleftharpoons G}\rightarrow{\rm AV}} / \overline{\overline{T}}^{\rm BS\rightarrow AV}$ and correspondingly $\bar{a} = \bar{a}^{\vec{J}^{\rm O\rightleftharpoons G}} / \bar{a}^{\vec{M}^{\rm O\rightleftharpoons G}} / \bar{a}^{\rm BS}$, and

$$\overline{\overline{T}}^{\vec{J}^{\rm O\rightleftharpoons G}\rightarrow{\rm AV}} \;=\; \left(\overline{\overline{\Psi}}_1\right)^{-1}\cdot\overline{\overline{\Psi}}_2 \tag{C-150a}$$

$$\overline{\overline{T}}^{\vec{M}^{\rm O\rightleftharpoons G}\rightarrow{\rm AV}} \;=\; \left(\overline{\overline{\Psi}}_3\right)^{-1}\cdot\overline{\overline{\Psi}}_4 \tag{C-150b}$$

and

$$\overline{\overline{T}}^{\rm BS\rightarrow AV} \;=\; {\rm nullspace}\left(\overline{\overline{\Psi}}_{\rm FCE}^{\rm DoJ/DoM}\right) \tag{C-151}$$

where

$$\overline{\overline{\Psi}}_1 = \begin{bmatrix} \overline{\overline{I}}^{\vec{J}^{\rm O\rightleftharpoons G}} & 0 & 0 & 0 & 0 \\ 0 & \overline{\overline{Z}}^{\vec{M}^{\rm O\rightleftharpoons G}\vec{J}^{\rm IV}} & \overline{\overline{Z}}^{\vec{M}^{\rm O\rightleftharpoons G}\vec{J}^{\rm IS}} & \overline{\overline{Z}}^{\vec{M}^{\rm O\rightleftharpoons G}\vec{M}^{\rm O\rightleftharpoons G}} & \overline{\overline{Z}}^{\vec{M}^{\rm O\rightleftharpoons G}\vec{M}^{\rm IV}} \\ 0 & \overline{\overline{Z}}^{\vec{J}^{\rm IV}\vec{J}^{\rm IV}} & \overline{\overline{Z}}^{\vec{J}^{\rm IV}\vec{J}^{\rm IS}} & \overline{\overline{Z}}^{\vec{J}^{\rm IV}\vec{M}^{\rm O\rightleftharpoons G}} & \overline{\overline{Z}}^{\vec{J}^{\rm IV}\vec{M}^{\rm IV}} \\ 0 & \overline{\overline{Z}}^{\vec{M}^{\rm IV}\vec{J}^{\rm IV}} & \overline{\overline{Z}}^{\vec{M}^{\rm IV}\vec{J}^{\rm IS}} & \overline{\overline{Z}}^{\vec{M}^{\rm IV}\vec{M}^{\rm O\rightleftharpoons G}} & \overline{\overline{Z}}^{\vec{M}^{\rm IV}\vec{M}^{\rm IV}} \\ 0 & \overline{\overline{Z}}^{\vec{J}^{\rm IS}\vec{J}^{\rm IV}} & \overline{\overline{Z}}^{\vec{J}^{\rm IS}\vec{J}^{\rm IS}} & \overline{\overline{Z}}^{\vec{J}^{\rm IS}\vec{M}^{\rm O\rightleftharpoons G}} & \overline{\overline{Z}}^{\vec{J}^{\rm IS}\vec{M}^{\rm IV}} \end{bmatrix} \tag{C-152a}$$

$$\overline{\overline{\Psi}}_2=\begin{bmatrix}\overline{\overline{I}}^{\vec{J}^{\mathrm{O}\rightleftharpoons\mathrm{G}}}\\-\overline{\overline{Z}}^{\vec{M}^{\mathrm{O}\rightleftharpoons\mathrm{G}}\vec{J}^{\mathrm{O}\rightleftharpoons\mathrm{G}}}\\-\overline{\overline{Z}}^{\vec{J}^{\mathrm{IV}}\vec{J}^{\mathrm{O}\rightleftharpoons\mathrm{G}}}\\-\overline{\overline{Z}}^{\vec{M}^{\mathrm{IV}}\vec{J}^{\mathrm{O}\rightleftharpoons\mathrm{G}}}\\-\overline{\overline{Z}}^{\vec{J}^{\mathrm{IS}}\vec{J}^{\mathrm{O}\rightleftharpoons\mathrm{G}}}\end{bmatrix}\tag{C-152b}$$

and

$$\overline{\overline{\Psi}}_3=\begin{bmatrix}0&0&0&\overline{\overline{I}}^{\vec{M}^{\mathrm{O}\rightleftharpoons\mathrm{G}}}&0\\\overline{\overline{Z}}^{\vec{J}^{\mathrm{O}\rightleftharpoons\mathrm{G}}\vec{J}^{\mathrm{O}\rightleftharpoons\mathrm{G}}}&\overline{\overline{Z}}^{\vec{J}^{\mathrm{O}\rightleftharpoons\mathrm{G}}\vec{J}^{\mathrm{IV}}}&\overline{\overline{Z}}^{\vec{J}^{\mathrm{O}\rightleftharpoons\mathrm{G}}\vec{J}^{\mathrm{IS}}}&0&\overline{\overline{Z}}^{\vec{J}^{\mathrm{O}\rightleftharpoons\mathrm{G}}\vec{M}^{\mathrm{IV}}}\\\overline{\overline{Z}}^{\vec{J}^{\mathrm{IV}}\vec{J}^{\mathrm{O}\rightleftharpoons\mathrm{G}}}&\overline{\overline{Z}}^{\vec{J}^{\mathrm{IV}}\vec{J}^{\mathrm{IV}}}&\overline{\overline{Z}}^{\vec{J}^{\mathrm{IV}}\vec{J}^{\mathrm{IS}}}&0&\overline{\overline{Z}}^{\vec{J}^{\mathrm{IV}}\vec{M}^{\mathrm{IV}}}\\\overline{\overline{Z}}^{\vec{M}^{\mathrm{IV}}\vec{J}^{\mathrm{O}\rightleftharpoons\mathrm{G}}}&\overline{\overline{Z}}^{\vec{M}^{\mathrm{IV}}\vec{J}^{\mathrm{IV}}}&\overline{\overline{Z}}^{\vec{M}^{\mathrm{IV}}\vec{J}^{\mathrm{IS}}}&0&\overline{\overline{Z}}^{\vec{M}^{\mathrm{IV}}\vec{M}^{\mathrm{IV}}}\\\overline{\overline{Z}}^{\vec{J}^{\mathrm{IS}}\vec{J}^{\mathrm{O}\rightleftharpoons\mathrm{G}}}&\overline{\overline{Z}}^{\vec{J}^{\mathrm{IS}}\vec{J}^{\mathrm{IV}}}&\overline{\overline{Z}}^{\vec{J}^{\mathrm{IS}}\vec{J}^{\mathrm{IS}}}&0&\overline{\overline{Z}}^{\vec{J}^{\mathrm{IS}}\vec{M}^{\mathrm{IV}}}\end{bmatrix}\tag{C-153a}$$

$$\overline{\overline{\Psi}}_4=\begin{bmatrix}\overline{\overline{I}}^{\vec{M}^{\mathrm{O}\rightleftharpoons\mathrm{G}}}\\-\overline{\overline{Z}}^{\vec{J}^{\mathrm{O}\rightleftharpoons\mathrm{G}}\vec{M}^{\mathrm{O}\rightleftharpoons\mathrm{G}}}\\-\overline{\overline{Z}}^{\vec{J}^{\mathrm{IV}}\vec{M}^{\mathrm{O}\rightleftharpoons\mathrm{G}}}\\-\overline{\overline{Z}}^{\vec{M}^{\mathrm{IV}}\vec{M}^{\mathrm{O}\rightleftharpoons\mathrm{G}}}\\-\overline{\overline{Z}}^{\vec{J}^{\mathrm{IS}}\vec{M}^{\mathrm{O}\rightleftharpoons\mathrm{G}}}\end{bmatrix}\tag{C-153b}$$

and

$$\overline{\overline{\Psi}}_{\mathrm{FCE}}^{\mathrm{DoJ}}=\begin{bmatrix}\overline{\overline{Z}}^{\vec{M}^{\mathrm{O}\rightleftharpoons\mathrm{G}}\vec{J}^{\mathrm{O}\rightleftharpoons\mathrm{G}}}&\overline{\overline{Z}}^{\vec{M}^{\mathrm{O}\rightleftharpoons\mathrm{G}}\vec{J}^{\mathrm{IV}}}&\overline{\overline{Z}}^{\vec{M}^{\mathrm{O}\rightleftharpoons\mathrm{G}}\vec{J}^{\mathrm{IS}}}&\overline{\overline{Z}}^{\vec{M}^{\mathrm{O}\rightleftharpoons\mathrm{G}}\vec{M}^{\mathrm{O}\rightleftharpoons\mathrm{G}}}&\overline{\overline{Z}}^{\vec{M}^{\mathrm{O}\rightleftharpoons\mathrm{G}}\vec{M}^{\mathrm{IV}}}\\\overline{\overline{Z}}^{\vec{J}^{\mathrm{IV}}\vec{J}^{\mathrm{O}\rightleftharpoons\mathrm{G}}}&\overline{\overline{Z}}^{\vec{J}^{\mathrm{IV}}\vec{J}^{\mathrm{IV}}}&\overline{\overline{Z}}^{\vec{J}^{\mathrm{IV}}\vec{J}^{\mathrm{IS}}}&\overline{\overline{Z}}^{\vec{J}^{\mathrm{IV}}\vec{M}^{\mathrm{O}\rightleftharpoons\mathrm{G}}}&\overline{\overline{Z}}^{\vec{J}^{\mathrm{IV}}\vec{M}^{\mathrm{IV}}}\\\overline{\overline{Z}}^{\vec{M}^{\mathrm{IV}}\vec{J}^{\mathrm{O}\rightleftharpoons\mathrm{G}}}&\overline{\overline{Z}}^{\vec{M}^{\mathrm{IV}}\vec{J}^{\mathrm{IV}}}&\overline{\overline{Z}}^{\vec{M}^{\mathrm{IV}}\vec{J}^{\mathrm{IS}}}&\overline{\overline{Z}}^{\vec{M}^{\mathrm{IV}}\vec{M}^{\mathrm{O}\rightleftharpoons\mathrm{G}}}&\overline{\overline{Z}}^{\vec{M}^{\mathrm{IV}}\vec{M}^{\mathrm{IV}}}\\\overline{\overline{Z}}^{\vec{J}^{\mathrm{IS}}\vec{J}^{\mathrm{O}\rightleftharpoons\mathrm{G}}}&\overline{\overline{Z}}^{\vec{J}^{\mathrm{IS}}\vec{J}^{\mathrm{IV}}}&\overline{\overline{Z}}^{\vec{J}^{\mathrm{IS}}\vec{J}^{\mathrm{IS}}}&\overline{\overline{Z}}^{\vec{J}^{\mathrm{IS}}\vec{M}^{\mathrm{O}\rightleftharpoons\mathrm{G}}}&\overline{\overline{Z}}^{\vec{J}^{\mathrm{IS}}\vec{M}^{\mathrm{IV}}}\end{bmatrix}\tag{C-154a}$$

$$\overline{\overline{\Psi}}_{\mathrm{FCE}}^{\mathrm{DoM}}=\begin{bmatrix}\overline{\overline{Z}}^{\vec{J}^{\mathrm{O}\rightleftharpoons\mathrm{G}}\vec{J}^{\mathrm{O}\rightleftharpoons\mathrm{G}}}&\overline{\overline{Z}}^{\vec{J}^{\mathrm{O}\rightleftharpoons\mathrm{G}}\vec{J}^{\mathrm{IV}}}&\overline{\overline{Z}}^{\vec{J}^{\mathrm{O}\rightleftharpoons\mathrm{G}}\vec{J}^{\mathrm{IS}}}&\overline{\overline{Z}}^{\vec{J}^{\mathrm{O}\rightleftharpoons\mathrm{G}}\vec{M}^{\mathrm{O}\rightleftharpoons\mathrm{G}}}&\overline{\overline{Z}}^{\vec{J}^{\mathrm{O}\rightleftharpoons\mathrm{G}}\vec{M}^{\mathrm{IV}}}\\\overline{\overline{Z}}^{\vec{J}^{\mathrm{IV}}\vec{J}^{\mathrm{O}\rightleftharpoons\mathrm{G}}}&\overline{\overline{Z}}^{\vec{J}^{\mathrm{IV}}\vec{J}^{\mathrm{IV}}}&\overline{\overline{Z}}^{\vec{J}^{\mathrm{IV}}\vec{J}^{\mathrm{IS}}}&\overline{\overline{Z}}^{\vec{J}^{\mathrm{IV}}\vec{M}^{\mathrm{O}\rightleftharpoons\mathrm{G}}}&\overline{\overline{Z}}^{\vec{J}^{\mathrm{IV}}\vec{M}^{\mathrm{IV}}}\\\overline{\overline{Z}}^{\vec{M}^{\mathrm{IV}}\vec{J}^{\mathrm{O}\rightleftharpoons\mathrm{G}}}&\overline{\overline{Z}}^{\vec{M}^{\mathrm{IV}}\vec{J}^{\mathrm{IV}}}&\overline{\overline{Z}}^{\vec{M}^{\mathrm{IV}}\vec{J}^{\mathrm{IS}}}&\overline{\overline{Z}}^{\vec{M}^{\mathrm{IV}}\vec{M}^{\mathrm{O}\rightleftharpoons\mathrm{G}}}&\overline{\overline{Z}}^{\vec{M}^{\mathrm{IV}}\vec{M}^{\mathrm{IV}}}\\\overline{\overline{Z}}^{\vec{J}^{\mathrm{IS}}\vec{J}^{\mathrm{O}\rightleftharpoons\mathrm{G}}}&\overline{\overline{Z}}^{\vec{J}^{\mathrm{IS}}\vec{J}^{\mathrm{IV}}}&\overline{\overline{Z}}^{\vec{J}^{\mathrm{IS}}\vec{J}^{\mathrm{IS}}}&\overline{\overline{Z}}^{\vec{J}^{\mathrm{IS}}\vec{M}^{\mathrm{O}\rightleftharpoons\mathrm{G}}}&\overline{\overline{Z}}^{\vec{J}^{\mathrm{IS}}\vec{M}^{\mathrm{IV}}}\end{bmatrix}\tag{C-154b}$$

Based on Eqs. (C-150) and (C-151), the IPO $P^{\mathrm{O}\rightleftharpoons\mathrm{G}}=(1/2)\iint_{\underset{\sim}{\mathbb{S}}^{\mathrm{O}\rightleftharpoons\mathrm{G}}}(\vec{E}\times\vec{H}^{\dagger})\cdot\hat{n}^{\rightarrow\mathrm{G}}dS$ can be rewritten as follows:

$$\begin{aligned}P^{\mathrm{O}\rightleftharpoons\mathrm{G}}=&\ (1/2)\left\langle\hat{n}^{\rightarrow\mathrm{G}}\times\vec{J}^{\mathrm{O}\rightleftharpoons\mathrm{G}},\vec{M}^{\mathrm{O}\rightleftharpoons\mathrm{G}}\right\rangle_{\underset{\sim}{\mathbb{S}}^{\mathrm{O}\rightleftharpoons\mathrm{G}}}\\=&-(1/2)\left\langle\vec{J}^{\mathrm{O}\rightleftharpoons\mathrm{G}},\mathcal{E}_0\left(\vec{J}^{\mathrm{O}\rightleftharpoons\mathrm{G}}+\vec{J}^{\mathrm{IV}}+\vec{J}^{\mathrm{IS}},\vec{M}^{\mathrm{O}\rightleftharpoons\mathrm{G}}+\vec{M}^{\mathrm{IV}}\right)\right\rangle_{\underset{\sim}{\mathbb{S}}^{\mathrm{O}\rightleftharpoons\mathrm{G}}}\\=&-(1/2)\left\langle\vec{M}^{\mathrm{O}\rightleftharpoons\mathrm{G}},\mathcal{H}_0\left(\vec{J}^{\mathrm{O}\rightleftharpoons\mathrm{G}}+\vec{J}^{\mathrm{IV}}+\vec{J}^{\mathrm{IS}},\vec{M}^{\mathrm{O}\rightleftharpoons\mathrm{G}}+\vec{M}^{\mathrm{IV}}\right)\right\rangle_{\underset{\sim}{\mathbb{S}}^{\mathrm{O}\rightleftharpoons\mathrm{G}}}^{\dagger}\end{aligned}\tag{C-155}$$

where integral surface $\underset{\sim}{\mathbb{S}}^{\mathrm{O}\rightleftharpoons\mathrm{G}}$ locates at the tra-guide side of $\mathbb{S}^{\mathrm{O}\rightleftharpoons\mathrm{G}}$. The IPO can be

discretized as follows:

$$P^{\mathrm{O}\rightleftharpoons\mathrm{G}} = \left(\bar{a}^{\mathrm{AV}}\right)^{\dagger}\cdot\overline{\overline{P}}{}^{\mathrm{O}\rightleftharpoons\mathrm{G}}_{\mathrm{curAV}}\cdot\bar{a}^{\mathrm{AV}} = \left(\bar{a}^{\mathrm{AV}}\right)^{\dagger}\cdot\overline{\overline{P}}{}^{\mathrm{O}\rightleftharpoons\mathrm{G}}_{\mathrm{intAV}}\cdot\bar{a}^{\mathrm{AV}} \tag{C-156}$$

in which

$$\overline{\overline{P}}{}^{\mathrm{O}\rightleftharpoons\mathrm{G}}_{\mathrm{curAV}} = \begin{bmatrix} 0 & 0 & 0 & \overline{\overline{C}}{}^{\vec{J}^{\mathrm{O}\rightleftharpoons\mathrm{G}}\vec{M}^{\mathrm{O}\rightleftharpoons\mathrm{G}}} & 0 \\ 0 & 0 & 0 & 0 & 0 \\ 0 & 0 & 0 & 0 & 0 \\ 0 & 0 & 0 & 0 & 0 \\ 0 & 0 & 0 & 0 & 0 \end{bmatrix} \tag{C-157}$$

corresponding to the first equality in Eq. (C-155), and

$$\overline{\overline{P}}{}^{\mathrm{O}\rightleftharpoons\mathrm{G}}_{\mathrm{intAV}} = \begin{bmatrix} \overline{\overline{P}}{}^{\vec{J}^{\mathrm{O}\rightleftharpoons\mathrm{G}}\vec{J}^{\mathrm{O}\rightleftharpoons\mathrm{G}}} & \overline{\overline{P}}{}^{\vec{J}^{\mathrm{O}\rightleftharpoons\mathrm{G}}\vec{J}^{\mathrm{IV}}} & \overline{\overline{P}}{}^{\vec{J}^{\mathrm{O}\rightleftharpoons\mathrm{G}}\vec{J}^{\mathrm{IS}}} & \overline{\overline{P}}{}^{\vec{J}^{\mathrm{O}\rightleftharpoons\mathrm{G}}\vec{M}^{\mathrm{O}\rightleftharpoons\mathrm{G}}} & \overline{\overline{P}}{}^{\vec{J}^{\mathrm{O}\rightleftharpoons\mathrm{G}}\vec{M}^{\mathrm{IV}}} \\ 0 & 0 & 0 & 0 & 0 \\ 0 & 0 & 0 & 0 & 0 \\ 0 & 0 & 0 & 0 & 0 \\ 0 & 0 & 0 & 0 & 0 \end{bmatrix} \tag{C-158a}$$

corresponding to the second equality in Eq. (C-155), and

$$\overline{\overline{P}}{}^{\mathrm{O}\rightleftharpoons\mathrm{G}}_{\mathrm{intAV}} = \begin{bmatrix} 0 & 0 & 0 & 0 & 0 \\ 0 & 0 & 0 & 0 & 0 \\ 0 & 0 & 0 & 0 & 0 \\ \overline{\overline{P}}{}^{\vec{M}^{\mathrm{O}\rightleftharpoons\mathrm{G}}\vec{J}^{\mathrm{O}\rightleftharpoons\mathrm{G}}} & \overline{\overline{P}}{}^{\vec{M}^{\mathrm{O}\rightleftharpoons\mathrm{G}}\vec{J}^{\mathrm{IV}}} & \overline{\overline{P}}{}^{\vec{M}^{\mathrm{O}\rightleftharpoons\mathrm{G}}\vec{J}^{\mathrm{IS}}} & \overline{\overline{P}}{}^{\vec{M}^{\mathrm{O}\rightleftharpoons\mathrm{G}}\vec{M}^{\mathrm{O}\rightleftharpoons\mathrm{G}}} & \overline{\overline{P}}{}^{\vec{M}^{\mathrm{O}\rightleftharpoons\mathrm{G}}\vec{M}^{\mathrm{IV}}} \\ 0 & 0 & 0 & 0 & 0 \end{bmatrix}^{\dagger} \tag{C-158b}$$

corresponding to the third equality in Eq. (C-155), where the elements of the submatrices are as follows:

$$c_{\xi\zeta}^{\vec{J}^{\mathrm{O}\rightleftharpoons\mathrm{G}}\vec{M}^{\mathrm{O}\rightleftharpoons\mathrm{G}}} = (1/2)\left\langle \hat{n}^{\rightarrow\mathrm{G}}\times\vec{b}_{\xi}^{\vec{J}^{\mathrm{O}\rightleftharpoons\mathrm{G}}}, \vec{b}_{\zeta}^{\vec{M}^{\mathrm{O}\rightleftharpoons\mathrm{G}}}\right\rangle_{\mathbb{S}^{\mathrm{O}\rightleftharpoons\mathrm{G}}} \tag{C-159}$$

and

$$p_{\xi\zeta}^{\vec{J}^{\mathrm{O}\rightleftharpoons\mathrm{G}}\vec{J}^{\mathrm{O}\rightleftharpoons\mathrm{G}}} = -(1/2)\left\langle \vec{b}_{\xi}^{\vec{J}^{\mathrm{O}\rightleftharpoons\mathrm{G}}}, -j\omega\mu_0\mathcal{L}_0\left(\vec{b}_{\zeta}^{\vec{J}^{\mathrm{O}\rightleftharpoons\mathrm{G}}}\right)\right\rangle_{\mathbb{S}^{\mathrm{O}\rightleftharpoons\mathrm{G}}} \tag{C-160a}$$

$$p_{\xi\zeta}^{\vec{J}^{\mathrm{O}\rightleftharpoons\mathrm{G}}\vec{J}^{\mathrm{IV}}} = -(1/2)\left\langle \vec{b}_{\xi}^{\vec{J}^{\mathrm{O}\rightleftharpoons\mathrm{G}}}, -j\omega\mu_0\mathcal{L}_0\left(\vec{b}_{\zeta}^{\vec{J}^{\mathrm{IV}}}\right)\right\rangle_{\mathbb{S}^{\mathrm{O}\rightleftharpoons\mathrm{G}}} \tag{C-160b}$$

$$p_{\xi\zeta}^{\vec{J}^{\mathrm{O}\rightleftharpoons\mathrm{G}}\vec{J}^{\mathrm{IS}}} = -(1/2)\left\langle \vec{b}_{\xi}^{\vec{J}^{\mathrm{O}\rightleftharpoons\mathrm{G}}}, -j\omega\mu_0\mathcal{L}_0\left(\vec{b}_{\zeta}^{\vec{J}^{\mathrm{IS}}}\right)\right\rangle_{\mathbb{S}^{\mathrm{O}\rightleftharpoons\mathrm{G}}} \tag{C-160c}$$

$$p_{\xi\zeta}^{\vec{J}^{\mathrm{O}\rightleftharpoons\mathrm{G}}\vec{M}^{\mathrm{O}\rightleftharpoons\mathrm{G}}} = -(1/2)\left\langle \vec{b}_{\xi}^{\vec{J}^{\mathrm{O}\rightleftharpoons\mathrm{G}}}, \hat{n}^{\rightarrow\mathrm{G}}\times\frac{1}{2}\vec{b}_{\zeta}^{\vec{M}^{\mathrm{O}\rightleftharpoons\mathrm{G}}} - \mathrm{P.V.}\,\mathcal{K}_0\left(\vec{b}_{\zeta}^{\vec{M}^{\mathrm{O}\rightleftharpoons\mathrm{G}}}\right)\right\rangle_{\mathbb{S}^{\mathrm{O}\rightleftharpoons\mathrm{G}}} \tag{C-160d}$$

$$p_{\xi\zeta}^{\vec{J}^{\mathrm{O}\rightleftharpoons\mathrm{G}}\vec{M}^{\mathrm{IV}}} = -(1/2)\left\langle \vec{b}_{\xi}^{\vec{J}^{\mathrm{O}\rightleftharpoons\mathrm{G}}}, -\mathcal{K}_0\left(\vec{b}_{\zeta}^{\vec{M}^{\mathrm{IV}}}\right)\right\rangle_{\mathbb{S}^{\mathrm{O}\rightleftharpoons\mathrm{G}}} \tag{C-160e}$$

and

$$p_{\xi\zeta}^{\vec{M}^{\mathrm{O}\rightleftharpoons\mathrm{G}}\vec{J}^{\mathrm{O}\rightleftharpoons\mathrm{G}}} = -(1/2)\left\langle \vec{b}_{\xi}^{\vec{M}^{\mathrm{O}\rightleftharpoons\mathrm{G}}}, \frac{1}{2}\vec{b}_{\zeta}^{\vec{J}^{\mathrm{O}\rightleftharpoons\mathrm{G}}} \times \hat{n}^{\rightarrow\mathrm{G}} + \mathrm{P.V.}\mathcal{K}_0\left(\vec{b}_{\zeta}^{\vec{J}^{\mathrm{O}\rightleftharpoons\mathrm{G}}}\right)\right\rangle_{\mathbb{S}^{\mathrm{O}\rightleftharpoons\mathrm{G}}} \tag{C-160f}$$

$$p_{\xi\zeta}^{\vec{M}^{\mathrm{O}\rightleftharpoons\mathrm{G}}\vec{J}^{\mathrm{IV}}} = -(1/2)\left\langle \vec{b}_{\xi}^{\vec{M}^{\mathrm{O}\rightleftharpoons\mathrm{G}}}, \mathcal{K}_0\left(\vec{b}_{\zeta}^{\vec{J}^{\mathrm{IV}}}\right)\right\rangle_{\mathbb{S}^{\mathrm{O}\rightleftharpoons\mathrm{G}}} \tag{C-160g}$$

$$p_{\xi\zeta}^{\vec{M}^{\mathrm{O}\rightleftharpoons\mathrm{G}}\vec{J}^{\mathrm{IS}}} = -(1/2)\left\langle \vec{b}_{\xi}^{\vec{M}^{\mathrm{O}\rightleftharpoons\mathrm{G}}}, \mathcal{K}_0\left(\vec{b}_{\zeta}^{\vec{J}^{\mathrm{IS}}}\right)\right\rangle_{\mathbb{S}^{\mathrm{O}\rightleftharpoons\mathrm{G}}} \tag{C-160h}$$

$$p_{\xi\zeta}^{\vec{M}^{\mathrm{O}\rightleftharpoons\mathrm{G}}\vec{M}^{\mathrm{O}\rightleftharpoons\mathrm{G}}} = -(1/2)\left\langle \vec{b}_{\xi}^{\vec{M}^{\mathrm{O}\rightleftharpoons\mathrm{G}}}, -j\omega\varepsilon_0\mathcal{L}_0\left(\vec{b}_{\zeta}^{\vec{M}^{\mathrm{O}\rightleftharpoons\mathrm{G}}}\right)\right\rangle_{\mathbb{S}^{\mathrm{O}\rightleftharpoons\mathrm{G}}} \tag{C-160i}$$

$$p_{\xi\zeta}^{\vec{M}^{\mathrm{O}\rightleftharpoons\mathrm{G}}\vec{M}^{\mathrm{IV}}} = -(1/2)\left\langle \vec{b}_{\xi}^{\vec{M}^{\mathrm{O}\rightleftharpoons\mathrm{G}}}, -j\omega\varepsilon_0\mathcal{L}_0\left(\vec{b}_{\zeta}^{\vec{M}^{\mathrm{IV}}}\right)\right\rangle_{\mathbb{S}^{\mathrm{O}\rightleftharpoons\mathrm{G}}} \tag{C-160j}$$

To obtain the IPO defined on modal space, we substitute Eq. (C-149) into the Eq. (C-156), and then we have that

$$P^{\mathrm{O}\rightleftharpoons\mathrm{G}} = \bar{a}^{\dagger}\cdot\underbrace{\left(\bar{\bar{T}}^{\dagger}\cdot\bar{\bar{P}}_{\mathrm{curAV}}^{\mathrm{O}\rightleftharpoons\mathrm{G}}\cdot\bar{\bar{T}}\right)}_{\bar{\bar{P}}_{\mathrm{cur}}^{\mathrm{O}\rightleftharpoons\mathrm{G}}}\cdot\bar{a} = \bar{a}^{\dagger}\cdot\underbrace{\left(\bar{\bar{T}}^{\dagger}\cdot\bar{\bar{P}}_{\mathrm{intAV}}^{\mathrm{O}\rightleftharpoons\mathrm{G}}\cdot\bar{\bar{T}}\right)}_{\bar{\bar{P}}_{\mathrm{int}}^{\mathrm{O}\rightleftharpoons\mathrm{G}}}\cdot\bar{a} \tag{C-161}$$

where subscripts "cur" and "int" are to emphasize that $\bar{\bar{P}}_{\mathrm{cur}}^{\mathrm{O}\rightleftharpoons\mathrm{G}}$ and $\bar{\bar{P}}_{\mathrm{int}}^{\mathrm{O}\rightleftharpoons\mathrm{G}}$ respectively originate from discretizing the current and interaction forms of IPO.

The process of constructing IP-DMs from orthogonalizing $\bar{\bar{P}}_{\mathrm{cur}}^{\mathrm{O}\rightleftharpoons\mathrm{G}}$ and $\bar{\bar{P}}_{\mathrm{int}}^{\mathrm{O}\rightleftharpoons\mathrm{G}}$ is completely the same as the one used in Sec. 8.2.

## C8 Surface-Volume Formulation of the PTT-Based DMT for the Augmented Tra-antenna-Rec-antenna System Discussed in Sec. 8.3

The method to establish the surface-volume formulation of the PTT-based DMT for the augmented tra-antenna-rec-antenna combined system discussed in Sec. 8.3 is similar to the method given in App. C7, and it will not be explicitly exhibited here for condensing the length of this report.

# Appendix D Some Detailed Formulations Related to This Report

In this App. D, some detailed formulations related to the contents in the main body of this report are provided for readers' reference.

## D1 Some Detailed Formulations Related to Sec. 3.4

The formulations used to calculate the elements of the matrices in Eqs. (3-126a)~(3-130b) are explicitly given as follows:

$$z_{\xi\zeta}^{\vec{M}_{\mathrm{in}}^{\mathrm{ES}}\vec{J}_{\mathrm{in}}^{\mathrm{ES}}} = \left\langle \vec{b}_{\xi}^{\vec{M}_{\mathrm{in}}^{\mathrm{ES}}}, \mathrm{P.V.}\mathcal{K}\left(\vec{b}_{\zeta}^{\vec{J}_{\mathrm{in}}^{\mathrm{ES}}}\right)\right\rangle_{\mathbb{S}_{\mathrm{in}}} - (1/2)\left\langle \vec{b}_{\xi}^{\vec{M}_{\mathrm{in}}^{\mathrm{ES}}}, \vec{b}_{\zeta}^{\vec{J}_{\mathrm{in}}^{\mathrm{ES}}} \times \hat{z}\right\rangle_{\mathbb{S}_{\mathrm{in}}} \tag{D-1a}$$

$$z_{\xi\zeta}^{\vec{M}_{\text{in}}^{\text{ES}}\vec{J}_{11}^{\text{ES}}} = \left\langle \vec{b}_{\xi}^{\vec{M}_{\text{in}}^{\text{ES}}}, \mathcal{K}\left(\vec{b}_{\zeta}^{\vec{J}_{11}^{\text{ES}}}\right)\right\rangle_{\mathbb{S}_{\text{in}}} \tag{D-1b}$$

$$z_{\xi\zeta}^{\vec{M}_{\text{in}}^{\text{ES}}\vec{J}_{10}^{\text{ES}}} = \left\langle \vec{b}_{\xi}^{\vec{M}_{\text{in}}^{\text{ES}}}, \mathcal{K}\left(\vec{b}_{\zeta}^{\vec{J}_{10}^{\text{ES}}}\right)\right\rangle_{\mathbb{S}_{\text{in}}} \tag{D-1c}$$

$$z_{\xi\zeta}^{\vec{M}_{\text{in}}^{\text{ES}}\vec{J}_{\text{out}}^{\text{ES}}} = \left\langle \vec{b}_{\xi}^{\vec{M}_{\text{in}}^{\text{ES}}}, \mathcal{K}\left(-\vec{b}_{\zeta}^{\vec{J}_{\text{out}}^{\text{ES}}}\right)\right\rangle_{\mathbb{S}_{\text{in}}} \tag{D-1d}$$

$$z_{\xi\zeta}^{\vec{M}_{\text{in}}^{\text{ES}}\vec{M}_{\text{in}}^{\text{ES}}} = \left\langle \vec{b}_{\xi}^{\vec{M}_{\text{in}}^{\text{ES}}}, -j\omega\varepsilon\mathcal{L}\left(\vec{b}_{\zeta}^{\vec{M}_{\text{in}}^{\text{ES}}}\right)\right\rangle_{\mathbb{S}_{\text{in}}} \tag{D-1e}$$

$$z_{\xi\zeta}^{\vec{M}_{\text{in}}^{\text{ES}}\vec{M}_{10}^{\text{ES}}} = \left\langle \vec{b}_{\xi}^{\vec{M}_{\text{in}}^{\text{ES}}}, -j\omega\varepsilon\mathcal{L}\left(\vec{b}_{\zeta}^{\vec{M}_{10}^{\text{ES}}}\right)\right\rangle_{\mathbb{S}_{\text{in}}} \tag{D-1f}$$

$$z_{\xi\zeta}^{\vec{M}_{\text{in}}^{\text{ES}}\vec{M}_{\text{out}}^{\text{ES}}} = \left\langle \vec{b}_{\xi}^{\vec{M}_{\text{in}}^{\text{ES}}}, -j\omega\varepsilon\mathcal{L}\left(-\vec{b}_{\zeta}^{\vec{M}_{\text{out}}^{\text{ES}}}\right)\right\rangle_{\mathbb{S}_{\text{in}}} \tag{D-1g}$$

and

$$z_{\xi\zeta}^{\vec{J}_{\text{in}}^{\text{ES}}\vec{J}_{\text{in}}^{\text{ES}}} = \left\langle \vec{b}_{\xi}^{\vec{J}_{\text{in}}^{\text{ES}}}, -j\omega\mu\mathcal{L}\left(\vec{b}_{\zeta}^{\vec{J}_{\text{in}}^{\text{ES}}}\right)\right\rangle_{\mathbb{S}_{\text{in}}} \tag{D-2a}$$

$$z_{\xi\zeta}^{\vec{J}_{\text{in}}^{\text{ES}}\vec{J}_{11}^{\text{ES}}} = \left\langle \vec{b}_{\xi}^{\vec{J}_{\text{in}}^{\text{ES}}}, -j\omega\mu\mathcal{L}\left(\vec{b}_{\zeta}^{\vec{J}_{11}^{\text{ES}}}\right)\right\rangle_{\mathbb{S}_{\text{in}}} \tag{D-2b}$$

$$z_{\xi\zeta}^{\vec{J}_{\text{in}}^{\text{ES}}\vec{J}_{10}^{\text{ES}}} = \left\langle \vec{b}_{\xi}^{\vec{J}_{\text{in}}^{\text{ES}}}, -j\omega\mu\mathcal{L}\left(\vec{b}_{\zeta}^{\vec{J}_{10}^{\text{ES}}}\right)\right\rangle_{\mathbb{S}_{\text{in}}} \tag{D-2c}$$

$$z_{\xi\zeta}^{\vec{J}_{\text{in}}^{\text{ES}}\vec{J}_{\text{out}}^{\text{ES}}} = \left\langle \vec{b}_{\xi}^{\vec{J}_{\text{in}}^{\text{ES}}}, -j\omega\mu\mathcal{L}\left(-\vec{b}_{\zeta}^{\vec{J}_{\text{out}}^{\text{ES}}}\right)\right\rangle_{\mathbb{S}_{\text{in}}} \tag{D-2d}$$

$$z_{\xi\zeta}^{\vec{J}_{\text{in}}^{\text{ES}}\vec{M}_{\text{in}}^{\text{ES}}} = \left\langle \vec{b}_{\xi}^{\vec{J}_{\text{in}}^{\text{ES}}}, -\text{P.V.}\mathcal{K}\left(\vec{b}_{\zeta}^{\vec{M}_{\text{in}}^{\text{ES}}}\right)\right\rangle_{\mathbb{S}_{\text{in}}} - (1/2)\left\langle \vec{b}_{\xi}^{\vec{J}_{\text{in}}^{\text{ES}}}, \hat{z}\times\vec{b}_{\zeta}^{\vec{M}_{\text{in}}^{\text{ES}}}\right\rangle_{\mathbb{S}_{\text{in}}} \tag{D-2e}$$

$$z_{\xi\zeta}^{\vec{J}_{\text{in}}^{\text{ES}}\vec{M}_{10}^{\text{ES}}} = \left\langle \vec{b}_{\xi}^{\vec{J}_{\text{in}}^{\text{ES}}}, -\mathcal{K}\left(\vec{b}_{\zeta}^{\vec{M}_{10}^{\text{ES}}}\right)\right\rangle_{\mathbb{S}_{\text{in}}} \tag{D-2f}$$

$$z_{\xi\zeta}^{\vec{J}_{\text{in}}^{\text{ES}}\vec{M}_{\text{out}}^{\text{ES}}} = \left\langle \vec{b}_{\xi}^{\vec{J}_{\text{in}}^{\text{ES}}}, -\mathcal{K}\left(-\vec{b}_{\zeta}^{\vec{M}_{\text{out}}^{\text{ES}}}\right)\right\rangle_{\mathbb{S}_{\text{in}}} \tag{D-2g}$$

and

$$z_{\xi\zeta}^{\vec{J}_{11}^{\text{ES}}\vec{J}_{\text{in}}^{\text{ES}}} = \left\langle \vec{b}_{\xi}^{\vec{J}_{11}^{\text{ES}}}, -j\omega\mu\mathcal{L}\left(\vec{b}_{\zeta}^{\vec{J}_{\text{in}}^{\text{ES}}}\right)\right\rangle_{\mathbb{S}_{11}} \tag{D-3a}$$

$$z_{\xi\zeta}^{\vec{J}_{11}^{\text{ES}}\vec{J}_{11}^{\text{ES}}} = \left\langle \vec{b}_{\xi}^{\vec{J}_{11}^{\text{ES}}}, -j\omega\mu\mathcal{L}\left(\vec{b}_{\zeta}^{\vec{J}_{11}^{\text{ES}}}\right)\right\rangle_{\mathbb{S}_{11}} \tag{D-3b}$$

$$z_{\xi\zeta}^{\vec{J}_{11}^{\text{ES}}\vec{J}_{10}^{\text{ES}}} = \left\langle \vec{b}_{\xi}^{\vec{J}_{11}^{\text{ES}}}, -j\omega\mu\mathcal{L}\left(\vec{b}_{\zeta}^{\vec{J}_{10}^{\text{ES}}}\right)\right\rangle_{\mathbb{S}_{11}} \tag{D-3c}$$

$$z_{\xi\zeta}^{\vec{J}_{11}^{\text{ES}}\vec{J}_{\text{out}}^{\text{ES}}} = \left\langle \vec{b}_{\xi}^{\vec{J}_{11}^{\text{ES}}}, -j\omega\mu\mathcal{L}\left(-\vec{b}_{\zeta}^{\vec{J}_{\text{out}}^{\text{ES}}}\right)\right\rangle_{\mathbb{S}_{11}} \tag{D-3d}$$

$$z_{\xi\zeta}^{\vec{J}_{11}^{\text{ES}}\vec{M}_{\text{in}}^{\text{ES}}} = \left\langle \vec{b}_{\xi}^{\vec{J}_{11}^{\text{ES}}}, -j\omega\mu\mathcal{L}\left(\vec{b}_{\zeta}^{\vec{M}_{\text{in}}^{\text{ES}}}\right)\right\rangle_{\mathbb{S}_{11}} \tag{D-3e}$$

$$z_{\xi\zeta}^{\vec{J}_{11}^{\text{ES}}\vec{M}_{10}^{\text{ES}}} = \left\langle \vec{b}_{\xi}^{\vec{J}_{11}^{\text{ES}}}, -j\omega\mu\mathcal{L}\left(\vec{b}_{\zeta}^{\vec{M}_{10}^{\text{ES}}}\right)\right\rangle_{\mathbb{S}_{11}} \tag{D-3f}$$

$$z_{\xi\zeta}^{\vec{J}_{11}^{\text{ES}}\vec{M}_{\text{out}}^{\text{ES}}} = \left\langle \vec{b}_{\xi}^{\vec{J}_{11}^{\text{ES}}}, -j\omega\mu\mathcal{L}\left(-\vec{b}_{\zeta}^{\vec{M}_{\text{out}}^{\text{ES}}}\right)\right\rangle_{\mathbb{S}_{11}} \tag{D-3g}$$

and

$$z_{\xi\zeta}^{\vec{J}_{00}^{\text{ES}}\vec{J}_{10}^{\text{ES}}} = \left\langle \vec{b}_{\xi}^{\vec{J}_{00}^{\text{ES}}}, -j\omega\mu_0\mathcal{L}_0\left(-\vec{b}_{\zeta}^{\vec{J}_{10}^{\text{ES}}}\right)\right\rangle_{\mathbb{S}_{00}} \tag{D-4a}$$

$$z_{\xi\zeta}^{\vec{J}_{00}^{\text{ES}}\vec{J}_{00}^{\text{ES}}} = \left\langle \vec{b}_{\xi}^{\vec{J}_{00}^{\text{ES}}}, -j\omega\mu_0\mathcal{L}_0\left(\vec{b}_{\zeta}^{\vec{J}_{00}^{\text{ES}}}\right)\right\rangle_{\mathbb{S}_{00}} \tag{D-4b}$$

$$z_{\xi\zeta}^{\vec{J}_{00}^{\text{ES}}\vec{M}_{10}^{\text{ES}}} = \left\langle \vec{b}_{\xi}^{\vec{J}_{00}^{\text{ES}}}, -\mathcal{K}_0\left(-\vec{b}_{\zeta}^{\vec{M}_{10}^{\text{ES}}}\right)\right\rangle_{\mathbb{S}_{00}} \tag{D-4c}$$

and

$$z_{\xi\zeta}^{\vec{J}_{10}^{\text{ES}}\vec{J}_{\text{in}}^{\text{ES}}} = \left\langle \vec{b}_{\xi}^{\vec{J}_{10}^{\text{ES}}}, -j\omega\mu\mathcal{L}\left(\vec{b}_{\zeta}^{\vec{J}_{\text{in}}^{\text{ES}}}\right)\right\rangle_{\mathbb{S}_{10}} \tag{D-5a}$$

$$z_{\xi\zeta}^{\vec{J}_{10}^{\text{ES}}\vec{J}_{11}^{\text{ES}}} = \left\langle \vec{b}_{\xi}^{\vec{J}_{10}^{\text{ES}}}, -j\omega\mu\mathcal{L}\left(\vec{b}_{\zeta}^{\vec{J}_{11}^{\text{ES}}}\right)\right\rangle_{\mathbb{S}_{10}} \tag{D-5b}$$

$$z_{\xi\zeta}^{\vec{J}_{10}^{\text{ES}}\vec{J}_{10}^{\text{ES}}} = \left\langle \vec{b}_{\xi}^{\vec{J}_{10}^{\text{ES}}}, -j\omega\mu\mathcal{L}\left(\vec{b}_{\zeta}^{\vec{J}_{10}^{\text{ES}}}\right)\right\rangle_{\mathbb{S}_{10}} - \left\langle \vec{b}_{\xi}^{\vec{J}_{10}^{\text{ES}}}, -j\omega\mu_0\mathcal{L}_0\left(-\vec{b}_{\zeta}^{\vec{J}_{10}^{\text{ES}}}\right)\right\rangle_{\mathbb{S}_{10}} \tag{D-5c}$$

$$z_{\xi\zeta}^{\vec{J}_{10}^{\text{ES}}\vec{J}_{00}^{\text{ES}}} = -\left\langle \vec{b}_{\xi}^{\vec{J}_{10}^{\text{ES}}}, -j\omega\mu_0\mathcal{L}_0\left(\vec{b}_{\zeta}^{\vec{J}_{00}^{\text{ES}}}\right)\right\rangle_{\mathbb{S}_{10}} \tag{D-5d}$$

$$z_{\xi\zeta}^{\vec{J}_{10}^{\text{ES}}\vec{J}_{\text{out}}^{\text{ES}}} = \left\langle \vec{b}_{\xi}^{\vec{J}_{10}^{\text{ES}}}, -j\omega\mu\mathcal{L}\left(-\vec{b}_{\zeta}^{\vec{J}_{\text{out}}^{\text{ES}}}\right)\right\rangle_{\mathbb{S}_{10}} \tag{D-5e}$$

$$z_{\xi\zeta}^{\vec{J}_{10}^{\text{ES}}\vec{M}_{\text{in}}^{\text{ES}}} = \left\langle \vec{b}_{\xi}^{\vec{J}_{10}^{\text{ES}}}, -\mathcal{K}\left(\vec{b}_{\zeta}^{\vec{M}_{\text{in}}^{\text{ES}}}\right)\right\rangle_{\mathbb{S}_{10}} \tag{D-5f}$$

$$z_{\xi\zeta}^{\vec{J}_{10}^{\text{ES}}\vec{M}_{10}^{\text{ES}}} = \left\langle \vec{b}_{\xi}^{\vec{J}_{10}^{\text{ES}}}, -\text{P.V.}\mathcal{K}\left(\vec{b}_{\zeta}^{\vec{M}_{10}^{\text{ES}}}\right)\right\rangle_{\mathbb{S}_{10}} - \left\langle \vec{b}_{\xi}^{\vec{J}_{10}^{\text{ES}}}, -\text{P.V.}\mathcal{K}_0\left(-\vec{b}_{\zeta}^{\vec{M}_{10}^{\text{ES}}}\right)\right\rangle_{\mathbb{S}_{10}} \tag{D-5g}$$

$$z_{\xi\zeta}^{\vec{J}_{10}^{\text{ES}}\vec{M}_{\text{out}}^{\text{ES}}} = \left\langle \vec{b}_{\xi}^{\vec{J}_{10}^{\text{ES}}}, -\mathcal{K}\left(-\vec{b}_{\zeta}^{\vec{M}_{\text{out}}^{\text{ES}}}\right)\right\rangle_{\mathbb{S}_{10}} \tag{D-5h}$$

and

$$z_{\xi\zeta}^{\vec{M}_{10}^{\text{ES}}\vec{J}_{\text{in}}^{\text{ES}}} = \left\langle \vec{b}_{\xi}^{\vec{M}_{10}^{\text{ES}}}, \mathcal{K}\left(\vec{b}_{\zeta}^{\vec{J}_{\text{in}}^{\text{ES}}}\right)\right\rangle_{\mathbb{S}_{10}} \tag{D-6a}$$

$$z_{\xi\zeta}^{\vec{M}_{10}^{\text{ES}}\vec{J}_{11}^{\text{ES}}} = \left\langle \vec{b}_{\xi}^{\vec{M}_{10}^{\text{ES}}}, \mathcal{K}\left(\vec{b}_{\zeta}^{\vec{J}_{11}^{\text{ES}}}\right)\right\rangle_{\mathbb{S}_{10}} \tag{D-6b}$$

$$z_{\xi\zeta}^{\vec{M}_{10}^{\text{ES}}\vec{J}_{10}^{\text{ES}}} = \left\langle \vec{b}_{\xi}^{\vec{M}_{10}^{\text{ES}}}, \text{P.V.}\mathcal{K}\left(\vec{b}_{\zeta}^{\vec{J}_{10}^{\text{ES}}}\right)\right\rangle_{\mathbb{S}_{10}} - \left\langle \vec{b}_{\xi}^{\vec{M}_{10}^{\text{ES}}}, \text{P.V.}\mathcal{K}_0\left(-\vec{b}_{\zeta}^{\vec{J}_{10}^{\text{ES}}}\right)\right\rangle_{\mathbb{S}_{10}} \tag{D-6c}$$

$$z_{\xi\zeta}^{\vec{M}_{10}^{\text{ES}}\vec{J}_{00}^{\text{ES}}} = -\left\langle \vec{b}_{\xi}^{\vec{M}_{10}^{\text{ES}}}, \mathcal{K}_0\left(\vec{b}_{\zeta}^{\vec{J}_{00}^{\text{ES}}}\right)\right\rangle_{\mathbb{S}_{10}} \tag{D-6d}$$

$$z_{\xi\zeta}^{\vec{M}_{10}^{\text{ES}}\vec{J}_{\text{out}}^{\text{ES}}} = \left\langle \vec{b}_{\xi}^{\vec{M}_{10}^{\text{ES}}}, \mathcal{K}\left(-\vec{b}_{\zeta}^{\vec{J}_{\text{out}}^{\text{ES}}}\right)\right\rangle_{\mathbb{S}_{10}} \tag{D-6e}$$

$$z_{\xi\zeta}^{\vec{M}_{10}^{\text{ES}}\vec{M}_{\text{in}}^{\text{ES}}} = \left\langle \vec{b}_{\xi}^{\vec{M}_{10}^{\text{ES}}}, -j\omega\varepsilon\mathcal{L}\left(\vec{b}_{\zeta}^{\vec{M}_{\text{in}}^{\text{ES}}}\right)\right\rangle_{\mathbb{S}_{10}} \tag{D-6f}$$

$$z_{\xi\zeta}^{\vec{M}_{10}^{\text{ES}}\vec{M}_{10}^{\text{ES}}} = \left\langle \vec{b}_{\xi}^{\vec{M}_{10}^{\text{ES}}}, -j\omega\varepsilon\mathcal{L}\left(\vec{b}_{\zeta}^{\vec{M}_{10}^{\text{ES}}}\right)\right\rangle_{\mathbb{S}_{10}} - \left\langle \vec{b}_{\xi}^{\vec{M}_{10}^{\text{ES}}}, -j\omega\varepsilon_0\mathcal{L}_0\left(-\vec{b}_{\zeta}^{\vec{M}_{10}^{\text{ES}}}\right)\right\rangle_{\mathbb{S}_{10}} \tag{D-6g}$$

$$z_{\xi\zeta}^{\vec{M}_{10}^{\text{ES}}\vec{M}_{\text{out}}^{\text{ES}}} = \left\langle \vec{b}_{\xi}^{\vec{M}_{10}^{\text{ES}}}, -j\omega\varepsilon\mathcal{L}\left(-\vec{b}_{\zeta}^{\vec{M}_{\text{out}}^{\text{ES}}}\right)\right\rangle_{\mathbb{S}_{10}} \tag{D-6h}$$

and

$$z_{\xi\zeta}^{\vec{J}_{\text{out}}^{\text{ES}}\vec{J}_{\text{in}}^{\text{ES}}} = \left\langle \vec{b}_{\xi}^{\vec{J}_{\text{out}}^{\text{ES}}}, -j\omega\mu\mathcal{L}\left(\vec{b}_{\zeta}^{\vec{J}_{\text{in}}^{\text{ES}}}\right)\right\rangle_{\mathbb{S}_{\text{out}}} \tag{D-7a}$$

$$z_{\xi\zeta}^{\vec{J}_{\text{out}}^{\text{ES}}\vec{J}_{11}^{\text{ES}}} = \left\langle \vec{b}_{\xi}^{\vec{J}_{\text{out}}^{\text{ES}}}, -j\omega\mu\mathcal{L}\left(\vec{b}_{\zeta}^{\vec{J}_{11}^{\text{ES}}}\right)\right\rangle_{\mathbb{S}_{\text{out}}} \tag{D-7b}$$

$$z_{\xi\zeta}^{\vec{J}_{\text{out}}^{\text{ES}}\vec{J}_{10}^{\text{ES}}} = \left\langle \vec{b}_{\xi}^{\vec{J}_{\text{out}}^{\text{ES}}}, -j\omega\mu\mathcal{L}\left(\vec{b}_{\zeta}^{\vec{J}_{10}^{\text{ES}}}\right)\right\rangle_{\mathbb{S}_{\text{out}}} \tag{D-7c}$$

$$z_{\xi\zeta}^{\vec{J}_{\text{out}}^{\text{ES}}\vec{J}_{\text{out}}^{\text{ES}}} = \left\langle \vec{b}_{\xi}^{\vec{J}_{\text{out}}^{\text{ES}}}, -j\omega\mu\mathcal{L}\left(-\vec{b}_{\zeta}^{\vec{J}_{\text{out}}^{\text{ES}}}\right)\right\rangle_{\mathbb{S}_{\text{out}}} - \left\langle \vec{b}_{\xi}^{\vec{J}_{\text{out}}^{\text{ES}}}, -j\omega\mu\mathcal{L}\left(\vec{b}_{\zeta}^{\vec{J}_{\text{out}}^{\text{ES}}}, 0\right)\right\rangle_{\mathbb{S}_{\text{out}}} \tag{D-7d}$$

$$z_{\xi\zeta}^{\vec{J}_{\text{out}}^{\text{ES}}\vec{M}_{\text{in}}^{\text{ES}}} = \left\langle \vec{b}_{\xi}^{\vec{J}_{\text{out}}^{\text{ES}}}, -\mathcal{K}\left(\vec{b}_{\zeta}^{\vec{M}_{\text{in}}^{\text{ES}}}\right)\right\rangle_{\mathbb{S}_{\text{out}}} \tag{D-7e}$$

$$z_{\xi\zeta}^{\vec{J}_{\text{out}}^{\text{ES}}\vec{M}_{10}^{\text{ES}}} = \left\langle \vec{b}_{\xi}^{\vec{J}_{\text{out}}^{\text{ES}}}, -\mathcal{K}\left(\vec{b}_{\zeta}^{\vec{M}_{10}^{\text{ES}}}\right)\right\rangle_{\mathbb{S}_{\text{out}}} \tag{D-7f}$$

$$z_{\xi\zeta}^{\vec{J}_{\text{out}}^{\text{ES}}\vec{M}_{\text{out}}^{\text{ES}}} = \left\langle \vec{b}_{\xi}^{\vec{J}_{\text{out}}^{\text{ES}}}, -\text{P.V.}\,\mathcal{K}\left(-\vec{b}_{\zeta}^{\vec{M}_{\text{out}}^{\text{ES}}}\right)\right\rangle_{\mathbb{S}_{\text{out}}} - \left\langle \vec{b}_{\xi}^{\vec{J}_{\text{out}}^{\text{ES}}}, -\text{P.V.}\,\mathcal{K}\left(0, \vec{b}_{\zeta}^{\vec{M}_{\text{out}}^{\text{ES}}}\right)\right\rangle_{\mathbb{S}_{\text{out}}} \tag{D-7g}$$

and

$$z_{\xi\zeta}^{\vec{M}_{\text{out}}^{\text{ES}}\vec{J}_{\text{in}}^{\text{ES}}} = \left\langle \vec{b}_{\xi}^{\vec{M}_{\text{out}}^{\text{ES}}}, \mathcal{K}\left(\vec{b}_{\zeta}^{\vec{J}_{\text{in}}^{\text{ES}}}\right)\right\rangle_{\mathbb{S}_{\text{out}}} \tag{D-8a}$$

$$z_{\xi\zeta}^{\vec{M}_{\text{out}}^{\text{ES}}\vec{J}_{11}^{\text{ES}}} = \left\langle \vec{b}_{\xi}^{\vec{M}_{\text{out}}^{\text{ES}}}, \mathcal{K}\left(\vec{b}_{\zeta}^{\vec{J}_{11}^{\text{ES}}}\right)\right\rangle_{\mathbb{S}_{\text{out}}} \tag{D-8b}$$

$$z_{\xi\zeta}^{\vec{M}_{\text{out}}^{\text{ES}}\vec{J}_{10}^{\text{ES}}} = \left\langle \vec{b}_{\xi}^{\vec{M}_{\text{out}}^{\text{ES}}}, \mathcal{K}\left(\vec{b}_{\zeta}^{\vec{J}_{10}^{\text{ES}}}\right)\right\rangle_{\mathbb{S}_{\text{out}}} \tag{D-8c}$$

$$z_{\xi\zeta}^{\vec{M}_{\text{out}}^{\text{ES}}\vec{J}_{\text{out}}^{\text{ES}}} = \left\langle \vec{b}_{\xi}^{\vec{M}_{\text{out}}^{\text{ES}}}, \text{P.V.}\,\mathcal{K}\left(-\vec{b}_{\zeta}^{\vec{J}_{\text{out}}^{\text{ES}}}\right)\right\rangle_{\mathbb{S}_{\text{out}}} - \left\langle \vec{b}_{\xi}^{\vec{M}_{\text{out}}^{\text{ES}}}, \text{P.V.}\,\mathcal{K}\left(\vec{b}_{\zeta}^{\vec{J}_{\text{out}}^{\text{ES}}}\right)\right\rangle_{\mathbb{S}_{\text{out}}} \tag{D-8d}$$

$$z_{\xi\zeta}^{\vec{M}_{\text{out}}^{\text{ES}}\vec{M}_{\text{in}}^{\text{ES}}} = \left\langle \vec{b}_{\xi}^{\vec{M}_{\text{out}}^{\text{ES}}}, -j\omega\varepsilon\mathcal{L}\left(\vec{b}_{\zeta}^{\vec{M}_{\text{in}}^{\text{ES}}}\right)\right\rangle_{\mathbb{S}_{\text{out}}} \tag{D-8e}$$

$$z_{\xi\zeta}^{\vec{M}_{\text{out}}^{\text{ES}}\vec{M}_{10}^{\text{ES}}} = \left\langle \vec{b}_{\xi}^{\vec{M}_{\text{out}}^{\text{ES}}}, -j\omega\varepsilon\mathcal{L}\left(\vec{b}_{\zeta}^{\vec{M}_{10}^{\text{ES}}}\right)\right\rangle_{\mathbb{S}_{\text{out}}} \tag{D-8f}$$

$$z_{\xi\zeta}^{\vec{M}_{\text{out}}^{\text{ES}}\vec{M}_{\text{out}}^{\text{ES}}} = \left\langle \vec{b}_{\xi}^{\vec{M}_{\text{out}}^{\text{ES}}}, -j\omega\varepsilon\mathcal{L}\left(-\vec{b}_{\zeta}^{\vec{M}_{\text{out}}^{\text{ES}}}\right)\right\rangle_{\mathbb{S}_{\text{out}}} - \left\langle \vec{b}_{\xi}^{\vec{M}_{\text{out}}^{\text{ES}}}, -j\omega\varepsilon\mathcal{L}\left(\vec{b}_{\zeta}^{\vec{M}_{\text{out}}^{\text{ES}}}\right)\right\rangle_{\mathbb{S}_{\text{out}}} \tag{D-8g}$$

The transformation matrix $\overline{\overline{T}}$ used in Eq. (3-131) is as follows:

$$\overline{\overline{T}} = \overline{\overline{T}}^{\vec{J}_{\text{in}}^{\text{ES}} \rightarrow \text{AV}} \text{ or } \overline{\overline{T}}^{\vec{M}_{\text{in}}^{\text{ES}} \rightarrow \text{AV}} \text{ or } \overline{\overline{T}}^{\text{BS} \rightarrow \text{AV}} \tag{D-9}$$

in which

$$\overline{\overline{T}}^{\vec{J}_{\text{in}}^{\text{ES}} \rightarrow \text{AV}} = \left(\overline{\overline{\Psi}}_1\right)^{-1} \cdot \overline{\overline{\Psi}}_2 \tag{D-10a}$$

$$\overline{\overline{T}}^{\vec{M}_{\text{in}}^{\text{ES}} \rightarrow \text{AV}} = \left(\overline{\overline{\Psi}}_3\right)^{-1} \cdot \overline{\overline{\Psi}}_4 \tag{D-10b}$$

$$\overline{\overline{T}}^{\text{BS} \rightarrow \text{AV}} = \text{nullspace}\left(\overline{\overline{\Psi}}_{\text{FCE}}^{\text{DoJ/DoM}}\right) \tag{D-11}$$

where

$$\bar{\bar{\Psi}}_1=\begin{bmatrix}
\bar{\bar{I}}^{\vec{J}_{\mathrm{in}}^{\mathrm{ES}}} & 0 & 0 & 0 & 0 & 0 & 0 & 0\\
0 & \bar{\bar{Z}}^{\vec{M}_{\mathrm{in}}^{\mathrm{ES}}\vec{J}_{11}^{\mathrm{ES}}} & \bar{\bar{Z}}^{\vec{M}_{\mathrm{in}}^{\mathrm{ES}}\vec{J}_{10}^{\mathrm{ES}}} & 0 & \bar{\bar{Z}}^{\vec{M}_{\mathrm{in}}^{\mathrm{ES}}\vec{J}_{\mathrm{out}}^{\mathrm{ES}}} & \bar{\bar{Z}}^{\vec{M}_{\mathrm{in}}^{\mathrm{ES}}\vec{M}_{\mathrm{in}}^{\mathrm{ES}}} & \bar{\bar{Z}}^{\vec{M}_{\mathrm{in}}^{\mathrm{ES}}\vec{M}_{10}^{\mathrm{ES}}} & \bar{\bar{Z}}^{\vec{M}_{\mathrm{in}}^{\mathrm{ES}}\vec{M}_{\mathrm{out}}^{\mathrm{ES}}}\\
0 & \bar{\bar{Z}}^{\vec{J}_{11}^{\mathrm{ES}}\vec{J}_{11}^{\mathrm{ES}}} & \bar{\bar{Z}}^{\vec{J}_{11}^{\mathrm{ES}}\vec{J}_{10}^{\mathrm{ES}}} & 0 & \bar{\bar{Z}}^{\vec{J}_{11}^{\mathrm{ES}}\vec{J}_{\mathrm{out}}^{\mathrm{ES}}} & \bar{\bar{Z}}^{\vec{J}_{11}^{\mathrm{ES}}\vec{M}_{\mathrm{in}}^{\mathrm{ES}}} & \bar{\bar{Z}}^{\vec{J}_{11}^{\mathrm{ES}}\vec{M}_{10}^{\mathrm{ES}}} & \bar{\bar{Z}}^{\vec{J}_{11}^{\mathrm{ES}}\vec{M}_{\mathrm{out}}^{\mathrm{ES}}}\\
0 & 0 & \bar{\bar{Z}}^{\vec{J}_{00}^{\mathrm{ES}}\vec{J}_{10}^{\mathrm{ES}}} & \bar{\bar{Z}}^{\vec{J}_{00}^{\mathrm{ES}}\vec{J}_{00}^{\mathrm{ES}}} & 0 & 0 & \bar{\bar{Z}}^{\vec{J}_{00}^{\mathrm{ES}}\vec{M}_{10}^{\mathrm{ES}}} & 0\\
0 & \bar{\bar{Z}}^{\vec{J}_{10}^{\mathrm{ES}}\vec{J}_{11}^{\mathrm{ES}}} & \bar{\bar{Z}}^{\vec{J}_{10}^{\mathrm{ES}}\vec{J}_{10}^{\mathrm{ES}}} & \bar{\bar{Z}}^{\vec{J}_{10}^{\mathrm{ES}}\vec{J}_{00}^{\mathrm{ES}}} & \bar{\bar{Z}}^{\vec{J}_{10}^{\mathrm{ES}}\vec{J}_{\mathrm{out}}^{\mathrm{ES}}} & \bar{\bar{Z}}^{\vec{J}_{10}^{\mathrm{ES}}\vec{M}_{\mathrm{in}}^{\mathrm{ES}}} & \bar{\bar{Z}}^{\vec{J}_{10}^{\mathrm{ES}}\vec{M}_{10}^{\mathrm{ES}}} & \bar{\bar{Z}}^{\vec{J}_{10}^{\mathrm{ES}}\vec{M}_{\mathrm{out}}^{\mathrm{ES}}}\\
0 & \bar{\bar{Z}}^{\vec{M}_{10}^{\mathrm{ES}}\vec{J}_{11}^{\mathrm{ES}}} & \bar{\bar{Z}}^{\vec{M}_{10}^{\mathrm{ES}}\vec{J}_{10}^{\mathrm{ES}}} & \bar{\bar{Z}}^{\vec{M}_{10}^{\mathrm{ES}}\vec{J}_{00}^{\mathrm{ES}}} & \bar{\bar{Z}}^{\vec{M}_{10}^{\mathrm{ES}}\vec{J}_{\mathrm{out}}^{\mathrm{ES}}} & \bar{\bar{Z}}^{\vec{M}_{10}^{\mathrm{ES}}\vec{M}_{\mathrm{in}}^{\mathrm{ES}}} & \bar{\bar{Z}}^{\vec{M}_{10}^{\mathrm{ES}}\vec{M}_{10}^{\mathrm{ES}}} & \bar{\bar{Z}}^{\vec{M}_{10}^{\mathrm{ES}}\vec{M}_{\mathrm{out}}^{\mathrm{ES}}}\\
0 & \bar{\bar{Z}}^{\vec{J}_{\mathrm{out}}^{\mathrm{ES}}\vec{J}_{11}^{\mathrm{ES}}} & \bar{\bar{Z}}^{\vec{J}_{\mathrm{out}}^{\mathrm{ES}}\vec{J}_{10}^{\mathrm{ES}}} & 0 & \bar{\bar{Z}}^{\vec{J}_{\mathrm{out}}^{\mathrm{ES}}\vec{J}_{\mathrm{out}}^{\mathrm{ES}}} & \bar{\bar{Z}}^{\vec{J}_{\mathrm{out}}^{\mathrm{ES}}\vec{M}_{\mathrm{in}}^{\mathrm{ES}}} & \bar{\bar{Z}}^{\vec{J}_{\mathrm{out}}^{\mathrm{ES}}\vec{M}_{10}^{\mathrm{ES}}} & \bar{\bar{Z}}^{\vec{J}_{\mathrm{out}}^{\mathrm{ES}}\vec{M}_{\mathrm{out}}^{\mathrm{ES}}}\\
0 & \bar{\bar{Z}}^{\vec{M}_{\mathrm{out}}^{\mathrm{ES}}\vec{J}_{11}^{\mathrm{ES}}} & \bar{\bar{Z}}^{\vec{M}_{\mathrm{out}}^{\mathrm{ES}}\vec{J}_{10}^{\mathrm{ES}}} & 0 & \bar{\bar{Z}}^{\vec{M}_{\mathrm{out}}^{\mathrm{ES}}\vec{J}_{\mathrm{out}}^{\mathrm{ES}}} & \bar{\bar{Z}}^{\vec{M}_{\mathrm{out}}^{\mathrm{ES}}\vec{M}_{\mathrm{in}}^{\mathrm{ES}}} & \bar{\bar{Z}}^{\vec{M}_{\mathrm{out}}^{\mathrm{ES}}\vec{M}_{10}^{\mathrm{ES}}} & \bar{\bar{Z}}^{\vec{M}_{\mathrm{out}}^{\mathrm{ES}}\vec{M}_{\mathrm{out}}^{\mathrm{ES}}}
\end{bmatrix} \tag{D-12a}$$

$$\bar{\bar{\Psi}}_2=\begin{bmatrix}
\bar{\bar{I}}^{\vec{J}_{\mathrm{in}}^{\mathrm{ES}}}\\
-\bar{\bar{Z}}^{\vec{M}_{\mathrm{in}}^{\mathrm{ES}}\vec{J}_{\mathrm{in}}^{\mathrm{ES}}}\\
-\bar{\bar{Z}}^{\vec{J}_{11}^{\mathrm{ES}}\vec{J}_{\mathrm{in}}^{\mathrm{ES}}}\\
0\\
-\bar{\bar{Z}}^{\vec{J}_{10}^{\mathrm{ES}}\vec{J}_{\mathrm{in}}^{\mathrm{ES}}}\\
-\bar{\bar{Z}}^{\vec{M}_{10}^{\mathrm{ES}}\vec{J}_{\mathrm{in}}^{\mathrm{ES}}}\\
-\bar{\bar{Z}}^{\vec{J}_{\mathrm{out}}^{\mathrm{ES}}\vec{J}_{\mathrm{in}}^{\mathrm{ES}}}\\
-\bar{\bar{Z}}^{\vec{M}_{\mathrm{out}}^{\mathrm{ES}}\vec{J}_{\mathrm{in}}^{\mathrm{ES}}}
\end{bmatrix} \tag{D-12b}$$

and

$$\bar{\bar{\Psi}}_3=\begin{bmatrix}
0 & 0 & 0 & 0 & 0 & \bar{\bar{I}}^{\vec{M}_{\mathrm{in}}^{\mathrm{ES}}} & 0 & 0\\
\bar{\bar{Z}}^{\vec{J}_{\mathrm{in}}^{\mathrm{ES}}\vec{J}_{\mathrm{in}}^{\mathrm{ES}}} & \bar{\bar{Z}}^{\vec{J}_{\mathrm{in}}^{\mathrm{ES}}\vec{J}_{11}^{\mathrm{ES}}} & \bar{\bar{Z}}^{\vec{J}_{\mathrm{in}}^{\mathrm{ES}}\vec{J}_{10}^{\mathrm{ES}}} & 0 & \bar{\bar{Z}}^{\vec{J}_{\mathrm{in}}^{\mathrm{ES}}\vec{J}_{\mathrm{out}}^{\mathrm{ES}}} & 0 & \bar{\bar{Z}}^{\vec{J}_{\mathrm{in}}^{\mathrm{ES}}\vec{M}_{10}^{\mathrm{ES}}} & \bar{\bar{Z}}^{\vec{J}_{\mathrm{in}}^{\mathrm{ES}}\vec{M}_{\mathrm{out}}^{\mathrm{ES}}}\\
\bar{\bar{Z}}^{\vec{J}_{11}^{\mathrm{ES}}\vec{J}_{\mathrm{in}}^{\mathrm{ES}}} & \bar{\bar{Z}}^{\vec{J}_{11}^{\mathrm{ES}}\vec{J}_{11}^{\mathrm{ES}}} & \bar{\bar{Z}}^{\vec{J}_{11}^{\mathrm{ES}}\vec{J}_{10}^{\mathrm{ES}}} & 0 & \bar{\bar{Z}}^{\vec{J}_{11}^{\mathrm{ES}}\vec{J}_{\mathrm{out}}^{\mathrm{ES}}} & 0 & \bar{\bar{Z}}^{\vec{J}_{11}^{\mathrm{ES}}\vec{M}_{10}^{\mathrm{ES}}} & \bar{\bar{Z}}^{\vec{J}_{11}^{\mathrm{ES}}\vec{M}_{\mathrm{out}}^{\mathrm{ES}}}\\
0 & 0 & \bar{\bar{Z}}^{\vec{J}_{00}^{\mathrm{ES}}\vec{J}_{10}^{\mathrm{ES}}} & \bar{\bar{Z}}^{\vec{J}_{00}^{\mathrm{ES}}\vec{J}_{00}^{\mathrm{ES}}} & 0 & 0 & \bar{\bar{Z}}^{\vec{J}_{00}^{\mathrm{ES}}\vec{M}_{10}^{\mathrm{ES}}} & 0\\
\bar{\bar{Z}}^{\vec{J}_{10}^{\mathrm{ES}}\vec{J}_{\mathrm{in}}^{\mathrm{ES}}} & \bar{\bar{Z}}^{\vec{J}_{10}^{\mathrm{ES}}\vec{J}_{11}^{\mathrm{ES}}} & \bar{\bar{Z}}^{\vec{J}_{10}^{\mathrm{ES}}\vec{J}_{10}^{\mathrm{ES}}} & \bar{\bar{Z}}^{\vec{J}_{10}^{\mathrm{ES}}\vec{J}_{00}^{\mathrm{ES}}} & \bar{\bar{Z}}^{\vec{J}_{10}^{\mathrm{ES}}\vec{J}_{\mathrm{out}}^{\mathrm{ES}}} & 0 & \bar{\bar{Z}}^{\vec{J}_{10}^{\mathrm{ES}}\vec{M}_{10}^{\mathrm{ES}}} & \bar{\bar{Z}}^{\vec{J}_{10}^{\mathrm{ES}}\vec{M}_{\mathrm{out}}^{\mathrm{ES}}}\\
\bar{\bar{Z}}^{\vec{M}_{10}^{\mathrm{ES}}\vec{J}_{\mathrm{in}}^{\mathrm{ES}}} & \bar{\bar{Z}}^{\vec{M}_{10}^{\mathrm{ES}}\vec{J}_{11}^{\mathrm{ES}}} & \bar{\bar{Z}}^{\vec{M}_{10}^{\mathrm{ES}}\vec{J}_{10}^{\mathrm{ES}}} & \bar{\bar{Z}}^{\vec{M}_{10}^{\mathrm{ES}}\vec{J}_{00}^{\mathrm{ES}}} & \bar{\bar{Z}}^{\vec{M}_{10}^{\mathrm{ES}}\vec{J}_{\mathrm{out}}^{\mathrm{ES}}} & 0 & \bar{\bar{Z}}^{\vec{M}_{10}^{\mathrm{ES}}\vec{M}_{10}^{\mathrm{ES}}} & \bar{\bar{Z}}^{\vec{M}_{10}^{\mathrm{ES}}\vec{M}_{\mathrm{out}}^{\mathrm{ES}}}\\
\bar{\bar{Z}}^{\vec{J}_{\mathrm{out}}^{\mathrm{ES}}\vec{J}_{\mathrm{in}}^{\mathrm{ES}}} & \bar{\bar{Z}}^{\vec{J}_{\mathrm{out}}^{\mathrm{ES}}\vec{J}_{11}^{\mathrm{ES}}} & \bar{\bar{Z}}^{\vec{J}_{\mathrm{out}}^{\mathrm{ES}}\vec{J}_{10}^{\mathrm{ES}}} & 0 & \bar{\bar{Z}}^{\vec{J}_{\mathrm{out}}^{\mathrm{ES}}\vec{J}_{\mathrm{out}}^{\mathrm{ES}}} & 0 & \bar{\bar{Z}}^{\vec{J}_{\mathrm{out}}^{\mathrm{ES}}\vec{M}_{10}^{\mathrm{ES}}} & \bar{\bar{Z}}^{\vec{J}_{\mathrm{out}}^{\mathrm{ES}}\vec{M}_{\mathrm{out}}^{\mathrm{ES}}}\\
\bar{\bar{Z}}^{\vec{M}_{\mathrm{out}}^{\mathrm{ES}}\vec{J}_{\mathrm{in}}^{\mathrm{ES}}} & \bar{\bar{Z}}^{\vec{M}_{\mathrm{out}}^{\mathrm{ES}}\vec{J}_{11}^{\mathrm{ES}}} & \bar{\bar{Z}}^{\vec{M}_{\mathrm{out}}^{\mathrm{ES}}\vec{J}_{10}^{\mathrm{ES}}} & 0 & \bar{\bar{Z}}^{\vec{M}_{\mathrm{out}}^{\mathrm{ES}}\vec{J}_{\mathrm{out}}^{\mathrm{ES}}} & 0 & \bar{\bar{Z}}^{\vec{M}_{\mathrm{out}}^{\mathrm{ES}}\vec{M}_{10}^{\mathrm{ES}}} & \bar{\bar{Z}}^{\vec{M}_{\mathrm{out}}^{\mathrm{ES}}\vec{M}_{\mathrm{out}}^{\mathrm{ES}}}
\end{bmatrix} \tag{D-13a}$$

$$\bar{\bar{\Psi}}_4=\begin{bmatrix}
\bar{\bar{I}}^{\vec{M}_{\mathrm{in}}^{\mathrm{ES}}}\\
-\bar{\bar{Z}}^{\vec{J}_{\mathrm{in}}^{\mathrm{ES}}\vec{M}_{\mathrm{in}}^{\mathrm{ES}}}\\
-\bar{\bar{Z}}^{\vec{J}_{11}^{\mathrm{ES}}\vec{M}_{\mathrm{in}}^{\mathrm{ES}}}\\
0\\
-\bar{\bar{Z}}^{\vec{J}_{10}^{\mathrm{ES}}\vec{M}_{\mathrm{in}}^{\mathrm{ES}}}\\
-\bar{\bar{Z}}^{\vec{M}_{10}^{\mathrm{ES}}\vec{M}_{\mathrm{in}}^{\mathrm{ES}}}\\
-\bar{\bar{Z}}^{\vec{J}_{\mathrm{out}}^{\mathrm{ES}}\vec{M}_{\mathrm{in}}^{\mathrm{ES}}}\\
-\bar{\bar{Z}}^{\vec{M}_{\mathrm{out}}^{\mathrm{ES}}\vec{M}_{\mathrm{in}}^{\mathrm{ES}}}
\end{bmatrix} \tag{D-13b}$$

and

$$\overline{\overline{\Psi}}{}^{\text{DoJ}}_{\text{FCE}}=\begin{bmatrix}
\overline{\overline{Z}}{}^{\vec{M}^{\text{ES}}_{\text{in}}\vec{J}^{\text{ES}}_{\text{in}}} & \overline{\overline{Z}}{}^{\vec{M}^{\text{ES}}_{\text{in}}\vec{J}^{\text{ES}}_{11}} & \overline{\overline{Z}}{}^{\vec{M}^{\text{ES}}_{\text{in}}\vec{J}^{\text{ES}}_{10}} & 0 & \overline{\overline{Z}}{}^{\vec{M}^{\text{ES}}_{\text{in}}\vec{J}^{\text{ES}}_{\text{out}}} & \overline{\overline{Z}}{}^{\vec{M}^{\text{ES}}_{\text{in}}\vec{M}^{\text{ES}}_{\text{in}}} & \overline{\overline{Z}}{}^{\vec{M}^{\text{ES}}_{\text{in}}\vec{M}^{\text{ES}}_{10}} & \overline{\overline{Z}}{}^{\vec{M}^{\text{ES}}_{\text{in}}\vec{M}^{\text{ES}}_{\text{out}}} \\
\overline{\overline{Z}}{}^{\vec{J}^{\text{ES}}_{11}\vec{J}^{\text{ES}}_{\text{in}}} & \overline{\overline{Z}}{}^{\vec{J}^{\text{ES}}_{11}\vec{J}^{\text{ES}}_{11}} & \overline{\overline{Z}}{}^{\vec{J}^{\text{ES}}_{11}\vec{J}^{\text{ES}}_{10}} & 0 & \overline{\overline{Z}}{}^{\vec{J}^{\text{ES}}_{11}\vec{J}^{\text{ES}}_{\text{out}}} & \overline{\overline{Z}}{}^{\vec{J}^{\text{ES}}_{11}\vec{M}^{\text{ES}}_{\text{in}}} & \overline{\overline{Z}}{}^{\vec{J}^{\text{ES}}_{11}\vec{M}^{\text{ES}}_{10}} & \overline{\overline{Z}}{}^{\vec{J}^{\text{ES}}_{11}\vec{M}^{\text{ES}}_{\text{out}}} \\
0 & 0 & \overline{\overline{Z}}{}^{\vec{J}^{\text{ES}}_{00}\vec{J}^{\text{ES}}_{10}} & \overline{\overline{Z}}{}^{\vec{J}^{\text{ES}}_{00}\vec{J}^{\text{ES}}_{00}} & 0 & 0 & \overline{\overline{Z}}{}^{\vec{J}^{\text{ES}}_{00}\vec{M}^{\text{ES}}_{10}} & 0 \\
\overline{\overline{Z}}{}^{\vec{J}^{\text{ES}}_{10}\vec{J}^{\text{ES}}_{\text{in}}} & \overline{\overline{Z}}{}^{\vec{J}^{\text{ES}}_{10}\vec{J}^{\text{ES}}_{11}} & \overline{\overline{Z}}{}^{\vec{J}^{\text{ES}}_{10}\vec{J}^{\text{ES}}_{10}} & \overline{\overline{Z}}{}^{\vec{J}^{\text{ES}}_{10}\vec{J}^{\text{ES}}_{00}} & \overline{\overline{Z}}{}^{\vec{J}^{\text{ES}}_{10}\vec{J}^{\text{ES}}_{\text{out}}} & \overline{\overline{Z}}{}^{\vec{J}^{\text{ES}}_{10}\vec{M}^{\text{ES}}_{\text{in}}} & \overline{\overline{Z}}{}^{\vec{J}^{\text{ES}}_{10}\vec{M}^{\text{ES}}_{10}} & \overline{\overline{Z}}{}^{\vec{J}^{\text{ES}}_{10}\vec{M}^{\text{ES}}_{\text{out}}} \\
\overline{\overline{Z}}{}^{\vec{M}^{\text{ES}}_{10}\vec{J}^{\text{ES}}_{\text{in}}} & \overline{\overline{Z}}{}^{\vec{M}^{\text{ES}}_{10}\vec{J}^{\text{ES}}_{11}} & \overline{\overline{Z}}{}^{\vec{M}^{\text{ES}}_{10}\vec{J}^{\text{ES}}_{10}} & \overline{\overline{Z}}{}^{\vec{M}^{\text{ES}}_{10}\vec{J}^{\text{ES}}_{00}} & \overline{\overline{Z}}{}^{\vec{M}^{\text{ES}}_{10}\vec{J}^{\text{ES}}_{\text{out}}} & \overline{\overline{Z}}{}^{\vec{M}^{\text{ES}}_{10}\vec{M}^{\text{ES}}_{\text{in}}} & \overline{\overline{Z}}{}^{\vec{M}^{\text{ES}}_{10}\vec{M}^{\text{ES}}_{10}} & \overline{\overline{Z}}{}^{\vec{M}^{\text{ES}}_{10}\vec{M}^{\text{ES}}_{\text{out}}} \\
\overline{\overline{Z}}{}^{\vec{J}^{\text{ES}}_{\text{out}}\vec{J}^{\text{ES}}_{\text{in}}} & \overline{\overline{Z}}{}^{\vec{J}^{\text{ES}}_{\text{out}}\vec{J}^{\text{ES}}_{11}} & \overline{\overline{Z}}{}^{\vec{J}^{\text{ES}}_{\text{out}}\vec{J}^{\text{ES}}_{10}} & 0 & \overline{\overline{Z}}{}^{\vec{J}^{\text{ES}}_{\text{out}}\vec{J}^{\text{ES}}_{\text{out}}} & \overline{\overline{Z}}{}^{\vec{J}^{\text{ES}}_{\text{out}}\vec{M}^{\text{ES}}_{\text{in}}} & \overline{\overline{Z}}{}^{\vec{J}^{\text{ES}}_{\text{out}}\vec{M}^{\text{ES}}_{10}} & \overline{\overline{Z}}{}^{\vec{J}^{\text{ES}}_{\text{out}}\vec{M}^{\text{ES}}_{\text{out}}} \\
\overline{\overline{Z}}{}^{\vec{M}^{\text{ES}}_{\text{out}}\vec{J}^{\text{ES}}_{\text{in}}} & \overline{\overline{Z}}{}^{\vec{M}^{\text{ES}}_{\text{out}}\vec{J}^{\text{ES}}_{11}} & \overline{\overline{Z}}{}^{\vec{M}^{\text{ES}}_{\text{out}}\vec{J}^{\text{ES}}_{10}} & 0 & \overline{\overline{Z}}{}^{\vec{M}^{\text{ES}}_{\text{out}}\vec{J}^{\text{ES}}_{\text{out}}} & \overline{\overline{Z}}{}^{\vec{M}^{\text{ES}}_{\text{out}}\vec{M}^{\text{ES}}_{\text{in}}} & \overline{\overline{Z}}{}^{\vec{M}^{\text{ES}}_{\text{out}}\vec{M}^{\text{ES}}_{10}} & \overline{\overline{Z}}{}^{\vec{M}^{\text{ES}}_{\text{out}}\vec{M}^{\text{ES}}_{\text{out}}}
\end{bmatrix} \tag{D-14a}$$

$$\overline{\overline{\Psi}}{}^{\text{DoM}}_{\text{FCE}}=\begin{bmatrix}
\overline{\overline{Z}}{}^{\vec{J}^{\text{ES}}_{\text{in}}\vec{J}^{\text{ES}}_{\text{in}}} & \overline{\overline{Z}}{}^{\vec{J}^{\text{ES}}_{\text{in}}\vec{J}^{\text{ES}}_{11}} & \overline{\overline{Z}}{}^{\vec{J}^{\text{ES}}_{\text{in}}\vec{J}^{\text{ES}}_{10}} & 0 & \overline{\overline{Z}}{}^{\vec{J}^{\text{ES}}_{\text{in}}\vec{J}^{\text{ES}}_{\text{out}}} & \overline{\overline{Z}}{}^{\vec{J}^{\text{ES}}_{\text{in}}\vec{M}^{\text{ES}}_{\text{in}}} & \overline{\overline{Z}}{}^{\vec{J}^{\text{ES}}_{\text{in}}\vec{M}^{\text{ES}}_{10}} & \overline{\overline{Z}}{}^{\vec{J}^{\text{ES}}_{\text{in}}\vec{M}^{\text{ES}}_{\text{out}}} \\
\overline{\overline{Z}}{}^{\vec{J}^{\text{ES}}_{11}\vec{J}^{\text{ES}}_{\text{in}}} & \overline{\overline{Z}}{}^{\vec{J}^{\text{ES}}_{11}\vec{J}^{\text{ES}}_{11}} & \overline{\overline{Z}}{}^{\vec{J}^{\text{ES}}_{11}\vec{J}^{\text{ES}}_{10}} & 0 & \overline{\overline{Z}}{}^{\vec{J}^{\text{ES}}_{11}\vec{J}^{\text{ES}}_{\text{out}}} & \overline{\overline{Z}}{}^{\vec{J}^{\text{ES}}_{11}\vec{M}^{\text{ES}}_{\text{in}}} & \overline{\overline{Z}}{}^{\vec{J}^{\text{ES}}_{11}\vec{M}^{\text{ES}}_{10}} & \overline{\overline{Z}}{}^{\vec{J}^{\text{ES}}_{11}\vec{M}^{\text{ES}}_{\text{out}}} \\
0 & 0 & \overline{\overline{Z}}{}^{\vec{J}^{\text{ES}}_{00}\vec{J}^{\text{ES}}_{10}} & \overline{\overline{Z}}{}^{\vec{J}^{\text{ES}}_{00}\vec{J}^{\text{ES}}_{00}} & 0 & 0 & \overline{\overline{Z}}{}^{\vec{J}^{\text{ES}}_{00}\vec{M}^{\text{ES}}_{10}} & 0 \\
\overline{\overline{Z}}{}^{\vec{J}^{\text{ES}}_{10}\vec{J}^{\text{ES}}_{\text{in}}} & \overline{\overline{Z}}{}^{\vec{J}^{\text{ES}}_{10}\vec{J}^{\text{ES}}_{11}} & \overline{\overline{Z}}{}^{\vec{J}^{\text{ES}}_{10}\vec{J}^{\text{ES}}_{10}} & \overline{\overline{Z}}{}^{\vec{J}^{\text{ES}}_{10}\vec{J}^{\text{ES}}_{00}} & \overline{\overline{Z}}{}^{\vec{J}^{\text{ES}}_{10}\vec{J}^{\text{ES}}_{\text{out}}} & \overline{\overline{Z}}{}^{\vec{J}^{\text{ES}}_{10}\vec{M}^{\text{ES}}_{\text{in}}} & \overline{\overline{Z}}{}^{\vec{J}^{\text{ES}}_{10}\vec{M}^{\text{ES}}_{10}} & \overline{\overline{Z}}{}^{\vec{J}^{\text{ES}}_{10}\vec{M}^{\text{ES}}_{\text{out}}} \\
\overline{\overline{Z}}{}^{\vec{M}^{\text{ES}}_{10}\vec{J}^{\text{ES}}_{\text{in}}} & \overline{\overline{Z}}{}^{\vec{M}^{\text{ES}}_{10}\vec{J}^{\text{ES}}_{11}} & \overline{\overline{Z}}{}^{\vec{M}^{\text{ES}}_{10}\vec{J}^{\text{ES}}_{10}} & \overline{\overline{Z}}{}^{\vec{M}^{\text{ES}}_{10}\vec{J}^{\text{ES}}_{00}} & \overline{\overline{Z}}{}^{\vec{M}^{\text{ES}}_{10}\vec{J}^{\text{ES}}_{\text{out}}} & \overline{\overline{Z}}{}^{\vec{M}^{\text{ES}}_{10}\vec{M}^{\text{ES}}_{\text{in}}} & \overline{\overline{Z}}{}^{\vec{M}^{\text{ES}}_{10}\vec{M}^{\text{ES}}_{10}} & \overline{\overline{Z}}{}^{\vec{M}^{\text{ES}}_{10}\vec{M}^{\text{ES}}_{\text{out}}} \\
\overline{\overline{Z}}{}^{\vec{J}^{\text{ES}}_{\text{out}}\vec{J}^{\text{ES}}_{\text{in}}} & \overline{\overline{Z}}{}^{\vec{J}^{\text{ES}}_{\text{out}}\vec{J}^{\text{ES}}_{11}} & \overline{\overline{Z}}{}^{\vec{J}^{\text{ES}}_{\text{out}}\vec{J}^{\text{ES}}_{10}} & 0 & \overline{\overline{Z}}{}^{\vec{J}^{\text{ES}}_{\text{out}}\vec{J}^{\text{ES}}_{\text{out}}} & \overline{\overline{Z}}{}^{\vec{J}^{\text{ES}}_{\text{out}}\vec{M}^{\text{ES}}_{\text{in}}} & \overline{\overline{Z}}{}^{\vec{J}^{\text{ES}}_{\text{out}}\vec{M}^{\text{ES}}_{10}} & \overline{\overline{Z}}{}^{\vec{J}^{\text{ES}}_{\text{out}}\vec{M}^{\text{ES}}_{\text{out}}} \\
\overline{\overline{Z}}{}^{\vec{M}^{\text{ES}}_{\text{out}}\vec{J}^{\text{ES}}_{\text{in}}} & \overline{\overline{Z}}{}^{\vec{M}^{\text{ES}}_{\text{out}}\vec{J}^{\text{ES}}_{11}} & \overline{\overline{Z}}{}^{\vec{M}^{\text{ES}}_{\text{out}}\vec{J}^{\text{ES}}_{10}} & 0 & \overline{\overline{Z}}{}^{\vec{M}^{\text{ES}}_{\text{out}}\vec{J}^{\text{ES}}_{\text{out}}} & \overline{\overline{Z}}{}^{\vec{M}^{\text{ES}}_{\text{out}}\vec{M}^{\text{ES}}_{\text{in}}} & \overline{\overline{Z}}{}^{\vec{M}^{\text{ES}}_{\text{out}}\vec{M}^{\text{ES}}_{10}} & \overline{\overline{Z}}{}^{\vec{M}^{\text{ES}}_{\text{out}}\vec{M}^{\text{ES}}_{\text{out}}}
\end{bmatrix} \tag{D-14b}$$

in which the various sub-matrices are just the ones used in Eqs. (3-126a)~(3-130b).

The power quadratic form matrix $\overline{\overline{P}}{}^{\text{in}}$ used in Eq. (3-133) is as follows:

$$\overline{\overline{P}}{}^{\text{in}} = \overline{\overline{P}}{}^{\text{in}}_{\text{curAV}} \text{ or } \overline{\overline{P}}{}^{\text{in}}_{\text{intAV}} \tag{D-15}$$

in which

$$\overline{\overline{P}}{}^{\text{in}}_{\text{curAV}}=\begin{bmatrix}
0 & 0 & 0 & 0 & 0 & \overline{\overline{C}}{}^{\vec{J}^{\text{ES}}_{\text{in}}\vec{M}^{\text{ES}}_{\text{in}}} & 0 & 0 \\
0 & 0 & 0 & 0 & 0 & 0 & 0 & 0 \\
0 & 0 & 0 & 0 & 0 & 0 & 0 & 0 \\
0 & 0 & 0 & 0 & 0 & 0 & 0 & 0 \\
0 & 0 & 0 & 0 & 0 & 0 & 0 & 0 \\
0 & 0 & 0 & 0 & 0 & 0 & 0 & 0 \\
0 & 0 & 0 & 0 & 0 & 0 & 0 & 0 \\
0 & 0 & 0 & 0 & 0 & 0 & 0 & 0
\end{bmatrix} \tag{D-16}$$

and

$$\overline{\overline{P}}{}^{\text{in}}_{\text{intAV}}=\begin{bmatrix}
\overline{\overline{P}}{}^{\vec{J}^{\text{ES}}_{\text{in}}\vec{J}^{\text{ES}}_{\text{in}}} & \overline{\overline{P}}{}^{\vec{J}^{\text{ES}}_{\text{in}}\vec{J}^{\text{ES}}_{11}} & \overline{\overline{P}}{}^{\vec{J}^{\text{ES}}_{\text{in}}\vec{J}^{\text{ES}}_{10}} & 0 & \overline{\overline{P}}{}^{\vec{J}^{\text{ES}}_{\text{in}}\vec{J}^{\text{ES}}_{\text{out}}} & \overline{\overline{P}}{}^{\vec{J}^{\text{ES}}_{\text{in}}\vec{M}^{\text{ES}}_{\text{in}}} & \overline{\overline{P}}{}^{\vec{J}^{\text{ES}}_{\text{in}}\vec{M}^{\text{ES}}_{10}} & \overline{\overline{P}}{}^{\vec{J}^{\text{ES}}_{\text{in}}\vec{M}^{\text{ES}}_{\text{out}}} \\
0 & 0 & 0 & 0 & 0 & 0 & 0 & 0 \\
0 & 0 & 0 & 0 & 0 & 0 & 0 & 0 \\
0 & 0 & 0 & 0 & 0 & 0 & 0 & 0 \\
0 & 0 & 0 & 0 & 0 & 0 & 0 & 0 \\
0 & 0 & 0 & 0 & 0 & 0 & 0 & 0 \\
0 & 0 & 0 & 0 & 0 & 0 & 0 & 0 \\
0 & 0 & 0 & 0 & 0 & 0 & 0 & 0
\end{bmatrix} \tag{D-17a}$$

for the third equality in Eq. (3-132), and

$$\overline{\overline{P}}_{\text{intAV}}^{\text{in}}=\begin{bmatrix}0&0&0&0&0&0&0&0\\0&0&0&0&0&0&0&0\\0&0&0&0&0&0&0&0\\0&0&0&0&0&0&0&0\\0&0&0&0&0&0&0&0\\\overline{\overline{P}}^{\vec{M}_{\text{in}}^{\text{ES}}\vec{J}_{\text{in}}^{\text{ES}}}&\overline{\overline{P}}^{\vec{M}_{\text{in}}^{\text{ES}}\vec{J}_{11}^{\text{ES}}}&\overline{\overline{P}}^{\vec{M}_{\text{in}}^{\text{ES}}\vec{J}_{10}^{\text{ES}}}&0&\overline{\overline{P}}^{\vec{M}_{\text{in}}^{\text{ES}}\vec{J}_{\text{out}}^{\text{ES}}}&\overline{\overline{P}}^{\vec{M}_{\text{in}}^{\text{ES}}\vec{M}_{\text{in}}^{\text{ES}}}&\overline{\overline{P}}^{\vec{M}_{\text{in}}^{\text{ES}}\vec{M}_{10}^{\text{ES}}}&\overline{\overline{P}}^{\vec{M}_{\text{in}}^{\text{ES}}\vec{M}_{\text{out}}^{\text{ES}}}\\0&0&0&0&0&0&0&0\\0&0&0&0&0&0&0&0\end{bmatrix}^{\dagger} \tag{D-17b}$$

for the fourth equality in Eq. (3-132), where the elements of the sub-matrices in above Eqs. (D-16) and (D-17) are as follows:

$$c_{\xi\zeta}^{\vec{J}_{\text{in}}^{\text{ES}}\vec{M}_{\text{in}}^{\text{ES}}} = (1/2)\left\langle \hat{z}\times\vec{b}_{\xi}^{\vec{J}_{\text{in}}^{\text{ES}}}, \vec{b}_{\zeta}^{\vec{M}_{\text{in}}^{\text{ES}}}\right\rangle_{\mathbb{S}_{\text{in}}} \tag{D-18}$$

and

$$p_{\xi\zeta}^{\vec{J}_{\text{in}}^{\text{ES}}\vec{J}_{\text{in}}^{\text{ES}}} = -(1/2)\left\langle \vec{b}_{\xi}^{\vec{J}_{\text{in}}^{\text{ES}}}, -j\omega\mu\mathcal{L}\left(\vec{b}_{\zeta}^{\vec{J}_{\text{in}}^{\text{ES}}}\right)\right\rangle_{\mathbb{S}_{\text{in}}} \tag{D-19a}$$

$$p_{\xi\zeta}^{\vec{J}_{\text{in}}^{\text{ES}}\vec{J}_{11}^{\text{ES}}} = -(1/2)\left\langle \vec{b}_{\xi}^{\vec{J}_{\text{in}}^{\text{ES}}}, -j\omega\mu\mathcal{L}\left(\vec{b}_{\zeta}^{\vec{J}_{11}^{\text{ES}}}\right)\right\rangle_{\mathbb{S}_{\text{in}}} \tag{D-19b}$$

$$p_{\xi\zeta}^{\vec{J}_{\text{in}}^{\text{ES}}\vec{J}_{10}^{\text{ES}}} = -(1/2)\left\langle \vec{b}_{\xi}^{\vec{J}_{\text{in}}^{\text{ES}}}, -j\omega\mu\mathcal{L}\left(\vec{b}_{\zeta}^{\vec{J}_{10}^{\text{ES}}}\right)\right\rangle_{\mathbb{S}_{\text{in}}} \tag{D-19c}$$

$$p_{\xi\zeta}^{\vec{J}_{\text{in}}^{\text{ES}}\vec{J}_{\text{out}}^{\text{ES}}} = -(1/2)\left\langle \vec{b}_{\xi}^{\vec{J}_{\text{in}}^{\text{ES}}}, -j\omega\mu\mathcal{L}\left(-\vec{b}_{\zeta}^{\vec{J}_{\text{out}}^{\text{ES}}}\right)\right\rangle_{\mathbb{S}_{\text{in}}} \tag{D-19d}$$

$$p_{\xi\zeta}^{\vec{J}_{\text{in}}^{\text{ES}}\vec{M}_{\text{in}}^{\text{ES}}} = -(1/2)\left\langle \vec{b}_{\xi}^{\vec{J}_{\text{in}}^{\text{ES}}}, \hat{z}\times\frac{1}{2}\vec{b}_{\zeta}^{\vec{M}_{\text{in}}^{\text{ES}}} - \text{P.V.}\,\mathcal{K}\left(\vec{b}_{\zeta}^{\vec{M}_{\text{in}}^{\text{ES}}}\right)\right\rangle_{\mathbb{S}_{\text{in}}} \tag{D-19e}$$

$$p_{\xi\zeta}^{\vec{J}_{\text{in}}^{\text{ES}}\vec{M}_{10}^{\text{ES}}} = -(1/2)\left\langle \vec{b}_{\xi}^{\vec{J}_{\text{in}}^{\text{ES}}}, -\mathcal{K}\left(\vec{b}_{\zeta}^{\vec{M}_{10}^{\text{ES}}}\right)\right\rangle_{\mathbb{S}_{\text{in}}} \tag{D-19f}$$

$$p_{\xi\zeta}^{\vec{J}_{\text{in}}^{\text{ES}}\vec{M}_{\text{out}}^{\text{ES}}} = -(1/2)\left\langle \vec{b}_{\xi}^{\vec{J}_{\text{in}}^{\text{ES}}}, -\mathcal{K}\left(-\vec{b}_{\zeta}^{\vec{M}_{\text{out}}^{\text{ES}}}\right)\right\rangle_{\mathbb{S}_{\text{in}}} \tag{D-19g}$$

and

$$p_{\xi\zeta}^{\vec{M}_{\text{in}}^{\text{ES}}\vec{J}_{\text{in}}^{\text{ES}}} = -(1/2)\left\langle \vec{b}_{\xi}^{\vec{M}_{\text{in}}^{\text{ES}}}, \frac{1}{2}\vec{b}_{\zeta}^{\vec{J}_{\text{in}}^{\text{ES}}}\times\hat{z} + \text{P.V.}\,\mathcal{K}\left(\vec{b}_{\zeta}^{\vec{J}_{\text{in}}^{\text{ES}}}\right)\right\rangle_{\mathbb{S}_{\text{in}}} \tag{D-19h}$$

$$p_{\xi\zeta}^{\vec{M}_{\text{in}}^{\text{ES}}\vec{J}_{11}^{\text{ES}}} = -(1/2)\left\langle \vec{b}_{\xi}^{\vec{M}_{\text{in}}^{\text{ES}}}, \mathcal{K}\left(\vec{b}_{\zeta}^{\vec{J}_{11}^{\text{ES}}}\right)\right\rangle_{\mathbb{S}_{\text{in}}} \tag{D-19i}$$

$$p_{\xi\zeta}^{\vec{M}_{\text{in}}^{\text{ES}}\vec{J}_{10}^{\text{ES}}} = -(1/2)\left\langle \vec{b}_{\xi}^{\vec{M}_{\text{in}}^{\text{ES}}}, \mathcal{K}\left(\vec{b}_{\zeta}^{\vec{J}_{10}^{\text{ES}}}\right)\right\rangle_{\mathbb{S}_{\text{in}}} \tag{D-19j}$$

$$p_{\xi\zeta}^{\vec{M}_{\text{in}}^{\text{ES}}\vec{J}_{\text{out}}^{\text{ES}}} = -(1/2)\left\langle \vec{b}_{\xi}^{\vec{M}_{\text{in}}^{\text{ES}}}, \mathcal{K}\left(-\vec{b}_{\zeta}^{\vec{J}_{\text{out}}^{\text{ES}}}\right)\right\rangle_{\mathbb{S}_{\text{in}}} \tag{D-19k}$$

$$p_{\xi\zeta}^{\vec{M}_{\text{in}}^{\text{ES}}\vec{M}_{\text{in}}^{\text{ES}}} = -(1/2)\left\langle \vec{b}_{\xi}^{\vec{M}_{\text{in}}^{\text{ES}}}, -j\omega\varepsilon\mathcal{L}\left(\vec{b}_{\zeta}^{\vec{M}_{\text{in}}^{\text{ES}}}\right)\right\rangle_{\mathbb{S}_{\text{in}}} \tag{D-19l}$$

$$p_{\xi\zeta}^{\vec{M}_{\text{in}}^{\text{ES}}\vec{M}_{10}^{\text{ES}}} = -(1/2)\left\langle \vec{b}_{\xi}^{\vec{M}_{\text{in}}^{\text{ES}}}, -j\omega\varepsilon\mathcal{L}\left(\vec{b}_{\zeta}^{\vec{M}_{10}^{\text{ES}}}\right)\right\rangle_{\mathbb{S}_{\text{in}}} \tag{D-19m}$$

$$p_{\xi\zeta}^{\vec{M}_{\text{in}}^{\text{ES}}\vec{M}_{\text{out}}^{\text{ES}}} = -(1/2)\left\langle \vec{b}_{\xi}^{\vec{M}_{\text{in}}^{\text{ES}}}, -j\omega\varepsilon\mathcal{L}\left(-\vec{b}_{\zeta}^{\vec{M}_{\text{out}}^{\text{ES}}}\right)\right\rangle_{\mathbb{S}_{\text{in}}} \tag{D-19n}$$

## D2 Some Detailed Formulations Related to Sec. 3.5

The formulations used to calculate the elements of the matrices in Eqs. (3-143a)~ (3-145b) are as follows:

$$z_{\xi\zeta}^{\vec{M}_{\rm in}^{\rm ES}\vec{J}_{\rm in}^{\rm ES}} = \left\langle \vec{b}_\xi^{\vec{M}_{\rm in}^{\rm ES}}, {\rm P.V.}\mathcal{K}_0\left(\vec{b}_\zeta^{\vec{J}_{\rm in}^{\rm ES}}\right)\right\rangle_{\mathbb{S}_{\rm in}} - (1/2)\left\langle \vec{b}_\xi^{\vec{M}_{\rm in}^{\rm ES}}, \vec{b}_\zeta^{\vec{J}_{\rm in}^{\rm ES}} \times \hat{n}_- \right\rangle_{\mathbb{S}_{\rm in}} \tag{D-20a}$$

$$z_{\xi\zeta}^{\vec{M}_{\rm in}^{\rm ES}\vec{J}^{\rm IS}} = \left\langle \vec{b}_\xi^{\vec{M}_{\rm in}^{\rm ES}}, \mathcal{K}\left(\vec{b}_\zeta^{\vec{J}^{\rm IS}}\right)\right\rangle_{\mathbb{S}_{\rm in}} \tag{D-20b}$$

$$z_{\xi\zeta}^{\vec{M}_{\rm in}^{\rm ES}\vec{J}_{\rm out}^{\rm ES}} = \left\langle \vec{b}_\xi^{\vec{M}_{\rm in}^{\rm ES}}, \mathcal{K}\left(-\vec{b}_\zeta^{\vec{J}_{\rm out}^{\rm ES}}\right)\right\rangle_{\mathbb{S}_{\rm in}} \tag{D-20c}$$

$$z_{\xi\zeta}^{\vec{M}_{\rm in}^{\rm ES}\vec{M}_{\rm in}^{\rm ES}} = \left\langle \vec{b}_\xi^{\vec{M}_{\rm in}^{\rm ES}}, -j\omega\varepsilon\mathcal{L}\left(\vec{b}_\zeta^{\vec{M}_{\rm in}^{\rm ES}}\right)\right\rangle_{\mathbb{S}_{\rm in}} \tag{D-20d}$$

$$z_{\xi\zeta}^{\vec{M}_{\rm in}^{\rm ES}\vec{M}_{\rm out}^{\rm ES}} = \left\langle \vec{b}_\xi^{\vec{M}_{\rm in}^{\rm ES}}, -j\omega\varepsilon\mathcal{L}\left(-\vec{b}_\zeta^{\vec{M}_{\rm out}^{\rm ES}}\right)\right\rangle_{\mathbb{S}_{\rm in}} \tag{D-20e}$$

and

$$z_{\xi\zeta}^{\vec{J}_{\rm in}^{\rm ES}\vec{J}_{\rm in}^{\rm ES}} = \left\langle \vec{b}_\xi^{\vec{J}_{\rm in}^{\rm ES}}, -j\omega\mu\mathcal{L}\left(\vec{b}_\zeta^{\vec{J}_{\rm in}^{\rm ES}}\right)\right\rangle_{\mathbb{S}_{\rm in}} \tag{D-21a}$$

$$z_{\xi\zeta}^{\vec{J}_{\rm in}^{\rm ES}\vec{J}^{\rm IS}} = \left\langle \vec{b}_\xi^{\vec{J}_{\rm in}^{\rm ES}}, -j\omega\mu\mathcal{L}\left(\vec{b}_\zeta^{\vec{J}^{\rm IS}}\right)\right\rangle_{\mathbb{S}_{\rm in}} \tag{D-21b}$$

$$z_{\xi\zeta}^{\vec{J}_{\rm in}^{\rm ES}\vec{J}_{\rm out}^{\rm ES}} = \left\langle \vec{b}_\xi^{\vec{J}_{\rm in}^{\rm ES}}, -j\omega\mu\mathcal{L}\left(-\vec{b}_\zeta^{\vec{J}_{\rm out}^{\rm ES}}\right)\right\rangle_{\mathbb{S}_{\rm in}} \tag{D-21c}$$

$$z_{\xi\zeta}^{\vec{J}_{\rm in}^{\rm ES}\vec{M}_{\rm in}^{\rm ES}} = \left\langle \vec{b}_\xi^{\vec{J}_{\rm in}^{\rm ES}}, -{\rm P.V.}\mathcal{K}\left(\vec{b}_\zeta^{\vec{M}_{\rm in}^{\rm ES}}\right)\right\rangle_{\mathbb{S}_{\rm in}} - (1/2)\left\langle \vec{b}_\xi^{\vec{J}_{\rm in}^{\rm ES}}, \hat{n}_- \times \vec{b}_\zeta^{\vec{M}_{\rm in}^{\rm ES}} \right\rangle_{\mathbb{S}_{\rm in}} \tag{D-21d}$$

$$z_{\xi\zeta}^{\vec{J}_{\rm in}^{\rm ES}\vec{M}_{\rm out}^{\rm ES}} = \left\langle \vec{b}_\xi^{\vec{J}_{\rm in}^{\rm ES}}, -\mathcal{K}\left(-\vec{b}_\zeta^{\vec{M}_{\rm out}^{\rm ES}}\right)\right\rangle_{\mathbb{S}_{\rm in}} \tag{D-21e}$$

and

$$z_{\xi\zeta}^{\vec{J}^{\rm IS}\vec{J}_{\rm in}^{\rm ES}} = \left\langle \vec{b}_\xi^{\vec{J}^{\rm IS}}, -j\omega\mu\mathcal{L}\left(\vec{b}_\zeta^{\vec{J}_{\rm in}^{\rm ES}}\right)\right\rangle_{\mathbb{S}_{\rm ele}} \tag{D-22a}$$

$$z_{\xi\zeta}^{\vec{J}^{\rm IS}\vec{J}^{\rm IS}} = \left\langle \vec{b}_\xi^{\vec{J}^{\rm IS}}, -j\omega\mu\mathcal{L}\left(\vec{b}_\zeta^{\vec{J}^{\rm IS}}\right)\right\rangle_{\mathbb{S}_{\rm ele}} \tag{D-22b}$$

$$z_{\xi\zeta}^{\vec{J}^{\rm IS}\vec{J}_{\rm out}^{\rm ES}} = \left\langle \vec{b}_\xi^{\vec{J}^{\rm IS}}, -j\omega\mu\mathcal{L}\left(-\vec{b}_\zeta^{\vec{J}_{\rm out}^{\rm ES}}\right)\right\rangle_{\mathbb{S}_{\rm ele}} \tag{D-22c}$$

$$z_{\xi\zeta}^{\vec{J}^{\rm IS}\vec{M}_{\rm in}^{\rm ES}} = \left\langle \vec{b}_\xi^{\vec{J}^{\rm IS}}, -\mathcal{K}\left(\vec{b}_\zeta^{\vec{M}_{\rm in}^{\rm ES}}\right)\right\rangle_{\mathbb{S}_{\rm ele}} \tag{D-22d}$$

$$z_{\xi\zeta}^{\vec{J}^{\rm IS}\vec{M}_{\rm out}^{\rm ES}} = \left\langle \vec{b}_\xi^{\vec{J}^{\rm IS}}, -\mathcal{K}\left(-\vec{b}_\zeta^{\vec{M}_{\rm out}^{\rm ES}}\right)\right\rangle_{\mathbb{S}_{\rm ele}} \tag{D-22e}$$

and

$$z_{\xi\zeta}^{\vec{J}_{\rm out}^{\rm ES}\vec{J}_{\rm in}^{\rm ES}} = \left\langle \vec{b}_\xi^{\vec{J}_{\rm out}^{\rm ES}}, -j\omega\mu\mathcal{L}\left(\vec{b}_\zeta^{\vec{J}_{\rm in}^{\rm ES}}\right)\right\rangle_{\mathbb{S}_{\rm out}} \tag{D-23a}$$

$$z_{\xi\zeta}^{\vec{J}_{\rm out}^{\rm ES}\vec{J}^{\rm IS}} = \left\langle \vec{b}_\xi^{\vec{J}_{\rm out}^{\rm ES}}, -j\omega\mu\mathcal{L}\left(\vec{b}_\zeta^{\vec{J}^{\rm IS}}\right)\right\rangle_{\mathbb{S}_{\rm out}} \tag{D-23b}$$

$$z_{\xi\zeta}^{\vec{J}_{\text{out}}^{\text{ES}}\vec{J}_{\text{out}}^{\text{ES}}} = \left\langle \vec{b}_{\xi}^{\vec{J}_{\text{out}}^{\text{ES}}}, -j\omega\mu\mathcal{L}\left(-\vec{b}_{\zeta}^{\vec{J}_{\text{out}}^{\text{ES}}}\right)\right\rangle_{\mathbb{S}_{\text{out}_-}} - \left\langle \vec{b}_{\xi}^{\vec{J}_{\text{out}}^{\text{ES}}}, -j\omega\mu\mathcal{L}\left(\vec{b}_{\zeta}^{\vec{J}_{\text{out}}^{\text{ES}}}\right)\right\rangle_{\mathbb{S}_{\text{out}_+}} \tag{D-23c}$$

$$z_{\xi\zeta}^{\vec{J}_{\text{out}}^{\text{ES}}\vec{M}_{\text{in}}^{\text{ES}}} = \left\langle \vec{b}_{\xi}^{\vec{J}_{\text{out}}^{\text{ES}}}, -\mathcal{K}\left(\vec{b}_{\zeta}^{\vec{M}_{\text{in}}^{\text{ES}}}\right)\right\rangle_{\mathbb{S}_{\text{out}}} \tag{D-23d}$$

$$z_{\xi\zeta}^{\vec{J}_{\text{out}}^{\text{ES}}\vec{M}_{\text{out}}^{\text{ES}}} = \left\langle \vec{b}_{\xi}^{\vec{J}_{\text{out}}^{\text{ES}}}, -\text{P.V.}\,\mathcal{K}\left(-\vec{b}_{\zeta}^{\vec{M}_{\text{out}}^{\text{ES}}}\right)\right\rangle_{\mathbb{S}_{\text{out}_-}} - \left\langle \vec{b}_{\xi}^{\vec{J}_{\text{out}}^{\text{ES}}}, -\text{P.V.}\,\mathcal{K}\left(\vec{b}_{\zeta}^{\vec{M}_{\text{out}}^{\text{ES}}}\right)\right\rangle_{\mathbb{S}_{\text{out}_+}} \tag{D-23e}$$

and

$$z_{\xi\zeta}^{\vec{M}_{\text{out}}^{\text{ES}}\vec{J}_{\text{in}}^{\text{ES}}} = \left\langle \vec{b}_{\xi}^{\vec{M}_{\text{out}}^{\text{ES}}}, \mathcal{K}\left(\vec{b}_{\zeta}^{\vec{J}_{\text{in}}^{\text{ES}}}\right)\right\rangle_{\mathbb{S}_{\text{out}}} \tag{D-24a}$$

$$z_{\xi\zeta}^{\vec{M}_{\text{out}}^{\text{ES}}\vec{J}^{\text{IS}}} = \left\langle \vec{b}_{\xi}^{\vec{M}_{\text{out}}^{\text{ES}}}, \mathcal{K}\left(\vec{b}_{\zeta}^{\vec{J}^{\text{IS}}}\right)\right\rangle_{\mathbb{S}_{\text{out}}} \tag{D-24b}$$

$$z_{\xi\zeta}^{\vec{M}_{\text{out}}^{\text{ES}}\vec{J}_{\text{out}}^{\text{ES}}} = \left\langle \vec{b}_{\xi}^{\vec{M}_{\text{out}}^{\text{ES}}}, \text{P.V.}\,\mathcal{K}\left(-\vec{b}_{\zeta}^{\vec{J}_{\text{out}}^{\text{ES}}}\right)\right\rangle_{\mathbb{S}_{\text{out}_-}} - \left\langle \vec{b}_{\xi}^{\vec{M}_{\text{out}}^{\text{ES}}}, \text{P.V.}\,\mathcal{K}\left(\vec{b}_{\zeta}^{\vec{J}_{\text{out}}^{\text{ES}}}\right)\right\rangle_{\mathbb{S}_{\text{out}_+}} \tag{D-24c}$$

$$z_{\xi\zeta}^{\vec{M}_{\text{out}}^{\text{ES}}\vec{M}_{\text{in}}^{\text{ES}}} = \left\langle \vec{b}_{\xi}^{\vec{M}_{\text{out}}^{\text{ES}}}, -j\omega\varepsilon\mathcal{L}\left(\vec{b}_{\zeta}^{\vec{M}_{\text{in}}^{\text{ES}}}\right)\right\rangle_{\mathbb{S}_{\text{out}}} \tag{D-24d}$$

$$z_{\xi\zeta}^{\vec{M}_{\text{out}}^{\text{ES}}\vec{M}_{\text{out}}^{\text{ES}}} = \left\langle \vec{b}_{\xi}^{\vec{M}_{\text{out}}^{\text{ES}}}, -j\omega\varepsilon\mathcal{L}\left(-\vec{b}_{\zeta}^{\vec{M}_{\text{out}}^{\text{ES}}}\right)\right\rangle_{\mathbb{S}_{\text{out}_-}} - \left\langle \vec{b}_{\xi}^{\vec{M}_{\text{out}}^{\text{ES}}}, -j\omega\varepsilon\mathcal{L}\left(\vec{b}_{\zeta}^{\vec{M}_{\text{out}}^{\text{ES}}}\right)\right\rangle_{\mathbb{S}_{\text{out}_+}} \tag{D-24e}$$

The transformation matrix $\overline{\overline{T}}$ used in Eq. (3-146) is $\overline{\overline{T}}^{\vec{J}_{\text{in}}^{\text{ES}}\to\text{AV}} / \overline{\overline{T}}^{\vec{M}_{\text{in}}^{\text{ES}}\to\text{AV}} / \overline{\overline{T}}^{\text{BS}\to\text{AV}}$, and the vector $\overline{a}$ in Eq. (3-146) is $\overline{a}^{\vec{J}_{\text{in}}^{\text{ES}}} / \overline{a}^{\vec{M}_{\text{in}}^{\text{ES}}} / \overline{a}^{\text{BS}}$ correspondingly. Here, the matrices $\overline{\overline{T}}^{\vec{J}_{\text{in}}^{\text{ES}}\to\text{AV}}$, $\overline{\overline{T}}^{\vec{M}_{\text{in}}^{\text{ES}}\to\text{AV}}$, and $\overline{\overline{T}}^{\text{BS}\to\text{AV}}$ are as follows:

$$\overline{\overline{T}}^{\vec{J}_{\text{in}}^{\text{ES}}\to\text{AV}} = \left(\overline{\overline{\Psi}}_1\right)^{-1}\cdot\overline{\overline{\Psi}}_2 \tag{D-25a}$$

$$\overline{\overline{T}}^{\vec{M}_{\text{in}}^{\text{ES}}\to\text{AV}} = \left(\overline{\overline{\Psi}}_3\right)^{-1}\cdot\overline{\overline{\Psi}}_4 \tag{D-25b}$$

and

$$\overline{\overline{T}}^{\text{BS}\to\text{AV}} = \text{nullspace}\left(\overline{\overline{\Psi}}^{\text{DoJ/DoM}}\right) \tag{D-26}$$

in which

$$\overline{\overline{\Psi}}_1 = \begin{bmatrix} \overline{\overline{I}}^{\vec{J}_{\text{in}}^{\text{ES}}} & 0 & 0 & 0 & 0 \\ 0 & \overline{\overline{Z}}^{\vec{M}_{\text{in}}^{\text{ES}}\vec{J}^{\text{IS}}} & \overline{\overline{Z}}^{\vec{M}_{\text{in}}^{\text{ES}}\vec{J}_{\text{out}}^{\text{ES}}} & \overline{\overline{Z}}^{\vec{M}_{\text{in}}^{\text{ES}}\vec{M}_{\text{in}}^{\text{ES}}} & \overline{\overline{Z}}^{\vec{M}_{\text{in}}^{\text{ES}}\vec{M}_{\text{out}}^{\text{ES}}} \\ 0 & \overline{\overline{Z}}^{\vec{J}^{\text{IS}}\vec{J}^{\text{IS}}} & \overline{\overline{Z}}^{\vec{J}^{\text{IS}}\vec{J}_{\text{out}}^{\text{ES}}} & \overline{\overline{Z}}^{\vec{J}^{\text{IS}}\vec{M}_{\text{in}}^{\text{ES}}} & \overline{\overline{Z}}^{\vec{J}^{\text{IS}}\vec{M}_{\text{out}}^{\text{ES}}} \\ 0 & \overline{\overline{Z}}^{\vec{J}_{\text{out}}^{\text{ES}}\vec{J}^{\text{IS}}} & \overline{\overline{Z}}^{\vec{J}_{\text{out}}^{\text{ES}}\vec{J}_{\text{out}}^{\text{ES}}} & \overline{\overline{Z}}^{\vec{J}_{\text{out}}^{\text{ES}}\vec{M}_{\text{in}}^{\text{ES}}} & \overline{\overline{Z}}^{\vec{J}_{\text{out}}^{\text{ES}}\vec{M}_{\text{out}}^{\text{ES}}} \\ 0 & \overline{\overline{Z}}^{\vec{M}_{\text{out}}^{\text{ES}}\vec{J}^{\text{IS}}} & \overline{\overline{Z}}^{\vec{M}_{\text{out}}^{\text{ES}}\vec{J}_{\text{out}}^{\text{ES}}} & \overline{\overline{Z}}^{\vec{M}_{\text{out}}^{\text{ES}}\vec{M}_{\text{in}}^{\text{ES}}} & \overline{\overline{Z}}^{\vec{M}_{\text{out}}^{\text{ES}}\vec{M}_{\text{out}}^{\text{ES}}} \end{bmatrix} \tag{D-27a}$$

$$\overline{\overline{\Psi}}_2 = \begin{bmatrix} \overline{\overline{I}}^{\vec{J}_{\text{in}}^{\text{ES}}} \\ -\overline{\overline{Z}}^{\vec{M}_{\text{in}}^{\text{ES}}\vec{J}_{\text{in}}^{\text{ES}}} \\ -\overline{\overline{Z}}^{\vec{J}^{\text{IS}}\vec{J}_{\text{in}}^{\text{ES}}} \\ -\overline{\overline{Z}}^{\vec{J}_{\text{out}}^{\text{ES}}\vec{J}_{\text{in}}^{\text{ES}}} \\ -\overline{\overline{Z}}^{\vec{M}_{\text{out}}^{\text{ES}}\vec{J}_{\text{in}}^{\text{ES}}} \end{bmatrix} \tag{D-27b}$$

and

$$\overline{\overline{\Psi}}_3 = \begin{bmatrix} 0 & 0 & 0 & \overline{\overline{I}}^{\vec{M}_{\mathrm{in}}^{\mathrm{ES}}} & 0 \\ \overline{\overline{Z}}^{\vec{J}_{\mathrm{in}}^{\mathrm{ES}}\vec{J}_{\mathrm{in}}^{\mathrm{ES}}} & \overline{\overline{Z}}^{\vec{J}_{\mathrm{in}}^{\mathrm{ES}}\vec{J}^{\mathrm{IS}}} & \overline{\overline{Z}}^{\vec{J}_{\mathrm{in}}^{\mathrm{ES}}\vec{J}_{\mathrm{out}}^{\mathrm{ES}}} & 0 & \overline{\overline{Z}}^{\vec{J}_{\mathrm{in}}^{\mathrm{ES}}\vec{M}_{\mathrm{out}}^{\mathrm{ES}}} \\ \overline{\overline{Z}}^{\vec{J}^{\mathrm{IS}}\vec{J}_{\mathrm{in}}^{\mathrm{ES}}} & \overline{\overline{Z}}^{\vec{J}^{\mathrm{IS}}\vec{J}^{\mathrm{IS}}} & \overline{\overline{Z}}^{\vec{J}^{\mathrm{IS}}\vec{J}_{\mathrm{out}}^{\mathrm{ES}}} & 0 & \overline{\overline{Z}}^{\vec{J}^{\mathrm{IS}}\vec{M}_{\mathrm{out}}^{\mathrm{ES}}} \\ \overline{\overline{Z}}^{\vec{J}_{\mathrm{out}}^{\mathrm{ES}}\vec{J}_{\mathrm{in}}^{\mathrm{ES}}} & \overline{\overline{Z}}^{\vec{J}_{\mathrm{out}}^{\mathrm{ES}}\vec{J}^{\mathrm{IS}}} & \overline{\overline{Z}}^{\vec{J}_{\mathrm{out}}^{\mathrm{ES}}\vec{J}_{\mathrm{out}}^{\mathrm{ES}}} & 0 & \overline{\overline{Z}}^{\vec{J}_{\mathrm{out}}^{\mathrm{ES}}\vec{M}_{\mathrm{out}}^{\mathrm{ES}}} \\ \overline{\overline{Z}}^{\vec{M}_{\mathrm{out}}^{\mathrm{ES}}\vec{J}_{\mathrm{in}}^{\mathrm{ES}}} & \overline{\overline{Z}}^{\vec{M}_{\mathrm{out}}^{\mathrm{ES}}\vec{J}^{\mathrm{IS}}} & \overline{\overline{Z}}^{\vec{M}_{\mathrm{out}}^{\mathrm{ES}}\vec{J}_{\mathrm{out}}^{\mathrm{ES}}} & 0 & \overline{\overline{Z}}^{\vec{M}_{\mathrm{out}}^{\mathrm{ES}}\vec{M}_{\mathrm{out}}^{\mathrm{ES}}} \end{bmatrix} \tag{D-28a}$$

$$\overline{\overline{\Psi}}_4 = \begin{bmatrix} \overline{\overline{I}}^{\vec{M}_{\mathrm{in}}^{\mathrm{ES}}} \\ -\overline{\overline{Z}}^{\vec{J}_{\mathrm{in}}^{\mathrm{ES}}\vec{M}_{\mathrm{in}}^{\mathrm{ES}}} \\ -\overline{\overline{Z}}^{\vec{J}^{\mathrm{IS}}\vec{M}_{\mathrm{in}}^{\mathrm{ES}}} \\ -\overline{\overline{Z}}^{\vec{J}_{\mathrm{out}}^{\mathrm{ES}}\vec{M}_{\mathrm{in}}^{\mathrm{ES}}} \\ -\overline{\overline{Z}}^{\vec{M}_{\mathrm{out}}^{\mathrm{ES}}\vec{M}_{\mathrm{in}}^{\mathrm{ES}}} \end{bmatrix} \tag{D-28b}$$

and

$$\overline{\overline{\Psi}}^{\mathrm{DoJ}} = \begin{bmatrix} \overline{\overline{Z}}^{\vec{M}_{\mathrm{in}}^{\mathrm{ES}}\vec{J}_{\mathrm{in}}^{\mathrm{ES}}} & \overline{\overline{Z}}^{\vec{M}_{\mathrm{in}}^{\mathrm{ES}}\vec{J}^{\mathrm{IS}}} & \overline{\overline{Z}}^{\vec{M}_{\mathrm{in}}^{\mathrm{ES}}\vec{J}_{\mathrm{out}}^{\mathrm{ES}}} & \overline{\overline{Z}}^{\vec{M}_{\mathrm{in}}^{\mathrm{ES}}\vec{M}_{\mathrm{in}}^{\mathrm{ES}}} & \overline{\overline{Z}}^{\vec{M}_{\mathrm{in}}^{\mathrm{ES}}\vec{M}_{\mathrm{out}}^{\mathrm{ES}}} \\ \overline{\overline{Z}}^{\vec{J}^{\mathrm{IS}}\vec{J}_{\mathrm{in}}^{\mathrm{ES}}} & \overline{\overline{Z}}^{\vec{J}^{\mathrm{IS}}\vec{J}^{\mathrm{IS}}} & \overline{\overline{Z}}^{\vec{J}^{\mathrm{IS}}\vec{J}_{\mathrm{out}}^{\mathrm{ES}}} & \overline{\overline{Z}}^{\vec{J}^{\mathrm{IS}}\vec{M}_{\mathrm{in}}^{\mathrm{ES}}} & \overline{\overline{Z}}^{\vec{J}^{\mathrm{IS}}\vec{M}_{\mathrm{out}}^{\mathrm{ES}}} \\ \overline{\overline{Z}}^{\vec{J}_{\mathrm{out}}^{\mathrm{ES}}\vec{J}_{\mathrm{in}}^{\mathrm{ES}}} & \overline{\overline{Z}}^{\vec{J}_{\mathrm{out}}^{\mathrm{ES}}\vec{J}^{\mathrm{IS}}} & \overline{\overline{Z}}^{\vec{J}_{\mathrm{out}}^{\mathrm{ES}}\vec{J}_{\mathrm{out}}^{\mathrm{ES}}} & \overline{\overline{Z}}^{\vec{J}_{\mathrm{out}}^{\mathrm{ES}}\vec{M}_{\mathrm{in}}^{\mathrm{ES}}} & \overline{\overline{Z}}^{\vec{J}_{\mathrm{out}}^{\mathrm{ES}}\vec{M}_{\mathrm{out}}^{\mathrm{ES}}} \\ \overline{\overline{Z}}^{\vec{M}_{\mathrm{out}}^{\mathrm{ES}}\vec{J}_{\mathrm{in}}^{\mathrm{ES}}} & \overline{\overline{Z}}^{\vec{M}_{\mathrm{out}}^{\mathrm{ES}}\vec{J}^{\mathrm{IS}}} & \overline{\overline{Z}}^{\vec{M}_{\mathrm{out}}^{\mathrm{ES}}\vec{J}_{\mathrm{out}}^{\mathrm{ES}}} & \overline{\overline{Z}}^{\vec{M}_{\mathrm{out}}^{\mathrm{ES}}\vec{M}_{\mathrm{in}}^{\mathrm{ES}}} & \overline{\overline{Z}}^{\vec{M}_{\mathrm{out}}^{\mathrm{ES}}\vec{M}_{\mathrm{out}}^{\mathrm{ES}}} \end{bmatrix} \tag{D-29a}$$

$$\overline{\overline{\Psi}}^{\mathrm{DoM}} = \begin{bmatrix} \overline{\overline{Z}}^{\vec{J}_{\mathrm{in}}^{\mathrm{ES}}\vec{J}_{\mathrm{in}}^{\mathrm{ES}}} & \overline{\overline{Z}}^{\vec{J}_{\mathrm{in}}^{\mathrm{ES}}\vec{J}^{\mathrm{IS}}} & \overline{\overline{Z}}^{\vec{J}_{\mathrm{in}}^{\mathrm{ES}}\vec{J}_{\mathrm{out}}^{\mathrm{ES}}} & \overline{\overline{Z}}^{\vec{J}_{\mathrm{in}}^{\mathrm{ES}}\vec{M}_{\mathrm{in}}^{\mathrm{ES}}} & \overline{\overline{Z}}^{\vec{J}_{\mathrm{in}}^{\mathrm{ES}}\vec{M}_{\mathrm{out}}^{\mathrm{ES}}} \\ \overline{\overline{Z}}^{\vec{J}^{\mathrm{IS}}\vec{J}_{\mathrm{in}}^{\mathrm{ES}}} & \overline{\overline{Z}}^{\vec{J}^{\mathrm{IS}}\vec{J}^{\mathrm{IS}}} & \overline{\overline{Z}}^{\vec{J}^{\mathrm{IS}}\vec{J}_{\mathrm{out}}^{\mathrm{ES}}} & \overline{\overline{Z}}^{\vec{J}^{\mathrm{IS}}\vec{M}_{\mathrm{in}}^{\mathrm{ES}}} & \overline{\overline{Z}}^{\vec{J}^{\mathrm{IS}}\vec{M}_{\mathrm{out}}^{\mathrm{ES}}} \\ \overline{\overline{Z}}^{\vec{J}_{\mathrm{out}}^{\mathrm{ES}}\vec{J}_{\mathrm{in}}^{\mathrm{ES}}} & \overline{\overline{Z}}^{\vec{J}_{\mathrm{out}}^{\mathrm{ES}}\vec{J}^{\mathrm{IS}}} & \overline{\overline{Z}}^{\vec{J}_{\mathrm{out}}^{\mathrm{ES}}\vec{J}_{\mathrm{out}}^{\mathrm{ES}}} & \overline{\overline{Z}}^{\vec{J}_{\mathrm{out}}^{\mathrm{ES}}\vec{M}_{\mathrm{in}}^{\mathrm{ES}}} & \overline{\overline{Z}}^{\vec{J}_{\mathrm{out}}^{\mathrm{ES}}\vec{M}_{\mathrm{out}}^{\mathrm{ES}}} \\ \overline{\overline{Z}}^{\vec{M}_{\mathrm{out}}^{\mathrm{ES}}\vec{J}_{\mathrm{in}}^{\mathrm{ES}}} & \overline{\overline{Z}}^{\vec{M}_{\mathrm{out}}^{\mathrm{ES}}\vec{J}^{\mathrm{IS}}} & \overline{\overline{Z}}^{\vec{M}_{\mathrm{out}}^{\mathrm{ES}}\vec{J}_{\mathrm{out}}^{\mathrm{ES}}} & \overline{\overline{Z}}^{\vec{M}_{\mathrm{out}}^{\mathrm{ES}}\vec{M}_{\mathrm{in}}^{\mathrm{ES}}} & \overline{\overline{Z}}^{\vec{M}_{\mathrm{out}}^{\mathrm{ES}}\vec{M}_{\mathrm{out}}^{\mathrm{ES}}} \end{bmatrix} \tag{D-29b}$$

The matrix $\overline{\overline{P}}^{\mathrm{in}}$ used in Eq. (3-148) is as follows:

$$\overline{\overline{P}}^{\mathrm{in}} = \begin{bmatrix} 0 & 0 & 0 & \overline{\overline{C}}^{\vec{J}_{\mathrm{in}}^{\mathrm{ES}}\vec{M}_{\mathrm{in}}^{\mathrm{ES}}} & 0 \\ 0 & 0 & 0 & 0 & 0 \\ 0 & 0 & 0 & 0 & 0 \\ 0 & 0 & 0 & 0 & 0 \\ 0 & 0 & 0 & 0 & 0 \end{bmatrix} \tag{D-30}$$

corresponding to the second equality in IPO (3-147), and

$$\overline{\overline{P}}^{\mathrm{in}} = \begin{bmatrix} \overline{\overline{P}}^{\vec{J}_{\mathrm{in}}^{\mathrm{ES}}\vec{J}_{\mathrm{in}}^{\mathrm{ES}}} & \overline{\overline{P}}^{\vec{J}_{\mathrm{in}}^{\mathrm{ES}}\vec{J}^{\mathrm{IS}}} & \overline{\overline{P}}^{\vec{J}_{\mathrm{in}}^{\mathrm{ES}}\vec{J}_{\mathrm{out}}^{\mathrm{ES}}} & \overline{\overline{P}}^{\vec{J}_{\mathrm{in}}^{\mathrm{ES}}\vec{M}_{\mathrm{in}}^{\mathrm{ES}}} & \overline{\overline{P}}^{\vec{J}_{\mathrm{in}}^{\mathrm{ES}}\vec{M}_{\mathrm{out}}^{\mathrm{ES}}} \\ 0 & 0 & 0 & 0 & 0 \\ 0 & 0 & 0 & 0 & 0 \\ 0 & 0 & 0 & 0 & 0 \\ 0 & 0 & 0 & 0 & 0 \end{bmatrix} \tag{D-31a}$$

corresponding to the third equality in IPO (3-147), and

$$\overline{\overline{P}}^{\mathrm{in}} = \begin{bmatrix} 0 & 0 & 0 & 0 & 0 \\ 0 & 0 & 0 & 0 & 0 \\ 0 & 0 & 0 & 0 & 0 \\ \overline{\overline{P}}^{\vec{M}_{\mathrm{in}}^{\mathrm{ES}}\vec{J}_{\mathrm{in}}^{\mathrm{ES}}} & \overline{\overline{P}}^{\vec{M}_{\mathrm{in}}^{\mathrm{ES}}\vec{J}^{\mathrm{IS}}} & \overline{\overline{P}}^{\vec{M}_{\mathrm{in}}^{\mathrm{ES}}\vec{J}_{\mathrm{out}}^{\mathrm{ES}}} & \overline{\overline{P}}^{\vec{M}_{\mathrm{in}}^{\mathrm{ES}}\vec{M}_{\mathrm{in}}^{\mathrm{ES}}} & \overline{\overline{P}}^{\vec{M}_{\mathrm{in}}^{\mathrm{ES}}\vec{M}_{\mathrm{out}}^{\mathrm{ES}}} \\ 0 & 0 & 0 & 0 & 0 \end{bmatrix}^{\dagger} \tag{D-31b}$$

corresponding to the fourth equality in IPO (3-147). The elements of the above various sub-matrices are as follows:

$$c_{\xi\zeta}^{\vec{J}_{\text{in}}^{\text{ES}}\vec{M}_{\text{in}}^{\text{ES}}} = (1/2)\left\langle \hat{z}\times\vec{b}_{\xi}^{\vec{J}_{\text{in}}^{\text{ES}}}, \vec{b}_{\zeta}^{\vec{M}_{\text{in}}^{\text{ES}}} \right\rangle_{\mathbb{S}_{\text{in}}} \tag{D-32}$$

and

$$p_{\xi\zeta}^{\vec{J}_{\text{in}}^{\text{ES}}\vec{J}_{\text{in}}^{\text{ES}}} = -(1/2)\left\langle \vec{b}_{\xi}^{\vec{J}_{\text{in}}^{\text{ES}}}, -j\omega\mu\mathcal{L}\left(\vec{b}_{\zeta}^{\vec{J}_{\text{in}}^{\text{ES}}}\right) \right\rangle_{\mathbb{S}_{\text{in}}} \tag{D-33a}$$

$$p_{\xi\zeta}^{\vec{J}_{\text{in}}^{\text{ES}}\vec{J}^{\text{IS}}} = -(1/2)\left\langle \vec{b}_{\xi}^{\vec{J}_{\text{in}}^{\text{ES}}}, -j\omega\mu\mathcal{L}\left(\vec{b}_{\zeta}^{\vec{J}^{\text{IS}}}\right) \right\rangle_{\mathbb{S}_{\text{in}}} \tag{D-33b}$$

$$p_{\xi\zeta}^{\vec{J}_{\text{in}}^{\text{ES}}\vec{J}_{\text{out}}^{\text{ES}}} = -(1/2)\left\langle \vec{b}_{\xi}^{\vec{J}_{\text{in}}^{\text{ES}}}, -j\omega\mu\mathcal{L}\left(\vec{b}_{\zeta}^{\vec{J}_{\text{out}}^{\text{ES}}}\right) \right\rangle_{\mathbb{S}_{\text{in}}} \tag{D-33c}$$

$$p_{\xi\zeta}^{\vec{J}_{\text{in}}^{\text{ES}}\vec{M}_{\text{in}}^{\text{ES}}} = -(1/2)\left\langle \vec{b}_{\xi}^{\vec{J}_{\text{in}}^{\text{ES}}}, \hat{z}\times\frac{1}{2}\vec{b}_{\zeta}^{\vec{M}_{\text{in}}^{\text{ES}}} - \text{P.V.}\,\mathcal{K}\left(\vec{b}_{\zeta}^{\vec{M}_{\text{in}}^{\text{ES}}}\right) \right\rangle_{\mathbb{S}_{\text{in}}} \tag{D-33d}$$

$$p_{\xi\zeta}^{\vec{J}_{\text{in}}^{\text{ES}}\vec{M}_{\text{out}}^{\text{ES}}} = -(1/2)\left\langle \vec{b}_{\xi}^{\vec{J}_{\text{in}}^{\text{ES}}}, -\mathcal{K}\left(\vec{b}_{\zeta}^{\vec{M}_{\text{out}}^{\text{ES}}}\right) \right\rangle_{\mathbb{S}_{\text{in}}} \tag{D-33e}$$

and

$$p_{\xi\zeta}^{\vec{M}_{\text{in}}^{\text{ES}}\vec{J}_{\text{in}}^{\text{ES}}} = -(1/2)\left\langle \vec{b}_{\xi}^{\vec{M}_{\text{in}}^{\text{ES}}}, \frac{1}{2}\vec{b}_{\zeta}^{\vec{J}_{\text{in}}^{\text{ES}}}\times\hat{z} + \text{P.V.}\,\mathcal{K}\left(\vec{b}_{\zeta}^{\vec{J}_{\text{in}}^{\text{ES}}}\right) \right\rangle_{\mathbb{S}_{\text{in}}} \tag{D-33f}$$

$$p_{\xi\zeta}^{\vec{M}_{\text{in}}^{\text{ES}}\vec{J}^{\text{IS}}} = -(1/2)\left\langle \vec{b}_{\xi}^{\vec{M}_{\text{in}}^{\text{ES}}}, \mathcal{K}\left(\vec{b}_{\zeta}^{\vec{J}^{\text{IS}}}\right) \right\rangle_{\mathbb{S}_{\text{in}}} \tag{D-33g}$$

$$p_{\xi\zeta}^{\vec{M}_{\text{in}}^{\text{ES}}\vec{J}_{\text{out}}^{\text{ES}}} = -(1/2)\left\langle \vec{b}_{\xi}^{\vec{M}_{\text{in}}^{\text{ES}}}, \mathcal{K}\left(\vec{b}_{\zeta}^{\vec{J}_{\text{out}}^{\text{ES}}}\right) \right\rangle_{\mathbb{S}_{\text{in}}} \tag{D-33h}$$

$$p_{\xi\zeta}^{\vec{M}_{\text{in}}^{\text{ES}}\vec{M}_{\text{in}}^{\text{ES}}} = -(1/2)\left\langle \vec{b}_{\xi}^{\vec{M}_{\text{in}}^{\text{ES}}}, -j\omega\varepsilon\mathcal{L}\left(\vec{b}_{\zeta}^{\vec{M}_{\text{in}}^{\text{ES}}}\right) \right\rangle_{\mathbb{S}_{\text{in}}} \tag{D-33i}$$

$$p_{\xi\zeta}^{\vec{M}_{\text{in}}^{\text{ES}}\vec{M}_{\text{out}}^{\text{ES}}} = -(1/2)\left\langle \vec{b}_{\xi}^{\vec{M}_{\text{in}}^{\text{ES}}}, -j\omega\varepsilon\mathcal{L}\left(\vec{b}_{\zeta}^{\vec{M}_{\text{out}}^{\text{ES}}}\right) \right\rangle_{\mathbb{S}_{\text{in}}} \tag{D-33j}$$

## D3 Some Detailed Formulations Related to Sec. 3.6

The formulations used to calculate the elements of the matrices in Eqs. (3-156a)~ (3-157) are as follows:

$$z_{\xi\zeta}^{\vec{M}_{\text{in}}^{\text{ES}}\vec{J}_{\text{in}}^{\text{ES}}} = \left\langle \vec{b}_{\xi}^{\vec{M}_{\text{in}}^{\text{ES}}}, \text{P.V.}\,\mathcal{K}_0\left(\vec{b}_{\zeta}^{\vec{J}_{\text{in}}^{\text{ES}}}\right) \right\rangle_{\mathbb{S}_{\text{in}}} - (1/2)\left\langle \vec{b}_{\xi}^{\vec{M}_{\text{in}}^{\text{ES}}}, \vec{b}_{\zeta}^{\vec{J}_{\text{in}}^{\text{ES}}}\times\hat{n} \right\rangle_{\mathbb{S}_{\text{in}}} \tag{D-34a}$$

$$z_{\xi\zeta}^{\vec{M}_{\text{in}}^{\text{ES}}\vec{J}^{\text{IS}}} = \left\langle \vec{b}_{\xi}^{\vec{M}_{\text{in}}^{\text{ES}}}, \mathcal{K}_0\left(\vec{b}_{\zeta}^{\vec{J}^{\text{IS}}}\right) \right\rangle_{\mathbb{S}_{\text{in}}} \tag{D-34b}$$

$$z_{\xi\zeta}^{\vec{M}_{\text{in}}^{\text{ES}}\vec{M}_{\text{in}}^{\text{ES}}} = \left\langle \vec{b}_{\xi}^{\vec{M}_{\text{in}}^{\text{ES}}}, -j\omega\varepsilon_0\mathcal{L}_0\left(\vec{b}_{\zeta}^{\vec{M}_{\text{in}}^{\text{ES}}}\right) \right\rangle_{\mathbb{S}_{\text{in}}} \tag{D-34c}$$

and

$$z_{\xi\zeta}^{\vec{J}_{\text{in}}^{\text{ES}}\vec{J}_{\text{in}}^{\text{ES}}} = \left\langle \vec{b}_{\xi}^{\vec{J}_{\text{in}}^{\text{ES}}}, -j\omega\mu_0\mathcal{L}_0\left(\vec{b}_{\zeta}^{\vec{J}_{\text{in}}^{\text{ES}}}\right)\right\rangle_{\mathbb{S}_{\text{in}}} \tag{D-35a}$$

$$z_{\xi\zeta}^{\vec{J}_{\text{in}}^{\text{ES}}\vec{J}^{\text{IS}}} = \left\langle \vec{b}_{\xi}^{\vec{J}_{\text{in}}^{\text{ES}}}, -j\omega\mu_0\mathcal{L}_0\left(\vec{b}_{\zeta}^{\vec{J}^{\text{IS}}}\right)\right\rangle_{\mathbb{S}_{\text{in}}} \tag{D-35b}$$

$$z_{\xi\zeta}^{\vec{J}_{\text{in}}^{\text{ES}}\vec{M}_{\text{in}}^{\text{ES}}} = \left\langle \vec{b}_{\xi}^{\vec{J}_{\text{in}}^{\text{ES}}}, -\text{P.V.}\,\mathcal{K}_0\left(\vec{b}_{\zeta}^{\vec{M}_{\text{in}}^{\text{ES}}}\right)\right\rangle_{\mathbb{S}_{\text{in}}} - (1/2)\left\langle \vec{b}_{\xi}^{\vec{J}_{\text{in}}^{\text{ES}}}, \hat{n}\times\vec{b}_{\zeta}^{\vec{M}_{\text{in}}^{\text{ES}}}\right\rangle_{\mathbb{S}_{\text{in}}} \tag{D-35c}$$

and

$$z_{\xi\zeta}^{\vec{J}^{\text{IS}}\vec{J}_{\text{in}}^{\text{ES}}} = \left\langle \vec{b}_{\xi}^{\vec{J}^{\text{IS}}}, -j\omega\mu_0\mathcal{L}_0\left(\vec{b}_{\zeta}^{\vec{J}_{\text{in}}^{\text{ES}}}\right)\right\rangle_{\mathbb{S}_{\text{ele}}} \tag{D-36a}$$

$$z_{\xi\zeta}^{\vec{J}^{\text{IS}}\vec{J}^{\text{IS}}} = \left\langle \vec{b}_{\xi}^{\vec{J}^{\text{IS}}}, -j\omega\mu_0\mathcal{L}_0\left(\vec{b}_{\zeta}^{\vec{J}^{\text{IS}}}\right)\right\rangle_{\mathbb{S}_{\text{ele}}} \tag{D-36b}$$

$$z_{\xi\zeta}^{\vec{J}^{\text{IS}}\vec{M}_{\text{in}}^{\text{ES}}} = \left\langle \vec{b}_{\xi}^{\vec{J}^{\text{IS}}}, -\mathcal{K}_0\left(\vec{b}_{\zeta}^{\vec{M}_{\text{in}}^{\text{ES}}}\right)\right\rangle_{\mathbb{S}_{\text{ele}}} \tag{D-36c}$$

The transformation matrix $\overline{\overline{T}}$ used in Eq. (3-158) is $\overline{\overline{T}}^{\vec{J}_{\text{in}}^{\text{ES}}\to\text{AV}} / \overline{\overline{T}}^{\vec{M}_{\text{in}}^{\text{ES}}\to\text{AV}} / \overline{\overline{T}}^{\text{BS}\to\text{AV}}$, and the vector $\bar{a}$ in Eq. (3-158) is $\bar{a}^{\vec{J}_{\text{in}}^{\text{ES}}} / \bar{a}^{\vec{M}_{\text{in}}^{\text{ES}}} / \bar{a}^{\text{BS}}$ correspondingly. Here, the matrices $\overline{\overline{T}}^{\vec{J}_{\text{in}}^{\text{ES}}\to\text{AV}}$, $\overline{\overline{T}}^{\vec{M}_{\text{in}}^{\text{ES}}\to\text{AV}}$, and $\overline{\overline{T}}^{\text{BS}\to\text{AV}}$ are as follows:

$$\overline{\overline{T}}^{\vec{J}_{\text{in}}^{\text{ES}}\to\text{AV}} = \left(\overline{\overline{\Psi}}_1\right)^{-1}\cdot\overline{\overline{\Psi}}_2 \tag{D-37a}$$

$$\overline{\overline{T}}^{\vec{M}_{\text{in}}^{\text{ES}}\to\text{AV}} = \left(\overline{\overline{\Psi}}_3\right)^{-1}\cdot\overline{\overline{\Psi}}_4 \tag{D-37b}$$

and

$$\overline{\overline{T}}^{\text{BS}\to\text{AV}} = \text{nullspace}\left(\overline{\overline{\Psi}}_{\text{FCE}}^{\text{DoJ/DoM}}\right) \tag{D-38}$$

in which

$$\overline{\overline{\Psi}}_1 = \begin{bmatrix} \overline{\overline{I}}^{\vec{J}_{\text{in}}^{\text{ES}}} & 0 & 0 \\ 0 & \overline{\overline{Z}}^{\vec{M}_{\text{in}}^{\text{ES}}\vec{J}^{\text{IS}}} & \overline{\overline{Z}}^{\vec{M}_{\text{in}}^{\text{ES}}\vec{M}_{\text{in}}^{\text{ES}}} \\ 0 & \overline{\overline{Z}}^{\vec{J}^{\text{IS}}\vec{J}^{\text{IS}}} & \overline{\overline{Z}}^{\vec{J}^{\text{IS}}\vec{M}_{\text{in}}^{\text{ES}}} \end{bmatrix} \tag{D-39a}$$

$$\overline{\overline{\Psi}}_2 = \begin{bmatrix} \overline{\overline{I}}^{\vec{J}_{\text{in}}^{\text{ES}}} \\ -\overline{\overline{Z}}^{\vec{M}_{\text{in}}^{\text{ES}}\vec{J}_{\text{in}}^{\text{ES}}} \\ -\overline{\overline{Z}}^{\vec{J}^{\text{IS}}\vec{J}_{\text{in}}^{\text{ES}}} \end{bmatrix} \tag{D-39b}$$

and

$$\overline{\overline{\Psi}}_3 = \begin{bmatrix} 0 & 0 & \overline{\overline{I}}^{\vec{M}_{\text{in}}^{\text{ES}}} \\ \overline{\overline{Z}}^{\vec{J}_{\text{in}}^{\text{ES}}\vec{J}_{\text{in}}^{\text{ES}}} & \overline{\overline{Z}}^{\vec{J}_{\text{in}}^{\text{ES}}\vec{J}^{\text{IS}}} & 0 \\ \overline{\overline{Z}}^{\vec{J}^{\text{IS}}\vec{J}_{\text{in}}^{\text{ES}}} & \overline{\overline{Z}}^{\vec{J}^{\text{IS}}\vec{J}^{\text{IS}}} & 0 \end{bmatrix} \tag{D-40a}$$

$$\overline{\overline{\Psi}}_4 = \begin{bmatrix} \overline{\overline{I}}^{\vec{M}_{\text{in}}^{\text{ES}}} \\ -\overline{\overline{Z}}^{\vec{J}_{\text{in}}^{\text{ES}}\vec{M}_{\text{in}}^{\text{ES}}} \\ -\overline{\overline{Z}}^{\vec{J}^{\text{IS}}\vec{M}_{\text{in}}^{\text{ES}}} \end{bmatrix} \tag{D-40b}$$

and

$$\bar{\bar{\Psi}}_{\text{FCE}}^{\text{DoJ}} = \begin{bmatrix} \bar{\bar{Z}}^{\vec{M}_{\text{in}}^{\text{ES}}\vec{J}_{\text{in}}^{\text{ES}}} & \bar{\bar{Z}}^{\vec{M}_{\text{in}}^{\text{ES}}\vec{J}^{\text{IS}}} & \bar{\bar{Z}}^{\vec{M}_{\text{in}}^{\text{ES}}\vec{M}_{\text{in}}^{\text{ES}}} \\ \bar{\bar{Z}}^{\vec{J}^{\text{IS}}\vec{J}_{\text{in}}^{\text{ES}}} & \bar{\bar{Z}}^{\vec{J}^{\text{IS}}\vec{J}^{\text{IS}}} & \bar{\bar{Z}}^{\vec{J}^{\text{IS}}\vec{M}_{\text{in}}^{\text{ES}}} \end{bmatrix} \tag{D-41a}$$

$$\bar{\bar{\Psi}}_{\text{FCE}}^{\text{DoM}} = \begin{bmatrix} \bar{\bar{Z}}^{\vec{J}_{\text{in}}^{\text{ES}}\vec{J}_{\text{in}}^{\text{ES}}} & \bar{\bar{Z}}^{\vec{J}_{\text{in}}^{\text{ES}}\vec{J}^{\text{IS}}} & \bar{\bar{Z}}^{\vec{J}_{\text{in}}^{\text{ES}}\vec{M}_{\text{in}}^{\text{ES}}} \\ \bar{\bar{Z}}^{\vec{J}^{\text{IS}}\vec{J}_{\text{in}}^{\text{ES}}} & \bar{\bar{Z}}^{\vec{J}^{\text{IS}}\vec{J}^{\text{IS}}} & \bar{\bar{Z}}^{\vec{J}^{\text{IS}}\vec{M}_{\text{in}}^{\text{ES}}} \end{bmatrix} \tag{D-41b}$$

The matrix $\bar{\bar{P}}^{\text{in}}$ used in Eq. (3-160) is as follows:

$$\bar{\bar{P}}^{\text{in}} = \begin{bmatrix} 0 & 0 & \bar{\bar{C}}^{\vec{J}_{\text{in}}^{\text{ES}}\vec{M}_{\text{in}}^{\text{ES}}} \\ 0 & 0 & 0 \\ 0 & 0 & 0 \end{bmatrix} \tag{D-42}$$

corresponding to the second equality in IPO (3-159), and

$$\bar{\bar{P}}^{\text{in}} = \begin{bmatrix} 0 & 0 & 0 \\ \bar{\bar{P}}^{\vec{J}_{\text{in}}^{\text{ES}}\vec{J}_{\text{in}}^{\text{ES}}} & \bar{\bar{P}}^{\vec{J}_{\text{in}}^{\text{ES}}\vec{J}^{\text{IS}}} & \bar{\bar{P}}^{\vec{J}_{\text{in}}^{\text{ES}}\vec{M}_{\text{in}}^{\text{ES}}} \\ 0 & 0 & 0 \end{bmatrix} \tag{D-43a}$$

corresponding to the third equality in IPO (3-159), and

$$\bar{\bar{P}}^{\text{in}} = \begin{bmatrix} 0 & 0 & 0 \\ 0 & 0 & 0 \\ \bar{\bar{P}}^{\vec{M}_{\text{in}}^{\text{ES}}\vec{J}_{\text{in}}^{\text{ES}}} & \bar{\bar{P}}^{\vec{M}_{\text{in}}^{\text{ES}}\vec{J}^{\text{IS}}} & \bar{\bar{P}}^{\vec{M}_{\text{in}}^{\text{ES}}\vec{M}_{\text{in}}^{\text{ES}}} \end{bmatrix} \tag{D-43b}$$

corresponding to the fourth equality in IPO (3-159). The elements of the above various sub-matrices are as follows:

$$c_{\xi\zeta}^{\vec{J}_{\text{in}}^{\text{ES}}\vec{M}_{\text{in}}^{\text{ES}}} = (1/2)\left\langle \hat{n}\times\vec{b}_{\xi}^{\vec{J}_{\text{in}}^{\text{ES}}}, \vec{b}_{\zeta}^{\vec{M}_{\text{in}}^{\text{ES}}} \right\rangle_{\mathbb{S}_{\text{in}}} \tag{D-44}$$

and

$$p_{\xi\zeta}^{\vec{J}_{\text{in}}^{\text{ES}}\vec{J}_{\text{in}}^{\text{ES}}} = -(1/2)\left\langle \vec{b}_{\xi}^{\vec{J}_{\text{in}}^{\text{ES}}}, -j\omega\mu_0\mathcal{L}_0\left(\vec{b}_{\zeta}^{\vec{J}_{\text{in}}^{\text{ES}}}\right) \right\rangle_{\mathbb{S}_{\text{in}}} \tag{D-45a}$$

$$p_{\xi\zeta}^{\vec{J}_{\text{in}}^{\text{ES}}\vec{J}^{\text{IS}}} = -(1/2)\left\langle \vec{b}_{\xi}^{\vec{J}_{\text{in}}^{\text{ES}}}, -j\omega\mu_0\mathcal{L}_0\left(\vec{b}_{\zeta}^{\vec{J}^{\text{IS}}}\right) \right\rangle_{\mathbb{S}_{\text{in}}} \tag{D-45b}$$

$$p_{\xi\zeta}^{\vec{J}_{\text{in}}^{\text{ES}}\vec{M}_{\text{in}}^{\text{ES}}} = -(1/2)\left\langle \vec{b}_{\xi}^{\vec{J}_{\text{in}}^{\text{ES}}}, \hat{n}\times\frac{1}{2}\vec{b}_{\zeta}^{\vec{M}_{\text{in}}^{\text{ES}}} - \text{P.V.}\,\mathcal{K}_0\left(\vec{b}_{\zeta}^{\vec{M}_{\text{in}}^{\text{ES}}}\right) \right\rangle_{\mathbb{S}_{\text{in}}} \tag{D-45c}$$

and

$$p_{\xi\zeta}^{\vec{M}_{\text{in}}^{\text{ES}}\vec{J}_{\text{in}}^{\text{ES}}} = -(1/2)\left\langle \vec{b}_{\xi}^{\vec{M}_{\text{in}}^{\text{ES}}}, \frac{1}{2}\vec{b}_{\zeta}^{\vec{J}_{\text{in}}^{\text{ES}}}\times\hat{n} + \text{P.V.}\,\mathcal{K}_0\left(\vec{b}_{\zeta}^{\vec{J}_{\text{in}}^{\text{ES}}}\right) \right\rangle_{\mathbb{S}_{\text{in}}} \tag{D-45d}$$

$$p_{\xi\zeta}^{\vec{M}_{\text{in}}^{\text{ES}}\vec{J}^{\text{IS}}} = -(1/2)\left\langle \vec{b}_{\xi}^{\vec{M}_{\text{in}}^{\text{ES}}}, \mathcal{K}_0\left(\vec{b}_{\zeta}^{\vec{J}^{\text{IS}}}\right) \right\rangle_{\mathbb{S}_{\text{in}}} \tag{D-45e}$$

$$p_{\xi\zeta}^{\vec{M}_{\text{in}}^{\text{ES}}\vec{M}_{\text{in}}^{\text{ES}}} = -(1/2)\left\langle \vec{b}_{\xi}^{\vec{M}_{\text{in}}^{\text{ES}}}, -j\omega\varepsilon_0\mathcal{L}_0\left(\vec{b}_{\zeta}^{\vec{M}_{\text{in}}^{\text{ES}}}\right) \right\rangle_{\mathbb{S}_{\text{in}}} \tag{D-45f}$$

## D4 Some Detailed Formulations Related to Sec. 4.3

The power quadratic form matrix $\overline{\overline{P}}^{\mathrm{A}\rightleftharpoons\mathrm{M}}$ used in Eq. (4-81) is as follows:

$$\overline{\overline{P}}^{\mathrm{A}\rightleftharpoons\mathrm{M}} = \overline{\overline{P}}_{\mathrm{curAV}}^{\mathrm{A}\rightleftharpoons\mathrm{M}} \text{ or } \overline{\overline{P}}_{\mathrm{intAV}}^{\mathrm{A}\rightleftharpoons\mathrm{M}} \tag{D-46}$$

in which

$$\overline{\overline{P}}_{\mathrm{curAV}}^{\mathrm{A}\rightleftharpoons\mathrm{M}} = \begin{bmatrix} 0 & 0 & 0 & 0 & 0 & \overline{\overline{C}}^{\vec{J}^{\mathrm{A}\rightleftharpoons\mathrm{M}}\vec{M}^{\mathrm{A}\rightleftharpoons\mathrm{M}}} & 0 & 0 \\ 0 & 0 & 0 & 0 & 0 & 0 & 0 & 0 \\ 0 & 0 & 0 & 0 & 0 & 0 & 0 & 0 \\ 0 & 0 & 0 & 0 & 0 & 0 & 0 & 0 \\ 0 & 0 & 0 & 0 & 0 & 0 & 0 & 0 \\ 0 & 0 & 0 & 0 & 0 & 0 & 0 & 0 \\ 0 & 0 & 0 & 0 & 0 & 0 & 0 & 0 \\ 0 & 0 & 0 & 0 & 0 & 0 & 0 & 0 \end{bmatrix} \tag{D-47}$$

for the first equality in Eq. (4-80), and

$$\overline{\overline{P}}_{\mathrm{intAV}}^{\mathrm{A}\rightleftharpoons\mathrm{M}} = \begin{bmatrix} \overline{\overline{P}}^{\vec{J}^{\mathrm{A}\rightleftharpoons\mathrm{M}}\vec{J}^{\mathrm{A}\rightleftharpoons\mathrm{M}}} & \overline{\overline{P}}^{\vec{J}^{\mathrm{A}\rightleftharpoons\mathrm{M}}\vec{J}^{\mathrm{M}}} & \overline{\overline{P}}^{\vec{J}^{\mathrm{A}\rightleftharpoons\mathrm{M}}\vec{J}} & \overline{\overline{P}}^{\vec{J}^{\mathrm{A}\rightleftharpoons\mathrm{M}}\vec{J}_{\mathrm{M}}} & \overline{\overline{P}}^{\vec{J}^{\mathrm{A}\rightleftharpoons\mathrm{M}}\vec{J}_{\mathrm{M}\rightleftharpoons\mathrm{A}}} & \overline{\overline{P}}^{\vec{J}^{\mathrm{A}\rightleftharpoons\mathrm{M}}\vec{M}^{\mathrm{A}\rightleftharpoons\mathrm{M}}} & \overline{\overline{P}}^{\vec{J}^{\mathrm{A}\rightleftharpoons\mathrm{M}}\vec{M}} & \overline{\overline{P}}^{\vec{J}^{\mathrm{A}\rightleftharpoons\mathrm{M}}\vec{M}_{\mathrm{M}\rightleftharpoons\mathrm{A}}} \\ 0 & 0 & 0 & 0 & 0 & 0 & 0 & 0 \\ 0 & 0 & 0 & 0 & 0 & 0 & 0 & 0 \\ 0 & 0 & 0 & 0 & 0 & 0 & 0 & 0 \\ 0 & 0 & 0 & 0 & 0 & 0 & 0 & 0 \\ 0 & 0 & 0 & 0 & 0 & 0 & 0 & 0 \\ 0 & 0 & 0 & 0 & 0 & 0 & 0 & 0 \\ 0 & 0 & 0 & 0 & 0 & 0 & 0 & 0 \end{bmatrix} \tag{D-48a}$$

for the second equality in Eq. (4-80), and

$$\overline{\overline{P}}_{\mathrm{intAV}}^{\mathrm{A}\rightleftharpoons\mathrm{M}} = \begin{bmatrix} 0 & 0 & 0 & 0 & 0 & 0 & 0 & 0 \\ 0 & 0 & 0 & 0 & 0 & 0 & 0 & 0 \\ 0 & 0 & 0 & 0 & 0 & 0 & 0 & 0 \\ 0 & 0 & 0 & 0 & 0 & 0 & 0 & 0 \\ 0 & 0 & 0 & 0 & 0 & 0 & 0 & 0 \\ \overline{\overline{P}}^{\vec{M}^{\mathrm{A}\rightleftharpoons\mathrm{M}}\vec{J}^{\mathrm{A}\rightleftharpoons\mathrm{M}}} & \overline{\overline{P}}^{\vec{M}^{\mathrm{A}\rightleftharpoons\mathrm{M}}\vec{J}^{\mathrm{M}}} & \overline{\overline{P}}^{\vec{M}^{\mathrm{A}\rightleftharpoons\mathrm{M}}\vec{J}} & \overline{\overline{P}}^{\vec{M}^{\mathrm{A}\rightleftharpoons\mathrm{M}}\vec{J}_{\mathrm{M}}} & \overline{\overline{P}}^{\vec{M}^{\mathrm{A}\rightleftharpoons\mathrm{M}}\vec{J}_{\mathrm{M}\rightleftharpoons\mathrm{A}}} & \overline{\overline{P}}^{\vec{M}^{\mathrm{A}\rightleftharpoons\mathrm{M}}\vec{M}^{\mathrm{A}\rightleftharpoons\mathrm{M}}} & \overline{\overline{P}}^{\vec{M}^{\mathrm{A}\rightleftharpoons\mathrm{M}}\vec{M}} & \overline{\overline{P}}^{\vec{M}^{\mathrm{A}\rightleftharpoons\mathrm{M}}\vec{M}_{\mathrm{M}\rightleftharpoons\mathrm{A}}} \\ 0 & 0 & 0 & 0 & 0 & 0 & 0 & 0 \\ 0 & 0 & 0 & 0 & 0 & 0 & 0 & 0 \end{bmatrix}^{\dagger} \tag{D-48b}$$

for the third equality in Eq. (4-80), where the elements of the sub-matrices in above Eqs. (D-47)~(D-48b) are as follows:

$$c_{\xi\zeta}^{\vec{J}^{\mathrm{A}\rightleftharpoons\mathrm{M}}\vec{M}^{\mathrm{A}\rightleftharpoons\mathrm{M}}} = (1/2)\left\langle \hat{n}^{\rightarrow\mathrm{M}} \times \vec{b}_{\xi}^{\vec{J}^{\mathrm{A}\rightleftharpoons\mathrm{M}}}, \vec{b}_{\zeta}^{\vec{M}^{\mathrm{A}\rightleftharpoons\mathrm{M}}} \right\rangle_{\mathbb{S}^{\mathrm{A}\rightleftharpoons\mathrm{M}}} \tag{D-49}$$

and

$$p_{\xi\zeta}^{\vec{J}^{\mathrm{A}\rightleftharpoons\mathrm{M}}\vec{J}^{\mathrm{A}\rightleftharpoons\mathrm{M}}} = -(1/2)\left\langle \vec{b}_{\xi}^{\vec{J}^{\mathrm{A}\rightleftharpoons\mathrm{M}}}, -j\omega\mu_0\mathcal{L}_0\left(\vec{b}_{\zeta}^{\vec{J}^{\mathrm{A}\rightleftharpoons\mathrm{M}}}\right) \right\rangle_{\mathbb{S}^{\mathrm{A}\rightleftharpoons\mathrm{M}}} \tag{D-50a}$$

$$p_{\xi\zeta}^{\vec{J}^{A\rightleftharpoons M}\vec{J}^{M}} = -(1/2)\left\langle \vec{b}_{\xi}^{\vec{J}^{A\rightleftharpoons M}}, -j\omega\mu_0\mathcal{L}_0\left(\vec{b}_{\zeta}^{\vec{J}^{M}}\right)\right\rangle_{\mathbb{S}^{A\rightleftharpoons M}} \tag{D-50b}$$

$$p_{\xi\zeta}^{\vec{J}^{A\rightleftharpoons M}\vec{J}} = -(1/2)\left\langle \vec{b}_{\xi}^{\vec{J}^{A\rightleftharpoons M}}, -j\omega\mu_0\mathcal{L}_0\left(\vec{b}_{\zeta}^{\vec{J}}\right)\right\rangle_{\mathbb{S}^{A\rightleftharpoons M}} \tag{D-50c}$$

$$p_{\xi\zeta}^{\vec{J}^{A\rightleftharpoons M}\vec{J}_{M}} = -(1/2)\left\langle \vec{b}_{\xi}^{\vec{J}^{A\rightleftharpoons M}}, -j\omega\mu_0\mathcal{L}_0\left(\vec{b}_{\zeta}^{\vec{J}_{M}}\right)\right\rangle_{\mathbb{S}^{A\rightleftharpoons M}} \tag{D-50d}$$

$$p_{\xi\zeta}^{\vec{J}^{A\rightleftharpoons M}\vec{J}_{M\rightleftharpoons A}} = -(1/2)\left\langle \vec{b}_{\xi}^{\vec{J}^{A\rightleftharpoons M}}, -j\omega\mu_0\mathcal{L}_0\left(\vec{b}_{\zeta}^{\vec{J}_{M\rightleftharpoons A}}\right)\right\rangle_{\mathbb{S}^{A\rightleftharpoons M}} \tag{D-50e}$$

$$p_{\xi\zeta}^{\vec{J}^{A\rightleftharpoons M}\vec{M}^{A\rightleftharpoons M}} = -(1/2)\left\langle \vec{b}_{\xi}^{\vec{J}^{A\rightleftharpoons M}}, -\text{P.V.}\,\mathcal{K}_0\left(\vec{b}_{\zeta}^{\vec{M}^{A\rightleftharpoons M}}\right) + \hat{n}^{\rightarrow M}\times\frac{1}{2}\vec{b}_{\zeta}^{\vec{M}^{A\rightleftharpoons M}}\right\rangle_{\mathbb{S}^{A\rightleftharpoons M}} \tag{D-50f}$$

$$p_{\xi\zeta}^{\vec{J}^{A\rightleftharpoons M}\vec{M}} = -(1/2)\left\langle \vec{b}_{\xi}^{\vec{J}^{A\rightleftharpoons M}}, -\mathcal{K}_0\left(\vec{b}_{\zeta}^{\vec{M}}\right)\right\rangle_{\mathbb{S}^{A\rightleftharpoons M}} \tag{D-50g}$$

$$p_{\xi\zeta}^{\vec{J}^{A\rightleftharpoons M}\vec{M}_{M\rightleftharpoons A}} = -(1/2)\left\langle \vec{b}_{\xi}^{\vec{J}^{A\rightleftharpoons M}}, -\mathcal{K}_0\left(\vec{b}_{\zeta}^{\vec{M}_{M\rightleftharpoons A}}\right)\right\rangle_{\mathbb{S}^{A\rightleftharpoons M}} \tag{D-50h}$$

and

$$p_{\xi\zeta}^{\vec{M}^{A\rightleftharpoons M}\vec{J}^{A\rightleftharpoons M}} = -(1/2)\left\langle \vec{b}_{\xi}^{\vec{M}^{A\rightleftharpoons M}}, \text{P.V.}\,\mathcal{K}_0\left(\vec{b}_{\zeta}^{\vec{J}^{A\rightleftharpoons M}}\right) + \frac{1}{2}\vec{b}_{\zeta}^{\vec{J}^{A\rightleftharpoons M}}\times\hat{n}^{\rightarrow M}\right\rangle_{\mathbb{S}^{A\rightleftharpoons M}} \tag{D-50i}$$

$$p_{\xi\zeta}^{\vec{M}^{A\rightleftharpoons M}\vec{J}^{M}} = -(1/2)\left\langle \vec{b}_{\xi}^{\vec{M}^{A\rightleftharpoons M}}, \mathcal{K}_0\left(\vec{b}_{\zeta}^{\vec{J}^{M}}\right)\right\rangle_{\mathbb{S}^{A\rightleftharpoons M}} \tag{D-50j}$$

$$p_{\xi\zeta}^{\vec{M}^{A\rightleftharpoons M}\vec{J}} = -(1/2)\left\langle \vec{b}_{\xi}^{\vec{M}^{A\rightleftharpoons M}}, \mathcal{K}_0\left(\vec{b}_{\zeta}^{\vec{J}}\right)\right\rangle_{\mathbb{S}^{A\rightleftharpoons M}} \tag{D-50k}$$

$$p_{\xi\zeta}^{\vec{M}^{A\rightleftharpoons M}\vec{J}_{M}} = -(1/2)\left\langle \vec{b}_{\xi}^{\vec{M}^{A\rightleftharpoons M}}, \mathcal{K}_0\left(\vec{b}_{\zeta}^{\vec{J}_{M}}\right)\right\rangle_{\mathbb{S}^{A\rightleftharpoons M}} \tag{D-50l}$$

$$p_{\xi\zeta}^{\vec{M}^{A\rightleftharpoons M}\vec{J}_{M\rightleftharpoons A}} = -(1/2)\left\langle \vec{b}_{\xi}^{\vec{M}^{A\rightleftharpoons M}}, \mathcal{K}_0\left(\vec{b}_{\zeta}^{\vec{J}_{M\rightleftharpoons A}}\right)\right\rangle_{\mathbb{S}^{A\rightleftharpoons M}} \tag{D-50m}$$

$$p_{\xi\zeta}^{\vec{M}^{A\rightleftharpoons M}\vec{M}^{A\rightleftharpoons M}} = -(1/2)\left\langle \vec{b}_{\xi}^{\vec{M}^{A\rightleftharpoons M}}, -j\omega\varepsilon_0\mathcal{L}_0\left(\vec{b}_{\zeta}^{\vec{M}^{A\rightleftharpoons M}}\right)\right\rangle_{\mathbb{S}^{A\rightleftharpoons M}} \tag{D-50n}$$

$$p_{\xi\zeta}^{\vec{M}^{A\rightleftharpoons M}\vec{M}} = -(1/2)\left\langle \vec{b}_{\xi}^{\vec{M}^{A\rightleftharpoons M}}, -j\omega\varepsilon_0\mathcal{L}_0\left(\vec{b}_{\zeta}^{\vec{M}}\right)\right\rangle_{\mathbb{S}^{A\rightleftharpoons M}} \tag{D-50o}$$

$$p_{\xi\zeta}^{\vec{M}^{A\rightleftharpoons M}\vec{M}_{M\rightleftharpoons A}} = -(1/2)\left\langle \vec{b}_{\xi}^{\vec{M}^{A\rightleftharpoons M}}, -j\omega\varepsilon_0\mathcal{L}_0\left(\vec{b}_{\zeta}^{\vec{M}_{M\rightleftharpoons A}}\right)\right\rangle_{\mathbb{S}^{A\rightleftharpoons M}} \tag{D-50p}$$

## D5 Some Detailed Formulations Related to Sec. 4.4

The formulations used to calculate the elements of the matrices in Eqs. (4-90a)~(4-92) are explicitly given as follows:

$$z_{\xi\zeta}^{\vec{M}_{M\rightleftharpoons A}\vec{J}_{M\rightleftharpoons A}} = \left\langle \vec{b}_{\xi}^{\vec{M}_{M\rightleftharpoons A}}, \mathcal{H}\left(\vec{b}_{\zeta}^{\vec{J}_{M\rightleftharpoons A}}, 0\right)\right\rangle_{\mathbb{S}_{M\rightleftharpoons A}} - \left\langle \vec{b}_{\xi}^{\vec{M}_{M\rightleftharpoons A}}, \vec{b}_{\zeta}^{\vec{J}_{M\rightleftharpoons A}}\times\hat{n}_{\rightarrow A}\right\rangle_{\mathbb{S}_{M\rightleftharpoons A}} \tag{D-51a}$$

$$z_{\xi\zeta}^{\vec{M}_{M\rightleftharpoons A}\vec{J}_{A\rightleftharpoons G}} = \left\langle \vec{b}_{\xi}^{\vec{M}_{M\rightleftharpoons A}}, \mathcal{H}\left(\vec{b}_{\zeta}^{\vec{J}_{A\rightleftharpoons G}}, 0\right)\right\rangle_{\mathbb{S}_{M\rightleftharpoons A}} \tag{D-51b}$$

$$z_{\xi\zeta}^{\vec{M}_{M\rightleftharpoons A}\vec{J}_{A}} = \left\langle \vec{b}_{\xi}^{\vec{M}_{M\rightleftharpoons A}}, \mathcal{H}\left(\vec{b}_{\zeta}^{\vec{J}_{A}}, 0\right)\right\rangle_{\mathbb{S}_{M\rightleftharpoons A}} \tag{D-51c}$$

$$z_{\xi\zeta}^{\vec{M}_{M\rightleftharpoons A}\vec{M}_{M\rightleftharpoons A}} = \left\langle \vec{b}_{\xi}^{\vec{M}_{M\rightleftharpoons A}}, \mathcal{H}\left(0, \vec{b}_{\zeta}^{\vec{M}_{M\rightleftharpoons A}}\right)\right\rangle_{\mathbb{S}_{M\rightleftharpoons A}} \tag{D-51d}$$

$$z_{\xi\zeta}^{\vec{M}_{\mathrm{M}\rightleftharpoons\mathrm{A}}\vec{M}_{\mathrm{A}\rightleftharpoons\mathrm{G}}} = \left\langle \vec{b}_{\xi}^{\vec{M}_{\mathrm{M}\rightleftharpoons\mathrm{A}}}, \mathcal{H}\left(0, \vec{b}_{\zeta}^{\vec{M}_{\mathrm{A}\rightleftharpoons\mathrm{G}}}\right)\right\rangle_{\underset{\sim}{\mathbb{S}}_{\mathrm{M}\rightleftharpoons\mathrm{A}}} \tag{D-51e}$$

and

$$z_{\xi\zeta}^{\vec{J}_{\mathrm{M}\rightleftharpoons\mathrm{A}}\vec{J}_{\mathrm{M}\rightleftharpoons\mathrm{A}}} = \left\langle \vec{b}_{\xi}^{\vec{J}_{\mathrm{M}\rightleftharpoons\mathrm{A}}}, \mathcal{E}\left(\vec{b}_{\zeta}^{\vec{J}_{\mathrm{M}\rightleftharpoons\mathrm{A}}}, 0\right)\right\rangle_{\underset{\sim}{\mathbb{S}}_{\mathrm{M}\rightleftharpoons\mathrm{A}}} \tag{D-52a}$$

$$z_{\xi\zeta}^{\vec{J}_{\mathrm{M}\rightleftharpoons\mathrm{A}}\vec{J}_{\mathrm{A}\rightleftharpoons\mathrm{G}}} = \left\langle \vec{b}_{\xi}^{\vec{J}_{\mathrm{M}\rightleftharpoons\mathrm{A}}}, \mathcal{E}\left(\vec{b}_{\zeta}^{\vec{J}_{\mathrm{A}\rightleftharpoons\mathrm{G}}}, 0\right)\right\rangle_{\underset{\sim}{\mathbb{S}}_{\mathrm{M}\rightleftharpoons\mathrm{A}}} \tag{D-52b}$$

$$z_{\xi\zeta}^{\vec{J}_{\mathrm{M}\rightleftharpoons\mathrm{A}}\vec{J}_{\mathrm{A}}} = \left\langle \vec{b}_{\xi}^{\vec{J}_{\mathrm{M}\rightleftharpoons\mathrm{A}}}, \mathcal{E}\left(\vec{b}_{\zeta}^{\vec{J}_{\mathrm{A}}}, 0\right)\right\rangle_{\underset{\sim}{\mathbb{S}}_{\mathrm{M}\rightleftharpoons\mathrm{A}}} \tag{D-52c}$$

$$z_{\xi\zeta}^{\vec{J}_{\mathrm{M}\rightleftharpoons\mathrm{A}}\vec{M}_{\mathrm{M}\rightleftharpoons\mathrm{A}}} = \left\langle \vec{b}_{\xi}^{\vec{J}_{\mathrm{M}\rightleftharpoons\mathrm{A}}}, \mathcal{E}\left(0, \vec{b}_{\zeta}^{\vec{M}_{\mathrm{M}\rightleftharpoons\mathrm{A}}}\right)\right\rangle_{\underset{\sim}{\mathbb{S}}_{\mathrm{M}\rightleftharpoons\mathrm{A}}} - \left\langle \vec{b}_{\xi}^{\vec{J}_{\mathrm{M}\rightleftharpoons\mathrm{A}}}, \hat{n}_{\rightarrow\mathrm{A}} \times \vec{b}_{\zeta}^{\vec{M}_{\mathrm{M}\rightleftharpoons\mathrm{A}}}\right\rangle_{\mathbb{S}_{\mathrm{M}\rightleftharpoons\mathrm{A}}} \tag{D-52d}$$

$$z_{\xi\zeta}^{\vec{J}_{\mathrm{M}\rightleftharpoons\mathrm{A}}\vec{M}_{\mathrm{A}\rightleftharpoons\mathrm{G}}} = \left\langle \vec{b}_{\xi}^{\vec{J}_{\mathrm{M}\rightleftharpoons\mathrm{A}}}, \mathcal{E}\left(0, \vec{b}_{\zeta}^{\vec{M}_{\mathrm{A}\rightleftharpoons\mathrm{G}}}\right)\right\rangle_{\underset{\sim}{\mathbb{S}}_{\mathrm{M}\rightleftharpoons\mathrm{A}}} \tag{D-52e}$$

and

$$z_{\xi\zeta}^{\vec{M}_{\mathrm{A}\rightleftharpoons\mathrm{G}}\vec{J}_{\mathrm{M}\rightleftharpoons\mathrm{A}}} = \left\langle \vec{b}_{\xi}^{\vec{M}_{\mathrm{A}\rightleftharpoons\mathrm{G}}}, \mathcal{H}\left(\vec{b}_{\zeta}^{\vec{J}_{\mathrm{M}\rightleftharpoons\mathrm{A}}}, 0\right)\right\rangle_{\tilde{\mathbb{S}}_{\mathrm{A}\rightleftharpoons\mathrm{G}}} \tag{D-53a}$$

$$z_{\xi\zeta}^{\vec{M}_{\mathrm{A}\rightleftharpoons\mathrm{G}}\vec{J}_{\mathrm{A}\rightleftharpoons\mathrm{G}}} = \left\langle \vec{b}_{\xi}^{\vec{M}_{\mathrm{A}\rightleftharpoons\mathrm{G}}}, \mathcal{H}\left(\vec{b}_{\zeta}^{\vec{J}_{\mathrm{A}\rightleftharpoons\mathrm{G}}}, 0\right)\right\rangle_{\tilde{\mathbb{S}}_{\mathrm{A}\rightleftharpoons\mathrm{G}}} - \left\langle \vec{b}_{\xi}^{\vec{M}_{\mathrm{A}\rightleftharpoons\mathrm{G}}}, \vec{b}_{\zeta}^{\vec{J}_{\mathrm{A}\rightleftharpoons\mathrm{G}}} \times \hat{n}_{\rightarrow\mathrm{A}}\right\rangle_{\mathbb{S}_{\mathrm{A}\rightleftharpoons\mathrm{G}}} \tag{D-53b}$$

$$z_{\xi\zeta}^{\vec{M}_{\mathrm{A}\rightleftharpoons\mathrm{G}}\vec{J}_{\mathrm{A}}} = \left\langle \vec{b}_{\xi}^{\vec{M}_{\mathrm{A}\rightleftharpoons\mathrm{G}}}, \mathcal{H}\left(\vec{b}_{\zeta}^{\vec{J}_{\mathrm{A}}}, 0\right)\right\rangle_{\tilde{\mathbb{S}}_{\mathrm{A}\rightleftharpoons\mathrm{G}}} \tag{D-53c}$$

$$z_{\xi\zeta}^{\vec{M}_{\mathrm{A}\rightleftharpoons\mathrm{G}}\vec{M}_{\mathrm{M}\rightleftharpoons\mathrm{A}}} = \left\langle \vec{b}_{\xi}^{\vec{M}_{\mathrm{A}\rightleftharpoons\mathrm{G}}}, \mathcal{H}\left(0, \vec{b}_{\zeta}^{\vec{M}_{\mathrm{M}\rightleftharpoons\mathrm{A}}}\right)\right\rangle_{\tilde{\mathbb{S}}_{\mathrm{A}\rightleftharpoons\mathrm{G}}} \tag{D-53d}$$

$$z_{\xi\zeta}^{\vec{M}_{\mathrm{A}\rightleftharpoons\mathrm{G}}\vec{M}_{\mathrm{A}\rightleftharpoons\mathrm{G}}} = \left\langle \vec{b}_{\xi}^{\vec{M}_{\mathrm{A}\rightleftharpoons\mathrm{G}}}, \mathcal{H}\left(0, \vec{b}_{\zeta}^{\vec{M}_{\mathrm{A}\rightleftharpoons\mathrm{G}}}\right)\right\rangle_{\tilde{\mathbb{S}}_{\mathrm{A}\rightleftharpoons\mathrm{G}}} \tag{D-53e}$$

and

$$z_{\xi\zeta}^{\vec{J}_{\mathrm{A}\rightleftharpoons\mathrm{G}}\vec{J}_{\mathrm{M}\rightleftharpoons\mathrm{A}}} = \left\langle \vec{b}_{\xi}^{\vec{J}_{\mathrm{A}\rightleftharpoons\mathrm{G}}}, \mathcal{E}\left(\vec{b}_{\zeta}^{\vec{J}_{\mathrm{M}\rightleftharpoons\mathrm{A}}}, 0\right)\right\rangle_{\tilde{\mathbb{S}}_{\mathrm{A}\rightleftharpoons\mathrm{G}}} \tag{D-54a}$$

$$z_{\xi\zeta}^{\vec{J}_{\mathrm{A}\rightleftharpoons\mathrm{G}}\vec{J}_{\mathrm{A}\rightleftharpoons\mathrm{G}}} = \left\langle \vec{b}_{\xi}^{\vec{J}_{\mathrm{A}\rightleftharpoons\mathrm{G}}}, \mathcal{E}\left(\vec{b}_{\zeta}^{\vec{J}_{\mathrm{A}\rightleftharpoons\mathrm{G}}}, 0\right)\right\rangle_{\tilde{\mathbb{S}}_{\mathrm{A}\rightleftharpoons\mathrm{G}}} \tag{D-54b}$$

$$z_{\xi\zeta}^{\vec{J}_{\mathrm{A}\rightleftharpoons\mathrm{G}}\vec{J}_{\mathrm{A}}} = \left\langle \vec{b}_{\xi}^{\vec{J}_{\mathrm{A}\rightleftharpoons\mathrm{G}}}, \mathcal{E}\left(\vec{b}_{\zeta}^{\vec{J}_{\mathrm{A}}}, 0\right)\right\rangle_{\tilde{\mathbb{S}}_{\mathrm{A}\rightleftharpoons\mathrm{G}}} \tag{D-54c}$$

$$z_{\xi\zeta}^{\vec{J}_{\mathrm{A}\rightleftharpoons\mathrm{G}}\vec{M}_{\mathrm{M}\rightleftharpoons\mathrm{A}}} = \left\langle \vec{b}_{\xi}^{\vec{J}_{\mathrm{A}\rightleftharpoons\mathrm{G}}}, \mathcal{E}\left(0, \vec{b}_{\zeta}^{\vec{M}_{\mathrm{M}\rightleftharpoons\mathrm{A}}}\right)\right\rangle_{\tilde{\mathbb{S}}_{\mathrm{A}\rightleftharpoons\mathrm{G}}} \tag{D-54d}$$

$$z_{\xi\zeta}^{\vec{J}_{\mathrm{A}\rightleftharpoons\mathrm{G}}\vec{M}_{\mathrm{A}\rightleftharpoons\mathrm{G}}} = \left\langle \vec{b}_{\xi}^{\vec{J}_{\mathrm{A}\rightleftharpoons\mathrm{G}}}, \mathcal{E}\left(0, \vec{b}_{\zeta}^{\vec{M}_{\mathrm{A}\rightleftharpoons\mathrm{G}}}\right)\right\rangle_{\tilde{\mathbb{S}}_{\mathrm{A}\rightleftharpoons\mathrm{G}}} - \left\langle \vec{b}_{\xi}^{\vec{J}_{\mathrm{A}\rightleftharpoons\mathrm{G}}}, \hat{n}_{\rightarrow\mathrm{A}} \times \vec{b}_{\zeta}^{\vec{M}_{\mathrm{A}\rightleftharpoons\mathrm{G}}}\right\rangle_{\mathbb{S}_{\mathrm{A}\rightleftharpoons\mathrm{G}}} \tag{D-54e}$$

and

$$z_{\xi\zeta}^{\vec{J}_{\mathrm{A}}\vec{J}_{\mathrm{M}\rightleftharpoons\mathrm{A}}} = \left\langle \vec{b}_{\xi}^{\vec{J}_{\mathrm{A}}}, \mathcal{E}\left(\vec{b}_{\zeta}^{\vec{J}_{\mathrm{M}\rightleftharpoons\mathrm{A}}}, 0\right)\right\rangle_{\tilde{\mathbb{S}}_{\mathrm{A}}} \tag{D-55a}$$

$$z_{\xi\zeta}^{\vec{J}_{\mathrm{A}}\vec{J}_{\mathrm{A}\rightleftharpoons\mathrm{G}}} = \left\langle \vec{b}_{\xi}^{\vec{J}_{\mathrm{A}}}, \mathcal{E}\left(\vec{b}_{\zeta}^{\vec{J}_{\mathrm{A}\rightleftharpoons\mathrm{G}}}, 0\right)\right\rangle_{\tilde{\mathbb{S}}_{\mathrm{A}}} \tag{D-55b}$$

$$z_{\xi\zeta}^{\vec{J}_{\mathrm{A}}\vec{J}_{\mathrm{A}}} = \left\langle \vec{b}_{\xi}^{\vec{J}_{\mathrm{A}}}, \mathcal{E}\left(\vec{b}_{\zeta}^{\vec{J}_{\mathrm{A}}}, 0\right)\right\rangle_{\tilde{\mathbb{S}}_{\mathrm{A}}} \tag{D-55c}$$

$$z_{\xi\zeta}^{\vec{J}_{\mathrm{A}}\vec{M}_{\mathrm{M}\rightleftharpoons\mathrm{A}}} = \left\langle \vec{b}_{\xi}^{\vec{J}_{\mathrm{A}}}, \mathcal{E}\left(0, \vec{b}_{\zeta}^{\vec{M}_{\mathrm{M}\rightleftharpoons\mathrm{A}}}\right)\right\rangle_{\tilde{\mathbb{S}}_{\mathrm{A}}} \tag{D-55d}$$

$$z_{\xi\zeta}^{\vec{J}_{\mathrm{A}}\vec{M}_{\mathrm{A}\rightleftharpoons\mathrm{G}}} = \left\langle \vec{b}_{\xi}^{\vec{J}_{\mathrm{A}}}, \mathcal{E}\left(0, \vec{b}_{\zeta}^{\vec{M}_{\mathrm{A}\rightleftharpoons\mathrm{G}}}\right)\right\rangle_{\tilde{\mathbb{S}}_{\mathrm{A}}} \tag{D-55e}$$

where the various integral surfaces with "~" are the ones located at the antenna side.

The transformation matrix $\overline{\overline{T}}$ used in Eq. (4-93) is as follows:

$$\overline{\overline{T}} = \overline{\overline{T}}^{\vec{J}_{\mathrm{M}\rightleftharpoons\mathrm{A}}\rightarrow\mathrm{AV}} \text{ or } \overline{\overline{T}}^{\vec{M}_{\mathrm{M}\rightleftharpoons\mathrm{A}}\rightarrow\mathrm{AV}} \text{ or } \overline{\overline{T}}^{\mathrm{BS}\rightarrow\mathrm{AV}} \tag{D-56}$$

in which

$$\overline{\overline{T}}^{\vec{J}_{\mathrm{M}\rightleftharpoons\mathrm{A}}\rightarrow\mathrm{AV}} = \left(\overline{\overline{\Psi}}_1\right)^{-1} \cdot \overline{\overline{\Psi}}_2 \tag{D-57a}$$

$$\overline{\overline{T}}^{\vec{M}_{\mathrm{M}\rightleftharpoons\mathrm{A}}\rightarrow\mathrm{AV}} = \left(\overline{\overline{\Psi}}_3\right)^{-1} \cdot \overline{\overline{\Psi}}_4 \tag{D-57b}$$

$$\overline{\overline{T}}^{\mathrm{BS}\rightarrow\mathrm{AV}} = \mathrm{nullspace}\left(\overline{\overline{\Psi}}_{\mathrm{FCE}}^{\mathrm{DoJ/DoM}}\right) \tag{D-58}$$

where

$$\overline{\overline{\Psi}}_1 = \begin{bmatrix} \overline{\overline{I}}^{\vec{J}_{\mathrm{M}\rightleftharpoons\mathrm{A}}} & 0 & 0 & 0 & 0 \\ 0 & \overline{\overline{Z}}^{\vec{M}_{\mathrm{M}\rightleftharpoons\mathrm{A}}\vec{J}_{\mathrm{A}\rightleftharpoons\mathrm{G}}} & \overline{\overline{Z}}^{\vec{M}_{\mathrm{M}\rightleftharpoons\mathrm{A}}\vec{J}_{\mathrm{A}}} & \overline{\overline{Z}}^{\vec{M}_{\mathrm{M}\rightleftharpoons\mathrm{A}}\vec{M}_{\mathrm{M}\rightleftharpoons\mathrm{A}}} & \overline{\overline{Z}}^{\vec{M}_{\mathrm{M}\rightleftharpoons\mathrm{A}}\vec{M}_{\mathrm{A}\rightleftharpoons\mathrm{G}}} \\ 0 & \overline{\overline{Z}}^{\vec{M}_{\mathrm{A}\rightleftharpoons\mathrm{G}}\vec{J}_{\mathrm{A}\rightleftharpoons\mathrm{G}}} & \overline{\overline{Z}}^{\vec{M}_{\mathrm{A}\rightleftharpoons\mathrm{G}}\vec{J}_{\mathrm{A}}} & \overline{\overline{Z}}^{\vec{M}_{\mathrm{A}\rightleftharpoons\mathrm{G}}\vec{M}_{\mathrm{M}\rightleftharpoons\mathrm{A}}} & \overline{\overline{Z}}^{\vec{M}_{\mathrm{A}\rightleftharpoons\mathrm{G}}\vec{M}_{\mathrm{A}\rightleftharpoons\mathrm{G}}} \\ 0 & \overline{\overline{Z}}^{\vec{J}_{\mathrm{A}\rightleftharpoons\mathrm{G}}\vec{J}_{\mathrm{A}\rightleftharpoons\mathrm{G}}} & \overline{\overline{Z}}^{\vec{J}_{\mathrm{A}\rightleftharpoons\mathrm{G}}\vec{J}_{\mathrm{A}}} & \overline{\overline{Z}}^{\vec{J}_{\mathrm{A}\rightleftharpoons\mathrm{G}}\vec{M}_{\mathrm{M}\rightleftharpoons\mathrm{A}}} & \overline{\overline{Z}}^{\vec{J}_{\mathrm{A}\rightleftharpoons\mathrm{G}}\vec{M}_{\mathrm{A}\rightleftharpoons\mathrm{G}}} \\ 0 & \overline{\overline{Z}}^{\vec{J}_{\mathrm{A}}\vec{J}_{\mathrm{A}\rightleftharpoons\mathrm{G}}} & \overline{\overline{Z}}^{\vec{J}_{\mathrm{A}}\vec{J}_{\mathrm{A}}} & \overline{\overline{Z}}^{\vec{J}_{\mathrm{A}}\vec{M}_{\mathrm{M}\rightleftharpoons\mathrm{A}}} & \overline{\overline{Z}}^{\vec{J}_{\mathrm{A}}\vec{M}_{\mathrm{A}\rightleftharpoons\mathrm{G}}} \end{bmatrix} \tag{D-59a}$$

$$\overline{\overline{\Psi}}_2 = \begin{bmatrix} \overline{\overline{I}}^{\vec{J}_{\mathrm{M}\rightleftharpoons\mathrm{A}}} \\ -\overline{\overline{Z}}^{\vec{M}_{\mathrm{M}\rightleftharpoons\mathrm{A}}\vec{J}_{\mathrm{M}\rightleftharpoons\mathrm{A}}} \\ -\overline{\overline{Z}}^{\vec{M}_{\mathrm{A}\rightleftharpoons\mathrm{G}}\vec{J}_{\mathrm{M}\rightleftharpoons\mathrm{A}}} \\ -\overline{\overline{Z}}^{\vec{J}_{\mathrm{A}\rightleftharpoons\mathrm{G}}\vec{J}_{\mathrm{M}\rightleftharpoons\mathrm{A}}} \\ -\overline{\overline{Z}}^{\vec{J}_{\mathrm{A}}\vec{J}_{\mathrm{M}\rightleftharpoons\mathrm{A}}} \end{bmatrix} \tag{D-59b}$$

and

$$\overline{\overline{\Psi}}_3 = \begin{bmatrix} 0 & 0 & 0 & \overline{\overline{I}}^{\vec{M}_{\mathrm{M}\rightleftharpoons\mathrm{A}}} & 0 \\ \overline{\overline{Z}}^{\vec{J}_{\mathrm{M}\rightleftharpoons\mathrm{A}}\vec{J}_{\mathrm{M}\rightleftharpoons\mathrm{A}}} & \overline{\overline{Z}}^{\vec{J}_{\mathrm{M}\rightleftharpoons\mathrm{A}}\vec{J}_{\mathrm{A}\rightleftharpoons\mathrm{G}}} & \overline{\overline{Z}}^{\vec{J}_{\mathrm{M}\rightleftharpoons\mathrm{A}}\vec{J}_{\mathrm{A}}} & 0 & \overline{\overline{Z}}^{\vec{J}_{\mathrm{M}\rightleftharpoons\mathrm{A}}\vec{M}_{\mathrm{A}\rightleftharpoons\mathrm{G}}} \\ \overline{\overline{Z}}^{\vec{M}_{\mathrm{A}\rightleftharpoons\mathrm{G}}\vec{J}_{\mathrm{M}\rightleftharpoons\mathrm{A}}} & \overline{\overline{Z}}^{\vec{M}_{\mathrm{A}\rightleftharpoons\mathrm{G}}\vec{J}_{\mathrm{A}\rightleftharpoons\mathrm{G}}} & \overline{\overline{Z}}^{\vec{M}_{\mathrm{A}\rightleftharpoons\mathrm{G}}\vec{J}_{\mathrm{A}}} & 0 & \overline{\overline{Z}}^{\vec{M}_{\mathrm{A}\rightleftharpoons\mathrm{G}}\vec{M}_{\mathrm{A}\rightleftharpoons\mathrm{G}}} \\ \overline{\overline{Z}}^{\vec{J}_{\mathrm{A}\rightleftharpoons\mathrm{G}}\vec{J}_{\mathrm{M}\rightleftharpoons\mathrm{A}}} & \overline{\overline{Z}}^{\vec{J}_{\mathrm{A}\rightleftharpoons\mathrm{G}}\vec{J}_{\mathrm{A}\rightleftharpoons\mathrm{G}}} & \overline{\overline{Z}}^{\vec{J}_{\mathrm{A}\rightleftharpoons\mathrm{G}}\vec{J}_{\mathrm{A}}} & 0 & \overline{\overline{Z}}^{\vec{J}_{\mathrm{A}\rightleftharpoons\mathrm{G}}\vec{M}_{\mathrm{A}\rightleftharpoons\mathrm{G}}} \\ \overline{\overline{Z}}^{\vec{J}_{\mathrm{A}}\vec{J}_{\mathrm{M}\rightleftharpoons\mathrm{A}}} & \overline{\overline{Z}}^{\vec{J}_{\mathrm{A}}\vec{J}_{\mathrm{A}\rightleftharpoons\mathrm{G}}} & \overline{\overline{Z}}^{\vec{J}_{\mathrm{A}}\vec{J}_{\mathrm{A}}} & 0 & \overline{\overline{Z}}^{\vec{J}_{\mathrm{A}}\vec{M}_{\mathrm{A}\rightleftharpoons\mathrm{G}}} \end{bmatrix} \tag{D-60a}$$

$$\overline{\overline{\Psi}}_4 = \begin{bmatrix} \overline{\overline{I}}^{\vec{M}_{\mathrm{M}\rightleftharpoons\mathrm{A}}} \\ -\overline{\overline{Z}}^{\vec{J}_{\mathrm{M}\rightleftharpoons\mathrm{A}}\vec{M}_{\mathrm{M}\rightleftharpoons\mathrm{A}}} \\ -\overline{\overline{Z}}^{\vec{M}_{\mathrm{A}\rightleftharpoons\mathrm{G}}\vec{M}_{\mathrm{M}\rightleftharpoons\mathrm{A}}} \\ -\overline{\overline{Z}}^{\vec{J}_{\mathrm{A}\rightleftharpoons\mathrm{G}}\vec{M}_{\mathrm{M}\rightleftharpoons\mathrm{A}}} \\ -\overline{\overline{Z}}^{\vec{J}_{\mathrm{A}}\vec{M}_{\mathrm{M}\rightleftharpoons\mathrm{A}}} \end{bmatrix} \tag{D-60b}$$

and

$$\overline{\overline{\Psi}}{}_{\mathrm{FCE}}^{\mathrm{DoJ}} = \begin{bmatrix} \overline{\overline{Z}}{}^{\vec{M}_{\mathrm{M\rightleftharpoons A}}\vec{J}_{\mathrm{M\rightleftharpoons A}}} & \overline{\overline{Z}}{}^{\vec{M}_{\mathrm{M\rightleftharpoons A}}\vec{J}_{\mathrm{A\rightleftharpoons G}}} & \overline{\overline{Z}}{}^{\vec{M}_{\mathrm{M\rightleftharpoons A}}\vec{J}_{\mathrm{A}}} & \overline{\overline{Z}}{}^{\vec{M}_{\mathrm{M\rightleftharpoons A}}\vec{M}_{\mathrm{M\rightleftharpoons A}}} & \overline{\overline{Z}}{}^{\vec{M}_{\mathrm{M\rightleftharpoons A}}\vec{M}_{\mathrm{A\rightleftharpoons G}}} \\ \overline{\overline{Z}}{}^{\vec{M}_{\mathrm{A\rightleftharpoons G}}\vec{J}_{\mathrm{M\rightleftharpoons A}}} & \overline{\overline{Z}}{}^{\vec{M}_{\mathrm{A\rightleftharpoons G}}\vec{J}_{\mathrm{A\rightleftharpoons G}}} & \overline{\overline{Z}}{}^{\vec{M}_{\mathrm{A\rightleftharpoons G}}\vec{J}_{\mathrm{A}}} & \overline{\overline{Z}}{}^{\vec{M}_{\mathrm{A\rightleftharpoons G}}\vec{M}_{\mathrm{M\rightleftharpoons A}}} & \overline{\overline{Z}}{}^{\vec{M}_{\mathrm{A\rightleftharpoons G}}\vec{M}_{\mathrm{A\rightleftharpoons G}}} \\ \overline{\overline{Z}}{}^{\vec{J}_{\mathrm{A\rightleftharpoons G}}\vec{J}_{\mathrm{M\rightleftharpoons A}}} & \overline{\overline{Z}}{}^{\vec{J}_{\mathrm{A\rightleftharpoons G}}\vec{J}_{\mathrm{A\rightleftharpoons G}}} & \overline{\overline{Z}}{}^{\vec{J}_{\mathrm{A\rightleftharpoons G}}\vec{J}_{\mathrm{A}}} & \overline{\overline{Z}}{}^{\vec{J}_{\mathrm{A\rightleftharpoons G}}\vec{M}_{\mathrm{M\rightleftharpoons A}}} & \overline{\overline{Z}}{}^{\vec{J}_{\mathrm{A\rightleftharpoons G}}\vec{M}_{\mathrm{A\rightleftharpoons G}}} \\ \overline{\overline{Z}}{}^{\vec{J}_{\mathrm{A}}\vec{J}_{\mathrm{M\rightleftharpoons A}}} & \overline{\overline{Z}}{}^{\vec{J}_{\mathrm{A}}\vec{J}_{\mathrm{A\rightleftharpoons G}}} & \overline{\overline{Z}}{}^{\vec{J}_{\mathrm{A}}\vec{J}_{\mathrm{A}}} & \overline{\overline{Z}}{}^{\vec{J}_{\mathrm{A}}\vec{M}_{\mathrm{M\rightleftharpoons A}}} & \overline{\overline{Z}}{}^{\vec{J}_{\mathrm{A}}\vec{M}_{\mathrm{A\rightleftharpoons G}}} \end{bmatrix} \tag{D-61a}$$

$$\overline{\overline{\Psi}}{}_{\mathrm{FCE}}^{\mathrm{DoM}} = \begin{bmatrix} \overline{\overline{Z}}{}^{\vec{J}_{\mathrm{M\rightleftharpoons A}}\vec{J}_{\mathrm{M\rightleftharpoons A}}} & \overline{\overline{Z}}{}^{\vec{J}_{\mathrm{M\rightleftharpoons A}}\vec{J}_{\mathrm{A\rightleftharpoons G}}} & \overline{\overline{Z}}{}^{\vec{J}_{\mathrm{M\rightleftharpoons A}}\vec{J}_{\mathrm{A}}} & \overline{\overline{Z}}{}^{\vec{J}_{\mathrm{M\rightleftharpoons A}}\vec{M}_{\mathrm{M\rightleftharpoons A}}} & \overline{\overline{Z}}{}^{\vec{J}_{\mathrm{M\rightleftharpoons A}}\vec{M}_{\mathrm{A\rightleftharpoons G}}} \\ \overline{\overline{Z}}{}^{\vec{M}_{\mathrm{A\rightleftharpoons G}}\vec{J}_{\mathrm{M\rightleftharpoons A}}} & \overline{\overline{Z}}{}^{\vec{M}_{\mathrm{A\rightleftharpoons G}}\vec{J}_{\mathrm{A\rightleftharpoons G}}} & \overline{\overline{Z}}{}^{\vec{M}_{\mathrm{A\rightleftharpoons G}}\vec{J}_{\mathrm{A}}} & \overline{\overline{Z}}{}^{\vec{M}_{\mathrm{A\rightleftharpoons G}}\vec{M}_{\mathrm{M\rightleftharpoons A}}} & \overline{\overline{Z}}{}^{\vec{M}_{\mathrm{A\rightleftharpoons G}}\vec{M}_{\mathrm{A\rightleftharpoons G}}} \\ \overline{\overline{Z}}{}^{\vec{J}_{\mathrm{A\rightleftharpoons G}}\vec{J}_{\mathrm{M\rightleftharpoons A}}} & \overline{\overline{Z}}{}^{\vec{J}_{\mathrm{A\rightleftharpoons G}}\vec{J}_{\mathrm{A\rightleftharpoons G}}} & \overline{\overline{Z}}{}^{\vec{J}_{\mathrm{A\rightleftharpoons G}}\vec{J}_{\mathrm{A}}} & \overline{\overline{Z}}{}^{\vec{J}_{\mathrm{A\rightleftharpoons G}}\vec{M}_{\mathrm{M\rightleftharpoons A}}} & \overline{\overline{Z}}{}^{\vec{J}_{\mathrm{A\rightleftharpoons G}}\vec{M}_{\mathrm{A\rightleftharpoons G}}} \\ \overline{\overline{Z}}{}^{\vec{J}_{\mathrm{A}}\vec{J}_{\mathrm{M\rightleftharpoons A}}} & \overline{\overline{Z}}{}^{\vec{J}_{\mathrm{A}}\vec{J}_{\mathrm{A\rightleftharpoons G}}} & \overline{\overline{Z}}{}^{\vec{J}_{\mathrm{A}}\vec{J}_{\mathrm{A}}} & \overline{\overline{Z}}{}^{\vec{J}_{\mathrm{A}}\vec{M}_{\mathrm{M\rightleftharpoons A}}} & \overline{\overline{Z}}{}^{\vec{J}_{\mathrm{A}}\vec{M}_{\mathrm{A\rightleftharpoons G}}} \end{bmatrix} \tag{D-61b}$$

The power quadratic form matrix $\overline{\overline{P}}_{\mathrm{M\rightleftharpoons A}}$ used in Eq. (4-97) is as follows:

$$\overline{\overline{P}}_{\mathrm{M\rightleftharpoons A}} = \left(\bar{a}^{\mathrm{AV}}\right)^{\dagger} \cdot \overline{\overline{P}}{}_{\mathrm{M\rightleftharpoons A}}^{\mathrm{curAV}} \cdot \bar{a}^{\mathrm{AV}} = \left(\bar{a}^{\mathrm{AV}}\right)^{\dagger} \cdot \overline{\overline{P}}{}_{\mathrm{M\rightleftharpoons A}}^{\mathrm{intAV}} \cdot \bar{a}^{\mathrm{AV}} \tag{D-62}$$

in which

$$\overline{\overline{P}}{}_{\mathrm{M\rightleftharpoons A}}^{\mathrm{curAV}} = \begin{bmatrix} 0 & 0 & 0 & \overline{\overline{C}}{}^{\vec{J}_{\mathrm{M\rightleftharpoons A}}\vec{M}_{\mathrm{M\rightleftharpoons A}}} & 0 \\ 0 & 0 & 0 & 0 & 0 \\ 0 & 0 & 0 & 0 & 0 \\ 0 & 0 & 0 & 0 & 0 \\ 0 & 0 & 0 & 0 & 0 \end{bmatrix} \tag{D-63}$$

$$\overline{\overline{P}}{}_{\mathrm{M\rightleftharpoons A}}^{\mathrm{intAV}} = \begin{cases} \begin{bmatrix} \overline{\overline{P}}{}^{\vec{J}_{\mathrm{M\rightleftharpoons A}}\vec{J}_{\mathrm{M\rightleftharpoons A}}} & \overline{\overline{P}}{}^{\vec{J}_{\mathrm{M\rightleftharpoons A}}\vec{J}_{\mathrm{A\rightleftharpoons G}}} & \overline{\overline{P}}{}^{\vec{J}_{\mathrm{M\rightleftharpoons A}}\vec{J}_{\mathrm{A}}} & \overline{\overline{P}}{}^{\vec{J}_{\mathrm{M\rightleftharpoons A}}\vec{M}_{\mathrm{M\rightleftharpoons A}}} & \overline{\overline{P}}{}^{\vec{J}_{\mathrm{M\rightleftharpoons A}}\vec{M}_{\mathrm{A\rightleftharpoons G}}} \\ 0 & 0 & 0 & 0 & 0 \\ 0 & 0 & 0 & 0 & 0 \\ 0 & 0 & 0 & 0 & 0 \\ 0 & 0 & 0 & 0 & 0 \end{bmatrix} & \text{for the 2nd equality in Eq. (4-96)} \\ \begin{bmatrix} 0 & 0 & 0 & 0 & 0 \\ 0 & 0 & 0 & 0 & 0 \\ 0 & 0 & 0 & 0 & 0 \\ \overline{\overline{P}}{}^{\vec{M}_{\mathrm{M\rightleftharpoons A}}\vec{J}_{\mathrm{M\rightleftharpoons A}}} & \overline{\overline{P}}{}^{\vec{M}_{\mathrm{M\rightleftharpoons A}}\vec{J}_{\mathrm{A\rightleftharpoons G}}} & \overline{\overline{P}}{}^{\vec{M}_{\mathrm{M\rightleftharpoons A}}\vec{J}_{\mathrm{A}}} & \overline{\overline{P}}{}^{\vec{M}_{\mathrm{M\rightleftharpoons A}}\vec{M}_{\mathrm{M\rightleftharpoons A}}} & \overline{\overline{P}}{}^{\vec{M}_{\mathrm{M\rightleftharpoons A}}\vec{M}_{\mathrm{A\rightleftharpoons G}}} \\ 0 & 0 & 0 & 0 & 0 \end{bmatrix}^{\dagger} & \text{for the 3rd equality in Eq. (4-96)} \end{cases} \tag{D-64}$$

where the elements of the sub-matrices in above Eqs. (D-63) and (D-64) are as follows:

$$c_{\xi\zeta}^{\vec{J}_{\mathrm{M\rightleftharpoons A}}\vec{M}_{\mathrm{M\rightleftharpoons A}}} = (1/2)\left\langle \hat{n}_{\to\mathrm{A}} \times \vec{b}_{\xi}^{\vec{J}_{\mathrm{M\rightleftharpoons A}}}, \vec{b}_{\zeta}^{\vec{M}_{\mathrm{M\rightleftharpoons A}}} \right\rangle_{\mathbb{S}_{\mathrm{M\rightleftharpoons A}}} \tag{D-65}$$

and

$$p_{\xi\zeta}^{\vec{J}_{\mathrm{M\rightleftharpoons A}}\vec{J}_{\mathrm{M\rightleftharpoons A}}} = -(1/2)\left\langle \vec{b}_{\xi}^{\vec{J}_{\mathrm{M\rightleftharpoons A}}}, \mathcal{E}\left(\vec{b}_{\zeta}^{\vec{J}_{\mathrm{M\rightleftharpoons A}}}, 0\right) \right\rangle_{\mathbb{S}_{\mathrm{M\rightleftharpoons A}}} \tag{D-66a}$$

$$p_{\xi\zeta}^{\vec{J}_{\mathrm{M\rightleftharpoons A}}\vec{J}_{\mathrm{A\rightleftharpoons G}}} = -(1/2)\left\langle \vec{b}_{\xi}^{\vec{J}_{\mathrm{M\rightleftharpoons A}}}, \mathcal{E}\left(\vec{b}_{\zeta}^{\vec{J}_{\mathrm{A\rightleftharpoons G}}}, 0\right) \right\rangle_{\mathbb{S}_{\mathrm{M\rightleftharpoons A}}} \tag{D-66b}$$

$$p_{\xi\zeta}^{\vec{J}_{\mathrm{M}\rightleftharpoons\mathrm{A}}\vec{J}_{\mathrm{A}}} = -(1/2)\left\langle \vec{b}_{\xi}^{\vec{J}_{\mathrm{M}\rightleftharpoons\mathrm{A}}}, \mathcal{E}\left(\vec{b}_{\zeta}^{\vec{J}_{\mathrm{A}}}, 0\right)\right\rangle_{\mathbb{S}_{\mathrm{M}\rightleftharpoons\mathrm{A}}} \tag{D-66c}$$

$$p_{\xi\zeta}^{\vec{J}_{\mathrm{M}\rightleftharpoons\mathrm{A}}\vec{M}_{\mathrm{M}\rightleftharpoons\mathrm{A}}} = -(1/2)\left\langle \vec{b}_{\xi}^{\vec{J}_{\mathrm{M}\rightleftharpoons\mathrm{A}}}, \mathcal{E}\left(0, \vec{b}_{\zeta}^{\vec{M}_{\mathrm{M}\rightleftharpoons\mathrm{A}}}\right)\right\rangle_{\mathbb{S}_{\mathrm{M}\rightleftharpoons\mathrm{A}}} \tag{D-66d}$$

$$p_{\xi\zeta}^{\vec{J}_{\mathrm{M}\rightleftharpoons\mathrm{A}}\vec{M}_{\mathrm{A}\rightleftharpoons\mathrm{G}}} = -(1/2)\left\langle \vec{b}_{\xi}^{\vec{J}_{\mathrm{M}\rightleftharpoons\mathrm{A}}}, \mathcal{E}\left(0, \vec{b}_{\zeta}^{\vec{M}_{\mathrm{A}\rightleftharpoons\mathrm{G}}}\right)\right\rangle_{\mathbb{S}_{\mathrm{M}\rightleftharpoons\mathrm{A}}} \tag{D-66e}$$

and

$$p_{\xi\zeta}^{\vec{M}_{\mathrm{M}\rightleftharpoons\mathrm{A}}\vec{J}_{\mathrm{M}\rightleftharpoons\mathrm{A}}} = -(1/2)\left\langle \vec{b}_{\xi}^{\vec{M}_{\mathrm{M}\rightleftharpoons\mathrm{A}}}, \mathcal{H}\left(\vec{b}_{\zeta}^{\vec{J}_{\mathrm{M}\rightleftharpoons\mathrm{A}}}, 0\right)\right\rangle_{\mathbb{S}_{\mathrm{M}\rightleftharpoons\mathrm{A}}} \tag{D-66f}$$

$$p_{\xi\zeta}^{\vec{M}_{\mathrm{M}\rightleftharpoons\mathrm{A}}\vec{J}_{\mathrm{A}\rightleftharpoons\mathrm{G}}} = -(1/2)\left\langle \vec{b}_{\xi}^{\vec{M}_{\mathrm{M}\rightleftharpoons\mathrm{A}}}, \mathcal{H}\left(\vec{b}_{\zeta}^{\vec{J}_{\mathrm{A}\rightleftharpoons\mathrm{G}}}, 0\right)\right\rangle_{\mathbb{S}_{\mathrm{M}\rightleftharpoons\mathrm{A}}} \tag{D-66g}$$

$$p_{\xi\zeta}^{\vec{M}_{\mathrm{M}\rightleftharpoons\mathrm{A}}\vec{J}_{\mathrm{A}}} = -(1/2)\left\langle \vec{b}_{\xi}^{\vec{M}_{\mathrm{M}\rightleftharpoons\mathrm{A}}}, \mathcal{H}\left(\vec{b}_{\zeta}^{\vec{J}_{\mathrm{A}}}, 0\right)\right\rangle_{\mathbb{S}_{\mathrm{M}\rightleftharpoons\mathrm{A}}} \tag{D-66h}$$

$$p_{\xi\zeta}^{\vec{M}_{\mathrm{M}\rightleftharpoons\mathrm{A}}\vec{M}_{\mathrm{M}\rightleftharpoons\mathrm{A}}} = -(1/2)\left\langle \vec{b}_{\xi}^{\vec{M}_{\mathrm{M}\rightleftharpoons\mathrm{A}}}, \mathcal{H}\left(0, \vec{b}_{\zeta}^{\vec{M}_{\mathrm{M}\rightleftharpoons\mathrm{A}}}\right)\right\rangle_{\mathbb{S}_{\mathrm{M}\rightleftharpoons\mathrm{A}}} \tag{D-66i}$$

$$p_{\xi\zeta}^{\vec{M}_{\mathrm{M}\rightleftharpoons\mathrm{A}}\vec{M}_{\mathrm{A}\rightleftharpoons\mathrm{G}}} = -(1/2)\left\langle \vec{b}_{\xi}^{\vec{M}_{\mathrm{M}\rightleftharpoons\mathrm{A}}}, \mathcal{H}\left(0, \vec{b}_{\zeta}^{\vec{M}_{\mathrm{A}\rightleftharpoons\mathrm{G}}}\right)\right\rangle_{\mathbb{S}_{\mathrm{M}\rightleftharpoons\mathrm{A}}} \tag{D-66j}$$

## D6 Some Detailed Formulations Related to Sec. 6.4

The formulations used to calculate the elements of the matrices in Eqs. (6-83a)~(6-89) are as follows:

$$z_{\xi\zeta}^{\vec{M}^{\mathrm{G}\rightleftharpoons\mathrm{A}}\vec{J}^{\mathrm{G}\rightleftharpoons\mathrm{A}}} = \left\langle \vec{b}_{\xi}^{\vec{M}^{\mathrm{G}\rightleftharpoons\mathrm{A}}}, \mathcal{H}_1\left(\vec{b}_{\zeta}^{\vec{J}^{\mathrm{G}\rightleftharpoons\mathrm{A}}}\right)\right\rangle_{\mathbb{S}^{\mathrm{G}\rightleftharpoons\mathrm{A}}} - \left\langle \vec{b}_{\xi}^{\vec{M}^{\mathrm{G}\rightleftharpoons\mathrm{A}}}, \vec{b}_{\zeta}^{\vec{J}^{\mathrm{G}\rightleftharpoons\mathrm{A}}} \times \hat{n}_1^{\rightarrow\mathrm{A}}\right\rangle_{\mathbb{S}^{\mathrm{G}\rightleftharpoons\mathrm{A}}} \tag{D-67a}$$

$$z_{\xi\zeta}^{\vec{M}^{\mathrm{G}\rightleftharpoons\mathrm{A}}\vec{J}_1^{\mathrm{A}\rightleftharpoons\mathrm{F}}} = \left\langle \vec{b}_{\xi}^{\vec{M}^{\mathrm{G}\rightleftharpoons\mathrm{A}}}, \mathcal{H}_1\left(\vec{b}_{\zeta}^{\vec{J}_1^{\mathrm{A}\rightleftharpoons\mathrm{F}}}\right)\right\rangle_{\mathbb{S}^{\mathrm{G}\rightleftharpoons\mathrm{A}}} \tag{D-67b}$$

$$z_{\xi\zeta}^{\vec{M}^{\mathrm{G}\rightleftharpoons\mathrm{A}}\vec{J}_{12}^{\mathrm{A}\rightleftharpoons\mathrm{A}}} = \left\langle \vec{b}_{\xi}^{\vec{M}^{\mathrm{G}\rightleftharpoons\mathrm{A}}}, \mathcal{H}_1\left(\vec{b}_{\zeta}^{\vec{J}_{12}^{\mathrm{A}\rightleftharpoons\mathrm{A}}}\right)\right\rangle_{\mathbb{S}^{\mathrm{G}\rightleftharpoons\mathrm{A}}} \tag{D-67c}$$

$$z_{\xi\zeta}^{\vec{M}^{\mathrm{G}\rightleftharpoons\mathrm{A}}\vec{J}_1^{\mathrm{A}}} = \left\langle \vec{b}_{\xi}^{\vec{M}^{\mathrm{G}\rightleftharpoons\mathrm{A}}}, \mathcal{H}_1\left(\vec{b}_{\zeta}^{\vec{J}_1^{\mathrm{A}}}\right)\right\rangle_{\mathbb{S}^{\mathrm{G}\rightleftharpoons\mathrm{A}}} \tag{D-67d}$$

$$z_{\xi\zeta}^{\vec{M}^{\mathrm{G}\rightleftharpoons\mathrm{A}}\vec{M}^{\mathrm{G}\rightleftharpoons\mathrm{A}}} = \left\langle \vec{b}_{\xi}^{\vec{M}^{\mathrm{G}\rightleftharpoons\mathrm{A}}}, \mathcal{H}_1\left(\vec{b}_{\zeta}^{\vec{M}^{\mathrm{G}\rightleftharpoons\mathrm{A}}}\right)\right\rangle_{\mathbb{S}^{\mathrm{G}\rightleftharpoons\mathrm{A}}} \tag{D-67e}$$

$$z_{\xi\zeta}^{\vec{M}^{\mathrm{G}\rightleftharpoons\mathrm{A}}\vec{M}_1^{\mathrm{A}\rightleftharpoons\mathrm{F}}} = \left\langle \vec{b}_{\xi}^{\vec{M}^{\mathrm{G}\rightleftharpoons\mathrm{A}}}, \mathcal{H}_1\left(\vec{b}_{\zeta}^{\vec{M}_1^{\mathrm{A}\rightleftharpoons\mathrm{F}}}\right)\right\rangle_{\mathbb{S}^{\mathrm{G}\rightleftharpoons\mathrm{A}}} \tag{D-67f}$$

$$z_{\xi\zeta}^{\vec{M}^{\mathrm{G}\rightleftharpoons\mathrm{A}}\vec{M}_{12}^{\mathrm{A}\rightleftharpoons\mathrm{A}}} = \left\langle \vec{b}_{\xi}^{\vec{M}^{\mathrm{G}\rightleftharpoons\mathrm{A}}}, \mathcal{H}_1\left(\vec{b}_{\zeta}^{\vec{M}_{12}^{\mathrm{A}\rightleftharpoons\mathrm{A}}}\right)\right\rangle_{\mathbb{S}^{\mathrm{G}\rightleftharpoons\mathrm{A}}} \tag{D-67g}$$

and

$$z_{\xi\zeta}^{\vec{J}^{\mathrm{G}\rightleftharpoons\mathrm{A}}\vec{J}^{\mathrm{G}\rightleftharpoons\mathrm{A}}} = \left\langle \vec{b}_{\xi}^{\vec{J}^{\mathrm{G}\rightleftharpoons\mathrm{A}}}, \mathcal{E}_1\left(\vec{b}_{\zeta}^{\vec{J}^{\mathrm{G}\rightleftharpoons\mathrm{A}}}\right)\right\rangle_{\mathbb{S}^{\mathrm{G}\rightleftharpoons\mathrm{A}}} \tag{D-68a}$$

$$z_{\xi\zeta}^{\vec{J}^{\mathrm{G}\rightleftharpoons\mathrm{A}}\vec{J}_1^{\mathrm{A}\rightleftharpoons\mathrm{F}}} = \left\langle \vec{b}_{\xi}^{\vec{J}^{\mathrm{G}\rightleftharpoons\mathrm{A}}}, \mathcal{E}_1\left(\vec{b}_{\zeta}^{\vec{J}_1^{\mathrm{A}\rightleftharpoons\mathrm{F}}}\right)\right\rangle_{\mathbb{S}^{\mathrm{G}\rightleftharpoons\mathrm{A}}} \tag{D-68b}$$

$$z_{\xi\zeta}^{\vec{J}^{\mathrm{G}\rightleftharpoons\mathrm{A}}\vec{J}_{12}^{\mathrm{A}\rightleftharpoons\mathrm{A}}} = \left\langle \vec{b}_{\xi}^{\vec{J}^{\mathrm{G}\rightleftharpoons\mathrm{A}}}, \mathcal{E}_1\left(\vec{b}_{\zeta}^{\vec{J}_{12}^{\mathrm{A}\rightleftharpoons\mathrm{A}}}\right)\right\rangle_{\mathbb{S}^{\mathrm{G}\rightleftharpoons\mathrm{A}}} \tag{D-68c}$$

$$z_{\xi\zeta}^{\vec{J}^{\mathrm{G}\rightleftharpoons\mathrm{A}}\vec{J}_{1}^{\mathrm{A}}} = \left\langle \vec{b}_{\xi}^{\vec{J}^{\mathrm{G}\rightleftharpoons\mathrm{A}}}, \mathcal{E}_1\left(\vec{b}_{\zeta}^{\vec{J}_{1}^{\mathrm{A}}}\right)\right\rangle_{\mathbb{S}^{\mathrm{G}\rightleftharpoons\mathrm{A}}} \tag{D-68d}$$

$$z_{\xi\zeta}^{\vec{J}^{\mathrm{G}\rightleftharpoons\mathrm{A}}\vec{M}^{\mathrm{G}\rightleftharpoons\mathrm{A}}} = \left\langle \vec{b}_{\xi}^{\vec{J}^{\mathrm{G}\rightleftharpoons\mathrm{A}}}, \mathcal{E}_1\left(\vec{b}_{\zeta}^{\vec{M}^{\mathrm{G}\rightleftharpoons\mathrm{A}}}\right)\right\rangle_{\mathbb{S}^{\mathrm{G}\rightleftharpoons\mathrm{A}}} - \left\langle \vec{b}_{\xi}^{\vec{J}^{\mathrm{G}\rightleftharpoons\mathrm{A}}}, \hat{n}_1^{\rightarrow\mathrm{A}} \times \vec{b}_{\zeta}^{\vec{M}^{\mathrm{G}\rightleftharpoons\mathrm{A}}}\right\rangle_{\mathbb{S}^{\mathrm{G}\rightleftharpoons\mathrm{A}}} \tag{D-68e}$$

$$z_{\xi\zeta}^{\vec{J}^{\mathrm{G}\rightleftharpoons\mathrm{A}}\vec{M}_{1}^{\mathrm{A}\rightleftharpoons\mathrm{F}}} = \left\langle \vec{b}_{\xi}^{\vec{J}^{\mathrm{G}\rightleftharpoons\mathrm{A}}}, \mathcal{E}_1\left(\vec{b}_{\zeta}^{\vec{M}_{1}^{\mathrm{A}\rightleftharpoons\mathrm{F}}}\right)\right\rangle_{\mathbb{S}^{\mathrm{G}\rightleftharpoons\mathrm{A}}} \tag{D-68f}$$

$$z_{\xi\zeta}^{\vec{J}^{\mathrm{G}\rightleftharpoons\mathrm{A}}\vec{M}_{12}^{\mathrm{A}\rightleftharpoons\mathrm{A}}} = \left\langle \vec{b}_{\xi}^{\vec{J}^{\mathrm{G}\rightleftharpoons\mathrm{A}}}, \mathcal{E}_1\left(\vec{b}_{\zeta}^{\vec{M}_{12}^{\mathrm{A}\rightleftharpoons\mathrm{A}}}\right)\right\rangle_{\mathbb{S}^{\mathrm{G}\rightleftharpoons\mathrm{A}}} \tag{D-68g}$$

and

$$z_{\xi\zeta}^{\vec{J}_1^{\mathrm{A}\rightleftharpoons\mathrm{F}}\vec{J}^{\mathrm{G}\rightleftharpoons\mathrm{A}}} = \left\langle \vec{b}_{\xi}^{\vec{J}_1^{\mathrm{A}\rightleftharpoons\mathrm{F}}}, \mathcal{E}_1\left(\vec{b}_{\zeta}^{\vec{J}^{\mathrm{G}\rightleftharpoons\mathrm{A}}}\right)\right\rangle_{\tilde{\mathbb{S}}_1^{\mathrm{A}\rightleftharpoons\mathrm{F}}} \tag{D-69a}$$

$$z_{\xi\zeta}^{\vec{J}_1^{\mathrm{A}\rightleftharpoons\mathrm{F}}\vec{J}_1^{\mathrm{A}\rightleftharpoons\mathrm{F}}} = \left\langle \vec{b}_{\xi}^{\vec{J}_1^{\mathrm{A}\rightleftharpoons\mathrm{F}}}, \mathcal{E}_1\left(\vec{b}_{\zeta}^{\vec{J}_1^{\mathrm{A}\rightleftharpoons\mathrm{F}}}\right)\right\rangle_{\tilde{\mathbb{S}}_1^{\mathrm{A}\rightleftharpoons\mathrm{F}}} - \left\langle \vec{b}_{\xi}^{\vec{J}_1^{\mathrm{A}\rightleftharpoons\mathrm{F}}}, -j\omega\mu_0\mathcal{L}_0\left(-\vec{b}_{\zeta}^{\vec{J}_1^{\mathrm{A}\rightleftharpoons\mathrm{F}}}\right)\right\rangle_{\mathbb{S}_1^{\mathrm{A}\rightleftharpoons\mathrm{F}}} \tag{D-69b}$$

$$z_{\xi\zeta}^{\vec{J}_1^{\mathrm{A}\rightleftharpoons\mathrm{F}}\vec{J}_{12}^{\mathrm{A}\rightleftharpoons\mathrm{A}}} = \left\langle \vec{b}_{\xi}^{\vec{J}_1^{\mathrm{A}\rightleftharpoons\mathrm{F}}}, \mathcal{E}_1\left(\vec{b}_{\zeta}^{\vec{J}_{12}^{\mathrm{A}\rightleftharpoons\mathrm{A}}}\right)\right\rangle_{\tilde{\mathbb{S}}_1^{\mathrm{A}\rightleftharpoons\mathrm{F}}} \tag{D-69c}$$

$$z_{\xi\zeta}^{\vec{J}_1^{\mathrm{A}\rightleftharpoons\mathrm{F}}\vec{J}_2^{\mathrm{A}\rightleftharpoons\mathrm{F}}} = -\left\langle \vec{b}_{\xi}^{\vec{J}_1^{\mathrm{A}\rightleftharpoons\mathrm{F}}}, -j\omega\mu_0\mathcal{L}_0\left(-\vec{b}_{\zeta}^{\vec{J}_2^{\mathrm{A}\rightleftharpoons\mathrm{F}}}\right)\right\rangle_{\mathbb{S}_1^{\mathrm{A}\rightleftharpoons\mathrm{F}}} \tag{D-69d}$$

$$z_{\xi\zeta}^{\vec{J}_1^{\mathrm{A}\rightleftharpoons\mathrm{F}}\vec{J}_1^{\mathrm{A}}} = \left\langle \vec{b}_{\xi}^{\vec{J}_1^{\mathrm{A}\rightleftharpoons\mathrm{F}}}, \mathcal{E}_1\left(\vec{b}_{\zeta}^{\vec{J}_1^{\mathrm{A}}}\right)\right\rangle_{\tilde{\mathbb{S}}_1^{\mathrm{A}\rightleftharpoons\mathrm{F}}} \tag{D-69e}$$

$$z_{\xi\zeta}^{\vec{J}_1^{\mathrm{A}\rightleftharpoons\mathrm{F}}\vec{J}^{\mathrm{F}}} = -\left\langle \vec{b}_{\xi}^{\vec{J}_1^{\mathrm{A}\rightleftharpoons\mathrm{F}}}, -j\omega\mu_0\mathcal{L}_0\left(\vec{b}_{\zeta}^{\vec{J}^{\mathrm{F}}}\right)\right\rangle_{\mathbb{S}_1^{\mathrm{A}\rightleftharpoons\mathrm{F}}} \tag{D-69f}$$

$$z_{\xi\zeta}^{\vec{J}_1^{\mathrm{A}\rightleftharpoons\mathrm{F}}\vec{M}^{\mathrm{G}\rightleftharpoons\mathrm{A}}} = \left\langle \vec{b}_{\xi}^{\vec{J}_1^{\mathrm{A}\rightleftharpoons\mathrm{F}}}, \mathcal{E}_1\left(\vec{b}_{\zeta}^{\vec{M}^{\mathrm{G}\rightleftharpoons\mathrm{A}}}\right)\right\rangle_{\tilde{\mathbb{S}}_1^{\mathrm{A}\rightleftharpoons\mathrm{F}}} \tag{D-69g}$$

$$z_{\xi\zeta}^{\vec{J}_1^{\mathrm{A}\rightleftharpoons\mathrm{F}}\vec{M}_1^{\mathrm{A}\rightleftharpoons\mathrm{F}}} = \left\langle \vec{b}_{\xi}^{\vec{J}_1^{\mathrm{A}\rightleftharpoons\mathrm{F}}}, \mathcal{E}_1\left(\vec{b}_{\zeta}^{\vec{M}_1^{\mathrm{A}\rightleftharpoons\mathrm{F}}}\right)\right\rangle_{\tilde{\mathbb{S}}_1^{\mathrm{A}\rightleftharpoons\mathrm{F}}} - \left\langle \vec{b}_{\xi}^{\vec{J}_1^{\mathrm{A}\rightleftharpoons\mathrm{F}}}, \frac{1}{2}\vec{b}_{\zeta}^{\vec{M}_1^{\mathrm{A}\rightleftharpoons\mathrm{F}}} \times \hat{n}^{\rightarrow\mathrm{F}} - \mathrm{P.V.}\,\mathcal{K}_0\left(-\vec{b}_{\zeta}^{\vec{M}_1^{\mathrm{A}\rightleftharpoons\mathrm{F}}}\right)\right\rangle_{\mathbb{S}_1^{\mathrm{A}\rightleftharpoons\mathrm{F}}} \tag{D-69h}$$

$$z_{\xi\zeta}^{\vec{J}_1^{\mathrm{A}\rightleftharpoons\mathrm{F}}\vec{M}_{12}^{\mathrm{A}\rightleftharpoons\mathrm{A}}} = \left\langle \vec{b}_{\xi}^{\vec{J}_1^{\mathrm{A}\rightleftharpoons\mathrm{F}}}, \mathcal{E}_1\left(\vec{b}_{\zeta}^{\vec{M}_{12}^{\mathrm{A}\rightleftharpoons\mathrm{A}}}\right)\right\rangle_{\tilde{\mathbb{S}}_1^{\mathrm{A}\rightleftharpoons\mathrm{F}}} \tag{D-69i}$$

$$z_{\xi\zeta}^{\vec{J}_1^{\mathrm{A}\rightleftharpoons\mathrm{F}}\vec{M}_2^{\mathrm{A}\rightleftharpoons\mathrm{F}}} = -\left\langle \vec{b}_{\xi}^{\vec{J}_1^{\mathrm{A}\rightleftharpoons\mathrm{F}}}, -\mathcal{K}_0\left(-\vec{b}_{\zeta}^{\vec{M}_2^{\mathrm{A}\rightleftharpoons\mathrm{F}}}\right)\right\rangle_{\mathbb{S}_1^{\mathrm{A}\rightleftharpoons\mathrm{F}}} \tag{D-69j}$$

and

$$z_{\xi\zeta}^{\vec{M}_1^{\mathrm{A}\rightleftharpoons\mathrm{F}}\vec{J}^{\mathrm{G}\rightleftharpoons\mathrm{A}}} = \left\langle \vec{b}_{\xi}^{\vec{M}_1^{\mathrm{A}\rightleftharpoons\mathrm{F}}}, \mathcal{H}_1\left(\vec{b}_{\zeta}^{\vec{J}^{\mathrm{G}\rightleftharpoons\mathrm{A}}}\right)\right\rangle_{\tilde{\mathbb{S}}_1^{\mathrm{A}\rightleftharpoons\mathrm{F}}} \tag{D-70a}$$

$$z_{\xi\zeta}^{\vec{M}_1^{\mathrm{A}\rightleftharpoons\mathrm{F}}\vec{J}_1^{\mathrm{A}\rightleftharpoons\mathrm{F}}} = \left\langle \vec{b}_{\xi}^{\vec{M}_1^{\mathrm{A}\rightleftharpoons\mathrm{F}}}, \mathcal{H}_1\left(\vec{b}_{\zeta}^{\vec{J}_1^{\mathrm{A}\rightleftharpoons\mathrm{F}}}\right)\right\rangle_{\tilde{\mathbb{S}}_1^{\mathrm{A}\rightleftharpoons\mathrm{F}}} - \left\langle \vec{b}_{\xi}^{\vec{M}_1^{\mathrm{A}\rightleftharpoons\mathrm{F}}}, \hat{n}^{\rightarrow\mathrm{F}} \times \frac{1}{2}\vec{b}_{\zeta}^{\vec{J}_1^{\mathrm{A}\rightleftharpoons\mathrm{F}}} + \mathrm{P.V.}\,\mathcal{K}_0\left(-\vec{b}_{\zeta}^{\vec{J}_1^{\mathrm{A}\rightleftharpoons\mathrm{F}}}\right)\right\rangle_{\mathbb{S}_1^{\mathrm{A}\rightleftharpoons\mathrm{F}}} \tag{D-70b}$$

$$z_{\xi\zeta}^{\vec{M}_1^{\mathrm{A}\rightleftharpoons\mathrm{F}}\vec{J}_{12}^{\mathrm{A}\rightleftharpoons\mathrm{A}}} = \left\langle \vec{b}_{\xi}^{\vec{M}_1^{\mathrm{A}\rightleftharpoons\mathrm{F}}}, \mathcal{H}_1\left(\vec{b}_{\zeta}^{\vec{J}_{12}^{\mathrm{A}\rightleftharpoons\mathrm{A}}}\right)\right\rangle_{\tilde{\mathbb{S}}_1^{\mathrm{A}\rightleftharpoons\mathrm{F}}} \tag{D-70c}$$

$$z_{\xi\zeta}^{\vec{M}_1^{\mathrm{A}\rightleftharpoons\mathrm{F}}\vec{J}_2^{\mathrm{A}\rightleftharpoons\mathrm{F}}} = -\left\langle \vec{b}_{\xi}^{\vec{M}_1^{\mathrm{A}\rightleftharpoons\mathrm{F}}}, \mathcal{K}_0\left(-\vec{b}_{\zeta}^{\vec{J}_2^{\mathrm{A}\rightleftharpoons\mathrm{F}}}\right)\right\rangle_{\mathbb{S}_1^{\mathrm{A}\rightleftharpoons\mathrm{F}}} \tag{D-70d}$$

$$z_{\xi\zeta}^{\vec{M}_1^{\mathrm{A}\rightleftharpoons\mathrm{F}}\vec{J}_1^{\mathrm{A}}} = \left\langle \vec{b}_\xi^{\vec{M}_1^{\mathrm{A}\rightleftharpoons\mathrm{F}}}, \mathcal{H}_1\left(\vec{b}_\zeta^{\vec{J}_1^{\mathrm{A}}}\right)\right\rangle_{\tilde{\mathbb{S}}_1^{\mathrm{A}\rightleftharpoons\mathrm{F}}} \tag{D-70e}$$

$$z_{\xi\zeta}^{\vec{M}_1^{\mathrm{A}\rightleftharpoons\mathrm{F}}\vec{J}^{\mathrm{F}}} = -\left\langle \vec{b}_\xi^{\vec{M}_1^{\mathrm{A}\rightleftharpoons\mathrm{F}}}, \mathcal{K}_0\left(\vec{b}_\zeta^{\vec{J}^{\mathrm{F}}}\right)\right\rangle_{\underset{\sim}{\mathbb{S}}_1^{\mathrm{A}\rightleftharpoons\mathrm{F}}} \tag{D-70f}$$

$$z_{\xi\zeta}^{\vec{M}_1^{\mathrm{A}\rightleftharpoons\mathrm{F}}\vec{M}^{\mathrm{G}\rightleftharpoons\mathrm{A}}} = \left\langle \vec{b}_\xi^{\vec{M}_1^{\mathrm{A}\rightleftharpoons\mathrm{F}}}, \mathcal{H}_1\left(\vec{b}_\zeta^{\vec{M}^{\mathrm{G}\rightleftharpoons\mathrm{A}}}\right)\right\rangle_{\tilde{\mathbb{S}}_1^{\mathrm{A}\rightleftharpoons\mathrm{F}}} \tag{D-70g}$$

$$z_{\xi\zeta}^{\vec{M}_1^{\mathrm{A}\rightleftharpoons\mathrm{F}}\vec{M}_1^{\mathrm{A}\rightleftharpoons\mathrm{F}}} = \left\langle \vec{b}_\xi^{\vec{M}_1^{\mathrm{A}\rightleftharpoons\mathrm{F}}}, \mathcal{H}_1\left(\vec{b}_\zeta^{\vec{M}_1^{\mathrm{A}\rightleftharpoons\mathrm{F}}}\right)\right\rangle_{\tilde{\mathbb{S}}_1^{\mathrm{A}\rightleftharpoons\mathrm{F}}} - \left\langle \vec{b}_\xi^{\vec{M}_1^{\mathrm{A}\rightleftharpoons\mathrm{F}}}, -j\omega\varepsilon_0\mathcal{L}_0\left(-\vec{b}_\zeta^{\vec{M}_1^{\mathrm{A}\rightleftharpoons\mathrm{F}}}\right)\right\rangle_{\underset{\sim}{\mathbb{S}}_1^{\mathrm{A}\rightleftharpoons\mathrm{F}}} \tag{D-70h}$$

$$z_{\xi\zeta}^{\vec{M}_1^{\mathrm{A}\rightleftharpoons\mathrm{F}}\vec{M}_{12}^{\mathrm{A}\rightleftharpoons\mathrm{A}}} = \left\langle \vec{b}_\xi^{\vec{M}_1^{\mathrm{A}\rightleftharpoons\mathrm{F}}}, \mathcal{H}_1\left(\vec{b}_\zeta^{\vec{M}_{12}^{\mathrm{A}\rightleftharpoons\mathrm{A}}}\right)\right\rangle_{\tilde{\mathbb{S}}_1^{\mathrm{A}\rightleftharpoons\mathrm{F}}} \tag{D-70i}$$

$$z_{\xi\zeta}^{\vec{M}_1^{\mathrm{A}\rightleftharpoons\mathrm{F}}\vec{M}_2^{\mathrm{A}\rightleftharpoons\mathrm{F}}} = -\left\langle \vec{b}_\xi^{\vec{M}_1^{\mathrm{A}\rightleftharpoons\mathrm{F}}}, -j\omega\varepsilon_0\mathcal{L}_0\left(-\vec{b}_\zeta^{\vec{M}_2^{\mathrm{A}\rightleftharpoons\mathrm{F}}}\right)\right\rangle_{\underset{\sim}{\mathbb{S}}_1^{\mathrm{A}\rightleftharpoons\mathrm{F}}} \tag{D-70j}$$

and

$$z_{\xi\zeta}^{\vec{J}_{12}^{\mathrm{A}\rightleftharpoons\mathrm{A}}\vec{J}^{\mathrm{G}\rightleftharpoons\mathrm{A}}} = \left\langle \vec{b}_\xi^{\vec{J}_{12}^{\mathrm{A}\rightleftharpoons\mathrm{A}}}, \mathcal{E}_1\left(\vec{b}_\zeta^{\vec{J}^{\mathrm{G}\rightleftharpoons\mathrm{A}}}\right)\right\rangle_{\tilde{\mathbb{S}}_{12}^{\mathrm{A}\rightleftharpoons\mathrm{A}}} \tag{D-71a}$$

$$z_{\xi\zeta}^{\vec{J}_{12}^{\mathrm{A}\rightleftharpoons\mathrm{A}}\vec{J}_1^{\mathrm{A}\rightleftharpoons\mathrm{F}}} = \left\langle \vec{b}_\xi^{\vec{J}_{12}^{\mathrm{A}\rightleftharpoons\mathrm{A}}}, \mathcal{E}_1\left(\vec{b}_\zeta^{\vec{J}_1^{\mathrm{A}\rightleftharpoons\mathrm{F}}}\right)\right\rangle_{\tilde{\mathbb{S}}_{12}^{\mathrm{A}\rightleftharpoons\mathrm{A}}} \tag{D-71b}$$

$$z_{\xi\zeta}^{\vec{J}_{12}^{\mathrm{A}\rightleftharpoons\mathrm{A}}\vec{J}_{12}^{\mathrm{A}\rightleftharpoons\mathrm{A}}} = \left\langle \vec{b}_\xi^{\vec{J}_{12}^{\mathrm{A}\rightleftharpoons\mathrm{A}}}, \mathcal{E}_1\left(\vec{b}_\zeta^{\vec{J}_{12}^{\mathrm{A}\rightleftharpoons\mathrm{A}}}\right)\right\rangle_{\tilde{\mathbb{S}}_{12}^{\mathrm{A}\rightleftharpoons\mathrm{A}}} - \left\langle \vec{b}_\xi^{\vec{J}_{12}^{\mathrm{A}\rightleftharpoons\mathrm{A}}}, \mathcal{E}_2\left(-\vec{b}_\zeta^{\vec{J}_{12}^{\mathrm{A}\rightleftharpoons\mathrm{A}}}\right)\right\rangle_{\underset{\sim}{\mathbb{S}}_{12}^{\mathrm{A}\rightleftharpoons\mathrm{A}}} \tag{D-71c}$$

$$z_{\xi\zeta}^{\vec{J}_{12}^{\mathrm{A}\rightleftharpoons\mathrm{A}}\vec{J}_2^{\mathrm{A}\rightleftharpoons\mathrm{F}}} = -\left\langle \vec{b}_\xi^{\vec{J}_{12}^{\mathrm{A}\rightleftharpoons\mathrm{A}}}, \mathcal{E}_2\left(\vec{b}_\zeta^{\vec{J}_2^{\mathrm{A}\rightleftharpoons\mathrm{F}}}\right)\right\rangle_{\underset{\sim}{\mathbb{S}}_{12}^{\mathrm{A}\rightleftharpoons\mathrm{A}}} \tag{D-71d}$$

$$z_{\xi\zeta}^{\vec{J}_{12}^{\mathrm{A}\rightleftharpoons\mathrm{A}}\vec{J}_1^{\mathrm{A}}} = \left\langle \vec{b}_\xi^{\vec{J}_{12}^{\mathrm{A}\rightleftharpoons\mathrm{A}}}, \mathcal{E}_1\left(\vec{b}_\zeta^{\vec{J}_1^{\mathrm{A}}}\right)\right\rangle_{\tilde{\mathbb{S}}_{12}^{\mathrm{A}\rightleftharpoons\mathrm{A}}} \tag{D-71e}$$

$$z_{\xi\zeta}^{\vec{J}_{12}^{\mathrm{A}\rightleftharpoons\mathrm{A}}\vec{J}_2^{\mathrm{A}}} = -\left\langle \vec{b}_\xi^{\vec{J}_{12}^{\mathrm{A}\rightleftharpoons\mathrm{A}}}, \mathcal{E}_2\left(\vec{b}_\zeta^{\vec{J}_2^{\mathrm{A}}}\right)\right\rangle_{\underset{\sim}{\mathbb{S}}_{12}^{\mathrm{A}\rightleftharpoons\mathrm{A}}} \tag{D-71f}$$

$$z_{\xi\zeta}^{\vec{J}_{12}^{\mathrm{A}\rightleftharpoons\mathrm{A}}\vec{M}^{\mathrm{G}\rightleftharpoons\mathrm{A}}} = \left\langle \vec{b}_\xi^{\vec{J}_{12}^{\mathrm{A}\rightleftharpoons\mathrm{A}}}, \mathcal{E}_1\left(\vec{b}_\zeta^{\vec{M}^{\mathrm{G}\rightleftharpoons\mathrm{A}}}\right)\right\rangle_{\tilde{\mathbb{S}}_{12}^{\mathrm{A}\rightleftharpoons\mathrm{A}}} \tag{D-71g}$$

$$z_{\xi\zeta}^{\vec{J}_{12}^{\mathrm{A}\rightleftharpoons\mathrm{A}}\vec{M}_1^{\mathrm{A}\rightleftharpoons\mathrm{F}}} = \left\langle \vec{b}_\xi^{\vec{J}_{12}^{\mathrm{A}\rightleftharpoons\mathrm{A}}}, \mathcal{E}_1\left(\vec{b}_\zeta^{\vec{M}_1^{\mathrm{A}\rightleftharpoons\mathrm{F}}}\right)\right\rangle_{\tilde{\mathbb{S}}_{12}^{\mathrm{A}\rightleftharpoons\mathrm{A}}} \tag{D-71h}$$

$$z_{\xi\zeta}^{\vec{J}_{12}^{\mathrm{A}\rightleftharpoons\mathrm{A}}\vec{M}_{12}^{\mathrm{A}\rightleftharpoons\mathrm{A}}} = \left\langle \vec{b}_\xi^{\vec{J}_{12}^{\mathrm{A}\rightleftharpoons\mathrm{A}}}, \mathcal{E}_1\left(\vec{b}_\zeta^{\vec{M}_{12}^{\mathrm{A}\rightleftharpoons\mathrm{A}}}\right)\right\rangle_{\tilde{\mathbb{S}}_{12}^{\mathrm{A}\rightleftharpoons\mathrm{A}}} - \left\langle \vec{b}_\xi^{\vec{J}_{12}^{\mathrm{A}\rightleftharpoons\mathrm{A}}}, \mathcal{E}_2\left(-\vec{b}_\zeta^{\vec{M}_{12}^{\mathrm{A}\rightleftharpoons\mathrm{A}}}\right)\right\rangle_{\underset{\sim}{\mathbb{S}}_{12}^{\mathrm{A}\rightleftharpoons\mathrm{A}}} \tag{D-71i}$$

$$z_{\xi\zeta}^{\vec{J}_{12}^{\mathrm{A}\rightleftharpoons\mathrm{A}}\vec{M}_2^{\mathrm{A}\rightleftharpoons\mathrm{F}}} = -\left\langle \vec{b}_\xi^{\vec{J}_{12}^{\mathrm{A}\rightleftharpoons\mathrm{A}}}, \mathcal{E}_2\left(\vec{b}_\zeta^{\vec{M}_2^{\mathrm{A}\rightleftharpoons\mathrm{F}}}\right)\right\rangle_{\underset{\sim}{\mathbb{S}}_{12}^{\mathrm{A}\rightleftharpoons\mathrm{A}}} \tag{D-71j}$$

and

$$z_{\xi\zeta}^{\vec{M}_{12}^{\mathrm{A}\rightleftharpoons\mathrm{A}}\vec{J}^{\mathrm{G}\rightleftharpoons\mathrm{A}}} = \left\langle \vec{b}_\xi^{\vec{M}_{12}^{\mathrm{A}\rightleftharpoons\mathrm{A}}}, \mathcal{H}_1\left(\vec{b}_\zeta^{\vec{J}^{\mathrm{G}\rightleftharpoons\mathrm{A}}}\right)\right\rangle_{\tilde{\mathbb{S}}_{12}^{\mathrm{A}\rightleftharpoons\mathrm{A}}} \tag{D-72a}$$

$$z_{\xi\zeta}^{\vec{M}_{12}^{\mathrm{A}\rightleftharpoons\mathrm{A}}\vec{J}_1^{\mathrm{A}\rightleftharpoons\mathrm{F}}} = \left\langle \vec{b}_\xi^{\vec{M}_{12}^{\mathrm{A}\rightleftharpoons\mathrm{A}}}, \mathcal{H}_1\left(\vec{b}_\zeta^{\vec{J}_1^{\mathrm{A}\rightleftharpoons\mathrm{F}}}\right)\right\rangle_{\tilde{\mathbb{S}}_{12}^{\mathrm{A}\rightleftharpoons\mathrm{A}}} \tag{D-72b}$$

$$z_{\xi\zeta}^{\vec{M}_{12}^{\mathrm{A}\rightleftharpoons\mathrm{A}}\vec{J}_{12}^{\mathrm{A}\rightleftharpoons\mathrm{A}}} = \left\langle \vec{b}_\xi^{\vec{M}_{12}^{\mathrm{A}\rightleftharpoons\mathrm{A}}}, \mathcal{H}_1\left(\vec{b}_\zeta^{\vec{J}_{12}^{\mathrm{A}\rightleftharpoons\mathrm{A}}}\right)\right\rangle_{\tilde{\mathbb{S}}_{12}^{\mathrm{A}\rightleftharpoons\mathrm{A}}} - \left\langle \vec{b}_\xi^{\vec{M}_{12}^{\mathrm{A}\rightleftharpoons\mathrm{A}}}, \mathcal{H}_2\left(-\vec{b}_\zeta^{\vec{J}_{12}^{\mathrm{A}\rightleftharpoons\mathrm{A}}}\right)\right\rangle_{\underset{\sim}{\mathbb{S}}_{12}^{\mathrm{A}\rightleftharpoons\mathrm{A}}} \tag{D-72c}$$

$$z_{\xi\zeta}^{\vec{M}_{12}^{\mathrm{A}\rightleftharpoons\mathrm{A}}\vec{J}_2^{\mathrm{A}\rightleftharpoons\mathrm{F}}} = -\left\langle \vec{b}_\xi^{\vec{M}_{12}^{\mathrm{A}\rightleftharpoons\mathrm{A}}}, \mathcal{H}_2\left(\vec{b}_\zeta^{\vec{J}_2^{\mathrm{A}\rightleftharpoons\mathrm{F}}}\right)\right\rangle_{\underset{\sim}{\mathbb{S}}_{12}^{\mathrm{A}\rightleftharpoons\mathrm{A}}} \tag{D-72d}$$

$$z_{\xi\zeta}^{\vec{M}_{12}^{\mathrm{A}\rightleftharpoons\mathrm{A}}\vec{J}_{1}^{\mathrm{A}}} = \left\langle \vec{b}_{\xi}^{\vec{M}_{12}^{\mathrm{A}\rightleftharpoons\mathrm{A}}}, \mathcal{H}_1\left(\vec{b}_{\zeta}^{\vec{J}_{1}^{\mathrm{A}}}\right)\right\rangle_{\tilde{\mathbb{S}}_{12}^{\mathrm{A}\rightleftharpoons\mathrm{A}}} \tag{D-72e}$$

$$z_{\xi\zeta}^{\vec{M}_{12}^{\mathrm{A}\rightleftharpoons\mathrm{A}}\vec{J}_{2}^{\mathrm{A}}} = -\left\langle \vec{b}_{\xi}^{\vec{M}_{12}^{\mathrm{A}\rightleftharpoons\mathrm{A}}}, \mathcal{H}_2\left(\vec{b}_{\zeta}^{\vec{J}_{2}^{\mathrm{A}}}\right)\right\rangle_{\mathbb{S}_{12}^{\mathrm{A}\rightleftharpoons\mathrm{A}}} \tag{D-72f}$$

$$z_{\xi\zeta}^{\vec{M}_{12}^{\mathrm{A}\rightleftharpoons\mathrm{A}}\vec{M}^{\mathrm{G}\rightleftharpoons\mathrm{A}}} = \left\langle \vec{b}_{\xi}^{\vec{M}_{12}^{\mathrm{A}\rightleftharpoons\mathrm{A}}}, \mathcal{H}_1\left(\vec{b}_{\zeta}^{\vec{M}^{\mathrm{G}\rightleftharpoons\mathrm{A}}}\right)\right\rangle_{\tilde{\mathbb{S}}_{12}^{\mathrm{A}\rightleftharpoons\mathrm{A}}} \tag{D-72g}$$

$$z_{\xi\zeta}^{\vec{M}_{12}^{\mathrm{A}\rightleftharpoons\mathrm{A}}\vec{M}_{1}^{\mathrm{A}\rightleftharpoons\mathrm{F}}} = \left\langle \vec{b}_{\xi}^{\vec{M}_{12}^{\mathrm{A}\rightleftharpoons\mathrm{A}}}, \mathcal{H}_1\left(\vec{b}_{\zeta}^{\vec{M}_{1}^{\mathrm{A}\rightleftharpoons\mathrm{F}}}\right)\right\rangle_{\tilde{\mathbb{S}}_{12}^{\mathrm{A}\rightleftharpoons\mathrm{A}}} \tag{D-72h}$$

$$z_{\xi\zeta}^{\vec{M}_{12}^{\mathrm{A}\rightleftharpoons\mathrm{A}}\vec{M}_{12}^{\mathrm{A}\rightleftharpoons\mathrm{A}}} = \left\langle \vec{b}_{\xi}^{\vec{M}_{12}^{\mathrm{A}\rightleftharpoons\mathrm{A}}}, \mathcal{H}_1\left(\vec{b}_{\zeta}^{\vec{M}_{12}^{\mathrm{A}\rightleftharpoons\mathrm{A}}}\right)\right\rangle_{\tilde{\mathbb{S}}_{12}^{\mathrm{A}\rightleftharpoons\mathrm{A}}} - \left\langle \vec{b}_{\xi}^{\vec{M}_{12}^{\mathrm{A}\rightleftharpoons\mathrm{A}}}, \mathcal{H}_2\left(-\vec{b}_{\zeta}^{\vec{M}_{12}^{\mathrm{A}\rightleftharpoons\mathrm{A}}}\right)\right\rangle_{\mathbb{S}_{12}^{\mathrm{A}\rightleftharpoons\mathrm{A}}} \tag{D-72i}$$

$$z_{\xi\zeta}^{\vec{M}_{12}^{\mathrm{A}\rightleftharpoons\mathrm{A}}\vec{M}_{2}^{\mathrm{A}\rightleftharpoons\mathrm{F}}} = -\left\langle \vec{b}_{\xi}^{\vec{M}_{12}^{\mathrm{A}\rightleftharpoons\mathrm{A}}}, \mathcal{H}_2\left(\vec{b}_{\zeta}^{\vec{M}_{2}^{\mathrm{A}\rightleftharpoons\mathrm{F}}}\right)\right\rangle_{\mathbb{S}_{12}^{\mathrm{A}\rightleftharpoons\mathrm{A}}} \tag{D-72j}$$

and

$$z_{\xi\zeta}^{\vec{J}_{2}^{\mathrm{A}\rightleftharpoons\mathrm{F}}\vec{J}_{1}^{\mathrm{A}\rightleftharpoons\mathrm{F}}} = -\left\langle \vec{b}_{\xi}^{\vec{J}_{2}^{\mathrm{A}\rightleftharpoons\mathrm{F}}}, -j\omega\mu_0\mathcal{L}_0\left(-\vec{b}_{\zeta}^{\vec{J}_{1}^{\mathrm{A}\rightleftharpoons\mathrm{F}}}\right)\right\rangle_{\mathbb{S}_{2}^{\mathrm{A}\rightleftharpoons\mathrm{F}}} \tag{D-73a}$$

$$z_{\xi\zeta}^{\vec{J}_{2}^{\mathrm{A}\rightleftharpoons\mathrm{F}}\vec{J}_{12}^{\mathrm{A}\rightleftharpoons\mathrm{A}}} = \left\langle \vec{b}_{\xi}^{\vec{J}_{2}^{\mathrm{A}\rightleftharpoons\mathrm{F}}}, \mathcal{E}_2\left(-\vec{b}_{\zeta}^{\vec{J}_{12}^{\mathrm{A}\rightleftharpoons\mathrm{A}}}\right)\right\rangle_{\tilde{\mathbb{S}}_{2}^{\mathrm{A}\rightleftharpoons\mathrm{F}}} \tag{D-73b}$$

$$z_{\xi\zeta}^{\vec{J}_{2}^{\mathrm{A}\rightleftharpoons\mathrm{F}}\vec{J}_{2}^{\mathrm{A}\rightleftharpoons\mathrm{F}}} = \left\langle \vec{b}_{\xi}^{\vec{J}_{2}^{\mathrm{A}\rightleftharpoons\mathrm{F}}}, \mathcal{E}_2\left(\vec{b}_{\zeta}^{\vec{J}_{2}^{\mathrm{A}\rightleftharpoons\mathrm{F}}}\right)\right\rangle_{\tilde{\mathbb{S}}_{2}^{\mathrm{A}\rightleftharpoons\mathrm{F}}} - \left\langle \vec{b}_{\xi}^{\vec{J}_{2}^{\mathrm{A}\rightleftharpoons\mathrm{F}}}, -j\omega\mu_0\mathcal{L}_0\left(-\vec{b}_{\zeta}^{\vec{J}_{2}^{\mathrm{A}\rightleftharpoons\mathrm{F}}}\right)\right\rangle_{\mathbb{S}_{2}^{\mathrm{A}\rightleftharpoons\mathrm{F}}} \tag{D-73c}$$

$$z_{\xi\zeta}^{\vec{J}_{2}^{\mathrm{A}\rightleftharpoons\mathrm{F}}\vec{J}_{2}^{\mathrm{A}}} = \left\langle \vec{b}_{\xi}^{\vec{J}_{2}^{\mathrm{A}\rightleftharpoons\mathrm{F}}}, \mathcal{E}_2\left(\vec{b}_{\zeta}^{\vec{J}_{2}^{\mathrm{A}}}\right)\right\rangle_{\tilde{\mathbb{S}}_{2}^{\mathrm{A}\rightleftharpoons\mathrm{F}}} \tag{D-73d}$$

$$z_{\xi\zeta}^{\vec{J}_{2}^{\mathrm{A}\rightleftharpoons\mathrm{F}}\vec{J}^{\mathrm{F}}} = -\left\langle \vec{b}_{\xi}^{\vec{J}_{2}^{\mathrm{A}\rightleftharpoons\mathrm{F}}}, -j\omega\mu_0\mathcal{L}_0\left(\vec{b}_{\zeta}^{\vec{J}^{\mathrm{F}}}\right)\right\rangle_{\mathbb{S}_{2}^{\mathrm{A}\rightleftharpoons\mathrm{F}}} \tag{D-73e}$$

$$z_{\xi\zeta}^{\vec{J}_{2}^{\mathrm{A}\rightleftharpoons\mathrm{F}}\vec{M}_{1}^{\mathrm{A}\rightleftharpoons\mathrm{F}}} = -\left\langle \vec{b}_{\xi}^{\vec{J}_{2}^{\mathrm{A}\rightleftharpoons\mathrm{F}}}, -\mathcal{K}_0\left(-\vec{b}_{\zeta}^{\vec{M}_{1}^{\mathrm{A}\rightleftharpoons\mathrm{F}}}\right)\right\rangle_{\mathbb{S}_{2}^{\mathrm{A}\rightleftharpoons\mathrm{F}}} \tag{D-73f}$$

$$z_{\xi\zeta}^{\vec{J}_{2}^{\mathrm{A}\rightleftharpoons\mathrm{F}}\vec{M}_{12}^{\mathrm{A}\rightleftharpoons\mathrm{A}}} = \left\langle \vec{b}_{\xi}^{\vec{J}_{2}^{\mathrm{A}\rightleftharpoons\mathrm{F}}}, \mathcal{E}_2\left(-\vec{b}_{\zeta}^{\vec{M}_{12}^{\mathrm{A}\rightleftharpoons\mathrm{A}}}\right)\right\rangle_{\tilde{\mathbb{S}}_{2}^{\mathrm{A}\rightleftharpoons\mathrm{F}}} \tag{D-73g}$$

$$z_{\xi\zeta}^{\vec{J}_{2}^{\mathrm{A}\rightleftharpoons\mathrm{F}}\vec{M}_{2}^{\mathrm{A}\rightleftharpoons\mathrm{F}}} = \left\langle \vec{b}_{\xi}^{\vec{J}_{2}^{\mathrm{A}\rightleftharpoons\mathrm{F}}}, \mathcal{E}_2\left(\vec{b}_{\zeta}^{\vec{M}_{2}^{\mathrm{A}\rightleftharpoons\mathrm{F}}}\right)\right\rangle_{\tilde{\mathbb{S}}_{2}^{\mathrm{A}\rightleftharpoons\mathrm{F}}} - \left\langle \vec{b}_{\xi}^{\vec{J}_{2}^{\mathrm{A}\rightleftharpoons\mathrm{F}}}, \frac{1}{2}\vec{b}_{\zeta}^{\vec{M}_{2}^{\mathrm{A}\rightleftharpoons\mathrm{F}}} \times \hat{n}^{\rightarrow\mathrm{F}} - \mathrm{P.V.}\,\mathcal{K}_0\left(-\vec{b}_{\zeta}^{\vec{M}_{2}^{\mathrm{A}\rightleftharpoons\mathrm{F}}}\right)\right\rangle_{\mathbb{S}_{2}^{\mathrm{A}\rightleftharpoons\mathrm{F}}} \tag{D-73h}$$

and

$$z_{\xi\zeta}^{\vec{M}_{2}^{\mathrm{A}\rightleftharpoons\mathrm{F}}\vec{J}_{1}^{\mathrm{A}\rightleftharpoons\mathrm{F}}} = -\left\langle \vec{b}_{\xi}^{\vec{M}_{2}^{\mathrm{A}\rightleftharpoons\mathrm{F}}}, \mathcal{K}_0\left(-\vec{b}_{\zeta}^{\vec{J}_{1}^{\mathrm{A}\rightleftharpoons\mathrm{F}}}\right)\right\rangle_{\mathbb{S}_{2}^{\mathrm{A}\rightleftharpoons\mathrm{F}}} \tag{D-74a}$$

$$z_{\xi\zeta}^{\vec{M}_{2}^{\mathrm{A}\rightleftharpoons\mathrm{F}}\vec{J}_{12}^{\mathrm{A}\rightleftharpoons\mathrm{A}}} = \left\langle \vec{b}_{\xi}^{\vec{M}_{2}^{\mathrm{A}\rightleftharpoons\mathrm{F}}}, \mathcal{H}_2\left(-\vec{b}_{\zeta}^{\vec{J}_{12}^{\mathrm{A}\rightleftharpoons\mathrm{A}}}\right)\right\rangle_{\tilde{\mathbb{S}}_{2}^{\mathrm{A}\rightleftharpoons\mathrm{F}}} \tag{D-74b}$$

$$z_{\xi\zeta}^{\vec{M}_{2}^{\mathrm{A}\rightleftharpoons\mathrm{F}}\vec{J}_{2}^{\mathrm{A}\rightleftharpoons\mathrm{F}}} = \left\langle \vec{b}_{\xi}^{\vec{M}_{2}^{\mathrm{A}\rightleftharpoons\mathrm{F}}}, \mathcal{H}_2\left(\vec{b}_{\zeta}^{\vec{J}_{2}^{\mathrm{A}\rightleftharpoons\mathrm{F}}}\right)\right\rangle_{\tilde{\mathbb{S}}_{2}^{\mathrm{A}\rightleftharpoons\mathrm{F}}} - \left\langle \vec{b}_{\xi}^{\vec{M}_{2}^{\mathrm{A}\rightleftharpoons\mathrm{F}}}, \hat{n}^{\rightarrow\mathrm{F}} \times \frac{1}{2}\vec{b}_{\zeta}^{\vec{J}_{2}^{\mathrm{A}\rightleftharpoons\mathrm{F}}} + \mathrm{P.V.}\,\mathcal{K}_0\left(-\vec{b}_{\zeta}^{\vec{J}_{2}^{\mathrm{A}\rightleftharpoons\mathrm{F}}}\right)\right\rangle_{\mathbb{S}_{2}^{\mathrm{A}\rightleftharpoons\mathrm{F}}} \tag{D-74c}$$

$$z_{\xi\zeta}^{\vec{M}_{2}^{\mathrm{A}\rightleftharpoons\mathrm{F}}\vec{J}_{2}^{\mathrm{A}}} = \left\langle \vec{b}_{\xi}^{\vec{M}_{2}^{\mathrm{A}\rightleftharpoons\mathrm{F}}}, \mathcal{H}_2\left(\vec{b}_{\zeta}^{\vec{J}_{2}^{\mathrm{A}}}\right)\right\rangle_{\tilde{\mathbb{S}}_{2}^{\mathrm{A}\rightleftharpoons\mathrm{F}}} \tag{D-74d}$$

$$z_{\xi\zeta}^{\vec{M}_{2}^{\mathrm{A}\rightleftharpoons\mathrm{F}}\vec{J}^{\mathrm{F}}} = -\left\langle \vec{b}_{\xi}^{\vec{M}_{2}^{\mathrm{A}\rightleftharpoons\mathrm{F}}}, \mathcal{K}_0\left(\vec{b}_{\zeta}^{\vec{J}^{\mathrm{F}}}\right)\right\rangle_{\mathbb{S}_{2}^{\mathrm{A}\rightleftharpoons\mathrm{F}}} \tag{D-74e}$$

$$z_{\xi\zeta}^{\vec{M}_2^{\mathrm{A}\rightleftharpoons\mathrm{F}}\vec{M}_1^{\mathrm{A}\rightleftharpoons\mathrm{F}}} = -\left\langle \vec{b}_\xi^{\vec{M}_2^{\mathrm{A}\rightleftharpoons\mathrm{F}}}, -j\omega\varepsilon_0\mathcal{L}_0\left(-\vec{b}_\zeta^{\vec{M}_1^{\mathrm{A}\rightleftharpoons\mathrm{F}}}\right)\right\rangle_{\mathbb{S}_2^{\mathrm{A}\rightleftharpoons\mathrm{F}}} \tag{D-74f}$$

$$z_{\xi\zeta}^{\vec{M}_2^{\mathrm{A}\rightleftharpoons\mathrm{F}}\vec{M}_{12}^{\mathrm{A}\rightleftharpoons\mathrm{A}}} = \left\langle \vec{b}_\xi^{\vec{M}_2^{\mathrm{A}\rightleftharpoons\mathrm{F}}}, \mathcal{H}_2\left(-\vec{b}_\zeta^{\vec{M}_{12}^{\mathrm{A}\rightleftharpoons\mathrm{A}}}\right)\right\rangle_{\tilde{\mathbb{S}}_2^{\mathrm{A}\rightleftharpoons\mathrm{F}}} \tag{D-74g}$$

$$z_{\xi\zeta}^{\vec{M}_2^{\mathrm{A}\rightleftharpoons\mathrm{F}}\vec{M}_2^{\mathrm{A}\rightleftharpoons\mathrm{F}}} = \left\langle \vec{b}_\xi^{\vec{M}_2^{\mathrm{A}\rightleftharpoons\mathrm{F}}}, \mathcal{H}_2\left(\vec{b}_\zeta^{\vec{M}_2^{\mathrm{A}\rightleftharpoons\mathrm{F}}}\right)\right\rangle_{\tilde{\mathbb{S}}_2^{\mathrm{A}\rightleftharpoons\mathrm{F}}} - \left\langle \vec{b}_\xi^{\vec{M}_2^{\mathrm{A}\rightleftharpoons\mathrm{F}}}, -j\omega\varepsilon_0\mathcal{L}_0\left(-\vec{b}_\zeta^{\vec{M}_2^{\mathrm{A}\rightleftharpoons\mathrm{F}}}\right)\right\rangle_{\mathbb{S}_2^{\mathrm{A}\rightleftharpoons\mathrm{F}}} \tag{D-74h}$$

and

$$z_{\xi\zeta}^{\vec{J}_1^{\mathrm{A}}\vec{J}^{\mathrm{G}\rightleftharpoons\mathrm{A}}} = \left\langle \vec{b}_\xi^{\vec{J}_1^{\mathrm{A}}}, \mathcal{E}_1\left(\vec{b}_\zeta^{\vec{J}^{\mathrm{G}\rightleftharpoons\mathrm{A}}}\right)\right\rangle_{\tilde{\mathbb{S}}_1^{\mathrm{A}}} \tag{D-75a}$$

$$z_{\xi\zeta}^{\vec{J}_1^{\mathrm{A}}\vec{J}_1^{\mathrm{A}\rightleftharpoons\mathrm{F}}} = \left\langle \vec{b}_\xi^{\vec{J}_1^{\mathrm{A}}}, \mathcal{E}_1\left(\vec{b}_\zeta^{\vec{J}_1^{\mathrm{A}\rightleftharpoons\mathrm{F}}}\right)\right\rangle_{\tilde{\mathbb{S}}_1^{\mathrm{A}}} \tag{D-75b}$$

$$z_{\xi\zeta}^{\vec{J}_1^{\mathrm{A}}\vec{J}_{12}^{\mathrm{A}\rightleftharpoons\mathrm{A}}} = \left\langle \vec{b}_\xi^{\vec{J}_1^{\mathrm{A}}}, \mathcal{E}_1\left(\vec{b}_\zeta^{\vec{J}_{12}^{\mathrm{A}\rightleftharpoons\mathrm{A}}}\right)\right\rangle_{\tilde{\mathbb{S}}_1^{\mathrm{A}}} \tag{D-75c}$$

$$z_{\xi\zeta}^{\vec{J}_1^{\mathrm{A}}\vec{J}_1^{\mathrm{A}}} = \left\langle \vec{b}_\xi^{\vec{J}_1^{\mathrm{A}}}, \mathcal{E}_1\left(\vec{b}_\zeta^{\vec{J}_1^{\mathrm{A}}}\right)\right\rangle_{\tilde{\mathbb{S}}_1^{\mathrm{A}}} \tag{D-75d}$$

$$z_{\xi\zeta}^{\vec{J}_1^{\mathrm{A}}\vec{M}^{\mathrm{G}\rightleftharpoons\mathrm{A}}} = \left\langle \vec{b}_\xi^{\vec{J}_1^{\mathrm{A}}}, \mathcal{E}_1\left(\vec{b}_\zeta^{\vec{M}^{\mathrm{G}\rightleftharpoons\mathrm{A}}}\right)\right\rangle_{\tilde{\mathbb{S}}_1^{\mathrm{A}}} \tag{D-75e}$$

$$z_{\xi\zeta}^{\vec{J}_1^{\mathrm{A}}\vec{M}_1^{\mathrm{A}\rightleftharpoons\mathrm{F}}} = \left\langle \vec{b}_\xi^{\vec{J}_1^{\mathrm{A}}}, \mathcal{E}_1\left(\vec{b}_\zeta^{\vec{M}_1^{\mathrm{A}\rightleftharpoons\mathrm{F}}}\right)\right\rangle_{\tilde{\mathbb{S}}_1^{\mathrm{A}}} \tag{D-75f}$$

$$z_{\xi\zeta}^{\vec{J}_1^{\mathrm{A}}\vec{M}_{12}^{\mathrm{A}\rightleftharpoons\mathrm{A}}} = \left\langle \vec{b}_\xi^{\vec{J}_1^{\mathrm{A}}}, \mathcal{E}_1\left(\vec{b}_\zeta^{\vec{M}_{12}^{\mathrm{A}\rightleftharpoons\mathrm{A}}}\right)\right\rangle_{\tilde{\mathbb{S}}_1^{\mathrm{A}}} \tag{D-75g}$$

and

$$z_{\xi\zeta}^{\vec{J}_2^{\mathrm{A}}\vec{J}_{12}^{\mathrm{A}\rightleftharpoons\mathrm{A}}} = \left\langle \vec{b}_\xi^{\vec{J}_2^{\mathrm{A}}}, \mathcal{E}_2\left(-\vec{b}_\zeta^{\vec{J}_{12}^{\mathrm{A}\rightleftharpoons\mathrm{A}}}\right)\right\rangle_{\tilde{\mathbb{S}}_2^{\mathrm{A}}} \tag{D-76a}$$

$$z_{\xi\zeta}^{\vec{J}_2^{\mathrm{A}}\vec{J}_2^{\mathrm{A}\rightleftharpoons\mathrm{F}}} = \left\langle \vec{b}_\xi^{\vec{J}_2^{\mathrm{A}}}, \mathcal{E}_2\left(\vec{b}_\zeta^{\vec{J}_2^{\mathrm{A}\rightleftharpoons\mathrm{F}}}\right)\right\rangle_{\tilde{\mathbb{S}}_2^{\mathrm{A}}} \tag{D-76b}$$

$$z_{\xi\zeta}^{\vec{J}_2^{\mathrm{A}}\vec{J}_2^{\mathrm{A}}} = \left\langle \vec{b}_\xi^{\vec{J}_2^{\mathrm{A}}}, \mathcal{E}_2\left(\vec{b}_\zeta^{\vec{J}_2^{\mathrm{A}}}\right)\right\rangle_{\tilde{\mathbb{S}}_2^{\mathrm{A}}} \tag{D-76c}$$

$$z_{\xi\zeta}^{\vec{J}_2^{\mathrm{A}}\vec{M}_{12}^{\mathrm{A}\rightleftharpoons\mathrm{A}}} = \left\langle \vec{b}_\xi^{\vec{J}_2^{\mathrm{A}}}, \mathcal{E}_2\left(-\vec{b}_\zeta^{\vec{M}_{12}^{\mathrm{A}\rightleftharpoons\mathrm{A}}}\right)\right\rangle_{\tilde{\mathbb{S}}_2^{\mathrm{A}}} \tag{D-76d}$$

$$z_{\xi\zeta}^{\vec{J}_2^{\mathrm{A}}\vec{M}_2^{\mathrm{A}\rightleftharpoons\mathrm{F}}} = \left\langle \vec{b}_\xi^{\vec{J}_2^{\mathrm{A}}}, \mathcal{E}_2\left(\vec{b}_\zeta^{\vec{M}_2^{\mathrm{A}\rightleftharpoons\mathrm{F}}}\right)\right\rangle_{\tilde{\mathbb{S}}_2^{\mathrm{A}}} \tag{D-76e}$$

and

$$z_{\xi\zeta}^{\vec{J}^{\mathrm{F}}\vec{J}_1^{\mathrm{A}\rightleftharpoons\mathrm{F}}} = \left\langle \vec{b}_\xi^{\vec{J}^{\mathrm{F}}}, -j\omega\mu_0\mathcal{L}_0\left(-\vec{b}_\zeta^{\vec{J}_1^{\mathrm{A}\rightleftharpoons\mathrm{F}}}\right)\right\rangle_{\mathbb{S}^{\mathrm{F}}} \tag{D-77a}$$

$$z_{\xi\zeta}^{\vec{J}^{\mathrm{F}}\vec{J}_2^{\mathrm{A}\rightleftharpoons\mathrm{F}}} = \left\langle \vec{b}_\xi^{\vec{J}^{\mathrm{F}}}, -j\omega\mu_0\mathcal{L}_0\left(-\vec{b}_\zeta^{\vec{J}_2^{\mathrm{A}\rightleftharpoons\mathrm{F}}}\right)\right\rangle_{\mathbb{S}^{\mathrm{F}}} \tag{D-77b}$$

$$z_{\xi\zeta}^{\vec{J}^{\mathrm{F}}\vec{J}^{\mathrm{F}}} = \left\langle \vec{b}_\xi^{\vec{J}^{\mathrm{F}}}, -j\omega\mu_0\mathcal{L}_0\left(\vec{b}_\zeta^{\vec{J}^{\mathrm{F}}}\right)\right\rangle_{\mathbb{S}^{\mathrm{F}}} \tag{D-77c}$$

$$z_{\xi\zeta}^{\vec{J}^{\mathrm{F}}\vec{M}_1^{\mathrm{A}\rightleftharpoons\mathrm{F}}} = \left\langle \vec{b}_\xi^{\vec{J}^{\mathrm{F}}}, -\mathcal{K}_0\left(-\vec{b}_\zeta^{\vec{M}_1^{\mathrm{A}\rightleftharpoons\mathrm{F}}}\right)\right\rangle_{\mathbb{S}^{\mathrm{F}}} \tag{D-77d}$$

$$z_{\xi\zeta}^{\vec{J}^{\mathrm{F}}\vec{M}_2^{\mathrm{A}\rightleftharpoons\mathrm{F}}} = \left\langle \vec{b}_\xi^{\vec{J}^{\mathrm{F}}}, -\mathcal{K}_0\left(-\vec{b}_\zeta^{\vec{M}_2^{\mathrm{A}\rightleftharpoons\mathrm{F}}}\right)\right\rangle_{\mathbb{S}^{\mathrm{F}}} \tag{D-77e}$$

where the various integral surfaces with "~" are the inner sub-boundaries of the related regions.

The transformation matrix $\overline{\overline{T}}$ used in Eq. (6-90) is as follows:

$$\overline{\overline{T}} = \overline{\overline{T}}^{\vec{J}^{\mathrm{G}\rightleftharpoons\mathrm{A}}\rightarrow\mathrm{AV}} \text{ or } \overline{\overline{T}}^{\vec{M}^{\mathrm{G}\rightleftharpoons\mathrm{A}}\rightarrow\mathrm{AV}} \text{ or } \overline{\overline{T}}^{\mathrm{BS}\rightarrow\mathrm{AV}} \tag{D-78}$$

in which

$$\overline{\overline{T}}^{\vec{J}^{\mathrm{G}\rightleftharpoons\mathrm{A}}\rightarrow\mathrm{AV}} = \left(\overline{\overline{\Psi}}_1\right)^{-1}\cdot\overline{\overline{\Psi}}_2 \tag{D-79a}$$

$$\overline{\overline{T}}^{\vec{M}^{\mathrm{G}\rightleftharpoons\mathrm{A}}\rightarrow\mathrm{AV}} = \left(\overline{\overline{\Psi}}_3\right)^{-1}\cdot\overline{\overline{\Psi}}_4 \tag{D-79b}$$

$$\overline{\overline{T}}^{\mathrm{BS}\rightarrow\mathrm{AV}} = \mathrm{nullspace}\left(\overline{\overline{\Psi}}_{\mathrm{FCE}}^{\mathrm{DoJ/DoM}}\right) \tag{D-80}$$

where

$$\overline{\overline{\Psi}}_1 = \begin{bmatrix}
\overline{\overline{I}}^{\vec{J}^{G\rightleftharpoons A}} & 0 & 0 & 0 & 0 & 0 & 0 & 0 & 0 & 0 & 0 \\
0 & \overline{\overline{Z}}^{\vec{M}^{G\rightleftharpoons A}\vec{J}_1^{A\rightleftharpoons F}} & \overline{\overline{Z}}^{\vec{M}^{G\rightleftharpoons A}\vec{J}_{12}^{A\rightleftharpoons A}} & 0 & \overline{\overline{Z}}^{\vec{M}^{G\rightleftharpoons A}\vec{J}_1^{A}} & 0 & 0 & \overline{\overline{Z}}^{\vec{M}^{G\rightleftharpoons A}\vec{M}^{G\rightleftharpoons A}} & \overline{\overline{Z}}^{\vec{M}^{G\rightleftharpoons A}\vec{M}_1^{A\rightleftharpoons F}} & \overline{\overline{Z}}^{\vec{M}^{G\rightleftharpoons A}\vec{M}_{12}^{A\rightleftharpoons A}} & 0 \\
0 & \overline{\overline{Z}}^{\vec{J}_1^{A\rightleftharpoons F}\vec{J}_1^{A\rightleftharpoons F}} & \overline{\overline{Z}}^{\vec{J}_1^{A\rightleftharpoons F}\vec{J}_{12}^{A\rightleftharpoons A}} & \overline{\overline{Z}}^{\vec{J}_1^{A\rightleftharpoons F}\vec{J}_2^{A\rightleftharpoons F}} & \overline{\overline{Z}}^{\vec{J}_1^{A\rightleftharpoons F}\vec{J}_1^{A}} & 0 & \overline{\overline{Z}}^{\vec{J}_1^{A\rightleftharpoons F}\vec{J}^{F}} & \overline{\overline{Z}}^{\vec{J}_1^{A\rightleftharpoons F}\vec{M}^{G\rightleftharpoons A}} & \overline{\overline{Z}}^{\vec{J}_1^{A\rightleftharpoons F}\vec{M}_1^{A\rightleftharpoons F}} & \overline{\overline{Z}}^{\vec{J}_1^{A\rightleftharpoons F}\vec{M}_{12}^{A\rightleftharpoons A}} & \overline{\overline{Z}}^{\vec{J}_1^{A\rightleftharpoons F}\vec{M}_2^{A\rightleftharpoons F}} \\
0 & \overline{\overline{Z}}^{\vec{M}_1^{A\rightleftharpoons F}\vec{J}_1^{A\rightleftharpoons F}} & \overline{\overline{Z}}^{\vec{M}_1^{A\rightleftharpoons F}\vec{J}_{12}^{A\rightleftharpoons A}} & \overline{\overline{Z}}^{\vec{M}_1^{A\rightleftharpoons F}\vec{J}_2^{A\rightleftharpoons F}} & \overline{\overline{Z}}^{\vec{M}_1^{A\rightleftharpoons F}\vec{J}_1^{A}} & 0 & \overline{\overline{Z}}^{\vec{M}_1^{A\rightleftharpoons F}\vec{J}^{F}} & \overline{\overline{Z}}^{\vec{M}_1^{A\rightleftharpoons F}\vec{M}^{G\rightleftharpoons A}} & \overline{\overline{Z}}^{\vec{M}_1^{A\rightleftharpoons F}\vec{M}_1^{A\rightleftharpoons F}} & \overline{\overline{Z}}^{\vec{M}_1^{A\rightleftharpoons F}\vec{M}_{12}^{A\rightleftharpoons A}} & \overline{\overline{Z}}^{\vec{M}_1^{A\rightleftharpoons F}\vec{M}_2^{A\rightleftharpoons F}} \\
0 & \overline{\overline{Z}}^{\vec{J}_{12}^{A\rightleftharpoons A}\vec{J}_1^{A\rightleftharpoons F}} & \overline{\overline{Z}}^{\vec{J}_{12}^{A\rightleftharpoons A}\vec{J}_{12}^{A\rightleftharpoons A}} & \overline{\overline{Z}}^{\vec{J}_{12}^{A\rightleftharpoons A}\vec{J}_2^{A\rightleftharpoons F}} & \overline{\overline{Z}}^{\vec{J}_{12}^{A\rightleftharpoons A}\vec{J}_1^{A}} & \overline{\overline{Z}}^{\vec{J}_{12}^{A\rightleftharpoons A}\vec{J}_2^{A}} & 0 & \overline{\overline{Z}}^{\vec{J}_{12}^{A\rightleftharpoons A}\vec{M}^{G\rightleftharpoons A}} & \overline{\overline{Z}}^{\vec{J}_{12}^{A\rightleftharpoons A}\vec{M}_1^{A\rightleftharpoons F}} & \overline{\overline{Z}}^{\vec{J}_{12}^{A\rightleftharpoons A}\vec{M}_{12}^{A\rightleftharpoons A}} & \overline{\overline{Z}}^{\vec{J}_{12}^{A\rightleftharpoons A}\vec{M}_2^{A\rightleftharpoons F}} \\
0 & \overline{\overline{Z}}^{\vec{M}_{12}^{A\rightleftharpoons A}\vec{J}_1^{A\rightleftharpoons F}} & \overline{\overline{Z}}^{\vec{M}_{12}^{A\rightleftharpoons A}\vec{J}_{12}^{A\rightleftharpoons A}} & \overline{\overline{Z}}^{\vec{M}_{12}^{A\rightleftharpoons A}\vec{J}_2^{A\rightleftharpoons F}} & \overline{\overline{Z}}^{\vec{M}_{12}^{A\rightleftharpoons A}\vec{J}_1^{A}} & \overline{\overline{Z}}^{\vec{M}_{12}^{A\rightleftharpoons A}\vec{J}_2^{A}} & 0 & \overline{\overline{Z}}^{\vec{M}_{12}^{A\rightleftharpoons A}\vec{M}^{G\rightleftharpoons A}} & \overline{\overline{Z}}^{\vec{M}_{12}^{A\rightleftharpoons A}\vec{M}_1^{A\rightleftharpoons F}} & \overline{\overline{Z}}^{\vec{M}_{12}^{A\rightleftharpoons A}\vec{M}_{12}^{A\rightleftharpoons A}} & \overline{\overline{Z}}^{\vec{M}_{12}^{A\rightleftharpoons A}\vec{M}_2^{A\rightleftharpoons F}} \\
0 & \overline{\overline{Z}}^{\vec{J}_2^{A\rightleftharpoons F}\vec{J}_1^{A\rightleftharpoons F}} & \overline{\overline{Z}}^{\vec{J}_2^{A\rightleftharpoons F}\vec{J}_{12}^{A\rightleftharpoons A}} & \overline{\overline{Z}}^{\vec{J}_2^{A\rightleftharpoons F}\vec{J}_2^{A\rightleftharpoons F}} & 0 & \overline{\overline{Z}}^{\vec{J}_2^{A\rightleftharpoons F}\vec{J}_2^{A}} & \overline{\overline{Z}}^{\vec{J}_2^{A\rightleftharpoons F}\vec{J}^{F}} & 0 & \overline{\overline{Z}}^{\vec{J}_2^{A\rightleftharpoons F}\vec{M}_1^{A\rightleftharpoons F}} & \overline{\overline{Z}}^{\vec{J}_2^{A\rightleftharpoons F}\vec{M}_{12}^{A\rightleftharpoons A}} & \overline{\overline{Z}}^{\vec{J}_2^{A\rightleftharpoons F}\vec{M}_2^{A\rightleftharpoons F}} \\
0 & \overline{\overline{Z}}^{\vec{M}_2^{A\rightleftharpoons F}\vec{J}_1^{A\rightleftharpoons F}} & \overline{\overline{Z}}^{\vec{M}_2^{A\rightleftharpoons F}\vec{J}_{12}^{A\rightleftharpoons A}} & \overline{\overline{Z}}^{\vec{M}_2^{A\rightleftharpoons F}\vec{J}_2^{A\rightleftharpoons F}} & 0 & \overline{\overline{Z}}^{\vec{M}_2^{A\rightleftharpoons F}\vec{J}_2^{A}} & \overline{\overline{Z}}^{\vec{M}_2^{A\rightleftharpoons F}\vec{J}^{F}} & 0 & \overline{\overline{Z}}^{\vec{M}_2^{A\rightleftharpoons F}\vec{M}_1^{A\rightleftharpoons F}} & \overline{\overline{Z}}^{\vec{M}_2^{A\rightleftharpoons F}\vec{M}_{12}^{A\rightleftharpoons A}} & \overline{\overline{Z}}^{\vec{M}_2^{A\rightleftharpoons F}\vec{M}_2^{A\rightleftharpoons F}} \\
0 & \overline{\overline{Z}}^{\vec{J}_1^{A}\vec{J}_1^{A\rightleftharpoons F}} & \overline{\overline{Z}}^{\vec{J}_1^{A}\vec{J}_{12}^{A\rightleftharpoons A}} & 0 & \overline{\overline{Z}}^{\vec{J}_1^{A}\vec{J}_1^{A}} & 0 & 0 & \overline{\overline{Z}}^{\vec{J}_1^{A}\vec{M}^{G\rightleftharpoons A}} & \overline{\overline{Z}}^{\vec{J}_1^{A}\vec{M}_1^{A\rightleftharpoons F}} & \overline{\overline{Z}}^{\vec{J}_1^{A}\vec{M}_{12}^{A\rightleftharpoons A}} & 0 \\
0 & 0 & \overline{\overline{Z}}^{\vec{J}_2^{A}\vec{J}_{12}^{A\rightleftharpoons A}} & \overline{\overline{Z}}^{\vec{J}_2^{A}\vec{J}_2^{A\rightleftharpoons F}} & 0 & \overline{\overline{Z}}^{\vec{J}_2^{A}\vec{J}_2^{A}} & 0 & 0 & 0 & \overline{\overline{Z}}^{\vec{J}_2^{A}\vec{M}_{12}^{A\rightleftharpoons A}} & \overline{\overline{Z}}^{\vec{J}_2^{A}\vec{M}_2^{A\rightleftharpoons F}} \\
0 & \overline{\overline{Z}}^{\vec{J}^{F}\vec{J}_1^{A\rightleftharpoons F}} & 0 & \overline{\overline{Z}}^{\vec{J}^{F}\vec{J}_2^{A\rightleftharpoons F}} & 0 & 0 & \overline{\overline{Z}}^{\vec{J}^{F}\vec{J}^{F}} & 0 & \overline{\overline{Z}}^{\vec{J}^{F}\vec{M}_1^{A\rightleftharpoons F}} & 0 & \overline{\overline{Z}}^{\vec{J}^{F}\vec{M}_2^{A\rightleftharpoons F}}
\end{bmatrix} \tag{D-81a}$$

$$\overline{\overline{\Psi}}_2 = \begin{bmatrix}
\overline{\overline{I}}^{\vec{J}^{G\rightleftharpoons A}} \\
-\overline{\overline{Z}}^{\vec{M}^{G\rightleftharpoons A}\vec{J}^{G\rightleftharpoons A}} \\
-\overline{\overline{Z}}^{\vec{J}_1^{A\rightleftharpoons F}\vec{J}^{G\rightleftharpoons A}} \\
-\overline{\overline{Z}}^{\vec{M}_1^{A\rightleftharpoons F}\vec{J}^{G\rightleftharpoons A}} \\
-\overline{\overline{Z}}^{\vec{J}_{12}^{A\rightleftharpoons A}\vec{J}^{G\rightleftharpoons A}} \\
-\overline{\overline{Z}}^{\vec{M}_{12}^{A\rightleftharpoons A}\vec{J}^{G\rightleftharpoons A}} \\
0 \\
0 \\
-\overline{\overline{Z}}^{\vec{J}_1^{A}\vec{J}^{G\rightleftharpoons A}} \\
0 \\
0
\end{bmatrix} \tag{D-81b}$$

and

$$\overline{\overline{\Psi}}_3 = \begin{bmatrix}
0 & 0 & 0 & 0 & 0 & 0 & 0 & \overline{\overline{I}}^{\vec{M}^{G\rightleftharpoons A}} & 0 & 0 & 0 \\
\overline{\overline{Z}}^{\vec{J}^{G\rightleftharpoons A}\vec{J}^{G\rightleftharpoons A}} & \overline{\overline{Z}}^{\vec{J}^{G\rightleftharpoons A}\vec{J}_1^{A\rightleftharpoons F}} & \overline{\overline{Z}}^{\vec{J}^{G\rightleftharpoons A}\vec{J}_{12}^{A\rightleftharpoons A}} & 0 & \overline{\overline{Z}}^{\vec{J}^{G\rightleftharpoons A}\vec{J}_1^{A}} & 0 & 0 & 0 & \overline{\overline{Z}}^{\vec{J}^{G\rightleftharpoons A}\vec{M}_1^{A\rightleftharpoons F}} & \overline{\overline{Z}}^{\vec{J}^{G\rightleftharpoons A}\vec{M}_{12}^{A\rightleftharpoons A}} & 0 \\
\overline{\overline{Z}}^{\vec{J}_1^{A\rightleftharpoons F}\vec{J}^{G\rightleftharpoons A}} & \overline{\overline{Z}}^{\vec{J}_1^{A\rightleftharpoons F}\vec{J}_1^{A\rightleftharpoons F}} & \overline{\overline{Z}}^{\vec{J}_1^{A\rightleftharpoons F}\vec{J}_{12}^{A\rightleftharpoons A}} & \overline{\overline{Z}}^{\vec{J}_1^{A\rightleftharpoons F}\vec{J}_2^{A\rightleftharpoons F}} & \overline{\overline{Z}}^{\vec{J}_1^{A\rightleftharpoons F}\vec{J}_1^{A}} & 0 & \overline{\overline{Z}}^{\vec{J}_1^{A\rightleftharpoons F}\vec{J}^{F}} & 0 & \overline{\overline{Z}}^{\vec{J}_1^{A\rightleftharpoons F}\vec{M}_1^{A\rightleftharpoons F}} & \overline{\overline{Z}}^{\vec{J}_1^{A\rightleftharpoons F}\vec{M}_{12}^{A\rightleftharpoons A}} & \overline{\overline{Z}}^{\vec{J}_1^{A\rightleftharpoons F}\vec{M}_2^{A\rightleftharpoons F}} \\
\overline{\overline{Z}}^{\vec{M}_1^{A\rightleftharpoons F}\vec{J}^{G\rightleftharpoons A}} & \overline{\overline{Z}}^{\vec{M}_1^{A\rightleftharpoons F}\vec{J}_1^{A\rightleftharpoons F}} & \overline{\overline{Z}}^{\vec{M}_1^{A\rightleftharpoons F}\vec{J}_{12}^{A\rightleftharpoons A}} & \overline{\overline{Z}}^{\vec{M}_1^{A\rightleftharpoons F}\vec{J}_2^{A\rightleftharpoons F}} & \overline{\overline{Z}}^{\vec{M}_1^{A\rightleftharpoons F}\vec{J}_1^{A}} & 0 & \overline{\overline{Z}}^{\vec{M}_1^{A\rightleftharpoons F}\vec{J}^{F}} & 0 & \overline{\overline{Z}}^{\vec{M}_1^{A\rightleftharpoons F}\vec{M}_1^{A\rightleftharpoons F}} & \overline{\overline{Z}}^{\vec{M}_1^{A\rightleftharpoons F}\vec{M}_{12}^{A\rightleftharpoons A}} & \overline{\overline{Z}}^{\vec{M}_1^{A\rightleftharpoons F}\vec{M}_2^{A\rightleftharpoons F}} \\
\overline{\overline{Z}}^{\vec{J}_{12}^{A\rightleftharpoons A}\vec{J}^{G\rightleftharpoons A}} & \overline{\overline{Z}}^{\vec{J}_{12}^{A\rightleftharpoons A}\vec{J}_1^{A\rightleftharpoons F}} & \overline{\overline{Z}}^{\vec{J}_{12}^{A\rightleftharpoons A}\vec{J}_{12}^{A\rightleftharpoons A}} & \overline{\overline{Z}}^{\vec{J}_{12}^{A\rightleftharpoons A}\vec{J}_2^{A\rightleftharpoons F}} & \overline{\overline{Z}}^{\vec{J}_{12}^{A\rightleftharpoons A}\vec{J}_1^{A}} & \overline{\overline{Z}}^{\vec{J}_{12}^{A\rightleftharpoons A}\vec{J}_2^{A}} & 0 & 0 & \overline{\overline{Z}}^{\vec{J}_{12}^{A\rightleftharpoons A}\vec{M}_1^{A\rightleftharpoons F}} & \overline{\overline{Z}}^{\vec{J}_{12}^{A\rightleftharpoons A}\vec{M}_{12}^{A\rightleftharpoons A}} & \overline{\overline{Z}}^{\vec{J}_{12}^{A\rightleftharpoons A}\vec{M}_2^{A\rightleftharpoons F}} \\
\overline{\overline{Z}}^{\vec{M}_{12}^{A\rightleftharpoons A}\vec{J}^{G\rightleftharpoons A}} & \overline{\overline{Z}}^{\vec{M}_{12}^{A\rightleftharpoons A}\vec{J}_1^{A\rightleftharpoons F}} & \overline{\overline{Z}}^{\vec{M}_{12}^{A\rightleftharpoons A}\vec{J}_{12}^{A\rightleftharpoons A}} & \overline{\overline{Z}}^{\vec{M}_{12}^{A\rightleftharpoons A}\vec{J}_2^{A\rightleftharpoons F}} & \overline{\overline{Z}}^{\vec{M}_{12}^{A\rightleftharpoons A}\vec{J}_1^{A}} & \overline{\overline{Z}}^{\vec{M}_{12}^{A\rightleftharpoons A}\vec{J}_2^{A}} & 0 & 0 & \overline{\overline{Z}}^{\vec{M}_{12}^{A\rightleftharpoons A}\vec{M}_1^{A\rightleftharpoons F}} & \overline{\overline{Z}}^{\vec{M}_{12}^{A\rightleftharpoons A}\vec{M}_{12}^{A\rightleftharpoons A}} & \overline{\overline{Z}}^{\vec{M}_{12}^{A\rightleftharpoons A}\vec{M}_2^{A\rightleftharpoons F}} \\
0 & \overline{\overline{Z}}^{\vec{J}_2^{A\rightleftharpoons F}\vec{J}_1^{A\rightleftharpoons F}} & \overline{\overline{Z}}^{\vec{J}_2^{A\rightleftharpoons F}\vec{J}_{12}^{A\rightleftharpoons A}} & \overline{\overline{Z}}^{\vec{J}_2^{A\rightleftharpoons F}\vec{J}_2^{A\rightleftharpoons F}} & 0 & \overline{\overline{Z}}^{\vec{J}_2^{A\rightleftharpoons F}\vec{J}_2^{A}} & \overline{\overline{Z}}^{\vec{J}_2^{A\rightleftharpoons F}\vec{J}^{F}} & 0 & \overline{\overline{Z}}^{\vec{J}_2^{A\rightleftharpoons F}\vec{M}_1^{A\rightleftharpoons F}} & \overline{\overline{Z}}^{\vec{J}_2^{A\rightleftharpoons F}\vec{M}_{12}^{A\rightleftharpoons A}} & \overline{\overline{Z}}^{\vec{J}_2^{A\rightleftharpoons F}\vec{M}_2^{A\rightleftharpoons F}} \\
0 & \overline{\overline{Z}}^{\vec{M}_2^{A\rightleftharpoons F}\vec{J}_1^{A\rightleftharpoons F}} & \overline{\overline{Z}}^{\vec{M}_2^{A\rightleftharpoons F}\vec{J}_{12}^{A\rightleftharpoons A}} & \overline{\overline{Z}}^{\vec{M}_2^{A\rightleftharpoons F}\vec{J}_2^{A\rightleftharpoons F}} & 0 & \overline{\overline{Z}}^{\vec{M}_2^{A\rightleftharpoons F}\vec{J}_2^{A}} & \overline{\overline{Z}}^{\vec{M}_2^{A\rightleftharpoons F}\vec{J}^{F}} & 0 & \overline{\overline{Z}}^{\vec{M}_2^{A\rightleftharpoons F}\vec{M}_1^{A\rightleftharpoons F}} & \overline{\overline{Z}}^{\vec{M}_2^{A\rightleftharpoons F}\vec{M}_{12}^{A\rightleftharpoons A}} & \overline{\overline{Z}}^{\vec{M}_2^{A\rightleftharpoons F}\vec{M}_2^{A\rightleftharpoons F}} \\
\overline{\overline{Z}}^{\vec{J}_1^{A}\vec{J}^{G\rightleftharpoons A}} & \overline{\overline{Z}}^{\vec{J}_1^{A}\vec{J}_1^{A\rightleftharpoons F}} & \overline{\overline{Z}}^{\vec{J}_1^{A}\vec{J}_{12}^{A\rightleftharpoons A}} & 0 & \overline{\overline{Z}}^{\vec{J}_1^{A}\vec{J}_1^{A}} & 0 & 0 & 0 & \overline{\overline{Z}}^{\vec{J}_1^{A}\vec{M}_1^{A\rightleftharpoons F}} & \overline{\overline{Z}}^{\vec{J}_1^{A}\vec{M}_{12}^{A\rightleftharpoons A}} & 0 \\
0 & 0 & \overline{\overline{Z}}^{\vec{J}_2^{A}\vec{J}_{12}^{A\rightleftharpoons A}} & \overline{\overline{Z}}^{\vec{J}_2^{A}\vec{J}_2^{A\rightleftharpoons F}} & 0 & \overline{\overline{Z}}^{\vec{J}_2^{A}\vec{J}_2^{A}} & 0 & 0 & 0 & \overline{\overline{Z}}^{\vec{J}_2^{A}\vec{M}_{12}^{A\rightleftharpoons A}} & \overline{\overline{Z}}^{\vec{J}_2^{A}\vec{M}_2^{A\rightleftharpoons F}} \\
0 & \overline{\overline{Z}}^{\vec{J}^{F}\vec{J}_1^{A\rightleftharpoons F}} & 0 & \overline{\overline{Z}}^{\vec{J}^{F}\vec{J}_2^{A\rightleftharpoons F}} & 0 & 0 & \overline{\overline{Z}}^{\vec{J}^{F}\vec{J}^{F}} & 0 & \overline{\overline{Z}}^{\vec{J}^{F}\vec{M}_1^{A\rightleftharpoons F}} & 0 & \overline{\overline{Z}}^{\vec{J}^{F}\vec{M}_2^{A\rightleftharpoons F}}
\end{bmatrix} \tag{D-82a}$$

$$\overline{\overline{\Psi}}_4 = \begin{bmatrix} \overline{\overline{I}}^{\vec{M}^{G\rightleftharpoons A}} \\ -\overline{\overline{Z}}^{\vec{J}^{G\rightleftharpoons A}\vec{M}^{G\rightleftharpoons A}} \\ -\overline{\overline{Z}}^{\vec{J}_1^{A\rightleftharpoons F}\vec{M}^{G\rightleftharpoons A}} \\ -\overline{\overline{Z}}^{\vec{M}_1^{A\rightleftharpoons F}\vec{M}^{G\rightleftharpoons A}} \\ -\overline{\overline{Z}}^{\vec{J}_{12}^{A\rightleftharpoons A}\vec{M}^{G\rightleftharpoons A}} \\ -\overline{\overline{Z}}^{\vec{M}_{12}^{A\rightleftharpoons A}\vec{M}^{G\rightleftharpoons A}} \\ 0 \\ 0 \\ -\overline{\overline{Z}}^{\vec{J}_1^{A}\vec{M}^{G\rightleftharpoons A}} \\ 0 \\ 0 \end{bmatrix} \tag{D-82b}$$

and

$$\overline{\overline{\Psi}}_{\text{FCE}}^{\text{DoJ}} = \begin{bmatrix}
\overline{\overline{Z}}^{\vec{M}^{G\rightleftharpoons A}\vec{J}^{G\rightleftharpoons A}} & \overline{\overline{Z}}^{\vec{M}^{G\rightleftharpoons A}\vec{J}_1^{A\rightleftharpoons F}} & \overline{\overline{Z}}^{\vec{M}^{G\rightleftharpoons A}\vec{J}_{12}^{A\rightleftharpoons A}} & 0 & \overline{\overline{Z}}^{\vec{M}^{G\rightleftharpoons A}\vec{J}_1^{A}} & 0 & 0 & \overline{\overline{Z}}^{\vec{M}^{G\rightleftharpoons A}\vec{M}^{G\rightleftharpoons A}} & \overline{\overline{Z}}^{\vec{M}^{G\rightleftharpoons A}\vec{M}_1^{A\rightleftharpoons F}} & \overline{\overline{Z}}^{\vec{M}^{G\rightleftharpoons A}\vec{M}_{12}^{A\rightleftharpoons A}} & 0 \\
\overline{\overline{Z}}^{\vec{J}_1^{A\rightleftharpoons F}\vec{J}^{G\rightleftharpoons A}} & \overline{\overline{Z}}^{\vec{J}_1^{A\rightleftharpoons F}\vec{J}_1^{A\rightleftharpoons F}} & \overline{\overline{Z}}^{\vec{J}_1^{A\rightleftharpoons F}\vec{J}_{12}^{A\rightleftharpoons A}} & \overline{\overline{Z}}^{\vec{J}_1^{A\rightleftharpoons F}\vec{J}_2^{A\rightleftharpoons F}} & \overline{\overline{Z}}^{\vec{J}_1^{A\rightleftharpoons F}\vec{J}_1^{A}} & 0 & \overline{\overline{Z}}^{\vec{J}_1^{A\rightleftharpoons F}\vec{J}^{F}} & \overline{\overline{Z}}^{\vec{J}_1^{A\rightleftharpoons F}\vec{M}^{G\rightleftharpoons A}} & \overline{\overline{Z}}^{\vec{J}_1^{A\rightleftharpoons F}\vec{M}_1^{A\rightleftharpoons F}} & \overline{\overline{Z}}^{\vec{J}_1^{A\rightleftharpoons F}\vec{M}_{12}^{A\rightleftharpoons A}} & \overline{\overline{Z}}^{\vec{J}_1^{A\rightleftharpoons F}\vec{M}_2^{A\rightleftharpoons F}} \\
\overline{\overline{Z}}^{\vec{M}_1^{A\rightleftharpoons F}\vec{J}^{G\rightleftharpoons A}} & \overline{\overline{Z}}^{\vec{M}_1^{A\rightleftharpoons F}\vec{J}_1^{A\rightleftharpoons F}} & \overline{\overline{Z}}^{\vec{M}_1^{A\rightleftharpoons F}\vec{J}_{12}^{A\rightleftharpoons A}} & \overline{\overline{Z}}^{\vec{M}_1^{A\rightleftharpoons F}\vec{J}_2^{A\rightleftharpoons F}} & \overline{\overline{Z}}^{\vec{M}_1^{A\rightleftharpoons F}\vec{J}_1^{A}} & 0 & \overline{\overline{Z}}^{\vec{M}_1^{A\rightleftharpoons F}\vec{J}^{F}} & \overline{\overline{Z}}^{\vec{M}_1^{A\rightleftharpoons F}\vec{M}^{G\rightleftharpoons A}} & \overline{\overline{Z}}^{\vec{M}_1^{A\rightleftharpoons F}\vec{M}_1^{A\rightleftharpoons F}} & \overline{\overline{Z}}^{\vec{M}_1^{A\rightleftharpoons F}\vec{M}_{12}^{A\rightleftharpoons A}} & \overline{\overline{Z}}^{\vec{M}_1^{A\rightleftharpoons F}\vec{M}_2^{A\rightleftharpoons F}} \\
\overline{\overline{Z}}^{\vec{J}_{12}^{A\rightleftharpoons A}\vec{J}^{G\rightleftharpoons A}} & \overline{\overline{Z}}^{\vec{J}_{12}^{A\rightleftharpoons A}\vec{J}_1^{A\rightleftharpoons F}} & \overline{\overline{Z}}^{\vec{J}_{12}^{A\rightleftharpoons A}\vec{J}_{12}^{A\rightleftharpoons A}} & \overline{\overline{Z}}^{\vec{J}_{12}^{A\rightleftharpoons A}\vec{J}_2^{A\rightleftharpoons F}} & \overline{\overline{Z}}^{\vec{J}_{12}^{A\rightleftharpoons A}\vec{J}_1^{A}} & \overline{\overline{Z}}^{\vec{J}_{12}^{A\rightleftharpoons A}\vec{J}_2^{A}} & 0 & \overline{\overline{Z}}^{\vec{J}_{12}^{A\rightleftharpoons A}\vec{M}^{G\rightleftharpoons A}} & \overline{\overline{Z}}^{\vec{J}_{12}^{A\rightleftharpoons A}\vec{M}_1^{A\rightleftharpoons F}} & \overline{\overline{Z}}^{\vec{J}_{12}^{A\rightleftharpoons A}\vec{M}_{12}^{A\rightleftharpoons A}} & \overline{\overline{Z}}^{\vec{J}_{12}^{A\rightleftharpoons A}\vec{M}_2^{A\rightleftharpoons F}} \\
\overline{\overline{Z}}^{\vec{M}_{12}^{A\rightleftharpoons A}\vec{J}^{G\rightleftharpoons A}} & \overline{\overline{Z}}^{\vec{M}_{12}^{A\rightleftharpoons A}\vec{J}_1^{A\rightleftharpoons F}} & \overline{\overline{Z}}^{\vec{M}_{12}^{A\rightleftharpoons A}\vec{J}_{12}^{A\rightleftharpoons A}} & \overline{\overline{Z}}^{\vec{M}_{12}^{A\rightleftharpoons A}\vec{J}_2^{A\rightleftharpoons F}} & \overline{\overline{Z}}^{\vec{M}_{12}^{A\rightleftharpoons A}\vec{J}_1^{A}} & \overline{\overline{Z}}^{\vec{M}_{12}^{A\rightleftharpoons A}\vec{J}_2^{A}} & 0 & \overline{\overline{Z}}^{\vec{M}_{12}^{A\rightleftharpoons A}\vec{M}^{G\rightleftharpoons A}} & \overline{\overline{Z}}^{\vec{M}_{12}^{A\rightleftharpoons A}\vec{M}_1^{A\rightleftharpoons F}} & \overline{\overline{Z}}^{\vec{M}_{12}^{A\rightleftharpoons A}\vec{M}_{12}^{A\rightleftharpoons A}} & \overline{\overline{Z}}^{\vec{M}_{12}^{A\rightleftharpoons A}\vec{M}_2^{A\rightleftharpoons F}} \\
0 & \overline{\overline{Z}}^{\vec{J}_2^{A\rightleftharpoons F}\vec{J}_1^{A\rightleftharpoons F}} & \overline{\overline{Z}}^{\vec{J}_2^{A\rightleftharpoons F}\vec{J}_{12}^{A\rightleftharpoons A}} & \overline{\overline{Z}}^{\vec{J}_2^{A\rightleftharpoons F}\vec{J}_2^{A\rightleftharpoons F}} & 0 & \overline{\overline{Z}}^{\vec{J}_2^{A\rightleftharpoons F}\vec{J}_2^{A}} & \overline{\overline{Z}}^{\vec{J}_2^{A\rightleftharpoons F}\vec{J}^{F}} & 0 & \overline{\overline{Z}}^{\vec{J}_2^{A\rightleftharpoons F}\vec{M}_1^{A\rightleftharpoons F}} & \overline{\overline{Z}}^{\vec{J}_2^{A\rightleftharpoons F}\vec{M}_{12}^{A\rightleftharpoons A}} & \overline{\overline{Z}}^{\vec{J}_2^{A\rightleftharpoons F}\vec{M}_2^{A\rightleftharpoons F}} \\
0 & \overline{\overline{Z}}^{\vec{M}_2^{A\rightleftharpoons F}\vec{J}_1^{A\rightleftharpoons F}} & \overline{\overline{Z}}^{\vec{M}_2^{A\rightleftharpoons F}\vec{J}_{12}^{A\rightleftharpoons A}} & \overline{\overline{Z}}^{\vec{M}_2^{A\rightleftharpoons F}\vec{J}_2^{A\rightleftharpoons F}} & 0 & \overline{\overline{Z}}^{\vec{M}_2^{A\rightleftharpoons F}\vec{J}_2^{A}} & \overline{\overline{Z}}^{\vec{M}_2^{A\rightleftharpoons F}\vec{J}^{F}} & 0 & \overline{\overline{Z}}^{\vec{M}_2^{A\rightleftharpoons F}\vec{M}_1^{A\rightleftharpoons F}} & \overline{\overline{Z}}^{\vec{M}_2^{A\rightleftharpoons F}\vec{M}_{12}^{A\rightleftharpoons A}} & \overline{\overline{Z}}^{\vec{M}_2^{A\rightleftharpoons F}\vec{M}_2^{A\rightleftharpoons F}} \\
\overline{\overline{Z}}^{\vec{J}_1^{A}\vec{J}^{G\rightleftharpoons A}} & \overline{\overline{Z}}^{\vec{J}_1^{A}\vec{J}_1^{A\rightleftharpoons F}} & \overline{\overline{Z}}^{\vec{J}_1^{A}\vec{J}_{12}^{A\rightleftharpoons A}} & 0 & \overline{\overline{Z}}^{\vec{J}_1^{A}\vec{J}_1^{A}} & 0 & 0 & \overline{\overline{Z}}^{\vec{J}_1^{A}\vec{M}^{G\rightleftharpoons A}} & \overline{\overline{Z}}^{\vec{J}_1^{A}\vec{M}_1^{A\rightleftharpoons F}} & \overline{\overline{Z}}^{\vec{J}_1^{A}\vec{M}_{12}^{A\rightleftharpoons A}} & 0 \\
0 & 0 & \overline{\overline{Z}}^{\vec{J}_2^{A}\vec{J}_{12}^{A\rightleftharpoons A}} & \overline{\overline{Z}}^{\vec{J}_2^{A}\vec{J}_2^{A\rightleftharpoons F}} & 0 & \overline{\overline{Z}}^{\vec{J}_2^{A}\vec{J}_2^{A}} & 0 & 0 & 0 & \overline{\overline{Z}}^{\vec{J}_2^{A}\vec{M}_{12}^{A\rightleftharpoons A}} & \overline{\overline{Z}}^{\vec{J}_2^{A}\vec{M}_2^{A\rightleftharpoons F}} \\
0 & \overline{\overline{Z}}^{\vec{J}^{F}\vec{J}_1^{A\rightleftharpoons F}} & 0 & \overline{\overline{Z}}^{\vec{J}^{F}\vec{J}_2^{A\rightleftharpoons F}} & 0 & 0 & \overline{\overline{Z}}^{\vec{J}^{F}\vec{J}^{F}} & 0 & \overline{\overline{Z}}^{\vec{J}^{F}\vec{M}_1^{A\rightleftharpoons F}} & 0 & \overline{\overline{Z}}^{\vec{J}^{F}\vec{M}_2^{A\rightleftharpoons F}}
\end{bmatrix} \tag{D-83a}$$

$$\overline{\overline{\Psi}}_{\text{FCE}}^{\text{DoM}} = \begin{bmatrix}
\overline{\overline{Z}}^{\vec{J}^{G\rightleftharpoons A}\vec{J}^{G\rightleftharpoons A}} & \overline{\overline{Z}}^{\vec{J}^{G\rightleftharpoons A}\vec{J}_1^{A\rightleftharpoons F}} & \overline{\overline{Z}}^{\vec{J}^{G\rightleftharpoons A}\vec{J}_{12}^{A\rightleftharpoons A}} & 0 & \overline{\overline{Z}}^{\vec{J}^{G\rightleftharpoons A}\vec{J}_1^{A}} & 0 & 0 & \overline{\overline{Z}}^{\vec{J}^{G\rightleftharpoons A}\vec{M}^{G\rightleftharpoons A}} & \overline{\overline{Z}}^{\vec{J}^{G\rightleftharpoons A}\vec{M}_1^{A\rightleftharpoons F}} & \overline{\overline{Z}}^{\vec{J}^{G\rightleftharpoons A}\vec{M}_{12}^{A\rightleftharpoons A}} & 0 \\
\overline{\overline{Z}}^{\vec{J}_1^{A\rightleftharpoons F}\vec{J}^{G\rightleftharpoons A}} & \overline{\overline{Z}}^{\vec{J}_1^{A\rightleftharpoons F}\vec{J}_1^{A\rightleftharpoons F}} & \overline{\overline{Z}}^{\vec{J}_1^{A\rightleftharpoons F}\vec{J}_{12}^{A\rightleftharpoons A}} & \overline{\overline{Z}}^{\vec{J}_1^{A\rightleftharpoons F}\vec{J}_2^{A\rightleftharpoons F}} & \overline{\overline{Z}}^{\vec{J}_1^{A\rightleftharpoons F}\vec{J}_1^{A}} & 0 & \overline{\overline{Z}}^{\vec{J}_1^{A\rightleftharpoons F}\vec{J}^{F}} & \overline{\overline{Z}}^{\vec{J}_1^{A\rightleftharpoons F}\vec{M}^{G\rightleftharpoons A}} & \overline{\overline{Z}}^{\vec{J}_1^{A\rightleftharpoons F}\vec{M}_1^{A\rightleftharpoons F}} & \overline{\overline{Z}}^{\vec{J}_1^{A\rightleftharpoons F}\vec{M}_{12}^{A\rightleftharpoons A}} & \overline{\overline{Z}}^{\vec{J}_1^{A\rightleftharpoons F}\vec{M}_2^{A\rightleftharpoons F}} \\
\overline{\overline{Z}}^{\vec{M}_1^{A\rightleftharpoons F}\vec{J}^{G\rightleftharpoons A}} & \overline{\overline{Z}}^{\vec{M}_1^{A\rightleftharpoons F}\vec{J}_1^{A\rightleftharpoons F}} & \overline{\overline{Z}}^{\vec{M}_1^{A\rightleftharpoons F}\vec{J}_{12}^{A\rightleftharpoons A}} & \overline{\overline{Z}}^{\vec{M}_1^{A\rightleftharpoons F}\vec{J}_2^{A\rightleftharpoons F}} & \overline{\overline{Z}}^{\vec{M}_1^{A\rightleftharpoons F}\vec{J}_1^{A}} & 0 & \overline{\overline{Z}}^{\vec{M}_1^{A\rightleftharpoons F}\vec{J}^{F}} & \overline{\overline{Z}}^{\vec{M}_1^{A\rightleftharpoons F}\vec{M}^{G\rightleftharpoons A}} & \overline{\overline{Z}}^{\vec{M}_1^{A\rightleftharpoons F}\vec{M}_1^{A\rightleftharpoons F}} & \overline{\overline{Z}}^{\vec{M}_1^{A\rightleftharpoons F}\vec{M}_{12}^{A\rightleftharpoons A}} & \overline{\overline{Z}}^{\vec{M}_1^{A\rightleftharpoons F}\vec{M}_2^{A\rightleftharpoons F}} \\
\overline{\overline{Z}}^{\vec{J}_{12}^{A\rightleftharpoons A}\vec{J}^{G\rightleftharpoons A}} & \overline{\overline{Z}}^{\vec{J}_{12}^{A\rightleftharpoons A}\vec{J}_1^{A\rightleftharpoons F}} & \overline{\overline{Z}}^{\vec{J}_{12}^{A\rightleftharpoons A}\vec{J}_{12}^{A\rightleftharpoons A}} & \overline{\overline{Z}}^{\vec{J}_{12}^{A\rightleftharpoons A}\vec{J}_2^{A\rightleftharpoons F}} & \overline{\overline{Z}}^{\vec{J}_{12}^{A\rightleftharpoons A}\vec{J}_1^{A}} & \overline{\overline{Z}}^{\vec{J}_{12}^{A\rightleftharpoons A}\vec{J}_2^{A}} & 0 & \overline{\overline{Z}}^{\vec{J}_{12}^{A\rightleftharpoons A}\vec{M}^{G\rightleftharpoons A}} & \overline{\overline{Z}}^{\vec{J}_{12}^{A\rightleftharpoons A}\vec{M}_1^{A\rightleftharpoons F}} & \overline{\overline{Z}}^{\vec{J}_{12}^{A\rightleftharpoons A}\vec{M}_{12}^{A\rightleftharpoons A}} & \overline{\overline{Z}}^{\vec{J}_{12}^{A\rightleftharpoons A}\vec{M}_2^{A\rightleftharpoons F}} \\
\overline{\overline{Z}}^{\vec{M}_{12}^{A\rightleftharpoons A}\vec{J}^{G\rightleftharpoons A}} & \overline{\overline{Z}}^{\vec{M}_{12}^{A\rightleftharpoons A}\vec{J}_1^{A\rightleftharpoons F}} & \overline{\overline{Z}}^{\vec{M}_{12}^{A\rightleftharpoons A}\vec{J}_{12}^{A\rightleftharpoons A}} & \overline{\overline{Z}}^{\vec{M}_{12}^{A\rightleftharpoons A}\vec{J}_2^{A\rightleftharpoons F}} & \overline{\overline{Z}}^{\vec{M}_{12}^{A\rightleftharpoons A}\vec{J}_1^{A}} & \overline{\overline{Z}}^{\vec{M}_{12}^{A\rightleftharpoons A}\vec{J}_2^{A}} & 0 & \overline{\overline{Z}}^{\vec{M}_{12}^{A\rightleftharpoons A}\vec{M}^{G\rightleftharpoons A}} & \overline{\overline{Z}}^{\vec{M}_{12}^{A\rightleftharpoons A}\vec{M}_1^{A\rightleftharpoons F}} & \overline{\overline{Z}}^{\vec{M}_{12}^{A\rightleftharpoons A}\vec{M}_{12}^{A\rightleftharpoons A}} & \overline{\overline{Z}}^{\vec{M}_{12}^{A\rightleftharpoons A}\vec{M}_2^{A\rightleftharpoons F}} \\
0 & \overline{\overline{Z}}^{\vec{J}_2^{A\rightleftharpoons F}\vec{J}_1^{A\rightleftharpoons F}} & \overline{\overline{Z}}^{\vec{J}_2^{A\rightleftharpoons F}\vec{J}_{12}^{A\rightleftharpoons A}} & \overline{\overline{Z}}^{\vec{J}_2^{A\rightleftharpoons F}\vec{J}_2^{A\rightleftharpoons F}} & 0 & \overline{\overline{Z}}^{\vec{J}_2^{A\rightleftharpoons F}\vec{J}_2^{A}} & \overline{\overline{Z}}^{\vec{J}_2^{A\rightleftharpoons F}\vec{J}^{F}} & 0 & \overline{\overline{Z}}^{\vec{J}_2^{A\rightleftharpoons F}\vec{M}_1^{A\rightleftharpoons F}} & \overline{\overline{Z}}^{\vec{J}_2^{A\rightleftharpoons F}\vec{M}_{12}^{A\rightleftharpoons A}} & \overline{\overline{Z}}^{\vec{J}_2^{A\rightleftharpoons F}\vec{M}_2^{A\rightleftharpoons F}} \\
0 & \overline{\overline{Z}}^{\vec{M}_2^{A\rightleftharpoons F}\vec{J}_1^{A\rightleftharpoons F}} & \overline{\overline{Z}}^{\vec{M}_2^{A\rightleftharpoons F}\vec{J}_{12}^{A\rightleftharpoons A}} & \overline{\overline{Z}}^{\vec{M}_2^{A\rightleftharpoons F}\vec{J}_2^{A\rightleftharpoons F}} & 0 & \overline{\overline{Z}}^{\vec{M}_2^{A\rightleftharpoons F}\vec{J}_2^{A}} & \overline{\overline{Z}}^{\vec{M}_2^{A\rightleftharpoons F}\vec{J}^{F}} & 0 & \overline{\overline{Z}}^{\vec{M}_2^{A\rightleftharpoons F}\vec{M}_1^{A\rightleftharpoons F}} & \overline{\overline{Z}}^{\vec{M}_2^{A\rightleftharpoons F}\vec{M}_{12}^{A\rightleftharpoons A}} & \overline{\overline{Z}}^{\vec{M}_2^{A\rightleftharpoons F}\vec{M}_2^{A\rightleftharpoons F}} \\
\overline{\overline{Z}}^{\vec{J}_1^{A}\vec{J}^{G\rightleftharpoons A}} & \overline{\overline{Z}}^{\vec{J}_1^{A}\vec{J}_1^{A\rightleftharpoons F}} & \overline{\overline{Z}}^{\vec{J}_1^{A}\vec{J}_{12}^{A\rightleftharpoons A}} & 0 & \overline{\overline{Z}}^{\vec{J}_1^{A}\vec{J}_1^{A}} & 0 & 0 & \overline{\overline{Z}}^{\vec{J}_1^{A}\vec{M}^{G\rightleftharpoons A}} & \overline{\overline{Z}}^{\vec{J}_1^{A}\vec{M}_1^{A\rightleftharpoons F}} & \overline{\overline{Z}}^{\vec{J}_1^{A}\vec{M}_{12}^{A\rightleftharpoons A}} & 0 \\
0 & 0 & \overline{\overline{Z}}^{\vec{J}_2^{A}\vec{J}_{12}^{A\rightleftharpoons A}} & \overline{\overline{Z}}^{\vec{J}_2^{A}\vec{J}_2^{A\rightleftharpoons F}} & 0 & \overline{\overline{Z}}^{\vec{J}_2^{A}\vec{J}_2^{A}} & 0 & 0 & 0 & \overline{\overline{Z}}^{\vec{J}_2^{A}\vec{M}_{12}^{A\rightleftharpoons A}} & \overline{\overline{Z}}^{\vec{J}_2^{A}\vec{M}_2^{A\rightleftharpoons F}} \\
0 & \overline{\overline{Z}}^{\vec{J}^{F}\vec{J}_1^{A\rightleftharpoons F}} & 0 & \overline{\overline{Z}}^{\vec{J}^{F}\vec{J}_2^{A\rightleftharpoons F}} & 0 & 0 & \overline{\overline{Z}}^{\vec{J}^{F}\vec{J}^{F}} & 0 & \overline{\overline{Z}}^{\vec{J}^{F}\vec{M}_1^{A\rightleftharpoons F}} & 0 & \overline{\overline{Z}}^{\vec{J}^{F}\vec{M}_2^{A\rightleftharpoons F}}
\end{bmatrix} \tag{D-83b}$$

The power quadratic form matrix $\overline{\overline{P}}^{G\rightleftharpoons A}$ used in Eq. (6-95) is as follows:

$$\overline{\overline{P}}^{G\rightleftharpoons A} = \overline{\overline{P}}_{\text{curAV}}^{G\rightleftharpoons A} \text{ or } \overline{\overline{P}}_{\text{intAV}}^{G\rightleftharpoons A} \tag{D-84}$$

in which

$$\overline{\overline{P}}_{\text{curAV}}^{G\rightleftharpoons A} = \begin{bmatrix}
0 & 0 & 0 & 0 & 0 & 0 & 0 & \overline{\overline{C}}^{\vec{J}^{G\rightleftharpoons A}\vec{M}^{G\rightleftharpoons A}} & 0 & 0 & 0 \\
0 & 0 & 0 & 0 & 0 & 0 & 0 & 0 & 0 & 0 & 0 \\
0 & 0 & 0 & 0 & 0 & 0 & 0 & 0 & 0 & 0 & 0 \\
0 & 0 & 0 & 0 & 0 & 0 & 0 & 0 & 0 & 0 & 0 \\
0 & 0 & 0 & 0 & 0 & 0 & 0 & 0 & 0 & 0 & 0 \\
0 & 0 & 0 & 0 & 0 & 0 & 0 & 0 & 0 & 0 & 0 \\
0 & 0 & 0 & 0 & 0 & 0 & 0 & 0 & 0 & 0 & 0 \\
0 & 0 & 0 & 0 & 0 & 0 & 0 & 0 & 0 & 0 & 0 \\
0 & 0 & 0 & 0 & 0 & 0 & 0 & 0 & 0 & 0 & 0 \\
0 & 0 & 0 & 0 & 0 & 0 & 0 & 0 & 0 & 0 & 0 \\
0 & 0 & 0 & 0 & 0 & 0 & 0 & 0 & 0 & 0 & 0
\end{bmatrix} \tag{D-85}$$

corresponding to the first equality in Eq. (6-94), and

$$\overline{\overline{P}}_{\text{intAV}}^{\text{G}\rightleftharpoons\text{A}}=\begin{bmatrix} \overline{\overline{P}}^{\vec{j}^{\text{G}\rightleftharpoons\text{A}}\vec{j}^{\text{G}\rightleftharpoons\text{A}}} & \overline{\overline{P}}^{\vec{j}^{\text{G}\rightleftharpoons\text{A}}\vec{J}_1^{\text{A}\rightleftharpoons\text{F}}} & \overline{\overline{P}}^{\vec{j}^{\text{G}\rightleftharpoons\text{A}}\vec{J}_{12}^{\text{A}\rightleftharpoons\text{A}}} & 0 & \overline{\overline{P}}^{\vec{j}^{\text{G}\rightleftharpoons\text{A}}\vec{J}_1^{\text{A}}} & 0 & 0 & \overline{\overline{P}}^{\vec{j}^{\text{G}\rightleftharpoons\text{A}}\vec{M}^{\text{G}\rightleftharpoons\text{A}}} & \overline{\overline{P}}^{\vec{j}^{\text{G}\rightleftharpoons\text{A}}\vec{M}_1^{\text{A}\rightleftharpoons\text{F}}} & \overline{\overline{P}}^{\vec{j}^{\text{G}\rightleftharpoons\text{A}}\vec{M}_{12}^{\text{A}\rightleftharpoons\text{A}}} & 0 \\ 0 & 0 & 0 & 0 & 0 & 0 & 0 & 0 & 0 & 0 & 0 \\ 0 & 0 & 0 & 0 & 0 & 0 & 0 & 0 & 0 & 0 & 0 \\ 0 & 0 & 0 & 0 & 0 & 0 & 0 & 0 & 0 & 0 & 0 \\ 0 & 0 & 0 & 0 & 0 & 0 & 0 & 0 & 0 & 0 & 0 \\ 0 & 0 & 0 & 0 & 0 & 0 & 0 & 0 & 0 & 0 & 0 \\ 0 & 0 & 0 & 0 & 0 & 0 & 0 & 0 & 0 & 0 & 0 \\ 0 & 0 & 0 & 0 & 0 & 0 & 0 & 0 & 0 & 0 & 0 \\ 0 & 0 & 0 & 0 & 0 & 0 & 0 & 0 & 0 & 0 & 0 \\ 0 & 0 & 0 & 0 & 0 & 0 & 0 & 0 & 0 & 0 & 0 \\ 0 & 0 & 0 & 0 & 0 & 0 & 0 & 0 & 0 & 0 & 0 \end{bmatrix} \tag{D-86a}$$

corresponding to the second equality in Eq. (6-94), and

$$\overline{\overline{P}}_{\text{intAV}}^{\text{G}\rightleftharpoons\text{A}}=\begin{bmatrix} 0 & 0 & 0 & 0 & 0 & 0 & 0 & 0 & 0 & 0 & 0 \\ 0 & 0 & 0 & 0 & 0 & 0 & 0 & 0 & 0 & 0 & 0 \\ 0 & 0 & 0 & 0 & 0 & 0 & 0 & 0 & 0 & 0 & 0 \\ 0 & 0 & 0 & 0 & 0 & 0 & 0 & 0 & 0 & 0 & 0 \\ 0 & 0 & 0 & 0 & 0 & 0 & 0 & 0 & 0 & 0 & 0 \\ 0 & 0 & 0 & 0 & 0 & 0 & 0 & 0 & 0 & 0 & 0 \\ 0 & 0 & 0 & 0 & 0 & 0 & 0 & 0 & 0 & 0 & 0 \\ \overline{\overline{P}}^{\vec{j}^{\text{G}\rightleftharpoons\text{A}}\vec{j}^{\text{G}\rightleftharpoons\text{A}}} & \overline{\overline{P}}^{\vec{j}^{\text{G}\rightleftharpoons\text{A}}\vec{J}_1^{\text{A}\rightleftharpoons\text{F}}} & \overline{\overline{P}}^{\vec{j}^{\text{G}\rightleftharpoons\text{A}}\vec{J}_{12}^{\text{A}\rightleftharpoons\text{A}}} & 0 & \overline{\overline{P}}^{\vec{j}^{\text{G}\rightleftharpoons\text{A}}\vec{J}_1^{\text{A}}} & 0 & 0 & \overline{\overline{P}}^{\vec{j}^{\text{G}\rightleftharpoons\text{A}}\vec{M}^{\text{G}\rightleftharpoons\text{A}}} & \overline{\overline{P}}^{\vec{j}^{\text{G}\rightleftharpoons\text{A}}\vec{M}_1^{\text{A}\rightleftharpoons\text{F}}} & \overline{\overline{P}}^{\vec{j}^{\text{G}\rightleftharpoons\text{A}}\vec{M}_{12}^{\text{A}\rightleftharpoons\text{A}}} & 0 \\ 0 & 0 & 0 & 0 & 0 & 0 & 0 & 0 & 0 & 0 & 0 \\ 0 & 0 & 0 & 0 & 0 & 0 & 0 & 0 & 0 & 0 & 0 \\ 0 & 0 & 0 & 0 & 0 & 0 & 0 & 0 & 0 & 0 & 0 \end{bmatrix}^{\dagger} \tag{D-86b}$$

corresponding to the third equality in Eq. (6-94), where the elements of the sub-matrices in the above Eqs. (D-85)~(D-86b) are as follows:

$$c_{\xi\zeta}^{\vec{j}^{\text{G}\rightleftharpoons\text{A}}\vec{M}^{\text{G}\rightleftharpoons\text{A}}} = (1/2)\left\langle \hat{n}^{\rightarrow\text{A}}\times\vec{b}_{\xi}^{\vec{j}^{\text{G}\rightleftharpoons\text{A}}},\vec{b}_{\zeta}^{\vec{M}^{\text{G}\rightleftharpoons\text{A}}}\right\rangle_{\mathbb{S}^{\text{G}\rightleftharpoons\text{A}}} \tag{D-87}$$

and

$$p_{\xi\zeta}^{\vec{j}^{\text{G}\rightleftharpoons\text{A}}\vec{j}^{\text{G}\rightleftharpoons\text{A}}} = -(1/2)\left\langle \vec{b}_{\xi}^{\vec{j}^{\text{G}\rightleftharpoons\text{A}}},\mathcal{E}_1\left(\vec{b}_{\zeta}^{\vec{j}^{\text{G}\rightleftharpoons\text{A}}}\right)\right\rangle_{\mathbb{S}_1^{\text{G}\rightleftharpoons\text{A}}} \tag{D-88a}$$

$$p_{\xi\zeta}^{\vec{j}^{\text{G}\rightleftharpoons\text{A}}\vec{J}_1^{\text{A}\rightleftharpoons\text{F}}} = -(1/2)\left\langle \vec{b}_{\xi}^{\vec{j}^{\text{G}\rightleftharpoons\text{A}}},\mathcal{E}_1\left(\vec{b}_{\zeta}^{\vec{J}_1^{\text{A}\rightleftharpoons\text{F}}}\right)\right\rangle_{\mathbb{S}_1^{\text{G}\rightleftharpoons\text{A}}} \tag{D-88b}$$

$$p_{\xi\zeta}^{\vec{j}^{\text{G}\rightleftharpoons\text{A}}\vec{J}_{12}^{\text{A}\rightleftharpoons\text{A}}} = -(1/2)\left\langle \vec{b}_{\xi}^{\vec{j}^{\text{G}\rightleftharpoons\text{A}}},\mathcal{E}_1\left(\vec{b}_{\zeta}^{\vec{J}_{12}^{\text{A}\rightleftharpoons\text{A}}}\right)\right\rangle_{\mathbb{S}_1^{\text{G}\rightleftharpoons\text{A}}} \tag{D-88c}$$

$$p_{\xi\zeta}^{\vec{j}^{\text{G}\rightleftharpoons\text{A}}\vec{J}_1^{\text{A}}} = -(1/2)\left\langle \vec{b}_{\xi}^{\vec{j}^{\text{G}\rightleftharpoons\text{A}}},\mathcal{E}_1\left(\vec{b}_{\zeta}^{\vec{J}_1^{\text{A}}}\right)\right\rangle_{\mathbb{S}_1^{\text{G}\rightleftharpoons\text{A}}} \tag{D-88d}$$

$$p_{\xi\zeta}^{\vec{j}^{\text{G}\rightleftharpoons\text{A}}\vec{M}^{\text{G}\rightleftharpoons\text{A}}} = -(1/2)\left\langle \vec{b}_{\xi}^{\vec{j}^{\text{G}\rightleftharpoons\text{A}}},\mathcal{E}_1\left(\vec{b}_{\zeta}^{\vec{M}^{\text{G}\rightleftharpoons\text{A}}}\right)\right\rangle_{\mathbb{S}_1^{\text{G}\rightleftharpoons\text{A}}} \tag{D-88e}$$

$$p_{\xi\zeta}^{\vec{j}^{\text{G}\rightleftharpoons\text{A}}\vec{M}_1^{\text{A}\rightleftharpoons\text{F}}} = -(1/2)\left\langle \vec{b}_{\xi}^{\vec{j}^{\text{G}\rightleftharpoons\text{A}}},\mathcal{E}_1\left(\vec{b}_{\zeta}^{\vec{M}_1^{\text{A}\rightleftharpoons\text{F}}}\right)\right\rangle_{\mathbb{S}_1^{\text{G}\rightleftharpoons\text{A}}} \tag{D-88f}$$

$$p_{\xi\zeta}^{\vec{j}^{\text{G}\rightleftharpoons\text{A}}\vec{M}_{12}^{\text{A}\rightleftharpoons\text{A}}} = -(1/2)\left\langle \vec{b}_{\xi}^{\vec{j}^{\text{G}\rightleftharpoons\text{A}}},\mathcal{E}_1\left(\vec{b}_{\zeta}^{\vec{M}_{12}^{\text{A}\rightleftharpoons\text{A}}}\right)\right\rangle_{\mathbb{S}_1^{\text{G}\rightleftharpoons\text{A}}} \tag{D-88g}$$

and

$$p_{\xi\zeta}^{\vec{M}^{\mathrm{G}\rightleftharpoons\mathrm{A}}\vec{J}^{\mathrm{G}\rightleftharpoons\mathrm{A}}} = -(1/2)\left\langle \vec{b}_{\xi}^{\vec{M}^{\mathrm{G}\rightleftharpoons\mathrm{A}}}, \mathcal{H}_1\left(\vec{b}_{\zeta}^{\vec{J}^{\mathrm{G}\rightleftharpoons\mathrm{A}}}\right)\right\rangle_{\mathbb{S}_1^{\mathrm{G}\rightleftharpoons\mathrm{A}}} \quad \text{(D-88h)}$$

$$p_{\xi\zeta}^{\vec{M}^{\mathrm{G}\rightleftharpoons\mathrm{A}}\vec{J}_1^{\mathrm{A}\rightleftharpoons\mathrm{F}}} = -(1/2)\left\langle \vec{b}_{\xi}^{\vec{M}^{\mathrm{G}\rightleftharpoons\mathrm{A}}}, \mathcal{H}_1\left(\vec{b}_{\zeta}^{\vec{J}_1^{\mathrm{A}\rightleftharpoons\mathrm{F}}}\right)\right\rangle_{\mathbb{S}_1^{\mathrm{G}\rightleftharpoons\mathrm{A}}} \quad \text{(D-88i)}$$

$$p_{\xi\zeta}^{\vec{M}^{\mathrm{G}\rightleftharpoons\mathrm{A}}\vec{J}_{12}^{\mathrm{A}\rightleftharpoons\mathrm{A}}} = -(1/2)\left\langle \vec{b}_{\xi}^{\vec{M}^{\mathrm{G}\rightleftharpoons\mathrm{A}}}, \mathcal{H}_1\left(\vec{b}_{\zeta}^{\vec{J}_{12}^{\mathrm{A}\rightleftharpoons\mathrm{A}}}\right)\right\rangle_{\mathbb{S}_1^{\mathrm{G}\rightleftharpoons\mathrm{A}}} \quad \text{(D-88j)}$$

$$p_{\xi\zeta}^{\vec{M}^{\mathrm{G}\rightleftharpoons\mathrm{A}}\vec{J}_1^{\mathrm{A}}} = -(1/2)\left\langle \vec{b}_{\xi}^{\vec{M}^{\mathrm{G}\rightleftharpoons\mathrm{A}}}, \mathcal{H}_1\left(\vec{b}_{\zeta}^{\vec{J}_1^{\mathrm{A}}}\right)\right\rangle_{\mathbb{S}_1^{\mathrm{G}\rightleftharpoons\mathrm{A}}} \quad \text{(D-88k)}$$

$$p_{\xi\zeta}^{\vec{M}^{\mathrm{G}\rightleftharpoons\mathrm{A}}\vec{M}^{\mathrm{G}\rightleftharpoons\mathrm{A}}} = -(1/2)\left\langle \vec{b}_{\xi}^{\vec{M}^{\mathrm{G}\rightleftharpoons\mathrm{A}}}, \mathcal{H}_1\left(\vec{b}_{\zeta}^{\vec{M}^{\mathrm{G}\rightleftharpoons\mathrm{A}}}\right)\right\rangle_{\mathbb{S}_1^{\mathrm{G}\rightleftharpoons\mathrm{A}}} \quad \text{(D-88l)}$$

$$p_{\xi\zeta}^{\vec{M}^{\mathrm{G}\rightleftharpoons\mathrm{A}}\vec{M}_1^{\mathrm{A}\rightleftharpoons\mathrm{F}}} = -(1/2)\left\langle \vec{b}_{\xi}^{\vec{M}^{\mathrm{G}\rightleftharpoons\mathrm{A}}}, \mathcal{H}_1\left(\vec{b}_{\zeta}^{\vec{M}_1^{\mathrm{A}\rightleftharpoons\mathrm{F}}}\right)\right\rangle_{\mathbb{S}_1^{\mathrm{G}\rightleftharpoons\mathrm{A}}} \quad \text{(D-88m)}$$

$$p_{\xi\zeta}^{\vec{M}^{\mathrm{G}\rightleftharpoons\mathrm{A}}\vec{M}_{12}^{\mathrm{A}\rightleftharpoons\mathrm{A}}} = -(1/2)\left\langle \vec{b}_{\xi}^{\vec{M}^{\mathrm{G}\rightleftharpoons\mathrm{A}}}, \mathcal{H}_1\left(\vec{b}_{\zeta}^{\vec{M}_{12}^{\mathrm{A}\rightleftharpoons\mathrm{A}}}\right)\right\rangle_{\mathbb{S}_1^{\mathrm{G}\rightleftharpoons\mathrm{A}}} \quad \text{(D-88n)}$$

## D7 Some Detailed Formulations Related to Sec. 6.6

The formulations used to calculate the elements of the matrices in Eqs. (6-111a)~(6-116) are as follows:

$$z_{\xi\zeta}^{\vec{M}^{\mathrm{G}\rightleftharpoons\mathrm{A}}\vec{J}^{\mathrm{G}\rightleftharpoons\mathrm{A}}} = \left\langle \vec{b}_{\xi}^{\vec{M}^{\mathrm{G}\rightleftharpoons\mathrm{A}}}, \hat{n}^{\rightarrow\mathrm{A}} \times \frac{1}{2}\vec{b}_{\zeta}^{\vec{J}^{\mathrm{G}\rightleftharpoons\mathrm{A}}} + \mathrm{P.V.}\,\mathcal{K}_0\left(\vec{b}_{\zeta}^{\vec{J}^{\mathrm{G}\rightleftharpoons\mathrm{A}}}\right)\right\rangle_{\mathbb{S}^{\mathrm{G}\rightleftharpoons\mathrm{A}}} \quad \text{(D-89a)}$$

$$z_{\xi\zeta}^{\vec{M}^{\mathrm{G}\rightleftharpoons\mathrm{A}}\vec{J}_1^{\mathrm{A}\rightleftharpoons\mathrm{F}}} = \left\langle \vec{b}_{\xi}^{\vec{M}^{\mathrm{G}\rightleftharpoons\mathrm{A}}}, \mathcal{K}_0\left(-\vec{b}_{\zeta}^{\vec{J}_1^{\mathrm{A}\rightleftharpoons\mathrm{F}}}\right)\right\rangle_{\mathbb{S}^{\mathrm{G}\rightleftharpoons\mathrm{A}}} \quad \text{(D-89b)}$$

$$z_{\xi\zeta}^{\vec{M}^{\mathrm{G}\rightleftharpoons\mathrm{A}}\vec{J}_2^{\mathrm{A}\rightleftharpoons\mathrm{F}}} = \left\langle \vec{b}_{\xi}^{\vec{M}^{\mathrm{G}\rightleftharpoons\mathrm{A}}}, \mathcal{K}_0\left(-\vec{b}_{\zeta}^{\vec{J}_2^{\mathrm{A}\rightleftharpoons\mathrm{F}}}\right)\right\rangle_{\mathbb{S}^{\mathrm{G}\rightleftharpoons\mathrm{A}}} \quad \text{(D-89c)}$$

$$z_{\xi\zeta}^{\vec{M}^{\mathrm{G}\rightleftharpoons\mathrm{A}}\vec{J}^{\mathrm{F}}} = \left\langle \vec{b}_{\xi}^{\vec{M}^{\mathrm{G}\rightleftharpoons\mathrm{A}}}, \mathcal{K}_0\left(\vec{b}_{\zeta}^{\vec{J}^{\mathrm{F}}}\right)\right\rangle_{\mathbb{S}^{\mathrm{G}\rightleftharpoons\mathrm{A}}} \quad \text{(D-89d)}$$

$$z_{\xi\zeta}^{\vec{M}^{\mathrm{G}\rightleftharpoons\mathrm{A}}\vec{M}^{\mathrm{G}\rightleftharpoons\mathrm{A}}} = \left\langle \vec{b}_{\xi}^{\vec{M}^{\mathrm{G}\rightleftharpoons\mathrm{A}}}, -j\omega\varepsilon_0\mathcal{L}_0\left(\vec{b}_{\zeta}^{\vec{M}^{\mathrm{G}\rightleftharpoons\mathrm{A}}}\right)\right\rangle_{\mathbb{S}^{\mathrm{G}\rightleftharpoons\mathrm{A}}} \quad \text{(D-89e)}$$

$$z_{\xi\zeta}^{\vec{M}^{\mathrm{G}\rightleftharpoons\mathrm{A}}\vec{M}_1^{\mathrm{A}\rightleftharpoons\mathrm{F}}} = \left\langle \vec{b}_{\xi}^{\vec{M}^{\mathrm{G}\rightleftharpoons\mathrm{A}}}, -j\omega\varepsilon_0\mathcal{L}_0\left(-\vec{b}_{\zeta}^{\vec{M}_1^{\mathrm{A}\rightleftharpoons\mathrm{F}}}\right)\right\rangle_{\mathbb{S}^{\mathrm{G}\rightleftharpoons\mathrm{A}}} \quad \text{(D-89f)}$$

$$z_{\xi\zeta}^{\vec{M}^{\mathrm{G}\rightleftharpoons\mathrm{A}}\vec{M}_2^{\mathrm{A}\rightleftharpoons\mathrm{F}}} = \left\langle \vec{b}_{\xi}^{\vec{M}^{\mathrm{G}\rightleftharpoons\mathrm{A}}}, -j\omega\varepsilon_0\mathcal{L}_0\left(-\vec{b}_{\zeta}^{\vec{M}_2^{\mathrm{A}\rightleftharpoons\mathrm{F}}}\right)\right\rangle_{\mathbb{S}^{\mathrm{G}\rightleftharpoons\mathrm{A}}} \quad \text{(D-89g)}$$

and

$$z_{\xi\zeta}^{\vec{J}^{\mathrm{G}\rightleftharpoons\mathrm{A}}\vec{J}^{\mathrm{G}\rightleftharpoons\mathrm{A}}} = \left\langle \vec{b}_{\xi}^{\vec{J}^{\mathrm{G}\rightleftharpoons\mathrm{A}}}, -j\omega\mu_0\mathcal{L}_0\left(\vec{b}_{\zeta}^{\vec{J}^{\mathrm{G}\rightleftharpoons\mathrm{A}}}\right)\right\rangle_{\mathbb{S}^{\mathrm{G}\rightleftharpoons\mathrm{A}}} \quad \text{(D-90a)}$$

$$z_{\xi\zeta}^{\vec{J}^{\mathrm{G}\rightleftharpoons\mathrm{A}}\vec{J}_1^{\mathrm{A}\rightleftharpoons\mathrm{F}}} = \left\langle \vec{b}_{\xi}^{\vec{J}^{\mathrm{G}\rightleftharpoons\mathrm{A}}}, -j\omega\mu_0\mathcal{L}_0\left(-\vec{b}_{\zeta}^{\vec{J}_1^{\mathrm{A}\rightleftharpoons\mathrm{F}}}\right)\right\rangle_{\mathbb{S}^{\mathrm{G}\rightleftharpoons\mathrm{A}}} \quad \text{(D-90b)}$$

$$z_{\xi\zeta}^{\vec{J}^{\mathrm{G}\rightleftharpoons\mathrm{A}}\vec{J}_2^{\mathrm{A}\rightleftharpoons\mathrm{F}}} = \left\langle \vec{b}_{\xi}^{\vec{J}^{\mathrm{G}\rightleftharpoons\mathrm{A}}}, -j\omega\mu_0\mathcal{L}_0\left(-\vec{b}_{\zeta}^{\vec{J}_2^{\mathrm{A}\rightleftharpoons\mathrm{F}}}\right)\right\rangle_{\mathbb{S}^{\mathrm{G}\rightleftharpoons\mathrm{A}}} \quad \text{(D-90c)}$$

$$z_{\xi\zeta}^{\vec{J}^{\mathrm{G}\rightleftharpoons\mathrm{A}}\vec{J}^{\mathrm{F}}} = \left\langle \vec{b}_{\xi}^{\vec{J}^{\mathrm{G}\rightleftharpoons\mathrm{A}}}, -j\omega\mu_0\mathcal{L}_0\left(\vec{b}_{\zeta}^{\vec{J}^{\mathrm{F}}}\right)\right\rangle_{\mathbb{S}^{\mathrm{G}\rightleftharpoons\mathrm{A}}} \quad \text{(D-90d)}$$

$$z_{\xi\zeta}^{\vec{J}^{\mathrm{G}\rightleftharpoons\mathrm{A}}\vec{M}^{\mathrm{G}\rightleftharpoons\mathrm{A}}} = \left\langle \vec{b}_{\xi}^{\vec{J}^{\mathrm{G}\rightleftharpoons\mathrm{A}}}, \frac{1}{2}\vec{b}_{\zeta}^{\vec{M}^{\mathrm{G}\rightleftharpoons\mathrm{A}}} \times \hat{n}^{\rightarrow\mathrm{A}} - \mathrm{P.V.}\,\mathcal{K}_0\left(\vec{b}_{\zeta}^{\vec{M}^{\mathrm{G}\rightleftharpoons\mathrm{A}}}\right)\right\rangle_{\mathbb{S}^{\mathrm{G}\rightleftharpoons\mathrm{A}}} \quad \text{(D-90e)}$$

$$z_{\xi\zeta}^{\vec{J}^{\mathrm{G}\rightleftharpoons\mathrm{A}}\vec{M}_1^{\mathrm{A}\rightleftharpoons\mathrm{F}}} = \left\langle \vec{b}_\xi^{\vec{J}^{\mathrm{G}\rightleftharpoons\mathrm{A}}}, -\mathcal{K}_0\left(-\vec{b}_\zeta^{\vec{M}_1^{\mathrm{A}\rightleftharpoons\mathrm{F}}}\right)\right\rangle_{\mathbb{S}^{\mathrm{G}\rightleftharpoons\mathrm{A}}} \tag{D-90f}$$

$$z_{\xi\zeta}^{\vec{J}^{\mathrm{G}\rightleftharpoons\mathrm{A}}\vec{M}_2^{\mathrm{A}\rightleftharpoons\mathrm{F}}} = \left\langle \vec{b}_\xi^{\vec{J}^{\mathrm{G}\rightleftharpoons\mathrm{A}}}, -\mathcal{K}_0\left(-\vec{b}_\zeta^{\vec{M}_2^{\mathrm{A}\rightleftharpoons\mathrm{F}}}\right)\right\rangle_{\mathbb{S}^{\mathrm{G}\rightleftharpoons\mathrm{A}}} \tag{D-90g}$$

and

$$z_{\xi\zeta}^{\vec{J}_1^{\mathrm{A}\rightleftharpoons\mathrm{F}}\vec{J}^{\mathrm{G}\rightleftharpoons\mathrm{A}}} = \left\langle \vec{b}_\xi^{\vec{J}_1^{\mathrm{A}\rightleftharpoons\mathrm{F}}}, -j\omega\mu_0\mathcal{L}_0\left(\vec{b}_\zeta^{\vec{J}^{\mathrm{G}\rightleftharpoons\mathrm{A}}}\right)\right\rangle_{\mathbb{S}_1^{\mathrm{A}\rightleftharpoons\mathrm{F}}} \tag{D-91a}$$

$$z_{\xi\zeta}^{\vec{J}_1^{\mathrm{A}\rightleftharpoons\mathrm{F}}\vec{J}_1^{\mathrm{A}\rightleftharpoons\mathrm{F}}} = \left\langle \vec{b}_\xi^{\vec{J}_1^{\mathrm{A}\rightleftharpoons\mathrm{F}}}, -j\omega\mu_0\mathcal{L}_0\left(-\vec{b}_\zeta^{\vec{J}_1^{\mathrm{A}\rightleftharpoons\mathrm{F}}}\right)\right\rangle_{\mathbb{S}_1^{\mathrm{A}\rightleftharpoons\mathrm{F}}} - \left\langle \vec{b}_\xi^{\vec{J}_1^{\mathrm{A}\rightleftharpoons\mathrm{F}}}, \mathcal{E}_1\left(\vec{b}_\zeta^{\vec{J}_1^{\mathrm{A}\rightleftharpoons\mathrm{F}}}\right)\right\rangle_{\tilde{\mathbb{S}}_1^{\mathrm{A}\rightleftharpoons\mathrm{F}}} \tag{D-91b}$$

$$z_{\xi\zeta}^{\vec{J}_1^{\mathrm{A}\rightleftharpoons\mathrm{F}}\vec{J}_2^{\mathrm{A}\rightleftharpoons\mathrm{F}}} = \left\langle \vec{b}_\xi^{\vec{J}_1^{\mathrm{A}\rightleftharpoons\mathrm{F}}}, -j\omega\mu_0\mathcal{L}_0\left(-\vec{b}_\zeta^{\vec{J}_2^{\mathrm{A}\rightleftharpoons\mathrm{F}}}\right)\right\rangle_{\mathbb{S}_1^{\mathrm{A}\rightleftharpoons\mathrm{F}}} \tag{D-91c}$$

$$z_{\xi\zeta}^{\vec{J}_1^{\mathrm{A}\rightleftharpoons\mathrm{F}}\vec{J}_1^{\mathrm{A}}} = -\left\langle \vec{b}_\xi^{\vec{J}_1^{\mathrm{A}\rightleftharpoons\mathrm{F}}}, \mathcal{E}_1\left(\vec{b}_\zeta^{\vec{J}_1^{\mathrm{A}}}\right)\right\rangle_{\tilde{\mathbb{S}}_1^{\mathrm{A}\rightleftharpoons\mathrm{F}}} \tag{D-91d}$$

$$z_{\xi\zeta}^{\vec{J}_1^{\mathrm{A}\rightleftharpoons\mathrm{F}}\vec{J}^{\mathrm{F}}} = \left\langle \vec{b}_\xi^{\vec{J}_1^{\mathrm{A}\rightleftharpoons\mathrm{F}}}, -j\omega\mu_0\mathcal{L}_0\left(\vec{b}_\zeta^{\vec{J}^{\mathrm{F}}}\right)\right\rangle_{\mathbb{S}_1^{\mathrm{A}\rightleftharpoons\mathrm{F}}} \tag{D-91e}$$

$$z_{\xi\zeta}^{\vec{J}_1^{\mathrm{A}\rightleftharpoons\mathrm{F}}\vec{M}^{\mathrm{G}\rightleftharpoons\mathrm{A}}} = \left\langle \vec{b}_\xi^{\vec{J}_1^{\mathrm{A}\rightleftharpoons\mathrm{F}}}, -\mathcal{K}_0\left(\vec{b}_\zeta^{\vec{M}^{\mathrm{G}\rightleftharpoons\mathrm{A}}}\right)\right\rangle_{\mathbb{S}_1^{\mathrm{A}\rightleftharpoons\mathrm{F}}} \tag{D-91f}$$

$$z_{\xi\zeta}^{\vec{J}_1^{\mathrm{A}\rightleftharpoons\mathrm{F}}\vec{M}_1^{\mathrm{A}\rightleftharpoons\mathrm{F}}} = \left\langle \vec{b}_\xi^{\vec{J}_1^{\mathrm{A}\rightleftharpoons\mathrm{F}}}, \frac{1}{2}\vec{b}_\zeta^{\vec{M}_1^{\mathrm{A}\rightleftharpoons\mathrm{F}}} \times \hat{n}^{\rightarrow\mathrm{F}} - \mathrm{P.V.}\,\mathcal{K}_0\left(-\vec{b}_\zeta^{\vec{M}_1^{\mathrm{A}\rightleftharpoons\mathrm{F}}}\right)\right\rangle_{\mathbb{S}_1^{\mathrm{A}\rightleftharpoons\mathrm{F}}} - \left\langle \vec{b}_\xi^{\vec{J}_1^{\mathrm{A}\rightleftharpoons\mathrm{F}}}, \mathcal{E}_1\left(\vec{b}_\zeta^{\vec{M}_1^{\mathrm{A}\rightleftharpoons\mathrm{F}}}\right)\right\rangle_{\tilde{\mathbb{S}}_1^{\mathrm{A}\rightleftharpoons\mathrm{F}}} \tag{D-91g}$$

$$z_{\xi\zeta}^{\vec{J}_1^{\mathrm{A}\rightleftharpoons\mathrm{F}}\vec{M}_2^{\mathrm{A}\rightleftharpoons\mathrm{F}}} = \left\langle \vec{b}_\xi^{\vec{J}_1^{\mathrm{A}\rightleftharpoons\mathrm{F}}}, -\mathcal{K}_0\left(-\vec{b}_\zeta^{\vec{M}_2^{\mathrm{A}\rightleftharpoons\mathrm{F}}}\right)\right\rangle_{\mathbb{S}_1^{\mathrm{A}\rightleftharpoons\mathrm{F}}} \tag{D-91h}$$

and

$$z_{\xi\zeta}^{\vec{M}_1^{\mathrm{A}\rightleftharpoons\mathrm{F}}\vec{J}^{\mathrm{G}\rightleftharpoons\mathrm{A}}} = \left\langle \vec{b}_\xi^{\vec{M}_1^{\mathrm{A}\rightleftharpoons\mathrm{F}}}, \mathcal{K}_0\left(\vec{b}_\zeta^{\vec{J}^{\mathrm{G}\rightleftharpoons\mathrm{A}}}\right)\right\rangle_{\mathbb{S}_1^{\mathrm{A}\rightleftharpoons\mathrm{F}}} \tag{D-92a}$$

$$z_{\xi\zeta}^{\vec{M}_1^{\mathrm{A}\rightleftharpoons\mathrm{F}}\vec{J}_1^{\mathrm{A}\rightleftharpoons\mathrm{F}}} = \left\langle \vec{b}_\xi^{\vec{M}_1^{\mathrm{A}\rightleftharpoons\mathrm{F}}}, \hat{n}^{\rightarrow\mathrm{F}} \times \frac{1}{2}\vec{b}_\zeta^{\vec{J}_1^{\mathrm{A}\rightleftharpoons\mathrm{F}}} + \mathrm{P.V.}\,\mathcal{K}_0\left(-\vec{b}_\zeta^{\vec{J}_1^{\mathrm{A}\rightleftharpoons\mathrm{F}}}\right)\right\rangle_{\mathbb{S}_1^{\mathrm{A}\rightleftharpoons\mathrm{F}}} - \left\langle \vec{b}_\xi^{\vec{M}_1^{\mathrm{A}\rightleftharpoons\mathrm{F}}}, \mathcal{H}_1\left(\vec{b}_\zeta^{\vec{J}_1^{\mathrm{A}\rightleftharpoons\mathrm{F}}}\right)\right\rangle_{\tilde{\mathbb{S}}_1^{\mathrm{A}\rightleftharpoons\mathrm{F}}} \tag{D-92b}$$

$$z_{\xi\zeta}^{\vec{M}_1^{\mathrm{A}\rightleftharpoons\mathrm{F}}\vec{J}_2^{\mathrm{A}\rightleftharpoons\mathrm{F}}} = \left\langle \vec{b}_\xi^{\vec{M}_1^{\mathrm{A}\rightleftharpoons\mathrm{F}}}, \mathcal{K}_0\left(-\vec{b}_\zeta^{\vec{J}_2^{\mathrm{A}\rightleftharpoons\mathrm{F}}}\right)\right\rangle_{\mathbb{S}_1^{\mathrm{A}\rightleftharpoons\mathrm{F}}} \tag{D-92c}$$

$$z_{\xi\zeta}^{\vec{M}_1^{\mathrm{A}\rightleftharpoons\mathrm{F}}\vec{J}_1^{\mathrm{A}}} = -\left\langle \vec{b}_\xi^{\vec{M}_1^{\mathrm{A}\rightleftharpoons\mathrm{F}}}, \mathcal{H}_1\left(\vec{b}_\zeta^{\vec{J}_1^{\mathrm{A}}}\right)\right\rangle_{\tilde{\mathbb{S}}_1^{\mathrm{A}\rightleftharpoons\mathrm{F}}} \tag{D-92d}$$

$$z_{\xi\zeta}^{\vec{M}_1^{\mathrm{A}\rightleftharpoons\mathrm{F}}\vec{J}^{\mathrm{F}}} = \left\langle \vec{b}_\xi^{\vec{M}_1^{\mathrm{A}\rightleftharpoons\mathrm{F}}}, \mathcal{K}_0\left(\vec{b}_\zeta^{\vec{J}^{\mathrm{F}}}\right)\right\rangle_{\mathbb{S}_1^{\mathrm{A}\rightleftharpoons\mathrm{F}}} \tag{D-92e}$$

$$z_{\xi\zeta}^{\vec{M}_1^{\mathrm{A}\rightleftharpoons\mathrm{F}}\vec{M}^{\mathrm{G}\rightleftharpoons\mathrm{A}}} = \left\langle \vec{b}_\xi^{\vec{M}_1^{\mathrm{A}\rightleftharpoons\mathrm{F}}}, -j\omega\varepsilon_0\mathcal{L}_0\left(\vec{b}_\zeta^{\vec{M}^{\mathrm{G}\rightleftharpoons\mathrm{A}}}\right)\right\rangle_{\mathbb{S}_1^{\mathrm{A}\rightleftharpoons\mathrm{F}}} \tag{D-92f}$$

$$z_{\xi\zeta}^{\vec{M}_1^{\mathrm{A}\rightleftharpoons\mathrm{F}}\vec{M}_1^{\mathrm{A}\rightleftharpoons\mathrm{F}}} = \left\langle \vec{b}_\xi^{\vec{M}_1^{\mathrm{A}\rightleftharpoons\mathrm{F}}}, -j\omega\varepsilon_0\mathcal{L}_0\left(-\vec{b}_\zeta^{\vec{M}_1^{\mathrm{A}\rightleftharpoons\mathrm{F}}}\right)\right\rangle_{\mathbb{S}_1^{\mathrm{A}\rightleftharpoons\mathrm{F}}} - \left\langle \vec{b}_\xi^{\vec{M}_1^{\mathrm{A}\rightleftharpoons\mathrm{F}}}, \mathcal{H}_1\left(\vec{b}_\zeta^{\vec{M}_1^{\mathrm{A}\rightleftharpoons\mathrm{F}}}\right)\right\rangle_{\tilde{\mathbb{S}}_1^{\mathrm{A}\rightleftharpoons\mathrm{F}}} \tag{D-92g}$$

$$z_{\xi\zeta}^{\vec{M}_1^{\mathrm{A}\rightleftharpoons\mathrm{F}}\vec{M}_2^{\mathrm{A}\rightleftharpoons\mathrm{F}}} = \left\langle \vec{b}_\xi^{\vec{M}_1^{\mathrm{A}\rightleftharpoons\mathrm{F}}}, -j\omega\varepsilon_0\mathcal{L}_0\left(-\vec{b}_\zeta^{\vec{M}_2^{\mathrm{A}\rightleftharpoons\mathrm{F}}}\right)\right\rangle_{\mathbb{S}_1^{\mathrm{A}\rightleftharpoons\mathrm{F}}} \tag{D-92h}$$

and

$$z_{\xi\zeta}^{\vec{J}_2^{\mathrm{A}\rightleftharpoons\mathrm{F}}\vec{J}^{\mathrm{G}\rightleftharpoons\mathrm{A}}} = \left\langle \vec{b}_\xi^{\vec{J}_2^{\mathrm{A}\rightleftharpoons\mathrm{F}}}, -j\omega\mu_0\mathcal{L}_0\left(\vec{b}_\zeta^{\vec{J}^{\mathrm{G}\rightleftharpoons\mathrm{A}}}\right)\right\rangle_{\mathbb{S}_2^{\mathrm{A}\rightleftharpoons\mathrm{F}}} \tag{D-93a}$$

$$z_{\xi\zeta}^{\vec{J}_2^{\mathrm{A}\rightleftharpoons\mathrm{F}}\vec{J}_1^{\mathrm{A}\rightleftharpoons\mathrm{F}}} = \left\langle \vec{b}_\xi^{\vec{J}_2^{\mathrm{A}\rightleftharpoons\mathrm{F}}}, -j\omega\mu_0\mathcal{L}_0\left(-\vec{b}_\zeta^{\vec{J}_1^{\mathrm{A}\rightleftharpoons\mathrm{F}}}\right)\right\rangle_{\mathbb{S}_2^{\mathrm{A}\rightleftharpoons\mathrm{F}}} \tag{D-93b}$$

$$z_{\xi\zeta}^{\vec{J}_2^{\mathrm{A}\rightleftharpoons\mathrm{F}}\vec{J}_2^{\mathrm{A}\rightleftharpoons\mathrm{F}}} = \left\langle \vec{b}_\xi^{\vec{J}_2^{\mathrm{A}\rightleftharpoons\mathrm{F}}}, -j\omega\mu_0\mathcal{L}_0\left(-\vec{b}_\zeta^{\vec{J}_2^{\mathrm{A}\rightleftharpoons\mathrm{F}}}\right)\right\rangle_{\mathbb{S}_2^{\mathrm{A}\rightleftharpoons\mathrm{F}}} - \left\langle \vec{b}_\xi^{\vec{J}_2^{\mathrm{A}\rightleftharpoons\mathrm{F}}}, \mathcal{E}_2\left(\vec{b}_\zeta^{\vec{J}_2^{\mathrm{A}\rightleftharpoons\mathrm{F}}}\right)\right\rangle_{\tilde{\mathbb{S}}_2^{\mathrm{A}\rightleftharpoons\mathrm{F}}} \tag{D-93c}$$

$$z_{\xi\zeta}^{\vec{J}_2^{\mathrm{A}\rightleftharpoons\mathrm{F}}\vec{J}_{23}^{\mathrm{A}\rightleftharpoons\mathrm{A}}} = -\left\langle \vec{b}_\xi^{\vec{J}_2^{\mathrm{A}\rightleftharpoons\mathrm{F}}}, \mathcal{E}_2\left(\vec{b}_\zeta^{\vec{J}_{23}^{\mathrm{A}\rightleftharpoons\mathrm{A}}}\right)\right\rangle_{\tilde{\mathbb{S}}_2^{\mathrm{A}\rightleftharpoons\mathrm{F}}} \tag{D-93d}$$

$$z_{\xi\zeta}^{\vec{J}_2^{\mathrm{A}\rightleftharpoons\mathrm{F}}\vec{J}^{\mathrm{F}}} = \left\langle \vec{b}_\xi^{\vec{J}_2^{\mathrm{A}\rightleftharpoons\mathrm{F}}}, -j\omega\mu_0\mathcal{L}_0\left(\vec{b}_\zeta^{\vec{J}^{\mathrm{F}}}\right)\right\rangle_{\mathbb{S}_2^{\mathrm{A}\rightleftharpoons\mathrm{F}}} \tag{D-93e}$$

$$z_{\xi\zeta}^{\vec{J}_2^{\mathrm{A}\rightleftharpoons\mathrm{F}}\vec{M}^{\mathrm{G}\rightleftharpoons\mathrm{A}}} = \left\langle \vec{b}_\xi^{\vec{J}_2^{\mathrm{A}\rightleftharpoons\mathrm{F}}}, -\mathcal{K}_0\left(\vec{b}_\zeta^{\vec{M}^{\mathrm{G}\rightleftharpoons\mathrm{A}}}\right)\right\rangle_{\mathbb{S}_2^{\mathrm{A}\rightleftharpoons\mathrm{F}}} \tag{D-93f}$$

$$z_{\xi\zeta}^{\vec{J}_2^{\mathrm{A}\rightleftharpoons\mathrm{F}}\vec{M}_1^{\mathrm{A}\rightleftharpoons\mathrm{F}}} = \left\langle \vec{b}_\xi^{\vec{J}_2^{\mathrm{A}\rightleftharpoons\mathrm{F}}}, -\mathcal{K}_0\left(-\vec{b}_\zeta^{\vec{M}_1^{\mathrm{A}\rightleftharpoons\mathrm{F}}}\right)\right\rangle_{\mathbb{S}_2^{\mathrm{A}\rightleftharpoons\mathrm{F}}} \tag{D-93g}$$

$$z_{\xi\zeta}^{\vec{J}_2^{\mathrm{A}\rightleftharpoons\mathrm{F}}\vec{M}_2^{\mathrm{A}\rightleftharpoons\mathrm{F}}} = \left\langle \vec{b}_\xi^{\vec{J}_2^{\mathrm{A}\rightleftharpoons\mathrm{F}}}, \frac{1}{2}\vec{b}_\zeta^{\vec{M}_2^{\mathrm{A}\rightleftharpoons\mathrm{F}}}\times\hat{n}^{\rightarrow\mathrm{F}} - \mathrm{P.V.}\,\mathcal{K}_0\left(-\vec{b}_\zeta^{\vec{M}_2^{\mathrm{A}\rightleftharpoons\mathrm{F}}}\right)\right\rangle_{\mathbb{S}_2^{\mathrm{A}\rightleftharpoons\mathrm{F}}} - \left\langle \vec{b}_\xi^{\vec{J}_2^{\mathrm{A}\rightleftharpoons\mathrm{F}}}, \mathcal{E}_2\left(\vec{b}_\zeta^{\vec{M}_2^{\mathrm{A}\rightleftharpoons\mathrm{F}}}\right)\right\rangle_{\tilde{\mathbb{S}}_2^{\mathrm{A}\rightleftharpoons\mathrm{F}}} \tag{D-93h}$$

$$z_{\xi\zeta}^{\vec{J}_2^{\mathrm{A}\rightleftharpoons\mathrm{F}}\vec{M}_{23}^{\mathrm{A}\rightleftharpoons\mathrm{A}}} = -\left\langle \vec{b}_\xi^{\vec{J}_2^{\mathrm{A}\rightleftharpoons\mathrm{F}}}, \mathcal{E}_2\left(\vec{b}_\zeta^{\vec{M}_{23}^{\mathrm{A}\rightleftharpoons\mathrm{A}}}\right)\right\rangle_{\tilde{\mathbb{S}}_2^{\mathrm{A}\rightleftharpoons\mathrm{F}}} \tag{D-93i}$$

and

$$z_{\xi\zeta}^{\vec{M}_2^{\mathrm{A}\rightleftharpoons\mathrm{F}}\vec{J}^{\mathrm{G}\rightleftharpoons\mathrm{A}}} = \left\langle \vec{b}_\xi^{\vec{M}_2^{\mathrm{A}\rightleftharpoons\mathrm{F}}}, \mathcal{K}_0\left(\vec{b}_\zeta^{\vec{J}^{\mathrm{G}\rightleftharpoons\mathrm{A}}}\right)\right\rangle_{\mathbb{S}_2^{\mathrm{A}\rightleftharpoons\mathrm{F}}} \tag{D-94a}$$

$$z_{\xi\zeta}^{\vec{M}_2^{\mathrm{A}\rightleftharpoons\mathrm{F}}\vec{J}_1^{\mathrm{A}\rightleftharpoons\mathrm{F}}} = \left\langle \vec{b}_\xi^{\vec{M}_2^{\mathrm{A}\rightleftharpoons\mathrm{F}}}, \mathcal{K}_0\left(-\vec{b}_\zeta^{\vec{J}_1^{\mathrm{A}\rightleftharpoons\mathrm{F}}}\right)\right\rangle_{\mathbb{S}_2^{\mathrm{A}\rightleftharpoons\mathrm{F}}} \tag{D-94b}$$

$$z_{\xi\zeta}^{\vec{M}_2^{\mathrm{A}\rightleftharpoons\mathrm{F}}\vec{J}_2^{\mathrm{A}\rightleftharpoons\mathrm{F}}} = \left\langle \vec{b}_\xi^{\vec{M}_2^{\mathrm{A}\rightleftharpoons\mathrm{F}}}, \hat{n}^{\rightarrow\mathrm{F}}\times\frac{1}{2}\vec{b}_\zeta^{\vec{J}_2^{\mathrm{A}\rightleftharpoons\mathrm{F}}} + \mathrm{P.V.}\,\mathcal{K}_0\left(-\vec{b}_\zeta^{\vec{J}_2^{\mathrm{A}\rightleftharpoons\mathrm{F}}}\right)\right\rangle_{\mathbb{S}_2^{\mathrm{A}\rightleftharpoons\mathrm{F}}} - \left\langle \vec{b}_\xi^{\vec{M}_2^{\mathrm{A}\rightleftharpoons\mathrm{F}}}, \mathcal{H}_2\left(\vec{b}_\zeta^{\vec{J}_2^{\mathrm{A}\rightleftharpoons\mathrm{F}}}\right)\right\rangle_{\tilde{\mathbb{S}}_2^{\mathrm{A}\rightleftharpoons\mathrm{F}}} \tag{D-94c}$$

$$z_{\xi\zeta}^{\vec{M}_2^{\mathrm{A}\rightleftharpoons\mathrm{F}}\vec{J}_{23}^{\mathrm{A}\rightleftharpoons\mathrm{A}}} = -\left\langle \vec{b}_\xi^{\vec{M}_2^{\mathrm{A}\rightleftharpoons\mathrm{F}}}, \mathcal{H}_2\left(\vec{b}_\zeta^{\vec{J}_{23}^{\mathrm{A}\rightleftharpoons\mathrm{A}}}\right)\right\rangle_{\tilde{\mathbb{S}}_2^{\mathrm{A}\rightleftharpoons\mathrm{F}}} \tag{D-94d}$$

$$z_{\xi\zeta}^{\vec{M}_2^{\mathrm{A}\rightleftharpoons\mathrm{F}}\vec{J}^{\mathrm{F}}} = \left\langle \vec{b}_\xi^{\vec{M}_2^{\mathrm{A}\rightleftharpoons\mathrm{F}}}, \mathcal{K}_0\left(\vec{b}_\zeta^{\vec{J}^{\mathrm{F}}}\right)\right\rangle_{\mathbb{S}_2^{\mathrm{A}\rightleftharpoons\mathrm{F}}} \tag{D-94e}$$

$$z_{\xi\zeta}^{\vec{M}_2^{\mathrm{A}\rightleftharpoons\mathrm{F}}\vec{M}^{\mathrm{G}\rightleftharpoons\mathrm{A}}} = \left\langle \vec{b}_\xi^{\vec{M}_2^{\mathrm{A}\rightleftharpoons\mathrm{F}}}, -j\omega\varepsilon_0\mathcal{L}_0\left(\vec{b}_\zeta^{\vec{M}^{\mathrm{G}\rightleftharpoons\mathrm{A}}}\right)\right\rangle_{\mathbb{S}_2^{\mathrm{A}\rightleftharpoons\mathrm{F}}} \tag{D-94f}$$

$$z_{\xi\zeta}^{\vec{M}_2^{\mathrm{A}\rightleftharpoons\mathrm{F}}\vec{M}_1^{\mathrm{A}\rightleftharpoons\mathrm{F}}} = \left\langle \vec{b}_\xi^{\vec{M}_2^{\mathrm{A}\rightleftharpoons\mathrm{F}}}, -j\omega\varepsilon_0\mathcal{L}_0\left(-\vec{b}_\zeta^{\vec{M}_1^{\mathrm{A}\rightleftharpoons\mathrm{F}}}\right)\right\rangle_{\mathbb{S}_2^{\mathrm{A}\rightleftharpoons\mathrm{F}}} \tag{D-94g}$$

$$z_{\xi\zeta}^{\vec{M}_2^{\mathrm{A}\rightleftharpoons\mathrm{F}}\vec{M}_2^{\mathrm{A}\rightleftharpoons\mathrm{F}}} = \left\langle \vec{b}_\xi^{\vec{M}_2^{\mathrm{A}\rightleftharpoons\mathrm{F}}}, -j\omega\varepsilon_0\mathcal{L}_0\left(-\vec{b}_\zeta^{\vec{M}_2^{\mathrm{A}\rightleftharpoons\mathrm{F}}}\right)\right\rangle_{\mathbb{S}_2^{\mathrm{A}\rightleftharpoons\mathrm{F}}} - \left\langle \vec{b}_\xi^{\vec{M}_2^{\mathrm{A}\rightleftharpoons\mathrm{F}}}, \mathcal{H}_2\left(\vec{b}_\zeta^{\vec{M}_2^{\mathrm{A}\rightleftharpoons\mathrm{F}}}\right)\right\rangle_{\tilde{\mathbb{S}}_2^{\mathrm{A}\rightleftharpoons\mathrm{F}}} \tag{D-94h}$$

$$z_{\xi\zeta}^{\vec{M}_2^{\mathrm{A}\rightleftharpoons\mathrm{F}}\vec{M}_{23}^{\mathrm{A}\rightleftharpoons\mathrm{A}}} = -\left\langle \vec{b}_\xi^{\vec{M}_2^{\mathrm{A}\rightleftharpoons\mathrm{F}}}, \mathcal{H}_2\left(\vec{b}_\zeta^{\vec{M}_{23}^{\mathrm{A}\rightleftharpoons\mathrm{A}}}\right)\right\rangle_{\tilde{\mathbb{S}}_2^{\mathrm{A}\rightleftharpoons\mathrm{F}}} \tag{D-94i}$$

and

$$z_{\xi\zeta}^{\vec{J}_{23}^{\mathrm{A}\rightleftharpoons\mathrm{A}}\vec{J}_2^{\mathrm{A}\rightleftharpoons\mathrm{F}}} = \left\langle \vec{b}_\xi^{\vec{J}_{23}^{\mathrm{A}\rightleftharpoons\mathrm{A}}}, \mathcal{E}_2\left(\vec{b}_\zeta^{\vec{J}_2^{\mathrm{A}\rightleftharpoons\mathrm{F}}}\right)\right\rangle_{\tilde{\mathbb{S}}_{23}^{\mathrm{A}\rightleftharpoons\mathrm{A}}} \tag{D-95a}$$

$$z_{\xi\zeta}^{\vec{J}_{23}^{\mathrm{A}\rightleftharpoons\mathrm{A}}\vec{J}_{23}^{\mathrm{A}\rightleftharpoons\mathrm{A}}} = \left\langle \vec{b}_{\xi}^{\vec{J}_{23}^{\mathrm{A}\rightleftharpoons\mathrm{A}}}, \mathcal{E}_2\left(\vec{b}_{\zeta}^{\vec{J}_{23}^{\mathrm{A}\rightleftharpoons\mathrm{A}}}\right)\right\rangle_{\tilde{\mathbb{S}}_{23}^{\mathrm{A}\rightleftharpoons\mathrm{A}}} - \left\langle \vec{b}_{\xi}^{\vec{J}_{23}^{\mathrm{A}\rightleftharpoons\mathrm{A}}}, \mathcal{E}_3\left(-\vec{b}_{\zeta}^{\vec{J}_{23}^{\mathrm{A}\rightleftharpoons\mathrm{A}}}\right)\right\rangle_{\mathbb{S}_{23}^{\mathrm{A}\rightleftharpoons\mathrm{A}}} \tag{D-95b}$$

$$z_{\xi\zeta}^{\vec{J}_{23}^{\mathrm{A}\rightleftharpoons\mathrm{A}}\vec{M}_{2}^{\mathrm{A}\rightleftharpoons\mathrm{F}}} = \left\langle \vec{b}_{\xi}^{\vec{J}_{23}^{\mathrm{A}\rightleftharpoons\mathrm{A}}}, \mathcal{E}_2\left(\vec{b}_{\zeta}^{\vec{M}_{2}^{\mathrm{A}\rightleftharpoons\mathrm{F}}}\right)\right\rangle_{\tilde{\mathbb{S}}_{23}^{\mathrm{A}\rightleftharpoons\mathrm{A}}} \tag{D-95c}$$

$$z_{\xi\zeta}^{\vec{J}_{23}^{\mathrm{A}\rightleftharpoons\mathrm{A}}\vec{M}_{23}^{\mathrm{A}\rightleftharpoons\mathrm{A}}} = \left\langle \vec{b}_{\xi}^{\vec{J}_{23}^{\mathrm{A}\rightleftharpoons\mathrm{A}}}, \mathcal{E}_2\left(\vec{b}_{\zeta}^{\vec{M}_{23}^{\mathrm{A}\rightleftharpoons\mathrm{A}}}\right)\right\rangle_{\tilde{\mathbb{S}}_{23}^{\mathrm{A}\rightleftharpoons\mathrm{A}}} - \left\langle \vec{b}_{\xi}^{\vec{J}_{23}^{\mathrm{A}\rightleftharpoons\mathrm{A}}}, \mathcal{E}_3\left(-\vec{b}_{\zeta}^{\vec{M}_{23}^{\mathrm{A}\rightleftharpoons\mathrm{A}}}\right)\right\rangle_{\mathbb{S}_{23}^{\mathrm{A}\rightleftharpoons\mathrm{A}}} \tag{D-95d}$$

and

$$z_{\xi\zeta}^{\vec{M}_{23}^{\mathrm{A}\rightleftharpoons\mathrm{A}}\vec{J}_{2}^{\mathrm{A}\rightleftharpoons\mathrm{F}}} = \left\langle \vec{b}_{\xi}^{\vec{M}_{23}^{\mathrm{A}\rightleftharpoons\mathrm{A}}}, \mathcal{H}_2\left(\vec{b}_{\zeta}^{\vec{J}_{2}^{\mathrm{A}\rightleftharpoons\mathrm{F}}}\right)\right\rangle_{\tilde{\mathbb{S}}_{23}^{\mathrm{A}\rightleftharpoons\mathrm{A}}} \tag{D-96a}$$

$$z_{\xi\zeta}^{\vec{M}_{23}^{\mathrm{A}\rightleftharpoons\mathrm{A}}\vec{J}_{23}^{\mathrm{A}\rightleftharpoons\mathrm{A}}} = \left\langle \vec{b}_{\xi}^{\vec{M}_{23}^{\mathrm{A}\rightleftharpoons\mathrm{A}}}, \mathcal{H}_2\left(\vec{b}_{\zeta}^{\vec{J}_{23}^{\mathrm{A}\rightleftharpoons\mathrm{A}}}\right)\right\rangle_{\tilde{\mathbb{S}}_{23}^{\mathrm{A}\rightleftharpoons\mathrm{A}}} - \left\langle \vec{b}_{\xi}^{\vec{M}_{23}^{\mathrm{A}\rightleftharpoons\mathrm{A}}}, \mathcal{H}_3\left(-\vec{b}_{\zeta}^{\vec{J}_{23}^{\mathrm{A}\rightleftharpoons\mathrm{A}}}\right)\right\rangle_{\mathbb{S}_{23}^{\mathrm{A}\rightleftharpoons\mathrm{A}}} \tag{D-96b}$$

$$z_{\xi\zeta}^{\vec{M}_{23}^{\mathrm{A}\rightleftharpoons\mathrm{A}}\vec{M}_{2}^{\mathrm{A}\rightleftharpoons\mathrm{F}}} = \left\langle \vec{b}_{\xi}^{\vec{M}_{23}^{\mathrm{A}\rightleftharpoons\mathrm{A}}}, \mathcal{H}_2\left(\vec{b}_{\zeta}^{\vec{M}_{2}^{\mathrm{A}\rightleftharpoons\mathrm{F}}}\right)\right\rangle_{\tilde{\mathbb{S}}_{23}^{\mathrm{A}\rightleftharpoons\mathrm{A}}} \tag{D-96c}$$

$$z_{\xi\zeta}^{\vec{M}_{23}^{\mathrm{A}\rightleftharpoons\mathrm{A}}\vec{M}_{23}^{\mathrm{A}\rightleftharpoons\mathrm{A}}} = \left\langle \vec{b}_{\xi}^{\vec{M}_{23}^{\mathrm{A}\rightleftharpoons\mathrm{A}}}, \mathcal{H}_2\left(\vec{b}_{\zeta}^{\vec{M}_{23}^{\mathrm{A}\rightleftharpoons\mathrm{A}}}\right)\right\rangle_{\tilde{\mathbb{S}}_{23}^{\mathrm{A}\rightleftharpoons\mathrm{A}}} - \left\langle \vec{b}_{\xi}^{\vec{M}_{23}^{\mathrm{A}\rightleftharpoons\mathrm{A}}}, \mathcal{H}_3\left(-\vec{b}_{\zeta}^{\vec{M}_{23}^{\mathrm{A}\rightleftharpoons\mathrm{A}}}\right)\right\rangle_{\mathbb{S}_{23}^{\mathrm{A}\rightleftharpoons\mathrm{A}}} \tag{D-96d}$$

and

$$z_{\xi\zeta}^{\vec{J}_{1}^{\mathrm{A}}\vec{J}_{1}^{\mathrm{A}\rightleftharpoons\mathrm{F}}} = \left\langle \vec{b}_{\xi}^{\vec{J}_{1}^{\mathrm{A}}}, \mathcal{E}_1\left(\vec{b}_{\zeta}^{\vec{J}_{1}^{\mathrm{A}\rightleftharpoons\mathrm{F}}}\right)\right\rangle_{\tilde{\mathbb{S}}_{1}^{\mathrm{A}}} \tag{D-97a}$$

$$z_{\xi\zeta}^{\vec{J}_{1}^{\mathrm{A}}\vec{J}_{1}^{\mathrm{A}}} = \left\langle \vec{b}_{\xi}^{\vec{J}_{1}^{\mathrm{A}}}, \mathcal{E}_1\left(\vec{b}_{\zeta}^{\vec{J}_{1}^{\mathrm{A}}}\right)\right\rangle_{\tilde{\mathbb{S}}_{1}^{\mathrm{A}}} \tag{D-97b}$$

$$z_{\xi\zeta}^{\vec{J}_{1}^{\mathrm{A}}\vec{M}_{1}^{\mathrm{A}\rightleftharpoons\mathrm{F}}} = \left\langle \vec{b}_{\xi}^{\vec{J}_{1}^{\mathrm{A}}}, \mathcal{E}_1\left(\vec{b}_{\zeta}^{\vec{M}_{1}^{\mathrm{A}\rightleftharpoons\mathrm{F}}}\right)\right\rangle_{\tilde{\mathbb{S}}_{1}^{\mathrm{A}}} \tag{D-97c}$$

and

$$z_{\xi\zeta}^{\vec{J}^{\mathrm{F}}\vec{J}^{\mathrm{G}\rightleftharpoons\mathrm{A}}} = \left\langle \vec{b}_{\xi}^{\vec{J}^{\mathrm{F}}}, -j\omega\mu_0\mathcal{L}_0\left(\vec{b}_{\zeta}^{\vec{J}^{\mathrm{G}\rightleftharpoons\mathrm{A}}}\right)\right\rangle_{\mathbb{S}^{\mathrm{F}}} \tag{D-98a}$$

$$z_{\xi\zeta}^{\vec{J}^{\mathrm{F}}\vec{J}_{1}^{\mathrm{A}\rightleftharpoons\mathrm{F}}} = \left\langle \vec{b}_{\xi}^{\vec{J}^{\mathrm{F}}}, -j\omega\mu_0\mathcal{L}_0\left(-\vec{b}_{\zeta}^{\vec{J}_{1}^{\mathrm{A}\rightleftharpoons\mathrm{F}}}\right)\right\rangle_{\mathbb{S}^{\mathrm{F}}} \tag{D-98b}$$

$$z_{\xi\zeta}^{\vec{J}^{\mathrm{F}}\vec{J}_{2}^{\mathrm{A}\rightleftharpoons\mathrm{F}}} = \left\langle \vec{b}_{\xi}^{\vec{J}^{\mathrm{F}}}, -j\omega\mu_0\mathcal{L}_0\left(-\vec{b}_{\zeta}^{\vec{J}_{2}^{\mathrm{A}\rightleftharpoons\mathrm{F}}}\right)\right\rangle_{\mathbb{S}^{\mathrm{F}}} \tag{D-98c}$$

$$z_{\xi\zeta}^{\vec{J}^{\mathrm{F}}\vec{J}^{\mathrm{F}}} = \left\langle \vec{b}_{\xi}^{\vec{J}^{\mathrm{F}}}, -j\omega\mu_0\mathcal{L}_0\left(\vec{b}_{\zeta}^{\vec{J}^{\mathrm{F}}}\right)\right\rangle_{\mathbb{S}^{\mathrm{F}}} \tag{D-98d}$$

$$z_{\xi\zeta}^{\vec{J}^{\mathrm{F}}\vec{M}^{\mathrm{G}\rightleftharpoons\mathrm{A}}} = \left\langle \vec{b}_{\xi}^{\vec{J}^{\mathrm{F}}}, -\mathcal{K}_0\left(\vec{b}_{\zeta}^{\vec{M}^{\mathrm{G}\rightleftharpoons\mathrm{A}}}\right)\right\rangle_{\mathbb{S}^{\mathrm{F}}} \tag{D-98e}$$

$$z_{\xi\zeta}^{\vec{J}^{\mathrm{F}}\vec{M}_{1}^{\mathrm{A}\rightleftharpoons\mathrm{F}}} = \left\langle \vec{b}_{\xi}^{\vec{J}^{\mathrm{F}}}, -\mathcal{K}_0\left(-\vec{b}_{\zeta}^{\vec{M}_{1}^{\mathrm{A}\rightleftharpoons\mathrm{F}}}\right)\right\rangle_{\mathbb{S}^{\mathrm{F}}} \tag{D-98f}$$

$$z_{\xi\zeta}^{\vec{J}^{\mathrm{F}}\vec{M}_{2}^{\mathrm{A}\rightleftharpoons\mathrm{F}}} = \left\langle \vec{b}_{\xi}^{\vec{J}^{\mathrm{F}}}, -\mathcal{K}_0\left(-\vec{b}_{\zeta}^{\vec{M}_{2}^{\mathrm{A}\rightleftharpoons\mathrm{F}}}\right)\right\rangle_{\mathbb{S}^{\mathrm{F}}} \tag{D-98g}$$

The transformation matrix $\overline{\overline{T}}$ used in Eq. (6-117) is as follows:

$$\overline{\overline{T}} = \overline{\overline{T}}^{\vec{J}^{\mathrm{G}\rightleftharpoons\mathrm{A}}\rightarrow\mathrm{AV}} \text{ or } \overline{\overline{T}}^{\vec{M}^{\mathrm{G}\rightleftharpoons\mathrm{A}}\rightarrow\mathrm{AV}} \text{ or } \overline{\overline{T}}^{\mathrm{BS}\rightarrow\mathrm{AV}} \tag{D-99}$$

in which

$$\bar{\bar{T}}^{\vec{J}^{G\rightleftharpoons A}\rightarrow AV} = \left(\bar{\bar{\Psi}}_1\right)^{-1}\cdot\bar{\bar{\Psi}}_2 \tag{D-100a}$$

$$\bar{\bar{T}}^{\vec{M}^{G\rightleftharpoons A}\rightarrow AV} = \left(\bar{\bar{\Psi}}_3\right)^{-1}\cdot\bar{\bar{\Psi}}_4 \tag{D-100b}$$

$$\bar{\bar{T}}^{BS\rightarrow AV} = \mathrm{nullspace}\left(\bar{\bar{\Psi}}_{FCE}^{DoJ/DoM}\right) \tag{D-101}$$

where

$$\bar{\bar{\Psi}}_1 = \begin{bmatrix}
\bar{\bar{I}}^{\vec{J}^{G\rightleftharpoons A}} & 0 & 0 & 0 & 0 & 0 & 0 & 0 & 0 & 0 \\
0 & \bar{\bar{Z}}^{\vec{M}^{G\rightleftharpoons A}\vec{J}_1^{A\rightleftharpoons F}} & \bar{\bar{Z}}^{\vec{M}^{G\rightleftharpoons A}\vec{J}_2^{A\rightleftharpoons F}} & 0 & 0 & \bar{\bar{Z}}^{\vec{M}^{G\rightleftharpoons A}\vec{J}^{F}} & \bar{\bar{Z}}^{\vec{M}^{G\rightleftharpoons A}\vec{M}^{G\rightleftharpoons A}} & \bar{\bar{Z}}^{\vec{M}^{G\rightleftharpoons A}\vec{M}_1^{A\rightleftharpoons F}} & \bar{\bar{Z}}^{\vec{M}^{G\rightleftharpoons A}\vec{M}_2^{A\rightleftharpoons F}} & 0 \\
0 & \bar{\bar{Z}}^{\vec{J}_1^{A\rightleftharpoons F}\vec{J}_1^{A\rightleftharpoons F}} & \bar{\bar{Z}}^{\vec{J}_1^{A\rightleftharpoons F}\vec{J}_2^{A\rightleftharpoons F}} & 0 & \bar{\bar{Z}}^{\vec{J}_1^{A\rightleftharpoons F}\vec{J}_1^{A}} & \bar{\bar{Z}}^{\vec{J}_1^{A\rightleftharpoons F}\vec{J}^{F}} & \bar{\bar{Z}}^{\vec{J}_1^{A\rightleftharpoons F}\vec{M}^{G\rightleftharpoons A}} & \bar{\bar{Z}}^{\vec{J}_1^{A\rightleftharpoons F}\vec{M}_1^{A\rightleftharpoons F}} & \bar{\bar{Z}}^{\vec{J}_1^{A\rightleftharpoons F}\vec{M}_2^{A\rightleftharpoons F}} & 0 \\
0 & \bar{\bar{Z}}^{\vec{M}_1^{A\rightleftharpoons F}\vec{J}_1^{A\rightleftharpoons F}} & \bar{\bar{Z}}^{\vec{M}_1^{A\rightleftharpoons F}\vec{J}_2^{A\rightleftharpoons F}} & 0 & \bar{\bar{Z}}^{\vec{M}_1^{A\rightleftharpoons F}\vec{J}_1^{A}} & \bar{\bar{Z}}^{\vec{M}_1^{A\rightleftharpoons F}\vec{J}^{F}} & \bar{\bar{Z}}^{\vec{M}_1^{A\rightleftharpoons F}\vec{M}^{G\rightleftharpoons A}} & \bar{\bar{Z}}^{\vec{M}_1^{A\rightleftharpoons F}\vec{M}_1^{A\rightleftharpoons F}} & \bar{\bar{Z}}^{\vec{M}_1^{A\rightleftharpoons F}\vec{M}_2^{A\rightleftharpoons F}} & 0 \\
0 & \bar{\bar{Z}}^{\vec{J}_2^{A\rightleftharpoons F}\vec{J}_1^{A\rightleftharpoons F}} & \bar{\bar{Z}}^{\vec{J}_2^{A\rightleftharpoons F}\vec{J}_2^{A\rightleftharpoons F}} & \bar{\bar{Z}}^{\vec{J}_2^{A\rightleftharpoons F}\vec{J}_{23}^{A\rightleftharpoons A}} & 0 & \bar{\bar{Z}}^{\vec{J}_2^{A\rightleftharpoons F}\vec{J}^{F}} & \bar{\bar{Z}}^{\vec{J}_2^{A\rightleftharpoons F}\vec{M}^{G\rightleftharpoons A}} & \bar{\bar{Z}}^{\vec{J}_2^{A\rightleftharpoons F}\vec{M}_1^{A\rightleftharpoons F}} & \bar{\bar{Z}}^{\vec{J}_2^{A\rightleftharpoons F}\vec{M}_2^{A\rightleftharpoons F}} & \bar{\bar{Z}}^{\vec{J}_2^{A\rightleftharpoons F}\vec{M}_{23}^{A\rightleftharpoons A}} \\
0 & \bar{\bar{Z}}^{\vec{M}_2^{A\rightleftharpoons F}\vec{J}_1^{A\rightleftharpoons F}} & \bar{\bar{Z}}^{\vec{M}_2^{A\rightleftharpoons F}\vec{J}_2^{A\rightleftharpoons F}} & \bar{\bar{Z}}^{\vec{M}_2^{A\rightleftharpoons F}\vec{J}_{23}^{A\rightleftharpoons A}} & 0 & \bar{\bar{Z}}^{\vec{M}_2^{A\rightleftharpoons F}\vec{J}^{F}} & \bar{\bar{Z}}^{\vec{M}_2^{A\rightleftharpoons F}\vec{M}^{G\rightleftharpoons A}} & \bar{\bar{Z}}^{\vec{M}_2^{A\rightleftharpoons F}\vec{M}_1^{A\rightleftharpoons F}} & \bar{\bar{Z}}^{\vec{M}_2^{A\rightleftharpoons F}\vec{M}_2^{A\rightleftharpoons F}} & \bar{\bar{Z}}^{\vec{M}_2^{A\rightleftharpoons F}\vec{M}_{23}^{A\rightleftharpoons A}} \\
0 & 0 & \bar{\bar{Z}}^{\vec{J}_{23}^{A\rightleftharpoons A}\vec{J}_2^{A\rightleftharpoons F}} & \bar{\bar{Z}}^{\vec{J}_{23}^{A\rightleftharpoons A}\vec{J}_{23}^{A\rightleftharpoons A}} & 0 & 0 & 0 & 0 & \bar{\bar{Z}}^{\vec{J}_{23}^{A\rightleftharpoons A}\vec{M}_2^{A\rightleftharpoons F}} & \bar{\bar{Z}}^{\vec{J}_{23}^{A\rightleftharpoons A}\vec{M}_{23}^{A\rightleftharpoons A}} \\
0 & 0 & \bar{\bar{Z}}^{\vec{M}_{23}^{A\rightleftharpoons A}\vec{J}_2^{A\rightleftharpoons F}} & \bar{\bar{Z}}^{\vec{M}_{23}^{A\rightleftharpoons A}\vec{J}_{23}^{A\rightleftharpoons A}} & 0 & 0 & 0 & 0 & \bar{\bar{Z}}^{\vec{M}_{23}^{A\rightleftharpoons A}\vec{M}_2^{A\rightleftharpoons F}} & \bar{\bar{Z}}^{\vec{M}_{23}^{A\rightleftharpoons A}\vec{M}_{23}^{A\rightleftharpoons A}} \\
0 & \bar{\bar{Z}}^{\vec{J}_1^{A}\vec{J}_1^{A\rightleftharpoons F}} & 0 & 0 & \bar{\bar{Z}}^{\vec{J}_1^{A}\vec{J}_1^{A}} & 0 & 0 & \bar{\bar{Z}}^{\vec{J}_1^{A}\vec{M}_1^{A\rightleftharpoons F}} & 0 & 0 \\
0 & \bar{\bar{Z}}^{\vec{J}^{F}\vec{J}_1^{A\rightleftharpoons F}} & \bar{\bar{Z}}^{\vec{J}^{F}\vec{J}_2^{A\rightleftharpoons F}} & 0 & 0 & \bar{\bar{Z}}^{\vec{J}^{F}\vec{J}^{F}} & \bar{\bar{Z}}^{\vec{J}^{F}\vec{M}^{G\rightleftharpoons A}} & \bar{\bar{Z}}^{\vec{J}^{F}\vec{M}_1^{A\rightleftharpoons F}} & \bar{\bar{Z}}^{\vec{J}^{F}\vec{M}_2^{A\rightleftharpoons F}} & 0
\end{bmatrix} \tag{D-102a}$$

$$\bar{\bar{\Psi}}_2 = \begin{bmatrix}
\bar{\bar{I}}^{\vec{J}^{G\rightleftharpoons A}} \\
-\bar{\bar{Z}}^{\vec{M}^{G\rightleftharpoons A}\vec{J}^{G\rightleftharpoons A}} \\
-\bar{\bar{Z}}^{\vec{J}_1^{A\rightleftharpoons F}\vec{J}^{G\rightleftharpoons A}} \\
-\bar{\bar{Z}}^{\vec{M}_1^{A\rightleftharpoons F}\vec{J}^{G\rightleftharpoons A}} \\
-\bar{\bar{Z}}^{\vec{J}_2^{A\rightleftharpoons F}\vec{J}^{G\rightleftharpoons A}} \\
-\bar{\bar{Z}}^{\vec{M}_2^{A\rightleftharpoons F}\vec{J}^{G\rightleftharpoons A}} \\
0 \\
0 \\
0 \\
-\bar{\bar{Z}}^{\vec{J}^{F}\vec{J}^{G\rightleftharpoons A}}
\end{bmatrix} \tag{D-102b}$$

and

$$\bar{\bar{\Psi}}_3 = \begin{bmatrix}
0 & 0 & 0 & 0 & 0 & 0 & \bar{\bar{I}}^{\vec{M}^{G\rightleftharpoons A}} & 0 & 0 & 0 \\
\bar{\bar{Z}}^{\vec{J}^{G\rightleftharpoons A}\vec{J}^{G\rightleftharpoons A}} & \bar{\bar{Z}}^{\vec{J}^{G\rightleftharpoons A}\vec{J}_1^{A\rightleftharpoons F}} & \bar{\bar{Z}}^{\vec{J}^{G\rightleftharpoons A}\vec{J}_2^{A\rightleftharpoons F}} & 0 & 0 & \bar{\bar{Z}}^{\vec{J}^{G\rightleftharpoons A}\vec{J}^{F}} & 0 & \bar{\bar{Z}}^{\vec{J}^{G\rightleftharpoons A}\vec{M}_1^{A\rightleftharpoons F}} & \bar{\bar{Z}}^{\vec{J}^{G\rightleftharpoons A}\vec{M}_2^{A\rightleftharpoons F}} & 0 \\
\bar{\bar{Z}}^{\vec{J}_1^{A\rightleftharpoons F}\vec{J}^{G\rightleftharpoons A}} & \bar{\bar{Z}}^{\vec{J}_1^{A\rightleftharpoons F}\vec{J}_1^{A\rightleftharpoons F}} & \bar{\bar{Z}}^{\vec{J}_1^{A\rightleftharpoons F}\vec{J}_2^{A\rightleftharpoons F}} & 0 & \bar{\bar{Z}}^{\vec{J}_1^{A\rightleftharpoons F}\vec{J}_1^{A}} & \bar{\bar{Z}}^{\vec{J}_1^{A\rightleftharpoons F}\vec{J}^{F}} & 0 & \bar{\bar{Z}}^{\vec{J}_1^{A\rightleftharpoons F}\vec{M}_1^{A\rightleftharpoons F}} & \bar{\bar{Z}}^{\vec{J}_1^{A\rightleftharpoons F}\vec{M}_2^{A\rightleftharpoons F}} & 0 \\
\bar{\bar{Z}}^{\vec{M}_1^{A\rightleftharpoons F}\vec{J}^{G\rightleftharpoons A}} & \bar{\bar{Z}}^{\vec{M}_1^{A\rightleftharpoons F}\vec{J}_1^{A\rightleftharpoons F}} & \bar{\bar{Z}}^{\vec{M}_1^{A\rightleftharpoons F}\vec{J}_2^{A\rightleftharpoons F}} & 0 & \bar{\bar{Z}}^{\vec{M}_1^{A\rightleftharpoons F}\vec{J}_1^{A}} & \bar{\bar{Z}}^{\vec{M}_1^{A\rightleftharpoons F}\vec{J}^{F}} & 0 & \bar{\bar{Z}}^{\vec{M}_1^{A\rightleftharpoons F}\vec{M}_1^{A\rightleftharpoons F}} & \bar{\bar{Z}}^{\vec{M}_1^{A\rightleftharpoons F}\vec{M}_2^{A\rightleftharpoons F}} & 0 \\
\bar{\bar{Z}}^{\vec{J}_2^{A\rightleftharpoons F}\vec{J}^{G\rightleftharpoons A}} & \bar{\bar{Z}}^{\vec{J}_2^{A\rightleftharpoons F}\vec{J}_1^{A\rightleftharpoons F}} & \bar{\bar{Z}}^{\vec{J}_2^{A\rightleftharpoons F}\vec{J}_2^{A\rightleftharpoons F}} & \bar{\bar{Z}}^{\vec{J}_2^{A\rightleftharpoons F}\vec{J}_{23}^{A\rightleftharpoons A}} & 0 & \bar{\bar{Z}}^{\vec{J}_2^{A\rightleftharpoons F}\vec{J}^{F}} & 0 & \bar{\bar{Z}}^{\vec{J}_2^{A\rightleftharpoons F}\vec{M}_1^{A\rightleftharpoons F}} & \bar{\bar{Z}}^{\vec{J}_2^{A\rightleftharpoons F}\vec{M}_2^{A\rightleftharpoons F}} & \bar{\bar{Z}}^{\vec{J}_2^{A\rightleftharpoons F}\vec{M}_{23}^{A\rightleftharpoons A}} \\
\bar{\bar{Z}}^{\vec{M}_2^{A\rightleftharpoons F}\vec{J}^{G\rightleftharpoons A}} & \bar{\bar{Z}}^{\vec{M}_2^{A\rightleftharpoons F}\vec{J}_1^{A\rightleftharpoons F}} & \bar{\bar{Z}}^{\vec{M}_2^{A\rightleftharpoons F}\vec{J}_2^{A\rightleftharpoons F}} & \bar{\bar{Z}}^{\vec{M}_2^{A\rightleftharpoons F}\vec{J}_{23}^{A\rightleftharpoons A}} & 0 & \bar{\bar{Z}}^{\vec{M}_2^{A\rightleftharpoons F}\vec{J}^{F}} & 0 & \bar{\bar{Z}}^{\vec{M}_2^{A\rightleftharpoons F}\vec{M}_1^{A\rightleftharpoons F}} & \bar{\bar{Z}}^{\vec{M}_2^{A\rightleftharpoons F}\vec{M}_2^{A\rightleftharpoons F}} & \bar{\bar{Z}}^{\vec{M}_2^{A\rightleftharpoons F}\vec{M}_{23}^{A\rightleftharpoons A}} \\
0 & 0 & \bar{\bar{Z}}^{\vec{J}_{23}^{A\rightleftharpoons A}\vec{J}_2^{A\rightleftharpoons F}} & \bar{\bar{Z}}^{\vec{J}_{23}^{A\rightleftharpoons A}\vec{J}_{23}^{A\rightleftharpoons A}} & 0 & 0 & 0 & 0 & \bar{\bar{Z}}^{\vec{J}_{23}^{A\rightleftharpoons A}\vec{M}_2^{A\rightleftharpoons F}} & \bar{\bar{Z}}^{\vec{J}_{23}^{A\rightleftharpoons A}\vec{M}_{23}^{A\rightleftharpoons A}} \\
0 & 0 & \bar{\bar{Z}}^{\vec{M}_{23}^{A\rightleftharpoons A}\vec{J}_2^{A\rightleftharpoons F}} & \bar{\bar{Z}}^{\vec{M}_{23}^{A\rightleftharpoons A}\vec{J}_{23}^{A\rightleftharpoons A}} & 0 & 0 & 0 & 0 & \bar{\bar{Z}}^{\vec{M}_{23}^{A\rightleftharpoons A}\vec{M}_2^{A\rightleftharpoons F}} & \bar{\bar{Z}}^{\vec{M}_{23}^{A\rightleftharpoons A}\vec{M}_{23}^{A\rightleftharpoons A}} \\
0 & \bar{\bar{Z}}^{\vec{J}_1^{A}\vec{J}_1^{A\rightleftharpoons F}} & 0 & 0 & \bar{\bar{Z}}^{\vec{J}_1^{A}\vec{J}_1^{A}} & 0 & 0 & \bar{\bar{Z}}^{\vec{J}_1^{A}\vec{M}_1^{A\rightleftharpoons F}} & 0 & 0 \\
\bar{\bar{Z}}^{\vec{J}^{F}\vec{J}^{G\rightleftharpoons A}} & \bar{\bar{Z}}^{\vec{J}^{F}\vec{J}_1^{A\rightleftharpoons F}} & \bar{\bar{Z}}^{\vec{J}^{F}\vec{J}_2^{A\rightleftharpoons F}} & 0 & 0 & \bar{\bar{Z}}^{\vec{J}^{F}\vec{J}^{F}} & 0 & \bar{\bar{Z}}^{\vec{J}^{F}\vec{M}_1^{A\rightleftharpoons F}} & \bar{\bar{Z}}^{\vec{J}^{F}\vec{M}_2^{A\rightleftharpoons F}} & 0
\end{bmatrix} \tag{D-103a}$$

$$\overline{\overline{\Psi}}_4 = \begin{bmatrix} \overline{\overline{I}}^{\vec{M}^{\mathrm{G\rightleftharpoons A}}} \\ -\overline{\overline{Z}}^{\vec{J}^{\mathrm{G\rightleftharpoons A}}\vec{M}^{\mathrm{G\rightleftharpoons A}}} \\ -\overline{\overline{Z}}^{\vec{J}_1^{\mathrm{A\rightleftharpoons F}}\vec{M}^{\mathrm{G\rightleftharpoons A}}} \\ -\overline{\overline{Z}}^{\vec{M}_1^{\mathrm{A\rightleftharpoons F}}\vec{M}^{\mathrm{G\rightleftharpoons A}}} \\ -\overline{\overline{Z}}^{\vec{J}_2^{\mathrm{A\rightleftharpoons F}}\vec{M}^{\mathrm{G\rightleftharpoons A}}} \\ -\overline{\overline{Z}}^{\vec{M}_2^{\mathrm{A\rightleftharpoons F}}\vec{M}^{\mathrm{G\rightleftharpoons A}}} \\ 0 \\ 0 \\ 0 \\ -\overline{\overline{Z}}^{\vec{J}^{\mathrm{F}}\vec{M}^{\mathrm{G\rightleftharpoons A}}} \end{bmatrix} \tag{D-103b}$$

and

$$\overline{\overline{\Psi}}_{\mathrm{FCE}}^{\mathrm{DoJ}} = \begin{bmatrix}
\overline{\overline{Z}}^{\vec{M}^{\mathrm{G\rightleftharpoons A}}\vec{J}^{\mathrm{G\rightleftharpoons A}}} & \overline{\overline{Z}}^{\vec{M}^{\mathrm{G\rightleftharpoons A}}\vec{J}_1^{\mathrm{A\rightleftharpoons F}}} & \overline{\overline{Z}}^{\vec{M}^{\mathrm{G\rightleftharpoons A}}\vec{J}_2^{\mathrm{A\rightleftharpoons F}}} & 0 & 0 & \overline{\overline{Z}}^{\vec{M}^{\mathrm{G\rightleftharpoons A}}\vec{J}^{\mathrm{F}}} & \overline{\overline{Z}}^{\vec{M}^{\mathrm{G\rightleftharpoons A}}\vec{M}^{\mathrm{G\rightleftharpoons A}}} & \overline{\overline{Z}}^{\vec{M}^{\mathrm{G\rightleftharpoons A}}\vec{M}_1^{\mathrm{A\rightleftharpoons F}}} & \overline{\overline{Z}}^{\vec{M}^{\mathrm{G\rightleftharpoons A}}\vec{M}_2^{\mathrm{A\rightleftharpoons F}}} & 0 \\
\overline{\overline{Z}}^{\vec{J}_1^{\mathrm{A\rightleftharpoons F}}\vec{J}^{\mathrm{G\rightleftharpoons A}}} & \overline{\overline{Z}}^{\vec{J}_1^{\mathrm{A\rightleftharpoons F}}\vec{J}_1^{\mathrm{A\rightleftharpoons F}}} & \overline{\overline{Z}}^{\vec{J}_1^{\mathrm{A\rightleftharpoons F}}\vec{J}_2^{\mathrm{A\rightleftharpoons F}}} & 0 & \overline{\overline{Z}}^{\vec{J}_1^{\mathrm{A\rightleftharpoons F}}\vec{J}_1^{\mathrm{A}}} & \overline{\overline{Z}}^{\vec{J}_1^{\mathrm{A\rightleftharpoons F}}\vec{J}^{\mathrm{F}}} & \overline{\overline{Z}}^{\vec{J}_1^{\mathrm{A\rightleftharpoons F}}\vec{M}^{\mathrm{G\rightleftharpoons A}}} & \overline{\overline{Z}}^{\vec{J}_1^{\mathrm{A\rightleftharpoons F}}\vec{M}_1^{\mathrm{A\rightleftharpoons F}}} & \overline{\overline{Z}}^{\vec{J}_1^{\mathrm{A\rightleftharpoons F}}\vec{M}_2^{\mathrm{A\rightleftharpoons F}}} & 0 \\
\overline{\overline{Z}}^{\vec{M}_1^{\mathrm{A\rightleftharpoons F}}\vec{J}^{\mathrm{G\rightleftharpoons A}}} & \overline{\overline{Z}}^{\vec{M}_1^{\mathrm{A\rightleftharpoons F}}\vec{J}_1^{\mathrm{A\rightleftharpoons F}}} & \overline{\overline{Z}}^{\vec{M}_1^{\mathrm{A\rightleftharpoons F}}\vec{J}_2^{\mathrm{A\rightleftharpoons F}}} & 0 & \overline{\overline{Z}}^{\vec{M}_1^{\mathrm{A\rightleftharpoons F}}\vec{J}_1^{\mathrm{A}}} & \overline{\overline{Z}}^{\vec{M}_1^{\mathrm{A\rightleftharpoons F}}\vec{J}^{\mathrm{F}}} & \overline{\overline{Z}}^{\vec{M}_1^{\mathrm{A\rightleftharpoons F}}\vec{M}^{\mathrm{G\rightleftharpoons A}}} & \overline{\overline{Z}}^{\vec{M}_1^{\mathrm{A\rightleftharpoons F}}\vec{M}_1^{\mathrm{A\rightleftharpoons F}}} & \overline{\overline{Z}}^{\vec{M}_1^{\mathrm{A\rightleftharpoons F}}\vec{M}_2^{\mathrm{A\rightleftharpoons F}}} & 0 \\
\overline{\overline{Z}}^{\vec{J}_2^{\mathrm{A\rightleftharpoons F}}\vec{J}^{\mathrm{G\rightleftharpoons A}}} & \overline{\overline{Z}}^{\vec{J}_2^{\mathrm{A\rightleftharpoons F}}\vec{J}_1^{\mathrm{A\rightleftharpoons F}}} & \overline{\overline{Z}}^{\vec{J}_2^{\mathrm{A\rightleftharpoons F}}\vec{J}_2^{\mathrm{A\rightleftharpoons F}}} & \overline{\overline{Z}}^{\vec{J}_2^{\mathrm{A\rightleftharpoons F}}\vec{J}_{23}^{\mathrm{A\rightleftharpoons A}}} & 0 & \overline{\overline{Z}}^{\vec{J}_2^{\mathrm{A\rightleftharpoons F}}\vec{J}^{\mathrm{F}}} & \overline{\overline{Z}}^{\vec{J}_2^{\mathrm{A\rightleftharpoons F}}\vec{M}^{\mathrm{G\rightleftharpoons A}}} & \overline{\overline{Z}}^{\vec{J}_2^{\mathrm{A\rightleftharpoons F}}\vec{M}_1^{\mathrm{A\rightleftharpoons F}}} & \overline{\overline{Z}}^{\vec{J}_2^{\mathrm{A\rightleftharpoons F}}\vec{M}_2^{\mathrm{A\rightleftharpoons F}}} & \overline{\overline{Z}}^{\vec{J}_2^{\mathrm{A\rightleftharpoons F}}\vec{M}_{23}^{\mathrm{A\rightleftharpoons A}}} \\
\overline{\overline{Z}}^{\vec{M}_2^{\mathrm{A\rightleftharpoons F}}\vec{J}^{\mathrm{G\rightleftharpoons A}}} & \overline{\overline{Z}}^{\vec{M}_2^{\mathrm{A\rightleftharpoons F}}\vec{J}_1^{\mathrm{A\rightleftharpoons F}}} & \overline{\overline{Z}}^{\vec{M}_2^{\mathrm{A\rightleftharpoons F}}\vec{J}_2^{\mathrm{A\rightleftharpoons F}}} & \overline{\overline{Z}}^{\vec{M}_2^{\mathrm{A\rightleftharpoons F}}\vec{J}_{23}^{\mathrm{A\rightleftharpoons A}}} & 0 & \overline{\overline{Z}}^{\vec{M}_2^{\mathrm{A\rightleftharpoons F}}\vec{J}^{\mathrm{F}}} & \overline{\overline{Z}}^{\vec{M}_2^{\mathrm{A\rightleftharpoons F}}\vec{M}^{\mathrm{G\rightleftharpoons A}}} & \overline{\overline{Z}}^{\vec{M}_2^{\mathrm{A\rightleftharpoons F}}\vec{M}_1^{\mathrm{A\rightleftharpoons F}}} & \overline{\overline{Z}}^{\vec{M}_2^{\mathrm{A\rightleftharpoons F}}\vec{M}_2^{\mathrm{A\rightleftharpoons F}}} & \overline{\overline{Z}}^{\vec{M}_2^{\mathrm{A\rightleftharpoons F}}\vec{M}_{23}^{\mathrm{A\rightleftharpoons A}}} \\
0 & 0 & \overline{\overline{Z}}^{\vec{J}_{23}^{\mathrm{A\rightleftharpoons A}}\vec{J}_2^{\mathrm{A\rightleftharpoons F}}} & \overline{\overline{Z}}^{\vec{J}_{23}^{\mathrm{A\rightleftharpoons A}}\vec{J}_{23}^{\mathrm{A\rightleftharpoons A}}} & 0 & 0 & 0 & 0 & \overline{\overline{Z}}^{\vec{J}_{23}^{\mathrm{A\rightleftharpoons A}}\vec{M}_2^{\mathrm{A\rightleftharpoons F}}} & \overline{\overline{Z}}^{\vec{J}_{23}^{\mathrm{A\rightleftharpoons A}}\vec{M}_{23}^{\mathrm{A\rightleftharpoons A}}} \\
0 & 0 & \overline{\overline{Z}}^{\vec{M}_{23}^{\mathrm{A\rightleftharpoons A}}\vec{J}_2^{\mathrm{A\rightleftharpoons F}}} & \overline{\overline{Z}}^{\vec{M}_{23}^{\mathrm{A\rightleftharpoons A}}\vec{J}_{23}^{\mathrm{A\rightleftharpoons A}}} & 0 & 0 & 0 & 0 & \overline{\overline{Z}}^{\vec{M}_{23}^{\mathrm{A\rightleftharpoons A}}\vec{M}_2^{\mathrm{A\rightleftharpoons F}}} & \overline{\overline{Z}}^{\vec{M}_{23}^{\mathrm{A\rightleftharpoons A}}\vec{M}_{23}^{\mathrm{A\rightleftharpoons A}}} \\
0 & \overline{\overline{Z}}^{\vec{J}_1^{\mathrm{A}}\vec{J}_1^{\mathrm{A\rightleftharpoons F}}} & 0 & 0 & \overline{\overline{Z}}^{\vec{J}_1^{\mathrm{A}}\vec{J}_1^{\mathrm{A}}} & 0 & 0 & \overline{\overline{Z}}^{\vec{J}_1^{\mathrm{A}}\vec{M}_1^{\mathrm{A\rightleftharpoons F}}} & 0 & 0 \\
\overline{\overline{Z}}^{\vec{J}^{\mathrm{F}}\vec{J}^{\mathrm{G\rightleftharpoons A}}} & \overline{\overline{Z}}^{\vec{J}^{\mathrm{F}}\vec{J}_1^{\mathrm{A\rightleftharpoons F}}} & \overline{\overline{Z}}^{\vec{J}^{\mathrm{F}}\vec{J}_2^{\mathrm{A\rightleftharpoons F}}} & 0 & 0 & \overline{\overline{Z}}^{\vec{J}^{\mathrm{F}}\vec{J}^{\mathrm{F}}} & \overline{\overline{Z}}^{\vec{J}^{\mathrm{F}}\vec{M}^{\mathrm{G\rightleftharpoons A}}} & \overline{\overline{Z}}^{\vec{J}^{\mathrm{F}}\vec{M}_1^{\mathrm{A\rightleftharpoons F}}} & \overline{\overline{Z}}^{\vec{J}^{\mathrm{F}}\vec{M}_2^{\mathrm{A\rightleftharpoons F}}} & 0
\end{bmatrix} \tag{D-104a}$$

$$\overline{\overline{\Psi}}_{\mathrm{FCE}}^{\mathrm{DoM}} = \begin{bmatrix}
\overline{\overline{Z}}^{\vec{J}^{\mathrm{G\rightleftharpoons A}}\vec{J}^{\mathrm{G\rightleftharpoons A}}} & \overline{\overline{Z}}^{\vec{J}^{\mathrm{G\rightleftharpoons A}}\vec{J}_1^{\mathrm{A\rightleftharpoons F}}} & \overline{\overline{Z}}^{\vec{J}^{\mathrm{G\rightleftharpoons A}}\vec{J}_2^{\mathrm{A\rightleftharpoons F}}} & 0 & 0 & \overline{\overline{Z}}^{\vec{J}^{\mathrm{G\rightleftharpoons A}}\vec{J}^{\mathrm{F}}} & \overline{\overline{Z}}^{\vec{J}^{\mathrm{G\rightleftharpoons A}}\vec{M}^{\mathrm{G\rightleftharpoons A}}} & \overline{\overline{Z}}^{\vec{J}^{\mathrm{G\rightleftharpoons A}}\vec{M}_1^{\mathrm{A\rightleftharpoons F}}} & \overline{\overline{Z}}^{\vec{J}^{\mathrm{G\rightleftharpoons A}}\vec{M}_2^{\mathrm{A\rightleftharpoons F}}} & 0 \\
\overline{\overline{Z}}^{\vec{J}_1^{\mathrm{A\rightleftharpoons F}}\vec{J}^{\mathrm{G\rightleftharpoons A}}} & \overline{\overline{Z}}^{\vec{J}_1^{\mathrm{A\rightleftharpoons F}}\vec{J}_1^{\mathrm{A\rightleftharpoons F}}} & \overline{\overline{Z}}^{\vec{J}_1^{\mathrm{A\rightleftharpoons F}}\vec{J}_2^{\mathrm{A\rightleftharpoons F}}} & 0 & \overline{\overline{Z}}^{\vec{J}_1^{\mathrm{A\rightleftharpoons F}}\vec{J}_1^{\mathrm{A}}} & \overline{\overline{Z}}^{\vec{J}_1^{\mathrm{A\rightleftharpoons F}}\vec{J}^{\mathrm{F}}} & \overline{\overline{Z}}^{\vec{J}_1^{\mathrm{A\rightleftharpoons F}}\vec{M}^{\mathrm{G\rightleftharpoons A}}} & \overline{\overline{Z}}^{\vec{J}_1^{\mathrm{A\rightleftharpoons F}}\vec{M}_1^{\mathrm{A\rightleftharpoons F}}} & \overline{\overline{Z}}^{\vec{J}_1^{\mathrm{A\rightleftharpoons F}}\vec{M}_2^{\mathrm{A\rightleftharpoons F}}} & 0 \\
\overline{\overline{Z}}^{\vec{M}_1^{\mathrm{A\rightleftharpoons F}}\vec{J}^{\mathrm{G\rightleftharpoons A}}} & \overline{\overline{Z}}^{\vec{M}_1^{\mathrm{A\rightleftharpoons F}}\vec{J}_1^{\mathrm{A\rightleftharpoons F}}} & \overline{\overline{Z}}^{\vec{M}_1^{\mathrm{A\rightleftharpoons F}}\vec{J}_2^{\mathrm{A\rightleftharpoons F}}} & 0 & \overline{\overline{Z}}^{\vec{M}_1^{\mathrm{A\rightleftharpoons F}}\vec{J}_1^{\mathrm{A}}} & \overline{\overline{Z}}^{\vec{M}_1^{\mathrm{A\rightleftharpoons F}}\vec{J}^{\mathrm{F}}} & \overline{\overline{Z}}^{\vec{M}_1^{\mathrm{A\rightleftharpoons F}}\vec{M}^{\mathrm{G\rightleftharpoons A}}} & \overline{\overline{Z}}^{\vec{M}_1^{\mathrm{A\rightleftharpoons F}}\vec{M}_1^{\mathrm{A\rightleftharpoons F}}} & \overline{\overline{Z}}^{\vec{M}_1^{\mathrm{A\rightleftharpoons F}}\vec{M}_2^{\mathrm{A\rightleftharpoons F}}} & 0 \\
\overline{\overline{Z}}^{\vec{J}_2^{\mathrm{A\rightleftharpoons F}}\vec{J}^{\mathrm{G\rightleftharpoons A}}} & \overline{\overline{Z}}^{\vec{J}_2^{\mathrm{A\rightleftharpoons F}}\vec{J}_1^{\mathrm{A\rightleftharpoons F}}} & \overline{\overline{Z}}^{\vec{J}_2^{\mathrm{A\rightleftharpoons F}}\vec{J}_2^{\mathrm{A\rightleftharpoons F}}} & \overline{\overline{Z}}^{\vec{J}_2^{\mathrm{A\rightleftharpoons F}}\vec{J}_{23}^{\mathrm{A\rightleftharpoons A}}} & 0 & \overline{\overline{Z}}^{\vec{J}_2^{\mathrm{A\rightleftharpoons F}}\vec{J}^{\mathrm{F}}} & \overline{\overline{Z}}^{\vec{J}_2^{\mathrm{A\rightleftharpoons F}}\vec{M}^{\mathrm{G\rightleftharpoons A}}} & \overline{\overline{Z}}^{\vec{J}_2^{\mathrm{A\rightleftharpoons F}}\vec{M}_1^{\mathrm{A\rightleftharpoons F}}} & \overline{\overline{Z}}^{\vec{J}_2^{\mathrm{A\rightleftharpoons F}}\vec{M}_2^{\mathrm{A\rightleftharpoons F}}} & \overline{\overline{Z}}^{\vec{J}_2^{\mathrm{A\rightleftharpoons F}}\vec{M}_{23}^{\mathrm{A\rightleftharpoons A}}} \\
\overline{\overline{Z}}^{\vec{M}_2^{\mathrm{A\rightleftharpoons F}}\vec{J}^{\mathrm{G\rightleftharpoons A}}} & \overline{\overline{Z}}^{\vec{M}_2^{\mathrm{A\rightleftharpoons F}}\vec{J}_1^{\mathrm{A\rightleftharpoons F}}} & \overline{\overline{Z}}^{\vec{M}_2^{\mathrm{A\rightleftharpoons F}}\vec{J}_2^{\mathrm{A\rightleftharpoons F}}} & \overline{\overline{Z}}^{\vec{M}_2^{\mathrm{A\rightleftharpoons F}}\vec{J}_{23}^{\mathrm{A\rightleftharpoons A}}} & 0 & \overline{\overline{Z}}^{\vec{M}_2^{\mathrm{A\rightleftharpoons F}}\vec{J}^{\mathrm{F}}} & \overline{\overline{Z}}^{\vec{M}_2^{\mathrm{A\rightleftharpoons F}}\vec{M}^{\mathrm{G\rightleftharpoons A}}} & \overline{\overline{Z}}^{\vec{M}_2^{\mathrm{A\rightleftharpoons F}}\vec{M}_1^{\mathrm{A\rightleftharpoons F}}} & \overline{\overline{Z}}^{\vec{M}_2^{\mathrm{A\rightleftharpoons F}}\vec{M}_2^{\mathrm{A\rightleftharpoons F}}} & \overline{\overline{Z}}^{\vec{M}_2^{\mathrm{A\rightleftharpoons F}}\vec{M}_{23}^{\mathrm{A\rightleftharpoons A}}} \\
0 & 0 & \overline{\overline{Z}}^{\vec{J}_{23}^{\mathrm{A\rightleftharpoons A}}\vec{J}_2^{\mathrm{A\rightleftharpoons F}}} & \overline{\overline{Z}}^{\vec{J}_{23}^{\mathrm{A\rightleftharpoons A}}\vec{J}_{23}^{\mathrm{A\rightleftharpoons A}}} & 0 & 0 & 0 & 0 & \overline{\overline{Z}}^{\vec{J}_{23}^{\mathrm{A\rightleftharpoons A}}\vec{M}_2^{\mathrm{A\rightleftharpoons F}}} & \overline{\overline{Z}}^{\vec{J}_{23}^{\mathrm{A\rightleftharpoons A}}\vec{M}_{23}^{\mathrm{A\rightleftharpoons A}}} \\
0 & 0 & \overline{\overline{Z}}^{\vec{M}_{23}^{\mathrm{A\rightleftharpoons A}}\vec{J}_2^{\mathrm{A\rightleftharpoons F}}} & \overline{\overline{Z}}^{\vec{M}_{23}^{\mathrm{A\rightleftharpoons A}}\vec{J}_{23}^{\mathrm{A\rightleftharpoons A}}} & 0 & 0 & 0 & 0 & \overline{\overline{Z}}^{\vec{M}_{23}^{\mathrm{A\rightleftharpoons A}}\vec{M}_2^{\mathrm{A\rightleftharpoons F}}} & \overline{\overline{Z}}^{\vec{M}_{23}^{\mathrm{A\rightleftharpoons A}}\vec{M}_{23}^{\mathrm{A\rightleftharpoons A}}} \\
0 & \overline{\overline{Z}}^{\vec{J}_1^{\mathrm{A}}\vec{J}_1^{\mathrm{A\rightleftharpoons F}}} & 0 & 0 & \overline{\overline{Z}}^{\vec{J}_1^{\mathrm{A}}\vec{J}_1^{\mathrm{A}}} & 0 & 0 & \overline{\overline{Z}}^{\vec{J}_1^{\mathrm{A}}\vec{M}_1^{\mathrm{A\rightleftharpoons F}}} & 0 & 0 \\
\overline{\overline{Z}}^{\vec{J}^{\mathrm{F}}\vec{J}^{\mathrm{G\rightleftharpoons A}}} & \overline{\overline{Z}}^{\vec{J}^{\mathrm{F}}\vec{J}_1^{\mathrm{A\rightleftharpoons F}}} & \overline{\overline{Z}}^{\vec{J}^{\mathrm{F}}\vec{J}_2^{\mathrm{A\rightleftharpoons F}}} & 0 & 0 & \overline{\overline{Z}}^{\vec{J}^{\mathrm{F}}\vec{J}^{\mathrm{F}}} & \overline{\overline{Z}}^{\vec{J}^{\mathrm{F}}\vec{M}^{\mathrm{G\rightleftharpoons A}}} & \overline{\overline{Z}}^{\vec{J}^{\mathrm{F}}\vec{M}_1^{\mathrm{A\rightleftharpoons F}}} & \overline{\overline{Z}}^{\vec{J}^{\mathrm{F}}\vec{M}_2^{\mathrm{A\rightleftharpoons F}}} & 0
\end{bmatrix} \tag{D-104b}$$

The power quadratic form matrix $\overline{\overline{P}}^{\mathrm{G\rightleftharpoons A}}$ used in Eq. (6-122) is as follows:

$$\overline{\overline{P}}^{\mathrm{G\rightleftharpoons A}} = \overline{\overline{P}}_{\mathrm{curAV}}^{\mathrm{G\rightleftharpoons A}} \text{ or } \overline{\overline{P}}_{\mathrm{intAV}}^{\mathrm{G\rightleftharpoons A}} \tag{D-105}$$

in which

$$\overline{\overline{P}}_{\mathrm{curAV}}^{\mathrm{G\rightleftharpoons A}} = \begin{bmatrix}
0 & 0 & 0 & 0 & 0 & 0 & \overline{\overline{C}}^{\vec{J}^{\mathrm{G\rightleftharpoons A}}\vec{M}^{\mathrm{G\rightleftharpoons A}}} & 0 & 0 & 0 \\
0 & 0 & 0 & 0 & 0 & 0 & 0 & 0 & 0 & 0 \\
0 & 0 & 0 & 0 & 0 & 0 & 0 & 0 & 0 & 0 \\
0 & 0 & 0 & 0 & 0 & 0 & 0 & 0 & 0 & 0 \\
0 & 0 & 0 & 0 & 0 & 0 & 0 & 0 & 0 & 0 \\
0 & 0 & 0 & 0 & 0 & 0 & 0 & 0 & 0 & 0 \\
0 & 0 & 0 & 0 & 0 & 0 & 0 & 0 & 0 & 0 \\
0 & 0 & 0 & 0 & 0 & 0 & 0 & 0 & 0 & 0 \\
0 & 0 & 0 & 0 & 0 & 0 & 0 & 0 & 0 & 0 \\
0 & 0 & 0 & 0 & 0 & 0 & 0 & 0 & 0 & 0
\end{bmatrix} \tag{D-106}$$

corresponding to the first equality in Eq. (6-121), and

$$\overline{\overline{P}}_{\text{intAV}}^{\text{G}\rightleftharpoons\text{A}}=\begin{bmatrix}\overline{\overline{P}}^{\vec{J}^{\text{G}\rightleftharpoons\text{A}}\vec{J}^{\text{G}\rightleftharpoons\text{A}}} & \overline{\overline{P}}^{\vec{J}^{\text{G}\rightleftharpoons\text{A}}\vec{J}_1^{\text{A}\rightleftharpoons\text{F}}} & \overline{\overline{P}}^{\vec{J}^{\text{G}\rightleftharpoons\text{A}}\vec{J}_2^{\text{A}\rightleftharpoons\text{F}}} & 0 & 0 & \overline{\overline{P}}^{\vec{J}^{\text{G}\rightleftharpoons\text{A}}\vec{J}^{\text{F}}} & \overline{\overline{P}}^{\vec{J}^{\text{G}\rightleftharpoons\text{A}}\vec{M}^{\text{G}\rightleftharpoons\text{A}}} & \overline{\overline{P}}^{\vec{J}^{\text{G}\rightleftharpoons\text{A}}\vec{M}_1^{\text{A}\rightleftharpoons\text{F}}} & \overline{\overline{P}}^{\vec{J}^{\text{G}\rightleftharpoons\text{A}}\vec{M}_2^{\text{A}\rightleftharpoons\text{F}}} & 0\\ 0&0&0&0&0&0&0&0&0&0\\ 0&0&0&0&0&0&0&0&0&0\\ 0&0&0&0&0&0&0&0&0&0\\ 0&0&0&0&0&0&0&0&0&0\\ 0&0&0&0&0&0&0&0&0&0\\ 0&0&0&0&0&0&0&0&0&0\\ 0&0&0&0&0&0&0&0&0&0\\ 0&0&0&0&0&0&0&0&0&0\\ 0&0&0&0&0&0&0&0&0&0\end{bmatrix}$$

(D-107a)

corresponding to the second equality in Eq. (6-121), and

$$\overline{\overline{P}}_{\text{intAV}}^{\text{G}\rightleftharpoons\text{A}}=\begin{bmatrix}0&0&0&0&0&0&0&0&0&0\\ 0&0&0&0&0&0&0&0&0&0\\ 0&0&0&0&0&0&0&0&0&0\\ 0&0&0&0&0&0&0&0&0&0\\ 0&0&0&0&0&0&0&0&0&0\\ 0&0&0&0&0&0&0&0&0&0\\ \overline{\overline{P}}^{\vec{M}^{\text{G}\rightleftharpoons\text{A}}\vec{J}^{\text{G}\rightleftharpoons\text{A}}} & \overline{\overline{P}}^{\vec{M}^{\text{G}\rightleftharpoons\text{A}}\vec{J}_1^{\text{A}\rightleftharpoons\text{F}}} & \overline{\overline{P}}^{\vec{M}^{\text{G}\rightleftharpoons\text{A}}\vec{J}_2^{\text{A}\rightleftharpoons\text{F}}} & 0 & 0 & \overline{\overline{P}}^{\vec{M}^{\text{G}\rightleftharpoons\text{A}}\vec{J}^{\text{F}}} & \overline{\overline{P}}^{\vec{M}^{\text{G}\rightleftharpoons\text{A}}\vec{M}^{\text{G}\rightleftharpoons\text{A}}} & \overline{\overline{P}}^{\vec{M}^{\text{G}\rightleftharpoons\text{A}}\vec{M}_1^{\text{A}\rightleftharpoons\text{F}}} & \overline{\overline{P}}^{\vec{M}^{\text{G}\rightleftharpoons\text{A}}\vec{M}_2^{\text{A}\rightleftharpoons\text{F}}} & 0\\ 0&0&0&0&0&0&0&0&0&0\\ 0&0&0&0&0&0&0&0&0&0\\ 0&0&0&0&0&0&0&0&0&0\end{bmatrix}^{\dagger}$$

(D-107b)

corresponding to the third equality in Eq. (6-121), where the elements of the sub-matrices in above Eqs. (D-106) and (D-107) are as follows:

$$c_{\xi\zeta}^{\vec{J}^{\text{G}\rightleftharpoons\text{A}}\vec{M}^{\text{G}\rightleftharpoons\text{A}}} = (1/2)\left\langle \hat{n}^{\rightarrow\text{A}}\times\vec{b}_{\xi}^{\vec{J}^{\text{G}\rightleftharpoons\text{A}}},\vec{b}_{\zeta}^{\vec{M}^{\text{G}\rightleftharpoons\text{A}}}\right\rangle_{\mathbb{S}^{\text{G}\rightleftharpoons\text{A}}} \tag{D-108}$$

and

$$p_{\xi\zeta}^{\vec{J}^{\text{G}\rightleftharpoons\text{A}}\vec{J}^{\text{G}\rightleftharpoons\text{A}}} = -(1/2)\left\langle \vec{b}_{\xi}^{\vec{J}^{\text{G}\rightleftharpoons\text{A}}},-j\omega\mu_0\mathcal{L}_0\left(\vec{b}_{\zeta}^{\vec{J}^{\text{G}\rightleftharpoons\text{A}}}\right)\right\rangle_{\mathbb{S}^{\text{G}\rightleftharpoons\text{A}}} \tag{D-109a}$$

$$p_{\xi\zeta}^{\vec{J}^{\text{G}\rightleftharpoons\text{A}}\vec{J}_1^{\text{A}\rightleftharpoons\text{F}}} = -(1/2)\left\langle \vec{b}_{\xi}^{\vec{J}^{\text{G}\rightleftharpoons\text{A}}},-j\omega\mu_0\mathcal{L}_0\left(-\vec{b}_{\zeta}^{\vec{J}_1^{\text{A}\rightleftharpoons\text{F}}}\right)\right\rangle_{\mathbb{S}^{\text{G}\rightleftharpoons\text{A}}} \tag{D-109b}$$

$$p_{\xi\zeta}^{\vec{J}^{\text{G}\rightleftharpoons\text{A}}\vec{J}_2^{\text{A}\rightleftharpoons\text{F}}} = -(1/2)\left\langle \vec{b}_{\xi}^{\vec{J}^{\text{G}\rightleftharpoons\text{A}}},-j\omega\mu_0\mathcal{L}_0\left(-\vec{b}_{\zeta}^{\vec{J}_2^{\text{A}\rightleftharpoons\text{F}}}\right)\right\rangle_{\mathbb{S}^{\text{G}\rightleftharpoons\text{A}}} \tag{D-109c}$$

$$p_{\xi\zeta}^{\vec{J}^{\text{G}\rightleftharpoons\text{A}}\vec{J}^{\text{F}}} = -(1/2)\left\langle \vec{b}_{\xi}^{\vec{J}^{\text{G}\rightleftharpoons\text{A}}},-j\omega\mu_0\mathcal{L}_0\left(\vec{b}_{\zeta}^{\vec{J}^{\text{F}}}\right)\right\rangle_{\mathbb{S}^{\text{G}\rightleftharpoons\text{A}}} \tag{D-109d}$$

$$p_{\xi\zeta}^{\vec{J}^{\text{G}\rightleftharpoons\text{A}}\vec{M}^{\text{G}\rightleftharpoons\text{A}}} = -(1/2)\left\langle \vec{b}_{\xi}^{\vec{J}^{\text{G}\rightleftharpoons\text{A}}},\hat{n}^{\rightarrow\text{A}}\times\frac{1}{2}\vec{b}_{\zeta}^{\vec{M}^{\text{G}\rightleftharpoons\text{A}}}-\text{P.V.}\,\mathcal{K}_0\left(\vec{b}_{\zeta}^{\vec{M}^{\text{G}\rightleftharpoons\text{A}}}\right)\right\rangle_{\mathbb{S}^{\text{G}\rightleftharpoons\text{A}}} \tag{D-109e}$$

$$p_{\xi\zeta}^{\vec{J}^{\text{G}\rightleftharpoons\text{A}}\vec{M}_1^{\text{A}\rightleftharpoons\text{F}}} = -(1/2)\left\langle \vec{b}_{\xi}^{\vec{J}^{\text{G}\rightleftharpoons\text{A}}},-\mathcal{K}_0\left(-\vec{b}_{\zeta}^{\vec{M}_1^{\text{A}\rightleftharpoons\text{F}}}\right)\right\rangle_{\mathbb{S}^{\text{G}\rightleftharpoons\text{A}}} \tag{D-109f}$$

$$p_{\xi\zeta}^{\vec{J}^{\text{G}\rightleftharpoons\text{A}}\vec{M}_2^{\text{A}\rightleftharpoons\text{F}}} = -(1/2)\left\langle \vec{b}_{\xi}^{\vec{J}^{\text{G}\rightleftharpoons\text{A}}},-\mathcal{K}_0\left(-\vec{b}_{\zeta}^{\vec{M}_2^{\text{A}\rightleftharpoons\text{F}}}\right)\right\rangle_{\mathbb{S}^{\text{G}\rightleftharpoons\text{A}}} \tag{D-109g}$$

and

$$p_{\xi\zeta}^{\vec{M}^{\text{G}\rightleftharpoons\text{A}}\vec{J}^{\text{G}\rightleftharpoons\text{A}}} = -(1/2)\left\langle \vec{b}_{\xi}^{\vec{M}^{\text{G}\rightleftharpoons\text{A}}},\frac{1}{2}\vec{b}_{\zeta}^{\vec{J}^{\text{G}\rightleftharpoons\text{A}}}\times\hat{n}^{\rightarrow\text{A}}+\text{P.V.}\,\mathcal{K}_0\left(\vec{b}_{\zeta}^{\vec{J}^{\text{G}\rightleftharpoons\text{A}}}\right)\right\rangle_{\mathbb{S}^{\text{G}\rightleftharpoons\text{A}}} \tag{D-109h}$$

$$p_{\xi\zeta}^{\vec{M}^{\mathrm{G}\rightleftharpoons\mathrm{A}}\vec{J}_1^{\mathrm{A}\rightleftharpoons\mathrm{F}}} = -(1/2)\left\langle \vec{b}_{\xi}^{\vec{M}^{\mathrm{G}\rightleftharpoons\mathrm{A}}}, \mathcal{K}_0\left(-\vec{b}_{\zeta}^{\vec{J}_1^{\mathrm{A}\rightleftharpoons\mathrm{F}}}\right)\right\rangle_{\mathbb{S}^{\mathrm{G}\rightleftharpoons\mathrm{A}}} \tag{D-109i}$$

$$p_{\xi\zeta}^{\vec{M}^{\mathrm{G}\rightleftharpoons\mathrm{A}}\vec{J}_2^{\mathrm{A}\rightleftharpoons\mathrm{F}}} = -(1/2)\left\langle \vec{b}_{\xi}^{\vec{M}^{\mathrm{G}\rightleftharpoons\mathrm{A}}}, \mathcal{K}_0\left(-\vec{b}_{\zeta}^{\vec{J}_2^{\mathrm{A}\rightleftharpoons\mathrm{F}}}\right)\right\rangle_{\mathbb{S}^{\mathrm{G}\rightleftharpoons\mathrm{A}}} \tag{D-109j}$$

$$p_{\xi\zeta}^{\vec{M}^{\mathrm{G}\rightleftharpoons\mathrm{A}}\vec{J}^{\mathrm{F}}} = -(1/2)\left\langle \vec{b}_{\xi}^{\vec{M}^{\mathrm{G}\rightleftharpoons\mathrm{A}}}, \mathcal{K}_0\left(\vec{b}_{\zeta}^{\vec{J}^{\mathrm{F}}}\right)\right\rangle_{\mathbb{S}^{\mathrm{G}\rightleftharpoons\mathrm{A}}} \tag{D-109k}$$

$$p_{\xi\zeta}^{\vec{M}^{\mathrm{G}\rightleftharpoons\mathrm{A}}\vec{M}^{\mathrm{G}\rightleftharpoons\mathrm{A}}} = -(1/2)\left\langle \vec{b}_{\xi}^{\vec{M}^{\mathrm{G}\rightleftharpoons\mathrm{A}}}, -j\omega\varepsilon_0\mathcal{L}_0\left(\vec{b}_{\zeta}^{\vec{M}^{\mathrm{G}\rightleftharpoons\mathrm{A}}}\right)\right\rangle_{\mathbb{S}^{\mathrm{G}\rightleftharpoons\mathrm{A}}} \tag{D-109l}$$

$$p_{\xi\zeta}^{\vec{M}^{\mathrm{G}\rightleftharpoons\mathrm{A}}\vec{M}_1^{\mathrm{A}\rightleftharpoons\mathrm{F}}} = -(1/2)\left\langle \vec{b}_{\xi}^{\vec{M}^{\mathrm{G}\rightleftharpoons\mathrm{A}}}, -j\omega\varepsilon_0\mathcal{L}_0\left(-\vec{b}_{\zeta}^{\vec{M}_1^{\mathrm{A}\rightleftharpoons\mathrm{F}}}\right)\right\rangle_{\mathbb{S}^{\mathrm{G}\rightleftharpoons\mathrm{A}}} \tag{D-109m}$$

$$p_{\xi\zeta}^{\vec{M}^{\mathrm{G}\rightleftharpoons\mathrm{A}}\vec{M}_2^{\mathrm{A}\rightleftharpoons\mathrm{F}}} = -(1/2)\left\langle \vec{b}_{\xi}^{\vec{M}^{\mathrm{G}\rightleftharpoons\mathrm{A}}}, -j\omega\varepsilon_0\mathcal{L}_0\left(-\vec{b}_{\zeta}^{\vec{M}_2^{\mathrm{A}\rightleftharpoons\mathrm{F}}}\right)\right\rangle_{\mathbb{S}^{\mathrm{G}\rightleftharpoons\mathrm{A}}} \tag{D-109n}$$

## D8 Some Detailed Formulations Related to Sec. 7.3

The formulations used to calculate the elements of the matrices in Eqs. (7-58a)~(7-60b) are as follows:

$$z_{\xi\zeta}^{\vec{M}_{\mathrm{aux}}\vec{J}_{\mathrm{aux}}} = \left\langle \vec{b}_{\xi}^{\vec{M}_{\mathrm{aux}}}, \hat{n}_{\rightarrow\mathrm{aux}}\times\frac{1}{2}\vec{b}_{\zeta}^{\vec{J}_{\mathrm{aux}}} + \mathrm{P.V.}\,\mathcal{K}_0\left(\vec{b}_{\zeta}^{\vec{J}_{\mathrm{aux}}}\right)\right\rangle_{\mathbb{S}_{\mathrm{aux}}} \tag{D-110a}$$

$$z_{\xi\zeta}^{\vec{M}_{\mathrm{aux}}\vec{J}_{\mathrm{M}}} = \left\langle \vec{b}_{\xi}^{\vec{M}_{\mathrm{aux}}}, \mathcal{K}_0\left(\vec{b}_{\zeta}^{\vec{J}_{\mathrm{M}}}\right)\right\rangle_{\mathbb{S}_{\mathrm{aux}}} \tag{D-110b}$$

$$z_{\xi\zeta}^{\vec{M}_{\mathrm{aux}}\vec{J}_{\mathrm{M}\rightleftharpoons\mathrm{A}}} = \left\langle \vec{b}_{\xi}^{\vec{M}_{\mathrm{aux}}}, \mathcal{K}_0\left(-\vec{b}_{\zeta}^{\vec{J}_{\mathrm{M}\rightleftharpoons\mathrm{A}}}\right)\right\rangle_{\mathbb{S}_{\mathrm{aux}}} \tag{D-110c}$$

$$z_{\xi\zeta}^{\vec{M}_{\mathrm{aux}}\vec{M}_{\mathrm{aux}}} = \left\langle \vec{b}_{\xi}^{\vec{M}_{\mathrm{aux}}}, -j\omega\varepsilon_0\mathcal{L}_0\left(\vec{b}_{\zeta}^{\vec{M}_{\mathrm{aux}}}\right)\right\rangle_{\mathbb{S}_{\mathrm{aux}}} \tag{D-110d}$$

$$z_{\xi\zeta}^{\vec{M}_{\mathrm{aux}}\vec{M}_{\mathrm{M}\rightleftharpoons\mathrm{A}}} = \left\langle \vec{b}_{\xi}^{\vec{M}_{\mathrm{aux}}}, -j\omega\varepsilon_0\mathcal{L}_0\left(-\vec{b}_{\zeta}^{\vec{M}_{\mathrm{M}\rightleftharpoons\mathrm{A}}}\right)\right\rangle_{\mathbb{S}_{\mathrm{aux}}} \tag{D-110e}$$

and

$$z_{\xi\zeta}^{\vec{J}_{\mathrm{aux}}\vec{J}_{\mathrm{aux}}} = \left\langle \vec{b}_{\xi}^{\vec{J}_{\mathrm{aux}}}, -j\omega\mu_0\mathcal{L}_0\left(\vec{b}_{\zeta}^{\vec{J}_{\mathrm{aux}}}\right)\right\rangle_{\mathbb{S}_{\mathrm{aux}}} \tag{D-111a}$$

$$z_{\xi\zeta}^{\vec{J}_{\mathrm{aux}}\vec{J}_{\mathrm{M}}} = \left\langle \vec{b}_{\xi}^{\vec{J}_{\mathrm{aux}}}, -j\omega\mu_0\mathcal{L}_0\left(\vec{b}_{\zeta}^{\vec{J}_{\mathrm{M}}}\right)\right\rangle_{\mathbb{S}_{\mathrm{aux}}} \tag{D-111b}$$

$$z_{\xi\zeta}^{\vec{J}_{\mathrm{aux}}\vec{J}_{\mathrm{M}\rightleftharpoons\mathrm{A}}} = \left\langle \vec{b}_{\xi}^{\vec{J}_{\mathrm{aux}}}, -j\omega\mu_0\mathcal{L}_0\left(-\vec{b}_{\zeta}^{\vec{J}_{\mathrm{M}\rightleftharpoons\mathrm{A}}}\right)\right\rangle_{\mathbb{S}_{\mathrm{aux}}} \tag{D-111c}$$

$$z_{\xi\zeta}^{\vec{J}_{\mathrm{aux}}\vec{M}_{\mathrm{aux}}} = \left\langle \vec{b}_{\xi}^{\vec{J}_{\mathrm{aux}}}, \frac{1}{2}\vec{b}_{\zeta}^{\vec{M}_{\mathrm{aux}}}\times\hat{n}_{\rightarrow\mathrm{aux}} - \mathrm{P.V.}\,\mathcal{K}_0\left(\vec{b}_{\zeta}^{\vec{M}_{\mathrm{aux}}}\right)\right\rangle_{\mathbb{S}_{\mathrm{aux}}} \tag{D-111d}$$

$$z_{\xi\zeta}^{\vec{J}_{\mathrm{aux}}\vec{M}_{\mathrm{M}\rightleftharpoons\mathrm{A}}} = \left\langle \vec{b}_{\xi}^{\vec{J}_{\mathrm{aux}}}, -\mathcal{K}_0\left(-\vec{b}_{\zeta}^{\vec{M}_{\mathrm{M}\rightleftharpoons\mathrm{A}}}\right)\right\rangle_{\mathbb{S}_{\mathrm{aux}}} \tag{D-111e}$$

and

$$z_{\xi\zeta}^{\vec{J}_{\mathrm{M}}\vec{J}_{\mathrm{aux}}} = \left\langle \vec{b}_{\xi}^{\vec{J}_{\mathrm{M}}}, -j\omega\mu_0\mathcal{L}_0\left(\vec{b}_{\zeta}^{\vec{J}_{\mathrm{aux}}}\right)\right\rangle_{\mathbb{S}_{\mathrm{M}}} \tag{D-112a}$$

$$z_{\xi\zeta}^{\vec{J}_{\mathrm{M}}\vec{J}_{\mathrm{M}}} = \left\langle \vec{b}_{\xi}^{\vec{J}_{\mathrm{M}}}, -j\omega\mu_0\mathcal{L}_0\left(\vec{b}_{\zeta}^{\vec{J}_{\mathrm{M}}}\right)\right\rangle_{\mathbb{S}_{\mathrm{M}}} \tag{D-112b}$$

$$z_{\xi\zeta}^{\vec{J}_{\text{M}}\vec{J}_{\text{M}\rightleftharpoons\text{A}}} = \left\langle \vec{b}_{\xi}^{\vec{J}_{\text{M}}}, -j\omega\mu_0\mathcal{L}_0\left(-\vec{b}_{\zeta}^{\vec{J}_{\text{M}\rightleftharpoons\text{A}}}\right)\right\rangle_{\mathbb{S}_{\text{M}}} \tag{D-112c}$$

$$z_{\xi\zeta}^{\vec{J}_{\text{M}}\vec{M}_{\text{aux}}} = \left\langle \vec{b}_{\xi}^{\vec{J}_{\text{M}}}, -\mathcal{K}_0\left(\vec{b}_{\zeta}^{\vec{M}_{\text{aux}}}\right)\right\rangle_{\mathbb{S}_{\text{M}}} \tag{D-112d}$$

$$z_{\xi\zeta}^{\vec{J}_{\text{M}}\vec{M}_{\text{M}\rightleftharpoons\text{A}}} = \left\langle \vec{b}_{\xi}^{\vec{J}_{\text{M}}}, -\mathcal{K}_0\left(-\vec{b}_{\zeta}^{\vec{M}_{\text{M}\rightleftharpoons\text{A}}}\right)\right\rangle_{\mathbb{S}_{\text{M}}} \tag{D-112e}$$

and

$$z_{\xi\zeta}^{\vec{J}_{\text{M}\rightleftharpoons\text{A}}\vec{J}_{\text{aux}}} = \left\langle \vec{b}_{\xi}^{\vec{J}_{\text{M}\rightleftharpoons\text{A}}}, -j\omega\mu_0\mathcal{L}_0\left(\vec{b}_{\zeta}^{\vec{J}_{\text{aux}}}\right)\right\rangle_{\mathbb{S}_{\text{M}\rightleftharpoons\text{A}}} \tag{D-113a}$$

$$z_{\xi\zeta}^{\vec{J}_{\text{M}\rightleftharpoons\text{A}}\vec{J}_{\text{M}}} = \left\langle \vec{b}_{\xi}^{\vec{J}_{\text{M}\rightleftharpoons\text{A}}}, -j\omega\mu_0\mathcal{L}_0\left(\vec{b}_{\zeta}^{\vec{J}_{\text{M}}}\right)\right\rangle_{\mathbb{S}_{\text{M}\rightleftharpoons\text{A}}} \tag{D-113b}$$

$$z_{\xi\zeta}^{\vec{J}_{\text{M}\rightleftharpoons\text{A}}\vec{J}_{\text{M}\rightleftharpoons\text{A}}} = \left\langle \vec{b}_{\xi}^{\vec{J}_{\text{M}\rightleftharpoons\text{A}}}, -j\omega\mu_0\mathcal{L}_0\left(-\vec{b}_{\zeta}^{\vec{J}_{\text{M}\rightleftharpoons\text{A}}}\right)\right\rangle_{\mathbb{S}_{\text{M}\rightleftharpoons\text{A}}} - \left\langle \vec{b}_{\xi}^{\vec{J}_{\text{M}\rightleftharpoons\text{A}}}, -j\omega\mu_0\mathcal{L}_0\left(\vec{b}_{\zeta}^{\vec{J}_{\text{M}\rightleftharpoons\text{A}}}\right)\right\rangle_{\mathbb{S}_{\text{M}\rightleftharpoons\text{A}}} \tag{D-113c}$$

$$z_{\xi\zeta}^{\vec{J}_{\text{M}\rightleftharpoons\text{A}}\vec{M}_{\text{aux}}} = \left\langle \vec{b}_{\xi}^{\vec{J}_{\text{M}\rightleftharpoons\text{A}}}, -\mathcal{K}_0\left(\vec{b}_{\zeta}^{\vec{M}_{\text{aux}}}\right)\right\rangle_{\mathbb{S}_{\text{M}\rightleftharpoons\text{A}}} \tag{D-113d}$$

$$z_{\xi\zeta}^{\vec{J}_{\text{M}\rightleftharpoons\text{A}}\vec{M}_{\text{M}\rightleftharpoons\text{A}}} = \left\langle \vec{b}_{\xi}^{\vec{J}_{\text{M}\rightleftharpoons\text{A}}}, -\text{P.V.}\mathcal{K}_0\left(-\vec{b}_{\zeta}^{\vec{M}_{\text{M}\rightleftharpoons\text{A}}}\right)\right\rangle_{\mathbb{S}_{\text{M}\rightleftharpoons\text{A}}} - \left\langle \vec{b}_{\xi}^{\vec{J}_{\text{M}\rightleftharpoons\text{A}}}, -\text{P.V.}\mathcal{K}_0\left(\vec{b}_{\zeta}^{\vec{M}_{\text{M}\rightleftharpoons\text{A}}}\right)\right\rangle_{\mathbb{S}_{\text{M}\rightleftharpoons\text{A}}} \tag{D-113e}$$

and

$$z_{\xi\zeta}^{\vec{M}_{\text{M}\rightleftharpoons\text{A}}\vec{J}_{\text{aux}}} = \left\langle \vec{b}_{\xi}^{\vec{M}_{\text{M}\rightleftharpoons\text{A}}}, \mathcal{K}_0\left(\vec{b}_{\zeta}^{\vec{J}_{\text{aux}}}\right)\right\rangle_{\mathbb{S}_{\text{M}\rightleftharpoons\text{A}}} \tag{D-114a}$$

$$z_{\xi\zeta}^{\vec{M}_{\text{M}\rightleftharpoons\text{A}}\vec{J}_{\text{M}}} = \left\langle \vec{b}_{\xi}^{\vec{M}_{\text{M}\rightleftharpoons\text{A}}}, \mathcal{K}_0\left(\vec{b}_{\zeta}^{\vec{J}_{\text{M}}}\right)\right\rangle_{\mathbb{S}_{\text{M}\rightleftharpoons\text{A}}} \tag{D-114b}$$

$$z_{\xi\zeta}^{\vec{M}_{\text{M}\rightleftharpoons\text{A}}\vec{J}_{\text{M}\rightleftharpoons\text{A}}} = \left\langle \vec{b}_{\xi}^{\vec{M}_{\text{M}\rightleftharpoons\text{A}}}, \text{P.V.}\mathcal{K}_0\left(-\vec{b}_{\zeta}^{\vec{J}_{\text{M}\rightleftharpoons\text{A}}}\right)\right\rangle_{\mathbb{S}_{\text{M}\rightleftharpoons\text{A}}} - \left\langle \vec{b}_{\xi}^{\vec{M}_{\text{M}\rightleftharpoons\text{A}}}, \text{P.V.}\mathcal{K}_0\left(\vec{b}_{\zeta}^{\vec{J}_{\text{M}\rightleftharpoons\text{A}}}\right)\right\rangle_{\mathbb{S}_{\text{M}\rightleftharpoons\text{A}}} \tag{D-114c}$$

$$z_{\xi\zeta}^{\vec{M}_{\text{M}\rightleftharpoons\text{A}}\vec{M}_{\text{aux}}} = \left\langle \vec{b}_{\xi}^{\vec{M}_{\text{M}\rightleftharpoons\text{A}}}, -j\omega\varepsilon_0\mathcal{L}_0\left(\vec{b}_{\zeta}^{\vec{M}_{\text{aux}}}\right)\right\rangle_{\mathbb{S}_{\text{M}\rightleftharpoons\text{A}}} \tag{D-114d}$$

$$z_{\xi\zeta}^{\vec{M}_{\text{M}\rightleftharpoons\text{A}}\vec{M}_{\text{M}\rightleftharpoons\text{A}}} = \left\langle \vec{b}_{\xi}^{\vec{M}_{\text{M}\rightleftharpoons\text{A}}}, -j\omega\varepsilon_0\mathcal{L}_0\left(-\vec{b}_{\zeta}^{\vec{M}_{\text{M}\rightleftharpoons\text{A}}}\right)\right\rangle_{\mathbb{S}_{\text{M}\rightleftharpoons\text{A}}} - \left\langle \vec{b}_{\xi}^{\vec{M}_{\text{M}\rightleftharpoons\text{A}}}, -j\omega\varepsilon_0\mathcal{L}_0\left(\vec{b}_{\zeta}^{\vec{M}_{\text{M}\rightleftharpoons\text{A}}}\right)\right\rangle_{\mathbb{S}_{\text{M}\rightleftharpoons\text{A}}} \tag{D-114e}$$

The transformation matrix $\overline{\overline{T}}$ used in Eq. (7-61) is given in the following Eq. (D-115)

$$\overline{\overline{T}} = \overline{\overline{T}}^{\vec{J}_{\text{aux}}\to\text{AV}} \text{ or } \overline{\overline{T}}^{\vec{M}_{\text{aux}}\to\text{AV}} \text{ or } \overline{\overline{T}}^{\text{BS}\to\text{AV}} \tag{D-115}$$

in which

$$\overline{\overline{T}}^{\vec{J}_{\text{aux}}\to\text{AV}} = \left(\overline{\overline{\Psi}}_1\right)^{-1}\cdot\overline{\overline{\Psi}}_2 \tag{D-116a}$$

$$\overline{\overline{T}}^{\vec{M}_{\text{aux}}\to\text{AV}} = \left(\overline{\overline{\Psi}}_3\right)^{-1}\cdot\overline{\overline{\Psi}}_4 \tag{D-116b}$$

$$\overline{\overline{T}}^{\text{BS}\to\text{AV}} = \text{null}\left(\overline{\overline{\Psi}}_{\text{FCE}}^{\text{DoJ/DoM}}\right) \tag{D-117}$$

where

$$\overline{\overline{\Psi}}_1=\begin{bmatrix}\overline{\overline{I}}^{\vec{J}_{\mathrm{aux}}} & 0 & 0 & 0 & 0\\ 0 & \overline{\overline{Z}}^{\vec{M}_{\mathrm{aux}}\vec{J}_{\mathrm{M}}} & \overline{\overline{Z}}^{\vec{M}_{\mathrm{aux}}\vec{J}_{\mathrm{M}\rightleftharpoons\mathrm{A}}} & \overline{\overline{Z}}^{\vec{M}_{\mathrm{aux}}\vec{M}_{\mathrm{aux}}} & \overline{\overline{Z}}^{\vec{M}_{\mathrm{aux}}\vec{M}_{\mathrm{M}\rightleftharpoons\mathrm{A}}}\\ 0 & \overline{\overline{Z}}^{\vec{J}_{\mathrm{M}}\vec{J}_{\mathrm{M}}} & \overline{\overline{Z}}^{\vec{J}_{\mathrm{M}}\vec{J}_{\mathrm{M}\rightleftharpoons\mathrm{A}}} & \overline{\overline{Z}}^{\vec{J}_{\mathrm{M}}\vec{M}_{\mathrm{aux}}} & \overline{\overline{Z}}^{\vec{J}_{\mathrm{M}}\vec{M}_{\mathrm{M}\rightleftharpoons\mathrm{A}}}\\ 0 & \overline{\overline{Z}}^{\vec{J}_{\mathrm{M}\rightleftharpoons\mathrm{A}}\vec{J}_{\mathrm{M}}} & \overline{\overline{Z}}^{\vec{J}_{\mathrm{M}\rightleftharpoons\mathrm{A}}\vec{J}_{\mathrm{M}\rightleftharpoons\mathrm{A}}} & \overline{\overline{Z}}^{\vec{J}_{\mathrm{M}\rightleftharpoons\mathrm{A}}\vec{M}_{\mathrm{aux}}} & \overline{\overline{Z}}^{\vec{J}_{\mathrm{M}\rightleftharpoons\mathrm{A}}\vec{M}_{\mathrm{M}\rightleftharpoons\mathrm{A}}}\\ 0 & \overline{\overline{Z}}^{\vec{M}_{\mathrm{M}\rightleftharpoons\mathrm{A}}\vec{J}_{\mathrm{M}}} & \overline{\overline{Z}}^{\vec{M}_{\mathrm{M}\rightleftharpoons\mathrm{A}}\vec{J}_{\mathrm{M}\rightleftharpoons\mathrm{A}}} & \overline{\overline{Z}}^{\vec{M}_{\mathrm{M}\rightleftharpoons\mathrm{A}}\vec{M}_{\mathrm{aux}}} & \overline{\overline{Z}}^{\vec{M}_{\mathrm{M}\rightleftharpoons\mathrm{A}}\vec{M}_{\mathrm{M}\rightleftharpoons\mathrm{A}}}\end{bmatrix} \tag{D-118a}$$

$$\overline{\overline{\Psi}}_2=\begin{bmatrix}\overline{\overline{I}}^{\vec{J}_{\mathrm{aux}}}\\ -\overline{\overline{Z}}^{\vec{M}_{\mathrm{aux}}\vec{J}_{\mathrm{aux}}}\\ -\overline{\overline{Z}}^{\vec{J}_{\mathrm{M}}\vec{J}_{\mathrm{aux}}}\\ -\overline{\overline{Z}}^{\vec{J}_{\mathrm{M}\rightleftharpoons\mathrm{A}}\vec{J}_{\mathrm{aux}}}\\ -\overline{\overline{Z}}^{\vec{M}_{\mathrm{M}\rightleftharpoons\mathrm{A}}\vec{J}_{\mathrm{aux}}}\end{bmatrix} \tag{D-118b}$$

and

$$\overline{\overline{\Psi}}_3=\begin{bmatrix}0 & 0 & 0 & \overline{\overline{I}}^{\vec{M}_{\mathrm{aux}}} & 0\\ \overline{\overline{Z}}^{\vec{J}_{\mathrm{aux}}\vec{J}_{\mathrm{aux}}} & \overline{\overline{Z}}^{\vec{J}_{\mathrm{aux}}\vec{J}_{\mathrm{M}}} & \overline{\overline{Z}}^{\vec{J}_{\mathrm{aux}}\vec{J}_{\mathrm{M}\rightleftharpoons\mathrm{A}}} & 0 & \overline{\overline{Z}}^{\vec{J}_{\mathrm{aux}}\vec{M}_{\mathrm{M}\rightleftharpoons\mathrm{A}}}\\ \overline{\overline{Z}}^{\vec{J}_{\mathrm{M}}\vec{J}_{\mathrm{aux}}} & \overline{\overline{Z}}^{\vec{J}_{\mathrm{M}}\vec{J}_{\mathrm{M}}} & \overline{\overline{Z}}^{\vec{J}_{\mathrm{M}}\vec{J}_{\mathrm{M}\rightleftharpoons\mathrm{A}}} & 0 & \overline{\overline{Z}}^{\vec{J}_{\mathrm{M}}\vec{M}_{\mathrm{M}\rightleftharpoons\mathrm{A}}}\\ \overline{\overline{Z}}^{\vec{J}_{\mathrm{M}\rightleftharpoons\mathrm{A}}\vec{J}_{\mathrm{aux}}} & \overline{\overline{Z}}^{\vec{J}_{\mathrm{M}\rightleftharpoons\mathrm{A}}\vec{J}_{\mathrm{M}}} & \overline{\overline{Z}}^{\vec{J}_{\mathrm{M}\rightleftharpoons\mathrm{A}}\vec{J}_{\mathrm{M}\rightleftharpoons\mathrm{A}}} & 0 & \overline{\overline{Z}}^{\vec{J}_{\mathrm{M}\rightleftharpoons\mathrm{A}}\vec{M}_{\mathrm{M}\rightleftharpoons\mathrm{A}}}\\ \overline{\overline{Z}}^{\vec{M}_{\mathrm{M}\rightleftharpoons\mathrm{A}}\vec{J}_{\mathrm{aux}}} & \overline{\overline{Z}}^{\vec{M}_{\mathrm{M}\rightleftharpoons\mathrm{A}}\vec{J}_{\mathrm{M}}} & \overline{\overline{Z}}^{\vec{M}_{\mathrm{M}\rightleftharpoons\mathrm{A}}\vec{J}_{\mathrm{M}\rightleftharpoons\mathrm{A}}} & 0 & \overline{\overline{Z}}^{\vec{M}_{\mathrm{M}\rightleftharpoons\mathrm{A}}\vec{M}_{\mathrm{M}\rightleftharpoons\mathrm{A}}}\end{bmatrix} \tag{D-119a}$$

$$\overline{\overline{\Psi}}_4=\begin{bmatrix}\overline{\overline{I}}^{\vec{M}_{\mathrm{aux}}}\\ -\overline{\overline{Z}}^{\vec{J}_{\mathrm{aux}}\vec{M}_{\mathrm{aux}}}\\ -\overline{\overline{Z}}^{\vec{J}_{\mathrm{M}}\vec{M}_{\mathrm{aux}}}\\ -\overline{\overline{Z}}^{\vec{J}_{\mathrm{M}\rightleftharpoons\mathrm{A}}\vec{M}_{\mathrm{aux}}}\\ -\overline{\overline{Z}}^{\vec{M}_{\mathrm{M}\rightleftharpoons\mathrm{A}}\vec{M}_{\mathrm{aux}}}\end{bmatrix} \tag{D-119b}$$

and

$$\overline{\overline{\Psi}}_{\mathrm{FCE}}^{\mathrm{DoJ}}=\begin{bmatrix}\overline{\overline{Z}}^{\vec{M}_{\mathrm{aux}}\vec{J}_{\mathrm{aux}}} & \overline{\overline{Z}}^{\vec{M}_{\mathrm{aux}}\vec{J}_{\mathrm{M}}} & \overline{\overline{Z}}^{\vec{M}_{\mathrm{aux}}\vec{J}_{\mathrm{M}\rightleftharpoons\mathrm{A}}} & \overline{\overline{Z}}^{\vec{M}_{\mathrm{aux}}\vec{M}_{\mathrm{aux}}} & \overline{\overline{Z}}^{\vec{M}_{\mathrm{aux}}\vec{M}_{\mathrm{M}\rightleftharpoons\mathrm{A}}}\\ \overline{\overline{Z}}^{\vec{J}_{\mathrm{M}}\vec{J}_{\mathrm{aux}}} & \overline{\overline{Z}}^{\vec{J}_{\mathrm{M}}\vec{J}_{\mathrm{M}}} & \overline{\overline{Z}}^{\vec{J}_{\mathrm{M}}\vec{J}_{\mathrm{M}\rightleftharpoons\mathrm{A}}} & \overline{\overline{Z}}^{\vec{J}_{\mathrm{M}}\vec{M}_{\mathrm{aux}}} & \overline{\overline{Z}}^{\vec{J}_{\mathrm{M}}\vec{M}_{\mathrm{M}\rightleftharpoons\mathrm{A}}}\\ \overline{\overline{Z}}^{\vec{J}_{\mathrm{M}\rightleftharpoons\mathrm{A}}\vec{J}_{\mathrm{aux}}} & \overline{\overline{Z}}^{\vec{J}_{\mathrm{M}\rightleftharpoons\mathrm{A}}\vec{J}_{\mathrm{M}}} & \overline{\overline{Z}}^{\vec{J}_{\mathrm{M}\rightleftharpoons\mathrm{A}}\vec{J}_{\mathrm{M}\rightleftharpoons\mathrm{A}}} & \overline{\overline{Z}}^{\vec{J}_{\mathrm{M}\rightleftharpoons\mathrm{A}}\vec{M}_{\mathrm{aux}}} & \overline{\overline{Z}}^{\vec{J}_{\mathrm{M}\rightleftharpoons\mathrm{A}}\vec{M}_{\mathrm{M}\rightleftharpoons\mathrm{A}}}\\ \overline{\overline{Z}}^{\vec{M}_{\mathrm{M}\rightleftharpoons\mathrm{A}}\vec{J}_{\mathrm{aux}}} & \overline{\overline{Z}}^{\vec{M}_{\mathrm{M}\rightleftharpoons\mathrm{A}}\vec{J}_{\mathrm{M}}} & \overline{\overline{Z}}^{\vec{M}_{\mathrm{M}\rightleftharpoons\mathrm{A}}\vec{J}_{\mathrm{M}\rightleftharpoons\mathrm{A}}} & \overline{\overline{Z}}^{\vec{M}_{\mathrm{M}\rightleftharpoons\mathrm{A}}\vec{M}_{\mathrm{aux}}} & \overline{\overline{Z}}^{\vec{M}_{\mathrm{M}\rightleftharpoons\mathrm{A}}\vec{M}_{\mathrm{M}\rightleftharpoons\mathrm{A}}}\end{bmatrix} \tag{D-120a}$$

$$\overline{\overline{\Psi}}_{\mathrm{FCE}}^{\mathrm{DoM}}=\begin{bmatrix}\overline{\overline{Z}}^{\vec{J}_{\mathrm{aux}}\vec{J}_{\mathrm{aux}}} & \overline{\overline{Z}}^{\vec{J}_{\mathrm{aux}}\vec{J}_{\mathrm{M}}} & \overline{\overline{Z}}^{\vec{J}_{\mathrm{aux}}\vec{J}_{\mathrm{M}\rightleftharpoons\mathrm{A}}} & \overline{\overline{Z}}^{\vec{J}_{\mathrm{aux}}\vec{M}_{\mathrm{aux}}} & \overline{\overline{Z}}^{\vec{J}_{\mathrm{aux}}\vec{M}_{\mathrm{M}\rightleftharpoons\mathrm{A}}}\\ \overline{\overline{Z}}^{\vec{J}_{\mathrm{M}}\vec{J}_{\mathrm{aux}}} & \overline{\overline{Z}}^{\vec{J}_{\mathrm{M}}\vec{J}_{\mathrm{M}}} & \overline{\overline{Z}}^{\vec{J}_{\mathrm{M}}\vec{J}_{\mathrm{M}\rightleftharpoons\mathrm{A}}} & \overline{\overline{Z}}^{\vec{J}_{\mathrm{M}}\vec{M}_{\mathrm{aux}}} & \overline{\overline{Z}}^{\vec{J}_{\mathrm{M}}\vec{M}_{\mathrm{M}\rightleftharpoons\mathrm{A}}}\\ \overline{\overline{Z}}^{\vec{J}_{\mathrm{M}\rightleftharpoons\mathrm{A}}\vec{J}_{\mathrm{aux}}} & \overline{\overline{Z}}^{\vec{J}_{\mathrm{M}\rightleftharpoons\mathrm{A}}\vec{J}_{\mathrm{M}}} & \overline{\overline{Z}}^{\vec{J}_{\mathrm{M}\rightleftharpoons\mathrm{A}}\vec{J}_{\mathrm{M}\rightleftharpoons\mathrm{A}}} & \overline{\overline{Z}}^{\vec{J}_{\mathrm{M}\rightleftharpoons\mathrm{A}}\vec{M}_{\mathrm{aux}}} & \overline{\overline{Z}}^{\vec{J}_{\mathrm{M}\rightleftharpoons\mathrm{A}}\vec{M}_{\mathrm{M}\rightleftharpoons\mathrm{A}}}\\ \overline{\overline{Z}}^{\vec{M}_{\mathrm{M}\rightleftharpoons\mathrm{A}}\vec{J}_{\mathrm{aux}}} & \overline{\overline{Z}}^{\vec{M}_{\mathrm{M}\rightleftharpoons\mathrm{A}}\vec{J}_{\mathrm{M}}} & \overline{\overline{Z}}^{\vec{M}_{\mathrm{M}\rightleftharpoons\mathrm{A}}\vec{J}_{\mathrm{M}\rightleftharpoons\mathrm{A}}} & \overline{\overline{Z}}^{\vec{M}_{\mathrm{M}\rightleftharpoons\mathrm{A}}\vec{M}_{\mathrm{aux}}} & \overline{\overline{Z}}^{\vec{M}_{\mathrm{M}\rightleftharpoons\mathrm{A}}\vec{M}_{\mathrm{M}\rightleftharpoons\mathrm{A}}}\end{bmatrix} \tag{D-120b}$$

The power quadratic form matrix $\overline{\overline{P}}_{\mathrm{M}\rightleftharpoons\mathrm{A}}$ used in Eq. (7-66) is as follows:

$$\overline{\overline{P}}_{\mathrm{M}\rightleftharpoons\mathrm{A}} = \overline{\overline{P}}_{\mathrm{M}\rightleftharpoons\mathrm{A}}^{\mathrm{curAV}} \text{ or } \overline{\overline{P}}_{\mathrm{M}\rightleftharpoons\mathrm{A}}^{\mathrm{intAV}} \tag{D-121}$$

in which

$$\overline{\overline{P}}_{\mathrm{M}\rightleftharpoons\mathrm{A}}^{\mathrm{curAV}} = \begin{bmatrix} 0 & 0 & 0 & 0 & 0 \\ 0 & 0 & 0 & 0 & 0 \\ 0 & 0 & 0 & 0 & \overline{\overline{C}}^{\vec{J}_{\mathrm{M}\rightleftharpoons\mathrm{A}}\vec{M}_{\mathrm{M}\rightleftharpoons\mathrm{A}}} \\ 0 & 0 & 0 & 0 & 0 \\ 0 & 0 & 0 & 0 & 0 \end{bmatrix} \tag{D-122}$$

corresponding to Eq. (7-64), and

$$\overline{\overline{P}}_{\mathrm{M}\rightleftharpoons\mathrm{A}}^{\mathrm{intAV}} = \begin{bmatrix} 0 & 0 & 0 & 0 & 0 \\ 0 & 0 & 0 & 0 & 0 \\ \overline{\overline{P}}^{\vec{J}_{\mathrm{M}\rightleftharpoons\mathrm{A}}\vec{J}_{\mathrm{aux}}} & \overline{\overline{P}}^{\vec{J}_{\mathrm{M}\rightleftharpoons\mathrm{A}}\vec{J}_{\mathrm{M}}} & \overline{\overline{P}}^{\vec{J}_{\mathrm{M}\rightleftharpoons\mathrm{A}}\vec{J}_{\mathrm{M}\rightleftharpoons\mathrm{A}}} & \overline{\overline{P}}^{\vec{J}_{\mathrm{M}\rightleftharpoons\mathrm{A}}\vec{M}_{\mathrm{aux}}} & \overline{\overline{P}}^{\vec{J}_{\mathrm{M}\rightleftharpoons\mathrm{A}}\vec{M}_{\mathrm{M}\rightleftharpoons\mathrm{A}}} \\ 0 & 0 & 0 & 0 & 0 \\ 0 & 0 & 0 & 0 & 0 \end{bmatrix} \tag{D-123a}$$

corresponding to Eq. (7-65a), and

$$\overline{\overline{P}}_{\mathrm{M}\rightleftharpoons\mathrm{A}}^{\mathrm{intAV}} = \begin{bmatrix} 0 & 0 & 0 & 0 & 0 \\ 0 & 0 & 0 & 0 & 0 \\ 0 & 0 & 0 & 0 & 0 \\ 0 & 0 & 0 & 0 & 0 \\ \overline{\overline{P}}^{\vec{M}_{\mathrm{M}\rightleftharpoons\mathrm{A}}\vec{J}_{\mathrm{aux}}} & \overline{\overline{P}}^{\vec{M}_{\mathrm{M}\rightleftharpoons\mathrm{A}}\vec{J}_{\mathrm{M}}} & \overline{\overline{P}}^{\vec{M}_{\mathrm{M}\rightleftharpoons\mathrm{A}}\vec{J}_{\mathrm{M}\rightleftharpoons\mathrm{A}}} & \overline{\overline{P}}^{\vec{M}_{\mathrm{M}\rightleftharpoons\mathrm{A}}\vec{M}_{\mathrm{aux}}} & \overline{\overline{P}}^{\vec{M}_{\mathrm{M}\rightleftharpoons\mathrm{A}}\vec{M}_{\mathrm{M}\rightleftharpoons\mathrm{A}}} \end{bmatrix}^{\dagger} \tag{D-123b}$$

corresponding to Eq. (7-65b), where the elements of the sub-matrices in the above Eqs. (D-122)~(D-123b) are as follows:

$$c_{\xi\zeta}^{\vec{J}_{\mathrm{M}\rightleftharpoons\mathrm{A}}\vec{M}_{\mathrm{M}\rightleftharpoons\mathrm{A}}} = (1/2)\left\langle \hat{n}_{\rightarrow\mathrm{PML}} \times \vec{b}_{\xi}^{\vec{J}_{\mathrm{M}\rightleftharpoons\mathrm{A}}}, \vec{b}_{\zeta}^{\vec{M}_{\mathrm{M}\rightleftharpoons\mathrm{A}}} \right\rangle_{\mathbb{S}_{\mathrm{M}\rightleftharpoons\mathrm{A}}} \tag{D-124}$$

and

$$p_{\xi\zeta}^{\vec{J}_{\mathrm{M}\rightleftharpoons\mathrm{A}}\vec{J}_{\mathrm{aux}}} = -(1/2)\left\langle \vec{b}_{\xi}^{\vec{J}_{\mathrm{M}\rightleftharpoons\mathrm{A}}}, -j\omega\mu_0\mathcal{L}_0\left(\vec{b}_{\zeta}^{\vec{J}_{\mathrm{aux}}}\right) \right\rangle_{\mathbb{S}_{\mathrm{M}\rightleftharpoons\mathrm{A}}} \tag{D-125a}$$

$$p_{\xi\zeta}^{\vec{J}_{\mathrm{M}\rightleftharpoons\mathrm{A}}\vec{J}_{\mathrm{M}}} = -(1/2)\left\langle \vec{b}_{\xi}^{\vec{J}_{\mathrm{M}\rightleftharpoons\mathrm{A}}}, -j\omega\mu_0\mathcal{L}_0\left(\vec{b}_{\zeta}^{\vec{J}_{\mathrm{M}}}\right) \right\rangle_{\mathbb{S}_{\mathrm{M}\rightleftharpoons\mathrm{A}}} \tag{D-125b}$$

$$p_{\xi\zeta}^{\vec{J}_{\mathrm{M}\rightleftharpoons\mathrm{A}}\vec{J}_{\mathrm{M}\rightleftharpoons\mathrm{A}}} = -(1/2)\left\langle \vec{b}_{\xi}^{\vec{J}_{\mathrm{M}\rightleftharpoons\mathrm{A}}}, -j\omega\mu_0\mathcal{L}_0\left(-\vec{b}_{\zeta}^{\vec{J}_{\mathrm{M}\rightleftharpoons\mathrm{A}}}\right) \right\rangle_{\mathbb{S}_{\mathrm{M}\rightleftharpoons\mathrm{A}}} \tag{D-125c}$$

$$p_{\xi\zeta}^{\vec{J}_{\mathrm{M}\rightleftharpoons\mathrm{A}}\vec{M}_{\mathrm{aux}}} = -(1/2)\left\langle \vec{b}_{\xi}^{\vec{J}_{\mathrm{M}\rightleftharpoons\mathrm{A}}}, -\mathcal{K}_0\left(\vec{b}_{\zeta}^{\vec{M}_{\mathrm{aux}}}\right) \right\rangle_{\mathbb{S}_{\mathrm{M}\rightleftharpoons\mathrm{A}}} \tag{D-125d}$$

$$p_{\xi\zeta}^{\vec{J}_{\mathrm{M}\rightleftharpoons\mathrm{A}}\vec{M}_{\mathrm{M}\rightleftharpoons\mathrm{A}}} = -(1/2)\left\langle \vec{b}_{\xi}^{\vec{J}_{\mathrm{M}\rightleftharpoons\mathrm{A}}}, \hat{n}_{\rightarrow\mathrm{PML}} \times \frac{1}{2}\vec{b}_{\zeta}^{\vec{M}_{\mathrm{M}\rightleftharpoons\mathrm{A}}} - \mathrm{P.V.}\,\mathcal{K}_0\left(-\vec{b}_{\zeta}^{\vec{M}_{\mathrm{M}\rightleftharpoons\mathrm{A}}}\right) \right\rangle_{\mathbb{S}_{\mathrm{M}\rightleftharpoons\mathrm{A}}} \tag{D-125e}$$

and

$$p_{\xi\zeta}^{\vec{M}_{\mathrm{M}\rightleftharpoons\mathrm{A}}\vec{J}_{\mathrm{aux}}} = -(1/2)\left\langle \vec{b}_{\xi}^{\vec{M}_{\mathrm{M}\rightleftharpoons\mathrm{A}}}, \mathcal{K}_0\left(\vec{b}_{\zeta}^{\vec{J}_{\mathrm{aux}}}\right)\right\rangle_{\mathbb{S}_{\mathrm{M}\rightleftharpoons\mathrm{A}}} \tag{D-125f}$$

$$p_{\xi\zeta}^{\vec{M}_{\mathrm{M}\rightleftharpoons\mathrm{A}}\vec{J}_{\mathrm{M}}} = -(1/2)\left\langle \vec{b}_{\xi}^{\vec{M}_{\mathrm{M}\rightleftharpoons\mathrm{A}}}, \mathcal{K}_0\left(\vec{b}_{\zeta}^{\vec{J}_{\mathrm{M}}}\right)\right\rangle_{\mathbb{S}_{\mathrm{M}\rightleftharpoons\mathrm{A}}} \tag{D-125g}$$

$$p_{\xi\zeta}^{\vec{M}_{\mathrm{M}\rightleftharpoons\mathrm{A}}\vec{J}_{\mathrm{M}\rightleftharpoons\mathrm{A}}} = -(1/2)\left\langle \vec{b}_{\xi}^{\vec{M}_{\mathrm{M}\rightleftharpoons\mathrm{A}}}, \frac{1}{2}\vec{b}_{\zeta}^{\vec{J}_{\mathrm{M}\rightleftharpoons\mathrm{A}}} \times \hat{n}_{\rightarrow\mathrm{PML}} + \mathrm{P.V.}\,\mathcal{K}_0\left(-\vec{b}_{\zeta}^{\vec{J}_{\mathrm{M}\rightleftharpoons\mathrm{A}}}\right)\right\rangle_{\mathbb{S}_{\mathrm{M}\rightleftharpoons\mathrm{A}}} \tag{D-125h}$$

$$p_{\xi\zeta}^{\vec{M}_{\mathrm{M}\rightleftharpoons\mathrm{A}}\vec{M}_{\mathrm{aux}}} = -(1/2)\left\langle \vec{b}_{\xi}^{\vec{M}_{\mathrm{M}\rightleftharpoons\mathrm{A}}}, -j\omega\varepsilon_0\mathcal{L}_0\left(\vec{b}_{\zeta}^{\vec{M}_{\mathrm{aux}}}\right)\right\rangle_{\mathbb{S}_{\mathrm{M}\rightleftharpoons\mathrm{A}}} \tag{D-125i}$$

$$p_{\xi\zeta}^{\vec{M}_{\mathrm{M}\rightleftharpoons\mathrm{A}}\vec{M}_{\mathrm{M}\rightleftharpoons\mathrm{A}}} = -(1/2)\left\langle \vec{b}_{\xi}^{\vec{M}_{\mathrm{M}\rightleftharpoons\mathrm{A}}}, -j\omega\varepsilon_0\mathcal{L}_0\left(-\vec{b}_{\zeta}^{\vec{M}_{\mathrm{M}\rightleftharpoons\mathrm{A}}}\right)\right\rangle_{\mathbb{S}_{\mathrm{M}\rightleftharpoons\mathrm{A}}} \tag{D-125j}$$

## D9 Some Detailed Formulations Related to Sec. 7.4

The formulations used to calculate the elements of the matrices in Eqs. (7-85a)~(7-92) are as follows:

$$z_{\xi\zeta}^{\vec{M}_{\mathrm{aux}}\vec{J}_{\mathrm{aux}}} = \left\langle \vec{b}_{\xi}^{\vec{M}_{\mathrm{aux}}}, \hat{n}_{\rightarrow\mathrm{aux}} \times \frac{1}{2}\vec{b}_{\zeta}^{\vec{J}_{\mathrm{aux}}} + \mathrm{P.V.}\,\mathcal{K}_0\left(\vec{b}_{\zeta}^{\vec{J}_{\mathrm{aux}}}\right)\right\rangle_{\mathbb{S}_{\mathrm{aux}}} \tag{D-126a}$$

$$z_{\xi\zeta}^{\vec{M}_{\mathrm{aux}}\vec{J}^{1}_{\mathrm{M}\rightleftharpoons\mathrm{A}}} = \left\langle \vec{b}_{\xi}^{\vec{M}_{\mathrm{aux}}}, \mathcal{K}_0\left(-\vec{b}_{\zeta}^{\vec{J}^{1}_{\mathrm{M}\rightleftharpoons\mathrm{A}}}\right)\right\rangle_{\mathbb{S}_{\mathrm{aux}}} \tag{D-126b}$$

$$z_{\xi\zeta}^{\vec{M}_{\mathrm{aux}}\vec{J}^{2}_{\mathrm{M}\rightleftharpoons\mathrm{A}}} = \left\langle \vec{b}_{\xi}^{\vec{M}_{\mathrm{aux}}}, \mathcal{K}_0\left(-\vec{b}_{\zeta}^{\vec{J}^{2}_{\mathrm{M}\rightleftharpoons\mathrm{A}}}\right)\right\rangle_{\mathbb{S}_{\mathrm{aux}}} \tag{D-126c}$$

$$z_{\xi\zeta}^{\vec{M}_{\mathrm{aux}}\vec{J}_{\mathrm{M}}} = \left\langle \vec{b}_{\xi}^{\vec{M}_{\mathrm{aux}}}, \mathcal{K}_0\left(\vec{b}_{\zeta}^{\vec{J}_{\mathrm{M}}}\right)\right\rangle_{\mathbb{S}_{\mathrm{aux}}} \tag{D-126d}$$

$$z_{\xi\zeta}^{\vec{M}_{\mathrm{aux}}\vec{M}_{\mathrm{aux}}} = \left\langle \vec{b}_{\xi}^{\vec{M}_{\mathrm{aux}}}, -j\omega\varepsilon_0\mathcal{K}_0\left(\vec{b}_{\zeta}^{\vec{M}_{\mathrm{aux}}}\right)\right\rangle_{\mathbb{S}_{\mathrm{aux}}} \tag{D-126e}$$

$$z_{\xi\zeta}^{\vec{M}_{\mathrm{aux}}\vec{M}^{1}_{\mathrm{M}\rightleftharpoons\mathrm{A}}} = \left\langle \vec{b}_{\xi}^{\vec{M}_{\mathrm{aux}}}, -j\omega\varepsilon_0\mathcal{K}_0\left(-\vec{b}_{\zeta}^{\vec{M}^{1}_{\mathrm{M}\rightleftharpoons\mathrm{A}}}\right)\right\rangle_{\mathbb{S}_{\mathrm{aux}}} \tag{D-126f}$$

$$z_{\xi\zeta}^{\vec{M}_{\mathrm{aux}}\vec{M}^{2}_{\mathrm{M}\rightleftharpoons\mathrm{A}}} = \left\langle \vec{b}_{\xi}^{\vec{M}_{\mathrm{aux}}}, -j\omega\varepsilon_0\mathcal{K}_0\left(-\vec{b}_{\zeta}^{\vec{M}^{2}_{\mathrm{M}\rightleftharpoons\mathrm{A}}}\right)\right\rangle_{\mathbb{S}_{\mathrm{aux}}} \tag{D-126g}$$

and

$$z_{\xi\zeta}^{\vec{J}_{\mathrm{aux}}\vec{J}_{\mathrm{aux}}} = \left\langle \vec{b}_{\xi}^{\vec{J}_{\mathrm{aux}}}, -j\omega\mu_0\mathcal{L}_0\left(\vec{b}_{\zeta}^{\vec{J}_{\mathrm{aux}}}\right)\right\rangle_{\mathbb{S}_{\mathrm{aux}}} \tag{D-127a}$$

$$z_{\xi\zeta}^{\vec{J}_{\mathrm{aux}}\vec{J}^{1}_{\mathrm{M}\rightleftharpoons\mathrm{A}}} = \left\langle \vec{b}_{\xi}^{\vec{J}_{\mathrm{aux}}}, -j\omega\mu_0\mathcal{L}_0\left(-\vec{b}_{\zeta}^{\vec{J}^{1}_{\mathrm{M}\rightleftharpoons\mathrm{A}}}\right)\right\rangle_{\mathbb{S}_{\mathrm{aux}}} \tag{D-127b}$$

$$z_{\xi\zeta}^{\vec{J}_{\mathrm{aux}}\vec{J}^{2}_{\mathrm{M}\rightleftharpoons\mathrm{A}}} = \left\langle \vec{b}_{\xi}^{\vec{J}_{\mathrm{aux}}}, -j\omega\mu_0\mathcal{L}_0\left(-\vec{b}_{\zeta}^{\vec{J}^{2}_{\mathrm{M}\rightleftharpoons\mathrm{A}}}\right)\right\rangle_{\mathbb{S}_{\mathrm{aux}}} \tag{D-127c}$$

$$z_{\xi\zeta}^{\vec{J}_{\mathrm{aux}}\vec{J}_{\mathrm{M}}} = \left\langle \vec{b}_{\xi}^{\vec{J}_{\mathrm{aux}}}, -j\omega\mu_0\mathcal{L}_0\left(\vec{b}_{\zeta}^{\vec{J}_{\mathrm{M}}}\right)\right\rangle_{\mathbb{S}_{\mathrm{aux}}} \tag{D-127d}$$

$$z_{\xi\zeta}^{\vec{J}_{\mathrm{aux}}\vec{M}_{\mathrm{aux}}} = \left\langle \vec{b}_{\xi}^{\vec{J}_{\mathrm{aux}}}, \frac{1}{2}\vec{b}_{\zeta}^{\vec{M}_{\mathrm{aux}}} \times \hat{n}_{\rightarrow\mathrm{aux}} - \mathrm{P.V.}\,\mathcal{K}_0\left(\vec{b}_{\zeta}^{\vec{M}_{\mathrm{aux}}}\right)\right\rangle_{\mathbb{S}_{\mathrm{aux}}} \tag{D-127e}$$

$$z_{\xi\zeta}^{\vec{J}_{\mathrm{aux}}\vec{M}^{1}_{\mathrm{M}\rightleftharpoons\mathrm{A}}} = \left\langle \vec{b}_{\xi}^{\vec{J}_{\mathrm{aux}}}, -\mathcal{K}_0\left(-\vec{b}_{\zeta}^{\vec{M}^{1}_{\mathrm{M}\rightleftharpoons\mathrm{A}}}\right)\right\rangle_{\mathbb{S}_{\mathrm{aux}}} \tag{D-127f}$$

$$z_{\xi\zeta}^{\vec{J}_{\mathrm{aux}}\vec{M}^{2}_{\mathrm{M}\rightleftharpoons\mathrm{A}}} = \left\langle \vec{b}_{\xi}^{\vec{J}_{\mathrm{aux}}}, -\mathcal{K}_0\left(-\vec{b}_{\zeta}^{\vec{M}^{2}_{\mathrm{M}\rightleftharpoons\mathrm{A}}}\right)\right\rangle_{\mathbb{S}_{\mathrm{aux}}} \qquad \text{(D-127g)}$$

and

$$z_{\xi\zeta}^{\vec{J}^{1}_{\mathrm{M}\rightleftharpoons\mathrm{A}}\vec{J}_{\mathrm{aux}}} = \left\langle \vec{b}_{\xi}^{\vec{J}^{1}_{\mathrm{M}\rightleftharpoons\mathrm{A}}}, -j\omega\mu_0\mathcal{L}_0\left(\vec{b}_{\zeta}^{\vec{J}_{\mathrm{aux}}}\right)\right\rangle_{\mathbb{S}^{1}_{\mathrm{M}\rightleftharpoons\mathrm{A}}} \qquad \text{(D-128a)}$$

$$z_{\xi\zeta}^{\vec{J}^{1}_{\mathrm{M}\rightleftharpoons\mathrm{A}}\vec{J}^{1}_{\mathrm{M}\rightleftharpoons\mathrm{A}}} = \left\langle \vec{b}_{\xi}^{\vec{J}^{1}_{\mathrm{M}\rightleftharpoons\mathrm{A}}}, -j\omega\mu_0\mathcal{L}_0\left(-\vec{b}_{\zeta}^{\vec{J}^{1}_{\mathrm{M}\rightleftharpoons\mathrm{A}}}\right)\right\rangle_{\mathbb{S}^{1}_{\mathrm{M}\rightleftharpoons\mathrm{A}}} - \left\langle \vec{b}_{\xi}^{\vec{J}^{1}_{\mathrm{M}\rightleftharpoons\mathrm{A}}}, \mathcal{E}_1\left(\vec{b}_{\zeta}^{\vec{J}^{1}_{\mathrm{M}\rightleftharpoons\mathrm{A}}}\right)\right\rangle_{\mathbb{S}^{1}_{\mathrm{M}\rightleftharpoons\mathrm{A}}} \qquad \text{(D-128b)}$$

$$z_{\xi\zeta}^{\vec{J}^{1}_{\mathrm{M}\rightleftharpoons\mathrm{A}}\vec{J}^{2}_{\mathrm{M}\rightleftharpoons\mathrm{A}}} = \left\langle \vec{b}_{\xi}^{\vec{J}^{1}_{\mathrm{M}\rightleftharpoons\mathrm{A}}}, -j\omega\mu_0\mathcal{L}_0\left(-\vec{b}_{\zeta}^{\vec{J}^{2}_{\mathrm{M}\rightleftharpoons\mathrm{A}}}\right)\right\rangle_{\mathbb{S}^{1}_{\mathrm{M}\rightleftharpoons\mathrm{A}}} \qquad \text{(D-128c)}$$

$$z_{\xi\zeta}^{\vec{J}^{1}_{\mathrm{M}\rightleftharpoons\mathrm{A}}\vec{J}^{12}_{\mathrm{A}\rightleftharpoons\mathrm{A}}} = -\left\langle \vec{b}_{\xi}^{\vec{J}^{1}_{\mathrm{M}\rightleftharpoons\mathrm{A}}}, \mathcal{E}_1\left(\vec{b}_{\zeta}^{\vec{J}^{12}_{\mathrm{A}\rightleftharpoons\mathrm{A}}}\right)\right\rangle_{\mathbb{S}^{1}_{\mathrm{M}\rightleftharpoons\mathrm{A}}} \qquad \text{(D-128d)}$$

$$z_{\xi\zeta}^{\vec{J}^{1}_{\mathrm{M}\rightleftharpoons\mathrm{A}}\vec{J}_{\mathrm{M}}} = \left\langle \vec{b}_{\xi}^{\vec{J}^{1}_{\mathrm{M}\rightleftharpoons\mathrm{A}}}, -j\omega\mu_0\mathcal{L}_0\left(\vec{b}_{\zeta}^{\vec{J}_{\mathrm{M}}}\right)\right\rangle_{\mathbb{S}^{1}_{\mathrm{M}\rightleftharpoons\mathrm{A}}} \qquad \text{(D-128e)}$$

$$z_{\xi\zeta}^{\vec{J}^{1}_{\mathrm{M}\rightleftharpoons\mathrm{A}}\vec{J}^{1}_{\mathrm{A}}} = -\left\langle \vec{b}_{\xi}^{\vec{J}^{1}_{\mathrm{M}\rightleftharpoons\mathrm{A}}}, \mathcal{E}_1\left(\vec{b}_{\zeta}^{\vec{J}^{1}_{\mathrm{A}}}\right)\right\rangle_{\mathbb{S}^{1}_{\mathrm{M}\rightleftharpoons\mathrm{A}}} \qquad \text{(D-128f)}$$

$$z_{\xi\zeta}^{\vec{J}^{1}_{\mathrm{M}\rightleftharpoons\mathrm{A}}\vec{M}_{\mathrm{aux}}} = \left\langle \vec{b}_{\xi}^{\vec{J}^{1}_{\mathrm{M}\rightleftharpoons\mathrm{A}}}, -\mathcal{K}_0\left(\vec{b}_{\zeta}^{\vec{M}_{\mathrm{aux}}}\right)\right\rangle_{\mathbb{S}^{1}_{\mathrm{M}\rightleftharpoons\mathrm{A}}} \qquad \text{(D-128g)}$$

$$z_{\xi\zeta}^{\vec{J}^{1}_{\mathrm{M}\rightleftharpoons\mathrm{A}}\vec{M}^{1}_{\mathrm{M}\rightleftharpoons\mathrm{A}}} = \left\langle \vec{b}_{\xi}^{\vec{J}^{1}_{\mathrm{M}\rightleftharpoons\mathrm{A}}}, \hat{n}^{1}_{\rightarrow\mathrm{A}} \times \frac{1}{2}\vec{b}_{\zeta}^{\vec{M}^{1}_{\mathrm{M}\rightleftharpoons\mathrm{A}}} - \mathrm{P.V.}\,\mathcal{K}_0\left(-\vec{b}_{\zeta}^{\vec{M}^{1}_{\mathrm{M}\rightleftharpoons\mathrm{A}}}\right)\right\rangle_{\mathbb{S}^{1}_{\mathrm{M}\rightleftharpoons\mathrm{A}}} - \left\langle \vec{b}_{\xi}^{\vec{J}^{1}_{\mathrm{M}\rightleftharpoons\mathrm{A}}}, \mathcal{E}_1\left(\vec{b}_{\zeta}^{\vec{M}^{1}_{\mathrm{M}\rightleftharpoons\mathrm{A}}}\right)\right\rangle_{\mathbb{S}^{1}_{\mathrm{M}\rightleftharpoons\mathrm{A}}} \qquad \text{(D-128h)}$$

$$z_{\xi\zeta}^{\vec{J}^{1}_{\mathrm{M}\rightleftharpoons\mathrm{A}}\vec{M}^{2}_{\mathrm{M}\rightleftharpoons\mathrm{A}}} = \left\langle \vec{b}_{\xi}^{\vec{J}^{1}_{\mathrm{M}\rightleftharpoons\mathrm{A}}}, -\mathcal{K}_0\left(-\vec{b}_{\zeta}^{\vec{M}^{2}_{\mathrm{M}\rightleftharpoons\mathrm{A}}}\right)\right\rangle_{\mathbb{S}^{1}_{\mathrm{M}\rightleftharpoons\mathrm{A}}} \qquad \text{(D-128i)}$$

$$z_{\xi\zeta}^{\vec{J}^{1}_{\mathrm{M}\rightleftharpoons\mathrm{A}}\vec{M}^{12}_{\mathrm{A}\rightleftharpoons\mathrm{A}}} = -\left\langle \vec{b}_{\xi}^{\vec{J}^{1}_{\mathrm{M}\rightleftharpoons\mathrm{A}}}, \mathcal{E}_1\left(\vec{b}_{\zeta}^{\vec{M}^{12}_{\mathrm{A}\rightleftharpoons\mathrm{A}}}\right)\right\rangle_{\mathbb{S}^{1}_{\mathrm{M}\rightleftharpoons\mathrm{A}}} \qquad \text{(D-128j)}$$

and

$$z_{\xi\zeta}^{\vec{M}^{1}_{\mathrm{M}\rightleftharpoons\mathrm{A}}\vec{J}_{\mathrm{aux}}} = \left\langle \vec{b}_{\xi}^{\vec{M}^{1}_{\mathrm{M}\rightleftharpoons\mathrm{A}}}, \mathcal{K}_0\left(\vec{b}_{\zeta}^{\vec{J}_{\mathrm{aux}}}\right)\right\rangle_{\mathbb{S}^{1}_{\mathrm{M}\rightleftharpoons\mathrm{A}}} \qquad \text{(D-129a)}$$

$$z_{\xi\zeta}^{\vec{M}^{1}_{\mathrm{M}\rightleftharpoons\mathrm{A}}\vec{J}^{1}_{\mathrm{M}\rightleftharpoons\mathrm{A}}} = \left\langle \vec{b}_{\xi}^{\vec{M}^{1}_{\mathrm{M}\rightleftharpoons\mathrm{A}}}, \frac{1}{2}\vec{b}_{\zeta}^{\vec{J}^{1}_{\mathrm{M}\rightleftharpoons\mathrm{A}}} \times \hat{n}^{1}_{\rightarrow\mathrm{A}} + \mathrm{P.V.}\,\mathcal{K}_0\left(-\vec{b}_{\zeta}^{\vec{J}^{1}_{\mathrm{M}\rightleftharpoons\mathrm{A}}}\right)\right\rangle_{\mathbb{S}^{1}_{\mathrm{M}\rightleftharpoons\mathrm{A}}} - \left\langle \vec{b}_{\xi}^{\vec{M}^{1}_{\mathrm{M}\rightleftharpoons\mathrm{A}}}, \mathcal{H}_1\left(\vec{b}_{\zeta}^{\vec{J}^{1}_{\mathrm{M}\rightleftharpoons\mathrm{A}}}\right)\right\rangle_{\mathbb{S}^{1}_{\mathrm{M}\rightleftharpoons\mathrm{A}}} \qquad \text{(D-129b)}$$

$$z_{\xi\zeta}^{\vec{M}^{1}_{\mathrm{M}\rightleftharpoons\mathrm{A}}\vec{J}^{2}_{\mathrm{M}\rightleftharpoons\mathrm{A}}} = \left\langle \vec{b}_{\xi}^{\vec{M}^{1}_{\mathrm{M}\rightleftharpoons\mathrm{A}}}, \mathcal{K}_0\left(-\vec{b}_{\zeta}^{\vec{J}^{2}_{\mathrm{M}\rightleftharpoons\mathrm{A}}}\right)\right\rangle_{\mathbb{S}^{1}_{\mathrm{M}\rightleftharpoons\mathrm{A}}} \qquad \text{(D-129c)}$$

$$z_{\xi\zeta}^{\vec{M}^{1}_{\mathrm{M}\rightleftharpoons\mathrm{A}}\vec{J}^{12}_{\mathrm{A}\rightleftharpoons\mathrm{A}}} = -\left\langle \vec{b}_{\xi}^{\vec{M}^{1}_{\mathrm{M}\rightleftharpoons\mathrm{A}}}, \mathcal{H}_1\left(\vec{b}_{\zeta}^{\vec{J}^{12}_{\mathrm{A}\rightleftharpoons\mathrm{A}}}\right)\right\rangle_{\mathbb{S}^{1}_{\mathrm{M}\rightleftharpoons\mathrm{A}}} \qquad \text{(D-129d)}$$

$$z_{\xi\zeta}^{\vec{M}^{1}_{\mathrm{M}\rightleftharpoons\mathrm{A}}\vec{J}_{\mathrm{M}}} = \left\langle \vec{b}_{\xi}^{\vec{M}^{1}_{\mathrm{M}\rightleftharpoons\mathrm{A}}}, \mathcal{K}_0\left(\vec{b}_{\zeta}^{\vec{J}_{\mathrm{M}}}\right)\right\rangle_{\mathbb{S}^{1}_{\mathrm{M}\rightleftharpoons\mathrm{A}}} \qquad \text{(D-129e)}$$

$$z_{\xi\zeta}^{\vec{M}^{1}_{\mathrm{M}\rightleftharpoons\mathrm{A}}\vec{J}^{1}_{\mathrm{A}}} = -\left\langle \vec{b}_{\xi}^{\vec{M}^{1}_{\mathrm{M}\rightleftharpoons\mathrm{A}}}, \mathcal{H}_1\left(\vec{b}_{\zeta}^{\vec{J}^{1}_{\mathrm{A}}}\right)\right\rangle_{\mathbb{S}^{1}_{\mathrm{M}\rightleftharpoons\mathrm{A}}} \qquad \text{(D-129f)}$$

$$z_{\xi\zeta}^{\vec{M}^{1}_{\mathrm{M}\rightleftharpoons\mathrm{A}}\vec{M}_{\mathrm{aux}}} = \left\langle \vec{b}_{\xi}^{\vec{M}^{1}_{\mathrm{M}\rightleftharpoons\mathrm{A}}}, -j\omega\varepsilon_0\mathcal{L}_0\left(\vec{b}_{\zeta}^{\vec{M}_{\mathrm{aux}}}\right)\right\rangle_{\mathbb{S}^{1}_{\mathrm{M}\rightleftharpoons\mathrm{A}}} \qquad \text{(D-129g)}$$

$$z_{\xi\zeta}^{\vec{M}^{1}_{\mathrm{M}\rightleftharpoons\mathrm{A}}\vec{M}^{1}_{\mathrm{M}\rightleftharpoons\mathrm{A}}} = \left\langle \vec{b}_{\xi}^{\vec{M}^{1}_{\mathrm{M}\rightleftharpoons\mathrm{A}}}, -j\omega\varepsilon_0\mathcal{L}_0\left(-\vec{b}_{\zeta}^{\vec{M}^{1}_{\mathrm{M}\rightleftharpoons\mathrm{A}}}\right)\right\rangle_{\mathbb{S}^{1}_{\mathrm{M}\rightleftharpoons\mathrm{A}}} - \left\langle \vec{b}_{\xi}^{\vec{M}^{1}_{\mathrm{M}\rightleftharpoons\mathrm{A}}}, \mathcal{H}_1\left(\vec{b}_{\zeta}^{\vec{M}^{1}_{\mathrm{M}\rightleftharpoons\mathrm{A}}}\right)\right\rangle_{\mathbb{S}^{1}_{\mathrm{M}\rightleftharpoons\mathrm{A}}} \qquad \text{(D-129h)}$$

$$z_{\xi\zeta}^{\vec{M}_{\mathrm{M}\rightleftharpoons\mathrm{A}}^{1}\vec{M}_{\mathrm{M}\rightleftharpoons\mathrm{A}}^{2}} = \left\langle \vec{b}_{\xi}^{\vec{M}_{\mathrm{M}\rightleftharpoons\mathrm{A}}^{1}}, -j\omega\varepsilon_0\mathcal{L}_0\left(-\vec{b}_{\zeta}^{\vec{M}_{\mathrm{M}\rightleftharpoons\mathrm{A}}^{2}}\right)\right\rangle_{\mathbb{S}_{\mathrm{M}\rightleftharpoons\mathrm{A}}^{1}} \tag{D-129i}$$

$$z_{\xi\zeta}^{\vec{M}_{\mathrm{M}\rightleftharpoons\mathrm{A}}^{1}\vec{M}_{\mathrm{A}\rightleftharpoons\mathrm{A}}^{12}} = -\left\langle \vec{b}_{\xi}^{\vec{M}_{\mathrm{M}\rightleftharpoons\mathrm{A}}^{1}}, \mathcal{H}_1\left(\vec{b}_{\zeta}^{\vec{M}_{\mathrm{A}\rightleftharpoons\mathrm{A}}^{12}}\right)\right\rangle_{\tilde{\mathbb{S}}_{\mathrm{M}\rightleftharpoons\mathrm{A}}^{1}} \tag{D-129j}$$

and

$$z_{\xi\zeta}^{\vec{J}_{\mathrm{M}\rightleftharpoons\mathrm{A}}^{2}\vec{J}_{\mathrm{aux}}} = \left\langle \vec{b}_{\xi}^{\vec{J}_{\mathrm{M}\rightleftharpoons\mathrm{A}}^{2}}, -j\omega\mu_0\mathcal{L}_0\left(\vec{b}_{\zeta}^{\vec{J}_{\mathrm{aux}}}\right)\right\rangle_{\mathbb{S}_{\mathrm{M}\rightleftharpoons\mathrm{A}}^{2}} \tag{D-130a}$$

$$z_{\xi\zeta}^{\vec{J}_{\mathrm{M}\rightleftharpoons\mathrm{A}}^{2}\vec{J}_{\mathrm{M}\rightleftharpoons\mathrm{A}}^{1}} = \left\langle \vec{b}_{\xi}^{\vec{J}_{\mathrm{M}\rightleftharpoons\mathrm{A}}^{2}}, -j\omega\mu_0\mathcal{L}_0\left(-\vec{b}_{\zeta}^{\vec{J}_{\mathrm{M}\rightleftharpoons\mathrm{A}}^{1}}\right)\right\rangle_{\mathbb{S}_{\mathrm{M}\rightleftharpoons\mathrm{A}}^{2}} \tag{D-130b}$$

$$z_{\xi\zeta}^{\vec{J}_{\mathrm{M}\rightleftharpoons\mathrm{A}}^{2}\vec{J}_{\mathrm{M}\rightleftharpoons\mathrm{A}}^{2}} = \left\langle \vec{b}_{\xi}^{\vec{J}_{\mathrm{M}\rightleftharpoons\mathrm{A}}^{2}}, -j\omega\mu_0\mathcal{L}_0\left(-\vec{b}_{\zeta}^{\vec{J}_{\mathrm{M}\rightleftharpoons\mathrm{A}}^{2}}\right)\right\rangle_{\mathbb{S}_{\mathrm{M}\rightleftharpoons\mathrm{A}}^{2}} - \left\langle \vec{b}_{\xi}^{\vec{J}_{\mathrm{M}\rightleftharpoons\mathrm{A}}^{2}}, \mathcal{E}_2\left(\vec{b}_{\zeta}^{\vec{J}_{\mathrm{M}\rightleftharpoons\mathrm{A}}^{2}}\right)\right\rangle_{\tilde{\mathbb{S}}_{\mathrm{M}\rightleftharpoons\mathrm{A}}^{2}} \tag{D-130c}$$

$$z_{\xi\zeta}^{\vec{J}_{\mathrm{M}\rightleftharpoons\mathrm{A}}^{2}\vec{J}_{\mathrm{A}\rightleftharpoons\mathrm{A}}^{12}} = -\left\langle \vec{b}_{\xi}^{\vec{J}_{\mathrm{M}\rightleftharpoons\mathrm{A}}^{2}}, \mathcal{E}_2\left(-\vec{b}_{\zeta}^{\vec{J}_{\mathrm{A}\rightleftharpoons\mathrm{A}}^{12}}\right)\right\rangle_{\tilde{\mathbb{S}}_{\mathrm{M}\rightleftharpoons\mathrm{A}}^{2}} \tag{D-130d}$$

$$z_{\xi\zeta}^{\vec{J}_{\mathrm{M}\rightleftharpoons\mathrm{A}}^{2}\vec{J}_{\mathrm{A}\rightleftharpoons\mathrm{G}}} = -\left\langle \vec{b}_{\xi}^{\vec{J}_{\mathrm{M}\rightleftharpoons\mathrm{A}}^{2}}, \mathcal{E}_2\left(\vec{b}_{\zeta}^{\vec{J}_{\mathrm{A}\rightleftharpoons\mathrm{G}}}\right)\right\rangle_{\tilde{\mathbb{S}}_{\mathrm{M}\rightleftharpoons\mathrm{A}}^{2}} \tag{D-130e}$$

$$z_{\xi\zeta}^{\vec{J}_{\mathrm{M}\rightleftharpoons\mathrm{A}}^{2}\vec{J}_{\mathrm{M}}} = \left\langle \vec{b}_{\xi}^{\vec{J}_{\mathrm{M}\rightleftharpoons\mathrm{A}}^{2}}, -j\omega\mu_0\mathcal{L}_0\left(\vec{b}_{\zeta}^{\vec{J}_{\mathrm{M}}}\right)\right\rangle_{\mathbb{S}_{\mathrm{M}\rightleftharpoons\mathrm{A}}^{2}} \tag{D-130f}$$

$$z_{\xi\zeta}^{\vec{J}_{\mathrm{M}\rightleftharpoons\mathrm{A}}^{2}\vec{J}_{\mathrm{A}}^{2}} = -\left\langle \vec{b}_{\xi}^{\vec{J}_{\mathrm{M}\rightleftharpoons\mathrm{A}}^{2}}, \mathcal{E}_2\left(\vec{b}_{\zeta}^{\vec{J}_{\mathrm{A}}^{2}}\right)\right\rangle_{\tilde{\mathbb{S}}_{\mathrm{M}\rightleftharpoons\mathrm{A}}^{2}} \tag{D-130g}$$

$$z_{\xi\zeta}^{\vec{J}_{\mathrm{M}\rightleftharpoons\mathrm{A}}^{2}\vec{M}_{\mathrm{aux}}} = \left\langle \vec{b}_{\xi}^{\vec{J}_{\mathrm{M}\rightleftharpoons\mathrm{A}}^{2}}, -\mathcal{K}_0\left(\vec{b}_{\zeta}^{\vec{M}_{\mathrm{aux}}}\right)\right\rangle_{\mathbb{S}_{\mathrm{M}\rightleftharpoons\mathrm{A}}^{2}} \tag{D-130h}$$

$$z_{\xi\zeta}^{\vec{J}_{\mathrm{M}\rightleftharpoons\mathrm{A}}^{2}\vec{M}_{\mathrm{M}\rightleftharpoons\mathrm{A}}^{1}} = \left\langle \vec{b}_{\xi}^{\vec{J}_{\mathrm{M}\rightleftharpoons\mathrm{A}}^{2}}, -\mathcal{K}_0\left(-\vec{b}_{\zeta}^{\vec{M}_{\mathrm{M}\rightleftharpoons\mathrm{A}}^{1}}\right)\right\rangle_{\mathbb{S}_{\mathrm{M}\rightleftharpoons\mathrm{A}}^{2}} \tag{D-130i}$$

$$z_{\xi\zeta}^{\vec{J}_{\mathrm{M}\rightleftharpoons\mathrm{A}}^{2}\vec{M}_{\mathrm{M}\rightleftharpoons\mathrm{A}}^{2}} = \left\langle \vec{b}_{\xi}^{\vec{J}_{\mathrm{M}\rightleftharpoons\mathrm{A}}^{2}}, \hat{n}_{\rightarrow\mathrm{A}}^{2}\times\frac{1}{2}\vec{b}_{\zeta}^{\vec{M}_{\mathrm{M}\rightleftharpoons\mathrm{A}}^{2}} - \mathrm{P.V.}\,\mathcal{K}_0\left(-\vec{b}_{\zeta}^{\vec{M}_{\mathrm{M}\rightleftharpoons\mathrm{A}}^{2}}\right)\right\rangle_{\mathbb{S}_{\mathrm{M}\rightleftharpoons\mathrm{A}}^{2}} - \left\langle \vec{b}_{\xi}^{\vec{J}_{\mathrm{M}\rightleftharpoons\mathrm{A}}^{2}}, \mathcal{E}_2\left(\vec{b}_{\zeta}^{\vec{M}_{\mathrm{M}\rightleftharpoons\mathrm{A}}^{2}}\right)\right\rangle_{\tilde{\mathbb{S}}_{\mathrm{M}\rightleftharpoons\mathrm{A}}^{2}} \tag{D-130j}$$

$$z_{\xi\zeta}^{\vec{J}_{\mathrm{M}\rightleftharpoons\mathrm{A}}^{2}\vec{M}_{\mathrm{A}\rightleftharpoons\mathrm{A}}^{12}} = -\left\langle \vec{b}_{\xi}^{\vec{J}_{\mathrm{M}\rightleftharpoons\mathrm{A}}^{2}}, \mathcal{E}_2\left(-\vec{b}_{\zeta}^{\vec{M}_{\mathrm{A}\rightleftharpoons\mathrm{A}}^{12}}\right)\right\rangle_{\tilde{\mathbb{S}}_{\mathrm{M}\rightleftharpoons\mathrm{A}}^{2}} \tag{D-130k}$$

$$z_{\xi\zeta}^{\vec{J}_{\mathrm{M}\rightleftharpoons\mathrm{A}}^{2}\vec{M}_{\mathrm{A}\rightleftharpoons\mathrm{G}}} = -\left\langle \vec{b}_{\xi}^{\vec{J}_{\mathrm{M}\rightleftharpoons\mathrm{A}}^{2}}, \mathcal{E}_2\left(\vec{b}_{\zeta}^{\vec{M}_{\mathrm{A}\rightleftharpoons\mathrm{G}}}\right)\right\rangle_{\tilde{\mathbb{S}}_{\mathrm{M}\rightleftharpoons\mathrm{A}}^{2}} \tag{D-130l}$$

and

$$z_{\xi\zeta}^{\vec{M}_{\mathrm{M}\rightleftharpoons\mathrm{A}}^{2}\vec{J}_{\mathrm{aux}}} = \left\langle \vec{b}_{\xi}^{\vec{M}_{\mathrm{M}\rightleftharpoons\mathrm{A}}^{2}}, \mathcal{K}_0\left(\vec{b}_{\zeta}^{\vec{J}_{\mathrm{aux}}}\right)\right\rangle_{\mathbb{S}_{\mathrm{M}\rightleftharpoons\mathrm{A}}^{2}} \tag{D-131a}$$

$$z_{\xi\zeta}^{\vec{M}_{\mathrm{M}\rightleftharpoons\mathrm{A}}^{2}\vec{J}_{\mathrm{M}\rightleftharpoons\mathrm{A}}^{1}} = \left\langle \vec{b}_{\xi}^{\vec{M}_{\mathrm{M}\rightleftharpoons\mathrm{A}}^{2}}, \mathcal{K}_0\left(-\vec{b}_{\zeta}^{\vec{J}_{\mathrm{M}\rightleftharpoons\mathrm{A}}^{1}}\right)\right\rangle_{\mathbb{S}_{\mathrm{M}\rightleftharpoons\mathrm{A}}^{2}} \tag{D-131b}$$

$$z_{\xi\zeta}^{\vec{M}_{\mathrm{M}\rightleftharpoons\mathrm{A}}^{2}\vec{J}_{\mathrm{M}\rightleftharpoons\mathrm{A}}^{2}} = \left\langle \vec{b}_{\xi}^{\vec{M}_{\mathrm{M}\rightleftharpoons\mathrm{A}}^{2}}, \frac{1}{2}\vec{b}_{\zeta}^{\vec{J}_{\mathrm{M}\rightleftharpoons\mathrm{A}}^{2}}\times\hat{n}_{\rightarrow\mathrm{A}}^{2} + \mathrm{P.V.}\,\mathcal{K}_0\left(-\vec{b}_{\zeta}^{\vec{J}_{\mathrm{M}\rightleftharpoons\mathrm{A}}^{2}}\right)\right\rangle_{\mathbb{S}_{\mathrm{M}\rightleftharpoons\mathrm{A}}^{2}} - \left\langle \vec{b}_{\xi}^{\vec{M}_{\mathrm{M}\rightleftharpoons\mathrm{A}}^{2}}, \mathcal{H}_2\left(\vec{b}_{\zeta}^{\vec{J}_{\mathrm{M}\rightleftharpoons\mathrm{A}}^{2}}\right)\right\rangle_{\tilde{\mathbb{S}}_{\mathrm{M}\rightleftharpoons\mathrm{A}}^{2}} \tag{D-131c}$$

$$z_{\xi\zeta}^{\vec{M}_{\mathrm{M}\rightleftharpoons\mathrm{A}}^{2}\vec{J}_{\mathrm{A}\rightleftharpoons\mathrm{A}}^{12}} = -\left\langle \vec{b}_{\xi}^{\vec{M}_{\mathrm{M}\rightleftharpoons\mathrm{A}}^{2}}, \mathcal{H}_2\left(-\vec{b}_{\zeta}^{\vec{J}_{\mathrm{A}\rightleftharpoons\mathrm{A}}^{12}}\right)\right\rangle_{\tilde{\mathbb{S}}_{\mathrm{M}\rightleftharpoons\mathrm{A}}^{2}} \tag{D-131d}$$

$$z_{\xi\zeta}^{\vec{M}^2_{\text{M}\rightleftharpoons\text{A}}\vec{J}_{\text{A}\rightleftharpoons\text{G}}} = -\left\langle \vec{b}_{\xi}^{\vec{M}^2_{\text{M}\rightleftharpoons\text{A}}}, \mathcal{H}_2\left(\vec{b}_{\zeta}^{\vec{J}_{\text{A}\rightleftharpoons\text{G}}}\right)\right\rangle_{\mathbb{S}^2_{\text{M}\rightleftharpoons\text{A}}} \tag{D-131e}$$

$$z_{\xi\zeta}^{\vec{M}^2_{\text{M}\rightleftharpoons\text{A}}\vec{J}_{\text{M}}} = \left\langle \vec{b}_{\xi}^{\vec{M}^2_{\text{M}\rightleftharpoons\text{A}}}, \mathcal{K}_0\left(\vec{b}_{\zeta}^{\vec{J}_{\text{M}}}\right)\right\rangle_{\mathbb{S}^2_{\text{M}\rightleftharpoons\text{A}}} \tag{D-131f}$$

$$z_{\xi\zeta}^{\vec{M}^2_{\text{M}\rightleftharpoons\text{A}}\vec{J}^2_{\text{A}}} = -\left\langle \vec{b}_{\xi}^{\vec{M}^2_{\text{M}\rightleftharpoons\text{A}}}, \mathcal{H}_2\left(\vec{b}_{\zeta}^{\vec{J}^2_{\text{A}}}\right)\right\rangle_{\mathbb{S}^2_{\text{M}\rightleftharpoons\text{A}}} \tag{D-131g}$$

$$z_{\xi\zeta}^{\vec{M}^2_{\text{M}\rightleftharpoons\text{A}}\vec{M}_{\text{aux}}} = \left\langle \vec{b}_{\xi}^{\vec{M}^2_{\text{M}\rightleftharpoons\text{A}}}, -j\omega\varepsilon_0\mathcal{L}_0\left(\vec{b}_{\zeta}^{\vec{M}_{\text{aux}}}\right)\right\rangle_{\mathbb{S}^2_{\text{M}\rightleftharpoons\text{A}}} \tag{D-131h}$$

$$z_{\xi\zeta}^{\vec{M}^2_{\text{M}\rightleftharpoons\text{A}}\vec{M}^1_{\text{M}\rightleftharpoons\text{A}}} = \left\langle \vec{b}_{\xi}^{\vec{M}^2_{\text{M}\rightleftharpoons\text{A}}}, -j\omega\varepsilon_0\mathcal{L}_0\left(-\vec{b}_{\zeta}^{\vec{M}^1_{\text{M}\rightleftharpoons\text{A}}}\right)\right\rangle_{\mathbb{S}^2_{\text{M}\rightleftharpoons\text{A}}} \tag{D-131i}$$

$$z_{\xi\zeta}^{\vec{M}^2_{\text{M}\rightleftharpoons\text{A}}\vec{M}^2_{\text{M}\rightleftharpoons\text{A}}} = \left\langle \vec{b}_{\xi}^{\vec{M}^2_{\text{M}\rightleftharpoons\text{A}}}, -j\omega\varepsilon_0\mathcal{L}_0\left(-\vec{b}_{\zeta}^{\vec{M}^2_{\text{M}\rightleftharpoons\text{A}}}\right)\right\rangle_{\mathbb{S}^2_{\text{M}\rightleftharpoons\text{A}}} - \left\langle \vec{b}_{\xi}^{\vec{M}^2_{\text{M}\rightleftharpoons\text{A}}}, \mathcal{H}_2\left(\vec{b}_{\zeta}^{\vec{M}^2_{\text{M}\rightleftharpoons\text{A}}}\right)\right\rangle_{\mathbb{S}^2_{\text{M}\rightleftharpoons\text{A}}} \tag{D-131j}$$

$$z_{\xi\zeta}^{\vec{M}^2_{\text{M}\rightleftharpoons\text{A}}\vec{M}^{12}_{\text{A}\rightleftharpoons\text{A}}} = -\left\langle \vec{b}_{\xi}^{\vec{M}^2_{\text{M}\rightleftharpoons\text{A}}}, \mathcal{H}_2\left(-\vec{b}_{\zeta}^{\vec{M}^{12}_{\text{A}\rightleftharpoons\text{A}}}\right)\right\rangle_{\mathbb{S}^2_{\text{M}\rightleftharpoons\text{A}}} \tag{D-131k}$$

$$z_{\xi\zeta}^{\vec{M}^2_{\text{M}\rightleftharpoons\text{A}}\vec{M}_{\text{A}\rightleftharpoons\text{G}}} = -\left\langle \vec{b}_{\xi}^{\vec{M}^2_{\text{M}\rightleftharpoons\text{A}}}, \mathcal{H}_2\left(\vec{b}_{\zeta}^{\vec{M}_{\text{A}\rightleftharpoons\text{G}}}\right)\right\rangle_{\mathbb{S}^2_{\text{M}\rightleftharpoons\text{A}}} \tag{D-131l}$$

and

$$z_{\xi\zeta}^{\vec{J}^{12}_{\text{A}\rightleftharpoons\text{A}}\vec{J}^1_{\text{M}\rightleftharpoons\text{A}}} = \left\langle \vec{b}_{\xi}^{\vec{J}^{12}_{\text{A}\rightleftharpoons\text{A}}}, \mathcal{E}_1\left(\vec{b}_{\zeta}^{\vec{J}^1_{\text{M}\rightleftharpoons\text{A}}}\right)\right\rangle_{\tilde{\mathbb{S}}^{12}_{\text{A}\rightleftharpoons\text{A}}} \tag{D-132a}$$

$$z_{\xi\zeta}^{\vec{J}^{12}_{\text{A}\rightleftharpoons\text{A}}\vec{J}^2_{\text{M}\rightleftharpoons\text{A}}} = -\left\langle \vec{b}_{\xi}^{\vec{J}^{12}_{\text{A}\rightleftharpoons\text{A}}}, \mathcal{E}_2\left(\vec{b}_{\zeta}^{\vec{J}^2_{\text{M}\rightleftharpoons\text{A}}}\right)\right\rangle_{\mathbb{S}^{12}_{\text{A}\rightleftharpoons\text{A}}} \tag{D-132b}$$

$$z_{\xi\zeta}^{\vec{J}^{12}_{\text{A}\rightleftharpoons\text{A}}\vec{J}^{12}_{\text{A}\rightleftharpoons\text{A}}} = \left\langle \vec{b}_{\xi}^{\vec{J}^{12}_{\text{A}\rightleftharpoons\text{A}}}, \mathcal{E}_1\left(\vec{b}_{\zeta}^{\vec{J}^{12}_{\text{A}\rightleftharpoons\text{A}}}\right)\right\rangle_{\tilde{\mathbb{S}}^{12}_{\text{A}\rightleftharpoons\text{A}}} - \left\langle \vec{b}_{\xi}^{\vec{J}^{12}_{\text{A}\rightleftharpoons\text{A}}}, \mathcal{E}_2\left(-\vec{b}_{\zeta}^{\vec{J}^{12}_{\text{A}\rightleftharpoons\text{A}}}\right)\right\rangle_{\mathbb{S}^{12}_{\text{A}\rightleftharpoons\text{A}}} \tag{D-132c}$$

$$z_{\xi\zeta}^{\vec{J}^{12}_{\text{A}\rightleftharpoons\text{A}}\vec{J}_{\text{A}\rightleftharpoons\text{G}}} = -\left\langle \vec{b}_{\xi}^{\vec{J}^{12}_{\text{A}\rightleftharpoons\text{A}}}, \mathcal{E}_2\left(\vec{b}_{\zeta}^{\vec{J}_{\text{A}\rightleftharpoons\text{G}}}\right)\right\rangle_{\mathbb{S}^{12}_{\text{A}\rightleftharpoons\text{A}}} \tag{D-132d}$$

$$z_{\xi\zeta}^{\vec{J}^{12}_{\text{A}\rightleftharpoons\text{A}}\vec{J}^1_{\text{A}}} = \left\langle \vec{b}_{\xi}^{\vec{J}^{12}_{\text{A}\rightleftharpoons\text{A}}}, \mathcal{E}_1\left(\vec{b}_{\zeta}^{\vec{J}^1_{\text{A}}}\right)\right\rangle_{\tilde{\mathbb{S}}^{12}_{\text{A}\rightleftharpoons\text{A}}} \tag{D-132e}$$

$$z_{\xi\zeta}^{\vec{J}^{12}_{\text{A}\rightleftharpoons\text{A}}\vec{J}^2_{\text{A}}} = -\left\langle \vec{b}_{\xi}^{\vec{J}^{12}_{\text{A}\rightleftharpoons\text{A}}}, \mathcal{E}_2\left(\vec{b}_{\zeta}^{\vec{J}^2_{\text{A}}}\right)\right\rangle_{\mathbb{S}^{12}_{\text{A}\rightleftharpoons\text{A}}} \tag{D-132f}$$

$$z_{\xi\zeta}^{\vec{J}^{12}_{\text{A}\rightleftharpoons\text{A}}\vec{M}^1_{\text{M}\rightleftharpoons\text{A}}} = \left\langle \vec{b}_{\xi}^{\vec{J}^{12}_{\text{A}\rightleftharpoons\text{A}}}, \mathcal{E}_1\left(\vec{b}_{\zeta}^{\vec{M}^1_{\text{M}\rightleftharpoons\text{A}}}\right)\right\rangle_{\tilde{\mathbb{S}}^{12}_{\text{A}\rightleftharpoons\text{A}}} \tag{D-132g}$$

$$z_{\xi\zeta}^{\vec{J}^{12}_{\text{A}\rightleftharpoons\text{A}}\vec{M}^2_{\text{M}\rightleftharpoons\text{A}}} = -\left\langle \vec{b}_{\xi}^{\vec{J}^{12}_{\text{A}\rightleftharpoons\text{A}}}, \mathcal{E}_2\left(\vec{b}_{\zeta}^{\vec{M}^2_{\text{M}\rightleftharpoons\text{A}}}\right)\right\rangle_{\mathbb{S}^{12}_{\text{A}\rightleftharpoons\text{A}}} \tag{D-132h}$$

$$z_{\xi\zeta}^{\vec{J}^{12}_{\text{A}\rightleftharpoons\text{A}}\vec{M}^{12}_{\text{A}\rightleftharpoons\text{A}}} = \left\langle \vec{b}_{\xi}^{\vec{J}^{12}_{\text{A}\rightleftharpoons\text{A}}}, \mathcal{E}_1\left(\vec{b}_{\zeta}^{\vec{M}^{12}_{\text{A}\rightleftharpoons\text{A}}}\right)\right\rangle_{\tilde{\mathbb{S}}^{12}_{\text{A}\rightleftharpoons\text{A}}} - \left\langle \vec{b}_{\xi}^{\vec{J}^{12}_{\text{A}\rightleftharpoons\text{A}}}, \mathcal{E}_2\left(-\vec{b}_{\zeta}^{\vec{M}^{12}_{\text{A}\rightleftharpoons\text{A}}}\right)\right\rangle_{\mathbb{S}^{12}_{\text{A}\rightleftharpoons\text{A}}} \tag{D-132i}$$

$$z_{\xi\zeta}^{\vec{J}^{12}_{\text{A}\rightleftharpoons\text{A}}\vec{M}_{\text{A}\rightleftharpoons\text{G}}} = -\left\langle \vec{b}_{\xi}^{\vec{J}^{12}_{\text{A}\rightleftharpoons\text{A}}}, \mathcal{E}_2\left(\vec{b}_{\zeta}^{\vec{M}_{\text{A}\rightleftharpoons\text{G}}}\right)\right\rangle_{\mathbb{S}^{12}_{\text{A}\rightleftharpoons\text{A}}} \tag{D-132j}$$

and

$$z_{\xi\zeta}^{\vec{M}^{12}_{\text{A}\rightleftharpoons\text{A}}\vec{J}^1_{\text{M}\rightleftharpoons\text{A}}} = \left\langle \vec{b}_{\xi}^{\vec{M}^{12}_{\text{A}\rightleftharpoons\text{A}}}, \mathcal{H}_1\left(\vec{b}_{\zeta}^{\vec{J}^1_{\text{M}\rightleftharpoons\text{A}}}\right)\right\rangle_{\tilde{\mathbb{S}}^{12}_{\text{A}\rightleftharpoons\text{A}}} \tag{D-133a}$$

$$z_{\xi\zeta}^{\vec{M}^{12}_{\text{A}\rightleftharpoons\text{A}}\vec{J}^2_{\text{M}\rightleftharpoons\text{A}}} = -\left\langle \vec{b}_{\xi}^{\vec{M}^{12}_{\text{A}\rightleftharpoons\text{A}}}, \mathcal{H}_2\left(\vec{b}_{\zeta}^{\vec{J}^2_{\text{M}\rightleftharpoons\text{A}}}\right)\right\rangle_{\mathbb{S}^{12}_{\text{A}\rightleftharpoons\text{A}}} \tag{D-133b}$$

$$z_{\xi\zeta}^{\vec{M}^{12}_{\mathrm{A}\rightleftharpoons\mathrm{A}}\vec{J}^{12}_{\mathrm{A}\rightleftharpoons\mathrm{A}}} = \left\langle \vec{b}_{\xi}^{\vec{M}^{12}_{\mathrm{A}\rightleftharpoons\mathrm{A}}}, \mathcal{H}_1\left(\vec{b}_{\zeta}^{\vec{J}^{12}_{\mathrm{A}\rightleftharpoons\mathrm{A}}}\right)\right\rangle_{\tilde{\mathbb{S}}^{12}_{\mathrm{A}\rightleftharpoons\mathrm{A}}} - \left\langle \vec{b}_{\xi}^{\vec{M}^{12}_{\mathrm{A}\rightleftharpoons\mathrm{A}}}, \mathcal{H}_2\left(-\vec{b}_{\zeta}^{\vec{J}^{12}_{\mathrm{A}\rightleftharpoons\mathrm{A}}}\right)\right\rangle_{\underset{\sim}{\mathbb{S}}^{12}_{\mathrm{A}\rightleftharpoons\mathrm{A}}} \quad \text{(D-133c)}$$

$$z_{\xi\zeta}^{\vec{M}^{12}_{\mathrm{A}\rightleftharpoons\mathrm{A}}\vec{J}_{\mathrm{A}\rightleftharpoons\mathrm{G}}} = -\left\langle \vec{b}_{\xi}^{\vec{M}^{12}_{\mathrm{A}\rightleftharpoons\mathrm{A}}}, \mathcal{H}_2\left(\vec{b}_{\zeta}^{\vec{J}_{\mathrm{A}\rightleftharpoons\mathrm{G}}}\right)\right\rangle_{\underset{\sim}{\mathbb{S}}^{12}_{\mathrm{A}\rightleftharpoons\mathrm{A}}} \quad \text{(D-133d)}$$

$$z_{\xi\zeta}^{\vec{M}^{12}_{\mathrm{A}\rightleftharpoons\mathrm{A}}\vec{J}^{1}_{\mathrm{A}}} = \left\langle \vec{b}_{\xi}^{\vec{M}^{12}_{\mathrm{A}\rightleftharpoons\mathrm{A}}}, \mathcal{H}_1\left(\vec{b}_{\zeta}^{\vec{J}^{1}_{\mathrm{A}}}\right)\right\rangle_{\tilde{\mathbb{S}}^{12}_{\mathrm{A}\rightleftharpoons\mathrm{A}}} \quad \text{(D-133e)}$$

$$z_{\xi\zeta}^{\vec{M}^{12}_{\mathrm{A}\rightleftharpoons\mathrm{A}}\vec{J}^{2}_{\mathrm{A}}} = -\left\langle \vec{b}_{\xi}^{\vec{M}^{12}_{\mathrm{A}\rightleftharpoons\mathrm{A}}}, \mathcal{H}_2\left(\vec{b}_{\zeta}^{\vec{J}^{2}_{\mathrm{A}}}\right)\right\rangle_{\underset{\sim}{\mathbb{S}}^{12}_{\mathrm{A}\rightleftharpoons\mathrm{A}}} \quad \text{(D-133f)}$$

$$z_{\xi\zeta}^{\vec{M}^{12}_{\mathrm{A}\rightleftharpoons\mathrm{A}}\vec{M}^{1}_{\mathrm{M}\rightleftharpoons\mathrm{A}}} = \left\langle \vec{b}_{\xi}^{\vec{M}^{12}_{\mathrm{A}\rightleftharpoons\mathrm{A}}}, \mathcal{H}_1\left(\vec{b}_{\zeta}^{\vec{M}^{1}_{\mathrm{M}\rightleftharpoons\mathrm{A}}}\right)\right\rangle_{\tilde{\mathbb{S}}^{12}_{\mathrm{A}\rightleftharpoons\mathrm{A}}} \quad \text{(D-133g)}$$

$$z_{\xi\zeta}^{\vec{M}^{12}_{\mathrm{A}\rightleftharpoons\mathrm{A}}\vec{M}^{2}_{\mathrm{M}\rightleftharpoons\mathrm{A}}} = -\left\langle \vec{b}_{\xi}^{\vec{M}^{12}_{\mathrm{A}\rightleftharpoons\mathrm{A}}}, \mathcal{H}_2\left(\vec{b}_{\zeta}^{\vec{M}^{2}_{\mathrm{M}\rightleftharpoons\mathrm{A}}}\right)\right\rangle_{\underset{\sim}{\mathbb{S}}^{12}_{\mathrm{A}\rightleftharpoons\mathrm{A}}} \quad \text{(D-133h)}$$

$$z_{\xi\zeta}^{\vec{M}^{12}_{\mathrm{A}\rightleftharpoons\mathrm{A}}\vec{M}^{12}_{\mathrm{A}\rightleftharpoons\mathrm{A}}} = \left\langle \vec{b}_{\xi}^{\vec{M}^{12}_{\mathrm{A}\rightleftharpoons\mathrm{A}}}, \mathcal{H}_1\left(\vec{b}_{\zeta}^{\vec{M}^{12}_{\mathrm{A}\rightleftharpoons\mathrm{A}}}\right)\right\rangle_{\tilde{\mathbb{S}}^{12}_{\mathrm{A}\rightleftharpoons\mathrm{A}}} - \left\langle \vec{b}_{\xi}^{\vec{M}^{12}_{\mathrm{A}\rightleftharpoons\mathrm{A}}}, \mathcal{H}_2\left(-\vec{b}_{\zeta}^{\vec{M}^{12}_{\mathrm{A}\rightleftharpoons\mathrm{A}}}\right)\right\rangle_{\underset{\sim}{\mathbb{S}}^{12}_{\mathrm{A}\rightleftharpoons\mathrm{A}}} \quad \text{(D-133i)}$$

$$z_{\xi\zeta}^{\vec{M}^{12}_{\mathrm{A}\rightleftharpoons\mathrm{A}}\vec{M}_{\mathrm{A}\rightleftharpoons\mathrm{G}}} = -\left\langle \vec{b}_{\xi}^{\vec{M}^{12}_{\mathrm{A}\rightleftharpoons\mathrm{A}}}, \mathcal{H}_2\left(\vec{b}_{\zeta}^{\vec{M}_{\mathrm{A}\rightleftharpoons\mathrm{G}}}\right)\right\rangle_{\underset{\sim}{\mathbb{S}}^{12}_{\mathrm{A}\rightleftharpoons\mathrm{A}}} \quad \text{(D-133j)}$$

and

$$z_{\xi\zeta}^{\vec{J}_{\mathrm{A}\rightleftharpoons\mathrm{G}}\vec{J}^{2}_{\mathrm{M}\rightleftharpoons\mathrm{A}}} = \left\langle \vec{b}_{\xi}^{\vec{J}_{\mathrm{A}\rightleftharpoons\mathrm{G}}}, \mathcal{E}_2\left(\vec{b}_{\zeta}^{\vec{J}^{2}_{\mathrm{M}\rightleftharpoons\mathrm{A}}}\right)\right\rangle_{\tilde{\mathbb{S}}_{\mathrm{A}\rightleftharpoons\mathrm{G}}} \quad \text{(D-134a)}$$

$$z_{\xi\zeta}^{\vec{J}_{\mathrm{A}\rightleftharpoons\mathrm{G}}\vec{J}^{12}_{\mathrm{A}\rightleftharpoons\mathrm{A}}} = \left\langle \vec{b}_{\xi}^{\vec{J}_{\mathrm{A}\rightleftharpoons\mathrm{G}}}, \mathcal{E}_2\left(-\vec{b}_{\zeta}^{\vec{J}^{12}_{\mathrm{A}\rightleftharpoons\mathrm{A}}}\right)\right\rangle_{\tilde{\mathbb{S}}_{\mathrm{A}\rightleftharpoons\mathrm{G}}} \quad \text{(D-134b)}$$

$$z_{\xi\zeta}^{\vec{J}_{\mathrm{A}\rightleftharpoons\mathrm{G}}\vec{J}_{\mathrm{A}\rightleftharpoons\mathrm{G}}} = \left\langle \vec{b}_{\xi}^{\vec{J}_{\mathrm{A}\rightleftharpoons\mathrm{G}}}, \mathcal{E}_2\left(\vec{b}_{\zeta}^{\vec{J}_{\mathrm{A}\rightleftharpoons\mathrm{G}}}\right)\right\rangle_{\tilde{\mathbb{S}}_{\mathrm{A}\rightleftharpoons\mathrm{G}}} - \left\langle \vec{b}_{\xi}^{\vec{J}_{\mathrm{A}\rightleftharpoons\mathrm{G}}}, \mathcal{E}_2\left(-\vec{b}_{\zeta}^{\vec{J}_{\mathrm{A}\rightleftharpoons\mathrm{G}}}\right)\right\rangle_{\underset{\sim}{\mathbb{S}}_{\mathrm{A}\rightleftharpoons\mathrm{G}}} \quad \text{(D-134c)}$$

$$z_{\xi\zeta}^{\vec{J}_{\mathrm{A}\rightleftharpoons\mathrm{G}}\vec{J}^{2}_{\mathrm{A}}} = \left\langle \vec{b}_{\xi}^{\vec{J}_{\mathrm{A}\rightleftharpoons\mathrm{G}}}, \mathcal{E}_2\left(\vec{b}_{\zeta}^{\vec{J}^{2}_{\mathrm{A}}}\right)\right\rangle_{\tilde{\mathbb{S}}_{\mathrm{A}\rightleftharpoons\mathrm{G}}} \quad \text{(D-134d)}$$

$$z_{\xi\zeta}^{\vec{J}_{\mathrm{A}\rightleftharpoons\mathrm{G}}\vec{M}^{2}_{\mathrm{M}\rightleftharpoons\mathrm{A}}} = \left\langle \vec{b}_{\xi}^{\vec{J}_{\mathrm{A}\rightleftharpoons\mathrm{G}}}, \mathcal{E}_2\left(\vec{b}_{\zeta}^{\vec{M}^{2}_{\mathrm{M}\rightleftharpoons\mathrm{A}}}\right)\right\rangle_{\tilde{\mathbb{S}}_{\mathrm{A}\rightleftharpoons\mathrm{G}}} \quad \text{(D-134e)}$$

$$z_{\xi\zeta}^{\vec{J}_{\mathrm{A}\rightleftharpoons\mathrm{G}}\vec{M}^{12}_{\mathrm{A}\rightleftharpoons\mathrm{A}}} = \left\langle \vec{b}_{\xi}^{\vec{J}_{\mathrm{A}\rightleftharpoons\mathrm{G}}}, \mathcal{E}_2\left(-\vec{b}_{\zeta}^{\vec{M}^{12}_{\mathrm{A}\rightleftharpoons\mathrm{A}}}\right)\right\rangle_{\tilde{\mathbb{S}}_{\mathrm{A}\rightleftharpoons\mathrm{G}}} \quad \text{(D-134f)}$$

$$z_{\xi\zeta}^{\vec{J}_{\mathrm{A}\rightleftharpoons\mathrm{G}}\vec{M}_{\mathrm{A}\rightleftharpoons\mathrm{G}}} = \left\langle \vec{b}_{\xi}^{\vec{J}_{\mathrm{A}\rightleftharpoons\mathrm{G}}}, \mathcal{E}_2\left(\vec{b}_{\zeta}^{\vec{M}_{\mathrm{A}\rightleftharpoons\mathrm{G}}}\right)\right\rangle_{\tilde{\mathbb{S}}_{\mathrm{A}\rightleftharpoons\mathrm{G}}} - \left\langle \vec{b}_{\xi}^{\vec{J}_{\mathrm{A}\rightleftharpoons\mathrm{G}}}, \mathcal{E}_2\left(-\vec{b}_{\zeta}^{\vec{M}_{\mathrm{A}\rightleftharpoons\mathrm{G}}}\right)\right\rangle_{\underset{\sim}{\mathbb{S}}_{\mathrm{A}\rightleftharpoons\mathrm{G}}} \quad \text{(D-134g)}$$

and

$$z_{\xi\zeta}^{\vec{M}_{\mathrm{A}\rightleftharpoons\mathrm{G}}\vec{J}^{2}_{\mathrm{M}\rightleftharpoons\mathrm{A}}} = \left\langle \vec{b}_{\xi}^{\vec{M}_{\mathrm{A}\rightleftharpoons\mathrm{G}}}, \mathcal{H}_2\left(\vec{b}_{\zeta}^{\vec{J}^{2}_{\mathrm{M}\rightleftharpoons\mathrm{A}}}\right)\right\rangle_{\tilde{\mathbb{S}}_{\mathrm{A}\rightleftharpoons\mathrm{G}}} \quad \text{(D-135a)}$$

$$z_{\xi\zeta}^{\vec{M}_{\mathrm{A}\rightleftharpoons\mathrm{G}}\vec{J}^{12}_{\mathrm{A}\rightleftharpoons\mathrm{A}}} = \left\langle \vec{b}_{\xi}^{\vec{M}_{\mathrm{A}\rightleftharpoons\mathrm{G}}}, \mathcal{H}_2\left(-\vec{b}_{\zeta}^{\vec{J}^{12}_{\mathrm{A}\rightleftharpoons\mathrm{A}}}\right)\right\rangle_{\tilde{\mathbb{S}}_{\mathrm{A}\rightleftharpoons\mathrm{G}}} \quad \text{(D-135b)}$$

$$z_{\xi\zeta}^{\vec{M}_{\mathrm{A}\rightleftharpoons\mathrm{G}}\vec{J}_{\mathrm{A}\rightleftharpoons\mathrm{G}}} = \left\langle \vec{b}_{\xi}^{\vec{M}_{\mathrm{A}\rightleftharpoons\mathrm{G}}}, \mathcal{H}_2\left(\vec{b}_{\zeta}^{\vec{J}_{\mathrm{A}\rightleftharpoons\mathrm{G}}}\right)\right\rangle_{\tilde{\mathbb{S}}_{\mathrm{A}\rightleftharpoons\mathrm{G}}} - \left\langle \vec{b}_{\xi}^{\vec{M}_{\mathrm{A}\rightleftharpoons\mathrm{G}}}, \mathcal{H}_2\left(-\vec{b}_{\zeta}^{\vec{J}_{\mathrm{A}\rightleftharpoons\mathrm{G}}}\right)\right\rangle_{\underset{\sim}{\mathbb{S}}_{\mathrm{A}\rightleftharpoons\mathrm{G}}} \quad \text{(D-135c)}$$

$$z_{\xi\zeta}^{\vec{M}_{\mathrm{A}\rightleftharpoons\mathrm{G}}\vec{J}^{2}_{\mathrm{A}}} = \left\langle \vec{b}_{\xi}^{\vec{M}_{\mathrm{A}\rightleftharpoons\mathrm{G}}}, \mathcal{H}_2\left(\vec{b}_{\zeta}^{\vec{J}^{2}_{\mathrm{A}}}\right)\right\rangle_{\tilde{\mathbb{S}}_{\mathrm{A}\rightleftharpoons\mathrm{G}}} \quad \text{(D-135d)}$$

$$z_{\xi\zeta}^{\vec{M}_{\mathrm{A}\rightleftharpoons\mathrm{G}}\vec{M}^{2}_{\mathrm{M}\rightleftharpoons\mathrm{A}}} = \left\langle \vec{b}_{\xi}^{\vec{M}_{\mathrm{A}\rightleftharpoons\mathrm{G}}}, \mathcal{H}_2\left(\vec{b}_{\zeta}^{\vec{M}^{2}_{\mathrm{M}\rightleftharpoons\mathrm{A}}}\right)\right\rangle_{\tilde{\mathbb{S}}_{\mathrm{A}\rightleftharpoons\mathrm{G}}} \quad \text{(D-135e)}$$

$$z_{\xi\zeta}^{\vec{M}_{A\rightleftharpoons G}\vec{M}_{A\rightleftharpoons A}^{12}} = \left\langle \vec{b}_{\xi}^{\vec{M}_{A\rightleftharpoons G}}, \mathcal{H}_2\left(-\vec{b}_{\zeta}^{\vec{M}_{A\rightleftharpoons A}^{12}}\right)\right\rangle_{\tilde{\mathbb{S}}_{A\rightleftharpoons G}} \tag{D-135f}$$

$$z_{\xi\zeta}^{\vec{M}_{A\rightleftharpoons G}\vec{M}_{A\rightleftharpoons G}} = \left\langle \vec{b}_{\xi}^{\vec{M}_{A\rightleftharpoons G}}, \mathcal{H}_2\left(\vec{b}_{\zeta}^{\vec{M}_{A\rightleftharpoons G}}\right)\right\rangle_{\tilde{\mathbb{S}}_{A\rightleftharpoons G}} - \left\langle \vec{b}_{\xi}^{\vec{M}_{A\rightleftharpoons G}}, \mathcal{H}_2\left(-\vec{b}_{\zeta}^{\vec{M}_{A\rightleftharpoons G}}\right)\right\rangle_{\underset{\sim}{\mathbb{S}}_{A\rightleftharpoons G}} \tag{D-135g}$$

and

$$z_{\xi\zeta}^{\vec{J}_M\vec{J}_{aux}} = \left\langle \vec{b}_{\xi}^{\vec{J}_M}, -j\omega\mu_0\mathcal{L}_0\left(\vec{b}_{\zeta}^{\vec{J}_{aux}}\right)\right\rangle_{\mathbb{S}_M} \tag{D-136a}$$

$$z_{\xi\zeta}^{\vec{J}_M\vec{J}_{M\rightleftharpoons A}^1} = \left\langle \vec{b}_{\xi}^{\vec{J}_M}, -j\omega\mu_0\mathcal{L}_0\left(-\vec{b}_{\zeta}^{\vec{J}_{M\rightleftharpoons A}^1}\right)\right\rangle_{\mathbb{S}_M} \tag{D-136b}$$

$$z_{\xi\zeta}^{\vec{J}_M\vec{J}_{M\rightleftharpoons A}^2} = \left\langle \vec{b}_{\xi}^{\vec{J}_M}, -j\omega\mu_0\mathcal{L}_0\left(-\vec{b}_{\zeta}^{\vec{J}_{M\rightleftharpoons A}^2}\right)\right\rangle_{\mathbb{S}_M} \tag{D-136c}$$

$$z_{\xi\zeta}^{\vec{J}_M\vec{J}_M} = \left\langle \vec{b}_{\xi}^{\vec{J}_M}, -j\omega\mu_0\mathcal{L}_0\left(\vec{b}_{\zeta}^{\vec{J}_M}\right)\right\rangle_{\mathbb{S}_M} \tag{D-136d}$$

$$z_{\xi\zeta}^{\vec{J}_M\vec{M}_{aux}} = \left\langle \vec{b}_{\xi}^{\vec{J}_M}, -\mathcal{K}_0\left(\vec{b}_{\zeta}^{\vec{M}_{aux}}\right)\right\rangle_{\mathbb{S}_M} \tag{D-136e}$$

$$z_{\xi\zeta}^{\vec{J}_M\vec{M}_{M\rightleftharpoons A}^1} = \left\langle \vec{b}_{\xi}^{\vec{J}_M}, -\mathcal{K}_0\left(-\vec{b}_{\zeta}^{\vec{M}_{M\rightleftharpoons A}^1}\right)\right\rangle_{\mathbb{S}_M} \tag{D-136f}$$

$$z_{\xi\zeta}^{\vec{J}_M\vec{M}_{M\rightleftharpoons A}^2} = \left\langle \vec{b}_{\xi}^{\vec{J}_M}, -\mathcal{K}_0\left(-\vec{b}_{\zeta}^{\vec{M}_{M\rightleftharpoons A}^2}\right)\right\rangle_{\mathbb{S}_M} \tag{D-136g}$$

and

$$z_{\xi\zeta}^{\vec{J}_A^1\vec{J}_{M\rightleftharpoons A}^1} = \left\langle \vec{b}_{\xi}^{\vec{J}_A^1}, \mathcal{E}_1\left(\vec{b}_{\zeta}^{\vec{J}_{M\rightleftharpoons A}^1}\right)\right\rangle_{\tilde{\mathbb{S}}_A^1} \tag{D-137a}$$

$$z_{\xi\zeta}^{\vec{J}_A^1\vec{J}_{A\rightleftharpoons A}^{12}} = \left\langle \vec{b}_{\xi}^{\vec{J}_A^1}, \mathcal{E}_1\left(\vec{b}_{\zeta}^{\vec{J}_{A\rightleftharpoons A}^{12}}\right)\right\rangle_{\tilde{\mathbb{S}}_A^1} \tag{D-137b}$$

$$z_{\xi\zeta}^{\vec{J}_A^1\vec{J}_A^1} = \left\langle \vec{b}_{\xi}^{\vec{J}_A^1}, \mathcal{E}_1\left(\vec{b}_{\zeta}^{\vec{J}_A^1}\right)\right\rangle_{\tilde{\mathbb{S}}_A^1} \tag{D-137c}$$

$$z_{\xi\zeta}^{\vec{J}_A^1\vec{M}_{M\rightleftharpoons A}^1} = \left\langle \vec{b}_{\xi}^{\vec{J}_A^1}, \mathcal{E}_1\left(\vec{b}_{\zeta}^{\vec{M}_{M\rightleftharpoons A}^1}\right)\right\rangle_{\tilde{\mathbb{S}}_A^1} \tag{D-137d}$$

$$z_{\xi\zeta}^{\vec{J}_A^1\vec{M}_{A\rightleftharpoons A}^{12}} = \left\langle \vec{b}_{\xi}^{\vec{J}_A^1}, \mathcal{E}_1\left(\vec{b}_{\zeta}^{\vec{M}_{A\rightleftharpoons A}^{12}}\right)\right\rangle_{\tilde{\mathbb{S}}_A^1} \tag{D-137e}$$

and

$$z_{\xi\zeta}^{\vec{J}_A^2\vec{J}_{M\rightleftharpoons A}^2} = \left\langle \vec{b}_{\xi}^{\vec{J}_A^2}, \mathcal{E}_2\left(\vec{b}_{\zeta}^{\vec{J}_{M\rightleftharpoons A}^2}\right)\right\rangle_{\tilde{\mathbb{S}}_A^2} \tag{D-138a}$$

$$z_{\xi\zeta}^{\vec{J}_A^2\vec{J}_{A\rightleftharpoons A}^{12}} = \left\langle \vec{b}_{\xi}^{\vec{J}_A^2}, \mathcal{E}_2\left(-\vec{b}_{\zeta}^{\vec{J}_{A\rightleftharpoons A}^{12}}\right)\right\rangle_{\tilde{\mathbb{S}}_A^2} \tag{D-138b}$$

$$z_{\xi\zeta}^{\vec{J}_A^2\vec{J}_{A\rightleftharpoons G}} = \left\langle \vec{b}_{\xi}^{\vec{J}_A^2}, \mathcal{E}_2\left(\vec{b}_{\zeta}^{\vec{J}_{A\rightleftharpoons G}}\right)\right\rangle_{\tilde{\mathbb{S}}_A^2} \tag{D-138c}$$

$$z_{\xi\zeta}^{\vec{J}_A^2\vec{J}_A^2} = \left\langle \vec{b}_{\xi}^{\vec{J}_A^2}, \mathcal{E}_2\left(\vec{b}_{\zeta}^{\vec{J}_A^2}\right)\right\rangle_{\tilde{\mathbb{S}}_A^2} \tag{D-138d}$$

$$z_{\xi\zeta}^{\vec{J}_A^2\vec{M}_{M\rightleftharpoons A}^2} = \left\langle \vec{b}_{\xi}^{\vec{J}_A^2}, \mathcal{E}_2\left(\vec{b}_{\zeta}^{\vec{M}_{M\rightleftharpoons A}^2}\right)\right\rangle_{\tilde{\mathbb{S}}_A^2} \tag{D-138e}$$

$$z_{\xi\zeta}^{\vec{J}_{\mathrm{A}}^{2}\vec{M}_{\mathrm{A\rightleftharpoons A}}^{12}} = \left\langle \vec{b}_{\xi}^{\vec{J}_{\mathrm{A}}^{2}}, \mathcal{E}_{2}\left(-\vec{b}_{\zeta}^{\vec{M}_{\mathrm{A\rightleftharpoons A}}^{12}}\right)\right\rangle_{\tilde{\mathbb{S}}_{\mathrm{A}}^{2}} \tag{D-138f}$$

$$z_{\xi\zeta}^{\vec{J}_{\mathrm{A}}^{2}\vec{M}_{\mathrm{A\rightleftharpoons G}}} = \left\langle \vec{b}_{\xi}^{\vec{J}_{\mathrm{A}}^{2}}, \mathcal{E}_{2}\left(\vec{b}_{\zeta}^{\vec{M}_{\mathrm{A\rightleftharpoons G}}}\right)\right\rangle_{\tilde{\mathbb{S}}_{\mathrm{A}}^{2}} \tag{D-138g}$$

The transformation matrix $\overline{\overline{T}}$ used in Eq. (7-93) is as follows:

$$\overline{\overline{T}} = \overline{\overline{T}}^{\vec{J}_{\mathrm{aux}}\to\mathrm{AV}} \text{ or } \overline{\overline{T}}^{\vec{M}_{\mathrm{aux}}\to\mathrm{AV}} \text{ or } \overline{\overline{T}}^{\mathrm{BS}\to\mathrm{AV}} \tag{D-139}$$

in which

$$\overline{\overline{T}}^{\vec{J}_{\mathrm{aux}}\to\mathrm{AV}} = \left(\overline{\overline{\Psi}}_{1}\right)^{-1}\cdot\overline{\overline{\Psi}}_{2} \tag{D-140a}$$

$$\overline{\overline{T}}^{\vec{M}_{\mathrm{aux}}\to\mathrm{AV}} = \left(\overline{\overline{\Psi}}_{3}\right)^{-1}\cdot\overline{\overline{\Psi}}_{4} \tag{D-140b}$$

$$\overline{\overline{T}}^{\mathrm{BS}\to\mathrm{AV}} = \mathrm{null}\left(\overline{\overline{\Psi}}_{\mathrm{FCE}}^{\mathrm{DoJ/DoM}}\right) \tag{D-141}$$

where

$$\overline{\overline{\Psi}}_{1} = \begin{bmatrix}
\overline{\overline{I}}^{\vec{J}_{\mathrm{aux}}} & 0 & 0 & 0 & 0 & 0 & 0 & 0 & 0 & 0 & 0 & 0 & 0 \\
0 & \overline{\overline{Z}}^{\vec{M}_{\mathrm{aux}}\vec{J}_{\mathrm{M\rightleftharpoons A}}^{1}} & \overline{\overline{Z}}^{\vec{M}_{\mathrm{aux}}\vec{J}_{\mathrm{M\rightleftharpoons A}}^{2}} & 0 & 0 & \overline{\overline{Z}}^{\vec{M}_{\mathrm{aux}}\vec{J}_{\mathrm{M}}} & 0 & 0 & \overline{\overline{Z}}^{\vec{M}_{\mathrm{aux}}\vec{M}_{\mathrm{aux}}} & \overline{\overline{Z}}^{\vec{M}_{\mathrm{aux}}\vec{M}_{\mathrm{M\rightleftharpoons A}}^{1}} & \overline{\overline{Z}}^{\vec{M}_{\mathrm{aux}}\vec{M}_{\mathrm{M\rightleftharpoons A}}^{2}} & 0 & 0 \\
0 & \overline{\overline{Z}}^{\vec{J}_{\mathrm{M\rightleftharpoons A}}^{1}\vec{J}_{\mathrm{M\rightleftharpoons A}}^{1}} & \overline{\overline{Z}}^{\vec{J}_{\mathrm{M\rightleftharpoons A}}^{1}\vec{J}_{\mathrm{M\rightleftharpoons A}}^{2}} & \overline{\overline{Z}}^{\vec{J}_{\mathrm{M\rightleftharpoons A}}^{1}\vec{J}_{\mathrm{A\rightleftharpoons A}}^{12}} & 0 & \overline{\overline{Z}}^{\vec{J}_{\mathrm{M\rightleftharpoons A}}^{1}\vec{J}_{\mathrm{M}}} & \overline{\overline{Z}}^{\vec{J}_{\mathrm{M\rightleftharpoons A}}^{1}\vec{J}_{\mathrm{A}}^{1}} & 0 & \overline{\overline{Z}}^{\vec{J}_{\mathrm{M\rightleftharpoons A}}^{1}\vec{M}_{\mathrm{aux}}} & \overline{\overline{Z}}^{\vec{J}_{\mathrm{M\rightleftharpoons A}}^{1}\vec{M}_{\mathrm{M\rightleftharpoons A}}^{1}} & \overline{\overline{Z}}^{\vec{J}_{\mathrm{M\rightleftharpoons A}}^{1}\vec{M}_{\mathrm{M\rightleftharpoons A}}^{2}} & \overline{\overline{Z}}^{\vec{J}_{\mathrm{M\rightleftharpoons A}}^{1}\vec{M}_{\mathrm{A\rightleftharpoons A}}^{12}} & 0 \\
0 & \overline{\overline{Z}}^{\vec{M}_{\mathrm{M\rightleftharpoons A}}^{1}\vec{J}_{\mathrm{M\rightleftharpoons A}}^{1}} & \overline{\overline{Z}}^{\vec{M}_{\mathrm{M\rightleftharpoons A}}^{1}\vec{J}_{\mathrm{M\rightleftharpoons A}}^{2}} & \overline{\overline{Z}}^{\vec{M}_{\mathrm{M\rightleftharpoons A}}^{1}\vec{J}_{\mathrm{A\rightleftharpoons A}}^{12}} & 0 & \overline{\overline{Z}}^{\vec{M}_{\mathrm{M\rightleftharpoons A}}^{1}\vec{J}_{\mathrm{M}}} & \overline{\overline{Z}}^{\vec{M}_{\mathrm{M\rightleftharpoons A}}^{1}\vec{J}_{\mathrm{A}}^{1}} & 0 & \overline{\overline{Z}}^{\vec{M}_{\mathrm{M\rightleftharpoons A}}^{1}\vec{M}_{\mathrm{aux}}} & \overline{\overline{Z}}^{\vec{M}_{\mathrm{M\rightleftharpoons A}}^{1}\vec{M}_{\mathrm{M\rightleftharpoons A}}^{1}} & \overline{\overline{Z}}^{\vec{M}_{\mathrm{M\rightleftharpoons A}}^{1}\vec{M}_{\mathrm{M\rightleftharpoons A}}^{2}} & \overline{\overline{Z}}^{\vec{M}_{\mathrm{M\rightleftharpoons A}}^{1}\vec{M}_{\mathrm{A\rightleftharpoons A}}^{12}} & 0 \\
0 & \overline{\overline{Z}}^{\vec{J}_{\mathrm{M\rightleftharpoons A}}^{2}\vec{J}_{\mathrm{M\rightleftharpoons A}}^{1}} & \overline{\overline{Z}}^{\vec{J}_{\mathrm{M\rightleftharpoons A}}^{2}\vec{J}_{\mathrm{M\rightleftharpoons A}}^{2}} & \overline{\overline{Z}}^{\vec{J}_{\mathrm{M\rightleftharpoons A}}^{2}\vec{J}_{\mathrm{A\rightleftharpoons A}}^{12}} & \overline{\overline{Z}}^{\vec{J}_{\mathrm{M\rightleftharpoons A}}^{2}\vec{J}_{\mathrm{A\rightleftharpoons G}}} & \overline{\overline{Z}}^{\vec{J}_{\mathrm{M\rightleftharpoons A}}^{2}\vec{J}_{\mathrm{M}}} & 0 & \overline{\overline{Z}}^{\vec{J}_{\mathrm{M\rightleftharpoons A}}^{2}\vec{J}_{\mathrm{A}}^{2}} & \overline{\overline{Z}}^{\vec{J}_{\mathrm{M\rightleftharpoons A}}^{2}\vec{M}_{\mathrm{aux}}} & \overline{\overline{Z}}^{\vec{J}_{\mathrm{M\rightleftharpoons A}}^{2}\vec{M}_{\mathrm{M\rightleftharpoons A}}^{1}} & \overline{\overline{Z}}^{\vec{J}_{\mathrm{M\rightleftharpoons A}}^{2}\vec{M}_{\mathrm{M\rightleftharpoons A}}^{2}} & \overline{\overline{Z}}^{\vec{J}_{\mathrm{M\rightleftharpoons A}}^{2}\vec{M}_{\mathrm{A\rightleftharpoons A}}^{12}} & \overline{\overline{Z}}^{\vec{J}_{\mathrm{M\rightleftharpoons A}}^{2}\vec{M}_{\mathrm{A\rightleftharpoons G}}} \\
0 & \overline{\overline{Z}}^{\vec{M}_{\mathrm{M\rightleftharpoons A}}^{2}\vec{J}_{\mathrm{M\rightleftharpoons A}}^{1}} & \overline{\overline{Z}}^{\vec{M}_{\mathrm{M\rightleftharpoons A}}^{2}\vec{J}_{\mathrm{M\rightleftharpoons A}}^{2}} & \overline{\overline{Z}}^{\vec{M}_{\mathrm{M\rightleftharpoons A}}^{2}\vec{J}_{\mathrm{A\rightleftharpoons A}}^{12}} & \overline{\overline{Z}}^{\vec{M}_{\mathrm{M\rightleftharpoons A}}^{2}\vec{J}_{\mathrm{A\rightleftharpoons G}}} & \overline{\overline{Z}}^{\vec{M}_{\mathrm{M\rightleftharpoons A}}^{2}\vec{J}_{\mathrm{M}}} & 0 & \overline{\overline{Z}}^{\vec{M}_{\mathrm{M\rightleftharpoons A}}^{2}\vec{J}_{\mathrm{A}}^{2}} & \overline{\overline{Z}}^{\vec{M}_{\mathrm{M\rightleftharpoons A}}^{2}\vec{M}_{\mathrm{aux}}} & \overline{\overline{Z}}^{\vec{M}_{\mathrm{M\rightleftharpoons A}}^{2}\vec{M}_{\mathrm{M\rightleftharpoons A}}^{1}} & \overline{\overline{Z}}^{\vec{M}_{\mathrm{M\rightleftharpoons A}}^{2}\vec{M}_{\mathrm{M\rightleftharpoons A}}^{2}} & \overline{\overline{Z}}^{\vec{M}_{\mathrm{M\rightleftharpoons A}}^{2}\vec{M}_{\mathrm{A\rightleftharpoons A}}^{12}} & \overline{\overline{Z}}^{\vec{M}_{\mathrm{M\rightleftharpoons A}}^{2}\vec{M}_{\mathrm{A\rightleftharpoons G}}} \\
0 & \overline{\overline{Z}}^{\vec{J}_{\mathrm{A\rightleftharpoons A}}^{12}\vec{J}_{\mathrm{M\rightleftharpoons A}}^{1}} & \overline{\overline{Z}}^{\vec{J}_{\mathrm{A\rightleftharpoons A}}^{12}\vec{J}_{\mathrm{M\rightleftharpoons A}}^{2}} & \overline{\overline{Z}}^{\vec{J}_{\mathrm{A\rightleftharpoons A}}^{12}\vec{J}_{\mathrm{A\rightleftharpoons A}}^{12}} & \overline{\overline{Z}}^{\vec{J}_{\mathrm{A\rightleftharpoons A}}^{12}\vec{J}_{\mathrm{A\rightleftharpoons G}}} & 0 & \overline{\overline{Z}}^{\vec{J}_{\mathrm{A\rightleftharpoons A}}^{12}\vec{J}_{\mathrm{A}}^{1}} & \overline{\overline{Z}}^{\vec{J}_{\mathrm{A\rightleftharpoons A}}^{12}\vec{J}_{\mathrm{A}}^{2}} & 0 & \overline{\overline{Z}}^{\vec{J}_{\mathrm{A\rightleftharpoons A}}^{12}\vec{M}_{\mathrm{M\rightleftharpoons A}}^{1}} & \overline{\overline{Z}}^{\vec{J}_{\mathrm{A\rightleftharpoons A}}^{12}\vec{M}_{\mathrm{M\rightleftharpoons A}}^{2}} & \overline{\overline{Z}}^{\vec{J}_{\mathrm{A\rightleftharpoons A}}^{12}\vec{M}_{\mathrm{A\rightleftharpoons A}}^{12}} & \overline{\overline{Z}}^{\vec{J}_{\mathrm{A\rightleftharpoons A}}^{12}\vec{M}_{\mathrm{A\rightleftharpoons G}}} \\
0 & \overline{\overline{Z}}^{\vec{M}_{\mathrm{A\rightleftharpoons A}}^{12}\vec{J}_{\mathrm{M\rightleftharpoons A}}^{1}} & \overline{\overline{Z}}^{\vec{M}_{\mathrm{A\rightleftharpoons A}}^{12}\vec{J}_{\mathrm{M\rightleftharpoons A}}^{2}} & \overline{\overline{Z}}^{\vec{M}_{\mathrm{A\rightleftharpoons A}}^{12}\vec{J}_{\mathrm{A\rightleftharpoons A}}^{12}} & \overline{\overline{Z}}^{\vec{M}_{\mathrm{A\rightleftharpoons A}}^{12}\vec{J}_{\mathrm{A\rightleftharpoons G}}} & 0 & \overline{\overline{Z}}^{\vec{M}_{\mathrm{A\rightleftharpoons A}}^{12}\vec{J}_{\mathrm{A}}^{1}} & \overline{\overline{Z}}^{\vec{M}_{\mathrm{A\rightleftharpoons A}}^{12}\vec{J}_{\mathrm{A}}^{2}} & 0 & \overline{\overline{Z}}^{\vec{M}_{\mathrm{A\rightleftharpoons A}}^{12}\vec{M}_{\mathrm{M\rightleftharpoons A}}^{1}} & \overline{\overline{Z}}^{\vec{M}_{\mathrm{A\rightleftharpoons A}}^{12}\vec{M}_{\mathrm{M\rightleftharpoons A}}^{2}} & \overline{\overline{Z}}^{\vec{M}_{\mathrm{A\rightleftharpoons A}}^{12}\vec{M}_{\mathrm{A\rightleftharpoons A}}^{12}} & \overline{\overline{Z}}^{\vec{M}_{\mathrm{A\rightleftharpoons A}}^{12}\vec{M}_{\mathrm{A\rightleftharpoons G}}} \\
0 & 0 & \overline{\overline{Z}}^{\vec{J}_{\mathrm{A\rightleftharpoons G}}\vec{J}_{\mathrm{M\rightleftharpoons A}}^{2}} & \overline{\overline{Z}}^{\vec{J}_{\mathrm{A\rightleftharpoons G}}\vec{J}_{\mathrm{A\rightleftharpoons A}}^{12}} & \overline{\overline{Z}}^{\vec{J}_{\mathrm{A\rightleftharpoons G}}\vec{J}_{\mathrm{A\rightleftharpoons G}}} & 0 & 0 & \overline{\overline{Z}}^{\vec{J}_{\mathrm{A\rightleftharpoons G}}\vec{J}_{\mathrm{A}}^{2}} & 0 & 0 & \overline{\overline{Z}}^{\vec{J}_{\mathrm{A\rightleftharpoons G}}\vec{M}_{\mathrm{M\rightleftharpoons A}}^{2}} & \overline{\overline{Z}}^{\vec{J}_{\mathrm{A\rightleftharpoons G}}\vec{M}_{\mathrm{A\rightleftharpoons A}}^{12}} & \overline{\overline{Z}}^{\vec{J}_{\mathrm{A\rightleftharpoons G}}\vec{M}_{\mathrm{A\rightleftharpoons G}}} \\
0 & 0 & \overline{\overline{Z}}^{\vec{M}_{\mathrm{A\rightleftharpoons G}}\vec{J}_{\mathrm{M\rightleftharpoons A}}^{2}} & \overline{\overline{Z}}^{\vec{M}_{\mathrm{A\rightleftharpoons G}}\vec{J}_{\mathrm{A\rightleftharpoons A}}^{12}} & \overline{\overline{Z}}^{\vec{M}_{\mathrm{A\rightleftharpoons G}}\vec{J}_{\mathrm{A\rightleftharpoons G}}} & 0 & 0 & \overline{\overline{Z}}^{\vec{M}_{\mathrm{A\rightleftharpoons G}}\vec{J}_{\mathrm{A}}^{2}} & 0 & 0 & \overline{\overline{Z}}^{\vec{M}_{\mathrm{A\rightleftharpoons G}}\vec{M}_{\mathrm{M\rightleftharpoons A}}^{2}} & \overline{\overline{Z}}^{\vec{M}_{\mathrm{A\rightleftharpoons G}}\vec{M}_{\mathrm{A\rightleftharpoons A}}^{12}} & \overline{\overline{Z}}^{\vec{M}_{\mathrm{A\rightleftharpoons G}}\vec{M}_{\mathrm{A\rightleftharpoons G}}} \\
0 & \overline{\overline{Z}}^{\vec{J}_{\mathrm{M}}\vec{J}_{\mathrm{M\rightleftharpoons A}}^{1}} & \overline{\overline{Z}}^{\vec{J}_{\mathrm{M}}\vec{J}_{\mathrm{M\rightleftharpoons A}}^{2}} & 0 & 0 & \overline{\overline{Z}}^{\vec{J}_{\mathrm{M}}\vec{J}_{\mathrm{M}}} & 0 & 0 & \overline{\overline{Z}}^{\vec{J}_{\mathrm{M}}\vec{M}_{\mathrm{aux}}} & \overline{\overline{Z}}^{\vec{J}_{\mathrm{M}}\vec{M}_{\mathrm{M\rightleftharpoons A}}^{1}} & \overline{\overline{Z}}^{\vec{J}_{\mathrm{M}}\vec{M}_{\mathrm{M\rightleftharpoons A}}^{2}} & 0 & 0 \\
0 & \overline{\overline{Z}}^{\vec{J}_{\mathrm{A}}^{1}\vec{J}_{\mathrm{M\rightleftharpoons A}}^{1}} & 0 & \overline{\overline{Z}}^{\vec{J}_{\mathrm{A}}^{1}\vec{J}_{\mathrm{A\rightleftharpoons A}}^{12}} & 0 & 0 & \overline{\overline{Z}}^{\vec{J}_{\mathrm{A}}^{1}\vec{J}_{\mathrm{A}}^{1}} & 0 & 0 & \overline{\overline{Z}}^{\vec{J}_{\mathrm{A}}^{1}\vec{M}_{\mathrm{M\rightleftharpoons A}}^{1}} & 0 & \overline{\overline{Z}}^{\vec{J}_{\mathrm{A}}^{1}\vec{M}_{\mathrm{A\rightleftharpoons A}}^{12}} & 0 \\
0 & 0 & \overline{\overline{Z}}^{\vec{J}_{\mathrm{A}}^{2}\vec{J}_{\mathrm{M\rightleftharpoons A}}^{2}} & \overline{\overline{Z}}^{\vec{J}_{\mathrm{A}}^{2}\vec{J}_{\mathrm{A\rightleftharpoons A}}^{12}} & \overline{\overline{Z}}^{\vec{J}_{\mathrm{A}}^{2}\vec{J}_{\mathrm{A\rightleftharpoons G}}} & 0 & 0 & \overline{\overline{Z}}^{\vec{J}_{\mathrm{A}}^{2}\vec{J}_{\mathrm{A}}^{2}} & 0 & 0 & \overline{\overline{Z}}^{\vec{J}_{\mathrm{A}}^{2}\vec{M}_{\mathrm{M\rightleftharpoons A}}^{2}} & \overline{\overline{Z}}^{\vec{J}_{\mathrm{A}}^{2}\vec{M}_{\mathrm{A\rightleftharpoons A}}^{12}} & \overline{\overline{Z}}^{\vec{J}_{\mathrm{A}}^{2}\vec{M}_{\mathrm{A\rightleftharpoons G}}}
\end{bmatrix} \tag{D-142a}$$

$$\overline{\overline{\Psi}}_{2} = \begin{bmatrix}
\overline{\overline{I}}^{\vec{J}_{\mathrm{aux}}} \\
-\overline{\overline{Z}}^{\vec{M}_{\mathrm{aux}}\vec{J}_{\mathrm{aux}}} \\
-\overline{\overline{Z}}^{\vec{J}_{\mathrm{M\rightleftharpoons A}}^{1}\vec{J}_{\mathrm{aux}}} \\
-\overline{\overline{Z}}^{\vec{M}_{\mathrm{M\rightleftharpoons A}}^{1}\vec{J}_{\mathrm{aux}}} \\
-\overline{\overline{Z}}^{\vec{J}_{\mathrm{M\rightleftharpoons A}}^{2}\vec{J}_{\mathrm{aux}}} \\
-\overline{\overline{Z}}^{\vec{M}_{\mathrm{M\rightleftharpoons A}}^{2}\vec{J}_{\mathrm{aux}}} \\
0 \\
0 \\
0 \\
0 \\
-\overline{\overline{Z}}^{\vec{J}_{\mathrm{M}}\vec{J}_{\mathrm{aux}}} \\
0 \\
0
\end{bmatrix} \tag{D-142b}$$

and

$$\overline{\overline{\Psi}}_{3}=\begin{bmatrix}
0 & 0 & 0 & 0 & 0 & 0 & 0 & 0 & \overline{\overline{I}}^{\vec{M}_{\mathrm{aux}}} & 0 & 0 & 0 & 0\\
\overline{\overline{Z}}^{\vec{J}_{\mathrm{aux}}\vec{J}_{\mathrm{aux}}} & \overline{\overline{Z}}^{\vec{J}_{\mathrm{aux}}\vec{J}^{1}_{\mathrm{M}\rightleftharpoons\mathrm{A}}} & \overline{\overline{Z}}^{\vec{J}_{\mathrm{aux}}\vec{J}^{2}_{\mathrm{M}\rightleftharpoons\mathrm{A}}} & 0 & 0 & \overline{\overline{Z}}^{\vec{J}_{\mathrm{aux}}\vec{J}_{\mathrm{M}}} & 0 & 0 & 0 & \overline{\overline{Z}}^{\vec{J}_{\mathrm{aux}}\vec{M}^{1}_{\mathrm{M}\rightleftharpoons\mathrm{A}}} & \overline{\overline{Z}}^{\vec{J}_{\mathrm{aux}}\vec{M}^{2}_{\mathrm{M}\rightleftharpoons\mathrm{A}}} & 0 & 0\\
\overline{\overline{Z}}^{\vec{J}^{1}_{\mathrm{M}\rightleftharpoons\mathrm{A}}\vec{J}_{\mathrm{aux}}} & \overline{\overline{Z}}^{\vec{J}^{1}_{\mathrm{M}\rightleftharpoons\mathrm{A}}\vec{J}^{1}_{\mathrm{M}\rightleftharpoons\mathrm{A}}} & \overline{\overline{Z}}^{\vec{J}^{1}_{\mathrm{M}\rightleftharpoons\mathrm{A}}\vec{J}^{2}_{\mathrm{M}\rightleftharpoons\mathrm{A}}} & \overline{\overline{Z}}^{\vec{J}^{1}_{\mathrm{M}\rightleftharpoons\mathrm{A}}\vec{J}^{12}_{\mathrm{A}\rightleftharpoons\mathrm{A}}} & 0 & \overline{\overline{Z}}^{\vec{J}^{1}_{\mathrm{M}\rightleftharpoons\mathrm{A}}\vec{J}_{\mathrm{M}}} & \overline{\overline{Z}}^{\vec{J}^{1}_{\mathrm{M}\rightleftharpoons\mathrm{A}}\vec{J}^{1}_{\mathrm{A}}} & 0 & 0 & \overline{\overline{Z}}^{\vec{J}^{1}_{\mathrm{M}\rightleftharpoons\mathrm{A}}\vec{M}^{1}_{\mathrm{M}\rightleftharpoons\mathrm{A}}} & \overline{\overline{Z}}^{\vec{J}^{1}_{\mathrm{M}\rightleftharpoons\mathrm{A}}\vec{M}^{2}_{\mathrm{M}\rightleftharpoons\mathrm{A}}} & \overline{\overline{Z}}^{\vec{J}^{1}_{\mathrm{M}\rightleftharpoons\mathrm{A}}\vec{M}^{12}_{\mathrm{A}\rightleftharpoons\mathrm{A}}} & 0\\
\overline{\overline{Z}}^{\vec{M}^{1}_{\mathrm{M}\rightleftharpoons\mathrm{A}}\vec{J}_{\mathrm{aux}}} & \overline{\overline{Z}}^{\vec{M}^{1}_{\mathrm{M}\rightleftharpoons\mathrm{A}}\vec{J}^{1}_{\mathrm{M}\rightleftharpoons\mathrm{A}}} & \overline{\overline{Z}}^{\vec{M}^{1}_{\mathrm{M}\rightleftharpoons\mathrm{A}}\vec{J}^{2}_{\mathrm{M}\rightleftharpoons\mathrm{A}}} & \overline{\overline{Z}}^{\vec{M}^{1}_{\mathrm{M}\rightleftharpoons\mathrm{A}}\vec{J}^{12}_{\mathrm{A}\rightleftharpoons\mathrm{A}}} & 0 & \overline{\overline{Z}}^{\vec{M}^{1}_{\mathrm{M}\rightleftharpoons\mathrm{A}}\vec{J}_{\mathrm{M}}} & \overline{\overline{Z}}^{\vec{M}^{1}_{\mathrm{M}\rightleftharpoons\mathrm{A}}\vec{J}^{1}_{\mathrm{A}}} & 0 & 0 & \overline{\overline{Z}}^{\vec{M}^{1}_{\mathrm{M}\rightleftharpoons\mathrm{A}}\vec{M}^{1}_{\mathrm{M}\rightleftharpoons\mathrm{A}}} & \overline{\overline{Z}}^{\vec{M}^{1}_{\mathrm{M}\rightleftharpoons\mathrm{A}}\vec{M}^{2}_{\mathrm{M}\rightleftharpoons\mathrm{A}}} & \overline{\overline{Z}}^{\vec{M}^{1}_{\mathrm{M}\rightleftharpoons\mathrm{A}}\vec{M}^{12}_{\mathrm{A}\rightleftharpoons\mathrm{A}}} & 0\\
\overline{\overline{Z}}^{\vec{J}^{2}_{\mathrm{M}\rightleftharpoons\mathrm{A}}\vec{J}_{\mathrm{aux}}} & \overline{\overline{Z}}^{\vec{J}^{2}_{\mathrm{M}\rightleftharpoons\mathrm{A}}\vec{J}^{1}_{\mathrm{M}\rightleftharpoons\mathrm{A}}} & \overline{\overline{Z}}^{\vec{J}^{2}_{\mathrm{M}\rightleftharpoons\mathrm{A}}\vec{J}^{2}_{\mathrm{M}\rightleftharpoons\mathrm{A}}} & \overline{\overline{Z}}^{\vec{J}^{2}_{\mathrm{M}\rightleftharpoons\mathrm{A}}\vec{J}^{12}_{\mathrm{A}\rightleftharpoons\mathrm{A}}} & \overline{\overline{Z}}^{\vec{J}^{2}_{\mathrm{M}\rightleftharpoons\mathrm{A}}\vec{J}_{\mathrm{A}\rightleftharpoons\mathrm{G}}} & \overline{\overline{Z}}^{\vec{J}^{2}_{\mathrm{M}\rightleftharpoons\mathrm{A}}\vec{J}_{\mathrm{M}}} & 0 & \overline{\overline{Z}}^{\vec{J}^{2}_{\mathrm{M}\rightleftharpoons\mathrm{A}}\vec{J}^{2}_{\mathrm{A}}} & 0 & \overline{\overline{Z}}^{\vec{J}^{2}_{\mathrm{M}\rightleftharpoons\mathrm{A}}\vec{M}^{1}_{\mathrm{M}\rightleftharpoons\mathrm{A}}} & \overline{\overline{Z}}^{\vec{J}^{2}_{\mathrm{M}\rightleftharpoons\mathrm{A}}\vec{M}^{2}_{\mathrm{M}\rightleftharpoons\mathrm{A}}} & \overline{\overline{Z}}^{\vec{J}^{2}_{\mathrm{M}\rightleftharpoons\mathrm{A}}\vec{M}^{12}_{\mathrm{A}\rightleftharpoons\mathrm{A}}} & \overline{\overline{Z}}^{\vec{J}^{2}_{\mathrm{M}\rightleftharpoons\mathrm{A}}\vec{M}_{\mathrm{A}\rightleftharpoons\mathrm{G}}}\\
\overline{\overline{Z}}^{\vec{M}^{2}_{\mathrm{M}\rightleftharpoons\mathrm{A}}\vec{J}_{\mathrm{aux}}} & \overline{\overline{Z}}^{\vec{M}^{2}_{\mathrm{M}\rightleftharpoons\mathrm{A}}\vec{J}^{1}_{\mathrm{M}\rightleftharpoons\mathrm{A}}} & \overline{\overline{Z}}^{\vec{M}^{2}_{\mathrm{M}\rightleftharpoons\mathrm{A}}\vec{J}^{2}_{\mathrm{M}\rightleftharpoons\mathrm{A}}} & \overline{\overline{Z}}^{\vec{M}^{2}_{\mathrm{M}\rightleftharpoons\mathrm{A}}\vec{J}^{12}_{\mathrm{A}\rightleftharpoons\mathrm{A}}} & \overline{\overline{Z}}^{\vec{M}^{2}_{\mathrm{M}\rightleftharpoons\mathrm{A}}\vec{J}_{\mathrm{A}\rightleftharpoons\mathrm{G}}} & \overline{\overline{Z}}^{\vec{M}^{2}_{\mathrm{M}\rightleftharpoons\mathrm{A}}\vec{J}_{\mathrm{M}}} & 0 & \overline{\overline{Z}}^{\vec{M}^{2}_{\mathrm{M}\rightleftharpoons\mathrm{A}}\vec{J}^{2}_{\mathrm{A}}} & 0 & \overline{\overline{Z}}^{\vec{M}^{2}_{\mathrm{M}\rightleftharpoons\mathrm{A}}\vec{M}^{1}_{\mathrm{M}\rightleftharpoons\mathrm{A}}} & \overline{\overline{Z}}^{\vec{M}^{2}_{\mathrm{M}\rightleftharpoons\mathrm{A}}\vec{M}^{2}_{\mathrm{M}\rightleftharpoons\mathrm{A}}} & \overline{\overline{Z}}^{\vec{M}^{2}_{\mathrm{M}\rightleftharpoons\mathrm{A}}\vec{M}^{12}_{\mathrm{A}\rightleftharpoons\mathrm{A}}} & \overline{\overline{Z}}^{\vec{M}^{2}_{\mathrm{M}\rightleftharpoons\mathrm{A}}\vec{M}_{\mathrm{A}\rightleftharpoons\mathrm{G}}}\\
0 & \overline{\overline{Z}}^{\vec{J}^{12}_{\mathrm{A}\rightleftharpoons\mathrm{A}}\vec{J}^{1}_{\mathrm{M}\rightleftharpoons\mathrm{A}}} & \overline{\overline{Z}}^{\vec{J}^{12}_{\mathrm{A}\rightleftharpoons\mathrm{A}}\vec{J}^{2}_{\mathrm{M}\rightleftharpoons\mathrm{A}}} & \overline{\overline{Z}}^{\vec{J}^{12}_{\mathrm{A}\rightleftharpoons\mathrm{A}}\vec{J}^{12}_{\mathrm{A}\rightleftharpoons\mathrm{A}}} & \overline{\overline{Z}}^{\vec{J}^{12}_{\mathrm{A}\rightleftharpoons\mathrm{A}}\vec{J}_{\mathrm{A}\rightleftharpoons\mathrm{G}}} & 0 & \overline{\overline{Z}}^{\vec{J}^{12}_{\mathrm{A}\rightleftharpoons\mathrm{A}}\vec{J}^{1}_{\mathrm{A}}} & \overline{\overline{Z}}^{\vec{J}^{12}_{\mathrm{A}\rightleftharpoons\mathrm{A}}\vec{J}^{2}_{\mathrm{A}}} & 0 & \overline{\overline{Z}}^{\vec{J}^{12}_{\mathrm{A}\rightleftharpoons\mathrm{A}}\vec{M}^{1}_{\mathrm{M}\rightleftharpoons\mathrm{A}}} & \overline{\overline{Z}}^{\vec{J}^{12}_{\mathrm{A}\rightleftharpoons\mathrm{A}}\vec{M}^{2}_{\mathrm{M}\rightleftharpoons\mathrm{A}}} & \overline{\overline{Z}}^{\vec{J}^{12}_{\mathrm{A}\rightleftharpoons\mathrm{A}}\vec{M}^{12}_{\mathrm{A}\rightleftharpoons\mathrm{A}}} & \overline{\overline{Z}}^{\vec{J}^{12}_{\mathrm{A}\rightleftharpoons\mathrm{A}}\vec{M}_{\mathrm{A}\rightleftharpoons\mathrm{G}}}\\
0 & \overline{\overline{Z}}^{\vec{M}^{12}_{\mathrm{A}\rightleftharpoons\mathrm{A}}\vec{J}^{1}_{\mathrm{M}\rightleftharpoons\mathrm{A}}} & \overline{\overline{Z}}^{\vec{M}^{12}_{\mathrm{A}\rightleftharpoons\mathrm{A}}\vec{J}^{2}_{\mathrm{M}\rightleftharpoons\mathrm{A}}} & \overline{\overline{Z}}^{\vec{M}^{12}_{\mathrm{A}\rightleftharpoons\mathrm{A}}\vec{J}^{12}_{\mathrm{A}\rightleftharpoons\mathrm{A}}} & \overline{\overline{Z}}^{\vec{M}^{12}_{\mathrm{A}\rightleftharpoons\mathrm{A}}\vec{J}_{\mathrm{A}\rightleftharpoons\mathrm{G}}} & 0 & \overline{\overline{Z}}^{\vec{M}^{12}_{\mathrm{A}\rightleftharpoons\mathrm{A}}\vec{J}^{1}_{\mathrm{A}}} & \overline{\overline{Z}}^{\vec{M}^{12}_{\mathrm{A}\rightleftharpoons\mathrm{A}}\vec{J}^{2}_{\mathrm{A}}} & 0 & \overline{\overline{Z}}^{\vec{M}^{12}_{\mathrm{A}\rightleftharpoons\mathrm{A}}\vec{M}^{1}_{\mathrm{M}\rightleftharpoons\mathrm{A}}} & \overline{\overline{Z}}^{\vec{M}^{12}_{\mathrm{A}\rightleftharpoons\mathrm{A}}\vec{M}^{2}_{\mathrm{M}\rightleftharpoons\mathrm{A}}} & \overline{\overline{Z}}^{\vec{M}^{12}_{\mathrm{A}\rightleftharpoons\mathrm{A}}\vec{M}^{12}_{\mathrm{A}\rightleftharpoons\mathrm{A}}} & \overline{\overline{Z}}^{\vec{M}^{12}_{\mathrm{A}\rightleftharpoons\mathrm{A}}\vec{M}_{\mathrm{A}\rightleftharpoons\mathrm{G}}}\\
0 & 0 & \overline{\overline{Z}}^{\vec{J}_{\mathrm{A}\rightleftharpoons\mathrm{G}}\vec{J}^{2}_{\mathrm{M}\rightleftharpoons\mathrm{A}}} & \overline{\overline{Z}}^{\vec{J}_{\mathrm{A}\rightleftharpoons\mathrm{G}}\vec{J}^{12}_{\mathrm{A}\rightleftharpoons\mathrm{A}}} & \overline{\overline{Z}}^{\vec{J}_{\mathrm{A}\rightleftharpoons\mathrm{G}}\vec{J}_{\mathrm{A}\rightleftharpoons\mathrm{G}}} & 0 & 0 & \overline{\overline{Z}}^{\vec{J}_{\mathrm{A}\rightleftharpoons\mathrm{G}}\vec{J}^{2}_{\mathrm{A}}} & 0 & 0 & \overline{\overline{Z}}^{\vec{J}_{\mathrm{A}\rightleftharpoons\mathrm{G}}\vec{M}^{2}_{\mathrm{M}\rightleftharpoons\mathrm{A}}} & \overline{\overline{Z}}^{\vec{J}_{\mathrm{A}\rightleftharpoons\mathrm{G}}\vec{M}^{12}_{\mathrm{A}\rightleftharpoons\mathrm{A}}} & \overline{\overline{Z}}^{\vec{J}_{\mathrm{A}\rightleftharpoons\mathrm{G}}\vec{M}_{\mathrm{A}\rightleftharpoons\mathrm{G}}}\\
0 & 0 & \overline{\overline{Z}}^{\vec{M}_{\mathrm{A}\rightleftharpoons\mathrm{G}}\vec{J}^{2}_{\mathrm{M}\rightleftharpoons\mathrm{A}}} & \overline{\overline{Z}}^{\vec{M}_{\mathrm{A}\rightleftharpoons\mathrm{G}}\vec{J}^{12}_{\mathrm{A}\rightleftharpoons\mathrm{A}}} & \overline{\overline{Z}}^{\vec{M}_{\mathrm{A}\rightleftharpoons\mathrm{G}}\vec{J}_{\mathrm{A}\rightleftharpoons\mathrm{G}}} & 0 & 0 & \overline{\overline{Z}}^{\vec{M}_{\mathrm{A}\rightleftharpoons\mathrm{G}}\vec{J}^{2}_{\mathrm{A}}} & 0 & 0 & \overline{\overline{Z}}^{\vec{M}_{\mathrm{A}\rightleftharpoons\mathrm{G}}\vec{M}^{2}_{\mathrm{M}\rightleftharpoons\mathrm{A}}} & \overline{\overline{Z}}^{\vec{M}_{\mathrm{A}\rightleftharpoons\mathrm{G}}\vec{M}^{12}_{\mathrm{A}\rightleftharpoons\mathrm{A}}} & \overline{\overline{Z}}^{\vec{M}_{\mathrm{A}\rightleftharpoons\mathrm{G}}\vec{M}_{\mathrm{A}\rightleftharpoons\mathrm{G}}}\\
\overline{\overline{Z}}^{\vec{J}_{\mathrm{M}}\vec{J}_{\mathrm{aux}}} & \overline{\overline{Z}}^{\vec{J}_{\mathrm{M}}\vec{J}^{1}_{\mathrm{M}\rightleftharpoons\mathrm{A}}} & \overline{\overline{Z}}^{\vec{J}_{\mathrm{M}}\vec{J}^{2}_{\mathrm{M}\rightleftharpoons\mathrm{A}}} & 0 & 0 & \overline{\overline{Z}}^{\vec{J}_{\mathrm{M}}\vec{J}_{\mathrm{M}}} & 0 & 0 & 0 & \overline{\overline{Z}}^{\vec{J}_{\mathrm{M}}\vec{M}^{1}_{\mathrm{M}\rightleftharpoons\mathrm{A}}} & \overline{\overline{Z}}^{\vec{J}_{\mathrm{M}}\vec{M}^{2}_{\mathrm{M}\rightleftharpoons\mathrm{A}}} & 0 & 0\\
0 & \overline{\overline{Z}}^{\vec{J}^{1}_{\mathrm{A}}\vec{J}^{1}_{\mathrm{M}\rightleftharpoons\mathrm{A}}} & 0 & \overline{\overline{Z}}^{\vec{J}^{1}_{\mathrm{A}}\vec{J}^{12}_{\mathrm{A}\rightleftharpoons\mathrm{A}}} & 0 & 0 & \overline{\overline{Z}}^{\vec{J}^{1}_{\mathrm{A}}\vec{J}^{1}_{\mathrm{A}}} & 0 & 0 & \overline{\overline{Z}}^{\vec{J}^{1}_{\mathrm{A}}\vec{M}^{1}_{\mathrm{M}\rightleftharpoons\mathrm{A}}} & 0 & \overline{\overline{Z}}^{\vec{J}^{1}_{\mathrm{A}}\vec{M}^{12}_{\mathrm{A}\rightleftharpoons\mathrm{A}}} & 0\\
0 & 0 & \overline{\overline{Z}}^{\vec{J}^{2}_{\mathrm{A}}\vec{J}^{2}_{\mathrm{M}\rightleftharpoons\mathrm{A}}} & \overline{\overline{Z}}^{\vec{J}^{2}_{\mathrm{A}}\vec{J}^{12}_{\mathrm{A}\rightleftharpoons\mathrm{A}}} & \overline{\overline{Z}}^{\vec{J}^{2}_{\mathrm{A}}\vec{J}_{\mathrm{A}\rightleftharpoons\mathrm{G}}} & 0 & 0 & \overline{\overline{Z}}^{\vec{J}^{2}_{\mathrm{A}}\vec{J}^{2}_{\mathrm{A}}} & 0 & 0 & \overline{\overline{Z}}^{\vec{J}^{2}_{\mathrm{A}}\vec{M}^{2}_{\mathrm{M}\rightleftharpoons\mathrm{A}}} & \overline{\overline{Z}}^{\vec{J}^{2}_{\mathrm{A}}\vec{M}^{12}_{\mathrm{A}\rightleftharpoons\mathrm{A}}} & \overline{\overline{Z}}^{\vec{J}^{2}_{\mathrm{A}}\vec{M}_{\mathrm{A}\rightleftharpoons\mathrm{G}}}
\end{bmatrix} \tag{D-143a}$$

$$\overline{\overline{\Psi}}_{4}=\begin{bmatrix}
\overline{\overline{I}}^{\vec{M}_{\mathrm{aux}}}\\
-\overline{\overline{Z}}^{\vec{J}_{\mathrm{aux}}\vec{M}_{\mathrm{aux}}}\\
-\overline{\overline{Z}}^{\vec{J}^{1}_{\mathrm{M}\rightleftharpoons\mathrm{A}}\vec{M}_{\mathrm{aux}}}\\
-\overline{\overline{Z}}^{\vec{M}^{1}_{\mathrm{M}\rightleftharpoons\mathrm{A}}\vec{M}_{\mathrm{aux}}}\\
-\overline{\overline{Z}}^{\vec{J}^{2}_{\mathrm{M}\rightleftharpoons\mathrm{A}}\vec{M}_{\mathrm{aux}}}\\
-\overline{\overline{Z}}^{\vec{M}^{2}_{\mathrm{M}\rightleftharpoons\mathrm{A}}\vec{M}_{\mathrm{aux}}}\\
0\\
0\\
0\\
0\\
-\overline{\overline{Z}}^{\vec{J}_{\mathrm{M}}\vec{M}_{\mathrm{aux}}}\\
0\\
0
\end{bmatrix} \tag{D-143b}$$

and

$$\overline{\overline{\Psi}}^{\mathrm{DoJ}}_{\mathrm{FCE}}=\begin{bmatrix}
\overline{\overline{Z}}^{\vec{M}_{\mathrm{aux}}\vec{J}_{\mathrm{aux}}} & \overline{\overline{Z}}^{\vec{M}_{\mathrm{aux}}\vec{J}^{1}_{\mathrm{M}\rightleftharpoons\mathrm{A}}} & \overline{\overline{Z}}^{\vec{M}_{\mathrm{aux}}\vec{J}^{2}_{\mathrm{M}\rightleftharpoons\mathrm{A}}} & 0 & 0 & \overline{\overline{Z}}^{\vec{M}_{\mathrm{aux}}\vec{J}_{\mathrm{M}}} & 0 & 0 & \overline{\overline{Z}}^{\vec{M}_{\mathrm{aux}}\vec{M}_{\mathrm{aux}}} & \overline{\overline{Z}}^{\vec{M}_{\mathrm{aux}}\vec{M}^{1}_{\mathrm{M}\rightleftharpoons\mathrm{A}}} & \overline{\overline{Z}}^{\vec{M}_{\mathrm{aux}}\vec{M}^{2}_{\mathrm{M}\rightleftharpoons\mathrm{A}}} & 0 & 0\\
\overline{\overline{Z}}^{\vec{J}^{1}_{\mathrm{M}\rightleftharpoons\mathrm{A}}\vec{J}_{\mathrm{aux}}} & \overline{\overline{Z}}^{\vec{J}^{1}_{\mathrm{M}\rightleftharpoons\mathrm{A}}\vec{J}^{1}_{\mathrm{M}\rightleftharpoons\mathrm{A}}} & \overline{\overline{Z}}^{\vec{J}^{1}_{\mathrm{M}\rightleftharpoons\mathrm{A}}\vec{J}^{2}_{\mathrm{M}\rightleftharpoons\mathrm{A}}} & \overline{\overline{Z}}^{\vec{J}^{1}_{\mathrm{M}\rightleftharpoons\mathrm{A}}\vec{J}^{12}_{\mathrm{A}\rightleftharpoons\mathrm{A}}} & 0 & \overline{\overline{Z}}^{\vec{J}^{1}_{\mathrm{M}\rightleftharpoons\mathrm{A}}\vec{J}_{\mathrm{M}}} & \overline{\overline{Z}}^{\vec{J}^{1}_{\mathrm{M}\rightleftharpoons\mathrm{A}}\vec{J}^{1}_{\mathrm{A}}} & 0 & \overline{\overline{Z}}^{\vec{J}^{1}_{\mathrm{M}\rightleftharpoons\mathrm{A}}\vec{M}_{\mathrm{aux}}} & \overline{\overline{Z}}^{\vec{J}^{1}_{\mathrm{M}\rightleftharpoons\mathrm{A}}\vec{M}^{1}_{\mathrm{M}\rightleftharpoons\mathrm{A}}} & \overline{\overline{Z}}^{\vec{J}^{1}_{\mathrm{M}\rightleftharpoons\mathrm{A}}\vec{M}^{2}_{\mathrm{M}\rightleftharpoons\mathrm{A}}} & \overline{\overline{Z}}^{\vec{J}^{1}_{\mathrm{M}\rightleftharpoons\mathrm{A}}\vec{M}^{12}_{\mathrm{A}\rightleftharpoons\mathrm{A}}} & 0\\
\overline{\overline{Z}}^{\vec{M}^{1}_{\mathrm{M}\rightleftharpoons\mathrm{A}}\vec{J}_{\mathrm{aux}}} & \overline{\overline{Z}}^{\vec{M}^{1}_{\mathrm{M}\rightleftharpoons\mathrm{A}}\vec{J}^{1}_{\mathrm{M}\rightleftharpoons\mathrm{A}}} & \overline{\overline{Z}}^{\vec{M}^{1}_{\mathrm{M}\rightleftharpoons\mathrm{A}}\vec{J}^{2}_{\mathrm{M}\rightleftharpoons\mathrm{A}}} & \overline{\overline{Z}}^{\vec{M}^{1}_{\mathrm{M}\rightleftharpoons\mathrm{A}}\vec{J}^{12}_{\mathrm{A}\rightleftharpoons\mathrm{A}}} & 0 & \overline{\overline{Z}}^{\vec{M}^{1}_{\mathrm{M}\rightleftharpoons\mathrm{A}}\vec{J}_{\mathrm{M}}} & \overline{\overline{Z}}^{\vec{M}^{1}_{\mathrm{M}\rightleftharpoons\mathrm{A}}\vec{J}^{1}_{\mathrm{A}}} & 0 & \overline{\overline{Z}}^{\vec{M}^{1}_{\mathrm{M}\rightleftharpoons\mathrm{A}}\vec{M}_{\mathrm{aux}}} & \overline{\overline{Z}}^{\vec{M}^{1}_{\mathrm{M}\rightleftharpoons\mathrm{A}}\vec{M}^{1}_{\mathrm{M}\rightleftharpoons\mathrm{A}}} & \overline{\overline{Z}}^{\vec{M}^{1}_{\mathrm{M}\rightleftharpoons\mathrm{A}}\vec{M}^{2}_{\mathrm{M}\rightleftharpoons\mathrm{A}}} & \overline{\overline{Z}}^{\vec{M}^{1}_{\mathrm{M}\rightleftharpoons\mathrm{A}}\vec{M}^{12}_{\mathrm{A}\rightleftharpoons\mathrm{A}}} & 0\\
\overline{\overline{Z}}^{\vec{J}^{2}_{\mathrm{M}\rightleftharpoons\mathrm{A}}\vec{J}_{\mathrm{aux}}} & \overline{\overline{Z}}^{\vec{J}^{2}_{\mathrm{M}\rightleftharpoons\mathrm{A}}\vec{J}^{1}_{\mathrm{M}\rightleftharpoons\mathrm{A}}} & \overline{\overline{Z}}^{\vec{J}^{2}_{\mathrm{M}\rightleftharpoons\mathrm{A}}\vec{J}^{2}_{\mathrm{M}\rightleftharpoons\mathrm{A}}} & \overline{\overline{Z}}^{\vec{J}^{2}_{\mathrm{M}\rightleftharpoons\mathrm{A}}\vec{J}^{12}_{\mathrm{A}\rightleftharpoons\mathrm{A}}} & \overline{\overline{Z}}^{\vec{J}^{2}_{\mathrm{M}\rightleftharpoons\mathrm{A}}\vec{J}_{\mathrm{A}\rightleftharpoons\mathrm{G}}} & \overline{\overline{Z}}^{\vec{J}^{2}_{\mathrm{M}\rightleftharpoons\mathrm{A}}\vec{J}_{\mathrm{M}}} & 0 & \overline{\overline{Z}}^{\vec{J}^{2}_{\mathrm{M}\rightleftharpoons\mathrm{A}}\vec{J}^{2}_{\mathrm{A}}} & \overline{\overline{Z}}^{\vec{J}^{2}_{\mathrm{M}\rightleftharpoons\mathrm{A}}\vec{M}_{\mathrm{aux}}} & \overline{\overline{Z}}^{\vec{J}^{2}_{\mathrm{M}\rightleftharpoons\mathrm{A}}\vec{M}^{1}_{\mathrm{M}\rightleftharpoons\mathrm{A}}} & \overline{\overline{Z}}^{\vec{J}^{2}_{\mathrm{M}\rightleftharpoons\mathrm{A}}\vec{M}^{2}_{\mathrm{M}\rightleftharpoons\mathrm{A}}} & \overline{\overline{Z}}^{\vec{J}^{2}_{\mathrm{M}\rightleftharpoons\mathrm{A}}\vec{M}^{12}_{\mathrm{A}\rightleftharpoons\mathrm{A}}} & \overline{\overline{Z}}^{\vec{J}^{2}_{\mathrm{M}\rightleftharpoons\mathrm{A}}\vec{M}_{\mathrm{A}\rightleftharpoons\mathrm{G}}}\\
\overline{\overline{Z}}^{\vec{M}^{2}_{\mathrm{M}\rightleftharpoons\mathrm{A}}\vec{J}_{\mathrm{aux}}} & \overline{\overline{Z}}^{\vec{M}^{2}_{\mathrm{M}\rightleftharpoons\mathrm{A}}\vec{J}^{1}_{\mathrm{M}\rightleftharpoons\mathrm{A}}} & \overline{\overline{Z}}^{\vec{M}^{2}_{\mathrm{M}\rightleftharpoons\mathrm{A}}\vec{J}^{2}_{\mathrm{M}\rightleftharpoons\mathrm{A}}} & \overline{\overline{Z}}^{\vec{M}^{2}_{\mathrm{M}\rightleftharpoons\mathrm{A}}\vec{J}^{12}_{\mathrm{A}\rightleftharpoons\mathrm{A}}} & \overline{\overline{Z}}^{\vec{M}^{2}_{\mathrm{M}\rightleftharpoons\mathrm{A}}\vec{J}_{\mathrm{A}\rightleftharpoons\mathrm{G}}} & \overline{\overline{Z}}^{\vec{M}^{2}_{\mathrm{M}\rightleftharpoons\mathrm{A}}\vec{J}_{\mathrm{M}}} & 0 & \overline{\overline{Z}}^{\vec{M}^{2}_{\mathrm{M}\rightleftharpoons\mathrm{A}}\vec{J}^{2}_{\mathrm{A}}} & \overline{\overline{Z}}^{\vec{M}^{2}_{\mathrm{M}\rightleftharpoons\mathrm{A}}\vec{M}_{\mathrm{aux}}} & \overline{\overline{Z}}^{\vec{M}^{2}_{\mathrm{M}\rightleftharpoons\mathrm{A}}\vec{M}^{1}_{\mathrm{M}\rightleftharpoons\mathrm{A}}} & \overline{\overline{Z}}^{\vec{M}^{2}_{\mathrm{M}\rightleftharpoons\mathrm{A}}\vec{M}^{2}_{\mathrm{M}\rightleftharpoons\mathrm{A}}} & \overline{\overline{Z}}^{\vec{M}^{2}_{\mathrm{M}\rightleftharpoons\mathrm{A}}\vec{M}^{12}_{\mathrm{A}\rightleftharpoons\mathrm{A}}} & \overline{\overline{Z}}^{\vec{M}^{2}_{\mathrm{M}\rightleftharpoons\mathrm{A}}\vec{M}_{\mathrm{A}\rightleftharpoons\mathrm{G}}}\\
0 & \overline{\overline{Z}}^{\vec{J}^{12}_{\mathrm{A}\rightleftharpoons\mathrm{A}}\vec{J}^{1}_{\mathrm{M}\rightleftharpoons\mathrm{A}}} & \overline{\overline{Z}}^{\vec{J}^{12}_{\mathrm{A}\rightleftharpoons\mathrm{A}}\vec{J}^{2}_{\mathrm{M}\rightleftharpoons\mathrm{A}}} & \overline{\overline{Z}}^{\vec{J}^{12}_{\mathrm{A}\rightleftharpoons\mathrm{A}}\vec{J}^{12}_{\mathrm{A}\rightleftharpoons\mathrm{A}}} & \overline{\overline{Z}}^{\vec{J}^{12}_{\mathrm{A}\rightleftharpoons\mathrm{A}}\vec{J}_{\mathrm{A}\rightleftharpoons\mathrm{G}}} & 0 & \overline{\overline{Z}}^{\vec{J}^{12}_{\mathrm{A}\rightleftharpoons\mathrm{A}}\vec{J}^{1}_{\mathrm{A}}} & \overline{\overline{Z}}^{\vec{J}^{12}_{\mathrm{A}\rightleftharpoons\mathrm{A}}\vec{J}^{2}_{\mathrm{A}}} & 0 & \overline{\overline{Z}}^{\vec{J}^{12}_{\mathrm{A}\rightleftharpoons\mathrm{A}}\vec{M}^{1}_{\mathrm{M}\rightleftharpoons\mathrm{A}}} & \overline{\overline{Z}}^{\vec{J}^{12}_{\mathrm{A}\rightleftharpoons\mathrm{A}}\vec{M}^{2}_{\mathrm{M}\rightleftharpoons\mathrm{A}}} & \overline{\overline{Z}}^{\vec{J}^{12}_{\mathrm{A}\rightleftharpoons\mathrm{A}}\vec{M}^{12}_{\mathrm{A}\rightleftharpoons\mathrm{A}}} & \overline{\overline{Z}}^{\vec{J}^{12}_{\mathrm{A}\rightleftharpoons\mathrm{A}}\vec{M}_{\mathrm{A}\rightleftharpoons\mathrm{G}}}\\
0 & \overline{\overline{Z}}^{\vec{M}^{12}_{\mathrm{A}\rightleftharpoons\mathrm{A}}\vec{J}^{1}_{\mathrm{M}\rightleftharpoons\mathrm{A}}} & \overline{\overline{Z}}^{\vec{M}^{12}_{\mathrm{A}\rightleftharpoons\mathrm{A}}\vec{J}^{2}_{\mathrm{M}\rightleftharpoons\mathrm{A}}} & \overline{\overline{Z}}^{\vec{M}^{12}_{\mathrm{A}\rightleftharpoons\mathrm{A}}\vec{J}^{12}_{\mathrm{A}\rightleftharpoons\mathrm{A}}} & \overline{\overline{Z}}^{\vec{M}^{12}_{\mathrm{A}\rightleftharpoons\mathrm{A}}\vec{J}_{\mathrm{A}\rightleftharpoons\mathrm{G}}} & 0 & \overline{\overline{Z}}^{\vec{M}^{12}_{\mathrm{A}\rightleftharpoons\mathrm{A}}\vec{J}^{1}_{\mathrm{A}}} & \overline{\overline{Z}}^{\vec{M}^{12}_{\mathrm{A}\rightleftharpoons\mathrm{A}}\vec{J}^{2}_{\mathrm{A}}} & 0 & \overline{\overline{Z}}^{\vec{M}^{12}_{\mathrm{A}\rightleftharpoons\mathrm{A}}\vec{M}^{1}_{\mathrm{M}\rightleftharpoons\mathrm{A}}} & \overline{\overline{Z}}^{\vec{M}^{12}_{\mathrm{A}\rightleftharpoons\mathrm{A}}\vec{M}^{2}_{\mathrm{M}\rightleftharpoons\mathrm{A}}} & \overline{\overline{Z}}^{\vec{M}^{12}_{\mathrm{A}\rightleftharpoons\mathrm{A}}\vec{M}^{12}_{\mathrm{A}\rightleftharpoons\mathrm{A}}} & \overline{\overline{Z}}^{\vec{M}^{12}_{\mathrm{A}\rightleftharpoons\mathrm{A}}\vec{M}_{\mathrm{A}\rightleftharpoons\mathrm{G}}}\\
0 & 0 & \overline{\overline{Z}}^{\vec{J}_{\mathrm{A}\rightleftharpoons\mathrm{G}}\vec{J}^{2}_{\mathrm{M}\rightleftharpoons\mathrm{A}}} & \overline{\overline{Z}}^{\vec{J}_{\mathrm{A}\rightleftharpoons\mathrm{G}}\vec{J}^{12}_{\mathrm{A}\rightleftharpoons\mathrm{A}}} & \overline{\overline{Z}}^{\vec{J}_{\mathrm{A}\rightleftharpoons\mathrm{G}}\vec{J}_{\mathrm{A}\rightleftharpoons\mathrm{G}}} & 0 & 0 & \overline{\overline{Z}}^{\vec{J}_{\mathrm{A}\rightleftharpoons\mathrm{G}}\vec{J}^{2}_{\mathrm{A}}} & 0 & 0 & \overline{\overline{Z}}^{\vec{J}_{\mathrm{A}\rightleftharpoons\mathrm{G}}\vec{M}^{2}_{\mathrm{M}\rightleftharpoons\mathrm{A}}} & \overline{\overline{Z}}^{\vec{J}_{\mathrm{A}\rightleftharpoons\mathrm{G}}\vec{M}^{12}_{\mathrm{A}\rightleftharpoons\mathrm{A}}} & \overline{\overline{Z}}^{\vec{J}_{\mathrm{A}\rightleftharpoons\mathrm{G}}\vec{M}_{\mathrm{A}\rightleftharpoons\mathrm{G}}}\\
0 & 0 & \overline{\overline{Z}}^{\vec{M}_{\mathrm{A}\rightleftharpoons\mathrm{G}}\vec{J}^{2}_{\mathrm{M}\rightleftharpoons\mathrm{A}}} & \overline{\overline{Z}}^{\vec{M}_{\mathrm{A}\rightleftharpoons\mathrm{G}}\vec{J}^{12}_{\mathrm{A}\rightleftharpoons\mathrm{A}}} & \overline{\overline{Z}}^{\vec{M}_{\mathrm{A}\rightleftharpoons\mathrm{G}}\vec{J}_{\mathrm{A}\rightleftharpoons\mathrm{G}}} & 0 & 0 & \overline{\overline{Z}}^{\vec{M}_{\mathrm{A}\rightleftharpoons\mathrm{G}}\vec{J}^{2}_{\mathrm{A}}} & 0 & 0 & \overline{\overline{Z}}^{\vec{M}_{\mathrm{A}\rightleftharpoons\mathrm{G}}\vec{M}^{2}_{\mathrm{M}\rightleftharpoons\mathrm{A}}} & \overline{\overline{Z}}^{\vec{M}_{\mathrm{A}\rightleftharpoons\mathrm{G}}\vec{M}^{12}_{\mathrm{A}\rightleftharpoons\mathrm{A}}} & \overline{\overline{Z}}^{\vec{M}_{\mathrm{A}\rightleftharpoons\mathrm{G}}\vec{M}_{\mathrm{A}\rightleftharpoons\mathrm{G}}}\\
\overline{\overline{Z}}^{\vec{J}_{\mathrm{M}}\vec{J}_{\mathrm{aux}}} & \overline{\overline{Z}}^{\vec{J}_{\mathrm{M}}\vec{J}^{1}_{\mathrm{M}\rightleftharpoons\mathrm{A}}} & \overline{\overline{Z}}^{\vec{J}_{\mathrm{M}}\vec{J}^{2}_{\mathrm{M}\rightleftharpoons\mathrm{A}}} & 0 & 0 & \overline{\overline{Z}}^{\vec{J}_{\mathrm{M}}\vec{J}_{\mathrm{M}}} & 0 & 0 & \overline{\overline{Z}}^{\vec{J}_{\mathrm{M}}\vec{M}_{\mathrm{aux}}} & \overline{\overline{Z}}^{\vec{J}_{\mathrm{M}}\vec{M}^{1}_{\mathrm{M}\rightleftharpoons\mathrm{A}}} & \overline{\overline{Z}}^{\vec{J}_{\mathrm{M}}\vec{M}^{2}_{\mathrm{M}\rightleftharpoons\mathrm{A}}} & 0 & 0\\
0 & \overline{\overline{Z}}^{\vec{J}^{1}_{\mathrm{A}}\vec{J}^{1}_{\mathrm{M}\rightleftharpoons\mathrm{A}}} & 0 & \overline{\overline{Z}}^{\vec{J}^{1}_{\mathrm{A}}\vec{J}^{12}_{\mathrm{A}\rightleftharpoons\mathrm{A}}} & 0 & 0 & \overline{\overline{Z}}^{\vec{J}^{1}_{\mathrm{A}}\vec{J}^{1}_{\mathrm{A}}} & 0 & 0 & \overline{\overline{Z}}^{\vec{J}^{1}_{\mathrm{A}}\vec{M}^{1}_{\mathrm{M}\rightleftharpoons\mathrm{A}}} & 0 & \overline{\overline{Z}}^{\vec{J}^{1}_{\mathrm{A}}\vec{M}^{12}_{\mathrm{A}\rightleftharpoons\mathrm{A}}} & 0\\
0 & 0 & \overline{\overline{Z}}^{\vec{J}^{2}_{\mathrm{A}}\vec{J}^{2}_{\mathrm{M}\rightleftharpoons\mathrm{A}}} & \overline{\overline{Z}}^{\vec{J}^{2}_{\mathrm{A}}\vec{J}^{12}_{\mathrm{A}\rightleftharpoons\mathrm{A}}} & \overline{\overline{Z}}^{\vec{J}^{2}_{\mathrm{A}}\vec{J}_{\mathrm{A}\rightleftharpoons\mathrm{G}}} & 0 & 0 & \overline{\overline{Z}}^{\vec{J}^{2}_{\mathrm{A}}\vec{J}^{2}_{\mathrm{A}}} & 0 & 0 & \overline{\overline{Z}}^{\vec{J}^{2}_{\mathrm{A}}\vec{M}^{2}_{\mathrm{M}\rightleftharpoons\mathrm{A}}} & \overline{\overline{Z}}^{\vec{J}^{2}_{\mathrm{A}}\vec{M}^{12}_{\mathrm{A}\rightleftharpoons\mathrm{A}}} & \overline{\overline{Z}}^{\vec{J}^{2}_{\mathrm{A}}\vec{M}_{\mathrm{A}\rightleftharpoons\mathrm{G}}}
\end{bmatrix} \tag{D-144a}$$

$$\overline{\overline{\Psi}}^{\mathrm{DoM}}_{\mathrm{FCE}}=\begin{bmatrix}
\overline{\overline{Z}}^{\vec{J}_{\mathrm{aux}}\vec{J}_{\mathrm{aux}}} & \overline{\overline{Z}}^{\vec{J}_{\mathrm{aux}}\vec{J}^{1}_{\mathrm{M}\rightleftharpoons\mathrm{A}}} & \overline{\overline{Z}}^{\vec{J}_{\mathrm{aux}}\vec{J}^{2}_{\mathrm{M}\rightleftharpoons\mathrm{A}}} & 0 & 0 & \overline{\overline{Z}}^{\vec{J}_{\mathrm{aux}}\vec{J}_{\mathrm{M}}} & 0 & 0 & \overline{\overline{Z}}^{\vec{J}_{\mathrm{aux}}\vec{M}_{\mathrm{aux}}} & \overline{\overline{Z}}^{\vec{J}_{\mathrm{aux}}\vec{M}^{1}_{\mathrm{M}\rightleftharpoons\mathrm{A}}} & \overline{\overline{Z}}^{\vec{J}_{\mathrm{aux}}\vec{M}^{2}_{\mathrm{M}\rightleftharpoons\mathrm{A}}} & 0 & 0\\
\overline{\overline{Z}}^{\vec{J}^{1}_{\mathrm{M}\rightleftharpoons\mathrm{A}}\vec{J}_{\mathrm{aux}}} & \overline{\overline{Z}}^{\vec{J}^{1}_{\mathrm{M}\rightleftharpoons\mathrm{A}}\vec{J}^{1}_{\mathrm{M}\rightleftharpoons\mathrm{A}}} & \overline{\overline{Z}}^{\vec{J}^{1}_{\mathrm{M}\rightleftharpoons\mathrm{A}}\vec{J}^{2}_{\mathrm{M}\rightleftharpoons\mathrm{A}}} & \overline{\overline{Z}}^{\vec{J}^{1}_{\mathrm{M}\rightleftharpoons\mathrm{A}}\vec{J}^{12}_{\mathrm{A}\rightleftharpoons\mathrm{A}}} & 0 & \overline{\overline{Z}}^{\vec{J}^{1}_{\mathrm{M}\rightleftharpoons\mathrm{A}}\vec{J}_{\mathrm{M}}} & \overline{\overline{Z}}^{\vec{J}^{1}_{\mathrm{M}\rightleftharpoons\mathrm{A}}\vec{J}^{1}_{\mathrm{A}}} & 0 & \overline{\overline{Z}}^{\vec{J}^{1}_{\mathrm{M}\rightleftharpoons\mathrm{A}}\vec{M}_{\mathrm{aux}}} & \overline{\overline{Z}}^{\vec{J}^{1}_{\mathrm{M}\rightleftharpoons\mathrm{A}}\vec{M}^{1}_{\mathrm{M}\rightleftharpoons\mathrm{A}}} & \overline{\overline{Z}}^{\vec{J}^{1}_{\mathrm{M}\rightleftharpoons\mathrm{A}}\vec{M}^{2}_{\mathrm{M}\rightleftharpoons\mathrm{A}}} & \overline{\overline{Z}}^{\vec{J}^{1}_{\mathrm{M}\rightleftharpoons\mathrm{A}}\vec{M}^{12}_{\mathrm{A}\rightleftharpoons\mathrm{A}}} & 0\\
\overline{\overline{Z}}^{\vec{M}^{1}_{\mathrm{M}\rightleftharpoons\mathrm{A}}\vec{J}_{\mathrm{aux}}} & \overline{\overline{Z}}^{\vec{M}^{1}_{\mathrm{M}\rightleftharpoons\mathrm{A}}\vec{J}^{1}_{\mathrm{M}\rightleftharpoons\mathrm{A}}} & \overline{\overline{Z}}^{\vec{M}^{1}_{\mathrm{M}\rightleftharpoons\mathrm{A}}\vec{J}^{2}_{\mathrm{M}\rightleftharpoons\mathrm{A}}} & \overline{\overline{Z}}^{\vec{M}^{1}_{\mathrm{M}\rightleftharpoons\mathrm{A}}\vec{J}^{12}_{\mathrm{A}\rightleftharpoons\mathrm{A}}} & 0 & \overline{\overline{Z}}^{\vec{M}^{1}_{\mathrm{M}\rightleftharpoons\mathrm{A}}\vec{J}_{\mathrm{M}}} & \overline{\overline{Z}}^{\vec{M}^{1}_{\mathrm{M}\rightleftharpoons\mathrm{A}}\vec{J}^{1}_{\mathrm{A}}} & 0 & \overline{\overline{Z}}^{\vec{M}^{1}_{\mathrm{M}\rightleftharpoons\mathrm{A}}\vec{M}_{\mathrm{aux}}} & \overline{\overline{Z}}^{\vec{M}^{1}_{\mathrm{M}\rightleftharpoons\mathrm{A}}\vec{M}^{1}_{\mathrm{M}\rightleftharpoons\mathrm{A}}} & \overline{\overline{Z}}^{\vec{M}^{1}_{\mathrm{M}\rightleftharpoons\mathrm{A}}\vec{M}^{2}_{\mathrm{M}\rightleftharpoons\mathrm{A}}} & \overline{\overline{Z}}^{\vec{M}^{1}_{\mathrm{M}\rightleftharpoons\mathrm{A}}\vec{M}^{12}_{\mathrm{A}\rightleftharpoons\mathrm{A}}} & 0\\
\overline{\overline{Z}}^{\vec{J}^{2}_{\mathrm{M}\rightleftharpoons\mathrm{A}}\vec{J}_{\mathrm{aux}}} & \overline{\overline{Z}}^{\vec{J}^{2}_{\mathrm{M}\rightleftharpoons\mathrm{A}}\vec{J}^{1}_{\mathrm{M}\rightleftharpoons\mathrm{A}}} & \overline{\overline{Z}}^{\vec{J}^{2}_{\mathrm{M}\rightleftharpoons\mathrm{A}}\vec{J}^{2}_{\mathrm{M}\rightleftharpoons\mathrm{A}}} & \overline{\overline{Z}}^{\vec{J}^{2}_{\mathrm{M}\rightleftharpoons\mathrm{A}}\vec{J}^{12}_{\mathrm{A}\rightleftharpoons\mathrm{A}}} & \overline{\overline{Z}}^{\vec{J}^{2}_{\mathrm{M}\rightleftharpoons\mathrm{A}}\vec{J}_{\mathrm{A}\rightleftharpoons\mathrm{G}}} & \overline{\overline{Z}}^{\vec{J}^{2}_{\mathrm{M}\rightleftharpoons\mathrm{A}}\vec{J}_{\mathrm{M}}} & 0 & \overline{\overline{Z}}^{\vec{J}^{2}_{\mathrm{M}\rightleftharpoons\mathrm{A}}\vec{J}^{2}_{\mathrm{A}}} & \overline{\overline{Z}}^{\vec{J}^{2}_{\mathrm{M}\rightleftharpoons\mathrm{A}}\vec{M}_{\mathrm{aux}}} & \overline{\overline{Z}}^{\vec{J}^{2}_{\mathrm{M}\rightleftharpoons\mathrm{A}}\vec{M}^{1}_{\mathrm{M}\rightleftharpoons\mathrm{A}}} & \overline{\overline{Z}}^{\vec{J}^{2}_{\mathrm{M}\rightleftharpoons\mathrm{A}}\vec{M}^{2}_{\mathrm{M}\rightleftharpoons\mathrm{A}}} & \overline{\overline{Z}}^{\vec{J}^{2}_{\mathrm{M}\rightleftharpoons\mathrm{A}}\vec{M}^{12}_{\mathrm{A}\rightleftharpoons\mathrm{A}}} & \overline{\overline{Z}}^{\vec{J}^{2}_{\mathrm{M}\rightleftharpoons\mathrm{A}}\vec{M}_{\mathrm{A}\rightleftharpoons\mathrm{G}}}\\
\overline{\overline{Z}}^{\vec{M}^{2}_{\mathrm{M}\rightleftharpoons\mathrm{A}}\vec{J}_{\mathrm{aux}}} & \overline{\overline{Z}}^{\vec{M}^{2}_{\mathrm{M}\rightleftharpoons\mathrm{A}}\vec{J}^{1}_{\mathrm{M}\rightleftharpoons\mathrm{A}}} & \overline{\overline{Z}}^{\vec{M}^{2}_{\mathrm{M}\rightleftharpoons\mathrm{A}}\vec{J}^{2}_{\mathrm{M}\rightleftharpoons\mathrm{A}}} & \overline{\overline{Z}}^{\vec{M}^{2}_{\mathrm{M}\rightleftharpoons\mathrm{A}}\vec{J}^{12}_{\mathrm{A}\rightleftharpoons\mathrm{A}}} & \overline{\overline{Z}}^{\vec{M}^{2}_{\mathrm{M}\rightleftharpoons\mathrm{A}}\vec{J}_{\mathrm{A}\rightleftharpoons\mathrm{G}}} & \overline{\overline{Z}}^{\vec{M}^{2}_{\mathrm{M}\rightleftharpoons\mathrm{A}}\vec{J}_{\mathrm{M}}} & 0 & \overline{\overline{Z}}^{\vec{M}^{2}_{\mathrm{M}\rightleftharpoons\mathrm{A}}\vec{J}^{2}_{\mathrm{A}}} & \overline{\overline{Z}}^{\vec{M}^{2}_{\mathrm{M}\rightleftharpoons\mathrm{A}}\vec{M}_{\mathrm{aux}}} & \overline{\overline{Z}}^{\vec{M}^{2}_{\mathrm{M}\rightleftharpoons\mathrm{A}}\vec{M}^{1}_{\mathrm{M}\rightleftharpoons\mathrm{A}}} & \overline{\overline{Z}}^{\vec{M}^{2}_{\mathrm{M}\rightleftharpoons\mathrm{A}}\vec{M}^{2}_{\mathrm{M}\rightleftharpoons\mathrm{A}}} & \overline{\overline{Z}}^{\vec{M}^{2}_{\mathrm{M}\rightleftharpoons\mathrm{A}}\vec{M}^{12}_{\mathrm{A}\rightleftharpoons\mathrm{A}}} & \overline{\overline{Z}}^{\vec{M}^{2}_{\mathrm{M}\rightleftharpoons\mathrm{A}}\vec{M}_{\mathrm{A}\rightleftharpoons\mathrm{G}}}\\
0 & \overline{\overline{Z}}^{\vec{J}^{12}_{\mathrm{A}\rightleftharpoons\mathrm{A}}\vec{J}^{1}_{\mathrm{M}\rightleftharpoons\mathrm{A}}} & \overline{\overline{Z}}^{\vec{J}^{12}_{\mathrm{A}\rightleftharpoons\mathrm{A}}\vec{J}^{2}_{\mathrm{M}\rightleftharpoons\mathrm{A}}} & \overline{\overline{Z}}^{\vec{J}^{12}_{\mathrm{A}\rightleftharpoons\mathrm{A}}\vec{J}^{12}_{\mathrm{A}\rightleftharpoons\mathrm{A}}} & \overline{\overline{Z}}^{\vec{J}^{12}_{\mathrm{A}\rightleftharpoons\mathrm{A}}\vec{J}_{\mathrm{A}\rightleftharpoons\mathrm{G}}} & 0 & \overline{\overline{Z}}^{\vec{J}^{12}_{\mathrm{A}\rightleftharpoons\mathrm{A}}\vec{J}^{1}_{\mathrm{A}}} & \overline{\overline{Z}}^{\vec{J}^{12}_{\mathrm{A}\rightleftharpoons\mathrm{A}}\vec{J}^{2}_{\mathrm{A}}} & 0 & \overline{\overline{Z}}^{\vec{J}^{12}_{\mathrm{A}\rightleftharpoons\mathrm{A}}\vec{M}^{1}_{\mathrm{M}\rightleftharpoons\mathrm{A}}} & \overline{\overline{Z}}^{\vec{J}^{12}_{\mathrm{A}\rightleftharpoons\mathrm{A}}\vec{M}^{2}_{\mathrm{M}\rightleftharpoons\mathrm{A}}} & \overline{\overline{Z}}^{\vec{J}^{12}_{\mathrm{A}\rightleftharpoons\mathrm{A}}\vec{M}^{12}_{\mathrm{A}\rightleftharpoons\mathrm{A}}} & \overline{\overline{Z}}^{\vec{J}^{12}_{\mathrm{A}\rightleftharpoons\mathrm{A}}\vec{M}_{\mathrm{A}\rightleftharpoons\mathrm{G}}}\\
0 & \overline{\overline{Z}}^{\vec{M}^{12}_{\mathrm{A}\rightleftharpoons\mathrm{A}}\vec{J}^{1}_{\mathrm{M}\rightleftharpoons\mathrm{A}}} & \overline{\overline{Z}}^{\vec{M}^{12}_{\mathrm{A}\rightleftharpoons\mathrm{A}}\vec{J}^{2}_{\mathrm{M}\rightleftharpoons\mathrm{A}}} & \overline{\overline{Z}}^{\vec{M}^{12}_{\mathrm{A}\rightleftharpoons\mathrm{A}}\vec{J}^{12}_{\mathrm{A}\rightleftharpoons\mathrm{A}}} & \overline{\overline{Z}}^{\vec{M}^{12}_{\mathrm{A}\rightleftharpoons\mathrm{A}}\vec{J}_{\mathrm{A}\rightleftharpoons\mathrm{G}}} & 0 & \overline{\overline{Z}}^{\vec{M}^{12}_{\mathrm{A}\rightleftharpoons\mathrm{A}}\vec{J}^{1}_{\mathrm{A}}} & \overline{\overline{Z}}^{\vec{M}^{12}_{\mathrm{A}\rightleftharpoons\mathrm{A}}\vec{J}^{2}_{\mathrm{A}}} & 0 & \overline{\overline{Z}}^{\vec{M}^{12}_{\mathrm{A}\rightleftharpoons\mathrm{A}}\vec{M}^{1}_{\mathrm{M}\rightleftharpoons\mathrm{A}}} & \overline{\overline{Z}}^{\vec{M}^{12}_{\mathrm{A}\rightleftharpoons\mathrm{A}}\vec{M}^{2}_{\mathrm{M}\rightleftharpoons\mathrm{A}}} & \overline{\overline{Z}}^{\vec{M}^{12}_{\mathrm{A}\rightleftharpoons\mathrm{A}}\vec{M}^{12}_{\mathrm{A}\rightleftharpoons\mathrm{A}}} & \overline{\overline{Z}}^{\vec{M}^{12}_{\mathrm{A}\rightleftharpoons\mathrm{A}}\vec{M}_{\mathrm{A}\rightleftharpoons\mathrm{G}}}\\
0 & 0 & \overline{\overline{Z}}^{\vec{J}_{\mathrm{A}\rightleftharpoons\mathrm{G}}\vec{J}^{2}_{\mathrm{M}\rightleftharpoons\mathrm{A}}} & \overline{\overline{Z}}^{\vec{J}_{\mathrm{A}\rightleftharpoons\mathrm{G}}\vec{J}^{12}_{\mathrm{A}\rightleftharpoons\mathrm{A}}} & \overline{\overline{Z}}^{\vec{J}_{\mathrm{A}\rightleftharpoons\mathrm{G}}\vec{J}_{\mathrm{A}\rightleftharpoons\mathrm{G}}} & 0 & 0 & \overline{\overline{Z}}^{\vec{J}_{\mathrm{A}\rightleftharpoons\mathrm{G}}\vec{J}^{2}_{\mathrm{A}}} & 0 & 0 & \overline{\overline{Z}}^{\vec{J}_{\mathrm{A}\rightleftharpoons\mathrm{G}}\vec{M}^{2}_{\mathrm{M}\rightleftharpoons\mathrm{A}}} & \overline{\overline{Z}}^{\vec{J}_{\mathrm{A}\rightleftharpoons\mathrm{G}}\vec{M}^{12}_{\mathrm{A}\rightleftharpoons\mathrm{A}}} & \overline{\overline{Z}}^{\vec{J}_{\mathrm{A}\rightleftharpoons\mathrm{G}}\vec{M}_{\mathrm{A}\rightleftharpoons\mathrm{G}}}\\
0 & 0 & \overline{\overline{Z}}^{\vec{M}_{\mathrm{A}\rightleftharpoons\mathrm{G}}\vec{J}^{2}_{\mathrm{M}\rightleftharpoons\mathrm{A}}} & \overline{\overline{Z}}^{\vec{M}_{\mathrm{A}\rightleftharpoons\mathrm{G}}\vec{J}^{12}_{\mathrm{A}\rightleftharpoons\mathrm{A}}} & \overline{\overline{Z}}^{\vec{M}_{\mathrm{A}\rightleftharpoons\mathrm{G}}\vec{J}_{\mathrm{A}\rightleftharpoons\mathrm{G}}} & 0 & 0 & \overline{\overline{Z}}^{\vec{M}_{\mathrm{A}\rightleftharpoons\mathrm{G}}\vec{J}^{2}_{\mathrm{A}}} & 0 & 0 & \overline{\overline{Z}}^{\vec{M}_{\mathrm{A}\rightleftharpoons\mathrm{G}}\vec{M}^{2}_{\mathrm{M}\rightleftharpoons\mathrm{A}}} & \overline{\overline{Z}}^{\vec{M}_{\mathrm{A}\rightleftharpoons\mathrm{G}}\vec{M}^{12}_{\mathrm{A}\rightleftharpoons\mathrm{A}}} & \overline{\overline{Z}}^{\vec{M}_{\mathrm{A}\rightleftharpoons\mathrm{G}}\vec{M}_{\mathrm{A}\rightleftharpoons\mathrm{G}}}\\
\overline{\overline{Z}}^{\vec{J}_{\mathrm{M}}\vec{J}_{\mathrm{aux}}} & \overline{\overline{Z}}^{\vec{J}_{\mathrm{M}}\vec{J}^{1}_{\mathrm{M}\rightleftharpoons\mathrm{A}}} & \overline{\overline{Z}}^{\vec{J}_{\mathrm{M}}\vec{J}^{2}_{\mathrm{M}\rightleftharpoons\mathrm{A}}} & 0 & 0 & \overline{\overline{Z}}^{\vec{J}_{\mathrm{M}}\vec{J}_{\mathrm{M}}} & 0 & 0 & \overline{\overline{Z}}^{\vec{J}_{\mathrm{M}}\vec{M}_{\mathrm{aux}}} & \overline{\overline{Z}}^{\vec{J}_{\mathrm{M}}\vec{M}^{1}_{\mathrm{M}\rightleftharpoons\mathrm{A}}} & \overline{\overline{Z}}^{\vec{J}_{\mathrm{M}}\vec{M}^{2}_{\mathrm{M}\rightleftharpoons\mathrm{A}}} & 0 & 0\\
0 & \overline{\overline{Z}}^{\vec{J}^{1}_{\mathrm{A}}\vec{J}^{1}_{\mathrm{M}\rightleftharpoons\mathrm{A}}} & 0 & \overline{\overline{Z}}^{\vec{J}^{1}_{\mathrm{A}}\vec{J}^{12}_{\mathrm{A}\rightleftharpoons\mathrm{A}}} & 0 & 0 & \overline{\overline{Z}}^{\vec{J}^{1}_{\mathrm{A}}\vec{J}^{1}_{\mathrm{A}}} & 0 & 0 & \overline{\overline{Z}}^{\vec{J}^{1}_{\mathrm{A}}\vec{M}^{1}_{\mathrm{M}\rightleftharpoons\mathrm{A}}} & 0 & \overline{\overline{Z}}^{\vec{J}^{1}_{\mathrm{A}}\vec{M}^{12}_{\mathrm{A}\rightleftharpoons\mathrm{A}}} & 0\\
0 & 0 & \overline{\overline{Z}}^{\vec{J}^{2}_{\mathrm{A}}\vec{J}^{2}_{\mathrm{M}\rightleftharpoons\mathrm{A}}} & \overline{\overline{Z}}^{\vec{J}^{2}_{\mathrm{A}}\vec{J}^{12}_{\mathrm{A}\rightleftharpoons\mathrm{A}}} & \overline{\overline{Z}}^{\vec{J}^{2}_{\mathrm{A}}\vec{J}_{\mathrm{A}\rightleftharpoons\mathrm{G}}} & 0 & 0 & \overline{\overline{Z}}^{\vec{J}^{2}_{\mathrm{A}}\vec{J}^{2}_{\mathrm{A}}} & 0 & 0 & \overline{\overline{Z}}^{\vec{J}^{2}_{\mathrm{A}}\vec{M}^{2}_{\mathrm{M}\rightleftharpoons\mathrm{A}}} & \overline{\overline{Z}}^{\vec{J}^{2}_{\mathrm{A}}\vec{M}^{12}_{\mathrm{A}\rightleftharpoons\mathrm{A}}} & \overline{\overline{Z}}^{\vec{J}^{2}_{\mathrm{A}}\vec{M}_{\mathrm{A}\rightleftharpoons\mathrm{G}}}
\end{bmatrix} \tag{D-144b}$$

The power quadratic form matrix $\overline{\overline{P}}^{\mathrm{G}\rightleftharpoons\mathrm{A}}$ used in Eq. (7-97) is as follows:

$$\overline{\overline{P}}_{\mathrm{M}\rightleftharpoons\mathrm{A}} = \overline{\overline{P}}^{\mathrm{curAV}}_{\mathrm{M}\rightleftharpoons\mathrm{A}} \text{ or } \overline{\overline{P}}^{\mathrm{intAV}}_{\mathrm{M}\rightleftharpoons\mathrm{A}} \tag{D-145}$$

in which

$$\overline{\overline{P}}^{\mathrm{curAV}}_{\mathrm{M}\rightleftharpoons\mathrm{A}} = \begin{bmatrix} 0 & 0 & 0 & 0 & 0 & 0 & 0 & 0 & 0 & 0 & 0 & 0 & 0 \\ 0 & 0 & 0 & 0 & 0 & 0 & 0 & 0 & 0 & \overline{\overline{C}}^{\vec{J}^{1}_{\mathrm{M}\rightleftharpoons\mathrm{A}}\vec{M}^{1}_{\mathrm{M}\rightleftharpoons\mathrm{A}}} & 0 & 0 & 0 \\ 0 & 0 & 0 & 0 & 0 & 0 & 0 & 0 & 0 & 0 & \overline{\overline{C}}^{\vec{J}^{2}_{\mathrm{M}\rightleftharpoons\mathrm{A}}\vec{M}^{2}_{\mathrm{M}\rightleftharpoons\mathrm{A}}} & 0 & 0 \\ 0 & 0 & 0 & 0 & 0 & 0 & 0 & 0 & 0 & 0 & 0 & 0 & 0 \\ 0 & 0 & 0 & 0 & 0 & 0 & 0 & 0 & 0 & 0 & 0 & 0 & 0 \\ 0 & 0 & 0 & 0 & 0 & 0 & 0 & 0 & 0 & 0 & 0 & 0 & 0 \\ 0 & 0 & 0 & 0 & 0 & 0 & 0 & 0 & 0 & 0 & 0 & 0 & 0 \\ 0 & 0 & 0 & 0 & 0 & 0 & 0 & 0 & 0 & 0 & 0 & 0 & 0 \\ 0 & 0 & 0 & 0 & 0 & 0 & 0 & 0 & 0 & 0 & 0 & 0 & 0 \\ 0 & 0 & 0 & 0 & 0 & 0 & 0 & 0 & 0 & 0 & 0 & 0 & 0 \\ 0 & 0 & 0 & 0 & 0 & 0 & 0 & 0 & 0 & 0 & 0 & 0 & 0 \\ 0 & 0 & 0 & 0 & 0 & 0 & 0 & 0 & 0 & 0 & 0 & 0 & 0 \\ 0 & 0 & 0 & 0 & 0 & 0 & 0 & 0 & 0 & 0 & 0 & 0 & 0 \end{bmatrix} \tag{D-146}$$

corresponding to the first equality in Eq. (7-96), and

$$\overline{\overline{P}}^{\mathrm{intAV}}_{\mathrm{M}\rightleftharpoons\mathrm{A}} = \begin{bmatrix} 0 & 0 & 0 & 0 & 0 & 0 & 0 & 0 & 0 & 0 & 0 & 0 & 0 \\ \overline{\overline{P}}^{\vec{J}^{1}_{\mathrm{M}\rightleftharpoons\mathrm{A}}\vec{J}_{\mathrm{aux}}} & \overline{\overline{P}}^{\vec{J}^{1}_{\mathrm{M}\rightleftharpoons\mathrm{A}}\vec{J}^{1}_{\mathrm{M}\rightleftharpoons\mathrm{A}}} & \overline{\overline{P}}^{\vec{J}^{1}_{\mathrm{M}\rightleftharpoons\mathrm{A}}\vec{J}^{2}_{\mathrm{M}\rightleftharpoons\mathrm{A}}} & 0 & 0 & \overline{\overline{P}}^{\vec{J}^{1}_{\mathrm{M}\rightleftharpoons\mathrm{A}}\vec{J}_{\mathrm{M}}} & 0 & 0 & \overline{\overline{P}}^{\vec{J}^{1}_{\mathrm{M}\rightleftharpoons\mathrm{A}}\vec{M}_{\mathrm{aux}}} & \overline{\overline{P}}^{\vec{J}^{1}_{\mathrm{M}\rightleftharpoons\mathrm{A}}\vec{M}^{1}_{\mathrm{M}\rightleftharpoons\mathrm{A}}} & \overline{\overline{P}}^{\vec{J}^{1}_{\mathrm{M}\rightleftharpoons\mathrm{A}}\vec{M}^{2}_{\mathrm{M}\rightleftharpoons\mathrm{A}}} & 0 & 0 \\ \overline{\overline{P}}^{\vec{J}^{2}_{\mathrm{M}\rightleftharpoons\mathrm{A}}\vec{J}_{\mathrm{aux}}} & \overline{\overline{P}}^{\vec{J}^{2}_{\mathrm{M}\rightleftharpoons\mathrm{A}}\vec{J}^{1}_{\mathrm{M}\rightleftharpoons\mathrm{A}}} & \overline{\overline{P}}^{\vec{J}^{2}_{\mathrm{M}\rightleftharpoons\mathrm{A}}\vec{J}^{2}_{\mathrm{M}\rightleftharpoons\mathrm{A}}} & 0 & 0 & \overline{\overline{P}}^{\vec{J}^{2}_{\mathrm{M}\rightleftharpoons\mathrm{A}}\vec{J}_{\mathrm{M}}} & 0 & 0 & \overline{\overline{P}}^{\vec{J}^{2}_{\mathrm{M}\rightleftharpoons\mathrm{A}}\vec{M}_{\mathrm{aux}}} & \overline{\overline{P}}^{\vec{J}^{2}_{\mathrm{M}\rightleftharpoons\mathrm{A}}\vec{M}^{1}_{\mathrm{M}\rightleftharpoons\mathrm{A}}} & \overline{\overline{P}}^{\vec{J}^{2}_{\mathrm{M}\rightleftharpoons\mathrm{A}}\vec{M}^{2}_{\mathrm{M}\rightleftharpoons\mathrm{A}}} & 0 & 0 \\ 0 & 0 & 0 & 0 & 0 & 0 & 0 & 0 & 0 & 0 & 0 & 0 & 0 \\ 0 & 0 & 0 & 0 & 0 & 0 & 0 & 0 & 0 & 0 & 0 & 0 & 0 \\ 0 & 0 & 0 & 0 & 0 & 0 & 0 & 0 & 0 & 0 & 0 & 0 & 0 \\ 0 & 0 & 0 & 0 & 0 & 0 & 0 & 0 & 0 & 0 & 0 & 0 & 0 \\ 0 & 0 & 0 & 0 & 0 & 0 & 0 & 0 & 0 & 0 & 0 & 0 & 0 \\ 0 & 0 & 0 & 0 & 0 & 0 & 0 & 0 & 0 & 0 & 0 & 0 & 0 \\ 0 & 0 & 0 & 0 & 0 & 0 & 0 & 0 & 0 & 0 & 0 & 0 & 0 \\ 0 & 0 & 0 & 0 & 0 & 0 & 0 & 0 & 0 & 0 & 0 & 0 & 0 \\ 0 & 0 & 0 & 0 & 0 & 0 & 0 & 0 & 0 & 0 & 0 & 0 & 0 \\ 0 & 0 & 0 & 0 & 0 & 0 & 0 & 0 & 0 & 0 & 0 & 0 & 0 \end{bmatrix} \tag{D-147a}$$

corresponding to the second equality in Eq. (7-96), and

$$\overline{\overline{P}}^{\mathrm{intAV}}_{\mathrm{M}\rightleftharpoons\mathrm{A}} = \begin{bmatrix} 0 & 0 & 0 & 0 & 0 & 0 & 0 & 0 & 0 & 0 & 0 & 0 & 0 \\ 0 & 0 & 0 & 0 & 0 & 0 & 0 & 0 & 0 & 0 & 0 & 0 & 0 \\ 0 & 0 & 0 & 0 & 0 & 0 & 0 & 0 & 0 & 0 & 0 & 0 & 0 \\ 0 & 0 & 0 & 0 & 0 & 0 & 0 & 0 & 0 & 0 & 0 & 0 & 0 \\ 0 & 0 & 0 & 0 & 0 & 0 & 0 & 0 & 0 & 0 & 0 & 0 & 0 \\ 0 & 0 & 0 & 0 & 0 & 0 & 0 & 0 & 0 & 0 & 0 & 0 & 0 \\ 0 & 0 & 0 & 0 & 0 & 0 & 0 & 0 & 0 & 0 & 0 & 0 & 0 \\ 0 & 0 & 0 & 0 & 0 & 0 & 0 & 0 & 0 & 0 & 0 & 0 & 0 \\ 0 & 0 & 0 & 0 & 0 & 0 & 0 & 0 & 0 & 0 & 0 & 0 & 0 \\ \overline{\overline{P}}^{\vec{M}^{1}_{\mathrm{M}\rightleftharpoons\mathrm{A}}\vec{J}_{\mathrm{aux}}} & \overline{\overline{P}}^{\vec{M}^{1}_{\mathrm{M}\rightleftharpoons\mathrm{A}}\vec{J}^{1}_{\mathrm{M}\rightleftharpoons\mathrm{A}}} & \overline{\overline{P}}^{\vec{M}^{1}_{\mathrm{M}\rightleftharpoons\mathrm{A}}\vec{J}^{2}_{\mathrm{M}\rightleftharpoons\mathrm{A}}} & 0 & 0 & \overline{\overline{P}}^{\vec{M}^{1}_{\mathrm{M}\rightleftharpoons\mathrm{A}}\vec{J}_{\mathrm{M}}} & 0 & 0 & \overline{\overline{P}}^{\vec{M}^{1}_{\mathrm{M}\rightleftharpoons\mathrm{A}}\vec{M}_{\mathrm{aux}}} & \overline{\overline{P}}^{\vec{M}^{1}_{\mathrm{M}\rightleftharpoons\mathrm{A}}\vec{M}^{1}_{\mathrm{M}\rightleftharpoons\mathrm{A}}} & \overline{\overline{P}}^{\vec{M}^{1}_{\mathrm{M}\rightleftharpoons\mathrm{A}}\vec{M}^{2}_{\mathrm{M}\rightleftharpoons\mathrm{A}}} & 0 & 0 \\ \overline{\overline{P}}^{\vec{M}^{2}_{\mathrm{M}\rightleftharpoons\mathrm{A}}\vec{J}_{\mathrm{aux}}} & \overline{\overline{P}}^{\vec{M}^{2}_{\mathrm{M}\rightleftharpoons\mathrm{A}}\vec{J}^{1}_{\mathrm{M}\rightleftharpoons\mathrm{A}}} & \overline{\overline{P}}^{\vec{M}^{2}_{\mathrm{M}\rightleftharpoons\mathrm{A}}\vec{J}^{2}_{\mathrm{M}\rightleftharpoons\mathrm{A}}} & 0 & 0 & \overline{\overline{P}}^{\vec{M}^{2}_{\mathrm{M}\rightleftharpoons\mathrm{A}}\vec{J}_{\mathrm{M}}} & 0 & 0 & \overline{\overline{P}}^{\vec{M}^{2}_{\mathrm{M}\rightleftharpoons\mathrm{A}}\vec{M}_{\mathrm{aux}}} & \overline{\overline{P}}^{\vec{M}^{2}_{\mathrm{M}\rightleftharpoons\mathrm{A}}\vec{M}^{1}_{\mathrm{M}\rightleftharpoons\mathrm{A}}} & \overline{\overline{P}}^{\vec{M}^{2}_{\mathrm{M}\rightleftharpoons\mathrm{A}}\vec{M}^{2}_{\mathrm{M}\rightleftharpoons\mathrm{A}}} & 0 & 0 \\ 0 & 0 & 0 & 0 & 0 & 0 & 0 & 0 & 0 & 0 & 0 & 0 & 0 \\ 0 & 0 & 0 & 0 & 0 & 0 & 0 & 0 & 0 & 0 & 0 & 0 & 0 \end{bmatrix}^{\dagger} \tag{D-147b}$$

corresponding to the third equality in Eq. (7-96), where the elements of the sub-matrices in above Eqs. (D-146)~(D-147b) are as follows:

$$c_{\xi\zeta}^{\vec{J}^1_{\mathrm{M}\rightleftharpoons\mathrm{A}}\vec{M}^1_{\mathrm{M}\rightleftharpoons\mathrm{A}}} = (1/2)\left\langle \hat{n}^1_{\to\mathrm{A}} \times \vec{b}_\xi^{\vec{J}^1_{\mathrm{M}\rightleftharpoons\mathrm{A}}}, \vec{b}_\zeta^{\vec{M}^1_{\mathrm{M}\rightleftharpoons\mathrm{A}}} \right\rangle_{\mathbb{S}^1_{\mathrm{M}\rightleftharpoons\mathrm{A}}} \tag{D-148a}$$

$$c_{\xi\zeta}^{\vec{J}^2_{\mathrm{M}\rightleftharpoons\mathrm{A}}\vec{M}^2_{\mathrm{M}\rightleftharpoons\mathrm{A}}} = (1/2)\left\langle \hat{n}^2_{\to\mathrm{A}} \times \vec{b}_\xi^{\vec{J}^2_{\mathrm{M}\rightleftharpoons\mathrm{A}}}, \vec{b}_\zeta^{\vec{M}^2_{\mathrm{M}\rightleftharpoons\mathrm{A}}} \right\rangle_{\mathbb{S}^2_{\mathrm{M}\rightleftharpoons\mathrm{A}}} \tag{D-148b}$$

and

$$p_{\xi\zeta}^{\vec{J}^1_{\mathrm{M}\rightleftharpoons\mathrm{A}}\vec{J}_{\mathrm{aux}}} = -(1/2)\left\langle \vec{b}_\xi^{\vec{J}^1_{\mathrm{M}\rightleftharpoons\mathrm{A}}}, -j\omega\mu_0\mathcal{L}_0\left(\vec{b}_\zeta^{\vec{J}_{\mathrm{aux}}}\right)\right\rangle_{\mathbb{S}^1_{\mathrm{M}\rightleftharpoons\mathrm{A}}} \tag{D-149a}$$

$$p_{\xi\zeta}^{\vec{J}^1_{\mathrm{M}\rightleftharpoons\mathrm{A}}\vec{J}^1_{\mathrm{M}\rightleftharpoons\mathrm{A}}} = -(1/2)\left\langle \vec{b}_\xi^{\vec{J}^1_{\mathrm{M}\rightleftharpoons\mathrm{A}}}, -j\omega\mu_0\mathcal{L}_0\left(-\vec{b}_\zeta^{\vec{J}^1_{\mathrm{M}\rightleftharpoons\mathrm{A}}}\right)\right\rangle_{\mathbb{S}^1_{\mathrm{M}\rightleftharpoons\mathrm{A}}} \tag{D-149b}$$

$$p_{\xi\zeta}^{\vec{J}^1_{\mathrm{M}\rightleftharpoons\mathrm{A}}\vec{J}^2_{\mathrm{M}\rightleftharpoons\mathrm{A}}} = -(1/2)\left\langle \vec{b}_\xi^{\vec{J}^1_{\mathrm{M}\rightleftharpoons\mathrm{A}}}, -j\omega\mu_0\mathcal{L}_0\left(-\vec{b}_\zeta^{\vec{J}^2_{\mathrm{M}\rightleftharpoons\mathrm{A}}}\right)\right\rangle_{\mathbb{S}^1_{\mathrm{M}\rightleftharpoons\mathrm{A}}} \tag{D-149c}$$

$$p_{\xi\zeta}^{\vec{J}^1_{\mathrm{M}\rightleftharpoons\mathrm{A}}\vec{J}_{\mathrm{M}}} = -(1/2)\left\langle \vec{b}_\xi^{\vec{J}^1_{\mathrm{M}\rightleftharpoons\mathrm{A}}}, -j\omega\mu_0\mathcal{L}_0\left(\vec{b}_\zeta^{\vec{J}_{\mathrm{M}}}\right)\right\rangle_{\mathbb{S}^1_{\mathrm{M}\rightleftharpoons\mathrm{A}}} \tag{D-149d}$$

$$p_{\xi\zeta}^{\vec{J}^1_{\mathrm{M}\rightleftharpoons\mathrm{A}}\vec{M}_{\mathrm{aux}}} = -(1/2)\left\langle \vec{b}_\xi^{\vec{J}^1_{\mathrm{M}\rightleftharpoons\mathrm{A}}}, -\mathcal{K}_0\left(\vec{b}_\zeta^{\vec{M}_{\mathrm{aux}}}\right)\right\rangle_{\mathbb{S}^1_{\mathrm{M}\rightleftharpoons\mathrm{A}}} \tag{D-149e}$$

$$p_{\xi\zeta}^{\vec{J}^1_{\mathrm{M}\rightleftharpoons\mathrm{A}}\vec{M}^1_{\mathrm{M}\rightleftharpoons\mathrm{A}}} = -(1/2)\left\langle \vec{b}_\xi^{\vec{J}^1_{\mathrm{M}\rightleftharpoons\mathrm{A}}}, \hat{n}^1_{\to\mathrm{A}} \times \frac{1}{2}\vec{b}_\zeta^{\vec{M}^1_{\mathrm{M}\rightleftharpoons\mathrm{A}}} - \mathrm{P.V.}\,\mathcal{K}_0\left(-\vec{b}_\zeta^{\vec{M}^1_{\mathrm{M}\rightleftharpoons\mathrm{A}}}\right)\right\rangle_{\mathbb{S}^1_{\mathrm{M}\rightleftharpoons\mathrm{A}}} \tag{D-149f}$$

$$p_{\xi\zeta}^{\vec{J}^1_{\mathrm{M}\rightleftharpoons\mathrm{A}}\vec{M}^2_{\mathrm{M}\rightleftharpoons\mathrm{A}}} = -(1/2)\left\langle \vec{b}_\xi^{\vec{J}^1_{\mathrm{M}\rightleftharpoons\mathrm{A}}}, -\mathcal{K}_0\left(-\vec{b}_\zeta^{\vec{M}^2_{\mathrm{M}\rightleftharpoons\mathrm{A}}}\right)\right\rangle_{\mathbb{S}^1_{\mathrm{M}\rightleftharpoons\mathrm{A}}} \tag{D-149g}$$

and

$$p_{\xi\zeta}^{\vec{M}^1_{\mathrm{M}\rightleftharpoons\mathrm{A}}\vec{J}_{\mathrm{aux}}} = -(1/2)\left\langle \vec{b}_\xi^{\vec{M}^1_{\mathrm{M}\rightleftharpoons\mathrm{A}}}, \mathcal{K}_0\left(\vec{b}_\zeta^{\vec{J}_{\mathrm{aux}}}\right)\right\rangle_{\mathbb{S}^1_{\mathrm{M}\rightleftharpoons\mathrm{A}}} \tag{D-149h}$$

$$p_{\xi\zeta}^{\vec{M}^1_{\mathrm{M}\rightleftharpoons\mathrm{A}}\vec{J}^1_{\mathrm{M}\rightleftharpoons\mathrm{A}}} = -(1/2)\left\langle \vec{b}_\xi^{\vec{M}^1_{\mathrm{M}\rightleftharpoons\mathrm{A}}}, \frac{1}{2}\vec{b}_\zeta^{\vec{J}^1_{\mathrm{M}\rightleftharpoons\mathrm{A}}} \times \hat{n}^1_{\to\mathrm{A}} + \mathrm{P.V.}\,\mathcal{K}_0\left(-\vec{b}_\zeta^{\vec{J}^1_{\mathrm{M}\rightleftharpoons\mathrm{A}}}\right)\right\rangle_{\mathbb{S}^1_{\mathrm{M}\rightleftharpoons\mathrm{A}}} \tag{D-149i}$$

$$p_{\xi\zeta}^{\vec{M}^1_{\mathrm{M}\rightleftharpoons\mathrm{A}}\vec{J}^2_{\mathrm{M}\rightleftharpoons\mathrm{A}}} = -(1/2)\left\langle \vec{b}_\xi^{\vec{M}^1_{\mathrm{M}\rightleftharpoons\mathrm{A}}}, \mathcal{K}_0\left(-\vec{b}_\zeta^{\vec{J}^2_{\mathrm{M}\rightleftharpoons\mathrm{A}}}\right)\right\rangle_{\mathbb{S}^1_{\mathrm{M}\rightleftharpoons\mathrm{A}}} \tag{D-149j}$$

$$p_{\xi\zeta}^{\vec{M}^1_{\mathrm{M}\rightleftharpoons\mathrm{A}}\vec{J}_{\mathrm{M}}} = -(1/2)\left\langle \vec{b}_\xi^{\vec{M}^1_{\mathrm{M}\rightleftharpoons\mathrm{A}}}, \mathcal{K}_0\left(\vec{b}_\zeta^{\vec{J}_{\mathrm{M}}}\right)\right\rangle_{\mathbb{S}^1_{\mathrm{M}\rightleftharpoons\mathrm{A}}} \tag{D-149k}$$

$$p_{\xi\zeta}^{\vec{M}^1_{\mathrm{M}\rightleftharpoons\mathrm{A}}\vec{M}_{\mathrm{aux}}} = -(1/2)\left\langle \vec{b}_\xi^{\vec{M}^1_{\mathrm{M}\rightleftharpoons\mathrm{A}}}, -j\omega\varepsilon_0\mathcal{L}_0\left(\vec{b}_\zeta^{\vec{M}_{\mathrm{aux}}}\right)\right\rangle_{\mathbb{S}^1_{\mathrm{M}\rightleftharpoons\mathrm{A}}} \tag{D-149l}$$

$$p_{\xi\zeta}^{\vec{M}^1_{\mathrm{M}\rightleftharpoons\mathrm{A}}\vec{M}^1_{\mathrm{M}\rightleftharpoons\mathrm{A}}} = -(1/2)\left\langle \vec{b}_\xi^{\vec{M}^1_{\mathrm{M}\rightleftharpoons\mathrm{A}}}, -j\omega\varepsilon_0\mathcal{L}_0\left(-\vec{b}_\zeta^{\vec{M}^1_{\mathrm{M}\rightleftharpoons\mathrm{A}}}\right)\right\rangle_{\mathbb{S}^1_{\mathrm{M}\rightleftharpoons\mathrm{A}}} \tag{D-149m}$$

$$p_{\xi\zeta}^{\vec{M}^1_{\mathrm{M}\rightleftharpoons\mathrm{A}}\vec{M}^2_{\mathrm{M}\rightleftharpoons\mathrm{A}}} = -(1/2)\left\langle \vec{b}_\xi^{\vec{M}^1_{\mathrm{M}\rightleftharpoons\mathrm{A}}}, -j\omega\varepsilon_0\mathcal{L}_0\left(-\vec{b}_\zeta^{\vec{M}^2_{\mathrm{M}\rightleftharpoons\mathrm{A}}}\right)\right\rangle_{\mathbb{S}^1_{\mathrm{M}\rightleftharpoons\mathrm{A}}} \tag{D-149n}$$

and

$$p_{\xi\zeta}^{\vec{J}^2_{\mathrm{M}\rightleftharpoons\mathrm{A}}\vec{J}_{\mathrm{aux}}} = -(1/2)\left\langle \vec{b}_\xi^{\vec{J}^2_{\mathrm{M}\rightleftharpoons\mathrm{A}}}, -j\omega\mu_0\mathcal{L}_0\left(\vec{b}_\zeta^{\vec{J}_{\mathrm{aux}}}\right)\right\rangle_{\mathbb{S}^2_{\mathrm{M}\rightleftharpoons\mathrm{A}}} \tag{D-149o}$$

$$p_{\xi\zeta}^{\vec{J}_{\mathrm{M}\rightleftharpoons\mathrm{A}}^{2}\vec{J}_{\mathrm{M}\rightleftharpoons\mathrm{A}}^{1}} = -(1/2)\left\langle \vec{b}_{\xi}^{\vec{J}_{\mathrm{M}\rightleftharpoons\mathrm{A}}^{2}}, -j\omega\mu_0\mathcal{L}_0\left(-\vec{b}_{\zeta}^{\vec{J}_{\mathrm{M}\rightleftharpoons\mathrm{A}}^{1}}\right)\right\rangle_{\mathbb{S}_{\mathrm{M}\rightleftharpoons\mathrm{A}}^{2}} \tag{D-149p}$$

$$p_{\xi\zeta}^{\vec{J}_{\mathrm{M}\rightleftharpoons\mathrm{A}}^{2}\vec{J}_{\mathrm{M}\rightleftharpoons\mathrm{A}}^{2}} = -(1/2)\left\langle \vec{b}_{\xi}^{\vec{J}_{\mathrm{M}\rightleftharpoons\mathrm{A}}^{2}}, -j\omega\mu_0\mathcal{L}_0\left(-\vec{b}_{\zeta}^{\vec{J}_{\mathrm{M}\rightleftharpoons\mathrm{A}}^{2}}\right)\right\rangle_{\mathbb{S}_{\mathrm{M}\rightleftharpoons\mathrm{A}}^{2}} \tag{D-149q}$$

$$p_{\xi\zeta}^{\vec{J}_{\mathrm{M}\rightleftharpoons\mathrm{A}}^{2}\vec{J}_{\mathrm{M}}} = -(1/2)\left\langle \vec{b}_{\xi}^{\vec{J}_{\mathrm{M}\rightleftharpoons\mathrm{A}}^{2}}, -j\omega\mu_0\mathcal{L}_0\left(\vec{b}_{\zeta}^{\vec{J}_{\mathrm{M}}}\right)\right\rangle_{\mathbb{S}_{\mathrm{M}\rightleftharpoons\mathrm{A}}^{2}} \tag{D-149r}$$

$$p_{\xi\zeta}^{\vec{J}_{\mathrm{M}\rightleftharpoons\mathrm{A}}^{2}\vec{M}_{\mathrm{aux}}} = -(1/2)\left\langle \vec{b}_{\xi}^{\vec{J}_{\mathrm{M}\rightleftharpoons\mathrm{A}}^{2}}, -\mathcal{K}_0\left(\vec{b}_{\zeta}^{\vec{M}_{\mathrm{aux}}}\right)\right\rangle_{\mathbb{S}_{\mathrm{M}\rightleftharpoons\mathrm{A}}^{2}} \tag{D-149s}$$

$$p_{\xi\zeta}^{\vec{J}_{\mathrm{M}\rightleftharpoons\mathrm{A}}^{2}\vec{M}_{\mathrm{M}\rightleftharpoons\mathrm{A}}^{1}} = -(1/2)\left\langle \vec{b}_{\xi}^{\vec{J}_{\mathrm{M}\rightleftharpoons\mathrm{A}}^{2}}, -\mathcal{K}_0\left(-\vec{b}_{\zeta}^{\vec{M}_{\mathrm{M}\rightleftharpoons\mathrm{A}}^{1}}\right)\right\rangle_{\mathbb{S}_{\mathrm{M}\rightleftharpoons\mathrm{A}}^{2}} \tag{D-149t}$$

$$p_{\xi\zeta}^{\vec{J}_{\mathrm{M}\rightleftharpoons\mathrm{A}}^{2}\vec{M}_{\mathrm{M}\rightleftharpoons\mathrm{A}}^{2}} = -(1/2)\left\langle \vec{b}_{\xi}^{\vec{J}_{\mathrm{M}\rightleftharpoons\mathrm{A}}^{2}}, \hat{n}_{\rightarrow\mathrm{A}}^{1}\times\frac{1}{2}\vec{b}_{\zeta}^{\vec{M}_{\mathrm{M}\rightleftharpoons\mathrm{A}}^{2}} - \mathrm{P.V.}\,\mathcal{K}_0\left(-\vec{b}_{\zeta}^{\vec{M}_{\mathrm{M}\rightleftharpoons\mathrm{A}}^{2}}\right)\right\rangle_{\mathbb{S}_{\mathrm{M}\rightleftharpoons\mathrm{A}}^{2}} \tag{D-149u}$$

and

$$p_{\xi\zeta}^{\vec{M}_{\mathrm{M}\rightleftharpoons\mathrm{A}}^{2}\vec{J}_{\mathrm{aux}}} = -(1/2)\left\langle \vec{b}_{\xi}^{\vec{M}_{\mathrm{M}\rightleftharpoons\mathrm{A}}^{2}}, \mathcal{K}_0\left(\vec{b}_{\zeta}^{\vec{J}_{\mathrm{aux}}}\right)\right\rangle_{\mathbb{S}_{\mathrm{M}\rightleftharpoons\mathrm{A}}^{2}} \tag{D-149v}$$

$$p_{\xi\zeta}^{\vec{M}_{\mathrm{M}\rightleftharpoons\mathrm{A}}^{2}\vec{J}_{\mathrm{M}\rightleftharpoons\mathrm{A}}^{1}} = -(1/2)\left\langle \vec{b}_{\xi}^{\vec{M}_{\mathrm{M}\rightleftharpoons\mathrm{A}}^{2}}, \mathcal{K}_0\left(-\vec{b}_{\zeta}^{\vec{J}_{\mathrm{M}\rightleftharpoons\mathrm{A}}^{1}}\right)\right\rangle_{\mathbb{S}_{\mathrm{M}\rightleftharpoons\mathrm{A}}^{2}} \tag{D-149w}$$

$$p_{\xi\zeta}^{\vec{M}_{\mathrm{M}\rightleftharpoons\mathrm{A}}^{2}\vec{J}_{\mathrm{M}\rightleftharpoons\mathrm{A}}^{2}} = -(1/2)\left\langle \vec{b}_{\xi}^{\vec{M}_{\mathrm{M}\rightleftharpoons\mathrm{A}}^{2}}, \frac{1}{2}\vec{b}_{\zeta}^{\vec{J}_{\mathrm{M}\rightleftharpoons\mathrm{A}}^{2}}\times\hat{n}_{\rightarrow\mathrm{A}}^{1} + \mathrm{P.V.}\,\mathcal{K}_0\left(-\vec{b}_{\zeta}^{\vec{J}_{\mathrm{M}\rightleftharpoons\mathrm{A}}^{2}}\right)\right\rangle_{\mathbb{S}_{\mathrm{M}\rightleftharpoons\mathrm{A}}^{2}} \tag{D-149x}$$

$$p_{\xi\zeta}^{\vec{M}_{\mathrm{M}\rightleftharpoons\mathrm{A}}^{2}\vec{J}_{\mathrm{M}}} = -(1/2)\left\langle \vec{b}_{\xi}^{\vec{M}_{\mathrm{M}\rightleftharpoons\mathrm{A}}^{2}}, \mathcal{K}_0\left(\vec{b}_{\zeta}^{\vec{J}_{\mathrm{M}}}\right)\right\rangle_{\mathbb{S}_{\mathrm{M}\rightleftharpoons\mathrm{A}}^{2}} \tag{D-149y}$$

$$p_{\xi\zeta}^{\vec{M}_{\mathrm{M}\rightleftharpoons\mathrm{A}}^{2}\vec{M}_{\mathrm{aux}}} = -(1/2)\left\langle \vec{b}_{\xi}^{\vec{M}_{\mathrm{M}\rightleftharpoons\mathrm{A}}^{2}}, -j\omega\varepsilon_0\mathcal{L}_0\left(\vec{b}_{\zeta}^{\vec{M}_{\mathrm{aux}}}\right)\right\rangle_{\mathbb{S}_{\mathrm{M}\rightleftharpoons\mathrm{A}}^{2}} \tag{D-149z}$$

$$p_{\xi\zeta}^{\vec{M}_{\mathrm{M}\rightleftharpoons\mathrm{A}}^{2}\vec{M}_{\mathrm{M}\rightleftharpoons\mathrm{A}}^{1}} = -(1/2)\left\langle \vec{b}_{\xi}^{\vec{M}_{\mathrm{M}\rightleftharpoons\mathrm{A}}^{2}}, -j\omega\varepsilon_0\mathcal{L}_0\left(-\vec{b}_{\zeta}^{\vec{M}_{\mathrm{M}\rightleftharpoons\mathrm{A}}^{1}}\right)\right\rangle_{\mathbb{S}_{\mathrm{M}\rightleftharpoons\mathrm{A}}^{2}} \tag{D-149α}$$

$$p_{\xi\zeta}^{\vec{M}_{\mathrm{M}\rightleftharpoons\mathrm{A}}^{2}\vec{M}_{\mathrm{M}\rightleftharpoons\mathrm{A}}^{2}} = -(1/2)\left\langle \vec{b}_{\xi}^{\vec{M}_{\mathrm{M}\rightleftharpoons\mathrm{A}}^{2}}, -j\omega\varepsilon_0\mathcal{L}_0\left(-\vec{b}_{\zeta}^{\vec{M}_{\mathrm{M}\rightleftharpoons\mathrm{A}}^{2}}\right)\right\rangle_{\mathbb{S}_{\mathrm{M}\rightleftharpoons\mathrm{A}}^{2}} \tag{D-149β}$$

# Appendix E Work-Energy Principle Based Characteristic Mode Theory with Solution Domain Compression for Material Scattering Systems

This App. E had been written as a journal paper by our research group (Ren-Zun Lian, Xing-Yue Guo and Ming-Yao Xia), and the original manuscript [AP2004-0708] entitled "Work-Energy Principe Based Characteristic Mode Theory with Solution Domain Compression for Material Scattering Systems"[18] was submitted to *IEEE Transactions on Antennas and Propagation* (*IEEE-TAP*) on 11-Apr-2020, and the revised manuscript [AP2004-0708.R1] was submitted to IEEE-TAP on 10-Nov-2020.

## E1 Abstract

Work-energy principle (WEP) framework is used to establish characteristic mode theory (CMT).

Under WEP framework, the physical purpose/picture of CMT is revealed, and that is to construct a set of steadily working energy-decoupled modes for scattering systems. Employing the physical picture, it is explained why the modal far fields of lossy scattering systems are not orthogonal; it is explained why the characteristic values of magnetodielectric scattering systems don't have clear physical meaning; the physical interpretation for normalizing modal real power to 1 is provided; the Parseval's identity related to CMT is derived.

Under WEP framework, driving power operator (DPO) is introduced as the generating operator of characteristic modes (CMs), and orthogonalizing DPO method is proposed to construct CMs. In the aspect of distinguishing independent variables from dependent variables, new DPO is more advantageous than traditional impedance matrix operator (IMO). In the aspect of constructing CMs, new orthogonalizing DPO method has a more satisfactory numerical performance than traditional orthogonalizing IMO method.

In addition, solution domain compression (SDC) scheme is developed to suppress spurious modes. New SDC scheme effectively avoids the matrix inversion process used in traditional dependent variable elimination (DVE) scheme.

## E2 Index Terms

Characteristic mode (CM), driving power operator (DPO), Parseval's identity, solution domain compression (SDC), work-energy principle (WEP).

## E3 Introduction

Under scattering matrix (SM) framework, Garbacz introduced the concept of characteristic mode (CM) in [5], and systematically established SM-based characteristic mode theory (CMT) in [6], and summarized his core ideology in [7]. The SM-based CMT (SM-CMT) has a very clear physical purpose (or physical picture), that is, for any pre-selected lossless objective scattering system (OSS) SM-CMT focuses on constructing a set of CMs with orthogonal modal far fields by orthogonalizing the perturbation matrix operator (PMO) obtained in SM framework.

Following Garbacz's pioneering works, Harrington *et al.*[8] alternatively established an integral equation (IE) based CMT, and published a series of papers[9,11,12]. By orthogonalizing the impedance matrix operator (IMO) obtained in IE framework, the IE-based CMT (IE-CMT) can also construct a set of CMs for any pre-selected lossless or lossy OSS.

Harrington's transformation for the carrying framework of CMT (from SM framework to IE framework, as shown in Tab. E-1) and transformation for the construction method of CMs (from orthogonalizing PMO method to orthogonalizing IMO method, as shown in Tab. E-1) are seminal works and of great significance for developing CMT, because the transformations greatly simplify the numerical computation of CMs. After half a century (from 1970 to now) development, IE-CMT has been greatly improved in many aspects[16], but still has some unsolved problems[17].

Table E-1 Evolutions of CMT and Comparisons From Aspects of Carrying Framework, Construction Method and Physical Picture/Purpose

| | **Carrying Framework** | **Construction Method** | **Physical Picture / Physical Purpose** |
|---|---|---|---|
| **SM-CMT (1960s)[5~7]** | scattering matrix (SM) framework | orthogonalizing perturbation matrix operator (PMO) method | to construct a set of working modes with orthogonal modal far fields |
| | | ↓ | |
| **IE-CMT (1970s)[8,9,11,12]** | integral equation (IE) framework | orthogonalizing impedance matrix operator (IMO) method | not clarified by its founders |
| | | ↓ | |
| **WEP-CMT (2019)[13]** | work-energy principle (WEP) framework | orthogonalizing driving power operator (DPO) method | to construct a set of steadily working modes without net energy exchange in any integral period |

Unlike SM-CMT[5~7], which had had a clear physical purpose/picture from the beginning of its establishment, IE-CMT[8,9,11,12] has been lacking of a clear physical picture since its establishment. Harrington *et al.*[9] provided a Poynting's theorem based physical interpretation to the IE-CMT for metallic OSSs, but [27] found out that the interpretation is not suitable for the IE-CMT of material OSSs. To provide a unified physical picture to both metallic and material IE-CMTs, [26,33] rebuilt Harrington's IE-CMT by using a new ideology. Recently, [13] further generalized the ideology used in [26,27,33] to work-energy principle (WEP) based CMT, and then realized the second transformation for the carrying framework of CMT — from IE framework to WEP framework (as shown in Tab. E-1).

The new WEP framework not only gives Harrington's IE-CMT a clear physical picture — to construct a set of steadily working energy-decoupled modes for scattering systems, but also leads to the second transformation for the construction method of CMs — from orthogonalizing IMO method to orthogonalizing driving power operator (DPO) method (as shown in Tab. E-1). Employing the physical picture, this paper will clarify some confusions existing in CMT for a long time. This paper will exhibit the fact that the new orthogonalizing DPO method has a more satisfactory numerical performance in the aspect of constructing CMs.

In addition, why the characteristic equation for complicated OSSs usually outputs some spurious modes and how to suppress the spurious modes are also the hot issues in recent years. Alroughani *et al.*[30] found out that the classical Poggio-Miller-Chang-Harrington-Wu-Tsai (PMCHWT) based surface CM formulation usually outputs some spurious modes. Alroughani *et al.*[30] and Miers *et al.*[253,254] proposed some methods for recognizing and filtering the spurious modes. Chen *et al.*[16,31] and some other researchers[26,33] attributed the spurious modes to overlooking the dependence relation between equivalent surface electric current $\boldsymbol{J}^{\text{ES}}$ and magnetic current $\boldsymbol{M}^{\text{ES}}$, and proposed some schemes to suppress the spurious modes by employing the transformations between $\boldsymbol{J}^{\text{ES}}$ and $\boldsymbol{M}^{\text{ES}}$, but the transformations proposed by Chen *et al.*[16,31] are essentially different from the ones proposed in [26,33]. In [26], the transformations were derived from the tangential continuation conditions of electromagnetic (EM) fields. In [33], the transformations were derived from the definitions for $\boldsymbol{J}^{\text{ES}}$ and $\boldsymbol{M}^{\text{ES}}$. Recently, [13] further proved that the transformations employed in [26] and [33] and then the suppression schemes proposed in [26] and [33] are essentially equivalent to each other. Guo and Xia *et al.*[34] also attributed the spurious modes to overlooking the dependence relation between $\boldsymbol{J}^{\text{ES}}$ and $\boldsymbol{M}^{\text{ES}}$, and proposed some alternative schemes for relating $\boldsymbol{J}^{\text{ES}}$ and $\boldsymbol{M}^{\text{ES}}$, by employing an intermediate variable — effective current, and the concept of effective current can be found in [255~258].

The spurious mode suppression schemes proposed in [13,26,33] need to inverse some full matrices related to first-kind[13,26,33] or second-kind[26] Fredholm's integral operator. But, the process to compute the inversion of full matrix requires a large CPU time or computer memory[259]. This paper proposes a new spurious mode suppression

scheme — solution domain compression (SDC) — which doesn't need to inverse any matrix, and the SDC scheme can be easily integrated into the WEP-based CMT (WEP-CMT).

This paper is organized as follows: Sec. E4 simply reviews the physical principle of WEP-CMT, where the physical picture of WEP-CMT is highlighted and some confusions on CMT are clarified; Secs. E5 and E6 propose a novel SDC scheme and its improved version for suppressing spurious modes, and the schemes don't need to inverse any matrix; Secs. E5 and E6 also do some necessary comparisons among different CM construction methods, and clearly exhibit the advantages of new orthogonalizing DPO method[13] over traditional orthogonalizing IMO method[8,9,11,12]; Sec. E7 concludes this paper.

In what follows, the $e^{j\omega t}$ convention is used throughout, and the time-domain quantities will be added time variable $t$ explicitly for example $q(t)$, but the frequency-domain quantities will not for example $q$. In addition, as known to all, for the linear quantities, we have $q(t) = \mathrm{Re}\{qe^{j\omega t}\}$; for the power-type quadratic quantities, we have $\mathrm{Re}\{q\} = (1/T)\int_0^T q(t)dt$, where $T$ is the period of the time-harmonic EM field[24,25].

## E4 Volume Formulation of the WEP-CMT for Material Scattering Systems

This section considers the EM scattering problem shown in Fig. E-1. In the figure, $V_{\rm OSS}$ is a material OSS with parameters $\boldsymbol{\mu}$ and $\boldsymbol{\varepsilon}^{\rm c} = \boldsymbol{\varepsilon} - j\boldsymbol{\sigma}/\omega$, where magnetic permeability $\boldsymbol{\mu}$, dielectric permittivity $\boldsymbol{\varepsilon}$, and electric conductivity $\boldsymbol{\sigma}$ are real and two-order symmetrical dyads depending on spatial variable $\boldsymbol{r}$ but independent of time variable $t$; $D_{\rm env}$ denotes the external environment surrounding $V_{\rm OSS}$, and it can be free space or not; $D_{\rm imp}$ is an externally impressed source and generates an EM field $\boldsymbol{F}^{\rm imp}$ in whole three-dimensional Euclidean space $\mathbb{E}^3$, where $\boldsymbol{F}^{\rm imp}$ is the abbreviated form of EM fields $(\boldsymbol{E}^{\rm imp}, \boldsymbol{H}^{\rm imp})$.

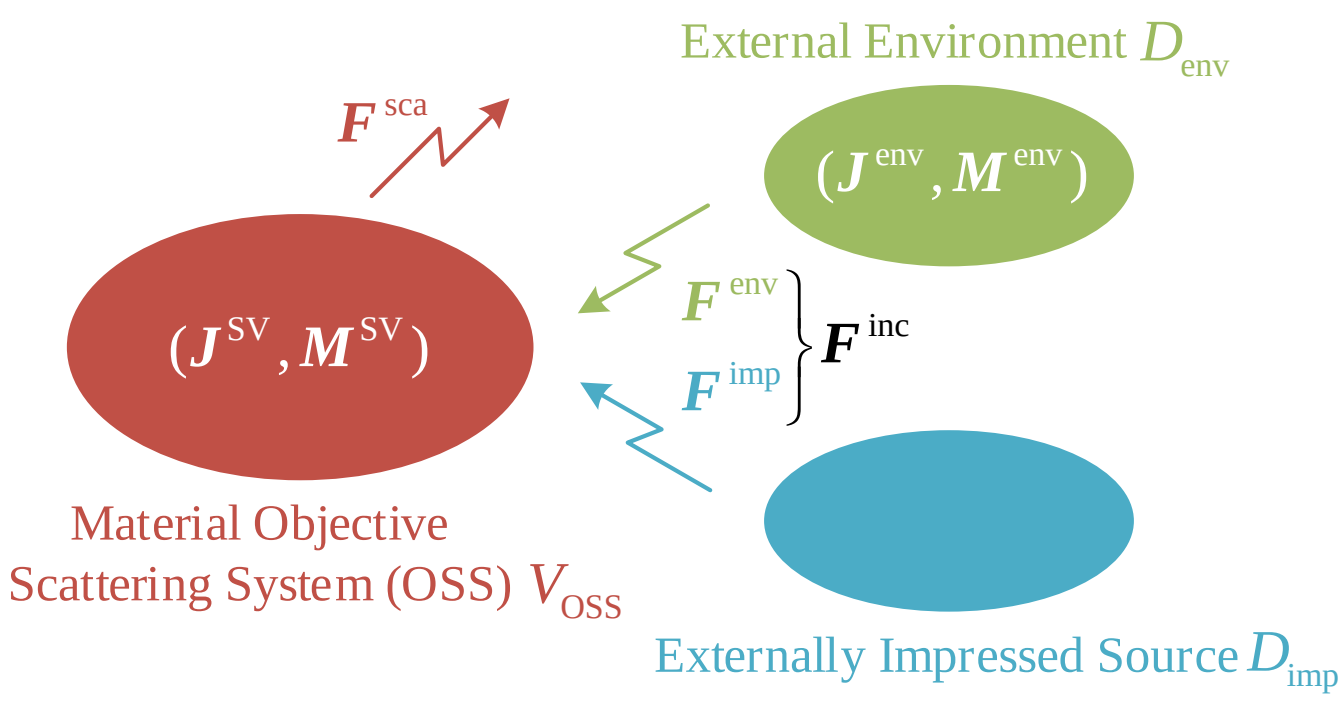

Figure E-1 EM scattering problem considered in Sec. E4

Due to the existence of $\boldsymbol{F}^{\text{imp}}$, a scattered volume current $\boldsymbol{C}^{\text{SV}}$ and a current $\boldsymbol{C}^{\text{env}}$ are induced on $V_{\text{OSS}}$ and $D_{\text{env}}$ respectively, and then a scattered field $\boldsymbol{F}^{\text{sca}}$ and an environment field $\boldsymbol{F}^{\text{env}}$ are generated in $\mathbb{E}^3$ by $\boldsymbol{C}^{\text{SV}}$ and $\boldsymbol{C}^{\text{env}}$ correspondingly, where $\boldsymbol{C}^{\text{SV/env}}$ is the abbreviated form of EM currents $(\boldsymbol{J}^{\text{SV/env}},\boldsymbol{M}^{\text{SV/env}})$. Based on superposition principle[24,260,261], $\boldsymbol{C}^{\text{SV}}$ and then $\boldsymbol{F}^{\text{sca}}$ can be viewed as the result of the excitation from two external fields $\boldsymbol{F}^{\text{imp}}$ and $\boldsymbol{F}^{\text{env}}$, so the two external fields are treated as a whole — incident field $\boldsymbol{F}^{\text{inc}}$ (or alternatively called externally resultant field) — in this paper, that is, $\boldsymbol{F}^{\text{inc}}=\boldsymbol{F}^{\text{imp}}+\boldsymbol{F}^{\text{env}}$. In addition, if the summation of $\boldsymbol{F}^{\text{inc}}$ and $\boldsymbol{F}^{\text{sca}}$ is denoted as total field $\boldsymbol{F}^{\text{tot}}$, that is, $\boldsymbol{F}^{\text{tot}}=\boldsymbol{F}^{\text{inc}}+\boldsymbol{F}^{\text{sca}}$, then it exists that $\boldsymbol{J}^{\text{SV}}=j\omega\Delta\boldsymbol{\varepsilon}^{\text{c}}\cdot\boldsymbol{E}^{\text{tot}}$ and $\boldsymbol{M}^{\text{SV}}=j\omega\Delta\boldsymbol{\mu}\cdot\boldsymbol{H}^{\text{tot}}$ due to volume equivalence principle[21,28], where $\Delta\boldsymbol{\varepsilon}^{\text{c}}=\boldsymbol{\varepsilon}^{\text{c}}-\mathbf{I}\varepsilon_0$ and $\Delta\boldsymbol{\mu}=\boldsymbol{\mu}-\mathbf{I}\mu_0$ and $\mathbf{I}$ is two-order unit dyad (that is, $\mathbf{I}=\hat{\boldsymbol{x}}\hat{\boldsymbol{x}}+\hat{\boldsymbol{y}}\hat{\boldsymbol{y}}+\hat{\boldsymbol{z}}\hat{\boldsymbol{z}}$).

In the following parts of this section, by focusing on the volume formulation of WEP-CMT, we will discuss the carrying framework of WEP-CMT, the generating operator of CMs, the construction method of CMs, the physical picture of WEP-CMT, some characteristic quantities of CM, and the concepts of physical modes and unphysical/spurious modes.

## E4.1 Carrying Framework of WEP-CMT

Based on superposition principle $\boldsymbol{F}^{\text{tot}}=\boldsymbol{F}^{\text{inc}}+\boldsymbol{F}^{\text{sca}}$, volume equivalence principle $\boldsymbol{J}^{\text{SV}}=j\omega\Delta\boldsymbol{\varepsilon}^{\text{c}}\cdot\boldsymbol{E}^{\text{tot}}$ & $\boldsymbol{M}^{\text{SV}}=j\omega\Delta\boldsymbol{\mu}\cdot\boldsymbol{H}^{\text{tot}}$, and Maxwell's equations $\nabla\times\boldsymbol{H}^{\text{sca}}=\boldsymbol{J}^{\text{SV}}+j\omega\varepsilon_0\boldsymbol{E}^{\text{sca}}$ & $\nabla\times\boldsymbol{E}^{\text{sca}}=-\boldsymbol{M}^{\text{SV}}-j\omega\mu_0\boldsymbol{H}^{\text{sca}}$ [21], it is not difficult to prove that

$$\mathcal{W}^{\text{Driv}}=\mathcal{E}^{\text{rad}}_{S_\infty}+\mathcal{E}^{\text{dis}}_{V_{\text{OSS}}}+\Delta\left(\mathcal{E}^{\text{mag}}_{\mathbb{E}^3}+\mathcal{E}^{\text{ele}}_{\mathbb{E}^3}\right)+\Delta\left(\mathcal{E}^{\text{mag}}_{V_{\text{OSS}}}+\mathcal{E}^{\text{pol}}_{V_{\text{OSS}}}\right)\tag{E-1}$$

The explicit expressions for the terms in (E-1) are given in (E-2)~(E-6).

$$\mathcal{W}^{\text{Driv}}=\int_{t_0}^{t_0+\Delta t}\left[\left\langle\boldsymbol{J}^{\text{SV}}(t),\boldsymbol{E}^{\text{inc}}(t)\right\rangle_{V_{\text{OSS}}}+\left\langle\boldsymbol{M}^{\text{SV}}(t),\boldsymbol{H}^{\text{inc}}(t)\right\rangle_{V_{\text{OSS}}}\right]dt\tag{E-2}$$

$$\mathcal{E}^{\text{rad}}_{S_\infty}=\int_{t_0}^{t_0+\Delta t}\left\{\oiint_{S_\infty}\left[\boldsymbol{E}^{\text{sca}}(t)\times\boldsymbol{H}^{\text{sca}}(t)\right]\cdot\hat{\boldsymbol{n}}^+_\infty dS\right\}dt\tag{E-3}$$

$$\mathcal{E}^{\text{dis}}_{V_{\text{OSS}}}=\int_{t_0}^{t_0+\Delta t}\left\langle\boldsymbol{\sigma}\cdot\boldsymbol{E}^{\text{tot}}(t),\boldsymbol{E}^{\text{tot}}(t)\right\rangle_{V_{\text{OSS}}}dt\tag{E-4}$$

$$\begin{aligned}\Delta\left(\mathcal{E}^{\text{mag}}_{\mathbb{E}^3}+\mathcal{E}^{\text{ele}}_{\mathbb{E}^3}\right)=&\left[\frac{1}{2}\left\langle\boldsymbol{H}^{\text{sca}}(t_0+\Delta t),\mu_0\boldsymbol{H}^{\text{sca}}(t_0+\Delta t)\right\rangle_{\mathbb{E}^3}+\frac{1}{2}\left\langle\varepsilon_0\boldsymbol{E}^{\text{sca}}(t_0+\Delta t),\boldsymbol{E}^{\text{sca}}(t_0+\Delta t)\right\rangle_{\mathbb{E}^3}\right]\\&-\left[\frac{1}{2}\left\langle\boldsymbol{H}^{\text{sca}}(t_0),\mu_0\boldsymbol{H}^{\text{sca}}(t_0)\right\rangle_{\mathbb{E}^3}+\frac{1}{2}\left\langle\varepsilon_0\boldsymbol{E}^{\text{sca}}(t_0),\boldsymbol{E}^{\text{sca}}(t_0)\right\rangle_{\mathbb{E}^3}\right]\end{aligned}\tag{E-5}$$

$$\Delta\left(\mathcal{E}_{V_{\rm OSS}}^{\rm mag}+\mathcal{E}_{V_{\rm OSS}}^{\rm pol}\right)= \left[\frac{1}{2}\left\langle \boldsymbol{H}^{\rm tot}(t_0+\Delta t),\Delta\boldsymbol{\mu}\cdot\boldsymbol{H}^{\rm tot}(t_0+\Delta t)\right\rangle_{V_{\rm OSS}}+\frac{1}{2}\left\langle\Delta\boldsymbol{\varepsilon}\cdot\boldsymbol{E}^{\rm tot}(t_0+\Delta t),\boldsymbol{E}^{\rm tot}(t_0+\Delta t)\right\rangle_{V_{\rm OSS}}\right]$$
$$-\left[\frac{1}{2}\left\langle \boldsymbol{H}^{\rm tot}(t_0),\Delta\boldsymbol{\mu}\cdot\boldsymbol{H}^{\rm tot}(t_0)\right\rangle_{V_{\rm OSS}}+\frac{1}{2}\left\langle\Delta\boldsymbol{\varepsilon}\cdot\boldsymbol{E}^{\rm tot}(t_0),\boldsymbol{E}^{\rm tot}(t_0)\right\rangle_{V_{\rm OSS}}\right] \quad \text{(E-6)}$$

In (E-2)~(E-6), time interval $\Delta t$ is a positive real number; the inner product is defined as that $<\boldsymbol{f},\boldsymbol{g}>_\Omega=\int_\Omega \boldsymbol{f}^*\cdot\boldsymbol{g}d\Omega$; $S_\infty$ is a spherical surface with infinite radius; $\hat{\boldsymbol{n}}_\infty^+$ is the outer normal direction of $S_\infty$.

Equation (E-1) has a very clear physical meaning: in time interval $t_0\sim t_0+\Delta t$, the work $\mathcal{W}^{\rm Driv}$ done by $\boldsymbol{F}^{\rm inc}$ on $\boldsymbol{C}^{\rm SV}$ is transformed into four parts — the radiated energy $\mathcal{E}_{S_\infty}^{\rm rad}$ passing through $S_\infty$, the Joule heating energy $\mathcal{E}_{V_{\rm OSS}}^{\rm dis}$ dissipated in $V_{\rm OSS}$, the increment of the magnetic energy $\mathcal{E}_{\mathbb{E}^3}^{\rm mag}$ and electric energy $\mathcal{E}_{\mathbb{E}^3}^{\rm ele}$ stored in $\mathbb{E}^3$, and the increment of the magnetization energy $\mathcal{E}_{V_{\rm OSS}}^{\rm mag}$ and polarization energy $\mathcal{E}_{V_{\rm OSS}}^{\rm pol}$ stored in $V_{\rm OSS}$. Thus, equation (E-1) is a quantitative expression for the transformation between work and energy, and it is very similar to the work-energy principle in mechanism[22,23], so this paper calls it the work-energy principle (WEP) in electromagnetism.

## E4.2 Generating Operator of CMs

The work term $\mathcal{W}^{\rm Driv}$ is just the source to sustain the work-energy transformation, and also the source to drive the steady working of the material OSS. Thus, $\mathcal{W}^{\rm Driv}$ is called driving work, and the associated power is correspondingly called driving power (DP) and denoted as $P^{\rm Driv}(t)$. Obviously, $P^{\rm Driv}(t)$ has operator expression

$$P^{\rm Driv}(t)=\left\langle\boldsymbol{J}^{\rm SV}(t),\boldsymbol{E}^{\rm inc}(t)\right\rangle_{V_{\rm OSS}}+\left\langle\boldsymbol{M}^{\rm SV}(t),\boldsymbol{H}^{\rm inc}(t)\right\rangle_{V_{\rm OSS}} \quad \text{(E-7)}$$

and the operator is accordingly called time-domain driving power operator (DPO).

Two different frequency-domain versions of $P^{\rm Driv}(t)$ are as follows:

$$P^{\rm driv}=(1/2)\left\langle\boldsymbol{J}^{\rm SV},\boldsymbol{E}^{\rm inc}\right\rangle_{V_{\rm OSS}}+(1/2)\left\langle\boldsymbol{M}^{\rm SV},\boldsymbol{H}^{\rm inc}\right\rangle_{V_{\rm OSS}} \quad \text{(E-8)}$$
$$P^{\rm DRIV}=(1/2)\left\langle\boldsymbol{J}^{\rm SV},\boldsymbol{E}^{\rm inc}\right\rangle_{V_{\rm OSS}}+(1/2)\left\langle\boldsymbol{H}^{\rm inc},\boldsymbol{M}^{\rm SV}\right\rangle_{V_{\rm OSS}} \quad \text{(E-9)}$$

where coefficient 1/2 originates from the time average for the power-type quadratic quantity of time-harmonic EM field[24,25]. It is not difficult to prove that $P^{\rm DRIV}$ can be alternatively expressed as

$$P^{\rm DRIV}=(1/2)\oint\!\!\!\oint_{S_\infty}\left[\boldsymbol{E}^{\rm sca}\times\left(\boldsymbol{H}^{\rm sca}\right)^*\right]\cdot\hat{\boldsymbol{n}}_\infty^+dS+(1/2)\left\langle\boldsymbol{\sigma}\cdot\boldsymbol{E}^{\rm tot},\boldsymbol{E}^{\rm tot}\right\rangle_{V_{\rm OSS}}$$
$$+j2\omega\left\{\left[(1/4)\left\langle\boldsymbol{H}^{\rm sca},\mu_0\boldsymbol{H}^{\rm sca}\right\rangle_{\mathbb{E}^3}-(1/4)\left\langle\varepsilon_0\boldsymbol{E}^{\rm sca},\boldsymbol{E}^{\rm sca}\right\rangle_{\mathbb{E}^3}\right]\right.$$
$$\left.+\left[(1/4)\left\langle\boldsymbol{H}^{\rm tot},\Delta\boldsymbol{\mu}\cdot\boldsymbol{H}^{\rm tot}\right\rangle_{V_{\rm OSS}}-(1/4)\left\langle\Delta\boldsymbol{\varepsilon}\cdot\boldsymbol{E}^{\rm tot},\boldsymbol{E}^{\rm tot}\right\rangle_{V_{\rm OSS}}\right]\right\} \quad \text{(E-10)}$$

but $P^{\text{driv}}$ doesn't have a similar expression. In addition, it is obvious that $\text{Re}\{P^{\text{driv}}\} = \text{Re}\{P^{\text{DRIV}}\}$, but $\text{Im}\{P^{\text{driv}}\} \neq \pm \text{Im}\{P^{\text{DRIV}}\}$ when $\boldsymbol{\varepsilon} \neq \mathbf{I}\varepsilon_0$ & $\boldsymbol{\mu} \neq \mathbf{I}\mu_0$, so $P^{\text{driv}} \neq P^{\text{DRIV}}, (P^{\text{DRIV}})^*$ if $\boldsymbol{\varepsilon} \neq \mathbf{I}\varepsilon_0$ & $\boldsymbol{\mu} \neq \mathbf{I}\mu_0$, and this is just the reason to use two different superscripts "driv" and "DRIV" on $P^{\text{driv}}$ and $P^{\text{DRIV}}$ respectively.

If $\boldsymbol{C}^{\text{SV}}$ is expanded in terms of some basis functions, $P^{\text{driv}}$ and $P^{\text{DRIV}}$ can be discretized into the following matrix forms

$$P^{\text{driv/DRIV}} = \underbrace{\begin{bmatrix} \mathbb{a}^{\boldsymbol{J}^{\text{SV}}} \\ \mathbb{a}^{\boldsymbol{M}^{\text{SV}}} \end{bmatrix}^H}_{\mathbb{a}^{\text{AV}}} \cdot \mathbb{P}^{\text{driv/DRIV}} \cdot \underbrace{\begin{bmatrix} \mathbb{a}^{\boldsymbol{J}^{\text{SV}}} \\ \mathbb{a}^{\boldsymbol{M}^{\text{SV}}} \end{bmatrix}}_{\mathbb{a}^{\text{AV}}} \tag{E-11}$$

by using superposition principle $\boldsymbol{F}^{\text{inc}} = \boldsymbol{F}^{\text{tot}} - \boldsymbol{F}^{\text{sca}}$, volume equivalence principle $\boldsymbol{J}^{\text{SV}} = j\omega\Delta\boldsymbol{\varepsilon}^{\text{c}} \cdot \boldsymbol{E}^{\text{tot}}$ & $\boldsymbol{M}^{\text{SV}} = j\omega\Delta\boldsymbol{\mu} \cdot \boldsymbol{H}^{\text{tot}}$, and source-field relation $\boldsymbol{F}^{\text{sca}} = \mathbf{G}_0^{JF} * \boldsymbol{J}^{\text{SV}} + \mathbf{G}_0^{MF} * \boldsymbol{M}^{\text{SV}}$ (here $\mathbf{G}_0^{JF}$ and $\mathbf{G}_0^{MF}$ are the free-space dyadic Green's functions[28], and the convolution integral operation "$*$" is defined as $\mathbf{G} * \boldsymbol{C} = \int_\Omega \mathbf{G}(\boldsymbol{r},\boldsymbol{r}') \cdot \boldsymbol{C}(\boldsymbol{r}') d\Omega'$). In (E-11), the superscript "$H$" represents the conjugate transpose operation for a matrix or vector; $\mathbb{a}^{\boldsymbol{J}^{\text{SV}}}$ and $\mathbb{a}^{\boldsymbol{M}^{\text{SV}}}$ are the column vectors constituted by the expansion coefficients of $\boldsymbol{J}^{\text{SV}}$ and $\boldsymbol{M}^{\text{SV}}$, and superscript "AV" is the acronym of "all variables ($\boldsymbol{J}^{\text{SV}}$ and $\boldsymbol{M}^{\text{SV}}$)"; the formulations for calculating the elements of $\mathbb{P}^{\text{driv}}$ and $\mathbb{P}^{\text{DRIV}}$ can be found in [11] and [13], and they are not repeated here.

## E4.3 Construction Method of CMs

There uniquely exists the following Toeplitz's decompositions for matrices $\mathbb{P}^{\text{driv}}$ and $\mathbb{P}^{\text{DRIV}}$ [46, Sec. 0.2.5].

$$\mathbb{P}^{\text{driv/DRIV}} = \mathbb{P}_+^{\text{driv/DRIV}} + j\,\mathbb{P}_-^{\text{driv/DRIV}} \tag{E-12}$$

where $\mathbb{P}_+^{\text{driv/DRIV}}$ and $\mathbb{P}_-^{\text{driv/DRIV}}$ are the positive and negative Hermitian parts of $\mathbb{P}^{\text{driv/DRIV}}$, and $\mathbb{P}_+^{\text{driv/DRIV}} = [\mathbb{P}^{\text{driv/DRIV}} + (\mathbb{P}^{\text{driv/DRIV}})^H]/2$ and $\mathbb{P}_-^{\text{driv/DRIV}} = [\mathbb{P}^{\text{driv/DRIV}} - (\mathbb{P}^{\text{driv/DRIV}})^H]/(2j)$.

Obviously, $\mathbb{P}_+^{\text{driv/DRIV}}$ and $\mathbb{P}_-^{\text{driv/DRIV}}$ are Hermitian. In addition, $\mathbb{P}_+^{\text{driv/DRIV}}$ is positive definite, since $(\mathbb{a}^{\text{AV}})^H \cdot \mathbb{P}_+^{\text{driv/DRIV}} \cdot \mathbb{a}^{\text{AV}}$ is equal to the summation of radiated and dissipated powers. Thus, there must exist a non-singular matrix $\mathbb{A}$, such that $\mathbb{A}^H \cdot \mathbb{P}_\pm^{\text{driv/DRIV}} \cdot \mathbb{A}$ is real diagonal matrix[46, Theorem 7.6.4]. Obviously, the $\mathbb{A}$ also diagonalizes $\mathbb{P}^{\text{driv/DRIV}}$ as follows:

$$\mathrm{A}^H \cdot \mathrm{P}^{\mathrm{driv/DRIV}} \cdot \mathrm{A} = \mathrm{diag}\left\{P_1^{\mathrm{driv/DRIV}}, P_2^{\mathrm{driv/DRIV}}, \cdots\right\} \tag{E-13}$$

The column vectors of the above $\mathrm{A}$ can be derived from solving the following characteristic equation[11~13]

$$\mathrm{P}_-^{\mathrm{driv/DRIV}} \cdot \mathrm{a}_\xi^{\mathrm{AV}} = \lambda_\xi \, \mathrm{P}_+^{\mathrm{driv/DRIV}} \cdot \mathrm{a}_\xi^{\mathrm{AV}} \tag{E-14}$$

If some derived modes $(\mathrm{a}_1^{\mathrm{AV}}, \mathrm{a}_2^{\mathrm{AV}}, \cdots, \mathrm{a}_d^{\mathrm{AV}})$ are $d$-order degenerate, then the following Gram-Schmidt orthogonalization process[46, Sec. 0.6.4] is necessary.

$$\left.\begin{aligned} \mathrm{a}_1^{\mathrm{AV}} &= \mathrm{a}_1^{\mathrm{AV}\prime} \\ \mathrm{a}_2^{\mathrm{AV}} - \chi_{12}\mathrm{a}_1^{\mathrm{AV}\prime} &= \mathrm{a}_2^{\mathrm{AV}\prime} \\ &\cdots \\ \mathrm{a}_d^{\mathrm{AV}} - \cdots - \chi_{2d}\mathrm{a}_2^{\mathrm{AV}\prime} - \chi_{1d}\mathrm{a}_1^{\mathrm{AV}\prime} &= \mathrm{a}_d^{\mathrm{AV}\prime} \end{aligned}\right\} \tag{E-15}$$

where the coefficients are calculated as follows:

$$\chi_{mn} = (\mathrm{a}_m^{\mathrm{AV}\prime})^H \cdot \mathrm{P}_+^{\mathrm{driv/DRIV}} \cdot \mathrm{a}_n^{\mathrm{AV}} \Big/ (\mathrm{a}_m^{\mathrm{AV}\prime})^H \cdot \mathrm{P}_+^{\mathrm{driv/DRIV}} \cdot \mathrm{a}_m^{\mathrm{AV}\prime} \tag{E-16}$$

For a lossless circular cylinder, whose radius and height are both 6mm and which is with $\boldsymbol{\mu} = \mathrm{I}2\mu_0$, $\boldsymbol{\varepsilon} = \mathrm{I}8\varepsilon_0$, and $\boldsymbol{\sigma} = \mathrm{I}0$, its characteristic values (CVs) in decibel (dB) and the associated modal significances (MSs) derived from orthogonalizing $\mathrm{P}^{\mathrm{driv}}$ and $\mathrm{P}^{\mathrm{DRIV}}$, that is, orthogonalizing $P^{\mathrm{driv}}$ and $P^{\mathrm{DRIV}}$, are shown in Fig. E-2 and Fig. E-3 respectively. Here, the MSs are calculated as that $\mathrm{MS}_\xi = 1/|1 + j\lambda_\xi|$. Obviously, different operators $P^{\mathrm{driv}}$ and $P^{\mathrm{DRIV}}$ generate different modes.

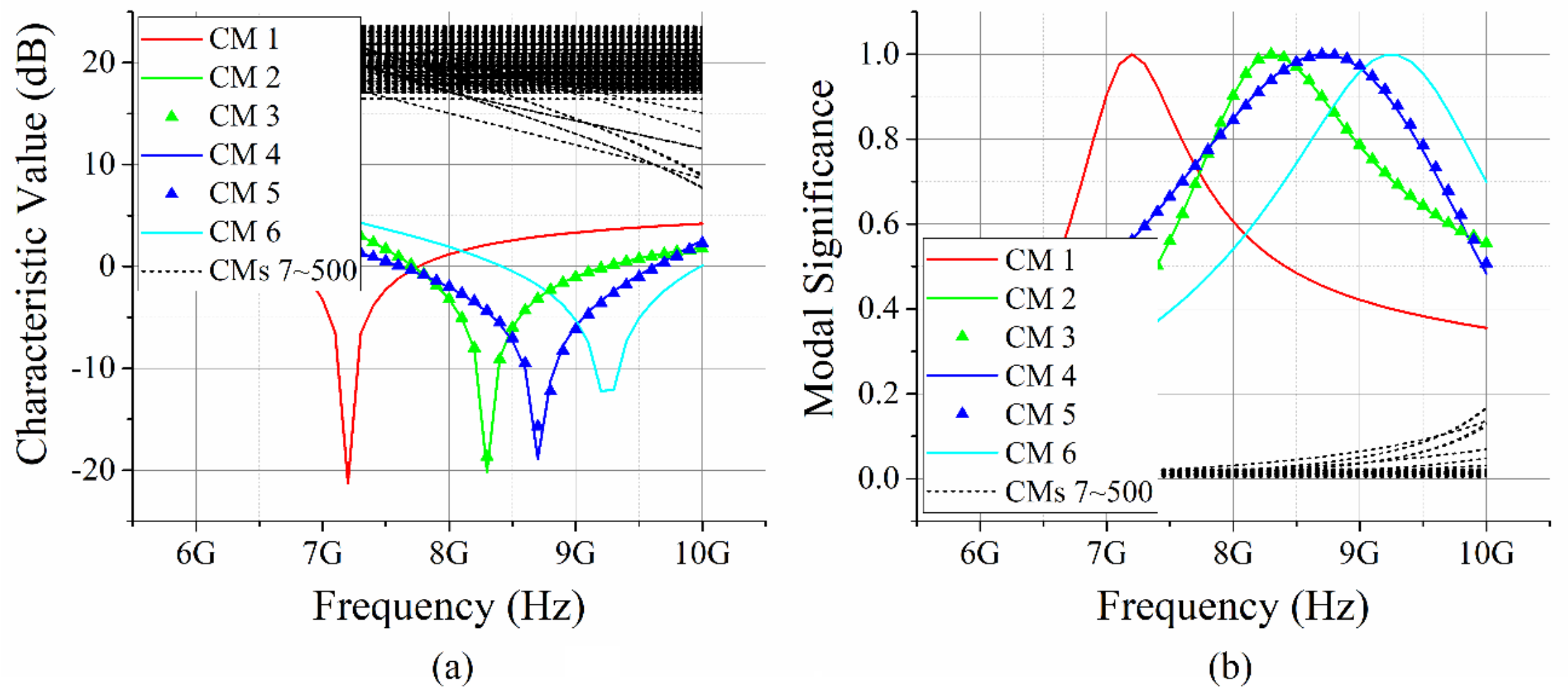

Figure E-2 $P^{\mathrm{driv}}$-based (a) CVs and (b) MSs of a lossless cylinder, whose radius and height are both 6mm and which is with $\boldsymbol{\mu} = \mathrm{I}2\mu_0$, $\boldsymbol{\varepsilon} = \mathrm{I}8\varepsilon_0$, and $\boldsymbol{\sigma} = \mathrm{I}0$

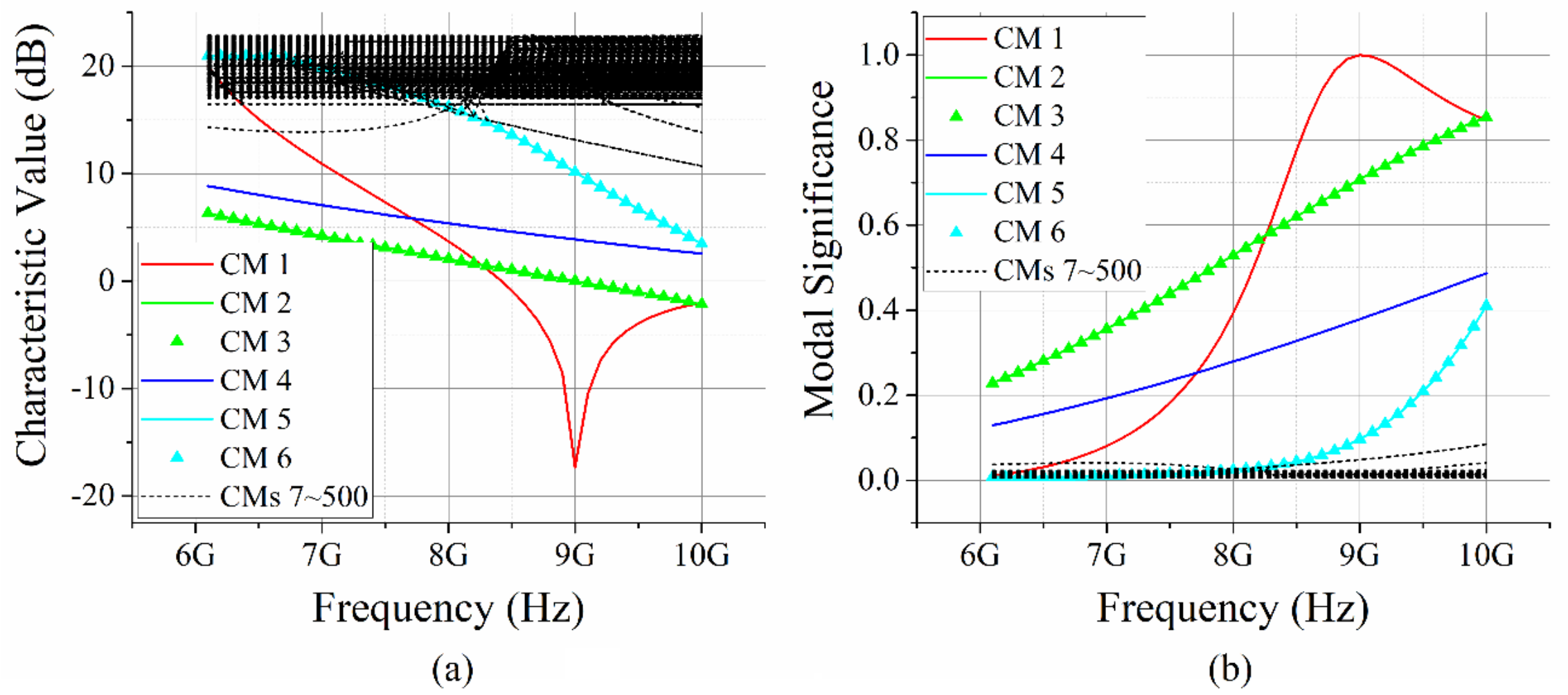

Figure E-3 $P^{\mathrm{DRIV}}$ -based (a) CVs and (b) MSs of a lossless cylinder, whose radius and height are both 6mm and which is with $\boldsymbol{\mu} = \mathbf{I}2\mu_0$, $\boldsymbol{\varepsilon} = \mathbf{I}8\varepsilon_0$, and $\boldsymbol{\sigma} = \mathbf{I}0$

## E4.4 Physical Picture of WEP-CMT

It is easy to prove that the CMs derived from orthogonalizing $\mathbb{P}^{\mathrm{driv}}$, that is, orthogonalizing $P^{\mathrm{driv}}$, satisfy orthogonality relation

$$P_\xi^{\mathrm{driv}} \delta_{\xi\zeta} = (1/2)\left\langle \boldsymbol{J}_\xi^{\mathrm{SV}}, \boldsymbol{E}_\zeta^{\mathrm{inc}} \right\rangle_{V_{\mathrm{OSS}}} + (1/2)\left\langle \boldsymbol{M}_\xi^{\mathrm{SV}}, \boldsymbol{H}_\zeta^{\mathrm{inc}} \right\rangle_{V_{\mathrm{OSS}}} \tag{E-17}$$

and the modes derived from orthogonalizing $\mathbb{P}^{\mathrm{DRIV}}$, that is, orthogonalizing $P^{\mathrm{DRIV}}$, satisfy orthogonality relation

$$P_\xi^{\mathrm{DRIV}} \delta_{\xi\zeta} = (1/2)\left\langle \boldsymbol{J}_\xi^{\mathrm{SV}}, \boldsymbol{E}_\zeta^{\mathrm{inc}} \right\rangle_{V_{\mathrm{OSS}}} + (1/2)\left\langle \boldsymbol{H}_\xi^{\mathrm{inc}}, \boldsymbol{M}_\zeta^{\mathrm{SV}} \right\rangle_{V_{\mathrm{OSS}}} \tag{E-18}$$

where $\delta_{\xi\zeta}$ is Kronecker's delta symbol.

Obviously, the CMs satisfying (E-17) are completely decoupled, that is, the action by the $\zeta$-th modal fields $(\boldsymbol{E}_\zeta^{\mathrm{inc}}, \boldsymbol{H}_\zeta^{\mathrm{inc}})$ on the $\xi$-th modal currents $(\boldsymbol{J}_\xi^{\mathrm{SV}}, \boldsymbol{M}_\xi^{\mathrm{SV}})$ is zero if $\xi \neq \zeta$. However, the modes satisfying (E-18) are not decoupled completely, as exhibited by the intertwining orthogonality (E-18) between $(\boldsymbol{E}_\zeta^{\mathrm{inc}}, \boldsymbol{M}_\zeta^{\mathrm{SV}})$ and $(\boldsymbol{J}_\xi^{\mathrm{SV}}, \boldsymbol{H}_\xi^{\mathrm{inc}})$ instead of between $(\boldsymbol{E}_\zeta^{\mathrm{inc}}, \boldsymbol{H}_\zeta^{\mathrm{inc}})$ and $(\boldsymbol{J}_\xi^{\mathrm{SV}}, \boldsymbol{M}_\xi^{\mathrm{SV}})$. Due to this reason, the following parts of this paper focus on orthogonalizing frequency-domain DPO $P^{\mathrm{driv}}$ rather than $P^{\mathrm{DRIV}}$.

In addition, the CMs satisfying frequency-domain orthogonality (E-17) also satisfy the following orthogonality relations

$$\mathrm{Re}\left\{P_\xi^{\mathrm{driv}}\right\} \delta_{\xi\zeta} = (1/T)\int_{t_0}^{t_0+T} \left[ \left\langle \boldsymbol{J}_\xi^{\mathrm{SV}}(t), \boldsymbol{E}_\zeta^{\mathrm{inc}}(t) \right\rangle_{V_{\mathrm{OSS}}} + \left\langle \boldsymbol{M}_\xi^{\mathrm{SV}}(t), \boldsymbol{H}_\zeta^{\mathrm{inc}}(t) \right\rangle_{V_{\mathrm{OSS}}} \right] dt \tag{E-19}$$

$$\text{Re}\left\{P_\xi^{\text{driv}}\right\}\delta_{\xi\zeta} = (1/2)\left\langle \frac{1}{\eta_0}\cdot \boldsymbol{E}_\xi^{\text{sca}}, \boldsymbol{E}_\zeta^{\text{sca}} \right\rangle_{S_\infty} + (1/2)\left\langle \boldsymbol{\sigma}\cdot \boldsymbol{E}_\xi^{\text{tot}}, \boldsymbol{E}_\zeta^{\text{tot}} \right\rangle_{V_{\text{OSS}}} \tag{E-20}$$

The time-domain orthogonality relation (E-19) has a very clear physical interpretation: in any integral period, the $\zeta$-th modal fields $(\boldsymbol{E}_\zeta^{\text{inc}}, \boldsymbol{H}_\zeta^{\text{inc}})$ don't supply net energy to the $\xi$-th modal currents $(\boldsymbol{J}_\xi^{\text{SV}}, \boldsymbol{M}_\xi^{\text{SV}})$, if $\xi \neq \zeta$. Then, it clearly reveals the physical picture/purpose of WEP-CMT — to construct a set of steadily working energy-decoupled modes for OSS. To visually exhibit that the $P^{\text{driv}}$-based CMs are time-averagely DP-decoupled but the $P^{\text{DRIV}}$-based modes are not, the time-average DP orthogonality matrices corresponding to the modes (at 9 GHz) in Figs. E-2 and E-3 are shown in Fig. E-4.

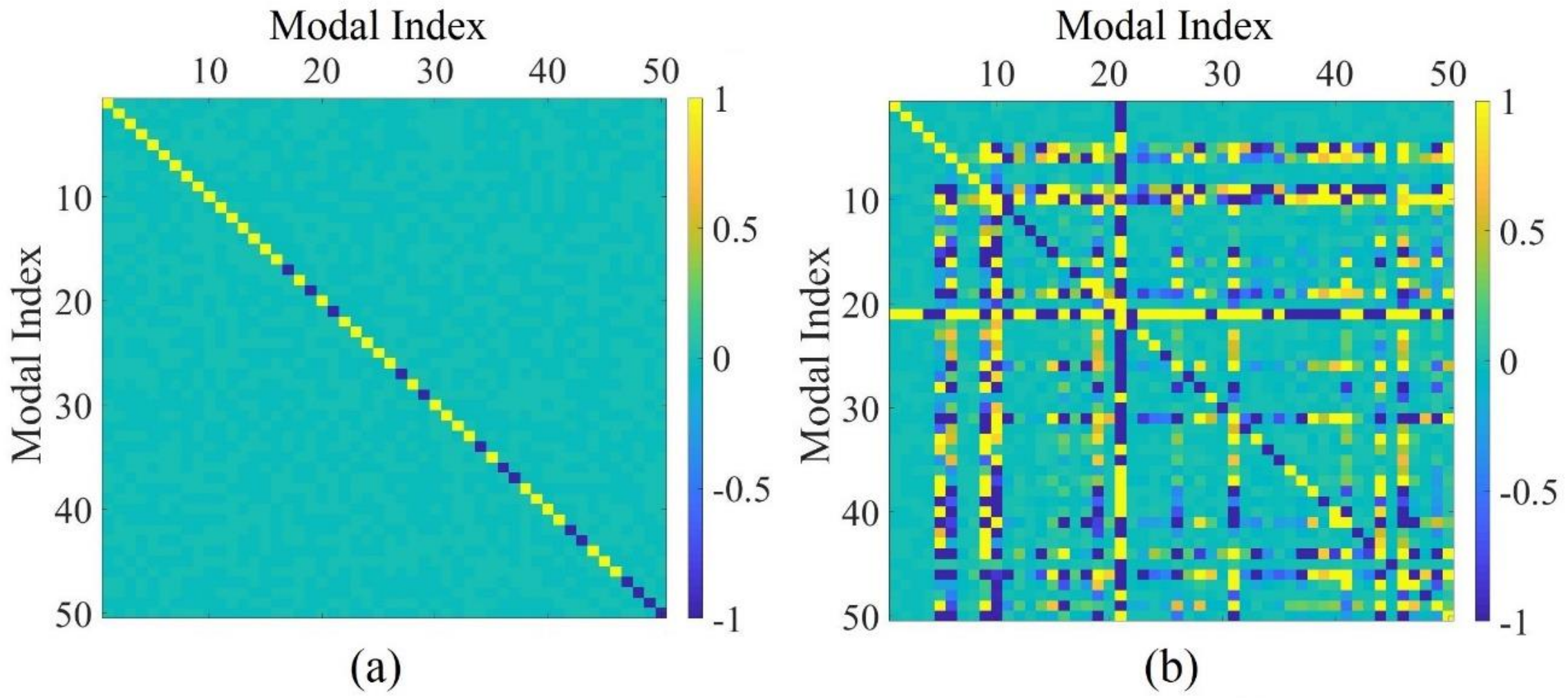

Figure E-4 Time-average DP orthogonality matrices about the modes derived from orthogonalizing frequency-domain DPOs (a) $P^{\text{driv}}$ and (b) $P^{\text{DRIV}}$, at 9 GHz

The frequency-domain orthogonality relation (E-20) clearly exhibits the fact that: when the OSS is lossy, that is, $\boldsymbol{\sigma} \neq \mathbf{I}0$, the modal far fields may not be orthogonal[13,17,26]. To visually exhibit this conclusion, we compute the modes of a lossy cylinder, whose radius and height are both 6mm and which is with $\boldsymbol{\mu} = \mathbf{I}2\mu_0$, $\boldsymbol{\varepsilon} = \mathbf{I}8\varepsilon_0$, and $\boldsymbol{\sigma} = \mathbf{I}1$, and show the corresponding CVs and MSs in Fig. E-5, and also show the time-average DP orthogonality matrix and far-field orthogonality matrix of the modes (at 9 GHz) in Fig. E-6. Obviously, the time-average DP orthogonality holds as shown in Fig. E-6(a), but the far-field orthogonality doesn't hold as shown in Fig. E-6(b). Here, we emphasize again that the physical purpose of WEP-CMT is to construct the modes without net energy exchange in integral period rather than the modes having orthogonal far fields. Two main reasons R1 and R2 leading to this conclusion are given as below.

R1. We consider two modes $(\boldsymbol{J}_1^{\text{SV}}, \boldsymbol{M}_1^{\text{SV}})$ and $(\boldsymbol{J}_2^{\text{SV}}, \boldsymbol{M}_2^{\text{SV}})$, which are driven by fields $(\boldsymbol{E}_1^{\text{inc}}, \boldsymbol{H}_1^{\text{inc}})$ and $(\boldsymbol{E}_2^{\text{inc}}, \boldsymbol{H}_2^{\text{inc}})$ respectively and generate fields $(\boldsymbol{E}_1^{\text{sca}}, \boldsymbol{H}_1^{\text{sca}})$ and $(\boldsymbol{E}_2^{\text{sca}}, \boldsymbol{H}_2^{\text{sca}})$ respectively. The modes have orthogonal far fields, but don't satisfy (E-19). Fields $(\boldsymbol{E}_1^{\text{inc}}, \boldsymbol{H}_1^{\text{inc}})$ carry information 1, and fields $(\boldsymbol{E}_2^{\text{inc}}, \boldsymbol{H}_2^{\text{inc}})$ carry information 2. Obviously, under the driving of $(\boldsymbol{E}_{1/2}^{\text{inc}}, \boldsymbol{H}_{1/2}^{\text{inc}})$, fields $(\boldsymbol{E}_{1/2}^{\text{sca}}, \boldsymbol{H}_{1/2}^{\text{sca}})$ carry the information 1 / information 2 from source zone to far zone. But, at the same time, fields $(\boldsymbol{E}_{1/2}^{\text{sca}}, \boldsymbol{H}_{1/2}^{\text{sca}})$ also carry the information 2 / information 1 from source zone to far zone, because $(\boldsymbol{E}_{1/2}^{\text{inc}}, \boldsymbol{H}_{1/2}^{\text{inc}})$ also provide energy to $(\boldsymbol{J}_{2/1}^{\text{SV}}, \boldsymbol{M}_{2/1}^{\text{SV}})$ due to the absence of (E-19). In other words, the modes not satisfying (E-19) cannot be driven independently, and then cannot achieve a real sense of decoupling for the information in far zone.

R2. Far-field orthogonality relation $< \boldsymbol{E}_\xi^{\text{sca}}, \boldsymbol{E}_\zeta^{\text{sca}} >_{S_\infty} = 0$ is a global relation rather than a local relation. Thus, it cannot be guaranteed that $(\boldsymbol{E}_\xi^{\text{sca}})^* \cdot \boldsymbol{E}_\zeta^{\text{sca}} = 0$ on the everywhere of $S_\infty$. This fact implies that: for any receiver not occupying whole $S_\infty$, relation $< \boldsymbol{E}_\xi^{\text{sca}} \cdot \boldsymbol{E}_\zeta^{\text{sca}} >_{\text{receiver}} = 0$ cannot be guaranteed. In other words, global orthogonality relation $< \boldsymbol{E}_\xi^{\text{sca}}, \boldsymbol{E}_\zeta^{\text{sca}} >_{S_\infty} = 0$ cannot provide a real sense of decoupling for the fields on any finite-sized receiver.

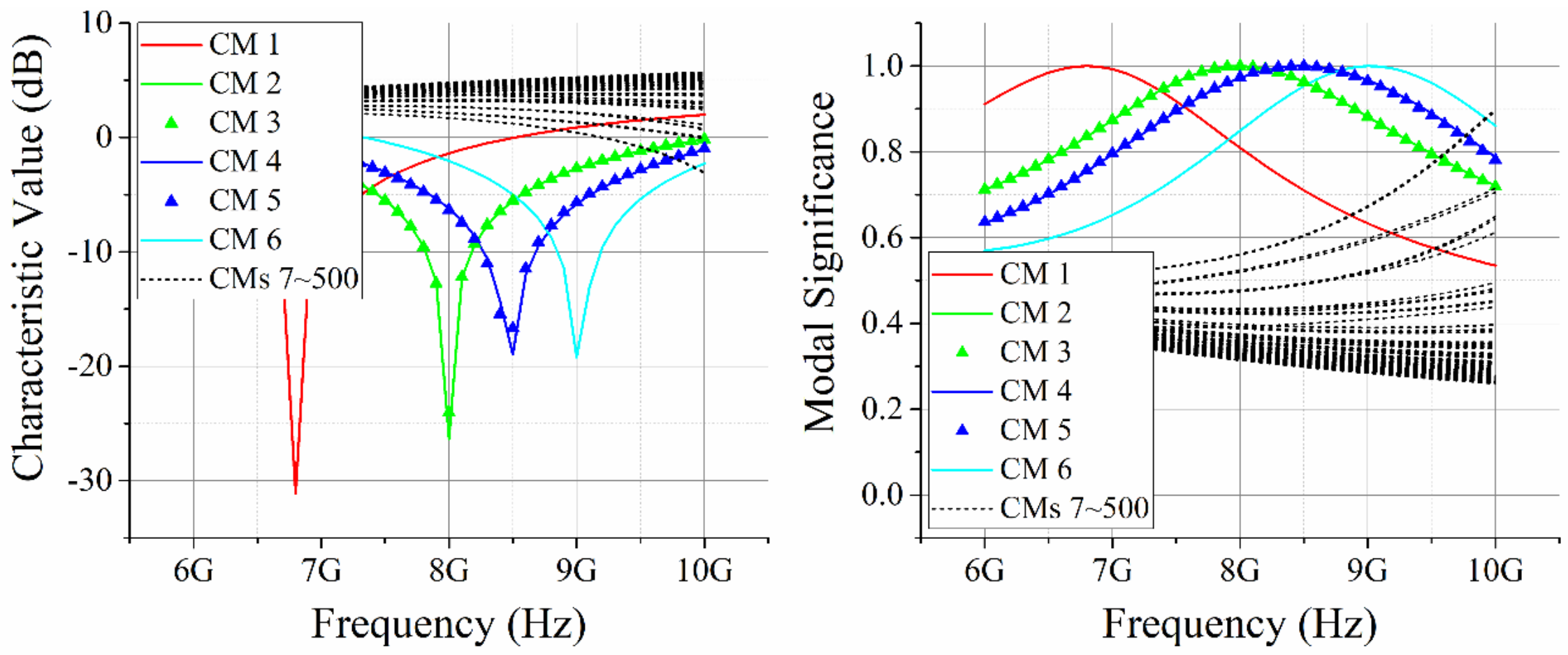

(a) (b)

Figure E-5 $P^{\text{driv}}$-based (a) CVs and (b) MSs of a lossy cylinder, whose radius and height are both 6mm and which is with $\boldsymbol{\mu} = \mathbf{I}2\mu_0$, $\boldsymbol{\varepsilon} = \mathbf{I}8\varepsilon_0$, and $\boldsymbol{\sigma} = \mathbf{I}1$

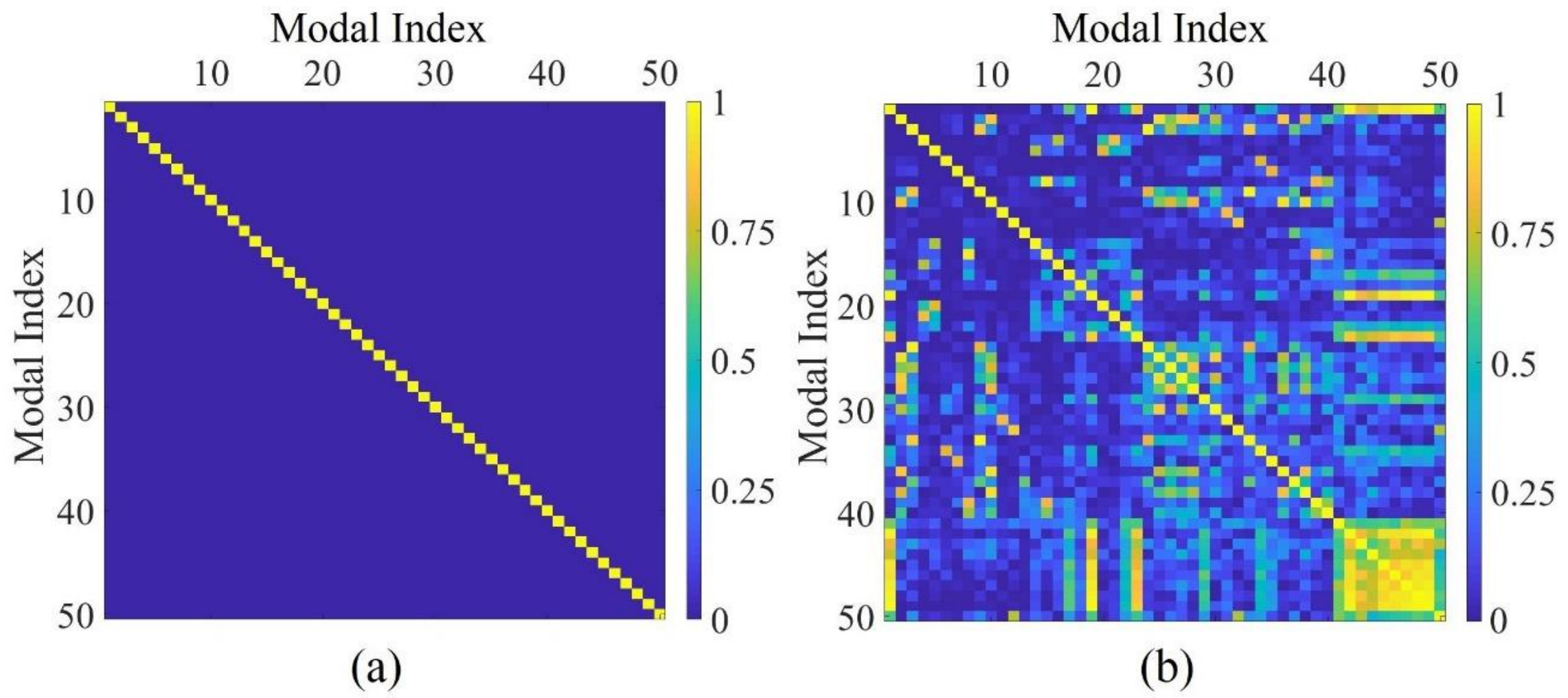

Figure E-6 (a) Time-average DP orthogonality matrix and (b) far-field orthogonality matrix of the modes (at 9 GHz) shown in Fig. E-5

## E4.5 Some Characteristic Quantities of CM

In this subsection, we focus on clarifying the concepts related to some important characteristic quantities of CM.

**Characteristic Value (CV)**

As everyone knows, CV $\lambda_\xi$ and modal DP $P_\xi^{\text{driv}}$ satisfy the following relation

$$\lambda_\xi = \text{Im}\left\{P_\xi^{\text{driv}}\right\} \big/ \text{Re}\left\{P_\xi^{\text{driv}}\right\} \tag{E-21}$$

Here, the physical meaning of $\text{Re}\{P_\xi^{\text{driv}}\}$ has been very clear — “the time average of modal driving power” / “the summation of modal radiated power and modal dissipated power”, but the physical meaning of $\text{Im}\{P_\xi^{\text{driv}}\}$ has not been clear (though the physical meaning of $\text{Im}\{P_\xi^{\text{DRIV}}\}$ is relatively clear as shown in (E-10)), and then the physical meaning of $\lambda_\xi$ has not been clear too. In addition, because $\text{Im}\{P_\xi^{\text{driv}}\} \neq \pm \text{Im}\{P_\xi^{\text{DRIV}}\}$ generally for magnetodielectric OSSs (which satisfy $\boldsymbol{\mu} \neq \mathbf{I}\mu_0$ & $\boldsymbol{\varepsilon} \neq \mathbf{I}\varepsilon_0$), thus the zero CVs don’t necessarily correspond to the resonant CMs, that is, $P^{\text{driv}}$-based $\lambda_\xi = 0$ doesn’t imply $\text{Im}\{P_\xi^{\text{DRIV}}\} = 0$, for magnetodielectric OSSs.

Very naturally, the above observations lead to the following questions Q1 and Q2.

Q1 What is the physical meaning of the $P^{\text{driv}}$-based CVs?

Q2 When the OSS is magnetodielectric, what do the $P^{\text{driv}}$-based zero CVs imply physically?

The answer to the question Q1 is as follows: in fact, this paper thinks that there is no need to expect $\lambda_\xi$ to have a clear physical meaning, because the CVs are only the byproducts during the process to construct CMs (the physical purpose of WEP-CMT is to construct the CMs, rather than to obtain the associated CVs). The answer to the question Q2 will be clear after our clarifying the physical meaning of characteristic quantity MS under WEP framework.

**Modal Normalization Factor**

Any working mode $(\mathbb{a},\boldsymbol{C}^{\rm SV},\boldsymbol{F}^{\rm inc})$ under the driving of a previously known incident field can be linearly expanded in terms of the CMs $\{\mathbb{a}_\xi,\boldsymbol{C}_\xi^{\rm SV},\boldsymbol{F}_\xi^{\rm inc}\}$ as follows:

$$\left.\begin{aligned} \mathbb{a} &= \sum\nolimits_\xi c_\xi \mathbb{a}_\xi \\ \boldsymbol{C}^{\rm SV} &= \sum\nolimits_\xi c_\xi \boldsymbol{C}_\xi^{\rm SV} \\ \boldsymbol{F}^{\rm inc} &= \sum\nolimits_\xi c_\xi \boldsymbol{F}_\xi^{\rm inc} \end{aligned}\right\} \tag{E-22}$$

with expansion coefficients

$$c_\xi = \frac{1}{P_\xi^{\rm driv}}\Big[(1/2)\big\langle \boldsymbol{J}_\xi^{\rm SV},\boldsymbol{E}^{\rm inc}\big\rangle_{V_{\rm OSS}} + (1/2)\big\langle \boldsymbol{M}_\xi^{\rm SV},\boldsymbol{H}^{\rm inc}\big\rangle_{V_{\rm OSS}}\Big] \tag{E-23}$$

because of the modal orthogonality (E-17) and completeness.

If we want $c_\xi$ to have ability to quantitatively reflect the weight of a CM in whole modal expansion formulation, it is necessary to appropriately normalize the corresponding CM. To find a reasonable normalization factor, we employ the time-domain orthogonality relation (E-19) due to its clear physical meaning. From (E-19), it is easy to derive the following time-average DP expansion formulation

$$\begin{aligned} &\frac{1}{T}\int_{t_0}^{t_0+T}\Big[\big\langle \boldsymbol{J}^{\rm SV}(t),\boldsymbol{E}^{\rm inc}(t)\big\rangle_{V_{\rm OSS}} + \big\langle \boldsymbol{M}^{\rm SV}(t),\boldsymbol{H}^{\rm inc}(t)\big\rangle_{V_{\rm OSS}}\Big]dt \\ &= {\rm Re}\big\{P^{\rm driv}\big\} \\ &= \sum\nolimits_\xi |c_\xi|^2\,{\rm Re}\big\{P_\xi^{\rm driv}\big\} \end{aligned} \tag{E-24}$$

where the first equality is evident[24,25], and the second equality is due to expansion formulation (E-22) and orthogonality relation (E-17). From time-average DP expansion formulation (E-24), it is easy to conclude that: a reasonable modal normalization way is to normalize ${\rm Re}\{P_\xi^{\rm driv}\}$ to 1, that is, a reasonable modal normalization factor is the square root of time-average modal DP ${\rm Re}\{P_\xi^{\rm driv}\}$.

By using $\sqrt{{\rm Re}\{P_\xi^{\rm driv}\}}$ as the modal normalization factor, it traditionally has that[8,9,11,12]

$$P_{\xi}^{\text{driv}} = \underbrace{\text{Re}\left\{P_{\xi}^{\text{driv}}\right\}}_{1} + j\,\underbrace{\text{Im}\left\{P_{\xi}^{\text{driv}}\right\}}_{\lambda_{\xi}} \tag{E-25}$$

and hence we immediately obtain the following famous Parseval's identity

$$\begin{aligned} &\frac{1}{T}\int_{t_0}^{t_0+T}\left[\left\langle \boldsymbol{J}^{\text{SV}}(t), \boldsymbol{E}^{\text{inc}}(t)\right\rangle_{V_{\text{OSS}}} + \left\langle \boldsymbol{M}^{\text{SV}}(t), \boldsymbol{H}^{\text{inc}}(t)\right\rangle_{V_{\text{OSS}}}\right]dt \\ &= \sum\nolimits_{\xi}\left|c_{\xi}\right|^{2} \end{aligned} \tag{E-26}$$

Substituting the above (E-25) into expansion coefficient (E-23), it is immediate to have that

$$c_{\xi} = \frac{1}{1+j\lambda_{\xi}}\left[(1/2)\left\langle \boldsymbol{J}_{\xi}^{\text{SV}}, \boldsymbol{E}^{\text{inc}}\right\rangle_{V_{\text{OSS}}} + (1/2)\left\langle \boldsymbol{M}_{\xi}^{\text{SV}}, \boldsymbol{H}^{\text{inc}}\right\rangle_{V_{\text{OSS}}}\right] \tag{E-27}$$

in which modal currents $\{\boldsymbol{J}_{\xi}^{\text{SV}}, \boldsymbol{M}_{\xi}^{\text{SV}}\}$ have been normalized such that $\text{Re}\{P_{\xi}^{\text{driv}}\} = 1$.

**Modal Significance (MS)**

Based on expansion coefficient formulation (E-27) and time-average DP expansion formulation (E-24), the following characteristic quantity — modal significance (MS) — is usually used to quantitatively describe the weight of a CM in whole modal expansion formulation[16].

$$\text{MS}_{\xi} = \frac{1}{\left|1+j\lambda_{\xi}\right|} = \frac{\text{Re}\left\{P_{\xi}^{\text{driv}}\right\}}{\left|P_{\xi}^{\text{driv}}\right|} \tag{E-28}$$

where the second equality is obvious because of (E-25). It is thus clear that, besides the above-mentioned physical meaning — modal weight of a CM in whole modal expansion formulation, $\text{MS}_{\xi}$ has another noteworthy physical meaning — weight of the $\text{Re}\{P_{\xi}^{\text{driv}}\}$ in whole modal DP $P_{\xi}^{\text{driv}}$.

In addition, it is evident that $\lambda_{\xi} = 0$ if and only if $\text{MS}_{\xi} = 1$ (because $\lambda_{\xi}$ is purely real). This implies that the CMs with $\lambda_{\xi} = 0$ have significant weights in whole expansion formulation. In fact, this is just the answer to the previous question Q2 "what do the $P^{\text{driv}}$-based zero CVs imply physically?".

Because of the clear physical meaning of MS, the following parts of this paper will always use MS rather than CV.

## E4.6 Physical Modes and Spurious Modes

The time average of modal DP is equal to the summation of modal radiated power and modal dissipated power as shown in (E-19) and (E-20), so it must be non-negative. However, there exist some negative elements distributing on the diagonal of the orthogonality matrix shown in Fig. E-4(a), that is, the negative elements correspond to the modes having negative time-average DP. This subsection focuses on explaining this phenomenon.

Due to volume equivalence principle $\boldsymbol{J}^{\text{SV}} = j\omega\Delta\boldsymbol{\varepsilon}^{\text{c}} \cdot \boldsymbol{E}^{\text{tot}}$ & $\boldsymbol{M}^{\text{SV}} = j\omega\Delta\boldsymbol{\mu}\cdot\boldsymbol{H}^{\text{tot}}$, and Maxwell's equations $\nabla\times\boldsymbol{H}^{\text{tot}} = j\omega\boldsymbol{\varepsilon}^{\text{c}}\cdot\boldsymbol{E}^{\text{tot}}$ & $\nabla\times\boldsymbol{E}^{\text{tot}} = -j\omega\boldsymbol{\mu}\cdot\boldsymbol{H}^{\text{tot}}$ [28], $\boldsymbol{J}^{\text{SV}}$ and $\boldsymbol{M}^{\text{SV}}$ satisfy the following dependence relations

$$\nabla\times\left[\left(j\omega\Delta\boldsymbol{\mu}\right)^{-1}\cdot\boldsymbol{M}^{\text{SV}}\right] = j\omega\boldsymbol{\varepsilon}\cdot\left[\left(j\omega\Delta\boldsymbol{\varepsilon}^{\text{c}}\right)^{-1}\cdot\boldsymbol{J}^{\text{SV}}\right] \tag{E-29}$$

$$\nabla\times\left[\left(j\omega\Delta\boldsymbol{\varepsilon}^{\text{c}}\right)^{-1}\cdot\boldsymbol{J}^{\text{SV}}\right] = -j\omega\boldsymbol{\mu}\cdot\left[\left(j\omega\Delta\boldsymbol{\mu}\right)^{-1}\cdot\boldsymbol{M}^{\text{SV}}\right] \tag{E-30}$$

so $\boldsymbol{J}^{\text{SV}}$ and $\boldsymbol{M}^{\text{SV}}$ are not independent of each other.

In the previous processes to construct CMs, the dependence relations (E-29) and (E-30) are not considered, such that some modes don't satisfy the relations (though some other modes indeed satisfy the relations automatically). The dependence relations (E-29) and (E-30) are essentially the Maxwell's equations, that is the physical law satisfied by $\boldsymbol{F}^{\text{tot}}$, so the modes satisfying them are called physical modes, while the modes not satisfying them are called unphysical modes (or more conventionally called spurious modes).

In the following parts of this paper, the physical modes and spurious modes are denoted as $\mathbb{a}^{\text{PhyAV}}$ and $\mathbb{a}^{\text{SpuAV}}$ respectively. Obviously, all the $\mathbb{a}^{\text{AV}}$ constitute a linear space $\{\mathbb{a}^{\text{AV}}\}$, and all the $\mathbb{a}^{\text{PhyAV}}$ also constitute a linear space — modal space $\{\mathbb{a}^{\text{PhyAV}}\}$. But, the set $\{\mathbb{a}^{\text{SpuAV}}\}$ constituted by all the $\mathbb{a}^{\text{SpuAV}}$ is not a linear space, because the set is not closed for addition, for example, both $\mathbb{a}^{\text{PhyAV}} + \mathbb{a}^{\text{SpuAV}}$ and $-\mathbb{a}^{\text{SpuAV}}$ are spurious but their summation $(\mathbb{a}^{\text{PhyAV}} + \mathbb{a}^{\text{SpuAV}}) + (-\mathbb{a}^{\text{SpuAV}})$ is evidently physical. In addition, it is obvious that $\{\mathbb{a}^{\text{AV}}\} = \{\mathbb{a}^{\text{PhyAV}}\} \cup \{\mathbb{a}^{\text{SpuAV}}\}$ as illustrated in Fig. E-7.

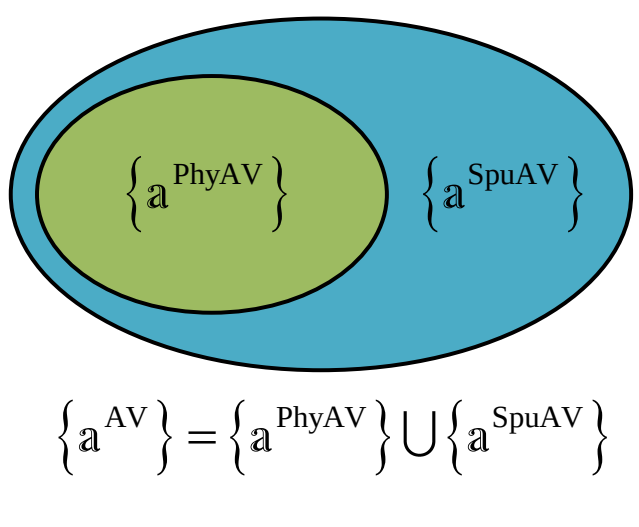

Figure E-7 Relations among space $\{\mathbb{a}^{\text{AV}}\}$, space $\{\mathbb{a}^{\text{PhyAV}}\}$, and set $\{\mathbb{a}^{\text{SpuAV}}\}$

Now, a very important question is how to effectively integrate the dependence relations between electric and magnetic currents into characteristic equation, such that all the modes outputted from the equation are physical? In the following section, we will carefully answer this question by focusing on the surface formulations of the WEP-CMT for material OSSs.

## E5 Surface Formulation of the WEP-CMT for Material Scattering Systems with SDC Scheme

This section considers the case that the OSS $V_{\rm OSS}$ is constituted by two material bodies $V_1$ and $V_2$ with parameters $(\boldsymbol{\mu}_1, \boldsymbol{\varepsilon}_1^{\rm c} = \boldsymbol{\varepsilon}_1 - j\boldsymbol{\sigma}_1/\omega)$ and $(\boldsymbol{\mu}_2, \boldsymbol{\varepsilon}_2^{\rm c} = \boldsymbol{\varepsilon}_2 - j\boldsymbol{\sigma}_2/\omega)$ respectively, as shown in Fig. E-8. The boundaries of $V_1$ and $V_2$ are denoted as $S_1$ and $S_2$ respectively, and $S_1 = S_{10} \cup S_{12}$ and $S_2 = S_{20} \cup S_{21}$. Here, $S_{10}/S_{20}$ is the interface between $V_1/V_2$ and environment, and $S_{12}/S_{21}$ is the interface between $V_1/V_2$ and $V_2/V_1$, and it is obvious that $S_{12} = S_{21}$.

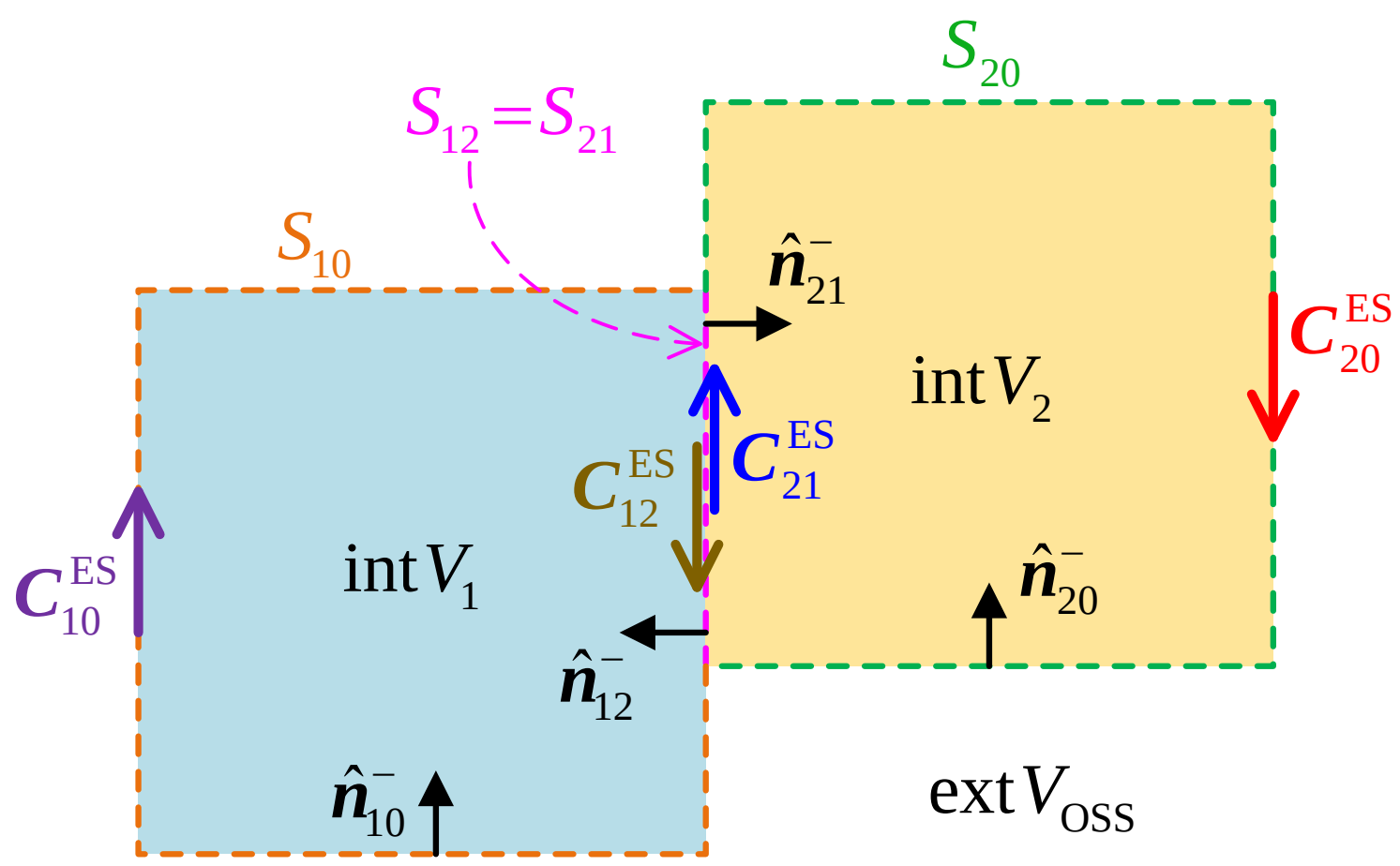

Figure E-8 Two-body material OSS considered in Sec. E5

Originating from famous Huygens-Fresnel principle[260], the following surface equivalence principles (SEPs) can be obtained

$$\left.\begin{array}{l}\text{ext}\,V_{\rm OSS}: \; -\boldsymbol{F}^{\rm sca} \\ \text{int}\,V_{\rm OSS}: \quad \boldsymbol{F}^{\rm inc}\end{array}\right\} = \mathcal{F}_0\left(\boldsymbol{J}_{10}^{\rm ES} + \boldsymbol{J}_{20}^{\rm ES}, \boldsymbol{M}_{10}^{\rm ES} + \boldsymbol{M}_{20}^{\rm ES}\right) \tag{E-31}$$

$$\text{int}\,V_1 \quad : \quad \boldsymbol{F}^{\rm tot} = \mathcal{F}_1\Big(\underbrace{\boldsymbol{J}_{10}^{\rm ES} + \boldsymbol{J}_{12}^{\rm ES}}_{\boldsymbol{J}_1^{\rm ES}}, \underbrace{\boldsymbol{M}_{10}^{\rm ES} + \boldsymbol{M}_{12}^{\rm ES}}_{\boldsymbol{M}_1^{\rm ES}}\Big) \tag{E-32}$$

$$\text{int}\,V_2 \quad : \quad \boldsymbol{F}^{\rm tot} = \mathcal{F}_2\Big(\underbrace{\boldsymbol{J}_{20}^{\rm ES} + \boldsymbol{J}_{21}^{\rm ES}}_{\boldsymbol{J}_2^{\rm ES}}, \underbrace{\boldsymbol{M}_{20}^{\rm ES} + \boldsymbol{M}_{21}^{\rm ES}}_{\boldsymbol{M}_2^{\rm ES}}\Big) \tag{E-33}$$

Here, $\text{ext}V_{\text{OSS}}$ and $\text{int}V_{\text{OSS}}$ denote the exterior and interior of $V_{\text{OSS}}$ respectively, and $\text{int}V_1$ and $\text{int}V_2$ are the interiors of $V_1$ and $V_2$ respectively; operator $\mathcal{F}_{0/1/2}$ is defined as that $\mathcal{F}_{0/1/2}\left(\boldsymbol{J},\boldsymbol{M}\right)=\mathbf{G}_{0/1/2}^{JF}*\boldsymbol{J}+\mathbf{G}_{0/1/2}^{MF}*\boldsymbol{M}$, where $\mathbf{G}_{0/1/2}^{JF}$ and $\mathbf{G}_{0/1/2}^{MF}$ are the dyadic Green's functions corresponding to material parameters $(\mu_0,\varepsilon_0)/(\boldsymbol{\mu}_1,\boldsymbol{\varepsilon}_1^{\text{c}})/(\boldsymbol{\mu}_2,\boldsymbol{\varepsilon}_2^{\text{c}})$ ; the equivalent surface electric current $\boldsymbol{J}_{10/12/21/20}^{\text{ES}}$ and equivalent surface magnetic current $\boldsymbol{M}_{10/12/21/20}^{\text{ES}}$ are defined as follows:

$$\boldsymbol{J}_{10/12/21/20}^{\text{ES}} = \hat{\boldsymbol{n}}_{10/12/21/20}^{-}\times\boldsymbol{H}_{-}^{\text{tot}} \tag{E-34}$$

$$\boldsymbol{M}_{10/12/21/20}^{\text{ES}} = \boldsymbol{E}_{-}^{\text{tot}}\times\hat{\boldsymbol{n}}_{10/12/21/20}^{-} \tag{E-35}$$

where $\hat{\boldsymbol{n}}_{10/12/21/20}^{-}$ is the inner normal direction of $S_{10/12/21/20}$ as shown in Fig. E-8, and $\boldsymbol{E}_{-}^{\text{tot}}$ and $\boldsymbol{H}_{-}^{\text{tot}}$ are the fields distributing on the inner surface of $S_{10/12/21/20}$.

In [13], it has been proved that the interaction between $(\boldsymbol{E}^{\text{inc}},\boldsymbol{H}^{\text{inc}})$ and $(\boldsymbol{J}_{1/2}^{\text{SV}},\boldsymbol{M}_{1/2}^{\text{SV}})$ and the interaction between $(\boldsymbol{E}^{\text{inc}},\boldsymbol{H}^{\text{inc}})$ and $(\boldsymbol{J}_{1/2}^{\text{ES}},\boldsymbol{M}_{1/2}^{\text{ES}})$ satisfy the following relation

$$\frac{1}{2}\left\langle\boldsymbol{J}_{1/2}^{\text{SV}},\boldsymbol{E}^{\text{inc}}\right\rangle_{V_{1/2}}+\frac{1}{2}\left\langle\boldsymbol{M}_{1/2}^{\text{SV}},\boldsymbol{H}^{\text{inc}}\right\rangle_{V_{1/2}}=-\frac{1}{2}\left\langle\boldsymbol{J}_{1/2}^{\text{ES}},\boldsymbol{E}^{\text{inc}}\right\rangle_{S_{1/2}}-\frac{1}{2}\left\langle\boldsymbol{M}_{1/2}^{\text{ES}},\boldsymbol{H}^{\text{inc}}\right\rangle_{S_{1/2}} \tag{E-36}$$

and $\boldsymbol{C}_{21}^{\text{ES}}=-\boldsymbol{C}_{12}^{\text{ES}}$, so

$$\begin{aligned}P^{\text{driv}} =& \left(1/2\right)\left\langle\boldsymbol{J}_1^{\text{SV}}+\boldsymbol{J}_2^{\text{SV}},\boldsymbol{E}^{\text{inc}}\right\rangle_{V_{\text{OSS}}}+\left(1/2\right)\left\langle\boldsymbol{M}_1^{\text{SV}}+\boldsymbol{M}_2^{\text{SV}},\boldsymbol{H}^{\text{inc}}\right\rangle_{V_{\text{OSS}}}\\ =& -\left(1/2\right)\left\langle\boldsymbol{J}_{10}^{\text{ES}}+\boldsymbol{J}_{20}^{\text{ES}},\boldsymbol{E}^{\text{inc}}\right\rangle_{S_{10}\cup S_{20}}-\left(1/2\right)\left\langle\boldsymbol{M}_{10}^{\text{ES}}+\boldsymbol{M}_{20}^{\text{ES}},\boldsymbol{H}^{\text{inc}}\right\rangle_{S_{10}\cup S_{20}}\end{aligned} \tag{E-37}$$

In fact, the above (E-37) is just the original surface formulation for the frequency-domain DPO of the material OSS $V_{\text{OSS}}$ shown in Fig. E-8.

In the following parts of this section, we will compare the numerical performances of two different surface formulations of DPO $P^{\text{driv}}$, which utilize different ways to express the tangential $\boldsymbol{F}^{\text{inc}}$ on $S_{10}\cup S_{20}$, and propose a new scheme for suppressing spurious modes.

## E5.1 PMCHWT-Based Surface CM Formulation with Two Different Spurious Mode Suppression Schemes

Because $\boldsymbol{F}^{\text{inc}}=\boldsymbol{F}^{\text{tot}}-\boldsymbol{F}^{\text{sca}}$, and the tangential components of $\boldsymbol{F}^{\text{tot}}$ and $\boldsymbol{F}^{\text{sca}}$ on $S_{10}\cup S_{20}$ are continuous, and the $\boldsymbol{F}^{\text{tot}}$ on $S_{10}^{-}\cup S_{20}^{-}$ (which is the internal surface of $S_{10}\cup S_{20}$) and the $\boldsymbol{F}^{\text{sca}}$ on $S_{10}^{+}\cup S_{20}^{+}$ (which is the external surface of $S_{10}\cup S_{20}$) can be expressed in terms of the equivalent surface currents as (E-31)~(E-33), thus

$$\begin{aligned}
P^{\text{driv}} &= -(1/2)\left\langle \boldsymbol{J}_{10}^{\text{ES}} + \boldsymbol{J}_{20}^{\text{ES}}, \boldsymbol{E}_{-}^{\text{tot}} - \boldsymbol{E}_{+}^{\text{sca}} \right\rangle_{S_{10} \cup S_{20}} \\
&\quad -(1/2)\left\langle \boldsymbol{M}_{10}^{\text{ES}} + \boldsymbol{M}_{20}^{\text{ES}}, \boldsymbol{H}_{-}^{\text{tot}} - \boldsymbol{H}_{+}^{\text{sca}} \right\rangle_{S_{10} \cup S_{20}} \\
&= -(1/2)\left\langle \boldsymbol{J}_{10}^{\text{ES}}, \mathcal{E}_1\left(\boldsymbol{J}_{10}^{\text{ES}} + \boldsymbol{J}_{12}^{\text{ES}}, \boldsymbol{M}_{10}^{\text{ES}} + \boldsymbol{M}_{12}^{\text{ES}}\right)\right\rangle_{S_{10}^{-}} \\
&\quad -(1/2)\left\langle \boldsymbol{J}_{20}^{\text{ES}}, \mathcal{E}_2\left(\boldsymbol{J}_{20}^{\text{ES}} - \boldsymbol{J}_{12}^{\text{ES}}, \boldsymbol{M}_{20}^{\text{ES}} - \boldsymbol{M}_{12}^{\text{ES}}\right)\right\rangle_{S_{20}^{-}} \\
&\quad -(1/2)\left\langle \boldsymbol{J}_{10}^{\text{ES}} + \boldsymbol{J}_{20}^{\text{ES}}, \mathcal{E}_0\left(\boldsymbol{J}_{10}^{\text{ES}} + \boldsymbol{J}_{20}^{\text{ES}}, \boldsymbol{M}_{10}^{\text{ES}} + \boldsymbol{M}_{20}^{\text{ES}}\right)\right\rangle_{S_{10}^{+} \cup S_{20}^{+}} \\
&\quad -(1/2)\left\langle \boldsymbol{M}_{10}^{\text{ES}}, \mathcal{H}_1\left(\boldsymbol{J}_{10}^{\text{ES}} + \boldsymbol{J}_{12}^{\text{ES}}, \boldsymbol{M}_{10}^{\text{ES}} + \boldsymbol{M}_{12}^{\text{ES}}\right)\right\rangle_{S_{10}^{-}} \\
&\quad -(1/2)\left\langle \boldsymbol{M}_{20}^{\text{ES}}, \mathcal{H}_2\left(\boldsymbol{J}_{20}^{\text{ES}} - \boldsymbol{J}_{12}^{\text{ES}}, \boldsymbol{M}_{20}^{\text{ES}} - \boldsymbol{M}_{12}^{\text{ES}}\right)\right\rangle_{S_{20}^{-}} \\
&\quad -(1/2)\left\langle \boldsymbol{M}_{10}^{\text{ES}} + \boldsymbol{M}_{20}^{\text{ES}}, \mathcal{H}_0\left(\boldsymbol{J}_{10}^{\text{ES}} + \boldsymbol{J}_{20}^{\text{ES}}, \boldsymbol{M}_{10}^{\text{ES}} + \boldsymbol{M}_{20}^{\text{ES}}\right)\right\rangle_{S_{10}^{+} \cup S_{20}^{+}} \qquad \text{(E-38)}
\end{aligned}$$

where relation $\boldsymbol{C}_{21}^{\text{ES}} = -\boldsymbol{C}_{12}^{\text{ES}}$ has been utilized.

If the related equivalent surface currents are expanded in terms of some basis functions, then the $P^{\text{driv}}$ given in (E-38) can be discretized into the following matrix form

$$P^{\text{driv}} = \underbrace{\begin{bmatrix} \mathbb{a}^{\boldsymbol{J}_{10}^{\text{ES}}} \\ \mathbb{a}^{\boldsymbol{J}_{12}^{\text{ES}}} \\ \mathbb{a}^{\boldsymbol{J}_{20}^{\text{ES}}} \\ \mathbb{a}^{\boldsymbol{M}_{10}^{\text{ES}}} \\ \mathbb{a}^{\boldsymbol{M}_{12}^{\text{ES}}} \\ \mathbb{a}^{\boldsymbol{M}_{20}^{\text{ES}}} \end{bmatrix}^H}_{\mathbb{a}^{\text{AV}}} \cdot \left[(1/2)\,\mathbb{Z}^{\text{PMCHWT}}\right] \cdot \underbrace{\begin{bmatrix} \mathbb{a}^{\boldsymbol{J}_{10}^{\text{ES}}} \\ \mathbb{a}^{\boldsymbol{J}_{12}^{\text{ES}}} \\ \mathbb{a}^{\boldsymbol{J}_{20}^{\text{ES}}} \\ \mathbb{a}^{\boldsymbol{M}_{10}^{\text{ES}}} \\ \mathbb{a}^{\boldsymbol{M}_{12}^{\text{ES}}} \\ \mathbb{a}^{\boldsymbol{M}_{20}^{\text{ES}}} \end{bmatrix}}_{\mathbb{a}^{\text{AV}}} \qquad \text{(E-39)}$$

Here, the sub-vectors of $\mathbb{a}^{\text{AV}}$ are constituted by the expansion coefficients of the corresponding currents; the formulations for calculating the elements of $\mathbb{Z}^{\text{PMCHWT}}$ can be found in [13].

If the dependence relations among the currents are ignored, then the modes of $V_{\text{OSS}}$ can be derived from directly solving characteristic equation $\mathbb{Z}_{-}^{\text{PMCHWT}} \cdot \mathbb{a}_{\xi}^{\text{AV}} = \lambda_{\xi}\, \mathbb{Z}_{+}^{\text{PMCHWT}} \cdot \mathbb{a}_{\xi}^{\text{AV}}$ where $\mathbb{Z}_{+}^{\text{PMCHWT}}$ and $\mathbb{Z}_{-}^{\text{PMCHWT}}$ are the positive and negative Hermitian parts of $\mathbb{Z}^{\text{PMCHWT}}$.

Now, we consider a two-body OSS shown in Fig. E-9, and the OSS is constituted by two circular cylinders, whose radiuses are both 6mm and heights are both 3mm. When the OSS is lossless and with parameters $(\boldsymbol{\mu}_1 = \mathbf{I}2\mu_0, \boldsymbol{\varepsilon}_1^{\text{c}} = \mathbf{I}8\varepsilon_0)$ and $(\boldsymbol{\mu}_2 = \mathbf{I}8\mu_0, \boldsymbol{\varepsilon}_2^{\text{c}} = \mathbf{I}2\varepsilon_0)$, the MSs of the first 500 CMs derived from volume formulation and the above PMCHWT-based surface formulation $\mathbb{Z}_{-}^{\text{PMCHWT}} \cdot \mathbb{a}_{\xi}^{\text{AV}} = \lambda_{\xi}\, \mathbb{Z}_{+}^{\text{PMCHWT}} \cdot \mathbb{a}_{\xi}^{\text{AV}}$ are shown in Fig. E-10(a) and Fig. E-10(c). When the OSS is lossy and with parameters $(\boldsymbol{\mu}_1 = \mathbf{I}2\mu_0, \boldsymbol{\varepsilon}_1^{\text{c}} = \mathbf{I}8\varepsilon_0 - j\mathbf{I}1/\omega)$ and $(\boldsymbol{\mu}_2 = \mathbf{I}8\mu_0, \boldsymbol{\varepsilon}_2^{\text{c}} = \mathbf{I}2\varepsilon_0)$, the MSs of the first 500 CMs derived from volume formulation and the above PMCHWT-based surface formulation $\mathbb{Z}_{-}^{\text{PMCHWT}} \cdot \mathrm{a}_{\xi}^{\text{AV}} = \lambda_{\xi} \, \mathbb{Z}_{+}^{\text{PMCHWT}} \cdot \mathrm{a}_{\xi}^{\text{AV}}$ are shown in Fig. E-10(b) and Fig. E-10(d). Obviously, the above PMCHWT-based surface formulation outputs many spurious modes.

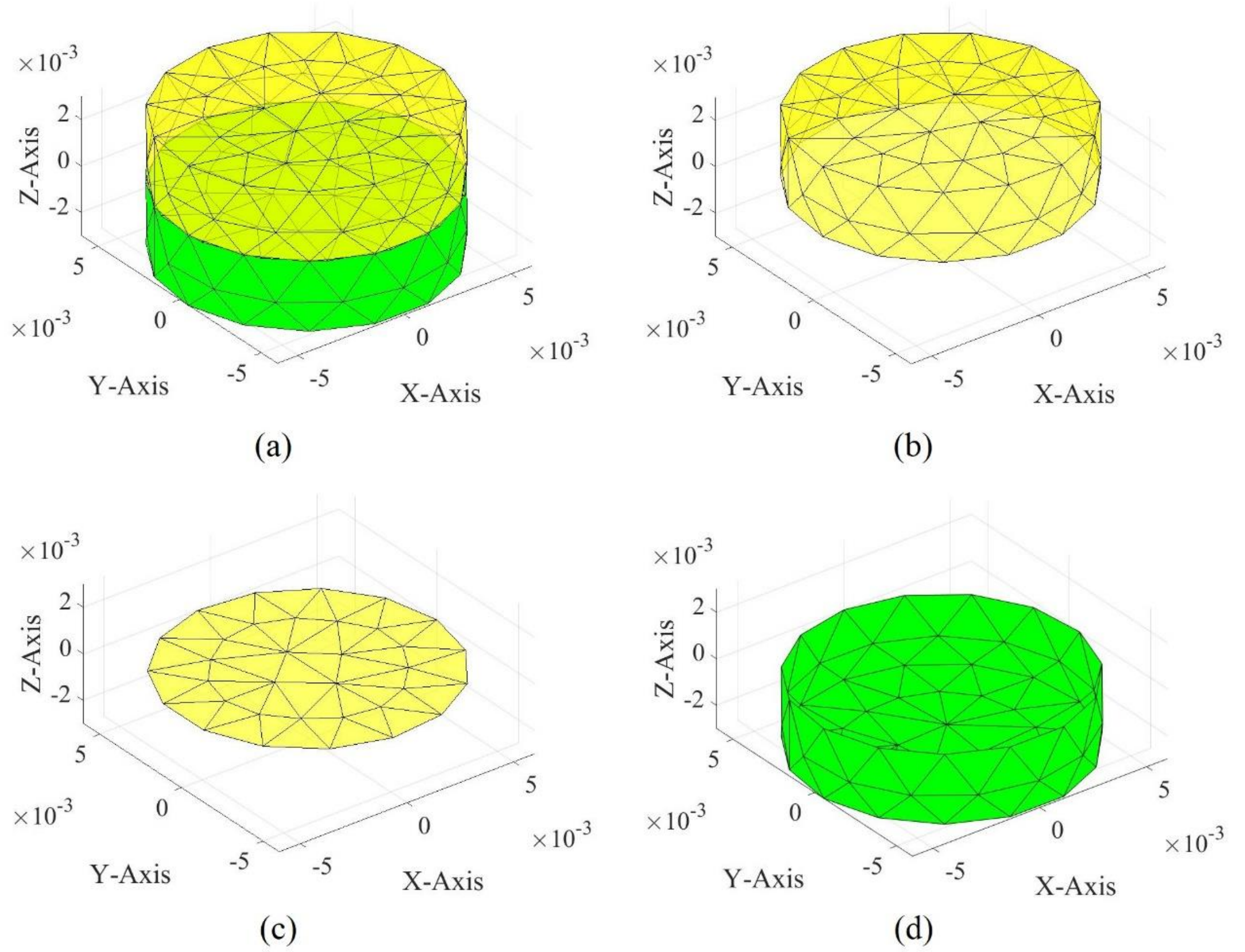

Figure E-9 (a) Topological structure and surface meshes of a two-body OSS; (b) $S_{10}$ of the OSS; (c) $S_{12}$ of the OSS; (d) $S_{20}$ of the OSS

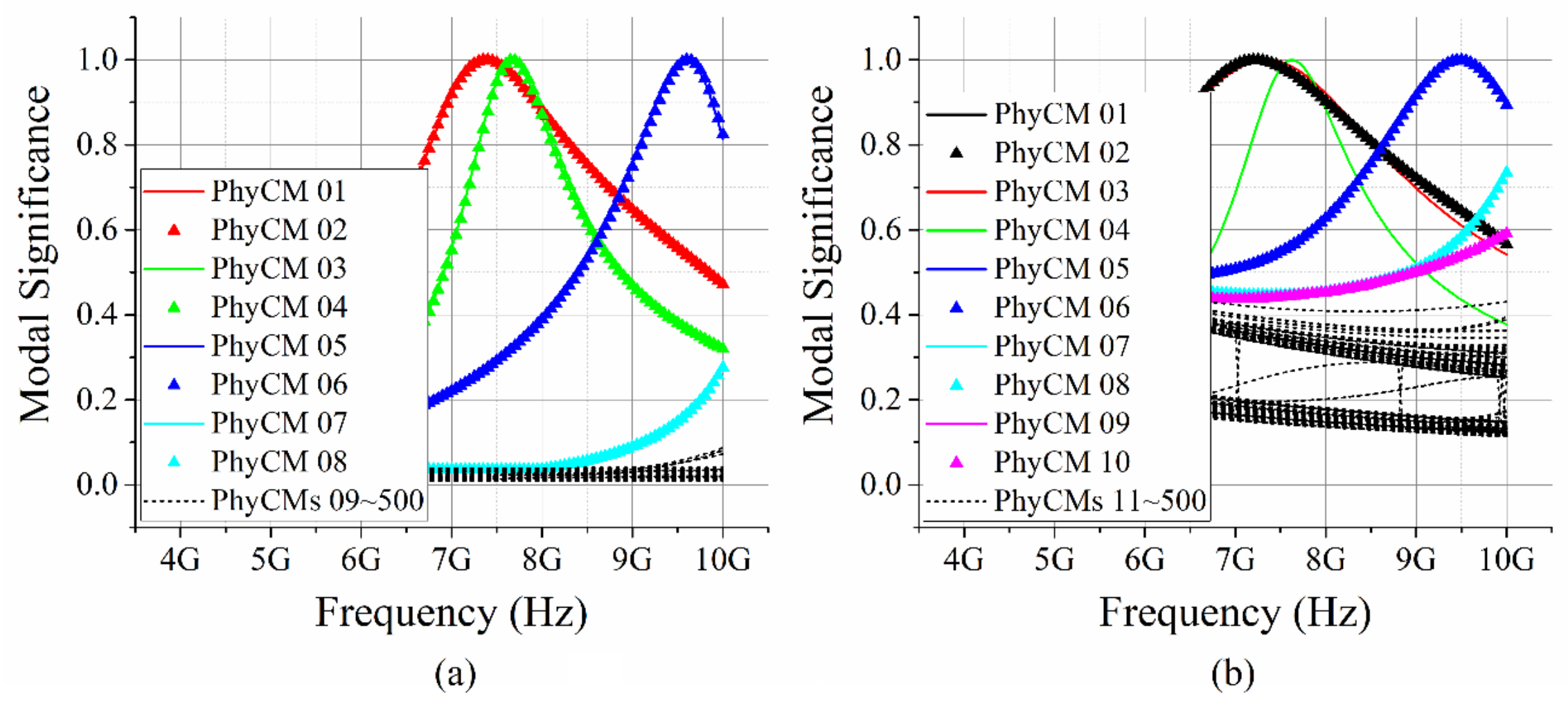

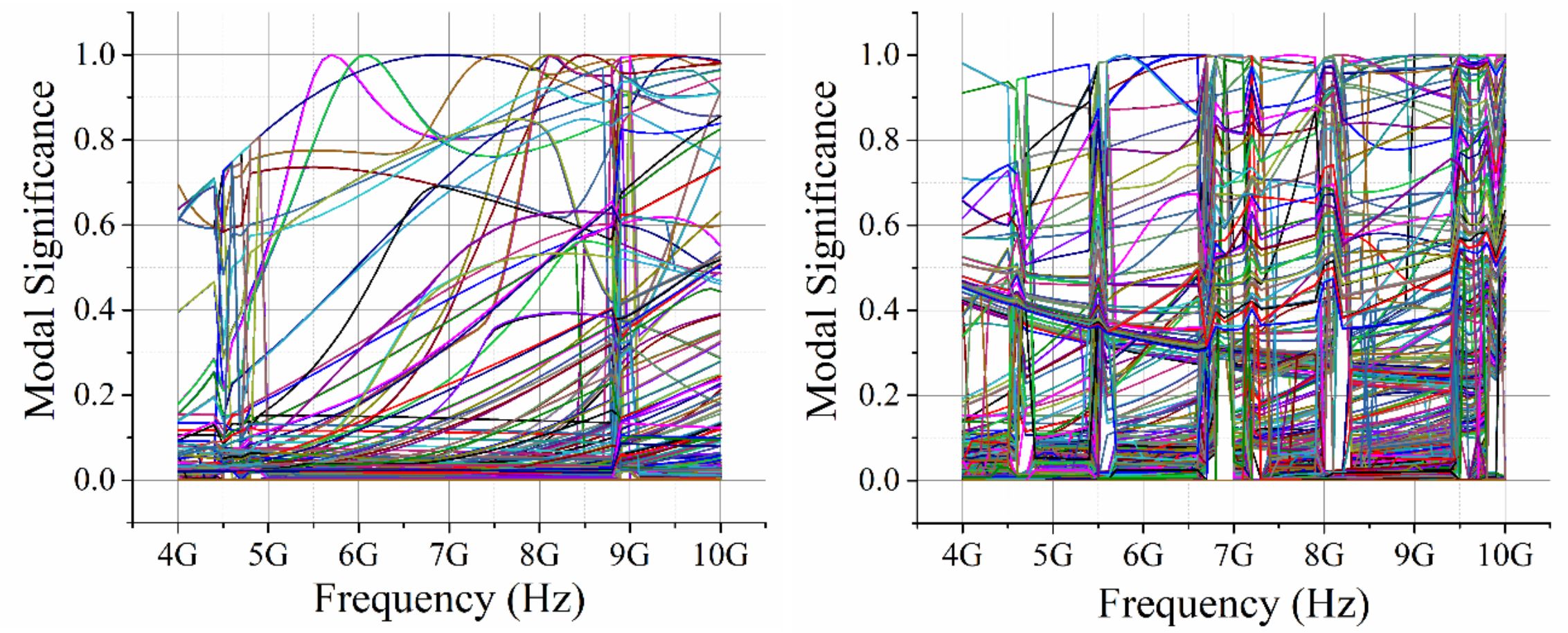

(c) (d)

Figure E-10 (a) (E-8)-based MSs of a lossless two-body OSS with topological structure Fig. E-9 and with material parameters $(\boldsymbol{\mu}_1 = \mathbf{I}2\mu_0, \boldsymbol{\varepsilon}_1^{\text{c}} = \mathbf{I}8\varepsilon_0)$ and $(\boldsymbol{\mu}_2 = \mathbf{I}8\mu_0, \boldsymbol{\varepsilon}_2^{\text{c}} = \mathbf{I}2\varepsilon_0)$; (b) (E-8)-based MSs of a lossy two-body OSS with topological structure Fig. E-9 and with material parameters $(\boldsymbol{\mu}_1 = \mathbf{I}2\mu_0, \boldsymbol{\varepsilon}_1^{\text{c}} = \mathbf{I}8\varepsilon_0 - j\mathbf{I}1/\omega)$ and $(\boldsymbol{\mu}_2 = \mathbf{I}8\mu_0, \boldsymbol{\varepsilon}_2^{\text{c}} = \mathbf{I}2\varepsilon_0)$; (c) (E-39)-based MSs of a lossless two-body OSS with topological structure Fig. E-9 and with material parameters $(\boldsymbol{\mu}_1 = \mathbf{I}2\mu_0, \boldsymbol{\varepsilon}_1^{\text{c}} = \mathbf{I}8\varepsilon_0)$ and $(\boldsymbol{\mu}_2 = \mathbf{I}8\mu_0, \boldsymbol{\varepsilon}_2^{\text{c}} = \mathbf{I}2\varepsilon_0)$; (d) (E-39)-based MSs of a lossy two-body OSS with topological structure Fig. E-9 and with material parameters $(\boldsymbol{\mu}_1 = \mathbf{I}2\mu_0, \boldsymbol{\varepsilon}_1^{\text{c}} = \mathbf{I}8\varepsilon_0 - j\mathbf{I}1/\omega)$ and $(\boldsymbol{\mu}_2 = \mathbf{I}8\mu_0, \boldsymbol{\varepsilon}_2^{\text{c}} = \mathbf{I}2\varepsilon_0)$

Below, we first simply review an existed scheme — dependent variable elimination (DVE)[13,26,33], and then propose a novel scheme — solution domain compression (SDC), for suppressing the spurious modes.

**PMCHWT-Based Surface CM Formulation with DVE**

The reason leading to the spurious modes is the overlooking of the dependence relations among the currents contained in DPO $P^{\text{driv}}$. To establish the dependence relations, [13] provided a series of integral equations as follows:

$$0 = \left[\mathcal{H}_1\left(\boldsymbol{J}_{10}^{\text{ES}} + \boldsymbol{J}_{12}^{\text{ES}}, \boldsymbol{M}_{10}^{\text{ES}} + \boldsymbol{M}_{12}^{\text{ES}}\right)\right]_{S_{10}^-}^{\tan} + \hat{\boldsymbol{n}}_{10}^- \times \boldsymbol{J}_{10}^{\text{ES}} \tag{E-40}$$

$$0 = \left[\mathcal{H}_2\left(\boldsymbol{J}_{20}^{\text{ES}} - \boldsymbol{J}_{12}^{\text{ES}}, \boldsymbol{M}_{20}^{\text{ES}} - \boldsymbol{M}_{12}^{\text{ES}}\right)\right]_{S_{20}^-}^{\tan} + \hat{\boldsymbol{n}}_{20}^- \times \boldsymbol{J}_{20}^{\text{ES}} \tag{E-41}$$

$$0 = \left[\mathcal{E}_1\left(\boldsymbol{J}_{10}^{\text{ES}} + \boldsymbol{J}_{12}^{\text{ES}}, \boldsymbol{M}_{10}^{\text{ES}} + \boldsymbol{M}_{12}^{\text{ES}}\right)\right]_{S_{10}^-}^{\tan} + \boldsymbol{M}_{10}^{\text{ES}} \times \hat{\boldsymbol{n}}_{10}^- \tag{E-42}$$

$$0 = \left[\mathcal{E}_2\left(\boldsymbol{J}_{20}^{\text{ES}} - \boldsymbol{J}_{12}^{\text{ES}}, \boldsymbol{M}_{20}^{\text{ES}} - \boldsymbol{M}_{12}^{\text{ES}}\right)\right]_{S_{20}^-}^{\tan} + \boldsymbol{M}_{20}^{\text{ES}} \times \hat{\boldsymbol{n}}_{20}^- \tag{E-43}$$

$$0 = \left[\mathcal{E}_1\left(\boldsymbol{J}_{10}^{\text{ES}} + \boldsymbol{J}_{12}^{\text{ES}}, \boldsymbol{M}_{10}^{\text{ES}} + \boldsymbol{M}_{12}^{\text{ES}}\right)\right]_{S_{12}^-}^{\tan} - \left[\mathcal{E}_2\left(\boldsymbol{J}_{20}^{\text{ES}} - \boldsymbol{J}_{12}^{\text{ES}}, \boldsymbol{M}_{20}^{\text{ES}} - \boldsymbol{M}_{12}^{\text{ES}}\right)\right]_{S_{12}^+}^{\tan} \tag{E-44}$$

$$0 = \left[\mathcal{H}_1\left(\boldsymbol{J}_{10}^{\text{ES}} + \boldsymbol{J}_{12}^{\text{ES}}, \boldsymbol{M}_{10}^{\text{ES}} + \boldsymbol{M}_{12}^{\text{ES}}\right)\right]_{S_{12}^-}^{\tan} - \left[\mathcal{H}_2\left(\boldsymbol{J}_{20}^{\text{ES}} - \boldsymbol{J}_{12}^{\text{ES}}, \boldsymbol{M}_{20}^{\text{ES}} - \boldsymbol{M}_{12}^{\text{ES}}\right)\right]_{S_{12}^+}^{\tan} \quad \text{(E-45)}$$

Here, integral equations (E-40)~(E-43) are based on SEPs (E-32)&(E-33) and definitions (E-34)&(E-35); integral equations (E-44) and (E-45) are based on SEPs (E-32)&(E-33), relation $\boldsymbol{C}_{21}^{\text{ES}} = -\boldsymbol{C}_{12}^{\text{ES}}$, and the tangential continuation condition of the $\boldsymbol{F}^{\text{tot}}$ on $S_{12}$.

By discretizing the integral equations (E-40)~(E-45) into matrix equations, the following transformation from basic variables (BVs, which are not only independent but also complete) to all variables (AVs, which consist of both the BVs and the other dependent variables) can be obtained.

$$\mathbb{a}^{\text{PhyAV}} = \mathbb{T} \cdot \mathbb{a}^{\text{BV}} \quad \text{(E-46)}$$

The formulations to calculate $\mathbb{T}$ can be found in [13], and they are not provided here, but we want to emphasize here that the process to obtain $\mathbb{T}$ involves calculating the inverses of some full matrices. Because the currents contained in $\mathbb{a}^{\text{PhyAV}}$ satisfy equations (E-40)~(E-45), thus $\mathbb{a}^{\text{PhyAV}}$ is physical automatically, and the superscript "PhyAV" is to emphasize this fact.

Substituting (E-46) into (E-39), we have the following matrix form of $P^{\text{driv}}$ with only BV $\mathbb{a}^{\text{BV}}$.

$$P^{\text{driv}} = \left(\mathbb{a}^{\text{BV}}\right)^H \cdot \Bigg( \frac{1}{2} \underbrace{\mathbb{T}^H \cdot \mathbb{Z}^{\text{PMCHWT}} \cdot \mathbb{T}}_{\underset{\sim}{\mathbb{Z}}^{\text{PMCHWT}}} \Bigg) \cdot \mathbb{a}^{\text{BV}} \quad \text{(E-47)}$$

In (E-47), all dependent variables have been eliminated (that is, expressed as the functions of BVs), and this is just the reason to call the scheme dependent variable elimination (DVE). Very naturally, we have the equation $\underset{\sim}{\mathbb{Z}}_{-}^{\text{PMCHWT}} \cdot \mathbb{a}_{\xi}^{\text{BV}} = \lambda_{\xi} \underset{\sim}{\mathbb{Z}}_{+}^{\text{PMCHWT}} \cdot \mathbb{a}_{\xi}^{\text{BV}}$ involving only BV, where $\underset{\sim}{\mathbb{Z}}_{+}^{\text{PMCHWT}}$ and $\underset{\sim}{\mathbb{Z}}_{-}^{\text{PMCHWT}}$ are the positive and negative Hermitian parts of $\underset{\sim}{\mathbb{Z}}^{\text{PMCHWT}}$. The corresponding physical CMs are $\mathbb{a}_{\xi}^{\text{PhyAV}} = \mathbb{T} \cdot \mathbb{a}_{\xi}^{\text{BV}}$.

**PMCHWT-Based Surface CM Formulation with SDC**

In fact, as the OSS becomes larger or more complicated, the computational complexity of the DVE scheme becomes worse, because of the step to inverse full matrices[259]. To resolve this problem, a novel spurious mode suppression scheme — solution domain compression (SDC) — is proposed as below.

By discretizing the integral equations (E-40), (E-41), (E-42), (E-43) and (E-44)&(E-45), we can obtain the following matrix equations

$$\mathbb{G}_{10}^{\text{DoJ}} \cdot \mathbb{a}^{\text{PhyAV}} = 0 \quad \text{(E-48)}$$

$$\mathbb{G}_{20}^{\text{DoJ}} \cdot \mathbb{a}^{\text{PhyAV}} = 0 \tag{E-49}$$
$$\mathbb{G}_{10}^{\text{DoM}} \cdot \mathbb{a}^{\text{PhyAV}} = 0 \tag{E-50}$$
$$\mathbb{G}_{20}^{\text{DoM}} \cdot \mathbb{a}^{\text{PhyAV}} = 0 \tag{E-51}$$
$$\mathbb{G}_{\text{FCE}} \cdot \mathbb{a}^{\text{PhyAV}} = 0 \tag{E-52}$$

where the superscripts "DoJ" and "DoM" are to emphasize the originations of $\mathbb{G}_{10/20}^{\text{DoJ}}$ (definition of $\boldsymbol{J}_{10/20}^{\text{ES}}$) and $\mathbb{G}_{10/20}^{\text{DoM}}$ (definition of $\boldsymbol{M}_{10/20}^{\text{ES}}$), and the subscript "FCE" is the acronym of "field continuation equation".

Based on the results obtained in [13,26,33], we can conclude that: equation (E-48) and equation (E-50) are equivalent to each other theoretically, and equation (E-49) and equation (E-51) are equivalent to each other theoretically, so if the matrices $\mathbb{G}_{10}^{\text{DoJ}}$, $\mathbb{G}_{20}^{\text{DoJ}}$, $\mathbb{G}_{10}^{\text{DoM}}$, $\mathbb{G}_{20}^{\text{DoM}}$, and $\mathbb{G}_{\text{FCE}}$ are assembled as

$$\mathbb{G}_{\text{FCE}}^{\text{DoJJ}} = \begin{bmatrix} \mathbb{G}_{10}^{\text{DoJ}} \\ \mathbb{G}_{20}^{\text{DoJ}} \\ \mathbb{G}_{\text{FCE}} \end{bmatrix},\ \mathbb{G}_{\text{FCE}}^{\text{DoMM}} = \begin{bmatrix} \mathbb{G}_{10}^{\text{DoM}} \\ \mathbb{G}_{20}^{\text{DoM}} \\ \mathbb{G}_{\text{FCE}} \end{bmatrix},\ \mathbb{G}_{\text{FCE}}^{\text{DoJM}} = \begin{bmatrix} \mathbb{G}_{10}^{\text{DoJ}} \\ \mathbb{G}_{20}^{\text{DoM}} \\ \mathbb{G}_{\text{FCE}} \end{bmatrix},\ \mathbb{G}_{\text{FCE}}^{\text{DoMJ}} = \begin{bmatrix} \mathbb{G}_{10}^{\text{DoM}} \\ \mathbb{G}_{20}^{\text{DoJ}} \\ \mathbb{G}_{\text{FCE}} \end{bmatrix} \tag{E-53}$$

then the following equations have the same solution space

$$\mathbb{G}_{\text{FCE}}^{\text{DoJJ/DoMM/DoJM/DoMJ}} \cdot \mathbb{a}^{\text{PhyAV}} = 0 \tag{E-54}$$

theoretically, and the solution space is just the null space of $\mathbb{G}_{\text{FCE}}^{\text{DoJJ/DoMM/DoJM/DoMJ}}$; a mode is physical, if and only if it satisfies (E-54), if and only if it belongs to the null space.

Thus, the null space is alternatively called physical modal space (or simply called modal space) in this paper, and any physical mode $\mathbb{a}^{\text{PhyAV}}$ can be expanded in terms of the basis $\{\mathbb{s}_1, \mathbb{s}_2, \cdots\}$ of the modal space as follows:

$$\mathbb{a}^{\text{PhyAV}} = \sum_i b_i \mathbb{s}_i = \underbrace{[\mathbb{s}_1, \mathbb{s}_2, \cdots]}_{\mathbb{S}} \cdot \underbrace{\begin{bmatrix} b_1 \\ b_2 \\ \vdots \end{bmatrix}}_{\mathbb{b}} \tag{E-55}$$

and vice versa. In (E-55), $\mathbb{b}$ is an arbitrary column vector, whose row number is equal to the dimension of the modal space.

In addition, it is obvious that

$$\begin{aligned} \text{Modal Space} &= \text{nullspace}\left(\mathbb{G}_{\text{FCE}}^{\text{DoJJ/DoMM/DoJM/DoMJ}}\right) \\ &= \text{span}\{\mathbb{s}_1, \mathbb{s}_2, \cdots\} \\ &= \text{range}(\mathbb{S}) \end{aligned} \tag{E-56}$$

where $\text{span}\{\mathbb{s}_1, \mathbb{s}_2, \cdots\}$ is the space spanned by $\{\mathbb{s}_1, \mathbb{s}_2, \cdots\}$, and $\text{range}(\mathbb{S})$ denotes the range of $\mathbb{S}$. Thus in this paper, the $\mathbb{G}_{\text{FCE}}^{\text{DoJJ/DoMM/DoJM/DoMJ}}$ and $\mathbb{S}$ are called the generating matrix/operator and spanning matrix/operator of the modal space.

Substituting (E-55) into (E-39), we have the following matrix form of $P^{\text{driv}}$ for and only for physical modes $\mathbb{a}^{\text{PhyAV}}$.

$$P^{\text{driv}} = \mathbb{b}^H \cdot \left( \frac{1}{2} \underbrace{\mathbb{S}^H \cdot \mathbb{Z}^{\text{PMCHWT}} \cdot \mathbb{S}}_{\tilde{\mathbb{Z}}^{\text{PMCHWT}}} \right) \cdot \mathbb{b} \tag{E-57}$$

and naturally we have the following characteristic equation $\tilde{\mathbb{Z}}_{-}^{\text{PMCHWT}} \cdot \mathbb{b}_{\xi} = \lambda_{\xi} \tilde{\mathbb{Z}}_{+}^{\text{PMCHWT}} \cdot \mathbb{b}_{\xi}$, where $\tilde{\mathbb{Z}}_{+}^{\text{PMCHWT}}$ and $\tilde{\mathbb{Z}}_{-}^{\text{PMCHWT}}$ are the positive and negative Hermitian parts of $\tilde{\mathbb{Z}}^{\text{PMCHWT}}$, and the associated CM $\mathbb{a}_{\xi}^{\text{PhyAV}}$ can be expressed as that $\mathbb{a}_{\xi}^{\text{PhyAV}} = \mathbb{S} \cdot \mathbb{b}_{\xi}$.

Obviously, the modal space is a subspace of the whole space constituted by all $\mathbb{a}^{\text{AV}}$ as shown in Fig. E-11, so the above SDC scheme is essentially to compress the solution domain, and this is just the reason to call it solution domain compression (SDC).

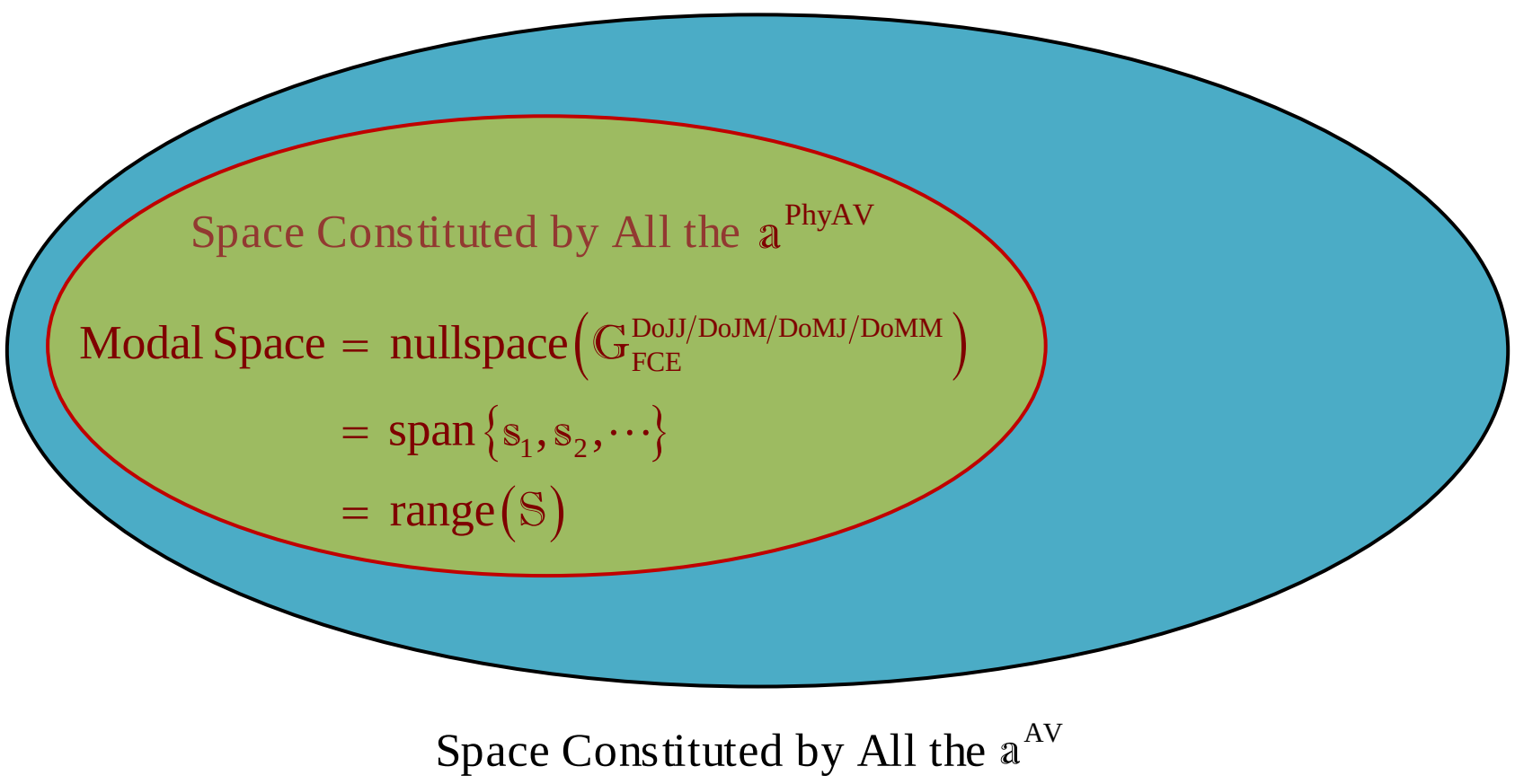

Figure E-11 Relations among the various spaces

For the lossless two-body OSS considered previously, its CMs derived from orthogonalizing the operator (E-57) with $\mathbb{G}_{\text{FCE}}^{\text{DoMJ}}$-based SDC are shown in Fig. E-12(a). Obviously, the SDC scheme effectively suppresses the spurious modes appeared in Fig. 10(c), but, at the same time, there also arise some other unwanted modes marked by the dotted lines. In fact, the unwanted modes originate from using PMCHWT operator rather than the SDC scheme as exhibited in the later Sec. E5.2.

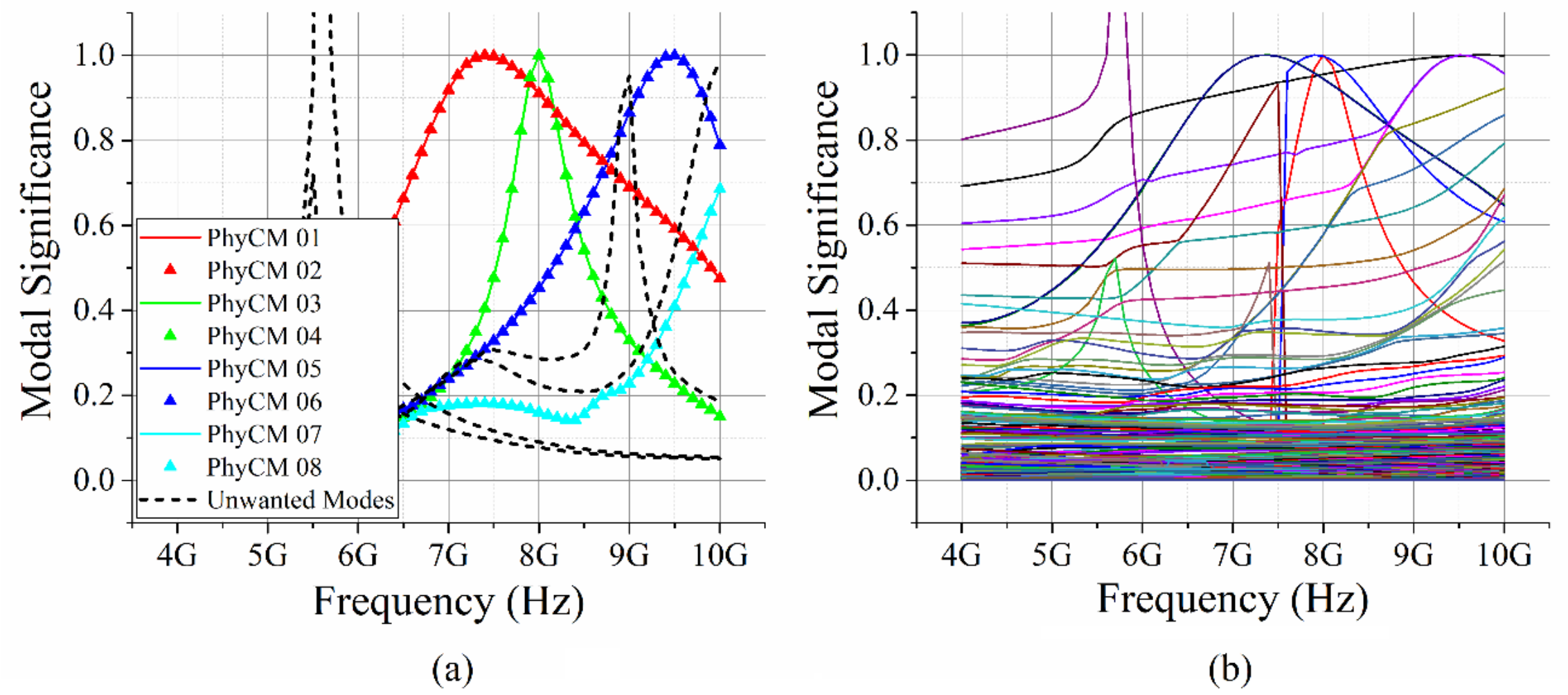

Figure E-12 (a) (E-57)-based MSs of a lossless two-body OSS with topological structure Fig. E-9 and with material parameters $(\boldsymbol{\mu}_1 = \mathbf{I}2\mu_0, \boldsymbol{\varepsilon}_1^{\rm c} = \mathbf{I}8\varepsilon_0)$ and $(\boldsymbol{\mu}_2 = \mathbf{I}8\mu_0, \boldsymbol{\varepsilon}_2^{\rm c} = \mathbf{I}2\varepsilon_0)$; (b) (E-57)-based MSs of a lossy two-body OSS with topological structure Fig. E-9 and with material parameters $(\boldsymbol{\mu}_1 = \mathbf{I}2\mu_0, \boldsymbol{\varepsilon}_1^{\rm c} = \mathbf{I}8\varepsilon_0 - j\mathbf{I}1/\omega)$ and $(\boldsymbol{\mu}_2 = \mathbf{I}8\mu_0, \boldsymbol{\varepsilon}_2^{\rm c} = \mathbf{I}2\varepsilon_0)$

For the lossy two-body OSS considered previously, we construct its CMs by orthogonalizing the operator (E-57) with $\mathrm{G}_{\rm FCE}^{\rm DoMJ}$ -based SDC, and the associated MSs are shown in Fig. E-12(b). Obviously, the results in Fig. E-12(b) are not satisfactory compared with the results in Fig. E-10(b). In fact, the reason leading to the invalidity of the operator (E-57) with $\mathrm{G}_{\rm FCE}^{\rm DoMJ}$ -based SDC for lossy OSSs originates from using PMCHWT operator rather than the SDC scheme as exhibited in the later Sec. E5.2.

Here, we emphasize again that the reason leading to the unwanted modes of the lossless OSS (shown in Fig. E-12 (a)) and the unsatisfactory results of the lossy OSS (shown in Fig. E-12 (b)) is using PMCHWT operator, rather than the new SDC scheme. In the following Sec. E5.2, we can effectively construct the CMs of the lossy OSS by employing another operator — driving power operator (DPO) — with the SDC scheme.

## E5.2 WEP-Based Surface CM Formulation with Two Different Spurious Mode Suppression Schemes

Because the tangential component of $\boldsymbol{F}^{\rm inc}$ is continuous on $S_{10} \cup S_{20}$, and the $\boldsymbol{F}^{\rm inc}$ on $S_{10}^- \cup S_{20}^-$ can be expressed in terms of the equivalent surface currents as SEP (E-31), thus DPO $P^{\rm driv}$ has surface formulation (E-58).

$$\begin{aligned}P^{\rm driv} = & -(1/2)\left\langle \boldsymbol{J}_{10}^{\rm ES} + \boldsymbol{J}_{20}^{\rm ES}, \mathcal{E}_0\left(\boldsymbol{J}_{10}^{\rm ES} + \boldsymbol{J}_{20}^{\rm ES}, \boldsymbol{M}_{10}^{\rm ES} + \boldsymbol{M}_{20}^{\rm ES}\right)\right\rangle_{S_{10}^- \cup S_{20}^-} \\ & -(1/2)\left\langle \boldsymbol{M}_{10}^{\rm ES} + \boldsymbol{M}_{20}^{\rm ES}, \mathcal{H}_0\left(\boldsymbol{J}_{10}^{\rm ES} + \boldsymbol{J}_{20}^{\rm ES}, \boldsymbol{M}_{10}^{\rm ES} + \boldsymbol{M}_{20}^{\rm ES}\right)\right\rangle_{S_{10}^- \cup S_{20}^-}\end{aligned}$$

$$\left.\begin{aligned} = & -(1/2)\left\langle \boldsymbol{J}_{10}^{\rm ES}+\boldsymbol{J}_{20}^{\rm ES}, -j\omega\mu_0\mathcal{L}_0\left(\boldsymbol{J}_{10}^{\rm ES}+\boldsymbol{J}_{20}^{\rm ES}\right)-\mathrm{P.V.}\mathcal{K}_0\left(\boldsymbol{M}_{10}^{\rm ES}+\boldsymbol{M}_{20}^{\rm ES}\right)\right\rangle_{S_{10}\cup S_{20}} \\ & -(1/2)\left\langle \boldsymbol{M}_{10}^{\rm ES}+\boldsymbol{M}_{20}^{\rm ES}, \mathrm{P.V.}\mathcal{K}_0\left(\boldsymbol{J}_{10}^{\rm ES}+\boldsymbol{J}_{20}^{\rm ES}\right)-j\omega\varepsilon_0\mathcal{L}_0\left(\boldsymbol{M}_{10}^{\rm ES}+\boldsymbol{M}_{20}^{\rm ES}\right)\right\rangle_{S_{10}\cup S_{20}} \end{aligned}\right\}\mathrm{PVT}$$
$$\left.\begin{aligned} & +(1/4)\left\langle \boldsymbol{J}_{10}^{\rm ES}+\boldsymbol{J}_{20}^{\rm ES}, \boldsymbol{M}_{10}^{\rm ES}\times\hat{\boldsymbol{n}}_{10}^{-}+\boldsymbol{M}_{20}^{\rm ES}\times\hat{\boldsymbol{n}}_{20}^{-}\right\rangle_{S_{10}\cup S_{20}} \\ & +(1/4)\left\langle \boldsymbol{M}_{10}^{\rm ES}+\boldsymbol{M}_{20}^{\rm ES}, \hat{\boldsymbol{n}}_{10}^{-}\times\boldsymbol{J}_{10}^{\rm ES}+\hat{\boldsymbol{n}}_{20}^{-}\times\boldsymbol{J}_{20}^{\rm ES}\right\rangle_{S_{10}\cup S_{20}} \end{aligned}\right\}\mathrm{SCT} \tag{E-58}$$

In (E-58), the operators $\mathcal{L}_0$ and $\mathcal{K}_0$ are defined as $\mathcal{L}_0(\boldsymbol{C}) = [1+(1/k_0^2)\nabla\nabla\cdot]\int_\Omega G_0(\boldsymbol{r},\boldsymbol{r}')\boldsymbol{C}(\boldsymbol{r}')d\Omega'$ and $\mathcal{K}_0(\boldsymbol{C}) = \nabla\times\int_\Omega G_0(\boldsymbol{r},\boldsymbol{r}')\boldsymbol{C}(\boldsymbol{r}')d\Omega'$ respectively; symbol "$\mathrm{P.V.}\mathcal{K}_0$" denotes the principal value of operator $\mathcal{K}_0$.

Similar to discretizing (E-38) into (E-39), the (E-58) can be discretized into the following matrix form

$$P^{\rm driv} = \underbrace{\begin{bmatrix} \mathrm{a}^{\boldsymbol{J}_{10}^{\rm ES}} \\ \mathrm{a}^{\boldsymbol{J}_{12}^{\rm ES}} \\ \mathrm{a}^{\boldsymbol{J}_{20}^{\rm ES}} \\ \mathrm{a}^{\boldsymbol{M}_{10}^{\rm ES}} \\ \mathrm{a}^{\boldsymbol{M}_{12}^{\rm ES}} \\ \mathrm{a}^{\boldsymbol{M}_{20}^{\rm ES}} \end{bmatrix}^H}_{\mathrm{a}^{\rm AV}} \cdot \underbrace{\left(\mathrm{P}_{\rm PVT}^{\rm driv}+\mathrm{P}_{\rm SCT}^{\rm driv}\right)}_{\mathrm{P}^{\rm driv}} \cdot \underbrace{\begin{bmatrix} \mathrm{a}^{\boldsymbol{J}_{10}^{\rm ES}} \\ \mathrm{a}^{\boldsymbol{J}_{12}^{\rm ES}} \\ \mathrm{a}^{\boldsymbol{J}_{20}^{\rm ES}} \\ \mathrm{a}^{\boldsymbol{M}_{10}^{\rm ES}} \\ \mathrm{a}^{\boldsymbol{M}_{12}^{\rm ES}} \\ \mathrm{a}^{\boldsymbol{M}_{20}^{\rm ES}} \end{bmatrix}}_{\mathrm{a}^{\rm AV}} \tag{E-59}$$

where subscripts "PVT" and "SCT" are the acronyms of "principal value term" and "singular current term" respectively.

**WEP-Based Surface CM Formulation with DVE**

Substituting (E-46) into (E-59), we have the matrix form with only BV $\mathrm{a}^{\rm BV}$ as follows:

$$P^{\rm driv} = \left(\mathrm{a}^{\rm BV}\right)^H \cdot \left(\overbrace{\underbrace{\mathrm{T}^H\cdot\mathrm{P}_{\rm PVT}^{\rm driv}\cdot\mathrm{T}}_{\underset{\sim}{\mathrm{P}}_{\rm PVT}^{\rm driv}} + \underbrace{\mathrm{T}^H\cdot\mathrm{P}_{\rm SCT}^{\rm driv}\cdot\mathrm{T}}_{\underset{\sim}{\mathrm{P}}_{\rm SCT}^{\rm driv}}}^{\underset{\sim}{\mathrm{P}}^{\rm driv}}\right)\cdot\mathrm{a}^{\rm BV} \tag{E-60}$$

and the corresponding characteristic equation as follows:

$$\underset{\sim}{\mathrm{P}}_{-}^{\rm driv}\cdot\mathrm{a}_\xi^{\rm BV} = \lambda_\xi\,\underset{\sim}{\mathrm{P}}_{+}^{\rm driv}\cdot\mathrm{a}_\xi^{\rm BV} \tag{E-61}$$

and the corresponding physical CMs as $\mathrm{a}_\xi^{\rm PhyAV} = \mathrm{T}\cdot\mathrm{a}_\xi^{\rm BV}$.

**WEP-Based Surface CM Formulation with SDC**

In modal space, we have the matrix form of $P^{\rm driv}$ as follows:

$$P^{\rm driv} = \mathrm{b}^H\cdot\left(\overbrace{\underbrace{\mathrm{S}^H\cdot\mathrm{P}_{\rm PVT}^{\rm driv}\cdot\mathrm{S}}_{\tilde{\mathrm{P}}_{\rm PVT}^{\rm driv}} + \underbrace{\mathrm{S}^H\cdot\mathrm{P}_{\rm SCT}^{\rm driv}\cdot\mathrm{S}}_{\tilde{\mathrm{P}}_{\rm SCT}^{\rm driv}}}^{\tilde{\mathrm{P}}^{\rm driv}}\right)\cdot\mathrm{b} \tag{E-62}$$

by substituting (E-55) into (E-59). The corresponding characteristic equation is as follows:

$$\tilde{\mathbb{P}}_{-}^{\text{driv}} \cdot \mathbb{b}_{\xi} = \lambda_{\xi}\, \tilde{\mathbb{P}}_{+}^{\text{driv}} \cdot \mathbb{b}_{\xi} \tag{E-63}$$

and the corresponding physical CMs are $\mathbb{a}_{\xi}^{\text{PhyAV}} = \mathbb{S} \cdot \mathbb{b}_{\xi}$.

For the lossless and lossy two-body OSSs considered previously, the MSs derived from (E-63) are shown in Fig. E-13. Obviously, the WEP-based DPO with SDC scheme successfully suppresses all the spurious modes, and, at the same time, doesn't introduce the unwanted modes shown in Figs. E-10(c), E-10(d), E-12(a) and E-12(b). In addition, Fig. E-13(b) implies that the reason leading to the invalidity of the operator (E-57) with $\mathbb{G}_{\text{FCE}}^{\text{DoMJ}}$-based SDC for lossy OSSs originates from the PMCHWT operator rather than the SDC scheme.

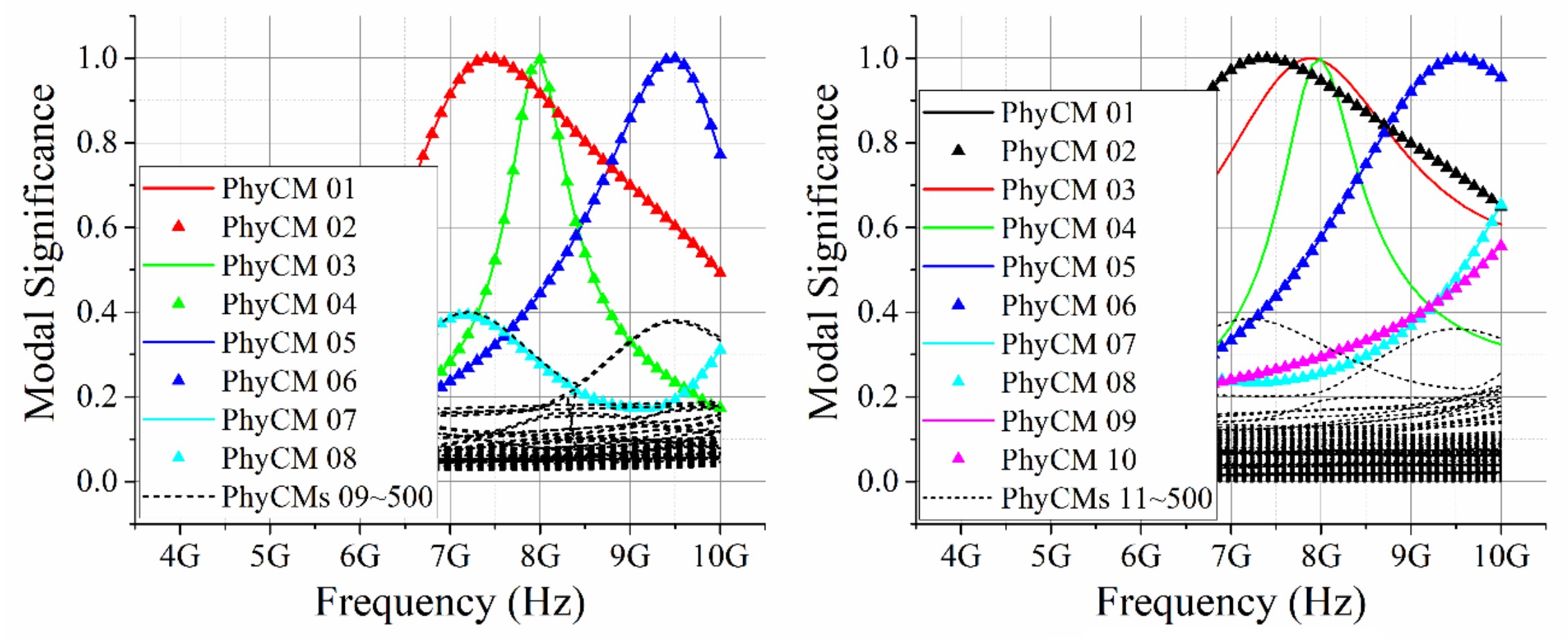

Figure E-13 (a) (E-63)-based MSs of a lossless two-body OSS with topological structure Fig. E-9 and with material parameters $(\boldsymbol{\mu}_1 = \mathbf{I}2\mu_0, \boldsymbol{\varepsilon}_1^{\text{c}} = \mathbf{I}8\varepsilon_0)$ and $(\boldsymbol{\mu}_2 = \mathbf{I}8\mu_0, \boldsymbol{\varepsilon}_2^{\text{c}} = \mathbf{I}2\varepsilon_0)$; (b) (E-63)-based MSs of a lossy two-body OSS with topological structure Fig. E-9 and with material parameters $(\boldsymbol{\mu}_1 = \mathbf{I}2\mu_0, \boldsymbol{\varepsilon}_1^{\text{c}} = \mathbf{I}8\varepsilon_0 - j\mathbf{I}1/\omega)$ and $(\boldsymbol{\mu}_2 = \mathbf{I}8\mu_0, \boldsymbol{\varepsilon}_2^{\text{c}} = \mathbf{I}2\varepsilon_0)$

### SCT and Its Physical Interpretation

It has been proved in [13] that $(\mathbb{a}^{\text{BV}})^H \cdot \mathbb{P}_{\text{SCT}}^{\text{driv}} \cdot \mathbb{a}^{\text{BV}}$ is always equal to the power dissipated in $V_{\text{OSS}}$, that is,

$$(\mathbb{a}^{\text{BV}})^H \cdot \mathbb{P}_{\text{SCT}}^{\text{driv}} \cdot \mathbb{a}^{\text{BV}} = (1/2)\langle \boldsymbol{\sigma}_1 \cdot \boldsymbol{E}^{\text{tot}}, \boldsymbol{E}^{\text{tot}} \rangle_{V_1} + (1/2)\langle \boldsymbol{\sigma}_2 \cdot \boldsymbol{E}^{\text{tot}}, \boldsymbol{E}^{\text{tot}} \rangle_{V_2} \tag{E-64}$$

and that

$$\boldsymbol{\sigma}_1 = \mathbf{I}0 = \boldsymbol{\sigma}_2. \Leftrightarrow V_{\text{OSS}} \text{ is lossless}. \Leftrightarrow \mathbb{P}_{\text{SCT}}^{\text{driv}} = 0. \tag{E-65}$$

Similarly, it can also be proved that

$$\text{b}^H \cdot \tilde{\text{P}}_{\text{SCT}}^{\text{driv}} \cdot \text{b} = (1/2)\langle \boldsymbol{\sigma}_1 \cdot \boldsymbol{E}^{\text{tot}}, \boldsymbol{E}^{\text{tot}} \rangle_{V_1} + (1/2)\langle \boldsymbol{\sigma}_2 \cdot \boldsymbol{E}^{\text{tot}}, \boldsymbol{E}^{\text{tot}} \rangle_{V_2} \tag{E-66}$$

and that

$$\boldsymbol{\sigma}_1 = \mathbf{I}0 = \boldsymbol{\sigma}_2 . \Leftrightarrow V_{\text{OSS}} \text{ is lossless} . \Leftrightarrow \tilde{\text{P}}_{\text{SCT}}^{\text{driv}} = 0. \tag{E-67}$$

Thus, when $V_{\text{OSS}}$ is lossless, we have that

$$P^{\text{driv}} \overset{\boldsymbol{\sigma}_1 = \mathbf{I}0 = \boldsymbol{\sigma}_2}{=\!=\!=\!=\!=} (\text{a}^{\text{BV}})^H \cdot \overbrace{(\text{T}^H \cdot \text{P}_{\text{PVT}}^{\text{driv}} \cdot \text{T})}^{\underset{\sim}{\text{P}}_{\text{PVT}}^{\text{driv}}} \cdot \text{a}^{\text{BV}} \tag{E-68}$$

$$P^{\text{driv}} \overset{\boldsymbol{\sigma}_1 = \mathbf{I}0 = \boldsymbol{\sigma}_2}{=\!=\!=\!=\!=} \text{b}^H \cdot \underbrace{(\text{S}^H \cdot \text{P}_{\text{PVT}}^{\text{driv}} \cdot \text{S})}_{\tilde{\text{P}}_{\text{PVT}}^{\text{driv}}} \cdot \text{b} \tag{E-69}$$

and then we have two alternative characteristic equations

$$\underset{\sim}{\text{P}}_{\text{PVT};-}^{\text{driv}} \cdot \text{a}_\xi^{\text{BV}} = \lambda_\xi \, \underset{\sim}{\text{P}}_{\text{PVT};+}^{\text{driv}} \cdot \text{a}_\xi^{\text{BV}} \tag{E-70}$$

$$\tilde{\text{P}}_{\text{PVT};-}^{\text{driv}} \cdot \text{b}_\xi = \lambda_\xi \, \tilde{\text{P}}_{\text{PVT};+}^{\text{driv}} \cdot \text{b}_\xi \tag{E-71}$$

for lossless $V_{\text{OSS}}$ only.

However, it must be emphasized here that: when $V_{\text{OSS}}$ is lossy, only matrix forms (E-60)&(E-62) and equations (E-61)&(E-63) are correct, while matrix forms (E-68)&(E-69) and equations (E-70)&(E-71) are incorrect as illustrated in the following example.

For the previous lossless two-body OSS, its CMs derived from the (E-71) with $\text{G}_{\text{FCE}}^{\text{DoMJ}}$-based SDC are shown in Fig. E-14. Obviously, the equation (E-71) is valid for constructing physical CMs and suppressing spurious modes, and, at the same time, has a more satisfactory numerical performance than characteristic equation (E-63).

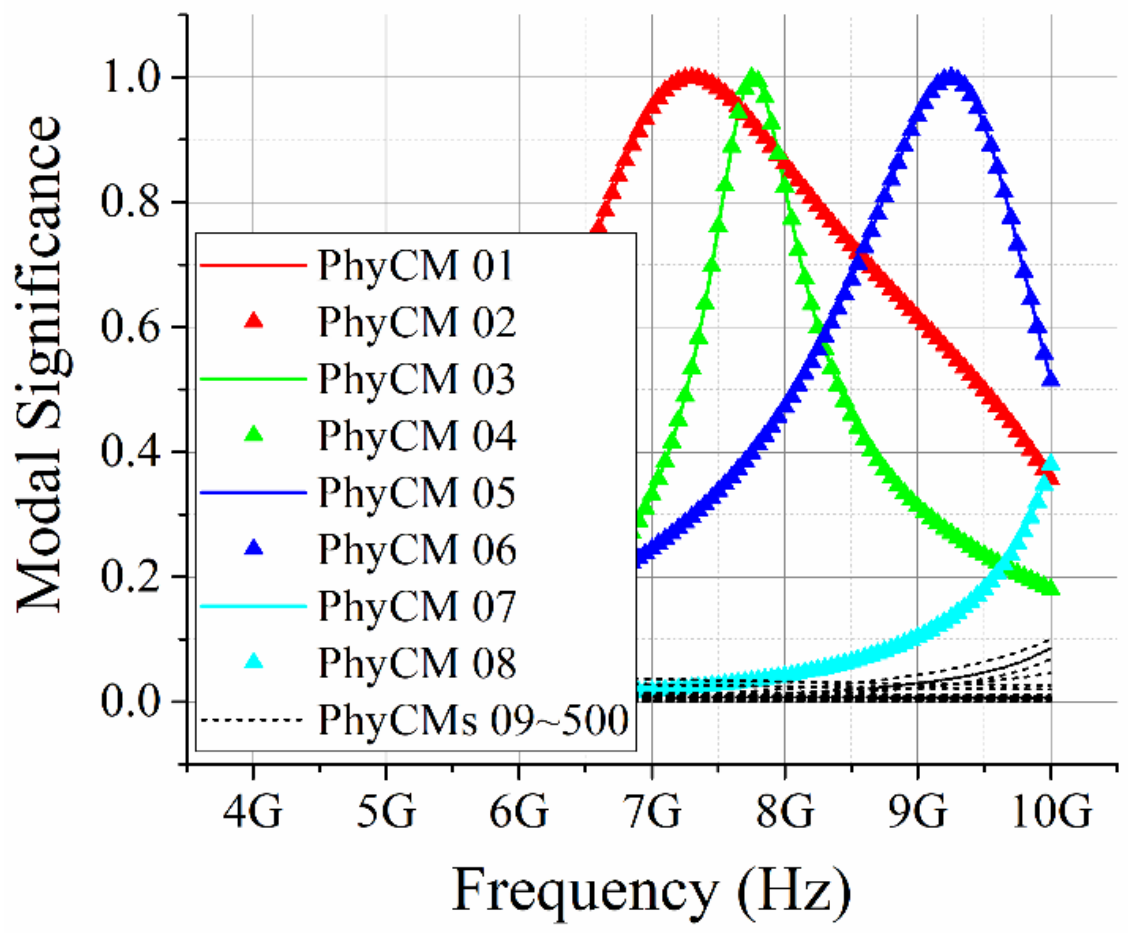

Figure E-14 (E-71)-based MSs of a lossless two-body OSS with topological structure Fig. E-9 and with material parameters $(\boldsymbol{\mu}_1 = \mathbf{I}2\mu_0, \boldsymbol{\varepsilon}_1^{\text{c}} = \mathbf{I}8\varepsilon_0)$ and $(\boldsymbol{\mu}_2 = \mathbf{I}8\mu_0, \boldsymbol{\varepsilon}_2^{\text{c}} = \mathbf{I}2\varepsilon_0)$

For the previous lossy two-body OSS, its CMs derived from the (E-71) with $\mathbb{G}_{\text{FCE}}^{\text{DoMJ}}$ -based SDC are shown in Fig. E-15. Obviously, the results are not consistent with the ones shown in Fig. E-10(b) and Fig. E-13(b), because the (E-71) doesn't include the SCT and the SCT is not zero in this lossy case. Thus, we want to emphasize here again that the matrix forms (E-68)&(E-69) and equations (E-70)&(E-71) are only correct for the lossless OSSs, but incorrect for the lossy OSSs.

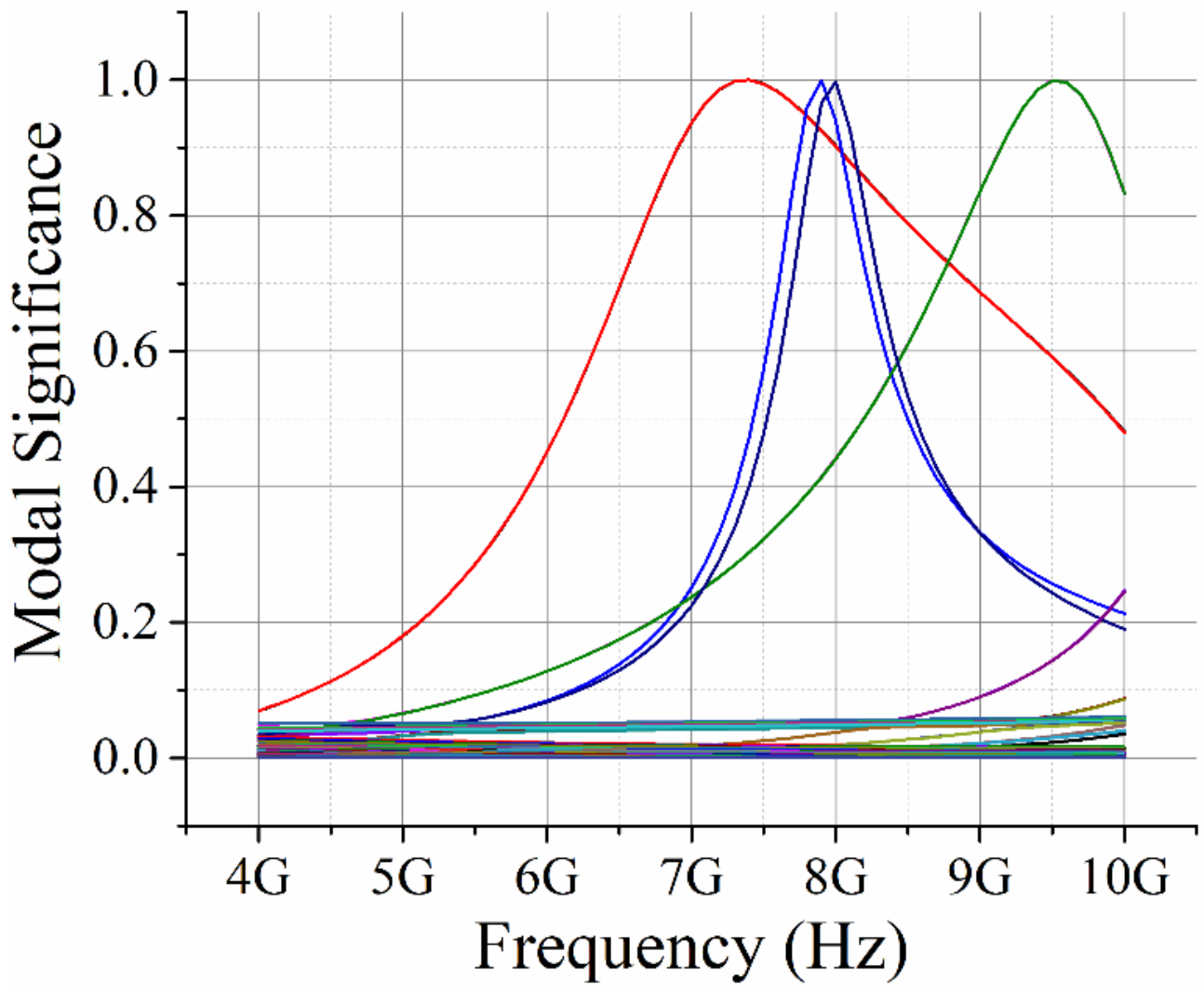

Figure E-15 (E-71)-based MSs of a lossy two-body OSS with topological structure Fig. E-9 and with material parameters $(\boldsymbol{\mu}_1 = \mathbf{I}2\mu_0, \boldsymbol{\varepsilon}_1^{\text{c}} = \mathbf{I}8\varepsilon_0 - j\mathbf{I}1/\omega)$ and $(\boldsymbol{\mu}_2 = \mathbf{I}8\mu_0, \boldsymbol{\varepsilon}_2^{\text{c}} = \mathbf{I}2\varepsilon_0)$

## E6 An Effective Method for Reducing the Computational Complexity of SDC Scheme

Compared with the conventional DVE scheme, the novel SDC scheme indeed avoids the matrix inversion process successfully, but, at the same time, it needs to calculate the basic solutions of equation (E-54). Unfortunately, the process to calculate the basic solutions is also computationally complicated. To resolve this problem, this section proposes an effective method for reducing the computational complexity of the original SDC scheme introduced in Sec. E5, and then obtains an improved SDC scheme with lower computational complexity.

For simplifying the symbolic system of the following discussions, we simply denote the matrix $\mathbb{G}_{\text{FCE}}^{\text{DoJJ/DoMM/DoJM/DoMJ}}$ and vector $\mathbb{a}^{\text{PhyAV}}$ used in (E-54) as $\mathbb{G}$ and $\mathbb{a}$ respectively. There exists the following equivalence relation

$$\mathbb{G} \cdot \mathbb{a} = 0 \;\Leftrightarrow\; \mathbb{a}^H \cdot \left(\mathbb{G}^H \cdot \mathbb{G}\right) \cdot \mathbb{a} = 0 \tag{E-72}$$

and the reasons are that: clearly $\mathbb{G}\cdot\mathbb{a}=0$ leads to $\mathbb{G}^H\cdot(\mathbb{G}\cdot\mathbb{a})=0$, so leads to $(\mathbb{G}^H\cdot\mathbb{G})\cdot\mathbb{a}=0$ due to the associativity for matrix multiplication, and then leads to $\mathbb{a}^H\cdot(\mathbb{G}^H\cdot\mathbb{G})\cdot\mathbb{a}=0$ ; obviously $0=\mathbb{a}^H\cdot(\mathbb{G}^H\cdot\mathbb{G})\cdot\mathbb{a}=(\mathbb{G}\cdot\mathbb{a})^H\cdot(\mathbb{G}\cdot\mathbb{a})$ due to the multiplication associativity and the relation $\mathbb{a}^H\cdot\mathbb{G}^H=(\mathbb{G}\cdot\mathbb{a})^H$, so $\mathbb{a}^H\cdot(\mathbb{G}^H\cdot\mathbb{G})\cdot\mathbb{a}=0$ implies $\mathbb{G}\cdot\mathbb{a}=0$ because $\mathbb{v}^H\cdot\mathbb{v}=0\Leftrightarrow\mathbb{v}=0$ for any complex vector $\mathbb{v}$.

Based on relation (E-72), we propose an alternative characteristic equation as follows:

$$\left(\mathbb{P}^{\text{driv}}_{\text{PVT};-}+\ell\cdot\mathbb{G}^H\cdot\mathbb{G}\right)\cdot\mathbb{a}_\xi=\lambda_\xi\,\mathbb{P}^{\text{driv}}_{\text{PVT};+}\cdot\mathbb{a}_\xi \tag{E-73}$$

for the lossless OSSs, and

$$\left(\mathbb{P}^{\text{driv}}_{-}+\ell\cdot\mathbb{G}^H\cdot\mathbb{G}\right)\cdot\mathbb{a}_\xi=\lambda_\xi\,\mathbb{P}^{\text{driv}}_{+}\cdot\mathbb{a}_\xi \tag{E-74}$$

for the lossy OSSs, where $\ell$ is an adjustable large real coefficient for example $\ell=10^{10}$.

Evidently, if we calculate the first several CMs which have relatively small $|\lambda_\xi|$ (the reason why the relatively small $|\lambda_\xi|$ are desirable had been carefully explained in Sec. E4), then the obtained CMs must satisfy equation $\mathbb{G}\cdot\mathbb{a}=0$, because

$$\lambda_\xi=\begin{cases}\dfrac{\mathbb{a}_\xi^H\cdot\mathbb{P}^{\text{driv}}_{\text{PVT};-}\cdot\mathbb{a}_\xi}{\mathbb{a}_\xi^H\cdot\mathbb{P}^{\text{driv}}_{\text{PVT};+}\cdot\mathbb{a}_\xi}+\ell\cdot\dfrac{\mathbb{a}_\xi^H\cdot\mathbb{G}^H\cdot\mathbb{G}\cdot\mathbb{a}_\xi}{\mathbb{a}_\xi^H\cdot\mathbb{P}^{\text{driv}}_{\text{PVT};+}\cdot\mathbb{a}_\xi} & \text{for lossless OSSs}\\[2ex] \dfrac{\mathbb{a}_\xi^H\cdot\mathbb{P}^{\text{driv}}_{-}\cdot\mathbb{a}_\xi}{\mathbb{a}_\xi^H\cdot\mathbb{P}^{\text{driv}}_{+}\cdot\mathbb{a}_\xi}+\ell\cdot\dfrac{\mathbb{a}_\xi^H\cdot\mathbb{G}^H\cdot\mathbb{G}\cdot\mathbb{a}_\xi}{\mathbb{a}_\xi^H\cdot\mathbb{P}^{\text{driv}}_{+}\cdot\mathbb{a}_\xi} & \text{for lossy OSSs}\end{cases} \tag{E-75}$$

and this implies that: if $\mathbb{G}\cdot\mathbb{a}\neq 0$, then the second term in the right-hand side of (E-75) is not zero due to relation (E-72), and then $|\lambda_\xi|$ will not be small due to the large coefficient $\ell$.

Obviously, the characteristic equations (E-73) and (E-74) involve neither calculating matrix inversion nor evaluating basic solutions.

To verify the validity of characteristic equations (E-73) and (E-74), we use them to calculate the CMs of the lossless and lossy two-body material OSSs considered in Secs. E4 and E5. The MSs corresponding to the first several CMs whose $|\lambda_\xi|$ are relatively small are shown in Fig. E-16. Clearly, the results shown in Figs. E-16(a) and E-16(b) are consistent with the ones shown in the Figs. E-10(a) and E-10(b), which are calculated from the volume formulation given in Sec. E4, and the ones shown in the Figs. E-14 and E-13(b), which are calculated from the surface formulation with the original SDC scheme proposed in Sec. E5.

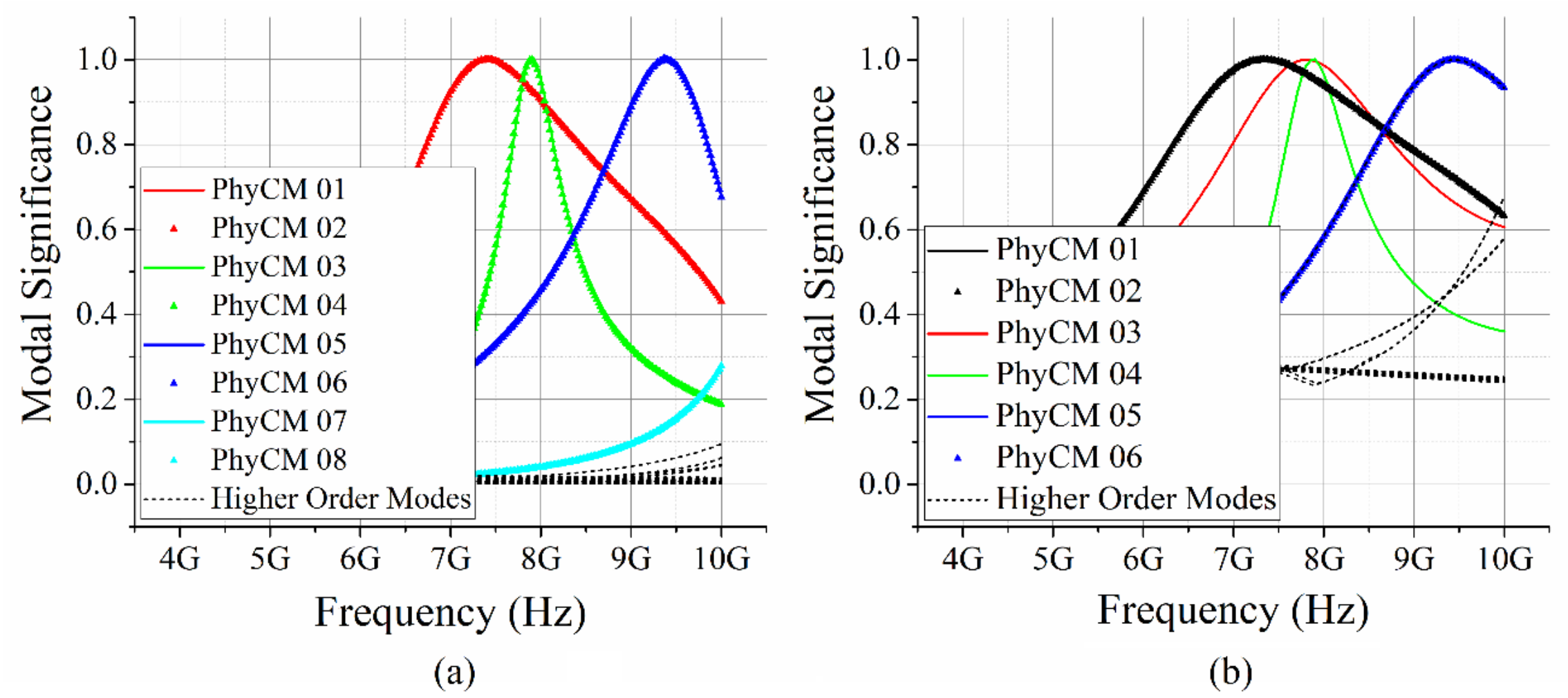

Figure E-16 (a) (E-73)-based MSs of a lossless two-body OSS with topological structure Fig. E-9 and with material parameters $(\boldsymbol{\mu}_1 = \mathbf{I}2\mu_0, \boldsymbol{\varepsilon}_1^{\rm c} = \mathbf{I}8\varepsilon_0)$ and $(\boldsymbol{\mu}_2 = \mathbf{I}8\mu_0, \boldsymbol{\varepsilon}_2^{\rm c} = \mathbf{I}2\varepsilon_0)$, where $\ell = 10^{10}$; (b) (E-74)-based MSs of a lossy two-body OSS with topological structure Fig. E-9 and with material parameters $(\boldsymbol{\mu}_1 = \mathbf{I}2\mu_0, \boldsymbol{\varepsilon}_1^{\rm c} = \mathbf{I}8\varepsilon_0 - j\mathbf{I}1/\omega)$ and $(\boldsymbol{\mu}_2 = \mathbf{I}8\mu_0, \boldsymbol{\varepsilon}_2^{\rm c} = \mathbf{I}2\varepsilon_0)$, where $\ell = 10^{10}$

## E7 Conclusions

For any pre-selected material objective scattering system (OSS), which can be either lossless or lossy, there exist some energy-decoupled working modes — characteristic modes (CMs). The CMs can span whole modal space, and don't exchange net energy in any integral period.

Traditionally, the CMs are constructed by orthogonalizing the impedance matrix operator (IMO) under integral equation (IE) framework. Alternatively, this paper effectively constructs the CMs by orthogonalizing the driving power operator (DPO) under work-energy principle (WEP) framework. This paper exhibits that the new WEP framework and new orthogonalizing DPO method have many advantages over the traditional IE framework and traditional orthogonalizing IMO method.

(1) The new WEP framework clearly reveals the physical picture/purpose of CMT.

(1.1) Based on the physical picture, the reason why the characteristic values (CVs) of magnetodielectric OSSs don't have clear physical meaning is clarified. The reason is that the CVs of magnetodielectric OSSs are only the byproducts during the process to construct CMs.

(1.2) Based on the physical picture, an improper physical interpretation for zero CV is corrected. The CMs with zero CVs have significant weights in whole modal expansion formulation, but they are not necessarily working at resonant states.

(1.3) Based on the physical picture, the reason why modal far fields of lossy OSSs are usually not orthogonal is answered. The answer is that the physical purpose of CMT is to construct the modes without net energy exchange in integral period rather than the modes with orthogonal far fields.

(1.4) Based on the physical picture, the reason why modal active powers are usually normalized to 1 is clarified. The reason is to let modal expansion coefficients have a clear physical meaning — the modal weight of the associated CM in whole modal expansion formulation, and to let the expansion formulation for time-average DP have the same mathematical form as the famous Parseval's identity.

(2) The new orthogonalizing DPO method has a more satisfactory numerical performance than the traditional orthogonalizing IMO method in the aspect of constructing CMs. A detailed and rigorous theoretical explanation for this phenomenon has been done by our group, and will be exhibited in our future article.

(3) By studying the physical meaning of the singular current term (SCT) contained in DPO, a further improvement for the numerical performance of DPO is proposed. The improvement scheme is to delete the SCT from DPO when the OSS is lossless.

(4) The variables contained in the new DPO are less than the ones contained in the traditional IMO, and this fact leads to the following conclusions.

(4.1) Discretizing DPO needs a smaller computational burden than discretizing IMO.

(4.2) From DPO, it is easier to distinguish the independent variables from the dependent variables of the related EM scattering problem, and then is easier to establish the transformation from the independent variables to the dependent variables and to compress solution domain from original space to modal space. (The transformation and compression can effectively suppress the spurious modes outputted from characteristic equation.)

Above these evidently exhibit the advantages of WEP framework and orthogonalizing DPO method.

Besides above these, this paper also proposes a novel spurious mode suppression scheme — solution domain compression (SDC). The novel SDC scheme is as effective as the traditional dependent variable elimination (DVE) scheme in the aspect of suppressing spurious modes, and, at the same time, it avoids the matrix inversion process used in the traditional DVE scheme successfully.

During the process to compress solution domain, the electric and magnetic field continuation conditions on the material-material interface are indispensable, but the one and only one of the electric and magnetic current definitions on the material-environment interface is necessary. In addition, we have the following further conclusions:

(i) when the material body is primarily electric, the definition of the magnetic current on the material-environment interface is more satisfactory for compressing solution domain;

(ii) when the material body is primarily magnetic, the definition of the electric current on the material-environment interface is more satisfactory for compressing solution domain.

A detailed and rigorous theoretical explanation for the above (i) and (ii) has been done by our group, and will be exhibited in our future article.

## Appendix F Work-Energy Principle Based Characteristic Mode Theory with Solution Domain Compression for Metal-Material Composite Scattering Systems

This App. F had been written as a journal paper by our research group (Ren-Zun Lian, Xing-Yue Guo and Ming-Yao Xia), and the original manuscript [AP2004-0709] entitled "Work-Energy Principe Based Characteristic Mode Theory with Solution Domain Compression for Metal-Material Composite Scattering Systems"[19] was submitted to IEEE-TAP (*IEEE Transactions on Antennas and Propagation*) on 12-Apr-2020, and the revised manuscript [AP2004-0709.R1] was submitted to IEEE-TAP on 10-Nov-2020.

### F1 Abstract

Characteristic mode theory (CMT) is established in a novel work-energy principle (WEP) framework. Under the framework, a novel driving power operator (DPO) is introduced as the generating operator for characteristic modes (CMs). Then, a novel orthogonalizing DPO method is proposed to construct the CMs. In addition, a novel solution domain compression (SDC) scheme is developed to suppress spurious modes.

Compared with the conventional integral equation (IE) framework, the novel WEP framework more clearly reveals the physical picture/purpose of CMT, and then easily derives the famous Parseval's identity. Compared with the conventional impedance matrix operator (IMO), the novel DPO is more advantageous in the aspect of distinguishing independent variables from dependent variables, and, at the same time, has a smaller computational burden. Compared with the conventional orthogonalizing IMO method, the novel orthogonalizing DPO method has a more acceptable numerical performance in the aspect of constructing CMs. Compared with the conventional dependent variable elimination (DVE) scheme, the novel SDC scheme doesn't need to inverse any matrix during suppressing spurious modes.

## F2 Index Terms

Characteristic mode (CM), driving power operator (DPO), Parseval's identity, solution domain compression (SDC), work-energy principle (WEP).

## F3 Introduction

In the later 1960s, Garbacz *et al.*[5~7] built a characteristic mode theory (CMT) in scattering matrix (SM) framework. By orthogonalizing perturbation matrix operator (PMO), the SM-based CMT (SM-CMT) can construct a set of orthogonal characteristic modes (CMs) for any pre-selected lossless objective scattering system (OSS). In the 1970s, Harrington *et al.*[8~12] established an alternative CMT under integral equation (IE) framework. By orthogonalizing impedance matrix operator (IMO), the IE-based CMT (IE-CMT) can construct a set of orthogonal CMs for any pre-selected lossless or lossy OSS.

Both Garbacz's CMs and Harrington's CMs depend only on the inherent physical properties (such as the topological structure and material parameters etc.) of the OSS, so both SM-CMT and IE-CMT are very valuable for analyzing and designing the inherent electromagnetic (EM) scattering characters of the OSS. To obtain IMO is easier than to obtain PMO, so IE-CMT has had a wider spread than SM-CMT in EM engineering society. A very comprehensive review for the various antenna applications of IE-CMT can be found in [14~16].

Under IE framework, electric field integral equation (EFIE) based IMO[8~10] is widely used to generate the CMs of metallic OSSs, and volume integral equation (VIE)[11] and Poggio-Miller-Chang-Harrington-Wu-Tsai (PMCHWT)[12] based IMOs are traditionally used to generate the CMs of material OSSs, and EFIE-VIE[38,39,262] and EFIE-PMCHWT[40~42] based IMOs are lately used to generate the CMs of metal-material composite OSSs. Very recently, EM-extinction theorem-based operator[33,34] is used to generate the CMs of material OSSs, and EFIE-EM-extinction-theorem[263] and EFIE-electric-extinction-theorem[43] based operators are used to generate the CMs of composite OSSs.

The EFIE-based IMO[8~10] is valid for generating metallic CMs, and has had many successful engineering applications[14~16]. The VIE and EFIE-VIE based IMOs are valid for generating material and composite CMs respectively, but they consume a large amount of computing resources[16]. For material and composite OSSs, the PMCHWT and EFIE-PMCHWT based IMOs require less computing resource, but they usually output some spurious modes[16,18,26,30,31,33,34,43,263]. The reasons leading to the spurious modes are mainly that[16,26,31,33,34,263]:

(i) the equivalent electric and magnetic currents $(\boldsymbol{J}^{\text{ES}},\boldsymbol{M}^{\text{ES}})$ distributing on the material boundaries are simultaneously included in the IMOs;

(ii) between $\boldsymbol{J}^{\text{ES}}$ and $\boldsymbol{M}^{\text{ES}}$, only one is independent, and the other one depends on the independent one;

(iii) the original PMCHWT and EFIE-PMCHWT based IMOs overlook the dependence relation between $\boldsymbol{J}^{\text{ES}}$ and $\boldsymbol{M}^{\text{ES}}$.

In [16,26,31,33,34,263], some different schemes were developed to eliminate the dependent variable in IMO, and the schemes are called dependent variable elimination (DVE).

During the process to realize DVE, it is necessary to inverse some full matrices related to first-kind[26,33,263] or second-kind[26,33] Fredholm's operator, and the matrix inverse process usually needs a large amount of computing resources[259]. To resolve this problem, [18] proposed a new scheme — solution domain compression (SDC) — to suppress the spurious modes, and the SDC scheme doesn't need to inverse any matrix. For lossless material OSSs, the IMO with DVE[26] and SDC[18] indeed has ability to suppress the spurious modes, but, at the same time, some extra unwanted modes will also be outputted[18]. For lossy material OSSs, the IMO with DVE and SDC cannot suppress the spurious modes effectively[18]. To resolve these problems, paper [18], under a new work-energy principle (WEP) framework rather than the traditional IE framework, proposed a new operator — driving power operator (DPO) — for generating CMs, and the DPO doesn't suffer from the problems mentioned above.

This paper focuses on generalizing the work done in [18] to metal-material composite OSSs, and is organized as follows: Sec. F4 does some necessary preparations for the subsequent sections; Sec. F5 provides a novel WEP framework for establishing

CMT, and, at the same time, introduces a novel DPO for generating CMs; Sec. F6 proposes a novel SDC scheme for suppressing spurious modes; Sec. F7 constructs CMs by orthogonalizing the DPO with SDC scheme, and derives famous Parseval's identity, and clearly reveals the fact that the physical purpose/picture of the WEP-based CMT (WEP-CMT) is to construct a set of energy-decoupled modes rather than a set of far-field-orthogonal modes for scattering systems; Sec. F8 gives some numerical examples to verify the theory and formulas established in this paper; Sec. F9 concludes this paper.

In what follows, the $e^{j\omega t}$ convention is used throughout, and the time-domain quantities will be added time variable $t$ explicitly for example $\boldsymbol{J}(t)$, but the frequency-domain quantities will not. Moreover, for the linear quantities, such as $\boldsymbol{E}$, we have that $\boldsymbol{E}(t) = \mathrm{Re}\{\boldsymbol{E}e^{j\omega t}\}$ ; for the power-type quadratic quantities, we have $\mathrm{Re}\{(1/2)\boldsymbol{J}^* \cdot \boldsymbol{E}\} = (1/T)\int_0^T \boldsymbol{J}(t)\cdot\boldsymbol{E}(t)dt$, where $T$ is the period of the time-harmonic EM field[24,25].

## F4 Preliminaries

In this paper, the metal-material composite OSS $\Sigma$ shown in Fig. F-1 is considered. The OSS $\Sigma$ is constituted by some metallic wires $L$, some metallic surfaces $S$, a metallic body $V$ and a material body $\Omega$. The permeability, permittivity and conductivity of $\Omega$ are denoted as $\boldsymbol{\mu}$, $\boldsymbol{\varepsilon}$ and $\boldsymbol{\sigma}$ respectively, and then the complex permittivity of $\Omega$ is $\boldsymbol{\varepsilon}_c = \boldsymbol{\varepsilon} - j\boldsymbol{\sigma}/\omega$. Here, $\boldsymbol{\mu}$, $\boldsymbol{\varepsilon}$ and $\boldsymbol{\sigma}$ are real and two-order symmetrical dyads.

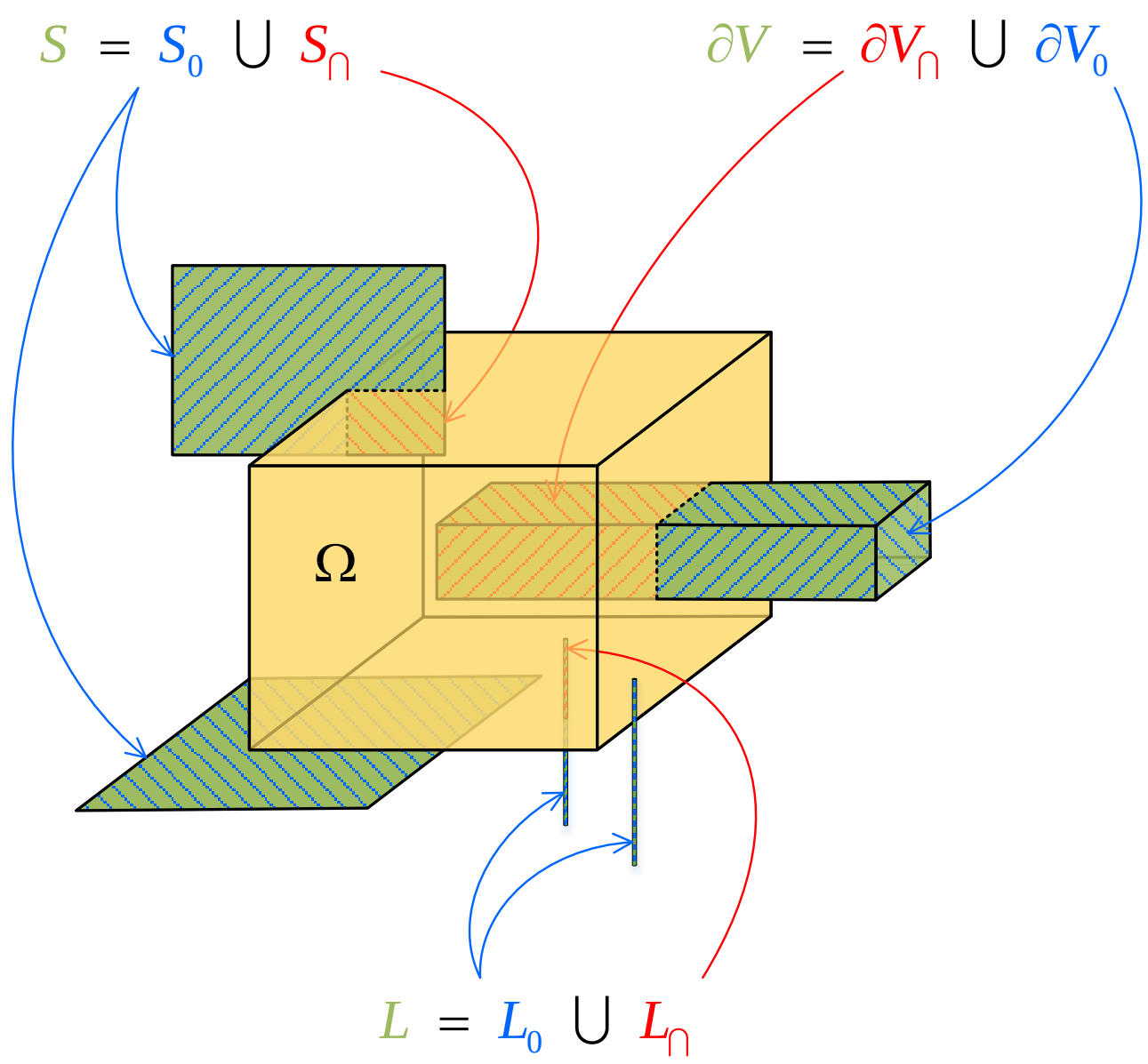

Figure F-1 Metal-material composite OSS considered in this paper

Obviously, the boundaries of $L$ and $S$ are just themselves[44]. The boundaries of $V$ and $\Omega$ are denoted as $\partial V$ and $\partial\Omega$ respectively. The interiors of $V$ and $\Omega$ are denoted as $\text{int}V$ and $\text{int}\,\Omega$ respectively. The exterior of $\Sigma$ is denoted as $\text{ext}\,\Sigma$.

When an incident field $\boldsymbol{F}^{\text{inc}}$, whose source distributes on $\text{ext}\,\Sigma$, excites the OSS, some scattered line electric current $\boldsymbol{J}^{\text{SL}}$ will be induced on $L$, and some scattered surface electric currents $\boldsymbol{J}^{\text{SS}}$ will be induced on $S \bigcup \partial V$, and some scattered volume electric and magnetic currents $(\boldsymbol{J}^{\text{SV}}, \boldsymbol{M}^{\text{SV}})$ will be induced on $\text{int}\,\Omega$[21,28]. The above scattered currents will generate a scattered field $\boldsymbol{F}^{\text{sca}}$ on whole three-dimensional Euclidean space $\mathbb{E}^3$. The summation of $\boldsymbol{F}^{\text{inc}}$ and $\boldsymbol{F}^{\text{sca}}$ is the total field, and the total field is denoted as $\boldsymbol{F}^{\text{tot}}$, that is, $\boldsymbol{F}^{\text{tot}} = \boldsymbol{F}^{\text{inc}} + \boldsymbol{F}^{\text{sca}}$.

Below, the whole $\mathbb{E}^3$ is decomposed into some sub-domains in Sec. F4.1 (domain decomposition); the metallic boundaries $L \bigcup S \bigcup \partial V$ and material boundary $\partial\Omega$ are decomposed into some sub-boundaries in Sec. F4.2 (boundary decomposition); the scattered currents on the metallic boundaries and the equivalent currents on the material boundary are decomposed into some sub-currents in Sec. F4.3 (current decomposition) according to the boundary decompositions given in Sec. F4.2; Sec. F4.4 (line-surface equivalence principle) expresses the fields on the sub-domains given in Sec. F4.1 in terms of the sub-currents obtained in Sec. F4.3.

## F4.1 Domain Decomposition

According to whether $\boldsymbol{F}^{\text{sca}}$ is continuous or not, whole $\mathbb{E}^3$ can be decomposed as follows:

$$\mathbb{E}^3 = \underbrace{\overbrace{L \cup S \cup \partial V}^{\text{metallic boundaries}} \cup \partial\Omega}_{\text{on which } \boldsymbol{F}^{\text{sca}} \text{ is not continuous}} \cup \underbrace{\text{int}\,\Omega \cup \text{int}V \cup \text{ext}\,\Sigma}_{\text{on which } \boldsymbol{F}^{\text{sca}} \text{ is continuous}} \tag{F-1}$$

On domains $\text{int}\,\Omega \bigcup \text{int}V \bigcup \text{ext}\,\Sigma$, field $\boldsymbol{F}^{\text{sca}}$ is continuous, and can be expressed in terms of the functions of the scattered currents $(\boldsymbol{J}^{\text{SL}}, \boldsymbol{J}^{\text{SS}})$ on metallic boundaries $L \bigcup S \bigcup \partial V$ and some equivalent currents $(\boldsymbol{J}^{\text{E}}, \boldsymbol{M}^{\text{E}})$ on material boundary $\partial\Omega$. However, field $\boldsymbol{F}^{\text{sca}}$ is not continuous on boundaries $L \bigcup S \bigcup \partial V \bigcup \partial\Omega$.

In fact, $(\boldsymbol{J}^{\text{SL}}, \boldsymbol{J}^{\text{SS}})$ and $(\boldsymbol{J}^{\text{E}}, \boldsymbol{M}^{\text{E}})$ are not independent of each other. To distinguish the independent ones from the dependent ones is indispensable for suppressing spurious modes[16,18,26,31,33,34,263]. To effectively distinguish them from each other, it is necessary to decompose the boundaries and currents as the following Sec. F4.2 and Sec. F4.3.

## F4.2 Boundary Decomposition

The metallic boundaries $L$, $S$ and $\partial V$ can be decomposed as follows:

$$L = L_{\cap} \cup L_0 \tag{F-2}$$

$$S = S_{\cap} \cup S_0 \tag{F-3}$$

$$\partial V = \partial V_{\cap} \cup \partial V_0 \tag{F-4}$$

In (F-2), $L_{\cap}$ is the part completely coated by $\Omega$, and the other part is defined as $L_0$ (that is, $L_0 = L \setminus L_{\cap}$). In (F-3), $S_{\cap}$ is the part completely coated by $\Omega$, and the other part is defined as $S_0$ (that is, $S_0 = S \setminus S_{\cap}$). In (F-4), $\partial V_{\cap}$ is the metal-material boundary, and $\partial V_0$ is the metal-environment boundary.

Obviously, $L_{\cap}$, $S_{\cap}$ and $\partial V_{\cap}$ belong to $\partial\Omega$. Thus, the material boundary $\partial\Omega$ can be decomposed as follows:

$$\partial\Omega = \underbrace{L_{\cap} \cup S_{\cap} \cup \partial V_{\cap}}_{\partial\Omega_{\cap}} \cup \partial\Omega_0 \tag{F-5}$$

where $L_{\cap}$, $S_{\cap}$ and $\partial V_{\cap}$ are collectively denoted as $\partial\Omega_{\cap}$ (that is, $\partial\Omega_{\cap} = L_{\cap} \cup S_{\cap} \cup \partial V_{\cap}$), and $\partial\Omega_0$ is defined as $\partial\Omega \setminus \partial\Omega_{\cap}$ (that is, $\partial\Omega_0 = \partial\Omega \setminus \partial\Omega_{\cap}$).

All the various sub-boundaries mentioned above are shown in Fig. F-1.

## F4.3 Current Decomposition

Based on the boundary decompositions (F-2)~(F-4), scattered electric currents $(\boldsymbol{J}^{\text{SL}}, \boldsymbol{J}^{\text{SS}})$ can be decomposed as follows:

$$\boldsymbol{J}^{\text{SL}} = \boldsymbol{J}_{\cap}^{\text{SL}} + \boldsymbol{J}_0^{\text{SL}} \tag{F-6}$$

$$\boldsymbol{J}^{\text{SS}} = \boldsymbol{J}_{\cap}^{\text{SS}} + \boldsymbol{J}_0^{\text{SS}} \tag{F-7}$$

In (F-6), $\boldsymbol{J}_{\cap}^{\text{SL}}$ and $\boldsymbol{J}_0^{\text{SL}}$ are the scattered currents distributing on $L_{\cap}$ and $L_0$ respectively. In (F-7), $\boldsymbol{J}_{\cap}^{\text{SS}}$ and $\boldsymbol{J}_0^{\text{SS}}$ are the scattered currents distributing on $S_{\cap} \cup \partial V_{\cap}$ and $S_0 \cup \partial V_0$ respectively.

Due to the homogeneous tangential electric field boundary condition on $L_{\cap} \cup S_{\cap} \cup \partial V_{\cap}$, there doesn't exist any equivalent magnetic current distributing on $\partial\Omega_{\cap}$, and, at the same time, the equivalent electric currents on $\partial\Omega_{\cap}$ are always equal to the scattered electric currents on $L_{\cap} \cup S_{\cap} \cup \partial V_{\cap}$, where $\partial\Omega_{\cap} = L_{\cap} \cup S_{\cap} \cup \partial V_{\cap}$ as shown in (F-5).

In addition, there exist some equivalent surface electric and magnetic currents $(\boldsymbol{J}_0^{\text{ES}}, \boldsymbol{M}_0^{\text{ES}})$ distributing on $\partial\Omega_0$, and they are defined as follows:

$$\boldsymbol{J}_0^{\text{ES}} = \hat{\boldsymbol{n}}_- \times \boldsymbol{H}_-^{\text{tot}} \quad , \quad \text{on } \partial\Omega_0 \tag{F-8}$$

$$\boldsymbol{M}_0^{\text{ES}} = \boldsymbol{E}_-^{\text{tot}} \times \hat{\boldsymbol{n}}_- \quad , \quad \text{on } \partial\Omega_0 \tag{F-9}$$

where $\boldsymbol{F}_-^{\text{tot}}$ is the total field distributing on the inner surface of $\partial\Omega_0$, and $\hat{\boldsymbol{n}}_-$ is the inner normal direction of $\partial\Omega_0$.

## F4.4 Line-Surface Equivalence Principle

Based on Huygens-Fresnel principle, the $\boldsymbol{F}^{\text{sca}}$ on $\text{ext}\,\Sigma$ and the $\boldsymbol{F}^{\text{inc}}$ on $\text{int}\,\Omega \bigcup \text{int}\,V$ can be expressed in terms of the functions of currents $(\boldsymbol{J}_0^{\text{SL}}, \boldsymbol{J}_0^{\text{SS}}, \boldsymbol{J}_0^{\text{ES}}, \boldsymbol{M}_0^{\text{ES}})$ as follows:

$$\left.\begin{aligned} \text{ext}\,\Sigma:\ & -\boldsymbol{F}^{\text{sca}} \\ \text{int}\,\Omega:\ & \ \ \boldsymbol{F}^{\text{inc}} \\ \text{int}\,V:\ & \ \ \boldsymbol{F}^{\text{inc}} \end{aligned}\right\} = \mathcal{F}_0\left(-\boldsymbol{J}_0^{\text{SL}} - \boldsymbol{J}_0^{\text{SS}} + \boldsymbol{J}_0^{\text{ES}}, \boldsymbol{M}_0^{\text{ES}}\right) \tag{F-10}$$

In above (F-10), $\boldsymbol{F} = \boldsymbol{E} / \boldsymbol{H}$ and correspondingly $\mathcal{F}_0 = \mathcal{E}_0 / \mathcal{H}_0$, and the operators $\mathcal{E}_0(\boldsymbol{J}, \boldsymbol{M})$ and $\mathcal{H}_0(\boldsymbol{J}, \boldsymbol{M})$ are defined as that $\mathcal{E}_0(\boldsymbol{J}, \boldsymbol{M}) = -j\omega\mu_0\mathcal{L}_0(\boldsymbol{J}) - \mathcal{K}_0(\boldsymbol{M})$ and $\mathcal{H}_0(\boldsymbol{J}, \boldsymbol{M}) = -j\omega\varepsilon_0\mathcal{L}_0(\boldsymbol{M}) + \mathcal{K}_0(\boldsymbol{J})$, where the operators $\mathcal{L}_0$ and $\mathcal{K}_0$ are defined as the ones given in [18] and [264].

Similarly, the $\boldsymbol{F}^{\text{tot}}$ on $\text{int}\,\Omega$ and $\text{int}\,V$ can be expressed in terms of the currents $(\boldsymbol{J}_\cap^{\text{SL}}, \boldsymbol{J}_\cap^{\text{SS}}, \boldsymbol{J}_0^{\text{ES}}, \boldsymbol{M}_0^{\text{ES}})$ as follows:

$$\text{int}\,\Omega:\ \boldsymbol{F}^{\text{tot}} = \mathcal{F}\left(\boldsymbol{J}_\cap^{\text{SL}} + \boldsymbol{J}_\cap^{\text{SS}} + \boldsymbol{J}_0^{\text{ES}}, \boldsymbol{M}_0^{\text{ES}}\right) \tag{F-11}$$

$$\text{int}\,V:\ \boldsymbol{F}^{\text{tot}} = 0 \tag{F-12}$$

In the above (F-11) and (F-12), $\boldsymbol{F} = \boldsymbol{E} / \boldsymbol{H}$ and correspondingly $\mathcal{F} = \mathcal{E} / \mathcal{H}$, and the operator $\mathcal{F}(\boldsymbol{J}, \boldsymbol{M})$ is defined as that $\mathcal{F}(\boldsymbol{J}, \boldsymbol{M}) = \mathbf{G}^{JF} * \boldsymbol{J} + \mathbf{G}^{MF} * \boldsymbol{M}$, where $\mathbf{G}^{JF}$ and $\mathbf{G}^{MF}$ are the dyadic Green's functions corresponding to material parameters $(\boldsymbol{\mu}, \boldsymbol{\varepsilon}_c)$, and the convolution integral operation "$*$" is defined as that $\mathbf{G} * \boldsymbol{C} = \int_\Gamma \mathbf{G}(\boldsymbol{r}, \boldsymbol{r}') \cdot \boldsymbol{C}(\boldsymbol{r}') d\Gamma'$.

Therefore, the $\boldsymbol{F}^{\text{sca}}$ on $\text{int}\,\Omega$ and $\text{int}\,V$ can be expressed in terms of the functions of currents $(\boldsymbol{J}_0^{\text{SL}}, \boldsymbol{J}_\cap^{\text{SL}}, \boldsymbol{J}_0^{\text{SS}}, \boldsymbol{J}_\cap^{\text{SS}}, \boldsymbol{J}_0^{\text{ES}}, \boldsymbol{M}_0^{\text{ES}})$ as follows:

$$\begin{aligned} \text{int}\,\Omega:\ \boldsymbol{F}^{\text{sca}} = \ & \mathcal{F}\left(\boldsymbol{J}_\cap^{\text{SL}} + \boldsymbol{J}_\cap^{\text{SS}} + \boldsymbol{J}_0^{\text{ES}}, \boldsymbol{M}_0^{\text{ES}}\right) \\ & - \mathcal{F}_0\left(-\boldsymbol{J}_0^{\text{SL}} - \boldsymbol{J}_0^{\text{SS}} + \boldsymbol{J}_0^{\text{ES}}, \boldsymbol{M}_0^{\text{ES}}\right) \end{aligned} \tag{F-13}$$

$$\text{int}\,V:\ \boldsymbol{F}^{\text{sca}} = -\mathcal{F}_0\left(-\boldsymbol{J}_0^{\text{SL}} - \boldsymbol{J}_0^{\text{SS}} + \boldsymbol{J}_0^{\text{ES}}, \boldsymbol{M}_0^{\text{ES}}\right) \tag{F-14}$$

because of (F-10)~(F-12) and that $\boldsymbol{F}^{\text{sca}} = \boldsymbol{F}^{\text{tot}} - \boldsymbol{F}^{\text{inc}}$.

The currents used in (F-10)~(F-14) are line-type or surface-type, so convolution integral formulations (F-10)~(F-14) are collectively referred to as line-surface equivalence principle (LSEP). Next, the LSEP (F-10) will be used to formulate the generating operator of CMs in the following Sec. F5, and the LSEP (F-11) will be used to compress the solution domain of characteristic equation in the future Sec. F6, and the LSEPs (F-10), (F-13) and (F-14) will be used to compute the modal scattering fields.

## F5 Work-Energy Principle and Driving Power Operator

In this section, the work-energy viewpoint is used to describe scattering problem, and some functions with power dimension are introduced to express the work-energy transformation process " $\boldsymbol{F}^{\text{inc}} \xrightarrow[\text{work}]{\text{do}} \text{OSS} \xrightarrow[\text{energy}]{\text{scatter}} \boldsymbol{F}^{\text{sca}} \xrightarrow[\text{energy}]{\text{carry}} \text{Infinity}$ " quantitatively. The work-energy viewpoint and power functions will be utilized to establish the WEP-based CMT for composite OSSs in the future Sec. F7.

### F5.1 Work-Energy Principle (WEP)

The conservation law of energy[260,261] tells us that the action of the incident field $\boldsymbol{F}^{\text{inc}}$ on the scattered currents $(\boldsymbol{J}^{\text{SL}}, \boldsymbol{J}^{\text{SS}}, \boldsymbol{J}^{\text{SV}}, \boldsymbol{M}^{\text{SV}})$ will give rise to a transformation between work and energy. The work-energy transformation can be quantitatively expressed as follows:

$$\mathcal{W}^{\text{Driv}} = \mathcal{E}_{S_\infty}^{\text{rad}} + \mathcal{E}_{\Omega}^{\text{dis}} + \Delta\mathcal{E}_{\mathbb{E}^3}^{\text{sto}} + \Delta\mathcal{E}_{\Omega}^{\text{sto}} \tag{F-15}$$

The mathematical expressions of the terms in (F-15) are given in (F-16)~(F-20).

$$\mathcal{W}^{\text{Driv}} = \int_{t_0}^{t_0+\Delta t} \Big[ \big\langle \boldsymbol{J}^{\text{SL}}(t) + \boldsymbol{J}^{\text{SS}}(t) + \boldsymbol{J}^{\text{SV}}(t), \boldsymbol{E}^{\text{inc}}(t) \big\rangle_{\Sigma} + \big\langle \boldsymbol{M}^{\text{SV}}(t), \boldsymbol{H}^{\text{inc}}(t) \big\rangle_{\Sigma} \Big] dt \tag{F-16}$$

$$\mathcal{E}_{S_\infty}^{\text{rad}} = \int_{t_0}^{t_0+\Delta t} \left\{ \oint\!\!\!\oint_{S_\infty} \big[ \boldsymbol{E}^{\text{sca}}(t) \times \boldsymbol{H}^{\text{sca}}(t) \big] \cdot d\boldsymbol{S} \right\} dt \tag{F-17}$$

$$\mathcal{E}_{\Omega}^{\text{dis}} = \int_{t_0}^{t_0+\Delta t} \big\langle \boldsymbol{\sigma} \cdot \boldsymbol{E}^{\text{tot}}(t), \boldsymbol{E}^{\text{tot}}(t) \big\rangle_{\Omega} dt \tag{F-18}$$

$$\begin{aligned} \Delta\mathcal{E}_{\mathbb{E}^3}^{\text{sto}} = & \left[ \frac{1}{2} \big\langle \boldsymbol{H}^{\text{sca}}(t_0+\Delta t), \mu_0 \boldsymbol{H}^{\text{sca}}(t_0+\Delta t) \big\rangle_{\mathbb{E}^3} + \frac{1}{2} \big\langle \varepsilon_0 \boldsymbol{E}^{\text{sca}}(t_0+\Delta t), \boldsymbol{E}^{\text{sca}}(t_0+\Delta t) \big\rangle_{\mathbb{E}^3} \right] \\ & - \left[ \frac{1}{2} \big\langle \boldsymbol{H}^{\text{sca}}(t_0), \mu_0 \boldsymbol{H}^{\text{sca}}(t_0) \big\rangle_{\mathbb{E}^3} + \frac{1}{2} \big\langle \varepsilon_0 \boldsymbol{E}^{\text{sca}}(t_0), \boldsymbol{E}^{\text{sca}}(t_0) \big\rangle_{\mathbb{E}^3} \right] \end{aligned} \tag{F-19}$$

$$\begin{aligned} \Delta\mathcal{E}_{\Omega}^{\text{sto}} = & \left[ \frac{1}{2} \big\langle \boldsymbol{H}^{\text{tot}}(t_0+\Delta t), \Delta\boldsymbol{\mu} \cdot \boldsymbol{H}^{\text{tot}}(t_0+\Delta t) \big\rangle_{\Omega} + \frac{1}{2} \big\langle \Delta\boldsymbol{\varepsilon} \cdot \boldsymbol{E}^{\text{tot}}(t_0+\Delta t), \boldsymbol{E}^{\text{tot}}(t_0+\Delta t) \big\rangle_{\Omega} \right] \\ & - \left[ \frac{1}{2} \big\langle \boldsymbol{H}^{\text{tot}}(t_0), \Delta\boldsymbol{\mu} \cdot \boldsymbol{H}^{\text{tot}}(t_0) \big\rangle_{\Omega} + \frac{1}{2} \big\langle \Delta\boldsymbol{\varepsilon} \cdot \boldsymbol{E}^{\text{tot}}(t_0), \boldsymbol{E}^{\text{tot}}(t_0) \big\rangle_{\Omega} \right] \end{aligned} \tag{F-20}$$

In (F-16)~(F-20), the time interval $\Delta t$ is a positive real number; $\Delta\boldsymbol{\mu} = \boldsymbol{\mu} - \mathbf{I}\mu_0$ and $\Delta\boldsymbol{\varepsilon} = \boldsymbol{\varepsilon} - \mathbf{I}\varepsilon_0$ , where $\mathbf{I}$ is the unit dyad; the inner product is defined as $< \boldsymbol{f}, \boldsymbol{g} >_\Gamma = \int_\Gamma \boldsymbol{f}^* \cdot \boldsymbol{g} d\Gamma$ ; $S_\infty$ is a spherical surface with infinite radius.

Evidently, equation (F-15) has a very clear physical explanation: in time interval $t_0 \sim t_0 + \Delta t$ , the work $\mathcal{W}^{\text{Driv}}$ done by fields $(\boldsymbol{E}^{\text{inc}}, \boldsymbol{H}^{\text{inc}})$ on currents $(\boldsymbol{J}^{\text{SL}}, \boldsymbol{J}^{\text{SS}}, \boldsymbol{J}^{\text{SV}}, \boldsymbol{M}^{\text{SV}})$ is transformed into four parts — the energy $\mathcal{E}_{S_\infty}^{\text{rad}}$ radiated to infinity by passing through $S_\infty$, the energy $\mathcal{E}_{\Omega}^{\text{dis}}$ dissipated in $\Omega$, the increase of the magnetic and electric field energies $\mathcal{E}_{\mathbb{E}^3}^{\text{sto}}$ stored in $\mathbb{E}^3$ , and the increase of the magnetization and polarization energies $\mathcal{E}_{\Omega}^{\text{sto}}$ stored in $\Omega$. Following the terminology used in mechanism[22,23], the above transformation relation between work and energy is called work-energy principle.

## F5.2 Driving Power Operator (DPO) — Line-Surface-Volume Formulation

Clearly, the work term $\mathcal{W}^{\text{Driv}}$ is just the source to sustain the work-energy transformation mentioned above, and it is also the source to drive a steady working of the composite OSS. Thus, $\mathcal{W}^{\text{Driv}}$ is called driving work, and the associated power is called driving power (DP) and denoted as $P^{\text{Driv}}(t)$. Evidently, $P^{\text{Driv}}(t)$ has the following time-domain operator expression

$$P^{\text{Driv}}(t) = \left\langle \boldsymbol{J}^{\text{SL}}(t) + \boldsymbol{J}^{\text{SS}}(t) + \boldsymbol{J}^{\text{SV}}(t), \boldsymbol{E}^{\text{inc}}(t) \right\rangle_\Sigma + \left\langle \boldsymbol{M}^{\text{SV}}(t), \boldsymbol{H}^{\text{inc}}(t) \right\rangle_\Sigma \tag{F-21}$$

and the operator is correspondingly called time-domain driving power operator (DPO).

The frequency-domain version of (F-21) is as follows:

$$P^{\text{driv}} = (1/2)\left\langle \boldsymbol{J}^{\text{SL}} + \boldsymbol{J}^{\text{SS}} + \boldsymbol{J}^{\text{SV}}, \boldsymbol{E}^{\text{inc}} \right\rangle_\Sigma + (1/2)\left\langle \boldsymbol{M}^{\text{SV}}, \boldsymbol{H}^{\text{inc}} \right\rangle_\Sigma \tag{F-22}$$

where coefficient 1/2 originates from the time average for the power-type quadratic quantity of time-harmonic EM field.

## F5.3 Driving Power Operator (DPO) — Line-Surface Formulation

Similarly to the conclusions given in [18,26,33,34], there exists the following relation

$$\frac{1}{2}\left\langle \boldsymbol{J}^{\text{SV}}, \boldsymbol{E}^{\text{inc}} \right\rangle_\Omega + \frac{1}{2}\left\langle \boldsymbol{M}^{\text{SV}}, \boldsymbol{H}^{\text{inc}} \right\rangle_\Omega = -\frac{1}{2}\left\langle \boldsymbol{J}_\cap^{\text{SL}} + \boldsymbol{J}_\cap^{\text{SS}} + \boldsymbol{J}_0^{\text{ES}}, \boldsymbol{E}^{\text{inc}} \right\rangle_{\partial\Omega} - \frac{1}{2}\left\langle \boldsymbol{M}_0^{\text{ES}}, \boldsymbol{H}^{\text{inc}} \right\rangle_{\partial\Omega} \tag{F-23}$$

Substituting the above (F-23) and the previous (F-2)~(F-7) into (F-22), we immediately obtain the following line-surface formulation

$$\begin{aligned} P^{\text{driv}} &= (1/2)\langle \boldsymbol{J}_0^{\text{SL}} + \boldsymbol{J}_0^{\text{SS}}, \boldsymbol{E}^{\text{inc}} \rangle_{L_0 \cup S_0 \cup \partial V_0} \\ &\quad - (1/2)\langle \boldsymbol{J}_0^{\text{ES}}, \boldsymbol{E}^{\text{inc}} \rangle_{\partial\Omega_0} - (1/2)\langle \boldsymbol{M}_0^{\text{ES}}, \boldsymbol{H}^{\text{inc}} \rangle_{\partial\Omega_0} \\ &= (1/2)\langle \boldsymbol{J}_0^{\text{SL}} + \boldsymbol{J}_0^{\text{SS}}, -\boldsymbol{E}^{\text{sca}} \rangle_{L_0 \cup S_0 \cup \partial V_0} \\ &\quad - (1/2)\langle \boldsymbol{J}_0^{\text{ES}}, \boldsymbol{E}^{\text{inc}} \rangle_{\partial\Omega_0} - (1/2)\langle \boldsymbol{M}_0^{\text{ES}}, \boldsymbol{H}^{\text{inc}} \rangle_{\partial\Omega_0} \end{aligned} \tag{F-24}$$

for frequency-domain DPO $P^{\text{driv}}$, where the second equality is because of the tangential electric field boundary condition $\boldsymbol{E}_{\text{tan}}^{\text{inc}} = -\boldsymbol{E}_{\text{tan}}^{\text{sca}}$ on $L_0 \cup S_0 \cup \partial V_0$. Applying LSEP (F-10) to (F-24), we have the $P^{\text{driv}}$ with only current variables as expressed in (F-25),

$$\begin{aligned} P^{\text{driv}} = &\left.\begin{aligned} &\frac{1}{2}\langle \boldsymbol{J}_0^{\text{SL}} + \boldsymbol{J}_0^{\text{SS}}, -j\omega\mu_0\mathcal{L}_0(-\boldsymbol{J}_0^{\text{SL}} - \boldsymbol{J}_0^{\text{SS}} + \boldsymbol{J}_0^{\text{ES}}) - \text{P.V.}\,\mathcal{K}_0(\boldsymbol{M}_0^{\text{ES}}) \rangle_{L_0 \cup S_0 \cup \partial V_0} \\ &- \frac{1}{2}\langle \boldsymbol{J}_0^{\text{ES}}, -j\omega\mu_0\mathcal{L}_0(-\boldsymbol{J}_0^{\text{SL}} - \boldsymbol{J}_0^{\text{SS}} + \boldsymbol{J}_0^{\text{ES}}) - \text{P.V.}\,\mathcal{K}_0(\boldsymbol{M}_0^{\text{ES}}) \rangle_{\partial\Omega_0} \\ &- \frac{1}{2}\langle \boldsymbol{M}_0^{\text{ES}}, \text{P.V.}\,\mathcal{K}_0(-\boldsymbol{J}_0^{\text{SL}} - \boldsymbol{J}_0^{\text{SS}} + \boldsymbol{J}_0^{\text{ES}}) - j\omega\varepsilon_0\mathcal{L}_0(\boldsymbol{M}_0^{\text{ES}}) \rangle_{\partial\Omega_0} \end{aligned}\right\} \text{PVT } P_{\text{PVT}}^{\text{driv}} \\ &\left.- \frac{1}{2}\langle \boldsymbol{J}_0^{\text{ES}}, \hat{\boldsymbol{n}}_- \times (1/2)\boldsymbol{M}_0^{\text{ES}} \rangle_{\partial\Omega_0} - (1/2)\langle \boldsymbol{M}_0^{\text{ES}}, (1/2)\boldsymbol{J}_0^{\text{ES}} \times \hat{\boldsymbol{n}}_- \rangle_{\partial\Omega_0} \right\} \text{SCT } P_{\text{SCT}}^{\text{driv}} \end{aligned} \tag{F-25}$$

where "$\text{P.V.}\,\mathcal{K}_0$" represents the principal value of operator $\mathcal{K}_0$ [264], and the subscripts "PVT" and "SCT" used in $P_{\text{PVT}}^{\text{driv}}$ and $P_{\text{SCT}}^{\text{driv}}$ are the acronyms of "principal value term" and "singular current term" respectively.

If the currents contained in (F-25) are expanded in terms of some proper basis functions, the (F-25) is immediately discretized into the following matrix form

$$P^{\text{driv}} = \mathrm{a}^H \cdot \underbrace{\left(\mathrm{P}_{\text{PVT}}^{\text{driv}} + \mathrm{P}_{\text{SCT}}^{\text{driv}}\right)}_{\mathrm{P}^{\text{driv}}} \cdot \mathrm{a} \tag{F-26}$$

in which the superscript "$H$" represents the conjugate transpose operation for a matrix or vector, and

$$\mathrm{P}_{\text{PVT}}^{\text{driv}} = \begin{bmatrix} \mathrm{P}_{\text{PVT}}^{\boldsymbol{J}_0^{\text{SL}}\boldsymbol{J}_0^{\text{SL}}} & 0 & \mathrm{P}_{\text{PVT}}^{\boldsymbol{J}_0^{\text{SL}}\boldsymbol{J}_0^{\text{SS}}} & 0 & \mathrm{P}_{\text{PVT}}^{\boldsymbol{J}_0^{\text{SL}}\boldsymbol{J}_0^{\text{ES}}} & \mathrm{P}_{\text{PVT}}^{\boldsymbol{J}_0^{\text{SL}}\boldsymbol{M}_0^{\text{ES}}} \\ 0 & 0 & 0 & 0 & 0 & 0 \\ \mathrm{P}_{\text{PVT}}^{\boldsymbol{J}_0^{\text{SS}}\boldsymbol{J}_0^{\text{SL}}} & 0 & \mathrm{P}_{\text{PVT}}^{\boldsymbol{J}_0^{\text{SS}}\boldsymbol{J}_0^{\text{SS}}} & 0 & \mathrm{P}_{\text{PVT}}^{\boldsymbol{J}_0^{\text{SS}}\boldsymbol{J}_0^{\text{ES}}} & \mathrm{P}_{\text{PVT}}^{\boldsymbol{J}_0^{\text{SS}}\boldsymbol{M}_0^{\text{ES}}} \\ 0 & 0 & 0 & 0 & 0 & 0 \\ \mathrm{P}_{\text{PVT}}^{\boldsymbol{J}_0^{\text{ES}}\boldsymbol{J}_0^{\text{SL}}} & 0 & \mathrm{P}_{\text{PVT}}^{\boldsymbol{J}_0^{\text{ES}}\boldsymbol{J}_0^{\text{SS}}} & 0 & \mathrm{P}_{\text{PVT}}^{\boldsymbol{J}_0^{\text{ES}}\boldsymbol{J}_0^{\text{ES}}} & \mathrm{P}_{\text{PVT}}^{\boldsymbol{J}_0^{\text{ES}}\boldsymbol{M}_0^{\text{ES}}} \\ \mathrm{P}_{\text{PVT}}^{\boldsymbol{M}_0^{\text{ES}}\boldsymbol{J}_0^{\text{SL}}} & 0 & \mathrm{P}_{\text{PVT}}^{\boldsymbol{M}_0^{\text{ES}}\boldsymbol{J}_0^{\text{SS}}} & 0 & \mathrm{P}_{\text{PVT}}^{\boldsymbol{M}_0^{\text{ES}}\boldsymbol{J}_0^{\text{ES}}} & \mathrm{P}_{\text{PVT}}^{\boldsymbol{M}_0^{\text{ES}}\boldsymbol{M}_0^{\text{ES}}} \end{bmatrix} \tag{F-27}$$

$$\mathrm{P}_{\text{SCT}}^{\text{driv}} = \begin{bmatrix} 0 & 0 & 0 & 0 & 0 & 0 \\ 0 & 0 & 0 & 0 & 0 & 0 \\ 0 & 0 & 0 & 0 & 0 & 0 \\ 0 & 0 & 0 & 0 & 0 & 0 \\ 0 & 0 & 0 & 0 & 0 & \mathrm{P}_{\text{SCT}}^{\boldsymbol{J}_0^{\text{ES}}\boldsymbol{M}_0^{\text{ES}}} \\ 0 & 0 & 0 & 0 & \mathrm{P}_{\text{SCT}}^{\boldsymbol{M}_0^{\text{ES}}\boldsymbol{J}_0^{\text{ES}}} & 0 \end{bmatrix} \tag{F-28}$$

$$\mathbf{a} = \begin{bmatrix} \mathbf{a}^{\boldsymbol{J}_0^{\text{SL}}} \\ \mathbf{a}^{\boldsymbol{J}_\cap^{\text{SL}}} \\ \mathbf{a}^{\boldsymbol{J}_0^{\text{SS}}} \\ \mathbf{a}^{\boldsymbol{J}_\cap^{\text{SS}}} \\ \mathbf{a}^{\boldsymbol{J}_0^{\text{ES}}} \\ \mathbf{a}^{\boldsymbol{M}_0^{\text{ES}}} \end{bmatrix} \tag{F-29}$$

In (F-29), $\mathbf{a}^{\boldsymbol{J}_0^{\text{SL}}}$, $\mathbf{a}^{\boldsymbol{J}_\cap^{\text{SL}}}$, $\mathbf{a}^{\boldsymbol{J}_0^{\text{SS}}}$, $\mathbf{a}^{\boldsymbol{J}_\cap^{\text{SS}}}$, $\mathbf{a}^{\boldsymbol{J}_0^{\text{ES}}}$ and $\mathbf{a}^{\boldsymbol{M}_0^{\text{ES}}}$ are the column vectors constituted by the expansion coefficients of $\boldsymbol{J}_0^{\text{SL}}$, $\boldsymbol{J}_\cap^{\text{SL}}$, $\boldsymbol{J}_0^{\text{SS}}$, $\boldsymbol{J}_\cap^{\text{SS}}$, $\boldsymbol{J}_0^{\text{ES}}$ and $\boldsymbol{M}_0^{\text{ES}}$ respectively; the 0s in (F-27) and (F-28) are the zero matrices with proper row and column numbers; the formulations for calculating the nonzero sub-matrices in (F-27) and (F-28) are trivial, and they are not explicitly listed here.

In the future Sec. F7, we will utilize the line-surface formulation of frequency-domain DPO to generate the CMs of composite OSSs, and employ the time-average version of DPO to derive famous Parseval's identity and to reveal the physical picture/purpose of CMT.

## F6 Solution Domain Compression and Its Variant

Before constructing CMs in Sec. F7, we first propose a preprocessing scheme — SDC — for the line-surface formulation of frequency-domain DPO in this section.

A directive usage of matrix operator $\mathbb{P}^{\text{driv}}$ gives characteristic equation $\mathbb{P}_-^{\text{driv}} \cdot \mathbf{a}_\xi = \lambda_\xi \mathbb{P}_+^{\text{driv}} \cdot \mathbf{a}_\xi$. Here, $\mathbb{P}_+^{\text{driv}}$ and $\mathbb{P}_-^{\text{driv}}$ are the positive and negative Hermitian parts of $\mathbb{P}^{\text{driv}}$, that is, $\mathbb{P}_+^{\text{driv}} = [\mathbb{P}^{\text{driv}} + (\mathbb{P}^{\text{driv}})^H]/2$ and $\mathbb{P}_-^{\text{driv}} = [\mathbb{P}^{\text{driv}} - (\mathbb{P}^{\text{driv}})^H]/(2j)$.

By solving the characteristic equation, we obtain the modes of the composite OSS shown in Fig. F-2, and the associated modal significances (MSs) are shown in Fig. F-3. The OSS is lossless, and with parameters $\boldsymbol{\mu} = \mathbf{I}4\mu_0$ and $\boldsymbol{\varepsilon}_{\text{c}} = \mathbf{I}4\varepsilon_0$. At the same time, we also calculate the CMs of the OSS by orthogonalizing the line-surface-volume formulation (F-22) of DPO (which is almost identical to the EFIE-VIE-based IMO except a coefficient 1/2), and show the associated MSs in Fig. F-4. Obviously, the results shown in Fig. F-3 are not consistent with the ones shown in Fig. F-4, and we provide a scheme for resolving the problem as below.

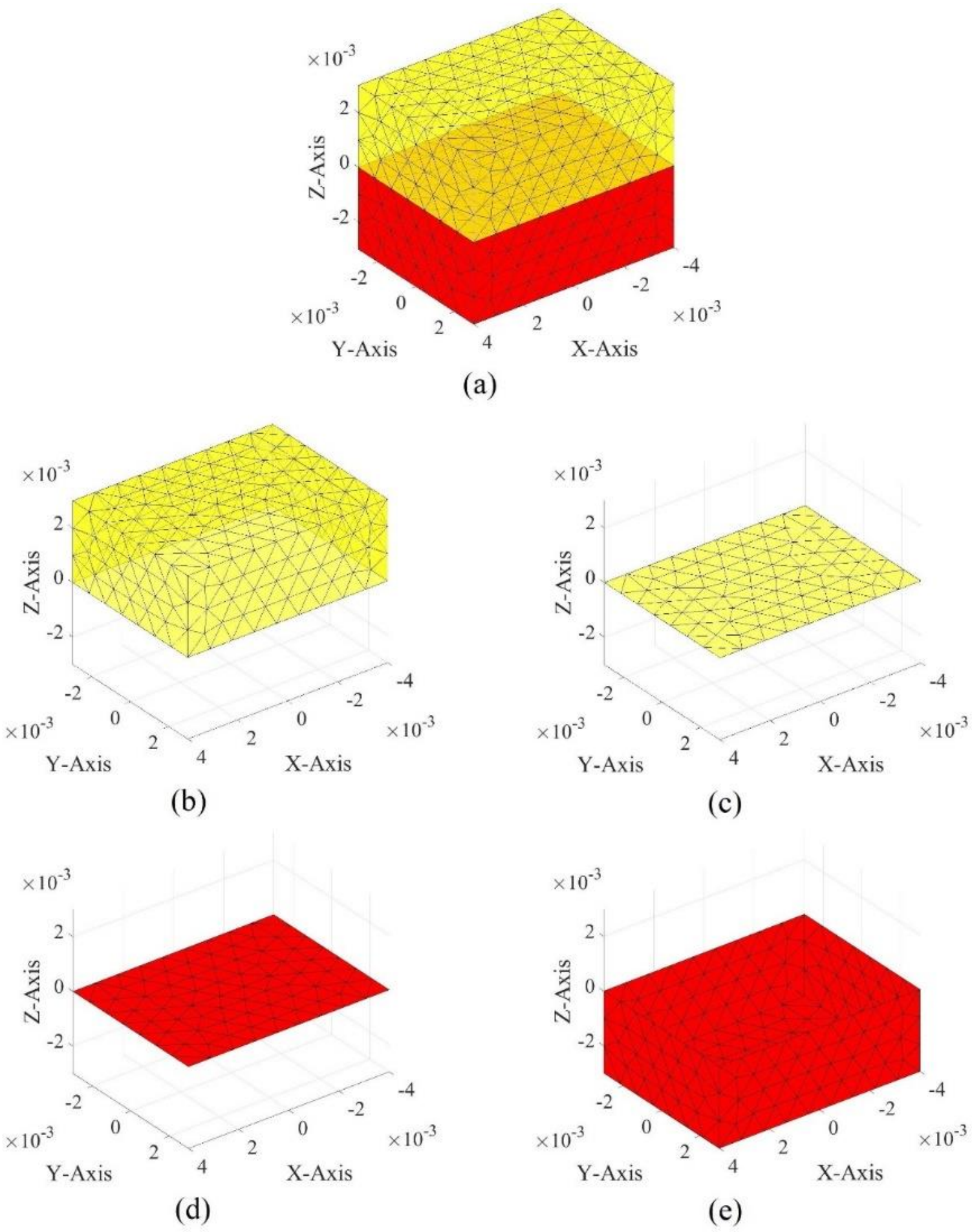

Figure F-2 Geometry of a composite OSS constituted by a material cube and a metallic cube, and the cubes have the same geometrical dimension $8\,\text{mm}\times 6\,\text{mm}\times 3\,\text{mm}$ . (a) Topological structure of whole OSS; (b) material-environment boundary $\partial\Omega_0$ ; (c) material-metal boundary $\partial\Omega_\cap$ ; (d) metal-material boundary $\partial V_\cap$ ; (e) metal-environment boundary $\partial V_0$

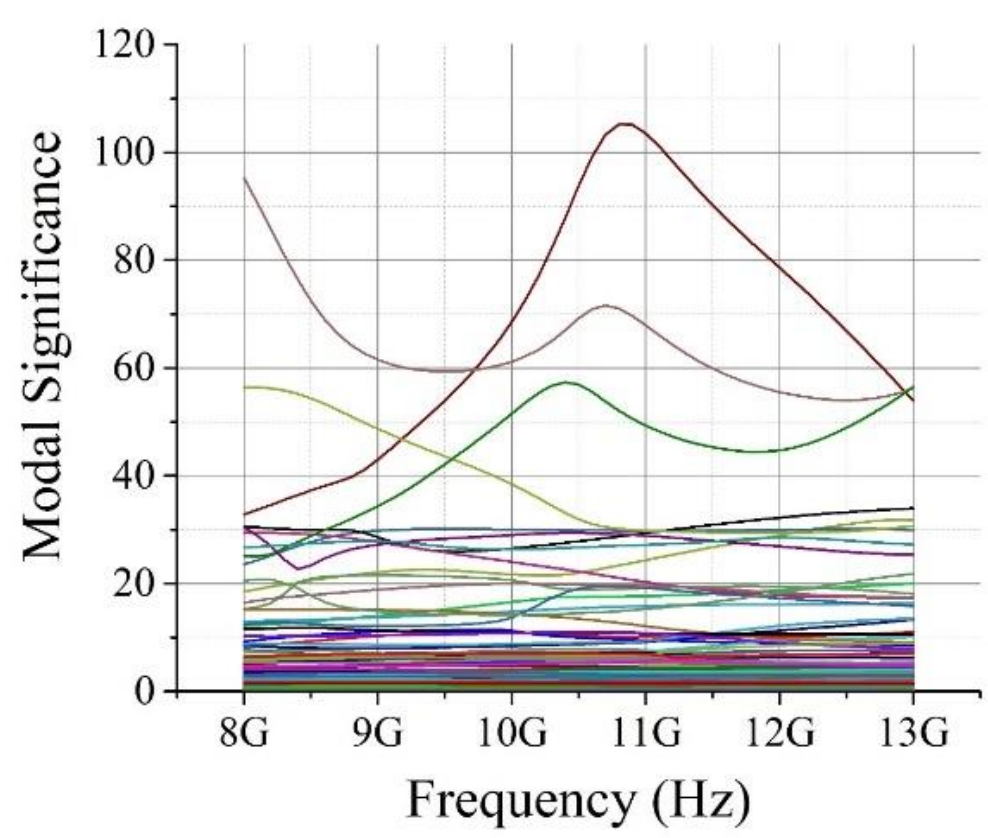

Figure F-3 MSs of some modes of a lossless composite OSS. The OSS is with parameters $\boldsymbol{\mu} = \mathbf{I}4\mu_0$ and $\boldsymbol{\varepsilon}_c = \mathbf{I}4\varepsilon_0$ , and has the topological structure shown in Fig. F-2. The modes shown in this figure are derived from solving characteristic equation $\mathrm{P}_-^{\text{driv}} \cdot \mathrm{a}_\xi = \lambda_\xi \mathrm{P}_+^{\text{driv}} \cdot \mathrm{a}_\xi$

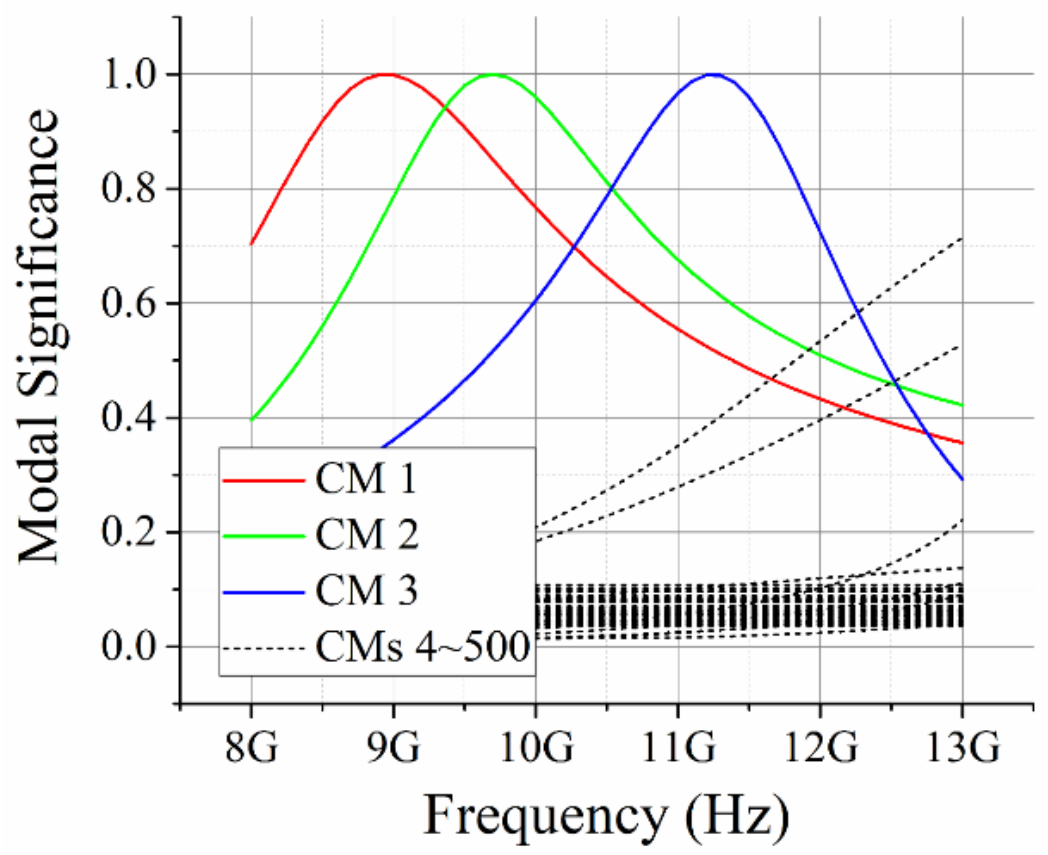

Figure F-4 MSs corresponding to the CMs of a lossless composite OSS. The OSS is with parameters $\boldsymbol{\mu} = \mathbf{I}4\mu_0$ and $\boldsymbol{\varepsilon}_c = \mathbf{I}4\varepsilon_0$, and has the topological structure shown in Fig. F-2. The modes shown in this figure are derived from orthogonalizing DPO (F-22)

The LSEP (F-11) and the current definitions (F-8) and (F-9) give the following integral equations

$$\left[\mathcal{H}\left(\boldsymbol{J}_{\cap}^{\text{SL}} + \boldsymbol{J}_{\cap}^{\text{SS}} + \boldsymbol{J}_{0}^{\text{ES}}, \boldsymbol{M}_{0}^{\text{ES}}\right)\right]_{\boldsymbol{r}'\to\boldsymbol{r}}^{\tan} + \hat{\boldsymbol{n}}_{-}\left(\boldsymbol{r}\right)\times\boldsymbol{J}_{0}^{\text{ES}}\left(\boldsymbol{r}\right) = 0 \tag{F-30}$$

$$\left[\mathcal{E}\left(\boldsymbol{J}_{\cap}^{\text{SL}} + \boldsymbol{J}_{\cap}^{\text{SS}} + \boldsymbol{J}_{0}^{\text{ES}}, \boldsymbol{M}_{0}^{\text{ES}}\right)\right]_{\boldsymbol{r}'\to\boldsymbol{r}}^{\tan} + \boldsymbol{M}_{0}^{\text{ES}}\left(\boldsymbol{r}\right)\times\hat{\boldsymbol{n}}_{-}\left(\boldsymbol{r}\right) = 0 \tag{F-31}$$

In the equations, $\boldsymbol{r}' \in \text{int}\,\Omega$ and $\boldsymbol{r} \in \partial\Omega_0$; subscript "$\boldsymbol{r}' \to \boldsymbol{r}$" means that $\boldsymbol{r}'$ approaches $\boldsymbol{r}$; superscript "tan" means that the equations are satisfied by tangential components.

The LSEP (F-11) and the tangential electric field continuation conditions on metallic boundaries $L_{\cap}$, $S_{\cap}$ and $\partial V_{\cap}$ give the following electric field integral equations

$$\left[\mathcal{E}\left(\boldsymbol{J}_{\cap}^{\text{SL}} + \boldsymbol{J}_{\cap}^{\text{SS}} + \boldsymbol{J}_{0}^{\text{ES}}, \boldsymbol{M}_{0}^{\text{ES}}\right)\right]_{\boldsymbol{r}'\to\boldsymbol{r}}^{\tan} = 0 \tag{F-32}$$

where $\boldsymbol{r}' \in \text{int}\,\Omega$ and $\boldsymbol{r} \in L_{\cap} \cup S_{\cap} \cup \partial V_{\cap}$.

By discretizing the above integral equations (F-30)&(F-31) and (F-32), we obtain the following matrix equations

$$\overbrace{\begin{bmatrix} 0 & \mathbb{G}^{\boldsymbol{M}_0^{\text{ES}}\boldsymbol{J}_{\cap}^{\text{SL}}} & 0 & \mathbb{G}^{\boldsymbol{M}_0^{\text{ES}}\boldsymbol{J}_{\cap}^{\text{SS}}} & \mathbb{G}^{\boldsymbol{M}_0^{\text{ES}}\boldsymbol{J}_0^{\text{ES}}} & \mathbb{G}^{\boldsymbol{M}_0^{\text{ES}}\boldsymbol{M}_0^{\text{ES}}} \end{bmatrix}}^{\mathbb{G}^{\text{DoJ}}} \cdot \mathbb{a} = 0 \tag{F-33}$$

$$\underbrace{\begin{bmatrix} 0 & \mathbb{G}^{\boldsymbol{J}_0^{\text{ES}}\boldsymbol{J}_{\cap}^{\text{SL}}} & 0 & \mathbb{G}^{\boldsymbol{J}_0^{\text{ES}}\boldsymbol{J}_{\cap}^{\text{SS}}} & \mathbb{G}^{\boldsymbol{J}_0^{\text{ES}}\boldsymbol{J}_0^{\text{ES}}} & \mathbb{G}^{\boldsymbol{J}_0^{\text{ES}}\boldsymbol{M}_0^{\text{ES}}} \end{bmatrix}}_{\mathbb{G}^{\text{DoM}}} \cdot \mathbb{a} = 0 \tag{F-34}$$

and

$$\underbrace{\begin{bmatrix} 0 & \mathbb{G}^{\boldsymbol{J}_{\cap}^{\text{SL}}\boldsymbol{J}_{\cap}^{\text{SL}}} & 0 & \mathbb{G}^{\boldsymbol{J}_{\cap}^{\text{SL}}\boldsymbol{J}_{\cap}^{\text{SS}}} & \mathbb{G}^{\boldsymbol{J}_{\cap}^{\text{SL}}\boldsymbol{J}_{0}^{\text{ES}}} & \mathbb{G}^{\boldsymbol{J}_{\cap}^{\text{SL}}\boldsymbol{M}_{0}^{\text{ES}}} \\ 0 & \mathbb{G}^{\boldsymbol{J}_{\cap}^{\text{SS}}\boldsymbol{J}_{\cap}^{\text{SL}}} & 0 & \mathbb{G}^{\boldsymbol{J}_{\cap}^{\text{SS}}\boldsymbol{J}_{\cap}^{\text{SS}}} & \mathbb{G}^{\boldsymbol{J}_{\cap}^{\text{SS}}\boldsymbol{J}_{0}^{\text{ES}}} & \mathbb{G}^{\boldsymbol{J}_{\cap}^{\text{SS}}\boldsymbol{M}_{0}^{\text{ES}}} \end{bmatrix}}_{\mathbb{G}_{\text{FCE}}} \cdot \mathbb{a} = 0 \tag{F-35}$$

satisfied by the physically realizable modes, in which the superscripts "DoJ" and "DoM" and subscript "FCE" are to emphasize the originations of matrices $\mathbb{G}^{\text{DoJ}}$ (from the definition of $\boldsymbol{J}_0^{\text{ES}}$), $\mathbb{G}^{\text{DoM}}$ (from the definition of $\boldsymbol{M}_0^{\text{ES}}$) and $\mathbb{G}_{\text{FCE}}$ (from field continuation equation).

By properly assembling the matrix equations, we have two theoretically equivalent matrix equations as follows:

$$\mathbb{G}_{\text{FCE}}^{\text{DoJ}} \cdot \mathbb{a} = 0 \quad \text{and} \quad \mathbb{G}_{\text{FCE}}^{\text{DoM}} \cdot \mathbb{a} = 0 \tag{F-36}$$

where

$$\mathbb{G}_{\text{FCE}}^{\text{DoJ/DoM}} = \begin{bmatrix} \mathbb{G}^{\text{DoJ/DoM}} \\ \mathbb{G}_{\text{FCE}} \end{bmatrix} \tag{F-37}$$

If the basic solutions of (F-36) are denoted as $\{\mathbb{s}_1, \mathbb{s}_2, \cdots\}$, then any physically realizable mode $\mathbb{a}$ can be uniquely expanded as that

$$\mathbb{a} = \sum_i b_i \mathbb{s}_i = \underbrace{\left[\mathbb{s}_1, \mathbb{s}_2, \cdots\right]}_{\mathbb{S}} \cdot \underbrace{\begin{bmatrix} b_1 \\ b_2 \\ \vdots \end{bmatrix}}_{\mathbb{b}} \tag{F-38}$$

Substituting the (F-38) into (F-26), we derive that

$$\begin{aligned} P^{\text{driv}} &= \underbrace{\left(\mathbb{S}\cdot\mathbb{b}\right)^H}_{\mathbb{a}} \cdot \underbrace{\left(\mathbb{P}_{\text{PVT}}^{\text{driv}} + \mathbb{P}_{\text{SCT}}^{\text{driv}}\right)}_{\mathbb{P}^{\text{driv}}} \cdot \underbrace{\left(\mathbb{S}\cdot\mathbb{b}\right)}_{\mathbb{a}} \\ &= \mathbb{b}^H \cdot \Big( \underbrace{\overbrace{\mathbb{S}^H \cdot \mathbb{P}_{\text{PVT}}^{\text{driv}} \cdot \mathbb{S}}^{\tilde{\mathbb{P}}_{\text{PVT}}^{\text{driv}}} + \overbrace{\mathbb{S}^H \cdot \mathbb{P}_{\text{SCT}}^{\text{driv}} \cdot \mathbb{S}}^{\tilde{\mathbb{P}}_{\text{SCT}}^{\text{driv}}}}_{\tilde{\mathbb{P}}^{\text{driv}}} \Big) \cdot \mathbb{b} \end{aligned} \tag{F-39}$$

It was pointed out in [18] and [34] that the SCT is always equal to the power dissipated by the material OSS considered in [18] and [34]. Similarly, it is not difficult to prove that the SCT is always equal to the power dissipated by the composite OSS considered in this paper, so there exist the following equivalence relations

$$\text{OSS is lossless.} \Leftrightarrow \mathbb{b}^H \cdot \tilde{\mathbb{P}}_{\text{SCT}}^{\text{driv}} \cdot \mathbb{b} = 0 \text{ for any } \mathbb{b}. \Leftrightarrow \tilde{\mathbb{P}}_{\text{SCT}}^{\text{driv}} = 0. \tag{F-40}$$

Thus, we have that

$$P^{\text{driv}} \overset{\text{OSS is lossless}}{=\!=\!=\!=\!=\!=\!=} \mathbb{b}^H \cdot \tilde{\mathbb{P}}_{\text{PVT}}^{\text{driv}} \cdot \mathbb{b} \tag{F-41}$$

for the lossless composite OSSs.

In the following Sec. F7, we will exhibit the fact that: unlike the modes (shown in Fig. F-3) directly derived from equation $\mathbb{P}_{-}^{\text{driv}} \cdot \mathbb{a}_{\xi} = \lambda_{\xi} \mathbb{P}_{+}^{\text{driv}} \cdot \mathbb{a}_{\xi}$, the modes derived from characteristic equations $\tilde{\mathbb{P}}_{-}^{\text{driv}} \cdot \mathbb{b}_{\xi} = \lambda_{\xi} \tilde{\mathbb{P}}_{+}^{\text{driv}} \cdot \mathbb{b}_{\xi}$ (for lossy OSSs) and $\tilde{\mathbb{P}}_{\text{PVT};-}^{\text{driv}} \cdot \mathbb{b}_{\xi} = \lambda_{\xi} \tilde{\mathbb{P}}_{\text{PVT};+}^{\text{driv}} \cdot \mathbb{b}_{\xi}$ (for lossless OSSs) are satisfactory.

## F7 Characteristic Modes, Parseval's Identity and Physical Picture

For a pre-selected composite OSS, there exists a set of inherently working modes — CMs. The CMs span whole modal space, and only depend on the inherent physical properties of the OSS. In this section, we, under WEP framework, construct the CMs by orthogonalizing frequency-domain DPO, and, at the same time, derive famous Parseval's identity, and then reveal the physical picture/purpose of CMT by employing the time-average version of modal orthogonality relations.

### F7.1 Characteristic Modes (CMs)

For square matrices $\tilde{\mathbb{P}}^{\text{driv}}$ and $\tilde{\mathbb{P}}_{\text{PVT}}^{\text{driv}}$, there must uniquely exist the following Toeplitz's decompositions[46, Sec. 0.2.5]

$$\tilde{\mathbb{P}}^{\text{driv}} = \tilde{\mathbb{P}}_{+}^{\text{driv}} + j\, \tilde{\mathbb{P}}_{-}^{\text{driv}} \tag{F-42}$$

$$\tilde{\mathbb{P}}_{\text{PVT}}^{\text{driv}} = \tilde{\mathbb{P}}_{\text{PVT};+}^{\text{driv}} + j\, \tilde{\mathbb{P}}_{\text{PVT};-}^{\text{driv}} \tag{F-43}$$

In (F-42), $\tilde{\mathbb{P}}_{+}^{\text{driv}}$ and $\tilde{\mathbb{P}}_{-}^{\text{driv}}$ are the positive and negative Hermitian parts of $\tilde{\mathbb{P}}^{\text{driv}}$, and $\tilde{\mathbb{P}}_{+}^{\text{driv}} = [\tilde{\mathbb{P}}^{\text{driv}} + (\tilde{\mathbb{P}}^{\text{driv}})^{H}]/2$ and $\tilde{\mathbb{P}}_{-}^{\text{driv}} = [\tilde{\mathbb{P}}^{\text{driv}} - (\tilde{\mathbb{P}}^{\text{driv}})^{H}]/(2j)$. In (F-43), $\tilde{\mathbb{P}}_{\text{PVT};+}^{\text{driv}}$ and $\tilde{\mathbb{P}}_{\text{PVT};-}^{\text{driv}}$ are the positive and negative Hermitian parts of $\tilde{\mathbb{P}}_{\text{PVT}}^{\text{driv}}$, and $\tilde{\mathbb{P}}_{\text{PVT};+}^{\text{driv}} = [\tilde{\mathbb{P}}_{\text{PVT}}^{\text{driv}} + (\tilde{\mathbb{P}}_{\text{PVT}}^{\text{driv}})^{H}]/2$ and $\tilde{\mathbb{P}}_{\text{PVT};-}^{\text{driv}} = [\tilde{\mathbb{P}}_{\text{PVT}}^{\text{driv}} - (\tilde{\mathbb{P}}_{\text{PVT}}^{\text{driv}})^{H}]/(2j)$.

In general, $\tilde{\mathbb{P}}_{+}^{\text{driv}}$ is positive definite, because $\mathbb{b}^{H} \cdot \tilde{\mathbb{P}}_{+}^{\text{driv}} \cdot \mathbb{b}$ is equal to the summation of the radiated and dissipated powers of any $\mathbb{b}$. Thus, there must exist a non-singular matrix having ability to simultaneously diagonalize $\tilde{\mathbb{P}}_{+}^{\text{driv}}$ and $\tilde{\mathbb{P}}_{-}^{\text{driv}}$ [46, Theorem 7.6.4]. Similarly, there also exist a non-singular matrix having ability to simultaneously diagonalize $\tilde{\mathbb{P}}_{\text{PVT};+}^{\text{driv}}$ and $\tilde{\mathbb{P}}_{\text{PVT};-}^{\text{driv}}$, if the OSS is lossless. The column vectors of the above-mentioned non-singular matrices can be obtained by solving the following characteristic equations

$$\tilde{\mathbb{P}}_{-}^{\text{driv}} \cdot \mathbb{b}_{\xi} = \lambda_{\xi}\, \tilde{\mathbb{P}}_{+}^{\text{driv}} \cdot \mathbb{b}_{\xi} \quad , \quad \text{for lossy OSSs} \tag{F-44}$$

$$\tilde{\mathbb{P}}_{\text{PVT};-}^{\text{driv}} \cdot \mathbb{b}_{\xi} = \lambda_{\xi}\, \tilde{\mathbb{P}}_{\text{PVT};+}^{\text{driv}} \cdot \mathbb{b}_{\xi} \quad , \quad \text{for lossless OSSs} \tag{F-45}$$

If the obtained CMs $(\mathrm{b}_1, \mathrm{b}_2, \cdots, \mathrm{b}_d)$ are $d$-order degenerate, then the following Gram-Schmidt orthogonalization process[46, Sec. 0.6.4] is necessary.

$$\left.\begin{aligned} \mathrm{b}_1 &= \mathrm{b}_1' \\ \mathrm{b}_2 - \chi_{12}\mathrm{b}_1' &= \mathrm{b}_2' \\ &\cdots \\ \mathrm{b}_d' - \cdots - \chi_{2d}\mathrm{b}_2' - \chi_{1d}\mathrm{b}_1' &= \mathrm{b}_d' \end{aligned}\right\} \tag{F-46}$$

where the coefficients are calculated as follows:

$$\chi_{mn} = \frac{(\mathrm{b}_m')^H \cdot \mathrm{P}_+^{\mathrm{driv}} \cdot \mathrm{b}_n}{(\mathrm{b}_m')^H \cdot \mathrm{P}_+^{\mathrm{driv}} \cdot \mathrm{b}_m'} \quad , \quad \text{for lossy OSSs} \tag{F-47}$$

$$\chi_{mn} = \frac{(\mathrm{b}_m')^H \cdot \mathrm{P}_{\mathrm{PVT};+}^{\mathrm{driv}} \cdot \mathrm{b}_n}{(\mathrm{b}_m')^H \cdot \mathrm{P}_{\mathrm{PVT};+}^{\mathrm{driv}} \cdot \mathrm{b}_m'} \quad , \quad \text{for lossless OSSs} \tag{F-48}$$

The current expansion vectors corresponding to the CMs can be calculated as that $\mathrm{a}_\xi = \mathrm{S} \cdot \mathrm{b}_\xi$.

Using (F-45) to calculate the CMs of the lossless OSS considered in Sec. F6, the associated MSs are shown in Fig. F-5. Clearly, the results are consistent with the ones shown in Fig. F-4. At the same time, we also show the MSs of the modes derived from orthogonalizing the traditional EFIE-PMCHW-based IMO with SDC scheme in Fig. F-6. Comparing the Figs. F-4, F-5 and F-6, it is not difficult to find out that the DPO is more satisfactory than the traditional IMO in the aspect of generating CMs.

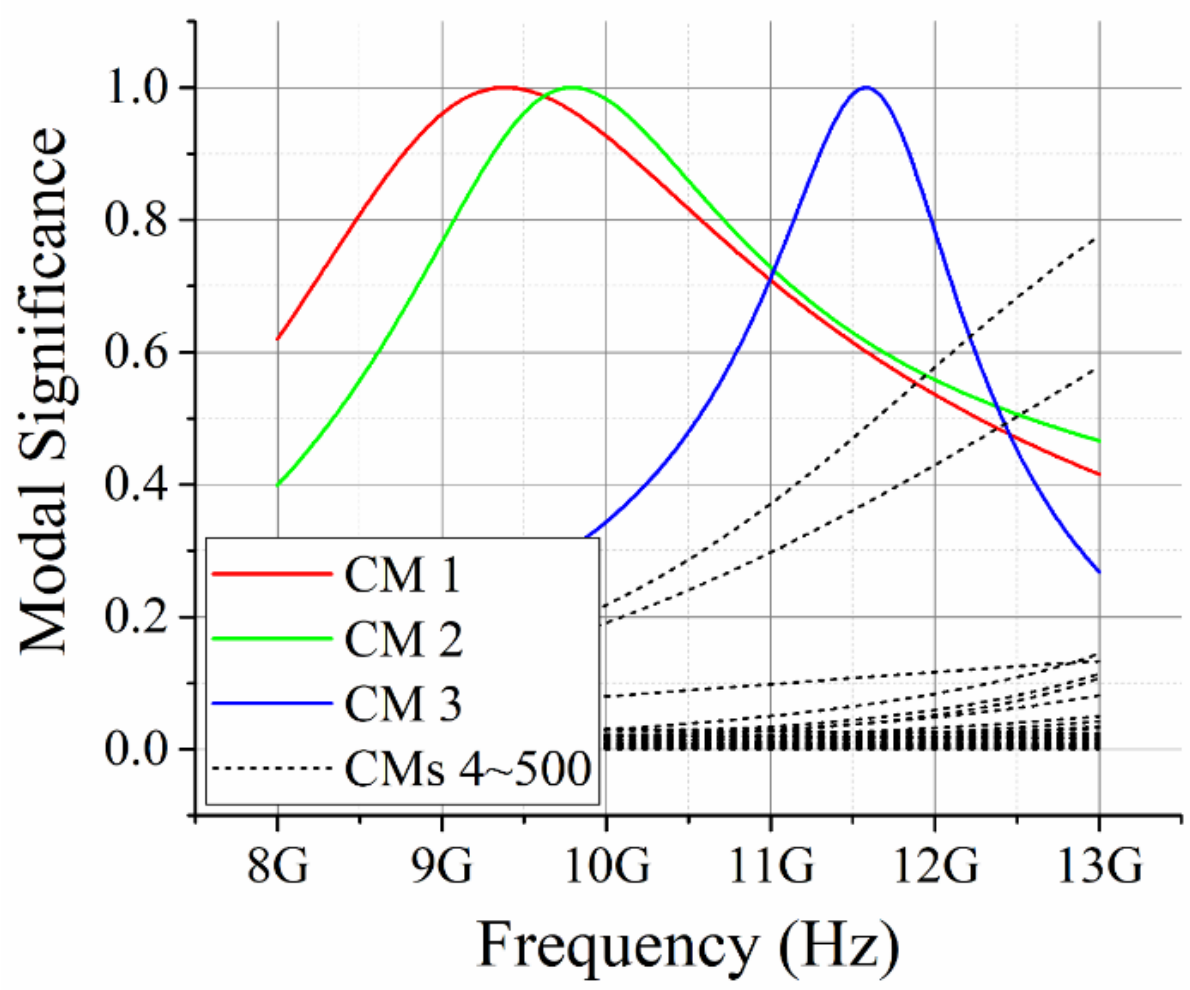

Figure F-5 MSs corresponding to the CMs of a lossless composite OSS. The OSS is with parameters $\boldsymbol{\mu} = \mathbf{I}4\mu_0$ and $\boldsymbol{\varepsilon}_c = \mathbf{I}4\varepsilon_0$, and has the topological structure shown in Fig. F-2. The modes shown in this figure are derived from solving characteristic equation (F-45)

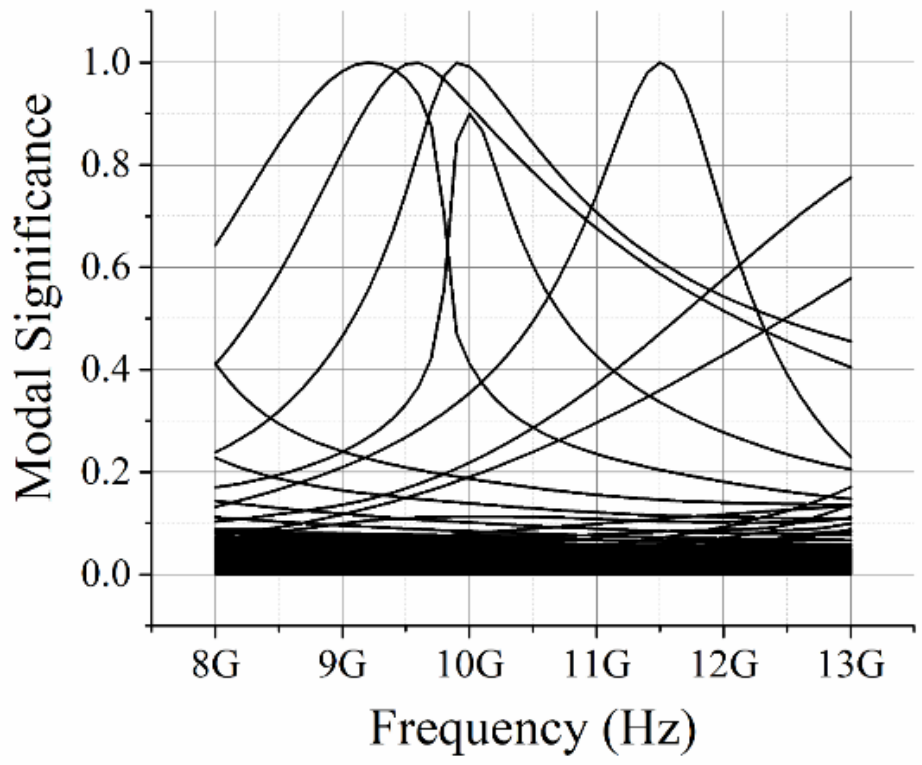

Figure F-6 MSs corresponding to the CMs of a lossless composite OSS. The OSS is with parameters $\boldsymbol{\mu} = \mathbf{I}4\mu_0$ and $\boldsymbol{\varepsilon}_c = \mathbf{I}4\varepsilon_0$, and has the topological structure shown in Fig. F-2. The modes shown in this figure are derived from orthogonalizing the traditional EFIE-PMCHWT-based IMO with SDC scheme

In addition, as pointed out in [18], to obtain the matrix $\mathbb{S}$ usually needs to consume a relatively large number of computational resources, and this problem can be effectively resolved by employing the following alternative characteristic equations

$$\left(\mathbb{P}_{-}^{\text{driv}} + \ell \cdot \mathbb{G}^{H} \cdot \mathbb{G}\right) \cdot \mathbf{a}_{\xi} = \lambda_{\xi} \, \mathbb{P}_{+}^{\text{driv}} \cdot \mathbf{a}_{\xi} \quad , \quad \text{for lossy OSSs} \qquad \text{(F-49)}$$

$$\left(\mathbb{P}_{\text{PVT};-}^{\text{driv}} + \ell \cdot \mathbb{G}^{H} \cdot \mathbb{G}\right) \cdot \mathbf{a}_{\xi} = \lambda_{\xi} \, \mathbb{P}_{\text{PVT};+}^{\text{driv}} \cdot \mathbf{a}_{\xi} \quad , \quad \text{for lossless OSSs} \qquad \text{(F-50)}$$

where $\mathbb{G} = \mathbb{G}_{\text{FCE}}^{\text{DoJ}} / \mathbb{G}_{\text{FCE}}^{\text{DoM}}$, and $\ell$ is an adjustable large real coefficient for example $\ell = 10^{10}$.

A rigorous theoretical explanation for the effectiveness of equations (F-49) and (F-50) had been given in [18]. Here, we emphasize that: only the modes with small or medium $|\lambda_{\xi}|$ are physical, but the ones with very large $|\lambda_{\xi}|$ are spurious. Based on this alternative scheme, we calculate the CMs of the previous lossless OSS, and show the associated MSs in Fig. F-7. Evidently, the results are consistent with the ones shown in Fig. F-4.

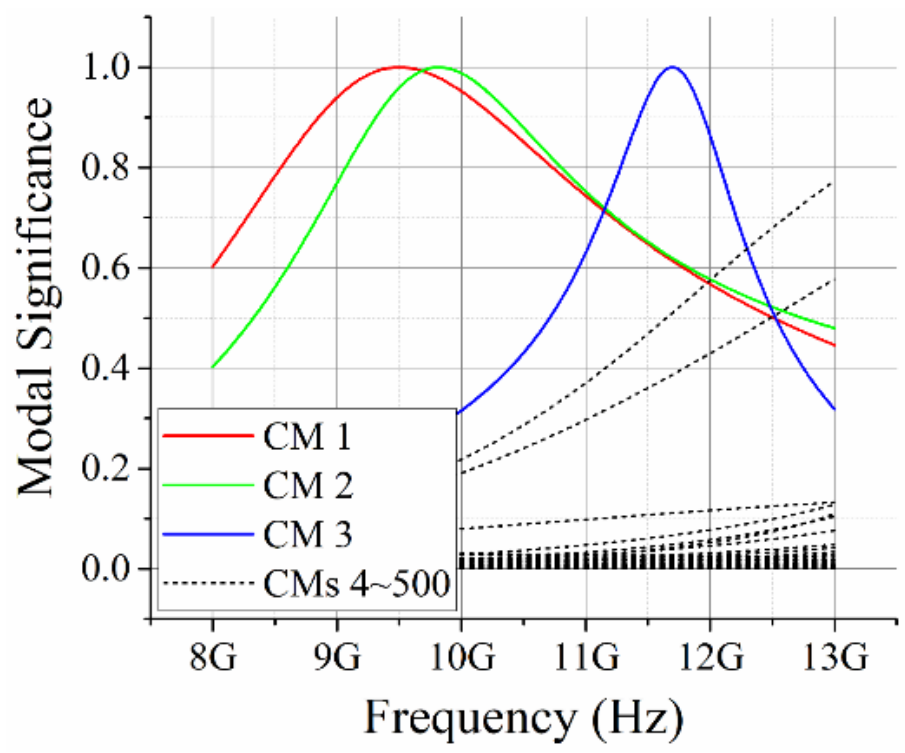

Figure F-7 MSs corresponding to the CMs of a lossless composite OSS. The OSS is with parameters $\boldsymbol{\mu} = \mathbf{I}4\mu_0$ and $\boldsymbol{\varepsilon}_c = \mathbf{I}4\varepsilon_0$, and has the topological structure shown in Fig. F-2. The modes shown in this figure are derived from solving characteristic equation (F-50)

## F7.2 Parseval's Identity

Similar to [18] and [26], it is easy to prove that the above-obtained CMs satisfy the following frequency-domain orthogonality relation

$$P_{\xi}^{\text{driv}}\delta_{\xi\zeta} = (1/2)\left\langle \boldsymbol{J}_{\xi}^{\text{SL}} + \boldsymbol{J}_{\xi}^{\text{SS}} + \boldsymbol{J}_{\xi}^{\text{SV}}, \boldsymbol{E}_{\zeta}^{\text{inc}} \right\rangle_{\Sigma} + (1/2)\left\langle \boldsymbol{M}_{\xi}^{\text{SV}}, \boldsymbol{H}_{\zeta}^{\text{inc}} \right\rangle_{\Sigma} \tag{F-51}$$

where $\delta_{\xi\zeta}$ is the Kronecker's delta symbol.

Because of the completeness of CMs, any working mode can be expressed in terms of the linear expansion of the CMs, and orthogonality relation (F-51) implies the following expansion coefficients

$$c_{\xi} = \frac{(1/2)\left\langle \boldsymbol{J}_{\xi}^{\text{SL}} + \boldsymbol{J}_{\xi}^{\text{SS}} + \boldsymbol{J}_{\xi}^{\text{SV}}, \boldsymbol{E}^{\text{inc}} \right\rangle_{\Sigma} + (1/2)\left\langle \boldsymbol{M}_{\xi}^{\text{SV}}, \boldsymbol{H}^{\text{inc}} \right\rangle_{\Sigma}}{1 + j\lambda_{\xi}} \tag{F-52}$$

Here the modal real power $\text{Re}\{P_{\xi}^{\text{driv}}\}$ has been normalized to 1.

By employing the above (F-51) and (F-52) and the previous (F-21) and (F-22), we can derive the following Parseval's identity

$$(1/T)\int_{t_0}^{t_0+T} P^{\text{Driv}}(t)\,dt = \sum_{\xi} \left| c_{\xi} \right|^2 \tag{F-53}$$

where $T$ is the time period of the time-harmonic EM field.

## F7.3 Physical Picture

Evidently, frequency-domain power-decoupling relation (F-51) implies the following time-domain energy-decoupling relation (or alternatively called time-average power-decoupling relation)

$$\underbrace{\text{Re}\left\{P_{\xi}^{\text{driv}}\right\}}_{1}\delta_{\xi\zeta} = \frac{1}{T}\int_{t_0}^{t_0+T}\left[\left\langle \boldsymbol{J}_{\xi}^{\text{SL}}(t) + \boldsymbol{J}_{\xi}^{\text{SS}}(t) + \boldsymbol{J}_{\xi}^{\text{SV}}(t), \boldsymbol{E}_{\zeta}^{\text{inc}}(t) \right\rangle_{\Sigma} + \left\langle \boldsymbol{M}_{\xi}^{\text{SV}}(t), \boldsymbol{H}_{\zeta}^{\text{inc}}(t) \right\rangle_{\Sigma}\right]dt \tag{F-54}$$

where $\text{Re}\{P_{\xi}^{\text{driv}}\} = 1$ as explained in [18]. Relation (F-54) has a very clear physical interpretation: in any integral period, if $\xi \neq \zeta$, the $\zeta$-th modal fields $(\boldsymbol{E}_{\zeta}^{\text{inc}}, \boldsymbol{H}_{\zeta}^{\text{inc}})$ don't supply net energy to the $\xi$-th modal currents $(\boldsymbol{J}_{\xi}^{\text{SL}}, \boldsymbol{J}_{\xi}^{\text{SS}}, \boldsymbol{J}_{\xi}^{\text{SV}}, \boldsymbol{M}_{\xi}^{\text{SV}})$. This physical interpretation gives CMT a very clear physical picture/purpose — to construct a set of steadily working energy-decoupled modes for scattering systems.

In addition, we can also prove the following frequency-domain equivalent of (F-54).

$$\underbrace{\text{Re}\left\{P_{\xi}^{\text{driv}}\right\}}_{1}\delta_{\xi\zeta} = \frac{1}{2\eta_0}\left\langle \boldsymbol{E}_{\xi}^{\text{sca}}, \boldsymbol{E}_{\zeta}^{\text{sca}} \right\rangle_{S_{\infty}} + (1/2)\left\langle \boldsymbol{\sigma} \cdot \boldsymbol{E}_{\xi}^{\text{tot}}, \boldsymbol{E}_{\zeta}^{\text{tot}} \right\rangle_{\Omega} \tag{F-55}$$

in which $\eta_0$ is the free-space wave impedance and $\eta_0 = \sqrt{\mu_0/\varepsilon_0}$ . Relation (F-55) clearly reveals the following important facts F1 and F2:

F1 when the OSS is lossless (that is, $\boldsymbol{\sigma} = 0$), the far-field orthogonality relation of CMs holds automatically, because the second term in the right-hand side is always zero;

F2 when the OSS is lossy (that is, $\boldsymbol{\sigma} \neq 0$), the far-field orthogonality relation of CMs cannot be guaranteed usually, because the second term in the right-hand side cannot be guaranteed to be zero.

Here, we emphasize again that the physical purpose/picture of CMT is to construct the modes without net energy exchange in integral period (that is, the modes satisfying (F-54)) rather than the modes with orthogonal modal far fields, though, in the lossless case, the formers indeed have orthogonal far fields automatically. The detailed explanations for this conclusion had been given in [18].

Using equations (F-44) and (F-49) to calculate the CMs of the lossy OSS whose topological structure is the same as the one shown in Fig. F-2 and material parameters are $\boldsymbol{\mu} = \mathbf{I}4\mu_0$ and $\boldsymbol{\varepsilon}_{\rm c} = \mathbf{I}4\varepsilon_0 - j\mathbf{I}0.5/\omega$, the associated MSs are illustrated in the following Fig. F-8. At the same time, the EFIE-VIE-based and EFIE-PMCHWT-based results are shown in Fig. F-9 and Fig. F-10 for comparisons. Obviously, the results shown in Fig. F-8 and Fig. F-9 are consistent with each other, but the they are not consistent with the ones shown in Fig. F-10.

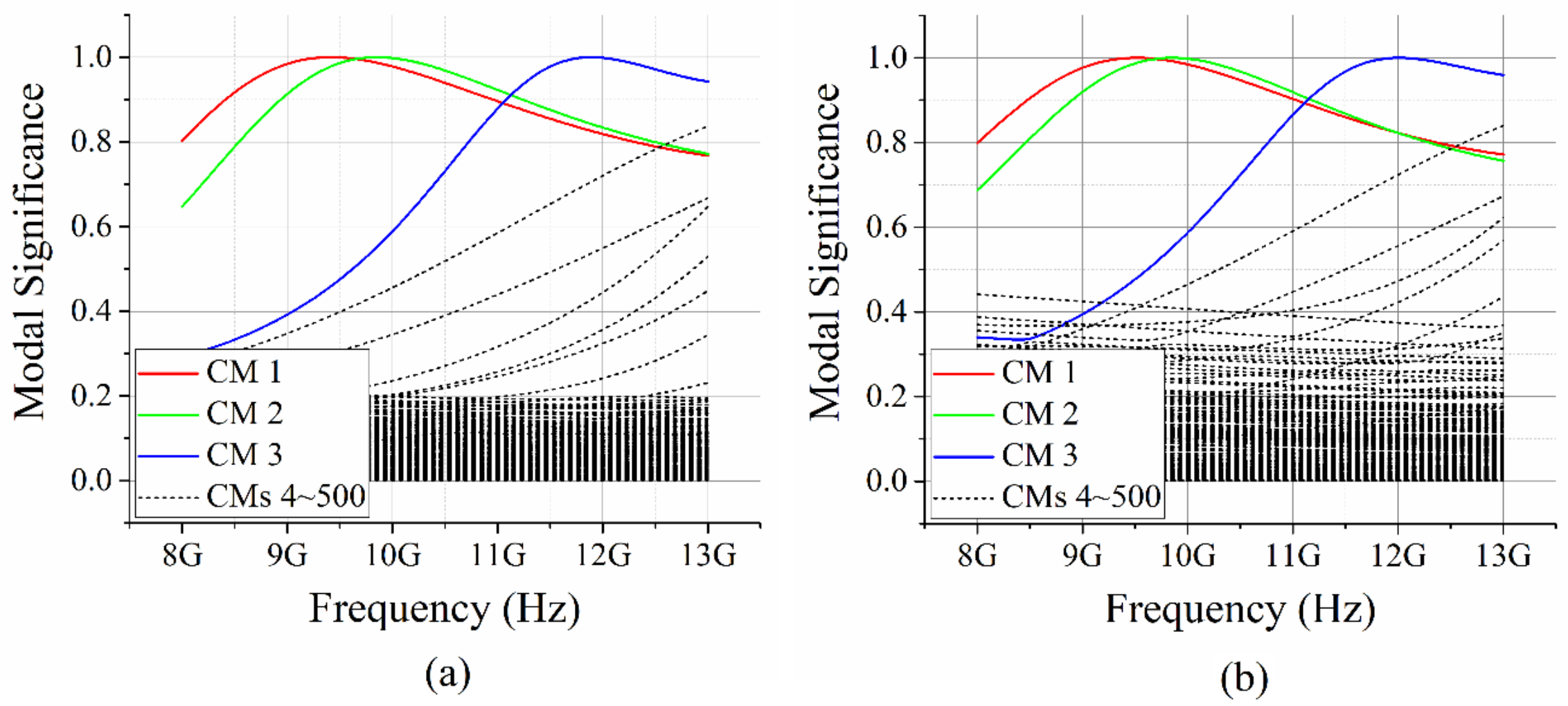

Figure F-8 MSs corresponding to the CMs of a lossy composite OSS. The OSS is with parameters $\boldsymbol{\mu} = \mathbf{I}4\mu_0$ and $\boldsymbol{\varepsilon}_{\rm c} = \mathbf{I}4\varepsilon_0 - j\mathbf{I}0.5/\omega$ , and has the topological structure shown in Fig. F-2. The CMs shown above are derived from solving (a) characteristic equation (F-44) and (b) characteristic equation (F-49) with $\ell = 10^{10}$

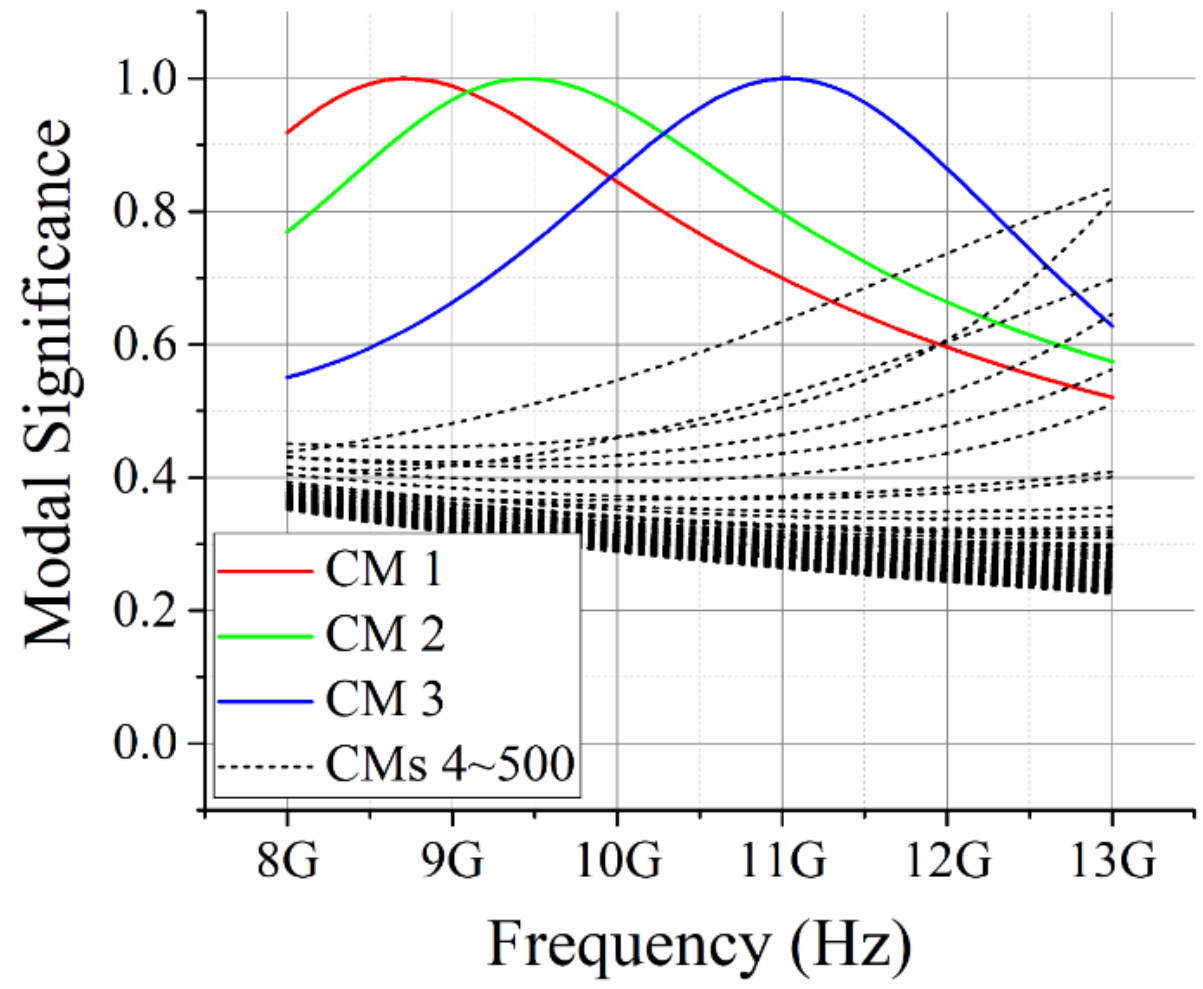

Figure F-9 MSs corresponding to the CMs of a lossy composite OSS. The OSS is with parameters $\boldsymbol{\mu} = \mathbf{I}4\mu_0$ and $\boldsymbol{\varepsilon}_c = \mathbf{I}4\varepsilon_0 - j\mathbf{I}0.5/\omega$, and has the topological structure shown in Fig. F-2. The CMs shown in this figure are derived from orthogonalizing DPO (F-22)

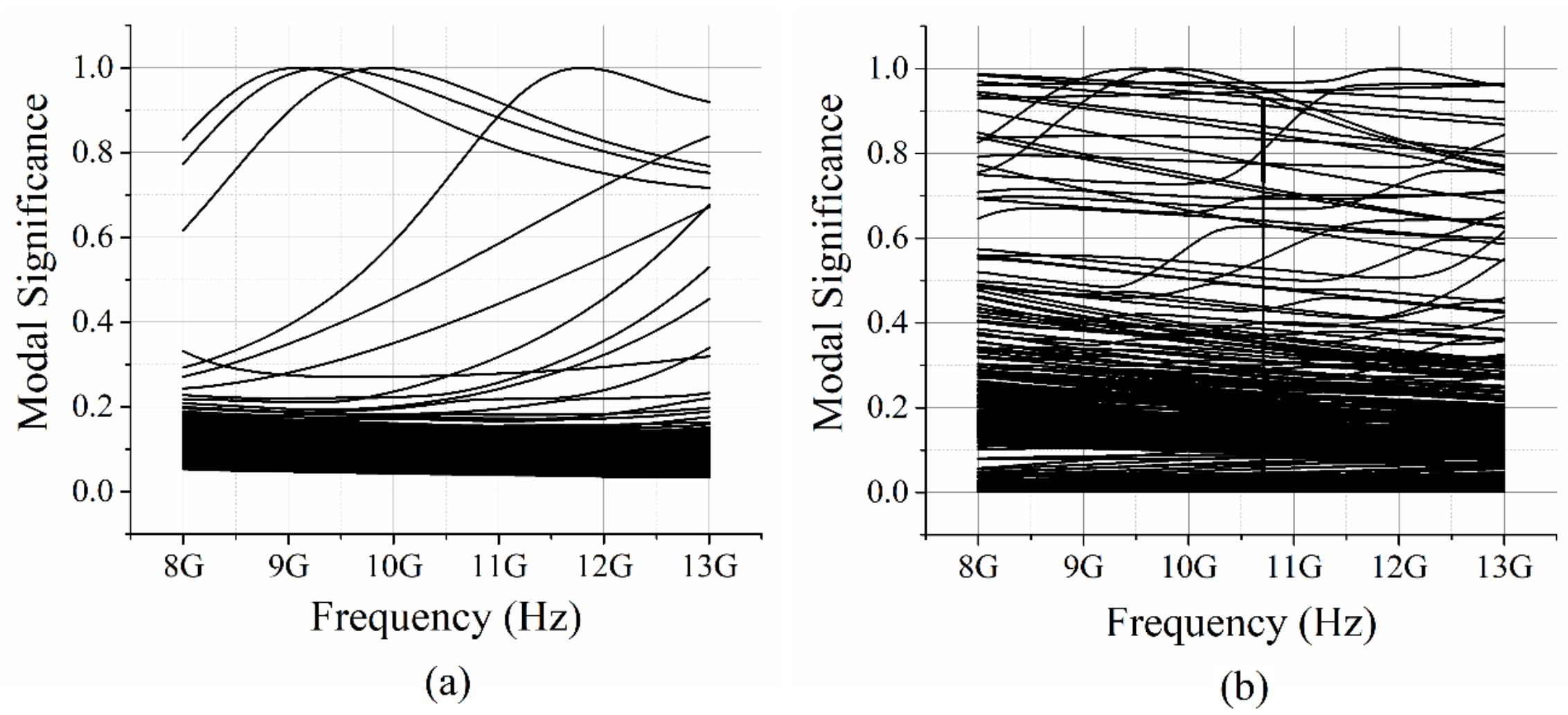

Figure F-10 MSs corresponding to the CMs of a lossy composite OSS. The OSS is with parameters $\boldsymbol{\mu} = \mathbf{I}4\mu_0$ and $\boldsymbol{\varepsilon}_c = \mathbf{I}4\varepsilon_0 - j\mathbf{I}0.5/\omega$, and has the topological structure shown in Fig. F-2. The CMs shown above are derived from solving EFIE-PMCHWT-based operator with (a) the original SDC scheme and (b) the weighted SDC scheme using $\ell = 10^{10}$

For the (F-44)-based CMs at 11 GHz, their time-average DP orthogonality matrix is shown in Fig. F-11(a), and their far-field orthogonality matrix is shown in Fig. F-11(b). Obviously, the modal orthogonality relations (F-54) and (F-55) indeed hold, but the modal far-field orthogonality relation doesn't.

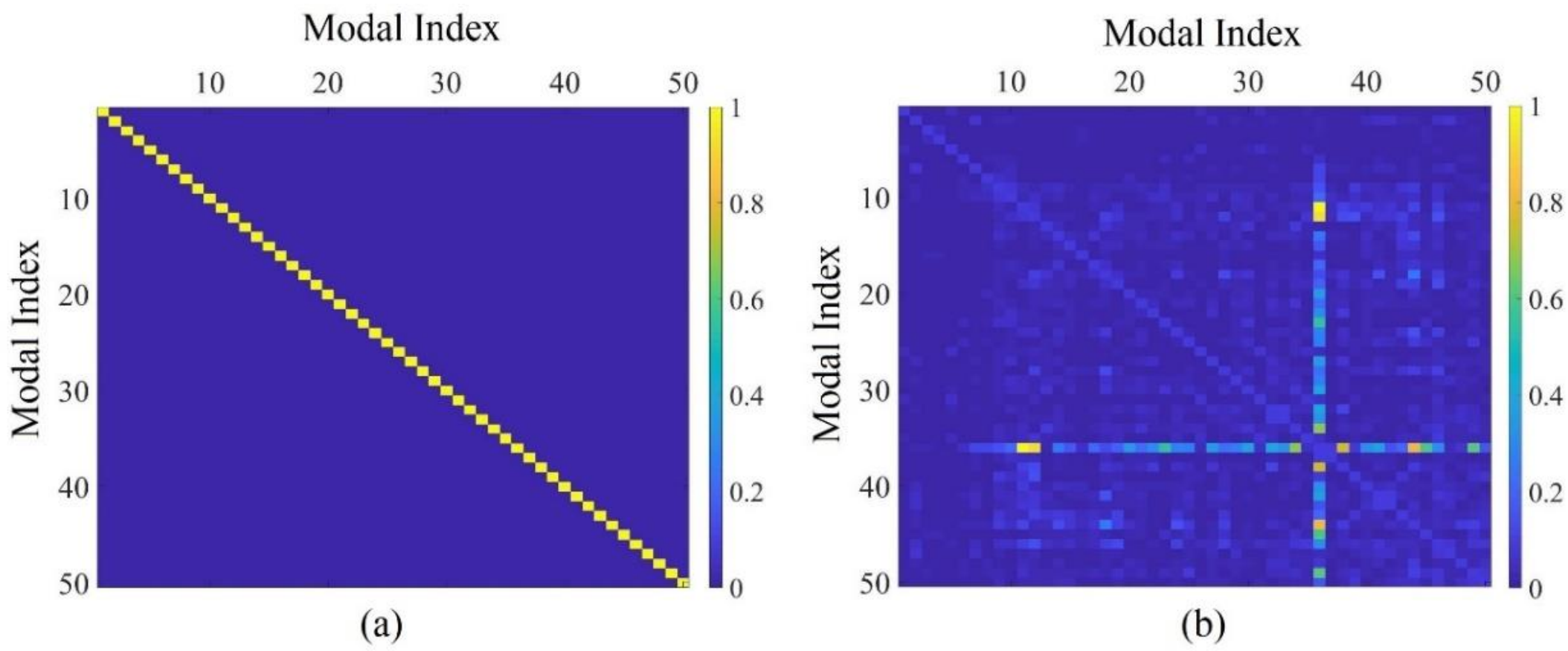

Figure F-11 (a) Time-average DP orthogonality matrix and (b) far-field orthogonality matrix about the CMs of a lossy composite OSS working at 11 GHz. The OSS is with parameters $\boldsymbol{\mu} = \mathbf{I}4\mu_0$ and $\boldsymbol{\varepsilon}_{\rm c} = \mathbf{I}4\varepsilon_0 - j\mathbf{I}0.5/\omega$, and has the topological structure shown in Fig. F-2. The CMs corresponding to this figure are derived from solving equation (F-44)

## F8 Numerical Examples

In the above Sec. F7, the CMs of a typical composite OSS — a metallic cube partially coated by a material cube — has been successfully constructed by orthogonalizing the frequency-domain DPO with SDC scheme. In this section, the CMs of some other typical composite OSSs are calculated in WEP framework to further verify the validity of the theory developed in this paper, and some comparisons are also done to exhibit the advantages of the WEP-CMT over the conventional IE-CMT.

### F8.1 A Metallic Core Completely Coated by a Material Shell

In this subsection, we consider the composite OSS constituted by a metallic cube and a material cubic shell, and the metallic cube is completely coated by the material cubic shell. The topological structure and the sub-boundaries of the composite OSS are shown in Fig. F-12.

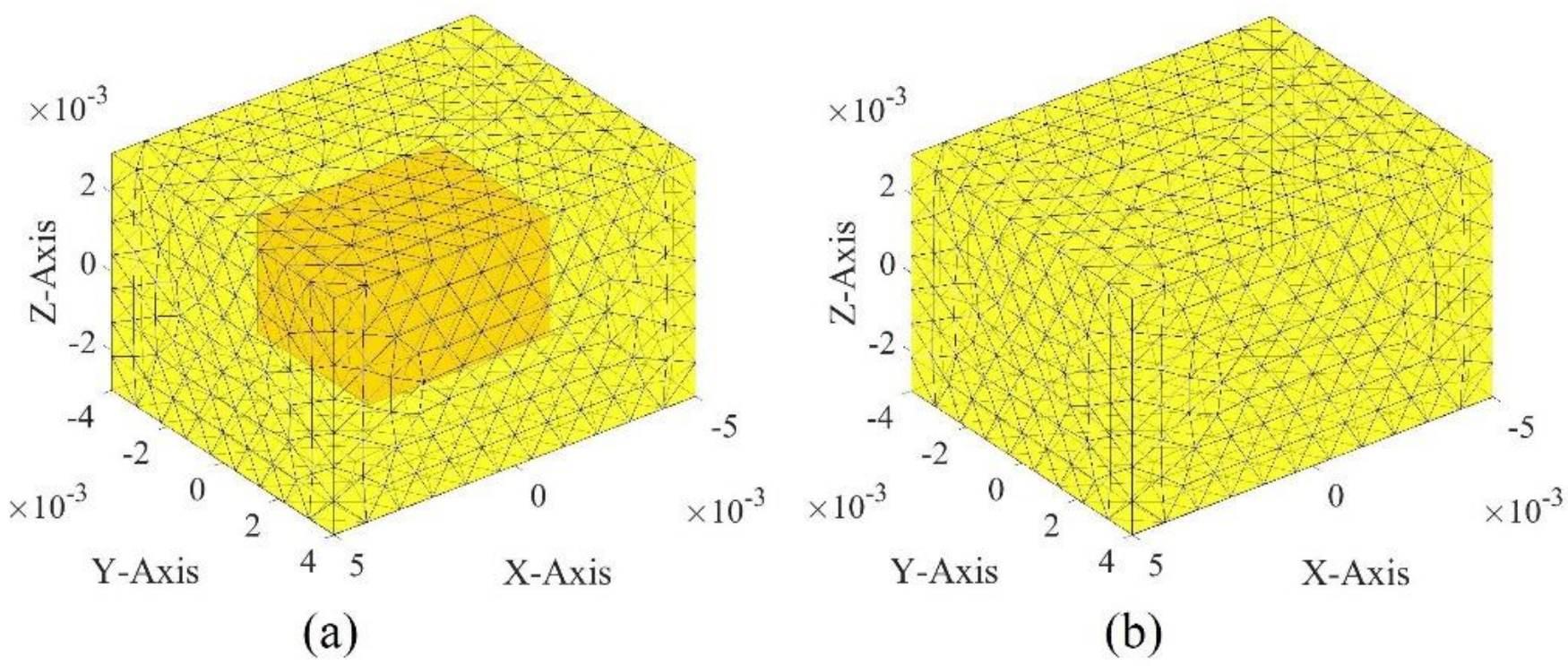

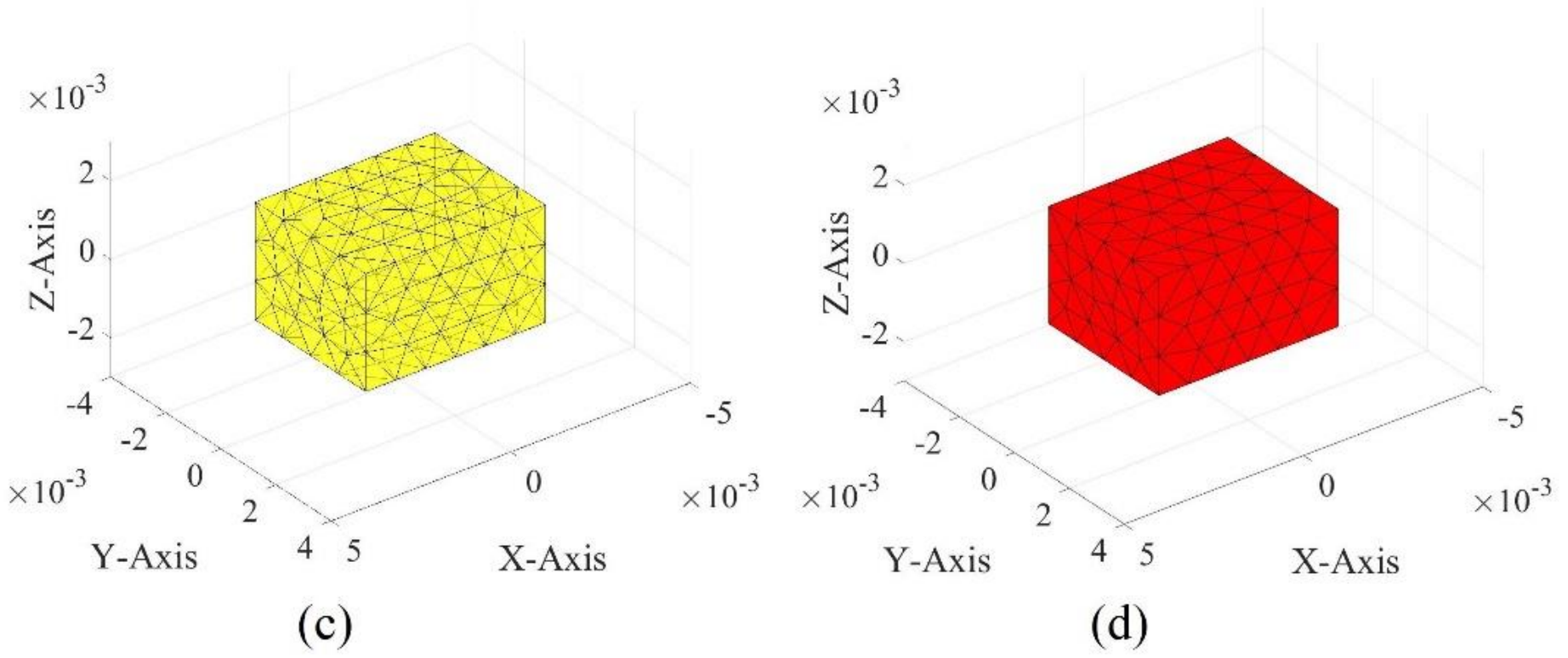

Figure F-12 Geometry of a composite OSS constituted by a metallic cube and a material shell, where the sizes of the metallic cube and the inner surface of the material shell are both $5\,\text{mm}\times 4\,\text{mm}\times 3\,\text{mm}$, and the size of the outer surface of the material shell is $10\,\text{mm}\times 8\,\text{mm}\times 6\,\text{mm}$. (a) Topological structure of whole OSS; (b) material-environment boundary $\partial\Omega_0$; (c) material-metal boundary $\partial\Omega_\cap$; (d) metal-material boundary $\partial V_\cap$ (which is just whole $\partial V$). Obviously, $\partial\Omega_\cap = \partial V_\cap$ for this composite OSS

When the composite OSS is lossless and with parameters $\boldsymbol{\mu} = \mathbf{I}4\mu_0$ and $\boldsymbol{\varepsilon}_c = \mathbf{I}4\varepsilon_0$, we use five different ways to construct its CMs, and show the obtained results in Figs. F-13(a) ~ F-13(e). In Fig. F-13(a), the CMs are derived from orthogonalizing EFIE-VIE-based IMO, that is DPO (F-22). In Figs. F-13(b) and F-13(c), the CMs are respectively derived from orthogonalizing the traditional EFIE-PMCHWT-based IMO with SDC scheme and weighted SDC scheme. In Fig. F-13(d), the CMs are derived from orthogonalizing the DPO with SDC scheme, that is, solving characteristic equation (F-45). In Fig. F-13(e), the CMs are derived from orthogonalizing the DPO with weighted SDC scheme, that is, solving characteristic equation (F-50).

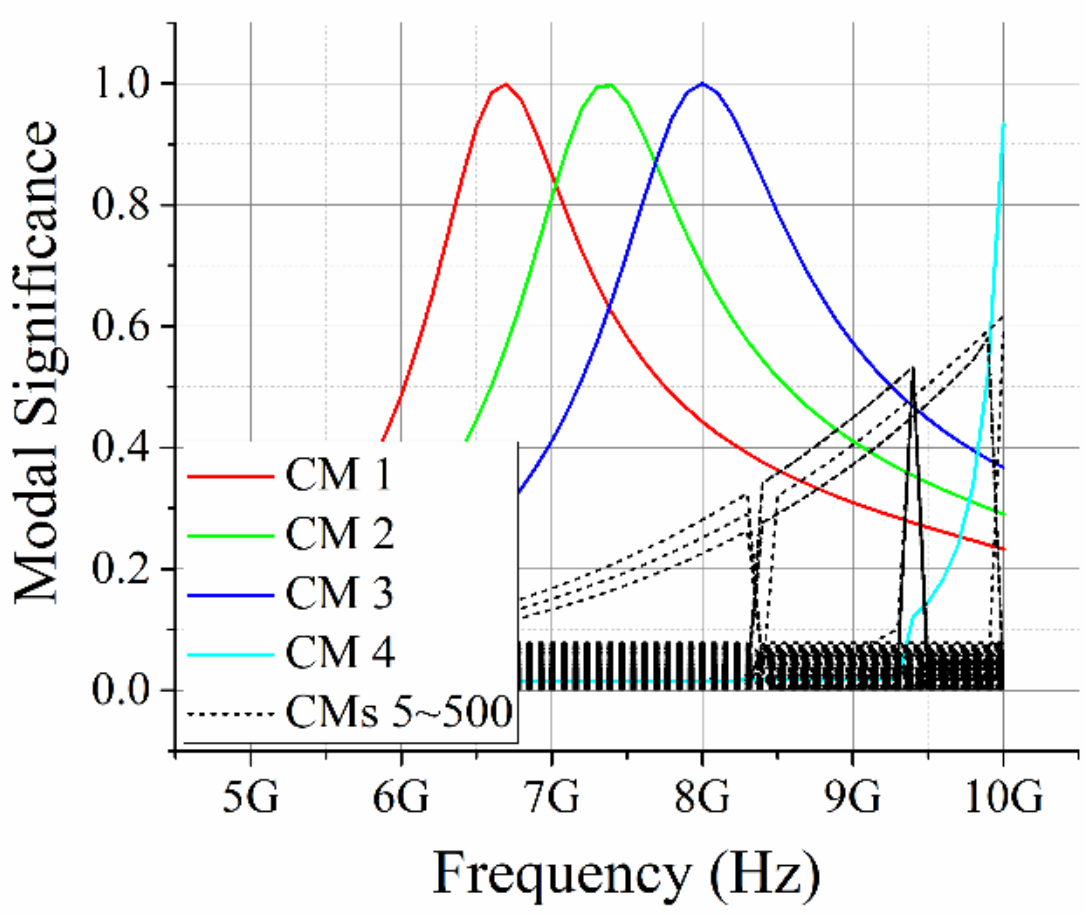

(a)

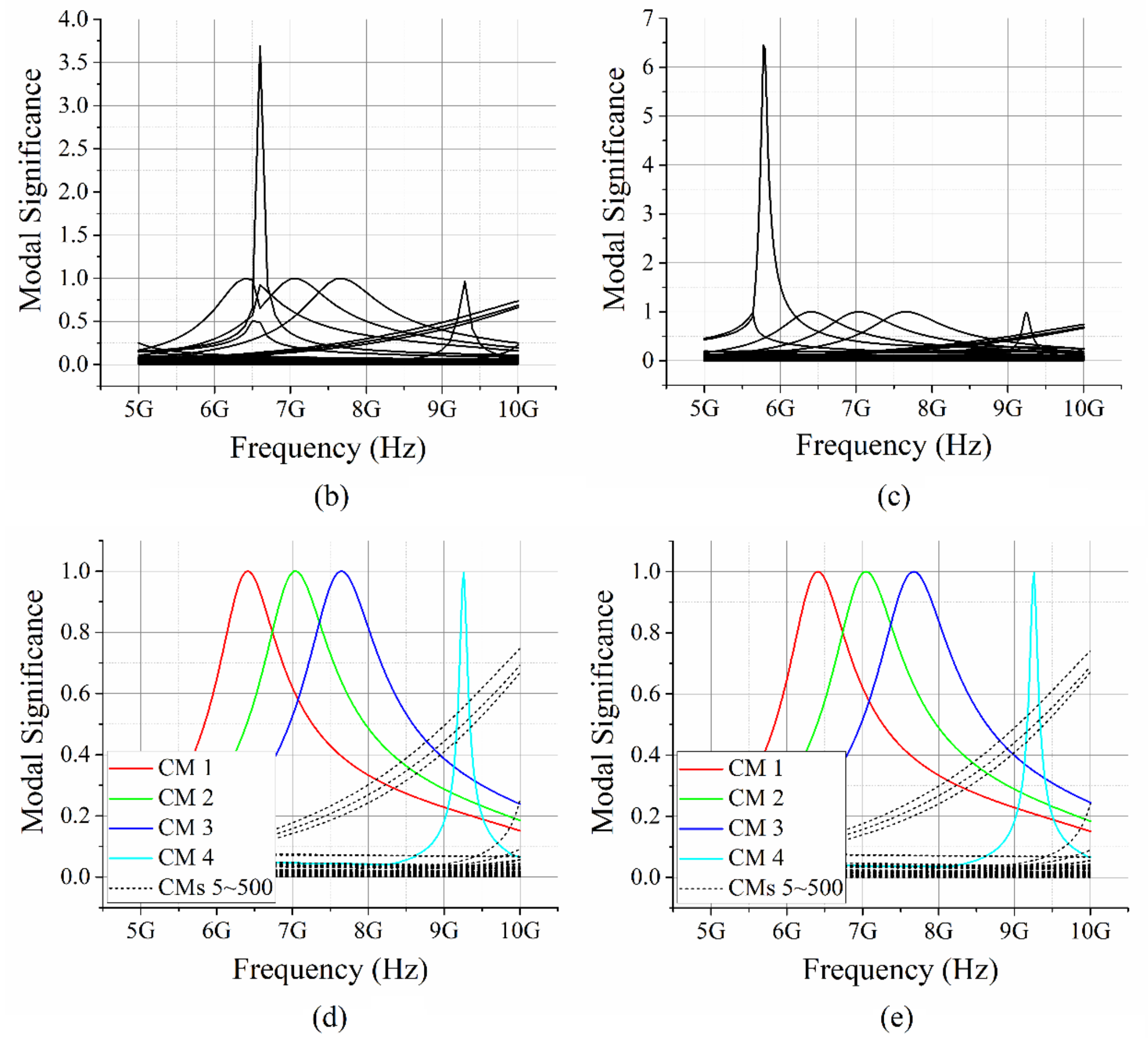

Figure F-13 MSs corresponding to the CMs of a lossless composite OSS. The OSS is with parameters $\boldsymbol{\mu} = \mathbf{I}4\mu_0$ and $\boldsymbol{\varepsilon}_{\rm c} = \mathbf{I}4\varepsilon_0$, and has the topological structure shown in Fig. F-12. The CMs are derived from (a) EFIE-VIE-based IMO, that is DPO (F-22), (b) the EFIE-PMCHWT-based IMO with SDC scheme, (c) the EFIE-PMCHWT-based IMO with weighted SDC scheme using $\ell = 10^{10}$, (d) characteristic equation (F-45), and (e) characteristic equation (F-50) with $\ell = 10^{10}$

When the composite OSS is lossy and with parameters $\boldsymbol{\mu} = \mathbf{I}4\mu_0$ and $\boldsymbol{\varepsilon}_{\rm c} = \mathbf{I}4\varepsilon_0 - j\mathbf{I}0.5/\omega$, we also use five different ways to construct its CMs, and show the obtained results in Figs. F-14(a) ~ F-14(e). In Fig. F-14(a), the CMs are derived from orthogonalizing EFIE-VIE-based IMO, that is DPO (F-22). In Figs. F-14(b) and F-14(c), the CMs are respectively derived from orthogonalizing the traditional EFIE-PMCHWT-based IMO with SDC scheme and weighted SDC scheme. In Fig. F-14(d), the CMs are derived from orthogonalizing the DPO with SDC scheme, that is, solving characteristic equation (F-44). In Fig. F-14(e), the CMs are derived from orthogonalizing the DPO with weighted SDC scheme, that is, solving characteristic equation (F-49).

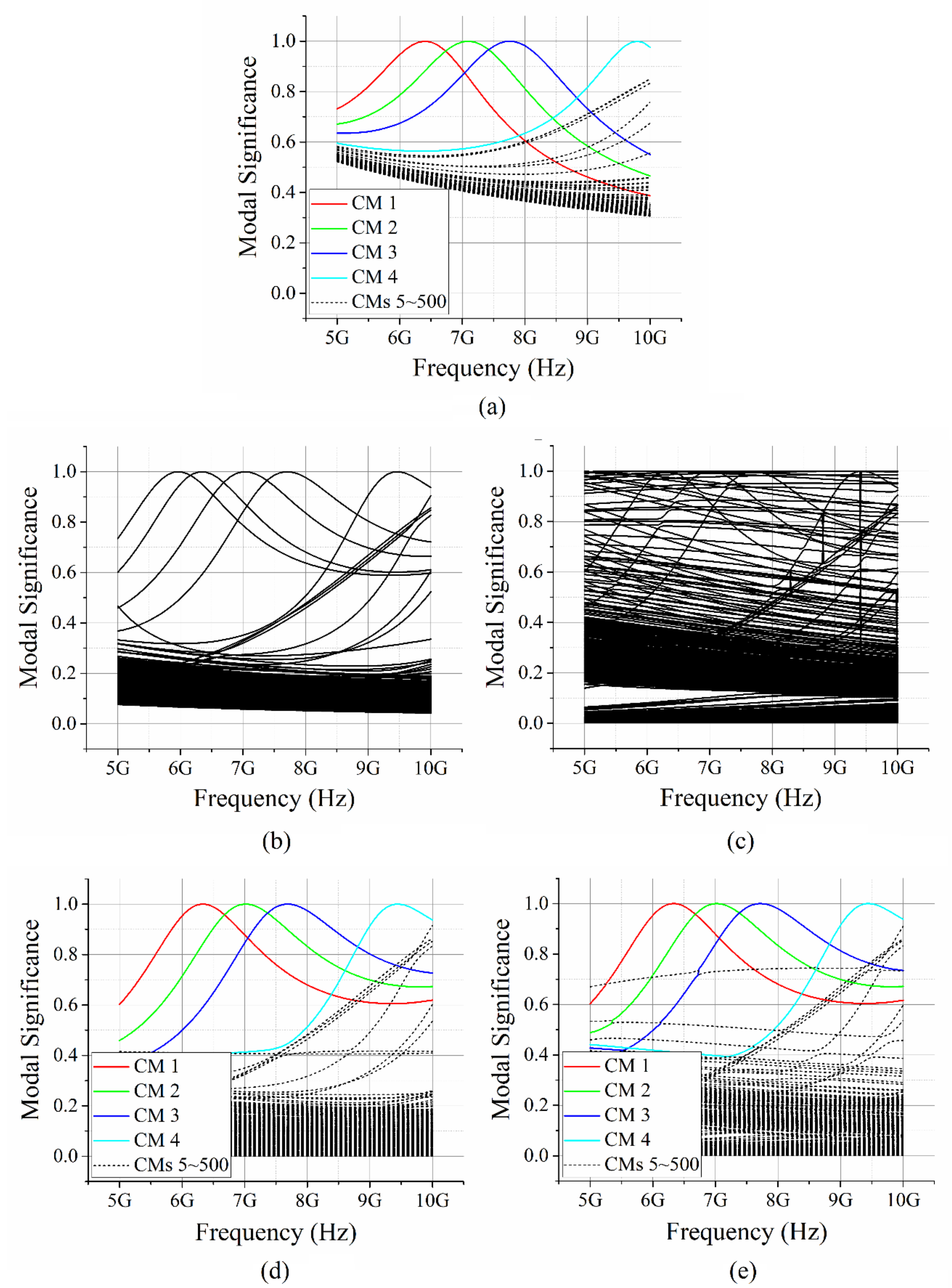

Figure F-14 MSs corresponding to the CMs of a lossy composite OSS. The OSS is with parameters $\boldsymbol{\mu} = \mathbf{I}4\mu_0$ and $\boldsymbol{\varepsilon}_c = \mathbf{I}4\varepsilon_0 - j\mathbf{I}0.5/\omega$, and has the topological structure shown in Fig. F-12. The CMs are derived from (a) EFIE-VIE-based IMO, that is DPO (F-22), (b) the EFIE-PMCHWT-based IMO with SDC scheme, (c) the EFIE-PMCHWT-based IMO with weighted SDC scheme using $\ell = 10^{10}$, (d) characteristic equation (F-44), and (e) characteristic equation (F-49) with $\ell = 10^{10}$

From comparing the results shown in Fig. F-13 and Fig. F-14, it is not difficult to conclude that: (i) the novel orthogonalizing DPO method is valid for constructing the CMs of both lossless and lossy composite OSSs; (ii) the novel DPO has a more satisfactory numerical performance than the traditional IMO in the aspect of generating CMs.

## F8.2 A Rectangular Metallic Patch Adhered to the Surface of a Cubic Material Body

In this subsection, we consider the composite OSS constituted by a metallic rectangular patch and a material cube, and the metallic patch is adhered to the material cube. The topological structure and the sub-boundaries of the composite OSS are shown in Fig. F-15.

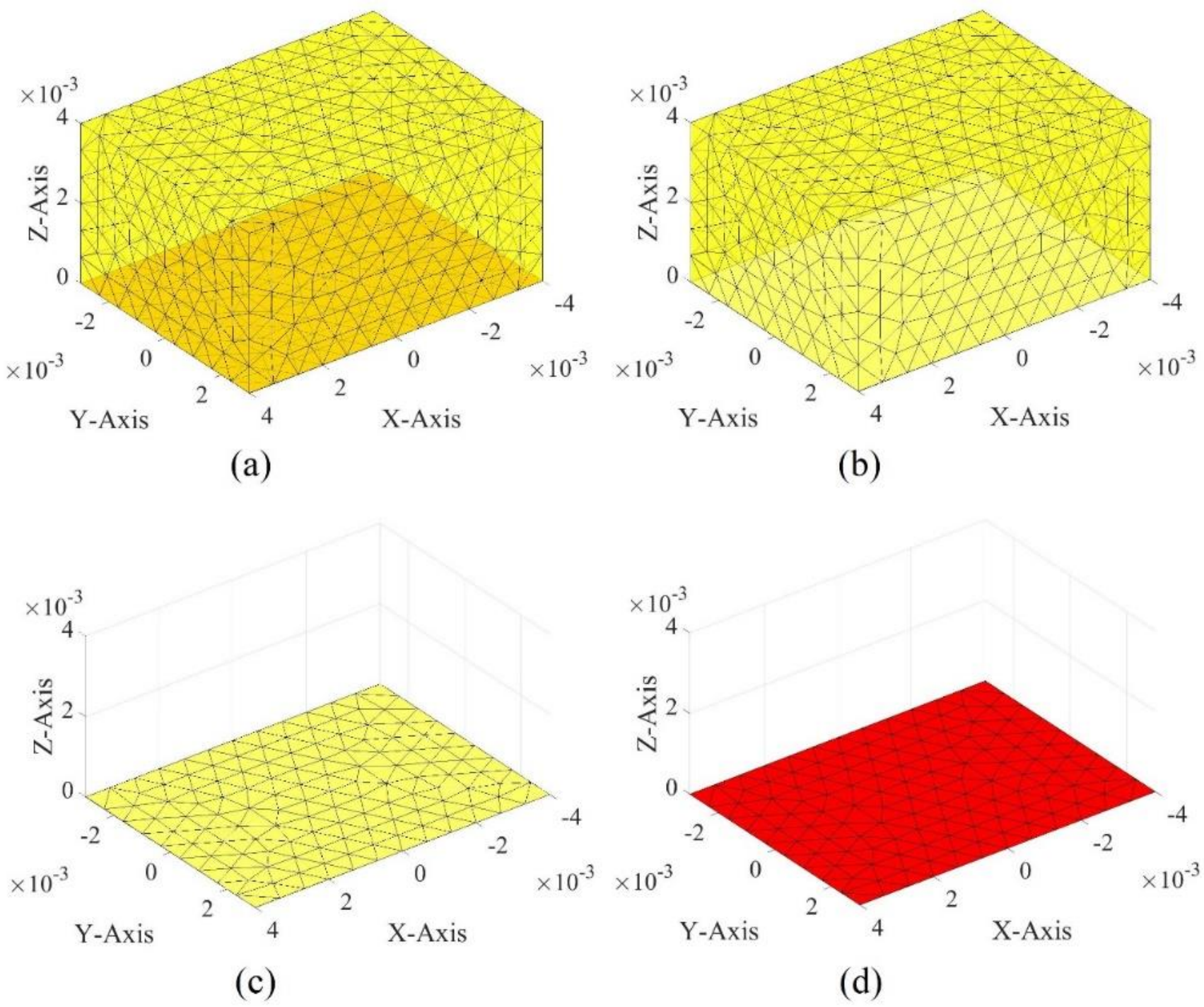

Figure F-15 Geometry of a composite OSS constituted by a rectangular metallic patch and a cubic material body, where the geometrical dimension of the metallic patch is $8\,\text{mm}\times 6\,\text{mm}$ , and the geometrical dimension of the material body is $8\,\text{mm}\times 6\,\text{mm}\times 4\,\text{mm}$ . (a) Topological structure of whole OSS; (b) material-environment boundary; (c) material-metal boundary; (d) metallic surface

When the composite OSS is lossless and with parameters $\boldsymbol{\mu} = \mathbf{I}4\mu_0$ and $\boldsymbol{\varepsilon}_c = \mathbf{I}4\varepsilon_0$, we use five different ways to construct its CMs, and show the obtained results in Figs. F-16(a) ~ F-16(e).

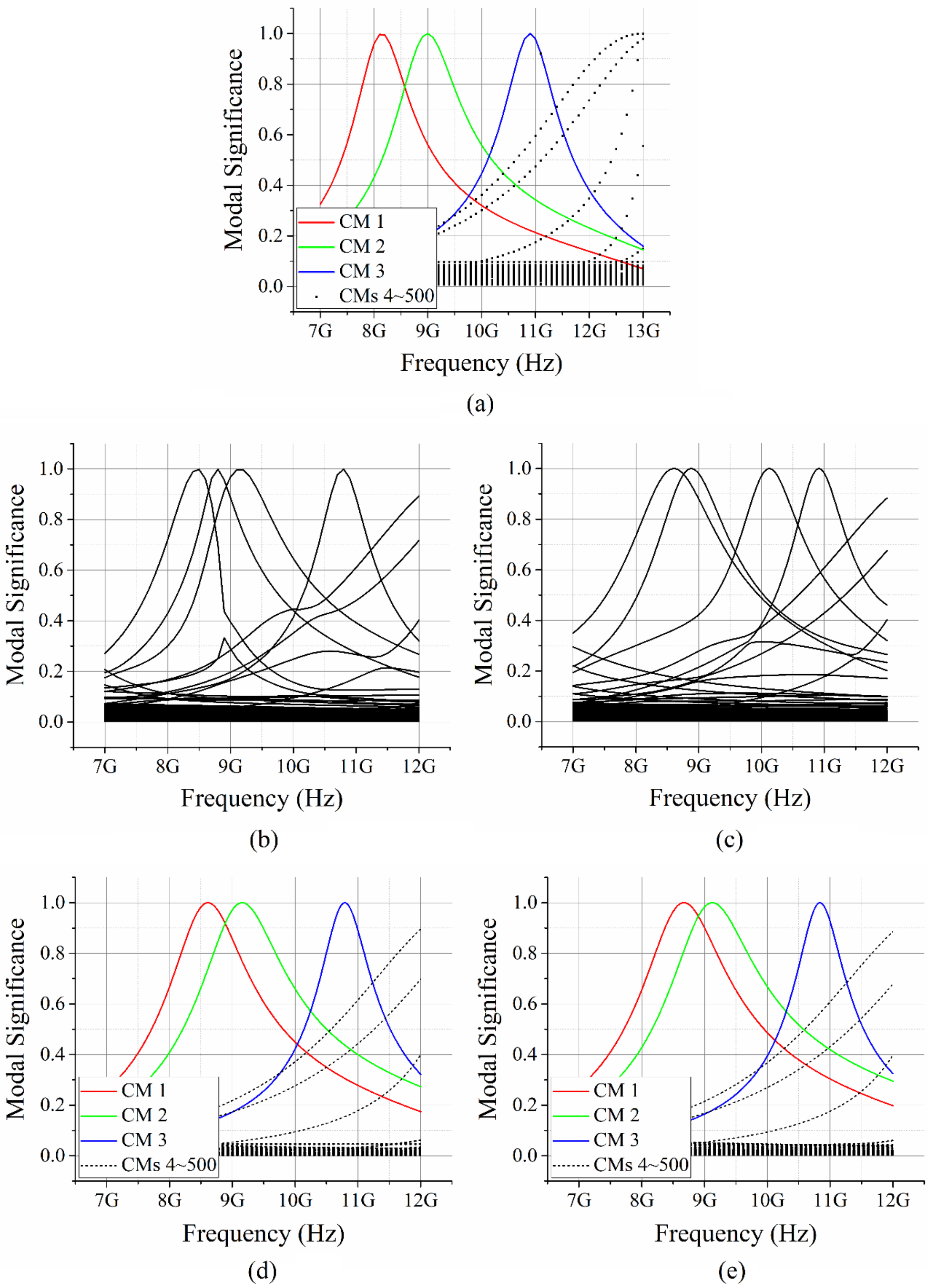

Figure F-16 MSs corresponding to the CMs of a lossless composite OSS. The OSS is with parameters $\boldsymbol{\mu} = \mathbf{I}4\mu_0$ and $\boldsymbol{\varepsilon}_c = \mathbf{I}4\varepsilon_0$, and has the topological structure shown in Fig. F-15. The CMs are derived from (a) EFIE-VIE-based IMO, that is DPO (F-22), (b) the EFIE-PMCHWT-based IMO with SDC scheme, (c) the EFIE-PMCHWT-based IMO with weighted SDC scheme using $\ell = 10^{10}$, (d) characteristic equation (F-45), and (e) characteristic equation (F-50) with $\ell = 10^{10}$

In Fig. F-16(a), the CMs are derived from orthogonalizing EFIE-VIE-based IMO, that is DPO (F-22). In Figs. F-16(b) and F-16(c), the CMs are respectively derived from orthogonalizing the traditional EFIE-PMCHWT-based IMO with SDC scheme and weighted SDC scheme. In Fig. F-16(d), the CMs are derived from orthogonalizing the DPO with SDC scheme, that is, solving characteristic equation (F-45). In Fig. F-16(e), the CMs are derived from orthogonalizing the DPO with weighted SDC scheme, that is, solving characteristic equation (F-50).

When the composite OSS is lossy and with parameters $\boldsymbol{\mu} = \mathbf{I}4\mu_0$ and $\boldsymbol{\varepsilon}_{\text{c}} = \mathbf{I}4\varepsilon_0 - j\mathbf{I}0.5/\omega$, we also use five different ways to construct its CMs, and show the obtained results in Figs. F-17(a) ~ F-17(e).

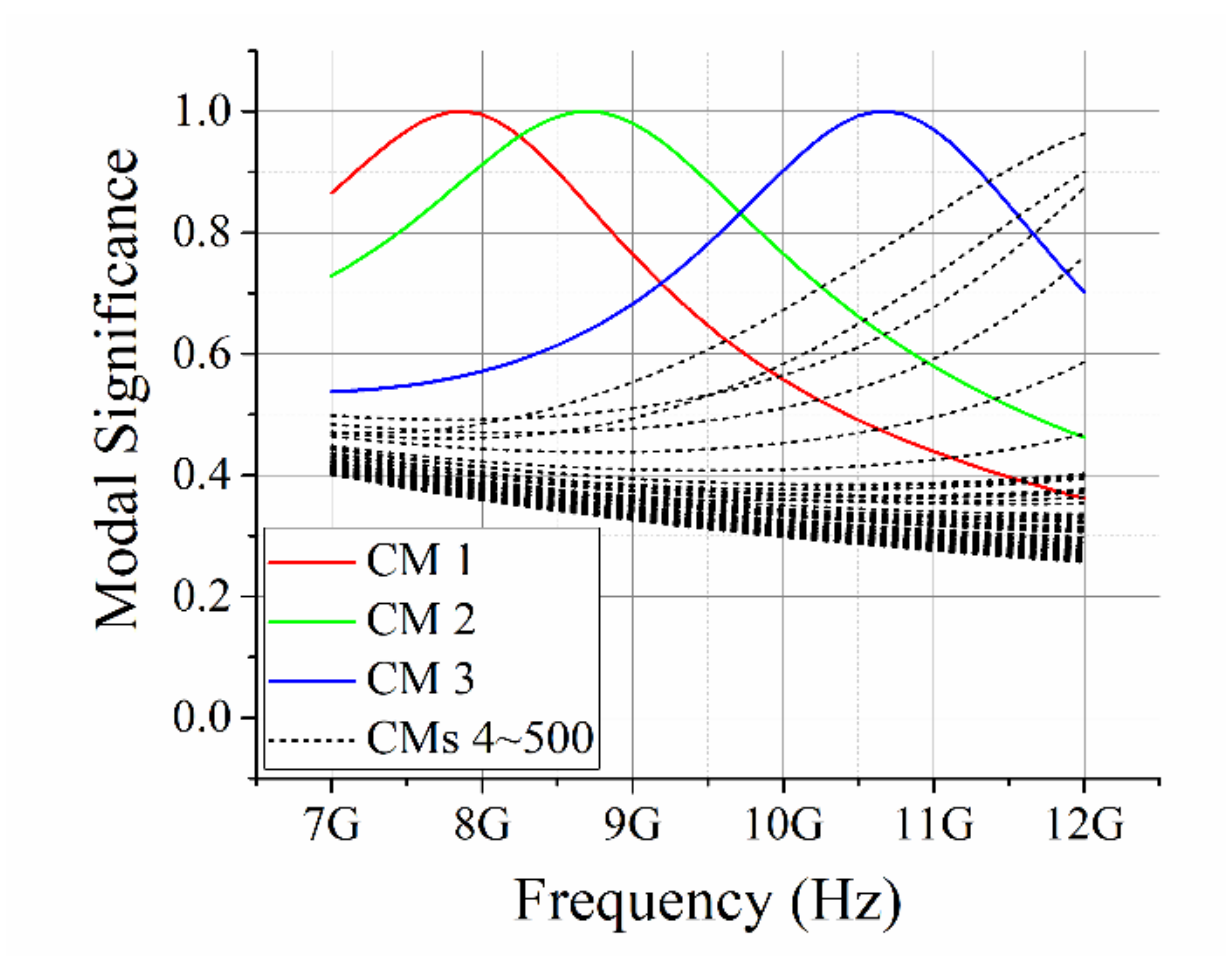

(a)

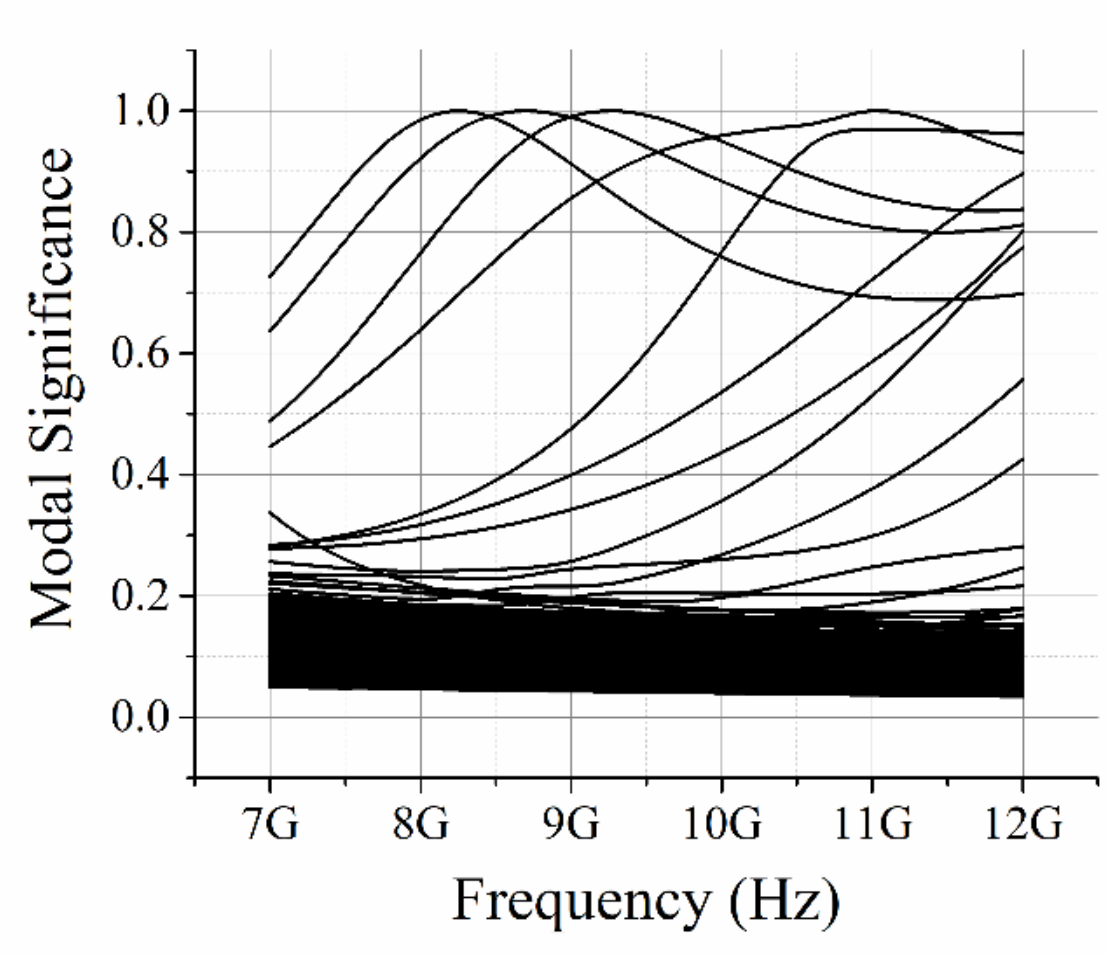

(b)

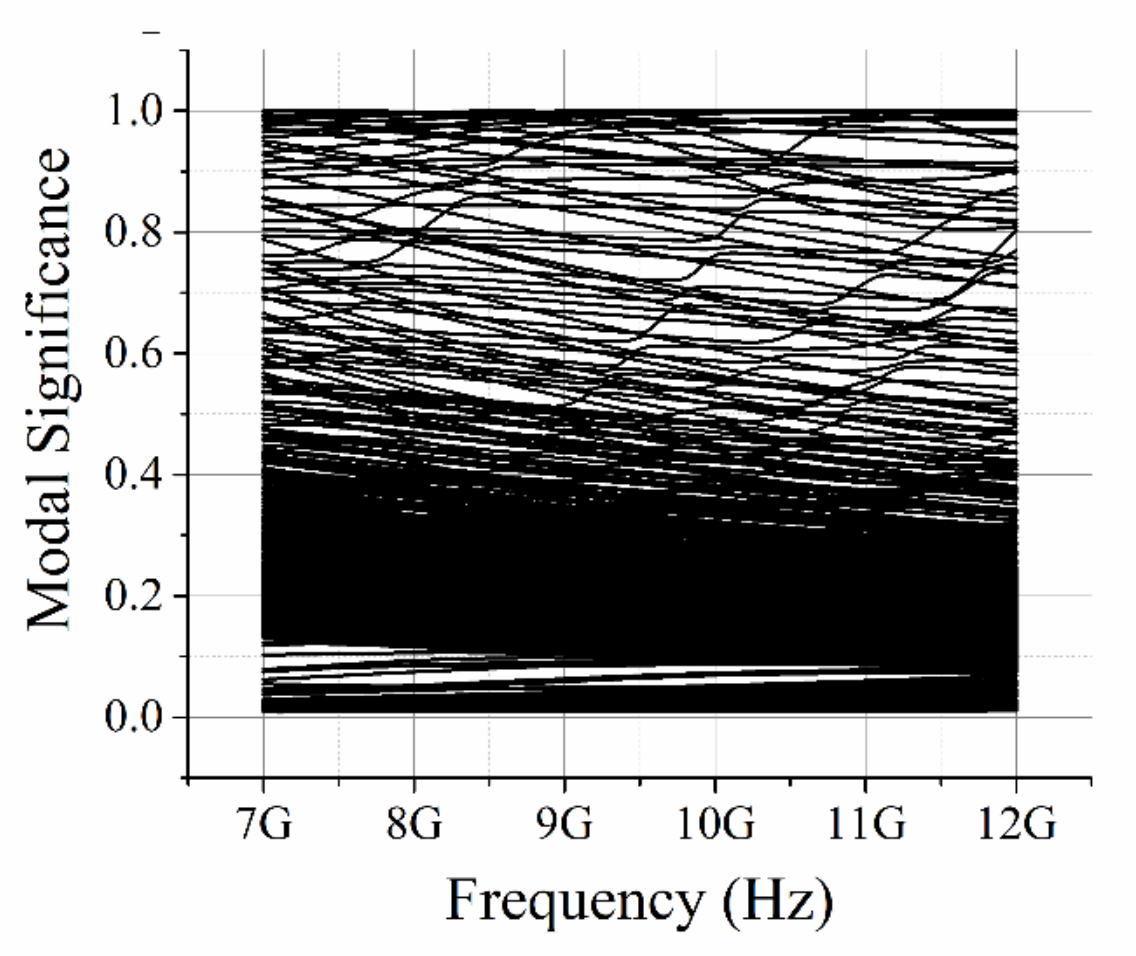

(c)

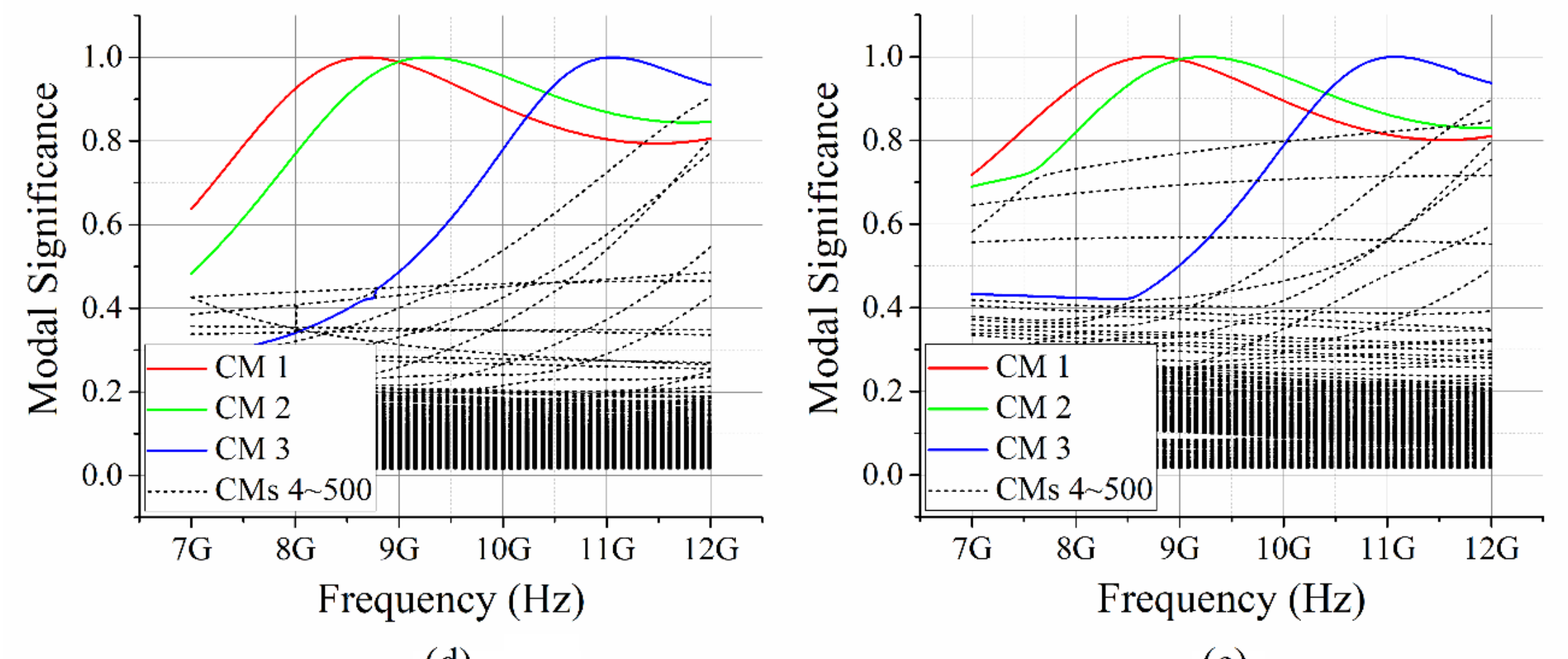

(d) (e)

Figure F-17 MSs corresponding to the CMs of a lossy composite OSS. The OSS is with parameters $\boldsymbol{\mu} = \mathbf{I}4\mu_0$ and $\boldsymbol{\varepsilon}_c = \mathbf{I}4\varepsilon_0 - j\mathbf{I}0.5/\omega$, and has the topological structure shown in Fig. F-15. The CMs are derived from (a) EFIE-VIE-based IMO, that is DPO (F-22), (b) the EFIE-PMCHWT-based IMO with SDC scheme, (c) the EFIE-PMCHWT-based IMO with weighted SDC scheme using $\ell = 10^{10}$, (d) characteristic equation (F-44), and (e) characteristic equation (F-49) with $\ell = 10^{10}$

In Fig. F-17(a), the CMs are derived from orthogonalizing EFIE-VIE-based IMO, that is DPO (F-22). In Figs. F-17(b) and F-17(c), the CMs are respectively derived from orthogonalizing the traditional EFIE-PMCHWT-based IMO with SDC scheme and weighted SDC scheme. In Fig. F-17(d), the CMs are derived from orthogonalizing the DPO with SDC scheme, that is, solving characteristic equation (F-44). In Fig. F-17(e), the CMs are derived from orthogonalizing the DPO with weighted SDC scheme, that is, solving characteristic equation (F-49).

## F9 Conclusions

Besides the traditional integral equation (IE) framework, work-energy principle (WEP) is also an effective framework for establishing characteristic mode theory (CMT). The novel WEP framework, in a clearer way, reveals the physical picture/purpose of CMT — to construct a set of steadily working energy-decoupled modes for scattering systems (rather than to construct the far-field-orthogonal modes). Under the novel WEP framework, a novel characteristic mode (CM) generating operator — driving power operator (DPO) — is introduced, and then a novel CM construction method — orthogonalizing DPO method — is proposed.

All the variables (except the equivalent electric or magnetic current on material-environment boundary) explicitly involved in the DPO are independent and complete. Using this observation and employing line-surface equivalence principle (LSEP), a novel scheme — solution domain compression (SDC) — and its alternative version are developed to suppress spurious modes. Unlike the traditional dependent variable elimination (DVE) scheme, the novel SDC scheme and its alternative version don't need to inverse any matrix during the process to suppress spurious modes.

Compared with the traditional IE-based orthogonalizing impedance matrix operator (IMO) method, the novel WEP-based orthogonalizing DPO method has a more desirable numerical performance in the aspect of constructing CMs. A detailed and rigorous theoretical explanation for this conclusion has been done by our group, and will be exhibited in our future article.

In addition, DPO always can be decomposed into two terms — principal value term (PVT) and singular current term (SCT). After employing the SDC scheme, the SCT is zero, if and only if the objective scattering system (OSS) is lossless. Based on this, this paper concludes that

(i) if the OSS is lossless, it is better to select the PVT (which is theoretically equal to the full DPO, but is numerically not) as CM generating operator;
(ii) if the OSS is lossy, it must select the full DPO (that is, PVT+SCT) as CM generating operator.

It is necessary to emphasize that the above conclusions (i) and (ii) hold under the condition that the SDC scheme has been utilized.

# Appendix G Work-Energy Principle Based Characteristic Mode Theory for Wireless Power Transfer Systems

This App. G had been written as a journal paper by our research group (Ren-Zun Lian, Ming-Yao Xia and Xing-Yue Guo), and the manuscript [AP2101-0035] entitled "Work-Energy Principe Based Characteristic Mode Theory for Wireless Power Transfer Systems"[349] was submitted to IEEE-TAP (*IEEE Transactions on Antennas and Propagation*) on 06-Jan-2021.

## G1 Abstract

Work-energy principle (WEP) governing wireless power transfer (WPT) process is derived. Driving power as the source to sustain a steady WPT is obtained. Transferring coefficient (TC) used to quantify power transfer efficiency is introduced.

WEP gives a clear physical picture to WPT process. The physical picture reveals the essential difference between transferring problem and scattering problem. The essential difference exposes the fact that the conventional characteristic mode theory (CMT) for scattering systems cannot be directly applied to transferring systems.

Under WEP framework, this paper establishes a CMT for transferring systems. By orthogonalizing driving power operator (DPO), the CMT can construct a set of energy-decoupled characteristic modes (CMs) for any pre-selected objective transferring system. It is proved that the obtained CM set contains the optimally transferring mode, which can maximize TC.

Employing the WEP-based CMT for transferring systems, this paper does the modal analysis for some typical two-coil transferring systems, and introduces the concepts of co-resonance and ci-resonance, and reveals some important differences and connections "between transferring problem and scattering problem", "between co-resonance phenomenon of transferring systems and external resonance phenomenon of scattering systems", and "between so-called magnetic resonance and classical electric-magnetic resonance".

## G2 Index Terms

Characteristic mode (CM), driving power operator (DPO), transferring coefficient (TC), wireless power transfer (WPT), work-energy principle (WEP).

## G3 Introduction

Wireless power transfer (WPT) system is a kind of electromagnetic (EM) device designed for wirelessly transferring EM power in an efficiency as high as possible and a distance as long as possible. The earliest researches on WPT can be dated back to the pioneers Hutin and Leblanc[265] and Tesla[266~268] etc., and more history on WPT could be found in [269]. According to the difference of working mechanism, WPT systems can be categorized into far-field/radiative WPT systems and near-field/nonradiative WPT systems[270]. The commonly used far-field WPT systems include microwave[271,272] and laser[273,274] WPT systems etc. The commonly used near-field WPT systems include inductive[275~277], capacitive[278,279], conductive[280,281], and magnetic resonance[282~287] WPT systems etc.

In the various WPT systems, the magnetic-resonance-based systems seem more desired in the applications of medium to high power levels, and more advantageous in the aspects of transferring capability, transferring efficiency, safety, and controllability[286,287]. The magnetic resonance WPT systems are focused on by this paper, and the principles and formulations obtained in this paper can be further generalized to the other kinds of WPT systems in the future. For convenience, the magnetic resonance WPT systems are simply called transferring systems in the following discussions.

A classical transferring system is shown in Fig. G-1. The transferring system is constituted by two metallic coils transmitting coil T and receiving coil R. Coil T is excited by a locally impressed driver (e.g. delta-gap source) used to inject power into coil T. Coil R is connected to a perfectly matched load (e.g. matched light bulb) used to extract power from coil R. The power is transmitted by coil T, and then wirelessly transferred from coil T to coil R by passing through the environment surrounding transferring system, and finally received by coil R.

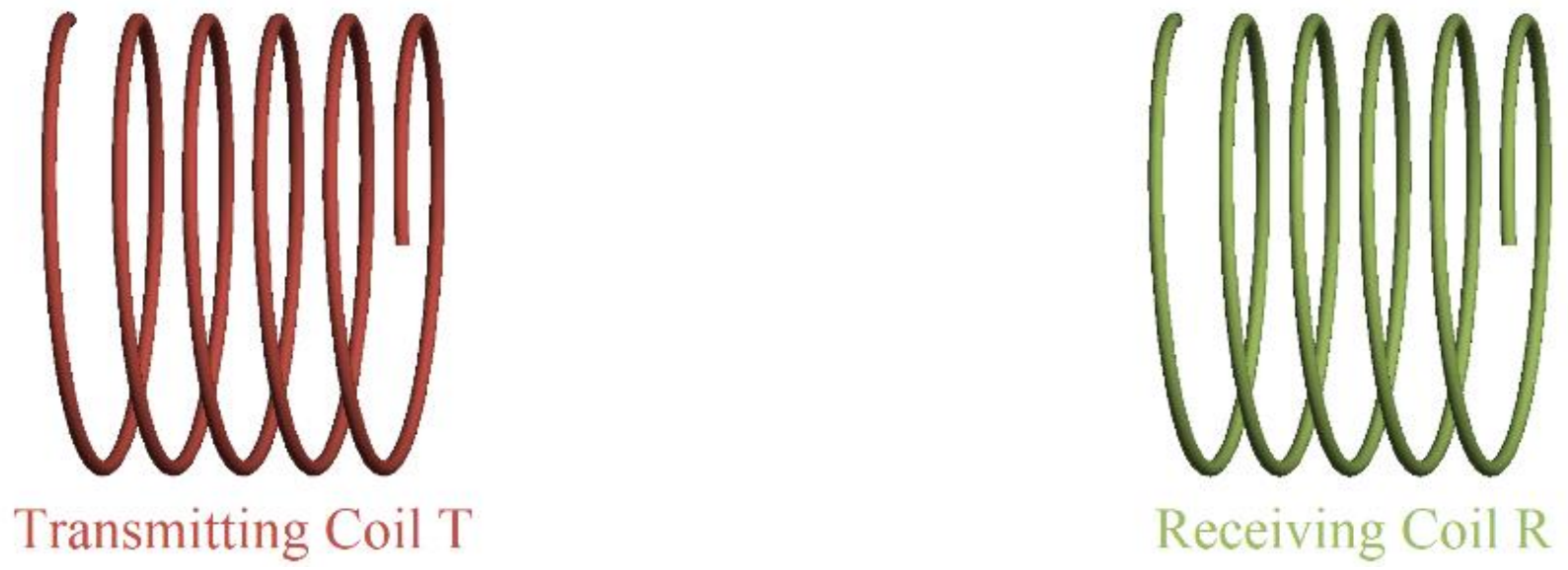

Figure G-1 A classical magnetic resonance WPT system constituted by a metallic transmitting coil T and a metallic receiving coil R

Inspired by the acoustic resonance between acoustic resonators, Tesla introduced the concept of magnetic resonance to the realm of WPT for the first time, and patented the well-known "Tesla coil"[288]. In 2007, Kurs *et al.*[282,283] employ the magnetic resonance between the coils T and R shown in Fig. G-1 to realize fully lighting up a 60-W light bulb from distances more than 2 m away. The time-average magnetic energy density distribution of the desired transferring mode is visualized in Fig. G-2. Evidently, the mode shown in Fig. G-2 indeed has ability to efficiently transfer power from coil T to coil R wirelessly. In recent years, the magnetic-resonance-based WPT technology has been widely applied in the realms of wirelessly charging consumer electronic products[289], electric vehicles (EV)[290,291], biomedical implants[292~294], underwater devices[295,296], Internet of things (IoT)[297,298], and industrial robots[299,300] etc.

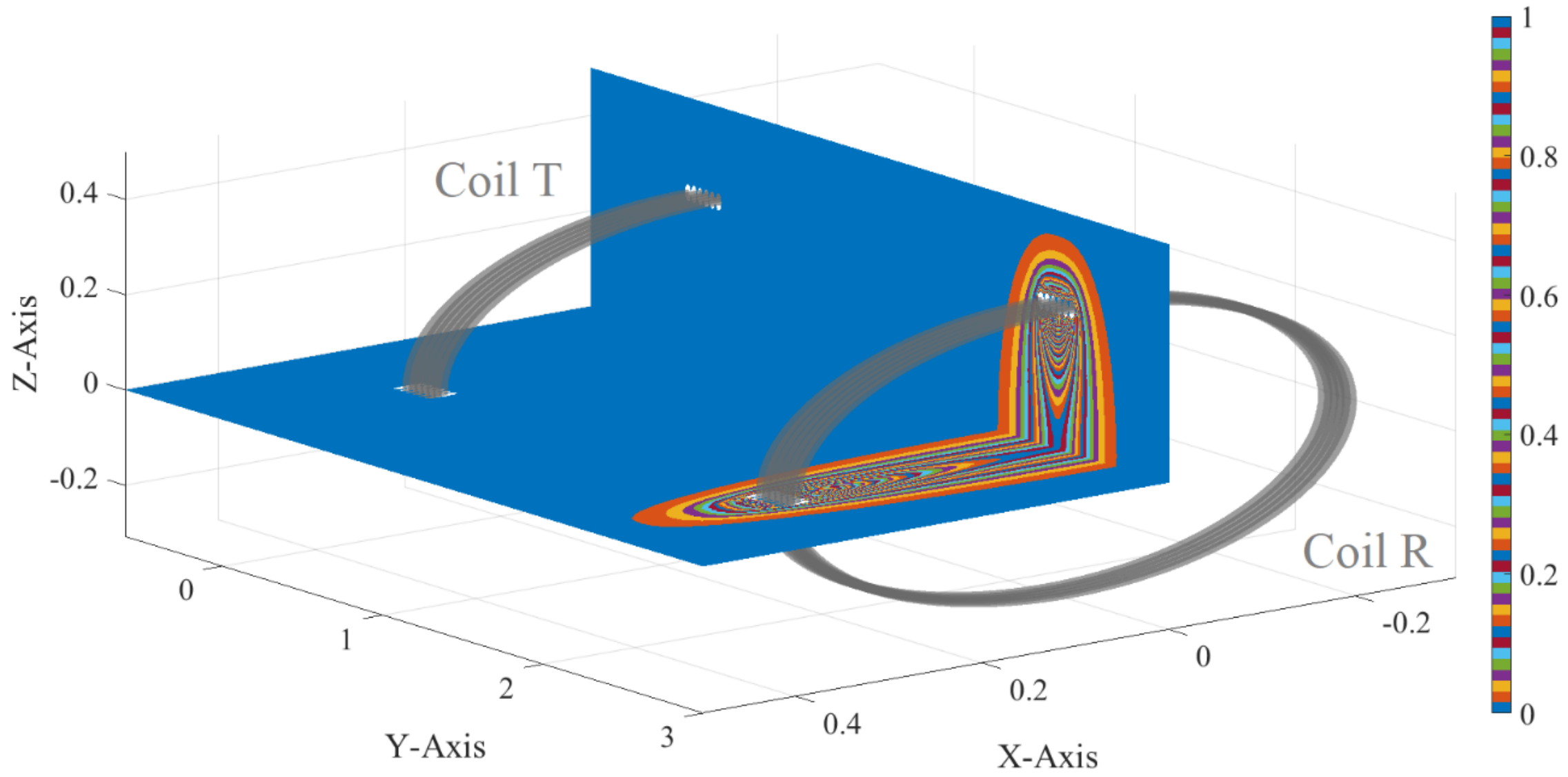

Figure G-2 Time-average magnetic energy density distribution of the desired transferring mode reported in [282]. The energy is almost completely transferred from coil T to coil R

Obviously, a systematical modal analysis method will significantly facilitate the theoretical analysis and engineering design for transferring systems. In fact, there have existed some different kinds of modal analysis methods for transferring systems, such as coupled-mode theory[282,283,301], classical circuit theory[302~304], and some other theories[305,306], and all of them are based on circuit models. However, the circuit-model-based modal analysis methods need to use some circuit-based concepts, such as scalar voltage, scalar current, self-inductance, mutual inductance, and capacitance etc., so they are some approximate but not rigorous methods. In addition, the employment for scalar voltage and scalar current implies that the circuit-model-based modal analysis methods are only applicable to the transferring systems working at low frequency and with simple geometrical structures (such as rectangular coils and circular coils[282~287] etc.). Thus, it is one of important challenges in the realm of magnetic-resonance-based WPT how to develop a rigorous, frequency-independent, and geometry-independent modal analysis method for transferring systems.

The conventional characteristic mode theory (CMT)[6~9,13,18] is just a rigorous, frequency-independent, and geometry-independent modal analysis method. However, the conventional CMT is a method for *scattering* systems as revealed by its physical picture[13,18], but cannot be directly applied to **transferring** systems (due to the different working mechanisms of *scattering* and **transferring** systems, for details please see Sec.

G4) as exhibited in the following example. A direct application of the conventional CMT to the transferring system shown in Fig. G-1 outputs some CMs, and the resonant CM has the time-average magnetic energy density distribution shown in Fig. G-3. Evidently, the resonant CM shown in Fig. G-3 doesn't realize the most efficient WPT from coil T to coil R like Fig. G-2.

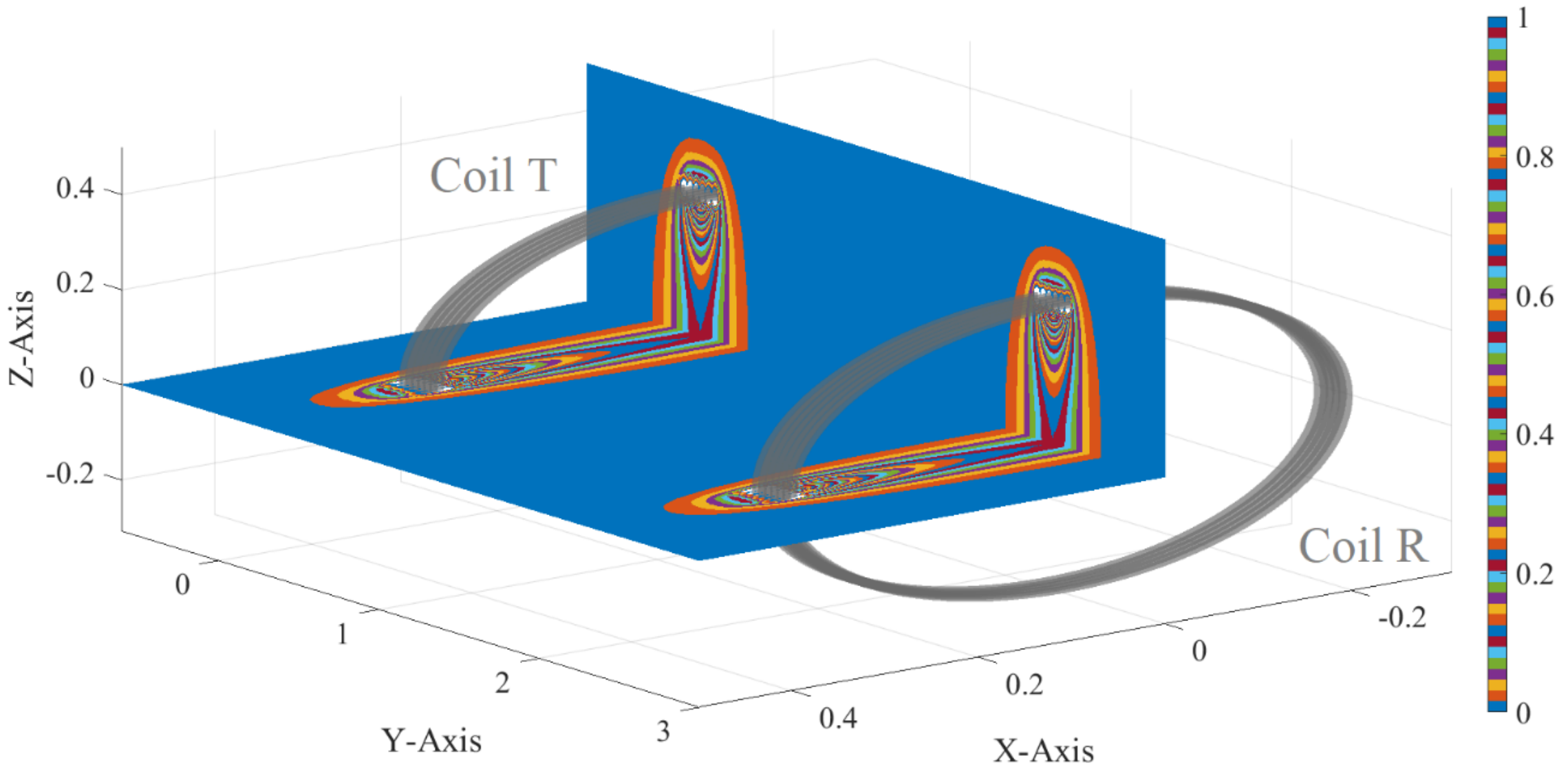

Figure G-3 Time-average magnetic energy density distribution of the resonant CM calculated from the conventional CMT for scattering systems[8,9,13]. The energy oscillates locally, but it is not transferred from coil T to coil R

The one shown in Fig. G-1 is a simplest and most classical metallic transferring system. Using it as a typical example, this paper is devoted to generalizing the conventional CMT for *scattering* systems (simply called *scattering* CMT) to a novel CMT for **transferring** systems (simply called **transferring** CMT), and this paper is organized as follows: Sec. G4 discusses the physical principle governing the working mechanism of transferring problem; Sec. G5 provides the mathematical formulas used to establish the transferring CMT; Sec. G6 employs the CMT to do some modal analysis for the classical transferring system shown in Fig. G-2 to exhibit the validity of the theory; Sec. G7 applies the transferring CMT to some typical variants of the classical transferring system for exhibiting the wide applicable range of the theory; Sec. G8 concludes this paper; Secs. G9.1~G9.4 provide some detailed formulations related to this paper.

In what follows, the $e^{j\omega t}$ convention and inner product $<\boldsymbol{f},\boldsymbol{g}>_{\Omega}=\int_{\Omega}\boldsymbol{f}^{\dagger}\cdot\boldsymbol{g}d\Omega$ are used throughout, where superscript "†" is the conjugate transpose operation for a scalar/vector/matrix. The environment surrounding transferring system is free space, and its permeability and permittivity are denoted as $\mu_0$ and $\varepsilon_0$ respectively. Time-domain and frequency-domain powers are denoted as $\mathcal{P}$ and $P$ respectively; time-domain and frequency-domain currents and fields are denoted as $(\boldsymbol{\mathcal{J}},\boldsymbol{\mathcal{F}})$ and $(\boldsymbol{J},\boldsymbol{F})$ respectively; frequency-domain power quadratic matrix and current expansion coefficient vector are denoted as $\mathbb{P}$ and $\mathbb{J}$ respectively. In addition, for the linear quantities (e.g. electric field intensity), we have that $\boldsymbol{\mathcal{E}} = \text{Re}\{\boldsymbol{E}e^{j\omega t}\}$; for the power-type quadratic quantity, we have that $\text{Re}\{(1/2)\boldsymbol{J}^{\dagger}\cdot\boldsymbol{E}\} = (1/T)\int_0^T \boldsymbol{\mathcal{J}}\cdot\boldsymbol{\mathcal{E}}dt$, where $T$ is the time period of the time-harmonic EM field.

## G4 Physical Principles

For the transferring system shown in Fig. G-1, a typical locally impressed driver is delta-gap source, and the source provides a voltage driving to coil T. The voltage driving has a field effect, i.e., the voltage driving can be equivalently viewed as a field driving. The voltage driving acts on coil T only (but doesn't act on coil R and load), so the equivalent field driving also acts on coil T only (but doesn't act on coil R and load), i.e., the driving field is localized/restricted in the region occupied by coil T, and this is just the reason to call it **locally** impressed driver.

The action of the driving field on coil T will induce a current on coil T, and the current will generate a field on surrounding environment. Similarly, the field generated by coil T will act on coil R, and the action will lead to an induced current on coil R, and the current will generate a field on surrounding environment. In addition, the fields generated by the coils act on the load as well.

In fact, there also exists a reaction from the field generated by coil R to the current distributing on coil T, and the reaction will affect the current distribution. However, there doesn't exist any reaction from the load to the coils, i.e., the field effect of the load is localized/restricted in the region occupied by the load itself, because the load is supposed as a perfectly **matched** one in this paper.

Through a complicated process, the above actions and reactions will reach a dynamic equilibrium finally, because the EM problem considered here is time-harmonic.

For the convenience of following discussions, the boundary surfaces of coil T and coil R are denoted as $S_{\text{t}}$ and $S_{\text{r}}$ respectively; the three-dimensional Euclidean space is denoted as $\mathbb{E}^3$; the boundary of $\mathbb{E}^3$ is denoted as $S_\infty$, which is a closed spherical surface with infinite radius. The driving field generated by locally impressed driver is denoted as $\mathcal{F}_{\text{driv}}$. At the state of dynamic equilibrium, the currents distributing on coil T and coil R are denoted as $\boldsymbol{J}_{\text{t}}$ and $\boldsymbol{J}_{\text{r}}$ respectively. The fields generated by $\boldsymbol{J}_{\text{t}}$ and $\boldsymbol{J}_{\text{r}}$ are denoted as $\mathcal{F}_{\text{t}}$ and $\mathcal{F}_{\text{r}}$ respectively.

The energy conservation law tells us that the above actions and reactions among driver, coil T, coil R, and load will result in a work-energy transformation, and the work-energy transformation can be quantitatively expressed as follows:

$$\mathcal{P}_{\text{driv}} = \mathcal{P}_{\text{sel}}^{\text{tt}} + \mathcal{P}_{\text{tra}} \tag{G-1}$$

called time-domain work-energy principle (WEP), where $\mathcal{P}_{\text{driv}} = <\boldsymbol{J}_{\text{t}}, \boldsymbol{\mathcal{E}}_{\text{driv}}>_{S_{\text{t}}}$ and $\mathcal{P}_{\text{sel}}^{\text{tt}} = -<\boldsymbol{J}_{\text{t}}, \boldsymbol{\mathcal{E}}_{\text{t}}>_{S_{\text{t}}}$ and $\mathcal{P}_{\text{tra}} = -<\boldsymbol{J}_{\text{t}}, \boldsymbol{\mathcal{E}}_{\text{r}}>_{S_{\text{t}}}$. A rigorous mathematical derivation for WEP (G-1) and a more detailed decomposition for $\mathcal{P}_{\text{driv}}$ are provided in Sec. G9.1.

Similar to the scattering problem discussed in [13,18], the current-field interaction $< \boldsymbol{J}_{\text{t}}, \boldsymbol{\mathcal{E}}_{\text{driv}} >_{S_{\text{t}}}$ is the source for driving a steady work-energy transformation, so $\mathcal{P}_{\text{driv}}$ is called driving power. Based on Maxwell's equations, the current-field interaction $- < \boldsymbol{J}_{\text{t}}, \boldsymbol{\mathcal{E}}_{\text{t}} >_{S_{\text{t}}}$ can be alternatively written as follows:

$$\begin{aligned} -\langle \boldsymbol{J}_{\text{t}}, \boldsymbol{\mathcal{E}}_{\text{t}} \rangle_{S_{\text{t}}} =& \oiint_{S_\infty} \left( \boldsymbol{\mathcal{E}}_{\text{t}} \times \boldsymbol{\mathcal{H}}_{\text{t}} \right) \cdot d\boldsymbol{S} \\ &+ \frac{d}{dt} \Big[ (1/2) \langle \boldsymbol{\mathcal{H}}_{\text{t}}, \boldsymbol{\mathcal{B}}_{\text{t}} \rangle_{\mathbb{E}^3} + (1/2) \langle \boldsymbol{\mathcal{D}}_{\text{t}}, \boldsymbol{\mathcal{E}}_{\text{t}} \rangle_{\mathbb{E}^3} \Big] \end{aligned} \tag{G-2}$$

called Poynting's theorem, where $\boldsymbol{\mathcal{D}}_{\text{t}} = \varepsilon_0 \boldsymbol{\mathcal{E}}_{\text{t}}$ and $\boldsymbol{\mathcal{B}}_{\text{t}} = \mu_0 \boldsymbol{\mathcal{H}}_{\text{t}}$. The theorem quantitatively governs the way how $\boldsymbol{J}_{\text{t}}$ provides EM power to $\mathcal{F}_{\text{t}}$, so $\mathcal{P}_{\text{sel}}^{\text{tt}}$ is called self-power of coil T. It has been explained previously that coil R is driven by coil T, and it will be further proven in Sec. G5 that $\boldsymbol{J}_{\text{r}}$ is uniquely determined by $\boldsymbol{J}_{\text{t}}$, so $\boldsymbol{J}_{\text{t}}$ can be viewed as the source for providing power to $\mathcal{F}_{\text{r}}$ and the power can be expressed as current-field interaction $- < \boldsymbol{J}_{\text{t}}, \boldsymbol{\mathcal{E}}_{\text{r}} >_{S_{\text{t}}}$ (which is similar to the current-field interaction $- < \boldsymbol{J}_{\text{t}}, \boldsymbol{\mathcal{E}}_{\text{t}} >_{S_{\text{t}}}$ of the coil T itself), and then $\mathcal{P}_{\text{tra}}$ is called transferred power from coil T to coil R.

The above these clearly reveal the physical picture of WPT process: driver drives coil T, and the driving power $\mathcal{P}_{\text{driv}}$ is transformed into two parts $\mathcal{P}_{\text{sel}}^{\text{tt}}$ and $\mathcal{P}_{\text{tra}}$ by the transferring system, where $\mathcal{P}_{\text{sel}}^{\text{tt}}$ is dissipated by coil T and $\mathcal{P}_{\text{tra}}$ is wirelessly transferred from coil T to coil R. The physical picture clearly reveals the fact that the working mechanism of **transferring** systems is different from the working mechanism of *scattering* systems (for details of the latter please see [13] and [18]). This is just the reason why the *scattering* CMT[6~9,13,18] cannot be directly applied to **transferring** systems.

For the WPT application, the transferred power $\mathcal{P}_{\text{tra}}$ is desired, and the dissipated power $\mathcal{P}_{\text{sel}}^{\text{tt}}$ is unwanted and expected to be as small as possible, so we introduce a novel concept of transferring coefficient (TC) as follows:

$$\text{TC} = \frac{\left(1/T\right)\int_0^T \mathcal{P}_{\text{tra}} dt}{\left(1/T\right)\int_0^T \mathcal{P}_{\text{driv}} dt} \tag{G-3}$$

to quantify the transferring efficiency of transferring system. From a relatively mathematical viewpoint, the central aim of designing transferring system is to search for a physically realizable working mode (or working state) such that TC is maximized. In this paper, the central aim is realized by applying a CMT-based modal analysis to the transferring system.

In the following Sec. G5, we provide the mathematical formulations used to establish the WEP-based CMT (WEP-CMT) for transferring systems. Because there exist relations $(1/T)\int_0^T \mathcal{P}_{\text{tra}} dt = \text{Re}\{P_{\text{tra}}\}$ and $(1/T)\int_0^T \mathcal{P}_{\text{driv}} dt = \text{Re}\{P_{\text{driv}}\}$, then the following Sec. G5 is discussed in frequency domain.

## G5 Mathematical Formulations

The EM field boundary condition on $S_{\text{t}}$ implies the relation that $[\boldsymbol{E}_{\text{driv}}]_{\text{tan}} = [-\boldsymbol{E}_{\text{t}} - \boldsymbol{E}_{\text{r}}]_{\text{tan}}$ on $S_{\text{t}}$, where subscript "tan" represents that the relation is satisfied by the tangential component. Electric fields $\boldsymbol{E}_{\text{t}}$ and $\boldsymbol{E}_{\text{r}}$ have operator expressions $\boldsymbol{E}_{\text{t}} = -j\omega\mu_0\mathcal{L}_0(\boldsymbol{J}_{\text{t}})$ and $\boldsymbol{E}_{\text{r}} = -j\omega\mu_0\mathcal{L}_0(\boldsymbol{J}_{\text{r}})$ respectively, in which $\mathcal{L}_0(\boldsymbol{J}_{\text{t/r}}) = [1 + (1/k_0^2)\nabla\nabla\cdot]\int_\Omega G_0(\boldsymbol{r},\boldsymbol{r}')\boldsymbol{J}_{\text{t/r}}(\boldsymbol{r}')d\Omega'$ where $G_0(\boldsymbol{r},\boldsymbol{r}') = e^{-jk_0|\boldsymbol{r}-\boldsymbol{r}'|}/4\pi|\boldsymbol{r}-\boldsymbol{r}'|$. Based on the above these, frequency-domain driving power $P_{\text{driv}}$ can be expressed as the following operator form

$$P_{\text{driv}} = -\left(1/2\right)\left\langle \boldsymbol{J}_{\text{t}}, -j\omega\mu_0\mathcal{L}_0\left(\boldsymbol{J}_{\text{t}} + \boldsymbol{J}_{\text{r}}\right)\right\rangle_{S_{\text{t}}} \tag{G-4}$$

called frequency-domain driving power operator (DPO).

In fact, the physically realizable $\boldsymbol{J}_{\text{t}}$ and $\boldsymbol{J}_{\text{r}}$ are not independent of each other, because they must satisfy the following electric field integral equation

$$\left[-j\omega\mu_0\mathcal{L}_0\left(\boldsymbol{J}_{\text{t}}\right) - j\omega\mu_0\mathcal{L}_0\left(\boldsymbol{J}_{\text{r}}\right)\right]_{\text{tan}} = 0 \text{ on } S_{\text{r}} \tag{G-5}$$

due to the homogeneous tangential electric field boundary condition $[\boldsymbol{E}_{\text{t}} + \boldsymbol{E}_{\text{r}}]_{\text{tan}} = 0$ on $S_{\text{r}}$. Applying the method of moments (MoM) to (G-5), the integral equation is immediately discretized into a matrix equation. By solving the matrix equation, the

following transformation

$$\mathrm{J}_{\mathrm{r}} = \mathrm{T} \cdot \mathrm{J}_{\mathrm{t}} \tag{G-6}$$

can be obtained, where $\mathrm{J}_{\mathrm{t}}$ and $\mathrm{J}_{\mathrm{r}}$ are the column vectors constituted by the expansion coefficients of $\boldsymbol{J}_{\mathrm{t}}$ and $\boldsymbol{J}_{\mathrm{r}}$ respectively. The detailed mathematical process for deriving (G-6) from (G-5) is provided in Sec. G9.2.

Similarly to discretizing integral equation (G-5), the DPO (G-4) can be discretized into its matrix form. Substituting transformation (G-6) into the matrix form, we immediately have that

$$P_{\mathrm{driv}} = \mathrm{J}_{\mathrm{t}}^{\dagger} \cdot \mathrm{P}_{\mathrm{driv}} \cdot \mathrm{J}_{\mathrm{t}} \tag{G-7}$$

The detailed mathematical formulations for calculating matrix $\mathrm{P}_{\mathrm{driv}}$ are listed in Sec. G9.3. In fact, the above process from (G-4) to (G-7) is just the dependent variable elimination (DVE) process used in the WEP-CMT for *scattering* systems[13,18].

WEP-CMT decomposes $\mathrm{P}_{\mathrm{driv}}$ in terms of its positive and negative Hermitian parts as that $\mathrm{P}_{\mathrm{driv}} = \mathrm{P}_{\mathrm{driv}}^{+} + j\mathrm{P}_{\mathrm{driv}}^{-}$ (where $\mathrm{P}_{\mathrm{driv}}^{+} = (\mathrm{P}_{\mathrm{driv}} + \mathrm{P}_{\mathrm{driv}}^{\dagger})/2$ and $\mathrm{P}_{\mathrm{driv}}^{-} = (\mathrm{P}_{\mathrm{driv}} - \mathrm{P}_{\mathrm{driv}}^{\dagger})/2j$), and constructs CMs by solving the following characteristic equation

$$\mathrm{P}_{\mathrm{driv}}^{-} \cdot \mathrm{J}_{\mathrm{t}} = \lambda\, \mathrm{P}_{\mathrm{driv}}^{+} \cdot \mathrm{J}_{\mathrm{t}} \tag{G-8}$$

where $\lambda$ and $\mathrm{J}_{\mathrm{t}}$ are the associated characteristic values and characteristic vectors respectively.

Using the above-obtained characteristic vector $\mathrm{J}_{\mathrm{t}}$, the TC of the CM can be calculated as follows:

$$\mathrm{TC} = \frac{\mathrm{J}_{\mathrm{t}}^{\dagger} \cdot \mathrm{P}_{\mathrm{tra}}^{+} \cdot \mathrm{J}_{\mathrm{t}}}{\mathrm{J}_{\mathrm{t}}^{\dagger} \cdot \mathrm{P}_{\mathrm{driv}}^{+} \cdot \mathrm{J}_{\mathrm{t}}} \tag{G-9}$$

where $\mathrm{P}_{\mathrm{tra}}^{+}$ is given in Sec. G9.3. By comparing the TCs of the CMs, the CM with the maximal TC can be easily found, and the CM is usually a desired selection for the WPT application.

In the following Sec. G6, we provide a WEP-CMT-based modal analysis for a classical example, which is just the transferring system reported in the seminal paper [282].

## G6 Numerical Verifications

The transferring system considered in [282] is constituted by two metallic coils as shown in Fig. G-1. The coils have the same radius 30 cm, height 20 cm, and turns 5.25. The coils are placed coaxially, and their distance is 2 m. The optimally transferring frequency (i.e. the working frequency of the optimally transferring mode) calculated from the coupled-mode theory used in [282] is $10.56 \pm 0.3$ MHz, and the optimally transferring frequency obtained from the measurement done in [282] is 9.90 MHz. The reason leading to a 5% discrepancy between the theoretical and measured values was explained in [282].

## G6.1 WEP-CMT-Based Modal Analysis

We use the WEP-CMT given in Secs. G4 and G5 to calculate the CMs of the transferring system, and show the TC curves of the first 5 CMs in Fig. G-4. Based on the analysis given in Secs. G4 and G5, it is not difficult to conclude that the CM 1 at 10.8988 MHz (which corresponds to the local maximum of the TC curve) works at the optimally transferring state. The coil current distribution and time-average magnetic energy density distribution of the optimally transferring mode are shown in Fig. G-5 and Fig. G-6 respectively.

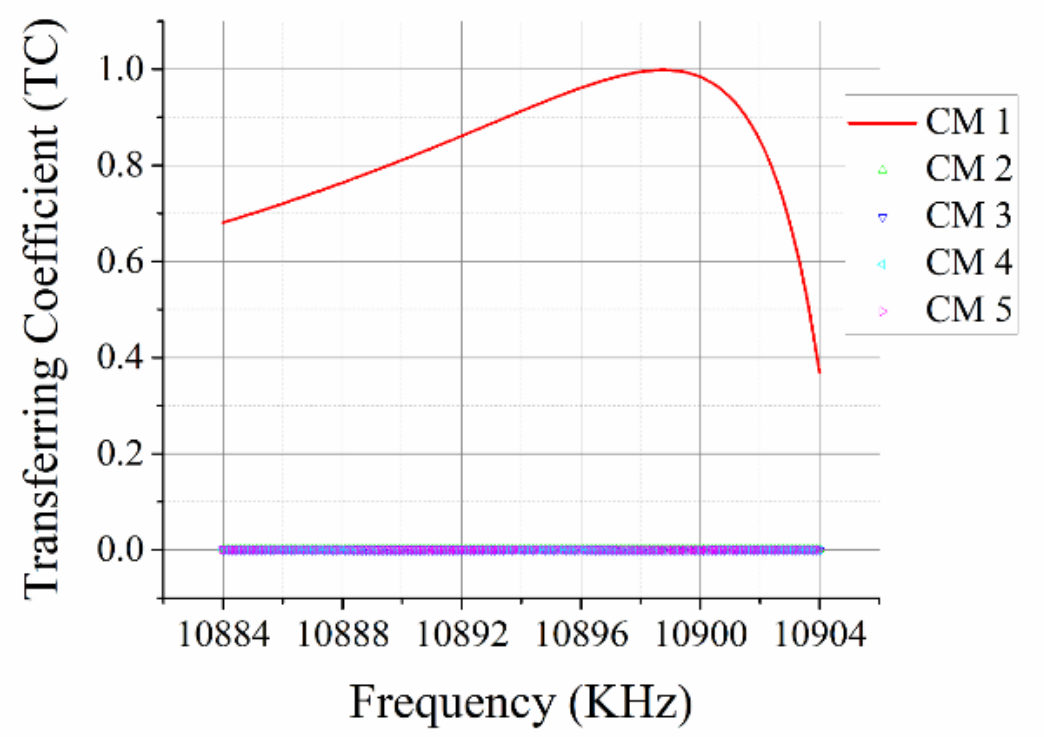

Figure G-4 TC curves of the first 5 low-order CMs calculated from the WEP-CMT established in this paper

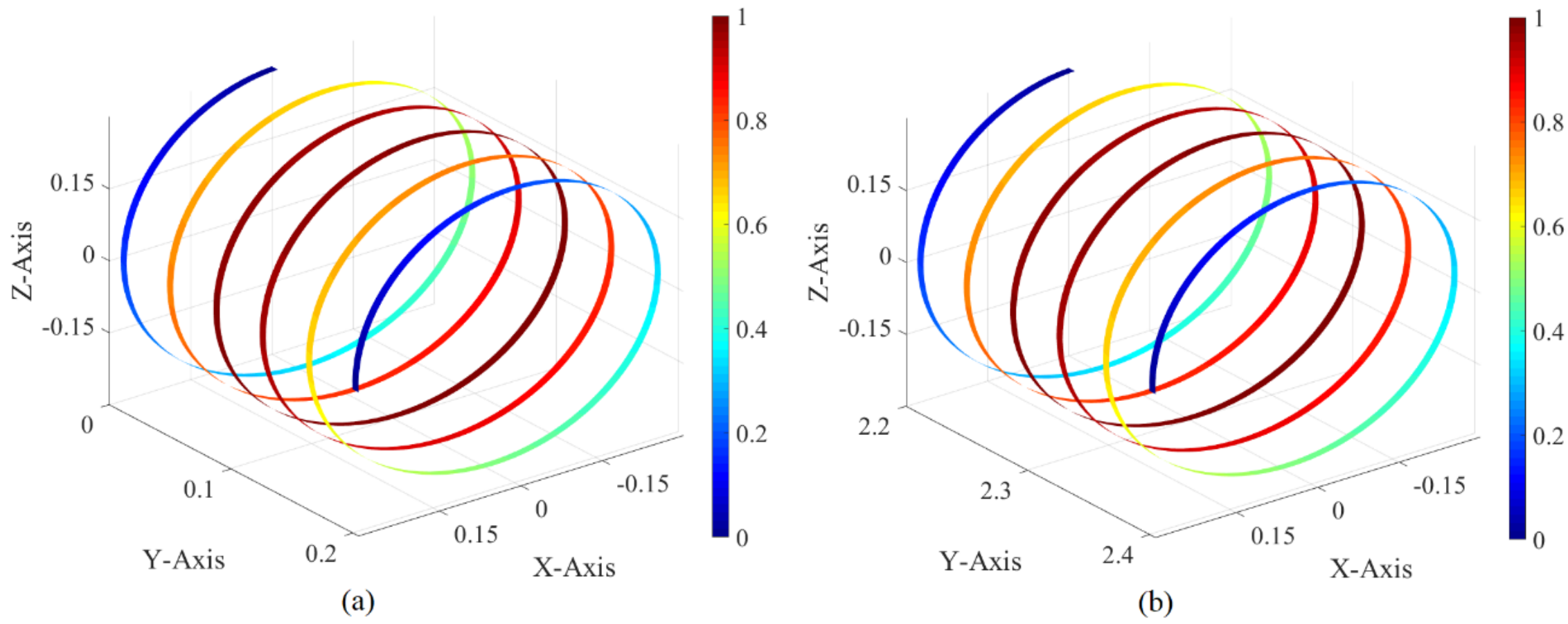

Figure G-5 For the CM 1 working at 10.8988 MHz, its current magnitudes distributing on (a) coil T and (b) coil R

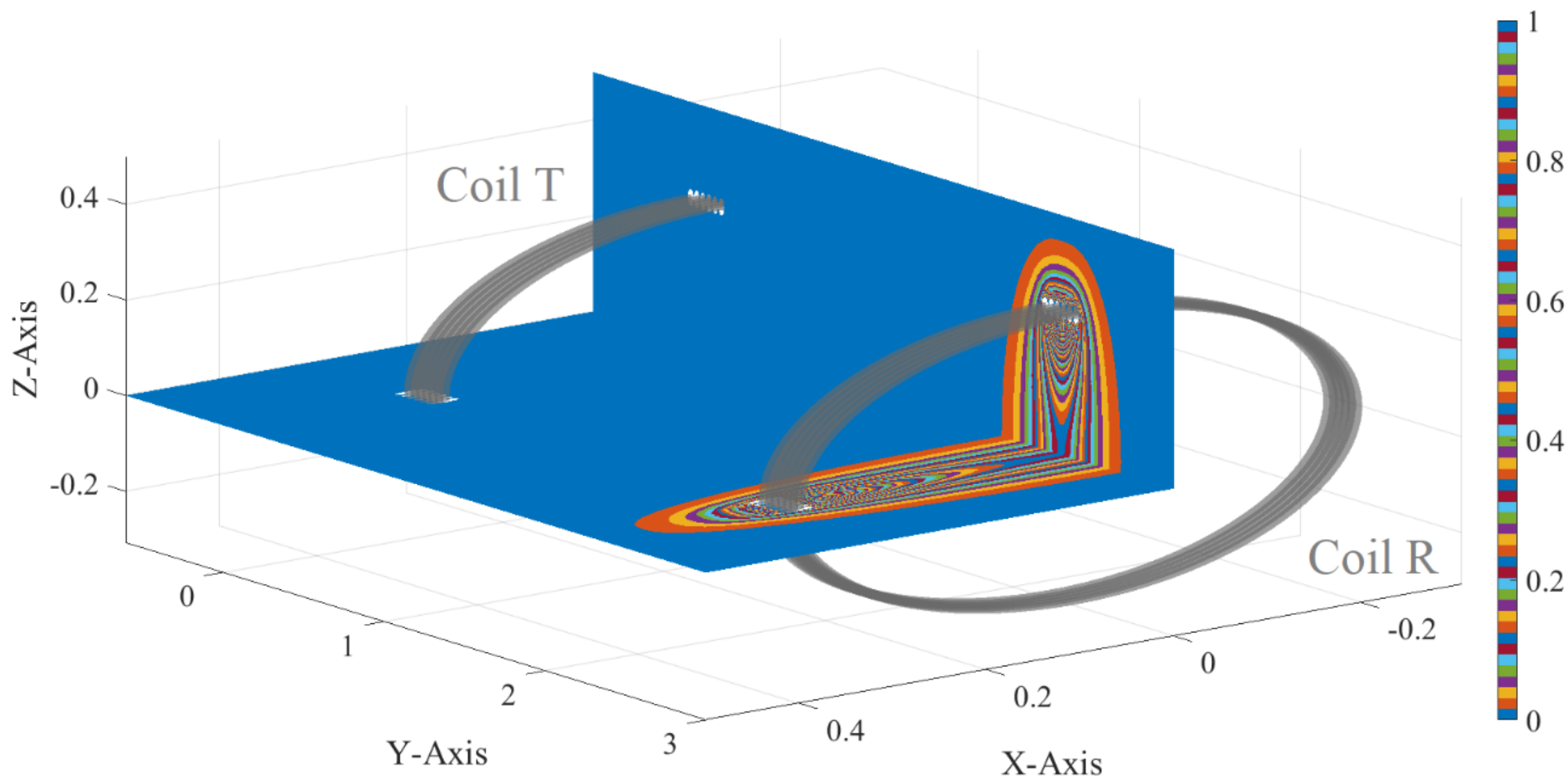

Figure G-6 For the CM 1 working at 10.8988 MHz, its time-average magnetic energy density distributing on xOy and yOz planes

Evidently, the CM 1 working at 10.8988 MHz corresponds to a half-wave current distribution for both the coil T and coil R as shown in Fig. G-5, and it indeed can efficiently transfer EM power from coil T to coil R in a wireless manner as shown in Fig. G-6.

In addition, the optimally transferring frequencies calculated from the classical coupled-mode theory ($10.56 \pm 0.3$ MHz) proposed in [282] and the WEP-CMT (10.8988 MHz) used in this paper are consistent with each other. The advantage of WEP-CMT over coupled-mode theory is reflected as follows:

A1 The WEP-CMT is a field-based modal analysis method, which is directly derived from Maxwell's equations and doesn't use any approximation; the coupled-mode theory is a circuit-model-based modal analysis method, which employs some circuit-model-based approximate quantities (such as scalar voltage, scalar current, effective inductance, and effective capacitance, etc.[282,283]).

A2 The WEP-CMT is applicable to the coils working at arbitrary frequency; the coupled-mode theory is only applicable to the coils working at low frequency at which the circuit model exists.

A3 The WEP-CMT is applicable to the coils with arbitrary topological structures; the coupled-mode theory is only applicable to the coils with simple topological structures, such that the coils can support sinusoidal scalar currents.

By employing an alternative field-based modal analysis, the following Sec. G6.2 verifies that the CM 1 at 10.8988 MHz is indeed the most efficient mode for WPT, and then exhibits that the optimally transferring mode is indeed included in the CM set constructed by the WEP-CMT.

## G6.2 An Alternative Modal Analysis

Because the matrices $\mathrm{P}_{\mathrm{tra}}^{+}$ and $\mathrm{P}_{\mathrm{driv}}^{+}$ in (G-9) are Hermitian, and the matrix $\mathrm{P}_{\mathrm{driv}}^{+}$ is positive definite, then the mode maximizing TC can be obtained from solving the following equation[307]

$$\mathrm{P}_{\mathrm{tra}}^{+} \cdot \mathrm{J}_{\mathrm{t}} = \tau \, \mathrm{P}_{\mathrm{driv}}^{+} \cdot \mathrm{J}_{\mathrm{t}} \tag{G-10}$$

Using the equation, we calculate the optimally transferring mode, and show the associated TC curve in Fig. G-7. Obviously, both the obtained optimally transferring frequency and optimally transferring coefficient are consistent with the ones obtained from the WEP-CMT-based modal analysis method.

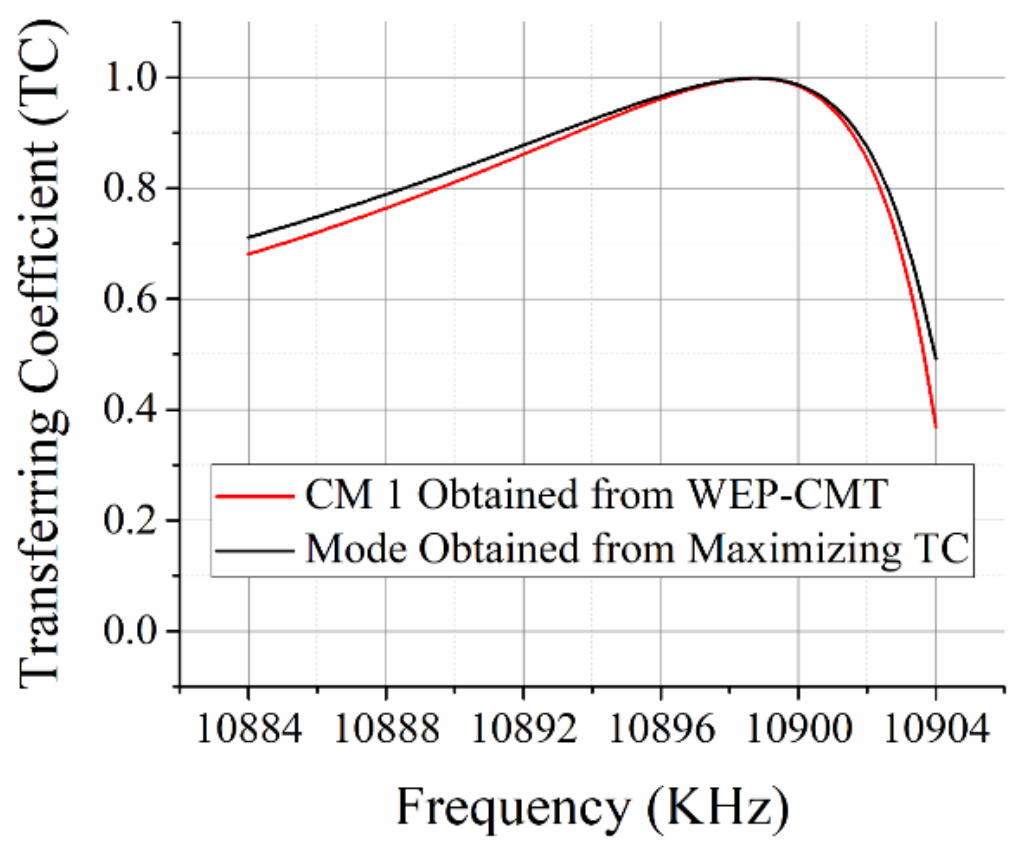

Figure G-7 TC curves of the optimally transferring modes obtained from two somewhat different modal analysis methods provided in this paper

## G6.3 Concepts of Co-resonance and Ci-resonance

In this sub-section, we, under the WEP framework, provide some further modal analysis to the CM 1 shown in Fig. G-4, and introduce the concepts of co-resonance and ci-resonance.

For the CM 1, its various resistance and reactance curves are shown in Fig. G-8. The definitions for the resistances and reactances are given in Sec. G9.4. From the Figs. G-8(a) and G-8(b), it is easy to observe the following facts:

F1 At optimally transferring frequency $f_{\mathrm{cor}} = 10.8988\,\mathrm{MHz}$, $R_{\mathrm{sel}}^{\mathrm{tt}}$ is very small; $R_{\mathrm{mut}}^{\mathrm{tr}}$ is equal to zero; $R_{\mathrm{driv}}$, $R_{\mathrm{sel}}^{\mathrm{rr}}$, and $R_{\mathrm{tra}}$ achieve their local maximums simultaneously, and the maximal values are the same.

F2 At frequency $f_{\text{cor}} = 10.8988\ \text{MHz}$, $X_{\text{sel}}^{\text{tt}}$, $X_{\text{mut}}^{\text{tr}}$, and $X_{\text{sel}}^{\text{rr}}$ are zero simultaneously, and then $X_{\text{driv}}$ and $X_{\text{tra}}$ are zero automatically because of that $X_{\text{driv}} = X_{\text{sel}}^{\text{tt}} + X_{\text{mut}}^{\text{tr}} + X_{\text{sel}}^{\text{rr}}$ and $X_{\text{tra}} = X_{\text{mut}}^{\text{tr}} + X_{\text{sel}}^{\text{rr}}$.

F3 At frequency $f_{\text{cir}} = 10.8873\ \text{MHz}$, $R_{\text{driv}}$, $R_{\text{sel}}^{\text{tt}}$, $R_{\text{mut}}^{\text{tr}}$, $R_{\text{sel}}^{\text{rr}}$, and $R_{\text{tra}}$ are very small.

F4 At frequency $f_{\text{cir}} = 10.8873\ \text{MHz}$, only $X_{\text{driv}}$ is zero, but the other reactances are not.

The subscripts "cor" and "cir" on $f_{\text{cor}}$ and $f_{\text{cir}}$ are explained as below.

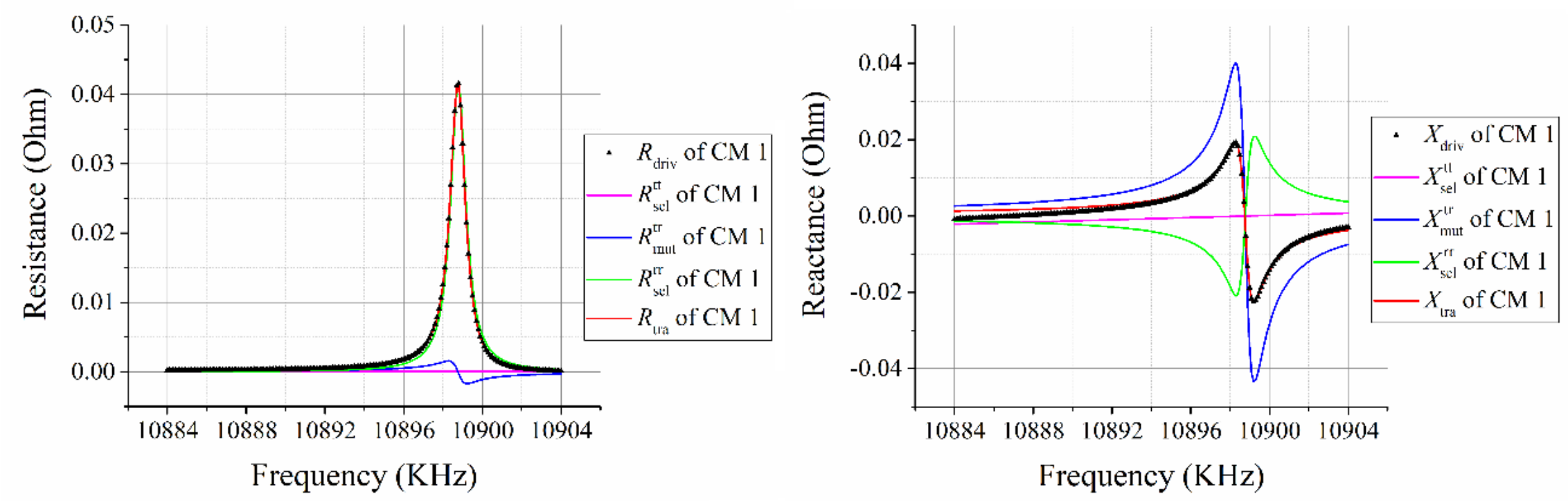

(a) (b)

Figure G-8 For the CM 1, its (a) resistance curves and (b) reactance curves

Based on the above observations, it is not difficult to conclude that:

**C1** At optimally transferring frequency $f_{\text{cor}}$, the driver provides power to coil T in a very efficient way, and the driving power is almost completely transferred from coil T to coil R except a very small part dissipated by coil T, and the transferred power is completely converted by coil R into its self-power but not into any mutual-power.

**C2** At optimally transferring frequency $f_{\text{cor}}$, coil T works at the state of self-resonance (because $X_{\text{sel}}^{\text{tt}} = 0$), and coil R works at the state of self-resonance as well (because $X_{\text{sel}}^{\text{rr}} = 0$), and, at the same time, the two coils also work at the state of mutual-resonance (because $X_{\text{mut}}^{\text{tr}} = 0$). The working state with $X_{\text{sel}}^{\text{tt}} = X_{\text{mut}}^{\text{tr}} = X_{\text{sel}}^{\text{rr}} = 0$ is particularly called co-resonance in this paper, and conventionally called magnetic-resonance in [282~287]. In fact, the co-resonance/magnetic-resonance automatically leads to that $X_{\text{tra}} = 0$ and $X_{\text{driv}} = 0$, because $X_{\text{tra}} = X_{\text{mut}}^{\text{tr}} + X_{\text{sel}}^{\text{rr}}$ and $X_{\text{driv}} = X_{\text{sel}}^{\text{tt}} + X_{\text{mut}}^{\text{tr}} + X_{\text{sel}}^{\text{rr}}$.

**C3** The co-resonance/magnetic-resonance satisfies the classical resonance condition that the reactance is zero, i.e., the stored electric energy is equal to the stored magnetic energy. Thus, the co-resonance/magnetic-resonance is a special case of the classical EM-resonance (electric-magnetic resonance).

**C4** At frequency $f_{\text{cir}}$, whole transferring system is resonant (because $X_{\text{driv}} = 0$), but neither coil T nor coil R is resonant (because $X_{\text{sel}}^{\text{tt}}, X_{\text{sel}}^{\text{rr}} \neq 0$). The resonance of whole transferring system originates from neutralizing the negative $X_{\text{sel}}^{\text{tt}} + X_{\text{sel}}^{\text{rr}}$ and the positive $X_{\text{mut}}^{\text{tr}}$. To distinguish this kind of resonance from the previous co-resonance, this kind of resonance is particularly called capacitive-inductive-neutralized resonance (simply denoted as ci-resonance) in this paper.

In addition, by observing Fig. G-4 and Fig. G-8(a), it is easy to find out that the co-resonant mode is more desired than the ci-resonant mode for WPT application, because the former corresponds to the local maximums of TC curve and $R_{\text{tra}}$ curve but the latter doesn't.

## G7 Typical Examples

To exhibit the values of WEP-CMT in the aspect of analyzing transferring systems and reveal the connections between transferring problem and scattering problem, this section employs WEP-CMT to do some valuable modal analysis for several typical variants of the classical two-coil transferring system considered in Sec. G6.

### G7.1 On Transferring Frequency — Part I

Now, we consider the *scattering* system only constituted by the coil T considered in Sec. G6, and we calculate its *scattering* CMs by using the WEP-CMT for *scattering* systems[8,9,13]. The modal significance (MS) curves associated with the first 5 CMs are shown in Fig. G-9. The resonant CM i with the local maximal MS has the same current distribution as the one shown in Fig. G-5(a).

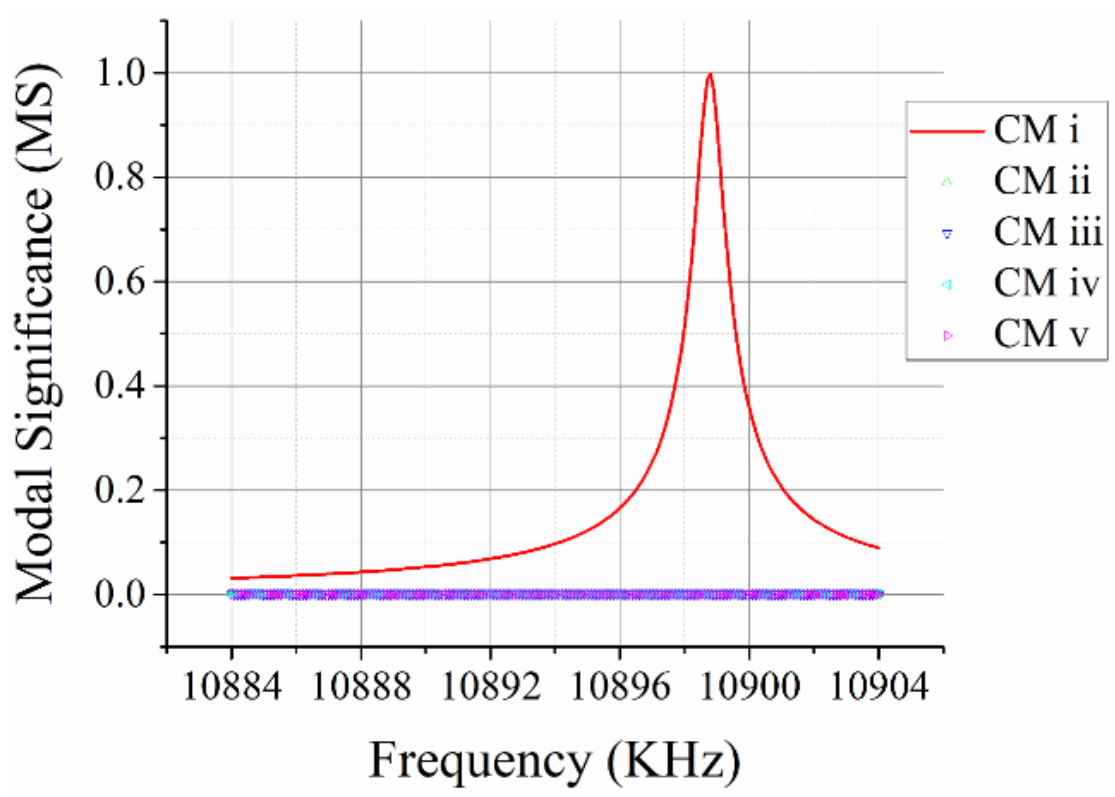

Figure G-9 MS curves of the first 5 CMs of the one-coil scattering system

Comparing the Fig. G-9 with the previous Figs. G-4, G-7 and G-8, it is evident that

**C5** The **co-resonance** frequency (i.e. the optimally transferring frequency) of the two-coil **transferring** system considered in Sec. G6 is equal to the conventional *external-resonance* frequency of the one-coil *scattering* system.

This conclusion effectively establishes a connection between the co-resonance-based/magnetic-resonance-based wireless power **transferring** problem and the conventional *scattering* problem.

## G7.2 On Transferring Frequency — Part II

In fact, the above-mentioned external-resonance frequency is the one-order external-resonance frequency of the one-coil scattering system, and the scattering system has some other higher-order external-resonance frequencies, such as the two-order external-resonance frequency 27.2646 MHz as shown in Figs. G-10 and G-11.

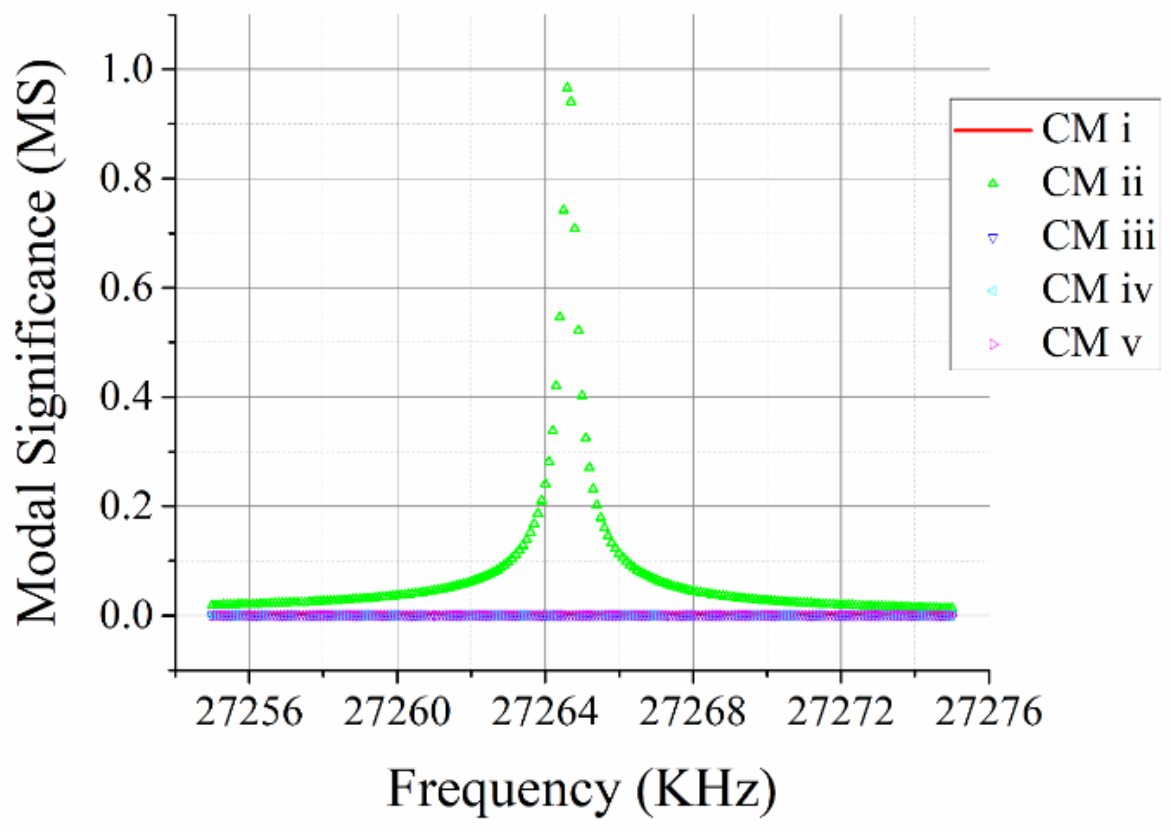

Figure G-10 MS curves of the first 5 CMs of the one-coil scattering system

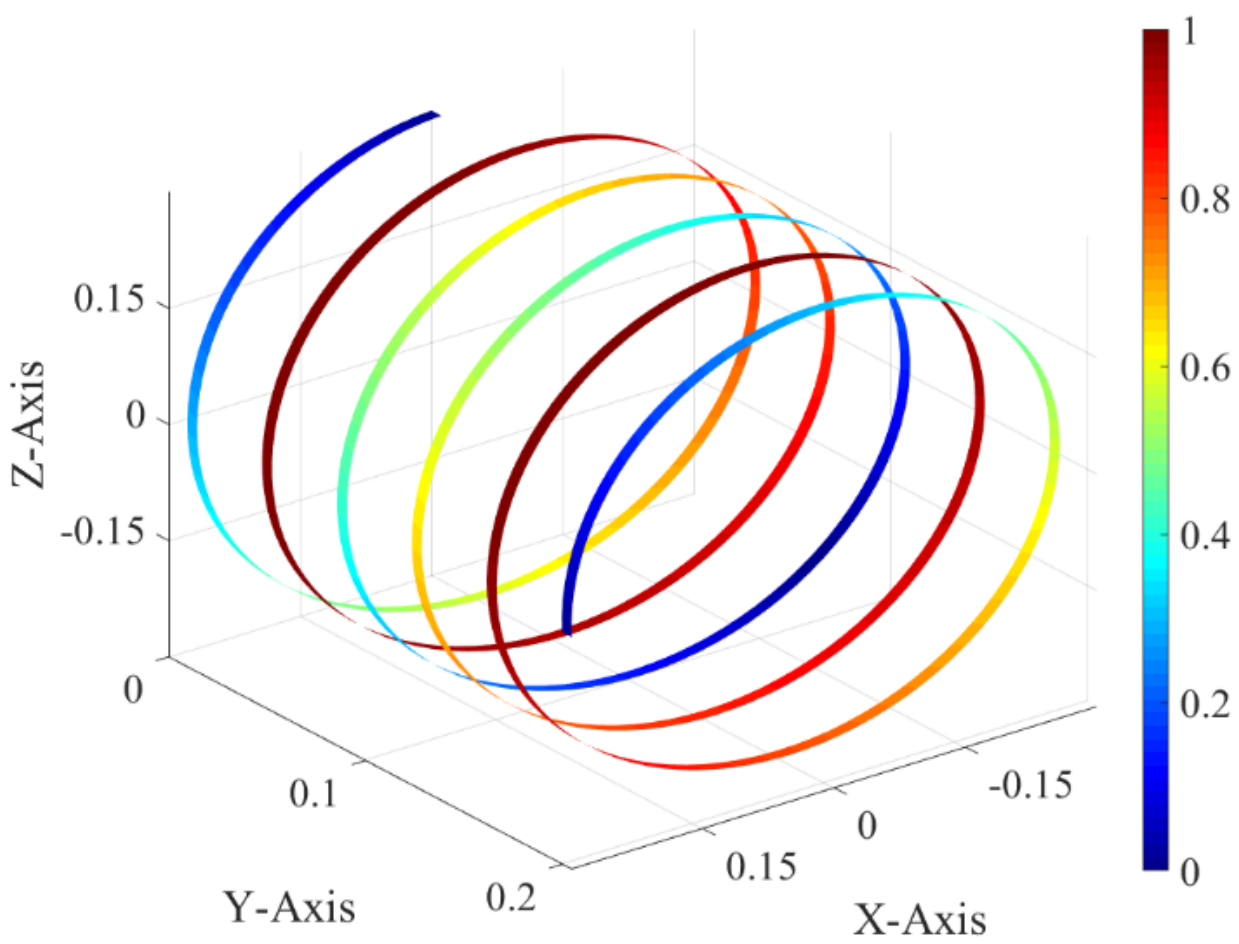

Figure G-11 Modal current magnitude distribution of the two-order externally resonant CM ii working at 27.2646 MHz

In the frequency band used in Fig. G-10, we calculate the CMs of the two-coil transferring system, and show the TC and reactance curves associated with CM 2 in Fig. G-12.

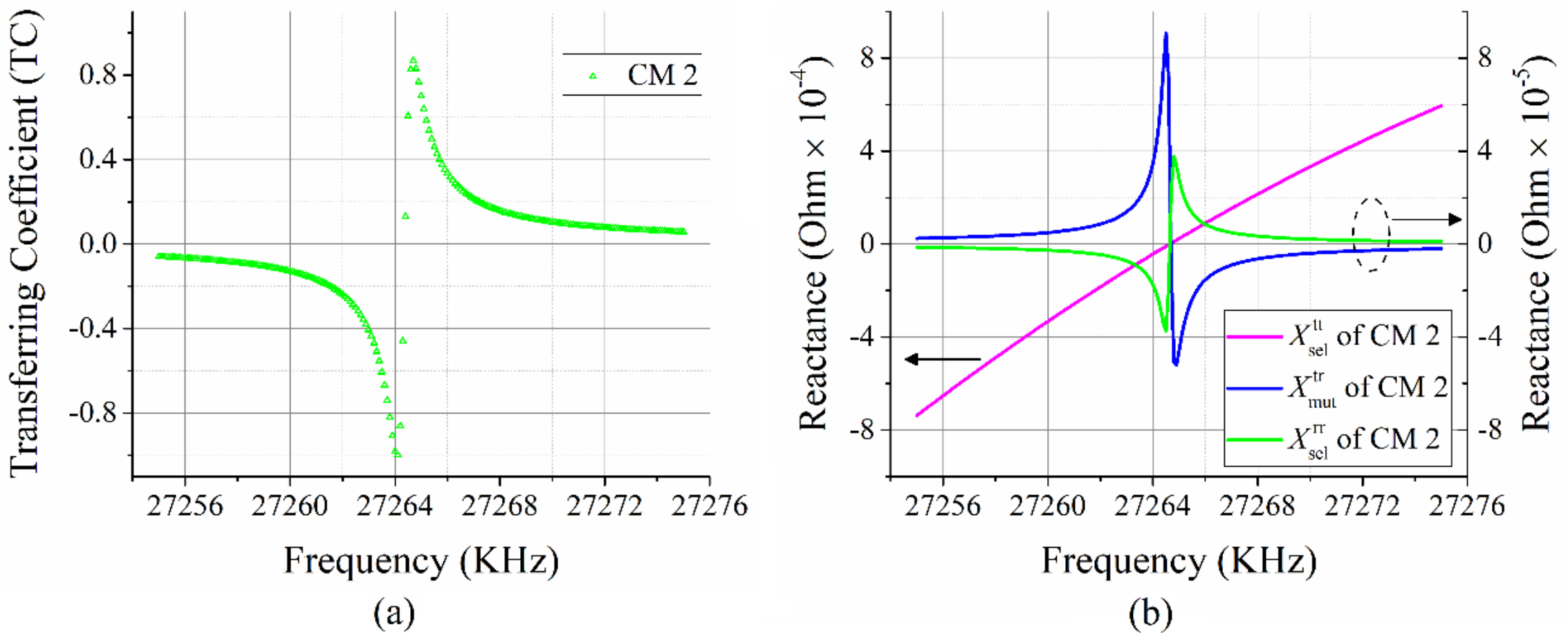

Figure G-12 (a) TC curve and (b) reactance curves associated with the CM 2 of the two-coil transferring system working in the frequency band used in Fig. G-10

The Fig. G-12(b) implies that the CM 2 is co-resonant at 27.2646 MHz. The time-average magnetic energy density distribution of the co-resonant CM 2 working at co-resonance frequency 27.2646 MHz is shown in Fig. G-13.

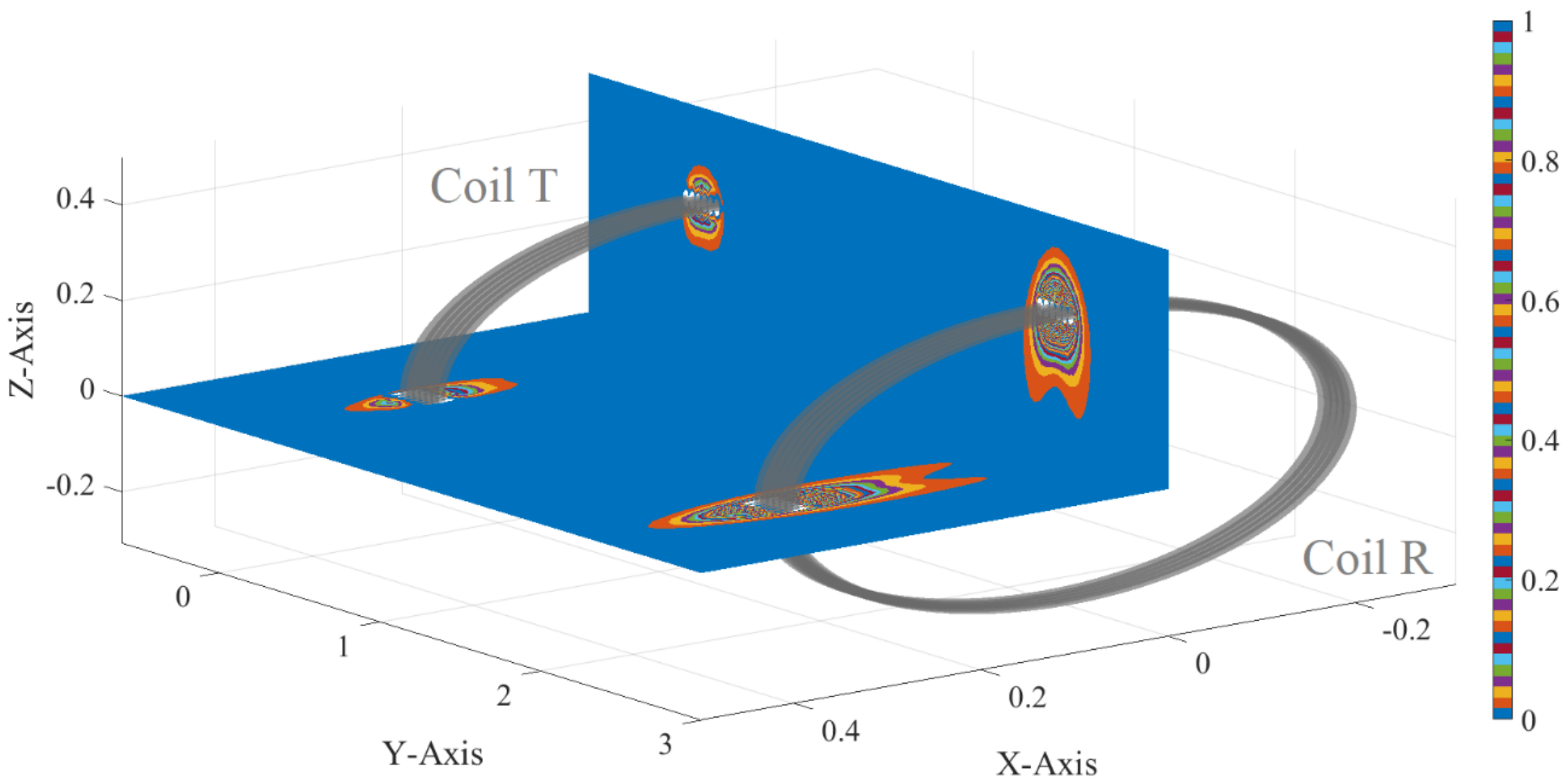

Figure G-13 For the co-resonant CM 2 working at 27.2646 MHz, its time-average magnetic energy density distributing on xOy and yOz planes

By comparing the Figs. G-12(a)&G-13 with the previous Figs. G-4&G-6, it is not difficult to find out that:

**C6** the one-order co-resonant mode has a higher wireless power transferring efficiency than the two-order co-resonant mode.

In fact, this conclusion can be generalized to the comparison between any lower-order and higher-order co-resonant modes.

## G7.3 On Transferring Frequency — Part III

The above-mentioned two-coil transferring system is constituted by two completely same coils, which have the same geometrical size and external-resonance frequencies 10.8988 MHz (one order) and 27.2646 MHz (two order) etc. Now, we replace the coil R with another different coil, whose geometrical size is "radius 30 cm, height 8.15 cm, and turns 2.138" and one-order external-resonance frequency is 27.2646 MHz and higher-order external resonance frequencies are larger than 27.2646 MHz. We calculate the CMs of the new two-coil transferring system working around frequencies 10.8988 MHz and 27.2646 MHz, and we show the corresponding TC and reactance curves in Fig. G-14 and Fig. G-15 respectively.

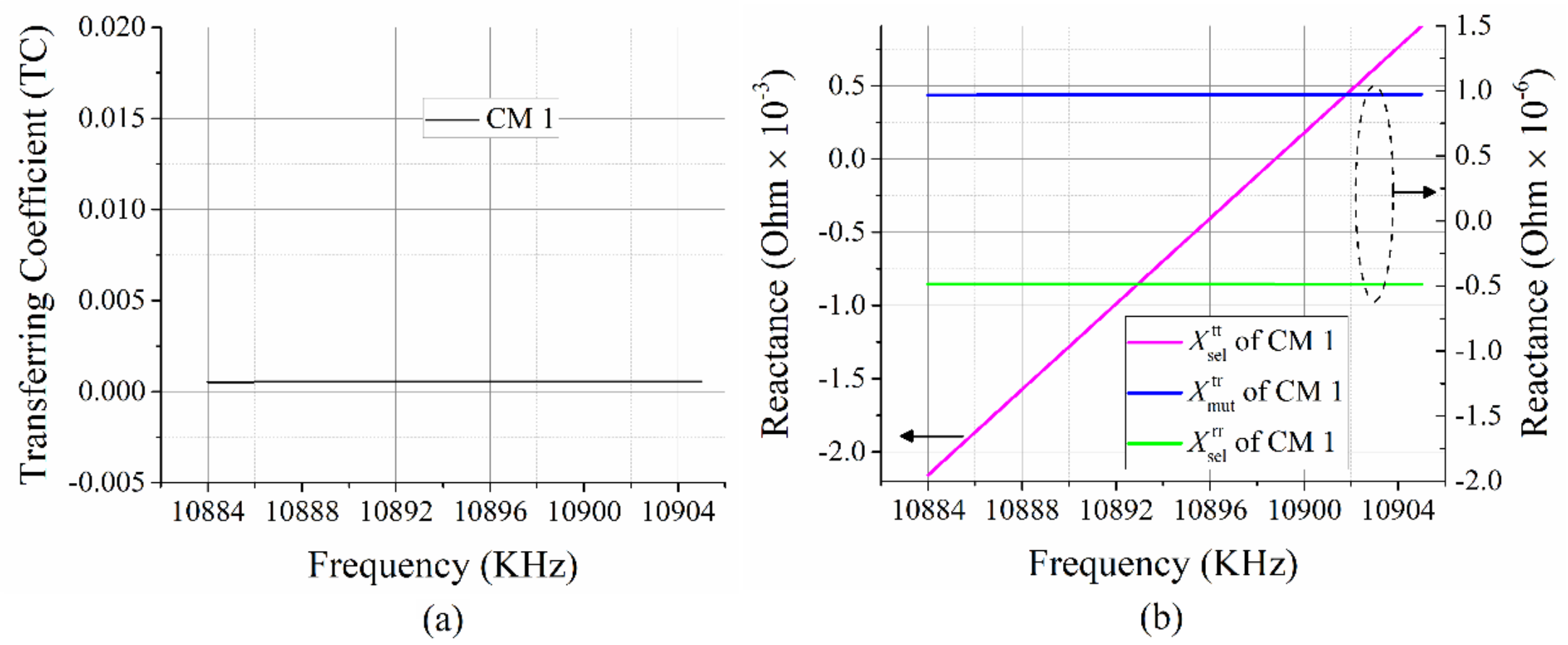

Figure G-14 (a) TC curve and (b) reactance curves associated with the CM 1 of the new two-coil transferring system working around 10.8988 MHz

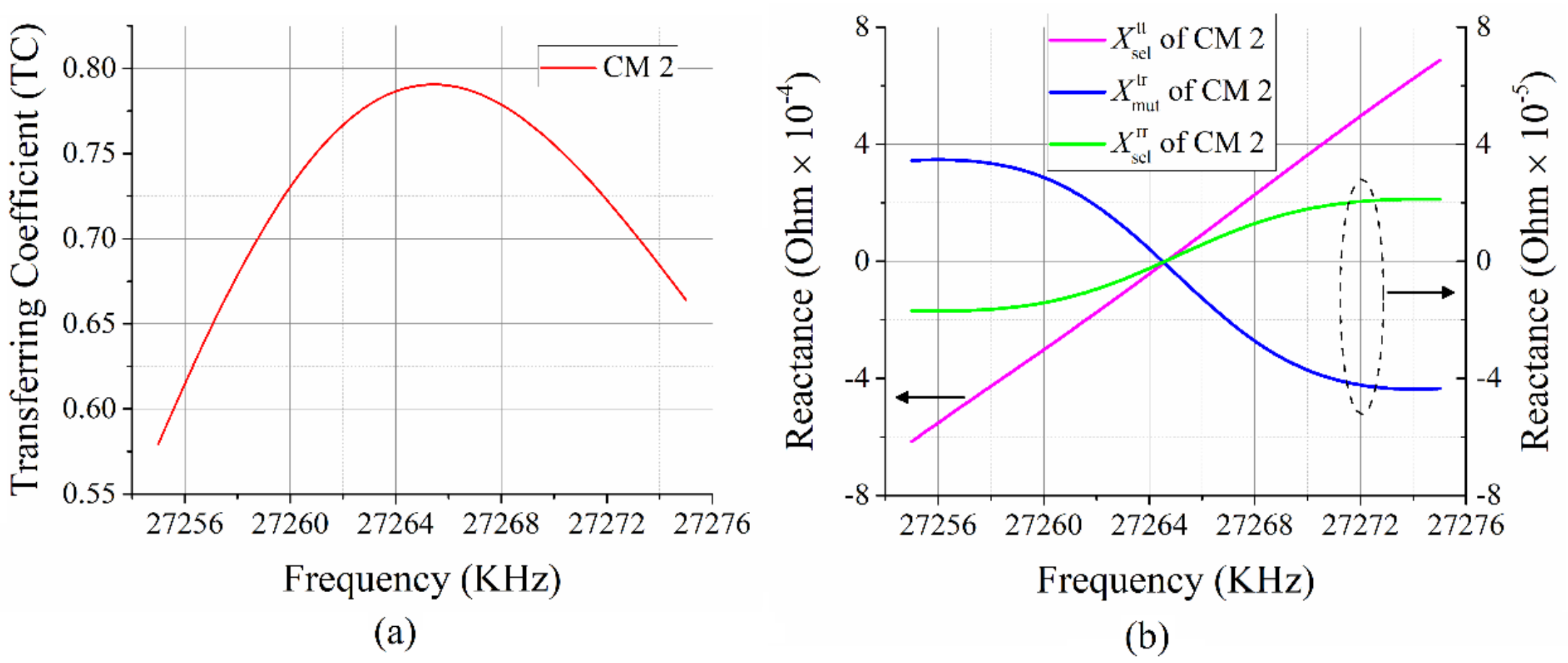

Figure G-15 (a) TC curve and (b) reactance curves associated with the CM 2 of the new two-coil transferring system working around 27.2646 MHz

From Figs. G-14 and G-15, it is easy to find out that: the coil T working at CM 1 is self-resonant at 10.8988 MHz ( $X_{\text{sel}}^{\text{tt}} = 0$ ), but 10.8988 MHz is not the co-resonance frequency of the new two-coil transferring system ( $X_{\text{mut}}^{\text{tr}}, X_{\text{sel}}^{\text{rr}} \neq 0$ ), and the TC is very small at 10.8988 MHz; 27.2646 MHz is the co-resonance frequency of the new two-coil transferring system, and the TC curve achieves its local maximum at 27.2646 MHz.

Based on the above observations and the comparisons among Figs. G-4, G-12(a), and G-15(a), we can conclude that:

**C7** When only one coil is self-resonant, the whole system cannot efficiently transfer power wirelessly.

**C8** When two coils have the same external-resonance frequency, and the whole two-coil transferring system works at the frequency, the system can transfer power wirelessly even if the orders of the frequencies are different.

**C9** The same-order co-resonance is more desired than the different-order co-resonance in WPT application.

To visualize the above conclusions, we show the corresponding time-average magnetic energy density distributions in Fig. G-16.

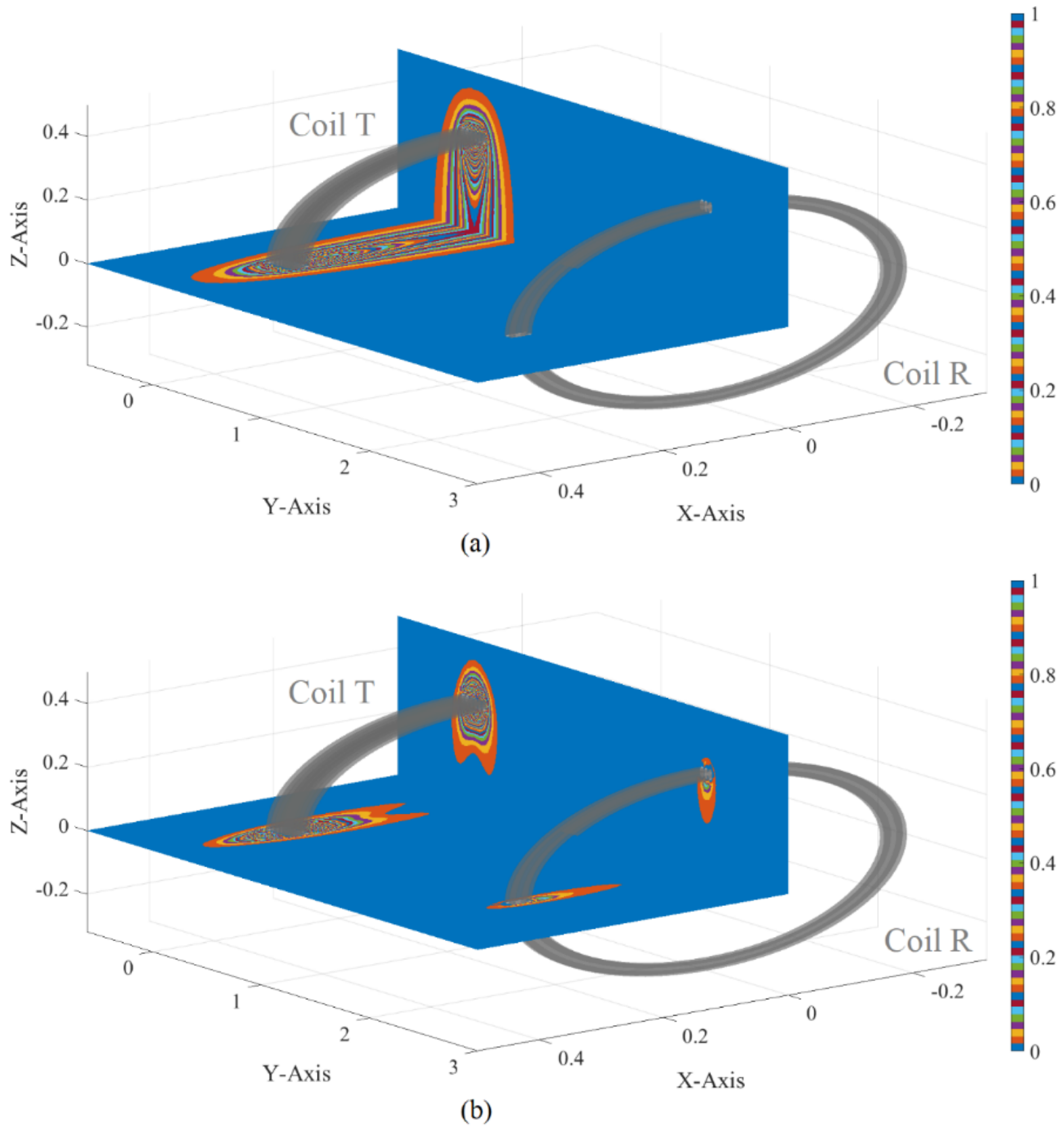

Figure G-16 Time-average magnetic energy density distributions of (a) the CM 1 working at 10.8988 MHz and (b) the CM 2 working at 27.1646 MHz

## G7.4 On Transferring Distance

In the above all discussions, the distance between coil T and coil R is 2 m. In this sub-section, we study how the distance influences transferring efficiency.

We use the WEP-CMT to calculate the two-coil transferring systems with a series of different coil-separation distances (where the coil T and coil R are the same as the ones considered in Sec. G6), and show the TC curves of the one-order CMs corresponding to the different coil-separation distances in Fig. G-17.

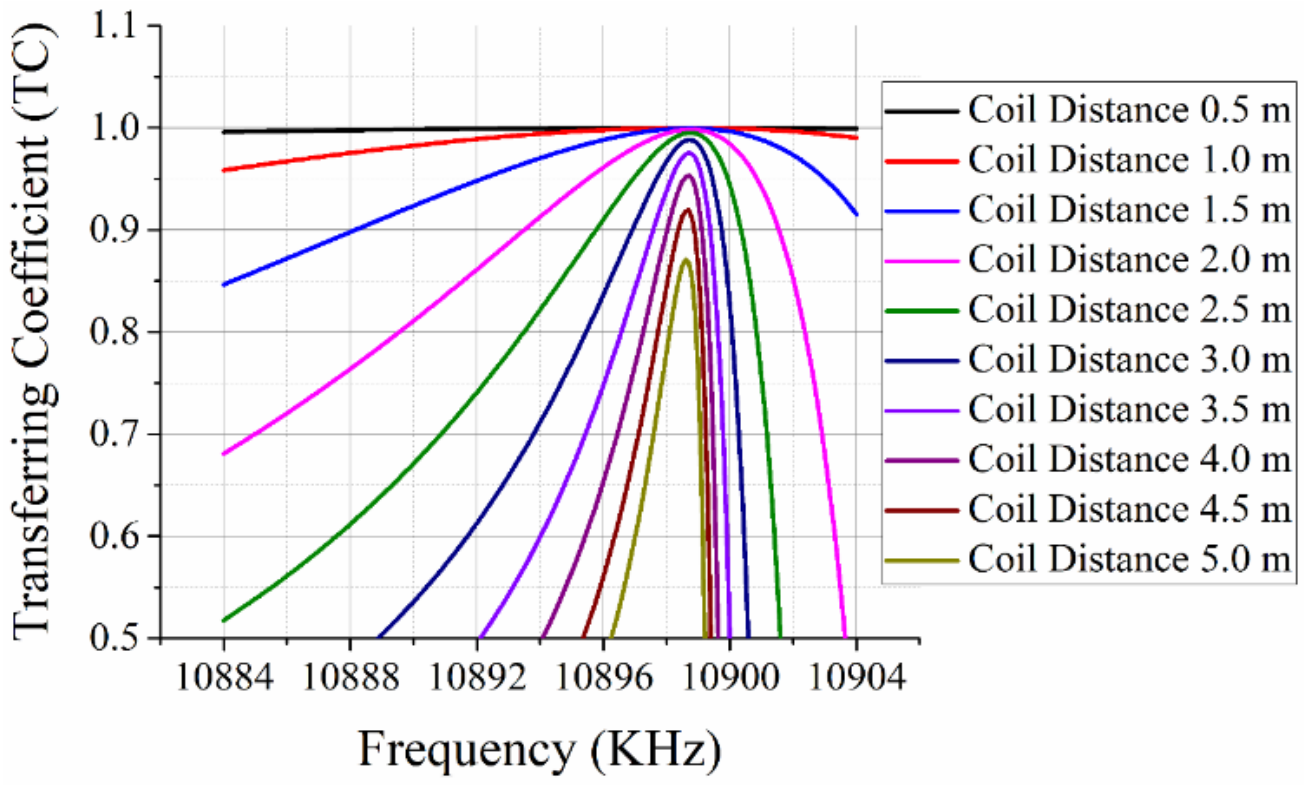

Figure G-17 TC curves of the one-order CMs corresponding to the different coil-separation distances

The Fig. G-17 clearly implies the following expected conclusion.

**C10** When two same coils are placed axially, the co-resonance frequency of the two-coil transferring system is almost independent of the distance between the coils, and the TC of the system is a monotonically decreasing function about the distance.

To visually exhibit the above conclusion, we show the time-average magnetic energy density distributions corresponding to distances 0.5 m, 2.5 m, and 5.0 m in Fig. G-18.

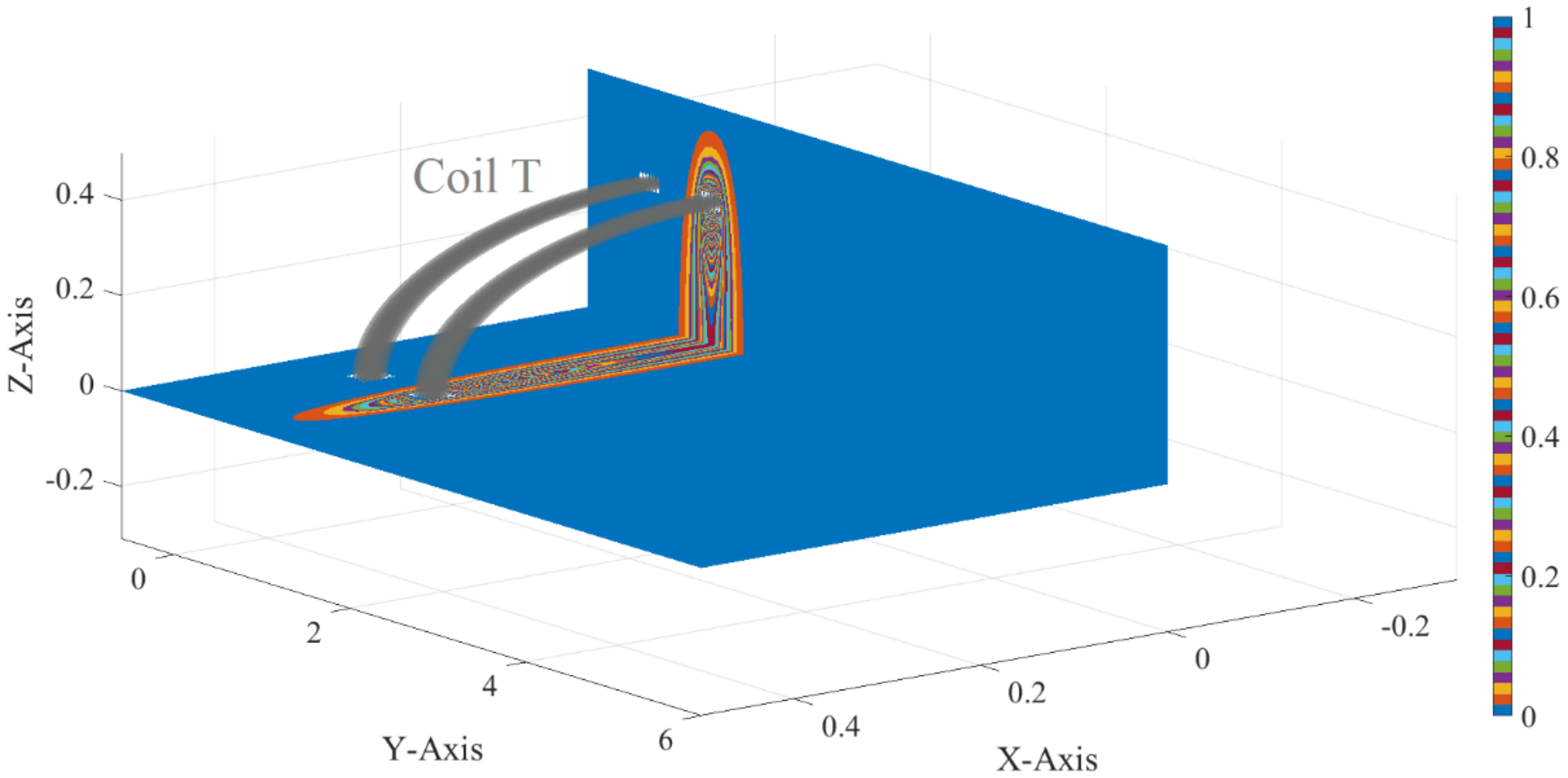

(a)

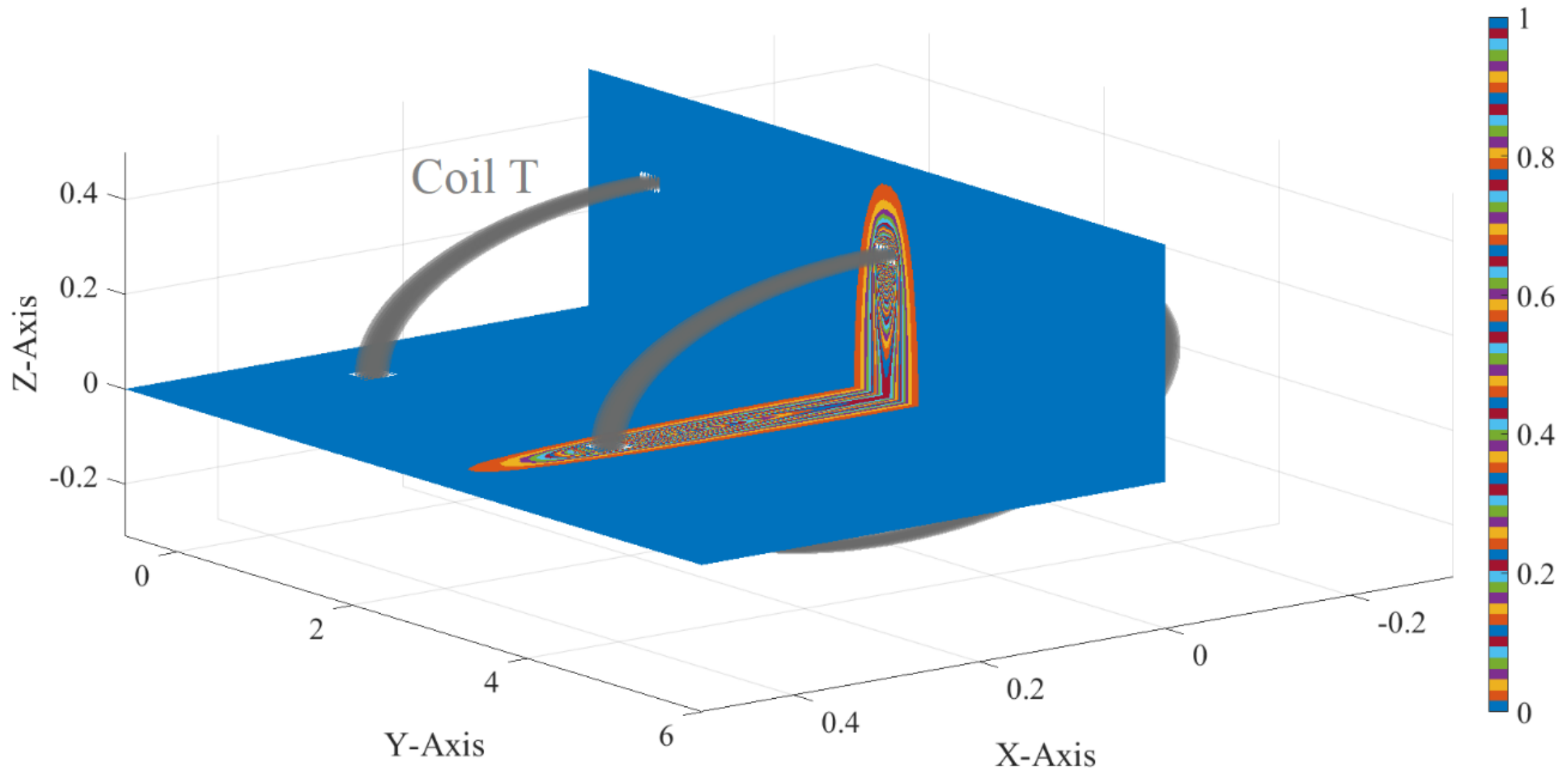

(b)

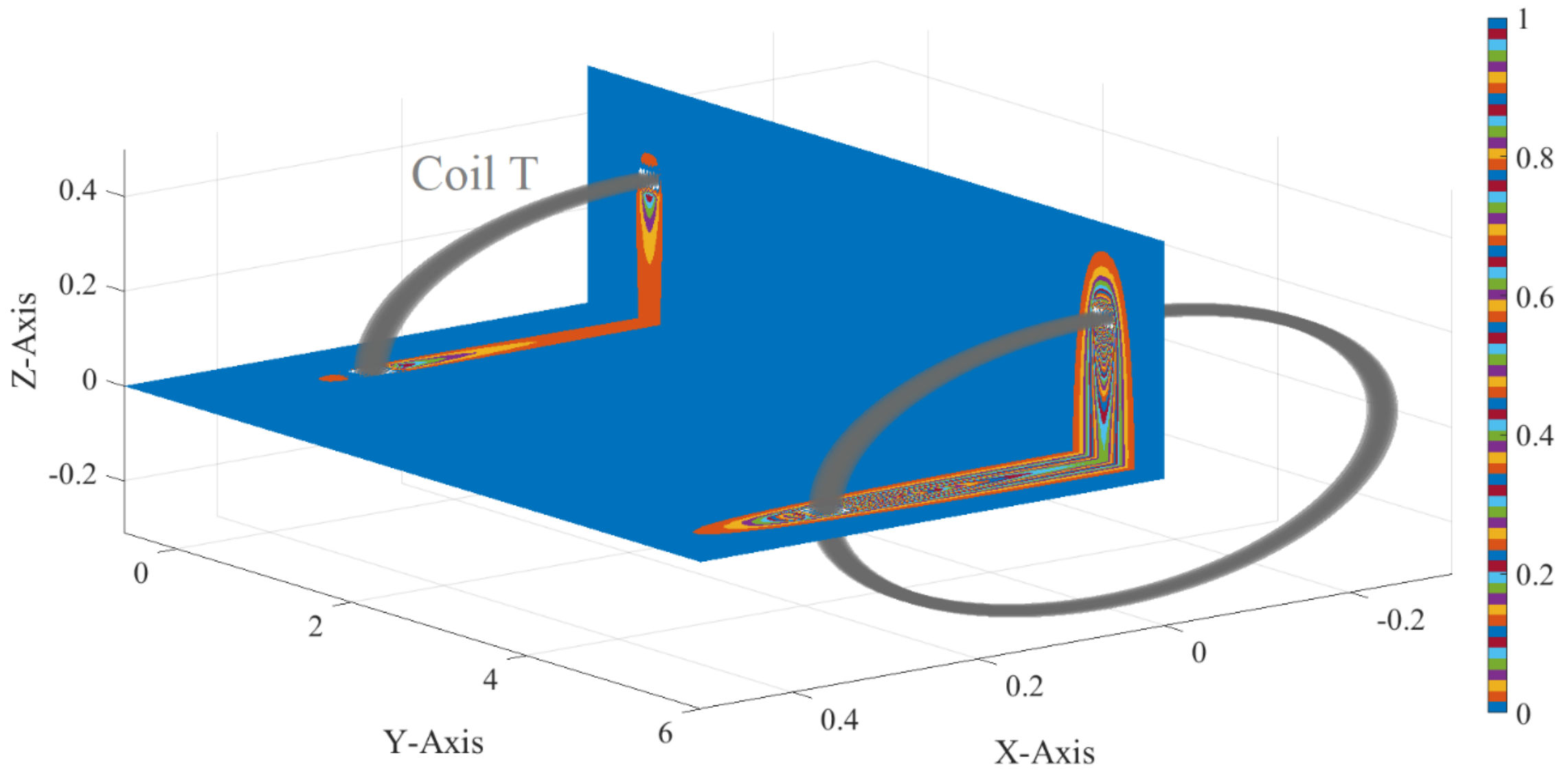

(c)

Figure G-18 Time-average magnetic energy density distributions corresponding to coil-separation distances (a) 0.5 m, (b) 2.5 m, and (c) 5.0 m

## G7.5 On Transferring Direction

In this sub-section, we employ the WEP-CMT to study how the relative direction between coil T and coil R influences the transferring efficiency. The angle between the axis of the two coils is denoted as $\alpha$ (in degree). We calculate the CMs of the transferring systems with $\alpha = 0°$, 10°, 20°, …, 90°, and show the corresponding TC curves in Fig. G-19.

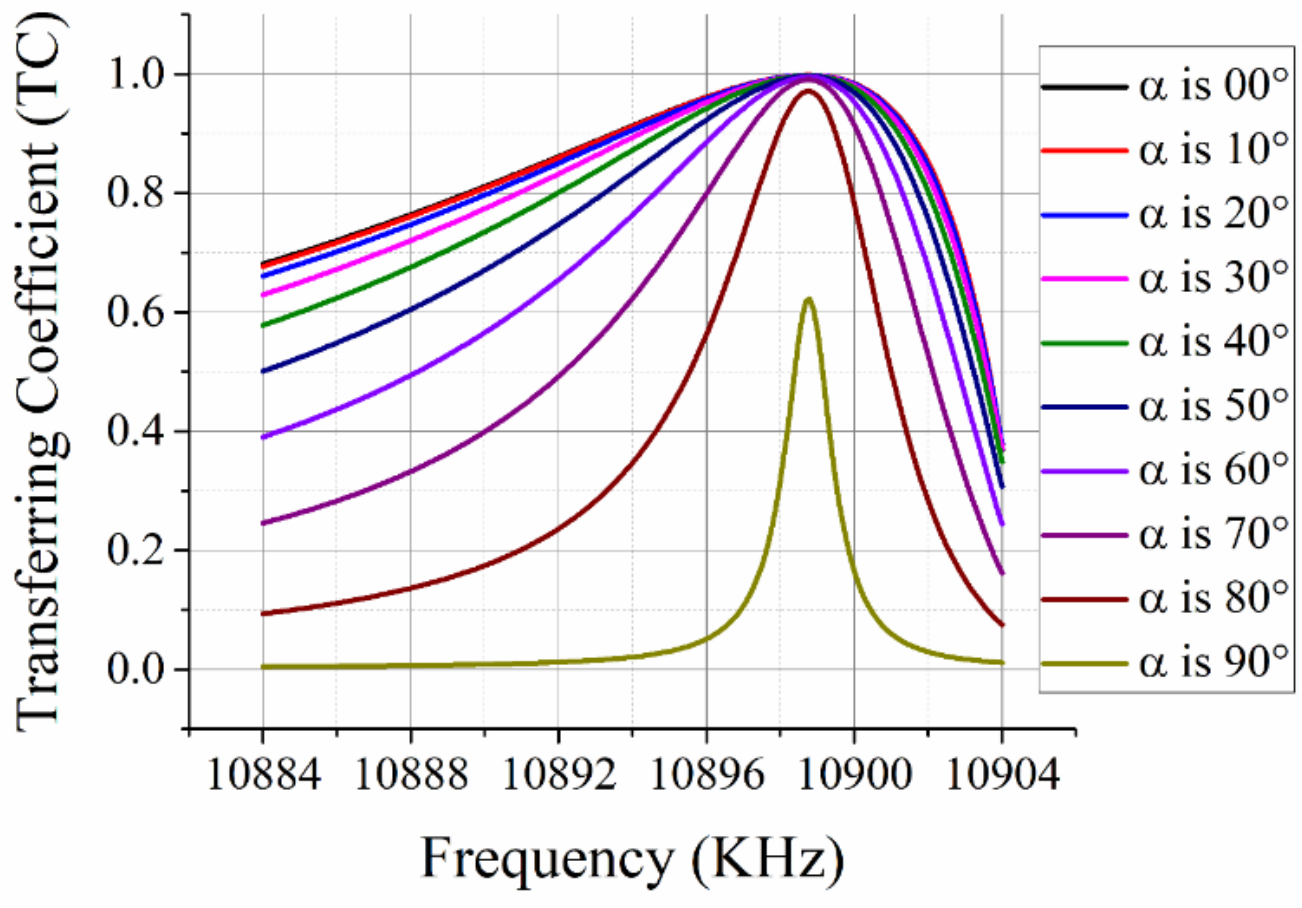

Figure G-19 TC curves of the one-order CMs corresponding to different coil-coil angle $\alpha$

Evidently, Fig. G-19 clearly implies the following expected conclusion

**C11** When two same coils are placed with an angle $\alpha$, the co-resonance frequency of the two-coil transferring system is almost independent of $\alpha$, and the TC of the system is a monotonically decreasing function about $\alpha$.

To visually exhibit the above conclusion, we show the time-average magnetic energy density distributions corresponding to $\alpha = 30°$, 60°, and 90° in Fig. G-20.

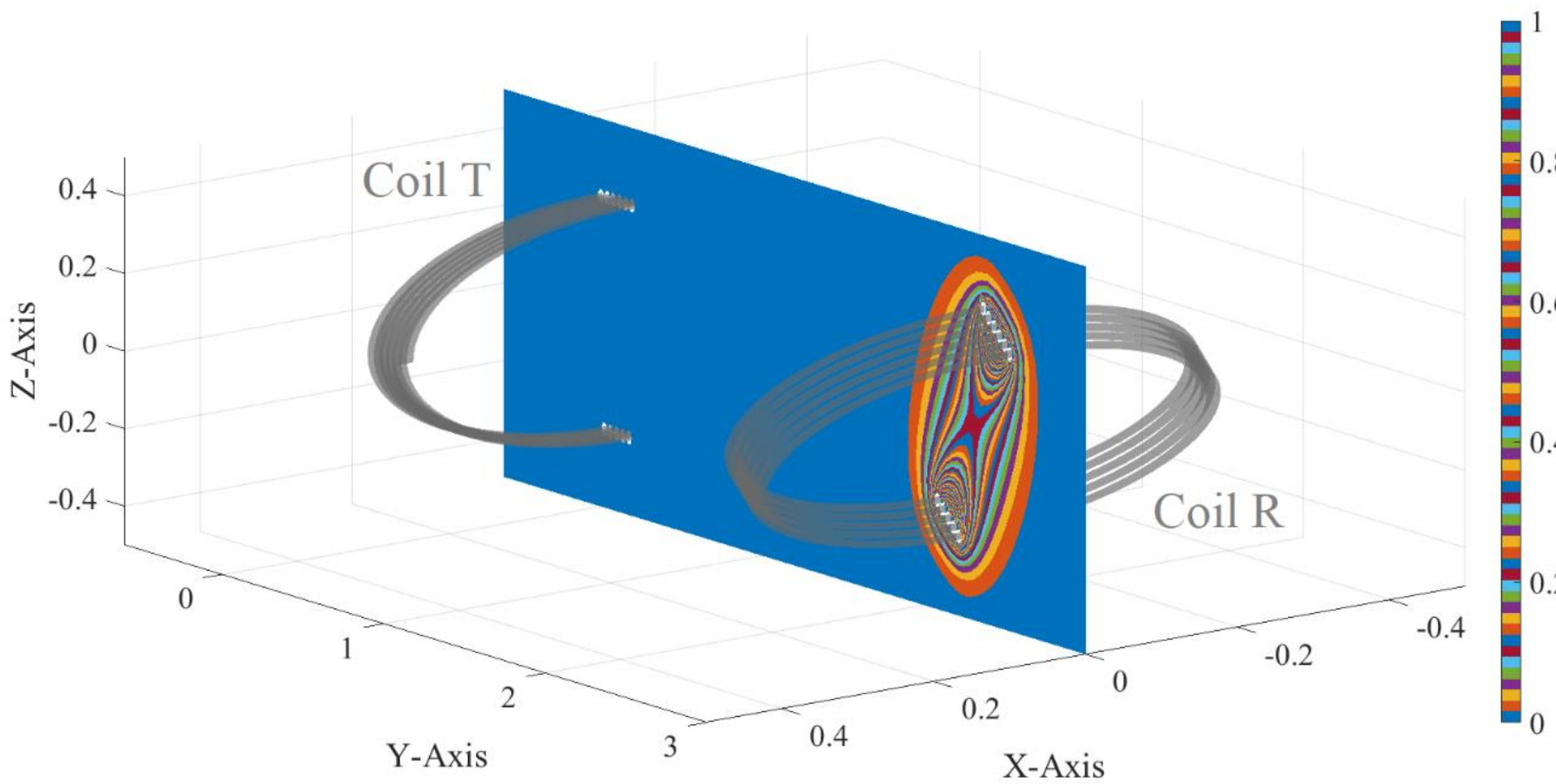

(a)

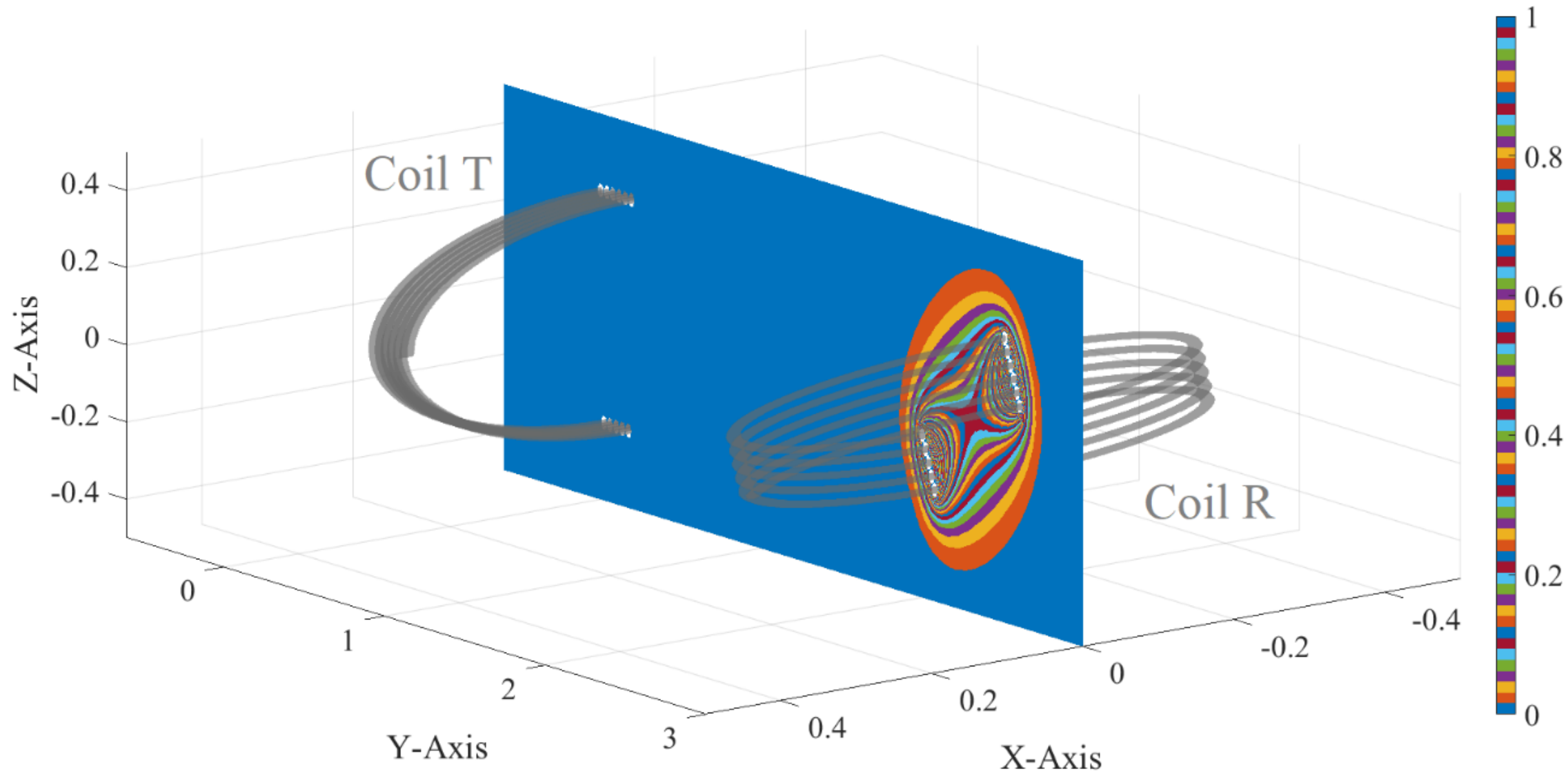

(b)

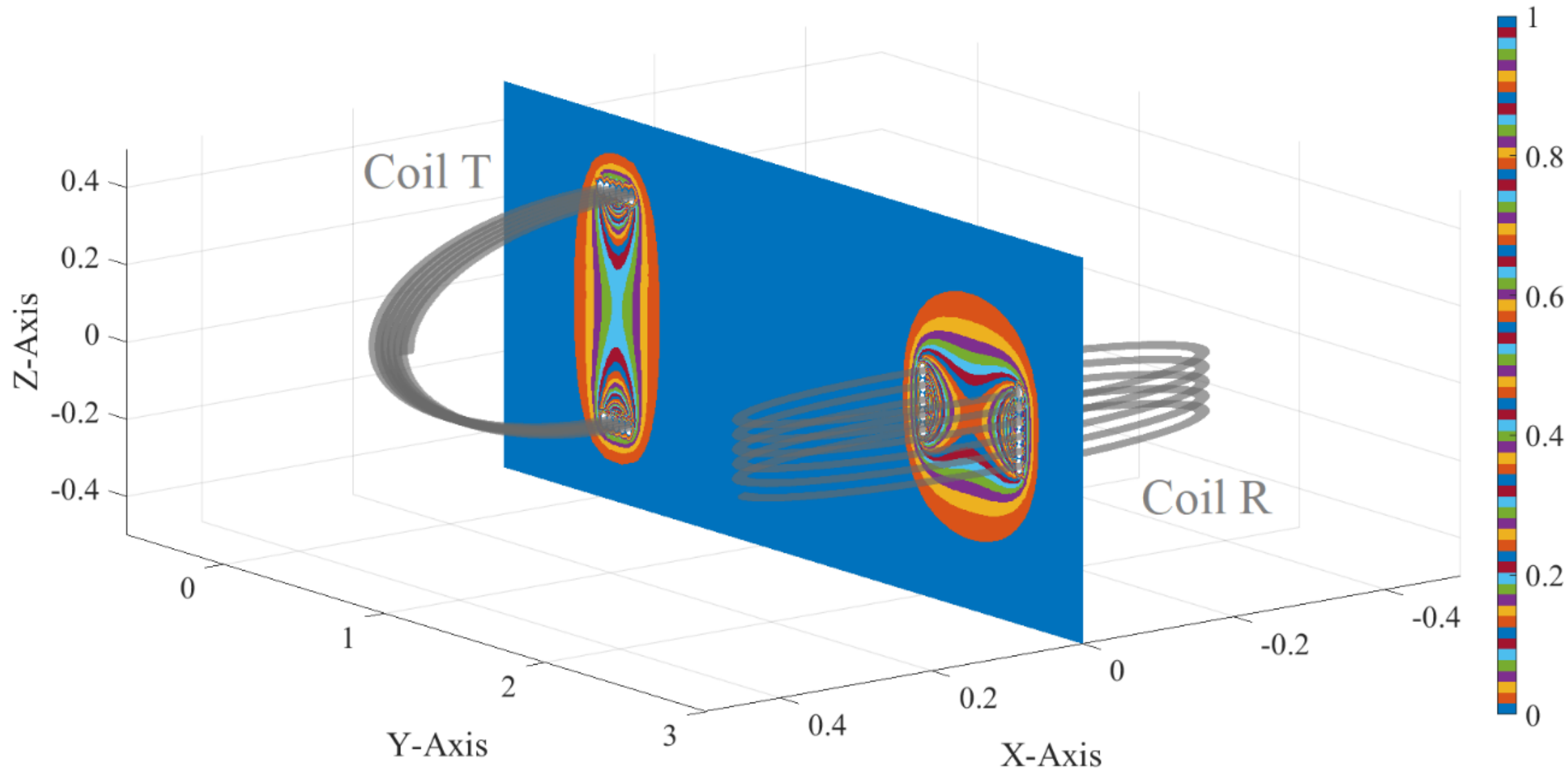

(c)

Figure G-20 Time-average magnetic energy density distributions corresponding to (a) $\alpha = 30°$ , (b) $\alpha = 60°$ , and (c) $\alpha = 90°$

## G7.6 On Multi-coil Transferring System

In the above discussions, only some typical two-coil transferring systems are analyzed. In this sub-section, we further apply the WEP-CMT to some typical multi-coil transferring systems.

Firstly, we consider the three-coil transferring system whose three coils have the same geometrical size and are coaxially placed with mutual distance 2 m. We use WEP-CMT to calculate the CMs of the transferring system, and show the TC curve corresponding to CM 1 in Fig. G-21. For the CM 1 working at the optimally transferring frequency 10.8872, its time-average magnetic energy density distribution is shown in Fig. G-22.

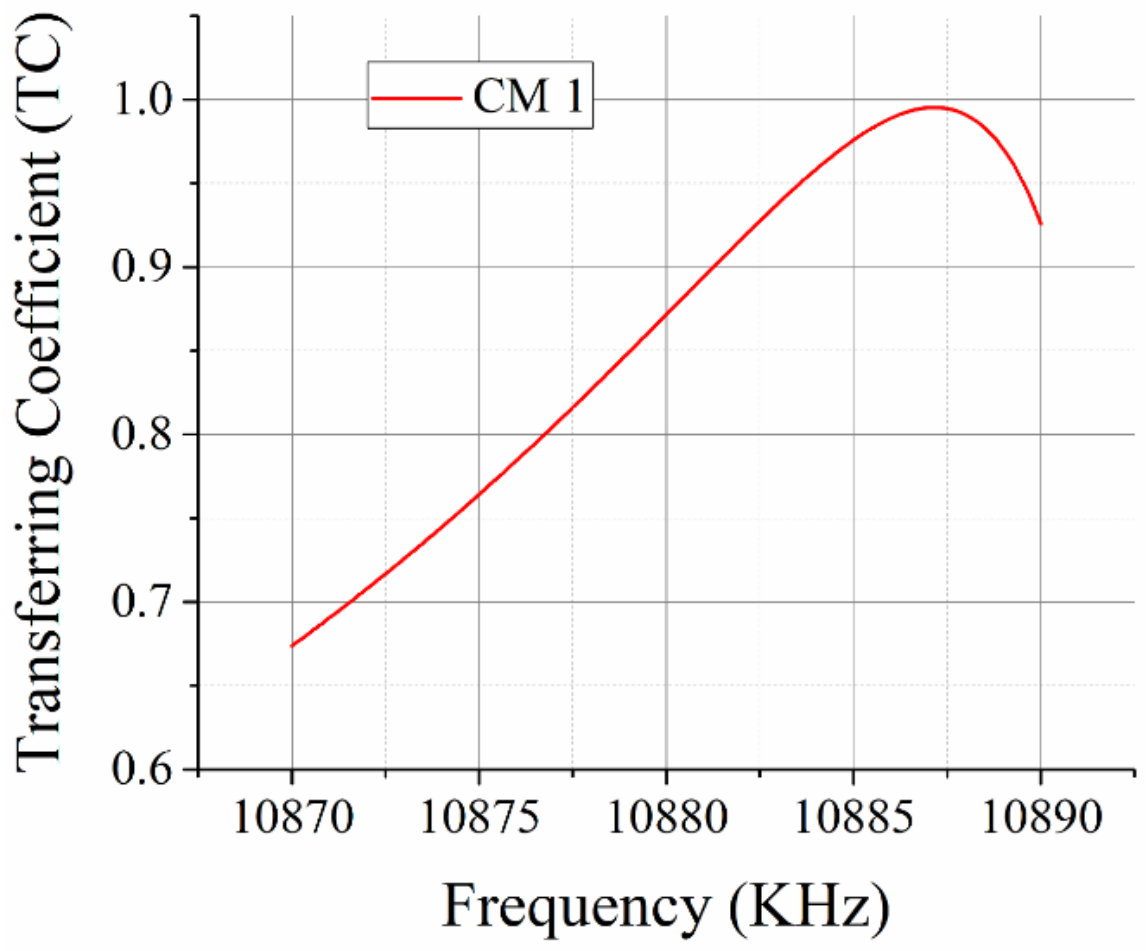

Figure G-21 TC curve associated with the CM 1 of the transferring system constituted by three coaxially placed coils

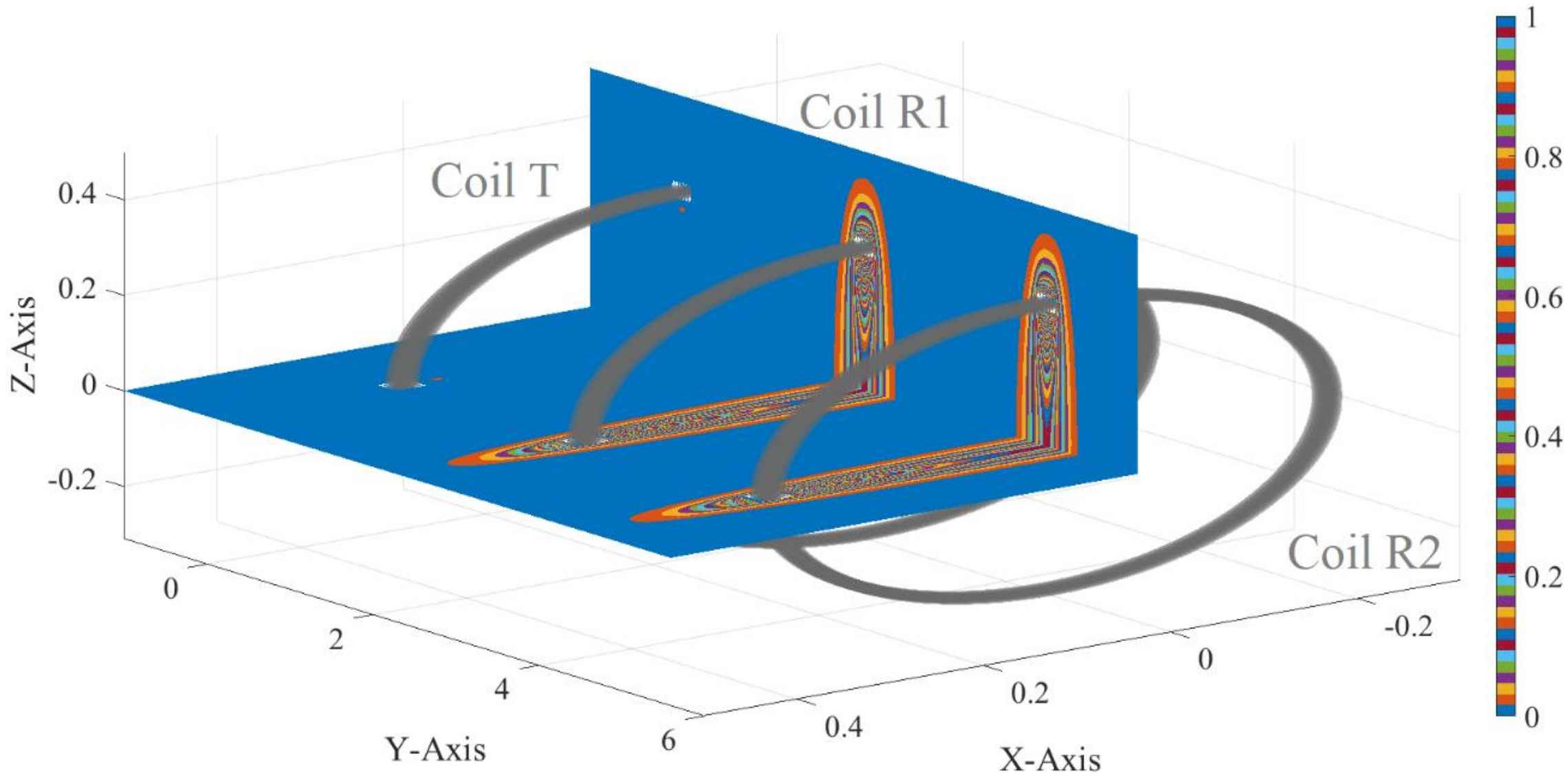

Figure G-22 For the CM 1 working at 10.8872 MHz (shown in Fig. G-21), its time-average magnetic energy density distributing on xOy and yOz planes

Secondly, we consider the three-coil transferring system whose three coils have the same geometrical size but are misaligned. We use WEP-CMT to calculate the CMs of the transferring system, and show the TC curve corresponding to CM 1 in Fig. G-23. For the CM 1 working at the optimally transferring frequency 10.8724, its time-average magnetic energy density distribution is shown in Fig. G-24.

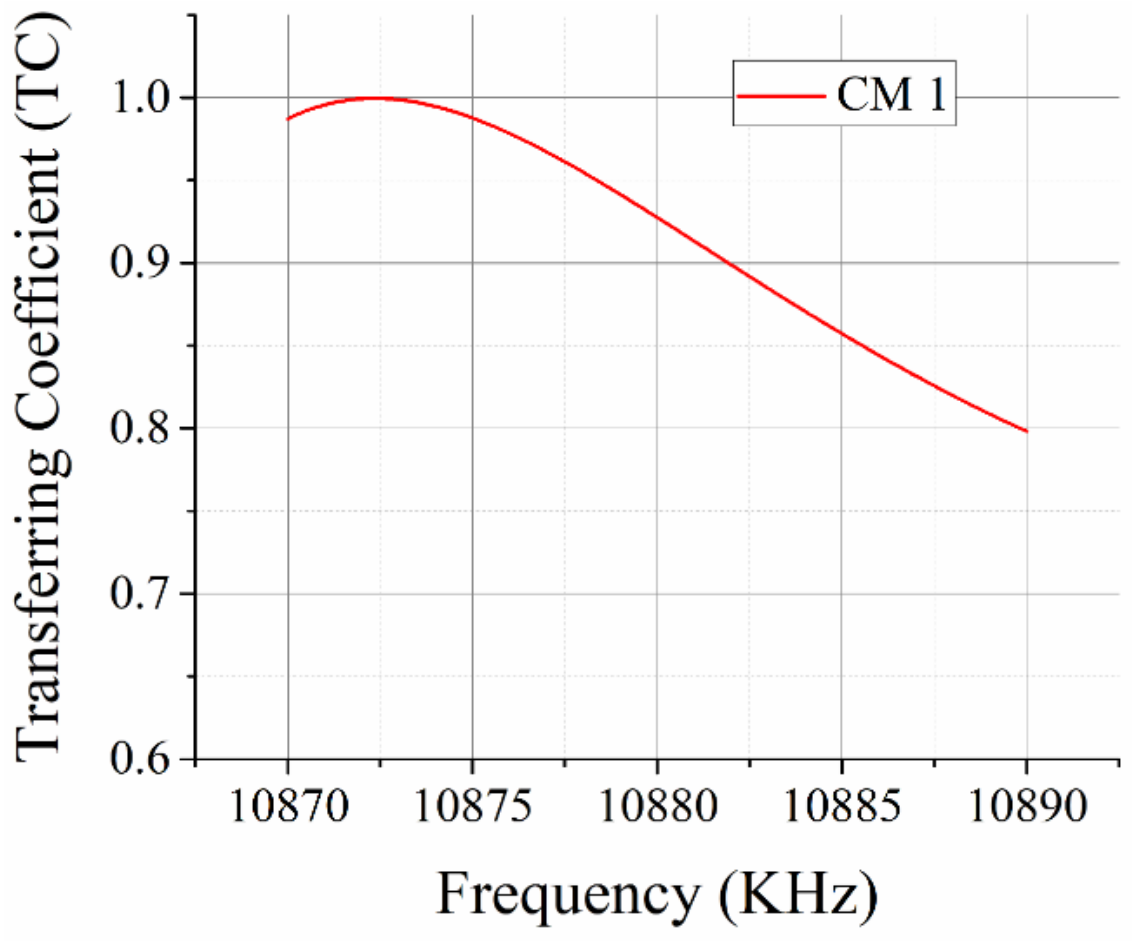

Figure G-23 TC curve associated with the CM 1 of the transferring system constituted by three misaligned coils

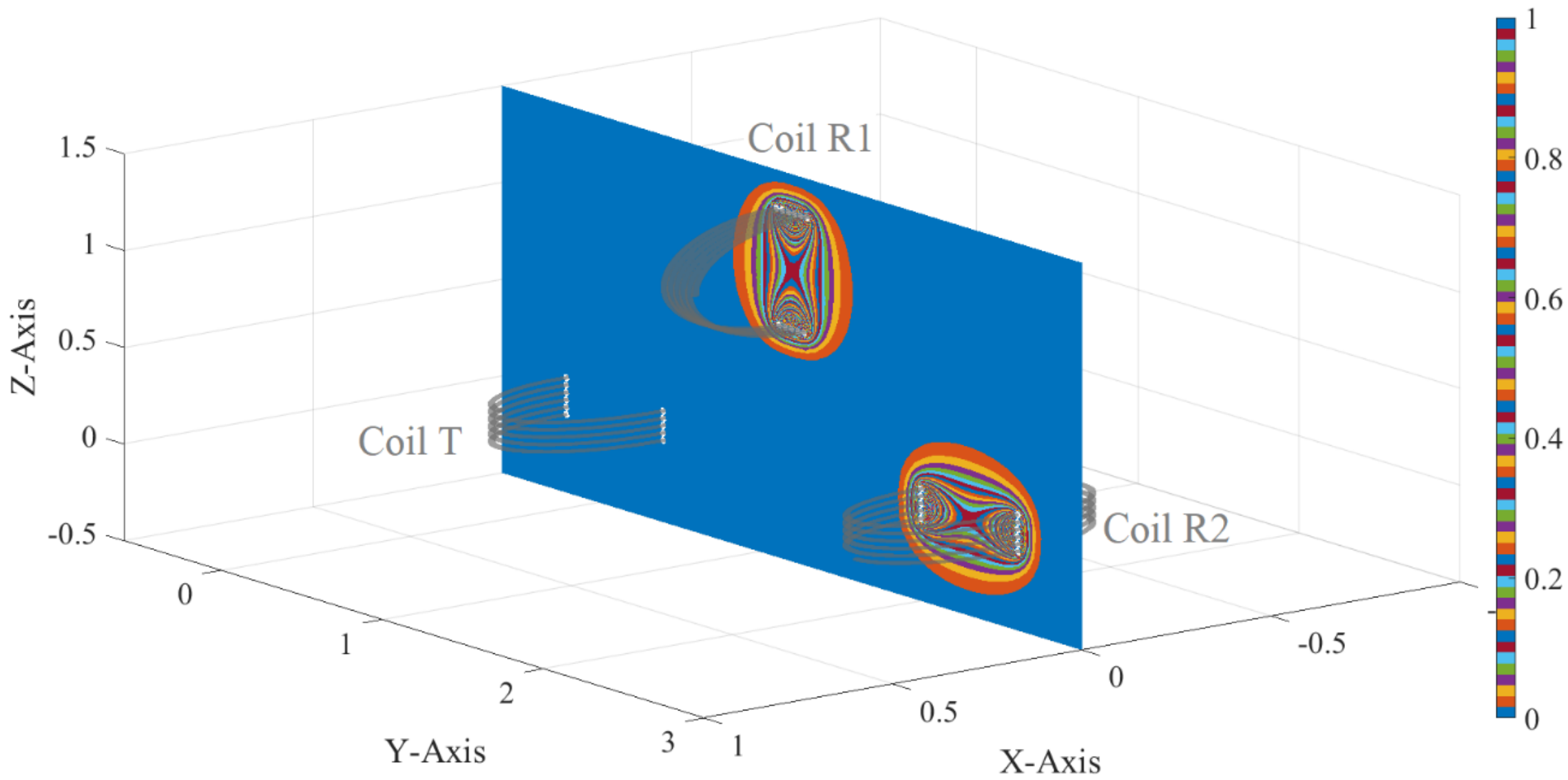

Figure G-24 For the CM 1 working at 10.8724 MHz (shown in Fig. G-23), its time-average magnetic energy density distributing on xOy and yOz planes

By comparing the Figs. G-21&G-23 with the previous Figs. G-4&G-17&G-19, it is not difficult to find out that:

**C12** The optimally transferring frequency of three-coil transferring system is usually different from the optimally transferring frequency of classical two-coil transferring system.

The reason leading to this conclusion is that the external-resonance frequency of scattering system R (discussed previously) is different from the external-resonance frequency of R1-R2 scattering system (discussed here). In fact, the conclusion is also applicable to the multi-coil transferring systems constituted by $N\ (\geq 3)$ coils.

## G7.7 On Surrounding Environment

In this sub-section, we use WEP-CMT to simply study how surrounding environment influences the WPT process.

We integrate a square thin metallic plane into the two-coil transferring system considered in Sec. G6. We use WEP-CMT to calculate the CMs of the transferring systems with three different planes having side-lengths 0.6 m, 1.2 m, and 2.4 m, and show the TC curves corresponding to the one-order CMs in Fig. G-25. We illustrate the time-average magnetic energy density distributions of the optimally transferring modes shown in Fig. G-25 as Fig. G-26.

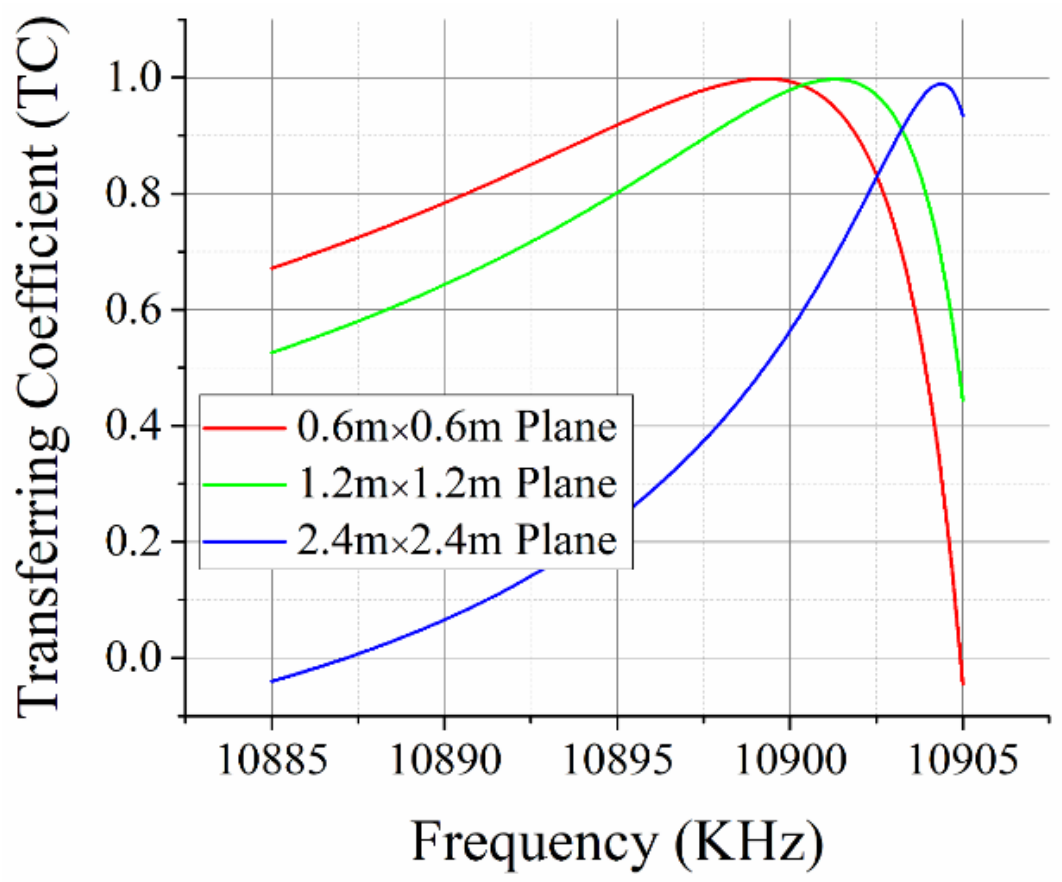

Figure G-25 TC curves of the one-order CMs corresponding to different plane sizes

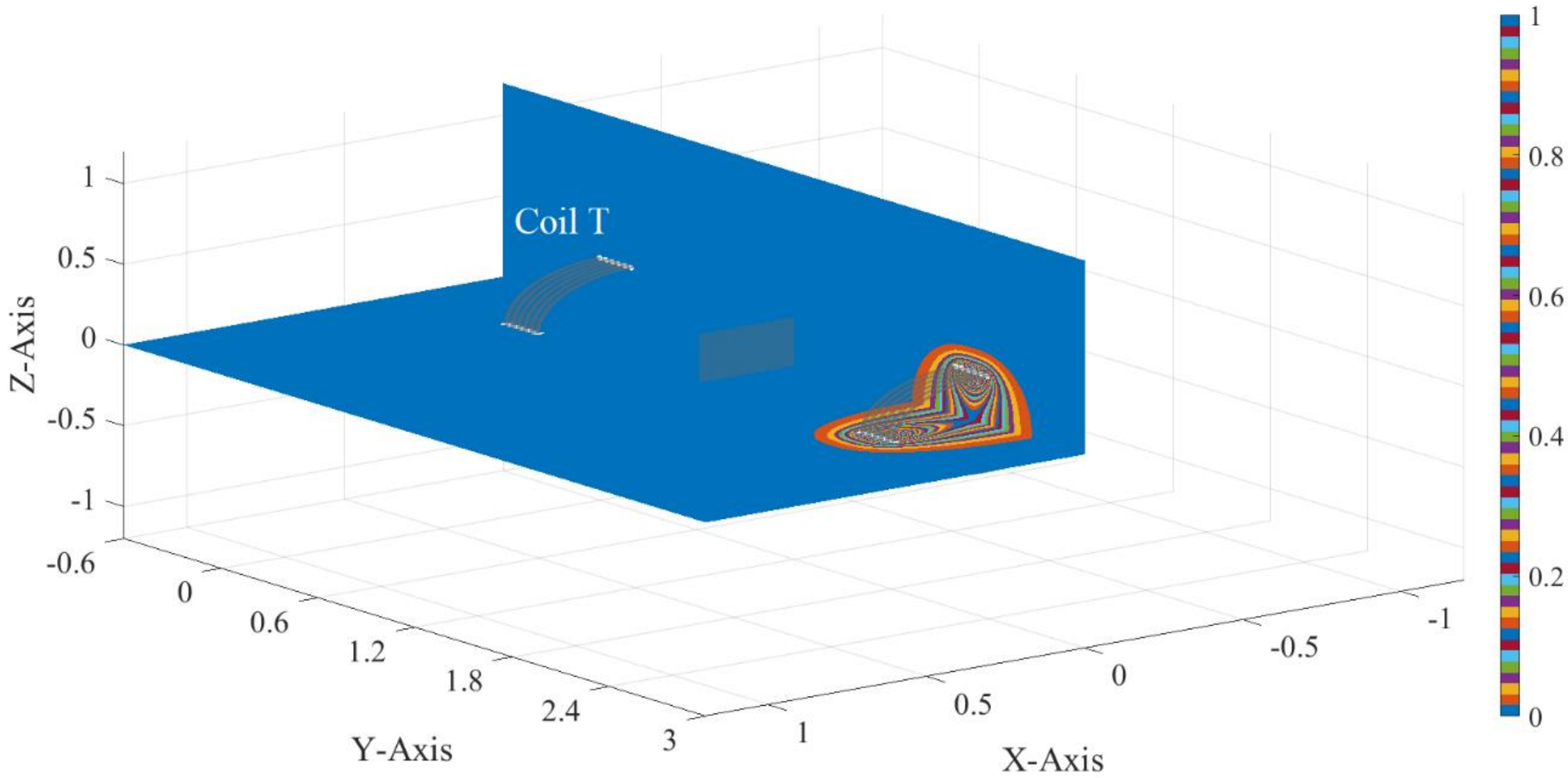

(a)

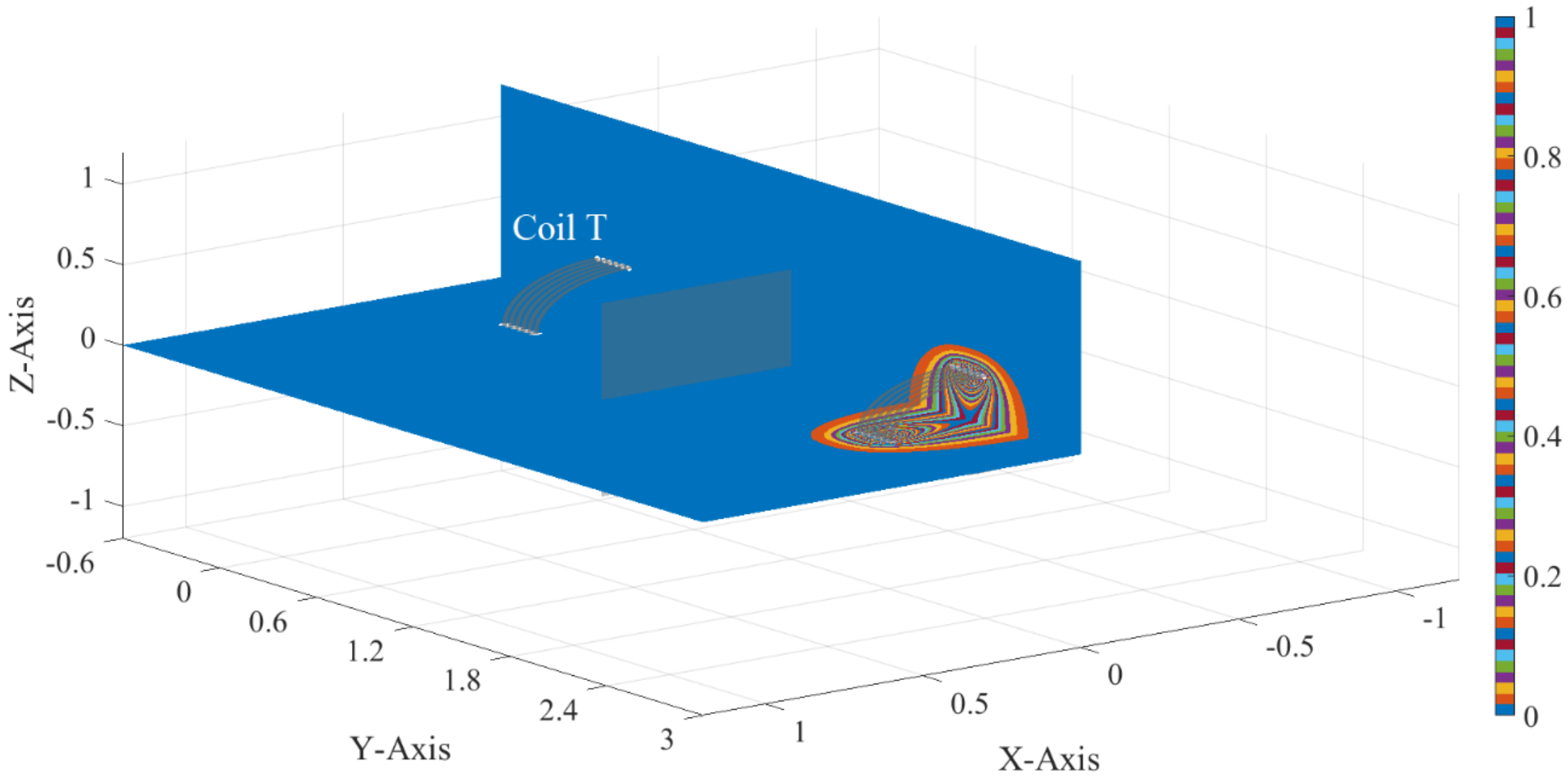

(b)

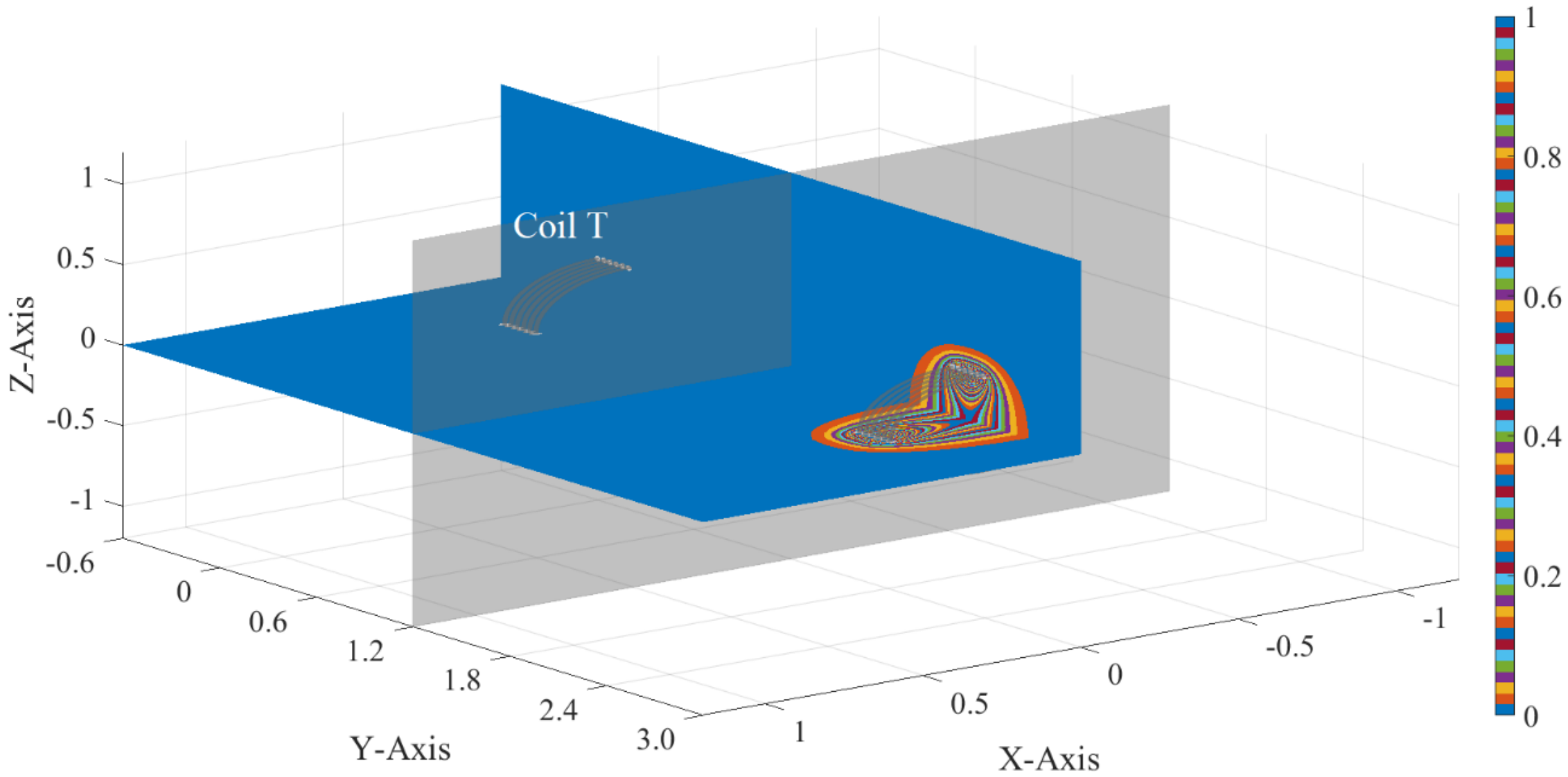

(c)

Figure G-26 Time-average magnetic energy density distributions of the optimally transferring modes with (a) $0.6\,\text{m}\times 0.6\,\text{m}$ plane, (b) $1.2\,\text{m}\times 1.2\,\text{m}$ plane, and (c) $2.4\,\text{m}\times 2.4\,\text{m}$ plane

From the above Figs. G-25 and G-26, it is not difficult to find out the following conclusion.

**C13** For the above-considered metallic planes placed between the coils, they almost don't hinder the power wirelessly transferred from coil T to coil R, but they indeed influence the optimally transferring frequency, and the altered optimally transferring frequency depends on the geometrical size of the plane.

The reason leading to the conclusion is that the external-resonance frequency of scattering system R is different from the external-resonance frequency of R-Plane scattering system.

In fact, the above conclusion can also be generalized to the other kinds of surrounding environments.

## G8 Conclusions

Focusing on the classical two-coil transferring system, which is constituted by a transmitting coil T (driven by locally impressed driver) and a receiving coil R (loaded by perfectly matched load), the WEP governing the WPT process of the transferring system is derived from Maxwell's vector field theory, and doesn't employ any scalar circuit model. The WEP gives the WPT process a very clear physical picture: during the WPT process, the driving power from driver to coil T acts as the source to sustain a steady WPT, and the driving power is finally transformed into two parts — a part is dissipated by coil T and the other part is wirelessly transferred from coil T to coil R. Based on the WEP, a novel concept of TC, which is equal to the ratio of time-average transferred power to time-average driving power, is introduced to quantitatively describe the transferring efficiency of the transferring system. From a relatively mathematical viewpoint, the design process for transferring system is to search for the physically realizable objective modes which can maximize TC, and the objective modes are called optimally transferring modes.

The WEP clearly reveals the fact that: the working mechanisms of *scattering* systems and **transferring** systems are different from each other. The fact clearly exposes the problem that: the conventional CMT for *scattering* systems cannot be directly applied to **transferring** systems. To resolve this problem, this paper, under WEP framework, generalizes the conventional *scattering* CMT to a novel **transferring** CMT. The WEP-based transferring CMT (transferring WEP-CMT) can be employed to search for the optimally transferring modes. Specifically, the transferring WEP-CMT constructs a set of energy-decoupled CMs by orthogonalizing DPO, and the optimally transferring modes can be found in the obtained CM set.

The validity of the transferring WEP-CMT is verified by applying it to the classical two-coil transferring system. After the verification, the WEP-CMT is further applied to the numerical experiments for some typical variants of the classical two-coil transferring system, and then some valuable conclusions are obtained as summarized below.

C1 The larger TC is, the more efficiently "driver drives coil T" and "power is transferred from coil T to coil R" and "coil R converts the transferred power into its self-power".

C2 When TC is maximized, coils T and R are not only self-resonant but also co-resonant. The co-resonant mode works at the optimally transferring state.

C3 Co-resonance is usually called magnetic resonance. The so-called magnetic resonance is a special case of classical electric-magnetic resonance.

C4 Ci-resonance is different from co-resonance. The ci-resonant mode cannot guarantee an efficient WPT from coil T to coil R.

C5 When coils T and R are the same, the **co-resonance** frequency of two-coil **transferring** system equals to the *external-resonance* frequency of one-coil *scattering* system.

C6 When coils T and R are the same, the lower-order co-resonant mode has a larger TC than the higher-order co-resonant mode.

C7 When coils T and R are different, and only one coil is resonant, the system usually cannot efficiently transfer power from coil T to coil R.

C8 When different coil T and coil R work at a same resonance frequency, the system can transfer power wirelessly even if the resonance orders of the coils are different.

C9 For a two-coil transferring system, the same-order co-resonance is usually more desired than the different-order co-resonance in WPT application.

C10 The co-resonance frequency of two same and aligned coils is almost independent of coil distance, and TC is a monotonically decreasing function about the distance.

C11 The co-resonance frequency of two same and misaligned coils is almost independent of the relative angle, and TC is a monotonically decreasing function about the angle.

C12 The optimally transferring frequency of $N$-coil ($N \geq 3$) system depends on the external-resonance frequency of the scattering system constituted by $N$-1 receiving coils.

C13 The optimally transferring frequency of the system in environment depends on the external-resonance frequency of the Environment-Receiving-coil scattering system.

In fact, the above conclusions also effectively establish some connections "between wireless power **transferring** problem and conventional *scattering* problem", "between **co-resonance** and conventional *external-resonance*", and "between so-called **magnetic resonance** and classical *electric-magnetic resonance*".

## G9 Appendices of AP2101-0035

In this section, some detailed derivations and formulations related to this paper are provided.

## G9.1 Rigorous Derivation for Work-Energy Principle (G-1) and More Detailed Decomposition for Driving Power

Based on the discussions in Sec. G4, the tangential components of electric fields $\boldsymbol{\mathcal{E}}_{\text{driv}} + \boldsymbol{\mathcal{E}}_{\text{t}} + \boldsymbol{\mathcal{E}}_{\text{r}}$ and $\boldsymbol{\mathcal{E}}_{\text{t}} + \boldsymbol{\mathcal{E}}_{\text{r}}$ are zero on $S_{\text{t}}$ and $S_{\text{r}}$ respectively, so

$$\begin{aligned}\overbrace{\langle \boldsymbol{J}_{\text{t}}, \boldsymbol{\mathcal{E}}_{\text{driv}} \rangle_{S_{\text{t}}}}^{\mathcal{P}_{\text{driv}}} &= \langle \boldsymbol{J}_{\text{t}}, -\boldsymbol{\mathcal{E}}_{\text{t}} - \boldsymbol{\mathcal{E}}_{\text{r}} \rangle_{S_{\text{t}}} + \overbrace{\langle \boldsymbol{J}_{\text{r}}, -\boldsymbol{\mathcal{E}}_{\text{t}} - \boldsymbol{\mathcal{E}}_{\text{r}} \rangle_{S_{\text{r}}}}^{0} \\ &= \underbrace{\langle \boldsymbol{J}_{\text{t}}, -\boldsymbol{\mathcal{E}}_{\text{t}} \rangle_{S_{\text{t}}}}_{\mathcal{P}_{\text{sel}}^{\text{tt}}} + \underbrace{\underbrace{\overbrace{\langle \boldsymbol{J}_{\text{t}}, -\boldsymbol{\mathcal{E}}_{\text{r}} \rangle_{S_{\text{t}}}}^{\mathcal{P}_{\text{tra}}} + \overbrace{\langle \boldsymbol{J}_{\text{r}}, -\boldsymbol{\mathcal{E}}_{\text{t}} \rangle_{S_{\text{r}}} + \langle \boldsymbol{J}_{\text{r}}, -\boldsymbol{\mathcal{E}}_{\text{r}} \rangle_{S_{\text{r}}}}^{0}}_{\mathcal{P}_{\text{mut}}^{\text{tr}} \;\; \cdots \;\; \mathcal{P}_{\text{sel}}^{\text{rr}}}}_{\mathcal{P}_{\text{tra}}}\end{aligned} \tag{G-11}$$

The physical meanings of $\mathcal{P}_{\text{sel}}^{\text{tt}}$ and $\mathcal{P}_{\text{tra}}$ had been carefully discussed in Sec. G4. The $\mathcal{P}_{\text{mut}}^{\text{tr}}$ and $\mathcal{P}_{\text{sel}}^{\text{rr}}$ can be similarly discussed, and they are respectively called mutual-power between coil T and coil R and self-power of coil R.

## G9.2 Deriving Transformation (G-6) from Electric Field Integral Equation (G-5)

Expanding the currents in (G-5) in terms of some proper basis functions and testing (G-5) with Galerkin's method, (G-5) is discretized as follows:

$$\mathbb{Z}_{\text{rt}} \cdot \mathbb{J}_{\text{t}} + \mathbb{Z}_{\text{rr}} \cdot \mathbb{J}_{\text{r}} = 0 \tag{G-12}$$

The elements of $\mathbb{Z}_{\text{rt}}$ are $[\mathbb{Z}_{\text{rt}}]_{\xi\zeta} = < \boldsymbol{b}_{\text{r};\xi}, -j\omega\mu_0 \mathcal{L}_0(\boldsymbol{b}_{\text{t};\zeta}) >_{S_{\text{r}}}$, and the elements of $\mathbb{Z}_{\text{rr}}$ are $[\mathbb{Z}_{\text{rr}}]_{\xi\zeta} = < \boldsymbol{b}_{\text{r};\xi}, -j\omega\mu_0 \mathcal{L}_0(\boldsymbol{b}_{\text{r};\zeta}) >_{S_{\text{r}}}$. By solving (G-12), the transformation $\mathbb{J}_{\text{r}} = \mathbb{T} \cdot \mathbb{J}_{\text{t}}$ given in (G-6) is immediately obtained, where $\mathbb{T} = -\mathbb{Z}_{\text{rr}}^{-1} \cdot \mathbb{Z}_{\text{rt}}$.

## G9.3 Deriving Matrix Quadratic Form (G-7) from Driving Power Operator (G-4)

By expanding the currents in (G-4), the (G-4) is discretized as follows:

$$P_{\text{driv}} = \begin{bmatrix} \mathbb{J}_{\text{t}} \\ \mathbb{J}_{\text{r}} \end{bmatrix}^{\dagger} \cdot \begin{bmatrix} \mathbb{P}_{\text{tt}} & \mathbb{P}_{\text{tr}} \\ \mathbb{O}_{\text{rt}} & \mathbb{O}_{\text{rr}} \end{bmatrix} \cdot \begin{bmatrix} \mathbb{J}_{\text{t}} \\ \mathbb{J}_{\text{r}} \end{bmatrix} \tag{G-13}$$

where the sub-matrices $\mathbb{O}_{\text{rt}}$ and $\mathbb{O}_{\text{rr}}$ are zero matrices with proper row number and column number, and the elements of sub-matrix $\mathbb{P}_{\text{tt}}$ are $(1/2) < \boldsymbol{b}_{\text{t};\xi}, j\omega\mu_0 \mathcal{L}_0(\boldsymbol{b}_{\text{t};\zeta}) >_{S_{\text{t}}}$, and the elements of sub-matrix $\mathbb{P}_{\text{tr}}$ are $(1/2) < \boldsymbol{b}_{\text{t};\xi}, j\omega\mu_0 \mathcal{L}_0(\boldsymbol{b}_{\text{r};\zeta}) >_{S_{\text{t}}}$.

Substituting transformation (G-6) into (G-13), matrix quadratic form (G-7) is immediately obtained, in which

$$\mathbb{P}_{\text{driv}} = \begin{bmatrix} \mathbb{I}_{\text{t}} \\ \mathbb{T} \end{bmatrix}^{\dagger} \cdot \begin{bmatrix} \mathbb{P}_{\text{tt}} & \mathbb{P}_{\text{tr}} \\ \mathbb{O}_{\text{rt}} & \mathbb{O}_{\text{rr}} \end{bmatrix} \cdot \begin{bmatrix} \mathbb{I}_{\text{t}} \\ \mathbb{T} \end{bmatrix} \tag{G-14}$$

where $\mathbb{I}_{\text{t}}$ is the unit matrix with the same dimension as the row number of $\mathbb{J}_{\text{t}}$.

In addition, the matrix $\mathbb{P}_{\text{tra}}^{+}$ used in (G-9) is as that $\mathbb{P}_{\text{tra}}^{+} = (\mathbb{P}_{\text{tra}} + \mathbb{P}_{\text{tra}}^{\dagger})/2$, in which

$$\mathbb{P}_{\text{tra}} = \begin{bmatrix} \mathbb{I}_{\text{t}} \\ \mathbb{T} \end{bmatrix}^{\dagger} \cdot \begin{bmatrix} \mathbb{O}_{\text{tt}} & \mathbb{P}_{\text{tr}} \\ \mathbb{O}_{\text{rt}} & \mathbb{O}_{\text{rr}} \end{bmatrix} \cdot \begin{bmatrix} \mathbb{I}_{\text{t}} \\ \mathbb{T} \end{bmatrix} \tag{G-15}$$

where sub-matrices $\mathbb{P}_{\text{tr}}$, $\mathbb{O}_{\text{rt}}$, and $\mathbb{O}_{\text{rr}}$ are the same as the ones used in (G-13) and (G-14), and submatrix $\mathbb{O}_{\text{tt}}$ is a zero matrix with proper row number and column number, and the derivation for (G-15) is similar to the derivation for (G-14).

## G9.4 Resistance and Reactance

For the convenience of Sec. G6 and using the convention of [308], we define resistances $R_{\text{driv}}$, $R_{\text{sel}}^{\text{tt}}$, $R_{\text{mut}}^{\text{tr}}$, $R_{\text{sel}}^{\text{rr}}$, and $R_{\text{tra}}$ as the powers $\text{Re}\{P_{\text{driv}}\}$, $\text{Re}\{P_{\text{sel}}^{\text{tt}}\}$, $\text{Re}\{P_{\text{mut}}^{\text{tr}}\}$, $\text{Re}\{P_{\text{sel}}^{\text{rr}}\}$, and $\text{Re}\{P_{\text{tra}}\}$ divided by $(1/2) < \boldsymbol{J}_{\text{t}}, \boldsymbol{J}_{\text{t}} >_{S_{\text{t}}}$, and we define reactances $X_{\text{driv}}$, $X_{\text{sel}}^{\text{tt}}$, $X_{\text{mut}}^{\text{tr}}$, $X_{\text{sel}}^{\text{rr}}$, and $X_{\text{tra}}$ as the powers $\text{Im}\{P_{\text{driv}}\}$, $\text{Im}\{P_{\text{sel}}^{\text{tt}}\}$, $\text{Im}\{P_{\text{mut}}^{\text{tr}}\}$, $\text{Im}\{P_{\text{sel}}^{\text{rr}}\}$, and $\text{Im}\{P_{\text{tra}}\}$ divided by $(1/2) < \boldsymbol{J}_{\text{t}}, \boldsymbol{J}_{\text{t}} >_{S_{\text{t}}}$.

# Appendix H Work-Energy Principle Based Characteristic Mode Analysis for Yagi-Uda Arrays

This App. H had been written as a journal paper by our research group (Ren-Zun Lian, Ming-Yao Xia and Xing-Yue Guo), and the manuscript [AP2101-0036] entitled “Work-Energy Principe Based Characteristic Mode Analysis for Yagi-Uda Arrays”[350] was submitted to IEEE-TAP (*IEEE Transactions on Antennas and Propagation*) on 06-Jan-2021.

## H1 Abstract

The difference between the working mechanism of scattering systems and the working mechanism of Yagi-Uda arrays is exposed. The similarity between the working mechanism of metallic wireless power transfer systems (simply called transferring systems) and the working mechanism of metallic Yagi-Uda arrays is revealed. The work-energy principle (WEP) based characteristic mode theory for metallic transferring systems is used to do the modal analysis for metallic Yagi-Uda arrays for the first time. The WEP-based characteristic mode analysis (CMA) for metallic Yagi-Uda arrays is further generalized to material Yagi-Uda arrays and metal-material composite Yagi-Uda arrays, and the WEP-based CMA (WEP-CMA) formulations for material and composite Yagi-Uda arrays are established for the first time. The validity of the WEP-CMA for various Yagi-Uda arrays is verified by comparing the WEP-CMA-based numerical results with some published simulation and measurement data.

## H2 Index Terms

Characteristic mode (CM), driving power operator (DPO), work-energy principle (WEP), Yagi-Uda array.

## H3 Introduction

Yagi-Uda antenna array was first studied by Uda and Yagi in the early 1920s, and publicly reported in the middle 1920s[309,310]. In 1984, the *Proceedings of the IEEE* reprinted several classical articles for celebrating the centennial year of IEEE (1884-1984), and Yagi's article [310] became the only reprinted one in the realm of electromagnetic (EM) antenna. This fact clearly illustrates the great significance of Yagi-Uda array in antennas and propagation society. Some histories about Yagi-Uda array can be found in [311~313].

A classical metallic Yagi-Uda array is shown in Fig. H-1, and it is constituted by a row of discrete metallic linear elements, one of which is driven by an external source while the others act as parasitic radiators whose currents are induced by near-field mutual coupling[74,75,77,78]. The linear metallic Yagi-Uda array is a typical discrete-element travelling-wave-type end-fire antenna, which usually works at HF (3-30 MHz), VHF (30-300 MHz) and UHF (300-3000 MHz) etc. bands[74,75,77,78].

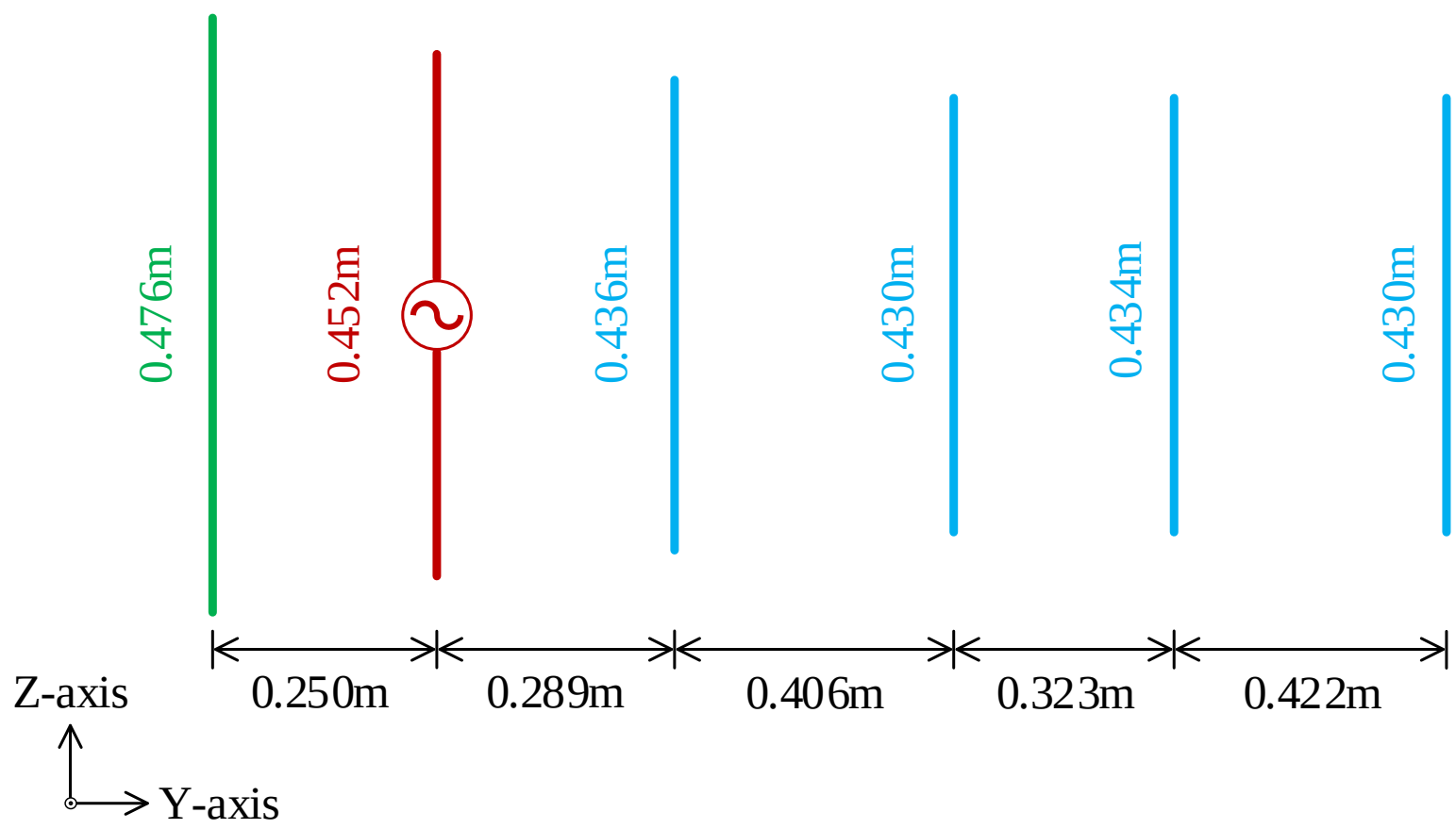

Figure H-1 Geometry and size of a typical 6-element linear metallic Yagi-Uda array reported in [344]

Besides the most classical linear metallic Yagi-Uda array shown in Fig. H-1, there also exist many different variants (sometimes called quasi Yagi-Uda arrays). The quasi Yagi-Uda arrays have been widely applied in the applications of frequency-modulated broadcast[314], domestic/mobile television signal transmission[315~317], point-to-point communication[318~322], long-distance communication[117,323~326], mobile communication[327~331], wireless local area network[332~336] and radio frequency identification[337~340] etc. due to their typical features of high radiation efficiency, high gain and directivity / narrow beamwidth, high front-to-back ratio, low level of minor lobes, good cross-polar discrimination, reasonable bandwidth, low cost and ease of fabrication etc.[74,75,77,78].

According to the difference of their constituent components, the various Yagi-Uda arrays can be categorized into three classes — metallic Yagi-Uda arrays[341~344], material Yagi-Uda arrays[345,346], and metal-material composite Yagi-Uda arrays[117,315~340]. A typical 6-element metallic Yagi-Uda array discussed in [344] is shown in Fig. H-1, and it has a dominant resonant mode working at 300 MHz (calculated from the formulation proposed in [344]), and the resonant mode has the far-field radiation pattern shown in Fig. H-2, which is end-fire.

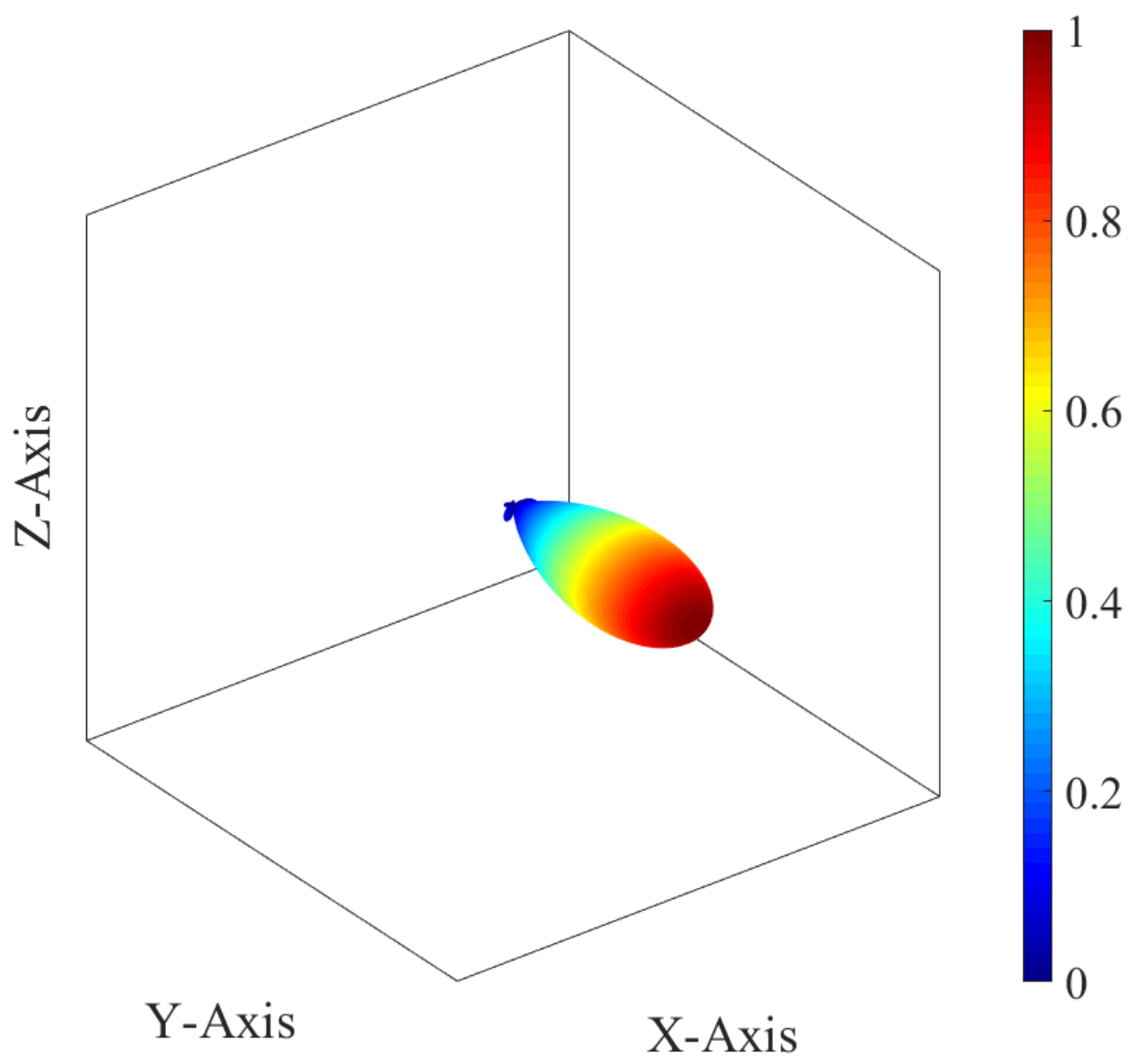

Figure H-2 Radiation pattern of the dominant resonant mode of the Yagi-Uda array shown in Fig. H-1

The analysis and design for resonant modes are the important topics in the realm of antenna engineering, and there have been some classical modal analysis and design theories, such as cavity model theory[171,175], dielectric waveguide model theory[130,131], eigen-mode theory[347,348], and characteristic mode theory (CMT)[6~9,13,16,349] etc. Among the various theories, the CMT has been attached great importance in the realm of antenna engineering[16] recently, because the CMT not only has a very wide applicable range but also is very easy for numerical realization. But unfortunately, the conventional CMT cannot be directly applied to doing the modal analysis for Yagi-Uda arrays as exhibited below. By directly applying the conventional CMT[8,9,13,16] to the Yagi-Uda array shown in Fig. H-1, the modal significances (MSs) associated to the obtained characteristic modes (CMs) are shown in Fig. H-3, and the radiation patterns of the resonant CMs are illustrated in Fig. H-4. Evidently, both the resonance frequencies and radiation patterns are incorrect.

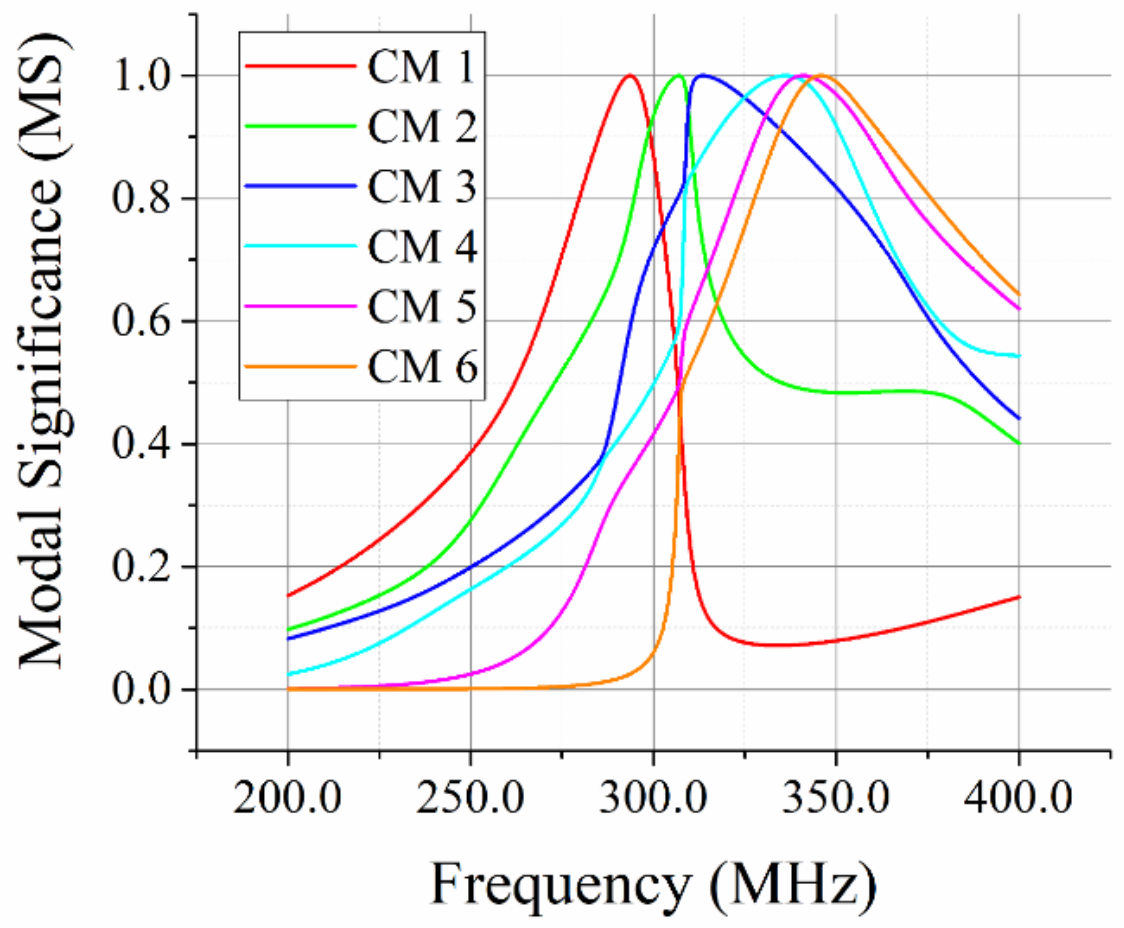

Figure H-3 MSs associated to the first six lower-order CMs calculated from the conventional CMT established by Harrington *et al.* in [8] and [9]

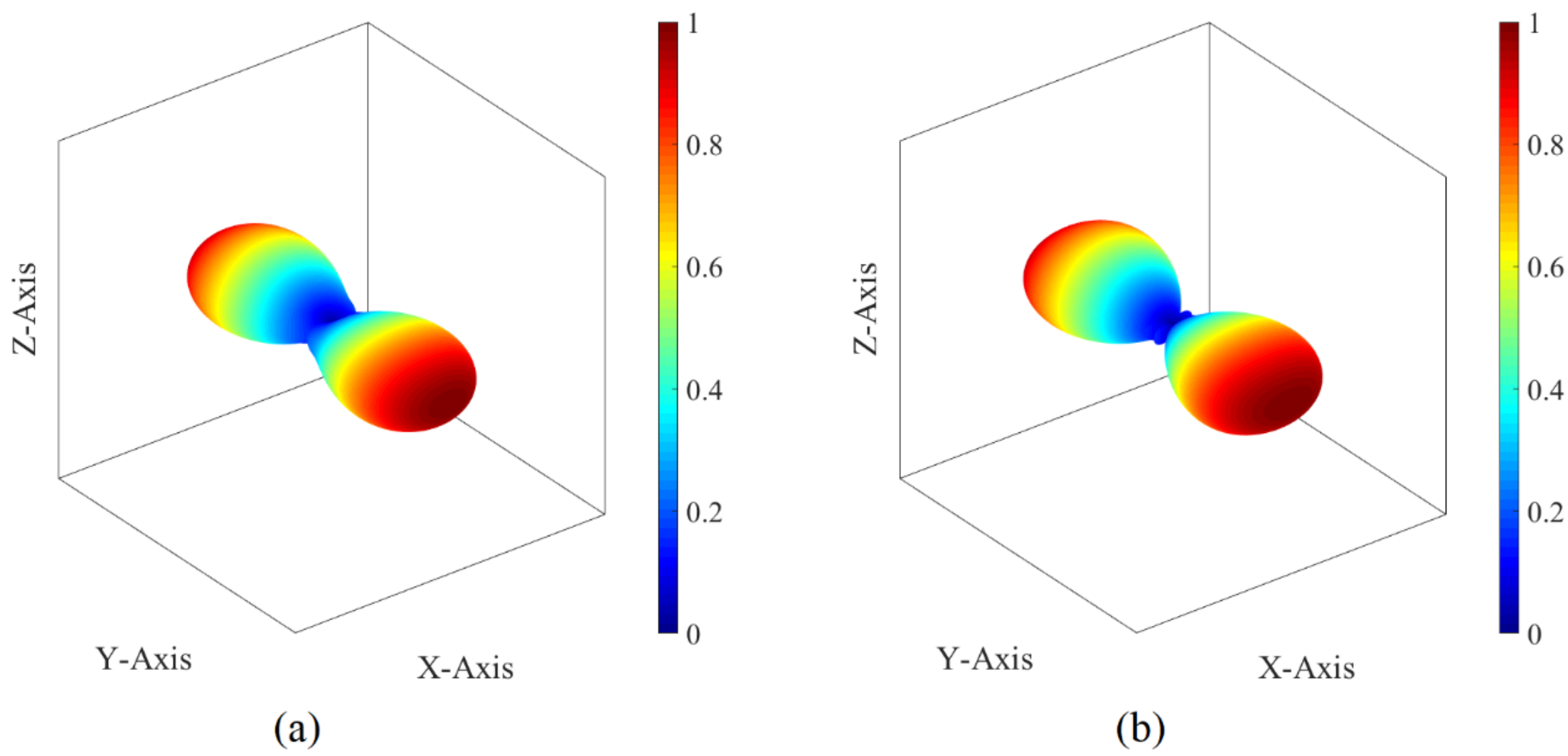

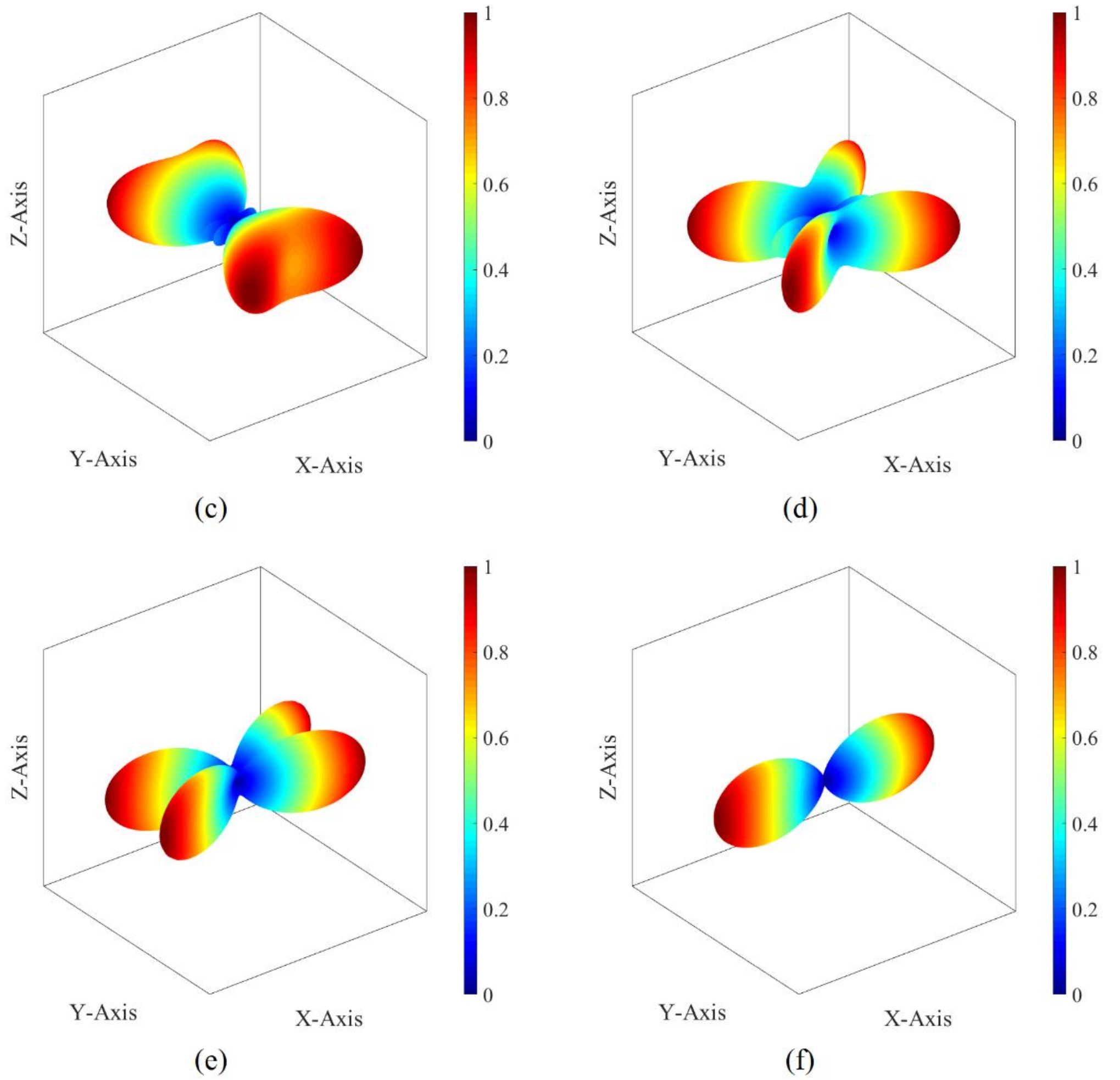

Figure H-4 Radiation patterns of the resonant (a) CM 1 working at 293.6 MHz, (b) CM 2 working at 307.0 MHz, (c) CM3 working at 313.6 MHz, (d) CM4 working at 336.2 MHz, (e) CM5 working at 340.9 MHz, and (f) CM6 at 345.9 MHz

The reasons leading to the above incorrect results mainly originate from the following two.

R1. The working mechanisms of scattering system and antenna system are different from each other. Scattering system is an object which is under the illumination of an externally incident field and generates a secondary scattered field. Antenna system is "*a device used for transmitting electromagnetic signals or power*"[37].

R2. The conventional CMT is a modal analysis method for scattering systems[13,349] rather than for antenna systems.

Recently, under working-energy principle (WEP) framework [349] established an alternative CMT which is for metallic wireless power transfer systems (simply called transferring systems) instead of for scattering systems. In this paper, it will be exhibited that the metallic Yagi-Uda arrays have a very similar working mechanism to the metallic transferring systems discussed in [349]. Then, this paper successfully applies the WEP-based CMT to doing the modal analysis for metallic Yagi-Uda arrays. In addition, this paper will further generalize the WEP-based characteristic mode analysis (CMA) method from metallic Yagi-Uda arrays to material Yagi-Uda arrays and then to metal-material composite Yagi-Uda arrays.

This paper is organized as follows: Sec. H4 discusses the WEP-based CMA (WEP-CMA) for metallic Yagi-Uda arrays; Sec. H5 generalizes the WEP-CMA to material Yagi-Uda arrays; Sec. H6 further generalizes the WEP-CMA to composite Yagi-Uda arrays; Sec. H7 concludes this paper; some detailed formulations related to this paper are provided in Sec. H8. The symbolic system of this paper is similar to [349], and will not be explained here. In addition, to distinguish surface electric current and volume electric current, the former is denoted as upper case letter $\boldsymbol{J}$, and the latter is denoted as lower case letter $\boldsymbol{j}$.

## H4 WEP-CMA for Metallic Yagi-Uda Arrays

Taking the one shown in Fig. H-1 as a typical example, this section proposes a WEP-CMA for metallic Yagi-Uda arrays.

### H4.1 Working Mechanism of Metallic Yagi-Uda Arrays

Traditionally, all the elements in a Yagi-Uda array are divided into three groups — feeding element, reflecting element, and directing elements[74,75,77,78]. In fact, all the elements can also be alternatively divided into two groups — active element and passive/parasitic elements[323], where the former is just the feeding element and the latter is the union of reflecting element and directing elements as shown in Fig. H-5.

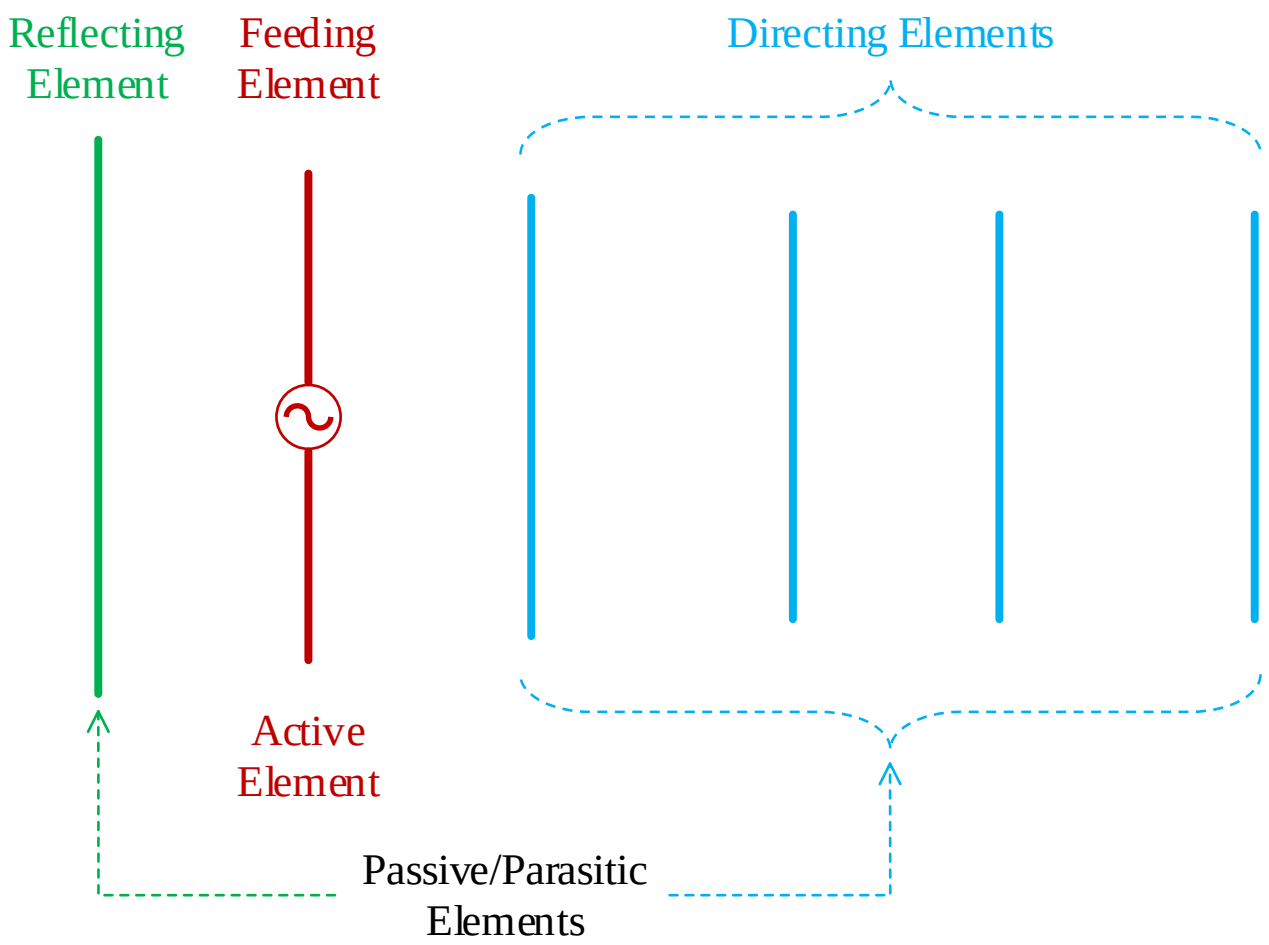

Figure H-5 Re-grouping for the elements used to constitute the Yagi-Uda array in Fig. H-1

When a localized driving field acts on the active element, an electric current will be induced on the active element. The field generated by the induced electric current on active element acts on the passive elements, and then induces some electric currents on the passive elements. In addition, the fields generated by the induced electric currents on passive elements also react on the active element.

For the time-harmonic EM problem, the above-mentioned action and reaction will reach a dynamic equilibrium finally. At the state of dynamic equilibrium, the driving field is denoted as $\boldsymbol{F}_{\text{driv}}$, and the induced electric currents on the active and passive elements are denoted as $\boldsymbol{J}_{\text{a}}$ and $\boldsymbol{J}_{\text{p}}$ respectively, and the fields generated by $\boldsymbol{J}_{\text{a}}$ and $\boldsymbol{J}_{\text{p}}$ are denoted as $\boldsymbol{F}_{\text{a}}$ and $\boldsymbol{F}_{\text{p}}$ respectively.

Obviously, the above-mentioned working mechanism of metallic Yagi-Uda arrays is almost identical to the working mechanism of the metallic transferring systems discussed in [349]. In the following Secs. H4.2 and H4.3, we will employ the WEP-based CMT established in [349] to do the modal analysis for metallic Yagi-Uda arrays.

## H4.2 Mathematical Formulations Related to the WEP-CMA for Metallic Yagi-Uda Arrays

The driving power $P_{\text{driv}} = (1/2) < \boldsymbol{J}_{\text{a}}, \boldsymbol{E}_{\text{driv}} >_{S_{\text{a}}}$ used to sustain a steady working of the metallic Yagi-Uda array has the following operator expression

$$P_{\text{driv}} = (1/2)\left\langle \boldsymbol{J}_{\text{a}} , j\omega\mu_0\mathcal{L}_0\left(\boldsymbol{J}_{\text{a}} + \boldsymbol{J}_{\text{p}}\right)\right\rangle_{S_{\text{a}}} \tag{H-1}$$

in which integral domain $S_{\text{a}}$ is the boundary surface of the active element, and integral operator $\mathcal{L}_0$ is defined as that $\mathcal{L}_0(\boldsymbol{X}) = [1 + (1/k_0^2)\nabla\nabla\cdot]\int_{\Omega} G_0(\boldsymbol{r},\boldsymbol{r}')\boldsymbol{X}(\boldsymbol{r}')d\Omega'$ where the scalar Green's function is as that $G_0(\boldsymbol{r},\boldsymbol{r}') = e^{-jk_0|\boldsymbol{r}-\boldsymbol{r}'|} / 4\pi |\boldsymbol{r}-\boldsymbol{r}'|$. In addition, the above integral form (H-1) of driving power operator (DPO) can be easily discretized into the following matrix form

$$P_{\text{driv}} = \mathrm{J}_{\text{a}}^{\dagger} \cdot \begin{bmatrix} \mathrm{P}_{\text{aa}} & \mathrm{P}_{\text{ap}} \end{bmatrix} \cdot \begin{bmatrix} \mathrm{J}_{\text{a}} \\ \mathrm{J}_{\text{p}} \end{bmatrix} \tag{H-2}$$

where $\mathrm{J}_{\text{a}}$ and $\mathrm{J}_{\text{p}}$ are the basis function expansion vectors of $\boldsymbol{J}_{\text{a}}$ and $\boldsymbol{J}_{\text{p}}$ respectively, and the formulations for calculating the sub-matrices are given in Sec. H8.1.

In fact, the $\mathrm{J}_{\text{a}}$ and $\mathrm{J}_{\text{p}}$ are not independent, and they satisfy the following transformation relation

$$\mathrm{J}_{\text{p}} = \mathrm{T} \cdot \mathrm{J}_{\text{a}} \tag{H-3}$$

which originates from the homogeneous tangential electric field boundary condition on $S_{\rm p}$. The formulations for calculating $\mathbb{T}$ are given in Sec. H8.1. Substituting (H-3) into (H-2), it is easy to derive the following

$$P_{\rm driv} = \mathbb{J}_{\rm a}^{\dagger} \cdot \mathbb{P}_{\rm driv} \cdot \mathbb{J}_{\rm a} \tag{H-4}$$

with independent variable $\mathbb{J}_{\rm a}$ only, where the formulations for calculating $\mathbb{P}_{\rm driv}$ are given in Sec. H8.1.

The energy-decoupled CMs of the metallic Yagi-Uda array can be derived from solving the following characteristic equation

$$\mathbb{P}_{\rm driv}^{-} \cdot \mathbb{J}_{\rm a} = \lambda\, \mathbb{P}_{\rm driv}^{+} \cdot \mathbb{J}_{\rm a} \tag{H-5}$$

where $\mathbb{P}_{\rm driv}^{+} = [\mathbb{P}_{\rm driv} + \mathbb{P}_{\rm driv}^{\dagger}]/2$ and $\mathbb{P}_{\rm driv}^{-} = [\mathbb{P}_{\rm driv} - \mathbb{P}_{\rm driv}^{\dagger}]/2j$. The above-obtained CMs satisfy frequency-domain power-decoupling relation $(1/2) < \boldsymbol{J}_{{\rm a};\xi}, \boldsymbol{E}_{{\rm driv};\zeta} >_{S_{\rm a}} = (1 + j\lambda)\delta_{\xi\zeta}$ where $\delta_{\xi\zeta}$ is Kronecker's delta symbol, and then satisfy the following energy-decoupling relation, i.e., time-average power-decoupling relation,

$$\left(1/T\right)\int_{t_0}^{t_0+T} \left\langle \boldsymbol{\mathcal{J}}_{{\rm a};\xi}, \boldsymbol{\mathcal{E}}_{{\rm driv};\zeta} \right\rangle dt = \delta_{\xi\zeta} \tag{H-6}$$

where the real parts of all modal complex powers have been normalized to 1 just like [9] did. Thus, there exists the following Parseval's identity

$$\left(1/T\right)\int_{t_0}^{t_0+T} \left\langle \boldsymbol{\mathcal{J}}_{\rm a}, \boldsymbol{\mathcal{E}}_{\rm driv} \right\rangle dt = \sum_{\xi} \left| c_{\xi} \right|^2 \tag{H-7}$$

in which $c_{\xi}$ is the CM-based modal expansion coefficient, and $c_{\xi} = (1/2) < \boldsymbol{J}_{{\rm a};\xi}, \boldsymbol{E}_{\rm driv} >_{S_{\rm a}} / (1 + j\lambda)$ where $\boldsymbol{E}_{\rm driv}$ is the frequency-domain version of a previously known time-domain driving field $\boldsymbol{\mathcal{E}}_{\rm driv}$. The above (H-6) and (H-7) have a very clear physical meaning: the CMs constructed above don't have net energy exchange in any integral period.

## H4.3 Numerical Verifications

In this sub-section, the WEP-CMA proposed above is applied to a specific metallic Yagi-Uda array to verify its validity. The geometry and size of the Yagi-Uda array are shown in Fig. H-1. The MSs associated to the first four lower-order CMs calculated from the WEP-CMA are shown in Fig. H-6.

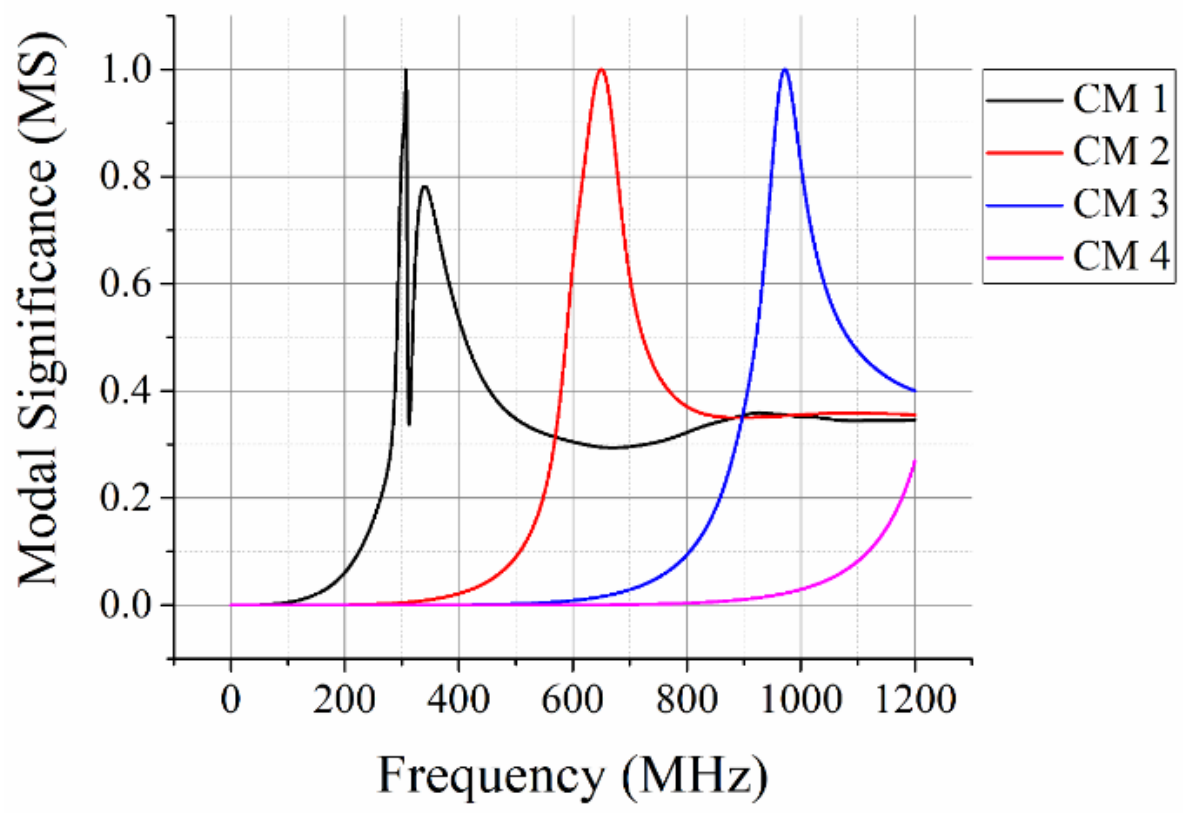

Figure H-6 MSs associated to the first four lower-order CMs calculated from the WEP-CMA proposed in this section

From Fig. H-6, it is easy to find out that the CM 1 is resonant at 307.3 MHz, and the resonance frequency is consistent with the one calculated from the formulation given in [344] except a 2% numerical error. The radiation pattern of the resonant CM 1 is shown in Fig. H-2. The distributions of the electric and magnetic fields of the resonant CM 1 are shown in Fig. H-7. Evidently, Fig. H-2 and Fig. H-7 satisfy the well-known features of linear metallic Yagi-Uda arrays — the radiated EM power propagates along the direction from reflecting element to directing elements[74,75,77,78]. In addition, the time-domain dynamic figures of the propagating electric and magnetic fields are also uploaded to IEEE manuscript system together with this paper.

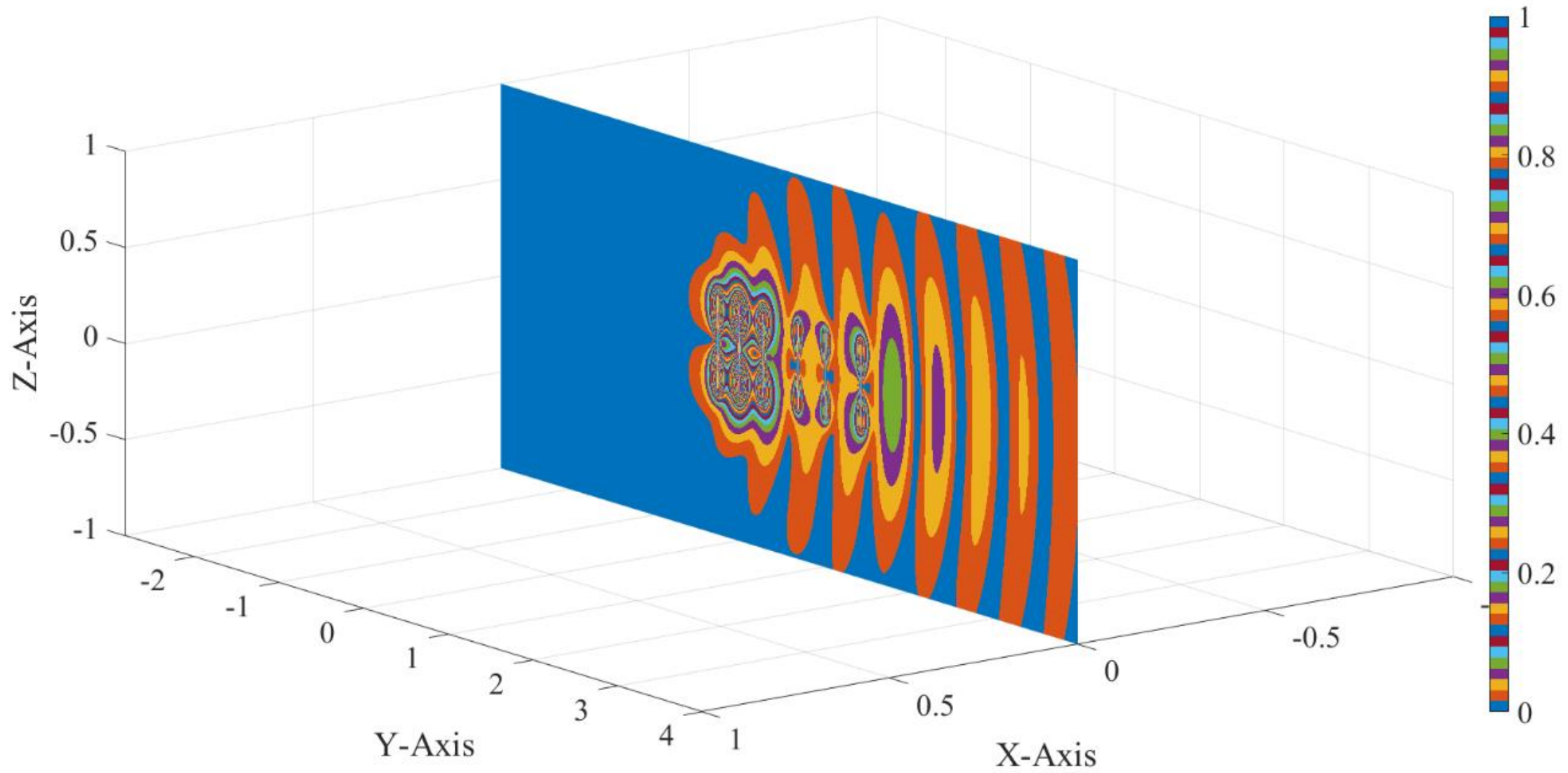

(a)

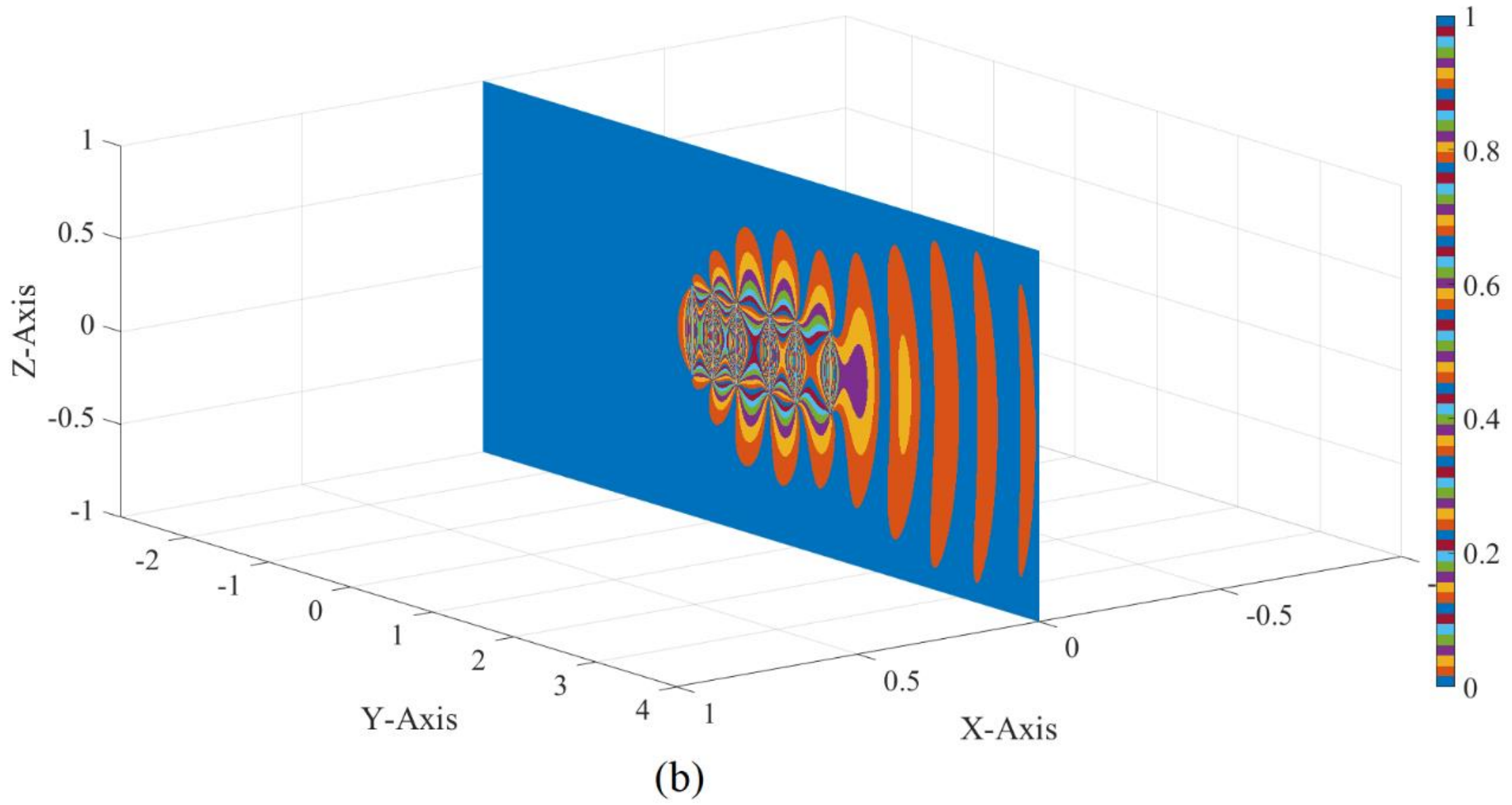

Figure H-7 Distributions of the (a) electric field and (b) magnetic field of the resonant CM 1 shown in Fig. H-6

In addition, the radiation patterns of the higher-order resonant CM 2 (working at 649.8 MHz) and CM 3 (working at 971.8 MHz) are shown in Fig. H-8. Evidently, both the resonant CM 2 and CM 3 don't work at the desired end-fire state. In fact, this is just the reason why "*higher resonances are available near lengths of* $\lambda$, $3\lambda/2$, *and so forth, but are seldom used*"[75].

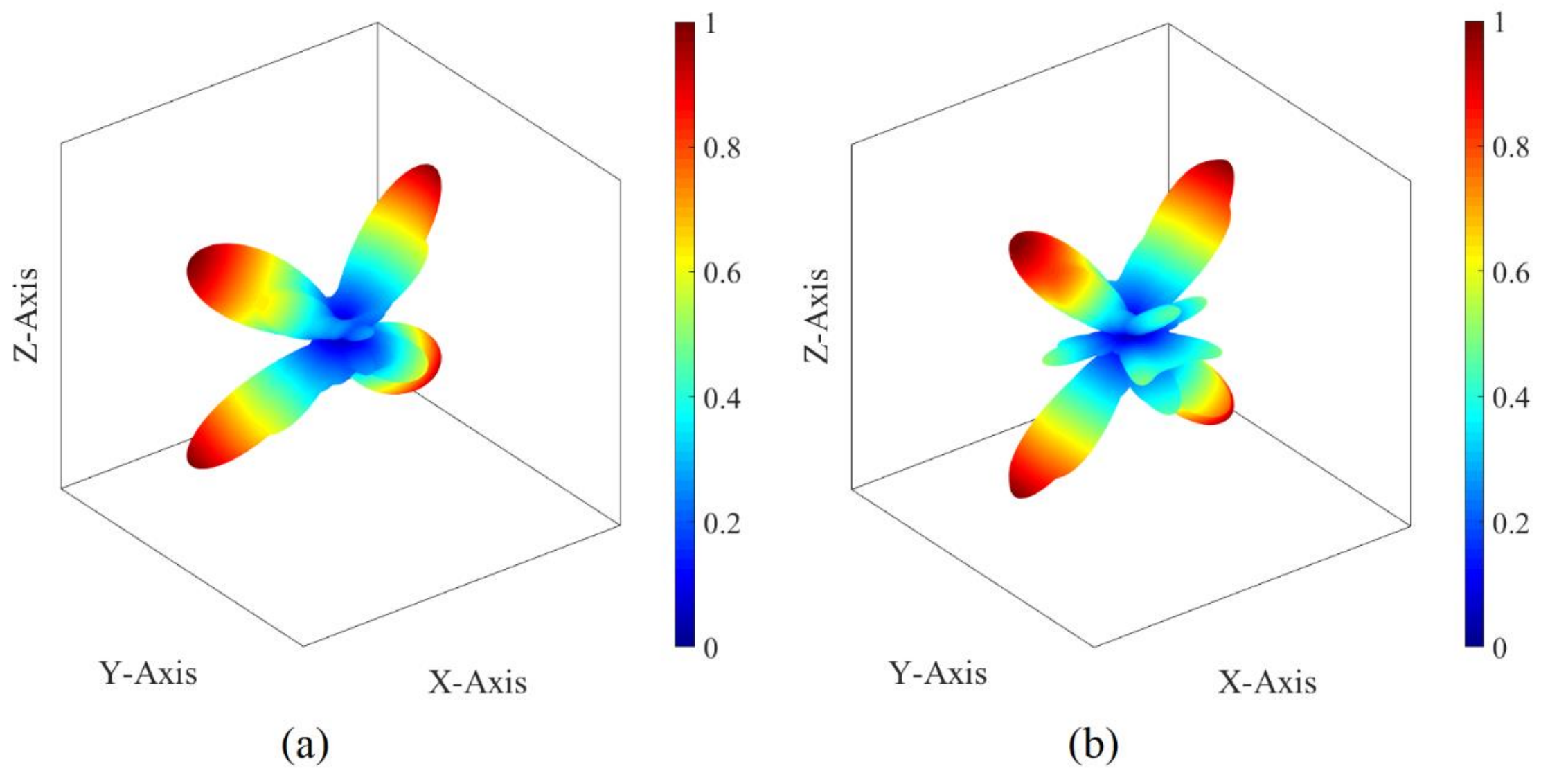

Figure H-8 Radiation patterns of the resonant (a) CM 2 working at 649.8 MHz and (b) CM 3 working at 971.8 MHz

## H5 WEP-CMA for Material Yagi-Uda Arrays

This section is devoted to generalizing the idea of Sec. H4 (which is for metallic Yagi-Uda arrays) to material Yagi-Uda arrays. A typical three-element material Yagi-Uda array reported in [345] is shown in Fig. H-9, and it has an active element $V_{\rm a}$ with boundary surface $S_{\rm a}$ and two passive elements $V_{\rm p1}$ and $V_{\rm p2}$ with boundary surfaces $S_{\rm p1}$ and $S_{\rm p2}$ respectively. For simplifying the following discussions, the elements are restricted to being non-magnetic in this section, and their complex permittivities are denoted as $\boldsymbol{\varepsilon}_{\rm a}^{\rm c}$, $\boldsymbol{\varepsilon}_{\rm p1}^{\rm c}$, and $\boldsymbol{\varepsilon}_{\rm p2}^{\rm c}$. The purely magnetic case and magneto-dielectric case can be similarly discussed.

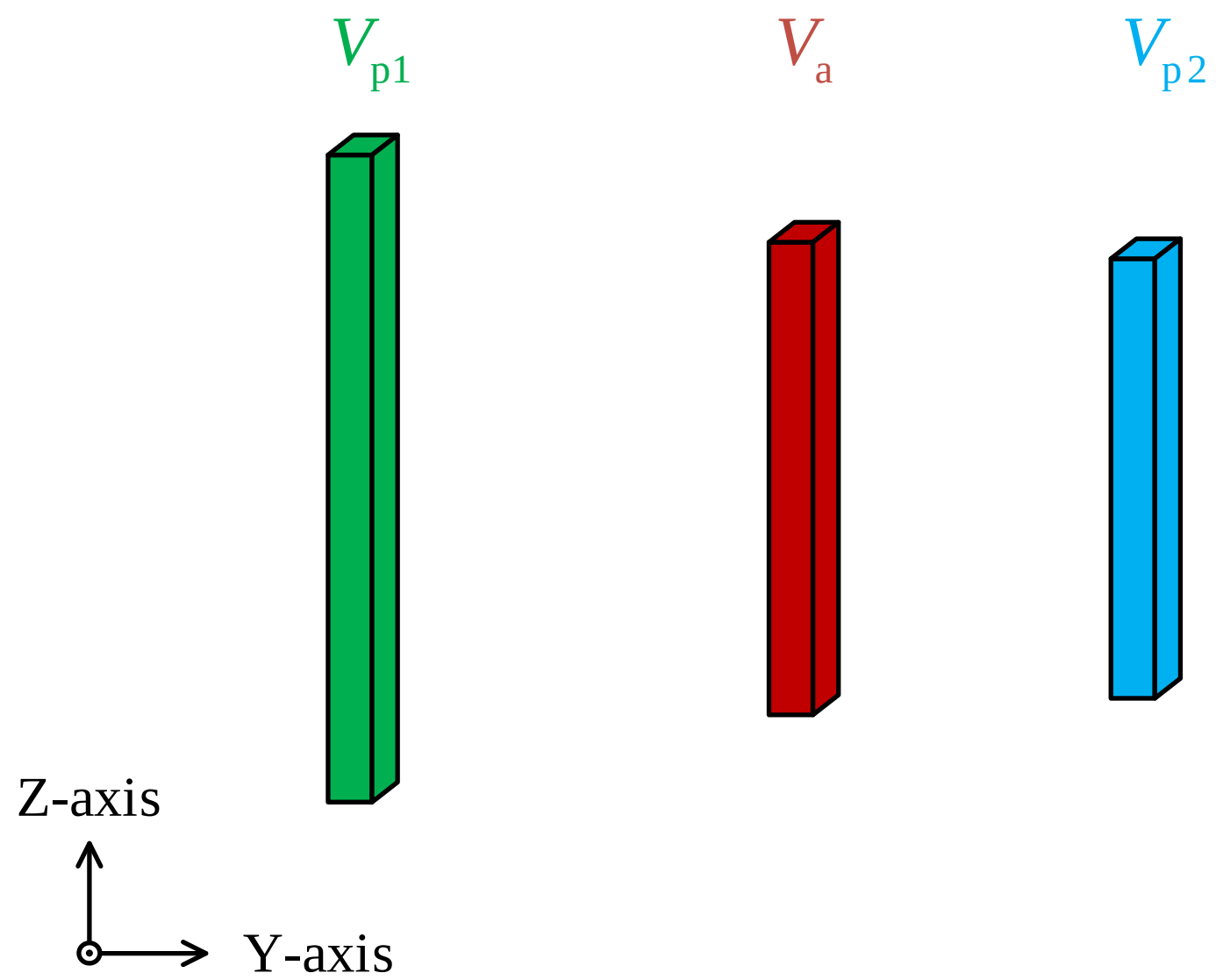

Figure H-9 Geometry of a typical 3-element linear material Yagi-Uda array reported in [345]

### H5.1 Volume Formulation of the WEP-CMA for Material Yagi-Uda Arrays

If the induced volume electric currents distributing on $V_{\rm a}$, $V_{\rm p1}$, and $V_{\rm p2}$ are denoted as $\boldsymbol{j}_{\rm a}$, $\boldsymbol{j}_{\rm p1}$, and $\boldsymbol{j}_{\rm p2}$ respectively, then the corresponding DPO is as follows:

$$P_{\rm driv} = (1/2)\left\langle \boldsymbol{j}_{\rm a}, \left(j\omega\Delta\boldsymbol{\varepsilon}_{\rm a}^{\rm c}\right)^{-1}\cdot\boldsymbol{j}_{\rm a} + j\omega\mu_0\mathcal{L}_0\left(\boldsymbol{j}_{\rm a}+\boldsymbol{j}_{\rm p1}+\boldsymbol{j}_{\rm p2}\right)\right\rangle_{V_{\rm a}} \tag{H-8}$$

where $\Delta\boldsymbol{\varepsilon}_{\rm a}^{\rm c} = \boldsymbol{\varepsilon}_{\rm a}^{\rm c} - \mathbf{I}\varepsilon_0$ and $\mathbf{I}$ is the unit dyad.

Similar to deriving (H-4) from (H-1), the following

$$P_{\rm driv} = \mathrm{j}_{\rm a}^{\dagger}\cdot\mathrm{P}_{\rm driv}\cdot\mathrm{j}_{\rm a} \tag{H-9}$$

with only independent variable $\mathrm{j}_{\rm a}$ can be derived from (H-8), and a detailed derivation process is given in Sec. H8.2. Here, $\mathrm{j}_{\mathrm{a}}$ is the basis function expansion vector of $\boldsymbol{j}_{\mathrm{a}}$.

Employing the $\mathrm{P}_{\mathrm{driv}}$ used in (H-9), the CMs of the material Yagi-Uda array can be calculated like Sec. H4.2.

The scheme provided in this sub-section is based on volume currents, and an alternative surface-current-based scheme will be given in the following sub-section.

## H5.2 Surface Formulation of the WEP-CMA for Material Yagi-Uda Arrays

If the equivalent surface currents distributing on $S_{\mathrm{a}}$, $S_{\mathrm{p1}}$, and $S_{\mathrm{p2}}$ are denoted as $(\boldsymbol{J}_{\mathrm{a}},\boldsymbol{M}_{\mathrm{a}})$, $(\boldsymbol{J}_{\mathrm{p1}},\boldsymbol{M}_{\mathrm{p1}})$, and $(\boldsymbol{J}_{\mathrm{p2}},\boldsymbol{M}_{\mathrm{p2}})$ respectively, then the volume-current version (H-8) of DPO can be alternatively written as the following surface-current version

$$P_{\mathrm{driv}} = -(1/2)\left\langle \boldsymbol{J}_{\mathrm{a}},\mathcal{E}_0\left(\boldsymbol{J}_{\mathrm{a}}+\boldsymbol{J}_{\mathrm{p1}}+\boldsymbol{J}_{\mathrm{p2}},\boldsymbol{M}_{\mathrm{a}}+\boldsymbol{M}_{\mathrm{p1}}+\boldsymbol{M}_{\mathrm{p2}}\right)\right\rangle_{S_{\mathrm{a}}^-} - (1/2)\left\langle \boldsymbol{M}_{\mathrm{a}},\mathcal{H}_0\left(\boldsymbol{J}_{\mathrm{a}}+\boldsymbol{J}_{\mathrm{p1}}+\boldsymbol{J}_{\mathrm{p2}},\boldsymbol{M}_{\mathrm{a}}+\boldsymbol{M}_{\mathrm{p1}}+\boldsymbol{M}_{\mathrm{p2}}\right)\right\rangle_{S_{\mathrm{a}}^-} \tag{H-10}$$

in which operator $\mathcal{E}_0$ is as $\mathcal{E}_0(\boldsymbol{J},\boldsymbol{M}) = -j\omega\mu_0\mathcal{L}_0(\boldsymbol{J}) - \mathcal{K}_0(\boldsymbol{M})$ and operator $\mathcal{H}_0$ is as $\mathcal{H}_0(\boldsymbol{J},\boldsymbol{M}) = \mathcal{K}_0(\boldsymbol{J}) - j\omega\varepsilon_0\mathcal{L}_0(\boldsymbol{M})$, where operator $\mathcal{L}_0$ is the same as the one used previously and operator $\mathcal{K}_0$ is defined as that $\mathcal{K}_0(\boldsymbol{X}) = \nabla\times\int_{\Omega} G_0(\boldsymbol{r},\boldsymbol{r}')\boldsymbol{X}(\boldsymbol{r}')d\Omega'$. In addition, the equivalent surface currents are defined by employing the inner normal directions of the boundaries of the array elements.

Similar to deriving (H-4) from (H-1), the following

$$P_{\mathrm{driv}} = \mathrm{M}_{\mathrm{a}}^{\dagger}\cdot\mathrm{P}_{\mathrm{driv}}\cdot\mathrm{M}_{\mathrm{a}} \tag{H-11}$$

with only independent variable $\mathrm{M}_{\mathrm{a}}$ can be derived from (H-10), and a detailed derivation process is given in Sec. H8.3. Here, $\mathrm{M}_{\mathrm{a}}$ is the basis function expansion vector of $\boldsymbol{M}_{\mathrm{a}}$. Theoretically, the independent variable can also be selected as the $\mathrm{J}_{\mathrm{a}}$, which is the basis function expansion vector of $\boldsymbol{J}_{\mathrm{a}}$. However, the numerical performances of the two selections are different, and it is more desirable to select $\mathrm{M}_{\mathrm{a}}$ as independent variable because the array elements are non-magnetic, and a similar explanation focusing on scattering systems can be found in [13].

Employing the $\mathrm{P}_{\mathrm{driv}}$ used in (H-11), the CMs of the material Yagi-Uda array can be calculated like Sec. H4.2.

## H5.3 Numerical Verifications

To verify the validities of the above volume and surface formulations of the WEP-CMA

for material Yagi-Uda arrays, the comparisons of the WEP-CMA-based numerical results with some published simulation and measurement data are provided in this sub-section. The specific material Yagi-Uda array analyzed here is the same as the one reported in [345], and its all elements are with $4.0 \text{ mm} \times 4.0 \text{ mm}$ cross section, and its elements $V_{\text{a}}$, $V_{\text{p1}}$, and $V_{\text{p2}}$ have lengths 46.35 mm, 77.6 mm, and 44.4 mm respectively, and the distance (side-to-side) between $V_{\text{a}}$ and $V_{\text{p1}}$ is 23.0 mm, and the distance (side-to-side) between $V_{\text{a}}$ and $V_{\text{p2}}$ is 10.7 mm.

The characteristic value (in decibel) of the dominant CM calculated from WEP-CMA and the modal $S_{11}$ parameter (in decibel) obtained from simulation and measurement published in [345] are shown in Fig. H-10 simultaneously. From the figure, it is easy to find out that the WEP-CMA-based resonance frequency is basically consistent with the data reported in [345], and the slight discrepancy is mainly originated from ignoring the feeding structure.

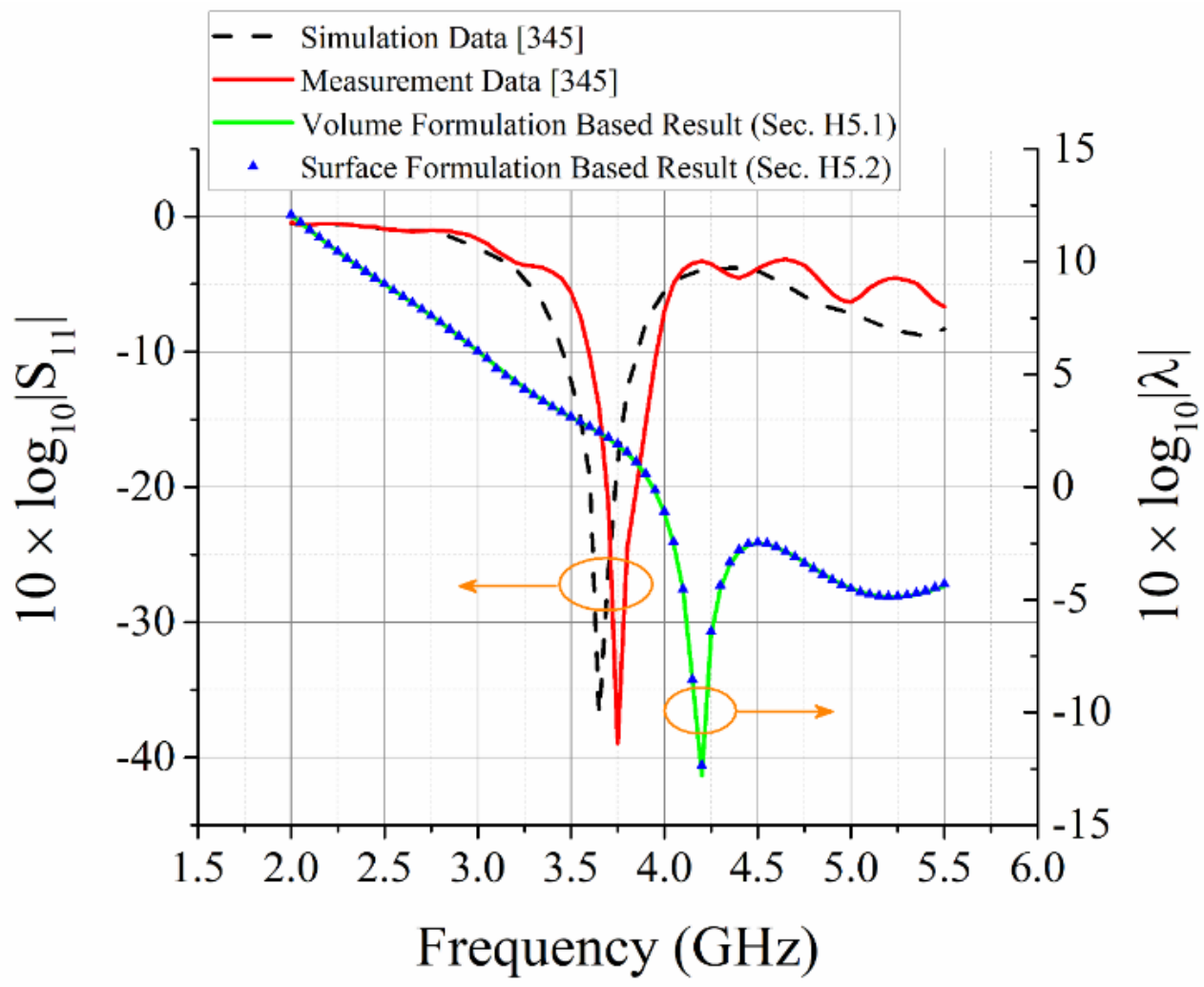

Figure H-10 Modal parameters of the dominant mode of the material Yagi-Uda array reported in [345]

For the resonant working state of the WEP-CMA-based CM shown in Fig. H-10, its radiation pattern is shown in Fig. H-11, and its electric and magnetic field distributions are shown in Fig. H-12. Evidently, Fig. H-11 and Fig. H-12 satisfy the well-known features of linear Yagi-Uda arrays — the radiated EM power propagates along the direction from reflecting element to directing elements. In addition, the time-domain dynamic figures of the propagating electric and magnetic fields are also uploaded to IEEE manuscript system together with this paper.

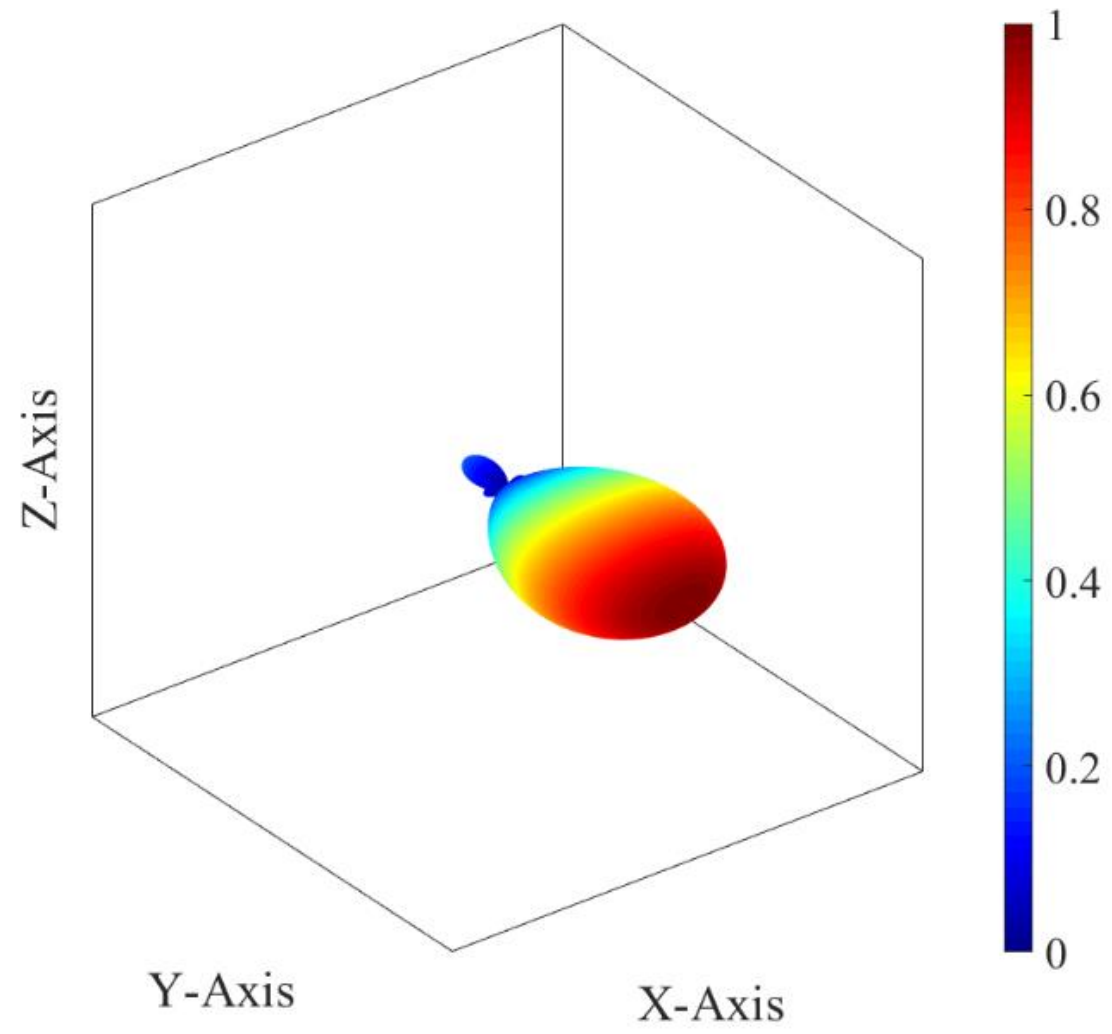

Figure H-11 Radiation pattern of the resonant state of the WEP-CMA-based dominant CM shown in Fig. H-10

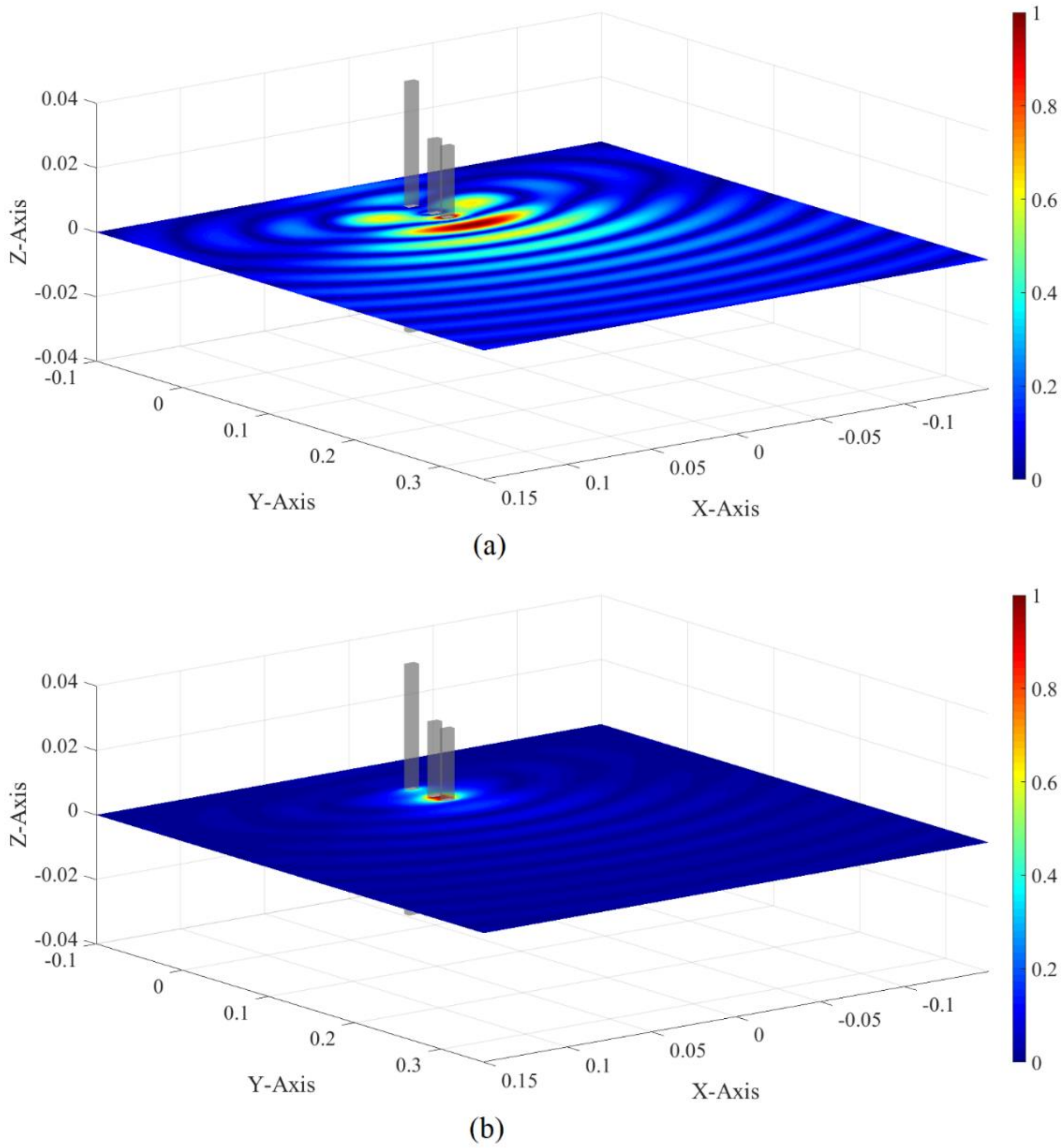

Figure H-12 Distributions of the (a) electric field and (b) magnetic field of the resonant state of the WEP-CMA-based CM shown in Fig. H-10

## H6 WEP-CMA for Composite Yagi-Uda Arrays

This section further generalizes the results given in Secs. H4 and H5 to the metal-material composite Yagi-Uda array shown in Fig. H-13, which was reported in [326]. The composite array is constituted by metallic active patch $S_{\rm a}$, metallic passive patches $S_{\rm p}$, metallic ground plane $V_{\rm g}$, and material substrate $V_{\rm s}$. The substrate $V_{\rm s}$ is restricted to being non-magnetic for simplifying the discussions, and its complex permittivity is $\boldsymbol{\varepsilon}_{\rm s}^{\rm c}$.

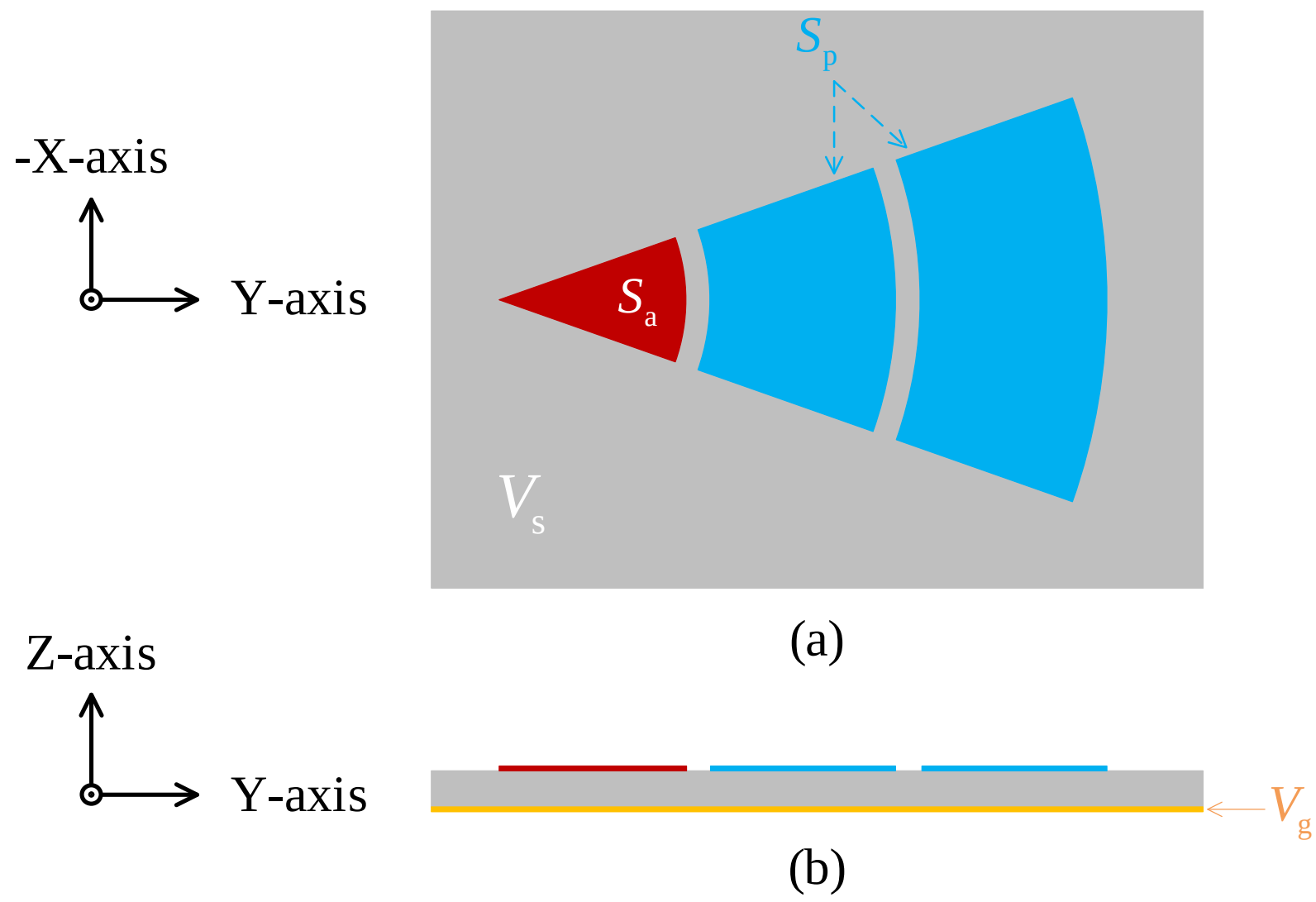

Figure H-13 Geometry of a 3-element composite Yagi-Uda array reported in [326]. (a) Top view; (b) lateral view

### H6.1 Volume-Surface Formulation of the WEP-CMA for Composite Yagi-Uda Arrays

If the induced surface electric currents on $S_{\rm a}$, $S_{\rm p}$, and the boundary of $V_{\rm g}$ are denoted as $\boldsymbol{J}_{\rm a}$, $\boldsymbol{J}_{\rm p}$, and $\boldsymbol{J}_{\rm g}$ respectively, and the induced volume electric current on $V_{\rm s}$ is denoted as $\boldsymbol{j}_{\rm s}$, then the corresponding DPO is as follows:

$$P_{\rm driv} = (1/2)\left\langle \boldsymbol{J}_{\rm a} \, , \, j\omega\mu_0\mathcal{L}_0\left(\boldsymbol{J}_{\rm a} + \boldsymbol{J}_{\rm p} + \boldsymbol{J}_{\rm g} \oplus \boldsymbol{j}_{\rm s}\right)\right\rangle_{S_{\rm a}} \tag{H-12}$$

Here, the utilization of symbol "$\oplus$" is to emphasize that the dimensions of surface current $\boldsymbol{J}_{\rm a} + \boldsymbol{J}_{\rm p} + \boldsymbol{J}_{\rm g}$ and volume current $\boldsymbol{j}_{\rm s}$ are different from each other.

Similar to deriving (H-4) from (H-1), the following

$$P_{\rm driv} = \mathrm{J}_{\rm a}^{\dagger} \cdot \mathrm{P}_{\rm driv} \cdot \mathrm{J}_{\rm a} \tag{H-13}$$

with only independent variable $\mathrm{J}_{\rm a}$ can be derived from (H-12), and a detailed derivation process is given in Sec. H8.4. Here, $\mathrm{J}_{\rm a}$ is the basis function expansion vector of $\boldsymbol{J}_{\rm a}$.

The formula provided in this sub-section is based on volume-surface currents, and an alternative surface-current-based formula will be given in the following sub-section.

## H6.2 Surface Formulation of the WEP-CMA for Composite Yagi-Uda Arrays

For the convenience of this sub-section, the interface between $V_{\rm g}$ and free space is denoted as $S_{\rm gf}$, and the interface between $V_{\rm s}$ and $S_{\rm a}$ is denoted as $S_{\rm sa}$, and the interface between $V_{\rm s}$ and $S_{\rm p}$ is denoted as $S_{\rm sp}$, and the interface between $V_{\rm s}$ and free space is denoted as $S_{\rm sf}$.

If the induced surface electric current on $S_{\rm gf}$ is denoted as $\boldsymbol{J}_{\rm gf}$, and the equivalent surface electric currents on $S_{\rm sa}$, $S_{\rm sp}$, and $S_{\rm sf}$ are denoted as $\boldsymbol{J}_{\rm sa}$, $\boldsymbol{J}_{\rm sp}$, and $\boldsymbol{J}_{\rm sf}$ respectively, and the equivalent surface magnetic current on $S_{\rm sf}$ is denoted as $\boldsymbol{M}_{\rm sf}$, then the volume-surface-current version (H-12) of DPO can be alternatively written as the following surface-current version

$$P_{\rm driv} = (1/2)\Big\langle \boldsymbol{J}_{\rm a}, -\mathcal{E}_0\big(\boldsymbol{J}_{\rm a} + \boldsymbol{J}_{\rm p} + \boldsymbol{J}_{\rm gf} - \boldsymbol{J}_{\rm sa} - \boldsymbol{J}_{\rm sp} - \boldsymbol{J}_{\rm sf}, -\boldsymbol{M}_{\rm sf}\big)\Big\rangle_{S_{\rm a}} \tag{H-14}$$

where the equivalent surface currents are defined by employing the inner normal direction of the boundary of the material substrate.

Similar to deriving (H-4) from (H-1), the following

$$P_{\rm driv} = \mathbb{J}_{\rm a}^{\dagger} \cdot \mathbb{P}_{\rm driv} \cdot \mathbb{J}_{\rm a} \tag{H-15}$$

with only independent variable $\mathbb{J}_{\rm a}$ can be derived from (H-14), and a detailed derivation process is given in Sec. H8.5.

Employing the $\mathbb{P}_{\rm driv}$ used in (H-13) or (H-15), the CMs of the composite Yagi-Uda array can be calculated like Sec. H4.2.

## H6.3 Numerical Verifications

This sub-section applies the WEP-CMA to the composite Yagi-Uda array reported in [326]. The geometry of the array is shown in Fig. H-13, and the size of the array is described in [326].

The characteristic value (in decibel) of the dominant CM calculated from WEP-CMA and the modal $S_{11}$ parameter (in decibel) obtained from simulation published in [326] are shown in Fig. H-14 simultaneously. The figure implies that the WEP-CMA-based resonance frequencies are basically consistent with the data reported in [326], and the slight discrepancy is mainly originated from ignoring the feeding structure.

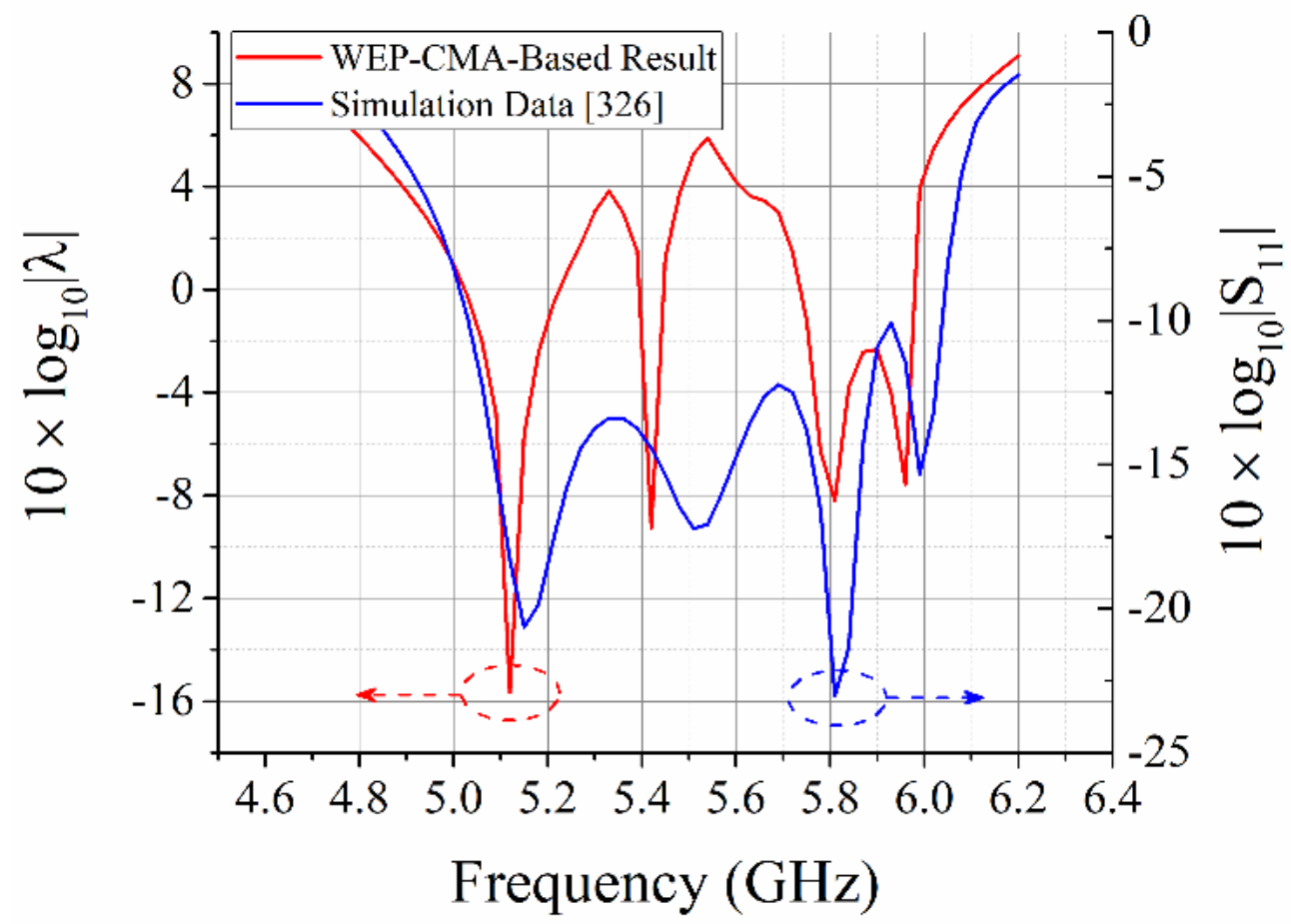

Figure H-14 Modal parameters of the dominant mode of the composite Yagi-Uda array reported in [326]

The radiation pattern of the CM working at resonance frequency 5.12 GHz is shown in Fig. H-15. The radiation pattern is consistent with the one reported in [326].

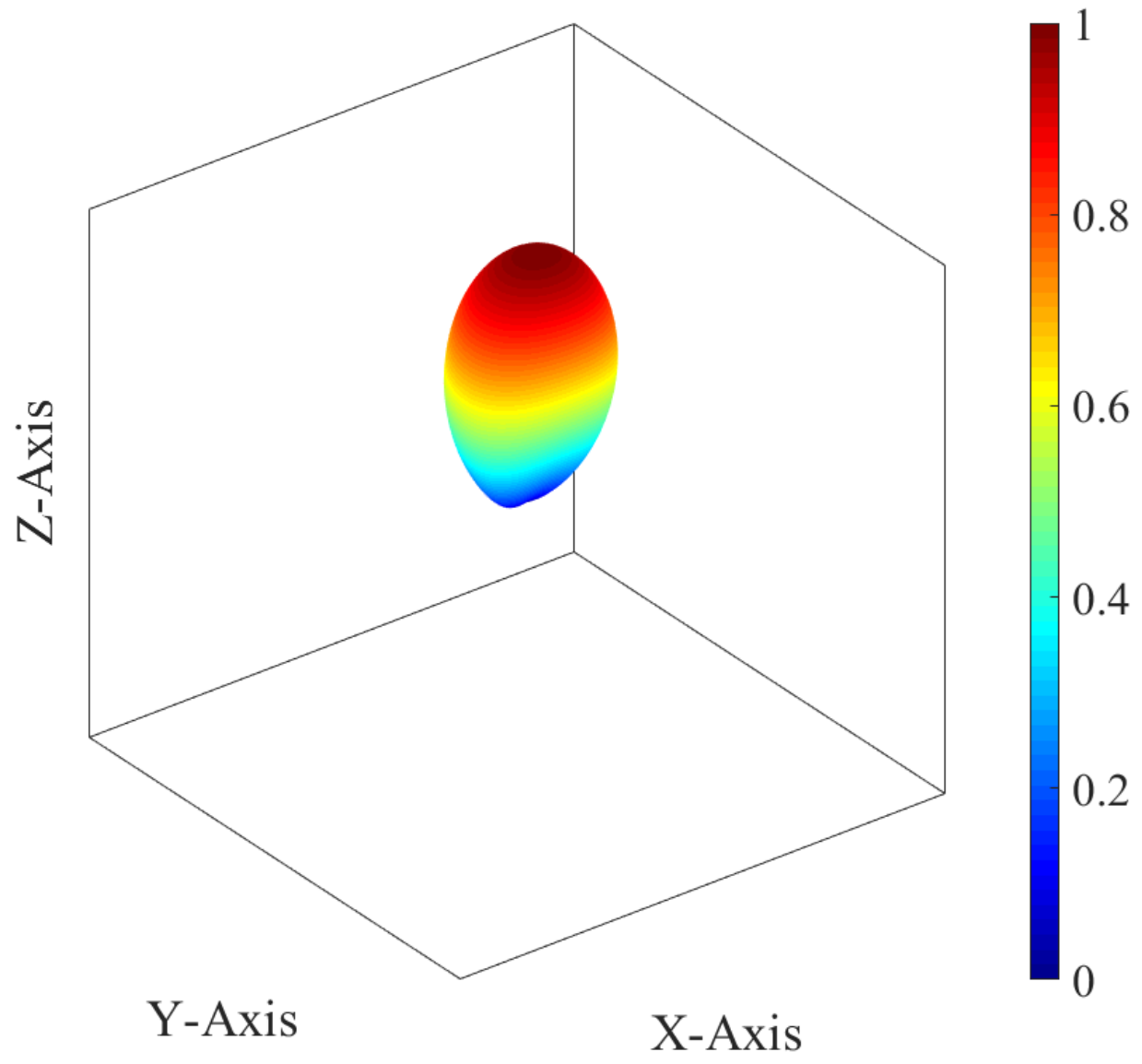

Figure H-15 Radiation pattern of the resonant state of the WEP-CMA-based CM shown in Fig. H-14

## H7 Conclusions

The same as the conclusions of our previous studies for metallic scattering and transferring systems, the WEP is also a quantitative depiction for the work-energy transformation process during the working of metallic Yagi-Uda arrays, and the driving power contained in WEP is just the source to sustain a steady work-energy transformation of the arrays. Employing the WEP, it is found out that the working mechanisms of metallic scattering systems and metallic Yagi-Uda arrays are different from each other, but the working mechanisms of metallic transferring systems and metallic Yagi-Uda arrays are similar to each other.

Based on the above these, the WEP-based CMT for metallic transferring systems can be successfully applied to doing the modal analysis for metallic Yagi-Uda arrays. By orthogonalizing DPO, the WEP-CMA can construct a set of energy-decoupled CMs for any previously selected objective metallic Yagi-Uda array. For a classical 6-element linear metallic Yagi-Uda array, the WEP-CMA implies that the dominant resonant CM is end-fire but the other higher-order resonant CMs are usually not, and then it is clearly explained why the higher-order resonant modes of the array are seldom used.

In addition, the WEP-CMA for metallic Yagi-Uda arrays can also be easily generalized to material Yagi-Uda arrays and composite Yagi-Uda arrays.

The validity and correctness of the WEP-CMA are verified by applying the WEP-CMA to the various Yagi-Uda arrays and comparing the WEP-CMA-based numerical results with the published simulation and measurement data.

## H8 Appendices of AP2101-0036

Some detailed formulations related to this paper are provided in the following appendices.

### H8.1 Detailed Formulations Related to Sec. H4.2

In (H-2), the elements of sub-matrix $\mathbb{P}_{\text{aa}}$ are calculated as that $[\mathbb{P}_{\text{aa}}]_{\xi\zeta} = (1/2) < \boldsymbol{b}_{\text{a};\xi}, j\omega\mu_0\mathcal{L}_0(\boldsymbol{b}_{\text{a};\zeta}) >_{S_{\text{a}}}$, and the elements of sub-matrix $\mathbb{P}_{\text{ap}}$ are calculated as that $[\mathbb{P}_{\text{ap}}]_{\xi\zeta} = (1/2) < \boldsymbol{b}_{\text{a};\xi}, j\omega\mu_0\mathcal{L}_0(\boldsymbol{b}_{\text{p};\zeta}) >_{S_{\text{a}}}$.

The $\mathbb{T}$ used in (H-3) is $\mathbb{T} = -\mathbb{Z}_{\text{pp}}^{-1} \cdot \mathbb{Z}_{\text{pa}}$. The elements of matrix $\mathbb{Z}_{\text{pp}}$ are calculated as that $[\mathbb{Z}_{\text{pp}}]_{\xi\zeta} =< \boldsymbol{b}_{\text{p};\xi}, -j\omega\mu_0\mathcal{L}_0(\boldsymbol{b}_{\text{p};\zeta}) >_{S_{\text{p}}}$, and the elements of matrix $\mathbb{Z}_{\text{pa}}$ are calculated as that $[\mathbb{Z}_{\text{pa}}]_{\xi\zeta} =< \boldsymbol{b}_{\text{p};\xi}, -j\omega\mu_0\mathcal{L}_0(\boldsymbol{b}_{\text{a};\zeta}) >_{S_{\text{p}}}$.

The $\mathbb{P}_{\text{driv}}$ used in (H-4) is as follows:

$$\mathbb{P}_{\text{driv}} = \begin{bmatrix} \mathbb{P}_{\text{aa}} & \mathbb{P}_{\text{ap}} \end{bmatrix} \cdot \begin{bmatrix} \mathbb{I} \\ \mathbb{T} \end{bmatrix} \tag{H-16}$$

where $\mathbb{I}$ is an unit matrix with proper order.

### H8.2 Detailed Formulations Related to Sec. H5.1

By expanding the currents in (H-8) in terms of some proper basis functions, the integral form (H-8) of DPO can be discretized into the following matrix form

$$P_{\text{driv}} = \mathrm{j}_{\text{a}}^{\dagger} \cdot \begin{bmatrix} \mathbb{P}_{\text{aa}} & \mathbb{P}_{\text{ap1}} & \mathbb{P}_{\text{ap2}} \end{bmatrix} \cdot \begin{bmatrix} \mathrm{j}_{\text{a}} \\ \mathrm{j}_{\text{p1}} \\ \mathrm{j}_{\text{p2}} \end{bmatrix} \tag{H-17}$$

where the elements of sub-matrix $\mathbb{P}_{\text{aa}}$ are calculated as that $[\mathbb{P}_{\text{aa}}]_{\xi\zeta} = (1/2) < \boldsymbol{b}_{\text{a};\xi}, (j\omega\Delta\boldsymbol{\varepsilon}_{\text{a}}^{\text{c}})^{-1} \cdot \boldsymbol{b}_{\text{a};\zeta} + j\omega\mu_0\mathcal{L}_0(\boldsymbol{b}_{\text{a};\zeta}) >_{V_{\text{a}}}$, and the elements of sub-matrix $\mathbb{P}_{\text{ap1/ap2}}$ are calculated as that $[\mathbb{P}_{\text{ap1/ap2}}]_{\xi\zeta} = (1/2) < \boldsymbol{b}_{\text{a};\xi}, j\omega\mu_0\mathcal{L}_0(\boldsymbol{b}_{\text{p1/p2};\zeta}) >_{V_{\text{a}}}$.

Because of volume equivalence principle, there exist the following integral equations

$$\boldsymbol{j}_{\text{p1}} = j\omega\Delta\boldsymbol{\varepsilon}_{\text{p1}}^{\text{c}} \cdot \left[ -j\omega\mu_0\mathcal{L}_0\left(\boldsymbol{j}_{\text{a}} + \boldsymbol{j}_{\text{p1}} + \boldsymbol{j}_{\text{p2}}\right)\right] \text{ on } V_{\text{p1}} \tag{H-18}$$

$$\boldsymbol{j}_{\text{p2}} = j\omega\Delta\boldsymbol{\varepsilon}_{\text{p2}}^{\text{c}} \cdot \left[ -j\omega\mu_0\mathcal{L}_0\left(\boldsymbol{j}_{\text{a}} + \boldsymbol{j}_{\text{p1}} + \boldsymbol{j}_{\text{p2}}\right)\right] \text{ on } V_{\text{p2}} \tag{H-19}$$

Applying the method of moments (MoM) to (H-18)~(H-19), the integral equations are immediately discretized into matrix equations. By solving the matrix equations, the following transformation

$$\begin{bmatrix} \mathrm{j}_{\text{p1}} \\ \mathrm{j}_{\text{p2}} \end{bmatrix} = \underbrace{\begin{bmatrix} \mathbb{Z}_{\text{p1p1}} & \mathbb{Z}_{\text{p1p2}} \\ \mathbb{Z}_{\text{p2p1}} & \mathbb{Z}_{\text{p2p2}} \end{bmatrix} \cdot \begin{bmatrix} \mathbb{Z}_{\text{p1a}} \\ \mathbb{Z}_{\text{p2a}} \end{bmatrix}}_{\mathbb{T}} \cdot \mathrm{j}_{\text{a}} \tag{H-20}$$

can be easily obtained, where the elements of $\mathbb{Z}_{\text{p1p1/p2p2}}$ are $< \boldsymbol{b}_{\text{p1/p2};\xi}, (j\omega\Delta\boldsymbol{\varepsilon}_{\text{p1/p2}}^{\text{c}})^{-1} \cdot \boldsymbol{b}_{\text{p1/p2};\zeta} + j\omega\mu_0\mathcal{L}_0(\boldsymbol{b}_{\text{p1/p2};\zeta}) >_{V_{\text{p1}}/V_{\text{p2}}}$, and the elements of sub-matrix $\mathbb{Z}_{\text{p1p2/p2p1}}$ are calculated as $< \boldsymbol{b}_{\text{p1/p2};\xi}, j\omega\mu_0\mathcal{L}_0(\boldsymbol{b}_{\text{p2/p1};\zeta}) >_{V_{\text{p1}}/V_{\text{p2}}}$, and the elements of $\mathbb{Z}_{\text{p1a/p2a}}$ are $< \boldsymbol{b}_{\text{p1/p2};\xi}, -j\omega\mu_0\mathcal{L}_0(\boldsymbol{b}_{\text{a};\zeta}) >_{V_{\text{p1}}/V_{\text{p2}}}$.

Substituting (H-20) into (H-17), we immediately have (H-9), where

$$\mathbb{P}_{\text{driv}} = \begin{bmatrix} \mathbb{P}_{\text{aa}} & \mathbb{P}_{\text{ap1}} & \mathbb{P}_{\text{ap2}} \end{bmatrix} \cdot \begin{bmatrix} \mathbb{I} \\ \mathbb{T} \end{bmatrix} \tag{H-21}$$

## H8.3 Detailed Formulations Related to Sec. H5.2

The integral form (H-10) of DPO can be discretized into the following matrix form

$$P_{\text{driv}} = \begin{bmatrix} \mathrm{J}_{\text{a}} \\ \mathrm{M}_{\text{a}} \end{bmatrix}^{\dagger} \cdot \begin{bmatrix} \mathbb{P}_{\text{aa}}^{\text{JJ}} & \mathbb{P}_{\text{ap1}}^{\text{JJ}} & \mathbb{P}_{\text{ap2}}^{\text{JJ}} & \mathbb{P}_{\text{aa}}^{\text{JM}} & \mathbb{P}_{\text{ap1}}^{\text{JM}} & \mathbb{P}_{\text{ap2}}^{\text{JM}} \\ \mathbb{P}_{\text{aa}}^{\text{MJ}} & \mathbb{P}_{\text{ap1}}^{\text{MJ}} & \mathbb{P}_{\text{ap2}}^{\text{MJ}} & \mathbb{P}_{\text{aa}}^{\text{MM}} & \mathbb{P}_{\text{ap1}}^{\text{MM}} & \mathbb{P}_{\text{ap2}}^{\text{MM}} \end{bmatrix} \cdot \begin{bmatrix} \mathrm{J}_{\text{a}} \\ \mathrm{J}_{\text{p1}} \\ \mathrm{J}_{\text{p2}} \\ \mathrm{M}_{\text{a}} \\ \mathrm{M}_{\text{p1}} \\ \mathrm{M}_{\text{p2}} \end{bmatrix} \tag{H-22}$$

by expanding the currents in terms of some proper basis functions. The formulations for calculating the elements of the various sub-matrices $\mathrm{P}$ are trivial, and they are not explicitly given here.

For the currents involved in (H-10), there exist the following integral equations

$$\left[\mathcal{E}_{\mathrm{a}}\left(\boldsymbol{J}_{\mathrm{a}},\boldsymbol{M}_{\mathrm{a}}\right)\right]_{S_{\mathrm{a}}^{-}}^{\tan} = \boldsymbol{n}_{\mathrm{a}}^{-}\times\boldsymbol{M}_{\mathrm{a}} \tag{H-23}$$

$$\left[\mathcal{E}_{\mathrm{p1}}\left(\boldsymbol{J}_{\mathrm{p1}},\boldsymbol{M}_{\mathrm{p1}}\right)\right]_{S_{\mathrm{p1}}^{-}}^{\tan} = -\left[\mathcal{E}_{0}\left(\boldsymbol{J}_{\mathrm{a}}+\boldsymbol{J}_{\mathrm{p1}}+\boldsymbol{J}_{\mathrm{p2}},\boldsymbol{M}_{\mathrm{a}}+\boldsymbol{M}_{\mathrm{p1}}+\boldsymbol{M}_{\mathrm{p2}}\right)\right]_{S_{\mathrm{p1}}^{+}}^{\tan} \tag{H-24}$$

$$\left[\mathcal{H}_{\mathrm{p1}}\left(\boldsymbol{J}_{\mathrm{p1}},\boldsymbol{M}_{\mathrm{p1}}\right)\right]_{S_{\mathrm{p1}}^{-}}^{\tan} = -\left[\mathcal{H}_{0}\left(\boldsymbol{J}_{\mathrm{a}}+\boldsymbol{J}_{\mathrm{p1}}+\boldsymbol{J}_{\mathrm{p2}},\boldsymbol{M}_{\mathrm{a}}+\boldsymbol{M}_{\mathrm{p1}}+\boldsymbol{M}_{\mathrm{p2}}\right)\right]_{S_{\mathrm{p1}}^{+}}^{\tan} \tag{H-25}$$

$$\left[\mathcal{E}_{\mathrm{p2}}\left(\boldsymbol{J}_{\mathrm{p2}},\boldsymbol{M}_{\mathrm{p2}}\right)\right]_{S_{\mathrm{p2}}^{-}}^{\tan} = -\left[\mathcal{E}_{0}\left(\boldsymbol{J}_{\mathrm{a}}+\boldsymbol{J}_{\mathrm{p1}}+\boldsymbol{J}_{\mathrm{p2}},\boldsymbol{M}_{\mathrm{a}}+\boldsymbol{M}_{\mathrm{p1}}+\boldsymbol{M}_{\mathrm{p2}}\right)\right]_{S_{\mathrm{p2}}^{+}}^{\tan} \tag{H-26}$$

$$\left[\mathcal{H}_{\mathrm{p2}}\left(\boldsymbol{J}_{\mathrm{p2}},\boldsymbol{M}_{\mathrm{p2}}\right)\right]_{S_{\mathrm{p2}}^{-}}^{\tan} = -\left[\mathcal{H}_{0}\left(\boldsymbol{J}_{\mathrm{a}}+\boldsymbol{J}_{\mathrm{p1}}+\boldsymbol{J}_{\mathrm{p2}},\boldsymbol{M}_{\mathrm{a}}+\boldsymbol{M}_{\mathrm{p1}}+\boldsymbol{M}_{\mathrm{p2}}\right)\right]_{S_{\mathrm{p2}}^{+}}^{\tan} \tag{H-27}$$

where operator $\mathcal{E}_{\mathrm{a}}$ is with parameters $(\boldsymbol{\varepsilon}_{\mathrm{s}}^{\mathrm{c}},\mu_0)$, and operators $\mathcal{E}_{\mathrm{p1/p2}}$ and $\mathcal{H}_{\mathrm{p1/p2}}$ are with parameters $(\boldsymbol{\varepsilon}_{\mathrm{p1/p2}}^{\mathrm{c}},\mu_0)$, and $S_{\mathrm{a/p1/p2}}^{-}$ is the inner surface of $S_{\mathrm{a/p1/p2}}$, and $S_{\mathrm{p1/p2}}^{+}$ is the outer surface of $S_{\mathrm{p1/p2}}$.

Similar to deriving (H-20) from (H-18) and (H-19), the following transformation

$$\begin{bmatrix}\mathrm{J}_{\mathrm{a}}\\ \mathrm{J}_{\mathrm{p1}}\\ \mathrm{J}_{\mathrm{p2}}\\ \mathrm{M}_{\mathrm{p1}}\\ \mathrm{M}_{\mathrm{p2}}\end{bmatrix}=\mathrm{T}\cdot\mathrm{M}_{\mathrm{a}} \tag{H-28}$$

can be derived from (H-23)~(H-27), and then we have the following sub-transformations

$$\mathrm{J}_{\mathrm{a}}=\mathrm{T}_1\cdot\mathrm{M}_{\mathrm{a}},\ \text{and}\ \begin{bmatrix}\mathrm{J}_{\mathrm{p1}}\\ \mathrm{J}_{\mathrm{p2}}\end{bmatrix}=\mathrm{T}_2\cdot\mathrm{M}_{\mathrm{a}},\ \text{and}\ \begin{bmatrix}\mathrm{M}_{\mathrm{p1}}\\ \mathrm{M}_{\mathrm{p2}}\end{bmatrix}=\mathrm{T}_3\cdot\mathrm{M}_{\mathrm{a}} \tag{H-29}$$

Substituting (H-29) into (H-22), we immediately have (H-11), where the $\mathrm{P}_{\mathrm{driv}}$ is as follows:

$$\mathrm{P}_{\mathrm{driv}}=\begin{bmatrix}\mathrm{T}_1\\ \mathrm{I}\end{bmatrix}^{\dagger}\cdot\begin{bmatrix}\mathrm{P}_{\mathrm{aa}}^{\mathrm{JJ}} & \mathrm{P}_{\mathrm{ap1}}^{\mathrm{JJ}} & \mathrm{P}_{\mathrm{ap2}}^{\mathrm{JJ}} & \mathrm{P}_{\mathrm{aa}}^{\mathrm{JM}} & \mathrm{P}_{\mathrm{ap1}}^{\mathrm{JM}} & \mathrm{P}_{\mathrm{ap2}}^{\mathrm{JM}}\\ \mathrm{P}_{\mathrm{aa}}^{\mathrm{MJ}} & \mathrm{P}_{\mathrm{ap1}}^{\mathrm{MJ}} & \mathrm{P}_{\mathrm{ap2}}^{\mathrm{MJ}} & \mathrm{P}_{\mathrm{aa}}^{\mathrm{MM}} & \mathrm{P}_{\mathrm{ap1}}^{\mathrm{MM}} & \mathrm{P}_{\mathrm{ap2}}^{\mathrm{MM}}\end{bmatrix}\cdot\begin{bmatrix}\mathrm{T}_1\\ \mathrm{T}_2\\ \mathrm{I}\\ \mathrm{T}_3\end{bmatrix} \tag{H-30}$$

## H8.4 Detailed Formulations Related to Sec. H6.1

Here, we only provide the integral equations used to establish the transformation from the independent variable to the dependent variables involved in (H-12) as follows:

$$0 = \left[ -j\omega\mu_0\mathcal{L}_0\left(\boldsymbol{J}_\mathrm{a} + \boldsymbol{J}_\mathrm{p} + \boldsymbol{J}_\mathrm{g} + \boldsymbol{j}_\mathrm{s}\right)\right]^{\tan} \quad \text{on } S_\mathrm{p} \cup S_\mathrm{g} \tag{H-31}$$

$$\boldsymbol{j}_\mathrm{s} = j\omega\Delta\boldsymbol{\varepsilon}_\mathrm{s}^\mathrm{c} \cdot \left[ -j\omega\mu_0\mathcal{L}_0\left(\boldsymbol{J}_\mathrm{a} + \boldsymbol{J}_\mathrm{p} + \boldsymbol{J}_\mathrm{g} + \boldsymbol{j}_\mathrm{s}\right)\right] \quad \text{on } V_\mathrm{s} \tag{H-32}$$

and the other detailed formulations related to Sec. H6.1 are similar to the ones used in Secs. H8.1~H8.3.

## H8.5 Detailed Formulations Related to Sec. H6.2

Here, we only provide the integral equations used to establish the transformation from the independent variable to the dependent variables involved in (H-14) as follows:

$$\begin{aligned}&\left[\mathcal{E}_0\left(\boldsymbol{J}_\mathrm{a} + \boldsymbol{J}_\mathrm{p} + \boldsymbol{J}_\mathrm{gf} - \boldsymbol{J}_\mathrm{sa} - \boldsymbol{J}_\mathrm{sp} - \boldsymbol{J}_\mathrm{sf}, -\boldsymbol{M}_\mathrm{sf}\right)\right]^{\tan} \\ &= 0 \qquad\qquad \text{on } S_\mathrm{p} \cup S_\mathrm{gf}\end{aligned} \tag{H-33}$$

$$\begin{aligned}&\left[\mathcal{E}_\mathrm{s}\left(\boldsymbol{J}_\mathrm{sa} + \boldsymbol{J}_\mathrm{sp} + \boldsymbol{J}_\mathrm{sg} + \boldsymbol{J}_\mathrm{sf}, \boldsymbol{M}_\mathrm{sf}\right)\right]^{\tan} \\ &= 0 \qquad\qquad \text{on } S_\mathrm{sa} \cup S_\mathrm{sp} \cup S_\mathrm{sg}\end{aligned} \tag{H-34}$$

$$\begin{aligned}&\left[\mathcal{E}_0\left(\boldsymbol{J}_\mathrm{a} + \boldsymbol{J}_\mathrm{p} + \boldsymbol{J}_\mathrm{gf} - \boldsymbol{J}_\mathrm{sa} - \boldsymbol{J}_\mathrm{sp} - \boldsymbol{J}_\mathrm{sf}, -\boldsymbol{M}_\mathrm{sf}\right)\right]_{S_\mathrm{sf}^+}^{\tan} \\ &= \left[\mathcal{E}_\mathrm{s}\left(\boldsymbol{J}_\mathrm{sa} + \boldsymbol{J}_\mathrm{sp} + \boldsymbol{J}_\mathrm{sg} + \boldsymbol{J}_\mathrm{sf}, \boldsymbol{M}_\mathrm{sf}\right)\right]_{S_\mathrm{sf}^-}^{\tan} \quad \text{on } S_\mathrm{sf}\end{aligned} \tag{H-35}$$

$$\begin{aligned}&\left[\mathcal{H}_0\left(\boldsymbol{J}_\mathrm{a} + \boldsymbol{J}_\mathrm{p} + \boldsymbol{J}_\mathrm{gf} - \boldsymbol{J}_\mathrm{sa} - \boldsymbol{J}_\mathrm{sp} - \boldsymbol{J}_\mathrm{sf}, -\boldsymbol{M}_\mathrm{sf}\right)\right]_{S_\mathrm{sf}^+}^{\tan} \\ &= \left[\mathcal{H}_\mathrm{s}\left(\boldsymbol{J}_\mathrm{sa} + \boldsymbol{J}_\mathrm{sp} + \boldsymbol{J}_\mathrm{sg} + \boldsymbol{J}_\mathrm{sf}, \boldsymbol{M}_\mathrm{sf}\right)\right]_{S_\mathrm{sf}^-}^{\tan} \quad \text{on } S_\mathrm{sf}\end{aligned} \tag{H-36}$$

where current $\boldsymbol{J}_\mathrm{sg}$ is the equivalent surface electric current on the interface between $V_\mathrm{s}$ and $V_\mathrm{g}$, operators $\mathcal{E}_\mathrm{s}$ and $\mathcal{H}_\mathrm{s}$ are with parameters $(\boldsymbol{\varepsilon}_\mathrm{s}^\mathrm{c}, \mu_0)$, and the other detailed formulations related to Sec. H6.2 are similar to the ones used in Secs. H8.1~H8.3.

# Acknowledgements

This work is dedicated to my mother for her constant understanding, support, and encouragement.

# Work-Energy Principle Based Characteristic Mode Theory with Solution Domain Compression for Material Scattering Systems

Ren-Zun Lian, Xing-Yue Guo, *Student Member, IEEE*, and Ming-Yao Xia, *Senior Member, IEEE*

***Abstract*—Work-energy principle (WEP) framework is used to establish characteristic mode theory (CMT).**

**Under WEP framework, the physical purpose/picture of CMT is revealed, and that is to construct a set of steadily working energy-decoupled modes for scattering systems. Employing the physical picture, it is explained why the modal far fields of lossy scattering systems are not orthogonal; it is explained why the characteristic values of magnetodielectric scattering systems don't have clear physical meaning; the physical interpretation for normalizing modal real power to 1 is provided; the Parseval's identity related to CMT is derived.**

**Under WEP framework, driving power operator (DPO) is introduced as the generating operator of characteristic modes (CMs), and orthogonalizing DPO method is proposed to construct CMs. In the aspect of distinguishing independent variables from dependent variables, new DPO is more advantageous than traditional impedance matrix operator (IMO). In the aspect of constructing CMs, new orthogonalizing DPO method has a more satisfactory numerical performance than traditional orthogonalizing IMO method.**

**In addition, solution domain compression (SDC) scheme is developed to suppress spurious modes. New SDC scheme effectively avoids the matrix inversion process used in traditional dependent variable elimination (DVE) scheme.**

## I. Introduction

UNDER scattering matrix (SM) framework, Garbacz introduced the concept of characteristic mode (CM) in [1], and systematically established SM-based characteristic mode theory (CMT) in [2], and summarized his core ideology in [3]. The SM-based CMT (SM-CMT) has a very clear physical purpose (or physical picture), that is, for any pre-selected lossless objective scattering system (OSS) SM-CMT focuses on constructing a set of CMs with orthogonal modal far fields by orthogonalizing the perturbation matrix operator (PMO) obtained in SM framework.

Following Garbacz's pioneering works, Harrington *et al.* [4] alternatively established an integral equation (IE) based CMT, and published a series of papers [5]-[7]. By orthogonalizing the impedance matrix operator (IMO) obtained in IE framework, the IE-based CMT (IE-CMT) can also construct a set of CMs for any pre-selected lossless or lossy OSS.

Harrington's transformation for the carrying framework of CMT (from SM framework to IE framework, as shown in Tab. I) and transformation for the construction method of CMs (from orthogonalizing PMO method to orthogonalizing IMO method, as shown in Tab. I) are seminal works and of great significance for developing CMT, because the transformations greatly simplify the numerical computation of CMs. After half a century (from 1970 to now) development, IE-CMT has been greatly improved in many aspects [8], but still has some unsolved problems [9].

Unlike SM-CMT [1]-[3], which had had a clear physical purpose/picture from the beginning of its establishment, IE-CMT [4]-[7] has been lacking of a clear physical picture since its establishment. Harrington *et al.* [5] provided a Poynting's theorem based physical interpretation to the IE-CMT for metallic OSSs, but [10] found out that the interpretation is not suitable for the IE-CMT of material OSSs. To provide a unified physical picture to both metallic and material IE-CMTs, [11]-[12] rebuilt Harrington's IE-CMT by using a new ideology. Recently, [13] further generalized the ideology used in [10]-[12] to work-energy principle (WEP) based CMT, and then realized the second transformation for the carrying framework of CMT — from IE framework to WEP framework (as shown in Tab. I).

The new WEP framework not only gives Harrington's IE-CMT a clear physical picture — to construct a set of steadily working energy-decoupled modes for scattering systems, but also leads to the second transformation for the construction method of CMs — from orthogonalizing IMO method to orthogonalizing driving power operator (DPO) method (as shown in Tab. I). Employing the physical picture, this paper will clarify some confusions existing in CMT for a long time. This paper will exhibit the fact that the new orthogonalizing DPO method has a more satisfactory numerical performance in the aspect of constructing CMs.

In addition, why the characteristic equation for complicated

[AP2004-0708] received April 11, 2020; [AP2004-0708.R1] revised November 10, 2020. This work was supported by XXXX under Project XXXXXXXX. (*Corresponding authors: Ren-Zun Lian; Ming-Yao Xia.*)

R. Z. Lian, X. Y. Guo, and M. Y. Xia are with the Department of Electronics, School of Electronics Engineering and Computer Science, Peking University, Beijing 100871, China. (E-mail: rzlian@vip.163.com; myxia@pku.edu.cn).

OSSs usually outputs some spurious modes and how to suppress the spurious modes are also the hot issues in recent years. Alroughani *et al.* [14] found out that the classical Poggio-Miller-Chang-Harrington-Wu-Tsai (PMCHWT) based surface CM formulation usually outputs some spurious modes. Alroughani *et al.* [14] and Miers *et al.* [15]-[16] proposed some methods for recognizing and filtering the spurious modes. Chen *et al.* [8], [17] and some other researchers [11]-[12] attributed the spurious modes to overlooking the dependence relation between equivalent surface electric current $\boldsymbol{J}^{\rm ES}$ and magnetic current $\boldsymbol{M}^{\rm ES}$, and proposed some schemes to suppress the spurious modes by employing the transformations between $\boldsymbol{J}^{\rm ES}$ and $\boldsymbol{M}^{\rm ES}$, but the transformations proposed by Chen *et al.* [8], [17] are essentially different from the ones proposed in [11] and [12]. In [11], the transformations were derived from the tangential continuation conditions of electromagnetic (EM) fields. In [12], the transformations were derived from the definitions for $\boldsymbol{J}^{\rm ES}$ and $\boldsymbol{M}^{\rm ES}$. Recently, [13] further proved that the transformations employed in [11] and [12] and then the suppression schemes proposed in [11] and [12] are essentially equivalent to each other. Guo and Xia *et al.* [18] also attributed the spurious modes to overlooking the dependence relation between $\boldsymbol{J}^{\rm ES}$ and $\boldsymbol{M}^{\rm ES}$, and proposed some alternative schemes for relating $\boldsymbol{J}^{\rm ES}$ and $\boldsymbol{M}^{\rm ES}$, by employing an intermediate variable — effective current, and the concept of effective current can be found in [19]-[22].

The spurious mode suppression schemes proposed in [11]-[13] need to inverse some full matrices related to first-kind [11]-[13] or second-kind [11] Fredholm's integral operator. But, the process to compute the inversion of full matrix requires a large CPU time or computer memory [23]. This paper proposes a new spurious mode suppression scheme — solution domain compression (SDC) — which doesn't need to inverse any matrix, and the SDC scheme can be easily integrated into the WEP-based CMT (WEP-CMT).

This paper is organized as follows: Sec. II simply reviews the physical principle of WEP-CMT, where the physical picture of WEP-CMT is highlighted and some confusions on CMT are clarified; Secs. III and IV propose a novel SDC scheme and its improved version for suppressing spurious modes, and the schemes don't need to inverse any matrix; Secs. III and IV also do some necessary comparisons among different CM construction methods, and clearly exhibit the advantages of new orthogonalizing DPO method [13] over traditional orthogonalizing IMO method [4]-[7]; Sec. V concludes this paper.

In what follows, the $e^{j\omega t}$ convention is used throughout, and the time-domain quantities will be added time variable $t$ explicitly for example $q(t)$, but the frequency-domain quantities will not for example $q$. In addition, as known to all, for the linear quantities, we have $q(t) = {\rm Re}\{qe^{j\omega t}\}$; for the power-type quadratic quantities, we have ${\rm Re}\{q\} = (1/T)\int_0^T q(t)dt$, where $T$ is the period of the time-harmonic EM field [24]-[25].

## II. Volume Formulation of the WEP-CMT for Material Scattering Systems

This section considers the EM scattering problem shown in Fig. 1. In the figure, $V_{\rm OSS}$ is a material OSS with parameters $\boldsymbol{\mu}$ and $\boldsymbol{\varepsilon}^{\rm c} = \boldsymbol{\varepsilon} - j\boldsymbol{\sigma}/\omega$, where magnetic permeability $\boldsymbol{\mu}$, dielectric permittivity $\boldsymbol{\varepsilon}$, and electric conductivity $\boldsymbol{\sigma}$ are real and two-order symmetrical dyads depending on spatial variable $\boldsymbol{r}$ but independent of time variable $t$; $D_{\rm env}$ denotes the external environment surrounding $V_{\rm OSS}$, and it can be free space or not; $D_{\rm imp}$ is an externally impressed source and generates an EM field $\boldsymbol{F}^{\rm imp}$ in whole three-dimensional Euclidean space $\mathbb{E}^3$, where $\boldsymbol{F}^{\rm imp}$ is the abbreviated form of EM fields $(\boldsymbol{E}^{\rm imp}, \boldsymbol{H}^{\rm imp})$.

Due to the existence of $\boldsymbol{F}^{\rm imp}$, a scattered volume current $\boldsymbol{C}^{\rm SV}$ and a current $\boldsymbol{C}^{\rm env}$ are induced on $V_{\rm OSS}$ and $D_{\rm env}$ respectively, and then a scattered field $\boldsymbol{F}^{\rm sca}$ and an environment field $\boldsymbol{F}^{\rm env}$ are generated in $\mathbb{E}^3$ by $\boldsymbol{C}^{\rm SV}$ and $\boldsymbol{C}^{\rm env}$ correspondingly, where $\boldsymbol{C}^{\rm SV/env}$ is the abbreviated form of EM currents $(\boldsymbol{J}^{\rm SV/env}, \boldsymbol{M}^{\rm SV/env})$. Based on superposition principle [25]-[27], $\boldsymbol{C}^{\rm SV}$ and then $\boldsymbol{F}^{\rm sca}$ can be viewed as the result of the excitation from two external fields $\boldsymbol{F}^{\rm imp}$ and $\boldsymbol{F}^{\rm env}$, so the two external fields are treated as a whole — incident field $\boldsymbol{F}^{\rm inc}$ (or alternatively called externally resultant field) — in this paper, that is, $\boldsymbol{F}^{\rm inc} = \boldsymbol{F}^{\rm imp} + \boldsymbol{F}^{\rm env}$. In addition, if the summation of $\boldsymbol{F}^{\rm inc}$ and $\boldsymbol{F}^{\rm sca}$ is denoted as total field $\boldsymbol{F}^{\rm tot}$, that is, $\boldsymbol{F}^{\rm tot} = \boldsymbol{F}^{\rm inc} + \boldsymbol{F}^{\rm sca}$, then it exists that $\boldsymbol{J}^{\text{SV}}=j\omega\Delta\boldsymbol{\varepsilon}^{\text{c}}\cdot\boldsymbol{E}^{\text{tot}}$ and $\boldsymbol{M}^{\text{SV}}=j\omega\Delta\boldsymbol{\mu}\cdot\boldsymbol{H}^{\text{tot}}$ due to volume equivalence principle [28]-[29], where $\Delta\boldsymbol{\varepsilon}^{\text{c}}=\boldsymbol{\varepsilon}^{\text{c}}-\mathbf{I}\varepsilon_0$ and $\Delta\boldsymbol{\mu}=\boldsymbol{\mu}-\mathbf{I}\mu_0$ and $\mathbf{I}$ is two-order unit dyad (that is, $\mathbf{I}=\hat{\boldsymbol{x}}\hat{\boldsymbol{x}}+\hat{\boldsymbol{y}}\hat{\boldsymbol{y}}+\hat{\boldsymbol{z}}\hat{\boldsymbol{z}}$).

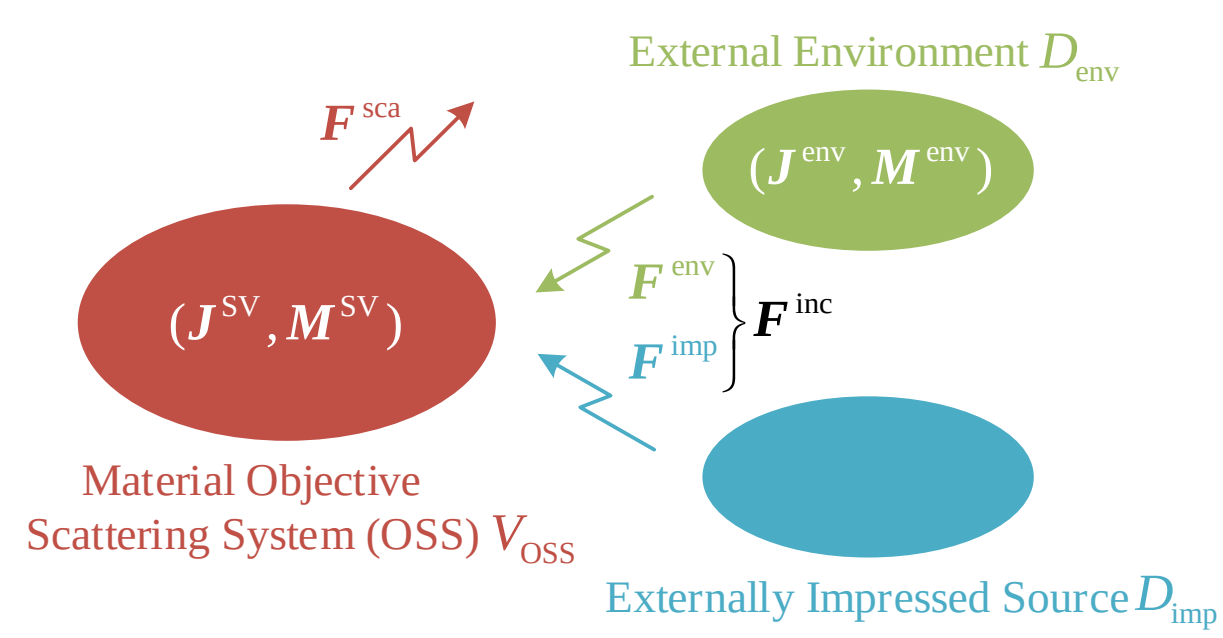

Fig. 1. EM scattering problem considered in Sec. II.

TABLE I Evolutions of CMT and Comparisons From Aspects of Carrying Framework, Construction Method and Physical Picture/Purpose

| | Carrying Framework | Construction Method | Physical Picture / Physical Purpose |
|---|---|---|---|
| **SM-CMT (1960s) [1]-[3]** | scattering matrix (SM) framework | orthogonalizing perturbation matrix operator (PMO) method | to construct a set of working modes with orthogonal modal far fields |
| | | ↓ | |
| **IE-CMT (1970s) [4]-[7]** | integral equation (IE) framework | orthogonalizing impedance matrix operator (IMO) method | not clarified by its founders |
| | | ↓ | |
| **WEP-CMT (2019) [13]** | work-energy principle (WEP) framework | orthogonalizing driving power operator (DPO) method | to construct a set of steadily working modes without net energy exchange in any integral period |

In the following parts of this section, by focusing on the volume formulation of WEP-CMT, we will discuss the carrying framework of WEP-CMT, the generating operator of CMs, the construction method of CMs, the physical picture of WEP-CMT, some characteristic quantities of CM, and the concepts of physical modes and unphysical/spurious modes.

### A. Carrying Framework of WEP-CMT

Based on superposition principle $\boldsymbol{F}^{\text{tot}}=\boldsymbol{F}^{\text{inc}}+\boldsymbol{F}^{\text{sca}}$, volume equivalence principle $\boldsymbol{J}^{\text{SV}}=j\omega\Delta\boldsymbol{\varepsilon}^{\text{c}}\cdot\boldsymbol{E}^{\text{tot}}$ & $\boldsymbol{M}^{\text{SV}}=j\omega\Delta\boldsymbol{\mu}\cdot\boldsymbol{H}^{\text{tot}}$, and Maxwell's equations $\nabla\times\boldsymbol{H}^{\text{sca}}=\boldsymbol{J}^{\text{SV}}+j\omega\varepsilon_0\boldsymbol{E}^{\text{sca}}$ & $\nabla\times\boldsymbol{E}^{\text{sca}}=-\boldsymbol{M}^{\text{SV}}-j\omega\mu_0\boldsymbol{H}^{\text{sca}}$ [28], it is not difficult to prove that

$$\mathcal{W}^{\text{Driv}}=\mathcal{E}^{\text{rad}}_{S_\infty}+\mathcal{E}^{\text{dis}}_{V_{\text{OSS}}}+\Delta\left(\mathcal{E}^{\text{mag}}_{\mathbb{E}^3}+\mathcal{E}^{\text{ele}}_{\mathbb{E}^3}\right)+\Delta\left(\mathcal{E}^{\text{mag}}_{V_{\text{OSS}}}+\mathcal{E}^{\text{pol}}_{V_{\text{OSS}}}\right)\tag{1}$$

The explicit expressions for the terms in (1) are given in (2)-(6). In (2)-(6), time interval $\Delta t$ is a positive real number; the inner product is defined as that $<\boldsymbol{f},\boldsymbol{g}>_\Omega=\int_\Omega \boldsymbol{f}^*\cdot\boldsymbol{g}\,d\Omega$; $S_\infty$ is a spherical surface with infinite radius; $\hat{\boldsymbol{n}}^+_\infty$ is the outer normal direction of $S_\infty$.

Equation (1) has a very clear physical meaning: in time interval $t_0\sim t_0+\Delta t$, the work $\mathcal{W}^{\text{Driv}}$ done by $\boldsymbol{F}^{\text{inc}}$ on $\boldsymbol{C}^{\text{SV}}$ is transformed into four parts — the radiated energy $\mathcal{E}^{\text{rad}}_{S_\infty}$ passing through $S_\infty$, the Joule heating energy $\mathcal{E}^{\text{dis}}_{V_{\text{OSS}}}$ dissipated in $V_{\text{OSS}}$, the increment of the magnetic energy $\mathcal{E}^{\text{mag}}_{\mathbb{E}^3}$ and electric energy $\mathcal{E}^{\text{ele}}_{\mathbb{E}^3}$ stored in $\mathbb{E}^3$, and the increment of the magnetization energy $\mathcal{E}^{\text{mag}}_{V_{\text{OSS}}}$ and polarization energy $\mathcal{E}^{\text{pol}}_{V_{\text{OSS}}}$ stored in $V_{\text{OSS}}$. Thus, equation (1) is a quantitative expression for the transformation between work and energy, and it is very similar to the work-energy principle in mechanism [30]-[31], so this paper calls it the work-energy principle (WEP) in electromagnetism.

### B. Generating Operator of CMs

The work term $\mathcal{W}^{\text{Driv}}$ is just the source to sustain the work-energy transformation, and also the source to drive the steady working of the material OSS. Thus, $\mathcal{W}^{\text{Driv}}$ is called driving work, and the associated power is correspondingly called driving power (DP) and denoted as $P^{\text{Driv}}(t)$. Obviously, $P^{\text{Driv}}(t)$ has operator expression

$$P^{\text{Driv}}(t)=\left\langle\boldsymbol{J}^{\text{SV}}(t),\boldsymbol{E}^{\text{inc}}(t)\right\rangle_{V_{\text{OSS}}}+\left\langle\boldsymbol{M}^{\text{SV}}(t),\boldsymbol{H}^{\text{inc}}(t)\right\rangle_{V_{\text{OSS}}}\tag{7}$$

and the operator is accordingly called time-domain driving power operator (DPO).

Two different frequency-domain versions of $P^{\text{Driv}}(t)$ are as follows:

$$P^{\text{driv}}=(1/2)\left\langle\boldsymbol{J}^{\text{SV}},\boldsymbol{E}^{\text{inc}}\right\rangle_{V_{\text{OSS}}}+(1/2)\left\langle\boldsymbol{M}^{\text{SV}},\boldsymbol{H}^{\text{inc}}\right\rangle_{V_{\text{OSS}}}\tag{8}$$

$$P^{\text{DRIV}}=(1/2)\left\langle\boldsymbol{J}^{\text{SV}},\boldsymbol{E}^{\text{inc}}\right\rangle_{V_{\text{OSS}}}+(1/2)\left\langle\boldsymbol{H}^{\text{inc}},\boldsymbol{M}^{\text{SV}}\right\rangle_{V_{\text{OSS}}}\tag{9}$$

where coefficient 1/2 originates from the time average for the power-type quadratic quantity of time-harmonic EM field [24]-[25]. It is not difficult to prove that $P^{\text{DRIV}}$ can be alternatively expressed as

$$\begin{aligned}P^{\text{DRIV}}=&(1/2)\oint\!\!\!\oint_{S_\infty}\left[\boldsymbol{E}^{\text{sca}}\times\left(\boldsymbol{H}^{\text{sca}}\right)^*\right]\cdot\hat{\boldsymbol{n}}^+_\infty dS+(1/2)\left\langle\boldsymbol{\sigma}\cdot\boldsymbol{E}^{\text{tot}},\boldsymbol{E}^{\text{tot}}\right\rangle_{V_{\text{OSS}}}\\&+j2\omega\left\{\left[\frac{1}{4}\left\langle\boldsymbol{H}^{\text{sca}},\mu_0\boldsymbol{H}^{\text{sca}}\right\rangle_{\mathbb{E}^3}-\frac{1}{4}\left\langle\varepsilon_0\boldsymbol{E}^{\text{sca}},\boldsymbol{E}^{\text{sca}}\right\rangle_{\mathbb{E}^3}\right]\right.\\&\left.+\left[\frac{1}{4}\left\langle\boldsymbol{H}^{\text{tot}},\Delta\boldsymbol{\mu}\cdot\boldsymbol{H}^{\text{tot}}\right\rangle_{V_{\text{OSS}}}-\frac{1}{4}\left\langle\Delta\boldsymbol{\varepsilon}\cdot\boldsymbol{E}^{\text{tot}},\boldsymbol{E}^{\text{tot}}\right\rangle_{V_{\text{OSS}}}\right]\right\}\end{aligned}\tag{10}$$

but $P^{\text{driv}}$ doesn't have a similar expression. In addition, it is obvious that $\text{Re}\{P^{\text{driv}}\}=\text{Re}\{P^{\text{DRIV}}\}$, but $\text{Im}\{P^{\text{driv}}\}\neq\pm\text{Im}\{P^{\text{DRIV}}\}$ when $\boldsymbol{\varepsilon}\neq\mathbf{I}\varepsilon_0$ & $\boldsymbol{\mu}\neq\mathbf{I}\mu_0$, so $P^{\text{driv}}\neq P^{\text{DRIV}},(P^{\text{DRIV}})^*$ if $\boldsymbol{\varepsilon}\neq\mathbf{I}\varepsilon_0$ & $\boldsymbol{\mu}\neq\mathbf{I}\mu_0$, and this is just the reason to use two different superscripts "driv" and "DRIV" on $P^{\text{driv}}$ and $P^{\text{DRIV}}$ respectively.

If $\boldsymbol{C}^{\text{SV}}$ is expanded in terms of some basis functions, $P^{\text{driv}}$ and $P^{\text{DRIV}}$ can be discretized into the following matrix forms

$$P^{\text{driv/DRIV}}=\underbrace{\begin{bmatrix}\text{a}^{\boldsymbol{J}^{\text{SV}}}\\\text{a}^{\boldsymbol{M}^{\text{SV}}}\end{bmatrix}}_{\text{a}^{\text{AV}}}{}^{H}\cdot\mathbb{P}^{\text{driv/DRIV}}\cdot\underbrace{\begin{bmatrix}\text{a}^{\boldsymbol{J}^{\text{SV}}}\\\text{a}^{\boldsymbol{M}^{\text{SV}}}\end{bmatrix}}_{\text{a}^{\text{AV}}}\tag{11}$$

by using superposition principle $\boldsymbol{F}^{\text{inc}}=\boldsymbol{F}^{\text{tot}}-\boldsymbol{F}^{\text{sca}}$, volume equivalence principle $\boldsymbol{J}^{\text{SV}}=j\omega\Delta\boldsymbol{\varepsilon}^{\text{c}}\cdot\boldsymbol{E}^{\text{tot}}$ & $\boldsymbol{M}^{\text{SV}}=j\omega\Delta\boldsymbol{\mu}\cdot\boldsymbol{H}^{\text{tot}}$, and source-field relation $\boldsymbol{F}^{\text{sca}}=\mathbf{G}^{JF}_0*\boldsymbol{J}^{\text{SV}}+\mathbf{G}^{MF}_0*\boldsymbol{M}^{\text{SV}}$ (here

$$\mathcal{W}^{\text{Driv}}=\int_{t_0}^{t_0+\Delta t}\left[\left\langle\boldsymbol{J}^{\text{SV}}(t),\boldsymbol{E}^{\text{inc}}(t)\right\rangle_{V_{\text{OSS}}}+\left\langle\boldsymbol{M}^{\text{SV}}(t),\boldsymbol{H}^{\text{inc}}(t)\right\rangle_{V_{\text{OSS}}}\right]dt\tag{2}$$

$$\mathcal{E}^{\text{rad}}_{S_\infty}=\int_{t_0}^{t_0+\Delta t}\left\{\oint\!\!\!\oint_{S_\infty}\left[\boldsymbol{E}^{\text{sca}}(t)\times\boldsymbol{H}^{\text{sca}}(t)\right]\cdot\hat{\boldsymbol{n}}^+_\infty dS\right\}dt\tag{3}$$

$$\mathcal{E}^{\text{dis}}_{V_{\text{OSS}}}=\int_{t_0}^{t_0+\Delta t}\left\langle\boldsymbol{\sigma}\cdot\boldsymbol{E}^{\text{tot}}(t),\boldsymbol{E}^{\text{tot}}(t)\right\rangle_{V_{\text{OSS}}}dt\tag{4}$$

$$\begin{aligned}\Delta\left(\mathcal{E}^{\text{mag}}_{\mathbb{E}^3}+\mathcal{E}^{\text{ele}}_{\mathbb{E}^3}\right)=&\left[(1/2)\left\langle\boldsymbol{H}^{\text{sca}}(t_0+\Delta t),\mu_0\boldsymbol{H}^{\text{sca}}(t_0+\Delta t)\right\rangle_{\mathbb{E}^3}+(1/2)\left\langle\varepsilon_0\boldsymbol{E}^{\text{sca}}(t_0+\Delta t),\boldsymbol{E}^{\text{sca}}(t_0+\Delta t)\right\rangle_{\mathbb{E}^3}\right]\\&-\left[(1/2)\left\langle\boldsymbol{H}^{\text{sca}}(t_0),\mu_0\boldsymbol{H}^{\text{sca}}(t_0)\right\rangle_{\mathbb{E}^3}+(1/2)\left\langle\varepsilon_0\boldsymbol{E}^{\text{sca}}(t_0),\boldsymbol{E}^{\text{sca}}(t_0)\right\rangle_{\mathbb{E}^3}\right]\end{aligned}\tag{5}$$

$$\begin{aligned}\Delta\left(\mathcal{E}^{\text{mag}}_{V_{\text{OSS}}}+\mathcal{E}^{\text{pol}}_{V_{\text{OSS}}}\right)=&\left[(1/2)\left\langle\boldsymbol{H}^{\text{tot}}(t_0+\Delta t),\Delta\boldsymbol{\mu}\cdot\boldsymbol{H}^{\text{tot}}(t_0+\Delta t)\right\rangle_{V_{\text{OSS}}}+(1/2)\left\langle\Delta\boldsymbol{\varepsilon}\cdot\boldsymbol{E}^{\text{tot}}(t_0+\Delta t),\boldsymbol{E}^{\text{tot}}(t_0+\Delta t)\right\rangle_{V_{\text{OSS}}}\right]\\&-\left[(1/2)\left\langle\boldsymbol{H}^{\text{tot}}(t_0),\Delta\boldsymbol{\mu}\cdot\boldsymbol{H}^{\text{tot}}(t_0)\right\rangle_{V_{\text{OSS}}}+(1/2)\left\langle\Delta\boldsymbol{\varepsilon}\cdot\boldsymbol{E}^{\text{tot}}(t_0),\boldsymbol{E}^{\text{tot}}(t_0)\right\rangle_{V_{\text{OSS}}}\right]\end{aligned}\tag{6}$$

$\mathbf{G}_0^{JF}$ and $\mathbf{G}_0^{MF}$ are the free-space dyadic Green's functions [29], and the convolution integral operation " $*$ " is defined as $\mathbf{G} * \boldsymbol{C} = \int_\Omega \mathbf{G}(\boldsymbol{r},\boldsymbol{r}') \cdot \boldsymbol{C}(\boldsymbol{r}') d\Omega'$ ). In (11), the superscript " $H$ " represents the conjugate transpose operation for a matrix or vector; $\mathrm{a}^{\boldsymbol{J}^{\mathrm{SV}}}$ and $\mathrm{a}^{\boldsymbol{M}^{\mathrm{SV}}}$ are the column vectors constituted by the expansion coefficients of $\boldsymbol{J}^{\mathrm{SV}}$ and $\boldsymbol{M}^{\mathrm{SV}}$, and superscript "AV" is the acronym of "all variables ( $\boldsymbol{J}^{\mathrm{SV}}$ and $\boldsymbol{M}^{\mathrm{SV}}$ )"; the formulations for calculating the elements of $\mathrm{P}^{\mathrm{driv}}$ and $\mathrm{P}^{\mathrm{DRIV}}$ can be found in [6] and [13], and they are not repeated here.

### *C. Construction Method of CMs*

There uniquely exists the following Toeplitz's decompositions for matrices $\mathrm{P}^{\mathrm{driv}}$ and $\mathrm{P}^{\mathrm{DRIV}}$ [32, Sec. 0.2.5].

$$\mathrm{P}^{\mathrm{driv/DRIV}} = \mathrm{P}_+^{\mathrm{driv/DRIV}} + j\, \mathrm{P}_-^{\mathrm{driv/DRIV}} \tag{12}$$

where $\mathrm{P}_+^{\mathrm{driv/DRIV}}$ and $\mathrm{P}_-^{\mathrm{driv/DRIV}}$ are the positive and negative Hermitian parts of $\mathrm{P}^{\mathrm{driv/DRIV}}$ , and $\mathrm{P}_+^{\mathrm{driv/DRIV}} = [\mathrm{P}^{\mathrm{driv/DRIV}} + (\mathrm{P}^{\mathrm{driv/DRIV}})^H]/2$ and $\mathrm{P}_-^{\mathrm{driv/DRIV}} = [\mathrm{P}^{\mathrm{driv/DRIV}} - (\mathrm{P}^{\mathrm{driv/DRIV}})^H]/(2j)$.

Obviously, $\mathrm{P}_+^{\mathrm{driv/DRIV}}$ and $\mathrm{P}_-^{\mathrm{driv/DRIV}}$ are Hermitian. In addition, $\mathrm{P}_+^{\mathrm{driv/DRIV}}$ is positive definite, since $(\mathrm{a}^{\mathrm{AV}})^H \cdot \mathrm{P}_+^{\mathrm{driv/DRIV}} \cdot \mathrm{a}^{\mathrm{AV}}$ is equal to the summation of radiated and dissipated powers. Thus, there must exist a non-singular matrix $\mathrm{A}$ , such that $\mathrm{A}^H \cdot \mathrm{P}_\pm^{\mathrm{driv/DRIV}} \cdot \mathrm{A}$ is real diagonal matrix [32, Theorem 7.6.4]. Obviously, the $\mathrm{A}$ also diagonalizes $\mathrm{P}^{\mathrm{driv/DRIV}}$ as follows:

$$\mathrm{A}^H \cdot \mathrm{P}^{\mathrm{driv/DRIV}} \cdot \mathrm{A} = \mathrm{diag}\left\{P_1^{\mathrm{driv/DRIV}}, P_2^{\mathrm{driv/DRIV}}, \cdots\right\} \tag{13}$$

The column vectors of the above $\mathrm{A}$ can be derived from solving the following characteristic equation [6]-[7], [13]

$$\mathrm{P}_-^{\mathrm{driv/DRIV}} \cdot \mathrm{a}_\xi^{\mathrm{AV}} = \lambda_\xi\, \mathrm{P}_+^{\mathrm{driv/DRIV}} \cdot \mathrm{a}_\xi^{\mathrm{AV}} \tag{14}$$

If some derived modes $(\mathrm{a}_1^{\mathrm{AV}}, \mathrm{a}_2^{\mathrm{AV}}, \cdots, \mathrm{a}_d^{\mathrm{AV}})$ are *d*-order degenerate, then the following Gram-Schmidt orthogonalization process [32, Sec. 0.6.4] is necessary.

$$\left.\begin{aligned} \mathrm{a}_1^{\mathrm{AV}} &= \mathrm{a}_1^{\mathrm{AV}\prime} \\ \mathrm{a}_2^{\mathrm{AV}} - \chi_{12}\mathrm{a}_1^{\mathrm{AV}\prime} &= \mathrm{a}_2^{\mathrm{AV}\prime} \\ &\cdots \\ \mathrm{a}_d^{\mathrm{AV}} - \cdots - \chi_{2d}\mathrm{a}_2^{\mathrm{AV}\prime} - \chi_{1d}\mathrm{a}_1^{\mathrm{AV}\prime} &= \mathrm{a}_d^{\mathrm{AV}\prime} \end{aligned}\right\} \tag{15}$$

where the coefficients are calculated as follows:

$$\chi_{mn} = (\mathrm{a}_m^{\mathrm{AV}\prime})^H \cdot \mathrm{P}_+^{\mathrm{driv/DRIV}} \cdot \mathrm{a}_n^{\mathrm{AV}} \Big/ (\mathrm{a}_m^{\mathrm{AV}\prime})^H \cdot \mathrm{P}_+^{\mathrm{driv/DRIV}} \cdot \mathrm{a}_m^{\mathrm{AV}\prime} \tag{16}$$

For a lossless circular cylinder, whose radius and height are both 6mm and which is with $\boldsymbol{\mu} = \mathbf{I}2\mu_0$ , $\boldsymbol{\varepsilon} = \mathbf{I}8\varepsilon_0$ , and $\boldsymbol{\sigma} = \mathbf{I}0$ , its characteristic values (CVs) in decibel (dB) and the associated modal significances (MSs) derived from orthogonalizing $\mathrm{P}^{\mathrm{driv}}$ and $\mathrm{P}^{\mathrm{DRIV}}$ , that is, orthogonalizing $P^{\mathrm{driv}}$ and $P^{\mathrm{DRIV}}$ , are shown in Fig. 2 and Fig. 3 respectively. Here, the MSs are calculated as that $\mathrm{MS}_\xi = 1/|1 + j\lambda_\xi|$ . Obviously, different operators $P^{\mathrm{driv}}$ and $P^{\mathrm{DRIV}}$ generate different modes.

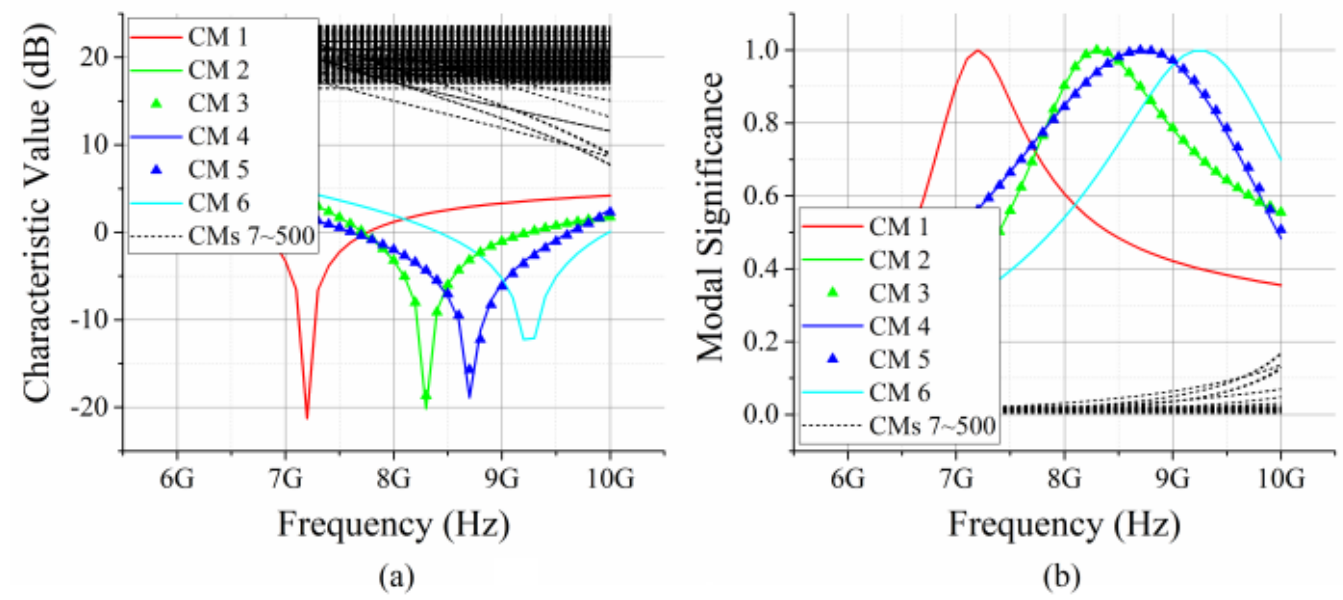

Fig. 2. $P^{\mathrm{driv}}$ -based (a) CVs and (b) MSs of a lossless cylinder, whose radius and height are both 6mm and which is with $\boldsymbol{\mu} = \mathbf{I}2\mu_0$ , $\boldsymbol{\varepsilon} = \mathbf{I}8\varepsilon_0$ , and $\boldsymbol{\sigma} = \mathbf{I}0$ .

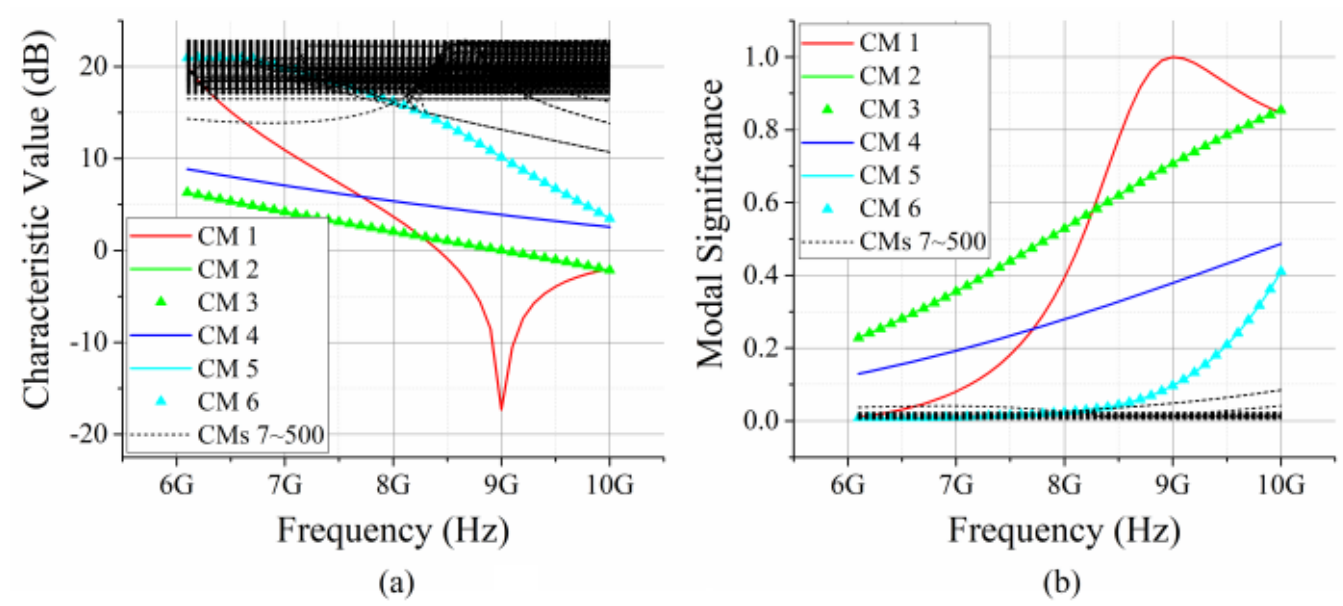

Fig. 3. $P^{\mathrm{DRIV}}$ -based (a) CVs and (b) MSs of a lossless cylinder, whose radius and height are both 6mm and which is with $\boldsymbol{\mu} = \mathbf{I}2\mu_0$ , $\boldsymbol{\varepsilon} = \mathbf{I}8\varepsilon_0$ , and $\boldsymbol{\sigma} = \mathbf{I}0$ .

### *D. Physical Picture of WEP-CMT*

It is easy to prove that the CMs derived from orthogonalizing $\mathrm{P}^{\mathrm{driv}}$ , that is, orthogonalizing $P^{\mathrm{driv}}$ , satisfy orthogonality relation

$$P_\xi^{\mathrm{driv}} \delta_{\xi\zeta} = (1/2)\left\langle \boldsymbol{J}_\xi^{\mathrm{SV}}, \boldsymbol{E}_\zeta^{\mathrm{inc}} \right\rangle_{V_{\mathrm{OSS}}} + (1/2)\left\langle \boldsymbol{M}_\xi^{\mathrm{SV}}, \boldsymbol{H}_\zeta^{\mathrm{inc}} \right\rangle_{V_{\mathrm{OSS}}} \tag{17}$$

and the modes derived from orthogonalizing $\mathrm{P}^{\mathrm{DRIV}}$ , that is, orthogonalizing $P^{\mathrm{DRIV}}$ , satisfy orthogonality relation

$$P_\xi^{\mathrm{DRIV}} \delta_{\xi\zeta} = (1/2)\left\langle \boldsymbol{J}_\xi^{\mathrm{SV}}, \boldsymbol{E}_\zeta^{\mathrm{inc}} \right\rangle_{V_{\mathrm{OSS}}} + (1/2)\left\langle \boldsymbol{H}_\xi^{\mathrm{inc}}, \boldsymbol{M}_\zeta^{\mathrm{SV}} \right\rangle_{V_{\mathrm{OSS}}} \tag{18}$$

where $\delta_{\xi\zeta}$ is Kronecker's delta symbol.

Obviously, the CMs satisfying (17) are completely decoupled, that is, the action by the $\zeta$ -th modal fields $(\boldsymbol{E}_\zeta^{\mathrm{inc}}, \boldsymbol{H}_\zeta^{\mathrm{inc}})$ on the $\xi$ -th modal currents $(\boldsymbol{J}_\xi^{\mathrm{SV}}, \boldsymbol{M}_\xi^{\mathrm{SV}})$ is zero if $\xi \neq \zeta$ . However, the modes satisfying (18) are not decoupled completely, as exhibited by the intertwining orthogonality (18) between $(\boldsymbol{E}_\zeta^{\mathrm{inc}}, \boldsymbol{M}_\zeta^{\mathrm{SV}})$ and $(\boldsymbol{J}_\xi^{\mathrm{SV}}, \boldsymbol{H}_\xi^{\mathrm{inc}})$ instead of between $(\boldsymbol{E}_\zeta^{\mathrm{inc}}, \boldsymbol{H}_\zeta^{\mathrm{inc}})$ and $(\boldsymbol{J}_\xi^{\mathrm{SV}}, \boldsymbol{M}_\xi^{\mathrm{SV}})$ . Due to this reason, the following parts of this paper focus on orthogonalizing frequency-domain DPO $P^{\mathrm{driv}}$ rather than $P^{\mathrm{DRIV}}$ .

In addition, the CMs satisfying frequency-domain orthogonality (17) also satisfy the following orthogonality relations

$$\mathrm{Re}\left\{P_\xi^{\mathrm{driv}}\right\} \delta_{\xi\zeta} = \frac{1}{T}\int_{t_0}^{t_0+T} \Big[ \left\langle \boldsymbol{J}_\xi^{\mathrm{SV}}(t), \boldsymbol{E}_\zeta^{\mathrm{inc}}(t) \right\rangle_{V_{\mathrm{OSS}}} + \left\langle \boldsymbol{M}_\xi^{\mathrm{SV}}(t), \boldsymbol{H}_\zeta^{\mathrm{inc}}(t) \right\rangle_{V_{\mathrm{OSS}}} \Big] dt \tag{19}$$

$$\mathrm{Re}\left\{P_\xi^{\mathrm{driv}}\right\} \delta_{\xi\zeta} = \frac{1}{2}\left\langle \frac{1}{\eta_0} \cdot \boldsymbol{E}_\xi^{\mathrm{sca}}, \boldsymbol{E}_\zeta^{\mathrm{sca}} \right\rangle_{S_\infty} + \frac{1}{2}\left\langle \boldsymbol{\sigma} \cdot \boldsymbol{E}_\xi^{\mathrm{tot}}, \boldsymbol{E}_\zeta^{\mathrm{tot}} \right\rangle_{V_{\mathrm{OSS}}} \tag{20}$$

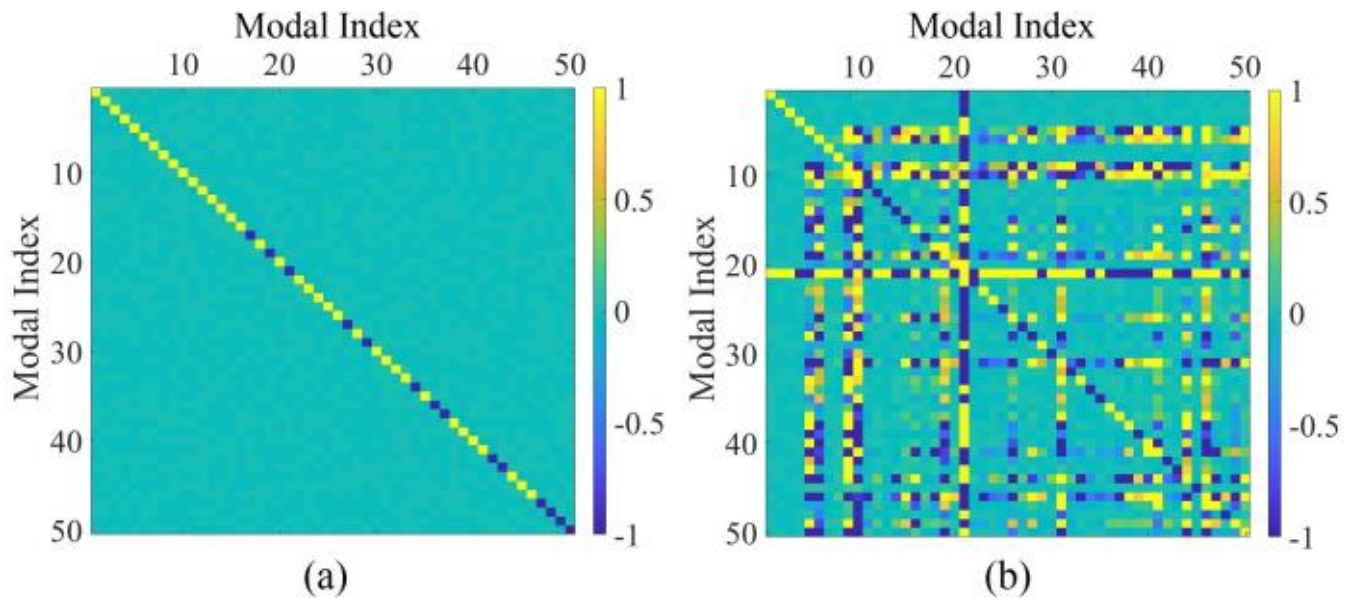

Fig. 4. Time-average DP orthogonality matrices about the modes derived from orthogonalizing frequency-domain DPOs (a) $P^{\rm driv}$ and (b) $P^{\rm DRIV}$, at 9 GHz.

The time-domain orthogonality relation (19) has a very clear physical interpretation: in any integral period, the $\zeta$-th modal fields $(\boldsymbol{E}_\zeta^{\rm inc}, \boldsymbol{H}_\zeta^{\rm inc})$ don't supply net energy to the $\xi$-th modal currents $(\boldsymbol{J}_\xi^{\rm SV}, \boldsymbol{M}_\xi^{\rm SV})$, if $\xi \neq \zeta$. Then, it clearly reveals the physical picture/purpose of WEP-CMT — to construct a set of steadily working energy-decoupled modes for OSS. To visually exhibit that the $P^{\rm driv}$-based CMs are time-averagely DP-decoupled but the $P^{\rm DRIV}$-based modes are not, the time-average DP orthogonality matrices corresponding to the modes (at 9 GHz) in Figs. 2 and 3 are shown in Fig. 4.

The frequency-domain orthogonality relation (20) clearly exhibits the fact that: when the OSS is lossy, that is, $\boldsymbol{\sigma} \neq \mathbf{I}0$, the modal far fields may not be orthogonal [9], [11], [13]. To visually exhibit this conclusion, we compute the modes of a lossy cylinder, whose radius and height are both 6mm and which is with $\boldsymbol{\mu} = \mathbf{I}2\mu_0$, $\boldsymbol{\varepsilon} = \mathbf{I}8\varepsilon_0$, and $\boldsymbol{\sigma} = \mathbf{I}1$, and show the corresponding CVs and MSs in Fig. 5, and also show the time-average DP orthogonality matrix and far-field orthogonality matrix of the modes (at 9 GHz) in Fig. 6. Obviously, the time-average DP orthogonality holds as shown in Fig. 6(a), but the far-field orthogonality doesn't hold as shown in Fig. 6(b). Here, we emphasize again that the physical purpose of WEP-CMT is to construct the modes without net energy exchange in integral period rather than the modes having orthogonal far fields. Two main reasons R1 and R2 leading to this conclusion are given as below.

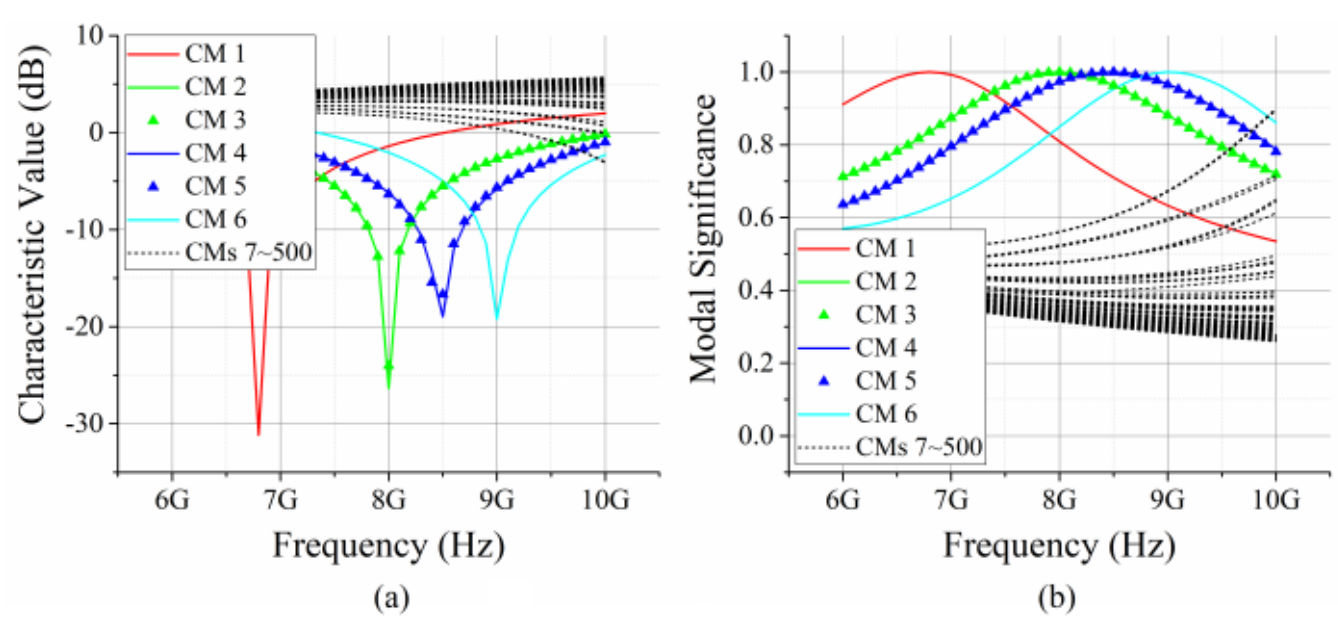

Fig. 5. $P^{\rm driv}$-based (a) CVs and (b) MSs of a lossy cylinder, whose radius and height are both 6mm and which is with $\boldsymbol{\mu} = \mathbf{I}2\mu_0$, $\boldsymbol{\varepsilon} = \mathbf{I}8\varepsilon_0$, and $\boldsymbol{\sigma} = \mathbf{I}1$.

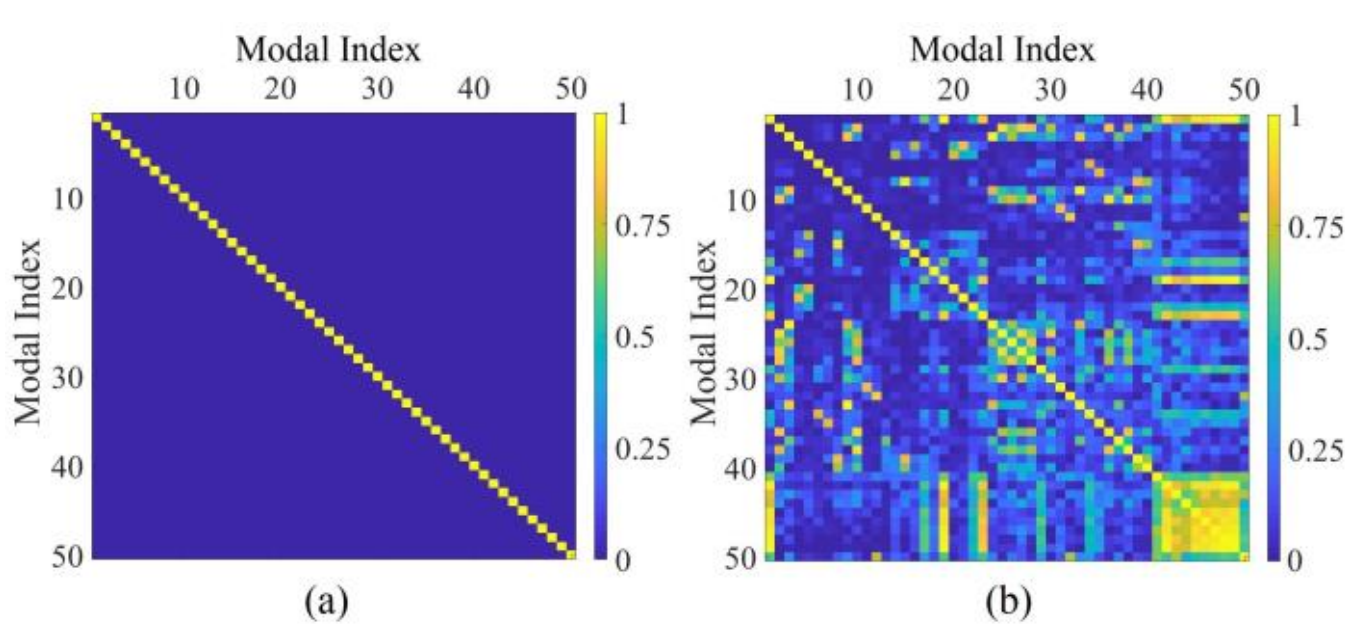

Fig. 6. (a) Time-average DP orthogonality matrix and (b) far-field orthogonality matrix of the modes (at 9 GHz) shown in Fig. 5.

R1. We consider two modes $(\boldsymbol{J}_1^{\rm SV}, \boldsymbol{M}_1^{\rm SV})$ and $(\boldsymbol{J}_2^{\rm SV}, \boldsymbol{M}_2^{\rm SV})$, which are driven by fields $(\boldsymbol{E}_1^{\rm inc}, \boldsymbol{H}_1^{\rm inc})$ and $(\boldsymbol{E}_2^{\rm inc}, \boldsymbol{H}_2^{\rm inc})$ respectively and generate fields $(\boldsymbol{E}_1^{\rm sca}, \boldsymbol{H}_1^{\rm sca})$ and $(\boldsymbol{E}_2^{\rm sca}, \boldsymbol{H}_2^{\rm sca})$ respectively. The modes have orthogonal far fields, but don't satisfy (19). Fields $(\boldsymbol{E}_1^{\rm inc}, \boldsymbol{H}_1^{\rm inc})$ carry information 1, and fields $(\boldsymbol{E}_2^{\rm inc}, \boldsymbol{H}_2^{\rm inc})$ carry information 2. Obviously, under the driving of $(\boldsymbol{E}_{1/2}^{\rm inc}, \boldsymbol{H}_{1/2}^{\rm inc})$, fields $(\boldsymbol{E}_{1/2}^{\rm sca}, \boldsymbol{H}_{1/2}^{\rm sca})$ carry the information 1 / information 2 from source zone to far zone. But, at the same time, fields $(\boldsymbol{E}_{1/2}^{\rm sca}, \boldsymbol{H}_{1/2}^{\rm sca})$ also carry the information 2 / information 1 from source zone to far zone, because $(\boldsymbol{E}_{1/2}^{\rm inc}, \boldsymbol{H}_{1/2}^{\rm inc})$ also provide energy to $(\boldsymbol{J}_{2/1}^{\rm SV}, \boldsymbol{M}_{2/1}^{\rm SV})$ due to the absence of (19). In other words, the modes not satisfying (19) cannot be driven independently, and then cannot achieve a real sense of decoupling for the information in far zone.

R2. Far-field orthogonality relation $< \boldsymbol{E}_\xi^{\rm sca}, \boldsymbol{E}_\zeta^{\rm sca} >_{S_\infty} = 0$ is a global relation rather than a local relation. Thus, it cannot be guaranteed that $(\boldsymbol{E}_\xi^{\rm sca})^* \cdot \boldsymbol{E}_\zeta^{\rm sca} = 0$ on the everywhere of $S_\infty$. This fact implies that: for any receiver not occupying whole $S_\infty$, relation $< \boldsymbol{E}_\xi^{\rm sca} \cdot \boldsymbol{E}_\zeta^{\rm sca} >_{\rm receiver} = 0$ cannot be guaranteed. In other words, global orthogonality relation $< \boldsymbol{E}_\xi^{\rm sca}, \boldsymbol{E}_\zeta^{\rm sca} >_{S_\infty} = 0$ cannot provide a real sense of decoupling for the fields on any finite-sized receiver.

### *E. Some Characteristic Quantities of CM*

In this subsection, we focus on clarifying the concepts related to some important characteristic quantities of CM.

**Characteristic Value (CV)**

As everyone knows, CV $\lambda_\xi$ and modal DP $P_\xi^{\rm driv}$ satisfy the following relation

$$\lambda_\xi = {\rm Im}\left\{P_\xi^{\rm driv}\right\} \Big/ {\rm Re}\left\{P_\xi^{\rm driv}\right\} \tag{21}$$

Here, the physical meaning of ${\rm Re}\{P_\xi^{\rm driv}\}$ has been very clear — "the time average of modal driving power" / "the summation of modal radiated power and modal dissipated power", but the physical meaning of ${\rm Im}\{P_\xi^{\rm driv}\}$ has not been clear (though the physical meaning of ${\rm Im}\{P_\xi^{\rm DRIV}\}$ is relatively clear as shown in (10)), and then the physical meaning of $\lambda_\xi$ has not been clear too. In addition, because ${\rm Im}\{P_\xi^{\rm driv}\} \neq \pm {\rm Im}\{P_\xi^{\rm DRIV}\}$ generally for magnetodielectric OSSs (which satisfy $\boldsymbol{\mu} \neq \mathbf{I}\mu_0$ & $\boldsymbol{\varepsilon} \neq \mathbf{I}\varepsilon_0$), thus the zero CVs don't necessarily correspond to the resonant CMs, that is, $P^{\rm driv}$-based $\lambda_\xi = 0$ doesn't imply ${\rm Im}\{P_\xi^{\rm DRIV}\} = 0$, for magnetodielectric OSSs.

Very naturally, the above observations lead to the following questions Q1 and Q2.

Q1 What is the physical meaning of the $P^{\rm driv}$-based CVs?

Q2 When the OSS is magnetodielectric, what do the $P^{\rm driv}$-based zero CVs imply physically?

The answer to the question Q1 is as follows: in fact, this paper thinks that there is no need to expect $\lambda_\xi$ to have a clear physical meaning, because the CVs are only the byproducts during the process to construct CMs (the physical purpose of WEP-CMT is to construct the CMs, rather than to obtain the associated CVs). The answer to the question Q2 will be clear after our clarifying the physical meaning of characteristic quantity MS under WEP framework.

**Modal Normalization Factor**

Any working mode $(\mathrm{a}, \boldsymbol{C}^{\text{SV}}, \boldsymbol{F}^{\text{inc}})$ under the driving of a previously known incident field can be linearly expanded in terms of the CMs $\{\mathrm{a}_\xi, \boldsymbol{C}_\xi^{\text{SV}}, \boldsymbol{F}_\xi^{\text{inc}}\}$ as follows:

$$\left.\begin{aligned} \mathrm{a} &= \sum\nolimits_\xi c_\xi \mathrm{a}_\xi \\ \boldsymbol{C}^{\text{SV}} &= \sum\nolimits_\xi c_\xi \boldsymbol{C}_\xi^{\text{SV}} \\ \boldsymbol{F}^{\text{inc}} &= \sum\nolimits_\xi c_\xi \boldsymbol{F}_\xi^{\text{inc}} \end{aligned}\right\} \tag{22}$$

with expansion coefficients

$$c_\xi = \frac{1}{P_\xi^{\text{driv}}}\Big[(1/2)\big\langle \boldsymbol{J}_\xi^{\text{SV}}, \boldsymbol{E}^{\text{inc}}\big\rangle_{V_{\text{OSS}}} + (1/2)\big\langle \boldsymbol{M}_\xi^{\text{SV}}, \boldsymbol{H}^{\text{inc}}\big\rangle_{V_{\text{OSS}}}\Big] \tag{23}$$

because of the modal orthogonality (17) and completeness.

If we want $c_\xi$ to have ability to quantitatively reflect the weight of a CM in whole modal expansion formulation, it is necessary to appropriately normalize the corresponding CM. To find a reasonable normalization factor, we employ the time-domain orthogonality relation (19) due to its clear physical meaning. From (19), it is easy to derive the following time-average DP expansion formulation

$$\begin{aligned} &\frac{1}{T}\int_{t_0}^{t_0+T}\Big[\big\langle \boldsymbol{J}^{\text{SV}}(t), \boldsymbol{E}^{\text{inc}}(t)\big\rangle_{V_{\text{OSS}}} + \big\langle \boldsymbol{M}^{\text{SV}}(t), \boldsymbol{H}^{\text{inc}}(t)\big\rangle_{V_{\text{OSS}}}\Big]dt \\ &= \operatorname{Re}\big\{P^{\text{driv}}\big\} \\ &= \sum\nolimits_\xi |c_\xi|^2 \operatorname{Re}\big\{P_\xi^{\text{driv}}\big\} \end{aligned} \tag{24}$$

where the first equality is evident [24]-[25], and the second equality is due to expansion formulation (22) and orthogonality relation (17). From time-average DP expansion formulation (24), it is easy to conclude that: a reasonable modal normalization way is to normalize $\operatorname{Re}\{P_\xi^{\text{driv}}\}$ to 1, that is, a reasonable modal normalization factor is the square root of time-average modal DP $\operatorname{Re}\{P_\xi^{\text{driv}}\}$.

By using $\sqrt{\operatorname{Re}\{P_\xi^{\text{driv}}\}}$ as the modal normalization factor, it traditionally has that [4]-[7]

$$P_\xi^{\text{driv}} = \underbrace{\operatorname{Re}\big\{P_\xi^{\text{driv}}\big\}}_{1} + j\,\underbrace{\operatorname{Im}\big\{P_\xi^{\text{driv}}\big\}}_{\lambda_\xi} \tag{25}$$

and hence we immediately obtain the following famous Parseval's identity

$$\begin{aligned} &\frac{1}{T}\int_{t_0}^{t_0+T}\Big[\big\langle \boldsymbol{J}^{\text{SV}}(t), \boldsymbol{E}^{\text{inc}}(t)\big\rangle_{V_{\text{OSS}}} + \big\langle \boldsymbol{M}^{\text{SV}}(t), \boldsymbol{H}^{\text{inc}}(t)\big\rangle_{V_{\text{OSS}}}\Big]dt \\ &= \sum\nolimits_\xi |c_\xi|^2 \end{aligned} \tag{26}$$

Substituting the above (25) into expansion coefficient (23), it is immediate to have that

$$c_\xi = \frac{1}{1+j\lambda_\xi}\Big[(1/2)\big\langle \boldsymbol{J}_\xi^{\text{SV}}, \boldsymbol{E}^{\text{inc}}\big\rangle_{V_{\text{OSS}}} + (1/2)\big\langle \boldsymbol{M}_\xi^{\text{SV}}, \boldsymbol{H}^{\text{inc}}\big\rangle_{V_{\text{OSS}}}\Big] \tag{27}$$

in which modal currents $\{\boldsymbol{J}_\xi^{\text{SV}}, \boldsymbol{M}_\xi^{\text{SV}}\}$ have been normalized such that $\operatorname{Re}\{P_\xi^{\text{driv}}\} = 1$.

**Modal Significance (MS)**

Based on expansion coefficient formulation (27) and time-average DP expansion formulation (24), the following characteristic quantity — modal significance (MS) — is usually used to quantitatively describe the weight of a CM in whole modal expansion formulation [8].

$$\text{MS}_\xi = \frac{1}{|1+j\lambda_\xi|} = \frac{\operatorname{Re}\big\{P_\xi^{\text{driv}}\big\}}{\big|P_\xi^{\text{driv}}\big|} \tag{28}$$

where the second equality is obvious because of (25). It is thus clear that, besides the above-mentioned physical meaning — modal weight of a CM in whole modal expansion formulation, $\text{MS}_\xi$ has another noteworthy physical meaning — weight of the $\operatorname{Re}\{P_\xi^{\text{driv}}\}$ in whole modal DP $P_\xi^{\text{driv}}$.

In addition, it is evident that $\lambda_\xi = 0$ if and only if $\text{MS}_\xi = 1$ (because $\lambda_\xi$ is purely real). This implies that the CMs with $\lambda_\xi = 0$ have significant weights in whole expansion formulation. In fact, this is just the answer to the previous question Q2 "what do the $P^{\text{driv}}$-based zero CVs imply physically?".

Because of the clear physical meaning of MS, the following parts of this paper will always use MS rather than CV.

### *F. Physical Modes and Spurious Modes*

The time average of modal DP is equal to the summation of modal radiated power and modal dissipated power as shown in (19) and (20), so it must be non-negative. However, there exist some negative elements distributing on the diagonal of the orthogonality matrix shown in Fig. 4(a), that is, the negative elements correspond to the modes having negative time-average DP. This subsection focuses on explaining this phenomenon.

Due to volume equivalence principle $\boldsymbol{J}^{\text{SV}} = j\omega\Delta\boldsymbol{\varepsilon}^{\text{c}}\cdot\boldsymbol{E}^{\text{tot}}$ & $\boldsymbol{M}^{\text{SV}} = j\omega\Delta\boldsymbol{\mu}\cdot\boldsymbol{H}^{\text{tot}}$, and Maxwell's equations $\nabla\times\boldsymbol{H}^{\text{tot}} = j\omega\boldsymbol{\varepsilon}^{\text{c}}\cdot\boldsymbol{E}^{\text{tot}}$ & $\nabla\times\boldsymbol{E}^{\text{tot}} = -j\omega\boldsymbol{\mu}\cdot\boldsymbol{H}^{\text{tot}}$ [29], $\boldsymbol{J}^{\text{SV}}$ and $\boldsymbol{M}^{\text{SV}}$ satisfy the following dependence relations

$$\nabla\times\Big[\big(j\omega\Delta\boldsymbol{\mu}\big)^{-1}\cdot\boldsymbol{M}^{\text{SV}}\Big] = \quad j\omega\boldsymbol{\varepsilon}\cdot\Big[\big(j\omega\Delta\boldsymbol{\varepsilon}^{\text{c}}\big)^{-1}\cdot\boldsymbol{J}^{\text{SV}}\Big] \tag{29}$$

$$\nabla\times\Big[\big(j\omega\Delta\boldsymbol{\varepsilon}^{\text{c}}\big)^{-1}\cdot\boldsymbol{J}^{\text{SV}}\Big] = -j\omega\boldsymbol{\mu}\cdot\Big[\big(j\omega\Delta\boldsymbol{\mu}\big)^{-1}\cdot\boldsymbol{M}^{\text{SV}}\Big] \tag{30}$$

so $\boldsymbol{J}^{\text{SV}}$ and $\boldsymbol{M}^{\text{SV}}$ are not independent of each other.

In the previous processes to construct CMs, the dependence relations (29) and (30) are not considered, such that some modes don't satisfy the relations (though some other modes indeed satisfy the relations automatically). The dependence relations (29) and (30) are essentially the Maxwell's equations,

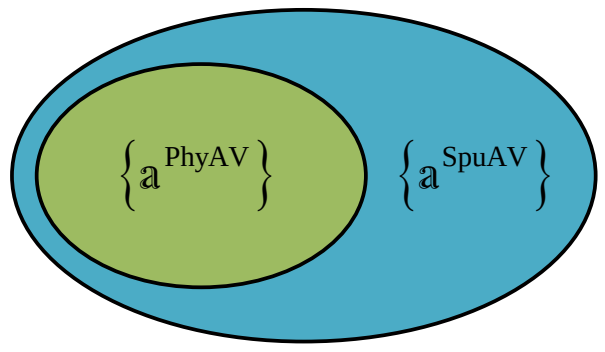

$$\{a^{\text{AV}}\}=\{a^{\text{PhyAV}}\}\bigcup\{a^{\text{SpuAV}}\}$$

Fig. 7. Relations among space $\{a^{\text{AV}}\}$, space $\{a^{\text{PhyAV}}\}$, and set $\{a^{\text{SpuAV}}\}$.

that is the physical law satisfied by $\boldsymbol{F}^{\text{tot}}$, so the modes satisfying them are called physical modes, while the modes not satisfying them are called unphysical modes (or more conventionally called spurious modes).

In the following parts of this paper, the physical modes and spurious modes are denoted as $a^{\text{PhyAV}}$ and $a^{\text{SpuAV}}$ respectively. Obviously, all the $a^{\text{AV}}$ constitute a linear space $\{a^{\text{AV}}\}$, and all the $a^{\text{PhyAV}}$ also constitute a linear space — modal space $\{a^{\text{PhyAV}}\}$. But, the set $\{a^{\text{SpuAV}}\}$ constituted by all the $a^{\text{SpuAV}}$ is not a linear space, because the set is not closed for addition, for example, both $a^{\text{PhyAV}}+a^{\text{SpuAV}}$ and $-a^{\text{SpuAV}}$ are spurious but their summation $(a^{\text{PhyAV}}+a^{\text{SpuAV}})+(-a^{\text{SpuAV}})$ is evidently physical. In addition, it is obvious that $\{a^{\text{AV}}\}=\{a^{\text{PhyAV}}\}\bigcup\{a^{\text{SpuAV}}\}$ as illustrated in Fig. 7.

Now, a very important question is how to effectively integrate the dependence relations between electric and magnetic currents into characteristic equation, such that all the modes outputted from the equation are physical? In the following section, we will carefully answer this question by focusing on the surface formulations of the WEP-CMT for material OSSs.

## III. Surface Formulation of the WEP-CMT for Material Scattering Systems with SDC Scheme

This section considers the case that the OSS $V_{\text{OSS}}$ is constituted by two material bodies $V_1$ and $V_2$ with parameters $(\boldsymbol{\mu}_1, \boldsymbol{\varepsilon}_1^{\text{c}}=\boldsymbol{\varepsilon}_1 - j\boldsymbol{\sigma}_1/\omega)$ and $(\boldsymbol{\mu}_2, \boldsymbol{\varepsilon}_2^{\text{c}}=\boldsymbol{\varepsilon}_2 - j\boldsymbol{\sigma}_2/\omega)$ respectively, as shown in Fig. 8. The boundaries of $V_1$ and $V_2$ are denoted as $S_1$ and $S_2$ respectively, and $S_1=S_{10}\bigcup S_{12}$ and $S_2=S_{20}\bigcup S_{21}$. Here, $S_{10}/S_{20}$ is the interface between $V_1/V_2$ and environment, and $S_{12}/S_{21}$ is the interface between $V_1/V_2$ and $V_2/V_1$, and it is obvious that $S_{12}=S_{21}$.

Originating from famous Huygens-Fresnel principle [26], the following surface equivalence principles (SEPs) can be obtained

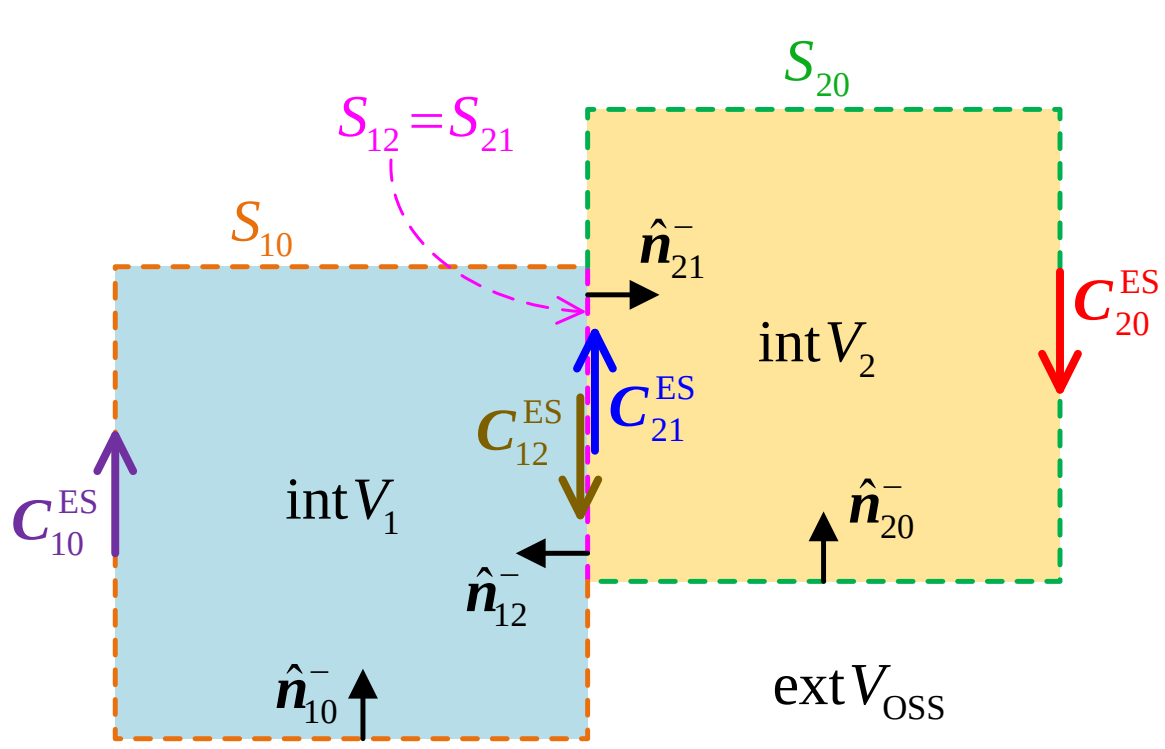

Fig. 8. Two-body material OSS considered in Sec. III.

$$\left.\begin{array}{l}\text{ext}\,V_{\text{OSS}}:-\boldsymbol{F}^{\text{sca}}\\ \text{int}\,V_{\text{OSS}}:\ \ \boldsymbol{F}^{\text{inc}}\end{array}\right\} = \mathcal{F}_0\left(\boldsymbol{J}_{10}^{\text{ES}}+\boldsymbol{J}_{20}^{\text{ES}},\boldsymbol{M}_{10}^{\text{ES}}+\boldsymbol{M}_{20}^{\text{ES}}\right) \tag{31}$$

$$\text{int}\,V_1 \ : \ \boldsymbol{F}^{\text{tot}} = \mathcal{F}_1\Big(\underbrace{\boldsymbol{J}_{10}^{\text{ES}}+\boldsymbol{J}_{12}^{\text{ES}}}_{\boldsymbol{J}_1^{\text{ES}}},\underbrace{\boldsymbol{M}_{10}^{\text{ES}}+\boldsymbol{M}_{12}^{\text{ES}}}_{\boldsymbol{M}_1^{\text{ES}}}\Big) \tag{32}$$

$$\text{int}\,V_2 \ : \ \boldsymbol{F}^{\text{tot}} = \mathcal{F}_2\Big(\underbrace{\boldsymbol{J}_{20}^{\text{ES}}+\boldsymbol{J}_{21}^{\text{ES}}}_{\boldsymbol{J}_2^{\text{ES}}},\underbrace{\boldsymbol{M}_{20}^{\text{ES}}+\boldsymbol{M}_{21}^{\text{ES}}}_{\boldsymbol{M}_2^{\text{ES}}}\Big) \tag{33}$$

Here, $\text{ext}\,V_{\text{OSS}}$ and $\text{int}\,V_{\text{OSS}}$ denote the exterior and interior of $V_{\text{OSS}}$ respectively, and $\text{int}\,V_1$ and $\text{int}\,V_2$ are the interiors of $V_1$ and $V_2$ respectively; operator $\mathcal{F}_{0/1/2}$ is defined as that $\mathcal{F}_{0/1/2}(\boldsymbol{J},\boldsymbol{M})=\mathbf{G}_{0/1/2}^{JF}*\boldsymbol{J}+\mathbf{G}_{0/1/2}^{MF}*\boldsymbol{M}$, where $\mathbf{G}_{0/1/2}^{JF}$ and $\mathbf{G}_{0/1/2}^{MF}$ are the dyadic Green's functions corresponding to material parameters $(\mu_0,\varepsilon_0)/(\boldsymbol{\mu}_1,\boldsymbol{\varepsilon}_1^{\text{c}})/(\boldsymbol{\mu}_2,\boldsymbol{\varepsilon}_2^{\text{c}})$; the equivalent surface electric current $\boldsymbol{J}_{10/12/21/20}^{\text{ES}}$ and equivalent surface magnetic current $\boldsymbol{M}_{10/12/21/20}^{\text{ES}}$ are defined as follows:

$$\boldsymbol{J}_{10/12/21/20}^{\text{ES}} = \hat{\boldsymbol{n}}_{10/12/21/20}^{-}\times\boldsymbol{H}_{-}^{\text{tot}} \tag{34}$$

$$\boldsymbol{M}_{10/12/21/20}^{\text{ES}} = \boldsymbol{E}_{-}^{\text{tot}}\times\hat{\boldsymbol{n}}_{10/12/21/20}^{-} \tag{35}$$

where $\hat{\boldsymbol{n}}_{10/12/21/20}^{-}$ is the inner normal direction of $S_{10/12/21/20}$ as shown in Fig. 8, and $\boldsymbol{E}_{-}^{\text{tot}}$ and $\boldsymbol{H}_{-}^{\text{tot}}$ are the fields distributing on the inner surface of $S_{10/12/21/20}$.

In [13], it has been proved that the interaction between $(\boldsymbol{E}^{\text{inc}},\boldsymbol{H}^{\text{inc}})$ and $(\boldsymbol{J}_{1/2}^{\text{SV}},\boldsymbol{M}_{1/2}^{\text{SV}})$ and the interaction between $(\boldsymbol{E}^{\text{inc}},\boldsymbol{H}^{\text{inc}})$ and $(\boldsymbol{J}_{1/2}^{\text{ES}},\boldsymbol{M}_{1/2}^{\text{ES}})$ satisfy the following relation

$$\begin{aligned}&(1/2)\left\langle\boldsymbol{J}_{1/2}^{\text{SV}},\boldsymbol{E}^{\text{inc}}\right\rangle_{V_{1/2}}+(1/2)\left\langle\boldsymbol{M}_{1/2}^{\text{SV}},\boldsymbol{H}^{\text{inc}}\right\rangle_{V_{1/2}}\\&=-(1/2)\left\langle\boldsymbol{J}_{1/2}^{\text{ES}},\boldsymbol{E}^{\text{inc}}\right\rangle_{S_{1/2}}-(1/2)\left\langle\boldsymbol{M}_{1/2}^{\text{ES}},\boldsymbol{H}^{\text{inc}}\right\rangle_{S_{1/2}}\end{aligned} \tag{36}$$

and $\boldsymbol{C}_{21}^{\text{ES}}=-\boldsymbol{C}_{12}^{\text{ES}}$, so

$$\begin{aligned}P^{\text{driv}}&=(1/2)\left\langle\boldsymbol{J}_1^{\text{SV}}+\boldsymbol{J}_2^{\text{SV}},\boldsymbol{E}^{\text{inc}}\right\rangle_{V_{\text{OSS}}}+(1/2)\left\langle\boldsymbol{M}_1^{\text{SV}}+\boldsymbol{M}_2^{\text{SV}},\boldsymbol{H}^{\text{inc}}\right\rangle_{V_{\text{OSS}}}\\&=-\frac{1}{2}\left\langle\boldsymbol{J}_{10}^{\text{ES}}+\boldsymbol{J}_{20}^{\text{ES}},\boldsymbol{E}^{\text{inc}}\right\rangle_{S_{10}\cup S_{20}}-\frac{1}{2}\left\langle\boldsymbol{M}_{10}^{\text{ES}}+\boldsymbol{M}_{20}^{\text{ES}},\boldsymbol{H}^{\text{inc}}\right\rangle_{S_{10}\cup S_{20}}\end{aligned} \tag{37}$$

In fact, the above (37) is just the original surface formulation for the frequency-domain DPO of the material OSS $V_{\text{OSS}}$ shown in Fig. 8.

In the following parts of this section, we will compare the numerical performances of two different surface formulations of DPO $P^{\text{driv}}$, which utilize different ways to express the tangential $\boldsymbol{F}^{\text{inc}}$ on $S_{10}\bigcup S_{20}$, and propose a new scheme for suppressing spurious modes.

### *A. PMCHWT-Based Surface CM Formulation with Two Different Spurious Mode Suppression Schemes*

Because $\boldsymbol{F}^{\text{inc}}=\boldsymbol{F}^{\text{tot}}-\boldsymbol{F}^{\text{sca}}$, and the tangential components of $\boldsymbol{F}^{\text{tot}}$ and $\boldsymbol{F}^{\text{sca}}$ on $S_{10}\bigcup S_{20}$ are continuous, and the $\boldsymbol{F}^{\text{tot}}$ on $S_{10}^{-}\bigcup S_{20}^{-}$ (which is the internal surface of $S_{10}\bigcup S_{20}$) and the $\boldsymbol{F}^{\text{sca}}$ on $S_{10}^{+}\bigcup S_{20}^{+}$ (which is the external surface of $S_{10}\bigcup S_{20}$) can be expressed in terms of the equivalent surface currents as (31)-(33), thus

$$\begin{aligned}P^{\text{driv}} =& -(1/2)\left\langle \boldsymbol{J}_{10}^{\text{ES}}+\boldsymbol{J}_{20}^{\text{ES}}, \boldsymbol{E}_{-}^{\text{tot}}-\boldsymbol{E}_{+}^{\text{sca}} \right\rangle_{S_{10}\cup S_{20}}\\
& -(1/2)\left\langle \boldsymbol{M}_{10}^{\text{ES}}+\boldsymbol{M}_{20}^{\text{ES}}, \boldsymbol{H}_{-}^{\text{tot}}-\boldsymbol{H}_{+}^{\text{sca}} \right\rangle_{S_{10}\cup S_{20}}\\
=& -(1/2)\left\langle \boldsymbol{J}_{10}^{\text{ES}}, \mathcal{E}_1\left(\boldsymbol{J}_{10}^{\text{ES}}+\boldsymbol{J}_{12}^{\text{ES}}, \boldsymbol{M}_{10}^{\text{ES}}+\boldsymbol{M}_{12}^{\text{ES}}\right)\right\rangle_{S_{10}^-}\\
& -(1/2)\left\langle \boldsymbol{J}_{20}^{\text{ES}}, \mathcal{E}_2\left(\boldsymbol{J}_{20}^{\text{ES}}-\boldsymbol{J}_{12}^{\text{ES}}, \boldsymbol{M}_{20}^{\text{ES}}-\boldsymbol{M}_{12}^{\text{ES}}\right)\right\rangle_{S_{20}^-}\\
& -(1/2)\left\langle \boldsymbol{J}_{10}^{\text{ES}}+\boldsymbol{J}_{20}^{\text{ES}}, \mathcal{E}_0\left(\boldsymbol{J}_{10}^{\text{ES}}+\boldsymbol{J}_{20}^{\text{ES}}, \boldsymbol{M}_{10}^{\text{ES}}+\boldsymbol{M}_{20}^{\text{ES}}\right)\right\rangle_{S_{10}^+\cup S_{20}^+}\\
& -(1/2)\left\langle \boldsymbol{M}_{10}^{\text{ES}}, \mathcal{H}_1\left(\boldsymbol{J}_{10}^{\text{ES}}+\boldsymbol{J}_{12}^{\text{ES}}, \boldsymbol{M}_{10}^{\text{ES}}+\boldsymbol{M}_{12}^{\text{ES}}\right)\right\rangle_{S_{10}^-}\\
& -(1/2)\left\langle \boldsymbol{M}_{20}^{\text{ES}}, \mathcal{H}_2\left(\boldsymbol{J}_{20}^{\text{ES}}-\boldsymbol{J}_{12}^{\text{ES}}, \boldsymbol{M}_{20}^{\text{ES}}-\boldsymbol{M}_{12}^{\text{ES}}\right)\right\rangle_{S_{20}^-}\\
& -(1/2)\left\langle \boldsymbol{M}_{10}^{\text{ES}}+\boldsymbol{M}_{20}^{\text{ES}}, \mathcal{H}_0\left(\boldsymbol{J}_{10}^{\text{ES}}+\boldsymbol{J}_{20}^{\text{ES}}, \boldsymbol{M}_{10}^{\text{ES}}+\boldsymbol{M}_{20}^{\text{ES}}\right)\right\rangle_{S_{10}^+\cup S_{20}^+} \quad (38)\end{aligned}$$

where relation $\boldsymbol{C}_{21}^{\text{ES}} = -\boldsymbol{C}_{12}^{\text{ES}}$ has been utilized.

If the related equivalent surface currents are expanded in terms of some basis functions, then the $P^{\text{driv}}$ given in (38) can be discretized into the following matrix form

$$P^{\text{driv}} = \underbrace{\begin{bmatrix} \mathrm{a}^{\boldsymbol{J}_{10}^{\text{ES}}} \\ \mathrm{a}^{\boldsymbol{J}_{12}^{\text{ES}}} \\ \mathrm{a}^{\boldsymbol{J}_{20}^{\text{ES}}} \\ \mathrm{a}^{\boldsymbol{M}_{10}^{\text{ES}}} \\ \mathrm{a}^{\boldsymbol{M}_{12}^{\text{ES}}} \\ \mathrm{a}^{\boldsymbol{M}_{20}^{\text{ES}}} \end{bmatrix}}_{\mathrm{a}^{\text{AV}}}^{H} \cdot \left[(1/2)\mathbb{Z}^{\text{PMCHWT}}\right] \cdot \underbrace{\begin{bmatrix} \mathrm{a}^{\boldsymbol{J}_{10}^{\text{ES}}} \\ \mathrm{a}^{\boldsymbol{J}_{12}^{\text{ES}}} \\ \mathrm{a}^{\boldsymbol{J}_{20}^{\text{ES}}} \\ \mathrm{a}^{\boldsymbol{M}_{10}^{\text{ES}}} \\ \mathrm{a}^{\boldsymbol{M}_{12}^{\text{ES}}} \\ \mathrm{a}^{\boldsymbol{M}_{20}^{\text{ES}}} \end{bmatrix}}_{\mathrm{a}^{\text{AV}}} \quad (39)$$

Here, the sub-vectors of $\mathrm{a}^{\text{AV}}$ are constituted by the expansion coefficients of the corresponding currents; the formulations for calculating the elements of $\mathbb{Z}^{\text{PMCHWT}}$ can be found in [13].

If the dependence relations among the currents are ignored, then the modes of $V_{\text{OSS}}$ can be derived from directly solving characteristic equation $\mathbb{Z}_{-}^{\text{PMCHWT}} \cdot \mathrm{a}_{\xi}^{\text{AV}} = \lambda_{\xi}\, \mathbb{Z}_{+}^{\text{PMCHWT}} \cdot \mathrm{a}_{\xi}^{\text{AV}}$ where $\mathbb{Z}_{+}^{\text{PMCHWT}}$ and $\mathbb{Z}_{-}^{\text{PMCHWT}}$ are the positive and negative Hermitian parts of $\mathbb{Z}^{\text{PMCHWT}}$.

Now, we consider a two-body OSS shown in Fig. 9, and the OSS is constituted by two circular cylinders, whose radiuses are both 6mm and heights are both 3mm. When the OSS is lossless and with parameters $(\boldsymbol{\mu}_1 = \mathbf{I}2\mu_0, \boldsymbol{\varepsilon}_1^{\text{c}} = \mathbf{I}8\varepsilon_0)$ and $(\boldsymbol{\mu}_2 = \mathbf{I}8\mu_0, \boldsymbol{\varepsilon}_2^{\text{c}} = \mathbf{I}2\varepsilon_0)$, the MSs of the first 500 CMs derived from volume formulation and the above PMCHWT-based surface formulation $\mathbb{Z}_{-}^{\text{PMCHWT}} \cdot \mathrm{a}_{\xi}^{\text{AV}} = \lambda_{\xi}\, \mathbb{Z}_{+}^{\text{PMCHWT}} \cdot \mathrm{a}_{\xi}^{\text{AV}}$ are shown in Fig. 10(a) and Fig. 10(c). When the OSS is lossy and with parameters $(\boldsymbol{\mu}_1 = \mathbf{I}2\mu_0, \boldsymbol{\varepsilon}_1^{\text{c}} = \mathbf{I}8\varepsilon_0 - j\mathbf{I}1/\omega)$ and $(\boldsymbol{\mu}_2 = \mathbf{I}8\mu_0, \boldsymbol{\varepsilon}_2^{\text{c}} = \mathbf{I}2\varepsilon_0)$, the MSs of the first 500 CMs derived from volume formulation and the above PMCHWT-based surface formulation $\mathbb{Z}_{-}^{\text{PMCHWT}} \cdot \mathrm{a}_{\xi}^{\text{AV}} = \lambda_{\xi}\, \mathbb{Z}_{+}^{\text{PMCHWT}} \cdot \mathrm{a}_{\xi}^{\text{AV}}$ are shown in Fig. 10(b) and Fig. 10(d). Obviously, the above PMCHWT-based surface formulation outputs many spurious modes.

Below, we first simply review an existed scheme — dependent variable elimination (DVE) [11]-[13], and then propose a novel scheme — solution domain compression (SDC), for suppressing the spurious modes.

### PMCHWT-Based Surface CM Formulation with DVE

The reason leading to the spurious modes is the overlooking of the dependence relations among the currents contained in DPO $P^{\text{driv}}$. To establish the dependence relations, [13] provided a series of integral equations as follows:

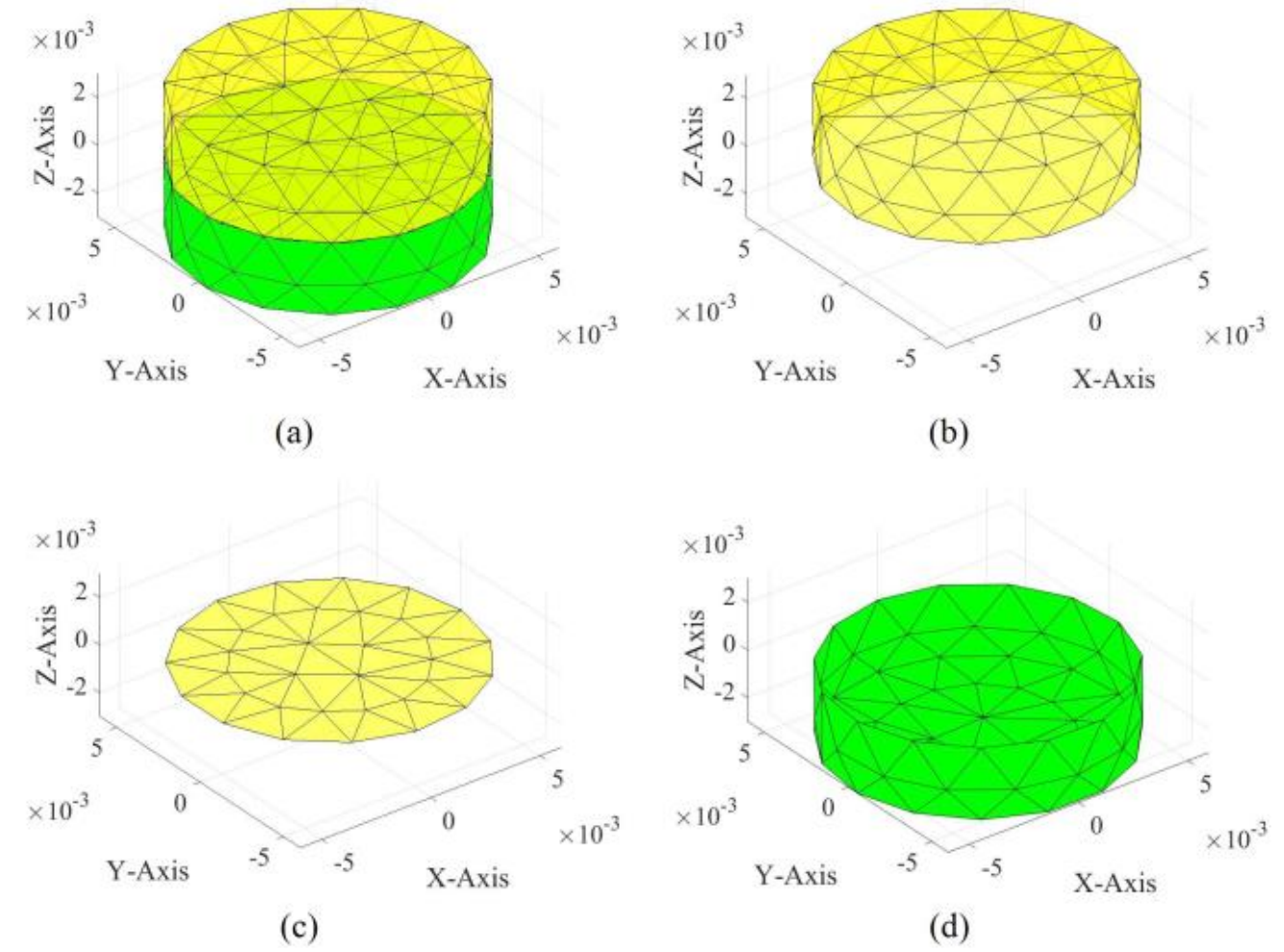

Fig. 9. (a) Topological structure and surface meshes of a two-body OSS; (b) $S_{10}$ of the OSS; (c) $S_{12}$ of the OSS; (d) $S_{20}$ of the OSS.

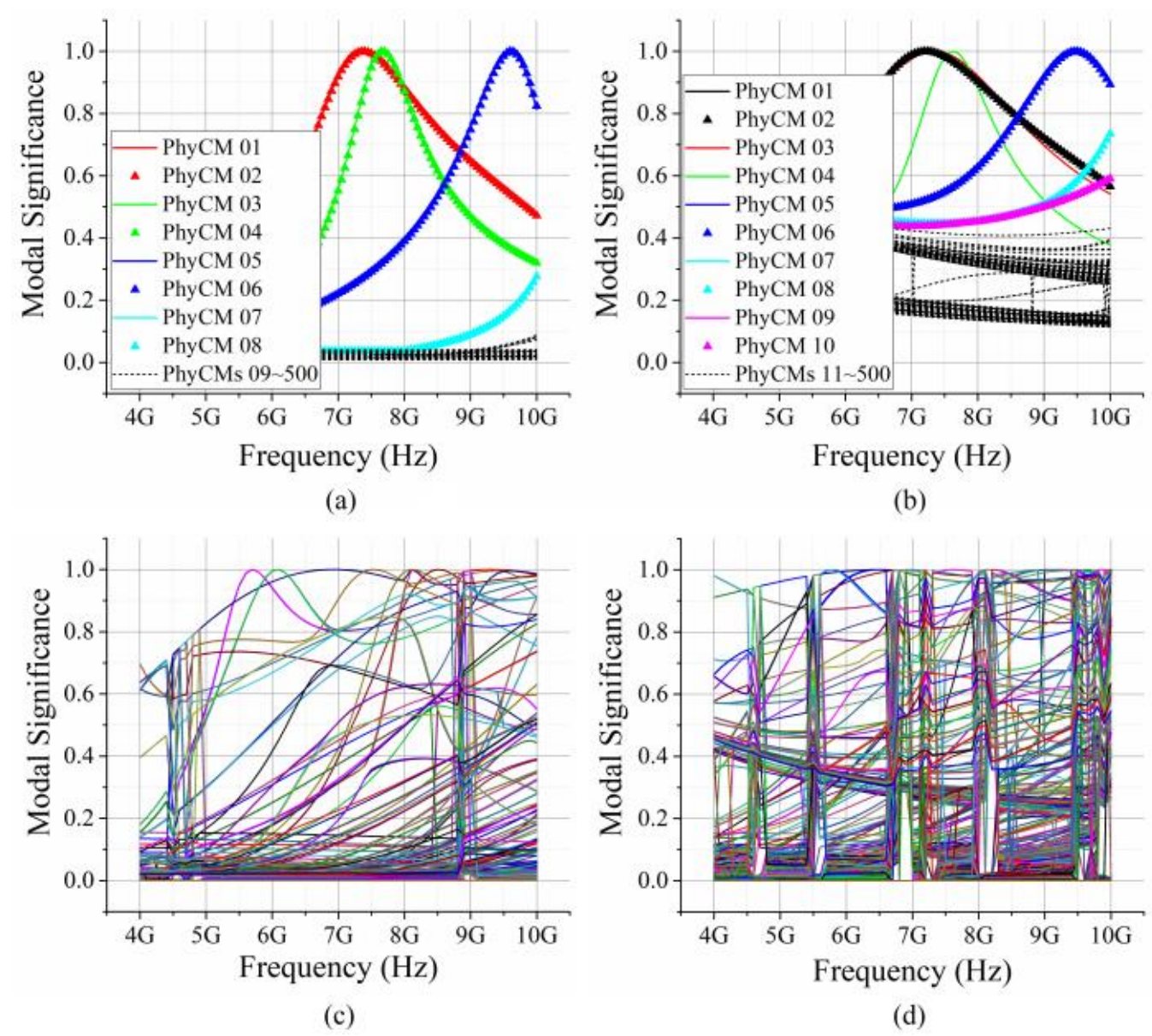

Fig. 10. (a) (8)-based MSs of a lossless two-body OSS with topological structure Fig. 9 and with material parameters $(\boldsymbol{\mu}_1 = \mathbf{I}2\mu_0, \boldsymbol{\varepsilon}_1^{\text{c}} = \mathbf{I}8\varepsilon_0)$ and $(\boldsymbol{\mu}_2 = \mathbf{I}8\mu_0, \boldsymbol{\varepsilon}_2^{\text{c}} = \mathbf{I}2\varepsilon_0)$; (b) (8)-based MSs of a lossy two-body OSS with topological structure Fig. 9 and with material parameters $(\boldsymbol{\mu}_1 = \mathbf{I}2\mu_0, \boldsymbol{\varepsilon}_1^{\text{c}} = \mathbf{I}8\varepsilon_0 - j\mathbf{I}1/\omega)$ and $(\boldsymbol{\mu}_2 = \mathbf{I}8\mu_0, \boldsymbol{\varepsilon}_2^{\text{c}} = \mathbf{I}2\varepsilon_0)$; (c) (39)-based MSs of a lossless two-body OSS with topological structure Fig. 9 and with material parameters $(\boldsymbol{\mu}_1 = \mathbf{I}2\mu_0, \boldsymbol{\varepsilon}_1^{\text{c}} = \mathbf{I}8\varepsilon_0)$ and $(\boldsymbol{\mu}_2 = \mathbf{I}8\mu_0, \boldsymbol{\varepsilon}_2^{\text{c}} = \mathbf{I}2\varepsilon_0)$; (d) (39)-based MSs of a lossy two-body OSS with topological structure Fig. 9 and with material parameters $(\boldsymbol{\mu}_1 = \mathbf{I}2\mu_0, \boldsymbol{\varepsilon}_1^{\text{c}} = \mathbf{I}8\varepsilon_0 - j\mathbf{I}1/\omega)$ and $(\boldsymbol{\mu}_2 = \mathbf{I}8\mu_0, \boldsymbol{\varepsilon}_2^{\text{c}} = \mathbf{I}2\varepsilon_0)$.

$$0 = \left[\mathcal{H}_1\left(\boldsymbol{J}_{10}^{\text{ES}}+\boldsymbol{J}_{12}^{\text{ES}}, \boldsymbol{M}_{10}^{\text{ES}}+\boldsymbol{M}_{12}^{\text{ES}}\right)\right]_{S_{10}^-}^{\tan} + \hat{\boldsymbol{n}}_{10}^{-}\times\boldsymbol{J}_{10}^{\text{ES}} \quad (40)$$

$$0 = \left[\mathcal{H}_2\left(\boldsymbol{J}_{20}^{\text{ES}}-\boldsymbol{J}_{12}^{\text{ES}}, \boldsymbol{M}_{20}^{\text{ES}}-\boldsymbol{M}_{12}^{\text{ES}}\right)\right]_{S_{20}^-}^{\tan} + \hat{\boldsymbol{n}}_{20}^{-}\times\boldsymbol{J}_{20}^{\text{ES}} \quad (41)$$

$$0 = \left[\mathcal{E}_1\left(\boldsymbol{J}_{10}^{\text{ES}}+\boldsymbol{J}_{12}^{\text{ES}}, \boldsymbol{M}_{10}^{\text{ES}}+\boldsymbol{M}_{12}^{\text{ES}}\right)\right]_{S_{10}^-}^{\tan} + \boldsymbol{M}_{10}^{\text{ES}}\times\hat{\boldsymbol{n}}_{10}^{-} \quad (42)$$

$$0 = \left[\mathcal{E}_2\left(\boldsymbol{J}_{20}^{\text{ES}}-\boldsymbol{J}_{12}^{\text{ES}}, \boldsymbol{M}_{20}^{\text{ES}}-\boldsymbol{M}_{12}^{\text{ES}}\right)\right]_{S_{20}^-}^{\tan} + \boldsymbol{M}_{20}^{\text{ES}}\times\hat{\boldsymbol{n}}_{20}^{-} \quad (43)$$

$$\begin{aligned}0 =& \left[\mathcal{E}_1\left(\boldsymbol{J}_{10}^{\text{ES}}+\boldsymbol{J}_{12}^{\text{ES}}, \boldsymbol{M}_{10}^{\text{ES}}+\boldsymbol{M}_{12}^{\text{ES}}\right)\right]_{S_{12}^-}^{\tan}\\
& -\left[\mathcal{E}_2\left(\boldsymbol{J}_{20}^{\text{ES}}-\boldsymbol{J}_{12}^{\text{ES}}, \boldsymbol{M}_{20}^{\text{ES}}-\boldsymbol{M}_{12}^{\text{ES}}\right)\right]_{S_{12}^+}^{\tan} \quad (44)\end{aligned}$$

$$
\begin{aligned}
0 \; = \; & \left[ \mathcal{H}_1 \left( \boldsymbol{J}_{10}^{\text{ES}} + \boldsymbol{J}_{12}^{\text{ES}}, \boldsymbol{M}_{10}^{\text{ES}} + \boldsymbol{M}_{12}^{\text{ES}} \right) \right]_{S_{12}^-}^{\tan} \\
& - \left[ \mathcal{H}_2 \left( \boldsymbol{J}_{20}^{\text{ES}} - \boldsymbol{J}_{12}^{\text{ES}}, \boldsymbol{M}_{20}^{\text{ES}} - \boldsymbol{M}_{12}^{\text{ES}} \right) \right]_{S_{12}^+}^{\tan} \qquad (45)
\end{aligned}
$$

Here, integral equations (40)-(43) are based on SEPs (32)&(33) and definitions (34)&(35); integral equations (44) and (45) are based on SEPs (32)&(33), relation $\boldsymbol{C}_{21}^{\text{ES}} = -\boldsymbol{C}_{12}^{\text{ES}}$, and the tangential continuation condition of the $\boldsymbol{F}^{\text{tot}}$ on $S_{12}$.

By discretizing the integral equations (40)-(45) into matrix equations, the following transformation from basic variables (BVs, which are not only independent but also complete) to all variables (AVs, which consist of both the BVs and the other dependent variables) can be obtained.

$$
\mathbb{a}^{\text{PhyAV}} = \mathbb{T} \cdot \mathbb{a}^{\text{BV}} \qquad (46)
$$

The formulations to calculate $\mathbb{T}$ can be found in [13], and they are not provided here, but we want to emphasize here that the process to obtain $\mathbb{T}$ involves calculating the inverses of some full matrices. Because the currents contained in $\mathbb{a}^{\text{PhyAV}}$ satisfy equations (40)-(45), thus $\mathbb{a}^{\text{PhyAV}}$ is physical automatically, and the superscript "PhyAV" is to emphasize this fact.

Substituting (46) into (39), we have the following matrix form of $P^{\text{driv}}$ with only BV $\mathbb{a}^{\text{BV}}$.

$$
P^{\text{driv}} = \left( \mathbb{a}^{\text{BV}} \right)^H \cdot \Bigg( \underbrace{\frac{1}{2} \mathbb{T}^H \cdot \mathbb{Z}^{\text{PMCHWT}} \cdot \mathbb{T}}_{\underset{\sim}{\mathbb{Z}}^{\text{PMCHWT}}} \Bigg) \cdot \mathbb{a}^{\text{BV}} \qquad (47)
$$

In (47), all dependent variables have been eliminated (that is, expressed as the functions of BVs), and this is just the reason to call the scheme dependent variable elimination (DVE). Very naturally, we have the equation $\underset{\sim}{\mathbb{Z}}_-^{\text{PMCHWT}} \cdot \mathbb{a}_\xi^{\text{BV}} = \lambda_\xi \, \underset{\sim}{\mathbb{Z}}_+^{\text{PMCHWT}} \cdot \mathbb{a}_\xi^{\text{BV}}$ involving only BV, where $\underset{\sim}{\mathbb{Z}}_+^{\text{PMCHWT}}$ and $\underset{\sim}{\mathbb{Z}}_-^{\text{PMCHWT}}$ are the positive and negative Hermitian parts of $\underset{\sim}{\mathbb{Z}}^{\text{PMCHWT}}$. The corresponding physical CMs are $\mathbb{a}_\xi^{\text{PhyAV}} = \mathbb{T} \cdot \mathbb{a}_\xi^{\text{BV}}$.

**PMCHWT-Based Surface CM Formulation with SDC**

In fact, as the OSS becomes larger or more complicated, the computational complexity of the DVE scheme becomes worse, because of the step to inverse full matrices [23]. To resolve this problem, a novel spurious mode suppression scheme — solution domain compression (SDC) — is proposed as below.

By discretizing the integral equations (40), (41), (42), (43) and (44)&(45), we can obtain the following matrix equations

$$
\begin{aligned}
\mathbb{G}_{10}^{\text{DoJ}} \cdot \mathbb{a}^{\text{PhyAV}} &= 0 \qquad (48) \\
\mathbb{G}_{20}^{\text{DoJ}} \cdot \mathbb{a}^{\text{PhyAV}} &= 0 \qquad (49) \\
\mathbb{G}_{10}^{\text{DoM}} \cdot \mathbb{a}^{\text{PhyAV}} &= 0 \qquad (50) \\
\mathbb{G}_{20}^{\text{DoM}} \cdot \mathbb{a}^{\text{PhyAV}} &= 0 \qquad (51) \\
\mathbb{G}_{\text{FCE}} \cdot \mathbb{a}^{\text{PhyAV}} &= 0 \qquad (52)
\end{aligned}
$$

where the superscripts "DoJ" and "DoM" are to emphasize the originations of $\mathbb{G}_{10/20}^{\text{DoJ}}$ (definition of $\boldsymbol{J}_{10/20}^{\text{ES}}$) and $\mathbb{G}_{10/20}^{\text{DoM}}$ (definition of $\boldsymbol{M}_{10/20}^{\text{ES}}$), and the subscript "FCE" is the acronym of "field continuation equation".

Based on the results obtained in [11]-[13], we can conclude that: equation (48) and equation (50) are equivalent to each other theoretically, and equation (49) and equation (51) are equivalent to each other theoretically, so if the matrices $\mathbb{G}_{10}^{\text{DoJ}}$, $\mathbb{G}_{20}^{\text{DoJ}}$, $\mathbb{G}_{10}^{\text{DoM}}$, $\mathbb{G}_{20}^{\text{DoM}}$, and $\mathbb{G}_{\text{FCE}}$ are assembled as

$$
\mathbb{G}_{\text{FCE}}^{\text{DoJJ}} = \begin{bmatrix} \mathbb{G}_{10}^{\text{DoJ}} \\ \mathbb{G}_{20}^{\text{DoJ}} \\ \mathbb{G}_{\text{FCE}} \end{bmatrix}, \mathbb{G}_{\text{FCE}}^{\text{DoMM}} = \begin{bmatrix} \mathbb{G}_{10}^{\text{DoM}} \\ \mathbb{G}_{20}^{\text{DoM}} \\ \mathbb{G}_{\text{FCE}} \end{bmatrix}, \mathbb{G}_{\text{FCE}}^{\text{DoJM}} = \begin{bmatrix} \mathbb{G}_{10}^{\text{DoJ}} \\ \mathbb{G}_{20}^{\text{DoM}} \\ \mathbb{G}_{\text{FCE}} \end{bmatrix}, \mathbb{G}_{\text{FCE}}^{\text{DoMJ}} = \begin{bmatrix} \mathbb{G}_{10}^{\text{DoM}} \\ \mathbb{G}_{20}^{\text{DoJ}} \\ \mathbb{G}_{\text{FCE}} \end{bmatrix} (53)
$$

then the following equations have the same solution space

$$
\mathbb{G}_{\text{FCE}}^{\text{DoJJ/DoMM/DoJM/DoMJ}} \cdot \mathbb{a}^{\text{PhyAV}} = 0 \qquad (54)
$$

theoretically, and the solution space is just the null space of $\mathbb{G}_{\text{FCE}}^{\text{DoJJ/DoMM/DoJM/DoMJ}}$; a mode is physical, if and only if it satisfies (54), if and only if it belongs to the null space.

Thus, the null space is alternatively called physical modal space (or simply called modal space) in this paper, and any physical mode $\mathbb{a}^{\text{PhyAV}}$ can be expanded in terms of the basis $\{\mathbb{s}_1, \mathbb{s}_2, \cdots\}$ of the modal space as follows:

$$
\mathbb{a}^{\text{PhyAV}} = \sum_i b_i \mathbb{s}_i = \underbrace{\left[ \mathbb{s}_1, \mathbb{s}_2, \cdots \right]}_{\mathbb{S}} \cdot \underbrace{\begin{bmatrix} b_1 \\ b_2 \\ \vdots \end{bmatrix}}_{\mathbb{b}} \qquad (55)
$$

and vice versa. In (55), $\mathbb{b}$ is an arbitrary column vector, whose row number is equal to the dimension of the modal space.

In addition, it is obvious that

$$
\begin{aligned}
\text{Modal Space} &= \text{nullspace}\left( \mathbb{G}_{\text{FCE}}^{\text{DoJJ/DoMM/DoJM/DoMJ}} \right) \\
&= \text{span}\left\{ \mathbb{s}_1, \mathbb{s}_2, \cdots \right\} \\
&= \text{range}\left( \mathbb{S} \right) \qquad (56)
\end{aligned}
$$

where $\text{span}\{\mathbb{s}_1, \mathbb{s}_2, \cdots\}$ is the space spanned by $\{\mathbb{s}_1, \mathbb{s}_2, \cdots\}$, and $\text{range}(\mathbb{S})$ denotes the range of $\mathbb{S}$. Thus in this paper, the $\mathbb{G}_{\text{FCE}}^{\text{DoJJ/DoMM/DoJM/DoMJ}}$ and $\mathbb{S}$ are called the generating matrix/operator and spanning matrix/operator of the modal space.

Substituting (55) into (39), we have the following matrix form of $P^{\text{driv}}$ for and only for physical modes $\mathbb{a}^{\text{PhyAV}}$.

$$
P^{\text{driv}} = \mathbb{b}^H \cdot \Bigg( \underbrace{\frac{1}{2} \mathbb{S}^H \cdot \mathbb{Z}^{\text{PMCHWT}} \cdot \mathbb{S}}_{\tilde{\mathbb{Z}}^{\text{PMCHWT}}} \Bigg) \cdot \mathbb{b} \qquad (57)
$$

and naturally we have the following characteristic equation $\tilde{\mathbb{Z}}_-^{\text{PMCHWT}} \cdot \mathbb{b}_\xi = \lambda_\xi \, \tilde{\mathbb{Z}}_+^{\text{PMCHWT}} \cdot \mathbb{b}_\xi$, where $\tilde{\mathbb{Z}}_+^{\text{PMCHWT}}$ and $\tilde{\mathbb{Z}}_-^{\text{PMCHWT}}$ are the positive and negative Hermitian parts of $\tilde{\mathbb{Z}}^{\text{PMCHWT}}$, and the associated CM $\mathbb{a}_\xi^{\text{PhyAV}}$ can be expressed as that $\mathbb{a}_\xi^{\text{PhyAV}} = \mathbb{S} \cdot \mathbb{b}_\xi$.

Obviously, the modal space is a subspace of the whole space constituted by all $\mathbb{a}^{\text{AV}}$ as shown in Fig. 11, so the above SDC scheme is essentially to compress the solution domain, and this is just the reason to call it solution domain compression (SDC).

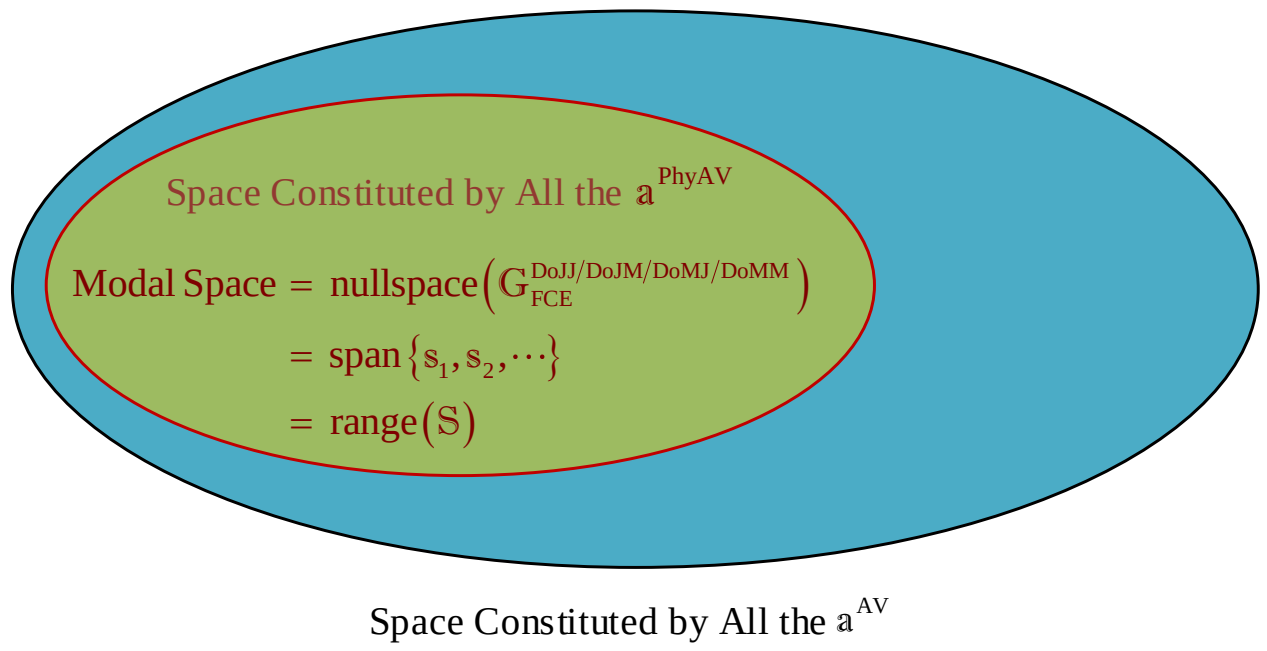

Fig. 11. Relations among the various spaces.

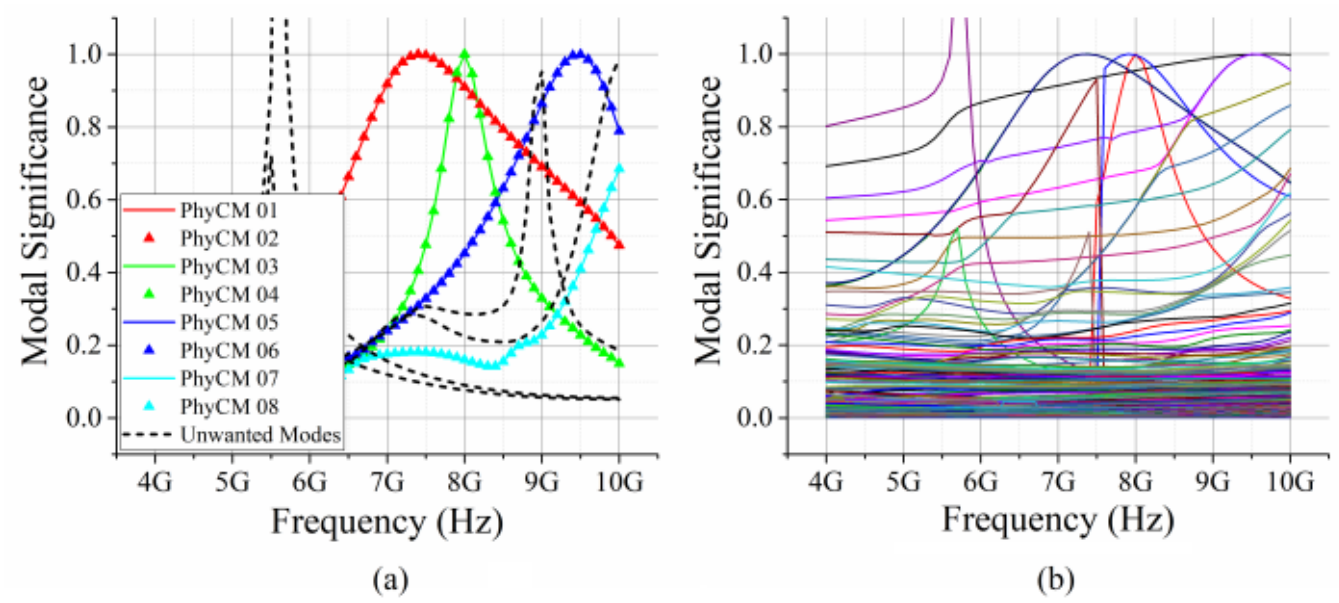

Fig. 12. (a) (57)-based MSs of a lossless two-body OSS with topological structure Fig. 9 and with material parameters $(\boldsymbol{\mu}_1=\mathbf{I}2\mu_0,\boldsymbol{\varepsilon}_1^{\mathrm{c}}=\mathbf{I}8\varepsilon_0)$ and $(\boldsymbol{\mu}_2=\mathbf{I}8\mu_0,\boldsymbol{\varepsilon}_2^{\mathrm{c}}=\mathbf{I}2\varepsilon_0)$; (b) (57)-based MSs of a lossy two-body OSS with topological structure Fig. 9 and with material parameters $(\boldsymbol{\mu}_1=\mathbf{I}2\mu_0,\boldsymbol{\varepsilon}_1^{\mathrm{c}}=\mathbf{I}8\varepsilon_0-j\mathbf{I}1/\omega)$ and $(\boldsymbol{\mu}_2=\mathbf{I}8\mu_0,\boldsymbol{\varepsilon}_2^{\mathrm{c}}=\mathbf{I}2\varepsilon_0)$.

For the lossless two-body OSS considered previously, its CMs derived from orthogonalizing the operator (57) with $\mathrm{G}_{\mathrm{FCE}}^{\mathrm{DoMJ}}$-based SDC are shown in Fig. 12(a). Obviously, the SDC scheme effectively suppresses the spurious modes appeared in Fig. 10(c), but, at the same time, there also arise some other unwanted modes marked by the dotted lines. In fact, the unwanted modes originate from using PMCHWT operator rather than the SDC scheme as exhibited in the later Sec. III-B.

For the lossy two-body OSS considered previously, we construct its CMs by orthogonalizing the operator (57) with $\mathrm{G}_{\mathrm{FCE}}^{\mathrm{DoMJ}}$-based SDC, and the associated MSs are shown in Fig. 12(b). Obviously, the results in Fig. 12(b) are not satisfactory compared with the results in Fig. 10(b). In fact, the reason leading to the invalidity of the operator (57) with $\mathrm{G}_{\mathrm{FCE}}^{\mathrm{DoMJ}}$-based SDC for lossy OSSs originates from using PMCHWT operator rather than the SDC scheme as exhibited in the later Sec. III-B.

Here, we emphasize again that the reason leading to the unwanted modes of the lossless OSS (shown in Fig. 12 (a)) and the unsatisfactory results of the lossy OSS (shown in Fig. 12 (b)) is using PMCHWT operator, rather than the new SDC scheme. In the following Sec. III-B, we can effectively construct the CMs of the lossy OSS by employing another operator — driving power operator (DPO) — with the SDC scheme.

### *B. WEP-Based Surface CM Formulation with Two Different Spurious Mode Suppression Schemes*

Because the tangential component of $\boldsymbol{F}^{\mathrm{inc}}$ is continuous on $S_{10}\bigcup S_{20}$, and the $\boldsymbol{F}^{\mathrm{inc}}$ on $S_{10}^-\bigcup S_{20}^-$ can be expressed in terms of the equivalent surface currents as SEP (31), thus DPO $P^{\mathrm{driv}}$ has surface formulation (58). In (58), the operators $\mathcal{L}_0$ and $\mathcal{K}_0$ are defined as $\mathcal{L}_0(\boldsymbol{C})=[1+(1/k_0^2)\nabla\nabla\cdot]\int_\Omega G_0(\boldsymbol{r},\boldsymbol{r}')\boldsymbol{C}(\boldsymbol{r}')d\Omega'$ and $\mathcal{K}_0(\boldsymbol{C})=\nabla\times\int_\Omega G_0(\boldsymbol{r},\boldsymbol{r}')\boldsymbol{C}(\boldsymbol{r}')d\Omega'$ respectively; symbol "$\mathrm{P.V.}\mathcal{K}_0$" denotes the principal value of operator $\mathcal{K}_0$.

Similar to discretizing (38) into (39), the (58) can be discretized into the following matrix form

$$P^{\mathrm{driv}}=\underbrace{\begin{bmatrix}\mathrm{a}^{\boldsymbol{J}_{10}^{\mathrm{ES}}}\\\mathrm{a}^{\boldsymbol{J}_{12}^{\mathrm{ES}}}\\\mathrm{a}^{\boldsymbol{J}_{20}^{\mathrm{ES}}}\\\mathrm{a}^{\boldsymbol{M}_{10}^{\mathrm{ES}}}\\\mathrm{a}^{\boldsymbol{M}_{12}^{\mathrm{ES}}}\\\mathrm{a}^{\boldsymbol{M}_{20}^{\mathrm{ES}}}\end{bmatrix}^H}_{\mathrm{a}^{\mathrm{AV}}}\cdot\underbrace{\left(\mathrm{P}_{\mathrm{PVT}}^{\mathrm{driv}}+\mathrm{P}_{\mathrm{SCT}}^{\mathrm{driv}}\right)}_{\mathrm{P}^{\mathrm{driv}}}\cdot\underbrace{\begin{bmatrix}\mathrm{a}^{\boldsymbol{J}_{10}^{\mathrm{ES}}}\\\mathrm{a}^{\boldsymbol{J}_{12}^{\mathrm{ES}}}\\\mathrm{a}^{\boldsymbol{J}_{20}^{\mathrm{ES}}}\\\mathrm{a}^{\boldsymbol{M}_{10}^{\mathrm{ES}}}\\\mathrm{a}^{\boldsymbol{M}_{12}^{\mathrm{ES}}}\\\mathrm{a}^{\boldsymbol{M}_{20}^{\mathrm{ES}}}\end{bmatrix}}_{\mathrm{a}^{\mathrm{AV}}}\tag{59}$$

where subscripts "PVT" and "SCT" are the acronyms of "principal value term" and "singular current term" respectively.

**WEP-Based Surface CM Formulation with DVE**

Substituting (46) into (59), we have the matrix form with only BV $\mathrm{a}^{\mathrm{BV}}$ as follows:

$$P^{\mathrm{driv}}=\left(\mathrm{a}^{\mathrm{BV}}\right)^H\cdot\Bigg(\overbrace{\underbrace{\mathrm{T}^H\cdot\mathrm{P}_{\mathrm{PVT}}^{\mathrm{driv}}\cdot\mathrm{T}}_{\underset{\sim}{\mathrm{P}}_{\mathrm{PVT}}^{\mathrm{driv}}}+\underbrace{\mathrm{T}^H\cdot\mathrm{P}_{\mathrm{SCT}}^{\mathrm{driv}}\cdot\mathrm{T}}_{\underset{\sim}{\mathrm{P}}_{\mathrm{SCT}}^{\mathrm{driv}}}}^{\underset{\sim}{\mathrm{P}}^{\mathrm{driv}}}\Bigg)\cdot\mathrm{a}^{\mathrm{BV}}\tag{60}$$

and the corresponding characteristic equation as follows:

$$\underset{\sim}{\mathrm{P}}_{-}^{\mathrm{driv}}\cdot\mathrm{a}_\xi^{\mathrm{BV}}=\lambda_\xi\,\underset{\sim}{\mathrm{P}}_{+}^{\mathrm{driv}}\cdot\mathrm{a}_\xi^{\mathrm{BV}}\tag{61}$$

and the corresponding physical CMs as $\mathrm{a}_\xi^{\mathrm{PhyAV}}=\mathrm{T}\cdot\mathrm{a}_\xi^{\mathrm{BV}}$.

**WEP-Based Surface CM Formulation with SDC**

In modal space, we have the matrix form of $P^{\mathrm{driv}}$ as follows:

$$P^{\mathrm{driv}}=\mathrm{b}^H\cdot\Bigg(\overbrace{\underbrace{\mathrm{S}^H\cdot\mathrm{P}_{\mathrm{PVT}}^{\mathrm{driv}}\cdot\mathrm{S}}_{\tilde{\mathrm{P}}_{\mathrm{PVT}}^{\mathrm{driv}}}+\underbrace{\mathrm{S}^H\cdot\mathrm{P}_{\mathrm{SCT}}^{\mathrm{driv}}\cdot\mathrm{S}}_{\tilde{\mathrm{P}}_{\mathrm{SCT}}^{\mathrm{driv}}}}^{\tilde{\mathrm{P}}^{\mathrm{driv}}}\Bigg)\cdot\mathrm{b}\tag{62}$$

by substituting (55) into (59). The corresponding characteristic equation is as follows:

$$\tilde{\mathrm{P}}_{-}^{\mathrm{driv}}\cdot\mathrm{b}_\xi=\lambda_\xi\,\tilde{\mathrm{P}}_{+}^{\mathrm{driv}}\cdot\mathrm{b}_\xi\tag{63}$$

and the corresponding physical CMs are $\mathrm{a}_\xi^{\mathrm{PhyAV}}=\mathrm{S}\cdot\mathrm{b}_\xi$.

$$\begin{aligned}P^{\mathrm{driv}}=&-(1/2)\left\langle\boldsymbol{J}_{10}^{\mathrm{ES}}+\boldsymbol{J}_{20}^{\mathrm{ES}},\mathcal{E}_0\left(\boldsymbol{J}_{10}^{\mathrm{ES}}+\boldsymbol{J}_{20}^{\mathrm{ES}},\boldsymbol{M}_{10}^{\mathrm{ES}}+\boldsymbol{M}_{20}^{\mathrm{ES}}\right)\right\rangle_{S_{10}^-\cup S_{20}^-}-(1/2)\left\langle\boldsymbol{M}_{10}^{\mathrm{ES}}+\boldsymbol{M}_{20}^{\mathrm{ES}},\mathcal{H}_0\left(\boldsymbol{J}_{10}^{\mathrm{ES}}+\boldsymbol{J}_{20}^{\mathrm{ES}},\boldsymbol{M}_{10}^{\mathrm{ES}}+\boldsymbol{M}_{20}^{\mathrm{ES}}\right)\right\rangle_{S_{10}^-\cup S_{20}^-}\\=&\underbrace{-\frac{1}{2}\left\langle\boldsymbol{J}_{10}^{\mathrm{ES}}+\boldsymbol{J}_{20}^{\mathrm{ES}},-j\omega\mu_0\mathcal{L}_0\left(\boldsymbol{J}_{10}^{\mathrm{ES}}+\boldsymbol{J}_{20}^{\mathrm{ES}}\right)-\mathrm{P.V.}\mathcal{K}_0\left(\boldsymbol{M}_{10}^{\mathrm{ES}}+\boldsymbol{M}_{20}^{\mathrm{ES}}\right)\right\rangle_{S_{10}\cup S_{20}}-\frac{1}{2}\left\langle\boldsymbol{M}_{10}^{\mathrm{ES}}+\boldsymbol{M}_{20}^{\mathrm{ES}},\mathrm{P.V.}\mathcal{K}_0\left(\boldsymbol{J}_{10}^{\mathrm{ES}}+\boldsymbol{J}_{20}^{\mathrm{ES}}\right)-j\omega\varepsilon_0\mathcal{L}_0\left(\boldsymbol{M}_{10}^{\mathrm{ES}}+\boldsymbol{M}_{20}^{\mathrm{ES}}\right)\right\rangle_{S_{10}\cup S_{20}}}_{\text{principal value term (PVT)}}\\&\underbrace{+\frac{1}{4}\left\langle\boldsymbol{J}_{10}^{\mathrm{ES}}+\boldsymbol{J}_{20}^{\mathrm{ES}},\boldsymbol{M}_{10}^{\mathrm{ES}}\times\hat{\boldsymbol{n}}_{10}^-+\boldsymbol{M}_{20}^{\mathrm{ES}}\times\hat{\boldsymbol{n}}_{20}^-\right\rangle_{S_{10}\cup S_{20}}+\frac{1}{4}\left\langle\boldsymbol{M}_{10}^{\mathrm{ES}}+\boldsymbol{M}_{20}^{\mathrm{ES}},\hat{\boldsymbol{n}}_{10}^-\times\boldsymbol{J}_{10}^{\mathrm{ES}}+\hat{\boldsymbol{n}}_{20}^-\times\boldsymbol{J}_{20}^{\mathrm{ES}}\right\rangle_{S_{10}\cup S_{20}}}_{\text{singular current term (SCT)}}\end{aligned}\tag{58}$$

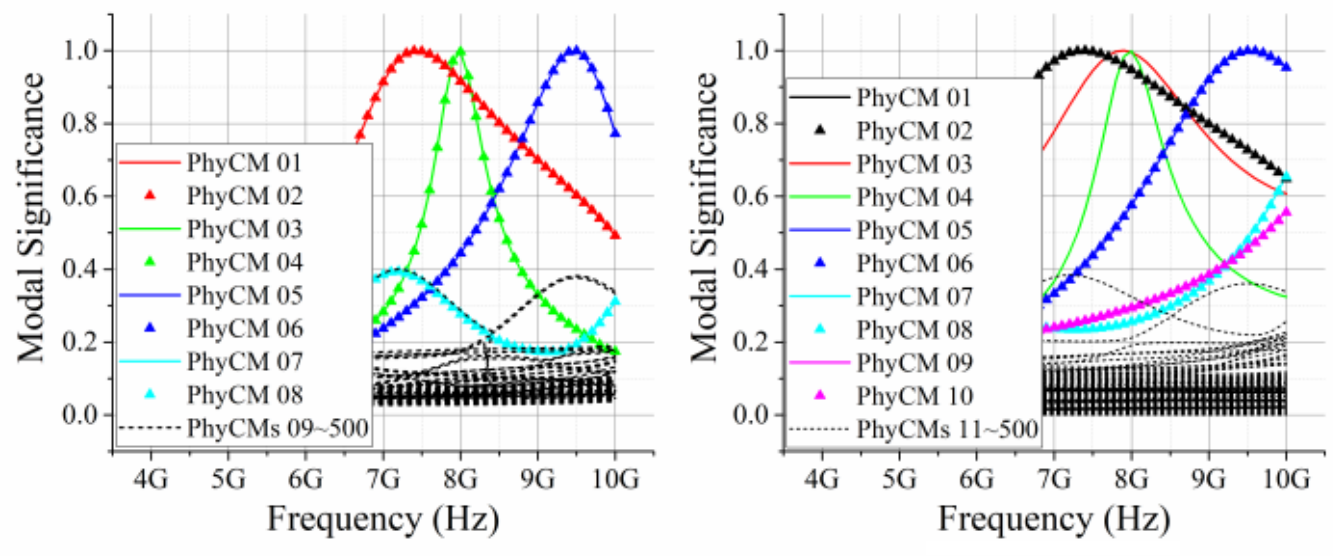

(a) (b)

Fig. 13. (a) (63)-based MSs of a lossless two-body OSS with topological structure Fig. 9 and with material parameters $(\boldsymbol{\mu}_1 = \mathbf{I}2\mu_0, \boldsymbol{\varepsilon}_1^{c} = \mathbf{I}8\varepsilon_0)$ and $(\boldsymbol{\mu}_2 = \mathbf{I}8\mu_0, \boldsymbol{\varepsilon}_2^{c} = \mathbf{I}2\varepsilon_0)$; (b) (63)-based MSs of a lossy two-body OSS with topological structure Fig. 9 and with material parameters $(\boldsymbol{\mu}_1 = \mathbf{I}2\mu_0, \boldsymbol{\varepsilon}_1^{c} = \mathbf{I}8\varepsilon_0 - j\mathbf{I}1/\omega)$ and $(\boldsymbol{\mu}_2 = \mathbf{I}8\mu_0, \boldsymbol{\varepsilon}_2^{c} = \mathbf{I}2\varepsilon_0)$.

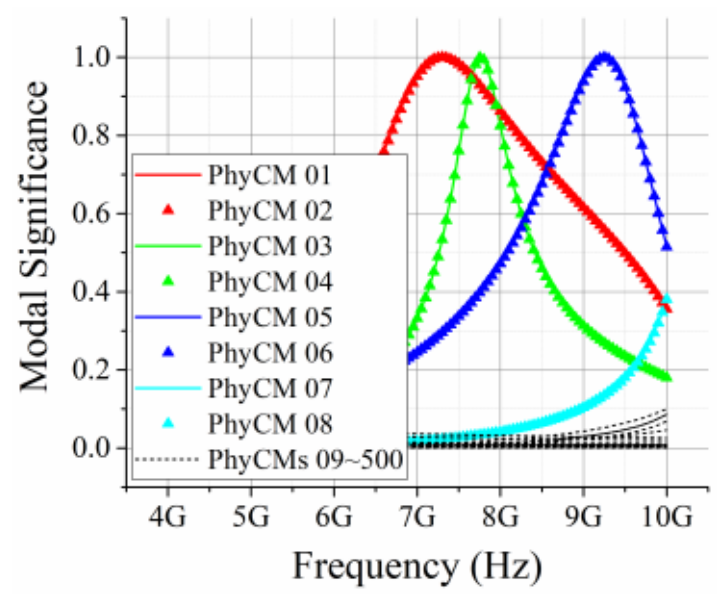

Fig. 14. (71)-based MSs of a lossless two-body OSS with topological structure Fig. 9 and with material parameters $(\boldsymbol{\mu}_1 = \mathbf{I}2\mu_0, \boldsymbol{\varepsilon}_1^{c} = \mathbf{I}8\varepsilon_0)$ and $(\boldsymbol{\mu}_2 = \mathbf{I}8\mu_0, \boldsymbol{\varepsilon}_2^{c} = \mathbf{I}2\varepsilon_0)$.

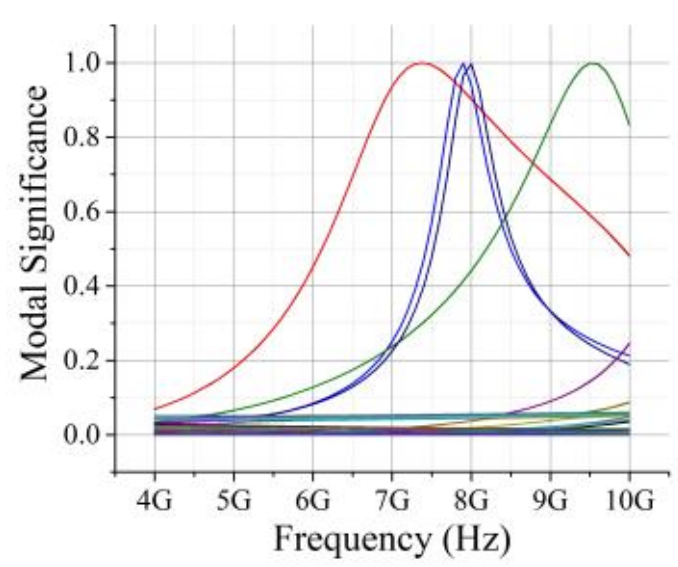

Fig. 15. (71)-based MSs of a lossy two-body OSS with topological structure Fig. 9 and with material parameters $(\boldsymbol{\mu}_1 = \mathbf{I}2\mu_0, \boldsymbol{\varepsilon}_1^{c} = \mathbf{I}8\varepsilon_0 - j\mathbf{I}1/\omega)$ and $(\boldsymbol{\mu}_2 = \mathbf{I}8\mu_0, \boldsymbol{\varepsilon}_2^{c} = \mathbf{I}2\varepsilon_0)$.

For the lossless and lossy two-body OSSs considered previously, the MSs derived from (63) are shown in Fig. 13. Obviously, the WEP-based DPO with SDC scheme successfully suppresses all the spurious modes, and, at the same time, doesn't introduce the unwanted modes shown in Figs. 10(c), 10(d), 12(a) and 12(b). In addition, Fig. 13(b) implies that the reason leading to the invalidity of the operator (57) with $\mathbb{G}_{\text{FCE}}^{\text{DoMJ}}$ -based SDC for lossy OSSs originates from the PMCHWT operator rather than the SDC scheme.

**SCT and Its Physical Interpretation**

It has been proved in [13] that $(\mathrm{a}^{\text{BV}})^{H} \cdot \underset{\sim}{\mathrm{P}}_{\text{SCT}}^{\text{driv}} \cdot \mathrm{a}^{\text{BV}}$ is always equal to the power dissipated in $V_{\text{OSS}}$, that is,

$$(\mathrm{a}^{\text{BV}})^{H} \cdot \underset{\sim}{\mathrm{P}}_{\text{SCT}}^{\text{driv}} \cdot \mathrm{a}^{\text{BV}} = \frac{1}{2}\left\langle \boldsymbol{\sigma}_1 \cdot \boldsymbol{E}^{\text{tot}}, \boldsymbol{E}^{\text{tot}} \right\rangle_{V_1} + \frac{1}{2}\left\langle \boldsymbol{\sigma}_2 \cdot \boldsymbol{E}^{\text{tot}}, \boldsymbol{E}^{\text{tot}} \right\rangle_{V_2} \quad (64)$$

and that

$$\boldsymbol{\sigma}_1 = \mathbf{I}0 = \boldsymbol{\sigma}_2. \Leftrightarrow V_{\text{OSS}} \text{ is lossless}. \Leftrightarrow \underset{\sim}{\mathrm{P}}_{\text{SCT}}^{\text{driv}} = 0. \quad (65)$$

Similarly, it can also be proved that

$$\mathrm{b}^{H} \cdot \tilde{\mathrm{P}}_{\text{SCT}}^{\text{driv}} \cdot \mathrm{b} = \frac{1}{2}\left\langle \boldsymbol{\sigma}_1 \cdot \boldsymbol{E}^{\text{tot}}, \boldsymbol{E}^{\text{tot}} \right\rangle_{V_1} + \frac{1}{2}\left\langle \boldsymbol{\sigma}_2 \cdot \boldsymbol{E}^{\text{tot}}, \boldsymbol{E}^{\text{tot}} \right\rangle_{V_2} \quad (66)$$

and that

$$\boldsymbol{\sigma}_1 = \mathbf{I}0 = \boldsymbol{\sigma}_2. \Leftrightarrow V_{\text{OSS}} \text{ is lossless}. \Leftrightarrow \tilde{\mathrm{P}}_{\text{SCT}}^{\text{driv}} = 0. \quad (67)$$

Thus, when $V_{\text{OSS}}$ is lossless, we have that

$$P^{\text{driv}} \overset{\boldsymbol{\sigma}_1 = \mathbf{I}0 = \boldsymbol{\sigma}_2}{=\!=\!=\!=} \left(\mathrm{a}^{\text{BV}}\right)^{H} \cdot \overbrace{\left(\mathrm{T}^{H} \cdot \mathrm{P}_{\text{PVT}}^{\text{driv}} \cdot \mathrm{T}\right)}^{\underset{\sim}{\mathrm{P}}_{\text{PVT}}^{\text{driv}}} \cdot \mathrm{a}^{\text{BV}} \quad (68)$$

$$P^{\text{driv}} \overset{\boldsymbol{\sigma}_1 = \mathbf{I}0 = \boldsymbol{\sigma}_2}{=\!=\!=\!=} \mathrm{b}^{H} \cdot \underbrace{\left(\mathrm{S}^{H} \cdot \mathrm{P}_{\text{PVT}}^{\text{driv}} \cdot \mathrm{S}\right)}_{\tilde{\mathrm{P}}_{\text{PVT}}^{\text{driv}}} \cdot \mathrm{b} \quad (69)$$

and then we have two alternative characteristic equations

$$\underset{\sim}{\mathrm{P}}_{\text{PVT};-}^{\text{driv}} \cdot \mathrm{a}_{\xi}^{\text{BV}} = \lambda_{\xi}\, \underset{\sim}{\mathrm{P}}_{\text{PVT};+}^{\text{driv}} \cdot \mathrm{a}_{\xi}^{\text{BV}} \quad (70)$$

$$\tilde{\mathrm{P}}_{\text{PVT};-}^{\text{driv}} \cdot \mathrm{b}_{\xi} = \lambda_{\xi}\, \tilde{\mathrm{P}}_{\text{PVT};+}^{\text{driv}} \cdot \mathrm{b}_{\xi} \quad (71)$$

for lossless $V_{\text{OSS}}$ only.

However, it must be emphasized here that: when $V_{\text{OSS}}$ is lossy, only matrix forms (60)&(62) and equations (61)&(63) are correct, while matrix forms (68)&(69) and equations (70)&(71) are incorrect as illustrated in the following example.

For the previous lossless two-body OSS, its CMs derived from the (71) with $\mathbb{G}_{\text{FCE}}^{\text{DoMJ}}$ -based SDC are shown in Fig. 14. Obviously, the equation (71) is valid for constructing physical CMs and suppressing spurious modes, and, at the same time, has a more satisfactory numerical performance than characteristic equation (63).

For the previous lossy two-body OSS, its CMs derived from the (71) with $\mathbb{G}_{\text{FCE}}^{\text{DoMJ}}$ -based SDC are shown in Fig. 15. Obviously, the results are not consistent with the ones shown in Fig. 10(b) and Fig. 13(b), because the (71) doesn't include the SCT and the SCT is not zero in this lossy case. Thus, we want to emphasize here again that the matrix forms (68)&(69) and equations (70)&(71) are only correct for the lossless OSSs, but incorrect for the lossy OSSs.

## IV. An Effective Method for Reducing the Computational Complexity of SDC Scheme

Compared with the conventional DVE scheme, the novel SDC scheme indeed avoids the matrix inversion process successfully, but, at the same time, it needs to calculate the basic solutions of equation (54). Unfortunately, the process to calculate the basic solutions is also computationally complicated. To resolve this problem, this section proposes an effective method for reducing the computational complexity of the original SDC scheme introduced in Sec. III, and then obtains an improved SDC scheme with lower computational complexity.

For simplifying the symbolic system of the following discussions, we simply denote the matrix $\mathrm{G}_{\mathrm{FCE}}^{\mathrm{DoJJ/DoMM/DoJM/DoMJ}}$ and vector $\mathrm{a}^{\mathrm{PhyAV}}$ used in (54) as $\mathrm{G}$ and $\mathrm{a}$ respectively. There exists the following equivalence relation

$$\mathrm{G}\cdot\mathrm{a}=0 \;\Leftrightarrow\; \mathrm{a}^{H}\cdot\left(\mathrm{G}^{H}\cdot\mathrm{G}\right)\cdot\mathrm{a}=0 \tag{72}$$

and the reasons are that: clearly $\mathrm{G}\cdot\mathrm{a}=0$ leads to $\mathrm{G}^{H}\cdot(\mathrm{G}\cdot\mathrm{a})=0$, so leads to $(\mathrm{G}^{H}\cdot\mathrm{G})\cdot\mathrm{a}=0$ due to the associativity for matrix multiplication, and then leads to $\mathrm{a}^{H}\cdot(\mathrm{G}^{H}\cdot\mathrm{G})\cdot\mathrm{a}=0$; obviously $0=\mathrm{a}^{H}\cdot(\mathrm{G}^{H}\cdot\mathrm{G})\cdot\mathrm{a}=(\mathrm{G}\cdot\mathrm{a})^{H}\cdot(\mathrm{G}\cdot\mathrm{a})$ due to the multiplication associativity and the relation $\mathrm{a}^{H}\cdot\mathrm{G}^{H}=(\mathrm{G}\cdot\mathrm{a})^{H}$, so $\mathrm{a}^{H}\cdot(\mathrm{G}^{H}\cdot\mathrm{G})\cdot\mathrm{a}=0$ implies $\mathrm{G}\cdot\mathrm{a}=0$ because $\mathrm{v}^{H}\cdot\mathrm{v}=0\Leftrightarrow\mathrm{v}=0$ for any complex vector $\mathrm{v}$.

Based on relation (72), we propose an alternative characteristic equation as follows:

$$\left(\mathrm{P}_{\mathrm{PVT};-}^{\mathrm{driv}}+\ell\cdot\mathrm{G}^{H}\cdot\mathrm{G}\right)\cdot\mathrm{a}_{\xi}=\lambda_{\xi}\,\mathrm{P}_{\mathrm{PVT};+}^{\mathrm{driv}}\cdot\mathrm{a}_{\xi} \tag{73}$$

for the lossless OSSs, and

$$\left(\mathrm{P}_{-}^{\mathrm{driv}}+\ell\cdot\mathrm{G}^{H}\cdot\mathrm{G}\right)\cdot\mathrm{a}_{\xi}=\lambda_{\xi}\,\mathrm{P}_{+}^{\mathrm{driv}}\cdot\mathrm{a}_{\xi} \tag{74}$$

for the lossy OSSs, where $\ell$ is an adjustable large real coefficient for example $\ell=10^{10}$.

Evidently, if we calculate the first several CMs which have relatively small $|\lambda_{\xi}|$ (the reason why the relatively small $|\lambda_{\xi}|$ are desirable had been carefully explained in Sec. II), then the obtained CMs must satisfy equation $\mathrm{G}\cdot\mathrm{a}=0$, because

$$\lambda_{\xi}=\begin{cases}\dfrac{\mathrm{a}_{\xi}^{H}\cdot\mathrm{P}_{\mathrm{PVT};-}^{\mathrm{driv}}\cdot\mathrm{a}_{\xi}}{\mathrm{a}_{\xi}^{H}\cdot\mathrm{P}_{\mathrm{PVT};+}^{\mathrm{driv}}\cdot\mathrm{a}_{\xi}}+\ell\cdot\dfrac{\mathrm{a}_{\xi}^{H}\cdot\mathrm{G}^{H}\cdot\mathrm{G}\cdot\mathrm{a}_{\xi}}{\mathrm{a}_{\xi}^{H}\cdot\mathrm{P}_{\mathrm{PVT};+}^{\mathrm{driv}}\cdot\mathrm{a}_{\xi}} & \text{for lossless OSSs}\\[2ex] \dfrac{\mathrm{a}_{\xi}^{H}\cdot\mathrm{P}_{-}^{\mathrm{driv}}\cdot\mathrm{a}_{\xi}}{\mathrm{a}_{\xi}^{H}\cdot\mathrm{P}_{+}^{\mathrm{driv}}\cdot\mathrm{a}_{\xi}}+\ell\cdot\dfrac{\mathrm{a}_{\xi}^{H}\cdot\mathrm{G}^{H}\cdot\mathrm{G}\cdot\mathrm{a}_{\xi}}{\mathrm{a}_{\xi}^{H}\cdot\mathrm{P}_{+}^{\mathrm{driv}}\cdot\mathrm{a}_{\xi}} & \text{for lossy OSSs}\end{cases} \tag{75}$$

and this implies that: if $\mathrm{G}\cdot\mathrm{a}\neq 0$, then the second term in the right-hand side of (75) is not zero due to relation (72), and then $|\lambda_{\xi}|$ will not be small due to the large coefficient $\ell$.

Obviously, the characteristic equations (73) and (74) involve neither calculating matrix inversion nor evaluating basic solutions.

To verify the validity of characteristic equations (73) and (74), we use them to calculate the CMs of the lossless and lossy two-body material OSSs considered in Secs. II and III. The MSs corresponding to the first several CMs whose $|\lambda_{\xi}|$ are relatively small are shown in Fig. 16. Clearly, the results shown in Figs. 16(a) and 16(b) are consistent with the ones shown in the Figs. 10(a) and 10(b), which are calculated from the volume formulation given in Sec. II, and the ones shown in the Figs. 14 and 13(b), which are calculated from the surface formulation with the original SDC scheme proposed in Sec. III.

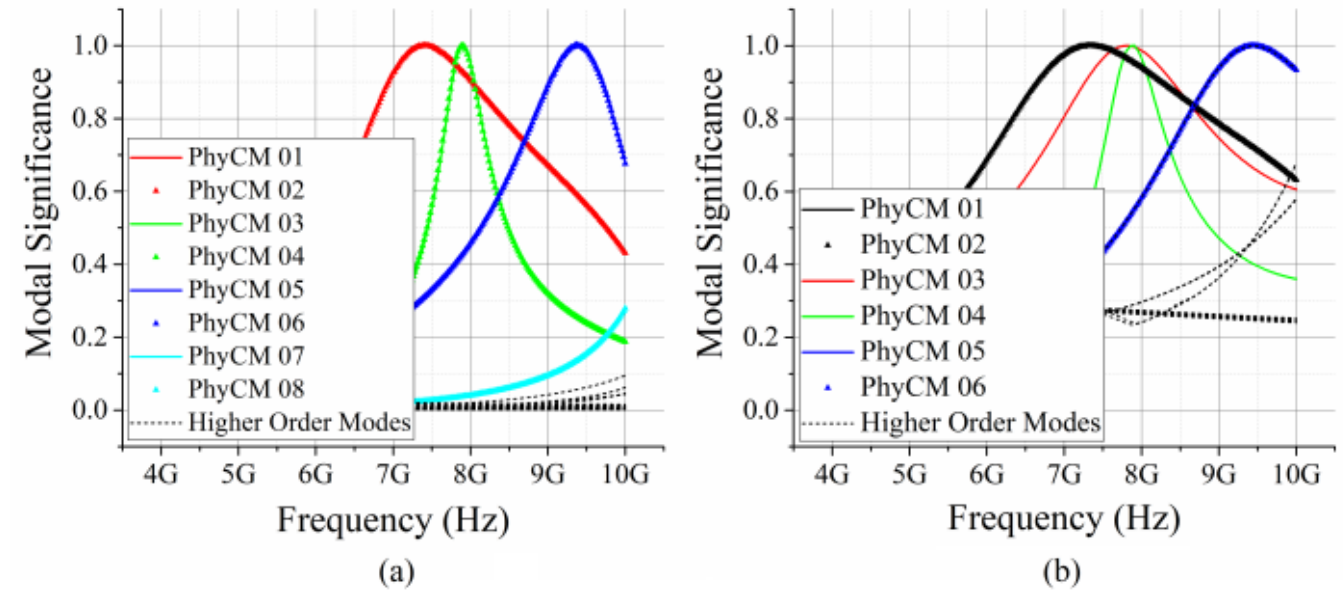

Fig. 16. (a) (73)-based MSs of a lossless two-body OSS with topological structure Fig. 9 and with material parameters $(\boldsymbol{\mu}_1=\mathbf{I}2\mu_0,\boldsymbol{\varepsilon}_1^{\mathrm{c}}=\mathbf{I}8\varepsilon_0)$ and $(\boldsymbol{\mu}_2=\mathbf{I}8\mu_0,\boldsymbol{\varepsilon}_2^{\mathrm{c}}=\mathbf{I}2\varepsilon_0)$, where $\ell=10^{10}$; (b) (74)-based MSs of a lossy two-body OSS with topological structure Fig. 9 and with material parameters $(\boldsymbol{\mu}_1=\mathbf{I}2\mu_0,\boldsymbol{\varepsilon}_1^{\mathrm{c}}=\mathbf{I}8\varepsilon_0-j\mathbf{I}1/\omega)$ and $(\boldsymbol{\mu}_2=\mathbf{I}8\mu_0,\boldsymbol{\varepsilon}_2^{\mathrm{c}}=\mathbf{I}2\varepsilon_0)$, where $\ell=10^{10}$.

## V. Conclusions

For any pre-selected material objective scattering system (OSS), which can be either lossless or lossy, there exist some energy-decoupled working modes — characteristic modes (CMs). The CMs can span whole modal space, and don't exchange net energy in any integral period.

Traditionally, the CMs are constructed by orthogonalizing the impedance matrix operator (IMO) under integral equation (IE) framework. Alternatively, this paper effectively constructs the CMs by orthogonalizing the driving power operator (DPO) under work-energy principle (WEP) framework. This paper exhibits that the new WEP framework and new orthogonalizing DPO method have many advantages over the traditional IE framework and traditional orthogonalizing IMO method.

(1) The new WEP framework clearly reveals the physical picture/purpose of CMT.

(1.1) Based on the physical picture, the reason why the characteristic values (CVs) of magnetodielectric OSSs don't have clear physical meaning is clarified. The reason is that the CVs of magnetodielectric OSSs are only the byproducts during the process to construct CMs.

(1.2) Based on the physical picture, an improper physical interpretation for zero CV is corrected. The CMs with zero CVs have significant weights in whole modal expansion formulation, but they are not necessarily working at resonant states.

(1.3) Based on the physical picture, the reason why modal far fields of lossy OSSs are usually not orthogonal is answered. The answer is that the physical purpose of CMT is to construct the modes without net energy exchange in integral period rather than the modes with orthogonal far fields.

(1.4) Based on the physical picture, the reason why modal active powers are usually normalized to 1 is clarified. The reason is to let modal expansion coefficients have a clear physical meaning — the modal weight of the associated CM in whole modal expansion formulation, and to let the expansion formulation for time-average DP have the same mathematical form as the famous Parseval's identity.

(2) The new orthogonalizing DPO method has a more satisfactory numerical performance than the traditional orthogonalizing IMO method in the aspect of constructing CMs. A detailed and rigorous theoretical explanation for this phenomenon has been done by our group, and will be exhibited in our future article.

(3) By studying the physical meaning of the singular current term (SCT) contained in DPO, a further improvement for the numerical performance of DPO is proposed. The improvement scheme is to delete the SCT from DPO when the OSS is lossless.

(4) The variables contained in the new DPO are less than the ones contained in the traditional IMO, and this fact leads to the following conclusions.

(4.1) Discretizing DPO needs a smaller computational burden than discretizing IMO.

(4.2) From DPO, it is easier to distinguish the independent variables from the dependent variables of the related EM scattering problem, and then is easier to establish the transformation from the independent variables to the dependent variables and to compress solution domain from original space to modal space. (The transformation and compression can effectively suppress the spurious modes outputted from characteristic equation.)

Above these evidently exhibit the advantages of WEP framework and orthogonalizing DPO method.

Besides above these, this paper also proposes a novel spurious mode suppression scheme — solution domain compression (SDC). The novel SDC scheme is as effective as the traditional dependent variable elimination (DVE) scheme in the aspect of suppressing spurious modes, and, at the same time, it avoids the matrix inversion process used in the traditional DVE scheme successfully.

During the process to compress solution domain, the electric and magnetic field continuation conditions on the material-material interface are indispensable, but the one and only one of the electric and magnetic current definitions on the material-environment interface is necessary. In addition, we have the following further conclusions:

(i) when the material body is primarily electric, the definition of the magnetic current on the material-environment interface is more satisfactory for compressing solution domain;

(ii) when the material body is primarily magnetic, the definition of the electric current on the material-environment interface is more satisfactory for compressing solution domain.

A detailed and rigorous theoretical explanation for the above (i) and (ii) has been done by our group, and will be exhibited in our future article.

## Acknowledgment

The authors would like to thank the reviewers and editors for their patient reviews, valuable comments, and selfless suggestions for improving this paper.

**Ren-Zun Lian** received the B.S. degree in optical engineering from the University of Electronic Science and Technology of China (UESTC), Chengdu, China, in 2011, and received the Ph.D. degree in electromagnetic field and microwave technology from UESTC, Chengdu, China, in 2019.

He is currently a postdoctoral researcher in Peking University (PKU), Beijing, China. His current research interests include mathematical physics, electromagnetic theory and computation, and antenna theory and design.

**Xing-Yue Guo** (S'17) received the B.S. degree in electronic information science and technology from the Southwest Jiaotong University, Chengdu, China, in 2011, and the M.E. degree from the University of Electronic Science and Technology of China, Chengdu, China, in 2014. From 2014 to 2017, she was a Research Assistant with the CAEP Software Center for High Performance Numerical Simulation, Beijing, China. Since 2017, she has been pursuing the Ph. D. degree at the School of Electronics Engineering and Computer Science, Peking University, Beijing, China. Her research interests include computational electromagnetics and applications.

**Ming-Yao Xia** (M'00-SM'03) received the Master and Ph. D degrees in electrical engineering from the Institute of Electronics, Chinese Academy of Sciences (IECAS), in 1988 and 1999, respectively. From 1988 to 2002, he was with IECAS as an Engineer and a Senior Engineer. He was a Visiting Scholar at the University of Oxford, U.K., from October 1995 to October 1996. From June 1999 to August 2000 and from January 2002 to June 2002, he was a Senior Research Assistant and a Research Fellow, respectively, with the City University of Hong Kong. He joined Peking University as an Associate Professor in 2002 and was promoted to Full Professor in 2004. He moved to the University of Electronic Science and Technology of China as a Chang-Jiang Professor nominated by the Ministry of Education of China in 2010. He returned to Peking University after finishing the appointment in 2013. He was a recipient of the Young Scientist Award of the URSI in 1993. He was awarded the first-class prize on Natural Science by the Chinese Academy of Sciences in 2001. He was the recipient of the Foundation for Outstanding Young Investigators presented by the National Natural Science Foundation of China in 2008. He served as an Associate Editor for the IEEE Transactions on Antennas and Propagation. His research interests include electromagnetic theory, numerical methods and applications, such as wave propagation and scattering, electromagnetic imaging and probing, microwave remote sensing, antennas and microwave components.

# Work-Energy Principle Based Characteristic Mode Theory with Solution Domain Compression for Metal-Material Composite Scattering Systems

Ren-Zun Lian, Xing-Yue Guo, *Student Member, IEEE*, and Ming-Yao Xia, *Senior Member, IEEE*

***Abstract*—Characteristic mode theory (CMT) is established in a novel work-energy principle (WEP) framework. Under the framework, a novel driving power operator (DPO) is introduced as the generating operator for characteristic modes (CMs). Then, a novel orthogonalizing DPO method is proposed to construct the CMs. In addition, a novel solution domain compression (SDC) scheme is developed to suppress spurious modes.**

**Compared with the conventional integral equation (IE) framework, the novel WEP framework more clearly reveals the physical picture/purpose of CMT, and then easily derives the famous Parseval's identity. Compared with the conventional impedance matrix operator (IMO), the novel DPO is more advantageous in the aspect of distinguishing independent variables from dependent variables, and, at the same time, has a smaller computational burden. Compared with the conventional orthogonalizing IMO method, the novel orthogonalizing DPO method has a more acceptable numerical performance in the aspect of constructing CMs. Compared with the conventional dependent variable elimination (DVE) scheme, the novel SDC scheme doesn't need to inverse any matrix during suppressing spurious modes.**

## I. Introduction

In the later 1960s, Garbacz *et al.* [1]-[3] built a characteristic mode theory (CMT) in scattering matrix (SM) framework. By orthogonalizing perturbation matrix operator (PMO), the SM-based CMT (SM-CMT) can construct a set of orthogonal characteristic modes (CMs) for any pre-selected lossless objective scattering system (OSS). In the 1970s, Harrington *et al.* [4]-[8] established an alternative CMT under integral equation (IE) framework. By orthogonalizing impedance matrix operator (IMO), the IE-based CMT (IE-CMT) can construct a set of orthogonal CMs for any pre-selected lossless or lossy OSS.

Both Garbacz's CMs and Harrington's CMs depend only on the inherent physical properties (such as the topological structure and material parameters etc.) of the OSS, so both SM-CMT and IE-CMT are very valuable for analyzing and designing the inherent electromagnetic (EM) scattering characters of the OSS. To obtain IMO is easier than to obtain PMO, so IE-CMT has had a wider spread than SM-CMT in EM engineering society. A very comprehensive review for the various antenna applications of IE-CMT can be found in [9]-[11].

Under IE framework, electric field integral equation (EFIE) based IMO [4]-[6] is widely used to generate the CMs of metallic OSSs, and volume integral equation (VIE) [7] and Poggio-Miller-Chang-Harrington-Wu-Tsai (PMCHWT) [8] based IMOs are traditionally used to generate the CMs of material OSSs, and EFIE-VIE [12]-[14] and EFIE-PMCHWT [15]-[17] based IMOs are lately used to generate the CMs of metal-material composite OSSs. Very recently, EM-extinction theorem-based operator [19]-[20] is used to generate the CMs of material OSSs, and EFIE-EM-extinction-theorem [21] and EFIE-electric-extinction-theorem [23] based operators are used to generate the CMs of composite OSSs.

The EFIE-based IMO [4]-[6] is valid for generating metallic CMs, and has had many successful engineering applications [9]-[11]. The VIE and EFIE-VIE based IMOs are valid for generating material and composite CMs respectively, but they consume a large amount of computing resources [11]. For material and composite OSSs, the PMCHWT and EFIE-PMCHWT based IMOs require less computing resource, but they usually output some spurious modes [11], [18]-[25]. The reasons leading to the spurious modes are mainly that [11], [18]-[21], [25]:

(i) the equivalent electric and magnetic currents $(\boldsymbol{J}^{\text{ES}}, \boldsymbol{M}^{\text{ES}})$ distributing on the material boundaries are simultaneously included in the IMOs;
(ii) between $\boldsymbol{J}^{\text{ES}}$ and $\boldsymbol{M}^{\text{ES}}$, only one is independent, and the other one depends on the independent one;
(iii) the original PMCHWT and EFIE-PMCHWT based IMOs overlook the dependence relation between $\boldsymbol{J}^{\text{ES}}$ and $\boldsymbol{M}^{\text{ES}}$.

In [11], [18]-[21] and [25], some different schemes were developed to eliminate the dependent variable in IMO, and the schemes are called dependent variable elimination (DVE).

During the process to realize DVE, it is necessary to inverse some full matrices related to first-kind [18]-[19], [21] or second-kind [18]-[19] Fredholm's operator, and the matrix inverse process usually needs a large amount of computing resources [26]. To resolve this problem, [22] proposed a new scheme —

[AP2004-0709] received April 12, 2020; [AP2004-0709.R1] revised November 10, 2020. This work was supported by XXXX under Project XXXXXXXX. (*Corresponding authors: Ren-Zun Lian; Ming-Yao Xia.*)

R. Z. Lian, X. Y. Guo and M. Y. Xia are with the Department of Electronics, School of Electronics Engineering and Computer Science, Peking University, Beijing 100871, China. (E-mail: rzlian@vip.163.com; myxia@pku.edu.cn).

solution domain compression (SDC) — to suppress the spurious modes, and the SDC scheme doesn't need to inverse any matrix. For lossless material OSSs, the IMO with DVE [18] and SDC [22] indeed has ability to suppress the spurious modes, but, at the same time, some extra unwanted modes will also be outputted [22]. For lossy material OSSs, the IMO with DVE and SDC cannot suppress the spurious modes effectively [22]. To resolve these problems, paper [22], under a new work-energy principle (WEP) framework rather than the traditional IE framework, proposed a new operator — driving power operator (DPO) — for generating CMs, and the DPO doesn't suffer from the problems mentioned above.

This paper focuses on generalizing the work done in [22] to metal-material composite OSSs, and is organized as follows: Sec. II does some necessary preparations for the subsequent sections; Sec. III provides a novel WEP framework for establishing CMT, and, at the same time, introduces a novel DPO for generating CMs; Sec. IV proposes a novel SDC scheme for suppressing spurious modes; Sec. V constructs CMs by orthogonalizing the DPO with SDC scheme, and derives famous Parseval's identity, and clearly reveals the fact that the physical purpose/picture of the WEP-based CMT (WEP-CMT) is to construct a set of energy-decoupled modes rather than a set of far-field-orthogonal modes for scattering systems; Sec. VI gives some numerical examples to verify the theory and formulas established in this paper; Sec. VII concludes this paper.

In what follows, the $e^{j\omega t}$ convention is used throughout, and the time-domain quantities will be added time variable $t$ explicitly for example $\boldsymbol{J}(t)$, but the frequency-domain quantities will not. Moreover, for the linear quantities, such as $\boldsymbol{E}$, we have that $\boldsymbol{E}(t) = \text{Re}\{\boldsymbol{E}e^{j\omega t}\}$; for the power-type quadratic quantities, we have $\text{Re}\{(1/2)\boldsymbol{J}^* \cdot \boldsymbol{E}\} = (1/T)\int_0^T \boldsymbol{J}(t)\cdot\boldsymbol{E}(t)dt$, where $T$ is the period of the time-harmonic EM field [27]-[28].

## II. PRELIMINARIES

In this paper, the metal-material composite OSS $\Sigma$ shown in Fig. 1 is considered. The OSS $\Sigma$ is constituted by some metallic wires $L$, some metallic surfaces $S$, a metallic body $V$ and a material body $\Omega$. The permeability, permittivity and conductivity of $\Omega$ are denoted as $\boldsymbol{\mu}$, $\boldsymbol{\varepsilon}$ and $\boldsymbol{\sigma}$ respectively, and then the complex permittivity of $\Omega$ is $\boldsymbol{\varepsilon}_c = \boldsymbol{\varepsilon} - j\boldsymbol{\sigma}/\omega$. Here, $\boldsymbol{\mu}$, $\boldsymbol{\varepsilon}$ and $\boldsymbol{\sigma}$ are real and two-order symmetrical dyads.

Obviously, the boundaries of $L$ and $S$ are just themselves [29]. The boundaries of $V$ and $\Omega$ are denoted as $\partial V$ and $\partial\Omega$ respectively. The interiors of $V$ and $\Omega$ are denoted as $\text{int}V$ and $\text{int}\Omega$ respectively. The exterior of $\Sigma$ is denoted as $\text{ext}\Sigma$.

When an incident field $\boldsymbol{F}^{\text{inc}}$, whose source distributes on $\text{ext}\Sigma$, excites the OSS, some scattered line electric current $\boldsymbol{J}^{\text{SL}}$ will be induced on $L$, and some scattered surface electric currents $\boldsymbol{J}^{\text{SS}}$ will be induced on $S \cup \partial V$, and some scattered volume electric and magnetic currents $(\boldsymbol{J}^{\text{SV}}, \boldsymbol{M}^{\text{SV}})$ will be induced on $\text{int}\Omega$ [30]-[31]. The above scattered currents will generate a scattered field $\boldsymbol{F}^{\text{sca}}$ on whole three-dimensional Euclidean space $\mathbb{E}^3$. The summation of $\boldsymbol{F}^{\text{inc}}$ and $\boldsymbol{F}^{\text{sca}}$ is the total field, and the total field is denoted as $\boldsymbol{F}^{\text{tot}}$, that is, $\boldsymbol{F}^{\text{tot}} = \boldsymbol{F}^{\text{inc}} + \boldsymbol{F}^{\text{sca}}$.

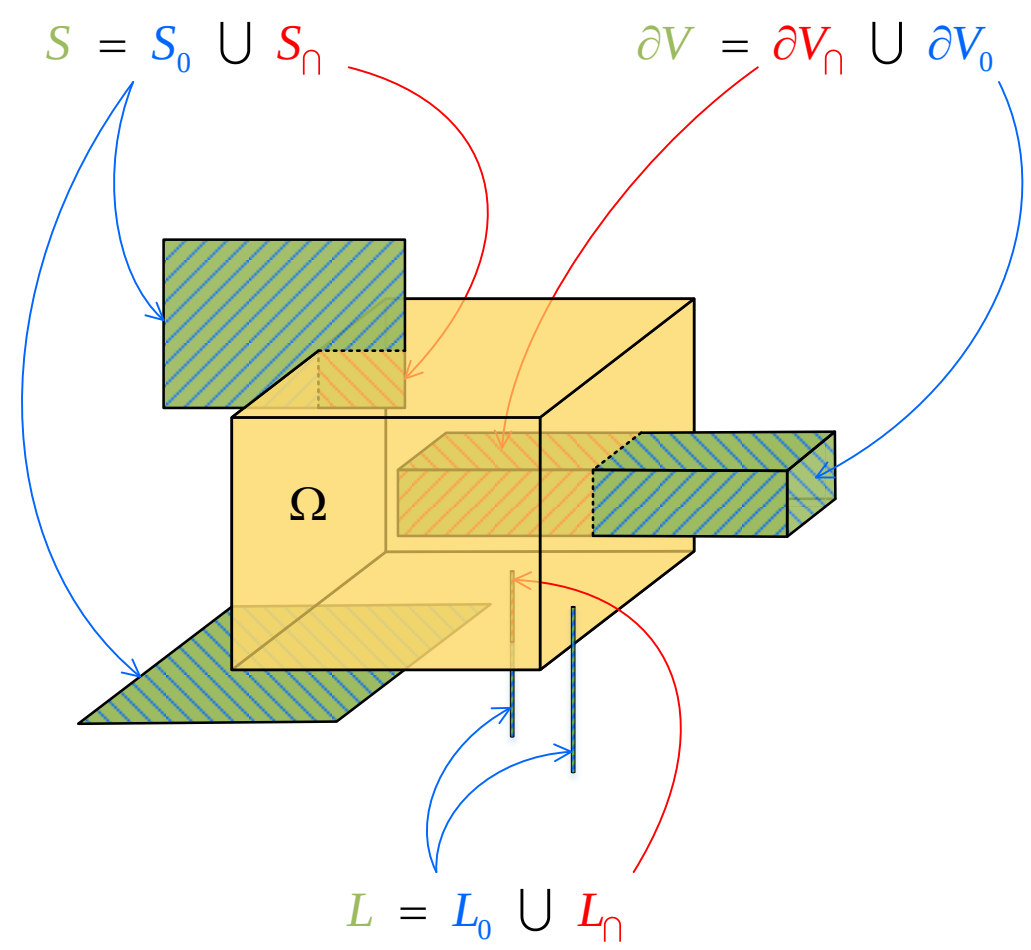

Fig. 1. Metal-material composite OSS considered in this paper.

Below, the whole $\mathbb{E}^3$ is decomposed into some sub-domains in Sec. II-A (domain decomposition); the metallic boundaries $L \cup S \cup \partial V$ and material boundary $\partial\Omega$ are decomposed into some sub-boundaries in Sec. II-B (boundary decomposition); the scattered currents on the metallic boundaries and the equivalent currents on the material boundary are decomposed into some sub-currents in Sec. II-C (current decomposition) according to the boundary decompositions given in Sec. II-B; Sec. II-D (line-surface equivalence principle) expresses the fields on the sub-domains given in Sec. II-A in terms of the sub-currents obtained in Sec. II-C.

### A. *Domain Decomposition*

According to whether $\boldsymbol{F}^{\text{sca}}$ is continuous or not, whole $\mathbb{E}^3$ can be decomposed as follows:

$$\mathbb{E}^3 = \underbrace{\overbrace{L \cup S \cup \partial V}^{\text{metallic boundaries}} \cup \partial\Omega}_{\text{on which } \boldsymbol{F}^{\text{sca}} \text{ is not continuous}} \cup \underbrace{\text{int}\Omega \cup \text{int}V \cup \text{ext}\Sigma}_{\text{on which } \boldsymbol{F}^{\text{sca}} \text{ is continuous}} \quad (1)$$

On domains $\text{int}\Omega \cup \text{int}V \cup \text{ext}\Sigma$, field $\boldsymbol{F}^{\text{sca}}$ is continuous, and can be expressed in terms of the functions of the scattered currents $(\boldsymbol{J}^{\text{SL}}, \boldsymbol{J}^{\text{SS}})$ on metallic boundaries $L \cup S \cup \partial V$ and some equivalent currents $(\boldsymbol{J}^{\text{E}}, \boldsymbol{M}^{\text{E}})$ on material boundary $\partial\Omega$. However, field $\boldsymbol{F}^{\text{sca}}$ is not continuous on boundaries $L \cup S \cup \partial V \cup \partial\Omega$.

In fact, $(\boldsymbol{J}^{\text{SL}}, \boldsymbol{J}^{\text{SS}})$ and $(\boldsymbol{J}^{\text{E}}, \boldsymbol{M}^{\text{E}})$ are not independent of each other. To distinguish the independent ones from the dependent ones is indispensable for suppressing spurious modes [11], [18]-[22], [25]. To effectively distinguish them from each other, it is necessary to decompose the boundaries and currents as the following Sec. II-B and Sec. II-C.

### B. *Boundary Decomposition*

The metallic boundaries $L$, $S$ and $\partial V$ can be decomposed as follows:

$$L = L_{\cap} \cup L_0 \quad (2)$$

$$S = S_{\cap} \cup S_0 \quad (3)$$

$$\partial V = \partial V_{\cap} \cup \partial V_0 \quad (4)$$

In (2), $L_{\cap}$ is the part completely coated by $\Omega$, and the other part is defined as $L_0$ (that is, $L_0 = L \setminus L_{\cap}$). In (3), $S_{\cap}$ is the part completely coated by $\Omega$, and the other part is defined as $S_0$ (that is, $S_0 = S \setminus S_{\cap}$). In (4), $\partial V_{\cap}$ is the metal-material boundary, and $\partial V_0$ is the metal-environment boundary.

Obviously, $L_{\cap}$, $S_{\cap}$ and $\partial V_{\cap}$ belong to $\partial\Omega$. Thus, the material boundary $\partial\Omega$ can be decomposed as follows:

$$\partial\Omega = \underbrace{L_{\cap} \bigcup S_{\cap} \bigcup \partial V_{\cap}}_{\partial\Omega_{\cap}} \bigcup \partial\Omega_0 \tag{5}$$

where $L_{\cap}$, $S_{\cap}$ and $\partial V_{\cap}$ are collectively denoted as $\partial\Omega_{\cap}$ (that is, $\partial\Omega_{\cap} = L_{\cap} \bigcup S_{\cap} \bigcup \partial V_{\cap}$), and $\partial\Omega_0$ is defined as $\partial\Omega \setminus \partial\Omega_{\cap}$ (that is, $\partial\Omega_0 = \partial\Omega \setminus \partial\Omega_{\cap}$).

All the various sub-boundaries mentioned above are shown in Fig. 1.

### *C. Current Decomposition*

Based on the boundary decompositions (2)-(4), scattered electric currents $(\boldsymbol{J}^{\text{SL}}, \boldsymbol{J}^{\text{SS}})$ can be decomposed as follows:

$$\boldsymbol{J}^{\text{SL}} = \boldsymbol{J}_{\cap}^{\text{SL}} + \boldsymbol{J}_0^{\text{SL}} \tag{6}$$
$$\boldsymbol{J}^{\text{SS}} = \boldsymbol{J}_{\cap}^{\text{SS}} + \boldsymbol{J}_0^{\text{SS}} \tag{7}$$

In (6), $\boldsymbol{J}_{\cap}^{\text{SL}}$ and $\boldsymbol{J}_0^{\text{SL}}$ are the scattered currents distributing on $L_{\cap}$ and $L_0$ respectively. In (7), $\boldsymbol{J}_{\cap}^{\text{SS}}$ and $\boldsymbol{J}_0^{\text{SS}}$ are the scattered currents distributing on $S_{\cap} \bigcup \partial V_{\cap}$ and $S_0 \bigcup \partial V_0$ respectively.

Due to the homogeneous tangential electric field boundary condition on $L_{\cap} \bigcup S_{\cap} \bigcup \partial V_{\cap}$, there doesn't exist any equivalent magnetic current distributing on $\partial\Omega_{\cap}$, and, at the same time, the equivalent electric currents on $\partial\Omega_{\cap}$ are always equal to the scattered electric currents on $L_{\cap} \bigcup S_{\cap} \bigcup \partial V_{\cap}$, where $\partial\Omega_{\cap} = L_{\cap} \bigcup S_{\cap} \bigcup \partial V_{\cap}$ as shown in (5).

In addition, there exist some equivalent surface electric and magnetic currents $(\boldsymbol{J}_0^{\text{ES}}, \boldsymbol{M}_0^{\text{ES}})$ distributing on $\partial\Omega_0$, and they are defined as follows:

$$\boldsymbol{J}_0^{\text{ES}} = \hat{\boldsymbol{n}}_- \times \boldsymbol{H}_-^{\text{tot}} \quad , \quad \text{on } \partial\Omega_0 \tag{8}$$
$$\boldsymbol{M}_0^{\text{ES}} = \boldsymbol{E}_-^{\text{tot}} \times \hat{\boldsymbol{n}}_- \quad , \quad \text{on } \partial\Omega_0 \tag{9}$$

where $\boldsymbol{F}_-^{\text{tot}}$ is the total field distributing on the inner surface of $\partial\Omega_0$, and $\hat{\boldsymbol{n}}_-$ is the inner normal direction of $\partial\Omega_0$.

### *D. Line-Surface Equivalence Principle*

Based on Huygens-Fresnel principle, the $\boldsymbol{F}^{\text{sca}}$ on $\text{ext}\,\Sigma$ and the $\boldsymbol{F}^{\text{inc}}$ on $\text{int}\,\Omega \bigcup \text{int}\,V$ can be expressed in terms of the functions of currents $(\boldsymbol{J}_0^{\text{SL}}, \boldsymbol{J}_0^{\text{SS}}, \boldsymbol{J}_0^{\text{ES}}, \boldsymbol{M}_0^{\text{ES}})$ as follows:

$$\left.\begin{aligned} \text{ext}\,\Sigma &: -\boldsymbol{F}^{\text{sca}} \\ \text{int}\,\Omega &: \quad \boldsymbol{F}^{\text{inc}} \\ \text{int}\,V &: \quad \boldsymbol{F}^{\text{inc}} \end{aligned}\right\} = \mathcal{F}_0\left(-\boldsymbol{J}_0^{\text{SL}} - \boldsymbol{J}_0^{\text{SS}} + \boldsymbol{J}_0^{\text{ES}}, \boldsymbol{M}_0^{\text{ES}}\right) \tag{10}$$

In above (10), $\boldsymbol{F} = \boldsymbol{E} / \boldsymbol{H}$ and correspondingly $\mathcal{F}_0 = \mathcal{E}_0 / \mathcal{H}_0$, and the operators $\mathcal{E}_0(\boldsymbol{J}, \boldsymbol{M})$ and $\mathcal{H}_0(\boldsymbol{J}, \boldsymbol{M})$ are defined as that $\mathcal{E}_0(\boldsymbol{J}, \boldsymbol{M}) = -j\omega\mu_0\mathcal{L}_0(\boldsymbol{J}) - \mathcal{K}_0(\boldsymbol{M})$ and $\mathcal{H}_0(\boldsymbol{J}, \boldsymbol{M}) = -j\omega\varepsilon_0\mathcal{L}_0(\boldsymbol{M}) + \mathcal{K}_0(\boldsymbol{J})$, where the operators $\mathcal{L}_0$ and $\mathcal{K}_0$ are defined as the ones given in [22] and [32].

Similarly, the $\boldsymbol{F}^{\text{tot}}$ on $\text{int}\,\Omega$ and $\text{int}\,V$ can be expressed in terms of the currents $(\boldsymbol{J}_{\cap}^{\text{SL}}, \boldsymbol{J}_{\cap}^{\text{SS}}, \boldsymbol{J}_0^{\text{ES}}, \boldsymbol{M}_0^{\text{ES}})$ as follows:

$$\text{int}\,\Omega : \ \boldsymbol{F}^{\text{tot}} = \mathcal{F}\left(\boldsymbol{J}_{\cap}^{\text{SL}} + \boldsymbol{J}_{\cap}^{\text{SS}} + \boldsymbol{J}_0^{\text{ES}}, \boldsymbol{M}_0^{\text{ES}}\right) \tag{11}$$
$$\text{int}\,V : \ \boldsymbol{F}^{\text{tot}} = 0 \tag{12}$$

In the above (11) and (12), $\boldsymbol{F} = \boldsymbol{E} / \boldsymbol{H}$ and correspondingly $\mathcal{F} = \mathcal{E} / \mathcal{H}$, and the operator $\mathcal{F}(\boldsymbol{J}, \boldsymbol{M})$ is defined as that $\mathcal{F}(\boldsymbol{J}, \boldsymbol{M}) = \mathbf{G}^{JF} * \boldsymbol{J} + \mathbf{G}^{MF} * \boldsymbol{M}$, where $\mathbf{G}^{JF}$ and $\mathbf{G}^{MF}$ are the dyadic Green's functions corresponding to material parameters $(\boldsymbol{\mu}, \boldsymbol{\varepsilon}_{\text{c}})$, and the convolution integral operation "$*$" is defined as that $\mathbf{G} * \boldsymbol{C} = \int_{\Gamma} \mathbf{G}(\boldsymbol{r}, \boldsymbol{r}') \cdot \boldsymbol{C}(\boldsymbol{r}') d\Gamma'$.

Therefore, the $\boldsymbol{F}^{\text{sca}}$ on $\text{int}\,\Omega$ and $\text{int}\,V$ can be expressed in terms of the functions of currents $(\boldsymbol{J}_0^{\text{SL}}, \boldsymbol{J}_{\cap}^{\text{SL}}, \boldsymbol{J}_0^{\text{SS}}, \boldsymbol{J}_{\cap}^{\text{SS}}, \boldsymbol{J}_0^{\text{ES}}, \boldsymbol{M}_0^{\text{ES}})$ as follows:

$$\begin{aligned} \text{int}\,\Omega : \ \boldsymbol{F}^{\text{sca}} = \ & \mathcal{F}\left(\boldsymbol{J}_{\cap}^{\text{SL}} + \boldsymbol{J}_{\cap}^{\text{SS}} + \boldsymbol{J}_0^{\text{ES}}, \boldsymbol{M}_0^{\text{ES}}\right) \\ & - \mathcal{F}_0\left(-\boldsymbol{J}_0^{\text{SL}} - \boldsymbol{J}_0^{\text{SS}} + \boldsymbol{J}_0^{\text{ES}}, \boldsymbol{M}_0^{\text{ES}}\right) \end{aligned} \tag{13}$$
$$\text{int}\,V : \ \boldsymbol{F}^{\text{sca}} = -\mathcal{F}_0\left(-\boldsymbol{J}_0^{\text{SL}} - \boldsymbol{J}_0^{\text{SS}} + \boldsymbol{J}_0^{\text{ES}}, \boldsymbol{M}_0^{\text{ES}}\right) \tag{14}$$

because of (10)-(12) and that $\boldsymbol{F}^{\text{sca}} = \boldsymbol{F}^{\text{tot}} - \boldsymbol{F}^{\text{inc}}$.

The currents used in (10)-(14) are line-type or surface-type, so convolution integral formulations (10)-(14) are collectively referred to as line-surface equivalence principle (LSEP). Next, the LSEP (10) will be used to formulate the generating operator of CMs in the following Sec. III, and the LSEP (11) will be used to compress the solution domain of characteristic equation in the future Sec. IV, and the LSEPs (10), (13) and (14) will be used to compute the modal scattering fields.

## III. Work-Energy Principle and Driving Power Operator

In this section, the work-energy viewpoint is used to describe scattering problem, and some functions with power dimension are introduced to express the work-energy transformation process "$\boldsymbol{F}^{\text{inc}} \xrightarrow[\text{work}]{\text{do}} \text{OSS} \xrightarrow[\text{energy}]{\text{scatter}} \boldsymbol{F}^{\text{sca}} \xrightarrow[\text{energy}]{\text{carry}} \text{Infinity}$" quantitatively. The work-energy viewpoint and power functions will be utilized to establish the WEP-based CMT for composite OSSs in the future Sec. V.

### *A. Work-Energy Principle (WEP)*

The conservation law of energy [33]-[34] tells us that the action of the incident field $\boldsymbol{F}^{\text{inc}}$ on the scattered currents $(\boldsymbol{J}^{\text{SL}}, \boldsymbol{J}^{\text{SS}}, \boldsymbol{J}^{\text{SV}}, \boldsymbol{M}^{\text{SV}})$ will give rise to a transformation between work and energy. The work-energy transformation can be quantitatively expressed as follows:

$$\boldsymbol{W}^{\text{Driv}} = \boldsymbol{\mathcal{E}}_{S_\infty}^{\text{rad}} + \boldsymbol{\mathcal{E}}_{\Omega}^{\text{dis}} + \Delta\boldsymbol{\mathcal{E}}_{\mathbb{E}^3}^{\text{sto}} + \Delta\boldsymbol{\mathcal{E}}_{\Omega}^{\text{sto}} \tag{15}$$

The mathematical expressions of the terms in (15) are given in (16)-(20). In (16)-(20), the time interval $\Delta t$ is a positive real number; $\Delta\boldsymbol{\mu} = \boldsymbol{\mu} - \mathbf{I}\mu_0$ and $\Delta\boldsymbol{\varepsilon} = \boldsymbol{\varepsilon} - \mathbf{I}\varepsilon_0$, where $\mathbf{I}$ is the unit dyad; the inner product is defined as $<\boldsymbol{f},\boldsymbol{g}>_\Gamma=\int_\Gamma \boldsymbol{f}^*\cdot\boldsymbol{g}d\Gamma$; $S_\infty$ is a spherical surface with infinite radius.

Evidently, equation (15) has a very clear physical explanation: in time interval $t_0 \sim t_0+\Delta t$, the work $\mathcal{W}^{\rm Driv}$ done by fields ($\boldsymbol{E}^{\rm inc},\boldsymbol{H}^{\rm inc}$) on currents ($\boldsymbol{J}^{\rm SL},\boldsymbol{J}^{\rm SS},\boldsymbol{J}^{\rm SV},\boldsymbol{M}^{\rm SV}$) is transformed into four parts — the energy $\mathcal{E}_{S_\infty}^{\rm rad}$ radiated to infinity by passing through $S_\infty$, the energy $\mathcal{E}_\Omega^{\rm dis}$ dissipated in $\Omega$, the increase of the magnetic and electric field energies $\mathcal{E}_{\mathbb{E}^3}^{\rm sto}$ stored in $\mathbb{E}^3$, and the increase of the magnetization and polarization energies $\mathcal{E}_\Omega^{\rm sto}$ stored in $\Omega$. Following the terminology used in mechanism [35]-[36], the above transformation relation between work and energy is called work-energy principle.

*B. Driving Power Operator (DPO) — Line-Surface-Volume Formulation*

Clearly, the work term $\mathcal{W}^{\rm Driv}$ is just the source to sustain the work-energy transformation mentioned above, and it is also the source to drive a steady working of the composite OSS. Thus, $\mathcal{W}^{\rm Driv}$ is called driving work, and the associated power is called driving power (DP) and denoted as $P^{\rm Driv}(t)$. Evidently, $P^{\rm Driv}(t)$ has the following time-domain operator expression

$$\begin{aligned} P^{\rm Driv}(t) = &\left\langle \boldsymbol{J}^{\rm SL}(t)+\boldsymbol{J}^{\rm SS}(t)+\boldsymbol{J}^{\rm SV}(t),\boldsymbol{E}^{\rm inc}(t)\right\rangle_\Sigma \\ &+\left\langle \boldsymbol{M}^{\rm SV}(t),\boldsymbol{H}^{\rm inc}(t)\right\rangle_\Sigma \end{aligned} \tag{21}$$

and the operator is correspondingly called time-domain driving power operator (DPO).

The frequency-domain version of (21) is as follows:

$$P^{\rm driv}=(1/2)\left\langle \boldsymbol{J}^{\rm SL}+\boldsymbol{J}^{\rm SS}+\boldsymbol{J}^{\rm SV},\boldsymbol{E}^{\rm inc}\right\rangle_\Sigma+(1/2)\left\langle \boldsymbol{M}^{\rm SV},\boldsymbol{H}^{\rm inc}\right\rangle_\Sigma \tag{22}$$

where coefficient 1/2 originates from the time average for the power-type quadratic quantity of time-harmonic EM field.

*C. Driving Power Operator (DPO) — Line-Surface Formulation*

Similarly to the conclusions given in [18]-[20] and [22], there exists the following relation

$$\begin{aligned} &(1/2)\left\langle \boldsymbol{J}^{\rm SV},\boldsymbol{E}^{\rm inc}\right\rangle_\Omega+(1/2)\left\langle \boldsymbol{M}^{\rm SV},\boldsymbol{H}^{\rm inc}\right\rangle_\Omega \\ &=-(1/2)\left\langle \boldsymbol{J}_\cap^{\rm SL}+\boldsymbol{J}_\cap^{\rm SS}+\boldsymbol{J}_0^{\rm ES},\boldsymbol{E}^{\rm inc}\right\rangle_{\partial\Omega}-(1/2)\left\langle \boldsymbol{M}_0^{\rm ES},\boldsymbol{H}^{\rm inc}\right\rangle_{\partial\Omega} \end{aligned} \tag{23}$$

Substituting the above (23) and the previous (2)-(7) into (22), we immediately obtain the following line-surface formulation

$$\begin{aligned} P^{\rm driv} = &(1/2)\left\langle \boldsymbol{J}_0^{\rm SL}+\boldsymbol{J}_0^{\rm SS},\boldsymbol{E}^{\rm inc}\right\rangle_{L_0\cup S_0\cup\partial V_0} \\ &-(1/2)\left\langle \boldsymbol{J}_0^{\rm ES},\boldsymbol{E}^{\rm inc}\right\rangle_{\partial\Omega_0}-(1/2)\left\langle \boldsymbol{M}_0^{\rm ES},\boldsymbol{H}^{\rm inc}\right\rangle_{\partial\Omega_0} \\ = &(1/2)\left\langle \boldsymbol{J}_0^{\rm SL}+\boldsymbol{J}_0^{\rm SS},-\boldsymbol{E}^{\rm sca}\right\rangle_{L_0\cup S_0\cup\partial V_0} \\ &-(1/2)\left\langle \boldsymbol{J}_0^{\rm ES},\boldsymbol{E}^{\rm inc}\right\rangle_{\partial\Omega_0}-(1/2)\left\langle \boldsymbol{M}_0^{\rm ES},\boldsymbol{H}^{\rm inc}\right\rangle_{\partial\Omega_0} \end{aligned} \tag{24}$$

for frequency-domain DPO $P^{\rm driv}$, where the second equality is because of the tangential electric field boundary condition $\boldsymbol{E}_{\rm tan}^{\rm inc}=-\boldsymbol{E}_{\rm tan}^{\rm sca}$ on $L_0\cup S_0\cup\partial V_0$. Applying LSEP (10) to (24), we have the $P^{\rm driv}$ with only current variables as expressed in (25), where "${\rm P.V.}\mathcal{K}_0$" represents the principal value of operator $\mathcal{K}_0$ [32], and the subscripts "PVT" and "SCT" used in $P_{\rm PVT}^{\rm driv}$ and $P_{\rm SCT}^{\rm driv}$ are the acronyms of "principal value term" and "singular current term" respectively.

If the currents contained in (25) are expanded in terms of some proper basis functions, the (25) is immediately discretized into the following matrix form

$$P^{\rm driv}=\mathbf{a}^H\cdot\underbrace{\left(\mathrm{P}_{\rm PVT}^{\rm driv}+\mathrm{P}_{\rm SCT}^{\rm driv}\right)}_{\mathrm{P}^{\rm driv}}\cdot\mathbf{a} \tag{26}$$

in which the superscript "$H$" represents the conjugate transpose operation for a matrix or vector, and

$$\mathrm{P}_{\rm PVT}^{\rm driv}=\begin{bmatrix} \mathrm{P}_{\rm PVT}^{\boldsymbol{J}_0^{\rm SL}\boldsymbol{J}_0^{\rm SL}} & 0 & \mathrm{P}_{\rm PVT}^{\boldsymbol{J}_0^{\rm SL}\boldsymbol{J}_0^{\rm SS}} & 0 & \mathrm{P}_{\rm PVT}^{\boldsymbol{J}_0^{\rm SL}\boldsymbol{J}_0^{\rm ES}} & \mathrm{P}_{\rm PVT}^{\boldsymbol{J}_0^{\rm SL}\boldsymbol{M}_0^{\rm ES}} \\ 0 & 0 & 0 & 0 & 0 & 0 \\ \mathrm{P}_{\rm PVT}^{\boldsymbol{J}_0^{\rm SS}\boldsymbol{J}_0^{\rm SL}} & 0 & \mathrm{P}_{\rm PVT}^{\boldsymbol{J}_0^{\rm SS}\boldsymbol{J}_0^{\rm SS}} & 0 & \mathrm{P}_{\rm PVT}^{\boldsymbol{J}_0^{\rm SS}\boldsymbol{J}_0^{\rm ES}} & \mathrm{P}_{\rm PVT}^{\boldsymbol{J}_0^{\rm SS}\boldsymbol{M}_0^{\rm ES}} \\ 0 & 0 & 0 & 0 & 0 & 0 \\ \mathrm{P}_{\rm PVT}^{\boldsymbol{J}_0^{\rm ES}\boldsymbol{J}_0^{\rm SL}} & 0 & \mathrm{P}_{\rm PVT}^{\boldsymbol{J}_0^{\rm ES}\boldsymbol{J}_0^{\rm SS}} & 0 & \mathrm{P}_{\rm PVT}^{\boldsymbol{J}_0^{\rm ES}\boldsymbol{J}_0^{\rm ES}} & \mathrm{P}_{\rm PVT}^{\boldsymbol{J}_0^{\rm ES}\boldsymbol{M}_0^{\rm ES}} \\ \mathrm{P}_{\rm PVT}^{\boldsymbol{M}_0^{\rm ES}\boldsymbol{J}_0^{\rm SL}} & 0 & \mathrm{P}_{\rm PVT}^{\boldsymbol{M}_0^{\rm ES}\boldsymbol{J}_0^{\rm SS}} & 0 & \mathrm{P}_{\rm PVT}^{\boldsymbol{M}_0^{\rm ES}\boldsymbol{J}_0^{\rm ES}} & \mathrm{P}_{\rm PVT}^{\boldsymbol{M}_0^{\rm ES}\boldsymbol{M}_0^{\rm ES}} \end{bmatrix} \tag{27}$$

$$\mathcal{W}^{\rm Driv} = \int_{t_0}^{t_0+\Delta t}\left[\left\langle \boldsymbol{J}^{\rm SL}(t)+\boldsymbol{J}^{\rm SS}(t)+\boldsymbol{J}^{\rm SV}(t),\boldsymbol{E}^{\rm inc}(t)\right\rangle_\Sigma+\left\langle \boldsymbol{M}^{\rm SV}(t),\boldsymbol{H}^{\rm inc}(t)\right\rangle_\Sigma\right]dt \tag{16}$$

$$\mathcal{E}_{S_\infty}^{\rm rad} = \int_{t_0}^{t_0+\Delta t}\left\{\oiint_{S_\infty}\left[\boldsymbol{E}^{\rm sca}(t)\times\boldsymbol{H}^{\rm sca}(t)\right]\cdot d\boldsymbol{S}\right\}dt \tag{17}$$

$$\mathcal{E}_\Omega^{\rm dis} = \int_{t_0}^{t_0+\Delta t}\left\langle \boldsymbol{\sigma}\cdot\boldsymbol{E}^{\rm tot}(t),\boldsymbol{E}^{\rm tot}(t)\right\rangle_\Omega dt \tag{18}$$

$$\begin{aligned} \Delta\mathcal{E}_{\mathbb{E}^3}^{\rm sto} = &\left[(1/2)\left\langle \boldsymbol{H}^{\rm sca}(t_0+\Delta t),\mu_0\boldsymbol{H}^{\rm sca}(t_0+\Delta t)\right\rangle_{\mathbb{E}^3}+(1/2)\left\langle \varepsilon_0\boldsymbol{E}^{\rm sca}(t_0+\Delta t),\boldsymbol{E}^{\rm sca}(t_0+\Delta t)\right\rangle_{\mathbb{E}^3}\right] \\ &-\left[(1/2)\left\langle \boldsymbol{H}^{\rm sca}(t_0),\mu_0\boldsymbol{H}^{\rm sca}(t_0)\right\rangle_{\mathbb{E}^3}+(1/2)\left\langle \varepsilon_0\boldsymbol{E}^{\rm sca}(t_0),\boldsymbol{E}^{\rm sca}(t_0)\right\rangle_{\mathbb{E}^3}\right] \end{aligned} \tag{19}$$

$$\begin{aligned} \Delta\mathcal{E}_\Omega^{\rm sto} = &\left[(1/2)\left\langle \boldsymbol{H}^{\rm tot}(t_0+\Delta t),\Delta\boldsymbol{\mu}\cdot\boldsymbol{H}^{\rm tot}(t_0+\Delta t)\right\rangle_\Omega+(1/2)\left\langle \Delta\boldsymbol{\varepsilon}\cdot\boldsymbol{E}^{\rm tot}(t_0+\Delta t),\boldsymbol{E}^{\rm tot}(t_0+\Delta t)\right\rangle_\Omega\right] \\ &-\left[(1/2)\left\langle \boldsymbol{H}^{\rm tot}(t_0),\Delta\boldsymbol{\mu}\cdot\boldsymbol{H}^{\rm tot}(t_0)\right\rangle_\Omega+(1/2)\left\langle \Delta\boldsymbol{\varepsilon}\cdot\boldsymbol{E}^{\rm tot}(t_0),\boldsymbol{E}^{\rm tot}(t_0)\right\rangle_\Omega\right] \end{aligned} \tag{20}$$

$$\begin{aligned} P^{\rm driv} = &\left.\begin{aligned} &(1/2)\left\langle \boldsymbol{J}_0^{\rm SL}+\boldsymbol{J}_0^{\rm SS},-j\omega\mu_0\mathcal{L}_0\left(-\boldsymbol{J}_0^{\rm SL}-\boldsymbol{J}_0^{\rm SS}+\boldsymbol{J}_0^{\rm ES}\right)-{\rm P.V.}\mathcal{K}_0\left(\boldsymbol{M}_0^{\rm ES}\right)\right\rangle_{L_0\cup S_0\cup\partial V_0} \\ &-(1/2)\left\langle \boldsymbol{J}_0^{\rm ES},-j\omega\mu_0\mathcal{L}_0\left(-\boldsymbol{J}_0^{\rm SL}-\boldsymbol{J}_0^{\rm SS}+\boldsymbol{J}_0^{\rm ES}\right)-{\rm P.V.}\mathcal{K}_0\left(\boldsymbol{M}_0^{\rm ES}\right)\right\rangle_{\partial\Omega_0} \\ &-(1/2)\left\langle \boldsymbol{M}_0^{\rm ES},{\rm P.V.}\mathcal{K}_0\left(-\boldsymbol{J}_0^{\rm SL}-\boldsymbol{J}_0^{\rm SS}+\boldsymbol{J}_0^{\rm ES}\right)-j\omega\varepsilon_0\mathcal{L}_0\left(\boldsymbol{M}_0^{\rm ES}\right)\right\rangle_{\partial\Omega_0} \end{aligned}\right\}\ \text{principal value term (PVT) } P_{\rm PVT}^{\rm driv} \\ &\left.-(1/2)\left\langle \boldsymbol{J}_0^{\rm ES},\hat{\boldsymbol{n}}_-\times(1/2)\boldsymbol{M}_0^{\rm ES}\right\rangle_{\partial\Omega_0}-(1/2)\left\langle \boldsymbol{M}_0^{\rm ES},(1/2)\boldsymbol{J}_0^{\rm ES}\times\hat{\boldsymbol{n}}_-\right\rangle_{\partial\Omega_0}\right\}\ \text{singular current term (SCT) } P_{\rm SCT}^{\rm driv} \end{aligned} \tag{25}$$

$$\mathrm{P}_{\mathrm{SCT}}^{\mathrm{driv}} = \begin{bmatrix} 0 & 0 & 0 & 0 & 0 & 0 \\ 0 & 0 & 0 & 0 & 0 & 0 \\ 0 & 0 & 0 & 0 & 0 & 0 \\ 0 & 0 & 0 & 0 & 0 & 0 \\ 0 & 0 & 0 & 0 & 0 & \mathrm{P}_{\mathrm{SCT}}^{\boldsymbol{J}_0^{\mathrm{ES}} \boldsymbol{M}_0^{\mathrm{ES}}} \\ 0 & 0 & 0 & 0 & \mathrm{P}_{\mathrm{SCT}}^{\boldsymbol{M}_0^{\mathrm{ES}} \boldsymbol{J}_0^{\mathrm{ES}}} & 0 \end{bmatrix} \tag{28}$$

$$\mathrm{a} = \begin{bmatrix} \mathrm{a}^{\boldsymbol{J}_0^{\mathrm{SL}}} \\ \mathrm{a}^{\boldsymbol{J}_\cap^{\mathrm{SL}}} \\ \mathrm{a}^{\boldsymbol{J}_0^{\mathrm{SS}}} \\ \mathrm{a}^{\boldsymbol{J}_\cap^{\mathrm{SS}}} \\ \mathrm{a}^{\boldsymbol{J}_0^{\mathrm{ES}}} \\ \mathrm{a}^{\boldsymbol{M}_0^{\mathrm{ES}}} \end{bmatrix} \tag{29}$$

In (29), $\mathrm{a}^{\boldsymbol{J}_0^{\mathrm{SL}}}$, $\mathrm{a}^{\boldsymbol{J}_\cap^{\mathrm{SL}}}$, $\mathrm{a}^{\boldsymbol{J}_0^{\mathrm{SS}}}$, $\mathrm{a}^{\boldsymbol{J}_\cap^{\mathrm{SS}}}$, $\mathrm{a}^{\boldsymbol{J}_0^{\mathrm{ES}}}$ and $\mathrm{a}^{\boldsymbol{M}_0^{\mathrm{ES}}}$ are the column vectors constituted by the expansion coefficients of $\boldsymbol{J}_0^{\mathrm{SL}}$, $\boldsymbol{J}_\cap^{\mathrm{SL}}$, $\boldsymbol{J}_0^{\mathrm{SS}}$, $\boldsymbol{J}_\cap^{\mathrm{SS}}$, $\boldsymbol{J}_0^{\mathrm{ES}}$ and $\boldsymbol{M}_0^{\mathrm{ES}}$ respectively; the 0s in (27) and (28) are the zero matrices with proper row and column numbers; the formulations for calculating the nonzero sub-matrices in (27) and (28) are trivial, and they are not explicitly listed here.

In the future Sec. V, we will utilize the line-surface formulation of frequency-domain DPO to generate the CMs of composite OSSs, and employ the time-average version of DPO to derive famous Parseval's identity and to reveal the physical picture/purpose of CMT.

## IV. Solution Domain Compression and Its Variant

Before constructing CMs in Sec. V, we first propose a pre-processing scheme — SDC — for the line-surface formulation of frequency-domain DPO in this section.

A directive usage of matrix operator $\mathrm{P}^{\mathrm{driv}}$ gives characteristic equation $\mathrm{P}_-^{\mathrm{driv}} \cdot \mathrm{a}_\xi = \lambda_\xi \mathrm{P}_+^{\mathrm{driv}} \cdot \mathrm{a}_\xi$. Here, $\mathrm{P}_+^{\mathrm{driv}}$ and $\mathrm{P}_-^{\mathrm{driv}}$ are the positive and negative Hermitian parts of $\mathrm{P}^{\mathrm{driv}}$, that is, $\mathrm{P}_+^{\mathrm{driv}} = [\mathrm{P}^{\mathrm{driv}} + (\mathrm{P}^{\mathrm{driv}})^H]/2$ and $\mathrm{P}_-^{\mathrm{driv}} = [\mathrm{P}^{\mathrm{driv}} - (\mathrm{P}^{\mathrm{driv}})^H]/(2j)$.

By solving the characteristic equation, we obtain the modes of the composite OSS shown in Fig. 2, and the associated modal significances (MSs) are shown in Fig. 3. The OSS is lossless, and with parameters $\boldsymbol{\mu} = \mathbf{I}4\mu_0$ and $\boldsymbol{\varepsilon}_{\mathrm{c}} = \mathbf{I}4\varepsilon_0$. At the same time, we also calculate the CMs of the OSS by orthogonalizing the line-surface-volume formulation (22) of DPO (which is almost identical to the EFIE-VIE-based IMO except a coefficient 1/2), and show the associated MSs in Fig. 4. Obviously, the results shown in Fig. 3 are not consistent with the ones shown in Fig. 4, and we provide a scheme for resolving the problem as below.

The LSEP (11) and the current definitions (8) and (9) give the following integral equations

$$\left[\mathcal{H}\left(\boldsymbol{J}_\cap^{\mathrm{SL}} + \boldsymbol{J}_\cap^{\mathrm{SS}} + \boldsymbol{J}_0^{\mathrm{ES}}, \boldsymbol{M}_0^{\mathrm{ES}}\right)\right]_{\boldsymbol{r}' \to \boldsymbol{r}}^{\tan} + \hat{\boldsymbol{n}}_-(\boldsymbol{r}) \times \boldsymbol{J}_0^{\mathrm{ES}}(\boldsymbol{r}) = 0 \tag{30}$$

$$\left[\mathcal{E}\left(\boldsymbol{J}_\cap^{\mathrm{SL}} + \boldsymbol{J}_\cap^{\mathrm{SS}} + \boldsymbol{J}_0^{\mathrm{ES}}, \boldsymbol{M}_0^{\mathrm{ES}}\right)\right]_{\boldsymbol{r}' \to \boldsymbol{r}}^{\tan} + \boldsymbol{M}_0^{\mathrm{ES}}(\boldsymbol{r}) \times \hat{\boldsymbol{n}}_-(\boldsymbol{r}) = 0 \tag{31}$$

In the equations, $\boldsymbol{r}' \in \mathrm{int}\,\Omega$ and $\boldsymbol{r} \in \partial\Omega_0$; subscript "$\boldsymbol{r}' \to \boldsymbol{r}$" means that $\boldsymbol{r}'$ approaches $\boldsymbol{r}$; superscript "tan" means that the equations are satisfied by tangential components.

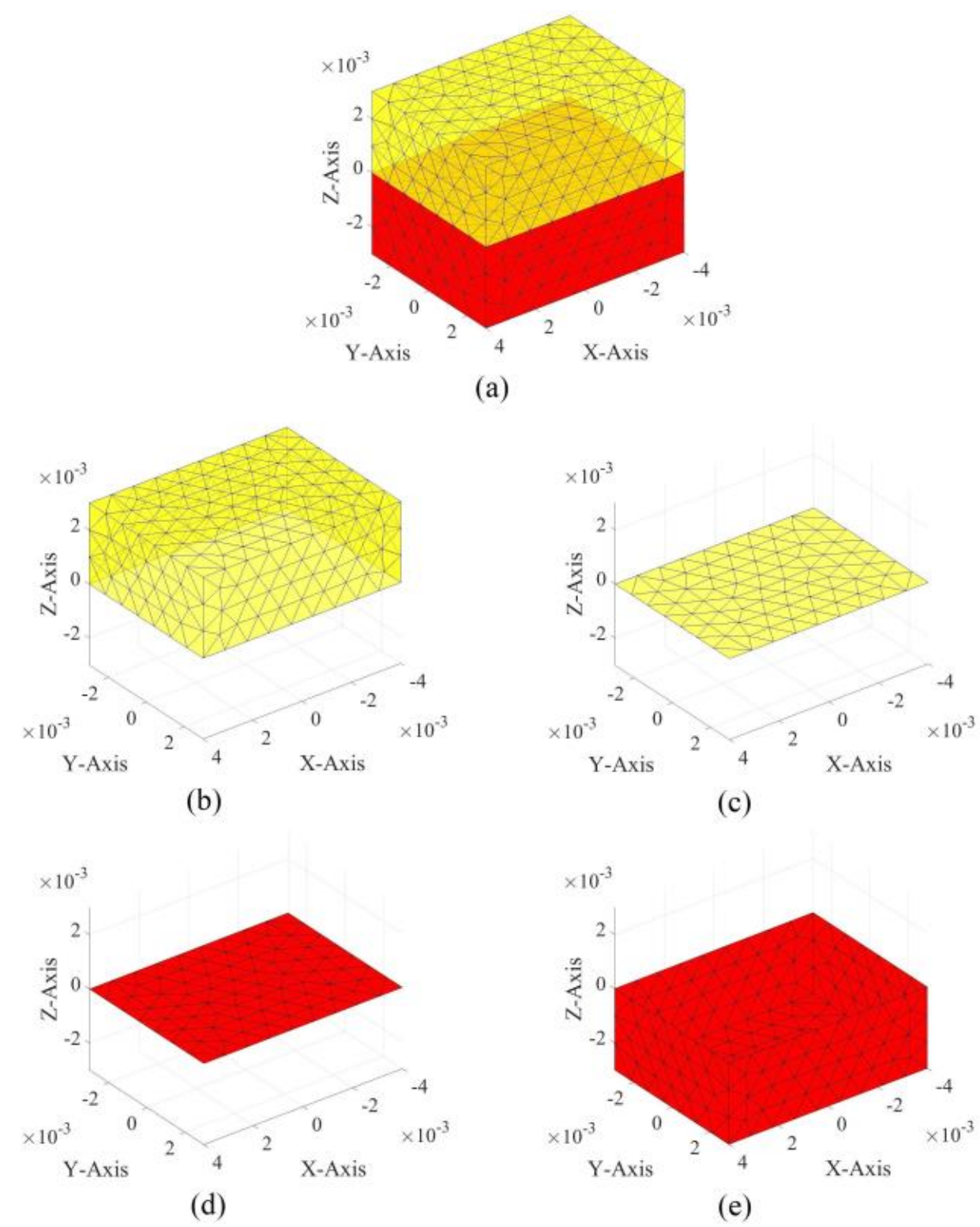

Fig. 2. Geometry of a composite OSS constituted by a material cube and a metallic cube, and the cubes have the same geometrical dimension $8\mathrm{mm} \times 6\mathrm{mm} \times 3\mathrm{mm}$. (a) Topological structure of whole OSS; (b) material-environment boundary $\partial\Omega_0$; (c) material-metal boundary $\partial\Omega_\cap$; (d) metal-material boundary $\partial V_\cap$; (e) metal-environment boundary $\partial V_0$.

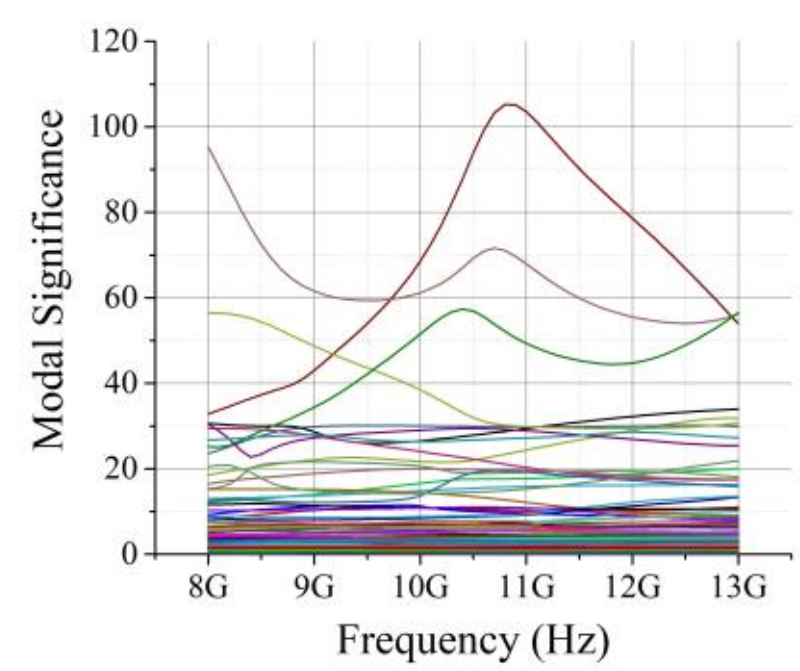

Fig. 3. MSs of some modes of a lossless composite OSS. The OSS is with parameters $\boldsymbol{\mu} = \mathbf{I}4\mu_0$ and $\boldsymbol{\varepsilon}_{\mathrm{c}} = \mathbf{I}4\varepsilon_0$, and has the topological structure shown in Fig. 2. The modes shown in this figure are derived from solving characteristic equation $\mathrm{P}_-^{\mathrm{driv}} \cdot \mathrm{a}_\xi = \lambda_\xi \mathrm{P}_+^{\mathrm{driv}} \cdot \mathrm{a}_\xi$.

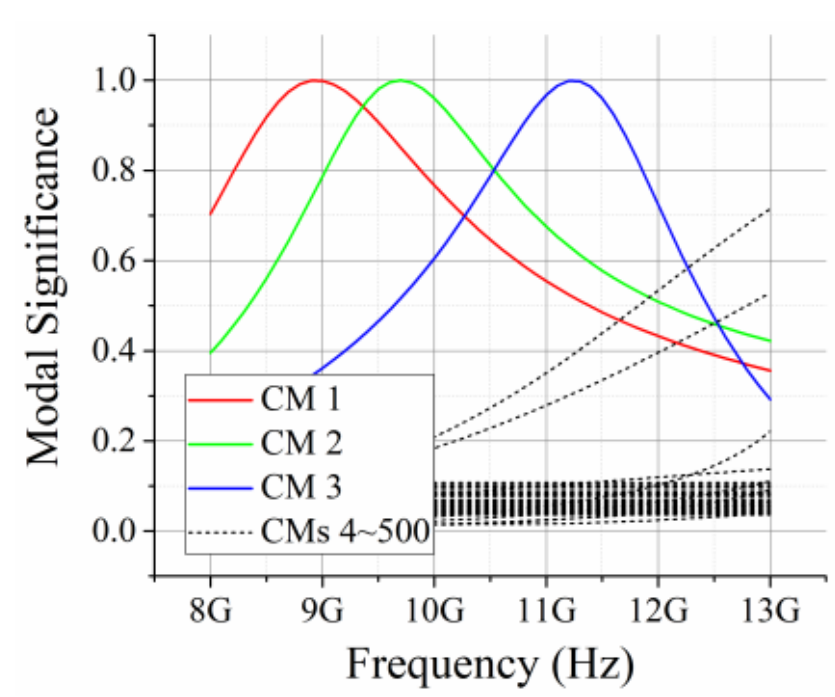

Fig. 4. MSs corresponding to the CMs of a lossless composite OSS. The OSS is with parameters $\boldsymbol{\mu} = \mathbf{I}4\mu_0$ and $\boldsymbol{\varepsilon}_{\mathrm{c}} = \mathbf{I}4\varepsilon_0$, and has the topological structure shown in Fig. 2. The modes shown in this figure are derived from orthogonalizing DPO (22).

The LSEP (11) and the tangential electric field continuation conditions on metallic boundaries $L_\cap$, $S_\cap$ and $\partial V_\cap$ give the following electric field integral equations

$$\left[\mathcal{E}\left(\boldsymbol{J}_\cap^{\rm SL}+\boldsymbol{J}_\cap^{\rm SS}+\boldsymbol{J}_0^{\rm ES},\boldsymbol{M}_0^{\rm ES}\right)\right]_{\boldsymbol{r}'\to\boldsymbol{r}}^{\tan}=0 \tag{32}$$

where $\boldsymbol{r}'\in{\rm int}\,\Omega$ and $\boldsymbol{r}\in L_\cap\cup S_\cap\cup\partial V_\cap$.

By discretizing the above integral equations (30)&(31) and (32), we obtain the following matrix equations

$$\overbrace{\begin{bmatrix}0 & \mathbb{G}^{\boldsymbol{M}_0^{\rm ES}\boldsymbol{J}_\cap^{\rm SL}} & 0 & \mathbb{G}^{\boldsymbol{M}_0^{\rm ES}\boldsymbol{J}_\cap^{\rm SS}} & \mathbb{G}^{\boldsymbol{M}_0^{\rm ES}\boldsymbol{J}_0^{\rm ES}} & \mathbb{G}^{\boldsymbol{M}_0^{\rm ES}\boldsymbol{M}_0^{\rm ES}}\end{bmatrix}}^{\mathbb{G}^{\rm DoJ}}\cdot\mathbb{a} = 0 \tag{33}$$

$$\underbrace{\begin{bmatrix}0 & \mathbb{G}^{\boldsymbol{J}_0^{\rm ES}\boldsymbol{J}_\cap^{\rm SL}} & 0 & \mathbb{G}^{\boldsymbol{J}_0^{\rm ES}\boldsymbol{J}_\cap^{\rm SS}} & \mathbb{G}^{\boldsymbol{J}_0^{\rm ES}\boldsymbol{J}_0^{\rm ES}} & \mathbb{G}^{\boldsymbol{J}_0^{\rm ES}\boldsymbol{M}_0^{\rm ES}}\end{bmatrix}}_{\mathbb{G}^{\rm DoM}}\cdot\mathbb{a} = 0 \tag{34}$$

and

$$\underbrace{\begin{bmatrix}0 & \mathbb{G}^{\boldsymbol{J}_\cap^{\rm SL}\boldsymbol{J}_\cap^{\rm SL}} & 0 & \mathbb{G}^{\boldsymbol{J}_\cap^{\rm SL}\boldsymbol{J}_\cap^{\rm SS}} & \mathbb{G}^{\boldsymbol{J}_\cap^{\rm SL}\boldsymbol{J}_0^{\rm ES}} & \mathbb{G}^{\boldsymbol{J}_\cap^{\rm SL}\boldsymbol{M}_0^{\rm ES}}\\ 0 & \mathbb{G}^{\boldsymbol{J}_\cap^{\rm SS}\boldsymbol{J}_\cap^{\rm SL}} & 0 & \mathbb{G}^{\boldsymbol{J}_\cap^{\rm SS}\boldsymbol{J}_\cap^{\rm SS}} & \mathbb{G}^{\boldsymbol{J}_\cap^{\rm SS}\boldsymbol{J}_0^{\rm ES}} & \mathbb{G}^{\boldsymbol{J}_\cap^{\rm SS}\boldsymbol{M}_0^{\rm ES}}\end{bmatrix}}_{\mathbb{G}_{\rm FCE}}\cdot\mathbb{a} = 0 \tag{35}$$

satisfied by the physically realizable modes, in which the superscripts "DoJ" and "DoM" and subscript "FCE" are to emphasize the originations of matrices $\mathbb{G}^{\rm DoJ}$ (from the definition of $\boldsymbol{J}_0^{\rm ES}$), $\mathbb{G}^{\rm DoM}$ (from the definition of $\boldsymbol{M}_0^{\rm ES}$) and $\mathbb{G}_{\rm FCE}$ (from field continuation equation).

By properly assembling the matrix equations, we have two theoretically equivalent matrix equations as follows:

$$\mathbb{G}_{\rm FCE}^{\rm DoJ}\cdot\mathbb{a}=0 \;\text{ and }\; \mathbb{G}_{\rm FCE}^{\rm DoM}\cdot\mathbb{a}=0 \tag{36}$$

where

$$\mathbb{G}_{\rm FCE}^{\rm DoJ/DoM}=\begin{bmatrix}\mathbb{G}^{\rm DoJ/DoM}\\ \mathbb{G}_{\rm FCE}\end{bmatrix} \tag{37}$$

If the basic solutions of (36) are denoted as $\{\mathbb{s}_1,\mathbb{s}_2,\cdots\}$, then any physically realizable mode $\mathbb{a}$ can be uniquely expanded as that

$$\mathbb{a}=\sum_i b_i\mathbb{s}_i=\underbrace{\left[\mathbb{s}_1,\mathbb{s}_2,\cdots\right]}_{\mathbb{S}}\cdot\underbrace{\begin{bmatrix}b_1\\ b_2\\ \vdots\end{bmatrix}}_{\mathbb{b}} \tag{38}$$

Substituting the (38) into (26), we derive that

$$\begin{aligned}P^{\rm driv} &= \underbrace{\left(\mathbb{S}\cdot\mathbb{b}\right)^H}_{\mathbb{a}}\cdot\underbrace{\left(\mathbb{P}_{\rm PVT}^{\rm driv}+\mathbb{P}_{\rm SCT}^{\rm driv}\right)}_{\mathbb{P}^{\rm driv}}\cdot\underbrace{\left(\mathbb{S}\cdot\mathbb{b}\right)}_{\mathbb{a}}\\ &= \mathbb{b}^H\cdot\Big(\underbrace{\overbrace{\mathbb{S}^H\cdot\mathbb{P}_{\rm PVT}^{\rm driv}\cdot\mathbb{S}}^{\tilde{\mathbb{P}}_{\rm PVT}^{\rm driv}}+\overbrace{\mathbb{S}^H\cdot\mathbb{P}_{\rm SCT}^{\rm driv}\cdot\mathbb{S}}^{\tilde{\mathbb{P}}_{\rm SCT}^{\rm driv}}}_{\tilde{\mathbb{P}}^{\rm driv}}\Big)\cdot\mathbb{b}\end{aligned} \tag{39}$$

It was pointed out in [20] and [22] that the SCT is always equal to the power dissipated by the material OSS considered in [20] and [22]. Similarly, it is not difficult to prove that the SCT is always equal to the power dissipated by the composite OSS considered in this paper, so there exist the following equivalence relations

$$\text{OSS is lossless.}\Leftrightarrow\mathbb{b}^H\cdot\tilde{\mathbb{P}}_{\rm SCT}^{\rm driv}\cdot\mathbb{b}=0\text{ for any }\mathbb{b}.\Leftrightarrow\tilde{\mathbb{P}}_{\rm SCT}^{\rm driv}=0. \tag{40}$$

Thus, we have that

$$P^{\rm driv}\overset{\text{OSS is lossless}}{=\!=\!=\!=\!=\!=\!=}\mathbb{b}^H\cdot\tilde{\mathbb{P}}_{\rm PVT}^{\rm driv}\cdot\mathbb{b} \tag{41}$$

for the lossless composite OSSs.

In the following Sec. V, we will exhibit the fact that: unlike the modes (shown in Fig. 3) directly derived from equation $\mathbb{P}_-^{\rm driv}\cdot\mathbb{a}_\xi=\lambda_\xi\mathbb{P}_+^{\rm driv}\cdot\mathbb{a}_\xi$, the modes derived from characteristic equations $\tilde{\mathbb{P}}_-^{\rm driv}\cdot\mathbb{b}_\xi=\lambda_\xi\tilde{\mathbb{P}}_+^{\rm driv}\cdot\mathbb{b}_\xi$ (for lossy OSSs) and $\tilde{\mathbb{P}}_{\rm PVT;-}^{\rm driv}\cdot\mathbb{b}_\xi=\lambda_\xi\tilde{\mathbb{P}}_{\rm PVT;+}^{\rm driv}\cdot\mathbb{b}_\xi$ (for lossless OSSs) are satisfactory.

## V. Characteristic Modes, Parseval's Identity and Physical Picture

For a pre-selected composite OSS, there exists a set of inherently working modes — CMs. The CMs span whole modal space, and only depend on the inherent physical properties of the OSS. In this section, we, under WEP framework, construct the CMs by orthogonalizing frequency-domain DPO, and, at the same time, derive famous Parseval's identity, and then reveal the physical picture/purpose of CMT by employing the time-average version of modal orthogonality relations.

### A. Characteristic Modes (CMs)

For square matrices $\tilde{\mathbb{P}}^{\rm driv}$ and $\tilde{\mathbb{P}}_{\rm PVT}^{\rm driv}$, there must uniquely exist the following Toeplitz's decompositions [37, Sec. 0.2.5]

$$\tilde{\mathbb{P}}^{\rm driv} = \tilde{\mathbb{P}}_+^{\rm driv}+j\,\tilde{\mathbb{P}}_-^{\rm driv} \tag{42}$$

$$\tilde{\mathbb{P}}_{\rm PVT}^{\rm driv} = \tilde{\mathbb{P}}_{\rm PVT;+}^{\rm driv}+j\,\tilde{\mathbb{P}}_{\rm PVT;-}^{\rm driv} \tag{43}$$

In (42), $\tilde{\mathbb{P}}_+^{\rm driv}$ and $\tilde{\mathbb{P}}_-^{\rm driv}$ are the positive and negative Hermitian parts of $\tilde{\mathbb{P}}^{\rm driv}$, and $\tilde{\mathbb{P}}_+^{\rm driv}=[\tilde{\mathbb{P}}^{\rm driv}+(\tilde{\mathbb{P}}^{\rm driv})^H]/2$ and $\tilde{\mathbb{P}}_-^{\rm driv}=[\tilde{\mathbb{P}}^{\rm driv}-(\tilde{\mathbb{P}}^{\rm driv})^H]/(2j)$. In (43), $\tilde{\mathbb{P}}_{\rm PVT;+}^{\rm driv}$ and $\tilde{\mathbb{P}}_{\rm PVT;-}^{\rm driv}$ are the positive and negative Hermitian parts of $\tilde{\mathbb{P}}_{\rm PVT}^{\rm driv}$, and $\tilde{\mathbb{P}}_{\rm PVT;+}^{\rm driv}=[\tilde{\mathbb{P}}_{\rm PVT}^{\rm driv}+(\tilde{\mathbb{P}}_{\rm PVT}^{\rm driv})^H]/2$ and $\tilde{\mathbb{P}}_{\rm PVT;-}^{\rm driv}=[\tilde{\mathbb{P}}_{\rm PVT}^{\rm driv}-(\tilde{\mathbb{P}}_{\rm PVT}^{\rm driv})^H]/(2j)$.

In general, $\tilde{\mathbb{P}}_+^{\rm driv}$ is positive definite, because $\mathbb{b}^H\cdot\tilde{\mathbb{P}}_+^{\rm driv}\cdot\mathbb{b}$ is equal to the summation of the radiated and dissipated powers of any $\mathbb{b}$. Thus, there must exist a non-singular matrix having ability to simultaneously diagonalize $\tilde{\mathbb{P}}_+^{\rm driv}$ and $\tilde{\mathbb{P}}_-^{\rm driv}$ [37, Theorem 7.6.4]. Similarly, there also exist a non-singular matrix having ability to simultaneously diagonalize $\tilde{\mathbb{P}}_{\rm PVT;+}^{\rm driv}$ and $\tilde{\mathbb{P}}_{\rm PVT;-}^{\rm driv}$, if the OSS is lossless. The column vectors of the above-mentioned non-singular matrices can be obtained by solving the following characteristic equations

$$\tilde{\mathbb{P}}_-^{\rm driv}\cdot\mathbb{b}_\xi=\lambda_\xi\,\tilde{\mathbb{P}}_+^{\rm driv}\cdot\mathbb{b}_\xi\quad,\quad\text{for lossy OSSs} \tag{44}$$

$$\tilde{\mathrm{P}}^{\mathrm{driv}}_{\mathrm{PVT};-}\cdot \mathrm{b}_{\xi} = \lambda_{\xi}\, \tilde{\mathrm{P}}^{\mathrm{driv}}_{\mathrm{PVT};+}\cdot \mathrm{b}_{\xi} \quad , \quad \text{for lossless OSSs} \tag{45}$$

If the obtained CMs $(\mathrm{b}_1, \mathrm{b}_2, \cdots, \mathrm{b}_d)$ are $d$-order degenerate, then the following Gram-Schmidt orthogonalization process [37, Sec. 0.6.4] is necessary.

$$\left.\begin{aligned} \mathrm{b}_1 &= \mathrm{b}_1' \\ \mathrm{b}_2 - \chi_{12}\mathrm{b}_1' &= \mathrm{b}_2' \\ &\cdots \\ \mathrm{b}_d' - \cdots - \chi_{2d}\mathrm{b}_2' - \chi_{1d}\mathrm{b}_1' &= \mathrm{b}_d' \end{aligned}\right\} \tag{46}$$

where the coefficients are calculated as follows:

$$\chi_{mn} = \frac{(\mathrm{b}_m')^H \cdot \mathrm{P}_+^{\mathrm{driv}} \cdot \mathrm{b}_n}{(\mathrm{b}_m')^H \cdot \mathrm{P}_+^{\mathrm{driv}} \cdot \mathrm{b}_m'} \quad , \quad \text{for lossy OSSs} \tag{47}$$

$$\chi_{mn} = \frac{(\mathrm{b}_m')^H \cdot \mathrm{P}_{\mathrm{PVT};+}^{\mathrm{driv}} \cdot \mathrm{b}_n}{(\mathrm{b}_m')^H \cdot \mathrm{P}_{\mathrm{PVT};+}^{\mathrm{driv}} \cdot \mathrm{b}_m'} \quad , \quad \text{for lossless OSSs} \tag{48}$$

The current expansion vectors corresponding to the CMs can be calculated as that $\mathrm{a}_{\xi} = \mathrm{S}\cdot\mathrm{b}_{\xi}$.

Using (45) to calculate the CMs of the lossless OSS considered in Sec. IV, the associated MSs are shown in Fig. 5. Clearly, the results are consistent with the ones shown in Fig. 4. At the same time, we also show the MSs of the modes derived from orthogonalizing the traditional EFIE-PMCHW-based IMO with SDC scheme in Fig. 6. Comparing the Figs. 4, 5 and 6, it is not difficult to find out that the DPO is more satisfactory than the traditional IMO in the aspect of generating CMs.

In addition, as pointed out in [22], to obtain the matrix $\mathrm{S}$ usually needs to consume a relatively large number of computational resources, and this problem can be effectively resolved by employing the following alternative characteristic equations

$$\left(\mathrm{P}_-^{\mathrm{driv}} + \ell\cdot \mathrm{G}^H\cdot\mathrm{G}\right)\cdot \mathrm{a}_{\xi} = \lambda_{\xi}\, \mathrm{P}_+^{\mathrm{driv}}\cdot \mathrm{a}_{\xi} \quad , \text{for lossy OSSs} \tag{49}$$

$$\left(\mathrm{P}_{\mathrm{PVT};-}^{\mathrm{driv}} + \ell\cdot \mathrm{G}^H\cdot\mathrm{G}\right)\cdot \mathrm{a}_{\xi} = \lambda_{\xi}\, \mathrm{P}_{\mathrm{PVT};+}^{\mathrm{driv}}\cdot \mathrm{a}_{\xi} \, , \text{for lossless OSSs} \tag{50}$$

where $\mathrm{G} = \mathrm{G}_{\mathrm{FCE}}^{\mathrm{DoJ}} / \mathrm{G}_{\mathrm{FCE}}^{\mathrm{DoM}}$, and $\ell$ is an adjustable large real coefficient for example $\ell = 10^{10}$.

A rigorous theoretical explanation for the effectiveness of equations (49) and (50) had been given in [22]. Here, we emphasize that: only the modes with small or medium $|\lambda_{\xi}|$ are physical, but the ones with very large $|\lambda_{\xi}|$ are spurious. Based on this alternative scheme, we calculate the CMs of the previous lossless OSS, and show the associated MSs in Fig. 7. Evidently, the results are consistent with the ones shown in Fig. 4.

### *B. Parseval's Identity*

Similar to [18] and [22], it is easy to prove that the above-obtained CMs satisfy the following frequency-domain orthogonality relation

$$P_{\xi}^{\mathrm{driv}}\delta_{\xi\zeta} = (1/2)\left\langle \boldsymbol{J}_{\xi}^{\mathrm{SL}} + \boldsymbol{J}_{\xi}^{\mathrm{SS}} + \boldsymbol{J}_{\xi}^{\mathrm{SV}}, \boldsymbol{E}_{\zeta}^{\mathrm{inc}}\right\rangle_{\Sigma} + (1/2)\left\langle \boldsymbol{M}_{\xi}^{\mathrm{SV}}, \boldsymbol{H}_{\zeta}^{\mathrm{inc}}\right\rangle_{\Sigma} \tag{51}$$

where $\delta_{\xi\zeta}$ is the Kronecker's delta symbol.

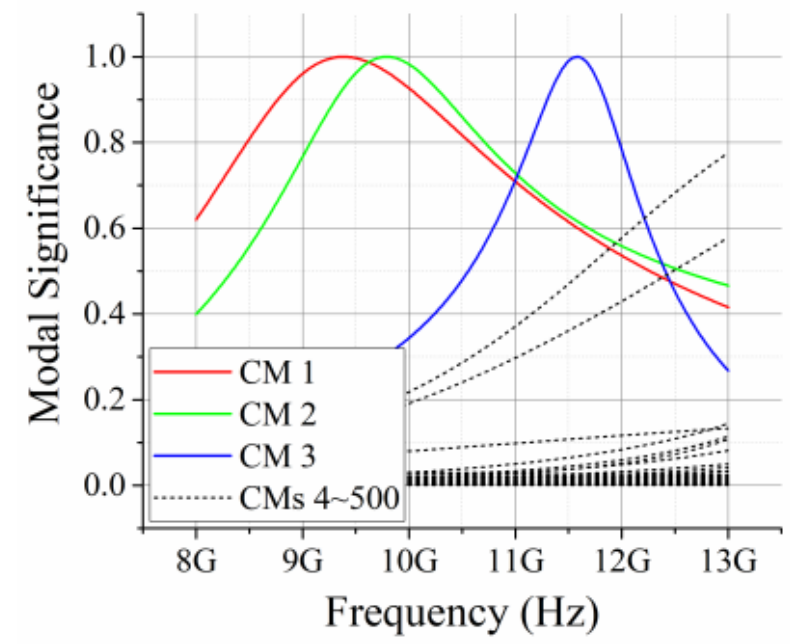

Fig. 5. MSs corresponding to the CMs of a lossless composite OSS. The OSS is with parameters $\boldsymbol{\mu} = \mathbf{I}4\mu_0$ and $\boldsymbol{\varepsilon}_{\mathrm{c}} = \mathbf{I}4\varepsilon_0$, and has the topological structure shown in Fig. 2. The modes shown in this figure are derived from solving characteristic equation (45).

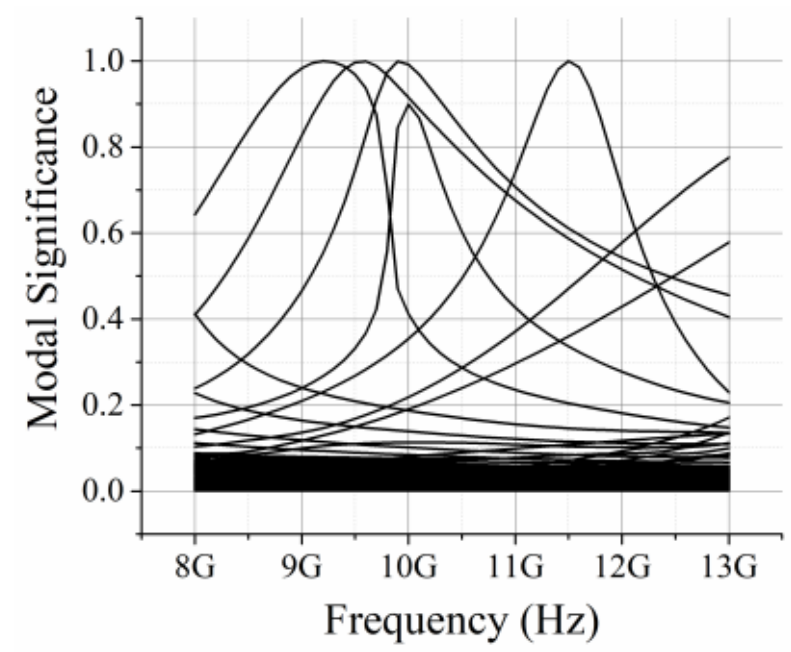

Fig. 6. MSs corresponding to the CMs of a lossless composite OSS. The OSS is with parameters $\boldsymbol{\mu} = \mathbf{I}4\mu_0$ and $\boldsymbol{\varepsilon}_{\mathrm{c}} = \mathbf{I}4\varepsilon_0$, and has the topological structure shown in Fig. 2. The modes shown in this figure are derived from orthogonalizing the traditional EFIE-PMCHWT-based IMO with SDC scheme.

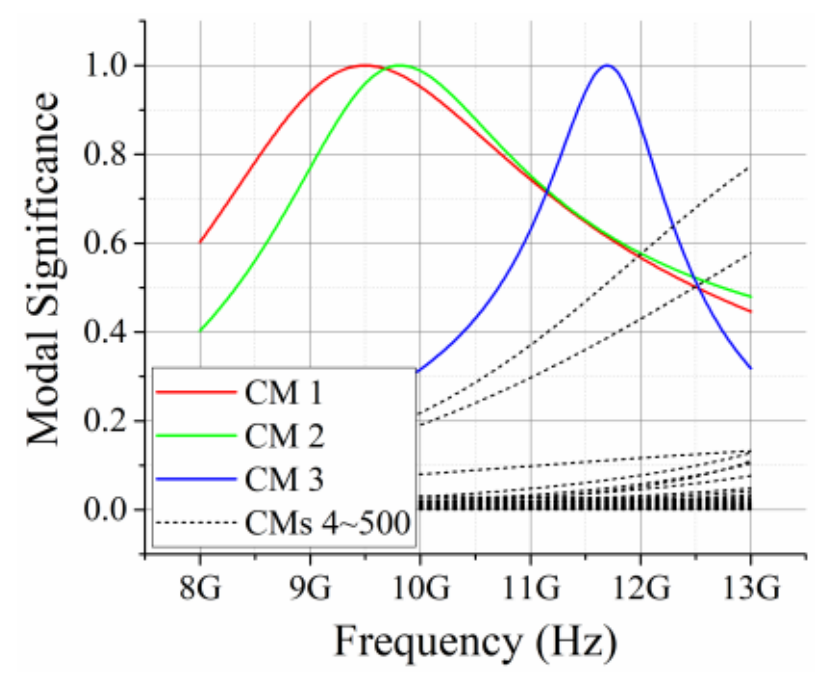

Fig. 7. MSs corresponding to the CMs of a lossless composite OSS. The OSS is with parameters $\boldsymbol{\mu} = \mathbf{I}4\mu_0$ and $\boldsymbol{\varepsilon}_{\mathrm{c}} = \mathbf{I}4\varepsilon_0$, and has the topological structure shown in Fig. 2. The modes shown in this figure are derived from solving characteristic equation (50).

Because of the completeness of CMs, any working mode can be expressed in terms of the linear expansion of the CMs, and orthogonality relation (51) implies the following expansion coefficients

$$c_{\xi} = \frac{(1/2)\left\langle \boldsymbol{J}_{\xi}^{\mathrm{SL}} + \boldsymbol{J}_{\xi}^{\mathrm{SS}} + \boldsymbol{J}_{\xi}^{\mathrm{SV}}, \boldsymbol{E}^{\mathrm{inc}}\right\rangle_{\Sigma} + (1/2)\left\langle \boldsymbol{M}_{\xi}^{\mathrm{SV}}, \boldsymbol{H}^{\mathrm{inc}}\right\rangle_{\Sigma}}{1 + j\lambda_{\xi}} \tag{52}$$

Here the modal real power $\mathrm{Re}\{P_{\xi}^{\mathrm{driv}}\}$ has been normalized to 1.

By employing the above (51) and (52) and the previous (21) and (22), we can derive the following Parseval's identity

$$\frac{1}{T}\int_{t_0}^{t_0+T} P^{\mathrm{Driv}}(t)\,dt = \sum_{\xi}\left|c_{\xi}\right|^2 \tag{53}$$

where $T$ is the time period of the time-harmonic EM field.

### C. Physical Picture

Evidently, frequency-domain power-decoupling relation (51) implies the following time-domain energy-decoupling relation (or alternatively called time-average power-decoupling relation)

$$\underbrace{\text{Re}\left\{P_{\xi}^{\text{driv}}\right\}}_{1}\delta_{\xi\zeta}=\frac{1}{T}\int_{t_0}^{t_0+T}\Big[\left\langle \boldsymbol{J}_{\xi}^{\text{SL}}(t)+\boldsymbol{J}_{\xi}^{\text{SS}}(t)+\boldsymbol{J}_{\xi}^{\text{SV}}(t),\boldsymbol{E}_{\zeta}^{\text{inc}}(t)\right\rangle_{\Sigma} +\left\langle \boldsymbol{M}_{\xi}^{\text{SV}}(t),\boldsymbol{H}_{\zeta}^{\text{inc}}(t)\right\rangle_{\Sigma}\Big]dt \qquad (54)$$

where $\text{Re}\{P_{\xi}^{\text{driv}}\}=1$ as explained in [22]. Relation (54) has a very clear physical interpretation: in any integral period, if $\xi\neq\zeta$, the $\zeta$-th modal fields $(\boldsymbol{E}_{\zeta}^{\text{inc}},\boldsymbol{H}_{\zeta}^{\text{inc}})$ don't supply net energy to the $\xi$-th modal currents $(\boldsymbol{J}_{\xi}^{\text{SL}},\boldsymbol{J}_{\xi}^{\text{SS}},\boldsymbol{J}_{\xi}^{\text{SV}},\boldsymbol{M}_{\xi}^{\text{SV}})$. This physical interpretation gives CMT a very clear physical picture/purpose — to construct a set of steadily working energy-decoupled modes for scattering systems.

In addition, we can also prove the following frequency-domain equivalent of (54).

$$\underbrace{\text{Re}\left\{P_{\xi}^{\text{driv}}\right\}}_{1}\delta_{\xi\zeta}=\frac{1}{2\eta_0}\left\langle \boldsymbol{E}_{\xi}^{\text{sca}},\boldsymbol{E}_{\zeta}^{\text{sca}}\right\rangle_{S_\infty}+(1/2)\left\langle \boldsymbol{\sigma}\cdot\boldsymbol{E}_{\xi}^{\text{tot}},\boldsymbol{E}_{\zeta}^{\text{tot}}\right\rangle_{\Omega} \qquad (55)$$

in which $\eta_0$ is the free-space wave impedance and $\eta_0=\sqrt{\mu_0/\varepsilon_0}$. Relation (55) clearly reveals the following important facts F1 and F2:

F1 when the OSS is lossless (that is, $\boldsymbol{\sigma}=0$), the far-field orthogonality relation of CMs holds automatically, because the second term in the right-hand side is always zero;

F2 when the OSS is lossy (that is, $\boldsymbol{\sigma}\neq 0$), the far-field orthogonality relation of CMs cannot be guaranteed usually, because the second term in the right-hand side cannot be guaranteed to be zero.

Here, we emphasize again that the physical purpose/picture of CMT is to construct the modes without net energy exchange in integral period (that is, the modes satisfying (54)) rather than the modes with orthogonal modal far fields, though, in the lossless case, the formers indeed have orthogonal far fields automatically. The detailed explanations for this conclusion had been given in [22].

Using equations (44) and (49) to calculate the CMs of the lossy OSS whose topological structure is the same as the one shown in Fig. 2 and material parameters are $\boldsymbol{\mu}=\mathbf{I}4\mu_0$ and $\boldsymbol{\varepsilon}_{\text{c}}=\mathbf{I}4\varepsilon_0-j\mathbf{I}0.5/\omega$, the associated MSs are illustrated in the following Fig. 8. At the same time, the EFIE-VIE-based and EFIE-PMCHWT-based results are shown in Fig. 9 and Fig. 10 for comparisons. Obviously, the results shown in Fig. 8 and Fig. 9 are consistent with each other, but the they are not consistent with the ones shown in Fig. 10.

For the (44)-based CMs at 11 GHz, their time-average DP orthogonality matrix is shown in Fig. 11(a), and their far-field orthogonality matrix is shown in Fig. 11(b). Obviously, the modal orthogonality relations (54) and (55) indeed hold, but the modal far-field orthogonality relation doesn't.

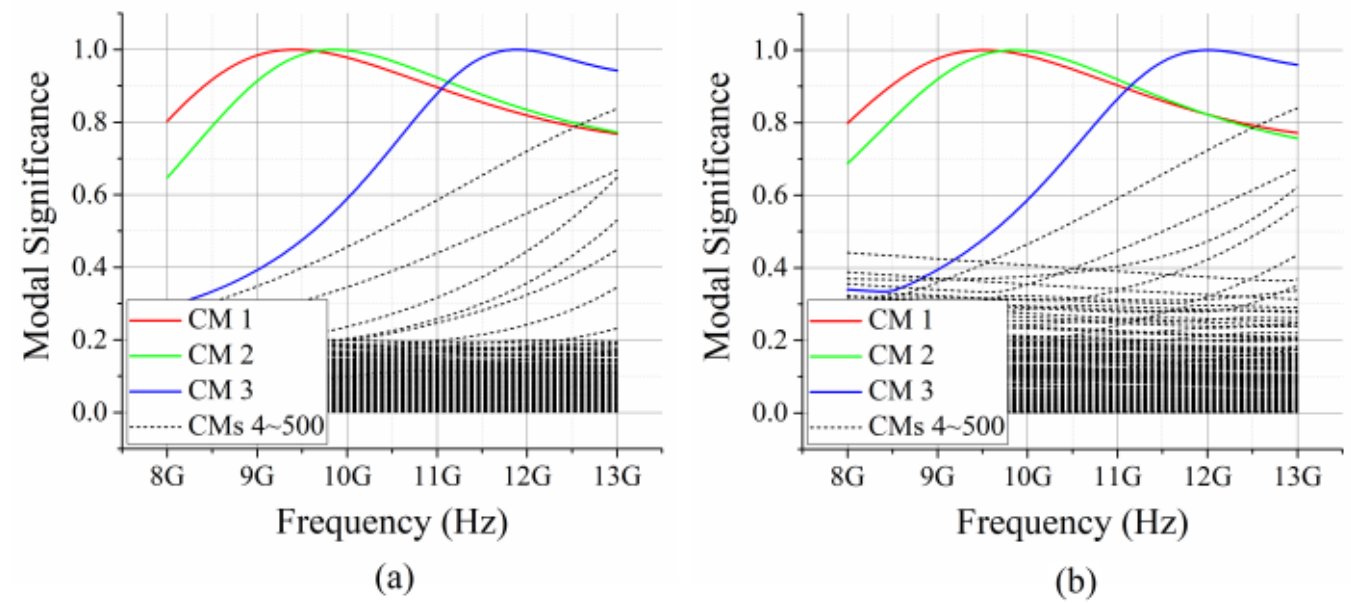

Fig. 8. MSs corresponding to the CMs of a lossy composite OSS. The OSS is with parameters $\boldsymbol{\mu}=\mathbf{I}4\mu_0$ and $\boldsymbol{\varepsilon}_{\text{c}}=\mathbf{I}4\varepsilon_0-j\mathbf{I}0.5/\omega$, and has the topological structure shown in Fig. 2. The CMs shown above are derived from solving (a) characteristic equation (44) and (b) characteristic equation (49) with $\ell=10^{10}$.

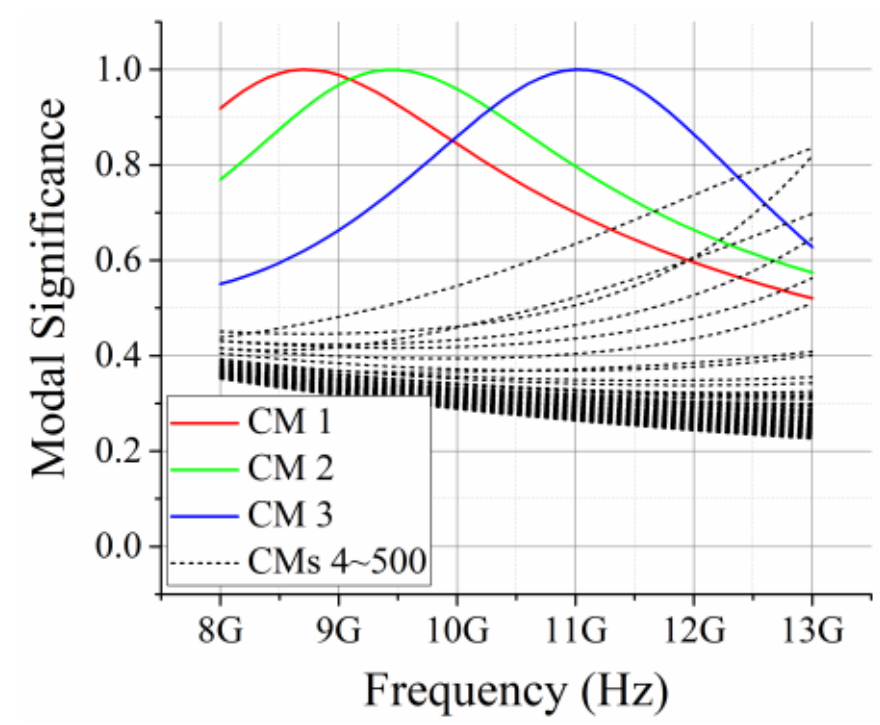

Fig. 9. MSs corresponding to the CMs of a lossy composite OSS. The OSS is with parameters $\boldsymbol{\mu}=\mathbf{I}4\mu_0$ and $\boldsymbol{\varepsilon}_{\text{c}}=\mathbf{I}4\varepsilon_0-j\mathbf{I}0.5/\omega$, and has the topological structure shown in Fig. 2. The CMs shown in this figure are derived from orthogonalizing DPO (22).

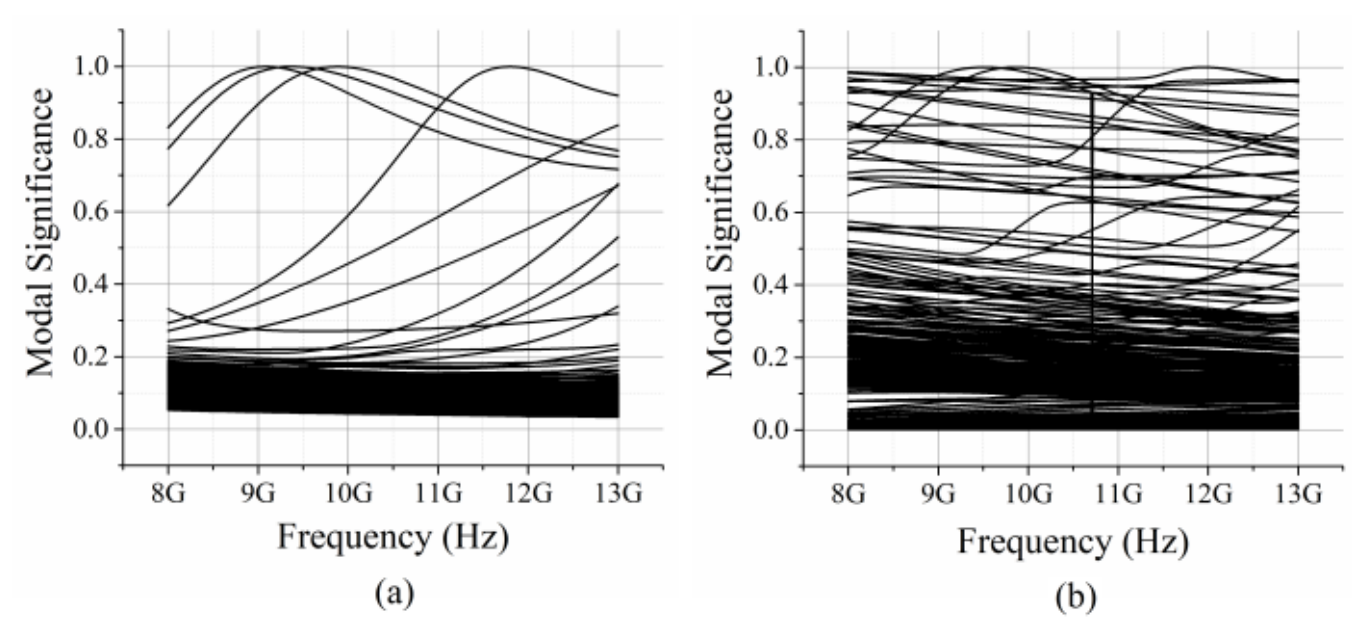

Fig. 10. MSs corresponding to the CMs of a lossy composite OSS. The OSS is with parameters $\boldsymbol{\mu}=\mathbf{I}4\mu_0$ and $\boldsymbol{\varepsilon}_{\text{c}}=\mathbf{I}4\varepsilon_0-j\mathbf{I}0.5/\omega$, and has the topological structure shown in Fig. 2. The CMs shown above are derived from solving EFIE-PMCHWT-based operator with (a) the original SDC scheme and (b) the weighted SDC scheme using $\ell=10^{10}$.

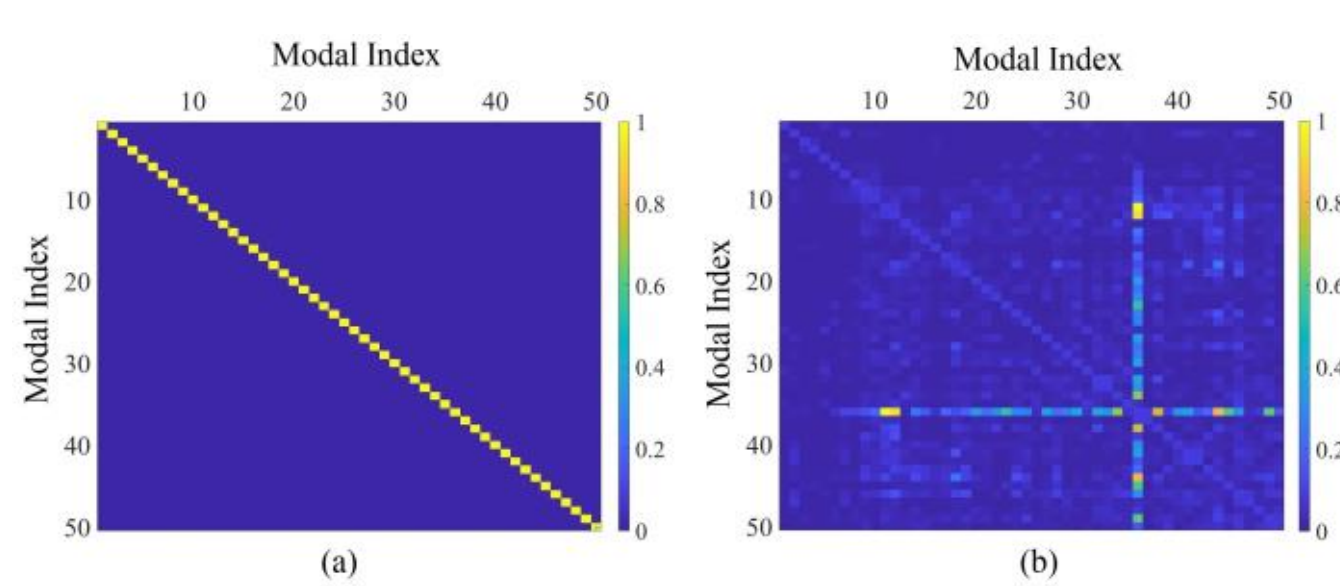

Fig. 11. (a) Time-average DP orthogonality matrix and (b) far-field orthogonality matrix about the CMs of a lossy composite OSS working at 11 GHz. The OSS is with parameters $\boldsymbol{\mu}=\mathbf{I}4\mu_0$ and $\boldsymbol{\varepsilon}_{\text{c}}=\mathbf{I}4\varepsilon_0-j\mathbf{I}0.5/\omega$, and has the topological structure shown in Fig. 2. The CMs corresponding to this figure are derived from solving equation (44).

## VI. Numerical Examples

In the above Sec. V, the CMs of a typical composite OSS — a metallic cube partially coated by a material cube — has been successfully constructed by orthogonalizing the frequency-domain DPO with SDC scheme. In this section, the CMs of some other typical composite OSSs are calculated in WEP framework to further verify the validity of the theory developed in this paper, and some comparisons are also done to exhibit the advantages of the WEP-CMT over the conventional IE-CMT.

### *A. A Metallic Core Completely Coated by a Material Shell*

In this subsection, we consider the composite OSS constituted by a metallic cube and a material cubic shell, and the metallic cube is completely coated by the material cubic shell. The topological structure and the sub-boundaries of the composite OSS are shown in Fig. 12.

When the composite OSS is lossless and with parameters $\boldsymbol{\mu} = \mathbf{I}4\mu_0$ and $\boldsymbol{\varepsilon}_c = \mathbf{I}4\varepsilon_0$, we use five different ways to construct its CMs, and show the obtained results in Figs. 13(a)-13(e). In Fig. 13(a), the CMs are derived from orthogonalizing EFIE-VIE-based IMO, that is DPO (22). In Figs. 13(b) and 13(c), the CMs are respectively derived from orthogonalizing the traditional EFIE-PMCHWT-based IMO with SDC scheme and weighted SDC scheme. In Fig. 13(d), the CMs are derived from orthogonalizing the DPO with SDC scheme, that is, solving characteristic equation (45). In Fig. 13(e), the CMs are derived from orthogonalizing the DPO with weighted SDC scheme, that is, solving characteristic equation (50).

When the composite OSS is lossy and with parameters $\boldsymbol{\mu} = \mathbf{I}4\mu_0$ and $\boldsymbol{\varepsilon}_c = \mathbf{I}4\varepsilon_0 - j\mathbf{I}0.5/\omega$, we also use five different ways to construct its CMs, and show the obtained results in Figs. 14(a)-14(e). In Fig. 14(a), the CMs are derived from orthogonalizing EFIE-VIE-based IMO, that is DPO (22). In Figs. 14(b) and 14(c), the CMs are respectively derived from orthogonalizing the traditional EFIE-PMCHWT-based IMO with SDC scheme and weighted SDC scheme. In Fig. 14(d), the CMs are derived from orthogonalizing the DPO with SDC scheme, that is, solving characteristic equation (44). In Fig. 14(e), the CMs are derived from orthogonalizing the DPO with weighted SDC scheme, that is, solving characteristic equation (49).

From comparing the results shown in Fig. 13 and Fig. 14, it is not difficult to conclude that: (i) the novel orthogonalizing DPO method is valid for constructing the CMs of both lossless and lossy composite OSSs; (ii) the novel DPO has a more satisfactory numerical performance than the traditional IMO in the aspect of generating CMs.

### *B. A Rectangular Metallic Patch Adhered to the Surface of a Cubic Material Body*

In this subsection, we consider the composite OSS constituted by a metallic rectangular patch and a material cube, and the metallic patch is adhered to the material cube. The topological structure and the sub-boundaries of the composite OSS are shown in Fig. 15.

When the composite OSS is lossless and with parameters $\boldsymbol{\mu} = \mathbf{I}4\mu_0$ and $\boldsymbol{\varepsilon}_c = \mathbf{I}4\varepsilon_0$, we use five different ways to construct its CMs, and show the obtained results in Figs. 16(a)-16(e).

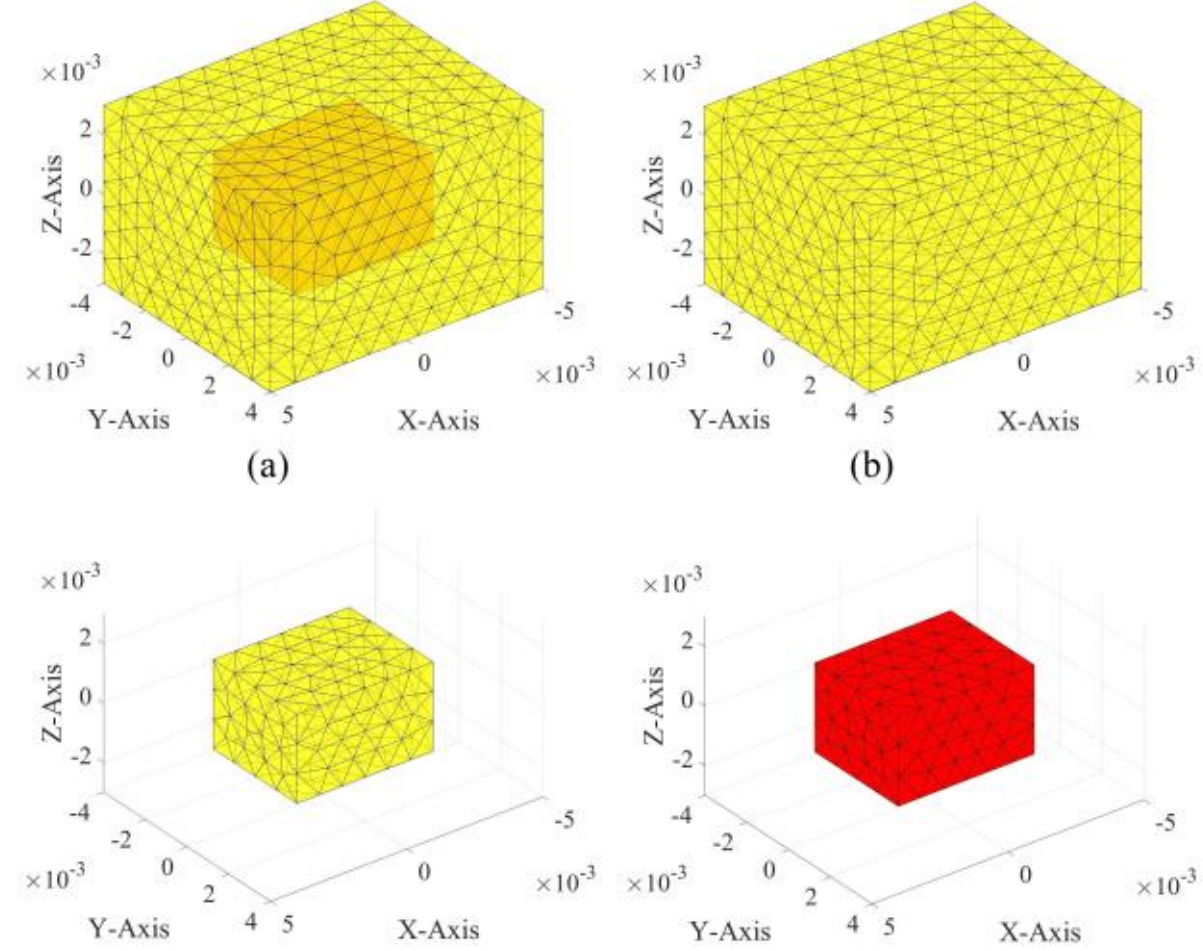

Fig. 12. Geometry of a composite OSS constituted by a metallic cube and a material shell, where the sizes of the metallic cube and the inner surface of the material shell are both $5\text{mm}\times4\text{mm}\times3\text{mm}$, and the size of the outer surface of the material shell is $10\text{mm}\times8\text{mm}\times6\text{mm}$. (a) Topological structure of whole OSS; (b) material-environment boundary $\partial\Omega_0$; (c) material-metal boundary $\partial\Omega_{\cap}$; (d) metal-material boundary $\partial V_{\cap}$ (which is just whole $\partial V$). Obviously, $\partial\Omega_{\cap} = \partial V_{\cap}$ for this composite OSS.

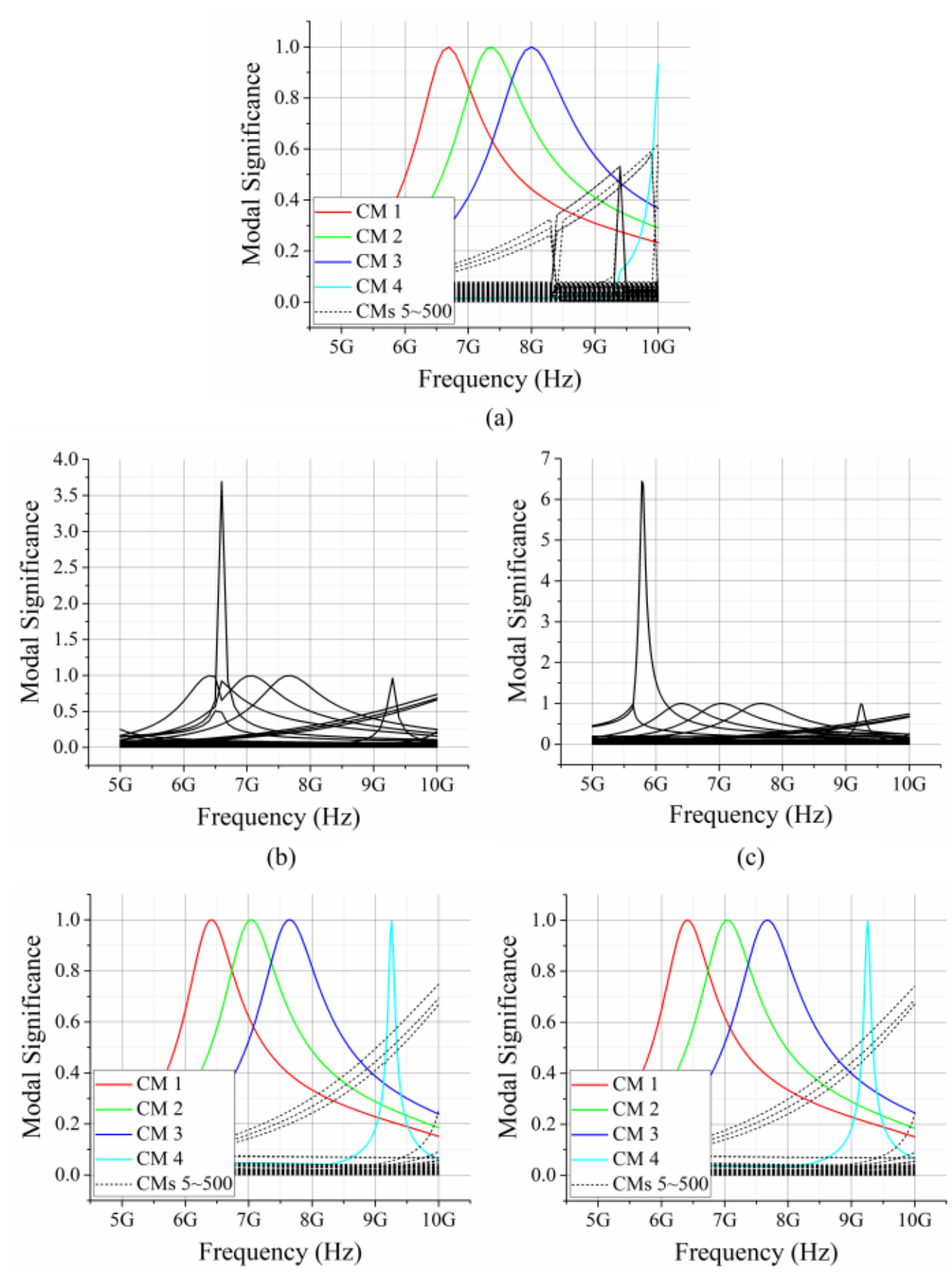

Fig. 13. MSs corresponding to the CMs of a lossless composite OSS. The OSS is with parameters $\boldsymbol{\mu} = \mathbf{I}4\mu_0$ and $\boldsymbol{\varepsilon}_c = \mathbf{I}4\varepsilon_0$, and has the topological structure shown in Fig. 12. The CMs are derived from (a) EFIE-VIE-based IMO, that is DPO (22), (b) the EFIE-PMCHWT-based IMO with SDC scheme, (c) the EFIE-PMCHWT-based IMO with weighted SDC scheme using $\ell = 10^{10}$, (d) characteristic equation (45), and (e) characteristic equation (50) with $\ell = 10^{10}$.

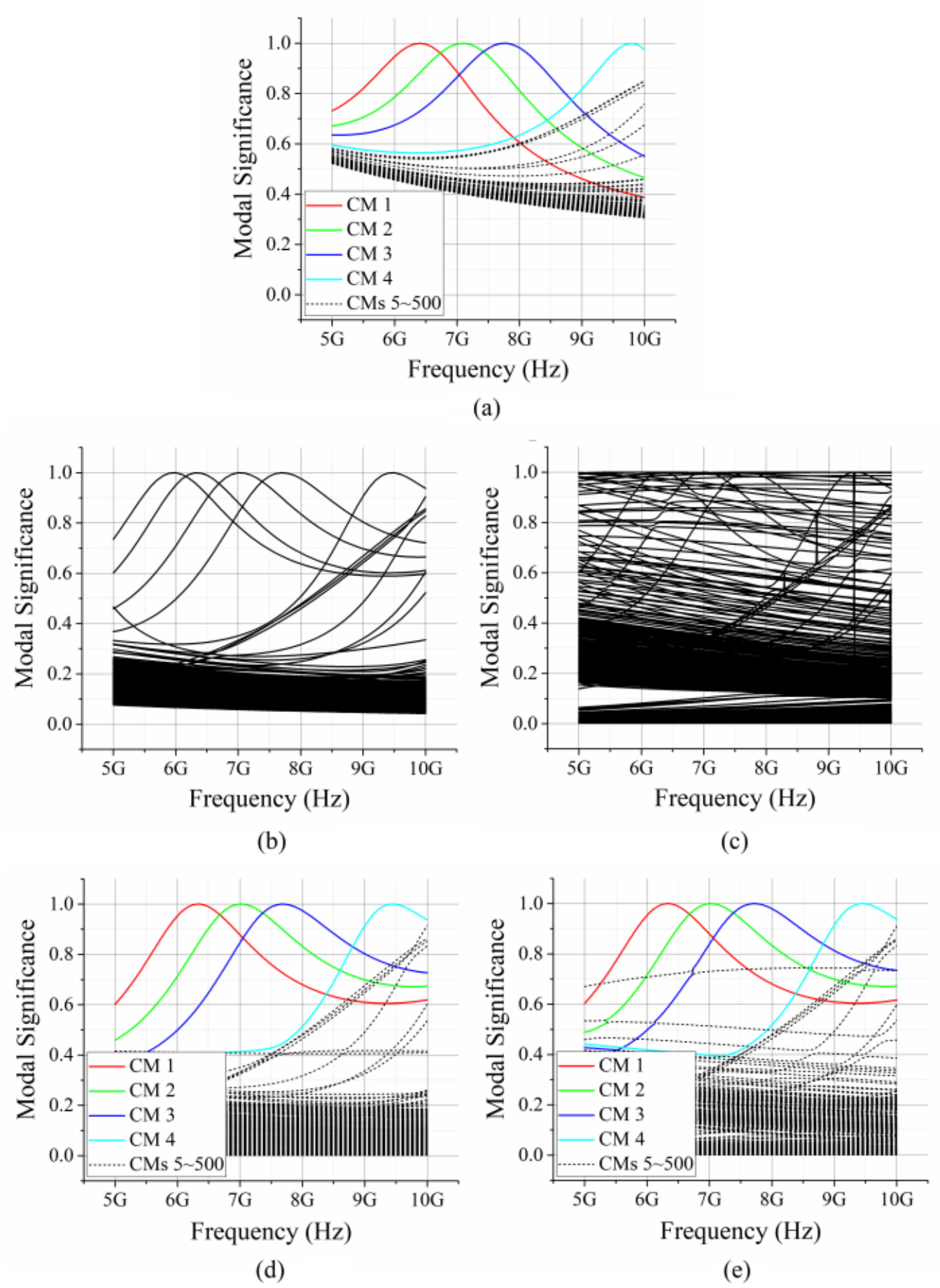

Fig. 14. MSs corresponding to the CMs of a lossy composite OSS. The OSS is with parameters $\boldsymbol{\mu} = \mathbf{I}4\mu_0$ and $\boldsymbol{\varepsilon}_c = \mathbf{I}4\varepsilon_0 - j\mathbf{I}0.5/\omega$, and has the topological structure shown in Fig. 12. The CMs are derived from (a) EFIE-VIE-based IMO, that is DPO (22), (b) the EFIE-PMCHWT-based IMO with SDC scheme, (c) the EFIE-PMCHWT-based IMO with weighted SDC scheme using $\ell = 10^{10}$, (d) characteristic equation (44), and (e) characteristic equation (49) with $\ell = 10^{10}$.

In Fig. 16(a), the CMs are derived from orthogonalizing EFIE-VIE-based IMO, that is DPO (22). In Figs. 16(b) and 16(c), the CMs are respectively derived from orthogonalizing the traditional EFIE-PMCHWT-based IMO with SDC scheme and weighted SDC scheme. In Fig. 16(d), the CMs are derived from orthogonalizing the DPO with SDC scheme, that is, solving characteristic equation (45). In Fig. 16(e), the CMs are derived from orthogonalizing the DPO with weighted SDC scheme, that is, solving characteristic equation (50).

When the composite OSS is lossy and with parameters $\boldsymbol{\mu} = \mathbf{I}4\mu_0$ and $\boldsymbol{\varepsilon}_c = \mathbf{I}4\varepsilon_0 - j\mathbf{I}0.5/\omega$, we also use five different ways to construct its CMs, and show the obtained results in Figs. 17(a)-17(e).

In Fig. 17(a), the CMs are derived from orthogonalizing EFIE-VIE-based IMO, that is DPO (22). In Figs. 17(b) and 17(c), the CMs are respectively derived from orthogonalizing the traditional EFIE-PMCHWT-based IMO with SDC scheme and weighted SDC scheme. In Fig. 17(d), the CMs are derived from orthogonalizing the DPO with SDC scheme, that is, solving characteristic equation (44). In Fig. 17(e), the CMs are derived from orthogonalizing the DPO with weighted SDC scheme, that is, solving characteristic equation (49).

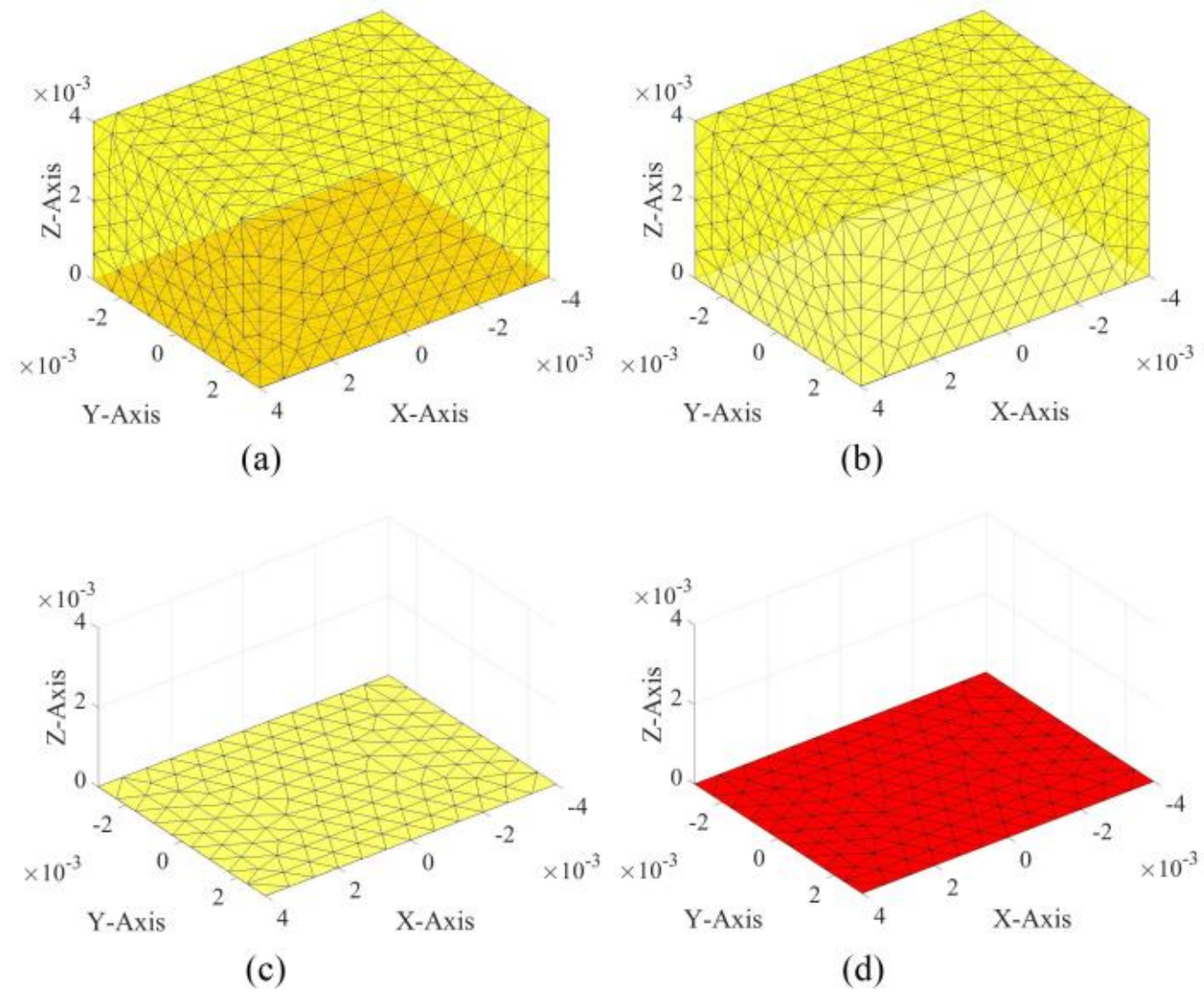

Fig. 15. Geometry of a composite OSS constituted by a rectangular metallic patch and a cubic material body, where the geometrical dimension of the metallic patch is $8\text{mm}\times 6\text{mm}$, and the geometrical dimension of the material body is $8\text{mm}\times 6\text{mm}\times 4\text{mm}$. (a) Topological structure of whole OSS; (b) material-environment boundary; (c) material-metal boundary; (d) metallic surface.

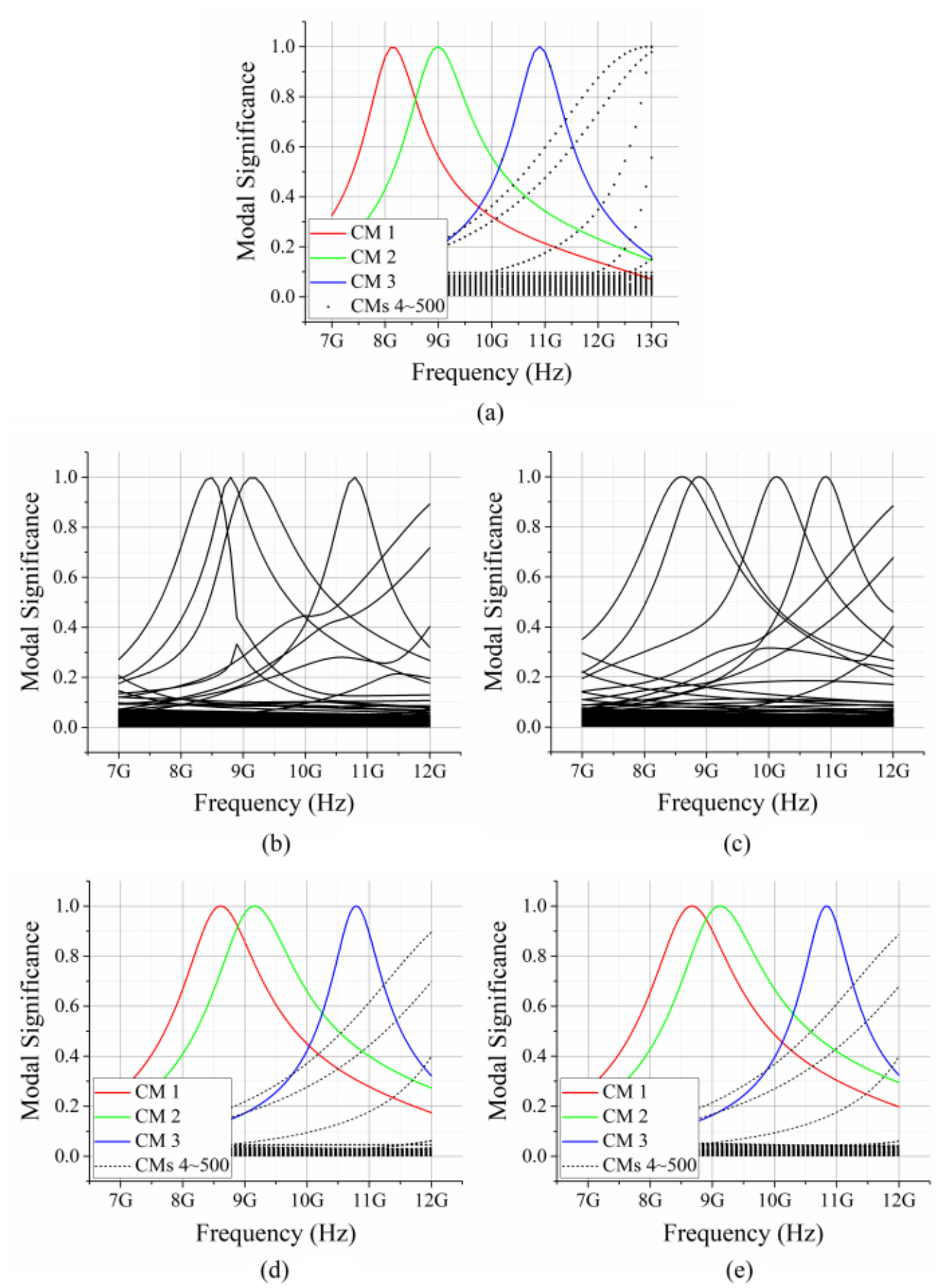

Fig. 16. MSs corresponding to the CMs of a lossless composite OSS. The OSS is with parameters $\boldsymbol{\mu} = \mathbf{I}4\mu_0$ and $\boldsymbol{\varepsilon}_c = \mathbf{I}4\varepsilon_0$, and has the topological structure shown in Fig. 15. The CMs are derived from (a) EFIE-VIE-based IMO, that is DPO (22), (b) the EFIE-PMCHWT-based IMO with SDC scheme, (c) the EFIE-PMCHWT-based IMO with weighted SDC scheme using $\ell = 10^{10}$, (d) characteristic equation (45), and (e) characteristic equation (50) with $\ell = 10^{10}$.

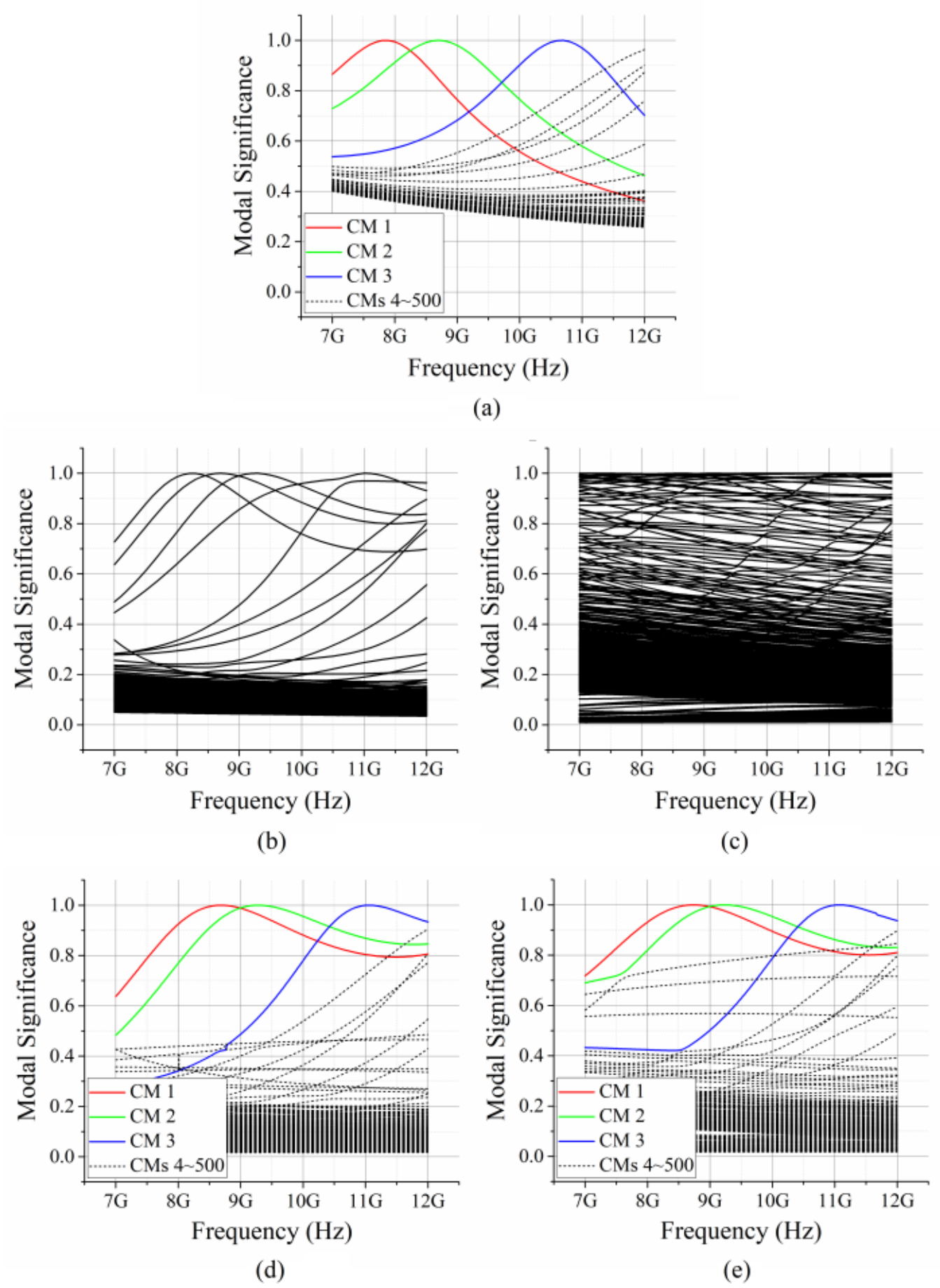

Fig. 17. MSs corresponding to the CMs of a lossy composite OSS. The OSS is with parameters $\boldsymbol{\mu} = \mathbf{I}4\mu_0$ and $\boldsymbol{\varepsilon}_c = \mathbf{I}4\varepsilon_0 - j\mathbf{I}0.5/\omega$, and has the topological structure shown in Fig. 15. The CMs are derived from (a) EFIE-VIE-based IMO, that is DPO (22), (b) the EFIE-PMCHWT-based IMO with SDC scheme, (c) the EFIE-PMCHWT-based IMO with weighted SDC scheme using $\ell = 10^{10}$, (d) characteristic equation (44), and (e) characteristic equation (49) with $\ell = 10^{10}$.

## VII. Conclusions

Besides the traditional integral equation (IE) framework, work-energy principle (WEP) is also an effective framework for establishing characteristic mode theory (CMT). The novel WEP framework, in a clearer way, reveals the physical picture/purpose of CMT — to construct a set of steadily working energy-decoupled modes for scattering systems (rather than to construct the far-field-orthogonal modes). Under the novel WEP framework, a novel characteristic mode (CM) generating operator — driving power operator (DPO) — is introduced, and then a novel CM construction method — orthogonalizing DPO method — is proposed.

All the variables (except the equivalent electric or magnetic current on material-environment boundary) explicitly involved in the DPO are independent and complete. Using this observation and employing line-surface equivalence principle (LSEP), a novel scheme — solution domain compression (SDC) — and its alternative version are developed to suppress spurious modes. Unlike the traditional dependent variable elimination (DVE) scheme, the novel SDC scheme and its alternative version don't need to inverse any matrix during the process to suppress spurious modes.

Compared with the traditional IE-based orthogonalizing impedance matrix operator (IMO) method, the novel WEP-based orthogonalizing DPO method has a more desirable numerical performance in the aspect of constructing CMs. A detailed and rigorous theoretical explanation for this conclusion has been done by our group, and will be exhibited in our future article.

In addition, DPO always can be decomposed into two terms — principal value term (PVT) and singular current term (SCT). After employing the SDC scheme, the SCT is zero, if and only if the objective scattering system (OSS) is lossless. Based on this, this paper concludes that

(i) if the OSS is lossless, it is better to select the PVT (which is theoretically equal to the full DPO, but is numerically not) as CM generating operator;

(ii) if the OSS is lossy, it must select the full DPO (that is, PVT+SCT) as CM generating operator.

It is necessary to emphasize that the above conclusions (i) and (ii) hold under the condition that the SDC scheme has been utilized.

## Acknowledgment

The authors would like to thank the reviewers and editors for their patient reviews, valuable comments, and selfless suggestions for improving this paper.

**Ren-Zun Lian** received the B.S. degree in optical engineering from the University of Electronic Science and Technology of China (UESTC), Chengdu, China, in 2011, and received the Ph.D. degree in electromagnetic field and microwave technology from UESTC, Chengdu, China, in 2019.

He is currently a postdoctoral researcher in Peking University (PKU), Beijing, China. His current research interests include mathematical physics, electromagnetic theory and computation, and antenna theory and design.

**Xing-Yue Guo** (S'17) received the B.S. degree in electronic information science and technology from the Southwest Jiaotong University, Chengdu, China, in 2011, and the M.E. degree from the University of Electronic Science and Technology of China, Chengdu, China, in 2014. From 2014 to 2017, she was a Research Assistant with the CAEP Software Center for High Performance Numerical Simulation, Beijing, China. Since 2017, she has been pursuing the Ph. D. degree at the School of Electronics Engineering and Computer Science, Peking University, Beijing, China. Her research interests include computational electromagnetics and applications.

**Ming-Yao Xia** (M'00-SM'03) received the Master and Ph. D degrees in electrical engineering from the Institute of Electronics, Chinese Academy of Sciences (IECAS), in 1988 and 1999, respectively. From 1988 to 2002, he was with IECAS as an Engineer and a Senior Engineer. He was a Visiting Scholar at the University of Oxford, U.K., from October 1995 to October 1996. From June 1999 to August 2000 and from January 2002 to June 2002, he was a Senior Research Assistant and a Research Fellow, respectively, with the City University of Hong Kong. He joined Peking University as an Associate Professor in 2002 and was promoted to Full Professor in 2004. He moved to the University of Electronic Science and Technology of China as a Chang-Jiang Professor nominated by the Ministry of Education of China in 2010. He returned to Peking University after finishing the appointment in 2013. He was a recipient of the Young Scientist Award of the URSI in 1993. He was awarded the first-class prize on Natural Science by the Chinese Academy of Sciences in 2001. He was the recipient of the Foundation for Outstanding Young Investigators presented by the National Natural Science Foundation of China in 2008. He served as an Associate Editor for the IEEE Transactions on Antennas and Propagation. His research interests include electromagnetic theory, numerical methods and applications, such as wave propagation and scattering, electromagnetic imaging and probing, microwave remote sensing, antennas and microwave components.

# Work-Energy Principle Based Characteristic Mode Theory for Wireless Power Transfer Systems

Ren-Zun Lian, Ming-Yao Xia, *Senior Member, IEEE*, and Xing-Yue Guo, *Student Member, IEEE*

***Abstract*—Work-energy principle (WEP) governing wireless power transfer (WPT) process is derived. Driving power as the source to sustain a steady WPT is obtained. Transferring coefficient (TC) used to quantify power transfer efficiency is introduced.**

**WEP gives a clear physical picture to WPT process. The physical picture reveals the essential difference between transferring problem and scattering problem. The essential difference exposes the fact that the conventional characteristic mode theory (CMT) for scattering systems cannot be directly applied to transferring systems.**

**Under WEP framework, this paper establishes a CMT for transferring systems. By orthogonalizing driving power operator (DPO), the CMT can construct a set of energy-decoupled characteristic modes (CMs) for any pre-selected objective transferring system. It is proved that the obtained CM set contains the optimally transferring mode, which can maximize TC.**

**Employing the WEP-based CMT for transferring systems, this paper does the modal analysis for some typical two-coil transferring systems, and introduces the concepts of co-resonance and ci-resonance, and reveals some important differences and connections "between transferring problem and scattering problem", "between co-resonance phenomenon of transferring systems and external resonance phenomenon of scattering systems", and "between so-called magnetic resonance and classical electric-magnetic resonance".**

## I. Introduction

WIRELESS power transfer (WPT) system is a kind of electromagnetic (EM) device designed for wirelessly transferring EM power in an efficiency as high as possible and a distance as long as possible. The earliest researches on WPT can be dated back to the pioneers Hutin and Leblanc [1] and Tesla [2]-[4] etc., and more history on WPT could be found in [5]. According to the difference of working mechanism, WPT systems can be categorized into far-field/radiative WPT systems and near-field/nonradiative WPT systems [6]. The commonly used far-field WPT systems include microwave [7]-[8] and laser [9]-[10] WPT systems etc. The commonly used near-field WPT systems include inductive [11]-[13], capacitive [14]-[15], conductive [16]-[17], and magnetic resonance [18]-[23] WPT systems etc.

In the various WPT systems, the magnetic-resonance-based systems seem more desired in the applications of medium to high power levels, and more advantageous in the aspects of transferring capability, transferring efficiency, safety, and controllability [22]-[23]. The magnetic resonance WPT systems are focused on by this paper, and the principles and formulations obtained in this paper can be further generalized to the other kinds of WPT systems in the future. For convenience, the magnetic resonance WPT systems are simply called transferring systems in the following discussions.

A classical transferring system is shown in Fig. 1. The transferring system is constituted by two metallic coils transmitting coil T and receiving coil R. Coil T is excited by a locally impressed driver (e.g. delta-gap source) used to inject power into coil T. Coil R is connected to a perfectly matched load (e.g. matched light bulb) used to extract power from coil R. The power is transmitted by coil T, and then wirelessly transferred from coil T to coil R by passing through the environment surrounding transferring system, and finally received by coil R.

Inspired by the acoustic resonance between acoustic resonators, Tesla introduced the concept of magnetic resonance to the realm of WPT for the first time, and patented the well-known "Tesla coil" [24]. In 2007, Kurs *et al.* [18]-[19] employ the magnetic resonance between the coils T and R shown in Fig. 1 to realize fully lighting up a 60-W light bulb from distances more than 2 m away. The time-average magnetic energy density distribution of the desired transferring mode is visualized in Fig. 2. Evidently, the mode shown in Fig. 2 indeed has ability to efficiently transfer power from coil T to coil R wirelessly. In recent years, the magnetic-resonance-based WPT technology has been widely applied in the realms of wirelessly charging consumer electronic products [25], electric vehicles (EV) [26]-[27], biomedical implants [28]-[30], underwater devices [31]-[32], Internet of things (IoT) [33]-[34], and industrial robots [35]-[36] etc.

Obviously, a systematical modal analysis method will significantly facilitate the theoretical analysis and engineering design for transferring systems. In fact, there have existed some different kinds of modal analysis methods for transferring systems, such as coupled-mode theory [18]-[19], [37], classical circuit theory [38]-[40], and some other theories [41]-[42], and

[AP2101-0035] received January 6, 2021. This work was supported by XXXX under Project XXXXXXXX. (*Corresponding authors: Ren-Zun Lian; Ming-Yao Xia.*)

R. Z. Lian, M. Y. Xia and X. Y. Guo are with the Department of Electronics, School of Electronics Engineering and Computer Science, Peking University, Beijing 100871, China. (E-mails: rzlian@vip.163.com; myxia@pku.edu.cn).

all of them are based on circuit models. However, the circuit-model-based modal analysis methods need to use some circuit-based concepts, such as scalar voltage, scalar current, self-inductance, mutual inductance, and capacitance etc., so they are some approximate but not rigorous methods. In addition, the employment for scalar voltage and scalar current implies that the circuit-model-based modal analysis methods are only applicable to the transferring systems working at low frequency and with simple geometrical structures (such as rectangular coils and circular coils [18]-[23] etc.). Thus, it is one of important challenges in the realm of magnetic-resonance-based WPT how to develop a rigorous, frequency-independent, and geometry-independent modal analysis method for transferring systems.

The conventional characteristic mode theory (CMT) [43]-[48] is just a rigorous, frequency-independent, and geometry-independent modal analysis method. However, the conventional CMT is a method for *scattering* systems as revealed by its physical picture [47]-[48], but cannot be directly applied to **transferring** systems (due to the different working mechanisms of *scattering* and **transferring** systems, for details please see Sec. II) as exhibited in the following example. A direct application of the conventional CMT to the transferring system shown in Fig. 1 outputs some CMs, and the resonant CM has the time-average magnetic energy density distribution shown in Fig. 3. Evidently, the resonant CM shown in Fig. 3 doesn't realize the most efficient WPT from coil T to coil R like Fig. 2.

The one shown in Fig. 1 is a simplest and most classical metallic transferring system. Using it as a typical example, this paper is devoted to generalizing the conventional CMT for *scattering* systems (simply called *scattering* CMT) to a novel CMT for **transferring** systems (simply called **transferring** CMT), and this paper is organized as follows: Sec. II discusses the physical principle governing the working mechanism of transferring problem; Sec. III provides the mathematical formulas used to establish the transferring CMT; Sec. IV employs the CMT to do some modal analysis for the classical transferring system shown in Fig. 2 to exhibit the validity of the theory; Sec. V applies the transferring CMT to some typical variants of the classical transferring system for exhibiting the wide applicable range of the theory; Sec. VI concludes this paper; Apps. A-D provide some detailed formulations related to this paper.

In what follows, the $e^{j\omega t}$ convention and inner product $<\boldsymbol{f},\boldsymbol{g}>_\Omega=\int_\Omega \boldsymbol{f}^\dagger\cdot\boldsymbol{g}d\Omega$ are used throughout, where superscript "$\dagger$" is the conjugate transpose operation for a scalar/vector/matrix. The environment surrounding transferring system is free space, and its permeability and permittivity are denoted as $\mu_0$ and $\varepsilon_0$ respectively. Time-domain and frequency-domain powers are denoted as $\mathcal{P}$ and $P$ respectively; time-domain and frequency-domain currents and fields are denoted as $(\boldsymbol{\mathcal{J}},\boldsymbol{\mathcal{F}})$ and $(\boldsymbol{J},\boldsymbol{F})$ respectively; frequency-domain power quadratic matrix and current expansion coefficient vector are denoted as $\mathbb{P}$ and $\mathbb{J}$ respectively. In addition, for the linear quantities (e.g. electric field intensity), we have that $\boldsymbol{\mathcal{E}}=\mathrm{Re}\{\boldsymbol{E}e^{j\omega t}\}$; for the power-type quadratic quantity, we have that $\mathrm{Re}\{(1/2)\boldsymbol{J}^\dagger\cdot\boldsymbol{E}\}=(1/T)\int_0^T\boldsymbol{\mathcal{J}}\cdot\boldsymbol{\mathcal{E}}dt$, where $T$ is the time period of the time-harmonic EM field.

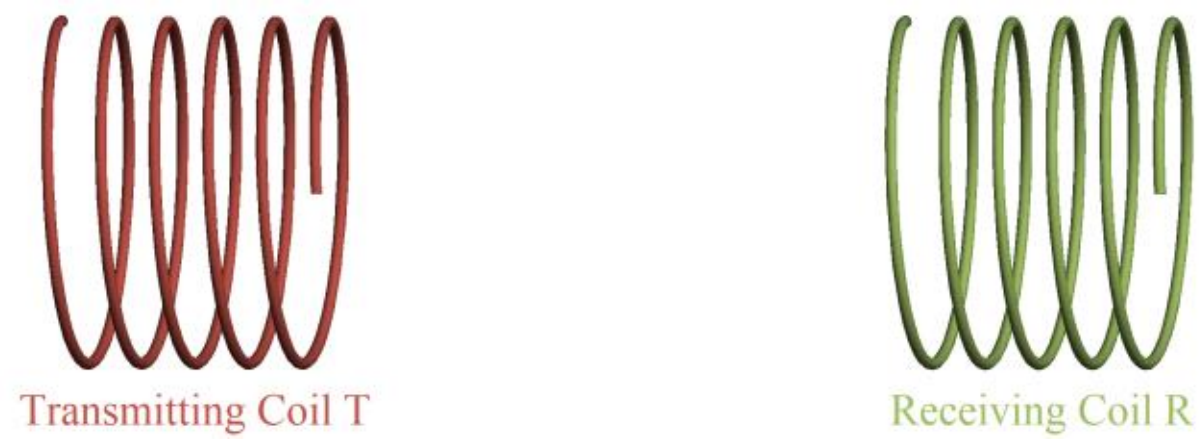

Fig. 1. A classical magnetic resonance WPT system constituted by a metallic transmitting coil T and a metallic receiving coil R.

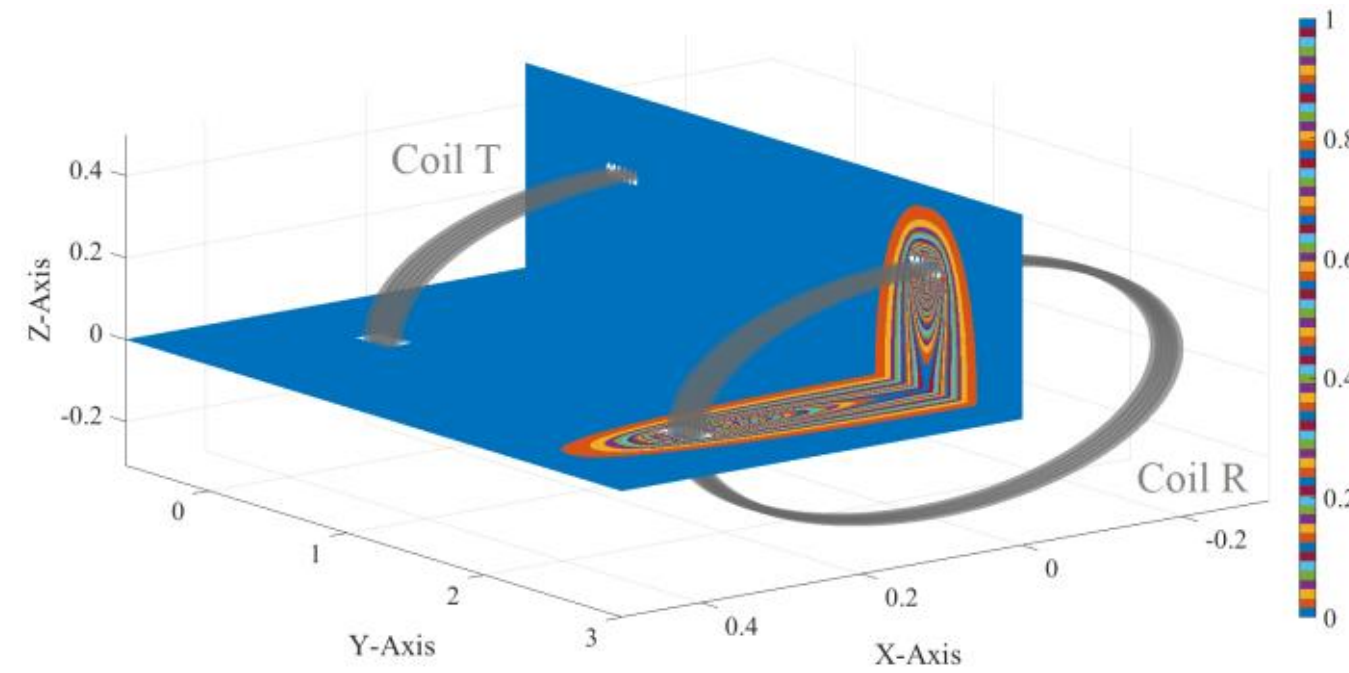

Fig. 2. Time-average magnetic energy density distribution of the desired transferring mode reported in [18]. The energy is almost completely transferred from coil T to coil R.

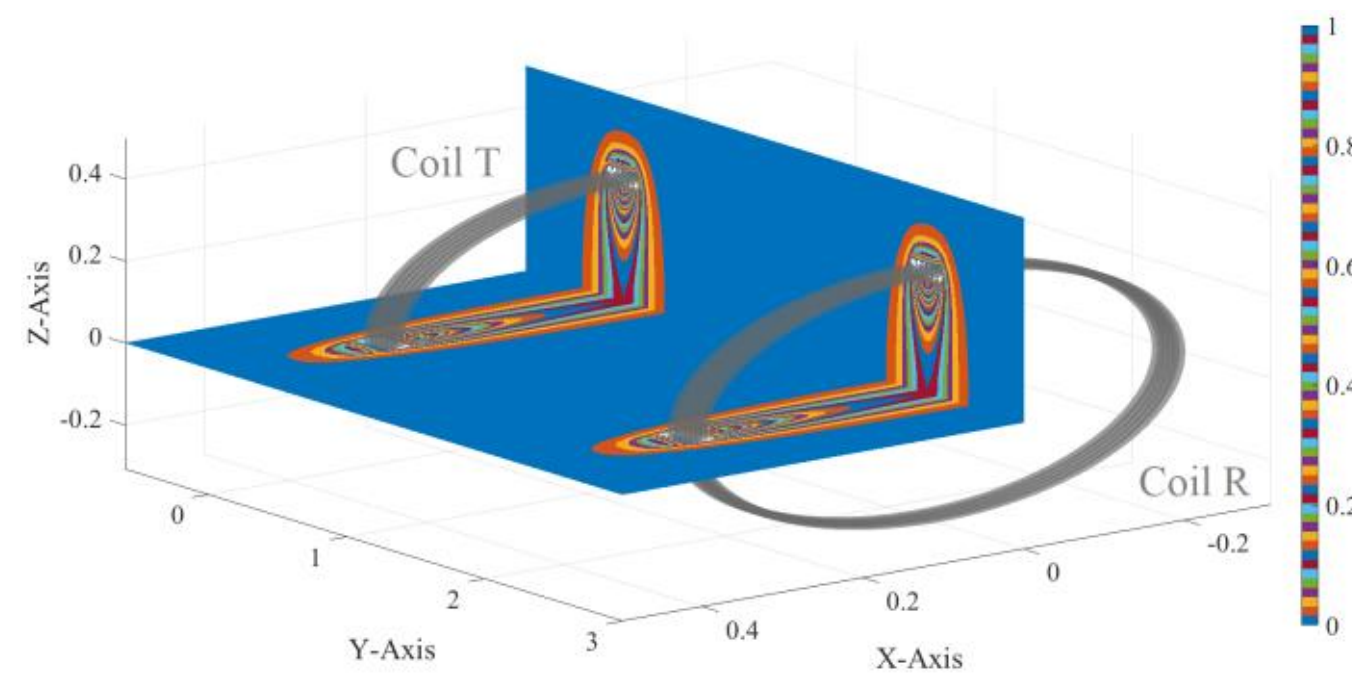

Fig. 3. Time-average magnetic energy density distribution of the resonant CM calculated from the conventional CMT for scattering systems [45]-[47]. The energy oscillates locally, but it is not transferred from coil T to coil R.

## II. Physical Principles

For the transferring system shown in Fig. 1, a typical locally impressed driver is delta-gap source, and the source provides a voltage driving to coil T. The voltage driving has a field effect, i.e., the voltage driving can be equivalently viewed as a field driving. The voltage driving acts on coil T only (but doesn't act on coil R and load), so the equivalent field driving also acts on coil T only (but doesn't act on coil R and load), i.e., the driving field is localized/restricted in the region occupied by coil T, and this is just the reason to call it **locally** impressed driver.

The action of the driving field on coil T will induce a current on coil T, and the current will generate a field on surrounding environment. Similarly, the field generated by coil T will act on coil R, and the action will lead to an induced current on coil R, and the current will generate a field on surrounding environment. In addition, the fields generated by the coils act on the load as well.

In fact, there also exists a reaction from the field generated by coil R to the current distributing on coil T, and the reaction will affect the current distribution. However, there doesn't exist any reaction from the load to the coils, i.e., the field effect of the load is localized/restricted in the region occupied by the load itself, because the load is supposed as a perfectly **matched** one in this paper.

Through a complicated process, the above actions and reactions will reach a dynamic equilibrium finally, because the EM problem considered here is time-harmonic.

For the convenience of following discussions, the boundary surfaces of coil T and coil R are denoted as $S_{\rm t}$ and $S_{\rm r}$ respectively; the three-dimensional Euclidean space is denoted as $\mathbb{E}^3$; the boundary of $\mathbb{E}^3$ is denoted as $S_\infty$, which is a closed spherical surface with infinite radius. The driving field generated by locally impressed driver is denoted as $\boldsymbol{\mathcal{F}}_{\rm driv}$. At the state of dynamic equilibrium, the currents distributing on coil T and coil R are denoted as $\boldsymbol{\mathcal{J}}_{\rm t}$ and $\boldsymbol{\mathcal{J}}_{\rm r}$ respectively. The fields generated by $\boldsymbol{\mathcal{J}}_{\rm t}$ and $\boldsymbol{\mathcal{J}}_{\rm r}$ are denoted as $\boldsymbol{\mathcal{F}}_{\rm t}$ and $\boldsymbol{\mathcal{F}}_{\rm r}$ respectively.

The energy conservation law tells us that the above actions and reactions among driver, coil T, coil R, and load will result in a work-energy transformation, and the work-energy transformation can be quantitatively expressed as follows:

$$\mathcal{P}_{\rm driv} = \mathcal{P}_{\rm sel}^{\rm tt} + \mathcal{P}_{\rm tra} \tag{1}$$

called time-domain work-energy principle (WEP), where $\mathcal{P}_{\rm driv} = <\boldsymbol{\mathcal{J}}_{\rm t}, \boldsymbol{\mathcal{E}}_{\rm driv}>_{S_{\rm t}}$ and $\mathcal{P}_{\rm sel}^{\rm tt} = -<\boldsymbol{\mathcal{J}}_{\rm t}, \boldsymbol{\mathcal{E}}_{\rm t}>_{S_{\rm t}}$ and $\mathcal{P}_{\rm tra} = -<\boldsymbol{\mathcal{J}}_{\rm t}, \boldsymbol{\mathcal{E}}_{\rm r}>_{S_{\rm t}}$. A rigorous mathematical derivation for WEP (1) and a more detailed decomposition for $\mathcal{P}_{\rm driv}$ are provided in App. A.

Similar to the scattering problem discussed in [47]-[48], the current-field interaction $<\boldsymbol{\mathcal{J}}_{\rm t}, \boldsymbol{\mathcal{E}}_{\rm driv}>_{S_{\rm t}}$ is the source for driving a steady work-energy transformation, so $\mathcal{P}_{\rm driv}$ is called driving power. Based on Maxwell's equations, the current-field interaction $-<\boldsymbol{\mathcal{J}}_{\rm t}, \boldsymbol{\mathcal{E}}_{\rm t}>_{S_{\rm t}}$ can be alternatively written as follows:

$$\begin{aligned} -\langle \boldsymbol{\mathcal{J}}_{\rm t}, \boldsymbol{\mathcal{E}}_{\rm t}\rangle_{S_{\rm t}} = & \oiint_{S_\infty} \left(\boldsymbol{\mathcal{E}}_{\rm t} \times \boldsymbol{\mathcal{H}}_{\rm t}\right)\cdot d\boldsymbol{S} \\ & + \frac{d}{dt}\left[(1/2)\langle \boldsymbol{\mathcal{H}}_{\rm t}, \boldsymbol{\mathcal{B}}_{\rm t}\rangle_{\mathbb{E}^3} + (1/2)\langle \boldsymbol{\mathcal{D}}_{\rm t}, \boldsymbol{\mathcal{E}}_{\rm t}\rangle_{\mathbb{E}^3}\right] \end{aligned} \tag{2}$$

called Poynting's theorem, where $\boldsymbol{\mathcal{D}}_{\rm t} = \varepsilon_0 \boldsymbol{\mathcal{E}}_{\rm t}$ and $\boldsymbol{\mathcal{B}}_{\rm t} = \mu_0 \boldsymbol{\mathcal{H}}_{\rm t}$. The theorem quantitatively governs the way how $\boldsymbol{\mathcal{J}}_{\rm t}$ provides EM power to $\boldsymbol{\mathcal{F}}_{\rm t}$, so $\mathcal{P}_{\rm sel}^{\rm tt}$ is called self-power of coil T. It has been explained previously that coil R is driven by coil T, and it will be further proven in Sec. III that $\boldsymbol{\mathcal{J}}_{\rm r}$ is uniquely determined by $\boldsymbol{\mathcal{J}}_{\rm t}$, so $\boldsymbol{\mathcal{J}}_{\rm t}$ can be viewed as the source for providing power to $\boldsymbol{\mathcal{F}}_{\rm r}$ and the power can be expressed as current-field interaction $-<\boldsymbol{\mathcal{J}}_{\rm t}, \boldsymbol{\mathcal{E}}_{\rm r}>_{S_{\rm t}}$ (which is similar to the current-field interaction $-<\boldsymbol{\mathcal{J}}_{\rm t}, \boldsymbol{\mathcal{E}}_{\rm t}>_{S_{\rm t}}$ of the coil T itself), and then $\mathcal{P}_{\rm tra}$ is called transferred power from coil T to coil R.

The above these clearly reveal the physical picture of WPT process: driver drives coil T, and the driving power $\mathcal{P}_{\rm driv}$ is transformed into two parts $\mathcal{P}_{\rm sel}^{\rm tt}$ and $\mathcal{P}_{\rm tra}$ by the transferring system, where $\mathcal{P}_{\rm sel}^{\rm tt}$ is dissipated by coil T and $\mathcal{P}_{\rm tra}$ is wirelessly transferred from coil T to coil R. The physical picture clearly reveals the fact that the working mechanism of **transferring** systems is different from the working mechanism of *scattering* systems (for details of the latter please see [47] and [48]). This is just the reason why the *scattering* CMT [43]-[48] cannot be directly applied to **transferring** systems.

For the WPT application, the transferred power $\mathcal{P}_{\rm tra}$ is desired, and the dissipated power $\mathcal{P}_{\rm sel}^{\rm tt}$ is unwanted and expected to be as small as possible, so we introduce a novel concept of transferring coefficient (TC) as follows:

$$\text{TC} = \frac{(1/T)\int_0^T \mathcal{P}_{\rm tra}\, dt}{(1/T)\int_0^T \mathcal{P}_{\rm driv}\, dt} \tag{3}$$

to quantify the transferring efficiency of transferring system. From a relatively mathematical viewpoint, the central aim of designing transferring system is to search for a physically realizable working mode (or working state) such that TC is maximized. In this paper, the central aim is realized by applying a CMT-based modal analysis to the transferring system.

In the following Sec. III, we provide the mathematical formulations used to establish the WEP-based CMT (WEP-CMT) for transferring systems. Because there exist relations $(1/T)\int_0^T \mathcal{P}_{\rm tra}\, dt = {\rm Re}\{P_{\rm tra}\}$ and $(1/T)\int_0^T \mathcal{P}_{\rm driv}\, dt = {\rm Re}\{P_{\rm driv}\}$, then the following Sec. III is discussed in frequency domain.

## III. Mathematical Formulations

The EM field boundary condition on $S_{\rm t}$ implies the relation that $[\boldsymbol{E}_{\rm driv}]_{\rm tan} = [-\boldsymbol{E}_{\rm t} - \boldsymbol{E}_{\rm r}]_{\rm tan}$ on $S_{\rm t}$, where subscript "tan" represents that the relation is satisfied by the tangential component. Electric fields $\boldsymbol{E}_{\rm t}$ and $\boldsymbol{E}_{\rm r}$ have operator expressions $\boldsymbol{E}_{\rm t} = -j\omega\mu_0 \mathcal{L}_0(\boldsymbol{J}_{\rm t})$ and $\boldsymbol{E}_{\rm r} = -j\omega\mu_0 \mathcal{L}_0(\boldsymbol{J}_{\rm r})$ respectively, in which $\mathcal{L}_0(\boldsymbol{J}_{\rm t/r}) = [1 + (1/k_0^2)\nabla\nabla\cdot]\int_\Omega G_0(\boldsymbol{r},\boldsymbol{r}')\boldsymbol{J}_{\rm t/r}(\boldsymbol{r}')d\Omega'$ where $G_0(\boldsymbol{r},\boldsymbol{r}') = e^{-jk_0|\boldsymbol{r}-\boldsymbol{r}'|}/4\pi|\boldsymbol{r}-\boldsymbol{r}'|$. Based on the above these, frequency-domain driving power $P_{\rm driv}$ can be expressed as the following operator form

$$P_{\rm driv} = -(1/2)\langle \boldsymbol{J}_{\rm t}, -j\omega\mu_0 \mathcal{L}_0(\boldsymbol{J}_{\rm t} + \boldsymbol{J}_{\rm r})\rangle_{S_{\rm t}} \tag{4}$$

called frequency-domain driving power operator (DPO).

In fact, the physically realizable $\boldsymbol{J}_{\rm t}$ and $\boldsymbol{J}_{\rm r}$ are not independent of each other, because they must satisfy the following electric field integral equation

$$\left[-j\omega\mu_0 \mathcal{L}_0(\boldsymbol{J}_{\rm t}) - j\omega\mu_0 \mathcal{L}_0(\boldsymbol{J}_{\rm r})\right]_{\rm tan} = 0 \ \text{ on } \ S_{\rm r} \tag{5}$$

due to the homogeneous tangential electric field boundary condition $[\boldsymbol{E}_{\rm t} + \boldsymbol{E}_{\rm r}]_{\rm tan} = 0$ on $S_{\rm r}$. Applying the method of moments (MoM) to (5), the integral equation is immediately discretized into a matrix equation. By solving the matrix equation, the following transformation

$$\mathbb{J}_{\rm r} = \mathbb{T}\cdot\mathbb{J}_{\rm t} \tag{6}$$

can be obtained, where $\mathbb{J}_{\rm t}$ and $\mathbb{J}_{\rm r}$ are the column vectors constituted by the expansion coefficients of $\boldsymbol{J}_{\rm t}$ and $\boldsymbol{J}_{\rm r}$ respectively. The detailed mathematical process for deriving (6) from

(5) is provided in App. B.

Similarly to discretizing integral equation (5), the DPO (4) can be discretized into its matrix form. Substituting transformation (6) into the matrix form, we immediately have that

$$P_{\text{driv}} = \mathrm{J}_{\text{t}}^{\dagger} \cdot \mathrm{P}_{\text{driv}} \cdot \mathrm{J}_{\text{t}} \tag{7}$$

The detailed mathematical formulations for calculating matrix $\mathrm{P}_{\text{driv}}$ are listed in App. C. In fact, the above process from (4) to (7) is just the dependent variable elimination (DVE) process used in the WEP-CMT for *scattering* systems [47]-[48].

WEP-CMT decomposes $\mathrm{P}_{\text{driv}}$ in terms of its positive and negative Hermitian parts as that $\mathrm{P}_{\text{driv}} = \mathrm{P}_{\text{driv}}^{+} + j\mathrm{P}_{\text{driv}}^{-}$ (where $\mathrm{P}_{\text{driv}}^{+} = (\mathrm{P}_{\text{driv}} + \mathrm{P}_{\text{driv}}^{\dagger})/2$ and $\mathrm{P}_{\text{driv}}^{-} = (\mathrm{P}_{\text{driv}} - \mathrm{P}_{\text{driv}}^{\dagger})/2j$), and constructs CMs by solving the following characteristic equation

$$\mathrm{P}_{\text{driv}}^{-} \cdot \mathrm{J}_{\text{t}} = \lambda\, \mathrm{P}_{\text{driv}}^{+} \cdot \mathrm{J}_{\text{t}} \tag{8}$$

where $\lambda$ and $\mathrm{J}_{\text{t}}$ are the associated characteristic values and characteristic vectors respectively.

Using the above-obtained characteristic vector $\mathrm{J}_{\text{t}}$, the TC of the CM can be calculated as follows:

$$\text{TC} = \frac{\mathrm{J}_{\text{t}}^{\dagger} \cdot \mathrm{P}_{\text{tra}}^{+} \cdot \mathrm{J}_{\text{t}}}{\mathrm{J}_{\text{t}}^{\dagger} \cdot \mathrm{P}_{\text{driv}}^{+} \cdot \mathrm{J}_{\text{t}}} \tag{9}$$

where $\mathrm{P}_{\text{tra}}^{+}$ is given in App. C. By comparing the TCs of the CMs, the CM with the maximal TC can be easily found, and the CM is usually a desired selection for the WPT application.

In the following Sec. IV, we provide a WEP-CMT-based modal analysis for a classical example, which is just the transferring system reported in the seminal paper [18].

## IV. Numerical Verifications

The transferring system considered in [18] is constituted by two metallic coils as shown in Fig. 1. The coils have the same radius 30 cm, height 20 cm, and turns 5.25. The coils are placed coaxially, and their distance is 2 m. The optimally transferring frequency (i.e. the working frequency of the optimally transferring mode) calculated from the coupled-mode theory used in [18] is $10.56 \pm 0.3\,\text{MHz}$, and the optimally transferring frequency obtained from the measurement done in [18] is 9.90 MHz. The reason leading to a 5% discrepancy between the theoretical and measured values was explained in [18].

### A. WEP-CMT-Based Modal Analysis

We use the WEP-CMT given in Secs. II and III to calculate the CMs of the transferring system, and show the TC curves of the first 5 CMs in Fig. 4. Based on the analysis given in Secs. II and III, it is not difficult to conclude that the CM 1 at 10.8988 MHz (which corresponds to the local maximum of the TC curve) works at the optimally transferring state. The coil current distribution and time-average magnetic energy density distribution of the optimally transferring mode are shown in Fig. 5 and Fig. 6 respectively.

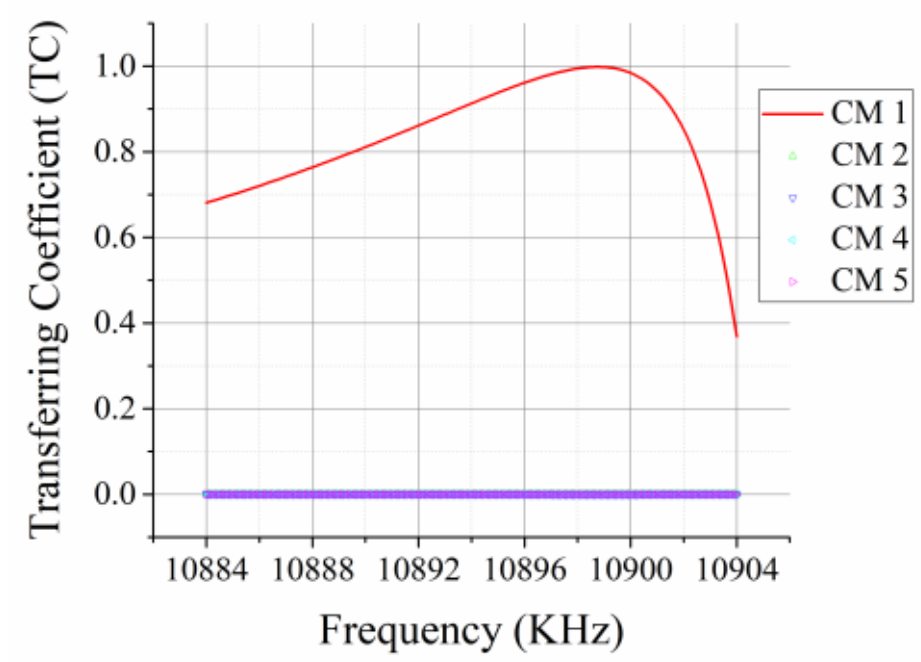

Fig. 4. TC curves of the first 5 low-order CMs calculated from the WEP-CMT established in this paper.

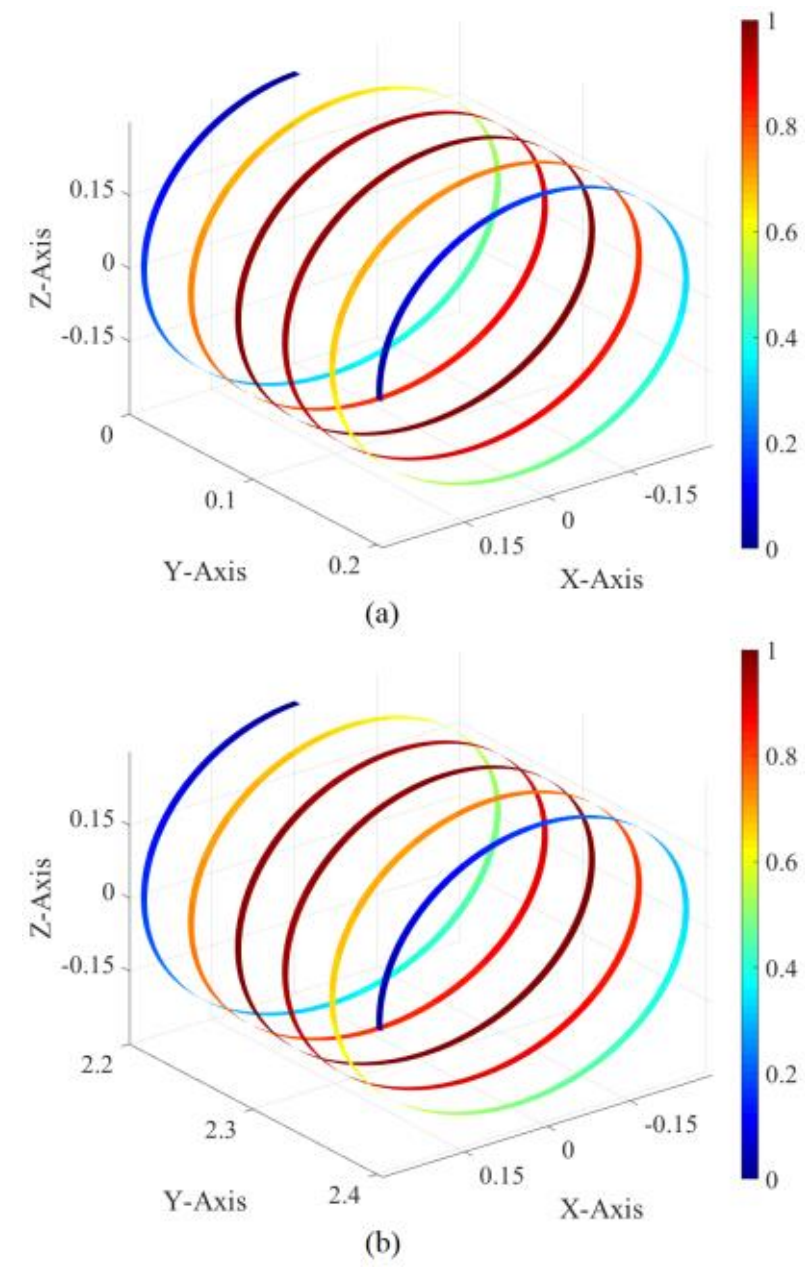

Fig. 5. For the CM 1 working at 10.8988 MHz, its current magnitudes distributing on (a) coil T and (b) coil R.

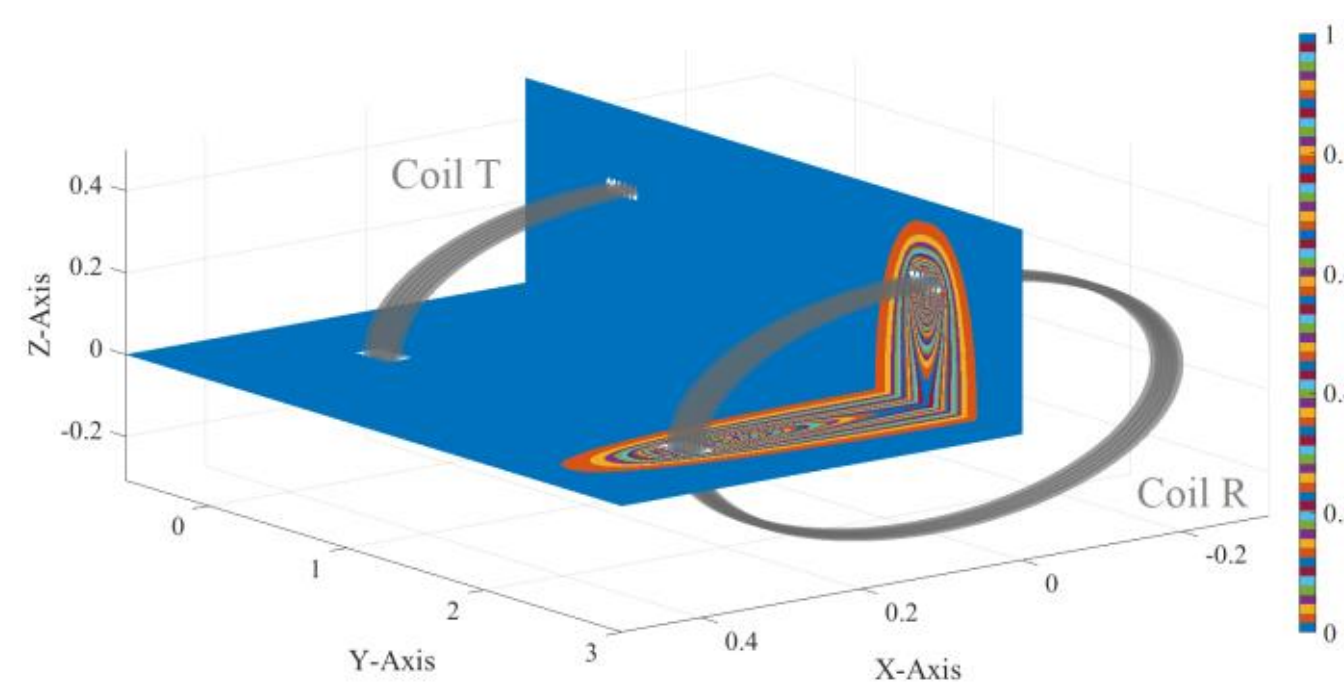

Fig. 6. For the CM 1 working at 10.8988 MHz, its time-average magnetic energy density distributing on xOy and yOz planes.

Evidently, the CM 1 working at 10.8988 MHz corresponds to a half-wave current distribution for both the coil T and coil R as shown in Fig. 5, and it indeed can efficiently transfer EM power from coil T to coil R in a wireless manner as shown in Fig. 6.

In addition, the optimally transferring frequencies calculated from the classical coupled-mode theory ($10.56 \pm 0.3\,\text{MHz}$) proposed in [18] and the WEP-CMT (10.8988 MHz) used in this paper are consistent with each other. The advantage of WEP-CMT over coupled-mode theory is reflected as follows:

A1 The WEP-CMT is a field-based modal analysis method, which is directly derived from Maxwell's equations and doesn't use any approximation; the coupled-mode theory is a circuit-model-based modal analysis method, which employs some circuit-model-based approximate quantities (such as scalar voltage, scalar current, effective inductance, and effective capacitance, etc. [18]-[19]).

A2 The WEP-CMT is applicable to the coils working at arbitrary frequency; the coupled-mode theory is only applicable to the coils working at low frequency at which the circuit model exists.

A3 The WEP-CMT is applicable to the coils with arbitrary topological structures; the coupled-mode theory is only applicable to the coils with simple topological structures, such that the coils can support sinusoidal scalar currents.

By employing an alternative field-based modal analysis, the following Sec. IV-B verifies that the CM 1 at 10.8988 MHz is indeed the most efficient mode for WPT, and then exhibits that the optimally transferring mode is indeed included in the CM set constructed by the WEP-CMT.

## B. *An Alternative Modal Analysis*

Because the matrices $\mathrm{P}_{\mathrm{tra}}^{+}$ and $\mathrm{P}_{\mathrm{driv}}^{+}$ in (9) are Hermitian, and the matrix $\mathrm{P}_{\mathrm{driv}}^{+}$ is positive definite, then the mode maximizing TC can be obtained from solving the following equation [49]

$$\mathrm{P}_{\mathrm{tra}}^{+} \cdot \mathrm{J}_{\mathrm{t}} = \tau\, \mathrm{P}_{\mathrm{driv}}^{+} \cdot \mathrm{J}_{\mathrm{t}} \tag{10}$$

Using the equation, we calculate the optimally transferring mode, and show the associated TC curve in Fig. 7. Obviously, both the obtained optimally transferring frequency and optimally transferring coefficient are consistent with the ones obtained from the WEP-CMT-based modal analysis method.

## C. *Concepts of Co-resonance and Ci-resonance*

In this sub-section, we, under the WEP framework, provide some further modal analysis to the CM 1 shown in Fig. 4, and introduce the concepts of co-resonance and ci-resonance.

For the CM 1, its various resistance and reactance curves are shown in Fig. 8. The definitions for the resistances and reactances are given in App. D. From the Figs. 8(a) and 8(b), it is easy to observe the following facts:

F1 At optimally transferring frequency $f_{\mathrm{cor}} = 10.8988\ \mathrm{MHz}$, $R_{\mathrm{sel}}^{\mathrm{tt}}$ is very small; $R_{\mathrm{mut}}^{\mathrm{tr}}$ is equal to zero; $R_{\mathrm{driv}}$, $R_{\mathrm{sel}}^{\mathrm{rr}}$, and $R_{\mathrm{tra}}$ achieve their local maximums simultaneously, and the maximal values are the same.

F2 At frequency $f_{\mathrm{cor}} = 10.8988\ \mathrm{MHz}$, $X_{\mathrm{sel}}^{\mathrm{tt}}$, $X_{\mathrm{mut}}^{\mathrm{tr}}$, and $X_{\mathrm{sel}}^{\mathrm{rr}}$ are zero simultaneously, and then $X_{\mathrm{driv}}$ and $X_{\mathrm{tra}}$ are zero automatically because of that $X_{\mathrm{driv}} = X_{\mathrm{sel}}^{\mathrm{tt}} + X_{\mathrm{mut}}^{\mathrm{tr}} + X_{\mathrm{sel}}^{\mathrm{rr}}$ and $X_{\mathrm{tra}} = X_{\mathrm{mut}}^{\mathrm{tr}} + X_{\mathrm{sel}}^{\mathrm{rr}}$.

F3 At frequency $f_{\mathrm{cir}} = 10.8873\ \mathrm{MHz}$, $R_{\mathrm{driv}}$, $R_{\mathrm{sel}}^{\mathrm{tt}}$, $R_{\mathrm{mut}}^{\mathrm{tr}}$, $R_{\mathrm{sel}}^{\mathrm{rr}}$, and $R_{\mathrm{tra}}$ are very small.

F4 At frequency $f_{\mathrm{cir}} = 10.8873\ \mathrm{MHz}$, only $X_{\mathrm{driv}}$ is zero, but the other reactances are not.

The subscripts "cor" and "cir" on $f_{\mathrm{cor}}$ and $f_{\mathrm{cir}}$ are explained as below.

Based on the above observations, it is not difficult to conclude that:

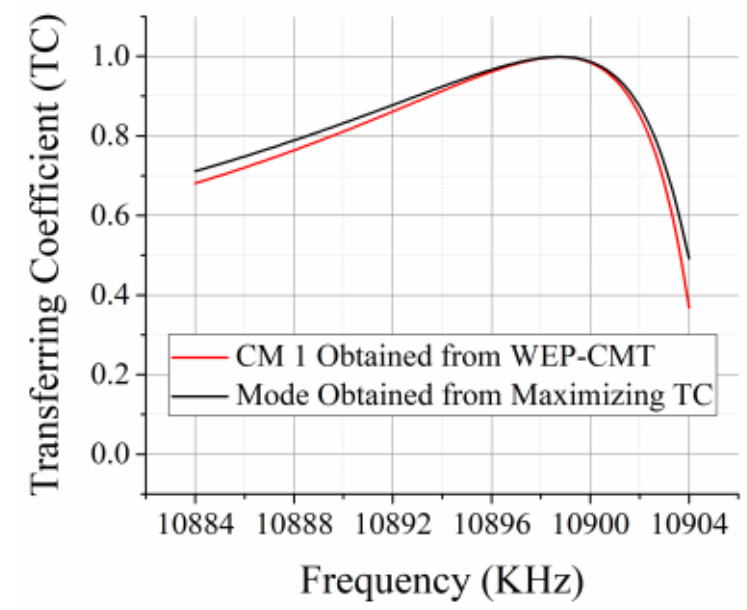

Fig. 7. TC curves of the optimally transferring modes obtained from two somewhat different modal analysis methods provided in this paper.

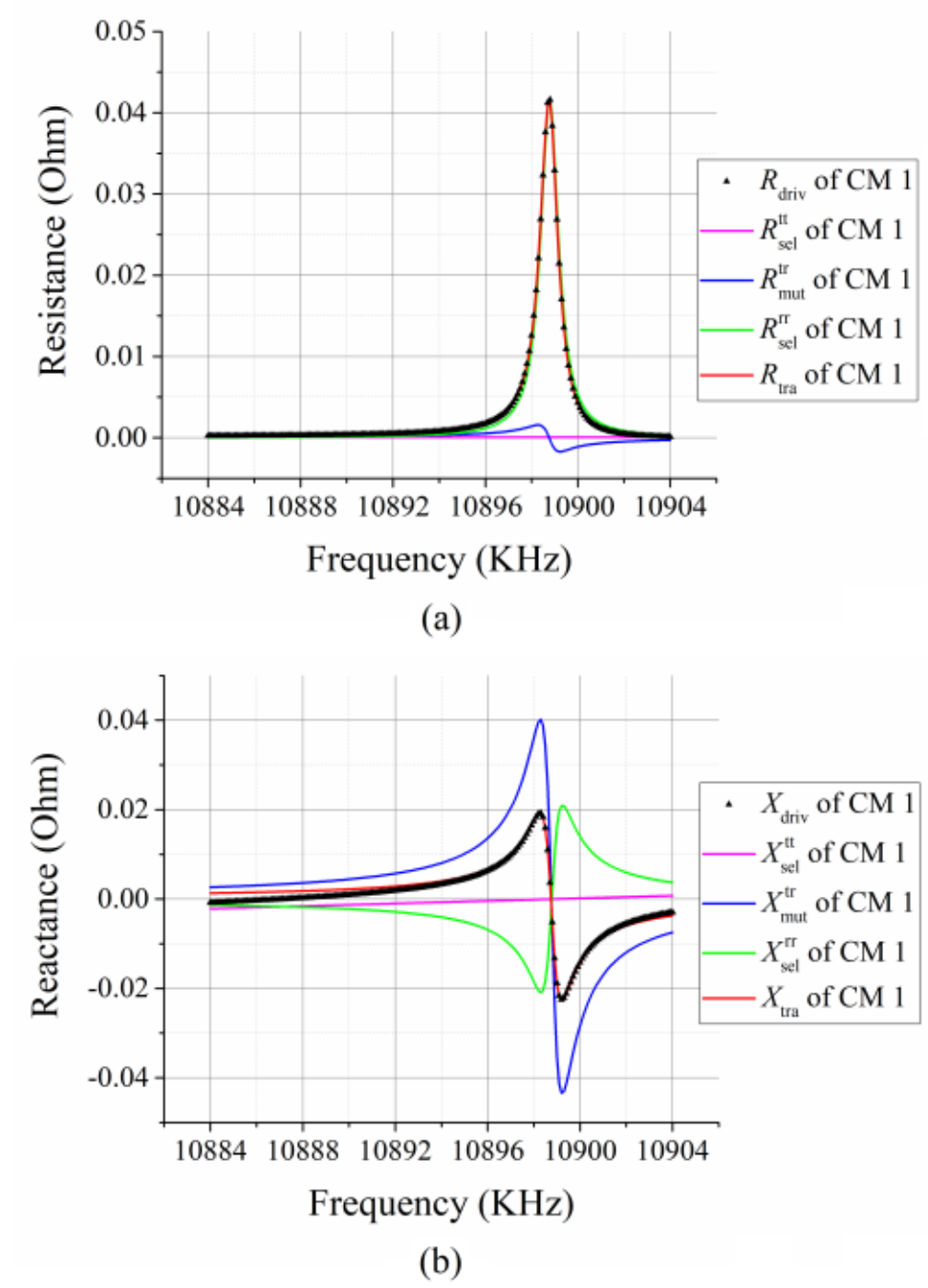

Fig. 8. For the CM 1, its (a) resistance curves and (b) reactance curves.

**C1** At optimally transferring frequency $f_{\mathrm{cor}}$, the driver provides power to coil T in a very efficient way, and the driving power is almost completely transferred from coil T to coil R except a very small part dissipated by coil T, and the transferred power is completely converted by coil R into its self-power but not into any mutual-power.

**C2** At optimally transferring frequency $f_{\mathrm{cor}}$, coil T works at the state of self-resonance (because $X_{\mathrm{sel}}^{\mathrm{tt}} = 0$), and coil R works at the state of self-resonance as well (because $X_{\mathrm{sel}}^{\mathrm{rr}} = 0$), and, at the same time, the two coils also work at the state of mutual-resonance (because $X_{\mathrm{mut}}^{\mathrm{tr}} = 0$). The working state with $X_{\mathrm{sel}}^{\mathrm{tt}} = X_{\mathrm{mut}}^{\mathrm{tr}} = X_{\mathrm{sel}}^{\mathrm{rr}} = 0$ is particularly called co-resonance in this paper, and conventionally called magnetic-resonance in [18]-[23]. In fact, the co-resonance/magnetic-resonance automatically leads to that $X_{\mathrm{tra}} = 0$ and $X_{\mathrm{driv}} = 0$, because $X_{\mathrm{tra}} = X_{\mathrm{mut}}^{\mathrm{tr}} + X_{\mathrm{sel}}^{\mathrm{rr}}$ and $X_{\mathrm{driv}} = X_{\mathrm{sel}}^{\mathrm{tt}} + X_{\mathrm{mut}}^{\mathrm{tr}} + X_{\mathrm{sel}}^{\mathrm{rr}}$.

**C3** The co-resonance/magnetic-resonance satisfies the classical resonance condition that the reactance is zero, i.e., the stored electric energy is equal to the stored magnetic energy. Thus, the co-resonance/magnetic-resonance is a special case of the classical EM-resonance (electric-magnetic resonance).

**C4** At frequency $f_{\text{cir}}$, whole transferring system is resonant (because $X_{\text{driv}} = 0$), but neither coil T nor coil R is resonant (because $X_{\text{sel}}^{\text{tt}}, X_{\text{sel}}^{\text{rr}} \neq 0$). The resonance of whole transferring system originates from neutralizing the negative $X_{\text{sel}}^{\text{tt}} + X_{\text{sel}}^{\text{rr}}$ and the positive $X_{\text{mut}}^{\text{tr}}$. To distinguish this kind of resonance from the previous co-resonance, this kind of resonance is particularly called capacitive-inductive-neutralized resonance (simply denoted as ci-resonance) in this paper.

In addition, by observing Fig. 4 and Fig. 8(a), it is easy to find out that the co-resonant mode is more desired than the ci-resonant mode for WPT application, because the former corresponds to the local maximums of TC curve and $R_{\text{tra}}$ curve but the latter doesn't.

## V. Typical Examples

To exhibit the values of WEP-CMT in the aspect of analyzing transferring systems and reveal the connections between transferring problem and scattering problem, this section employs WEP-CMT to do some valuable modal analysis for several typical variants of the classical two-coil transferring system considered in Sec. IV.

### *A. On Transferring Frequency — Part I*

Now, we consider the *scattering* system only constituted by the coil T considered in Sec. IV, and we calculate its *scattering* CMs by using the WEP-CMT for *scattering* systems [45]-[47]. The modal significance (MS) curves associated with the first 5 CMs are shown in Fig. 9. The resonant CM i with the local maximal MS has the same current distribution as the one shown in Fig. 5(a).

Comparing the Fig. 9 with the previous Figs. 4, 7 and 8, it is evident that

**C5** The **co-resonance** frequency (i.e. the optimally transferring frequency) of the two-coil **transferring** system considered in Sec. IV is equal to the conventional *external-resonance* frequency of the one-coil *scattering* system.

This conclusion effectively establishes a connection between the co-resonance-based/magnetic-resonance-based wireless power **transferring** problem and the conventional *scattering* problem.

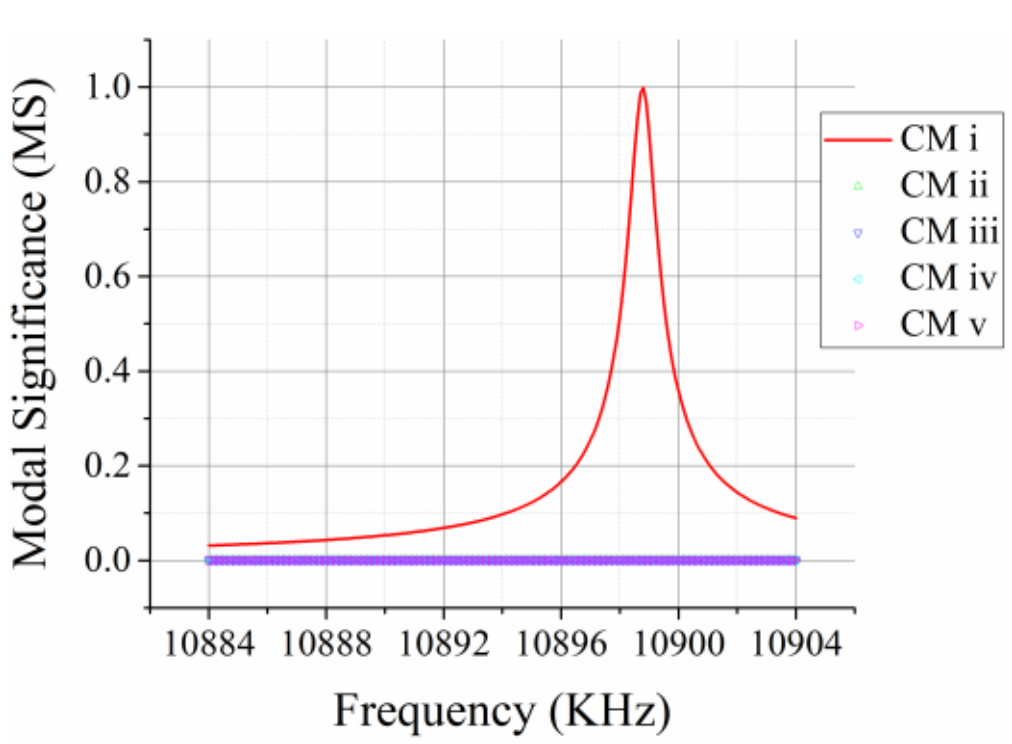

Fig. 9. MS curves of the first 5 CMs of the one-coil scattering system.

### *B. On Transferring Frequency — Part II*

In fact, the above-mentioned external-resonance frequency is the one-order external-resonance frequency of the one-coil scattering system, and the scattering system has some other higher-order external-resonance frequencies, such as the two-order external-resonance frequency 27.2646 MHz as shown in Figs. 10 and 11.

In the frequency band used in Fig. 10, we calculate the CMs of the two-coil transferring system, and show the TC and reactance curves associated with CM 2 in Fig. 12.

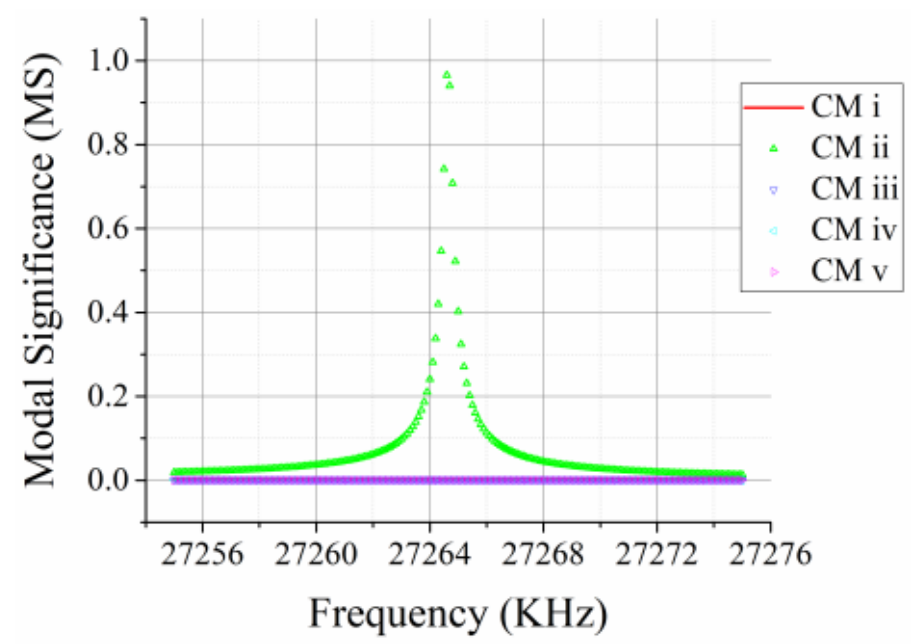

Fig. 10. MS curves of the first 5 CMs of the one-coil scattering system.

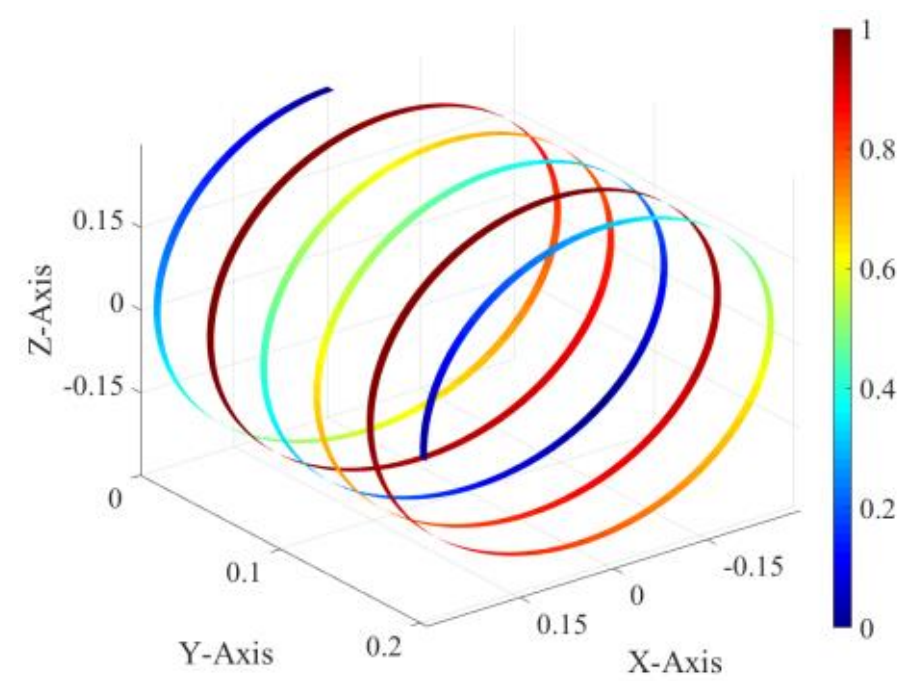

Fig. 11. Modal current magnitude distribution of the two-order externally resonant CM ii working at 27.2646 MHz.

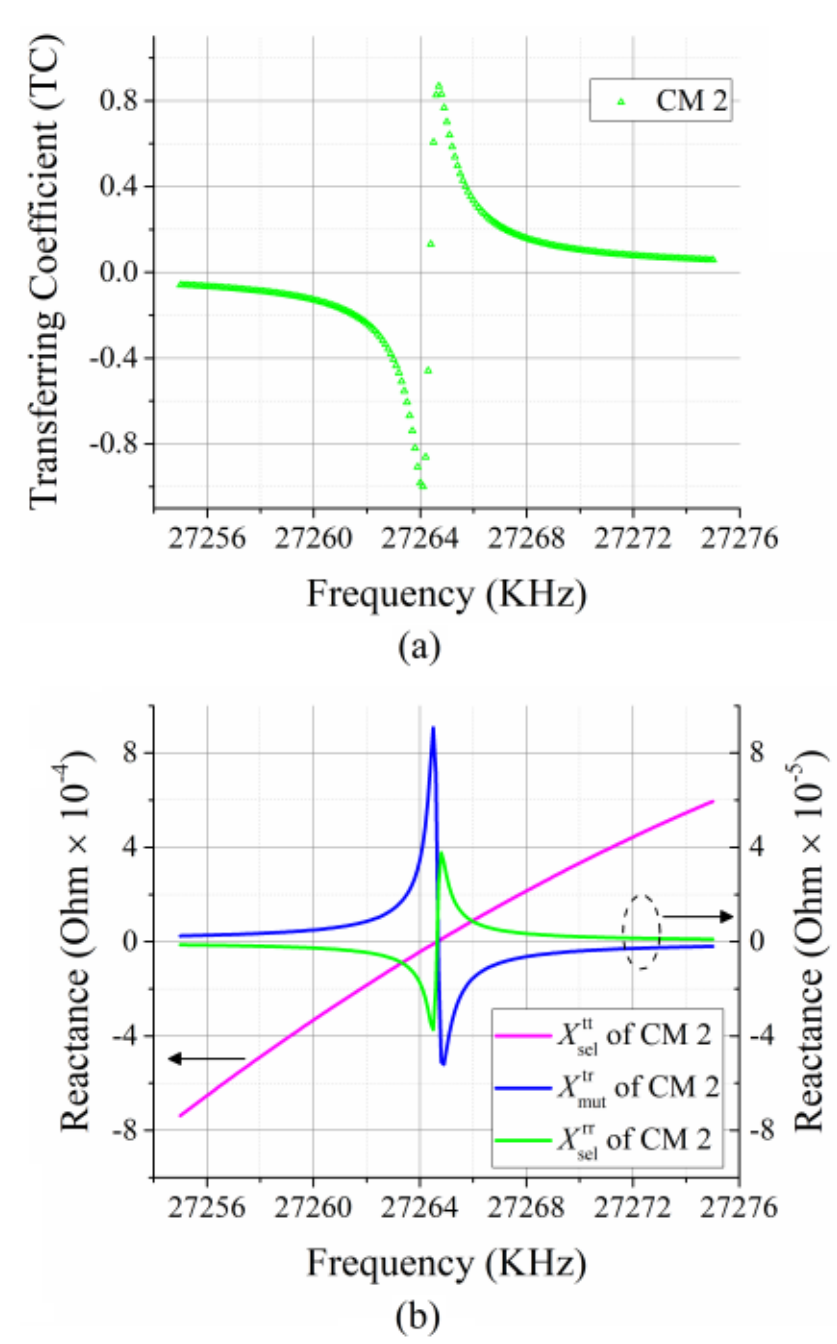

Fig. 12. (a) TC curve and (b) reactance curves associated with the CM 2 of the two-coil transferring system working in the frequency band used in Fig. 10.

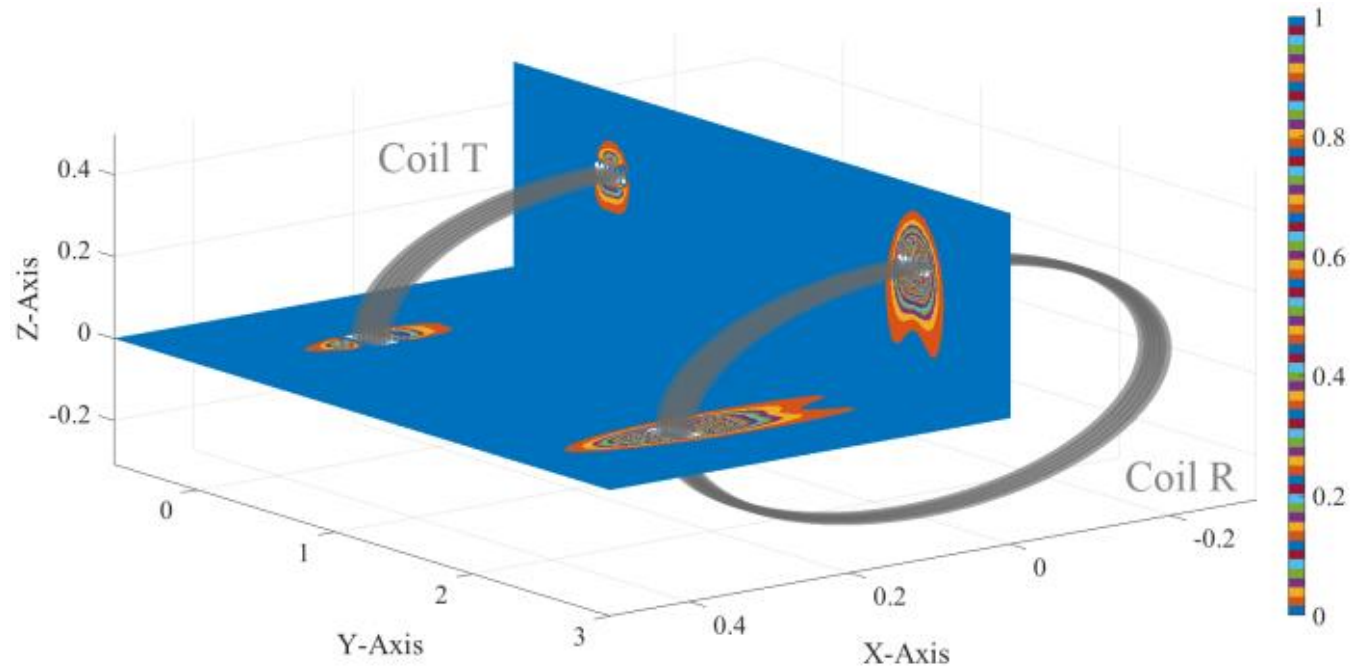

Fig. 13. For the co-resonant CM 2 working at 27.2646 MHz, its time-average magnetic energy density distributing on xOy and yOz planes.

The Fig. 12(b) implies that the CM 2 is co-resonant at 27.2646 MHz. The time-average magnetic energy density distribution of the co-resonant CM 2 working at co-resonance frequency 27.2646 MHz is shown in Fig. 13.

By comparing the Figs. 12(a)&13 with the previous Figs. 4&6, it is not difficult to find out that:

**C6** the one-order co-resonant mode has a higher wireless power transferring efficiency than the two-order co-resonant mode.

In fact, this conclusion can be generalized to the comparison between any lower-order and higher-order co-resonant modes.

### *C. On Transferring Frequency — Part III*

The above-mentioned two-coil transferring system is constituted by two completely same coils, which have the same geometrical size and external-resonance frequencies 10.8988 MHz (one order) and 27.2646 MHz (two order) etc. Now, we replace the coil R with another different coil, whose geometrical size is "radius 30 cm, height 8.15 cm, and turns 2.138" and one-order external-resonance frequency is 27.2646 MHz and higher-order external resonance frequencies are larger than 27.2646 MHz. We calculate the CMs of the new two-coil transferring system working around frequencies 10.8988 MHz and 27.2646 MHz, and we show the corresponding TC and reactance curves in Fig. 14 and Fig. 15 respectively.

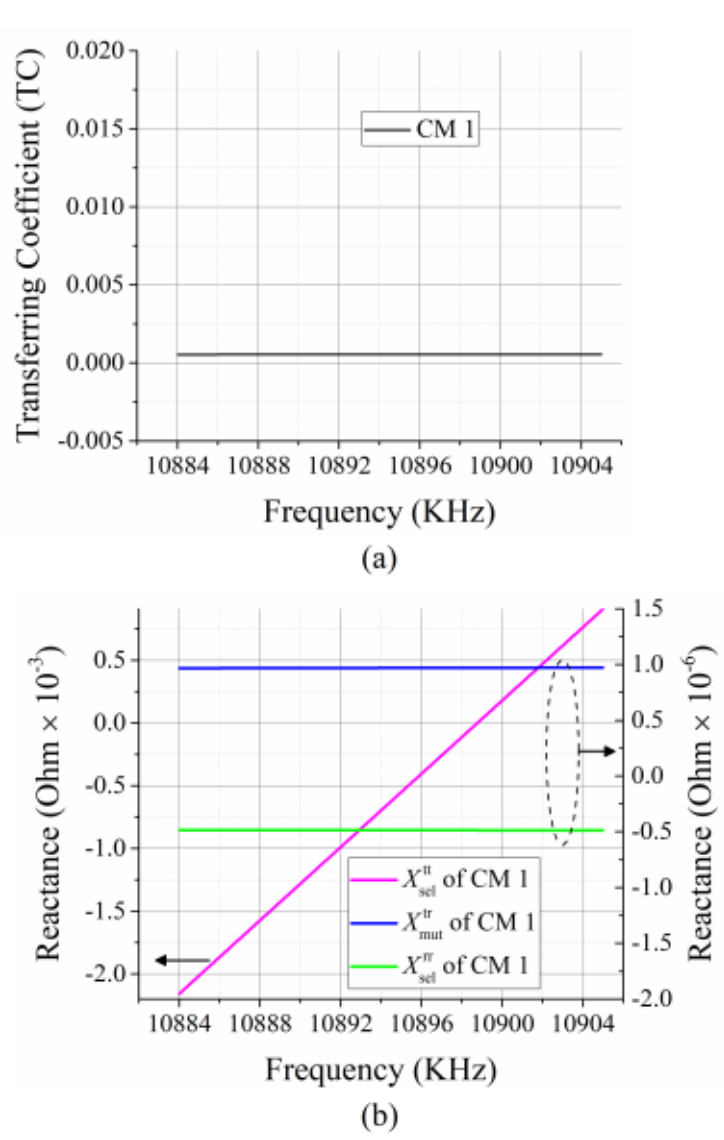

Fig. 14. (a) TC curve and (b) reactance curves associated with the CM 1 of the new two-coil transferring system working around 10.8988 MHz.

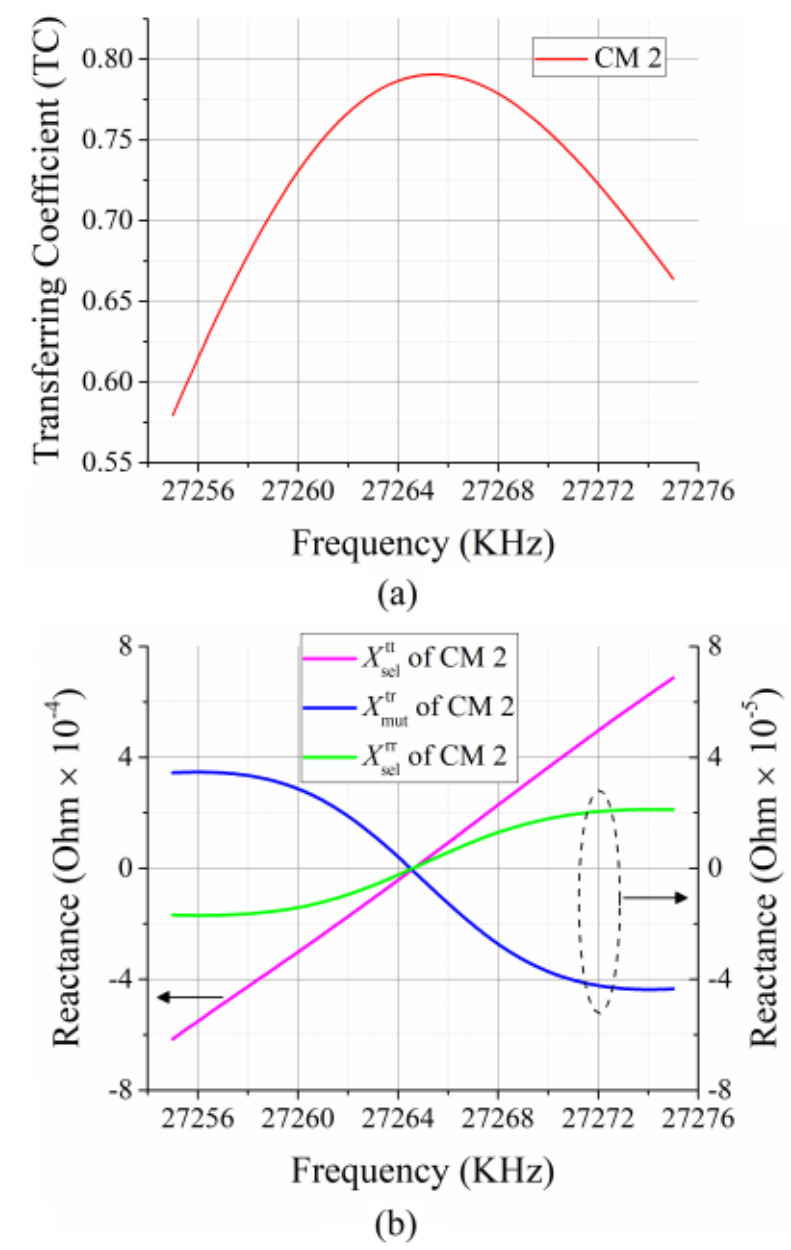

Fig. 15. (a) TC curve and (b) reactance curves associated with the CM 2 of the new two-coil transferring system working around 27.2646 MHz.

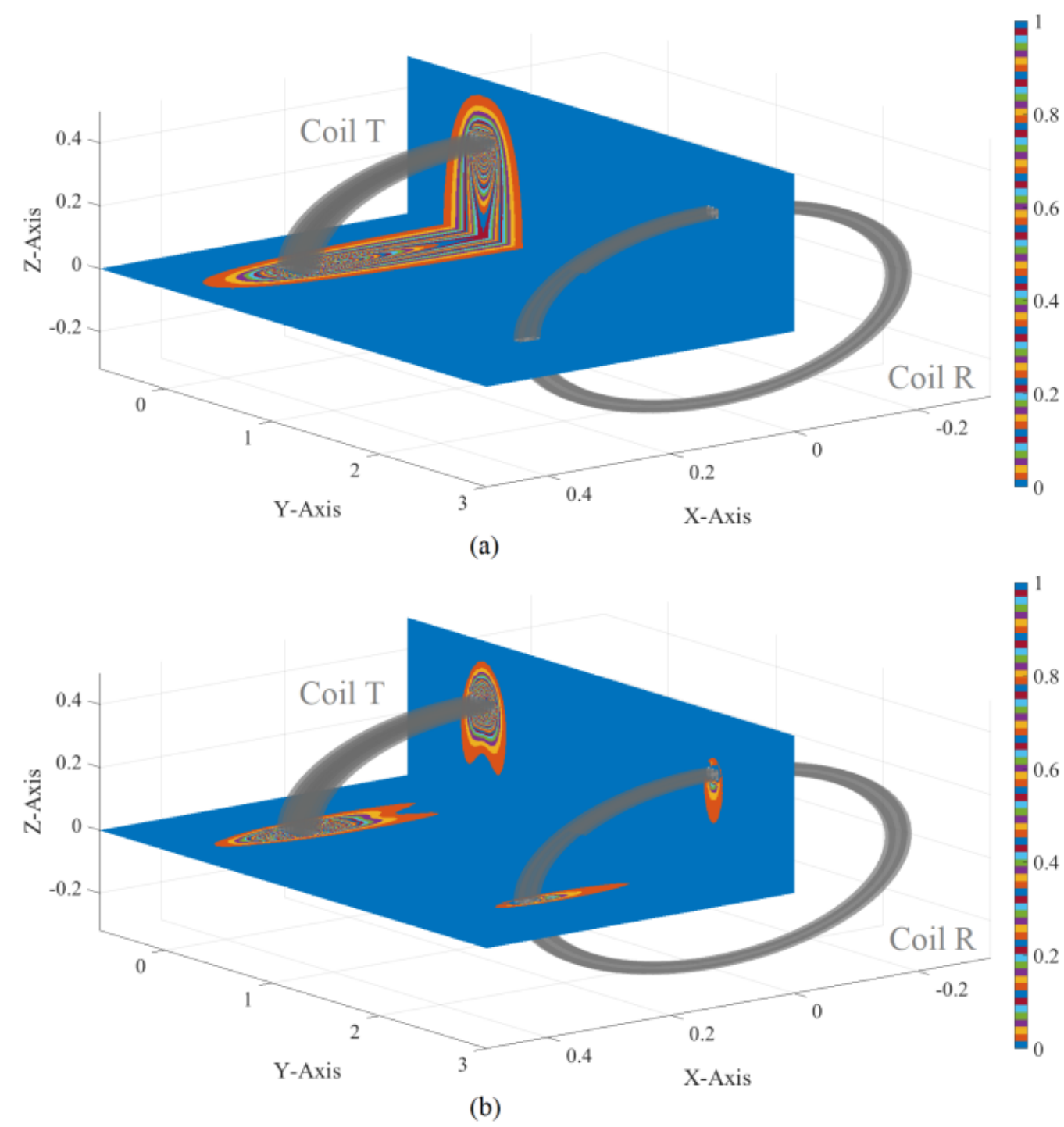

Fig. 16. Time-average magnetic energy density distributions of (a) the CM 1 working at 10.8988 MHz and (b) the CM 2 working at 27.2646 MHz.

From Figs. 14 and 15, it is easy to find out that: the coil T working at CM 1 is self-resonant at 10.8988 MHz ( $X_{sel}^{tt}=0$ ), but 10.8988 MHz is not the co-resonance frequency of the new two-coil transferring system ( $X_{mut}^{tr}, X_{sel}^{rr} \neq 0$ ), and the TC is very small at 10.8988 MHz; 27.2646 MHz is the co-resonance frequency of the new two-coil transferring system, and the TC curve achieves its local maximum at 27.2646 MHz.

Based on the above observations and the comparisons among Figs. 4, 12(a), and 15(a), we can conclude that:

**C7** When only one coil is self-resonant, the whole system cannot efficiently transfer power wirelessly.

**C8** When two coils have the same external-resonance frequency, and the whole two-coil transferring system works at the frequency, the system can transfer power wirelessly even if the orders of the frequencies are different.

**C9** The same-order co-resonance is more desired than the different-order co-resonance in WPT application.

To visualize the above conclusions, we show the corresponding time-average magnetic energy density distributions in Fig. 16.

### *D. On Transferring Distance*

In the above all discussions, the distance between coil T and coil R is 2 m. In this sub-section, we study how the distance influences transferring efficiency.

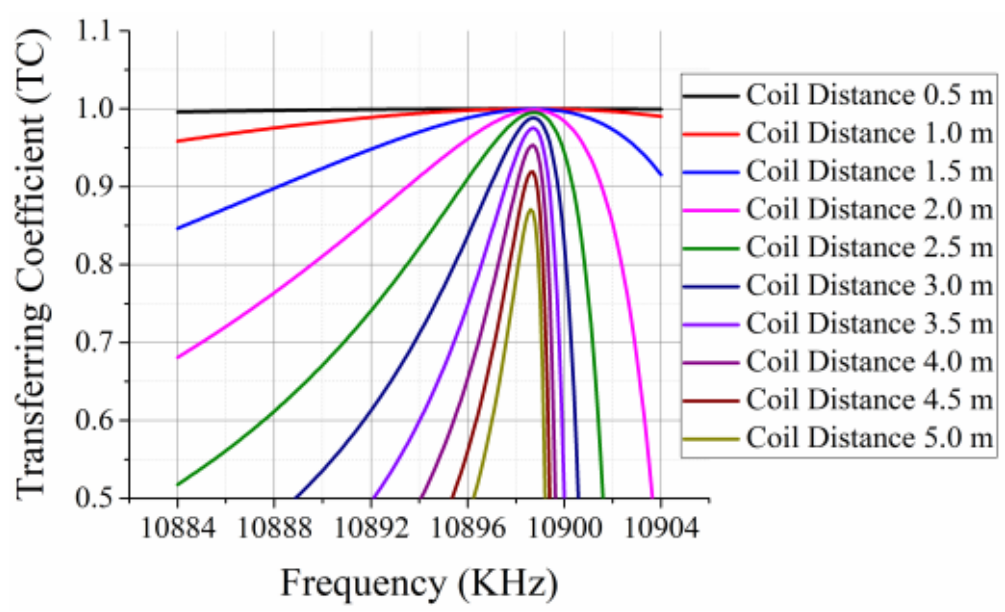

Fig. 17. TC curves of the one-order CMs corresponding to the different coil-separation distances.

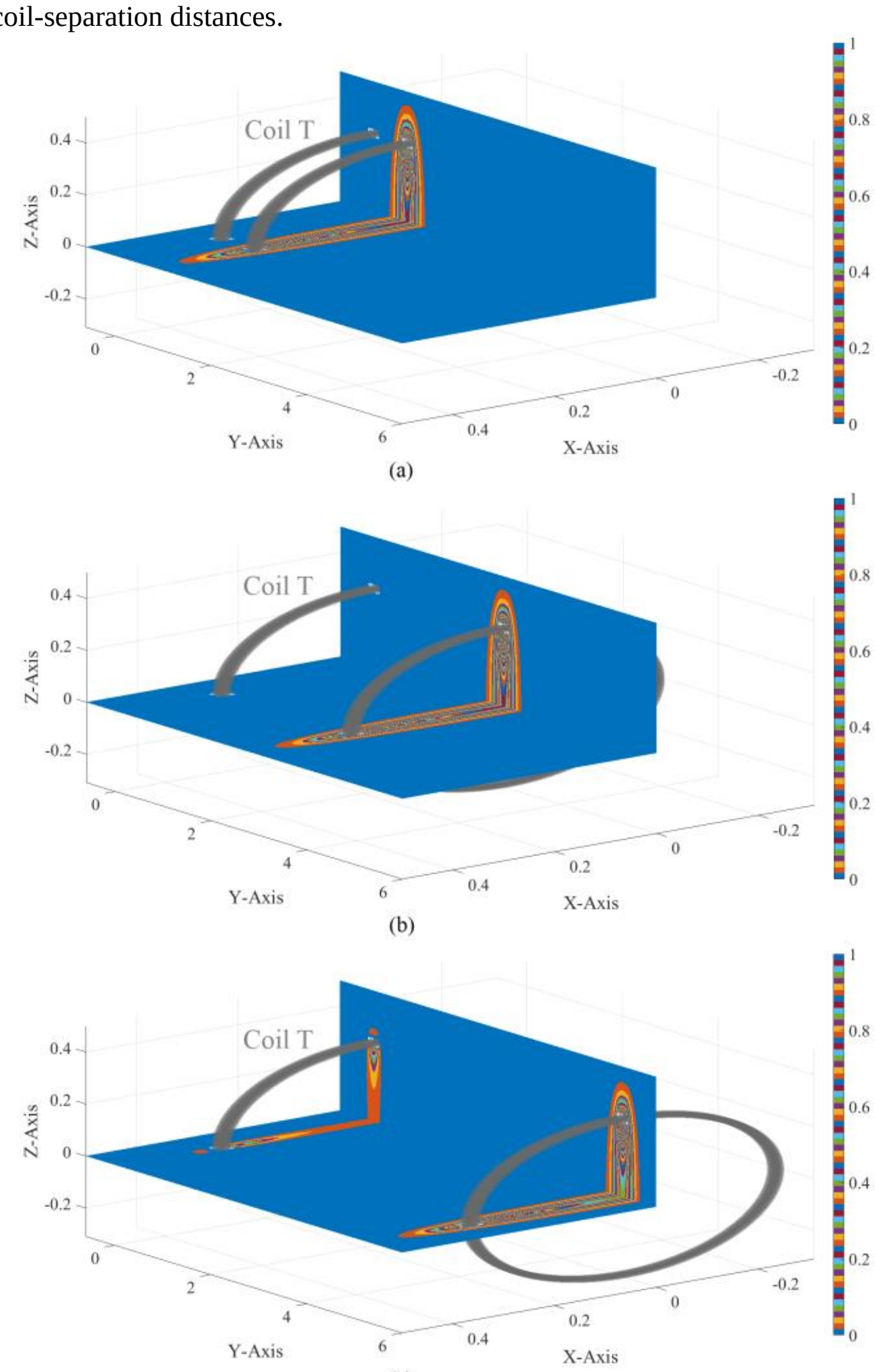

Fig. 18. Time-average magnetic energy density distributions corresponding to coil-separation distances (a) 0.5 m, (b) 2.5 m, and (c) 5.0 m.

We use the WEP-CMT to calculate the two-coil transferring systems with a series of different coil-separation distances (where the coil T and coil R are the same as the ones considered in Sec. IV), and show the TC curves of the one-order CMs corresponding to the different coil-separation distances in Fig. 17.

The Fig. 17 clearly implies the following expected conclusion.

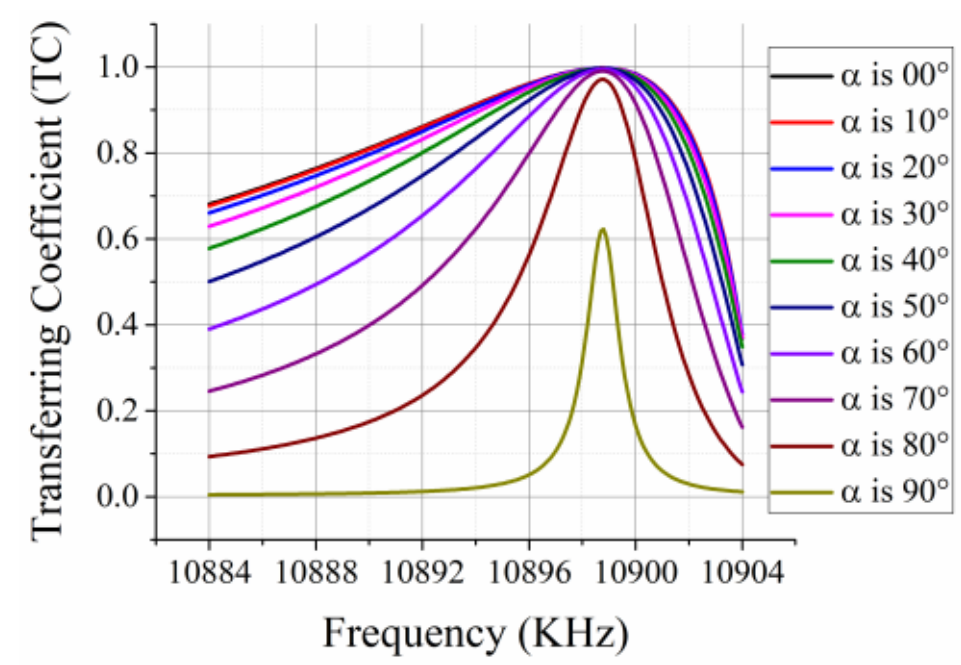

Fig. 19. TC curves of the one-order CMs corresponding to different coil-coil angle $\alpha$.

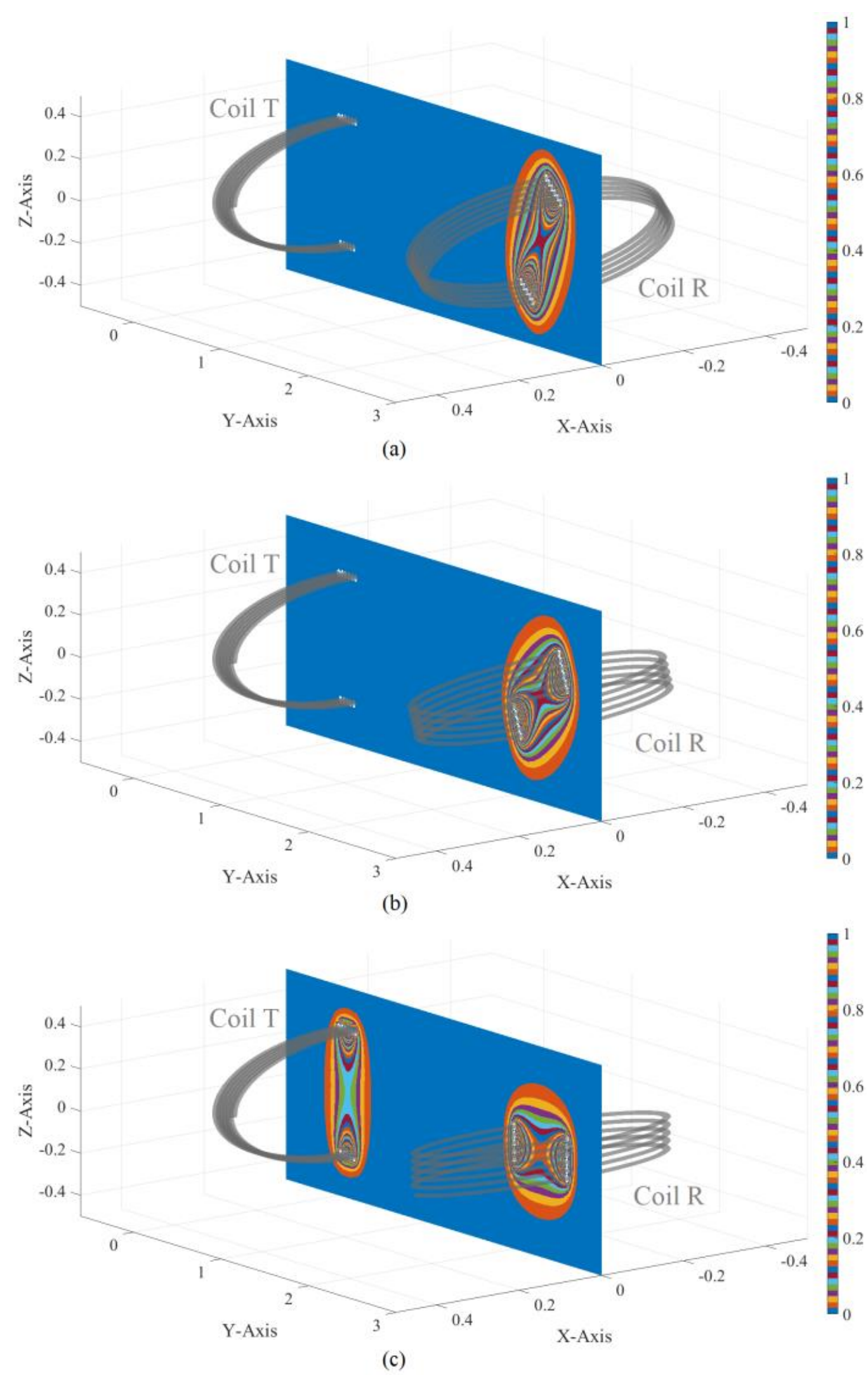

Fig. 20. Time-average magnetic energy density distributions corresponding to (a) $\alpha = 30°$, (b) $\alpha = 60°$, and (c) $\alpha = 90°$.

**C10** When two same coils are placed axially, the co-resonance frequency of the two-coil transferring system is almost independent of the distance between the coils, and the TC of the system is a monotonically decreasing function about the distance.

To visually exhibit the above conclusion, we show the time-average magnetic energy density distributions corresponding to distances 0.5 m, 2.5 m, and 5.0 m in Fig. 18.

### *E. On Transferring Direction*

In this sub-section, we employ the WEP-CMT to study how the relative direction between coil T and coil R influences the transferring efficiency. The angle between the axis of the two coils is denoted as $\alpha$ (in degree). We calculate the CMs of the transferring systems with $\alpha = 0°$, 10°, 20°, …, 90°, and show the corresponding TC curves in Fig. 19.

Evidently, Fig. 19 clearly implies the following expected conclusion

**C11** When two same coils are placed with an angle $\alpha$, the co-resonance frequency of the two-coil transferring system is almost independent of $\alpha$, and the TC of the system is a monotonically decreasing function about $\alpha$.

To visually exhibit the above conclusion, we show the time-average magnetic energy density distributions corresponding to $\alpha = 30°$, 60°, and 90° in Fig. 20.

### *F. On Multi-coil Transferring System*

In the above discussions, only some typical two-coil transferring systems are analyzed. In this sub-section, we further apply the WEP-CMT to some typical multi-coil transferring systems.

Firstly, we consider the three-coil transferring system whose three coils have the same geometrical size and are coaxially placed with mutual distance 2 m. We use WEP-CMT to calculate the CMs of the transferring system, and show the TC curve corresponding to CM 1 in Fig. 21. For the CM 1 working at the optimally transferring frequency 10.8872, its time-average magnetic energy density distribution is shown in Fig. 22.

Secondly, we consider the three-coil transferring system whose three coils have the same geometrical size but are misaligned. We use WEP-CMT to calculate the CMs of the transferring system, and show the TC curve corresponding to CM 1 in Fig. 23. For the CM 1 working at the optimally transferring frequency 10.8724, its time-average magnetic energy density distribution is shown in Fig. 24.

By comparing the Figs. 21&23 with the previous Figs. 4&17&19, it is not difficult to find out that:

**C12** The optimally transferring frequency of three-coil transferring system is usually different from the optimally transferring frequency of classical two-coil transferring system.

The reason leading to this conclusion is that the external-resonance frequency of scattering system R (discussed previously) is different from the external-resonance frequency of R1-R2 scattering system (discussed here). In fact, the conclusion is also applicable to the multi-coil transferring systems constituted by $N\,(\geq 3)$ coils.

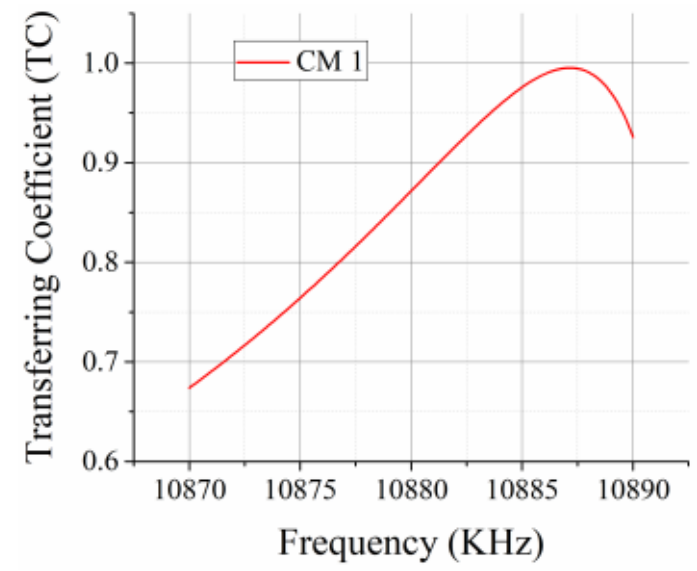

Fig. 21. TC curve associated with the CM 1 of the transferring system constituted by three coaxially placed coils.

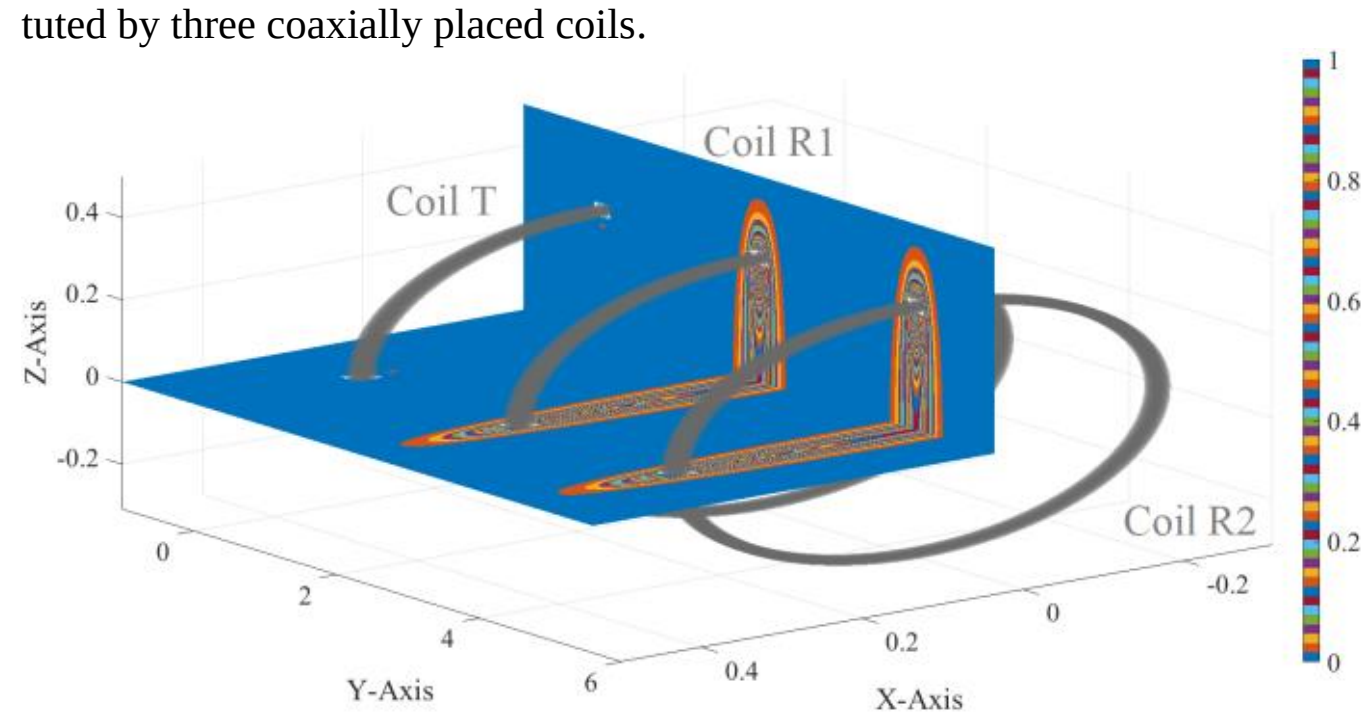

Fig. 22. For the CM 1 working at 10.8872 MHz (shown in Fig. 21), its time-average magnetic energy density distributing on xOy and yOz planes.

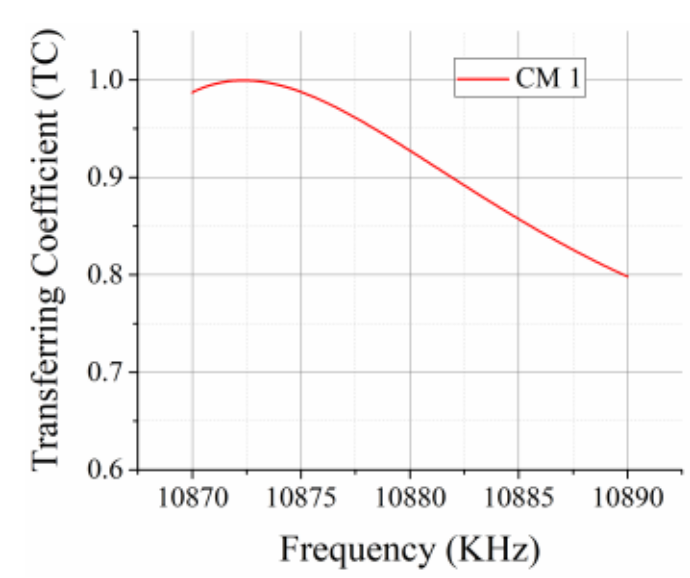

Fig. 23. TC curve associated with the CM 1 of the transferring system constituted by three misaligned coils.

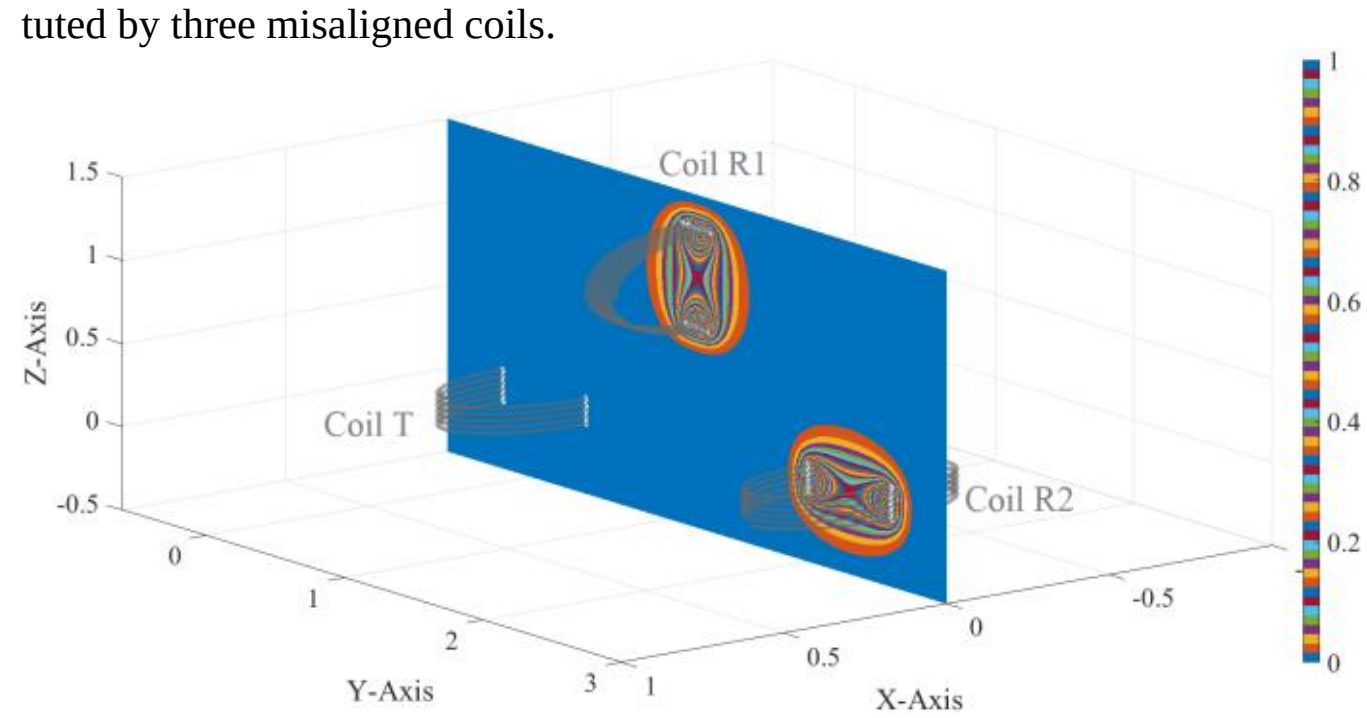

Fig. 24. For the CM 1 working at 10.8724 MHz (shown in Fig. 23), its time-average magnetic energy density distributing on xOy and yOz planes.

### *G. On Surrounding Environment*

In this sub-section, we use WEP-CMT to simply study how surrounding environment influences the WPT process.

We integrate a square thin metallic plane into the two-coil transferring system considered in Sec. IV. We use WEP-CMT to calculate the CMs of the transferring systems with three different planes having side-lengths 0.6 m, 1.2 m, and 2.4 m, and show the TC curves corresponding to the one-order CMs in Fig. 25. We illustrate the time-average magnetic energy density distributions of the optimally transferring modes shown in Fig. 25 as Fig. 26.

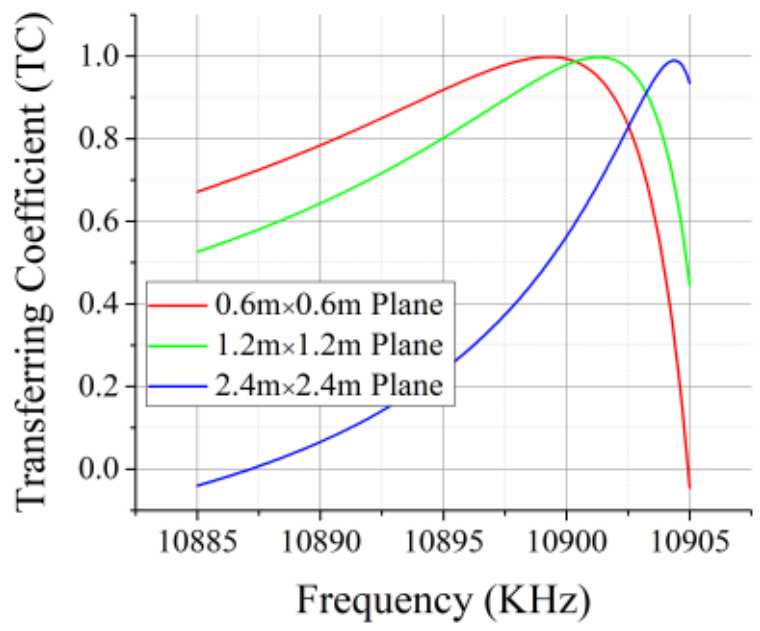

Fig. 25. TC curves of the one-order CMs corresponding to different plane sizes.

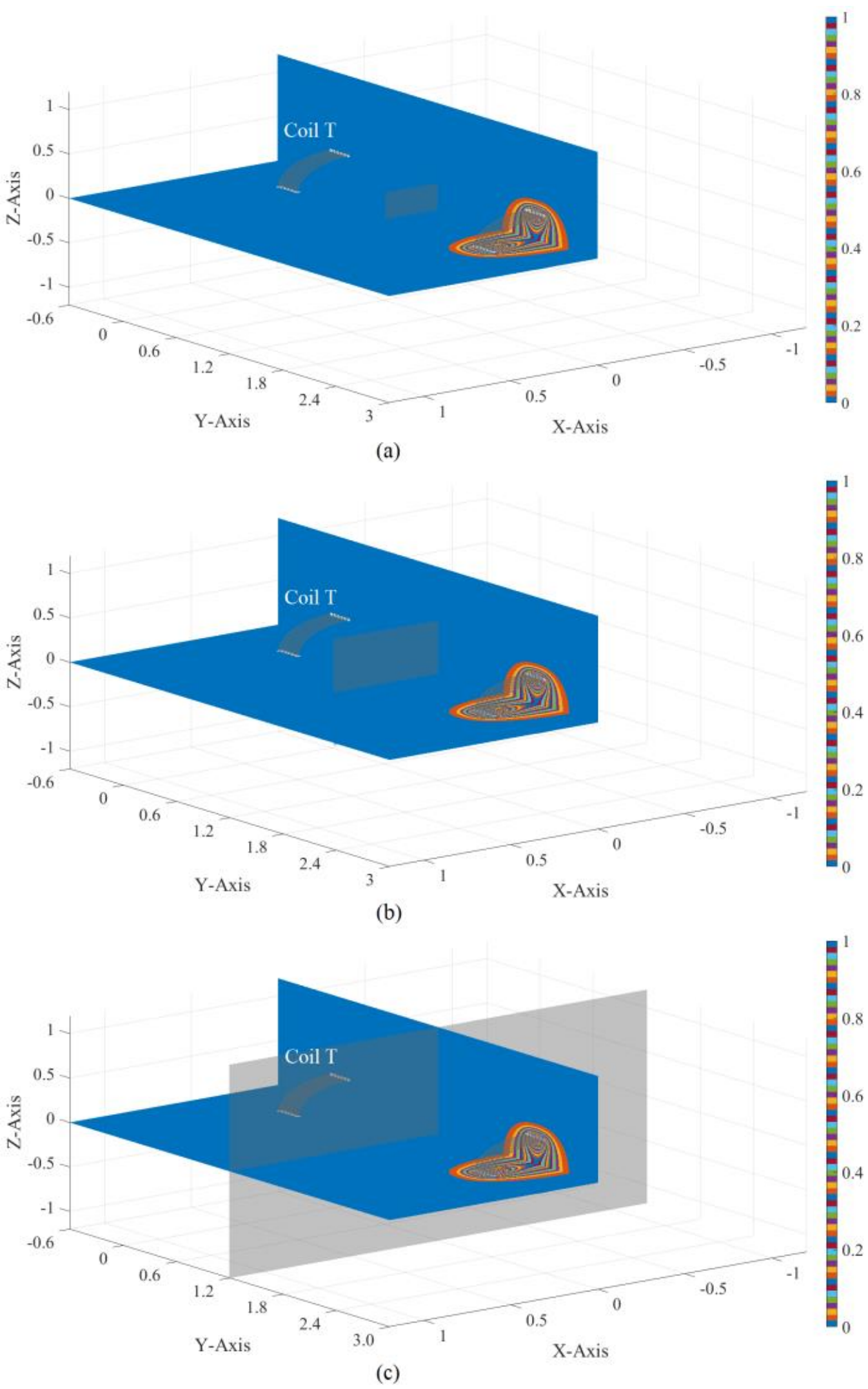

Fig. 26. Time-average magnetic energy density distributions of the optimally transferring modes with (a) 0.6m×0.6m plane, (b) 1.2m×1.2m plane, and (c) 2.4m×2.4m plane.

From the above Figs. 25 and 26, it is not difficult to find out the following conclusion.

**C13** For the above-considered metallic planes placed between the coils, they almost don't hinder the power wirelessly transferred from coil T to coil R, but they indeed influence the optimally transferring frequency, and the altered optimally transferring frequency depends on the geometrical size of the plane.

The reason leading to the conclusion is that the external-resonance frequency of scattering system R is different from the external-resonance frequency of R-Plane scattering system. In fact, the above conclusion can also be generalized to the other kinds of surrounding environments.

## VI. Conclusions

Focusing on the classical two-coil transferring system, which is constituted by a transmitting coil T (driven by locally impressed driver) and a receiving coil R (loaded by perfectly matched load), the WEP governing the WPT process of the transferring system is derived from Maxwell's vector field theory, and doesn't employ any scalar circuit model. The WEP gives the WPT process a very clear physical picture: during the WPT process, the driving power from driver to coil T acts as the source to sustain a steady WPT, and the driving power is finally transformed into two parts — a part is dissipated by coil T and the other part is wirelessly transferred from coil T to coil R. Based on the WEP, a novel concept of TC, which is equal to the ratio of time-average transferred power to time-average driving power, is introduced to quantitatively describe the transferring efficiency of the transferring system. From a relatively mathematical viewpoint, the design process for transferring system is to search for the physically realizable objective modes which can maximize TC, and the objective modes are called optimally transferring modes.

The WEP clearly reveals the fact that: the working mechanisms of *scattering* systems and **transferring** systems are different from each other. The fact clearly exposes the problem that: the conventional CMT for *scattering* systems cannot be directly applied to **transferring** systems. To resolve this problem, this paper, under WEP framework, generalizes the conventional *scattering* CMT to a novel **transferring** CMT. The WEP-based transferring CMT (transferring WEP-CMT) can be employed to search for the optimally transferring modes. Specifically, the transferring WEP-CMT constructs a set of energy-decoupled CMs by orthogonalizing DPO, and the optimally transferring modes can be found in the obtained CM set.

The validity of the transferring WEP-CMT is verified by applying it to the classical two-coil transferring system. After the verification, the WEP-CMT is further applied to the numerical experiments for some typical variants of the classical two-coil transferring system, and then some valuable conclusions are obtained as summarized below.

C1 The larger TC is, the more efficiently "driver drives coil T" and "power is transferred from coil T to coil R" and "coil R converts the transferred power into its self-power".

C2 When TC is maximized, coils T and R are not only self-resonant but also co-resonant. The co-resonant mode works at the optimally transferring state.

C3 Co-resonance is usually called magnetic resonance. The so-called magnetic resonance is a special case of classical electric-magnetic resonance.

C4 Ci-resonance is different from co-resonance. The ci-resonant mode cannot guarantee an efficient WPT from coil T to coil R.

C5 When coils T and R are the same, the **co-resonance** frequency of two-coil **transferring** system equals to the *external-resonance* frequency of one-coil *scattering* system.
C6 When coils T and R are the same, the lower-order co-resonant mode has a larger TC than the higher-order co-resonant mode.
C7 When coils T and R are different, and only one coil is resonant, the system usually cannot efficiently transfer power from coil T to coil R.
C8 When different coil T and coil R work at a same resonance frequency, the system can transfer power wirelessly even if the resonance orders of the coils are different.
C9 For a two-coil transferring system, the same-order co-resonance is usually more desired than the different-order co-resonance in WPT application.
C10 The co-resonance frequency of two same and aligned coils is almost independent of coil distance, and TC is a monotonically decreasing function about the distance.
C11 The co-resonance frequency of two same and misaligned coils is almost independent of the relative angle, and TC is a monotonically decreasing function about the angle.
C12 The optimally transferring frequency of *N*-coil (*N*≥3) system depends on the external-resonance frequency of the scattering system constituted by *N*-1 receiving coils.
C13 The optimally transferring frequency of the system in environment depends on the external-resonance frequency of the Environment-Receiving-coil scattering system.

In fact, the above conclusions also effectively establish some connections “between wireless power **transferring** problem and conventional *scattering* problem”, “between **co-resonance** and conventional *external-resonance*”, and “between so-called **magnetic resonance** and classical *electric-magnetic resonance*”.

## APPENDICES

In this section, some detailed derivations and formulations related to this paper are provided.

### A. Rigorous Derivation for Work-Energy Principle (1) and More Detailed Decomposition for Driving Power

Based on the discussions in Sec. II, the tangential components of electric fields $\boldsymbol{\mathcal{E}}_{\text{driv}} + \boldsymbol{\mathcal{E}}_{\text{t}} + \boldsymbol{\mathcal{E}}_{\text{r}}$ and $\boldsymbol{\mathcal{E}}_{\text{t}} + \boldsymbol{\mathcal{E}}_{\text{r}}$ are zero on $S_{\text{t}}$ and $S_{\text{r}}$ respectively, so

$$\begin{aligned}\overbrace{\langle \boldsymbol{J}_{\text{t}}, \boldsymbol{\mathcal{E}}_{\text{driv}} \rangle_{S_{\text{t}}}}^{\mathcal{P}_{\text{driv}}} &= \langle \boldsymbol{J}_{\text{t}}, -\boldsymbol{\mathcal{E}}_{\text{t}} - \boldsymbol{\mathcal{E}}_{\text{r}} \rangle_{S_{\text{t}}} + \overbrace{\langle \boldsymbol{J}_{\text{r}}, -\boldsymbol{\mathcal{E}}_{\text{t}} - \boldsymbol{\mathcal{E}}_{\text{r}} \rangle_{S_{\text{r}}}}^{0} \\ &= \underbrace{\langle \boldsymbol{J}_{\text{t}}, -\boldsymbol{\mathcal{E}}_{\text{t}} \rangle_{S_{\text{t}}}}_{\mathcal{P}_{\text{sel}}^{\text{tt}}} + \underbrace{\overbrace{\underbrace{\langle \boldsymbol{J}_{\text{t}}, -\boldsymbol{\mathcal{E}}_{\text{r}} \rangle_{S_{\text{t}}} + \langle \boldsymbol{J}_{\text{r}}, -\boldsymbol{\mathcal{E}}_{\text{t}} \rangle_{S_{\text{r}}}}_{\mathcal{P}_{\text{mut}}^{\text{tr}}}}^{\mathcal{P}_{\text{tra}}} + \underbrace{\langle \boldsymbol{J}_{\text{r}}, -\boldsymbol{\mathcal{E}}_{\text{r}} \rangle_{S_{\text{r}}}}_{\mathcal{P}_{\text{sel}}^{\text{rr}}}}_{\mathcal{P}_{\text{tra}}}\end{aligned} \tag{11}$$

The physical meanings of $\mathcal{P}_{\text{sel}}^{\text{tt}}$ and $\mathcal{P}_{\text{tra}}$ had been carefully discussed in Sec. II. The $\mathcal{P}_{\text{mut}}^{\text{tr}}$ and $\mathcal{P}_{\text{sel}}^{\text{rr}}$ can be similarly discussed, and they are respectively called mutual-power between coil T and coil R and self-power of coil R.

### B. Deriving Transformation (6) from Electric Field Integral Equation (5)

Expanding the currents in (5) in terms of some proper basis functions and testing (5) with Galerkin’s method, (5) is discretized as follows:

$$\mathbb{Z}_{\text{rt}} \cdot \mathbb{J}_{\text{t}} + \mathbb{Z}_{\text{rr}} \cdot \mathbb{J}_{\text{r}} = 0 \tag{12}$$

The elements of $\mathbb{Z}_{\text{rt}}$ are $[\mathbb{Z}_{\text{rt}}]_{\xi\zeta} = <\boldsymbol{b}_{\text{r};\xi}, -j\omega\mu_0\mathcal{L}_0(\boldsymbol{b}_{\text{t};\zeta})>_{S_{\text{r}}}$, and the elements of $\mathbb{Z}_{\text{rr}}$ are $[\mathbb{Z}_{\text{rr}}]_{\xi\zeta} = <\boldsymbol{b}_{\text{r};\xi}, -j\omega\mu_0\mathcal{L}_0(\boldsymbol{b}_{\text{r};\zeta})>_{S_{\text{r}}}$. By solving (12), the transformation $\mathbb{J}_{\text{r}} = \mathbb{T} \cdot \mathbb{J}_{\text{t}}$ given in (6) is immediately obtained, where $\mathbb{T} = -\mathbb{Z}_{\text{rr}}^{-1} \cdot \mathbb{Z}_{\text{rt}}$.

### C. Deriving Matrix Quadratic Form (7) from Driving Power Operator (4)

By expanding the currents in (4), the (4) is discretized as follows:

$$P_{\text{driv}} = \begin{bmatrix} \mathbb{J}_{\text{t}} \\ \mathbb{J}_{\text{r}} \end{bmatrix}^{\dagger} \cdot \begin{bmatrix} \mathbb{P}_{\text{tt}} & \mathbb{P}_{\text{tr}} \\ \mathbb{O}_{\text{rt}} & \mathbb{O}_{\text{rr}} \end{bmatrix} \cdot \begin{bmatrix} \mathbb{J}_{\text{t}} \\ \mathbb{J}_{\text{r}} \end{bmatrix} \tag{13}$$

where the sub-matrices $\mathbb{O}_{\text{rt}}$ and $\mathbb{O}_{\text{rr}}$ are zero matrices with proper row number and column number, and the elements of sub-matrix $\mathbb{P}_{\text{tt}}$ are $(1/2) < \boldsymbol{b}_{\text{t};\xi}, j\omega\mu_0\mathcal{L}_0(\boldsymbol{b}_{\text{t};\zeta}) >_{S_{\text{t}}}$, and the elements of sub-matrix $\mathbb{P}_{\text{tr}}$ are $(1/2) < \boldsymbol{b}_{\text{t};\xi}, j\omega\mu_0\mathcal{L}_0(\boldsymbol{b}_{\text{r};\zeta}) >_{S_{\text{t}}}$.

Substituting transformation (6) into (13), matrix quadratic form (7) is immediately obtained, in which

$$\mathbb{P}_{\text{driv}} = \begin{bmatrix} \mathbb{I}_{\text{t}} \\ \mathbb{T} \end{bmatrix}^{\dagger} \cdot \begin{bmatrix} \mathbb{P}_{\text{tt}} & \mathbb{P}_{\text{tr}} \\ \mathbb{O}_{\text{rt}} & \mathbb{O}_{\text{rr}} \end{bmatrix} \cdot \begin{bmatrix} \mathbb{I}_{\text{t}} \\ \mathbb{T} \end{bmatrix} \tag{14}$$

where $\mathbb{I}_{\text{t}}$ is the unit matrix with the same dimension as the row number of $\mathbb{J}_{\text{t}}$.

In addition, the matrix $\mathbb{P}_{\text{tra}}^{+}$ used in (9) is as that $\mathbb{P}_{\text{tra}}^{+} = (\mathbb{P}_{\text{tra}} + \mathbb{P}_{\text{tra}}^{\dagger})/2$, in which

$$\mathbb{P}_{\text{tra}} = \begin{bmatrix} \mathbb{I}_{\text{t}} \\ \mathbb{T} \end{bmatrix}^{\dagger} \cdot \begin{bmatrix} \mathbb{O}_{\text{tt}} & \mathbb{P}_{\text{tr}} \\ \mathbb{O}_{\text{rt}} & \mathbb{O}_{\text{rr}} \end{bmatrix} \cdot \begin{bmatrix} \mathbb{I}_{\text{t}} \\ \mathbb{T} \end{bmatrix} \tag{15}$$

where sub-matrices $\mathbb{P}_{\text{tr}}$, $\mathbb{O}_{\text{rt}}$, and $\mathbb{O}_{\text{rr}}$ are the same as the ones used in (13) and (14), and submatrix $\mathbb{O}_{\text{tt}}$ is a zero matrix with proper row number and column number, and the derivation for (15) is similar to the derivation for (14).

### D. Resistance and Reactance

For the convenience of Sec. IV and using the convention of [50], we define resistances $R_{\text{driv}}$, $R_{\text{sel}}^{\text{tt}}$, $R_{\text{mut}}^{\text{tr}}$, $R_{\text{sel}}^{\text{rr}}$, and $R_{\text{tra}}$ as the powers $\text{Re}\{P_{\text{driv}}\}$, $\text{Re}\{P_{\text{sel}}^{\text{tt}}\}$, $\text{Re}\{P_{\text{mut}}^{\text{tr}}\}$, $\text{Re}\{P_{\text{sel}}^{\text{rr}}\}$, and $\text{Re}\{P_{\text{tra}}\}$ divided by $(1/2) < \boldsymbol{J}_{\text{t}}, \boldsymbol{J}_{\text{t}} >_{S_{\text{t}}}$, and we define reactances $X_{\text{driv}}$, $X_{\text{sel}}^{\text{tt}}$, $X_{\text{mut}}^{\text{tr}}$, $X_{\text{sel}}^{\text{rr}}$, and $X_{\text{tra}}$ as the powers $\text{Im}\{P_{\text{driv}}\}$, $\text{Im}\{P_{\text{sel}}^{\text{tt}}\}$, $\text{Im}\{P_{\text{mut}}^{\text{tr}}\}$, $\text{Im}\{P_{\text{sel}}^{\text{rr}}\}$, and $\text{Im}\{P_{\text{tra}}\}$ divided by $(1/2) < \boldsymbol{J}_{\text{t}}, \boldsymbol{J}_{\text{t}} >_{S_{\text{t}}}$.

**Ren-Zun Lian** received the B.S. degree in optical engineering from the University of Electronic Science and Technology of China (UESTC), Chengdu, China, in 2011, and received the Ph.D. degree in electromagnetic field and microwave technology from UESTC, Chengdu, China, in 2019.

He is currently a postdoctoral researcher in Peking University (PKU), Beijing, China. His current research interests include mathematical physics, electromagnetic theory and computation, and antenna theory and design.

**Ming-Yao Xia** (M’00-SM’03) received the Master and Ph. D degrees in electrical engineering from the Institute of Electronics, Chinese Academy of Sciences (IECAS), in 1988 and 1999, respectively. From 1988 to 2002, he was with IECAS as an Engineer and a Senior Engineer. He was a Visiting Scholar at the University of Oxford, U.K., from October 1995 to October 1996. From June 1999 to August 2000 and from January 2002 to June 2002, he was a Senior Research Assistant and a Research Fellow, respectively, with the City University of Hong Kong. He joined Peking University as an Associate Professor in 2002 and was promoted to Full Professor in 2004. He moved to the University of Electronic Science and Technology of China as a Chang-Jiang Professor nominated by the Ministry of Education of China in 2010. He returned to Peking University after finishing the appointment in 2013. He was a recipient of the Young Scientist Award of the URSI in 1993. He was awarded the first-class prize on Natural Science by the Chinese Academy of Sciences in 2001. He was the recipient of the Foundation for Outstanding Young Investigators presented by the National Natural Science Foundation of China in 2008. He served as an Associate Editor for the IEEE Transactions on Antennas and Propagation. His research interests include electromagnetic theory, numerical methods and applications, such as wave propagation and scattering, electromagnetic imaging and probing, microwave remote sensing, antennas and microwave components.

**Xing-Yue Guo** (S’17) received the B.S. degree in electronic information science and technology from the Southwest Jiaotong University, Chengdu, China, in 2011, and the M.E. degree from the University of Electronic Science and Technology of China, Chengdu, China, in 2014. From 2014 to 2017, she was a Research Assistant with the CAEP Software Center for High Performance Numerical Simulation, Beijing, China. Since 2017, she has been pursuing the Ph. D. degree at the School of Electronics Engineering and Computer Science, Peking University, Beijing, China. Her research interests include computational electromagnetics and applications.

# Work-Energy Principle Based Characteristic Mode Analysis for Yagi-Uda Arrays

Ren-Zun Lian, Ming-Yao Xia, *Senior Member, IEEE*, and Xing-Yue Guo, *Student Member, IEEE*

***Abstract*—The difference between the working mechanism of scattering systems and the working mechanism of Yagi-Uda arrays is exposed. The similarity between the working mechanism of metallic wireless power transfer systems (simply called transferring systems) and the working mechanism of metallic Yagi-Uda arrays is revealed. The work-energy principle (WEP) based characteristic mode theory for metallic transferring systems is used to do the modal analysis for metallic Yagi-Uda arrays for the first time. The WEP-based characteristic mode analysis (CMA) for metallic Yagi-Uda arrays is further generalized to material Yagi-Uda arrays and metal-material composite Yagi-Uda arrays, and the WEP-based CMA (WEP-CMA) formulations for material and composite Yagi-Uda arrays are established for the first time. The validity of the WEP-CMA for various Yagi-Uda arrays is verified by comparing the WEP-CMA-based numerical results with some published simulation and measurement data.**

## I. Introduction

YAGI-UDA antenna array was first studied by Uda and Yagi in the early 1920s, and publicly reported in the middle 1920s [1]-[2]. In 1984, the *Proceedings of the IEEE* reprinted several classical articles for celebrating the centennial year of IEEE (1884-1984), and Yagi's article [2] became the only reprinted one in the realm of electromagnetic (EM) antenna. This fact clearly illustrates the great significance of Yagi-Uda array in antennas and propagation society. Some histories about Yagi-Uda array can be found in [3]-[5].

A classical metallic Yagi-Uda array is shown in Fig. 1, and it is constituted by a row of discrete metallic linear elements, one of which is driven by an external source while the others act as parasitic radiators whose currents are induced by near-field mutual coupling [6]-[9]. The linear metallic Yagi-Uda array is a typical discrete-element travelling-wave-type end-fire antenna, which usually works at HF (3-30 MHz), VHF (30-300 MHz) and UHF (300-3000 MHz) etc. bands [6]-[9].

Besides the most classical linear metallic Yagi-Uda array shown in Fig. 1, there also exist many different variants (sometimes called quasi Yagi-Uda arrays). The quasi Yagi-Uda arrays have been widely applied in the applications of frequency-modulated broadcast [10], domestic/mobile television signal transmission [11]-[13], point-to-point communication [14]-[18], long-distance communication [19]-[23], mobile communication [24]-[28], wireless local area network [29]-[33] and radio frequency identification [34]-[37] etc. due to their typical features of high radiation efficiency, high gain and directivity / narrow beamwidth, high front-to-back ratio, low level of minor lobes, good cross-polar discrimination, reasonable bandwidth, low cost and ease of fabrication etc. [6]-[9].

According to the difference of their constituent components, the various Yagi-Uda arrays can be categorized into three classes — metallic Yagi-Uda arrays [38]-[41], material Yagi-Uda arrays [42]-[43], and metal-material composite Yagi-Uda arrays [11]-[37]. A typical 6-element metallic Yagi-Uda array discussed in [41] is shown in Fig. 1, and it has a dominant resonant mode working at 300 MHz (calculated from the formulation proposed in [41]), and the resonant mode has the far-field radiation pattern shown in Fig. 2, which is end-fire.

The analysis and design for resonant modes are the important topics in the realm of antenna engineering, and there have been some classical modal analysis and design theories, such as cavity model theory [44]-[45], dielectric waveguide model theory [46]-[47], eigen-mode theory [48]-[49], and characteristic mode theory (CMT) [50]-[56] etc. Among the various theories, the CMT has been attached great importance in the realm of antenna engineering [54] recently, because the CMT not only has a very wide applicable range but also is very easy for numerical realization. But unfortunately, the conventional CMT cannot be directly applied to doing the modal analysis for Yagi-Uda arrays as exhibited below. By directly applying the conventional CMT [52]-[55] to the Yagi-Uda array shown in Fig. 1, the modal significances (MSs) associated to the obtained characteristic modes (CMs) are shown in Fig. 3, and the radiation patterns of the resonant CMs are illustrated in Fig. 4. Evidently, both the resonance frequencies and radiation patterns are incorrect.

The reasons leading to the above incorrect results mainly originate from the following two.

R1. The working mechanisms of scattering system and antenna system are different from each other. Scattering system is an object which is under the illumination of an externally incident field and generates a secondary scattered field. Antenna system is "*a device used for trans-*

[AP2101-0036] received January 6, 2021. This work was supported by XXXX under Project XXXXXXXX. (*Corresponding authors: Ren-Zun Lian; Ming-Yao Xia.*)

R. Z. Lian, M. Y. Xia and X. Y. Guo are with the Department of Electronics, School of Electronics Engineering and Computer Science, Peking University, Beijing 100871, China. (E-mails: rzlian@vip.163.com; myxia@pku.edu.cn).

*mitting electromagnetic signals or power*” [57].

R2. The conventional CMT is a modal analysis method for scattering systems [55]-[56] rather than for antenna systems.

Recently, under working-energy principle (WEP) framework [56] established an alternative CMT which is for metallic wireless power transfer systems (simply called transferring systems) instead of for scattering systems. In this paper, it will be exhibited that the metallic Yagi-Uda arrays have a very similar working mechanism to the metallic transferring systems discussed in [56]. Then, this paper successfully applies the WEP-based CMT to doing the modal analysis for metallic Yagi-Uda arrays. In addition, this paper will further generalize the WEP-based characteristic mode analysis (CMA) method from metallic Yagi-Uda arrays to material Yagi-Uda arrays and then to metal-material composite Yagi-Uda arrays.

This paper is organized as follows: Sec. II discusses the WEP-based CMA (WEP-CMA) for metallic Yagi-Uda arrays; Sec. III generalizes the WEP-CMA to material Yagi-Uda arrays; Sec. IV further generalizes the WEP-CMA to composite Yagi-Uda arrays; Sec. V concludes this paper; some detailed formulations related to this paper are provided in the appendices. The symbolic system of this paper is similar to [56], and will not be explained here. In addition, to distinguish surface electric current and volume electric current, the former is denoted as upper case letter $\boldsymbol{J}$, and the latter is denoted as lower case letter $\boldsymbol{j}$.

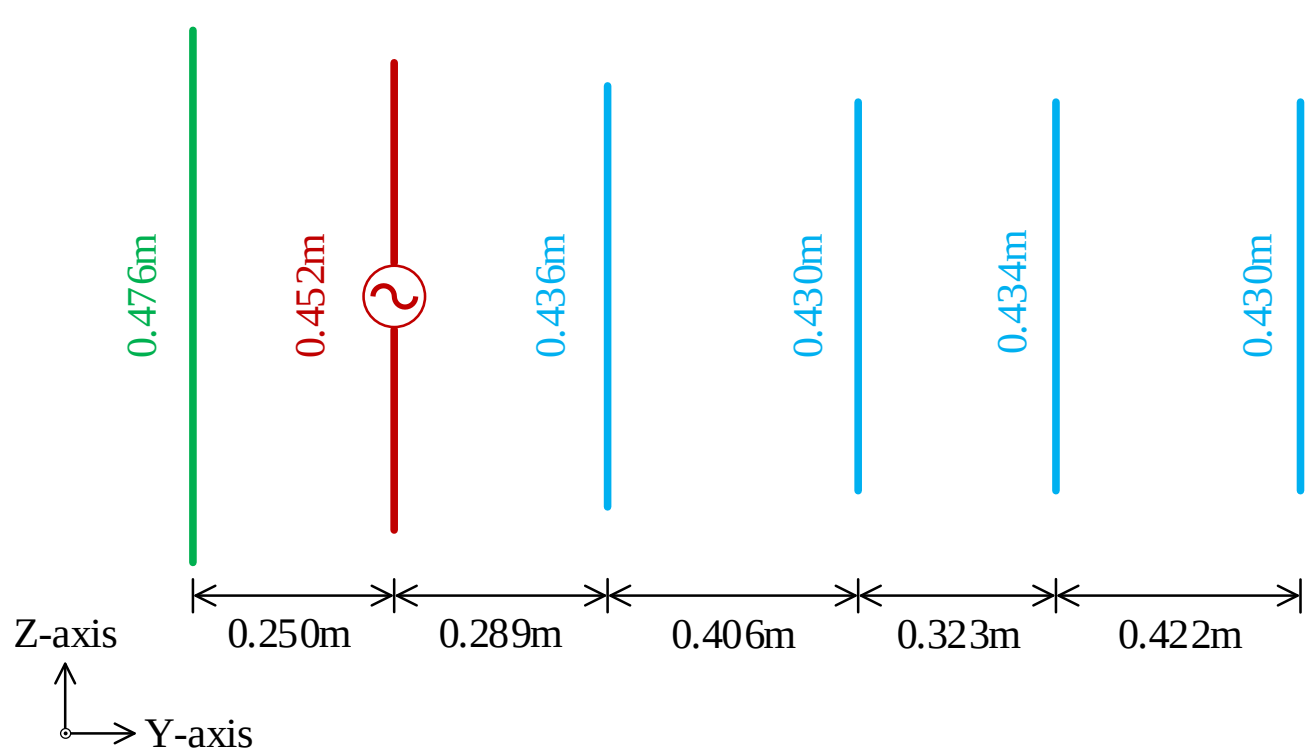

Fig. 1. Geometry and size of a typical 6-element linear metallic Yagi-Uda array reported in [41].

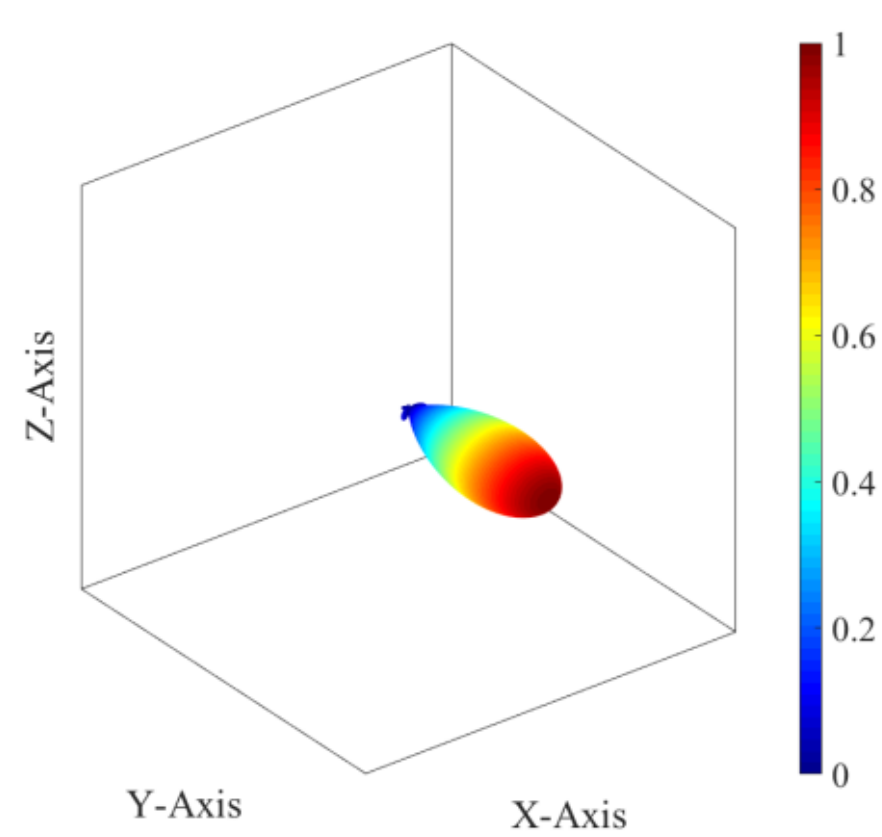

Fig. 2. Radiation pattern of the dominant resonant mode of the Yagi-Uda array shown in Fig. 1.

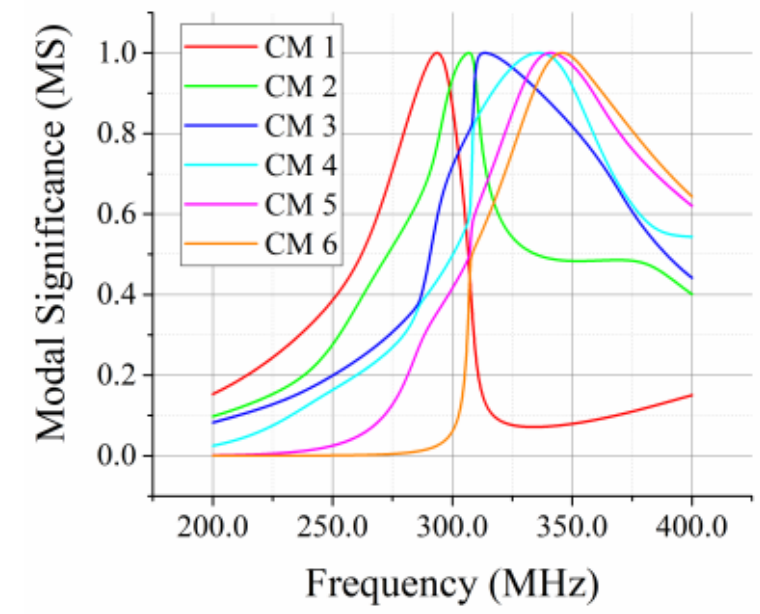

Fig. 3. MSs associated to the first six lower-order CMs calculated from the conventional CMT established by Harrington *et al.* in [52] and [53].

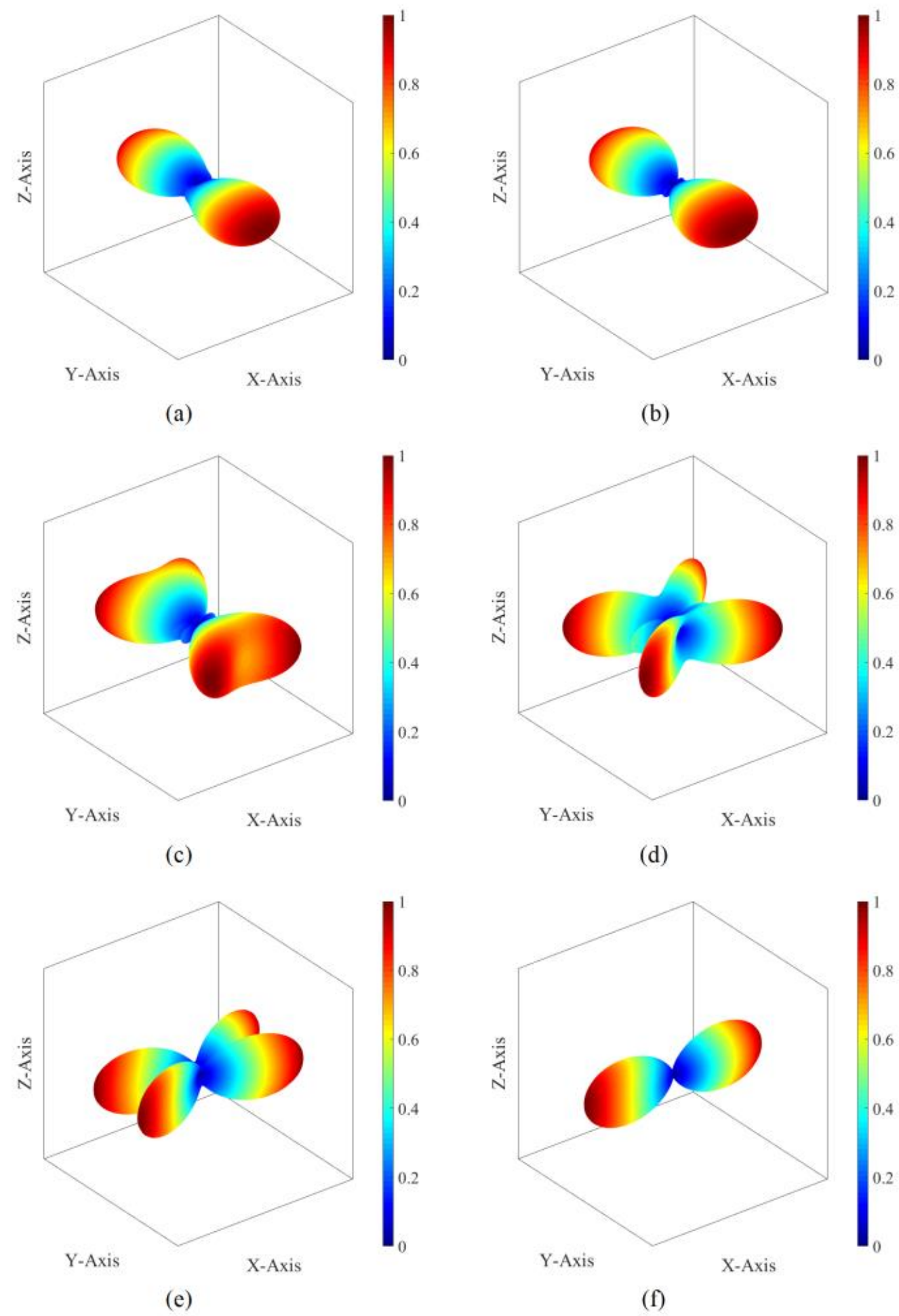

Fig. 4. Radiation patterns of the resonant (a) CM 1 working at 293.6 MHz, (b) CM 2 working at 307.0 MHz, (c) CM3 working at 313.6 MHz, (d) CM4 working at 336.2 MHz, (e) CM5 working at 340.9 MHz, and (f) CM6 working at 345.9 MHz.

## II. WEP-CMA FOR METALLIC YAGI-UDA ARRAYS

Taking the one shown in Fig. 1 as a typical example, this section proposes a WEP-CMA for metallic Yagi-Uda arrays.

### A. *Working Mechanism of Metallic Yagi-Uda Arrays*

Traditionally, all the elements in a Yagi-Uda array are divided into three groups — feeding element, reflecting element, and directing elements [6]-[9]. In fact, all the elements can also be alternatively divided into two groups — active element and passive/parasitic elements [19], where the former is just the feeding element and the latter is the union of reflecting element and directing elements as shown in Fig. 5.

When a localized driving field acts on the active element, an electric current will be induced on the active element. The field generated by the induced electric current on active element acts on the passive elements, and then induces some electric currents on the passive elements. In addition, the fields generated by the induced electric currents on passive elements also react on the active element.

For the time-harmonic EM problem, the above-mentioned action and reaction will reach a dynamic equilibrium finally. At the state of dynamic equilibrium, the driving field is denoted as $\boldsymbol{F}_{\rm driv}$, and the induced electric currents on the active and passive elements are denoted as $\boldsymbol{J}_{\rm a}$ and $\boldsymbol{J}_{\rm p}$ respectively, and the fields generated by $\boldsymbol{J}_{\rm a}$ and $\boldsymbol{J}_{\rm p}$ are denoted as $\boldsymbol{F}_{\rm a}$ and $\boldsymbol{F}_{\rm p}$ respectively.

Obviously, the above-mentioned working mechanism of metallic Yagi-Uda arrays is almost identical to the working mechanism of the metallic transferring systems discussed in [56]. In the following Secs. II-B and II-C, we will employ the WEP-based CMT established in [56] to do the modal analysis for metallic Yagi-Uda arrays.

### *B. Mathematical Formulations Related to the WEP-CMA for Metallic Yagi-Uda Arrays*

The driving power $P_{\rm driv} = (1/2) < \boldsymbol{J}_{\rm a}, \boldsymbol{E}_{\rm driv} >_{S_{\rm a}}$ used to sustain a steady working of the metallic Yagi-Uda array has the following operator expression

$$P_{\rm driv} = (1/2)\left\langle \boldsymbol{J}_{\rm a} \, , \, j\omega\mu_0\mathcal{L}_0\left(\boldsymbol{J}_{\rm a} + \boldsymbol{J}_{\rm p}\right)\right\rangle_{S_{\rm a}} \tag{1}$$

in which integral domain $S_{\rm a}$ is the boundary surface of the active element, and integral operator $\mathcal{L}_0$ is defined as that $\mathcal{L}_0(\boldsymbol{X}) = [1 + (1/k_0^2)\nabla\nabla\cdot]\int_\Omega G_0(\boldsymbol{r},\boldsymbol{r}')\boldsymbol{X}(\boldsymbol{r}')d\Omega'$ where the scalar Green's function is as that $G_0(\boldsymbol{r},\boldsymbol{r}') = e^{-jk_0|\boldsymbol{r}-\boldsymbol{r}'|} / 4\pi\,|\,\boldsymbol{r}-\boldsymbol{r}'\,|$. In addition, the above integral form (1) of driving power operator (DPO) can be easily discretized into the following matrix form

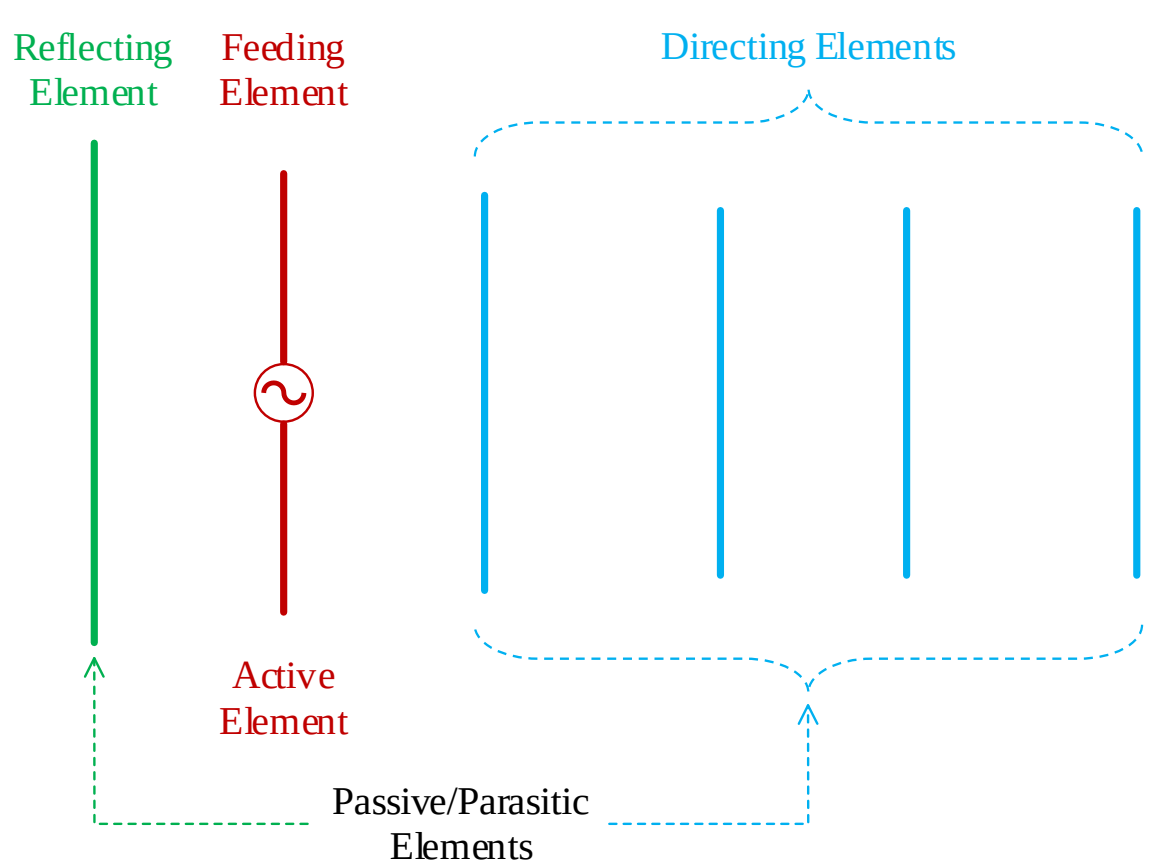

Fig. 5. A re-grouping for the elements used to constitute the Yagi-Uda array shown in Fig. 1.

$$P_{\rm driv} = \mathrm{J}_{\rm a}^\dagger \cdot \begin{bmatrix} \mathrm{P}_{\rm aa} & \mathrm{P}_{\rm ap} \end{bmatrix} \cdot \begin{bmatrix} \mathrm{J}_{\rm a} \\ \mathrm{J}_{\rm p} \end{bmatrix} \tag{2}$$

where $\mathrm{J}_{\rm a}$ and $\mathrm{J}_{\rm p}$ are the basis function expansion vectors of $\boldsymbol{J}_{\rm a}$ and $\boldsymbol{J}_{\rm p}$ respectively, and the formulations for calculating the sub-matrices are given in App. A.

In fact, the $\mathrm{J}_{\rm a}$ and $\mathrm{J}_{\rm p}$ are not independent, and they satisfy the following transformation relation

$$\mathrm{J}_{\rm p} = \mathrm{T} \cdot \mathrm{J}_{\rm a} \tag{3}$$

which originates from the homogeneous tangential electric field boundary condition on $S_{\rm p}$. The formulations for calculating $\mathrm{T}$ are given in App. A. Substituting (3) into (2), it is easy to derive the following

$$P_{\rm driv} = \mathrm{J}_{\rm a}^\dagger \cdot \mathrm{P}_{\rm driv} \cdot \mathrm{J}_{\rm a} \tag{4}$$

with independent variable $\mathrm{J}_{\rm a}$ only, where the formulations for calculating $\mathrm{P}_{\rm driv}$ are given in App. A.

The energy-decoupled CMs of the metallic Yagi-Uda array can be derived from solving the following characteristic equation

$$\mathrm{P}_{\rm driv}^- \cdot \mathrm{J}_{\rm a} = \lambda\, \mathrm{P}_{\rm driv}^+ \cdot \mathrm{J}_{\rm a} \tag{5}$$

where $\mathrm{P}_{\rm driv}^+ = [\mathrm{P}_{\rm driv} + \mathrm{P}_{\rm driv}^\dagger]/2$ and $\mathrm{P}_{\rm driv}^- = [\mathrm{P}_{\rm driv} - \mathrm{P}_{\rm driv}^\dagger]/2j$. The above-obtained CMs satisfy frequency-domain power-decoupling relation $(1/2) < \boldsymbol{J}_{{\rm a};\xi}, \boldsymbol{E}_{{\rm driv};\zeta} >_{S_{\rm a}} = (1 + j\lambda)\delta_{\xi\zeta}$ where $\delta_{\xi\zeta}$ is Kronecker's delta symbol, and then satisfy the following energy-decoupling relation, i.e., time-average power-decoupling relation,

$$(1/T)\int_{t_0}^{t_0+T} \left\langle \mathcal{J}_{{\rm a};\xi}, \mathcal{E}_{{\rm driv};\zeta} \right\rangle dt = \delta_{\xi\zeta} \tag{6}$$

where the real parts of all modal complex powers have been normalized to 1 just like [53] did. Thus, there exists the following Parseval's identity

$$(1/T)\int_{t_0}^{t_0+T} \left\langle \mathcal{J}_{\rm a}, \mathcal{E}_{\rm driv} \right\rangle dt = \sum\nolimits_\xi |c_\xi|^2 \tag{7}$$

in which $c_\xi$ is the CM-based modal expansion coefficient, and $c_\xi = (1/2) < \boldsymbol{J}_{{\rm a};\xi}, \boldsymbol{E}_{\rm driv} >_{S_{\rm a}} / (1 + j\lambda)$ where $\boldsymbol{E}_{\rm driv}$ is the frequency-domain version of a previously known time-domain driving field $\mathcal{E}_{\rm driv}$. The above (6) and (7) have a very clear physical meaning: the CMs constructed above don't have net energy exchange in any integral period.

### *C. Numerical Verifications*

In this sub-section, the WEP-CMA proposed above is applied to a specific metallic Yagi-Uda array to verify its validity. The geometry and size of the Yagi-Uda array are shown in Fig. 1. The MSs associated to the first four lower-order CMs calculated from the WEP-CMA are shown in Fig. 6.

From Fig. 6, it is easy to find out that the CM 1 is resonant at 307.3 MHz, and the resonance frequency is consistent with the one calculated from the formulation given in [41] except a 2% numerical error. The radiation pattern of the resonant CM 1 is shown in Fig. 2. The distributions of the electric and magnetic fields of the resonant CM 1 are shown in Fig. 7. Evidently, Fig. 2 and Fig. 7 satisfy the well-known features of linear metallic Yagi-Uda arrays — the radiated EM power propagates along the direction from reflecting element to directing elements [6]-[9]. In addition, the time-domain dynamic figures of the propagating electric and magnetic fields are also uploaded to IEEE manuscript system together with this paper.

In addition, the radiation patterns of the higher-order resonant CM 2 (working at 649.8 MHz) and CM 3 (working at 971.8 MHz) are shown in Fig. 8. Evidently, both the resonant CM 2 and CM 3 don't work at the desired end-fire state. In fact, this is just the reason why "*higher resonances are available near lengths of* $\lambda$, $3\lambda/2$, *and so forth, but are seldom used*" [9].

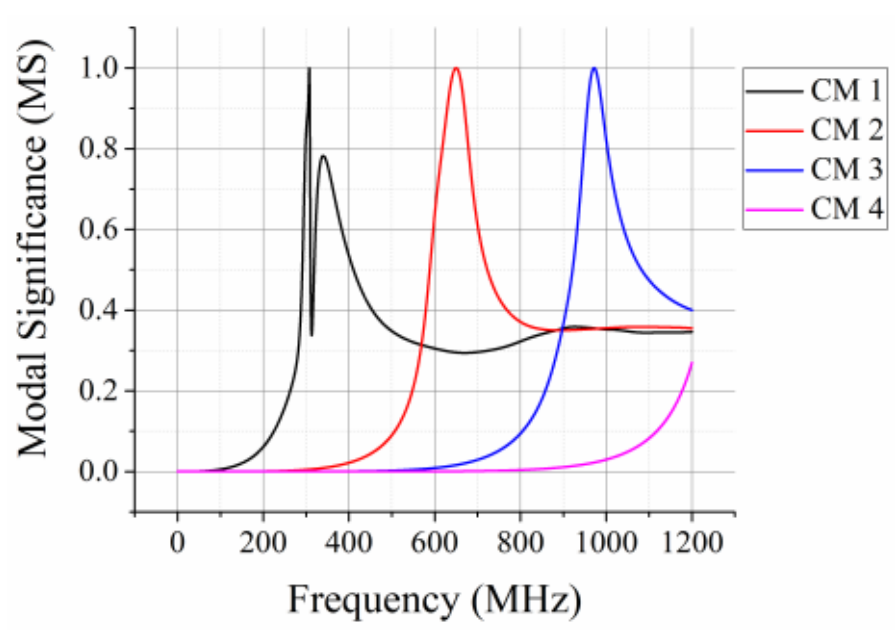

Fig. 6. MSs associated to the first four lower-order CMs calculated from the WEP-CMA proposed in this section.

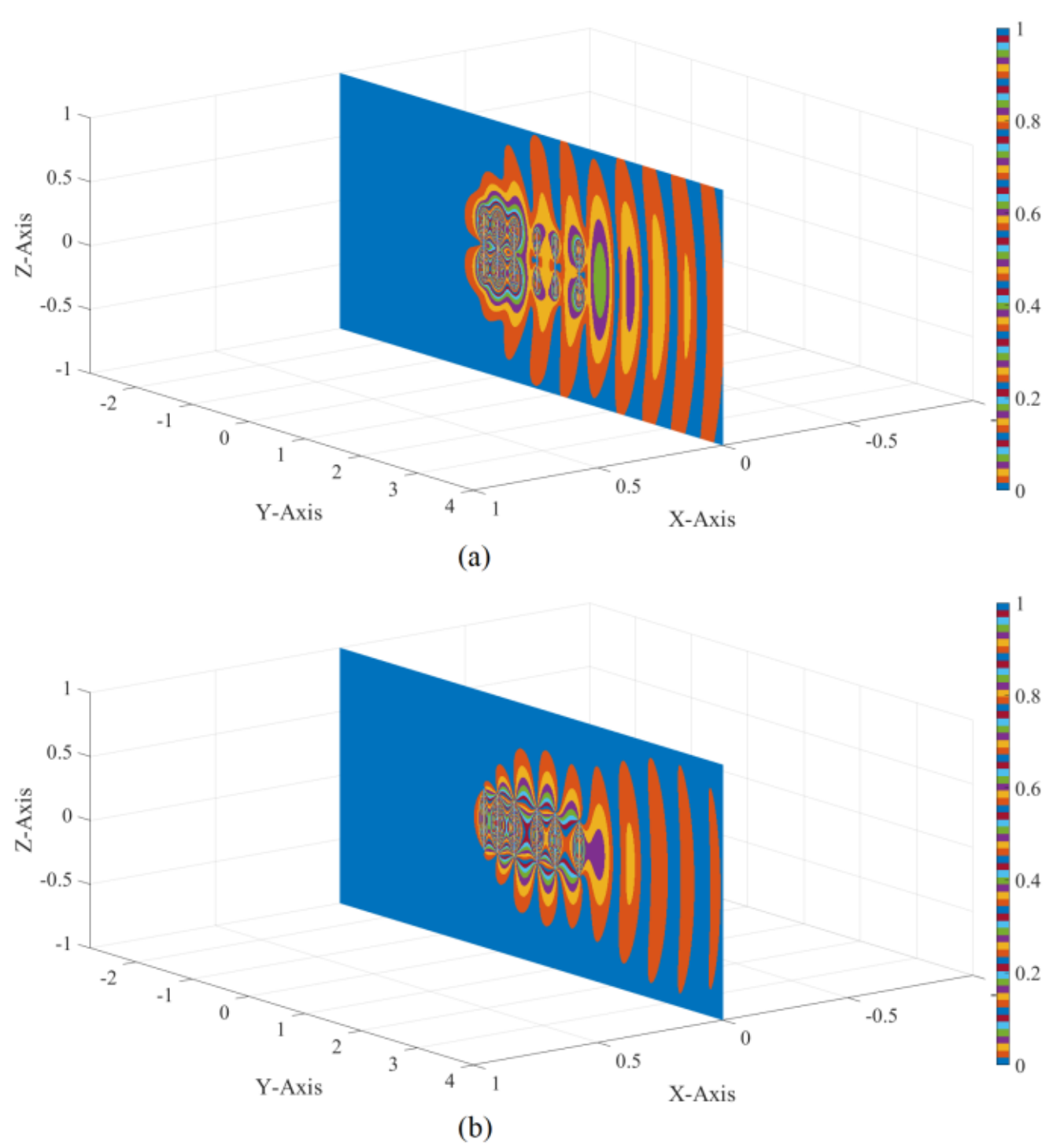

Fig. 7. Distributions of the (a) electric field and (b) magnetic field of the resonant CM 1 shown in Fig. 6.

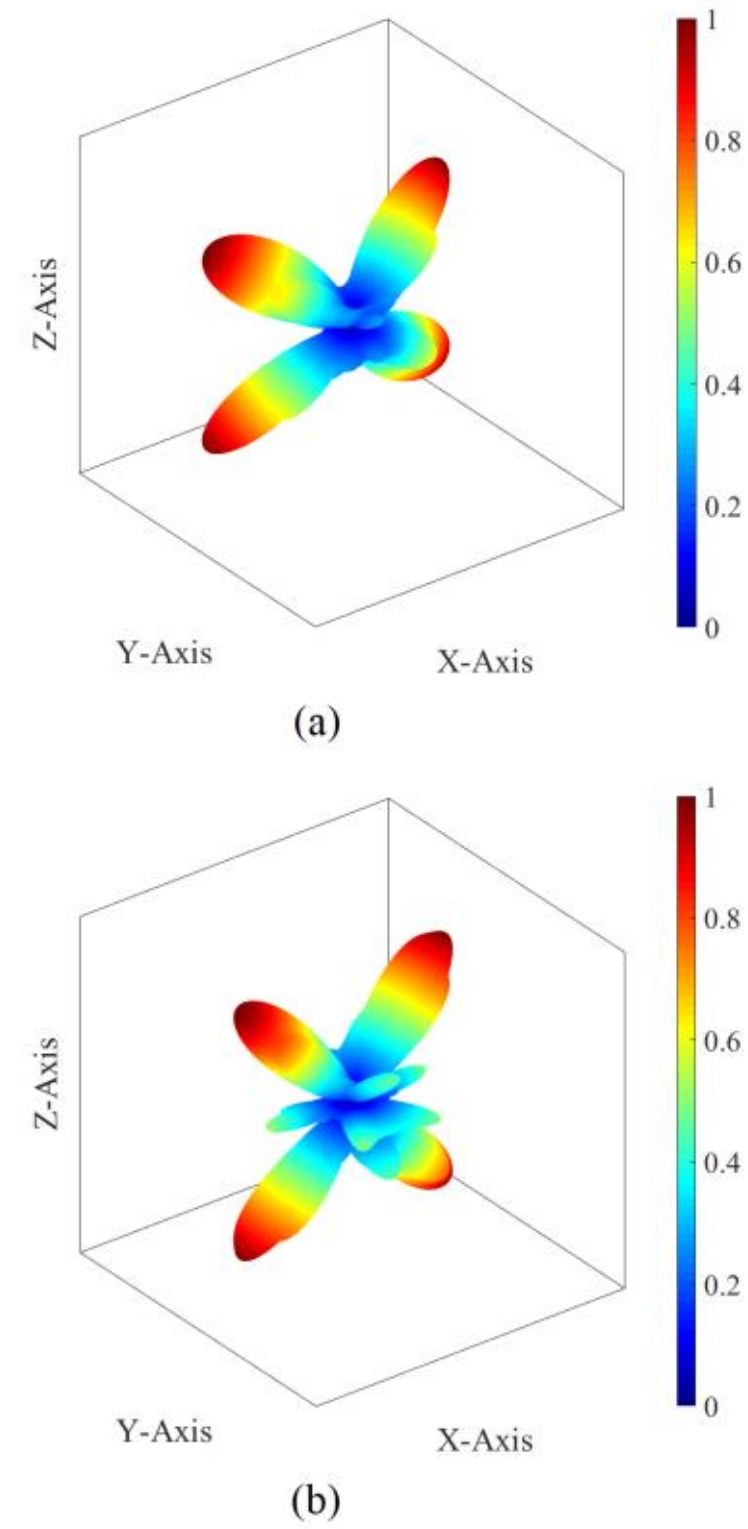

Fig. 8. Radiation patterns of the resonant (a) CM 2 working at 649.8 MHz and (b) CM 3 working at 971.8 MHz.

## III. WEP-CMA for Material Yagi-Uda Arrays

This section is devoted to generalizing the idea of Sec. II (which is for metallic Yagi-Uda arrays) to material Yagi-Uda arrays. A typical three-element material Yagi-Uda array reported in [42] is shown in Fig. 9, and it has an active element $V_a$ with boundary surface $S_a$ and two passive elements $V_{p1}$ and $V_{p2}$ with boundary surfaces $S_{p1}$ and $S_{p2}$ respectively. For simplifying the following discussions, the elements are restricted to being non-magnetic in this section, and their complex permittivities are denoted as $\boldsymbol{\varepsilon}_a^c$, $\boldsymbol{\varepsilon}_{p1}^c$, and $\boldsymbol{\varepsilon}_{p2}^c$. The purely magnetic case and magneto-dielectric case can be similarly discussed.

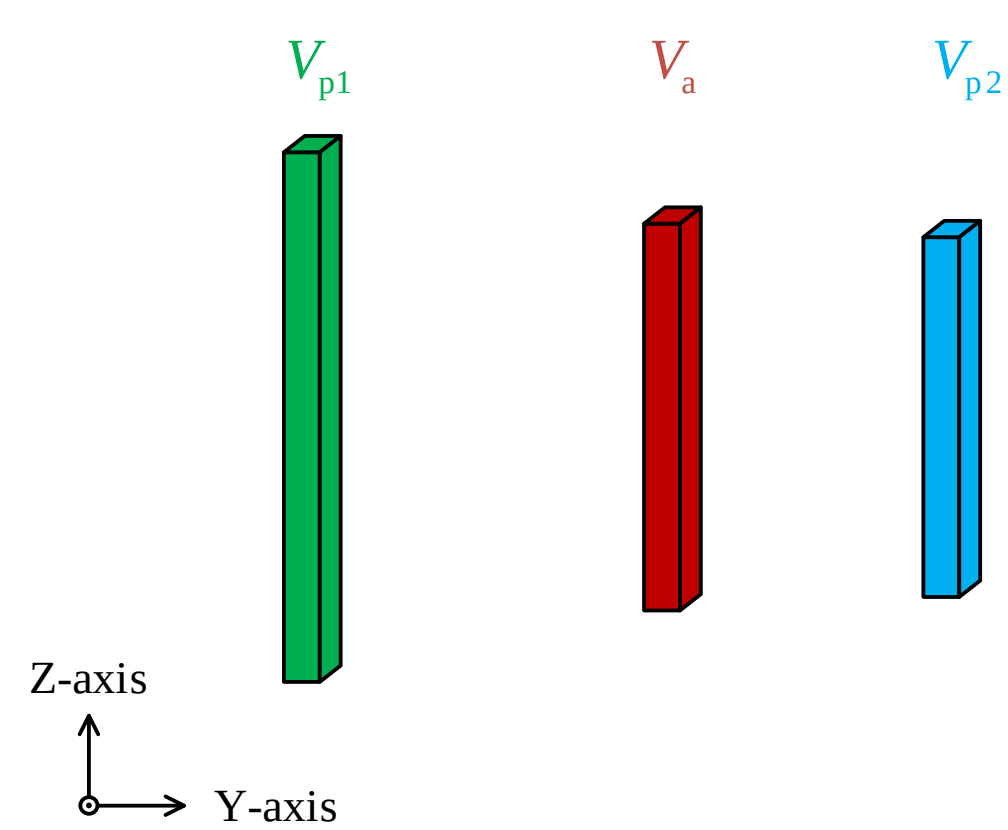

Fig. 9. Geometry of a typical 3-element linear material Yagi-Uda array reported in [42].

### A. Volume Formulation of the WEP-CMA for Material Yagi-Uda Arrays

If the induced volume electric currents distributing on $V_{\text{a}}$, $V_{\text{p1}}$, and $V_{\text{p2}}$ are denoted as $\boldsymbol{j}_{\text{a}}$, $\boldsymbol{j}_{\text{p1}}$, and $\boldsymbol{j}_{\text{p2}}$ respectively, then the corresponding DPO is as follows:

$$P_{\text{driv}} = \frac{1}{2}\left\langle \boldsymbol{j}_{\text{a}}, \left(j\omega\Delta\boldsymbol{\varepsilon}_{\text{a}}^{\text{c}}\right)^{-1}\cdot\boldsymbol{j}_{\text{a}} + j\omega\mu_0\mathcal{L}_0\left(\boldsymbol{j}_{\text{a}}+\boldsymbol{j}_{\text{p1}}+\boldsymbol{j}_{\text{p2}}\right)\right\rangle_{V_{\text{a}}} \quad (8)$$

where $\Delta\boldsymbol{\varepsilon}_{\text{a}}^{\text{c}} = \boldsymbol{\varepsilon}_{\text{a}}^{\text{c}} - \mathbf{I}\varepsilon_0$ and $\mathbf{I}$ is the unit dyad.

Similar to deriving (4) from (1), the following

$$P_{\text{driv}} = \mathrm{j}_{\text{a}}^{\dagger}\cdot\mathrm{P}_{\text{driv}}\cdot\mathrm{j}_{\text{a}} \quad (9)$$

with only independent variable $\mathrm{j}_{\text{a}}$ can be derived from (8), and a detailed derivation process is given in App. B. Here, $\mathrm{j}_{\text{a}}$ is the basis function expansion vector of $\boldsymbol{j}_{\text{a}}$.

Employing the $\mathrm{P}_{\text{driv}}$ used in (9), the CMs of the material Yagi-Uda array can be calculated like Sec. II-B.

The scheme provided in this sub-section is based on volume currents, and an alternative surface-current-based scheme will be given in the following sub-section.

### B. Surface Formulation of the WEP-CMA for Material Yagi-Uda Arrays

If the equivalent surface currents distributing on $S_{\text{a}}$, $S_{\text{p1}}$, and $S_{\text{p2}}$ are denoted as $(\boldsymbol{J}_{\text{a}},\boldsymbol{M}_{\text{a}})$, $(\boldsymbol{J}_{\text{p1}},\boldsymbol{M}_{\text{p1}})$, and $(\boldsymbol{J}_{\text{p2}},\boldsymbol{M}_{\text{p2}})$ respectively, then the volume-current version (8) of DPO can be alternatively written as the following surface-current version

$$\begin{aligned} P_{\text{driv}} = &-(1/2)\left\langle \boldsymbol{J}_{\text{a}}, \mathcal{E}_0\left(\boldsymbol{J}_{\text{a}}+\boldsymbol{J}_{\text{p1}}+\boldsymbol{J}_{\text{p2}}, \boldsymbol{M}_{\text{a}}+\boldsymbol{M}_{\text{p1}}+\boldsymbol{M}_{\text{p2}}\right)\right\rangle_{S_{\text{a}}^-} \\ &-(1/2)\left\langle \boldsymbol{M}_{\text{a}}, \mathcal{H}_0\left(\boldsymbol{J}_{\text{a}}+\boldsymbol{J}_{\text{p1}}+\boldsymbol{J}_{\text{p2}}, \boldsymbol{M}_{\text{a}}+\boldsymbol{M}_{\text{p1}}+\boldsymbol{M}_{\text{p2}}\right)\right\rangle_{S_{\text{a}}^-} \end{aligned} \quad (10)$$

in which operator $\mathcal{E}_0$ is as $\mathcal{E}_0(\boldsymbol{J},\boldsymbol{M}) = -j\omega\mu_0\mathcal{L}_0(\boldsymbol{J}) - \mathcal{K}_0(\boldsymbol{M})$ and operator $\mathcal{H}_0$ is as $\mathcal{H}_0(\boldsymbol{J},\boldsymbol{M}) = \mathcal{K}_0(\boldsymbol{J}) - j\omega\varepsilon_0\mathcal{L}_0(\boldsymbol{M})$, where operator $\mathcal{L}_0$ is the same as the one used previously and operator $\mathcal{K}_0$ is defined as that $\mathcal{K}_0(\boldsymbol{X}) = \nabla\times\int_{\Omega} G_0(\boldsymbol{r},\boldsymbol{r}')\boldsymbol{X}(\boldsymbol{r}')d\Omega'$. In addition, the equivalent surface currents are defined by employing the inner normal directions of the boundaries of the array elements.

Similar to deriving (4) from (1), the following

$$P_{\text{driv}} = \mathrm{M}_{\text{a}}^{\dagger}\cdot\mathrm{P}_{\text{driv}}\cdot\mathrm{M}_{\text{a}} \quad (11)$$

with only independent variable $\mathrm{M}_{\text{a}}$ can be derived from (10), and a detailed derivation process is given in App. C. Here, $\mathrm{M}_{\text{a}}$ is the basis function expansion vector of $\boldsymbol{M}_{\text{a}}$. Theoretically, the independent variable can also be selected as the $\mathrm{J}_{\text{a}}$, which is the basis function expansion vector of $\boldsymbol{J}_{\text{a}}$. However, the numerical performances of the two selections are different, and it is more desirable to select $\mathrm{M}_{\text{a}}$ as independent variable because the array elements are non-magnetic, and a similar explanation focusing on scattering systems can be found in [55].

Employing the $\mathrm{P}_{\text{driv}}$ used in (11), the CMs of the material Yagi-Uda array can be calculated like Sec. II-B.

### C. Numerical Verifications

To verify the validities of the above volume and surface formulations of the WEP-CMA for material Yagi-Uda arrays, the comparisons of the WEP-CMA-based numerical results with some published simulation and measurement data are provided in this sub-section. The specific material Yagi-Uda array analyzed here is the same as the one reported in [42], and its all elements are with $4.0\,\text{mm}\times 4.0\,\text{mm}$ cross section, and its elements $V_{\text{a}}$, $V_{\text{p1}}$, and $V_{\text{p2}}$ have lengths 46.35 mm, 77.6 mm, and 44.4 mm respectively, and the distance (side-to-side) between $V_{\text{a}}$ and $V_{\text{p1}}$ is 23.0 mm, and the distance (side-to-side) between $V_{\text{a}}$ and $V_{\text{p2}}$ is 10.7 mm.

The characteristic value (in decibel) of the dominant CM calculated from WEP-CMA and the modal $S_{11}$ parameter (in decibel) obtained from simulation and measurement published in [42] are shown in Fig. 10 simultaneously. From the figure, it is easy to find out that the WEP-CMA-based resonance frequency is basically consistent with the data reported in [42], and the slight discrepancy is mainly originated from ignoring the feeding structure.

For the resonant working state of the WEP-CMA-based CM shown in Fig. 10, its radiation pattern is shown in Fig. 11, and its electric and magnetic field distributions are shown in Fig. 12. Evidently, Fig. 11 and Fig. 12 satisfy the well-known features of linear Yagi-Uda arrays — the radiated EM power propagates along the direction from reflecting element to directing elements. In addition, the time-domain dynamic figures of the propagating electric and magnetic fields are also uploaded to IEEE manuscript system together with this paper.

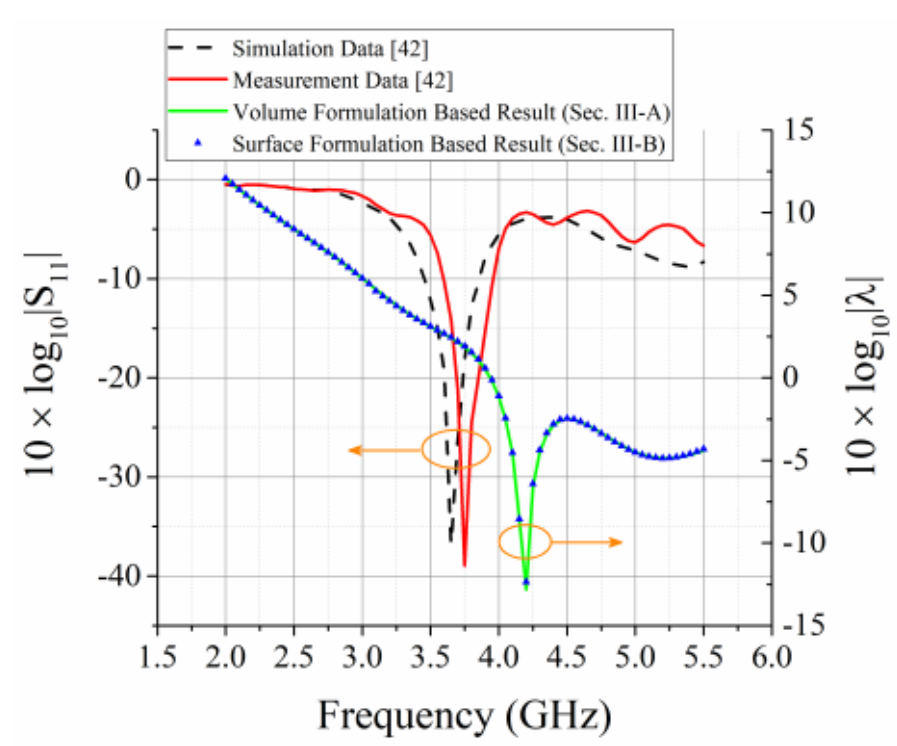

Fig. 10. Modal parameters of the dominant mode of the material Yagi-Uda array reported in [42].

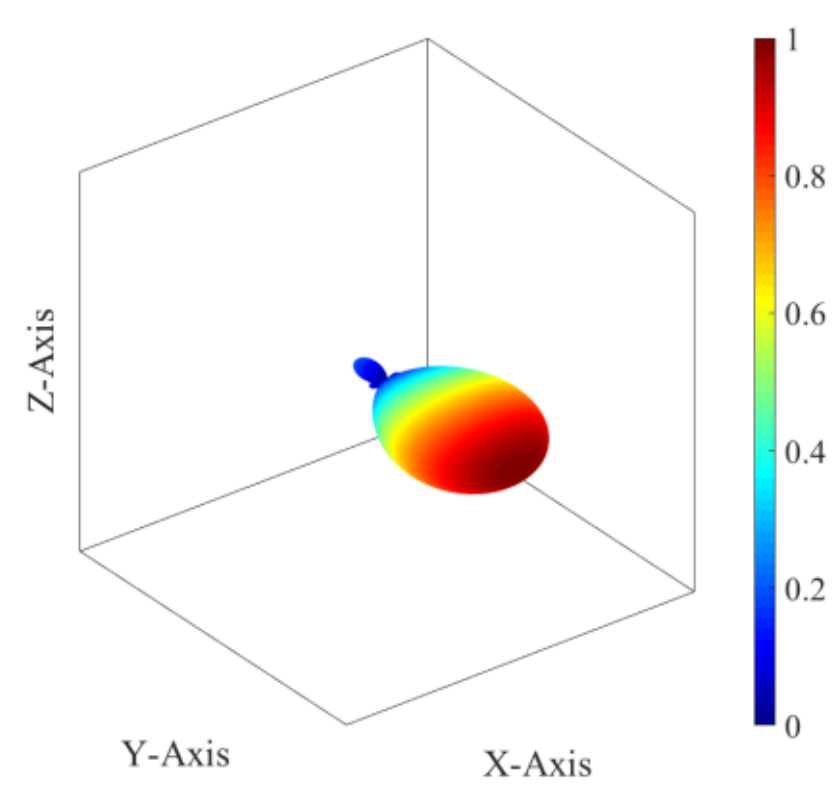

Fig. 11. Radiation pattern of the resonant state of the WEP-CMA-based dominant CM shown in Fig. 10.

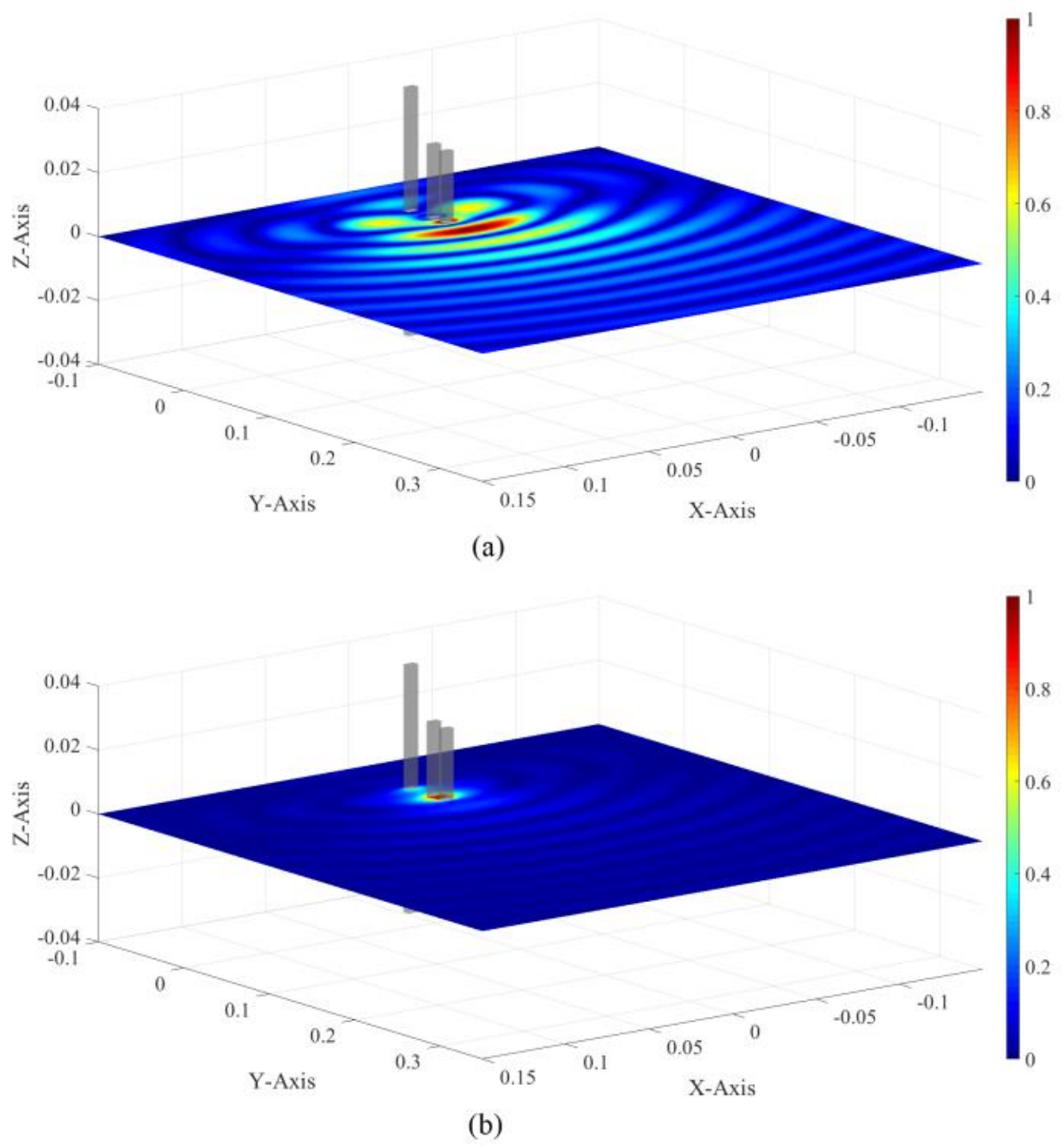

Fig. 12. Distributions of the (a) electric field and (b) magnetic field of the resonant state of the WEP-CMA-based CM shown in Fig. 10.

## IV. WEP-CMA for Composite Yagi-Uda Arrays

This section further generalizes the results given in Secs. II and III to the metal-material composite Yagi-Uda array shown in Fig. 13, which was reported in [22]. The composite array is constituted by metallic active patch $S_{\rm a}$, metallic passive patches $S_{\rm p}$, metallic ground plane $V_{\rm g}$, and material substrate $V_{\rm s}$. The substrate $V_{\rm s}$ is restricted to being non-magnetic for simplifying the discussions, and its complex permittivity is $\boldsymbol{\varepsilon}_{\rm s}^{\rm c}$.

### A. Volume-Surface Formulation of the WEP-CMA for Composite Yagi-Uda Arrays

If the induced surface electric currents on $S_{\rm a}$, $S_{\rm p}$, and the boundary of $V_{\rm g}$ are denoted as $\boldsymbol{J}_{\rm a}$, $\boldsymbol{J}_{\rm p}$, and $\boldsymbol{J}_{\rm g}$ respectively, and the induced volume electric current on $V_{\rm s}$ is denoted as $\boldsymbol{j}_{\rm s}$, then the corresponding DPO is as follows:

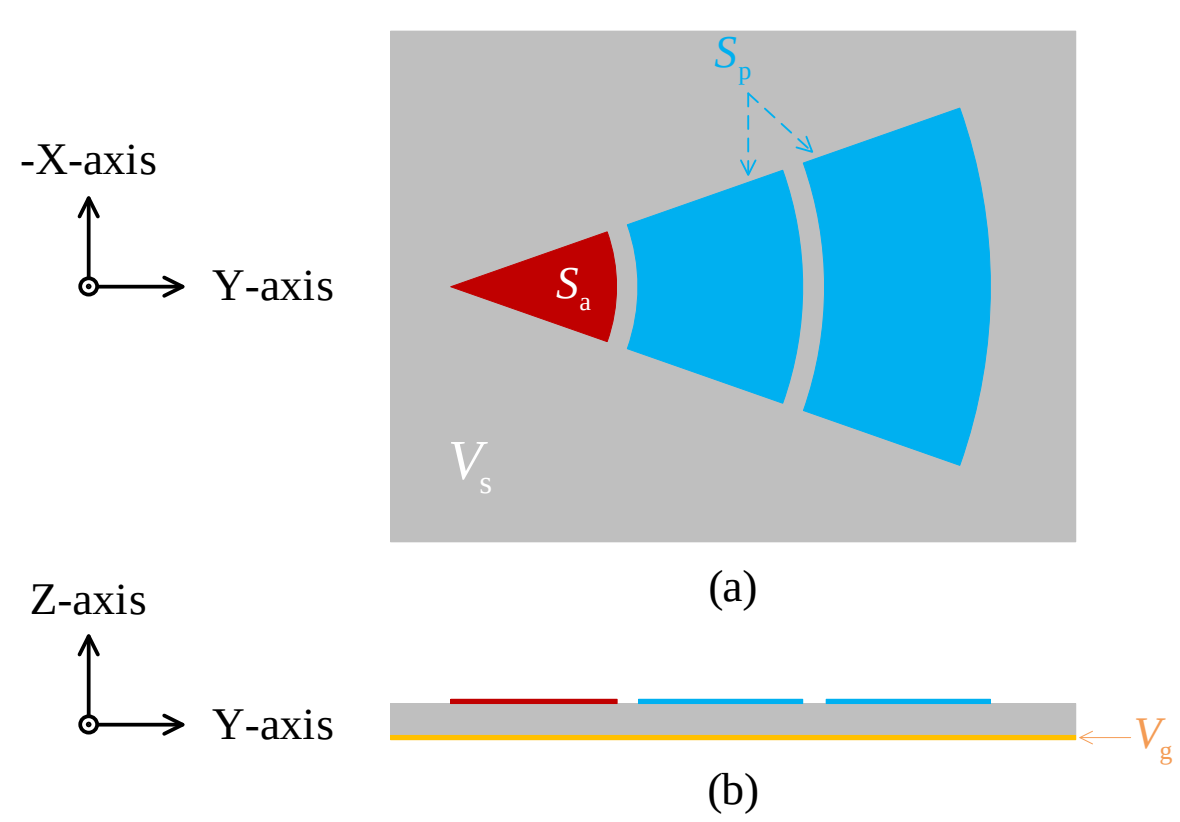

Fig. 13. Geometry of a 3-element composite Yagi-Uda array reported in [22]. (a) Top view; (b) lateral view.

$$P_{\rm driv} = (1/2)\left\langle \boldsymbol{J}_{\rm a} \, , \, j\omega\mu_0\mathcal{L}_0\left(\boldsymbol{J}_{\rm a}+\boldsymbol{J}_{\rm p}+\boldsymbol{J}_{\rm g}\oplus\boldsymbol{j}_{\rm s}\right)\right\rangle_{S_{\rm a}} \tag{12}$$

Here, the utilization of symbol "$\oplus$" is to emphasize that the dimensions of surface current $\boldsymbol{J}_{\rm a}+\boldsymbol{J}_{\rm p}+\boldsymbol{J}_{\rm g}$ and volume current $\boldsymbol{j}_{\rm s}$ are different from each other.

Similar to deriving (4) from (1), the following

$$P_{\rm driv} = \mathbb{J}_{\rm a}^{\dagger}\cdot\mathbb{P}_{\rm driv}\cdot\mathbb{J}_{\rm a} \tag{13}$$

with only independent variable $\mathbb{J}_{\rm a}$ can be derived from (12), and a detailed derivation process is given in App. D. Here, $\mathbb{J}_{\rm a}$ is the basis function expansion vector of $\boldsymbol{J}_{\rm a}$.

The formula provided in this sub-section is based on volume-surface currents, and an alternative surface-current-based formula will be given in the following sub-section.

### B. Surface Formulation of the WEP-CMA for Composite Yagi-Uda Arrays

For the convenience of this sub-section, the interface between $V_{\rm g}$ and free space is denoted as $S_{\rm gf}$, and the interface between $V_{\rm s}$ and $S_{\rm a}$ is denoted as $S_{\rm sa}$, and the interface between $V_{\rm s}$ and $S_{\rm p}$ is denoted as $S_{\rm sp}$, and the interface between $V_{\rm s}$ and free space is denoted as $S_{\rm sf}$.

If the induced surface electric current on $S_{\rm gf}$ is denoted as $\boldsymbol{J}_{\rm gf}$, and the equivalent surface electric currents on $S_{\rm sa}$, $S_{\rm sp}$, and $S_{\rm sf}$ are denoted as $\boldsymbol{J}_{\rm sa}$, $\boldsymbol{J}_{\rm sp}$, and $\boldsymbol{J}_{\rm sf}$ respectively, and the equivalent surface magnetic current on $S_{\rm sf}$ is denoted as $\boldsymbol{M}_{\rm sf}$, then the volume-surface-current version (12) of DPO can be alternatively written as the following surface-current version

$$P_{\rm driv} = \frac{1}{2}\left\langle \boldsymbol{J}_{\rm a} \, , -\mathcal{E}_0\left(\boldsymbol{J}_{\rm a}+\boldsymbol{J}_{\rm p}+\boldsymbol{J}_{\rm gf}-\boldsymbol{J}_{\rm sa}-\boldsymbol{J}_{\rm sp}-\boldsymbol{J}_{\rm sf}, -\boldsymbol{M}_{\rm sf}\right)\right\rangle_{S_{\rm a}} \tag{14}$$

where the equivalent surface currents are defined by employing the inner normal direction of the boundary of the material substrate.

Similar to deriving (4) from (1), the following

$$P_{\rm driv} = \mathbb{J}_{\rm a}^{\dagger}\cdot\mathbb{P}_{\rm driv}\cdot\mathbb{J}_{\rm a} \tag{15}$$

with only independent variable $\mathbb{J}_{\rm a}$ can be derived from (14), and a detailed derivation process is given in App. E.

Employing the $\mathbb{P}_{\rm driv}$ used in (13) or (15), the CMs of the composite Yagi-Uda array can be calculated like Sec. II-B.

### C. Numerical Verifications

This sub-section applies the WEP-CMA to the composite Yagi-Uda array reported in [22]. The geometry of the array is shown in Fig. 13, and the size of the array is described in [22].

The characteristic value (in decibel) of the dominant CM calculated from WEP-CMA and the modal $S_{11}$ parameter (in decibel) obtained from simulation published in [22] are shown in Fig. 14 simultaneously. The figure implies that the WEP-CMA-based resonance frequencies are basically consistent with the data reported in [22], and the slight discrepancy is mainly originated from ignoring the feeding structure.

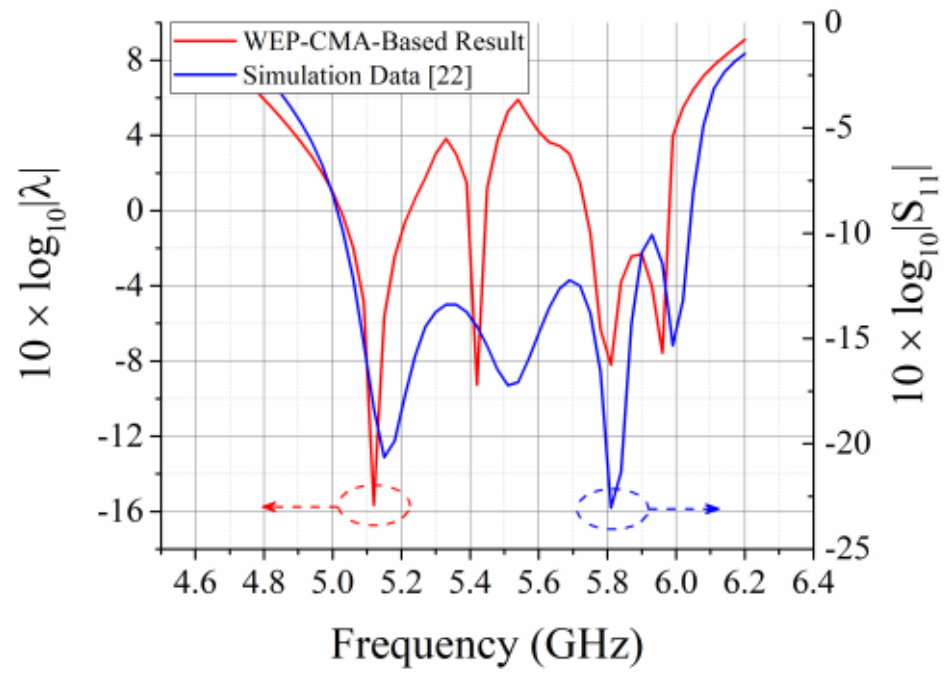

Fig. 14. Modal parameters of the dominant mode of the composite Yagi-Uda array reported in [22].

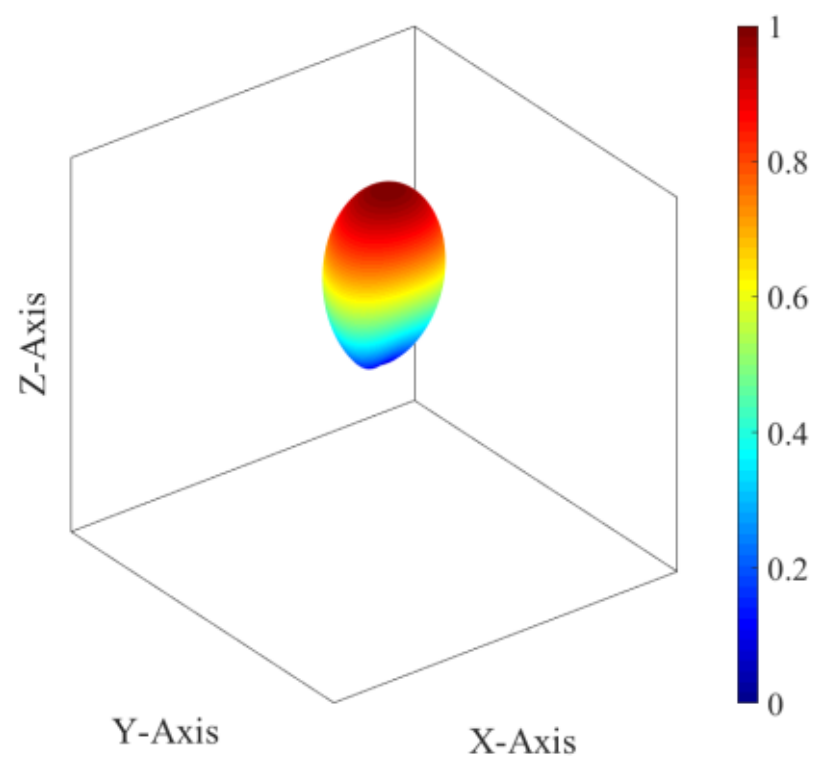

Fig. 15. Radiation pattern of the resonant state of the WEP-CMA-based CM shown in Fig. 14.

The radiation pattern of the CM working at resonance frequency 5.12 GHz is shown in Fig. 15. The radiation pattern is consistent with the one reported in [22].

## V. Conclusions

The same as the conclusions of our previous studies for metallic scattering and transferring systems, the WEP is also a quantitative depiction for the work-energy transformation process during the working of metallic Yagi-Uda arrays, and the driving power contained in WEP is just the source to sustain a steady work-energy transformation of the arrays. Employing the WEP, it is found out that the working mechanisms of metallic scattering systems and metallic Yagi-Uda arrays are different from each other, but the working mechanisms of metallic transferring systems and metallic Yagi-Uda arrays are similar to each other.

Based on the above these, the WEP-based CMT for metallic transferring systems can be successfully applied to doing the modal analysis for metallic Yagi-Uda arrays. By orthogonalizing DPO, the WEP-CMA can construct a set of energy-decoupled CMs for any previously selected objective metallic Yagi-Uda array. For a classical 6-element linear metallic Yagi-Uda array, the WEP-CMA implies that the dominant resonant CM is end-fire but the other higher-order resonant CMs are usually not, and then it is clearly explained why the higher-order resonant modes of the array are seldom used.

In addition, the WEP-CMA for metallic Yagi-Uda arrays can also be easily generalized to material Yagi-Uda arrays and composite Yagi-Uda arrays.

The validity and correctness of the WEP-CMA are verified by applying the WEP-CMA to the various Yagi-Uda arrays and comparing the WEP-CMA-based numerical results with the published simulation and measurement data.

## Appendices

Some detailed formulations related to this paper are provided in the following appendices.

### A. *Detailed Formulations Related to Sec. II-B*

In (2), the elements of sub-matrix $\mathbb{P}_{\text{aa}}$ are calculated as that $[\mathbb{P}_{\text{aa}}]_{\xi\zeta} = (1/2)\langle \boldsymbol{b}_{\text{a};\xi}, j\omega\mu_0\mathcal{L}_0(\boldsymbol{b}_{\text{a};\zeta})\rangle_{S_\text{a}}$, and the elements of sub-matrix $\mathbb{P}_{\text{ap}}$ are calculated as that $[\mathbb{P}_{\text{ap}}]_{\xi\zeta} = (1/2)\langle \boldsymbol{b}_{\text{a};\xi}, j\omega\mu_0\mathcal{L}_0(\boldsymbol{b}_{\text{p};\zeta})\rangle_{S_\text{a}}$.

The $\mathbb{T}$ used in (3) is $\mathbb{T} = -\mathbb{Z}_{\text{pp}}^{-1}\cdot\mathbb{Z}_{\text{pa}}$. The elements of matrix $\mathbb{Z}_{\text{pp}}$ are calculated as that $[\mathbb{Z}_{\text{pp}}]_{\xi\zeta} = \langle \boldsymbol{b}_{\text{p};\xi}, -j\omega\mu_0\mathcal{L}_0(\boldsymbol{b}_{\text{p};\zeta})\rangle_{S_\text{p}}$, and the elements of matrix $\mathbb{Z}_{\text{pa}}$ are calculated as that $[\mathbb{Z}_{\text{pa}}]_{\xi\zeta} = \langle \boldsymbol{b}_{\text{p};\xi}, -j\omega\mu_0\mathcal{L}_0(\boldsymbol{b}_{\text{a};\zeta})\rangle_{S_\text{p}}$.

The $\mathbb{P}_{\text{driv}}$ used in (4) is as follows:

$$\mathbb{P}_{\text{driv}} = \begin{bmatrix}\mathbb{P}_{\text{aa}} & \mathbb{P}_{\text{ap}}\end{bmatrix}\cdot\begin{bmatrix}\mathbb{I}\\ \mathbb{T}\end{bmatrix} \tag{16}$$

where $\mathbb{I}$ is an unit matrix with proper order.

### B. *Detailed Formulations Related to Sec. III-A*

By expanding the currents in (8) in terms of some proper basis functions, the integral form (8) of DPO can be discretized into the following matrix form

$$P_{\text{driv}} = \mathbf{j}_\text{a}^\dagger\cdot\begin{bmatrix}\mathbb{P}_{\text{aa}} & \mathbb{P}_{\text{ap1}} & \mathbb{P}_{\text{ap2}}\end{bmatrix}\cdot\begin{bmatrix}\mathbf{j}_\text{a}\\ \mathbf{j}_{\text{p1}}\\ \mathbf{j}_{\text{p2}}\end{bmatrix} \tag{17}$$

where the elements of sub-matrix $\mathbb{P}_{\text{aa}}$ are calculated as that $[\mathbb{P}_{\text{aa}}]_{\xi\zeta} = (1/2)\langle \boldsymbol{b}_{\text{a};\xi}, (j\omega\Delta\boldsymbol{\varepsilon}_\text{a}^\text{c})^{-1}\cdot\boldsymbol{b}_{\text{a};\zeta} + j\omega\mu_0\mathcal{L}_0(\boldsymbol{b}_{\text{a};\zeta})\rangle_{V_\text{a}}$, and the elements of sub-matrix $\mathbb{P}_{\text{ap1/ap2}}$ are calculated as that $[\mathbb{P}_{\text{ap1/ap2}}]_{\xi\zeta} = (1/2)\langle \boldsymbol{b}_{\text{a};\xi}, j\omega\mu_0\mathcal{L}_0(\boldsymbol{b}_{\text{p1/p2};\zeta})\rangle_{V_\text{a}}$.

Because of volume equivalence principle, there exist the following integral equations

$$\boldsymbol{j}_{\text{p1}} = j\omega\Delta\boldsymbol{\varepsilon}_{\text{p1}}^\text{c}\cdot\left[-j\omega\mu_0\mathcal{L}_0\left(\boldsymbol{j}_\text{a}+\boldsymbol{j}_{\text{p1}}+\boldsymbol{j}_{\text{p2}}\right)\right] \text{ on } V_{\text{p1}} \tag{18}$$

$$\boldsymbol{j}_{\text{p2}} = j\omega\Delta\boldsymbol{\varepsilon}_{\text{p2}}^\text{c}\cdot\left[-j\omega\mu_0\mathcal{L}_0\left(\boldsymbol{j}_\text{a}+\boldsymbol{j}_{\text{p1}}+\boldsymbol{j}_{\text{p2}}\right)\right] \text{ on } V_{\text{p2}} \tag{19}$$

Applying the method of moments (MoM) to (18)-(19), the integral equations are immediately discretized into matrix equations. By solving the matrix equations, the following transformation

$$\begin{bmatrix}\mathbf{j}_{\text{p1}}\\ \mathbf{j}_{\text{p2}}\end{bmatrix} = \underbrace{\begin{bmatrix}\mathbb{Z}_{\text{p1p1}} & \mathbb{Z}_{\text{p1p2}}\\ \mathbb{Z}_{\text{p2p1}} & \mathbb{Z}_{\text{p2p2}}\end{bmatrix}\cdot\begin{bmatrix}\mathbb{Z}_{\text{p1a}}\\ \mathbb{Z}_{\text{p2a}}\end{bmatrix}}_{\mathbb{T}}\cdot\mathbf{j}_\text{a} \tag{20}$$

can be easily obtained, where the elements of $\mathbb{Z}_{\mathrm{p1p1/p2p2}}$ are $<\boldsymbol{b}_{\mathrm{p1/p2};\xi}, (j\omega\Delta\boldsymbol{\varepsilon}^{\mathrm{c}}_{\mathrm{p1/p2}})^{-1}\cdot\boldsymbol{b}_{\mathrm{p1/p2};\zeta} + j\omega\mu_0\mathcal{L}_0(\boldsymbol{b}_{\mathrm{p1/p2};\zeta})>_{V_{\mathrm{p1}}/V_{\mathrm{p2}}}$, and the elements of sub-matrix $\mathbb{Z}_{\mathrm{p1p2/p2p1}}$ are calculated as $<\boldsymbol{b}_{\mathrm{p1/p2};\xi}, j\omega\mu_0\mathcal{L}_0(\boldsymbol{b}_{\mathrm{p2/p1};\zeta})>_{V_{\mathrm{p1}}/V_{\mathrm{p2}}}$, and the elements of $\mathbb{Z}_{\mathrm{p1a/p2a}}$ are $<\boldsymbol{b}_{\mathrm{p1/p2};\xi}, -j\omega\mu_0\mathcal{L}_0(\boldsymbol{b}_{\mathrm{a};\zeta})>_{V_{\mathrm{p1}}/V_{\mathrm{p2}}}$.

Substituting (20) into (17), we immediately have (9), where

$$\mathrm{P}_{\mathrm{driv}} = \begin{bmatrix} \mathrm{P}_{\mathrm{aa}} & \mathrm{P}_{\mathrm{ap1}} & \mathrm{P}_{\mathrm{ap2}} \end{bmatrix} \cdot \begin{bmatrix} \mathrm{I} \\ \mathrm{T} \end{bmatrix} \tag{21}$$

### C. Detailed Formulations Related to Sec. III-B

The integral form (10) of DPO can be discretized into the following matrix form

$$P_{\mathrm{driv}} = \begin{bmatrix} \mathrm{J}_{\mathrm{a}} \\ \mathrm{M}_{\mathrm{a}} \end{bmatrix}^{\dagger} \cdot \begin{bmatrix} \mathrm{P}^{\mathrm{JJ}}_{\mathrm{aa}} & \mathrm{P}^{\mathrm{JJ}}_{\mathrm{ap1}} & \mathrm{P}^{\mathrm{JJ}}_{\mathrm{ap2}} & \mathrm{P}^{\mathrm{JM}}_{\mathrm{aa}} & \mathrm{P}^{\mathrm{JM}}_{\mathrm{ap1}} & \mathrm{P}^{\mathrm{JM}}_{\mathrm{ap2}} \\ \mathrm{P}^{\mathrm{MJ}}_{\mathrm{aa}} & \mathrm{P}^{\mathrm{MJ}}_{\mathrm{ap1}} & \mathrm{P}^{\mathrm{MJ}}_{\mathrm{ap2}} & \mathrm{P}^{\mathrm{MM}}_{\mathrm{aa}} & \mathrm{P}^{\mathrm{MM}}_{\mathrm{ap1}} & \mathrm{P}^{\mathrm{MM}}_{\mathrm{ap2}} \end{bmatrix} \cdot \begin{bmatrix} \mathrm{J}_{\mathrm{a}} \\ \mathrm{J}_{\mathrm{p1}} \\ \mathrm{J}_{\mathrm{p2}} \\ \mathrm{M}_{\mathrm{a}} \\ \mathrm{M}_{\mathrm{p1}} \\ \mathrm{M}_{\mathrm{p2}} \end{bmatrix} \tag{22}$$

by expanding the currents in terms of some proper basis functions. The formulations for calculating the elements of the various sub-matrices $\mathbb{P}$ are trivial, and they are not explicitly given here.

For the currents involved in (10), there exist the following integral equations

$$\left[\mathcal{E}_{\mathrm{a}}\left(\boldsymbol{J}_{\mathrm{a}}, \boldsymbol{M}_{\mathrm{a}}\right)\right]^{\mathrm{tan}}_{S^-_{\mathrm{a}}} = \boldsymbol{n}^-_{\mathrm{a}} \times \boldsymbol{M}_{\mathrm{a}} \tag{23}$$

$$\begin{aligned} &\left[\mathcal{E}_{\mathrm{p1}}\left(\boldsymbol{J}_{\mathrm{p1}}, \boldsymbol{M}_{\mathrm{p1}}\right)\right]^{\mathrm{tan}}_{S^-_{\mathrm{p1}}} \\ &= -\left[\mathcal{E}_0\left(\boldsymbol{J}_{\mathrm{a}} + \boldsymbol{J}_{\mathrm{p1}} + \boldsymbol{J}_{\mathrm{p2}}, \boldsymbol{M}_{\mathrm{a}} + \boldsymbol{M}_{\mathrm{p1}} + \boldsymbol{M}_{\mathrm{p2}}\right)\right]^{\mathrm{tan}}_{S^+_{\mathrm{p1}}} \end{aligned} \tag{24}$$

$$\begin{aligned} &\left[\mathcal{H}_{\mathrm{p1}}\left(\boldsymbol{J}_{\mathrm{p1}}, \boldsymbol{M}_{\mathrm{p1}}\right)\right]^{\mathrm{tan}}_{S^-_{\mathrm{p1}}} \\ &= -\left[\mathcal{H}_0\left(\boldsymbol{J}_{\mathrm{a}} + \boldsymbol{J}_{\mathrm{p1}} + \boldsymbol{J}_{\mathrm{p2}}, \boldsymbol{M}_{\mathrm{a}} + \boldsymbol{M}_{\mathrm{p1}} + \boldsymbol{M}_{\mathrm{p2}}\right)\right]^{\mathrm{tan}}_{S^+_{\mathrm{p1}}} \end{aligned} \tag{25}$$

$$\begin{aligned} &\left[\mathcal{E}_{\mathrm{p2}}\left(\boldsymbol{J}_{\mathrm{p2}}, \boldsymbol{M}_{\mathrm{p2}}\right)\right]^{\mathrm{tan}}_{S^-_{\mathrm{p2}}} \\ &= -\left[\mathcal{E}_0\left(\boldsymbol{J}_{\mathrm{a}} + \boldsymbol{J}_{\mathrm{p1}} + \boldsymbol{J}_{\mathrm{p2}}, \boldsymbol{M}_{\mathrm{a}} + \boldsymbol{M}_{\mathrm{p1}} + \boldsymbol{M}_{\mathrm{p2}}\right)\right]^{\mathrm{tan}}_{S^+_{\mathrm{p2}}} \end{aligned} \tag{26}$$

$$\begin{aligned} &\left[\mathcal{H}_{\mathrm{p2}}\left(\boldsymbol{J}_{\mathrm{p2}}, \boldsymbol{M}_{\mathrm{p2}}\right)\right]^{\mathrm{tan}}_{S^-_{\mathrm{p2}}} \\ &= -\left[\mathcal{H}_0\left(\boldsymbol{J}_{\mathrm{a}} + \boldsymbol{J}_{\mathrm{p1}} + \boldsymbol{J}_{\mathrm{p2}}, \boldsymbol{M}_{\mathrm{a}} + \boldsymbol{M}_{\mathrm{p1}} + \boldsymbol{M}_{\mathrm{p2}}\right)\right]^{\mathrm{tan}}_{S^+_{\mathrm{p2}}} \end{aligned} \tag{27}$$

where operator $\mathcal{E}_{\mathrm{a}}$ is with parameters $(\boldsymbol{\varepsilon}^{\mathrm{c}}_{\mathrm{s}}, \mu_0)$, and operators $\mathcal{E}_{\mathrm{p1/p2}}$ and $\mathcal{H}_{\mathrm{p1/p2}}$ are with parameters $(\boldsymbol{\varepsilon}^{\mathrm{c}}_{\mathrm{p1/p2}}, \mu_0)$, and $S^-_{\mathrm{a/p1/p2}}$ is the inner surface of $S_{\mathrm{a/p1/p2}}$, and $S^+_{\mathrm{p1/p2}}$ is the outer surface of $S_{\mathrm{p1/p2}}$.

Similar to deriving (20) from (18) and (19), the following transformation

$$\begin{bmatrix} \mathrm{J}_{\mathrm{a}} \\ \mathrm{J}_{\mathrm{p1}} \\ \mathrm{J}_{\mathrm{p2}} \\ \mathrm{M}_{\mathrm{p1}} \\ \mathrm{M}_{\mathrm{p2}} \end{bmatrix} = \mathrm{T} \cdot \mathrm{M}_{\mathrm{a}} \tag{28}$$

can be derived from (23)-(27), and then we have the following sub-transformations

$$\mathrm{J}_{\mathrm{a}} = \mathrm{T}_1 \cdot \mathrm{M}_{\mathrm{a}}, \text{ and } \begin{bmatrix} \mathrm{J}_{\mathrm{p1}} \\ \mathrm{J}_{\mathrm{p2}} \end{bmatrix} = \mathrm{T}_2 \cdot \mathrm{M}_{\mathrm{a}}, \text{ and } \begin{bmatrix} \mathrm{M}_{\mathrm{p1}} \\ \mathrm{M}_{\mathrm{p2}} \end{bmatrix} = \mathrm{T}_3 \cdot \mathrm{M}_{\mathrm{a}} \tag{29}$$

Substituting (29) into (22), we immediately have (11), where the $\mathrm{P}_{\mathrm{driv}}$ is as follows:

$$\mathrm{P}_{\mathrm{driv}} = \begin{bmatrix} \mathrm{T}_1 \\ \mathrm{I} \end{bmatrix}^{\dagger} \cdot \begin{bmatrix} \mathrm{P}^{\mathrm{JJ}}_{\mathrm{aa}} & \mathrm{P}^{\mathrm{JJ}}_{\mathrm{ap1}} & \mathrm{P}^{\mathrm{JJ}}_{\mathrm{ap2}} & \mathrm{P}^{\mathrm{JM}}_{\mathrm{aa}} & \mathrm{P}^{\mathrm{JM}}_{\mathrm{ap1}} & \mathrm{P}^{\mathrm{JM}}_{\mathrm{ap2}} \\ \mathrm{P}^{\mathrm{MJ}}_{\mathrm{aa}} & \mathrm{P}^{\mathrm{MJ}}_{\mathrm{ap1}} & \mathrm{P}^{\mathrm{MJ}}_{\mathrm{ap2}} & \mathrm{P}^{\mathrm{MM}}_{\mathrm{aa}} & \mathrm{P}^{\mathrm{MM}}_{\mathrm{ap1}} & \mathrm{P}^{\mathrm{MM}}_{\mathrm{ap2}} \end{bmatrix} \cdot \begin{bmatrix} \mathrm{T}_1 \\ \mathrm{T}_2 \\ \mathrm{I} \\ \mathrm{T}_3 \end{bmatrix} \tag{30}$$

### D. Detailed Formulations Related to Sec. IV-A

Here, we only provide the integral equations used to establish the transformation from the independent variable to the dependent variables involved in (12) as follows:

$$0 = \left[-j\omega\mu_0\mathcal{L}_0\left(\boldsymbol{J}_{\mathrm{a}} + \boldsymbol{J}_{\mathrm{p}} + \boldsymbol{J}_{\mathrm{g}} + \boldsymbol{j}_{\mathrm{s}}\right)\right]^{\mathrm{tan}} \quad \text{on } S_{\mathrm{p}} \cup S_{\mathrm{g}} \tag{31}$$

$$\boldsymbol{j}_{\mathrm{s}} = j\omega\Delta\boldsymbol{\varepsilon}^{\mathrm{c}}_{\mathrm{s}} \cdot \left[-j\omega\mu_0\mathcal{L}_0\left(\boldsymbol{J}_{\mathrm{a}} + \boldsymbol{J}_{\mathrm{p}} + \boldsymbol{J}_{\mathrm{g}} + \boldsymbol{j}_{\mathrm{s}}\right)\right] \text{ on } V_{\mathrm{s}} \tag{32}$$

and the other detailed formulations related to Sec. IV-A are similar to the ones used in Apps. A-C.

### E. Detailed Formulations Related to Sec. IV-B

Here, we only provide the integral equations used to establish the transformation from the independent variable to the dependent variables involved in (14) as follows:

$$\begin{aligned} &\left[\mathcal{E}_0\left(\boldsymbol{J}_{\mathrm{a}} + \boldsymbol{J}_{\mathrm{p}} + \boldsymbol{J}_{\mathrm{gf}} - \boldsymbol{J}_{\mathrm{sa}} - \boldsymbol{J}_{\mathrm{sp}} - \boldsymbol{J}_{\mathrm{sf}}, -\boldsymbol{M}_{\mathrm{sf}}\right)\right]^{\mathrm{tan}} \\ &= 0 \qquad \text{on } S_{\mathrm{p}} \cup S_{\mathrm{gf}} \end{aligned} \tag{33}$$

$$\begin{aligned} &\left[\mathcal{E}_{\mathrm{s}}\left(\boldsymbol{J}_{\mathrm{sa}} + \boldsymbol{J}_{\mathrm{sp}} + \boldsymbol{J}_{\mathrm{sg}} + \boldsymbol{J}_{\mathrm{sf}}, \boldsymbol{M}_{\mathrm{sf}}\right)\right]^{\mathrm{tan}} \\ &= 0 \qquad \text{on } S_{\mathrm{sa}} \cup S_{\mathrm{sp}} \cup S_{\mathrm{sg}} \end{aligned} \tag{34}$$

$$\begin{aligned} &\left[\mathcal{E}_0\left(\boldsymbol{J}_{\mathrm{a}} + \boldsymbol{J}_{\mathrm{p}} + \boldsymbol{J}_{\mathrm{gf}} - \boldsymbol{J}_{\mathrm{sa}} - \boldsymbol{J}_{\mathrm{sp}} - \boldsymbol{J}_{\mathrm{sf}}, -\boldsymbol{M}_{\mathrm{sf}}\right)\right]^{\mathrm{tan}}_{S^+_{\mathrm{sf}}} \\ &= \left[\mathcal{E}_{\mathrm{s}}\left(\boldsymbol{J}_{\mathrm{sa}} + \boldsymbol{J}_{\mathrm{sp}} + \boldsymbol{J}_{\mathrm{sg}} + \boldsymbol{J}_{\mathrm{sf}}, \boldsymbol{M}_{\mathrm{sf}}\right)\right]^{\mathrm{tan}}_{S^-_{\mathrm{sf}}} \quad \text{on } S_{\mathrm{sf}} \end{aligned} \tag{35}$$

$$\begin{aligned} &\left[\mathcal{H}_0\left(\boldsymbol{J}_{\mathrm{a}} + \boldsymbol{J}_{\mathrm{p}} + \boldsymbol{J}_{\mathrm{gf}} - \boldsymbol{J}_{\mathrm{sa}} - \boldsymbol{J}_{\mathrm{sp}} - \boldsymbol{J}_{\mathrm{sf}}, -\boldsymbol{M}_{\mathrm{sf}}\right)\right]^{\mathrm{tan}}_{S^+_{\mathrm{sf}}} \\ &= \left[\mathcal{H}_{\mathrm{s}}\left(\boldsymbol{J}_{\mathrm{sa}} + \boldsymbol{J}_{\mathrm{sp}} + \boldsymbol{J}_{\mathrm{sg}} + \boldsymbol{J}_{\mathrm{sf}}, \boldsymbol{M}_{\mathrm{sf}}\right)\right]^{\mathrm{tan}}_{S^-_{\mathrm{sf}}} \quad \text{on } S_{\mathrm{sf}} \end{aligned} \tag{36}$$

where current $\boldsymbol{J}_{\mathrm{sg}}$ is the equivalent surface electric current on the interface between $V_{\mathrm{s}}$ and $V_{\mathrm{g}}$, operators $\mathcal{E}_{\mathrm{s}}$ and $\mathcal{H}_{\mathrm{s}}$ are with parameters $(\boldsymbol{\varepsilon}_{\mathrm{s}}^{\mathrm{c}}, \mu_0)$, and the other detailed formulations related to Sec. IV-B are similar to the ones used in Apps. A-C.

**Ren-Zun Lian** received the B.S. degree in optical engineering from the University of Electronic Science and Technology of China (UESTC), Chengdu, China, in 2011, and received the Ph.D. degree in electromagnetic field and microwave technology from UESTC, Chengdu, China, in 2019.

He is currently a postdoctoral researcher in Peking University (PKU), Beijing, China. His current research interests include mathematical physics, electromagnetic theory and computation, and antenna theory and design.

**Ming-Yao Xia** (M'00-SM'03) received the Master and Ph. D degrees in electrical engineering from the Institute of Electronics, Chinese Academy of Sciences (IECAS), in 1988 and 1999, respectively. From 1988 to 2002, he was with IECAS as an Engineer and a Senior Engineer. He was a Visiting Scholar at the University of Oxford, U.K., from October 1995 to October 1996. From June 1999 to August 2000 and from January 2002 to June 2002, he was a Senior Research Assistant and a Research Fellow, respectively, with the City University of Hong Kong. He joined Peking University as an Associate Professor in 2002 and was promoted to Full Professor in 2004. He moved to the University of Electronic Science and Technology of China as a Chang-Jiang Professor nominated by the Ministry of Education of China in 2010. He returned to Peking University after finishing the appointment in 2013. He was a recipient of the Young Scientist Award of the URSI in 1993. He was awarded the first-class prize on Natural Science by the Chinese Academy of Sciences in 2001. He was the recipient of the Foundation for Outstanding Young Investigators presented by the National Natural Science Foundation of China in 2008. He served as an Associate Editor for the IEEE Transactions on Antennas and Propagation. His research interests include electromagnetic theory, numerical methods and applications, such as wave propagation and scattering, electromagnetic imaging and probing, microwave remote sensing, antennas and microwave components.

**Xing-Yue Guo** (S'17) received the B.S. degree in electronic information science and technology from the Southwest Jiaotong University, Chengdu, China, in 2011, and the M.E. degree from the University of Electronic Science and Technology of China, Chengdu, China, in 2014. From 2014 to 2017, she was a Research Assistant with the CAEP Software Center for High Performance Numerical Simulation, Beijing, China. Since 2017, she has been pursuing the Ph. D. degree at the School of Electronics Engineering and Computer Science, Peking University, Beijing, China. Her research interests include computational electromagnetics and applications.